\documentclass[12pt]{article}

\usepackage{latexsym}
\usepackage{amssymb,amsfonts,amsmath}
\usepackage{graphicx} 
\usepackage{indentfirst}
\usepackage{bbm}
\usepackage{amssymb}
\usepackage{verbatim}
\usepackage{amsmath, amsthm,amssymb}
\usepackage{mathrsfs}
\usepackage{hyperref}
\usepackage{amsfonts}
\usepackage{dsfont}
\usepackage{cite}
\usepackage{xcolor}
\usepackage{slashed}
\usepackage{booktabs}
\usepackage{changepage}
\usepackage{setspace}

\PassOptionsToPackage{linecolor=blue,backgroundcolor=blue!25,bordercolor=blue,textsize=scriptsize}{todonotes}
\usepackage{todonotes}

\parskip=\medskipamount

\usepackage[mathscr]{euscript}

\newcommand{\de}{{\nabla}}

\numberwithin{equation}{section}

\newcommand {\cB}{{\cal B}}
\newcommand {\cC}{{\cal C}}
\newcommand {\cD}{{\cal D}}

\newcommand {\cF}{{\cal F}}
\newcommand {\cG}{{\cal G}}
\newcommand {\cH}{{\cal H}}

\newcommand {\cJ}{{\cal J}}
\newcommand {\cK}{{\cal K}}
\newcommand {\cL}{{\cal L}}

\newcommand {\cN}{{\cal N}}
\newcommand {\cO}{{\cal O}}

\newcommand {\cR}{{\cal R}}

\newcommand {\cU}{{\cal U}}
\newcommand {\cV}{{\cal V}}

\newcommand{\scT}{{\mathscr{T}}}
\newcommand{\bbD}{\mathbb D}

\newcommand{\RM}{R(M)}
\newcommand{\RD}{R(\mathbb D)}

\newcommand{\RS}{R(S)}
\newcommand{\RK}{R(K)}

\def\a{\alpha}
\def\b{\beta}

\def\d{\delta}

\def\g{\gamma}
\def\G{\Gamma}

\def\k{\kappa}
\def\l{\lambda}

\def\o{\omega}

\def\q{\theta}

\def\s{\sigma}

\def\D{\Delta}
\def\F{\Phi}

\def\L{\Lambda}
\def\O{\Omega}

\def\S{\Sigma}

\def\ri{{\rm i}}

\newcommand{\rd}{\mathrm d}

\newcommand{\hm}{{{m}}}
\newcommand{\hn}{{{n}}}
\newcommand{\hp}{{{p}}}
\newcommand{\hq}{{{q}}}
\newcommand{\ha}{{{a}}}
\newcommand{\hb}{{{b}}}
\newcommand{\hc}{{{c}}}
\newcommand{\hd}{{{d}}}
\newcommand{\he}{{{e}}}
\newcommand{\hM}{{{M}}}
\newcommand{\hN}{{{N}}}
\newcommand{\hA}{{{A}}}
\newcommand{\hB}{{{B}}}
\newcommand{\hC}{{{C}}}
\newcommand{\hal}{{{\a}}}
\newcommand{\hbe}{{{\b}}}
\newcommand{\hga}{{{\g}}}
\newcommand{\hde}{{{\d}}}
\newcommand{\hrh}{{{\rho}}}

\newcommand{\ve}{\varepsilon}

\newcommand{\pa}{\partial}
\newcommand{\hf}{\frac12}

\newcommand{\vf}{\varphi}

\newcommand{\be}{\begin{equation}}
\newcommand{\ee}{\end{equation}}
\newcommand{\bea}{\begin{eqnarray}}
\newcommand{\eea}{\end{eqnarray}}
\newcommand{\non}{\nonumber}
\newcommand{\ba}{\begin{array}}
\newcommand{\ea}{\end{array}}

\newcommand{\bsubeq}{\begin{subequations}}
\newcommand{\esubeq}{\end{subequations}}

\newcommand{\gD}{{\mathbb D}}

\newcommand{\bm}[1]{\mbox{\boldmath$#1$}}

\def\double #1{#1{\hbox{\kern-2pt $#1$}}}

\newcommand{\eps}{\varepsilon}

\newcommand{\eol}{\notag \\}

\newcommand{\loco}{\vert}
\newcommand{\doubar}{{{\loco}\!{\loco}}}
\newcommand{\veps}{\varepsilon}

\expandafter\ifx\csname url\endcsname\relax
  \def\url#1{\texttt{#1}}\fi
\expandafter\ifx\csname urlprefix\endcsname\relax\fi
\providecommand{\eprint}[2][]{\url{#2}}

\begin{document}

\begin{titlepage}
\begin{flushright}
November, 2023
\end{flushright}
\vspace{5mm}

\begin{center}
{\Large \bf 
Components of curvature-squared invariants of minimal supergravity in five dimensions}
\end{center}

\begin{center}

{\bf
Gregory Gold,
Jessica Hutomo,
Saurish Khandelwal,\\
and Gabriele Tartaglino-Mazzucchelli
} \\
\vspace{5mm}

\footnotesize{
{\it 
School of Mathematics and Physics, University of Queensland,
\\
 St Lucia, Brisbane, Queensland 4072, Australia}
}
\vspace{2mm}
~\\
\texttt{g.gold@uq.edu.au; 
j.hutomo@uq.edu.au; 
s.khandelwal@uq.edu.au;
\\ g.tartaglino-mazzucchelli@uq.edu.au}\\
\vspace{2mm}

\end{center}

\begin{abstract}
\baselineskip=14pt

We present for the first time the component structure of the supersymmetric completions for all curvature-squared invariants of five-dimensional, off-shell (gauged) minimal supergravity, including all fermions. This is achieved by using an interplay between superspace and superconformal tensor calculus techniques, and by employing results from arXiv:1410.8682 and arXiv:2302.14295. Our analysis is based on using a standard Weyl multiplet of conformal supergravity coupled to a vector and a linear multiplet compensator to engineer off-shell Poincar\'e supergravity.
We compute all the descendants of the composite linear multiplets that describe gauged supergravity together with the three independent four-derivative invariants. These are the building blocks of the locally superconformal invariant actions. A derivation of the primary equations of motion for minimal gauged off-shell supergravity deformed by an arbitrary combination of these three locally superconformal invariants, is then provided. Finally, all the covariant descendants in the multiplets of equations of motion are obtained by applying a series of $Q$-supersymmetry transformations, equivalent to successively applying  superspace spinor derivatives to the primary equations of motion.

\end{abstract}
\vspace{5mm}

\vfill
\end{titlepage}


\newpage
\renewcommand{\thefootnote}{\arabic{footnote}}
\setcounter{footnote}{0}

\tableofcontents{}
\vspace{1cm}
\bigskip\hrule

\allowdisplaybreaks
\section{Introduction}

Even though the first (two-derivative) supergravity was constructed (for $\cN=1$ supersymmetry in four dimensions) almost five decades ago \cite{FvNF,DZ} (see also \cite{Townsend,Breitenlohner,Siegel77-80,SG,Siegel77-77,WZ,old1,old2,SohniusW1,SohniusW3} and the pedagogical reviews \cite{SUPERSPACE,WB,BK,Freedman:2012zz,Kuzenko:2022skv}), higher-order locally supersymmetric invariants are still largely unknown.
However, in an effective field theory approach, quantum corrections in string theory take the form of an infinite series of (supersymmetric) higher-derivative terms; see, e.\,g., \cite{Sethi:2017phn,Becker:1996gj,Antoniadis:1997eg,Antoniadis:2003sw,Liu:2013dna} and references therein. Many open problems in string theory, for example its vacua structure, are unresolved due to the lack of information about the full quantum corrected supergravity effective action. 
More complexity arises due to the fact that the purely gravitational higher-curvature terms are related by supersymmetry to contributions depending on $p$-forms, which describe part of the string spectrum. These terms, which have not yet been fully understood, play an important role in studying, for example, the moduli space in compactified string theory and the low-energy description of string dualities. Even the first $\alpha^\prime$ corrections, associated with curvature-squared terms, have not been completely understood to date. Our paper represents an important step forward in this direction.

A remarkable success of supergravity in the last 25 years has been its role in the development of holographic correspondences such as the AdS/CFT \cite{Maldacena:1997re,Witten:1998qj,Gubser:1998bc} (see also the classic reviews \cite{Aharony:1999ti,DHoker:2002nbb}). 
The amount of evidence that planar (large-$N$, at first order in the $1/N$ expansion) field theory can be described by a dual theory of (two-derivative) gravity, and vice versa, is absolutely outstanding. These results have revolutionised our understanding of both quantum field and gravity theories and have shed new light in several fields of research such as black hole physics, condensed matter and integrable systems, and quantum information theory, just to mention a few. One aspect to stress is that most of these developments were based on the leading-order effective field theory description of string theory. It is of fundamental importance for holography to advance to higher orders so that we may assess the validity of the correspondence. In fact, precision tests beyond the leading order have become increasingly important in the last few years. The reason is that, on the field theory side, a series of breakthroughs based on integrability and localisation techniques has allowed several observables in superconformal field theories (SCFT) to be computed exactly. Higher-order (in $1/N$) corrections in quantum field theories translate into higher-curvature terms on the gravity side, calling for new supergravity higher-derivative analyses. 

When {\emph{off-shell}} techniques are available for supergravity,\footnote{Off-shell means that the symmetry algebra closes without using equations of motion; see, e.g., \cite{SUPERSPACE,WB,BK,Freedman:2012zz}.} systematic approaches exist to construct locally supersymmetric higher-derivative invariants.
In ${\rm D}\leq 6$ space-time dimensions, off-shell techniques are now thoroughly developed and understood for up to eight real supercharges -- see \cite{Freedman:2012zz,Lauria:2020rhc,SUPERSPACE,WB,BK, Kuzenko:2022skv,Kuzenko:2022ajd} for reviews. By using these approaches, exact, off-shell higher-derivative supergravity models have been constructed, see for example, the following list of references
\cite{HOT,BSS1,LopesCardoso:1998tkj,Mohaupt:2000mj,BRS,CVanP,Bergshoeff:2012ax,Butter:2013rba,Butter:2013lta,Kuzenko:2013vha,OP1,OP2,OzkanThesis,BKNT-M14,Kuzenko:2015jxa,BKNT16,Butter:2016mtk,BNT-M17,NOPT-M17,Butter:2018wss,Butter:2019edc,Hegde:2019ioy,Mishra:2020jlc,deWS,Ozkan:2016csy,deWit:2010za,Butter:2014iwa,GHKT-M,Gold:2023ymc}. These results have recently enabled new pioneering works towards obtaining precision, higher-order, tests in AdS/CFT from the gravity side -- see for example \cite{Baggio:2014hua,Bobev:2020zov,Bobev:2020egg,Bobev:2021oku,Bobev:2021qxx,Liu:2022sew,Hristov:2021qsw,Hristov:2022lcw,Bobev:2022bjm,Cassani:2022lrk,Gold:2023ymc} and references therein. We underline, by extending results obtained in \cite{BKNT-M14} and more recently in \cite{GHKT-M}, that in \cite{Gold:2023ymc} and in the current paper, the construction of a complete basis to study (minimal) gauged supergravity in five dimensions modified by the three possible curvature-squared terms has been completed. As such, we have control of all first $\alpha^\prime$ correction of a universal sector of string compactifications to five dimensions preserving at least eight real supercharges. This then allows inspection from the gravity side of new first order tests of the AdS$_5$/CFT$_4$.

The playground of our paper is minimal five-dimensional (5D) supergravity. 
On-shell, this was introduced four decades ago in \cite{Cremmer,CN}, and the first off-shell description was given in \cite{Howe5Dsugra} by the use of superspace techniques. The matter couplings in 5D minimal supergravity have since been extensively studied at the component level, both  by using on-shell \cite{GST1,GST2,GZ,CD} and off-shell approaches \cite{Zucker1,Zucker2,Zucker3,Ohashi1,Ohashi2,Ohashi3,Ohashi4,Bergshoeff1,Bergshoeff2,Bergshoeff3}. The superspace approach to general off-shell 5D $\cN=1$ supergravity-matter systems was then developed in \cite{KT-M_5D2,KT-M_5D3,KT-M08,BKNT-M14,Howe:2020hxi}. In our paper we will employ an interplay between component techniques based on the superconformal tensor calculus and superspace -- see \cite{Freedman:2012zz,Lauria:2020rhc,Kuzenko:2022skv,Kuzenko:2022ajd} for reviews -- the merging of which goes under the name of {\emph{conformal superspace}.}
Conformal superspace was originally introduced by D.\,Butter for 4D $\cN=1$ supergravity in a seminal work \cite{Butter4DN=1} and then extended to other space-time dimensions $2\leq {\rm D} \leq 6$ for various amount of supersymmetry in \cite{Butter4DN=2,BKNT-M1,BKNT-M14,BKNT16,BNT-M17,Kuzenko:2022qnb} -- see \cite{Kuzenko:2022skv,Kuzenko:2022ajd} for recent reviews. 

In superconformal approaches to supergravity, the key idea is to enlarge the supergravity gauge group to be described by local superconformal transformations and, potentially, internal symmetries. Local Poincar\'e supersymmetry is then recovered by making appropriate choices of gauge fixing conditions for the non-physical symmetries within the conformal algebra implemented through the use of compensating multiplets. The extra symmetries make some of the local supersymmetry  of the gravitational sector (both the transformations and symmetry algebra) more manageable and allow for increased freedom to appropriately fix gauges and frame choices in the final steps of various analyses -- see \cite{Freedman:2012zz,Lauria:2020rhc,Kuzenko:2022skv,Kuzenko:2022ajd} for pedagogical reviews. To achieve minimal 5D supergravity off-shell from superconformal techniques one can couple the standard Weyl multiplet of 5D conformal supergravity to two off-shell conformal compensators: a vector multiplet and a linear multiplet \cite{Zucker1,Zucker2,Zucker3,Ohashi1,Ohashi2,Ohashi3,Ohashi4,Bergshoeff1,Bergshoeff2,Bergshoeff3,KT-M_5D2,KT-M_5D3,KT-M08,BKNT-M14}. These will be the off-shell multiplets employed in our paper. 

Within the framework of these superconformal approaches, the minimal five dimensional two-derivative (gauged) supergravity theory is obtained by the combination of the locally superconformal two-derivative theories for the vector and linear multiplets together with a BF coupling of the two superconformal compensators that leads to a supersymmetric cosmological constant. Using this setup, locally supersymmetric completions of the Weyl tensor squared and the scalar curvature squared were constructed for the first time, respectively, in \cite{HOT} and \cite{OP2} by using component field techniques. 

The third independent, locally superconformal invariant, which includes a Ricci tensor-squared term, was constructed in superspace in \cite{BKNT-M14} by using a 5D analog of the ``Log multiplet'' construction in 4D $\cN=2$ supergravity of \cite{Butter:2013lta}. However, due to the computational complexity associated to the construction of this invariant, it took almost 10 years to obtain the results starting from the very simple building block in superspace. For example, one must compute up to eight local $Q$-supersymmetry transformations on the logarithm of the primary field $W$ of the vector multiplet compensator to obtain the component action of the 5D Log invariant. In our paper we present for the first time the component structure of this invariant in a standard Weyl multiplet background. See \cite{Gold:2023ymc} for the bosonic terms based on using a dilaton Weyl multiplet instead of the standard Weyl multiplet of conformal supergravity. Moreover, by employing the techniques mentioned above, and by a substantial use of the computer algebra program {\it Cadabra} \cite{Cadabra-1,Cadabra-2}, in our work we have also obtained all the fermionic contributions for all the three curvature-squared invariant for the first time. The reader who is only interested in the bosonic results of all the curvature squared actions in a standard Weyl background can look directly  to section \ref{section-5} of our paper, specifically: equations \eqref{Weyl-action-full} and \eqref{Weyl-action-degauged} for the Weyl squared invariant; equations \eqref{Log-action-0} and \eqref{log-action-degauged} for the Log invariant; and equations \eqref{action-R2-bos} and \eqref{desc-W-R2_000} for the scalar curvature-squared invariant. Due to the size of the fermionic structures arising for the building blocks of the invariants, much of the fermionic contributions are given in the (more than 200 pages long) supplementary file accompanying our paper.

Having obtained all the building blocks for minimal gauged supergravity with the addition of the three curvature-squared invariants, we are in the position to obtain for the first time the multiplets of equations of motions (EOMs) for this five-parameter class of theories. Because these models are obtained within a superconformal tensor calculus approach, all the EOMs turn out to be organised in terms of superconformal multiplets described by three main primary (super)fields: a scalar primary supercurrent operator $J$ which arises from the variation of the fields of the standard Weyl multiplet; a composite vector multiplet of equations of motion which arises from the variation of the linear multiplet compensator; and a composite linear multiplet of equations of motion which arises from the variation of the vector multiplet compensator. The primary (super)fields associated with these three multiplets of EOMs were presented in \cite{GHKT-M}. In this paper we will present a derivation of the primary EOMs based on an interplay of superspace and component techniques together with all their superconformal descendants. An advantage in obtaining the equations of motion by this approach is that covariance will be manifest in every result, which is nontrivial because the various component actions have Chern-Simons terms and naked gravitini terms. Similar to the other building blocks of the curvature-squared invariants, the equations of motion have remarkable length and complexity. For this reason, once more, we have decided to relegate most of the complete fermionic results of the EOM analysis in the supplementary file associated to this paper. It is our hope that these results will be later used to advance our understanding of the on-shell structure of these models.

Our paper is organised as follows. In section \ref{section-2} we review the structure of the 5D, $\cN=1$ superconformal multiplets that will be used in this work. We do present results both in conformal superspace and in components in the so called traceless frame employed in \cite{BKNT-M14}. It is useful to mention that our notation and conventions correspond to that of \cite{BKNT-M14} (see also \cite{Hutomo:2022hdi}, where some typos from \cite{BKNT-M14} were fixed).
In section \ref{section-3} we review the definition of the BF action principle which is the building block for all the invariants studied in this paper. We then review the construction of all the two-derivative invariants that define minimal gauged supergravity in five dimensions. 
In section \ref{section-4}, elaborating on the results of \cite{GHKT-M}, we describe the component structure of all these composite primary multiplets (including the fermionic terms that can be found in the supplementary file) which are used to construct the three independent four-derivative invariants studied in our paper.
By using all the building blocks of sections \ref{section-2}--\ref{section-4}, in section \ref{section-5} we present the bosonic part of the three curvature-squared invariants, including the ``Log invariant" which is presented for the first time in subsection \ref{sc-logW-action} in a standard Weyl basis.\footnote{The Log invariant in the gauged dilaton Weyl basis has been obtained recently in \cite{Gold:2023ymc}.}
Sections \ref{section-6} and \ref{section-7} are respectively devoted to firstly provide a derivation of the primary equations of motion which first appeared in \cite{GHKT-M}, and secondly to analyse all the descendants of the equations of motion.
We conclude our paper in section \ref{Conclusion} where we comment on possible developments and applications of our results.
We accompany our paper with three appendices and a long supplementary file.
Appendix \ref{App-A} collects results about conformal superspace in the traceless frame of \cite{BKNT-M14,Hutomo:2022hdi}.
In appendix \ref{Building-Blocks} we present several equations that we have used to turn the result we have obtained in \textit{Cadabra} for the $\cH^{a}_{\log}$ descendant of the Log multiplet to its simplified form given in equations \eqref{Ha-log-comp} and \eqref{Ha-log-comp_b}. 
Finally, in appendix \ref{app:Conventions} we explain how to map our notations and conventions to the ones of some other groups.


\section{Superconformal multiplets}
\label{section-2}

This section is devoted to a review of several superconformal multiplets in five dimensions which are the building blocks for the various supersymmetric curvature-squared invariants discussed in this paper. We will first describe the standard Weyl multiplet and then move on to the off-shell Abelian vector and linear multiplets. For each of the previously mentioned multiplets, we will present both their formulations in superspace  and components for completeness. Our analysis follows the superspace and components notation and results of \cite{BKNT-M14,Hutomo:2022hdi,GHKT-M}.\footnote{For various discussions on off-shell multiplets in five dimensions, see also other works based on superspace, e.g., \cite{KL,KT-M5D1,Howe5Dsugra,HL,Howe,KT-M_5D2,KT-M_5D3,KT-M08,K06}, and component approaches, e.g., \cite{GST1,GST2,GZ,CD,Zucker1,Zucker2,Zucker3,Ohashi1,Ohashi2,Ohashi3,Ohashi4,Bergshoeff1,Bergshoeff2,Bergshoeff3}. }

\subsection{The standard Weyl multiplet}
\label{SWM}
We begin by describing the standard Weyl multiplet in conformal superspace and components. 

\subsubsection{The standard Weyl multiplet in superspace}
\label{SWM-superspace}
In five dimensions, the standard Weyl multiplet \cite{Bergshoeff1} is associated with the gauging of the superconformal algebra $\rm F^2(4)$.  The multiplet contains $32+32$ physical components described by a set of independent gauge fields: the vielbein $e_\hm{}^\ha$,
the gravitino $\psi_\hm{}_\hal^i$,
the ${\rm SU(2)}_R$ gauge field $\phi_\hm{}^{ij}$, and a dilatation gauge field $b_\hm$. The other fields associated with the remaining gauge symmetries: the spin connection  $\omega_\hm{}^{\ha\hb}$, the $S$-supersymmetry connection
$\phi_\hm{}_\hal^i$, and the special conformal connection
$\mathfrak{f}_\hm{}^\ha$ are composite, meaning that they are determined in terms of the other fields by imposing some appropriate
curvature constraints. The standard Weyl multiplet also contains a set of matter fields: a real antisymmetric
tensor $w_{\ha\hb}$, a fermion $\chi_\hal^i$, and a real auxiliary scalar $D$. 

The $\cN=1$ conformal superspace is parametrised by
local bosonic $(x^{\hm})$ and fermionic $(\theta_i)$ coordinates 
$z^{\hM} = (x^{\hm},\q^{{\mu}}_i)$, 
where $\hm = 0, 1, 2,3, 4$, ${\mu} = 1, \cdots, 4$, and $i = 1, 2$.
To gauge the
superconformal algebra, one introduces
covariant derivatives $ {\nabla}_{\hA} = (\nabla_{\ha} , \nabla_{\hal}^i)$ which have the form
\bsubeq
\bea\label{eq:covD}
\de_{\hA} 
= E_{\hA} - \o_{\hA}{}^{\underline{b}}X_{\underline{b}} 
&=& E_{\hA} - \hf \O_{\hA}{}^{\ha \hb} M_{\ha \hb} - \Phi_{\hA}{}^{ij} J_{ij} - B_{\hA} \mathbb{D} - \mathfrak{F}_{\hA}{}^{\hB} K_{\hB}~,
\\ &=& E_{\hA} - \hf \O_{\hA}{}^{\ha \hb} M_{\ha \hb} - \Phi_{\hA}{}^{ij} J_{ij} - B_{\hA} \mathbb{D} - \mathfrak{F}_{\hA}{}^{\a i} S_{\a i} - \mathfrak{F}_{\hA}{}^{a} K_{a} ~.~~~~
\eea
\esubeq
Here $E_{\hA} = E_{\hA}{}^{\hM} \partial_{\hM}$ is the inverse super-vielbein,
$M_{\ha \hb}$ are the Lorentz generators, $J_{ij}$ are generators of the
${\rm SU(2)}_R$ $R$-symmetry group,
$\mathbb D$ is the dilatation generator, and $K_{\hA} = (K_{\ha}, S_{\hal i})$ are the special superconformal
generators.
The super-vielbein one-form is $E^{\hA} =\rd z^{\hM} E_{\hM}{}^{\hA}$ with $E_{\hM}{}^{\hA} E_{\hA}{}^{\hN} =\d_{\hM}^{\hN}$ and
 $E_{\hA}{}^{\hM} E_{\hM}{}^{\hB}=\d_{\hA}^{\hB}$.
Associated with each generator $X_{\underline{a}} = (M_{\ha\hb},J_{ij},\bbD, S_{\hal i}, K_\ha)$ is the connection super one-form 
$\omega^{\underline{a}} = (\O^{\ha\hb},\F^{ij},B,\mathfrak{F}^{\hal i},\mathfrak{F}^{\ha})= \rd z^\hM \omega_\hM{}^{\underline{a}} = E^{\hA} \o_{\hA}{}^{\underline{a}}$.

The algebra of covariant derivatives
\begin{align}
[ \nabla_\hA , \nabla_\hB \}
	&= -\mathscr{T}_{\hA\hB}{}^\hC \nabla_\hC
	- \frac{1}{2} {\mathscr{R}(M)}_{\hA\hB}{}^{\hc\hd} M_{\hc\hd}
	- {\mathscr{R}(J)}_{\hA\hB}{}^{kl} J_{kl}
	\non \\ & \quad
	- {\mathscr{R}(\mathbb{D})}_{\hA\hB} \mathbb D
	- {\mathscr{R}(S)}_{\hA\hB}{}^{\hga k} S_{\hga k}
	- {\mathscr{R}(K)}_{\hA\hB}{}^\hc K_\hc~,
	\label{nablanabla}
\end{align}
is constrained to be  expressed  in terms of a single primary superfield, the super-Weyl tensor $W_{\hal \hbe}$,\footnote{Here and in what follows, an antisymmetric rank-2 tensor $T_{a b} = -T_{b a}$ can equivalently be written as: $T_{a b} = (\Sigma_{a b}){}^{\a \b} T_{\a \b}$ and $T_{\hal \hbe} = 1/2 \,({\S}^{\ha \hb})_{\hal \hbe}T_{\ha \hb}$.} which has the following properties
\be
W_{\hal \hbe} = W_{\hbe \hal} \ , \quad
K_{\hA} W_{\hal \hbe} = 0 \ ,
 \quad \mathbb D W_{\hal \hbe} = W_{\hal \hbe} \ ,
\ee
and satisfies the Bianchi identity
\be 
\nabla_\hga^k W_{\hal \hbe} = \nabla_{(\hal}^k W_{\hbe \hga )} + \frac{2}{5} \eps_{\hga (\hal} \nabla^{\hde k} W_{\hbe ) \hde} \ . \label{WBI}
\ee
In  \eqref{nablanabla},
$ \mathscr{T}_{\hA \hB}{}^C$ is the torsion, while $ \mathscr{R}(M)_{\hA \hB}{}^{\hc \hd}$,
$ \mathscr{R}(J)_{\hA \hB}{}^{kl}$, $ \mathscr{R}(\mathbb D)_{\hA \hB}$, $ \mathscr{R}(S)_{\hA \hB}{}^{ {{\g}}k}$, and $ \mathscr{R}(K)_{\hA \hB}{}^{\hc}$
are the curvatures associated with Lorentz, ${\rm SU(2)}_R$,
dilatation, $S$-supersymmetry, and special conformal
boosts, respectively.

In this paper we make use of the ``traceless'' frame conventional constraints for the conformal superspace algebra employed in appendix C of \cite{BKNT-M14} as well as in \cite{Hutomo:2022hdi}.
The full algebra of covariant derivatives \eqref{nablanabla} (including the explicit expressions for the torsion and curvature tensors) are given in appendix \ref{App-A} of our paper.

Let us introduce the dimension-3/2 superfields
\bsubeq \label{descendantsW-5d}
\begin{gather}
W_{\hal \hbe \hga}{}^k := \nabla_{(\hal}^k W_{\hbe \hga )} \ , \quad X_\hal^i := \frac{2}{5} \nabla^{\hbe i} W_{\hbe\hal} \ ,
\end{gather}
and the following dimension-2 descendant superfields 
\bea
W_{\hal \hbe \hga \hde} := \nabla_{(\hal}^k W_{\hbe \hga \hde) k} \ , \quad 
X_{\hal \hbe}{}^{ij} := \nabla_{(\hal}^{(i} X_{\hbe)}^{j)} 
\ , \quad
Y := \ri \nabla^{\hga k} X_{\hga k} \ .
\eea
\esubeq
It can be checked that only the superfields \eqref{descendantsW-5d} and their vector derivatives appear upon taking successive spinor derivatives of $W_{\a \b}$. Specific relations that connect the various descendant fields which we will need later are given below:
\bsubeq \label{eq:Wdervs}
\bea
\nabla_{\hga}^k W_{\hal\hbe} 
&=& W_{\hal\hbe\hga}{}^k + \eps_{\hga (\hal} X_{\hbe)}^k \ ,  \label{Wdervs-a}
\\
\nabla^{i}_{\a}{X^{j}_{{\beta}}} 
&=& 
X_{{\alpha} {\beta}}\,^{i j}
+\frac{\ri}{8} \eps^{i j} \eps_{{\alpha} {\beta}} Y
-\frac{3\ri}{2} \eps^{i j} (\Gamma^{{a}})_{{\alpha}}{}^{ \rho} {\nabla}_{{a}}{W_{{\beta} {\rho}}}
-2\ri \eps^{i j} W_{{\alpha}}{}^{{\rho}} W_{{\beta} {\rho}}
\non\\
&&
+\frac{\ri}{2} \eps^{i j} \eps_{{\alpha} {\beta}} W^{{\gamma} {\delta}} W_{{\gamma} {\delta}}
-\frac{\ri}{2}\eps^{i j} (\Gamma^{{a}})_{{\beta}}{}^{{\rho}} {\nabla}_{{a}}{W_{{\alpha} {\rho}}}
~,
\\
\nabla^{i}_{{\alpha}}{W_{{\beta} {\gamma} {\lambda}}{}^{j}}
&=& -  \frac{1}{2} \eps^{i j} \Big( W_{{\alpha} {\beta} {\gamma} {\lambda}} + 3 \ri  (\Gamma_{{a}})_{{\alpha} ({\beta}}  {\nabla}^{{a}}{W_{{\gamma} {\lambda})}} + 3 \ri \eps_{{\alpha} ({\beta}} (\Gamma_{{a}})_{{\gamma}}{}^{{\tau}}   {\nabla}^{{a}}{W_{{\lambda}) {\tau}}}\Big) \non \\&&- \frac{3}{2}\eps_{{\alpha} ({\beta}} X_{{\gamma} {\lambda})}\,^{i j} \ , 
\\ 
\nabla^{i}_{{\alpha}}{W_{{\beta} {\gamma} {\lambda} {\rho}}} 
&=& - 4 \ri (\Gamma_{{a}})_{{\alpha} ({\beta}} {\nabla}^{{a}}{W_{{\gamma} {\lambda} {\rho})}\,^{i}}- 6 \ri  W_{{\alpha} ({\beta} {\gamma}}\,^{i} W_{{\lambda} {\rho})}+6 \ri W_{{\alpha} ({\beta}} W_{{\gamma} {\lambda} {\rho})}\,^{i}\non \\&& +6 \ri \eps_{{\alpha} ({\beta}} \Big( W_{{\gamma} {\lambda}} X^{i}_{{\rho})}-2 (\Gamma_{{a}})_{{\gamma}}{}^{{\tau}}   {\nabla}^{{a}}{W_{{\lambda} {\rho}) {\tau}}\,^{i}}-  W_{{\gamma}}{}^{{\tau}} W_{{\lambda} {\rho} ){\tau}}\,^{i}\Big) \ , \\
\nabla^{i}_{{\alpha}}{X_{{\beta} {\gamma}}\,^{j k}} 
&=& \ri  \eps^{i (j} \Big( -3   W_{({\beta}}{}^ {\lambda} W_{ {\gamma}) {\alpha}\lambda}\,^{ k)} -  \eps_{{\alpha} ({\beta}}  W^{{\rho} {\tau}} W_{{\gamma}) {\rho} {\tau}}\,^{k)}  -  W_{{\alpha} {\lambda}} W_{{\beta} {\gamma} }\,^{k){\lambda}}  - \frac{3}{2} W_{{\beta} {\gamma}} X^{k)}_{{\alpha}}  
\non\\&& \qquad ~
+\frac{1}{2} W_{{\alpha} ({\beta}} X^{k)}_{{\gamma})} +\frac{3}{2} \eps_{{\alpha} ({\beta}}  W_{{\gamma}) {\lambda}} X^{k) {\lambda}}  + 2 (\Gamma^{{a}})_{{\alpha}}{}^{{\rho}}  {\nabla}_{{a}}{W_{{\beta} {\gamma} {\rho}}\,^{k)}} 
\non\\&& \qquad  ~
+ 2  (\Gamma^{{a}})_{({\beta} }{}^{{\rho}}   {\nabla}_{{a}}{W_{ {\gamma}) {\alpha} {\rho}}\,^{k)}}  -  (\Gamma^{{a}})_{{\alpha} ({\beta}}  {\nabla}_{{a}}{X^{k)}_{{\gamma})}}
+  \eps_{{\alpha} ({\beta}} (\Gamma^{{a}})_{{\gamma}) {\lambda}}  {\nabla}_{{a}}{X^{k) {\lambda}}} \Big) 
~,~~~~~~ \\
\nabla^{i}_{{\alpha}} {Y} 
&=& 8(\Gamma^{{a}})_{{\alpha}}{}^{{\beta}}  {\nabla}_{{a}}{X^{i}_{ {\beta}}} + 8 W_{{\alpha}}{}^{{\beta}} X^{i}_{{\beta}}
~.
\eea
\esubeq
As implied by \eqref{WBI}, the  dimension-2 superfields $X_{\a\b}{}^{ij}$ and $W_{\a\b\g\d}$ obey the following Bianchi identities in the traceless frame:
\bsubeq\label{WeylmultipletBI}
\bea
  \nabla_{(\a}{}^{\g} X_{\b)\g}{}^{ij} &=& -\frac{1}{2} X^{\g (i} W_{\a \b \g}{}^{j)}
  ~,\\
   \nabla_{(\a}{}^{\l} W_{\b \g \tau)\l} &=& 3 \ri  \nabla_{(\a}{}^{\l} \Big(W_{\b \g} W_{\tau)\l}\Big)
   ~.
 \eea
\esubeq
The independent descendant superfields of $W_{\a\b}$, specifically $W_{\hal \hbe \hga}{}^k,\, X_\hal^i,\,W_{\hal \hbe \hga \hde},\,X_{\hal \hbe}{}^{ij},\,Y$, are all annihilated by $K_{\ha}$. However, under $S$-supersymmetry, they transform as follows:
\bsubeq 
\bea 
S_{\hal i} W_{\hbe\hga\hde}{}^j 
&=& 
6 \d^j_i \eps_{\hal (\hbe} W_{\hga \hde)} \ , \qquad
S_{\hal i} X_\hbe^j = 4 \d_i^j W_{\hal\hbe}~, 
\\
S_{\hal i} W_{\hbe\hga\hde\hrh} &=& 24 \eps_{\hal (\hbe} W_{\hga\hde \hrh)}{}_i \ , \qquad
S_{\hal i} Y = 8 \ri X_{\hal i} 
~, 
\\
S_{\hal i} X_{\hbe\hga}{}^{jk} &=& - 4 \d_i^{(j} W_{\hal\hbe\hga}{}^{k)} + 4 \d_i^{(j} \eps_{\hal (\hbe} X_{\hga)}^{k)} \ .\label{S-on-X_Y-a}
\eea 
\esubeq
We also stress that in the paper, and in the supplementary file, we will often use the following notation:
\bea
\Phi_{a b}{}^{ij}
:=
{\mathscr{R}(J)}_{a b}{}^{ij} 
= -\frac{3 \ri}{4} X_{\ha \hb}{}^{ij}
~.
\eea

The gauge group of conformal supergravity is denoted by $\cG$. It is generated by
{\it covariant general coordinate transformations},
$\delta_{\rm cgct}$, associated with a local superdiffeomorphism parameter $\xi^{\hA}$ and
{\it standard superconformal transformations},
$\delta_{\cH}$, associated with the local superfield parameters:
the dilatation $\s$, Lorentz $\L^{\ha \hb}=-\L^{\hb \ha}$,  ${\rm SU(2)}_R$ $\L^{ij}=\L^{ji}$,
 and special conformal (bosonic and fermionic) transformations $\L^{\hA}=(\eta^{\hal i},\L^{\ha}_{K})$.
The covariant derivatives transform as
\bsubeq
\label{sugra-group}
\bea
\d_\cG \nabla_{\hA} &=& [\cK , \nabla_{\hA}] \ ,
\label{TransCD}
\eea
with
\bea
\cK = \xi^{\hC}  {\nabla}_{\hC} + \hf  {\L}^{\ha \hb} M_{\ha \hb} +  {\L}^{ij} J_{ij} +  \s \mathbb D +  {\L}^{\hA} K_{\hA} ~.
\eea
\esubeq
A covariant (or tensor) superfield $U$ transforms as
\be
\d_{\cG} U =
(\d_{\rm cgct}
+\d_{\cH}) U =
 \cK U
 ~. \label{trans-primary}
\ee
The superfield $U$ is said to be \emph{superconformal primary} of dimension $\D$ if $K_{\hA} U = 0$ (it suffices to require that $S_{\a i} U = 0$) and $\mathbb D U = \D U$.

\subsubsection{The standard Weyl multiplet in components}
Let us now explain how to systematically obtain various component fields of the standard Weyl multiplet from the superspace geometry described in the subsection \ref{SWM-superspace}. The vielbein $(e_{\hm}{}^{\ha})$ and gravitini $(\psi_{\hm}{}^{i}_{ \hal})$ appear as the coefficients of $\rd x^\hm$ of the super-vielbein $E^\hA = (E^\ha, E^\hal_i) = \rd z^\hM\, E_\hM{}^\hA$,
\bea\label{singlebar}
e_{\hm}{}^{\ha} (x):= E_{\hm}{}^{\ha}(z)|~, \qquad \psi_{\hm}{}^{i}_{ \hal}(x) := 2 E_{\hm}{}^{i}_{\hal} (z)|~,
\eea
where a single vertical line next to a superfield denotes the usual component projection to $\q = 0$, i.e., $V(z) | := V(z) |_{\q=0}$. This operation can be written in a coordinate-independent way using the so-called double-bar projection \cite{Baulieu:1986dp,Binetruy:2000zx}
\begin{align}\label{doublebar}
e{}^{\ha} = \rd x^{\hm} e_{\hm}{}^{\ha} = E^{\ha}\doubar~,~~~~~~
\psi_{\hal}^{i} = \rd x^{\hm} \psi_{\hm}{}^{i}_{ \hal} =  2 E_{\hal}^{i} \doubar ~,
\end{align}
where the double-bar denotes setting $\q = \rd \q = 0$.
The remaining fundamental and composite one-forms are also obtained by taking the projections of the corresponding superspace one-forms,
\begin{align}
	\phi^{ij} := \Phi{}^{ij} \doubar~, \quad
	b := B\doubar ~, \quad
\omega^{\ha \hb} := \Omega{}^{\ha \hb} \doubar ~, \quad  
\phi^{\hal i} := 2 \,\mathfrak F{}^{\hal i}\doubar~, \quad
\mathfrak{f}{}^{\ha} := \mathfrak{F}{}^{\ha}\doubar~. 
\end{align}

The covariant matter fields are contained within the super-Weyl tensor $W_{\hal \hbe}$ and its independent descendants, 
\bsubeq \label{comps-W}
\bea
W_{\hal \hbe} &:=&  W_{\hal \hbe}\loco \ ,
\\
\chi_{\hal}^{i} &:=& \frac{3 \ri}{32}X_{\hal}^{i} \loco = \frac{3\ri}{80}\de^{\hbe i} W_{\hbe \hal}\loco~,
\\
D&:=& -\frac{3}{128}Y \loco=
-\frac{3\ri}{320}\de_{\hal}^k\de_{\hbe k}W^{\hal \hbe}\loco
~.
\eea
\esubeq
The other components of the super Weyl tensor are given by $W_{\ha \hb}{}_{\hal}^{i} := W_{\ha \hb}{}_{\hal}^{i} \loco$, $W_{\hal \hbe \hga \hde} := W_{\hal \hbe \hga \hde}  \loco$ and $\Phi_{a b}{}^{ij} := \Phi_{a b}{}^{ij} \loco = -\frac{3 \ri}{4} X_{\ha \hb}{}^{ij} \loco= -\frac{3 \ri}{4} (\S_{\ha\hb})^{\hal\hbe}X_{\hal \hbe}{}^{ij} \loco$.\footnote{In the case of $W_{\a\b}$, $W_{\ha \hb}{}_{\hal}^{i}$, $\Phi_{a b}{}^{i j}$, and $W_{\hal \hbe \hga \hde}$, unless specified, it should be clear from the context if we refer to the superfields or their component projections.}
These will turn out to be composite and expressed in terms of the component curvatures.

Taking the double-bar projection of $\de = E^{\hA} \de_{\hA}$, the component vector covariant derivative $\de_{\ha}$ is defined to coincide with the projection of the superspace derivative $\de_{\ha} \loco$,
\begin{align}
e_{\hm}{}^{\ha} \nabla_{\ha} = \pa_{\hm}
	- \frac{1}{2} \psi_{\hm}{}^{\hal}_i \nabla_{\hal}^i \loco
	- \frac{1}{2} \omega_{\hm}{}^{\ha \hb} M_{\ha \hb}
	- b_{\hm} \mathbb D
	- \phi_{\hm}{}^{ij} J_{ij}
	- \frac{1}{2} \phi_{\hm}{}^{\hal i} S_{ \hal i}
	- {\mathfrak f}_{\hm}{}^{\ha} K_{\ha}
	~.
\end{align}
Here, the projected spinor covariant derivative $\nabla_{\hal}^i\loco$
corresponds to the generator of $Q$-supersymmetry. It is defined such that
if $\cU = U\vert$, then $Q_{\hal}^i \cU:=\nabla_{\hal}^i| \cU := (\nabla_{\hal}^i U) \loco$.
For the other generators, e.g.,
$M_{\ha \hb}\cU =(M_{\ha \hb} U)\loco$, there is no ambiguity in identifying the bar projection; hence, local diffeomorphisms,
$Q$-supersymmetry transformations, and so forth descend naturally
from their corresponding rule in superspace.

The component supercovariant curvature tensors are given by
\begin{align}
[\nabla_\ha, \nabla_\hb]
	&= - R(P)_{\ha \hb}{}^\hc \nabla_{\hc}
	- R(Q)_{\ha \hb}{}^{\hal}_i \nabla_\hal^i\loco
	- \frac{1}{2} \RM_{\ha \hb}{}^{\hc \hd} M_{\hc \hd}
	- R(J)_{\ha\hb}{}^{ij} J_{ij}
	\eol & \quad
	- \RD_{\ha\hb} \gD 
	- \RS_{\ha\hb}{}^{\hga k} S_{\hga k}
	- \RK_{\ha\hb}{}^\hc K_\hc~.
\end{align}
We have introduced $R(P)_{\ha \hb}{}^{\hc} = \scT_{\ha \hb}{}^\hc \loco$ and 
$R(Q)_{\ha \hb}{}_\hal^i = \scT_{\ha\hb}{}_\hal^i \loco$.
$ {R}(M)_{\ha \hb}{}^{\hc \hd}$, $ {R}(J)_{\ha \hb}{}^{ij}$, $ {R}(\mathbb D)_{\ha \hb}$, $ {R}(S)_{\ha \hb}{}^{\hga k}$, and $ {R}(K)_{\ha \hb}{}^{\hc}$ coincide with the lowest components of the corresponding superspace curvature tensors given in appendix \ref{App-A}.

The constraints on the superspace curvatures determine how to supercovariantise
a given component curvature by taking the double-bar projection of the
superspace torsion and each of the curvature two forms. One then finds \cite{BKNT-M14}\footnote{When defining products of spinors, the spinor indices are sometimes suppressed. The spinor indices should be viewed as contracted from top left to bottom right.}
\begin{subequations}\label{eq:hatCurvs}
\begin{align}
R(P)_{\ha\hb}{}^\hc &= 2 \,e_\ha{}^\hm e_\hb{}^\hn \cD_{[\hm} e_{\hn]}{}^\hc
	- \frac{\ri}{2} \psi_{\ha  j} \G^\hc \psi_\hb{}^j\ , \\
R(Q)_{\ha \hb}{}_{\hal}^i &=
	e_\ha{}^\hm e_\hb{}^\hn \cD_{[\hm} \psi_{\hn]}{}_\hal^i
	+ \ri (\G_{[\ha } \phi_{\hb]}{}^i)_\hal
	\eol & \quad
	+ \frac{1}{8} W_{\hc\hd} \Big(
		3 (\S^{\hc\hd} \G_{[\ha })_\hal{}^\hbe - (\G_{[\ha } \S^{\hc\hd})_\hal{}^\hbe
		\Big) \psi_{\hb]}{}_\hbe^i~, \\
R(M)_{\ha\hb}{}^{\hc\hd} &= \cR(\omega)_{\ha\hb}{}^{\hc\hd}
	+ 8 \,\delta_{[\ha }{}^{[\hc} {\mathfrak f}_{\hb]}{}^{\hd]}
	- 2 \,\psi_{[\ha  j} \Sigma^{\hc\hd} \phi_{\hb]}{}^{j}
	\eol & \quad
	+ \frac{16\ri}{3} \delta_{[\ha }{}^{[\hc}  \psi_{\hb]i} \G^{\hd]} \chi^i
	- \ri \psi_{[\ha  i} \Big(\G_{\hb]} R(Q)^{\hc\hd  i} +
		2\G^{[\hc} R(Q)_{\hb]}{}^{\hd] i} \Big)
	\eol & \quad
	+ \frac{\ri}{2} \psi_{\ha j} \psi_\hb{}^j W^{\hc\hd}
	- \frac{\ri}{4} (\psi_{\ha  j} \G_\he \psi_{b}{}^j) \tilde W^{\hc\hd\he} \ , \\
R(J)_{\ha\hb}{}^{ij} &= \cR(\phi)_{\ha\hb}{}^{ij}
	- 3 \,\psi_{[\ha }^{(i} \phi_{\hb]}{}^{j)}
	- 8\ri \,\psi_{[\ha }^{(i} \G_{\hb]} \chi^{j)} \ , \\
R(\bbD)_{\ha\hb} &= 2\, e_\ha{}^\hm e_\hb{}^\hn \pa_{[\hm} b_{\hn]}
	+ 4 \,{\mathfrak f}_{[\ha\hb]}
	+ \psi_{[\ha  j} \phi_{\hb]}{}^j
	+ \frac{8\ri}{3} \,\psi_{[\ha  j} \G_{\hb]} \chi^j
	 \ ,
\end{align}
\end{subequations}
where we have also defined
\bea
\tilde{W}_{abc} = \hf \ve_{abcde} W^{de}~.
\eea
In  \eqref{eq:hatCurvs}, we have introduced the spin, dilatation, and
${\rm SU}(2)_R$ covariant derivative
\begin{align}
\cD_\hm = \pa_\hm
	- \frac{1}{2} \omega_\hm{}^{\hb \hc} M_{\hb \hc}
	- b_\hm \mathbb D
	- \phi_\hm{}^{ij} J_{ij}~,  \qquad
\cD_{\ha} = e_{\ha}{}^{\hm} \cD_{\hm}~.
\label{cD-def}
\end{align}
Note that the covariant derivative $\cD_{a}$ obeys the following commutation relation
\bea \label{degauged-vec-algebra}
    [\cD_{a}, \cD_{b}] = - \frac{1}{2} \cR(\o)_{a b}{}^{c d} M_{c d} - \cR (\phi)_{ab}{}^{ij} J_{i j}~,
\eea
where the curvatures takes the form
\bsubeq
\begin{align}
\cR(\omega)_{\ha\hb}{}^{\hc\hd} &:= 2 e_\ha{}^\hm e_\hb{}^\hn \Big(\partial_{[\hm} \omega_{\hn]}{}^{\hc\hd} - \omega_{[\hm}{}^{\hc\he} \omega_{\hn]}{}_\he{}^\hd\Big)  \ , \\
\cR(\phi)_{\ha\hb}{}^{ij} &:= 2\,e_\ha{}^\hm e_\hb{}^\hn \Big( \pa_{[\hm} \phi_{\hn]}{}^{ij}
	+ \phi_{[\hm}{}^{k (i} {\phi}^{j)}{}_{\hn]k }\Big)~.
\end{align}
\esubeq
The component curvatures turn out to obey ``traceless'' conventional constraints \cite{BKNT-M14}
\begin{align} \label{componentConstTRv2}
R(P)_{\ha\hb}{}^\hc &= 0~, \qquad
(\G^\ha)_\hal{}^\hbe R(Q)_{\ha \hb}{}_{\hbe}^i = 0~, \qquad
R(M)_{\ha\hb}{}^{\hc\hb} = 0~,
\end{align}
thus allowing us to solve for the composite connections as follows:
\begin{subequations} \label{componentComposite}
\begin{align}
\omega_{\ha\hb\hc} &= 
	\omega(e)_{\ha\hb\hc} + \frac{\ri}{4} (
	\psi_{\ha k} \G_\hc \psi_\hb{}^k
	+ \psi_{\hc k} \G_\hb \psi_\ha{}^k
	- \psi_{\hb k} \G_\ha \psi_\hc{}^k)
	+ 2 b_{[\hb} \eta_{\hc ] \ha} ~, \\
\ri \,\phi_{\hm}{}^i
	&= \frac{2}{3} (\G^{[\hp} \delta_\hm{}^{\hq]} + \frac{1}{4} \G_\hm \S^{\hp\hq})
		\Big(
		\cD_{[\hp} \psi_{\hq]}{}^i
		+ \frac{1}{8} W_{\hc\hd}
			\big(3 \S^{\hc\hd} \G_{[\hp} \psi_{\hq]}{}^i
			- \G_{[\hp} \S^{\hc\hd} \psi_{\hq]}{}^i
		\big)\Big)~, \\
{\mathfrak f}_\ha{}^\hb &=
	- \frac{1}{6}\cR(\omega)_{\ha \hc}{}^{\hb\hc}
	+ \frac{1}{48} \delta_\ha{}^\hb \cR(\omega)_{\hc\hd}{}^{\hc\hd}
	- \frac{\ri}{6} \psi_{\hc j} \G^{[\hb} R(Q)_\ha{}^{\hc]j}
	- \frac{\ri}{12} \psi_{\hc j} \G_\ha R(Q)^{\hb\hc j}
	\eol & \quad
	+ \frac{1}{3} \psi_{[\ha  j} \S^{\hb\hd} \phi_{\hd]}{}^j
	- \frac{1}{24} \delta_\ha{}^\hb (\psi_{\hc j} \S^{\hc\hd}  \phi_\hd{}^j)
	- \frac{2\ri}{3} (\psi_{\ha  j} \G^\hb \chi^j)
	\eol & \quad
	- \frac{\ri}{12} \psi_{\ha j} \psi_\hc{}^j W^{\hb\hc}
	+ \frac{\ri}{24} (\psi_{\ha j} \G_\he \psi_\hd{}^j) \tilde W^{\hb\hd\he}
	\eol & \quad
	+ \frac{\ri}{192} \delta_\ha{}^\hb
		\Big(2 (\psi_{\hc j} \psi_\hd{}^j) W^{\hc\hd} -
		(\psi_{\hc j} \G_\he \psi_\hd{}^j) \tilde W^{\hc\hd \he}\Big)~.
\end{align}
\end{subequations}
Here $\o(e)_{\ha \hb \hc} = -\hf (\cC_{\ha \hb \hc} + \cC_{\hc \ha \hb } -\cC_{\hb \hc \ha})$ is the usual torsion-free spin connection expressed in terms of the anholonomy coefficient $\cC_{\hm \hn}{}^{\ha} := 2 \pa_{[\hm} e_{\hn]}{}^{\ha}$. The Riemann tensor is denoted by $\cR(\omega)_{a b}{}^{c d}=\cR_{ab}{}^{cd}$, whereas $\cR_{a}{}^{b} = \cR_{ac}{}^{b c}$ and $\cR=\cR_{ab}{}^{ab}$ correspond to Ricci tensor and Ricci scalar, respectively.
The dilatation curvature $R(\mathbb{D})_{\ha \hb}$ now vanishes due to eqs.~\eqref{componentComposite}.  
It is worth noting that in the traceless frame $\phi_m{}^i$ has no dependence upon the matter field $\chi^i$, and $\mathfrak{f}_a{}^b$ has no dependence upon $D$. This choice minimises the dependence of the covariant derivatives upon some matter ``auxiliary'' fields and will simplify part of the analysis in the coming sections.

The supersymmetry transformations of the fundamental gauge connections of the Weyl
multiplet can be derived directly from the transformations of their corresponding
superspace one-forms.
We restrict
to all local superconformal transformations except local 
translations (covariant general coordinate transformations).
We denote such transformations by $\d = \d_{Q} + \d_{\cH}$ and define the operator
\bea
\delta
= 
\xi^{\hal}_iQ_{\hal}^i
+\hf \l^{\ha \hb}M_{\ha \hb} 
+ \l^{ij}J_{ij} 
+ \l_{\mathbb{D}}\mathbb{D} 
+ \l^{\ha} K_{\ha}
+ \eta^{\hal i} S_{\hal i}
~.
\eea
Here, the local component parameters are respectively defined as the $\q=0$ components of the corresponding superfield parameters, $\xi^{\hal}_i := \xi^{\hal}_i \loco$, $\l^{\ha \hb} := \L^{\ha \hb} \loco$, $\l^{ij} := \L^{ij} \loco$, $\l_{\mathbb{D}} := \sigma \loco$, $\l^{\ha} := \L^{\ha}_K \loco$, and $\eta^{\hal i} := \eta^{\hal i} \loco$.
The local superconformal transformations of 
the independent connection fields of the standard Weyl multiplet
are given by \cite{BKNT-M14}
\bsubeq\label{transf-standard-Weyl}
\bea
\delta e_{\hm}{}^{\ha} 
&=& 
\ri\, (\xi_i\G^{\ha}{\psi}_{\hm}{}^i)
- \lambda_{\mathbb{D}}{e}_{\hm}{}^{\ha}
+ \lambda^{\ha}{}_{\hb}e_{\hm}{}^{\hb}
~,
\label{d-vielbein}
\\
\delta \psi_\hm{}_\hal^i &=&
	2 \cD_\hm \xi_\hal^i
	- \frac{1}{4} W_{\hc\hd} \Big(
		(\G_\hm \S^{\hc\hd})_\hal{}^\hbe 
		- 3 (\Sigma^{\hc\hd} \G_\hm)_\hal{}^\hbe \Big) \xi_\hbe^i
	+ 2 \ri \,(\G_\hm \eta^i)_\hal \non\\
&&+\frac{1}{2}\lambda^{\ha \hb}(\S_{\ha \hb} \psi_{\hm}{}^{i})_{\hal}
+ \lambda^{i}{}_{j} \,\psi_{\hm}{}_{\hal}^{j}
- \frac{1}{2}\lambda_{\mathbb{D}}\,{\psi}_{\hm}{}^{i}_{\hal}
~,
\label{d-gravitino}
\\
 \delta  \phi_{\hm}{}^{ij} &=& \pa_{\hm}\l^{ij}
-2 \phi_{\hm}{}^{(i}{}_{k} \l^{j) k}
+3 \xi^{(i} \phi_{\hm}{}^{j)} 
-3 \eta^{(i} \psi_{\hm}{}^{j)}
+8 \ri \xi^{(i} \G_{\hm} \chi^{j)}~,
\label{d-SU2}
\\
\delta b_{\hm}
&=& \partial_{\hm} \lambda_{\mathbb{D}} 
-\frac{8\ri}{3}\,\xi_i \G_{\hm} \chi^{i} 
-\xi_i   \phi_m{}^i
-\psi_{\hm}{}^i\eta_i  
- 2\lambda_{\hm}
~.
\label{d-dilatation}
\eea
In like fashion, one can derive the transformations $\d \o_{\hm}{}^{\ha \hb}, \d \phi_{\hm \hal}{}^i$, and $\d \mathfrak{f}_{\hm \ha}$, which we omit since these fields are composite. For the covariant matter fields, the transformations are given by \cite{BKNT-M14}
\bea
 \delta W_{\ha \hb} &=& 
2 \ri \, \xi_{i} R(Q)_{\ha \hb}{}^{i} 
- \frac{32 \ri}{3} \xi_{i} \S_{\ha \hb} \chi^i
- 2 \l_{[\ha}{}^{\hc} W_{\hb] \hc} 
+ \l_{\mathbb{D}} W_{\ha \hb}~,
  \label{d-W}
\\
\delta  \chi^{\hal i}
&=& \hf \xi^{\hal i} D
-\frac{1}{16} (\xi_{j} \S^{\ha \hb})^{\hal} R(J)_{\ha \hb}{}^{ij}
-\frac{3}{128} (\de_{\ha} W_{\hb \hc}) \Big( 3 (\xi^{i} \G^{\ha} \S^{\hb \hc})^{\hal} + (\xi^{i} \S^{\hb \hc} \G^{\ha})^{\hal} \Big)\non\\
&&+ \frac{3}{256} W_{\ha \hb} W_{\hc \hd} \ve^{\ha \hb \hc \hd \he} (\xi^i \G_{\he})^{\hal}
+ \frac{3\ri}{16} (\eta^{i} \S^{\ha \hb})^{\hal} W_{\ha \hb}
 \non\\
&&-\hf \l^{\ha \hb}(\chi^{i}\S_{\ha \hb})^{\hal}
+\l^{i}{}_{j}\chi^{\hal j}
+\frac{3}{2} \l_{\mathbb{D}} \chi^{\hal i}
~,
\label{d-chi}
\\
\delta D
&=& 2 \ri \, \xi_{i}\G^{\ha}\de_{\ha} \chi^{i} +  \ri W_{\ha \hb} (\xi_{i} \S^{\ha \hb} \chi^i)
+2 \eta_{i} \chi^{i}
+2 \l_{\mathbb{D} }D
~,
\label{d-D}
\eea
\esubeq
where
\bsubeq\label{nabla-on-W-and-Chi}
\bea
\de_{\ha}W_{\hb \hc} &=& 
\cD_{\ha}W_{\hb \hc}
-\ri \psi_{\ha i} R(Q)_{\hb \hc}{}^{i}
+ \frac{16 \ri}{3} \psi_{\ha i} \S_{\hb \hc} \chi^i~,\\
\de_{\ha} \chi^{\hal i}
&=& 
\cD_{\ha}\chi^{\hal i} 
-\frac{1}{4} \psi_{\ha}{}^{\hal i} D 
-\frac{3\ri}{32} (\phi_{\ha}{}^i \S^{\hb \hc})^{\hal}W_{\hb \hc}
+\frac{1}{32} (\psi_{\ha j} \S^{\hb \hc})^{\hal} R(J)_{\hb \hc}{}^{ij}
\non\\
&&+ \frac{3}{256} (\de_{\hb} W_{\hc \hd}) \Big( 3 (\psi_{\ha}{}^{i} \G^{\hb} \S^{\hc \hd})^{\hal}+ (\psi_{\ha}{}^{i} \S^{\hc \hd} \G^{\hb})^{\hal} \Big)
\non\\
&&
-\frac{3}{512} W_{\hb \hc} W_{\hd \he} \ve^{\hb \hc \hd \he {f}} (\psi_{\ha}{}^i \G_{{f}})^{\hal}~.~~~~~~~~~~
\eea
For completeness, we give the following projection rules of the vector covariant derivative action on the descendant component fields
\bea
\nabla_{a} W_{\b \g }{}_{\l}^{i} &=& \cD_{a} W_{\b \g }{}_{\l}^{i} -  \frac{1}{2} \psi_{a}{}^{i \a}  \Big( W_{{\alpha} {\beta} {\gamma} {\lambda}} + 3 \ri  (\Gamma_{{a}})_{{\alpha} ({\beta}}  {\nabla}^{{a}}{W_{{\gamma} {\lambda})}} + 3 \ri \eps_{{\alpha} ({\beta}} (\Gamma_{{a}})_{{\gamma}}{}^{{\tau}}   {\nabla}^{{a}}{W_{{\lambda}) {\tau}}}\Big)
\non \\
&& 
+ \ri \psi_{a}{}_{j (\b}   \Phi_{{\gamma} {\lambda})}\,^{i j} + 3 \phi_{a}{}^{i}_{(\b}   W_{\g \l)}~,
\\ 
\nabla_{a} D &=& 
\cD_{a} D - \ri \psi_{a}{}^\a_i (\Gamma^{{a}})_{{\alpha}}{}^{{\beta}}  {\nabla}_{{a}}{\chi^{i}_{ {\beta}}} - \ri \psi_{a}{}^\a_i W_{{\alpha}}{}^{{\beta}} \chi^{i}_{{\beta}} +  \phi_{a}{}^{i \a}  \chi_{\hal i} ~,
\\
\nabla_{a} \Phi_{\b \g}{}^{j k} &=&~
\cD_{a} \Phi_{\b \g}{}^{j k} + \frac{3}{8} \psi_{a}{}^{\a (j} \Big( -3   W_{({\beta}}{}^ {\lambda} W_{ {\gamma}) {\alpha}\lambda}\,^{ k)} -  \eps_{{\alpha} ({\beta}}  W^{{\rho} {\tau}} W_{{\gamma}) {\rho} {\tau}}\,^{k)}  -  W_{{\alpha} {\lambda}} W_{{\beta} {\gamma} }\,^{k){\lambda}}  
\non\\&& \qquad ~
+ 16 \ri W_{{\beta} {\gamma}} \chi^{k)}_{{\alpha}} - \frac{16 \ri}{3} W_{{\alpha} ({\beta}} \chi^{k)}_{{\gamma})} - 16 \ri \eps_{{\alpha} ({\beta}}  W_{{\gamma}) {\lambda}} \chi^{k) {\lambda}}  + 2 (\Gamma^{{a}})_{{\alpha}}{}^{{\rho}}  {\nabla}_{{a}}{W_{{\beta} {\gamma} {\rho}}\,^{k)}} 
\non\\&& \qquad  ~
+ 2  (\Gamma^{{a}})_{({\beta} }{}^{{\rho}}   {\nabla}_{{a}}{W_{ {\gamma}) {\alpha} {\rho}}\,^{k)}}  + \frac{32 \ri}{3}  (\Gamma^{{a}})_{{\alpha} ({\beta}}  {\nabla}_{{a}}{\chi^{k)}_{{\gamma})}}
- \frac{32 \ri}{3}  \eps_{{\alpha} ({\beta}} (\Gamma^{{a}})_{{\gamma}) {\lambda}}  {\nabla}_{{a}}{\chi^{k) {\lambda}}} \Big) 
\non\\
&& 
- \frac{3 \ri}{2} \phi_{a}{}^{\a (j} W_{\hal\hbe\hga}{}^{k)} - 16 \phi_{a}{}^{(j}_{(\b} \chi_{\hga)}^{k)}~, 
 \\
\nabla_{a} W_{\b \g \l \rho} &=& 
\cD_{a} W_{\b \g \l \rho} + \ri \psi_{a}{}^{\a}_{i}\Big( 2  (\Gamma_{{a}})_{{\alpha} ({\beta}} {\nabla}^{{a}}{W_{{\gamma} {\lambda}}{}_{{\rho})}^{i}}+ 3   W_{{\alpha} ({\beta}}{}_{{\gamma}}^{i} W_{{\lambda} {\rho})}- 3 W_{{\alpha} ({\beta}} W_{{\gamma} {\lambda}}{}_{{\rho})}^{i} \Big)
\non \\&&
+ 3 \ri \psi_{a}{}^{\a}_{i (\b}  \Big( - \frac{32 \ri}{3}W_{{\gamma} {\lambda}} \chi^{i}_{{\rho})}-2 (\Gamma_{{a}})_{{\gamma}}{}^{{\tau}}   {\nabla}^{{a}}{W_{{\lambda} {\rho}) {\tau}}\,^{i}}-  W_{{\gamma}}{}^{{\tau}} W_{{\lambda} {\rho} ){\tau}}\,^{i}\Big) \non \\&& 
+ 12 \phi_{a}{}^{i}_{(\b} W_{\hga\hde \hrh)}{}_i~.
\eea
\esubeq
%


\subsection{The Abelian vector multiplet}
Following the presentation and conventions of \cite{Hutomo:2022hdi, GHKT-M}, let us turn to the description of an off-shell Abelian vector multiplet. 

\subsubsection{The Abelian vector multiplet in superspace}
\label{vector-superspace}

To describe the Abelian vector multiplet \cite{HL, Zupnik:1999iy} in conformal superspace \cite{BKNT-M14}, we introduce a real primary superfield $W$ of dimension 1,
\bsubeq\label{vector-defs}
\bea
(W)^* = W~, \qquad K_A W=0~, \qquad \bbD W= W~.
\eea
The superfield $W$ is subject to the Bianchi identity
\bea 
\nabla_{{\a}}^{(i} \nabla_{{\b}}^{j)} W 
= \frac{1}{4} \eps_{{\a} {\b}} \nabla^{{\g} (i} \nabla_{{\g}}^{j)} 
W
~.
\label{vector-Bianchi}
\eea
\esubeq
Acting with spinor covariant derivatives on $W$ gives the following descendants:
 \bsubeq \label{components-vect}
\begin{align}
\l_\hal^i := - \ri\nabla_\hal^i W \ , \qquad
X^{ij} := \frac{\ri}{4} \nabla^{\hal (i} \nabla_\hal^{j)} W 
= - \frac{1}{4} \nabla^{\hal (i}  \l_\hal^{j)}~.
\end{align}
These superfields, along with
\be 
F_{\hal\hbe} := - \frac{\ri}{4} \nabla^k_{(\hal} \nabla_{\hbe) k}  W - W_{\a \b} W
= \frac{1}{4} \nabla_{(\hal}^k \l_{\hbe) k} 
- W_{\a \b} W
~,
\ee
\esubeq
satisfy the identities:
\bsubeq \label{VMIdentities}
\begin{align}
\nabla_\hal^i \l_\hbe^j 
&= - 2 \eps^{ij} \big(F_{\hal \hbe} + W_{\hal\hbe}  W\big) 
- \eps_{\hal\hbe} X^{ij} - \eps^{ij} \nabla_{\hal\hbe} W \ , \\
\nabla_\hal^i F_{\hbe\hga} 
&=
 - \ri \nabla_{\hal (\hbe} \l_{\hga)}^i 
 - \ri \eps_{\hal (\hbe} \nabla_{\hga )}{}^\hde \l_\hde^i 
- \frac{3\ri}{2} W_{\hbe\hga} \l_\hal^i - W_{\hal\hbe \hga}{}^i  W \non\\
&~~~+ \frac{\ri}{2} W_{\a (\b} \l_{\g)}^i
-\frac{3 \ri}{2} \ve_{\a (\b} W_{\g)}{}^{\d} \l_{\d}^i\ , \\
\nabla_\hal^i X^{jk} &= 2 \ri \eps^{i (j} 
\Big(
\nabla_\hal{}^\hbe \l_\hbe^{k)} 
- \hf W_{\hal\hbe} \l^{\hbe k)} 
+ \frac{3\ri}{4} X_\hal^{k)} W
\Big) 
\ .
\end{align}
\esubeq
 Due to \eqref{vector-Bianchi}, a dimension-2 superfield of a vector multiplet in the traceless frame obeys the following Bianchi identity:
\begin{align}
         \nabla_{(\a}{}^{\g} F_{\b)\g} = \frac{1}{2} \l^{\g k} W_{\a \b \g k}
         ~.
         \label{F-bianchi}
\end{align}
It is useful to note that the $S$-supersymmetry generator acts on the descendants as
\begin{align}
S_\hal^i \l_\hbe^j &= - 2 \ri \eps_{\hal\hbe} \eps^{ij}  W \ , \qquad
S_\hal^i F_{\hbe\hga} = 4 \eps_{\hal (\hbe}  \l_{\hga)}^i \ , \qquad
S_\hal^i X^{jk} = - 2 \eps^{i (j}  \l_\hal^{k)} \ ,
\label{S-on-W}
\end{align}
while all the superfields are annihilated by $K_a$.

It is worth noting that there exists a prepotential formulation for the Abelian vector multiplet which was developed in \cite{BKNT-M14}, see also \cite{K06,KL,KT-M08,KT-M_5D3} for related works in other superspaces. The formulation of \cite{BKNT-M14} is based on a real primary superfield $V_{ij}$ of dimension $-2$, i.e., $\mathbb{D} V_{ij} = -2 V_{ij}$. Here $V_{ij}$ transforms as an isovector under ${\rm SU}(2)_R$ transformations and is the 5D analogue of Mezincescu's prepotential \cite{Mezincescu, HST, BK11} for the 4D $\cN=2$ Abelian vector multiplet. This then allows us to represent the field strength $W$ as
\bea
W= -\frac{3\ri}{40} \de_{ij} \D^{ijkl} V_{kl}~,
\label{mezincescu}
\eea
where we have defined the operators
 \bsubeq
 \bea
 \D^{ijkl} &:=& 
 -\frac{1}{96} \ve^{\a \b \g \d} \de_{\a}^{(i} \de_{\b}^j \de_{\g}^k \de_{\d}^{l)} = -\frac{1}{32} \de^{(ij} \de^{kl)} = \D^{(ijkl)}~, \\
 \de^{ij} &:=& \de^{\a (i} \de_{\a}^{j)}~.
 \eea
 \esubeq
Let us also point out that $V_{ij}$ in \eqref{mezincescu} is defined modulo gauge transformations of the form 
\bea
\d V_{kl} = \de_{\a}^p \L^{\a}{}_{klp}~, \qquad \L^{\a}{}_{klp} = \L^{\a}{}_{(klp)}~,
\label{gauge-vector}
\eea
with the gauge parameter $\L^{\a}{}_{klp}$ being a dimension $-5/2$ primary superfield,
\bea
S_{\a}^i \L^{\b}{}_{jkl}=0~, \qquad  \mathbb{D} \L^{\b}{}_{jkl} = -\frac{5}{2} \L^{\b}{}_{jkl}~.
\eea

\subsubsection{The Abelian vector multiplet in components}

The component structure of the vector multiplet follows directly from the superfield definitions \eqref{components-vect}. It contains a real scalar field $W := W \loco$,
gaugini $\lambda_{\hal}^i:= \lambda_{\hal}^i \loco$, a triplet of auxiliary fields $X^{ij} := X^{ij} \loco$,
and a real Abelian gauge connection $v_{\hm} := V_m \loco$ or, equivalently,
its real field strength $f_{mn} :=  F_{mn}\loco = 2 \pa_{[m} v_{n]}$.  The  field strength $f_{mn}$ may be expressed in terms of the covariant field strength $F_{ab}:=F_{ab} \loco$ via the relation
\bea
F_{ab} = f_{ab} + \ri (\G_{[a})_{\a}{}^{\b} \psi_{b]}{}^{\a}_{k} \l_{\b}^k  
+ \frac{\ri}{2} \psi_{[a}{}^{\g}_{k} \psi_{b]}{}^{k}_{\g} W~, \qquad f_{ab}:= e_{a}{}^{m} e_{b}{}^{n} f_{mn}~.
\label{cov-Fab-grav}
\eea
The dilatation weights of the vector multiplet fields are summarised in Table
\ref{weights-vector}.
\begin{table}[hbt!]
\begin{center}
\begin{tabular}{ |c| c| c| c| c| c|} 
 \hline
& $W$ 
& $\lambda_{\hal}^i$
& $X^{ij}$
& $F_{\ha \hb}$
& $v_{\hm}$
\\ 
\hline
 \hline
$\mathbb{D}$
&$1$
&$3/2$
&$2$
&$2$
&$0$
\\
\hline
\end{tabular}
\caption{\footnotesize{Dilatation weights of the
Abelian vector multiplet.}\label{weights-vector}}
\end{center}
\end{table}

The transformations of the component fields in a standard Weyl multiplet background can be obtained from the corresponding superfields. They read
\bsubeq
\bea
\d W
&=&
\ri \xi_i \l^i + \l_\mathbb{D} W~,
\\ 
\d\lambda_{\hal}^i
&=&
- (\S^{\ha \hb} \xi^i)_{\hal} F_{\ha \hb} - (\S^{\ha \hb} \xi^i)_{\hal} W_{\ha \hb} W
+ \xi_{\hal}{}_j X^{ij} + (\G^{\ha} {\xi}^i)_{\hal} \de_{\ha} W \nonumber \\
&&+ \frac{1}{2}\l^{\ha \hb}(\S_{\ha \hb} \l^i)_{\hal}  + \l^i{}_j \l^j_{\hal} + \frac{3}{2} \l_{\mathbb{D}} \l^i_{\hal} + 2 \ri \eta_{\hal}{}^i W~,
\label{transf-lambda}
\\
\d X^{ij}
&=& -2 \ri \xi^{(i} \G^{\ha} \de_{\ha} {\l}^{j)} 
 - \frac{ \ri}{2} \xi^{(i} \S^{\ha \hb} \l^{j)} W_{\ha \hb} 
-16 \ri  (\xi^{(i} \chi^{j)}) W
\non\\
&&+ 2 \l^{(i}{}_k X^{j)k} 
-2 \eta^{(i} \l^{j)}
+ 2 \l_{\mathbb{D}} X^{ij}~,
\\
\d v_{\hm}
&=& \ri (\xi^i \psi_{\hm i}) W
- \ri (\xi^{i} \G_{\hm} \l_{i}  )
+\partial_{\hm} \l_V
~,
\label{dvm}
\eea
\esubeq
where
\bsubeq
\bea
\de_{\ha} W
&=&
\cD_{\ha} W - \frac{\ri}{2}\psi_{\ha}{}_i \l^i
~,
\\
\de_{\ha} \l^i_{\hal}
&=&
\cD_{\ha} \l^i_{\hal} 
+ \hf (\S^{\hb \hc}\psi_{\ha}{}^{i})_{\hal} \Big( F_{\hb \hc}
+ W_{\hb \hc} W \Big)
-\hf \psi_{\ha}{}_{\hal}{}_j X^{ij} 
\non\\
&&
-\hf (\G^{\hb} {\psi}_{\ha}{}^i)_{\hal} \de_{\hb} W 
- \ri \phi_{\ha}{}_{\hal}{}^i W
~.
\eea
For completeness, we include
\bea
    \de_{a} X^{j k} &=& \cD_{a} X^{j k} + \ri \psi_{a}{}^\a_i \eps^{i (j} \Big( \nabla_\hal{}^\hbe \l_\hbe^{k)} - \hf W_{\hal\hbe} \l^{\hbe k)} + \frac{3\ri}{4} X_\hal^{k)} W \Big) + \phi_{a}^{\a (j}   \l_\hal^{k)}
    ~,\\
    \de_{a} F_{\b \g}  &=& 
    \cD_{a} F_{\b \g} + \frac{\ri}{2} \psi_{a}{}^\a_i \bigg(
 \nabla_{\hal (\hbe} \l_{\hga)}^i 
+\eps_{\hal (\hbe} \nabla_{\hga )}{}^\hde \l_\hde^i 
+ \frac{3}{2} W_{\hbe\hga} \l_\hal^i - \ri W_{\hal\hbe \hga}{}^i  W \non\\
&&~~~- \frac{1}{2} W_{\a (\b} \l_{\g)}^i
+\frac{3}{2} \ve_{\a (\b} W_{\g)}{}^{\d} \l_{\d}^i \bigg) + 2 \phi_{a}{}^{i}_{(\b}  \l_{\hga) i}~.
\eea
\esubeq
Note that we have also included in \eqref{dvm}
the gauge field transformation
parametrised by the local real parameter $\l_V$.

\subsection{The linear multiplet}
We now turn to the discussion of the off-shell linear multiplet coupled to conformal supergravity.

\subsubsection{The linear multiplet in superspace}
The linear multiplet \cite{FS2,deWit:1980gt,deWit:1980lyi,deWit:1983xhu,N=2tensor,Siegel:1978yi,Siegel80,SSW,deWit:1982na,KLR,LR3}, 
 or $\cO(2)$ multiplet, can be described in terms of the dimension three  primary superfield $G^{ij}= G^{ji}$ with the properties
\bsubeq\label{linear-constr}
\bea
\de_{\a}^{(i} G^{jk)} &=& 0~, 
\label{O(2)constraints}\\
K_A G^{ij} &=& 0~, \qquad \mathbb{D} G^{ij} = 3 G^{ij}~,
\eea
\esubeq
where $G^{ij}$ is assumed to be real, $(G^{ij})^{*}= \ve_{ik} \ve_{jl} G^{kl}$.

The following tower of descending identities is useful to elaborate on the component structure of the superfield $G^{ij}$:
\bsubeq \label{O2spinorderivs}
\begin{align}
\nabla_\hal^i G^{jk} &= 2 \eps^{i(j} \varphi_\hal^{k)} \ , \\
\nabla_\hal^i \varphi_\hbe^j &= - \frac{\ri}{2} \eps^{ij} \eps_{\hal\hbe} F + \frac{\ri}{2} \eps^{ij} \cH_{\hal\hbe} + \ri \nabla_{\hal\hbe} G^{ij} \ , \\
\nabla_\hal^i F &= - 2 \nabla_\hal{}^\hbe \varphi_\hbe^i - 3 W_{\hal\hbe} \varphi^{\hbe i} - \frac{3}{2} X_{\hal j} G^{ij} \ , \\
\nabla_\hal^i \cH_{\ha} &= 4 (\S_{\ha\hb})_\hal{}^\hbe \nabla^\hb \varphi_\hbe^i - \frac{3}{2} (\G_\ha)_\hal{}^\hbe W_{\hbe \hga} \varphi^{\hga i}
-\frac{1}{2} (\G_\ha)_\hga{}^\hbe W_{\hbe \hal} \varphi^{\hga i} \ ,
\end{align}
\esubeq
with the independent descendant superfields being defined as
\bsubeq \label{O2superfieldComps}
\begin{align}
\varphi_\hal^i &:= \frac{1}{3} \nabla_{\hal j} G^{ij} \ , \\
F &:= \frac{\ri}{12} \nabla^{\hga i} \nabla_\hga^j G_{ij} = - \frac{\ri}{4} \nabla^{\hga k} \varphi_{\hga k} \ ,\\
\cH_{abcd} &:= \frac{\ri}{12} \eps_{abcde} (\G^{e})^{\a \b} \nabla_{\a}^i \nabla_{\b}^j G_{ij} 
\equiv \eps_{\ha\hb\hc\hd\he} \cH^\he \ .
\end{align}
\esubeq
Here $\cH^{a}$ obeys the differential condition 
\be 
\nabla_\ha \cH^\ha = 0~, \qquad {\cH}^{\ha}:= -\frac{1}{4!}\ve^{\ha \hb \hc \hd \he} \cH_{\hb \hc \hd \he}~.
\ee
The descendants \eqref{O2superfieldComps} are all annihilated by $K_a$. Under the action of $S$-supersymmetry, they transform as follows:
\begin{align} 
S_\hal^i \varphi_\hbe^j = - 6 \eps_{\hal \hbe} G^{ij} \ , ~~~~~~
S_\hal^i F = 6 \ri \varphi_\hal^i \ , ~~~~~~
S_\hal^i \cH_\hb &= - 8 \ri (\G_\hb)_\hal{}^\hbe \varphi_\hbe^i \ .
\label{O2-S-actions}
\end{align}
We refer the reader to \cite{BKNT-M14} for a superform description of the linear multiplet. 

There exists a prepotential formulation for the linear multiplet in superspace \cite{BKNT-M14}. The constraints \eqref{linear-constr} may be solved in terms of an arbitrary primary real dimensionless scalar prepotential $\O$, 
\bea
S_\hal^i\O=0~,~~~~~~
\bbD\O=0
~,
\eea
which leads to
\bea
G^{ij}
&=&
-\frac{3\ri}{40}\D^{ijkl}\de_{kl}\O ~.
\label{def-G-0-b}
\eea
An important property of $G^{ij}$ defined by 
\eqref{def-G-0-b} is that it is invariant under  gauge transformations 
of $\O$
of the form 
\bea
\d\Omega=-\frac{\ri}{2}(\G^\ha)^{\hal\hbe}\de_\hal^i \de_\hbe^jB_\ha{}_{ij}
~,
\label{gauge-O2-0}
\eea
where the gauge parameter is assumed to have the properties
\bea
B_{\ha}{}^{ij}=B_{\ha}{}^{ji}
~,~~~~~~
S_{\hal}^{i} B_\ha{}^{jk}=0
~,~~~~~~
\bbD B_\ha{}^{ij}=-B_\ha{}^{ij}
~,
\label{gauge-inv-O2-1}
\eea
and is otherwise arbitrary.

It is useful to note that given a system of $n$ Abelian vector multiplets $W^I$, with $I=1,2, \dots n$,  all satisfying \eqref{vector-defs}, we can construct the following composite linear multiplet and its descendants \cite{BKNT-M14}:
\bsubeq \label{O2composite-N}
\bea
H^{ij}
&=& C_{JK} \Big\{ \,2 W^J X^{ij\, K}
- \ri \l^{\a J\,(i } \l_{\a}^{j) K}\Big\}
~,\\
\vf_{\a\, }^i &=& C_{JK} \bigg\{\,
\ri X^{ij\, J} \l_{\a j}^K 
-2 \ri F_{\a \b}^{J} \l^{\b i K}
- \frac{3}{2} X_{\a}^i W^J W^K
-2 \ri W^J \de_{\a \b} \l^{\b i K}\non\\
&&~~~~~~~~
- \ri (\de_{\a \b} W^J) \l^{\b i K}
- 3 \ri W_{\a \b} W^J \l^{\b i K}
\bigg\}~, \\
F &=& C_{JK} \bigg\{\,
X^{ij J}X_{ij}^{K}
- F^{ab J} F_{ab}^{K}
+ 4 W^J \Box W^K
+ 2 (\de^a W^J) \de_{a} W^K \non\\
&&~~~~~~~~
+ 2 \ri (\de_{\a}{}^{\b} \l_{\b}^{iJ}) \l^{\a K}_{i}{} -6 W^{ab} F_{ab}^J W^K 
-\frac{39}{8} W^{ab} W_{ab} W^J W^K \label{composite_FVMfermionic}
 \non\\
&&~~~~~~~~
+ \frac{3}{8} Y W^J W^K
+ 6 X^{\a i}\l_{\a i}^J W^K
-3 \ri W_{\a \b} \l^{\a i J} \l^{\b K}_{i}
\bigg\}
~, \\
\cH_{a} &=& C_{JK} \bigg\{
-\hf \ve_{abcde} F^{bc\, J} F^{de \,K}
+ 4 \de^{b} \Big( W^J F_{ba}^K + \frac{3}{2} W_{ba} W^J W^K\Big)\non\\
&&~~~~~~~~~
+ 2 \ri (\S_{ba})^{\a \b} \de^{b} (\l_{\a}^{iJ} \l_{\b i}^K)
\bigg\}
~,
\eea
\esubeq
where $\Box := \de^{a} \de_{a}$ and $C_{JK} = C_{(JK)}$ is a constant symmetric in $J$ and $K$.
Equation \eqref{O2composite-N} is the superspace analogue of the composite linear multiplet constructed for the first time in \cite{Bergshoeff1}. Note that it is possible to create a composite linear multiplet from a single Abelian vector multiplet by setting $C_{JK} = 1$ when $J = K = 1$ and zero otherwise. This case will play an important role.

\subsubsection{The linear multiplet in components}
The components of the linear multiplet follows directly from the superfield description:
an SU$(2)_R$ triplet of Lorentz scalar fields $G^{ij} = G^{ij} \loco$\,; 
a spinor field $\varphi_{\hal i}  = \varphi_{\hal i} \loco$\,;
a scalar field $F  = F \loco$;
and a covariant closed anti-symmetric four-form field strength $\cH_{abcd} := \cH_{abcd} \loco$. 
The latter is equivalent to a conserved dual vector 
$\cH^{\ha}:= -1/{4!}\,\ve^{\ha \hb \hc \hd \he}\cH_{\hb \hc \hd \he}$.\footnote{The Levi-Civita tensor with world indices 
is defined as $\ve^{mnpqr} := \ve^{\ha\hb\hc\hd\he} e_\ha{}^{m} e_\hb{}^{n} e_\hc{}^{p} e_\hd{}^{q} e_\he{}^{r}$, such that
$\ve_{abcde}$ and $\ve^{abcde}$ are normalised as $\ve_{01234} = -\ve^{01234}= 1$.}
It holds that
\bea
\cH^a = h^a 
+ 2 (\S^{ab})_\a{}^\b\psi_{b}{}^\a_i\vf_\b^i
- \frac{\ri}{2} \ve^{abcde} (\S_{bc})_{\a\b}\psi_{d}{}^\a_{i}\psi_{e}{}^\b_{j} G^{ij} ~.
\eea
The covariant conservation equation for ${\cH}_{\ha}$ is
\bea
\de^{\ha} {\cH}_{\ha} 
&=& 
0~.
\label{covariant-current}
\eea
The constraint implies that there exists a gauge three-form potential, $b_{\hm \hn \hp}$, 
and its exterior derivative, such that $h_{\hm \hn \hp \hq}:= 4 \pa_{[\hm}b_{\hn \hp \hq]}$, with $b_{mnp} := \cB_{mnp} \loco$.

The local superconformal transformations of the covariant fields can be derived using \eqref{O2spinorderivs} and \eqref{O2-S-actions}, which give
\bsubeq
\label{linear-multiplet}
\bea
\d G_{ij}
&=&
-2 \xi_{(i} \varphi_{j)}
- 2\lambda_{(i}{}^{k}  G_{j) k} 
+3 \lambda_{\mathbb{D}} G_{ i j}
~,
\\
\d \varphi_{\hal i}
&=&
-\frac{\ri}{2}\xi_{\hal i} F 
-\frac{\ri}{2}{\cH}_{\ha} (\G^{\ha} {\xi}_{i})_{\hal} 
- \ri   
(\G^{\ha} {\xi}^{j})_{\hal} 
\de_{\ha} G_{i j}
+6 \eta_{\hal}{}^{j}   G_{ij}
\nonumber\\ 
&& 
+\frac{1}{2} \lambda^{\ha \hb}
(\S_{\ha \hb}\varphi_{i})_{\hal} 
-\lambda_i{}^j \varphi_{j \hal}
+\frac{7}{2}\lambda_{\mathbb{D}}\varphi_{\hal i}~,
\\
\d F
&=&
2 {\xi}^{i} \G^{\ha} \de_{\ha} \varphi_i
- \frac{3}{2}({\xi}^{i} \S^{\ha \hb} {\varphi}_i)
W_{\ha \hb}
+16 \ri ({\xi}^{i} {\chi}^{ j}) G_{i j} 
\non\\
&&+6 \ri\, \eta^{i} \varphi_{i} +4 \lambda_{\mathbb{D}} F ~,\\
\d {\cH}_{\ha}
&=&
-4 \xi^{i} \S_{\ha \hb} \de^{\hb} \varphi_{ i}
-\frac{1}{2} (\xi^{i} \G^{\hb}{\varphi}_{i})W_{\ha \hb}
+ \hf \ve_{\ha \hb \hc \hd \he} W^{\hb \hc} (\xi^{i} \S^{\hd \he}{\varphi}_{i})
\non\\
&&
+ \lambda_{\ha}{}^{\hb}\,{\cH}_{\hb}
+4 \lambda_{\mathbb{D}} \,{\cH}_{\ha}
-8 \ri
\eta^{i}\G_{\ha} {\varphi}_i
~,
\eea
\esubeq
where
\bsubeq \label{eq:degaugeLinear1}
\bea
\de_{\ha} G_{ij} &=& \cD_{\ha} G_{ij} +  \psi_{\ha(i} \vf_{j)}~, \\
\de_{\ha} \vf_{{\a} i} &=& \cD_{\ha} \vf_{{\a} i} 
+\frac{\ri}{4} \psi_{\ha {\a} i } F 
+\frac{\ri}{4} (\G^{\hb} \psi_{\ha i})_{{\a}} \cH_{\hb}
+\frac{\ri}{2}  (\G^{\hb} \psi_{\ha}{}^{j})_{{\a}} \de_{\hb} G_{ij} - 3 \phi_{\ha {\a}}{}^{j} G_{ij}~.
\eea
For completeness, we also have
\bea
    \nabla_a F &=& 
    \cD_a F - (\Gamma_b \psi_{a i})_\a  \nabla^b \varphi^{i \a} + \frac{3}{2}\psi_{a i}{}^\b W_{\a \b}\varphi^{i \a} - 8 \ri  G^{i j} \psi_{a j}{}^\a \chi_{i \a} 
 + 3 \ri \phi_{a i}{}^{\a}\varphi^{i}_{\a} ~,
    \\
    \nabla_a \cH_b &=& 
    \cD_a \cH_b - 2 (\Sigma_{b c} \psi_{a i})_\a \nabla^c \varphi^{i \a} + \frac{3}{4} (\Gamma_{b} \psi_{a i})_\a W^{\a \b}\varphi^i_\b + \frac{1}{4} \psi_{a i}{}^{\b} W_{\a \b} (\Gamma_b \varphi^i)^\a
    \non \\ 
    &&
    + 4 \ri (\Gamma_{b}\phi_{a i})_\a\varphi^{i \a} ~.
\eea
\esubeq
The locally superconformal transformations of $b_{\hm \hn \hp}$ are 
\bea
\d b_{\hm \hn \hp}
&=&
2 \, \ve_{\ha \hb \hc \hd \he} e_{\hm} {}^{\ha} e_{\hn}{}^{\hb} e_{\hp}{}^{\hc} (\xi_i\S^{\hd \he} \varphi^i)
- 12 \ri (\psi_{[\hm}{}^{i} \S_{\hn \hp]} \xi^j) G_{ij}
+3 \pa_{[\hm}l_{\hn \hp]}~,
\label{d-bmn}
\eea
where we have also included 
the gauge transformation
$\d_l b_{\hm \hn \hp}= 3 \pa_{[\hm}l_{\hn \hp]}$
leaving 
$h_{\hm \hn \hp \hq}$
and
${\cH}^{\ha}$ invariant.
The dilatation weights
of the $\cO(2)$ multiplet are summarised in Table 
\ref{chiral-dilatation-weights-LINEAR}.
\begin{table}[hbt!]
\begin{center}
\begin{tabular}{ |c||c| c| c| c| c| c| c|} 
 \hline
& $G_{ij}$ 
& $\varphi_{\hal i}$
& $F$
& ${\cH}_{\ha}$
&$b_{\hm \hn \hp}$
\\ 
\hline
 \hline
$\mathbb{D}$
&$3$
&$7/2$
&$4$
&$4$
&$0$
\\
\hline
\end{tabular}
\caption{\footnotesize{Dilatation weights of the off-shell $\cO (2)$ multiplet.}
\label{chiral-dilatation-weights-LINEAR}}
\end{center}
\end{table}

\section{Superconformal actions}
\label{section-3}

In this section we review a main action principle that has played an important role in constructing various locally superconformal invariants including the two-derivative gauged supergravity action. The reader should refer to \cite{BKNT-M14} for a more complete analysis.

\subsection{BF action}
Consider a full superspace integral
\begin{align}
S[\cL] = \int\rd^{5|8}z\,  E\, \cL~,\qquad\rd^{5|8}z:=\rd^5x\,\rd^8\q~, \qquad
E := {\rm Ber}(E_\hM{}^\hA)~,
\end{align}
where the Lagrangian $\cL$ is a conformal primary superfield of dimension $+1$, i.e, $\mathbb{D} \cL= \cL$. The above action is locally superconformal, that is, invariant under the supergravity gauge transformations \eqref{sugra-group}.

One of the main building blocks for the construction of general supergravity-matter couplings \cite{Ohashi1,Ohashi2,Ohashi3,Ohashi4,Bergshoeff1,Bergshoeff2,Bergshoeff3} and higher-derivative invariants in \cite{BKNT-M14} is the so-called BF action principle. It is based on an appropriate product of a linear multiplet with an Abelian vector multiplet, which may be described by the following form in superspace
\bsubeq\label{BF-action-0}
\bea
S_{\rm{BF}}
=
\int\rd^{5|8}z\, E\, 
 \O W
=
 \int \rd^{5|8}z\, E\, 
 G^{ij} V_{ij}
 ~.
  \label{SGV}
\eea
As shown above, the BF action can be written in different ways, see \cite{BKNT-M14} for even more variants. In the first form in \eqref{SGV}, it involves the field strength of the vector multiplet, 
$W$, and the prepotential of the linear multiplet, $\O$.  
Equivalently, the BF action can be written in terms of the Mezincescu's prepotential $V_{ij}$ and the field strength $G^{ij}$ of the linear multiplet described by the right-hand side of \eqref{SGV}. 
In addition,  the  functionals $\int \rd^{5|8}z\, E\, 
 \O W$ and 
 $\int \rd^{5|8}z\, E\, 
 G^{ij} V_{ij}$
are, respectively, invariant under the gauge transformations \eqref{gauge-O2-0} and \eqref{gauge-vector}, thanks to the defining differential constraints satisfied by $W$ and $G^{ij}$, \eqref{vector-Bianchi} and \eqref{O(2)constraints}.

In components, and in our notation, the BF action takes the form \cite{BKNT-M14} 
\bea
    S_{\rm BF} &=& - \int \rd^5 x \, e \Big( \,v_{{a}} \cH^{{a}} + W F + X_{i j} G^{i j} + 2 \l^{\a k} \vf_{\a k} 
    \non\\
    &&- \psi_{{a }}{}^{\a}_i(\Gamma^{{a}})_{\a}{}^{\b} \vf_{\b}^i  W - \ri \psi_{{a}}{}^{\a}_{i} (\Gamma^{{a}})_{\a}{}^{\b} \l_{\b j} G^{i j} +  \ri \psi_{{a}}{}^{\a}_{i} (\Sigma^{{a} {b}}){}_{\a}{}^{\b} \psi_{{b} \b j} W G^{i j} \Big)
    \label{BF-Scomp}\\
     &=& 
    - \int \rd^5 x \, e \Big( -\frac{1}{12} \ve^{abcde} f_{ab} b_{cde} + W F + X_{i j} G^{i j} + 2 \l^{\a k} \vf_{\a k} 
     \non\\
    &&- \psi_{{a }}{}^{\a}_i(\Gamma^{{a}})_{\a}{}^{\b} \vf_{\b}^i  W - \ri \psi_{{a}}{}^{\a}_{i} (\Gamma^{{a}})_{\a}{}^{\b} \l_{\b j} G^{i j} +  \ri \psi_{{a}}{}^{\a}_{i} (\Sigma^{{a} {b}}){}_{\a}{}^{\b} \psi_{{b} \b j} W G^{i j} \Big)~,
    \label{BF-Scomp-3form}
\eea
\esubeq 
where it should be noted that sometimes we associate the same symbol for the covariant component fields and the corresponding superfields, when the interpretation is clear from the context.

\subsection{Vector multiplet compensator}

The two-derivative action for the vector multiplet compensator can be constructed via the BF action principle \eqref{SGV}, with the linear multiplet being a composite superfield.  We denote by $H^{ij}_{\rm VM}$ the composite linear multiplet \eqref{O2composite-N},
\bea 
H^{ij}_{\rm{VM}} &=& \ri (\de^{\a(i} W) \de_{\a}^{j)} W + \frac{\ri}{2} W \de^{\a(i} \de_{\a}^{j)} W \non\\
&=& -\ri {\lambda}^{\a i} {\l}_{\a}^j +2 {W} {X}^{ij}~.\label{HijVM}
\eea
One can then check that $H_{\rm{VM}}^{ij}$  is a primary superfield, $S_{\a}^k H^{ij}_{\rm{VM}}= 0$ and has dimension 3. Thanks to the Bianchi identity \eqref{vector-Bianchi}, it also satisfies the analyticity constraint
\bea
\de^{(i}_{\a} H^{jk)}_{\rm{VM}}=0~.
\eea

The vector multiplet action may then be written as an integral over the full superspace,
\bea
S_{\rm{VM}}&=&
\frac{1}{4} \int\rd^{5|8}z\, E\,   V_{ij} { H }_{\rm VM}^{ij} ~.~~~
\label{vm-action}
\eea
This invariant also admits another representation, which can be obtained by applying \eqref{SGV} once more
\bsubeq \label{prep-vm-action}
\bea
S_{\rm VM}= 
\frac{1}{4}
 \int \rd^{5|8}z\, E\, 
 {\bm \O}_{\rm VM} W~,
\eea
where we have introduced the primary superfield \cite{BKNT-M14}
\bea
 {\bm \O}_{\rm VM} &=& 
\frac{\ri}{4}\Big(
 W\de^{ij} V_{ij}
-2  (\de^{\hal i}V_{ij})\de_\hal^{j}W
-2 V_{ij}\de^{ij} W \Big)~.
\eea
\esubeq
It can be shown that, by making use of  \eqref{mezincescu}, \eqref{def-G-0-b}, \eqref{prep-vm-action}, and integration by parts, we obtain the following relation
\bsubeq
\bea
\frac{1}{4} \int\rd^{5|8}z\, E\,     V_{ij} \, \d{ H }_{\rm VM}^{ij} &=& \frac{1}{2} \int\rd^{5|8}z\, E\,    \d W {\bm \O}_{\rm VM} \\
&=& \frac{1}{2} \Big(-\frac{3\ri}{40}\Big) \int\rd^{5|8}z\, E\,   {\bm \O}_{\rm VM} \de_{ij} \Delta^{ijkl} \d V_{kl}  \\
&=& \frac{1}{2}  \int\rd^{5|8}z\, E\,   \d V_{ij} H^{ij}_{\rm VM}~. 
\eea
\esubeq
Making use of the representations \eqref{vm-action} and \eqref{prep-vm-action}, it is seen that the variation of  $S_{\rm VM}$, induced by an arbitrary variation of the Mezincescu's prepotential $V_{ij}$ reads
\bea \label{var-SVM}
\d S_{\rm{VM}}
&=&
\frac{3}{4}\int \rd^{5|8}z\, E\, \d V_{ij}
{ H }_{\rm VM}^{ij} ~.
\eea
The above variation vanishes when $\d V_{ij}$  is a gauge transformation \eqref{gauge-vector}. This implies that
\bea
\int \rd^{5|8}z\, E\, \L^{\a}{}_{ijk} \de_{\a}^{(k} H_{\rm VM}^{ij)} = 0~,
\eea
that is, $\de_{\a}^{(i} H_{\rm VM}^{jk)} = 0$.
This result holds for any dynamical system involving an Abelian vector multiplet \cite{BKNT-M14}. The variation with respect to the prepotential $V_{ij}$ couples to a composite linear multiplet which depends on the specific form of the associated action principle -- let us call this, in general, $\mathbf{H}^{ij}$ which satisfies by construction the constraints \eqref{linear-constr}.
The equation of motion (EOM) for a vector multiplet is then $\mathbf{H}^{ij}=0$.
In the case of eq.~\eqref{vm-action}, the EOM for the vector multiplet compensator is $H^{ij}_{\rm VM}=0$.

Let us now work out the component form of $S_{\rm VM}$. The bosonic part of such an action takes the form
\bea
S_{\rm VM} 
&=&
-\frac{1}{4}  \int \rd^5 x\, e\, \bigg\{ 
v^{a} \cH_{a}{}_{\rm VM}+ W F_{\rm VM} + X_{i j} H^{i j}_{\rm VM}
\bigg\}~.
\label{BF}
\eea
This amounts to taking the bosonic sector\footnote{Here and in what follows, whenever we only explicitly present the bosonic sectors, any supercovariant field strength (which contains fermionic terms) can also be replaced with its purely bosonic analogue, e.g., $\cH^a$ with $h_a$, $F_{ab}$ with $f_{ab}$ and so on, as this will only modify the suppressed fermionic terms.}  of the bar projection of eqs.~\eqref{O2composite-N} for a single Abelian vector multiplet:
\bsubeq \label{composite-O2-BF}
\bea 
H^{i j}_{\rm VM}|_{bosonic}  
&=& 2 W X^{ij}~, \\
F_{\rm VM}|_{bosonic} &=& X^{ij} X_{ij} - f^{\ha\hb} f_{\ha\hb} 
+ 4W \nabla^\ha \nabla_\ha W + 2 (\nabla^\ha W )\nabla_\ha W  \non\\
&&- 6 W \,W^{\ha\hb} f_{\ha\hb} - \frac{39}{8} W^2 W^{\ha\hb} W_{\ha\hb}
-16 W^2 D~, \label{Composite_Fvm} \\
\cH_\ha{}_{\rm VM}|_{bosonic} &=& - \frac{1}{2} \eps_{\ha\hb\hc\hd\he} f^{\hb\hc} f^{\hd\he}
+ 4 \nabla^\hb (W f_{\hb\ha} + \frac{3}{2} W^2 W_{\hb\ha})~,
\eea
\esubeq
and plugging \eqref{composite-O2-BF} back into \eqref{BF}.
This procedure results in 
\bea
S_{\rm VM}
&=& 
 \int \rd^5 x\, e\, \bigg\{ 
- \frac{1}{2} W ({\de}^{\ha} W) {\de}_{\ha} W
- W^2 \de^a \de_a W
- \frac{3}{4} W X^{ij} X_{ij} 
+\frac{1}{8} \ve_{\ha \hb \hc \hd \he}v^{\ha} f^{\hb \hc} f^{\hd \he} \non\\
&&+ \frac{3}{4}W f^{\ha \hb} f_{\ha \hb}
+ \frac{9}{4} W^2 W^{\ha \hb} f_{\ha \hb}
+ \frac{39}{32}W^3 W^{ab} W_{ab}
+ 4 W^3 D 
\bigg\}~.
\label{BF-2}
\eea
The expression \eqref{BF-2} can be written in terms of the degauged covariant derivative $\cD_a$ \eqref{cD-def}. For the main body of this paper, whenever we apply the degauging procedure, we are effectively replacing the superconformal vector covariant derivative $\de_a$ with the following relation
\bea\label{bosonic-degauging}
\de_a = \cD_a -\mathfrak{f}_{a}{}^{b} K_{b}~,
\eea
since we only focus on the bosonic part. For results including fermionic terms, we refer to the supplementary file of our paper. 
As such, we have that
\bea
\de^a \de_a W = \cD^a \cD_a W
+\frac{1}{8} W \cR~. 
\eea
After performing integration by parts, one then arrives at
\bea
S_{\rm VM}
&=& 
 \int \rd^5 x\, e\, \bigg\{ 
- \frac{1}{8} W^3 {\cR}
+ \frac{3}{2} W ({\cD}^{\ha} W) {\cD}_{\ha} W
- \frac{3}{4} W X^{ij} X_{ij}
+\frac{1}{8} \ve_{\ha \hb \hc \hd \he}v^{\ha} f^{\hb \hc} f^{\hd \he} 
\non\\
&&+ \frac{3}{4} W f^{\ha \hb} f_{\ha \hb}
+ \frac{9}{4} W^2 W^{\ha \hb} f_{\ha \hb}
+ \frac{39}{32} W^3 W^{\ha \hb}W_{\ha \hb} + 4 W^3 D
\bigg\}~.
\label{bosonic-swm}
\eea
Upon gauge fixing dilatation by imposing $W=1$, the first term gives a scalar curvature term, $\cR$.
The gauge fixing $W=1$ can be achieved by requiring $W\ne 0$ meaning that the vector multiplet is a conformal compensator.

\subsection{Linear multiplet compensator}\label{subsection Linear multiplet compensator}

Within the BF action principle \eqref{SGV}, the dynamical part of the linear multiplet action is described by a vector multiplet built out of the linear multiplet. 
We denote by $\mathbf{W}$ the composite vector multiplet field strength:
\bea
\mathbf{W} := 
 \frac{\ri}{16} G \nabla^{\hal i} \nabla_\hal^j \Big(\frac{G_{ij}}{G^2}\Big) = \frac{1}{4}F G^{-1} - \frac{\ri}{8} G_{i j} \varphi^{i \a} \varphi^{j}_{\a} G^{-3}~,
 \label{comp-W}
\eea
with
\bea
G := \sqrt{\hf G^{ij} G_{ij}}
\eea
being nowhere vanishing, $G \neq 0$. At the component level, the vector multiplet \eqref{comp-W} was first derived by Zucker
\cite{Zucker:2000ks} as a 5D analogue of the improved 4D $\cN=2$ tensor multiplet \cite{deWit:1982na}. 
The field strength $\mathbf{W}$ obeys 
the constraints \eqref{vector-defs}.

The component fields of the above composite vector multiplet in our notation have been worked out in \cite{GHKT-M}, see \cite{Zucker:2000ks} for the first analysis in components. 
They consist of the $\q =0$ projection of $\mathbf{W}$, along with the descendant components of $\mathbf{W}$: $\bm{\l}_{\a}^i= -\ri \de_{\a}^i \mathbf{W} |$, $ \mathbf{X}^{ij} 
= \frac{\ri}{4} \nabla^{\hal (i} \nabla_\hal^{j)} \mathbf{W} |$ and $\mathbf{F}_{{a} {b}} = \frac{1}{4} (\Sigma_{{a} {b}})^{\a \b} \de^{k}_{(\a} \bm{\l}_{\b) k} | - W_{{a} {b}} \mathbf{W} |$~. Their explicit expressions are
\bsubeq \label{desc-W-R2}
\bea
\bm{\l}_{\a}^i 
&=& ~~ G^{-1} \Bigg\{ -\frac{\ri}{2} \de_{\a \b} \vf^{\b i}
+ \frac{3\ri}{4} W_{\a \b} \vf^{\b i} + 4 G^{ij} \chi_{\a j}\Bigg\}
\non\\
&&
+ G^{-3} \Bigg\{ -\frac{\ri}{8} F G^{ij} \vf_{\a j} - \frac{\ri}{8} G^{ij} \cH_{\a \b} \vf^{\b}_{j}
+ \frac{\ri}{4} G_{jk} \vf^{\b k} \de_{\a \b} G^{ij}
+ \frac{1}{4} \vf^{\b i} \vf_{\b}^j \vf_{\a j} \Bigg\}
\non\\
&&
+  G^{-5} \Bigg\{ -\frac{3}{8} G^{ij} G_{kl} \vf^{\b k} \vf_{\b}^l \vf_{\a j}\Bigg\}~,
\\
\mathbf{X}^{ij} 
&=& ~~ G^{-1} \Bigg\{\, \hf \Box G^{ij} 
+ \frac{3}{64} W^{ab} W_{ab} G^{ij} +2 D G^{ij}
+ 8 \chi^{\a (i} \vf_{\a}^{j)}
\Bigg\}\non\\
&&+ G^{-3} \Bigg\{ -\frac{1}{16} F^2 G^{ij} - \frac{1}{16} \cH^{a} \cH_{a} G^{ij} + \frac{1}{4} \cH^{a} G^{k(i} \de_{a} G_k{}^{j)}-\frac{1}{4} G_{kl} \big(\de^{a}   G^{k(i}\big) \de_{a} G^{j)l}
\non\\
&&~~~~~~~~~~
- 4 G^{ij} G_{kl} \chi^{\a k} \vf_{\a}^l
-\frac{\ri}{8} F \vf^{\a(i} \vf_{\a}^{j)}
+\frac{3\ri}{8} W^{\a \b} G^{ij} \vf^{k}_{\a} \vf_{\b k}
\non\\
&&~~~~~~~~~~
+ \frac{\ri}{16}(\G^a)^{\a \b} \Big(\cH_{a} \vf^{(i}_{\a} \vf^{j)}_{\b} 
+ 8 G^{k(i}  \big(\de_{a} \vf^{j)}_{\a} \big) \vf_{\b k} + 2 \vf_{\a}^{(i}   \big(\de_{a} G^{j)k} \big) \vf_{\b k} \Big)
\Bigg\}\non\\
&&+ G^{-5} \Bigg\{\,\frac{3\ri}{16} F G^{ij} G_{kl} \vf^{\a k} \vf_{\a}^l+ \frac{3\ri}{16} G^{k(i} G^{j)l} (\G^a)_{\a \b} \cH_{a} \vf_{k}^{\a} \vf_{l}^{\b}
-\frac{3\ri}{8}G^{mn}G^{k(i}  \big(\de_{\a \b} G_m{}^{j)} \big) \vf^{\a}_{k} \vf^{\b}_{n}
\non\\
&&~~~~~~~~~
+ \frac{3}{8} G^{k(i} \vf^{j)}_{\a}  \vf^{\a l} \vf^{\b}_{k} \vf_{\b l} -\frac{3}{8} G^{kl} \vf^{\a(i} \vf_{\a}^{j)} \vf_{k}^{\b} \vf_{\b l}
\Bigg\}
\non\\
&&
+  G^{-7} \Bigg\{ \frac{15}{32} G^{ij}G_{kl} G_{mn} \vf^{\a k} \vf_{\a}^{l} \vf^{\b m} \vf_{\b}^n \Bigg\} ~, \label{compositeXijlinear}
\\
\mathbf{F}_{{a} {b}} 
&=& ~~ G^{-1} \Bigg\{ \hf \de_{[a} \cH_{b]} + \frac{1}{2} G_{ij}\Phi_{ab}{}^{ij} + \frac{\ri}{4} W_{ab \a}{}^i\vf^{\a}_i
\Bigg\} \non\\
&&+ G^{-3} \Bigg\{ \,\frac{1}{4} G_{ij} \cH_{[a} \de_{b]} G^{ij}
-\frac{1}{4} G_{ij}  \big(\de_{[a} G^{ik} \big) \de_{b]} G_k{}^{j} + \frac{\ri}{2} G_{ij} (\G_{[a})^{\a \b} (\de_{b]} \vf_{\a}^i) \vf_{\b}^{j} 
\non\\
&&
~~~~~~~~~
- \frac{\ri}{8} (\G_{[a})^{\a \b}  \big(\de_{b]} G_{ij} \big) \vf^{i}_{\a} \vf^{j}_{\b}
\Bigg\}
\non\\
&&
+ G^{-5}  \Bigg\{ -\frac{3\ri}{8} G^{k}{}_{(i} G^{l}{}_{j)} (\G_{[a})^{\a \b}  \big(\de_{b]} G_{kl} \big) \vf_{\a}^{i} \vf^{j}_{\b} \Bigg\}
~.
\eea
\esubeq

The action for the linear multiplet compensator may be expressed as 
\bea \label{S_boldW}
S_{\rm L} =  \int \rd^{5|8}z\, E\,  \O \mathbf{W} ~.
\eea
Varying the prepotential $\O$ leads to
\bea
\d S_{\rm L}
= 
 \int \rd^{5|8}z\, E\, \d \O \,\mathbf{W} ~.
\label{rep8.11}
\eea
The previous form holds for the first-order variation of a matter system which includes a linear multiplet with respect to its prepotential $\O$. In particular, the variation must vanish if $\d \O$ is given by \eqref{gauge-O2-0}. This holds provided $\mathbf{W}$ satisfies the Bianchi identity \eqref{vector-Bianchi}. 
In general, any dynamical system involving a linear multiplet possesses a composite vector multiplet $\mathbf{W}$. The EOM for the linear multiplet is $\mathbf{W}=0$, and for the linear multiplet action of eq.~\eqref{S_boldW}, this is given by $\mathbf{W}$ defined in \eqref{comp-W}.

In components, the linear multiplet action, $S_{\rm L}$, is obtained by making use of the component BF action \eqref{BF-Scomp-3form}, with the fields of the vector multiplet now being composite. That is, we substitute the expressions \eqref{desc-W-R2} into \eqref{BF-Scomp-3form}. After degauging \eqref{bosonic-degauging}, the bosonic part of $S_{\rm L}$ is given by \cite{BKNT-M14}
\bea
\label{eq:TensorComp}
S_{\rm L} 
&=& 
\int \rd^{5}x\, e\, 
\bigg\{ 
- \frac{3}{8} G\cR 
-4 D G
 - \frac{1}{8G} F^2
- \frac{3}{32} W^{\ha\hb} W_{\ha\hb} G
+\frac{1}{4} G^{-1}( \cD_\ha G^{ij} )\cD^\ha G_{ij}
\eol &&
- \frac{1}{8} G^{-1} \cH^\ha \cH_\ha
 + \frac{1}{24} \ve^{\ha\hb\hc\hd\he} b_{\hc\hd\he} \Big(
\frac{1}{2} G^{-3} (\cD_\ha G_{ik}) (\cD_\hb G_j{}^k) G^{ij}
+ G^{-1} \Phi_{\ha\hb}{}^{ij} G_{ij}
\Big)
\bigg\}~.~~~~~~~~~
\eea

\subsection{Gauged supergravity action}
An off-shell formulation for 5D minimal supergravity can be constructed by coupling the standard Weyl multiplet to two off-shell compensators: vector and linear multiplets \cite{Zucker1,Zucker2,Zucker3,Ohashi1,Ohashi2,Ohashi3,Ohashi4,Bergshoeff1,Bergshoeff2,Bergshoeff3,BKNT-M14}.
The complete (gauged) supergravity action, $S_{\rm gSG}$, is given by the following two-derivative action \cite{BKNT-M14}:
\begin{subequations} \label{gsg-action}
\bea
S_{\rm gSG}&=&  S_{\rm VM} + S_{\rm L} +\k \, S_{\rm BF} =
 \int \rd^{5|8}z\, E\, \Big\{ 
\frac{1}{4} V_{ij} { H }_{\rm VM}^{ij} 
+\O \mathbf{ W } 
+\k V_{ij} G^{ij} \Big\}\label{rep8.13a} \\
&=&
 \int \rd^{5|8}z\, E\,  \Big\{ 
\frac{1}{4} V_{ij} { H }_{\rm VM}^{ij} 
+ \O \mathbf{ W}  
+\k \O W \Big\}~. \label{rep8.13b}
\eea
\end{subequations}
The BF action $\k S_{\rm BF}$ describes a supersymmetric cosmological term. The $\k=0$ case corresponds to Poincar\'e supergravity, while $\k \neq 0$ leads to gauged or anti-de Sitter supergravity. In components, it reads
\bea \label{BF:cosmological}
\k S_{\rm BF} =  -\k \int \rd^{5}x\, e\, 
\Big\{  WF+ X^{ij} G_{ij} + v_a \cH^a \Big\}
+{\rm fermions}~.
\eea
Upon gauge fixing dilatation and superconformal symmetries (dilatation, $S$ and $K$) and integrating out the various auxiliary fields, one obtains the on-shell Poincare supergravity action of \cite{Cremmer, CN}. The contributions from the scalar curvature terms in \eqref{bosonic-swm} and \eqref{eq:TensorComp} combine to give the normalised Einstein-Hilbert term $-\hf \cR$ plus a cosmological constant, see, e.g., \cite{BKNT-M14} for details.

\section{Curvature-squared multiplets in components}\label{section-4}

In \cite{BKNT-M14}, three independent
curvature-squared invariants \cite{BKNT-M14, HOT, BRS, OP1, OP2} were defined in superspace in the standard Weyl multiplet background. The full expressions of all the composite primary superfields generating these invariants were presented recently in \cite{GHKT-M}. In this section we describe the component structure of all these composite primary multiplets (including the fermionic terms which can be found in the supplementary file), thus elaborating the results of \cite{GHKT-M}.

\subsection{Weyl-squared}
In superspace, a supersymmetric completion of a Weyl-squared term can be generated using a composite linear multiplet $H^{ij}_{\rm Weyl}$ built out of the super Weyl tensor and its descendants \cite{BKNT-M14}. The appropriate composite linear multiplet is described by the primary superfield
\begin{equation}
    H^{i j}_{\textrm{Weyl}} := - \frac{\ri}{2}   W^{{\alpha} {\beta} {\gamma} i} W_{{\alpha} {\beta} {\gamma}}\,^{j}+\frac{3\ri}{2}   W^{{\alpha} {\beta}} X_{{\alpha} {\beta}}\,^{i j} - \frac{3 \ri}{4}   X^{{\alpha} i} X^{j}_{{\alpha}}~.
    \label{H-Weyl}
\end{equation}
Here $H^{ij}_{\rm Weyl}$ satisfies the constraints \eqref{linear-constr}, and corresponds to the composite linear multiplet which was first constructed in components by Hanaki, Ohashi, and Tachikawa in \cite{HOT}.

With the aid of relations \eqref{eq:Wdervs}, along with the definition of the Weyl multiplet components given in \eqref{comps-W}, the components of $H^{ij}_{\rm Weyl}$ are straightforward to compute. They include the $\q =0$ projection of $H^{ij}_{\rm Weyl}$,  together with the descendant components:
\bsubeq \label{desc-H-Weyl}
\begin{align}
\varphi_{\rm Weyl}^{\hal\, i} &= \frac{1}{3} \nabla^{\hal}_{j} H^{ij}_{\rm Weyl} \loco~, \\
 F_{{\rm Weyl}} &=  \frac{\ri}{12} \de^{\a}_{i} \de_{\a j}H^{i j}_{\rm{Weyl}}\loco~, \\
\cH^{{a}}_{{\rm Weyl}} &= \frac{\ri}{12} (\Gamma^{{a}})^{\a \b} \de_{\a i} \de_{\b j} H^{i j}_{\rm{Weyl}} \loco~.
\end{align}
\esubeq
The bosonic part of the
descendants \eqref{desc-H-Weyl} was given in \cite{HOT}, while its complete structure including fermions is given for the first time in this paper and can be found in the supplementary file. Below we have omitted the fermionic contributions in the expressions for $F_{\rm Weyl}$ and $\cH^{a}_{\rm Weyl}$.
\bsubeq \label{desc-H-Weyl-full}
\bea
     H^{i j}_{\textrm{Weyl}} 
     &=& - \frac{\ri}{2}   W^{{\alpha} {\beta} {\gamma} i} W_{{\alpha} {\beta} {\gamma}}\,^{j}-   W^{{a} {b}} \Phi_{{a}{b}}{}^{i j} + \frac{256 \ri}{3}   \chi^{{\alpha} i} \chi^{j}_{{\alpha}}~, \\
    \varphi^{i}_{{\alpha} \, {\rm Weyl}}
    &=&{} C_{{\alpha} {\beta}{\gamma} {\lambda}}  W^{{\beta} {\gamma} {\lambda} i}+\frac{3}{4} W_{{\alpha} {\beta}} W_{{\gamma} {\lambda}} W^{{\beta} {\gamma} {\lambda} i} - \frac{3}{4}  W_{{\beta} {\gamma} {\lambda} }\,^{i} {\nabla}_{{\alpha}}{}^{ {\beta}}{W^{{\gamma} {\lambda}}}
    \non \\&&
    +\frac{3}{4} W_{{\alpha} {\beta} {\lambda}}\,^{i} {\nabla}^{{\beta} {\gamma} }{W_{{\gamma}}{}^{{\lambda}}} 
    -   \Phi^{{\beta} {\gamma} i j} W_{{\alpha} {\beta} {\gamma}j} - \frac{32 \ri}{3}\Phi_{{\alpha} {\beta}}\,^{i j} \chi_{j}^{{\beta}} - \frac{3}{2}   {W^{{\gamma} {\lambda}}} {\nabla}_{{\alpha}}{}^{ {\beta}}W_{{\beta} {\gamma} {\lambda} }\,^{i} 
    \non \\&&
    -8 \ri W^{{\beta} {\gamma}} W_{{\beta} {\gamma}} \chi^{i}_{{\alpha}}
    - \frac{3}{2} W_{{\beta} {\lambda}} {\nabla}^{{\beta} {\gamma}}{W_{{\alpha} {\gamma}}{}^{{\lambda} i}}-8\ri  W_{{\gamma} {\lambda}} {\nabla}_{{\alpha}}{}^{ {\gamma}}{\chi^{{\lambda} i}} +8\ri W_{{\alpha} {\beta}} {\nabla}^{{\beta} {\rho}}{\chi^{i}_{{\rho}}} 
    \non \\&&
    -\frac{128 \ri}{3}D \chi^{i}_{{\alpha}} - 12 \ri  \chi^{{\lambda} i} {\nabla}_{{\alpha}}{}^{ {\gamma}}{W_{{\gamma} {\lambda}}} +  4 \ri \chi^{i}_{{\beta}} {\nabla}^{{\beta} {\rho}}{W_{{\alpha} {\rho}}}~,\\
 F_{{\rm Weyl}} &=\, & \frac{128}{3}{D}^{2}  + 8 W^{{a} {b}} W_{{a} {b}} D  -\frac{2}{3}\Phi^{{a} {b} i j} \Phi_{{a} {b} i j}+  \frac{1}{4} C^{{a} {b} {c} {d}} C_{{a} {b} {c} {d}} + \frac{15}{8} C^{{a} {b} {c} {d}} W_{{a} {b}} W_{{c} {d}}  \non \\&& + \frac{81}{32} W^{{a} {b}} W_{{b} {c}} W^{{c} {d}} W_{{d} {a}}  -3 W_{{c} {d}} {{\nabla}_{{b}}{\nabla}^{{c}}{W^{{b} {d}}}} - \frac{33}{128}  W^{{a} {b}} W_{{a} {b}} W^{{c} {d}} W_{{c} {d}} \non \\&& +\frac{3}{2} {\nabla}^{{a}}{W^{{c} {e}}} {\nabla}_{{c}}{W_{{a} {e}}} - \frac{3}{2} {\nabla}^{{a}}{W^{{c} {e}}} {\nabla}_{{a}}{W_{{c} {e}}}   +\frac{9}{16}\ve^{{a} {b} {c} {d} {e}} W_{{a} {b}} W_{{c} {d}} {\nabla}^{{f}}{W_{{e} {f}}}~,\\
 \cH^{{a}}_{{\rm Weyl}} &=&~ \frac{1}{8}\ve^{{a} {b} {c} {d} {e}}C_{{b} {c}}\,^{{e_{1}} {e_{2}}} C_{{d} {e} {e_{1}} {e_{2}} }-\frac{1}{6}\ve^{{a} {b} {c} {d} {e}} \Phi_{{b} {c}}\,^{i j} \Phi_{{d} {e} i j} + {\nabla}_{{b}} \tilde{\cH}^{{a} b}_{{\rm Weyl}}~,
 \eea
 where
 \bea
     \tilde{\cH}^{{a} b}_{{\rm Weyl}} &=& 8W^{{a} {b}} D   + \frac{3}{2} C^{{a}{b}{c}{d}} {W_{{c} {d}}}+ \frac{3}{16}  W^{{a} {b}}W^{{c} {d}} {W_{{c} {d}}} - \frac{9}{4} W^{{a} {c}} {W_{{c} {d}}} W^{{d} {b}}
   \non \\&&
     -\frac{9}{4}\ve^{{a} {b} {c} {d} {e}} {W_{{c} {e_{1}}}} {\nabla}_{{d}}{W_{{e}}{}^{{e_{1}}}}- \frac{3}{2}\ve^{{a} {b} {c} {d} {e}}  {W_{{c} {e_{1}}}} {\nabla}^{{e_{1}}}{W_{{d}{e}}} 
   ~.\label{desc-H-Weyl-full-d}
\eea
\esubeq
Here the Weyl tensor $C_{abcd}$ is defined through
\bea
  C_{abcd} := {\mathscr{R}}(M)_{{a}{b}{c}{d}}  &=&  - \frac{\ri}{4} W_{{a} {b} {c} {d}}  - \frac{1}{2}  W_{{a} {b}}W_{{c} {d}} - \frac{1}{8} \eta_{{a} [ {c}} \eta_{{d}] {b}} W^{{e} {f}} W_{{e} {f}} \non\\
   &&+ \frac{1}{2} W_{{a} [{c}} W_{{d}] {b}} -   W_{{e} [ {a}} \eta_{{b}] [{c}} W^{{e}}{}_{ {d}]}~.
\eea
We also note the relation
\bea
\nabla_{a} \nabla_{b} W_{cd} 
&=& 
\cD_{a} \cD_{b} W_{c d} - 2 \mathfrak{f}_{a b} W_{c d} + 4 \mathfrak{f}_{a [c} W_{d] b} - 4 \mathfrak{f}_{a f} \eta_{b [c} W_{d]}{}^{f}
+\cdots
~,
\eea
with 
\bea
\mathfrak{f}_{ab} = -\frac{1}{6} \cR_{ab} + \frac{1}{48} \eta_{a b} \cR + \dots ~, \qquad \mathfrak{f}_{a}{}^a = -\frac{1}{16} \cR + \dots~,
\eea
and the ellipsis corresponding to omitted fermionic contributions. 
Having obtained the full component expressions, the first non-trivial check we have done is to show that the divergence of $\cH^a_{\rm Weyl}$ vanishes up to fermions, that is $\de_a \cH^{a}_{\rm Weyl} = 0$. 
One can show that this holds with the use of the Bianchi identities on $C_{abcd}$ and $\Phi_{ab}{}^{ij}$, along with the fact that $\de_{a} \de_{b} \tilde{\cH}^{{a} b}_{{\rm Weyl}} = 0$ up to fermions as $\tilde{\cH}^{{a} b}_{{\rm Weyl}}$ is a $K$-invariant, $\rm SU(2)$-singlet, antisymmetric tensor. 
In addition, we have checked explicitly that all the component fields described in \eqref{desc-H-Weyl-full} are $K$-primary.

\subsection{Log}\label{subsection Log descendant}
As an extension of the higher-derivative chiral invariant in 4D $\cN=2$ supergravity \cite{Butter:2013lta}, Ref. \cite{BKNT-M14} proposed a composite linear superfield which describes a supersymmetric Ricci tensor-squared term. This primary superfield, $H^{ij}_{\log}$, makes use of the standard Weyl multiplet coupled to the off-shell
vector multiplet compensator. It is given by\footnote{Note the overall minus sign difference between the definition of the $\log{W}$ invariant in this paper and the one of \cite{BKNT-M14}.}
\begin{equation}
    H^{i j}_{{\log}} = - \frac{3 \ri}{40} \Delta^{i j k l}\de_{k l} \log{W} =  \frac{3 \ri}{1280} \de^{(i j} \de^{k l)} \de_{k l} \log{W}~.
    \label{logW-0}
\end{equation}
Due to the complexity in computing the action of up to eight spinor derivatives on the ``log multiplet", the component structure of $H^{ij}_{{\log}}$ has just been studied recently.
With the aid of the \textit{Cadabra} software \cite{Cadabra-1,Cadabra-2}, the full expression of $H^{ij}_{\log}$ in its expanded form in terms of the descendants of $W$ and $W_{\a\b}$ appeared for the first time in \cite{GHKT-M}. Its lowest component, $H^{ij}_{\log} \loco$, is given by
\bsubeq \label{desc-H-logW-full}
\bea  \label{HlogW-full}
    H^{i j}_{\log}
    &=&
    {}  \frac{1}{2}  W_{{a} {b}}   \Phi^{a b i j}  - \frac{272\ri}{3} \chi^{{\alpha} i} \chi^{j}_{{\alpha}} 
    \non\\
    &&
    +{{W}}^{-1} \Bigg\{\,-6X^{i j} D + \frac{1}{2}  F_{{a} {b}} \Phi^{{a}{b} i j}  - \frac{1}{2}  \Box{{X^{i j}}} - \frac{9}{64}X^{i j} W^{{a} {b}} W_{{a} {b}} + \frac{1}{4} W^{{a}{b}} W_{{a}{b}}{}^{\alpha (i} \lambda^{j)}_{\alpha} \non\\
    &&~~~~~~~~~~
    + 8 \ri (\Gamma^{{a}})^{{\alpha} {\beta}}   \chi^{(i}_{{\alpha}}  {\nabla}_{{a}}{\lambda^{j)}_{{\beta}}} - 8 \ri (\Gamma^{{a}})^{{\alpha} {\beta}}   \lambda^{(i}_{{\alpha}} {\nabla}_{{a}}{\chi^{j)}_{{\beta}}}\Bigg\}
    \non\\
    &&
    + {{W}}^{-2} \Bigg\{\,\frac{1}{2} X^{i j}  \Box{{{W}}}+\frac{1}{2}  \big({\nabla}^{{a}}{{W}} \big) {\nabla}_{{a}}{X^{i j}} 
+ \frac{1}{4} F^{{a}{b}} W_{{a}{b}}{}^{\alpha (i} \lambda^{j)}_{\alpha}-\frac{\ri}{2}  \lambda^{{\alpha}(j}  \Box{{\lambda^{i)}_{{\alpha}}}} \non\\
&&~~~~~~~~~~~
-\frac{\ri}{4}    \big({\nabla}^{{a}}{\lambda^{{\alpha} i}} \big) {\nabla}_{{a}}{\lambda^{j}_{{\alpha}}} - \frac{3 \ri}{16} \ve^{{a} {b} {c} {d} {e}} (\Sigma_{{a} {b}})^{{\alpha} {\beta}}   W_{{d} {e}} \lambda^{(i}_{{\alpha}}  {\nabla}_{{c}}{\lambda^{j)}_{{\beta}}} - \frac{3 \ri}{8} (\Gamma^{{a}})^{{\alpha} {\beta}}  \lambda^{i}_{{\alpha}} \lambda^{j}_{{\beta}}  {\nabla}^{{c}}{W_{{a} {c}}} \non\\
&&~~~~~~~~~~~
 - 2 \ri  D \lambda^{{\alpha} i} \lambda^{j}_{{\alpha}}
- 4 \ri (\Sigma_{{a} {b}})^{{\alpha} {\beta}}  F^{{a} {b}} \chi^{(i}_{{\alpha}} \lambda^{j)}_{{\beta}}  
+ \frac{3 \ri}{128} W^{{a} {b}} W_{{a} {b}} \lambda^{{\alpha} i} \lambda^{j}_{{\alpha}} 
\non\\
&&~~~~~~~~~~~
+\frac{9 \ri}{256} \ve^{{a} {b} {c} {d} {e}} (\Gamma_{a})^{{\alpha} {\beta}}   W_{{b} {c}} W_{{d} {e}} \lambda^{i}_{{\alpha}} \lambda^{j}_{{\beta}} 
+ 4 \ri X^{i j} \chi^{{\alpha} k} \lambda_{{\alpha} k}
\Bigg\} 
\non\\
&&
+ {{W}}^{-3} \Bigg\{\,\frac{1}{8}X^{i j} F^{{a} {b}} F_{{a} {b}}  - \frac{1}{8} X^{i j} X^{k l} X_{k l} - \frac{1}{4} X^{i j}   \big({\nabla}^{{a}}{{W}} \big) {\nabla}_{{a}}{{W}}  \non\\
&&~~~~~~~~~~
- \frac{\ri}{8} \ve^{{a} {b} {c} {d} {e}} (\Sigma_{{a} {b}})^{{\alpha} {\beta}}   F_{{d} {e}} \lambda^{(i}_{{\alpha}}  {\nabla}_{c}{\lambda^{j)}_{{\beta}}} - \frac{\ri}{4} (\Gamma^{{a}})^{{\alpha} {\beta}}  F_{{a} {b}} \lambda^{(i}_{{\alpha}}  {\nabla}^{{b}}{\lambda^{j)}_{{\beta}}}  
\non\\
&&~~~~~~~~~~
- \frac{\ri}{4} (\Gamma^{{a}})^{{\alpha} {\beta}}   \lambda^{i}_{{\alpha}} \lambda^{j}_{{\beta}}  {\nabla}^{{c}}{F_{{a} {c}}} 
+\frac{\ri}{4} (\Gamma^{{a}})^{{\alpha} {\beta}}  X^{i j} \lambda^{k}_{{\alpha}} {\nabla}_{{a}}{\lambda_{{\beta} k}}+\frac{\ri}{4} (\Gamma^{{a}})^{{\alpha} {\beta}}  X^{k (i} \lambda^{j)}_{{\alpha}}  {\nabla}_{{a}}{\lambda_{{\beta} k}} 
\non\\
&&~~~~~~~~~~
 - \frac{\ri}{4} (\Gamma^{{a}})^{{\alpha} {\beta}}  \lambda^{(i}_{{\alpha}}    \big({\nabla}_{{a}}{X^{j) l}} \big) \lambda_{{\beta} l}
 + \frac{\ri}{4}   \lambda^{{\alpha} i} \lambda^{j}_{{\alpha}}  \Box{{{W}}} + \frac{3 \ri}{4}   \big({\nabla}^{{a}}{{W}} \big) \lambda^{{\alpha} (i}   {\nabla}_{{a}}{\lambda^{j)}_{{\a}}} 
 \non\\
&&~~~~~~~~~~
- \frac{\ri}{2} (\Sigma_{{a} {b}})^{{\alpha} {\beta}}  \big({\nabla}^{{a}}{{W}}  \big)\lambda^{(i}_{{\alpha}}   {\nabla}^{{b}}{\lambda^{j)}_{{\beta}}}  
 - \frac{3 \ri}{16} (\Gamma^{{a}})^{{\alpha} {\beta}}  W_{{a} {b}} \lambda^{i}_{{\alpha}} \lambda^{j}_{{\beta}}  {\nabla}^{{b}}{{W}} + \frac{3 \ri}{32} W_{{a} {b}} F^{{a} {b}} \lambda^{{\alpha} i} \lambda^{j}_{{\alpha}} 
 \non\\
 &&~~~~~~~~~~
+\frac{9 \ri}{64} \ve^{{a} {b} {c} {d} {e}} (\Gamma_{{a}})^{{\alpha} {\beta}}   W_{{b} {c}} F_{{d} {e}} \lambda^{i}_{{\alpha}} \lambda^{j}_{{\beta}}  - \frac{3 \ri}{32}  (\Sigma_{{a} {b}})^{{\alpha} {\beta}}  X^{i j} W^{{a} {b}} \lambda^{k}_{{\alpha}} \lambda_{{\beta}k}  
 \non\\
 &&~~~~~~~~~~
- 4  \chi^{{\alpha} k} \lambda^{(i}_{{\alpha}} \lambda^{{\beta} j)} \lambda_{{\beta} k}-4   \chi^{{\alpha} k} \lambda^{{\beta} i} \lambda^{j}_{{\beta}} \lambda_{{\alpha} k} \Bigg \}
 \non\\
 &&
 + {{W}}^{-4}\Bigg\{  -\frac{3 \ri}{16}   \lambda^{{\alpha} i} \lambda^{j}_{{\alpha}}   \big({\nabla}^{{a}}{{W}} \big) {\nabla}_{{a}}{{W}} 
 - \frac{3 \ri}{8} (\Gamma^{{a}})^{{\alpha} {\beta}}   X^{ k (i} \lambda^{j)}_{{\alpha}} \lambda_{{\beta}k}  {\nabla}_{{a}}{{W}} \non\\
 &&~~~~~~~~~~~
 +\frac{3\ri}{8} (\Gamma^{{a}})^{{\alpha} {\beta}}  F_{{a} {b}} \lambda^{i}_{{\alpha}} \lambda^{j}_{{\beta}}  {\nabla}^{{b}}{{W}} 
 + \frac{3 \ri}{32} F^{{a} {b}} F_{{a} {b}} \lambda^{{\alpha} i} \lambda^{j}_{{\alpha}} +\frac{3\ri }{64} \ve^{{a} {b} {c} {d} {e}} (\Gamma_{{a}})^{{\alpha} {\beta}}   F_{{b} {c}} F_{{d} {e}} \lambda^{i}_{{\alpha}} \lambda^{j}_{{\beta}}  \non\\
 &&~~~~~~~~~~~
 - \frac{3 \ri}{16} (\Sigma_{{a} {b}})^{{\alpha} {\beta}}  X^{i j} F^{{a} {b}} \lambda^{k}_{{\alpha}} \lambda_{{\beta} k}    -\frac{3 \ri}{16}  X^{i j} X^{k l} \lambda^{\alpha}_{k} \lambda_{{\alpha} l}   -\frac{3 \ri}{32}  X^{k l} X_{ k l} \lambda^{{\alpha} i} \lambda^{j}_{{\alpha}}  \non\\
 &&~~~~~~~~~~~
 -\frac{15}{64} (\Gamma^{{a}})^{{\alpha} {\beta}}   \lambda^{i}_{{\alpha}} \lambda^{j}_{{\beta}} \lambda^{{\gamma} k}  {\nabla}_{{a}}{\lambda_{{\gamma} k}}-\frac{9}{32} (\Gamma^{{a}})^{{\alpha} {\beta}}    \lambda^{(i}_{{\alpha}} \lambda^{{\rho} j)} \lambda^{k}_{{\beta}}  {\nabla}_{{a}}{\lambda_{{\rho} k}}
 \non\\
 &&~~~~~~~~~~~
 -\frac{15}{32}(\Gamma^{{a}})^{{\alpha} {\beta}}   \lambda^{(i}_{{\alpha}} \lambda^{{\rho} j)} \lambda^{k}_{{\rho}}  {\nabla}_{{a}}{\lambda_{{\beta} k}}
   -\frac{15}{64} (\Gamma^{{a}})^{{\alpha} {\beta}}   \lambda^{{\rho} i} \lambda^{j}_{{\rho}} \lambda^{k}_{{\alpha}}  {\nabla}_{{a}}{\lambda_{{\beta} k}}  
      \non\\
   &&~~~~~~~~~~~
+ \frac{9}{32} (\Sigma_{{a} {b}})^{{\alpha} {\beta}}   W^{{a} {b}} \lambda^{(i}_{{\alpha}} \lambda^{{\rho} j)} \lambda^{k}_{{\beta}} \lambda_{{\rho} k} 
+ \frac{9}{64} (\Sigma_{{a} {b}})^{{\alpha} {\beta}}  W^{{a} {b}} \lambda^{{\rho} i} \lambda^{j}_{{\rho}} \lambda^{k}_{{\alpha}} \lambda_{{\beta} k} \Bigg\} 
   \non\\
   &&
   + {{W}}^{-5} \Bigg\{\, \frac{3}{8} (\Gamma^{{a}})^{{\alpha} {\beta}}    \lambda^{(i}_{{\alpha}} \lambda^{{\rho} j)} \lambda^{k}_{{\beta}} \lambda_{{\rho} k}  {\nabla}_{{a}}{{W}}  + \frac{3}{8} (\Sigma_{{a} {b}})^{{\alpha} {\beta}}   F^{{a} {b}} \lambda^{{\rho} i} \lambda^{j}_{{\rho}} \lambda^{k}_{{\alpha}} \lambda_{{\beta} k}    \non\\
   &&~~~~~~~~~~
  +\frac{3}{8}  X^{k l} \lambda^{{\alpha} i} \lambda^{j}_{{\alpha}} \lambda^{{\beta}}_{k} \lambda_{{\beta} l} 
  \Bigg\} \non\\
  &&
  + {W}^{-6}\Bigg\{\,\frac{3 \ri}{32}   \lambda^{{\alpha} i} \lambda^{j}_{{\alpha}} \lambda^{{\beta} k} \lambda^{l}_{{\beta}} \lambda_{k}^{{\gamma}} \lambda_{{\gamma} l} 
  +\frac{3 \ri}{16}   \lambda^{{\alpha} i} 
  \lambda^{k}_{{\alpha}}  
  \lambda^{{\beta} j}
  \lambda^{l}_{{\beta}} 
  \lambda_{k}^{{\gamma}}
  \lambda_{{\gamma} l} \Bigg\} 
~.
\eea
The full expressions of the descendant component fields of $H^{ij}_{\log}$, including the fermionic terms, are very involved and can be found in the supplementary file. In this subsection, we will only present the bosonic sectors. The component $F_{\log} = \frac{\ri}{12} \de^{\a}_{i} \de_{\a j}H^{i j}_{\log} \loco$ takes the form
\bea
    F_{\log} 
    &=&  
    -\frac{1}{2}W^{{a} {b}} W^{{c} {d}} C_{{a}{b}{c}{d}} + \frac{8}{3}{D}^{2} + 4 \Box{D} + \frac{368}{3}W^{{a} {b}} W_{{a} {b}} D -\frac{1}{6}\Phi^{{a}{b} i j} \Phi_{{a}{b} i j} 
    \non\\
    &&
    + 2 W_{{c} {d}} {\nabla}^{{c}}{{\nabla}_{{a}}{W^{{a} {d}}}} - \frac{1}{2} W_{{c} {d}} {\nabla}_{{a}}{{\nabla}^{{c}}{W^{{a} {d}}}} + \frac{39}{16} W_{{c} {d}}\Box{W^{{c} {d}}}   - \frac{1005}{2048}W^{{a} {b}}  W_{{a} {b}} W^{{c} {d}} W_{{c} {d}} 
    \non\\
    &&
    - \frac{9}{4} {\nabla}^{{a}}{W_{{a} {b}}} {\nabla}_{{c}}{W^{{b} {{c}}}} - \frac{3}{4} {\nabla}^{{a}}{W^{{b} {c}}} {\nabla}_{{b}}{W_{{a} {c}}}+\frac{33}{16}{\nabla}^{{a}}{W^{{b} {c}}}{\nabla}_{{a}}{W_{{b} {c}}}+\frac{9}{16}\ve^{{a} {b} {c} {d} {e}} W_{{b} {c}} W_{{d} {e}} {\nabla}^{f}{W_{{a} f}}  
    \non\\
    &&
    + {{W}}^{-1} \Bigg\{\, - F_{{c} {d}} {\nabla}^{{c}}{{\nabla}_{{a}}{W^{{a} {d}}}} - \frac{1}{2} F_{{c} {d}} {\nabla}_{{a}}{{\nabla}^{{c}}{W^{{a} {d}}}} + \frac{3}{2} F_{{c} {d}} \Box{W^{{c} {d}}} + 2 W_{{c} {d}} {\nabla}^{{c}}{{\nabla}_{{a}}{F^{{a} {d}}}}
    \non\\
    &&~~~~~~~~~~~
    + \frac{1}{2} W_{{c} {d}} \Box{F^{{c} {d}}} - 4 D\Box{W}-\Box\Box{W} - 4 {\nabla}^{{a}}{W} {\nabla}_{{a}}{D} - \frac{1}{4} W^{{a} {b}} F^{{c} {d}} C_{{a} {b} {c} {d}}
    \non\\
    &&~~~~~~~~~~~
    - \frac{9}{2} W_{{a}}{}^{{c}} W_{{c} {b}} {\nabla}^{{a}}{{\nabla}^{{b}}{W}} + \frac{9}{2} W_{{c} {d}} {\nabla}_{{a}}{W} {\nabla}^{{c}}{W^{{a} {d}}} - \frac{3}{16} W_{{c} {d}} {\nabla}_{{a}}{W} {\nabla}^{{a}}{W^{{c} {d}}}
    \non\\
    &&~~~~~~~~~~~
    + \frac{9}{2} W_{{c} {d}} {\nabla}^{{c}}{W} {\nabla}_{{a}}{W^{{a} {d}}} - \frac{3}{32} W^{{c} {d}} W_{{c} {d}} \Box{W} - \frac{3}{2}{\nabla}^{{a}}{F_{{a} {b}}} {\nabla}_{{c}}{W^{{b} {{c}}}} 
    \non\\
    &&~~~~~~~~~~~
    +\frac{3}{2}{\nabla}^{{a}}{F^{{b} {c}}}{\nabla}_{{a}}{W_{{b} {c}}} + \frac{9}{16} \ve^{{a} {b} {c} {d} {e}} \bigg(2 W_{{c} {d}} F_{{e} f} {\nabla}^{f}W_{{a} {b}} +  W_{{b} {c}} W_{{d} {e}} {\nabla}^{f}{F_{{a} f}} 
    \bigg) \Bigg\}
    \non\\
    &&
    + {{W}}^{-2} \Bigg\{\, - F^{{a} {b}} F_{{a} {b}} D - \frac{1}{2} F_{{c} {d}}\Box{F^{{c} {d}}} + \frac{1}{4} F^{{a} {b}} F^{{c} {d}} C_{{a} {b} {c} {d}}  - \frac{75}{128}W^{{a} {b}} W_{{a} {b}} F^{{c} {d}} F_{{c} {d}} 
    \non\\
    &&~~~~~~~~~~~
    + \frac{9}{4}W^{{a} {b}} W_{{b} {c}} F^{{c} {d}} F_{{d} {a}} - \frac{3}{2} F_{{c} {d}} {\nabla}_{{a}}{W} {\nabla}^{{a}}{W^{{c} {d}}} - \frac{3}{2} W_{{c} {d}} {\nabla}_{{a}}{W} {\nabla}^{{a}}{F^{{c} {d}}} 
    \non\\
    &&~~~~~~~~~~~
    - \frac{3}{2} W^{{c} {d}} F_{{c} {d}} \Box{W} + 2 D {\nabla}^{{a}}{W} {\nabla}_{{a}}{W} + 2{\nabla}^{{a}}{W} {\nabla}_{{a}}{\Box{W}}+\Box{W} \Box{W}
     \non\\
    &&~~~~~~~~~~~
    + \frac{1}{2} {\nabla}^{{a}}{{\nabla}^{{b}}{W}} {\nabla}_{{a}}{{\nabla}_{{b}}{W}} + 3 X^{i j} X_{i j} D + \frac{1}{2} X^{i j} \Box{X_{i j }}+\frac{1}{4} {\nabla}^{{a}}{X^{i j}} {\nabla}_{{a}}{X_{i j}}
    \non\\
    &&~~~~~~~~~~~
    + \frac{9}{4} W_{{a}}{}^{{c}} W_{{c} {b}} {\nabla}^{{a}}{W} {\nabla}^{{b}}{W} + \frac{3}{64} W^{{c} {d}} W_{{c} {d}} {\nabla}^{{a}}{W} {\nabla}_{{a}}{W} - \frac{1}{4} {\nabla}^{{a}}{F^{{b} {c}}} {\nabla}_{{a}}{F_{{b} {c}}}
    \non\\
    &&~~~~~~~~~~~
    -\frac{1}{2}  X^{i j} F^{{a} {b}} \Phi_{{a} {b} i j } +\frac{9}{128} X^{i j} X_{i j} W^{{a} {b}} W_{{a} {b}} + \frac{3}{8}\ve^{{a} {b} {c} {d} {e}}  W_{{c} {d}} F_{{e} f} {\nabla}^{f}{F_{{a} {b}}}
    \non\\
    &&~~~~~~~~~~~
    - \frac{3}{8}\ve^{{a} {b} {c} {d} {e}} F_{{b} {c}} F_{{d} {e}} {\nabla}^{f}{W_{{a} f}} 
    \Bigg\}
    \non\\
    &&
    +{{W}}^{-3} \Bigg\{\, - \frac{3}{4}W^{{a} {b}} F^{{c} {d}} F_{{a} {b}} F_{{c} {d}}+\frac{3}{2}W^{{a} {b}} F^{{c} {d}} F_{{a} {c}} F_{{b} {d}}- F_{{a}}{}^{{c}} F_{{c} {b}} {\nabla}^{{a}}{{\nabla}^{{b}}{W}}
    \non\\
    &&~~~~~~~~~~~
    + F_{{c} {d}} {\nabla}^{{a}}{W} {\nabla}_{{a}}{F^{{c} {d}}} + F_{{c} {d}} {\nabla}^{{c}}{W} {\nabla}_{{a}}{F^{{a} {d}}} + \frac{1}{4} F^{{c} {d}} F_{{c} {d}} \Box{W} 
    \non\\
    &&~~~~~~~~~~~
    + \frac{3}{2} W^{{c} {d}} F_{{c} {d}} {\nabla}^{{a}}{W} {\nabla}_{{a}}{W} - \frac{3}{2} {\nabla}^{{a}}{W} {\nabla}_{{a}}{W}\Box{W} - {\nabla}^{{a}}{W} {\nabla}^{{b}}{W} {\nabla}_{{a}}{{\nabla}_{{b}}{W}} 
    \non\\
    &&~~~~~~~~~~~
    - \frac{3}{4} X^{i j} X_{i j } \Box{W} - \frac{3}{2} X^{i j} {\nabla}^{{a}}{W} {\nabla}_{{a}}{X_{i j}} - \frac{1}{8}\ve^{{a} {b} {c} {d} {e}} F_{{b} {c}} F_{{d} {e}} {\nabla}^{f}{F_{{a} f}} \Bigg\}
    \non\\
    &&
    +{{W}}^{-4} \Bigg\{\,- \frac{3}{32}F^{{a} {b}}F_{{a} {b}} F^{{c} {d}}  F_{{c} {d}}+\frac{3}{8}F^{{a} {b}} F_{{b} {c}} F^{{c} {d}} F_{{d} {a}} +\frac{3}{2}F_{{a}}{}^{{c}} F_{{c} {b}} {\nabla}^{{a}}{W} {\nabla}^{{b}}{W} {\nabla}_{{a}}{W} 
    \non\\
    &&~~~~~~~~~~~
    - \frac{3}{8} F^{{c} {d}} F_{{c} {d}} {\nabla}^{{a}}{W} - \frac{3}{16} X^{i j} X_{i j} F^{{a} {b}} F_{{a} {b}}+\frac{3}{8} ({\nabla}^{{a}}{W} {\nabla}_{{a}}{W})^2 
    \non\\
    &&~~~~~~~~~~~
    +\frac{9}{8}  X^{i j} X_{i j} {\nabla}^{{a}}{W} {\nabla}_{{a}}{W}+\frac{3}{32} X^{i j} X_{i j} X^{k l} X_{k l} \Bigg\}  
    \non\\
    &&
    + \, {\rm fermions}~.
\eea

The initial expression for the top component, $\cH^{{a}}_{\log} = \frac{\ri}{12} (\Gamma^{{a}})^{\a \b} \de_{\a i} \de_{\b j} H^{i j}_{\log} \loco $, obtained from the computer algebra program \textit{Cadabra}, is quite extensive and contains 261 terms in the bosonic part alone. For completeness, this expression is given in appendix \ref{Building-Blocks}. However, there are ample opportunities for simplification by utilising the algebra of commutators \eqref{vect-vect commutation}, symmetry properties of the Weyl tensor and the Bianchi identities \eqref{WeylmultipletBI}, \eqref{F-bianchi}. Additionally, many terms in the expression can be further combined as they exhibit similar structures and are linearly dependent, see appendix \ref{Building-Blocks}.

The ultimate goal is to express $\cH^{a}_{\log}$ in terms of total derivatives as much as possible, that is, $\cH^{a}_{\log} = \de_{b} \tilde{\cH}^{ab}_{\log} + \dots$. The reason is that, once coupled to the vector potential $v_a$ in the component BF action \eqref{BF-Scomp}, this term will be proportional to $F_{ab} \tilde{\cH}^{ab}_{\log}$ upon integration by parts. This task, however, proves to be challenging, as determining the most suitable combinations of terms to combine into total derivatives is not obvious beforehand. Despite these difficulties, the expression provided by \textit{Cadabra} remains approachable, and progress has been made by manually combining terms to reach a final expression of the form 
\bea
    \cH^{{a}}_{\log} &=& \frac{1}{12}\ve^{{a} {b} {c} {d} {e}} \Phi_{{b} {c}}\,^{i j} \Phi_{{d} {e} i j} - \frac{3}{8} {\nabla}_{{b}}{C^{{a} {b} {c} {d}}} W_{{c} {d}}
    \non \\
    &&
    - \frac{3}{8}  W^{-1} {\nabla}_{{b}}{C^{{a} {b} {c} {d}}} F_{{c} {d}}  + \nabla_b \tilde{\cH}^{a b}_{\log} + {\rm fermions}~,~~~~~~~~~
    \label{Ha-log-comp}
\eea
where $\tilde{\cH}^{a b}_{\log}$ is antisymmetric and is given by
\bea
    \tilde{\cH}^{a b}_{\log} &=& 
    - 4 W^{{a} {b}} D + \frac{9}{4}\ve^{{a} {b} {c} {d} {e}} W_{{c} {d}} {\nabla}^f {W_{{e} f}} + \frac{3}{4} \ve^{{a} {b} {c} {d} {e}} W_{{c} f} {\nabla}^{f}{W_{{d} {e}}} - \frac{3}{32}W^{{a} {b}} W^{{c} {d}} W_{{c} {d}} 
    \non\\
    &&
    - \frac{9}{2}W^{{a} {c}} W_{{c} {d}} W^{{b} {d}} - \frac{3}{2}C^{{a} {b} {c} {d}} W_{{c} {d}}+\frac{3}{2} \Box{W^{{a} {b}}}   \non\\
    && 
    + W^{-1} \Bigg\{\, 4 F^{{a} {b}} D + \frac{3}{4}\ve^{{a} {b} {c} {d} {e}} F_{{c} f} {\nabla}^{f}{W_{{d} {e}}} + \frac{3}{2}\ve^{{a} {b} {c} {d} {e}} F_{{d} {e}} {\nabla}^{f}{W_{{c} f}} + \frac{3}{2}\ve^{{a} {b} {c} {d} {e}} W_{{d} {e}} {\nabla}^{f}{F_{{c} f}}
    \non\\
    &&~~~~~~~~~~~
    +\frac{3}{4}\ve^{{a} {b} {c} {d} {e}}  W_{{e} f} {\nabla}^{f}{F_{{c} {d}}} + \frac{9}{8} \ve^{{c} {d} {e} f [{b}} {\nabla}^{{a}]}{W} W_{{c} {d}} W_{{e} f} + 3 {\nabla}^{{c}}{W^{[{b}}{}_{{c}}} {\nabla}^{{a}]}{W} 
    \non\\
    &&~~~~~~~~~~~
    + 6  {\nabla}^{[{a}}{W_{{c}}{}^{{b}]}} {\nabla}^{{c}}{W} + 3 W^{{a} {b}} \Box{W}+ \Box{F^{{a} {b}}}+\frac{75}{32} W^{{c} {d}} W_{{c} {d}} F^{{a} {b}}
    \non\\
    &&~~~~~~~~~~~
    + 9 W^{{c} [{a}} W_{{c} {d}} F^{{b}] {d}} + \Phi^{{a} {b}}{}_{i j} X^{i j}- C^{{a} {b} {c} {d}} F_{{c} {d}}\Bigg\}
    \non\\
    &&
    +W^{-2}\Bigg\{\, \frac{1}{4} \ve^{{a} {b} {c} {d} {e}} F_{{c} f} {\nabla}^{f}{F_{{d} {e}}} + \frac{1}{2} \ve^{{a} {b} {c} {d} {e}} F_{{c} {d}} {\nabla}^{f}{F_{{e} f}} - \frac{3}{8} \ve^{{a} {b} {c} {d} {e}} W_{{c} {d}} F_{{e} f} {\nabla}^{f}{W}
    \non\\
    &&~~~~~~~~~~~
    +\frac{3}{8} \ve^{{a} {b} {c} {d} {e}} W_{{c} f} F_{{d} {e}} {\nabla}^{f}{W} + \frac{3}{8} \ve^{ {c} {d} {e} f [{a}} W_{{c} {d}} F_{{e} f} {\nabla}^{{b}]}{W} - \frac{1}{2} F^{{a} {b}}  \Box{W}
    \non\\
    &&~~~~~~~~~~~
    + F^{{c} [{a}} {\nabla}^{{b}]}{{\nabla}_{{c}}{W}} +  {\nabla}^{[{a}}{W} {\nabla}_{{c}}{F^{{b}] {c}}} - \frac{3}{2} W^{{a} {b}} {\nabla}^{{c}}{W}  {\nabla}_{{c}}{W} + 3  W^{{c} [{a}} F_{{c} {d}} F^{{b}] {d}}
    \non\\
    &&~~~~~~~~~~~
    + \frac{3}{4} W^{{a} {b}} F^{{c} {d}} F_{{c} {d}} + \frac{3}{2} F^{{c} [{a}} W_{{c} {d}} F^{{b}] {d}} - \frac{1}{2}{\nabla}^{{c}}{W}  {\nabla}_{{c}}{F^{{a} {b}}} + \frac{3}{2} F^{{a} {b}} W^{{c} {d}} F_{{c} {d}}\Bigg\}
    \non\\
    &&
    + W^{-3} \Bigg\{ \frac{1}{4}  \ve^{{c} {d} {e} f [{a}} F_{{c} {d}} F_{{e} f} {\nabla}^{{b}]}{W} + \frac{1}{2} F^{{a} {b}} {\nabla}^{{c}}{W} {\nabla}_{{c}}{W}
    \non\\
    &&~~~~~~~~~~~~~
    +\frac{1}{4} F^{{a} {b}} F^{{c} {d}} F_{{c} {d}}+ F^{{a} {c}} F_{{c} {d}} F^{{d} {b}}+\frac{1}{4} X^{i j} X_{i j} F^{{a} {b}}\Bigg\}~.
    \label{Ha-log-comp_b}
\eea
\esubeq

We also note the following useful relations (up to fermions)
\bsubeq \label{degauged-vec-Weyl}
\bea
\nabla_{a} \nabla_{b} W &=& \cD_{a} \cD_{b} W - 2 \mathfrak{f}_{a b} W + \dots~, \\
\nabla_{a} \nabla_{b} X^{ij} &=& \cD_{a} \cD_{b} X^{i j} - 4 \mathfrak{f}_{a b} X^{i j}+ \dots~,\\
\nabla_{a} \nabla_{b} F_{c d} &=& \cD_{a} \cD_{b} F_{c d} - 4 \mathfrak{f}_{a b} F_{c d} + 4 \mathfrak{f}_{a [c} F_{d] b} - 4 \mathfrak{f}_{a f} \eta_{b [c} F_{d]}{}^{f}+ \dots~, \\
\de_{a}  \Box W &=& 
\cD_{a} \cD_{b} \cD^{b} W 
-2\cD_{a}( \mathfrak{f}_{b}{}^b W) 
+ 2 \mathfrak{f}_{a b} \cD^{b} W+ \dots~, \\
\Box \Box W &=& 
\cD_{a} \cD^{a} \cD_{b} \cD^{b}W 
- 2 \cD_{a}\cD^{a}( \mathfrak{f}_{b}{}^{b} W ) + 4 \mathfrak{f}_{a b} \cD^{a} \cD^{b}W  \non  \\
&& + 2 \cD^{a}\mathfrak{f}_{a b} \cD^{b} W - 6 \mathfrak{f}_{a}{}^{a} \cD_{b} \cD^{b} W  - 4 \mathfrak{f}_{a b} \mathfrak{f}^{a b} W + 12 \mathfrak{f}_{a}{}^{a} \mathfrak{f}_{b}{}^{b} W + \dots ~, \\
\nabla_{a} \nabla_{b} D &=&  \cD_{a} \cD_{b} D - 4 \mathfrak{f}_{a b} D + \dots~.
\eea
\esubeq

It is useful to express the component field $\cH^a_{\log}$ in terms of the degauged covariant derivative $\cD_a$. Degauging all the covariant derivative $\de_a$ in \eqref{Ha-log-comp} but not those implicit in ${\tilde{\cH}^{a b}_{\log}}$ results in the following expression:
\bsubeq
\bea
    \cH^{a}_{\log} &=&  \frac{1}{12}\ve^{{a} {b} {c} {d} {e}} \Phi_{{b} {c}}\,^{i j} \Phi_{{d} {e} i j} + \cD_{b}{\tilde{\cH}^{a b}_{\log}} - \mathfrak{f}_{b c} K^{c}\tilde{\cH}^{a b}_{\log}
    \non \\
    && - \frac{3}{8} \cD_{{b}}{C^{{a} {b} {c} {d}}} W_{{c} {d}}  - \frac{3}{8}  W^{-1} \cD_{{b}}{C^{{a} {b} {c} {d}}} F_{{c} {d}}~.
\eea
This can equivalently be written as
\bea\label{HalogWdegauged}
    \cH^{a}_{\log} &=&  \frac{1}{12}\ve^{{a} {b} {c} {d} {e}} \Phi_{{b} {c}}\,^{i j} \Phi_{{d} {e} i j} + \cD_{b}({\tilde{\cH}^{a b}_{\log}} + \hat{\cH}^{a b}_{\log})~,
\eea
where $\hat{\cH}^{a b}_{\log}$ is antisymmetric and is explicitly defined as
\bea \label{non-covariant terms}
    \hat{\cH}^{a b}_{\log}
    &=&  \frac{5}{16} \cR W^{ab} + \cR^{c [a} W^{b]}{}_{c} + \frac{5}{16} \cR F^{ab} W^{-1} + \cR^{c [a} F^{b]}{}_{c} W^{-1}.
\eea
\esubeq
In order to derive \eqref{HalogWdegauged}, we have made use of the following Riemann tensor properties
\bsubeq\label{Riemann tensor properties}
\bea
    C^{abcd} &=& \cR^{abcd} + 4 \eta^{c[a} \mathfrak{f}^{b]d} - 4 \eta^{d[a} \mathfrak{f}^{b]c}~,\\
    \cD_{[a}\cR_{bc]de} &=& 0~,\\
    \cD_{d}\cR^{abcd} &=& -2 \cD^{[a}\cR^{b]c}~,\\
    \cD_{a}\cR^{ab} &=& \frac{1}{2}\cD^{a} \cR~.
\eea
\esubeq
Having obtained the full component expressions, the first non-trivial check we have done is to show that the divergence of $\cH^a_{\rm log}$ vanishes up to fermions, that is $\de_a \cH^{a}_{\rm log} = 0$. Unlike the Weyl case, $\tilde{\cH}^{a b}_{\log}$ is not $K$-invariant. Therefore,
\bea
    \de_{a} \de_{b} \tilde{\cH}^{a b}_{\log} &=& -\frac{1}{2} {R}(K)_{\ha\hb}{}^\hc K_{c} \tilde{\cH}^{a b}_{\log} \non \\
    &=& \frac{3}{8} \de^{b}C_{a b c d} \left(\de^{a} W^{c d} +  W^{-1}  \de^{a} F^{c d} -  W^{-2} F^{c d} \de^{a}W\right)~.
\eea
The above identity, along with the following relation
\bea
    \de_{a} \de_{b}C^{a b c d} &=& 0~,\\
    \de_{[a} \Phi_{b c]}{}^{ij} &=& 0~,
\eea
implies $\de_a \cH^{a}_{\rm log} = 0$.
In addition, we have checked explicitly the $K$-invariance of all the composite descendant components of $H^{ij}_{\rm log}$ including fermions.

\subsection{Scalar curvature squared}
Following the component field construction of \cite{OP2}, given a composite vector multiplet \eqref{comp-W} and its corresponding descendants \eqref{desc-W-R2}, one can then use \eqref{O2composite-N} to construct a composite primary superfield based on a single composite vector multiplet \cite{BKNT-M14}
\bea
H^{ij}_{R^2}:=H^{ij}_{\rm{VM}}[\mathbf{W}] &=& \ri (\de^{\a(i} \mathbf{W}) \de_{\a}^{j)} \mathbf{W} + \frac{\ri}{2} \mathbf{W} \de^{\a(i} \de_{\a}^{j)} \mathbf{W} \non\\
&=& -\ri \bm{\lambda}^{\a i} \bm{\l}_{\a}^j +2 \mathbf{W} \mathbf{X}^{ij}~.
\label{Hij-R2}
\eea
Its component structure can be compactly described in terms of \eqref{desc-W-R2}. They read
\bsubeq \label{Hij-R2-desc}
\bea
 H^{i j}_{R^2} &=& 2 \mathbf{W} \mathbf{X}^{i j} - \ri \bm{\l}^{\a (i} \bm{\l}^{j)}_{\a}
 ~, \\
 \varphi^{i}_{\a R^2} &=& \ri \mathbf{X}^{i j} \bm{\l}_{\a j} - 2 \ri \mathbf{F}_{\a \b}\bm{\l}^{\b i} + 16 \ri \mathbf{W}^2 \chi^{i}_{\a} - 2 \ri \mathbf{W} \de_{\a \b}\bm{\l}^{\b i}
 \non\\
 &&
 - \ri \bm{\l}^{\b i} \de_{\a \b}\mathbf{W} - 3 \ri W_{\a \b}\mathbf{W} \bm{\l}^{\b i}
 ~, \\
 F_{R^2} &=& \mathbf{X}^{i j} \mathbf{X}_{i j} - \mathbf{F}^{{a} {b}} \mathbf{F}_{{a} {b}} + 4 \mathbf{W} \de^{{a}}\de_{{a}}\mathbf{W} + 2 (\de^{{a}}\mathbf{W})(\de_{{a}}\mathbf{W}) - 2 \ri (\de_{\a \b} \bm{\l}^{\b i}) \bm{\l}^{\a}_{i}
  \non\\
 &&
 - 6 W^{{a} {b}}\mathbf{F}_{{a} {b}} \mathbf{W} - \frac{39}{8}W^{{a} {b}}W_{{a} {b}}\mathbf{W}^2 - 16 D \mathbf{W}^2 + 64 \ri \chi^{\a}_{i} \bm{\l}^{i}_{\a} \mathbf{W} - 3 \ri W_{\a \b} \bm{\l}^{\a i} \bm{\l}^{\b}_{i}
 ~,~~~~~~~\\
 \cH^{{a}}_{R^2} &=& - \frac{1}{2} \ve^{{a} {b} {c} {d} {e}} \mathbf{F}_{{b} {c}} \mathbf{F}_{{d} {e}} - \de_{{b}}\left( 4\mathbf{W} \mathbf{F}^{{a} {b}} + 6 W^{{a} {b}} \mathbf{W}^2 + 2 \ri (\Sigma^{{a} {b}})^{\a \b} \bm{\l}^{i}_{\a} \bm{\l}_{\b i} \right)
 ~.
\eea
\esubeq
In addition to eqs. \eqref{eq:degaugeLinear1}, the following identities are useful in which fermions\footnote{See the supplementary file for the results including fermions.} are suppressed at two or more covariant vector derivatives:
\bsubeq
\bea
    \nabla_a \nabla_b G^{ij} &=& \cD_a \cD_b G^{ij} - 6 G^{i j} \mathfrak{f}_{a b}~,
    \\
    \nabla_a \nabla_b F &=& \cD_a \cD_b F - 8 F \mathfrak{f}_{a b}~,
    \\
    \nabla_a \nabla_b \cH_c &=&  \cD_a \cD_b \cH_c - 8 \cH_c \mathfrak{f}_{a b} - 2 \cH_b \mathfrak{f}_{a c} + 2 \eta_{b c} \cH_{d} \mathfrak{f}_{a}{}^{d}~,
    \\
    \nabla_a \Box G^{ij} &=& 
    \cD_a \cD_b\cD^b G^{ij} 
    -6\cD_a(\mathfrak{f}_b{}^b G^{ij} )
    - 6 \mathfrak{f}_{a b} \cD^{b}G^{i j}~.
\eea
\esubeq
The composite multiplet \eqref{Hij-R2-desc} will be used in section \ref{Scalar-curvature-squared} to describe the supersymmetric completion of the scalar curvature-squared invariant.

Having obtained the full component expressions, the first non-trivial check we have done is to show that the divergence of $\cH^a_{R^2}$ vanishes up to fermions, that is, $\de_a \cH^{a}_{R^2} = 0$.
Another consistency check is the $K$-invariance for the component fields \eqref{Hij-R2-desc}. Both of these can be shown in a straighforward way because the composite vector multiplet fields are all primary.

\section{Curvature-squared actions in a standard Weyl background}
\label{section-5}

Within the superconformal approach, supersymmetric completions of the Weyl tensor squared and the scalar curvature squared were constructed for the first time, respectively in \cite{HOT} and \cite{OP2} by using component field techniques. 
The third independent invariant necessary to obtain all the curvature-squared models in five dimensions includes the Ricci tensor-squared term. In the standard Weyl multiplet background, this invariant was defined in superspace in \cite{BKNT-M14}.

In section \ref{section-4}, we have presented the bosonic component structure (and their fermionic counterparts in the supplementary file) of all the composite primary linear multiplets generating the three curvature-squared terms for 5D minimal supergravity based on the standard Weyl multiplet: Weyl, Ricci tensor, and scalar curvature squared. In this section we will use these multiplets and the BF action principle to reproduce the results of \cite{HOT} and \cite{OP2}. In particular, by exploiting the computer algebra program \textit{Cadabra} \cite{Cadabra-1,Cadabra-2}, the full bosonic part of the new ``Log invariant" will be presented for the first time in subsection \ref{sc-logW-action}.\footnote{The Log invariant in the gauged dilaton Weyl basis has been obtained recently in \cite{Gold:2023ymc}.}

\subsection{Weyl-squared}
The off-shell Weyl-squared action in the standard Weyl multiplet background was first constructed in \cite{HOT}. In our notation, such an invariant can be constructed using the BF action principle
\bsubeq \label{S-Weyl-0}
\bea 
S_{\rm{Weyl}} 
&=& \int \rd^{5|8} z\, E\, V_{ij} H^{ij}_{\rm Weyl} \\
&=&  \int \rd^5 x \, e \, \cL_{\rm Weyl}~, 
\eea
where
\bea
\cL_{\rm Weyl} &=& -\Big( v_{{a}} \cH^{{a}}_{\rm Weyl} + W F_{\rm Weyl} + X_{i j} H^{i j}_{\rm Weyl} 
- 2 \l_{\a}^k \varphi^{\a}_{k \, \rm Weyl} \non\\
    &&
    - \psi^{\a}_{{a} i}( \Gamma^{{a}})_{\a}{}^{\b} \varphi_{\b\, \rm Weyl}^i W - \ri \psi^{\a}_{{a} i} (\Gamma^{{a}})_{\a}{}^{\b} \l_{\b j} H^{i j}_{\rm Weyl} +  \ri \psi^{\a}_{{a} i} (\Sigma^{{a} {b}})_{\a }{}^{\b} \psi_{\b {b} j} W H^{i j}_{\rm Weyl}
   \Big)~.~~~~~~~~
   \label{S-Weyl-b}
\eea
\end{subequations}
Note that the fields of the linear multiplet are composed in terms of the super Weyl tensor and its descendants. Upon plugging the composite expressions \eqref{desc-H-Weyl-full} into \eqref{S-Weyl-b} and disregarding total derivatives, the bosonic part of the invariant reads
\bea
    \cL_{\rm Weyl} &=&  -\frac{1}{8}\varepsilon^{{a} {b} {c} {d} {e}}v_{{a}}C_{{b} {c}}\,^{{e} {f}} C_{{d} {e} {e} {f} } + \frac{1}{6}\varepsilon^{{a} {b} {c} {d} {e}} v_{{a}} \Phi_{{b} {c}}\,^{i j} \Phi_{{d} {e} i j}  
    - 4F_{a b} W^{{a} {b}} {D}   - \frac{3}{4} F_{a b} C^{{a}{b}{c}{d}} {W_{{c} {d}}} 
     \non \\
    && 
   - \frac{3}{32} F_{a b}  W^{{a} {b}}W^{{c} {d}} {W_{{c} {d}}}  + \frac{9}{8} F_{a b} W^{{a} {c}} {W_{{c} {d}}} W^{{d} {b}} 
    + \frac{9}{8} \varepsilon^{{a} {b} {c} {d} {e}} F_{a b}  {W_{{c} f}} {\nabla}_{{d}}{W_{{e}}{}^{f}}
    \non \\
    && 
    +\frac{3}{4}\varepsilon^{{a} {b} {c} {d} {e}}  F_{a b}  {W_{{c} f}} {\nabla}^{f}{W_{{d}{e}}}  - \frac{9}{16}W \varepsilon^{{a} {b} {c} {d} {e}} W_{{a} {b}} W_{{c} {d}} {\nabla}^{{f}}{W_{{e} {f}}}
    - \frac{128}{3} W {D}^{2}  - 8 W W^{{a} {b}} W_{{a} {b}} D  
    \non\\
    &&
  + \frac{2}{3} W \Phi^{{a} {b} i j} \Phi_{{a} {b} i j} -  \frac{1}{4} W C^{{a} {b} {c} {d}} C_{{a} {b} {c} {d}} 
    - \frac{3}{8} W C^{{a} {b} {c} {d}} W_{{a} {b}} W_{{c} {d}} - \frac{81}{32} W W^{{a} {b}} W_{{b} {c}} W^{{c} {d}} W_{{d} {a}}  
    \non \\
    && 
  + \frac{33}{128}  W W^{{a} {b}} W_{{a} {b}} W^{{c} {d}} W_{{c} {d}} 
    + 3 W W_{{c} {d}} {\nabla}^{{c}}{{\nabla}_{{b}}{W^{{b} {d}}}}-\frac{3}{2} W {\nabla}^{{a}}{W^{{c} {e}}} {\nabla}_{{c}}{W_{{a} {e}}}
      \non\\
    && 
  + \frac{3}{2} W {\nabla}^{{a}}{W^{{c} {e}}} {\nabla}_{{a}}{W_{{c} {e}}}  + X_{i j} W^{a b} \Phi_{a b}{}^{ i j}~.~~~~~~
\label{Weyl-action-full}
\eea
 In obtaining \eqref{Weyl-action-full}, we have performed integration by parts on $v_a$ dependent terms. It is useful to look at these terms more closely. Substituting \eqref{desc-H-Weyl-full-d} and turning off the fermions, we get
\bea
\int \rd^5 x \, e \, v_{{a}} \cH^{{a}}_{\rm Weyl} 
&=&
\int \rd^5 x \, e \, v_{{a}} \bigg\{ 
\frac{1}{8}\varepsilon^{{a} {b} {c} {d} {e}}C_{{b} {c}}\,^{f g} C_{{d} {e} f g }
-\frac{1}{6}\varepsilon^{{a} {b} {c} {d} {e}} \Phi_{{b} {c}}\,^{i j} \Phi_{{d} {e} i j}  + \de_{b} \tilde{\cH}^{a b}_{\rm Weyl} \bigg\}~,~~~~~~~~
\eea
with 
\bea
    \tilde{\cH}^{a b}_{\rm Weyl} 
    &=&
    8 W^{{a} {b}} D   + \frac{3}{2} C^{{a}{b}{c}{d}} {W_{{c} {d}}}+ \frac{3}{16}  W^{{a} {b}}W^{{c} {d}} {W_{{c} {d}}}  - \frac{9}{4} W^{{a} {c}} {W_{{c} {d}}} W^{{d} {b}}  
    \non\\
    &&
    -\frac{9}{4}\varepsilon^{{a} {b} {c} {d} {e}} {W_{{c} f}} {\nabla}_{{d}}{W_{{e}}{}^{f}} - \frac{3}{2}\varepsilon^{{a} {b} {c} {d} {e}}  {W_{{c} f}} {\nabla}^{f}{W_{{d}{e}}}~.
\eea
Writing the superconformally covariant vector derivative $\de_{a}$ in terms of the degauged covariant derivative $\cD_a$ gives
\bea
\de_{b} \tilde{\cH}^{a b}_{\rm Weyl} &=& \cD_{b} \tilde{\cH}^{a b}_{\rm Weyl}- \mathfrak{f}_{b}{}^c K_{c} \tilde{\cH}^{a b}_{\rm Weyl} + \text{gravitini terms} ~.
\eea
One can explicitly check that $K_{c} \tilde{\cH}^{ab}_{\rm Weyl} = 0$ and $\de_{b} \tilde{\cH}^{a b}_{\rm Weyl} = \cD_{b} \tilde{\cH}^{a b}_{\rm Weyl} + {\rm fermions}$. Thus, 
\bea \label{Weyl-action-cov}
\int \rd^5 x \, e \, v_{{a}} \cH^{{a}}_{\rm Weyl} 
&=& 
\int \rd^5 x \, e \,  \bigg\{ 
\frac{1}{8}\varepsilon^{{a} {b} {c} {d} {e}} v_a C_{{b} {c}}\,^{f g} C_{{d} {e} f g }
-\frac{1}{6}\varepsilon^{{a} {b} {c} {d} {e}} v_a \Phi_{{b} {c}}\,^{i j} \Phi_{{d} {e} i j}  + \hf F_{a b} \tilde{\cH}^{a b}_{\rm Weyl} \bigg\}~.~~~~~~~~
\eea

Finally, in order to extract the $K$-connections, which encode the Ricci tensor contributions, we degauge the action \eqref{Weyl-action-cov}. Making use of \eqref{nabla-on-W-and-Chi} and \eqref{degauged-vec-Weyl}, and setting $W=1$, the bosonic part of the Weyl-squared Lagrangian reads
\bea
    \cL_{\rm Weyl} &=&  -\frac{1}{8}\ve^{{a} {b} {c} {d} {e}}v_{{a}} \cR_{{b} {c}}\,^{ {e} {f}} \cR_{{d} {e}  {e} {f}} + \frac{1}{6}\ve^{{a} {b} {c} {d} {e}} v_{{a}} \Phi_{{b} {c}}\,^{i j} \Phi_{{d} {e} i j}  
    \non \\
    && 
    - 4F_{a b} W^{{a} {b}} {D} - \frac{3}{4} F_{a b} \cR^{{a}{b}{c}{d}} {W_{{c} {d}}} - \frac{1}{8} \cR F_{a b} W^{a b} -  F_{a b} \cR^{[a}{}_{c} W^{b] c} - \frac{3}{32}  F_{a b} W^{{a} {b}}W^{{c} {d}} {W_{{c} {d}}} 
    \non \\
    && 
    + \frac{9}{8} F_{a b} W^{{a} {c}} {W_{{c} {d}}} W^{{d} {b}}  +\frac{9}{8}\ve^{{a} {b} {c} {d} {e}} F_{a b} {W_{{c} f}} \cD_{{d}}{W_{{e}}{}^{f}}+ \frac{3}{4}\ve^{{a} {b} {c} {d} {e}}  F_{a b} {W_{{c} f}} \cD^{f}{W_{{d}{e}}} -\frac{128}{3}{D}^{2} 
    \non \\
    &&   
    - 8 W^{{a} {b}} W_{{a} {b}} D  +\frac{2}{3}\Phi^{{a} {b} i j} \Phi_{{a} {b} i j} -  \frac{1}{4} \cR^{{a} {b} {c} {d}} \cR_{{a} {b} {c} {d}} + \frac{1}{3} \cR_{a b} \cR^{a b} - \frac{1}{24} \cR^2 - \frac{3}{8} \cR^{{a} {b} {c} {d}} W_{{a} {b}} W_{{c} {d}} 
    \non \\
    && 
    - \frac{81}{32} W^{{a} {b}} W_{{b} {c}} W^{{c} {d}} W_{{d} {a}}  +3 W_{{c} {d}} \cD^{{c}}{\cD_{{b}}{W^{{b} {d}}}} - \frac{3}{2} \cR_{a b} W^{b d} W^{a}{}_{d} + \frac{3}{16} \cR W^{a b} W_{a b} 
    \non \\
    && 
    + \frac{33}{128}  W^{{a} {b}} W_{{a} {b}} W^{{c} {d}} W_{{c} {d}} - \frac{3}{2} \cD^{{a}}{W^{{c} {e}}} \cD_{{c}}{W_{{a} {e}}} + \frac{3}{2} \cD^{{a}}{W^{{c} {e}}} \cD_{{a}}{W_{{c} {e}}} 
    \non \\
    && 
    - \frac{9}{16}\ve^{{a} {b} {c} {d} {e}} W_{{a} {b}} W_{{c} {d}} \cD^{{f}}{W_{{e} {f}}} + X_{i j} W^{a b} \Phi_{a b}{}^{ i j}~. \label{Weyl-action-degauged}
\eea
Note that we have also used
\bea
C_{abcd} &=&  
\cR_{{a} {b} {c} {d}} + \frac{1}{6} \eta_{{c} [ {a}} \eta_{{b} ] {d}} \cR - \frac{2}{3} \eta_{{c} [ {a}} \cR_{{b} ] {d}} + \frac{2}{3} \eta_{{d} [ {a}} \cR_{{b} ] {c}}~.
\eea
The action \eqref{Weyl-action-degauged}, up to a change of notations, coincides with the result of \cite{HOT}.

\subsection{Log}
\label{sc-logW-action}

Making use of the BF action principle, eqs.~\eqref{SGV} and \eqref{BF-Scomp}, the primary superfield $H^{ij}_{\log }$ can be used to define the following locally superconformal invariant in a standard Weyl multiplet background
\begin{subequations} 
\label{S-log-0}
\bea
S_{\log} &=& \int\rd^{5|8}z\, E\,   V_{ij} { H }_{{\log}}^{ij} \label{S-log} \\
&=&  - \int \rd^5 x \, e \, \bigg( v_{{a}} \cH^{{a}}_{{\log}} + W F_{{\log}} + X_{i j} H^{i j}_{{\log}} + 2 \l^k \varphi_{k \, {\log}}  \non\\
    &&- \psi_{{a} i} \Gamma^{{a}} \varphi_{{\log}}^i W - \ri \psi_{{a} i} \Gamma^{{a}} \l_{j} H^{i j}_{{\log}} +  \ri \psi_{{a} i} \Sigma^{{a} {b}} \psi_{{b} j} W H^{i j}_{{\log}} \bigg)~. \label{S-log-b}
\eea
\end{subequations}

As can be seen from the expressions of the descendants of $H^{ij}_{\log }$ given in \eqref{desc-H-logW-full} and the supplementary file for fermionic counterpart, the component form of \eqref{S-log-b} is fairly involved. Let us now focus on the bosonic sector of \eqref{S-log-b}:
\be
   \cL_{\log} = - \Big( v_{a} \cH^{a}_{\log}  + W F_{\log} + X_{i j} H^{i j}_{\log} \Big).
\ee
The relevant term in the Lagrangian can then be integrated by parts:
\bea
   \cL_{\log} &=& - \frac{1}{12}v_{a}\ve^{{a} {b} {c} {d} {e}} \Phi_{{b} {c}}\,^{i j} \Phi_{{d} {e} i j} - v_{a}\cD_{b}{(\hat{H}^{a b}_{\log } + \tilde{H}^{a b}_{\log })}  - W F_{\log } - X_{i j} H^{i j}_{\log } \non\\
   &=& - \frac{1}{12}v_{a}\ve^{{a} {b} {c} {d} {e}} \Phi_{{b} {c}}\,^{i j} \Phi_{{d} {e} i j} - \frac{1}{2}F_{a b}{(\hat{H}^{a b}_{\log } + \tilde{H}^{a b}_{\log })}  - W F_{\log } - X_{i j} H^{i j}_{\log }.~~~~~
\eea
Inserting the composite expressions \eqref{desc-H-logW-full} and \eqref{HalogWdegauged} into the above action yields
\bsubeq\label{Log-action-0}
\bea
    \cL_{\log} &=& \cL_{\log}^{cov} + \cL_{\log}^{\prime}~,
\eea
where we have defined
\bea \label{Log-action}
     \cL_{\log}^{cov}
    &=& 
    W \Bigg\{\, \frac{3}{4} C_{{a} {b} {c} {d}} W^{{a} {b}} W^{{c} {d}} - \frac{8}{3} {D}^{2} -4 \Box{D} + \frac{23}{8} W^{{a} {b}} W_{{a} {b}} D + \frac{1}{6}\Phi^{{a} {b} i j} \Phi_{{a} {b}  i j}  
    \non \\
    && ~~~~~~
    - \frac{3}{2} W_{{c} {d}} {\nabla}^{{c}}{{\nabla}_{{a}}{W^{{a} {d}}}} - \frac{39}{16}  W_{{c} {d}} \Box{W^{{c} {d}}} + \frac{1005}{2048} W^{{a} {b}} W^{{c} {d}} W_{{a} {b}} W_{{c} {d}} 
    \non \\
    && ~~~~~~
    - \frac{9}{4}  {\nabla}^{{c}}{W_{{c} {d}}} {\nabla}_{{a}}{W^{{a} {d}}} + \frac{3}{4}  {\nabla}^{{c}}{W^{{a} {d}}} {\nabla}_{{a}}{W_{{c} {d}}} - \frac{33}{16}  {\nabla}^{{a}}{W^{{c} {d}}} {\nabla}_{{a}}{W_{{c} {d}}} 
    \non \\
    && ~~~~~~
    - \frac{9}{16}\ve^{{a} {b} {c} {d} {e}}  W_{{a} {b}} W_{{c} {d}} {\nabla}^{{f}}{W_{{e} {f}}} \Bigg\} 
    \non \\
    && 
    + \Bigg\{ - \frac{1}{12}\ve^{{a} {b} {c} {d} {e}} v_{{a}} \Phi_{{b} {c}}\,^{i j} \Phi_{{d} {e} i j} 
    - \frac{1}{2}  X^{i j} W_{{a} {b}} \Phi^{{a} {b}}{}_{i j} +  \frac{5}{4} C^{{a} {b} {c} {d}}  F_{{a} {b}} W_{{c} {d}} + 2 W^{{a} {b}} F_{{a} {b}} D
    \non \\
    && ~~~~~
     + \frac{3}{64} W^{{a} {b}} W^{{c} {d}} W_{{a} {b}} F_{{c} {d}} + \frac{9}{4} W^{{a} {b}} F^{{c} {d}} W_{{a} {c}} W_{{b} {d}}- \frac{9}{8}\ve^{{a} {b} {c} {d} {e}} W_{{a} {b}} F_{{c} {d}} {\nabla}^{{f}}{W_{{e} {f}}} 
    \non \\
    && ~~~~~
     - \frac{3}{8}\ve^{{a} {b} {c} {d} {e}} W_{{a} {f}} F_{{b} {c}} {\nabla}^{{f}}{W_{{d} {e}}} - \frac{9}{4} F_{{c} {d}} \Box{W^{{c} {d}}} +  \frac{3}{2} F_{{c} {d}} {\nabla}^{{c}}{{\nabla}_{{a}}{W^{{a} {d}}}} 
    \non \\
    && ~~~~~
    - 2  W_{{c} {d}} {\nabla}^{{c}}{{\nabla}_{{a}}{F^{{a} {d}}}} - \frac{1}{2} W_{{c} {d}} \Box{F^{{c} {d}}} + 4 D \Box{W}  + 4 {\nabla}^{{a}}{W} {\nabla}_{{a}}{D} 
    \non \\
    && ~~~~~
    + \frac{9}{2} W^{{a} {c}} W_{{c} {d}} {\nabla}_{{a}}{{\nabla}^{{d}}{W}} - \frac{9}{2} W_{{c} {d}} {\nabla}_{{a}}{W} {\nabla}^{{c}}{W^{{a} {d}}} + \frac{3}{16} W_{{c} {d}} {\nabla}^{{a}}{W} {\nabla}_{{a}}{W^{{c} {d}}} 
    \non \\
    && ~~~~~
    - \frac{9}{2} W_{{c} {d}} {\nabla}^{{c}}{W} {\nabla}_{{a}}{W^{{a} {d}}} + \frac{3}{32} W^{{c} {d}} W_{{c} {d}} \Box{W} - \frac{3}{2} {\nabla}^{{c}}{F_{{c} {d}}} {\nabla}_{{a}}{W^{{a} {d}}} 
    \non \\
    && ~~~~~
    - \frac{3}{2} {\nabla}^{{a}}{F^{{c} {d}}} {\nabla}_{{a}}{W_{{c} {d}}}  - \frac{9}{16}\ve^{{a} {b} {c} {d} {e} } {\de}^{{f}} \left( W_{{a} {b}} W_{{c} {d}} F_{{e} {f}} \right)  + \Box \Box {W}\Bigg\} 
    \non \\
    &&
    + W^{-1}\Bigg\{- \frac{1}{2} X^{i j} F_{{a} {b}} \Phi^{{a} {b}}{}_{i j} + \frac{1}{4}C^{{a} {b} {c} {d}} F_{{a} {b}} F_{{c} {d}} + 3 X^{i j} X_{i j} D - F^{{a} {b}} F_{{a} {b}} D 
    \non \\
    && ~~~~~~~~~
    + \frac{9}{128} X^{i j} X_{i j} W^{{a} {b}} W_{{a} {b}} - \frac{75}{128}W^{{a} {b}} W_{{a} {b}} F^{{c} {d}} F_{{c} {d}} + \frac{9}{4}W^{{a} {b}} W_{{a} {c}} F^{{c} {d}} F_{{b} {d}} 
    \non \\
    && ~~~~~~~~~
    - \frac{3}{8}\ve^{{a} {b} {c} {d} {e}} F_{{a} {b}} F_{{c} {f}}  {\nabla}^{{f}}{W_{{d} {e}}}- \frac{3}{8}\ve^{{a} {b} {c} {d} {e}} F_{{a} {b}} F_{{c} {d}}  {\nabla}^{{f}}{W_{{e} {f}}} - \frac{3}{4}\ve^{{a} {b} {c} {d} {e}} W_{{a} {b}} F_{{c} {d}}  {\nabla}^{{f}}{F_{{e} {f}}} 
    \non \\
    && ~~~~~~~~~
    - \frac{3}{8}\ve^{{a} {b} {c} {d} {e}} W_{{a} {f}} F_{{b} {c}} {\nabla}^{{f}}{F_{{d} {e}}} - \frac{3}{8}\ve^{{a} {b} {c} {d} {e}} W_{{a} {b}} F_{{c} {f}}  {\nabla}^{{f}}{F_{{d} {e}}} 
    \non \\
    && ~~~~~~~~~ + \frac{9}{16}\ve^{{a} {b} {c} {d} {e}} W_{{a} {b}} W_{{c} {d}} F_{{e} {f}}  {\nabla}^{{f}}{W}
    - \frac{3}{2} F_{{c} {d}}  {\nabla}_{{a}}{W^{{a} {c}}} {\nabla}^{{d}}{W} 
    \non \\
    && ~~~~~~~~~ 
    - 3 F_{{a} {c}}  {\nabla}^{{c}}{W^{{a}}{}_{ {d}}} {\nabla}^{{d}}{W} + \frac{3}{2} F_{{c} {d}}  {\nabla}^{{a}}{W} {\nabla}_{{a}}{W^{{c} {d}}}  + \frac{3}{2} W_{{c} {d}}  {\nabla}^{{a}}{W} {\nabla}_{{a}}{F^{{c} {d}}}
    \non \\
    && ~~~~~~~~~
    - 2 D  {\nabla}^{{a}}{W} {\nabla}_{{a}}{W}    - \frac{1}{4} {\nabla}^{{a}}{X^{i j}} {\nabla}_{{a}}{X_{i j}} + \frac{9}{4} W_{{a} {c}} W^{{a}}{}_{{d}}  {\nabla}^{{c}}{W} {\nabla}^{{d}}{W} 
    \non \\
    && ~~~~~~~~~ 
    - \frac{3}{64} W^{{c} {d}} W_{{c} {d}}  {\nabla}^{{a}}{W} {\nabla}_{{a}}{W} + \frac{1}{4} {\nabla}^{{a}}{F^{{c} {d}}} {\nabla}_{{a}}{F_{{c} {d}}} - 2  {\nabla}_{{a}}{W} {\nabla}^{{a}} \Box{W} 
    \non \\
    && ~~~~~~~~~ 
    - \Box{W} \Box{W}  - \frac{1}{2} {\nabla}_{{a}}{{\nabla}_{{b}}{W}} {\nabla}^{{a}}{{\nabla}^{{b}}{W}}  \Bigg\} \non \\
    && 
    + W^{-2}\Bigg\{\, 
     - \frac{3}{8}W^{{a} {b}} F^{{c} {d}} F_{{a} {b}} F_{{c} {d}} + \frac{3}{4}W_{{a} {b}} F^{{a} {c}} F^{{b} {d}} F_{{c} {d}} - \frac{1}{8}\ve^{{a} {b} {c} {d} {e}} F_{{a} {b}} F_{{c} {f}}  {\nabla}^{{f}}{F_{{d} {e}}} 
     \non \\
    && ~~~~~~~~~ - \frac{1}{8}\ve^{{a} {b} {c} {d} {e}} F_{{a} {b}} F_{{c} {d}} {\nabla}^{{f}}{F_{{e} {f}}} - \frac{3}{16}\ve^{{a} {b} {c} {d} {e}} W_{{a} {f}} F_{{b} {c}} F_{{d} {e}} {\nabla}^{{f}}{W} \non \\
    && ~~~~~~~~~
    + \frac{1}{2} F_{{a}}{}^{{b}} F_{{b} {c}}  {\nabla}^{{a}}{{\nabla}^{{c}}{W}} - \frac{1}{2} F_{{c} {d}}  {\nabla}^{{c}}{W} {\nabla}_{{a}}{F^{{a} {d}}}  - \frac{3}{4} W^{{c} {d}} F_{{c} {d}}  {\nabla}^{{a}}{W} {\nabla}_{{a}}{W}
    \non \\
    && ~~~~~~~~~
    - \frac{3}{4} F_{{c} {d}} {\nabla}^{{a}}{W} {\nabla}_{{a}}{F^{{c} {d}}} 
    + \frac{3}{2} {\nabla}^{{a}}{W} {\nabla}_{{a}}{W} \Box{W} + {\nabla}^{{a}}{W} {\nabla}^{{c}}{W} {\nabla}_{{a}}{{\nabla}_{{c}}{W}} 
    \non \\
    && ~~~~~~~~~
    + \frac{1}{4} X^{i j} X_{i j}  \Box{W} +  X^{i j} {\nabla}^{{a}}{W} {\nabla}_{{a}}{X_{i j}}\Bigg\}  
    \non \\
    &&
    + W^{-3}\Bigg\{ - \frac{1}{32}F^{{a} {b}} F^{{c} {d}} F_{{a} {b}} F_{{c} {d}}+ \frac{1}{8}F^{{a} {b}} F^{{c} {d}} F_{{a} {c}} F_{{b} {d}} - \frac{1}{16} X^{i j} X_{i j} F^{{a} {b}} F_{{a} {b}}    
    \non \\
    && ~~~~~~~~~
    + \frac{1}{32}   X^{i j} X_{i j} X^{k l} X_{k l} - \frac{1}{8}\ve^{{a} {b} {c} {d} {e}} F_{{a} {b}} F_{{c} {d}} F_{{e} {f}} {\nabla}^{{f}}{W} 
    \non \\
    && ~~~~~~~~~
    + \frac{1}{8} F^{{c} {d}} F_{{c} {d}}  {\nabla}^{{a}}{W} {\nabla}_{{a}}{W}  + \frac{3}{2} F_{{a} {c}} F^{{a}}{}_{{d}}  {\nabla}^{{c}}{W} {\nabla}^{{d}}{W}  
    \non \\
    && ~~~~~~~~~
    - \frac{7}{8} X^{i j} X_{i j}  {\nabla}^{{a}}{W} {\nabla}_{{a}}{W}  - \frac{3}{8} {\nabla}^{{a}}{W} {\nabla}_{{a}}{W} {\nabla}^{{c}}{W} {\nabla}_{{c}}{W}
    \Bigg\}~, \\
    \cL_{\log}^{\prime} &=& - \frac{5}{32} \cR W^{{a} {b}} F_{{a} {b}}+\frac{1}{2}\cR_{{a} {b}}  W^{a}{}_{ {d}} F^{{b} {d}} - \frac{5}{32} \cR F^{{a} {b}} F_{{a} {b}} {W}^{-1}+\frac{1}{2} \cR_{{a} {b}} F^{a}{}_{{d}} F^{{b} {d}}  {W}^{-1}~.
\eea
\esubeq

Here, $\cL_{\log}^{\prime}$ emerges from \eqref{non-covariant terms} and can be expressed in a superconformal covariant form as ${\nabla}^{{a}}X_{{a}}$ up to total derivative terms, i.e.,
\begin{align}
    \cL_{\log}^{\prime} = {\nabla}^{{a}}X_{{a}} - {\mathcal{D}^{{a}}}X_{{a}} =  - \mathfrak{f}^{{a} {b}} K_{{b}}X_{{a}}~, 
\end{align} 
where, inside the integral, ${\mathcal{D}^{{a}}}X_{{a}}$ vanishes. Therefore, we are seeking an $X_{{a}}$ such that $- \mathfrak{f}^{{a} {b}} K_{{b}}X_{{a}}$ coincides with $\cL_{\log}^{\prime}$, and consequently allows the action to be expressed as:
\bea
    \cL_{\log} = \cL_{\log}^{cov} + {\nabla}^{{a}}X_{{a}}~.
\eea
We proceed by writing the most general ansatz for $X_{{a}}$ that generates the desired curvature terms:
\bea 
    X_{{a}} 
    &=& 
    x_1 W_{{b} {c}}  {\nabla}_{{a}}{F^{{b} {c}}} + x_2 W_{{a} {c}}  {\nabla}_{{b}}{F^{{b} {c}}} + x_3 F_{{b} {c}}  {\nabla}_{{a}}{W^{{b} {c}}} + x_4 F_{{a} {c}}  {\nabla}_{{b}}{W^{{b} {c}}} + x_5 F^{{b} {c}}  {\nabla}_{{b}}{W_{{a} {c}}} 
    \non \\
    &&
    + x_6 {W}^{-1} {\nabla}_{{a}}{W} F_{{b} {c}}  {W^{{b} {c}}} + x_7 {W}^{-1}{\nabla}_{{b}}{W} F_{{a} {c}}  {W^{{b} {c}}} + x_8 {W}^{-1} {\nabla}_{{b}}{W} F^{{b} {c}} {W_{{a} {c}}}
    \non \\
    && 
    + x_9 {W}^{-1}{\nabla}_{{a}}(F_{{b} {c}} {F^{{b} {c}}})+x_{10}{W}^{-1} {\nabla}_{{b}}( F_{{a} {c}} {F^{{b} {c}}})~,
\eea
and demand that
\bea
    - \mathfrak{f}^{{a} {b}} K_{{b}}X_{{a}}
    &=& 
    - \frac{5}{32} \cR W^{{a} {b}} F_{{a} {b}} + \frac{1}{2} \cR_{{a} {b}} \eta^{{a} {c}} W_{{c} {d}} F^{{b} {d}} 
    \non \\
    && 
    - \frac{5}{32} \cR F^{{a} {b}} F_{{a} {b}} {W}^{-1}+\frac{1}{2} \cR_{{a} {b}} \eta^{{a} {c}} F^{{b} {d}} F_{{c} {d}} {W}^{-1}~,
    \non \\
    K^{a} \cL_{\log} &=& 0~.
\eea
The second constraint ensures the $K$-invariance of the Log action. Although these constraints result in five linearly independent equations, they are insufficient to uniquely fix $X_a$. One interesting choice of solution that we opted for in this case is
\bea
    X_{{a}}&=&  ~~ \frac{3}{16} {\nabla}_{{a}}(W_{{b} {c}} {F^{{b} {c}}})-\frac{3}{2} {\nabla}_{{b}}( F_{{a} {c}} {W^{{b} {c}}}) + \frac{3}{16}  {W}^{-1} ({\nabla}_{{a}}{W}) W_{{b} {c}} {F^{{b} {c}}} 
    \non \\
    && 
    - \frac{3}{2}  {W}^{-1} ({\nabla}_{{b}}{W}) F_{{a} {c}} {W^{{b} {c}}} + \frac{3}{16} {W}^{-1} {\nabla}_{{a}}(F_{{b} {c}} {F^{{b} {c}}}) - \frac{3}{2} {W}^{-1} {\nabla}_{{b}}(  F_{{a} {c}} {F^{{b} {c}}})~,
\eea
which then leads to the final expression
\bea \label{log-action-covariant}
   && \cL_{\log}  
    = \cL_{\log}^{cov} 
    +\de^{{a}}\Bigg\{~\frac{3}{16} {\nabla}_{{a}}(W_{{b} {c}} {F^{{b} {c}}})-\frac{3}{2} {\nabla}_{{b}}( F_{{a} {c}} {W^{{b} {c}}}) + \frac{3}{16}  {W}^{-1} ({\nabla}_{{a}}{W}) W_{{b} {c}} {F^{{b} {c}}} 
    \non \\
    && ~~~~~~~~~~~~
    -\frac{3}{2}  {W}^{-1} ({\nabla}_{{b}}{W}) F_{{a} {c}} {W^{{b} {c}}}
    +\frac{3}{16} {W}^{-1} {\nabla}_{{a}}(F_{{b} {c}} {F^{{b} {c}}})-\frac{3}{2} {W}^{-1} {\nabla}_{{b}}(  F_{{a} {c}} {F^{{b} {c}}}) 
    \Bigg\}
    ~.~~~~~~~~~
\eea
Note that the resulting component action \eqref{log-action-covariant} includes, for example, a $(\Box\Box W)$ term, which upon using the degauge identity \eqref{degauged-vec-Weyl}, gives rise to a Ricci tensor squared combination. 


Unlike the Weyl-squared case, manually degauging the Log action \eqref{log-action-covariant} proves to be technically involved, primarily due to its  considerable length. Consequently, we made use of \textit{Cadabra} to carry out the degauging procedure and then applied various simplification techniques to refine the initial expression. These include integration by parts as well as the utilisation of degauged commutator algebra \eqref{degauged-vec-algebra}, symmetries of the Riemann tensor, and degauged Bianchi identities $\cD_{[a}F_{b c]}=0~, \cD_{[a}\Phi_{b c]}^{ij}=0$ (up to fermions).
Furthermore, the following identity is also useful for simplification at final stages
\bea
\ve^{{a} {b} {c} {d} {e}} {W_{{a} {b}}} W_{{c} f} \mathcal{D}^{f}F_{{d} {e}} =  - \frac{1}{2}\ve^{{a} {b} {c} {d} {e}} W_{{a} {b}} W_{{c} {d}} \mathcal{D}^{f}{F_{{e} f}}~.
\eea
Setting $W=1$, the bosonic sector of the Log invariant reads
\bea
    \cL_{\log} 
    &=& 
    -\frac{1}{12} \ve^{{a} {b} {c} {d} {e}} v_{{a}} \Phi_{{b} {c}}\,^{i j} \Phi_{{d} {e} i j} 
    +\frac{1}{4} \cR_{{a} {b} {c} {d}}\big(  F^{{a} {b}} F^{{c} {d}}  + W^{{a} {b}} W^{{c} {d}} \big) 
    + \frac{1}{4} \cR_{{a} {b}}\big( 4 W^{{a} {c}} F^{{b}}{}_{{c}} - 3 W^{{a} {c}} W^{{b}}{}_{{c}}\big)
    \non \\
    &&
    - \frac{1}{256} \cR \big( 24 F^{{a} {b}} F_{{a} {b}}  + 80 W^{{a} {b}} F_{{a} {b}}+ 27 W^{{a} {b}} W_{{a} {b}} + 128 D -8  X^{i j} X_{i j}\big)
    \non\\
    && 
    - \frac{1}{6} \cR_{{a} {b}} \cR^{{a} {b}} + \frac{23}{384}{\cR}^{2} - \frac{8}{3}{D}^{2} - \frac{1}{2}  X^{i j} \big(W^{{a} {b}} \Phi_{{a} {b} i j} + F^{{a} {b}} \Phi_{{a} {b} i j}\big) 
    \non \\
    &&
    + \frac{1}{128} X^{i j} X_{i j}\big( 9  W^{{a} {b}} W_{{a} {b}}-  8 F^{{a} {b}} F_{{a} {b}} + 4  X^{k l} X_{k l}\big)
    \non \\
    &&
    + \frac{1}{8} D \big(16 W^{{a} {b}} F_{{a} {b}} - 8 F^{{a} {b}} F_{{a} {b}} + 23 W^{{a} {b}} W_{{a} {b}} + 24 X^{i j} X_{i j} \big)
    \non \\
    &&
    + W^{{a} {b}}F^{{c} {d}}\bigg(\frac{3}{64}  W_{{a} {b}}W_{{c} {d}} +\frac{9}{4} W_{{a} {c}} W_{{b} {d}}  - \frac{75}{128} W_{{a} {b}}  F_{{c} {d}}+\frac{9}{4} W_{{a} {c}}  F_{{b} {d}} 
    - \frac{3}{8} F_{{a} {b}} F_{{c} {d}}+\frac{3}{4} F_{{a} {c}} F_{{b} {d}} \bigg) 
    \non \\
    &&
    -\frac{1}{32} F^{{a} {b}} F^{{c} {d}} \big(  F_{{a} {b}} F_{{c} {d}}- 4  F_{{a} {c}} F_{{b} {d}} \big) +\frac{1005}{2048}W^{{a} {b}} W^{{c} {d}} W_{{a} {b}} W_{{c} {d}}
    + \frac{1}{6}\Phi_{{a} {b}}\,^{i j} \Phi^{{a} {b}}{}_{i j}
    \non \\
    && 
    - \frac{3}{16}\ve^{{a} {b} {c} {d} {e}}(\mathcal{D}^{f}{W_{{a} f}}) \big( 4 {W_{{b} {c}}}  F_{{d} {e}} +  F_{{b} {c}} F_{{d} {e}} + 3 W_{{b} {c}} W_{{d} {e}} \big) \non \\
    &&
    - \frac{1}{16}\ve^{{a} {b} {c} {d} {e}} (\mathcal{D}^{f}{F_{{a} f}} )\big( 6 W_{{b} {c}}  F_{{d} {e}}+  F_{{b} {c}} F_{{d} {e}}  + 3 W_{{b} {c}} W_{{d} {e}}\big)  \non \\
    &&
    - \frac{3}{8} W^{{c} {d}} \mathcal{D}^{{a}}{\mathcal{D}_{{a}}{W_{{c} {d}}}} 
     -  \frac{3}{2}F^{{a}{c}}  \mathcal{D}_{{c}}{\mathcal{D}^{{d}}{W_{{a} {d}}}} 
    - \frac{1}{4} (\mathcal{D}^{{a}}{X^{i j}}) \mathcal{D}_{{a}}{X_{i j}} 
    + \frac{1}{4} (\mathcal{D}^{{a}}{F^{{b} {c}}} )\mathcal{D}_{{a}}{F_{{b} {c}}}~.\label{log-action-degauged}
\eea
Note that the degauged action \eqref{log-action-degauged} contains both the supersymmetric Ricci tensor and Ricci scalar terms.

\subsection{Scalar curvature squared}\label{Scalar-curvature-squared}
We now construct an action describing the supersymmetric completion of the Ricci scalar squared in the standard Weyl multiplet. It is obtained by plugging the composite linear multiplet \eqref{Hij-R2-desc} into the following BF action
\bsubeq
\bea
    S_{R^2} &=& \int\rd^{5|8}z\, E\,   V_{ij} { H }_{R^2}^{ij} \label{action-R2-a} \\
    &=& - \int \rd^5 x \, e \Big( \,v_{{a}} \cH^{{a}}_{R^2} + W F_{R^2} + X_{i j} H^{i j}_{R^2} - 2 \l^{\a}_{i} \vf_{\a \, R^2}^i 
    \non\\
    &&- \psi_{{a }}{}^{\a}_i(\Gamma^{{a}})_{\a}{}^{\b} \vf_{\b \, R^2}^i  W - \ri \psi_{{a}}{}^{\a}_{i} (\Gamma^{{a}})_{\a}{}^{\b} \l_{\b j} H^{i j}_{R^2} +  \ri \psi_{{a}}{}^{\a}_{i} (\Sigma^{{a} {b}}){}_{\a}{}^{\b} \psi_{{b} \b j} W H^{i j}_{R^2} \Big)~.~~~~
    \label{action-R2}
\eea
\esubeq
The bosonic part of such an action reads
\bsubeq \label{action-R2-bos}
\bea
S_{R^2} 
&=&  - \int \rd^5 x \, e \, \bigg(  - \frac{1}{2}  \ve_{{a} {b} {c} {d} {e}} v^{{a}} \mathbf{F}^{{b} {c}} \mathbf{F}^{{d} {e}} - v^{{a}} \de^{{b}}\left( 4\mathbf{W} \mathbf{F}_{{a} {b}} + 6 W_{{a} {b}} \mathbf{W}^2 \right) \non\\
    && +  W \mathbf{X}^{i j} \mathbf{X}_{i j} - W \mathbf{F}^{{a} {b}} \mathbf{F}_{{a} {b}} + 4 W \mathbf{W} \de^{{a}}\de_{{a}}\mathbf{W} + 2 W (\de^{{a}}\mathbf{W})(\de_{{a}}\mathbf{W}) 
  \non\\
 &&
 - 6 W W^{{a} {b}}\mathbf{F}_{{a} {b}} \mathbf{W} - \frac{39}{8} W W^{{a} {b}}W_{{a} {b}}\mathbf{W}^2 - 16 W D \mathbf{W}^2 + 2 X_{i j}  \mathbf{W} \mathbf{X}^{i j} \bigg)~,
\eea
which, upon integrating by parts, gives
\bea
    S_{R^2} 
    &=& 
    \int \rd^5 x \, e \, \bigg(   \frac{1}{2}  \ve_{{a} {b} {c} {d} {e}} v^{{a}} \mathbf{F}^{{b} {c}} \mathbf{F}^{{d} {e}} +F^{{a} {b}} \left( 2\mathbf{W} \mathbf{F}_{{a} {b}} + 3 W_{{a} {b}} \mathbf{W}^2 \right) \non \\
    &&
    - W \mathbf{X}^{i j} \mathbf{X}_{i j} + W \mathbf{F}^{{a} {b}} \mathbf{F}_{{a} {b}} - 4 W \mathbf{W} \de^{{a}}\de_{{a}}\mathbf{W} - 2 W (\de^{{a}}\mathbf{W})\de_{{a}}\mathbf{W} 
    \non\\
    &&
    + 6 W W^{{a} {b}}\mathbf{F}_{{a} {b}} \mathbf{W} +  \frac{39}{8}W W^{{a} {b}}W_{{a} {b}}\mathbf{W}^2 + 16 D W \mathbf{W}^2 - 2 X_{i j}  \mathbf{W} \mathbf{X}^{i j} \bigg)~. \label{action-R2-b}
\eea
\esubeq
Here we note that the composite expressions for the vector multiplet are given explicitly in \eqref{comp-W} and \eqref{desc-W-R2}.  After degauging and omitting the fermionic terms, the composite vector multiplet now takes the form
\bsubeq \label{desc-W-R2_000}
\bea
\mathbf{W} &=& 
\frac{1}{4}F G^{-1} ~,
 \\
\mathbf{X}^{ij} 
&=& ~~ G^{-1} \bigg\{\, \hf \cD^a \cD_a G^{ij} + \frac{3}{16} \cR  G^{ij} 
+ \frac{3}{64} W^{ab} W_{ab} G^{ij} +2 D G^{ij}
\bigg\}\non\\
&&+ G^{-3} \bigg\{ -\frac{1}{16} F^2 G^{ij} - \frac{1}{16} \cH^{a} \cH_{a} G^{ij} + \frac{1}{4} \cH^{a} G^{k(i} \cD_{a} G_k{}^{j)}
\non\\
&&~~~~~~~~~~
-\frac{1}{4} G_{kl} \big(\cD^{a}   G^{k(i}\big) \cD_{a} G^{j)l}
\bigg\}~,
\\
\mathbf{F}_{{a} {b}} 
&=& ~~ G^{-1} \Bigg\{ \hf \cD_{[a} \cH_{b]} -\frac{3\ri}{8} G_{ij}X_{ab}{}^{ij} 
\Bigg\} \non\\
&&+ G^{-3} \Bigg\{ \,\frac{1}{4} G_{ij} \cH_{[a} \cD_{b]} G^{ij}
-\frac{1}{4} G_{ij}  \big(\cD_{[a} G^{ik} \big) \cD_{b]} G_k{}^{j} 
\Bigg\}
~.
\eea
\esubeq
Inserting the above expressions into \eqref{action-R2-b} and imposing the gauge fixing condition $W=1$ leads to the supersymmetric extension of the Ricci scalar squared action \cite{OP2}.

Let us point out that while the component actions presented in this section do not include fermionic terms, we can, in principle, construct the complete actions for all the three invariants, using the composite expressions given in the supplementary file. In the remainder of this paper, we will require some of these fermionic terms to derive multiplet of EOMs.

\section{Derivation of superconformal equations of motion}\label{section-6}

The goal of this section is to obtain superconformal primary equations of motion in superspace that describe minimal 5D gauged supergravity based on a standard Weyl multiplet and deformed by an arbitrary combination of the three curvature-squared invariants described by the action  
\bea
S_{\rm HD}&=&  S_{\rm gSG} + \a \, S_{\rm Weyl} +\b  \, S_{{\log}} + \g \, S_{R^2}
~,
\label{HD-action}
\eea
where the three independent curvature-squared invariants are described by eqs. \eqref{S-Weyl-0}, \eqref{S-log-0}, and \eqref{action-R2}. 
The results in this section were first presented in \cite{GHKT-M} but with no derivation.

In principle, one can obtain these equations of motion by varying the superspace action \eqref{HD-action} with respect to the superfield prepotentials of the standard Weyl multiplet ($\mathfrak U$), the vector multiplet compensator ($V_{ij}$), and the linear multiplet compensator ($\Omega$). These variations lead to the supercurrent superfield $\cJ$, the linear multiplet of the EOM of $V_{ij}$, and the vector multiplet of the EOM of $\Omega$, respectively. The result would be manifestly covariant, as is well known from the very well developed results for theories with four supercharges, see \cite{SUPERSPACE,BK}. Alternatively, we can reduce \eqref{HD-action} to components and vary it with respect to the highest dimension independent component fields ($D$, $X_{i j}$, and $F$) of the corresponding multiplets. Because the prepotential formulation of the five-dimensional standard Weyl multiplet in superspace has not been developed to date, we will use mostly a component approach, in particular to obtain $\cJ$. We also stress that we are interested in obtaining the full equations of motions, including all the potential fermionic terms.

 It is worth pointing out that in superspace the equations of motion derived by varying prepotentials would be manifestly covariant. Hence, one expects the same to be true once the superspace results are reduced to component fields. However, in the component approach of finding the EOMs, the component action computed from \eqref{HD-action} includes thousands of terms when fermions are considered, and it is not manifestly covariant due to the presence of naked gravitini and Chern-Simons terms. Although the action lacks manifest covariance, it has recently been explicitly demonstrated in components that for any supergravity theory, there exist covariant equations of motion that are equivalent to the regular field equations \cite{Ferrara:2017yhz,Vanhecke:2017chr}. These covariant equations are obtained by covariantising the regular field equations, resulting in a multiplet of field equations \cite{Ferrara:2017yhz,Vanhecke:2017chr}. 

Adopting this approach, in this section we derive the covariant field equations for the Weyl multiplet, linear multiplet, and vector multiplet for the minimal 5D gauged supergravity deformed by an arbitrary combination of the three curvature-squared invariants described by the action $S_{\rm HD}$. The above argument suggests that to obtain these equations of motion, it is sufficient to initially vary the gravitini-independent part (or by setting gravitini to zero), in a degauged action, to derive the regular field equations. Hence, we begin with 
\bea\label{BF-action-including-fermion}
    S_{\rm BF} &=& - \int \rd^5 x \, e \Big( \,v_{{a}} \cH^{{a}} + W F + X_{i j} G^{i j} + 2 \l^{\a k} \vf_{\a k}  \Big)~,
\eea
which, by suitably choosing primary composite linear or vector multiplets (e.g., eqs. \eqref{desc-W-R2}, \eqref{desc-H-Weyl-full}, and \eqref{desc-H-logW-full}), becomes the building block in finding EOMs for various two- and four-derivative invariants. Subsequently, we vary the degauged action with respect to the highest dimension auxiliary fields $D$, $X_{i j}$, and $F$ to derive the regular field equations for Weyl, vector, and linear multiplets, respectively. To obtain covariant EOMs, we then appropriately replace the degauged derivative to the conformal covariant derivative and translate component fields into superfields. This uplifting process effectively eliminates the curvature terms present in the degauged EOMs. The resulting covariant EOMs then describe the primary fields, i.e., the top components of the multiplets of the equations of motion that arise from the variation of the full superfields. It is then straightforward to reinterpret them as the locally superconformal multiplet of field equations. Once this result is obtained, it is then possible to systematically produce the entire multiplet of equations of motion by the successive action of superspace spinor derivatives (equivalently $Q$-supersymmetry transformations) on the primary EOMs.
Once again, it is important to emphasise that in our analysis, we can carefully disregard the influence of gravitini in both the degauging process and the subsequent transformation of regular EOMs into superconformally covariant EOMs. We then run several consistency checks to ensure the results are correct.

\subsection{Two-derivative EOMs}
\label{EOM:EH}
The on-shell structure of the two-derivative gauged supergravity action, $S_{\rm gSG}$, has been discussed in \cite{BKNT-M14}. Recall that
\begin{subequations} 
\bea
\label{gauged_sugra}
S_{\rm gSG}&=&  S_{\rm VM} + S_{\rm L} +\k \, S_{\rm BF} =
 \int \rd^{5|8}z\, E\, \Big\{ 
\frac{1}{4} V_{ij} { H }_{\rm VM}^{ij} 
+\O \mathbf{ W } 
+\k V_{ij} G^{ij} \Big\} \\
&=&
 \int \rd^{5|8}z\, E\,  \Big\{ 
\frac{1}{4} V_{ij} { H }_{\rm VM}^{ij} 
+ \O \mathbf{ W}  
+\k \O W \Big\}~. 
\eea
\end{subequations}
Also, the BF action is
\bea
    S_{\rm BF} &=& - \int \rd^5 x \, e \Big( \,v_{{a}} \cH^{{a}} + W F + X_{i j} G^{i j} + 2 \l^{\a k} \vf_{\a k} 
    \non\\
    &&- \psi_{{a }}{}^{\a}_i(\Gamma^{{a}})_{\a}{}^{\b} \vf_{\b}^i  W - \ri \psi_{{a}}{}^{\a}_{i} (\Gamma^{{a}})_{\a}{}^{\b} \l_{\b j} G^{i j} +  \ri \psi_{{a}}{}^{\a}_{i} (\Sigma^{{a} {b}}){}_{\a}{}^{\b} \psi_{{b} \b j} W G^{i j} \Big)~.
\eea
Here we elaborate more on the derivation of the EOMs using the component form of $S_{\rm gSG}$. While only the bosonic part of the component actions \eqref{bosonic-swm}, \eqref{eq:TensorComp}, and \eqref{BF:cosmological} for the gauged supergravity \eqref{gauged_sugra} are provided, it is possible to derive the full expressions including fermionic terms, by directly examining the composite fields \eqref{O2composite-N} and \eqref{desc-W-R2} from which the action is derived. 
\vspace{0.3cm} \\
\textbf{\underline{Standard Weyl multiplet}}
\vspace{0.3cm} \\
The conformal supergravity equation of motion is obtained by varying the superspace form of $S_{\rm gSG}$ with respect to the standard Weyl multiplet prepotential superfield $\mathfrak U$ or equivalently the component form of $S_{\rm gSG}$ with respect to the auxiliary field $D$. To derive the equation of motion for the Weyl multiplet from the component action, the $D$-dependencies arise from the composite fields \eqref{composite_FVMfermionic} and \eqref{compositeXijlinear}. Consequently, the equation of motion can be expressed as follows (note that no fermions are involved):
\bea
0 = J_{EH} &=&  -\frac{1}{4} W \frac{\delta F_{\rm VM}}{\delta D} - \frac{\delta \mathbf{X}^{ij} }{\delta D} G_{ij}
~,
\non \\
0 = J_{EH} &=& 4 (W^3-G)~.
\eea
Note that here and in what follows, we have neglected the gravitini-dependent terms throughout because of the argument given in the beginning of this section. It can be proven explicitly that their contribution either cancels out or adds up to make some of the fields in the equations of motion manifestly superconformally covariant.
\vspace{0.3cm} \\
\textbf{\underline{Vector multiplet}} 
\vspace{0.3cm} \\
For the two-derivative gauged supergravity action \eqref{rep8.13a}, the EOM for the vector multiplet was given in \cite{BKNT-M14}. In superspace, it was obtained by varying $S_{\rm gSG}$ with respect to the Mezincescu's superfield prepotential $V_{ij}$.  An arbitrary variation of $V_{ij}$ leads to
\bea
\d S_{\rm gSG}&=& \d S_{\rm VM} + \k \,\d S_{\rm BF}
\non\\
&=& \frac{1}{4} \int\rd^{5|8}z\, E\,   \Big[  \d V_{ij} { H }_{\rm VM}^{ij} + V_{ij} \d { H }_{\rm VM}^{ij} \Big] + \k \int\rd^{5|8}z\, E\,    \d V_{ij} G^{ij}~.~~~
\eea
Making use of the relation \eqref{var-SVM}, and various integration by parts of spinor covariant derivatives,\footnote{See comments in \cite{BKNT-M14}, and also the recent reviews \cite{Kuzenko:2022skv,Kuzenko:2022ajd}, for subtleties in performing (5D) conformal superspace integration by parts.} the total variation of  $S_{\rm gSG}$ induced by an arbitrary variation of the vector multiplet prepotential $V_{ij}$ is given by
\bea
\d S_{\rm gSG}
&=&
\frac{3}{4}\int \rd^{5|8}z\, E\, \d V_{ij}
{ H }_{\rm VM}^{ij} + \k \int\rd^{5|8}z\, E\,    \d V_{ij} G^{ij}~,
\eea
and, hence, 
\bea
\frac{\d S_{\rm gSG}}{\d V_{ij}}=0 \qquad \Longrightarrow \qquad \frac{3}{4} H^{ij}_{\rm VM} + \k G^{ij} = 0~.
\label{2deriv-VM}
\eea

To obtain the EOM for the vector multiplet from the component action, we begin by taking the variation of the component gauged supergravity action with respect to the auxiliary field $X^{ij}$. Upon examining the composite fields \eqref{O2composite-N} and \eqref{desc-W-R2} for their dependence on $X^{ij}$, the resulting EOM can be expressed as follows:
\bea
0 &=& - \frac{1}{4}W \frac{\d F_{\rm VM}}{\d X^{ij}} -   \frac{1}{4} H^{i j}_{\rm VM} -  \frac{1}{4} X_{i j} \frac{\d H^{i j}_{\rm VM}}{\d X^{i j}}-  \frac{1}{2} \l^{\a k} \frac{\d \vf_{\a k}{}_{\rm VM}}{\d X^{i j}} - \k G^{ij} 
~,
\non \\
\implies 0 &=&-  \frac{3}{4} H^{i j}_{\rm VM} - \k G^{ij} ~,\label{2deriv-comp-VM}
\eea
where we have neglected the gravitini-dependent terms in our analysis. This, as expected, corresponds to the lowest component of the superfield EOM \eqref{2deriv-VM}. Note the relative sign difference between the superspace equation \eqref{2deriv-VM} and the component equation \eqref{2deriv-comp-VM}. This relative sign arises when considering the variation of the action with respect to either the superfield prepotential or its corresponding auxiliary field. A similar relative sign difference between the superspace and component EOMs will also exist for other EOMs.
\vspace{0.3cm}\\
\textbf{\underline{Linear multiplet}} 
\vspace{0.3cm} \\
Lastly, we consider the EOM for the linear multiplet compensator. In the superspace approach, this can be done by varying with respect to the superfield prepotential $\Omega$; hence, at the two-derivative level, from \eqref{rep8.13b} we find that
\bea\label{liner-2der-EOM-superspace}
\mathbf{W} + \k W = 0~.
\eea
From the component action side, the linear multiplet EOM is obtained by varying the action with respect to auxiliary field $F$. As no composite fields in \eqref{O2composite-N} and \eqref{desc-W-R2} have dependence on $F$, the EOM is given by:
\bea
    0 &= - \mathbf{W} - \k W~,
\eea
where we have neglected the gravitini dependent terms in our analysis for the EOM.

\subsection{Four-derivative: Standard Weyl multiplet EOMs}\label{JEOM:StandardWeyl}

We now proceed in examining the $D$-dependent terms in each curvature-squared invariant.

\subsubsection{Weyl-squared}
\label{JEOM:Weyl2}
As can be seen from the explicit expressions \eqref{desc-H-Weyl-full}, each of the descendant components of the composite $H^{ij}_{\rm Weyl}$ carry $D$-dependence. Suppressing the explicit gravitini terms, the $D$-dependent terms of the Weyl-squared Lagrangian \eqref{S-Weyl-b} are
\bsubeq  \label{LWeyl-eom-1}
\bea
    \mathcal{L}_{{\rm Weyl} {\rm , D}} &=& -\Big( v_{a} \cH_{{\rm Weyl}}^{a} + W F_{{\rm Weyl}} - 2 \l_{i}^{\a} \varphi^{i}_{\a \, \rm Weyl}  \Big) 
    \\
    &=&  -8 v_{a} \cD_{b} (D W^{ab} ) - W \Big( \frac{128}{3} D^2 + 8 W^{a b}W_{a b} D \Big) + 2 \l_{i}^{\a} \Big(-\frac{128}{3} \ri D \chi^{i}_{\a} \Big)~~~~~~~~\\
    &=& - \frac{128}{3} W D^2 -8 W W^{a b}W_{a b} D -8 v_a  {\cD}_{b}\left(D W^{a b} \right)  - \frac{256}{3} \ri \l_{i}^{\a} \chi^{i}_{\a} D~. \label{L-Weyl-D}
\eea
\esubeq
Upon integrating by parts, an arbitrary variation of $D$ leads to the following EOM
\bea\label{Weyl-D-EOM}
\frac{\pa}{\pa D} \cL_{\rm Weyl, D} = -\frac{256}{3} W D - 8 W W^{ab} W_{a b}-4 F^{ab} W_{ab} + \frac{256}{3} \ri \l^{\a i} \chi_{\a i} = 0~.
\eea
We can uplift the above EOM to superspace \cite{Ferrara:2017yhz,Vanhecke:2017chr}, by promoting every component field to superfield. This amounts to replacing $D \rightarrow -\frac{3}{128} Y$ and $\chi_{\a i} \rightarrow \frac{3}{32} \ri X_{\a i}$ while all other fields should now be understood as superfields. The covariant superconformal primary EOM is then
\bea  
0= J_{{\rm Weyl}} 
=
- \frac{3}{64} W Y + \frac{3}{16} W W^{{a} {b}}W_{{a} {b}}  + \frac{3}{32}  F_{{a} {b}} W^{{a} {b}}   - \frac{3}{16} \l_{i}^{\a} X^{i}_{\a}  
  ~.
\eea
Note the extra factor of $-\frac{3}{128}$ in the above equation. This arises when we replace $D$ with $Y$ on the left-hand side of \eqref{Weyl-D-EOM}.

\subsubsection{Log} \label{JEOM:logW}
Next, we examine the Log invariant and explicitly write out the $D$-dependent contributions. Suppressing the gravitini terms, upon substituting the composite fields \eqref{desc-H-logW-full} (of which their explicit fermionic dependence given in subsection 2.2 of the supplementary file)  into the BF Lagrangian \eqref{BF-action-including-fermion}, the relevant $D$-dependent terms are given by
\bea
   \mathcal{L}_{\log, \, D} &=& -v_{{a}} \cH_{\log}^{{a}} - {W} F_{\log} - X_{i j} H^{i j}_{\log} + 2 \l_{i}^{\a} \varphi^{i}_{\a\,\log}
   ~,\\
    &=& 2 v_{{a}}{\cD}_{{b}}\Big( 2 W^{{a} {b}} D -  2 W^{-1}  F^{{a} {b}} D 
    +  \ri W^{-2} (\Sigma^{{a} {b}})^{\a \b}   \l^{i}_{\a} \l_{\b i} D\Big) 
    \non \\
    &&-W \bigg\{\frac{8}{3} D^2 -  \frac{23}{8} W^{{a} {b}} W_{{a} {b}} D - W^{-2} F^{{a} {b}} F_{{a} {b}} D   +3 W^{-2}  X^{i j} X_{i j} D 
    \non \\
    &&
    -  4 \ri W^{-2} (\Gamma^{{a}})^{\a \b} \l^{i}_{\a}  ({\cD}_{{a}}\l_{\b i}) D -4 W^{-1} D  {\cD}^{{a}} {\cD}_{{a}} W +2 W^{-2} D  ({\cD}^{{a}}W) {\cD}_{{a}} W 
    \non \\
    &&
    +2  \ri W^{-3}  (\Sigma^{{a} {b}})^{\a \b}  F_{{a} {b}} \l^{i}_{\a} \l_{\b i} D 
    +4 \ri  W^{-3} X^{i j} \l^{\a}_{i} \l_{\a j} D
    \non \\
    &&
    - \frac{3}{2} W^{-4} \l^{\a i} \l_{\a}^j \l^{\b}_{i} \l_{\b j} D -4 W^{-1} ({\cD}_{{a}} W) {\cD}^{{a}}D +4  {\cD}^{{a}}{\cD}_{{a}}D + \hf \cR D
\bigg\} 
    \non \\
    &&- X_{i j}\bigg\{-6 W^{-1} X^{i j} D  - 2 \ri  W^{-2} \l^{\a i} \l^{j}_{\a} D\bigg\} 
    \non \\
    &&
    + 2 \l^{\a}_{j}\bigg\{- \frac{8}{3} \ri \chi^{j}_{\a} D + 4 \ri (\Gamma_{ {b} })_{\a \b} W^{-1}   ({\cD}^{{b}} \l^{\b j}) D - \ri W^{-2} (\Sigma^{{a} {b}})_{\a \b}  F_{{a} {b}} \l^{\b j}  D
    \non \\
    &&
    +2 \ri W^{-2}  X^{i j} \l_{\a i} D + 2\ri (\Gamma_{{b}})_{\a \b} \l^{\b j}  {\cD}^{{b}}( W^{-1}D )
    - W^{-3}\l^{\b j} \l_{\a}^i \l_{\b i} D \bigg\}~.
\eea
Upon integrating by parts, an arbitrary variation of $D$ leads to the following EOM:
\bea\label{log-D-EOM}
0 = \frac{\pa \cL_{\log , D}}{\pa D} &=& 
-\frac{16}{3} W D
+  \frac{23}{8} W^{{a} {b}} W_{{a} {b}}  W 
-  4 {\cD}_{{a}}{\cD}^{{a}}W - \hf \cR W
+2 F_{{a}{b}} W^{{a} {b}}
- \frac{16 \ri}{3} \chi^{\a i} \l_{\a i}
\non\\
&&
- F^{{a} {b}} F_{{a} {b}}  W^{-1}
+3  X^{i j} X_{i j}  W^{-1} 
-4 \ri  (\Gamma^{{a}})^{\a \b} W^{-1}  \l^{i}_{\a} {\cD}_{{a}}\l_{\b i}   
\non\\
&&
-2  W^{-1} ({\cD}^{{a}}W) {\cD}_{{a}}W
+ \ri  (\Sigma^{{a} {b}})^{\a \b}  F_{{a}{b}} \l^{i}_{\a} \l_{\b i} W^{-2}  
\non\\
&&
+ 2 \ri  X^{i j}  \l^{\a}_{i} \l_{\a j}W^{-2}  
+\frac{1}{2} \l^{\a i} \l^{\b}_i \l_{\a}^{j} \l_{\b j} W^{-3} ~.
\eea
We can uplift this EOM to superspace to get the covariant superconformal primary EOM
\bea\label{log-Y-EOM}
0=
J_{\log}
&=& 
-\frac{3}{1024} W Y 
-  \frac{69}{1024} W^{{a} {b}} W_{{a} {b}}  W 
+  \frac{3}{32}   {\de}_{{a}}{\de}^{{a}}W 
-  \frac{3}{64} F_{{a}{b}} W^{{a} {b}}
- \frac{3}{256}\l^{\a}_{j} X^{j}_{\a}   
\non\\
&&
+ \frac{3}{128} F^{{a} {b}} F_{{a} {b}}  W^{-1}
-  \frac{9}{128}  X^{i j} X_{i j}  W^{-1} 
+  \frac{3 \ri}{32} (\Gamma^{{a}})^{\a \b} W^{-1}  \l^{i}_{\a} {\de}_{{a}}\l_{\b i}   
\non\\
&&
+  \frac{3}{64}  W^{-1} ({\de}^{{a}}W) {\de}_{{a}}W
- \frac{3 \ri}{128}  (\Sigma^{{a} {b}})^{\a \b}  F_{{a}{b}} \l^{i}_{\a} \l_{\b i} W^{-2}  
\non\\
&&
-  \frac{3 \ri}{64}  X^{i j}  \l^{\a}_{i} \l_{\a j}W^{-2}  
- \frac{3}{256} \l^{\a i} \l^{\b}_i \l_{\a}^{j} \l_{\b j} W^{-3} 
~,~~~~~~~
\eea
where, in the uplifting process the curvature term has been incorporated into the degauged d'Alembertian operator making it superconformal, i.e., $\cD^{a} \cD_{a} W \rightarrow \de^{a} \de_{a} W$. All the fields should now be understood as superfields. Note the additional factor of $-\frac{3}{128}$ in the above equation as a result of replacing $D$ with $Y$ on the left-hand side of \eqref{log-D-EOM}.

\subsubsection{Scalar curvature squared} \label{JEOM:R2}
As shown in subsection \ref{subsection Linear multiplet compensator}, the scalar curvature-squared action can be expressed more compactly in terms of the composite vector multiplet fields \eqref{desc-W-R2}. 
From \eqref{desc-W-R2} we also see that only $\mathbf{X}_{ij}$ has a $D$-dependent term, $\frac{\d}{\d D}\mathbf{X}^{ij} = 2 G^{-1} G^{ij}$. As such, to derive the EOM for $D$, the relevant terms in the Lagrangian are given by
\bea
\cL_{R^2, D} &=& -v_a \cH^a_{R^2} -W F_{R^2}- X_{ij} H^{ij}_{R^2} + 2 \l^{\a}_{i} \vf^i_{\a R^2} \non\\
&=& -W \mathbf{X}^{ij} \mathbf{X}_{ij} + 16 W D \mathbf{W}^2 -2 X_{ij} \mathbf{W} \mathbf{X}^{ij} + 2 \ri \l^{\a}_{i} \mathbf{X}^{ij} \bm{\l_{\a j}}~.
\eea
In the above we have neglected the gravitini-dependent terms due to a similar reasoning as in the previous subsections. An arbitrary variation with respect to $D$ leads to the following EOM:
\bea\label{Scalar curvature squared-D-EOM}
0 = \frac{\pa \cL_{R^2 , D}}{\pa D} &=& 
 16 W \mathbf{W}^2 -4 G^{-1}\Big(W G^{ij} \mathbf{X}_{ij}  + X_{ij} \mathbf{W} G^{ij} - \ri \l^{\a}_{i}  G^{ij} \bm{\l_{\a j}}\Big)~.
\eea
We can uplift this EOM to superspace. Once again recall the overall factor of $-\frac{3}{128}$ as we replace $D$ with $Y$ on the left-hand side of \eqref{Scalar curvature squared-D-EOM}. The covariant superconformal primary EOM for $Y$ then reads
\bea
0=
J_{R^2} &=& -\frac{3}{8} W \mathbf{W}^2  
+\frac{3}{32} G^{-1} \Big(\,
W G^{ij} \mathbf{X}_{ij} 
+ G^{ij}X_{ij} \mathbf{W}
-\ri G^{ij} \lambda^{\a}_{i} \bm{\l}_{\a j}
\Big)~.
\label{Y-eom-comp}
\eea

\subsection{Four-derivative: Vector multiplet EOMs} \label{JEOM:Weyl}

In the Weyl-squared case, the composite primary superfield \eqref{H-Weyl}
\begin{equation}
    H^{i j}_{\textrm{Weyl}} := - \frac{\ri}{2}   W^{{\alpha} {\beta} {\gamma} i} W_{{\alpha} {\beta} {\gamma}}\,^{j}+\frac{3\ri}{2}   W^{{\alpha} {\beta}} X_{{\alpha} {\beta}}\,^{i j} - \frac{3 \ri}{4}   X^{{\alpha} i} X^{j}_{{\alpha}}~,
\end{equation}
and subsequently its descendants \eqref{desc-H-Weyl-full} are constructed solely out of the Weyl multiplet. Therefore, 
\bea
\frac{\d S_{\rm Weyl}}{\d V_{ij}}= H^{ij}_{\rm Weyl}~, \qquad \frac{\d S_{\rm Weyl}}{\d X_{ij}}= -H^{ij}_{\rm Weyl} 
\eea
and then the component equation of motion is simply given by
\bea
-H^{ij}_{\rm Weyl} = 0~.
\eea

Similar arguments hold for the scalar curvature-squared invariant. Here the generating primary superfield $H^{ij}_{R^2}$ 
is given by \eqref{Hij-R2}
\bea
H^{ij}_{R^2} &=& -\ri \bm{\lambda}^{\a i} \bm{\l}_{\a}^j +2 \mathbf{W} \mathbf{X}^{ij}~,
\eea
and subsequently its descendants \eqref{desc-W-R2} are constructed solely out of the linear multiplet. This again implies that
\bea
\frac{\d S_{R^2}}{\d V_{ij}}= H^{ij}_{R^2}~, \qquad \frac{\d S_{R^2}}{\d X_{ij}}= -H^{ij}_{R^2} ~,
\eea
and then the component equation of motion is simply
\bea
-H^{ij}_{R^2} = 0~.
\eea

As for the Log invariant, it requires more explanations on the variation of the primary superfield $H^{ij}_{\log}$ with respect to the $V_{kl}$ prepotential. For this we recall that
\bea
W=-\frac{3\ri}{40}\de_{ij}\D^{ijkl}V_{kl}
~,~~~
\mathbb{D}V_{ij}=-2V_{ij}
~,~~~
S^\g_k V_{ij}=0
~.
\eea
while
\bea
H^{ij}_{\log}
=
-\frac{3\ri}{40}\D^{ijkl}\de_{kl}\log W~.
\eea
The BF action can then be written as
\bea
S_{\log}
=\int\rd^{{5|8}}z \, E \, V_{ij}H^{ij}_{\log}~.
\eea
If we vary with respect to $V_{kl}$ the above  we obtain
\bea
\d S_{\log}
=\int\rd^{{5|8}}z \, E \, \Big\{
\d V_{ij}H^{ij}_{\log}
+V_{ij}\d H^{ij}_{\log}
\Big\}
~,
\eea
and then
\bea
\d S_{\log}
=\int\rd^{{5|8}}z \, E \, \Big\{
\d V_{ij}H^{ij}_{\log}
+V_{ij}\Big{[}-\frac{3\ri}{40}\D^{ijkl}\de_{kl}W^{-1}\d W\Big{]}
\Big\}~.
\eea
After integration by parts of the last term, it follows
\bea
\d S_{\log}
=\int\rd^{{5|8}}z \, E \, \Big\{
\d V_{ij}H^{ij}_{\log}
+\d W
\Big\}
=\int\rd^{{5|8}}z \, E \, \Big\{
\d V_{ij}H^{ij}_{\log} 
\Big\}
~.
\eea
Here we have used the fact that the full conformal superspace integral of a vector multiplet is (expected to be) zero as a total derivative. This is clearly true in flat superspace. Moreover, we know the same holds for a tensor multiplet in 6D $\cN=(1,0)$ and a vector multiplet in 4D $\cN=2$ conformal superspace \cite{BKNT16, BNT-M17, Arias:2014ona}, and we expect the same in 5D. 
We can prove it by using the $A$-action principle in 5D conformal superspace, a result that we plan to present elsewhere.

To obtain the EOM from the component action, we begin by taking the variation of the Log supergravity action with respect to the auxiliary field $X^{ij}$. As can be seen from the explicit expressions \eqref{desc-H-logW-full}, each of the descendant components of the composite $H^{ij}_{\log}$ carry $X^{ij}$ dependence. The resulting EOM is given by:
\bea
- H^{i j}_{{\log}}- v_{{a}} \frac{\d \cH^{{a}}_{{\log}}}{\d X^{ij}} - W \frac{\d F_{{\log}}}{\d X^{ij}} - X_{i j} \frac{\d H^{i j}_{{\log}}}{\d X^{ij}}  - 2 \l^{\a k} \frac{\d \varphi_{\a k \, {\log}}}{\d X^{ij}} = 0~.
\eea
Next, we will demonstrate that the bosonic contribution arising from the $X^{ij}$ dependencies found in $H^{ij}_{\log}$ and its descendants does indeed sum up to zero within the EOM, i.e.,
\bea\label{vanishingcombination}
 - v_{{a}} \frac{\d \cH^{{a}}_{{\log}}}{\d X^{ij}} - W \frac{\d F_{{\log}}}{\d X^{ij}} - X_{i j} \frac{\d H^{i j}_{{\log}}}{\d X^{ij}}  
\eea
(at least explicitly up to fermions) vanishes. We expect that this will hold even when we extend our analysis to include fermionic terms. This aligns with the expectations derived from the superspace argument that the full superspace integral of a vector multiplet is  zero as a total derivative. To see it explicitly, note that
\bsubeq
\bea
   X_{i j} \frac{\d H^{i j}_{{\log}}}{\d X^{ij}} &=& -\frac{1}{16} W^{-1} \cR X_{ij} -6  W^{-1} X_{i j} D   - \frac{1}{2}  \cD^{a} \cD_{a}(W^{-1} X^{i j})  - \frac{9}{64}  W^{-1} X_{i j} W^{{a} {b}} W_{{a} {b}}
    \non\\
    &&
    + \frac{1}{2} X_{i j} {W}^{-2}  \cD^{a}\cD_{a}(W)
    -\frac{1}{2}  {\cD}_{{a}}\big({W}^{-2} {X_{i j}} ({\cD}^{{a}}{{W}}) \big)  +  \frac{1}{8}X_{i j}  {{W}}^{-3} F^{{a} {b}} F_{{a} {b}}
\non\\
&&
  - \frac{3}{8} X_{i j} X^{k l} X_{k l} {{W}}^{-3} - \frac{1}{4} X_{i j}   {{W}}^{-3} \big({\cD}^{{a}}{{W}} \big) {\cD}_{{a}}{{W}}  ~,
  \\
   W \frac{\d F_{{\log}}}{\d X^{ij}} &=&  ~ \frac{1}{16} W^{-1} \cR X_{ij} + 6  W^{-1}  X_{i j} D + \frac{1}{2}  W^{-1} \cD^{a}\cD_{a}{X_{i j }} + \frac{1}{2}  \cD^{a}\cD_{a}(W^{-1} X_{i j}) 
    \non\\
    &&
    -\frac{1}{2} {\cD}^{{a}}( W^{-1} {\cD}_{{a}}{X_{i j}}) -\frac{1}{2} W^{-1}  F^{{a} {b}} \Phi_{{a} {b} i j } +\frac{9}{64} W^{-1} X_{i j} W^{{a} {b}} W_{{a} {b}}
    \non\\
    &&
    - \frac{3}{2} X_{i j } W^{-2} \cD^{a}\cD_{a}(W) - \frac{3}{2} W^{-2} {\cD}^{{a}}{W} {\cD}_{{a}}{X_{i j}}  + \frac{3}{2} {\cD}_{{a}}(W^{-2} X_{i j} {\cD}^{{a}}{W}) 
    \non\\
    &&
    - \frac{3}{8} W^{-3} X_{i j} F^{{a} {b}} F_{{a} {b}}
    +\frac{9}{4}  W^{-3} X_{i j} {\cD}^{{a}}{W} {\cD}_{{a}}{W}+\frac{3}{8} W^{-3} X_{i j} X^{k l} X_{k l} ~,
    \\
    v_{{a}} \frac{\d \cH^{{a}}_{{\log}}}{\d X^{ij}} &=&  v_{a} \cD_{b}(W^{-1} \Phi^{{a} {b}}{}_{i j} X^{i j} +\frac{1}{4} W^{-3} X^{i j} X_{i j} F^{{a} {b}}) 
    ~,
    \non \\
     &=&   \frac{1}{2} W^{-1} F_{a b} \Phi^{{a} {b}}{}_{i j}  +\frac{1}{4} W^{-3}  X_{i j} F_{a b} F^{{a} {b}}
     +{\rm total~derivative}~.
\eea
\esubeq
Substituting the above expressions in \eqref{vanishingcombination} all terms cancel out implying that it holds
\bea
    - v_{{a}} \frac{\d \cH^{{a}}_{{\log}}}{\d X^{ij}} - W \frac{\d F_{{\log}}}{\d X^{ij}} - X_{i j} \frac{\d H^{i j}_{{\log}}}{\d X^{ij}} = 0~.
\eea
The vector multiplet EOM for the Log-invariant then reads
\bea
    - H^{i j}_{{\log}} = 0~.
\eea

\subsection{Four-derivative: Linear multiplet EOMs} \label{LEOM:Weyl}
Both the Weyl-squared and Log component actions do not have $F$-dependent terms, so we only have to vary the scalar curvature-squared action \eqref{action-R2}. The relevant terms are given by
\bsubeq
\bea
\cL_{R^2, F} &=& - \Big( v_a \cH^a_{R^2} + W F_{R^2}+ X_{ij} H^{ij}_{R^2} - 2 \l^{\a}_{i} \vf^i_{\a R^2} \Big)~.
\eea
After substituting $H^{ij}_{R^2}, \vf^i_{\a R^2}, F_{R^2}$, and $\cH^a_{R^2}$ according to \eqref{Hij-R2-desc} and performing integration by parts, we find that 
\bea
\cL_{R^2,\, F} &=&
  F^{a b} \left( 2\mathbf{W} \mathbf{F}_{a b} + 3 W_{a b} \mathbf{W}^2 \right) 
    - W \mathbf{X}^{i j} \mathbf{X}_{i j} - 4 W \mathbf{W} \cD^{{a}}\cD_{{a}}\mathbf{W}
    -\hf \cR W \mathbf{W}^2 
  \non\\
 &&
 - 2 W (\cD^{{a}}\mathbf{W})\cD_{{a}}\mathbf{W} + 6 W W^{a b}\mathbf{F}_{a b} \mathbf{W} +  \frac{39}{8}W W^{a b}W_{a b}\mathbf{W}^2 +16 D W \mathbf{W}^2 
 \non\\
 && 
 - 2 X_{i j}  \mathbf{W} \mathbf{X}^{i j} 
 + \ri (\S^{ab})^{\a \b} F_{ab} \bm{\l}_{\a}^i \bm{\l}_{\b i}
 -2\ri W (\cD_{\a}{}^{\b} \bm{\l}_{\b}^i) \bm{\l}^{\a}_i 
 +64 W \chi^{\a i} \bm{\l}_{\a i} \mathbf{W}
 \non\\
 &&
  + 3\ri W W_{\a \b} \bm{\l}^{\a i} \bm{\l}^{\b}_i
  + \ri X_{ij} \bm{\l}^{\a i} \bm{\l}_{\a}^j 
  \non\\
 &&+2 {\l}^\a_i \Big( ~\ri \mathbf{X}^{ij} \bm{\l}_{\a j} -2 \ri \mathbf{F}_{\a \b} \bm{\l}^{\b i} +16 \ri \chi_{\a}^i\mathbf{W}^2 - 2 \ri \mathbf{W} \cD_{\a \b} \bm{\l}^{\b i}  - \ri (\cD_{\a \b} \mathbf{W}) \bm{\l}^{\b i} 
 \non\\
 &&~~~~~~~~~
 -3 \ri W_{\a \b} \mathbf{W} \bm{\l}^{\b i} \Big)
 ~. 
\eea
\esubeq
Since the above is expressed in terms of the composite vector multiplet fields, we make use of \eqref{desc-W-R2} to work out the $F$-dependence. Specifically, 
\bsubeq
\bea
\frac{\d}{\d F} \mathbf{W} &=& \frac{1}{4}G^{-1}~, \\
\frac{\d}{\d F} \bm{\l}_{\a}^i &=& 
-\frac{\ri}{8}G^{-3} G^{ij} \vf_{\a j}~, \\
\frac{\d}{\d F} \mathbf{X}^{ij} &=& 
-\frac{1}{8} G^{-3} F G^{ij}
-\frac{\ri}{8} G^{-3} \vf^{\a i} \vf_{\a}{}^{j}
+\frac{3 \ri}{16}G^{-5} G^{ij} G^{kl} \vf^{\a}_k \vf_{\a l}~.
\eea
\esubeq
Varying $\cL_{R^2, F}$ with respect to $F$ yields
\bea
0&=&
-G^{-1} \bigg[~
\hf X^{ij} \mathbf{X}_{ij}
-\frac{1}{2} F^{\ha \hb} \mathbf{F}_{\ha \hb}
+W \cD^a \cD_a \mathbf{W} + \mathbf{W} \cD^a \cD_a W+  \big(\cD^{\ha} W \big) \cD_{\ha} \mathbf{W} 
\non\\
&&~~~~~~~
+\frac{1}{4} \cR W \mathbf{W}
-8D W \mathbf{W}
-\frac{3}{2} W^{\ha \hb} \big(F_{\ha \hb} \mathbf{W} + \mathbf{F}_{\ha \hb} W \big)
-\frac{39}{16} W^{\ha \hb} W_{\ha \hb} W \mathbf{W} 
\non\\
&&~~~~~~~
+ \frac{\ri}{2} \l^{\a i} \cD_{\a}{}^{\b} \bm{\l}_{\b i}
+ \frac{\ri}{2} \bm{\l}^{\a i} \cD_{\a}{}^{\b} \l_{\b i}
-16 \ri \chi^{\a i} \big( W \bm{\l}_{\a i} + \bm{W} \l_{\a i}\big) 
-\frac{3\ri}{2} W^{\a \b} \l_{\a}^i \bm{\l}_{\b i} 
\bigg]
\non\\
&&
-G^{-3} \bigg[~
\frac{1}{4} G^{ij} \vf_{\b i} \big( \l_{\a j} \cD^{\a \b} \mathbf{W} + \bm{\l}_{\a j} \cD^{\a \b} W
\big)+ \frac{1}{2} G^{ij} \vf_{\b i} \big( W \cD^{\a \b} \bm{\l}_{\a j} + \mathbf{W} \cD^{\a \b} \l_{\a j}
\big)
\non\\
&&~~~~~~~~
-\frac{1}{2}G^{ij} \vf_{\a i} \big( F^{\a \b} \bm{\l}_{\b j} + \mathbf{F}^{\a \b} \l_{\b j} \big)
-\frac{1}{4}G^{ij} F \big( W \mathbf{X}_{ij} + \mathbf{W} X_{i j} \big)
\non\\
&&~~~~~~~~
-\frac{3}{4}G^{ij} W^{\a \b} \vf_{\a i} \big(\mathbf{W} \l_{\b j} + W \bm{\l}_{\b j} \big)
+\frac{\ri}{4}F G^{ij}\l^{\a}_{i} \bm{\l}_{\a j}
+ 8 G_{ij} \chi^{\a i} \vf_{\a}^j W \mathbf{W} 
\non\\
&&~~~~~~~~
+\frac{1}{4} G_{ij} \vf^{\a i} \big( X^{jk} \bm{\l}_{\a k} + \mathbf{X}^{jk} \l_{\a k} \big)
-\frac{\ri}{4} \vf^{\a i} \vf^{j}_\a \big( X_{ij} \mathbf{W} + \mathbf{X}_{ij} W \big)
- \frac{1}{4} \vf^{\a i} \vf^{j}_{\a} \l^{\b}_i \bm{\l}_{\b j}
\bigg]
\non\\
&&
-G^{-5} \bigg[~
\frac{3\ri}{8} G^{ij} G^{kl} \vf^{\a}_{k} \vf_{\a l}
\Big( 
X_{ij} \mathbf{W} + \mathbf{X}_{ij} W 
-\ri \l^{\b}_i \bm{\l}_{\b j}
\Big) \bigg] 
~.
\eea
After covariantising the previous result, one obtains the following superconformal primary EOM \cite{GHKT-M}:
\bea
0= -W_{R^2}
~,
\eea
with
\bea
W_{R^2}&= & 
G^{-1} \bigg[~
\hf X^{ij} \mathbf{X}_{ij}
-\frac{1}{2} F^{\ha \hb} \mathbf{F}_{\ha \hb}
+W \Box \mathbf{W} + \mathbf{W} \Box W+  \big(\de^{\ha} W \big) \de_{\ha} \mathbf{W}
\non\\
&&~~~~~~~
+ \frac{3}{16} Y W \mathbf{W}
-\frac{3}{2} W^{\ha \hb} \big(F_{\ha \hb} \mathbf{W} + \mathbf{F}_{\ha \hb} W \big)
-\frac{39}{16} W^{\ha \hb} W_{\ha \hb} W \mathbf{W} 
\non\\
&&~~~~~~~
+ \frac{\ri}{2} \l^{\a i} \de_{\a}{}^{\b} \bm{\l}_{\b i}
+ \frac{\ri}{2} \bm{\l}^{\a i} \de_{\a}{}^{\b} \l_{\b i}
+ \frac{3}{2} X^{\a i} \big( W \bm{\l}_{\a i} + \bm{W} \l_{\a i}\big) 
-\frac{3\ri}{2} W^{\a \b} \l_{\a}^i \bm{\l}_{\b i} 
\bigg]
\non\\
&&
+G^{-3} \bigg[~
\frac{1}{4} G^{ij} \vf_{\b i} \big( \l_{\a j} \de^{\a \b} \mathbf{W} + \bm{\l}_{\a j} \de^{\a \b} W
\big)+ \frac{1}{2} G^{ij} \vf_{\b i} \big( W \de^{\a \b} \bm{\l}_{\a j} + \mathbf{W} \de^{\a \b} \l_{\a j}
\big)
\non\\
&&~~~~~~~~
-\frac{1}{2}G^{ij} \vf_{\a i} \big( F^{\a \b} \bm{\l}_{\b j} + \mathbf{F}^{\a \b} \l_{\b j} \big)
-\frac{1}{4}G^{ij} F \big( W \mathbf{X}_{ij} + \mathbf{W} X_{i j} \big)
\non\\
&&~~~~~~~~
-\frac{3}{4}G^{ij} W^{\a \b} \vf_{\a i} \big(\mathbf{W} \l_{\b j} + W \bm{\l}_{\b j} \big)
+\frac{\ri}{4}F G^{ij}\l^{\a}_{i} \bm{\l}_{\a j}
+ \frac{3\ri}{4} G_{ij} X^{\a i} \vf_{\a}^j W \mathbf{W} 
\non\\
&&~~~~~~~~
+\frac{1}{4} G_{ij} \vf^{\a i} \big( X^{jk} \bm{\l}_{\a k} + \mathbf{X}^{jk} \l_{\a k} \big)
-\frac{\ri}{4} \vf^{\a i} \vf^{j}_\a \big( X_{ij} \mathbf{W} + \mathbf{X}_{ij} W -\ri\l^{\b}_i \bm{\l}_{\b j}\big)
\bigg]
\non\\
&&
+G^{-5} \bigg[~
\frac{3\ri}{8} G^{ij} G^{kl} \vf^{\a}_{k} \vf_{\a l}
\Big( 
X_{ij} \mathbf{W} + \mathbf{X}_{ij} W 
-\ri \l^{\b}_i \bm{\l}_{\b j}
\Big) \bigg] 
~.
\label{F-eom-comp}
\eea

\subsection{Consistency checks}
This subsection is devoted to the various consistency checks that have been performed on the vector, linear, and Weyl multiplet EOMs. As we will comment more in detail soon, and as expected, these EOMs are primary superfields. Furthermore, it has been checked that the vector multiplet EOMs satisfy the linear multiplet constraint \eqref{linear-constr}, while the linear multiplet EOM adheres to the vector multiplet constraint \eqref{vector-defs}. The Weyl multiplet EOM is found to satisfy the conformal supercurrent conservation equation \cite{HL}, which we will elaborate further.

The vector multiplet EOM for the higher-derivative action $S_{\rm HD}$ \eqref{HD-action}  is
\bea\label{VM-eom}
0 = \frac{3}{4} H^{ij}_{\rm VM} + \k G^{ij} + \a H^{ij}_{\rm Weyl} + \b H^{ij}_{\log}+ \g H^{ij}_{R^2}~.
\eea
Here the first two terms correspond to the EOM for the vector multiplet in the two-derivative supergravity theory $S_{\rm gSG}$ while the remaining three terms describe the contribution coming from the three independent curvature-squared invariants. It is clear that, as expected, the right-hand side of \eqref{VM-eom} is a primary superfield satisfying the linear multiplet constraint \eqref{linear-constr}. 

Next, the resulting linear multiplet EOM in the higher-derivative action $S_{\rm HD}$ \eqref{HD-action} reads
\be
 0=\mathbf{W} + \kappa W + \gamma W_{R^2}~.
 \label{LM-eom}
\ee
It is possible to check from the explicit form \eqref{F-eom-comp} that $W_{R^2}$ is primary, $S_{\a}^i W_{R^2} = 0$. Here we have made use of the relations \eqref{S-on-W} and \eqref{O2-S-actions} as well as the analogous identities for the bold superfields
\bsubeq
\bea
S_{\a}^i \bm{\l}_{\b}^j &=& -2\ri \ve_{\a \b} \ve^{ij} \bm{W}~, \\
S_{\a}^i \bm{F}_{\b \g} &=& 4 \ve_{\a (\b} \bm{\l}_{\g)}{}^i~, \\
S_{\a}^i \bm{X}^{jk} &=& -2\ri  \ve^{i(j} \bm{\l}_{\a}^{k)}~, \\
S_{\a}^i  \de_{\b}{}^{\g}\bm{\l}^{j}_{\g} &=& 5\ri \ve_{\a \b} \bm{X}^{i j} + 2\ri \ve^{i j} \bm{F}_{\a\b} -\ri \ve^{i j} \de_{\a \b} \bm{W}~, \\
S_{\a}^i \Box \bm{W} &=& -\frac{9}{2} W_{\a \b} \bm{\l}^{\b i} +2 \de_{\a \b} \bm{\l}^{\b i}~.
\eea
\esubeq
Alternatively, it is useful to note that $W_{R^2}$ can be expressed as
\bsubeq\label{W-R2-composite}
\begin{equation} \label{WR2_as_W1}
    W_{R^2} = \frac{\ri}{32} G \de_{i j} \mathcal{R}^{i j}_{1}~,
\end{equation}
where
\begin{equation}
    \mathcal{R}^{i j}_{1} =G^{-2} \left(\d^i_k \d^j_l - \frac{1}{2 G^2} G^{i j}G_{k l}\right) H^{k l}_{\rm bilinear}
    ~,
\end{equation}
and
\begin{equation}
    H^{k l}_{\rm bilinear} = 2 W \mathbf{X}^{k l} + 2 \mathbf{W} X^{k l} - 2 \l^{\a k} \bm{\l}^{l}_{\a}~.
\end{equation}
\esubeq 
This remarkably simple form of \eqref{W-R2-composite} guarantees that the right-hand side of eq.~\eqref{LM-eom}, and in particular \eqref{F-eom-comp}, is a primary superfield satisfying the vector multiplet constraints \eqref{vector-defs}, as expected. 

Finally, the Weyl multiplet EOM for the higher-derivative action is
\bsubeq \label{Y-eom}
\bea
0&=& \cJ=J_{\rm EH}
+\alpha J_{\rm Weyl}
+\beta J_{\log} + \gamma J_{{R^2}}~,
\label{Y-eom-0}
\eea
with
\bea
J_{\rm EH}
&=&
\frac{3}{32}(G- W^{3}) 
~,
\label{Y-eom-1}
\\
J_{\rm Weyl}
&=&
- \frac{3}{64} W Y + \frac{3}{16} W W^{{a} {b}}W_{{a} {b}}  + \frac{3}{32}  F_{{a} {b}} W^{{a} {b}}   - \frac{3}{16} \l_{i}^{\alpha} X^{i}_{\alpha}
~,
\\
J_{\log}
&=&
-\frac{3}{1024}W Y 
-  \frac{69}{1024} W^{{a} {b}} W_{{a} {b}}  W 
+  \frac{3}{32}  \Box W 
-  \frac{3}{64} F_{{a}{b}} W^{{a} {b}}
- \frac{3}{256}\l^{\alpha}_{j} X^{j}_{\alpha} 
\non\\
&&+ \frac{3}{128} F^{{a} {b}} F_{{a} {b}}  W^{-1}  
-  \frac{9}{128}  X^{i j} X_{i j}  W^{-1} 
+  \frac{3\ri}{32}  (\Gamma^{{a}})^{\alpha \beta} W^{-1}  \l^{i}_{\alpha} {\de}_{{a}}\l_{\beta i} 
\non\\
&&+  \frac{3}{64}  W^{-1} ({\de}^{{a}}W) {\de}_{{a}}W  
- \frac{3 \ri }{128} (\Sigma^{{a} {b}})^{\alpha \beta}  F_{{a}{b}} \l^{i}_{\alpha} \l_{\beta i} W^{-2}   
-\frac{3\ri}{64}    X^{i j}  \l^{\alpha}_{i} \l_{ \alpha j} W^{-2}  
\non\\
&&- \frac{3}{256}     \l^{\alpha i} \l^{\beta}_{i} \l^{j}_{\alpha} \l_{{\beta} j} W^{-3}
~,
\label{Y-eom-2}
\\
J_{{R^2}}
&=&
-\frac{3}{8} W \mathbf{W}^2  
+\frac{3}{32} G^{-1} \Big(\,
W G^{ij} \mathbf{X}_{ij} 
+ G^{ij}X_{ij} \mathbf{W}
-\ri G^{ij} \lambda^{\a}_{i} \bm{\l}_{\a j}
\Big)
~.
\label{Y-eom-3}
\eea
\esubeq
Here $J_{\rm EH}$ is the EOM from the two-derivative gauged supergravity action, $S_{\rm gSG}$, which does not have any contribution from the cosmological constant term $\kappa$.

Let us comment on the Weyl multiplet EOM \eqref{Y-eom}. As previously explained in \cite{BKNT-M14, GHKT-M}, the 5D Weyl multiplet may be described by a single unconstrained real prepotential $\mathfrak U$, in complete analogy with the case of 4D $\cN=2$ supergravity \cite{KT}. Given a system of matter superfields $\vf^i$, a Noether coupling between $\mathfrak U$ and the matter supercurrent $\cJ$ can be constructed
\bea
S[\vf^i] &=& \int\rd^{5|8}z\, E\,   {\mathfrak U} \cJ = \int \rd^5 x \, e \, \Big( D J + \cdots \Big)~,
\eea
where $J = \cJ \loco$. 
The Weyl multiplet EOM \eqref{Y-eom} is obtained by varying the supergravity action with respect to $\mathfrak U$
\bea
\frac{\d S[\vf^i]}{\d \mathfrak U} = \cJ = 0~.
\eea
In five dimensions, the $\cN=1$ and $\cN=2$ supercurrent multiplets were introduced by Howe and Lindstr\"om \cite{HL}, see also \cite{Howe}. The conformal supercurrent, $\cJ$, is a dimension-3 primary real scalar superfield which satisfies the conservation equation \cite{Kuzenko:2014eqa}
\bea
\de^{ij} \cJ = 0~,
\label{supercurrent}
\eea
provided the dynamical matter superfields obey their EOMs, ${\d S[\vf^i]}/{\d \vf^i} = 0$.
Hence, we shall prove that the expression $\cJ$ in \eqref{Y-eom} satisfies the conservation constraint \eqref{supercurrent}. It has been shown in \cite{BKNT-M14} that this constraint holds for $J_{\rm EH}$. 
For each invariant, it has indeed been verified in \cite{GHKT-M} that the corresponding $J$ is a primary superfield of dimension 3. Furthermore,  a very non-trivial consistency check of \eqref{Y-eom-1}--\eqref{Y-eom-3} is to verify that $\de^{ij}J=0$,
provided the vector and linear multiplets equations of motion of eqs.~\eqref{VM-eom} and \eqref{LM-eom}, respectively, are imposed. Using \textit{Cadabra} an explicit calculation shows that, off-shell, it holds
\bsubeq \label{supercurrent_equation}
\bea
\de^{ij}J_{\rm Weyl}&=&\frac{3 \ri}{4} W H^{ij}_{\rm Weyl}
~,\\
\de^{ij}J_{\log}&=&\frac{3 \ri}{4} W H^{ij}_{\log}
 ~,\\
\de^{ij}J_{R^2}&=& \frac{3 \ri}{4}  W H^{ij}_{R^2}
- \frac{3 \ri}{4} G^{ij}W_{R^2}
~.
\eea
\esubeq 
The right-hand sides of \eqref{supercurrent_equation} are all proportional to the composite vector and linear multiplets appearing in \eqref{VM-eom} and \eqref{LM-eom}. Consequently, the supercurrent conservation equation \eqref{supercurrent} is satisfied once the EOMs for the compensators are used.

\section{EOM descendants}
\label{section-7}

In section \ref{section-6}, we derived the manifestly covariant superconformal primary superfield equations of motions (EOMs) for the vector, linear, and Weyl multiplet. In this section, we will find the descendants of these EOMs, which are obtained by successive applications of superspace spinor derivatives or, equivalently, successive application of $Q$-supersymmetry transformations. Computing the descendants is of importance. It suffices to mention that the EOM resulting from the variaton of the vielbein in components is at the bottom of the multiplet of EOMs. One advantage of deriving all the EOMs as descendants of the primary ones is that covariance is manifest at every stage. This fact would be very non-trivial if one attempted to obtain the EOMs directly by varying the component actions, including all fermions.

The component structure of the vector multiplet compensator EOM is straightforward to obtain, as it naturally forms a (composite) linear multiplet. More specifically, in the two-derivative gauge supergravity $S_{\rm gSG}$, the EOM \eqref{2deriv-VM} leads to a combination of $G^{ij}$ and $H^{ij}_{\rm VM}$, whose descendants can be found using \eqref{O2superfieldComps} and \eqref{O2composite-N}. In the four-derivative case, the EOMs in the Weyl-squared, Log, and scalar curvature-squared invariants are described by $H^{ij}_{\rm Weyl}$,  $H^{ij}_{\rm log}$, and $H^{ij}_{R^2}$, respectively. Their corresponding descendants have already been presented in eqs.~\eqref{desc-H-Weyl-full}, \eqref{desc-H-logW-full}, and \eqref{Hij-R2-desc}.  

The component structure of the linear multiplet compensator EOM is straightforward to obtain, as it naturally forms a (composite) vector multiplet. More specifically, in $S_{\rm gSG}$, the EOM \eqref{liner-2der-EOM-superspace} leads to a combination of $W$ and  $\mathbf{W}$, whose descendants can be found using \eqref{components-vect} and \eqref{desc-W-R2}. In the four-derivative case, 
since both the Weyl-squared and Log invariants do not have any dependence on the linear multiplet (e.g., there are no $F$-dependent terms), 
the only contribution to the EOM comes from the scalar curvature-squared invariant which leads to the non-trivial structure $W_{R^2}$,  eq.~\eqref{F-eom-comp}. We will present its new component structure in subsection \ref{WR2-desc}.

To our knowledge, the component structure for the Weyl multiplet EOM, specifically the supercurrent, was not known before. Hence, it is the subject of the next subsection to derive the independent components of the supercurrent multiplet and use them to compute its descendants for all two and four-derivative invariants.

\subsection{Component structure of the conformal supercurrent}
The 5D $\mathcal{N} = 1$ conformal supercurrent is described by a real primary scalar superfield $J$ of dimension 3, which obeys the conservation equation
\bsubeq \label{supercurrent_constraint}
\be 
    \de^{2}_{ij}J = 0~,
\ee
where we have defined\footnote{Note that in this section, our $\de^2_{ij} = \hf \de_{ij}$. We insert the superscript ``2" to easily keep track the number of spinor derivatives acting on $J$.}
\be
\de^{2}_{ij} = \frac{1}{2} \de_{(i}\de_{j)}~.
\ee
\esubeq

To define the independent components of the conformal supercurrent $J$, one can systematically act with spinor derivatives on $J$. At dim-$\frac{7}{2}$, we introduce
\bea
J^{1}_{\a i} = \de_{\a i} J~.
\eea

At dim-$4$, we start by applying two spinor derivatives on $J$. A generic two-spinor-derivative combination can be expanded as
\be
\de_{\a i} \de_{\b j}  = \ri \ve_{i j} (\G^a)_{\a \b} \de_a  - \frac{1}{2} \ve_{i j} (\Sigma^{ab})_{\a \b} \de^{2}_{ab} + \frac{1}{2} \ve_{\a \b} \de^{2}_{i j} +\frac{1}{2} (\G^{a})_{\a \b} \de^{2}_{a i j} + l.d. ~,
\ee
where $l.d.$ contains terms related to lower-dimensional supercurrent descendants and Weyl multiplet. Furthermore, the irreducible $\de^{2}$ operators are defined as
\be
\de^{2}_{a i j} = \frac{1}{2}\de_{(i} \G_{a} \de_{j)}
~, \qquad \text{and}\qquad    \de^{2}_{ab} = \frac{1}{2}\de^{i} \Sigma_{a b} \de_{i}~.
\ee
It follows from the constraint \eqref{supercurrent_constraint} that the independent descendants at dim-4 are 
\bea
J^{2}_{ab} = \de^{2}_{ab}J~, \quad J^{2}_{a i j} = \de^{2}_{a i j}J~.
\eea
Note that, given the superconformal primary supercurrent $J$, dim-$\frac{7}{2}$ and dim-$4$ descendants prove to be $K$-primary.

At dim-$\frac{9}{2}$, following the discussion 
of \cite{Gates:2014cqa}, there exist three irreducible $\de^{3}$ operators, specifically
\bea\label{irreducible-three-derivative}
{\de}^{3}_{\a i} = \{\de_{\a}^j, \de^{2}_{ij}\}~,~~~{\de}^{3}_{a \a i} = \{\de_{\a}^j, \de^{2}_{a ij}\}~,~~~{\de}^{3}_{\a i j k} = \{\de_{\a (i}, \de^{2}_{ j k)}\} &\cong& 2 \de_{\a (i} \de^{2}_{ j k)}~,
\eea
where we have omitted writing $l.d.$ and $\cong$ means that the equality holds only up to $l.d.$ terms. It is easy to verify that a generic object with three spinor derivatives can be expanded in terms of these three irreducible $\de^{3}$ operators \eqref{irreducible-three-derivative}, since
\bea
    \de \de \de \propto \de \de^2 = \hf \{\de, \de^2\} + \hf [ \de , \de^{2} ] &\cong& \hf \{\de, \de^2\}~,
\eea
where the anticommutator can be expanded in \eqref{irreducible-three-derivative} according to eq.~(A.13) of \cite{Gates:2014cqa}. However, it is important to note that eq.~(A.13) there was given in flat space. When uplifted to curved space, one then picks up additional contributions to $l.d.$ terms.   
Furthermore, when acting on $J$, the following two $\de^{3}$ operators do not lead to independent components as a result of the conservation constraint \eqref{supercurrent_constraint}:
\bsubeq
\bea
{\de}^{3}_{\a i j k}J &\cong& 2 \de_{\a (i} \de^{2}_{ j k)}J = 0~, \\
{\de}^{3}_{\a i}J &=& \{\de_{\a}^j, \de^{2}_{ij}\}J =  \de^{2}_{ij} \de_{\a}^j J = [\de^{2}_{ij}, \de_{\a}^j] J \propto \de_{a} J^1_{\a i} + l.d.~.
\eea
\esubeq
Therefore, the only independent component at dim-$\frac{9}{2}$ is ${\de}^{3}_{a \a i}J$. Now, using the fact that
\be
(\G^{a})_{\a}{}^{\b} {\de}^{3}_{a \b i} = -{\de}^{3}_{\a i}~,
\ee
this implies that the gamma trace part of  ${\de}^{3}_{a \a i}J$ is a total derivative, i.e.,
\be
(\G^{a})_{\a}{}^{\b} {\de}^{3}_{a \b i}J = -{\de}^{3}_{\a i}J \propto - \de_{a} J^1_{\a i}~.
\ee
Hence, we can subtract the trace by modifying the above as follows
\be
{\de}^{3}_{a \a i}J +\frac{1}{5} (\G_{a})_{\a}{^{\b}} (\G^{b})_{\b}{^{\g}} {\de}^{3}_{b \g i}J~.
\ee
Lastly, we modify the descendant to make it $K$-primary. This amounts to adding a total derivative term to obtain the dim-$\frac{9}{2}$ component
\bea
 J^{3}_{a \a i} =  \left( \d_{a}^{b} \d^{\g}_{\a} + \frac{1}{5}  (\Gamma_{a})_{\a}{}^{\b} (\Gamma^{b})_{\b}{}^{\g} \right)\left({\de}^{3}_{b \g i}J - 3 \ri \de_{b} J^{1}_{\g i} \right)~.
\eea

Next, we move on to dim-5 descendants. All the independent components at this level will come from
\bea \label{general_dim_5}
\de_{\b j} {\de}^{3}_{a \a i} J &=& 2 \de_{\b j} \de_{\a}^k \de^{2}_{a ik} J + \de_{\b j} [\de^{2}_{a ik},\de_{\a}^k]  J~ 
\non \\&\cong&  (\Sigma^{cd})_{ \b \a} \de^{2}_{cd}\de^{2}_{a ij}J- (\G^{c})_{ \b \a} \de^{2}_{c j k}\de^{2}_{a i}{}^{k}J~.
\eea
Let us consider the $\S$ trace part:
\bea
\de_{(\b (i} {\de}^{3}_{|a| \a) j)} J &=& (\Sigma^{cd})_{ \b \a} \de^{2}_{cd}\de^{2}_{a ij}J   \non \\
&=&  - \de_{(\a}^{k}  \de_{\b)k} \de^{2}_{a ij}J \non
\\
&=&  - \frac{1}{2} \de_{(\a}^{k}  \{ \de_{\b)k} , \de^{2}_{a ij} \} J - \frac{1}{2} \de_{(\a}^{k}  [ \de_{\b)k} , \de^{2}_{a ij} ] J \non
\\
&\cong&  \frac{1}{3} \de_{(\a (i}  {\de}^{3}_{|a| \b)j)} J ~.
\eea
The above equation implies that the $\S$-trace part is a $l.d.$ term and hence, does not give any independent component. We are only left with the $\G$-trace part, which is given by:
\bea 
\de_{[\b j} {\de}^{3}_{a \a] i}J &\cong&   - (\G^{c})_{ \b \a} \de^{2}_{c j k}\de^{2}_{a i}{}^{k}J   \non
\\&\cong&      \frac{2}{3} \de_{[\b j} {\de}^{3}_{a \a] i}J + \frac{1}{6} \ve_{j i} \de_{[\b}{}^{k} {\de}^{3}_{a \a] k}J~. 
\eea
The symmetric part in $(i j)$ of the above equation implies that $\de_{[\b (j} {\de}^{3}_{a \a] i)}J$ is again a $l.d.$ term, while the antisymmetric part is unfixed. Therefore, the only independent part left of $\de_{\b j} {\de}^{3}_{a \a i} J $ is the following:
\bea
\de_{[\b [j} {\de}^{3}_{a \a] i]} J  &\cong&   - (\G^{c})_{ \b \a} \de^{2}_{c [j |k|}\de^{2}_{a i]}{}^{k}J  \non 
\\ &\cong&   \frac{1}{2} (\G^{c})_{ \b \a} \ve_{j i}\de^{2}_{(c}{}^{ k l}\de^{2}_{a) k l}J
~.
\eea
The independent component is thus $\de^{2}_{(c}{}^{ k l}\de^{2}_{a) k l}J$. Once again the $\eta$-trace part of this component is a total derivative and can be removed as follows 
\bea
 \de^{2}_{(a}{}^{ k l}\de^{2}_{b) k l}J - \frac{1}{5} \eta_{ab} \de^{2 c k l}\de^{2}_{c k l}J
 ~.
\eea
Lastly, we modify the descendant to make it $K$-primary. This amounts to adding a total derivative term to the above descendant to obtain the dim-$5$ component
\bea
    J^{4}_{ab} =  \left( \d^{c}_{(a} \d^{d}_{b)}- \frac{1}{5} \eta_{a b} \eta^{c d}\right) \left(\de^{2}_{c}{}^{ k l}\de^{2}_{d k l}J - 6\de_{c}\de_{d}J\right)~.
\eea

In summary, we have derived the independent components of the conformal supercurrent multiplet $J$. They are defined as follows:
\bsubeq \label{eq:Jcomponents}
\bea
\text{dim-3} \quad && J  ~,
\\
\text{dim-$\frac{7}{2}$} \quad && J^{1}_{\a i} = \de_{\a i} J  ~, 
\\
\text{dim-4} \quad && J^{2}_{ab} = \de^{2}_{ab}J ~, \qquad J^{2}_{a i j} = \de^{2}_{a i j}J ~,
\\
\text{dim-$\frac{9}{2}$} \quad && J^{3}_{a \a i} =  \left( \d_{a}^{b} \d^{\g}_{\a} + \frac{1}{5}  (\Gamma_{a})_{\a}{}^{\b} (\Gamma^{b})_{\b}{}^{\g} \right)\left({\de}^{3}_{b \g i}J  - 3 \ri \de_{b} J^{1}_{\g i}  \right)~, 
\\
\text{dim-5} \quad && J^{4}_{ab} =  \left( \d^{c}_{(a} \d^{d}_{b)}- \frac{1}{5} \eta_{a b} \eta^{c d}\right) \left(\de^{2}_{c}{}^{ k l}\de^{2}_{d k l}J  - 6\de_{c}\de_{d}J \right)~.
\eea
\esubeq
The component fields in the supercurrent multiplet will be defined by using the same names as the superfields, i.e., $J:=J\loco$, $J^1_{\a i}:=J^1_{\a i}\loco$, and so forth. As previously discussed by Howe and Lindstr\"{o}m in \cite{HL,Howe}, associated to each component of the above current multiplet, we may then identify a corresponding field in the Weyl multiplet whose variation leads to the various $J$s in eq.~\eqref{eq:Jcomponents}. This is summarised in Table \ref{weights-vector_0} below:
\begin{table}[hbt!]
\begin{center}
\begin{tabular}{ |c| c| c| c| c| c| c|} 
 \hline
Current multiplet & $J$ 
& $J^{1}_{\a i}$
& $J^{2}_{ab}$
& $J^{2}_{a ij}$
& $J^{3}_{a \a i}$
& $J^{4}_{ab}$
\\ 
\hline
 \hline
Fields
& $D$
& $\chi^{\a i}$
& $W_{ab}$
& $V_m^{ij}$
& $\hat{\psi}_m{}^{\a}_{i}$
& $\hat{g}_{mn}$
\\
\hline
\end{tabular}
\caption{\footnotesize{Current multiplet and the corresponding fields to which they couple via a Noether coupling. With $\hat{\psi}_m{}^{\a}_{i}$ and
$\hat{g}_{mn}$ we denote the gamma traceless part of the gravitini and the traceless part of the metric, respectively.}\label{weights-vector_0}}
\end{center}
\end{table}

\subsection{Einstein--Hilbert}

In this subsection we derive the EOM descendants of the Weyl multiplet in the two-derivative gauged supergravity action, $S_{\rm gSG}$. The descendants of the vector compensator EOM can be readily obtained from \eqref{O2superfieldComps} and \eqref{O2composite-N}. To compute the descendants of the linear compensator EOM, the relevant equations are given in \eqref{components-vect} and \eqref{desc-W-R2}.

Recall the Weyl multiplet EOM for the action $S_{\rm gSG}$ given in subsection \ref{EOM:EH}:
\be
    J_{\rm EH} = \frac{3}{32}\left(G - W^3\right)
    ~.
\ee
We present here the results for each component descendants of $J_{\rm EH}$ based on the structures found in eqs.~\eqref{eq:Jcomponents}: 
\bsubeq 
\bea
    J^{1}_{ \a i, {\rm EH}} &=& \frac{3}{32} G_{i j} \varphi^{j}_{\alpha} {G}^{-1} - \frac{9}{32} \ri  \lambda_{i {\alpha}} ~,
    \\
    J^{2}_{ a b, {\rm EH}} &=& \frac{9}{16} \ri F_{a b} W^{2} + \frac{9}{16} \ri W_{a b} W^{3} - \frac{3}{32}(\Sigma_{a b})^{\alpha \beta}  \varphi^{i}_{\alpha} \varphi_{i \beta} {G}^{-1} - \frac{9}{32} (\Sigma_{a b})^{\alpha \beta} \lambda^{i}_{\alpha} \lambda_{i \beta} W ~,
    \\
    J^{2}_{a i j, {\rm EH}} &=& \frac{3}{64}  (\Gamma_{a})_{\alpha \beta} \varphi_{i}^{\alpha} \varphi_{j}^{\beta} {G}^{-1} + \frac{3}{32} \ri \cH_{a} G_{i j} {G}^{-1} - \frac{3}{16} \ri {G}^{-1}   G^k{}_{(i} \nabla_{a}{G_{j) k}} \non \\
    && + \frac{3}{64} (\Gamma_{\a})^{\alpha \beta} G_{i k} G_{j l} {G}^{-3} \varphi^{k}_{\alpha} \varphi^{l}_{\beta} - \frac{9}{32}(\Gamma_{a})_{\alpha \beta}  W \lambda_{i}^{\alpha} \lambda_{j}^{\beta} ~,
    \\
    J^{3}_{a \a i, {\rm EH}} &=& ~~
    G^{-1} \bigg\{  \frac{9}{20} \ri \left( \d_a^b \d_\alpha^\beta - \frac{1}{2} (\Sigma_a{}^b)_\a{}^\b \right)\left( \frac{1}{2}\nabla_{b}G_{i j} \varphi^{j}_{\beta} - \frac{3}{2} G_{i j} \nabla_b \varphi^j_\beta - \cH_b \varphi_{i \b} \right) 
    \non \\
    && ~~~~~~~~~~~~
    - \frac{9 \ri}{320} \left( \ve_{a b c d e} (\Sigma^{d e})_\a{}^\b + 3 \eta_{a b} (\Gamma_c)_\a{}^\b \right) G_{i j} W^{b c} \varphi^j_\b \bigg\}
    \non \\
    &&
    + G^{-3} \bigg\{ \frac{27}{80}{\rm i} G_{i j} G_{k l}\left(\d_a^b \d_\a^\b - \frac{1}{2} (\Sigma_{a}{}^b)_\a{}^\b\right)  \nabla_{b}{G^{k l}} \varphi^{j}_{\b} 
    \non \\
    && ~~~~~~~~~~~~
    + \frac{3}{16}\left(\d_\a^\l (\Gamma_a)^{\b \rho} - \frac{1}{5}\ve_{\b \rho} (\Gamma_a)_\a{}^{\lambda} \right) G_{i j}\varphi^k_{\l} \varphi^j_\b \varphi_{k \rho} \bigg\}
    \non \\
    && 
    + ~~~~~ \bigg\{ \frac{9 \ri}{16}  \left(\d_\a^\l(\Gamma_a)^{\b \rho} - \frac{1}{5} \ve^{\b \rho} (\Gamma_a)_\a{}^{\l}\right) \l_{j \l}\l_{i \b} \l^j_\rho \bigg\}
    \non \\
    && 
    + ~ W  \bigg\{ \frac{27}{40}\left( \ve_{a b c d e} (\Sigma^{d e})_{\alpha \beta} + 3 \eta_{a b} (\Gamma_c)_{ \a \b}\right)F^{b c} \l_i^\b  - \frac{27}{20} \left( \d_a^b \d^\b_\a - \frac{1}{2} (\Sigma_a{}^b)_\a{}^\b\right) \l_{i \b} \nabla_b W \bigg\}
    \non \\
    &&
    +  W^2 \bigg\{  \frac{189}{320} \left(\ve_{a b c d e} (\Sigma^{d e})_{\alpha \beta} + 3 \eta_{a b} (\Gamma_c)_{\a \b}\right) W^{b c} \l^\b_i 
    \non \\
    && ~~~~~~~~~~~~
    + \frac{27}{40} \left( \d_a^b \d^\b_\a - \frac{1}{2} (\Sigma_a{}^b)_\a{}^\b\right) \nabla_{b}\l_{i \b}  \bigg\} ~,
    \\
    J^{4}_{a b, {\rm EH}} &=& ~~~
    \left(\d_{(a}^c\d_{b)}^d - \frac{1}{5}\eta_{a b}\eta^{c d}\right) \times \bigg[
    \non \\
    && ~~~~~~~~
    G \bigg\{ \frac{27}{8}W_{c}{}^e W_{d e} \bigg\}
    \non \\
    && ~
    + G^{-1} \bigg\{- \frac{9}{16}\cH_{c}\cH_{d} - \frac{9}{32}\nabla_{c}{G_{i j}} \nabla_{d}{G^{i j}} + \frac{27}{32}G_{i j} \nabla_{c}{\nabla_{d}{G^{i j}}} - \frac{9 \ri}{8} (\Gamma_c)^{\a \b} \nabla_{d} \varphi^i_\a \varphi_{i \b} 
    \non \\
    && ~~~~~~~~~~~~~~~
    - \frac{9 \ri}{16} (\Sigma_{c}{}^{e})^{\a \b} W_{d e} \varphi^i_\a \varphi_{i \b}\bigg\}
     \non \\
    &&  ~
    + G^{-3} \bigg\{  - \frac{27}{64}G_{i j} G_{k l} \nabla_{c}{G^{i j}} \nabla_{d}{G^{k l}} + \frac{9}{32}{\rm i} \cH_{c} G_{i j} (\Gamma_{d})^{\alpha \beta}  \varphi^{i}_{\alpha} \varphi^{j}_{\beta}
    \non \\
    && ~~~~~~~~~~~~~~~
    - \frac{9}{16}{\rm i} G_{i j} (\Gamma_{c})^{\alpha \beta} \nabla_{d}{G^{i}\,_{k}} \varphi^{j}_{\alpha} \varphi^{k}_{\beta}  \bigg\}
    \non \\
    && ~
    + ~~~~~~ \bigg\{ \frac{27 \ri}{8} (\Sigma_{c e})^{\a \b}F_{d}{}^e \lambda_{i \alpha} \lambda^{i}_\beta \bigg\}
    \non \\
    &&  ~  
    + ~~~ W \bigg\{ - \frac{27}{4} F_{c}\,^{e} F_{d e}  - \frac{27}{8} \nabla_{c}{W} \nabla_{d}{W}  + \frac{27 \ri}{16} (\Sigma_{c}{}^{\, e})^{\alpha \beta}  W_{d e} \lambda_{i \alpha} \lambda^{i}_{\beta} 
    \non \\
    && ~~~~~~~~~~~~~~~
    + \frac{27 \ri}{8} (\Gamma_{c})^{\alpha \beta}  \lambda_{i \alpha} \nabla_{d}{\lambda^{i}_{\beta}} \bigg\}
    \non \\
    && ~
    + W^2 ~ \bigg\{- \frac{27}{2}W_{c}\,^{e} F_{d e}   + \frac{27}{16} \nabla_{c}{\nabla_{d}{W}}  \bigg\}
    \non \\
    && ~
    + W^3 ~ \bigg\{- \frac{27}{4}W_{c}\,^{e} W_{d e} \bigg\} ~~ \bigg]~.
    \label{J4-EH}
\eea
\esubeq
Note that, we have proven each to be explicitly $K$-primary.

\subsection{Weyl-squared}
  As previously discussed, the Weyl-squared invariant does not contribute to the linear multiplet EOM, while the descendants of the vector multiplet EOM have been described earlier in eq.~\eqref{desc-H-Weyl-full}. Thus, in this subsection we are mainly focusing on deriving the descendants of the Weyl multiplet EOM.

In superspace, the Weyl multiplet EOM  was presented in subsection \ref{JEOM:Weyl2}. It defines a conformal supercurrent multiplet
\be
    J_{\rm Weyl} = - \frac{3}{64} W Y + \frac{3}{16} W W^{{a} {b}}W_{{a} {b}}  + \frac{3}{32}  F_{{a} {b}} W^{{a} {b}}   - \frac{3}{16} \l_{i}^{\alpha} X^{i}_{\alpha}
    ~.
\ee
We present here the results for one- and two-derivative descendants of $J_{\rm Weyl}$ based on the structures found in eqs.~\eqref{eq:Jcomponents}. Note that, we have explicitly proven each to be $K$-primary.  In subsection 4.2 of the supplementary file, the reader can find the complete expressions of $J^3_{a \a i, {\rm Weyl}}$, along with the degauged, gauged fixed $(W=1)$, bosonic part of $J^4_{a b, {\rm Weyl}}$. It holds that
\bsubeq
\bea
    J^{1}_{ \a i, {\rm Weyl}} &=& - \frac{3}{128} \ri Y \lambda_{i \alpha} - \frac{3}{8} (\Gamma_{a})_{\alpha}{}^{\beta}  W \nabla^{a}{X_{i \beta}} + \frac{3}{32} \ri  W_{a b} W^{a b} \lambda_{i \alpha} + \frac{9}{128}  W W_{a b} W^{a b}{}_{ \alpha i} \non \\
    && + \frac{9}{256} \ve^{a b c d e} (\Gamma_{e})_{\alpha}{}^{\beta}  W W_{a b} W_{c d \beta i} - \frac{9}{32} (\Sigma^{a c})_{\alpha}{}^{\beta}  W W_{a}{}^{b} W_{c b \beta i} + \frac{3}{128} F^{a b} W_{a b \alpha i}  \non \\
    && + \frac{3}{256} \ve^{a b c d e} (\Gamma_{e})_{\alpha}{}^{\beta}  F_{a b} W_{c d \beta i} - \frac{3}{32} (\Sigma^{a c})_{\alpha}{}^{ \beta} F_{a}{}^{b} W_{c b \beta i} + \frac{3}{32} (\Sigma_{a b})_{\alpha \beta}  F^{a b} X_{i}^{ \beta}  
    \non \\
    && 
    + \frac{3}{16} \ri (\Gamma^{a})_{\alpha \beta}  W_{a b} {\nabla}^{b}{\lambda_{i}^{\beta}} + \frac{1}{8} \ri \Phi_{a b}\,_{i j}   (\Sigma^{a b})_{\alpha}{}^{\beta} \lambda^{j}_{\beta}  +\frac{3}{32} \ri  \ve^{a b c d e } (\Sigma_{a b})_{\alpha \beta}   \lambda_{i}^{\beta} \nabla_{e} {W_{c d}} \non \\
    && + \frac{3}{32} \ri (\Gamma^{a})_{\alpha \beta}  \lambda_{i}^{\beta} \nabla^{b}{W_{a b}} - \frac{3}{16} X_{i j} X^{j}_{\alpha} + \frac{3}{16}(\Gamma^{a})_{\alpha \beta}   X_{i}^{\beta} \nabla_{a}{W} ~,
    \\
    J^2_{a i j, {\rm Weyl}} &=&  
    - \frac{1}{16} \ri \ve_{a b c d e} F^{b c} \Phi^{d e}{}_{i j} - \frac{1}{2}  \ri  \nabla^{b}( W \Phi_{a b i j}) - \frac{3}{64} \ri (\Gamma_{a})_{\alpha \beta} W^{c d} W_{c d}{}^{\alpha}{}_{(i} \lambda_{j)}^{\beta} 
    \non \\ 
    &&
    + \frac{9}{32} (\Gamma_{a})_{\alpha \beta} W X_{(i}^{\alpha} X_{j)}^{\beta} - \frac{9}{32} \ri (\Gamma^{b})_{\alpha \beta}  W_{a b} X_{(i}^{\alpha} \lambda_{j)}^{\beta} + \frac{3}{64} \ve_{a b c d e} W^{b c}{}_{\alpha (i} W^{d e \alpha}{}_{j)} W 
    \non \\ 
    &&
    + \frac{3}{16} \ri W_{[a}{}^{c} W_{b]  c}{}_{\alpha (i} (\Gamma^{b})^{\alpha \rho} \lambda_{j) \rho} + \frac{3}{128} \ri \ve_{a b c d e} W^{b c} W^{d e}{}_{\alpha (i}  \lambda_{j)}^{\alpha} 
    \non \\
    &&
    - \frac{3}{64} \ri (\Sigma_{a b})_{\alpha \beta} \ve^{b c d e f} W_{c d} W_{e f}{}^{ \alpha}\,_{(i} \lambda_{j)}^{\beta} - \frac{3}{32} \ri  \ve_{a b c d e} (\Sigma^{d e})^{\alpha}{}_{\lambda} W^{b f} W^{c}{}_{ f \alpha (i} \lambda_{j)}^{\lambda} 
    \non \\
    &&
    - \frac{3}{8} \ri  \nabla^{b}(X_{i j} W_{a b}) - \frac{3}{8} \ri (\Sigma_{a b})_{\alpha \beta}   \nabla^{b}(\lambda_{(i}^{\alpha} X_{j)}^{\beta})  - \frac{3}{8} \ri \nabla^{b}(W_{a b \alpha (i} \lambda_{j)}^{\alpha})  ~,
    \\
    J^2_{a b, {\rm Weyl}} &=& 
    \frac{3}{128}{\rm i} F_{a b} Y + \frac{9}{128}{\rm i} W W_{a b} Y + \frac{1}{4}{\rm i} \Phi_{a b i j} X^{i j} - \frac{9}{16}{\rm i} W C_{a b c d} W^{c d} - \frac{3}{16}{\rm i} C_{a b c d} F^{c d}
    \non\\
    &&
    + \frac{81}{32}{\rm i} W W_{c [a} W_{b] d} W^{c d} + \frac{45}{128}{\rm i} W W_{a b} W^{c d} W_{c d} + \frac{9}{16}{\rm i} W_{c [a} F_{b] d}W^{c d} 
    \non\\
    &&
    + \frac{9}{32}{\rm i} W_{c [a} W_{b] d} F^{c d} - \frac{3}{128}{\rm i} F_{a b} W^{c d} W_{c d} + \frac{3}{16}{\rm i} \ve_{a b c d e} W^{e (f} \nabla_{f}{F^{d) c}}
    \non\\
    &&    
    + \frac{9}{64}{\rm i} \ve_{a b c d e} F^{c d} \nabla_{f}{W^{f e}} + \frac{3}{64}{\rm i} \ve_{a b c d e} F^{c f} \nabla_{f}{W^{d e}} + \frac{3}{32}{\rm i} \ve_{c d e f [a} F^{c d} \nabla^{f}{W_{b]}{}^{e}} 
    \non\\
    &&    
    - \frac{3}{16}{\rm i} \ve_{ c d e f [a} \nabla^{f}{F^{c d}} W_{b]}{}^{e} - \frac{15}{32}{\rm i} \ve_{c d e f [a} F_{b]}{}^{c} \nabla^{f}{W^{d e}} - \frac{3}{16}{\rm i} \ve_{c d e f [a} \nabla^{f}{F_{b]}{}^{c}} W^{d e}  
    \non\\
    &&
    - \frac{9}{64}{\rm i} \ve_{a b c d e} W W^{c f} \nabla_{f}{W^{d e}} - \frac{27}{64}{\rm i} \ve_{a b c d e} W W^{d e} \nabla_{f}{W^{c f}} + \frac{9}{32}{\rm i} \ve_{c d e f [a} W \nabla^{c}(W^{d e} W_{b]}{}^{f})  
    \non\\
    &&
    - \frac{9}{64}{\rm i} \ve_{a b c d e} W^{c d} W^{e f} \nabla_{f}{W} - \frac{9}{32}{\rm i} \ve_{c d e f [a} W_{b]}{}^{c} W^{d e} \nabla^{f}{W} - \frac{3}{2}{\rm i} W \nabla^{c}{\nabla_{[a}{W_{b] c}}} 
    \non\\
    &&
    - \frac{3}{4}{\rm i} W \nabla_{[a}{\nabla^{c}{W_{b] c}}} - \frac{3}{2}{\rm i} \nabla^{c}{W} \nabla_{[a}{W_{b] c}} - \frac{3}{4}{\rm i} \nabla_{[a}{W} \nabla^{c}{W_{b] c}} - \frac{3}{4}{\rm i} W_{c [a} \nabla^{c}{\nabla_{b]}{W}}
    \non\\
    &&
    - \frac{3}{4}{\rm i} \nabla_{c}{W} \nabla^{c}{W_{a b}} - \frac{3}{4}{\rm i} W \nabla_{c}{\nabla^{c}{W_{a b}}} + \frac{15}{32} W_{a b}\,_{\alpha}^{i} X_{i}^{\alpha} W + \frac{9}{32}{\rm i} W_{a b} \lambda_{\alpha}^{i} X_{i}^{\alpha} 
    \non\\
    &&
    + \frac{3}{16}W_{[a}\,^{c \alpha i} W_{b] c \alpha i} W  + \frac{9}{32}{\rm i} W_{[a}\,^{c} W_{b] c}\,^{\alpha}\,_{i} \lambda^{i}_{\alpha} - \frac{3}{16}{\rm i} (\Sigma_{c [a})^{\alpha \beta} W^{c d} W_{b] d \alpha i} \lambda^i_\beta 
    \non\\
    &&
    + \frac{3}{16} \ri (\Sigma_{c [a})^{\alpha \beta} W_{b] d} W^{c d}\,_{\alpha i}  \lambda^{i}_{\beta} + \frac{3}{16}{\rm i} (\Sigma^{c d})^{\alpha \beta} W_{c [a} W_{b] d \alpha i} \lambda^{i}_{\beta} + \frac{3}{64}{\rm i}(\Sigma_{a b})^{\alpha \beta}  W^{c d} W_{c d}\,_{\alpha i} \lambda^{i}_{\beta}
    \non\\
    &&
     + \frac{3}{64}{\rm i} (\Sigma_{c d})^{\alpha \beta} W^{c d} W_{a b \alpha i} \lambda^{i}_{\beta}  + \frac{3}{64}{\rm i} (\Sigma_{c d})^{\alpha \beta} W_{a b} W^{c d}\,_{\alpha i}  \lambda^{i}_{\beta}
    \non\\
    &&
    - \frac{3}{128}{\rm i} \ve_{a b c d e} (\Gamma_{f})^{\alpha \beta} W^{c f} W^{d e}\,_{\alpha i}  \lambda^{i}_{\beta} + \frac{3}{64}{\rm i} \ve_{c d e f [a} (\Gamma^{c})^{\alpha \beta} W^{d e} W_{b]}{}^{f}{}_{\alpha i} \lambda^{i}_{\beta}   
    \non\\
    && 
    - \frac{3}{16}{\rm i} \ve_{a b c d e}  \lambda_{i}^{\alpha}(\Sigma^{c d})_\alpha{}^{\beta} \nabla^{e}{X^{i}_{\beta}}  - \frac{3}{16}{\rm i} \ve_{a b c d e}  X_{i}^{\alpha} (\Sigma^{c d})_\alpha{}^{ \beta} \nabla^{e}{\lambda^{i}_{\beta}} - \frac{3}{16}{\rm i}  \lambda_{i \alpha} (\Gamma_{[a})^{\alpha \beta} \nabla_{b]}{X^{i}_{\beta}}    
    \non\\
    &&
    - \frac{3}{16}{\rm i}  X_{i \alpha} (\Gamma_{[a})^{\alpha \beta} \nabla_{b]}{\lambda^{i}_{\beta}} - \frac{3}{64}{\rm i} W_{a b}\,^{\alpha}\,_{i} (\Gamma_{c})_{\alpha}{}^{\beta} \nabla^{c}{\lambda^{i}_{\beta}} + \frac{3}{32}{\rm i} W_{c [a}\,^{\alpha i} (\Gamma^{c})_{|\alpha |}{}^{\beta} \nabla_{b]}{\lambda_{i \beta}}
    \non\\
    &&
    - \frac{3}{32}{\rm i} W_{c [a}\,^{\alpha i} (\Gamma_{b]})_{\alpha}{}^{\beta} \nabla^{c}{\lambda_{i \beta}} + \frac{3}{64}{\rm i} \ve_{a b c d e} W^{c d}\,^{\alpha}\,_{i}  (\Sigma^{e f})_{\alpha}{}^{\beta} \nabla_{f}{\lambda^{i}_{\beta}}
    \non\\
    &&
    - \frac{3}{32}{\rm i} \ve_{c d e f [a} W_{b]}\,^{c \alpha}\,_{i}  (\Sigma^{d e})_{\alpha}{}^{\beta} \nabla^{f}{\lambda^{i}_{\beta}} - \frac{9}{128}{\rm i} \ve_{a b c d e} W^{c d}\,^{\alpha}\,_{i}  \nabla^{e}{\lambda^{i}_{\alpha}}  + \frac{39}{128}{\rm i} \ve_{a b c d e} \lambda^{\alpha}_{i} \nabla^{e}{W^{c d}{}_{\alpha}\,^{i}}
    \non\\
    &&
    - \frac{3}{64}{\rm i} \ve_{a b c d e}  \lambda_{i}^{ \alpha} (\Sigma^{e f})_{\alpha}{}^{\beta} \nabla_{f}{W^{c d}{}_{\beta}\,^{i}} + \frac{3}{32}{\rm i} \ve_{c d e f [a} \lambda_{i}^{\alpha} (\Sigma^{d e})_{|\alpha|}{}^{\beta} \nabla^{f}{W_{b]}{}^{c}{}_{\beta}\,^{i}} 
    \non\\
    &&
    - \frac{3}{32}{\rm i}  \lambda_{i}^{\alpha} (\Gamma^{c})_{\alpha}{}^{\beta} \nabla_{[a}{W_{b] c \beta}\,^{i}} + \frac{3}{32}{\rm i}  \lambda_{i}^{\alpha} (\Gamma_{[a})_{\alpha}{}^{\beta} \nabla^{c}{W_{b] c \beta}\,^{i}} 
    \non\\
    &&
    - \frac{3}{64}{\rm i}  \lambda_{i}^{\alpha} (\Gamma_{c})_{\alpha}{}^{\beta} \nabla^{c}{W_{a b \beta}\,^{i}} 
    ~.
\eea
\esubeq

\subsection{Log}

The Log invariant does not contribute to the linear multiplet EOM and the descendants of the vector multiplet EOM have been described in \eqref{desc-H-logW-full}. Our main focus in this subsection is thus to derive the descendants of the Weyl multiplet EOM.

The Weyl multiplet EOM in superspace was derived in subsection \ref{JEOM:logW}. It takes the following form
\bea
    J_{\log}
&=&
-\frac{3}{1024}W Y 
-  \frac{69}{1024} W^{{a} {b}} W_{{a} {b}}  W 
+  \frac{3}{32}  \Box W 
-  \frac{3}{64} F_{{a}{b}} W^{{a} {b}}
- \frac{3}{256}\l^{\alpha}_{j} X^{j}_{\alpha} 
\non\\
&&+ \frac{3}{128} F^{{a} {b}} F_{{a} {b}}  W^{-1}  
-  \frac{9}{128}  X^{i j} X_{i j}  W^{-1} 
+  \frac{3\ri}{32}  (\Gamma^{{a}})^{\alpha \beta} W^{-1}  \l^{i}_{\alpha} {\de}_{{a}}\l_{\beta i} 
\non\\
&&+  \frac{3}{64}  W^{-1} ({\de}^{{a}}W) {\de}_{{a}}W  
- \frac{3 \ri }{128} (\Sigma^{{a} {b}})^{\alpha \beta}  F_{{a}{b}} \l^{i}_{\alpha} \l_{\beta i} W^{-2}   
-\frac{3\ri}{64}    X^{i j}  \l^{\alpha}_{i} \l_{ \alpha j} W^{-2}  
\non\\
&&-  \frac{3 \ri}{32}  (\Gamma_{{b}})^{\alpha \beta}  \  \l_{\alpha}^{j} \l_{{\beta} j} W^{-2} {\de}^{{b}} W  
- \frac{3}{256}     \l^{\alpha i} \l^{\beta}_{i} \l^{j}_{\alpha} \l_{{\beta} j} W^{-3}
~.
\eea
We present here the results for one and two-derivative descendants of $J_{\log}$ based on the structures found in ~\eqref{eq:Jcomponents}. Note that, we have proven each to be explicitly $K$-primary. Due to their size, only the bosonic parts of $J^2_{a i j, \log }$ and $J^2_{a b, \log}$ are provided below. 
In subsection 4.3 of the supplementary file, the reader can find the complete expressions of $J^2_{a i j, \log }$, $J^2_{a b, \log}$, $J^3_{a \a i, {\log}}$, along with the degauged, gauged fixed $(W=1)$, bosonic part of $J^4_{a b, {\log}}$. It holds:
\bsubeq
\bea
     J^{1}_{\a i, \log} &=& ~~~~
     W ~\bigg\{ \frac{3}{256}\ve_{a b c d e} (\Gamma^{a})_{\alpha \beta}  W^{b c} W^{d e}{}^{\beta}{}_{i} + \frac{3}{32}(\Sigma^{a b})_{\alpha \beta}  W_{a}{}^{c} W_{c b}{}^{\beta}{}_{i} - \frac{3}{128} W_{a b} W^{a b}{}_{\alpha i} 
     \non\\
     && ~~~~~~~~
     + \frac{81}{512} (\Sigma_{a b})_{\alpha \beta} W^{a b} X_{i}^{\beta} + \frac{3}{128}(\Gamma^{a})_{\alpha \beta} \nabla_{a}{X_{i}^{\beta}} \bigg\}
     \non\\
     &&
     + ~~~~~~ \bigg\{ \frac{1}{16}{\rm i} (\Sigma_{a b})_{\alpha \beta} \Phi^{a b}\,_{i j}  \lambda^{j \beta} - \frac{3}{2048}{\rm i} Y \lambda_{i \alpha}   - \frac{231}{2048}{\rm i} W^{a b} W_{a b} \lambda_{i \alpha}  + \frac{3}{32}{\rm i} \Box\lambda_{i \alpha}
     \non \\
     && ~~~~~~~~
     - \frac{27}{2048}{\rm i} \ve_{a b c d e} (\Gamma^{a})_{\alpha \beta} W^{b c} W^{d e} \lambda_{i}^{\beta} + \frac{3}{256}(\Gamma^{a})_{\alpha \beta} X_{i}^{\beta} \nabla_{a}{W} 
     \non \\
     && ~~~~~~~~
     + \frac{3}{256}{\rm i} \ve_{a b c d e} (\Sigma^{a b})_{\alpha}{}^{\beta} \lambda_{i \beta} \nabla^{c}{W^{d e}}    + \frac{9}{128}{\rm i} \ve_{a b c d e} (\Sigma^{a b})_{\alpha \beta} W^{c d} \nabla^{e}{\lambda_{i}^{\beta}}
     \non\\
     && ~~~~~~~~
    + \frac{21}{128}{\rm i} (\Gamma^{a})_{\alpha \beta} \lambda_{i}^{\beta} \nabla^{b}{W_{a b}} - \frac{3}{32}{\rm i} (\Gamma^{a})_{\alpha \beta} W_{a b} \nabla^{b}{\lambda_{i}^{\beta}} 
     \non\\
     && ~~~~~~~~
     - \frac{3}{256}\ve_{a b c d e} (\Gamma^{a})_{\alpha}{}^{\beta} F^{b c} W^{d e}{}_{\beta i} - \frac{3}{128}F^{a b} W_{a b \alpha i} - \frac{3}{32}(\Sigma_{a b})_{\alpha}{}^{\beta} F^{a c} W_{c}{}^{b}{}_{\beta i}
     \non\\
     && ~~~~~~~~
     + \frac{15}{256}(\Sigma_{a b})_{\alpha \beta} F^{a b} X_{i}^{\beta} + \frac{51}{256}X_{i j} X^{j}_{\alpha} \bigg\}
     \non\\
     &&
     + W^{-1}\bigg\{ \frac{3}{64}{\rm i} \ve^{a b c d e} (\Sigma_{a b})_{\alpha \beta} F_{c d} \nabla_{e}{\lambda_{i}^{\beta}} + \frac{27}{512}{\rm i} \ve_{a b c d e} (\Gamma^{a})_{\alpha}{}^{\beta} W_{b c} F_{d e} \lambda_{i \beta} 
    \non\\
     && ~~~~~~~~
     - \frac{27}{256}{\rm i} W^{a b} F_{a b} \lambda_{i \alpha} - \frac{3}{16}{\rm i} (\Gamma_{a})_{\alpha}{}^{\beta} X_{i j} \nabla^{a}{\lambda^{j}_{\beta}} + \frac{9}{64}{\rm i} \lambda^{\beta}_{i} \lambda_{j \beta} X^{j}_{\alpha} 
     \non\\
     && ~~~~~~~~
     + \frac{3}{16}{\rm i} (\Sigma_{a b})_{\alpha \beta} \nabla^{a}{W} \nabla^{b}{\lambda_{i}^{\beta}} + \frac{3}{64}{\rm i} \ve_{a b c d e} (\Sigma^{a b})_{\alpha}{}^{\beta} \lambda_{i \beta} \nabla^{c}{F^{d e}} 
     \non\\
     && ~~~~~~~~
     + \frac{3}{32}{\rm i} (\Gamma^{a})_{\alpha \beta} \lambda_{i}^{\beta} \nabla^{b}{F_{a b}} + \frac{27}{128}{\rm i} (\Gamma^{a})_{\alpha \beta} W_{a b} \lambda_{i}^{\beta}  \nabla^{b}{W} - \frac{3}{32}{\rm i} (\Gamma_{a})_{\alpha}{}^{\beta} \lambda^j_{\beta}  \nabla^{a}{X_{i j}} 
     \non\\
     && ~~~~~~~~
     + \frac{3}{32}{\rm i} \lambda_{i \alpha} \Box W + \frac{3}{16}{\rm i} (\Sigma_{a b})_{\alpha \beta} \lambda_{i}^{\beta} \nabla^{a}{\nabla^{b}{W}} + \frac{9}{32}{\rm i} \lambda_{\alpha (i} \lambda^{\beta}_{j)} X^{j}_{\beta} \bigg\}
    \non\\
     &&
     + W^{-2}\bigg\{\frac{9}{128}{\rm i} \lambda_{i \alpha} X_{j k} X^{j k}  - \frac{3}{32}(\Gamma_{a})^{\beta \rho} \lambda_{i \alpha} \lambda_{j \beta} \nabla^{a}{\lambda^{j}_{\rho}} - \frac{3}{64}{\rm i} \lambda_{i \alpha} \nabla_{a}{W} \nabla^{a}{W}
     \non\\
     && ~~~~~~~~
     + \frac{3}{64}(\Gamma_{a})_{\alpha \beta} \lambda_{j}^{\beta} \lambda^{j \rho} \nabla^{a}{\lambda_{i \rho}} + \frac{9}{256}(\Sigma_{a b})^{\beta \rho} W^{a b} \lambda_{i \alpha} \lambda_{j \beta} \lambda^{j}_{\rho}
     \non\\
     && ~~~~~~~~
     - \frac{3}{64}(\Gamma_{a})^{\beta \rho} \lambda_{j \alpha} \lambda^{j}_{\beta} \nabla^{a}{\lambda_{i \rho}} + \frac{9}{256}(\Sigma_{a b})^{\beta \rho} W^{a b} \lambda_{j \alpha} \lambda_{i \beta} \lambda^{j}_{\rho}
     \non\\
     && ~~~~~~~~
     - \frac{3}{256}{\rm i} \ve_{a b c d e} (\Gamma^{a})_{\alpha \beta} F_{b c} F_{d e} \lambda_{i}^{\beta} + \frac{3}{64}{\rm i} (\Sigma_{a b})_{\alpha}{}^{\beta} X_{i j} F^{a b} \lambda^{j}_{\beta}
     \non\\
     && ~~~~~~~~
     + \frac{3}{128}{\rm i} \ve_{a b c d e} (\Sigma^{a b})_{\alpha}{}^{\beta} F_{c d} \lambda_{i \beta} \nabla^{e}{W} - \frac{3}{64}{\rm i} (\Gamma^{a})_{\alpha \beta} F_{a b} \lambda_{i}^{\beta} \nabla^{b}{W} 
     \non\\
     && ~~~~~~~~
     + \frac{3}{32}(\Gamma_{a})_{\alpha}{}^{\beta} \lambda^{\rho}_{i} \lambda_{j \rho} \nabla^{a}{\lambda^{j}_{\beta}} + \frac{3}{32}{\rm i} X_{i j} X^{j}\,_{k} \lambda^{k}_{\alpha}
     + \frac{3}{32}{\rm i} (\Gamma_{a})_{\alpha}{}^{\beta} X_{i j} \lambda^{j}_{\beta} \nabla^{a}{W}  
     \non\\
     && ~~~~~~~~
     + \frac{9}{256}(\Sigma_{a b})_{\alpha \beta} W^{a b} \lambda^{\rho}_{i} \lambda_{j}^{\beta} \lambda^{j}_{\rho} \bigg\}
     \non\\
     &&
     + W^{-3}\bigg\{\frac{3}{64}(\Sigma_{a b})^{\beta \rho} F^{a b} \lambda_{i \alpha} \lambda_{j \beta} \lambda^{j}_{\rho} - \frac{3}{32}X_{j k} \lambda_{i \alpha} \lambda^{j \beta} \lambda^{k}_{\beta} + \frac{3}{64}(\Sigma_{a b})_{\alpha \beta} F^{a b} \lambda^{\rho}_{i} \lambda_{j}^{\beta} \lambda^{j}_{\rho} 
     \non\\
     && ~~~~~~~~
     +\frac{3}{64} X_{i j} \lambda_{k \alpha} \lambda^{j \beta} \lambda^{k}_{\beta}  + \frac{3}{64}(\Gamma_{a})_{\alpha \beta} \lambda^{\rho}_{i} \lambda_{j}^{\beta} \lambda^{j}_{\rho}  \nabla^{a}{W} \bigg\}
     \non\\
     &&
     + W^{-4} \bigg\{\frac{9}{256}{\rm i} \lambda_{i \alpha} \lambda^{\beta j} \lambda^{\rho}_{j} \lambda^k_{\beta} \lambda_{k \rho} \bigg\} ~,
     \\
    J^2_{a i j, \log} &=& - \frac{1}{8}{\rm i} W \nabla^{b}{\Phi_{a b i j}} - \frac{1}{8}{\rm i} \Phi_{a b i j} \nabla^{b}{W} + \frac{1}{32}{\rm i} \ve_{a}{}^{b c d e} \Phi_{d e i j} F_{b c} +\frac{3}{16}{\rm i} X_{i j} \nabla^{b}{W_{a b}} 
    \non\\
    &&
     + \frac{3}{16}{\rm i} W_{a b} \nabla^{b}{X_{i j}} + \frac{3}{16}{\rm i} F_{a b} {W}^{-1} \nabla^{b}{X_{i j}} + \frac{3}{16}{\rm i} X_{i j} {W}^{-1} \nabla^{b}{F_{a b}} 
    \non \\
    &&
    + \frac{3}{16}{\rm i} X_{(i }{}^{k} {W}^{-1} \nabla_{a}{X_{j) k}} - \frac{3}{16}{\rm i} X_{i j} F_{a b} {W}^{-2} \nabla^{b}{W} + {\rm fermions} ~,
    \\
    J^{2}_{a b, \log}&=& ~~~~
    W ~ \bigg\{ \frac{3}{16}{\rm i} C_{a b c d} W^{c d} - \frac{9}{256}{\rm i} W_{a b} Y + \frac{117}{256}{\rm i} W_{a b} W^{c d} W_{c d} - \frac{3}{16}{\rm i} W^{c d} W_{c [a} W_{b] d} 
    \non\\
    &&  ~~~~~~~~
    + \frac{3}{16}{\rm i} W_{c [a} W_{b] d} W^{c d} - \frac{27}{64}{\rm i} \ve_{a b c d e} W^{c d} \nabla_{f}{W^{e f}} - \frac{9}{32}{\rm i} \ve_{c d e f [a} W_{b]}{}^{c} \nabla^{d}{W^{e f}} 
    \non\\
    && ~~~~~~~~
    - \frac{9}{64}{\rm i} \ve_{a b c d e} W^{c f} \nabla_{f}{W^{d e}} + \frac{9}{32}{\rm i} \ve_{c d e f}{}_{[a} W^{c d} \nabla^{e}{W_{b]}{}^{f}} - \frac{3}{16}{\rm i} \Box W_{a b} \bigg\}
    \non\\
    &&
    + ~~~~~~ \bigg\{\frac{3}{16}{\rm i} C_{a b c d} F^{c d} - \frac{3}{256}{\rm i} F_{a b} Y - \frac{1}{8}{\rm i} \Phi_{a b}{}^{i j} X_{i j} + \frac{3}{256}{\rm i} F_{a b} W_{c d} W^{c d} 
    \non\\
    && ~~~~~~~~
    - \frac{3}{32}{\rm i} W_{a b} F_{c d} W^{c d} - \frac{21}{16}{\rm i} W_{c [a} F_{b] d} W^{c d} - \frac{9}{16}{\rm i} W_{c [a} W_{b] d} F^{c d}  
    \non\\
    && ~~~~~~~~
    + \frac{3}{32}{\rm i} W_{a b} W_{c d} F^{c d} + \frac{3}{16}{\rm i} W_{c [a} F_{b] d} W^{c d} - \frac{9}{64}{\rm i} \ve_{a b c d e} W^{c d} W^{e f} \nabla_{f}{W}
    \non\\
    && ~~~~~~~~
    - \frac{9}{32}{\rm i} \ve_{c d e f [a} W_{b] c} W_{d e} \nabla^{f}{W} - \frac{3}{16}{\rm i} \nabla_{c}{W} \nabla^{c}{W_{a b}} - \frac{3}{4}{\rm i} \nabla_{[a}{W} \nabla^{c}{W_{b] c}} 
    \non\\
    && ~~~~~~~~
    + \frac{3}{4}{\rm i} \nabla^{c}{W} \nabla_{[a}{W_{b] c}} - \frac{9}{32}{\rm i} \ve_{a b c d e} F^{c d} \nabla_{f}{W^{e f}} - \frac{3}{16}{\rm i} \ve_{c d e f [a} F_{b]}{}^{c} \nabla^{d}{W_{e f}} 
    \non\\
    && ~~~~~~~~
    - \frac{3}{32}{\rm i} \ve_{a b c d e} W^{c f} \nabla_{f}{F^{d e}} + \frac{3}{16}{\rm i} \ve_{c d e f [a} W^{c d} \nabla^{e}{F_{b]}{}^{f}} + \frac{9}{8}{\rm i} W_{c [a} \nabla^{c}{\nabla_{b]}{W}} 
    \non\\
    && ~~~~~~~~
    - \frac{3}{4}{\rm i} W_{c [a} \nabla_{b]}{\nabla^{c}{W}} - \frac{3}{16}{\rm i} \ve_{a b c d e} W^{c d} \nabla_{f}{F^{e f}} + \frac{3}{8}{\rm i} \ve_{c d e f [a} W_{b]}{}^{c} \nabla^{d}{F^{e f}}
    \non\\
    && ~~~~~~~~
    - \frac{3}{32}{\rm i} \ve_{a b c d e} F^{c f} \nabla_{f}{W^{d e}} + \frac{3}{16}{\rm i} \ve_{c d e f [a} F^{c d} \nabla^{e}{W_{b]}{}^{f}} - \frac{3}{16}{\rm i} \Box F_{a b} - \frac{9}{16}{\rm i} W_{a b} \Box W \bigg\}
    \non\\
     &&
     + W^{-1} \bigg\{ - \frac{9}{8}{\rm i} W_{c [a} F_{b] d} F^{c d} - \frac{9}{32}{\rm i} W_{a b} F_{c d} F^{c d} - \frac{3}{64}{\rm i} \ve_{a b c d e} F^{c f} \nabla_{f}{F^{d e}} 
     \non\\
     && ~~~~~~~~
     + \frac{3}{32}{\rm i} \ve_{c d e f [a} F^{c d} \nabla^{e}F_{b]}{}^{f} - \frac{9}{64}{\rm i} \ve_{a b c d e} F^{c d} \nabla_{f}{F^{e f}} + \frac{9}{32}{\rm i} \ve_{c d e f [a} F_{b]}{}^{c} \nabla^{d}{F^{e f}}
     \non\\
     && ~~~~~~~~
     + \frac{9}{32}{\rm i} \ve_{a b c d e} W^{c d} F^{e f} \nabla_{f}{W} - \frac{3}{8}{\rm i} F_{a b} \Box W + \frac{3}{8}{\rm i} F_{c [a} \nabla^{c}{\nabla_{b]}{W}} - \frac{3}{4}{\rm i} F_{c [a} \nabla_{b]}{\nabla^{c}{W}} 
     \non\\
     && ~~~~~~~~
     + \frac{3}{16}{\rm i} \nabla_{c}{W} \nabla^{c}{F_{a b}} - \frac{3}{4}{\rm i} \nabla_{[a}{W} \nabla^{c}{F_{b] c}} + \frac{3}{4}{\rm i} \nabla^{c}{W} \nabla_{[a}{F_{b] c}} 
     \non\\
     && ~~~~~~~~
     + \frac{3}{8}{\rm i} \ve_{a b c d e} \nabla^{c}{W} \nabla^{d}{\nabla^{e}{W}} - \frac{9}{8}{\rm i} W_{c [a} \nabla_{b]}{W} \nabla^{c}{W} \bigg\}
     \non\\
     &&
     + W^{-2} \bigg\{ - \frac{3}{32}{\rm i} F_{a b} F_{c d} F^{c d} - \frac{3}{16}{\rm i} F_{c [a} F_{b] d} F^{c d} + \frac{3}{32}{\rm i} \ve_{a b c d e} F^{c d} F^{e f} \nabla_{f}{W} 
    \non \\
     && ~~~~~~~~
     + \frac{3}{16}{\rm i} \ve_{c d e f [a} F_{b]}{}^{c} F^{d e} \nabla^{f}{W} + \frac{3}{8}{\rm i} F_{c [a}\nabla_{b]}{W} \nabla^{c}{W} 
     \non\\
     && ~~~~~~~~
     + \frac{3}{16}{\rm i} F_{a b} \nabla_{c}{W} \nabla^{c}{W}\bigg\} 
     + {\rm fermions}~.
\eea
\esubeq

\subsection{Scalar curvature squared}

In this subsection, we will first describe the descendants of the supercurrent multiplet associated to the scalar curvature squared $J_{R^2}$. Unlike the Weyl-squared and Log invariants, the EOM for the linear multiplet compensator presented in subsection \ref{LEOM:Weyl} leads to new nontrivial descendants which we will compute in subsection \ref{WR2-desc}. The EOM for the vector multiplet compensator gives rise to the composite linear multiplet superfield $H^{ij}_{R^2}$, whose descendants have been given earlier in eq.~\eqref{Hij-R2-desc}.

\subsubsection{Standard Weyl multiplet}

The Weyl multiplet EOM for the curvature-squared invariant was given in \ref{JEOM:R2}. It defines a conformal supercurrent multiplet
\bea
    J_{{R^2}}
    &=&
    -\frac{3}{8} W \mathbf{W}^2  +\frac{3}{32} G^{-1} \Big(\, W G^{ij} \mathbf{X}_{ij} + G^{ij}X_{ij} \mathbf{W} -\ri G^{ij} \lambda^{\a}_{i} \bm{\l}_{\a j} \Big) \non \\
    &=&
    \frac{3}{32} G^{-1} \Big(\, G^{ij} W \mathbf{X}_{ij} + G^{ij}X_{ij} \mathbf{W} -\ri G^{ij} \lambda^{\a}_{i} \bm{\l}_{\a j}  - W\mathbf{W} F \Big) 
    \non \\
    &&
    + \frac{3 \ri}{64} G^{-3} W \mathbf{W} G_{i j} \varphi^{i \a} \varphi^j_\a 
    ~.
\eea
We present here the results for one and two-derivative descendants of $J_{R^2}$ based on the structures found in eqs.~\eqref{eq:Jcomponents}. Note that, we have proven each to be explicitly $K$-primary. Due to their size, only the bosonic parts of $J^2_{a i j, R^2}$ and $J^2_{a b, R^2}$ are provided below. In subsection 4.4 of the supplementary file, the reader can find the complete expressions of $J^2_{a i j, R^2}$, $J^2_{a b, R^2}$, $J^3_{a \a i, R^2}$, along with the degauged, gauged fixed $(W=1)$, bosonic part of $J^4_{a b, R^2}$. It holds that
\bsubeq
\bea
     J^{1}_{\a i, R^2} 
     &=& 
     ~~
     G^{-1}\bigg\{ ~
     W \boldsymbol{W}  \left(- \frac{3}{16}(\Gamma_{a})_{\alpha \beta}  \nabla^{a}{\varphi_{i}^{\beta}} + \frac{9}{64}(\Sigma_{a b})_{\alpha \beta} W^{a b} \varphi_{i}^{\beta} - \frac{27}{64}G_{i j} X^{j}_{\alpha} \right)
     \non \\
     && ~~~~~~~~
     - \frac{3}{32}{\rm i} F \left(\boldsymbol{W} \lambda_{i \alpha} + W \boldsymbol{\lambda}_{i \alpha} \right)+ \frac{3}{32}{\rm i} G_{j k} \left(\mathbf{X}^{j k} \lambda_{i \alpha} + X^{j k} \boldsymbol{\lambda}_{i \alpha}\right)
     \non \\
     && ~~~~~~~~
     + \frac{3}{16}\left( W \mathbf{X}_{i j} + \boldsymbol{W} X_{i j}\right) \varphi^{j}_{\alpha} + \frac{3}{16}{\rm i} G_{i j} (\Gamma_{a})_{\alpha}{}^{\beta} \left( W  \nabla^{a}{\boldsymbol{\lambda}^{j}_{\beta}} + \boldsymbol{W} \nabla^{a}{\lambda^{j}_{\beta}} \right) 
     \non \\
     && ~~~~~~~~
     + \frac{9}{64}{\rm i} G_{i j} (\Sigma_{a b})_{\alpha}{}^{\beta} W^{a b} \left(W \boldsymbol{\lambda}^{j}_{\beta}  + \boldsymbol{W} \lambda^{j}_{\beta} \right)- \frac{3}{32}{\rm i} \left( \lambda_i^{\beta} \boldsymbol{\lambda}_{j \beta} + \boldsymbol{\lambda}_i^{ \beta} \lambda_{j \beta}\right) \varphi^{j}_{\alpha}
     \non \\
     && ~~~~~~~~
      + \frac{3}{32}{\rm i} G_{i j} (\Sigma_{a b})_{\alpha}{}^{\beta} \left(F^{a b} \boldsymbol{\lambda}^{j}_{\beta} + \mathbf{F}^{a b} \lambda^{j}_{\beta} \right) - \frac{3}{32}{\rm i} G_{j k} \left(X_{i}{}^{j} \boldsymbol{\lambda}^{k}_{\alpha} + \mathbf{X}_i{}^{j} \lambda^{k}_{\alpha}\right)
     \non \\
     && ~~~~~~~~
     +\frac{3}{32}{\rm i} G_{i j} (\Gamma_{a})_{\alpha}{}^{\beta} \left(\lambda^{j}_{\beta}  \nabla^{a}{\boldsymbol{W}} + \boldsymbol{\lambda}^{j}_{\beta} \nabla^{a}{W} \right)\bigg\}
    \non \\
     &&
     + G^{-3} \bigg\{~ \frac{3}{64}F G_{i j} W \boldsymbol{W} \varphi^{j}_{\alpha} - \frac{3}{64}G_{j k} \left(\boldsymbol{W} \lambda_{i \alpha} + W \boldsymbol{\lambda}_{i \alpha} \right) \varphi^{j \beta} \varphi^{k}_{\beta} + \frac{3}{32}{\rm i} W \boldsymbol{W}  \varphi_i^{\beta} \varphi_{j \alpha} \varphi^{j}_{\beta}
    \non \\
     && ~~~~~~~~
     - \frac{3}{64}\cH_{a} G_{i j} (\Gamma^{a})_{\alpha}{}^{\beta} W \boldsymbol{W}  \varphi^{j}_{\beta} - \frac{3}{32}G_{j k} (\Gamma_{a})_{\alpha}{}^{\beta} W \boldsymbol{W} \nabla^{a}{G_i{}^{j}} \varphi^{k}_{\beta}
     \non \\
     && ~~~~~~~~
     - \frac{3}{32}G_{i j} G_{k l} \left(W \mathbf{X}^{k l} + \boldsymbol{W} X^{k l} \right) \varphi^{j}_{\alpha} + \frac{3}{32}{\rm i} G_{i j} G_{k l} \lambda^{k \beta} \boldsymbol{\lambda}^{l}_{\beta}  \varphi^{j}_{\alpha} \bigg\}
    \non \\
     &&
     + G^{-5} \bigg\{ - \frac{9}{64}{\rm i} G_{i j} G_{k l} W \boldsymbol{W} \varphi^{j}_{\alpha} \varphi^{k \beta} \varphi^{l}_{\beta} \bigg\}~,
    \\   
    J^{2}_{a i j, R^2}&=& ~~
    G^{-1} \bigg\{ ~ \frac{3}{16}{\rm i} \cH_{a} \left(W \mathbf{X}_{i j} + \boldsymbol{W} X_{i j}\right)  - \frac{3}{32}{\rm i} \ve_{a}\,^{b c d e} G_{i j} F_{b c} \mathbf{F}_{d e}
    \non \\
    && ~~~~~~~~~~~
    - \frac{9}{8}{\rm i} G_{i j} \nabla^{b}\left( W \boldsymbol{W} W_{a b}\right) + \frac{3}{8}{\rm i} \left(W \mathbf{X}_{k (i} + \boldsymbol{W} X_{k (i} \right) \nabla_{a}{G_{j)}\,^{k}} 
    \non \\
    && ~~~~~~~~~~~
    - \frac{3}{8}{\rm i} G_{i j} \nabla^{b}\left(W \mathbf{F}_{a b} + \boldsymbol{W} F_{a b} \right) - \frac{3}{8}{\rm i} G^k{}_{(i}  \nabla_{a} \left(W  \mathbf{X}_{j) k} + \boldsymbol{W} X_{j) k} \right)  \bigg\}
    \non \\
    &&
    + G^{-3} \bigg\{  - \frac{3}{32}{\rm i} \cH_{a} G_{i j} G_{k l} \left( W \mathbf{X}^{k l} + \boldsymbol{W} X^{k l}\right)
    \non \\
    && ~~~~~~~~~~~
    - \frac{3}{16}{\rm i} G_{l m} \left(W \mathbf{X}^{l m}  + \boldsymbol{W} X^{l m} \right) G_{k (i} \nabla_{a}{G_{j)}\,^{k}} \bigg\} 
+ {\rm fermions}~,
    \\ 
    J^{2}_{a b, R^2} &=& ~~ G^{-1} \bigg\{~
    \frac{3}{16}{\rm i} F \left(\boldsymbol{W} F_{a b} + W \mathbf{F}_{a b}\right) + \frac{9}{16}{\rm i} F W \boldsymbol{W} W_{a b} + \frac{3}{8}{\rm i} W \boldsymbol{W}  \nabla_{[a}{\cH_{b]}} 
    \non \\
    && ~~~~~~~~~~~
    + \frac{3}{8}{\rm i} \Phi_{a b i j} G^{i j} W \boldsymbol{W} {G}^{-1} \bigg\} 
    \non \\
    &&
     + G^{-3} \bigg\{~
     \frac{3}{16}{\rm i} G_{i j} W \boldsymbol{W}  \cH_{[a} \nabla_{b]}{G^{i j}} - \frac{3}{16}{\rm i} G_{i j} W \boldsymbol{W}  \nabla_{[a}{G^{i k}} \nabla_{b]}{G^{j}{}_{k}}\bigg\} 
     \non\\
     &&
     + {\rm fermions}~.
\eea
\esubeq

\subsubsection{Linear multiplet}
\label{WR2-desc}

Recall that the EOM for the auxiliary field $F$ for the curvature-squared invariant was given in subsection \ref{LEOM:Weyl}. It leads to the expression $W_{R^2}$, eq.~\eqref{F-eom-comp}, which defines a composite vector multiplet. 
We present here, for the first time, the results for the bosonic parts of the component descendants of $W_{R^2}$ based on the structures defined in eqs.~\eqref{components-vect}. Note that, we have proven the bosons for each to be explicitly $K$-primary. Due to their size, the entire component results with fermions of $X^{i j}_{R^2}$ and $F_{a b, R^2}$ as well as the fermionic descendant $\l^i_{\a, R^2}$ are given in subsection 4.4 of the corresponding supplementary file. It holds:
\bsubeq
\bea
    X^{i j}_{R^2} 
    &=& ~~
    G^{-1} \bigg\{ \Box(
    W \mathbf{X}^{i j} + \mathbf{W} X^{i j} )   + \frac{3}{32}\left( W^{{a} {b}} W_{{a} {b}} - Y\right)\left(W \mathbf{X}^{i j} + \mathbf{W} X^{i j}\right)     \bigg\}
    \non \\
    && 
    +G^{-3} \bigg\{ \left(\frac{1}{2} \cH_{{a}} G^{k (i} -  G^{k l} {\nabla}_{{a}}{G^{(i}{}_{l}}\right){\nabla}^{{a}}\left(\mathbf{X}^{j)}{}_{k} W + X^{j)}{}_{k} \mathbf{W}\right)
    \non \\
    &&~~~~~~~~ 
    + \left(\frac{1}{2}\cH_{{a}} G^{i j} - G^{k (i} {\nabla}_{{a}}{G^{j)}{}_{k}}\right)\bigg[{\nabla}_{{b}}{(W \mathbf{F}^{{a} {b}}}) + {\nabla}_{{b}}({\mathbf{W} F^{{a} {b}}}) + 3{\nabla}_{{b}}\left( W \mathbf{W} W^{{a} {b}}\right) 
    \non \\
    &&~~~~~~~~ ~~~~~~
    + \frac{1}{4} \ve^{{a} {b} {c} {d} {e}} F_{{b} {c}} \mathbf{F}_{{d} {e}}\bigg] + \frac{3}{8} F G^{k (i} \left( X^{j) l} \mathbf{X}_{k l} + \mathbf{X}^{j) l} X_{k l}\right)
    \non \\
    &&~~~~~~~~ 
    - \frac{1}{2} F G^{i j}  \left(W \Box{\mathbf{W}} + \mathbf{W} \Box{W}\right)+ \frac{3}{4} F G^{i j} W_{{a} {b}} \left( W \mathbf{F}^{{a} {b}} + \mathbf{W} F^{{a} {b}}\right)  
    \non \\
    &&~~~~~~~~ 
    +  \frac{3}{32}\left(13 W^{{a} {b}} W_{{a} {b}} - Y \right)F G^{i j} W \mathbf{W}  - \frac{1}{8} \left( {F}^{2} + \cH_{{a}} \cH^{{a}}\right) \left(W \mathbf{X}^{i j} + \mathbf{W} X^{i j} \right) 
    \non \\
    &&~~~~~~~~
    + \bigg[- \frac{1}{2}{\nabla}^{{a}}{G^{k i}} {\nabla}_{{a}}{G^{j l}}  - \frac{1}{2} G^{k l} \Box{G^{i j}}  
    \non \\
    &&~~~~~~~~ ~~~~~~
    +\frac{3}{64} G^{i j} G^{k l}(Y - W^{{a} {b}} W_{{a} {b}}) \bigg] \left(W \mathbf{X}_{k l} + \mathbf{W} X_{k l}\right)
    \non \\
    &&~~~~~~~~
    - \frac{5}{8} F G^{k l} \left(X^{i j} \mathbf{X}_{k l} + \mathbf{X}^{i j} X_{k l}\right) + \frac{1}{2} \cH_{{a}} \left(\mathbf{W} X^{k (i} + W \mathbf{X}^{k (i}\right){\nabla}^{{a}}{G^{j)}{}_{k}}\non \\
    &&~~~~~~~~
    + \frac{1}{4}F G^{i j} \mathbf{F}^{{a} {b}} F_{{a} {b}}  - \frac{1}{2}F G^{i j} {\nabla}^{{a}}{W} {\nabla}_{{a}}{\mathbf{W}} + \frac{5}{4} F G^{k l} X^{(i}{}_{k} \mathbf{X}^{j)}{}_{l}     \bigg\}
    \non \\
    && 
    +G^{-5} \bigg\{ \bigg[\frac{3}{4} G_{k l} {\nabla}^{{a}}{G^{i k}} {\nabla}_{{a}}{G^{j l}} + \frac{3}{16} \left(\cH_{{a}} \cH^{{a}} + F^{2}\right) G^{i j}  
    \non \\
    &&~~~~~~~~ ~~~~~~
    - \frac{3}{4} \cH_{{a}}  G^{ k (i}{\nabla}^{{a}}{G^{j)}{}_{k}}\bigg] \left(W \mathbf{X}_{m n}G^{m n} + \mathbf{W} X_{m n}G^{m n}\right) \bigg\} 
    \non\\
    &&
    + {\rm fermions}~,
    \\
    F_{{a} {b}, R^2} 
    &=&    
    6 {\nabla}_{[{b}}\Big( G^{-1} W_{{a}]}{}^{{d}} \left(W {{\nabla}_{{d}}}\mathbf{W} + \mathbf{W} {{\nabla}_{{d}}}{W}\right)\Big) + 2 {\nabla}_{[{b}}\Big( G^{-1}   {\nabla}^{{e}} \left({W}{\mathbf{F}_{{a}] {e}}} + {\mathbf{W}} {{F}_{{a}] {e}}} \right) \Big)
    \non \\
    &&
    -6 {\nabla}_{[{a}} \left( G^{-1} \mathbf{W} {W}  {\nabla}^{{e}}{W_{{b}] {e}}}\right) - \frac{1}{4}\ve_{a b c d e}    \nabla_f \Big( G^{-1}\left(F^{c d} \mathbf{F}^{e f} + \mathbf{F}^{c d} F^{e f}\right) \Big) 
    \non \\
   &&
   -\frac{1}{2}  \ve_{ {c} {d} {e} {f} [{a}}  {\nabla}^{{f}} \Big( G^{-1} \left( \mathbf{F}_{{b}]}{}^{{c}} F^{{d}  {e}}\,+{F}_{{b}]}{}^{{c}} \mathbf{F}^{{d}  {e}}\,\right) \Big) 
   \non \\
   &&
   - \frac{1}{2}{\nabla}_{[{a}}\Big( G^{-3} G^{i j} \cH_{{b}]}   \left( {W} \mathbf{X}_{i j}+ {\mathbf{W}} {X}_{i j}\right)\Big)
   \non \\
   &&
   + G^{-3} {\nabla}_{[{a}}\Big(G_{i}{}^{k}\left(\mathbf{X}^{ i j }{W}+ {X}^{i j } {\mathbf{W}} \right) {\nabla}_{{b}]}{G_{j k}}\Big)
   \non \\
   &&
   - \frac{G^{-3}}{2}   \left(W \mathbf{X}^{i j}+ \mathbf{W} {X}^{i j}\right) {\nabla}_{{a}}{G_{i}{}^{k}} {\nabla}_{{b}}{G_{j  k}} \non \\
   &&
   +\frac{3}{4} G^{-5} G_{i j} G^{k l}  \left(W \mathbf{X}^{i j} + \mathbf{W} {X}^{i j}\right) {\nabla}_{{a}}{G_{k}{}^{m}} {\nabla}_{{b}}{G_{l m}} \label{FabR2}
   \non\\
   &&
   + {\rm fermions}
   ~.
\eea
\esubeq

Having described in detail all the descendants of the EOMs for the Weyl multiplet and compensators, let us end this section by commenting on the physical relevance of our results; in particular, how these might be useful for understanding higher-order corrections to Einstein's field equations. Recall that, after gauge fixing, the Poincar\'e supergravity multiplet contains the metric $g_{mn}$, gravitini $\psi_{m}{}_{\a}^{i}$, graviphoton $A_{m}$, along with the matter fields $D$, $\chi_{\a}^{i}$, and $W_{ab}$. As indicated in Table 3 and looking at the descendants of the Weyl multiplet EOM, the component current $J^4_{m n} = e_m{}^a \, e_n{}^b J^4_{a b}$ is associated to the variation of the traceless part of the metric, $\d \hat{g}_{mn}$. Hence, to compute higher-order corrections to the traceless part of Einstein's equations, one essentially needs to consider $J^4_{ab}$ from all the three four-derivative invariants: $J^4_{ab, \rm Weyl}$, $J^4_{ab, \rm log}$, and $J^4_{ab, R^2}$. The latter are given in the auxiliary file, specifically in subsections 4.2.5, 4.3.5 and 4.4.5, respectively. On the other hand, the trace part of Einstein's equations comes from taking a linear combination of the components $F_{\rm Weyl}$ \eqref{desc-H-Weyl-full} and $F_{\rm log}$ \eqref{desc-H-logW-full}.

The component current $J^3_{a \a i}$ is associated to variation of the gamma-traceless part of the gravitini, $\d \hat{\psi}_{m}{}_{i}^{\a}$, while the variation with respect to the trace part is given by the fermionic descendant of the vector compensator EOM \eqref{VM-eom}. The composite field $\cH_{a}$ coming from \eqref{VM-eom} is associated to the variation of graviphoton of the Poincar\'e supergravity multiplet. Finally, the composite field strength $F_{ab}$ coming from \eqref{LM-eom} (see also \eqref{FabR2}) is associated to the variation of the Hodge dual of the three-form potential $b_{mnp}$.

\section{Conclusion}
\label{Conclusion}

By using an interplay between superspace and superconformal tensor calculus techniques, in our paper we have presented for the first time a comprehensive description of the component structure of the supersymmetric completions for all curvature-squared invariants of five-dimensional, off-shell (gauged) minimal supergravity. Our results include both bosonic as well as fermionic contributions for the five-parameter family of theories described by the action
\bea
S_{\rm HD}&=&  S_{\rm gSG} + \a \, S_{\rm Weyl} +\b  \, S_{{\log}} + \g \, S_{R^2}
~,
\label{HD-action-2}
\eea
where $S_{\rm gSG}$ describes the supersymmetric two-derivative (gauged) minimal Poincar\'e supergravity with a cosmological constant, while the three independent curvature-squared invariants are described by eqs.~\eqref{S-Weyl-0}, \eqref{S-log-0}, \eqref{action-R2}, and further results in section \ref{section-5}. As a first application of our results, we have also analysed the structure of the multiplets of equations of motion for the theory defined by \eqref{HD-action-2}. It did take more than fifteen years from the first papers on curvature-squared supergravity in five dimensions \cite{HOT,OP2,OP1,BKNT-M14} to complete the detailed construction of these invariants. We expect that the results in our paper will find several developments and applications. We comment here on a few directions.

A key motivation of our analysis was the recent developments of \cite{Bobev:2020zov,Bobev:2020egg,Bobev:2021oku,Bobev:2021qxx,Liu:2022sew,Hristov:2021qsw,Hristov:2022lcw,Bobev:2022bjm,Cassani:2022lrk} studying $\alpha^\prime$ corrected gravity for precision tests of the AdS/CFT correspondence. 
In \cite{Bobev:2021qxx,Bobev:2022bjm,Cassani:2022lrk} the formulation of minimal gauged supergravity in 5D based on the standard Weyl multiplet was used, as in our paper. 
Only the Weyl-squared and scalar curvature-squared invariants were employed in these works, since the Log (Ricci squared) invariant had not been constructed by using component fields. Arguments were given to justify why only two invariants might suffice to compute (BPS) on-shell observables (as for example the entropy of five-dimensional black holes), but it remains an open question to understand to which extent only a subset of invariants suffices in general. We expect that, for generic observables and calculations in quantum corrected minimal gauged supergravity, the knowledge of all invariants could play an important role. 

In \cite{Liu:2022sew} three invariants were considered in a dilaton Weyl background to engineer Poincar\'e supergravity. However, this analysis did assume that the gauged two-derivative supergravity could be added to the three ungauged curvature-squared invariants of \cite{BRS,OP2} to obtain a five-parameter family of locally supersymmetric invariants. A description of the invariants in a gauged dilaton Weyl background was given in \cite{Gold:2023ymc}, see also \cite{GJSGMY-2023} for an extended upcoming analysis.
We have discussed off-shell formulations based on the standard Weyl multiplet and the dilaton Weyl multiplet. The dilaton Weyl multiplet mentioned above is the one first introduced in \cite{Ohashi3, Bergshoeff1} and extended in \cite{CO} to a version suitable for gauged supergravity. This is defined as an on-shell vector multiplet coupled to the standard Weyl multiplet -- for this reason we will denote it as vector-dilaton Weyl multiplet. Recently, it was noticed in \cite{Gold:2022bdk,Hutomo:2022hdi} that there exist even more variant versions of Weyl multiplets. In fact, by considering an on-shell hypermultiplet in a standard Weyl multiplet background one can obtain another variant Weyl multiplet of off-shell conformal supergravity which was referred to as hyper-dilaton Weyl. Thanks to the results in our paper, we can straightforwardly obtain a basis of independent curvature-squared invariants in all the dilaton Weyl backgrounds. The vector-dilaton Weyl multiplet has been largely used to engineer general matter couplings in Poincar\'e supergravity \cite{Ohashi3, Bergshoeff1}, including the recent curvature-squared analysis in \cite{Gold:2023ymc,GJSGMY-2023}. By using another new independent off-shell scalar curvature-squared invariant that we constructed in \cite{GHKT-M}, it would be possible to engineer general off-shell supergravities based on the hyper-dilaton Weyl multiplet of \cite{Gold:2022bdk,Hutomo:2022hdi}. An interesting feature of this formulation, once coupled to a system of vector multiplets, is how it is straightforward to describe general gaugings off-shell. It is in fact a natural avenue of future investigations to extend the results in this paper to general off-shell matter-coupled supergravity, at least for arbitrary couplings with physical vector multiplets. These models would be important for applications involving $\alpha^\prime$ corrected string compactifications to five dimensions where physical scalar multiples would describe coordinates of the internal geometry and their moduli.

 Another natural question concerning curvature-squared invariants is how the results we have obtained here and in \cite{Gold:2023ymc,GJSGMY-2023} connects with the known invariants in lower and higher dimensions. In \cite{Banerjee:2011ts} an off-shell dimensional reduction approach was developed. There it was noticed how the Weyl-squared invariant in 5D, once reduced to 4D, leads to a linear combination of the 4D Weyl squared and the 4D Log invariants. It would be interesting to extend this analysis by using all the independent 5D invariants that are now available from this year. Knowing the precise connection between 4D and 5D might help to answer questions in AdS/CFT realising the roles that half BPS (e.g., 4D $F$-term) and $D$-term invariants can generally have in both dimensions. The classification of all curvature-squared invariants for six-dimensional $\cN=(1,0)$ minimal Poincar\'e supergravity was obtained six years ago in \cite{NOPT-M17,Butter:2018wss}. One could develop an off-shell 5D/6D map and identify how the invariants in different dimensions are related -- see \cite{BRS} for the 5D/6D Riemann squared based on a dilaton Weyl multiplet. One wonders if understanding off-shell relations between higher-derivative invariants in different dimensions might shed some light to better understand higher-derivative supergravities in dimensions higher than six where superconformal techniques are not (directly) applicable.

 It is known that among all the curvature-squared terms one predominant role is played by the Gauss-Bonnet combination. This is a topological term in four dimensions. In ${\rm D}>4$ supersymmetric Gauss-Bonnet invariants are expected to play a key role in the description of the first $\alpha^\prime$ corrections to compactified string theory \cite{Zwiebach:1985uq,Deser:1986xr}. A reason being that, though higher derivative, the Gauss-Bonnet combination,
 \bea
    \cL_{\rm GB} &=& \
    -\frac{1}{4}\cR^{a b c d} \cR_{a b c d} + \cR^{a b} \cR_{a b} -\frac{1}{4} \cR^2
    ~,
    \label{GB-0}
\eea
leads to the same physical degrees of freedom as standard massless gravity with a cosmological constant when studying fluctuations around maximally symmetric backgrounds. In five dimensions, one remarkable property of $\alpha^\prime$ corrected minimal gauged supergravity is that, up to perturbatively integrating out auxiliary fields and making curvature dependent field redefinitions, the complicated Lagrangian in \eqref{HD-action-2} simplifies to the following one \cite{Gold:2023ymc}
\bea
e^{-1}{\cal L}_{2\partial+4\partial}
=
a_0 \cR
+a_1\kappa^2 
+a_2 f^{ab} f_{ab}
+a_3\varepsilon^{abcde}v_{a}f_{bc} f_{de}
 +\alpha {\cal L}_{\rm GB}^{\rm on-shell}
 ~.
 \label{on-shell-24}
\eea
Here, the various constants $a_i$ are functions of the five independent constants in \eqref{HD-action-2}, and ${\cal L}_{\rm GB}^{\rm on-shell}$ is an appropriate linear combination of \eqref{GB-0} and the following terms: 
$f^{ab} f_{ab}$,
$C_{abcd}f^{ab}f^{cd}$, and $\varepsilon^{abcde}v_{a}\cR_{bc}{}^{fg}\cR_{defg}$; see \cite{Gold:2023ymc} for more detail.\footnote{Note that the result obtained in \cite{Gold:2023ymc} is based on the gauged dilaton Weyl multiplet. However, the on-shell action obtained by perturbatively integrating out auxiliary fields and performing field redefinitions would agree with the one obtained by starting from the standard Weyl multiplet invariants of our current paper. We also underline that the result in equation \eqref{on-shell-24} in a standard Weyl multiplet and in gauged supergravity was first obtained in \cite{Cassani:2022lrk} -- see also \cite{Bobev:2021qxx,Bobev:2022bjm} -- by using only the Weyl-squared and the scalar curvature-squared invariants.}

One might ask what is an off-shell extension of ${\cal L}_{\rm GB}|_{\rm onshell}$. By looking only at the curvature-squared invariants that we constructed in this paper, it is straightforward to realise that combinations of the independent curvature-squared invariants with proper coefficients leads to off-shell supersymmetric completions of the Gauss-Bonnet combination.
One can show that this can be obtained by using just two curvature-squared invariants: the Weyl squared and Log. It is simple to show that the combination
\bea
\cL^{\rm off-shell}_{\rm GB} 
&=& 
\cL_{\rm Weyl} 
- 4 \cL_{\log }  
~,
\label{GB-offshell}
\eea
is such that the kinetic terms for the SU(2)$_R$ connection $\phi_{m}{}^{ij}$ cancel 
between the Weyl squared \eqref{Weyl-action-degauged} and the Log \eqref{log-action-degauged} invariants. These observations indicate that there is no massive graviton nor dynamical massive vector generated by \eqref{GB-offshell}, as expected from a Gauss-Bonnet invariant. The equation of motion for the field $D$ are algebraic, as it is clear from the $D$ dependent terms in the bosonic part of $\cL^{\rm off-shell}_{\rm GB}$,
\bea
    \cL^{\rm off-shell}_{\rm GB,D} &=& 
      - 32{D}^{2} + 2 \cR D - 12 F_{a b} W^{{a} {b}} {D}     
     + 4 D F^{{a} {b}} F_{{a} {b}}
     \non \\
     &&  - \frac{39}{2} D   W^{{a} {b}} W_{{a} {b}} - 12 D   X^{i j} X_{i j} ~.  
\eea
It is then possible to eliminate $D$ by substituting its equation of motion in the action. One can easily check that the final result is $\cL_{\rm GB}$ plus terms that leads to a supersymmetric completions. Moreover, one can prove that in an ungauged vector-dilaton Weyl background, the combination above leads to the off-shell Gauss-Bonnet invariant first constructed in \cite{OP1,OP2}.
The simplifications arising with the 5D Gauss-Bonnet invariant make easier the study of solutions of gauged supergravity modified with four-derivative terms, which are known to be difficult to analyse in general. One might now construct new classes of solutions and prove if and how solutions of the two-derivative theory are modified.

So far we have commented mostly on four-derivative corrections. However, the building blocks of our paper allow to construct invariants with arbitrary high orders in $\alpha^\prime$. For example, now that we have obtained all the information about the three independent composite linear multiplets associated to the Weyl squared, Log, and scalar curvature-squared invariants, we might construct an arbitrary (appropriately constrained) function of these invariants as for the system of linear multiplets engineered in \cite{Ozkan:2016csy}.\footnote{The results of \cite{Ozkan:2016csy} extend to five dimension the four dimensional construction of \cite{deWS}. See \cite{BK11}, and \cite{BKNT-M14} in 5D, for their superspace formulation and for other (infinite series of) higher-derivative terms.} To be more precise, we label $G_I^{ij}=(G^{ij},H^{ij}_{\rm VM},H^{ij}_{\rm Weyl},H^{ij}_{\log},H^{ij}_{R^2})$ with $I=0,1,2,3,4$ as the different linear multiplets used in our paper, the compensator, and the various composite ones. Then, in a covariant projective superspace approach,\footnote{We refer the reader to \cite{BK11,BKNT-M14}, and the recent review \cite{Kuzenko:2022ajd} and references therein, for detail about the covariant projective formalism and its application to higher derivative supergravity. More in general, projective superspace \cite{Karlhede:1984vr,Lindstrom:1987ks,Lindstrom:1989ne}, together with the related harmonic superspace \cite{Galperin:1984av,Galperin:2001seg}, are powerful techniques to formulate general off-shell matter couplings in supersymmetric models with eight real supercharges in ${\rm D}\leq 6$ dimensions where infinite number of auxiliary/matter fields might be necessary. See  \cite{Lindstrom:2009afn,K06,KT-M_5D2,KT-M_5D3,KT-M08,Kuzenko:2008ep,Kuzenko:2009zu,BKNT-M14,Linch:2012zh,Butter:2014gha,Butter:2014xua,Butter:2015nza,Kuzenko:2022ajd} for further references in the curved (supergravity) case.} one can formulate a half BPS invariant built out of the linear multiplets with the projective superspace Lagrangian given by
\bea
\cL^{(2)}
=
G_I^{(2)}\cF^I[G_J^{(2)}]
~,~~~~~~
G_I^{(2)}
=
v_iv_jG_I^{ij}
~.
\label{L-projective}
\eea 
Here $v_i$ are constant isotwistor (twistor-like) coordinates describing an auxiliary $\mathbb{C}P^1$ manifold which is used to engineer a 5D locally supersymmetric invariant associated to the projective Lagrangian $\cL^{(2)}$, see \cite{BKNT-M14}. The functions $\cF^I[G_J^{(2)}]$ are only required to be homogeneous of degree zero in their argument: 
\bea
G_K^{(2)}\frac{\pa\cF^I[G_J^{(2)}]}{\pa G_K^{(2)}}=0
~.
\eea 
One could also consider a single function $\mathbb{F}[G_I^{(2)}]$ homogenous of degree one in $G_I^{(2)}$, which is locally equivalent to the description above.
 The details are not important here, but what we want to stress is the existence of a function that can lead to arbitrarily high curvature corrections in 5D minimal supergravity. In four dimensions, it is known how for a large class of observables and applications to AdS/CFT an $F$-term constructed out of an holomorphic function with arguments given by the composite vector multiplets describing the 4D curvature squared invariants of $\cN=2$ supergravity \cite{LopesCardoso:1998tkj,Mohaupt:2000mj,deWit:2010za,Butter:2013lta,Butter:2014iwa,Kuzenko:2015jxa} suffices to computing observables based on the on-shell action as for example the entropy of (supersymmetric) black holes -- see \cite{Hristov:2021qsw,Hristov:2022lcw}, and references therein, for inspiring recent results in the four dimensional case. Along this line, it would be interesting to investigate if and how the model described by \eqref{L-projective} encodes enough information to study general higher-derivative correction in minimal supergravity in the AdS/CFT context.

\vspace{0.3cm}
\noindent
{\bf Acknowledgements:}\\
We are grateful to M.\,Ozkan and Y.\,Pang for discussions and collaboration related to this work.
G.T.-M. thanks N.\,Bobev, and K.\,Hristov, for recent discussions. G.T.-M. is also grateful to D.\,Butter, S.\,M.\,Kuzenko, and J.\,Novak for the joint collaboration on \cite{BKNT-M14} that led to several of the structures used in this paper.
This work is supported by the Australian Research Council (ARC) Future Fellowship FT180100353 and by the Capacity Building Package of the University of Queensland.
G.G. and S.K. are supported by the postgraduate scholarships at the University of Queensland.

\appendix

\section{Conformal superspace summary}
\label{App-A}

Here we collect results about conformal superspace in the traceless frame of \cite{BKNT-M14} focusing on the ingredients relevant to our discussion in the paper. We adhere with the notations and conventions of \cite{BKNT-M14}.
The reader should also look at \cite{Hutomo:2022hdi}, where some typos from \cite{BKNT-M14} were fixed.

The Lorentz generators act on the superspace covariant derivatives $\de_{\hA} = (\de_{\ha}, \de_{\hal}^{i})$ in the following way
\bsubeq \label{SCA}
\begin{align} 
[M_{\ha \hb} , M_{\hc \hd}] &= 2 \eta_{\hc [\ha} M_{\hb] \hd} - 2 \eta_{{d} [ \ha} M_{\hb ] \hc}~, \\
[M_{\ha \hb}, \de_\hc] &= 2 \eta_{\hc [\ha} \de_{\hb]}~, \\ 
[M_{\hal \hbe} , \de_\hga^i ] &= \eps_{\hga (\hal} \de_{\hbe)}^i~, 
\end{align}
where $M_{\hal \hbe} = 1/2 (\S^{\ha \hb})_{\a\b} M_{\ha \hb}$. 
The $\rm SU(2)_{R}$ and dilatation generators satisfy
\begin{align}
[J^{ij} , J^{kl}] &= \eps^{k(i} J^{j) l} + \eps^{l (i} J^{j) k}~,
\quad 
[J^{ij} , \de_\hal^k ] = \eps^{k (i} \de_\hal^{j)}~\, \\
[\mathbb{D} , \de_\ha] &= \de_\ha~, 
\quad 
[\gD , \de_\hal^i ] = \hf \de_\hal^i~.
\end{align}
The Lorentz and $\rm SU(2)_R$ generators act on the special conformal generators $K_\hA = (K_\ha , S_{\hal i})$ according to the rules
\begin{align} 
[M_{\ha \hb} , K_\hc] &= 2 \eta_{\hc [\ha} K_{\hb]} \ , 
\quad
[M_{\hal \hbe} , S_\hga^i ] = \eps_{\hga (\hal} S_{\hbe)}^i \,
\quad
[J^{ij} , S_\hal^k ] = \eps^{k (i} S_\hal^{j)}~, 
\end{align}
while the dilatation generator acts on $K_{\hA}$ as
\begin{align} 
[\mathbb{D} , K_\ha] = - K_{\ha}~, 
\quad 
[\gD , S_{\hal i} ] = - \hf S_{\hal i}~.
\end{align}
The generators $K_{\hA}$ obey the only nontrivial anti-commutation relation
\begin{align} 
\{ S_\hal^i , S_\hbe^j \} &= - 2 \ri \,\eps^{ij} (\G^{\hc})_{\hal\hbe} K_{\hc}~.
\end{align}
The algebra of $K_{\hA}$ with $\de_{\hA}$ is given by
\begin{align}
[K_\ha , \de_\hb] &= 2 \eta_{\ha \hb} \gD + 2 M_{\ha\hb}~, \\
{[} K_\ha , \de_\hal^i {]} &=  \ri (\G_\ha)_\hal{}^\hbe S_{\hbe}^i~, \\
\{ S_{\hal i} , \de_\hbe^j \} &= 2 \eps_{\hal \hbe} \d_i^j \gD - 4 \d_i^j M_{\hal\hbe} + 6 \eps_{\hal\hbe} J_i{}^j~, \label{alg-S-spinor} \\
[S_{\hbe i}, \nabla_\ha] &= 
	\ri (\Gamma_\ha)_\hbe{}^\hal \nabla_{\hal i}
	- \frac{1}{4} W_\ha{}^\hb (\Gamma_\hb)_\hbe{}^\hal S_{\hal i}
	+ \frac{\ri}{8} (\G_\ha \G^\hb)_\hbe{}^\hga X_{\hga i} K_\hb
	- \frac{\ri}{4} W_{\ha\hb\hbe i} K^\hb~.
\end{align}
\esubeq

The anticommutator of two spinor derivatives,
$\{{\de}_{\hal}^i,{\de}_{\hbe}^j\}$, has the following non-zero
 torsion and curvatures
\begin{subequations}  \label{new-frame_spinor-spinor}
\begin{align}
\scT_{\hal}^i{}_\hbe^j{}^\hc &= 2 \ri \ve^{ij} (\Gamma^\hc)_{\hal\hbe}~, \\
{\mathscr{R}}(M)_{\hal}^i{}_\hbe^j{}^{\hc\hd}
	&= 2 \ri \ve^{ij} \ve_{\hal\hbe} W^{\hc\hd}
		+ \ri \ve^{ij} (\Gamma_\hb)_{\hal\hbe} \tilde W^{\hb\hc\hd} ~, \\
{\mathscr{R}}(S)_{\hal}^i{}_\hbe^j{}^{\hga k}
	&= \frac{3\ri}{4} \ve^{ij} \ve_{\hal\hbe} X^{\hga k}
		+ \ri \ve^{ij} \delta^\hga_{[\hal} X_{\hbe]}^k ~, \\
{\mathscr{R}}(K)_{\hal}^i{}_\hbe^j{}^{\hc}
	&= - \frac{\ri}{2} \eps^{ij} \eps_{\hal\hbe} \nabla^\hb W_\hb{}^\hc
	+ \frac{\ri}{2} \eps^{ij} (\Gamma^\ha)_{\hal\hbe} \nabla^\hd \tilde W_{\hd\ha}{}^\hc
	- \frac{\ri}{32}  \eps^{ij} (\Gamma^\hc)_{\hal\hbe} Y
	\non\\ & \quad
	+ \frac{\ri}{4} \eps^{ij} \eps_{\hal\hbe} \tilde W_\hc{}^{\hd\he} W_{\hd\he}
	+ \frac{\ri}{2} \eps^{ij} (\Gamma^\ha)_{\hal\hbe}
	\Big(W_{\ha\hd} W^{\hc\hd} 
	- \frac{3}{16} W^{\hb\hd}W_{\hb\hd} \delta_\ha{}^\hc	
	\Big)~.
\end{align}
\end{subequations}
The non-vanishing torsion and curvatures in the spinor-vector commutator $[\de_{\hb}, \de_{\hal}^i]$ are:
\begin{subequations} \label{new-vectspinor}
\begin{align}
\scT_{\hb}{}_{\hal}^i{}^{{\g}}_k
	&= \frac{1}{4} \delta^i_k \Big(
		3(\Gamma_\hb)_\hal{}^\hbe W_\hbe{}^{{\g}}
		- W_\hal{}^\hbe (\Gamma_\hb)_\hbe{}^{{\g}}
	\Big)~, \\
{\mathscr{R}}(D)_{\hb}{}_{\hal}^i &= -\frac{1}{4} (\Gamma_\hb)_\hal{}^\hga X_\hga^i~, \\
{\mathscr{R}}(J)_{\hb}{}_\hal^i{}^{jk} &= -\frac{3}{4}(\Gamma_\hb)_\hal{}^\hga \eps^{i(j} X_\hga^{k)}~, \\
{\mathscr{R}}(M)_\hb{}_\hal^i{}^{\hc\hd} 
	&= -(\Gamma_\hb)_\hal{}^\hga W^{\hc\hd}{}_\hga^i 
	- \frac{1}{4} \ve_\hb{}^{\hc\hd\he{f}} W_{\he{f}}{}_\hal^i 
	+ \hf \delta_\hb^{[\hc} (\Gamma^{\hd]})_\hal{}^\hga X_\hga^i ~,\\
{\mathscr{R}}(S)_{\hb}{}_{ \hal}^i{}^{\hga j}
	&=
	\frac{1}{16} X_{\hc\hd}{}^{ij} (\S^{\hc\hd} \G_\hb - 2 \G_\hb \S^{\hc\hd})_\hal{}^\hga
	\non\\ & \quad
	- \frac{3\ri}{8} \ve^{ij}\, \nabla_{[\hb} W_{\hc\hd]} (\S^{\hc\hd})_\hal{}^\hga 
	- \frac{\ri}{8} \ve^{ij}\,\nabla_\hd W^{\hd\hc} (\S_{\hc\hb})_\hal{}^\hga 
	\non\\ & \quad
	+ \frac{3\ri}{16} \ve^{ij} \, \nabla^\hd W_{\hd\hb} \,\delta_\hal^\hga
	- \frac{\ri}{8} \ve^{ij}\, \nabla^\hc \tilde W_{\hc\hb}{}^\hd (\G_\hd)_\hal{}^\hga
	\non\\ & \quad
	+ \frac{\ri}{16} \ve^{ij} \tilde W^{\hc\hd\he} W_{\hd\he} (\S_{\hc\hb})_\hal{}^\hga
	- \frac{3\ri}{32} \ve^{ij} \tilde W_{\hb\hd\he} W^{\hd\he} \delta_\hal{}^\hga
	\non\\ & \quad
	+ \frac{\ri}{4} \ve^{ij}\, W_{\hb\hd} W^{\hc\hd} (\G_\hc)_\hal{}^\hga 
	- \frac{3\ri}{64} \ve^{ij}\,W^{\hc\hd}W_{\hc\hd} (\G_\hb)_\hal{}^\hga ~, \\
{\mathscr{R}}(K)_{\hb}{}_{ \hal}^i{}^\hc &=
	\frac{1}{6} (\G^\hc)_\hal{}^\hbe \nabla^\hd W_{\hd\hb}{}_\hbe^i
	+ \frac{1}{12} (\G_\hb)_\hal{}^\hbe \nabla^\hd W{}_\hd{}^{\hc}{}_\hbe^i
	+ \frac{1}{6} {\nabla}_\hal{}^\hbe W_\hb{}^\hc{}_\hbe^i
	- \frac{1}{24} \ve_\hb{}^{\hc\hd\he f} \nabla_\hd W_{\he{f}}{}_\hal^i
	\non\\ & \quad
	+ \frac{1}{8} (\G^\hc)_\hal{}^\hbe \nabla_\hb X^i_\hbe
	+ \frac{1}{64} W^{\hd\he} (3 \G_\hb \S_{\hd\he} \G^\hc - \S_{\hd\he} \G_\hb \G^\hc)_\hal{}^\hbe X_\hbe^i
	\non\\ & \quad
	- \frac{1}{48} \tilde W_{\hb\hd\he} (\G^\hc)_\hal{}^\hbe W^{\hd\he}{}_\hbe^i
	+ \frac{1}{8} \delta_\hb{}^\hc W^{\hd\he} W_{\hd\he}{}_\hal^i
	\non\\ & \quad
	+ \frac{1}{12} (\S_\hb{}^\hc)_\hal{}^\hbe W_{\hd\he}{}_\hbe^i W^{\hd\he}
	- \frac{1}{12} W^{\hd\he} (\S_{\hd\he})_\hal{}^\hbe W_{\hb}{}^{\hc}{}_\hbe{}^i
	\non\\ & \quad
	+ \frac{13}{48} W_{\hb\hd} W^{\hd\hc}{}_\hal^i
	+ \frac{11}{48} W_{\hb\hd}{}_\hal^i W^{\hd\hc}
	- \frac{13}{96} (\G_\hb)_\hal{}^\hbe W_{\hd\he}{}_\hbe^i \, \tilde W^{\hd\he\hc}
	~.
\end{align}
\end{subequations}
The commutator of two vector derivatives $[\de_{\ha}, \de_{\hb}]$ has the following non-zero torsion and curvatures:
\begin{subequations}\label{vect-vect commutation}
\begin{align}
\scT_{\ha\hb}{}^\hal_i &= -\frac{\ri}{2} W_{\ha\hb}{}^\hal_i~, \\
{\mathscr{R}}(J)_{\ha\hb}{}^{ij} &= -\frac{3\ri}{4} X_{\ha\hb}{}^{ij}~, \\
{\mathscr{R}}(M)_{\ha\hb}{}^{\hc\hd} &=
	-\frac{1}{4} (\Sigma_{\ha\hb})^{\hal\hbe}
	(\Sigma^{\hc\hd})^{\hga \hde}
	\Big( \ri W_{\hal\hbe\hga\hde}
		+ 3 W_{(\hal \hbe} W_{\hga \hde)}
	\Big)~, \label{eq:newRMab} \\
{\mathscr{R}}(S)_{\ha\hb}{}_\hal^i
	&=
	- \frac{1}{2} {\nabla}_\hal{}^\hbe W_{\ha\hb\hbe}{}^i
	- \frac{1}{2} (\G_{[\ha })_\hal{}^\hbe \nabla^\hc W_{\hb]\hc \hbe}{}^i
	\non\\ & \quad
	- \frac{1}{8} W_\hal{}^\hbe W_{\ha\hb\hbe}{}^i
	+ \frac{1}{16} (\S_{\ha\hb})_\hal{}^\hbe W^{\hc\hd} W_{\hc\hd\hbe}{}^i
	+ \frac{3}{8} W^\hc{}_{[\ha } W_{\hb]\hc\hal}{}^i~, \\
{\mathscr{R}}(K)_{\ha\hb}{}^\hc &=
	\frac{1}{4} \nabla_d  {\mathscr{R}}(M)_{\ha\hb}{}^{\hc\hd}
	- \frac{\ri}{16} W_{\ha b}{}^\hal_j (\G^\hc)_\hal{}^\hbe X_\hbe^j
	- \frac{\ri}{8} W_{\hd [\ha}{}^\hal_j (\G_{\hb]})_\hal{}^\hbe W^{\hc\hd}_\hbe{}^j
	\non\\ & \quad
	+ \frac{\ri}{8} W_{\ha \hd }{}^\hal_i (\G^\hc)_\hal{}^\hbe
		W_\hb{}^{\hd}{}_\hbe{}^i~.
\end{align}
\end{subequations}


\section{Building blocks}
\label{Building-Blocks}

The following is the bosonic $\cH^{a}_{\log }$ expression obtained from Cadabra:
\begin{adjustwidth}{0cm}{3cm}
\begin{spacing}{1.75} 
$\cH^{a}_{\log} = \frac{3}{32}W^{a}\,_{b} \nabla^{b}{Y}+\frac{3}{32}Y \nabla_{b}{W^{a b}}+\frac{149}{256}\ve_{{e_{1}}}\,^{d e b c} W_{b c} \nabla^{{e_{1}}}{\nabla^{a}{W_{d e}}} - \frac{25}{512}\ve_{{e_{1}}}\,^{d e b c} W_{b c} \nabla^{a}{\nabla^{{e_{1}}}{W_{d e}}}+\frac{4203}{8192}\ve_{b c {e_{1}} d e} \nabla^{a}{W^{b c}} \nabla^{{e_{1}}}{W^{d e}}+\frac{133}{256}\ve^{a}\,_{{e_{1}}}\,^{d b c} W_{b c} \nabla^{{e_{1}}}{\nabla^{e}{W_{d e}}}+\frac{43}{256}\ve^{a}\,_{{e_{1}}}\,^{d e b} W_{b c} \nabla^{c}{\nabla^{{e_{1}}}{W_{d e}}} - \frac{73}{128}\ve_{e {e_{1}}}\,^{c d b} W_{b}\,^{a} \nabla^{e}{\nabla^{{e_{1}}}{W_{c d}}} - \frac{273}{512}\ve^{a d e b c} W_{b c} \nabla_{{e_{1}}}{\nabla^{{e_{1}}}{W_{d e}}} - \frac{33}{128}\ve^{a}\,_{e {e_{1}} d}\,^{b} W_{b c} \nabla^{e}{\nabla^{{e_{1}}}{W^{d c}}} - \frac{31}{64}\ve^{a}\,_{{e_{1}}}\,^{d e b} W_{b c} \nabla^{{e_{1}}}{\nabla^{c}{W_{d e}}}+\frac{85}{128}\ve^{a}\,_{{e_{1}}}\,^{d b c} W_{b c} \nabla^{e}{\nabla^{{e_{1}}}{W_{d e}}}+\frac{81}{256}\ve_{e {e_{1}} d}\,^{b c} W_{b c} \nabla^{e}{\nabla^{{e_{1}}}{W^{a d}}}+\frac{5211}{4096}\ve^{a}\,_{{e_{1}}}\,^{b c d} \nabla^{{e_{1}}}{W_{b c}} \nabla^{e}{W_{d e}} - \frac{99}{2048}\ve_{e b {e_{1}} c d} \nabla^{e}{W^{a b}} \nabla^{{e_{1}}}{W^{c d}} - \frac{4203}{8192}\ve^{a b c d e} \nabla_{{e_{1}}}{W_{b c}} \nabla^{{e_{1}}}{W_{d e}} - \frac{1329}{4096}\ve^{a}\,_{{e_{1}}}\,^{b d e} \nabla^{{e_{1}}}{W_{b c}} \nabla^{c}{W_{d e}} - \frac{99}{1024}\ve^{a}\,_{e}\,^{b}\,_{{e_{1}} d} \nabla^{e}{W_{b c}} \nabla^{{e_{1}}}{W^{d c}} - \frac{3}{32}F^{a}\,_{b} {W}^{-1} \nabla^{b}{Y}%
 - \frac{3}{32}Y {W}^{-1} \nabla_{b}{F^{a b}}+\frac{137}{256}\ve_{{e_{1}}}\,^{d e b c} F_{b c} {W}^{-1} \nabla^{{e_{1}}}{\nabla^{a}{W_{d e}}} - \frac{25}{512}\ve_{{e_{1}}}\,^{d e b c} F_{b c} {W}^{-1} \nabla^{a}{\nabla^{{e_{1}}}{W_{d e}}} - \frac{89}{256}\ve^{a}\,_{{e_{1}}}\,^{d e b} F_{b c} {W}^{-1} \nabla^{c}{\nabla^{{e_{1}}}{W_{d e}}} - \frac{249}{512}\ve^{a d e b c} F_{b c} {W}^{-1} \nabla_{{e_{1}}}{\nabla^{{e_{1}}}{W_{d e}}} - \frac{15}{256}\ve_{e {e_{1}} d}\,^{b c} F_{b c} {W}^{-1} \nabla^{e}{\nabla^{{e_{1}}}{W^{a d}}}+\frac{1}{8}\ve^{a}\,_{{e_{1}}}\,^{d e b} F_{b c} {W}^{-1} \nabla^{{e_{1}}}{\nabla^{c}{W_{d e}}}+\frac{61}{256}\ve^{a}\,_{{e_{1}}}\,^{d b c} F_{b c} {W}^{-1} \nabla^{{e_{1}}}{\nabla^{e}{W_{d e}}} - \frac{7}{128}\ve_{e {e_{1}}}\,^{c d b} F_{b}\,^{a} {W}^{-1} \nabla^{e}{\nabla^{{e_{1}}}{W_{c d}}}+\frac{37}{128}\ve^{a}\,_{{e_{1}}}\,^{d b c} F_{b c} {W}^{-1} \nabla^{e}{\nabla^{{e_{1}}}{W_{d e}}}+\frac{3}{128}\ve^{a}\,_{e {e_{1}} d}\,^{b} F_{b c} {W}^{-1} \nabla^{e}{\nabla^{{e_{1}}}{W^{d c}}} - \frac{13}{32}\ve^{a}\,_{{e_{1}}}\,^{b c d} W_{d e} {W}^{-1} \nabla^{{e_{1}}}{\nabla^{e}{F_{b c}}} - \frac{17}{32}\ve^{a b c d e} W_{d e} {W}^{-1} \nabla_{{e_{1}}}{\nabla^{{e_{1}}}{F_{b c}}}+\frac{1}{4}\ve_{e {e_{1}} b}\,^{c d} W_{c d} {W}^{-1} \nabla^{e}{\nabla^{{e_{1}}}{F^{a b}}}+\frac{11}{32}\ve^{a}\,_{{e_{1}}}\,^{b c d} W_{d e} {W}^{-1} \nabla^{e}{\nabla^{{e_{1}}}{F_{b c}}}+\frac{1}{4}\ve^{a}\,_{{e_{1}}}\,^{b d e} W_{d e} {W}^{-1} \nabla^{c}{\nabla^{{e_{1}}}{F_{b c}}} - \frac{11}{32}\ve_{e {e_{1}}}\,^{b c d} W_{d}\,^{a} {W}^{-1} \nabla^{e}{\nabla^{{e_{1}}}{F_{b c}}}+\frac{3}{16}\ve^{a}\,_{{e_{1}}}\,^{b d e} W_{d e} {W}^{-1} \nabla^{{e_{1}}}{\nabla^{c}{F_{b c}}} - \frac{3}{16}\ve^{a}\,_{e {e_{1}} b}\,^{d} W_{d c} {W}^{-1} \nabla^{e}{\nabla^{{e_{1}}}{F^{b c}}}+\frac{27}{128}\ve_{b c {e_{1}} d e} {W}^{-1} \nabla^{a}{F^{b c}} \nabla^{{e_{1}}}{W^{d e}}%
+\frac{15}{64}\ve^{a}\,_{{e_{1}}}\,^{b c d} {W}^{-1} \nabla^{{e_{1}}}{F_{b c}} \nabla^{e}{W_{d e}} - \frac{33}{128}\ve^{a b}\,_{{e_{1}}}\,^{d e} {W}^{-1} \nabla^{c}{F_{b c}} \nabla^{{e_{1}}}{W_{d e}}+\frac{105}{128}\ve_{e b {e_{1}} c d} {W}^{-1} \nabla^{e}{F^{a b}} \nabla^{{e_{1}}}{W^{c d}} - \frac{15}{16}\ve^{a b c d e} {W}^{-1} \nabla_{{e_{1}}}{F_{b c}} \nabla^{{e_{1}}}{W_{d e}} - \frac{51}{64}\ve^{a b c}\,_{{e_{1}}}\,^{d} {W}^{-1} \nabla^{e}{F_{b c}} \nabla^{{e_{1}}}{W_{d e}} - \frac{27}{128}\ve_{{e_{1}} b c d e} {W}^{-1} \nabla^{{e_{1}}}{F^{b c}} \nabla^{a}{W^{d e}}+\frac{15}{128}\ve^{a}\,_{{e_{1}}}\,^{b d e} {W}^{-1} \nabla^{{e_{1}}}{F_{b c}} \nabla^{c}{W_{d e}} - \frac{39}{64}\ve^{a}\,_{e}\,^{b}\,_{{e_{1}} d} {W}^{-1} \nabla^{e}{F_{b c}} \nabla^{{e_{1}}}{W^{d c}}+\frac{1}{32}\ve_{{e_{1}}}\,^{b c d e} W_{d e} {W}^{-1} \nabla^{{e_{1}}}{\nabla^{a}{F_{b c}}}+\frac{3}{32}F^{a}\,_{b} Y {W}^{-2} \nabla^{b}{W}+\frac{1}{16}{\rm i} W_{b c} \nabla_{d}{W^{a d b c}}+\frac{1}{16}{\rm i} W_{b c} \nabla_{d}{W^{b c a d}} - \frac{1}{16}{\rm i} W_{b c} \nabla_{d}{W^{d b a c}}+\frac{1}{16}{\rm i} W_{b c} \nabla_{d}{W^{a b d c}}+\frac{3}{2}\nabla_{c}{\nabla^{c}{\nabla_{b}{W^{a b}}}} - \frac{15}{8}{W}^{-1} \nabla^{b}{W_{b c}} \nabla^{c}{\nabla^{a}{W}}+2W^{a}\,_{b} {W}^{-1} \nabla^{b}{\nabla_{c}{\nabla^{c}{W}}}+\frac{3}{4}{W}^{-1} \nabla^{a}{W_{b c}} \nabla^{b}{\nabla^{c}{W}}+\frac{3}{2}{W}^{-1} \nabla_{b}{W^{a b}} \nabla_{c}{\nabla^{c}{W}}+\frac{3}{4}{W}^{-1} \nabla_{c}{W^{a}\,_{b}} \nabla^{b}{\nabla^{c}{W}}%
+\frac{9}{4}{W}^{-1} \nabla_{c}{W^{a}\,_{b}} \nabla^{c}{\nabla^{b}{W}}+\frac{3}{8}{W}^{-1} \nabla^{b}{W_{b c}} \nabla^{a}{\nabla^{c}{W}}+3{W}^{-1} \nabla_{b}{W} \nabla_{c}{\nabla^{c}{W^{a b}}}-2{W}^{-1} \nabla_{c}{W} \nabla_{b}{\nabla^{c}{W^{a b}}}+2{W}^{-1} \nabla^{b}{W} \nabla^{c}{\nabla^{a}{W_{b c}}}+{W}^{-1} \nabla^{b}{W} \nabla^{a}{\nabla^{c}{W_{b c}}}+\frac{1}{2}{W}^{-1} \nabla_{c}{W} \nabla^{c}{\nabla_{b}{W^{a b}}} - \frac{103}{512}\ve_{{e_{1}}}\,^{d e b c} F_{b c} {W}^{-2} \nabla^{a}{W} \nabla^{{e_{1}}}{W_{d e}}+\frac{73}{64}\ve_{{e_{1}}}\,^{d e b c} W_{b c} {W}^{-1} \nabla^{a}{W} \nabla^{{e_{1}}}{W_{d e}} - \frac{155}{512}\ve_{{e_{1}}}\,^{d e b c} W_{b c} {W}^{-1} \nabla^{{e_{1}}}{W} \nabla^{a}{W_{d e}} - \frac{19}{512}\ve^{a}\,_{{e_{1}}}\,^{d e b} F_{b c} {W}^{-2} \nabla^{{e_{1}}}{W} \nabla^{c}{W_{d e}}+\frac{33}{64}\ve^{a d e b c} F_{b c} {W}^{-2} \nabla_{{e_{1}}}{W} \nabla^{{e_{1}}}{W_{d e}} - \frac{63}{256}\ve_{e {e_{1}} d}\,^{b c} F_{b c} {W}^{-2} \nabla^{e}{W} \nabla^{{e_{1}}}{W^{a d}} - \frac{221}{512}\ve^{a}\,_{{e_{1}}}\,^{d e b} F_{b c} {W}^{-2} \nabla^{c}{W} \nabla^{{e_{1}}}{W_{d e}} - \frac{161}{512}\ve_{{e_{1}}}\,^{d e b c} F_{b c} {W}^{-2} \nabla^{{e_{1}}}{W} \nabla^{a}{W_{d e}}+\frac{5}{32}\ve^{a}\,_{{e_{1}}}\,^{d b c} F_{b c} {W}^{-2} \nabla^{e}{W} \nabla^{{e_{1}}}{W_{d e}}+\frac{43}{512}\ve_{e {e_{1}}}\,^{c d b} F_{b}\,^{a} {W}^{-2} \nabla^{e}{W} \nabla^{{e_{1}}}{W_{c d}} - \frac{5}{8}\ve^{a}\,_{{e_{1}}}\,^{d b c} F_{b c} {W}^{-2} \nabla^{{e_{1}}}{W} \nabla^{e}{W_{d e}} - \frac{83}{256}\ve^{a}\,_{e {e_{1}} d}\,^{b} F_{b c} {W}^{-2} \nabla^{e}{W} \nabla^{{e_{1}}}{W^{d c}}+\frac{3}{16}\ve_{{e_{1}}}\,^{d e b c} W_{d e} F_{b c} {W}^{-2} \nabla^{a}{\nabla^{{e_{1}}}{W}}%
 - \frac{3}{8}\ve_{{e_{1}}}\,^{d e b c} W_{d e} F_{b c} {W}^{-2} \nabla^{{e_{1}}}{\nabla^{a}{W}}+\frac{3}{32}\ve_{{e_{1}}}\,^{b c d e} W_{b c} W_{d e} {W}^{-1} \nabla^{a}{\nabla^{{e_{1}}}{W}}+\frac{15}{32}\ve_{{e_{1}}}\,^{b c d e} W_{b c} W_{d e} {W}^{-1} \nabla^{{e_{1}}}{\nabla^{a}{W}} - \frac{9}{16}\ve^{a b c d e} W_{b c} W_{d e} {W}^{-1} \nabla_{{e_{1}}}{\nabla^{{e_{1}}}{W}}+\frac{25}{256}\ve_{{e_{1}}}\,^{b c d e} W_{d e} {W}^{-2} \nabla^{a}{W} \nabla^{{e_{1}}}{F_{b c}}+\frac{11}{32}\ve^{a}\,_{{e_{1}}}\,^{b c d} W_{d e} {W}^{-2} \nabla^{{e_{1}}}{W} \nabla^{e}{F_{b c}}+\frac{153}{256}\ve^{a b c d e} W_{d e} {W}^{-2} \nabla_{{e_{1}}}{W} \nabla^{{e_{1}}}{F_{b c}} - \frac{81}{128}\ve_{e {e_{1}} b}\,^{c d} W_{c d} {W}^{-2} \nabla^{e}{W} \nabla^{{e_{1}}}{F^{a b}} - \frac{19}{128}\ve^{a}\,_{{e_{1}}}\,^{b c d} W_{d e} {W}^{-2} \nabla^{e}{W} \nabla^{{e_{1}}}{F_{b c}} - \frac{25}{128}\ve_{{e_{1}}}\,^{b c d e} W_{d e} {W}^{-2} \nabla^{{e_{1}}}{W} \nabla^{a}{F_{b c}} - \frac{37}{64}\ve^{a}\,_{{e_{1}}}\,^{b d e} W_{d e} {W}^{-2} \nabla^{c}{W} \nabla^{{e_{1}}}{F_{b c}} - \frac{3}{64}\ve_{e {e_{1}}}\,^{b c d} W_{d}\,^{a} {W}^{-2} \nabla^{e}{W} \nabla^{{e_{1}}}{F_{b c}} - \frac{61}{128}\ve^{a}\,_{{e_{1}}}\,^{b d e} W_{d e} {W}^{-2} \nabla^{{e_{1}}}{W} \nabla^{c}{F_{b c}}+\frac{9}{64}\ve^{a}\,_{e {e_{1}} b}\,^{d} W_{d c} {W}^{-2} \nabla^{e}{W} \nabla^{{e_{1}}}{F^{b c}}+\frac{137}{512}\ve^{a}\,_{{e_{1}}}\,^{d e b} W_{b c} {W}^{-1} \nabla^{{e_{1}}}{W} \nabla^{c}{W_{d e}} - \frac{429}{512}\ve^{a d e b c} W_{b c} {W}^{-1} \nabla_{{e_{1}}}{W} \nabla^{{e_{1}}}{W_{d e}}+\frac{39}{128}\ve_{e {e_{1}} d}\,^{b c} W_{b c} {W}^{-1} \nabla^{e}{W} \nabla^{{e_{1}}}{W^{a d}}+\frac{157}{512}\ve^{a}\,_{{e_{1}}}\,^{d e b} W_{b c} {W}^{-1} \nabla^{c}{W} \nabla^{{e_{1}}}{W_{d e}}+\frac{35}{128}\ve^{a}\,_{{e_{1}}}\,^{d b c} W_{b c} {W}^{-1} \nabla^{e}{W} \nabla^{{e_{1}}}{W_{d e}} - \frac{173}{512}\ve_{e {e_{1}}}\,^{c d b} W_{b}\,^{a} {W}^{-1} \nabla^{e}{W} \nabla^{{e_{1}}}{W_{c d}}%
+\frac{77}{256}\ve^{a}\,_{{e_{1}}}\,^{d b c} W_{b c} {W}^{-1} \nabla^{{e_{1}}}{W} \nabla^{e}{W_{d e}} - \frac{17}{256}\ve^{a}\,_{e {e_{1}} d}\,^{b} W_{b c} {W}^{-1} \nabla^{e}{W} \nabla^{{e_{1}}}{W^{d c}} - \frac{1}{8}\ve_{{e_{1}}}\,^{b c d e} F_{b c} F_{d e} {W}^{-3} \nabla^{{e_{1}}}{\nabla^{a}{W}}+\frac{3}{16}\ve^{a}\,_{{e_{1}}}\,^{d e b} F_{b c} {W}^{-3} \nabla^{{e_{1}}}{W} \nabla^{c}{F_{d e}}+\frac{37}{64}\ve^{a d e b c} F_{b c} {W}^{-3} \nabla_{{e_{1}}}{W} \nabla^{{e_{1}}}{F_{d e}} - \frac{9}{32}\ve_{e {e_{1}} d}\,^{b c} F_{b c} {W}^{-3} \nabla^{e}{W} \nabla^{{e_{1}}}{F^{a d}} - \frac{1}{32}\ve^{a}\,_{{e_{1}}}\,^{d e b} F_{b c} {W}^{-3} \nabla^{c}{W} \nabla^{{e_{1}}}{F_{d e}} - \frac{7}{32}\ve_{{e_{1}}}\,^{d e b c} F_{b c} {W}^{-3} \nabla^{{e_{1}}}{W} \nabla^{a}{F_{d e}} - \frac{1}{16}\ve^{a}\,_{{e_{1}}}\,^{d b c} F_{b c} {W}^{-3} \nabla^{e}{W} \nabla^{{e_{1}}}{F_{d e}} - \frac{1}{4}\ve_{e {e_{1}}}\,^{c d b} F_{b}\,^{a} {W}^{-3} \nabla^{e}{W} \nabla^{{e_{1}}}{F_{c d}}+\frac{9}{64}\ve_{{e_{1}}}\,^{d e b c} F_{b c} {W}^{-3} \nabla^{a}{W} \nabla^{{e_{1}}}{F_{d e}} - \frac{9}{32}\ve^{a}\,_{{e_{1}}}\,^{d b c} F_{b c} {W}^{-3} \nabla^{{e_{1}}}{W} \nabla^{e}{F_{d e}}-\ve^{a}\,_{e {e_{1}}}\,^{d b} W_{d}\,^{c} F_{b c} {W}^{-2} \nabla^{e}{\nabla^{{e_{1}}}{W}} - \frac{1}{4}\ve^{a}\,_{{e_{1}}}\,^{d e b} W_{d e} F_{b c} {W}^{-2} \nabla^{{e_{1}}}{\nabla^{c}{W}} - \frac{1}{8}\ve^{a}\,_{{e_{1}}}\,^{d b c} W_{d e} F_{b c} {W}^{-2} \nabla^{e}{\nabla^{{e_{1}}}{W}}+\frac{1}{4}\ve_{e {e_{1}}}\,^{d b c} W_{d}\,^{a} F_{b c} {W}^{-2} \nabla^{e}{\nabla^{{e_{1}}}{W}} - \frac{1}{8}\ve^{a}\,_{{e_{1}}}\,^{d e b} W_{d e} F_{b c} {W}^{-2} \nabla^{c}{\nabla^{{e_{1}}}{W}}+\frac{1}{2}\ve^{a}\,_{{e_{1}}}\,^{d b c} W_{d e} F_{b c} {W}^{-2} \nabla^{{e_{1}}}{\nabla^{e}{W}}+\frac{3}{8}W^{a}\,_{b} W_{c d} \nabla^{b}{W^{c d}}+\frac{9}{32}W^{b c} W_{b c} \nabla_{d}{W^{a d}}%
 - \frac{21}{4}W^{a}\,_{b} W_{c d} \nabla^{c}{W^{b d}} - \frac{9}{2}W^{b}\,_{c} W_{b d} \nabla^{c}{W^{a d}} - \frac{9}{2}W^{a b} W_{b c} \nabla_{d}{W^{d c}}+\frac{1}{12}\ve^{a b c d e} \Phi_{b c i j} \Phi_{d e}\,^{i j}+\frac{1}{16}{\rm i} F_{b c} {W}^{-1} \nabla_{d}{W^{a d b c}}+\frac{1}{16}{\rm i} F_{b c} {W}^{-1} \nabla_{d}{W^{b c a d}} - \frac{1}{16}{\rm i} F_{b c} {W}^{-1} \nabla_{d}{W^{d b a c}}+\frac{1}{16}{\rm i} F_{b c} {W}^{-1} \nabla_{d}{W^{a b d c}}+X_{i j} {W}^{-1} \nabla_{b}{\Phi^{a b i j}}+\Phi^{a}\,_{b i j} {W}^{-1} \nabla^{b}{X^{i j}}+{W}^{-1} \nabla_{c}{\nabla^{c}{\nabla_{b}{F^{a b}}}} - \frac{3}{2}{W}^{-2} \nabla^{a}{W} \nabla^{b}{W} \nabla^{c}{W_{b c}}+F^{a}\,_{b} {W}^{-3} \nabla^{b}{W} \nabla_{c}{\nabla^{c}{W}}+2F^{a}\,_{b} {W}^{-3} \nabla_{c}{W} \nabla^{c}{\nabla^{b}{W}}-F_{b c} {W}^{-3} \nabla^{b}{W} \nabla^{c}{\nabla^{a}{W}}-3W^{a}\,_{b} {W}^{-2} \nabla^{b}{W} \nabla_{c}{\nabla^{c}{W}}-W^{a}\,_{b} {W}^{-2} \nabla_{c}{W} \nabla^{c}{\nabla^{b}{W}}-2W^{a}\,_{b} {W}^{-2} \nabla_{c}{W} \nabla^{b}{\nabla^{c}{W}}+\frac{3}{2}{W}^{-3} \nabla_{c}{W} \nabla^{c}{W} \nabla_{b}{F^{a b}}-{W}^{-3} \nabla^{a}{W} \nabla^{b}{W} \nabla^{c}{F_{b c}}%
+\frac{3}{8}\ve_{{e_{1}}}\,^{d e b c} W_{d e} F_{b c} {W}^{-3} \nabla^{a}{W} \nabla^{{e_{1}}}{W} - \frac{9}{16}\ve_{{e_{1}}}\,^{b c d e} W_{b c} W_{d e} {W}^{-2} \nabla^{a}{W} \nabla^{{e_{1}}}{W}+\frac{3}{8}\ve_{{e_{1}}}\,^{b c d e} F_{b c} F_{d e} {W}^{-4} \nabla^{a}{W} \nabla^{{e_{1}}}{W} - \frac{3}{8}\ve^{a b c d e} F_{b c} F_{d e} {W}^{-4} \nabla_{{e_{1}}}{W} \nabla^{{e_{1}}}{W} - \frac{3}{8}\ve^{a d e b c} W_{d e} F_{b c} {W}^{-3} \nabla_{{e_{1}}}{W} \nabla^{{e_{1}}}{W}+\frac{9}{16}\ve^{a b c d e} W_{b c} W_{d e} {W}^{-2} \nabla_{{e_{1}}}{W} \nabla^{{e_{1}}}{W}+\frac{3}{4}\ve^{a}\,_{{e_{1}}}\,^{d e b} W_{d e} F_{b c} {W}^{-3} \nabla^{{e_{1}}}{W} \nabla^{c}{W} - \frac{3}{4}\ve^{a}\,_{{e_{1}}}\,^{d b c} W_{d e} F_{b c} {W}^{-3} \nabla^{{e_{1}}}{W} \nabla^{e}{W}+\frac{3}{2}F^{a}\,_{b} F_{c d} {W}^{-2} \nabla^{b}{W^{c d}}+\frac{3}{4}F^{b c} F_{b c} {W}^{-2} \nabla_{d}{W^{a d}} - \frac{3}{2}F^{a}\,_{b} F_{c d} {W}^{-2} \nabla^{c}{W^{b d}} - \frac{3}{2}F^{b}\,_{c} F_{b d} {W}^{-2} \nabla^{c}{W^{a d}} - \frac{3}{2}F^{a b} F_{b c} {W}^{-2} \nabla_{d}{W^{d c}}+\frac{3}{8}W^{a}\,_{d} F_{b c} {W}^{-1} \nabla^{d}{W^{b c}} - \frac{3}{8}W^{a}\,_{d} W_{b c} {W}^{-1} \nabla^{d}{F^{b c}} - \frac{1}{2}W^{b}\,_{d} F_{b c} {W}^{-2} \nabla^{a}{F^{d c}}+\frac{1}{2}W^{a}\,_{d} F_{b c} {W}^{-2} \nabla^{d}{F^{b c}}+\frac{3}{2}W^{b c} F_{b c} {W}^{-2} \nabla_{d}{F^{a d}}+\frac{1}{2}W^{a}\,_{d} F_{b c} {W}^{-2} \nabla^{b}{F^{d c}}+\frac{1}{4}W_{c d} F^{a}\,_{b} {W}^{-2} \nabla^{b}{F^{c d}}%
-W_{d c} F^{a}\,_{b} {W}^{-2} \nabla^{d}{F^{c b}} - \frac{3}{2}W^{a b} F_{b c} {W}^{-2} \nabla_{d}{F^{d c}} - \frac{3}{2}W_{b d} F^{a b} {W}^{-2} \nabla_{c}{F^{c d}}-W^{b}\,_{d} F_{b c} {W}^{-2} \nabla^{d}{F^{a c}}-2W^{b}\,_{d} F_{b c} {W}^{-2} \nabla^{c}{F^{a d}}+\frac{3}{8}W^{b c} F_{b c} {W}^{-1} \nabla_{d}{W^{a d}} - \frac{39}{8}W^{a}\,_{d} F_{b c} {W}^{-1} \nabla^{b}{W^{d c}}+\frac{39}{8}W_{c d} F^{a}\,_{b} {W}^{-1} \nabla^{b}{W^{c d}}+\frac{39}{8}W_{c d} F^{a}\,_{b} {W}^{-1} \nabla^{c}{W^{d b}}+\frac{3}{8}W^{a b} F_{b c} {W}^{-1} \nabla_{d}{W^{d c}} - \frac{39}{8}W_{b c} F^{a b} {W}^{-1} \nabla_{d}{W^{d c}}+\frac{3}{8}W^{b}\,_{d} F_{b c} {W}^{-1} \nabla^{d}{W^{a c}} - \frac{39}{8}W^{b}\,_{d} F_{b c} {W}^{-1} \nabla^{c}{W^{a d}}+\frac{75}{32}W^{c d} W_{c d} {W}^{-1} \nabla_{b}{F^{a b}}+\frac{3}{4}W^{a}\,_{b} W_{d c} {W}^{-1} \nabla^{d}{F^{b c}} - \frac{9}{2}W^{c}\,_{d} W_{c b} {W}^{-1} \nabla^{d}{F^{a b}} - \frac{9}{2}W^{a d} W_{d c} {W}^{-1} \nabla_{b}{F^{b c}}+\frac{1}{4}F^{b c} F_{b c} {W}^{-3} \nabla_{d}{F^{a d}}-F^{a b} F_{b c} {W}^{-3} \nabla_{d}{F^{d c}}-F^{b}\,_{c} F_{b d} {W}^{-3} \nabla^{c}{F^{a d}}%
+\frac{1}{2}X_{i j} F^{a}\,_{b} {W}^{-3} \nabla^{b}{X^{i j}}+\frac{1}{4}X_{i j} X^{i j} {W}^{-3} \nabla_{b}{F^{a b}} - \frac{1}{24}{\rm i} F^{b c} W^{a}\,_{d b c} {W}^{-2} \nabla^{d}{W} - \frac{1}{24}{\rm i} F^{b c} W_{b c}\,^{a}\,_{d} {W}^{-2} \nabla^{d}{W} - \frac{1}{24}{\rm i} F^{b c} W_{b d}\,^{a}\,_{c} {W}^{-2} \nabla^{d}{W}+\frac{1}{24}{\rm i} F^{b c} W^{a}\,_{b c d} {W}^{-2} \nabla^{d}{W} - \frac{1}{24}{\rm i} W^{b c} W^{a}\,_{d b c} {W}^{-1} \nabla^{d}{W} - \frac{1}{24}{\rm i} W^{b c} W_{b c}\,^{a}\,_{d} {W}^{-1} \nabla^{d}{W} - \frac{1}{24}{\rm i} W^{b c} W_{b d}\,^{a}\,_{c} {W}^{-1} \nabla^{d}{W}+\frac{1}{24}{\rm i} W^{b c} W^{a}\,_{b c d} {W}^{-1} \nabla^{d}{W} - \frac{5}{192}{\rm i} \ve^{a d e {e_{1}} {e_{2}}} W^{b c} W_{d e} W_{{e_{1}} {e_{2}} b c} - \frac{5}{192}{\rm i} \ve^{a d e {e_{1}} {e_{2}}} W^{b c} W_{d e} W_{b c {e_{1}} {e_{2}}} - \frac{5}{192}{\rm i} \ve^{a d e {e_{1}} {e_{2}}} W^{b c} W_{d e} W_{{e_{1}} b {e_{2}} c}+\frac{5}{384}{\rm i} \ve^{b c d {e_{1}} {e_{2}}} W_{b c} W_{d}\,^{e} W^{a}\,_{e {e_{1}} {e_{2}}} - \frac{5}{96}{\rm i} \ve^{a b d {e_{1}} {e_{2}}} W_{b}\,^{c} W_{d}\,^{e} W_{{e_{1}} c {e_{2}} e}-\Phi^{a}\,_{b i j} X^{i j} {W}^{-2} \nabla^{b}{W}+W^{a}\,_{b} {W}^{-1} \nabla_{c}{\nabla^{c}{\nabla^{b}{W}}}-F^{a}\,_{b} {W}^{-2} \nabla_{c}{\nabla^{c}{\nabla^{b}{W}}}-{W}^{-2} \nabla_{b}{F^{a b}} \nabla_{c}{\nabla^{c}{W}}-{W}^{-2} \nabla_{c}{F^{a}\,_{b}} \nabla^{c}{\nabla^{b}{W}}%
-{W}^{-2} \nabla_{b}{W} \nabla_{c}{\nabla^{c}{F^{a b}}}-{W}^{-2} \nabla_{c}{W} \nabla^{c}{\nabla_{b}{F^{a b}}} - \frac{3}{2}F^{a}\,_{b} {W}^{-4} \nabla_{c}{W} \nabla^{c}{W} \nabla^{b}{W}+3W^{a}\,_{b} {W}^{-3} \nabla_{c}{W} \nabla^{c}{W} \nabla^{b}{W} - \frac{1}{16}\ve^{a}\,_{e {e_{1}} d}\,^{b} F_{b c} {W}^{-3} \nabla^{e}{W} \nabla^{{e_{1}}}{F^{d c}}+\frac{3}{16}\ve^{a d e b c} W_{d e} F_{b c} {W}^{-2} \nabla_{{e_{1}}}{\nabla^{{e_{1}}}{W}} - \frac{3}{4}F^{b c} F^{a}\,_{d} F_{b c} {W}^{-4} \nabla^{d}{W} - \frac{21}{8}F^{a b} F^{c}\,_{b} F_{c d} {W}^{-4} \nabla^{d}{W} - \frac{45}{8}F^{b c} F^{a}\,_{b} F_{c d} {W}^{-4} \nabla^{d}{W} - \frac{1}{4}W^{b c} W^{a}\,_{d} F_{b c} {W}^{-2} \nabla^{d}{W} - \frac{5}{16}W^{b c} W^{a}\,_{d} W_{b c} {W}^{-1} \nabla^{d}{W}-3W^{c d} F^{a}\,_{b} F_{c d} {W}^{-3} \nabla^{b}{W} - \frac{9}{2}W^{c}\,_{d} F^{a b} F_{c b} {W}^{-3} \nabla^{d}{W}+3W^{a c} F^{b}\,_{c} F_{b d} {W}^{-3} \nabla^{d}{W} - \frac{3}{2}W^{a}\,_{d} F^{b c} F_{b c} {W}^{-3} \nabla^{d}{W}+\frac{15}{2}W_{b d} F^{b c} F^{a}\,_{c} {W}^{-3} \nabla^{d}{W}-3W^{b c} F^{a}\,_{b} F_{c d} {W}^{-3} \nabla^{d}{W} - \frac{77}{32}W^{c d} W_{c d} F^{a}\,_{b} {W}^{-2} \nabla^{b}{W} - \frac{1}{4}W^{a c} W^{b}\,_{d} F_{b c} {W}^{-2} \nabla^{d}{W}+\frac{19}{4}W^{c b} W_{c d} F^{a}\,_{b} {W}^{-2} \nabla^{d}{W}%
+\frac{213}{64}W^{a d} W^{b}\,_{d} F_{b c} {W}^{-2} \nabla^{c}{W} - \frac{91}{64}W^{d b} W^{a}\,_{d} F_{b c} {W}^{-2} \nabla^{c}{W} - \frac{37}{8}W^{d c} W^{b}\,_{d} F_{b c} {W}^{-2} \nabla^{a}{W} - \frac{7}{32}W^{a b} W^{c}\,_{b} W_{c d} {W}^{-1} \nabla^{d}{W} - \frac{15}{32}W^{b c} W^{a}\,_{b} W_{c d} {W}^{-1} \nabla^{d}{W} - \frac{5}{32}\ve^{a d e {e_{1}} {e_{2}}} W^{b c} W_{d e} W_{{e_{1}} {e_{2}}} W_{b c} - \frac{3}{4}X_{i j} X^{i j} F^{a}\,_{b} {W}^{-4} \nabla^{b}{W} - \frac{1}{96}{\rm i} \ve^{a d e {e_{1}} {e_{2}}} F^{b c} F_{d e} W_{{e_{1}} {e_{2}} b c} {W}^{-2} - \frac{1}{96}{\rm i} \ve^{a d e {e_{1}} {e_{2}}} F^{b c} F_{d e} W_{b c {e_{1}} {e_{2}}} {W}^{-2} - \frac{1}{96}{\rm i} \ve^{a d e {e_{1}} {e_{2}}} F^{b c} F_{d e} W_{{e_{1}} b {e_{2}} c} {W}^{-2}+\frac{1}{192}{\rm i} \ve^{b c d {e_{1}} {e_{2}}} F_{b c} F_{d}\,^{e} W^{a}\,_{e {e_{1}} {e_{2}}} {W}^{-2} - \frac{1}{48}{\rm i} \ve^{a b d {e_{1}} {e_{2}}} F_{b}\,^{c} F_{d}\,^{e} W_{{e_{1}} c {e_{2}} e} {W}^{-2} - \frac{5}{192}{\rm i} \ve^{a d e {e_{1}} {e_{2}}} W_{d e} F^{b c} W_{{e_{1}} {e_{2}} b c} {W}^{-1} - \frac{5}{192}{\rm i} \ve^{a d e {e_{1}} {e_{2}}} W_{d e} F^{b c} W_{b c {e_{1}} {e_{2}}} {W}^{-1} - \frac{5}{512}{\rm i} \ve^{a b c {e_{1}} {e_{2}}} W^{d e} F_{b c} W_{{e_{1}} {e_{2}} d e} {W}^{-1}+\frac{13}{768}{\rm i} \ve^{d b c {e_{1}} {e_{2}}} W_{d}\,^{e} F_{b c} W_{{e_{1}} {e_{2}}}\,^{a}\,_{e} {W}^{-1} - \frac{13}{384}{\rm i} \ve^{a d e {e_{1}} {e_{2}}} W_{d e} F^{b c} W_{{e_{1}} b {e_{2}} c} {W}^{-1} - \frac{17}{1536}{\rm i} \ve^{a b c {e_{1}} {e_{2}}} W^{d e} F_{b c} W_{d e {e_{1}} {e_{2}}} {W}^{-1} - \frac{11}{768}{\rm i} \ve^{d e b {e_{1}} {e_{2}}} W_{d e} F_{b}\,^{c} W_{{e_{1}} {e_{2}}}\,^{a}\,_{c} {W}^{-1}+\frac{1}{64}{\rm i} \ve^{d b c {e_{1}} {e_{2}}} W_{d}\,^{e} F_{b c} W^{a}\,_{e {e_{1}} {e_{2}}} {W}^{-1}%
 - \frac{7}{96}{\rm i} \ve^{a d b {e_{1}} {e_{2}}} W_{d}\,^{e} F_{b}\,^{c} W_{{e_{1}} e {e_{2}} c} {W}^{-1} - \frac{1}{384}{\rm i} \ve^{a b c {e_{1}} {e_{2}}} W^{d e} F_{b c} W_{{e_{1}} d {e_{2}} e} {W}^{-1}+{W}^{-3} \nabla_{c}{W} \nabla_{b}{W} \nabla^{c}{F^{a b}}-3{W}^{-2} \nabla_{c}{W} \nabla_{b}{W} \nabla^{c}{W^{a b}} - \frac{1}{4}\ve^{a d e b c} F_{b c} {W}^{-2} \nabla_{{e_{1}}}{\nabla^{{e_{1}}}{F_{d e}}} - \frac{1}{8}\ve^{a b c d e} {W}^{-2} \nabla_{{e_{1}}}{F_{b c}} \nabla^{{e_{1}}}{F_{d e}} - \frac{5}{32}\ve^{a d e {e_{1}} {e_{2}}} W^{b c} W_{d e} W_{{e_{1}} {e_{2}}} F_{b c} {W}^{-1} - \frac{7}{32}\ve^{a {e_{1}} {e_{2}} b c} W^{d e} W_{{e_{1}} {e_{2}}} F_{b c} F_{d e} {W}^{-2}+\frac{7}{128}\ve^{a d e {e_{1}} {e_{2}}} W_{d e} W_{{e_{1}} {e_{2}}} F^{b c} F_{b c} {W}^{-2}+\frac{11}{32}\ve^{a {e_{1}} {e_{2}} b d} W^{c e} W_{{e_{1}} {e_{2}}} F_{b c} F_{d e} {W}^{-2} - \frac{3}{64}\ve^{{e_{2}} b c d e} W^{a {e_{1}}} W_{{e_{2}} {e_{1}}} F_{b c} F_{d e} {W}^{-2} - \frac{1}{4}\ve^{a e {e_{1}} {e_{2}} d} W_{e {e_{1}}} W_{{e_{2}} b} F^{b c} F_{d c} {W}^{-2}+\frac{7}{16}\ve^{{e_{1}} {e_{2}} b c d} W_{{e_{1}}}\,^{a} W_{{e_{2}}}\,^{e} F_{b c} F_{d e} {W}^{-2}+\frac{3}{32}\ve^{e {e_{1}} {e_{2}} c d} W_{e {e_{1}}} W_{{e_{2}} b} F^{a b} F_{c d} {W}^{-2} - \frac{3}{16}\ve^{a {e_{1}} {e_{2}} b d} W_{{e_{1}}}\,^{c} W_{{e_{2}}}\,^{e} F_{b c} F_{d e} {W}^{-2}+\frac{5}{64}\ve^{a b c d e} W^{{e_{1}} {e_{2}}} W_{{e_{1}} {e_{2}}} F_{b c} F_{d e} {W}^{-2} - \frac{1}{16}\ve^{a {e_{1}} {e_{2}} b c} W^{d e} W_{{e_{1}} {e_{2}}} W_{d e} F_{b c} {W}^{-1}+\frac{3}{32}\ve^{a e {e_{1}} {e_{2}} b} W^{d c} W_{e {e_{1}}} W_{{e_{2}} d} F_{b c} {W}^{-1}+\frac{3}{64}\ve^{e {e_{1}} {e_{2}} b c} W^{a d} W_{e {e_{1}}} W_{{e_{2}} d} F_{b c} {W}^{-1} - \frac{3}{32}\ve^{a d e {e_{1}} {e_{2}}} W_{d e} W_{{e_{1}} b} W_{{e_{2}} c} F^{b c} {W}^{-1}%
 - \frac{3}{64}\ve^{d e {e_{1}} {e_{2}} b} W^{a c} W_{d e} W_{{e_{1}} {e_{2}}} F_{b c} {W}^{-1}+\frac{1}{8}\ve^{a b c d e} F_{b c} F_{d e} {W}^{-3} \nabla_{{e_{1}}}{\nabla^{{e_{1}}}{W}}$
 \end{spacing}
 \end{adjustwidth}
 The expression for $\cH^{a}_{\log}$ presented in subsection \ref{subsection Log descendant} serves as one of the key outcomes of this paper. In this section, we have gathered a set of simplifications that hold up to fermions and have been applied to the above expression to obtain equations \eqref{Ha-log-comp} and \eqref{Ha-log-comp_b}. It is important to note that while some of these identities could be made even simpler using algebra and the Bianchi identity, our choice of simplification aims to express $\cH^{a}_{\log}$ in the form of total derivatives to the greatest extent possible. It holds that
\bsubeq
\bea
    (1)
    &=&
    \ve^{{a} {b} {c} {d} {e}} \Bigg(\frac{133}{256}  W_{{d} {e}} {\nabla}_{b}{{\nabla}^{f}}{W_{c f}} +\frac{43}{256} W_{{e} f} {\nabla}^{f}{{\nabla}_{b}{W_{c d}}}      - \frac{273}{512}  W_{{d} {e}} {\nabla}^{f}{{\nabla}_{f}{W_{b c}}}  
    \non \\&&~~~~~~~~~~~
    - \frac{33}{128}  W_{{e} f} {\nabla}_{b}{{\nabla}_{c}{W_{d}{}^{f}}}  - \frac{31}{64} W_{{e} f} {\nabla}_{b}{{\nabla}^{f}{W_{c d}}}  +\frac{85}{128}  W_{{d} {e}} {\nabla}^{f}{{\nabla}_{b}{W_{c f}}} \Bigg) 
    \non \\&&
    +\ve^{{b} {c} {d} {e} f}\bigg(\frac{149}{256} W_{e f} {\nabla}_{b}{{\nabla}^{{a}}{W_{cd}}} - \frac{25}{512} W_{{e} f} {\nabla}^{{a}}{{\nabla}_{b}{W_{cd}}} - \frac{73}{128}  \eta^{{a} {e_{6}}} W_{f {e_{6}}} {\nabla}_{b}{\nabla}_{c}{W_{d e}} 
    \non \\&&~~~~~~~~~~~~~~~
    +\frac{81}{256} W_{{e} f} {\nabla}_{b}{{\nabla}_{c}{W^{{a}}{}_{ d}}} \Bigg) 
    \non \\
    &=& \ve^{{a} {b} {c} {d} {e}}\Bigg( \frac{9}{4}  W_{{d} {e}} {\nabla}_{b}{{\nabla}^{f}}{W_{c f}} + \frac{3}{4} W_{{e} f} {\nabla}_{b}{{\nabla}^{f}{W_{c d}}}  + \frac{5\ri}{64} W^{f g} W_{{b} {c}} W_{{d} {e} f g} + \frac{5}{32} W^{f g} W_{{b} {c}} W_{{d} {e}} W_{f g} \Bigg)~, 
    \non\\&& \\
    (2)
    &=&
    \ve^{{a} {b} {c} {d} {e}} \Bigg(\frac{5211}{4096} {\nabla}_{b}{W_{c d}} {\nabla}^{f}{W_{e f}}  - \frac{4203}{8192}{\nabla}^{f}{W_{b c}} {\nabla}_{f}{W_{d e}} - \frac{1329}{4096}{\nabla}_{b}{W_{c f}} {\nabla}^{f}{W_{d e}} 
    \non \\&&~~~~~~~~~~~
    - \frac{99}{1024} {\nabla}_{b}{W_{c}{}^{f}} {\nabla}_{d}{W_{e f}} \Bigg) 
    \non \\&&+\ve^{{b} {c} {d} {e} f} \Bigg( \frac{4203}{8192}  {\nabla}^{{a}}{W_{b c}} {\nabla}_{d}{W_{e f}}- \frac{99}{2048} {\nabla}_{b}{W^{{a}}{}_{c}} {\nabla}_{d}{W_{e f}} \Bigg) \non \\
    \non \\&=&  \frac{3}{4}\ve^{{a} {b} {c} {d} {e}} \Big( {\nabla}_{b}{W_{c f}} {\nabla}^{f}{W_{d e}} +3 {\nabla}_{b}{W_{c d}} {\nabla}^{f}{W_{e f}} \Big)~, \non \\ && \\
    (3)
    &=&
    \frac{3}{2} {\nabla}^{{b}}{{\nabla}_{{b}}{{\nabla}_{{c}}{W^{{a} {c}}}}} = \frac{3}{2} {{\nabla}_{{c}}{\nabla}^{{b}}{{\nabla}_{{b}}{W^{{a} {c}}}}} - \frac{3}{2} \nabla_{b}(C^{abcd} W_{cd}) + \frac{3}{8} \nabla_{b}(C^{abcd}) W_{cd} ~, \non \\ && \\
    (4)&=& \ve^{{b} {c} {d} {e} {f}}\Bigg( \frac{137}{256}  F_{{e} {f}} {\nabla}_{b}{{\nabla}^{{a}}{W_{c d}}} - \frac{25}{512} F_{{e} {f}} {\nabla}^{{a}}{\nabla}_{b}{W_{c d}} - \frac{15}{256}  F_{{e} {f}} {\nabla}_{b}{\nabla}_{c}{W^{{a}}{}_{d}} 
    \non \\&&~~~~~~~~~~~
    - \frac{7}{128}  \eta^{{a} {e_{6}}} F_{{f} {e_{6}}} {\nabla}_{b}{{\nabla}_{c}{W_{d e}}} \Bigg)\non \\ 
    \non \\&&
    -\ve^{{a} {b} {c} {d} {e}} \Bigg( \frac{89}{256}  F_{{e} f} {\nabla}^{f}{{\nabla}_{b}{W_{c d}}} - \frac{249}{512}  F_{{d} {e}} {\nabla}^{f}{{\nabla}_{f}{W_{b c}}} +\frac{1}{8}F_{{e} f} {\nabla}_{b}{{\nabla}^{f}{W_{c d}}}+\frac{61}{256}  F_{{d} {e}} {\nabla}_{b}{{\nabla}^{f}}{W_{c f}} \non \\ 
    \non \\&&~~~~~~~~~~~~~~~
    +\frac{37}{128} F_{{d} {e}} {\nabla}^{f}{{\nabla}_{b}{W_{c f}}}+\frac{3}{128}F_{{e} f} {\nabla}_{b}{{\nabla}_{c}{W_{d}{}^{ f}}} \Bigg)\non \\ 
    \non \\&=& \ve^{{a} {b} {c} {d} {e}} \Bigg(
    \frac{3}{4} F_{{e} f} {\nabla}_{b}{{\nabla}^{f}{W_{c d}}}+\frac{5}{4} F_{{d} {e}} {\nabla}_{b}{{\nabla}^{f}}{W_{c f}} +\frac{1}{4}F_{{d} {e}} {\nabla}^{f}{{\nabla}_{b}{W_{c f}}}-\frac{3}{4}F_{{e} f} {\nabla}_{b}{{\nabla}_{c}{W_{d}{}^{ f}}} \Bigg)~, \non \\&& \\
    (5)&=&\ve^{{a} {b} {c} {d} {e}} \Bigg(- \frac{13}{32}  W_{{e} f} {\nabla}_{b}{{\nabla}^{f}{F_{c d}}} - \frac{17}{32} W_{{d} {e}} {\nabla}^{f}{{\nabla}_{f}{F_{b c}}} +\frac{11}{32}  W_{{e} {f}} {\nabla}^{{f}}{{\nabla}_{b}{F_{c d}}}
    \non \\&&~~~~~~~~~~~
    +\frac{1}{4} W_{{d} {e}} {\nabla}^{f}{{\nabla}_{b}{F_{c f}}} +\frac{3}{16} W_{{d} {e}} {\nabla}_{b}{{\nabla}^{f}{F_{c f}}} 
    - \frac{3}{16} W_{{e} f} {\nabla}_{b}{{\nabla}_{c}{F_{d}{}^{ f}}} \Bigg)  
    \non \\&&
    +\ve^{{b} {c} {d} {e} {e_{1}}} \Bigg(\frac{1}{4}  W_{{e} {e_{1}}} {\nabla}_{b}{{\nabla}_{c}{F^{{a}}{}_{d}}} - \frac{11}{32} \eta^{{a} {e_{6}}} W_{{e_{1}} {e_{6}}} {\nabla}_{b}{{\nabla}_{c}{F_{d e}}}+\frac{1}{32}  W_{{e} {e_{1}}} {\nabla}_{b}{{\nabla}^{{a}}{F_{c d}}} \Bigg) 
    \non \\
    &=&
    \ve^{{a} {b} {c} {d} {e}} \Bigg( - \frac{1}{4} W_{{e} {f}} {\nabla}_{b}{{\nabla}^{f}{F_{c d}}} - \frac{1}{2} W_{{d} {e}} {\nabla}^{f}{{\nabla}_{f}{F_{b c}}} \Bigg)~, \non \\ && 
    \\ 
    (6)&=& \ve^{{a} {b} {c} {d} {e}} \Bigg(- \frac{33}{128}   {\nabla}^{{e_{5}}}{F_{b {e_{5}}}} {\nabla}_{c}{W_{d e}}- \frac{15}{16} {\nabla}^{{e_{5}}}{F_{b c}} {\nabla}_{{e_{5}}}{W_{d e}} - \frac{51}{64}   {\nabla}^{{e_{5}}}{F_{b c}} {\nabla}_{d}{W_{{e} {e_{5}}}} 
    \non \\&&~~~~~~~~~~~
    +\frac{15}{128}  {\nabla}_{{b}}{F_{c {e_{5}}}} {\nabla}^{{e_{5}}}{W_{d e}} - \frac{39}{64} {\nabla}_{{b}}{F_{c}{}^{ {e_{5}}}} {\nabla}_{d}{W_{e {e_{5}}}} \Bigg) 
    \non \\&&
    + \ve^{{b} {c} {d} {e} {f}} \Bigg( \frac{27}{128}  {\nabla}^{{a}}{F_{b c}} {\nabla}_{d}{W_{ e f}} +\frac{105}{128}   {\nabla}_{b}{F^{{a}}{}_{ c}} {\nabla}_{d}{W_{e f}} \Bigg)
    \non \\
    &=&
    \ve^{{a} {b} {c} {d} {e}}\Bigg(- \frac{3}{2} {\nabla}^{{e_{5}}}{F_{b {e_{5}}}} {\nabla}_{c}{W_{d e}} +\frac{3}{4}{\nabla}_{{b}}{F_{c {e_{5}}}} {\nabla}^{{e_{5}}}{W_{d e}}- \frac{3}{2}{\nabla}_{{b}}{F_{c}{}^{ {e_{5}}}} {\nabla}_{d}{W_{e {e_{5}}}} \Bigg)~, \non \\ && \\
    (7)
    &=&
    \ve^{{a} {b} {c} {d} {e}} \Bigg(\frac{137}{512}  W_{{e} {e_{4}}}  {\nabla}_{{b}}{W} {\nabla}^{{e_{4}}}{W_{cd}} - \frac{429}{512}W_{{d} {e}}  {\nabla}^{{e_{4}}}{W} {\nabla}_{{e_{4}}}{W_{b c}}+\frac{157}{512}  W_{{e} {e_{4}}}  {\nabla}^{{e_{4}}}{W} {\nabla}_{{b}}{W_{c d}}
    \non \\&&~~~~~~~~~~~
    +\frac{35}{128} W_{{d} {e}}  {\nabla}^{{e_{4}}}{W} {\nabla}_{{b}}{W_{c {e_{4}}}} +\frac{77}{256}  W_{{d} {e}}  {\nabla}_{{b}}{W} {\nabla}^{{e_{4}}}{W_{c {e_{4}}}} - \frac{17}{256}W_{{e} {e_{4}}}  {\nabla}_{{b}}{W} {\nabla}_{c}{W_{d}{}^{ {e_{4}}}} \Bigg)
    \non \\&&
    +\ve^{{b} {c} {d} {e} {f}} \Bigg(\frac{73}{64} W_{{e} {f}}  {\nabla}^{{a}}{W} {\nabla}_{b}{W_{c d}} - \frac{155}{512} W_{{e} {f}}  {\nabla}_{b}{W} {\nabla}^{{a}}{W_{c d}} +\frac{39}{128}  W_{{e} {f}}  {\nabla}_{b}{W} {\nabla}_{c}{W^{{a}}{}_{ d}}
    \non \\&&~~~~~~~~~~~~~~~
    - \frac{173}{512} W_{{f}}{}^{a}  {\nabla}_{b}{W} {\nabla}_{c}{W_{d e}}\Bigg)
    \non \\
    &=&
    \frac{9}{4}\ve^{{a} {b} {c} {d} {e}} \Bigg(  W_{{e} {e_{4}}}  {\nabla}^{{e_{4}}}{W} {\nabla}_{{b}}{W_{cd}}+ W_{{d} {e}}  {\nabla}^{{e_{4}}}{W} {\nabla}_{{b}}{W_{c {e_{4}}}}\Bigg)~, 
    \non \\ && \\
    (8)
    &=&
    \ve^{{a} {b} {c} {d} {e}}\Bigg(- \frac{5}{192}  W_{{b} {c}} F^{{f} {e_{2}}} W_{{d} {e} {f} {e_{2}}}  - \frac{5}{192}  W_{{b} {c}} F^{{f} {e_{2}}} W_{{f} {e_{2}} {d} {e}}  - \frac{5}{512}  W^{{f} {e_{2}}} F_{{b} {c}} W_{{d} {e} {f} {e_{2}}} \non \non \\&&~~~~~~~~~~~
    - \frac{13}{384}  W_{{b} {c}} F^{{f} {e_{2}}} W_{{d} {f} {e} {e_{2}}} 
    - \frac{17}{1536}  W^{{f} {e_{2}}} F_{{b} {c}} W_{{f} {e_{2}} {d} {e}} 
    - \frac{7}{96}  \eta^{{f} {e_{2}}} \eta^{{e_{3}} {e_{4}}} W_{{b} {f}} F_{{c} {e_{3}}} W_{{d} {e_{2}} {e} {e_{4}}} 
    \non \\&&~~~~~~~~~~~
    - \frac{1}{384} W^{{f} {e_{2}}} F_{{b} {c}} W_{{d} {f} {e} {e_{2}}} \Bigg) 
    \non \\&&
    +\ve^{{b} {c} {d} {e} {f}} \Bigg(\frac{13}{768}  \eta^{{a} {e_{2}}} \eta^{{e_{3}} {e_{4}}} W_{{b} {e_{3}}} F_{{c} {d}} W_{{e} {f} {e_{2}} {e_{4}}}   - \frac{11}{768}  \eta^{{a} {e_{2}}} \eta^{{e_{3}} {e_{4}}} W_{{b} {c}} F_{{d} {e_{3}}} W_{{e} {f} {e_{2}} {e_{4}}}
    \non \\&&~~~~~~~~~~~~~~~
    +\frac{1}{64}  \eta^{{a} {e_{2}}} \eta^{{e_{3}} {e_{4}}} W_{{b} {e_{3}}} F_{{c} {d}} W_{{e_{2}} {e_{4}} {e} {f}} \Bigg) 
    \non \\
    &= &
    -\frac{1}{8} \ve^{{a} {b} {c} {d} {e}} \Bigg( W_{{b} {c}} F^{{f} {e_{2}}} W_{{d} {e} {f} {e_{2}}}  + \frac{1}{4}   W^{{f} {e_{2}}} F_{{b} {c}} W_{{d} {e} {f} {e_{2}}}  - \frac{3}{4}  \eta^{{f} {e_{2}}} \eta^{{e_{3}} {e_{4}}} W_{{b} {f}} F_{{c} {e_{3}}} W_{{d}  {e} {e_{2}} {e_{4}}}\Bigg)  ~, 
    \non \\ && \\
    (9)
    &=&
    \ve^{{a} {b} {c} {d} {e}} \Bigg(\frac{3}{16} F_{{e} {e_{4}}}  {\nabla}_{{b}}{W} {\nabla}^{{e_{4}}}{F_{c d}}+\frac{37}{64}  F_{{d} {e}}  {\nabla}^{{e_{4}}}{W} {\nabla}_{{e_{4}}}{F_{b c}} - \frac{1}{16}  F_{{d} {e}}  {\nabla}^{{e_{4}}}{W} {\nabla}_{b}{F_{c {e_{4}}}}  
    \non \\&&~~~~~~~~~~~
    - \frac{9}{32}  F_{{d} {e}}  {\nabla}_{{b}}{W} {\nabla}^{{e_{4}}}{F_{c {e_{4}}}} - \frac{1}{16} F_{{e} {e_{4}}}  {\nabla}_{{b}}{W} {\nabla}_{c}{F_{d}{}^{{e_{4}}}}\Bigg)  
    \non \\&&
    +\ve^{{b} {c} {d} {e} {f}} \Bigg(- \frac{9}{32} F_{{e} {f}}  {\nabla}_{b}{W} {\nabla}_{c}{F^{{a}}{}_{d}} - \frac{7}{32} F_{{e} {f}}  {\nabla}_{b}{W} {\nabla}^{{a}}{F_{cd}} \Bigg) 
    \non \\
    &=&
    \ve^{{a} {b} {c} {d} {e}} \Bigg(- \frac{1}{2} F_{{e} {e_{4}}}  {\nabla}_{{b}}{W} {\nabla}^{{e_{4}}}{F_{c d}} +\frac{1}{4}F_{{d} {e}}  {\nabla}^{{e_{4}}}{W} {\nabla}_{{e_{4}}}{F_{b c}} - F_{{d} {e}}  {\nabla}_{{b}}{W} {\nabla}^{{e_{4}}}{F_{c {e_{4}}}}  \Bigg)~,
    \non \\ && \\
    (10)
    &=&
    \ve^{{b} {c} {d} {e} {f}} \Bigg(- \frac{103}{512}  F_{{e} {f}}  {\nabla}^{{a}}{W} {\nabla}_{b}{W_{c d}} - \frac{63}{256}  F_{{e} {f}}  {\nabla}_{b}{W} {\nabla}_{c}{W^{{a}}{}_{d} } - \frac{161}{512}  F_{{e} {f}}  {\nabla}_{b}{W} {\nabla}^{{a}}{W_{c d}} 
    \non \\&&~~~~~~~~~~~
    +\frac{43}{512} F_{{f}}{}^{a}  {\nabla}_{b}{W} {\nabla}_{c}{W_{d e}} \Bigg) 
    \non \\&&
    + \ve^{{a} {b} {c} {d} {e}}\Bigg(- \frac{19}{512}  F_{{e} {e_{4}}}  {\nabla}_{{b}}{W} {\nabla}^{{e_{4}}}{W_{c d}} +\frac{33}{64}  F_{{d} {e}}  {\nabla}^{{e_{4}}}{W} {\nabla}_{{e_{4}}}{W_{b c}}  - \frac{221}{512}  F_{{e} {e_{4}}}  {\nabla}^{{e_{4}}}{W} {\nabla}_{{b}}{W_{c d}}  
    \non \\&&~~~~~~~~~~~~~~~
    +\frac{5}{32} F_{{d} {e}}  {\nabla}^{{e_{4}}}{W} {\nabla}_{{b}}{W_{c {e_{4}}}} - \frac{5}{8}  F_{{d} {e}}  {\nabla}_{{b}}{W} {\nabla}^{{e_{4}}}{W_{c {e_{4}}}} - \frac{83}{256} F_{{e} {e_{4}}}  {\nabla}_{{b}}{W} {\nabla}_{c}{W_{d}{}^{{e_{4}}}} \Bigg)  
    \non \\
    &=&
    \ve^{{a} {b} {c} {d} {e}} (- \frac{3}{4} F_{{e} {e_{4}}}  {\nabla}_{{b}}{W} {\nabla}^{{e_{4}}}{W_{c d}} - \frac{3}{4} F_{{e} {e_{4}}}  {\nabla}^{{e_{4}}}{W} {\nabla}_{{b}}{W_{c d}} - \frac{3}{2} F_{{d} {e}}  {\nabla}_{{b}}{W} {\nabla}^{{e_{4}}}{W_{c {e_{4}}}} \Bigg)~, 
    \non \\ && \\
    (11)
    &=&
    \ve^{{a} {b} {c} {d} {e}} \Bigg(\frac{11}{32}  W_{{e} {e_{4}}}  {\nabla}_{{b}}{W} {\nabla}^{{e_{4}}}{F_{c d}} +\frac{153}{256}  W_{{d} {e}}  {\nabla}^{{e_{4}}}{W} {\nabla}_{{e_{4}}}{F_{{b} c}} - \frac{19}{128}  W_{{e} {e_{4}}}  {\nabla}^{{e_{4}}}{W} {\nabla}_{{b}}{F_{c d}}  
    \non \\&&~~~~~~~~~~~
    - \frac{37}{64}  W_{{d} {e}}  {\nabla}^{{e_{4}}}{W} {\nabla}_{{b}}{F_{c {e_{4}}}} - \frac{61}{128} W_{{d} {e}}  {\nabla}_{{b}}{W} {\nabla}^{{e_{4}}}{F_{c {e_{4}}}}+\frac{9}{64} W_{{e} {e_{4}}}  {\nabla}_{{b}}{W} {\nabla}_{c}{F_{d}{}^{ {e_{4}}}} \Bigg)
    \non \\&&
    + \ve^{{b} {c} {d} {e} {f}} \Bigg(- \frac{81}{128} W_{{e} {f}}  {\nabla}_{b}{W} {\nabla}_{c}{F^{{a}}{}_{d}}  - \frac{25}{128} W_{{e} {f}}  {\nabla}_{b}{W} {\nabla}^{{a}}{F_{c d}} \Bigg) 
    \non \\
    &=&
    \ve^{{a} {b} {c} {d} {e}}  \Bigg(-\frac{3}{4} W_{{e} {e_{4}}}  {\nabla}_{{b}}{W} {\nabla}^{{e_{4}}}{F_{c d}} + \frac{3}{8}W_{{d} {e}}  {\nabla}^{{e_{4}}}{W} {\nabla}_{{e_{4}}}{F_{{b} c}} 
    - \frac{3}{2}W_{{d} {e}}  {\nabla}_{{b}}{W} {\nabla}^{{e_{4}}}{F_{c {e_{4}}}} \Bigg)~. \non \\ &&
\eea
\esubeq

\section{Different 5D $\cN=1$ notation and conventions}
\label{app:Conventions}

Table \ref{tab:ConvComp} provides
a brief translation scheme between our conventions and other groups \cite{Ohashi3, Bergshoeff1, deWit:2009de} that worked on the superconformal tensor calculus in five dimensions.

\begin{table*}[!htb]
\centering
\renewcommand{\arraystretch}{1.4}
\resizebox{16cm}{!}{
\begin{tabular}{@{}cccc@{}} \toprule
Our conventions & de Wit and Katmadas & Bergshoeff et al. & Fujita et al. \\ \midrule
$\eta^{\ha\hb}$ & $\eta^{ab}$ & $\eta^{ab}$ & $-\eta^{ab}$ \\
$\Gamma^\ha$ & $-\ri \gamma^a$ & $\ri \gamma^a$ & $\gamma^a$ \\
$\S^{\ha\hb}$ & $\frac{1}{2} \gamma^{ab}$ & $\frac{1}{2} \gamma^{ab}$ & $-\frac{1}{2} \gamma^{ab}$ \\
$\veps^{\ha\hb\hc\hd\he}$ & $-\ri\veps^{abcde}$ & $-\veps^{abcde}$ & $\veps^{abcde}$ \\
\midrule
$\psi_\hm{}^i$ & $\psi_\mu{}^i$ & $\psi_\mu{}^i$ & $2 \,\psi_\mu{}^i$ \\
$\cV_\hm{}^i{}_j$ & $-\frac{1}{2} V_\mu{}_j{}^i$ & $-V_\mu{}^i{}_j$ & $V_\mu{}^i{}_j$ \\
$\omega_\hm{}^{\ha\hb}$ & $\omega_\mu{}^{ab}$ & $-\omega_\mu{}^{ab}$ & $-\omega_\mu{}^{ab}$ \\
$\ri \,\phi_\hm{}^i$ & $\phi_\mu{}^i$ &
	$-\phi_\mu{}^i + \frac{1}{3} T_{ab} \gamma^{ab} \psi_\mu{}^i$ &
	$2 \,\phi_\mu{}^i - \frac{2}{3} v_{ab} \gamma^{ab} \psi_\mu{}^i$ \\
$f_\hm{}^\ha$ & $-f_\mu{}^a + \frac{1}{3} \psi_{\mu i} \gamma^a \chi^i$ &
	$-f_\mu{}^a + \frac{1}{3} \psi_{\mu}{}^i \gamma^a \chi_i$ &
	$-f_\mu{}^a + \frac{\ri}{24} \psi_\mu{}^i \gamma^a \chi_i$ \\ 
\midrule
$w^{\ha\hb}$ & $-4 T^{ab}$ & $\frac{16}{3} T^{ab}$ & $\frac{4}{3} v^{ab}$ \\
$\chi^i$ & $\chi^i$ & $\chi^i$ & $\frac{1}{32} \chi^i$ \\
$D$ & $D$ & $D$ & $\frac{1}{16} (D - \frac{8}{3} v^{ab} v_{ab})$ \\
\midrule
$R(Q)_{\ha\hb}{}^i$ & $\frac{1}{2} R(Q)_{ab}{}^i$ & 
	$\frac{1}{2}  R(Q)_{ab}{}^i$ &
	$R(Q)_{ab}{}^i$ \\
$R(M)_{\ha\hb}{}^{\hc\hd}$ &
	$R(M)_{ab}{}^{cd}$ &
	$- R(M)_{ab}{}^{cd}$ &
	$- R(M)_{ab}{}^{cd}$ \\
$R(J)_{\ha\hb}{}^i{}_j$ & 
	$-\frac{1}{2} R(\cV)_{ab}{}_j{}^i$ & 
	$-R(V)_{ab}{}^i{}_j$ & 
	$R(U)_{ab}{}^i{}_j$ \\
$\ri\, R(S)_{\ha\hb}{}^i$ & $\frac{1}{2} R(S)_{ab}{}^i$ &
	$-\frac{1}{2} R(S)_{ab}{}^i + \frac{1}{6} T_{cd} \gamma^{cd} R(Q)_{ab}{}^i$ &
	$R(S)_{ab}{}^i - \frac{1}{3} v_{cd} \gamma^{cd}  R(Q)_{ab}{}^i$\\
$R(K)_{\ha\hb}{}^\hc$ & $-R(K)_{ab}{}^c + \frac{1}{3} R(Q)_{ab i} \gamma^c \chi^i$
	& $-R(K)_{ab}{}^c + \frac{1}{3} R(Q)_{ab}{}^i \gamma^c \chi_i$
	& $-R(K)_{ab}{}^c + \frac{\ri}{24} R(Q)_{ab}{}^i \gamma^c \chi_i$ \\
\bottomrule
\end{tabular}}
\caption{Conventions for Weyl multiplet}\label{tab:ConvComp}
\end{table*}

It is important to note that the definitions of supersymmetry are
different between the various groups. Here the differences amount to
not only to normalisations, but also to additional field-dependent $S$
and $K$ transformations in the definition of $\delta_Q$. That is,
given a transformation $\delta_Q + \delta_S + \delta_K$ in our conventions
with respective parameters $\xi_\hal^i$, $\eta_\hal^i$, and $\L_K^\ha$,
we will find a transformation $\delta'_Q + \delta'_S + \delta'_K$ with
new parameters $\eps^i$, $\eta'^i$, and $\L_K'^a$ given in 
Table \ref{tab:N2SusyConventions}.

\begin{table*}[!hbt]
\centering
\renewcommand{\arraystretch}{1.4}
\begin{tabular}{@{}cccc@{}} \toprule
de Wit and Katmadas & Bergshoeff et al. & Fujita et al. \\ \midrule
$\eps^i = 2\,\xi^i$ & $\eps^i = 2\,\xi^i$ & $\eps^i =\xi^i$ \\
$\eta'^i = 2\ri \eta^i$
	& $\eta'^i = -2\ri \eta^i + \frac{2}{3} T_{ab} \gamma^{ab} \xi^i$ 
	& $\eta'^i = \ri\eta^i + \frac{1}{3} v_{ab} \gamma^{ab} \xi^i$ \\
$\L_K'^a = -\L_K^a + \frac{2}{3} \xi_i \gamma^a \chi^i$
	& $\L_K'^a = -\L_K^a + \frac{2}{3} \xi^i \gamma^a \chi_i$
	& $\L_K'^a = -\L_K^a + \frac{\ri}{24} \xi^i \gamma^a \chi_i$ \\
\bottomrule
\end{tabular}
\caption{Conventions for $\delta_Q+\delta_S+\delta_K$}\label{tab:N2SusyConventions}
\end{table*}

We also emphasise that each group uses the same
vector derivative $D_a$, corresponding to our $\nabla_\ha$, modulo
differing overall normalisations of the superconformal generators.
The additional gravitino-dependent terms in the $S$-supersymmetry
and special conformal connections in Table \ref{tab:ConvComp}
cancel against additional terms found within $\delta_Q$, so that the
vector derivative is unchanged.

For completeness, we also give in Table \ref{tab:ConvVect}
the relation between our conventions for the vector multiplet and the other
groups.
\begin{table*}[!hbt]
\centering
\renewcommand{\arraystretch}{1.4}
\begin{tabular}{@{}cccc@{}} \toprule
Our conventions & de Wit and Katmadas & Bergshoeff et al. & Fujita et al. \\ \midrule
$W$ & $\sigma$ & $\sigma$ & $M$ \\
$\l^i$ & $\Omega^i$ & $-\psi^i$ & $-2\Omega^i$ \\
$X^{ij}$ & $2 Y^{ij}$ & $2 Y^{ij}$ & $2Y^{ij}$ \\
\bottomrule
\end{tabular}
\caption{Conventions for vector multiplet}\label{tab:ConvVect}
\end{table*}
%



\begin{footnotesize}

\end{footnotesize}

\end{document}



\renewcommand{\thefootnote}{\arabic{footnote}}
\setcounter{footnote}{0}

\begin{flushright}
November, 2023
\end{flushright}
\vspace{5mm}

\begin{center}
{\Large \bf 
Supplementary file for\\ ``Components of curvature-squared invariants of minimal supergravity in five dimensions''
}
\end{center}

\begin{center}

{\bf
Gregory Gold,
Jessica Hutomo,
Saurish Khandelwal,\\
and Gabriele Tartaglino-Mazzucchelli
} \\
\vspace{5mm}

\footnotesize{
{\it 
School of Mathematics and Physics, University of Queensland,
\\
 St Lucia, Brisbane, Queensland 4072, Australia}
}
\vspace{2mm}
~\\
\texttt{g.gold@uq.edu.au; 
j.hutomo@uq.edu.au; 
s.khandelwal@uq.edu.au;
\\ g.tartaglino-mazzucchelli@uq.edu.au}\\
\vspace{2mm}

\end{center}


\bigskip\hrule
\tableofcontents{}
\vspace{1cm}
\bigskip\hrule

\allowdisplaybreaks

\section{Overview}

This file serves as a supplementary file to the primary paper \cite{GGJS2023-3}. We present here the complete superconformal component results of various identities, superfields, and actions computed directly by \emph{Cadabra} including all bosonic and fermionic contributions. The only exceptions where we present only the  bosonic contribution are the four (fermionic) derivative supercurrent descendants of sections \ref{J4WeylComplete}, \ref{J4logWComplete}, and \ref{J4R2Complete} as well as the BF action for the log invariant given in \ref{BFlogWDegauged} in which we have written the superconformally covariant vector derivative $\nabla_a$ in terms of thee degauged covariant derivative $\cD_a$, gauged fixed $W=1$, and removed fermions.

Due to the computational source of these results, a unique symmetric and antisymmetric index notation is used denoted by underlining and hatting indices, respectively. For example
\bsubeq
\bea
    \l^{\a}_{\underline{i}} \l^{\b}_{\underline{j}} &:=& \l^{\a}_{(i} \l^{\b}_{j)} = \hf \l^{\a}_{i} \l^{\b}_{j} + \hf \l^{\a}_{j} \l^{\b}_{i} ~, \\
    W_{c \hat{a}} W_{\hat{b} d} &:=&  W_{c [a} W_{b] d} = \hf W_{c a} W_{b d} - \hf W_{c b} W_{a d} ~.
\eea
\esubeq
These are employed only when a field has explicit symmetric or antisymmetric indices by construction. Note also that two descendants of the super-Weyl tensor $W_{\a \b} = \hf (\Sigma^{a b})_{\a \b} W_{a b}$ are written here in their form prior to component projection being $Y$ and $X^i_\a$. These projections are trivial however.
\bsubeq
\bea
    \chi^i_\a &:=& \frac{3 \ri}{32} X^i_\a \loco~, \\
    D &:=& -\frac{3}{128} Y \loco~.
\eea
\esubeq
Also recall $\Phi_{a b}{}^{i j} = - \frac{3 \ri}{4} X_{a b}{}^{i j}$. These redefinitions and projections should be applied depending on the context of the results (either looking at genuine superfield dependants or at component results). We trust the reader will have enough information to decide depending on the context and their need. Also keep in mind the difference in notation between this file and the main paper of the Riemann tensor and its contractions ($R \rightarrow \cR$) as well as the composite connection ($f \rightarrow \mathfrak{f}$).

Each subsection of this document is titled after a single field or object, and the contents therein include its component structure. Be aware that in most cases the bosonic parts of the results are larger than and may not appear at a glance to match those of the main paper \cite{GGJS2023-3}. This is because there are various simplifications (e.g., algebraic manipulations, sometimes integration by parts) that are performed by hand on the results given in \cite{GGJS2023-3}. 

This file is organised as follows. In sections \ref{sec:Weyl2} and \ref{sec:logW}, we give results for the descendants of the composite primary superfields, $H^{i j}_{\rm Weyl}$ and $H^{i j}_{\log W}$, respectively, both of which were introduced in section 3 of the main paper \cite{GGJS2023-3}. In section \ref{BFlogWDegauged} of the supplementary file, we give the raw bosonic results which were simplified and given in subsection 5.2.1 of the main paper. In sections \ref{SupercurrentWeylComplete}, \ref{SupercurrentlogWComplete}, and \ref{SupercurrentR2Complete}, we give results for the descendants of the supercurrent equations of motion (or $Y$ equations of motion) for each invariant discussed in subsections 7.3, 7.4, and 7.5.1, respectively, of the main paper. In section \ref{SupercurrentR2Complete}, we also give results for the descendants of the linear multiplet compensator equations of motion (or $F$ equations of motion) discussed in subsection 7.5.2 of the main paper. Finally, in section \ref{DegaugedIdentities}, we give the degauged expressions of various vector covariant derivatives acting on multiplet fields that are applicable throughout this file and the main paper.

\section{Curvature-squared multiplets in components} \label{sec:curvaturesquared}

\subsection{Weyl squared} \label{sec:Weyl2}

\subsubsection{$\varphi^i_{\a \, {\rm Weyl}}$}

\begin{adjustwidth}{0cm}{2cm}
\doublespacedmathbegin
-C_{\alpha}\,^{\beta \rho \lambda} W_{\beta \rho \lambda}\,^{i} - \frac{3}{4}W_{\alpha}\,^{\beta} W^{\rho \lambda} W_{\beta \rho \lambda}\,^{i} - \frac{3}{4}W^{\beta \rho}\,_{\lambda}\,^{i} \nabla_{\alpha}\,^{\lambda}{W_{\beta \rho}} - \frac{3}{4}W_{\alpha}\,^{\beta}\,_{\rho}\,^{i} \nabla^{\rho \lambda}{W_{\beta \lambda}}+\Phi^{\beta \rho i}\,_{j} W_{\alpha \beta \rho}\,^{j}+\Phi_{\alpha}\,^{\beta i}\,_{j} X^{j}_{\beta} - \frac{3}{2}W^{\beta \rho} \nabla_{\alpha}\,^{\lambda}{W_{\beta \rho \lambda}\,^{i}}+\frac{3}{4}W^{\beta \rho} W_{\beta \rho} X^{i}_{\alpha}+\frac{3}{2}W^{\beta}\,_{\rho} \nabla^{\rho \lambda}{W_{\alpha \beta \lambda}\,^{i}} - \frac{3}{4}W^{\beta}\,_{\rho} \nabla_{\alpha}\,^{\rho}{X^{i}_{\beta}} - \frac{3}{4}W_{\alpha \beta} \nabla^{\beta \rho}{X^{i}_{\rho}} - \frac{3}{32}Y X^{i}_{\alpha}+\frac{9}{8}X^{i \beta} \nabla_{\alpha}\,^{\rho}{W_{\beta \rho}} - \frac{3}{8}X^{i}_{\rho} \nabla^{\rho \beta}{W_{\alpha \beta}}
\doublespacedmathend
\end{adjustwidth}

\subsubsection{$F_{{\rm Weyl}}$}

\begin{adjustwidth}{0cm}{2cm}
\doublespacedmathbegin
{}\frac{3}{4}{\rm i} W_{a b}\,^{\alpha}\,_{i} \epsilon^{a b}\,_{e}\,^{c d} \nabla^{e}{W_{c d \alpha}\,^{i}}+\frac{3}{4}{\rm i} W^{a b} W_{a}\,^{c \alpha}\,_{i} W_{b c \alpha}\,^{i}+\frac{21}{8}{\rm i} W^{a b} W_{a b}\,^{\alpha}\,_{i} X^{i}_{\alpha}+\frac{1}{12}C_{a b c d} C^{a b c d}+\frac{5}{4}C_{a b c d} W^{a b} W^{c d}+\frac{5}{4}C_{a b c d} W^{a c} W^{b d} - \frac{33}{128}W^{a b} W^{c d} W_{a b} W_{c d}+\frac{17}{4}W^{a b} W^{c d} W_{a c} W_{b d} - \frac{55}{32}W_{a}\,^{b} W^{a c} W^{d}\,_{b} W_{d c}+\frac{1}{12}C_{a b c d} C^{c d a b}+\frac{1}{6}C_{a b c d} C^{a c b d} - \frac{3}{2}\nabla_{c}{W_{a b}} \nabla^{c}{W^{a b}}+\frac{3}{2}\nabla_{c}{W_{a b}} \nabla^{a}{W^{c b}} - \frac{2}{3}\Phi^{a b}\,_{i j} \Phi_{a b}\,^{i j} - \frac{3}{2}{\rm i} (\Gamma_{a})^{\alpha \beta} X_{i \alpha} \nabla^{a}{X^{i}_{\beta}}-3W_{a b} \nabla_{c}{\nabla^{a}{W^{c b}}}+\frac{9}{16}\epsilon^{e a b c d} W_{a b} W_{c d} \nabla^{{e_{1}}}{W_{e {e_{1}}}} - \frac{3}{16}W^{a b} W_{a b} Y+\frac{3}{128}{Y}^{2}
\doublespacedmathend
\end{adjustwidth}

\subsubsection{$\cH^{a}_{{\rm Weyl}}$}

\begin{adjustwidth}{0cm}{4cm}
\doublespacedmathbegin
-{\rm i} W^{b c \alpha}\,_{i} \nabla^{a}{W_{b c \alpha}\,^{i}}-3{\rm i} W^{b}\,_{c}\,^{\alpha}\,_{i} \nabla^{c}{W^{a}\,_{b \alpha}\,^{i}}+\frac{9}{16}{\rm i} \epsilon^{a b c d {e_{1}}} W_{b c} W_{d}\,^{e \alpha}\,_{i} W_{{e_{1}} e \alpha}\,^{i} - \frac{9}{64}{\rm i} (\Gamma^{a})^{\alpha \beta} W^{b c} W_{b c \alpha i} X^{i}_{\beta}+\frac{27}{128}{\rm i} \epsilon^{a b c d e} W_{b c} W_{d e}\,^{\alpha}\,_{i} X^{i}_{\alpha}+\frac{9}{64}{\rm i} \epsilon^{{e_{1}} b c d e} (\Sigma_{{e_{1}}}{}^{\, a})^{\alpha \beta} W_{b c} W_{d e \alpha i} X^{i}_{\beta}+\frac{9}{32}{\rm i} W^{b}\,_{c} W_{b d}\,^{\alpha}\,_{i} \epsilon^{a d e {e_{1}} c} (\Sigma_{e {e_{1}}})_{\alpha}{}^{\beta} X^{i}_{\beta}+\frac{9}{32}{\rm i} W^{a b} W_{b c}\,^{\alpha}\,_{i} (\Gamma^{c})_{\alpha}{}^{\beta} X^{i}_{\beta}+\frac{9}{32}{\rm i} W_{b}\,^{c} W^{a}\,_{c}\,^{\alpha}\,_{i} (\Gamma^{b})_{\alpha}{}^{\beta} X^{i}_{\beta} - \frac{9}{32}{\rm i} \epsilon^{b c d e {e_{1}}} W_{b c} W_{d e}\,^{\alpha}\,_{i} W_{{e_{1}}}\,^{a}\,_{\alpha}\,^{i}+\frac{9}{16}{\rm i} \epsilon^{a b d e {e_{1}}} W_{b}\,^{c} W_{d e}\,^{\alpha}\,_{i} W_{{e_{1}} c \alpha}\,^{i}+\frac{5}{192}\epsilon^{a b c {e_{1}} {e_{2}}} C_{b c}\,^{d e} C_{{e_{1}} {e_{2}} d e} - \frac{5}{64}\epsilon^{a b c {e_{1}} {e_{2}}} C_{b c d e} W^{d e} W_{{e_{1}} {e_{2}}}+\frac{3}{64}\epsilon^{a b c {e_{1}} {e_{2}}} C_{b c}\,^{d e} W_{{e_{1}} d} W_{{e_{2}} e} - \frac{1}{32}\epsilon^{a b c d {e_{2}}} C_{b c d e} W^{e {e_{1}}} W_{{e_{2}} {e_{1}}} - \frac{19}{64}\epsilon^{a d e {e_{1}} {e_{2}}} W^{b c} W_{d e} W_{{e_{1}} {e_{2}}} W_{b c}+\frac{15}{16}\epsilon^{a d e {e_{1}} {e_{2}}} W^{b c} W_{d e} W_{{e_{1}} b} W_{{e_{2}} c} - \frac{1}{4}\epsilon^{a}\,_{b}\,^{d e {e_{1}}} W^{b c} W_{d e} W_{{e_{1}}}\,^{{e_{2}}} W_{{e_{2}} c} - \frac{1}{16}\epsilon^{a}\,_{b d}\,^{{e_{1}} {e_{2}}} W^{b c} W^{d e} W_{{e_{1}} c} W_{{e_{2}} e}%
+\frac{1}{32}\epsilon^{a b c {e_{1}} {e_{2}}} C_{b c}\,^{d e} C_{d e {e_{1}} {e_{2}}} - \frac{5}{64}\epsilon^{a d e {e_{1}} {e_{2}}} C_{b c d e} W^{b c} W_{{e_{1}} {e_{2}}}+\frac{3}{64}\epsilon^{a d e {e_{1}} {e_{2}}} C^{b c}\,_{d e} W_{{e_{1}} b} W_{{e_{2}} c} - \frac{1}{32}\epsilon^{a b d e {e_{2}}} C_{b c d e} W^{c {e_{1}}} W_{{e_{2}} {e_{1}}}+\frac{1}{48}\epsilon^{a b c {e_{1}} {e_{2}}} C_{b c}\,^{d e} C_{{e_{1}} d {e_{2}} e} - \frac{1}{16}\epsilon^{a b d {e_{1}} {e_{2}}} C_{b c d e} W^{c e} W_{{e_{1}} {e_{2}}} - \frac{3}{16}\epsilon^{a b d {e_{1}} {e_{2}}} C_{b}\,^{c}\,_{d}\,^{e} W_{{e_{1}} c} W_{{e_{2}} e}+\frac{5}{192}\epsilon^{a d e {e_{1}} {e_{2}}} C^{b c}\,_{d e} C_{b c {e_{1}} {e_{2}}}+\frac{1}{12}\epsilon^{a b d {e_{1}} {e_{2}}} C_{b}\,^{c}\,_{d}\,^{e} C_{{e_{1}} c {e_{2}} e}+\frac{1}{48}\epsilon^{a b d {e_{1}} {e_{2}}} C_{b}\,^{c}\,_{d}\,^{e} C_{c e {e_{1}} {e_{2}}}+\frac{1}{2}C^{a}\,_{b c d} \nabla^{b}{W^{c d}} - \frac{9}{16}W^{a}\,_{b} W_{c d} \nabla^{b}{W^{c d}} - \frac{9}{16}W^{b c} W_{b c} \nabla_{d}{W^{a d}}+\frac{15}{4}W^{a}\,_{b} W_{c d} \nabla^{c}{W^{b d}}+\frac{9}{4}W^{a b} W_{b c} \nabla_{d}{W^{d c}}+\frac{9}{4}W^{b}\,_{c} W_{b d} \nabla^{c}{W^{a d}}+\frac{1}{2}C_{b c}\,^{a}\,_{d} \nabla^{d}{W^{b c}} - \frac{1}{2}C_{b c}\,^{a}\,_{d} \nabla^{b}{W^{c d}}+\frac{1}{2}C^{a}\,_{b c d} \nabla^{c}{W^{b d}} - \frac{3}{64}\epsilon^{b c}\,_{{e_{1}}}\,^{d e} \nabla^{a}{W_{b c}} \nabla^{{e_{1}}}{W_{d e}}%
+\frac{3}{64}\epsilon^{a b c d e} \nabla_{{e_{1}}}{W_{b c}} \nabla^{{e_{1}}}{W_{d e}}+\frac{3}{8}\epsilon^{a}\,_{e}\,^{b}\,_{{e_{1}} d} \nabla^{e}{W_{b c}} \nabla^{{e_{1}}}{W^{d c}} - \frac{39}{32}\epsilon^{a}\,_{{e_{1}}}\,^{b c d} \nabla^{{e_{1}}}{W_{b c}} \nabla^{e}{W_{d e}}+\frac{21}{16}\epsilon_{e b {e_{1}}}\,^{c d} \nabla^{e}{W^{a b}} \nabla^{{e_{1}}}{W_{c d}}+\frac{3}{2}{\rm i} W^{a}\,_{b}\,^{\alpha}\,_{i} \nabla^{b}{X^{i}_{\alpha}} - \frac{3}{32}\epsilon^{a}\,_{{e_{1}}}\,^{b d e} \nabla^{{e_{1}}}{W_{b c}} \nabla^{c}{W_{d e}} - \frac{1}{6}\epsilon^{a b c d e} \Phi_{b c i j} \Phi_{d e}\,^{i j} - \frac{3}{2}{\rm i} X_{i}^{\alpha} \nabla^{b}{W^{a}\,_{b \alpha}\,^{i}} - \frac{3}{2}{\rm i} (\Sigma^{a}{}_{\, b})^{\alpha \beta} X_{i \alpha} \nabla^{b}{X^{i}_{\beta}}+\frac{1}{2}W^{c d} \nabla^{b}{C^{a}\,_{b c d}}+\frac{3}{4}W^{b c} W_{b c} \nabla^{d}{W^{a}\,_{d}}+\frac{15}{16}W^{a}\,_{b} W^{c d} \nabla^{b}{W_{c d}}+\frac{3}{4}W^{b c} W_{b d} \nabla^{d}{W^{a}\,_{c}} - \frac{3}{4}W^{b c} W^{a}\,_{b} \nabla^{d}{W_{c d}} - \frac{3}{4}W_{b}\,^{c} W_{d c} \nabla^{b}{W^{a d}} - \frac{3}{2}W^{a b} W_{c}\,^{d} \nabla^{c}{W_{b d}} - \frac{3}{4}W^{a b} W^{c}\,_{b} \nabla^{d}{W_{c d}}+\frac{1}{2}W^{b c} \nabla^{d}{C_{b c}\,^{a}\,_{d}}+\frac{1}{2}W^{b d} \nabla^{c}{C_{b c}\,^{a}\,_{d}} - \frac{1}{2}W^{b c} \nabla^{d}{C^{a}\,_{b c d}}%
+\frac{3}{8}\epsilon_{{e_{1}}}\,^{d e b c} W_{b c} \nabla^{a}{\nabla^{{e_{1}}}{W_{d e}}} - \frac{3}{4}\epsilon^{a}\,_{{e_{1}}}\,^{d e b} W_{b c} \nabla^{{e_{1}}}{\nabla^{c}{W_{d e}}}+\frac{3}{16}\epsilon^{a}\,_{{e_{1}}}\,^{d b c} W_{b c} \nabla^{e}{\nabla^{{e_{1}}}{W_{d e}}} - \frac{9}{16}\epsilon_{e {e_{1}} d}\,^{b c} W_{b c} \nabla^{e}{\nabla^{{e_{1}}}{W^{a d}}} - \frac{3}{16}\epsilon^{a}\,_{{e_{1}}}\,^{d b c} W_{b c} \nabla^{{e_{1}}}{\nabla^{e}{W_{d e}}} - \frac{3}{4}\epsilon^{a}\,_{{e_{1}}}\,^{d e b} W_{b c} \nabla^{c}{\nabla^{{e_{1}}}{W_{d e}}}+\frac{1}{16}\epsilon^{c d e {e_{1}} {e_{2}}} C^{a b}\,_{c d} W_{e {e_{1}}} W_{{e_{2}} b}+\frac{5}{8}\epsilon^{c d e {e_{1}} {e_{2}}} W^{a b} W_{c d} W_{e {e_{1}}} W_{{e_{2}} b} - \frac{3}{8}\epsilon_{{e_{1}}}\,^{d e b c} W_{b c} \nabla^{{e_{1}}}{\nabla^{a}{W_{d e}}}+\frac{3}{8}\epsilon^{a}\,_{e {e_{1}} d}\,^{b} W_{b c} \nabla^{e}{\nabla^{{e_{1}}}{W^{d c}}} - \frac{3}{16}W^{a}\,_{b} \nabla^{b}{Y} - \frac{3}{16}Y \nabla_{b}{W^{a b}}
\doublespacedmathend
\end{adjustwidth}

\subsection{Log} \label{sec:logW}

\subsubsection{$\varphi^i_{\a \, \log}$}

\begin{adjustwidth}{0cm}{3cm}
\doublespacedmathbegin
{} - \frac{3}{512}Y X^{j}_{\alpha} - \frac{3}{16}W_{\alpha \beta} \nabla^{\beta \rho}{X^{j}_{\rho}} - \frac{3}{8}X^{j}_{\rho} \nabla^{\rho \beta}{W_{\alpha \beta}} - \frac{3}{4}W^{\beta}\,_{\rho} \nabla^{\rho \lambda}{W_{\alpha \beta \lambda}\,^{j}} - \frac{3}{8}F_{\alpha \beta} {W}^{-1} \nabla^{\beta \rho}{X^{j}_{\rho}} - \frac{3}{4}X^{j}_{\rho} {W}^{-1} \nabla^{\rho \beta}{F_{\alpha \beta}} - \frac{1}{2}\Phi_{\alpha}\,^{\beta j}\,_{i} X^{i}_{\beta} - \frac{3}{32}\nabla_{\beta \rho}{\nabla^{\beta \rho}{X^{j}_{\alpha}}}+\frac{15}{16}W^{\beta}\,_{\rho} \nabla_{\alpha}\,^{\rho}{X^{j}_{\beta}} - \frac{27}{16}X^{j \beta} \nabla_{\alpha}\,^{\rho}{W_{\beta \rho}} - \frac{3}{4}F^{\beta}\,_{\rho} {W}^{-1} \nabla^{\rho \lambda}{W_{\alpha \beta \lambda}\,^{j}}+\frac{27}{32}W_{\alpha \beta} X^{j}_{\rho} {W}^{-1} \nabla^{\beta \rho}{W} - \frac{27}{32}W_{\alpha}\,^{\beta} W_{\beta}\,^{\rho} X^{j}_{\rho} - \frac{285}{256}W^{\beta \rho} W_{\beta \rho} X^{j}_{\alpha} - \frac{1}{2}{\rm i} \Phi_{\alpha \beta}\,^{j}\,_{i} {W}^{-1} \nabla^{\beta \rho}{\lambda^{i}_{\rho}}+\frac{1}{9}{\rm i} \lambda_{i \rho} {W}^{-1} \nabla^{\rho \beta}{\Phi_{\alpha \beta}\,^{j i}}+\frac{3}{32}{\rm i} Y {W}^{-1} \nabla_{\alpha}\,^{\beta}{\lambda^{j}_{\beta}}+\frac{3}{64}{\rm i} \lambda^{j}_{\beta} {W}^{-1} \nabla_{\alpha}\,^{\beta}{Y} - \frac{3}{64}{\rm i} F_{\alpha}\,^{\beta} Y \lambda^{j}_{\beta} {W}^{-2}%
+\frac{3}{64}{\rm i} X^{j}\,_{i} Y \lambda^{i}_{\alpha} {W}^{-2}+\frac{3}{8}F^{\beta}\,_{\rho} {W}^{-1} \nabla_{\alpha}\,^{\rho}{X^{j}_{\beta}}+\frac{3}{4}W^{\beta \rho} \nabla_{\alpha}\,^{\lambda}{W_{\beta \rho \lambda}\,^{j}} - \frac{3}{4}F_{\alpha}\,^{\beta} F_{\beta}\,^{\rho} X^{j}_{\rho} {W}^{-2} - \frac{3}{16}F^{\beta \rho} F_{\beta \rho} X^{j}_{\alpha} {W}^{-2} - \frac{27}{16}W_{\alpha}\,^{\beta} F_{\beta}\,^{\rho} X^{j}_{\rho} {W}^{-1}+\frac{27}{16}W^{\beta \rho} F_{\alpha \beta} X^{j}_{\rho} {W}^{-1} - \frac{27}{16}W^{\beta \rho} F_{\beta \rho} X^{j}_{\alpha} {W}^{-1} - \frac{1}{2}W_{\alpha}\,^{\beta} W^{\rho \lambda} W_{\beta \rho \lambda}\,^{j}+\frac{3}{8}X_{i \beta} {W}^{-1} \nabla_{\alpha}\,^{\beta}{X^{j i}} - \frac{3}{4}X^{j}\,_{i} {W}^{-1} \nabla_{\alpha}\,^{\beta}{X^{i}_{\beta}}+\frac{3}{8}X^{j}\,_{i} F_{\alpha}\,^{\beta} X^{i}_{\beta} {W}^{-2}+\frac{9}{16}\lambda_{i \alpha} X^{j \beta} X^{i}_{\beta} {W}^{-1} - \frac{3}{4}\lambda^{\beta}_{i} X^{j}_{\alpha} X^{i}_{\beta} {W}^{-1}+\frac{3}{8}\lambda^{\beta}_{i} X^{j}_{\beta} X^{i}_{\alpha} {W}^{-1}+\frac{3}{16}\lambda^{j \beta} X_{i \alpha} X^{i}_{\beta} {W}^{-1}+\frac{3}{32}X_{i k} X^{i k} X^{j}_{\alpha} {W}^{-2}+\frac{3}{4}{\rm i} W_{\beta \rho} {W}^{-1} \nabla_{\alpha}\,^{\beta}{\nabla^{\rho \lambda}{\lambda^{j}_{\lambda}}}+\frac{19}{32}{\rm i} \lambda^{j}_{\lambda} {W}^{-1} \nabla^{\lambda \beta}{\nabla_{\alpha}\,^{\rho}{W_{\beta \rho}}}+\frac{5}{32}{\rm i} \lambda^{j}_{\lambda} {W}^{-1} \nabla_{\alpha}\,^{\beta}{\nabla^{\lambda \rho}{W_{\beta \rho}}}%
+\frac{63}{64}{\rm i} {W}^{-1} \nabla_{\alpha}\,^{\beta}{W_{\beta \rho}} \nabla^{\rho \lambda}{\lambda^{j}_{\lambda}} - \frac{15}{64}{\rm i} {W}^{-1} \nabla^{\beta \lambda}{W_{\beta \rho}} \nabla_{\alpha}\,^{\rho}{\lambda^{j}_{\lambda}} - \frac{1}{4}{\rm i} W_{\alpha \beta} {W}^{-1} \nabla_{\lambda}\,^{\beta}{\nabla^{\lambda \rho}{\lambda^{j}_{\rho}}}+\frac{3}{32}{\rm i} \lambda^{j}_{\alpha} {W}^{-1} \nabla_{\lambda}\,^{\beta}{\nabla^{\lambda \rho}{W_{\beta \rho}}}+\frac{1}{16}{\rm i} \lambda^{j}_{\rho} {W}^{-1} \nabla_{\lambda}\,^{\beta}{\nabla^{\lambda \rho}{W_{\alpha \beta}}}+\frac{3}{32}{\rm i} \lambda^{j}_{\rho} {W}^{-1} \nabla_{\lambda}\,^{\rho}{\nabla^{\lambda \beta}{W_{\alpha \beta}}}+\frac{15}{64}{\rm i} {W}^{-1} \nabla_{\lambda}\,^{\beta}{W_{\alpha \beta}} \nabla^{\lambda \rho}{\lambda^{j}_{\rho}} - \frac{3}{32}{\rm i} {W}^{-1} \nabla_{\lambda}\,^{\rho}{W_{\alpha \beta}} \nabla^{\lambda \beta}{\lambda^{j}_{\rho}} - \frac{9}{64}{\rm i} {W}^{-1} \nabla_{\lambda}\,^{\beta}{W_{\beta \rho}} \nabla^{\lambda \rho}{\lambda^{j}_{\alpha}} - \frac{1}{2}{\rm i} F_{\alpha \beta} F^{\rho}\,_{\lambda} {W}^{-3} \nabla^{\beta \lambda}{\lambda^{j}_{\rho}}+\frac{1}{2}{\rm i} F_{\alpha \beta} \lambda^{j \rho} {W}^{-3} \nabla^{\beta \lambda}{F_{\rho \lambda}}+\frac{3}{8}{\rm i} F_{\alpha \beta} \lambda^{j \rho} {W}^{-2} \nabla^{\beta \lambda}{W_{\rho \lambda}} - \frac{1}{2}{\rm i} F_{\alpha}\,^{\beta} \lambda^{j}_{\lambda} {W}^{-3} \nabla^{\lambda \rho}{F_{\beta \rho}} - \frac{13}{16}{\rm i} F_{\alpha}\,^{\beta} \lambda^{j}_{\lambda} {W}^{-2} \nabla^{\lambda \rho}{W_{\beta \rho}}+\frac{1}{16}{\rm i} F^{\beta}\,_{\rho} \lambda^{j}_{\alpha} {W}^{-2} \nabla^{\rho \lambda}{W_{\beta \lambda}} - \frac{1}{2}{\rm i} F^{\beta}\,_{\rho} \lambda^{j}_{\lambda} {W}^{-3} \nabla^{\rho \lambda}{F_{\alpha \beta}} - \frac{7}{16}{\rm i} F^{\beta}\,_{\rho} \lambda^{j}_{\lambda} {W}^{-2} \nabla^{\rho \lambda}{W_{\alpha \beta}} - \frac{3}{8}{\rm i} W_{\alpha \lambda} F^{\beta}\,_{\rho} {W}^{-2} \nabla^{\lambda \rho}{\lambda^{j}_{\beta}} - \frac{9}{8}{\rm i} W_{\alpha \beta} W^{\rho}\,_{\lambda} {W}^{-1} \nabla^{\beta \lambda}{\lambda^{j}_{\rho}}+\frac{1}{4}{\rm i} W_{\alpha \lambda} \lambda^{j \beta} {W}^{-2} \nabla^{\lambda \rho}{F_{\beta \rho}}%
+\frac{9}{16}{\rm i} W_{\alpha \beta} \lambda^{j \rho} {W}^{-1} \nabla^{\beta \lambda}{W_{\rho \lambda}}+\frac{9}{16}{\rm i} W_{\alpha}\,^{\beta} W_{\beta \rho} {W}^{-1} \nabla^{\rho \lambda}{\lambda^{j}_{\lambda}} - \frac{1}{2}{\rm i} W_{\alpha}\,^{\beta} \lambda^{j}_{\lambda} {W}^{-2} \nabla^{\lambda \rho}{F_{\beta \rho}}+\frac{23}{64}{\rm i} W_{\alpha}\,^{\beta} \lambda^{j}_{\lambda} {W}^{-1} \nabla^{\lambda \rho}{W_{\beta \rho}}+\frac{3}{8}{\rm i} W^{\rho}\,_{\lambda} F_{\alpha \beta} {W}^{-2} \nabla^{\lambda \beta}{\lambda^{j}_{\rho}}+\frac{3}{8}{\rm i} W^{\beta}\,_{\rho} F_{\alpha \beta} {W}^{-2} \nabla^{\rho \lambda}{\lambda^{j}_{\lambda}} - \frac{9}{32}{\rm i} W^{\beta}\,_{\rho} W_{\beta \lambda} {W}^{-1} \nabla^{\rho \lambda}{\lambda^{j}_{\alpha}} - \frac{59}{64}{\rm i} W^{\beta}\,_{\rho} \lambda^{j}_{\alpha} {W}^{-1} \nabla^{\rho \lambda}{W_{\beta \lambda}} - \frac{3}{8}{\rm i} W^{\beta}\,_{\rho} \lambda^{j}_{\lambda} {W}^{-2} \nabla^{\rho \lambda}{F_{\alpha \beta}}+\frac{23}{64}{\rm i} W^{\beta}\,_{\rho} \lambda^{j}_{\lambda} {W}^{-1} \nabla^{\rho \lambda}{W_{\alpha \beta}} - \frac{3}{8}{\rm i} W^{\rho}\,_{\lambda} \lambda^{j}_{\rho} {W}^{-2} \nabla^{\lambda \beta}{F_{\alpha \beta}}+\frac{9}{16}{\rm i} W^{\beta}\,_{\rho} \lambda^{j}_{\beta} {W}^{-1} \nabla^{\rho \lambda}{W_{\alpha \lambda}} - \frac{3}{64}{\rm i} Y \lambda^{j}_{\beta} {W}^{-2} \nabla_{\alpha}\,^{\beta}{W} - \frac{1}{4}{\rm i} F_{\alpha \beta} \lambda_{i \rho} {W}^{-3} \nabla^{\beta \rho}{X^{j i}} - \frac{3}{16}{\rm i} W_{\alpha \beta} \lambda_{i \rho} {W}^{-2} \nabla^{\beta \rho}{X^{j i}}+\frac{1}{4}{\rm i} X^{j}\,_{i} F_{\alpha \beta} {W}^{-3} \nabla^{\beta \rho}{\lambda^{i}_{\rho}}+\frac{3}{8}{\rm i} X^{j}\,_{i} W_{\alpha \beta} {W}^{-2} \nabla^{\beta \rho}{\lambda^{i}_{\rho}}+\frac{1}{6}{\rm i} X^{j}\,_{i} \lambda^{i}_{\rho} {W}^{-3} \nabla^{\rho \beta}{F_{\alpha \beta}}+\frac{3}{16}{\rm i} X^{j}\,_{i} \lambda^{i}_{\rho} {W}^{-2} \nabla^{\rho \beta}{W_{\alpha \beta}}+\frac{3}{16}{\rm i} \lambda_{i \alpha} X^{j}_{\beta} {W}^{-2} \nabla^{\beta \rho}{\lambda^{i}_{\rho}}%
+\frac{5}{32}{\rm i} \lambda_{i \alpha} X^{i}_{\beta} {W}^{-2} \nabla^{\beta \rho}{\lambda^{j}_{\rho}} - \frac{1}{8}{\rm i} \lambda_{i \alpha} \lambda^{i}_{\rho} {W}^{-2} \nabla^{\rho \beta}{X^{j}_{\beta}} - \frac{5}{32}{\rm i} \lambda_{i \beta} X^{j}_{\alpha} {W}^{-2} \nabla^{\beta \rho}{\lambda^{i}_{\rho}} - \frac{1}{32}{\rm i} \lambda_{i \rho} X^{i}_{\beta} {W}^{-2} \nabla^{\rho \beta}{\lambda^{j}_{\alpha}}+\frac{1}{32}{\rm i} \lambda^{j}_{\alpha} X_{i \beta} {W}^{-2} \nabla^{\beta \rho}{\lambda^{i}_{\rho}} - \frac{1}{16}{\rm i} \lambda^{j}_{\alpha} \lambda_{i \rho} {W}^{-2} \nabla^{\rho \beta}{X^{i}_{\beta}} - \frac{1}{16}{\rm i} \lambda^{j}_{\beta} X_{i \alpha} {W}^{-2} \nabla^{\beta \rho}{\lambda^{i}_{\rho}}+\frac{1}{32}{\rm i} \lambda^{j}_{\rho} X_{i \beta} {W}^{-2} \nabla^{\rho \beta}{\lambda^{i}_{\alpha}}+\frac{3}{16}{\rm i} \lambda^{j}_{\rho} \lambda_{i \alpha} {W}^{-2} \nabla^{\rho \beta}{X^{i}_{\beta}} - \frac{1}{8}{\rm i} \lambda^{j}_{\beta} \lambda_{i \rho} {W}^{-2} \nabla^{\beta \rho}{X^{i}_{\alpha}} - \frac{3}{8}X^{j}_{\beta} {W}^{-1} \nabla_{\rho}\,^{\beta}{\nabla_{\alpha}\,^{\rho}{W}} - \frac{3}{8}{W}^{-1} \nabla_{\alpha \rho}{W} \nabla^{\rho \beta}{X^{j}_{\beta}} - \frac{3}{32}{W}^{-1} \nabla_{\beta \rho}{W} \nabla^{\beta \rho}{X^{j}_{\alpha}} - \frac{3}{8}F^{\beta}\,_{\rho} X^{j}_{\beta} {W}^{-2} \nabla_{\alpha}\,^{\rho}{W}+\frac{27}{32}W^{\beta}\,_{\rho} X^{j}_{\beta} {W}^{-1} \nabla_{\alpha}\,^{\rho}{W}+\frac{3}{4}F^{\beta \rho} {W}^{-1} \nabla_{\alpha}\,^{\lambda}{W_{\beta \rho \lambda}\,^{j}} - \frac{1}{2}F_{\alpha}\,^{\beta} F^{\rho \lambda} W_{\beta \rho \lambda}\,^{j} {W}^{-2} - \frac{1}{2}{\rm i} C_{\alpha}\,^{\beta \rho \lambda} F_{\beta \rho} \lambda^{j}_{\lambda} {W}^{-2}+\frac{9}{4}{\rm i} W_{\alpha}\,^{\beta} W^{\rho \lambda} F_{\beta \rho} \lambda^{j}_{\lambda} {W}^{-2} - \frac{9}{4}{\rm i} W_{\alpha}\,^{\lambda} W^{\beta \rho} F_{\beta \rho} \lambda^{j}_{\lambda} {W}^{-2}%
 - \frac{1}{8}W_{\alpha}\,^{\lambda} F^{\beta \rho} W_{\lambda \beta \rho}\,^{j} {W}^{-1} - \frac{7}{8}W^{\rho \lambda} F_{\alpha}\,^{\beta} W_{\rho \lambda \beta}\,^{j} {W}^{-1}+\frac{9}{8}W^{\beta \lambda} F_{\beta}\,^{\rho} W_{\alpha \lambda \rho}\,^{j} {W}^{-1} - \frac{1}{2}{\rm i} C_{\alpha}\,^{\beta \rho \lambda} W_{\beta \rho} \lambda^{j}_{\lambda} {W}^{-1} - \frac{81}{16}{\rm i} W_{\alpha}\,^{\beta} W_{\beta}\,^{\rho} W_{\rho}\,^{\lambda} \lambda^{j}_{\lambda} {W}^{-1} - \frac{81}{32}{\rm i} W_{\alpha}\,^{\beta} W^{\rho \lambda} W_{\rho \lambda} \lambda^{j}_{\beta} {W}^{-1}+\frac{1}{2}{\rm i} \Phi_{\beta}\,^{\rho j}\,_{i} {W}^{-1} \nabla^{\beta}\,_{\alpha}{\lambda^{i}_{\rho}} - \frac{1}{9}{\rm i} \lambda^{\rho}_{i} {W}^{-1} \nabla_{\alpha}\,^{\beta}{\Phi_{\beta \rho}\,^{j i}}+\frac{3}{8}X^{j}\,_{i} X^{i}_{\beta} {W}^{-2} \nabla_{\alpha}\,^{\beta}{W}+\frac{1}{2}{\rm i} \Phi^{\beta \rho j}\,_{i} F_{\beta \rho} \lambda^{i}_{\alpha} {W}^{-2} - \frac{1}{4}{\rm i} \Phi^{\beta \rho j}\,_{i} W_{\alpha \beta} \lambda^{i}_{\rho} {W}^{-1}+\frac{3}{4}{\rm i} \Phi_{\alpha}\,^{\beta j}\,_{i} W_{\beta}\,^{\rho} \lambda^{i}_{\rho} {W}^{-1}+\frac{3}{4}{\rm i} \Phi^{\beta \rho j}\,_{i} W_{\beta \rho} \lambda^{i}_{\alpha} {W}^{-1}+\frac{1}{4}X^{j}\,_{i} F^{\beta \rho} W_{\alpha \beta \rho}\,^{i} {W}^{-2}+\frac{1}{4}X^{j}\,_{i} W^{\beta \rho} W_{\alpha \beta \rho}\,^{i} {W}^{-1} - \frac{1}{12}\lambda^{\beta}_{i} X^{i \rho} W_{\alpha \beta \rho}\,^{j} {W}^{-1}+\frac{1}{12}\lambda^{j \beta} X_{i}^{\rho} W_{\alpha \beta \rho}\,^{i} {W}^{-1} - \frac{3}{128}Y \lambda^{j \beta} \lambda_{i \alpha} \lambda^{i}_{\beta} {W}^{-3}+\frac{1}{4}{\rm i} F_{\beta \rho} \lambda^{j}_{\lambda} {W}^{-3} \nabla^{\beta \lambda}{\nabla_{\alpha}\,^{\rho}{W}}+\frac{3}{16}{\rm i} W_{\beta \rho} \lambda^{j}_{\lambda} {W}^{-2} \nabla^{\beta \lambda}{\nabla_{\alpha}\,^{\rho}{W}}%
 - \frac{3}{8}{\rm i} W_{\beta \rho} \lambda^{j}_{\lambda} {W}^{-2} \nabla_{\alpha}\,^{\beta}{\nabla^{\rho \lambda}{W}} - \frac{9}{16}{\rm i} W_{\beta \rho} {W}^{-2} \nabla_{\alpha}\,^{\beta}{W} \nabla^{\rho \lambda}{\lambda^{j}_{\lambda}} - \frac{3}{16}{\rm i} W_{\beta \rho} {W}^{-2} \nabla^{\beta \lambda}{W} \nabla_{\alpha}\,^{\rho}{\lambda^{j}_{\lambda}} - \frac{1}{4}{\rm i} \lambda^{j}_{\lambda} {W}^{-3} \nabla_{\alpha}\,^{\beta}{W} \nabla^{\lambda \rho}{F_{\beta \rho}} - \frac{49}{128}{\rm i} \lambda^{j}_{\lambda} {W}^{-2} \nabla_{\alpha}\,^{\beta}{W} \nabla^{\lambda \rho}{W_{\beta \rho}}+\frac{25}{128}{\rm i} \lambda^{j}_{\lambda} {W}^{-2} \nabla^{\lambda \beta}{W} \nabla_{\alpha}\,^{\rho}{W_{\beta \rho}} - \frac{1}{8}{\rm i} F_{\alpha \beta} {W}^{-3} \nabla_{\lambda}\,^{\beta}{W} \nabla^{\lambda \rho}{\lambda^{j}_{\rho}} - \frac{1}{8}{\rm i} F_{\alpha \beta} {W}^{-3} \nabla_{\lambda}\,^{\rho}{W} \nabla^{\lambda \beta}{\lambda^{j}_{\rho}}+\frac{1}{8}{\rm i} W_{\alpha \beta} \lambda^{j}_{\rho} {W}^{-2} \nabla_{\lambda}\,^{\beta}{\nabla^{\lambda \rho}{W}} - \frac{3}{16}{\rm i} W_{\alpha \beta} {W}^{-2} \nabla_{\lambda}\,^{\rho}{W} \nabla^{\lambda \beta}{\lambda^{j}_{\rho}}+\frac{1}{128}{\rm i} \lambda^{j}_{\alpha} {W}^{-2} \nabla_{\lambda}\,^{\beta}{W} \nabla^{\lambda \rho}{W_{\beta \rho}}+\frac{1}{8}{\rm i} \lambda^{j}_{\rho} {W}^{-3} \nabla_{\lambda}\,^{\rho}{W} \nabla^{\lambda \beta}{F_{\alpha \beta}}+\frac{1}{8}{\rm i} \lambda^{j}_{\rho} {W}^{-3} \nabla_{\lambda}\,^{\beta}{W} \nabla^{\lambda \rho}{F_{\alpha \beta}}+\frac{15}{128}{\rm i} \lambda^{j}_{\rho} {W}^{-2} \nabla_{\lambda}\,^{\rho}{W} \nabla^{\lambda \beta}{W_{\alpha \beta}}+\frac{1}{16}{\rm i} \lambda^{j}_{\rho} {W}^{-2} \nabla_{\lambda}\,^{\beta}{W} \nabla^{\lambda \rho}{W_{\alpha \beta}}+\frac{3}{4}{\rm i} F_{\alpha \beta} F^{\rho}\,_{\lambda} \lambda^{j}_{\rho} {W}^{-4} \nabla^{\beta \lambda}{W} - \frac{3}{4}{\rm i} F_{\alpha}\,^{\beta} F_{\beta \rho} \lambda^{j}_{\lambda} {W}^{-4} \nabla^{\rho \lambda}{W}+\frac{3}{4}{\rm i} W_{\alpha \lambda} F^{\beta}\,_{\rho} \lambda^{j}_{\beta} {W}^{-3} \nabla^{\lambda \rho}{W}+\frac{9}{16}{\rm i} W_{\alpha \beta} W^{\rho}\,_{\lambda} \lambda^{j}_{\rho} {W}^{-2} \nabla^{\beta \lambda}{W} - \frac{3}{16}{\rm i} W_{\alpha}\,^{\beta} F_{\beta \rho} \lambda^{j}_{\lambda} {W}^{-3} \nabla^{\rho \lambda}{W}%
+\frac{9}{16}{\rm i} W_{\alpha}\,^{\beta} W_{\beta \rho} \lambda^{j}_{\lambda} {W}^{-2} \nabla^{\rho \lambda}{W} - \frac{3}{16}{\rm i} W^{\beta}\,_{\rho} F_{\alpha \beta} \lambda^{j}_{\lambda} {W}^{-3} \nabla^{\rho \lambda}{W}+\frac{9}{16}{\rm i} W^{\beta}\,_{\lambda} F_{\beta \rho} \lambda^{j}_{\alpha} {W}^{-3} \nabla^{\lambda \rho}{W}+\frac{27}{64}{\rm i} W^{\beta}\,_{\rho} W_{\beta \lambda} \lambda^{j}_{\alpha} {W}^{-2} \nabla^{\rho \lambda}{W} - \frac{1}{12}{\rm i} \lambda^{\beta}_{i} W_{\alpha \beta \rho}\,^{j} {W}^{-2} \nabla^{\rho \lambda}{\lambda^{i}_{\lambda}}+\frac{1}{12}{\rm i} \lambda^{\beta}_{i} W_{\alpha \beta \rho}\,^{i} {W}^{-2} \nabla^{\rho \lambda}{\lambda^{j}_{\lambda}}+\frac{1}{24}{\rm i} \lambda^{j}_{\lambda} \lambda^{\beta}_{i} {W}^{-2} \nabla^{\lambda \rho}{W_{\alpha \beta \rho}\,^{i}} - \frac{1}{6}{\rm i} \lambda^{j \beta} W_{\alpha \beta \rho i} {W}^{-2} \nabla^{\rho \lambda}{\lambda^{i}_{\lambda}} - \frac{1}{24}{\rm i} \lambda^{j \beta} \lambda_{i \lambda} {W}^{-2} \nabla^{\lambda \rho}{W_{\alpha \beta \rho}\,^{i}} - \frac{3}{8}{\rm i} X^{j}\,_{i} F_{\alpha \beta} \lambda^{i}_{\rho} {W}^{-4} \nabla^{\beta \rho}{W} - \frac{9}{16}{\rm i} X^{j}\,_{i} W_{\alpha \beta} \lambda^{i}_{\rho} {W}^{-3} \nabla^{\beta \rho}{W}+\frac{1}{32}{\rm i} \lambda_{i \alpha} \lambda^{i}_{\rho} X^{j}_{\beta} {W}^{-3} \nabla^{\rho \beta}{W} - \frac{11}{96}{\rm i} \lambda^{j}_{\alpha} \lambda_{i \rho} X^{i}_{\beta} {W}^{-3} \nabla^{\rho \beta}{W} - \frac{1}{96}{\rm i} \lambda^{j}_{\rho} \lambda_{i \alpha} X^{i}_{\beta} {W}^{-3} \nabla^{\rho \beta}{W}+\frac{1}{24}{\rm i} \lambda^{j}_{\beta} \lambda_{i \rho} X^{i}_{\alpha} {W}^{-3} \nabla^{\beta \rho}{W}+\frac{3}{64}X^{j}_{\alpha} {W}^{-2} \nabla_{\beta \rho}{W} \nabla^{\beta \rho}{W} - \frac{1}{4}F^{\beta \rho} W_{\beta \rho \lambda}\,^{j} {W}^{-2} \nabla_{\alpha}\,^{\lambda}{W} - \frac{1}{4}W^{\beta \rho} W_{\beta \rho \lambda}\,^{j} {W}^{-1} \nabla_{\alpha}\,^{\lambda}{W}+\frac{1}{24}\lambda_{i \beta} {W}^{-3} \nabla_{\alpha}\,^{\beta}{\lambda^{j}_{\rho}} \nabla^{\rho \lambda}{\lambda^{i}_{\lambda}} - \frac{7}{24}\lambda_{i \beta} {W}^{-3} \nabla_{\alpha}\,^{\rho}{\lambda^{j}_{\rho}} \nabla^{\beta \lambda}{\lambda^{i}_{\lambda}}%
 - \frac{1}{24}\lambda_{i \beta} {W}^{-3} \nabla^{\beta \lambda}{\lambda^{j}_{\rho}} \nabla_{\alpha}\,^{\rho}{\lambda^{i}_{\lambda}}+\frac{1}{24}\lambda^{j}_{\beta} \lambda_{i \rho} {W}^{-3} \nabla^{\rho \lambda}{\nabla_{\alpha}\,^{\beta}{\lambda^{i}_{\lambda}}}+\frac{1}{24}\lambda^{j}_{\beta} \lambda_{i \rho} {W}^{-3} \nabla_{\alpha}\,^{\beta}{\nabla^{\rho \lambda}{\lambda^{i}_{\lambda}}}+\frac{1}{24}\lambda^{j}_{\beta} {W}^{-3} \nabla_{\alpha}\,^{\beta}{\lambda_{i \rho}} \nabla^{\rho \lambda}{\lambda^{i}_{\lambda}} - \frac{1}{24}\lambda^{j}_{\beta} \lambda_{i \rho} {W}^{-3} \nabla^{\beta \lambda}{\nabla_{\alpha}\,^{\rho}{\lambda^{i}_{\lambda}}} - \frac{1}{24}\lambda^{j}_{\beta} \lambda_{i \rho} {W}^{-3} \nabla_{\alpha}\,^{\rho}{\nabla^{\beta \lambda}{\lambda^{i}_{\lambda}}} - \frac{1}{6}\lambda^{j}_{\beta} \lambda_{i \rho} {W}^{-3} \nabla^{\beta \rho}{\nabla_{\alpha}\,^{\lambda}{\lambda^{i}_{\lambda}}} - \frac{1}{6}\lambda^{j}_{\beta} {W}^{-3} \nabla_{\alpha}\,^{\rho}{\lambda_{i \rho}} \nabla^{\beta \lambda}{\lambda^{i}_{\lambda}} - \frac{1}{24}\lambda^{j}_{\beta} {W}^{-3} \nabla_{\alpha}\,^{\lambda}{\lambda_{i \rho}} \nabla^{\beta \rho}{\lambda^{i}_{\lambda}}+\frac{1}{24}\lambda_{i \beta} {W}^{-3} \nabla_{\lambda}\,^{\beta}{\lambda^{j}_{\alpha}} \nabla^{\lambda \rho}{\lambda^{i}_{\rho}} - \frac{1}{48}\lambda_{i \beta} {W}^{-3} \nabla_{\lambda}\,^{\beta}{\lambda^{j}_{\rho}} \nabla^{\lambda \rho}{\lambda^{i}_{\alpha}} - \frac{1}{48}\lambda_{i \beta} {W}^{-3} \nabla_{\lambda}\,^{\rho}{\lambda^{j}_{\rho}} \nabla^{\lambda \beta}{\lambda^{i}_{\alpha}} - \frac{1}{24}\lambda^{j}_{\alpha} \lambda_{i \beta} {W}^{-3} \nabla_{\lambda}\,^{\beta}{\nabla^{\lambda \rho}{\lambda^{i}_{\rho}}}+\frac{1}{24}\lambda^{j}_{\beta} \lambda_{i \alpha} {W}^{-3} \nabla_{\lambda}\,^{\beta}{\nabla^{\lambda \rho}{\lambda^{i}_{\rho}}}+\frac{1}{16}\lambda^{j}_{\beta} {W}^{-3} \nabla_{\lambda}\,^{\beta}{\lambda_{i \alpha}} \nabla^{\lambda \rho}{\lambda^{i}_{\rho}}+\frac{1}{48}\lambda^{j}_{\beta} {W}^{-3} \nabla_{\lambda}\,^{\rho}{\lambda_{i \alpha}} \nabla^{\lambda \beta}{\lambda^{i}_{\rho}} - \frac{3}{8}F_{\alpha \beta} \lambda^{j}_{\rho} \lambda^{\lambda}_{i} {W}^{-4} \nabla^{\beta \rho}{\lambda^{i}_{\lambda}} - \frac{3}{8}F_{\alpha \beta} \lambda^{j \rho} \lambda_{i \lambda} {W}^{-4} \nabla^{\beta \lambda}{\lambda^{i}_{\rho}} - \frac{3}{8}F_{\alpha \beta} \lambda^{j \rho} \lambda_{i \rho} {W}^{-4} \nabla^{\beta \lambda}{\lambda^{i}_{\lambda}}+\frac{1}{8}F_{\alpha}\,^{\beta} \lambda^{j}_{\rho} \lambda_{i \lambda} {W}^{-4} \nabla^{\rho \lambda}{\lambda^{i}_{\beta}}%
+\frac{1}{2}F_{\alpha}\,^{\beta} \lambda^{j}_{\rho} \lambda_{i \beta} {W}^{-4} \nabla^{\rho \lambda}{\lambda^{i}_{\lambda}} - \frac{1}{2}F_{\alpha}\,^{\beta} \lambda^{j}_{\beta} \lambda_{i \rho} {W}^{-4} \nabla^{\rho \lambda}{\lambda^{i}_{\lambda}} - \frac{1}{4}F^{\beta}\,_{\rho} \lambda_{i \alpha} \lambda^{i}_{\lambda} {W}^{-4} \nabla^{\rho \lambda}{\lambda^{j}_{\beta}} - \frac{1}{8}F^{\beta}\,_{\rho} \lambda^{j}_{\alpha} \lambda_{i \beta} {W}^{-4} \nabla^{\rho \lambda}{\lambda^{i}_{\lambda}}+\frac{1}{4}F^{\beta}\,_{\rho} \lambda^{j}_{\lambda} \lambda_{i \alpha} {W}^{-4} \nabla^{\rho \lambda}{\lambda^{i}_{\beta}}+\frac{1}{8}F^{\beta}\,_{\rho} \lambda^{j}_{\lambda} \lambda_{i \beta} {W}^{-4} \nabla^{\rho \lambda}{\lambda^{i}_{\alpha}}+\frac{1}{8}F^{\beta}\,_{\rho} \lambda^{j}_{\beta} \lambda_{i \alpha} {W}^{-4} \nabla^{\rho \lambda}{\lambda^{i}_{\lambda}} - \frac{1}{8}F^{\beta}\,_{\rho} \lambda^{j}_{\beta} \lambda_{i \lambda} {W}^{-4} \nabla^{\rho \lambda}{\lambda^{i}_{\alpha}} - \frac{35}{96}W_{\alpha \beta} \lambda^{j}_{\rho} \lambda^{\lambda}_{i} {W}^{-3} \nabla^{\beta \rho}{\lambda^{i}_{\lambda}} - \frac{37}{96}W_{\alpha \beta} \lambda^{j \rho} \lambda_{i \lambda} {W}^{-3} \nabla^{\beta \lambda}{\lambda^{i}_{\rho}} - \frac{35}{96}W_{\alpha \beta} \lambda^{j \rho} \lambda_{i \rho} {W}^{-3} \nabla^{\beta \lambda}{\lambda^{i}_{\lambda}} - \frac{1}{96}W_{\alpha}\,^{\beta} \lambda_{i \beta} \lambda^{i}_{\rho} {W}^{-3} \nabla^{\rho \lambda}{\lambda^{j}_{\lambda}}+\frac{1}{96}W_{\alpha}\,^{\beta} \lambda^{j}_{\rho} \lambda_{i \lambda} {W}^{-3} \nabla^{\rho \lambda}{\lambda^{i}_{\beta}}+\frac{7}{24}W_{\alpha}\,^{\beta} \lambda^{j}_{\rho} \lambda_{i \beta} {W}^{-3} \nabla^{\rho \lambda}{\lambda^{i}_{\lambda}} - \frac{7}{32}W_{\alpha}\,^{\beta} \lambda^{j}_{\beta} \lambda_{i \rho} {W}^{-3} \nabla^{\rho \lambda}{\lambda^{i}_{\lambda}} - \frac{1}{4}W^{\beta}\,_{\rho} \lambda_{i \alpha} \lambda^{i}_{\lambda} {W}^{-3} \nabla^{\rho \lambda}{\lambda^{j}_{\beta}} - \frac{3}{32}W^{\beta}\,_{\rho} \lambda_{i \alpha} \lambda^{i}_{\beta} {W}^{-3} \nabla^{\rho \lambda}{\lambda^{j}_{\lambda}}+\frac{3}{32}W^{\beta}\,_{\rho} \lambda_{i \beta} \lambda^{i}_{\lambda} {W}^{-3} \nabla^{\rho \lambda}{\lambda^{j}_{\alpha}}+\frac{1}{8}W^{\beta}\,_{\rho} \lambda^{j}_{\alpha} \lambda_{i \lambda} {W}^{-3} \nabla^{\rho \lambda}{\lambda^{i}_{\beta}} - \frac{1}{32}W^{\beta}\,_{\rho} \lambda^{j}_{\alpha} \lambda_{i \beta} {W}^{-3} \nabla^{\rho \lambda}{\lambda^{i}_{\lambda}}%
+\frac{1}{8}W^{\beta}\,_{\rho} \lambda^{j}_{\lambda} \lambda_{i \alpha} {W}^{-3} \nabla^{\rho \lambda}{\lambda^{i}_{\beta}}+\frac{1}{32}W^{\beta}\,_{\rho} \lambda^{j}_{\lambda} \lambda_{i \beta} {W}^{-3} \nabla^{\rho \lambda}{\lambda^{i}_{\alpha}}+\frac{1}{8}W^{\beta}\,_{\rho} \lambda^{j}_{\beta} \lambda_{i \alpha} {W}^{-3} \nabla^{\rho \lambda}{\lambda^{i}_{\lambda}} - \frac{1}{8}W^{\beta}\,_{\rho} \lambda^{j}_{\beta} \lambda_{i \lambda} {W}^{-3} \nabla^{\rho \lambda}{\lambda^{i}_{\alpha}} - \frac{1}{32}\lambda^{j}_{\alpha} \lambda^{\beta}_{i} \lambda^{i}_{\lambda} {W}^{-3} \nabla^{\lambda \rho}{W_{\beta \rho}}+\frac{1}{4}\lambda^{j}_{\lambda} \lambda_{i \alpha} \lambda^{i \beta} {W}^{-4} \nabla^{\lambda \rho}{F_{\beta \rho}}+\frac{9}{32}\lambda^{j}_{\lambda} \lambda_{i \alpha} \lambda^{i \beta} {W}^{-3} \nabla^{\lambda \rho}{W_{\beta \rho}} - \frac{1}{4}\lambda^{j}_{\rho} \lambda^{\beta}_{i} \lambda^{i}_{\lambda} {W}^{-4} \nabla^{\rho \lambda}{F_{\alpha \beta}} - \frac{5}{32}\lambda^{j}_{\rho} \lambda^{\beta}_{i} \lambda^{i}_{\lambda} {W}^{-3} \nabla^{\rho \lambda}{W_{\alpha \beta}} - \frac{1}{4}\lambda^{j \beta} \lambda_{i \alpha} \lambda^{i}_{\lambda} {W}^{-4} \nabla^{\lambda \rho}{F_{\beta \rho}} - \frac{1}{4}\lambda^{j \beta} \lambda_{i \alpha} \lambda^{i}_{\lambda} {W}^{-3} \nabla^{\lambda \rho}{W_{\beta \rho}}+\frac{5}{64}X_{i k} \lambda^{j}_{\alpha} \lambda^{i}_{\beta} {W}^{-4} \nabla^{\beta \rho}{\lambda^{k}_{\rho}} - \frac{13}{64}X_{i k} \lambda^{j}_{\beta} \lambda^{i}_{\alpha} {W}^{-4} \nabla^{\beta \rho}{\lambda^{k}_{\rho}}+\frac{9}{64}X_{i k} \lambda^{j}_{\beta} \lambda^{i}_{\rho} {W}^{-4} \nabla^{\beta \rho}{\lambda^{k}_{\alpha}}+\frac{21}{64}X^{j}\,_{i} \lambda^{i}_{\alpha} \lambda_{k \beta} {W}^{-4} \nabla^{\beta \rho}{\lambda^{k}_{\rho}}+\frac{1}{64}X^{j}\,_{i} \lambda^{i}_{\beta} \lambda_{k \rho} {W}^{-4} \nabla^{\beta \rho}{\lambda^{k}_{\alpha}}+\frac{5}{64}X^{j}\,_{i} \lambda_{k \alpha} \lambda^{i}_{\beta} {W}^{-4} \nabla^{\beta \rho}{\lambda^{k}_{\rho}}+\frac{1}{8}\lambda_{k \alpha} \lambda^{k}_{\beta} \lambda_{i \rho} {W}^{-4} \nabla^{\beta \rho}{X^{j i}}+\frac{3}{64}\lambda^{j}_{\alpha} \lambda_{i \beta} \lambda_{k \rho} {W}^{-4} \nabla^{\beta \rho}{X^{i k}} - \frac{3}{32}\lambda^{j}_{\beta} \lambda_{i \alpha} \lambda_{k \rho} {W}^{-4} \nabla^{\beta \rho}{X^{i k}}%
+\frac{1}{8}{\rm i} {W}^{-1} \nabla_{\rho \lambda}{\nabla^{\rho \lambda}{\nabla_{\alpha}\,^{\beta}{\lambda^{j}_{\beta}}}} - \frac{3}{8}{\rm i} F_{\beta \rho} \lambda^{j}_{\lambda} {W}^{-4} \nabla_{\alpha}\,^{\beta}{W} \nabla^{\rho \lambda}{W}+\frac{9}{16}{\rm i} W_{\beta \rho} \lambda^{j}_{\lambda} {W}^{-3} \nabla_{\alpha}\,^{\beta}{W} \nabla^{\rho \lambda}{W}+\frac{1}{16}{\rm i} \lambda^{j \beta} {W}^{-1} \nabla_{\lambda}\,^{\rho}{\nabla_{\alpha}\,^{\lambda}{W_{\beta \rho}}}+\frac{1}{32}{\rm i} \lambda^{j \beta} {W}^{-1} \nabla_{\alpha \lambda}{\nabla^{\lambda \rho}{W_{\beta \rho}}}+\frac{3}{32}{\rm i} {W}^{-1} \nabla_{\alpha \lambda}{W^{\beta}\,_{\rho}} \nabla^{\lambda \rho}{\lambda^{j}_{\beta}}+\frac{3}{64}{\rm i} {W}^{-1} \nabla_{\lambda}\,^{\rho}{W^{\beta}\,_{\rho}} \nabla_{\alpha}\,^{\lambda}{\lambda^{j}_{\beta}} - \frac{3}{32}{\rm i} W_{\alpha \beta} \lambda^{j}_{\rho} {W}^{-3} \nabla_{\lambda}\,^{\beta}{W} \nabla^{\lambda \rho}{W} - \frac{1}{8}{\rm i} F_{\alpha}\,^{\beta} {W}^{-2} \nabla_{\rho \lambda}{\nabla^{\rho \lambda}{\lambda^{j}_{\beta}}} - \frac{1}{16}{\rm i} W_{\alpha}\,^{\beta} {W}^{-1} \nabla_{\rho \lambda}{\nabla^{\rho \lambda}{\lambda^{j}_{\beta}}}+\frac{1}{8}{\rm i} \lambda^{j \beta} {W}^{-2} \nabla_{\rho \lambda}{\nabla^{\rho \lambda}{F_{\alpha \beta}}} - \frac{1}{8}{\rm i} {W}^{-2} \nabla_{\rho \lambda}{F_{\alpha}\,^{\beta}} \nabla^{\rho \lambda}{\lambda^{j}_{\beta}}+\frac{3}{32}{\rm i} {W}^{-1} \nabla_{\rho \lambda}{W_{\alpha}\,^{\beta}} \nabla^{\rho \lambda}{\lambda^{j}_{\beta}} - \frac{5}{16}{\rm i} F^{\beta}\,_{\rho} \lambda^{j \lambda} {W}^{-2} \nabla_{\alpha}\,^{\rho}{W_{\beta \lambda}}+\frac{1}{4}{\rm i} F^{\beta \rho} F_{\beta \rho} {W}^{-3} \nabla_{\alpha}\,^{\lambda}{\lambda^{j}_{\lambda}} - \frac{1}{16}{\rm i} F^{\beta \rho} \lambda^{j}_{\lambda} {W}^{-2} \nabla_{\alpha}\,^{\lambda}{W_{\beta \rho}} - \frac{17}{16}{\rm i} F^{\beta \rho} \lambda^{j}_{\beta} {W}^{-2} \nabla_{\alpha}\,^{\lambda}{W_{\rho \lambda}}+\frac{3}{8}{\rm i} W^{\beta}\,_{\lambda} F_{\beta}\,^{\rho} {W}^{-2} \nabla_{\alpha}\,^{\lambda}{\lambda^{j}_{\rho}}+\frac{9}{16}{\rm i} W^{\beta \rho} W_{\beta \lambda} {W}^{-1} \nabla_{\alpha}\,^{\lambda}{\lambda^{j}_{\rho}} - \frac{3}{4}{\rm i} W^{\beta}\,_{\lambda} \lambda^{j \rho} {W}^{-2} \nabla_{\alpha}\,^{\lambda}{F_{\beta \rho}}%
 - \frac{23}{64}{\rm i} W^{\beta}\,_{\rho} \lambda^{j \lambda} {W}^{-1} \nabla_{\alpha}\,^{\rho}{W_{\beta \lambda}} - \frac{15}{32}{\rm i} W^{\beta \rho} W_{\beta \rho} {W}^{-1} \nabla_{\alpha}\,^{\lambda}{\lambda^{j}_{\lambda}}+\frac{11}{64}{\rm i} W^{\beta \rho} \lambda^{j}_{\lambda} {W}^{-1} \nabla_{\alpha}\,^{\lambda}{W_{\beta \rho}} - \frac{23}{64}{\rm i} W^{\beta \rho} \lambda^{j}_{\beta} {W}^{-1} \nabla_{\alpha}\,^{\lambda}{W_{\rho \lambda}} - \frac{3}{2}{\rm i} F_{\alpha}\,^{\beta} F_{\beta}\,^{\rho} F_{\rho}\,^{\lambda} \lambda^{j}_{\lambda} {W}^{-4} - \frac{3}{4}{\rm i} F_{\alpha}\,^{\beta} F^{\rho \lambda} F_{\rho \lambda} \lambda^{j}_{\beta} {W}^{-4} - \frac{3}{2}{\rm i} W_{\alpha}\,^{\beta} F_{\beta}\,^{\rho} F_{\rho}\,^{\lambda} \lambda^{j}_{\lambda} {W}^{-3} - \frac{3}{8}{\rm i} W_{\alpha}\,^{\lambda} F^{\beta \rho} F_{\beta \rho} \lambda^{j}_{\lambda} {W}^{-3} - \frac{27}{16}{\rm i} W_{\alpha}\,^{\lambda} W_{\lambda}\,^{\beta} F_{\beta}\,^{\rho} \lambda^{j}_{\rho} {W}^{-2}+\frac{9}{4}{\rm i} W^{\beta \rho} F_{\alpha \beta} F_{\rho}\,^{\lambda} \lambda^{j}_{\lambda} {W}^{-3} - \frac{3}{2}{\rm i} W^{\rho \lambda} F_{\alpha}\,^{\beta} F_{\rho \lambda} \lambda^{j}_{\beta} {W}^{-3}+\frac{3}{2}{\rm i} W^{\rho \lambda} F_{\alpha}\,^{\beta} F_{\rho \beta} \lambda^{j}_{\lambda} {W}^{-3} - \frac{15}{32}{\rm i} W^{\rho \lambda} W_{\rho \lambda} F_{\alpha}\,^{\beta} \lambda^{j}_{\beta} {W}^{-2}+\frac{27}{16}{\rm i} W^{\rho \beta} W_{\rho}\,^{\lambda} F_{\alpha \beta} \lambda^{j}_{\lambda} {W}^{-2} - \frac{1}{8}{\rm i} X^{j}\,_{i} X^{i}\,_{k} X^{k}\,_{l} \lambda^{l}_{\alpha} {W}^{-4} - \frac{5}{32}{\rm i} X^{j}\,_{i} X_{k l} X^{k l} \lambda^{i}_{\alpha} {W}^{-4}+\frac{1}{16}{\rm i} X^{j}\,_{i} {W}^{-2} \nabla_{\beta \rho}{\nabla^{\beta \rho}{\lambda^{i}_{\alpha}}} - \frac{1}{16}{\rm i} \lambda_{i \alpha} {W}^{-2} \nabla_{\beta \rho}{\nabla^{\beta \rho}{X^{j i}}}+\frac{1}{16}{\rm i} {W}^{-2} \nabla_{\beta \rho}{X^{j}\,_{i}} \nabla^{\beta \rho}{\lambda^{i}_{\alpha}}+\frac{3}{16}{\rm i} W^{\beta}\,_{\rho} \lambda_{i \beta} {W}^{-2} \nabla_{\alpha}\,^{\rho}{X^{j i}}%
 - \frac{1}{4}{\rm i} X^{j}\,_{i} F^{\beta}\,_{\rho} {W}^{-3} \nabla_{\alpha}\,^{\rho}{\lambda^{i}_{\beta}} - \frac{3}{8}{\rm i} X^{j}\,_{i} W^{\beta}\,_{\rho} {W}^{-2} \nabla_{\alpha}\,^{\rho}{\lambda^{i}_{\beta}}+\frac{1}{12}{\rm i} X^{j}\,_{i} \lambda^{i \beta} {W}^{-3} \nabla_{\alpha}\,^{\rho}{F_{\beta \rho}}+\frac{9}{16}{\rm i} X^{j}\,_{i} \lambda^{i \beta} {W}^{-2} \nabla_{\alpha}\,^{\rho}{W_{\beta \rho}}+\frac{1}{16}{\rm i} \lambda_{i \rho} X^{j \beta} {W}^{-2} \nabla_{\alpha}\,^{\rho}{\lambda^{i}_{\beta}} - \frac{1}{32}{\rm i} \lambda_{i \rho} X^{i \beta} {W}^{-2} \nabla_{\alpha}\,^{\rho}{\lambda^{j}_{\beta}}+\frac{1}{32}{\rm i} \lambda^{\rho}_{i} X^{j}_{\beta} {W}^{-2} \nabla_{\alpha}\,^{\beta}{\lambda^{i}_{\rho}}+\frac{1}{8}{\rm i} \lambda^{\rho}_{i} X^{i}_{\beta} {W}^{-2} \nabla_{\alpha}\,^{\beta}{\lambda^{j}_{\rho}} - \frac{11}{32}{\rm i} \lambda^{\beta}_{i} X^{i}_{\beta} {W}^{-2} \nabla_{\alpha}\,^{\rho}{\lambda^{j}_{\rho}}+\frac{3}{32}{\rm i} \lambda^{j}_{\rho} X_{i}^{\beta} {W}^{-2} \nabla_{\alpha}\,^{\rho}{\lambda^{i}_{\beta}} - \frac{1}{16}{\rm i} \lambda^{j}_{\rho} \lambda^{\beta}_{i} {W}^{-2} \nabla_{\alpha}\,^{\rho}{X^{i}_{\beta}} - \frac{3}{16}{\rm i} \lambda^{j \rho} X_{i \beta} {W}^{-2} \nabla_{\alpha}\,^{\beta}{\lambda^{i}_{\rho}} - \frac{1}{32}{\rm i} \lambda^{j \beta} X_{i \beta} {W}^{-2} \nabla_{\alpha}\,^{\rho}{\lambda^{i}_{\rho}}+\frac{1}{16}{\rm i} \lambda^{j \beta} \lambda_{i \rho} {W}^{-2} \nabla_{\alpha}\,^{\rho}{X^{i}_{\beta}} - \frac{1}{4}{\rm i} \lambda^{j \rho} \lambda_{i \rho} {W}^{-2} \nabla_{\alpha}\,^{\beta}{X^{i}_{\beta}}+\frac{1}{12}{\rm i} \lambda^{j}_{\lambda} \lambda^{\beta}_{i} W_{\alpha \beta \rho}\,^{i} {W}^{-3} \nabla^{\lambda \rho}{W} - \frac{1}{12}{\rm i} \lambda^{j \beta} \lambda_{i \lambda} W_{\alpha \beta \rho}\,^{i} {W}^{-3} \nabla^{\lambda \rho}{W} - \frac{1}{16}{\rm i} F_{\alpha}\,^{\beta} \lambda_{i \beta} \lambda^{i \rho} X^{j}_{\rho} {W}^{-3}+\frac{11}{16}{\rm i} F_{\alpha}\,^{\beta} \lambda^{j}_{\beta} \lambda^{\rho}_{i} X^{i}_{\rho} {W}^{-3} - \frac{1}{16}{\rm i} F_{\alpha}\,^{\beta} \lambda^{j \rho} \lambda_{i \beta} X^{i}_{\rho} {W}^{-3}%
+\frac{3}{8}{\rm i} F_{\alpha}\,^{\beta} \lambda^{j \rho} \lambda_{i \rho} X^{i}_{\beta} {W}^{-3}+\frac{5}{16}{\rm i} F^{\beta \rho} \lambda_{i \alpha} \lambda^{i}_{\beta} X^{j}_{\rho} {W}^{-3} - \frac{5}{16}{\rm i} F^{\beta \rho} \lambda_{i \beta} \lambda^{i}_{\rho} X^{j}_{\alpha} {W}^{-3} - \frac{1}{16}{\rm i} F^{\beta \rho} \lambda^{j}_{\alpha} \lambda_{i \beta} X^{i}_{\rho} {W}^{-3} - \frac{7}{16}{\rm i} F^{\beta \rho} \lambda^{j}_{\beta} \lambda_{i \alpha} X^{i}_{\rho} {W}^{-3}+\frac{1}{8}{\rm i} F^{\beta \rho} \lambda^{j}_{\beta} \lambda_{i \rho} X^{i}_{\alpha} {W}^{-3} - \frac{1}{8}{\rm i} W_{\alpha}\,^{\beta} \lambda_{i \beta} \lambda^{i \rho} X^{j}_{\rho} {W}^{-2}+\frac{23}{32}{\rm i} W_{\alpha}\,^{\beta} \lambda^{j}_{\beta} \lambda^{\rho}_{i} X^{i}_{\rho} {W}^{-2} - \frac{1}{8}{\rm i} W_{\alpha}\,^{\beta} \lambda^{j \rho} \lambda_{i \beta} X^{i}_{\rho} {W}^{-2}+\frac{9}{32}{\rm i} W_{\alpha}\,^{\beta} \lambda^{j \rho} \lambda_{i \rho} X^{i}_{\beta} {W}^{-2}+\frac{3}{16}{\rm i} W^{\beta \rho} \lambda_{i \alpha} \lambda^{i}_{\beta} X^{j}_{\rho} {W}^{-2} - \frac{9}{32}{\rm i} W^{\beta \rho} \lambda_{i \beta} \lambda^{i}_{\rho} X^{j}_{\alpha} {W}^{-2} - \frac{3}{32}{\rm i} W^{\beta \rho} \lambda^{j}_{\alpha} \lambda_{i \beta} X^{i}_{\rho} {W}^{-2} - \frac{3}{8}{\rm i} W^{\beta \rho} \lambda^{j}_{\beta} \lambda_{i \alpha} X^{i}_{\rho} {W}^{-2}+\frac{9}{32}{\rm i} W^{\beta \rho} \lambda^{j}_{\beta} \lambda_{i \rho} X^{i}_{\alpha} {W}^{-2}+\frac{3}{4}{\rm i} X^{j}\,_{i} F_{\alpha}\,^{\beta} F_{\beta}\,^{\rho} \lambda^{i}_{\rho} {W}^{-4}+\frac{3}{8}{\rm i} X^{j}\,_{i} F^{\beta \rho} F_{\beta \rho} \lambda^{i}_{\alpha} {W}^{-4}+\frac{9}{8}{\rm i} X^{j}\,_{i} W_{\alpha}\,^{\beta} F_{\beta}\,^{\rho} \lambda^{i}_{\rho} {W}^{-3}+\frac{9}{16}{\rm i} X^{j}\,_{i} W_{\alpha}\,^{\beta} W_{\beta}\,^{\rho} \lambda^{i}_{\rho} {W}^{-2} - \frac{9}{8}{\rm i} X^{j}\,_{i} W^{\beta \rho} F_{\alpha \beta} \lambda^{i}_{\rho} {W}^{-3}%
+\frac{3}{4}{\rm i} X^{j}\,_{i} W^{\beta \rho} F_{\beta \rho} \lambda^{i}_{\alpha} {W}^{-3}+\frac{3}{16}{\rm i} X^{j}\,_{i} W^{\beta \rho} W_{\beta \rho} \lambda^{i}_{\alpha} {W}^{-2} - \frac{1}{8}{\rm i} X_{i k} X^{i k} {W}^{-3} \nabla_{\alpha}\,^{\beta}{\lambda^{j}_{\beta}} - \frac{1}{12}{\rm i} X_{i k} \lambda^{j}_{\beta} {W}^{-3} \nabla_{\alpha}\,^{\beta}{X^{i k}} - \frac{1}{24}{\rm i} X_{i k} \lambda^{i}_{\beta} {W}^{-3} \nabla_{\alpha}\,^{\beta}{X^{j k}}+\frac{1}{8}{\rm i} X^{j}\,_{i} X^{i}\,_{k} {W}^{-3} \nabla_{\alpha}\,^{\beta}{\lambda^{k}_{\beta}} - \frac{1}{6}{\rm i} X^{j}\,_{i} \lambda_{k \beta} {W}^{-3} \nabla_{\alpha}\,^{\beta}{X^{i k}}+\frac{3}{16}{\rm i} X_{i k} X^{i k} F_{\alpha}\,^{\beta} \lambda^{j}_{\beta} {W}^{-4}+\frac{5}{48}{\rm i} X_{i k} X^{i k} W_{\alpha}\,^{\beta} \lambda^{j}_{\beta} {W}^{-3}+\frac{1}{16}{\rm i} X_{i k} \lambda^{j}_{\alpha} \lambda^{i \beta} X^{k}_{\beta} {W}^{-3} - \frac{1}{16}{\rm i} X_{i k} \lambda^{j \beta} \lambda^{i}_{\alpha} X^{k}_{\beta} {W}^{-3}+\frac{1}{32}{\rm i} X_{i k} \lambda^{j \beta} \lambda^{i}_{\beta} X^{k}_{\alpha} {W}^{-3}+\frac{1}{16}{\rm i} X_{i k} \lambda^{i}_{\alpha} \lambda^{k \beta} X^{j}_{\beta} {W}^{-3}+\frac{5}{32}{\rm i} X_{i k} \lambda^{i \beta} \lambda^{k}_{\beta} X^{j}_{\alpha} {W}^{-3}+\frac{1}{48}{\rm i} X^{j}\,_{i} X^{i}\,_{k} W_{\alpha}\,^{\beta} \lambda^{k}_{\beta} {W}^{-3} - \frac{1}{2}{\rm i} X^{j}\,_{i} \lambda^{i}_{\alpha} \lambda^{\beta}_{k} X^{k}_{\beta} {W}^{-3} - \frac{7}{32}{\rm i} X^{j}\,_{i} \lambda^{i \beta} \lambda_{k \beta} X^{k}_{\alpha} {W}^{-3} - \frac{3}{16}{\rm i} X^{j}\,_{i} \lambda_{k \alpha} \lambda^{i \beta} X^{k}_{\beta} {W}^{-3} - \frac{1}{16}{\rm i} X^{j}\,_{i} \lambda_{k \alpha} \lambda^{k \beta} X^{i}_{\beta} {W}^{-3} - \frac{3}{8}\lambda^{j}_{\beta} \lambda_{i \rho} {W}^{-4} \nabla_{\alpha}\,^{\beta}{W} \nabla^{\rho \lambda}{\lambda^{i}_{\lambda}}%
 - \frac{1}{16}\lambda^{j}_{\beta} \lambda_{i \rho} {W}^{-4} \nabla^{\rho \lambda}{W} \nabla_{\alpha}\,^{\beta}{\lambda^{i}_{\lambda}}+\frac{1}{8}\lambda^{j}_{\beta} \lambda_{i \rho} \lambda^{i}_{\lambda} {W}^{-4} \nabla^{\beta \rho}{\nabla_{\alpha}\,^{\lambda}{W}}+\frac{3}{8}\lambda^{j}_{\beta} \lambda_{i \rho} {W}^{-4} \nabla_{\alpha}\,^{\rho}{W} \nabla^{\beta \lambda}{\lambda^{i}_{\lambda}}+\frac{1}{16}\lambda^{j}_{\beta} \lambda_{i \rho} {W}^{-4} \nabla^{\beta \lambda}{W} \nabla_{\alpha}\,^{\rho}{\lambda^{i}_{\lambda}}+\frac{1}{16}\lambda^{j}_{\beta} \lambda_{i \rho} {W}^{-4} \nabla_{\alpha}\,^{\lambda}{W} \nabla^{\beta \rho}{\lambda^{i}_{\lambda}}+\frac{1}{4}\lambda^{j}_{\beta} \lambda_{i \rho} {W}^{-4} \nabla^{\beta \rho}{W} \nabla_{\alpha}\,^{\lambda}{\lambda^{i}_{\lambda}}+\frac{1}{16}\lambda_{i \alpha} \lambda^{i}_{\beta} {W}^{-4} \nabla_{\lambda}\,^{\beta}{W} \nabla^{\lambda \rho}{\lambda^{j}_{\rho}}+\frac{1}{16}\lambda_{i \alpha} \lambda^{i}_{\beta} {W}^{-4} \nabla_{\lambda}\,^{\rho}{W} \nabla^{\lambda \beta}{\lambda^{j}_{\rho}}+\frac{1}{16}\lambda^{j}_{\alpha} \lambda_{i \beta} {W}^{-4} \nabla_{\lambda}\,^{\beta}{W} \nabla^{\lambda \rho}{\lambda^{i}_{\rho}} - \frac{1}{8}\lambda^{j}_{\beta} \lambda_{i \alpha} {W}^{-4} \nabla_{\lambda}\,^{\beta}{W} \nabla^{\lambda \rho}{\lambda^{i}_{\rho}} - \frac{1}{16}\lambda^{j}_{\beta} \lambda_{i \alpha} {W}^{-4} \nabla_{\lambda}\,^{\rho}{W} \nabla^{\lambda \beta}{\lambda^{i}_{\rho}}+\frac{1}{2}F_{\alpha \beta} \lambda^{j \rho} \lambda_{i \rho} \lambda^{i}_{\lambda} {W}^{-5} \nabla^{\beta \lambda}{W} - \frac{1}{2}F_{\alpha}\,^{\beta} \lambda^{j}_{\rho} \lambda_{i \beta} \lambda^{i}_{\lambda} {W}^{-5} \nabla^{\rho \lambda}{W} - \frac{1}{2}F^{\beta}\,_{\rho} \lambda^{j}_{\lambda} \lambda_{i \alpha} \lambda^{i}_{\beta} {W}^{-5} \nabla^{\rho \lambda}{W}+\frac{1}{2}F^{\beta}\,_{\rho} \lambda^{j}_{\beta} \lambda_{i \alpha} \lambda^{i}_{\lambda} {W}^{-5} \nabla^{\rho \lambda}{W}+\frac{9}{16}W_{\alpha \beta} \lambda^{j \rho} \lambda_{i \rho} \lambda^{i}_{\lambda} {W}^{-4} \nabla^{\beta \lambda}{W} - \frac{3}{32}W_{\alpha}\,^{\beta} \lambda^{j}_{\rho} \lambda_{i \beta} \lambda^{i}_{\lambda} {W}^{-4} \nabla^{\rho \lambda}{W} - \frac{9}{32}W^{\beta}\,_{\rho} \lambda^{j}_{\alpha} \lambda_{i \beta} \lambda^{i}_{\lambda} {W}^{-4} \nabla^{\rho \lambda}{W}+\frac{3}{32}W^{\beta}\,_{\rho} \lambda^{j}_{\lambda} \lambda_{i \alpha} \lambda^{i}_{\beta} {W}^{-4} \nabla^{\rho \lambda}{W}+\frac{3}{16}W^{\beta}\,_{\rho} \lambda^{j}_{\beta} \lambda_{i \alpha} \lambda^{i}_{\lambda} {W}^{-4} \nabla^{\rho \lambda}{W}%
 - \frac{1}{16}X_{i k} \lambda^{j}_{\alpha} \lambda^{i}_{\beta} \lambda^{k}_{\rho} {W}^{-5} \nabla^{\beta \rho}{W}+\frac{5}{16}X_{i k} \lambda^{j}_{\beta} \lambda^{i}_{\alpha} \lambda^{k}_{\rho} {W}^{-5} \nabla^{\beta \rho}{W}+\frac{3}{16}X^{j}\,_{i} \lambda_{k \alpha} \lambda^{i}_{\beta} \lambda^{k}_{\rho} {W}^{-5} \nabla^{\beta \rho}{W} - \frac{1}{16}{\rm i} \lambda^{j}_{\beta} {W}^{-2} \nabla_{\rho \lambda}{\nabla^{\rho \lambda}{\nabla_{\alpha}\,^{\beta}{W}}} - \frac{1}{16}{\rm i} {W}^{-2} \nabla_{\alpha}\,^{\beta}{W} \nabla_{\rho \lambda}{\nabla^{\rho \lambda}{\lambda^{j}_{\beta}}} - \frac{1}{8}{\rm i} {W}^{-2} \nabla_{\rho \lambda}{W} \nabla^{\rho \lambda}{\nabla_{\alpha}\,^{\beta}{\lambda^{j}_{\beta}}} - \frac{1}{8}{\rm i} {W}^{-2} \nabla_{\alpha}\,^{\beta}{\lambda^{j}_{\beta}} \nabla_{\rho \lambda}{\nabla^{\rho \lambda}{W}} - \frac{1}{16}{\rm i} {W}^{-2} \nabla_{\rho \lambda}{\lambda^{j}_{\beta}} \nabla^{\rho \lambda}{\nabla_{\alpha}\,^{\beta}{W}} - \frac{1}{4}{\rm i} F^{\beta}\,_{\rho} {W}^{-3} \nabla_{\alpha \lambda}{W} \nabla^{\lambda \rho}{\lambda^{j}_{\beta}} - \frac{3}{16}{\rm i} W^{\beta}\,_{\rho} \lambda^{j}_{\beta} {W}^{-2} \nabla_{\lambda}\,^{\rho}{\nabla_{\alpha}\,^{\lambda}{W}} - \frac{3}{16}{\rm i} W^{\beta}\,_{\rho} {W}^{-2} \nabla_{\alpha \lambda}{W} \nabla^{\lambda \rho}{\lambda^{j}_{\beta}}+\frac{1}{4}{\rm i} \lambda^{j \beta} {W}^{-3} \nabla_{\alpha \lambda}{W} \nabla^{\lambda \rho}{F_{\beta \rho}}+\frac{49}{128}{\rm i} \lambda^{j \beta} {W}^{-2} \nabla_{\alpha \lambda}{W} \nabla^{\lambda \rho}{W_{\beta \rho}}+\frac{1}{8}{\rm i} F_{\alpha}\,^{\beta} \lambda^{j}_{\beta} {W}^{-3} \nabla_{\rho \lambda}{\nabla^{\rho \lambda}{W}}+\frac{3}{16}{\rm i} F_{\alpha}\,^{\beta} {W}^{-3} \nabla_{\rho \lambda}{W} \nabla^{\rho \lambda}{\lambda^{j}_{\beta}} - \frac{1}{16}{\rm i} W_{\alpha}\,^{\beta} \lambda^{j}_{\beta} {W}^{-2} \nabla_{\rho \lambda}{\nabla^{\rho \lambda}{W}} - \frac{3}{16}{\rm i} \lambda^{j \beta} {W}^{-3} \nabla_{\rho \lambda}{W} \nabla^{\rho \lambda}{F_{\alpha \beta}}+\frac{5}{32}{\rm i} \lambda^{j \beta} {W}^{-2} \nabla_{\rho \lambda}{W} \nabla^{\rho \lambda}{W_{\alpha \beta}}+\frac{3}{4}{\rm i} F^{\beta \rho} F_{\beta \lambda} \lambda^{j}_{\rho} {W}^{-4} \nabla_{\alpha}\,^{\lambda}{W} - \frac{3}{8}{\rm i} F^{\beta \rho} F_{\beta \rho} \lambda^{j}_{\lambda} {W}^{-4} \nabla_{\alpha}\,^{\lambda}{W}%
 - \frac{27}{16}{\rm i} W^{\beta}\,_{\lambda} F_{\beta}\,^{\rho} \lambda^{j}_{\rho} {W}^{-3} \nabla_{\alpha}\,^{\lambda}{W}+\frac{21}{16}{\rm i} W^{\beta \lambda} F_{\beta \rho} \lambda^{j}_{\lambda} {W}^{-3} \nabla_{\alpha}\,^{\rho}{W} - \frac{9}{16}{\rm i} W^{\beta \rho} F_{\beta \rho} \lambda^{j}_{\lambda} {W}^{-3} \nabla_{\alpha}\,^{\lambda}{W} - \frac{3}{64}{\rm i} W^{\beta \rho} W_{\beta \rho} \lambda^{j}_{\lambda} {W}^{-2} \nabla_{\alpha}\,^{\lambda}{W}+\frac{1}{24}{\rm i} X^{j}\,_{i} \lambda^{i}_{\beta} {W}^{-3} \nabla_{\rho}\,^{\beta}{\nabla_{\alpha}\,^{\rho}{W}}+\frac{11}{48}{\rm i} X^{j}\,_{i} {W}^{-3} \nabla_{\alpha \rho}{W} \nabla^{\rho \beta}{\lambda^{i}_{\beta}}+\frac{1}{48}{\rm i} X^{j}\,_{i} {W}^{-3} \nabla_{\rho}\,^{\beta}{W} \nabla_{\alpha}\,^{\rho}{\lambda^{i}_{\beta}} - \frac{7}{48}{\rm i} \lambda_{i \beta} {W}^{-3} \nabla_{\alpha \rho}{W} \nabla^{\rho \beta}{X^{j i}}+\frac{1}{48}{\rm i} \lambda_{i \beta} {W}^{-3} \nabla_{\rho}\,^{\beta}{W} \nabla_{\alpha}\,^{\rho}{X^{j i}} - \frac{5}{48}{\rm i} X^{j}\,_{i} \lambda^{i}_{\alpha} {W}^{-3} \nabla_{\beta \rho}{\nabla^{\beta \rho}{W}} - \frac{7}{96}{\rm i} X^{j}\,_{i} {W}^{-3} \nabla_{\beta \rho}{W} \nabla^{\beta \rho}{\lambda^{i}_{\alpha}}+\frac{11}{96}{\rm i} \lambda_{i \alpha} {W}^{-3} \nabla_{\beta \rho}{W} \nabla^{\beta \rho}{X^{j i}}+\frac{3}{8}{\rm i} X^{j}\,_{i} F^{\beta}\,_{\rho} \lambda^{i}_{\beta} {W}^{-4} \nabla_{\alpha}\,^{\rho}{W}+\frac{3}{16}{\rm i} X^{j}\,_{i} W^{\beta}\,_{\rho} \lambda^{i}_{\beta} {W}^{-3} \nabla_{\alpha}\,^{\rho}{W} - \frac{1}{12}{\rm i} \lambda^{\beta}_{i} W_{\beta}\,^{\rho}\,_{\lambda}\,^{j} {W}^{-2} \nabla_{\alpha}\,^{\lambda}{\lambda^{i}_{\rho}}+\frac{1}{12}{\rm i} \lambda^{\beta}_{i} W_{\beta}\,^{\rho}\,_{\lambda}\,^{i} {W}^{-2} \nabla_{\alpha}\,^{\lambda}{\lambda^{j}_{\rho}} - \frac{1}{6}{\rm i} \lambda^{j \beta} W_{\beta}\,^{\rho}\,_{\lambda i} {W}^{-2} \nabla_{\alpha}\,^{\lambda}{\lambda^{i}_{\rho}}+\frac{1}{32}{\rm i} \lambda^{\beta}_{i} \lambda^{i}_{\rho} X^{j}_{\beta} {W}^{-3} \nabla_{\alpha}\,^{\rho}{W}+\frac{35}{96}{\rm i} \lambda^{j}_{\rho} \lambda^{\beta}_{i} X^{i}_{\beta} {W}^{-3} \nabla_{\alpha}\,^{\rho}{W} - \frac{29}{96}{\rm i} \lambda^{j \beta} \lambda_{i \rho} X^{i}_{\beta} {W}^{-3} \nabla_{\alpha}\,^{\rho}{W}%
+\frac{25}{48}{\rm i} \lambda^{j \rho} \lambda_{i \rho} X^{i}_{\beta} {W}^{-3} \nabla_{\alpha}\,^{\beta}{W} - \frac{1}{6}{\rm i} F^{\beta \rho} \lambda_{i \alpha} \lambda^{i \lambda} W_{\beta \rho \lambda}\,^{j} {W}^{-3} - \frac{1}{6}{\rm i} F^{\beta \rho} \lambda^{j}_{\beta} \lambda^{\lambda}_{i} W_{\alpha \rho \lambda}\,^{i} {W}^{-3}+\frac{1}{6}{\rm i} F^{\beta \rho} \lambda^{j \lambda} \lambda_{i \alpha} W_{\beta \rho \lambda}\,^{i} {W}^{-3}+\frac{1}{6}{\rm i} F^{\beta \rho} \lambda^{j \lambda} \lambda_{i \beta} W_{\alpha \rho \lambda}\,^{i} {W}^{-3}+\frac{1}{24}{\rm i} W_{\alpha}\,^{\beta} \lambda^{\rho}_{i} \lambda^{i \lambda} W_{\beta \rho \lambda}\,^{j} {W}^{-2} - \frac{1}{24}{\rm i} W_{\alpha}\,^{\beta} \lambda^{j \rho} \lambda^{\lambda}_{i} W_{\beta \rho \lambda}\,^{i} {W}^{-2} - \frac{5}{24}{\rm i} W^{\beta \rho} \lambda_{i \alpha} \lambda^{i \lambda} W_{\beta \rho \lambda}\,^{j} {W}^{-2}+\frac{1}{8}{\rm i} W^{\beta \rho} \lambda_{i \beta} \lambda^{i \lambda} W_{\alpha \rho \lambda}\,^{j} {W}^{-2} - \frac{1}{16}{\rm i} W^{\beta \rho} \lambda^{j}_{\alpha} \lambda^{\lambda}_{i} W_{\beta \rho \lambda}\,^{i} {W}^{-2}+\frac{1}{16}{\rm i} W^{\beta \rho} \lambda^{j}_{\beta} \lambda^{\lambda}_{i} W_{\alpha \rho \lambda}\,^{i} {W}^{-2}+\frac{13}{48}{\rm i} W^{\beta \rho} \lambda^{j \lambda} \lambda_{i \alpha} W_{\beta \rho \lambda}\,^{i} {W}^{-2} - \frac{3}{16}{\rm i} W^{\beta \rho} \lambda^{j \lambda} \lambda_{i \beta} W_{\alpha \rho \lambda}\,^{i} {W}^{-2}+\frac{1}{24}{\rm i} W^{\beta \rho} \lambda^{j \lambda} \lambda_{i \lambda} W_{\alpha \beta \rho}\,^{i} {W}^{-2}+\frac{5}{32}{\rm i} X_{i k} X^{i k} \lambda^{j}_{\beta} {W}^{-4} \nabla_{\alpha}\,^{\beta}{W} - \frac{1}{4}{\rm i} X^{j}\,_{i} X^{i}\,_{k} \lambda^{k}_{\beta} {W}^{-4} \nabla_{\alpha}\,^{\beta}{W} - \frac{1}{48}{\rm i} X_{k i} \lambda^{j \beta} \lambda^{k \rho} W_{\alpha \beta \rho}\,^{i} {W}^{-3}+\frac{1}{48}{\rm i} X^{j}\,_{k} \lambda^{k \beta} \lambda^{\rho}_{i} W_{\alpha \beta \rho}\,^{i} {W}^{-3}+\frac{1}{48}{\rm i} X^{j}\,_{i} \lambda^{\beta}_{k} \lambda^{k \rho} W_{\alpha \beta \rho}\,^{i} {W}^{-3} - \frac{7}{48}\Phi_{\alpha}\,^{\beta}\,_{i k} \lambda^{j}_{\beta} \lambda^{i \rho} \lambda^{k}_{\rho} {W}^{-3}%
+\frac{3}{16}\Phi_{\alpha}\,^{\beta}\,_{i k} \lambda^{j \rho} \lambda^{i}_{\beta} \lambda^{k}_{\rho} {W}^{-3}+\frac{1}{48}\Phi_{\alpha}\,^{\beta j}\,_{i} \lambda^{i \rho} \lambda_{k \beta} \lambda^{k}_{\rho} {W}^{-3}+\frac{1}{12}\Phi^{\beta \rho}\,_{i k} \lambda^{j}_{\beta} \lambda^{i}_{\alpha} \lambda^{k}_{\rho} {W}^{-3} - \frac{1}{36}\Phi^{\beta \rho j}\,_{i} \lambda^{i}_{\alpha} \lambda_{k \beta} \lambda^{k}_{\rho} {W}^{-3} - \frac{1}{36}\Phi^{\beta \rho j}\,_{i} \lambda_{k \alpha} \lambda^{i}_{\beta} \lambda^{k}_{\rho} {W}^{-3}+\frac{1}{8}X_{i k} X^{i k} \lambda^{j \beta} \lambda_{l \alpha} \lambda^{l}_{\beta} {W}^{-5} - \frac{1}{4}X_{i k} X^{i}\,_{l} \lambda^{j \beta} \lambda^{k}_{\alpha} \lambda^{l}_{\beta} {W}^{-5}+\frac{3}{8}X^{j}\,_{i} X_{k l} \lambda^{i}_{\alpha} \lambda^{k \beta} \lambda^{l}_{\beta} {W}^{-5}+\frac{1}{24}\lambda_{i \beta} {W}^{-3} \nabla_{\lambda}\,^{\beta}{\lambda^{j \rho}} \nabla_{\alpha}\,^{\lambda}{\lambda^{i}_{\rho}}+\frac{1}{24}\lambda^{j}_{\beta} \lambda^{\rho}_{i} {W}^{-3} \nabla_{\lambda}\,^{\beta}{\nabla_{\alpha}\,^{\lambda}{\lambda^{i}_{\rho}}}+\frac{1}{24}\lambda^{j}_{\beta} {W}^{-3} \nabla_{\alpha \lambda}{\lambda^{\rho}_{i}} \nabla^{\lambda \beta}{\lambda^{i}_{\rho}} - \frac{1}{24}\lambda^{j \beta} \lambda_{i \rho} {W}^{-3} \nabla_{\lambda}\,^{\rho}{\nabla_{\alpha}\,^{\lambda}{\lambda^{i}_{\beta}}} - \frac{1}{4}\lambda^{j}_{\beta} \lambda_{i \rho} \lambda^{i}_{\lambda} {W}^{-5} \nabla_{\alpha}\,^{\rho}{W} \nabla^{\beta \lambda}{W}+\frac{1}{24}\lambda_{i \alpha} \lambda^{i \beta} {W}^{-3} \nabla_{\rho \lambda}{\nabla^{\rho \lambda}{\lambda^{j}_{\beta}}}+\frac{1}{24}\lambda_{i \alpha} {W}^{-3} \nabla_{\rho \lambda}{\lambda^{j \beta}} \nabla^{\rho \lambda}{\lambda^{i}_{\beta}} - \frac{1}{32}\lambda^{\beta}_{i} {W}^{-3} \nabla_{\rho \lambda}{\lambda^{j}_{\beta}} \nabla^{\rho \lambda}{\lambda^{i}_{\alpha}} - \frac{1}{24}\lambda^{j \beta} \lambda_{i \alpha} {W}^{-3} \nabla_{\rho \lambda}{\nabla^{\rho \lambda}{\lambda^{i}_{\beta}}} - \frac{1}{48}\lambda^{j \beta} \lambda_{i \beta} {W}^{-3} \nabla_{\rho \lambda}{\nabla^{\rho \lambda}{\lambda^{i}_{\alpha}}} - \frac{1}{32}\lambda^{j \beta} {W}^{-3} \nabla_{\rho \lambda}{\lambda_{i \alpha}} \nabla^{\rho \lambda}{\lambda^{i}_{\beta}}+\frac{1}{8}F^{\beta}\,_{\rho} \lambda^{j}_{\beta} \lambda^{\lambda}_{i} {W}^{-4} \nabla_{\alpha}\,^{\rho}{\lambda^{i}_{\lambda}}%
 - \frac{1}{8}F^{\beta}\,_{\rho} \lambda^{j \lambda} \lambda_{i \beta} {W}^{-4} \nabla_{\alpha}\,^{\rho}{\lambda^{i}_{\lambda}}+\frac{1}{4}F^{\beta}\,_{\rho} \lambda^{j \lambda} \lambda_{i \lambda} {W}^{-4} \nabla_{\alpha}\,^{\rho}{\lambda^{i}_{\beta}} - \frac{3}{8}F^{\beta \rho} \lambda_{i \beta} \lambda^{i}_{\rho} {W}^{-4} \nabla_{\alpha}\,^{\lambda}{\lambda^{j}_{\lambda}} - \frac{1}{8}F^{\beta \rho} \lambda^{j}_{\lambda} \lambda_{i \beta} {W}^{-4} \nabla_{\alpha}\,^{\lambda}{\lambda^{i}_{\rho}}+\frac{1}{8}F^{\beta \rho} \lambda^{j}_{\beta} \lambda_{i \lambda} {W}^{-4} \nabla_{\alpha}\,^{\lambda}{\lambda^{i}_{\rho}} - \frac{3}{32}W^{\beta}\,_{\rho} \lambda_{i \beta} \lambda^{i \lambda} {W}^{-3} \nabla_{\alpha}\,^{\rho}{\lambda^{j}_{\lambda}}+\frac{1}{8}W^{\beta}\,_{\rho} \lambda^{j}_{\beta} \lambda^{\lambda}_{i} {W}^{-3} \nabla_{\alpha}\,^{\rho}{\lambda^{i}_{\lambda}} - \frac{1}{32}W^{\beta}\,_{\rho} \lambda^{j \lambda} \lambda_{i \beta} {W}^{-3} \nabla_{\alpha}\,^{\rho}{\lambda^{i}_{\lambda}}+\frac{3}{8}W^{\beta}\,_{\rho} \lambda^{j \lambda} \lambda_{i \lambda} {W}^{-3} \nabla_{\alpha}\,^{\rho}{\lambda^{i}_{\beta}} - \frac{1}{32}W^{\beta \rho} \lambda_{i \beta} \lambda^{i}_{\lambda} {W}^{-3} \nabla_{\alpha}\,^{\lambda}{\lambda^{j}_{\rho}} - \frac{5}{32}W^{\beta \rho} \lambda_{i \beta} \lambda^{i}_{\rho} {W}^{-3} \nabla_{\alpha}\,^{\lambda}{\lambda^{j}_{\lambda}} - \frac{3}{32}W^{\beta \rho} \lambda^{j}_{\lambda} \lambda_{i \beta} {W}^{-3} \nabla_{\alpha}\,^{\lambda}{\lambda^{i}_{\rho}}+\frac{1}{8}W^{\beta \rho} \lambda^{j}_{\beta} \lambda_{i \lambda} {W}^{-3} \nabla_{\alpha}\,^{\lambda}{\lambda^{i}_{\rho}} - \frac{1}{32}W^{\beta \rho} \lambda^{j}_{\beta} \lambda_{i \rho} {W}^{-3} \nabla_{\alpha}\,^{\lambda}{\lambda^{i}_{\lambda}}+\frac{1}{32}\lambda^{j}_{\lambda} \lambda^{\beta}_{i} \lambda^{i \rho} {W}^{-3} \nabla_{\alpha}\,^{\lambda}{W_{\beta \rho}} - \frac{1}{32}\lambda^{j \beta} \lambda^{\rho}_{i} \lambda^{i}_{\lambda} {W}^{-3} \nabla_{\alpha}\,^{\lambda}{W_{\beta \rho}} - \frac{7}{32}\lambda^{j \lambda} \lambda_{i \lambda} \lambda^{i \beta} {W}^{-3} \nabla_{\alpha}\,^{\rho}{W_{\beta \rho}}-F_{\alpha}\,^{\beta} F_{\beta}\,^{\rho} \lambda^{j \lambda} \lambda_{i \rho} \lambda^{i}_{\lambda} {W}^{-5}+\frac{3}{2}F_{\alpha}\,^{\beta} F^{\rho \lambda} \lambda^{j}_{\beta} \lambda_{i \rho} \lambda^{i}_{\lambda} {W}^{-5} - \frac{1}{2}F^{\beta \rho} F_{\beta \rho} \lambda^{j \lambda} \lambda_{i \alpha} \lambda^{i}_{\lambda} {W}^{-5}%
+F^{\beta \rho} F_{\beta}\,^{\lambda} \lambda^{j}_{\rho} \lambda_{i \alpha} \lambda^{i}_{\lambda} {W}^{-5} - \frac{9}{8}W_{\alpha}\,^{\beta} F_{\beta}\,^{\rho} \lambda^{j \lambda} \lambda_{i \rho} \lambda^{i}_{\lambda} {W}^{-4}+\frac{19}{16}W_{\alpha}\,^{\lambda} F^{\beta \rho} \lambda^{j}_{\lambda} \lambda_{i \beta} \lambda^{i}_{\rho} {W}^{-4}+\frac{1}{8}W_{\alpha}\,^{\lambda} F^{\beta \rho} \lambda^{j}_{\beta} \lambda_{i \lambda} \lambda^{i}_{\rho} {W}^{-4} - \frac{9}{8}W_{\alpha}\,^{\beta} W_{\beta}\,^{\rho} \lambda^{j \lambda} \lambda_{i \rho} \lambda^{i}_{\lambda} {W}^{-3}+\frac{35}{32}W_{\alpha}\,^{\beta} W^{\rho \lambda} \lambda^{j}_{\beta} \lambda_{i \rho} \lambda^{i}_{\lambda} {W}^{-3} - \frac{19}{16}W_{\alpha}\,^{\beta} W^{\rho \lambda} \lambda^{j}_{\rho} \lambda_{i \beta} \lambda^{i}_{\lambda} {W}^{-3}+\frac{9}{8}W^{\beta \rho} F_{\alpha \beta} \lambda^{j \lambda} \lambda_{i \rho} \lambda^{i}_{\lambda} {W}^{-4} - \frac{3}{4}W^{\rho \lambda} F_{\alpha}\,^{\beta} \lambda^{j}_{\rho} \lambda_{i \lambda} \lambda^{i}_{\beta} {W}^{-4}+\frac{3}{4}W^{\rho \lambda} F_{\alpha}\,^{\beta} \lambda^{j}_{\beta} \lambda_{i \rho} \lambda^{i}_{\lambda} {W}^{-4} - \frac{9}{8}W^{\beta \rho} F_{\beta \rho} \lambda^{j \lambda} \lambda_{i \alpha} \lambda^{i}_{\lambda} {W}^{-4}+\frac{9}{8}W^{\beta \lambda} F_{\beta}\,^{\rho} \lambda^{j}_{\lambda} \lambda_{i \alpha} \lambda^{i}_{\rho} {W}^{-4} - \frac{9}{8}W^{\beta \lambda} F_{\beta}\,^{\rho} \lambda^{j}_{\rho} \lambda_{i \alpha} \lambda^{i}_{\lambda} {W}^{-4} - \frac{33}{64}W^{\beta \rho} W_{\beta \rho} \lambda^{j \lambda} \lambda_{i \alpha} \lambda^{i}_{\lambda} {W}^{-3}+\frac{9}{8}W^{\beta \rho} W_{\beta}\,^{\lambda} \lambda^{j}_{\rho} \lambda_{i \alpha} \lambda^{i}_{\lambda} {W}^{-3}+\frac{3}{64}X_{i k} \lambda^{j}_{\beta} \lambda^{i \rho} {W}^{-4} \nabla_{\alpha}\,^{\beta}{\lambda^{k}_{\rho}} - \frac{3}{64}X_{i k} \lambda^{j \beta} \lambda^{i}_{\rho} {W}^{-4} \nabla_{\alpha}\,^{\rho}{\lambda^{k}_{\beta}}+\frac{3}{64}X_{i k} \lambda^{j \beta} \lambda^{i}_{\beta} {W}^{-4} \nabla_{\alpha}\,^{\rho}{\lambda^{k}_{\rho}}+\frac{3}{16}X_{i k} \lambda^{i \beta} \lambda^{k}_{\beta} {W}^{-4} \nabla_{\alpha}\,^{\rho}{\lambda^{j}_{\rho}} - \frac{5}{64}X^{j}\,_{i} \lambda^{i}_{\beta} \lambda^{\rho}_{k} {W}^{-4} \nabla_{\alpha}\,^{\beta}{\lambda^{k}_{\rho}}%
 - \frac{3}{64}X^{j}\,_{i} \lambda^{i \beta} \lambda_{k \rho} {W}^{-4} \nabla_{\alpha}\,^{\rho}{\lambda^{k}_{\beta}} - \frac{13}{64}X^{j}\,_{i} \lambda^{i \beta} \lambda_{k \beta} {W}^{-4} \nabla_{\alpha}\,^{\rho}{\lambda^{k}_{\rho}}+\frac{1}{8}X^{j}\,_{i} \lambda^{\beta}_{k} \lambda^{k}_{\rho} {W}^{-4} \nabla_{\alpha}\,^{\rho}{\lambda^{i}_{\beta}}+\frac{5}{64}\lambda^{j}_{\beta} \lambda^{\rho}_{i} \lambda_{k \rho} {W}^{-4} \nabla_{\alpha}\,^{\beta}{X^{i k}}+\frac{5}{32}\lambda^{j \beta} \lambda_{i \beta} \lambda_{k \rho} {W}^{-4} \nabla_{\alpha}\,^{\rho}{X^{i k}} - \frac{3}{4}X_{i k} F_{\alpha}\,^{\beta} \lambda^{j}_{\beta} \lambda^{i \rho} \lambda^{k}_{\rho} {W}^{-5} - \frac{17}{32}X_{i k} W_{\alpha}\,^{\beta} \lambda^{j}_{\beta} \lambda^{i \rho} \lambda^{k}_{\rho} {W}^{-4}+\frac{1}{32}X_{i k} W_{\alpha}\,^{\beta} \lambda^{j \rho} \lambda^{i}_{\beta} \lambda^{k}_{\rho} {W}^{-4}+\frac{3}{16}X_{i k} W^{\beta \rho} \lambda^{j}_{\beta} \lambda^{i}_{\alpha} \lambda^{k}_{\rho} {W}^{-4} - \frac{3}{4}X^{j}\,_{i} F^{\beta \rho} \lambda^{i}_{\alpha} \lambda_{k \beta} \lambda^{k}_{\rho} {W}^{-5}+\frac{1}{32}X^{j}\,_{i} W_{\alpha}\,^{\beta} \lambda^{i \rho} \lambda_{k \beta} \lambda^{k}_{\rho} {W}^{-4} - \frac{3}{8}X^{j}\,_{i} W^{\beta \rho} \lambda^{i}_{\alpha} \lambda_{k \beta} \lambda^{k}_{\rho} {W}^{-4} - \frac{3}{16}X^{j}\,_{i} W^{\beta \rho} \lambda_{k \alpha} \lambda^{i}_{\beta} \lambda^{k}_{\rho} {W}^{-4}+\frac{3}{64}\lambda^{j}_{\alpha} \lambda^{\rho}_{k} \lambda^{k \beta} \lambda_{i \rho} X^{i}_{\beta} {W}^{-4}+\frac{27}{64}\lambda^{j \rho} \lambda_{k \alpha} \lambda^{k}_{\rho} \lambda^{\beta}_{i} X^{i}_{\beta} {W}^{-4} - \frac{3}{32}\lambda^{j \rho} \lambda_{k \alpha} \lambda^{k \beta} \lambda_{i \rho} X^{i}_{\beta} {W}^{-4}+\frac{3}{64}\lambda^{j \beta} \lambda_{k \alpha} \lambda^{k \rho} \lambda_{i \rho} X^{i}_{\beta} {W}^{-4} - \frac{9}{64}\lambda^{j \rho} \lambda_{i \alpha} \lambda_{k \rho} \lambda^{k \beta} X^{i}_{\beta} {W}^{-4}+\frac{9}{32}\lambda^{j \beta} \lambda_{k \beta} \lambda^{k \rho} \lambda_{i \rho} X^{i}_{\alpha} {W}^{-4}+\frac{1}{16}{\rm i} \lambda^{j}_{\beta} {W}^{-3} \nabla_{\rho \lambda}{W} \nabla^{\lambda \beta}{\nabla_{\alpha}\,^{\rho}{W}}%
+\frac{1}{16}{\rm i} \lambda^{j}_{\beta} {W}^{-3} \nabla_{\rho}\,^{\beta}{W} \nabla_{\lambda}\,^{\rho}{\nabla_{\alpha}\,^{\lambda}{W}}+\frac{1}{16}{\rm i} {W}^{-3} \nabla_{\alpha \rho}{W} \nabla^{\rho}\,_{\lambda}{W} \nabla^{\lambda \beta}{\lambda^{j}_{\beta}}+\frac{1}{16}{\rm i} {W}^{-3} \nabla_{\alpha \rho}{W} \nabla_{\lambda}\,^{\beta}{W} \nabla^{\rho \lambda}{\lambda^{j}_{\beta}}+\frac{1}{16}{\rm i} \lambda^{j}_{\beta} {W}^{-3} \nabla_{\alpha}\,^{\beta}{W} \nabla_{\rho \lambda}{\nabla^{\rho \lambda}{W}}+\frac{3}{32}{\rm i} \lambda^{j}_{\beta} {W}^{-3} \nabla_{\rho \lambda}{W} \nabla^{\rho \lambda}{\nabla_{\alpha}\,^{\beta}{W}}+\frac{3}{32}{\rm i} {W}^{-3} \nabla_{\alpha}\,^{\beta}{W} \nabla_{\rho \lambda}{W} \nabla^{\rho \lambda}{\lambda^{j}_{\beta}}+\frac{1}{16}{\rm i} {W}^{-3} \nabla_{\rho \lambda}{W} \nabla^{\rho \lambda}{W} \nabla_{\alpha}\,^{\beta}{\lambda^{j}_{\beta}}+\frac{3}{8}{\rm i} F^{\beta}\,_{\rho} \lambda^{j}_{\beta} {W}^{-4} \nabla_{\alpha \lambda}{W} \nabla^{\lambda \rho}{W} - \frac{9}{32}{\rm i} W^{\beta}\,_{\rho} \lambda^{j}_{\beta} {W}^{-3} \nabla_{\alpha \lambda}{W} \nabla^{\lambda \rho}{W} - \frac{3}{32}{\rm i} F_{\alpha}\,^{\beta} \lambda^{j}_{\beta} {W}^{-4} \nabla_{\rho \lambda}{W} \nabla^{\rho \lambda}{W}+\frac{3}{32}{\rm i} W_{\alpha}\,^{\beta} \lambda^{j}_{\beta} {W}^{-3} \nabla_{\rho \lambda}{W} \nabla^{\rho \lambda}{W} - \frac{1}{4}{\rm i} X^{j}\,_{i} \lambda^{i}_{\beta} {W}^{-4} \nabla_{\alpha \rho}{W} \nabla^{\rho \beta}{W}+\frac{5}{64}{\rm i} X^{j}\,_{i} \lambda^{i}_{\alpha} {W}^{-4} \nabla_{\beta \rho}{W} \nabla^{\beta \rho}{W}+\frac{5}{16}{\rm i} \lambda^{j}_{\beta} \lambda_{i \alpha} \lambda^{i}_{\rho} \lambda^{\lambda}_{k} {W}^{-5} \nabla^{\beta \rho}{\lambda^{k}_{\lambda}}+\frac{3}{16}{\rm i} \lambda^{j}_{\beta} \lambda_{i \alpha} \lambda^{i \rho} \lambda_{k \lambda} {W}^{-5} \nabla^{\beta \lambda}{\lambda^{k}_{\rho}}+\frac{5}{16}{\rm i} \lambda^{j}_{\beta} \lambda_{i \alpha} \lambda^{i \rho} \lambda_{k \rho} {W}^{-5} \nabla^{\beta \lambda}{\lambda^{k}_{\lambda}} - \frac{1}{16}{\rm i} \lambda^{j}_{\beta} \lambda^{\rho}_{i} \lambda^{i}_{\lambda} \lambda_{k \rho} {W}^{-5} \nabla^{\beta \lambda}{\lambda^{k}_{\alpha}} - \frac{3}{16}{\rm i} \lambda^{j \beta} \lambda_{i \alpha} \lambda^{i}_{\rho} \lambda_{k \lambda} {W}^{-5} \nabla^{\rho \lambda}{\lambda^{k}_{\beta}} - \frac{1}{16}{\rm i} \lambda^{j \beta} \lambda_{i \alpha} \lambda^{i}_{\rho} \lambda_{k \beta} {W}^{-5} \nabla^{\rho \lambda}{\lambda^{k}_{\lambda}}+\frac{5}{16}{\rm i} \lambda^{j \beta} \lambda_{i \alpha} \lambda^{i}_{\beta} \lambda_{k \rho} {W}^{-5} \nabla^{\rho \lambda}{\lambda^{k}_{\lambda}}%
+\frac{1}{16}{\rm i} \lambda^{j \beta} \lambda_{i \beta} \lambda^{i}_{\rho} \lambda_{k \lambda} {W}^{-5} \nabla^{\rho \lambda}{\lambda^{k}_{\alpha}} - \frac{3}{16}\lambda^{j}_{\beta} \lambda^{\rho}_{i} {W}^{-4} \nabla_{\alpha \lambda}{W} \nabla^{\lambda \beta}{\lambda^{i}_{\rho}} - \frac{1}{16}\lambda^{j}_{\beta} \lambda^{\rho}_{i} {W}^{-4} \nabla_{\lambda}\,^{\beta}{W} \nabla_{\alpha}\,^{\lambda}{\lambda^{i}_{\rho}} - \frac{3}{16}\lambda^{j \beta} \lambda_{i \rho} {W}^{-4} \nabla_{\alpha \lambda}{W} \nabla^{\lambda \rho}{\lambda^{i}_{\beta}}+\frac{1}{16}\lambda^{j \beta} \lambda_{i \rho} {W}^{-4} \nabla_{\lambda}\,^{\rho}{W} \nabla_{\alpha}\,^{\lambda}{\lambda^{i}_{\beta}} - \frac{5}{16}\lambda^{j \beta} \lambda_{i \beta} {W}^{-4} \nabla_{\alpha \lambda}{W} \nabla^{\lambda \rho}{\lambda^{i}_{\rho}} - \frac{3}{32}\lambda_{i \alpha} \lambda^{i \beta} {W}^{-4} \nabla_{\rho \lambda}{W} \nabla^{\rho \lambda}{\lambda^{j}_{\beta}}+\frac{1}{16}\lambda^{j \beta} \lambda_{i \alpha} \lambda^{i}_{\beta} {W}^{-4} \nabla_{\rho \lambda}{\nabla^{\rho \lambda}{W}}+\frac{3}{32}\lambda^{j \beta} \lambda_{i \alpha} {W}^{-4} \nabla_{\rho \lambda}{W} \nabla^{\rho \lambda}{\lambda^{i}_{\beta}}+\frac{1}{32}\lambda^{j \beta} \lambda_{i \beta} {W}^{-4} \nabla_{\rho \lambda}{W} \nabla^{\rho \lambda}{\lambda^{i}_{\alpha}} - \frac{1}{2}F^{\beta}\,_{\rho} \lambda^{j \lambda} \lambda_{i \beta} \lambda^{i}_{\lambda} {W}^{-5} \nabla_{\alpha}\,^{\rho}{W}+\frac{3}{4}F^{\beta \rho} \lambda^{j}_{\lambda} \lambda_{i \beta} \lambda^{i}_{\rho} {W}^{-5} \nabla_{\alpha}\,^{\lambda}{W} - \frac{15}{32}W^{\beta}\,_{\rho} \lambda^{j \lambda} \lambda_{i \beta} \lambda^{i}_{\lambda} {W}^{-4} \nabla_{\alpha}\,^{\rho}{W}+\frac{9}{32}W^{\beta \rho} \lambda^{j}_{\lambda} \lambda_{i \beta} \lambda^{i}_{\rho} {W}^{-4} \nabla_{\alpha}\,^{\lambda}{W} - \frac{9}{32}W^{\beta \rho} \lambda^{j}_{\beta} \lambda_{i \rho} \lambda^{i}_{\lambda} {W}^{-4} \nabla_{\alpha}\,^{\lambda}{W} - \frac{7}{16}X_{i k} \lambda^{j}_{\beta} \lambda^{i \rho} \lambda^{k}_{\rho} {W}^{-5} \nabla_{\alpha}\,^{\beta}{W} - \frac{5}{16}X_{i k} \lambda^{j \beta} \lambda^{i}_{\beta} \lambda^{k}_{\rho} {W}^{-5} \nabla_{\alpha}\,^{\rho}{W}+\frac{1}{16}X^{j}\,_{i} \lambda^{i \beta} \lambda_{k \beta} \lambda^{k}_{\rho} {W}^{-5} \nabla_{\alpha}\,^{\rho}{W}+\frac{1}{4}\lambda^{j}_{\alpha} \lambda^{\beta}_{k} \lambda^{k \rho} \lambda^{\lambda}_{i} W_{\beta \rho \lambda}\,^{i} {W}^{-4} - \frac{1}{4}\lambda^{j \beta} \lambda_{i \alpha} \lambda^{\rho}_{k} \lambda^{k \lambda} W_{\beta \rho \lambda}\,^{i} {W}^{-4}%
 - \frac{1}{32}\lambda^{j \lambda} \lambda_{k \lambda} \lambda^{k \beta} \lambda^{\rho}_{i} W_{\alpha \beta \rho}\,^{i} {W}^{-4} - \frac{1}{32}\lambda^{j \lambda} \lambda_{i \lambda} \lambda^{\beta}_{k} \lambda^{k \rho} W_{\alpha \beta \rho}\,^{i} {W}^{-4} - \frac{1}{32}\lambda^{j \beta} \lambda^{\lambda}_{k} \lambda^{k \rho} \lambda_{i \lambda} W_{\alpha \beta \rho}\,^{i} {W}^{-4} - \frac{3}{64}{\rm i} \lambda^{j}_{\beta} {W}^{-4} \nabla_{\alpha}\,^{\beta}{W} \nabla_{\rho \lambda}{W} \nabla^{\rho \lambda}{W} - \frac{5}{16}{\rm i} \lambda^{j}_{\beta} \lambda_{i \alpha} \lambda^{i \rho} \lambda_{k \rho} \lambda^{k}_{\lambda} {W}^{-6} \nabla^{\beta \lambda}{W}+\frac{5}{16}{\rm i} \lambda^{j \beta} \lambda_{i \alpha} \lambda^{i}_{\rho} \lambda_{k \beta} \lambda^{k}_{\lambda} {W}^{-6} \nabla^{\rho \lambda}{W}+\frac{1}{4}\lambda^{j \beta} \lambda_{i \beta} \lambda^{i}_{\rho} {W}^{-5} \nabla_{\alpha \lambda}{W} \nabla^{\lambda \rho}{W} - \frac{1}{16}\lambda^{j \beta} \lambda_{i \alpha} \lambda^{i}_{\beta} {W}^{-5} \nabla_{\rho \lambda}{W} \nabla^{\rho \lambda}{W}+\frac{1}{16}{\rm i} X_{i k} \lambda^{j \beta} \lambda^{i}_{\alpha} \lambda^{k \rho} \lambda_{l \beta} \lambda^{l}_{\rho} {W}^{-6} - \frac{3}{16}{\rm i} X_{i k} \lambda^{j \beta} \lambda_{l \alpha} \lambda^{i}_{\beta} \lambda^{k \rho} \lambda^{l}_{\rho} {W}^{-6}+\frac{9}{16}{\rm i} X_{i k} \lambda^{j \beta} \lambda_{l \alpha} \lambda^{i \rho} \lambda^{k}_{\rho} \lambda^{l}_{\beta} {W}^{-6} - \frac{1}{32}{\rm i} X^{j}\,_{i} \lambda^{i}_{\alpha} \lambda^{\beta}_{k} \lambda^{k \rho} \lambda_{l \beta} \lambda^{l}_{\rho} {W}^{-6} - \frac{1}{16}{\rm i} X^{j}\,_{i} \lambda_{k \alpha} \lambda^{i \beta} \lambda^{k \rho} \lambda_{l \beta} \lambda^{l}_{\rho} {W}^{-6}+\frac{1}{4}{\rm i} \lambda^{j \beta} \lambda_{i \beta} \lambda^{\rho}_{k} \lambda^{k}_{\lambda} {W}^{-5} \nabla_{\alpha}\,^{\lambda}{\lambda^{i}_{\rho}} - \frac{1}{4}{\rm i} \lambda^{j \beta} \lambda_{i \beta} \lambda^{i \rho} \lambda_{k \rho} {W}^{-5} \nabla_{\alpha}\,^{\lambda}{\lambda^{k}_{\lambda}}+\frac{5}{16}{\rm i} F_{\alpha}\,^{\beta} \lambda^{j}_{\beta} \lambda^{\rho}_{i} \lambda^{i \lambda} \lambda_{k \rho} \lambda^{k}_{\lambda} {W}^{-6} - \frac{5}{8}{\rm i} F_{\alpha}\,^{\beta} \lambda^{j \rho} \lambda_{i \beta} \lambda^{i \lambda} \lambda_{k \rho} \lambda^{k}_{\lambda} {W}^{-6} - \frac{5}{4}{\rm i} F^{\beta \rho} \lambda^{j \lambda} \lambda_{i \alpha} \lambda^{i}_{\lambda} \lambda_{k \beta} \lambda^{k}_{\rho} {W}^{-6}+\frac{9}{32}{\rm i} W_{\alpha}\,^{\beta} \lambda^{j}_{\beta} \lambda^{\rho}_{i} \lambda^{i \lambda} \lambda_{k \rho} \lambda^{k}_{\lambda} {W}^{-5} - \frac{9}{16}{\rm i} W_{\alpha}\,^{\beta} \lambda^{j \rho} \lambda_{i \beta} \lambda^{i \lambda} \lambda_{k \rho} \lambda^{k}_{\lambda} {W}^{-5}%
 - \frac{9}{32}{\rm i} W^{\beta \rho} \lambda^{j}_{\beta} \lambda_{i \alpha} \lambda^{i \lambda} \lambda_{k \rho} \lambda^{k}_{\lambda} {W}^{-5}+\frac{21}{32}{\rm i} W^{\beta \rho} \lambda^{j \lambda} \lambda_{i \alpha} \lambda^{i}_{\beta} \lambda_{k \rho} \lambda^{k}_{\lambda} {W}^{-5} - \frac{21}{32}{\rm i} W^{\beta \rho} \lambda^{j \lambda} \lambda_{i \alpha} \lambda^{i}_{\lambda} \lambda_{k \beta} \lambda^{k}_{\rho} {W}^{-5}+\frac{5}{32}{\rm i} \lambda^{j}_{\beta} \lambda^{\rho}_{i} \lambda^{i \lambda} \lambda_{k \rho} \lambda^{k}_{\lambda} {W}^{-6} \nabla_{\alpha}\,^{\beta}{W}+\frac{5}{16}{\rm i} \lambda^{j \beta} \lambda_{i \beta} \lambda^{i \rho} \lambda_{k \rho} \lambda^{k}_{\lambda} {W}^{-6} \nabla_{\alpha}\,^{\lambda}{W}+\frac{3}{16}\lambda^{j \beta} \lambda_{i \alpha} \lambda^{i}_{\beta} \lambda^{\rho}_{k} \lambda^{k \lambda} \lambda_{l \rho} \lambda^{l}_{\lambda} {W}^{-7} - \frac{3}{8}\lambda^{j \beta} \lambda_{i \alpha} \lambda^{i \rho} \lambda_{k \beta} \lambda^{k \lambda} \lambda_{l \rho} \lambda^{l}_{\lambda} {W}^{-7}
\doublespacedmathend
\end{adjustwidth}

\subsubsection{$F_{\log}$}

\begin{adjustwidth}{0cm}{5cm}
\doublespacedmathbegin
- \frac{3}{32}{\rm i} (\Gamma_{a})^{\alpha \beta} X_{i \alpha} \nabla^{a}{X^{i}_{\beta}} - \frac{81}{128}{\rm i} (\Sigma_{a b})^{\alpha \beta} W^{a b} X_{i \alpha} X^{i}_{\beta}+\frac{3}{2048}{Y}^{2}+\frac{69}{1024}W^{a b} W_{a b} Y+\frac{39}{16}W_{a b} \nabla_{c}{\nabla^{c}{W^{a b}}} - \frac{1}{2}W_{a b} \nabla_{c}{\nabla^{a}{W^{c b}}}+2W_{a b} \nabla^{a}{\nabla_{c}{W^{c b}}}+\frac{9}{16}\epsilon^{e a b c d} W_{a b} W_{c d} \nabla^{{e_{1}}}{W_{e {e_{1}}}} - \frac{1005}{2048}W^{a b} W^{c d} W_{a b} W_{c d} - \frac{1}{2}W^{a b} W^{c d} W_{a c} W_{b d} - \frac{11}{16}{\rm i} W^{a b} W_{a b}\,^{\alpha}\,_{i} X^{i}_{\alpha}+\frac{33}{16}\nabla_{c}{W_{a b}} \nabla^{c}{W^{a b}} - \frac{9}{4}\nabla^{a}{W_{a b}} \nabla_{c}{W^{b c}} - \frac{3}{4}\nabla_{c}{W_{a b}} \nabla^{a}{W^{c b}} - \frac{1}{3}C_{a b c d} W^{a b} W^{c d}+\frac{1}{2}W_{a}\,^{b} W^{a c} W^{d}\,_{b} W_{d c} - \frac{1}{3}C_{a b c d} W^{a c} W^{b d} - \frac{3}{4}{\rm i} W^{a b} W_{a}\,^{c \alpha}\,_{i} W_{b c \alpha}\,^{i} - \frac{3}{2}{W}^{-1} \nabla_{a}{\lambda^{\alpha}_{i}} \nabla^{a}{X^{i}_{\alpha}}%
+\frac{3}{2}(\Sigma_{a b})^{\beta \alpha} {W}^{-1} \nabla^{a}{\lambda_{i \beta}} \nabla^{b}{X^{i}_{\alpha}}+\frac{27}{16}(\Gamma^{a})^{\beta \alpha} W_{a b} \lambda_{i \beta} {W}^{-1} \nabla^{b}{X^{i}_{\alpha}}+\frac{3}{16}\epsilon^{c d}\,_{e}\,^{a b} (\Sigma_{c d})^{\beta \alpha} F_{a b} \lambda_{i \beta} {W}^{-2} \nabla^{e}{X^{i}_{\alpha}} - \frac{3}{8}(\Gamma^{a})^{\beta \alpha} F_{a b} \lambda_{i \beta} {W}^{-2} \nabla^{b}{X^{i}_{\alpha}}+\frac{3}{2}F_{a b} {W}^{-1} \nabla_{c}{\nabla^{c}{W^{a b}}} - \frac{1}{2}F_{a b} {W}^{-1} \nabla_{c}{\nabla^{a}{W^{c b}}}-F_{a b} {W}^{-1} \nabla^{a}{\nabla_{c}{W^{c b}}}+\frac{9}{16}\epsilon^{e {e_{1}} c d a} W_{c d} F_{a b} {W}^{-1} \nabla^{b}{W_{e {e_{1}}}}+\frac{9}{8}\epsilon_{{e_{1}} e}\,^{c a b} W_{c d} F_{a b} {W}^{-1} \nabla^{{e_{1}}}{W^{e d}} - \frac{9}{16}\epsilon^{e c d a b} W_{c d} F_{a b} {W}^{-1} \nabla^{{e_{1}}}{W_{e {e_{1}}}} - \frac{9}{8}\epsilon_{{e_{1}}}\,^{d e c a} W_{c}\,^{b} F_{a b} {W}^{-1} \nabla^{{e_{1}}}{W_{d e}}+\frac{5}{32}W^{c d} W^{a b} W_{c d} F_{a b} {W}^{-1} - \frac{1}{8}W^{c a} W^{d b} W_{c d} F_{a b} {W}^{-1} - \frac{5}{16}{\rm i} F^{a b} W_{a b}\,^{\alpha}\,_{i} X^{i}_{\alpha} {W}^{-1}+\frac{9}{8}{W}^{-1} \nabla_{c}{F_{a b}} \nabla^{c}{W^{a b}} - \frac{3}{2}{W}^{-1} \nabla^{a}{F_{a b}} \nabla_{c}{W^{b c}}+\frac{3}{4}{W}^{-1} \nabla_{c}{F_{a b}} \nabla^{a}{W^{c b}}+\frac{9}{16}\epsilon^{a c d e {e_{1}}} W_{c d} W_{e {e_{1}}} {W}^{-1} \nabla^{b}{F_{a b}}+\frac{1}{32}\epsilon^{c d}\,_{e}\,^{a b} (\Sigma_{c d})^{\beta \alpha} \lambda_{i \beta} X^{i}_{\alpha} {W}^{-2} \nabla^{e}{F_{a b}}+\frac{3}{8}(\Gamma^{a})^{\beta \alpha} \lambda_{i \beta} X^{i}_{\alpha} {W}^{-2} \nabla^{b}{F_{a b}}%
+\frac{3}{4}X_{i}^{\alpha} {W}^{-1} \nabla_{a}{\nabla^{a}{\lambda^{i}_{\alpha}}}-(\Sigma_{a b})^{\alpha \beta} X_{i \alpha} {W}^{-1} \nabla^{a}{\nabla^{b}{\lambda^{i}_{\beta}}}+\frac{27}{16}(\Gamma^{a})^{\beta \alpha} \lambda_{i \beta} X^{i}_{\alpha} {W}^{-1} \nabla^{b}{W_{a b}} - \frac{27}{16}(\Gamma^{a})^{\alpha \beta} W_{a b} X_{i \alpha} {W}^{-1} \nabla^{b}{\lambda^{i}_{\beta}} - \frac{1}{8}\Phi^{a b}\,_{j i} (\Sigma_{a b})^{\beta \alpha} \lambda^{j}_{\beta} X^{i}_{\alpha} {W}^{-1} - \frac{1}{6}\Phi^{a b}\,_{i j} \Phi_{a b}\,^{i j} - \frac{3}{32}\nabla_{a}{\nabla^{a}{Y}}+\frac{3}{4}{W}^{-1} \nabla^{a}{\lambda^{\alpha}_{i}} \nabla^{b}{W_{a b \alpha}\,^{i}} - \frac{53}{256}\epsilon_{e}\,^{a b c d} W_{a b} \lambda^{\alpha}_{i} {W}^{-1} \nabla^{e}{W_{c d \alpha}\,^{i}} - \frac{9}{128}(\Gamma_{c})^{\beta \alpha} W^{a b} \lambda_{i \beta} {W}^{-1} \nabla^{c}{W_{a b \alpha}\,^{i}} - \frac{9}{128}\epsilon^{e a b c d} (\Sigma_{e {e_{1}}})^{\beta \alpha} W_{a b} \lambda_{i \beta} {W}^{-1} \nabla^{{e_{1}}}{W_{c d \alpha}\,^{i}} - \frac{9}{64}\epsilon^{d e}\,_{{e_{1}}}\,^{a c} (\Sigma_{d e})^{\beta \alpha} W_{a}\,^{b} \lambda_{i \beta} {W}^{-1} \nabla^{{e_{1}}}{W_{c b \alpha}\,^{i}} - \frac{9}{64}(\Gamma^{c})^{\beta \alpha} W^{a}\,_{b} \lambda_{i \beta} {W}^{-1} \nabla^{b}{W_{c a \alpha}\,^{i}}+\frac{9}{64}(\Gamma^{a})^{\beta \alpha} W_{a}\,^{b} \lambda_{i \beta} {W}^{-1} \nabla^{c}{W_{b c \alpha}\,^{i}} - \frac{27}{128}\epsilon_{e}\,^{a b c d} F_{a b} \lambda^{\alpha}_{i} {W}^{-2} \nabla^{e}{W_{c d \alpha}\,^{i}} - \frac{1}{12}C_{a b c d} W^{a b} F^{c d} {W}^{-1} - \frac{5}{32}W^{c d} W_{c d} W_{a b} F^{a b} {W}^{-1} - \frac{1}{8}W^{c d} W_{c a} W_{d b} F^{a b} {W}^{-1}+\frac{1}{4}W_{a}\,^{c} W^{d}\,_{c} W_{d b} F^{a b} {W}^{-1} - \frac{1}{12}C_{a b c d} W^{c d} F^{a b} {W}^{-1}%
 - \frac{1}{6}C_{a b c d} W^{a c} F^{b d} {W}^{-1} - \frac{3}{4}{\rm i} F^{a b} W_{a}\,^{c \alpha}\,_{i} W_{b c \alpha}\,^{i} {W}^{-1} - \frac{3}{16}W_{a b} {W}^{-1} \nabla_{c}{W} \nabla^{c}{W^{a b}}+\frac{9}{2}W_{a b} {W}^{-1} \nabla_{c}{W} \nabla^{a}{W^{c b}}+\frac{9}{2}W_{a b} {W}^{-1} \nabla^{a}{W} \nabla_{c}{W^{c b}} - \frac{3}{128}W^{a b} W_{a b}\,^{\alpha}\,_{i} (\Gamma_{c})_{\alpha}{}^{\beta} {W}^{-1} \nabla^{c}{\lambda^{i}_{\beta}} - \frac{25}{256}\epsilon^{a b c d}\,_{e} W_{a b} W_{c d}\,^{\alpha}\,_{i} {W}^{-1} \nabla^{e}{\lambda^{i}_{\alpha}} - \frac{3}{128}\epsilon^{e a b c d} (\Sigma_{e {e_{1}}})^{\alpha \beta} W_{a b} W_{c d \alpha i} {W}^{-1} \nabla^{{e_{1}}}{\lambda^{i}_{\beta}} - \frac{3}{64}\epsilon^{d e a c}\,_{{e_{1}}} (\Sigma_{d e})^{\alpha \beta} W_{a}\,^{b} W_{c b \alpha i} {W}^{-1} \nabla^{{e_{1}}}{\lambda^{i}_{\beta}}+\frac{3}{64}(\Gamma^{c})^{\alpha \beta} W^{a}\,_{b} W_{c a \alpha i} {W}^{-1} \nabla^{b}{\lambda^{i}_{\beta}} - \frac{3}{64}(\Gamma^{a})^{\alpha \beta} W_{a}\,^{b} W_{b c \alpha i} {W}^{-1} \nabla^{c}{\lambda^{i}_{\beta}}+\frac{1}{2}{\rm i} \Phi_{a b i j} (\Gamma^{a})^{\alpha \beta} \lambda^{i}_{\alpha} {W}^{-2} \nabla^{b}{\lambda^{j}_{\beta}} - \frac{1}{6}W_{a b}\,^{\alpha}\,_{i} \Phi^{a b i}\,_{j} \lambda^{j}_{\alpha} {W}^{-1}+\frac{3}{32}W_{a b}\,^{\alpha}\,_{i} (\Gamma_{c})_{\alpha}{}^{\beta} \lambda^{i}_{\beta} {W}^{-1} \nabla^{c}{W^{a b}} - \frac{3}{64}W_{a b}\,^{\alpha}\,_{i} \epsilon^{a b}\,_{e}\,^{c d} \lambda^{i}_{\alpha} {W}^{-1} \nabla^{e}{W_{c d}}+\frac{1}{128}W_{a b}\,^{\alpha}\,_{i} \epsilon^{a b e c d} (\Sigma_{e {e_{1}}})_{\alpha}{}^{\beta} \lambda^{i}_{\beta} {W}^{-1} \nabla^{{e_{1}}}{W_{c d}} - \frac{1}{64}W_{a b}\,^{\alpha}\,_{i} \epsilon^{a d e}\,_{{e_{1}} c} (\Sigma_{d e})_{\alpha}{}^{\beta} \lambda^{i}_{\beta} {W}^{-1} \nabla^{{e_{1}}}{W^{b c}}+\frac{7}{32}W_{a b}\,^{\alpha}\,_{i} (\Gamma_{c})_{\alpha}{}^{\beta} \lambda^{i}_{\beta} {W}^{-1} \nabla^{a}{W^{b c}}+\frac{3}{16}W_{a b}\,^{\alpha}\,_{i} (\Gamma^{a})_{\alpha}{}^{\beta} \lambda^{i}_{\beta} {W}^{-1} \nabla_{c}{W^{b c}} - \frac{3}{4}\lambda^{\alpha}_{i} {W}^{-1} \nabla_{a}{\nabla^{a}{X^{i}_{\alpha}}}%
+\frac{3}{4}(\Sigma_{a b})^{\beta \alpha} \lambda_{i \beta} {W}^{-1} \nabla^{a}{\nabla^{b}{X^{i}_{\alpha}}} - \frac{1}{16}(\Sigma_{a b})^{\alpha \beta} W^{a b} W^{c d} W_{c d \alpha i} \lambda^{i}_{\beta} {W}^{-1}+\frac{1}{16}W^{a b} W_{a b} W^{c d \alpha}\,_{i} (\Sigma_{c d})_{\alpha}{}^{\beta} \lambda^{i}_{\beta} {W}^{-1} - \frac{1}{4}W^{a b} W_{a}\,^{c} W_{b}\,^{d \alpha}\,_{i} (\Sigma_{c d})_{\alpha}{}^{\beta} \lambda^{i}_{\beta} {W}^{-1}+\frac{1}{16}W^{a b} W^{c d} W_{a b}\,^{\alpha}\,_{i} (\Sigma_{c d})_{\alpha}{}^{\beta} \lambda^{i}_{\beta} {W}^{-1} - \frac{1}{32}\epsilon^{a b c e {e_{1}}} W_{a b} W_{c d} W_{e {e_{1}}}\,^{\alpha}\,_{i} (\Gamma^{d})_{\alpha}{}^{\beta} \lambda^{i}_{\beta} {W}^{-1}+\frac{1}{8}W^{a b} W_{a}\,^{c} W_{b c}\,^{\alpha}\,_{i} \lambda^{i}_{\alpha} {W}^{-1} - \frac{1}{16}\epsilon^{a b c e}\,_{{e_{1}}} (\Gamma^{{e_{1}}})^{\alpha \beta} W_{a b} W_{c}\,^{d} W_{e d \alpha i} \lambda^{i}_{\beta} {W}^{-1}+\frac{1}{4}(\Sigma^{c d})^{\alpha \beta} W^{a b} W_{a c} W_{b d \alpha i} \lambda^{i}_{\beta} {W}^{-1} - \frac{1}{4}(\Sigma_{a c})^{\alpha \beta} W^{a b} W^{c d} W_{b d \alpha i} \lambda^{i}_{\beta} {W}^{-1} - \frac{13}{64}W_{a b}\,^{\alpha}\,_{i} \epsilon^{a b e}\,_{{e_{1}} d} (\Sigma_{e c})_{\alpha}{}^{\beta} \lambda^{i}_{\beta} {W}^{-1} \nabla^{{e_{1}}}{W^{c d}} - \frac{13}{128}W_{a b}\,^{\alpha}\,_{i} \epsilon^{a e {e_{1}} c d} (\Sigma_{e {e_{1}}})_{\alpha}{}^{\beta} \lambda^{i}_{\beta} {W}^{-1} \nabla^{b}{W_{c d}} - \frac{3}{32}{\rm i} (\Gamma_{a})^{\alpha \beta} Y \lambda_{i \alpha} {W}^{-2} \nabla^{a}{\lambda^{i}_{\beta}}+\frac{3}{32}Y {W}^{-1} \nabla_{a}{\nabla^{a}{W}}+\frac{3}{32}{W}^{-1} \nabla_{a}{W} \nabla^{a}{Y} - \frac{27}{32}W^{a b} F_{a b} \lambda^{\alpha}_{i} X^{i}_{\alpha} {W}^{-2} - \frac{27}{64}\epsilon_{e}\,^{c d a b} (\Gamma^{e})^{\beta \alpha} W_{c d} F_{a b} \lambda_{i \beta} X^{i}_{\alpha} {W}^{-2}+\frac{3}{128}F^{a b} F_{a b} Y {W}^{-2}+\frac{3}{64}{\rm i} (\Sigma_{a b})^{\alpha \beta} F^{a b} Y \lambda_{i \alpha} \lambda^{i}_{\beta} {W}^{-3} - \frac{3}{4}(\Gamma_{a})^{\beta \alpha} X_{i j} \lambda^{i}_{\beta} {W}^{-2} \nabla^{a}{X^{j}_{\alpha}}%
 - \frac{9}{128}X_{i j} X^{i j} Y {W}^{-2} - \frac{3}{32}{\rm i} X_{i j} Y \lambda^{i \alpha} \lambda^{j}_{\alpha} {W}^{-3} - \frac{3}{16}\epsilon^{c d}\,_{e}\,^{a b} (\Sigma_{c d})^{\alpha \beta} F_{a b} X_{i \alpha} {W}^{-2} \nabla^{e}{\lambda^{i}_{\beta}}+\frac{3}{8}(\Gamma^{a})^{\alpha \beta} F_{a b} X_{i \alpha} {W}^{-2} \nabla^{b}{\lambda^{i}_{\beta}} - \frac{3}{8}\epsilon^{e a b c d} F_{a b} F_{c d} {W}^{-2} \nabla^{{e_{1}}}{W_{e {e_{1}}}} - \frac{75}{128}W^{c d} W_{c d} F^{a b} F_{a b} {W}^{-2} - \frac{1}{8}W^{a b} W^{c d} F_{a b} F_{c d} {W}^{-2} - \frac{1}{8}W^{a c} W^{b d} F_{a b} F_{c d} {W}^{-2} - \frac{3}{16}\epsilon_{e}\,^{a b c d} (\Gamma^{e})^{\beta \alpha} F_{a b} F_{c d} \lambda_{i \beta} X^{i}_{\alpha} {W}^{-3}+\frac{3}{4}(\Gamma_{a})^{\alpha \beta} X_{i j} X^{i}_{\alpha} {W}^{-2} \nabla^{a}{\lambda^{j}_{\beta}} - \frac{1}{2}\Phi_{a b i j} X^{i j} F^{a b} {W}^{-2}+\frac{3}{4}(\Sigma_{a b})^{\beta \alpha} X_{i j} F^{a b} \lambda^{i}_{\beta} X^{j}_{\alpha} {W}^{-3}+\frac{9}{16}\lambda^{\beta}_{i} \lambda_{j \beta} X^{i \alpha} X^{j}_{\alpha} {W}^{-2} - \frac{11}{16}\lambda^{\alpha}_{i} \lambda^{\beta}_{j} X^{i}_{\alpha} X^{j}_{\beta} {W}^{-2}+\frac{7}{16}\lambda^{\beta}_{i} \lambda^{\alpha}_{j} X^{i}_{\alpha} X^{j}_{\beta} {W}^{-2}+\frac{1}{8}\lambda^{\alpha}_{j} \lambda^{j \beta} X_{i \alpha} X^{i}_{\beta} {W}^{-2}+\frac{9}{128}X_{i j} X^{i j} W^{a b} W_{a b} {W}^{-2}+\frac{167}{384}X_{i j} X^{i j} \lambda^{\alpha}_{k} X^{k}_{\alpha} {W}^{-3} - \frac{1}{4}{\rm i} (\Sigma^{a}{}_{\, c})^{\alpha \beta} W_{a b} \lambda_{i \alpha} {W}^{-2} \nabla^{c}{\nabla^{b}{\lambda^{i}_{\beta}}}+\frac{3}{8}{\rm i} (\Sigma_{a b})^{\alpha \beta} W^{a b} \lambda_{i \alpha} {W}^{-2} \nabla_{c}{\nabla^{c}{\lambda^{i}_{\beta}}}%
+\frac{1}{16}{\rm i} \epsilon_{c d e}\,^{a b} (\Gamma^{c})^{\alpha \beta} W_{a b} \lambda_{i \alpha} {W}^{-2} \nabla^{d}{\nabla^{e}{\lambda^{i}_{\beta}}} - \frac{1}{4}{\rm i} W_{a b} \lambda^{\alpha}_{i} {W}^{-2} \nabla^{a}{\nabla^{b}{\lambda^{i}_{\alpha}}}+{\rm i} (\Sigma^{a}{}_{\, c})^{\alpha \beta} W_{a b} \lambda_{i \alpha} {W}^{-2} \nabla^{b}{\nabla^{c}{\lambda^{i}_{\beta}}}+\frac{1}{2}W_{a b} {W}^{-1} \nabla_{c}{\nabla^{c}{F^{a b}}}+2W_{c b} {W}^{-1} \nabla^{c}{\nabla_{a}{F^{a b}}} - \frac{3}{32}W^{a b} W_{a b} {W}^{-1} \nabla_{c}{\nabla^{c}{W}}+\frac{9}{2}W^{a}\,_{b} W_{a c} {W}^{-1} \nabla^{b}{\nabla^{c}{W}} - \frac{1}{2}\epsilon_{c d e}\,^{a b} {W}^{-1} \nabla^{c}{W} \nabla^{d}{\nabla^{e}{W_{a b}}}+\frac{3}{8}{\rm i} (\Sigma_{a b})^{\alpha \beta} \lambda_{i \alpha} \lambda^{i}_{\beta} {W}^{-2} \nabla_{c}{\nabla^{c}{W^{a b}}}+\frac{1}{4}{\rm i} (\Sigma_{c a})^{\alpha \beta} \lambda_{i \alpha} \lambda^{i}_{\beta} {W}^{-2} \nabla^{c}{\nabla_{b}{W^{a b}}}+\frac{1}{2}{\rm i} (\Sigma_{c a})^{\alpha \beta} \lambda_{i \alpha} \lambda^{i}_{\beta} {W}^{-2} \nabla_{b}{\nabla^{c}{W^{a b}}}+\frac{9}{4}{\rm i} (\Sigma_{a c})^{\alpha \beta} \lambda_{i \alpha} {W}^{-2} \nabla_{b}{W^{a b}} \nabla^{c}{\lambda^{i}_{\beta}}+\frac{3}{8}{\rm i} (\Sigma_{a b})^{\alpha \beta} \lambda_{i \alpha} {W}^{-2} \nabla_{c}{W^{a b}} \nabla^{c}{\lambda^{i}_{\beta}}+\frac{3}{16}{\rm i} \epsilon_{c d}\,^{a b}\,_{e} (\Gamma^{c})^{\alpha \beta} \lambda_{i \alpha} {W}^{-2} \nabla^{d}{W_{a b}} \nabla^{e}{\lambda^{i}_{\beta}}+\frac{3}{4}{\rm i} \lambda^{\alpha}_{i} {W}^{-2} \nabla^{a}{W_{a b}} \nabla^{b}{\lambda^{i}_{\alpha}} - \frac{9}{8}\epsilon_{c}\,^{a b}\,_{d e} {W}^{-1} \nabla^{c}{W_{a b}} \nabla^{d}{\nabla^{e}{W}}+\frac{1}{4}\lambda^{\alpha}_{i} {W}^{-1} \nabla^{a}{\nabla^{b}{W_{a b \alpha}\,^{i}}}-{\rm i} (\Sigma^{a}{}_{\, c})^{\alpha \beta} F_{a b} {W}^{-3} \nabla^{c}{\lambda_{i \alpha}} \nabla^{b}{\lambda^{i}_{\beta}}+\frac{1}{4}{\rm i} (\Sigma_{a b})^{\alpha \beta} F^{a b} {W}^{-3} \nabla_{c}{\lambda_{i \alpha}} \nabla^{c}{\lambda^{i}_{\beta}} - \frac{1}{8}{\rm i} \epsilon_{c d e}\,^{a b} (\Gamma^{c})^{\alpha \beta} F_{a b} {W}^{-3} \nabla^{d}{\lambda_{i \alpha}} \nabla^{e}{\lambda^{i}_{\beta}}%
+\frac{1}{4}{\rm i} F_{a b} {W}^{-3} \nabla^{a}{\lambda^{\alpha}_{i}} \nabla^{b}{\lambda^{i}_{\alpha}}+\frac{3}{16}{\rm i} \epsilon_{e}\,^{c d a b} W_{c d} F_{a b} \lambda^{\alpha}_{i} {W}^{-3} \nabla^{e}{\lambda^{i}_{\alpha}}+\frac{3}{8}{\rm i} (\Gamma_{c})^{\alpha \beta} W^{a b} F_{a b} \lambda_{i \alpha} {W}^{-3} \nabla^{c}{\lambda^{i}_{\beta}}+\frac{3}{4}{\rm i} (\Gamma^{c})^{\alpha \beta} W_{c}\,^{a} F_{a b} \lambda_{i \alpha} {W}^{-3} \nabla^{b}{\lambda^{i}_{\beta}}+\frac{3}{8}{\rm i} \epsilon^{e}\,_{{e_{1}}}\,^{c d a} (\Sigma_{e}{}^{\, b})^{\alpha \beta} W_{c d} F_{a b} \lambda_{i \alpha} {W}^{-3} \nabla^{{e_{1}}}{\lambda^{i}_{\beta}} - \frac{3}{16}{\rm i} \epsilon^{e {e_{1}} c a b} (\Sigma_{e {e_{1}}})^{\alpha \beta} W_{c d} F_{a b} \lambda_{i \alpha} {W}^{-3} \nabla^{d}{\lambda^{i}_{\beta}} - \frac{3}{4}{\rm i} \epsilon^{e c d a b} (\Sigma_{e {e_{1}}})^{\alpha \beta} W_{c d} F_{a b} \lambda_{i \alpha} {W}^{-3} \nabla^{{e_{1}}}{\lambda^{i}_{\beta}}+\frac{1}{16}\epsilon^{a b c d}\,_{e} F_{a b} W_{c d}\,^{\alpha}\,_{i} {W}^{-2} \nabla^{e}{\lambda^{i}_{\alpha}}+\frac{3}{32}{\rm i} \epsilon_{e}\,^{a b c d} F_{a b} F_{c d} \lambda^{\alpha}_{i} {W}^{-4} \nabla^{e}{\lambda^{i}_{\alpha}} - \frac{3}{16}{\rm i} (\Gamma_{c})^{\alpha \beta} F^{a b} F_{a b} \lambda_{i \alpha} {W}^{-4} \nabla^{c}{\lambda^{i}_{\beta}}+\frac{3}{4}{\rm i} (\Gamma^{a})^{\alpha \beta} F_{a}\,^{b} F_{b c} \lambda_{i \alpha} {W}^{-4} \nabla^{c}{\lambda^{i}_{\beta}} - \frac{3}{8}{\rm i} \epsilon^{e}\,_{{e_{1}}}\,^{a b c} (\Sigma_{e}{}^{\, d})^{\alpha \beta} F_{a b} F_{c d} \lambda_{i \alpha} {W}^{-4} \nabla^{{e_{1}}}{\lambda^{i}_{\beta}}+\frac{3}{16}{\rm i} \epsilon^{e {e_{1}} a b c} (\Sigma_{e {e_{1}}})^{\alpha \beta} F_{a b} F_{c d} \lambda_{i \alpha} {W}^{-4} \nabla^{d}{\lambda^{i}_{\beta}}+\frac{1}{4}F^{a b} F_{a b} {W}^{-3} \nabla_{c}{\nabla^{c}{W}}+F^{a}\,_{b} F_{a c} {W}^{-3} \nabla^{b}{\nabla^{c}{W}} - \frac{3}{4}W^{c d} F^{a b} F_{c d} F_{a b} {W}^{-3}+\frac{3}{2}W^{c d} F^{a b} F_{c a} F_{d b} {W}^{-3}+\frac{5}{2}W^{d c} W_{d a} F^{a b} F_{c b} {W}^{-2} - \frac{1}{2}{\rm i} (\Sigma_{a b})^{\alpha \beta} \lambda_{i \alpha} {W}^{-3} \nabla_{c}{F^{a b}} \nabla^{c}{\lambda^{i}_{\beta}}+\frac{1}{4}{\rm i} \lambda^{\alpha}_{i} {W}^{-3} \nabla^{a}{F_{a b}} \nabla^{b}{\lambda^{i}_{\alpha}}%
+\frac{1}{2}{\rm i} (\Sigma_{a c})^{\alpha \beta} \lambda_{i \alpha} {W}^{-3} \nabla_{b}{F^{a b}} \nabla^{c}{\lambda^{i}_{\beta}} - \frac{1}{32}{\rm i} \epsilon^{e}\,_{{e_{1}} b}\,^{c d} (\Sigma_{e a})^{\alpha \beta} W_{c d} \lambda_{i \alpha} \lambda^{i}_{\beta} {W}^{-3} \nabla^{{e_{1}}}{F^{a b}} - \frac{1}{64}{\rm i} \epsilon^{e {e_{1}} a b c} (\Sigma_{e {e_{1}}})^{\alpha \beta} W_{c d} \lambda_{i \alpha} \lambda^{i}_{\beta} {W}^{-3} \nabla^{d}{F_{a b}} - \frac{1}{8}\epsilon^{e a b c d} F_{a b} F_{c d} {W}^{-3} \nabla^{{e_{1}}}{F_{e {e_{1}}}} - \frac{3}{16}\epsilon^{c e {e_{1}} a b} W_{e {e_{1}}} F_{a b} {W}^{-2} \nabla^{d}{F_{c d}} - \frac{1}{8}\epsilon_{{e_{1}}}\,^{c d e a} W_{e}\,^{b} F_{a b} {W}^{-2} \nabla^{{e_{1}}}{F_{c d}}+F_{a b} {W}^{-3} \nabla_{c}{W} \nabla^{c}{F^{a b}}+F_{a b} {W}^{-3} \nabla^{a}{W} \nabla_{c}{F^{c b}}+\frac{3}{16}{\rm i} \epsilon^{e}\,_{{e_{1}} d}\,^{a b} (\Sigma_{e c})^{\alpha \beta} F_{a b} \lambda_{i \alpha} \lambda^{i}_{\beta} {W}^{-4} \nabla^{{e_{1}}}{F^{c d}}+\frac{3}{32}{\rm i} \epsilon^{e {e_{1}} c d a} (\Sigma_{e {e_{1}}})^{\alpha \beta} F_{a b} \lambda_{i \alpha} \lambda^{i}_{\beta} {W}^{-4} \nabla^{b}{F_{c d}} - \frac{1}{2}{\rm i} (\Sigma^{a}{}_{\, c})^{\alpha \beta} F_{a b} \lambda_{i \alpha} {W}^{-3} \nabla^{c}{\nabla^{b}{\lambda^{i}_{\beta}}} - \frac{1}{4}{\rm i} (\Sigma_{a b})^{\alpha \beta} F^{a b} \lambda_{i \alpha} {W}^{-3} \nabla_{c}{\nabla^{c}{\lambda^{i}_{\beta}}} - \frac{1}{2}{\rm i} (\Sigma^{a}{}_{\, c})^{\alpha \beta} F_{a b} \lambda_{i \alpha} {W}^{-3} \nabla^{b}{\nabla^{c}{\lambda^{i}_{\beta}}}+\frac{3}{8}{\rm i} \epsilon^{e}\,_{{e_{1}} d}\,^{a b} (\Sigma_{e c})^{\alpha \beta} F_{a b} \lambda_{i \alpha} \lambda^{i}_{\beta} {W}^{-3} \nabla^{{e_{1}}}{W^{c d}}+\frac{3}{16}{\rm i} \epsilon^{e {e_{1}} c d a} (\Sigma_{e {e_{1}}})^{\alpha \beta} F_{a b} \lambda_{i \alpha} \lambda^{i}_{\beta} {W}^{-3} \nabla^{b}{W_{c d}} - \frac{11}{32}{\rm i} (\Sigma_{c d})^{\alpha \beta} W^{c d} W^{a b} F_{a b} \lambda_{i \alpha} \lambda^{i}_{\beta} {W}^{-3} - \frac{15}{32}{\rm i} (\Sigma_{a b})^{\alpha \beta} W^{c d} W_{c d} F^{a b} \lambda_{i \alpha} \lambda^{i}_{\beta} {W}^{-3}+\frac{19}{8}{\rm i} (\Sigma^{d a})^{\alpha \beta} W^{c b} W_{d c} F_{a b} \lambda_{i \alpha} \lambda^{i}_{\beta} {W}^{-3}+\frac{1}{2}{\rm i} (\Sigma_{c d})^{\alpha \beta} W^{c a} W^{d b} F_{a b} \lambda_{i \alpha} \lambda^{i}_{\beta} {W}^{-3}+\frac{81}{128}W^{a c} F_{a}\,^{b} W_{c b}\,^{\alpha}\,_{i} \lambda^{i}_{\alpha} {W}^{-2}%
+\frac{1}{16}(\Sigma_{a b})^{\alpha \beta} F^{a b} F^{c d} W_{c d \alpha i} \lambda^{i}_{\beta} {W}^{-3} - \frac{1}{16}F^{a b} F_{a b} W^{c d \alpha}\,_{i} (\Sigma_{c d})_{\alpha}{}^{\beta} \lambda^{i}_{\beta} {W}^{-3}+\frac{1}{4}F^{a b} F_{a}\,^{c} W_{b}\,^{d \alpha}\,_{i} (\Sigma_{c d})_{\alpha}{}^{\beta} \lambda^{i}_{\beta} {W}^{-3} - \frac{1}{16}F^{a b} F^{c d} W_{a b}\,^{\alpha}\,_{i} (\Sigma_{c d})_{\alpha}{}^{\beta} \lambda^{i}_{\beta} {W}^{-3}+\frac{1}{32}\epsilon^{a b c e {e_{1}}} F_{a b} F_{c d} W_{e {e_{1}}}\,^{\alpha}\,_{i} (\Gamma^{d})_{\alpha}{}^{\beta} \lambda^{i}_{\beta} {W}^{-3} - \frac{1}{8}F^{a b} F_{a}\,^{c} W_{b c}\,^{\alpha}\,_{i} \lambda^{i}_{\alpha} {W}^{-3}+\frac{1}{16}\epsilon^{a b c e}\,_{{e_{1}}} (\Gamma^{{e_{1}}})^{\alpha \beta} F_{a b} F_{c}\,^{d} W_{e d \alpha i} \lambda^{i}_{\beta} {W}^{-3} - \frac{1}{4}(\Sigma^{c d})^{\alpha \beta} F^{a b} F_{a c} W_{b d \alpha i} \lambda^{i}_{\beta} {W}^{-3}+\frac{1}{4}(\Sigma_{a c})^{\alpha \beta} F^{a b} F^{c d} W_{b d \alpha i} \lambda^{i}_{\beta} {W}^{-3} - \frac{3}{8}{\rm i} \Phi_{a b i j} F^{a b} \lambda^{i \alpha} \lambda^{j}_{\alpha} {W}^{-3} - \frac{1}{16}{\rm i} \epsilon^{d e}\,_{c}\,^{a b} \Phi_{d e i j} (\Gamma^{c})^{\alpha \beta} F_{a b} \lambda^{i}_{\alpha} \lambda^{j}_{\beta} {W}^{-3}+\frac{9}{32}{\rm i} \epsilon^{e c d a b} (\Sigma_{e {e_{1}}})^{\alpha \beta} F_{a b} \lambda_{i \alpha} \lambda^{i}_{\beta} {W}^{-3} \nabla^{{e_{1}}}{W_{c d}} - \frac{9}{16}{\rm i} \epsilon^{d e}\,_{{e_{1}} c}\,^{a} (\Sigma_{d e})^{\alpha \beta} F_{a b} \lambda_{i \alpha} \lambda^{i}_{\beta} {W}^{-3} \nabla^{{e_{1}}}{W^{c b}}+\frac{25}{256}(\Sigma_{a b})^{\alpha \beta} W^{c d} F^{a b} W_{c d \alpha i} \lambda^{i}_{\beta} {W}^{-2}+\frac{27}{256}W^{a b} F_{a b} W^{c d \alpha}\,_{i} (\Sigma_{c d})_{\alpha}{}^{\beta} \lambda^{i}_{\beta} {W}^{-2} - \frac{27}{64}W^{a c} F_{a}\,^{b} W_{b}\,^{d \alpha}\,_{i} (\Sigma_{c d})_{\alpha}{}^{\beta} \lambda^{i}_{\beta} {W}^{-2}+\frac{27}{256}W^{c d} F^{a b} W_{a b}\,^{\alpha}\,_{i} (\Sigma_{c d})_{\alpha}{}^{\beta} \lambda^{i}_{\beta} {W}^{-2} - \frac{27}{512}\epsilon^{c d a e {e_{1}}} W_{c d} F_{a b} W_{e {e_{1}}}\,^{\alpha}\,_{i} (\Gamma^{b})_{\alpha}{}^{\beta} \lambda^{i}_{\beta} {W}^{-2} - \frac{27}{256}\epsilon^{c a b e}\,_{{e_{1}}} (\Gamma^{{e_{1}}})^{\alpha \beta} W_{c}\,^{d} F_{a b} W_{e d \alpha i} \lambda^{i}_{\beta} {W}^{-2}+\frac{27}{64}(\Sigma^{b d})^{\alpha \beta} W^{a c} F_{a b} W_{c d \alpha i} \lambda^{i}_{\beta} {W}^{-2}%
 - \frac{27}{64}(\Sigma_{c a})^{\alpha \beta} W^{c d} F^{a b} W_{d b \alpha i} \lambda^{i}_{\beta} {W}^{-2}+\frac{3}{32}{\rm i} \epsilon^{e c d a b} (\Sigma_{e {e_{1}}})^{\alpha \beta} F_{a b} \lambda_{i \alpha} \lambda^{i}_{\beta} {W}^{-4} \nabla^{{e_{1}}}{F_{c d}} - \frac{3}{16}{\rm i} \epsilon^{d e}\,_{{e_{1}} c}\,^{a} (\Sigma_{d e})^{\alpha \beta} F_{a b} \lambda_{i \alpha} \lambda^{i}_{\beta} {W}^{-4} \nabla^{{e_{1}}}{F^{c b}} - \frac{7}{64}(\Gamma_{c})^{\beta \alpha} F^{a b} \lambda_{i \beta} {W}^{-2} \nabla^{c}{W_{a b \alpha}\,^{i}} - \frac{7}{64}\epsilon^{e a b c d} (\Sigma_{e {e_{1}}})^{\beta \alpha} F_{a b} \lambda_{i \beta} {W}^{-2} \nabla^{{e_{1}}}{W_{c d \alpha}\,^{i}} - \frac{7}{32}\epsilon^{d e}\,_{{e_{1}}}\,^{a c} (\Sigma_{d e})^{\beta \alpha} F_{a}\,^{b} \lambda_{i \beta} {W}^{-2} \nabla^{{e_{1}}}{W_{c b \alpha}\,^{i}} - \frac{7}{32}(\Gamma^{c})^{\beta \alpha} F^{a}\,_{b} \lambda_{i \beta} {W}^{-2} \nabla^{b}{W_{c a \alpha}\,^{i}}+\frac{7}{32}(\Gamma^{a})^{\beta \alpha} F_{a}\,^{b} \lambda_{i \beta} {W}^{-2} \nabla^{c}{W_{b c \alpha}\,^{i}}+\frac{9}{64}{\rm i} \epsilon^{e c d a b} (\Sigma_{e {e_{1}}})^{\alpha \beta} W_{a b} \lambda_{i \alpha} \lambda^{i}_{\beta} {W}^{-2} \nabla^{{e_{1}}}{W_{c d}}+\frac{7}{64}{\rm i} \epsilon^{e a b c d} (\Sigma_{e {e_{1}}})^{\alpha \beta} W_{c d} \lambda_{i \alpha} \lambda^{i}_{\beta} {W}^{-3} \nabla^{{e_{1}}}{F_{a b}}+\frac{5}{32}{\rm i} \epsilon^{d e}\,_{{e_{1}} a}\,^{c} (\Sigma_{d e})^{\alpha \beta} W_{c b} \lambda_{i \alpha} \lambda^{i}_{\beta} {W}^{-3} \nabla^{{e_{1}}}{F^{a b}}+\frac{3}{16}\epsilon^{c d e {e_{1}} a} W_{e {e_{1}}} F_{a b} {W}^{-2} \nabla^{b}{F_{c d}}+\frac{3}{8}\epsilon_{{e_{1}} c}\,^{e a b} W_{e d} F_{a b} {W}^{-2} \nabla^{{e_{1}}}{F^{c d}} - \frac{9}{128}{\rm i} \epsilon_{e}\,^{a b c d} W_{a b} W_{c d} \lambda^{\alpha}_{i} {W}^{-2} \nabla^{e}{\lambda^{i}_{\alpha}}+\frac{15}{64}{\rm i} (\Gamma_{c})^{\alpha \beta} W^{a b} W_{a b} \lambda_{i \alpha} {W}^{-2} \nabla^{c}{\lambda^{i}_{\beta}}+\frac{9}{16}{\rm i} (\Gamma^{a})^{\alpha \beta} W_{a}\,^{b} W_{b c} \lambda_{i \alpha} {W}^{-2} \nabla^{c}{\lambda^{i}_{\beta}} - \frac{9}{32}{\rm i} \epsilon^{e}\,_{{e_{1}}}\,^{a b c} (\Sigma_{e}{}^{\, d})^{\alpha \beta} W_{a b} W_{c d} \lambda_{i \alpha} {W}^{-2} \nabla^{{e_{1}}}{\lambda^{i}_{\beta}}+\frac{9}{64}{\rm i} \epsilon^{e {e_{1}} a b c} (\Sigma_{e {e_{1}}})^{\alpha \beta} W_{a b} W_{c d} \lambda_{i \alpha} {W}^{-2} \nabla^{d}{\lambda^{i}_{\beta}}+\frac{9}{64}{\rm i} \epsilon^{e a b c d} (\Sigma_{e {e_{1}}})^{\alpha \beta} W_{a b} W_{c d} \lambda_{i \alpha} {W}^{-2} \nabla^{{e_{1}}}{\lambda^{i}_{\beta}} - \frac{3}{4}{\rm i} (\Sigma^{a}{}_{\, c})^{\alpha \beta} W_{a b} {W}^{-2} \nabla^{c}{\lambda_{i \alpha}} \nabla^{b}{\lambda^{i}_{\beta}}%
+\frac{3}{4}{\rm i} (\Sigma_{a b})^{\alpha \beta} W^{a b} {W}^{-2} \nabla_{c}{\lambda_{i \alpha}} \nabla^{c}{\lambda^{i}_{\beta}} - \frac{3}{16}{\rm i} \epsilon_{c d e}\,^{a b} (\Gamma^{c})^{\alpha \beta} W_{a b} {W}^{-2} \nabla^{d}{\lambda_{i \alpha}} \nabla^{e}{\lambda^{i}_{\beta}} - \frac{1}{2}W_{a b} {W}^{-2} \nabla_{c}{W} \nabla^{c}{F^{a b}}-2W_{c b} {W}^{-2} \nabla_{a}{W} \nabla^{c}{F^{a b}} - \frac{23}{512}\epsilon^{a b c d}\,_{e} W_{a b} W_{c d}\,^{\alpha}\,_{i} \lambda^{i}_{\alpha} {W}^{-2} \nabla^{e}{W} - \frac{81}{256}{\rm i} (\Sigma_{a b})^{\alpha \beta} W^{a b} W^{c d} W_{c d} \lambda_{i \alpha} \lambda^{i}_{\beta} {W}^{-2}+\frac{89}{64}{\rm i} (\Sigma^{c d})^{\alpha \beta} W^{a b} W_{c a} W_{d b} \lambda_{i \alpha} \lambda^{i}_{\beta} {W}^{-2}+\frac{1}{8}{\rm i} \Phi_{a b i j} W^{a b} \lambda^{i \alpha} \lambda^{j}_{\alpha} {W}^{-2} - \frac{1}{8}{\rm i} \epsilon^{d e}\,_{c}\,^{a b} \Phi_{d e i j} (\Gamma^{c})^{\alpha \beta} W_{a b} \lambda^{i}_{\alpha} \lambda^{j}_{\beta} {W}^{-2}+\frac{19}{256}W^{a b} W_{a b}\,^{\alpha}\,_{i} (\Gamma_{c})_{\alpha}{}^{\beta} \lambda^{i}_{\beta} {W}^{-2} \nabla^{c}{W}+\frac{19}{256}\epsilon^{e a b c d} (\Sigma_{e {e_{1}}})^{\alpha \beta} W_{a b} W_{c d \alpha i} \lambda^{i}_{\beta} {W}^{-2} \nabla^{{e_{1}}}{W}+\frac{19}{128}\epsilon^{d e a c}\,_{{e_{1}}} (\Sigma_{d e})^{\alpha \beta} W_{a}\,^{b} W_{c b \alpha i} \lambda^{i}_{\beta} {W}^{-2} \nabla^{{e_{1}}}{W} - \frac{19}{128}(\Gamma^{c})^{\alpha \beta} W^{a}\,_{b} W_{c a \alpha i} \lambda^{i}_{\beta} {W}^{-2} \nabla^{b}{W}+\frac{19}{128}(\Gamma^{a})^{\alpha \beta} W_{a}\,^{b} W_{b c \alpha i} \lambda^{i}_{\beta} {W}^{-2} \nabla^{c}{W}+\frac{1}{8}F^{a b} W_{a b}\,^{\alpha}\,_{i} (\Gamma_{c})_{\alpha}{}^{\beta} {W}^{-2} \nabla^{c}{\lambda^{i}_{\beta}}+\frac{1}{8}\epsilon^{e a b c d} (\Sigma_{e {e_{1}}})^{\alpha \beta} F_{a b} W_{c d \alpha i} {W}^{-2} \nabla^{{e_{1}}}{\lambda^{i}_{\beta}}+\frac{1}{4}\epsilon^{d e a c}\,_{{e_{1}}} (\Sigma_{d e})^{\alpha \beta} F_{a}\,^{b} W_{c b \alpha i} {W}^{-2} \nabla^{{e_{1}}}{\lambda^{i}_{\beta}} - \frac{1}{4}(\Gamma^{c})^{\alpha \beta} F^{a}\,_{b} W_{c a \alpha i} {W}^{-2} \nabla^{b}{\lambda^{i}_{\beta}}+\frac{1}{4}(\Gamma^{a})^{\alpha \beta} F_{a}\,^{b} W_{b c \alpha i} {W}^{-2} \nabla^{c}{\lambda^{i}_{\beta}} - \frac{3}{2}W^{a b} F_{a b} {W}^{-2} \nabla_{c}{\nabla^{c}{W}}%
-W^{a}\,_{c} F_{a b} {W}^{-2} \nabla^{c}{\nabla^{b}{W}}+W^{a}\,_{c} F_{a b} {W}^{-2} \nabla^{b}{\nabla^{c}{W}}+\frac{3}{4}\lambda^{\alpha}_{i} {W}^{-2} \nabla_{a}{W} \nabla^{a}{X^{i}_{\alpha}} - \frac{3}{64}Y {W}^{-2} \nabla_{a}{W} \nabla^{a}{W} - \frac{1}{4}{\rm i} \lambda^{\alpha}_{i} {W}^{-3} \nabla_{a}{X^{i}\,_{j}} \nabla^{a}{\lambda^{j}_{\alpha}} - \frac{1}{2}{\rm i} (\Sigma_{a b})^{\alpha \beta} \lambda_{i \alpha} {W}^{-3} \nabla^{a}{X^{i}\,_{j}} \nabla^{b}{\lambda^{j}_{\beta}}+\frac{3}{16}{\rm i} (\Gamma^{a})^{\alpha \beta} W_{a b} \lambda_{i \alpha} \lambda_{j \beta} {W}^{-3} \nabla^{b}{X^{i j}} - \frac{3}{8}{\rm i} (\Gamma^{a})^{\alpha \beta} F_{a b} \lambda_{i \alpha} \lambda_{j \beta} {W}^{-4} \nabla^{b}{X^{i j}}+\frac{1}{4}{\rm i} X_{i j} {W}^{-3} \nabla_{a}{\lambda^{i \alpha}} \nabla^{a}{\lambda^{j}_{\alpha}}+\frac{1}{2}{\rm i} (\Sigma_{a b})^{\alpha \beta} X_{i j} {W}^{-3} \nabla^{a}{\lambda^{i}_{\alpha}} \nabla^{b}{\lambda^{j}_{\beta}}+\frac{3}{8}{\rm i} \epsilon^{c d}\,_{e}\,^{a b} (\Sigma_{c d})^{\alpha \beta} X_{i j} W_{a b} \lambda^{i}_{\alpha} {W}^{-3} \nabla^{e}{\lambda^{j}_{\beta}}+\frac{3}{8}{\rm i} (\Gamma^{a})^{\alpha \beta} X_{i j} W_{a b} \lambda^{i}_{\alpha} {W}^{-3} \nabla^{b}{\lambda^{j}_{\beta}}+\frac{3}{8}{\rm i} \epsilon^{c d}\,_{e}\,^{a b} (\Sigma_{c d})^{\alpha \beta} X_{i j} F_{a b} \lambda^{i}_{\alpha} {W}^{-4} \nabla^{e}{\lambda^{j}_{\beta}} - \frac{1}{4}X_{i j} F^{a b} W_{a b}\,^{\alpha i} \lambda^{j}_{\alpha} {W}^{-3} - \frac{1}{4}X_{i j} W^{a b} W_{a b}\,^{\alpha i} \lambda^{j}_{\alpha} {W}^{-2}+\frac{3}{8}{\rm i} (\Gamma^{a})^{\alpha \beta} X_{i j} \lambda^{i}_{\alpha} \lambda^{j}_{\beta} {W}^{-4} \nabla^{b}{F_{a b}}+\frac{3}{4}{\rm i} X_{i j} \lambda^{i \alpha} {W}^{-3} \nabla_{a}{\nabla^{a}{\lambda^{j}_{\alpha}}}+\frac{15}{16}{\rm i} (\Gamma^{a})^{\alpha \beta} X_{i j} \lambda^{i}_{\alpha} \lambda^{j}_{\beta} {W}^{-3} \nabla^{b}{W_{a b}} - \frac{3}{64}{\rm i} X_{i j} W^{a b} W_{a b} \lambda^{i \alpha} \lambda^{j}_{\alpha} {W}^{-3} - \frac{9}{128}{\rm i} \epsilon_{e}\,^{a b c d} (\Gamma^{e})^{\alpha \beta} X_{i j} W_{a b} W_{c d} \lambda^{i}_{\alpha} \lambda^{j}_{\beta} {W}^{-3}%
 - \frac{11}{192}{\rm i} \Phi^{a b}\,_{i k} (\Sigma_{a b})^{\alpha \beta} X^{i}\,_{j} \lambda^{k}_{\alpha} \lambda^{j}_{\beta} {W}^{-3}+{\rm i} (\Gamma_{a})^{\alpha \beta} \lambda^{\rho}_{i} \lambda_{j \rho} X^{i}_{\alpha} {W}^{-3} \nabla^{a}{\lambda^{j}_{\beta}}+\frac{3}{4}\lambda^{\alpha}_{i} X^{i}_{\alpha} {W}^{-2} \nabla_{a}{\nabla^{a}{W}}+\frac{7}{8}(\Sigma_{a b})^{\beta \alpha} \lambda_{i \beta} X^{i}_{\alpha} {W}^{-2} \nabla^{a}{\nabla^{b}{W}} - \frac{31}{384}W^{a b \alpha}\,_{i} (\Sigma_{a b})^{\beta \rho} \lambda_{j \beta} \lambda^{j}_{\rho} X^{i}_{\alpha} {W}^{-2} - \frac{7}{192}W^{a b \alpha}\,_{i} (\Sigma_{a b})^{\beta \rho} \lambda^{i}_{\beta} \lambda_{j \rho} X^{j}_{\alpha} {W}^{-2}+\frac{3}{8}{\rm i} (\Gamma_{a})^{\beta \alpha} \lambda_{j \beta} \lambda^{j \rho} \lambda_{i \rho} {W}^{-3} \nabla^{a}{X^{i}_{\alpha}}+\frac{5}{6}{\rm i} (\Gamma_{a})^{\beta \rho} \lambda_{j \beta} \lambda^{\alpha}_{i} X^{i}_{\alpha} {W}^{-3} \nabla^{a}{\lambda^{j}_{\rho}} - \frac{3}{4}X_{i}^{\alpha} {W}^{-2} \nabla_{a}{W} \nabla^{a}{\lambda^{i}_{\alpha}} - \frac{11}{32}{\rm i} (\Gamma_{a})^{\beta \alpha} \lambda_{i \beta} \lambda^{\rho}_{j} X^{i}_{\alpha} {W}^{-3} \nabla^{a}{\lambda^{j}_{\rho}}+\frac{5}{24}{\rm i} (\Gamma_{a})^{\beta \rho} \lambda_{j \beta} \lambda^{j \alpha} X_{i \alpha} {W}^{-3} \nabla^{a}{\lambda^{i}_{\rho}}+\frac{3}{32}{\rm i} (\Gamma_{a})^{\beta \alpha} \lambda_{j \beta} \lambda^{j \rho} X_{i \alpha} {W}^{-3} \nabla^{a}{\lambda^{i}_{\rho}}+\frac{1}{6}C_{a b c d} F^{a b} F^{c d} {W}^{-2}+\frac{1}{8}W_{a b} W_{c d} F^{a b} F^{c d} {W}^{-2}+\frac{1}{8}W_{a c} W_{b d} F^{a b} F^{c d} {W}^{-2} - \frac{1}{4}W^{d}\,_{b} W_{d c} F_{a}\,^{b} F^{a c} {W}^{-2}+\frac{1}{6}C_{a b c d} F^{a c} F^{b d} {W}^{-2}+\frac{1}{24}{\rm i} (\Sigma^{c d})^{\alpha \beta} C_{a b c d} W^{a b} \lambda_{i \alpha} \lambda^{i}_{\beta} {W}^{-2} - \frac{1}{8}{\rm i} (\Sigma_{a c})^{\alpha \beta} W^{a b} W^{c d} W_{b d} \lambda_{i \alpha} \lambda^{i}_{\beta} {W}^{-2}+\frac{1}{24}{\rm i} (\Sigma^{a b})^{\alpha \beta} C_{a b c d} W^{c d} \lambda_{i \alpha} \lambda^{i}_{\beta} {W}^{-2}%
+\frac{1}{12}{\rm i} (\Sigma^{a c})^{\alpha \beta} C_{a b c d} W^{b d} \lambda_{i \alpha} \lambda^{i}_{\beta} {W}^{-2}+\frac{1}{24}{\rm i} (\Sigma^{c d})^{\alpha \beta} C_{a b c d} F^{a b} \lambda_{i \alpha} \lambda^{i}_{\beta} {W}^{-3}+\frac{1}{16}{\rm i} (\Sigma_{c d})^{\alpha \beta} W^{c d} W_{a b} F^{a b} \lambda_{i \alpha} \lambda^{i}_{\beta} {W}^{-3}+\frac{1}{16}{\rm i} (\Sigma^{c d})^{\alpha \beta} W_{c a} W_{d b} F^{a b} \lambda_{i \alpha} \lambda^{i}_{\beta} {W}^{-3} - \frac{1}{8}{\rm i} (\Sigma_{c a})^{\alpha \beta} W^{c d} W_{d b} F^{a b} \lambda_{i \alpha} \lambda^{i}_{\beta} {W}^{-3}+\frac{1}{24}{\rm i} (\Sigma^{a b})^{\alpha \beta} C_{a b c d} F^{c d} \lambda_{i \alpha} \lambda^{i}_{\beta} {W}^{-3}+\frac{1}{12}{\rm i} (\Sigma^{a c})^{\alpha \beta} C_{a b c d} F^{b d} \lambda_{i \alpha} \lambda^{i}_{\beta} {W}^{-3} - \frac{9}{16}{\rm i} (\Sigma_{a b})^{\beta \alpha} W^{a b} \lambda_{j \beta} \lambda^{j \rho} \lambda_{i \rho} X^{i}_{\alpha} {W}^{-3} - \frac{9}{256}Y \lambda^{\alpha}_{i} \lambda^{i \beta} \lambda_{j \alpha} \lambda^{j}_{\beta} {W}^{-4} - \frac{1}{4}{\rm i} (\Gamma_{a})^{\alpha \beta} \lambda_{i \alpha} {W}^{-3} \nabla_{b}{\lambda^{i}_{\beta}} \nabla^{a}{\nabla^{b}{W}} - \frac{1}{4}{\rm i} \epsilon^{a b}\,_{c d e} (\Sigma_{a b})^{\alpha \beta} \lambda_{i \alpha} {W}^{-3} \nabla^{c}{\lambda^{i}_{\beta}} \nabla^{d}{\nabla^{e}{W}} - \frac{1}{4}{\rm i} (\Gamma_{a})^{\alpha \beta} \lambda_{i \alpha} {W}^{-3} \nabla_{b}{\lambda^{i}_{\beta}} \nabla^{b}{\nabla^{a}{W}} - \frac{5}{4}{\rm i} (\Gamma_{a})^{\alpha \beta} \lambda_{i \alpha} {W}^{-3} \nabla^{a}{\lambda^{i}_{\beta}} \nabla_{b}{\nabla^{b}{W}}+\frac{7}{8}{\rm i} (\Sigma^{a}{}_{\, c})^{\alpha \beta} W_{a b} \lambda_{i \alpha} \lambda^{i}_{\beta} {W}^{-3} \nabla^{c}{\nabla^{b}{W}} - \frac{9}{16}{\rm i} (\Sigma_{a b})^{\alpha \beta} W^{a b} \lambda_{i \alpha} \lambda^{i}_{\beta} {W}^{-3} \nabla_{c}{\nabla^{c}{W}} - \frac{1}{8}{\rm i} (\Sigma^{a}{}_{\, c})^{\alpha \beta} W_{a b} \lambda_{i \alpha} \lambda^{i}_{\beta} {W}^{-3} \nabla^{b}{\nabla^{c}{W}}+\frac{1}{2}\epsilon_{c d e}\,^{a b} F_{a b} {W}^{-3} \nabla^{c}{W} \nabla^{d}{\nabla^{e}{W}}+\frac{3}{4}{\rm i} (\Sigma^{a}{}_{\, c})^{\alpha \beta} F_{a b} \lambda_{i \alpha} \lambda^{i}_{\beta} {W}^{-4} \nabla^{c}{\nabla^{b}{W}}+\frac{3}{8}{\rm i} (\Sigma_{a b})^{\alpha \beta} F^{a b} \lambda_{i \alpha} \lambda^{i}_{\beta} {W}^{-4} \nabla_{c}{\nabla^{c}{W}}+\frac{3}{4}{\rm i} (\Sigma^{a}{}_{\, c})^{\alpha \beta} F_{a b} \lambda_{i \alpha} \lambda^{i}_{\beta} {W}^{-4} \nabla^{b}{\nabla^{c}{W}}%
+\frac{3}{4}\epsilon_{c d e}\,^{a b} W_{a b} {W}^{-2} \nabla^{c}{W} \nabla^{d}{\nabla^{e}{W}}+\frac{3}{4}{\rm i} (\Sigma^{a}{}_{\, c})^{\alpha \beta} W_{a b} \lambda_{i \alpha} {W}^{-3} \nabla^{c}{W} \nabla^{b}{\lambda^{i}_{\beta}} - \frac{9}{8}{\rm i} (\Sigma_{a b})^{\alpha \beta} W^{a b} \lambda_{i \alpha} {W}^{-3} \nabla_{c}{W} \nabla^{c}{\lambda^{i}_{\beta}}+\frac{3}{4}{\rm i} (\Sigma^{a}{}_{\, c})^{\alpha \beta} W_{a b} \lambda_{i \alpha} {W}^{-3} \nabla^{b}{W} \nabla^{c}{\lambda^{i}_{\beta}}+\frac{3}{64}W^{a b} W_{a b} {W}^{-2} \nabla_{c}{W} \nabla^{c}{W} - \frac{9}{4}W^{a}\,_{b} W_{a c} {W}^{-2} \nabla^{b}{W} \nabla^{c}{W}+\frac{3}{4}{\rm i} (\Sigma_{a b})^{\alpha \beta} \lambda_{i \alpha} \lambda^{i}_{\beta} {W}^{-4} \nabla_{c}{W} \nabla^{c}{F^{a b}} - \frac{3}{4}{\rm i} (\Sigma_{c a})^{\alpha \beta} \lambda_{i \alpha} \lambda^{i}_{\beta} {W}^{-4} \nabla^{c}{W} \nabla_{b}{F^{a b}} - \frac{1}{4}{\rm i} (\Gamma_{a})^{\alpha \beta} \lambda_{i \alpha} {W}^{-3} \nabla_{b}{W} \nabla^{a}{\nabla^{b}{\lambda^{i}_{\beta}}} - \frac{5}{4}{\rm i} (\Gamma_{a})^{\alpha \beta} \lambda_{i \alpha} {W}^{-3} \nabla_{b}{W} \nabla^{b}{\nabla^{a}{\lambda^{i}_{\beta}}} - \frac{1}{4}{\rm i} (\Gamma_{a})^{\alpha \beta} \lambda_{i \alpha} {W}^{-3} \nabla^{a}{W} \nabla_{b}{\nabla^{b}{\lambda^{i}_{\beta}}} - \frac{9}{16}{\rm i} (\Sigma_{a b})^{\alpha \beta} \lambda_{i \alpha} \lambda^{i}_{\beta} {W}^{-3} \nabla_{c}{W} \nabla^{c}{W^{a b}} - \frac{3}{8}{\rm i} (\Sigma_{c a})^{\alpha \beta} \lambda_{i \alpha} \lambda^{i}_{\beta} {W}^{-3} \nabla^{c}{W} \nabla_{b}{W^{a b}} - \frac{3}{8}{\rm i} (\Sigma_{c a})^{\alpha \beta} \lambda_{i \alpha} \lambda^{i}_{\beta} {W}^{-3} \nabla_{b}{W} \nabla^{c}{W^{a b}}+\frac{3}{2}{\rm i} (\Sigma^{a}{}_{\, c})^{\alpha \beta} F_{a b} \lambda_{i \alpha} {W}^{-4} \nabla^{c}{W} \nabla^{b}{\lambda^{i}_{\beta}}+\frac{3}{4}{\rm i} (\Sigma_{a b})^{\alpha \beta} F^{a b} \lambda_{i \alpha} {W}^{-4} \nabla_{c}{W} \nabla^{c}{\lambda^{i}_{\beta}}+\frac{3}{2}{\rm i} (\Sigma^{a}{}_{\, c})^{\alpha \beta} F_{a b} \lambda_{i \alpha} {W}^{-4} \nabla^{b}{W} \nabla^{c}{\lambda^{i}_{\beta}}+\frac{3}{2}W^{a b} F_{a b} {W}^{-3} \nabla_{c}{W} \nabla^{c}{W}+{\rm i} (\Gamma_{a})^{\alpha \beta} {W}^{-3} \nabla_{b}{W} \nabla^{a}{\lambda_{i \alpha}} \nabla^{b}{\lambda^{i}_{\beta}} - \frac{3}{2}F_{a b} {W}^{-2} \nabla_{c}{W} \nabla^{c}{W^{a b}}%
+\frac{1}{8}\lambda^{\alpha}_{i} {W}^{-2} \nabla^{a}{W} \nabla^{b}{W_{a b \alpha}\,^{i}} - \frac{3}{8}F^{a b} F_{a b} {W}^{-4} \nabla_{c}{W} \nabla^{c}{W} - \frac{3}{2}F^{a}\,_{b} F_{a c} {W}^{-4} \nabla^{b}{W} \nabla^{c}{W} - \frac{27}{32}{\rm i} (\Sigma_{a b})^{\alpha \beta} W^{c d} F^{a b} F_{c d} \lambda_{i \alpha} \lambda^{i}_{\beta} {W}^{-4}+\frac{27}{8}{\rm i} (\Sigma_{d a})^{\alpha \beta} W^{d c} F^{a b} F_{c b} \lambda_{i \alpha} \lambda^{i}_{\beta} {W}^{-4} - \frac{45}{64}{\rm i} (\Sigma_{c d})^{\alpha \beta} W^{c d} F^{a b} F_{a b} \lambda_{i \alpha} \lambda^{i}_{\beta} {W}^{-4}+\frac{9}{16}{\rm i} (\Sigma^{a c})^{\alpha \beta} W^{b d} F_{a b} F_{c d} \lambda_{i \alpha} \lambda^{i}_{\beta} {W}^{-4} - \frac{3}{8}{\rm i} \epsilon^{e a b c d} (\Sigma_{e {e_{1}}})^{\alpha \beta} F_{a b} F_{c d} \lambda_{i \alpha} {W}^{-4} \nabla^{{e_{1}}}{\lambda^{i}_{\beta}}+\frac{1}{32}{\rm i} W_{a b}\,^{\alpha}\,_{i} \epsilon^{a b c d}\,_{e} (\Sigma_{c d})^{\beta \rho} \lambda^{i}_{\beta} \lambda_{j \alpha} {W}^{-3} \nabla^{e}{\lambda^{j}_{\rho}} - \frac{1}{16}{\rm i} W_{a b}\,^{\alpha}\,_{i} (\Gamma^{a})^{\beta \rho} \lambda^{i}_{\beta} \lambda_{j \alpha} {W}^{-3} \nabla^{b}{\lambda^{j}_{\rho}} - \frac{1}{2}W_{a b}\,^{\alpha}\,_{i} \lambda^{i}_{\alpha} {W}^{-2} \nabla^{a}{\nabla^{b}{W}}+\frac{1}{24}W^{a b \alpha}\,_{i} W_{a}\,^{c}\,_{\alpha}\,^{i} (\Sigma_{b c})^{\beta \rho} \lambda_{j \beta} \lambda^{j}_{\rho} {W}^{-2}+\frac{1}{32}W^{a b \alpha}\,_{i} W_{a b \alpha j} \lambda^{i \beta} \lambda^{j}_{\beta} {W}^{-2} - \frac{1}{64}W_{a b}\,^{\alpha}\,_{i} W_{c d \alpha j} \epsilon^{a b c d}\,_{e} (\Gamma^{e})^{\beta \rho} \lambda^{i}_{\beta} \lambda^{j}_{\rho} {W}^{-2}+\frac{1}{24}W^{a b \alpha}\,_{i} W_{a}\,^{c}\,_{\alpha j} (\Sigma_{b c})^{\beta \rho} \lambda^{i}_{\beta} \lambda^{j}_{\rho} {W}^{-2}+\frac{3}{64}{\rm i} \epsilon^{c d}\,_{e}\,^{a b} (\Sigma_{c d})^{\beta \rho} \lambda_{j \beta} \lambda^{j}_{\rho} \lambda^{\alpha}_{i} {W}^{-3} \nabla^{e}{W_{a b \alpha}\,^{i}} - \frac{3}{64}{\rm i} W_{a b}\,^{\alpha}\,_{i} \epsilon^{a b c d}\,_{e} (\Sigma_{c d})^{\beta \rho} \lambda_{j \alpha} \lambda^{j}_{\beta} {W}^{-3} \nabla^{e}{\lambda^{i}_{\rho}}+\frac{3}{32}{\rm i} W_{a b}\,^{\alpha}\,_{i} (\Gamma^{a})^{\beta \rho} \lambda_{j \alpha} \lambda^{j}_{\beta} {W}^{-3} \nabla^{b}{\lambda^{i}_{\rho}} - \frac{3}{64}{\rm i} \epsilon^{c d}\,_{e}\,^{a b} (\Sigma_{c d})^{\beta \rho} \lambda_{j \beta} \lambda^{j \alpha} \lambda_{i \rho} {W}^{-3} \nabla^{e}{W_{a b \alpha}\,^{i}}+\frac{3}{32}{\rm i} (\Gamma^{a})^{\beta \rho} \lambda_{j \beta} \lambda^{j \alpha} \lambda_{i \rho} {W}^{-3} \nabla^{b}{W_{a b \alpha}\,^{i}}%
 - \frac{9}{4}{\rm i} X_{i j} \lambda^{i \alpha} {W}^{-4} \nabla_{a}{W} \nabla^{a}{\lambda^{j}_{\alpha}} - \frac{9}{32}{\rm i} X_{i j} W^{a b} F_{a b} \lambda^{i \alpha} \lambda^{j}_{\alpha} {W}^{-4} - \frac{27}{64}{\rm i} \epsilon_{e}\,^{c d a b} (\Gamma^{e})^{\alpha \beta} X_{i j} W_{c d} F_{a b} \lambda^{i}_{\alpha} \lambda^{j}_{\beta} {W}^{-4}+\frac{77}{128}{\rm i} (\Sigma_{a b})^{\beta \rho} W^{a b} \lambda_{j \beta} \lambda^{j \alpha} \lambda_{i \rho} X^{i}_{\alpha} {W}^{-3} - \frac{139}{256}{\rm i} (\Sigma_{a b})^{\beta \rho} W^{a b} \lambda_{j \beta} \lambda^{j}_{\rho} \lambda^{\alpha}_{i} X^{i}_{\alpha} {W}^{-3} - \frac{3}{4}\lambda^{\alpha}_{i} X^{i}_{\alpha} {W}^{-3} \nabla_{a}{W} \nabla^{a}{W}+\frac{7}{32}{\rm i} (\Gamma_{a})^{\beta \alpha} \lambda_{j \beta} \lambda^{\rho}_{i} X^{i}_{\alpha} {W}^{-3} \nabla^{a}{\lambda^{j}_{\rho}} - \frac{1}{8}{\rm i} (\Gamma_{a})^{\beta \rho} \lambda_{i \beta} \lambda_{j \rho} X^{i \alpha} {W}^{-3} \nabla^{a}{\lambda^{j}_{\alpha}} - \frac{1}{2}W_{a b}\,^{\alpha}\,_{i} {W}^{-2} \nabla^{a}{W} \nabla^{b}{\lambda^{i}_{\alpha}} - \frac{1}{16}(\Gamma_{a})^{\alpha \beta} (\Gamma_{b})^{\rho \lambda} \lambda_{i \alpha} \lambda_{j \beta} {W}^{-4} \nabla^{a}{\lambda^{i}_{\rho}} \nabla^{b}{\lambda^{j}_{\lambda}}+\frac{5}{12}{\rm i} (\Gamma_{a})^{\beta \rho} \lambda_{i \beta} \lambda^{\alpha}_{j} X^{i}_{\alpha} {W}^{-3} \nabla^{a}{\lambda^{j}_{\rho}}+\frac{1}{48}{\rm i} W_{a b}\,^{\alpha}\,_{i} \epsilon^{a b c d}\,_{e} (\Sigma_{c d})^{\beta \rho} \lambda_{j \beta} \lambda^{j}_{\rho} {W}^{-3} \nabla^{e}{\lambda^{i}_{\alpha}}+\frac{21}{32}(\Gamma_{a})^{\alpha \beta} (\Gamma_{b})^{\rho \lambda} \lambda_{i \alpha} \lambda_{j \rho} {W}^{-4} \nabla^{a}{\lambda^{i}_{\beta}} \nabla^{b}{\lambda^{j}_{\lambda}} - \frac{1}{32}(\Gamma_{a})^{\alpha \beta} (\Gamma_{b})^{\rho \lambda} \lambda_{i \alpha} \lambda_{j \rho} {W}^{-4} \nabla^{a}{\lambda^{j}_{\lambda}} \nabla^{b}{\lambda^{i}_{\beta}} - \frac{1}{48}{\rm i} W_{a b}\,^{\alpha}\,_{i} \epsilon^{a b c d}\,_{e} (\Sigma_{c d})^{\beta \rho} \lambda^{i}_{\beta} \lambda_{j \rho} {W}^{-3} \nabla^{e}{\lambda^{j}_{\alpha}}+\frac{1}{8}{\rm i} W_{a b}\,^{\alpha}\,_{i} (\Gamma^{a})^{\beta \rho} \lambda^{i}_{\beta} \lambda_{j \rho} {W}^{-3} \nabla^{b}{\lambda^{j}_{\alpha}}+\frac{7}{48}{\rm i} W^{a b} W_{a b}\,^{\alpha}\,_{i} \lambda^{i \beta} \lambda_{j \alpha} \lambda^{j}_{\beta} {W}^{-3}+\frac{83}{192}{\rm i} (\Sigma_{a}{}^{\, c})^{\beta \rho} W^{a b} W_{b c}\,^{\alpha}\,_{i} \lambda^{i}_{\alpha} \lambda_{j \beta} \lambda^{j}_{\rho} {W}^{-3} - \frac{1}{384}{\rm i} \epsilon^{a b c d}\,_{e} (\Gamma^{e})^{\beta \rho} W_{a b} W_{c d}\,^{\alpha}\,_{i} \lambda^{i}_{\beta} \lambda_{j \rho} \lambda^{j}_{\alpha} {W}^{-3}+\frac{1}{48}{\rm i} (\Sigma_{a}{}^{\, c})^{\beta \rho} W^{a b} W_{b c}\,^{\alpha}\,_{i} \lambda^{i}_{\beta} \lambda_{j \rho} \lambda^{j}_{\alpha} {W}^{-3}%
+\frac{1}{16}(\Gamma_{a})^{\alpha \beta} (\Gamma_{b})^{\rho \lambda} \lambda_{i \alpha} \lambda^{i}_{\rho} \lambda_{j \beta} {W}^{-4} \nabla^{b}{\nabla^{a}{\lambda^{j}_{\lambda}}}+\frac{3}{16}(\Gamma_{a})^{\alpha \beta} (\Gamma_{b})^{\rho \lambda} \lambda_{i \alpha} \lambda^{i}_{\rho} \lambda_{j \beta} {W}^{-4} \nabla^{a}{\nabla^{b}{\lambda^{j}_{\lambda}}}+\frac{1}{16}(\Gamma_{a})^{\alpha \beta} (\Gamma_{b})^{\rho \lambda} \lambda_{i \alpha} \lambda^{i}_{\rho} {W}^{-4} \nabla^{a}{\lambda_{j \beta}} \nabla^{b}{\lambda^{j}_{\lambda}}+\frac{1}{16}(\Gamma_{a})^{\alpha \beta} (\Gamma_{b})^{\rho \lambda} \lambda_{i \alpha} \lambda^{i}_{\rho} {W}^{-4} \nabla^{a}{\lambda_{j \lambda}} \nabla^{b}{\lambda^{j}_{\beta}} - \frac{1}{8}\lambda^{\alpha}_{i} \lambda^{\beta}_{j} {W}^{-4} \nabla_{a}{\lambda^{i}_{\alpha}} \nabla^{a}{\lambda^{j}_{\beta}} - \frac{3}{8}(\Sigma_{a b})^{\alpha \beta} \lambda_{i \alpha} \lambda^{\rho}_{j} {W}^{-4} \nabla^{a}{\lambda^{i}_{\beta}} \nabla^{b}{\lambda^{j}_{\rho}}+\frac{1}{8}(\Sigma_{a b})^{\alpha \beta} \lambda_{i \alpha} \lambda^{\rho}_{j} {W}^{-4} \nabla^{a}{\lambda^{i}_{\rho}} \nabla^{b}{\lambda^{j}_{\beta}}+\frac{1}{2}\lambda^{\alpha}_{i} \lambda^{i \beta} \lambda_{j \alpha} {W}^{-4} \nabla_{a}{\nabla^{a}{\lambda^{j}_{\beta}}} - \frac{1}{8}(\Sigma_{a b})^{\alpha \beta} \lambda_{i \alpha} \lambda^{i \rho} \lambda_{j \rho} {W}^{-4} \nabla^{a}{\nabla^{b}{\lambda^{j}_{\beta}}}+\frac{1}{16}\lambda^{\alpha}_{i} \lambda^{i \beta} {W}^{-4} \nabla_{a}{\lambda_{j \alpha}} \nabla^{a}{\lambda^{j}_{\beta}} - \frac{1}{8}(\Sigma_{a b})^{\alpha \beta} \lambda_{i \alpha} \lambda^{i \rho} {W}^{-4} \nabla^{a}{\lambda_{j \beta}} \nabla^{b}{\lambda^{j}_{\rho}}+\frac{27}{256}\epsilon^{c d}\,_{e}\,^{a b} (\Sigma_{c d})^{\alpha \beta} W_{a b} \lambda_{i \alpha} \lambda^{i}_{\beta} \lambda^{\rho}_{j} {W}^{-4} \nabla^{e}{\lambda^{j}_{\rho}}+\frac{1}{4}\epsilon^{c d}\,_{e}\,^{a b} (\Sigma_{c d})^{\alpha \beta} F_{a b} \lambda_{i \alpha} \lambda^{i}_{\beta} \lambda^{\rho}_{j} {W}^{-5} \nabla^{e}{\lambda^{j}_{\rho}}+\frac{13}{128}{\rm i} F^{a b} W_{a b}\,^{\alpha}\,_{i} \lambda^{i \beta} \lambda_{j \alpha} \lambda^{j}_{\beta} {W}^{-4} - \frac{13}{256}{\rm i} \epsilon^{a b c d}\,_{e} (\Gamma^{e})^{\beta \rho} F_{a b} W_{c d}\,^{\alpha}\,_{i} \lambda^{i}_{\beta} \lambda_{j \rho} \lambda^{j}_{\alpha} {W}^{-4}+\frac{13}{32}{\rm i} (\Sigma_{a}{}^{\, c})^{\beta \rho} F^{a b} W_{b c}\,^{\alpha}\,_{i} \lambda^{i}_{\beta} \lambda_{j \rho} \lambda^{j}_{\alpha} {W}^{-4}+\frac{1}{16}{\rm i} (\Sigma_{a}{}^{\, c})^{\beta \rho} F^{a b} W_{b c}\,^{\alpha}\,_{i} \lambda^{i}_{\alpha} \lambda_{j \beta} \lambda^{j}_{\rho} {W}^{-4} - \frac{137}{128}{\rm i} (\Sigma_{a b})^{\beta \rho} F^{a b} \lambda_{j \beta} \lambda^{j}_{\rho} \lambda^{\alpha}_{i} X^{i}_{\alpha} {W}^{-4} - \frac{9}{8}{\rm i} (\Sigma_{a b})^{\beta \alpha} F^{a b} \lambda_{j \beta} \lambda^{j \rho} \lambda_{i \rho} X^{i}_{\alpha} {W}^{-4}+\frac{7}{64}{\rm i} (\Sigma_{a b})^{\beta \rho} F^{a b} \lambda_{j \beta} \lambda^{j \alpha} \lambda_{i \rho} X^{i}_{\alpha} {W}^{-4}%
 - \frac{33}{256}\epsilon^{c d}\,_{e}\,^{a b} (\Sigma_{c d})^{\alpha \beta} W_{a b} \lambda_{i \alpha} \lambda^{i \rho} \lambda_{j \beta} {W}^{-4} \nabla^{e}{\lambda^{j}_{\rho}} - \frac{19}{128}(\Gamma^{a})^{\alpha \beta} W_{a b} \lambda_{i \alpha} \lambda^{i \rho} \lambda_{j \beta} {W}^{-4} \nabla^{b}{\lambda^{j}_{\rho}} - \frac{1}{16}\epsilon^{c d}\,_{e}\,^{a b} (\Sigma_{c d})^{\alpha \beta} F_{a b} \lambda_{i \alpha} \lambda^{i \rho} \lambda_{j \beta} {W}^{-5} \nabla^{e}{\lambda^{j}_{\rho}} - \frac{5}{8}(\Gamma^{a})^{\alpha \beta} F_{a b} \lambda_{i \alpha} \lambda^{i \rho} \lambda_{j \beta} {W}^{-5} \nabla^{b}{\lambda^{j}_{\rho}}+\frac{1}{16}\lambda^{\alpha}_{i} \lambda_{j \alpha} {W}^{-4} \nabla_{a}{\lambda^{i \beta}} \nabla^{a}{\lambda^{j}_{\beta}} - \frac{7}{8}(\Sigma_{a b})^{\alpha \beta} \lambda^{\rho}_{i} \lambda_{j \rho} {W}^{-4} \nabla^{a}{\lambda^{i}_{\alpha}} \nabla^{b}{\lambda^{j}_{\beta}}+\frac{125}{256}\epsilon^{c d}\,_{e}\,^{a b} (\Sigma_{c d})^{\alpha \beta} W_{a b} \lambda_{i \alpha} \lambda^{i \rho} \lambda_{j \rho} {W}^{-4} \nabla^{e}{\lambda^{j}_{\beta}}+\frac{19}{128}(\Gamma^{a})^{\alpha \beta} W_{a b} \lambda_{i \alpha} \lambda^{i \rho} \lambda_{j \rho} {W}^{-4} \nabla^{b}{\lambda^{j}_{\beta}}+\frac{11}{16}\epsilon^{c d}\,_{e}\,^{a b} (\Sigma_{c d})^{\alpha \beta} F_{a b} \lambda_{i \alpha} \lambda^{i \rho} \lambda_{j \rho} {W}^{-5} \nabla^{e}{\lambda^{j}_{\beta}} - \frac{3}{8}(\Gamma^{a})^{\alpha \beta} F_{a b} \lambda_{i \alpha} \lambda^{i \rho} \lambda_{j \rho} {W}^{-5} \nabla^{b}{\lambda^{j}_{\beta}} - \frac{19}{128}(\Sigma_{a b})^{\rho \lambda} (\Gamma_{c})^{\alpha \beta} W^{a b} \lambda_{i \rho} \lambda^{i}_{\alpha} \lambda_{j \beta} {W}^{-4} \nabla^{c}{\lambda^{j}_{\lambda}} - \frac{1}{8}(\Sigma_{a b})^{\rho \lambda} (\Gamma_{c})^{\alpha \beta} F^{a b} \lambda_{i \rho} \lambda^{i}_{\alpha} \lambda_{j \beta} {W}^{-5} \nabla^{c}{\lambda^{j}_{\lambda}}+\frac{7}{32}(\Gamma_{a})^{\alpha \beta} (\Gamma_{b})^{\rho \lambda} \lambda_{i \alpha} \lambda_{j \rho} {W}^{-4} \nabla^{a}{\lambda^{j}_{\beta}} \nabla^{b}{\lambda^{i}_{\lambda}}+\frac{7}{128}(\Sigma_{a b})^{\rho \lambda} (\Gamma_{c})^{\alpha \beta} W^{a b} \lambda_{i \rho} \lambda^{i}_{\alpha} \lambda_{j \lambda} {W}^{-4} \nabla^{c}{\lambda^{j}_{\beta}} - \frac{1}{8}(\Sigma_{a b})^{\rho \lambda} (\Gamma_{c})^{\alpha \beta} F^{a b} \lambda_{i \rho} \lambda^{i}_{\alpha} \lambda_{j \lambda} {W}^{-5} \nabla^{c}{\lambda^{j}_{\beta}}+\frac{47}{128}(\Sigma_{a b})^{\rho \lambda} (\Gamma_{c})^{\alpha \beta} W^{a b} \lambda_{i \rho} \lambda^{i}_{\lambda} \lambda_{j \alpha} {W}^{-4} \nabla^{c}{\lambda^{j}_{\beta}}+\frac{5}{4}(\Sigma_{a b})^{\rho \lambda} (\Gamma_{c})^{\alpha \beta} F^{a b} \lambda_{i \rho} \lambda^{i}_{\lambda} \lambda_{j \alpha} {W}^{-5} \nabla^{c}{\lambda^{j}_{\beta}} - \frac{1}{8}(\Sigma_{a b})^{\alpha \beta} \lambda_{i \alpha} \lambda^{i}_{\beta} {W}^{-4} \nabla^{a}{\lambda^{\rho}_{j}} \nabla^{b}{\lambda^{j}_{\rho}} - \frac{1}{8}(\Sigma_{a b})^{\alpha \beta} \lambda_{i \alpha} \lambda_{j \beta} {W}^{-4} \nabla^{a}{\lambda^{i \rho}} \nabla^{b}{\lambda^{j}_{\rho}}+\frac{1}{32}(\Gamma_{a})^{\alpha \beta} (\Gamma_{b})^{\rho \lambda} \lambda_{i \alpha} \lambda_{j \rho} {W}^{-4} \nabla^{a}{\lambda^{i}_{\lambda}} \nabla^{b}{\lambda^{j}_{\beta}}%
+\frac{1}{4}(\Gamma^{a})^{\alpha \beta} \lambda_{i \alpha} \lambda^{i \rho} \lambda_{j \beta} \lambda^{j}_{\rho} {W}^{-5} \nabla^{b}{F_{a b}}+\frac{1}{8}(\Sigma_{a b})^{\alpha \beta} \lambda_{i \alpha} \lambda^{i \rho} \lambda_{j \beta} {W}^{-4} \nabla^{a}{\nabla^{b}{\lambda^{j}_{\rho}}}+\frac{9}{16}(\Gamma^{a})^{\alpha \beta} \lambda_{i \alpha} \lambda^{i \rho} \lambda_{j \beta} \lambda^{j}_{\rho} {W}^{-4} \nabla^{b}{W_{a b}} - \frac{3}{128}{\rm i} W_{a b}\,^{\alpha}\,_{i} \epsilon^{a b c d}\,_{e} (\Sigma_{c d})^{\beta \rho} \lambda^{i}_{\beta} \lambda_{j \alpha} \lambda^{j}_{\rho} {W}^{-4} \nabla^{e}{W} - \frac{3}{64}{\rm i} W_{a b}\,^{\alpha}\,_{i} (\Gamma^{a})^{\beta \rho} \lambda^{i}_{\beta} \lambda_{j \alpha} \lambda^{j}_{\rho} {W}^{-4} \nabla^{b}{W}+\frac{9}{256}W^{a b} W_{a b} \lambda^{\alpha}_{i} \lambda^{i \beta} \lambda_{j \alpha} \lambda^{j}_{\beta} {W}^{-4} - \frac{27}{128}\epsilon_{e}\,^{a b c d} (\Gamma^{e})^{\alpha \beta} W_{a b} W_{c d} \lambda_{i \alpha} \lambda^{i \rho} \lambda_{j \beta} \lambda^{j}_{\rho} {W}^{-4} - \frac{105}{256}(\Sigma_{a b})^{\alpha \beta} (\Sigma_{c d})^{\rho \lambda} W^{a b} W^{c d} \lambda_{i \alpha} \lambda^{i}_{\beta} \lambda_{j \rho} \lambda^{j}_{\lambda} {W}^{-4}+\frac{57}{128}(\Sigma_{a b})^{\alpha \beta} (\Sigma_{c d})^{\rho \lambda} W^{a b} W^{c d} \lambda_{i \alpha} \lambda^{i}_{\rho} \lambda_{j \beta} \lambda^{j}_{\lambda} {W}^{-4}+\frac{25}{192}\Phi^{a b}\,_{i j} (\Sigma_{a b})^{\alpha \beta} \lambda^{i}_{\alpha} \lambda^{j \rho} \lambda_{k \beta} \lambda^{k}_{\rho} {W}^{-4}+\frac{119}{384}\Phi^{a b}\,_{i j} (\Sigma_{a b})^{\alpha \beta} \lambda^{i \rho} \lambda^{j}_{\rho} \lambda_{k \alpha} \lambda^{k}_{\beta} {W}^{-4} - \frac{1}{8}(\Sigma_{a b})^{\alpha \beta} \lambda_{i \alpha} \lambda^{i}_{\beta} \lambda^{\rho}_{j} {W}^{-4} \nabla^{a}{\nabla^{b}{\lambda^{j}_{\rho}}} - \frac{3}{64}{\rm i} W_{a b}\,^{\alpha}\,_{i} \epsilon^{a b c d}\,_{e} (\Sigma_{c d})^{\beta \rho} \lambda^{i}_{\alpha} \lambda_{j \beta} \lambda^{j}_{\rho} {W}^{-4} \nabla^{e}{W} - \frac{3}{8}{\rm i} (\Gamma_{a})^{\alpha \beta} X_{i j} \lambda^{i}_{\alpha} \lambda_{k \beta} {W}^{-4} \nabla^{a}{X^{j k}} - \frac{9}{8}{\rm i} X_{i j} \lambda^{i \alpha} \lambda^{j}_{\alpha} {W}^{-4} \nabla_{a}{\nabla^{a}{W}}+\frac{9}{64}{\rm i} (\Sigma_{a b})^{\alpha \beta} X_{i j} X^{i}\,_{k} W^{a b} \lambda^{j}_{\alpha} \lambda^{k}_{\beta} {W}^{-4} - \frac{61}{64}{\rm i} X_{i j} \lambda^{i \beta} \lambda^{j \alpha} \lambda_{k \beta} X^{k}_{\alpha} {W}^{-4}+\frac{83}{64}{\rm i} X_{i j} \lambda^{i \beta} \lambda^{j}_{\beta} \lambda^{\alpha}_{k} X^{k}_{\alpha} {W}^{-4}+\frac{11}{64}{\rm i} X_{i j} \lambda^{i \beta} \lambda_{k \beta} \lambda^{k \alpha} X^{j}_{\alpha} {W}^{-4} - \frac{1}{32}(\Gamma_{a})^{\alpha \beta} X_{i j} \lambda^{i \rho} \lambda_{k \alpha} \lambda^{k}_{\rho} {W}^{-5} \nabla^{a}{\lambda^{j}_{\beta}}%
+\frac{1}{32}(\Gamma_{a})^{\alpha \beta} X_{i j} \lambda^{i}_{\alpha} \lambda_{k \beta} \lambda^{k \rho} {W}^{-5} \nabla^{a}{\lambda^{j}_{\rho}}+\frac{1}{2}{\rm i} (\Gamma_{a})^{\alpha \beta} X_{i j} X^{i j} \lambda_{k \alpha} {W}^{-4} \nabla^{a}{\lambda^{k}_{\beta}} - \frac{21}{16}(\Gamma_{a})^{\alpha \beta} X_{i j} \lambda^{i \rho} \lambda^{j}_{\rho} \lambda_{k \alpha} {W}^{-5} \nabla^{a}{\lambda^{k}_{\beta}} - \frac{13}{32}(\Gamma_{a})^{\alpha \beta} X_{i j} \lambda^{i}_{\alpha} \lambda^{j \rho} \lambda_{k \beta} {W}^{-5} \nabla^{a}{\lambda^{k}_{\rho}}+\frac{1}{8}{\rm i} (\Gamma_{a})^{\alpha \beta} X_{i j} X^{i}\,_{k} \lambda^{j}_{\alpha} {W}^{-4} \nabla^{a}{\lambda^{k}_{\beta}} - \frac{33}{32}(\Gamma_{a})^{\alpha \beta} X_{i j} \lambda^{i}_{\alpha} \lambda^{j \rho} \lambda_{k \rho} {W}^{-5} \nabla^{a}{\lambda^{k}_{\beta}} - \frac{9}{8}{\rm i} \lambda^{\alpha}_{i} \lambda_{j \alpha} {W}^{-4} \nabla_{a}{W} \nabla^{a}{X^{i j}}+\frac{3}{8}(\Gamma_{a})^{\alpha \beta} \lambda_{k \alpha} \lambda^{k \rho} \lambda_{i \beta} \lambda_{j \rho} {W}^{-5} \nabla^{a}{X^{i j}}+\frac{1}{2}{\rm i} (\Gamma_{a})^{\alpha \beta} \lambda_{i \alpha} {W}^{-2} \nabla_{b}{\nabla^{b}{\nabla^{a}{\lambda^{i}_{\beta}}}}-{W}^{-1} \nabla_{a}{\nabla^{a}{\nabla_{b}{\nabla^{b}{W}}}}+\frac{3}{4}{\rm i} (\Gamma_{a})^{\alpha \beta} \lambda_{i \alpha} {W}^{-4} \nabla^{a}{W} \nabla_{b}{W} \nabla^{b}{\lambda^{i}_{\beta}} - \frac{9}{8}{\rm i} (\Sigma^{a}{}_{\, c})^{\alpha \beta} W_{a b} \lambda_{i \alpha} \lambda^{i}_{\beta} {W}^{-4} \nabla^{c}{W} \nabla^{b}{W}+\frac{27}{32}{\rm i} (\Sigma_{a b})^{\alpha \beta} W^{a b} \lambda_{i \alpha} \lambda^{i}_{\beta} {W}^{-4} \nabla_{c}{W} \nabla^{c}{W}-3{\rm i} (\Sigma^{a}{}_{\, c})^{\alpha \beta} F_{a b} \lambda_{i \alpha} \lambda^{i}_{\beta} {W}^{-5} \nabla^{c}{W} \nabla^{b}{W} - \frac{3}{4}{\rm i} (\Sigma_{a b})^{\alpha \beta} F^{a b} \lambda_{i \alpha} \lambda^{i}_{\beta} {W}^{-5} \nabla_{c}{W} \nabla^{c}{W} - \frac{1}{2}{\rm i} (\Gamma_{a})^{\alpha \beta} {W}^{-2} \nabla^{a}{\lambda_{i \alpha}} \nabla_{b}{\nabla^{b}{\lambda^{i}_{\beta}}} - \frac{1}{2}F_{a b} {W}^{-2} \nabla_{c}{\nabla^{c}{F^{a b}}} - \frac{1}{4}{\rm i} (\Sigma_{a b})^{\alpha \beta} \lambda_{i \alpha} \lambda^{i}_{\beta} {W}^{-3} \nabla_{c}{\nabla^{c}{F^{a b}}}+\frac{1}{2}{\rm i} (\Gamma_{a})^{\alpha \beta} {W}^{-2} \nabla_{b}{\lambda_{i \alpha}} \nabla^{b}{\nabla^{a}{\lambda^{i}_{\beta}}} - \frac{1}{4}{W}^{-2} \nabla_{c}{F_{a b}} \nabla^{c}{F^{a b}}%
 - \frac{3}{32}F^{a b} F^{c d} F_{a b} F_{c d} {W}^{-4}+\frac{3}{8}F^{a b} F^{c d} F_{a c} F_{b d} {W}^{-4} - \frac{3}{8}{\rm i} (\Sigma_{a b})^{\alpha \beta} F^{a b} F^{c d} F_{c d} \lambda_{i \alpha} \lambda^{i}_{\beta} {W}^{-5}+\frac{3}{2}{\rm i} (\Sigma^{c d})^{\alpha \beta} F^{a b} F_{c a} F_{d b} \lambda_{i \alpha} \lambda^{i}_{\beta} {W}^{-5} - \frac{23}{192}X_{i j} X^{i}\,_{k} \lambda^{j \alpha} X^{k}_{\alpha} {W}^{-3} - \frac{27}{128}{\rm i} (\Sigma_{a b})^{\alpha \beta} X_{i j} X^{i j} W^{a b} \lambda_{k \alpha} \lambda^{k}_{\beta} {W}^{-4} - \frac{1}{8}X_{i j} X^{i}\,_{k} X^{j}\,_{l} X^{k l} {W}^{-4} - \frac{5}{8}{\rm i} X_{i j} X^{i}\,_{k} X^{j}\,_{l} \lambda^{k \alpha} \lambda^{l}_{\alpha} {W}^{-5}+\frac{5}{32}X_{i j} X^{i j} X_{k l} X^{k l} {W}^{-4}+\frac{11}{16}{\rm i} X_{i j} X^{i j} X_{k l} \lambda^{k \alpha} \lambda^{l}_{\alpha} {W}^{-5}+\frac{1}{2}X_{i j} {W}^{-2} \nabla_{a}{\nabla^{a}{X^{i j}}}+\frac{1}{4}{\rm i} \lambda^{\alpha}_{i} \lambda_{j \alpha} {W}^{-3} \nabla_{a}{\nabla^{a}{X^{i j}}}+\frac{1}{4}{W}^{-2} \nabla_{a}{X_{i j}} \nabla^{a}{X^{i j}} - \frac{3}{16}X_{i j} X^{i j} F^{a b} F_{a b} {W}^{-4} - \frac{3}{8}{\rm i} X_{i j} F^{a b} F_{a b} \lambda^{i \alpha} \lambda^{j}_{\alpha} {W}^{-5} - \frac{3}{16}{\rm i} \epsilon_{e}\,^{a b c d} (\Gamma^{e})^{\alpha \beta} X_{i j} F_{a b} F_{c d} \lambda^{i}_{\alpha} \lambda^{j}_{\beta} {W}^{-5} - \frac{3}{4}X_{i j} X^{i j} {W}^{-3} \nabla_{a}{\nabla^{a}{W}} - \frac{3}{2}X_{i j} {W}^{-3} \nabla_{a}{W} \nabla^{a}{X^{i j}} - \frac{37}{384}{\rm i} \Phi^{a b}\,_{i j} (\Sigma_{a b})^{\alpha \beta} X^{i j} \lambda_{k \alpha} \lambda^{k}_{\beta} {W}^{-3} - \frac{9}{16}{\rm i} (\Sigma_{a b})^{\alpha \beta} X_{i j} X^{i j} F^{a b} \lambda_{k \alpha} \lambda^{k}_{\beta} {W}^{-5}%
+\frac{15}{8}{\rm i} (\Gamma_{a})^{\alpha \beta} \lambda_{i \alpha} {W}^{-4} \nabla_{b}{W} \nabla^{b}{W} \nabla^{a}{\lambda^{i}_{\beta}}+\frac{3}{8}{\rm i} \epsilon^{a b}\,_{c d e} (\Sigma_{a b})^{\alpha \beta} \lambda_{i \alpha} \lambda^{i}_{\beta} {W}^{-4} \nabla^{c}{W} \nabla^{d}{\nabla^{e}{W}} - \frac{1}{4}(\Gamma_{a})^{\alpha \beta} (\Gamma_{b})^{\rho \lambda} \lambda_{i \alpha} \lambda^{i}_{\rho} \lambda_{j \beta} \lambda^{j}_{\lambda} {W}^{-5} \nabla^{a}{\nabla^{b}{W}} - \frac{5}{8}(\Gamma_{a})^{\alpha \beta} (\Gamma_{b})^{\rho \lambda} \lambda_{i \alpha} \lambda^{i}_{\rho} \lambda_{j \beta} {W}^{-5} \nabla^{a}{W} \nabla^{b}{\lambda^{j}_{\lambda}} - \frac{3}{8}(\Gamma_{a})^{\alpha \beta} (\Gamma_{b})^{\rho \lambda} \lambda_{i \alpha} \lambda^{i}_{\rho} \lambda_{j \beta} {W}^{-5} \nabla^{b}{W} \nabla^{a}{\lambda^{j}_{\lambda}}-2\lambda^{\alpha}_{i} \lambda^{i \beta} \lambda_{j \alpha} {W}^{-5} \nabla_{a}{W} \nabla^{a}{\lambda^{j}_{\beta}}+\frac{1}{4}(\Sigma_{a b})^{\alpha \beta} \lambda_{i \alpha} \lambda^{i \rho} \lambda_{j \rho} {W}^{-5} \nabla^{a}{W} \nabla^{b}{\lambda^{j}_{\beta}} - \frac{9}{16}\epsilon_{e}\,^{c d a b} (\Gamma^{e})^{\alpha \beta} W_{c d} F_{a b} \lambda_{i \alpha} \lambda^{i \rho} \lambda_{j \beta} \lambda^{j}_{\rho} {W}^{-5}+\frac{1}{2}(\Sigma_{c d})^{\alpha \beta} (\Sigma_{a b})^{\rho \lambda} W^{c d} F^{a b} \lambda_{i \alpha} \lambda^{i}_{\rho} \lambda_{j \beta} \lambda^{j}_{\lambda} {W}^{-5} - \frac{7}{8}(\Sigma_{c d})^{\alpha \beta} (\Sigma_{a b})^{\rho \lambda} W^{c d} F^{a b} \lambda_{i \alpha} \lambda^{i}_{\beta} \lambda_{j \rho} \lambda^{j}_{\lambda} {W}^{-5}+\frac{5}{4}(\Sigma_{a b})^{\alpha \beta} \lambda_{i \alpha} \lambda^{i \rho} \lambda_{j \beta} {W}^{-5} \nabla^{a}{W} \nabla^{b}{\lambda^{j}_{\rho}}+(\Sigma_{a b})^{\alpha \beta} \lambda_{i \alpha} \lambda^{i}_{\beta} \lambda^{\rho}_{j} {W}^{-5} \nabla^{a}{W} \nabla^{b}{\lambda^{j}_{\rho}}+\frac{9}{4}{\rm i} X_{i j} \lambda^{i \alpha} \lambda^{j}_{\alpha} {W}^{-5} \nabla_{a}{W} \nabla^{a}{W} - \frac{1}{2}(\Gamma_{a})^{\alpha \beta} X_{i j} \lambda^{i}_{\alpha} \lambda^{j}_{\beta} \lambda^{\rho}_{k} {W}^{-5} \nabla^{a}{\lambda^{k}_{\rho}}+\frac{55}{64}(\Sigma_{a b})^{\alpha \beta} X_{i j} W^{a b} \lambda^{i \rho} \lambda^{j}_{\rho} \lambda_{k \alpha} \lambda^{k}_{\beta} {W}^{-5}+\frac{17}{32}(\Sigma_{a b})^{\alpha \beta} X_{i j} W^{a b} \lambda^{i}_{\alpha} \lambda^{j \rho} \lambda_{k \beta} \lambda^{k}_{\rho} {W}^{-5}+{W}^{-2} \nabla_{a}{W} \nabla_{b}{\nabla^{b}{\nabla^{a}{W}}}+{W}^{-2} \nabla_{a}{W} \nabla^{a}{\nabla_{b}{\nabla^{b}{W}}}+{W}^{-2} \nabla_{a}{\nabla^{a}{W}} \nabla_{b}{\nabla^{b}{W}}+\frac{1}{2}{W}^{-2} \nabla_{a}{\nabla_{b}{W}} \nabla^{a}{\nabla^{b}{W}}%
+\frac{9}{8}X_{i j} X^{i j} {W}^{-4} \nabla_{a}{W} \nabla^{a}{W}+\frac{11}{64}X_{i j} X^{i j} \lambda^{\alpha}_{k} \lambda^{k \beta} \lambda_{l \alpha} \lambda^{l}_{\beta} {W}^{-6}+\frac{3}{8}{\rm i} (\Sigma_{a b})^{\alpha \beta} X_{i j} X^{i}\,_{k} F^{a b} \lambda^{j}_{\alpha} \lambda^{k}_{\beta} {W}^{-5} - \frac{9}{16}X_{i j} X^{i}\,_{k} \lambda^{j \alpha} \lambda^{k \beta} \lambda_{l \alpha} \lambda^{l}_{\beta} {W}^{-6} - \frac{37}{32}X_{i j} X_{k l} \lambda^{i \alpha} \lambda^{j}_{\alpha} \lambda^{k \beta} \lambda^{l}_{\beta} {W}^{-6}+\frac{5}{8}(\Gamma_{a})^{\alpha \beta} (\Gamma_{b})^{\rho \lambda} \lambda_{i \alpha} \lambda^{i}_{\rho} \lambda_{j \beta} \lambda^{j}_{\lambda} {W}^{-6} \nabla^{a}{W} \nabla^{b}{W} - \frac{5}{16}\epsilon_{e}\,^{a b c d} (\Gamma^{e})^{\alpha \beta} F_{a b} F_{c d} \lambda_{i \alpha} \lambda^{i \rho} \lambda_{j \beta} \lambda^{j}_{\rho} {W}^{-6} - \frac{15}{16}(\Sigma_{a b})^{\alpha \beta} (\Sigma_{c d})^{\rho \lambda} F^{a b} F^{c d} \lambda_{i \alpha} \lambda^{i}_{\beta} \lambda_{j \rho} \lambda^{j}_{\lambda} {W}^{-6}+\frac{65}{32}(\Sigma_{a b})^{\alpha \beta} X_{i j} F^{a b} \lambda^{i \rho} \lambda^{j}_{\rho} \lambda_{k \alpha} \lambda^{k}_{\beta} {W}^{-6}+\frac{11}{64}{\rm i} W^{a b \alpha}\,_{i} (\Sigma_{a b})^{\beta \rho} X^{i}\,_{j} \lambda^{j}_{\beta} \lambda_{k \alpha} \lambda^{k}_{\rho} {W}^{-4}+\frac{39}{64}\lambda^{\beta}_{j} \lambda^{j \rho} \lambda_{k \beta} \lambda^{k}_{\rho} \lambda^{\alpha}_{i} X^{i}_{\alpha} {W}^{-5} - \frac{33}{32}\lambda^{\beta}_{j} \lambda^{j \rho} \lambda_{k \beta} \lambda^{k \alpha} \lambda_{i \rho} X^{i}_{\alpha} {W}^{-5}-{W}^{-3} \nabla_{a}{W} \nabla_{b}{W} \nabla^{a}{\nabla^{b}{W}} - \frac{3}{2}{W}^{-3} \nabla_{a}{W} \nabla^{a}{W} \nabla_{b}{\nabla^{b}{W}}+\frac{25}{64}{\rm i} (\Gamma_{a})^{\alpha \beta} \lambda_{i \alpha} \lambda^{i \rho} \lambda_{j \beta} \lambda^{j}_{\rho} \lambda^{\lambda}_{k} {W}^{-6} \nabla^{a}{\lambda^{k}_{\lambda}} - \frac{3}{16}W^{a b \alpha}\,_{i} (\Sigma_{a b})^{\beta \rho} \lambda^{i \lambda} \lambda_{j \alpha} \lambda^{j}_{\beta} \lambda_{k \rho} \lambda^{k}_{\lambda} {W}^{-5}+\frac{5}{16}{\rm i} (\Gamma_{a})^{\alpha \beta} \lambda_{i \alpha} \lambda^{i \rho} \lambda_{j \beta} \lambda_{k \rho} \lambda^{k \lambda} {W}^{-6} \nabla^{a}{\lambda^{j}_{\lambda}}+\frac{35}{32}{\rm i} (\Gamma_{a})^{\alpha \beta} \lambda_{i \alpha} \lambda^{i \rho} \lambda_{j \rho} \lambda^{j \lambda} \lambda_{k \lambda} {W}^{-6} \nabla^{a}{\lambda^{k}_{\beta}} - \frac{15}{32}{\rm i} (\Gamma_{a})^{\alpha \beta} \lambda_{i \alpha} \lambda^{i \rho} \lambda_{j \beta} \lambda^{j \lambda} \lambda_{k \rho} {W}^{-6} \nabla^{a}{\lambda^{k}_{\lambda}}+\frac{35}{64}{\rm i} (\Gamma_{a})^{\alpha \beta} \lambda_{i \alpha} \lambda^{\rho}_{j} \lambda^{j \lambda} \lambda_{k \rho} \lambda^{k}_{\lambda} {W}^{-6} \nabla^{a}{\lambda^{i}_{\beta}}%
 - \frac{1}{2}\lambda^{\alpha}_{i} \lambda^{i \beta} \lambda_{j \alpha} \lambda^{j}_{\beta} {W}^{-5} \nabla_{a}{\nabla^{a}{W}}+\frac{3}{8}{W}^{-4} \nabla_{a}{W} \nabla^{a}{W} \nabla_{b}{W} \nabla^{b}{W}+\frac{5}{4}\lambda^{\alpha}_{i} \lambda^{i \beta} \lambda_{j \alpha} \lambda^{j}_{\beta} {W}^{-6} \nabla_{a}{W} \nabla^{a}{W}+\frac{75}{64}{\rm i} (\Sigma_{a b})^{\alpha \beta} W^{a b} \lambda_{i \alpha} \lambda^{i \rho} \lambda_{j \beta} \lambda^{j \lambda} \lambda_{k \rho} \lambda^{k}_{\lambda} {W}^{-6} - \frac{105}{128}{\rm i} (\Sigma_{a b})^{\alpha \beta} W^{a b} \lambda_{i \alpha} \lambda^{i}_{\beta} \lambda^{\rho}_{j} \lambda^{j \lambda} \lambda_{k \rho} \lambda^{k}_{\lambda} {W}^{-6} - \frac{5}{16}(\Sigma_{a b})^{\alpha \beta} X_{i j} F^{a b} \lambda^{i}_{\alpha} \lambda^{j \rho} \lambda_{k \beta} \lambda^{k}_{\rho} {W}^{-6} - \frac{7}{32}X_{i j} X_{k l} \lambda^{i \alpha} \lambda^{j \beta} \lambda^{k}_{\alpha} \lambda^{l}_{\beta} {W}^{-6} - \frac{3}{16}{\rm i} X_{i j} \lambda^{i \alpha} \lambda^{j \beta} \lambda_{k \alpha} \lambda^{k \rho} \lambda_{l \beta} \lambda^{l}_{\rho} {W}^{-7}+\frac{33}{32}{\rm i} X_{i j} \lambda^{i \alpha} \lambda^{j}_{\alpha} \lambda^{\beta}_{k} \lambda^{k \rho} \lambda_{l \beta} \lambda^{l}_{\rho} {W}^{-7} - \frac{45}{32}{\rm i} (\Sigma_{a b})^{\alpha \beta} F^{a b} \lambda_{i \alpha} \lambda^{i}_{\beta} \lambda^{\rho}_{j} \lambda^{j \lambda} \lambda_{k \rho} \lambda^{k}_{\lambda} {W}^{-7}+\frac{15}{16}{\rm i} (\Sigma_{a b})^{\alpha \beta} F^{a b} \lambda_{i \alpha} \lambda^{i \rho} \lambda_{j \beta} \lambda^{j \lambda} \lambda_{k \rho} \lambda^{k}_{\lambda} {W}^{-7}+\frac{21}{64}\lambda^{\alpha}_{i} \lambda^{i \beta} \lambda_{j \alpha} \lambda^{j}_{\beta} \lambda^{\rho}_{k} \lambda^{k \lambda} \lambda_{l \rho} \lambda^{l}_{\lambda} {W}^{-8} - \frac{21}{32}\lambda^{\alpha}_{i} \lambda^{i \beta} \lambda_{j \alpha} \lambda^{j \rho} \lambda_{k \beta} \lambda^{k \lambda} \lambda_{l \rho} \lambda^{l}_{\lambda} {W}^{-8}
\doublespacedmathend
\end{adjustwidth}

\subsubsection{$\cH^{a}_{\log}$}

\begin{adjustwidth}{0cm}{5cm}
\doublespacedmathbegin
{}\frac{3}{32}W^{a}\,_{b} \nabla^{b}{Y}+\frac{3}{32}Y \nabla_{b}{W^{a b}}+\frac{23}{64}\epsilon_{{e_{1}}}\,^{d e b c} W_{b c} \nabla^{{e_{1}}}{\nabla^{a}{W_{d e}}} - \frac{3}{32}\epsilon_{{e_{1}}}\,^{d e b c} W_{b c} \nabla^{a}{\nabla^{{e_{1}}}{W_{d e}}}+\frac{567}{2048}\epsilon^{b c}\,_{{e_{1}}}\,^{d e} \nabla^{a}{W_{b c}} \nabla^{{e_{1}}}{W_{d e}}+\frac{3}{8}\epsilon^{a}\,_{{e_{1}}}\,^{d e b} W_{b c} \nabla^{{e_{1}}}{\nabla^{c}{W_{d e}}}+\frac{3}{8}\epsilon^{a}\,_{{e_{1}}}\,^{d b c} W_{b c} \nabla^{e}{\nabla^{{e_{1}}}{W_{d e}}} - \frac{1}{16}\epsilon_{e {e_{1}} d}\,^{b c} W_{b c} \nabla^{e}{\nabla^{{e_{1}}}{W^{a d}}}+\frac{43}{32}\epsilon^{a}\,_{{e_{1}}}\,^{d b c} W_{b c} \nabla^{{e_{1}}}{\nabla^{e}{W_{d e}}} - \frac{5}{32}\epsilon^{a}\,_{{e_{1}}}\,^{d e b} W_{b c} \nabla^{c}{\nabla^{{e_{1}}}{W_{d e}}} - \frac{5}{32}\epsilon_{e {e_{1}}}\,^{c d b} W_{b}\,^{a} \nabla^{e}{\nabla^{{e_{1}}}{W_{c d}}}+\frac{1743}{1024}\epsilon^{a}\,_{{e_{1}}}\,^{b c d} \nabla^{{e_{1}}}{W_{b c}} \nabla^{e}{W_{d e}} - \frac{3}{512}\epsilon_{e b {e_{1}}}\,^{c d} \nabla^{e}{W^{a b}} \nabla^{{e_{1}}}{W_{c d}} - \frac{567}{2048}\epsilon^{a b c d e} \nabla_{{e_{1}}}{W_{b c}} \nabla^{{e_{1}}}{W_{d e}} - \frac{3}{256}\epsilon^{a}\,_{e}\,^{b}\,_{{e_{1}} d} \nabla^{e}{W_{b c}} \nabla^{{e_{1}}}{W^{d c}} - \frac{3}{32}F^{a}\,_{b} {W}^{-1} \nabla^{b}{Y} - \frac{3}{32}Y {W}^{-1} \nabla_{b}{F^{a b}} - \frac{15}{16}{\rm i} (\Sigma^{a}{}_{\, b})^{\alpha \beta} X_{i \alpha} \nabla^{b}{X^{i}_{\beta}}+\frac{29}{64}\epsilon_{{e_{1}}}\,^{d e b c} F_{b c} {W}^{-1} \nabla^{{e_{1}}}{\nabla^{a}{W_{d e}}}%
 - \frac{3}{32}\epsilon_{{e_{1}}}\,^{d e b c} F_{b c} {W}^{-1} \nabla^{a}{\nabla^{{e_{1}}}{W_{d e}}}+\frac{9}{16}\epsilon^{a}\,_{{e_{1}}}\,^{d b c} F_{b c} {W}^{-1} \nabla^{e}{\nabla^{{e_{1}}}{W_{d e}}}+\frac{1}{8}\epsilon_{e {e_{1}} d}\,^{b c} F_{b c} {W}^{-1} \nabla^{e}{\nabla^{{e_{1}}}{W^{a d}}}+\frac{7}{32}\epsilon^{a}\,_{{e_{1}}}\,^{d b c} F_{b c} {W}^{-1} \nabla^{{e_{1}}}{\nabla^{e}{W_{d e}}}+\frac{1}{32}\epsilon^{a}\,_{{e_{1}}}\,^{d e b} F_{b c} {W}^{-1} \nabla^{c}{\nabla^{{e_{1}}}{W_{d e}}} - \frac{11}{32}\epsilon_{e {e_{1}}}\,^{c d b} F_{b}\,^{a} {W}^{-1} \nabla^{e}{\nabla^{{e_{1}}}{W_{c d}}}+\frac{3}{4}\epsilon^{a}\,_{{e_{1}}}\,^{b d e} W_{d e} {W}^{-1} \nabla^{{e_{1}}}{\nabla^{c}{F_{b c}}} - \frac{1}{8}\epsilon^{a}\,_{{e_{1}}}\,^{b c d} W_{d e} {W}^{-1} \nabla^{e}{\nabla^{{e_{1}}}{F_{b c}}}+\frac{15}{64}\epsilon^{b c}\,_{{e_{1}}}\,^{d e} {W}^{-1} \nabla^{a}{F_{b c}} \nabla^{{e_{1}}}{W_{d e}}+\frac{3}{4}\epsilon^{a}\,_{{e_{1}}}\,^{b c d} {W}^{-1} \nabla^{{e_{1}}}{F_{b c}} \nabla^{e}{W_{d e}}+\frac{3}{32}\epsilon^{a b}\,_{{e_{1}}}\,^{d e} {W}^{-1} \nabla^{c}{F_{b c}} \nabla^{{e_{1}}}{W_{d e}}+\frac{9}{8}\epsilon_{e b {e_{1}}}\,^{c d} {W}^{-1} \nabla^{e}{F^{a b}} \nabla^{{e_{1}}}{W_{c d}} - \frac{1}{16}\epsilon_{{e_{1}}}\,^{b c d e} W_{d e} {W}^{-1} \nabla^{{e_{1}}}{\nabla^{a}{F_{b c}}} - \frac{1}{8}\epsilon^{a}\,_{{e_{1}}}\,^{b d e} W_{d e} {W}^{-1} \nabla^{c}{\nabla^{{e_{1}}}{F_{b c}}}+\frac{1}{4}\epsilon^{a}\,_{{e_{1}}}\,^{b c d} W_{d e} {W}^{-1} \nabla^{{e_{1}}}{\nabla^{e}{F_{b c}}}+\frac{1}{8}\epsilon_{e {e_{1}}}\,^{b c d} W_{d}\,^{a} {W}^{-1} \nabla^{e}{\nabla^{{e_{1}}}{F_{b c}}} - \frac{15}{32}\epsilon_{{e_{1}}}\,^{b c d e} {W}^{-1} \nabla^{{e_{1}}}{F_{b c}} \nabla^{a}{W_{d e}} - \frac{27}{32}\epsilon^{a b c}\,_{{e_{1}}}\,^{d} {W}^{-1} \nabla^{e}{F_{b c}} \nabla^{{e_{1}}}{W_{d e}}+\frac{15}{16}\epsilon^{a}\,_{{e_{1}}}\,^{b d e} {W}^{-1} \nabla^{{e_{1}}}{F_{b c}} \nabla^{c}{W_{d e}}+\frac{3}{32}F^{a}\,_{b} Y {W}^{-2} \nabla^{b}{W}%
+(\Gamma_{b})^{\alpha \beta} X_{i \alpha} {W}^{-1} \nabla^{b}{\nabla^{a}{\lambda^{i}_{\beta}}} - \frac{1}{4}(\Gamma_{b})^{\alpha \beta} X_{i \alpha} {W}^{-1} \nabla^{a}{\nabla^{b}{\lambda^{i}_{\beta}}} - \frac{3}{8}(\Gamma_{b})^{\beta \alpha} \lambda_{i \beta} {W}^{-1} \nabla^{b}{\nabla^{a}{X^{i}_{\alpha}}} - \frac{3}{8}(\Gamma_{b})^{\beta \alpha} \lambda_{i \beta} {W}^{-1} \nabla^{a}{\nabla^{b}{X^{i}_{\alpha}}} - \frac{3}{4}(\Gamma_{b})^{\beta \alpha} {W}^{-1} \nabla^{a}{\lambda_{i \beta}} \nabla^{b}{X^{i}_{\alpha}} - \frac{3}{4}(\Gamma_{b})^{\beta \alpha} {W}^{-1} \nabla^{b}{\lambda_{i \beta}} \nabla^{a}{X^{i}_{\alpha}}+\frac{1}{4}\epsilon^{a b c}\,_{d e} (\Sigma_{b c})^{\alpha \beta} X_{i \alpha} {W}^{-1} \nabla^{d}{\nabla^{e}{\lambda^{i}_{\beta}}} - \frac{3}{4}(\Gamma^{a})^{\alpha \beta} X_{i \alpha} {W}^{-1} \nabla_{b}{\nabla^{b}{\lambda^{i}_{\beta}}} - \frac{1}{8}\epsilon^{a b c}\,_{d e} (\Sigma_{b c})^{\beta \alpha} \lambda_{i \beta} {W}^{-1} \nabla^{d}{\nabla^{e}{X^{i}_{\alpha}}}+\frac{3}{4}(\Gamma^{a})^{\beta \alpha} \lambda_{i \beta} {W}^{-1} \nabla_{b}{\nabla^{b}{X^{i}_{\alpha}}} - \frac{3}{4}\epsilon^{a b c}\,_{d e} (\Sigma_{b c})^{\beta \alpha} {W}^{-1} \nabla^{d}{\lambda_{i \beta}} \nabla^{e}{X^{i}_{\alpha}}+\frac{3}{2}(\Gamma^{a})^{\beta \alpha} {W}^{-1} \nabla_{b}{\lambda_{i \beta}} \nabla^{b}{X^{i}_{\alpha}} - \frac{1}{4}W^{c d} \nabla^{b}{C^{a}\,_{b c d}} - \frac{3}{8}W^{b c} W_{b c} \nabla^{d}{W^{a}\,_{d}} - \frac{15}{32}W^{a}\,_{b} W^{c d} \nabla^{b}{W_{c d}}+\frac{9}{32}W^{a}\,_{b} W_{c d} \nabla^{b}{W^{c d}} - \frac{3}{8}W^{b c} W_{b d} \nabla^{d}{W^{a}\,_{c}}+\frac{3}{8}W^{b c} W^{a}\,_{b} \nabla^{d}{W_{c d}}+\frac{3}{8}W_{b}\,^{c} W_{d c} \nabla^{b}{W^{a d}}+\frac{3}{4}W^{a b} W_{c}\,^{d} \nabla^{c}{W_{b d}}%
+\frac{3}{8}W^{a b} W^{c}\,_{b} \nabla^{d}{W_{c d}} - \frac{1}{4}W^{b c} \nabla^{d}{C_{b c}\,^{a}\,_{d}} - \frac{1}{4}W^{b d} \nabla^{c}{C_{b c}\,^{a}\,_{d}}+\frac{1}{4}W^{b c} \nabla^{d}{C^{a}\,_{b c d}} - \frac{3}{4}{\rm i} W^{a}\,_{b}\,^{\alpha}\,_{i} \nabla^{b}{X^{i}_{\alpha}}+\frac{27}{16}{\rm i} X_{i}^{\alpha} \nabla^{b}{W^{a}\,_{b \alpha}\,^{i}}+\frac{3}{2}\nabla_{c}{\nabla^{c}{\nabla_{b}{W^{a b}}}} - \frac{15}{8}{W}^{-1} \nabla^{b}{W_{b c}} \nabla^{c}{\nabla^{a}{W}}+2W^{a}\,_{b} {W}^{-1} \nabla^{b}{\nabla_{c}{\nabla^{c}{W}}}+\frac{3}{4}{W}^{-1} \nabla^{a}{W_{b c}} \nabla^{b}{\nabla^{c}{W}}+\frac{3}{2}{W}^{-1} \nabla_{b}{W^{a b}} \nabla_{c}{\nabla^{c}{W}}+\frac{3}{4}{W}^{-1} \nabla_{c}{W^{a}\,_{b}} \nabla^{b}{\nabla^{c}{W}}+\frac{9}{4}{W}^{-1} \nabla_{c}{W^{a}\,_{b}} \nabla^{c}{\nabla^{b}{W}}+\frac{3}{8}{W}^{-1} \nabla^{b}{W_{b c}} \nabla^{a}{\nabla^{c}{W}}+3{W}^{-1} \nabla_{b}{W} \nabla_{c}{\nabla^{c}{W^{a b}}}-2{W}^{-1} \nabla_{c}{W} \nabla_{b}{\nabla^{c}{W^{a b}}}+2{W}^{-1} \nabla^{b}{W} \nabla^{c}{\nabla^{a}{W_{b c}}}+{W}^{-1} \nabla^{b}{W} \nabla^{a}{\nabla^{c}{W_{b c}}}+\frac{1}{2}{W}^{-1} \nabla_{c}{W} \nabla^{c}{\nabla_{b}{W^{a b}}} - \frac{3}{16}\epsilon_{{e_{1}}}\,^{d e b c} F_{b c} {W}^{-2} \nabla^{a}{W} \nabla^{{e_{1}}}{W_{d e}}%
+\frac{273}{256}\epsilon_{{e_{1}}}\,^{d e b c} W_{b c} {W}^{-1} \nabla^{a}{W} \nabla^{{e_{1}}}{W_{d e}} - \frac{39}{64}\epsilon_{{e_{1}}}\,^{d e b c} W_{b c} {W}^{-1} \nabla^{{e_{1}}}{W} \nabla^{a}{W_{d e}} - \frac{73}{64}\epsilon^{a}\,_{{e_{1}}}\,^{d b c} F_{b c} {W}^{-2} \nabla^{{e_{1}}}{W} \nabla^{e}{W_{d e}} - \frac{119}{128}\epsilon^{a}\,_{{e_{1}}}\,^{d e b} F_{b c} {W}^{-2} \nabla^{c}{W} \nabla^{{e_{1}}}{W_{d e}}+\frac{71}{128}\epsilon_{e {e_{1}}}\,^{c d b} F_{b}\,^{a} {W}^{-2} \nabla^{e}{W} \nabla^{{e_{1}}}{W_{c d}} - \frac{11}{64}\epsilon^{a}\,_{{e_{1}}}\,^{d e b} F_{b c} {W}^{-2} \nabla^{{e_{1}}}{W} \nabla^{c}{W_{d e}}+\frac{5}{128}\epsilon^{a}\,_{{e_{1}}}\,^{d b c} F_{b c} {W}^{-2} \nabla^{e}{W} \nabla^{{e_{1}}}{W_{d e}} - \frac{43}{128}\epsilon_{e {e_{1}} d}\,^{b c} F_{b c} {W}^{-2} \nabla^{e}{W} \nabla^{{e_{1}}}{W^{a d}}+\frac{3}{16}\epsilon_{{e_{1}}}\,^{d e b c} W_{d e} F_{b c} {W}^{-2} \nabla^{a}{\nabla^{{e_{1}}}{W}} - \frac{3}{8}\epsilon_{{e_{1}}}\,^{d e b c} W_{d e} F_{b c} {W}^{-2} \nabla^{{e_{1}}}{\nabla^{a}{W}}+\frac{3}{32}\epsilon_{{e_{1}}}\,^{b c d e} W_{b c} W_{d e} {W}^{-1} \nabla^{a}{\nabla^{{e_{1}}}{W}}+\frac{15}{32}\epsilon_{{e_{1}}}\,^{b c d e} W_{b c} W_{d e} {W}^{-1} \nabla^{{e_{1}}}{\nabla^{a}{W}} - \frac{9}{16}\epsilon^{a b c d e} W_{b c} W_{d e} {W}^{-1} \nabla_{{e_{1}}}{\nabla^{{e_{1}}}{W}}+\frac{1}{16}\epsilon_{{e_{1}}}\,^{b c d e} W_{d e} {W}^{-2} \nabla^{a}{W} \nabla^{{e_{1}}}{F_{b c}} - \frac{15}{16}\epsilon^{a}\,_{{e_{1}}}\,^{b d e} W_{d e} {W}^{-2} \nabla^{{e_{1}}}{W} \nabla^{c}{F_{b c}} - \frac{3}{16}\epsilon^{a}\,_{{e_{1}}}\,^{b c d} W_{d e} {W}^{-2} \nabla^{e}{W} \nabla^{{e_{1}}}{F_{b c}}+\frac{1}{16}\epsilon_{e {e_{1}}}\,^{b c d} W_{d}\,^{a} {W}^{-2} \nabla^{e}{W} \nabla^{{e_{1}}}{F_{b c}} - \frac{1}{2}\epsilon^{a}\,_{{e_{1}}}\,^{b d e} W_{d e} {W}^{-2} \nabla^{c}{W} \nabla^{{e_{1}}}{F_{b c}} - \frac{5}{8}\epsilon_{e {e_{1}} b}\,^{c d} W_{c d} {W}^{-2} \nabla^{e}{W} \nabla^{{e_{1}}}{F^{a b}}+\frac{59}{64}\epsilon^{a}\,_{{e_{1}}}\,^{d b c} W_{b c} {W}^{-1} \nabla^{{e_{1}}}{W} \nabla^{e}{W_{d e}}%
+\frac{85}{128}\epsilon^{a}\,_{{e_{1}}}\,^{d e b} W_{b c} {W}^{-1} \nabla^{c}{W} \nabla^{{e_{1}}}{W_{d e}} - \frac{35}{64}\epsilon_{e {e_{1}}}\,^{c d b} W_{b}\,^{a} {W}^{-1} \nabla^{e}{W} \nabla^{{e_{1}}}{W_{c d}}+\frac{43}{64}\epsilon^{a}\,_{{e_{1}}}\,^{d e b} W_{b c} {W}^{-1} \nabla^{{e_{1}}}{W} \nabla^{c}{W_{d e}}+\frac{53}{128}\epsilon^{a}\,_{{e_{1}}}\,^{d b c} W_{b c} {W}^{-1} \nabla^{e}{W} \nabla^{{e_{1}}}{W_{d e}}+\frac{19}{64}\epsilon_{e {e_{1}} d}\,^{b c} W_{b c} {W}^{-1} \nabla^{e}{W} \nabla^{{e_{1}}}{W^{a d}} - \frac{17}{64}\epsilon^{a d e b c} W_{b c} \nabla_{{e_{1}}}{\nabla^{{e_{1}}}{W_{d e}}} - \frac{3}{16}\epsilon^{a}\,_{e {e_{1}} d}\,^{b} W_{b c} \nabla^{e}{\nabla^{{e_{1}}}{W^{d c}}}+\frac{195}{1024}\epsilon^{a}\,_{{e_{1}}}\,^{b d e} \nabla^{{e_{1}}}{W_{b c}} \nabla^{c}{W_{d e}} - \frac{1}{8}\epsilon_{{e_{1}}}\,^{b c d e} F_{b c} F_{d e} {W}^{-3} \nabla^{{e_{1}}}{\nabla^{a}{W}} - \frac{1}{2}\epsilon^{a}\,_{{e_{1}}}\,^{d b c} F_{b c} {W}^{-3} \nabla^{{e_{1}}}{W} \nabla^{e}{F_{d e}} - \frac{3}{256}\epsilon_{{e_{1}}}\,^{d e b c} F_{b c} {W}^{-2} \nabla^{{e_{1}}}{W} \nabla^{a}{W_{d e}} - \frac{1}{4}\epsilon^{a}\,_{{e_{1}}}\,^{d e b} W_{d e} F_{b c} {W}^{-2} \nabla^{{e_{1}}}{\nabla^{c}{W}}+\frac{1}{2}\epsilon^{a}\,_{{e_{1}}}\,^{d b c} W_{d e} F_{b c} {W}^{-2} \nabla^{e}{\nabla^{{e_{1}}}{W}} - \frac{3}{8}\epsilon_{e {e_{1}}}\,^{d b c} W_{d}\,^{a} F_{b c} {W}^{-2} \nabla^{e}{\nabla^{{e_{1}}}{W}}+\frac{1}{32}\epsilon_{{e_{1}}}\,^{b c d e} W_{d e} {W}^{-2} \nabla^{{e_{1}}}{W} \nabla^{a}{F_{b c}} - \frac{9}{2}W^{a b} W_{b c} \nabla_{d}{W^{d c}}+\frac{9}{32}W^{b c} W_{b c} \nabla_{d}{W^{a d}} - \frac{9}{2}W^{b}\,_{c} W_{b d} \nabla^{c}{W^{a d}} - \frac{21}{4}W^{a}\,_{b} W_{c d} \nabla^{c}{W^{b d}} - \frac{21}{256}\epsilon^{a}\,_{d e}\,^{b c} \lambda^{\alpha}_{i} {W}^{-1} \nabla^{d}{\nabla^{e}{W_{b c \alpha}\,^{i}}}%
 - \frac{255}{512}\epsilon^{a}\,_{d e}\,^{b c} {W}^{-1} \nabla^{d}{\lambda^{\alpha}_{i}} \nabla^{e}{W_{b c \alpha}\,^{i}} - \frac{1}{8}W_{b c}\,^{\alpha}\,_{i} \epsilon^{b c d e}\,_{{e_{1}}} (\Sigma_{d e})_{\alpha}{}^{\beta} {W}^{-1} \nabla^{a}{\nabla^{{e_{1}}}{\lambda^{i}_{\beta}}}+\frac{1}{4}W_{b c}\,^{\alpha}\,_{i} (\Gamma^{b})_{\alpha}{}^{\beta} {W}^{-1} \nabla^{a}{\nabla^{c}{\lambda^{i}_{\beta}}} - \frac{1}{4}W^{a}\,_{b}\,^{\alpha}\,_{i} (\Gamma_{c})_{\alpha}{}^{\beta} {W}^{-1} \nabla^{b}{\nabla^{c}{\lambda^{i}_{\beta}}}+\frac{1}{8}W_{b c}\,^{\alpha}\,_{i} \epsilon^{a b c}\,_{d e} {W}^{-1} \nabla^{d}{\nabla^{e}{\lambda^{i}_{\alpha}}} - \frac{1}{4}W_{b c}\,^{\alpha}\,_{i} \epsilon^{a b c d}\,_{{e_{1}}} (\Sigma_{d e})_{\alpha}{}^{\beta} {W}^{-1} \nabla^{{e_{1}}}{\nabla^{e}{\lambda^{i}_{\beta}}}+\frac{1}{4}W_{b c}\,^{\alpha}\,_{i} \epsilon^{a b d e}\,_{{e_{1}}} (\Sigma_{d e})_{\alpha}{}^{\beta} {W}^{-1} \nabla^{c}{\nabla^{{e_{1}}}{\lambda^{i}_{\beta}}}+\frac{1}{4}W_{b c}\,^{\alpha}\,_{i} (\Gamma^{b})_{\alpha}{}^{\beta} {W}^{-1} \nabla^{c}{\nabla^{a}{\lambda^{i}_{\beta}}}+\frac{1}{4}W_{b c}\,^{\alpha}\,_{i} (\Gamma^{a})_{\alpha}{}^{\beta} {W}^{-1} \nabla^{b}{\nabla^{c}{\lambda^{i}_{\beta}}}+\frac{1}{4}W_{b}\,^{a \alpha}\,_{i} \epsilon^{b c d}\,_{e {e_{1}}} (\Sigma_{c d})_{\alpha}{}^{\beta} {W}^{-1} \nabla^{e}{\nabla^{{e_{1}}}{\lambda^{i}_{\beta}}} - \frac{1}{4}W_{b}\,^{a \alpha}\,_{i} (\Gamma^{b})_{\alpha}{}^{\beta} {W}^{-1} \nabla_{c}{\nabla^{c}{\lambda^{i}_{\beta}}} - \frac{1}{4}W^{a}\,_{b}\,^{\alpha}\,_{i} (\Gamma_{c})_{\alpha}{}^{\beta} {W}^{-1} \nabla^{c}{\nabla^{b}{\lambda^{i}_{\beta}}} - \frac{27}{512}\epsilon^{d e}\,_{{e_{1}}}\,^{b c} (\Sigma_{d e})^{\beta \alpha} \lambda_{i \beta} {W}^{-1} \nabla^{{e_{1}}}{\nabla^{a}{W_{b c \alpha}\,^{i}}} - \frac{1}{8}(\Gamma^{b})^{\beta \alpha} \lambda_{i \beta} {W}^{-1} \nabla^{c}{\nabla^{a}{W_{b c \alpha}\,^{i}}}+\frac{1}{8}(\Gamma_{c})^{\beta \alpha} \lambda_{i \beta} {W}^{-1} \nabla^{c}{\nabla^{b}{W^{a}\,_{b \alpha}\,^{i}}} - \frac{27}{256}\epsilon^{a d}\,_{{e_{1}}}\,^{b c} (\Sigma_{d e})^{\beta \alpha} \lambda_{i \beta} {W}^{-1} \nabla^{e}{\nabla^{{e_{1}}}{W_{b c \alpha}\,^{i}}} - \frac{27}{256}\epsilon^{a d e}\,_{{e_{1}}}\,^{b} (\Sigma_{d e})^{\beta \alpha} \lambda_{i \beta} {W}^{-1} \nabla^{{e_{1}}}{\nabla^{c}{W_{b c \alpha}\,^{i}}} - \frac{11}{128}(\Gamma^{a})^{\beta \alpha} \lambda_{i \beta} {W}^{-1} \nabla^{b}{\nabla^{c}{W_{b c \alpha}\,^{i}}} - \frac{1}{8}(\Gamma^{b})^{\beta \alpha} \lambda_{i \beta} {W}^{-1} \nabla^{a}{\nabla^{c}{W_{b c \alpha}\,^{i}}} - \frac{11}{128}\epsilon^{c d}\,_{e {e_{1}}}\,^{b} (\Sigma_{c d})^{\beta \alpha} \lambda_{i \beta} {W}^{-1} \nabla^{e}{\nabla^{{e_{1}}}{W_{b}\,^{a}\,_{\alpha}\,^{i}}}%
+\frac{1}{8}(\Gamma^{b})^{\beta \alpha} \lambda_{i \beta} {W}^{-1} \nabla_{c}{\nabla^{c}{W_{b}\,^{a}\,_{\alpha}\,^{i}}}+\frac{1}{8}(\Gamma_{c})^{\beta \alpha} \lambda_{i \beta} {W}^{-1} \nabla^{b}{\nabla^{c}{W^{a}\,_{b \alpha}\,^{i}}} - \frac{5}{512}\epsilon^{d e}\,_{{e_{1}}}\,^{b c} (\Sigma_{d e})^{\beta \alpha} \lambda_{i \beta} {W}^{-1} \nabla^{a}{\nabla^{{e_{1}}}{W_{b c \alpha}\,^{i}}} - \frac{5}{256}\epsilon^{a d}\,_{{e_{1}}}\,^{b c} (\Sigma_{d e})^{\beta \alpha} \lambda_{i \beta} {W}^{-1} \nabla^{{e_{1}}}{\nabla^{e}{W_{b c \alpha}\,^{i}}} - \frac{5}{256}\epsilon^{a d e}\,_{{e_{1}}}\,^{b} (\Sigma_{d e})^{\beta \alpha} \lambda_{i \beta} {W}^{-1} \nabla^{c}{\nabla^{{e_{1}}}{W_{b c \alpha}\,^{i}}}+\frac{33}{1024}\epsilon^{d e}\,_{{e_{1}}}\,^{b c} (\Sigma_{d e})^{\beta \alpha} {W}^{-1} \nabla^{a}{\lambda_{i \beta}} \nabla^{{e_{1}}}{W_{b c \alpha}\,^{i}}+\frac{3}{8}(\Gamma^{b})^{\beta \alpha} {W}^{-1} \nabla^{a}{\lambda_{i \beta}} \nabla^{c}{W_{b c \alpha}\,^{i}} - \frac{3}{8}(\Gamma_{c})^{\beta \alpha} {W}^{-1} \nabla^{b}{\lambda_{i \beta}} \nabla^{c}{W^{a}\,_{b \alpha}\,^{i}}+\frac{33}{512}\epsilon^{a d}\,_{{e_{1}}}\,^{b c} (\Sigma_{d e})^{\beta \alpha} {W}^{-1} \nabla^{{e_{1}}}{\lambda_{i \beta}} \nabla^{e}{W_{b c \alpha}\,^{i}}+\frac{33}{512}\epsilon^{a d e}\,_{{e_{1}}}\,^{b} (\Sigma_{d e})^{\beta \alpha} {W}^{-1} \nabla^{c}{\lambda_{i \beta}} \nabla^{{e_{1}}}{W_{b c \alpha}\,^{i}} - \frac{129}{256}(\Gamma^{a})^{\beta \alpha} {W}^{-1} \nabla^{b}{\lambda_{i \beta}} \nabla^{c}{W_{b c \alpha}\,^{i}}+\frac{3}{8}(\Gamma^{b})^{\beta \alpha} {W}^{-1} \nabla^{c}{\lambda_{i \beta}} \nabla^{a}{W_{b c \alpha}\,^{i}}+\frac{63}{256}\epsilon^{c d}\,_{e {e_{1}}}\,^{b} (\Sigma_{c d})^{\beta \alpha} {W}^{-1} \nabla^{e}{\lambda_{i \beta}} \nabla^{{e_{1}}}{W_{b}\,^{a}\,_{\alpha}\,^{i}} - \frac{3}{8}(\Gamma^{b})^{\beta \alpha} {W}^{-1} \nabla_{c}{\lambda_{i \beta}} \nabla^{c}{W_{b}\,^{a}\,_{\alpha}\,^{i}} - \frac{3}{8}(\Gamma_{c})^{\beta \alpha} {W}^{-1} \nabla^{c}{\lambda_{i \beta}} \nabla^{b}{W^{a}\,_{b \alpha}\,^{i}}+\frac{159}{1024}\epsilon^{d e}\,_{{e_{1}}}\,^{b c} (\Sigma_{d e})^{\beta \alpha} {W}^{-1} \nabla^{{e_{1}}}{\lambda_{i \beta}} \nabla^{a}{W_{b c \alpha}\,^{i}}+\frac{159}{512}\epsilon^{a d}\,_{{e_{1}}}\,^{b c} (\Sigma_{d e})^{\beta \alpha} {W}^{-1} \nabla^{e}{\lambda_{i \beta}} \nabla^{{e_{1}}}{W_{b c \alpha}\,^{i}}+\frac{159}{512}\epsilon^{a d e}\,_{{e_{1}}}\,^{b} (\Sigma_{d e})^{\beta \alpha} {W}^{-1} \nabla^{{e_{1}}}{\lambda_{i \beta}} \nabla^{c}{W_{b c \alpha}\,^{i}} - \frac{3}{8}(\Gamma_{b})^{\alpha \beta} X_{i \alpha} {W}^{-2} \nabla^{a}{W} \nabla^{b}{\lambda^{i}_{\beta}} - \frac{3}{4}(\Gamma_{b})^{\alpha \beta} X_{i \alpha} {W}^{-2} \nabla^{b}{W} \nabla^{a}{\lambda^{i}_{\beta}}%
+\frac{11}{16}(\Gamma_{b})^{\beta \alpha} \lambda_{i \beta} X^{i}_{\alpha} {W}^{-2} \nabla^{b}{\nabla^{a}{W}} - \frac{5}{16}(\Gamma_{b})^{\beta \alpha} \lambda_{i \beta} X^{i}_{\alpha} {W}^{-2} \nabla^{a}{\nabla^{b}{W}}+\frac{3}{8}(\Gamma_{b})^{\beta \alpha} \lambda_{i \beta} {W}^{-2} \nabla^{a}{W} \nabla^{b}{X^{i}_{\alpha}}+\frac{3}{4}(\Gamma_{b})^{\beta \alpha} \lambda_{i \beta} {W}^{-2} \nabla^{b}{W} \nabla^{a}{X^{i}_{\alpha}} - \frac{3}{8}\epsilon^{a b c}\,_{d e} (\Sigma_{b c})^{\alpha \beta} X_{i \alpha} {W}^{-2} \nabla^{d}{W} \nabla^{e}{\lambda^{i}_{\beta}}+\frac{9}{8}(\Gamma^{a})^{\alpha \beta} X_{i \alpha} {W}^{-2} \nabla_{b}{W} \nabla^{b}{\lambda^{i}_{\beta}}+\frac{1}{16}\epsilon^{a b c}\,_{d e} (\Sigma_{b c})^{\beta \alpha} \lambda_{i \beta} X^{i}_{\alpha} {W}^{-2} \nabla^{d}{\nabla^{e}{W}} - \frac{3}{8}(\Gamma^{a})^{\beta \alpha} \lambda_{i \beta} X^{i}_{\alpha} {W}^{-2} \nabla_{b}{\nabla^{b}{W}}+\frac{3}{8}\epsilon^{a b c}\,_{d e} (\Sigma_{b c})^{\beta \alpha} \lambda_{i \beta} {W}^{-2} \nabla^{d}{W} \nabla^{e}{X^{i}_{\alpha}} - \frac{9}{8}(\Gamma^{a})^{\beta \alpha} \lambda_{i \beta} {W}^{-2} \nabla_{b}{W} \nabla^{b}{X^{i}_{\alpha}}+\frac{1}{12}\epsilon^{a b c d e} \Phi_{b c i j} \Phi_{d e}\,^{i j} - \frac{1}{4}F^{c d} {W}^{-1} \nabla^{b}{C^{a}\,_{b c d}} - \frac{3}{8}W^{b c} F_{b c} {W}^{-1} \nabla^{d}{W^{a}\,_{d}} - \frac{3}{8}W^{a}\,_{d} F^{b c} {W}^{-1} \nabla^{d}{W_{b c}}+\frac{153}{32}W_{c d} F^{a}\,_{b} {W}^{-1} \nabla^{b}{W^{c d}} - \frac{3}{32}W^{c d} F^{a}\,_{b} {W}^{-1} \nabla^{b}{W_{c d}} - \frac{3}{8}W_{d}\,^{c} F_{b c} {W}^{-1} \nabla^{d}{W^{a b}} - \frac{3}{8}W^{a c} F^{b}\,_{c} {W}^{-1} \nabla^{d}{W_{b d}}+\frac{3}{8}W^{b d} F_{b c} {W}^{-1} \nabla^{c}{W^{a}\,_{d}}+\frac{3}{8}W^{a d} F_{b}\,^{c} {W}^{-1} \nabla^{b}{W_{d c}}%
 - \frac{3}{8}W^{b c} F^{a}\,_{b} {W}^{-1} \nabla^{d}{W_{c d}} - \frac{3}{8}W_{c}\,^{d} F^{a b} {W}^{-1} \nabla^{c}{W_{d b}} - \frac{1}{4}F^{b c} {W}^{-1} \nabla^{d}{C_{b c}\,^{a}\,_{d}} - \frac{1}{4}F^{b d} {W}^{-1} \nabla^{c}{C_{b c}\,^{a}\,_{d}}+\frac{1}{4}F^{b c} {W}^{-1} \nabla^{d}{C^{a}\,_{b c d}} - \frac{3}{32}{\rm i} (\Sigma^{a}{}_{\, b})^{\alpha \beta} Y \lambda_{i \alpha} {W}^{-2} \nabla^{b}{\lambda^{i}_{\beta}} - \frac{3}{64}{\rm i} (\Sigma^{a}{}_{\, b})^{\alpha \beta} \lambda_{i \alpha} \lambda^{i}_{\beta} {W}^{-2} \nabla^{b}{Y}+X_{i j} {W}^{-1} \nabla_{b}{\Phi^{a b i j}}+\Phi^{a}\,_{b i j} {W}^{-1} \nabla^{b}{X^{i j}}+{W}^{-1} \nabla_{c}{\nabla^{c}{\nabla_{b}{F^{a b}}}} - \frac{3}{2}{W}^{-2} \nabla^{a}{W} \nabla^{b}{W} \nabla^{c}{W_{b c}}+F^{a}\,_{b} {W}^{-3} \nabla^{b}{W} \nabla_{c}{\nabla^{c}{W}}+2F^{a}\,_{b} {W}^{-3} \nabla_{c}{W} \nabla^{c}{\nabla^{b}{W}}-F_{b c} {W}^{-3} \nabla^{b}{W} \nabla^{c}{\nabla^{a}{W}}-3W^{a}\,_{b} {W}^{-2} \nabla^{b}{W} \nabla_{c}{\nabla^{c}{W}}-W^{a}\,_{b} {W}^{-2} \nabla_{c}{W} \nabla^{c}{\nabla^{b}{W}}-2W^{a}\,_{b} {W}^{-2} \nabla_{c}{W} \nabla^{b}{\nabla^{c}{W}}+\frac{3}{2}{W}^{-3} \nabla_{c}{W} \nabla^{c}{W} \nabla_{b}{F^{a b}}-{W}^{-3} \nabla^{a}{W} \nabla^{b}{W} \nabla^{c}{F_{b c}}+\frac{3}{8}\epsilon_{{e_{1}}}\,^{d e b c} W_{d e} F_{b c} {W}^{-3} \nabla^{a}{W} \nabla^{{e_{1}}}{W}%
 - \frac{9}{16}\epsilon_{{e_{1}}}\,^{b c d e} W_{b c} W_{d e} {W}^{-2} \nabla^{a}{W} \nabla^{{e_{1}}}{W}+\frac{3}{8}\epsilon_{{e_{1}}}\,^{b c d e} F_{b c} F_{d e} {W}^{-4} \nabla^{a}{W} \nabla^{{e_{1}}}{W} - \frac{3}{8}\epsilon^{a b c d e} F_{b c} F_{d e} {W}^{-4} \nabla_{{e_{1}}}{W} \nabla^{{e_{1}}}{W} - \frac{3}{8}\epsilon^{a d e b c} W_{d e} F_{b c} {W}^{-3} \nabla_{{e_{1}}}{W} \nabla^{{e_{1}}}{W}+\frac{9}{16}\epsilon^{a b c d e} W_{b c} W_{d e} {W}^{-2} \nabla_{{e_{1}}}{W} \nabla^{{e_{1}}}{W} - \frac{23}{64}\epsilon^{a d e b c} F_{b c} {W}^{-1} \nabla_{{e_{1}}}{\nabla^{{e_{1}}}{W_{d e}}} - \frac{3}{16}\epsilon^{a}\,_{e {e_{1}} d}\,^{b} F_{b c} {W}^{-1} \nabla^{e}{\nabla^{{e_{1}}}{W^{d c}}}+\frac{3}{4}\epsilon^{a}\,_{{e_{1}}}\,^{d e b} W_{d e} F_{b c} {W}^{-3} \nabla^{{e_{1}}}{W} \nabla^{c}{W} - \frac{3}{4}\epsilon^{a}\,_{{e_{1}}}\,^{d b c} W_{d e} F_{b c} {W}^{-3} \nabla^{{e_{1}}}{W} \nabla^{e}{W} - \frac{7}{16}\epsilon^{a b c d e} W_{d e} {W}^{-1} \nabla_{{e_{1}}}{\nabla^{{e_{1}}}{F_{b c}}} - \frac{1}{8}\epsilon_{e {e_{1}} b}\,^{c d} W_{c d} {W}^{-1} \nabla^{e}{\nabla^{{e_{1}}}{F^{a b}}} - \frac{45}{64}\epsilon^{a b c d e} {W}^{-1} \nabla_{{e_{1}}}{F_{b c}} \nabla^{{e_{1}}}{W_{d e}}+\frac{3}{2}F^{a}\,_{b} F_{c d} {W}^{-2} \nabla^{b}{W^{c d}} - \frac{3}{2}F^{a b} F_{b c} {W}^{-2} \nabla_{d}{W^{d c}}+\frac{3}{4}F^{b c} F_{b c} {W}^{-2} \nabla_{d}{W^{a d}} - \frac{3}{2}F^{b}\,_{c} F_{b d} {W}^{-2} \nabla^{c}{W^{a d}} - \frac{3}{2}F^{a}\,_{b} F_{c d} {W}^{-2} \nabla^{c}{W^{b d}}+\frac{1}{4}F^{b c} F_{b c} {W}^{-3} \nabla_{d}{F^{a d}}+\frac{3}{8}W^{a}\,_{d} F_{b c} {W}^{-1} \nabla^{d}{W^{b c}} - \frac{3}{8}W^{a}\,_{d} W_{b c} {W}^{-1} \nabla^{d}{F^{b c}}%
 - \frac{1}{2}W^{b}\,_{d} F_{b c} {W}^{-2} \nabla^{a}{F^{d c}}+\frac{1}{2}W^{a}\,_{d} F_{b c} {W}^{-2} \nabla^{d}{F^{b c}} - \frac{3}{2}W_{b d} F^{a b} {W}^{-2} \nabla_{c}{F^{c d}}+\frac{1}{4}W_{c d} F^{a}\,_{b} {W}^{-2} \nabla^{b}{F^{c d}}+\frac{3}{2}W^{b c} F_{b c} {W}^{-2} \nabla_{d}{F^{a d}}-2W^{b}\,_{d} F_{b c} {W}^{-2} \nabla^{c}{F^{a d}}-W^{b}\,_{d} F_{b c} {W}^{-2} \nabla^{d}{F^{a c}} - \frac{3}{2}W^{a b} F_{b c} {W}^{-2} \nabla_{d}{F^{d c}}+\frac{1}{2}W^{a}\,_{d} F_{b c} {W}^{-2} \nabla^{b}{F^{d c}} - \frac{39}{8}W_{b c} F^{a b} {W}^{-1} \nabla_{d}{W^{d c}}+\frac{3}{8}W^{b c} F_{b c} {W}^{-1} \nabla_{d}{W^{a d}} - \frac{39}{8}W^{b}\,_{d} F_{b c} {W}^{-1} \nabla^{c}{W^{a d}}+\frac{3}{8}W^{b}\,_{d} F_{b c} {W}^{-1} \nabla^{d}{W^{a c}}+\frac{39}{8}W_{c d} F^{a}\,_{b} {W}^{-1} \nabla^{c}{W^{d b}}+\frac{3}{8}W^{a b} F_{b c} {W}^{-1} \nabla_{d}{W^{d c}} - \frac{39}{8}W^{a}\,_{d} F_{b c} {W}^{-1} \nabla^{b}{W^{d c}}-W_{d c} F^{a}\,_{b} {W}^{-2} \nabla^{d}{F^{c b}} - \frac{9}{2}W^{a d} W_{d c} {W}^{-1} \nabla_{b}{F^{b c}}+\frac{75}{32}W^{c d} W_{c d} {W}^{-1} \nabla_{b}{F^{a b}} - \frac{9}{2}W^{c}\,_{d} W_{c b} {W}^{-1} \nabla^{d}{F^{a b}}%
+\frac{3}{4}W^{a}\,_{b} W_{d c} {W}^{-1} \nabla^{d}{F^{b c}}-F^{b}\,_{c} F_{b d} {W}^{-3} \nabla^{c}{F^{a d}}-F^{a b} F_{b c} {W}^{-3} \nabla_{d}{F^{d c}}+\frac{5}{32}W_{b c}\,^{\alpha}\,_{i} \epsilon^{a b c}\,_{d e} {W}^{-2} \nabla^{d}{W} \nabla^{e}{\lambda^{i}_{\alpha}} - \frac{1}{16}W_{b c}\,^{\alpha}\,_{i} \epsilon^{a b c}\,_{d e} \lambda^{i}_{\alpha} {W}^{-2} \nabla^{d}{\nabla^{e}{W}}+\frac{1}{8}W_{b c}\,^{\alpha}\,_{i} \epsilon^{b c d e}\,_{{e_{1}}} (\Sigma_{d e})_{\alpha}{}^{\beta} {W}^{-2} \nabla^{a}{W} \nabla^{{e_{1}}}{\lambda^{i}_{\beta}} - \frac{5}{16}W_{b c}\,^{\alpha}\,_{i} (\Gamma^{b})_{\alpha}{}^{\beta} {W}^{-2} \nabla^{a}{W} \nabla^{c}{\lambda^{i}_{\beta}}+\frac{5}{16}W^{a}\,_{b}\,^{\alpha}\,_{i} (\Gamma_{c})_{\alpha}{}^{\beta} {W}^{-2} \nabla^{b}{W} \nabla^{c}{\lambda^{i}_{\beta}}+\frac{1}{4}W_{b c}\,^{\alpha}\,_{i} \epsilon^{a b c d}\,_{{e_{1}}} (\Sigma_{d e})_{\alpha}{}^{\beta} {W}^{-2} \nabla^{{e_{1}}}{W} \nabla^{e}{\lambda^{i}_{\beta}} - \frac{1}{4}W_{b c}\,^{\alpha}\,_{i} \epsilon^{a b d e}\,_{{e_{1}}} (\Sigma_{d e})_{\alpha}{}^{\beta} {W}^{-2} \nabla^{c}{W} \nabla^{{e_{1}}}{\lambda^{i}_{\beta}} - \frac{5}{16}W_{b c}\,^{\alpha}\,_{i} (\Gamma^{b})_{\alpha}{}^{\beta} {W}^{-2} \nabla^{c}{W} \nabla^{a}{\lambda^{i}_{\beta}} - \frac{3}{16}W_{b c}\,^{\alpha}\,_{i} (\Gamma^{a})_{\alpha}{}^{\beta} {W}^{-2} \nabla^{b}{W} \nabla^{c}{\lambda^{i}_{\beta}} - \frac{3}{16}W_{b}\,^{a \alpha}\,_{i} \epsilon^{b c d}\,_{e {e_{1}}} (\Sigma_{c d})_{\alpha}{}^{\beta} {W}^{-2} \nabla^{e}{W} \nabla^{{e_{1}}}{\lambda^{i}_{\beta}}+\frac{5}{16}W_{b}\,^{a \alpha}\,_{i} (\Gamma^{b})_{\alpha}{}^{\beta} {W}^{-2} \nabla_{c}{W} \nabla^{c}{\lambda^{i}_{\beta}}+\frac{5}{16}W^{a}\,_{b}\,^{\alpha}\,_{i} (\Gamma_{c})_{\alpha}{}^{\beta} {W}^{-2} \nabla^{c}{W} \nabla^{b}{\lambda^{i}_{\beta}}+\frac{1}{32}W_{b c}\,^{\alpha}\,_{i} \epsilon^{b c d e}\,_{{e_{1}}} (\Sigma_{d e})_{\alpha}{}^{\beta} {W}^{-2} \nabla^{{e_{1}}}{W} \nabla^{a}{\lambda^{i}_{\beta}}+\frac{1}{16}W_{b c}\,^{\alpha}\,_{i} \epsilon^{a b c d}\,_{{e_{1}}} (\Sigma_{d e})_{\alpha}{}^{\beta} {W}^{-2} \nabla^{e}{W} \nabla^{{e_{1}}}{\lambda^{i}_{\beta}} - \frac{1}{16}W_{b c}\,^{\alpha}\,_{i} \epsilon^{a b d e}\,_{{e_{1}}} (\Sigma_{d e})_{\alpha}{}^{\beta} {W}^{-2} \nabla^{{e_{1}}}{W} \nabla^{c}{\lambda^{i}_{\beta}}+\frac{1}{16}W_{b c}\,^{\alpha}\,_{i} \epsilon^{b c d e}\,_{{e_{1}}} (\Sigma_{d e})_{\alpha}{}^{\beta} \lambda^{i}_{\beta} {W}^{-2} \nabla^{a}{\nabla^{{e_{1}}}{W}} - \frac{1}{8}W_{b c}\,^{\alpha}\,_{i} (\Gamma^{b})_{\alpha}{}^{\beta} \lambda^{i}_{\beta} {W}^{-2} \nabla^{a}{\nabla^{c}{W}}%
+\frac{1}{8}W^{a}\,_{b}\,^{\alpha}\,_{i} (\Gamma_{c})_{\alpha}{}^{\beta} \lambda^{i}_{\beta} {W}^{-2} \nabla^{b}{\nabla^{c}{W}}+\frac{1}{8}W_{b c}\,^{\alpha}\,_{i} \epsilon^{a b c d}\,_{{e_{1}}} (\Sigma_{d e})_{\alpha}{}^{\beta} \lambda^{i}_{\beta} {W}^{-2} \nabla^{{e_{1}}}{\nabla^{e}{W}} - \frac{1}{8}W_{b c}\,^{\alpha}\,_{i} \epsilon^{a b d e}\,_{{e_{1}}} (\Sigma_{d e})_{\alpha}{}^{\beta} \lambda^{i}_{\beta} {W}^{-2} \nabla^{c}{\nabla^{{e_{1}}}{W}} - \frac{1}{8}W_{b c}\,^{\alpha}\,_{i} (\Gamma^{b})_{\alpha}{}^{\beta} \lambda^{i}_{\beta} {W}^{-2} \nabla^{c}{\nabla^{a}{W}} - \frac{1}{8}W_{b c}\,^{\alpha}\,_{i} (\Gamma^{a})_{\alpha}{}^{\beta} \lambda^{i}_{\beta} {W}^{-2} \nabla^{b}{\nabla^{c}{W}} - \frac{1}{8}W_{b}\,^{a \alpha}\,_{i} \epsilon^{b c d}\,_{e {e_{1}}} (\Sigma_{c d})_{\alpha}{}^{\beta} \lambda^{i}_{\beta} {W}^{-2} \nabla^{e}{\nabla^{{e_{1}}}{W}}+\frac{1}{8}W_{b}\,^{a \alpha}\,_{i} (\Gamma^{b})_{\alpha}{}^{\beta} \lambda^{i}_{\beta} {W}^{-2} \nabla_{c}{\nabla^{c}{W}}+\frac{1}{8}W^{a}\,_{b}\,^{\alpha}\,_{i} (\Gamma_{c})_{\alpha}{}^{\beta} \lambda^{i}_{\beta} {W}^{-2} \nabla^{c}{\nabla^{b}{W}}+\frac{83}{6144}\epsilon^{d e}\,_{{e_{1}}}\,^{b c} (\Sigma_{d e})^{\beta \alpha} \lambda_{i \beta} {W}^{-2} \nabla^{a}{W} \nabla^{{e_{1}}}{W_{b c \alpha}\,^{i}} - \frac{1}{32}(\Gamma^{b})^{\beta \alpha} \lambda_{i \beta} {W}^{-2} \nabla^{a}{W} \nabla^{c}{W_{b c \alpha}\,^{i}}+\frac{1}{32}(\Gamma_{c})^{\beta \alpha} \lambda_{i \beta} {W}^{-2} \nabla^{b}{W} \nabla^{c}{W^{a}\,_{b \alpha}\,^{i}}+\frac{35}{3072}\epsilon^{a}\,_{d e}\,^{b c} \lambda^{\alpha}_{i} {W}^{-2} \nabla^{d}{W} \nabla^{e}{W_{b c \alpha}\,^{i}}+\frac{83}{3072}\epsilon^{a d}\,_{{e_{1}}}\,^{b c} (\Sigma_{d e})^{\beta \alpha} \lambda_{i \beta} {W}^{-2} \nabla^{{e_{1}}}{W} \nabla^{e}{W_{b c \alpha}\,^{i}}+\frac{83}{3072}\epsilon^{a d e}\,_{{e_{1}}}\,^{b} (\Sigma_{d e})^{\beta \alpha} \lambda_{i \beta} {W}^{-2} \nabla^{c}{W} \nabla^{{e_{1}}}{W_{b c \alpha}\,^{i}}+\frac{157}{1536}(\Gamma^{a})^{\beta \alpha} \lambda_{i \beta} {W}^{-2} \nabla^{b}{W} \nabla^{c}{W_{b c \alpha}\,^{i}} - \frac{1}{32}(\Gamma^{b})^{\beta \alpha} \lambda_{i \beta} {W}^{-2} \nabla^{c}{W} \nabla^{a}{W_{b c \alpha}\,^{i}} - \frac{131}{1536}\epsilon^{c d}\,_{e {e_{1}}}\,^{b} (\Sigma_{c d})^{\beta \alpha} \lambda_{i \beta} {W}^{-2} \nabla^{e}{W} \nabla^{{e_{1}}}{W_{b}\,^{a}\,_{\alpha}\,^{i}}+\frac{1}{32}(\Gamma^{b})^{\beta \alpha} \lambda_{i \beta} {W}^{-2} \nabla_{c}{W} \nabla^{c}{W_{b}\,^{a}\,_{\alpha}\,^{i}}+\frac{1}{32}(\Gamma_{c})^{\beta \alpha} \lambda_{i \beta} {W}^{-2} \nabla^{c}{W} \nabla^{b}{W^{a}\,_{b \alpha}\,^{i}} - \frac{179}{6144}\epsilon^{d e}\,_{{e_{1}}}\,^{b c} (\Sigma_{d e})^{\beta \alpha} \lambda_{i \beta} {W}^{-2} \nabla^{{e_{1}}}{W} \nabla^{a}{W_{b c \alpha}\,^{i}}%
 - \frac{179}{3072}\epsilon^{a d}\,_{{e_{1}}}\,^{b c} (\Sigma_{d e})^{\beta \alpha} \lambda_{i \beta} {W}^{-2} \nabla^{e}{W} \nabla^{{e_{1}}}{W_{b c \alpha}\,^{i}} - \frac{179}{3072}\epsilon^{a d e}\,_{{e_{1}}}\,^{b} (\Sigma_{d e})^{\beta \alpha} \lambda_{i \beta} {W}^{-2} \nabla^{{e_{1}}}{W} \nabla^{c}{W_{b c \alpha}\,^{i}} - \frac{3}{4}(\Gamma_{b})^{\beta \alpha} \lambda_{i \beta} X^{i}_{\alpha} {W}^{-3} \nabla^{a}{W} \nabla^{b}{W}+\frac{3}{4}(\Gamma^{a})^{\beta \alpha} \lambda_{i \beta} X^{i}_{\alpha} {W}^{-3} \nabla_{b}{W} \nabla^{b}{W}+\frac{5}{16}(\Sigma_{b c})^{\beta \alpha} \lambda_{i \beta} X^{i}_{\alpha} {W}^{-2} \nabla^{a}{F^{b c}}+\frac{3}{8}\epsilon^{a}\,_{d e}\,^{b c} (\Gamma^{d})^{\alpha \beta} F_{b c} X_{i \alpha} {W}^{-2} \nabla^{e}{\lambda^{i}_{\beta}} - \frac{3}{8}\epsilon^{a}\,_{d e}\,^{b c} (\Gamma^{d})^{\beta \alpha} F_{b c} \lambda_{i \beta} {W}^{-2} \nabla^{e}{X^{i}_{\alpha}} - \frac{27}{16}W^{a}\,_{b} X_{i}^{\alpha} {W}^{-1} \nabla^{b}{\lambda^{i}_{\alpha}}+\frac{27}{16}W^{a}\,_{b} \lambda^{\alpha}_{i} {W}^{-1} \nabla^{b}{X^{i}_{\alpha}}+\frac{27}{32}\epsilon^{a}\,_{d e}\,^{b c} (\Gamma^{d})^{\alpha \beta} W_{b c} X_{i \alpha} {W}^{-1} \nabla^{e}{\lambda^{i}_{\beta}} - \frac{27}{32}\epsilon^{a}\,_{d e}\,^{b c} (\Gamma^{d})^{\beta \alpha} W_{b c} \lambda_{i \beta} {W}^{-1} \nabla^{e}{X^{i}_{\alpha}}+\frac{1}{32}\epsilon^{a}\,_{d e}\,^{b c} (\Gamma^{d})^{\beta \alpha} \lambda_{i \beta} X^{i}_{\alpha} {W}^{-2} \nabla^{e}{F_{b c}} - \frac{5}{8}(\Sigma_{c b})^{\beta \alpha} \lambda_{i \beta} X^{i}_{\alpha} {W}^{-2} \nabla^{c}{F^{a b}}+\frac{27}{16}\lambda^{\alpha}_{i} X^{i}_{\alpha} {W}^{-1} \nabla_{b}{W^{a b}} - \frac{27}{32}\epsilon^{a}\,_{d e}\,^{b c} (\Gamma^{d})^{\beta \alpha} \lambda_{i \beta} X^{i}_{\alpha} {W}^{-1} \nabla^{e}{W_{b c}}+\frac{1}{2}X_{i j} F^{a}\,_{b} {W}^{-3} \nabla^{b}{X^{i j}}+\frac{3}{4}(\Sigma^{a}{}_{\, b})^{\alpha \beta} X_{i j} X^{i}_{\alpha} {W}^{-2} \nabla^{b}{\lambda^{j}_{\beta}}+\frac{1}{4}X_{i j} X^{i j} {W}^{-3} \nabla_{b}{F^{a b}} - \frac{3}{4}(\Sigma^{a}{}_{\, b})^{\beta \alpha} X_{i j} \lambda^{i}_{\beta} {W}^{-2} \nabla^{b}{X^{j}_{\alpha}} - \frac{3}{4}(\Sigma^{a}{}_{\, b})^{\beta \alpha} \lambda_{j \beta} X_{i \alpha} {W}^{-2} \nabla^{b}{X^{j i}}%
+\frac{1}{6}C^{a}\,_{b c d} F^{c d} {W}^{-2} \nabla^{b}{W}+\frac{1}{4}W^{a}\,_{d} W_{b c} F^{b c} {W}^{-2} \nabla^{d}{W} - \frac{75}{32}W^{c d} W_{c d} F^{a}\,_{b} {W}^{-2} \nabla^{b}{W} - \frac{1}{4}W^{a}\,_{b} W_{c d} F^{b c} {W}^{-2} \nabla^{d}{W} - \frac{1}{4}W^{a d} W_{d c} F_{b}\,^{c} {W}^{-2} \nabla^{b}{W} - \frac{1}{4}W^{c}\,_{b} W_{c d} F^{a b} {W}^{-2} \nabla^{d}{W}+\frac{1}{6}C_{b c}\,^{a}\,_{d} F^{b c} {W}^{-2} \nabla^{d}{W}+\frac{1}{6}C_{b c}\,^{a}\,_{d} F^{b d} {W}^{-2} \nabla^{c}{W} - \frac{1}{6}C^{a}\,_{b c d} F^{b c} {W}^{-2} \nabla^{d}{W}+\frac{1}{6}C^{a}\,_{b c d} W^{c d} {W}^{-1} \nabla^{b}{W} - \frac{1}{4}W^{b c} W^{a}\,_{b} W_{c d} {W}^{-1} \nabla^{d}{W} - \frac{1}{4}W^{a b} W_{c}\,^{d} W_{b d} {W}^{-1} \nabla^{c}{W}+\frac{1}{6}C_{b c}\,^{a}\,_{d} W^{b c} {W}^{-1} \nabla^{d}{W}+\frac{1}{6}C_{b c}\,^{a}\,_{d} W^{b d} {W}^{-1} \nabla^{c}{W} - \frac{1}{6}C^{a}\,_{b c d} W^{b c} {W}^{-1} \nabla^{d}{W}+\frac{5}{48}\epsilon^{a b c {e_{1}} {e_{2}}} C_{b c d e} W^{d e} W_{{e_{1}} {e_{2}}} - \frac{121}{1024}\epsilon^{a d e {e_{1}} {e_{2}}} W^{b c} W_{d e} W_{{e_{1}} {e_{2}}} W_{b c}+\frac{563}{768}\epsilon^{a d e {e_{1}} {e_{2}}} W^{b c} W_{d e} W_{{e_{1}} b} W_{{e_{2}} c}+\frac{25}{96}\epsilon^{a}\,_{b}\,^{d e {e_{1}}} W^{b c} W_{d e} W_{{e_{1}}}\,^{{e_{2}}} W_{{e_{2}} c}+\frac{5}{48}\epsilon^{a d e {e_{1}} {e_{2}}} C_{b c d e} W^{b c} W_{{e_{1}} {e_{2}}}%
+\frac{5}{48}\epsilon^{a b d {e_{1}} {e_{2}}} C_{b c d e} W^{c e} W_{{e_{1}} {e_{2}}} - \frac{5}{96}\epsilon^{c d e {e_{1}} {e_{2}}} C^{a b}\,_{c d} W_{e {e_{1}}} W_{{e_{2}} b}+\frac{283}{1536}\epsilon^{c d e {e_{1}} {e_{2}}} W^{a b} W_{c d} W_{e {e_{1}}} W_{{e_{2}} b}+\frac{5}{24}\epsilon^{a b d {e_{1}} {e_{2}}} C_{b}\,^{c}\,_{d}\,^{e} W_{{e_{1}} c} W_{{e_{2}} e}+\frac{3}{16}{\rm i} \epsilon^{d e}\,_{{e_{1}}}\,^{b c} (\Sigma_{d e})^{\alpha \beta} F_{b c} \lambda_{i \alpha} {W}^{-3} \nabla^{{e_{1}}}{\nabla^{a}{\lambda^{i}_{\beta}}}+\frac{1}{4}{\rm i} (\Gamma^{b})^{\alpha \beta} F_{b c} \lambda_{i \alpha} {W}^{-3} \nabla^{c}{\nabla^{a}{\lambda^{i}_{\beta}}}+\frac{1}{8}{\rm i} \epsilon^{d e}\,_{{e_{1}}}\,^{b c} (\Sigma_{d e})^{\alpha \beta} F_{b c} {W}^{-3} \nabla^{a}{\lambda_{i \alpha}} \nabla^{{e_{1}}}{\lambda^{i}_{\beta}} - \frac{1}{4}{\rm i} (\Gamma^{b})^{\alpha \beta} F_{b c} {W}^{-3} \nabla^{a}{\lambda_{i \alpha}} \nabla^{c}{\lambda^{i}_{\beta}}+\frac{25}{128}{\rm i} \epsilon^{d e}\,_{{e_{1}}}\,^{b c} (\Sigma_{d e})^{\alpha \beta} W_{b c} \lambda_{i \alpha} {W}^{-2} \nabla^{{e_{1}}}{\nabla^{a}{\lambda^{i}_{\beta}}}+\frac{1}{4}{\rm i} (\Gamma^{b})^{\alpha \beta} W_{b c} \lambda_{i \alpha} {W}^{-2} \nabla^{c}{\nabla^{a}{\lambda^{i}_{\beta}}}+\frac{3}{32}{\rm i} \epsilon^{d e}\,_{{e_{1}}}\,^{b c} (\Sigma_{d e})^{\alpha \beta} W_{b c} {W}^{-2} \nabla^{a}{\lambda_{i \alpha}} \nabla^{{e_{1}}}{\lambda^{i}_{\beta}} - \frac{3}{8}{\rm i} (\Gamma^{b})^{\alpha \beta} W_{b c} {W}^{-2} \nabla^{a}{\lambda_{i \alpha}} \nabla^{c}{\lambda^{i}_{\beta}} - \frac{5}{64}{\rm i} \epsilon^{d e}\,_{{e_{1}}}\,^{b c} (\Sigma_{d e})^{\alpha \beta} \lambda_{i \alpha} {W}^{-3} \nabla^{{e_{1}}}{F_{b c}} \nabla^{a}{\lambda^{i}_{\beta}}+\frac{1}{4}{\rm i} (\Gamma^{b})^{\alpha \beta} \lambda_{i \alpha} {W}^{-3} \nabla^{c}{F_{b c}} \nabla^{a}{\lambda^{i}_{\beta}}+\frac{11}{256}{\rm i} \epsilon^{d e}\,_{{e_{1}}}\,^{b c} (\Sigma_{d e})^{\alpha \beta} \lambda_{i \alpha} {W}^{-2} \nabla^{{e_{1}}}{W_{b c}} \nabla^{a}{\lambda^{i}_{\beta}}+\frac{3}{8}{\rm i} (\Gamma^{b})^{\alpha \beta} \lambda_{i \alpha} {W}^{-2} \nabla^{c}{W_{b c}} \nabla^{a}{\lambda^{i}_{\beta}}+\frac{139}{1024}{\rm i} \epsilon^{d e b c}\,_{{e_{1}}} (\Sigma_{d e})^{\alpha \beta} \lambda_{i \alpha} {W}^{-2} \nabla^{a}{W_{b c}} \nabla^{{e_{1}}}{\lambda^{i}_{\beta}} - \frac{1}{8}{\rm i} \epsilon^{a}\,_{d e}\,^{b c} F_{b c} {W}^{-3} \nabla^{d}{\lambda^{\alpha}_{i}} \nabla^{e}{\lambda^{i}_{\alpha}} - \frac{1}{2}{\rm i} \epsilon^{a d}\,_{{e_{1}}}\,^{b c} (\Sigma_{d e})^{\alpha \beta} F_{b c} {W}^{-3} \nabla^{e}{\lambda_{i \alpha}} \nabla^{{e_{1}}}{\lambda^{i}_{\beta}} - \frac{1}{4}{\rm i} (\Gamma^{a})^{\alpha \beta} F_{b c} {W}^{-3} \nabla^{b}{\lambda_{i \alpha}} \nabla^{c}{\lambda^{i}_{\beta}}%
+\frac{1}{16}{\rm i} \epsilon^{a}\,_{d e}\,^{b c} W_{b c} \lambda^{\alpha}_{i} {W}^{-2} \nabla^{d}{\nabla^{e}{\lambda^{i}_{\alpha}}} - \frac{3}{16}{\rm i} \epsilon^{a}\,_{d e}\,^{b c} W_{b c} {W}^{-2} \nabla^{d}{\lambda^{\alpha}_{i}} \nabla^{e}{\lambda^{i}_{\alpha}} - \frac{5}{128}{\rm i} \epsilon^{d e}\,_{{e_{1}}}\,^{b c} (\Sigma_{d e})^{\alpha \beta} W_{b c} \lambda_{i \alpha} {W}^{-2} \nabla^{a}{\nabla^{{e_{1}}}{\lambda^{i}_{\beta}}}+\frac{1}{8}{\rm i} (\Gamma^{b})^{\alpha \beta} W_{b c} \lambda_{i \alpha} {W}^{-2} \nabla^{a}{\nabla^{c}{\lambda^{i}_{\beta}}} - \frac{1}{2}{\rm i} (\Gamma_{c})^{\alpha \beta} W^{a}\,_{b} \lambda_{i \alpha} {W}^{-2} \nabla^{b}{\nabla^{c}{\lambda^{i}_{\beta}}} - \frac{5}{8}{\rm i} \epsilon^{a d}\,_{{e_{1}}}\,^{b c} (\Sigma_{d e})^{\alpha \beta} W_{b c} \lambda_{i \alpha} {W}^{-2} \nabla^{{e_{1}}}{\nabla^{e}{\lambda^{i}_{\beta}}} - \frac{5}{16}{\rm i} \epsilon^{a d e}\,_{{e_{1}}}\,^{b} (\Sigma_{d e})^{\alpha \beta} W_{b c} \lambda_{i \alpha} {W}^{-2} \nabla^{c}{\nabla^{{e_{1}}}{\lambda^{i}_{\beta}}}+\frac{1}{4}{\rm i} (\Gamma^{a})^{\alpha \beta} W_{b c} \lambda_{i \alpha} {W}^{-2} \nabla^{b}{\nabla^{c}{\lambda^{i}_{\beta}}}+\frac{9}{64}{\rm i} \epsilon^{c d}\,_{e {e_{1}}}\,^{b} (\Sigma_{c d})^{\alpha \beta} W_{b}\,^{a} \lambda_{i \alpha} {W}^{-2} \nabla^{e}{\nabla^{{e_{1}}}{\lambda^{i}_{\beta}}}+\frac{1}{8}{\rm i} (\Gamma_{c})^{\alpha \beta} W^{a}\,_{b} \lambda_{i \alpha} {W}^{-2} \nabla^{c}{\nabla^{b}{\lambda^{i}_{\beta}}} - \frac{3}{8}{\rm i} (\Gamma^{b})^{\alpha \beta} W_{b}\,^{a} \lambda_{i \alpha} {W}^{-2} \nabla_{c}{\nabla^{c}{\lambda^{i}_{\beta}}}+\frac{3}{8}{\rm i} (\Gamma_{c})^{\alpha \beta} W^{a}\,_{b} {W}^{-2} \nabla^{c}{\lambda_{i \alpha}} \nabla^{b}{\lambda^{i}_{\beta}} - \frac{15}{16}{\rm i} \epsilon^{a d}\,_{{e_{1}}}\,^{b c} (\Sigma_{d e})^{\alpha \beta} W_{b c} {W}^{-2} \nabla^{e}{\lambda_{i \alpha}} \nabla^{{e_{1}}}{\lambda^{i}_{\beta}} - \frac{3}{16}{\rm i} \epsilon^{a d e}\,_{{e_{1}}}\,^{b} (\Sigma_{d e})^{\alpha \beta} W_{b c} {W}^{-2} \nabla^{{e_{1}}}{\lambda_{i \alpha}} \nabla^{c}{\lambda^{i}_{\beta}} - \frac{3}{16}{\rm i} \epsilon^{a}\,_{d}\,^{b c}\,_{e} \lambda^{\alpha}_{i} {W}^{-2} \nabla^{d}{W_{b c}} \nabla^{e}{\lambda^{i}_{\alpha}} - \frac{3}{8}{\rm i} (\Gamma_{c})^{\alpha \beta} \lambda_{i \alpha} {W}^{-2} \nabla_{b}{W^{a b}} \nabla^{c}{\lambda^{i}_{\beta}}+\frac{7}{128}{\rm i} \epsilon^{a d}\,_{e c {e_{1}}} (\Sigma_{d b})^{\alpha \beta} \lambda_{i \alpha} {W}^{-2} \nabla^{e}{W^{b c}} \nabla^{{e_{1}}}{\lambda^{i}_{\beta}} - \frac{183}{1024}{\rm i} \epsilon^{a d e b c} (\Sigma_{d e})^{\alpha \beta} \lambda_{i \alpha} {W}^{-2} \nabla_{{e_{1}}}{W_{b c}} \nabla^{{e_{1}}}{\lambda^{i}_{\beta}}+\frac{3}{8}{\rm i} (\Gamma_{b})^{\alpha \beta} \lambda_{i \alpha} {W}^{-2} \nabla_{c}{W^{a b}} \nabla^{c}{\lambda^{i}_{\beta}} - \frac{129}{128}{\rm i} \epsilon^{a d}\,_{{e_{1}}}\,^{b c} (\Sigma_{d e})^{\alpha \beta} \lambda_{i \alpha} {W}^{-2} \nabla^{{e_{1}}}{W_{b c}} \nabla^{e}{\lambda^{i}_{\beta}}%
 - \frac{55}{512}{\rm i} \epsilon^{a d e b}\,_{{e_{1}}} (\Sigma_{d e})^{\alpha \beta} \lambda_{i \alpha} {W}^{-2} \nabla^{c}{W_{b c}} \nabla^{{e_{1}}}{\lambda^{i}_{\beta}}+\frac{95}{256}{\rm i} \epsilon^{c d}\,_{e b {e_{1}}} (\Sigma_{c d})^{\alpha \beta} \lambda_{i \alpha} {W}^{-2} \nabla^{e}{W^{a b}} \nabla^{{e_{1}}}{\lambda^{i}_{\beta}}+\frac{1}{16}{\rm i} \epsilon^{d e b c}\,_{{e_{1}}} (\Sigma_{d e})^{\alpha \beta} \lambda_{i \alpha} {W}^{-3} \nabla^{a}{F_{b c}} \nabla^{{e_{1}}}{\lambda^{i}_{\beta}}+\frac{1}{4}{\rm i} (\Gamma_{c})^{\alpha \beta} \lambda_{i \alpha} {W}^{-3} \nabla_{b}{F^{a b}} \nabla^{c}{\lambda^{i}_{\beta}} - \frac{3}{16}{\rm i} \epsilon^{a d}\,_{{e_{1}}}\,^{b c} (\Sigma_{d e})^{\alpha \beta} \lambda_{i \alpha} {W}^{-3} \nabla^{{e_{1}}}{F_{b c}} \nabla^{e}{\lambda^{i}_{\beta}} - \frac{1}{4}{\rm i} (\Gamma^{a})^{\alpha \beta} \lambda_{i \alpha} {W}^{-3} \nabla^{b}{F_{b c}} \nabla^{c}{\lambda^{i}_{\beta}}+\frac{9}{32}{\rm i} \epsilon^{c d}\,_{e b {e_{1}}} (\Sigma_{c d})^{\alpha \beta} \lambda_{i \alpha} {W}^{-3} \nabla^{e}{F^{a b}} \nabla^{{e_{1}}}{\lambda^{i}_{\beta}}+\frac{1}{4}{\rm i} (\Gamma_{b})^{\alpha \beta} \lambda_{i \alpha} {W}^{-3} \nabla_{c}{F^{a b}} \nabla^{c}{\lambda^{i}_{\beta}}+\frac{1}{4}{\rm i} (\Gamma_{c})^{\alpha \beta} F^{a}\,_{b} \lambda_{i \alpha} {W}^{-3} \nabla^{c}{\nabla^{b}{\lambda^{i}_{\beta}}} - \frac{1}{4}{\rm i} \epsilon^{a d}\,_{{e_{1}}}\,^{b c} (\Sigma_{d e})^{\alpha \beta} F_{b c} \lambda_{i \alpha} {W}^{-3} \nabla^{e}{\nabla^{{e_{1}}}{\lambda^{i}_{\beta}}} - \frac{3}{16}{\rm i} \epsilon^{a d e}\,_{{e_{1}}}\,^{b} (\Sigma_{d e})^{\alpha \beta} F_{b c} \lambda_{i \alpha} {W}^{-3} \nabla^{{e_{1}}}{\nabla^{c}{\lambda^{i}_{\beta}}} - \frac{1}{16}{\rm i} \epsilon^{c d}\,_{e {e_{1}}}\,^{b} (\Sigma_{c d})^{\alpha \beta} F_{b}\,^{a} \lambda_{i \alpha} {W}^{-3} \nabla^{e}{\nabla^{{e_{1}}}{\lambda^{i}_{\beta}}} - \frac{1}{4}{\rm i} (\Gamma^{b})^{\alpha \beta} F_{b}\,^{a} \lambda_{i \alpha} {W}^{-3} \nabla_{c}{\nabla^{c}{\lambda^{i}_{\beta}}}+\frac{1}{4}{\rm i} (\Gamma_{c})^{\alpha \beta} F^{a}\,_{b} \lambda_{i \alpha} {W}^{-3} \nabla^{b}{\nabla^{c}{\lambda^{i}_{\beta}}}+\frac{1}{32}{\rm i} \epsilon^{d e}\,_{{e_{1}}}\,^{b c} (\Sigma_{d e})^{\alpha \beta} F_{b c} \lambda_{i \alpha} {W}^{-3} \nabla^{a}{\nabla^{{e_{1}}}{\lambda^{i}_{\beta}}} - \frac{1}{16}{\rm i} \epsilon^{a d}\,_{{e_{1}}}\,^{b c} (\Sigma_{d e})^{\alpha \beta} F_{b c} \lambda_{i \alpha} {W}^{-3} \nabla^{{e_{1}}}{\nabla^{e}{\lambda^{i}_{\beta}}} - \frac{3}{16}{\rm i} \epsilon^{a d}\,_{{e_{1}}}\,^{b c} (\Sigma_{d e})^{\alpha \beta} W_{b c} \lambda_{i \alpha} {W}^{-2} \nabla^{e}{\nabla^{{e_{1}}}{\lambda^{i}_{\beta}}}+\frac{1}{16}{\rm i} \epsilon^{d e}\,_{{e_{1}}}\,^{b c} (\Sigma_{d e})^{\alpha \beta} \lambda_{i \alpha} \lambda^{i}_{\beta} {W}^{-2} \nabla^{{e_{1}}}{\nabla^{a}{W_{b c}}}+\frac{19}{128}{\rm i} \epsilon^{a d}\,_{{e_{1}}}\,^{b c} (\Sigma_{d e})^{\alpha \beta} \lambda_{i \alpha} \lambda^{i}_{\beta} {W}^{-2} \nabla^{e}{\nabla^{{e_{1}}}{W_{b c}}}+\frac{7}{64}{\rm i} \epsilon^{a d e}\,_{{e_{1}}}\,^{b} (\Sigma_{d e})^{\alpha \beta} \lambda_{i \alpha} \lambda^{i}_{\beta} {W}^{-2} \nabla^{{e_{1}}}{\nabla^{c}{W_{b c}}}%
 - \frac{7}{64}{\rm i} \epsilon^{c d}\,_{e {e_{1}} b} (\Sigma_{c d})^{\alpha \beta} \lambda_{i \alpha} \lambda^{i}_{\beta} {W}^{-2} \nabla^{e}{\nabla^{{e_{1}}}{W^{a b}}}+\frac{1}{32}{\rm i} \epsilon^{d e}\,_{{e_{1}}}\,^{b c} (\Sigma_{d e})^{\alpha \beta} \lambda_{i \alpha} \lambda^{i}_{\beta} {W}^{-2} \nabla^{a}{\nabla^{{e_{1}}}{W_{b c}}}+\frac{5}{128}{\rm i} \epsilon^{a d}\,_{{e_{1}}}\,^{b c} (\Sigma_{d e})^{\alpha \beta} \lambda_{i \alpha} \lambda^{i}_{\beta} {W}^{-2} \nabla^{{e_{1}}}{\nabla^{e}{W_{b c}}}+\frac{5}{64}{\rm i} \epsilon^{a d e}\,_{{e_{1}}}\,^{b} (\Sigma_{d e})^{\alpha \beta} \lambda_{i \alpha} \lambda^{i}_{\beta} {W}^{-2} \nabla^{c}{\nabla^{{e_{1}}}{W_{b c}}}+\frac{1}{4}{\rm i} (\Gamma_{c})^{\alpha \beta} \lambda_{i \alpha} {W}^{-3} \nabla^{c}{F^{a}\,_{b}} \nabla^{b}{\lambda^{i}_{\beta}}+\frac{5}{32}{\rm i} \epsilon^{a d b c}\,_{{e_{1}}} (\Sigma_{d e})^{\alpha \beta} \lambda_{i \alpha} {W}^{-3} \nabla^{e}{F_{b c}} \nabla^{{e_{1}}}{\lambda^{i}_{\beta}}+\frac{3}{16}{\rm i} \epsilon^{a d e}\,_{{e_{1}}}\,^{b} (\Sigma_{d e})^{\alpha \beta} \lambda_{i \alpha} {W}^{-3} \nabla^{{e_{1}}}{F_{b c}} \nabla^{c}{\lambda^{i}_{\beta}}+\frac{123}{512}{\rm i} \epsilon^{a d b c}\,_{{e_{1}}} (\Sigma_{d e})^{\alpha \beta} \lambda_{i \alpha} {W}^{-2} \nabla^{e}{W_{b c}} \nabla^{{e_{1}}}{\lambda^{i}_{\beta}}+\frac{73}{256}{\rm i} \epsilon^{a d e}\,_{{e_{1}}}\,^{b} (\Sigma_{d e})^{\alpha \beta} \lambda_{i \alpha} {W}^{-2} \nabla^{{e_{1}}}{W_{b c}} \nabla^{c}{\lambda^{i}_{\beta}} - \frac{7}{8}{\rm i} W^{a}\,_{b}\,^{\alpha}\,_{i} X^{i}_{\alpha} {W}^{-1} \nabla^{b}{W}+\frac{3}{32}{\rm i} (\Sigma^{a}{}_{\, b})^{\alpha \beta} Y \lambda_{i \alpha} \lambda^{i}_{\beta} {W}^{-3} \nabla^{b}{W}+\frac{23}{1024}{\rm i} \epsilon^{a b c d e} W_{b c} W_{d e}\,^{\alpha}\,_{i} X^{i}_{\alpha}+\frac{19}{512}{\rm i} (\Gamma^{a})^{\alpha \beta} W^{b c} W_{b c \alpha i} X^{i}_{\beta} - \frac{19}{512}{\rm i} \epsilon^{{e_{1}} b c d e} (\Sigma_{{e_{1}}}{}^{\, a})^{\alpha \beta} W_{b c} W_{d e \alpha i} X^{i}_{\beta}+\frac{19}{256}{\rm i} W_{b}\,^{c} W_{d c}\,^{\alpha}\,_{i} \epsilon^{a b d e {e_{1}}} (\Sigma_{e {e_{1}}})_{\alpha}{}^{\beta} X^{i}_{\beta} - \frac{19}{256}{\rm i} W^{a b} W_{b c}\,^{\alpha}\,_{i} (\Gamma^{c})_{\alpha}{}^{\beta} X^{i}_{\beta} - \frac{19}{256}{\rm i} W_{b}\,^{c} W^{a}\,_{c}\,^{\alpha}\,_{i} (\Gamma^{b})_{\alpha}{}^{\beta} X^{i}_{\beta} - \frac{1}{6}{\rm i} (\Gamma_{b})^{\alpha \beta} X_{i j} \lambda^{i}_{\alpha} {W}^{-3} \nabla^{b}{\nabla^{a}{\lambda^{j}_{\beta}}} - \frac{1}{12}{\rm i} (\Gamma_{b})^{\alpha \beta} X_{i j} \lambda^{i}_{\alpha} {W}^{-3} \nabla^{a}{\nabla^{b}{\lambda^{j}_{\beta}}} - \frac{1}{4}{\rm i} (\Gamma_{b})^{\alpha \beta} X_{i j} {W}^{-3} \nabla^{a}{\lambda^{i}_{\alpha}} \nabla^{b}{\lambda^{j}_{\beta}}%
 - \frac{1}{6}{\rm i} (\Gamma_{b})^{\alpha \beta} \lambda_{i \alpha} \lambda_{j \beta} {W}^{-3} \nabla^{b}{\nabla^{a}{X^{i j}}}+\frac{1}{2}{\rm i} (\Gamma_{b})^{\alpha \beta} \lambda_{i \alpha} {W}^{-3} \nabla^{a}{X^{i}\,_{j}} \nabla^{b}{\lambda^{j}_{\beta}}+\frac{1}{4}{\rm i} (\Gamma_{b})^{\alpha \beta} \lambda_{i \alpha} {W}^{-3} \nabla^{b}{X^{i}\,_{j}} \nabla^{a}{\lambda^{j}_{\beta}}+\frac{1}{4}{\rm i} (\Gamma^{a})^{\alpha \beta} X_{i j} \lambda^{i}_{\alpha} {W}^{-3} \nabla_{b}{\nabla^{b}{\lambda^{j}_{\beta}}} - \frac{1}{4}{\rm i} \epsilon^{a b c}\,_{d e} (\Sigma_{b c})^{\alpha \beta} X_{i j} {W}^{-3} \nabla^{d}{\lambda^{i}_{\alpha}} \nabla^{e}{\lambda^{j}_{\beta}}+\frac{1}{4}{\rm i} (\Gamma^{a})^{\alpha \beta} X_{i j} {W}^{-3} \nabla_{b}{\lambda^{i}_{\alpha}} \nabla^{b}{\lambda^{j}_{\beta}} - \frac{1}{12}{\rm i} (\Gamma_{b})^{\alpha \beta} \lambda_{i \alpha} \lambda_{j \beta} {W}^{-3} \nabla^{a}{\nabla^{b}{X^{i j}}}+\frac{1}{4}{\rm i} (\Gamma^{a})^{\alpha \beta} \lambda_{i \alpha} \lambda_{j \beta} {W}^{-3} \nabla_{b}{\nabla^{b}{X^{i j}}}+\frac{1}{4}{\rm i} \epsilon^{a b c}\,_{d e} (\Sigma_{b c})^{\alpha \beta} \lambda_{i \alpha} {W}^{-3} \nabla^{d}{X^{i}\,_{j}} \nabla^{e}{\lambda^{j}_{\beta}} - \frac{3}{4}{\rm i} (\Gamma^{a})^{\alpha \beta} \lambda_{i \alpha} {W}^{-3} \nabla_{b}{X^{i}\,_{j}} \nabla^{b}{\lambda^{j}_{\beta}}-\Phi^{a}\,_{b i j} X^{i j} {W}^{-2} \nabla^{b}{W}+W^{a}\,_{b} {W}^{-1} \nabla_{c}{\nabla^{c}{\nabla^{b}{W}}}-F^{a}\,_{b} {W}^{-2} \nabla_{c}{\nabla^{c}{\nabla^{b}{W}}}-{W}^{-2} \nabla_{b}{F^{a b}} \nabla_{c}{\nabla^{c}{W}}-{W}^{-2} \nabla_{c}{F^{a}\,_{b}} \nabla^{c}{\nabla^{b}{W}}-{W}^{-2} \nabla_{b}{W} \nabla_{c}{\nabla^{c}{F^{a b}}}-{W}^{-2} \nabla_{c}{W} \nabla^{c}{\nabla_{b}{F^{a b}}} - \frac{3}{2}F^{a}\,_{b} {W}^{-4} \nabla_{c}{W} \nabla^{c}{W} \nabla^{b}{W}+3W^{a}\,_{b} {W}^{-3} \nabla_{c}{W} \nabla^{c}{W} \nabla^{b}{W}+\frac{1}{8}\epsilon_{{e_{1}}}\,^{d e b c} F_{b c} {W}^{-3} \nabla^{a}{W} \nabla^{{e_{1}}}{F_{d e}}%
 - \frac{1}{8}\epsilon_{{e_{1}}}\,^{d e b c} F_{b c} {W}^{-3} \nabla^{{e_{1}}}{W} \nabla^{a}{F_{d e}}+\frac{1}{2}\epsilon^{a d e b c} F_{b c} {W}^{-3} \nabla_{{e_{1}}}{W} \nabla^{{e_{1}}}{F_{d e}}+\frac{1}{2}\epsilon^{a}\,_{e {e_{1}} d}\,^{b} F_{b c} {W}^{-3} \nabla^{e}{W} \nabla^{{e_{1}}}{F^{d c}}+\frac{51}{256}\epsilon^{a d e b c} F_{b c} {W}^{-2} \nabla_{{e_{1}}}{W} \nabla^{{e_{1}}}{W_{d e}}+\frac{7}{16}\epsilon^{a}\,_{e {e_{1}} d}\,^{b} F_{b c} {W}^{-2} \nabla^{e}{W} \nabla^{{e_{1}}}{W^{d c}}+\frac{13}{32}\epsilon^{a b c d e} W_{d e} {W}^{-2} \nabla_{{e_{1}}}{W} \nabla^{{e_{1}}}{F_{b c}}+\frac{3}{8}\epsilon^{a}\,_{e {e_{1}} b}\,^{d} W_{d c} {W}^{-2} \nabla^{e}{W} \nabla^{{e_{1}}}{F^{b c}} - \frac{117}{256}\epsilon^{a d e b c} W_{b c} {W}^{-1} \nabla_{{e_{1}}}{W} \nabla^{{e_{1}}}{W_{d e}} - \frac{1}{2}\epsilon^{a}\,_{e {e_{1}} d}\,^{b} W_{b c} {W}^{-1} \nabla^{e}{W} \nabla^{{e_{1}}}{W^{d c}}+\frac{1}{4}\epsilon^{a}\,_{{e_{1}}}\,^{d e b} F_{b c} {W}^{-3} \nabla^{{e_{1}}}{W} \nabla^{c}{F_{d e}} - \frac{1}{4}\epsilon_{e {e_{1}} d}\,^{b c} F_{b c} {W}^{-3} \nabla^{e}{W} \nabla^{{e_{1}}}{F^{a d}} - \frac{1}{4}\epsilon^{a}\,_{{e_{1}}}\,^{d e b} F_{b c} {W}^{-3} \nabla^{c}{W} \nabla^{{e_{1}}}{F_{d e}} - \frac{1}{8}\epsilon^{a}\,_{{e_{1}}}\,^{d b c} W_{d e} F_{b c} {W}^{-2} \nabla^{{e_{1}}}{\nabla^{e}{W}}+\frac{3}{16}\epsilon^{a d e b c} W_{d e} F_{b c} {W}^{-2} \nabla_{{e_{1}}}{\nabla^{{e_{1}}}{W}}+\frac{1}{4}\epsilon^{a}\,_{e {e_{1}}}\,^{d b} W_{d}\,^{c} F_{b c} {W}^{-2} \nabla^{e}{\nabla^{{e_{1}}}{W}} - \frac{1}{8}\epsilon^{a}\,_{{e_{1}}}\,^{d e b} W_{d e} F_{b c} {W}^{-2} \nabla^{c}{\nabla^{{e_{1}}}{W}} - \frac{3}{4}F^{b c} F^{a}\,_{d} F_{b c} {W}^{-4} \nabla^{d}{W}+3F^{a b} F^{c}\,_{b} F_{c d} {W}^{-4} \nabla^{d}{W} - \frac{1}{4}W^{b c} W^{a}\,_{d} F_{b c} {W}^{-2} \nabla^{d}{W}-3W^{c d} F^{a}\,_{b} F_{c d} {W}^{-3} \nabla^{b}{W}%
+6W_{b d} F^{b c} F^{a}\,_{c} {W}^{-3} \nabla^{d}{W}-3W^{c}\,_{d} F^{a b} F_{c b} {W}^{-3} \nabla^{d}{W}+3W^{a c} F^{b}\,_{c} F_{b d} {W}^{-3} \nabla^{d}{W} - \frac{3}{2}W^{a}\,_{d} F^{b c} F_{b c} {W}^{-3} \nabla^{d}{W}-3W^{b c} F^{a}\,_{b} F_{c d} {W}^{-3} \nabla^{d}{W}+\frac{19}{4}W^{c b} W_{c d} F^{a}\,_{b} {W}^{-2} \nabla^{d}{W}+\frac{19}{4}W^{a d} W^{b}\,_{d} F_{b c} {W}^{-2} \nabla^{c}{W} - \frac{1}{4}W^{a c} W^{b}\,_{d} F_{b c} {W}^{-2} \nabla^{d}{W} - \frac{9}{8}W^{d c} W^{b}\,_{d} F_{b c} {W}^{-2} \nabla^{a}{W} - \frac{3}{32}W_{b c}\,^{\alpha}\,_{i} \epsilon^{b c d e}\,_{{e_{1}}} (\Sigma_{d e})_{\alpha}{}^{\beta} \lambda^{i}_{\beta} {W}^{-3} \nabla^{a}{W} \nabla^{{e_{1}}}{W}+\frac{3}{8}W_{b c}\,^{\alpha}\,_{i} (\Gamma^{b})_{\alpha}{}^{\beta} \lambda^{i}_{\beta} {W}^{-3} \nabla^{a}{W} \nabla^{c}{W} - \frac{3}{8}W^{a}\,_{b}\,^{\alpha}\,_{i} (\Gamma_{c})_{\alpha}{}^{\beta} \lambda^{i}_{\beta} {W}^{-3} \nabla^{b}{W} \nabla^{c}{W} - \frac{3}{16}W_{b c}\,^{\alpha}\,_{i} \epsilon^{a b c d}\,_{{e_{1}}} (\Sigma_{d e})_{\alpha}{}^{\beta} \lambda^{i}_{\beta} {W}^{-3} \nabla^{e}{W} \nabla^{{e_{1}}}{W}+\frac{3}{16}W_{b c}\,^{\alpha}\,_{i} \epsilon^{a b d e}\,_{{e_{1}}} (\Sigma_{d e})_{\alpha}{}^{\beta} \lambda^{i}_{\beta} {W}^{-3} \nabla^{c}{W} \nabla^{{e_{1}}}{W} - \frac{3}{16}W_{b}\,^{a \alpha}\,_{i} (\Gamma^{b})_{\alpha}{}^{\beta} \lambda^{i}_{\beta} {W}^{-3} \nabla_{c}{W} \nabla^{c}{W} - \frac{9}{32}F^{b c} W_{b c}\,^{\alpha}\,_{i} {W}^{-2} \nabla^{a}{\lambda^{i}_{\alpha}}+\frac{53}{512}F^{b c} \lambda^{\alpha}_{i} {W}^{-2} \nabla^{a}{W_{b c \alpha}\,^{i}} - \frac{991}{4096}W^{b c} W_{b c}\,^{\alpha}\,_{i} {W}^{-1} \nabla^{a}{\lambda^{i}_{\alpha}} - \frac{107}{4096}W^{b c} \lambda^{\alpha}_{i} {W}^{-1} \nabla^{a}{W_{b c \alpha}\,^{i}}+\frac{1}{32}W_{b c}\,^{\alpha}\,_{i} \lambda^{i}_{\alpha} {W}^{-1} \nabla^{a}{W^{b c}}%
+\frac{213}{4096}W_{b c}\,^{\alpha}\,_{i} \epsilon^{b c}\,_{{e_{1}}}\,^{d e} (\Gamma^{{e_{1}}})_{\alpha}{}^{\beta} \lambda^{i}_{\beta} {W}^{-1} \nabla^{a}{W_{d e}}+\frac{1}{8}W^{b}\,_{c}\,^{\alpha}\,_{i} (\Sigma_{b d})_{\alpha}{}^{\beta} \lambda^{i}_{\beta} {W}^{-1} \nabla^{a}{W^{c d}}+\frac{163}{256}F^{b}\,_{c} \lambda^{\alpha}_{i} {W}^{-2} \nabla^{c}{W^{a}\,_{b \alpha}\,^{i}} - \frac{19}{256}F^{a b} \lambda^{\alpha}_{i} {W}^{-2} \nabla^{c}{W_{b c \alpha}\,^{i}} - \frac{1}{2}F^{b}\,_{c} W^{a}\,_{b}\,^{\alpha}\,_{i} {W}^{-2} \nabla^{c}{\lambda^{i}_{\alpha}}+\frac{3}{4}F^{a b} W_{b c}\,^{\alpha}\,_{i} {W}^{-2} \nabla^{c}{\lambda^{i}_{\alpha}} - \frac{1}{8}(\Sigma_{b c})^{\alpha \beta} F^{b c} W^{a}\,_{d \alpha i} {W}^{-2} \nabla^{d}{\lambda^{i}_{\beta}} - \frac{3}{64}\epsilon^{b c d e}\,_{{e_{1}}} (\Gamma^{{e_{1}}})^{\alpha \beta} F_{b c} W_{d e \alpha i} {W}^{-2} \nabla^{a}{\lambda^{i}_{\beta}}+\frac{3}{8}(\Sigma_{b}{}^{\, d})^{\alpha \beta} F^{b c} W_{c d \alpha i} {W}^{-2} \nabla^{a}{\lambda^{i}_{\beta}} - \frac{1}{4}(\Sigma^{a}{}_{\, d})^{\alpha \beta} F^{b c} W_{b c \alpha i} {W}^{-2} \nabla^{d}{\lambda^{i}_{\beta}}+\frac{1}{8}F^{a}\,_{b} W^{c d \alpha}\,_{i} (\Sigma_{c d})_{\alpha}{}^{\beta} {W}^{-2} \nabla^{b}{\lambda^{i}_{\beta}}+\frac{1}{4}F^{a b} W^{c}\,_{d}\,^{\alpha}\,_{i} (\Sigma_{b c})_{\alpha}{}^{\beta} {W}^{-2} \nabla^{d}{\lambda^{i}_{\beta}} - \frac{1}{4}F^{b}\,_{c} W^{a d \alpha}\,_{i} (\Sigma_{b d})_{\alpha}{}^{\beta} {W}^{-2} \nabla^{c}{\lambda^{i}_{\beta}}+\frac{1}{8}F^{b c} W^{a}\,_{d}\,^{\alpha}\,_{i} (\Sigma_{b c})_{\alpha}{}^{\beta} {W}^{-2} \nabla^{d}{\lambda^{i}_{\beta}}+\frac{1}{32}\epsilon^{a b c d e} F_{b c} W_{d e}\,^{\alpha}\,_{i} (\Gamma_{{e_{1}}})_{\alpha}{}^{\beta} {W}^{-2} \nabla^{{e_{1}}}{\lambda^{i}_{\beta}} - \frac{1}{32}\epsilon^{b c d e}\,_{{e_{1}}} F_{b c} W_{d e}\,^{\alpha}\,_{i} (\Gamma^{a})_{\alpha}{}^{\beta} {W}^{-2} \nabla^{{e_{1}}}{\lambda^{i}_{\beta}} - \frac{1}{8}\epsilon^{a b d}\,_{e {e_{1}}} (\Gamma^{e})^{\alpha \beta} F_{b}\,^{c} W_{d c \alpha i} {W}^{-2} \nabla^{{e_{1}}}{\lambda^{i}_{\beta}} - \frac{1}{2}(\Sigma^{a d})^{\alpha \beta} F^{b}\,_{c} W_{d b \alpha i} {W}^{-2} \nabla^{c}{\lambda^{i}_{\beta}} - \frac{1}{2}(\Sigma_{b}{}^{\, a})^{\alpha \beta} F^{b c} W_{c d \alpha i} {W}^{-2} \nabla^{d}{\lambda^{i}_{\beta}} - \frac{1}{4}(\Sigma^{c}{}_{\, d})^{\alpha \beta} F^{a b} W_{b c \alpha i} {W}^{-2} \nabla^{d}{\lambda^{i}_{\beta}}%
 - \frac{83}{3072}\epsilon_{{e_{1}}}\,^{b c d e} (\Gamma^{{e_{1}}})^{\beta \alpha} F_{b c} \lambda_{i \beta} {W}^{-2} \nabla^{a}{W_{d e \alpha}\,^{i}} - \frac{203}{384}(\Sigma^{d}{}_{\, b})^{\beta \alpha} F^{b c} \lambda_{i \beta} {W}^{-2} \nabla^{a}{W_{d c \alpha}\,^{i}} - \frac{163}{768}(\Sigma^{a}{}_{\, d})^{\beta \alpha} F^{b c} \lambda_{i \beta} {W}^{-2} \nabla^{d}{W_{b c \alpha}\,^{i}} - \frac{163}{768}(\Sigma^{c d})^{\beta \alpha} F^{a}\,_{b} \lambda_{i \beta} {W}^{-2} \nabla^{b}{W_{c d \alpha}\,^{i}}+\frac{221}{384}(\Sigma^{b d})^{\beta \alpha} F_{b c} \lambda_{i \beta} {W}^{-2} \nabla^{c}{W^{a}\,_{d \alpha}\,^{i}} - \frac{19}{384}(\Sigma^{c}{}_{\, b})^{\beta \alpha} F^{a b} \lambda_{i \beta} {W}^{-2} \nabla^{d}{W_{c d \alpha}\,^{i}} - \frac{403}{768}(\Sigma_{b c})^{\beta \alpha} F^{b c} \lambda_{i \beta} {W}^{-2} \nabla^{d}{W^{a}\,_{d \alpha}\,^{i}}+\frac{3}{128}\epsilon^{a b c d e} (\Gamma_{{e_{1}}})^{\beta \alpha} F_{b c} \lambda_{i \beta} {W}^{-2} \nabla^{{e_{1}}}{W_{d e \alpha}\,^{i}} - \frac{1}{16}\epsilon_{{e_{1}}}\,^{b c d e} (\Gamma^{a})^{\beta \alpha} F_{b c} \lambda_{i \beta} {W}^{-2} \nabla^{{e_{1}}}{W_{d e \alpha}\,^{i}}+\frac{101}{768}\epsilon^{a}\,_{e {e_{1}}}\,^{b d} (\Gamma^{e})^{\beta \alpha} F_{b}\,^{c} \lambda_{i \beta} {W}^{-2} \nabla^{{e_{1}}}{W_{d c \alpha}\,^{i}} - \frac{19}{384}(\Sigma^{a}{}_{\, b})^{\beta \alpha} F^{b c} \lambda_{i \beta} {W}^{-2} \nabla^{d}{W_{c d \alpha}\,^{i}} - \frac{221}{384}(\Sigma_{b d})^{\beta \alpha} F^{b c} \lambda_{i \beta} {W}^{-2} \nabla^{d}{W^{a}\,_{c \alpha}\,^{i}}+\frac{163}{384}(\Sigma^{c}{}_{\, d})^{\beta \alpha} F^{a b} \lambda_{i \beta} {W}^{-2} \nabla^{d}{W_{c b \alpha}\,^{i}} - \frac{163}{384}(\Sigma^{a d})^{\beta \alpha} F^{b}\,_{c} \lambda_{i \beta} {W}^{-2} \nabla^{c}{W_{d b \alpha}\,^{i}} - \frac{1021}{2048}W^{b}\,_{c} W^{a}\,_{b}\,^{\alpha}\,_{i} {W}^{-1} \nabla^{c}{\lambda^{i}_{\alpha}}+\frac{893}{2048}W^{a b} W_{b c}\,^{\alpha}\,_{i} {W}^{-1} \nabla^{c}{\lambda^{i}_{\alpha}}+\frac{581}{2048}W^{b}\,_{c} \lambda^{\alpha}_{i} {W}^{-1} \nabla^{c}{W^{a}\,_{b \alpha}\,^{i}} - \frac{69}{2048}W^{a b} \lambda^{\alpha}_{i} {W}^{-1} \nabla^{c}{W_{b c \alpha}\,^{i}}+\frac{1283}{2048}(\Sigma_{b c})^{\alpha \beta} W^{b c} W^{a}\,_{d \alpha i} {W}^{-1} \nabla^{d}{\lambda^{i}_{\beta}} - \frac{13}{2048}\epsilon^{b c d e}\,_{{e_{1}}} (\Gamma^{{e_{1}}})^{\alpha \beta} W_{b c} W_{d e \alpha i} {W}^{-1} \nabla^{a}{\lambda^{i}_{\beta}}%
+\frac{13}{256}(\Sigma_{b}{}^{\, d})^{\alpha \beta} W^{b c} W_{c d \alpha i} {W}^{-1} \nabla^{a}{\lambda^{i}_{\beta}} - \frac{511}{2048}(\Sigma^{a}{}_{\, d})^{\alpha \beta} W^{b c} W_{b c \alpha i} {W}^{-1} \nabla^{d}{\lambda^{i}_{\beta}}+\frac{1}{4}W^{a}\,_{b} W^{c d \alpha}\,_{i} (\Sigma_{c d})_{\alpha}{}^{\beta} {W}^{-1} \nabla^{b}{\lambda^{i}_{\beta}}+\frac{193}{512}W^{a b} W^{c}\,_{d}\,^{\alpha}\,_{i} (\Sigma_{b c})_{\alpha}{}^{\beta} {W}^{-1} \nabla^{d}{\lambda^{i}_{\beta}} - \frac{1}{2}W^{b}\,_{c} W^{a d \alpha}\,_{i} (\Sigma_{b d})_{\alpha}{}^{\beta} {W}^{-1} \nabla^{c}{\lambda^{i}_{\beta}}+\frac{193}{1024}W^{b c} W^{a}\,_{d}\,^{\alpha}\,_{i} (\Sigma_{b c})_{\alpha}{}^{\beta} {W}^{-1} \nabla^{d}{\lambda^{i}_{\beta}}+\frac{449}{8192}\epsilon^{a b c d e} W_{b c} W_{d e}\,^{\alpha}\,_{i} (\Gamma_{{e_{1}}})_{\alpha}{}^{\beta} {W}^{-1} \nabla^{{e_{1}}}{\lambda^{i}_{\beta}} - \frac{449}{8192}\epsilon^{b c d e}\,_{{e_{1}}} W_{b c} W_{d e}\,^{\alpha}\,_{i} (\Gamma^{a})_{\alpha}{}^{\beta} {W}^{-1} \nabla^{{e_{1}}}{\lambda^{i}_{\beta}} - \frac{449}{2048}\epsilon^{a b d}\,_{e {e_{1}}} (\Gamma^{e})^{\alpha \beta} W_{b}\,^{c} W_{d c \alpha i} {W}^{-1} \nabla^{{e_{1}}}{\lambda^{i}_{\beta}} - \frac{511}{1024}(\Sigma^{a d})^{\alpha \beta} W^{b}\,_{c} W_{d b \alpha i} {W}^{-1} \nabla^{c}{\lambda^{i}_{\beta}} - \frac{385}{1024}(\Sigma_{b}{}^{\, a})^{\alpha \beta} W^{b c} W_{c d \alpha i} {W}^{-1} \nabla^{d}{\lambda^{i}_{\beta}}+\frac{225}{512}(\Sigma_{b d})^{\alpha \beta} W^{b c} W_{c}\,^{a}\,_{\alpha i} {W}^{-1} \nabla^{d}{\lambda^{i}_{\beta}} - \frac{449}{1024}(\Sigma^{c}{}_{\, d})^{\alpha \beta} W^{a b} W_{b c \alpha i} {W}^{-1} \nabla^{d}{\lambda^{i}_{\beta}} - \frac{875}{24576}\epsilon_{{e_{1}}}\,^{b c d e} (\Gamma^{{e_{1}}})^{\beta \alpha} W_{b c} \lambda_{i \beta} {W}^{-1} \nabla^{a}{W_{d e \alpha}\,^{i}} - \frac{1643}{3072}(\Sigma^{d}{}_{\, b})^{\beta \alpha} W^{b c} \lambda_{i \beta} {W}^{-1} \nabla^{a}{W_{d c \alpha}\,^{i}} - \frac{1047}{2048}(\Sigma^{a}{}_{\, d})^{\beta \alpha} W^{b c} \lambda_{i \beta} {W}^{-1} \nabla^{d}{W_{b c \alpha}\,^{i}} - \frac{1047}{2048}(\Sigma^{c d})^{\beta \alpha} W^{a}\,_{b} \lambda_{i \beta} {W}^{-1} \nabla^{b}{W_{c d \alpha}\,^{i}}+\frac{489}{1024}(\Sigma^{b d})^{\beta \alpha} W_{b c} \lambda_{i \beta} {W}^{-1} \nabla^{c}{W^{a}\,_{d \alpha}\,^{i}} - \frac{23}{1024}(\Sigma^{c}{}_{\, b})^{\beta \alpha} W^{a b} \lambda_{i \beta} {W}^{-1} \nabla^{d}{W_{c d \alpha}\,^{i}}+\frac{489}{2048}(\Sigma_{b c})^{\beta \alpha} W^{b c} \lambda_{i \beta} {W}^{-1} \nabla^{d}{W^{a}\,_{d \alpha}\,^{i}}%
+\frac{1}{16}\epsilon^{a b c d e} (\Gamma_{{e_{1}}})^{\beta \alpha} W_{b c} \lambda_{i \beta} {W}^{-1} \nabla^{{e_{1}}}{W_{d e \alpha}\,^{i}} - \frac{3}{32}\epsilon_{{e_{1}}}\,^{b c d e} (\Gamma^{a})^{\beta \alpha} W_{b c} \lambda_{i \beta} {W}^{-1} \nabla^{{e_{1}}}{W_{d e \alpha}\,^{i}}+\frac{233}{2048}\epsilon^{a}\,_{e {e_{1}}}\,^{b d} (\Gamma^{e})^{\beta \alpha} W_{b}\,^{c} \lambda_{i \beta} {W}^{-1} \nabla^{{e_{1}}}{W_{d c \alpha}\,^{i}} - \frac{23}{1024}(\Sigma^{a}{}_{\, b})^{\beta \alpha} W^{b c} \lambda_{i \beta} {W}^{-1} \nabla^{d}{W_{c d \alpha}\,^{i}} - \frac{489}{1024}(\Sigma_{b d})^{\beta \alpha} W^{b c} \lambda_{i \beta} {W}^{-1} \nabla^{d}{W^{a}\,_{c \alpha}\,^{i}}+\frac{1047}{1024}(\Sigma^{c}{}_{\, d})^{\beta \alpha} W^{a b} \lambda_{i \beta} {W}^{-1} \nabla^{d}{W_{c b \alpha}\,^{i}} - \frac{1047}{1024}(\Sigma^{a d})^{\beta \alpha} W^{b}\,_{c} \lambda_{i \beta} {W}^{-1} \nabla^{c}{W_{d b \alpha}\,^{i}}+\frac{1}{12}{\rm i} (\Sigma^{c d})^{\alpha \beta} C^{a}\,_{b c d} \lambda_{i \alpha} {W}^{-2} \nabla^{b}{\lambda^{i}_{\beta}}+\frac{9}{16}{\rm i} (\Sigma_{b c})^{\alpha \beta} W^{b c} W^{a}\,_{d} \lambda_{i \alpha} {W}^{-2} \nabla^{d}{\lambda^{i}_{\beta}}+\frac{15}{16}{\rm i} (\Sigma^{a}{}_{\, d})^{\alpha \beta} W^{b c} W_{b c} \lambda_{i \alpha} {W}^{-2} \nabla^{d}{\lambda^{i}_{\beta}} - \frac{9}{8}{\rm i} (\Sigma^{c}{}_{\, b})^{\alpha \beta} W^{a b} W_{c d} \lambda_{i \alpha} {W}^{-2} \nabla^{d}{\lambda^{i}_{\beta}} - \frac{259}{256}{\rm i} (\Sigma_{b d})^{\alpha \beta} W^{b c} W^{a}\,_{c} \lambda_{i \alpha} {W}^{-2} \nabla^{d}{\lambda^{i}_{\beta}}+\frac{9}{4}{\rm i} (\Sigma^{a}{}_{\, b})^{\alpha \beta} W^{b c} W_{c d} \lambda_{i \alpha} {W}^{-2} \nabla^{d}{\lambda^{i}_{\beta}}+\frac{1}{12}{\rm i} (\Sigma^{b c})^{\alpha \beta} C_{b c}\,^{a}\,_{d} \lambda_{i \alpha} {W}^{-2} \nabla^{d}{\lambda^{i}_{\beta}} - \frac{317}{256}{\rm i} (\Sigma^{c}{}_{\, d})^{\alpha \beta} W^{a b} W_{c b} \lambda_{i \alpha} {W}^{-2} \nabla^{d}{\lambda^{i}_{\beta}}+\frac{1}{12}{\rm i} (\Sigma^{b d})^{\alpha \beta} C_{b c}\,^{a}\,_{d} \lambda_{i \alpha} {W}^{-2} \nabla^{c}{\lambda^{i}_{\beta}} - \frac{1}{12}{\rm i} (\Sigma^{b c})^{\alpha \beta} C^{a}\,_{b c d} \lambda_{i \alpha} {W}^{-2} \nabla^{d}{\lambda^{i}_{\beta}} - \frac{5}{32}W^{a}\,_{b}\,^{\alpha}\,_{i} (\Sigma_{c d})_{\alpha}{}^{\beta} \lambda^{i}_{\beta} {W}^{-1} \nabla^{b}{W^{c d}} - \frac{5}{8}W^{b}\,_{c}\,^{\alpha}\,_{i} (\Sigma_{b d})_{\alpha}{}^{\beta} \lambda^{i}_{\beta} {W}^{-1} \nabla^{d}{W^{a c}}+\frac{5}{16}W^{b c \alpha}\,_{i} (\Sigma_{b c})_{\alpha}{}^{\beta} \lambda^{i}_{\beta} {W}^{-1} \nabla_{d}{W^{a d}}%
+\frac{5}{16}W_{b c}\,^{\alpha}\,_{i} (\Sigma^{a}{}_{\, d})_{\alpha}{}^{\beta} \lambda^{i}_{\beta} {W}^{-1} \nabla^{d}{W^{b c}}+\frac{5}{8}W^{b}\,_{c}\,^{\alpha}\,_{i} (\Sigma^{a}{}_{\, b})_{\alpha}{}^{\beta} \lambda^{i}_{\beta} {W}^{-1} \nabla_{d}{W^{c d}} - \frac{5}{32}W_{b c}\,^{\alpha}\,_{i} \epsilon^{a b c}\,_{{e_{1}}}\,^{d} (\Gamma^{e})_{\alpha}{}^{\beta} \lambda^{i}_{\beta} {W}^{-1} \nabla^{{e_{1}}}{W_{d e}}+\frac{3}{32}W_{b c}\,^{\alpha}\,_{i} \lambda^{i}_{\alpha} {W}^{-1} \nabla^{b}{W^{a c}} - \frac{3}{64}W_{b c}\,^{\alpha}\,_{i} \epsilon^{a b}\,_{{e_{1}}}\,^{d e} (\Gamma^{{e_{1}}})_{\alpha}{}^{\beta} \lambda^{i}_{\beta} {W}^{-1} \nabla^{c}{W_{d e}}+\frac{3}{16}W^{b}\,_{c}\,^{\alpha}\,_{i} (\Sigma_{b d})_{\alpha}{}^{\beta} \lambda^{i}_{\beta} {W}^{-1} \nabla^{c}{W^{a d}}+\frac{3}{16}W_{b c}\,^{\alpha}\,_{i} (\Sigma^{a}{}_{\, d})_{\alpha}{}^{\beta} \lambda^{i}_{\beta} {W}^{-1} \nabla^{b}{W^{c d}}+\frac{149}{2048}W_{b}\,^{a \alpha}\,_{i} \epsilon^{b}\,_{e {e_{1}}}\,^{c d} (\Gamma^{e})_{\alpha}{}^{\beta} \lambda^{i}_{\beta} {W}^{-1} \nabla^{{e_{1}}}{W_{c d}}+\frac{9}{16}W^{a}\,_{d}\,^{\alpha}\,_{i} (\Sigma_{b c})_{\alpha}{}^{\beta} \lambda^{i}_{\beta} {W}^{-2} \nabla^{d}{F^{b c}} - \frac{3}{32}W_{b c}\,^{\alpha}\,_{i} \lambda^{i}_{\alpha} {W}^{-2} \nabla^{a}{F^{b c}}+\frac{1}{64}W_{d e}\,^{\alpha}\,_{i} \epsilon^{d e}\,_{{e_{1}}}\,^{b c} (\Gamma^{{e_{1}}})_{\alpha}{}^{\beta} \lambda^{i}_{\beta} {W}^{-2} \nabla^{a}{F_{b c}}+\frac{1}{8}W^{d}\,_{b}\,^{\alpha}\,_{i} (\Sigma_{d c})_{\alpha}{}^{\beta} \lambda^{i}_{\beta} {W}^{-2} \nabla^{a}{F^{b c}} - \frac{1}{8}W^{c}\,_{b}\,^{\alpha}\,_{i} (\Sigma_{c d})_{\alpha}{}^{\beta} \lambda^{i}_{\beta} {W}^{-2} \nabla^{d}{F^{a b}}+\frac{1}{16}W^{c d \alpha}\,_{i} (\Sigma_{c d})_{\alpha}{}^{\beta} \lambda^{i}_{\beta} {W}^{-2} \nabla_{b}{F^{a b}}+\frac{1}{16}W_{b c}\,^{\alpha}\,_{i} (\Sigma^{a}{}_{\, d})_{\alpha}{}^{\beta} \lambda^{i}_{\beta} {W}^{-2} \nabla^{d}{F^{b c}}+\frac{1}{8}W^{d}\,_{b}\,^{\alpha}\,_{i} (\Sigma^{a}{}_{\, d})_{\alpha}{}^{\beta} \lambda^{i}_{\beta} {W}^{-2} \nabla_{c}{F^{b c}} - \frac{1}{32}W_{d e}\,^{\alpha}\,_{i} \epsilon^{a d e}\,_{{e_{1}}}\,^{b} (\Gamma^{c})_{\alpha}{}^{\beta} \lambda^{i}_{\beta} {W}^{-2} \nabla^{{e_{1}}}{F_{b c}}+\frac{5}{16}W_{c b}\,^{\alpha}\,_{i} \lambda^{i}_{\alpha} {W}^{-2} \nabla^{c}{F^{a b}} - \frac{1}{32}W_{d e}\,^{\alpha}\,_{i} \epsilon^{a d}\,_{{e_{1}}}\,^{b c} (\Gamma^{{e_{1}}})_{\alpha}{}^{\beta} \lambda^{i}_{\beta} {W}^{-2} \nabla^{e}{F_{b c}}+\frac{1}{8}W^{c}\,_{d}\,^{\alpha}\,_{i} (\Sigma_{c b})_{\alpha}{}^{\beta} \lambda^{i}_{\beta} {W}^{-2} \nabla^{d}{F^{a b}}%
+\frac{1}{8}W_{d b}\,^{\alpha}\,_{i} (\Sigma^{a}{}_{\, c})_{\alpha}{}^{\beta} \lambda^{i}_{\beta} {W}^{-2} \nabla^{d}{F^{b c}} - \frac{3}{16}W^{a}\,_{b}\,^{\alpha}\,_{i} \lambda^{i}_{\alpha} {W}^{-2} \nabla_{c}{F^{b c}}+\frac{1}{32}W_{d}\,^{a \alpha}\,_{i} \epsilon^{d}\,_{e {e_{1}}}\,^{b c} (\Gamma^{e})_{\alpha}{}^{\beta} \lambda^{i}_{\beta} {W}^{-2} \nabla^{{e_{1}}}{F_{b c}} - \frac{1}{8}W^{a d \alpha}\,_{i} (\Sigma_{d b})_{\alpha}{}^{\beta} \lambda^{i}_{\beta} {W}^{-2} \nabla_{c}{F^{b c}} - \frac{1}{8}W^{a}\,_{b}\,^{\alpha}\,_{i} (\Sigma_{d c})_{\alpha}{}^{\beta} \lambda^{i}_{\beta} {W}^{-2} \nabla^{d}{F^{b c}}+\frac{3}{4}\epsilon^{a}\,_{d e}\,^{b c} (\Gamma^{d})^{\beta \alpha} F_{b c} \lambda_{i \beta} X^{i}_{\alpha} {W}^{-3} \nabla^{e}{W} - \frac{27}{16}W^{a}\,_{b} \lambda^{\alpha}_{i} X^{i}_{\alpha} {W}^{-2} \nabla^{b}{W}+\frac{27}{32}\epsilon^{a}\,_{d e}\,^{b c} (\Gamma^{d})^{\beta \alpha} W_{b c} \lambda_{i \beta} X^{i}_{\alpha} {W}^{-2} \nabla^{e}{W} - \frac{31}{768}\epsilon^{a}\,_{{e_{1}}}\,^{b d e} (\Gamma^{c})^{\beta \alpha} F_{b c} \lambda_{i \beta} {W}^{-2} \nabla^{{e_{1}}}{W_{d e \alpha}\,^{i}}+\frac{91}{1536}\epsilon_{e {e_{1}}}\,^{b c d} (\Gamma^{e})^{\beta \alpha} F_{b}\,^{a} \lambda_{i \beta} {W}^{-2} \nabla^{{e_{1}}}{W_{c d \alpha}\,^{i}}+\frac{91}{1536}\epsilon^{a}\,_{{e_{1}}}\,^{b c d} (\Gamma^{{e_{1}}})^{\beta \alpha} F_{b c} \lambda_{i \beta} {W}^{-2} \nabla^{e}{W_{d e \alpha}\,^{i}} - \frac{1}{16}\epsilon^{b c d}\,_{e {e_{1}}} (\Gamma^{e})^{\alpha \beta} F_{b c} W_{d}\,^{a}\,_{\alpha i} {W}^{-2} \nabla^{{e_{1}}}{\lambda^{i}_{\beta}} - \frac{1}{4}(\Sigma^{b d})^{\alpha \beta} F_{b c} W^{a}\,_{d \alpha i} {W}^{-2} \nabla^{c}{\lambda^{i}_{\beta}}+\frac{1}{4}(\Sigma^{c}{}_{\, d})^{\alpha \beta} F^{a b} W_{c b \alpha i} {W}^{-2} \nabla^{d}{\lambda^{i}_{\beta}} - \frac{1}{8}(\Sigma^{c d})^{\alpha \beta} F^{a}\,_{b} W_{c d \alpha i} {W}^{-2} \nabla^{b}{\lambda^{i}_{\beta}} - \frac{1}{16}\epsilon^{a b d e}\,_{{e_{1}}} (\Gamma^{c})^{\alpha \beta} F_{b c} W_{d e \alpha i} {W}^{-2} \nabla^{{e_{1}}}{\lambda^{i}_{\beta}} - \frac{1}{16}\epsilon^{a b c d}\,_{{e_{1}}} (\Gamma^{{e_{1}}})^{\alpha \beta} F_{b c} W_{d e \alpha i} {W}^{-2} \nabla^{e}{\lambda^{i}_{\beta}} - \frac{1}{4}(\Sigma_{b}{}^{\, c})^{\alpha \beta} F^{a b} W_{c d \alpha i} {W}^{-2} \nabla^{d}{\lambda^{i}_{\beta}} - \frac{29}{1536}\epsilon^{a}\,_{{e_{1}}}\,^{b d e} (\Gamma^{{e_{1}}})^{\beta \alpha} F_{b c} \lambda_{i \beta} {W}^{-2} \nabla^{c}{W_{d e \alpha}\,^{i}} - \frac{29}{1536}\epsilon_{e {e_{1}}}\,^{b c d} (\Gamma^{e})^{\beta \alpha} F_{b c} \lambda_{i \beta} {W}^{-2} \nabla^{{e_{1}}}{W_{d}\,^{a}\,_{\alpha}\,^{i}}%
+\frac{63}{8192}\epsilon^{b c d e}\,_{{e_{1}}} (\Gamma^{a})^{\alpha \beta} W_{b c} W_{d e \alpha i} {W}^{-1} \nabla^{{e_{1}}}{\lambda^{i}_{\beta}}+\frac{63}{1024}W^{a b} W_{b}\,^{c \alpha}\,_{i} (\Sigma_{c d})_{\alpha}{}^{\beta} {W}^{-1} \nabla^{d}{\lambda^{i}_{\beta}}+\frac{63}{1024}W^{b c} W_{b}\,^{a \alpha}\,_{i} (\Sigma_{c d})_{\alpha}{}^{\beta} {W}^{-1} \nabla^{d}{\lambda^{i}_{\beta}} - \frac{63}{1024}W^{b c} W_{b}\,^{d \alpha}\,_{i} (\Sigma_{c d})_{\alpha}{}^{\beta} {W}^{-1} \nabla^{a}{\lambda^{i}_{\beta}} - \frac{63}{4096}\epsilon^{a b c d}\,_{{e_{1}}} W_{b c} W_{d e}\,^{\alpha}\,_{i} (\Gamma^{e})_{\alpha}{}^{\beta} {W}^{-1} \nabla^{{e_{1}}}{\lambda^{i}_{\beta}} - \frac{63}{4096}\epsilon^{b c d}\,_{e {e_{1}}} W_{b c} W_{d}\,^{a \alpha}\,_{i} (\Gamma^{e})_{\alpha}{}^{\beta} {W}^{-1} \nabla^{{e_{1}}}{\lambda^{i}_{\beta}}+\frac{63}{4096}\epsilon^{a b d e}\,_{{e_{1}}} W_{b c} W_{d e}\,^{\alpha}\,_{i} (\Gamma^{{e_{1}}})_{\alpha}{}^{\beta} {W}^{-1} \nabla^{c}{\lambda^{i}_{\beta}} - \frac{407}{2048}\epsilon^{a}\,_{{e_{1}}}\,^{b d e} (\Gamma^{c})^{\beta \alpha} W_{b c} \lambda_{i \beta} {W}^{-1} \nabla^{{e_{1}}}{W_{d e \alpha}\,^{i}}+\frac{535}{4096}\epsilon_{e {e_{1}}}\,^{b c d} (\Gamma^{e})^{\beta \alpha} W_{b}\,^{a} \lambda_{i \beta} {W}^{-1} \nabla^{{e_{1}}}{W_{c d \alpha}\,^{i}}+\frac{535}{4096}\epsilon^{a}\,_{{e_{1}}}\,^{b c d} (\Gamma^{{e_{1}}})^{\beta \alpha} W_{b c} \lambda_{i \beta} {W}^{-1} \nabla^{e}{W_{d e \alpha}\,^{i}}+\frac{1}{4096}\epsilon^{b c d}\,_{e {e_{1}}} (\Gamma^{e})^{\alpha \beta} W_{b c} W_{d}\,^{a}\,_{\alpha i} {W}^{-1} \nabla^{{e_{1}}}{\lambda^{i}_{\beta}}+\frac{1}{1024}(\Sigma^{b d})^{\alpha \beta} W_{b c} W^{a}\,_{d \alpha i} {W}^{-1} \nabla^{c}{\lambda^{i}_{\beta}} - \frac{1}{1024}(\Sigma^{c}{}_{\, d})^{\alpha \beta} W^{a b} W_{c b \alpha i} {W}^{-1} \nabla^{d}{\lambda^{i}_{\beta}}+\frac{1}{2048}(\Sigma^{c d})^{\alpha \beta} W^{a}\,_{b} W_{c d \alpha i} {W}^{-1} \nabla^{b}{\lambda^{i}_{\beta}}+\frac{1}{4096}\epsilon^{a b d e}\,_{{e_{1}}} (\Gamma^{c})^{\alpha \beta} W_{b c} W_{d e \alpha i} {W}^{-1} \nabla^{{e_{1}}}{\lambda^{i}_{\beta}}+\frac{1}{4096}\epsilon^{a b c d}\,_{{e_{1}}} (\Gamma^{{e_{1}}})^{\alpha \beta} W_{b c} W_{d e \alpha i} {W}^{-1} \nabla^{e}{\lambda^{i}_{\beta}}+\frac{1}{1024}(\Sigma_{b}{}^{\, c})^{\alpha \beta} W^{a b} W_{c d \alpha i} {W}^{-1} \nabla^{d}{\lambda^{i}_{\beta}}+\frac{279}{4096}\epsilon^{a}\,_{{e_{1}}}\,^{b d e} (\Gamma^{{e_{1}}})^{\beta \alpha} W_{b c} \lambda_{i \beta} {W}^{-1} \nabla^{c}{W_{d e \alpha}\,^{i}}+\frac{279}{4096}\epsilon_{e {e_{1}}}\,^{b c d} (\Gamma^{e})^{\beta \alpha} W_{b c} \lambda_{i \beta} {W}^{-1} \nabla^{{e_{1}}}{W_{d}\,^{a}\,_{\alpha}\,^{i}} - \frac{149}{4096}W_{b c}\,^{\alpha}\,_{i} \epsilon^{b c}\,_{{e_{1}}}\,^{d e} (\Gamma^{a})_{\alpha}{}^{\beta} \lambda^{i}_{\beta} {W}^{-1} \nabla^{{e_{1}}}{W_{d e}}%
 - \frac{149}{2048}W_{b c}\,^{\alpha}\,_{i} \epsilon^{b c}\,_{e {e_{1}} d} (\Gamma^{e})_{\alpha}{}^{\beta} \lambda^{i}_{\beta} {W}^{-1} \nabla^{{e_{1}}}{W^{a d}} - \frac{1}{4}W^{a}\,_{b}\,^{\alpha}\,_{i} X^{i}\,_{j} {W}^{-2} \nabla^{b}{\lambda^{j}_{\alpha}}+\frac{11}{24}X_{j i} \lambda^{j \alpha} {W}^{-2} \nabla^{b}{W^{a}\,_{b \alpha}\,^{i}}+{\rm i} \Phi^{a}\,_{b i j} \lambda^{i \alpha} {W}^{-2} \nabla^{b}{\lambda^{j}_{\alpha}}+{\rm i} \Phi^{b}\,_{c i j} (\Sigma^{a}{}_{\, b})^{\alpha \beta} \lambda^{i}_{\alpha} {W}^{-2} \nabla^{c}{\lambda^{j}_{\beta}}+\frac{1}{2}{\rm i} \lambda^{\alpha}_{i} \lambda_{j \alpha} {W}^{-2} \nabla_{b}{\Phi^{a b i j}} - \frac{5}{72}{\rm i} \epsilon^{a}\,_{b e c d} (\Gamma^{b})^{\alpha \beta} \lambda_{i \alpha} \lambda_{j \beta} {W}^{-2} \nabla^{e}{\Phi^{c d i j}} - \frac{3}{4}X_{i j} X^{i j} F^{a}\,_{b} {W}^{-4} \nabla^{b}{W}+\frac{3}{2}(\Sigma^{a}{}_{\, b})^{\beta \alpha} X_{i j} \lambda^{i}_{\beta} X^{j}_{\alpha} {W}^{-3} \nabla^{b}{W}+\frac{1}{24}\epsilon^{a b c {e_{1}} {e_{2}}} C_{b c d e} F^{d e} F_{{e_{1}} {e_{2}}} {W}^{-2}+\frac{5}{96}\epsilon^{a {e_{1}} {e_{2}} d e} W_{{e_{1}} {e_{2}}} W_{b c} F^{b c} F_{d e} {W}^{-2}+\frac{31}{384}\epsilon^{a b c d e} W^{{e_{1}} {e_{2}}} W_{{e_{1}} {e_{2}}} F_{b c} F_{d e} {W}^{-2}+\frac{5}{96}\epsilon^{a {e_{1}} {e_{2}} d e} W_{{e_{1}} b} W_{{e_{2}} c} F^{b c} F_{d e} {W}^{-2}+\frac{5}{48}\epsilon^{a}\,_{b}\,^{{e_{1}} d e} W_{{e_{1}}}\,^{{e_{2}}} W_{{e_{2}} c} F^{b c} F_{d e} {W}^{-2}+\frac{1}{24}\epsilon^{a d e {e_{1}} {e_{2}}} C_{b c d e} F^{b c} F_{{e_{1}} {e_{2}}} {W}^{-2}+\frac{1}{24}\epsilon^{a b d {e_{1}} {e_{2}}} C_{b c d e} F^{c e} F_{{e_{1}} {e_{2}}} {W}^{-2} - \frac{1}{48}\epsilon^{c d e {e_{1}} {e_{2}}} C^{a b}\,_{c d} F_{e {e_{1}}} F_{{e_{2}} b} {W}^{-2}+\frac{139}{192}\epsilon^{{e_{1}} {e_{2}} b c d} W^{a e} W_{{e_{1}} {e_{2}}} F_{b c} F_{d e} {W}^{-2} - \frac{17}{48}\epsilon^{{e_{1}} {e_{2}} b c d} W_{{e_{1}}}\,^{a} W_{{e_{2}}}\,^{e} F_{b c} F_{d e} {W}^{-2}+\frac{25}{192}\epsilon^{{e_{2}} b c d e} W^{a {e_{1}}} W_{{e_{2}} {e_{1}}} F_{b c} F_{d e} {W}^{-2}%
+\frac{1}{32}\epsilon^{a {e_{2}} b c d} W^{{e_{1}} e} W_{{e_{2}} {e_{1}}} F_{b c} F_{d e} {W}^{-2}+\frac{1}{12}\epsilon^{a b d {e_{1}} {e_{2}}} C_{b}\,^{c}\,_{d}\,^{e} F_{{e_{1}} c} F_{{e_{2}} e} {W}^{-2}+\frac{7}{12}\epsilon^{a {e_{1}} {e_{2}} b d} W_{{e_{1}}}\,^{c} W_{{e_{2}}}\,^{e} F_{b c} F_{d e} {W}^{-2} - \frac{1}{24}\epsilon^{a {e_{1}} {e_{2}} b d} W^{c e} W_{{e_{1}} {e_{2}}} F_{b c} F_{d e} {W}^{-2}+\frac{5}{48}\epsilon^{a b c {e_{1}} {e_{2}}} C_{b c d e} W_{{e_{1}} {e_{2}}} F^{d e} {W}^{-1}+\frac{53}{384}\epsilon^{a d e {e_{1}} {e_{2}}} W_{d e} W_{{e_{1}} {e_{2}}} W_{b c} F^{b c} {W}^{-1}+\frac{7}{384}\epsilon^{a {e_{1}} {e_{2}} b c} W^{d e} W_{{e_{1}} {e_{2}}} W_{d e} F_{b c} {W}^{-1}+\frac{221}{384}\epsilon^{a d e {e_{1}} {e_{2}}} W_{d e} W_{{e_{1}} b} W_{{e_{2}} c} F^{b c} {W}^{-1}+\frac{53}{192}\epsilon^{a}\,_{b}\,^{d e {e_{1}}} W_{d e} W_{{e_{1}}}\,^{{e_{2}}} W_{{e_{2}} c} F^{b c} {W}^{-1}+\frac{5}{48}\epsilon^{a d e {e_{1}} {e_{2}}} C_{b c d e} W_{{e_{1}} {e_{2}}} F^{b c} {W}^{-1}+\frac{5}{128}\epsilon^{a b c {e_{1}} {e_{2}}} C_{b c d e} W^{d e} F_{{e_{1}} {e_{2}}} {W}^{-1}+\frac{5}{96}\epsilon^{a {e_{1}} {e_{2}} b c} W^{d e} W_{{e_{1}} d} W_{{e_{2}} e} F_{b c} {W}^{-1}+\frac{17}{192}\epsilon^{a}\,_{d}\,^{{e_{1}} b c} W^{d e} W_{{e_{1}}}\,^{{e_{2}}} W_{{e_{2}} e} F_{b c} {W}^{-1} - \frac{13}{192}\epsilon^{b c {e_{2}} e {e_{1}}} C_{b c}\,^{a d} W_{{e_{2}} d} F_{e {e_{1}}} {W}^{-1}+\frac{27}{128}\epsilon^{e {e_{1}} {e_{2}} b c} W^{a d} W_{e {e_{1}}} W_{{e_{2}} d} F_{b c} {W}^{-1}+\frac{13}{96}\epsilon^{a b d {e_{1}} {e_{2}}} C_{b c d e} W_{{e_{1}} {e_{2}}} F^{c e} {W}^{-1}+\frac{17}{384}\epsilon^{a d e {e_{1}} {e_{2}}} C_{b c d e} W^{b c} F_{{e_{1}} {e_{2}}} {W}^{-1}+\frac{11}{192}\epsilon^{b c {e_{1}} {e_{2}} e} C_{b c}\,^{a d} W_{{e_{1}} {e_{2}}} F_{e d} {W}^{-1}+\frac{95}{384}\epsilon^{d e {e_{1}} {e_{2}} b} W^{a c} W_{d e} W_{{e_{1}} {e_{2}}} F_{b c} {W}^{-1}+\frac{11}{384}\epsilon^{d e {e_{1}} {e_{2}} b} W_{d e} W_{{e_{1}}}\,^{a} W_{{e_{2}}}\,^{c} F_{b c} {W}^{-1}%
+\frac{307}{384}\epsilon^{a e {e_{1}} {e_{2}} b} W^{d c} W_{e {e_{1}}} W_{{e_{2}} d} F_{b c} {W}^{-1} - \frac{1}{16}\epsilon^{c d {e_{2}} e {e_{1}}} C^{a b}\,_{c d} W_{{e_{2}} b} F_{e {e_{1}}} {W}^{-1}+\frac{7}{24}\epsilon^{a b d {e_{2}} {e_{1}}} C_{b}\,^{c}\,_{d}\,^{e} W_{{e_{2}} c} F_{{e_{1}} e} {W}^{-1}+\frac{1}{96}\epsilon^{a b d {e_{1}} {e_{2}}} C_{b c d e} W^{c e} F_{{e_{1}} {e_{2}}} {W}^{-1} - \frac{1}{4}{\rm i} \lambda^{\alpha}_{i} {W}^{-3} \nabla^{a}{W} \nabla_{b}{\nabla^{b}{\lambda^{i}_{\alpha}}}+\frac{1}{4}{\rm i} \lambda^{\alpha}_{i} {W}^{-3} \nabla_{b}{W} \nabla^{b}{\nabla^{a}{\lambda^{i}_{\alpha}}}+{\rm i} (\Sigma_{b c})^{\alpha \beta} \lambda_{i \alpha} {W}^{-3} \nabla^{b}{W} \nabla^{c}{\nabla^{a}{\lambda^{i}_{\beta}}} - \frac{1}{4}{\rm i} \lambda^{\alpha}_{i} {W}^{-3} \nabla_{b}{\lambda^{i}_{\alpha}} \nabla^{b}{\nabla^{a}{W}} - \frac{1}{2}{\rm i} (\Sigma_{b c})^{\alpha \beta} \lambda_{i \alpha} {W}^{-3} \nabla^{b}{\lambda^{i}_{\beta}} \nabla^{c}{\nabla^{a}{W}} - \frac{1}{4}{\rm i} {W}^{-3} \nabla_{b}{W} \nabla^{a}{\lambda^{\alpha}_{i}} \nabla^{b}{\lambda^{i}_{\alpha}}+\frac{3}{2}{\rm i} (\Sigma_{b c})^{\alpha \beta} {W}^{-3} \nabla^{b}{W} \nabla^{a}{\lambda_{i \alpha}} \nabla^{c}{\lambda^{i}_{\beta}} - \frac{1}{2}{\rm i} (\Sigma^{a}{}_{\, b})^{\alpha \beta} \lambda_{i \alpha} {W}^{-3} \nabla_{c}{W} \nabla^{b}{\nabla^{c}{\lambda^{i}_{\beta}}} - \frac{1}{2}{\rm i} (\Sigma^{a}{}_{\, b})^{\alpha \beta} \lambda_{i \alpha} {W}^{-3} \nabla^{b}{W} \nabla_{c}{\nabla^{c}{\lambda^{i}_{\beta}}} - \frac{5}{2}{\rm i} (\Sigma^{a}{}_{\, b})^{\alpha \beta} \lambda_{i \alpha} {W}^{-3} \nabla_{c}{W} \nabla^{c}{\nabla^{b}{\lambda^{i}_{\beta}}}+\frac{1}{2}{\rm i} (\Sigma_{b c})^{\alpha \beta} \lambda_{i \alpha} {W}^{-3} \nabla^{b}{W} \nabla^{a}{\nabla^{c}{\lambda^{i}_{\beta}}} - \frac{3}{2}{\rm i} (\Sigma^{a}{}_{\, b})^{\alpha \beta} \lambda_{i \alpha} {W}^{-3} \nabla^{b}{\lambda^{i}_{\beta}} \nabla_{c}{\nabla^{c}{W}}+\frac{1}{4}{\rm i} \lambda^{\alpha}_{i} {W}^{-3} \nabla^{a}{\lambda^{i}_{\alpha}} \nabla_{b}{\nabla^{b}{W}}-{\rm i} (\Sigma^{a}{}_{\, b})^{\alpha \beta} \lambda_{i \alpha} {W}^{-3} \nabla_{c}{\lambda^{i}_{\beta}} \nabla^{c}{\nabla^{b}{W}}-3{\rm i} (\Sigma^{a}{}_{\, b})^{\alpha \beta} {W}^{-3} \nabla_{c}{W} \nabla^{b}{\lambda_{i \alpha}} \nabla^{c}{\lambda^{i}_{\beta}}+\frac{1}{2}{\rm i} (\Sigma^{a}{}_{\, b})^{\alpha \beta} {W}^{-3} \nabla^{b}{W} \nabla_{c}{\lambda_{i \alpha}} \nabla^{c}{\lambda^{i}_{\beta}}%
+\frac{1}{4}{\rm i} \epsilon^{a}\,_{b c d e} (\Gamma^{b})^{\alpha \beta} {W}^{-3} \nabla^{c}{W} \nabla^{d}{\lambda_{i \alpha}} \nabla^{e}{\lambda^{i}_{\beta}} - \frac{9}{64}{\rm i} \epsilon^{d e}\,_{{e_{1}}}\,^{b c} (\Sigma_{d e})^{\alpha \beta} F_{b c} \lambda_{i \alpha} {W}^{-4} \nabla^{{e_{1}}}{W} \nabla^{a}{\lambda^{i}_{\beta}} - \frac{3}{4}{\rm i} (\Gamma^{b})^{\alpha \beta} F_{b c} \lambda_{i \alpha} {W}^{-4} \nabla^{c}{W} \nabla^{a}{\lambda^{i}_{\beta}} - \frac{9}{32}{\rm i} \epsilon^{d e}\,_{{e_{1}}}\,^{b c} (\Sigma_{d e})^{\alpha \beta} F_{b c} \lambda_{i \alpha} {W}^{-4} \nabla^{a}{W} \nabla^{{e_{1}}}{\lambda^{i}_{\beta}} - \frac{15}{256}{\rm i} \epsilon^{d e}\,_{{e_{1}}}\,^{b c} (\Sigma_{d e})^{\alpha \beta} W_{b c} \lambda_{i \alpha} \lambda^{i}_{\beta} {W}^{-3} \nabla^{{e_{1}}}{\nabla^{a}{W}} - \frac{9}{256}{\rm i} \epsilon^{d e}\,_{{e_{1}}}\,^{b c} (\Sigma_{d e})^{\alpha \beta} W_{b c} \lambda_{i \alpha} {W}^{-3} \nabla^{a}{W} \nabla^{{e_{1}}}{\lambda^{i}_{\beta}} - \frac{21}{128}{\rm i} \epsilon^{d e}\,_{{e_{1}}}\,^{b c} (\Sigma_{d e})^{\alpha \beta} W_{b c} \lambda_{i \alpha} {W}^{-3} \nabla^{{e_{1}}}{W} \nabla^{a}{\lambda^{i}_{\beta}} - \frac{3}{4}{\rm i} (\Gamma^{b})^{\alpha \beta} W_{b c} \lambda_{i \alpha} {W}^{-3} \nabla^{c}{W} \nabla^{a}{\lambda^{i}_{\beta}} - \frac{3}{4}{\rm i} (\Gamma_{c})^{\alpha \beta} F^{a}\,_{b} \lambda_{i \alpha} {W}^{-4} \nabla^{b}{W} \nabla^{c}{\lambda^{i}_{\beta}} - \frac{3}{4}{\rm i} \epsilon^{a d e}\,_{{e_{1}}}\,^{b} (\Sigma_{d e})^{\alpha \beta} F_{b c} \lambda_{i \alpha} {W}^{-4} \nabla^{c}{W} \nabla^{{e_{1}}}{\lambda^{i}_{\beta}}+\frac{9}{16}{\rm i} \epsilon^{c d}\,_{e {e_{1}}}\,^{b} (\Sigma_{c d})^{\alpha \beta} F_{b}\,^{a} \lambda_{i \alpha} {W}^{-4} \nabla^{e}{W} \nabla^{{e_{1}}}{\lambda^{i}_{\beta}}+\frac{3}{4}{\rm i} (\Gamma^{b})^{\alpha \beta} F_{b}\,^{a} \lambda_{i \alpha} {W}^{-4} \nabla_{c}{W} \nabla^{c}{\lambda^{i}_{\beta}}+\frac{3}{8}{\rm i} \epsilon^{a}\,_{d e}\,^{b c} F_{b c} \lambda^{\alpha}_{i} {W}^{-4} \nabla^{d}{W} \nabla^{e}{\lambda^{i}_{\alpha}}+\frac{9}{16}{\rm i} \epsilon^{d}\,_{e {e_{1}}}\,^{b c} (\Sigma_{d}{}^{\, a})^{\alpha \beta} F_{b c} \lambda_{i \alpha} {W}^{-4} \nabla^{e}{W} \nabla^{{e_{1}}}{\lambda^{i}_{\beta}}+\frac{3}{4}{\rm i} \epsilon^{a d}\,_{{e_{1}}}\,^{b c} (\Sigma_{d e})^{\alpha \beta} F_{b c} \lambda_{i \alpha} {W}^{-4} \nabla^{e}{W} \nabla^{{e_{1}}}{\lambda^{i}_{\beta}}+\frac{3}{32}{\rm i} \epsilon^{a d e}\,_{{e_{1}}}\,^{b} (\Sigma_{d e})^{\alpha \beta} F_{b c} \lambda_{i \alpha} {W}^{-4} \nabla^{{e_{1}}}{W} \nabla^{c}{\lambda^{i}_{\beta}} - \frac{3}{4}{\rm i} (\Gamma_{c})^{\alpha \beta} F^{a}\,_{b} \lambda_{i \alpha} {W}^{-4} \nabla^{c}{W} \nabla^{b}{\lambda^{i}_{\beta}}+\frac{3}{4}{\rm i} (\Gamma^{a})^{\alpha \beta} F_{b c} \lambda_{i \alpha} {W}^{-4} \nabla^{b}{W} \nabla^{c}{\lambda^{i}_{\beta}}+\frac{45}{32}{\rm i} \epsilon^{a d}\,_{{e_{1}}}\,^{b c} (\Sigma_{d e})^{\alpha \beta} F_{b c} \lambda_{i \alpha} {W}^{-4} \nabla^{{e_{1}}}{W} \nabla^{e}{\lambda^{i}_{\beta}}+\frac{3}{8}{\rm i} \epsilon^{a}\,_{d e}\,^{b c} W_{b c} \lambda^{\alpha}_{i} {W}^{-3} \nabla^{d}{W} \nabla^{e}{\lambda^{i}_{\alpha}}%
+\frac{27}{256}{\rm i} \epsilon^{d e}\,_{{e_{1}}}\,^{b c} (\Sigma_{d e})^{\alpha \beta} W_{b c} \lambda_{i \alpha} \lambda^{i}_{\beta} {W}^{-3} \nabla^{a}{\nabla^{{e_{1}}}{W}}+\frac{3}{32}{\rm i} \epsilon^{a d}\,_{e {e_{1}}}\,^{b} (\Sigma_{d}{}^{\, c})^{\alpha \beta} W_{b c} \lambda_{i \alpha} \lambda^{i}_{\beta} {W}^{-3} \nabla^{e}{\nabla^{{e_{1}}}{W}} - \frac{3}{64}{\rm i} \epsilon^{a d e b c} (\Sigma_{d e})^{\alpha \beta} W_{b c} \lambda_{i \alpha} \lambda^{i}_{\beta} {W}^{-3} \nabla_{{e_{1}}}{\nabla^{{e_{1}}}{W}}+\frac{33}{128}{\rm i} \epsilon^{a d}\,_{{e_{1}}}\,^{b c} (\Sigma_{d e})^{\alpha \beta} W_{b c} \lambda_{i \alpha} \lambda^{i}_{\beta} {W}^{-3} \nabla^{{e_{1}}}{\nabla^{e}{W}}+\frac{23}{128}{\rm i} \epsilon^{a d e}\,_{{e_{1}}}\,^{b} (\Sigma_{d e})^{\alpha \beta} W_{b c} \lambda_{i \alpha} \lambda^{i}_{\beta} {W}^{-3} \nabla^{c}{\nabla^{{e_{1}}}{W}} - \frac{9}{64}{\rm i} \epsilon^{c d}\,_{e {e_{1}}}\,^{b} (\Sigma_{c d})^{\alpha \beta} W_{b}\,^{a} \lambda_{i \alpha} \lambda^{i}_{\beta} {W}^{-3} \nabla^{e}{\nabla^{{e_{1}}}{W}}+\frac{3}{4}{\rm i} (\Gamma_{c})^{\alpha \beta} W^{a}\,_{b} \lambda_{i \alpha} {W}^{-3} \nabla^{b}{W} \nabla^{c}{\lambda^{i}_{\beta}}+\frac{165}{128}{\rm i} \epsilon^{a d}\,_{{e_{1}}}\,^{b c} (\Sigma_{d e})^{\alpha \beta} W_{b c} \lambda_{i \alpha} {W}^{-3} \nabla^{{e_{1}}}{W} \nabla^{e}{\lambda^{i}_{\beta}}+\frac{3}{16}{\rm i} \epsilon^{a d e}\,_{{e_{1}}}\,^{b} (\Sigma_{d e})^{\alpha \beta} W_{b c} \lambda_{i \alpha} {W}^{-3} \nabla^{c}{W} \nabla^{{e_{1}}}{\lambda^{i}_{\beta}}+\frac{33}{128}{\rm i} \epsilon^{c d}\,_{e {e_{1}}}\,^{b} (\Sigma_{c d})^{\alpha \beta} W_{b}\,^{a} \lambda_{i \alpha} {W}^{-3} \nabla^{e}{W} \nabla^{{e_{1}}}{\lambda^{i}_{\beta}}+\frac{3}{4}{\rm i} (\Gamma^{b})^{\alpha \beta} W_{b}\,^{a} \lambda_{i \alpha} {W}^{-3} \nabla_{c}{W} \nabla^{c}{\lambda^{i}_{\beta}}+\frac{81}{128}{\rm i} \epsilon^{d}\,_{e {e_{1}}}\,^{b c} (\Sigma_{d}{}^{\, a})^{\alpha \beta} W_{b c} \lambda_{i \alpha} {W}^{-3} \nabla^{e}{W} \nabla^{{e_{1}}}{\lambda^{i}_{\beta}}+\frac{15}{16}{\rm i} \epsilon^{a d}\,_{{e_{1}}}\,^{b c} (\Sigma_{d e})^{\alpha \beta} W_{b c} \lambda_{i \alpha} {W}^{-3} \nabla^{e}{W} \nabla^{{e_{1}}}{\lambda^{i}_{\beta}}+\frac{75}{128}{\rm i} \epsilon^{a d e}\,_{{e_{1}}}\,^{b} (\Sigma_{d e})^{\alpha \beta} W_{b c} \lambda_{i \alpha} {W}^{-3} \nabla^{{e_{1}}}{W} \nabla^{c}{\lambda^{i}_{\beta}} - \frac{41}{64}{\rm i} \epsilon^{a d}\,_{e {e_{1}} c} (\Sigma_{d b})^{\alpha \beta} \lambda_{i \alpha} \lambda^{i}_{\beta} {W}^{-3} \nabla^{e}{W} \nabla^{{e_{1}}}{W^{b c}}+\frac{9}{64}{\rm i} \epsilon^{a d e b c} (\Sigma_{d e})^{\alpha \beta} \lambda_{i \alpha} \lambda^{i}_{\beta} {W}^{-3} \nabla_{{e_{1}}}{W} \nabla^{{e_{1}}}{W_{b c}} - \frac{9}{64}{\rm i} \epsilon^{d e}\,_{{e_{1}}}\,^{b c} (\Sigma_{d e})^{\alpha \beta} \lambda_{i \alpha} \lambda^{i}_{\beta} {W}^{-3} \nabla^{{e_{1}}}{W} \nabla^{a}{W_{b c}} - \frac{17}{128}{\rm i} \epsilon^{a d}\,_{{e_{1}}}\,^{b c} (\Sigma_{d e})^{\alpha \beta} \lambda_{i \alpha} \lambda^{i}_{\beta} {W}^{-3} \nabla^{{e_{1}}}{W} \nabla^{e}{W_{b c}} - \frac{11}{64}{\rm i} \epsilon^{a d e}\,_{{e_{1}}}\,^{b} (\Sigma_{d e})^{\alpha \beta} \lambda_{i \alpha} \lambda^{i}_{\beta} {W}^{-3} \nabla^{c}{W} \nabla^{{e_{1}}}{W_{b c}} - \frac{23}{64}{\rm i} \epsilon^{c d}\,_{e {e_{1}} b} (\Sigma_{c d})^{\alpha \beta} \lambda_{i \alpha} \lambda^{i}_{\beta} {W}^{-3} \nabla^{e}{W} \nabla^{{e_{1}}}{W^{a b}}%
 - \frac{3}{16}{\rm i} \epsilon^{d e}\,_{{e_{1}}}\,^{b c} (\Sigma_{d e})^{\alpha \beta} F_{b c} \lambda_{i \alpha} \lambda^{i}_{\beta} {W}^{-4} \nabla^{{e_{1}}}{\nabla^{a}{W}}+\frac{3}{32}{\rm i} \epsilon^{d e}\,_{{e_{1}}}\,^{b c} (\Sigma_{d e})^{\alpha \beta} \lambda_{i \alpha} \lambda^{i}_{\beta} {W}^{-4} \nabla^{a}{W} \nabla^{{e_{1}}}{F_{b c}} - \frac{3}{16}{\rm i} \epsilon^{a d e}\,_{{e_{1}}}\,^{b} (\Sigma_{d e})^{\alpha \beta} \lambda_{i \alpha} \lambda^{i}_{\beta} {W}^{-4} \nabla^{c}{W} \nabla^{{e_{1}}}{F_{b c}} - \frac{3}{8}{\rm i} \epsilon^{c d}\,_{e {e_{1}} b} (\Sigma_{c d})^{\alpha \beta} \lambda_{i \alpha} \lambda^{i}_{\beta} {W}^{-4} \nabla^{e}{W} \nabla^{{e_{1}}}{F^{a b}}+\frac{5}{128}{\rm i} \epsilon^{a d}\,_{{e_{1}}}\,^{b c} (\Sigma_{d e})^{\alpha \beta} \lambda_{i \alpha} \lambda^{i}_{\beta} {W}^{-3} \nabla^{e}{W} \nabla^{{e_{1}}}{W_{b c}} - \frac{7}{64}{\rm i} \epsilon^{a d e}\,_{{e_{1}}}\,^{b} (\Sigma_{d e})^{\alpha \beta} \lambda_{i \alpha} \lambda^{i}_{\beta} {W}^{-3} \nabla^{{e_{1}}}{W} \nabla^{c}{W_{b c}} - \frac{3}{16}{\rm i} \epsilon^{a b c d {e_{1}}} W_{b c} W_{d}\,^{e \alpha}\,_{i} W_{{e_{1}} e \alpha}\,^{i}+\frac{9}{32}{\rm i} \epsilon^{b c d e {e_{1}}} W_{b c} W_{d e}\,^{\alpha}\,_{i} W_{{e_{1}}}\,^{a}\,_{\alpha}\,^{i} - \frac{9}{16}{\rm i} \epsilon^{a b d e {e_{1}}} W_{b}\,^{c} W_{d e}\,^{\alpha}\,_{i} W_{{e_{1}} c \alpha}\,^{i}+\frac{77}{256}{\rm i} \epsilon^{a b c d e} F_{b c} W_{d e}\,^{\alpha}\,_{i} X^{i}_{\alpha} {W}^{-1}+\frac{1}{128}{\rm i} (\Gamma^{a})^{\alpha \beta} F^{b c} W_{b c \alpha i} X^{i}_{\beta} {W}^{-1} - \frac{1}{128}{\rm i} \epsilon^{{e_{1}} b c d e} (\Sigma_{{e_{1}}}{}^{\, a})^{\alpha \beta} F_{b c} W_{d e \alpha i} X^{i}_{\beta} {W}^{-1}+\frac{1}{64}{\rm i} F_{b}\,^{c} W_{d c}\,^{\alpha}\,_{i} \epsilon^{a b d e {e_{1}}} (\Sigma_{e {e_{1}}})_{\alpha}{}^{\beta} X^{i}_{\beta} {W}^{-1} - \frac{1}{64}{\rm i} F^{a b} W_{b c}\,^{\alpha}\,_{i} (\Gamma^{c})_{\alpha}{}^{\beta} X^{i}_{\beta} {W}^{-1} - \frac{1}{64}{\rm i} F_{b}\,^{c} W^{a}\,_{c}\,^{\alpha}\,_{i} (\Gamma^{b})_{\alpha}{}^{\beta} X^{i}_{\beta} {W}^{-1}+\frac{3}{8}{\rm i} (\Gamma_{b})^{\alpha \beta} X_{i j} \lambda^{i}_{\alpha} \lambda^{j}_{\beta} {W}^{-4} \nabla^{b}{\nabla^{a}{W}}+\frac{3}{4}{\rm i} (\Gamma_{b})^{\alpha \beta} X_{i j} \lambda^{i}_{\alpha} {W}^{-4} \nabla^{a}{W} \nabla^{b}{\lambda^{j}_{\beta}}+\frac{3}{4}{\rm i} (\Gamma_{b})^{\alpha \beta} X_{i j} \lambda^{i}_{\alpha} {W}^{-4} \nabla^{b}{W} \nabla^{a}{\lambda^{j}_{\beta}}+\frac{3}{8}{\rm i} (\Gamma_{b})^{\alpha \beta} \lambda_{i \alpha} \lambda_{j \beta} {W}^{-4} \nabla^{a}{W} \nabla^{b}{X^{i j}}+\frac{3}{4}{\rm i} (\Gamma_{b})^{\alpha \beta} \lambda_{i \alpha} \lambda_{j \beta} {W}^{-4} \nabla^{b}{W} \nabla^{a}{X^{i j}}%
 - \frac{3}{8}{\rm i} (\Gamma^{a})^{\alpha \beta} X_{i j} \lambda^{i}_{\alpha} \lambda^{j}_{\beta} {W}^{-4} \nabla_{b}{\nabla^{b}{W}}+\frac{3}{4}{\rm i} \epsilon^{a b c}\,_{d e} (\Sigma_{b c})^{\alpha \beta} X_{i j} \lambda^{i}_{\alpha} {W}^{-4} \nabla^{d}{W} \nabla^{e}{\lambda^{j}_{\beta}} - \frac{3}{2}{\rm i} (\Gamma^{a})^{\alpha \beta} X_{i j} \lambda^{i}_{\alpha} {W}^{-4} \nabla_{b}{W} \nabla^{b}{\lambda^{j}_{\beta}} - \frac{9}{8}{\rm i} (\Gamma^{a})^{\alpha \beta} \lambda_{i \alpha} \lambda_{j \beta} {W}^{-4} \nabla_{b}{W} \nabla^{b}{X^{i j}} - \frac{3}{16}\epsilon^{a b c d e} \Phi_{b c j i} (\Sigma_{d e})^{\beta \alpha} \lambda^{j}_{\beta} X^{i}_{\alpha} {W}^{-1} - \frac{5}{8}\Phi_{b}\,^{a}\,_{j i} (\Gamma^{b})^{\beta \alpha} \lambda^{j}_{\beta} X^{i}_{\alpha} {W}^{-1}+{W}^{-3} \nabla_{c}{W} \nabla_{b}{W} \nabla^{c}{F^{a b}}-3{W}^{-2} \nabla_{c}{W} \nabla_{b}{W} \nabla^{c}{W^{a b}} - \frac{1}{4}\epsilon^{a d e b c} F_{b c} {W}^{-2} \nabla_{{e_{1}}}{\nabla^{{e_{1}}}{F_{d e}}} - \frac{1}{8}\epsilon^{a b c d e} {W}^{-2} \nabla_{{e_{1}}}{F_{b c}} \nabla^{{e_{1}}}{F_{d e}} - \frac{3}{64}\epsilon^{a d e {e_{1}} {e_{2}}} F^{b c} F_{d e} F_{{e_{1}} {e_{2}}} F_{b c} {W}^{-4}+\frac{3}{16}\epsilon^{a d e {e_{1}} {e_{2}}} F^{b c} F_{d e} F_{{e_{1}} b} F_{{e_{2}} c} {W}^{-4}+\frac{3}{32}\epsilon^{c d e {e_{1}} {e_{2}}} F^{a b} F_{c d} F_{e {e_{1}}} F_{{e_{2}} b} {W}^{-4} - \frac{3}{16}\epsilon^{a {e_{1}} {e_{2}} d e} W_{{e_{1}} {e_{2}}} F^{b c} F_{d e} F_{b c} {W}^{-3}+\frac{3}{8}\epsilon^{a {e_{1}} {e_{2}} d e} W_{{e_{1}} {e_{2}}} F^{b c} F_{d b} F_{e c} {W}^{-3}+\frac{3}{8}\epsilon^{b c d e {e_{1}}} W^{a {e_{2}}} F_{b c} F_{d e} F_{{e_{1}} {e_{2}}} {W}^{-3} - \frac{27}{64}\epsilon^{a d e {e_{1}} {e_{2}}} W^{b c} W_{d e} W_{{e_{1}} {e_{2}}} F_{b c} {W}^{-1} - \frac{13}{64}\epsilon^{a {e_{1}} {e_{2}} b c} W^{d e} W_{{e_{1}} {e_{2}}} F_{b c} F_{d e} {W}^{-2}+\frac{5}{4}\epsilon^{a e {e_{1}} {e_{2}} d} W_{e {e_{1}}} W_{{e_{2}} b} F^{b c} F_{d c} {W}^{-2} - \frac{29}{128}\epsilon^{a d e {e_{1}} {e_{2}}} W_{d e} W_{{e_{1}} {e_{2}}} F^{b c} F_{b c} {W}^{-2}%
+\frac{3}{32}\epsilon^{a b c d e} W^{{e_{1}} {e_{2}}} F_{b c} F_{d e} F_{{e_{1}} {e_{2}}} {W}^{-3} - \frac{3}{16}\epsilon^{{e_{2}} c d e {e_{1}}} W_{{e_{2}} b} F^{a b} F_{c d} F_{e {e_{1}}} {W}^{-3} - \frac{9}{32}\epsilon^{e {e_{1}} {e_{2}} c d} W_{e {e_{1}}} W_{{e_{2}} b} F^{a b} F_{c d} {W}^{-2}+\frac{419}{8192}W^{b c} W_{b c}\,^{\alpha}\,_{i} \lambda^{i}_{\alpha} {W}^{-2} \nabla^{a}{W} - \frac{9}{64}F^{b c} W_{b c}\,^{\alpha}\,_{i} \lambda^{i}_{\alpha} {W}^{-3} \nabla^{a}{W}+\frac{17}{32}F^{b}\,_{c} W^{a}\,_{b}\,^{\alpha}\,_{i} \lambda^{i}_{\alpha} {W}^{-3} \nabla^{c}{W} - \frac{13}{32}F^{a b} W_{b c}\,^{\alpha}\,_{i} \lambda^{i}_{\alpha} {W}^{-3} \nabla^{c}{W}+\frac{7}{32}(\Sigma^{a}{}_{\, d})^{\alpha \beta} F^{b c} W_{b c \alpha i} \lambda^{i}_{\beta} {W}^{-3} \nabla^{d}{W} - \frac{3}{16}F^{a}\,_{b} W^{c d \alpha}\,_{i} (\Sigma_{c d})_{\alpha}{}^{\beta} \lambda^{i}_{\beta} {W}^{-3} \nabla^{b}{W}+\frac{3}{8}F^{b}\,_{c} W^{a d \alpha}\,_{i} (\Sigma_{b d})_{\alpha}{}^{\beta} \lambda^{i}_{\beta} {W}^{-3} \nabla^{c}{W} - \frac{3}{128}\epsilon^{a b c d e} F_{b c} W_{d e}\,^{\alpha}\,_{i} (\Gamma_{{e_{1}}})_{\alpha}{}^{\beta} \lambda^{i}_{\beta} {W}^{-3} \nabla^{{e_{1}}}{W}+\frac{3}{128}\epsilon^{b c d e}\,_{{e_{1}}} F_{b c} W_{d e}\,^{\alpha}\,_{i} (\Gamma^{a})_{\alpha}{}^{\beta} \lambda^{i}_{\beta} {W}^{-3} \nabla^{{e_{1}}}{W}+\frac{3}{32}\epsilon^{a b d}\,_{e {e_{1}}} (\Gamma^{e})^{\alpha \beta} F_{b}\,^{c} W_{d c \alpha i} \lambda^{i}_{\beta} {W}^{-3} \nabla^{{e_{1}}}{W}+\frac{7}{16}(\Sigma^{a d})^{\alpha \beta} F^{b}\,_{c} W_{d b \alpha i} \lambda^{i}_{\beta} {W}^{-3} \nabla^{c}{W}+\frac{1}{16}(\Sigma_{b}{}^{\, a})^{\alpha \beta} F^{b c} W_{c d \alpha i} \lambda^{i}_{\beta} {W}^{-3} \nabla^{d}{W} - \frac{1}{8}(\Sigma_{b d})^{\alpha \beta} F^{b c} W_{c}\,^{a}\,_{\alpha i} \lambda^{i}_{\beta} {W}^{-3} \nabla^{d}{W}+\frac{3}{16}(\Sigma^{c}{}_{\, d})^{\alpha \beta} F^{a b} W_{b c \alpha i} \lambda^{i}_{\beta} {W}^{-3} \nabla^{d}{W}+\frac{19}{32}(\Sigma_{b c})^{\alpha \beta} F^{b c} W^{a}\,_{d \alpha i} \lambda^{i}_{\beta} {W}^{-3} \nabla^{d}{W}+\frac{961}{4096}W^{b}\,_{c} W^{a}\,_{b}\,^{\alpha}\,_{i} \lambda^{i}_{\alpha} {W}^{-2} \nabla^{c}{W} - \frac{705}{4096}W^{a b} W_{b c}\,^{\alpha}\,_{i} \lambda^{i}_{\alpha} {W}^{-2} \nabla^{c}{W}%
+\frac{65}{12288}\epsilon^{b c d e}\,_{{e_{1}}} (\Gamma^{{e_{1}}})^{\alpha \beta} W_{b c} W_{d e \alpha i} \lambda^{i}_{\beta} {W}^{-2} \nabla^{a}{W} - \frac{65}{1536}(\Sigma_{b}{}^{\, d})^{\alpha \beta} W^{b c} W_{c d \alpha i} \lambda^{i}_{\beta} {W}^{-2} \nabla^{a}{W}+\frac{3841}{12288}(\Sigma^{a}{}_{\, d})^{\alpha \beta} W^{b c} W_{b c \alpha i} \lambda^{i}_{\beta} {W}^{-2} \nabla^{d}{W} - \frac{27}{128}W^{a}\,_{b} W^{c d \alpha}\,_{i} (\Sigma_{c d})_{\alpha}{}^{\beta} \lambda^{i}_{\beta} {W}^{-2} \nabla^{b}{W} - \frac{431}{3072}W^{a b} W^{c}\,_{d}\,^{\alpha}\,_{i} (\Sigma_{b c})_{\alpha}{}^{\beta} \lambda^{i}_{\beta} {W}^{-2} \nabla^{d}{W}+\frac{27}{64}W^{b}\,_{c} W^{a d \alpha}\,_{i} (\Sigma_{b d})_{\alpha}{}^{\beta} \lambda^{i}_{\beta} {W}^{-2} \nabla^{c}{W} - \frac{431}{6144}W^{b c} W^{a}\,_{d}\,^{\alpha}\,_{i} (\Sigma_{b c})_{\alpha}{}^{\beta} \lambda^{i}_{\beta} {W}^{-2} \nabla^{d}{W} - \frac{1727}{49152}\epsilon^{a b c d e} W_{b c} W_{d e}\,^{\alpha}\,_{i} (\Gamma_{{e_{1}}})_{\alpha}{}^{\beta} \lambda^{i}_{\beta} {W}^{-2} \nabla^{{e_{1}}}{W}+\frac{1727}{49152}\epsilon^{b c d e}\,_{{e_{1}}} W_{b c} W_{d e}\,^{\alpha}\,_{i} (\Gamma^{a})_{\alpha}{}^{\beta} \lambda^{i}_{\beta} {W}^{-2} \nabla^{{e_{1}}}{W}+\frac{1727}{12288}\epsilon^{a b d}\,_{e {e_{1}}} (\Gamma^{e})^{\alpha \beta} W_{b}\,^{c} W_{d c \alpha i} \lambda^{i}_{\beta} {W}^{-2} \nabla^{{e_{1}}}{W}+\frac{3841}{6144}(\Sigma^{a d})^{\alpha \beta} W^{b}\,_{c} W_{d b \alpha i} \lambda^{i}_{\beta} {W}^{-2} \nabla^{c}{W}+\frac{2111}{6144}(\Sigma_{b}{}^{\, a})^{\alpha \beta} W^{b c} W_{c d \alpha i} \lambda^{i}_{\beta} {W}^{-2} \nabla^{d}{W} - \frac{239}{3072}(\Sigma_{b d})^{\alpha \beta} W^{b c} W_{c}\,^{a}\,_{\alpha i} \lambda^{i}_{\beta} {W}^{-2} \nabla^{d}{W}+\frac{1727}{6144}(\Sigma^{c}{}_{\, d})^{\alpha \beta} W^{a b} W_{b c \alpha i} \lambda^{i}_{\beta} {W}^{-2} \nabla^{d}{W} - \frac{3487}{4096}(\Sigma_{b c})^{\alpha \beta} W^{b c} W^{a}\,_{d \alpha i} \lambda^{i}_{\beta} {W}^{-2} \nabla^{d}{W} - \frac{3}{128}\epsilon^{b c d e}\,_{{e_{1}}} (\Gamma^{a})^{\alpha \beta} F_{b c} W_{d e \alpha i} \lambda^{i}_{\beta} {W}^{-3} \nabla^{{e_{1}}}{W} - \frac{3}{16}F^{a b} W_{b}\,^{c \alpha}\,_{i} (\Sigma_{c d})_{\alpha}{}^{\beta} \lambda^{i}_{\beta} {W}^{-3} \nabla^{d}{W} - \frac{3}{16}F^{b c} W_{b}\,^{a \alpha}\,_{i} (\Sigma_{c d})_{\alpha}{}^{\beta} \lambda^{i}_{\beta} {W}^{-3} \nabla^{d}{W}+\frac{3}{16}F^{b c} W_{b}\,^{d \alpha}\,_{i} (\Sigma_{c d})_{\alpha}{}^{\beta} \lambda^{i}_{\beta} {W}^{-3} \nabla^{a}{W}+\frac{3}{64}\epsilon^{a b c d}\,_{{e_{1}}} F_{b c} W_{d e}\,^{\alpha}\,_{i} (\Gamma^{e})_{\alpha}{}^{\beta} \lambda^{i}_{\beta} {W}^{-3} \nabla^{{e_{1}}}{W}%
+\frac{3}{64}\epsilon^{b c d}\,_{e {e_{1}}} F_{b c} W_{d}\,^{a \alpha}\,_{i} (\Gamma^{e})_{\alpha}{}^{\beta} \lambda^{i}_{\beta} {W}^{-3} \nabla^{{e_{1}}}{W} - \frac{3}{64}\epsilon^{a b d e}\,_{{e_{1}}} F_{b c} W_{d e}\,^{\alpha}\,_{i} (\Gamma^{{e_{1}}})_{\alpha}{}^{\beta} \lambda^{i}_{\beta} {W}^{-3} \nabla^{c}{W}+\frac{1}{64}\epsilon^{b c d}\,_{e {e_{1}}} (\Gamma^{e})^{\alpha \beta} F_{b c} W_{d}\,^{a}\,_{\alpha i} \lambda^{i}_{\beta} {W}^{-3} \nabla^{{e_{1}}}{W}+\frac{1}{16}(\Sigma^{b d})^{\alpha \beta} F_{b c} W^{a}\,_{d \alpha i} \lambda^{i}_{\beta} {W}^{-3} \nabla^{c}{W} - \frac{1}{16}(\Sigma^{c}{}_{\, d})^{\alpha \beta} F^{a b} W_{c b \alpha i} \lambda^{i}_{\beta} {W}^{-3} \nabla^{d}{W}+\frac{1}{32}(\Sigma^{c d})^{\alpha \beta} F^{a}\,_{b} W_{c d \alpha i} \lambda^{i}_{\beta} {W}^{-3} \nabla^{b}{W}+\frac{1}{64}\epsilon^{a b d e}\,_{{e_{1}}} (\Gamma^{c})^{\alpha \beta} F_{b c} W_{d e \alpha i} \lambda^{i}_{\beta} {W}^{-3} \nabla^{{e_{1}}}{W}+\frac{1}{64}\epsilon^{a b c d}\,_{{e_{1}}} (\Gamma^{{e_{1}}})^{\alpha \beta} F_{b c} W_{d e \alpha i} \lambda^{i}_{\beta} {W}^{-3} \nabla^{e}{W}+\frac{1}{16}(\Sigma_{b}{}^{\, c})^{\alpha \beta} F^{a b} W_{c d \alpha i} \lambda^{i}_{\beta} {W}^{-3} \nabla^{d}{W} - \frac{865}{49152}\epsilon^{b c d e}\,_{{e_{1}}} (\Gamma^{a})^{\alpha \beta} W_{b c} W_{d e \alpha i} \lambda^{i}_{\beta} {W}^{-2} \nabla^{{e_{1}}}{W} - \frac{865}{6144}W^{a b} W_{b}\,^{c \alpha}\,_{i} (\Sigma_{c d})_{\alpha}{}^{\beta} \lambda^{i}_{\beta} {W}^{-2} \nabla^{d}{W} - \frac{865}{6144}W^{b c} W_{b}\,^{a \alpha}\,_{i} (\Sigma_{c d})_{\alpha}{}^{\beta} \lambda^{i}_{\beta} {W}^{-2} \nabla^{d}{W}+\frac{865}{6144}W^{b c} W_{b}\,^{d \alpha}\,_{i} (\Sigma_{c d})_{\alpha}{}^{\beta} \lambda^{i}_{\beta} {W}^{-2} \nabla^{a}{W}+\frac{865}{24576}\epsilon^{a b c d}\,_{{e_{1}}} W_{b c} W_{d e}\,^{\alpha}\,_{i} (\Gamma^{e})_{\alpha}{}^{\beta} \lambda^{i}_{\beta} {W}^{-2} \nabla^{{e_{1}}}{W}+\frac{865}{24576}\epsilon^{b c d}\,_{e {e_{1}}} W_{b c} W_{d}\,^{a \alpha}\,_{i} (\Gamma^{e})_{\alpha}{}^{\beta} \lambda^{i}_{\beta} {W}^{-2} \nabla^{{e_{1}}}{W} - \frac{865}{24576}\epsilon^{a b d e}\,_{{e_{1}}} W_{b c} W_{d e}\,^{\alpha}\,_{i} (\Gamma^{{e_{1}}})_{\alpha}{}^{\beta} \lambda^{i}_{\beta} {W}^{-2} \nabla^{c}{W}+\frac{1249}{24576}\epsilon^{b c d}\,_{e {e_{1}}} (\Gamma^{e})^{\alpha \beta} W_{b c} W_{d}\,^{a}\,_{\alpha i} \lambda^{i}_{\beta} {W}^{-2} \nabla^{{e_{1}}}{W}+\frac{1249}{6144}(\Sigma^{b d})^{\alpha \beta} W_{b c} W^{a}\,_{d \alpha i} \lambda^{i}_{\beta} {W}^{-2} \nabla^{c}{W} - \frac{1249}{6144}(\Sigma^{c}{}_{\, d})^{\alpha \beta} W^{a b} W_{c b \alpha i} \lambda^{i}_{\beta} {W}^{-2} \nabla^{d}{W}+\frac{1249}{12288}(\Sigma^{c d})^{\alpha \beta} W^{a}\,_{b} W_{c d \alpha i} \lambda^{i}_{\beta} {W}^{-2} \nabla^{b}{W}%
+\frac{1249}{24576}\epsilon^{a b d e}\,_{{e_{1}}} (\Gamma^{c})^{\alpha \beta} W_{b c} W_{d e \alpha i} \lambda^{i}_{\beta} {W}^{-2} \nabla^{{e_{1}}}{W}+\frac{1249}{24576}\epsilon^{a b c d}\,_{{e_{1}}} (\Gamma^{{e_{1}}})^{\alpha \beta} W_{b c} W_{d e \alpha i} \lambda^{i}_{\beta} {W}^{-2} \nabla^{e}{W}+\frac{1249}{6144}(\Sigma_{b}{}^{\, c})^{\alpha \beta} W^{a b} W_{c d \alpha i} \lambda^{i}_{\beta} {W}^{-2} \nabla^{d}{W} - \frac{5}{16}(\Gamma_{b})^{\alpha \beta} \lambda_{i \alpha} \lambda^{i \rho} \lambda_{j \beta} {W}^{-4} \nabla^{b}{\nabla^{a}{\lambda^{j}_{\rho}}}+\frac{1}{4}(\Gamma_{b})^{\alpha \beta} \lambda_{i \alpha} \lambda_{j \beta} {W}^{-4} \nabla^{a}{\lambda^{i \rho}} \nabla^{b}{\lambda^{j}_{\rho}}+\frac{1}{4}(\Gamma_{b})^{\alpha \beta} \lambda_{i \alpha} \lambda^{\rho}_{j} {W}^{-4} \nabla^{a}{\lambda^{i}_{\beta}} \nabla^{b}{\lambda^{j}_{\rho}} - \frac{3}{8}(\Gamma_{b})^{\alpha \beta} \lambda_{i \alpha} \lambda^{\rho}_{j} {W}^{-4} \nabla^{a}{\lambda^{i}_{\rho}} \nabla^{b}{\lambda^{j}_{\beta}} - \frac{1}{16}(\Gamma_{b})^{\alpha \beta} \lambda_{i \alpha} \lambda^{\rho}_{j} {W}^{-4} \nabla^{a}{\lambda^{j}_{\beta}} \nabla^{b}{\lambda^{i}_{\rho}}+\frac{7}{16}(\Gamma_{b})^{\alpha \beta} \lambda_{i \alpha} \lambda^{\rho}_{j} {W}^{-4} \nabla^{a}{\lambda^{j}_{\rho}} \nabla^{b}{\lambda^{i}_{\beta}}+\frac{7}{16}(\Gamma_{b})^{\alpha \beta} \lambda^{\rho}_{i} \lambda_{j \rho} {W}^{-4} \nabla^{a}{\lambda^{i}_{\alpha}} \nabla^{b}{\lambda^{j}_{\beta}} - \frac{1}{4}(\Gamma_{b})^{\alpha \beta} \lambda_{i \alpha} \lambda^{i \rho} {W}^{-4} \nabla^{a}{\lambda_{j \rho}} \nabla^{b}{\lambda^{j}_{\beta}}+\frac{3}{16}\epsilon^{a b c}\,_{d e} (\Sigma_{b c})^{\alpha \beta} \lambda_{i \alpha} \lambda^{i \rho} {W}^{-4} \nabla^{d}{\lambda_{j \beta}} \nabla^{e}{\lambda^{j}_{\rho}}+\frac{7}{16}(\Gamma^{a})^{\alpha \beta} \lambda_{i \alpha} \lambda^{i \rho} {W}^{-4} \nabla_{b}{\lambda_{j \beta}} \nabla^{b}{\lambda^{j}_{\rho}} - \frac{1}{4}(\Gamma_{b})^{\alpha \beta} \lambda_{i \alpha} \lambda^{i \rho} {W}^{-4} \nabla^{a}{\lambda_{j \beta}} \nabla^{b}{\lambda^{j}_{\rho}} - \frac{3}{16}(\Gamma^{a})^{\alpha \beta} \lambda_{i \alpha} \lambda_{j \beta} {W}^{-4} \nabla_{b}{\lambda^{i \rho}} \nabla^{b}{\lambda^{j}_{\rho}}+\frac{7}{32}\epsilon^{a b c}\,_{d e} (\Sigma_{b c})^{\alpha \beta} \lambda_{i \alpha} \lambda^{\rho}_{j} {W}^{-4} \nabla^{d}{\lambda^{i}_{\beta}} \nabla^{e}{\lambda^{j}_{\rho}}+\frac{1}{8}(\Gamma^{a})^{\alpha \beta} \lambda_{i \alpha} \lambda^{\rho}_{j} {W}^{-4} \nabla_{b}{\lambda^{i}_{\beta}} \nabla^{b}{\lambda^{j}_{\rho}} - \frac{5}{32}\epsilon^{a b c}\,_{d e} (\Sigma_{b c})^{\alpha \beta} \lambda_{i \alpha} \lambda^{\rho}_{j} {W}^{-4} \nabla^{d}{\lambda^{i}_{\rho}} \nabla^{e}{\lambda^{j}_{\beta}}+\frac{1}{2}(\Gamma^{a})^{\alpha \beta} \lambda_{i \alpha} \lambda^{\rho}_{j} {W}^{-4} \nabla_{b}{\lambda^{i}_{\rho}} \nabla^{b}{\lambda^{j}_{\beta}}+\frac{13}{32}\epsilon^{a b c}\,_{d e} (\Sigma_{b c})^{\alpha \beta} \lambda^{\rho}_{i} \lambda_{j \rho} {W}^{-4} \nabla^{d}{\lambda^{i}_{\alpha}} \nabla^{e}{\lambda^{j}_{\beta}}%
 - \frac{3}{8}(\Gamma^{a})^{\alpha \beta} \lambda^{\rho}_{i} \lambda_{j \rho} {W}^{-4} \nabla_{b}{\lambda^{i}_{\alpha}} \nabla^{b}{\lambda^{j}_{\beta}}+\frac{1}{4}(\Sigma^{a}{}_{\, c})^{\rho \lambda} (\Gamma_{b})^{\alpha \beta} \lambda_{i \rho} \lambda^{i}_{\alpha} \lambda_{j \beta} {W}^{-4} \nabla^{b}{\nabla^{c}{\lambda^{j}_{\lambda}}} - \frac{1}{8}(\Sigma^{a}{}_{\, c})^{\rho \lambda} (\Gamma_{b})^{\alpha \beta} \lambda_{i \rho} \lambda^{i}_{\lambda} \lambda_{j \alpha} {W}^{-4} \nabla^{b}{\nabla^{c}{\lambda^{j}_{\beta}}} - \frac{1}{8}(\Sigma^{a}{}_{\, c})^{\rho \lambda} (\Gamma_{b})^{\alpha \beta} \lambda_{i \rho} \lambda^{i}_{\lambda} \lambda_{j \alpha} {W}^{-4} \nabla^{c}{\nabla^{b}{\lambda^{j}_{\beta}}} - \frac{1}{8}(\Sigma^{a}{}_{\, c})^{\rho \lambda} (\Gamma_{b})^{\alpha \beta} \lambda_{i \rho} \lambda^{i}_{\lambda} {W}^{-4} \nabla^{c}{\lambda_{j \alpha}} \nabla^{b}{\lambda^{j}_{\beta}} - \frac{1}{16}(\Gamma_{b})^{\alpha \beta} \lambda_{i \alpha} \lambda^{i \rho} \lambda_{j \rho} {W}^{-4} \nabla^{b}{\nabla^{a}{\lambda^{j}_{\beta}}}+\frac{1}{8}(\Sigma^{a}{}_{\, c})^{\rho \lambda} (\Gamma_{b})^{\alpha \beta} \lambda_{i \rho} \lambda^{i}_{\alpha} \lambda_{j \lambda} {W}^{-4} \nabla^{b}{\nabla^{c}{\lambda^{j}_{\beta}}} - \frac{1}{16}(\Gamma_{b})^{\alpha \beta} \lambda_{i \alpha} \lambda^{i \rho} \lambda_{j \rho} {W}^{-4} \nabla^{a}{\nabla^{b}{\lambda^{j}_{\beta}}}+\frac{1}{8}(\Sigma^{a}{}_{\, c})^{\rho \lambda} (\Gamma_{b})^{\alpha \beta} \lambda_{i \rho} \lambda^{i}_{\alpha} \lambda_{j \lambda} {W}^{-4} \nabla^{c}{\nabla^{b}{\lambda^{j}_{\beta}}} - \frac{1}{16}(\Sigma^{a}{}_{\, c})^{\rho \lambda} (\Gamma_{b})^{\alpha \beta} \lambda_{i \rho} \lambda_{j \alpha} {W}^{-4} \nabla^{c}{\lambda^{i}_{\beta}} \nabla^{b}{\lambda^{j}_{\lambda}}+\frac{23}{16}(\Sigma^{a}{}_{\, c})^{\rho \lambda} (\Gamma_{b})^{\alpha \beta} \lambda_{i \rho} \lambda_{j \alpha} {W}^{-4} \nabla^{c}{\lambda^{i}_{\lambda}} \nabla^{b}{\lambda^{j}_{\beta}} - \frac{3}{16}(\Sigma^{a}{}_{\, c})^{\rho \lambda} (\Gamma_{b})^{\alpha \beta} \lambda_{i \rho} \lambda_{j \alpha} {W}^{-4} \nabla^{c}{\lambda^{j}_{\beta}} \nabla^{b}{\lambda^{i}_{\lambda}}+\frac{9}{16}(\Sigma^{a}{}_{\, c})^{\rho \lambda} (\Gamma_{b})^{\alpha \beta} \lambda_{i \rho} \lambda_{j \alpha} {W}^{-4} \nabla^{c}{\lambda^{j}_{\lambda}} \nabla^{b}{\lambda^{i}_{\beta}} - \frac{1}{8}(\Sigma^{a}{}_{\, c})^{\rho \lambda} (\Gamma_{b})^{\alpha \beta} \lambda_{i \rho} \lambda_{j \lambda} {W}^{-4} \nabla^{c}{\lambda^{i}_{\alpha}} \nabla^{b}{\lambda^{j}_{\beta}}+\frac{1}{8}(\Sigma^{a}{}_{\, c})^{\rho \lambda} (\Gamma_{b})^{\alpha \beta} \lambda_{i \rho} \lambda^{i}_{\alpha} {W}^{-4} \nabla^{c}{\lambda_{j \lambda}} \nabla^{b}{\lambda^{j}_{\beta}}+\frac{1}{8}(\Sigma^{a}{}_{\, c})^{\rho \lambda} (\Gamma_{b})^{\alpha \beta} \lambda_{i \rho} \lambda^{i}_{\alpha} {W}^{-4} \nabla^{c}{\lambda_{j \beta}} \nabla^{b}{\lambda^{j}_{\lambda}}+\frac{1}{8}(\Sigma^{a}{}_{\, c})^{\rho \lambda} (\Gamma_{b})^{\alpha \beta} \lambda_{i \alpha} \lambda_{j \beta} {W}^{-4} \nabla^{c}{\lambda^{i}_{\rho}} \nabla^{b}{\lambda^{j}_{\lambda}}+\frac{1}{12}W^{a}\,_{b}\,^{\alpha}\,_{i} X^{i}\,_{j} \lambda^{j}_{\alpha} {W}^{-3} \nabla^{b}{W}-{\rm i} \Phi^{a}\,_{b i j} \lambda^{i \alpha} \lambda^{j}_{\alpha} {W}^{-3} \nabla^{b}{W}+\frac{1}{8}{\rm i} \epsilon^{a c d}\,_{b e} \Phi_{c d i j} (\Gamma^{b})^{\alpha \beta} \lambda^{i}_{\alpha} \lambda^{j}_{\beta} {W}^{-3} \nabla^{e}{W}%
+\frac{7}{32}(\Gamma^{a})^{\rho \alpha} \lambda_{j \rho} \lambda^{j \beta} X_{i \alpha} X^{i}_{\beta} {W}^{-2}+\frac{121}{384}(\Gamma^{a})^{\beta \rho} \lambda_{i \beta} \lambda_{j \rho} X^{i \alpha} X^{j}_{\alpha} {W}^{-2} - \frac{163}{192}(\Gamma^{a})^{\rho \alpha} \lambda_{i \rho} \lambda^{\beta}_{j} X^{i}_{\alpha} X^{j}_{\beta} {W}^{-2}+\frac{79}{192}(\Gamma^{a})^{\rho \beta} \lambda_{i \rho} \lambda^{\alpha}_{j} X^{i}_{\alpha} X^{j}_{\beta} {W}^{-2}+\frac{121}{384}(\Gamma^{a})^{\alpha \beta} \lambda^{\rho}_{i} \lambda_{j \rho} X^{i}_{\alpha} X^{j}_{\beta} {W}^{-2}+\frac{41}{384}(\Gamma^{a})^{\beta \alpha} X_{i j} X^{i j} \lambda_{k \beta} X^{k}_{\alpha} {W}^{-3} - \frac{41}{192}(\Gamma^{a})^{\beta \alpha} X_{i j} X^{i}\,_{k} \lambda^{j}_{\beta} X^{k}_{\alpha} {W}^{-3}+{\rm i} (\Sigma^{a}{}_{\, b})^{\alpha \beta} \lambda_{i \alpha} {W}^{-2} \nabla_{c}{\nabla^{c}{\nabla^{b}{\lambda^{i}_{\beta}}}}+{\rm i} (\Sigma^{a}{}_{\, b})^{\alpha \beta} {W}^{-2} \nabla^{b}{\lambda_{i \alpha}} \nabla_{c}{\nabla^{c}{\lambda^{i}_{\beta}}}+{\rm i} (\Sigma^{a}{}_{\, b})^{\alpha \beta} {W}^{-2} \nabla_{c}{\lambda_{i \alpha}} \nabla^{c}{\nabla^{b}{\lambda^{i}_{\beta}}}+\frac{3}{4}{\rm i} \lambda^{\alpha}_{i} {W}^{-4} \nabla^{a}{W} \nabla_{b}{W} \nabla^{b}{\lambda^{i}_{\alpha}}-3{\rm i} (\Sigma_{b c})^{\alpha \beta} \lambda_{i \alpha} {W}^{-4} \nabla^{a}{W} \nabla^{b}{W} \nabla^{c}{\lambda^{i}_{\beta}} - \frac{3}{4}{\rm i} (\Sigma_{b c})^{\alpha \beta} \lambda_{i \alpha} \lambda^{i}_{\beta} {W}^{-4} \nabla^{b}{W} \nabla^{c}{\nabla^{a}{W}}+\frac{3}{4}{\rm i} (\Sigma^{a}{}_{\, b})^{\alpha \beta} \lambda_{i \alpha} \lambda^{i}_{\beta} {W}^{-4} \nabla^{b}{W} \nabla_{c}{\nabla^{c}{W}}+\frac{3}{2}{\rm i} (\Sigma^{a}{}_{\, b})^{\alpha \beta} \lambda_{i \alpha} \lambda^{i}_{\beta} {W}^{-4} \nabla_{c}{W} \nabla^{c}{\nabla^{b}{W}} - \frac{3}{4}{\rm i} \lambda^{\alpha}_{i} {W}^{-4} \nabla_{b}{W} \nabla^{b}{W} \nabla^{a}{\lambda^{i}_{\alpha}}+\frac{15}{4}{\rm i} (\Sigma^{a}{}_{\, b})^{\alpha \beta} \lambda_{i \alpha} {W}^{-4} \nabla_{c}{W} \nabla^{c}{W} \nabla^{b}{\lambda^{i}_{\beta}}+\frac{3}{2}{\rm i} (\Sigma^{a}{}_{\, b})^{\alpha \beta} \lambda_{i \alpha} {W}^{-4} \nabla^{b}{W} \nabla_{c}{W} \nabla^{c}{\lambda^{i}_{\beta}} - \frac{45}{256}{\rm i} \epsilon^{d e}\,_{{e_{1}}}\,^{b c} (\Sigma_{d e})^{\alpha \beta} W_{b c} \lambda_{i \alpha} \lambda^{i}_{\beta} {W}^{-4} \nabla^{a}{W} \nabla^{{e_{1}}}{W}+\frac{3}{8}{\rm i} \epsilon^{d e}\,_{{e_{1}}}\,^{b c} (\Sigma_{d e})^{\alpha \beta} F_{b c} \lambda_{i \alpha} \lambda^{i}_{\beta} {W}^{-5} \nabla^{a}{W} \nabla^{{e_{1}}}{W}%
 - \frac{3}{4}{\rm i} \epsilon^{a d}\,_{{e_{1}}}\,^{b c} (\Sigma_{d e})^{\alpha \beta} F_{b c} \lambda_{i \alpha} \lambda^{i}_{\beta} {W}^{-5} \nabla^{e}{W} \nabla^{{e_{1}}}{W}+\frac{3}{4}{\rm i} \epsilon^{a d e}\,_{{e_{1}}}\,^{b} (\Sigma_{d e})^{\alpha \beta} F_{b c} \lambda_{i \alpha} \lambda^{i}_{\beta} {W}^{-5} \nabla^{{e_{1}}}{W} \nabla^{c}{W} - \frac{117}{128}{\rm i} \epsilon^{a d}\,_{{e_{1}}}\,^{b c} (\Sigma_{d e})^{\alpha \beta} W_{b c} \lambda_{i \alpha} \lambda^{i}_{\beta} {W}^{-4} \nabla^{e}{W} \nabla^{{e_{1}}}{W} - \frac{27}{128}{\rm i} \epsilon^{a d e}\,_{{e_{1}}}\,^{b} (\Sigma_{d e})^{\alpha \beta} W_{b c} \lambda_{i \alpha} \lambda^{i}_{\beta} {W}^{-4} \nabla^{{e_{1}}}{W} \nabla^{c}{W}+\frac{1}{64}{\rm i} \epsilon^{d}\,_{e {e_{1}}}\,^{b c} (\Sigma_{d}{}^{\, a})^{\alpha \beta} W_{b c} \lambda_{i \alpha} {W}^{-2} \nabla^{e}{\nabla^{{e_{1}}}{\lambda^{i}_{\beta}}} - \frac{3}{64}{\rm i} \epsilon^{a d}\,_{e {e_{1}} c} (\Sigma_{d b})^{\alpha \beta} \lambda_{i \alpha} \lambda^{i}_{\beta} {W}^{-2} \nabla^{e}{\nabla^{{e_{1}}}{W^{b c}}} - \frac{3}{32}{\rm i} \epsilon^{a d e b c} (\Sigma_{d e})^{\alpha \beta} \lambda_{i \alpha} \lambda^{i}_{\beta} {W}^{-2} \nabla_{{e_{1}}}{\nabla^{{e_{1}}}{W_{b c}}}+\frac{1}{32}{\rm i} \epsilon^{d}\,_{e}\,^{b c}\,_{{e_{1}}} (\Sigma_{d}{}^{\, a})^{\alpha \beta} \lambda_{i \alpha} {W}^{-3} \nabla^{e}{F_{b c}} \nabla^{{e_{1}}}{\lambda^{i}_{\beta}} - \frac{3}{32}{\rm i} \epsilon^{a d e b}\,_{{e_{1}}} (\Sigma_{d e})^{\alpha \beta} \lambda_{i \alpha} {W}^{-3} \nabla^{c}{F_{b c}} \nabla^{{e_{1}}}{\lambda^{i}_{\beta}} - \frac{1}{32}{\rm i} \epsilon^{d}\,_{e}\,^{b c}\,_{{e_{1}}} (\Sigma_{d}{}^{\, a})^{\alpha \beta} \lambda_{i \alpha} {W}^{-2} \nabla^{e}{W_{b c}} \nabla^{{e_{1}}}{\lambda^{i}_{\beta}} - \frac{7}{32}{\rm i} \epsilon^{a d e b c} (\Sigma_{d e})^{\alpha \beta} F_{b c} \lambda_{i \alpha} {W}^{-3} \nabla_{{e_{1}}}{\nabla^{{e_{1}}}{\lambda^{i}_{\beta}}} - \frac{1}{16}{\rm i} \epsilon^{d}\,_{e {e_{1}}}\,^{b c} (\Sigma_{d}{}^{\, a})^{\alpha \beta} F_{b c} \lambda_{i \alpha} {W}^{-3} \nabla^{e}{\nabla^{{e_{1}}}{\lambda^{i}_{\beta}}} - \frac{1}{8}{\rm i} \epsilon^{a d e b c} (\Sigma_{d e})^{\alpha \beta} F_{b c} {W}^{-3} \nabla_{{e_{1}}}{\lambda_{i \alpha}} \nabla^{{e_{1}}}{\lambda^{i}_{\beta}} - \frac{5}{32}{\rm i} \epsilon^{a d e b c} (\Sigma_{d e})^{\alpha \beta} W_{b c} \lambda_{i \alpha} {W}^{-2} \nabla_{{e_{1}}}{\nabla^{{e_{1}}}{\lambda^{i}_{\beta}}} - \frac{5}{128}{\rm i} \epsilon^{d}\,_{e {e_{1}}}\,^{b c} (\Sigma_{d}{}^{\, a})^{\alpha \beta} \lambda_{i \alpha} \lambda^{i}_{\beta} {W}^{-2} \nabla^{e}{\nabla^{{e_{1}}}{W_{b c}}} - \frac{15}{64}{\rm i} \epsilon^{a d e b c} (\Sigma_{d e})^{\alpha \beta} \lambda_{i \alpha} {W}^{-3} \nabla_{{e_{1}}}{F_{b c}} \nabla^{{e_{1}}}{\lambda^{i}_{\beta}}+\frac{3}{32}{\rm i} (\Sigma^{d}{}_{\, b})^{\alpha \beta} F^{b c} F_{d c} \lambda_{i \alpha} {W}^{-4} \nabla^{a}{\lambda^{i}_{\beta}}+\frac{49}{256}{\rm i} (\Sigma^{d}{}_{\, b})^{\alpha \beta} W^{b c} W_{d c} \lambda_{i \alpha} {W}^{-2} \nabla^{a}{\lambda^{i}_{\beta}}+\frac{3}{4}{\rm i} (\Sigma_{b c})^{\alpha \beta} F^{b c} F^{a}\,_{d} \lambda_{i \alpha} {W}^{-4} \nabla^{d}{\lambda^{i}_{\beta}}+\frac{15}{16}{\rm i} (\Sigma_{c d})^{\alpha \beta} F^{a}\,_{b} \lambda_{i \alpha} \lambda^{i}_{\beta} {W}^{-3} \nabla^{b}{W^{c d}}%
+\frac{3}{2}{\rm i} (\Sigma^{a}{}_{\, b})^{\alpha \beta} F^{b c} F_{c d} \lambda_{i \alpha} {W}^{-4} \nabla^{d}{\lambda^{i}_{\beta}}+\frac{3}{8}{\rm i} (\Sigma^{a}{}_{\, d})^{\alpha \beta} F^{b c} F_{b c} \lambda_{i \alpha} {W}^{-4} \nabla^{d}{\lambda^{i}_{\beta}}+\frac{9}{64}{\rm i} \epsilon^{a}\,_{{e_{1}}}\,^{b c d} (\Gamma^{e})^{\alpha \beta} F_{b c} F_{d e} \lambda_{i \alpha} {W}^{-4} \nabla^{{e_{1}}}{\lambda^{i}_{\beta}} - \frac{15}{64}{\rm i} \epsilon^{a}\,_{{e_{1}}}\,^{b c d} (\Gamma^{{e_{1}}})^{\alpha \beta} F_{b c} F_{d e} \lambda_{i \alpha} {W}^{-4} \nabla^{e}{\lambda^{i}_{\beta}} - \frac{3}{2}{\rm i} (\Sigma^{c}{}_{\, b})^{\alpha \beta} F^{a b} F_{c d} \lambda_{i \alpha} {W}^{-4} \nabla^{d}{\lambda^{i}_{\beta}} - \frac{9}{64}{\rm i} \epsilon_{e {e_{1}}}\,^{b c d} (\Gamma^{e})^{\alpha \beta} F_{b c} F_{d}\,^{a} \lambda_{i \alpha} {W}^{-4} \nabla^{{e_{1}}}{\lambda^{i}_{\beta}} - \frac{15}{32}{\rm i} (\Sigma^{c}{}_{\, d})^{\alpha \beta} F^{a b} F_{c b} \lambda_{i \alpha} {W}^{-4} \nabla^{d}{\lambda^{i}_{\beta}} - \frac{3}{128}{\rm i} \epsilon^{a b c d e} (\Gamma_{{e_{1}}})^{\alpha \beta} F_{b c} F_{d e} \lambda_{i \alpha} {W}^{-4} \nabla^{{e_{1}}}{\lambda^{i}_{\beta}} - \frac{33}{32}{\rm i} (\Sigma_{b d})^{\alpha \beta} F^{b c} F^{a}\,_{c} \lambda_{i \alpha} {W}^{-4} \nabla^{d}{\lambda^{i}_{\beta}}+\frac{9}{16}{\rm i} (\Sigma_{b c})^{\alpha \beta} F^{b c} \lambda_{i \alpha} \lambda^{i}_{\beta} {W}^{-3} \nabla_{d}{W^{a d}} - \frac{3}{8}{\rm i} (\Sigma^{b}{}_{\, d})^{\alpha \beta} F_{b c} \lambda_{i \alpha} \lambda^{i}_{\beta} {W}^{-3} \nabla^{d}{W^{a c}} - \frac{3}{8}{\rm i} (\Sigma^{a b})^{\alpha \beta} F_{b c} \lambda_{i \alpha} \lambda^{i}_{\beta} {W}^{-3} \nabla_{d}{W^{d c}}+\frac{9}{16}{\rm i} (\Sigma^{a}{}_{\, d})^{\alpha \beta} F_{b c} \lambda_{i \alpha} \lambda^{i}_{\beta} {W}^{-3} \nabla^{d}{W^{b c}} - \frac{9}{8}{\rm i} (\Sigma^{b}{}_{\, d})^{\alpha \beta} F_{b c} \lambda_{i \alpha} \lambda^{i}_{\beta} {W}^{-3} \nabla^{c}{W^{a d}} - \frac{9}{8}{\rm i} (\Sigma^{a}{}_{\, d})^{\alpha \beta} F_{b c} \lambda_{i \alpha} \lambda^{i}_{\beta} {W}^{-3} \nabla^{b}{W^{d c}}+\frac{9}{8}{\rm i} (\Sigma_{d c})^{\alpha \beta} F^{a}\,_{b} \lambda_{i \alpha} \lambda^{i}_{\beta} {W}^{-3} \nabla^{d}{W^{c b}}+\frac{9}{8}{\rm i} (\Sigma_{b c})^{\alpha \beta} F^{a b} \lambda_{i \alpha} \lambda^{i}_{\beta} {W}^{-3} \nabla_{d}{W^{c d}}+\frac{9}{8}{\rm i} (\Sigma_{b c})^{\alpha \beta} W^{a}\,_{d} F^{b c} \lambda_{i \alpha} {W}^{-3} \nabla^{d}{\lambda^{i}_{\beta}}+\frac{9}{32}{\rm i} (\Sigma_{c d})^{\alpha \beta} W^{a}\,_{b} \lambda_{i \alpha} \lambda^{i}_{\beta} {W}^{-2} \nabla^{b}{W^{c d}}+\frac{15}{8}{\rm i} (\Sigma_{c d})^{\alpha \beta} W^{c d} F^{a}\,_{b} \lambda_{i \alpha} {W}^{-3} \nabla^{b}{\lambda^{i}_{\beta}}%
 - \frac{9}{4}{\rm i} (\Sigma_{c d})^{\alpha \beta} W^{c b} F^{a}\,_{b} \lambda_{i \alpha} {W}^{-3} \nabla^{d}{\lambda^{i}_{\beta}}+\frac{9}{4}{\rm i} (\Sigma^{a}{}_{\, d})^{\alpha \beta} W^{d b} F_{b c} \lambda_{i \alpha} {W}^{-3} \nabla^{c}{\lambda^{i}_{\beta}}+\frac{9}{8}{\rm i} (\Sigma^{a}{}_{\, d})^{\alpha \beta} W^{b c} F_{b c} \lambda_{i \alpha} {W}^{-3} \nabla^{d}{\lambda^{i}_{\beta}}+\frac{7}{128}{\rm i} \epsilon^{a}\,_{{e_{1}}}\,^{d e b} (\Gamma^{c})^{\alpha \beta} W_{d e} F_{b c} \lambda_{i \alpha} {W}^{-3} \nabla^{{e_{1}}}{\lambda^{i}_{\beta}} - \frac{17}{128}{\rm i} \epsilon^{a}\,_{{e_{1}}}\,^{d b c} (\Gamma^{{e_{1}}})^{\alpha \beta} W_{d e} F_{b c} \lambda_{i \alpha} {W}^{-3} \nabla^{e}{\lambda^{i}_{\beta}} - \frac{9}{4}{\rm i} (\Sigma^{c}{}_{\, b})^{\alpha \beta} W_{c d} F^{a b} \lambda_{i \alpha} {W}^{-3} \nabla^{d}{\lambda^{i}_{\beta}}+\frac{3}{4}{\rm i} (\Sigma^{a}{}_{\, b})^{\alpha \beta} W_{c d} F^{b c} \lambda_{i \alpha} {W}^{-3} \nabla^{d}{\lambda^{i}_{\beta}} - \frac{5}{256}{\rm i} \epsilon_{e {e_{1}}}\,^{d b c} (\Gamma^{e})^{\alpha \beta} W_{d}\,^{a} F_{b c} \lambda_{i \alpha} {W}^{-3} \nabla^{{e_{1}}}{\lambda^{i}_{\beta}} - \frac{3}{4}{\rm i} (\Sigma^{b}{}_{\, d})^{\alpha \beta} W^{a c} F_{b c} \lambda_{i \alpha} {W}^{-3} \nabla^{d}{\lambda^{i}_{\beta}} - \frac{9}{4}{\rm i} (\Sigma^{b}{}_{\, d})^{\alpha \beta} W^{a d} F_{b c} \lambda_{i \alpha} {W}^{-3} \nabla^{c}{\lambda^{i}_{\beta}} - \frac{29}{512}{\rm i} \epsilon^{a d e b c} (\Gamma_{{e_{1}}})^{\alpha \beta} W_{d e} F_{b c} \lambda_{i \alpha} {W}^{-3} \nabla^{{e_{1}}}{\lambda^{i}_{\beta}} - \frac{9}{128}{\rm i} \epsilon^{a}\,_{e {e_{1}}}\,^{d b} (\Gamma^{e})^{\alpha \beta} W_{d}\,^{c} F_{b c} \lambda_{i \alpha} {W}^{-3} \nabla^{{e_{1}}}{\lambda^{i}_{\beta}}+\frac{69}{256}{\rm i} \epsilon^{a}\,_{{e_{1}}}\,^{b c d} (\Gamma^{e})^{\alpha \beta} W_{b c} W_{d e} \lambda_{i \alpha} {W}^{-2} \nabla^{{e_{1}}}{\lambda^{i}_{\beta}} - \frac{67}{256}{\rm i} \epsilon^{a}\,_{{e_{1}}}\,^{b c d} (\Gamma^{{e_{1}}})^{\alpha \beta} W_{b c} W_{d e} \lambda_{i \alpha} {W}^{-2} \nabla^{e}{\lambda^{i}_{\beta}} - \frac{69}{256}{\rm i} \epsilon_{e {e_{1}}}\,^{b c d} (\Gamma^{e})^{\alpha \beta} W_{b c} W_{d}\,^{a} \lambda_{i \alpha} {W}^{-2} \nabla^{{e_{1}}}{\lambda^{i}_{\beta}}+\frac{1}{512}{\rm i} \epsilon^{a b c d e} (\Gamma_{{e_{1}}})^{\alpha \beta} W_{b c} W_{d e} \lambda_{i \alpha} {W}^{-2} \nabla^{{e_{1}}}{\lambda^{i}_{\beta}}+\frac{1}{4}{\rm i} (\Sigma^{d}{}_{\, b})^{\alpha \beta} W_{d c} \lambda_{i \alpha} \lambda^{i}_{\beta} {W}^{-3} \nabla^{a}{F^{b c}}+\frac{1}{4}{\rm i} (\Sigma_{b c})^{\alpha \beta} W^{a}\,_{d} \lambda_{i \alpha} \lambda^{i}_{\beta} {W}^{-3} \nabla^{d}{F^{b c}}+\frac{15}{16}{\rm i} (\Sigma_{c d})^{\alpha \beta} W^{c d} \lambda_{i \alpha} \lambda^{i}_{\beta} {W}^{-3} \nabla_{b}{F^{a b}} - \frac{11}{8}{\rm i} (\Sigma^{c}{}_{\, d})^{\alpha \beta} W_{c b} \lambda_{i \alpha} \lambda^{i}_{\beta} {W}^{-3} \nabla^{d}{F^{a b}}%
 - \frac{9}{8}{\rm i} (\Sigma^{a d})^{\alpha \beta} W_{d c} \lambda_{i \alpha} \lambda^{i}_{\beta} {W}^{-3} \nabla_{b}{F^{b c}}+\frac{1}{8}{\rm i} (\Sigma^{a}{}_{\, d})^{\alpha \beta} W_{b c} \lambda_{i \alpha} \lambda^{i}_{\beta} {W}^{-3} \nabla^{d}{F^{b c}} - \frac{7}{8}{\rm i} (\Sigma^{c}{}_{\, b})^{\alpha \beta} W_{c d} \lambda_{i \alpha} \lambda^{i}_{\beta} {W}^{-3} \nabla^{d}{F^{a b}}+\frac{1}{2}{\rm i} (\Sigma^{a}{}_{\, b})^{\alpha \beta} W_{d c} \lambda_{i \alpha} \lambda^{i}_{\beta} {W}^{-3} \nabla^{d}{F^{b c}} - \frac{1}{4}{\rm i} (\Sigma_{d b})^{\alpha \beta} W^{a}\,_{c} \lambda_{i \alpha} \lambda^{i}_{\beta} {W}^{-3} \nabla^{d}{F^{b c}}+\frac{9}{8}{\rm i} (\Sigma_{d b})^{\alpha \beta} W^{a d} \lambda_{i \alpha} \lambda^{i}_{\beta} {W}^{-3} \nabla_{c}{F^{b c}}+\frac{9}{32}{\rm i} (\Sigma_{b c})^{\alpha \beta} W^{b c} \lambda_{i \alpha} \lambda^{i}_{\beta} {W}^{-2} \nabla_{d}{W^{a d}} - \frac{9}{8}{\rm i} (\Sigma^{b}{}_{\, d})^{\alpha \beta} W_{b c} \lambda_{i \alpha} \lambda^{i}_{\beta} {W}^{-2} \nabla^{d}{W^{a c}} - \frac{9}{8}{\rm i} (\Sigma^{a b})^{\alpha \beta} W_{b c} \lambda_{i \alpha} \lambda^{i}_{\beta} {W}^{-2} \nabla_{d}{W^{d c}}+\frac{15}{16}{\rm i} (\Sigma^{a}{}_{\, d})^{\alpha \beta} W_{b c} \lambda_{i \alpha} \lambda^{i}_{\beta} {W}^{-2} \nabla^{d}{W^{b c}} - \frac{9}{16}{\rm i} (\Sigma^{b}{}_{\, d})^{\alpha \beta} W_{b c} \lambda_{i \alpha} \lambda^{i}_{\beta} {W}^{-2} \nabla^{c}{W^{a d}} - \frac{9}{8}{\rm i} (\Sigma^{a}{}_{\, d})^{\alpha \beta} W_{b c} \lambda_{i \alpha} \lambda^{i}_{\beta} {W}^{-2} \nabla^{b}{W^{d c}}+\frac{9}{8}{\rm i} (\Sigma_{d c})^{\alpha \beta} W^{a}\,_{b} \lambda_{i \alpha} \lambda^{i}_{\beta} {W}^{-2} \nabla^{d}{W^{c b}}+\frac{9}{16}{\rm i} (\Sigma_{b c})^{\alpha \beta} W^{a b} \lambda_{i \alpha} \lambda^{i}_{\beta} {W}^{-2} \nabla_{d}{W^{c d}}+\frac{3}{128}{\rm i} \epsilon_{{e_{1}}}\,^{b c d e} (\Gamma^{{e_{1}}})^{\alpha \beta} F_{b c} F_{d e} \lambda_{i \alpha} {W}^{-4} \nabla^{a}{\lambda^{i}_{\beta}}+\frac{3}{4}{\rm i} (\Sigma_{b c})^{\alpha \beta} F^{a b} \lambda_{i \alpha} \lambda^{i}_{\beta} {W}^{-4} \nabla_{d}{F^{c d}} - \frac{3}{4}{\rm i} (\Sigma^{b}{}_{\, d})^{\alpha \beta} F_{b c} \lambda_{i \alpha} \lambda^{i}_{\beta} {W}^{-4} \nabla^{d}{F^{a c}}+\frac{3}{8}{\rm i} (\Sigma_{b c})^{\alpha \beta} F^{b c} \lambda_{i \alpha} \lambda^{i}_{\beta} {W}^{-4} \nabla_{d}{F^{a d}} - \frac{3}{4}{\rm i} (\Sigma^{a b})^{\alpha \beta} F_{b c} \lambda_{i \alpha} \lambda^{i}_{\beta} {W}^{-4} \nabla_{d}{F^{d c}} - \frac{3}{4}{\rm i} (\Sigma^{b}{}_{\, d})^{\alpha \beta} F_{b c} \lambda_{i \alpha} \lambda^{i}_{\beta} {W}^{-4} \nabla^{c}{F^{a d}}%
 - \frac{7}{128}{\rm i} \epsilon_{e {e_{1}}}\,^{c d b} (\Gamma^{e})^{\alpha \beta} W_{c d} F_{b}\,^{a} \lambda_{i \alpha} {W}^{-3} \nabla^{{e_{1}}}{\lambda^{i}_{\beta}}+\frac{5}{256}{\rm i} \epsilon^{a}\,_{{e_{1}}}\,^{d b c} (\Gamma^{e})^{\alpha \beta} W_{d e} F_{b c} \lambda_{i \alpha} {W}^{-3} \nabla^{{e_{1}}}{\lambda^{i}_{\beta}} - \frac{43}{256}{\rm i} \epsilon^{a}\,_{{e_{1}}}\,^{d e b} (\Gamma^{{e_{1}}})^{\alpha \beta} W_{d e} F_{b c} \lambda_{i \alpha} {W}^{-3} \nabla^{c}{\lambda^{i}_{\beta}}+\frac{29}{512}{\rm i} \epsilon_{{e_{1}}}\,^{d e b c} (\Gamma^{{e_{1}}})^{\alpha \beta} W_{d e} F_{b c} \lambda_{i \alpha} {W}^{-3} \nabla^{a}{\lambda^{i}_{\beta}} - \frac{1}{512}{\rm i} \epsilon_{{e_{1}}}\,^{b c d e} (\Gamma^{{e_{1}}})^{\alpha \beta} W_{b c} W_{d e} \lambda_{i \alpha} {W}^{-2} \nabla^{a}{\lambda^{i}_{\beta}} - \frac{3}{16}{\rm i} \epsilon^{a b c d {e_{1}}} F_{b c} W_{d}\,^{e \alpha}\,_{i} W_{{e_{1}} e \alpha}\,^{i} {W}^{-1}+\frac{9}{32}{\rm i} \epsilon^{b c d e {e_{1}}} F_{b c} W_{d e}\,^{\alpha}\,_{i} W_{{e_{1}}}\,^{a}\,_{\alpha}\,^{i} {W}^{-1} - \frac{9}{16}{\rm i} \epsilon^{a b d e {e_{1}}} F_{b}\,^{c} W_{d e}\,^{\alpha}\,_{i} W_{{e_{1}} c \alpha}\,^{i} {W}^{-1} - \frac{3}{2}{\rm i} (\Gamma_{b})^{\alpha \beta} X_{i j} \lambda^{i}_{\alpha} \lambda^{j}_{\beta} {W}^{-5} \nabla^{a}{W} \nabla^{b}{W}+\frac{3}{2}{\rm i} (\Gamma^{a})^{\alpha \beta} X_{i j} \lambda^{i}_{\alpha} \lambda^{j}_{\beta} {W}^{-5} \nabla_{b}{W} \nabla^{b}{W} - \frac{7}{24}{\rm i} \lambda^{\beta}_{i} \lambda_{j \beta} X^{i \alpha} {W}^{-3} \nabla^{a}{\lambda^{j}_{\alpha}}+\frac{37}{96}{\rm i} \lambda^{\alpha}_{i} \lambda^{\beta}_{j} X^{i}_{\alpha} {W}^{-3} \nabla^{a}{\lambda^{j}_{\beta}} - \frac{19}{96}{\rm i} \lambda^{\beta}_{i} \lambda^{\alpha}_{j} X^{i}_{\alpha} {W}^{-3} \nabla^{a}{\lambda^{j}_{\beta}}+\frac{3}{8}{\rm i} F^{a}\,_{b} \lambda^{\alpha}_{i} \lambda_{j \alpha} {W}^{-4} \nabla^{b}{X^{i j}} - \frac{3}{16}{\rm i} \epsilon^{a}\,_{d e}\,^{b c} (\Gamma^{d})^{\alpha \beta} F_{b c} \lambda_{i \alpha} \lambda_{j \beta} {W}^{-4} \nabla^{e}{X^{i j}}+\frac{3}{16}{\rm i} W^{a}\,_{b} \lambda^{\alpha}_{i} \lambda_{j \alpha} {W}^{-3} \nabla^{b}{X^{i j}} - \frac{9}{32}{\rm i} \epsilon^{a}\,_{d e}\,^{b c} (\Gamma^{d})^{\alpha \beta} W_{b c} \lambda_{i \alpha} \lambda_{j \beta} {W}^{-3} \nabla^{e}{X^{i j}}+\frac{3}{4}{\rm i} X_{i j} F^{a}\,_{b} \lambda^{i \alpha} {W}^{-4} \nabla^{b}{\lambda^{j}_{\alpha}} - \frac{3}{8}{\rm i} \epsilon^{a}\,_{d e}\,^{b c} (\Gamma^{d})^{\alpha \beta} X_{i j} F_{b c} \lambda^{i}_{\alpha} {W}^{-4} \nabla^{e}{\lambda^{j}_{\beta}}+\frac{3}{8}{\rm i} X_{i j} W^{a}\,_{b} \lambda^{i \alpha} {W}^{-3} \nabla^{b}{\lambda^{j}_{\alpha}}%
 - \frac{9}{16}{\rm i} \epsilon^{a}\,_{d e}\,^{b c} (\Gamma^{d})^{\alpha \beta} X_{i j} W_{b c} \lambda^{i}_{\alpha} {W}^{-3} \nabla^{e}{\lambda^{j}_{\beta}}+\frac{3}{8}{\rm i} X_{i j} \lambda^{i \alpha} \lambda^{j}_{\alpha} {W}^{-4} \nabla_{b}{F^{a b}}+\frac{3}{16}{\rm i} X_{i j} \lambda^{i \alpha} \lambda^{j}_{\alpha} {W}^{-3} \nabla_{b}{W^{a b}} - \frac{9}{32}{\rm i} \epsilon^{a}\,_{d e}\,^{b c} (\Gamma^{d})^{\alpha \beta} X_{i j} \lambda^{i}_{\alpha} \lambda^{j}_{\beta} {W}^{-3} \nabla^{e}{W_{b c}} - \frac{13}{48}{\rm i} (\Sigma^{a}{}_{\, b})^{\beta \rho} \lambda_{j \beta} \lambda^{j}_{\rho} X_{i}^{\alpha} {W}^{-3} \nabla^{b}{\lambda^{i}_{\alpha}}+\frac{21}{128}{\rm i} (\Sigma^{a}{}_{\, b})^{\beta \rho} \lambda_{j \beta} \lambda^{j}_{\rho} \lambda^{\alpha}_{i} {W}^{-3} \nabla^{b}{X^{i}_{\alpha}} - \frac{3}{32}{\rm i} \lambda^{\alpha}_{j} \lambda^{j \beta} X_{i \alpha} {W}^{-3} \nabla^{a}{\lambda^{i}_{\beta}}+\frac{31}{48}{\rm i} (\Sigma^{a}{}_{\, b})^{\beta \alpha} \lambda_{j \beta} \lambda^{j \rho} X_{i \alpha} {W}^{-3} \nabla^{b}{\lambda^{i}_{\rho}}+\frac{1}{6}{\rm i} (\Sigma^{a}{}_{\, b})^{\beta \rho} \lambda_{j \beta} \lambda^{j \alpha} X_{i \alpha} {W}^{-3} \nabla^{b}{\lambda^{i}_{\rho}} - \frac{21}{128}{\rm i} (\Sigma^{a}{}_{\, b})^{\beta \rho} \lambda_{j \beta} \lambda^{j \alpha} \lambda_{i \rho} {W}^{-3} \nabla^{b}{X^{i}_{\alpha}}+\frac{53}{128}{\rm i} (\Sigma^{a}{}_{\, b})^{\beta \alpha} \lambda_{j \beta} \lambda^{j \rho} \lambda_{i \rho} {W}^{-3} \nabla^{b}{X^{i}_{\alpha}} - \frac{35}{96}{\rm i} (\Sigma^{a}{}_{\, b})^{\beta \rho} \lambda_{i \beta} \lambda_{j \rho} X^{i \alpha} {W}^{-3} \nabla^{b}{\lambda^{j}_{\alpha}}+\frac{1}{48}{\rm i} (\Sigma^{a}{}_{\, b})^{\beta \alpha} \lambda_{i \beta} \lambda^{\rho}_{j} X^{i}_{\alpha} {W}^{-3} \nabla^{b}{\lambda^{j}_{\rho}}+\frac{17}{32}{\rm i} (\Sigma^{a}{}_{\, b})^{\beta \rho} \lambda_{i \beta} \lambda^{\alpha}_{j} X^{i}_{\alpha} {W}^{-3} \nabla^{b}{\lambda^{j}_{\rho}} - \frac{11}{32}{\rm i} (\Sigma^{a}{}_{\, b})^{\beta \alpha} \lambda_{j \beta} \lambda^{\rho}_{i} X^{i}_{\alpha} {W}^{-3} \nabla^{b}{\lambda^{j}_{\rho}}+\frac{85}{48}{\rm i} (\Sigma^{a}{}_{\, b})^{\beta \rho} \lambda_{j \beta} \lambda^{\alpha}_{i} X^{i}_{\alpha} {W}^{-3} \nabla^{b}{\lambda^{j}_{\rho}}+\frac{131}{96}{\rm i} (\Sigma^{a}{}_{\, b})^{\alpha \beta} \lambda^{\rho}_{i} \lambda_{j \rho} X^{i}_{\alpha} {W}^{-3} \nabla^{b}{\lambda^{j}_{\beta}}+\frac{1}{12}{\rm i} (\Gamma^{a})^{\beta \rho} (\Gamma_{b})^{\lambda \alpha} \lambda_{j \beta} \lambda^{j}_{\lambda} X_{i \alpha} {W}^{-3} \nabla^{b}{\lambda^{i}_{\rho}}+\frac{1}{32}{\rm i} (\Gamma^{a})^{\beta \alpha} (\Gamma_{b})^{\rho \lambda} \lambda_{j \beta} \lambda^{j}_{\rho} X_{i \alpha} {W}^{-3} \nabla^{b}{\lambda^{i}_{\lambda}} - \frac{43}{256}{\rm i} (\Gamma^{a})^{\beta \alpha} (\Gamma_{b})^{\rho \lambda} \lambda_{j \beta} \lambda^{j}_{\rho} \lambda_{i \lambda} {W}^{-3} \nabla^{b}{X^{i}_{\alpha}}%
+\frac{43}{256}{\rm i} (\Gamma^{a})^{\beta \rho} (\Gamma_{b})^{\lambda \alpha} \lambda_{j \beta} \lambda^{j}_{\lambda} \lambda_{i \rho} {W}^{-3} \nabla^{b}{X^{i}_{\alpha}}+\frac{19}{192}{\rm i} (\Gamma^{a})^{\alpha \beta} (\Gamma_{b})^{\rho \lambda} \lambda_{i \rho} \lambda_{j \lambda} X^{i}_{\alpha} {W}^{-3} \nabla^{b}{\lambda^{j}_{\beta}} - \frac{7}{96}{\rm i} (\Gamma^{a})^{\beta \rho} (\Gamma_{b})^{\lambda \alpha} \lambda_{j \beta} \lambda_{i \lambda} X^{i}_{\alpha} {W}^{-3} \nabla^{b}{\lambda^{j}_{\rho}} - \frac{23}{192}{\rm i} (\Gamma^{a})^{\beta \alpha} (\Gamma_{b})^{\rho \lambda} \lambda_{j \beta} \lambda_{i \rho} X^{i}_{\alpha} {W}^{-3} \nabla^{b}{\lambda^{j}_{\lambda}}+\frac{11}{192}{\rm i} (\Gamma^{a})^{\beta \rho} (\Gamma_{b})^{\lambda \alpha} \lambda_{i \beta} \lambda_{j \lambda} X^{i}_{\alpha} {W}^{-3} \nabla^{b}{\lambda^{j}_{\rho}} - \frac{23}{48}{\rm i} (\Gamma^{a})^{\beta \alpha} (\Gamma_{b})^{\rho \lambda} \lambda_{i \beta} \lambda_{j \rho} X^{i}_{\alpha} {W}^{-3} \nabla^{b}{\lambda^{j}_{\lambda}}+\frac{37}{192}{\rm i} (\Gamma^{a})^{\beta \rho} (\Gamma_{b})^{\alpha \lambda} \lambda_{i \beta} \lambda_{j \rho} X^{i}_{\alpha} {W}^{-3} \nabla^{b}{\lambda^{j}_{\lambda}}+\frac{1}{12}W_{b c}\,^{\alpha}\,_{i} \epsilon^{a b c d e} \Phi_{d e}\,^{i}\,_{j} \lambda^{j}_{\alpha} {W}^{-1} - \frac{1}{16}{\rm i} X_{i j} X^{i j} \lambda^{\alpha}_{k} {W}^{-4} \nabla^{a}{\lambda^{k}_{\alpha}}+\frac{1}{8}{\rm i} X_{i j} X^{i}\,_{k} \lambda^{j \alpha} {W}^{-4} \nabla^{a}{\lambda^{k}_{\alpha}}+\frac{3}{8}{\rm i} (\Sigma^{a}{}_{\, b})^{\alpha \beta} X_{i j} X^{i j} \lambda_{k \alpha} {W}^{-4} \nabla^{b}{\lambda^{k}_{\beta}} - \frac{1}{2}{\rm i} (\Sigma^{a}{}_{\, b})^{\alpha \beta} X_{i j} \lambda^{i}_{\alpha} \lambda_{k \beta} {W}^{-4} \nabla^{b}{X^{j k}}+\frac{1}{8}{\rm i} (\Sigma^{a}{}_{\, b})^{\alpha \beta} X_{i j} \lambda_{k \alpha} \lambda^{k}_{\beta} {W}^{-4} \nabla^{b}{X^{i j}}+\frac{1}{8}\epsilon^{a b c d e} F_{b c} F_{d e} {W}^{-3} \nabla_{{e_{1}}}{\nabla^{{e_{1}}}{W}} - \frac{1}{64}F^{b c} F_{d e} W_{b c}\,^{\alpha}\,_{i} \epsilon^{a d e {e_{1}} {e_{2}}} (\Sigma_{{e_{1}} {e_{2}}})_{\alpha}{}^{\beta} \lambda^{i}_{\beta} {W}^{-3}+\frac{1}{32}F^{b c} F_{d}\,^{a} W_{b c}\,^{\alpha}\,_{i} (\Gamma^{d})_{\alpha}{}^{\beta} \lambda^{i}_{\beta} {W}^{-3}+\frac{5}{64}\epsilon^{a d e {e_{1}} {e_{2}}} (\Sigma_{b c})^{\alpha \beta} F^{b c} F_{d e} W_{{e_{1}} {e_{2}} \alpha i} \lambda^{i}_{\beta} {W}^{-3}+\frac{1}{64}\epsilon^{b c d e {e_{1}}} F_{b c} F_{d}\,^{a} W_{e {e_{1}}}\,^{\alpha}\,_{i} \lambda^{i}_{\alpha} {W}^{-3}+\frac{1}{16}(\Gamma^{d})^{\alpha \beta} F^{b c} F_{b}\,^{a} W_{d c \alpha i} \lambda^{i}_{\beta} {W}^{-3} - \frac{1}{32}(\Gamma^{d})^{\alpha \beta} F^{b c} F_{d}\,^{a} W_{b c \alpha i} \lambda^{i}_{\beta} {W}^{-3}%
 - \frac{1}{32}(\Gamma^{d})^{\alpha \beta} F^{b c} F_{b c} W_{d}\,^{a}\,_{\alpha i} \lambda^{i}_{\beta} {W}^{-3} - \frac{1}{16}(\Gamma^{d})^{\alpha \beta} F^{b c} F_{d b} W_{c}\,^{a}\,_{\alpha i} \lambda^{i}_{\beta} {W}^{-3}+\frac{1}{32}\epsilon^{b c d {e_{1}} {e_{2}}} (\Sigma^{a e})^{\alpha \beta} F_{b c} F_{d e} W_{{e_{1}} {e_{2}} \alpha i} \lambda^{i}_{\beta} {W}^{-3} - \frac{1}{32}\epsilon^{{e_{2}} c d e {e_{1}}} (\Sigma_{{e_{2}} b})^{\alpha \beta} F^{a b} F_{c d} W_{e {e_{1}} \alpha i} \lambda^{i}_{\beta} {W}^{-3} - \frac{1}{16}(\Gamma^{a})^{\alpha \beta} F^{b c} F_{b}\,^{d} W_{c d \alpha i} \lambda^{i}_{\beta} {W}^{-3} - \frac{1}{32}\epsilon^{a b c d {e_{1}}} F_{b c} F_{d}\,^{e} W_{{e_{1}} e}\,^{\alpha}\,_{i} \lambda^{i}_{\alpha} {W}^{-3}+\frac{1}{16}\epsilon^{{e_{2}} b c d {e_{1}}} (\Sigma_{{e_{2}}}{}^{\, a})^{\alpha \beta} F_{b c} F_{d}\,^{e} W_{{e_{1}} e \alpha i} \lambda^{i}_{\beta} {W}^{-3}+\frac{1}{16}F^{b c} F_{b d} W_{c e}\,^{\alpha}\,_{i} \epsilon^{a d e {e_{1}} {e_{2}}} (\Sigma_{{e_{1}} {e_{2}}})_{\alpha}{}^{\beta} \lambda^{i}_{\beta} {W}^{-3}+\frac{1}{16}F^{b c} F_{b}\,^{a} W_{c d}\,^{\alpha}\,_{i} (\Gamma^{d})_{\alpha}{}^{\beta} \lambda^{i}_{\beta} {W}^{-3} - \frac{1}{16}F^{b c} F_{b d} W_{c}\,^{a \alpha}\,_{i} (\Gamma^{d})_{\alpha}{}^{\beta} \lambda^{i}_{\beta} {W}^{-3} - \frac{1}{16}F_{b}\,^{c} F_{d}\,^{e} W_{c e}\,^{\alpha}\,_{i} \epsilon^{a b d {e_{1}} {e_{2}}} (\Sigma_{{e_{1}} {e_{2}}})_{\alpha}{}^{\beta} \lambda^{i}_{\beta} {W}^{-3} - \frac{1}{8}F^{a b} F_{c}\,^{d} W_{b d}\,^{\alpha}\,_{i} (\Gamma^{c})_{\alpha}{}^{\beta} \lambda^{i}_{\beta} {W}^{-3}+\frac{1}{48}{\rm i} \epsilon^{a {e_{1}} {e_{2}} d e} (\Sigma_{{e_{1}} {e_{2}}})^{\alpha \beta} C_{b c d e} F^{b c} \lambda_{i \alpha} \lambda^{i}_{\beta} {W}^{-3}+\frac{11}{384}{\rm i} \epsilon^{a {e_{1}} {e_{2}} d e} (\Sigma_{{e_{1}} {e_{2}}})^{\alpha \beta} W_{d e} W_{b c} F^{b c} \lambda_{i \alpha} \lambda^{i}_{\beta} {W}^{-3}+\frac{3}{256}{\rm i} \epsilon^{a {e_{1}} {e_{2}} b c} (\Sigma_{{e_{1}} {e_{2}}})^{\alpha \beta} W^{d e} W_{d e} F_{b c} \lambda_{i \alpha} \lambda^{i}_{\beta} {W}^{-3} - \frac{1}{384}{\rm i} \epsilon^{a {e_{1}} {e_{2}} d e} (\Sigma_{{e_{1}} {e_{2}}})^{\alpha \beta} W_{d b} W_{e c} F^{b c} \lambda_{i \alpha} \lambda^{i}_{\beta} {W}^{-3}+\frac{1}{16}{\rm i} \epsilon^{a {e_{1}} {e_{2}}}\,_{b}\,^{d} (\Sigma_{{e_{1}} {e_{2}}})^{\alpha \beta} W_{d}\,^{e} W_{e c} F^{b c} \lambda_{i \alpha} \lambda^{i}_{\beta} {W}^{-3}+\frac{1}{192}{\rm i} \epsilon^{a b c {e_{1}} {e_{2}}} (\Sigma^{d e})^{\alpha \beta} C_{b c d e} F_{{e_{1}} {e_{2}}} \lambda_{i \alpha} \lambda^{i}_{\beta} {W}^{-3} - \frac{1}{192}{\rm i} \epsilon^{a {e_{1}} {e_{2}} b c} (\Sigma_{d e})^{\alpha \beta} W^{d e} W_{{e_{1}} {e_{2}}} F_{b c} \lambda_{i \alpha} \lambda^{i}_{\beta} {W}^{-3}+\frac{1}{384}{\rm i} \epsilon^{a d {e_{1}} b c} (\Sigma^{e {e_{2}}})^{\alpha \beta} W_{d e} W_{{e_{1}} {e_{2}}} F_{b c} \lambda_{i \alpha} \lambda^{i}_{\beta} {W}^{-3}%
 - \frac{1}{64}{\rm i} \epsilon^{a {e_{2}} {e_{1}} b c} (\Sigma_{{e_{2}} d})^{\alpha \beta} W^{d e} W_{{e_{1}} e} F_{b c} \lambda_{i \alpha} \lambda^{i}_{\beta} {W}^{-3} - \frac{1}{48}{\rm i} \epsilon^{{e_{2}} b c e {e_{1}}} (\Sigma_{{e_{2}}}{}^{\, d})^{\alpha \beta} C_{b c}\,^{a}\,_{d} F_{e {e_{1}}} \lambda_{i \alpha} \lambda^{i}_{\beta} {W}^{-3}+\frac{1}{192}{\rm i} \epsilon^{{e_{2}} e {e_{1}} b c} (\Sigma_{{e_{2}} d})^{\alpha \beta} W^{a d} W_{e {e_{1}}} F_{b c} \lambda_{i \alpha} \lambda^{i}_{\beta} {W}^{-3} - \frac{1}{96}{\rm i} \epsilon^{{e_{2}} d e b c} (\Sigma_{{e_{2}}}{}^{\, {e_{1}}})^{\alpha \beta} W_{d}\,^{a} W_{e {e_{1}}} F_{b c} \lambda_{i \alpha} \lambda^{i}_{\beta} {W}^{-3} - \frac{1}{64}{\rm i} \epsilon^{{e_{1}} {e_{2}} e b c} (\Sigma_{{e_{1}} {e_{2}}})^{\alpha \beta} W^{a d} W_{e d} F_{b c} \lambda_{i \alpha} \lambda^{i}_{\beta} {W}^{-3}+\frac{1}{24}{\rm i} \epsilon^{a {e_{1}} {e_{2}} b d} (\Sigma_{{e_{1}} {e_{2}}})^{\alpha \beta} C_{b c d e} F^{c e} \lambda_{i \alpha} \lambda^{i}_{\beta} {W}^{-3}+\frac{1}{64}{\rm i} \epsilon^{a {e_{1}} {e_{2}} b c} (\Sigma_{{e_{1}} {e_{2}}})^{\alpha \beta} C_{b c d e} F^{d e} \lambda_{i \alpha} \lambda^{i}_{\beta} {W}^{-3} - \frac{1}{48}{\rm i} \epsilon^{{e_{1}} {e_{2}}}\,_{e}\,^{b c} (\Sigma_{{e_{1}} {e_{2}}})^{\alpha \beta} C_{b}\,^{a}\,_{c d} F^{e d} \lambda_{i \alpha} \lambda^{i}_{\beta} {W}^{-3} - \frac{1}{192}{\rm i} \epsilon^{{e_{1}} {e_{2}}}\,_{b}\,^{d e} (\Sigma_{{e_{1}} {e_{2}}})^{\alpha \beta} W_{d}\,^{a} W_{e c} F^{b c} \lambda_{i \alpha} \lambda^{i}_{\beta} {W}^{-3} - \frac{1}{192}{\rm i} \epsilon^{{e_{1}} {e_{2}}}\,_{b}\,^{d e} (\Sigma_{{e_{1}} {e_{2}}})^{\alpha \beta} W_{d e} W^{a}\,_{c} F^{b c} \lambda_{i \alpha} \lambda^{i}_{\beta} {W}^{-3} - \frac{67}{1536}\epsilon^{d b c e {e_{1}}} W_{d}\,^{a} F_{b c} W_{e {e_{1}}}\,^{\alpha}\,_{i} \lambda^{i}_{\alpha} {W}^{-2}+\frac{97}{768}\epsilon^{a d e b {e_{1}}} W_{d e} F_{b}\,^{c} W_{{e_{1}} c}\,^{\alpha}\,_{i} \lambda^{i}_{\alpha} {W}^{-2} - \frac{805}{6144}W^{d e} F_{b c} W_{d e}\,^{\alpha}\,_{i} \epsilon^{a b c {e_{1}} {e_{2}}} (\Sigma_{{e_{1}} {e_{2}}})_{\alpha}{}^{\beta} \lambda^{i}_{\beta} {W}^{-2}+\frac{805}{3072}W^{c d} F_{b}\,^{a} W_{c d}\,^{\alpha}\,_{i} (\Gamma^{b})_{\alpha}{}^{\beta} \lambda^{i}_{\beta} {W}^{-2}+\frac{469}{6144}\epsilon^{a d e {e_{1}} {e_{2}}} (\Sigma_{b c})^{\alpha \beta} W_{d e} F^{b c} W_{{e_{1}} {e_{2}} \alpha i} \lambda^{i}_{\beta} {W}^{-2} - \frac{277}{6144}\epsilon^{c d b e {e_{1}}} W_{c d} F_{b}\,^{a} W_{e {e_{1}}}\,^{\alpha}\,_{i} \lambda^{i}_{\alpha} {W}^{-2}+\frac{61}{1536}(\Gamma^{d})^{\alpha \beta} W^{a b} F_{b}\,^{c} W_{d c \alpha i} \lambda^{i}_{\beta} {W}^{-2}+\frac{133}{3072}(\Gamma^{d})^{\alpha \beta} W_{d}\,^{a} F^{b c} W_{b c \alpha i} \lambda^{i}_{\beta} {W}^{-2}+\frac{205}{3072}(\Gamma^{d})^{\alpha \beta} W^{b c} F_{b c} W_{d}\,^{a}\,_{\alpha i} \lambda^{i}_{\beta} {W}^{-2}+\frac{205}{1536}(\Gamma^{d})^{\alpha \beta} W_{d b} F^{b c} W_{c}\,^{a}\,_{\alpha i} \lambda^{i}_{\beta} {W}^{-2}%
 - \frac{133}{3072}\epsilon^{d e b {e_{1}} {e_{2}}} (\Sigma^{a c})^{\alpha \beta} W_{d e} F_{b c} W_{{e_{1}} {e_{2}} \alpha i} \lambda^{i}_{\beta} {W}^{-2}+\frac{133}{3072}\epsilon^{{e_{2}} c d e {e_{1}}} (\Sigma_{{e_{2}} b})^{\alpha \beta} W_{c d} F^{a b} W_{e {e_{1}} \alpha i} \lambda^{i}_{\beta} {W}^{-2}+\frac{155}{1536}(\Gamma^{a})^{\alpha \beta} W^{b d} F_{b}\,^{c} W_{d c \alpha i} \lambda^{i}_{\beta} {W}^{-2} - \frac{15}{128}\epsilon^{a d b c {e_{1}}} W_{d}\,^{e} F_{b c} W_{{e_{1}} e}\,^{\alpha}\,_{i} \lambda^{i}_{\alpha} {W}^{-2}+\frac{9}{64}\epsilon^{{e_{2}} d b c {e_{1}}} (\Sigma_{{e_{2}}}{}^{\, a})^{\alpha \beta} W_{d}\,^{e} F_{b c} W_{{e_{1}} e \alpha i} \lambda^{i}_{\beta} {W}^{-2}+\frac{9}{64}W^{b d} F_{b c} W_{d e}\,^{\alpha}\,_{i} \epsilon^{a c e {e_{1}} {e_{2}}} (\Sigma_{{e_{1}} {e_{2}}})_{\alpha}{}^{\beta} \lambda^{i}_{\beta} {W}^{-2}+\frac{9}{64}W^{b c} F_{b}\,^{a} W_{c d}\,^{\alpha}\,_{i} (\Gamma^{d})_{\alpha}{}^{\beta} \lambda^{i}_{\beta} {W}^{-2} - \frac{9}{64}W^{b d} F_{b c} W_{d}\,^{a \alpha}\,_{i} (\Gamma^{c})_{\alpha}{}^{\beta} \lambda^{i}_{\beta} {W}^{-2} - \frac{9}{64}W_{d}\,^{e} F_{b}\,^{c} W_{e c}\,^{\alpha}\,_{i} \epsilon^{a d b {e_{1}} {e_{2}}} (\Sigma_{{e_{1}} {e_{2}}})_{\alpha}{}^{\beta} \lambda^{i}_{\beta} {W}^{-2} - \frac{9}{64}W^{a d} F_{b}\,^{c} W_{d c}\,^{\alpha}\,_{i} (\Gamma^{b})_{\alpha}{}^{\beta} \lambda^{i}_{\beta} {W}^{-2} - \frac{9}{64}W_{c}\,^{d} F^{a b} W_{b d}\,^{\alpha}\,_{i} (\Gamma^{c})_{\alpha}{}^{\beta} \lambda^{i}_{\beta} {W}^{-2}+\frac{3}{256}W_{d e} F^{b c} W_{b c}\,^{\alpha}\,_{i} \epsilon^{a d e {e_{1}} {e_{2}}} (\Sigma_{{e_{1}} {e_{2}}})_{\alpha}{}^{\beta} \lambda^{i}_{\beta} {W}^{-2}+\frac{3}{128}W_{d}\,^{a} F^{b c} W_{b c}\,^{\alpha}\,_{i} (\Gamma^{d})_{\alpha}{}^{\beta} \lambda^{i}_{\beta} {W}^{-2} - \frac{3}{256}\epsilon^{a b c {e_{1}} {e_{2}}} (\Sigma_{d e})^{\alpha \beta} W^{d e} F_{b c} W_{{e_{1}} {e_{2}} \alpha i} \lambda^{i}_{\beta} {W}^{-2} - \frac{61}{1536}(\Gamma^{d})^{\alpha \beta} W^{b c} F_{b}\,^{a} W_{d c \alpha i} \lambda^{i}_{\beta} {W}^{-2}+\frac{83}{1536}(\Gamma^{b})^{\alpha \beta} W^{c d} F_{b c} W_{d}\,^{a}\,_{\alpha i} \lambda^{i}_{\beta} {W}^{-2}+\frac{3}{128}(\Gamma^{b})^{\alpha \beta} W^{c d} F_{b}\,^{a} W_{c d \alpha i} \lambda^{i}_{\beta} {W}^{-2}+\frac{3}{128}\epsilon^{d b c {e_{1}} {e_{2}}} (\Sigma^{a e})^{\alpha \beta} W_{d e} F_{b c} W_{{e_{1}} {e_{2}} \alpha i} \lambda^{i}_{\beta} {W}^{-2} - \frac{3}{128}\epsilon^{{e_{2}} b c e {e_{1}}} (\Sigma_{{e_{2}} d})^{\alpha \beta} W^{a d} F_{b c} W_{e {e_{1}} \alpha i} \lambda^{i}_{\beta} {W}^{-2} - \frac{3}{128}W_{d e} F_{b}\,^{c} W_{{e_{1}} c}\,^{\alpha}\,_{i} \epsilon^{a d e b {e_{1}}} \lambda^{i}_{\alpha} {W}^{-2}%
 - \frac{61}{1536}(\Gamma^{c})^{\alpha \beta} W_{c}\,^{d} F^{a b} W_{b d \alpha i} \lambda^{i}_{\beta} {W}^{-2}+\frac{3}{64}(\Sigma_{{e_{2}}}{}^{\, {e_{1}}})^{\alpha \beta} W_{d e} F_{b}\,^{c} W_{{e_{1}} c \alpha i} \epsilon^{a {e_{2}} d e b} \lambda^{i}_{\beta} {W}^{-2} - \frac{3}{64}(\Sigma_{b {e_{2}}})^{\alpha \beta} W_{d e} F^{b c} W_{c {e_{1}} \alpha i} \epsilon^{a {e_{2}} d e {e_{1}}} \lambda^{i}_{\beta} {W}^{-2}+\frac{3}{64}\epsilon^{{e_{1}} {e_{2}} d b e} (\Sigma_{{e_{1}} {e_{2}}})^{\alpha \beta} W_{d}\,^{a} F_{b}\,^{c} W_{e c \alpha i} \lambda^{i}_{\beta} {W}^{-2}+\frac{61}{1536}(\Gamma^{b})^{\alpha \beta} W^{a d} F_{b}\,^{c} W_{d c \alpha i} \lambda^{i}_{\beta} {W}^{-2} - \frac{5}{256}W^{b c} W_{d e} W_{b c}\,^{\alpha}\,_{i} \epsilon^{a d e {e_{1}} {e_{2}}} (\Sigma_{{e_{1}} {e_{2}}})_{\alpha}{}^{\beta} \lambda^{i}_{\beta} {W}^{-1}+\frac{3}{32}W^{b c} W_{d}\,^{a} W_{b c}\,^{\alpha}\,_{i} (\Gamma^{d})_{\alpha}{}^{\beta} \lambda^{i}_{\beta} {W}^{-1}+\frac{23}{256}\epsilon^{a d e {e_{1}} {e_{2}}} (\Sigma_{b c})^{\alpha \beta} W^{b c} W_{d e} W_{{e_{1}} {e_{2}} \alpha i} \lambda^{i}_{\beta} {W}^{-1}+\frac{5}{256}\epsilon^{b c d e {e_{1}}} W_{b c} W_{d}\,^{a} W_{e {e_{1}}}\,^{\alpha}\,_{i} \lambda^{i}_{\alpha} {W}^{-1}+\frac{17}{128}(\Gamma^{d})^{\alpha \beta} W^{b c} W_{b}\,^{a} W_{d c \alpha i} \lambda^{i}_{\beta} {W}^{-1} - \frac{5}{128}(\Gamma^{d})^{\alpha \beta} W^{b c} W_{d}\,^{a} W_{b c \alpha i} \lambda^{i}_{\beta} {W}^{-1} - \frac{5}{128}(\Gamma^{d})^{\alpha \beta} W^{b c} W_{b c} W_{d}\,^{a}\,_{\alpha i} \lambda^{i}_{\beta} {W}^{-1} - \frac{3}{128}(\Gamma^{d})^{\alpha \beta} W^{b c} W_{d b} W_{c}\,^{a}\,_{\alpha i} \lambda^{i}_{\beta} {W}^{-1}+\frac{3}{32}\epsilon^{b c d {e_{1}} {e_{2}}} (\Sigma^{a e})^{\alpha \beta} W_{b c} W_{d e} W_{{e_{1}} {e_{2}} \alpha i} \lambda^{i}_{\beta} {W}^{-1} - \frac{3}{32}\epsilon^{{e_{2}} c d e {e_{1}}} (\Sigma_{{e_{2}} b})^{\alpha \beta} W^{a b} W_{c d} W_{e {e_{1}} \alpha i} \lambda^{i}_{\beta} {W}^{-1} - \frac{3}{16}(\Gamma^{a})^{\alpha \beta} W^{b c} W_{b}\,^{d} W_{c d \alpha i} \lambda^{i}_{\beta} {W}^{-1} - \frac{17}{256}\epsilon^{a b c d {e_{1}}} W_{b c} W_{d}\,^{e} W_{{e_{1}} e}\,^{\alpha}\,_{i} \lambda^{i}_{\alpha} {W}^{-1}+\frac{17}{128}\epsilon^{{e_{2}} b c d {e_{1}}} (\Sigma_{{e_{2}}}{}^{\, a})^{\alpha \beta} W_{b c} W_{d}\,^{e} W_{{e_{1}} e \alpha i} \lambda^{i}_{\beta} {W}^{-1}+\frac{17}{128}W^{b c} W_{b d} W_{c e}\,^{\alpha}\,_{i} \epsilon^{a d e {e_{1}} {e_{2}}} (\Sigma_{{e_{1}} {e_{2}}})_{\alpha}{}^{\beta} \lambda^{i}_{\beta} {W}^{-1}+\frac{17}{128}W^{b c} W_{b}\,^{a} W_{c d}\,^{\alpha}\,_{i} (\Gamma^{d})_{\alpha}{}^{\beta} \lambda^{i}_{\beta} {W}^{-1}%
 - \frac{17}{128}W^{b c} W_{b d} W_{c}\,^{a \alpha}\,_{i} (\Gamma^{d})_{\alpha}{}^{\beta} \lambda^{i}_{\beta} {W}^{-1} - \frac{17}{128}W_{b}\,^{c} W_{d}\,^{e} W_{c e}\,^{\alpha}\,_{i} \epsilon^{a b d {e_{1}} {e_{2}}} (\Sigma_{{e_{1}} {e_{2}}})_{\alpha}{}^{\beta} \lambda^{i}_{\beta} {W}^{-1} - \frac{17}{64}W^{a b} W_{c}\,^{d} W_{b d}\,^{\alpha}\,_{i} (\Gamma^{c})_{\alpha}{}^{\beta} \lambda^{i}_{\beta} {W}^{-1} - \frac{7}{256}W_{b c} W_{d}\,^{e} W_{{e_{1}} e}\,^{\alpha}\,_{i} \epsilon^{a b c d {e_{1}}} \lambda^{i}_{\alpha} {W}^{-1}+\frac{7}{128}(\Sigma_{{e_{2}}}{}^{\, {e_{1}}})^{\alpha \beta} W_{b c} W_{d}\,^{e} W_{{e_{1}} e \alpha i} \epsilon^{a {e_{2}} b c d} \lambda^{i}_{\beta} {W}^{-1} - \frac{7}{128}(\Sigma_{b {e_{2}}})^{\alpha \beta} W^{b c} W_{d e} W_{c {e_{1}} \alpha i} \epsilon^{a {e_{2}} d e {e_{1}}} \lambda^{i}_{\beta} {W}^{-1}+\frac{7}{128}\epsilon^{{e_{1}} {e_{2}} b c e} (\Sigma_{{e_{1}} {e_{2}}})^{\alpha \beta} W_{b}\,^{a} W_{c}\,^{d} W_{e d \alpha i} \lambda^{i}_{\beta} {W}^{-1}+\frac{1}{384}{\rm i} \epsilon^{a b c {e_{1}} {e_{2}}} (\Sigma^{d e})^{\alpha \beta} C_{b c d e} W_{{e_{1}} {e_{2}}} \lambda_{i \alpha} \lambda^{i}_{\beta} {W}^{-2} - \frac{7}{1536}{\rm i} \epsilon^{a d e {e_{1}} {e_{2}}} (\Sigma_{b c})^{\alpha \beta} W^{b c} W_{d e} W_{{e_{1}} {e_{2}}} \lambda_{i \alpha} \lambda^{i}_{\beta} {W}^{-2}+\frac{67}{3072}{\rm i} \epsilon^{a {e_{1}} {e_{2}} d e} (\Sigma_{{e_{1}} {e_{2}}})^{\alpha \beta} W^{b c} W_{d e} W_{b c} \lambda_{i \alpha} \lambda^{i}_{\beta} {W}^{-2} - \frac{1}{1536}{\rm i} \epsilon^{a b c d {e_{1}}} (\Sigma^{e {e_{2}}})^{\alpha \beta} W_{b c} W_{d e} W_{{e_{1}} {e_{2}}} \lambda_{i \alpha} \lambda^{i}_{\beta} {W}^{-2} - \frac{19}{768}{\rm i} \epsilon^{a {e_{2}} d e {e_{1}}} (\Sigma_{{e_{2}} b})^{\alpha \beta} W^{b c} W_{d e} W_{{e_{1}} c} \lambda_{i \alpha} \lambda^{i}_{\beta} {W}^{-2}+\frac{7}{384}{\rm i} \epsilon^{a {e_{1}} {e_{2}} b c} (\Sigma_{{e_{1}} {e_{2}}})^{\alpha \beta} C_{b c d e} W^{d e} \lambda_{i \alpha} \lambda^{i}_{\beta} {W}^{-2}+\frac{49}{1536}{\rm i} \epsilon^{a {e_{1}} {e_{2}} d e} (\Sigma_{{e_{1}} {e_{2}}})^{\alpha \beta} W^{b c} W_{d b} W_{e c} \lambda_{i \alpha} \lambda^{i}_{\beta} {W}^{-2}+\frac{29}{384}{\rm i} \epsilon^{a {e_{1}} {e_{2}}}\,_{b}\,^{d} (\Sigma_{{e_{1}} {e_{2}}})^{\alpha \beta} W^{b c} W_{d}\,^{e} W_{e c} \lambda_{i \alpha} \lambda^{i}_{\beta} {W}^{-2} - \frac{1}{256}{\rm i} \epsilon^{a d e {e_{1}} {e_{2}}} (\Sigma^{b c})^{\alpha \beta} C_{b c d e} W_{{e_{1}} {e_{2}}} \lambda_{i \alpha} \lambda^{i}_{\beta} {W}^{-2} - \frac{1}{384}{\rm i} \epsilon^{{e_{2}} c d e {e_{1}}} (\Sigma_{{e_{2}}}{}^{\, b})^{\alpha \beta} C^{a}\,_{b c d} W_{e {e_{1}}} \lambda_{i \alpha} \lambda^{i}_{\beta} {W}^{-2}+\frac{1}{128}{\rm i} \epsilon^{{e_{2}} c d e {e_{1}}} (\Sigma_{{e_{2}} b})^{\alpha \beta} W^{a b} W_{c d} W_{e {e_{1}}} \lambda_{i \alpha} \lambda^{i}_{\beta} {W}^{-2} - \frac{1}{64}{\rm i} \epsilon^{{e_{2}} b c d e} (\Sigma_{{e_{2}}}{}^{\, {e_{1}}})^{\alpha \beta} W_{b c} W_{d}\,^{a} W_{e {e_{1}}} \lambda_{i \alpha} \lambda^{i}_{\beta} {W}^{-2} - \frac{13}{768}{\rm i} \epsilon^{{e_{1}} {e_{2}} c d e} (\Sigma_{{e_{1}} {e_{2}}})^{\alpha \beta} W^{a b} W_{c d} W_{e b} \lambda_{i \alpha} \lambda^{i}_{\beta} {W}^{-2}%
 - \frac{1}{192}{\rm i} \epsilon^{a {e_{2}} b c {e_{1}}} (\Sigma_{{e_{2}}}{}^{\, d})^{\alpha \beta} C_{b c d}\,^{e} W_{{e_{1}} e} \lambda_{i \alpha} \lambda^{i}_{\beta} {W}^{-2}+\frac{1}{192}{\rm i} \epsilon^{{e_{1}} {e_{2}} b c e} (\Sigma_{{e_{1}} {e_{2}}})^{\alpha \beta} C_{b}\,^{a}\,_{c}\,^{d} W_{e d} \lambda_{i \alpha} \lambda^{i}_{\beta} {W}^{-2}+\frac{19}{768}{\rm i} \epsilon^{a {e_{1}} {e_{2}} d e} (\Sigma_{{e_{1}} {e_{2}}})^{\alpha \beta} C_{b c d e} W^{b c} \lambda_{i \alpha} \lambda^{i}_{\beta} {W}^{-2} - \frac{11}{384}{\rm i} \epsilon^{{e_{2}} b c e {e_{1}}} (\Sigma_{{e_{2}}}{}^{\, d})^{\alpha \beta} C_{b c}\,^{a}\,_{d} W_{e {e_{1}}} \lambda_{i \alpha} \lambda^{i}_{\beta} {W}^{-2} - \frac{1}{192}{\rm i} \epsilon^{a {e_{2}} d e {e_{1}}} (\Sigma_{{e_{2}}}{}^{\, b})^{\alpha \beta} C_{b}\,^{c}\,_{d e} W_{{e_{1}} c} \lambda_{i \alpha} \lambda^{i}_{\beta} {W}^{-2} - \frac{1}{192}{\rm i} \epsilon^{{e_{1}} {e_{2}} b d e} (\Sigma_{{e_{1}} {e_{2}}})^{\alpha \beta} C_{b}\,^{c}\,_{d}\,^{a} W_{e c} \lambda_{i \alpha} \lambda^{i}_{\beta} {W}^{-2} - \frac{11}{3072}W_{d}\,^{e} F_{b c} W_{{e_{1}} e}\,^{\alpha}\,_{i} \epsilon^{a d b c {e_{1}}} \lambda^{i}_{\alpha} {W}^{-2} - \frac{11}{768}(\Sigma_{{e_{2}}}{}^{\, c})^{\alpha \beta} W_{d}\,^{e} F_{b c} W_{{e_{1}} e \alpha i} \epsilon^{a {e_{2}} d b {e_{1}}} \lambda^{i}_{\beta} {W}^{-2} - \frac{11}{3072}\epsilon^{{e_{1}} {e_{2}} b c e} (\Sigma_{{e_{1}} {e_{2}}})^{\alpha \beta} W^{a d} F_{b c} W_{e d \alpha i} \lambda^{i}_{\beta} {W}^{-2}+\frac{11}{3072}\epsilon^{{e_{1}} {e_{2}} d b c} (\Sigma_{{e_{1}} {e_{2}}})^{\alpha \beta} W_{d}\,^{e} F_{b c} W^{a}\,_{e \alpha i} \lambda^{i}_{\beta} {W}^{-2}+\frac{5}{96}{\rm i} \epsilon^{a {e_{1}} {e_{2}} b d} (\Sigma_{{e_{1}} {e_{2}}})^{\alpha \beta} C_{b c d e} W^{c e} \lambda_{i \alpha} \lambda^{i}_{\beta} {W}^{-2} - \frac{5}{192}{\rm i} \epsilon^{{e_{1}} {e_{2}}}\,_{e}\,^{b c} (\Sigma_{{e_{1}} {e_{2}}})^{\alpha \beta} C_{b}\,^{a}\,_{c d} W^{e d} \lambda_{i \alpha} \lambda^{i}_{\beta} {W}^{-2} - \frac{5}{768}{\rm i} \epsilon^{{e_{1}} {e_{2}}}\,_{b}\,^{d e} (\Sigma_{{e_{1}} {e_{2}}})^{\alpha \beta} W^{b c} W_{d}\,^{a} W_{e c} \lambda_{i \alpha} \lambda^{i}_{\beta} {W}^{-2} - \frac{5}{768}{\rm i} \epsilon^{{e_{1}} {e_{2}}}\,_{b}\,^{d e} (\Sigma_{{e_{1}} {e_{2}}})^{\alpha \beta} W^{b c} W_{d e} W^{a}\,_{c} \lambda_{i \alpha} \lambda^{i}_{\beta} {W}^{-2}+\frac{11}{8}(\Gamma_{b})^{\alpha \beta} \lambda_{i \alpha} \lambda^{i \rho} \lambda_{j \beta} {W}^{-5} \nabla^{b}{W} \nabla^{a}{\lambda^{j}_{\rho}}+\frac{5}{8}(\Gamma_{b})^{\alpha \beta} \lambda_{i \alpha} \lambda^{i \rho} \lambda_{j \rho} {W}^{-5} \nabla^{a}{W} \nabla^{b}{\lambda^{j}_{\beta}}+\frac{5}{8}(\Gamma_{b})^{\alpha \beta} \lambda_{i \alpha} \lambda^{i \rho} \lambda_{j \rho} {W}^{-5} \nabla^{b}{W} \nabla^{a}{\lambda^{j}_{\beta}}+\frac{3}{8}(\Gamma_{b})^{\alpha \beta} \lambda_{i \alpha} \lambda^{i \rho} \lambda_{j \beta} {W}^{-5} \nabla^{a}{W} \nabla^{b}{\lambda^{j}_{\rho}} - \frac{1}{8}\epsilon^{a b c}\,_{d e} (\Sigma_{b c})^{\alpha \beta} \lambda_{i \alpha} \lambda^{i \rho} \lambda_{j \beta} {W}^{-5} \nabla^{d}{W} \nabla^{e}{\lambda^{j}_{\rho}} - \frac{7}{4}(\Gamma^{a})^{\alpha \beta} \lambda_{i \alpha} \lambda^{i \rho} \lambda_{j \beta} {W}^{-5} \nabla_{b}{W} \nabla^{b}{\lambda^{j}_{\rho}}%
+\epsilon^{a b c}\,_{d e} (\Sigma_{b c})^{\alpha \beta} \lambda_{i \alpha} \lambda^{i \rho} \lambda_{j \rho} {W}^{-5} \nabla^{d}{W} \nabla^{e}{\lambda^{j}_{\beta}} - \frac{5}{4}(\Gamma^{a})^{\alpha \beta} \lambda_{i \alpha} \lambda^{i \rho} \lambda_{j \rho} {W}^{-5} \nabla_{b}{W} \nabla^{b}{\lambda^{j}_{\beta}}+\frac{1}{4}(\Gamma_{b})^{\alpha \beta} \lambda_{i \alpha} \lambda^{i \rho} \lambda_{j \beta} \lambda^{j}_{\rho} {W}^{-5} \nabla^{b}{\nabla^{a}{W}} - \frac{1}{2}(\Sigma^{a}{}_{\, c})^{\rho \lambda} (\Gamma_{b})^{\alpha \beta} \lambda_{i \rho} \lambda^{i}_{\alpha} \lambda_{j \beta} {W}^{-5} \nabla^{b}{W} \nabla^{c}{\lambda^{j}_{\lambda}}+\frac{3}{2}(\Sigma^{a}{}_{\, c})^{\rho \lambda} (\Gamma_{b})^{\alpha \beta} \lambda_{i \rho} \lambda^{i}_{\lambda} \lambda_{j \alpha} {W}^{-5} \nabla^{c}{W} \nabla^{b}{\lambda^{j}_{\beta}}+\frac{1}{4}(\Sigma^{a}{}_{\, c})^{\rho \lambda} (\Gamma_{b})^{\alpha \beta} \lambda_{i \rho} \lambda^{i}_{\lambda} \lambda_{j \alpha} {W}^{-5} \nabla^{b}{W} \nabla^{c}{\lambda^{j}_{\beta}} - \frac{3}{4}(\Sigma^{a}{}_{\, c})^{\rho \lambda} (\Gamma_{b})^{\alpha \beta} \lambda_{i \rho} \lambda^{i}_{\alpha} \lambda_{j \lambda} {W}^{-5} \nabla^{c}{W} \nabla^{b}{\lambda^{j}_{\beta}} - \frac{1}{4}(\Sigma^{a}{}_{\, c})^{\rho \lambda} (\Gamma_{b})^{\alpha \beta} \lambda_{i \rho} \lambda^{i}_{\alpha} \lambda_{j \lambda} {W}^{-5} \nabla^{b}{W} \nabla^{c}{\lambda^{j}_{\beta}} - \frac{1}{16}{\rm i} \epsilon^{a d e b c} \Phi_{d e i j} F_{b c} \lambda^{i \alpha} \lambda^{j}_{\alpha} {W}^{-3} - \frac{1}{8}{\rm i} \Phi_{b c i j} (\Gamma^{a})^{\alpha \beta} F^{b c} \lambda^{i}_{\alpha} \lambda^{j}_{\beta} {W}^{-3} - \frac{1}{8}{\rm i} \epsilon^{a d e b c} \Phi_{d e i j} W_{b c} \lambda^{i \alpha} \lambda^{j}_{\alpha} {W}^{-2}+\frac{1}{8}{\rm i} \Phi^{a b}\,_{i j} (\Gamma^{c})^{\alpha \beta} W_{b c} \lambda^{i}_{\alpha} \lambda^{j}_{\beta} {W}^{-2} - \frac{1}{8}{\rm i} \Phi_{b c i j} (\Gamma^{a})^{\alpha \beta} W^{b c} \lambda^{i}_{\alpha} \lambda^{j}_{\beta} {W}^{-2} - \frac{1}{8}{\rm i} \Phi_{c b i j} (\Gamma^{c})^{\alpha \beta} W^{a b} \lambda^{i}_{\alpha} \lambda^{j}_{\beta} {W}^{-2} - \frac{7}{64}\epsilon^{a b c d e} X_{i j} F_{b c} W_{d e}\,^{\alpha i} \lambda^{j}_{\alpha} {W}^{-3} - \frac{1}{96}(\Gamma^{a})^{\alpha \beta} X_{i j} F^{b c} W_{b c \alpha}\,^{i} \lambda^{j}_{\beta} {W}^{-3}+\frac{1}{96}\epsilon^{{e_{1}} b c d e} (\Sigma_{{e_{1}}}{}^{\, a})^{\alpha \beta} X_{i j} F_{b c} W_{d e \alpha}\,^{i} \lambda^{j}_{\beta} {W}^{-3} - \frac{1}{48}X_{i j} F_{b}\,^{c} W_{d c}\,^{\alpha i} \epsilon^{a b d e {e_{1}}} (\Sigma_{e {e_{1}}})_{\alpha}{}^{\beta} \lambda^{j}_{\beta} {W}^{-3}+\frac{1}{48}X_{i j} F^{a b} W_{b c}\,^{\alpha i} (\Gamma^{c})_{\alpha}{}^{\beta} \lambda^{j}_{\beta} {W}^{-3}+\frac{1}{48}X_{i j} F_{b}\,^{c} W^{a}\,_{c}\,^{\alpha i} (\Gamma^{b})_{\alpha}{}^{\beta} \lambda^{j}_{\beta} {W}^{-3}%
+\frac{23}{1536}\epsilon^{a b c d e} X_{i j} W_{b c} W_{d e}\,^{\alpha i} \lambda^{j}_{\alpha} {W}^{-2} - \frac{13}{768}(\Gamma^{a})^{\alpha \beta} X_{i j} W^{b c} W_{b c \alpha}\,^{i} \lambda^{j}_{\beta} {W}^{-2}+\frac{13}{768}\epsilon^{{e_{1}} b c d e} (\Sigma_{{e_{1}}}{}^{\, a})^{\alpha \beta} X_{i j} W_{b c} W_{d e \alpha}\,^{i} \lambda^{j}_{\beta} {W}^{-2} - \frac{13}{384}X_{i j} W_{b}\,^{c} W_{d c}\,^{\alpha i} \epsilon^{a b d e {e_{1}}} (\Sigma_{e {e_{1}}})_{\alpha}{}^{\beta} \lambda^{j}_{\beta} {W}^{-2}+\frac{13}{384}X_{i j} W^{a b} W_{b c}\,^{\alpha i} (\Gamma^{c})_{\alpha}{}^{\beta} \lambda^{j}_{\beta} {W}^{-2}+\frac{13}{384}X_{i j} W_{b}\,^{c} W^{a}\,_{c}\,^{\alpha i} (\Gamma^{b})_{\alpha}{}^{\beta} \lambda^{j}_{\beta} {W}^{-2} - \frac{7}{384}W_{b c}\,^{\alpha}\,_{i} \epsilon^{a b c d e} (\Sigma_{d e})^{\beta \rho} \lambda_{j \beta} \lambda^{j}_{\rho} X^{i}_{\alpha} {W}^{-2}+\frac{1}{24}W_{b c}\,^{\alpha}\,_{i} \epsilon^{a b c d e} (\Sigma_{d e})^{\beta \rho} \lambda^{i}_{\beta} \lambda_{j \rho} X^{j}_{\alpha} {W}^{-2}+\frac{1}{12}W_{b}\,^{a \alpha}\,_{i} (\Gamma^{b})^{\beta \rho} \lambda^{i}_{\beta} \lambda_{j \rho} X^{j}_{\alpha} {W}^{-2}+\frac{7}{768}W_{b c}\,^{\alpha}\,_{i} \epsilon^{a b c d e} (\Sigma_{d e})^{\rho \beta} \lambda_{j \alpha} \lambda^{j}_{\rho} X^{i}_{\beta} {W}^{-2} - \frac{7}{384}W_{b}\,^{a \alpha}\,_{i} (\Gamma^{b})^{\rho \beta} \lambda_{j \alpha} \lambda^{j}_{\rho} X^{i}_{\beta} {W}^{-2}+\frac{1}{384}W_{b c}\,^{\alpha}\,_{i} \epsilon^{a b c d e} (\Sigma_{d e})^{\rho \beta} \lambda^{i}_{\alpha} \lambda_{j \rho} X^{j}_{\beta} {W}^{-2} - \frac{1}{192}W_{b}\,^{a \alpha}\,_{i} (\Gamma^{b})^{\rho \beta} \lambda^{i}_{\alpha} \lambda_{j \rho} X^{j}_{\beta} {W}^{-2}+\frac{73}{2304}{\rm i} \epsilon^{a b c d e} \Phi_{b c i j} (\Sigma_{d e})^{\alpha \beta} X^{i j} \lambda_{k \alpha} \lambda^{k}_{\beta} {W}^{-3}+\frac{71}{1152}{\rm i} \epsilon^{a b c d e} \Phi_{b c i k} (\Sigma_{d e})^{\alpha \beta} X^{i}\,_{j} \lambda^{k}_{\alpha} \lambda^{j}_{\beta} {W}^{-3}+\frac{1}{12}{\rm i} \Phi_{b}\,^{a}\,_{i k} (\Gamma^{b})^{\alpha \beta} X^{i}\,_{j} \lambda^{k}_{\alpha} \lambda^{j}_{\beta} {W}^{-3} - \frac{1}{2}{\rm i} (\Sigma^{a}{}_{\, b})^{\alpha \beta} \lambda_{i \alpha} \lambda^{i}_{\beta} {W}^{-3} \nabla_{c}{\nabla^{c}{\nabla^{b}{W}}} - \frac{3}{2}{\rm i} (\Sigma^{a}{}_{\, b})^{\alpha \beta} \lambda_{i \alpha} \lambda^{i}_{\beta} {W}^{-5} \nabla^{b}{W} \nabla_{c}{W} \nabla^{c}{W}+\frac{3}{128}{\rm i} \epsilon^{a d}\,_{{e_{1}}}\,^{b c} (\Sigma_{d e})^{\alpha \beta} W_{b c} \lambda_{i \alpha} \lambda^{i}_{\beta} {W}^{-3} \nabla^{e}{\nabla^{{e_{1}}}{W}} - \frac{11}{128}{\rm i} \epsilon^{a d e}\,_{{e_{1}}}\,^{b} (\Sigma_{d e})^{\alpha \beta} W_{b c} \lambda_{i \alpha} \lambda^{i}_{\beta} {W}^{-3} \nabla^{{e_{1}}}{\nabla^{c}{W}}%
 - \frac{3}{8}{\rm i} \epsilon^{a d}\,_{e {e_{1}} c} (\Sigma_{d b})^{\alpha \beta} \lambda_{i \alpha} \lambda^{i}_{\beta} {W}^{-4} \nabla^{e}{W} \nabla^{{e_{1}}}{F^{b c}}+\frac{3}{8}{\rm i} \epsilon^{a d e b c} (\Sigma_{d e})^{\alpha \beta} \lambda_{i \alpha} \lambda^{i}_{\beta} {W}^{-4} \nabla_{{e_{1}}}{W} \nabla^{{e_{1}}}{F_{b c}} - \frac{3}{32}{\rm i} \epsilon^{d e}\,_{{e_{1}}}\,^{b c} (\Sigma_{d e})^{\alpha \beta} \lambda_{i \alpha} \lambda^{i}_{\beta} {W}^{-4} \nabla^{{e_{1}}}{W} \nabla^{a}{F_{b c}}+\frac{27}{64}{\rm i} \epsilon^{a d e b c} (\Sigma_{d e})^{\alpha \beta} F_{b c} \lambda_{i \alpha} {W}^{-4} \nabla_{{e_{1}}}{W} \nabla^{{e_{1}}}{\lambda^{i}_{\beta}} - \frac{1}{16}{\rm i} \epsilon^{d}\,_{e {e_{1}}}\,^{b c} (\Sigma_{d}{}^{\, a})^{\alpha \beta} W_{b c} \lambda_{i \alpha} \lambda^{i}_{\beta} {W}^{-3} \nabla^{e}{\nabla^{{e_{1}}}{W}}+\frac{51}{256}{\rm i} \epsilon^{a d e b c} (\Sigma_{d e})^{\alpha \beta} W_{b c} \lambda_{i \alpha} {W}^{-3} \nabla_{{e_{1}}}{W} \nabla^{{e_{1}}}{\lambda^{i}_{\beta}} - \frac{3}{16}{\rm i} \epsilon^{d}\,_{e {e_{1}}}\,^{b c} (\Sigma_{d}{}^{\, a})^{\alpha \beta} \lambda_{i \alpha} \lambda^{i}_{\beta} {W}^{-4} \nabla^{e}{W} \nabla^{{e_{1}}}{F_{b c}} - \frac{3}{16}{\rm i} \epsilon^{a d e}\,_{{e_{1}}}\,^{b} (\Sigma_{d e})^{\alpha \beta} \lambda_{i \alpha} \lambda^{i}_{\beta} {W}^{-4} \nabla^{{e_{1}}}{W} \nabla^{c}{F_{b c}} - \frac{19}{128}{\rm i} \epsilon^{d}\,_{e {e_{1}}}\,^{b c} (\Sigma_{d}{}^{\, a})^{\alpha \beta} \lambda_{i \alpha} \lambda^{i}_{\beta} {W}^{-3} \nabla^{e}{W} \nabla^{{e_{1}}}{W_{b c}} - \frac{3}{2}{\rm i} (\Sigma_{b c})^{\alpha \beta} F^{b c} F^{a}\,_{d} \lambda_{i \alpha} \lambda^{i}_{\beta} {W}^{-5} \nabla^{d}{W}+\frac{3}{2}{\rm i} (\Sigma^{c}{}_{\, d})^{\alpha \beta} F^{a b} F_{c b} \lambda_{i \alpha} \lambda^{i}_{\beta} {W}^{-5} \nabla^{d}{W}-3{\rm i} (\Sigma^{a}{}_{\, b})^{\alpha \beta} F^{b c} F_{c d} \lambda_{i \alpha} \lambda^{i}_{\beta} {W}^{-5} \nabla^{d}{W} - \frac{3}{4}{\rm i} (\Sigma^{a}{}_{\, d})^{\alpha \beta} F^{b c} F_{b c} \lambda_{i \alpha} \lambda^{i}_{\beta} {W}^{-5} \nabla^{d}{W}+3{\rm i} (\Sigma^{c}{}_{\, b})^{\alpha \beta} F^{a b} F_{c d} \lambda_{i \alpha} \lambda^{i}_{\beta} {W}^{-5} \nabla^{d}{W}+\frac{3}{2}{\rm i} (\Sigma_{b d})^{\alpha \beta} F^{b c} F^{a}\,_{c} \lambda_{i \alpha} \lambda^{i}_{\beta} {W}^{-5} \nabla^{d}{W} - \frac{27}{16}{\rm i} (\Sigma_{b c})^{\alpha \beta} W^{a}\,_{d} F^{b c} \lambda_{i \alpha} \lambda^{i}_{\beta} {W}^{-4} \nabla^{d}{W} - \frac{9}{16}{\rm i} (\Sigma_{b c})^{\alpha \beta} W^{b c} W^{a}\,_{d} \lambda_{i \alpha} \lambda^{i}_{\beta} {W}^{-3} \nabla^{d}{W} - \frac{45}{16}{\rm i} (\Sigma_{c d})^{\alpha \beta} W^{c d} F^{a}\,_{b} \lambda_{i \alpha} \lambda^{i}_{\beta} {W}^{-4} \nabla^{b}{W}+\frac{27}{8}{\rm i} (\Sigma_{c d})^{\alpha \beta} W^{c b} F^{a}\,_{b} \lambda_{i \alpha} \lambda^{i}_{\beta} {W}^{-4} \nabla^{d}{W} - \frac{27}{8}{\rm i} (\Sigma^{a}{}_{\, d})^{\alpha \beta} W^{d b} F_{b c} \lambda_{i \alpha} \lambda^{i}_{\beta} {W}^{-4} \nabla^{c}{W}%
 - \frac{27}{16}{\rm i} (\Sigma^{a}{}_{\, d})^{\alpha \beta} W^{b c} F_{b c} \lambda_{i \alpha} \lambda^{i}_{\beta} {W}^{-4} \nabla^{d}{W}+\frac{27}{8}{\rm i} (\Sigma^{c}{}_{\, b})^{\alpha \beta} W_{c d} F^{a b} \lambda_{i \alpha} \lambda^{i}_{\beta} {W}^{-4} \nabla^{d}{W}+\frac{9}{8}{\rm i} (\Sigma^{b}{}_{\, d})^{\alpha \beta} W^{a c} F_{b c} \lambda_{i \alpha} \lambda^{i}_{\beta} {W}^{-4} \nabla^{d}{W}+\frac{27}{8}{\rm i} (\Sigma^{b}{}_{\, d})^{\alpha \beta} W^{a d} F_{b c} \lambda_{i \alpha} \lambda^{i}_{\beta} {W}^{-4} \nabla^{c}{W}+\frac{315}{256}{\rm i} (\Sigma^{c}{}_{\, d})^{\alpha \beta} W^{a b} W_{c b} \lambda_{i \alpha} \lambda^{i}_{\beta} {W}^{-3} \nabla^{d}{W} - \frac{9}{4}{\rm i} (\Sigma^{a}{}_{\, b})^{\alpha \beta} W^{b c} W_{c d} \lambda_{i \alpha} \lambda^{i}_{\beta} {W}^{-3} \nabla^{d}{W} - \frac{15}{16}{\rm i} (\Sigma^{a}{}_{\, d})^{\alpha \beta} W^{b c} W_{b c} \lambda_{i \alpha} \lambda^{i}_{\beta} {W}^{-3} \nabla^{d}{W}+\frac{9}{8}{\rm i} (\Sigma^{c}{}_{\, b})^{\alpha \beta} W^{a b} W_{c d} \lambda_{i \alpha} \lambda^{i}_{\beta} {W}^{-3} \nabla^{d}{W}+\frac{261}{256}{\rm i} (\Sigma_{b d})^{\alpha \beta} W^{b c} W^{a}\,_{c} \lambda_{i \alpha} \lambda^{i}_{\beta} {W}^{-3} \nabla^{d}{W}+\frac{3}{4}{\rm i} (\Sigma^{d}{}_{\, b})^{\alpha \beta} F^{b c} F_{d c} \lambda_{i \alpha} \lambda^{i}_{\beta} {W}^{-5} \nabla^{a}{W} - \frac{9}{8}{\rm i} (\Sigma^{a}{}_{\, b})^{\alpha \beta} W_{c d} F^{b c} \lambda_{i \alpha} \lambda^{i}_{\beta} {W}^{-4} \nabla^{d}{W}+\frac{27}{128}{\rm i} (\Sigma^{d}{}_{\, b})^{\alpha \beta} W^{b c} W_{d c} \lambda_{i \alpha} \lambda^{i}_{\beta} {W}^{-3} \nabla^{a}{W} - \frac{3}{32}{\rm i} W^{b c \alpha}\,_{i} (\Sigma_{b c})^{\beta \rho} \lambda_{j \beta} \lambda^{j}_{\rho} {W}^{-3} \nabla^{a}{\lambda^{i}_{\alpha}}+\frac{7}{48}{\rm i} W^{b}\,_{c}\,^{\alpha}\,_{i} (\Sigma^{a}{}_{\, b})^{\beta \rho} \lambda_{j \beta} \lambda^{j}_{\rho} {W}^{-3} \nabla^{c}{\lambda^{i}_{\alpha}} - \frac{7}{48}{\rm i} W^{a b \alpha}\,_{i} (\Sigma_{b c})^{\beta \rho} \lambda_{j \beta} \lambda^{j}_{\rho} {W}^{-3} \nabla^{c}{\lambda^{i}_{\alpha}}+\frac{9}{512}{\rm i} (\Sigma^{b c})^{\beta \rho} \lambda_{j \beta} \lambda^{j}_{\rho} \lambda^{\alpha}_{i} {W}^{-3} \nabla^{a}{W_{b c \alpha}\,^{i}}+\frac{19}{256}{\rm i} (\Sigma^{a b})^{\beta \rho} \lambda_{j \beta} \lambda^{j}_{\rho} \lambda^{\alpha}_{i} {W}^{-3} \nabla^{c}{W_{b c \alpha}\,^{i}} - \frac{25}{768}{\rm i} (\Sigma^{b}{}_{\, c})^{\beta \rho} \lambda_{j \beta} \lambda^{j}_{\rho} \lambda^{\alpha}_{i} {W}^{-3} \nabla^{c}{W^{a}\,_{b \alpha}\,^{i}}+\frac{3}{32}{\rm i} W^{b c \alpha}\,_{i} (\Sigma_{b c})^{\beta \rho} \lambda^{i}_{\beta} \lambda_{j \rho} {W}^{-3} \nabla^{a}{\lambda^{j}_{\alpha}} - \frac{3}{32}{\rm i} W^{a}\,_{b}\,^{\alpha}\,_{i} \lambda^{i \beta} \lambda_{j \beta} {W}^{-3} \nabla^{b}{\lambda^{j}_{\alpha}}%
 - \frac{3}{64}{\rm i} W_{b c}\,^{\alpha}\,_{i} \epsilon^{a b c}\,_{d e} (\Gamma^{d})^{\beta \rho} \lambda^{i}_{\beta} \lambda_{j \rho} {W}^{-3} \nabla^{e}{\lambda^{j}_{\alpha}} - \frac{7}{48}{\rm i} W^{b}\,_{c}\,^{\alpha}\,_{i} (\Sigma^{a}{}_{\, b})^{\beta \rho} \lambda^{i}_{\beta} \lambda_{j \rho} {W}^{-3} \nabla^{c}{\lambda^{j}_{\alpha}}+\frac{7}{48}{\rm i} W^{a b \alpha}\,_{i} (\Sigma_{b c})^{\beta \rho} \lambda^{i}_{\beta} \lambda_{j \rho} {W}^{-3} \nabla^{c}{\lambda^{j}_{\alpha}} - \frac{9}{512}{\rm i} (\Sigma^{b c})^{\beta \rho} \lambda_{j \beta} \lambda^{j \alpha} \lambda_{i \rho} {W}^{-3} \nabla^{a}{W_{b c \alpha}\,^{i}} - \frac{263}{1536}{\rm i} \lambda^{\beta}_{j} \lambda^{j \alpha} \lambda_{i \beta} {W}^{-3} \nabla^{b}{W^{a}\,_{b \alpha}\,^{i}} - \frac{103}{3072}{\rm i} \epsilon^{a}\,_{d e}\,^{b c} (\Gamma^{d})^{\beta \rho} \lambda_{j \beta} \lambda^{j \alpha} \lambda_{i \rho} {W}^{-3} \nabla^{e}{W_{b c \alpha}\,^{i}} - \frac{19}{256}{\rm i} (\Sigma^{a b})^{\beta \rho} \lambda_{j \beta} \lambda^{j \alpha} \lambda_{i \rho} {W}^{-3} \nabla^{c}{W_{b c \alpha}\,^{i}}+\frac{25}{768}{\rm i} (\Sigma^{b}{}_{\, c})^{\beta \rho} \lambda_{j \beta} \lambda^{j \alpha} \lambda_{i \rho} {W}^{-3} \nabla^{c}{W^{a}\,_{b \alpha}\,^{i}}+\frac{1}{24}{\rm i} W^{a}\,_{b}\,^{\alpha}\,_{i} \lambda_{j \alpha} \lambda^{j \beta} {W}^{-3} \nabla^{b}{\lambda^{i}_{\beta}}+\frac{1}{16}{\rm i} W^{b c \alpha}\,_{i} (\Sigma_{b c})^{\beta \rho} \lambda_{j \alpha} \lambda^{j}_{\beta} {W}^{-3} \nabla^{a}{\lambda^{i}_{\rho}}+\frac{1}{16}{\rm i} W_{b c}\,^{\alpha}\,_{i} \epsilon^{a b c}\,_{d e} (\Gamma^{d})^{\beta \rho} \lambda_{j \alpha} \lambda^{j}_{\beta} {W}^{-3} \nabla^{e}{\lambda^{i}_{\rho}}+\frac{1}{8}{\rm i} W^{b}\,_{c}\,^{\alpha}\,_{i} (\Sigma^{a}{}_{\, b})^{\beta \rho} \lambda_{j \alpha} \lambda^{j}_{\beta} {W}^{-3} \nabla^{c}{\lambda^{i}_{\rho}}+\frac{1}{24}{\rm i} W^{a b \alpha}\,_{i} (\Sigma_{b c})^{\beta \rho} \lambda_{j \alpha} \lambda^{j}_{\beta} {W}^{-3} \nabla^{c}{\lambda^{i}_{\rho}} - \frac{1}{96}{\rm i} W^{a}\,_{b}\,^{\alpha}\,_{i} \lambda^{i \beta} \lambda_{j \alpha} {W}^{-3} \nabla^{b}{\lambda^{j}_{\beta}} - \frac{1}{32}{\rm i} W^{b c \alpha}\,_{i} (\Sigma_{b c})^{\beta \rho} \lambda^{i}_{\beta} \lambda_{j \alpha} {W}^{-3} \nabla^{a}{\lambda^{j}_{\rho}} - \frac{3}{64}{\rm i} W_{b c}\,^{\alpha}\,_{i} \epsilon^{a b c}\,_{d e} (\Gamma^{d})^{\beta \rho} \lambda^{i}_{\beta} \lambda_{j \alpha} {W}^{-3} \nabla^{e}{\lambda^{j}_{\rho}} - \frac{1}{16}{\rm i} W^{b}\,_{c}\,^{\alpha}\,_{i} (\Sigma^{a}{}_{\, b})^{\beta \rho} \lambda^{i}_{\beta} \lambda_{j \alpha} {W}^{-3} \nabla^{c}{\lambda^{j}_{\rho}} - \frac{5}{48}{\rm i} W^{a b \alpha}\,_{i} (\Sigma_{b c})^{\beta \rho} \lambda^{i}_{\beta} \lambda_{j \alpha} {W}^{-3} \nabla^{c}{\lambda^{j}_{\rho}} - \frac{3}{2}{\rm i} X_{i j} F^{a}\,_{b} \lambda^{i \alpha} \lambda^{j}_{\alpha} {W}^{-5} \nabla^{b}{W}+\frac{3}{4}{\rm i} \epsilon^{a}\,_{d e}\,^{b c} (\Gamma^{d})^{\alpha \beta} X_{i j} F_{b c} \lambda^{i}_{\alpha} \lambda^{j}_{\beta} {W}^{-5} \nabla^{e}{W}%
 - \frac{9}{16}{\rm i} X_{i j} W^{a}\,_{b} \lambda^{i \alpha} \lambda^{j}_{\alpha} {W}^{-4} \nabla^{b}{W}+\frac{27}{32}{\rm i} \epsilon^{a}\,_{d e}\,^{b c} (\Gamma^{d})^{\alpha \beta} X_{i j} W_{b c} \lambda^{i}_{\alpha} \lambda^{j}_{\beta} {W}^{-4} \nabla^{e}{W} - \frac{5}{4}{\rm i} (\Sigma^{a}{}_{\, b})^{\beta \rho} \lambda_{j \beta} \lambda^{j}_{\rho} \lambda^{\alpha}_{i} X^{i}_{\alpha} {W}^{-4} \nabla^{b}{W}+\frac{31}{32}{\rm i} (\Sigma^{a}{}_{\, b})^{\beta \rho} \lambda_{j \beta} \lambda^{j \alpha} \lambda_{i \rho} X^{i}_{\alpha} {W}^{-4} \nabla^{b}{W} - \frac{61}{32}{\rm i} (\Sigma^{a}{}_{\, b})^{\beta \alpha} \lambda_{j \beta} \lambda^{j \rho} \lambda_{i \rho} X^{i}_{\alpha} {W}^{-4} \nabla^{b}{W}+\frac{11}{64}{\rm i} (\Gamma^{a})^{\beta \alpha} (\Gamma_{b})^{\rho \lambda} \lambda_{j \beta} \lambda^{j}_{\rho} \lambda_{i \lambda} X^{i}_{\alpha} {W}^{-4} \nabla^{b}{W} - \frac{11}{64}{\rm i} (\Gamma^{a})^{\beta \rho} (\Gamma_{b})^{\lambda \alpha} \lambda_{j \beta} \lambda^{j}_{\lambda} \lambda_{i \rho} X^{i}_{\alpha} {W}^{-4} \nabla^{b}{W} - \frac{9}{16}{\rm i} (\Sigma^{a}{}_{\, b})^{\alpha \beta} X_{i j} X^{i j} \lambda_{k \alpha} \lambda^{k}_{\beta} {W}^{-5} \nabla^{b}{W} - \frac{3}{8}{\rm i} (\Sigma^{a}{}_{\, b})^{\alpha \beta} X_{i j} X^{i}\,_{k} \lambda^{j}_{\alpha} \lambda^{k}_{\beta} {W}^{-5} \nabla^{b}{W} - \frac{5}{4}(\Gamma_{b})^{\alpha \beta} \lambda_{i \alpha} \lambda^{i \rho} \lambda_{j \beta} \lambda^{j}_{\rho} {W}^{-6} \nabla^{a}{W} \nabla^{b}{W}+\frac{5}{4}(\Gamma^{a})^{\alpha \beta} \lambda_{i \alpha} \lambda^{i \rho} \lambda_{j \beta} \lambda^{j}_{\rho} {W}^{-6} \nabla_{b}{W} \nabla^{b}{W} - \frac{1}{16}\epsilon^{a b c}\,_{d e} (\Sigma_{b c})^{\alpha \beta} \lambda_{i \alpha} \lambda^{i}_{\beta} \lambda^{\rho}_{j} {W}^{-4} \nabla^{d}{\nabla^{e}{\lambda^{j}_{\rho}}} - \frac{1}{16}\epsilon^{a b c}\,_{d e} (\Sigma_{b c})^{\alpha \beta} \lambda_{i \alpha} \lambda^{i}_{\beta} {W}^{-4} \nabla^{d}{\lambda^{\rho}_{j}} \nabla^{e}{\lambda^{j}_{\rho}}+\frac{1}{16}\epsilon^{a b c}\,_{d e} (\Sigma_{b c})^{\alpha \beta} \lambda_{i \alpha} \lambda^{i \rho} \lambda_{j \beta} {W}^{-4} \nabla^{d}{\nabla^{e}{\lambda^{j}_{\rho}}}+\frac{3}{8}(\Gamma^{a})^{\alpha \beta} \lambda_{i \alpha} \lambda^{i \rho} \lambda_{j \beta} {W}^{-4} \nabla_{b}{\nabla^{b}{\lambda^{j}_{\rho}}} - \frac{1}{16}(\Gamma_{b})^{\alpha \beta} \lambda_{i \alpha} \lambda^{i \rho} \lambda_{j \beta} {W}^{-4} \nabla^{a}{\nabla^{b}{\lambda^{j}_{\rho}}} - \frac{1}{16}\epsilon^{a b c}\,_{d e} (\Sigma_{b c})^{\alpha \beta} \lambda_{i \alpha} \lambda_{j \beta} {W}^{-4} \nabla^{d}{\lambda^{i \rho}} \nabla^{e}{\lambda^{j}_{\rho}} - \frac{1}{8}(\Sigma_{b c})^{\rho \lambda} (\Gamma^{a})^{\alpha \beta} \lambda_{i \rho} \lambda^{i}_{\alpha} \lambda_{j \beta} {W}^{-4} \nabla^{b}{\nabla^{c}{\lambda^{j}_{\lambda}}}+\frac{1}{4}(\Sigma_{b c})^{\rho \lambda} (\Gamma^{a})^{\alpha \beta} \lambda_{i \rho} \lambda^{i}_{\alpha} {W}^{-4} \nabla^{b}{\lambda_{j \lambda}} \nabla^{c}{\lambda^{j}_{\beta}} - \frac{3}{16}(\Sigma_{b c})^{\rho \lambda} (\Gamma^{a})^{\alpha \beta} \lambda_{i \rho} \lambda_{j \alpha} {W}^{-4} \nabla^{b}{\lambda^{i}_{\beta}} \nabla^{c}{\lambda^{j}_{\lambda}}%
+\frac{1}{16}(\Sigma_{b c})^{\rho \lambda} (\Gamma^{a})^{\alpha \beta} \lambda_{i \rho} \lambda_{j \alpha} {W}^{-4} \nabla^{b}{\lambda^{i}_{\lambda}} \nabla^{c}{\lambda^{j}_{\beta}} - \frac{1}{16}(\Sigma_{b c})^{\rho \lambda} (\Gamma^{a})^{\alpha \beta} \lambda_{i \alpha} \lambda_{j \beta} {W}^{-4} \nabla^{b}{\lambda^{i}_{\rho}} \nabla^{c}{\lambda^{j}_{\lambda}} - \frac{1}{2}(\Sigma_{b c})^{\alpha \beta} F^{b c} \lambda_{i \alpha} \lambda^{i}_{\beta} \lambda^{\rho}_{j} {W}^{-5} \nabla^{a}{\lambda^{j}_{\rho}} - \frac{3}{8}(\Sigma_{b c})^{\alpha \beta} F^{b c} \lambda_{i \alpha} \lambda^{i \rho} \lambda_{j \rho} {W}^{-5} \nabla^{a}{\lambda^{j}_{\beta}} - \frac{69}{128}(\Sigma_{b c})^{\alpha \beta} W^{b c} \lambda_{i \alpha} \lambda^{i}_{\beta} \lambda^{\rho}_{j} {W}^{-4} \nabla^{a}{\lambda^{j}_{\rho}} - \frac{15}{128}(\Sigma_{b c})^{\alpha \beta} W^{b c} \lambda_{i \alpha} \lambda^{i \rho} \lambda_{j \beta} {W}^{-4} \nabla^{a}{\lambda^{j}_{\rho}} - \frac{41}{128}(\Sigma_{b c})^{\alpha \beta} W^{b c} \lambda_{i \alpha} \lambda^{i \rho} \lambda_{j \rho} {W}^{-4} \nabla^{a}{\lambda^{j}_{\beta}}+\frac{3}{8}F^{a}\,_{b} \lambda^{\alpha}_{i} \lambda^{i \beta} \lambda_{j \alpha} {W}^{-5} \nabla^{b}{\lambda^{j}_{\beta}} - \frac{1}{2}(\Sigma^{a b})^{\alpha \beta} F_{b c} \lambda_{i \alpha} \lambda^{i}_{\beta} \lambda^{\rho}_{j} {W}^{-5} \nabla^{c}{\lambda^{j}_{\rho}}-(\Sigma_{b c})^{\alpha \beta} F^{a b} \lambda_{i \alpha} \lambda^{i}_{\beta} \lambda^{\rho}_{j} {W}^{-5} \nabla^{c}{\lambda^{j}_{\rho}}+\frac{1}{8}(\Sigma_{b c})^{\alpha \beta} F^{b c} \lambda_{i \alpha} \lambda^{i \rho} \lambda_{j \beta} {W}^{-5} \nabla^{a}{\lambda^{j}_{\rho}} - \frac{5}{16}\epsilon^{a}\,_{d e}\,^{b c} (\Gamma^{d})^{\alpha \beta} F_{b c} \lambda_{i \alpha} \lambda^{i \rho} \lambda_{j \beta} {W}^{-5} \nabla^{e}{\lambda^{j}_{\rho}} - \frac{1}{4}(\Sigma^{a b})^{\alpha \beta} F_{b c} \lambda_{i \alpha} \lambda^{i \rho} \lambda_{j \beta} {W}^{-5} \nabla^{c}{\lambda^{j}_{\rho}} - \frac{5}{4}(\Sigma_{b c})^{\alpha \beta} F^{a b} \lambda_{i \alpha} \lambda^{i \rho} \lambda_{j \beta} {W}^{-5} \nabla^{c}{\lambda^{j}_{\rho}} - \frac{1}{2}\epsilon^{a}\,_{d e}\,^{b c} (\Gamma^{d})^{\alpha \beta} F_{b c} \lambda_{i \alpha} \lambda^{i \rho} \lambda_{j \rho} {W}^{-5} \nabla^{e}{\lambda^{j}_{\beta}}-(\Sigma^{a b})^{\alpha \beta} F_{b c} \lambda_{i \alpha} \lambda^{i \rho} \lambda_{j \rho} {W}^{-5} \nabla^{c}{\lambda^{j}_{\beta}}+\frac{1}{2}(\Sigma_{b c})^{\alpha \beta} F^{a b} \lambda_{i \alpha} \lambda^{i \rho} \lambda_{j \rho} {W}^{-5} \nabla^{c}{\lambda^{j}_{\beta}}+\frac{1}{4}(\Sigma^{a}{}_{\, d})^{\alpha \beta} (\Sigma_{b c})^{\rho \lambda} F^{b c} \lambda_{i \alpha} \lambda^{i}_{\beta} \lambda_{j \rho} {W}^{-5} \nabla^{d}{\lambda^{j}_{\lambda}} - \frac{1}{4}(\Sigma^{a}{}_{\, d})^{\alpha \beta} (\Sigma_{b c})^{\rho \lambda} F^{b c} \lambda_{i \alpha} \lambda^{i}_{\rho} \lambda_{j \beta} {W}^{-5} \nabla^{d}{\lambda^{j}_{\lambda}}+\frac{3}{2}(\Sigma^{a}{}_{\, d})^{\alpha \beta} (\Sigma_{b c})^{\rho \lambda} F^{b c} \lambda_{i \alpha} \lambda_{j \rho} \lambda^{j}_{\lambda} {W}^{-5} \nabla^{d}{\lambda^{i}_{\beta}}%
+\frac{45}{128}W^{a}\,_{b} \lambda^{\alpha}_{i} \lambda^{i \beta} \lambda_{j \alpha} {W}^{-4} \nabla^{b}{\lambda^{j}_{\beta}} - \frac{39}{64}(\Sigma^{a b})^{\alpha \beta} W_{b c} \lambda_{i \alpha} \lambda^{i}_{\beta} \lambda^{\rho}_{j} {W}^{-4} \nabla^{c}{\lambda^{j}_{\rho}} - \frac{67}{64}(\Sigma_{b c})^{\alpha \beta} W^{a b} \lambda_{i \alpha} \lambda^{i}_{\beta} \lambda^{\rho}_{j} {W}^{-4} \nabla^{c}{\lambda^{j}_{\rho}} - \frac{97}{256}\epsilon^{a}\,_{d e}\,^{b c} (\Gamma^{d})^{\alpha \beta} W_{b c} \lambda_{i \alpha} \lambda^{i \rho} \lambda_{j \beta} {W}^{-4} \nabla^{e}{\lambda^{j}_{\rho}} - \frac{27}{64}(\Sigma^{a b})^{\alpha \beta} W_{b c} \lambda_{i \alpha} \lambda^{i \rho} \lambda_{j \beta} {W}^{-4} \nabla^{c}{\lambda^{j}_{\rho}} - \frac{83}{64}(\Sigma_{b c})^{\alpha \beta} W^{a b} \lambda_{i \alpha} \lambda^{i \rho} \lambda_{j \beta} {W}^{-4} \nabla^{c}{\lambda^{j}_{\rho}} - \frac{73}{128}\epsilon^{a}\,_{d e}\,^{b c} (\Gamma^{d})^{\alpha \beta} W_{b c} \lambda_{i \alpha} \lambda^{i \rho} \lambda_{j \rho} {W}^{-4} \nabla^{e}{\lambda^{j}_{\beta}} - \frac{25}{32}(\Sigma^{a b})^{\alpha \beta} W_{b c} \lambda_{i \alpha} \lambda^{i \rho} \lambda_{j \rho} {W}^{-4} \nabla^{c}{\lambda^{j}_{\beta}}+\frac{1}{4}(\Sigma_{b c})^{\alpha \beta} W^{a b} \lambda_{i \alpha} \lambda^{i \rho} \lambda_{j \rho} {W}^{-4} \nabla^{c}{\lambda^{j}_{\beta}}+\frac{9}{64}(\Sigma^{a}{}_{\, d})^{\alpha \beta} (\Sigma_{b c})^{\rho \lambda} W^{b c} \lambda_{i \alpha} \lambda^{i}_{\beta} \lambda_{j \rho} {W}^{-4} \nabla^{d}{\lambda^{j}_{\lambda}} - \frac{9}{64}(\Sigma^{a}{}_{\, d})^{\alpha \beta} (\Sigma_{b c})^{\rho \lambda} W^{b c} \lambda_{i \alpha} \lambda^{i}_{\rho} \lambda_{j \beta} {W}^{-4} \nabla^{d}{\lambda^{j}_{\lambda}} - \frac{15}{32}(\Sigma^{a}{}_{\, d})^{\alpha \beta} (\Sigma_{b c})^{\rho \lambda} W^{b c} \lambda_{i \alpha} \lambda^{i}_{\rho} \lambda_{j \lambda} {W}^{-4} \nabla^{d}{\lambda^{j}_{\beta}}+\frac{3}{4}(\Sigma^{a}{}_{\, d})^{\alpha \beta} (\Sigma_{b c})^{\rho \lambda} W^{b c} \lambda_{i \alpha} \lambda_{j \rho} \lambda^{j}_{\lambda} {W}^{-4} \nabla^{d}{\lambda^{i}_{\beta}} - \frac{3}{64}(\Sigma^{a}{}_{\, d})^{\alpha \beta} (\Sigma_{b c})^{\rho \lambda} \lambda_{i \alpha} \lambda^{i}_{\beta} \lambda_{j \rho} \lambda^{j}_{\lambda} {W}^{-4} \nabla^{d}{W^{b c}}+\frac{3}{64}(\Sigma^{a}{}_{\, d})^{\alpha \beta} (\Sigma_{b c})^{\rho \lambda} \lambda_{i \alpha} \lambda^{i}_{\rho} \lambda_{j \beta} \lambda^{j}_{\lambda} {W}^{-4} \nabla^{d}{W^{b c}}+\frac{27}{128}\lambda^{\alpha}_{i} \lambda^{i \beta} \lambda_{j \alpha} \lambda^{j}_{\beta} {W}^{-4} \nabla_{b}{W^{a b}} - \frac{45}{256}\epsilon^{a}\,_{d e}\,^{b c} (\Gamma^{d})^{\alpha \beta} \lambda_{i \alpha} \lambda^{i \rho} \lambda_{j \beta} \lambda^{j}_{\rho} {W}^{-4} \nabla^{e}{W_{b c}}+\frac{1}{4}\lambda^{\alpha}_{i} \lambda^{i \beta} \lambda_{j \alpha} \lambda^{j}_{\beta} {W}^{-5} \nabla_{b}{F^{a b}}+\frac{3}{16}\epsilon^{d e}\,_{{e_{1}}}\,^{b c} (\Sigma_{d e})^{\rho \lambda} (\Gamma^{a})^{\alpha \beta} F_{b c} \lambda_{i \rho} \lambda^{i}_{\alpha} \lambda_{j \beta} {W}^{-5} \nabla^{{e_{1}}}{\lambda^{j}_{\lambda}}+\frac{1}{8}(\Gamma^{a})^{\alpha \beta} (\Gamma^{b})^{\rho \lambda} F_{b c} \lambda_{i \alpha} \lambda^{i}_{\rho} \lambda_{j \beta} {W}^{-5} \nabla^{c}{\lambda^{j}_{\lambda}}%
+\frac{1}{4}(\Gamma^{a})^{\alpha \beta} (\Gamma^{b})^{\rho \lambda} F_{b c} \lambda_{i \alpha} \lambda^{i}_{\rho} \lambda_{j \lambda} {W}^{-5} \nabla^{c}{\lambda^{j}_{\beta}} - \frac{1}{4}(\Gamma_{c})^{\alpha \beta} (\Gamma^{b})^{\rho \lambda} F_{b}\,^{a} \lambda_{i \alpha} \lambda^{i}_{\rho} \lambda_{j \beta} {W}^{-5} \nabla^{c}{\lambda^{j}_{\lambda}} - \frac{1}{16}\epsilon^{a e {e_{1}} b c} (\Sigma_{e {e_{1}}})^{\rho \lambda} (\Gamma_{d})^{\alpha \beta} F_{b c} \lambda_{i \rho} \lambda^{i}_{\alpha} \lambda_{j \lambda} {W}^{-5} \nabla^{d}{\lambda^{j}_{\beta}} - \frac{1}{8}(\Gamma_{c})^{\alpha \beta} (\Gamma^{b})^{\rho \lambda} F_{b}\,^{a} \lambda_{i \alpha} \lambda^{i}_{\rho} \lambda_{j \lambda} {W}^{-5} \nabla^{c}{\lambda^{j}_{\beta}}+\frac{1}{4}\epsilon^{a e {e_{1}} b c} (\Sigma_{e {e_{1}}})^{\rho \lambda} (\Gamma_{d})^{\alpha \beta} F_{b c} \lambda_{i \rho} \lambda^{i}_{\lambda} \lambda_{j \alpha} {W}^{-5} \nabla^{d}{\lambda^{j}_{\beta}}+\frac{33}{256}\epsilon^{d e}\,_{{e_{1}}}\,^{b c} (\Sigma_{d e})^{\rho \lambda} (\Gamma^{a})^{\alpha \beta} W_{b c} \lambda_{i \rho} \lambda^{i}_{\alpha} \lambda_{j \beta} {W}^{-4} \nabla^{{e_{1}}}{\lambda^{j}_{\lambda}}+\frac{15}{128}(\Gamma^{a})^{\alpha \beta} (\Gamma^{b})^{\rho \lambda} W_{b c} \lambda_{i \alpha} \lambda^{i}_{\rho} \lambda_{j \beta} {W}^{-4} \nabla^{c}{\lambda^{j}_{\lambda}}+\frac{15}{64}(\Gamma^{a})^{\alpha \beta} (\Gamma^{b})^{\rho \lambda} W_{b c} \lambda_{i \alpha} \lambda^{i}_{\rho} \lambda_{j \lambda} {W}^{-4} \nabla^{c}{\lambda^{j}_{\beta}}+\frac{1}{32}\epsilon^{a e {e_{1}} b c} (\Sigma_{e {e_{1}}})^{\rho \lambda} (\Gamma_{d})^{\alpha \beta} W_{b c} \lambda_{i \rho} \lambda^{i}_{\alpha} \lambda_{j \beta} {W}^{-4} \nabla^{d}{\lambda^{j}_{\lambda}} - \frac{11}{64}(\Gamma_{c})^{\alpha \beta} (\Gamma^{b})^{\rho \lambda} W_{b}\,^{a} \lambda_{i \alpha} \lambda^{i}_{\rho} \lambda_{j \beta} {W}^{-4} \nabla^{c}{\lambda^{j}_{\lambda}} - \frac{17}{256}\epsilon^{a e {e_{1}} b c} (\Sigma_{e {e_{1}}})^{\rho \lambda} (\Gamma_{d})^{\alpha \beta} W_{b c} \lambda_{i \rho} \lambda^{i}_{\alpha} \lambda_{j \lambda} {W}^{-4} \nabla^{d}{\lambda^{j}_{\beta}} - \frac{23}{128}(\Gamma_{c})^{\alpha \beta} (\Gamma^{b})^{\rho \lambda} W_{b}\,^{a} \lambda_{i \alpha} \lambda^{i}_{\rho} \lambda_{j \lambda} {W}^{-4} \nabla^{c}{\lambda^{j}_{\beta}}+\frac{53}{256}\epsilon^{a e {e_{1}} b c} (\Sigma_{e {e_{1}}})^{\rho \lambda} (\Gamma_{d})^{\alpha \beta} W_{b c} \lambda_{i \rho} \lambda^{i}_{\lambda} \lambda_{j \alpha} {W}^{-4} \nabla^{d}{\lambda^{j}_{\beta}}+\frac{1}{4}(\Gamma_{c})^{\alpha \beta} (\Gamma_{b})^{\rho \lambda} \lambda_{i \alpha} \lambda^{i}_{\rho} \lambda_{j \beta} \lambda^{j}_{\lambda} {W}^{-5} \nabla^{c}{F^{a b}}+\frac{27}{128}(\Gamma_{c})^{\alpha \beta} (\Gamma_{b})^{\rho \lambda} \lambda_{i \alpha} \lambda^{i}_{\rho} \lambda_{j \beta} \lambda^{j}_{\lambda} {W}^{-4} \nabla^{c}{W^{a b}}+\frac{1}{4}(\Gamma^{a})^{\alpha \beta} (\Gamma^{b})^{\rho \lambda} \lambda_{i \alpha} \lambda^{i}_{\rho} \lambda_{j \beta} \lambda^{j}_{\lambda} {W}^{-5} \nabla^{c}{F_{b c}}+\frac{27}{128}(\Gamma^{a})^{\alpha \beta} (\Gamma^{b})^{\rho \lambda} \lambda_{i \alpha} \lambda^{i}_{\rho} \lambda_{j \beta} \lambda^{j}_{\lambda} {W}^{-4} \nabla^{c}{W_{b c}} - \frac{1}{48}W_{b c}\,^{\alpha}\,_{i} W_{d}\,^{e}\,_{\alpha}\,^{i} \epsilon^{a b c d {e_{1}}} (\Sigma_{{e_{1}} e})^{\beta \rho} \lambda_{j \beta} \lambda^{j}_{\rho} {W}^{-2} - \frac{1}{96}W_{b c}\,^{\alpha}\,_{i} W_{d}\,^{a}\,_{\alpha}\,^{i} \epsilon^{b c d e {e_{1}}} (\Sigma_{e {e_{1}}})^{\beta \rho} \lambda_{j \beta} \lambda^{j}_{\rho} {W}^{-2} - \frac{1}{64}W_{b c}\,^{\alpha}\,_{i} W_{d e \alpha j} \epsilon^{a b c d e} \lambda^{i \beta} \lambda^{j}_{\beta} {W}^{-2}%
+\frac{1}{8}W_{b}\,^{c \alpha}\,_{i} W^{a}\,_{c \alpha j} (\Gamma^{b})^{\beta \rho} \lambda^{i}_{\beta} \lambda^{j}_{\rho} {W}^{-2} - \frac{1}{32}W^{b c \alpha}\,_{i} W_{b c \alpha j} (\Gamma^{a})^{\beta \rho} \lambda^{i}_{\beta} \lambda^{j}_{\rho} {W}^{-2} - \frac{1}{48}W_{b c}\,^{\alpha}\,_{i} W_{d}\,^{e}\,_{\alpha j} \epsilon^{a b c d {e_{1}}} (\Sigma_{{e_{1}} e})^{\beta \rho} \lambda^{i}_{\beta} \lambda^{j}_{\rho} {W}^{-2} - \frac{1}{96}W_{b c}\,^{\alpha}\,_{i} W_{d}\,^{a}\,_{\alpha j} \epsilon^{b c d e {e_{1}}} (\Sigma_{e {e_{1}}})^{\beta \rho} \lambda^{i}_{\beta} \lambda^{j}_{\rho} {W}^{-2} - \frac{21}{32}X_{i j} \lambda^{i \alpha} \lambda^{j \beta} \lambda_{k \alpha} {W}^{-5} \nabla^{a}{\lambda^{k}_{\beta}} - \frac{9}{8}(\Sigma^{a}{}_{\, b})^{\alpha \beta} X_{i j} \lambda^{i}_{\alpha} \lambda^{j \rho} \lambda_{k \rho} {W}^{-5} \nabla^{b}{\lambda^{k}_{\beta}} - \frac{7}{32}X_{i j} \lambda^{i \alpha} \lambda_{k \alpha} \lambda^{k \beta} {W}^{-5} \nabla^{a}{\lambda^{j}_{\beta}}+\frac{15}{16}(\Sigma^{a}{}_{\, b})^{\alpha \beta} X_{i j} \lambda^{i}_{\alpha} \lambda_{k \beta} \lambda^{k \rho} {W}^{-5} \nabla^{b}{\lambda^{j}_{\rho}} - \frac{3}{16}(\Sigma^{a}{}_{\, b})^{\alpha \beta} X_{i j} \lambda^{i}_{\alpha} \lambda^{j \rho} \lambda_{k \beta} {W}^{-5} \nabla^{b}{\lambda^{k}_{\rho}} - \frac{21}{16}(\Sigma^{a}{}_{\, b})^{\alpha \beta} X_{i j} \lambda^{i \rho} \lambda^{j}_{\rho} \lambda_{k \alpha} {W}^{-5} \nabla^{b}{\lambda^{k}_{\beta}} - \frac{3}{16}(\Sigma^{a}{}_{\, b})^{\alpha \beta} X_{i j} \lambda^{i \rho} \lambda_{k \alpha} \lambda^{k}_{\beta} {W}^{-5} \nabla^{b}{\lambda^{j}_{\rho}} - \frac{1}{16}(\Sigma^{a}{}_{\, b})^{\alpha \beta} \lambda_{k \alpha} \lambda^{k}_{\beta} \lambda^{\rho}_{i} \lambda_{j \rho} {W}^{-5} \nabla^{b}{X^{i j}}+\frac{5}{8}(\Sigma^{a}{}_{\, b})^{\alpha \beta} \lambda_{k \alpha} \lambda^{k \rho} \lambda_{i \beta} \lambda_{j \rho} {W}^{-5} \nabla^{b}{X^{i j}} - \frac{1}{32}(\Gamma^{a})^{\alpha \beta} (\Gamma_{b})^{\rho \lambda} X_{i j} \lambda^{i}_{\rho} \lambda^{j}_{\lambda} \lambda_{k \alpha} {W}^{-5} \nabla^{b}{\lambda^{k}_{\beta}} - \frac{1}{16}(\Gamma^{a})^{\alpha \beta} (\Gamma_{b})^{\rho \lambda} X_{i j} \lambda^{i}_{\alpha} \lambda^{j}_{\rho} \lambda_{k \lambda} {W}^{-5} \nabla^{b}{\lambda^{k}_{\beta}}+\frac{5}{32}(\Gamma^{a})^{\alpha \beta} (\Gamma_{b})^{\rho \lambda} X_{i j} \lambda^{i}_{\alpha} \lambda^{j}_{\rho} \lambda_{k \beta} {W}^{-5} \nabla^{b}{\lambda^{k}_{\lambda}} - \frac{1}{4}(\Gamma^{a})^{\alpha \beta} (\Gamma_{b})^{\rho \lambda} X_{i j} \lambda^{i}_{\rho} \lambda_{k \alpha} \lambda^{k}_{\lambda} {W}^{-5} \nabla^{b}{\lambda^{j}_{\beta}} - \frac{13}{32}(\Gamma^{a})^{\alpha \beta} (\Gamma_{b})^{\rho \lambda} X_{i j} \lambda^{i}_{\alpha} \lambda^{j}_{\beta} \lambda_{k \rho} {W}^{-5} \nabla^{b}{\lambda^{k}_{\lambda}}+\frac{1}{32}(\Gamma^{a})^{\alpha \beta} (\Gamma_{b})^{\rho \lambda} X_{i j} \lambda^{i}_{\alpha} \lambda_{k \beta} \lambda^{k}_{\rho} {W}^{-5} \nabla^{b}{\lambda^{j}_{\lambda}} - \frac{3}{16}(\Gamma^{a})^{\alpha \beta} (\Gamma_{b})^{\rho \lambda} \lambda_{k \alpha} \lambda^{k}_{\rho} \lambda_{i \beta} \lambda_{j \lambda} {W}^{-5} \nabla^{b}{X^{i j}}%
+\frac{3}{16}{\rm i} (\Gamma^{a})^{\alpha \beta} X_{i j} X^{i j} X_{k l} \lambda^{k}_{\alpha} \lambda^{l}_{\beta} {W}^{-5} - \frac{3}{8}{\rm i} (\Gamma^{a})^{\alpha \beta} X_{i j} X^{i}\,_{k} X^{j}\,_{l} \lambda^{k}_{\alpha} \lambda^{l}_{\beta} {W}^{-5} - \frac{3}{8}{\rm i} \epsilon^{a d e b c} (\Sigma_{d e})^{\alpha \beta} F_{b c} \lambda_{i \alpha} \lambda^{i}_{\beta} {W}^{-5} \nabla_{{e_{1}}}{W} \nabla^{{e_{1}}}{W}+\frac{45}{256}{\rm i} \epsilon^{a d e b c} (\Sigma_{d e})^{\alpha \beta} W_{b c} \lambda_{i \alpha} \lambda^{i}_{\beta} {W}^{-4} \nabla_{{e_{1}}}{W} \nabla^{{e_{1}}}{W} - \frac{3}{32}{\rm i} \epsilon^{a d e b c} (\Sigma_{d e})^{\alpha \beta} W_{b c} {W}^{-2} \nabla_{{e_{1}}}{\lambda_{i \alpha}} \nabla^{{e_{1}}}{\lambda^{i}_{\beta}} - \frac{1}{8}{\rm i} \epsilon^{a d e b c} (\Sigma_{d e})^{\alpha \beta} \lambda_{i \alpha} \lambda^{i}_{\beta} {W}^{-3} \nabla_{{e_{1}}}{\nabla^{{e_{1}}}{F_{b c}}} - \frac{9}{64}{\rm i} \epsilon^{a {e_{1}} {e_{2}} b c} (\Sigma_{{e_{1}} {e_{2}}})^{\alpha \beta} W^{d e} F_{b c} F_{d e} \lambda_{i \alpha} \lambda^{i}_{\beta} {W}^{-4}+\frac{9}{128}{\rm i} \epsilon^{a b c d e} (\Sigma_{{e_{1}} {e_{2}}})^{\alpha \beta} W^{{e_{1}} {e_{2}}} F_{b c} F_{d e} \lambda_{i \alpha} \lambda^{i}_{\beta} {W}^{-4}+\frac{9}{64}{\rm i} \epsilon^{{e_{1}} b c d e} (\Sigma^{a {e_{2}}})^{\alpha \beta} W_{{e_{1}} {e_{2}}} F_{b c} F_{d e} \lambda_{i \alpha} \lambda^{i}_{\beta} {W}^{-4} - \frac{9}{64}{\rm i} \epsilon^{{e_{2}} b c d e} (\Sigma_{{e_{2}} {e_{1}}})^{\alpha \beta} W^{a {e_{1}}} F_{b c} F_{d e} \lambda_{i \alpha} \lambda^{i}_{\beta} {W}^{-4} - \frac{3}{128}{\rm i} \epsilon^{a {e_{1}} {e_{2}} d e} (\Sigma_{{e_{1}} {e_{2}}})^{\alpha \beta} W^{b c} W_{d e} F_{b c} \lambda_{i \alpha} \lambda^{i}_{\beta} {W}^{-3} - \frac{1}{64}{\rm i} \epsilon^{d e {e_{1}} b c} (\Sigma^{a {e_{2}}})^{\alpha \beta} W_{d e} W_{{e_{1}} {e_{2}}} F_{b c} \lambda_{i \alpha} \lambda^{i}_{\beta} {W}^{-3} - \frac{3}{128}{\rm i} \epsilon^{b c d e {e_{1}}} (\Sigma^{a {e_{2}}})^{\alpha \beta} W_{b c} W_{d e} W_{{e_{1}} {e_{2}}} \lambda_{i \alpha} \lambda^{i}_{\beta} {W}^{-2} - \frac{1}{32}{\rm i} \epsilon^{{e_{2}} d e {e_{1}} b} (\Sigma_{{e_{2}}}{}^{\, a})^{\alpha \beta} W_{d e} W_{{e_{1}}}\,^{c} F_{b c} \lambda_{i \alpha} \lambda^{i}_{\beta} {W}^{-3}+\frac{1}{32}{\rm i} \epsilon^{a {e_{1}} {e_{2}} e b} (\Sigma_{{e_{1}} {e_{2}}})^{\alpha \beta} W^{d c} W_{e d} F_{b c} \lambda_{i \alpha} \lambda^{i}_{\beta} {W}^{-3}+\frac{9}{64}{\rm i} \epsilon^{{e_{1}} {e_{2}} b c d} (\Sigma^{a e})^{\alpha \beta} W_{{e_{1}} {e_{2}}} F_{b c} F_{d e} \lambda_{i \alpha} \lambda^{i}_{\beta} {W}^{-4}+\frac{9}{32}{\rm i} \epsilon^{{e_{2}} {e_{1}} b c d} (\Sigma_{{e_{2}}}{}^{\, a})^{\alpha \beta} W_{{e_{1}}}\,^{e} F_{b c} F_{d e} \lambda_{i \alpha} \lambda^{i}_{\beta} {W}^{-4}+\frac{9}{32}{\rm i} \epsilon^{a {e_{1}} {e_{2}} b d} (\Sigma_{{e_{1}} {e_{2}}})^{\alpha \beta} W^{c e} F_{b c} F_{d e} \lambda_{i \alpha} \lambda^{i}_{\beta} {W}^{-4} - \frac{5}{8}{\rm i} W^{b c \alpha}\,_{i} (\Sigma_{b c})^{\beta \rho} \lambda^{i}_{\alpha} \lambda_{j \beta} \lambda^{j}_{\rho} {W}^{-4} \nabla^{a}{W}+\frac{1}{64}{\rm i} W^{b c \alpha}\,_{i} (\Sigma_{b c})^{\beta \rho} \lambda^{i}_{\beta} \lambda_{j \alpha} \lambda^{j}_{\rho} {W}^{-4} \nabla^{a}{W}%
+\frac{3}{64}{\rm i} W^{a}\,_{b}\,^{\alpha}\,_{i} \lambda^{i \beta} \lambda_{j \alpha} \lambda^{j}_{\beta} {W}^{-4} \nabla^{b}{W}+\frac{19}{128}{\rm i} W_{b c}\,^{\alpha}\,_{i} \epsilon^{a b c}\,_{d e} (\Gamma^{d})^{\beta \rho} \lambda^{i}_{\beta} \lambda_{j \alpha} \lambda^{j}_{\rho} {W}^{-4} \nabla^{e}{W} - \frac{7}{32}{\rm i} W^{b}\,_{c}\,^{\alpha}\,_{i} (\Sigma^{a}{}_{\, b})^{\beta \rho} \lambda^{i}_{\beta} \lambda_{j \alpha} \lambda^{j}_{\rho} {W}^{-4} \nabla^{c}{W}+\frac{7}{32}{\rm i} W^{a b \alpha}\,_{i} (\Sigma_{b c})^{\beta \rho} \lambda^{i}_{\beta} \lambda_{j \alpha} \lambda^{j}_{\rho} {W}^{-4} \nabla^{c}{W} - \frac{1}{4}{\rm i} W^{b}\,_{c}\,^{\alpha}\,_{i} (\Sigma^{a}{}_{\, b})^{\beta \rho} \lambda^{i}_{\alpha} \lambda_{j \beta} \lambda^{j}_{\rho} {W}^{-4} \nabla^{c}{W}+\frac{1}{4}{\rm i} W^{a b \alpha}\,_{i} (\Sigma_{b c})^{\beta \rho} \lambda^{i}_{\alpha} \lambda_{j \beta} \lambda^{j}_{\rho} {W}^{-4} \nabla^{c}{W} - \frac{33}{128}{\rm i} \epsilon^{a d e b c} (\Sigma_{d e})^{\beta \rho} F_{b c} \lambda_{j \beta} \lambda^{j}_{\rho} \lambda^{\alpha}_{i} X^{i}_{\alpha} {W}^{-4}+\frac{3}{64}{\rm i} \epsilon^{a d e b c} (\Sigma_{d e})^{\beta \rho} F_{b c} \lambda_{j \beta} \lambda^{j \alpha} \lambda_{i \rho} X^{i}_{\alpha} {W}^{-4} - \frac{3}{8}{\rm i} (\Gamma^{b})^{\beta \rho} F_{b}\,^{a} \lambda_{j \beta} \lambda^{j \alpha} \lambda_{i \rho} X^{i}_{\alpha} {W}^{-4} - \frac{3}{16}{\rm i} \epsilon^{a d e b c} (\Sigma_{d e})^{\beta \alpha} F_{b c} \lambda_{j \beta} \lambda^{j \rho} \lambda_{i \rho} X^{i}_{\alpha} {W}^{-4}+\frac{3}{8}{\rm i} (\Gamma^{b})^{\beta \alpha} F_{b}\,^{a} \lambda_{j \beta} \lambda^{j \rho} \lambda_{i \rho} X^{i}_{\alpha} {W}^{-4} - \frac{3}{8}{\rm i} (\Sigma_{b c})^{\lambda \alpha} (\Gamma^{a})^{\beta \rho} F^{b c} \lambda_{j \lambda} \lambda^{j}_{\beta} \lambda_{i \rho} X^{i}_{\alpha} {W}^{-4}+\frac{7}{64}{\rm i} (\Sigma_{b c})^{\rho \lambda} (\Gamma^{a})^{\beta \alpha} F^{b c} \lambda_{j \rho} \lambda^{j}_{\beta} \lambda_{i \lambda} X^{i}_{\alpha} {W}^{-4} - \frac{65}{128}{\rm i} (\Sigma_{b c})^{\rho \lambda} (\Gamma^{a})^{\beta \alpha} F^{b c} \lambda_{j \rho} \lambda^{j}_{\lambda} \lambda_{i \beta} X^{i}_{\alpha} {W}^{-4} - \frac{1567}{6144}{\rm i} \epsilon^{a d e b c} (\Sigma_{d e})^{\beta \rho} W_{b c} \lambda_{j \beta} \lambda^{j}_{\rho} \lambda^{\alpha}_{i} X^{i}_{\alpha} {W}^{-3}+\frac{487}{6144}{\rm i} \epsilon^{a d e b c} (\Sigma_{d e})^{\beta \rho} W_{b c} \lambda_{j \beta} \lambda^{j \alpha} \lambda_{i \rho} X^{i}_{\alpha} {W}^{-3} - \frac{1207}{3072}{\rm i} (\Gamma^{b})^{\beta \rho} W_{b}\,^{a} \lambda_{j \beta} \lambda^{j \alpha} \lambda_{i \rho} X^{i}_{\alpha} {W}^{-3} - \frac{1207}{6144}{\rm i} \epsilon^{a d e b c} (\Sigma_{d e})^{\beta \alpha} W_{b c} \lambda_{j \beta} \lambda^{j \rho} \lambda_{i \rho} X^{i}_{\alpha} {W}^{-3}+\frac{1207}{3072}{\rm i} (\Gamma^{b})^{\beta \alpha} W_{b}\,^{a} \lambda_{j \beta} \lambda^{j \rho} \lambda_{i \rho} X^{i}_{\alpha} {W}^{-3} - \frac{1207}{3072}{\rm i} (\Sigma_{b c})^{\lambda \alpha} (\Gamma^{a})^{\beta \rho} W^{b c} \lambda_{j \lambda} \lambda^{j}_{\beta} \lambda_{i \rho} X^{i}_{\alpha} {W}^{-3}%
+\frac{1039}{3072}{\rm i} (\Sigma_{b c})^{\rho \lambda} (\Gamma^{a})^{\beta \alpha} W^{b c} \lambda_{j \rho} \lambda^{j}_{\beta} \lambda_{i \lambda} X^{i}_{\alpha} {W}^{-3} - \frac{1291}{3072}{\rm i} (\Sigma_{b c})^{\rho \lambda} (\Gamma^{a})^{\beta \alpha} W^{b c} \lambda_{j \rho} \lambda^{j}_{\lambda} \lambda_{i \beta} X^{i}_{\alpha} {W}^{-3} - \frac{3}{32}{\rm i} \epsilon^{a d e b c} (\Sigma_{d e})^{\alpha \beta} X_{i j} X^{i j} F_{b c} \lambda_{k \alpha} \lambda^{k}_{\beta} {W}^{-5} - \frac{15}{256}{\rm i} \epsilon^{a d e b c} (\Sigma_{d e})^{\alpha \beta} X_{i j} X^{i j} W_{b c} \lambda_{k \alpha} \lambda^{k}_{\beta} {W}^{-4}+\frac{3}{16}{\rm i} \epsilon^{a d e b c} (\Sigma_{d e})^{\alpha \beta} X_{i j} X^{i}\,_{k} F_{b c} \lambda^{j}_{\alpha} \lambda^{k}_{\beta} {W}^{-5}+\frac{15}{128}{\rm i} \epsilon^{a d e b c} (\Sigma_{d e})^{\alpha \beta} X_{i j} X^{i}\,_{k} W_{b c} \lambda^{j}_{\alpha} \lambda^{k}_{\beta} {W}^{-4}+\frac{73}{128}{\rm i} (\Gamma^{a})^{\beta \rho} X_{i j} \lambda^{i}_{\beta} \lambda^{j}_{\rho} \lambda^{\alpha}_{k} X^{k}_{\alpha} {W}^{-4} - \frac{47}{128}{\rm i} (\Gamma^{a})^{\beta \rho} X_{i j} \lambda^{i}_{\beta} \lambda^{j \alpha} \lambda_{k \rho} X^{k}_{\alpha} {W}^{-4}+\frac{73}{128}{\rm i} (\Gamma^{a})^{\beta \alpha} X_{i j} \lambda^{i}_{\beta} \lambda^{j \rho} \lambda_{k \rho} X^{k}_{\alpha} {W}^{-4}+\frac{33}{128}{\rm i} (\Gamma^{a})^{\beta \rho} X_{i j} \lambda^{i}_{\beta} \lambda_{k \rho} \lambda^{k \alpha} X^{j}_{\alpha} {W}^{-4}+\frac{47}{128}{\rm i} (\Gamma^{a})^{\beta \alpha} X_{i j} \lambda^{i \rho} \lambda^{j}_{\rho} \lambda_{k \beta} X^{k}_{\alpha} {W}^{-4}+\frac{7}{128}{\rm i} (\Gamma^{a})^{\beta \alpha} X_{i j} \lambda^{i \rho} \lambda_{k \beta} \lambda^{k}_{\rho} X^{j}_{\alpha} {W}^{-4}+\frac{1}{2}\epsilon^{a b c}\,_{d e} (\Sigma_{b c})^{\alpha \beta} \lambda_{i \alpha} \lambda^{i}_{\beta} \lambda^{\rho}_{j} {W}^{-5} \nabla^{d}{W} \nabla^{e}{\lambda^{j}_{\rho}}+\frac{3}{4}(\Sigma_{b c})^{\rho \lambda} (\Gamma^{a})^{\alpha \beta} \lambda_{i \rho} \lambda^{i}_{\alpha} \lambda_{j \beta} {W}^{-5} \nabla^{b}{W} \nabla^{c}{\lambda^{j}_{\lambda}}+\frac{15}{16}\epsilon^{a}\,_{d e}\,^{b c} (\Gamma^{d})^{\alpha \beta} F_{b c} \lambda_{i \alpha} \lambda^{i \rho} \lambda_{j \beta} \lambda^{j}_{\rho} {W}^{-6} \nabla^{e}{W} - \frac{15}{8}(\Sigma^{a}{}_{\, d})^{\alpha \beta} (\Sigma_{b c})^{\rho \lambda} F^{b c} \lambda_{i \alpha} \lambda^{i}_{\beta} \lambda_{j \rho} \lambda^{j}_{\lambda} {W}^{-6} \nabla^{d}{W} - \frac{9}{32}W^{a}\,_{b} \lambda^{\alpha}_{i} \lambda^{i \beta} \lambda_{j \alpha} \lambda^{j}_{\beta} {W}^{-5} \nabla^{b}{W}+\frac{63}{64}\epsilon^{a}\,_{d e}\,^{b c} (\Gamma^{d})^{\alpha \beta} W_{b c} \lambda_{i \alpha} \lambda^{i \rho} \lambda_{j \beta} \lambda^{j}_{\rho} {W}^{-5} \nabla^{e}{W} - \frac{15}{16}(\Sigma^{a}{}_{\, d})^{\alpha \beta} (\Sigma_{b c})^{\rho \lambda} W^{b c} \lambda_{i \alpha} \lambda^{i}_{\beta} \lambda_{j \rho} \lambda^{j}_{\lambda} {W}^{-5} \nabla^{d}{W}+\frac{15}{16}(\Sigma^{a}{}_{\, d})^{\alpha \beta} (\Sigma_{b c})^{\rho \lambda} W^{b c} \lambda_{i \alpha} \lambda^{i}_{\rho} \lambda_{j \beta} \lambda^{j}_{\lambda} {W}^{-5} \nabla^{d}{W}%
 - \frac{5}{8}F^{a}\,_{b} \lambda^{\alpha}_{i} \lambda^{i \beta} \lambda_{j \alpha} \lambda^{j}_{\beta} {W}^{-6} \nabla^{b}{W} - \frac{5}{8}(\Gamma^{a})^{\alpha \beta} (\Gamma^{b})^{\rho \lambda} F_{b c} \lambda_{i \alpha} \lambda^{i}_{\rho} \lambda_{j \beta} \lambda^{j}_{\lambda} {W}^{-6} \nabla^{c}{W}+\frac{5}{8}(\Gamma_{c})^{\alpha \beta} (\Gamma^{b})^{\rho \lambda} F_{b}\,^{a} \lambda_{i \alpha} \lambda^{i}_{\rho} \lambda_{j \beta} \lambda^{j}_{\lambda} {W}^{-6} \nabla^{c}{W} - \frac{9}{32}(\Gamma^{a})^{\alpha \beta} (\Gamma^{b})^{\rho \lambda} W_{b c} \lambda_{i \alpha} \lambda^{i}_{\rho} \lambda_{j \beta} \lambda^{j}_{\lambda} {W}^{-5} \nabla^{c}{W}+\frac{9}{32}(\Gamma_{c})^{\alpha \beta} (\Gamma^{b})^{\rho \lambda} W_{b}\,^{a} \lambda_{i \alpha} \lambda^{i}_{\rho} \lambda_{j \beta} \lambda^{j}_{\lambda} {W}^{-5} \nabla^{c}{W}+\frac{5}{4}(\Sigma^{a}{}_{\, b})^{\alpha \beta} X_{i j} \lambda^{i}_{\alpha} \lambda^{j \rho} \lambda_{k \beta} \lambda^{k}_{\rho} {W}^{-6} \nabla^{b}{W}+\frac{15}{8}(\Sigma^{a}{}_{\, b})^{\alpha \beta} X_{i j} \lambda^{i \rho} \lambda^{j}_{\rho} \lambda_{k \alpha} \lambda^{k}_{\beta} {W}^{-6} \nabla^{b}{W}+\frac{5}{8}(\Gamma^{a})^{\alpha \beta} (\Gamma_{b})^{\rho \lambda} X_{i j} \lambda^{i}_{\alpha} \lambda^{j}_{\rho} \lambda_{k \beta} \lambda^{k}_{\lambda} {W}^{-6} \nabla^{b}{W}+\frac{3}{16}{\rm i} \epsilon^{a d e b c} (\Sigma_{d e})^{\alpha \beta} F_{b c} \lambda_{i \alpha} \lambda^{i}_{\beta} {W}^{-4} \nabla_{{e_{1}}}{\nabla^{{e_{1}}}{W}}+\frac{1}{16}{\rm i} \epsilon^{{e_{1}} b c d e} (\Sigma_{{e_{1}}}{}^{\, a})^{\beta \rho} F_{b c} W_{d e}\,^{\alpha}\,_{i} \lambda^{i}_{\alpha} \lambda_{j \beta} \lambda^{j}_{\rho} {W}^{-4} - \frac{1}{8}{\rm i} F_{b}\,^{c} W_{d c}\,^{\alpha}\,_{i} \epsilon^{a b d e {e_{1}}} (\Sigma_{e {e_{1}}})^{\beta \rho} \lambda^{i}_{\alpha} \lambda_{j \beta} \lambda^{j}_{\rho} {W}^{-4}+\frac{21}{128}{\rm i} (\Gamma^{a})^{\beta \rho} F^{b c} W_{b c}\,^{\alpha}\,_{i} \lambda^{i}_{\beta} \lambda_{j \rho} \lambda^{j}_{\alpha} {W}^{-4}+\frac{5}{64}{\rm i} \epsilon^{a b c d e} F_{b c} W_{d e}\,^{\alpha}\,_{i} \lambda^{i \beta} \lambda_{j \alpha} \lambda^{j}_{\beta} {W}^{-4}+\frac{1}{32}{\rm i} \epsilon^{{e_{1}} b c d e} (\Sigma_{{e_{1}}}{}^{\, a})^{\beta \rho} F_{b c} W_{d e}\,^{\alpha}\,_{i} \lambda^{i}_{\beta} \lambda_{j \rho} \lambda^{j}_{\alpha} {W}^{-4} - \frac{1}{16}{\rm i} F_{b}\,^{c} W_{d c}\,^{\alpha}\,_{i} \epsilon^{a b d e {e_{1}}} (\Sigma_{e {e_{1}}})^{\beta \rho} \lambda^{i}_{\beta} \lambda_{j \alpha} \lambda^{j}_{\rho} {W}^{-4} - \frac{1}{16}{\rm i} F^{a b} W_{b c}\,^{\alpha}\,_{i} (\Gamma^{c})^{\beta \rho} \lambda^{i}_{\beta} \lambda_{j \alpha} \lambda^{j}_{\rho} {W}^{-4} - \frac{1}{16}{\rm i} F_{b}\,^{c} W^{a}\,_{c}\,^{\alpha}\,_{i} (\Gamma^{b})^{\beta \rho} \lambda^{i}_{\beta} \lambda_{j \alpha} \lambda^{j}_{\rho} {W}^{-4}+\frac{1}{32}{\rm i} (\Sigma_{{e_{1}}}{}^{\, e})^{\beta \rho} F_{b c} W_{d e}\,^{\alpha}\,_{i} \epsilon^{a {e_{1}} b c d} \lambda^{i}_{\alpha} \lambda_{j \beta} \lambda^{j}_{\rho} {W}^{-4}+\frac{1}{64}{\rm i} \epsilon^{e {e_{1}} b c d} (\Sigma_{e {e_{1}}})^{\beta \rho} F_{b}\,^{a} W_{c d}\,^{\alpha}\,_{i} \lambda^{i}_{\alpha} \lambda_{j \beta} \lambda^{j}_{\rho} {W}^{-4}+\frac{1}{256}{\rm i} F_{b c} W_{d e}\,^{\alpha}\,_{i} \epsilon^{a b c d e} \lambda^{i \beta} \lambda_{j \alpha} \lambda^{j}_{\beta} {W}^{-4}%
 - \frac{1}{64}{\rm i} (\Gamma^{b})^{\beta \rho} F_{b}\,^{c} W^{a}\,_{c}\,^{\alpha}\,_{i} \lambda^{i}_{\beta} \lambda_{j \rho} \lambda^{j}_{\alpha} {W}^{-4}+\frac{1}{64}{\rm i} (\Sigma_{{e_{1}}}{}^{\, e})^{\beta \rho} F_{b c} W_{d e}\,^{\alpha}\,_{i} \epsilon^{a {e_{1}} b c d} \lambda^{i}_{\beta} \lambda_{j \rho} \lambda^{j}_{\alpha} {W}^{-4}+\frac{1}{128}{\rm i} \epsilon^{e {e_{1}} b c d} (\Sigma_{e {e_{1}}})^{\beta \rho} F_{b}\,^{a} W_{c d}\,^{\alpha}\,_{i} \lambda^{i}_{\beta} \lambda_{j \rho} \lambda^{j}_{\alpha} {W}^{-4} - \frac{1}{64}{\rm i} (\Gamma^{c})^{\beta \rho} F^{a b} W_{c b}\,^{\alpha}\,_{i} \lambda^{i}_{\beta} \lambda_{j \rho} \lambda^{j}_{\alpha} {W}^{-4}+\frac{1171}{12288}{\rm i} \epsilon^{{e_{1}} b c d e} (\Sigma_{{e_{1}}}{}^{\, a})^{\beta \rho} W_{b c} W_{d e}\,^{\alpha}\,_{i} \lambda^{i}_{\alpha} \lambda_{j \beta} \lambda^{j}_{\rho} {W}^{-3} - \frac{1267}{6144}{\rm i} W_{b}\,^{c} W_{d c}\,^{\alpha}\,_{i} \epsilon^{a b d e {e_{1}}} (\Sigma_{e {e_{1}}})^{\beta \rho} \lambda^{i}_{\alpha} \lambda_{j \beta} \lambda^{j}_{\rho} {W}^{-3}+\frac{865}{4096}{\rm i} (\Gamma^{a})^{\beta \rho} W^{b c} W_{b c}\,^{\alpha}\,_{i} \lambda^{i}_{\beta} \lambda_{j \rho} \lambda^{j}_{\alpha} {W}^{-3}+\frac{289}{8192}{\rm i} \epsilon^{a b c d e} W_{b c} W_{d e}\,^{\alpha}\,_{i} \lambda^{i \beta} \lambda_{j \alpha} \lambda^{j}_{\beta} {W}^{-3} - \frac{19}{12288}{\rm i} \epsilon^{{e_{1}} b c d e} (\Sigma_{{e_{1}}}{}^{\, a})^{\beta \rho} W_{b c} W_{d e}\,^{\alpha}\,_{i} \lambda^{i}_{\beta} \lambda_{j \rho} \lambda^{j}_{\alpha} {W}^{-3}+\frac{115}{6144}{\rm i} W_{b}\,^{c} W_{d c}\,^{\alpha}\,_{i} \epsilon^{a b d e {e_{1}}} (\Sigma_{e {e_{1}}})^{\beta \rho} \lambda^{i}_{\beta} \lambda_{j \alpha} \lambda^{j}_{\rho} {W}^{-3} - \frac{1139}{6144}{\rm i} W^{a b} W_{b c}\,^{\alpha}\,_{i} (\Gamma^{c})^{\beta \rho} \lambda^{i}_{\beta} \lambda_{j \alpha} \lambda^{j}_{\rho} {W}^{-3} - \frac{1139}{6144}{\rm i} W_{b}\,^{c} W^{a}\,_{c}\,^{\alpha}\,_{i} (\Gamma^{b})^{\beta \rho} \lambda^{i}_{\beta} \lambda_{j \alpha} \lambda^{j}_{\rho} {W}^{-3}+\frac{5}{192}{\rm i} (\Sigma_{{e_{1}}}{}^{\, e})^{\beta \rho} W_{b c} W_{d e}\,^{\alpha}\,_{i} \epsilon^{a {e_{1}} b c d} \lambda^{i}_{\alpha} \lambda_{j \beta} \lambda^{j}_{\rho} {W}^{-3}+\frac{5}{384}{\rm i} \epsilon^{e {e_{1}} b c d} (\Sigma_{e {e_{1}}})^{\beta \rho} W_{b}\,^{a} W_{c d}\,^{\alpha}\,_{i} \lambda^{i}_{\alpha} \lambda_{j \beta} \lambda^{j}_{\rho} {W}^{-3}+\frac{5}{1536}{\rm i} W_{b c} W_{d e}\,^{\alpha}\,_{i} \epsilon^{a b c d e} \lambda^{i \beta} \lambda_{j \alpha} \lambda^{j}_{\beta} {W}^{-3} - \frac{5}{384}{\rm i} (\Gamma^{b})^{\beta \rho} W_{b}\,^{c} W^{a}\,_{c}\,^{\alpha}\,_{i} \lambda^{i}_{\beta} \lambda_{j \rho} \lambda^{j}_{\alpha} {W}^{-3}+\frac{5}{384}{\rm i} (\Sigma_{{e_{1}}}{}^{\, e})^{\beta \rho} W_{b c} W_{d e}\,^{\alpha}\,_{i} \epsilon^{a {e_{1}} b c d} \lambda^{i}_{\beta} \lambda_{j \rho} \lambda^{j}_{\alpha} {W}^{-3}+\frac{5}{768}{\rm i} \epsilon^{e {e_{1}} b c d} (\Sigma_{e {e_{1}}})^{\beta \rho} W_{b}\,^{a} W_{c d}\,^{\alpha}\,_{i} \lambda^{i}_{\beta} \lambda_{j \rho} \lambda^{j}_{\alpha} {W}^{-3} - \frac{5}{384}{\rm i} (\Gamma^{c})^{\beta \rho} W^{a b} W_{c b}\,^{\alpha}\,_{i} \lambda^{i}_{\beta} \lambda_{j \rho} \lambda^{j}_{\alpha} {W}^{-3}+\frac{1}{768}{\rm i} (\Gamma^{a})^{\alpha \beta} W^{b c} W_{b c \alpha i} \lambda^{i \rho} \lambda_{j \beta} \lambda^{j}_{\rho} {W}^{-3}%
 - \frac{1}{768}{\rm i} \epsilon^{{e_{1}} b c d e} (\Sigma_{{e_{1}}}{}^{\, a})^{\alpha \beta} W_{b c} W_{d e \alpha i} \lambda^{i \rho} \lambda_{j \beta} \lambda^{j}_{\rho} {W}^{-3}+\frac{1}{384}{\rm i} W_{b}\,^{c} W_{d c}\,^{\alpha}\,_{i} \epsilon^{a b d e {e_{1}}} (\Sigma_{e {e_{1}}})_{\alpha}{}^{\beta} \lambda^{i \rho} \lambda_{j \beta} \lambda^{j}_{\rho} {W}^{-3} - \frac{1}{384}{\rm i} W^{a b} W_{b c}\,^{\alpha}\,_{i} (\Gamma^{c})_{\alpha}{}^{\beta} \lambda^{i \rho} \lambda_{j \beta} \lambda^{j}_{\rho} {W}^{-3} - \frac{1}{384}{\rm i} W_{b}\,^{c} W^{a}\,_{c}\,^{\alpha}\,_{i} (\Gamma^{b})_{\alpha}{}^{\beta} \lambda^{i \rho} \lambda_{j \beta} \lambda^{j}_{\rho} {W}^{-3} - \frac{9}{128}{\rm i} W_{b}\,^{a \alpha}\,_{i} (\Gamma^{b})^{\beta \rho} X_{j k} \lambda^{i}_{\alpha} \lambda^{j}_{\beta} \lambda^{k}_{\rho} {W}^{-4}+\frac{3}{128}{\rm i} W_{b c}\,^{\alpha}\,_{i} \epsilon^{a b c d e} (\Sigma_{d e})^{\beta \rho} X^{i}\,_{j} \lambda^{j}_{\beta} \lambda_{k \alpha} \lambda^{k}_{\rho} {W}^{-4} - \frac{1}{16}{\rm i} W_{b}\,^{a \alpha}\,_{i} (\Gamma^{b})^{\beta \rho} X^{i}\,_{j} \lambda^{j}_{\beta} \lambda_{k \alpha} \lambda^{k}_{\rho} {W}^{-4}+\frac{7}{384}\epsilon^{a b c d e} \Phi_{b c i j} (\Sigma_{d e})^{\alpha \beta} \lambda^{i}_{\alpha} \lambda^{j \rho} \lambda_{k \beta} \lambda^{k}_{\rho} {W}^{-4} - \frac{1}{96}\Phi_{b}\,^{a}\,_{i j} (\Gamma^{b})^{\alpha \beta} \lambda^{i}_{\alpha} \lambda^{j \rho} \lambda_{k \beta} \lambda^{k}_{\rho} {W}^{-4}+\frac{13}{768}\epsilon^{a b c d e} \Phi_{b c i j} (\Sigma_{d e})^{\alpha \beta} \lambda^{i \rho} \lambda^{j}_{\rho} \lambda_{k \alpha} \lambda^{k}_{\beta} {W}^{-4}+\frac{3}{64}\Phi^{b c}\,_{i j} (\Sigma_{b c})^{\rho \lambda} (\Gamma^{a})^{\alpha \beta} \lambda^{i}_{\alpha} \lambda^{j}_{\beta} \lambda_{k \rho} \lambda^{k}_{\lambda} {W}^{-4}+\frac{1}{96}\Phi^{b c}\,_{i j} (\Sigma_{b c})^{\rho \lambda} (\Gamma^{a})^{\alpha \beta} \lambda^{i}_{\rho} \lambda^{j}_{\alpha} \lambda_{k \lambda} \lambda^{k}_{\beta} {W}^{-4}+\frac{9}{64}(\Gamma^{a})^{\alpha \beta} X_{i j} X^{i j} \lambda_{k \alpha} \lambda^{k \rho} \lambda_{l \beta} \lambda^{l}_{\rho} {W}^{-6} - \frac{11}{16}(\Gamma^{a})^{\alpha \beta} X_{i j} X^{i}\,_{k} \lambda^{j}_{\alpha} \lambda^{k \rho} \lambda_{l \beta} \lambda^{l}_{\rho} {W}^{-6} - \frac{19}{32}(\Gamma^{a})^{\alpha \beta} X_{i j} X_{k l} \lambda^{i}_{\alpha} \lambda^{j}_{\beta} \lambda^{k \rho} \lambda^{l}_{\rho} {W}^{-6} - \frac{1}{32}(\Gamma^{a})^{\alpha \beta} X_{i j} X_{k l} \lambda^{i}_{\alpha} \lambda^{j \rho} \lambda^{k}_{\beta} \lambda^{l}_{\rho} {W}^{-6}+\frac{1}{8}(\Gamma^{a})^{\alpha \beta} \lambda_{i \alpha} \lambda^{i \rho} \lambda_{j \rho} {W}^{-4} \nabla_{b}{\nabla^{b}{\lambda^{j}_{\beta}}}+\frac{5}{64}\epsilon^{a b c d e} F_{b c} F_{d e} \lambda^{\alpha}_{i} \lambda^{i \beta} \lambda_{j \alpha} \lambda^{j}_{\beta} {W}^{-6} - \frac{5}{8}(\Gamma^{c})^{\alpha \beta} F^{a b} F_{c b} \lambda_{i \alpha} \lambda^{i \rho} \lambda_{j \beta} \lambda^{j}_{\rho} {W}^{-6}+\frac{5}{16}(\Gamma^{a})^{\alpha \beta} F^{b c} F_{b c} \lambda_{i \alpha} \lambda^{i \rho} \lambda_{j \beta} \lambda^{j}_{\rho} {W}^{-6}%
 - \frac{15}{32}\epsilon^{a {e_{1}} {e_{2}} d e} (\Sigma_{{e_{1}} {e_{2}}})^{\alpha \beta} (\Sigma_{b c})^{\rho \lambda} F^{b c} F_{d e} \lambda_{i \alpha} \lambda^{i}_{\beta} \lambda_{j \rho} \lambda^{j}_{\lambda} {W}^{-6} - \frac{5}{64}\epsilon_{{e_{1}}}\,^{b c d e} (\Gamma^{a})^{\alpha \beta} (\Gamma^{{e_{1}}})^{\rho \lambda} F_{b c} F_{d e} \lambda_{i \alpha} \lambda^{i}_{\rho} \lambda_{j \beta} \lambda^{j}_{\lambda} {W}^{-6}+\frac{9}{64}\epsilon^{a d e b c} W_{d e} F_{b c} \lambda^{\alpha}_{i} \lambda^{i \beta} \lambda_{j \alpha} \lambda^{j}_{\beta} {W}^{-5} - \frac{9}{16}(\Gamma^{c})^{\alpha \beta} W_{c b} F^{a b} \lambda_{i \alpha} \lambda^{i \rho} \lambda_{j \beta} \lambda^{j}_{\rho} {W}^{-5}+\frac{9}{16}(\Gamma^{a})^{\alpha \beta} W^{b c} F_{b c} \lambda_{i \alpha} \lambda^{i \rho} \lambda_{j \beta} \lambda^{j}_{\rho} {W}^{-5} - \frac{9}{16}(\Gamma^{b})^{\alpha \beta} W^{a c} F_{b c} \lambda_{i \alpha} \lambda^{i \rho} \lambda_{j \beta} \lambda^{j}_{\rho} {W}^{-5} - \frac{7}{16}\epsilon^{a {e_{1}} {e_{2}} d e} (\Sigma_{{e_{1}} {e_{2}}})^{\alpha \beta} (\Sigma_{b c})^{\rho \lambda} W_{d e} F^{b c} \lambda_{i \alpha} \lambda^{i}_{\beta} \lambda_{j \rho} \lambda^{j}_{\lambda} {W}^{-5} - \frac{1}{32}\epsilon^{a {e_{1}} {e_{2}} d e} (\Sigma_{{e_{1}} {e_{2}}})^{\alpha \beta} (\Sigma_{b c})^{\rho \lambda} W_{d e} F^{b c} \lambda_{i \alpha} \lambda^{i}_{\rho} \lambda_{j \beta} \lambda^{j}_{\lambda} {W}^{-5}+\frac{171}{2048}\epsilon^{a b c d e} W_{b c} W_{d e} \lambda^{\alpha}_{i} \lambda^{i \beta} \lambda_{j \alpha} \lambda^{j}_{\beta} {W}^{-4} - \frac{171}{256}(\Gamma^{c})^{\alpha \beta} W^{a b} W_{c b} \lambda_{i \alpha} \lambda^{i \rho} \lambda_{j \beta} \lambda^{j}_{\rho} {W}^{-4}+\frac{171}{512}(\Gamma^{a})^{\alpha \beta} W^{b c} W_{b c} \lambda_{i \alpha} \lambda^{i \rho} \lambda_{j \beta} \lambda^{j}_{\rho} {W}^{-4} - \frac{21}{64}\epsilon^{a {e_{1}} {e_{2}} d e} (\Sigma_{{e_{1}} {e_{2}}})^{\alpha \beta} (\Sigma_{b c})^{\rho \lambda} W^{b c} W_{d e} \lambda_{i \alpha} \lambda^{i}_{\beta} \lambda_{j \rho} \lambda^{j}_{\lambda} {W}^{-4}+\frac{177}{512}\epsilon^{a {e_{1}} {e_{2}} d e} (\Sigma_{{e_{1}} {e_{2}}})^{\alpha \beta} (\Sigma_{b c})^{\rho \lambda} W^{b c} W_{d e} \lambda_{i \alpha} \lambda^{i}_{\rho} \lambda_{j \beta} \lambda^{j}_{\lambda} {W}^{-4}+\frac{9}{32}\epsilon^{a {e_{1}} {e_{2}} b c} (\Sigma_{{e_{1}} {e_{2}}})^{\alpha \beta} (\Sigma_{d e})^{\rho \lambda} W^{d e} F_{b c} \lambda_{i \alpha} \lambda^{i}_{\rho} \lambda_{j \beta} \lambda^{j}_{\lambda} {W}^{-5} - \frac{9}{32}\epsilon^{a {e_{1}} {e_{2}} b c} (\Sigma_{{e_{1}} {e_{2}}})^{\alpha \beta} (\Sigma_{d e})^{\rho \lambda} W^{d e} F_{b c} \lambda_{i \alpha} \lambda^{i}_{\beta} \lambda_{j \rho} \lambda^{j}_{\lambda} {W}^{-5} - \frac{9}{64}\epsilon_{{e_{1}}}\,^{d e b c} (\Gamma^{a})^{\alpha \beta} (\Gamma^{{e_{1}}})^{\rho \lambda} W_{d e} F_{b c} \lambda_{i \alpha} \lambda^{i}_{\rho} \lambda_{j \beta} \lambda^{j}_{\lambda} {W}^{-5} - \frac{171}{2048}\epsilon_{{e_{1}}}\,^{b c d e} (\Gamma^{a})^{\alpha \beta} (\Gamma^{{e_{1}}})^{\rho \lambda} W_{b c} W_{d e} \lambda_{i \alpha} \lambda^{i}_{\rho} \lambda_{j \beta} \lambda^{j}_{\lambda} {W}^{-4} - \frac{5}{32}\epsilon^{a d e b c} (\Sigma_{d e})^{\alpha \beta} X_{i j} F_{b c} \lambda^{i}_{\alpha} \lambda^{j \rho} \lambda_{k \beta} \lambda^{k}_{\rho} {W}^{-6}+\frac{5}{8}(\Gamma^{b})^{\alpha \beta} X_{i j} F_{b}\,^{a} \lambda^{i}_{\alpha} \lambda^{j \rho} \lambda_{k \beta} \lambda^{k}_{\rho} {W}^{-6}+\frac{25}{64}\epsilon^{a d e b c} (\Sigma_{d e})^{\alpha \beta} X_{i j} F_{b c} \lambda^{i \rho} \lambda^{j}_{\rho} \lambda_{k \alpha} \lambda^{k}_{\beta} {W}^{-6}%
+\frac{5}{8}(\Sigma_{b c})^{\rho \lambda} (\Gamma^{a})^{\alpha \beta} X_{i j} F^{b c} \lambda^{i}_{\alpha} \lambda^{j}_{\beta} \lambda_{k \rho} \lambda^{k}_{\lambda} {W}^{-6} - \frac{5}{128}\epsilon^{a d e b c} (\Sigma_{d e})^{\alpha \beta} X_{i j} W_{b c} \lambda^{i}_{\alpha} \lambda^{j \rho} \lambda_{k \beta} \lambda^{k}_{\rho} {W}^{-5}+\frac{9}{16}(\Gamma^{b})^{\alpha \beta} X_{i j} W_{b}\,^{a} \lambda^{i}_{\alpha} \lambda^{j \rho} \lambda_{k \beta} \lambda^{k}_{\rho} {W}^{-5}+\frac{77}{256}\epsilon^{a d e b c} (\Sigma_{d e})^{\alpha \beta} X_{i j} W_{b c} \lambda^{i \rho} \lambda^{j}_{\rho} \lambda_{k \alpha} \lambda^{k}_{\beta} {W}^{-5}+\frac{45}{128}(\Sigma_{b c})^{\rho \lambda} (\Gamma^{a})^{\alpha \beta} X_{i j} W^{b c} \lambda^{i}_{\alpha} \lambda^{j}_{\beta} \lambda_{k \rho} \lambda^{k}_{\lambda} {W}^{-5}+\frac{27}{64}(\Sigma_{b c})^{\rho \lambda} (\Gamma^{a})^{\alpha \beta} X_{i j} W^{b c} \lambda^{i}_{\rho} \lambda^{j}_{\alpha} \lambda_{k \lambda} \lambda^{k}_{\beta} {W}^{-5}+\frac{31}{64}(\Gamma^{a})^{\beta \rho} \lambda_{j \beta} \lambda^{j \lambda} \lambda_{k \rho} \lambda^{k}_{\lambda} \lambda^{\alpha}_{i} X^{i}_{\alpha} {W}^{-5}+\frac{15}{64}(\Gamma^{a})^{\beta \rho} \lambda_{j \beta} \lambda^{j \lambda} \lambda_{k \rho} \lambda^{k \alpha} \lambda_{i \lambda} X^{i}_{\alpha} {W}^{-5} - \frac{17}{64}(\Gamma^{a})^{\beta \rho} \lambda_{j \beta} \lambda^{j \lambda} \lambda_{i \rho} \lambda_{k \lambda} \lambda^{k \alpha} X^{i}_{\alpha} {W}^{-5}+\frac{25}{32}(\Gamma^{a})^{\beta \alpha} \lambda_{j \beta} \lambda^{j \rho} \lambda_{k \rho} \lambda^{k \lambda} \lambda_{i \lambda} X^{i}_{\alpha} {W}^{-5}+\frac{7}{64}(\Gamma^{a})^{\beta \alpha} \lambda_{i \beta} \lambda^{\rho}_{j} \lambda^{j \lambda} \lambda_{k \rho} \lambda^{k}_{\lambda} X^{i}_{\alpha} {W}^{-5}+\frac{5}{32}{\rm i} \lambda^{\alpha}_{i} \lambda^{i \beta} \lambda_{j \alpha} \lambda^{j}_{\beta} \lambda^{\rho}_{k} {W}^{-6} \nabla^{a}{\lambda^{k}_{\rho}}+\frac{15}{16}{\rm i} \lambda^{\alpha}_{i} \lambda^{i \beta} \lambda_{j \alpha} \lambda^{j \rho} \lambda_{k \beta} {W}^{-6} \nabla^{a}{\lambda^{k}_{\rho}} - \frac{5}{4}{\rm i} (\Sigma^{a}{}_{\, b})^{\alpha \beta} \lambda_{i \alpha} \lambda^{i \rho} \lambda_{j \beta} \lambda^{j \lambda} \lambda_{k \rho} {W}^{-6} \nabla^{b}{\lambda^{k}_{\lambda}}+\frac{15}{8}{\rm i} (\Sigma^{a}{}_{\, b})^{\alpha \beta} \lambda_{i \alpha} \lambda^{i \rho} \lambda_{j \rho} \lambda^{j \lambda} \lambda_{k \lambda} {W}^{-6} \nabla^{b}{\lambda^{k}_{\beta}}+\frac{5}{16}{\rm i} (\Sigma^{a}{}_{\, b})^{\alpha \beta} \lambda_{i \alpha} \lambda^{\rho}_{j} \lambda^{j \lambda} \lambda_{k \rho} \lambda^{k}_{\lambda} {W}^{-6} \nabla^{b}{\lambda^{i}_{\beta}}+\frac{25}{64}{\rm i} (\Gamma^{a})^{\alpha \beta} (\Gamma_{b})^{\rho \lambda} \lambda_{i \alpha} \lambda^{i}_{\rho} \lambda_{j \beta} \lambda^{j}_{\lambda} \lambda^{\gamma}_{k} {W}^{-6} \nabla^{b}{\lambda^{k}_{\gamma}} - \frac{5}{32}{\rm i} (\Gamma^{a})^{\alpha \beta} (\Gamma_{b})^{\rho \lambda} \lambda_{i \alpha} \lambda^{i}_{\rho} \lambda_{j \lambda} \lambda^{j \gamma} \lambda_{k \gamma} {W}^{-6} \nabla^{b}{\lambda^{k}_{\beta}}+\frac{15}{32}{\rm i} (\Gamma^{a})^{\alpha \beta} (\Gamma_{b})^{\rho \lambda} \lambda_{i \alpha} \lambda^{i}_{\rho} \lambda_{j \beta} \lambda^{j \gamma} \lambda_{k \lambda} {W}^{-6} \nabla^{b}{\lambda^{k}_{\gamma}}+\frac{5}{32}{\rm i} (\Gamma^{a})^{\alpha \beta} (\Gamma_{b})^{\rho \lambda} \lambda_{i \alpha} \lambda^{i}_{\rho} \lambda_{j \beta} \lambda^{j \gamma} \lambda_{k \gamma} {W}^{-6} \nabla^{b}{\lambda^{k}_{\lambda}}%
+\frac{5}{32}{\rm i} (\Gamma^{a})^{\alpha \beta} (\Gamma_{b})^{\rho \lambda} \lambda_{i \alpha} \lambda^{i \gamma} \lambda_{j \rho} \lambda^{j}_{\gamma} \lambda_{k \lambda} {W}^{-6} \nabla^{b}{\lambda^{k}_{\beta}}+\frac{25}{64}{\rm i} (\Gamma^{a})^{\alpha \beta} (\Gamma_{b})^{\rho \lambda} \lambda_{i \alpha} \lambda^{i \gamma} \lambda_{j \beta} \lambda^{j}_{\gamma} \lambda_{k \rho} {W}^{-6} \nabla^{b}{\lambda^{k}_{\lambda}} - \frac{1}{4}(\Gamma^{a})^{\alpha \beta} \lambda_{i \alpha} \lambda^{i \rho} \lambda_{j \beta} \lambda^{j}_{\rho} {W}^{-5} \nabla_{b}{\nabla^{b}{W}} - \frac{1}{4}W^{b c \alpha}\,_{i} (\Sigma_{b c})^{\lambda \gamma} (\Gamma^{a})^{\beta \rho} \lambda^{i}_{\beta} \lambda_{j \alpha} \lambda^{j}_{\lambda} \lambda_{k \gamma} \lambda^{k}_{\rho} {W}^{-5} - \frac{1}{32}W_{b c}\,^{\alpha}\,_{i} \epsilon^{a b c d e} (\Sigma_{d e})^{\beta \rho} \lambda^{i \lambda} \lambda_{j \alpha} \lambda^{j}_{\beta} \lambda_{k \rho} \lambda^{k}_{\lambda} {W}^{-5}+\frac{1}{16}W_{b}\,^{a \alpha}\,_{i} (\Gamma^{b})^{\beta \rho} \lambda^{i \lambda} \lambda_{j \alpha} \lambda^{j}_{\beta} \lambda_{k \rho} \lambda^{k}_{\lambda} {W}^{-5} - \frac{1}{16}W_{b c}\,^{\alpha}\,_{i} \epsilon^{a b c d e} (\Sigma_{d e})^{\beta \rho} \lambda^{i \lambda} \lambda_{j \alpha} \lambda^{j}_{\lambda} \lambda_{k \beta} \lambda^{k}_{\rho} {W}^{-5} - \frac{15}{16}{\rm i} (\Sigma^{a}{}_{\, b})^{\alpha \beta} \lambda_{i \alpha} \lambda^{i}_{\beta} \lambda^{\rho}_{j} \lambda^{j \lambda} \lambda_{k \rho} \lambda^{k}_{\lambda} {W}^{-7} \nabla^{b}{W}+\frac{15}{8}{\rm i} (\Sigma^{a}{}_{\, b})^{\alpha \beta} \lambda_{i \alpha} \lambda^{i \rho} \lambda_{j \beta} \lambda^{j \lambda} \lambda_{k \rho} \lambda^{k}_{\lambda} {W}^{-7} \nabla^{b}{W} - \frac{15}{16}{\rm i} (\Gamma^{a})^{\alpha \beta} (\Gamma_{b})^{\rho \lambda} \lambda_{i \alpha} \lambda^{i}_{\rho} \lambda_{j \beta} \lambda^{j \gamma} \lambda_{k \lambda} \lambda^{k}_{\gamma} {W}^{-7} \nabla^{b}{W}+\frac{3}{32}{\rm i} (\Gamma^{a})^{\alpha \beta} X_{i j} \lambda^{i}_{\alpha} \lambda^{j}_{\beta} \lambda^{\rho}_{k} \lambda^{k \lambda} \lambda_{l \rho} \lambda^{l}_{\lambda} {W}^{-7} - \frac{3}{8}{\rm i} (\Gamma^{a})^{\alpha \beta} X_{i j} \lambda^{i}_{\alpha} \lambda^{j \rho} \lambda_{k \beta} \lambda^{k \lambda} \lambda_{l \rho} \lambda^{l}_{\lambda} {W}^{-7}+\frac{3}{4}{\rm i} (\Gamma^{a})^{\alpha \beta} X_{i j} \lambda^{i \rho} \lambda^{j}_{\rho} \lambda_{k \alpha} \lambda^{k \lambda} \lambda_{l \beta} \lambda^{l}_{\lambda} {W}^{-7}+\frac{9}{16}{\rm i} (\Gamma^{a})^{\alpha \beta} X_{i j} \lambda^{i \rho} \lambda^{j \lambda} \lambda_{k \alpha} \lambda^{k}_{\rho} \lambda_{l \beta} \lambda^{l}_{\lambda} {W}^{-7} - \frac{15}{64}{\rm i} \epsilon^{a d e b c} (\Sigma_{d e})^{\alpha \beta} F_{b c} \lambda_{i \alpha} \lambda^{i}_{\beta} \lambda^{\rho}_{j} \lambda^{j \lambda} \lambda_{k \rho} \lambda^{k}_{\lambda} {W}^{-7}+\frac{15}{32}{\rm i} \epsilon^{a d e b c} (\Sigma_{d e})^{\alpha \beta} F_{b c} \lambda_{i \alpha} \lambda^{i \rho} \lambda_{j \beta} \lambda^{j \lambda} \lambda_{k \rho} \lambda^{k}_{\lambda} {W}^{-7} - \frac{15}{16}{\rm i} (\Sigma_{b c})^{\rho \lambda} (\Gamma^{a})^{\alpha \beta} F^{b c} \lambda_{i \rho} \lambda^{i}_{\lambda} \lambda_{j \alpha} \lambda^{j \gamma} \lambda_{k \beta} \lambda^{k}_{\gamma} {W}^{-7} - \frac{15}{64}{\rm i} \epsilon^{a d e b c} (\Sigma_{d e})^{\alpha \beta} W_{b c} \lambda_{i \alpha} \lambda^{i}_{\beta} \lambda^{\rho}_{j} \lambda^{j \lambda} \lambda_{k \rho} \lambda^{k}_{\lambda} {W}^{-6}+\frac{15}{32}{\rm i} \epsilon^{a d e b c} (\Sigma_{d e})^{\alpha \beta} W_{b c} \lambda_{i \alpha} \lambda^{i \rho} \lambda_{j \beta} \lambda^{j \lambda} \lambda_{k \rho} \lambda^{k}_{\lambda} {W}^{-6}+\frac{45}{64}{\rm i} (\Sigma_{b c})^{\rho \lambda} (\Gamma^{a})^{\alpha \beta} W^{b c} \lambda_{i \rho} \lambda^{i}_{\alpha} \lambda_{j \lambda} \lambda^{j \gamma} \lambda_{k \beta} \lambda^{k}_{\gamma} {W}^{-6}%
 - \frac{75}{128}{\rm i} (\Sigma_{b c})^{\rho \lambda} (\Gamma^{a})^{\alpha \beta} W^{b c} \lambda_{i \rho} \lambda^{i}_{\lambda} \lambda_{j \alpha} \lambda^{j \gamma} \lambda_{k \beta} \lambda^{k}_{\gamma} {W}^{-6}+\frac{21}{64}(\Gamma^{a})^{\alpha \beta} \lambda_{i \alpha} \lambda^{i \rho} \lambda_{j \beta} \lambda^{j}_{\rho} \lambda^{\lambda}_{k} \lambda^{k \gamma} \lambda_{l \lambda} \lambda^{l}_{\gamma} {W}^{-8} - \frac{21}{32}(\Gamma^{a})^{\alpha \beta} \lambda_{i \alpha} \lambda^{i \rho} \lambda_{j \beta} \lambda^{j \lambda} \lambda_{k \rho} \lambda^{k \gamma} \lambda_{l \lambda} \lambda^{l}_{\gamma} {W}^{-8}
\doublespacedmathend
\end{adjustwidth}

\section{Curvature-squared actions} 

\subsection{Log Degauged and Gauge-fixed bosonic Lagrangian} \label{BFlogWDegauged}

\begin{adjustwidth}{0em}{3cm}
\doublespacedmathbegin
- \frac{1}{12}\nu_{e} \epsilon^{e a b c d} \Phi_{a b i j} \Phi_{c d}\,^{i j} - \frac{5}{16}R W^{a b} F_{a b}+\frac{1}{3}R^{a}\,_{c} W^{c b} F_{a b} - \frac{3}{32}R F^{a b} F_{a b}+\frac{1}{3}R^{c}\,_{a} F^{a b} F_{c b} - \frac{3}{64}W^{a b} F_{a b} Y+\frac{3}{4}\epsilon^{c d e a b} \mathcal{D}^{{e_{1}}}{W_{c d}} W_{e {e_{1}}} F_{a b}+\frac{3}{64}W^{c d} W^{a b} W_{c d} F_{a b}+\frac{9}{4}W^{c d} W_{c a} W_{d b} F^{a b}+\frac{11}{12}R_{c d a b} W^{c d} F^{a b} - \frac{4}{3}R_{c a} W^{c}\,_{b} F^{a b} - \frac{9}{4}\mathcal{D}_{c}{\mathcal{D}^{c}{W_{a b}}} F^{a b}+\frac{3}{128}F^{a b} F_{a b} Y - \frac{3}{16}\epsilon^{e a b c d} \mathcal{D}^{{e_{1}}}{W_{e {e_{1}}}} F_{a b} F_{c d} - \frac{3}{8}\epsilon^{e {e_{1}} a c d} W_{e {e_{1}}} \mathcal{D}^{b}{F_{a b}} F_{c d} - \frac{75}{128}W^{c d} W_{c d} F^{a b} F_{a b}+\frac{9}{4}W^{d c} W_{d a} F^{a b} F_{c b} - \frac{1}{2}\Phi_{a b i j} X^{i j} F^{a b}+\frac{1}{3}R_{a b c d} F^{a b} F^{c d}%
 - \frac{1}{3}R_{a c} F^{a}\,_{b} F^{c b} - \frac{1}{16}\epsilon^{a c d e {e_{1}}} \mathcal{D}^{b}{F_{a b}} F_{c d} F_{e {e_{1}}} - \frac{3}{8}W^{c d} F^{a b} F_{c d} F_{a b}+\frac{3}{4}W^{c d} F^{a b} F_{c a} F_{d b} - \frac{1}{32}F^{a b} F^{c d} F_{a b} F_{c d}+\frac{1}{8}F^{a b} F^{c d} F_{a c} F_{b d} - \frac{1}{16}X_{i j} X^{i j} F^{a b} F_{a b}+\frac{1}{3}R_{a b c d} W^{a b} W^{c d} - \frac{2}{3}R_{a b} W^{a}\,_{c} W^{b c} - \frac{27}{256}R W^{a b} W_{a b}+\frac{1}{3}R_{a b c d} W^{a c} W^{b d} - \frac{3}{2048}{Y}^{2}+\frac{3}{32}\mathcal{D}_{a}{\mathcal{D}^{a}{Y}}+\frac{3}{256}R Y - \frac{69}{1024}W^{a b} W_{a b} Y+\frac{1}{6}\Phi_{a b i j} \Phi^{a b i j} - \frac{39}{16}W^{a b} \mathcal{D}_{c}{\mathcal{D}^{c}{W_{a b}}}+\frac{1}{2}\mathcal{D}^{b}{\mathcal{D}^{c}{W^{a}\,_{b}}} W_{a c}-2\mathcal{D}^{c}{\mathcal{D}^{b}{W^{a}\,_{b}}} W_{a c} - \frac{1}{3}R^{a}\,_{b} W^{b c} W_{a c}%
+\frac{1005}{2048}W^{a b} W^{c d} W_{a b} W_{c d} - \frac{33}{16}\mathcal{D}_{c}{W^{a b}} \mathcal{D}^{c}{W_{a b}} - \frac{9}{4}\mathcal{D}^{b}{W^{a}\,_{b}} \mathcal{D}^{c}{W_{a c}}+\frac{3}{4}\mathcal{D}^{c}{W^{a}\,_{b}} \mathcal{D}^{b}{W_{a c}} - \frac{9}{16}\epsilon^{a c d e {e_{1}}} \mathcal{D}^{b}{W_{a b}} W_{c d} W_{e {e_{1}}}+\mathcal{D}^{b}{\mathcal{D}^{c}{W^{a}\,_{c}}} F_{a b}+\frac{1}{2}\mathcal{D}^{c}{\mathcal{D}^{b}{W^{a}\,_{c}}} F_{a b} - \frac{1}{2}W^{a b} \mathcal{D}_{c}{\mathcal{D}^{c}{F_{a b}}}-2W^{a}\,_{c} \mathcal{D}^{c}{\mathcal{D}^{b}{F_{a b}}}+\frac{1}{3}\mathcal{D}_{b}{\mathcal{D}^{b}{R^{a}\,_{a}}} - \frac{5}{24}\mathcal{D}_{a}{\mathcal{D}^{a}{R}} - \frac{1}{6}R^{a b} R_{a b}+\frac{23}{384}{R}^{2}+\frac{1}{6}R_{c a d b} W^{c d} F^{a b} - \frac{3}{2}\mathcal{D}_{c}{W^{a b}} \mathcal{D}^{c}{F_{a b}} - \frac{3}{2}\mathcal{D}^{c}{W^{a}\,_{c}} \mathcal{D}^{b}{F_{a b}} - \frac{9}{16}\epsilon^{c d e {e_{1}} a} W_{c d} W_{e {e_{1}}} \mathcal{D}^{b}{F_{a b}} - \frac{1}{6}R_{a c b d} F^{a b} F^{c d} - \frac{9}{128}X_{i j} X^{i j} Y - \frac{1}{4}\mathcal{D}_{a}{X_{i j}} \mathcal{D}^{a}{X^{i j}}%
+\frac{1}{4}\mathcal{D}_{c}{F^{a b}} \mathcal{D}^{c}{F_{a b}}+\frac{9}{128}X_{i j} X^{i j} W^{a b} W_{a b}+\frac{1}{32}R X_{i j} X^{i j}+\frac{1}{32}X_{i j} X^{i j} X_{k l} X^{k l} - \frac{1}{2}\Phi_{a b i j} X^{i j} W^{a b}
\doublespacedmathend
\end{adjustwidth}

\section{EOM Descendants} \label{sec:supercurrents}

\subsection{Einstein--Hilbert} \label{SupercurrentEHComplete}

\subsubsection{$J^3_{a i \alpha, {\rm EH}}$} \label{J3EHComplete}

\begin{adjustwidth}{0cm}{5cm}
\doublespacedmathbegin
\frac{3}{32}G_{j k} (\Gamma_{a})^{\beta \rho} {G}^{-3} \varphi^{j}_{\alpha} \varphi_{i \beta} \varphi^{k}_{\rho} - \frac{9}{20}{\rm i} \mathcal{H}_{a} {G}^{-1} \varphi_{i \alpha}+\frac{9}{40}{\rm i} \mathcal{H}^{b} (\Sigma_{a b})_{\alpha}{}^{\beta} {G}^{-1} \varphi_{i \beta}+\frac{9}{20}{\rm i} {G}^{-1} \nabla_{a}{G_{i j}} \varphi^{j}_{\alpha} - \frac{9}{40}{\rm i} (\Sigma_{a b})_{\alpha}{}^{\beta} {G}^{-1} \nabla^{b}{G_{i j}} \varphi^{j}_{\beta} - \frac{3}{32}G_{i j} (\Gamma_{a})^{\beta \rho} {G}^{-3} \varphi_{k \alpha} \varphi^{j}_{\beta} \varphi^{k}_{\rho} - \frac{3}{32}G_{j k} (\Gamma_{a})^{\beta \rho} {G}^{-3} \varphi_{i \alpha} \varphi^{j}_{\beta} \varphi^{k}_{\rho}+\frac{9}{40}{\rm i} G_{i j} G_{k l} {G}^{-3} \nabla_{a}{G^{j k}} \varphi^{l}_{\alpha} - \frac{9}{80}{\rm i} G_{i j} G_{k l} (\Sigma_{a b})_{\alpha}{}^{\beta} {G}^{-3} \nabla^{b}{G^{j k}} \varphi^{l}_{\beta} - \frac{27}{40}{\rm i} G_{i j} {G}^{-1} \nabla_{a}{\varphi^{j}_{\alpha}}+\frac{27}{80}{\rm i} G_{i j} (\Sigma_{a b})_{\alpha}{}^{\beta} {G}^{-1} \nabla^{b}{\varphi^{j}_{\beta}} - \frac{9}{320}{\rm i} \epsilon^{d e}\,_{a}\,^{b c} G_{i j} (\Sigma_{d e})_{\alpha}{}^{\beta} W_{b c} {G}^{-1} \varphi^{j}_{\beta} - \frac{27}{320}{\rm i} G_{i j} (\Gamma^{b})_{\alpha}{}^{\beta} W_{a b} {G}^{-1} \varphi^{j}_{\beta}+\frac{9}{16}{\rm i} (\Gamma_{a})^{\beta \rho} \lambda_{j \alpha} \lambda_{i \beta} \lambda^{j}_{\rho} - \frac{27}{40}\epsilon^{d e}\,_{a}\,^{b c} (\Sigma_{d e})_{\alpha}{}^{\beta} W F_{b c} \lambda_{i \beta} - \frac{81}{40}(\Gamma^{b})_{\alpha}{}^{\beta} W F_{a b} \lambda_{i \beta} - \frac{189}{320}\epsilon^{d e}\,_{a}\,^{b c} (\Sigma_{d e})_{\alpha}{}^{\beta} W_{b c} \lambda_{i \beta} {W}^{2} - \frac{567}{320}(\Gamma^{b})_{\alpha}{}^{\beta} W_{a b} \lambda_{i \beta} {W}^{2}+\frac{27}{40}{W}^{2} \nabla_{a}{\lambda_{i \alpha}}%
 - \frac{27}{80}(\Sigma_{a b})_{\alpha}{}^{\beta} {W}^{2} \nabla^{b}{\lambda_{i \beta}} - \frac{3}{160}G_{j k} (\Gamma_{a})_{\alpha}{}^{\beta} {G}^{-3} \varphi_{i}^{\rho} \varphi^{j}_{\beta} \varphi^{k}_{\rho} - \frac{3}{160}G_{j k} (\Gamma_{a})_{\alpha}{}^{\beta} {G}^{-3} \varphi_{i \beta} \varphi^{j \rho} \varphi^{k}_{\rho}+\frac{3}{160}G_{i j} (\Gamma_{a})_{\alpha}{}^{\beta} {G}^{-3} \varphi^{j \rho} \varphi_{k \beta} \varphi^{k}_{\rho} - \frac{9}{80}{\rm i} (\Gamma_{a})_{\alpha}{}^{\beta} \lambda^{\rho}_{i} \lambda_{j \beta} \lambda^{j}_{\rho}+\frac{9}{80}{\rm i} G_{i j} G_{k l} {G}^{-3} \nabla_{a}{G^{k l}} \varphi^{j}_{\alpha} - \frac{27}{20}W \lambda_{i \alpha} \nabla_{a}{W} - \frac{9}{160}{\rm i} G_{i j} G_{k l} (\Sigma_{a b})_{\alpha}{}^{\beta} {G}^{-3} \nabla^{b}{G^{k l}} \varphi^{j}_{\beta}+\frac{27}{40}(\Sigma_{a b})_{\alpha}{}^{\beta} W \lambda_{i \beta} \nabla^{b}{W}
\doublespacedmathend
\end{adjustwidth}
 
\subsubsection{$J^4_{a b, {\rm EH}}$} \label{J4EHComplete}

\begin{adjustwidth}{0cm}{5cm}
\doublespacedmathbegin
{}\frac{9}{128}(\Gamma_{\underline{a}})^{\alpha \beta} (\Gamma_{\underline{b}})^{\rho \lambda} {G}^{-3} \varphi_{i \alpha} \varphi^{i}_{\rho} \varphi_{j \beta} \varphi^{j}_{\lambda}+\frac{9}{128}G_{i j} G_{k l} (\Gamma_{\underline{a}})^{\alpha \beta} (\Gamma_{\underline{b}})^{\rho \lambda} {G}^{-5} \varphi^{i}_{\alpha} \varphi^{j}_{\rho} \varphi^{k}_{\beta} \varphi^{l}_{\lambda}+\frac{9}{32}{\rm i} \mathcal{H}_{\underline{a}} G_{i j} (\Gamma_{\underline{b}})^{\alpha \beta} {G}^{-3} \varphi^{i}_{\alpha} \varphi^{j}_{\beta} - \frac{9}{160}{\rm i} \mathcal{H}_{c} G_{i j} (\Gamma^{c})^{\alpha \beta} \eta_{\underline{a} \underline{b}} {G}^{-3} \varphi^{i}_{\alpha} \varphi^{j}_{\beta} - \frac{9}{16}{\rm i} G_{i j} (\Gamma_{\underline{a}})^{\alpha \beta} {G}^{-3} \nabla_{\underline{b}}{G^{i}\,_{k}} \varphi^{j}_{\alpha} \varphi^{k}_{\beta}+\frac{9}{80}{\rm i} G_{i j} (\Gamma_{c})^{\alpha \beta} \eta_{\underline{a} \underline{b}} {G}^{-3} \nabla^{c}{G^{i}\,_{k}} \varphi^{j}_{\alpha} \varphi^{k}_{\beta} - \frac{9}{40}{\rm i} (\Gamma_{c})^{\alpha \beta} \eta_{\underline{a} \underline{b}} {G}^{-1} \nabla^{c}{\varphi_{i \alpha}} \varphi^{i}_{\beta} - \frac{9}{80}{\rm i} (\Sigma_{c d})^{\alpha \beta} \eta_{\underline{a} \underline{b}} W^{c d} {G}^{-1} \varphi_{i \alpha} \varphi^{i}_{\beta}+\frac{9}{8}{\rm i} (\Gamma_{\underline{a}})^{\alpha \beta} {G}^{-1} \nabla_{\underline{b}}{\varphi_{i \alpha}} \varphi^{i}_{\beta}+\frac{9}{16}{\rm i} (\Sigma_{\underline{a}}{}^{\, c})^{\alpha \beta} W_{\underline{b} c} {G}^{-1} \varphi_{i \alpha} \varphi^{i}_{\beta} - \frac{9}{16}\mathcal{H}_{\underline{a}} \mathcal{H}_{\underline{b}} {G}^{-1}+\frac{9}{80}\mathcal{H}^{c} \mathcal{H}_{c} \eta_{\underline{a} \underline{b}} {G}^{-1} - \frac{21}{32}{G}^{-1} \nabla_{\underline{a}}{G_{i j}} \nabla_{\underline{b}}{G^{i j}}+\frac{21}{160}\eta_{\underline{a} \underline{b}} {G}^{-1} \nabla_{c}{G_{i j}} \nabla^{c}{G^{i j}}+\frac{9}{128}G_{i j} G_{k l} (\Gamma_{\underline{a}})^{\alpha \beta} (\Gamma_{\underline{b}})^{\rho \lambda} {G}^{-5} \varphi^{i}_{\alpha} \varphi^{j}_{\beta} \varphi^{k}_{\rho} \varphi^{l}_{\lambda} - \frac{3}{8}G_{i j} G_{k l} {G}^{-3} \nabla_{\underline{a}}{G^{i k}} \nabla_{\underline{b}}{G^{j l}}+\frac{3}{40}G_{i j} G_{k l} \eta_{\underline{a} \underline{b}} {G}^{-3} \nabla_{c}{G^{i k}} \nabla^{c}{G^{j l}} - \frac{3}{64}G_{i j} G_{k l} {G}^{-3} \nabla_{\underline{a}}{G^{i j}} \nabla_{\underline{b}}{G^{k l}}+\frac{3}{320}G_{i j} G_{k l} \eta_{\underline{a} \underline{b}} {G}^{-3} \nabla_{c}{G^{i j}} \nabla^{c}{G^{k l}}%
+\frac{27}{32}G_{i j} {G}^{-1} \nabla_{\underline{a}}{\nabla_{\underline{b}}{G^{i j}}} - \frac{27}{160}G_{i j} \eta_{\underline{a} \underline{b}} {G}^{-1} \nabla_{c}{\nabla^{c}{G^{i j}}} - \frac{27}{40}G \eta_{\underline{a} \underline{b}} W^{c d} W_{c d}+\frac{27}{8}G W_{\underline{a}}\,^{c} W_{\underline{b} c} - \frac{27}{40}{\rm i} (\Sigma_{c d})^{\alpha \beta} \eta_{\underline{a} \underline{b}} F^{c d} \lambda_{i \alpha} \lambda^{i}_{\beta}+\frac{27}{8}{\rm i} (\Sigma_{\underline{a}}{}^{\, c})^{\alpha \beta} F_{\underline{b} c} \lambda_{i \alpha} \lambda^{i}_{\beta} - \frac{27}{80}{\rm i} (\Sigma_{c d})^{\alpha \beta} \eta_{\underline{a} \underline{b}} W W^{c d} \lambda_{i \alpha} \lambda^{i}_{\beta}+\frac{27}{16}{\rm i} (\Sigma_{\underline{a}}{}^{\, c})^{\alpha \beta} W W_{\underline{b} c} \lambda_{i \alpha} \lambda^{i}_{\beta}+\frac{27}{8}{\rm i} (\Gamma_{\underline{a}})^{\alpha \beta} W \lambda_{i \alpha} \nabla_{\underline{b}}{\lambda^{i}_{\beta}} - \frac{27}{40}{\rm i} (\Gamma_{c})^{\alpha \beta} \eta_{\underline{a} \underline{b}} W \lambda_{i \alpha} \nabla^{c}{\lambda^{i}_{\beta}} - \frac{27}{4}W F_{\underline{a}}\,^{c} F_{\underline{b} c}+\frac{27}{20}\eta_{\underline{a} \underline{b}} W F^{c d} F_{c d} - \frac{27}{2}W_{\underline{a}}\,^{c} F_{\underline{b} c} {W}^{2}+\frac{27}{10}\eta_{\underline{a} \underline{b}} W^{c d} F_{c d} {W}^{2} - \frac{27}{4}W_{\underline{a}}\,^{c} W_{\underline{b} c} {W}^{3}+\frac{27}{20}\eta_{\underline{a} \underline{b}} W^{c d} W_{c d} {W}^{3}+\frac{27}{40}\eta_{\underline{a} \underline{b}} W \nabla_{c}{W} \nabla^{c}{W} - \frac{27}{8}W \nabla_{\underline{a}}{W} \nabla_{\underline{b}}{W}+\frac{9}{640}\eta_{\underline{a} \underline{b}} {G}^{-3} \varphi_{i}^{\alpha} \varphi^{i \beta} \varphi_{j \alpha} \varphi^{j}_{\beta}+\frac{9}{640}G_{i j} G_{k l} \eta_{\underline{a} \underline{b}} {G}^{-5} \varphi^{i \alpha} \varphi^{j \beta} \varphi^{k}_{\alpha} \varphi^{l}_{\beta}%
+\frac{9}{640}G_{i j} G_{k l} \eta_{\underline{a} \underline{b}} {G}^{-5} \varphi^{i \alpha} \varphi^{j}_{\alpha} \varphi^{k \beta} \varphi^{l}_{\beta}+\frac{27}{16}{W}^{2} \nabla_{\underline{a}}{\nabla_{\underline{b}}{W}} - \frac{27}{80}\eta_{\underline{a} \underline{b}} {W}^{2} \nabla_{c}{\nabla^{c}{W}}
\doublespacedmathend
\end{adjustwidth}

\subsection{Weyl squared} \label{SupercurrentWeylComplete}

\subsubsection{$J^1_{\alpha i, {\rm Weyl}}$} \label{J1WeylComplete}

\begin{adjustwidth}{0cm}{0cm}
\doublespacedmathbegin
- \frac{3}{128}{\rm i} Y \lambda_{i \alpha} - \frac{3}{8}(\Gamma_{a})_{\alpha}{}^{\beta} W \nabla^{a}{X_{i \beta}}+\frac{3}{32}{\rm i} W^{a b} W_{a b} \lambda_{i \alpha}+\frac{9}{128}W W^{a b} W_{a b \alpha i}+\frac{9}{256}\epsilon_{e}\,^{a b c d} (\Gamma^{e})_{\alpha}{}^{\beta} W W_{a b} W_{c d \beta i}+\frac{9}{32}(\Sigma^{c}{}_{\, a})_{\alpha}{}^{\beta} W W^{a b} W_{c b \beta i}+\frac{3}{128}F^{a b} W_{a b \alpha i}+\frac{3}{256}\epsilon_{e}\,^{a b c d} (\Gamma^{e})_{\alpha}{}^{\beta} F_{a b} W_{c d \beta i}+\frac{3}{32}(\Sigma^{c}{}_{\, a})_{\alpha}{}^{\beta} F^{a b} W_{c b \beta i} - \frac{3}{32}(\Sigma_{a b})_{\alpha}{}^{\beta} F^{a b} X_{i \beta} - \frac{3}{16}{\rm i} (\Gamma^{a})_{\alpha}{}^{\beta} W_{a b} \nabla^{b}{\lambda_{i \beta}} - \frac{3}{16}X_{i j} X^{j}_{\alpha} - \frac{3}{16}(\Gamma_{a})_{\alpha}{}^{\beta} X_{i \beta} \nabla^{a}{W}+\frac{1}{8}{\rm i} \Phi^{a b}\,_{i j} (\Sigma_{a b})_{\alpha}{}^{\beta} \lambda^{j}_{\beta} - \frac{3}{32}{\rm i} \epsilon^{c d}\,_{e}\,^{a b} (\Sigma_{c d})_{\alpha}{}^{\beta} \lambda_{i \beta} \nabla^{e}{W_{a b}} - \frac{3}{32}{\rm i} (\Gamma^{a})_{\alpha}{}^{\beta} \lambda_{i \beta} \nabla^{b}{W_{a b}}
\doublespacedmathend
\end{adjustwidth}

\subsubsection{$J^2_{a i j,  {\rm Weyl}}$} \label{J2aijWeylComplete}

\begin{adjustwidth}{0cm}{2cm}
\doublespacedmathbegin
{} - \frac{3}{8}{\rm i} (\Sigma_{a b})^{\beta \alpha} \lambda_{\underline{i} \beta} \nabla^{b}{X_{\underline{j} \alpha}} - \frac{1}{2}{\rm i} W \nabla^{b}{\Phi_{a b \underline{i} \underline{j}}}+\frac{9}{32}(\Gamma_{a})^{\alpha \beta} W X_{\underline{i} \alpha} X_{\underline{j} \beta} - \frac{3}{64}{\rm i} (\Gamma_{a})^{\beta \alpha} W^{b c} \lambda_{\underline{i} \beta} W_{b c \alpha \underline{j}} - \frac{3}{128}{\rm i} \epsilon_{a}\,^{b c d e} W_{b c} \lambda^{\alpha}_{\underline{i}} W_{d e \alpha \underline{j}}+\frac{3}{64}{\rm i} \epsilon^{{e_{1}} b c d e} (\Sigma_{a {e_{1}}})^{\beta \alpha} W_{b c} \lambda_{\underline{i} \beta} W_{d e \alpha \underline{j}} - \frac{3}{32}{\rm i} \epsilon^{e {e_{1}}}\,_{a}\,^{b d} (\Sigma_{e {e_{1}}})^{\beta \alpha} W_{b}\,^{c} \lambda_{\underline{i} \beta} W_{d c \alpha \underline{j}}+\frac{3}{32}{\rm i} (\Gamma^{c})^{\beta \alpha} W_{a}\,^{b} \lambda_{\underline{i} \beta} W_{c b \alpha \underline{j}} - \frac{3}{32}{\rm i} (\Gamma^{b})^{\beta \alpha} W_{b}\,^{c} \lambda_{\underline{i} \beta} W_{a c \alpha \underline{j}} - \frac{9}{32}{\rm i} (\Gamma^{b})^{\beta \alpha} W_{a b} \lambda_{\underline{i} \beta} X_{\underline{j} \alpha} - \frac{3}{64}\epsilon_{a}\,^{b c d e} W W_{b c}\,^{\alpha}\,_{\underline{i}} W_{d e \alpha \underline{j}}+\frac{3}{8}{\rm i} W_{a b}\,^{\alpha}\,_{\underline{i}} \nabla^{b}{\lambda_{\underline{j} \alpha}}+\frac{3}{8}{\rm i} (\Sigma_{a b})^{\alpha \beta} X_{\underline{i} \alpha} \nabla^{b}{\lambda_{\underline{j} \beta}} - \frac{3}{8}{\rm i} W_{a b} \nabla^{b}{X_{\underline{i} \underline{j}}} - \frac{3}{8}{\rm i} X_{\underline{i} \underline{j}} \nabla^{b}{W_{a b}} - \frac{1}{2}{\rm i} \Phi_{a b \underline{i} \underline{j}} \nabla^{b}{W}+\frac{3}{8}{\rm i} \lambda^{\alpha}_{\underline{i}} \nabla^{b}{W_{a b \alpha \underline{j}}} - \frac{1}{16}{\rm i} \epsilon^{d e}\,_{a}\,^{b c} \Phi_{d e \underline{i} \underline{j}} F_{b c}
\doublespacedmathend
\end{adjustwidth}

\subsubsection{$J^2_{a b, {\rm Weyl}}$} \label{J2abWeylComplete}

\begin{adjustwidth}{0cm}{4cm}
\doublespacedmathbegin
{}\frac{3}{16}{\rm i} \epsilon^{c d}\,_{\hat{a} \hat{b} e} (\Sigma_{c d})^{\beta \alpha} \lambda_{i \beta} \nabla^{e}{X^{i}_{\alpha}} - \frac{3}{16}{\rm i} (\Gamma_{\hat{a}})^{\beta \alpha} \lambda_{i \beta} \nabla_{\hat{b}}{X^{i}_{\alpha}}+\frac{3}{128}{\rm i} F_{\hat{a} \hat{b}} Y+\frac{9}{128}{\rm i} W W_{\hat{a} \hat{b}} Y - \frac{3}{4}{\rm i} W \nabla_{c}{\nabla^{c}{W_{\hat{a} \hat{b}}}} - \frac{3}{2}{\rm i} W \nabla^{c}{\nabla_{\hat{a}}{W_{\hat{b} c}}} - \frac{3}{4}{\rm i} W \nabla_{\hat{a}}{\nabla^{c}{W_{\hat{b} c}}} - \frac{9}{64}{\rm i} \epsilon_{\hat{a} \hat{b}}\,^{e {e_{1}} c} W W_{c d} \nabla^{d}{W_{e {e_{1}}}}+\frac{9}{32}{\rm i} \epsilon_{\hat{a} {e_{1}}}\,^{e c d} W W_{c d} \nabla^{{e_{1}}}{W_{\hat{b} e}} - \frac{27}{64}{\rm i} \epsilon_{\hat{a} \hat{b}}\,^{e c d} W W_{c d} \nabla^{{e_{1}}}{W_{e {e_{1}}}}+\frac{9}{32}{\rm i} \epsilon_{\hat{a} {e_{1}}}\,^{d e c} W W_{\hat{b} c} \nabla^{{e_{1}}}{W_{d e}}+\frac{45}{128}{\rm i} W W^{c d} W_{\hat{a} \hat{b}} W_{c d} - \frac{99}{32}{\rm i} W W^{c d} W_{\hat{a} c} W_{\hat{b} d} - \frac{15}{32}W X_{i}^{\alpha} W_{\hat{a} \hat{b} \alpha}\,^{i}+\frac{3}{64}{\rm i} (\Sigma_{c d})^{\beta \alpha} W^{c d} \lambda_{i \beta} W_{\hat{a} \hat{b} \alpha}\,^{i}+\frac{3}{64}{\rm i} (\Sigma_{\hat{a} \hat{b}})^{\beta \alpha} W^{c d} \lambda_{i \beta} W_{c d \alpha}\,^{i} - \frac{3}{16}{\rm i} (\Sigma_{\hat{a}}{}^{\, d})^{\beta \alpha} W_{\hat{b}}\,^{c} \lambda_{i \beta} W_{d c \alpha}\,^{i}+\frac{3}{64}{\rm i} (\Sigma^{c d})^{\beta \alpha} W_{\hat{a} \hat{b}} \lambda_{i \beta} W_{c d \alpha}\,^{i}+\frac{3}{128}{\rm i} \epsilon_{\hat{a} \hat{b}}\,^{c e {e_{1}}} (\Gamma^{d})^{\beta \alpha} W_{c d} \lambda_{i \beta} W_{e {e_{1}} \alpha}\,^{i}%
 - \frac{9}{32}{\rm i} W_{\hat{a}}\,^{c} \lambda^{\alpha}_{i} W_{\hat{b} c \alpha}\,^{i} - \frac{3}{64}{\rm i} \epsilon_{\hat{a} {e_{1}}}\,^{c d e} (\Gamma^{{e_{1}}})^{\beta \alpha} W_{c d} \lambda_{i \beta} W_{\hat{b} e \alpha}\,^{i} - \frac{3}{16}{\rm i} (\Sigma^{c d})^{\beta \alpha} W_{\hat{a} c} \lambda_{i \beta} W_{\hat{b} d \alpha}\,^{i}+\frac{3}{16}{\rm i} (\Sigma_{\hat{a} c})^{\beta \alpha} W^{c d} \lambda_{i \beta} W_{\hat{b} d \alpha}\,^{i} - \frac{3}{16}W W_{\hat{a}}\,^{c \alpha}\,_{i} W_{\hat{b} c \alpha}\,^{i} - \frac{3}{16}{\rm i} W C_{\hat{a} \hat{b} c d} W^{c d} - \frac{9}{16}{\rm i} W W_{\hat{a}}\,^{c} W_{\hat{b}}\,^{d} W_{d c} - \frac{3}{16}{\rm i} W C_{c d \hat{a} \hat{b}} W^{c d} - \frac{3}{8}{\rm i} W C_{\hat{a} c \hat{b} d} W^{c d} - \frac{9}{128}{\rm i} \epsilon_{\hat{a} \hat{b} e}\,^{c d} W_{c d}\,^{\alpha}\,_{i} \nabla^{e}{\lambda^{i}_{\alpha}}+\frac{3}{64}{\rm i} (\Gamma_{c})^{\alpha \beta} W_{\hat{a} \hat{b} \alpha i} \nabla^{c}{\lambda^{i}_{\beta}} - \frac{3}{64}{\rm i} \epsilon^{e}\,_{\hat{a} \hat{b}}\,^{c d} (\Sigma_{e {e_{1}}})^{\alpha \beta} W_{c d \alpha i} \nabla^{{e_{1}}}{\lambda^{i}_{\beta}} - \frac{3}{32}{\rm i} \epsilon^{d e}\,_{\hat{a} {e_{1}}}\,^{c} (\Sigma_{d e})^{\alpha \beta} W_{\hat{b} c \alpha i} \nabla^{{e_{1}}}{\lambda^{i}_{\beta}}+\frac{3}{32}{\rm i} (\Gamma^{c})^{\alpha \beta} W_{\hat{b} c \alpha i} \nabla_{\hat{a}}{\lambda^{i}_{\beta}} - \frac{3}{32}{\rm i} (\Gamma_{\hat{a}})^{\alpha \beta} W_{\hat{b} c \alpha i} \nabla^{c}{\lambda^{i}_{\beta}} - \frac{1}{16}{\rm i} C_{\hat{a} \hat{b} c d} F^{c d} - \frac{3}{32}{\rm i} W_{\hat{a} \hat{b}} W_{c d} F^{c d} - \frac{3}{128}{\rm i} W^{c d} W_{c d} F_{\hat{a} \hat{b}} - \frac{9}{32}{\rm i} W_{\hat{a} c} W_{\hat{b} d} F^{c d} - \frac{3}{16}{\rm i} W_{\hat{b}}\,^{d} W_{d c} F_{\hat{a}}\,^{c}%
 - \frac{1}{16}{\rm i} C_{c d \hat{a} \hat{b}} F^{c d} - \frac{1}{8}{\rm i} C_{\hat{a} c \hat{b} d} F^{c d}+\frac{3}{64}{\rm i} \epsilon_{\hat{a} \hat{b}}\,^{e {e_{1}} c} F_{c d} \nabla^{d}{W_{e {e_{1}}}} - \frac{3}{32}{\rm i} \epsilon_{\hat{a} {e_{1}}}\,^{e c d} F_{c d} \nabla^{{e_{1}}}{W_{\hat{b} e}} - \frac{9}{64}{\rm i} \epsilon_{\hat{a} \hat{b}}\,^{e c d} F_{c d} \nabla^{{e_{1}}}{W_{e {e_{1}}}}+\frac{15}{32}{\rm i} \epsilon_{\hat{a} {e_{1}}}\,^{d e c} F_{\hat{b} c} \nabla^{{e_{1}}}{W_{d e}}+\frac{3}{16}{\rm i} \epsilon^{c d}\,_{\hat{a} \hat{b} e} (\Sigma_{c d})^{\alpha \beta} X_{i \alpha} \nabla^{e}{\lambda^{i}_{\beta}} - \frac{3}{16}{\rm i} (\Gamma_{\hat{a}})^{\alpha \beta} X_{i \alpha} \nabla_{\hat{b}}{\lambda^{i}_{\beta}}+\frac{9}{32}{\rm i} W_{\hat{a} \hat{b}} \lambda^{\alpha}_{i} X^{i}_{\alpha}+\frac{3}{32}{\rm i} W^{c d} W_{\hat{a} \hat{b}} F_{c d} - \frac{3}{4}{\rm i} W^{d c} W_{\hat{a} d} F_{\hat{b} c} - \frac{3}{32}{\rm i} \epsilon_{\hat{a} \hat{b}}\,^{c d e} W_{e {e_{1}}} \nabla^{{e_{1}}}{F_{c d}}+\frac{3}{16}{\rm i} \epsilon_{\hat{a} {e_{1}}}\,^{c d e} W_{d e} \nabla^{{e_{1}}}{F_{\hat{b} c}} - \frac{9}{64}{\rm i} \epsilon_{\hat{a} \hat{b}}\,^{c d e} W_{c d} W_{e {e_{1}}} \nabla^{{e_{1}}}{W}+\frac{9}{32}{\rm i} \epsilon_{\hat{a} {e_{1}}}\,^{c d e} W_{\hat{b} c} W_{d e} \nabla^{{e_{1}}}{W} - \frac{3}{4}{\rm i} W_{\hat{b} c} \nabla^{c}{\nabla_{\hat{a}}{W}}+\frac{3}{32}{\rm i} \epsilon_{\hat{a} \hat{b}}\,^{c e {e_{1}}} W_{e {e_{1}}} \nabla^{d}{F_{c d}}+\frac{3}{16}{\rm i} \epsilon_{\hat{a} {e_{1}}}\,^{c d e} W_{\hat{b} e} \nabla^{{e_{1}}}{F_{c d}}+\frac{1}{4}{\rm i} \Phi_{\hat{a} \hat{b} i j} X^{i j} - \frac{3}{4}{\rm i} \nabla_{c}{W} \nabla^{c}{W_{\hat{a} \hat{b}}}%
 - \frac{3}{2}{\rm i} \nabla^{c}{W} \nabla_{\hat{a}}{W_{\hat{b} c}} - \frac{3}{4}{\rm i} \nabla_{\hat{a}}{W} \nabla^{c}{W_{\hat{b} c}}+\frac{39}{128}{\rm i} \epsilon_{\hat{a} \hat{b} e}\,^{c d} \lambda^{\alpha}_{i} \nabla^{e}{W_{c d \alpha}\,^{i}}+\frac{3}{64}{\rm i} (\Gamma_{c})^{\beta \alpha} \lambda_{i \beta} \nabla^{c}{W_{\hat{a} \hat{b} \alpha}\,^{i}}+\frac{3}{64}{\rm i} \epsilon^{e}\,_{\hat{a} \hat{b}}\,^{c d} (\Sigma_{e {e_{1}}})^{\beta \alpha} \lambda_{i \beta} \nabla^{{e_{1}}}{W_{c d \alpha}\,^{i}}+\frac{3}{32}{\rm i} \epsilon^{d e}\,_{\hat{a} {e_{1}}}\,^{c} (\Sigma_{d e})^{\beta \alpha} \lambda_{i \beta} \nabla^{{e_{1}}}{W_{\hat{b} c \alpha}\,^{i}}+\frac{3}{32}{\rm i} (\Gamma^{c})^{\beta \alpha} \lambda_{i \beta} \nabla_{\hat{a}}{W_{\hat{b} c \alpha}\,^{i}} - \frac{3}{32}{\rm i} (\Gamma_{\hat{a}})^{\beta \alpha} \lambda_{i \beta} \nabla^{c}{W_{\hat{b} c \alpha}\,^{i}}
 \doublespacedmathend
 \end{adjustwidth}

\subsubsection{$J^3_{a i \alpha, {\rm Weyl}}$} \label{J3WeylComplete}

\begin{adjustwidth}{0cm}{5cm}
\doublespacedmathbegin
{} - \frac{45}{64}{\rm i} (\Gamma_{a})^{\beta}{}_{\lambda} F_{\alpha \beta} \nabla^{\lambda \rho}{X_{i \rho}} - \frac{873}{512}{\rm i} (\Gamma_{a})^{\beta}{}_{\lambda} W W_{\alpha \beta} \nabla^{\lambda \rho}{X_{i \rho}}+\frac{9}{64}{\rm i} (\Gamma_{a})_{\rho \lambda} \nabla^{\rho}\,_{\alpha}{W} \nabla^{\lambda \beta}{X_{i \beta}} - \frac{9}{512}(\Gamma_{a})^{\beta}{}_{\rho} \lambda_{i \beta} \nabla^{\rho}\,_{\alpha}{Y} - \frac{45}{128}(\Gamma_{a})^{\lambda}{}_{\gamma} \lambda_{i \lambda} \nabla^{\gamma \beta}{\nabla_{\alpha}\,^{\rho}{W_{\beta \rho}}} - \frac{9}{128}(\Gamma_{a})^{\lambda}{}_{\gamma} W^{\beta}\,_{\rho} \lambda_{i \lambda} \nabla^{\gamma \rho}{W_{\alpha \beta}}+\frac{9}{32}(\Gamma_{a})^{\lambda}{}_{\gamma} W_{\alpha}\,^{\beta} \lambda_{i \lambda} \nabla^{\gamma \rho}{W_{\beta \rho}} - \frac{27}{128}(\Gamma_{a})^{\lambda}{}_{\gamma} W^{\beta \rho} \lambda_{i \lambda} \nabla^{\gamma}\,_{\alpha}{W_{\beta \rho}}+\frac{9}{128}(\Gamma_{a})^{\rho}{}_{\lambda} \lambda_{i \rho} \nabla^{\lambda}\,_{\gamma}{\nabla^{\gamma \beta}{W_{\alpha \beta}}} - \frac{33}{32}\Phi^{\rho \beta}\,_{i j} (\Gamma_{a})_{\rho}{}^{\lambda} W_{\alpha \beta} \lambda^{j}_{\lambda} - \frac{45}{512}(\Gamma_{a})^{\beta \rho} W_{\alpha \beta} Y \lambda_{i \rho}+\frac{45}{32}(\Gamma_{a})^{\gamma \rho} W_{\alpha \beta} \lambda_{i \gamma} \nabla^{\beta \lambda}{W_{\rho \lambda}} - \frac{81}{16}(\Gamma_{a})^{\rho \gamma} W_{\alpha}\,^{\beta} W_{\rho}\,^{\lambda} W_{\beta \lambda} \lambda_{i \gamma}+\frac{207}{64}(\Gamma_{a})^{\beta \gamma} W_{\alpha \beta} W^{\rho \lambda} W_{\rho \lambda} \lambda_{i \gamma}+\frac{39}{160}\Phi^{\beta \rho}\,_{i j} (\Gamma_{a})_{\alpha}{}^{\lambda} W_{\beta \rho} \lambda^{j}_{\lambda}+\frac{63}{160}(\Gamma_{a})_{\alpha}{}^{\gamma} W^{\beta}\,_{\rho} \lambda_{i \gamma} \nabla^{\rho \lambda}{W_{\beta \lambda}}+\frac{21}{32}\Phi_{\alpha}\,^{\beta}\,_{i j} (\Gamma_{a})^{\rho \lambda} W_{\beta \rho} \lambda^{j}_{\lambda} - \frac{261}{256}(\Gamma_{a})^{\beta \gamma} W_{\beta \rho} \lambda_{i \gamma} \nabla^{\rho \lambda}{W_{\alpha \lambda}}+\frac{9}{256}(\Gamma_{a})^{\beta \gamma} W_{\beta}\,^{\rho} \lambda_{i \gamma} \nabla_{\alpha}\,^{\lambda}{W_{\rho \lambda}}%
+\frac{45}{128}{\rm i} (\Gamma_{a})^{\lambda \beta} \lambda_{j \lambda} X_{i}^{\rho} W_{\alpha \beta \rho}\,^{j}+\frac{39}{128}{\rm i} (\Gamma_{a})^{\rho \beta} \lambda_{j \rho} X_{i \alpha} X^{j}_{\beta} - \frac{15}{128}{\rm i} (\Gamma_{a})^{\rho \beta} \lambda_{j \rho} X^{j}_{\alpha} X_{i \beta}+\frac{87}{640}{\rm i} (\Gamma_{a})_{\alpha}{}^{\rho} \lambda_{j \rho} X_{i}^{\beta} X^{j}_{\beta}+\frac{27}{64}{\rm i} (\Gamma_{a})^{\rho \beta} \lambda_{i \rho} X_{j \alpha} X^{j}_{\beta} - \frac{9}{128}(\Gamma_{a})^{\gamma \lambda} W^{\beta}\,_{\rho} \lambda_{i \gamma} \nabla_{\alpha}\,^{\rho}{W_{\lambda \beta}} - \frac{3}{32}{\rm i} (\Gamma_{a})_{\alpha \rho} X_{i j} \nabla^{\rho \beta}{X^{j}_{\beta}}+\frac{69}{128}{\rm i} (\Gamma_{a})^{\lambda \beta} \lambda_{i \lambda} X_{j}^{\rho} W_{\alpha \beta \rho}\,^{j} - \frac{7}{16}(\Gamma_{a})^{\beta}{}_{\rho} \lambda_{j \alpha} \nabla^{\rho \lambda}{\Phi_{\lambda \beta i}\,^{j}}+\frac{9}{16}{\rm i} (\Gamma_{a})^{\beta}{}_{\gamma} W \nabla^{\gamma \rho}{\nabla_{\alpha}\,^{\lambda}{W_{\beta \rho \lambda i}}} - \frac{45}{32}{\rm i} (\Gamma_{a})^{\beta}{}_{\gamma} W W_{\alpha \beta}\,^{\rho}\,_{i} \nabla^{\gamma \lambda}{W_{\rho \lambda}}+\frac{135}{256}{\rm i} (\Gamma_{a})^{\lambda}{}_{\gamma} W W^{\beta}\,_{\rho} \nabla^{\gamma \rho}{W_{\alpha \lambda \beta i}}+\frac{27}{64}{\rm i} (\Gamma_{a})^{\lambda}{}_{\gamma} W W_{\alpha}\,^{\beta}\,_{\rho i} \nabla^{\gamma \rho}{W_{\lambda \beta}}+\frac{9}{32}{\rm i} (\Gamma_{a})^{\beta}{}_{\gamma} W W_{\beta}\,^{\rho \lambda}\,_{i} \nabla^{\gamma}\,_{\alpha}{W_{\rho \lambda}} - \frac{45}{256}{\rm i} (\Gamma_{a})^{\lambda}{}_{\gamma} W W^{\beta \rho} \nabla^{\gamma}\,_{\alpha}{W_{\lambda \beta \rho i}}+\frac{81}{320}{\rm i} (\Gamma_{a})_{\alpha \gamma} W W^{\beta \rho}\,_{\lambda i} \nabla^{\gamma \lambda}{W_{\beta \rho}}+\frac{297}{320}{\rm i} (\Gamma_{a})_{\alpha \gamma} W W^{\beta \rho} \nabla^{\gamma \lambda}{W_{\beta \rho \lambda i}} - \frac{9}{64}{\rm i} (\Gamma_{a})^{\beta}{}_{\gamma} W W_{\beta}\,^{\rho}\,_{\lambda i} \nabla^{\gamma \lambda}{W_{\alpha \rho}}+\frac{9}{256}{\rm i} (\Gamma_{a})^{\rho}{}_{\gamma} W W_{\alpha}\,^{\beta} \nabla^{\gamma \lambda}{W_{\rho \beta \lambda i}} - \frac{369}{256}{\rm i} (\Gamma_{a})^{\beta}{}_{\lambda} W X_{i \alpha} \nabla^{\lambda \rho}{W_{\beta \rho}}%
 - \frac{459}{512}{\rm i} (\Gamma_{a})^{\beta}{}_{\lambda} W W_{\beta \rho} \nabla^{\lambda \rho}{X_{i \alpha}} - \frac{9}{32}{\rm i} (\Gamma_{a})^{\beta}{}_{\lambda} W \nabla^{\lambda}\,_{\gamma}{\nabla^{\gamma \rho}{W_{\alpha \beta \rho i}}}+\frac{9}{32}{\rm i} (\Gamma_{a})_{\lambda \gamma} W \nabla^{\lambda \beta}{\nabla^{\gamma \rho}{W_{\alpha \beta \rho i}}}+\frac{9}{64}{\rm i} (\Gamma_{a})^{\beta}{}_{\rho} W \nabla^{\rho}\,_{\lambda}{\nabla_{\alpha}\,^{\lambda}{X_{i \beta}}}+\frac{9}{64}{\rm i} (\Gamma_{a})_{\rho \lambda} W \nabla^{\rho \beta}{\nabla^{\lambda}\,_{\alpha}{X_{i \beta}}}+\frac{45}{256}{\rm i} (\Gamma_{a})^{\rho}{}_{\lambda} W X_{i \rho} \nabla^{\lambda \beta}{W_{\alpha \beta}}+\frac{45}{128}{\rm i} (\Gamma_{a})^{\rho}{}_{\lambda} W W_{\alpha \beta} \nabla^{\lambda \beta}{X_{i \rho}} - \frac{27}{64}{\rm i} (\Gamma_{a})^{\beta}{}_{\lambda} W X_{i \rho} \nabla^{\lambda \rho}{W_{\alpha \beta}}+\frac{27}{64}{\rm i} (\Gamma_{a})_{\rho \lambda} W \nabla^{\rho}\,_{\alpha}{\nabla^{\lambda \beta}{X_{i \beta}}}+\frac{171}{320}{\rm i} (\Gamma_{a})_{\alpha \rho} W \nabla^{\rho}\,_{\lambda}{\nabla^{\lambda \beta}{X_{i \beta}}} - \frac{27}{32}{\rm i} (\Gamma_{a})^{\beta}{}_{\lambda} W X_{i}^{\rho} \nabla^{\lambda}\,_{\alpha}{W_{\beta \rho}}+\frac{675}{512}{\rm i} (\Gamma_{a})^{\beta}{}_{\lambda} W W_{\beta}\,^{\rho} \nabla^{\lambda}\,_{\alpha}{X_{i \rho}}+\frac{1197}{1280}{\rm i} (\Gamma_{a})_{\alpha \lambda} W X_{i}^{\beta} \nabla^{\lambda \rho}{W_{\beta \rho}} - \frac{1323}{1280}{\rm i} (\Gamma_{a})_{\alpha \lambda} W W^{\beta}\,_{\rho} \nabla^{\lambda \rho}{X_{i \beta}} - \frac{207}{32}{\rm i} (\Gamma_{a})^{\beta \gamma} W W_{\alpha \beta} W^{\rho \lambda} W_{\gamma \rho \lambda i} - \frac{405}{64}{\rm i} (\Gamma_{a})^{\beta \rho} W W_{\alpha \beta} W_{\rho}\,^{\lambda} X_{i \lambda}+\frac{135}{32}{\rm i} (\Gamma_{a})^{\rho \gamma} W W_{\alpha}\,^{\beta} W_{\rho}\,^{\lambda} W_{\gamma \beta \lambda i} - \frac{963}{1280}{\rm i} (\Gamma_{a})_{\alpha}{}^{\lambda} W W^{\beta}\,_{\rho} \nabla^{\rho \gamma}{W_{\lambda \beta \gamma i}} - \frac{9}{20}{\rm i} (\Gamma_{a})_{\alpha}{}^{\beta} W W_{\beta}\,^{\rho} W^{\lambda \gamma} W_{\rho \lambda \gamma i}+\frac{135}{256}{\rm i} (\Gamma_{a})_{\alpha}{}^{\beta} W W_{\beta}\,^{\rho} W_{\rho}\,^{\lambda} X_{i \lambda}%
+\frac{135}{512}{\rm i} (\Gamma_{a})_{\alpha}{}^{\lambda} W W^{\beta \rho} W_{\beta \rho} X_{i \lambda} - \frac{1449}{2560}{\rm i} (\Gamma_{a})_{\alpha}{}^{\beta} W W_{\beta \rho} \nabla^{\rho \lambda}{X_{i \lambda}}+\frac{27}{16}{\rm i} (\Gamma_{a})^{\beta \lambda} W W_{\beta \rho} \nabla^{\rho \gamma}{W_{\alpha \lambda \gamma i}} - \frac{243}{32}{\rm i} (\Gamma_{a})^{\beta \gamma} W W_{\beta}\,^{\rho} W_{\rho}\,^{\lambda} W_{\alpha \gamma \lambda i}+\frac{567}{256}{\rm i} (\Gamma_{a})^{\beta \lambda} W W_{\beta}\,^{\rho} \nabla_{\alpha}\,^{\gamma}{W_{\lambda \rho \gamma i}}+\frac{837}{512}{\rm i} (\Gamma_{a})^{\beta \lambda} W W_{\beta \rho} \nabla_{\alpha}\,^{\rho}{X_{i \lambda}}+\frac{189}{256}{\rm i} (\Gamma_{a})^{\rho \lambda} W W_{\alpha}\,^{\beta} W_{\rho \beta} X_{i \lambda} - \frac{1431}{512}{\rm i} (\Gamma_{a})^{\beta \lambda} W W_{\beta}\,^{\rho} W_{\lambda \rho} X_{i \alpha} - \frac{45}{512}{\rm i} (\Gamma_{a})_{\rho \lambda} W W_{\alpha}\,^{\beta} \nabla^{\rho \lambda}{X_{i \beta}}+\frac{3}{10}{\rm i} \Phi^{\beta \rho}\,_{i j} (\Gamma_{a})_{\alpha}{}^{\lambda} W W_{\beta \rho \lambda}\,^{j} - \frac{9}{4}{\rm i} \Phi_{\alpha}\,^{\rho}\,_{i j} (\Gamma_{a})_{\rho}{}^{\beta} W X^{j}_{\beta} - \frac{9}{20}{\rm i} \Phi^{\rho \beta}\,_{i j} (\Gamma_{a})_{\alpha \rho} W X^{j}_{\beta} - \frac{927}{256}{\rm i} (\Gamma_{a})^{\lambda \beta} W X_{i \lambda} \nabla_{\alpha}\,^{\rho}{W_{\beta \rho}}+\frac{99}{64}{\rm i} (\Gamma_{a})^{\beta \gamma} W W_{\beta}\,^{\rho}\,_{\lambda i} \nabla_{\alpha}\,^{\lambda}{W_{\gamma \rho}} - \frac{351}{320}{\rm i} (\Gamma_{a})_{\alpha}{}^{\beta} W W_{\beta}\,^{\rho}\,_{\lambda i} \nabla^{\lambda \gamma}{W_{\rho \gamma}} - \frac{27}{16}{\rm i} (\Gamma_{a})^{\beta \lambda} W W_{\alpha \beta \rho i} \nabla^{\rho \gamma}{W_{\lambda \gamma}}+\frac{81}{128}{\rm i} (\Gamma_{a})_{\rho \lambda} W X_{i}^{\beta} \nabla^{\rho \lambda}{W_{\alpha \beta}}+\frac{1179}{1280}{\rm i} (\Gamma_{a})_{\alpha}{}^{\beta} W X_{i \lambda} \nabla^{\lambda \rho}{W_{\beta \rho}} - \frac{39}{128}{\rm i} (\Gamma_{a})^{\beta \rho} \lambda_{j \alpha} X_{i \beta} X^{j}_{\rho}+\frac{9}{16}{\rm i} (\Gamma_{a})^{\gamma \lambda} \lambda_{j \gamma} W_{\alpha}\,^{\beta \rho j} W_{\lambda \beta \rho i}%
 - \frac{33}{128}{\rm i} (\Gamma_{a})^{\lambda \beta} \lambda_{j \lambda} X^{j \rho} W_{\alpha \beta \rho i} - \frac{81}{32}{\rm i} (\Gamma_{a})^{\beta \gamma} W^{\rho \lambda} F_{\alpha \beta} W_{\gamma \rho \lambda i}+\frac{27}{16}(\Gamma_{a})^{\beta \gamma} C_{\alpha \beta}\,^{\rho \lambda} W_{\rho \lambda} \lambda_{i \gamma}+\frac{135}{64}{\rm i} (\Gamma_{a})^{\rho \beta} W_{\rho}\,^{\lambda} F_{\alpha \beta} X_{i \lambda}+\frac{27}{32}{\rm i} (\Gamma_{a})^{\beta}{}_{\lambda} W_{\beta}\,^{\rho} X_{i \rho} \nabla^{\lambda}\,_{\alpha}{W} - \frac{15}{32}{\rm i} (\Gamma_{a})^{\gamma \lambda} \lambda_{i \gamma} W_{\alpha}\,^{\beta \rho}\,_{j} W_{\lambda \beta \rho}\,^{j}+\frac{9}{160}{\rm i} (\Gamma_{a})_{\alpha}{}^{\lambda} X_{i j} W^{\beta \rho} W_{\lambda \beta \rho}\,^{j}+\frac{9}{64}{\rm i} (\Gamma_{a})_{\alpha}{}^{\beta} X_{i j} W_{\beta}\,^{\rho} X^{j}_{\rho}+\frac{3}{8}{\rm i} (\Gamma_{a})^{\beta \gamma} \lambda_{j \alpha} W_{\beta}\,^{\rho \lambda}\,_{i} W_{\gamma \rho \lambda}\,^{j} - \frac{9}{4}{\rm i} (\Gamma_{a})^{\beta \gamma} W C_{\alpha \beta}\,^{\rho \lambda} W_{\gamma \rho \lambda i} - \frac{9}{8}(\Gamma_{a})^{\beta}{}_{\gamma} C_{\alpha \beta}\,^{\rho}\,_{\lambda} \nabla^{\gamma \lambda}{\lambda_{i \rho}}+\frac{9}{32}(\Gamma_{a})^{\rho}{}_{\gamma} W_{\alpha \beta} W_{\rho}\,^{\lambda} \nabla^{\gamma \beta}{\lambda_{i \lambda}} - \frac{45}{32}(\Gamma_{a})^{\beta}{}_{\gamma} W_{\alpha \beta} W^{\rho}\,_{\lambda} \nabla^{\gamma \lambda}{\lambda_{i \rho}} - \frac{9}{64}(\Gamma_{a})^{\beta}{}_{\lambda} \nabla_{\alpha \gamma}{W_{\beta}\,^{\rho}} \nabla^{\lambda \gamma}{\lambda_{i \rho}} - \frac{9}{64}(\Gamma_{a})^{\beta}{}_{\lambda} \nabla_{\alpha}\,^{\gamma}{W_{\beta \rho}} \nabla^{\lambda \rho}{\lambda_{i \gamma}} - \frac{9}{64}(\Gamma_{a})_{\lambda \gamma} \nabla^{\lambda \rho}{W^{\beta}\,_{\rho}} \nabla^{\gamma}\,_{\alpha}{\lambda_{i \beta}} - \frac{9}{320}(\Gamma_{a})_{\alpha \lambda} \nabla^{\beta \gamma}{W_{\beta \rho}} \nabla^{\lambda \rho}{\lambda_{i \gamma}} - \frac{9}{80}(\Gamma_{a})_{\alpha \lambda} \nabla_{\gamma}\,^{\rho}{W^{\beta}\,_{\rho}} \nabla^{\lambda \gamma}{\lambda_{i \beta}}+\frac{9}{64}(\Gamma_{a})_{\lambda \gamma} \nabla^{\lambda \beta}{W_{\beta \rho}} \nabla^{\gamma \rho}{\lambda_{i \alpha}} - \frac{9}{16}{\rm i} (\Gamma_{a})^{\rho}{}_{\gamma} W_{\rho}\,^{\beta}\,_{\lambda i} \nabla^{\gamma \lambda}{F_{\alpha \beta}}%
+\frac{9}{16}{\rm i} (\Gamma_{a})^{\beta}{}_{\lambda} W_{\alpha \beta \rho j} \nabla^{\lambda \rho}{X_{i}\,^{j}}+\frac{9}{32}{\rm i} (\Gamma_{a})^{\beta}{}_{\gamma} W_{\beta \rho \lambda i} \nabla^{\gamma \rho}{\nabla_{\alpha}\,^{\lambda}{W}} - \frac{153}{320}{\rm i} (\Gamma_{a})_{\alpha}{}^{\gamma} W^{\beta \lambda} F_{\beta}\,^{\rho} W_{\gamma \lambda \rho i} - \frac{9}{20}{\rm i} (\Gamma_{a})_{\alpha}{}^{\lambda} W^{\beta}\,_{\rho} W_{\lambda \beta \gamma i} \nabla^{\rho \gamma}{W} - \frac{45}{16}{\rm i} (\Gamma_{a})^{\lambda \gamma} W_{\lambda}\,^{\beta} F_{\beta}\,^{\rho} W_{\alpha \gamma \rho i}+\frac{9}{32}{\rm i} (\Gamma_{a})^{\beta \lambda} X_{i j} W_{\beta}\,^{\rho} W_{\alpha \lambda \rho}\,^{j}+\frac{9}{32}{\rm i} (\Gamma_{a})^{\beta \lambda} W_{\beta \rho} W_{\alpha \lambda \gamma i} \nabla^{\rho \gamma}{W}+\frac{81}{64}{\rm i} (\Gamma_{a})^{\beta \gamma} W_{\alpha}\,^{\lambda} F_{\beta}\,^{\rho} W_{\gamma \lambda \rho i} - \frac{45}{32}{\rm i} (\Gamma_{a})^{\lambda \gamma} W_{\alpha \lambda} F^{\beta \rho} W_{\gamma \beta \rho i} - \frac{3}{160}{\rm i} (\Gamma_{a})_{\alpha}{}^{\beta} \lambda^{\gamma}_{i} W_{\beta}\,^{\rho \lambda}\,_{j} W_{\gamma \rho \lambda}\,^{j}+\frac{45}{64}{\rm i} (\Gamma_{a})^{\lambda \beta} \lambda^{\rho}_{i} X_{j \lambda} W_{\alpha \beta \rho}\,^{j}+\frac{159}{640}{\rm i} (\Gamma_{a})_{\alpha}{}^{\beta} \lambda^{\rho}_{i} X_{j}^{\lambda} W_{\beta \rho \lambda}\,^{j} - \frac{123}{640}{\rm i} (\Gamma_{a})_{\alpha}{}^{\beta} \lambda^{\rho}_{j} X^{j \lambda} W_{\beta \rho \lambda i} - \frac{45}{64}{\rm i} (\Gamma_{a})^{\lambda \beta} \lambda^{\rho}_{j} X^{j}_{\lambda} W_{\alpha \beta \rho i} - \frac{9}{640}{\rm i} (\Gamma_{a})_{\alpha}{}^{\beta} \lambda^{\rho}_{j} X_{i}^{\lambda} W_{\beta \rho \lambda}\,^{j} - \frac{27}{64}{\rm i} (\Gamma_{a})^{\lambda \beta} \lambda^{\rho}_{j} X_{i \lambda} W_{\alpha \beta \rho}\,^{j} - \frac{3}{2}{\rm i} \Phi^{\lambda \beta}\,_{i j} (\Gamma_{a})_{\lambda}{}^{\rho} W W_{\alpha \beta \rho}\,^{j}+\frac{1}{8}\Phi_{\rho}\,^{\lambda}\,_{i j} (\Gamma_{a})_{\alpha \beta} \nabla^{\rho \beta}{\lambda^{j}_{\lambda}}+\frac{1}{4}\Phi_{\lambda}\,^{\beta}\,_{i j} (\Gamma_{a})_{\beta \rho} \nabla^{\lambda \rho}{\lambda^{j}_{\alpha}} - \frac{9}{256}(\Gamma_{a})^{\beta}{}_{\rho} Y \nabla^{\rho}\,_{\alpha}{\lambda_{i \beta}}%
 - \frac{9}{16}(\Gamma_{a})^{\lambda}{}_{\gamma} \nabla_{\alpha}\,^{\beta}{W_{\beta \rho}} \nabla^{\gamma \rho}{\lambda_{i \lambda}} - \frac{9}{32}(\Gamma_{a})^{\lambda}{}_{\gamma} W_{\alpha}\,^{\beta} W_{\beta \rho} \nabla^{\gamma \rho}{\lambda_{i \lambda}}+\frac{9}{32}(\Gamma_{a})^{\lambda}{}_{\gamma} W^{\beta \rho} W_{\beta \rho} \nabla^{\gamma}\,_{\alpha}{\lambda_{i \lambda}} - \frac{9}{64}(\Gamma_{a})^{\rho}{}_{\lambda} \nabla_{\gamma}\,^{\beta}{W_{\alpha \beta}} \nabla^{\lambda \gamma}{\lambda_{i \rho}} - \frac{9}{32}{\rm i} (\Gamma_{a})^{\beta}{}_{\lambda} X_{i \rho} \nabla^{\lambda \rho}{F_{\alpha \beta}} - \frac{45}{64}{\rm i} (\Gamma_{a})^{\beta}{}_{\lambda} W_{\alpha \beta} X_{i \rho} \nabla^{\lambda \rho}{W}+\frac{9}{64}{\rm i} (\Gamma_{a})_{\rho \lambda} X_{i \beta} \nabla^{\rho \beta}{\nabla^{\lambda}\,_{\alpha}{W}} - \frac{9}{128}{\rm i} (\Gamma_{a})^{\beta \lambda} W_{\alpha}\,^{\rho} F_{\beta \rho} X_{i \lambda}+\frac{9}{64}{\rm i} (\Gamma_{a})^{\rho}{}_{\lambda} W_{\alpha \beta} X_{i \rho} \nabla^{\lambda \beta}{W}+\frac{99}{128}{\rm i} (\Gamma_{a})_{\alpha}{}^{\beta} W^{\rho \lambda} F_{\beta \rho} X_{i \lambda} - \frac{63}{64}{\rm i} (\Gamma_{a})_{\alpha \lambda} W^{\beta}\,_{\rho} X_{i \beta} \nabla^{\lambda \rho}{W} - \frac{117}{64}{\rm i} (\Gamma_{a})^{\beta}{}_{\lambda} W_{\beta \rho} X_{i \alpha} \nabla^{\lambda \rho}{W} - \frac{9}{64}{\rm i} (\Gamma_{a})^{\lambda \beta} W_{\alpha \lambda} F_{\beta}\,^{\rho} X_{i \rho}+\frac{27}{320}{\rm i} (\Gamma_{a})_{\alpha}{}^{\beta} \lambda^{\rho}_{i} X_{j \beta} X^{j}_{\rho} - \frac{153}{128}{\rm i} (\Gamma_{a})^{\rho \lambda} W_{\rho}\,^{\beta} F_{\alpha \beta} X_{i \lambda}+\frac{45}{64}{\rm i} (\Gamma_{a})^{\beta \rho} X_{i j} W_{\alpha \beta} X^{j}_{\rho}+\frac{99}{64}{\rm i} (\Gamma_{a})^{\beta \lambda} W_{\beta \rho} X_{i \lambda} \nabla_{\alpha}\,^{\rho}{W} - \frac{9}{512}(\Gamma_{a})_{\alpha}{}^{\beta} W_{\beta}\,^{\rho} Y \lambda_{i \rho} - \frac{27}{16}(\Gamma_{a})^{\beta \lambda} W_{\beta}\,^{\rho} \lambda_{i \rho} \nabla_{\alpha}\,^{\gamma}{W_{\lambda \gamma}}+\frac{27}{128}(\Gamma_{a})^{\beta}{}_{\gamma} W_{\beta}\,^{\rho} \lambda_{i \rho} \nabla^{\gamma \lambda}{W_{\alpha \lambda}}%
 - \frac{1}{8}\Phi_{\alpha \lambda i j} (\Gamma_{a})^{\beta}{}_{\rho} \nabla^{\lambda \rho}{\lambda^{j}_{\beta}}+\frac{141}{640}{\rm i} (\Gamma_{a})_{\alpha}{}^{\rho} \lambda^{\beta}_{j} X_{i \beta} X^{j}_{\rho} - \frac{87}{640}{\rm i} (\Gamma_{a})_{\alpha}{}^{\beta} \lambda^{\rho}_{j} X_{i \beta} X^{j}_{\rho} - \frac{27}{16}\Phi_{\alpha}\,^{\lambda}\,_{i j} (\Gamma_{a})_{\lambda}{}^{\beta} W_{\beta}\,^{\rho} \lambda^{j}_{\rho} - \frac{9}{16}(\Gamma_{a})^{\beta}{}_{\lambda} W_{\beta \rho} \nabla^{\lambda \rho}{\nabla_{\alpha}\,^{\gamma}{\lambda_{i \gamma}}}+\frac{63}{64}(\Gamma_{a})^{\beta}{}_{\gamma} W_{\beta \rho} \lambda^{\lambda}_{i} \nabla^{\gamma \rho}{W_{\alpha \lambda}}+\frac{351}{320}(\Gamma_{a})_{\alpha}{}^{\beta} W_{\beta}\,^{\rho} W_{\rho \lambda} \nabla^{\lambda \gamma}{\lambda_{i \gamma}}+\frac{81}{64}(\Gamma_{a})_{\alpha}{}^{\beta} W_{\beta}\,^{\rho} W_{\rho}\,^{\lambda} W_{\lambda}\,^{\gamma} \lambda_{i \gamma}+\frac{45}{16}(\Gamma_{a})^{\beta \rho} W_{\alpha \beta} W_{\rho \lambda} \nabla^{\lambda \gamma}{\lambda_{i \gamma}} - \frac{81}{64}(\Gamma_{a})^{\beta \rho} W_{\alpha \beta} W_{\rho}\,^{\lambda} W_{\lambda}\,^{\gamma} \lambda_{i \gamma} - \frac{63}{64}(\Gamma_{a})^{\rho}{}_{\lambda} W_{\alpha}\,^{\beta} W_{\rho \beta} \nabla^{\lambda \gamma}{\lambda_{i \gamma}} - \frac{51}{160}\Phi^{\beta \lambda}\,_{i j} (\Gamma_{a})_{\alpha}{}^{\rho} W_{\beta \rho} \lambda^{j}_{\lambda}+\frac{15}{16}\Phi^{\rho \lambda}\,_{i j} (\Gamma_{a})_{\rho}{}^{\beta} W_{\alpha \beta} \lambda^{j}_{\lambda} - \frac{9}{16}(\Gamma_{a})^{\beta \lambda} W_{\beta \rho} \lambda^{\gamma}_{i} \nabla_{\alpha}\,^{\rho}{W_{\lambda \gamma}} - \frac{81}{256}(\Gamma_{a})^{\beta \rho} W_{\alpha \beta} \lambda_{i \gamma} \nabla^{\gamma \lambda}{W_{\rho \lambda}} - \frac{9}{16}(\Gamma_{a})^{\beta \lambda} W_{\beta \rho} \lambda_{i \gamma} \nabla^{\rho \gamma}{W_{\alpha \lambda}}+\frac{171}{256}(\Gamma_{a})^{\beta}{}_{\gamma} W_{\alpha \beta} \lambda^{\rho}_{i} \nabla^{\gamma \lambda}{W_{\rho \lambda}}+\frac{9}{256}(\Gamma_{a})_{\alpha}{}^{\beta} W_{\beta \rho} \lambda^{\lambda}_{i} \nabla^{\rho \gamma}{W_{\lambda \gamma}} - \frac{3}{4}\Phi_{\alpha}\,^{\beta}\,_{i j} (\Gamma_{a})_{\beta \rho} \nabla^{\rho \lambda}{\lambda^{j}_{\lambda}}+\frac{9}{256}(\Gamma_{a})_{\alpha \beta} Y \nabla^{\beta \rho}{\lambda_{i \rho}}%
 - \frac{27}{64}(\Gamma_{a})^{\beta}{}_{\lambda} \nabla_{\alpha}\,^{\rho}{W_{\beta \rho}} \nabla^{\lambda \gamma}{\lambda_{i \gamma}}+\frac{63}{320}(\Gamma_{a})_{\alpha \lambda} W^{\beta \rho} W_{\beta \rho} \nabla^{\lambda \gamma}{\lambda_{i \gamma}} - \frac{9}{32}(\Gamma_{a})_{\rho \lambda} \nabla^{\rho \beta}{W_{\alpha \beta}} \nabla^{\lambda \gamma}{\lambda_{i \gamma}} - \frac{9}{32}{\rm i} (\Gamma_{a})^{\rho}{}_{\lambda} X_{i \rho} \nabla^{\lambda \beta}{F_{\alpha \beta}}+\frac{9}{64}{\rm i} (\Gamma_{a})^{\beta}{}_{\rho} X_{i \beta} \nabla^{\rho}\,_{\lambda}{\nabla_{\alpha}\,^{\lambda}{W}} - \frac{333}{640}{\rm i} (\Gamma_{a})_{\alpha}{}^{\lambda} W^{\beta \rho} F_{\beta \rho} X_{i \lambda} - \frac{45}{64}(\Gamma_{a})^{\beta}{}_{\lambda} \nabla^{\lambda \rho}{W_{\beta \rho}} \nabla_{\alpha}\,^{\gamma}{\lambda_{i \gamma}} - \frac{207}{256}(\Gamma_{a})^{\rho}{}_{\gamma} W_{\alpha}\,^{\beta} \lambda_{i \beta} \nabla^{\gamma \lambda}{W_{\rho \lambda}} - \frac{9}{16}{\rm i} (\Gamma_{a})^{\beta}{}_{\lambda} X_{i j} \nabla^{\lambda \rho}{W_{\alpha \beta \rho}\,^{j}}+\frac{3}{32}{\rm i} (\Gamma_{a})^{\beta}{}_{\rho} X_{i j} \nabla^{\rho}\,_{\alpha}{X^{j}_{\beta}}+\frac{9}{8}{\rm i} (\Gamma_{a})^{\beta}{}_{\gamma} \nabla^{\gamma \rho}{W} \nabla_{\alpha}\,^{\lambda}{W_{\beta \rho \lambda i}} - \frac{9}{32}{\rm i} (\Gamma_{a})^{\beta}{}_{\gamma} W_{\beta}\,^{\rho} W_{\alpha \rho \lambda i} \nabla^{\gamma \lambda}{W}+\frac{81}{160}{\rm i} (\Gamma_{a})_{\alpha \gamma} W^{\beta \rho} W_{\beta \rho \lambda i} \nabla^{\gamma \lambda}{W} - \frac{9}{16}{\rm i} (\Gamma_{a})^{\beta}{}_{\lambda} \nabla^{\lambda}\,_{\gamma}{W} \nabla^{\gamma \rho}{W_{\alpha \beta \rho i}}+\frac{9}{16}{\rm i} (\Gamma_{a})_{\lambda \gamma} \nabla^{\lambda \beta}{W} \nabla^{\gamma \rho}{W_{\alpha \beta \rho i}}+\frac{9}{64}{\rm i} (\Gamma_{a})^{\beta}{}_{\rho} \nabla^{\rho}\,_{\lambda}{W} \nabla_{\alpha}\,^{\lambda}{X_{i \beta}} - \frac{9}{64}{\rm i} (\Gamma_{a})_{\rho \lambda} \nabla^{\rho \beta}{W} \nabla^{\lambda}\,_{\alpha}{X_{i \beta}}+\frac{63}{320}{\rm i} (\Gamma_{a})_{\alpha \rho} \nabla^{\rho}\,_{\lambda}{W} \nabla^{\lambda \beta}{X_{i \beta}} - \frac{129}{160}\Phi^{\lambda \beta}\,_{i j} (\Gamma_{a})_{\alpha \lambda} W_{\beta}\,^{\rho} \lambda^{j}_{\rho}+\frac{27}{32}{\rm i} (\Gamma_{a})^{\rho}{}_{\gamma} F_{\alpha}\,^{\beta} \nabla^{\gamma \lambda}{W_{\rho \beta \lambda i}}%
+\frac{9}{16}{\rm i} (\Gamma_{a})^{\beta}{}_{\gamma} \nabla_{\alpha}\,^{\rho}{W} \nabla^{\gamma \lambda}{W_{\beta \rho \lambda i}}+\frac{9}{8}(\Gamma_{a})^{\beta}{}_{\gamma} \lambda^{\rho}_{i} \nabla^{\gamma \lambda}{C_{\alpha \beta \rho \lambda}} - \frac{9}{128}(\Gamma_{a})^{\rho}{}_{\gamma} W_{\alpha \beta} \lambda^{\lambda}_{i} \nabla^{\gamma \beta}{W_{\rho \lambda}}+\frac{9}{128}(\Gamma_{a})^{\lambda}{}_{\gamma} W^{\beta}\,_{\rho} \lambda_{i \beta} \nabla^{\gamma \rho}{W_{\alpha \lambda}} - \frac{9}{64}(\Gamma_{a})^{\beta}{}_{\lambda} \lambda^{\rho}_{i} \nabla^{\lambda}\,_{\gamma}{\nabla_{\alpha}\,^{\gamma}{W_{\beta \rho}}} - \frac{9}{64}(\Gamma_{a})^{\beta}{}_{\lambda} \lambda_{i \gamma} \nabla^{\lambda \rho}{\nabla_{\alpha}\,^{\gamma}{W_{\beta \rho}}} - \frac{9}{128}(\Gamma_{a})^{\beta}{}_{\lambda} \lambda_{i \gamma} \nabla^{\lambda}\,_{\alpha}{\nabla^{\gamma \rho}{W_{\beta \rho}}} - \frac{27}{320}(\Gamma_{a})_{\alpha \lambda} \lambda_{i \gamma} \nabla^{\lambda \beta}{\nabla^{\gamma \rho}{W_{\beta \rho}}} - \frac{99}{320}(\Gamma_{a})_{\alpha \lambda} \lambda^{\beta}_{i} \nabla^{\lambda}\,_{\gamma}{\nabla^{\gamma \rho}{W_{\beta \rho}}} - \frac{9}{128}(\Gamma_{a})^{\beta}{}_{\lambda} \lambda_{i \alpha} \nabla^{\lambda}\,_{\gamma}{\nabla^{\gamma \rho}{W_{\beta \rho}}}+\frac{9}{128}(\Gamma_{a})_{\lambda \gamma} \lambda_{i \alpha} \nabla^{\lambda \beta}{\nabla^{\gamma \rho}{W_{\beta \rho}}}+\frac{9}{10}(\Gamma_{a})_{\alpha}{}^{\beta} C_{\beta}\,^{\rho \lambda \gamma} W_{\rho \lambda} \lambda_{i \gamma}+\frac{99}{128}(\Gamma_{a})_{\alpha}{}^{\beta} W_{\beta}\,^{\rho} W^{\lambda \gamma} W_{\lambda \gamma} \lambda_{i \rho}+\frac{333}{640}(\Gamma_{a})_{\alpha \gamma} W^{\beta}\,_{\rho} \lambda^{\lambda}_{i} \nabla^{\gamma \rho}{W_{\beta \lambda}} - \frac{207}{640}(\Gamma_{a})_{\alpha}{}^{\lambda} W^{\beta}\,_{\rho} \lambda_{i \gamma} \nabla^{\rho \gamma}{W_{\lambda \beta}} - \frac{27}{256}(\Gamma_{a})_{\alpha}{}^{\lambda} W^{\beta}\,_{\rho} \lambda_{i \beta} \nabla^{\rho \gamma}{W_{\lambda \gamma}}+\frac{351}{1280}(\Gamma_{a})_{\alpha \gamma} W^{\beta \rho} \lambda_{i \beta} \nabla^{\gamma \lambda}{W_{\rho \lambda}}+\frac{45}{16}(\Gamma_{a})^{\beta \gamma} C_{\alpha \beta}\,^{\rho \lambda} W_{\gamma \rho} \lambda_{i \lambda} - \frac{81}{128}(\Gamma_{a})^{\rho \gamma} W_{\alpha}\,^{\beta} W_{\rho}\,^{\lambda} W_{\gamma \lambda} \lambda_{i \beta}+\frac{27}{128}(\Gamma_{a})_{\lambda \gamma} W_{\alpha}\,^{\beta} \lambda^{\rho}_{i} \nabla^{\lambda \gamma}{W_{\beta \rho}}%
 - \frac{45}{128}(\Gamma_{a})^{\rho}{}_{\lambda} W_{\alpha}\,^{\beta} \lambda_{i \gamma} \nabla^{\lambda \gamma}{W_{\rho \beta}}+\frac{63}{128}(\Gamma_{a})_{\lambda \gamma} W^{\beta \rho} \lambda_{i \beta} \nabla^{\lambda \gamma}{W_{\alpha \rho}}+\frac{9}{128}{\rm i} (\Gamma_{a})_{\rho \lambda} F_{\alpha}\,^{\beta} \nabla^{\rho \lambda}{X_{i \beta}} - \frac{3}{64}{\rm i} (\Gamma_{a})_{\beta \rho} X_{i j} \nabla^{\beta \rho}{X^{j}_{\alpha}}+\frac{9}{1024}(\Gamma_{a})_{\beta \rho} \lambda_{i \alpha} \nabla^{\beta \rho}{Y}+\frac{99}{256}(\Gamma_{a})_{\lambda \gamma} \lambda^{\beta}_{i} \nabla^{\lambda \gamma}{\nabla_{\alpha}\,^{\rho}{W_{\beta \rho}}} - \frac{45}{128}(\Gamma_{a})_{\lambda \gamma} W^{\beta \rho} \lambda_{i \alpha} \nabla^{\lambda \gamma}{W_{\beta \rho}} - \frac{45}{256}(\Gamma_{a})_{\rho \lambda} \lambda_{i \gamma} \nabla^{\rho \lambda}{\nabla^{\gamma \beta}{W_{\alpha \beta}}}+\frac{1}{4}(\Gamma_{a})^{\beta}{}_{\rho} \lambda^{\lambda}_{j} \nabla^{\rho}\,_{\alpha}{\Phi_{\beta \lambda i}\,^{j}} - \frac{7}{80}(\Gamma_{a})_{\alpha \beta} \lambda^{\lambda}_{j} \nabla^{\beta \rho}{\Phi_{\rho \lambda i}\,^{j}} - \frac{3}{80}{\rm i} (\Gamma_{a})_{\alpha}{}^{\beta} \lambda^{\gamma}_{j} W_{\beta}\,^{\rho \lambda j} W_{\gamma \rho \lambda i} - \frac{1}{32}(\Gamma_{a})_{\beta \rho} \lambda^{\lambda}_{j} \nabla^{\beta \rho}{\Phi_{\alpha \lambda i}\,^{j}} - \frac{1}{8}\Phi_{\lambda}\,^{\beta}\,_{i j} (\Gamma_{a})_{\beta}{}^{\rho} \nabla^{\lambda}\,_{\alpha}{\lambda^{j}_{\rho}}+\frac{15}{32}\Phi^{\lambda \beta}\,_{i j} (\Gamma_{a})_{\lambda}{}^{\rho} W_{\beta \rho} \lambda^{j}_{\alpha} - \frac{1}{8}\Phi_{\rho}\,^{\beta}\,_{i j} (\Gamma_{a})_{\alpha \beta} \nabla^{\rho \lambda}{\lambda^{j}_{\lambda}}+\frac{45}{32}{\rm i} (\Gamma_{a})^{\beta \lambda} F_{\beta}\,^{\rho} \nabla_{\alpha}\,^{\gamma}{W_{\lambda \rho \gamma i}} - \frac{9}{64}{\rm i} (\Gamma_{a})^{\lambda \beta} W_{\lambda}\,^{\gamma} F_{\beta}\,^{\rho} W_{\alpha \gamma \rho i}+\frac{27}{32}{\rm i} (\Gamma_{a})^{\beta \gamma} W^{\rho \lambda} F_{\beta \rho} W_{\alpha \gamma \lambda i} - \frac{81}{320}{\rm i} (\Gamma_{a})_{\alpha}{}^{\beta} W^{\lambda \gamma} F_{\beta}\,^{\rho} W_{\lambda \gamma \rho i} - \frac{9}{16}{\rm i} (\Gamma_{a})^{\beta}{}_{\gamma} F_{\beta}\,^{\rho} \nabla^{\gamma \lambda}{W_{\alpha \rho \lambda i}}%
 - \frac{9}{16}{\rm i} (\Gamma_{a})^{\beta \lambda} F_{\beta \rho} \nabla^{\rho \gamma}{W_{\alpha \lambda \gamma i}}+\frac{9}{32}{\rm i} (\Gamma_{a})^{\beta}{}_{\lambda} F_{\beta}\,^{\rho} \nabla^{\lambda}\,_{\alpha}{X_{i \rho}}+\frac{9}{64}{\rm i} (\Gamma_{a})^{\beta \lambda} F_{\beta \rho} \nabla_{\alpha}\,^{\rho}{X_{i \lambda}} - \frac{9}{64}{\rm i} (\Gamma_{a})_{\alpha}{}^{\beta} F_{\beta \rho} \nabla^{\rho \lambda}{X_{i \lambda}} - \frac{27}{256}(\Gamma_{a})^{\beta \lambda} W_{\beta \rho} \lambda_{i \alpha} \nabla^{\rho \gamma}{W_{\lambda \gamma}} - \frac{81}{256}(\Gamma_{a})^{\beta}{}_{\gamma} W_{\beta}\,^{\rho} \lambda_{i \alpha} \nabla^{\gamma \lambda}{W_{\rho \lambda}} - \frac{7}{16}(\Gamma_{a})^{\beta \rho} \lambda_{j \beta} \nabla_{\alpha}\,^{\lambda}{\Phi_{\lambda \rho i}\,^{j}} - \frac{45}{64}(\Gamma_{a})^{\lambda}{}_{\gamma} \lambda_{i \lambda} \nabla_{\alpha}\,^{\beta}{\nabla^{\gamma \rho}{W_{\beta \rho}}} - \frac{27}{128}(\Gamma_{a})^{\lambda \beta} \lambda_{i \lambda} \nabla_{\alpha \gamma}{\nabla^{\gamma \rho}{W_{\beta \rho}}}+\frac{9}{16}{\rm i} (\Gamma_{a})^{\beta}{}_{\gamma} W \nabla_{\alpha}\,^{\rho}{\nabla^{\gamma \lambda}{W_{\beta \rho \lambda i}}}+\frac{9}{64}{\rm i} (\Gamma_{a})_{\rho \lambda} W \nabla_{\alpha}\,^{\beta}{\nabla^{\rho \lambda}{X_{i \beta}}}+\frac{9}{16}{\rm i} (\Gamma_{a})^{\beta}{}_{\gamma} W W_{\beta}\,^{\rho} \nabla^{\gamma \lambda}{W_{\alpha \rho \lambda i}}+\frac{9}{128}(\Gamma_{a})_{\lambda \gamma} W^{\beta \rho} W_{\beta \rho} \nabla^{\lambda \gamma}{\lambda_{i \alpha}}+\frac{27}{128}{\rm i} (\Gamma_{a})_{\lambda \gamma} W W^{\beta \rho} \nabla^{\lambda \gamma}{W_{\alpha \beta \rho i}}+\frac{9}{64}(\Gamma_{a})_{\rho \lambda} \nabla^{\rho}\,_{\gamma}{W_{\alpha}\,^{\beta}} \nabla^{\lambda \gamma}{\lambda_{i \beta}}+\frac{9}{64}(\Gamma_{a})_{\rho \lambda} \nabla^{\rho \gamma}{W_{\alpha \beta}} \nabla^{\lambda \beta}{\lambda_{i \gamma}} - \frac{9}{128}(\Gamma_{a})_{\lambda \gamma} \nabla_{\alpha}\,^{\rho}{W^{\beta}\,_{\rho}} \nabla^{\lambda \gamma}{\lambda_{i \beta}} - \frac{3}{4}{\rm i} (\Gamma_{a})^{\beta \lambda} \lambda^{\gamma}_{j} W_{\alpha \beta}\,^{\rho j} W_{\lambda \gamma \rho i}+\frac{3}{8}{\rm i} (\Gamma_{a})^{\beta \lambda} \lambda^{\gamma}_{i} W_{\alpha \beta}\,^{\rho}\,_{j} W_{\lambda \gamma \rho}\,^{j}+\frac{9}{64}(\Gamma_{a})^{\beta}{}_{\lambda} \nabla^{\lambda}\,_{\alpha}{W_{\beta \rho}} \nabla^{\rho \gamma}{\lambda_{i \gamma}}%
+\frac{9}{64}(\Gamma_{a})^{\beta}{}_{\rho} \nabla^{\rho}\,_{\gamma}{W_{\alpha \beta}} \nabla^{\gamma \lambda}{\lambda_{i \lambda}} - \frac{9}{80}(\Gamma_{a})_{\alpha \lambda} \nabla^{\lambda \beta}{W_{\beta \rho}} \nabla^{\rho \gamma}{\lambda_{i \gamma}}+\frac{9}{16}{\rm i} (\Gamma_{a})^{\lambda \beta} W_{\alpha \lambda \gamma i} \nabla^{\gamma \rho}{F_{\beta \rho}} - \frac{9}{32}{\rm i} (\Gamma_{a})^{\beta}{}_{\lambda} W_{\alpha \beta \rho i} \nabla_{\gamma}\,^{\rho}{\nabla^{\lambda \gamma}{W}} - \frac{9}{64}{\rm i} (\Gamma_{a})^{\lambda}{}_{\gamma} F^{\beta \rho} \nabla^{\gamma}\,_{\alpha}{W_{\lambda \beta \rho i}}+\frac{9}{128}{\rm i} (\Gamma_{a})_{\lambda \gamma} F^{\beta \rho} \nabla^{\lambda \gamma}{W_{\alpha \beta \rho i}}+\frac{9}{32}{\rm i} (\Gamma_{a})^{\lambda}{}_{\gamma} F^{\beta}\,_{\rho} \nabla^{\gamma \rho}{W_{\alpha \lambda \beta i}} - \frac{81}{128}{\rm i} (\Gamma_{a})_{\alpha}{}^{\lambda} W_{\lambda}\,^{\beta} F_{\beta}\,^{\rho} X_{i \rho}+\frac{45}{128}{\rm i} (\Gamma_{a})^{\lambda \beta} W_{\lambda}\,^{\rho} F_{\beta \rho} X_{i \alpha} - \frac{9}{64}{\rm i} (\Gamma_{a})_{\alpha \gamma} F^{\beta \rho} \nabla^{\gamma \lambda}{W_{\beta \rho \lambda i}} - \frac{9}{32}{\rm i} (\Gamma_{a})_{\alpha}{}^{\lambda} F^{\beta}\,_{\rho} \nabla^{\rho \gamma}{W_{\lambda \beta \gamma i}} - \frac{9}{64}(\Gamma_{a})^{\beta}{}_{\gamma} W_{\beta}\,^{\rho} \lambda^{\lambda}_{i} \nabla^{\gamma}\,_{\alpha}{W_{\rho \lambda}} - \frac{9}{128}(\Gamma_{a})^{\lambda}{}_{\gamma} W^{\beta \rho} \lambda_{i \beta} \nabla^{\gamma}\,_{\alpha}{W_{\lambda \rho}}+\frac{9}{64}(\Gamma_{a})^{\beta}{}_{\lambda} W_{\beta}\,^{\rho} \lambda_{i \gamma} \nabla^{\lambda \gamma}{W_{\alpha \rho}} - \frac{9}{64}{\rm i} (\Gamma_{a})_{\alpha \lambda} F^{\beta}\,_{\rho} \nabla^{\lambda \rho}{X_{i \beta}} - \frac{9}{64}{\rm i} (\Gamma_{a})^{\beta}{}_{\lambda} F_{\beta \rho} \nabla^{\lambda \rho}{X_{i \alpha}} - \frac{459}{1280}(\Gamma_{a})_{\alpha}{}^{\beta} W_{\beta}\,^{\rho} \lambda_{i \gamma} \nabla^{\gamma \lambda}{W_{\rho \lambda}} - \frac{27}{320}{\rm i} (\Gamma_{a})_{\alpha}{}^{\lambda} W_{\lambda}\,^{\gamma} F^{\beta \rho} W_{\gamma \beta \rho i}+\frac{27}{64}(\Gamma_{a})^{\beta}{}_{\gamma} W_{\beta}\,^{\rho} W_{\rho}\,^{\lambda} \nabla^{\gamma}\,_{\alpha}{\lambda_{i \lambda}} - \frac{9}{32}{\rm i} (\Gamma_{a})^{\beta}{}_{\lambda} X_{i}^{\rho} \nabla^{\lambda}\,_{\alpha}{F_{\beta \rho}}%
+\frac{9}{64}{\rm i} (\Gamma_{a})_{\rho \lambda} X_{i \beta} \nabla^{\rho}\,_{\alpha}{\nabla^{\lambda \beta}{W}} - \frac{27}{64}(\Gamma_{a})^{\beta}{}_{\gamma} W_{\beta}\,^{\rho} W_{\rho \lambda} \nabla^{\gamma \lambda}{\lambda_{i \alpha}}+\frac{9}{40}{\rm i} (\Gamma_{a})_{\alpha \rho} X_{i \beta} \nabla_{\lambda}\,^{\beta}{\nabla^{\rho \lambda}{W}} - \frac{27}{64}(\Gamma_{a})^{\beta \lambda} W_{\beta}\,^{\rho} W_{\lambda \rho} \nabla_{\alpha}\,^{\gamma}{\lambda_{i \gamma}}+\frac{63}{64}{\rm i} (\Gamma_{a})^{\rho \gamma} W_{\rho}\,^{\lambda} F_{\alpha}\,^{\beta} W_{\gamma \lambda \beta i}+\frac{9}{16}{\rm i} (\Gamma_{a})^{\lambda \beta} W_{\lambda}\,^{\rho}\,_{\gamma i} \nabla_{\alpha}\,^{\gamma}{F_{\beta \rho}}+\frac{45}{32}{\rm i} (\Gamma_{a})^{\beta \lambda} W_{\beta}\,^{\rho} W_{\lambda \rho \gamma i} \nabla_{\alpha}\,^{\gamma}{W}+\frac{9}{32}{\rm i} (\Gamma_{a})^{\beta}{}_{\gamma} W_{\beta \rho \lambda i} \nabla_{\alpha}\,^{\rho}{\nabla^{\gamma \lambda}{W}}+\frac{9}{40}{\rm i} (\Gamma_{a})_{\alpha}{}^{\beta} \lambda^{\gamma}_{j} W_{\beta}\,^{\rho \lambda}\,_{i} W_{\gamma \rho \lambda}\,^{j} - \frac{27}{64}(\Gamma_{a})^{\lambda}{}_{\gamma} \nabla^{\gamma \beta}{W_{\beta \rho}} \nabla_{\alpha}\,^{\rho}{\lambda_{i \lambda}} - \frac{27}{64}(\Gamma_{a})^{\beta \gamma} W_{\beta}\,^{\rho} W_{\rho \lambda} \nabla_{\alpha}\,^{\lambda}{\lambda_{i \gamma}} - \frac{9}{64}(\Gamma_{a})^{\beta \lambda} \nabla_{\gamma}\,^{\rho}{W_{\beta \rho}} \nabla_{\alpha}\,^{\gamma}{\lambda_{i \lambda}}+\frac{9}{64}{\rm i} (\Gamma_{a})_{\rho \lambda} X_{i \beta} \nabla_{\alpha}\,^{\beta}{\nabla^{\rho \lambda}{W}}+\frac{9}{64}(\Gamma_{a})_{\lambda \gamma} W^{\beta}\,_{\rho} \nabla_{\alpha}\,^{\rho}{\nabla^{\lambda \gamma}{\lambda_{i \beta}}} - \frac{9}{32}(\Gamma_{a})^{\lambda}{}_{\gamma} W_{\beta \rho} \nabla_{\alpha}\,^{\beta}{\nabla^{\gamma \rho}{\lambda_{i \lambda}}} - \frac{9}{32}(\Gamma_{a})^{\beta}{}_{\lambda} W_{\beta \rho} \nabla_{\alpha}\,^{\rho}{\nabla^{\lambda \gamma}{\lambda_{i \gamma}}} - \frac{27}{32}(\Gamma_{a})^{\rho \gamma} W_{\alpha \beta} W_{\rho \lambda} \nabla^{\beta \lambda}{\lambda_{i \gamma}}+\frac{153}{320}(\Gamma_{a})_{\alpha}{}^{\gamma} W^{\beta}\,_{\rho} W_{\beta \lambda} \nabla^{\rho \lambda}{\lambda_{i \gamma}}+\frac{21}{16}{\rm i} (\Gamma_{a})^{\beta \lambda} \lambda^{\gamma}_{j} W_{\alpha \beta}\,^{\rho}\,_{i} W_{\lambda \gamma \rho}\,^{j}+\frac{27}{640}(\Gamma_{a})_{\alpha \lambda} W^{\beta \rho} \lambda_{i \gamma} \nabla^{\lambda \gamma}{W_{\beta \rho}}%
 - \frac{9}{32}{\rm i} (\Gamma_{a})^{\beta}{}_{\lambda} X_{i \alpha} \nabla^{\lambda \rho}{F_{\beta \rho}}+\frac{9}{64}{\rm i} (\Gamma_{a})_{\beta \rho} X_{i \alpha} \nabla^{\beta}\,_{\lambda}{\nabla^{\rho \lambda}{W}}+\frac{9}{32}(\Gamma_{a})^{\rho}{}_{\lambda} W_{\alpha \beta} \nabla_{\gamma}\,^{\beta}{\nabla^{\lambda \gamma}{\lambda_{i \rho}}}+\frac{9}{64}(\Gamma_{a})_{\rho \lambda} W_{\alpha \beta} \nabla^{\beta \gamma}{\nabla^{\rho \lambda}{\lambda_{i \gamma}}} - \frac{9}{32}(\Gamma_{a})_{\rho \lambda} W_{\alpha \beta} \nabla^{\rho \beta}{\nabla^{\lambda \gamma}{\lambda_{i \gamma}}}+\frac{27}{64}(\Gamma_{a})_{\lambda \gamma} W_{\alpha}\,^{\beta} W_{\beta}\,^{\rho} \nabla^{\lambda \gamma}{\lambda_{i \rho}} - \frac{9}{64}(\Gamma_{a})_{\rho \lambda} \lambda^{\beta}_{i} \nabla^{\rho}\,_{\gamma}{\nabla^{\lambda \gamma}{W_{\alpha \beta}}} - \frac{9}{64}(\Gamma_{a})_{\rho \lambda} \lambda_{i \gamma} \nabla^{\rho \beta}{\nabla^{\lambda \gamma}{W_{\alpha \beta}}}+\frac{9}{128}(\Gamma_{a})^{\beta}{}_{\lambda} \lambda^{\rho}_{i} \nabla_{\alpha \gamma}{\nabla^{\lambda \gamma}{W_{\beta \rho}}} - \frac{9}{128}(\Gamma_{a})_{\lambda \gamma} \lambda^{\beta}_{i} \nabla_{\alpha}\,^{\rho}{\nabla^{\lambda \gamma}{W_{\beta \rho}}}+\frac{9}{128}(\Gamma_{a})^{\beta}{}_{\lambda} \lambda_{i \gamma} \nabla_{\alpha}\,^{\rho}{\nabla^{\lambda \gamma}{W_{\beta \rho}}}+\frac{9}{128}(\Gamma_{a})^{\beta}{}_{\lambda} \lambda_{i \gamma} \nabla^{\gamma \rho}{\nabla^{\lambda}\,_{\alpha}{W_{\beta \rho}}}+\frac{9}{128}(\Gamma_{a})^{\beta}{}_{\rho} \lambda_{i \lambda} \nabla_{\gamma}\,^{\lambda}{\nabla^{\rho \gamma}{W_{\alpha \beta}}}+\frac{9}{128}(\Gamma_{a})_{\rho \lambda} \lambda_{i \gamma} \nabla^{\gamma \beta}{\nabla^{\rho \lambda}{W_{\alpha \beta}}} - \frac{27}{640}(\Gamma_{a})_{\alpha}{}^{\beta} \lambda_{i \lambda} \nabla_{\gamma}\,^{\lambda}{\nabla^{\gamma \rho}{W_{\beta \rho}}}+\frac{9}{40}(\Gamma_{a})_{\alpha \lambda} \lambda_{i \gamma} \nabla^{\gamma \beta}{\nabla^{\lambda \rho}{W_{\beta \rho}}}+\frac{9}{128}(\Gamma_{a})_{\rho \lambda} \lambda_{i \gamma} \nabla^{\rho \gamma}{\nabla^{\lambda \beta}{W_{\alpha \beta}}}+\frac{3}{80}(\Gamma_{a})_{\alpha}{}^{\beta} \lambda_{j \lambda} \nabla^{\lambda \rho}{\Phi_{\rho \beta i}\,^{j}} - \frac{3}{16}(\Gamma_{a})^{\beta}{}_{\rho} \lambda_{j \lambda} \nabla^{\rho \lambda}{\Phi_{\alpha \beta i}\,^{j}} - \frac{3}{16}{\rm i} (\Gamma_{a})^{\gamma \lambda} \lambda_{j \gamma} W_{\alpha}\,^{\beta \rho}\,_{i} W_{\lambda \beta \rho}\,^{j}%
+\frac{9}{64}{\rm i} (\Gamma_{a})_{\beta \rho} \nabla^{\beta}\,_{\lambda}{W} \nabla^{\rho \lambda}{X_{i \alpha}}+\frac{9}{512}(\Gamma_{a})_{\alpha \beta} \lambda_{i \rho} \nabla^{\beta \rho}{Y} - \frac{9}{64}{\rm i} (\Gamma_{a})_{\rho \lambda} \nabla^{\rho \lambda}{W} \nabla_{\alpha}\,^{\beta}{X_{i \beta}} - \frac{27}{320}(\Gamma_{a})_{\alpha}{}^{\lambda} \lambda_{i \lambda} \nabla_{\gamma}\,^{\beta}{\nabla^{\gamma \rho}{W_{\beta \rho}}}+\frac{9}{80}{\rm i} (\Gamma_{a})_{\alpha}{}^{\beta} W \nabla_{\gamma}\,^{\rho}{\nabla^{\gamma \lambda}{W_{\beta \rho \lambda i}}}+\frac{27}{160}{\rm i} (\Gamma_{a})_{\alpha \rho} W \nabla_{\lambda}\,^{\beta}{\nabla^{\rho \lambda}{X_{i \beta}}} - \frac{9}{160}{\rm i} (\Gamma_{a})_{\alpha}{}^{\beta} W \nabla_{\rho \lambda}{\nabla^{\rho \lambda}{X_{i \beta}}}+\frac{3}{40}{\rm i} (\Gamma_{a})_{\alpha}{}^{\gamma} \lambda_{j \gamma} W^{\beta \rho \lambda}\,_{i} W_{\beta \rho \lambda}\,^{j} - \frac{9}{20}{\rm i} (\Gamma_{a})_{\alpha}{}^{\beta} W C_{\beta}\,^{\rho \lambda \gamma} W_{\rho \lambda \gamma i}+\frac{27}{320}(\Gamma_{a})_{\alpha \lambda} \nabla^{\lambda}\,_{\gamma}{W^{\beta}\,_{\rho}} \nabla^{\gamma \rho}{\lambda_{i \beta}} - \frac{9}{320}(\Gamma_{a})_{\alpha}{}^{\beta} \nabla_{\lambda \gamma}{W_{\beta}\,^{\rho}} \nabla^{\lambda \gamma}{\lambda_{i \rho}} - \frac{9}{320}(\Gamma_{a})_{\alpha}{}^{\beta} \nabla_{\gamma}\,^{\lambda}{W_{\beta \rho}} \nabla^{\gamma \rho}{\lambda_{i \lambda}} - \frac{9}{320}(\Gamma_{a})_{\alpha}{}^{\lambda} \nabla_{\gamma}\,^{\beta}{W_{\beta \rho}} \nabla^{\gamma \rho}{\lambda_{i \lambda}}+\frac{9}{80}{\rm i} (\Gamma_{a})_{\alpha}{}^{\lambda} W_{\lambda}\,^{\beta}\,_{\gamma i} \nabla^{\gamma \rho}{F_{\beta \rho}}+\frac{9}{160}{\rm i} (\Gamma_{a})_{\alpha}{}^{\beta} W_{\beta \rho \lambda i} \nabla_{\gamma}\,^{\rho}{\nabla^{\gamma \lambda}{W}} - \frac{9}{80}(\Gamma_{a})_{\alpha \gamma} W^{\beta \rho} W_{\beta \lambda} \nabla^{\gamma \lambda}{\lambda_{i \rho}}+\frac{9}{80}(\Gamma_{a})_{\alpha}{}^{\beta} W_{\beta \rho} \nabla_{\gamma}\,^{\rho}{\nabla^{\gamma \lambda}{\lambda_{i \lambda}}}+\frac{9}{32}(\Gamma_{a})_{\alpha}{}^{\beta} W_{\beta \rho} W^{\lambda}\,_{\gamma} \nabla^{\rho \gamma}{\lambda_{i \lambda}}+\frac{9}{160}(\Gamma_{a})_{\alpha}{}^{\beta} \nabla_{\gamma}\,^{\rho}{W_{\beta \rho}} \nabla^{\gamma \lambda}{\lambda_{i \lambda}}+\frac{9}{32}{\rm i} (\Gamma_{a})_{\alpha \lambda} X_{i}^{\beta} \nabla^{\lambda \rho}{F_{\beta \rho}}%
+\frac{63}{320}{\rm i} (\Gamma_{a})_{\alpha \rho} X_{i \beta} \nabla^{\rho}\,_{\lambda}{\nabla^{\lambda \beta}{W}} - \frac{9}{320}{\rm i} (\Gamma_{a})_{\alpha}{}^{\beta} X_{i \beta} \nabla_{\rho \lambda}{\nabla^{\rho \lambda}{W}} - \frac{9}{160}{\rm i} (\Gamma_{a})_{\alpha \rho} \nabla_{\lambda}\,^{\beta}{W} \nabla^{\rho \lambda}{X_{i \beta}} - \frac{9}{320}{\rm i} (\Gamma_{a})_{\alpha}{}^{\beta} \nabla_{\rho \lambda}{W} \nabla^{\rho \lambda}{X_{i \beta}} - \frac{9}{160}(\Gamma_{a})_{\alpha \lambda} \lambda^{\beta}_{i} \nabla_{\gamma}\,^{\rho}{\nabla^{\lambda \gamma}{W_{\beta \rho}}}+\frac{9}{160}(\Gamma_{a})_{\alpha}{}^{\beta} \lambda^{\rho}_{i} \nabla_{\lambda \gamma}{\nabla^{\lambda \gamma}{W_{\beta \rho}}}+\frac{9}{160}(\Gamma_{a})_{\alpha}{}^{\beta} \lambda_{i \lambda} \nabla_{\gamma}\,^{\rho}{\nabla^{\gamma \lambda}{W_{\beta \rho}}} - \frac{9}{40}(\Gamma_{a})_{\alpha}{}^{\lambda} W_{\beta \rho} \nabla_{\gamma}\,^{\beta}{\nabla^{\gamma \rho}{\lambda_{i \lambda}}}+\frac{9}{512}(\Gamma_{a})_{\beta \rho} Y \nabla^{\beta \rho}{\lambda_{i \alpha}} - \frac{9}{32}{\rm i} (\Gamma_{a})_{\rho \lambda} W \nabla^{\rho \lambda}{\nabla_{\alpha}\,^{\beta}{X_{i \beta}}}+\frac{27}{64}{\rm i} (\Gamma_{a})_{\lambda \gamma} W^{\beta \rho} W_{\alpha \beta \rho i} \nabla^{\lambda \gamma}{W}+\frac{27}{64}{\rm i} (\Gamma_{a})_{\lambda \gamma} W W_{\alpha}\,^{\beta \rho}\,_{i} \nabla^{\lambda \gamma}{W_{\beta \rho}}+\frac{9}{64}{\rm i} (\Gamma_{a})_{\lambda \gamma} W_{\alpha}\,^{\beta \rho}\,_{i} \nabla^{\lambda \gamma}{F_{\beta \rho}}+\frac{9}{64}{\rm i} (\Gamma_{a})_{\rho \lambda} X_{i}^{\beta} \nabla^{\rho \lambda}{F_{\alpha \beta}} - \frac{9}{64}(\Gamma_{a})_{\lambda \gamma} \nabla^{\lambda \gamma}{W^{\beta}\,_{\rho}} \nabla_{\alpha}\,^{\rho}{\lambda_{i \beta}} - \frac{9}{64}(\Gamma_{a})_{\lambda \gamma} W^{\beta}\,_{\rho} \nabla^{\lambda \gamma}{\nabla_{\alpha}\,^{\rho}{\lambda_{i \beta}}} - \frac{9}{64}(\Gamma_{a})_{\rho \lambda} W_{\alpha \beta} \nabla^{\rho \lambda}{\nabla^{\beta \gamma}{\lambda_{i \gamma}}} - \frac{9}{64}{\rm i} (\Gamma_{a})_{\rho \lambda} X_{i \beta} \nabla^{\rho \lambda}{\nabla_{\alpha}\,^{\beta}{W}} - \frac{3}{16}\Phi_{\alpha}\,^{\lambda}\,_{i j} (\Gamma_{a})_{\beta \rho} \nabla^{\beta \rho}{\lambda^{j}_{\lambda}}+\frac{9}{128}(\Gamma_{a})_{\rho \lambda} \nabla^{\beta \gamma}{W_{\alpha \beta}} \nabla^{\rho \lambda}{\lambda_{i \gamma}}%
+\frac{9}{80}(\Gamma_{a})_{\alpha \lambda} W^{\beta}\,_{\rho} \nabla^{\lambda}\,_{\gamma}{\nabla^{\gamma \rho}{\lambda_{i \beta}}}
\doublespacedmathend
\end{adjustwidth}

\subsubsection{$J^4_{a b, {\rm Weyl}}$ Degauged and Gauge Fixed Bosons} \label{J4WeylComplete}

\begin{adjustwidth}{0cm}{5cm}
\doublespacedmathbegin
{} - \frac{9}{8}\mathcal{D}^{d}{\mathcal{D}_{\underline{a}}{W_{\underline{b}}\,^{c}}} F_{c d}+\frac{9}{8}\mathcal{D}^{d}{\mathcal{D}_{\underline{a}}{W^{c}\,_{d}}} F_{\underline{b} c}+\frac{27}{8}\mathcal{D}_{\underline{a}}{\mathcal{D}^{d}{W_{\underline{b}}\,^{c}}} F_{c d}+\frac{9}{8}\mathcal{D}_{\underline{a}}{\mathcal{D}^{d}{W^{c}\,_{d}}} F_{\underline{b} c} - \frac{9}{8}\mathcal{D}^{c}{\mathcal{D}^{d}{W_{\underline{a} d}}} F_{\underline{b} c}+\frac{81}{40}\eta_{\underline{a} \underline{b}} \mathcal{D}^{e}{\mathcal{D}^{d}{W^{c}\,_{e}}} F_{c d} - \frac{9}{16}\eta_{\underline{a} \underline{b}} \mathcal{D}_{e}{\mathcal{D}^{e}{W_{c d}}} F^{c d}+\frac{9}{16}\mathcal{D}_{\underline{a}}{\mathcal{D}_{\underline{b}}{W_{c d}}} F^{c d}+\frac{9}{4}\mathcal{D}_{d}{\mathcal{D}^{d}{W_{\underline{a}}\,^{c}}} F_{\underline{b} c} - \frac{45}{8}\mathcal{D}^{d}{\mathcal{D}^{c}{W_{\underline{a} d}}} F_{\underline{b} c}+\frac{9}{40}\eta_{\underline{a} \underline{b}} \mathcal{D}^{d}{\mathcal{D}^{e}{W^{c}\,_{e}}} F_{c d}+\frac{9}{16}R_{\underline{a} \underline{b}} W^{c d} F_{c d} - \frac{21}{80}R \eta_{\underline{a} \underline{b}} W^{c d} F_{c d} - \frac{9}{8}R_{\underline{a} c} W_{\underline{b} d} F^{c d} - \frac{3}{16}R W_{\underline{b} c} F_{\underline{a}}\,^{c} - \frac{27}{8}R_{\underline{a} d} W^{d c} F_{\underline{b} c}+\frac{15}{16}R W_{\underline{a}}\,^{c} F_{\underline{b} c}+3R^{d c} W_{\underline{a} d} F_{\underline{b} c}+\frac{3}{8}R^{c}\,_{e} \eta_{\underline{a} \underline{b}} W^{e d} F_{c d}%
+\frac{1665}{1024}\epsilon^{e {e_{1}} {e_{2}} c d} \mathcal{D}_{\underline{a}}{W_{e {e_{1}}}} W_{\underline{b} {e_{2}}} F_{c d}+\frac{99}{512}\epsilon_{\underline{a}}\,^{e {e_{1}} {e_{2}} c} \mathcal{D}_{\underline{b}}{W_{e}\,^{d}} W_{{e_{1}} {e_{2}}} F_{c d}+\frac{585}{1024}\epsilon_{\underline{a}}\,^{d e {e_{1}} {e_{2}}} \mathcal{D}^{c}{W_{d e}} W_{{e_{1}} {e_{2}}} F_{\underline{b} c}+\frac{891}{512}\epsilon_{\underline{a} {e_{2}}}\,^{e c d} \mathcal{D}^{{e_{2}}}{W_{e}\,^{{e_{1}}}} W_{\underline{b} {e_{1}}} F_{c d} - \frac{135}{128}\epsilon_{\underline{a} {e_{2}}}\,^{{e_{1}} c d} \mathcal{D}^{{e_{2}}}{W_{\underline{b}}\,^{e}} W_{{e_{1}} e} F_{c d} - \frac{495}{256}\epsilon_{{e_{2}}}\,^{d e {e_{1}} c} \mathcal{D}^{{e_{2}}}{W_{d e}} W_{\underline{a} {e_{1}}} F_{\underline{b} c} - \frac{153}{256}\epsilon_{\underline{a}}\,^{d {e_{1}} {e_{2}} c} \mathcal{D}^{e}{W_{d e}} W_{{e_{1}} {e_{2}}} F_{\underline{b} c}+\frac{981}{512}\epsilon_{\underline{a} {e_{2}}}\,^{d e {e_{1}}} \mathcal{D}^{{e_{2}}}{W_{d}\,^{c}} W_{e {e_{1}}} F_{\underline{b} c} - \frac{891}{5120}\epsilon^{e {e_{1}} {e_{2}} {e_{3}} c} \eta_{\underline{a} \underline{b}} \mathcal{D}^{d}{W_{e {e_{1}}}} W_{{e_{2}} {e_{3}}} F_{c d}+\frac{45}{256}\epsilon^{e {e_{1}} {e_{2}} c d} \mathcal{D}_{\underline{a}}{W_{\underline{b} e}} W_{{e_{1}} {e_{2}}} F_{c d} - \frac{153}{128}\epsilon_{\underline{a}}\,^{e {e_{1}} {e_{2}} c} \mathcal{D}_{\underline{b}}{W_{e {e_{1}}}} W_{{e_{2}}}\,^{d} F_{c d}+\frac{63}{1024}\epsilon^{d e {e_{1}} {e_{2}} c} \mathcal{D}_{\underline{a}}{W_{d e}} W_{{e_{1}} {e_{2}}} F_{\underline{b} c} - \frac{495}{256}\epsilon_{\underline{a} {e_{2}}}\,^{d {e_{1}} c} \mathcal{D}^{{e_{2}}}{W_{d}\,^{e}} W_{{e_{1}} e} F_{\underline{b} c} - \frac{9}{16}W_{\underline{a}}\,^{c} F_{\underline{b} c} Y+\frac{9}{80}\eta_{\underline{a} \underline{b}} W^{c d} F_{c d} Y+\frac{27}{128}W^{d e} W_{\underline{a}}\,^{c} W_{d e} F_{\underline{b} c} - \frac{63}{80}\eta_{\underline{a} \underline{b}} W^{e {e_{1}}} W^{c d} W_{e {e_{1}}} F_{c d} - \frac{2979}{512}\eta_{\underline{a} \underline{b}} W^{e c} W^{{e_{1}} d} W_{e {e_{1}}} F_{c d}+\frac{225}{32}W^{d e} W_{\underline{a} d} W^{c}\,_{e} F_{\underline{b} c}+\frac{153}{32}W^{e d} W_{\underline{a}}\,^{c} W_{\underline{b} e} F_{c d}%
+\frac{9711}{2560}\eta_{\underline{a} \underline{b}} W^{e {e_{1}}} W_{e c} W_{{e_{1}} d} F^{c d}+\frac{9}{128}\epsilon_{\underline{a}}\,^{e {e_{1}} c d} \mathcal{D}^{{e_{2}}}{W_{e {e_{1}}}} W_{\underline{b} {e_{2}}} F_{c d} - \frac{1143}{512}\epsilon_{\underline{a} {e_{2}}}\,^{e {e_{1}} c} \mathcal{D}^{{e_{2}}}{W_{\underline{b}}\,^{d}} W_{e {e_{1}}} F_{c d}+\frac{9}{32}\epsilon_{\underline{a} {e_{2}} c}\,^{e {e_{1}}} \mathcal{D}^{{e_{2}}}{W_{e {e_{1}}}} W_{\underline{b} d} F^{c d}+\frac{99}{256}\epsilon_{{e_{2}}}\,^{e {e_{1}} c d} \mathcal{D}^{{e_{2}}}{W_{\underline{a} e}} W_{\underline{b} {e_{1}}} F_{c d} - \frac{9}{16}\epsilon_{e {e_{3}}}\,^{{e_{1}} {e_{2}} c} \eta_{\underline{a} \underline{b}} W^{e d} \mathcal{D}^{{e_{3}}}{W_{{e_{1}} {e_{2}}}} F_{c d} - \frac{2439}{2560}\epsilon_{{e_{3}}}\,^{e {e_{2}} c d} \eta_{\underline{a} \underline{b}} \mathcal{D}^{{e_{3}}}{W_{e}\,^{{e_{1}}}} W_{{e_{2}} {e_{1}}} F_{c d}+\frac{963}{512}\epsilon_{\underline{a} d {e_{2}}}\,^{e {e_{1}}} W^{d c} \mathcal{D}^{{e_{2}}}{W_{e {e_{1}}}} F_{\underline{b} c} - \frac{27}{256}\epsilon_{\underline{a} e {e_{2}}}\,^{{e_{1}} c} W^{e d} \mathcal{D}^{{e_{2}}}{W_{\underline{b} {e_{1}}}} F_{c d}+\frac{99}{512}\epsilon_{\underline{a}}\,^{e {e_{2}} c d} \mathcal{D}^{{e_{1}}}{W_{e {e_{1}}}} W_{\underline{b} {e_{2}}} F_{c d}+\frac{351}{1024}\epsilon_{\underline{a}}\,^{{e_{1}} {e_{2}} c d} \mathcal{D}^{e}{W_{\underline{b} e}} W_{{e_{1}} {e_{2}}} F_{c d}+\frac{405}{256}\epsilon_{{e_{2}}}\,^{d e {e_{1}} c} \mathcal{D}^{{e_{2}}}{W_{\underline{a} d}} W_{e {e_{1}}} F_{\underline{b} c} - \frac{729}{512}\epsilon_{{e_{3}}}\,^{e {e_{1}} {e_{2}} c} \eta_{\underline{a} \underline{b}} \mathcal{D}^{{e_{3}}}{W_{e}\,^{d}} W_{{e_{1}} {e_{2}}} F_{c d}+\frac{99}{128}\epsilon_{\underline{a} {e_{2}}}\,^{e {e_{1}} c} \mathcal{D}^{{e_{2}}}{W_{e}\,^{d}} W_{\underline{b} {e_{1}}} F_{c d}+\frac{459}{512}\epsilon_{\underline{a}}\,^{e {e_{1}} c d} \mathcal{D}^{{e_{2}}}{W_{\underline{b} e}} W_{{e_{1}} {e_{2}}} F_{c d} - \frac{27}{16}W^{c d} \mathcal{D}_{\underline{a}}{\mathcal{D}_{\underline{b}}{W_{c d}}} - \frac{63}{16}\mathcal{D}^{d}{\mathcal{D}_{\underline{a}}{W^{c}\,_{d}}} W_{\underline{b} c}+\frac{45}{16}\mathcal{D}_{\underline{a}}{\mathcal{D}^{d}{W^{c}\,_{d}}} W_{\underline{b} c} - \frac{27}{16}\mathcal{D}_{\underline{a}}{\mathcal{D}^{d}{W_{\underline{b}}\,^{c}}} W_{c d} - \frac{63}{16}\mathcal{D}^{c}{\mathcal{D}^{d}{W_{\underline{a} c}}} W_{\underline{b} d}%
 - \frac{207}{16}\mathcal{D}^{d}{\mathcal{D}^{c}{W_{\underline{a} c}}} W_{\underline{b} d} - \frac{81}{8}\mathcal{D}_{d}{\mathcal{D}^{d}{W_{\underline{a}}\,^{c}}} W_{\underline{b} c}+\frac{261}{80}\eta_{\underline{a} \underline{b}} \mathcal{D}^{e}{\mathcal{D}^{d}{W^{c}\,_{d}}} W_{c e}+\frac{171}{80}\eta_{\underline{a} \underline{b}} W^{c d} \mathcal{D}_{e}{\mathcal{D}^{e}{W_{c d}}} - \frac{27}{80}\eta_{\underline{a} \underline{b}} \mathcal{D}^{d}{\mathcal{D}^{e}{W^{c}\,_{d}}} W_{c e} - \frac{27}{32}R_{\underline{a} \underline{b}} W^{c d} W_{c d}+\frac{9}{32}R \eta_{\underline{a} \underline{b}} W^{c d} W_{c d} - \frac{45}{8}R_{\underline{a} c} W^{c d} W_{\underline{b} d} - \frac{9}{16}R W_{\underline{a}}\,^{c} W_{\underline{b} c}+\frac{33}{8}R^{c d} W_{\underline{a} c} W_{\underline{b} d} - \frac{33}{20}R^{c}\,_{d} \eta_{\underline{a} \underline{b}} W^{d e} W_{c e} - \frac{4005}{1024}\epsilon_{\underline{a}}\,^{c d {e_{1}} {e_{2}}} \mathcal{D}^{e}{W_{c d}} W_{\underline{b} e} W_{{e_{1}} {e_{2}}} - \frac{1017}{1024}\epsilon_{\underline{a} {e_{2}}}\,^{c e {e_{1}}} \mathcal{D}^{{e_{2}}}{W_{c}\,^{d}} W_{\underline{b} d} W_{e {e_{1}}} - \frac{1179}{512}\epsilon_{\underline{a} {e_{2}}}\,^{d e {e_{1}}} \mathcal{D}^{{e_{2}}}{W_{\underline{b}}\,^{c}} W_{d e} W_{{e_{1}} c} - \frac{585}{1024}\epsilon_{\underline{a}}\,^{c e {e_{1}} {e_{2}}} \mathcal{D}^{d}{W_{c d}} W_{\underline{b} e} W_{{e_{1}} {e_{2}}}+\frac{1575}{2048}\epsilon^{c d e {e_{1}} {e_{2}}} \eta_{\underline{a} \underline{b}} \mathcal{D}^{{e_{3}}}{W_{c d}} W_{e {e_{1}}} W_{{e_{2}} {e_{3}}}+\frac{5553}{2048}\epsilon^{c d e {e_{1}} {e_{2}}} \mathcal{D}_{\underline{a}}{W_{c d}} W_{\underline{b} e} W_{{e_{1}} {e_{2}}}+\frac{3879}{1024}\epsilon_{\underline{a}}\,^{c e {e_{1}} {e_{2}}} \mathcal{D}_{\underline{b}}{W_{c}\,^{d}} W_{e {e_{1}}} W_{{e_{2}} d} - \frac{189}{128}\epsilon_{\underline{a} {e_{2}}}\,^{c e {e_{1}}} \mathcal{D}^{{e_{2}}}{W_{c}\,^{d}} W_{\underline{b} e} W_{{e_{1}} d} - \frac{135}{64}W_{\underline{a}}\,^{c} W_{\underline{b} c} Y%
+\frac{27}{64}\eta_{\underline{a} \underline{b}} W^{c d} W_{c d} Y - \frac{2619}{1024}\epsilon_{\underline{a}}\,^{d e {e_{1}} {e_{2}}} \mathcal{D}^{c}{W_{\underline{b} c}} W_{d e} W_{{e_{1}} {e_{2}}} - \frac{2529}{1024}\epsilon_{{e_{2}}}\,^{c d e {e_{1}}} \mathcal{D}^{{e_{2}}}{W_{\underline{a} c}} W_{\underline{b} d} W_{e {e_{1}}}+\frac{99}{64}\epsilon_{\underline{a}}\,^{c d e {e_{1}}} \mathcal{D}^{{e_{2}}}{W_{\underline{b} c}} W_{d e} W_{{e_{1}} {e_{2}}} - \frac{261}{1024}\epsilon_{\underline{a}}\,^{c d e {e_{1}}} \mathcal{D}^{{e_{2}}}{W_{c d}} W_{\underline{b} e} W_{{e_{1}} {e_{2}}}+\frac{315}{1024}\epsilon_{\underline{a} {e_{2}}}\,^{c d {e_{1}}} \mathcal{D}^{{e_{2}}}{W_{c d}} W_{\underline{b}}\,^{e} W_{{e_{1}} e}+\frac{603}{128}W^{c d} W_{\underline{a}}\,^{e} W_{\underline{b} e} W_{c d} - \frac{297}{320}\eta_{\underline{a} \underline{b}} W^{c d} W^{e {e_{1}}} W_{c d} W_{e {e_{1}}} - \frac{243}{80}\eta_{\underline{a} \underline{b}} W^{c d} W^{e {e_{1}}} W_{c e} W_{d {e_{1}}}+\frac{495}{32}W^{c d} W_{\underline{a}}\,^{e} W_{\underline{b} c} W_{e d}+\frac{6}{5}\Phi_{c d i j} \eta_{\underline{a} \underline{b}} X^{i j} W^{c d} - \frac{9}{2}\mathcal{D}^{c}{\mathcal{D}^{d}{R_{\underline{a} c \underline{b} d}}} - \frac{135}{8}\mathcal{D}^{c}{W_{\underline{a} c}} \mathcal{D}^{d}{W_{\underline{b} d}}+\frac{27}{8}\mathcal{D}^{d}{W_{\underline{a} c}} \mathcal{D}^{c}{W_{\underline{b} d}} - \frac{27}{16}W_{\underline{a} c} \mathcal{D}^{d}{\mathcal{D}^{c}{W_{\underline{b} d}}} - \frac{27}{16}W_{\underline{a} c} \mathcal{D}^{c}{\mathcal{D}^{d}{W_{\underline{b} d}}}-27\mathcal{D}^{d}{W_{\underline{b}}\,^{c}} \mathcal{D}_{\underline{a}}{W_{c d}}+\frac{27}{8}\eta_{\underline{a} \underline{b}} \mathcal{D}^{d}{W^{c}\,_{d}} \mathcal{D}^{e}{W_{c e}} - \frac{243}{40}\eta_{\underline{a} \underline{b}} \mathcal{D}^{e}{W^{c}\,_{d}} \mathcal{D}^{d}{W_{c e}} - \frac{27}{2}\mathcal{D}_{d}{W_{\underline{a}}\,^{c}} \mathcal{D}^{d}{W_{\underline{b} c}}%
+\frac{3}{2}\mathcal{D}^{c}{\mathcal{D}^{e}{R_{c}\,^{d}\,_{e d}}} \eta_{\underline{a} \underline{b}}+\frac{9}{16}\eta_{\underline{a} \underline{b}} W^{c}\,_{d} \mathcal{D}^{d}{\mathcal{D}^{e}{W_{c e}}} - \frac{3}{8}\mathcal{D}_{e}{\mathcal{D}^{e}{R^{c d}\,_{c d}}} \eta_{\underline{a} \underline{b}}+\frac{351}{80}\eta_{\underline{a} \underline{b}} \mathcal{D}_{e}{W^{c d}} \mathcal{D}^{e}{W_{c d}} - \frac{3}{2}\mathcal{D}_{\underline{b}}{\mathcal{D}^{d}{R_{\underline{a}}\,^{c}\,_{d c}}}+\frac{9}{16}W_{\underline{b}}\,^{c} \mathcal{D}_{\underline{a}}{\mathcal{D}^{d}{W_{c d}}} - \frac{135}{16}\mathcal{D}_{\underline{a}}{W^{c d}} \mathcal{D}_{\underline{b}}{W_{c d}}+\frac{3}{8}\mathcal{D}_{\underline{a}}{\mathcal{D}_{\underline{b}}{R^{c d}\,_{c d}}}+\frac{3}{2}\mathcal{D}_{d}{\mathcal{D}^{d}{R_{\underline{a}}\,^{c}\,_{\underline{b} c}}}+\frac{9}{8}W_{\underline{a}}\,^{c} \mathcal{D}_{d}{\mathcal{D}^{d}{W_{\underline{b} c}}} - \frac{3}{2}\mathcal{D}^{d}{\mathcal{D}_{\underline{b}}{R_{\underline{a}}\,^{c}\,_{d c}}}+\frac{9}{16}W^{c}\,_{d} \mathcal{D}^{d}{\mathcal{D}_{\underline{a}}{W_{\underline{b} c}}}+\frac{9}{16}\mathcal{D}^{d}{\mathcal{D}_{\underline{a}}{W_{\underline{b}}\,^{c}}} W_{c d}+\frac{9}{16}W_{\underline{b}}\,^{c} \mathcal{D}^{d}{\mathcal{D}_{\underline{a}}{W_{c d}}}+\frac{9}{16}\eta_{\underline{a} \underline{b}} W^{c}\,_{d} \mathcal{D}^{e}{\mathcal{D}^{d}{W_{c e}}}+\frac{9}{16}W^{c}\,_{d} \mathcal{D}_{\underline{a}}{\mathcal{D}^{d}{W_{\underline{b} c}}}+\frac{1}{2}R_{\underline{a} \underline{b}} R+\frac{3}{5}R^{c d} R_{c d} \eta_{\underline{a} \underline{b}} - \frac{3}{40}\eta_{\underline{a} \underline{b}} {R}^{2} - \frac{3}{2}R_{\underline{a}}\,^{c} R_{\underline{b} c}%
 - \frac{1}{8}R R_{\underline{a} \underline{b}}+\frac{15}{8}\epsilon^{c}\,_{\underline{b} {e_{1}}}\,^{d e} \mathcal{D}^{{e_{1}}}{R_{\underline{a} c}} W_{d e} - \frac{9}{32}\epsilon_{\underline{a} e {e_{1}}}\,^{c d} \mathcal{D}^{e}{\mathcal{D}_{\underline{b}}{\mathcal{D}^{{e_{1}}}{W_{c d}}}}+\frac{81}{32}\epsilon_{\underline{a} e {e_{1}}}\,^{c d} \mathcal{D}^{e}{\mathcal{D}^{{e_{1}}}{\mathcal{D}_{\underline{b}}{W_{c d}}}}+\frac{27}{16}\epsilon_{\underline{a} d e {e_{1}}}\,^{c} \mathcal{D}^{d}{\mathcal{D}^{e}{\mathcal{D}^{{e_{1}}}{W_{\underline{b} c}}}}+\frac{9}{32}\epsilon_{\underline{a} e {e_{1}}}\,^{c d} \mathcal{D}_{\underline{b}}{\mathcal{D}^{e}{\mathcal{D}^{{e_{1}}}{W_{c d}}}} - \frac{9}{32}\epsilon_{e {e_{1}} {e_{2}}}\,^{c d} \eta_{\underline{a} \underline{b}} \mathcal{D}^{e}{\mathcal{D}^{{e_{1}}}{\mathcal{D}^{{e_{2}}}{W_{c d}}}}+\frac{9}{8}\epsilon^{d e}\,_{\underline{b}}\,^{{e_{1}} {e_{2}}} R_{\underline{a} c d e} \mathcal{D}^{c}{W_{{e_{1}} {e_{2}}}} - \frac{9}{16}\epsilon^{d e}\,_{\underline{b} {e_{2}}}\,^{{e_{1}}} R_{\underline{a}}\,^{c}\,_{d e} \mathcal{D}^{{e_{2}}}{W_{{e_{1}} c}}+\frac{9}{8}\epsilon^{c d}\,_{\underline{b} {e_{2}}}\,^{{e_{1}}} R_{\underline{a} c d}\,^{e} \mathcal{D}^{{e_{2}}}{W_{{e_{1}} e}}+\frac{3}{4}\epsilon^{d e}\,_{\underline{b}}\,^{{e_{1}} {e_{2}}} \mathcal{D}^{c}{R_{\underline{a} c d e}} W_{{e_{1}} {e_{2}}}+\frac{15}{64}\epsilon_{\underline{a}}\,^{c d {e_{1}} {e_{2}}} W_{c d} \mathcal{D}^{e}{W_{\underline{b} e}} W_{{e_{1}} {e_{2}}}+\frac{15}{64}\epsilon_{\underline{a}}\,^{d e {e_{1}} {e_{2}}} W_{\underline{b} c} W_{d e} \mathcal{D}^{c}{W_{{e_{1}} {e_{2}}}} - \frac{9}{32}\epsilon_{\underline{a}}\,^{c d e {e_{1}}} W_{c d} \mathcal{D}^{{e_{2}}}{W_{\underline{b} e}} W_{{e_{1}} {e_{2}}} - \frac{9}{32}\epsilon_{\underline{a}}\,^{c d e {e_{1}}} W_{\underline{b} c} W_{d e} \mathcal{D}^{{e_{2}}}{W_{{e_{1}} {e_{2}}}}+\frac{3}{8}\epsilon^{c d e}\,_{\underline{a} {e_{2}}} \mathcal{D}^{{e_{2}}}{R_{c d e}\,^{{e_{1}}}} W_{\underline{b} {e_{1}}}+\frac{3}{32}\epsilon_{\underline{a} {e_{2}}}\,^{d e {e_{1}}} W_{\underline{b}}\,^{c} W_{d e} \mathcal{D}^{{e_{2}}}{W_{{e_{1}} c}}+\frac{3}{16}\epsilon_{\underline{a} {e_{2}}}\,^{d e {e_{1}}} W_{\underline{b}}\,^{c} \mathcal{D}^{{e_{2}}}{W_{d c}} W_{e {e_{1}}} - \frac{3}{8}\epsilon^{d e}\,_{\underline{b} {e_{2}}}\,^{{e_{1}}} \mathcal{D}^{{e_{2}}}{R_{\underline{a}}\,^{c}\,_{d e}} W_{{e_{1}} c} - \frac{3}{16}\epsilon_{\underline{a} {e_{2}}}\,^{d e {e_{1}}} W_{\underline{b}}\,^{c} W_{d c} \mathcal{D}^{{e_{2}}}{W_{e {e_{1}}}}%
 - \frac{3}{16}\epsilon_{\underline{a} c {e_{2}}}\,^{e {e_{1}}} W^{c d} W_{e d} \mathcal{D}^{{e_{2}}}{W_{\underline{b} {e_{1}}}} - \frac{9}{64}\epsilon_{\underline{a} {e_{2}}}\,^{c d {e_{1}}} W_{\underline{b} c} W_{d}\,^{e} \mathcal{D}^{{e_{2}}}{W_{{e_{1}} e}}+\frac{3}{64}\epsilon_{\underline{a} {e_{2}}}\,^{c d {e_{1}}} W_{\underline{b} c} \mathcal{D}^{{e_{2}}}{W_{d}\,^{e}} W_{{e_{1}} e} - \frac{3}{8}\epsilon^{c d e}\,_{{e_{2}}}\,^{{e_{1}}} \mathcal{D}^{{e_{2}}}{R_{\underline{a} c d e}} W_{\underline{b} {e_{1}}}+\frac{3}{64}\epsilon_{{e_{2}}}\,^{c d e {e_{1}}} W_{\underline{a} c} \mathcal{D}^{{e_{2}}}{W_{\underline{b} d}} W_{e {e_{1}}} - \frac{3}{8}\epsilon^{c d}\,_{\underline{b}}\,^{{e_{1}} {e_{2}}} \mathcal{D}^{e}{R_{\underline{a} c d e}} W_{{e_{1}} {e_{2}}}+\frac{3}{64}\epsilon_{\underline{a} {e_{2}}}\,^{c d {e_{1}}} W_{c d} \mathcal{D}^{{e_{2}}}{W_{\underline{b}}\,^{e}} W_{{e_{1}} e} - \frac{3}{16}\epsilon^{c d e {e_{2}} {e_{3}}} \mathcal{D}^{{e_{1}}}{R_{c d e {e_{1}}}} \eta_{\underline{a} \underline{b}} W_{{e_{2}} {e_{3}}} - \frac{3}{128}\epsilon^{c d e {e_{2}} {e_{3}}} \eta_{\underline{a} \underline{b}} W_{c d} \mathcal{D}^{{e_{1}}}{W_{e {e_{1}}}} W_{{e_{2}} {e_{3}}} - \frac{3}{128}\epsilon^{c d e {e_{2}} {e_{3}}} \eta_{\underline{a} \underline{b}} W_{c d} W_{e {e_{1}}} \mathcal{D}^{{e_{1}}}{W_{{e_{2}} {e_{3}}}}+\frac{3}{128}\epsilon^{c d e {e_{1}} {e_{2}}} \eta_{\underline{a} \underline{b}} W_{c d} \mathcal{D}^{{e_{3}}}{W_{e {e_{1}}}} W_{{e_{2}} {e_{3}}}+\frac{3}{128}\epsilon^{c d e {e_{1}} {e_{2}}} \eta_{\underline{a} \underline{b}} W_{c d} W_{e {e_{1}}} \mathcal{D}^{{e_{3}}}{W_{{e_{2}} {e_{3}}}}+\frac{15}{64}\epsilon_{\underline{a}}\,^{d e {e_{1}} {e_{2}}} W_{\underline{b} c} \mathcal{D}^{c}{W_{d e}} W_{{e_{1}} {e_{2}}}+\frac{15}{64}\epsilon_{\underline{a}}\,^{c d e {e_{1}}} W_{c d} W_{e {e_{1}}} \mathcal{D}^{{e_{2}}}{W_{\underline{b} {e_{2}}}} - \frac{3}{16}\epsilon_{\underline{a}}\,^{c d {e_{1}} {e_{2}}} W_{\underline{b} c} \mathcal{D}^{e}{W_{d e}} W_{{e_{1}} {e_{2}}}+\frac{3}{16}\epsilon_{\underline{a}}\,^{c d e {e_{2}}} W_{c d} W_{e {e_{1}}} \mathcal{D}^{{e_{1}}}{W_{\underline{b} {e_{2}}}} - \frac{3}{64}\epsilon_{{e_{2}}}\,^{c d e {e_{1}}} W_{\underline{a} c} W_{d e} \mathcal{D}^{{e_{2}}}{W_{\underline{b} {e_{1}}}}+\frac{3}{64}\epsilon_{\underline{a} {e_{2}}}\,^{c d e} W_{c d} W_{e}\,^{{e_{1}}} \mathcal{D}^{{e_{2}}}{W_{\underline{b} {e_{1}}}}+\frac{2349}{10240}\epsilon^{c e {e_{1}} {e_{2}} {e_{3}}} \eta_{\underline{a} \underline{b}} \mathcal{D}^{d}{W_{c d}} W_{e {e_{1}}} W_{{e_{2}} {e_{3}}} - \frac{261}{2048}\epsilon^{c d e {e_{1}} {e_{2}}} \mathcal{D}_{\underline{a}}{W_{\underline{b} c}} W_{d e} W_{{e_{1}} {e_{2}}}%
 - \frac{261}{1024}\epsilon_{{e_{3}}}\,^{c e {e_{1}} {e_{2}}} \eta_{\underline{a} \underline{b}} \mathcal{D}^{{e_{3}}}{W_{c}\,^{d}} W_{e {e_{1}}} W_{{e_{2}} d}+\frac{69}{8}R_{\underline{a}}\,^{c}\,_{d e} W^{d e} W_{\underline{b} c}+\frac{69}{8}R^{c}\,_{d} W^{d}\,_{\underline{a}} W_{\underline{b} c} - \frac{69}{40}R_{c d e {e_{1}}} \eta_{\underline{a} \underline{b}} W^{c d} W^{e {e_{1}}}+\frac{69}{20}R_{c d} \eta_{\underline{a} \underline{b}} W^{c}\,_{e} W^{d e}+\frac{9}{8}R_{\underline{a}}\,^{c}\,_{\underline{b} d} W^{d e} W_{c e} - \frac{9}{128}W_{\underline{a}}\,^{c} W^{d e} W_{\underline{b} c} W_{d e}+\frac{69}{8}R_{\underline{a} c}\,^{d}\,_{e} W^{c e} W_{\underline{b} d} - \frac{69}{40}R_{c d e {e_{1}}} \eta_{\underline{a} \underline{b}} W^{c e} W^{d {e_{1}}}+\frac{6}{5}\Phi_{c d i j} \Phi^{c d i j} \eta_{\underline{a} \underline{b}}-6\Phi_{\underline{a}}\,^{c}\,_{i j} \Phi_{\underline{b} c}\,^{i j}+\frac{9}{2}R_{\underline{a}}\,^{c}\,_{d e} W^{d e} F_{\underline{b} c}+\frac{9}{2}R^{c}\,_{d} W^{d}\,_{\underline{a}} F_{\underline{b} c} - \frac{57}{40}R_{e {e_{1}} c d} \eta_{\underline{a} \underline{b}} W^{e {e_{1}}} F^{c d}+\frac{57}{20}R_{e c} \eta_{\underline{a} \underline{b}} W^{e}\,_{d} F^{c d}+\frac{9}{32}W^{d c} W_{\underline{a}}\,^{e} W_{e d} F_{\underline{b} c} - \frac{45}{8}R_{\underline{a}}\,^{c}\,_{\underline{b} e} W^{e d} F_{c d} - \frac{15}{8}R_{\underline{a}}\,^{c} W_{\underline{b}}\,^{d} F_{c d}+\frac{45}{128}W_{\underline{a}}\,^{c} W^{d e} W_{d e} F_{\underline{b} c}+\frac{27}{8}W^{c d} W_{\underline{a}}\,^{e} W_{\underline{b} e} F_{c d}%
 - \frac{45}{32}W_{\underline{a}}\,^{c} W^{e d} W_{\underline{b} e} F_{c d}+\frac{21}{8}R_{\underline{a}}\,^{e}\,_{c d} W_{\underline{b} e} F^{c d}+\frac{21}{8}R^{d}\,_{c} W_{\underline{b} d} F^{c}\,_{\underline{a}}+\frac{9}{2}R_{\underline{a} d}\,^{c}\,_{e} W^{d e} F_{\underline{b} c} - \frac{57}{40}R_{e c {e_{1}} d} \eta_{\underline{a} \underline{b}} W^{e {e_{1}}} F^{c d}+\frac{3}{2}R_{\underline{a}}\,^{c d e} R_{\underline{b} c d e} - \frac{3}{2}R^{c d} R_{\underline{a} c \underline{b} d} - \frac{9}{32}W_{c}\,^{d} W^{c e} W_{\underline{a} d} W_{\underline{b} e} - \frac{3}{10}R^{c d e {e_{1}}} R_{c d e {e_{1}}} \eta_{\underline{a} \underline{b}}+\frac{3}{2}R_{\underline{a}}\,^{c d e} R_{\underline{b} d c e} - \frac{3}{10}R^{c d e {e_{1}}} R_{c e d {e_{1}}} \eta_{\underline{a} \underline{b}} - \frac{3}{4}\epsilon^{{e_{1}} {e_{2}}}\,_{\underline{b}}\,^{c d} R_{\underline{a} e {e_{1}} {e_{2}}} \mathcal{D}^{e}{F_{c d}}+\frac{33}{32}\epsilon_{\underline{a}}\,^{{e_{1}} {e_{2}} c d} W_{\underline{b} e} W_{{e_{1}} {e_{2}}} \mathcal{D}^{e}{F_{c d}}+\frac{21}{32}\epsilon_{\underline{a}}\,^{e {e_{1}} c d} W_{\underline{b} e} W_{{e_{1}} {e_{2}}} \mathcal{D}^{{e_{2}}}{F_{c d}}+\frac{3}{8}\epsilon^{e {e_{1}}}\,_{\underline{b} {e_{2}}}\,^{c} R_{\underline{a}}\,^{d}\,_{e {e_{1}}} \mathcal{D}^{{e_{2}}}{F_{c d}}+\frac{21}{32}\epsilon_{\underline{a} {e_{2}}}\,^{e {e_{1}} c} W_{\underline{b}}\,^{d} W_{e {e_{1}}} \mathcal{D}^{{e_{2}}}{F_{c d}} - \frac{3}{8}\epsilon_{\underline{a} {e_{2}}}\,^{e {e_{1}} c} W_{\underline{b} e} W_{{e_{1}}}\,^{d} \mathcal{D}^{{e_{2}}}{F_{c d}} - \frac{3}{4}\epsilon^{e {e_{1}}}\,_{\underline{b} {e_{2}}}\,^{c} R_{\underline{a} e {e_{1}}}\,^{d} \mathcal{D}^{{e_{2}}}{F_{c d}}+\frac{21}{8}R_{\underline{a} c}\,^{e}\,_{d} W_{\underline{b} e} F^{c d}+\frac{9}{8}\mathcal{D}_{\underline{a}}{W^{c d}} \mathcal{D}_{\underline{b}}{F_{c d}}%
+\frac{9}{4}\mathcal{D}_{\underline{a}}{W_{\underline{b}}\,^{c}} \mathcal{D}^{d}{F_{c d}}+\frac{9}{4}\mathcal{D}_{\underline{a}}{W^{c}\,_{d}} \mathcal{D}^{d}{F_{\underline{b} c}}+\frac{9}{4}\mathcal{D}^{d}{W_{\underline{b}}\,^{c}} \mathcal{D}_{\underline{a}}{F_{c d}}+\frac{9}{4}\mathcal{D}^{d}{W^{c}\,_{d}} \mathcal{D}_{\underline{a}}{F_{\underline{b} c}} - \frac{9}{8}\eta_{\underline{a} \underline{b}} \mathcal{D}_{e}{W^{c d}} \mathcal{D}^{e}{F_{c d}} - \frac{27}{4}\mathcal{D}^{c}{W_{\underline{a} d}} \mathcal{D}^{d}{F_{\underline{b} c}} - \frac{27}{4}\mathcal{D}^{d}{W_{\underline{a} d}} \mathcal{D}^{c}{F_{\underline{b} c}}+\frac{9}{4}\eta_{\underline{a} \underline{b}} \mathcal{D}^{e}{W^{c}\,_{e}} \mathcal{D}^{d}{F_{c d}}+\frac{9}{4}\eta_{\underline{a} \underline{b}} \mathcal{D}^{d}{W^{c}\,_{e}} \mathcal{D}^{e}{F_{c d}}+\frac{9}{2}\mathcal{D}_{d}{W_{\underline{a}}\,^{c}} \mathcal{D}^{d}{F_{\underline{b} c}}+\frac{1233}{5120}\epsilon^{e {e_{1}} {e_{2}} c d} \eta_{\underline{a} \underline{b}} \mathcal{D}^{{e_{3}}}{W_{e {e_{1}}}} W_{{e_{2}} {e_{3}}} F_{c d} - \frac{9}{256}\epsilon^{e {e_{2}} {e_{3}} c d} \eta_{\underline{a} \underline{b}} \mathcal{D}^{{e_{1}}}{W_{e {e_{1}}}} W_{{e_{2}} {e_{3}}} F_{c d}+\frac{9}{256}\epsilon_{{e_{3}}}\,^{e {e_{1}} {e_{2}} c} \eta_{\underline{a} \underline{b}} \mathcal{D}^{{e_{3}}}{W_{e {e_{1}}}} W_{{e_{2}}}\,^{d} F_{c d} - \frac{9}{512}\epsilon_{\underline{a}}\,^{e {e_{2}} c d} \mathcal{D}_{\underline{b}}{W_{e}\,^{{e_{1}}}} W_{{e_{2}} {e_{1}}} F_{c d} - \frac{765}{512}\epsilon_{\underline{a} {e_{2}}}\,^{e {e_{1}} c} \mathcal{D}^{{e_{2}}}{W_{e {e_{1}}}} W_{\underline{b}}\,^{d} F_{c d}+\frac{765}{512}\epsilon_{\underline{a} {e_{2}}}\,^{d e {e_{1}}} \mathcal{D}^{{e_{2}}}{W_{d e}} W_{{e_{1}}}\,^{c} F_{\underline{b} c}+\frac{81}{128}\epsilon_{\underline{a}}\,^{e {e_{1}} {e_{2}} c} \mathcal{D}^{d}{W_{\underline{b} e}} W_{{e_{1}} {e_{2}}} F_{c d} - \frac{81}{512}\epsilon_{\underline{a}}\,^{e {e_{1}} {e_{2}} c} \mathcal{D}^{d}{W_{e {e_{1}}}} W_{\underline{b} {e_{2}}} F_{c d}+\frac{81}{64}\epsilon_{\underline{a} {e_{2}}}\,^{e {e_{1}} c} \mathcal{D}^{{e_{2}}}{W_{\underline{b} e}} W_{{e_{1}}}\,^{d} F_{c d}-6\Phi_{\underline{a}}\,^{c}\,_{i j} X^{i j} W_{\underline{b} c}%
 - \frac{9}{16}\epsilon_{\underline{a}}\,^{d e {e_{1}} {e_{2}}} W_{d e} W_{{e_{1}} {e_{2}}} \mathcal{D}^{c}{F_{\underline{b} c}}+\frac{9}{8}\epsilon_{{e_{2}}}\,^{d e {e_{1}} c} W_{\underline{a} d} W_{e {e_{1}}} \mathcal{D}^{{e_{2}}}{F_{\underline{b} c}} - \frac{9}{32}\epsilon_{\underline{a}}\,^{d e {e_{1}} c} W_{d e} W_{{e_{1}} {e_{2}}} \mathcal{D}^{{e_{2}}}{F_{\underline{b} c}} - \frac{27}{32}\epsilon_{\underline{a} {e_{2}}}\,^{d e {e_{1}}} W_{d e} W_{{e_{1}}}\,^{c} \mathcal{D}^{{e_{2}}}{F_{\underline{b} c}}+\frac{9}{40}\epsilon^{e {e_{1}} {e_{2}} {e_{3}} c} \eta_{\underline{a} \underline{b}} W_{e {e_{1}}} W_{{e_{2}} {e_{3}}} \mathcal{D}^{d}{F_{c d}}+\frac{9}{32}\epsilon_{\underline{a}}\,^{e {e_{1}} {e_{2}} c} W_{e {e_{1}}} W_{{e_{2}}}\,^{d} \mathcal{D}_{\underline{b}}{F_{c d}} - \frac{9}{2}W_{\underline{a} d} \mathcal{D}^{d}{\mathcal{D}^{c}{F_{\underline{b} c}}} - \frac{9}{32}\epsilon^{e {e_{1}} {e_{2}} c d} \eta_{\underline{a} \underline{b}} W_{e {e_{1}}} W_{{e_{2}} {e_{3}}} \mathcal{D}^{{e_{3}}}{F_{c d}} - \frac{9}{32}\epsilon_{\underline{a}}\,^{e {e_{1}} {e_{2}} c} W_{\underline{b} e} W_{{e_{1}} {e_{2}}} \mathcal{D}^{d}{F_{c d}}+\frac{9}{64}\epsilon^{e {e_{1}} {e_{2}} c d} W_{\underline{b} e} W_{{e_{1}} {e_{2}}} \mathcal{D}_{\underline{a}}{F_{c d}}+\frac{99}{512}\epsilon_{\underline{a}}\,^{d e {e_{1}} c} \mathcal{D}^{{e_{2}}}{W_{d e}} W_{{e_{1}} {e_{2}}} F_{\underline{b} c}+\frac{9}{32}\epsilon_{\underline{a} {e_{2}}}\,^{{e_{1}} c d} W_{\underline{b}}\,^{e} W_{{e_{1}} e} \mathcal{D}^{{e_{2}}}{F_{c d}} - \frac{9}{8}\epsilon^{{e_{1}} {e_{2}}}\,_{\underline{b}}\,^{c d} \mathcal{D}^{e}{R_{\underline{a} e {e_{1}} {e_{2}}}} F_{c d} - \frac{45}{128}\epsilon_{\underline{a}}\,^{{e_{1}} {e_{2}} c d} W_{\underline{b} e} \mathcal{D}^{e}{W_{{e_{1}} {e_{2}}}} F_{c d}+\frac{27}{64}\epsilon_{\underline{a}}\,^{e {e_{1}} c d} W_{\underline{b} e} \mathcal{D}^{{e_{2}}}{W_{{e_{1}} {e_{2}}}} F_{c d} - \frac{9}{16}\epsilon^{d e {e_{1}}}\,_{\underline{a} {e_{2}}} \mathcal{D}^{{e_{2}}}{R_{d e {e_{1}}}\,^{c}} F_{\underline{b} c}+\frac{9}{128}\epsilon_{\underline{a} {e_{2}}}\,^{d e {e_{1}}} W_{d e} \mathcal{D}^{{e_{2}}}{W_{{e_{1}}}\,^{c}} F_{\underline{b} c}+\frac{9}{16}\epsilon^{e {e_{1}}}\,_{\underline{b} {e_{2}}}\,^{c} \mathcal{D}^{{e_{2}}}{R_{\underline{a}}\,^{d}\,_{e {e_{1}}}} F_{c d}+\frac{9}{32}\epsilon_{\underline{a} {e_{2}}}\,^{e {e_{1}} c} W_{\underline{b}}\,^{d} \mathcal{D}^{{e_{2}}}{W_{e {e_{1}}}} F_{c d}+\frac{9}{64}\epsilon_{\underline{a} {e_{2}}}\,^{e {e_{1}} c} W_{e {e_{1}}} \mathcal{D}^{{e_{2}}}{W_{\underline{b}}\,^{d}} F_{c d}%
 - \frac{27}{128}\epsilon_{\underline{a} {e_{2}}}\,^{e {e_{1}} c} W_{\underline{b} e} \mathcal{D}^{{e_{2}}}{W_{{e_{1}}}\,^{d}} F_{c d}+\frac{9}{16}\epsilon^{d e {e_{1}}}\,_{{e_{2}}}\,^{c} \mathcal{D}^{{e_{2}}}{R_{\underline{a} d e {e_{1}}}} F_{\underline{b} c}+\frac{9}{128}\epsilon_{{e_{2}}}\,^{d e {e_{1}} c} W_{\underline{a} d} \mathcal{D}^{{e_{2}}}{W_{e {e_{1}}}} F_{\underline{b} c}+\frac{9}{16}\epsilon^{e {e_{1}}}\,_{\underline{b}}\,^{c d} \mathcal{D}^{{e_{2}}}{R_{\underline{a} e {e_{1}} {e_{2}}}} F_{c d} - \frac{9}{16}\epsilon^{e}\,_{\underline{b} {e_{1}}}\,^{c d} \mathcal{D}^{{e_{1}}}{R_{\underline{a} e}} F_{c d} - \frac{27}{128}\epsilon_{\underline{a} {e_{2}}}\,^{{e_{1}} c d} W_{\underline{b}}\,^{e} \mathcal{D}^{{e_{2}}}{W_{{e_{1}} e}} F_{c d}+\frac{9}{32}\epsilon^{e {e_{1}} {e_{2}} c d} \mathcal{D}^{{e_{3}}}{R_{e {e_{1}} {e_{2}} {e_{3}}}} \eta_{\underline{a} \underline{b}} F_{c d}+\frac{9}{256}\epsilon^{e {e_{2}} {e_{3}} c d} \eta_{\underline{a} \underline{b}} W_{e {e_{1}}} \mathcal{D}^{{e_{1}}}{W_{{e_{2}} {e_{3}}}} F_{c d} - \frac{9}{256}\epsilon^{e {e_{1}} {e_{2}} c d} \eta_{\underline{a} \underline{b}} W_{e {e_{1}}} \mathcal{D}^{{e_{3}}}{W_{{e_{2}} {e_{3}}}} F_{c d} - \frac{45}{128}\epsilon_{\underline{a}}\,^{e {e_{1}} c d} W_{e {e_{1}}} \mathcal{D}^{{e_{2}}}{W_{\underline{b} {e_{2}}}} F_{c d} - \frac{9}{32}\epsilon_{\underline{a}}\,^{e {e_{2}} c d} W_{e {e_{1}}} \mathcal{D}^{{e_{1}}}{W_{\underline{b} {e_{2}}}} F_{c d} - \frac{9}{128}\epsilon_{{e_{2}}}\,^{d e {e_{1}} c} W_{d e} \mathcal{D}^{{e_{2}}}{W_{\underline{a} {e_{1}}}} F_{\underline{b} c} - \frac{27}{128}\epsilon_{\underline{a} {e_{2}}}\,^{e c d} W_{e}\,^{{e_{1}}} \mathcal{D}^{{e_{2}}}{W_{\underline{b} {e_{1}}}} F_{c d}+\frac{9}{80}\eta_{\underline{a} \underline{b}} W^{c d} \mathcal{D}_{e}{\mathcal{D}^{e}{F_{c d}}}+\frac{9}{10}\eta_{\underline{a} \underline{b}} W^{c}\,_{e} \mathcal{D}^{e}{\mathcal{D}^{d}{F_{c d}}} - \frac{9}{16}W^{c d} \mathcal{D}_{\underline{a}}{\mathcal{D}_{\underline{b}}{F_{c d}}}
 \doublespacedmathend
 \end{adjustwidth}
 
\subsection{Log} \label{SupercurrentlogWComplete}

\subsubsection{$J^1_{\alpha i, \log}$} \label{J1logWComplete}

\begin{adjustwidth}{0cm}{5cm}
\doublespacedmathbegin
- \frac{3}{2048}{\rm i} Y \lambda_{i \alpha} - \frac{3}{128}(\Gamma_{a})_{\alpha}{}^{\beta} W \nabla^{a}{X_{i \beta}} - \frac{81}{512}(\Sigma_{a b})_{\alpha}{}^{\beta} W W^{a b} X_{i \beta} - \frac{231}{2048}{\rm i} W^{a b} W_{a b} \lambda_{i \alpha} - \frac{3}{128}W W^{a b} W_{a b \alpha i} - \frac{3}{256}\epsilon_{e}\,^{a b c d} (\Gamma^{e})_{\alpha}{}^{\beta} W W_{a b} W_{c d \beta i} - \frac{3}{32}(\Sigma^{c}{}_{\, a})_{\alpha}{}^{\beta} W W^{a b} W_{c b \beta i}+\frac{27}{2048}{\rm i} \epsilon_{e}\,^{a b c d} (\Gamma^{e})_{\alpha}{}^{\beta} W_{a b} W_{c d} \lambda_{i \beta} - \frac{3}{256}(\Gamma_{a})_{\alpha}{}^{\beta} X_{i \beta} \nabla^{a}{W} - \frac{1}{16}{\rm i} \Phi^{a b}\,_{i j} (\Sigma_{a b})_{\alpha}{}^{\beta} \lambda^{j}_{\beta}+\frac{3}{256}{\rm i} \epsilon^{c d}\,_{e}\,^{a b} (\Sigma_{c d})_{\alpha}{}^{\beta} \lambda_{i \beta} \nabla^{e}{W_{a b}} - \frac{21}{128}{\rm i} (\Gamma^{a})_{\alpha}{}^{\beta} \lambda_{i \beta} \nabla^{b}{W_{a b}}+\frac{3}{32}{\rm i} \nabla_{a}{\nabla^{a}{\lambda_{i \alpha}}} - \frac{9}{128}{\rm i} \epsilon^{c d}\,_{e}\,^{a b} (\Sigma_{c d})_{\alpha}{}^{\beta} W_{a b} \nabla^{e}{\lambda_{i \beta}}+\frac{3}{32}{\rm i} (\Gamma^{a})_{\alpha}{}^{\beta} W_{a b} \nabla^{b}{\lambda_{i \beta}} - \frac{3}{128}F^{a b} W_{a b \alpha i} - \frac{3}{256}\epsilon_{e}\,^{a b c d} (\Gamma^{e})_{\alpha}{}^{\beta} F_{a b} W_{c d \beta i} - \frac{3}{32}(\Sigma^{c}{}_{\, a})_{\alpha}{}^{\beta} F^{a b} W_{c b \beta i} - \frac{15}{256}(\Sigma_{a b})_{\alpha}{}^{\beta} F^{a b} X_{i \beta}%
+\frac{51}{256}X_{i j} X^{j}_{\alpha} - \frac{3}{64}{\rm i} \epsilon^{c d}\,_{e}\,^{a b} (\Sigma_{c d})_{\alpha}{}^{\beta} F_{a b} {W}^{-1} \nabla^{e}{\lambda_{i \beta}} - \frac{27}{256}{\rm i} W^{a b} F_{a b} \lambda_{i \alpha} {W}^{-1}+\frac{27}{512}{\rm i} \epsilon_{e}\,^{c d a b} (\Gamma^{e})_{\alpha}{}^{\beta} W_{c d} F_{a b} \lambda_{i \beta} {W}^{-1} - \frac{3}{16}{\rm i} (\Gamma_{a})_{\alpha}{}^{\beta} X_{i j} {W}^{-1} \nabla^{a}{\lambda^{j}_{\beta}}+\frac{9}{128}{\rm i} X_{j k} X^{j k} \lambda_{i \alpha} {W}^{-2}+\frac{9}{64}{\rm i} \lambda^{\beta}_{i} \lambda_{j \beta} X^{j}_{\alpha} {W}^{-1} - \frac{3}{16}{\rm i} (\Sigma_{a b})_{\alpha}{}^{\beta} {W}^{-1} \nabla^{a}{W} \nabla^{b}{\lambda_{i \beta}} - \frac{3}{32}(\Gamma_{a})^{\beta \rho} \lambda_{i \alpha} \lambda_{j \beta} {W}^{-2} \nabla^{a}{\lambda^{j}_{\rho}}+\frac{3}{64}{\rm i} \epsilon^{c d}\,_{e}\,^{a b} (\Sigma_{c d})_{\alpha}{}^{\beta} \lambda_{i \beta} {W}^{-1} \nabla^{e}{F_{a b}} - \frac{3}{32}{\rm i} (\Gamma^{a})_{\alpha}{}^{\beta} \lambda_{i \beta} {W}^{-1} \nabla^{b}{F_{a b}} - \frac{27}{128}{\rm i} (\Gamma^{a})_{\alpha}{}^{\beta} W_{a b} \lambda_{i \beta} {W}^{-1} \nabla^{b}{W}+\frac{3}{32}{\rm i} (\Gamma_{a})_{\alpha}{}^{\beta} \lambda_{j \beta} {W}^{-1} \nabla^{a}{X_{i}\,^{j}}+\frac{3}{32}{\rm i} \lambda_{i \alpha} {W}^{-1} \nabla_{a}{\nabla^{a}{W}} - \frac{3}{16}{\rm i} (\Sigma_{a b})_{\alpha}{}^{\beta} \lambda_{i \beta} {W}^{-1} \nabla^{a}{\nabla^{b}{W}}+\frac{9}{64}{\rm i} \lambda_{i \alpha} \lambda^{\beta}_{j} X^{j}_{\beta} {W}^{-1}+\frac{9}{64}{\rm i} \lambda_{j \alpha} \lambda^{\beta}_{i} X^{j}_{\beta} {W}^{-1} - \frac{3}{64}{\rm i} \lambda_{i \alpha} {W}^{-2} \nabla_{a}{W} \nabla^{a}{W} - \frac{3}{64}(\Gamma_{a})_{\alpha}{}^{\beta} \lambda_{j \beta} \lambda^{j \rho} {W}^{-2} \nabla^{a}{\lambda_{i \rho}}+\frac{9}{256}(\Sigma_{a b})^{\beta \rho} W^{a b} \lambda_{i \alpha} \lambda_{j \beta} \lambda^{j}_{\rho} {W}^{-2}%
 - \frac{3}{64}(\Gamma_{a})^{\beta \rho} \lambda_{j \alpha} \lambda^{j}_{\beta} {W}^{-2} \nabla^{a}{\lambda_{i \rho}}+\frac{9}{256}(\Sigma_{a b})^{\beta \rho} W^{a b} \lambda_{j \alpha} \lambda_{i \beta} \lambda^{j}_{\rho} {W}^{-2}+\frac{3}{256}{\rm i} \epsilon_{e}\,^{a b c d} (\Gamma^{e})_{\alpha}{}^{\beta} F_{a b} F_{c d} \lambda_{i \beta} {W}^{-2}+\frac{3}{64}{\rm i} (\Sigma_{a b})_{\alpha}{}^{\beta} X_{i j} F^{a b} \lambda^{j}_{\beta} {W}^{-2}+\frac{3}{128}{\rm i} \epsilon^{c d}\,_{e}\,^{a b} (\Sigma_{c d})_{\alpha}{}^{\beta} F_{a b} \lambda_{i \beta} {W}^{-2} \nabla^{e}{W}+\frac{3}{64}{\rm i} (\Gamma^{a})_{\alpha}{}^{\beta} F_{a b} \lambda_{i \beta} {W}^{-2} \nabla^{b}{W}+\frac{3}{64}(\Sigma_{a b})^{\beta \rho} F^{a b} \lambda_{i \alpha} \lambda_{j \beta} \lambda^{j}_{\rho} {W}^{-3}+\frac{3}{32}(\Gamma_{a})_{\alpha}{}^{\beta} \lambda^{\rho}_{i} \lambda_{j \rho} {W}^{-2} \nabla^{a}{\lambda^{j}_{\beta}}+\frac{3}{32}{\rm i} X_{i j} X^{j}\,_{k} \lambda^{k}_{\alpha} {W}^{-2}+\frac{3}{32}{\rm i} (\Gamma_{a})_{\alpha}{}^{\beta} X_{i j} \lambda^{j}_{\beta} {W}^{-2} \nabla^{a}{W} - \frac{3}{32}X_{j k} \lambda_{i \alpha} \lambda^{j \beta} \lambda^{k}_{\beta} {W}^{-3} - \frac{3}{64}(\Sigma_{a b})_{\alpha}{}^{\beta} F^{a b} \lambda^{\rho}_{i} \lambda_{j \beta} \lambda^{j}_{\rho} {W}^{-3}+\frac{3}{64}X_{i j} \lambda_{k \alpha} \lambda^{j \beta} \lambda^{k}_{\beta} {W}^{-3} - \frac{3}{64}(\Gamma_{a})_{\alpha}{}^{\beta} \lambda^{\rho}_{i} \lambda_{j \beta} \lambda^{j}_{\rho} {W}^{-3} \nabla^{a}{W} - \frac{9}{256}(\Sigma_{a b})_{\alpha}{}^{\beta} W^{a b} \lambda^{\rho}_{i} \lambda_{j \beta} \lambda^{j}_{\rho} {W}^{-2}+\frac{9}{256}{\rm i} \lambda_{i \alpha} \lambda^{\beta}_{j} \lambda^{j \rho} \lambda_{k \beta} \lambda^{k}_{\rho} {W}^{-4}
\doublespacedmathend
\end{adjustwidth}
 
\subsubsection{$J^2_{a b, \log}$} \label{J2ablogWComplete}

\begin{adjustwidth}{0cm}{5cm}
\doublespacedmathbegin
- \frac{3}{128}{\rm i} \epsilon^{c d}\,_{\hat{a} \hat{b} e} (\Sigma_{c d})^{\beta \alpha} \lambda_{i \beta} \nabla^{e}{X^{i}_{\alpha}} - \frac{21}{64}{\rm i} (\Gamma_{\hat{a}})^{\beta \alpha} \lambda_{i \beta} \nabla_{\hat{b}}{X^{i}_{\alpha}} - \frac{117}{256}{\rm i} W_{\hat{a} \hat{b}} \lambda^{\alpha}_{i} X^{i}_{\alpha} - \frac{27}{512}{\rm i} \epsilon_{\hat{a} \hat{b} e}\,^{c d} (\Gamma^{e})^{\beta \alpha} W_{c d} \lambda_{i \beta} X^{i}_{\alpha} - \frac{3}{256}{\rm i} F_{\hat{a} \hat{b}} Y - \frac{9}{256}{\rm i} W W_{\hat{a} \hat{b}} Y - \frac{27}{64}{\rm i} \epsilon_{\hat{a} \hat{b}}\,^{e c d} W W_{c d} \nabla^{{e_{1}}}{W_{e {e_{1}}}}+\frac{9}{32}{\rm i} \epsilon_{\hat{a} {e_{1}}}\,^{d e c} W W_{\hat{b} c} \nabla^{{e_{1}}}{W_{d e}} - \frac{3}{16}{\rm i} W \nabla_{c}{\nabla^{c}{W_{\hat{a} \hat{b}}}}+\frac{117}{256}{\rm i} W W^{c d} W_{\hat{a} \hat{b}} W_{c d}+\frac{3}{16}{\rm i} W W^{c d} W_{\hat{a} c} W_{\hat{b} d} - \frac{3}{16}W_{\hat{a} \hat{b}}\,^{\alpha}\,_{i} W X^{i}_{\alpha} - \frac{81}{512}(\Sigma_{\hat{a} \hat{b}})^{\alpha \beta} W X_{i \alpha} X^{i}_{\beta} - \frac{263}{4096}{\rm i} (\Sigma_{c d})^{\alpha \beta} W^{c d} W_{\hat{a} \hat{b} \alpha i} \lambda^{i}_{\beta} - \frac{31}{4096}{\rm i} W^{c d} W_{c d}\,^{\alpha}\,_{i} (\Sigma_{\hat{a} \hat{b}})_{\alpha}{}^{\beta} \lambda^{i}_{\beta}+\frac{31}{1024}{\rm i} W_{\hat{a}}\,^{c} W_{c}\,^{d \alpha}\,_{i} (\Sigma_{\hat{b} d})_{\alpha}{}^{\beta} \lambda^{i}_{\beta} - \frac{31}{4096}{\rm i} W_{\hat{a} \hat{b}} W^{c d \alpha}\,_{i} (\Sigma_{c d})_{\alpha}{}^{\beta} \lambda^{i}_{\beta}+\frac{25}{8192}{\rm i} \epsilon^{c e {e_{1}}}\,_{\hat{a} \hat{b}} W_{c d} W_{e {e_{1}}}\,^{\alpha}\,_{i} (\Gamma^{d})_{\alpha}{}^{\beta} \lambda^{i}_{\beta} - \frac{261}{2048}{\rm i} W_{\hat{a}}\,^{c} W_{\hat{b} c}\,^{\alpha}\,_{i} \lambda^{i}_{\alpha}%
 - \frac{23}{4096}{\rm i} \epsilon^{c d e}\,_{\hat{b} {e_{1}}} (\Gamma^{{e_{1}}})^{\alpha \beta} W_{c d} W_{\hat{a} e \alpha i} \lambda^{i}_{\beta} - \frac{23}{1024}{\rm i} (\Sigma^{c d})^{\alpha \beta} W_{\hat{a} c} W_{\hat{b} d \alpha i} \lambda^{i}_{\beta}+\frac{23}{1024}{\rm i} (\Sigma_{\hat{a} c})^{\alpha \beta} W^{c d} W_{\hat{b} d \alpha i} \lambda^{i}_{\beta}+\frac{3}{256}{\rm i} W^{c d} W_{c d} F_{\hat{a} \hat{b}}+\frac{1}{16}{\rm i} W C_{\hat{a} \hat{b} c d} W^{c d}+\frac{3}{16}{\rm i} W W_{\hat{a}}\,^{c} W_{\hat{b}}\,^{d} W_{d c}+\frac{1}{16}{\rm i} W C_{c d \hat{a} \hat{b}} W^{c d}+\frac{1}{8}{\rm i} W C_{\hat{a} c \hat{b} d} W^{c d} - \frac{9}{64}{\rm i} \epsilon_{\hat{a} \hat{b}}\,^{e {e_{1}} c} W W_{c d} \nabla^{d}{W_{e {e_{1}}}}+\frac{9}{32}{\rm i} \epsilon_{\hat{a} {e_{1}}}\,^{e c d} W W_{c d} \nabla^{{e_{1}}}{W_{\hat{b} e}} - \frac{3}{32}{\rm i} W^{c d} W_{\hat{a} \hat{b}} F_{c d}+\frac{21}{16}{\rm i} W^{d c} W_{\hat{a} d} F_{\hat{b} c}+\frac{9}{16}{\rm i} W_{\hat{a} c} W_{\hat{b} d} F^{c d} - \frac{9}{64}{\rm i} \epsilon_{\hat{a} \hat{b}}\,^{c d e} W_{c d} W_{e {e_{1}}} \nabla^{{e_{1}}}{W}+\frac{9}{32}{\rm i} \epsilon_{\hat{a} {e_{1}}}\,^{c d e} W_{\hat{b} c} W_{d e} \nabla^{{e_{1}}}{W} - \frac{3}{16}{\rm i} \nabla_{c}{W} \nabla^{c}{W_{\hat{a} \hat{b}}} - \frac{3}{4}{\rm i} \nabla_{\hat{a}}{W} \nabla^{c}{W_{\hat{b} c}}+\frac{3}{4}{\rm i} \nabla^{c}{W} \nabla_{\hat{a}}{W_{\hat{b} c}} - \frac{21}{128}{\rm i} \epsilon^{c d}\,_{\hat{a} \hat{b} e} (\Sigma_{c d})^{\alpha \beta} X_{i \alpha} \nabla^{e}{\lambda^{i}_{\beta}}+\frac{33}{64}{\rm i} (\Gamma_{\hat{a}})^{\alpha \beta} X_{i \alpha} \nabla_{\hat{b}}{\lambda^{i}_{\beta}}%
 - \frac{303}{2048}{\rm i} \epsilon_{\hat{a} \hat{b} e}\,^{c d} \lambda^{\alpha}_{i} \nabla^{e}{W_{c d \alpha}\,^{i}} - \frac{31}{1024}{\rm i} (\Gamma_{c})^{\beta \alpha} \lambda_{i \beta} \nabla^{c}{W_{\hat{a} \hat{b} \alpha}\,^{i}} - \frac{7}{256}{\rm i} \epsilon^{e}\,_{\hat{a} \hat{b}}\,^{c d} (\Sigma_{e {e_{1}}})^{\beta \alpha} \lambda_{i \beta} \nabla^{{e_{1}}}{W_{c d \alpha}\,^{i}} - \frac{19}{512}{\rm i} \epsilon^{d e}\,_{\hat{a} {e_{1}}}\,^{c} (\Sigma_{d e})^{\beta \alpha} \lambda_{i \beta} \nabla^{{e_{1}}}{W_{\hat{b} c \alpha}\,^{i}} - \frac{31}{512}{\rm i} (\Gamma^{c})^{\beta \alpha} \lambda_{i \beta} \nabla_{\hat{a}}{W_{\hat{b} c \alpha}\,^{i}}+\frac{19}{512}{\rm i} (\Gamma_{\hat{a}})^{\beta \alpha} \lambda_{i \beta} \nabla^{c}{W_{\hat{b} c \alpha}\,^{i}} - \frac{1}{8}{\rm i} \Phi_{\hat{a} \hat{b} i j} X^{i j} - \frac{9}{32}{\rm i} \epsilon_{\hat{a} \hat{b}}\,^{e c d} F_{c d} \nabla^{{e_{1}}}{W_{e {e_{1}}}}+\frac{3}{16}{\rm i} \epsilon_{\hat{a} {e_{1}}}\,^{d e c} F_{\hat{b} c} \nabla^{{e_{1}}}{W_{d e}} - \frac{3}{16}{\rm i} \nabla_{c}{\nabla^{c}{F_{\hat{a} \hat{b}}}} - \frac{9}{16}{\rm i} W_{\hat{a} \hat{b}} \nabla_{c}{\nabla^{c}{W}} - \frac{3}{32}{\rm i} \epsilon_{\hat{a} \hat{b}}\,^{c d e} W_{e {e_{1}}} \nabla^{{e_{1}}}{F_{c d}}+\frac{3}{16}{\rm i} \epsilon_{\hat{a} {e_{1}}}\,^{c d e} W_{d e} \nabla^{{e_{1}}}{F_{\hat{b} c}}+\frac{9}{8}{\rm i} W_{\hat{b} c} \nabla^{c}{\nabla_{\hat{a}}{W}} - \frac{3}{4}{\rm i} W_{\hat{b} c} \nabla_{\hat{a}}{\nabla^{c}{W}} - \frac{3}{16}{\rm i} \epsilon_{\hat{a} \hat{b}}\,^{c e {e_{1}}} W_{e {e_{1}}} \nabla^{d}{F_{c d}} - \frac{3}{8}{\rm i} \epsilon_{\hat{a} {e_{1}}}\,^{c d e} W_{\hat{b} e} \nabla^{{e_{1}}}{F_{c d}}+\frac{9}{128}{\rm i} W_{c d}\,^{\alpha}\,_{i} \epsilon^{c d}\,_{\hat{a} \hat{b} e} \nabla^{e}{\lambda^{i}_{\alpha}}+\frac{3}{64}{\rm i} W_{\hat{a} \hat{b}}\,^{\alpha}\,_{i} (\Gamma_{c})_{\alpha}{}^{\beta} \nabla^{c}{\lambda^{i}_{\beta}} - \frac{3}{64}{\rm i} W_{c d}\,^{\alpha}\,_{i} \epsilon^{c d e}\,_{\hat{a} \hat{b}} (\Sigma_{e {e_{1}}})_{\alpha}{}^{\beta} \nabla^{{e_{1}}}{\lambda^{i}_{\beta}}%
+\frac{3}{32}{\rm i} W_{\hat{a} c}\,^{\alpha}\,_{i} \epsilon^{c d e}\,_{\hat{b} {e_{1}}} (\Sigma_{d e})_{\alpha}{}^{\beta} \nabla^{{e_{1}}}{\lambda^{i}_{\beta}} - \frac{3}{32}{\rm i} W_{\hat{a} c}\,^{\alpha}\,_{i} (\Gamma^{c})_{\alpha}{}^{\beta} \nabla_{\hat{b}}{\lambda^{i}_{\beta}}+\frac{3}{32}{\rm i} W_{\hat{a} c}\,^{\alpha}\,_{i} (\Gamma_{\hat{b}})_{\alpha}{}^{\beta} \nabla^{c}{\lambda^{i}_{\beta}} - \frac{3}{256}{\rm i} \epsilon^{e}\,_{\hat{a} \hat{b} {e_{1}}}\,^{c} (\Sigma_{e}{}^{\, d})^{\beta \alpha} \lambda_{i \beta} \nabla^{{e_{1}}}{W_{c d \alpha}\,^{i}} - \frac{1}{1024}{\rm i} \epsilon^{e {e_{1}}}\,_{\hat{a} \hat{b}}\,^{c} (\Sigma_{e {e_{1}}})^{\beta \alpha} \lambda_{i \beta} \nabla^{d}{W_{c d \alpha}\,^{i}} - \frac{3}{1024}{\rm i} \epsilon^{e {e_{1}}}\,_{\hat{a}}\,^{c d} (\Sigma_{e {e_{1}}})^{\beta \alpha} \lambda_{i \beta} \nabla_{\hat{b}}{W_{c d \alpha}\,^{i}}+\frac{3}{4096}{\rm i} \epsilon_{\hat{a} \hat{b}}\,^{c e {e_{1}}} W_{c d} W_{e {e_{1}}}\,^{\alpha}\,_{i} (\Gamma^{d})_{\alpha}{}^{\beta} \lambda^{i}_{\beta} - \frac{1}{2048}{\rm i} \epsilon_{\hat{a} \hat{b} {e_{1}}}\,^{c e} (\Gamma^{{e_{1}}})^{\alpha \beta} W_{c}\,^{d} W_{e d \alpha i} \lambda^{i}_{\beta} - \frac{1}{2048}{\rm i} (\Sigma^{c d})^{\alpha \beta} W_{\hat{a} \hat{b}} W_{c d \alpha i} \lambda^{i}_{\beta}+\frac{1}{256}{\rm i} (\Sigma_{\hat{a}}{}^{\, d})^{\alpha \beta} W_{\hat{b}}\,^{c} W_{d c \alpha i} \lambda^{i}_{\beta} - \frac{3}{2048}{\rm i} (\Sigma_{\hat{a} \hat{b}})^{\alpha \beta} W^{c d} W_{c d \alpha i} \lambda^{i}_{\beta} - \frac{1}{4096}{\rm i} \epsilon_{\hat{a} \hat{b}}\,^{c d e} (\Gamma^{{e_{1}}})^{\alpha \beta} W_{c d} W_{e {e_{1}} \alpha i} \lambda^{i}_{\beta} - \frac{1}{4096}{\rm i} \epsilon_{\hat{a} \hat{b}}\,^{c e {e_{1}}} (\Gamma^{d})^{\alpha \beta} W_{c d} W_{e {e_{1}} \alpha i} \lambda^{i}_{\beta}+\frac{1}{4096}{\rm i} (\Gamma_{\hat{a}})^{\alpha \beta} W_{c d} W_{e {e_{1}} \alpha i} \epsilon^{c d e {e_{1}}}\,_{\hat{b}} \lambda^{i}_{\beta}+\frac{1}{2048}{\rm i} (\Gamma^{c})^{\alpha \beta} W_{c d} W_{e {e_{1}} \alpha i} \epsilon_{\hat{a}}\,^{d e {e_{1}}}\,_{\hat{b}} \lambda^{i}_{\beta}+\frac{1}{2048}{\rm i} (\Gamma_{{e_{1}}})^{\alpha \beta} W_{\hat{b} c} W_{d e \alpha i} \epsilon_{\hat{a}}\,^{{e_{1}} c d e} \lambda^{i}_{\beta}+\frac{1}{2048}{\rm i} \epsilon^{c d e}\,_{\hat{b} {e_{1}}} (\Gamma^{{e_{1}}})^{\alpha \beta} W_{\hat{a} c} W_{d e \alpha i} \lambda^{i}_{\beta} - \frac{1}{1024}{\rm i} \epsilon^{e}\,_{\hat{a} \hat{b} {e_{1}} d} (\Gamma^{{e_{1}}})^{\alpha \beta} W^{c d} W_{e c \alpha i} \lambda^{i}_{\beta}+\frac{1}{2048}{\rm i} (\Gamma^{c})^{\alpha \beta} W_{\hat{b} c} W_{d e \alpha i} \epsilon_{\hat{a}}\,^{d e {e_{1}} {e_{2}}} (\Sigma_{{e_{1}} {e_{2}}})_{\beta}{}^{\rho} \lambda^{i}_{\rho}+\frac{1}{16}{\rm i} C_{\hat{a} \hat{b} c d} F^{c d}%
+\frac{3}{32}{\rm i} W_{\hat{a} \hat{b}} W_{c d} F^{c d}+\frac{3}{16}{\rm i} W_{\hat{b}}\,^{d} W_{d c} F_{\hat{a}}\,^{c}+\frac{1}{16}{\rm i} C_{c d \hat{a} \hat{b}} F^{c d}+\frac{1}{8}{\rm i} C_{\hat{a} c \hat{b} d} F^{c d} - \frac{3}{32}{\rm i} \epsilon_{\hat{a} \hat{b}}\,^{e {e_{1}} c} F_{c d} \nabla^{d}{W_{e {e_{1}}}}+\frac{3}{16}{\rm i} \epsilon_{\hat{a} {e_{1}}}\,^{e c d} F_{c d} \nabla^{{e_{1}}}{W_{\hat{b} e}}+\frac{3}{16}{W}^{-1} \nabla_{\hat{a}}{\lambda^{\alpha}_{i}} \nabla_{\hat{b}}{\lambda^{i}_{\alpha}}+\frac{9}{8}(\Sigma_{\hat{a} c})^{\alpha \beta} {W}^{-1} \nabla^{c}{\lambda_{i \alpha}} \nabla_{\hat{b}}{\lambda^{i}_{\beta}} - \frac{9}{32}(\Sigma_{\hat{a} \hat{b}})^{\alpha \beta} {W}^{-1} \nabla_{c}{\lambda_{i \alpha}} \nabla^{c}{\lambda^{i}_{\beta}}+\frac{3}{32}\epsilon_{\hat{a} \hat{b} c d e} (\Gamma^{c})^{\alpha \beta} {W}^{-1} \nabla^{d}{\lambda_{i \alpha}} \nabla^{e}{\lambda^{i}_{\beta}}+\frac{9}{256}\epsilon_{\hat{a} \hat{b} e}\,^{c d} W_{c d} \lambda^{\alpha}_{i} {W}^{-1} \nabla^{e}{\lambda^{i}_{\alpha}} - \frac{9}{128}(\Gamma_{c})^{\alpha \beta} W_{\hat{a} \hat{b}} \lambda_{i \alpha} {W}^{-1} \nabla^{c}{\lambda^{i}_{\beta}} - \frac{9}{64}(\Gamma_{\hat{a}})^{\alpha \beta} W_{\hat{b} c} \lambda_{i \alpha} {W}^{-1} \nabla^{c}{\lambda^{i}_{\beta}} - \frac{9}{64}(\Gamma^{c})^{\alpha \beta} W_{\hat{b} c} \lambda_{i \alpha} {W}^{-1} \nabla_{\hat{a}}{\lambda^{i}_{\beta}} - \frac{3}{128}\epsilon_{\hat{a} \hat{b} e}\,^{c d} F_{c d} \lambda^{\alpha}_{i} {W}^{-2} \nabla^{e}{\lambda^{i}_{\alpha}} - \frac{3}{64}(\Gamma_{c})^{\alpha \beta} F_{\hat{a} \hat{b}} \lambda_{i \alpha} {W}^{-2} \nabla^{c}{\lambda^{i}_{\beta}} - \frac{3}{32}(\Gamma_{\hat{a}})^{\alpha \beta} F_{\hat{b} c} \lambda_{i \alpha} {W}^{-2} \nabla^{c}{\lambda^{i}_{\beta}}+\frac{3}{64}\epsilon^{e}\,_{\hat{a} \hat{b} {e_{1}}}\,^{c} (\Sigma_{e}{}^{\, d})^{\alpha \beta} F_{c d} \lambda_{i \alpha} {W}^{-2} \nabla^{{e_{1}}}{\lambda^{i}_{\beta}} - \frac{3}{128}\epsilon^{e {e_{1}}}\,_{\hat{a}}\,^{c d} (\Sigma_{e {e_{1}}})^{\alpha \beta} F_{c d} \lambda_{i \alpha} {W}^{-2} \nabla_{\hat{b}}{\lambda^{i}_{\beta}} - \frac{3}{64}{\rm i} \epsilon_{\hat{a} \hat{b}}\,^{e {e_{1}} c} F_{c d} {W}^{-1} \nabla^{d}{F_{e {e_{1}}}}%
+\frac{3}{32}{\rm i} \epsilon_{\hat{a} {e_{1}}}\,^{e c d} F_{c d} {W}^{-1} \nabla^{{e_{1}}}{F_{\hat{b} e}} - \frac{3}{8}{\rm i} F_{\hat{a} \hat{b}} {W}^{-1} \nabla_{c}{\nabla^{c}{W}}+\frac{3}{8}{\rm i} F_{\hat{b} c} {W}^{-1} \nabla^{c}{\nabla_{\hat{a}}{W}} - \frac{3}{4}{\rm i} F_{\hat{b} c} {W}^{-1} \nabla_{\hat{a}}{\nabla^{c}{W}} - \frac{9}{32}{\rm i} W_{\hat{a} \hat{b}} F^{c d} F_{c d} {W}^{-1}+\frac{9}{32}{\rm i} \epsilon_{\hat{a} \hat{b}}\,^{e {e_{1}} c} W_{e {e_{1}}} F_{c d} {W}^{-1} \nabla^{d}{W} - \frac{27}{128}{\rm i} F_{\hat{a} \hat{b}} \lambda^{\alpha}_{i} X^{i}_{\alpha} {W}^{-1} - \frac{27}{256}{\rm i} \epsilon_{\hat{a} \hat{b} e}\,^{c d} (\Gamma^{e})^{\beta \alpha} F_{c d} \lambda_{i \beta} X^{i}_{\alpha} {W}^{-1}+\frac{81}{1024}(\Sigma_{\hat{a} \hat{b}})^{\alpha \beta} W^{c d} W_{c d} \lambda_{i \alpha} \lambda^{i}_{\beta} {W}^{-1}+\frac{81}{512}(\Sigma_{c d})^{\alpha \beta} W^{c d} W_{\hat{a} \hat{b}} \lambda_{i \alpha} \lambda^{i}_{\beta} {W}^{-1} - \frac{81}{128}(\Sigma_{\hat{a} c})^{\alpha \beta} W^{c d} W_{\hat{b} d} \lambda_{i \alpha} \lambda^{i}_{\beta} {W}^{-1} - \frac{81}{256}(\Sigma^{c d})^{\alpha \beta} W_{\hat{a} c} W_{\hat{b} d} \lambda_{i \alpha} \lambda^{i}_{\beta} {W}^{-1}+\frac{9}{256}(\Sigma_{\hat{a} \hat{b}})^{\alpha \beta} W^{c d} F_{c d} \lambda_{i \alpha} \lambda^{i}_{\beta} {W}^{-2}+\frac{27}{256}(\Sigma_{c d})^{\alpha \beta} W_{\hat{a} \hat{b}} F^{c d} \lambda_{i \alpha} \lambda^{i}_{\beta} {W}^{-2}+\frac{9}{256}(\Sigma_{c d})^{\alpha \beta} W^{c d} F_{\hat{a} \hat{b}} \lambda_{i \alpha} \lambda^{i}_{\beta} {W}^{-2} - \frac{9}{64}(\Sigma_{\hat{a} d})^{\alpha \beta} W^{d c} F_{\hat{b} c} \lambda_{i \alpha} \lambda^{i}_{\beta} {W}^{-2}+\frac{21}{128}\epsilon^{e}\,_{\hat{a} \hat{b}}\,^{c d} (\Sigma_{e {e_{1}}})^{\alpha \beta} F_{c d} \lambda_{i \alpha} {W}^{-2} \nabla^{{e_{1}}}{\lambda^{i}_{\beta}} - \frac{9}{64}\epsilon^{d e}\,_{\hat{a} {e_{1}}}\,^{c} (\Sigma_{d e})^{\alpha \beta} F_{\hat{b} c} \lambda_{i \alpha} {W}^{-2} \nabla^{{e_{1}}}{\lambda^{i}_{\beta}} - \frac{9}{64}{\rm i} \epsilon_{\hat{a} \hat{b}}\,^{e c d} F_{c d} {W}^{-1} \nabla^{{e_{1}}}{F_{e {e_{1}}}} - \frac{9}{32}{\rm i} \epsilon_{\hat{a} {e_{1}}}\,^{d e c} F_{\hat{b} c} {W}^{-1} \nabla^{{e_{1}}}{F_{d e}}%
+\frac{9}{8}{\rm i} W_{\hat{a} c} F^{c d} F_{\hat{b} d} {W}^{-1} - \frac{9}{32}(\Sigma_{\hat{a} c})^{\alpha \beta} W_{\hat{b} d} F^{c d} \lambda_{i \alpha} \lambda^{i}_{\beta} {W}^{-2} - \frac{9}{32}(\Sigma^{d c})^{\alpha \beta} W_{\hat{a} d} F_{\hat{b} c} \lambda_{i \alpha} \lambda^{i}_{\beta} {W}^{-2}+\frac{15}{256}(\Sigma_{\hat{a} \hat{b}})^{\alpha \beta} F^{c d} F_{c d} \lambda_{i \alpha} \lambda^{i}_{\beta} {W}^{-3} - \frac{3}{64}\epsilon^{c d}\,_{\hat{a} \hat{b} e} (\Sigma_{c d})^{\alpha \beta} X_{i j} \lambda^{i}_{\alpha} {W}^{-2} \nabla^{e}{\lambda^{j}_{\beta}}+\frac{3}{16}(\Gamma_{\hat{a}})^{\alpha \beta} X_{i j} \lambda^{i}_{\alpha} {W}^{-2} \nabla_{\hat{b}}{\lambda^{j}_{\beta}}+\frac{9}{256}X_{i j} W_{\hat{a} \hat{b}} \lambda^{i \alpha} \lambda^{j}_{\alpha} {W}^{-2}+\frac{9}{512}\epsilon_{\hat{a} \hat{b} e}\,^{c d} (\Gamma^{e})^{\alpha \beta} X_{i j} W_{c d} \lambda^{i}_{\alpha} \lambda^{j}_{\beta} {W}^{-2}+\frac{3}{32}(\Sigma_{\hat{a} \hat{b}})^{\alpha \beta} X_{i j} X^{i j} \lambda_{k \alpha} \lambda^{k}_{\beta} {W}^{-3}+\frac{27}{64}{\rm i} (\Gamma_{\hat{a}})^{\beta \alpha} \lambda_{i \beta} X^{i}_{\alpha} {W}^{-1} \nabla_{\hat{b}}{W}+\frac{9}{128}(\Sigma_{\hat{a} \hat{b}})^{\beta \alpha} \lambda_{j \beta} \lambda^{j \rho} \lambda_{i \rho} X^{i}_{\alpha} {W}^{-2} - \frac{3}{8}(\Sigma_{\hat{a} c})^{\alpha \beta} \lambda_{i \alpha} {W}^{-2} \nabla^{c}{W} \nabla_{\hat{b}}{\lambda^{i}_{\beta}}+\frac{9}{32}(\Sigma_{\hat{a} \hat{b}})^{\alpha \beta} \lambda_{i \alpha} {W}^{-2} \nabla_{c}{W} \nabla^{c}{\lambda^{i}_{\beta}} - \frac{3}{64}\epsilon_{\hat{a} \hat{b} c d e} (\Gamma^{c})^{\alpha \beta} \lambda_{i \alpha} {W}^{-2} \nabla^{d}{W} \nabla^{e}{\lambda^{i}_{\beta}} - \frac{3}{16}\lambda^{\alpha}_{i} {W}^{-2} \nabla_{\hat{a}}{W} \nabla_{\hat{b}}{\lambda^{i}_{\alpha}} - \frac{15}{16}(\Sigma_{\hat{a} c})^{\alpha \beta} \lambda_{i \alpha} {W}^{-2} \nabla_{\hat{b}}{W} \nabla^{c}{\lambda^{i}_{\beta}}+\frac{3}{16}{\rm i} {W}^{-1} \nabla_{c}{W} \nabla^{c}{F_{\hat{a} \hat{b}}} - \frac{3}{4}{\rm i} {W}^{-1} \nabla_{\hat{a}}{W} \nabla^{c}{F_{\hat{b} c}}+\frac{3}{4}{\rm i} {W}^{-1} \nabla^{c}{W} \nabla_{\hat{a}}{F_{\hat{b} c}}+\frac{3}{8}{\rm i} \epsilon_{\hat{a} \hat{b} c d e} {W}^{-1} \nabla^{c}{W} \nabla^{d}{\nabla^{e}{W}}%
 - \frac{9}{8}{\rm i} W_{\hat{b} c} {W}^{-1} \nabla_{\hat{a}}{W} \nabla^{c}{W}+\frac{3}{32}{\rm i} (\Sigma_{\hat{a} \hat{b}})^{\rho \lambda} (\Gamma_{c})^{\alpha \beta} \lambda_{i \rho} \lambda^{i}_{\lambda} \lambda_{j \alpha} {W}^{-3} \nabla^{c}{\lambda^{j}_{\beta}}+\frac{3}{32}(\Gamma_{\hat{a}})^{\alpha \beta} \lambda_{i \alpha} \lambda_{j \beta} {W}^{-2} \nabla_{\hat{b}}{X^{i j}} - \frac{9}{256}\epsilon^{e}\,_{\hat{a} \hat{b}}\,^{c d} (\Sigma_{e {e_{1}}})^{\alpha \beta} W_{c d} \lambda_{i \alpha} \lambda^{i}_{\beta} {W}^{-2} \nabla^{{e_{1}}}{W} - \frac{9}{128}(\Sigma_{\hat{a} \hat{b}})^{\beta \rho} \lambda_{j \beta} \lambda^{j \alpha} \lambda_{i \rho} X^{i}_{\alpha} {W}^{-2}+\frac{9}{128}(\Sigma_{\hat{a} \hat{b}})^{\beta \rho} \lambda_{j \beta} \lambda^{j}_{\rho} \lambda^{\alpha}_{i} X^{i}_{\alpha} {W}^{-2} - \frac{3}{32}\epsilon^{d e}\,_{\hat{a} {e_{1}}}\,^{c} (\Sigma_{d e})^{\alpha \beta} \lambda_{i \alpha} \lambda^{i}_{\beta} {W}^{-2} \nabla^{{e_{1}}}{F_{\hat{b} c}}+\frac{3}{8}(\Sigma_{\hat{a} c})^{\alpha \beta} \lambda_{i \alpha} {W}^{-1} \nabla^{c}{\nabla_{\hat{b}}{\lambda^{i}_{\beta}}} - \frac{3}{16}(\Sigma_{\hat{a} \hat{b}})^{\alpha \beta} \lambda_{i \alpha} {W}^{-1} \nabla_{c}{\nabla^{c}{\lambda^{i}_{\beta}}}+\frac{3}{32}\epsilon_{\hat{a} \hat{b} c d e} (\Gamma^{c})^{\alpha \beta} \lambda_{i \alpha} {W}^{-1} \nabla^{d}{\nabla^{e}{\lambda^{i}_{\beta}}}+\frac{3}{16}\lambda^{\alpha}_{i} {W}^{-1} \nabla_{\hat{a}}{\nabla_{\hat{b}}{\lambda^{i}_{\alpha}}}+\frac{3}{8}(\Sigma_{\hat{a} c})^{\alpha \beta} \lambda_{i \alpha} {W}^{-1} \nabla_{\hat{b}}{\nabla^{c}{\lambda^{i}_{\beta}}} - \frac{3}{16}(\Sigma_{\hat{a} c})^{\alpha \beta} \lambda_{i \alpha} \lambda^{i}_{\beta} {W}^{-2} \nabla_{\hat{b}}{\nabla^{c}{W}}+\frac{3}{32}(\Sigma_{\hat{a} \hat{b}})^{\alpha \beta} \lambda_{i \alpha} \lambda^{i}_{\beta} {W}^{-2} \nabla_{c}{\nabla^{c}{W}}+\frac{3}{16}{\rm i} F_{\hat{a} \hat{b}} {W}^{-2} \nabla_{c}{W} \nabla^{c}{W} - \frac{15}{128}(\Sigma_{\hat{a} \hat{b}})^{\alpha \beta} \lambda_{i \alpha} \lambda^{i}_{\beta} {W}^{-3} \nabla_{c}{W} \nabla^{c}{W} - \frac{3}{64}{\rm i} \epsilon^{c d}\,_{\hat{a} \hat{b} e} (\Sigma_{c d})^{\alpha \beta} \lambda_{i \alpha} \lambda^{i \rho} \lambda_{j \beta} {W}^{-3} \nabla^{e}{\lambda^{j}_{\rho}} - \frac{3}{32}{\rm i} (\Gamma_{\hat{a}})^{\alpha \beta} \lambda_{i \alpha} \lambda^{i \rho} \lambda_{j \beta} {W}^{-3} \nabla_{\hat{b}}{\lambda^{j}_{\rho}} - \frac{9}{128}{\rm i} (\Sigma_{\hat{a} \hat{b}})^{\alpha \beta} (\Sigma_{c d})^{\rho \lambda} W^{c d} \lambda_{i \alpha} \lambda^{i}_{\beta} \lambda_{j \rho} \lambda^{j}_{\lambda} {W}^{-3}+\frac{3}{32}{\rm i} (\Sigma_{\hat{a} \hat{b}})^{\rho \lambda} (\Gamma_{c})^{\alpha \beta} \lambda_{i \rho} \lambda^{i}_{\alpha} \lambda_{j \lambda} {W}^{-3} \nabla^{c}{\lambda^{j}_{\beta}}%
+\frac{9}{128}{\rm i} (\Sigma_{\hat{a} \hat{b}})^{\alpha \beta} (\Sigma_{c d})^{\rho \lambda} W^{c d} \lambda_{i \alpha} \lambda^{i}_{\rho} \lambda_{j \beta} \lambda^{j}_{\lambda} {W}^{-3} - \frac{3}{32}{\rm i} F^{c d} F_{\hat{a} \hat{b}} F_{c d} {W}^{-2}+\frac{3}{16}{\rm i} F^{c d} F_{\hat{a} c} F_{\hat{b} d} {W}^{-2}+\frac{3}{32}{\rm i} \epsilon_{\hat{a} \hat{b}}\,^{c d e} F_{c d} F_{e {e_{1}}} {W}^{-2} \nabla^{{e_{1}}}{W} - \frac{3}{16}{\rm i} \epsilon_{\hat{a} {e_{1}}}\,^{c d e} F_{\hat{b} c} F_{d e} {W}^{-2} \nabla^{{e_{1}}}{W}+\frac{9}{128}(\Sigma_{c d})^{\alpha \beta} F^{c d} F_{\hat{a} \hat{b}} \lambda_{i \alpha} \lambda^{i}_{\beta} {W}^{-3} - \frac{9}{32}(\Sigma_{\hat{a} c})^{\alpha \beta} F^{c d} F_{\hat{b} d} \lambda_{i \alpha} \lambda^{i}_{\beta} {W}^{-3} - \frac{3}{64}(\Sigma^{c d})^{\alpha \beta} F_{\hat{a} c} F_{\hat{b} d} \lambda_{i \alpha} \lambda^{i}_{\beta} {W}^{-3}+\frac{3}{128}X_{i j} F_{\hat{a} \hat{b}} \lambda^{i \alpha} \lambda^{j}_{\alpha} {W}^{-3}+\frac{9}{256}\epsilon_{\hat{a} \hat{b} e}\,^{c d} (\Gamma^{e})^{\alpha \beta} X_{i j} F_{c d} \lambda^{i}_{\alpha} \lambda^{j}_{\beta} {W}^{-3}+\frac{3}{8}{\rm i} F_{\hat{b} c} {W}^{-2} \nabla_{\hat{a}}{W} \nabla^{c}{W} - \frac{3}{32}\epsilon^{e}\,_{\hat{b} {e_{1}}}\,^{c d} (\Sigma_{\hat{a} e})^{\alpha \beta} F_{c d} \lambda_{i \alpha} \lambda^{i}_{\beta} {W}^{-3} \nabla^{{e_{1}}}{W} - \frac{3}{64}\epsilon^{e {e_{1}}}\,_{\hat{a} \hat{b}}\,^{c} (\Sigma_{e {e_{1}}})^{\alpha \beta} F_{c d} \lambda_{i \alpha} \lambda^{i}_{\beta} {W}^{-3} \nabla^{d}{W} - \frac{9}{256}{\rm i} \epsilon_{\hat{a} \hat{b} e}\,^{c d} (\Gamma^{e})^{\alpha \beta} W_{c d} \lambda_{i \alpha} \lambda^{i \rho} \lambda_{j \beta} \lambda^{j}_{\rho} {W}^{-3} - \frac{9}{128}{\rm i} (\Sigma_{\hat{a} \hat{b}})^{\alpha \beta} (\Sigma_{c d})^{\rho \lambda} F^{c d} \lambda_{i \alpha} \lambda^{i}_{\beta} \lambda_{j \rho} \lambda^{j}_{\lambda} {W}^{-4}+\frac{3}{64}{\rm i} \epsilon^{c d}\,_{\hat{a} \hat{b} e} (\Sigma_{c d})^{\alpha \beta} \lambda_{i \alpha} \lambda^{i \rho} \lambda_{j \rho} {W}^{-3} \nabla^{e}{\lambda^{j}_{\beta}} - \frac{3}{32}{\rm i} (\Gamma_{\hat{a}})^{\alpha \beta} \lambda_{i \alpha} \lambda^{i \rho} \lambda_{j \rho} {W}^{-3} \nabla_{\hat{b}}{\lambda^{j}_{\beta}} - \frac{21}{128}(\Sigma_{\hat{a} \hat{b}})^{\alpha \beta} X_{i j} X^{i}\,_{k} \lambda^{j}_{\alpha} \lambda^{k}_{\beta} {W}^{-3} - \frac{15}{64}(\Gamma_{\hat{a}})^{\alpha \beta} X_{i j} \lambda^{i}_{\alpha} \lambda^{j}_{\beta} {W}^{-3} \nabla_{\hat{b}}{W}+\frac{9}{64}{\rm i} (\Sigma_{\hat{a} \hat{b}})^{\alpha \beta} X_{i j} \lambda^{i \rho} \lambda^{j}_{\rho} \lambda_{k \alpha} \lambda^{k}_{\beta} {W}^{-4}%
 - \frac{9}{128}\epsilon^{e}\,_{\hat{a} \hat{b}}\,^{c d} (\Sigma_{e {e_{1}}})^{\alpha \beta} F_{c d} \lambda_{i \alpha} \lambda^{i}_{\beta} {W}^{-3} \nabla^{{e_{1}}}{W}+\frac{9}{64}\epsilon^{d e}\,_{\hat{a} {e_{1}}}\,^{c} (\Sigma_{d e})^{\alpha \beta} F_{\hat{b} c} \lambda_{i \alpha} \lambda^{i}_{\beta} {W}^{-3} \nabla^{{e_{1}}}{W} - \frac{9}{256}{\rm i} \epsilon_{\hat{a} \hat{b} e}\,^{c d} (\Gamma^{e})^{\alpha \beta} F_{c d} \lambda_{i \alpha} \lambda^{i \rho} \lambda_{j \beta} \lambda^{j}_{\rho} {W}^{-4} - \frac{9}{64}{\rm i} (\Sigma_{\hat{a} \hat{b}})^{\alpha \beta} X_{i j} \lambda^{i}_{\alpha} \lambda^{j \rho} \lambda_{k \beta} \lambda^{k}_{\rho} {W}^{-4}+\frac{9}{32}(\Sigma_{\hat{a} c})^{\alpha \beta} \lambda_{i \alpha} \lambda^{i}_{\beta} {W}^{-3} \nabla^{c}{W} \nabla_{\hat{b}}{W}+\frac{9}{64}{\rm i} (\Gamma_{\hat{a}})^{\alpha \beta} \lambda_{i \alpha} \lambda^{i \rho} \lambda_{j \beta} \lambda^{j}_{\rho} {W}^{-4} \nabla_{\hat{b}}{W}+\frac{9}{128}(\Sigma_{\hat{a} \hat{b}})^{\alpha \beta} \lambda_{i \alpha} \lambda^{i}_{\beta} \lambda^{\rho}_{j} \lambda^{j \lambda} \lambda_{k \rho} \lambda^{k}_{\lambda} {W}^{-5}
 \doublespacedmathend
 \end{adjustwidth}
 
\subsubsection{$J^{2}_{a i j, \log}$} \label{J2aijlogWComplete}

\begin{adjustwidth}{0cm}{5cm}
\doublespacedmathbegin
{} - \frac{9}{64}{\rm i} \lambda^{\alpha}_{\underline{i}} \nabla_{a}{X_{\underline{j} \alpha}}+\frac{3}{16}{\rm i} (\Sigma_{a b})^{\beta \alpha} \lambda_{\underline{i} \beta} \nabla^{b}{X_{\underline{j} \alpha}} - \frac{27}{256}{\rm i} \epsilon^{d e}\,_{a}\,^{b c} (\Sigma_{d e})^{\beta \alpha} W_{b c} \lambda_{\underline{i} \beta} X_{\underline{j} \alpha}+\frac{9}{64}{\rm i} (\Gamma^{b})^{\beta \alpha} W_{a b} \lambda_{\underline{i} \beta} X_{\underline{j} \alpha} - \frac{1}{8}{\rm i} W \nabla^{b}{\Phi_{a b \underline{i} \underline{j}}}+\frac{9}{512}(\Gamma_{a})^{\alpha \beta} W X_{\underline{i} \alpha} X_{\underline{j} \beta}+\frac{39}{1024}{\rm i} (\Gamma_{a})^{\alpha \beta} W^{b c} W_{b c \alpha \underline{i}} \lambda_{\underline{j} \beta} - \frac{69}{2048}{\rm i} \epsilon^{b c d e}\,_{a} W_{b c} W_{d e}\,^{\alpha}\,_{\underline{i}} \lambda_{\underline{j} \alpha}+\frac{39}{1024}{\rm i} \epsilon^{{e_{1}} b c d e} (\Sigma_{a {e_{1}}})^{\alpha \beta} W_{b c} W_{d e \alpha \underline{i}} \lambda_{\underline{j} \beta}+\frac{39}{512}{\rm i} W_{b}\,^{c} W_{d c}\,^{\alpha}\,_{\underline{i}} \epsilon^{b d e {e_{1}}}\,_{a} (\Sigma_{e {e_{1}}})_{\alpha}{}^{\beta} \lambda_{\underline{j} \beta}+\frac{39}{512}{\rm i} W_{a}\,^{b} W_{c b}\,^{\alpha}\,_{\underline{i}} (\Gamma^{c})_{\alpha}{}^{\beta} \lambda_{\underline{j} \beta} - \frac{39}{512}{\rm i} W_{b}\,^{c} W_{a c}\,^{\alpha}\,_{\underline{i}} (\Gamma^{b})_{\alpha}{}^{\beta} \lambda_{\underline{j} \beta} - \frac{1}{8}{\rm i} \Phi_{a b \underline{i} \underline{j}} \nabla^{b}{W}+\frac{9}{64}{\rm i} X_{\underline{i}}^{\alpha} \nabla_{a}{\lambda_{\underline{j} \alpha}} - \frac{3}{16}{\rm i} (\Sigma_{a b})^{\alpha \beta} X_{\underline{i} \alpha} \nabla^{b}{\lambda_{\underline{j} \beta}}+\frac{9}{32}{\rm i} \lambda^{\alpha}_{\underline{i}} \nabla^{b}{W_{a b \alpha \underline{j}}}+\frac{1}{32}{\rm i} \epsilon^{d e}\,_{a}\,^{b c} \Phi_{d e \underline{i} \underline{j}} F_{b c}+\frac{3}{16}{\rm i} X_{\underline{i} \underline{j}} \nabla^{b}{W_{a b}}+\frac{3}{16}{\rm i} W_{a b} \nabla^{b}{X_{\underline{i} \underline{j}}}%
 - \frac{3}{16}(\Gamma_{b})^{\alpha \beta} {W}^{-1} \nabla_{a}{\lambda_{\underline{i} \alpha}} \nabla^{b}{\lambda_{\underline{j} \beta}}+\frac{3}{32}(\Gamma_{a})^{\alpha \beta} {W}^{-1} \nabla_{b}{\lambda_{\underline{i} \alpha}} \nabla^{b}{\lambda_{\underline{j} \beta}} - \frac{3}{32}(\Sigma_{b c})^{\alpha \beta} F^{b c} \lambda_{\underline{i} \alpha} {W}^{-2} \nabla_{a}{\lambda_{\underline{j} \beta}}+\frac{3}{16}{\rm i} F_{a b} {W}^{-1} \nabla^{b}{X_{\underline{i} \underline{j}}} - \frac{15}{128}{\rm i} \epsilon^{b c d e}\,_{a} F_{b c} W_{d e}\,^{\alpha}\,_{\underline{i}} \lambda_{\underline{j} \alpha} {W}^{-1}+\frac{3}{64}{\rm i} (\Gamma_{a})^{\alpha \beta} F^{b c} W_{b c \alpha \underline{i}} \lambda_{\underline{j} \beta} {W}^{-1}+\frac{3}{64}{\rm i} \epsilon^{{e_{1}} b c d e} (\Sigma_{a {e_{1}}})^{\alpha \beta} F_{b c} W_{d e \alpha \underline{i}} \lambda_{\underline{j} \beta} {W}^{-1}+\frac{3}{32}{\rm i} F_{b}\,^{c} W_{d c}\,^{\alpha}\,_{\underline{i}} \epsilon^{b d e {e_{1}}}\,_{a} (\Sigma_{e {e_{1}}})_{\alpha}{}^{\beta} \lambda_{\underline{j} \beta} {W}^{-1}+\frac{3}{32}{\rm i} F_{a}\,^{b} W_{c b}\,^{\alpha}\,_{\underline{i}} (\Gamma^{c})_{\alpha}{}^{\beta} \lambda_{\underline{j} \beta} {W}^{-1} - \frac{3}{32}{\rm i} F_{b}\,^{c} W_{a c}\,^{\alpha}\,_{\underline{i}} (\Gamma^{b})_{\alpha}{}^{\beta} \lambda_{\underline{j} \beta} {W}^{-1} - \frac{9}{128}{\rm i} \epsilon^{d e}\,_{a}\,^{b c} (\Sigma_{d e})^{\beta \alpha} F_{b c} \lambda_{\underline{i} \beta} X_{\underline{j} \alpha} {W}^{-1}+\frac{45}{1024}(\Gamma_{a})^{\alpha \beta} W^{b c} W_{b c} \lambda_{\underline{i} \alpha} \lambda_{\underline{j} \beta} {W}^{-1} - \frac{27}{2048}\epsilon_{a}\,^{b c d e} W_{b c} W_{d e} \lambda^{\alpha}_{\underline{i}} \lambda_{\underline{j} \alpha} {W}^{-1} - \frac{9}{64}(\Sigma_{b c})^{\alpha \beta} W^{b c} \lambda_{\underline{i} \alpha} {W}^{-1} \nabla_{a}{\lambda_{\underline{j} \beta}} - \frac{9}{32}(\Sigma^{b}{}_{\, c})^{\alpha \beta} W_{a b} \lambda_{\underline{i} \alpha} {W}^{-1} \nabla^{c}{\lambda_{\underline{j} \beta}} - \frac{9}{32}(\Sigma_{a}{}^{\, b})^{\alpha \beta} W_{b c} \lambda_{\underline{i} \alpha} {W}^{-1} \nabla^{c}{\lambda_{\underline{j} \beta}}+\frac{9}{256}(\Gamma_{a})^{\alpha \beta} W^{b c} F_{b c} \lambda_{\underline{i} \alpha} \lambda_{\underline{j} \beta} {W}^{-2} - \frac{27}{256}(\Gamma^{c})^{\alpha \beta} W_{a}\,^{b} W_{c b} \lambda_{\underline{i} \alpha} \lambda_{\underline{j} \beta} {W}^{-1}+\frac{9}{512}\epsilon_{a}\,^{d e b c} W_{d e} F_{b c} \lambda^{\alpha}_{\underline{i}} \lambda_{\underline{j} \alpha} {W}^{-2} - \frac{9}{128}(\Gamma^{c})^{\alpha \beta} W_{c}\,^{b} F_{a b} \lambda_{\underline{i} \alpha} \lambda_{\underline{j} \beta} {W}^{-2}%
 - \frac{3}{256}(\Gamma_{a})^{\alpha \beta} F^{b c} F_{b c} \lambda_{\underline{i} \alpha} \lambda_{\underline{j} \beta} {W}^{-3}+\frac{3}{32}X_{\underline{i} k} \lambda^{\alpha}_{\underline{j}} {W}^{-2} \nabla_{a}{\lambda^{k}_{\alpha}} - \frac{3}{16}(\Sigma_{a b})^{\alpha \beta} X_{\underline{i} k} \lambda_{\underline{j} \alpha} {W}^{-2} \nabla^{b}{\lambda^{k}_{\beta}}+\frac{3}{16}{\rm i} X_{\underline{i} \underline{j}} {W}^{-1} \nabla^{b}{F_{a b}} - \frac{3}{16}{\rm i} X_{\underline{i} k} {W}^{-1} \nabla_{a}{X_{\underline{j}}\,^{k}}+\frac{9}{512}{\rm i} (\Gamma_{a})^{\beta \alpha} X_{\underline{i} k} \lambda^{k}_{\beta} X_{\underline{j} \alpha} {W}^{-1} - \frac{63}{512}{\rm i} (\Gamma_{a})^{\beta \alpha} X_{\underline{i} \underline{j}} \lambda_{k \beta} X^{k}_{\alpha} {W}^{-1} - \frac{9}{512}{\rm i} (\Gamma_{a})^{\beta \alpha} X_{\underline{i} k} \lambda_{\underline{j} \beta} X^{k}_{\alpha} {W}^{-1} - \frac{9}{128}(\Gamma_{a})^{\alpha \beta} X_{k l} X^{k l} \lambda_{\underline{i} \alpha} \lambda_{\underline{j} \beta} {W}^{-3}+\frac{9}{32}{\rm i} (\Sigma_{a b})^{\beta \alpha} \lambda_{\underline{i} \beta} X_{\underline{j} \alpha} {W}^{-1} \nabla^{b}{W} - \frac{9}{64}\lambda^{\alpha}_{\underline{i}} \lambda_{\underline{j} \alpha} {W}^{-1} \nabla^{b}{W_{a b}} - \frac{117}{512}(\Gamma_{a})^{\beta \alpha} \lambda_{\underline{i} \beta} \lambda^{\rho}_{\underline{j}} \lambda_{k \rho} X^{k}_{\alpha} {W}^{-2}+\frac{3}{32}(\Gamma_{b})^{\alpha \beta} \lambda_{\underline{i} \alpha} {W}^{-2} \nabla^{b}{W} \nabla_{a}{\lambda_{\underline{j} \beta}} - \frac{3}{16}(\Gamma_{a})^{\alpha \beta} \lambda_{\underline{i} \alpha} {W}^{-2} \nabla_{b}{W} \nabla^{b}{\lambda_{\underline{j} \beta}}+\frac{9}{16}{\rm i} W_{a b}\,^{\alpha}\,_{\underline{i}} \lambda_{\underline{j} \alpha} {W}^{-1} \nabla^{b}{W} - \frac{3}{32}{\rm i} (\Gamma_{a})^{\alpha \beta} (\Gamma_{b})^{\rho \lambda} \lambda_{\underline{i} \alpha} \lambda_{\underline{j} \beta} \lambda_{k \rho} {W}^{-3} \nabla^{b}{\lambda^{k}_{\lambda}} - \frac{3}{64}\epsilon_{a d e}\,^{b c} (\Gamma^{d})^{\alpha \beta} \lambda_{\underline{i} \alpha} \lambda_{\underline{j} \beta} {W}^{-2} \nabla^{e}{F_{b c}} - \frac{3}{32}\lambda^{\alpha}_{\underline{i}} \lambda_{\underline{j} \alpha} {W}^{-2} \nabla^{b}{F_{a b}} - \frac{9}{256}\epsilon_{a d e}\,^{b c} (\Gamma^{d})^{\alpha \beta} W_{b c} \lambda_{\underline{i} \alpha} \lambda_{\underline{j} \beta} {W}^{-2} \nabla^{e}{W}+\frac{9}{128}\epsilon_{a d e}\,^{b c} (\Gamma^{d})^{\alpha \beta} \lambda_{\underline{i} \alpha} \lambda_{\underline{j} \beta} {W}^{-1} \nabla^{e}{W_{b c}}%
+\frac{3}{32}\lambda^{\alpha}_{\underline{i}} \lambda_{k \alpha} {W}^{-2} \nabla_{a}{X_{\underline{j}}\,^{k}}+\frac{3}{16}(\Sigma_{a b})^{\alpha \beta} \lambda_{\underline{i} \alpha} \lambda_{k \beta} {W}^{-2} \nabla^{b}{X_{\underline{j}}\,^{k}} - \frac{3}{32}(\Gamma_{b})^{\alpha \beta} \lambda_{\underline{i} \alpha} \lambda_{\underline{j} \beta} {W}^{-2} \nabla^{b}{\nabla_{a}{W}}+\frac{3}{32}(\Gamma_{b})^{\alpha \beta} \lambda_{\underline{i} \alpha} \lambda_{\underline{j} \beta} {W}^{-2} \nabla_{a}{\nabla^{b}{W}} - \frac{3}{32}(\Gamma_{a})^{\alpha \beta} \lambda_{\underline{i} \alpha} \lambda_{\underline{j} \beta} {W}^{-2} \nabla_{b}{\nabla^{b}{W}}+\frac{9}{256}\epsilon^{d e}\,_{a}\,^{b c} (\Sigma_{d e})^{\alpha \beta} X_{\underline{i} k} W_{b c} \lambda_{\underline{j} \alpha} \lambda^{k}_{\beta} {W}^{-2} - \frac{9}{128}(\Gamma_{a})^{\beta \rho} \lambda_{\underline{i} \beta} \lambda_{k \rho} \lambda^{k \alpha} X_{\underline{j} \alpha} {W}^{-2} - \frac{9}{64}(\Gamma_{a})^{\beta \rho} \lambda_{\underline{i} \beta} \lambda_{\underline{j} \rho} \lambda^{\alpha}_{k} X^{k}_{\alpha} {W}^{-2}+\frac{9}{64}(\Gamma_{a})^{\beta \rho} \lambda_{\underline{i} \beta} \lambda^{\alpha}_{\underline{j}} \lambda_{k \rho} X^{k}_{\alpha} {W}^{-2} - \frac{13}{256}\epsilon^{b c d e}\,_{a} \Phi_{b c \underline{i} k} (\Sigma_{d e})^{\alpha \beta} \lambda_{\underline{j} \alpha} \lambda^{k}_{\beta} {W}^{-1} - \frac{3}{8}\Phi_{a b \underline{i} k} (\Gamma^{b})^{\alpha \beta} \lambda_{\underline{j} \alpha} \lambda^{k}_{\beta} {W}^{-1}+\frac{3}{16}(\Gamma_{b})^{\alpha \beta} \lambda_{\underline{i} \alpha} {W}^{-1} \nabla^{b}{\nabla_{a}{\lambda_{\underline{j} \beta}}} - \frac{3}{16}\epsilon^{b c}\,_{a d e} (\Sigma_{b c})^{\alpha \beta} \lambda_{\underline{i} \alpha} {W}^{-1} \nabla^{d}{\nabla^{e}{\lambda_{\underline{j} \beta}}} - \frac{3}{16}(\Gamma_{b})^{\alpha \beta} \lambda_{\underline{i} \alpha} {W}^{-1} \nabla_{a}{\nabla^{b}{\lambda_{\underline{j} \beta}}}+\frac{3}{16}(\Gamma_{a})^{\alpha \beta} \lambda_{\underline{i} \alpha} {W}^{-1} \nabla_{b}{\nabla^{b}{\lambda_{\underline{j} \beta}}} - \frac{11}{512}\epsilon^{b c d e}\,_{a} \Phi_{b c \underline{i} \underline{j}} (\Sigma_{d e})^{\alpha \beta} \lambda_{k \alpha} \lambda^{k}_{\beta} {W}^{-1} - \frac{9}{512}(\Gamma_{a})^{\alpha \beta} Y \lambda_{\underline{i} \alpha} \lambda_{\underline{j} \beta} {W}^{-1}+\frac{9}{128}(\Gamma_{a})^{\alpha \beta} \lambda_{\underline{i} \alpha} \lambda_{\underline{j} \beta} {W}^{-3} \nabla_{b}{W} \nabla^{b}{W} - \frac{3}{16}F_{a b} \lambda^{\alpha}_{\underline{i}} {W}^{-2} \nabla^{b}{\lambda_{\underline{j} \alpha}}+\frac{3}{16}(\Sigma_{a b})^{\alpha \beta} X_{\underline{i} k} \lambda^{k}_{\alpha} {W}^{-2} \nabla^{b}{\lambda_{\underline{j} \beta}}%
+\frac{3}{16}{\rm i} (\Sigma_{a b})^{\alpha \beta} \lambda_{\underline{i} \alpha} \lambda_{k \beta} \lambda^{k \rho} {W}^{-3} \nabla^{b}{\lambda_{\underline{j} \rho}}+\frac{3}{32}(\Sigma_{a b})^{\alpha \beta} \lambda_{k \alpha} \lambda^{k}_{\beta} {W}^{-2} \nabla^{b}{X_{\underline{i} \underline{j}}} - \frac{9}{256}\epsilon^{d e}\,_{a}\,^{b c} (\Sigma_{d e})^{\alpha \beta} X_{\underline{i} \underline{j}} W_{b c} \lambda_{k \alpha} \lambda^{k}_{\beta} {W}^{-2}+\frac{3}{128}W_{b c}\,^{\alpha}\,_{\underline{i}} \epsilon^{b c d e}\,_{a} (\Sigma_{d e})^{\beta \rho} \lambda_{\underline{j} \beta} \lambda_{k \alpha} \lambda^{k}_{\rho} {W}^{-2}+\frac{3}{64}W_{a b}\,^{\alpha}\,_{\underline{i}} (\Gamma^{b})^{\beta \rho} \lambda_{\underline{j} \beta} \lambda_{k \alpha} \lambda^{k}_{\rho} {W}^{-2}+\frac{9}{512}(\Gamma_{a})^{\beta \alpha} \lambda^{\rho}_{\underline{i}} \lambda_{k \beta} \lambda^{k}_{\rho} X_{\underline{j} \alpha} {W}^{-2}+\frac{9}{128}{\rm i} (\Sigma_{b c})^{\rho \lambda} (\Gamma_{a})^{\alpha \beta} W^{b c} \lambda_{\underline{i} \alpha} \lambda_{\underline{j} \beta} \lambda_{k \rho} \lambda^{k}_{\lambda} {W}^{-3} - \frac{3}{32}{\rm i} (\Gamma_{a})^{\alpha \beta} (\Gamma_{b})^{\rho \lambda} \lambda_{\underline{i} \alpha} \lambda_{k \beta} \lambda^{k}_{\rho} {W}^{-3} \nabla^{b}{\lambda_{\underline{j} \lambda}}+\frac{9}{128}{\rm i} (\Sigma_{b c})^{\rho \lambda} (\Gamma_{a})^{\alpha \beta} W^{b c} \lambda_{\underline{i} \rho} \lambda_{\underline{j} \alpha} \lambda_{k \lambda} \lambda^{k}_{\beta} {W}^{-3} - \frac{3}{16}{\rm i} X_{\underline{i} \underline{j}} F_{a b} {W}^{-2} \nabla^{b}{W}+\frac{3}{128}\epsilon^{d e}\,_{a}\,^{b c} (\Sigma_{d e})^{\alpha \beta} X_{\underline{i} k} F_{b c} \lambda_{\underline{j} \alpha} \lambda^{k}_{\beta} {W}^{-3} - \frac{3}{64}(\Gamma^{b})^{\alpha \beta} X_{\underline{i} k} F_{a b} \lambda_{\underline{j} \alpha} \lambda^{k}_{\beta} {W}^{-3}+\frac{9}{64}F_{a b} \lambda^{\alpha}_{\underline{i}} \lambda_{\underline{j} \alpha} {W}^{-3} \nabla^{b}{W} - \frac{3}{128}\epsilon_{a d e}\,^{b c} (\Gamma^{d})^{\alpha \beta} F_{b c} \lambda_{\underline{i} \alpha} \lambda_{\underline{j} \beta} {W}^{-3} \nabla^{e}{W}+\frac{9}{128}{\rm i} \epsilon^{d e}\,_{a}\,^{b c} (\Sigma_{d e})^{\alpha \beta} W_{b c} \lambda_{\underline{i} \alpha} \lambda^{\rho}_{\underline{j}} \lambda_{k \beta} \lambda^{k}_{\rho} {W}^{-3} - \frac{9}{64}{\rm i} (\Gamma^{b})^{\alpha \beta} W_{a b} \lambda_{\underline{i} \alpha} \lambda^{\rho}_{\underline{j}} \lambda_{k \beta} \lambda^{k}_{\rho} {W}^{-3}+\frac{9}{128}{\rm i} (\Sigma_{b c})^{\rho \lambda} (\Gamma_{a})^{\alpha \beta} F^{b c} \lambda_{\underline{i} \alpha} \lambda_{\underline{j} \beta} \lambda_{k \rho} \lambda^{k}_{\lambda} {W}^{-4} - \frac{3}{32}X_{\underline{i} \underline{j}} \lambda^{\alpha}_{k} {W}^{-2} \nabla_{a}{\lambda^{k}_{\alpha}}+\frac{3}{16}(\Sigma_{a b})^{\alpha \beta} X_{\underline{i} \underline{j}} \lambda_{k \alpha} {W}^{-2} \nabla^{b}{\lambda^{k}_{\beta}} - \frac{3}{16}{\rm i} \lambda^{\alpha}_{\underline{i}} \lambda^{\beta}_{\underline{j}} \lambda_{k \alpha} {W}^{-3} \nabla_{a}{\lambda^{k}_{\beta}}%
 - \frac{3}{8}{\rm i} (\Sigma_{a b})^{\alpha \beta} \lambda_{\underline{i} \alpha} \lambda^{\rho}_{\underline{j}} \lambda_{k \rho} {W}^{-3} \nabla^{b}{\lambda^{k}_{\beta}} - \frac{27}{512}(\Gamma_{a})^{\beta \alpha} \lambda^{\rho}_{\underline{i}} \lambda_{\underline{j} \rho} \lambda_{k \beta} X^{k}_{\alpha} {W}^{-2} - \frac{3}{16}(\Gamma_{a})^{\alpha \beta} X_{\underline{i} k} X^{k}\,_{l} \lambda_{\underline{j} \alpha} \lambda^{l}_{\beta} {W}^{-3}+\frac{3}{64}X_{\underline{i} k} \lambda^{\alpha}_{\underline{j}} \lambda^{k}_{\alpha} {W}^{-3} \nabla_{a}{W}+\frac{9}{32}(\Sigma_{a b})^{\alpha \beta} X_{\underline{i} k} \lambda_{\underline{j} \alpha} \lambda^{k}_{\beta} {W}^{-3} \nabla^{b}{W} - \frac{9}{64}{\rm i} (\Gamma_{a})^{\alpha \beta} X_{k l} \lambda_{\underline{i} \alpha} \lambda_{\underline{j} \beta} \lambda^{k \rho} \lambda^{l}_{\rho} {W}^{-4}+\frac{3}{16}{\rm i} (\Sigma_{a b})^{\alpha \beta} \lambda^{\rho}_{\underline{i}} \lambda_{k \alpha} \lambda^{k}_{\rho} {W}^{-3} \nabla^{b}{\lambda_{\underline{j} \beta}} - \frac{3}{128}\epsilon^{d e}\,_{a}\,^{b c} (\Sigma_{d e})^{\alpha \beta} X_{\underline{i} \underline{j}} F_{b c} \lambda_{k \alpha} \lambda^{k}_{\beta} {W}^{-3}+\frac{3}{512}\epsilon_{a}\,^{b c d e} F_{b c} F_{d e} \lambda^{\alpha}_{\underline{i}} \lambda_{\underline{j} \alpha} {W}^{-3} - \frac{3}{64}(\Gamma^{c})^{\alpha \beta} F_{a}\,^{b} F_{c b} \lambda_{\underline{i} \alpha} \lambda_{\underline{j} \beta} {W}^{-3}+\frac{9}{128}{\rm i} \epsilon^{d e}\,_{a}\,^{b c} (\Sigma_{d e})^{\alpha \beta} F_{b c} \lambda_{\underline{i} \alpha} \lambda^{\rho}_{\underline{j}} \lambda_{k \beta} \lambda^{k}_{\rho} {W}^{-4} - \frac{9}{64}{\rm i} (\Gamma^{b})^{\alpha \beta} F_{a b} \lambda_{\underline{i} \alpha} \lambda^{\rho}_{\underline{j}} \lambda_{k \beta} \lambda^{k}_{\rho} {W}^{-4} - \frac{3}{32}(\Sigma_{a b})^{\alpha \beta} X_{\underline{i} \underline{j}} \lambda_{k \alpha} \lambda^{k}_{\beta} {W}^{-3} \nabla^{b}{W}+\frac{3}{128}(\Gamma_{a})^{\alpha \beta} X_{\underline{i} k} X_{\underline{j} l} \lambda^{k}_{\alpha} \lambda^{l}_{\beta} {W}^{-3} - \frac{9}{64}{\rm i} (\Gamma_{a})^{\alpha \beta} X_{\underline{i} k} \lambda_{\underline{j} \alpha} \lambda^{k \rho} \lambda_{l \beta} \lambda^{l}_{\rho} {W}^{-4} - \frac{3}{64}(\Gamma_{b})^{\alpha \beta} \lambda_{\underline{i} \alpha} \lambda_{\underline{j} \beta} {W}^{-3} \nabla_{a}{W} \nabla^{b}{W}+\frac{9}{32}{\rm i} (\Sigma_{a b})^{\alpha \beta} \lambda_{\underline{i} \alpha} \lambda^{\rho}_{\underline{j}} \lambda_{k \beta} \lambda^{k}_{\rho} {W}^{-4} \nabla^{b}{W} - \frac{9}{128}(\Gamma_{a})^{\alpha \beta} \lambda_{\underline{i} \alpha} \lambda_{\underline{j} \beta} \lambda^{\rho}_{k} \lambda^{k \lambda} \lambda_{l \rho} \lambda^{l}_{\lambda} {W}^{-5}
 \doublespacedmathend
 \end{adjustwidth}

\subsubsection{$J^3_{a i \alpha, \log}$} \label{J3logWComplete}

\begin{adjustwidth}{0cm}{5cm}
\doublespacedmathbegin
{}\frac{9}{256}{\rm i} (\Gamma_{a})^{\beta}{}_{\lambda} F_{\alpha \beta} \nabla^{\lambda \rho}{X_{i \rho}}+\frac{2943}{8192}{\rm i} (\Gamma_{a})^{\beta}{}_{\lambda} W W_{\alpha \beta} \nabla^{\lambda \rho}{X_{i \rho}}+\frac{189}{4096}{\rm i} (\Gamma_{a})_{\rho \lambda} \nabla^{\rho}\,_{\alpha}{W} \nabla^{\lambda \beta}{X_{i \beta}} - \frac{101}{512}(\Gamma_{a})^{\beta}{}_{\rho} \lambda_{j \beta} \nabla^{\rho \lambda}{\Phi_{\alpha \lambda i}\,^{j}} - \frac{243}{16384}(\Gamma_{a})^{\beta}{}_{\rho} \lambda_{i \beta} \nabla^{\rho}\,_{\alpha}{Y}+\frac{117}{4096}(\Gamma_{a})^{\lambda}{}_{\gamma} \lambda_{i \lambda} \nabla^{\gamma \beta}{\nabla_{\alpha}\,^{\rho}{W_{\beta \rho}}} - \frac{6615}{4096}(\Gamma_{a})^{\lambda}{}_{\gamma} W^{\beta}\,_{\rho} \lambda_{i \lambda} \nabla^{\gamma \rho}{W_{\alpha \beta}}+\frac{5013}{8192}(\Gamma_{a})^{\lambda}{}_{\gamma} W_{\alpha}\,^{\beta} \lambda_{i \lambda} \nabla^{\gamma \rho}{W_{\beta \rho}}+\frac{639}{4096}(\Gamma_{a})^{\lambda}{}_{\gamma} W^{\beta \rho} \lambda_{i \lambda} \nabla^{\gamma}\,_{\alpha}{W_{\beta \rho}}+\frac{81}{4096}(\Gamma_{a})^{\rho}{}_{\lambda} \lambda_{i \rho} \nabla^{\lambda}\,_{\gamma}{\nabla^{\gamma \beta}{W_{\alpha \beta}}}+\frac{3243}{1024}\Phi^{\rho \beta}\,_{i j} (\Gamma_{a})_{\rho}{}^{\lambda} W_{\alpha \beta} \lambda^{j}_{\lambda} - \frac{225}{4096}(\Gamma_{a})^{\beta \rho} W_{\alpha \beta} Y \lambda_{i \rho}+\frac{1845}{8192}(\Gamma_{a})^{\gamma \rho} W_{\alpha \beta} \lambda_{i \gamma} \nabla^{\beta \lambda}{W_{\rho \lambda}} - \frac{22689}{2048}(\Gamma_{a})^{\rho \gamma} W_{\alpha}\,^{\beta} W_{\rho}\,^{\lambda} W_{\beta \lambda} \lambda_{i \gamma}+\frac{18819}{4096}(\Gamma_{a})^{\beta \gamma} W_{\alpha \beta} W^{\rho \lambda} W_{\rho \lambda} \lambda_{i \gamma} - \frac{1527}{5120}\Phi^{\beta \rho}\,_{i j} (\Gamma_{a})_{\alpha}{}^{\lambda} W_{\beta \rho} \lambda^{j}_{\lambda}+\frac{16677}{20480}(\Gamma_{a})_{\alpha}{}^{\gamma} W^{\beta}\,_{\rho} \lambda_{i \gamma} \nabla^{\rho \lambda}{W_{\beta \lambda}} - \frac{1653}{1024}\Phi_{\alpha}\,^{\beta}\,_{i j} (\Gamma_{a})^{\rho \lambda} W_{\beta \rho} \lambda^{j}_{\lambda} - \frac{2043}{4096}(\Gamma_{a})^{\beta \gamma} W_{\beta \rho} \lambda_{i \gamma} \nabla^{\rho \lambda}{W_{\alpha \lambda}}%
 - \frac{1341}{2048}(\Gamma_{a})^{\beta \gamma} W_{\beta}\,^{\rho} \lambda_{i \gamma} \nabla_{\alpha}\,^{\lambda}{W_{\rho \lambda}} - \frac{3}{512}{\rm i} (\Gamma_{a})^{\lambda \beta} \lambda_{j \lambda} X_{i}^{\rho} W_{\alpha \beta \rho}\,^{j}+\frac{33}{4096}{\rm i} (\Gamma_{a})^{\rho \beta} \lambda_{j \rho} X_{i \alpha} X^{j}_{\beta}+\frac{321}{4096}{\rm i} (\Gamma_{a})^{\rho \beta} \lambda_{j \rho} X^{j}_{\alpha} X_{i \beta}+\frac{1101}{20480}{\rm i} (\Gamma_{a})_{\alpha}{}^{\rho} \lambda_{j \rho} X_{i}^{\beta} X^{j}_{\beta} - \frac{63}{512}{\rm i} (\Gamma_{a})^{\rho \beta} \lambda_{i \rho} X_{j \alpha} X^{j}_{\beta} - \frac{603}{4096}(\Gamma_{a})^{\gamma \lambda} W^{\beta}\,_{\rho} \lambda_{i \gamma} \nabla_{\alpha}\,^{\rho}{W_{\lambda \beta}} - \frac{189}{64}{\rm i} (\Gamma_{a})^{\rho \beta} W_{\rho}\,^{\lambda} F_{\alpha \beta} X_{i \lambda}+\frac{891}{512}{\rm i} (\Gamma_{a})^{\beta \rho} W W_{\alpha \beta} W_{\rho}\,^{\lambda} X_{i \lambda} - \frac{2439}{8192}{\rm i} (\Gamma_{a})^{\beta}{}_{\lambda} W_{\beta}\,^{\rho} X_{i \rho} \nabla^{\lambda}\,_{\alpha}{W}+\frac{2121}{10240}{\rm i} (\Gamma_{a})_{\alpha \rho} X_{i j} \nabla^{\rho \beta}{X^{j}_{\beta}} - \frac{81}{256}{\rm i} (\Gamma_{a})^{\lambda \beta} \lambda_{i \lambda} X_{j}^{\rho} W_{\alpha \beta \rho}\,^{j} - \frac{117}{512}{\rm i} (\Gamma_{a})_{\alpha}{}^{\beta} X_{i j} W_{\beta}\,^{\rho} X^{j}_{\rho} - \frac{3}{8}(\Gamma_{a})^{\beta}{}_{\rho} \lambda_{j \alpha} \nabla^{\rho \lambda}{\Phi_{\lambda \beta i}\,^{j}}+\frac{243}{2048}{\rm i} (\Gamma_{a})^{\beta}{}_{\gamma} W \nabla^{\gamma \rho}{\nabla_{\alpha}\,^{\lambda}{W_{\beta \rho \lambda i}}}+\frac{2361}{4096}{\rm i} (\Gamma_{a})^{\beta}{}_{\gamma} W W_{\alpha \beta}\,^{\rho}\,_{i} \nabla^{\gamma \lambda}{W_{\rho \lambda}} - \frac{1731}{4096}{\rm i} (\Gamma_{a})^{\lambda}{}_{\gamma} W W^{\beta}\,_{\rho} \nabla^{\gamma \rho}{W_{\alpha \lambda \beta i}} - \frac{543}{1024}{\rm i} (\Gamma_{a})^{\lambda}{}_{\gamma} W W_{\alpha}\,^{\beta}\,_{\rho i} \nabla^{\gamma \rho}{W_{\lambda \beta}}+\frac{129}{4096}{\rm i} (\Gamma_{a})^{\beta}{}_{\gamma} W W_{\beta}\,^{\rho \lambda}\,_{i} \nabla^{\gamma}\,_{\alpha}{W_{\rho \lambda}}+\frac{579}{4096}{\rm i} (\Gamma_{a})^{\lambda}{}_{\gamma} W W^{\beta \rho} \nabla^{\gamma}\,_{\alpha}{W_{\lambda \beta \rho i}}%
 - \frac{651}{10240}{\rm i} (\Gamma_{a})_{\alpha \gamma} W W^{\beta \rho}\,_{\lambda i} \nabla^{\gamma \lambda}{W_{\beta \rho}} - \frac{7251}{40960}{\rm i} (\Gamma_{a})_{\alpha \gamma} W W^{\beta \rho} \nabla^{\gamma \lambda}{W_{\beta \rho \lambda i}} - \frac{1917}{4096}{\rm i} (\Gamma_{a})^{\beta}{}_{\gamma} W W_{\beta}\,^{\rho}\,_{\lambda i} \nabla^{\gamma \lambda}{W_{\alpha \rho}}+\frac{9969}{8192}{\rm i} (\Gamma_{a})^{\rho}{}_{\gamma} W W_{\alpha}\,^{\beta} \nabla^{\gamma \lambda}{W_{\rho \beta \lambda i}}+\frac{1899}{8192}{\rm i} (\Gamma_{a})^{\beta}{}_{\lambda} W X_{i \alpha} \nabla^{\lambda \rho}{W_{\beta \rho}} - \frac{9}{8192}{\rm i} (\Gamma_{a})^{\beta}{}_{\lambda} W W_{\beta \rho} \nabla^{\lambda \rho}{X_{i \alpha}} - \frac{159}{2048}{\rm i} (\Gamma_{a})^{\beta}{}_{\lambda} W \nabla^{\lambda}\,_{\gamma}{\nabla^{\gamma \rho}{W_{\alpha \beta \rho i}}}+\frac{3}{32}{\rm i} (\Gamma_{a})_{\lambda \gamma} W \nabla^{\lambda \beta}{\nabla^{\gamma \rho}{W_{\alpha \beta \rho i}}}+\frac{81}{2048}{\rm i} (\Gamma_{a})^{\beta}{}_{\rho} W \nabla^{\rho}\,_{\lambda}{\nabla_{\alpha}\,^{\lambda}{X_{i \beta}}}+\frac{153}{4096}{\rm i} (\Gamma_{a})_{\rho \lambda} W \nabla^{\rho \beta}{\nabla^{\lambda}\,_{\alpha}{X_{i \beta}}}+\frac{2169}{4096}{\rm i} (\Gamma_{a})^{\rho}{}_{\lambda} W X_{i \rho} \nabla^{\lambda \beta}{W_{\alpha \beta}}+\frac{1845}{4096}{\rm i} (\Gamma_{a})^{\rho}{}_{\lambda} W W_{\alpha \beta} \nabla^{\lambda \beta}{X_{i \rho}}+\frac{675}{8192}{\rm i} (\Gamma_{a})^{\beta}{}_{\lambda} W X_{i \rho} \nabla^{\lambda \rho}{W_{\alpha \beta}}+\frac{9}{4096}{\rm i} (\Gamma_{a})_{\rho \lambda} W \nabla^{\rho}\,_{\alpha}{\nabla^{\lambda \beta}{X_{i \beta}}}+\frac{441}{10240}{\rm i} (\Gamma_{a})_{\alpha \rho} W \nabla^{\rho}\,_{\lambda}{\nabla^{\lambda \beta}{X_{i \beta}}} - \frac{2547}{8192}{\rm i} (\Gamma_{a})^{\beta}{}_{\lambda} W X_{i}^{\rho} \nabla^{\lambda}\,_{\alpha}{W_{\beta \rho}}+\frac{3087}{8192}{\rm i} (\Gamma_{a})^{\beta}{}_{\lambda} W W_{\beta}\,^{\rho} \nabla^{\lambda}\,_{\alpha}{X_{i \rho}}+\frac{1593}{5120}{\rm i} (\Gamma_{a})_{\alpha \lambda} W X_{i}^{\beta} \nabla^{\lambda \rho}{W_{\beta \rho}} - \frac{279}{1024}{\rm i} (\Gamma_{a})_{\alpha \lambda} W W^{\beta}\,_{\rho} \nabla^{\lambda \rho}{X_{i \beta}}+\frac{8949}{40960}{\rm i} (\Gamma_{a})_{\alpha}{}^{\lambda} W W^{\beta}\,_{\rho} \nabla^{\rho \gamma}{W_{\lambda \beta \gamma i}}%
+\frac{63}{256}{\rm i} (\Gamma_{a})_{\alpha}{}^{\beta} W W_{\beta}\,^{\rho} W^{\lambda \gamma} W_{\rho \lambda \gamma i}+\frac{16893}{40960}{\rm i} (\Gamma_{a})_{\alpha}{}^{\beta} W W_{\beta}\,^{\rho} W_{\rho}\,^{\lambda} X_{i \lambda} - \frac{18099}{81920}{\rm i} (\Gamma_{a})_{\alpha}{}^{\lambda} W W^{\beta \rho} W_{\beta \rho} X_{i \lambda}+\frac{6669}{40960}{\rm i} (\Gamma_{a})_{\alpha}{}^{\beta} W W_{\beta \rho} \nabla^{\rho \lambda}{X_{i \lambda}}+\frac{45}{128}{\rm i} (\Gamma_{a})^{\beta \lambda} W W_{\beta \rho} \nabla^{\rho \gamma}{W_{\alpha \lambda \gamma i}} - \frac{135}{256}{\rm i} (\Gamma_{a})^{\rho \gamma} W W_{\alpha}\,^{\beta} W_{\rho}\,^{\lambda} W_{\gamma \beta \lambda i}+\frac{225}{128}{\rm i} (\Gamma_{a})^{\beta \gamma} W W_{\alpha \beta} W^{\rho \lambda} W_{\gamma \rho \lambda i} - \frac{513}{256}{\rm i} (\Gamma_{a})^{\beta \gamma} W W_{\beta}\,^{\rho} W_{\rho}\,^{\lambda} W_{\alpha \gamma \lambda i}+\frac{5007}{8192}{\rm i} (\Gamma_{a})^{\beta \lambda} W W_{\beta}\,^{\rho} \nabla_{\alpha}\,^{\gamma}{W_{\lambda \rho \gamma i}} - \frac{9}{8192}{\rm i} (\Gamma_{a})^{\beta \lambda} W W_{\beta \rho} \nabla_{\alpha}\,^{\rho}{X_{i \lambda}}+\frac{7767}{8192}{\rm i} (\Gamma_{a})^{\rho \lambda} W W_{\alpha}\,^{\beta} W_{\rho \beta} X_{i \lambda} - \frac{12969}{16384}{\rm i} (\Gamma_{a})^{\beta \lambda} W W_{\beta}\,^{\rho} W_{\lambda \rho} X_{i \alpha}+\frac{1377}{8192}{\rm i} (\Gamma_{a})_{\rho \lambda} W W_{\alpha}\,^{\beta} \nabla^{\rho \lambda}{X_{i \beta}} - \frac{9}{80}{\rm i} \Phi^{\beta \rho}\,_{i j} (\Gamma_{a})_{\alpha}{}^{\lambda} W W_{\beta \rho \lambda}\,^{j} - \frac{27}{64}{\rm i} \Phi_{\alpha}\,^{\rho}\,_{i j} (\Gamma_{a})_{\rho}{}^{\beta} W X^{j}_{\beta} - \frac{27}{320}{\rm i} \Phi^{\rho \beta}\,_{i j} (\Gamma_{a})_{\alpha \rho} W X^{j}_{\beta}+\frac{9}{16}{\rm i} \Phi^{\lambda \beta}\,_{i j} (\Gamma_{a})_{\lambda}{}^{\rho} W W_{\alpha \beta \rho}\,^{j} - \frac{1899}{8192}{\rm i} (\Gamma_{a})^{\lambda \beta} W X_{i \lambda} \nabla_{\alpha}\,^{\rho}{W_{\beta \rho}}+\frac{675}{4096}{\rm i} (\Gamma_{a})^{\beta \gamma} W W_{\beta}\,^{\rho}\,_{\lambda i} \nabla_{\alpha}\,^{\lambda}{W_{\gamma \rho}}+\frac{189}{4096}{\rm i} (\Gamma_{a})_{\alpha}{}^{\beta} W W_{\beta}\,^{\rho}\,_{\lambda i} \nabla^{\lambda \gamma}{W_{\rho \gamma}}%
 - \frac{27}{64}{\rm i} (\Gamma_{a})^{\beta \lambda} W W_{\alpha \beta \rho i} \nabla^{\rho \gamma}{W_{\lambda \gamma}}+\frac{531}{8192}{\rm i} (\Gamma_{a})_{\rho \lambda} W X_{i}^{\beta} \nabla^{\rho \lambda}{W_{\alpha \beta}}+\frac{2583}{40960}{\rm i} (\Gamma_{a})_{\alpha}{}^{\beta} W X_{i \lambda} \nabla^{\lambda \rho}{W_{\beta \rho}} - \frac{393}{4096}{\rm i} (\Gamma_{a})^{\beta \rho} \lambda_{j \alpha} X_{i \beta} X^{j}_{\rho}+\frac{231}{512}{\rm i} (\Gamma_{a})^{\lambda \beta} \lambda_{j \lambda} X^{j \rho} W_{\alpha \beta \rho i}+\frac{333}{128}{\rm i} (\Gamma_{a})^{\beta \gamma} W^{\rho \lambda} F_{\alpha \beta} W_{\gamma \rho \lambda i} - \frac{2061}{4096}{\rm i} (\Gamma_{a})^{\lambda}{}_{\gamma} W^{\beta \rho} W_{\lambda \beta \rho i} \nabla^{\gamma}\,_{\alpha}{W} - \frac{63}{128}(\Gamma_{a})^{\beta \gamma} C_{\alpha \beta}\,^{\rho \lambda} W_{\rho \lambda} \lambda_{i \gamma} - \frac{81}{1280}{\rm i} (\Gamma_{a})_{\alpha}{}^{\lambda} X_{i j} W^{\beta \rho} W_{\lambda \beta \rho}\,^{j}+\frac{39}{512}{\rm i} (\Gamma_{a})^{\lambda \beta} \lambda^{\rho}_{j} X_{i \lambda} W_{\alpha \beta \rho}\,^{j} - \frac{1461}{20480}{\rm i} (\Gamma_{a})_{\alpha}{}^{\beta} \lambda^{\rho}_{j} X_{i \beta} X^{j}_{\rho} - \frac{441}{1024}{\rm i} (\Gamma_{a})^{\rho \lambda} W_{\rho}\,^{\beta} F_{\alpha \beta} X_{i \lambda} - \frac{1701}{8192}{\rm i} (\Gamma_{a})^{\beta \lambda} W_{\beta \rho} X_{i \lambda} \nabla_{\alpha}\,^{\rho}{W} - \frac{27}{32}\Phi_{\alpha}\,^{\lambda}\,_{i j} (\Gamma_{a})_{\lambda}{}^{\beta} W_{\beta}\,^{\rho} \lambda^{j}_{\rho} - \frac{45}{4096}(\Gamma_{a})_{\alpha}{}^{\beta} W_{\beta}\,^{\rho} Y \lambda_{i \rho} - \frac{639}{1024}(\Gamma_{a})^{\beta \lambda} W_{\beta}\,^{\rho} \lambda_{i \rho} \nabla_{\alpha}\,^{\gamma}{W_{\lambda \gamma}} - \frac{243}{64}(\Gamma_{a})^{\rho \lambda} W_{\alpha}\,^{\beta} W_{\rho \beta} W_{\lambda}\,^{\gamma} \lambda_{i \gamma}+\frac{16209}{10240}(\Gamma_{a})_{\alpha}{}^{\beta} W_{\beta}\,^{\rho} W^{\lambda \gamma} W_{\lambda \gamma} \lambda_{i \rho}+\frac{297}{1024}(\Gamma_{a})^{\beta}{}_{\gamma} W_{\beta}\,^{\rho} \lambda_{i \rho} \nabla^{\gamma \lambda}{W_{\alpha \lambda}}+\frac{15}{512}{\rm i} (\Gamma_{a})^{\lambda \beta} \lambda^{\rho}_{i} X_{j \lambda} W_{\alpha \beta \rho}\,^{j}%
 - \frac{63}{2560}{\rm i} (\Gamma_{a})_{\alpha}{}^{\beta} \lambda^{\rho}_{i} X_{j \beta} X^{j}_{\rho} - \frac{585}{512}{\rm i} (\Gamma_{a})^{\beta \rho} X_{i j} W_{\alpha \beta} X^{j}_{\rho}+\frac{3}{64}{\rm i} (\Gamma_{a})^{\beta \lambda} \lambda^{\gamma}_{j} W_{\alpha \beta}\,^{\rho j} W_{\lambda \gamma \rho i}+\frac{3}{64}{\rm i} (\Gamma_{a})_{\alpha}{}^{\beta} \lambda^{\rho}_{j} X^{j \lambda} W_{\beta \rho \lambda i} - \frac{111}{512}{\rm i} (\Gamma_{a})^{\lambda \beta} \lambda^{\rho}_{j} X^{j}_{\lambda} W_{\alpha \beta \rho i} - \frac{819}{256}{\rm i} (\Gamma_{a})^{\rho \gamma} W_{\rho}\,^{\lambda} F_{\alpha}\,^{\beta} W_{\gamma \lambda \beta i}+\frac{2109}{4096}{\rm i} (\Gamma_{a})^{\beta \lambda} W_{\beta}\,^{\rho} W_{\lambda \rho \gamma i} \nabla_{\alpha}\,^{\gamma}{W}+\frac{9}{64}(\Gamma_{a})^{\beta \gamma} C_{\alpha \beta}\,^{\rho \lambda} W_{\gamma \rho} \lambda_{i \lambda}+\frac{28521}{2048}(\Gamma_{a})^{\beta \rho} W_{\alpha \beta} W_{\rho}\,^{\lambda} W_{\lambda}\,^{\gamma} \lambda_{i \gamma}+\frac{17505}{4096}(\Gamma_{a})^{\rho \gamma} W_{\alpha}\,^{\beta} W_{\rho}\,^{\lambda} W_{\gamma \lambda} \lambda_{i \beta} - \frac{981}{4096}(\Gamma_{a})^{\beta}{}_{\gamma} W_{\beta}\,^{\rho} \lambda^{\lambda}_{i} \nabla^{\gamma}\,_{\alpha}{W_{\rho \lambda}}+\frac{45}{512}(\Gamma_{a})^{\beta \lambda} W_{\beta \rho} \lambda^{\gamma}_{i} \nabla_{\alpha}\,^{\rho}{W_{\lambda \gamma}} - \frac{4365}{4096}(\Gamma_{a})^{\beta \lambda} W_{\beta}\,^{\rho} \lambda_{i \gamma} \nabla_{\alpha}\,^{\gamma}{W_{\lambda \rho}} - \frac{9477}{10240}(\Gamma_{a})_{\alpha}{}^{\beta} W_{\beta \rho} \lambda^{\lambda}_{i} \nabla^{\rho \gamma}{W_{\lambda \gamma}}+\frac{837}{20480}(\Gamma_{a})_{\alpha}{}^{\beta} W_{\beta}\,^{\rho} \lambda_{i \gamma} \nabla^{\gamma \lambda}{W_{\rho \lambda}} - \frac{4545}{8192}(\Gamma_{a})^{\beta \rho} W_{\alpha \beta} \lambda_{i \gamma} \nabla^{\gamma \lambda}{W_{\rho \lambda}} - \frac{16191}{8192}(\Gamma_{a})^{\beta}{}_{\gamma} W_{\alpha \beta} \lambda^{\rho}_{i} \nabla^{\gamma \lambda}{W_{\rho \lambda}}+\frac{4401}{4096}(\Gamma_{a})^{\beta}{}_{\gamma} W_{\beta}\,^{\rho} \lambda_{i \alpha} \nabla^{\gamma \lambda}{W_{\rho \lambda}} - \frac{765}{2048}(\Gamma_{a})^{\beta \lambda} W_{\beta \rho} \lambda_{i \alpha} \nabla^{\rho \gamma}{W_{\lambda \gamma}} - \frac{1161}{5120}\Phi^{\beta \lambda}\,_{i j} (\Gamma_{a})_{\alpha}{}^{\rho} W_{\beta \rho} \lambda^{j}_{\lambda}%
 - \frac{69}{32}\Phi^{\rho \lambda}\,_{i j} (\Gamma_{a})_{\rho}{}^{\beta} W_{\alpha \beta} \lambda^{j}_{\lambda}+\frac{1653}{1024}\Phi^{\lambda \beta}\,_{i j} (\Gamma_{a})_{\lambda}{}^{\rho} W_{\beta \rho} \lambda^{j}_{\alpha}+\frac{3}{16}{\rm i} (\Gamma_{a})^{\beta \lambda} \lambda^{\gamma}_{i} W_{\alpha \beta}\,^{\rho}\,_{j} W_{\lambda \gamma \rho}\,^{j} - \frac{147}{2560}{\rm i} (\Gamma_{a})_{\alpha}{}^{\beta} \lambda^{\rho}_{i} X_{j}^{\lambda} W_{\beta \rho \lambda}\,^{j} - \frac{81}{256}{\rm i} (\Gamma_{a})^{\beta \lambda} X_{i j} W_{\beta}\,^{\rho} W_{\alpha \lambda \rho}\,^{j}+\frac{357}{2048}{\rm i} (\Gamma_{a})^{\beta}{}_{\gamma} \nabla^{\gamma \rho}{W} \nabla_{\alpha}\,^{\lambda}{W_{\beta \rho \lambda i}} - \frac{3171}{4096}{\rm i} (\Gamma_{a})^{\lambda}{}_{\gamma} W^{\beta}\,_{\rho} W_{\alpha \lambda \beta i} \nabla^{\gamma \rho}{W}+\frac{555}{256}{\rm i} (\Gamma_{a})^{\beta}{}_{\gamma} W_{\beta}\,^{\rho} W_{\alpha \rho \lambda i} \nabla^{\gamma \lambda}{W} - \frac{951}{10240}{\rm i} (\Gamma_{a})_{\alpha \gamma} W^{\beta \rho} W_{\beta \rho \lambda i} \nabla^{\gamma \lambda}{W}+\frac{8115}{4096}{\rm i} (\Gamma_{a})^{\rho}{}_{\gamma} W_{\alpha}\,^{\beta} W_{\rho \beta \lambda i} \nabla^{\gamma \lambda}{W}+\frac{1755}{8192}{\rm i} (\Gamma_{a})^{\beta}{}_{\lambda} W_{\beta \rho} X_{i \alpha} \nabla^{\lambda \rho}{W}+\frac{75}{2048}{\rm i} (\Gamma_{a})^{\beta}{}_{\lambda} \nabla^{\lambda}\,_{\gamma}{W} \nabla^{\gamma \rho}{W_{\alpha \beta \rho i}} - \frac{201}{512}{\rm i} (\Gamma_{a})_{\lambda \gamma} \nabla^{\lambda \beta}{W} \nabla^{\gamma \rho}{W_{\alpha \beta \rho i}} - \frac{9}{128}{\rm i} (\Gamma_{a})^{\beta}{}_{\rho} \nabla^{\rho}\,_{\lambda}{W} \nabla_{\alpha}\,^{\lambda}{X_{i \beta}} - \frac{153}{4096}{\rm i} (\Gamma_{a})_{\rho \lambda} \nabla^{\rho \beta}{W} \nabla^{\lambda}\,_{\alpha}{X_{i \beta}}+\frac{63}{2048}{\rm i} (\Gamma_{a})^{\rho}{}_{\lambda} W_{\alpha \beta} X_{i \rho} \nabla^{\lambda \beta}{W}+\frac{11673}{8192}{\rm i} (\Gamma_{a})^{\beta}{}_{\lambda} W_{\alpha \beta} X_{i \rho} \nabla^{\lambda \rho}{W}+\frac{207}{5120}{\rm i} (\Gamma_{a})_{\alpha \rho} \nabla^{\rho}\,_{\lambda}{W} \nabla^{\lambda \beta}{X_{i \beta}} - \frac{8091}{20480}{\rm i} (\Gamma_{a})_{\alpha \lambda} W^{\beta}\,_{\rho} X_{i \beta} \nabla^{\lambda \rho}{W}+\frac{93}{128}\Phi_{\lambda}\,^{\beta}\,_{i j} (\Gamma_{a})_{\beta \rho} \nabla^{\lambda \rho}{\lambda^{j}_{\alpha}}%
+\frac{663}{5120}\Phi^{\lambda \beta}\,_{i j} (\Gamma_{a})_{\alpha \lambda} W_{\beta}\,^{\rho} \lambda^{j}_{\rho}+\frac{9}{1024}(\Gamma_{a})^{\beta}{}_{\rho} Y \nabla^{\rho}\,_{\alpha}{\lambda_{i \beta}} - \frac{189}{1024}(\Gamma_{a})^{\lambda}{}_{\gamma} \nabla_{\alpha}\,^{\beta}{W_{\beta \rho}} \nabla^{\gamma \rho}{\lambda_{i \lambda}}+\frac{63}{64}(\Gamma_{a})^{\lambda}{}_{\gamma} W_{\alpha}\,^{\beta} W_{\beta \rho} \nabla^{\gamma \rho}{\lambda_{i \lambda}} - \frac{1197}{2048}(\Gamma_{a})^{\lambda}{}_{\gamma} W^{\beta \rho} W_{\beta \rho} \nabla^{\gamma}\,_{\alpha}{\lambda_{i \lambda}}+\frac{297}{1024}(\Gamma_{a})^{\rho}{}_{\lambda} \nabla_{\gamma}\,^{\beta}{W_{\alpha \beta}} \nabla^{\lambda \gamma}{\lambda_{i \rho}} - \frac{45}{256}{\rm i} (\Gamma_{a})^{\beta}{}_{\lambda} X_{i \rho} \nabla^{\lambda \rho}{F_{\alpha \beta}} - \frac{207}{2560}{\rm i} (\Gamma_{a})_{\alpha \rho} X_{j \beta} \nabla^{\rho \beta}{X_{i}\,^{j}} - \frac{27}{512}{\rm i} (\Gamma_{a})_{\rho \lambda} X_{i \beta} \nabla^{\rho \beta}{\nabla^{\lambda}\,_{\alpha}{W}}+\frac{441}{1024}{\rm i} (\Gamma_{a})^{\beta \lambda} W_{\alpha}\,^{\rho} F_{\beta \rho} X_{i \lambda} - \frac{2547}{5120}{\rm i} (\Gamma_{a})_{\alpha}{}^{\beta} W^{\rho \lambda} F_{\beta \rho} X_{i \lambda} - \frac{423}{1024}{\rm i} (\Gamma_{a})^{\lambda \beta} W_{\lambda}\,^{\rho} F_{\beta \rho} X_{i \alpha} - \frac{6201}{8192}{\rm i} (\Gamma_{a})_{\rho \lambda} W_{\alpha}\,^{\beta} X_{i \beta} \nabla^{\rho \lambda}{W}+\frac{261}{256}{\rm i} (\Gamma_{a})^{\lambda \beta} W_{\alpha \lambda} F_{\beta}\,^{\rho} X_{i \rho}+\frac{1173}{20480}{\rm i} (\Gamma_{a})_{\alpha}{}^{\rho} \lambda^{\beta}_{j} X_{i \beta} X^{j}_{\rho}+\frac{459}{512}{\rm i} (\Gamma_{a})^{\rho}{}_{\gamma} F_{\alpha}\,^{\beta} \nabla^{\gamma \lambda}{W_{\rho \beta \lambda i}}+\frac{27}{128}{\rm i} (\Gamma_{a})^{\beta}{}_{\lambda} X_{i j} \nabla^{\lambda \rho}{W_{\alpha \beta \rho}\,^{j}}+\frac{75}{2048}{\rm i} (\Gamma_{a})^{\beta}{}_{\gamma} \nabla_{\alpha}\,^{\rho}{W} \nabla^{\gamma \lambda}{W_{\beta \rho \lambda i}}+\frac{9}{64}(\Gamma_{a})^{\beta}{}_{\gamma} \lambda^{\rho}_{i} \nabla^{\gamma \lambda}{C_{\alpha \beta \rho \lambda}}+\frac{135}{512}(\Gamma_{a})^{\rho}{}_{\gamma} W_{\alpha \beta} \lambda^{\lambda}_{i} \nabla^{\gamma \beta}{W_{\rho \lambda}}%
+\frac{3285}{4096}(\Gamma_{a})^{\lambda}{}_{\gamma} W^{\beta}\,_{\rho} \lambda_{i \beta} \nabla^{\gamma \rho}{W_{\alpha \lambda}}+\frac{5661}{4096}(\Gamma_{a})^{\beta}{}_{\gamma} W_{\beta \rho} \lambda^{\lambda}_{i} \nabla^{\gamma \rho}{W_{\alpha \lambda}} - \frac{6471}{8192}(\Gamma_{a})^{\rho}{}_{\gamma} W_{\alpha}\,^{\beta} \lambda_{i \beta} \nabla^{\gamma \lambda}{W_{\rho \lambda}} - \frac{27}{256}(\Gamma_{a})^{\beta}{}_{\lambda} \lambda^{\rho}_{i} \nabla^{\lambda}\,_{\gamma}{\nabla_{\alpha}\,^{\gamma}{W_{\beta \rho}}} - \frac{45}{256}(\Gamma_{a})_{\lambda \gamma} \lambda^{\beta}_{i} \nabla^{\lambda \rho}{\nabla^{\gamma}\,_{\alpha}{W_{\beta \rho}}} - \frac{27}{256}(\Gamma_{a})^{\beta}{}_{\lambda} \lambda_{i \gamma} \nabla^{\lambda \rho}{\nabla_{\alpha}\,^{\gamma}{W_{\beta \rho}}}+\frac{45}{512}(\Gamma_{a})_{\lambda \gamma} \lambda^{\beta}_{i} \nabla^{\lambda}\,_{\alpha}{\nabla^{\gamma \rho}{W_{\beta \rho}}} - \frac{27}{512}(\Gamma_{a})^{\beta}{}_{\lambda} \lambda_{i \gamma} \nabla^{\lambda}\,_{\alpha}{\nabla^{\gamma \rho}{W_{\beta \rho}}}+\frac{297}{2560}(\Gamma_{a})_{\alpha \lambda} \lambda_{i \gamma} \nabla^{\lambda \beta}{\nabla^{\gamma \rho}{W_{\beta \rho}}} - \frac{153}{1280}(\Gamma_{a})_{\alpha \lambda} \lambda^{\beta}_{i} \nabla^{\lambda}\,_{\gamma}{\nabla^{\gamma \rho}{W_{\beta \rho}}} - \frac{135}{1024}(\Gamma_{a})^{\beta}{}_{\lambda} \lambda_{i \alpha} \nabla^{\lambda}\,_{\gamma}{\nabla^{\gamma \rho}{W_{\beta \rho}}}+\frac{189}{1024}(\Gamma_{a})_{\lambda \gamma} \lambda_{i \alpha} \nabla^{\lambda \beta}{\nabla^{\gamma \rho}{W_{\beta \rho}}} - \frac{9}{128}(\Gamma_{a})_{\alpha}{}^{\beta} C_{\beta}\,^{\rho \lambda \gamma} W_{\rho \lambda} \lambda_{i \gamma}+\frac{4941}{1280}(\Gamma_{a})_{\alpha}{}^{\beta} W_{\beta}\,^{\rho} W_{\rho}\,^{\lambda} W_{\lambda}\,^{\gamma} \lambda_{i \gamma}+\frac{207}{1280}(\Gamma_{a})_{\alpha \gamma} W^{\beta}\,_{\rho} \lambda^{\lambda}_{i} \nabla^{\gamma \rho}{W_{\beta \lambda}}+\frac{171}{640}(\Gamma_{a})_{\alpha}{}^{\lambda} W^{\beta}\,_{\rho} \lambda_{i \gamma} \nabla^{\rho \gamma}{W_{\lambda \beta}} - \frac{4347}{10240}(\Gamma_{a})_{\alpha}{}^{\lambda} W^{\beta}\,_{\rho} \lambda_{i \beta} \nabla^{\rho \gamma}{W_{\lambda \gamma}}+\frac{3537}{20480}(\Gamma_{a})_{\alpha \gamma} W^{\beta \rho} \lambda_{i \beta} \nabla^{\gamma \lambda}{W_{\rho \lambda}}+\frac{1845}{4096}(\Gamma_{a})^{\beta \lambda} W_{\beta \rho} \lambda_{i \gamma} \nabla^{\rho \gamma}{W_{\alpha \lambda}} - \frac{171}{2048}(\Gamma_{a})_{\lambda \gamma} W_{\alpha}\,^{\beta} \lambda^{\rho}_{i} \nabla^{\lambda \gamma}{W_{\beta \rho}}%
 - \frac{9}{1024}(\Gamma_{a})^{\rho}{}_{\lambda} W_{\alpha}\,^{\beta} \lambda_{i \gamma} \nabla^{\lambda \gamma}{W_{\rho \beta}}+\frac{3}{64}{\rm i} (\Gamma_{a})_{\alpha}{}^{\beta} \lambda^{\gamma}_{i} W_{\beta}\,^{\rho \lambda}\,_{j} W_{\gamma \rho \lambda}\,^{j} - \frac{2565}{4096}(\Gamma_{a})^{\lambda}{}_{\gamma} W^{\beta \rho} \lambda_{i \beta} \nabla^{\gamma}\,_{\alpha}{W_{\lambda \rho}} - \frac{2889}{4096}(\Gamma_{a})_{\lambda \gamma} W^{\beta \rho} \lambda_{i \beta} \nabla^{\lambda \gamma}{W_{\alpha \rho}}+\frac{1629}{2048}(\Gamma_{a})^{\beta}{}_{\lambda} W_{\beta}\,^{\rho} \lambda_{i \gamma} \nabla^{\lambda \gamma}{W_{\alpha \rho}}+\frac{549}{2048}{\rm i} (\Gamma_{a})_{\rho \lambda} F_{\alpha}\,^{\beta} \nabla^{\rho \lambda}{X_{i \beta}}+\frac{489}{4096}{\rm i} (\Gamma_{a})_{\beta \rho} X_{i j} \nabla^{\beta \rho}{X^{j}_{\alpha}}+\frac{63}{2048}{\rm i} (\Gamma_{a})_{\rho \lambda} \nabla_{\alpha}\,^{\beta}{W} \nabla^{\rho \lambda}{X_{i \beta}} - \frac{513}{32768}(\Gamma_{a})_{\beta \rho} \lambda_{i \alpha} \nabla^{\beta \rho}{Y}+\frac{1575}{8192}(\Gamma_{a})_{\lambda \gamma} \lambda^{\beta}_{i} \nabla^{\lambda \gamma}{\nabla_{\alpha}\,^{\rho}{W_{\beta \rho}}}+\frac{1701}{4096}(\Gamma_{a})_{\lambda \gamma} W^{\beta \rho} \lambda_{i \alpha} \nabla^{\lambda \gamma}{W_{\beta \rho}}+\frac{1107}{8192}(\Gamma_{a})_{\rho \lambda} \lambda_{i \gamma} \nabla^{\rho \lambda}{\nabla^{\gamma \beta}{W_{\alpha \beta}}}+\frac{117}{256}{\rm i} (\Gamma_{a})^{\rho}{}_{\lambda} F_{\alpha \beta} \nabla^{\lambda \beta}{X_{i \rho}} - \frac{381}{2048}{\rm i} (\Gamma_{a})^{\beta}{}_{\rho} X_{i j} \nabla^{\rho}\,_{\alpha}{X^{j}_{\beta}}+\frac{909}{4096}{\rm i} (\Gamma_{a})^{\beta}{}_{\rho} \nabla_{\alpha \lambda}{W} \nabla^{\rho \lambda}{X_{i \beta}} - \frac{297}{81920}(\Gamma_{a})_{\alpha \beta} \lambda_{i \rho} \nabla^{\beta \rho}{Y}+\frac{999}{4096}(\Gamma_{a})^{\beta}{}_{\lambda} \lambda_{i \gamma} \nabla^{\lambda \gamma}{\nabla_{\alpha}\,^{\rho}{W_{\beta \rho}}}+\frac{711}{2048}(\Gamma_{a})_{\alpha \lambda} W^{\beta \rho} \lambda_{i \gamma} \nabla^{\lambda \gamma}{W_{\beta \rho}} - \frac{963}{4096}(\Gamma_{a})_{\rho \lambda} \lambda_{i \gamma} \nabla^{\rho \gamma}{\nabla^{\lambda \beta}{W_{\alpha \beta}}} - \frac{1}{32}(\Gamma_{a})^{\beta}{}_{\rho} \lambda^{\lambda}_{j} \nabla^{\rho}\,_{\alpha}{\Phi_{\beta \lambda i}\,^{j}}%
+\frac{9}{160}(\Gamma_{a})_{\alpha \beta} \lambda^{\lambda}_{j} \nabla^{\beta \rho}{\Phi_{\rho \lambda i}\,^{j}} - \frac{33}{320}{\rm i} (\Gamma_{a})_{\alpha}{}^{\beta} \lambda^{\gamma}_{j} W_{\beta}\,^{\rho \lambda}\,_{i} W_{\gamma \rho \lambda}\,^{j} - \frac{39}{64}{\rm i} (\Gamma_{a})^{\beta \lambda} \lambda^{\gamma}_{j} W_{\alpha \beta}\,^{\rho}\,_{i} W_{\lambda \gamma \rho}\,^{j} - \frac{3}{320}{\rm i} (\Gamma_{a})_{\alpha}{}^{\beta} \lambda^{\gamma}_{j} W_{\beta}\,^{\rho \lambda j} W_{\gamma \rho \lambda i} - \frac{11}{1024}(\Gamma_{a})_{\beta \rho} \lambda^{\lambda}_{j} \nabla^{\beta \rho}{\Phi_{\alpha \lambda i}\,^{j}}+\frac{69}{512}(\Gamma_{a})^{\beta}{}_{\rho} \lambda_{j \lambda} \nabla^{\rho \lambda}{\Phi_{\alpha \beta i}\,^{j}} - \frac{5}{128}\Phi^{\beta \lambda}\,_{i j} (\Gamma_{a})_{\beta \rho} \nabla^{\rho}\,_{\alpha}{\lambda^{j}_{\lambda}} - \frac{27}{128}\Phi_{\lambda}\,^{\beta}\,_{i j} (\Gamma_{a})_{\beta}{}^{\rho} \nabla^{\lambda}\,_{\alpha}{\lambda^{j}_{\rho}}+\frac{153}{640}\Phi_{\rho}\,^{\beta}\,_{i j} (\Gamma_{a})_{\alpha \beta} \nabla^{\rho \lambda}{\lambda^{j}_{\lambda}}+\frac{21}{64}\Phi_{\alpha}\,^{\beta}\,_{i j} (\Gamma_{a})_{\beta \rho} \nabla^{\rho \lambda}{\lambda^{j}_{\lambda}}+\frac{45}{512}{\rm i} (\Gamma_{a})^{\beta \lambda} F_{\beta}\,^{\rho} \nabla_{\alpha}\,^{\gamma}{W_{\lambda \rho \gamma i}}+\frac{297}{256}{\rm i} (\Gamma_{a})^{\lambda \beta} W_{\lambda}\,^{\gamma} F_{\beta}\,^{\rho} W_{\alpha \gamma \rho i}+\frac{81}{1280}{\rm i} (\Gamma_{a})_{\alpha}{}^{\beta} W^{\lambda \gamma} F_{\beta}\,^{\rho} W_{\lambda \gamma \rho i} - \frac{999}{256}{\rm i} (\Gamma_{a})^{\beta \gamma} W_{\alpha}\,^{\lambda} F_{\beta}\,^{\rho} W_{\gamma \lambda \rho i}+\frac{243}{256}{\rm i} (\Gamma_{a})^{\beta}{}_{\gamma} F_{\beta}\,^{\rho} \nabla^{\gamma \lambda}{W_{\alpha \rho \lambda i}}+\frac{315}{1024}{\rm i} (\Gamma_{a})^{\beta}{}_{\lambda} F_{\beta}\,^{\rho} \nabla^{\lambda}\,_{\alpha}{X_{i \rho}} - \frac{27}{128}{\rm i} (\Gamma_{a})^{\beta \lambda} F_{\beta \rho} \nabla_{\alpha}\,^{\rho}{X_{i \lambda}}+\frac{27}{640}{\rm i} (\Gamma_{a})_{\alpha}{}^{\beta} F_{\beta \rho} \nabla^{\rho \lambda}{X_{i \lambda}}+\frac{333}{512}(\Gamma_{a})^{\beta}{}_{\lambda} \nabla^{\lambda \rho}{W_{\beta \rho}} \nabla_{\alpha}\,^{\gamma}{\lambda_{i \gamma}}+\frac{117}{512}{\rm i} (\Gamma_{a})^{\beta}{}_{\rho} X_{j \beta} \nabla^{\rho}\,_{\alpha}{X_{i}\,^{j}}%
+\frac{9}{32}(\Gamma_{a})^{\beta}{}_{\lambda} W_{\beta \rho} \nabla^{\lambda \rho}{\nabla_{\alpha}\,^{\gamma}{\lambda_{i \gamma}}} - \frac{9}{4}(\Gamma_{a})^{\rho}{}_{\gamma} W_{\alpha}\,^{\beta} W_{\rho \lambda} \nabla^{\gamma \lambda}{\lambda_{i \beta}} - \frac{1161}{1280}(\Gamma_{a})_{\alpha}{}^{\beta} W_{\beta}\,^{\rho} W_{\rho \lambda} \nabla^{\lambda \gamma}{\lambda_{i \gamma}} - \frac{513}{128}(\Gamma_{a})^{\beta \rho} W_{\alpha \beta} W_{\rho \lambda} \nabla^{\lambda \gamma}{\lambda_{i \gamma}}+\frac{81}{256}(\Gamma_{a})^{\rho}{}_{\lambda} W_{\alpha}\,^{\beta} W_{\rho \beta} \nabla^{\lambda \gamma}{\lambda_{i \gamma}}+\frac{13}{256}\Phi_{\alpha}\,^{\lambda}\,_{i j} (\Gamma_{a})_{\beta \rho} \nabla^{\beta \rho}{\lambda^{j}_{\lambda}} - \frac{9}{8192}(\Gamma_{a})_{\beta \rho} Y \nabla^{\beta \rho}{\lambda_{i \alpha}}+\frac{153}{2048}(\Gamma_{a})_{\lambda \gamma} \nabla_{\alpha}\,^{\rho}{W^{\beta}\,_{\rho}} \nabla^{\lambda \gamma}{\lambda_{i \beta}} - \frac{639}{512}(\Gamma_{a})_{\lambda \gamma} W_{\alpha}\,^{\beta} W_{\beta}\,^{\rho} \nabla^{\lambda \gamma}{\lambda_{i \rho}} - \frac{819}{2048}(\Gamma_{a})_{\lambda \gamma} W^{\beta \rho} W_{\beta \rho} \nabla^{\lambda \gamma}{\lambda_{i \alpha}} - \frac{243}{2048}(\Gamma_{a})_{\rho \lambda} \nabla^{\beta \gamma}{W_{\alpha \beta}} \nabla^{\rho \lambda}{\lambda_{i \gamma}} - \frac{9}{256}{\rm i} (\Gamma_{a})_{\rho \lambda} X_{i}^{\beta} \nabla^{\rho \lambda}{F_{\alpha \beta}} - \frac{45}{512}{\rm i} (\Gamma_{a})_{\rho \lambda} X_{i \beta} \nabla^{\rho \lambda}{\nabla_{\alpha}\,^{\beta}{W}}+\frac{567}{5120}{\rm i} (\Gamma_{a})_{\alpha}{}^{\lambda} W_{\lambda}\,^{\beta} F_{\beta}\,^{\rho} X_{i \rho}+\frac{11817}{40960}{\rm i} (\Gamma_{a})_{\alpha}{}^{\beta} W_{\beta \rho} X_{i \lambda} \nabla^{\rho \lambda}{W} - \frac{63}{10240}(\Gamma_{a})_{\alpha \beta} Y \nabla^{\beta \rho}{\lambda_{i \rho}}+\frac{405}{1024}(\Gamma_{a})^{\beta}{}_{\lambda} \nabla_{\alpha}\,^{\rho}{W_{\beta \rho}} \nabla^{\lambda \gamma}{\lambda_{i \gamma}}+\frac{27}{10240}(\Gamma_{a})_{\alpha \lambda} W^{\beta \rho} W_{\beta \rho} \nabla^{\lambda \gamma}{\lambda_{i \gamma}}+\frac{585}{1024}(\Gamma_{a})_{\rho \lambda} \nabla^{\rho \beta}{W_{\alpha \beta}} \nabla^{\lambda \gamma}{\lambda_{i \gamma}}+\frac{45}{256}{\rm i} (\Gamma_{a})^{\rho}{}_{\lambda} X_{i \rho} \nabla^{\lambda \beta}{F_{\alpha \beta}}%
 - \frac{45}{512}{\rm i} (\Gamma_{a})^{\beta}{}_{\rho} X_{i \beta} \nabla^{\rho}\,_{\lambda}{\nabla_{\alpha}\,^{\lambda}{W}} - \frac{1341}{5120}{\rm i} (\Gamma_{a})_{\alpha}{}^{\lambda} W^{\beta \rho} F_{\beta \rho} X_{i \lambda}+\frac{9}{128}{\rm i} (\Gamma_{a})_{\beta \rho} X_{j \alpha} \nabla^{\beta \rho}{X_{i}\,^{j}} - \frac{3}{64}{\rm i} (\Gamma_{a})_{\beta \rho} \lambda_{j \alpha} {W}^{-2} \nabla^{\beta}\,_{\gamma}{\lambda^{\lambda}_{i}} \nabla^{\rho \gamma}{\lambda^{j}_{\lambda}} - \frac{9}{64}(\Gamma_{a})_{\rho \lambda} {W}^{-1} \nabla^{\rho}\,_{\gamma}{F_{\alpha}\,^{\beta}} \nabla^{\lambda \gamma}{\lambda_{i \beta}}+\frac{9}{64}(\Gamma_{a})_{\rho \lambda} W_{\alpha}\,^{\beta} {W}^{-1} \nabla^{\rho}\,_{\gamma}{W} \nabla^{\lambda \gamma}{\lambda_{i \beta}} - \frac{9}{32}(\Gamma_{a})_{\rho \lambda} \nabla^{\rho}\,_{\gamma}{W_{\alpha}\,^{\beta}} \nabla^{\lambda \gamma}{\lambda_{i \beta}}+\frac{3}{32}(\Gamma_{a})_{\beta \rho} {W}^{-1} \nabla^{\beta}\,_{\lambda}{X_{i j}} \nabla^{\rho \lambda}{\lambda^{j}_{\alpha}} - \frac{9}{128}(\Gamma_{a})_{\beta \rho} {W}^{-1} \nabla^{\beta}\,_{\gamma}{\lambda_{i \lambda}} \nabla^{\rho \gamma}{\nabla_{\alpha}\,^{\lambda}{W}}+\frac{63}{256}(\Gamma_{a})_{\rho \lambda} X_{i j} W_{\alpha}\,^{\beta} {W}^{-1} \nabla^{\rho \lambda}{\lambda^{j}_{\beta}} - \frac{9}{128}(\Gamma_{a})_{\rho \lambda} W_{\alpha \beta} {W}^{-1} \nabla^{\beta \gamma}{W} \nabla^{\rho \lambda}{\lambda_{i \gamma}}+\frac{63}{640}(\Gamma_{a})_{\alpha \gamma} W^{\beta}\,_{\lambda} F_{\beta}\,^{\rho} {W}^{-1} \nabla^{\gamma \lambda}{\lambda_{i \rho}}+\frac{171}{320}(\Gamma_{a})_{\alpha \lambda} X_{i j} W^{\beta}\,_{\rho} {W}^{-1} \nabla^{\lambda \rho}{\lambda^{j}_{\beta}} - \frac{99}{640}(\Gamma_{a})_{\alpha \lambda} W_{\beta \rho} {W}^{-1} \nabla^{\beta \gamma}{W} \nabla^{\lambda \rho}{\lambda_{i \gamma}} - \frac{27}{128}(\Gamma_{a})^{\lambda}{}_{\gamma} W_{\lambda}\,^{\beta} F_{\beta}\,^{\rho} {W}^{-1} \nabla^{\gamma}\,_{\alpha}{\lambda_{i \rho}} - \frac{9}{32}(\Gamma_{a})^{\beta}{}_{\lambda} X_{i j} W_{\beta}\,^{\rho} {W}^{-1} \nabla^{\lambda}\,_{\alpha}{\lambda^{j}_{\rho}}+\frac{27}{256}(\Gamma_{a})^{\beta}{}_{\lambda} W_{\beta \rho} {W}^{-1} \nabla^{\rho \gamma}{W} \nabla^{\lambda}\,_{\alpha}{\lambda_{i \gamma}} - \frac{81}{128}(\Gamma_{a})^{\beta}{}_{\gamma} W_{\alpha \lambda} F_{\beta}\,^{\rho} {W}^{-1} \nabla^{\gamma \lambda}{\lambda_{i \rho}} - \frac{99}{64}(\Gamma_{a})^{\rho}{}_{\gamma} W_{\alpha \beta} W_{\rho}\,^{\lambda} \nabla^{\gamma \beta}{\lambda_{i \lambda}}+\frac{9}{32}(\Gamma_{a})_{\rho \lambda} W_{\alpha \beta} {W}^{-1} \nabla^{\rho \gamma}{W} \nabla^{\lambda \beta}{\lambda_{i \gamma}}%
+\frac{9}{64}(\Gamma_{a})^{\lambda}{}_{\gamma} W_{\alpha \lambda} F^{\beta}\,_{\rho} {W}^{-1} \nabla^{\gamma \rho}{\lambda_{i \beta}}+\frac{531}{128}(\Gamma_{a})^{\beta}{}_{\gamma} W_{\alpha \beta} W^{\rho}\,_{\lambda} \nabla^{\gamma \lambda}{\lambda_{i \rho}} - \frac{9}{64}(\Gamma_{a})_{\lambda \gamma} \lambda^{\beta}_{i} W_{\alpha \beta}\,^{\rho}\,_{j} {W}^{-1} \nabla^{\lambda \gamma}{\lambda^{j}_{\rho}} - \frac{3}{80}(\Gamma_{a})_{\alpha \gamma} \lambda^{\beta}_{i} W_{\beta}\,^{\rho}\,_{\lambda j} {W}^{-1} \nabla^{\gamma \lambda}{\lambda^{j}_{\rho}}+\frac{21}{64}(\Gamma_{a})^{\beta}{}_{\lambda} \lambda^{\rho}_{i} W_{\alpha \beta \rho j} {W}^{-1} \nabla^{\lambda \gamma}{\lambda^{j}_{\gamma}}+\frac{3}{16}(\Gamma_{a})^{\lambda}{}_{\gamma} \lambda_{i \lambda} W_{\alpha}\,^{\beta}\,_{\rho j} {W}^{-1} \nabla^{\gamma \rho}{\lambda^{j}_{\beta}} - \frac{39}{512}(\Gamma_{a})^{\beta}{}_{\rho} \lambda_{i \alpha} X_{j \beta} {W}^{-1} \nabla^{\rho \lambda}{\lambda^{j}_{\lambda}} - \frac{45}{256}(\Gamma_{a})^{\rho}{}_{\lambda} \lambda_{i \alpha} X_{j \beta} {W}^{-1} \nabla^{\lambda \beta}{\lambda^{j}_{\rho}}+\frac{93}{512}(\Gamma_{a})^{\beta}{}_{\rho} \lambda_{i \lambda} X_{j \beta} {W}^{-1} \nabla^{\rho \lambda}{\lambda^{j}_{\alpha}}+\frac{207}{1280}(\Gamma_{a})_{\alpha \rho} \lambda^{\beta}_{i} X_{j \beta} {W}^{-1} \nabla^{\rho \lambda}{\lambda^{j}_{\lambda}}+\frac{3}{80}(\Gamma_{a})_{\alpha \rho} \lambda_{i \lambda} X_{j}^{\beta} {W}^{-1} \nabla^{\rho \lambda}{\lambda^{j}_{\beta}}+\frac{9}{512}(\Gamma_{a})^{\rho}{}_{\lambda} \lambda_{i \rho} X_{j}^{\beta} {W}^{-1} \nabla^{\lambda}\,_{\alpha}{\lambda^{j}_{\beta}} - \frac{81}{512}(\Gamma_{a})_{\beta \rho} \lambda^{\lambda}_{i} X_{j \alpha} {W}^{-1} \nabla^{\beta \rho}{\lambda^{j}_{\lambda}} - \frac{123}{1280}(\Gamma_{a})_{\alpha \rho} \lambda^{\lambda}_{i} X_{j \beta} {W}^{-1} \nabla^{\rho \beta}{\lambda^{j}_{\lambda}} - \frac{15}{512}(\Gamma_{a})^{\beta}{}_{\rho} \lambda^{\lambda}_{i} X_{j \beta} {W}^{-1} \nabla^{\rho}\,_{\alpha}{\lambda^{j}_{\lambda}}+\frac{27}{512}(\Gamma_{a})_{\beta \rho} \lambda^{\lambda}_{j} X^{j}_{\alpha} {W}^{-1} \nabla^{\beta \rho}{\lambda_{i \lambda}}+\frac{147}{1280}(\Gamma_{a})_{\alpha \rho} \lambda^{\lambda}_{j} X^{j}_{\beta} {W}^{-1} \nabla^{\rho \beta}{\lambda_{i \lambda}} - \frac{15}{512}(\Gamma_{a})^{\beta}{}_{\rho} \lambda^{\lambda}_{j} X^{j}_{\beta} {W}^{-1} \nabla^{\rho}\,_{\alpha}{\lambda_{i \lambda}} - \frac{3}{128}(\Gamma_{a})_{\alpha \rho} \lambda^{\lambda}_{j} X_{i \beta} {W}^{-1} \nabla^{\rho \beta}{\lambda^{j}_{\lambda}} - \frac{3}{512}(\Gamma_{a})^{\beta}{}_{\rho} \lambda^{\lambda}_{j} X_{i \beta} {W}^{-1} \nabla^{\rho}\,_{\alpha}{\lambda^{j}_{\lambda}}%
 - \frac{1}{5}\Phi_{\rho}\,^{\lambda}\,_{i j} (\Gamma_{a})_{\alpha \beta} \nabla^{\rho \beta}{\lambda^{j}_{\lambda}} - \frac{9}{64}(\Gamma_{a})_{\lambda \gamma} \nabla^{\lambda}\,_{\alpha}{W^{\beta}\,_{\rho}} \nabla^{\gamma \rho}{\lambda_{i \beta}}+\frac{9}{64}(\Gamma_{a})^{\beta}{}_{\lambda} \nabla_{\alpha \gamma}{W_{\beta}\,^{\rho}} \nabla^{\lambda \gamma}{\lambda_{i \rho}} - \frac{441}{2560}(\Gamma_{a})_{\alpha \lambda} \nabla^{\beta \gamma}{W_{\beta \rho}} \nabla^{\lambda \rho}{\lambda_{i \gamma}}+\frac{9}{128}(\Gamma_{a})^{\beta}{}_{\lambda} \nabla^{\rho \gamma}{W_{\beta \rho}} \nabla^{\lambda}\,_{\alpha}{\lambda_{i \gamma}} - \frac{9}{32}(\Gamma_{a})_{\rho \lambda} \nabla^{\rho \gamma}{W_{\alpha \beta}} \nabla^{\lambda \beta}{\lambda_{i \gamma}} - \frac{9}{32}(\Gamma_{a})_{\lambda \gamma} \nabla^{\lambda \rho}{W^{\beta}\,_{\rho}} \nabla^{\gamma}\,_{\alpha}{\lambda_{i \beta}} - \frac{81}{512}(\Gamma_{a})_{\alpha \lambda} \nabla_{\gamma}\,^{\rho}{W^{\beta}\,_{\rho}} \nabla^{\lambda \gamma}{\lambda_{i \beta}}+\frac{9}{512}(\Gamma_{a})_{\lambda \gamma} \nabla^{\lambda \beta}{W_{\beta \rho}} \nabla^{\gamma \rho}{\lambda_{i \alpha}}+\frac{117}{512}(\Gamma_{a})^{\beta}{}_{\lambda} \nabla_{\gamma}\,^{\rho}{W_{\beta \rho}} \nabla^{\lambda \gamma}{\lambda_{i \alpha}}+\frac{3}{32}(\Gamma_{a})^{\beta}{}_{\gamma} \lambda^{\rho}_{j} W_{\beta \rho}\,^{\lambda j} {W}^{-1} \nabla^{\gamma}\,_{\alpha}{\lambda_{i \lambda}} - \frac{3}{32}(\Gamma_{a})^{\beta}{}_{\lambda} \lambda^{\rho}_{j} W_{\alpha \beta \rho}\,^{j} {W}^{-1} \nabla^{\lambda \gamma}{\lambda_{i \gamma}}+\frac{21}{512}(\Gamma_{a})^{\beta}{}_{\rho} \lambda_{j \alpha} X^{j}_{\beta} {W}^{-1} \nabla^{\rho \lambda}{\lambda_{i \lambda}} - \frac{45}{256}(\Gamma_{a})^{\rho}{}_{\lambda} \lambda_{j \alpha} X^{j}_{\beta} {W}^{-1} \nabla^{\lambda \beta}{\lambda_{i \rho}}+\frac{9}{512}(\Gamma_{a})^{\beta}{}_{\rho} \lambda_{j \lambda} X^{j}_{\beta} {W}^{-1} \nabla^{\rho \lambda}{\lambda_{i \alpha}} - \frac{213}{1280}(\Gamma_{a})_{\alpha \rho} \lambda^{\beta}_{j} X^{j}_{\beta} {W}^{-1} \nabla^{\rho \lambda}{\lambda_{i \lambda}}+\frac{45}{256}(\Gamma_{a})^{\rho}{}_{\lambda} \lambda^{\beta}_{j} X^{j}_{\beta} {W}^{-1} \nabla^{\lambda}\,_{\alpha}{\lambda_{i \rho}} - \frac{9}{80}(\Gamma_{a})_{\alpha \rho} \lambda_{j \lambda} X^{j \beta} {W}^{-1} \nabla^{\rho \lambda}{\lambda_{i \beta}} - \frac{3}{64}{\rm i} (\Gamma_{a})_{\beta \rho} \lambda_{j \alpha} {W}^{-2} \nabla^{\beta \gamma}{\lambda_{i \lambda}} \nabla^{\rho \lambda}{\lambda^{j}_{\gamma}} - \frac{9}{64}(\Gamma_{a})_{\rho \lambda} {W}^{-1} \nabla^{\rho \gamma}{F_{\alpha \beta}} \nabla^{\lambda \beta}{\lambda_{i \gamma}}%
 - \frac{9}{128}(\Gamma_{a})_{\beta \rho} {W}^{-1} \nabla^{\beta}\,_{\gamma}{\lambda_{i \lambda}} \nabla^{\rho \lambda}{\nabla_{\alpha}\,^{\gamma}{W}}+\frac{117}{128}(\Gamma_{a})^{\lambda}{}_{\gamma} W_{\alpha}\,^{\beta} F_{\beta \rho} {W}^{-1} \nabla^{\gamma \rho}{\lambda_{i \lambda}} - \frac{27}{128}(\Gamma_{a})^{\rho}{}_{\lambda} X_{i j} W_{\alpha \beta} {W}^{-1} \nabla^{\lambda \beta}{\lambda^{j}_{\rho}}+\frac{99}{256}(\Gamma_{a})^{\rho}{}_{\lambda} W_{\alpha \beta} {W}^{-1} \nabla_{\gamma}\,^{\beta}{W} \nabla^{\lambda \gamma}{\lambda_{i \rho}} - \frac{45}{128}(\Gamma_{a})_{\alpha \gamma} W^{\beta \lambda} F_{\beta \rho} {W}^{-1} \nabla^{\gamma \rho}{\lambda_{i \lambda}} - \frac{45}{128}(\Gamma_{a})_{\alpha \lambda} W^{\beta}\,_{\rho} {W}^{-1} \nabla_{\gamma}\,^{\rho}{W} \nabla^{\lambda \gamma}{\lambda_{i \beta}}+\frac{27}{128}(\Gamma_{a})^{\lambda}{}_{\gamma} W_{\lambda}\,^{\beta} F_{\beta \rho} {W}^{-1} \nabla^{\gamma \rho}{\lambda_{i \alpha}}+\frac{27}{64}(\Gamma_{a})^{\beta}{}_{\lambda} X_{i j} W_{\beta \rho} {W}^{-1} \nabla^{\lambda \rho}{\lambda^{j}_{\alpha}}+\frac{27}{256}(\Gamma_{a})^{\beta}{}_{\lambda} W_{\beta \rho} {W}^{-1} \nabla_{\gamma}\,^{\rho}{W} \nabla^{\lambda \gamma}{\lambda_{i \alpha}} - \frac{63}{128}(\Gamma_{a})^{\beta}{}_{\gamma} W_{\alpha}\,^{\lambda} F_{\beta \rho} {W}^{-1} \nabla^{\gamma \rho}{\lambda_{i \lambda}} - \frac{21}{64}(\Gamma_{a})^{\beta}{}_{\gamma} \lambda^{\rho}_{i} W_{\beta \rho \lambda j} {W}^{-1} \nabla^{\gamma \lambda}{\lambda^{j}_{\alpha}}+\frac{63}{512}(\Gamma_{a})_{\rho \lambda} \lambda_{i \alpha} X_{j}^{\beta} {W}^{-1} \nabla^{\rho \lambda}{\lambda^{j}_{\beta}}+\frac{15}{64}(\Gamma_{a})_{\rho \lambda} \lambda^{\beta}_{i} X_{j \beta} {W}^{-1} \nabla^{\rho \lambda}{\lambda^{j}_{\alpha}}+\frac{15}{128}(\Gamma_{a})^{\beta}{}_{\rho} \lambda_{i \lambda} X_{j \alpha} {W}^{-1} \nabla^{\rho \lambda}{\lambda^{j}_{\beta}}+\frac{9}{32}(\Gamma_{a})^{\beta}{}_{\rho} \lambda_{j \lambda} X^{j}_{\alpha} {W}^{-1} \nabla^{\rho \lambda}{\lambda_{i \beta}}+\frac{21}{256}(\Gamma_{a})^{\beta}{}_{\rho} \lambda_{j \lambda} X_{i \alpha} {W}^{-1} \nabla^{\rho \lambda}{\lambda^{j}_{\beta}} - \frac{21}{256}(\Gamma_{a})_{\alpha \rho} \lambda_{j \lambda} X_{i}^{\beta} {W}^{-1} \nabla^{\rho \lambda}{\lambda^{j}_{\beta}}+\frac{21}{512}(\Gamma_{a})^{\beta}{}_{\rho} \lambda_{j \lambda} X_{i \beta} {W}^{-1} \nabla^{\rho \lambda}{\lambda^{j}_{\alpha}}+\frac{9}{64}(\Gamma_{a})^{\beta}{}_{\lambda} \nabla_{\alpha}\,^{\gamma}{W_{\beta \rho}} \nabla^{\lambda \rho}{\lambda_{i \gamma}}+\frac{3}{32}(\Gamma_{a})^{\beta}{}_{\gamma} \lambda^{\rho}_{j} W_{\beta \rho \lambda}\,^{j} {W}^{-1} \nabla^{\gamma \lambda}{\lambda_{i \alpha}}%
 - \frac{9}{512}(\Gamma_{a})_{\rho \lambda} \lambda_{j \alpha} X^{j \beta} {W}^{-1} \nabla^{\rho \lambda}{\lambda_{i \beta}}+\frac{3}{128}(\Gamma_{a})_{\rho \lambda} \lambda^{\beta}_{j} X^{j}_{\beta} {W}^{-1} \nabla^{\rho \lambda}{\lambda_{i \alpha}} - \frac{105}{512}(\Gamma_{a})^{\rho}{}_{\lambda} \lambda_{i \rho} X_{j \beta} {W}^{-1} \nabla^{\lambda \beta}{\lambda^{j}_{\alpha}}+\frac{189}{64}(\Gamma_{a})^{\beta}{}_{\gamma} W^{\rho}\,_{\lambda} F_{\alpha \beta} {W}^{-1} \nabla^{\gamma \lambda}{\lambda_{i \rho}}+\frac{27}{64}(\Gamma_{a})_{\lambda \gamma} W^{\beta}\,_{\rho} {W}^{-1} \nabla^{\lambda}\,_{\alpha}{W} \nabla^{\gamma \rho}{\lambda_{i \beta}}+\frac{9}{16}{\rm i} (\Gamma_{a})^{\lambda}{}_{\gamma} W^{\beta}\,_{\rho} \lambda_{j \alpha} \lambda_{i \lambda} {W}^{-2} \nabla^{\gamma \rho}{\lambda^{j}_{\beta}} - \frac{81}{64}(\Gamma_{a})^{\lambda}{}_{\gamma} W^{\beta}\,_{\rho} \lambda_{i \lambda} {W}^{-1} \nabla^{\gamma \rho}{F_{\alpha \beta}}+\frac{9}{64}(\Gamma_{a})^{\lambda}{}_{\gamma} W_{\alpha}\,^{\beta} W_{\beta \rho} \lambda_{i \lambda} {W}^{-1} \nabla^{\gamma \rho}{W}+\frac{81}{128}(\Gamma_{a})^{\lambda}{}_{\gamma} W_{\beta \rho} \lambda_{i \lambda} {W}^{-1} \nabla^{\gamma \beta}{\nabla_{\alpha}\,^{\rho}{W}} - \frac{63}{32}(\Gamma_{a})^{\lambda \gamma} W_{\alpha}\,^{\beta} W_{\lambda}\,^{\rho} F_{\beta \rho} \lambda_{i \gamma} {W}^{-1} - \frac{63}{256}(\Gamma_{a})^{\rho \gamma} W_{\alpha \beta} W_{\rho \lambda} \lambda_{i \gamma} {W}^{-1} \nabla^{\beta \lambda}{W}+\frac{1521}{1280}(\Gamma_{a})_{\alpha}{}^{\gamma} W^{\beta}\,_{\rho} W_{\beta \lambda} \lambda_{i \gamma} {W}^{-1} \nabla^{\rho \lambda}{W}+\frac{171}{32}(\Gamma_{a})^{\beta \gamma} W_{\alpha}\,^{\lambda} W_{\lambda}\,^{\rho} F_{\beta \rho} \lambda_{i \gamma} {W}^{-1}+\frac{171}{128}(\Gamma_{a})^{\lambda \gamma} W_{\alpha \lambda} W^{\beta \rho} F_{\beta \rho} \lambda_{i \gamma} {W}^{-1}+\frac{1593}{2048}(\Gamma_{a})^{\beta \gamma} W_{\beta}\,^{\rho} \lambda_{i \gamma} \lambda^{\lambda}_{j} W_{\alpha \rho \lambda}\,^{j} {W}^{-1}+\frac{4821}{10240}(\Gamma_{a})_{\alpha}{}^{\gamma} W^{\beta \rho} \lambda_{i \gamma} \lambda^{\lambda}_{j} W_{\beta \rho \lambda}\,^{j} {W}^{-1}+\frac{151}{512}(\Gamma_{a})^{\gamma \rho} W_{\alpha}\,^{\beta} \lambda_{i \gamma} \lambda^{\lambda}_{j} W_{\rho \beta \lambda}\,^{j} {W}^{-1} - \frac{437}{2048}(\Gamma_{a})^{\gamma \lambda} W^{\beta \rho} \lambda_{i \gamma} \lambda_{j \beta} W_{\alpha \lambda \rho}\,^{j} {W}^{-1}+\frac{27}{128}(\Gamma_{a})^{\lambda \gamma} W^{\beta \rho} \lambda_{i \lambda} \lambda_{j \gamma} W_{\alpha \beta \rho}\,^{j} {W}^{-1}+\frac{981}{4096}(\Gamma_{a})^{\beta \lambda} W_{\beta}\,^{\rho} \lambda_{j \alpha} \lambda_{i \lambda} X^{j}_{\rho} {W}^{-1}%
 - \frac{111}{2048}(\Gamma_{a})^{\lambda \rho} W_{\alpha}\,^{\beta} \lambda_{i \lambda} \lambda_{j \beta} X^{j}_{\rho} {W}^{-1} - \frac{405}{2048}(\Gamma_{a})^{\rho \lambda} W_{\alpha}\,^{\beta} \lambda_{i \rho} \lambda_{j \lambda} X^{j}_{\beta} {W}^{-1} - \frac{699}{4096}(\Gamma_{a})^{\beta \lambda} W_{\alpha \beta} \lambda_{i \lambda} \lambda^{\rho}_{j} X^{j}_{\rho} {W}^{-1} - \frac{1407}{2560}(\Gamma_{a})_{\alpha}{}^{\lambda} W^{\beta \rho} \lambda_{i \lambda} \lambda_{j \beta} X^{j}_{\rho} {W}^{-1} - \frac{2793}{4096}(\Gamma_{a})^{\beta \lambda} W_{\beta}\,^{\rho} \lambda_{i \lambda} \lambda_{j \rho} X^{j}_{\alpha} {W}^{-1} - \frac{135}{128}(\Gamma_{a})^{\rho}{}_{\gamma} W_{\rho \lambda} F_{\alpha}\,^{\beta} {W}^{-1} \nabla^{\gamma \lambda}{\lambda_{i \beta}} - \frac{27}{64}(\Gamma_{a})^{\beta}{}_{\lambda} W_{\beta \rho} {W}^{-1} \nabla_{\alpha}\,^{\gamma}{W} \nabla^{\lambda \rho}{\lambda_{i \gamma}} - \frac{21}{64}{\rm i} (\Gamma_{a})^{\beta}{}_{\lambda} W_{\beta \rho} \lambda_{j \alpha} \lambda^{\gamma}_{i} {W}^{-2} \nabla^{\lambda \rho}{\lambda^{j}_{\gamma}}+\frac{81}{128}(\Gamma_{a})^{\rho}{}_{\gamma} W_{\rho \lambda} \lambda^{\beta}_{i} {W}^{-1} \nabla^{\gamma \lambda}{F_{\alpha \beta}}+\frac{9}{256}(\Gamma_{a})^{\rho}{}_{\gamma} W_{\alpha}\,^{\beta} W_{\rho \lambda} \lambda_{i \beta} {W}^{-1} \nabla^{\gamma \lambda}{W} - \frac{81}{256}(\Gamma_{a})^{\beta}{}_{\lambda} W_{\beta \rho} \lambda_{i \gamma} {W}^{-1} \nabla^{\lambda \rho}{\nabla_{\alpha}\,^{\gamma}{W}}+\frac{1161}{640}(\Gamma_{a})_{\alpha}{}^{\lambda} W_{\lambda}\,^{\gamma} W_{\gamma}\,^{\beta} F_{\beta}\,^{\rho} \lambda_{i \rho} {W}^{-1}+\frac{333}{128}(\Gamma_{a})^{\lambda \gamma} W_{\alpha \lambda} W_{\gamma}\,^{\beta} F_{\beta}\,^{\rho} \lambda_{i \rho} {W}^{-1} - \frac{45}{64}(\Gamma_{a})^{\beta \rho} W_{\alpha \beta} W_{\rho \lambda} \lambda_{i \gamma} {W}^{-1} \nabla^{\lambda \gamma}{W} - \frac{81}{32}(\Gamma_{a})^{\gamma \beta} W_{\alpha}\,^{\lambda} W_{\gamma \lambda} F_{\beta}\,^{\rho} \lambda_{i \rho} {W}^{-1}+\frac{9}{64}(\Gamma_{a})^{\rho}{}_{\lambda} W_{\alpha}\,^{\beta} W_{\rho \beta} \lambda_{i \gamma} {W}^{-1} \nabla^{\lambda \gamma}{W}+\frac{1665}{4096}(\Gamma_{a})^{\beta \lambda} W_{\beta}\,^{\rho} \lambda_{j \alpha} \lambda_{i \rho} X^{j}_{\lambda} {W}^{-1}+\frac{3327}{4096}(\Gamma_{a})^{\beta \lambda} W_{\beta}\,^{\rho} \lambda_{i \alpha} \lambda_{j \rho} X^{j}_{\lambda} {W}^{-1} - \frac{1683}{4096}(\Gamma_{a})^{\beta \lambda} W_{\beta}\,^{\rho} \lambda_{i \alpha} \lambda_{j \lambda} X^{j}_{\rho} {W}^{-1} - \frac{2187}{20480}(\Gamma_{a})_{\alpha}{}^{\beta} W_{\beta}\,^{\rho} \lambda^{\lambda}_{i} \lambda_{j \lambda} X^{j}_{\rho} {W}^{-1}%
 - \frac{1593}{4096}(\Gamma_{a})^{\beta \rho} W_{\alpha \beta} \lambda^{\lambda}_{i} \lambda_{j \lambda} X^{j}_{\rho} {W}^{-1} - \frac{135}{128}(\Gamma_{a})^{\rho}{}_{\gamma} W_{\rho}\,^{\lambda} F_{\alpha \beta} {W}^{-1} \nabla^{\gamma \beta}{\lambda_{i \lambda}} - \frac{27}{64}(\Gamma_{a})^{\beta}{}_{\lambda} W_{\beta}\,^{\rho} {W}^{-1} \nabla_{\alpha \gamma}{W} \nabla^{\lambda \gamma}{\lambda_{i \rho}} - \frac{21}{64}{\rm i} (\Gamma_{a})^{\beta}{}_{\lambda} W_{\beta}\,^{\rho} \lambda_{j \alpha} \lambda_{i \gamma} {W}^{-2} \nabla^{\lambda \gamma}{\lambda^{j}_{\rho}}+\frac{135}{128}(\Gamma_{a})^{\rho}{}_{\lambda} W_{\rho}\,^{\beta} \lambda_{i \gamma} {W}^{-1} \nabla^{\lambda \gamma}{F_{\alpha \beta}} - \frac{27}{256}(\Gamma_{a})^{\beta}{}_{\lambda} W_{\beta \rho} \lambda_{i \gamma} {W}^{-1} \nabla^{\lambda \gamma}{\nabla_{\alpha}\,^{\rho}{W}} - \frac{783}{640}(\Gamma_{a})_{\alpha}{}^{\lambda} W_{\lambda}\,^{\beta} W^{\rho \gamma} F_{\beta \rho} \lambda_{i \gamma} {W}^{-1}+\frac{1161}{1280}(\Gamma_{a})_{\alpha}{}^{\beta} W_{\beta \rho} W^{\lambda}\,_{\gamma} \lambda_{i \lambda} {W}^{-1} \nabla^{\rho \gamma}{W} - \frac{243}{256}(\Gamma_{a})^{\beta \lambda} W_{\beta \rho} W_{\lambda \gamma} \lambda_{i \alpha} {W}^{-1} \nabla^{\rho \gamma}{W}+\frac{117}{32}(\Gamma_{a})^{\gamma \beta} W_{\alpha}\,^{\lambda} W_{\gamma}\,^{\rho} F_{\beta \rho} \lambda_{i \lambda} {W}^{-1} - \frac{59}{1024}(\Gamma_{a})^{\beta \lambda} W_{\beta}\,^{\rho} \lambda_{i \alpha} \lambda^{\gamma}_{j} W_{\lambda \rho \gamma}\,^{j} {W}^{-1}+\frac{419}{2048}(\Gamma_{a})^{\beta \lambda} W_{\beta}\,^{\rho} \lambda^{\gamma}_{i} \lambda_{j \gamma} W_{\alpha \lambda \rho}\,^{j} {W}^{-1} - \frac{129}{128}(\Gamma_{a})^{\beta \gamma} W_{\beta}\,^{\rho} \lambda^{\lambda}_{i} \lambda_{j \gamma} W_{\alpha \rho \lambda}\,^{j} {W}^{-1} - \frac{105}{4096}(\Gamma_{a})^{\beta \lambda} W_{\alpha \beta} \lambda^{\rho}_{i} \lambda_{j \lambda} X^{j}_{\rho} {W}^{-1} - \frac{3099}{20480}(\Gamma_{a})_{\alpha}{}^{\beta} W_{\beta}\,^{\rho} \lambda_{i \rho} \lambda^{\lambda}_{j} X^{j}_{\lambda} {W}^{-1}+\frac{12357}{20480}(\Gamma_{a})_{\alpha}{}^{\beta} W_{\beta}\,^{\rho} \lambda^{\lambda}_{i} \lambda_{j \rho} X^{j}_{\lambda} {W}^{-1}+\frac{3}{64}{\rm i} (\Gamma_{a})^{\beta}{}_{\rho} \lambda_{j \alpha} {W}^{-2} \nabla^{\rho}\,_{\gamma}{\lambda^{j}_{\beta}} \nabla^{\gamma \lambda}{\lambda_{i \lambda}}+\frac{9}{64}(\Gamma_{a})^{\beta}{}_{\rho} {W}^{-1} \nabla^{\rho}\,_{\gamma}{F_{\alpha \beta}} \nabla^{\gamma \lambda}{\lambda_{i \lambda}}+\frac{9}{64}(\Gamma_{a})^{\beta}{}_{\rho} W_{\alpha \beta} {W}^{-1} \nabla^{\rho}\,_{\gamma}{W} \nabla^{\gamma \lambda}{\lambda_{i \lambda}} - \frac{9}{128}(\Gamma_{a})^{\beta}{}_{\rho} \nabla^{\rho}\,_{\gamma}{W_{\alpha \beta}} \nabla^{\gamma \lambda}{\lambda_{i \lambda}}%
+\frac{9}{128}(\Gamma_{a})_{\beta \rho} {W}^{-1} \nabla_{\gamma}\,^{\lambda}{\lambda_{i \lambda}} \nabla^{\beta \gamma}{\nabla^{\rho}\,_{\alpha}{W}} - \frac{81}{64}(\Gamma_{a})^{\beta}{}_{\lambda} W_{\alpha}\,^{\rho} F_{\beta \rho} {W}^{-1} \nabla^{\lambda \gamma}{\lambda_{i \gamma}}+\frac{27}{128}(\Gamma_{a})_{\rho \lambda} W_{\alpha \beta} {W}^{-1} \nabla^{\rho \beta}{W} \nabla^{\lambda \gamma}{\lambda_{i \gamma}}+\frac{207}{640}(\Gamma_{a})_{\alpha}{}^{\beta} W^{\rho}\,_{\lambda} F_{\beta \rho} {W}^{-1} \nabla^{\lambda \gamma}{\lambda_{i \gamma}}+\frac{441}{1280}(\Gamma_{a})_{\alpha \lambda} W_{\beta \rho} {W}^{-1} \nabla^{\lambda \beta}{W} \nabla^{\rho \gamma}{\lambda_{i \gamma}} - \frac{45}{128}(\Gamma_{a})^{\lambda \beta} W_{\lambda}\,^{\rho} F_{\beta \rho} {W}^{-1} \nabla_{\alpha}\,^{\gamma}{\lambda_{i \gamma}}+\frac{63}{128}(\Gamma_{a})^{\beta \lambda} W_{\beta}\,^{\rho} W_{\lambda \rho} \nabla_{\alpha}\,^{\gamma}{\lambda_{i \gamma}}+\frac{9}{16}(\Gamma_{a})^{\beta}{}_{\lambda} W_{\beta \rho} {W}^{-1} \nabla^{\lambda \rho}{W} \nabla_{\alpha}\,^{\gamma}{\lambda_{i \gamma}}+\frac{81}{128}(\Gamma_{a})_{\rho \lambda} W_{\alpha \beta} {W}^{-1} \nabla^{\rho \lambda}{W} \nabla^{\beta \gamma}{\lambda_{i \gamma}} - \frac{9}{128}(\Gamma_{a})^{\lambda \beta} W_{\alpha \lambda} F_{\beta \rho} {W}^{-1} \nabla^{\rho \gamma}{\lambda_{i \gamma}}+\frac{3}{32}(\Gamma_{a})^{\lambda \beta} \lambda_{j \lambda} W_{\alpha \beta \rho}\,^{j} {W}^{-1} \nabla^{\rho \gamma}{\lambda_{i \gamma}} - \frac{309}{2560}(\Gamma_{a})_{\alpha}{}^{\rho} \lambda_{j \rho} X^{j}_{\beta} {W}^{-1} \nabla^{\beta \lambda}{\lambda_{i \lambda}} - \frac{15}{64}(\Gamma_{a})^{\rho \beta} \lambda_{j \rho} X^{j}_{\beta} {W}^{-1} \nabla_{\alpha}\,^{\lambda}{\lambda_{i \lambda}} - \frac{21}{512}(\Gamma_{a})^{\beta}{}_{\rho} \lambda_{j \beta} X^{j}_{\alpha} {W}^{-1} \nabla^{\rho \lambda}{\lambda_{i \lambda}} - \frac{27}{128}(\Gamma_{a})^{\beta}{}_{\lambda} \nabla^{\lambda}\,_{\alpha}{W_{\beta \rho}} \nabla^{\rho \gamma}{\lambda_{i \gamma}}+\frac{9}{32}(\Gamma_{a})_{\alpha \lambda} \nabla^{\lambda \beta}{W_{\beta \rho}} \nabla^{\rho \gamma}{\lambda_{i \gamma}}+\frac{9}{256}(\Gamma_{a})_{\rho \lambda} \nabla^{\rho \lambda}{W_{\alpha \beta}} \nabla^{\beta \gamma}{\lambda_{i \gamma}} - \frac{9}{128}(\Gamma_{a})_{\alpha}{}^{\beta} \nabla_{\gamma}\,^{\rho}{W_{\beta \rho}} \nabla^{\gamma \lambda}{\lambda_{i \lambda}}+\frac{27}{64}(\Gamma_{a})^{\rho}{}_{\lambda} {W}^{-1} \nabla_{\gamma}\,^{\beta}{F_{\alpha \beta}} \nabla^{\lambda \gamma}{\lambda_{i \rho}} - \frac{3}{64}(\Gamma_{a})^{\beta}{}_{\rho} {W}^{-1} \nabla_{\alpha \lambda}{X_{i j}} \nabla^{\rho \lambda}{\lambda^{j}_{\beta}}%
+\frac{27}{128}(\Gamma_{a})^{\beta}{}_{\rho} {W}^{-1} \nabla^{\rho}\,_{\lambda}{\lambda_{i \beta}} \nabla^{\lambda}\,_{\gamma}{\nabla_{\alpha}\,^{\gamma}{W}} - \frac{81}{128}(\Gamma_{a})^{\lambda}{}_{\gamma} W^{\beta \rho} F_{\beta \rho} {W}^{-1} \nabla^{\gamma}\,_{\alpha}{\lambda_{i \lambda}} - \frac{81}{128}(\Gamma_{a})^{\lambda}{}_{\gamma} W^{\beta}\,_{\rho} F_{\alpha \beta} {W}^{-1} \nabla^{\gamma \rho}{\lambda_{i \lambda}} - \frac{81}{256}(\Gamma_{a})^{\lambda}{}_{\gamma} W_{\beta \rho} {W}^{-1} \nabla_{\alpha}\,^{\beta}{W} \nabla^{\gamma \rho}{\lambda_{i \lambda}}+\frac{45}{256}(\Gamma_{a})^{\rho}{}_{\lambda} \lambda^{\beta}_{i} X_{j \beta} {W}^{-1} \nabla^{\lambda}\,_{\alpha}{\lambda^{j}_{\rho}}+\frac{9}{256}(\Gamma_{a})^{\rho}{}_{\lambda} \lambda^{\beta}_{j} X_{i \beta} {W}^{-1} \nabla^{\lambda}\,_{\alpha}{\lambda^{j}_{\rho}} - \frac{9}{256}(\Gamma_{a})^{\rho}{}_{\lambda} \lambda_{j \alpha} X_{i \beta} {W}^{-1} \nabla^{\lambda \beta}{\lambda^{j}_{\rho}} - \frac{21}{128}\Phi_{\alpha \lambda i j} (\Gamma_{a})^{\beta}{}_{\rho} \nabla^{\lambda \rho}{\lambda^{j}_{\beta}}+\frac{27}{32}(\Gamma_{a})_{\lambda \gamma} W^{\beta \rho} F_{\alpha \beta} {W}^{-1} \nabla^{\lambda \gamma}{\lambda_{i \rho}} - \frac{135}{256}(\Gamma_{a})_{\lambda \gamma} W^{\beta}\,_{\rho} {W}^{-1} \nabla_{\alpha}\,^{\rho}{W} \nabla^{\lambda \gamma}{\lambda_{i \beta}}+\frac{171}{512}{\rm i} (\Gamma_{a})_{\lambda \gamma} W^{\beta \rho} \lambda_{j \alpha} \lambda_{i \beta} {W}^{-2} \nabla^{\lambda \gamma}{\lambda^{j}_{\rho}} - \frac{297}{256}(\Gamma_{a})_{\lambda \gamma} W^{\rho \beta} \lambda_{i \rho} {W}^{-1} \nabla^{\lambda \gamma}{F_{\alpha \beta}}+\frac{135}{128}(\Gamma_{a})_{\lambda \gamma} W_{\alpha}\,^{\beta} W_{\beta}\,^{\rho} \lambda_{i \rho} {W}^{-1} \nabla^{\lambda \gamma}{W}+\frac{27}{512}(\Gamma_{a})_{\lambda \gamma} W^{\beta}\,_{\rho} \lambda_{i \beta} {W}^{-1} \nabla^{\lambda \gamma}{\nabla_{\alpha}\,^{\rho}{W}} - \frac{657}{128}(\Gamma_{a})^{\lambda \beta} W_{\alpha \lambda} W^{\rho \gamma} F_{\beta \rho} \lambda_{i \gamma} {W}^{-1} - \frac{333}{128}(\Gamma_{a})^{\beta}{}_{\gamma} W_{\alpha \beta} W^{\rho}\,_{\lambda} \lambda_{i \rho} {W}^{-1} \nabla^{\gamma \lambda}{W} - \frac{1163}{2560}(\Gamma_{a})_{\alpha}{}^{\lambda} W^{\beta \rho} \lambda_{i \beta} \lambda^{\gamma}_{j} W_{\lambda \rho \gamma}\,^{j} {W}^{-1} - \frac{1039}{2048}(\Gamma_{a})^{\gamma \lambda} W^{\beta \rho} \lambda_{i \beta} \lambda_{j \gamma} W_{\alpha \lambda \rho}\,^{j} {W}^{-1} - \frac{9}{256}(\Gamma_{a})^{\lambda \rho} W_{\alpha}\,^{\beta} \lambda_{i \beta} \lambda_{j \lambda} X^{j}_{\rho} {W}^{-1} - \frac{87}{10240}(\Gamma_{a})_{\alpha}{}^{\lambda} W^{\beta \rho} \lambda_{i \beta} \lambda_{j \lambda} X^{j}_{\rho} {W}^{-1}%
+\frac{3621}{10240}(\Gamma_{a})_{\alpha}{}^{\lambda} W^{\beta \rho} \lambda_{i \beta} \lambda_{j \rho} X^{j}_{\lambda} {W}^{-1}+\frac{45}{512}(\Gamma_{a})^{\rho}{}_{\lambda} \lambda_{j \rho} X^{j \beta} {W}^{-1} \nabla^{\lambda}\,_{\alpha}{\lambda_{i \beta}}+\frac{3}{512}(\Gamma_{a})^{\rho}{}_{\lambda} \lambda_{j \rho} X^{j}_{\beta} {W}^{-1} \nabla^{\lambda \beta}{\lambda_{i \alpha}}+\frac{63}{128}{\rm i} (\Gamma_{a})^{\lambda}{}_{\gamma} W^{\beta}\,_{\rho} \lambda_{j \alpha} \lambda^{j}_{\lambda} {W}^{-2} \nabla^{\gamma \rho}{\lambda_{i \beta}}+\frac{27}{64}(\Gamma_{a})^{\rho \lambda} X_{i j} W_{\alpha}\,^{\beta} W_{\rho \beta} \lambda^{j}_{\lambda} {W}^{-1} - \frac{27}{80}(\Gamma_{a})_{\alpha}{}^{\lambda} X_{i j} W^{\beta \rho} W_{\beta \rho} \lambda^{j}_{\lambda} {W}^{-1} - \frac{63}{256}(\Gamma_{a})_{\alpha}{}^{\gamma} W^{\beta \rho} \lambda^{\lambda}_{i} \lambda_{j \gamma} W_{\beta \rho \lambda}\,^{j} {W}^{-1} - \frac{775}{2048}(\Gamma_{a})^{\gamma \rho} W_{\alpha}\,^{\beta} \lambda^{\lambda}_{i} \lambda_{j \gamma} W_{\rho \beta \lambda}\,^{j} {W}^{-1} - \frac{393}{4096}(\Gamma_{a})^{\beta \lambda} W_{\beta}\,^{\rho} \lambda_{i \rho} \lambda_{j \lambda} X^{j}_{\alpha} {W}^{-1} - \frac{1791}{2048}(\Gamma_{a})^{\beta \lambda} W_{\beta}\,^{\rho} \lambda_{j \lambda} \lambda^{j}_{\rho} X_{i \alpha} {W}^{-1} - \frac{27}{5120}(\Gamma_{a})_{\alpha}{}^{\lambda} W^{\beta \rho} \lambda_{j \lambda} \lambda^{j}_{\beta} X_{i \rho} {W}^{-1}+\frac{429}{1024}(\Gamma_{a})^{\lambda \rho} W_{\alpha}\,^{\beta} \lambda_{j \lambda} \lambda^{j}_{\beta} X_{i \rho} {W}^{-1} - \frac{69}{256}{\rm i} (\Gamma_{a})^{\beta}{}_{\lambda} W_{\beta \rho} \lambda_{j \alpha} \lambda^{j \gamma} {W}^{-2} \nabla^{\lambda \rho}{\lambda_{i \gamma}} - \frac{27}{128}(\Gamma_{a})^{\beta}{}_{\lambda} W_{\beta \rho} \lambda_{j \alpha} {W}^{-1} \nabla^{\lambda \rho}{X_{i}\,^{j}} - \frac{27}{40}(\Gamma_{a})_{\alpha}{}^{\beta} X_{i j} W_{\beta}\,^{\rho} W_{\rho}\,^{\lambda} \lambda^{j}_{\lambda} {W}^{-1}+\frac{81}{64}(\Gamma_{a})^{\beta \rho} X_{i j} W_{\alpha \beta} W_{\rho}\,^{\lambda} \lambda^{j}_{\lambda} {W}^{-1} - \frac{69}{256}{\rm i} (\Gamma_{a})^{\beta}{}_{\lambda} W_{\beta}\,^{\rho} \lambda_{j \alpha} \lambda^{j}_{\gamma} {W}^{-2} \nabla^{\lambda \gamma}{\lambda_{i \rho}}+\frac{9}{128}(\Gamma_{a})^{\beta}{}_{\rho} W_{\alpha \beta} \lambda_{j \lambda} {W}^{-1} \nabla^{\rho \lambda}{X_{i}\,^{j}}+\frac{27}{32}(\Gamma_{a})^{\beta \lambda} X_{i j} W_{\beta}\,^{\rho} W_{\lambda \rho} \lambda^{j}_{\alpha} {W}^{-1}+\frac{1907}{2048}(\Gamma_{a})^{\beta \lambda} W_{\beta}\,^{\rho} \lambda_{j \alpha} \lambda^{\gamma}_{i} W_{\lambda \rho \gamma}\,^{j} {W}^{-1}%
+\frac{3}{32}{\rm i} (\Gamma_{a})^{\beta}{}_{\rho} \lambda_{j \alpha} {W}^{-2} \nabla^{\rho}\,_{\gamma}{\lambda_{i \beta}} \nabla^{\gamma \lambda}{\lambda^{j}_{\lambda}} - \frac{9}{80}(\Gamma_{a})_{\alpha \beta} {W}^{-1} \nabla^{\beta}\,_{\lambda}{X_{i j}} \nabla^{\lambda \rho}{\lambda^{j}_{\rho}} - \frac{9}{320}(\Gamma_{a})_{\alpha}{}^{\beta} X_{i j} W_{\beta \rho} {W}^{-1} \nabla^{\rho \lambda}{\lambda^{j}_{\lambda}} - \frac{9}{320}(\Gamma_{a})_{\alpha}{}^{\beta} \lambda^{\rho}_{i} W_{\beta \rho \lambda j} {W}^{-1} \nabla^{\lambda \gamma}{\lambda^{j}_{\gamma}} - \frac{15}{32}(\Gamma_{a})^{\lambda \beta} \lambda_{i \lambda} W_{\alpha \beta \rho j} {W}^{-1} \nabla^{\rho \gamma}{\lambda^{j}_{\gamma}}+\frac{75}{512}(\Gamma_{a})_{\alpha}{}^{\beta} \lambda_{i \rho} X_{j \beta} {W}^{-1} \nabla^{\rho \lambda}{\lambda^{j}_{\lambda}} - \frac{9}{2560}(\Gamma_{a})_{\alpha}{}^{\rho} \lambda_{i \rho} X_{j \beta} {W}^{-1} \nabla^{\beta \lambda}{\lambda^{j}_{\lambda}} - \frac{21}{64}(\Gamma_{a})^{\rho \beta} \lambda_{i \rho} X_{j \beta} {W}^{-1} \nabla_{\alpha}\,^{\lambda}{\lambda^{j}_{\lambda}}+\frac{147}{512}(\Gamma_{a})^{\beta}{}_{\rho} \lambda_{i \beta} X_{j \alpha} {W}^{-1} \nabla^{\rho \lambda}{\lambda^{j}_{\lambda}} - \frac{81}{512}(\Gamma_{a})^{\beta}{}_{\rho} \lambda_{j \beta} X_{i \alpha} {W}^{-1} \nabla^{\rho \lambda}{\lambda^{j}_{\lambda}} - \frac{273}{2560}(\Gamma_{a})_{\alpha}{}^{\rho} \lambda_{j \rho} X_{i \beta} {W}^{-1} \nabla^{\beta \lambda}{\lambda^{j}_{\lambda}} - \frac{27}{256}(\Gamma_{a})^{\rho \beta} \lambda_{j \rho} X_{i \beta} {W}^{-1} \nabla_{\alpha}\,^{\lambda}{\lambda^{j}_{\lambda}}+\frac{9}{64}{\rm i} (\Gamma_{a})_{\lambda \gamma} W^{\beta \rho} \lambda_{j \alpha} \lambda^{j}_{\beta} {W}^{-2} \nabla^{\lambda \gamma}{\lambda_{i \rho}} - \frac{45}{256}(\Gamma_{a})_{\rho \lambda} W_{\alpha}\,^{\beta} \lambda_{j \beta} {W}^{-1} \nabla^{\rho \lambda}{X_{i}\,^{j}}+\frac{1379}{10240}(\Gamma_{a})_{\alpha}{}^{\lambda} W^{\beta \rho} \lambda^{\gamma}_{i} \lambda_{j \beta} W_{\lambda \rho \gamma}\,^{j} {W}^{-1} - \frac{3}{64}{\rm i} (\Gamma_{a})^{\beta}{}_{\rho} \lambda_{i \beta} {W}^{-2} \nabla^{\rho}\,_{\gamma}{\lambda^{\lambda}_{j}} \nabla_{\alpha}\,^{\gamma}{\lambda^{j}_{\lambda}} - \frac{3}{64}{\rm i} (\Gamma_{a})^{\beta}{}_{\rho} \lambda_{i \beta} {W}^{-2} \nabla^{\rho \gamma}{\lambda_{j \lambda}} \nabla_{\alpha}\,^{\lambda}{\lambda^{j}_{\gamma}}+\frac{15}{64}{\rm i} (\Gamma_{a})^{\rho}{}_{\lambda} W_{\alpha \beta} \lambda_{i \rho} \lambda^{\gamma}_{j} {W}^{-2} \nabla^{\lambda \beta}{\lambda^{j}_{\gamma}}+\frac{15}{128}{\rm i} (\Gamma_{a})^{\rho}{}_{\lambda} W_{\alpha}\,^{\beta} \lambda_{i \rho} \lambda_{j \gamma} {W}^{-2} \nabla^{\lambda \gamma}{\lambda^{j}_{\beta}}+\frac{3}{128}{\rm i} (\Gamma_{a})^{\beta}{}_{\rho} \lambda_{i \beta} {W}^{-2} \nabla^{\rho}\,_{\alpha}{\lambda_{j \lambda}} \nabla^{\lambda \gamma}{\lambda^{j}_{\gamma}}%
 - \frac{3}{128}{\rm i} (\Gamma_{a})^{\beta}{}_{\rho} \lambda_{i \beta} {W}^{-2} \nabla^{\rho}\,_{\gamma}{\lambda_{j \alpha}} \nabla^{\gamma \lambda}{\lambda^{j}_{\lambda}}+\frac{9}{256}{\rm i} (\Gamma_{a})^{\lambda}{}_{\gamma} W^{\beta \rho} \lambda_{i \lambda} \lambda_{j \beta} {W}^{-2} \nabla^{\gamma}\,_{\alpha}{\lambda^{j}_{\rho}}+\frac{27}{256}{\rm i} (\Gamma_{a})^{\lambda}{}_{\gamma} W^{\beta}\,_{\rho} \lambda_{i \lambda} \lambda_{j \beta} {W}^{-2} \nabla^{\gamma \rho}{\lambda^{j}_{\alpha}}+\frac{9}{16}(\Gamma_{a})^{\beta}{}_{\gamma} F_{\alpha \beta} F^{\rho}\,_{\lambda} {W}^{-2} \nabla^{\gamma \lambda}{\lambda_{i \rho}} - \frac{3}{160}(\Gamma_{a})_{\alpha \lambda} X_{i j} F^{\beta}\,_{\rho} {W}^{-2} \nabla^{\lambda \rho}{\lambda^{j}_{\beta}}+\frac{3}{16}{\rm i} (\Gamma_{a})^{\lambda}{}_{\gamma} F^{\beta}\,_{\rho} \lambda_{j \alpha} \lambda_{i \lambda} {W}^{-3} \nabla^{\gamma \rho}{\lambda^{j}_{\beta}} - \frac{9}{32}(\Gamma_{a})^{\lambda}{}_{\gamma} F^{\beta}\,_{\rho} \lambda_{i \lambda} {W}^{-2} \nabla^{\gamma \rho}{F_{\alpha \beta}} - \frac{81}{64}(\Gamma_{a})^{\lambda}{}_{\gamma} W_{\alpha}\,^{\beta} F_{\beta \rho} \lambda_{i \lambda} {W}^{-2} \nabla^{\gamma \rho}{W} - \frac{63}{256}(\Gamma_{a})^{\lambda}{}_{\gamma} F^{\beta}\,_{\rho} \lambda_{i \lambda} {W}^{-1} \nabla^{\gamma \rho}{W_{\alpha \beta}}+\frac{9}{64}(\Gamma_{a})^{\lambda}{}_{\gamma} F_{\beta \rho} \lambda_{i \lambda} {W}^{-2} \nabla^{\gamma \beta}{\nabla_{\alpha}\,^{\rho}{W}} - \frac{81}{32}(\Gamma_{a})^{\beta \gamma} W_{\alpha}\,^{\lambda} F_{\beta}\,^{\rho} F_{\lambda \rho} \lambda_{i \gamma} {W}^{-2} - \frac{27}{64}(\Gamma_{a})^{\beta \gamma} W_{\alpha \lambda} F_{\beta \rho} \lambda_{i \gamma} {W}^{-2} \nabla^{\lambda \rho}{W} - \frac{27}{8}(\Gamma_{a})^{\lambda \gamma} W_{\lambda}\,^{\rho} F_{\alpha}\,^{\beta} F_{\rho \beta} \lambda_{i \gamma} {W}^{-2} - \frac{27}{32}(\Gamma_{a})^{\rho \gamma} W_{\rho \lambda} F_{\alpha \beta} \lambda_{i \gamma} {W}^{-2} \nabla^{\lambda \beta}{W}+\frac{45}{128}(\Gamma_{a})^{\lambda \gamma} W_{\alpha \lambda} F^{\beta \rho} F_{\beta \rho} \lambda_{i \gamma} {W}^{-2}+\frac{3}{16}(\Gamma_{a})^{\lambda \gamma} F^{\beta \rho} \lambda_{i \lambda} \lambda_{j \gamma} W_{\alpha \beta \rho}\,^{j} {W}^{-2} - \frac{99}{256}(\Gamma_{a})^{\beta \lambda} F_{\beta}\,^{\rho} \lambda_{j \alpha} \lambda_{i \lambda} X^{j}_{\rho} {W}^{-2}+\frac{33}{32}(\Gamma_{a})^{\lambda \rho} F_{\alpha}\,^{\beta} \lambda_{i \lambda} \lambda_{j \beta} X^{j}_{\rho} {W}^{-2} - \frac{369}{256}(\Gamma_{a})^{\beta \lambda} F_{\alpha \beta} \lambda_{i \lambda} \lambda^{\rho}_{j} X^{j}_{\rho} {W}^{-2} - \frac{9}{1280}(\Gamma_{a})_{\alpha}{}^{\lambda} F^{\beta \rho} \lambda_{i \lambda} \lambda_{j \beta} X^{j}_{\rho} {W}^{-2}%
 - \frac{171}{256}(\Gamma_{a})^{\beta \lambda} F_{\beta}\,^{\rho} \lambda_{i \lambda} \lambda_{j \rho} X^{j}_{\alpha} {W}^{-2}+\frac{45}{512}(\Gamma_{a})^{\beta \gamma} F_{\beta}\,^{\rho} \lambda_{i \gamma} {W}^{-1} \nabla_{\alpha}\,^{\lambda}{W_{\rho \lambda}}+\frac{81}{256}(\Gamma_{a})^{\lambda}{}_{\gamma} F^{\beta \rho} \lambda_{i \lambda} {W}^{-1} \nabla^{\gamma}\,_{\alpha}{W_{\beta \rho}} - \frac{27}{256}(\Gamma_{a})^{\gamma \lambda} F^{\beta}\,_{\rho} \lambda_{i \gamma} {W}^{-1} \nabla_{\alpha}\,^{\rho}{W_{\lambda \beta}}+\frac{63}{320}(\Gamma_{a})_{\alpha}{}^{\gamma} F^{\beta}\,_{\rho} \lambda_{i \gamma} {W}^{-1} \nabla^{\rho \lambda}{W_{\beta \lambda}} - \frac{315}{512}(\Gamma_{a})^{\gamma \rho} F_{\alpha \beta} \lambda_{i \gamma} {W}^{-1} \nabla^{\beta \lambda}{W_{\rho \lambda}}+\frac{189}{512}(\Gamma_{a})^{\beta \gamma} F_{\beta \rho} \lambda_{i \gamma} {W}^{-1} \nabla^{\rho \lambda}{W_{\alpha \lambda}}+\frac{513}{512}(\Gamma_{a})^{\lambda}{}_{\gamma} F_{\alpha}\,^{\beta} \lambda_{i \lambda} {W}^{-1} \nabla^{\gamma \rho}{W_{\beta \rho}}+\frac{297}{64}(\Gamma_{a})^{\beta \gamma} W^{\rho \lambda} W_{\rho \lambda} F_{\alpha \beta} \lambda_{i \gamma} {W}^{-1} - \frac{3}{64}{\rm i} (\Gamma_{a})^{\beta}{}_{\rho} \lambda_{j \beta} {W}^{-2} \nabla^{\rho}\,_{\gamma}{\lambda^{\lambda}_{i}} \nabla_{\alpha}\,^{\gamma}{\lambda^{j}_{\lambda}} - \frac{3}{64}{\rm i} (\Gamma_{a})^{\beta}{}_{\rho} \lambda_{j \beta} {W}^{-2} \nabla^{\rho \gamma}{\lambda_{i \lambda}} \nabla_{\alpha}\,^{\lambda}{\lambda^{j}_{\gamma}}+\frac{3}{128}{\rm i} (\Gamma_{a})^{\beta}{}_{\rho} \lambda_{j \beta} {W}^{-2} \nabla^{\rho}\,_{\alpha}{\lambda_{i \lambda}} \nabla^{\lambda \gamma}{\lambda^{j}_{\gamma}}+\frac{3}{128}{\rm i} (\Gamma_{a})^{\beta}{}_{\rho} \lambda_{j \beta} {W}^{-2} \nabla^{\rho}\,_{\gamma}{\lambda_{i \alpha}} \nabla^{\gamma \lambda}{\lambda^{j}_{\lambda}}+\frac{63}{256}{\rm i} (\Gamma_{a})^{\lambda}{}_{\gamma} W^{\beta \rho} \lambda_{j \lambda} \lambda^{j}_{\beta} {W}^{-2} \nabla^{\gamma}\,_{\alpha}{\lambda_{i \rho}} - \frac{27}{256}{\rm i} (\Gamma_{a})^{\lambda}{}_{\gamma} W^{\beta}\,_{\rho} \lambda_{j \lambda} \lambda^{j}_{\beta} {W}^{-2} \nabla^{\gamma \rho}{\lambda_{i \alpha}}+\frac{3}{16}{\rm i} (\Gamma_{a})^{\lambda}{}_{\gamma} F^{\beta}\,_{\rho} \lambda_{j \alpha} \lambda^{j}_{\lambda} {W}^{-3} \nabla^{\gamma \rho}{\lambda_{i \beta}} - \frac{9}{32}(\Gamma_{a})^{\rho}{}_{\lambda} F_{\alpha \beta} \lambda_{j \rho} {W}^{-2} \nabla^{\lambda \beta}{X_{i}\,^{j}}+\frac{9}{32}(\Gamma_{a})^{\beta \lambda} X_{i j} W_{\alpha}\,^{\rho} F_{\beta \rho} \lambda^{j}_{\lambda} {W}^{-2} - \frac{63}{320}(\Gamma_{a})_{\alpha}{}^{\lambda} X_{i j} W^{\beta \rho} F_{\beta \rho} \lambda^{j}_{\lambda} {W}^{-2} - \frac{9}{32}(\Gamma_{a})^{\rho \lambda} X_{i j} W_{\rho}\,^{\beta} F_{\alpha \beta} \lambda^{j}_{\lambda} {W}^{-2}%
 - \frac{3}{8}(\Gamma_{a})^{\beta \gamma} F_{\beta}\,^{\rho} \lambda^{\lambda}_{i} \lambda_{j \gamma} W_{\alpha \rho \lambda}\,^{j} {W}^{-2} - \frac{3}{160}(\Gamma_{a})_{\alpha}{}^{\gamma} F^{\beta \rho} \lambda^{\lambda}_{i} \lambda_{j \gamma} W_{\beta \rho \lambda}\,^{j} {W}^{-2} - \frac{3}{32}(\Gamma_{a})^{\gamma \rho} F_{\alpha}\,^{\beta} \lambda^{\lambda}_{i} \lambda_{j \gamma} W_{\rho \beta \lambda}\,^{j} {W}^{-2} - \frac{81}{256}(\Gamma_{a})^{\beta \lambda} F_{\beta}\,^{\rho} \lambda_{i \alpha} \lambda_{j \lambda} X^{j}_{\rho} {W}^{-2} - \frac{69}{128}(\Gamma_{a})^{\lambda \rho} F_{\alpha}\,^{\beta} \lambda_{i \beta} \lambda_{j \lambda} X^{j}_{\rho} {W}^{-2}+\frac{171}{256}(\Gamma_{a})^{\beta \lambda} F_{\alpha \beta} \lambda^{\rho}_{i} \lambda_{j \lambda} X^{j}_{\rho} {W}^{-2}+\frac{21}{256}(\Gamma_{a})_{\alpha}{}^{\lambda} F^{\beta \rho} \lambda_{i \beta} \lambda_{j \lambda} X^{j}_{\rho} {W}^{-2}+\frac{75}{256}(\Gamma_{a})^{\beta \lambda} F_{\beta}\,^{\rho} \lambda_{i \rho} \lambda_{j \lambda} X^{j}_{\alpha} {W}^{-2} - \frac{25}{16}\Phi_{\alpha}\,^{\beta}\,_{i j} (\Gamma_{a})^{\rho \lambda} F_{\beta \rho} \lambda^{j}_{\lambda} {W}^{-1} - \frac{1}{80}\Phi^{\beta \rho}\,_{i j} (\Gamma_{a})_{\alpha}{}^{\lambda} F_{\beta \rho} \lambda^{j}_{\lambda} {W}^{-1}+\frac{19}{8}\Phi^{\rho \beta}\,_{i j} (\Gamma_{a})_{\rho}{}^{\lambda} F_{\alpha \beta} \lambda^{j}_{\lambda} {W}^{-1}+\frac{3}{64}(\Gamma_{a})^{\beta}{}_{\rho} {W}^{-1} \nabla^{\rho}\,_{\lambda}{X_{i j}} \nabla_{\alpha}\,^{\lambda}{\lambda^{j}_{\beta}} - \frac{21}{64}(\Gamma_{a})_{\beta \rho} {W}^{-1} \nabla^{\beta}\,_{\alpha}{X_{i j}} \nabla^{\rho \lambda}{\lambda^{j}_{\lambda}}+\frac{9}{64}(\Gamma_{a})^{\beta}{}_{\lambda} W_{\beta}\,^{\rho} \lambda_{j \rho} {W}^{-1} \nabla^{\lambda}\,_{\alpha}{X_{i}\,^{j}} - \frac{171}{640}(\Gamma_{a})_{\alpha \lambda} W^{\beta}\,_{\rho} \lambda_{j \beta} {W}^{-1} \nabla^{\lambda \rho}{X_{i}\,^{j}} - \frac{9}{32}(\Gamma_{a})^{\beta}{}_{\lambda} F_{\beta \rho} \lambda_{j \alpha} {W}^{-2} \nabla^{\lambda \rho}{X_{i}\,^{j}}+\frac{9}{16}(\Gamma_{a})^{\beta}{}_{\lambda} F_{\beta \rho} {W}^{-1} \nabla^{\lambda \rho}{\nabla_{\alpha}\,^{\gamma}{\lambda_{i \gamma}}}+\frac{9}{64}(\Gamma_{a})^{\beta}{}_{\gamma} F_{\beta \rho} \lambda^{\lambda}_{i} {W}^{-1} \nabla^{\gamma \rho}{W_{\alpha \lambda}}+\frac{45}{128}{\rm i} (\Gamma_{a})^{\beta}{}_{\lambda} F_{\beta \rho} \nabla^{\lambda \rho}{X_{i \alpha}} - \frac{783}{320}(\Gamma_{a})_{\alpha}{}^{\beta} W^{\lambda \rho} W_{\lambda}\,^{\gamma} F_{\beta \rho} \lambda_{i \gamma} {W}^{-1}%
+\frac{387}{128}(\Gamma_{a})^{\rho \beta} W_{\rho \lambda} F_{\alpha \beta} {W}^{-1} \nabla^{\lambda \gamma}{\lambda_{i \gamma}} - \frac{1107}{128}(\Gamma_{a})^{\rho \beta} W_{\rho}\,^{\lambda} W_{\lambda}\,^{\gamma} F_{\alpha \beta} \lambda_{i \gamma} {W}^{-1}+\frac{135}{64}(\Gamma_{a})^{\lambda \beta} W_{\alpha}\,^{\rho} W_{\lambda}\,^{\gamma} F_{\beta \rho} \lambda_{i \gamma} {W}^{-1} - \frac{81}{320}{\rm i} (\Gamma_{a})_{\alpha}{}^{\beta} X_{i j} F_{\beta}\,^{\rho} X^{j}_{\rho} {W}^{-1} - \frac{81}{64}{\rm i} (\Gamma_{a})^{\beta \rho} X_{i j} F_{\alpha \beta} X^{j}_{\rho} {W}^{-1} - \frac{7}{16}\Phi^{\beta \lambda}\,_{i j} (\Gamma_{a})_{\alpha}{}^{\rho} F_{\beta \rho} \lambda^{j}_{\lambda} {W}^{-1} - \frac{27}{16}\Phi^{\rho \lambda}\,_{i j} (\Gamma_{a})_{\rho}{}^{\beta} F_{\alpha \beta} \lambda^{j}_{\lambda} {W}^{-1} - \frac{423}{512}(\Gamma_{a})^{\beta \lambda} F_{\beta}\,^{\rho} \lambda_{i \rho} {W}^{-1} \nabla_{\alpha}\,^{\gamma}{W_{\lambda \gamma}} - \frac{99}{256}(\Gamma_{a})^{\beta}{}_{\gamma} F_{\beta}\,^{\rho} \lambda^{\lambda}_{i} {W}^{-1} \nabla^{\gamma}\,_{\alpha}{W_{\rho \lambda}}+\frac{135}{256}(\Gamma_{a})^{\beta \lambda} F_{\beta \rho} \lambda^{\gamma}_{i} {W}^{-1} \nabla_{\alpha}\,^{\rho}{W_{\lambda \gamma}}+\frac{153}{512}(\Gamma_{a})_{\alpha}{}^{\beta} F_{\beta}\,^{\rho} \lambda_{i \gamma} {W}^{-1} \nabla^{\gamma \lambda}{W_{\rho \lambda}} - \frac{63}{256}(\Gamma_{a})^{\beta \rho} F_{\alpha \beta} \lambda_{i \gamma} {W}^{-1} \nabla^{\gamma \lambda}{W_{\rho \lambda}} - \frac{99}{256}(\Gamma_{a})^{\beta \lambda} F_{\beta \rho} \lambda_{i \gamma} {W}^{-1} \nabla^{\rho \gamma}{W_{\alpha \lambda}} - \frac{297}{512}(\Gamma_{a})^{\beta}{}_{\gamma} F_{\beta}\,^{\rho} \lambda_{i \rho} {W}^{-1} \nabla^{\gamma \lambda}{W_{\alpha \lambda}} - \frac{63}{32}(\Gamma_{a})^{\beta}{}_{\gamma} F_{\alpha \beta} \lambda^{\rho}_{i} {W}^{-1} \nabla^{\gamma \lambda}{W_{\rho \lambda}} - \frac{1737}{2560}(\Gamma_{a})_{\alpha}{}^{\beta} F_{\beta \rho} \lambda^{\lambda}_{i} {W}^{-1} \nabla^{\rho \gamma}{W_{\lambda \gamma}}+\frac{531}{512}(\Gamma_{a})^{\beta}{}_{\gamma} F_{\beta}\,^{\rho} \lambda_{i \alpha} {W}^{-1} \nabla^{\gamma \lambda}{W_{\rho \lambda}}+\frac{9}{512}(\Gamma_{a})^{\beta \lambda} F_{\beta \rho} \lambda_{i \alpha} {W}^{-1} \nabla^{\rho \gamma}{W_{\lambda \gamma}}+\frac{405}{256}(\Gamma_{a})_{\alpha}{}^{\beta} W^{\lambda \gamma} W_{\lambda \gamma} F_{\beta}\,^{\rho} \lambda_{i \rho} {W}^{-1} - \frac{27}{64}(\Gamma_{a})^{\lambda \beta} W_{\lambda}\,^{\gamma} W_{\gamma}\,^{\rho} F_{\beta \rho} \lambda_{i \alpha} {W}^{-1}%
+\frac{3}{64}(\Gamma_{a})^{\beta}{}_{\gamma} \lambda^{\rho}_{j} W_{\beta \rho}\,^{\lambda}\,_{i} {W}^{-1} \nabla^{\gamma}\,_{\alpha}{\lambda^{j}_{\lambda}}+\frac{15}{64}(\Gamma_{a})^{\beta \gamma} \lambda^{\rho}_{j} W_{\beta \rho \lambda i} {W}^{-1} \nabla_{\alpha}\,^{\lambda}{\lambda^{j}_{\gamma}}+\frac{2523}{2048}(\Gamma_{a})^{\beta \lambda} W_{\beta}\,^{\rho} \lambda_{j \alpha} \lambda^{j \gamma} W_{\lambda \rho \gamma i} {W}^{-1} - \frac{33}{64}(\Gamma_{a})^{\beta \rho} W_{\alpha \beta} \lambda^{\lambda}_{j} \lambda^{j \gamma} W_{\rho \lambda \gamma i} {W}^{-1}+\frac{843}{2048}(\Gamma_{a})^{\gamma \rho} W_{\alpha}\,^{\beta} \lambda_{j \gamma} \lambda^{j \lambda} W_{\rho \beta \lambda i} {W}^{-1} - \frac{27}{320}(\Gamma_{a})_{\alpha}{}^{\beta} \lambda^{\rho}_{j} W_{\beta \rho \lambda i} {W}^{-1} \nabla^{\lambda \gamma}{\lambda^{j}_{\gamma}}+\frac{15}{64}(\Gamma_{a})^{\beta}{}_{\lambda} \lambda^{\rho}_{j} W_{\alpha \beta \rho i} {W}^{-1} \nabla^{\lambda \gamma}{\lambda^{j}_{\gamma}} - \frac{45}{64}(\Gamma_{a})^{\beta \lambda} W_{\beta}\,^{\rho} \lambda_{j \rho} \lambda^{j \gamma} W_{\alpha \lambda \gamma i} {W}^{-1} - \frac{63}{16}{\rm i} (\Gamma_{a})^{\rho \gamma} F_{\alpha}\,^{\beta} F_{\rho}\,^{\lambda} W_{\gamma \beta \lambda i} {W}^{-1} - \frac{9}{16}{\rm i} (\Gamma_{a})^{\beta \lambda} F_{\beta}\,^{\rho} W_{\lambda \rho \gamma i} {W}^{-1} \nabla_{\alpha}\,^{\gamma}{W}+\frac{27}{16}(\Gamma_{a})^{\beta \gamma} C_{\alpha \beta}\,^{\rho \lambda} F_{\gamma \rho} \lambda_{i \lambda} {W}^{-1} - \frac{117}{128}(\Gamma_{a})^{\beta \lambda} F_{\beta}\,^{\rho} \lambda_{i \gamma} {W}^{-1} \nabla_{\alpha}\,^{\gamma}{W_{\lambda \rho}}+\frac{25}{16}\Phi^{\lambda \beta}\,_{i j} (\Gamma_{a})_{\lambda}{}^{\rho} F_{\beta \rho} \lambda^{j}_{\alpha} {W}^{-1}+\frac{15}{32}(\Gamma_{a})^{\beta \lambda} F_{\beta}\,^{\rho} \lambda_{j \alpha} \lambda^{j \gamma} W_{\lambda \rho \gamma i} {W}^{-2} - \frac{309}{512}(\Gamma_{a})^{\gamma \lambda} W^{\beta \rho} \lambda_{j \alpha} \lambda^{j}_{\gamma} W_{\lambda \beta \rho i} {W}^{-1} - \frac{21}{32}(\Gamma_{a})^{\lambda \beta} \lambda_{j \lambda} W_{\alpha \beta \rho i} {W}^{-1} \nabla^{\rho \gamma}{\lambda^{j}_{\gamma}}+\frac{927}{1024}(\Gamma_{a})^{\gamma \lambda} W^{\beta \rho} \lambda_{j \gamma} \lambda^{j}_{\beta} W_{\alpha \lambda \rho i} {W}^{-1}+\frac{45}{16}{\rm i} (\Gamma_{a})^{\beta \gamma} F_{\alpha \beta} F^{\rho \lambda} W_{\gamma \rho \lambda i} {W}^{-1}+\frac{27}{8}{\rm i} (\Gamma_{a})^{\lambda \gamma} W_{\alpha \lambda} F^{\beta \rho} W_{\gamma \beta \rho i} - \frac{9}{16}{\rm i} (\Gamma_{a})^{\lambda}{}_{\gamma} F^{\beta \rho} W_{\lambda \beta \rho i} {W}^{-1} \nabla^{\gamma}\,_{\alpha}{W}%
 - \frac{9}{8}(\Gamma_{a})^{\beta \gamma} C_{\alpha \beta}\,^{\rho \lambda} F_{\rho \lambda} \lambda_{i \gamma} {W}^{-1} - \frac{9}{32}(\Gamma_{a})^{\gamma \lambda} F^{\beta \rho} \lambda_{j \alpha} \lambda^{j}_{\gamma} W_{\lambda \beta \rho i} {W}^{-2}+\frac{27}{512}(\Gamma_{a})^{\rho}{}_{\lambda} \lambda_{j \rho} X_{i}^{\beta} {W}^{-1} \nabla^{\lambda}\,_{\alpha}{\lambda^{j}_{\beta}}+\frac{45}{512}(\Gamma_{a})^{\rho \lambda} \lambda_{j \rho} X_{i \beta} {W}^{-1} \nabla_{\alpha}\,^{\beta}{\lambda^{j}_{\lambda}}+\frac{639}{1024}(\Gamma_{a})^{\beta \lambda} W_{\beta}\,^{\rho} \lambda_{j \alpha} \lambda^{j}_{\lambda} X_{i \rho} {W}^{-1} - \frac{2145}{2048}(\Gamma_{a})^{\beta \lambda} W_{\alpha \beta} \lambda_{j \lambda} \lambda^{j \rho} X_{i \rho} {W}^{-1}+\frac{27}{32}{\rm i} (\Gamma_{a})^{\beta \rho} F_{\alpha \beta} F_{\rho}\,^{\lambda} X_{i \lambda} {W}^{-1} - \frac{27}{64}{\rm i} (\Gamma_{a})^{\beta}{}_{\lambda} F_{\beta}\,^{\rho} X_{i \rho} {W}^{-1} \nabla^{\lambda}\,_{\alpha}{W} - \frac{81}{512}(\Gamma_{a})^{\beta \rho} F_{\alpha \beta} Y \lambda_{i \rho} {W}^{-1}+\frac{63}{256}(\Gamma_{a})^{\beta \lambda} F_{\beta}\,^{\rho} \lambda_{j \alpha} \lambda^{j}_{\lambda} X_{i \rho} {W}^{-2}+\frac{237}{2560}(\Gamma_{a})_{\alpha}{}^{\beta} \lambda_{j \rho} X_{i \beta} {W}^{-1} \nabla^{\rho \lambda}{\lambda^{j}_{\lambda}}+\frac{45}{512}(\Gamma_{a})^{\beta}{}_{\rho} \lambda_{j \alpha} X_{i \beta} {W}^{-1} \nabla^{\rho \lambda}{\lambda^{j}_{\lambda}}+\frac{9}{16}{\rm i} (\Gamma_{a})^{\rho \lambda} F_{\alpha}\,^{\beta} F_{\rho \beta} X_{i \lambda} {W}^{-1}+\frac{9}{64}{\rm i} (\Gamma_{a})^{\beta \lambda} F_{\beta \rho} X_{i \lambda} {W}^{-1} \nabla_{\alpha}\,^{\rho}{W} - \frac{81}{2560}(\Gamma_{a})_{\alpha}{}^{\beta} F_{\beta}\,^{\rho} Y \lambda_{i \rho} {W}^{-1}+\frac{9}{128}(\Gamma_{a})^{\beta \lambda} F_{\beta}\,^{\rho} \lambda_{j \alpha} \lambda^{j}_{\rho} X_{i \lambda} {W}^{-2} - \frac{195}{512}(\Gamma_{a})^{\beta \rho} \lambda_{i \lambda} X_{j \beta} {W}^{-1} \nabla_{\alpha}\,^{\lambda}{\lambda^{j}_{\rho}}+\frac{33}{64}(\Gamma_{a})^{\beta \lambda} F_{\beta}\,^{\rho} \lambda_{j \alpha} \lambda_{i \rho} X^{j}_{\lambda} {W}^{-2}+\frac{21}{64}(\Gamma_{a})^{\beta \gamma} \lambda^{\rho}_{i} W_{\beta \rho \lambda j} {W}^{-1} \nabla_{\alpha}\,^{\lambda}{\lambda^{j}_{\gamma}}+\frac{33}{64}(\Gamma_{a})^{\beta \rho} W_{\alpha \beta} \lambda^{\lambda}_{i} \lambda^{\gamma}_{j} W_{\rho \lambda \gamma}\,^{j} {W}^{-1}%
+\frac{63}{64}(\Gamma_{a})^{\beta \lambda} W_{\beta}\,^{\rho} \lambda^{\gamma}_{i} \lambda_{j \rho} W_{\alpha \lambda \gamma}\,^{j} {W}^{-1} - \frac{9}{32}{\rm i} (\Gamma_{a})^{\beta \lambda} X_{i j} F_{\beta}\,^{\rho} W_{\alpha \lambda \rho}\,^{j} {W}^{-1}+\frac{15}{32}(\Gamma_{a})^{\beta \lambda} F_{\beta}\,^{\rho} \lambda_{j \alpha} \lambda^{\gamma}_{i} W_{\lambda \rho \gamma}\,^{j} {W}^{-2} - \frac{3}{16}(\Gamma_{a})^{\gamma \beta} \lambda_{i \gamma} W_{\beta}\,^{\rho}\,_{\lambda j} {W}^{-1} \nabla_{\alpha}\,^{\lambda}{\lambda^{j}_{\rho}} - \frac{911}{2048}(\Gamma_{a})^{\gamma \lambda} W^{\beta \rho} \lambda_{j \alpha} \lambda_{i \gamma} W_{\lambda \beta \rho}\,^{j} {W}^{-1} - \frac{9}{160}{\rm i} (\Gamma_{a})_{\alpha}{}^{\lambda} X_{i j} F^{\beta \rho} W_{\lambda \beta \rho}\,^{j} {W}^{-1}+\frac{63}{512}(\Gamma_{a})^{\rho \lambda} \lambda_{i \rho} X_{j \beta} {W}^{-1} \nabla_{\alpha}\,^{\beta}{\lambda^{j}_{\lambda}} - \frac{189}{512}(\Gamma_{a})^{\lambda}{}_{\gamma} W^{\beta \rho} W_{\beta \rho} \lambda_{i \lambda} {W}^{-1} \nabla^{\gamma}\,_{\alpha}{W}+\frac{27}{16}{\rm i} (\Gamma_{a})^{\lambda \gamma} W^{\beta \rho} W_{\beta \rho} \lambda_{j \alpha} \lambda_{i \lambda} \lambda^{j}_{\gamma} {W}^{-2}+\frac{675}{128}(\Gamma_{a})^{\rho \gamma} W_{\rho}\,^{\lambda} W_{\lambda}\,^{\beta} F_{\alpha \beta} \lambda_{i \gamma} {W}^{-1} - \frac{513}{256}(\Gamma_{a})^{\beta \gamma} W_{\beta}\,^{\rho} W_{\rho \lambda} \lambda_{i \gamma} {W}^{-1} \nabla_{\alpha}\,^{\lambda}{W}+\frac{27}{128}(\Gamma_{a})^{\beta}{}_{\gamma} W_{\beta}\,^{\rho} W_{\rho}\,^{\lambda} \lambda_{i \lambda} {W}^{-1} \nabla^{\gamma}\,_{\alpha}{W}+\frac{747}{256}{\rm i} (\Gamma_{a})^{\beta \gamma} W_{\beta}\,^{\rho} W_{\rho}\,^{\lambda} \lambda_{j \alpha} \lambda_{i \lambda} \lambda^{j}_{\gamma} {W}^{-2}+\frac{9}{32}(\Gamma_{a})^{\beta}{}_{\lambda} W_{\beta \rho} {W}^{-1} \nabla^{\lambda}\,_{\alpha}{W} \nabla^{\rho \gamma}{\lambda_{i \gamma}}+\frac{33}{256}{\rm i} (\Gamma_{a})^{\beta \lambda} W_{\beta \rho} \lambda_{j \alpha} \lambda^{j}_{\lambda} {W}^{-2} \nabla^{\rho \gamma}{\lambda_{i \gamma}} - \frac{27}{128}(\Gamma_{a})^{\rho \gamma} W_{\rho \lambda} \lambda_{i \gamma} {W}^{-1} \nabla^{\lambda \beta}{F_{\alpha \beta}}+\frac{27}{128}(\Gamma_{a})^{\beta \lambda} W_{\beta \rho} \lambda_{j \lambda} {W}^{-1} \nabla_{\alpha}\,^{\rho}{X_{i}\,^{j}} - \frac{27}{256}(\Gamma_{a})^{\beta \lambda} W_{\beta \rho} \lambda_{i \lambda} {W}^{-1} \nabla_{\gamma}\,^{\rho}{\nabla_{\alpha}\,^{\gamma}{W}} - \frac{225}{64}{\rm i} (\Gamma_{a})^{\beta \gamma} W_{\beta}\,^{\rho} W_{\rho}\,^{\lambda} \lambda_{j \alpha} \lambda_{i \gamma} \lambda^{j}_{\lambda} {W}^{-2}+\frac{51}{128}{\rm i} (\Gamma_{a})^{\beta \lambda} W_{\beta \rho} \lambda_{j \alpha} \lambda_{i \lambda} {W}^{-2} \nabla^{\rho \gamma}{\lambda^{j}_{\gamma}}%
 - \frac{45}{64}(\Gamma_{a})^{\rho \lambda} F_{\alpha}\,^{\beta} \lambda_{i \rho} \lambda_{j \lambda} X^{j}_{\beta} {W}^{-2}+\frac{27}{64}{\rm i} (\Gamma_{a})^{\lambda \gamma} W^{\beta}\,_{\rho} \lambda_{i \lambda} \lambda_{j \gamma} {W}^{-2} \nabla_{\alpha}\,^{\rho}{\lambda^{j}_{\beta}} - \frac{45}{16}{\rm i} (\Gamma_{a})^{\lambda \gamma} W_{\alpha}\,^{\beta} W_{\beta}\,^{\rho} \lambda_{i \lambda} \lambda_{j \gamma} \lambda^{j}_{\rho} {W}^{-2}+\frac{9}{128}{\rm i} (\Gamma_{a})^{\rho \lambda} W_{\alpha \beta} \lambda_{i \rho} \lambda_{j \lambda} {W}^{-2} \nabla^{\beta \gamma}{\lambda^{j}_{\gamma}}+\frac{81}{16}(\Gamma_{a})^{\beta \gamma} W^{\rho \lambda} F_{\alpha \beta} F_{\rho \lambda} \lambda_{i \gamma} {W}^{-2} - \frac{27}{64}(\Gamma_{a})^{\lambda}{}_{\gamma} W^{\beta \rho} F_{\beta \rho} \lambda_{i \lambda} {W}^{-2} \nabla^{\gamma}\,_{\alpha}{W}+\frac{27}{16}{\rm i} (\Gamma_{a})^{\lambda \gamma} W^{\beta \rho} F_{\beta \rho} \lambda_{j \alpha} \lambda_{i \lambda} \lambda^{j}_{\gamma} {W}^{-3}+\frac{459}{256}(\Gamma_{a})^{\rho \gamma} W_{\rho}\,^{\lambda} W_{\gamma \lambda} F_{\alpha}\,^{\beta} \lambda_{i \beta} {W}^{-1}+\frac{27}{64}(\Gamma_{a})^{\beta \lambda} W_{\beta}\,^{\rho} W_{\lambda \rho} \lambda_{i \gamma} {W}^{-1} \nabla_{\alpha}\,^{\gamma}{W}+\frac{513}{512}{\rm i} (\Gamma_{a})^{\beta \lambda} W_{\beta}\,^{\rho} W_{\lambda \rho} \lambda_{j \alpha} \lambda^{\gamma}_{i} \lambda^{j}_{\gamma} {W}^{-2} - \frac{9}{32}(\Gamma_{a})^{\beta \lambda} W_{\beta}\,^{\rho} \lambda_{i \rho} \lambda^{\gamma}_{j} W_{\alpha \lambda \gamma}\,^{j} {W}^{-1}+\frac{135}{64}(\Gamma_{a})^{\rho \lambda} W_{\rho}\,^{\beta} W_{\lambda}\,^{\gamma} F_{\alpha \beta} \lambda_{i \gamma} {W}^{-1}+\frac{27}{128}(\Gamma_{a})^{\beta \lambda} W_{\beta}\,^{\rho} W_{\lambda \gamma} \lambda_{i \rho} {W}^{-1} \nabla_{\alpha}\,^{\gamma}{W}+\frac{27}{32}{\rm i} (\Gamma_{a})^{\beta \lambda} W_{\beta}\,^{\rho} W_{\lambda}\,^{\gamma} \lambda_{j \alpha} \lambda_{i \rho} \lambda^{j}_{\gamma} {W}^{-2}+\frac{33}{128}{\rm i} (\Gamma_{a})^{\beta \lambda} W_{\beta \rho} \lambda_{i \lambda} \lambda^{\gamma}_{j} {W}^{-2} \nabla_{\alpha}\,^{\rho}{\lambda^{j}_{\gamma}}+\frac{9}{64}{\rm i} (\Gamma_{a})^{\beta \lambda} W_{\beta}\,^{\rho} \lambda_{i \lambda} \lambda_{j \gamma} {W}^{-2} \nabla_{\alpha}\,^{\gamma}{\lambda^{j}_{\rho}} - \frac{1713}{512}{\rm i} (\Gamma_{a})^{\rho \gamma} W_{\alpha}\,^{\beta} W_{\rho}\,^{\lambda} \lambda_{i \gamma} \lambda_{j \beta} \lambda^{j}_{\lambda} {W}^{-2} - \frac{69}{128}{\rm i} (\Gamma_{a})^{\beta \rho} W_{\alpha \beta} \lambda_{i \rho} \lambda_{j \lambda} {W}^{-2} \nabla^{\lambda \gamma}{\lambda^{j}_{\gamma}}+\frac{843}{256}{\rm i} (\Gamma_{a})^{\beta \gamma} W_{\alpha \beta} W^{\rho \lambda} \lambda_{i \gamma} \lambda_{j \rho} \lambda^{j}_{\lambda} {W}^{-2} - \frac{27}{8}(\Gamma_{a})^{\gamma \beta} W_{\gamma}\,^{\rho} F_{\alpha \beta} F_{\rho}\,^{\lambda} \lambda_{i \lambda} {W}^{-2}%
 - \frac{27}{320}(\Gamma_{a})_{\alpha}{}^{\lambda} X_{i j} W_{\lambda}\,^{\beta} F_{\beta}\,^{\rho} \lambda^{j}_{\rho} {W}^{-2} - \frac{27}{64}(\Gamma_{a})^{\lambda}{}_{\gamma} W_{\lambda}\,^{\beta} F_{\beta}\,^{\rho} \lambda_{i \rho} {W}^{-2} \nabla^{\gamma}\,_{\alpha}{W} - \frac{27}{16}(\Gamma_{a})^{\lambda \gamma} W_{\lambda}\,^{\beta} F_{\beta \rho} \lambda_{i \gamma} {W}^{-2} \nabla_{\alpha}\,^{\rho}{W} - \frac{39}{32}{\rm i} (\Gamma_{a})^{\lambda \gamma} W_{\lambda}\,^{\beta} F_{\beta}\,^{\rho} \lambda_{j \alpha} \lambda_{i \gamma} \lambda^{j}_{\rho} {W}^{-3} - \frac{39}{128}{\rm i} (\Gamma_{a})^{\beta \lambda} W_{\beta \rho} \lambda^{\gamma}_{i} \lambda_{j \lambda} {W}^{-2} \nabla_{\alpha}\,^{\rho}{\lambda^{j}_{\gamma}} - \frac{9}{32}{\rm i} (\Gamma_{a})^{\beta \lambda} W_{\beta}\,^{\rho} \lambda_{i \gamma} \lambda_{j \lambda} {W}^{-2} \nabla_{\alpha}\,^{\gamma}{\lambda^{j}_{\rho}}+\frac{81}{64}{\rm i} (\Gamma_{a})^{\rho \lambda} W_{\alpha}\,^{\beta} W_{\rho \beta} \lambda^{\gamma}_{i} \lambda_{j \lambda} \lambda^{j}_{\gamma} {W}^{-2}+\frac{627}{512}{\rm i} (\Gamma_{a})^{\rho \gamma} W_{\alpha}\,^{\beta} W_{\rho}\,^{\lambda} \lambda_{i \beta} \lambda_{j \gamma} \lambda^{j}_{\lambda} {W}^{-2} - \frac{21}{64}{\rm i} (\Gamma_{a})^{\beta \rho} W_{\alpha \beta} \lambda_{i \lambda} \lambda_{j \rho} {W}^{-2} \nabla^{\lambda \gamma}{\lambda^{j}_{\gamma}} - \frac{21}{64}{\rm i} (\Gamma_{a})^{\beta \lambda} W_{\beta \rho} \lambda_{i \alpha} \lambda_{j \lambda} {W}^{-2} \nabla^{\rho \gamma}{\lambda^{j}_{\gamma}} - \frac{39}{32}{\rm i} (\Gamma_{a})^{\beta \gamma} W_{\alpha \beta} W^{\rho \lambda} \lambda_{i \rho} \lambda_{j \gamma} \lambda^{j}_{\lambda} {W}^{-2}+\frac{117}{256}{\rm i} (\Gamma_{a})^{\beta \gamma} W_{\beta}\,^{\rho} W_{\rho}\,^{\lambda} \lambda_{i \alpha} \lambda_{j \gamma} \lambda^{j}_{\lambda} {W}^{-2}+\frac{15}{8}{\rm i} (\Gamma_{a})^{\lambda \gamma} W_{\lambda}\,^{\beta} F_{\beta}\,^{\rho} \lambda_{j \alpha} \lambda_{i \rho} \lambda^{j}_{\gamma} {W}^{-3}+\frac{3}{64}{\rm i} (\Gamma_{a})_{\beta \rho} \lambda_{j \alpha} {W}^{-2} \nabla^{\beta \lambda}{\lambda_{i \lambda}} \nabla^{\rho \gamma}{\lambda^{j}_{\gamma}}+\frac{9}{64}(\Gamma_{a})_{\rho \lambda} {W}^{-1} \nabla^{\rho \beta}{F_{\alpha \beta}} \nabla^{\lambda \gamma}{\lambda_{i \gamma}}+\frac{9}{128}(\Gamma_{a})_{\beta \rho} {W}^{-1} \nabla^{\beta \lambda}{\lambda_{i \lambda}} \nabla^{\rho}\,_{\gamma}{\nabla_{\alpha}\,^{\gamma}{W}}+\frac{9}{16}(\Gamma_{a})_{\alpha \lambda} W^{\beta \rho} F_{\beta \rho} {W}^{-1} \nabla^{\lambda \gamma}{\lambda_{i \gamma}}+\frac{27}{32}(\Gamma_{a})^{\beta}{}_{\lambda} W_{\beta \rho} {W}^{-1} \nabla_{\alpha}\,^{\rho}{W} \nabla^{\lambda \gamma}{\lambda_{i \gamma}}+\frac{99}{128}(\Gamma_{a})^{\rho}{}_{\gamma} W_{\rho}\,^{\lambda} \lambda_{i \lambda} {W}^{-1} \nabla^{\gamma \beta}{F_{\alpha \beta}}+\frac{9}{4}(\Gamma_{a})^{\rho}{}_{\gamma} W_{\alpha \beta} W_{\rho}\,^{\lambda} \lambda_{i \lambda} {W}^{-1} \nabla^{\gamma \beta}{W}%
 - \frac{99}{256}(\Gamma_{a})^{\beta}{}_{\lambda} W_{\beta}\,^{\rho} \lambda_{i \rho} {W}^{-1} \nabla^{\lambda}\,_{\gamma}{\nabla_{\alpha}\,^{\gamma}{W}}+\frac{783}{640}(\Gamma_{a})_{\alpha}{}^{\lambda} W_{\lambda}\,^{\gamma} W^{\beta \rho} F_{\beta \rho} \lambda_{i \gamma} {W}^{-1} - \frac{45}{256}{\rm i} (\Gamma_{a})^{\lambda \gamma} W^{\beta}\,_{\rho} \lambda_{i \lambda} \lambda_{j \beta} {W}^{-2} \nabla_{\alpha}\,^{\rho}{\lambda^{j}_{\gamma}} - \frac{57}{1280}{\rm i} (\Gamma_{a})_{\alpha}{}^{\lambda} W^{\beta}\,_{\rho} \lambda_{i \lambda} \lambda_{j \beta} {W}^{-2} \nabla^{\rho \gamma}{\lambda^{j}_{\gamma}}+\frac{159}{256}{\rm i} (\Gamma_{a})^{\rho}{}_{\lambda} W_{\alpha}\,^{\beta} \lambda_{i \rho} \lambda_{j \beta} {W}^{-2} \nabla^{\lambda \gamma}{\lambda^{j}_{\gamma}} - \frac{81}{16}(\Gamma_{a})^{\beta \rho} W^{\lambda \gamma} F_{\alpha \beta} F_{\rho \lambda} \lambda_{i \gamma} {W}^{-2} - \frac{9}{160}(\Gamma_{a})_{\alpha}{}^{\beta} X_{i j} W^{\rho \lambda} F_{\beta \rho} \lambda^{j}_{\lambda} {W}^{-2}+\frac{27}{64}(\Gamma_{a})^{\beta}{}_{\gamma} W^{\rho \lambda} F_{\beta \rho} \lambda_{i \lambda} {W}^{-2} \nabla^{\gamma}\,_{\alpha}{W} - \frac{81}{16}(\Gamma_{a})^{\rho \gamma} W^{\beta \lambda} F_{\alpha \beta} F_{\rho \lambda} \lambda_{i \gamma} {W}^{-2}+\frac{81}{64}(\Gamma_{a})^{\beta \gamma} W^{\rho}\,_{\lambda} F_{\beta \rho} \lambda_{i \gamma} {W}^{-2} \nabla_{\alpha}\,^{\lambda}{W}+\frac{63}{32}{\rm i} (\Gamma_{a})^{\beta \gamma} W^{\rho \lambda} F_{\beta \rho} \lambda_{j \alpha} \lambda_{i \gamma} \lambda^{j}_{\lambda} {W}^{-3} - \frac{9}{32}{\rm i} (\Gamma_{a})^{\lambda}{}_{\gamma} W^{\beta \rho} \lambda_{i \beta} \lambda_{j \lambda} {W}^{-2} \nabla^{\gamma}\,_{\alpha}{\lambda^{j}_{\rho}}+\frac{9}{64}{\rm i} (\Gamma_{a})^{\lambda \gamma} W^{\beta}\,_{\rho} \lambda_{i \beta} \lambda_{j \lambda} {W}^{-2} \nabla_{\alpha}\,^{\rho}{\lambda^{j}_{\gamma}} - \frac{39}{320}{\rm i} (\Gamma_{a})_{\alpha}{}^{\lambda} W^{\beta}\,_{\rho} \lambda_{i \beta} \lambda_{j \lambda} {W}^{-2} \nabla^{\rho \gamma}{\lambda^{j}_{\gamma}}+\frac{3}{128}{\rm i} (\Gamma_{a})^{\rho}{}_{\lambda} W_{\alpha}\,^{\beta} \lambda_{i \beta} \lambda_{j \rho} {W}^{-2} \nabla^{\lambda \gamma}{\lambda^{j}_{\gamma}}+\frac{243}{256}{\rm i} (\Gamma_{a})_{\alpha}{}^{\gamma} W^{\beta \rho} W_{\beta}\,^{\lambda} \lambda_{i \rho} \lambda_{j \gamma} \lambda^{j}_{\lambda} {W}^{-2} - \frac{9}{4}{\rm i} (\Gamma_{a})^{\beta \gamma} W^{\rho \lambda} F_{\beta \rho} \lambda_{j \alpha} \lambda_{i \lambda} \lambda^{j}_{\gamma} {W}^{-3}+\frac{3}{16}{\rm i} (\Gamma_{a})^{\lambda \gamma} F^{\beta}\,_{\rho} \lambda_{i \lambda} \lambda_{j \gamma} {W}^{-3} \nabla_{\alpha}\,^{\rho}{\lambda^{j}_{\beta}} - \frac{9}{8}{\rm i} (\Gamma_{a})^{\lambda \gamma} W_{\alpha}\,^{\beta} F_{\beta}\,^{\rho} \lambda_{i \lambda} \lambda_{j \gamma} \lambda^{j}_{\rho} {W}^{-3} - \frac{3}{8}{\rm i} (\Gamma_{a})^{\rho \lambda} F_{\alpha \beta} \lambda_{i \rho} \lambda_{j \lambda} {W}^{-3} \nabla^{\beta \gamma}{\lambda^{j}_{\gamma}}%
+\frac{27}{16}{\rm i} (\Gamma_{a})^{\lambda \gamma} W^{\beta \rho} F_{\alpha \beta} \lambda_{i \lambda} \lambda_{j \gamma} \lambda^{j}_{\rho} {W}^{-3}+\frac{9}{16}(\Gamma_{a})^{\beta \gamma} F_{\alpha \beta} F^{\rho \lambda} F_{\rho \lambda} \lambda_{i \gamma} {W}^{-3} - \frac{3}{80}(\Gamma_{a})_{\alpha}{}^{\lambda} X_{i j} F^{\beta \rho} F_{\beta \rho} \lambda^{j}_{\lambda} {W}^{-3}+\frac{9}{32}{\rm i} (\Gamma_{a})^{\lambda \gamma} F^{\beta \rho} F_{\beta \rho} \lambda_{j \alpha} \lambda_{i \lambda} \lambda^{j}_{\gamma} {W}^{-4} - \frac{33}{64}{\rm i} (\Gamma_{a})^{\beta}{}_{\rho} \lambda_{i \beta} {W}^{-2} \nabla^{\rho \lambda}{\lambda_{j \lambda}} \nabla_{\alpha}\,^{\gamma}{\lambda^{j}_{\gamma}} - \frac{3}{64}(\Gamma_{a})_{\alpha \beta} X_{i j} X^{j}\,_{k} {W}^{-2} \nabla^{\beta \rho}{\lambda^{k}_{\rho}}+\frac{3}{16}(\Gamma_{a})_{\beta \rho} X_{i j} {W}^{-2} \nabla^{\beta}\,_{\alpha}{W} \nabla^{\rho \lambda}{\lambda^{j}_{\lambda}}+\frac{3}{64}{\rm i} (\Gamma_{a})^{\beta}{}_{\rho} X_{j k} \lambda^{j}_{\alpha} \lambda_{i \beta} {W}^{-3} \nabla^{\rho \lambda}{\lambda^{k}_{\lambda}}+\frac{3}{64}(\Gamma_{a})^{\beta}{}_{\rho} X_{j k} \lambda_{i \beta} {W}^{-2} \nabla^{\rho}\,_{\alpha}{X^{j k}} - \frac{105}{256}(\Gamma_{a})^{\beta \rho} X_{j k} X^{j k} W_{\alpha \beta} \lambda_{i \rho} {W}^{-2} - \frac{15}{64}(\Gamma_{a})^{\rho \beta} X_{j k} \lambda^{j}_{\alpha} \lambda_{i \rho} X^{k}_{\beta} {W}^{-2} - \frac{9}{512}(\Gamma_{a})^{\beta \rho} X_{j k} \lambda_{i \beta} \lambda^{j}_{\rho} X^{k}_{\alpha} {W}^{-2}+\frac{171}{2560}(\Gamma_{a})_{\alpha}{}^{\rho} X_{j k} \lambda_{i \rho} \lambda^{j \beta} X^{k}_{\beta} {W}^{-2} - \frac{53}{128}\Phi_{\alpha}\,^{\beta}\,_{j k} (\Gamma_{a})_{\beta}{}^{\rho} X^{j k} \lambda_{i \rho} {W}^{-1} - \frac{27}{128}{\rm i} (\Gamma_{a})^{\beta}{}_{\rho} \lambda_{j \beta} {W}^{-2} \nabla^{\rho \lambda}{\lambda^{j}_{\lambda}} \nabla_{\alpha}\,^{\gamma}{\lambda_{i \gamma}} - \frac{51}{128}{\rm i} (\Gamma_{a})^{\beta}{}_{\rho} X_{i j} \lambda_{k \alpha} \lambda^{k}_{\beta} {W}^{-3} \nabla^{\rho \lambda}{\lambda^{j}_{\lambda}}+\frac{3}{64}(\Gamma_{a})^{\rho}{}_{\lambda} X_{i j} \lambda^{j}_{\rho} {W}^{-2} \nabla^{\lambda \beta}{F_{\alpha \beta}}+\frac{9}{64}(\Gamma_{a})^{\rho}{}_{\lambda} X_{i j} W_{\alpha \beta} \lambda^{j}_{\rho} {W}^{-2} \nabla^{\lambda \beta}{W} - \frac{9}{256}(\Gamma_{a})^{\rho}{}_{\lambda} X_{i j} \lambda^{j}_{\rho} {W}^{-1} \nabla^{\lambda \beta}{W_{\alpha \beta}} - \frac{3}{128}(\Gamma_{a})^{\beta}{}_{\rho} X_{i j} \lambda^{j}_{\beta} {W}^{-2} \nabla^{\rho}\,_{\lambda}{\nabla_{\alpha}\,^{\lambda}{W}}%
 - \frac{141}{128}(\Gamma_{a})^{\beta \rho} X_{i j} X^{j}\,_{k} W_{\alpha \beta} \lambda^{k}_{\rho} {W}^{-2}+\frac{27}{128}(\Gamma_{a})^{\beta \lambda} X_{i j} W_{\beta \rho} \lambda^{j}_{\lambda} {W}^{-2} \nabla_{\alpha}\,^{\rho}{W} - \frac{57}{128}(\Gamma_{a})^{\rho \beta} X_{i j} \lambda^{j}_{\alpha} \lambda_{k \rho} X^{k}_{\beta} {W}^{-2} - \frac{15}{512}(\Gamma_{a})^{\beta \rho} X_{i j} \lambda^{j}_{\beta} \lambda_{k \rho} X^{k}_{\alpha} {W}^{-2} - \frac{381}{2560}(\Gamma_{a})_{\alpha}{}^{\rho} X_{i j} \lambda^{j \beta} \lambda_{k \rho} X^{k}_{\beta} {W}^{-2}+\frac{3}{160}(\Gamma_{a})_{\alpha}{}^{\rho} X_{i j} \lambda^{j}_{\rho} \lambda^{\beta}_{k} X^{k}_{\beta} {W}^{-2}+\frac{105}{512}(\Gamma_{a})^{\rho \beta} X_{i j} \lambda_{k \alpha} \lambda^{j}_{\rho} X^{k}_{\beta} {W}^{-2} - \frac{27}{320}(\Gamma_{a})_{\alpha}{}^{\rho} X_{i j} \lambda_{k \rho} \lambda^{k \beta} X^{j}_{\beta} {W}^{-2} - \frac{27}{64}(\Gamma_{a})^{\rho \beta} X_{i j} \lambda_{k \alpha} \lambda^{k}_{\rho} X^{j}_{\beta} {W}^{-2}+\frac{53}{128}\Phi_{\alpha}\,^{\beta}\,_{j k} (\Gamma_{a})_{\beta}{}^{\rho} X_{i}\,^{j} \lambda^{k}_{\rho} {W}^{-1}+\frac{27}{32}(\Gamma_{a})^{\lambda \beta} X_{i j} \lambda^{j}_{\lambda} {W}^{-1} \nabla_{\alpha}\,^{\rho}{W_{\beta \rho}}+\frac{45}{64}(\Gamma_{a})^{\beta}{}_{\lambda} {W}^{-1} \nabla^{\lambda \rho}{F_{\beta \rho}} \nabla_{\alpha}\,^{\gamma}{\lambda_{i \gamma}}+\frac{9}{128}(\Gamma_{a})^{\beta}{}_{\gamma} W_{\alpha}\,^{\lambda} \lambda_{i \lambda} {W}^{-1} \nabla^{\gamma \rho}{F_{\beta \rho}}+\frac{135}{256}{\rm i} (\Gamma_{a})^{\beta}{}_{\lambda} X_{i \alpha} \nabla^{\lambda \rho}{F_{\beta \rho}}+\frac{15}{64}(\Gamma_{a})^{\beta}{}_{\lambda} X_{i j} \lambda^{j}_{\alpha} {W}^{-2} \nabla^{\lambda \rho}{F_{\beta \rho}} - \frac{3}{32}(\Gamma_{a})^{\beta}{}_{\rho} X_{i j} {W}^{-1} \nabla^{\rho}\,_{\lambda}{\nabla_{\alpha}\,^{\lambda}{\lambda^{j}_{\beta}}} - \frac{3}{32}(\Gamma_{a})_{\beta \rho} X_{i j} {W}^{-1} \nabla^{\beta \lambda}{\nabla^{\rho}\,_{\alpha}{\lambda^{j}_{\lambda}}}+\frac{27}{64}(\Gamma_{a})^{\beta}{}_{\lambda} X_{i j} \lambda^{j}_{\alpha} {W}^{-1} \nabla^{\lambda \rho}{W_{\beta \rho}} - \frac{99}{256}(\Gamma_{a})^{\beta}{}_{\rho} X_{i j} \lambda^{j}_{\lambda} {W}^{-1} \nabla^{\rho \lambda}{W_{\alpha \beta}} - \frac{15}{64}(\Gamma_{a})_{\beta \rho} X_{i j} {W}^{-1} \nabla^{\beta}\,_{\alpha}{\nabla^{\rho \lambda}{\lambda^{j}_{\lambda}}}%
 - \frac{57}{160}(\Gamma_{a})_{\alpha \beta} X_{i j} {W}^{-1} \nabla^{\beta}\,_{\lambda}{\nabla^{\lambda \rho}{\lambda^{j}_{\rho}}}+\frac{27}{256}(\Gamma_{a})^{\beta}{}_{\lambda} X_{i j} \lambda^{j \rho} {W}^{-1} \nabla^{\lambda}\,_{\alpha}{W_{\beta \rho}} - \frac{351}{1280}(\Gamma_{a})_{\alpha \lambda} X_{i j} \lambda^{j \beta} {W}^{-1} \nabla^{\lambda \rho}{W_{\beta \rho}}+\frac{9}{64}{\rm i} (\Gamma_{a})^{\beta}{}_{\lambda} X_{i j} W_{\alpha \beta \rho}\,^{j} {W}^{-1} \nabla^{\lambda \rho}{W} - \frac{27}{64}(\Gamma_{a})^{\beta \lambda} X_{i j} W_{\beta \rho} {W}^{-1} \nabla_{\alpha}\,^{\rho}{\lambda^{j}_{\lambda}}+\frac{53}{640}\Phi^{\beta \rho}\,_{j k} (\Gamma_{a})_{\alpha \beta} X_{i}\,^{j} \lambda^{k}_{\rho} {W}^{-1} - \frac{63}{256}(\Gamma_{a})_{\rho \lambda} X_{i j} \lambda^{j \beta} {W}^{-1} \nabla^{\rho \lambda}{W_{\alpha \beta}} - \frac{15}{128}(\Gamma_{a})_{\beta \rho} {W}^{-1} \nabla^{\beta \rho}{X_{i j}} \nabla_{\alpha}\,^{\lambda}{\lambda^{j}_{\lambda}}+\frac{3}{32}(\Gamma_{a})_{\beta \rho} X_{j k} \lambda^{j}_{\alpha} {W}^{-2} \nabla^{\beta \rho}{X_{i}\,^{k}}+\frac{3}{16}(\Gamma_{a})_{\beta \rho} X_{i j} {W}^{-1} \nabla^{\beta \rho}{\nabla_{\alpha}\,^{\lambda}{\lambda^{j}_{\lambda}}}+\frac{45}{128}{\rm i} (\Gamma_{a})_{\beta \rho} X_{i j} X^{j}_{\alpha} {W}^{-1} \nabla^{\beta \rho}{W}+\frac{91}{128}\Phi_{\alpha}\,^{\beta}\,_{i j} (\Gamma_{a})_{\beta}{}^{\rho} X^{j}\,_{k} \lambda^{k}_{\rho} {W}^{-1}+\frac{91}{640}\Phi^{\beta \rho}\,_{i j} (\Gamma_{a})_{\alpha \beta} X^{j}\,_{k} \lambda^{k}_{\rho} {W}^{-1} - \frac{53}{640}\Phi^{\beta \rho}\,_{j k} (\Gamma_{a})_{\alpha \beta} X^{j k} \lambda_{i \rho} {W}^{-1} - \frac{45}{128}(\Gamma_{a})_{\beta \rho} {W}^{-1} \nabla_{\alpha}\,^{\lambda}{\lambda_{i \lambda}} \nabla^{\beta}\,_{\gamma}{\nabla^{\rho \gamma}{W}}+\frac{27}{256}(\Gamma_{a})_{\rho \lambda} W_{\alpha}\,^{\beta} \lambda_{i \beta} {W}^{-1} \nabla^{\rho}\,_{\gamma}{\nabla^{\lambda \gamma}{W}} - \frac{135}{512}{\rm i} (\Gamma_{a})_{\beta \rho} X_{i \alpha} \nabla^{\beta}\,_{\lambda}{\nabla^{\rho \lambda}{W}} - \frac{15}{128}(\Gamma_{a})_{\beta \rho} X_{i j} \lambda^{j}_{\alpha} {W}^{-2} \nabla^{\beta}\,_{\lambda}{\nabla^{\rho \lambda}{W}}+\frac{3}{32}(\Gamma_{a})_{\beta \rho} X_{i j} {W}^{-1} \nabla^{\beta}\,_{\lambda}{\nabla^{\rho \lambda}{\lambda^{j}_{\alpha}}}+\frac{207}{640}{\rm i} (\Gamma_{a})_{\alpha \rho} X_{i j} X^{j}_{\beta} {W}^{-1} \nabla^{\rho \beta}{W}%
 - \frac{9}{128}{\rm i} (\Gamma_{a})^{\beta}{}_{\rho} X_{i j} X^{j}_{\beta} {W}^{-1} \nabla^{\rho}\,_{\alpha}{W} - \frac{27}{512}(\Gamma_{a})^{\rho \beta} X_{j k} \lambda^{j}_{\alpha} \lambda^{k}_{\rho} X_{i \beta} {W}^{-2}+\frac{3}{16}{\rm i} (\Gamma_{a})^{\beta \rho} X_{j k} \lambda_{i \beta} \lambda^{j}_{\rho} {W}^{-3} \nabla_{\alpha}\,^{\lambda}{\lambda^{k}_{\lambda}} - \frac{165}{256}{\rm i} (\Gamma_{a})^{\rho \lambda} X_{j k} W_{\alpha}\,^{\beta} \lambda_{i \rho} \lambda^{j}_{\lambda} \lambda^{k}_{\beta} {W}^{-3} - \frac{3}{8}(\Gamma_{a})^{\beta \rho} X_{j k} X^{j k} F_{\alpha \beta} \lambda_{i \rho} {W}^{-3} - \frac{3}{64}(\Gamma_{a})^{\beta}{}_{\rho} X_{j k} X^{j k} \lambda_{i \beta} {W}^{-3} \nabla^{\rho}\,_{\alpha}{W} - \frac{9}{16}{\rm i} (\Gamma_{a})^{\beta \rho} X_{j k} X^{j k} \lambda_{l \alpha} \lambda_{i \beta} \lambda^{l}_{\rho} {W}^{-4} - \frac{9}{32}{\rm i} (\Gamma_{a})^{\rho}{}_{\lambda} F_{\alpha \beta} X_{i \rho} {W}^{-1} \nabla^{\lambda \beta}{W}+\frac{1}{32}\Phi_{\alpha}\,^{\beta}\,_{i j} (\Gamma_{a})_{\beta \rho} \lambda^{j}_{\lambda} {W}^{-1} \nabla^{\rho \lambda}{W}+\frac{189}{10240}(\Gamma_{a})_{\alpha \beta} Y \lambda_{i \rho} {W}^{-1} \nabla^{\beta \rho}{W} - \frac{171}{512}(\Gamma_{a})^{\beta}{}_{\lambda} \lambda_{i \gamma} {W}^{-1} \nabla^{\lambda \gamma}{W} \nabla_{\alpha}\,^{\rho}{W_{\beta \rho}}+\frac{909}{2560}(\Gamma_{a})_{\alpha \lambda} W^{\beta \rho} W_{\beta \rho} \lambda_{i \gamma} {W}^{-1} \nabla^{\lambda \gamma}{W}+\frac{45}{512}(\Gamma_{a})^{\beta}{}_{\rho} \lambda_{j \alpha} \lambda^{j}_{\lambda} X_{i \beta} {W}^{-2} \nabla^{\rho \lambda}{W}+\frac{171}{512}(\Gamma_{a})^{\beta}{}_{\rho} \lambda_{j \alpha} \lambda_{i \lambda} X^{j}_{\beta} {W}^{-2} \nabla^{\rho \lambda}{W} - \frac{747}{512}(\Gamma_{a})^{\rho}{}_{\gamma} F_{\alpha}\,^{\beta} \lambda_{i \beta} {W}^{-1} \nabla^{\gamma \lambda}{W_{\rho \lambda}} - \frac{117}{256}(\Gamma_{a})^{\beta}{}_{\lambda} \lambda_{i \gamma} {W}^{-1} \nabla_{\alpha}\,^{\gamma}{W} \nabla^{\lambda \rho}{W_{\beta \rho}} - \frac{369}{1024}{\rm i} (\Gamma_{a})^{\beta}{}_{\lambda} \lambda_{j \alpha} \lambda^{\gamma}_{i} \lambda^{j}_{\gamma} {W}^{-2} \nabla^{\lambda \rho}{W_{\beta \rho}} - \frac{109}{1024}(\Gamma_{a})^{\beta}{}_{\lambda} \lambda^{\gamma}_{i} \lambda_{j \gamma} {W}^{-1} \nabla^{\lambda \rho}{W_{\alpha \beta \rho}\,^{j}} - \frac{117}{2048}(\Gamma_{a})^{\beta}{}_{\rho} \lambda^{\lambda}_{i} \lambda_{j \lambda} {W}^{-1} \nabla^{\rho}\,_{\alpha}{X^{j}_{\beta}} - \frac{423}{10240}(\Gamma_{a})_{\alpha \rho} \lambda^{\lambda}_{i} \lambda_{j \lambda} {W}^{-1} \nabla^{\rho \beta}{X^{j}_{\beta}}%
+\frac{511}{10240}(\Gamma_{a})_{\alpha}{}^{\lambda} W^{\beta \rho} \lambda^{\gamma}_{i} \lambda_{j \gamma} W_{\lambda \beta \rho}\,^{j} {W}^{-1} - \frac{27}{32}(\Gamma_{a})^{\beta \rho} F_{\alpha \beta} \lambda^{\lambda}_{i} \lambda_{j \lambda} X^{j}_{\rho} {W}^{-2}+\frac{261}{2560}(\Gamma_{a})_{\alpha}{}^{\beta} X_{i j} \lambda^{j \rho} \lambda_{k \rho} X^{k}_{\beta} {W}^{-2}+\frac{15}{64}(\Gamma_{a})^{\beta}{}_{\rho} \lambda^{\lambda}_{i} \lambda_{j \lambda} X^{j}_{\beta} {W}^{-2} \nabla^{\rho}\,_{\alpha}{W}+\frac{333}{512}(\Gamma_{a})^{\rho \beta} \lambda_{i \rho} \lambda_{j \lambda} X^{j}_{\beta} {W}^{-2} \nabla_{\alpha}\,^{\lambda}{W} - \frac{3}{256}{\rm i} \Phi_{\alpha}\,^{\beta}\,_{j k} (\Gamma_{a})_{\beta}{}^{\rho} \lambda_{i \rho} \lambda^{j \lambda} \lambda^{k}_{\lambda} {W}^{-2} - \frac{57}{512}{\rm i} (\Gamma_{a})^{\rho \beta} \lambda_{k \alpha} \lambda_{i \rho} \lambda^{k \lambda} \lambda_{j \lambda} X^{j}_{\beta} {W}^{-3}+\frac{87}{512}(\Gamma_{a})^{\rho \beta} X_{j k} \lambda_{i \alpha} \lambda^{j}_{\rho} X^{k}_{\beta} {W}^{-2} - \frac{243}{512}(\Gamma_{a})^{\rho \beta} \lambda_{i \lambda} \lambda_{j \rho} X^{j}_{\beta} {W}^{-2} \nabla_{\alpha}\,^{\lambda}{W} - \frac{99}{256}{\rm i} \Phi_{\alpha}\,^{\beta}\,_{j k} (\Gamma_{a})_{\beta}{}^{\rho} \lambda^{\lambda}_{i} \lambda^{j}_{\rho} \lambda^{k}_{\lambda} {W}^{-2}+\frac{27}{5120}{\rm i} (\Gamma_{a})_{\alpha}{}^{\beta} Y \lambda^{\rho}_{i} \lambda_{j \beta} \lambda^{j}_{\rho} {W}^{-2}+\frac{333}{1024}{\rm i} (\Gamma_{a})^{\lambda \beta} \lambda^{\gamma}_{i} \lambda_{j \lambda} \lambda^{j}_{\gamma} {W}^{-2} \nabla_{\alpha}\,^{\rho}{W_{\beta \rho}} - \frac{1647}{2560}{\rm i} (\Gamma_{a})_{\alpha}{}^{\lambda} W^{\beta \rho} W_{\beta \rho} \lambda^{\gamma}_{i} \lambda_{j \lambda} \lambda^{j}_{\gamma} {W}^{-2}+\frac{135}{1024}{\rm i} (\Gamma_{a})^{\rho}{}_{\lambda} \lambda^{\gamma}_{i} \lambda_{j \rho} \lambda^{j}_{\gamma} {W}^{-2} \nabla^{\lambda \beta}{W_{\alpha \beta}}+\frac{153}{256}{\rm i} (\Gamma_{a})^{\rho \beta} \lambda_{k \alpha} \lambda^{\lambda}_{i} \lambda^{k}_{\rho} \lambda_{j \lambda} X^{j}_{\beta} {W}^{-3} - \frac{9}{32}(\Gamma_{a})^{\beta}{}_{\rho} F_{\alpha \beta} {W}^{-2} \nabla^{\rho}\,_{\gamma}{W} \nabla^{\gamma \lambda}{\lambda_{i \lambda}}+\frac{27}{320}(\Gamma_{a})_{\alpha \beta} X_{i j} {W}^{-2} \nabla^{\beta}\,_{\lambda}{W} \nabla^{\lambda \rho}{\lambda^{j}_{\rho}}+\frac{9}{64}(\Gamma_{a})_{\beta \rho} {W}^{-2} \nabla^{\beta}\,_{\alpha}{W} \nabla^{\rho}\,_{\gamma}{W} \nabla^{\gamma \lambda}{\lambda_{i \lambda}} - \frac{3}{16}{\rm i} (\Gamma_{a})^{\beta}{}_{\rho} \lambda_{j \alpha} \lambda_{i \beta} {W}^{-3} \nabla^{\rho}\,_{\gamma}{W} \nabla^{\gamma \lambda}{\lambda^{j}_{\lambda}}+\frac{75}{128}{\rm i} (\Gamma_{a})^{\beta \lambda} W_{\beta}\,^{\rho} \lambda_{i \lambda} \lambda_{j \rho} {W}^{-2} \nabla_{\alpha}\,^{\gamma}{\lambda^{j}_{\gamma}}%
+\frac{27}{64}(\Gamma_{a})^{\rho}{}_{\lambda} \lambda_{i \rho} {W}^{-2} \nabla^{\lambda}\,_{\gamma}{W} \nabla^{\gamma \beta}{F_{\alpha \beta}}+\frac{45}{128}(\Gamma_{a})^{\rho}{}_{\lambda} W_{\alpha \beta} \lambda_{i \rho} {W}^{-2} \nabla^{\lambda}\,_{\gamma}{W} \nabla^{\gamma \beta}{W}+\frac{135}{1024}(\Gamma_{a})^{\rho}{}_{\lambda} \lambda_{i \rho} {W}^{-1} \nabla^{\lambda}\,_{\gamma}{W} \nabla^{\gamma \beta}{W_{\alpha \beta}} - \frac{27}{128}(\Gamma_{a})^{\beta}{}_{\rho} \lambda_{i \beta} {W}^{-2} \nabla^{\rho}\,_{\lambda}{W} \nabla^{\lambda}\,_{\gamma}{\nabla_{\alpha}\,^{\gamma}{W}}+\frac{27}{32}(\Gamma_{a})^{\lambda}{}_{\gamma} W^{\beta}\,_{\rho} F_{\alpha \beta} \lambda_{i \lambda} {W}^{-2} \nabla^{\gamma \rho}{W} - \frac{27}{64}(\Gamma_{a})^{\beta}{}_{\rho} \lambda_{i \beta} \lambda_{j \lambda} X^{j}_{\alpha} {W}^{-2} \nabla^{\rho \lambda}{W}+\frac{81}{512}(\Gamma_{a})^{\rho}{}_{\lambda} \lambda_{i \rho} \lambda^{\beta}_{j} X^{j}_{\beta} {W}^{-2} \nabla^{\lambda}\,_{\alpha}{W} - \frac{243}{512}(\Gamma_{a})^{\rho}{}_{\lambda} \lambda_{j \alpha} \lambda_{i \rho} X^{j}_{\beta} {W}^{-2} \nabla^{\lambda \beta}{W}+\frac{729}{1024}(\Gamma_{a})^{\lambda}{}_{\gamma} \lambda_{i \lambda} {W}^{-1} \nabla^{\gamma \beta}{W} \nabla_{\alpha}\,^{\rho}{W_{\beta \rho}} - \frac{3}{32}{\rm i} (\Gamma_{a})^{\beta}{}_{\rho} \lambda_{j \alpha} \lambda^{j}_{\beta} {W}^{-3} \nabla^{\rho}\,_{\gamma}{W} \nabla^{\gamma \lambda}{\lambda_{i \lambda}}+\frac{3}{64}{\rm i} (\Gamma_{a})^{\beta}{}_{\rho} \lambda_{j \beta} {W}^{-2} \nabla^{\rho}\,_{\gamma}{\lambda^{j}_{\alpha}} \nabla^{\gamma \lambda}{\lambda_{i \lambda}}+\frac{51}{256}{\rm i} (\Gamma_{a})^{\rho}{}_{\lambda} W_{\alpha}\,^{\beta} \lambda_{j \rho} \lambda^{j}_{\beta} {W}^{-2} \nabla^{\lambda \gamma}{\lambda_{i \gamma}}+\frac{33}{64}{\rm i} (\Gamma_{a})^{\beta \lambda} W_{\beta}\,^{\rho} \lambda_{j \lambda} \lambda^{j}_{\rho} {W}^{-2} \nabla_{\alpha}\,^{\gamma}{\lambda_{i \gamma}} - \frac{147}{256}{\rm i} (\Gamma_{a})^{\beta \rho} W_{\alpha \beta} \lambda_{j \rho} \lambda^{j}_{\lambda} {W}^{-2} \nabla^{\lambda \gamma}{\lambda_{i \gamma}}+\frac{3}{64}(\Gamma_{a})^{\beta}{}_{\rho} \lambda_{j \beta} {W}^{-2} \nabla^{\rho}\,_{\lambda}{W} \nabla_{\alpha}\,^{\lambda}{X_{i}\,^{j}}+\frac{9}{64}(\Gamma_{a})^{\beta}{}_{\rho} \lambda_{i \lambda} \lambda_{j \beta} X^{j}_{\alpha} {W}^{-2} \nabla^{\rho \lambda}{W} - \frac{99}{512}(\Gamma_{a})^{\rho}{}_{\lambda} \lambda^{\beta}_{i} \lambda_{j \rho} X^{j}_{\beta} {W}^{-2} \nabla^{\lambda}\,_{\alpha}{W}+\frac{45}{512}(\Gamma_{a})^{\rho}{}_{\lambda} \lambda_{j \rho} \lambda^{j \beta} X_{i \beta} {W}^{-2} \nabla^{\lambda}\,_{\alpha}{W} - \frac{81}{512}(\Gamma_{a})^{\rho}{}_{\lambda} \lambda_{j \alpha} \lambda^{j}_{\rho} X_{i \beta} {W}^{-2} \nabla^{\lambda \beta}{W} - \frac{63}{64}\Phi_{\alpha \lambda i j} (\Gamma_{a})^{\beta}{}_{\rho} \lambda^{j}_{\beta} {W}^{-1} \nabla^{\lambda \rho}{W}%
+\frac{15}{64}(\Gamma_{a})_{\beta \rho} \lambda_{j \alpha} {W}^{-2} \nabla^{\beta}\,_{\lambda}{W} \nabla^{\rho \lambda}{X_{i}\,^{j}} - \frac{27}{64}(\Gamma_{a})_{\beta \rho} {W}^{-1} \nabla^{\beta}\,_{\gamma}{W} \nabla^{\rho \gamma}{\nabla_{\alpha}\,^{\lambda}{\lambda_{i \lambda}}}+\frac{27}{256}(\Gamma_{a})_{\rho \lambda} \lambda^{\beta}_{i} {W}^{-1} \nabla^{\rho}\,_{\gamma}{W} \nabla^{\lambda \gamma}{W_{\alpha \beta}} - \frac{1359}{4096}{\rm i} (\Gamma_{a})_{\beta \rho} \nabla^{\beta}\,_{\lambda}{W} \nabla^{\rho \lambda}{X_{i \alpha}} - \frac{423}{1280}(\Gamma_{a})_{\alpha \gamma} W^{\beta \rho} W_{\beta \lambda} \lambda_{i \rho} {W}^{-1} \nabla^{\gamma \lambda}{W}+\frac{59}{128}\Phi_{\alpha}\,^{\lambda}\,_{i j} (\Gamma_{a})_{\beta \rho} \lambda^{j}_{\lambda} {W}^{-1} \nabla^{\beta \rho}{W}+\frac{7}{40}\Phi_{\rho}\,^{\lambda}\,_{i j} (\Gamma_{a})_{\alpha \beta} \lambda^{j}_{\lambda} {W}^{-1} \nabla^{\rho \beta}{W} - \frac{15}{64}\Phi^{\beta \lambda}\,_{i j} (\Gamma_{a})_{\beta \rho} \lambda^{j}_{\lambda} {W}^{-1} \nabla^{\rho}\,_{\alpha}{W} - \frac{63}{256}(\Gamma_{a})_{\lambda \gamma} \lambda^{\beta}_{i} {W}^{-1} \nabla^{\lambda \rho}{W} \nabla^{\gamma}\,_{\alpha}{W_{\beta \rho}} - \frac{9}{64}(\Gamma_{a})^{\beta}{}_{\lambda} \lambda^{\rho}_{i} {W}^{-1} \nabla^{\lambda}\,_{\gamma}{W} \nabla_{\alpha}\,^{\gamma}{W_{\beta \rho}}+\frac{729}{5120}(\Gamma_{a})_{\alpha \lambda} \lambda_{i \gamma} {W}^{-1} \nabla^{\lambda \beta}{W} \nabla^{\gamma \rho}{W_{\beta \rho}} - \frac{9}{32}(\Gamma_{a})^{\beta}{}_{\lambda} \lambda_{i \gamma} {W}^{-1} \nabla^{\lambda}\,_{\alpha}{W} \nabla^{\gamma \rho}{W_{\beta \rho}} - \frac{9}{128}(\Gamma_{a})_{\rho \lambda} \lambda_{i \gamma} {W}^{-1} \nabla^{\rho \beta}{W} \nabla^{\lambda \gamma}{W_{\alpha \beta}} - \frac{117}{256}(\Gamma_{a})^{\beta}{}_{\rho} \lambda_{i \lambda} {W}^{-1} \nabla^{\rho}\,_{\gamma}{W} \nabla^{\gamma \lambda}{W_{\alpha \beta}} - \frac{9}{32}(\Gamma_{a})_{\lambda \gamma} \lambda^{\beta}_{i} {W}^{-1} \nabla^{\lambda}\,_{\alpha}{W} \nabla^{\gamma \rho}{W_{\beta \rho}}+\frac{621}{5120}(\Gamma_{a})_{\alpha \lambda} \lambda^{\beta}_{i} {W}^{-1} \nabla^{\lambda}\,_{\gamma}{W} \nabla^{\gamma \rho}{W_{\beta \rho}} - \frac{135}{1024}(\Gamma_{a})_{\lambda \gamma} \lambda_{i \alpha} {W}^{-1} \nabla^{\lambda \beta}{W} \nabla^{\gamma \rho}{W_{\beta \rho}} - \frac{243}{1024}(\Gamma_{a})^{\beta}{}_{\lambda} \lambda_{i \alpha} {W}^{-1} \nabla^{\lambda}\,_{\gamma}{W} \nabla^{\gamma \rho}{W_{\beta \rho}}+\frac{135}{1024}(\Gamma_{a})_{\lambda \gamma} W^{\beta \rho} W_{\beta \rho} \lambda_{i \alpha} {W}^{-1} \nabla^{\lambda \gamma}{W} - \frac{27}{128}(\Gamma_{a})^{\beta}{}_{\gamma} W_{\beta}\,^{\rho} W_{\rho \lambda} \lambda_{i \alpha} {W}^{-1} \nabla^{\gamma \lambda}{W}%
+\frac{9}{16}{\rm i} (\Gamma_{a})^{\rho}{}_{\gamma} F_{\alpha}\,^{\beta} W_{\rho \beta \lambda i} {W}^{-1} \nabla^{\gamma \lambda}{W}+\frac{63}{64}{\rm i} (\Gamma_{a})^{\beta}{}_{\gamma} W_{\beta \rho \lambda i} {W}^{-1} \nabla^{\gamma \rho}{W} \nabla_{\alpha}\,^{\lambda}{W} - \frac{27}{32}(\Gamma_{a})^{\beta}{}_{\gamma} C_{\alpha \beta}\,^{\rho}\,_{\lambda} \lambda_{i \rho} {W}^{-1} \nabla^{\gamma \lambda}{W}+\frac{99}{256}(\Gamma_{a})^{\beta}{}_{\lambda} \lambda_{i \gamma} {W}^{-1} \nabla^{\lambda \rho}{W} \nabla_{\alpha}\,^{\gamma}{W_{\beta \rho}} - \frac{43}{64}\Phi_{\lambda}\,^{\beta}\,_{i j} (\Gamma_{a})_{\beta \rho} \lambda^{j}_{\alpha} {W}^{-1} \nabla^{\lambda \rho}{W} - \frac{3}{64}(\Gamma_{a})^{\beta}{}_{\gamma} \lambda_{j \alpha} \lambda^{j \rho} W_{\beta \rho \lambda i} {W}^{-2} \nabla^{\gamma \lambda}{W} - \frac{15}{64}(\Gamma_{a})^{\beta}{}_{\gamma} \lambda^{\rho}_{j} W_{\beta \rho \lambda i} {W}^{-1} \nabla^{\gamma \lambda}{\lambda^{j}_{\alpha}}+\frac{45}{32}{\rm i} (\Gamma_{a})^{\beta}{}_{\lambda} F_{\alpha \beta} X_{i \rho} {W}^{-1} \nabla^{\lambda \rho}{W}+\frac{9}{64}{\rm i} (\Gamma_{a})_{\rho \lambda} X_{i \beta} {W}^{-1} \nabla^{\rho}\,_{\alpha}{W} \nabla^{\lambda \beta}{W}+\frac{27}{2048}(\Gamma_{a})^{\beta}{}_{\rho} Y \lambda_{i \beta} {W}^{-1} \nabla^{\rho}\,_{\alpha}{W} - \frac{75}{512}(\Gamma_{a})^{\rho}{}_{\lambda} \lambda_{j \rho} X_{i \beta} {W}^{-1} \nabla^{\lambda \beta}{\lambda^{j}_{\alpha}} - \frac{63}{128}{\rm i} (\Gamma_{a})_{\rho \lambda} F_{\alpha}\,^{\beta} X_{i \beta} {W}^{-1} \nabla^{\rho \lambda}{W}+\frac{9}{256}{\rm i} (\Gamma_{a})_{\rho \lambda} X_{i \beta} {W}^{-1} \nabla^{\rho \lambda}{W} \nabla_{\alpha}\,^{\beta}{W}+\frac{135}{4096}(\Gamma_{a})_{\beta \rho} Y \lambda_{i \alpha} {W}^{-1} \nabla^{\beta \rho}{W} - \frac{405}{1024}(\Gamma_{a})_{\lambda \gamma} \lambda^{\beta}_{i} {W}^{-1} \nabla^{\lambda \gamma}{W} \nabla_{\alpha}\,^{\rho}{W_{\beta \rho}}+\frac{27}{1024}(\Gamma_{a})_{\rho \lambda} \lambda_{i \gamma} {W}^{-1} \nabla^{\rho \lambda}{W} \nabla^{\gamma \beta}{W_{\alpha \beta}}+\frac{63}{512}(\Gamma_{a})_{\rho \lambda} \lambda_{j \alpha} \lambda^{j \beta} X_{i \beta} {W}^{-2} \nabla^{\rho \lambda}{W}+\frac{3}{64}(\Gamma_{a})_{\rho \lambda} \lambda^{\beta}_{j} X_{i \beta} {W}^{-1} \nabla^{\rho \lambda}{\lambda^{j}_{\alpha}}+\frac{153}{512}(\Gamma_{a})_{\rho \lambda} \lambda_{j \alpha} \lambda^{\beta}_{i} X^{j}_{\beta} {W}^{-2} \nabla^{\rho \lambda}{W} - \frac{21}{64}(\Gamma_{a})^{\beta}{}_{\gamma} \lambda_{j \alpha} \lambda^{\rho}_{i} W_{\beta \rho \lambda}\,^{j} {W}^{-2} \nabla^{\gamma \lambda}{W}%
 - \frac{3}{4}{\rm i} (\Gamma_{a})^{\beta \rho} F_{\alpha \beta} \lambda_{i \rho} \lambda_{j \lambda} {W}^{-3} \nabla^{\lambda \gamma}{\lambda^{j}_{\gamma}} - \frac{9}{128}{\rm i} (\Gamma_{a})_{\alpha}{}^{\beta} X_{i j} \lambda^{j}_{\beta} \lambda_{k \rho} {W}^{-3} \nabla^{\rho \lambda}{\lambda^{k}_{\lambda}} - \frac{9}{32}{\rm i} (\Gamma_{a})^{\beta \rho} \lambda_{i \beta} \lambda_{j \rho} {W}^{-3} \nabla_{\alpha \gamma}{W} \nabla^{\gamma \lambda}{\lambda^{j}_{\lambda}}+\frac{9}{16}(\Gamma_{a})^{\beta \rho} \lambda_{j \alpha} \lambda_{i \beta} \lambda^{j}_{\rho} \lambda_{k \lambda} {W}^{-4} \nabla^{\lambda \gamma}{\lambda^{k}_{\gamma}} - \frac{3}{32}{\rm i} (\Gamma_{a})^{\rho \lambda} \lambda_{i \rho} \lambda_{j \lambda} \lambda^{j}_{\gamma} {W}^{-3} \nabla^{\gamma \beta}{F_{\alpha \beta}}+\frac{45}{1024}{\rm i} (\Gamma_{a})^{\rho \lambda} \lambda_{i \rho} \lambda_{j \lambda} \lambda^{j}_{\gamma} {W}^{-2} \nabla^{\gamma \beta}{W_{\alpha \beta}}+\frac{3}{16}{\rm i} (\Gamma_{a})^{\beta \rho} \lambda_{i \beta} \lambda_{j \rho} \lambda_{k \lambda} {W}^{-3} \nabla_{\alpha}\,^{\lambda}{X^{j k}} - \frac{3}{64}{\rm i} (\Gamma_{a})^{\beta \rho} \lambda_{i \beta} \lambda_{j \rho} \lambda^{j}_{\lambda} {W}^{-3} \nabla_{\gamma}\,^{\lambda}{\nabla_{\alpha}\,^{\gamma}{W}} - \frac{27}{64}{\rm i} (\Gamma_{a})^{\lambda \gamma} W^{\beta}\,_{\rho} \lambda_{i \lambda} \lambda_{j \gamma} \lambda^{j}_{\beta} {W}^{-3} \nabla_{\alpha}\,^{\rho}{W}+\frac{63}{512}{\rm i} (\Gamma_{a})^{\beta \rho} \lambda_{i \beta} \lambda_{k \rho} \lambda^{k \lambda} \lambda_{j \lambda} X^{j}_{\alpha} {W}^{-3} - \frac{15}{128}{\rm i} \Phi_{\alpha}\,^{\lambda}\,_{j k} (\Gamma_{a})^{\beta \rho} \lambda_{i \beta} \lambda^{j}_{\rho} \lambda^{k}_{\lambda} {W}^{-2}+\frac{9}{1024}{\rm i} (\Gamma_{a})^{\lambda \gamma} \lambda_{i \lambda} \lambda_{j \gamma} \lambda^{j \beta} {W}^{-2} \nabla_{\alpha}\,^{\rho}{W_{\beta \rho}} - \frac{27}{32}(\Gamma_{a})^{\gamma \rho} F_{\alpha \beta} \lambda_{i \gamma} {W}^{-2} \nabla^{\beta \lambda}{F_{\rho \lambda}} - \frac{9}{16}(\Gamma_{a})^{\gamma \beta} W_{\alpha \lambda} \lambda_{i \gamma} {W}^{-1} \nabla^{\lambda \rho}{F_{\beta \rho}}+\frac{3}{16}(\Gamma_{a})^{\lambda \beta} X_{i j} \lambda^{j}_{\lambda} {W}^{-2} \nabla_{\alpha}\,^{\rho}{F_{\beta \rho}} - \frac{9}{16}(\Gamma_{a})^{\lambda \beta} \lambda_{i \lambda} {W}^{-2} \nabla_{\alpha \gamma}{W} \nabla^{\gamma \rho}{F_{\beta \rho}}+\frac{27}{32}(\Gamma_{a})^{\beta \rho} F_{\alpha \beta} \lambda_{i \gamma} {W}^{-2} \nabla^{\gamma \lambda}{F_{\rho \lambda}}+\frac{45}{128}(\Gamma_{a})^{\lambda \beta} W_{\alpha \lambda} \lambda_{i \gamma} {W}^{-1} \nabla^{\gamma \rho}{F_{\beta \rho}} - \frac{9}{64}(\Gamma_{a})^{\beta}{}_{\lambda} \lambda_{i \gamma} {W}^{-2} \nabla^{\lambda}\,_{\alpha}{W} \nabla^{\gamma \rho}{F_{\beta \rho}} - \frac{3}{8}{\rm i} (\Gamma_{a})^{\lambda \beta} \lambda_{j \alpha} \lambda_{i \gamma} \lambda^{j}_{\lambda} {W}^{-3} \nabla^{\gamma \rho}{F_{\beta \rho}}%
+\frac{3}{64}{\rm i} (\Gamma_{a})^{\beta \rho} \lambda_{i \lambda} \lambda_{j \beta} {W}^{-2} \nabla_{\gamma}\,^{\lambda}{\nabla_{\alpha}\,^{\gamma}{\lambda^{j}_{\rho}}}+\frac{3}{64}{\rm i} (\Gamma_{a})^{\beta}{}_{\rho} \lambda_{i \lambda} \lambda_{j \beta} {W}^{-2} \nabla^{\lambda \gamma}{\nabla^{\rho}\,_{\alpha}{\lambda^{j}_{\gamma}}} - \frac{345}{1024}{\rm i} (\Gamma_{a})^{\lambda \beta} \lambda_{j \alpha} \lambda_{i \gamma} \lambda^{j}_{\lambda} {W}^{-2} \nabla^{\gamma \rho}{W_{\beta \rho}}+\frac{3}{128}{\rm i} (\Gamma_{a})^{\beta \lambda} W_{\beta \rho} \lambda_{i \gamma} \lambda_{j \lambda} {W}^{-2} \nabla^{\rho \gamma}{\lambda^{j}_{\alpha}}+\frac{3}{64}{\rm i} (\Gamma_{a})^{\rho \lambda} W_{\alpha \beta} \lambda_{i \gamma} \lambda_{j \rho} {W}^{-2} \nabla^{\beta \gamma}{\lambda^{j}_{\lambda}}+\frac{3}{32}{\rm i} (\Gamma_{a})^{\beta}{}_{\rho} \lambda_{i \lambda} \lambda_{j \beta} {W}^{-2} \nabla_{\alpha}\,^{\lambda}{\nabla^{\rho \gamma}{\lambda^{j}_{\gamma}}}+\frac{3}{32}{\rm i} (\Gamma_{a})^{\lambda \beta} \lambda_{i \gamma} \lambda_{j \lambda} \lambda^{j \rho} {W}^{-2} \nabla_{\alpha}\,^{\gamma}{W_{\beta \rho}}+\frac{75}{1024}{\rm i} (\Gamma_{a})_{\alpha}{}^{\lambda} \lambda_{i \gamma} \lambda_{j \lambda} \lambda^{j \beta} {W}^{-2} \nabla^{\gamma \rho}{W_{\beta \rho}}+\frac{123}{640}{\rm i} (\Gamma_{a})_{\alpha}{}^{\lambda} W^{\beta}\,_{\rho} \lambda_{i \gamma} \lambda_{j \lambda} {W}^{-2} \nabla^{\rho \gamma}{\lambda^{j}_{\beta}}+\frac{473}{1024}(\Gamma_{a})^{\lambda \beta} \lambda_{i \gamma} \lambda_{j \lambda} {W}^{-1} \nabla^{\gamma \rho}{W_{\alpha \beta \rho}\,^{j}}+\frac{3}{128}{\rm i} (\Gamma_{a})^{\rho}{}_{\lambda} W_{\alpha \beta} \lambda^{\gamma}_{i} \lambda_{j \rho} {W}^{-2} \nabla^{\lambda \beta}{\lambda^{j}_{\gamma}}+\frac{105}{512}{\rm i} (\Gamma_{a})^{\rho \gamma} W_{\alpha}\,^{\beta} W_{\rho}\,^{\lambda} \lambda_{i \lambda} \lambda_{j \gamma} \lambda^{j}_{\beta} {W}^{-2} - \frac{9}{64}{\rm i} (\Gamma_{a})^{\lambda}{}_{\gamma} W^{\beta}\,_{\rho} \lambda_{i \alpha} \lambda_{j \lambda} {W}^{-2} \nabla^{\gamma \rho}{\lambda^{j}_{\beta}}+\frac{649}{2048}(\Gamma_{a})^{\gamma \lambda} W^{\beta \rho} \lambda_{i \alpha} \lambda_{j \gamma} W_{\lambda \beta \rho}\,^{j} {W}^{-1}+\frac{3}{32}{\rm i} (\Gamma_{a})^{\beta \lambda} W_{\beta}\,^{\rho} \lambda_{i \rho} \lambda_{j \lambda} {W}^{-2} \nabla_{\alpha}\,^{\gamma}{\lambda^{j}_{\gamma}}+\frac{3}{64}{\rm i} (\Gamma_{a})^{\rho}{}_{\lambda} W_{\alpha}\,^{\beta} \lambda_{i \gamma} \lambda_{j \rho} {W}^{-2} \nabla^{\lambda \gamma}{\lambda^{j}_{\beta}}+\frac{3}{16}(\Gamma_{a})^{\gamma \lambda} F^{\beta \rho} \lambda_{i \alpha} \lambda_{j \gamma} W_{\lambda \beta \rho}\,^{j} {W}^{-2} - \frac{33}{128}{\rm i} \Phi_{\alpha}\,^{\lambda}\,_{j k} (\Gamma_{a})^{\beta \rho} \lambda_{i \lambda} \lambda^{j}_{\beta} \lambda^{k}_{\rho} {W}^{-2} - \frac{5}{32}{\rm i} \Phi^{\beta \lambda}\,_{j k} (\Gamma_{a})_{\beta}{}^{\rho} \lambda_{i \alpha} \lambda^{j}_{\rho} \lambda^{k}_{\lambda} {W}^{-2} - \frac{33}{128}{\rm i} \Phi^{\beta \lambda}\,_{j k} (\Gamma_{a})_{\beta}{}^{\rho} \lambda^{j}_{\alpha} \lambda_{i \lambda} \lambda^{k}_{\rho} {W}^{-2}%
 - \frac{9}{512}{\rm i} (\Gamma_{a})^{\rho \beta} \lambda_{i \lambda} \lambda_{j \rho} \lambda^{j}_{\gamma} {W}^{-2} \nabla^{\lambda \gamma}{W_{\alpha \beta}}+\frac{57}{512}{\rm i} (\Gamma_{a})^{\lambda \beta} \lambda^{\rho}_{i} \lambda_{j \lambda} \lambda^{j}_{\gamma} {W}^{-2} \nabla_{\alpha}\,^{\gamma}{W_{\beta \rho}}+\frac{87}{512}{\rm i} (\Gamma_{a})^{\lambda}{}_{\gamma} \lambda_{i \alpha} \lambda_{j \lambda} \lambda^{j \beta} {W}^{-2} \nabla^{\gamma \rho}{W_{\beta \rho}}+\frac{375}{1024}{\rm i} (\Gamma_{a})^{\lambda}{}_{\gamma} \lambda_{j \alpha} \lambda^{\beta}_{i} \lambda^{j}_{\lambda} {W}^{-2} \nabla^{\gamma \rho}{W_{\beta \rho}} - \frac{21}{512}{\rm i} (\Gamma_{a})^{\rho}{}_{\lambda} \lambda_{i \gamma} \lambda_{j \rho} \lambda^{j \beta} {W}^{-2} \nabla^{\lambda \gamma}{W_{\alpha \beta}}+\frac{21}{256}{\rm i} (\Gamma_{a})^{\rho}{}_{\lambda} \lambda^{\beta}_{i} \lambda_{j \rho} \lambda^{j}_{\gamma} {W}^{-2} \nabla^{\lambda \gamma}{W_{\alpha \beta}}+\frac{51}{512}{\rm i} (\Gamma_{a})^{\lambda \beta} \lambda_{i \alpha} \lambda_{j \lambda} \lambda^{j}_{\gamma} {W}^{-2} \nabla^{\gamma \rho}{W_{\beta \rho}} - \frac{687}{5120}{\rm i} (\Gamma_{a})_{\alpha}{}^{\lambda} \lambda^{\beta}_{i} \lambda_{j \lambda} \lambda^{j}_{\gamma} {W}^{-2} \nabla^{\gamma \rho}{W_{\beta \rho}}+\frac{207}{1280}(\Gamma_{a})_{\alpha}{}^{\rho} \lambda_{i \lambda} \lambda_{j \rho} X^{j}_{\beta} {W}^{-2} \nabla^{\lambda \beta}{W}+\frac{9}{16}(\Gamma_{a})^{\beta \lambda} W_{\beta \rho} \lambda_{i \lambda} {W}^{-2} \nabla_{\alpha \gamma}{W} \nabla^{\gamma \rho}{W}+\frac{27}{32}(\Gamma_{a})^{\rho \beta} W_{\rho \lambda} F_{\alpha \beta} \lambda_{i \gamma} {W}^{-2} \nabla^{\lambda \gamma}{W}+\frac{27}{128}(\Gamma_{a})^{\beta}{}_{\lambda} W_{\beta \rho} \lambda_{i \gamma} {W}^{-2} \nabla^{\lambda}\,_{\alpha}{W} \nabla^{\rho \gamma}{W} - \frac{27}{64}{\rm i} (\Gamma_{a})^{\beta \lambda} W_{\beta \rho} \lambda_{j \alpha} \lambda_{i \gamma} \lambda^{j}_{\lambda} {W}^{-3} \nabla^{\rho \gamma}{W} - \frac{387}{1024}(\Gamma_{a})^{\lambda \beta} \lambda_{i \lambda} {W}^{-1} \nabla_{\alpha \gamma}{W} \nabla^{\gamma \rho}{W_{\beta \rho}}+\frac{375}{2048}(\Gamma_{a})^{\rho \beta} \lambda_{i \lambda} \lambda_{j \rho} {W}^{-1} \nabla_{\alpha}\,^{\lambda}{X^{j}_{\beta}} - \frac{903}{10240}(\Gamma_{a})_{\alpha}{}^{\rho} \lambda_{i \lambda} \lambda_{j \rho} {W}^{-1} \nabla^{\lambda \beta}{X^{j}_{\beta}}+\frac{9}{32}(\Gamma_{a})^{\beta}{}_{\rho} F_{\alpha \beta} \lambda_{j \lambda} {W}^{-2} \nabla^{\rho \lambda}{X_{i}\,^{j}} - \frac{3}{64}(\Gamma_{a})_{\beta \rho} \lambda_{j \lambda} {W}^{-2} \nabla^{\beta}\,_{\alpha}{W} \nabla^{\rho \lambda}{X_{i}\,^{j}}+\frac{3}{32}(\Gamma_{a})^{\beta}{}_{\rho} X_{j k} \lambda^{j}_{\beta} {W}^{-2} \nabla^{\rho}\,_{\alpha}{X_{i}\,^{k}} - \frac{3}{16}(\Gamma_{a})^{\beta}{}_{\rho} \lambda_{j \beta} {W}^{-2} \nabla_{\alpha \lambda}{W} \nabla^{\rho \lambda}{X_{i}\,^{j}}%
+\frac{3}{16}{\rm i} (\Gamma_{a})^{\beta}{}_{\rho} \lambda_{k \alpha} \lambda^{k}_{\beta} \lambda_{j \lambda} {W}^{-3} \nabla^{\rho \lambda}{X_{i}\,^{j}} - \frac{15}{64}{\rm i} (\Gamma_{a})^{\beta}{}_{\rho} \lambda_{i \beta} \lambda_{j \lambda} {W}^{-2} \nabla^{\rho \lambda}{\nabla_{\alpha}\,^{\gamma}{\lambda^{j}_{\gamma}}} - \frac{3}{64}{\rm i} (\Gamma_{a})^{\beta}{}_{\rho} \lambda_{j \beta} \lambda^{j}_{\lambda} {W}^{-2} \nabla^{\rho \lambda}{\nabla_{\alpha}\,^{\gamma}{\lambda_{i \gamma}}} - \frac{75}{512}{\rm i} (\Gamma_{a})^{\rho}{}_{\lambda} \lambda_{i \rho} \lambda^{\beta}_{j} \lambda^{j}_{\gamma} {W}^{-2} \nabla^{\lambda \gamma}{W_{\alpha \beta}} - \frac{51}{256}{\rm i} (\Gamma_{a})^{\rho}{}_{\lambda} W_{\alpha}\,^{\beta} \lambda_{j \rho} \lambda^{j}_{\gamma} {W}^{-2} \nabla^{\lambda \gamma}{\lambda_{i \beta}}+\frac{207}{1024}(\Gamma_{a})^{\beta}{}_{\rho} \lambda_{i \beta} \lambda_{j \lambda} {W}^{-1} \nabla^{\rho \lambda}{X^{j}_{\alpha}}+\frac{63}{1024}(\Gamma_{a})^{\beta}{}_{\rho} \lambda_{j \beta} \lambda^{j}_{\lambda} {W}^{-1} \nabla^{\rho \lambda}{X_{i \alpha}}+\frac{33}{512}(\Gamma_{a})^{\beta \rho} X_{j k} \lambda^{j}_{\beta} \lambda^{k}_{\rho} X_{i \alpha} {W}^{-2}+\frac{35}{128}{\rm i} \Phi_{\alpha}\,^{\lambda}\,_{i j} (\Gamma_{a})^{\beta \rho} \lambda^{j}_{\beta} \lambda_{k \lambda} \lambda^{k}_{\rho} {W}^{-2}+\frac{41}{640}{\rm i} \Phi^{\rho \lambda}\,_{i j} (\Gamma_{a})_{\alpha}{}^{\beta} \lambda^{j}_{\beta} \lambda_{k \rho} \lambda^{k}_{\lambda} {W}^{-2} - \frac{9}{64}{\rm i} \Phi^{\beta \lambda}\,_{i j} (\Gamma_{a})_{\beta}{}^{\rho} \lambda_{k \alpha} \lambda^{j}_{\rho} \lambda^{k}_{\lambda} {W}^{-2}+\frac{1}{320}{\rm i} \Phi^{\rho \lambda}\,_{j k} (\Gamma_{a})_{\alpha}{}^{\beta} \lambda_{i \rho} \lambda^{j}_{\beta} \lambda^{k}_{\lambda} {W}^{-2}+\frac{9}{512}{\rm i} (\Gamma_{a})^{\lambda}{}_{\gamma} \lambda_{i \lambda} \lambda^{\beta}_{j} \lambda^{j \rho} {W}^{-2} \nabla^{\gamma}\,_{\alpha}{W_{\beta \rho}}+\frac{129}{512}{\rm i} (\Gamma_{a})^{\lambda \beta} \lambda_{i \lambda} \lambda^{\rho}_{j} \lambda^{j}_{\gamma} {W}^{-2} \nabla_{\alpha}\,^{\gamma}{W_{\beta \rho}} - \frac{27}{512}{\rm i} (\Gamma_{a})^{\lambda}{}_{\gamma} \lambda^{\beta}_{i} \lambda_{j \lambda} \lambda^{j \rho} {W}^{-2} \nabla^{\gamma}\,_{\alpha}{W_{\beta \rho}}+\frac{3}{64}{\rm i} (\Gamma_{a})_{\alpha}{}^{\lambda} \lambda_{i \lambda} \lambda^{\beta}_{j} \lambda^{j}_{\gamma} {W}^{-2} \nabla^{\gamma \rho}{W_{\beta \rho}} - \frac{105}{512}{\rm i} (\Gamma_{a})^{\lambda \beta} \lambda_{j \alpha} \lambda_{i \lambda} \lambda^{j}_{\gamma} {W}^{-2} \nabla^{\gamma \rho}{W_{\beta \rho}}+\frac{27}{64}(\Gamma_{a})^{\rho}{}_{\lambda} F_{\alpha \beta} \lambda_{i \rho} {W}^{-2} \nabla_{\gamma}\,^{\beta}{\nabla^{\lambda \gamma}{W}}+\frac{9}{32}(\Gamma_{a})^{\rho}{}_{\lambda} W_{\alpha \beta} \lambda_{i \rho} {W}^{-1} \nabla_{\gamma}\,^{\beta}{\nabla^{\lambda \gamma}{W}}+\frac{3}{32}(\Gamma_{a})^{\beta}{}_{\rho} X_{i j} \lambda^{j}_{\beta} {W}^{-2} \nabla_{\alpha \lambda}{\nabla^{\rho \lambda}{W}}%
+\frac{9}{32}(\Gamma_{a})^{\beta}{}_{\rho} \lambda_{i \beta} {W}^{-2} \nabla_{\alpha \lambda}{W} \nabla_{\gamma}\,^{\lambda}{\nabla^{\rho \gamma}{W}} - \frac{27}{64}(\Gamma_{a})^{\beta}{}_{\rho} F_{\alpha \beta} \lambda_{i \lambda} {W}^{-2} \nabla_{\gamma}\,^{\lambda}{\nabla^{\rho \gamma}{W}} - \frac{9}{128}(\Gamma_{a})_{\beta \rho} \lambda_{i \lambda} {W}^{-2} \nabla^{\beta}\,_{\alpha}{W} \nabla_{\gamma}\,^{\lambda}{\nabla^{\rho \gamma}{W}}+\frac{3}{16}{\rm i} (\Gamma_{a})^{\beta}{}_{\rho} \lambda_{j \alpha} \lambda_{i \lambda} \lambda^{j}_{\beta} {W}^{-3} \nabla_{\gamma}\,^{\lambda}{\nabla^{\rho \gamma}{W}}+\frac{3}{64}{\rm i} (\Gamma_{a})^{\beta}{}_{\rho} \lambda_{i \lambda} \lambda_{j \beta} {W}^{-2} \nabla_{\gamma}\,^{\lambda}{\nabla^{\rho \gamma}{\lambda^{j}_{\alpha}}}+\frac{15}{64}(\Gamma_{a})^{\gamma \beta} \lambda^{\rho}_{i} \lambda_{j \gamma} W_{\beta \rho \lambda}\,^{j} {W}^{-2} \nabla_{\alpha}\,^{\lambda}{W} - \frac{63}{128}(\Gamma_{a})^{\rho \lambda} \lambda_{i \rho} \lambda_{j \lambda} X^{j}_{\beta} {W}^{-2} \nabla_{\alpha}\,^{\beta}{W} - \frac{9}{512}(\Gamma_{a})^{\rho}{}_{\lambda} \lambda_{i \alpha} \lambda_{j \rho} X^{j}_{\beta} {W}^{-2} \nabla^{\lambda \beta}{W}+\frac{285}{512}(\Gamma_{a})^{\lambda}{}_{\gamma} \lambda^{\beta}_{i} \lambda_{j \lambda} {W}^{-1} \nabla^{\gamma \rho}{W_{\alpha \beta \rho}\,^{j}} - \frac{491}{512}(\Gamma_{a})^{\gamma \beta} \lambda^{\rho}_{i} \lambda_{j \gamma} {W}^{-1} \nabla_{\alpha}\,^{\lambda}{W_{\beta \rho \lambda}\,^{j}} - \frac{27}{32}(\Gamma_{a})^{\rho \beta} X_{i j} W_{\rho}\,^{\lambda} F_{\alpha \beta} \lambda^{j}_{\lambda} {W}^{-2} - \frac{9}{128}(\Gamma_{a})^{\beta}{}_{\lambda} X_{i j} W_{\beta}\,^{\rho} \lambda^{j}_{\rho} {W}^{-2} \nabla^{\lambda}\,_{\alpha}{W} - \frac{21}{128}{\rm i} (\Gamma_{a})^{\beta \lambda} X_{i j} W_{\beta}\,^{\rho} \lambda_{k \alpha} \lambda^{j}_{\rho} \lambda^{k}_{\lambda} {W}^{-3} - \frac{153}{256}(\Gamma_{a})^{\beta \lambda} F_{\alpha \beta} \lambda_{j \lambda} \lambda^{j \rho} X_{i \rho} {W}^{-2} - \frac{27}{1024}{\rm i} (\Gamma_{a})^{\beta \rho} Y \lambda_{j \alpha} \lambda_{i \beta} \lambda^{j}_{\rho} {W}^{-2} - \frac{81}{256}{\rm i} (\Gamma_{a})^{\rho \lambda} \lambda_{j \alpha} \lambda^{j}_{\rho} \lambda_{k \lambda} \lambda^{k \beta} X_{i \beta} {W}^{-3} - \frac{63}{128}{\rm i} (\Gamma_{a})^{\rho \lambda} \lambda_{k \alpha} \lambda_{i \rho} \lambda^{k}_{\lambda} \lambda^{\beta}_{j} X^{j}_{\beta} {W}^{-3} - \frac{27}{128}{\rm i} (\Gamma_{a})^{\rho \lambda} \lambda_{k \alpha} \lambda^{\beta}_{i} \lambda^{k}_{\rho} \lambda_{j \lambda} X^{j}_{\beta} {W}^{-3}+\frac{625}{1024}(\Gamma_{a})^{\gamma \beta} \lambda_{i \gamma} \lambda^{\rho}_{j} {W}^{-1} \nabla_{\alpha}\,^{\lambda}{W_{\beta \rho \lambda}\,^{j}}+\frac{435}{1024}(\Gamma_{a})^{\gamma \beta} \lambda_{j \gamma} \lambda^{j \rho} {W}^{-1} \nabla_{\alpha}\,^{\lambda}{W_{\beta \rho \lambda i}}%
 - \frac{747}{2048}(\Gamma_{a})^{\beta \gamma} W_{\beta}\,^{\rho} \lambda_{j \gamma} \lambda^{j \lambda} W_{\alpha \rho \lambda i} {W}^{-1} - \frac{3567}{10240}(\Gamma_{a})_{\alpha}{}^{\gamma} W^{\beta \rho} \lambda_{j \gamma} \lambda^{j \lambda} W_{\beta \rho \lambda i} {W}^{-1} - \frac{405}{1024}(\Gamma_{a})^{\lambda}{}_{\gamma} \lambda_{i \lambda} \lambda^{\beta}_{j} {W}^{-1} \nabla^{\gamma \rho}{W_{\alpha \beta \rho}\,^{j}} - \frac{195}{1024}(\Gamma_{a})^{\lambda}{}_{\gamma} \lambda_{j \lambda} \lambda^{j \beta} {W}^{-1} \nabla^{\gamma \rho}{W_{\alpha \beta \rho i}} - \frac{365}{1024}(\Gamma_{a})^{\lambda \beta} \lambda_{i \lambda} \lambda_{j \gamma} {W}^{-1} \nabla^{\gamma \rho}{W_{\alpha \beta \rho}\,^{j}}+\frac{27}{512}(\Gamma_{a})^{\lambda \beta} \lambda_{j \lambda} \lambda^{j}_{\gamma} {W}^{-1} \nabla^{\gamma \rho}{W_{\alpha \beta \rho i}} - \frac{99}{512}(\Gamma_{a})^{\rho}{}_{\lambda} \lambda_{i \rho} \lambda^{\beta}_{j} {W}^{-1} \nabla^{\lambda}\,_{\alpha}{X^{j}_{\beta}} - \frac{9}{1024}(\Gamma_{a})^{\rho}{}_{\lambda} \lambda_{j \rho} \lambda^{j \beta} {W}^{-1} \nabla^{\lambda}\,_{\alpha}{X_{i \beta}} - \frac{219}{2048}(\Gamma_{a})^{\rho \beta} \lambda_{i \rho} \lambda_{j \lambda} {W}^{-1} \nabla_{\alpha}\,^{\lambda}{X^{j}_{\beta}}+\frac{3}{256}(\Gamma_{a})^{\rho \beta} \lambda_{j \rho} \lambda^{j}_{\lambda} {W}^{-1} \nabla_{\alpha}\,^{\lambda}{X_{i \beta}}+\frac{51}{10240}(\Gamma_{a})_{\alpha}{}^{\rho} \lambda_{i \rho} \lambda_{j \lambda} {W}^{-1} \nabla^{\lambda \beta}{X^{j}_{\beta}} - \frac{39}{2560}(\Gamma_{a})_{\alpha}{}^{\rho} \lambda_{j \rho} \lambda^{j}_{\lambda} {W}^{-1} \nabla^{\lambda \beta}{X_{i \beta}} - \frac{45}{512}(\Gamma_{a})^{\rho}{}_{\lambda} \lambda_{j \alpha} \lambda_{i \rho} {W}^{-1} \nabla^{\lambda \beta}{X^{j}_{\beta}}+\frac{81}{1024}(\Gamma_{a})^{\rho}{}_{\lambda} \lambda_{j \alpha} \lambda^{j}_{\rho} {W}^{-1} \nabla^{\lambda \beta}{X_{i \beta}}+\frac{83}{64}\Phi_{\lambda}\,^{\beta}\,_{i j} (\Gamma_{a})_{\beta}{}^{\rho} \lambda^{j}_{\rho} {W}^{-1} \nabla^{\lambda}\,_{\alpha}{W}+\frac{7}{64}{\rm i} \Phi^{\beta \lambda}\,_{i j} (\Gamma_{a})_{\beta}{}^{\rho} \lambda_{k \alpha} \lambda^{j}_{\lambda} \lambda^{k}_{\rho} {W}^{-2}+\frac{765}{1024}(\Gamma_{a})^{\lambda}{}_{\gamma} \lambda_{i \lambda} {W}^{-1} \nabla_{\alpha}\,^{\beta}{W} \nabla^{\gamma \rho}{W_{\beta \rho}}+\frac{261}{1024}(\Gamma_{a})^{\rho}{}_{\lambda} \lambda^{\beta}_{i} \lambda_{j \rho} {W}^{-1} \nabla^{\lambda}\,_{\alpha}{X^{j}_{\beta}} - \frac{171}{1024}(\Gamma_{a})^{\rho}{}_{\lambda} \lambda_{i \alpha} \lambda_{j \rho} {W}^{-1} \nabla^{\lambda \beta}{X^{j}_{\beta}}+\frac{3}{80}(\Gamma_{a})_{\alpha}{}^{\beta} X_{i j} \lambda^{j}_{\lambda} {W}^{-2} \nabla^{\lambda \rho}{F_{\beta \rho}}%
+\frac{3}{32}{\rm i} (\Gamma_{a})^{\lambda \beta} \lambda_{j \alpha} \lambda_{i \lambda} \lambda^{j}_{\gamma} {W}^{-3} \nabla^{\gamma \rho}{F_{\beta \rho}}+\frac{3}{128}{\rm i} (\Gamma_{a})^{\beta \lambda} W_{\beta \rho} \lambda_{i \lambda} \lambda_{j \gamma} {W}^{-2} \nabla^{\rho \gamma}{\lambda^{j}_{\alpha}} - \frac{3}{128}{\rm i} (\Gamma_{a})^{\rho \lambda} W_{\alpha \beta} \lambda_{i \rho} \lambda_{j \gamma} {W}^{-2} \nabla^{\beta \gamma}{\lambda^{j}_{\lambda}} - \frac{15}{64}{\rm i} (\Gamma_{a})^{\beta}{}_{\rho} \lambda_{i \beta} \lambda_{j \lambda} {W}^{-2} \nabla_{\alpha}\,^{\lambda}{\nabla^{\rho \gamma}{\lambda^{j}_{\gamma}}}+\frac{35}{128}{\rm i} \Phi^{\beta \lambda}\,_{j k} (\Gamma_{a})_{\beta}{}^{\rho} \lambda^{j}_{\alpha} \lambda_{i \rho} \lambda^{k}_{\lambda} {W}^{-2}+\frac{117}{640}(\Gamma_{a})_{\alpha}{}^{\beta} X_{i j} W_{\beta \rho} \lambda^{j}_{\lambda} {W}^{-2} \nabla^{\rho \lambda}{W}+\frac{9}{160}(\Gamma_{a})_{\alpha \beta} X_{i j} \lambda_{k \rho} {W}^{-2} \nabla^{\beta \rho}{X^{j k}}+\frac{3}{16}{\rm i} (\Gamma_{a})^{\beta}{}_{\rho} \lambda_{j \alpha} \lambda_{i \beta} \lambda_{k \lambda} {W}^{-3} \nabla^{\rho \lambda}{X^{j k}}+\frac{3}{32}(\Gamma_{a})_{\alpha \beta} X_{i j} \lambda^{j}_{\rho} {W}^{-2} \nabla_{\lambda}\,^{\rho}{\nabla^{\beta \lambda}{W}} - \frac{3}{64}{\rm i} (\Gamma_{a})^{\beta}{}_{\rho} \lambda_{j \alpha} \lambda_{i \beta} \lambda^{j}_{\lambda} {W}^{-3} \nabla_{\gamma}\,^{\lambda}{\nabla^{\rho \gamma}{W}} - \frac{3}{16}{\rm i} (\Gamma_{a})^{\lambda}{}_{\gamma} \lambda_{j \alpha} \lambda_{i \lambda} \lambda^{j \beta} {W}^{-2} \nabla^{\gamma \rho}{W_{\beta \rho}} - \frac{141}{640}(\Gamma_{a})_{\alpha}{}^{\beta} X_{i j} X^{j}\,_{k} W_{\beta}\,^{\rho} \lambda^{k}_{\rho} {W}^{-2} - \frac{123}{256}{\rm i} (\Gamma_{a})^{\beta \lambda} X_{j k} W_{\beta}\,^{\rho} \lambda^{j}_{\alpha} \lambda_{i \lambda} \lambda^{k}_{\rho} {W}^{-3} - \frac{9}{256}{\rm i} (\Gamma_{a})^{\rho \lambda} \lambda_{j \alpha} \lambda_{i \rho} \lambda_{k \lambda} \lambda^{k \beta} X^{j}_{\beta} {W}^{-3} - \frac{27}{64}(\Gamma_{a})^{\rho}{}_{\lambda} F_{\alpha \beta} {W}^{-1} \nabla_{\gamma}\,^{\beta}{\nabla^{\lambda \gamma}{\lambda_{i \rho}}} - \frac{81}{128}(\Gamma_{a})^{\rho}{}_{\lambda} W_{\alpha \beta} \nabla_{\gamma}\,^{\beta}{\nabla^{\lambda \gamma}{\lambda_{i \rho}}} - \frac{9}{32}(\Gamma_{a})^{\beta}{}_{\rho} {W}^{-1} \nabla_{\alpha \lambda}{W} \nabla_{\gamma}\,^{\lambda}{\nabla^{\rho \gamma}{\lambda_{i \beta}}} - \frac{3}{128}{\rm i} (\Gamma_{a})^{\beta}{}_{\rho} \lambda_{j \alpha} \lambda^{j}_{\lambda} {W}^{-2} \nabla_{\gamma}\,^{\lambda}{\nabla^{\rho \gamma}{\lambda_{i \beta}}}+\frac{9}{32}(\Gamma_{a})^{\beta}{}_{\rho} \lambda_{i \lambda} {W}^{-1} \nabla_{\gamma}\,^{\lambda}{\nabla^{\rho \gamma}{F_{\alpha \beta}}}+\frac{27}{128}(\Gamma_{a})^{\beta}{}_{\rho} \lambda_{i \lambda} {W}^{-1} \nabla_{\gamma}\,^{\lambda}{W} \nabla^{\rho \gamma}{W_{\alpha \beta}}%
+\frac{45}{256}(\Gamma_{a})^{\beta}{}_{\rho} \lambda_{i \lambda} \nabla_{\gamma}\,^{\lambda}{\nabla^{\rho \gamma}{W_{\alpha \beta}}}+\frac{21}{320}(\Gamma_{a})_{\alpha \beta} \lambda_{j \rho} {W}^{-1} \nabla_{\lambda}\,^{\rho}{\nabla^{\beta \lambda}{X_{i}\,^{j}}} - \frac{9}{64}(\Gamma_{a})_{\beta \rho} \lambda_{i \lambda} {W}^{-1} \nabla_{\gamma}\,^{\lambda}{\nabla^{\beta \gamma}{\nabla^{\rho}\,_{\alpha}{W}}}+\frac{117}{256}(\Gamma_{a})^{\beta}{}_{\lambda} F_{\beta}\,^{\rho} \lambda_{i \gamma} {W}^{-1} \nabla^{\lambda \gamma}{W_{\alpha \rho}} - \frac{9}{64}(\Gamma_{a})_{\rho \lambda} W_{\alpha \beta} \lambda_{i \gamma} {W}^{-1} \nabla^{\rho \gamma}{\nabla^{\lambda \beta}{W}}+\frac{279}{640}(\Gamma_{a})_{\alpha}{}^{\beta} W^{\rho}\,_{\lambda} \lambda_{i \gamma} {W}^{-1} \nabla^{\lambda \gamma}{F_{\beta \rho}}+\frac{9}{640}(\Gamma_{a})_{\alpha}{}^{\beta} W_{\beta \rho} \lambda_{j \lambda} {W}^{-1} \nabla^{\rho \lambda}{X_{i}\,^{j}} - \frac{747}{1280}(\Gamma_{a})_{\alpha \lambda} W_{\beta \rho} \lambda_{i \gamma} {W}^{-1} \nabla^{\beta \gamma}{\nabla^{\lambda \rho}{W}} - \frac{171}{128}(\Gamma_{a})^{\lambda \beta} W_{\lambda}\,^{\rho} \lambda_{i \gamma} {W}^{-1} \nabla_{\alpha}\,^{\gamma}{F_{\beta \rho}}+\frac{153}{256}(\Gamma_{a})^{\beta}{}_{\lambda} W_{\beta \rho} \lambda_{i \gamma} {W}^{-1} \nabla_{\alpha}\,^{\gamma}{\nabla^{\lambda \rho}{W}} - \frac{27}{128}(\Gamma_{a})_{\rho \lambda} W_{\alpha \beta} \lambda_{i \gamma} {W}^{-1} \nabla^{\beta \gamma}{\nabla^{\rho \lambda}{W}}+\frac{463}{2560}(\Gamma_{a})_{\alpha}{}^{\beta} \lambda^{\rho}_{i} \lambda_{j \gamma} {W}^{-1} \nabla^{\gamma \lambda}{W_{\beta \rho \lambda}\,^{j}}+\frac{153}{512}(\Gamma_{a})^{\beta \rho} \lambda_{j \lambda} X^{j}_{\beta} {W}^{-1} \nabla_{\alpha}\,^{\lambda}{\lambda_{i \rho}} - \frac{33}{1280}(\Gamma_{a})_{\alpha}{}^{\rho} \lambda_{j \lambda} X^{j}_{\beta} {W}^{-1} \nabla^{\lambda \beta}{\lambda_{i \rho}} - \frac{309}{1280}(\Gamma_{a})_{\alpha}{}^{\rho} \lambda_{i \lambda} X_{j \beta} {W}^{-1} \nabla^{\lambda \beta}{\lambda^{j}_{\rho}} - \frac{9}{320}(\Gamma_{a})_{\alpha}{}^{\rho} \lambda_{j \lambda} X_{i \beta} {W}^{-1} \nabla^{\lambda \beta}{\lambda^{j}_{\rho}} - \frac{7}{80}(\Gamma_{a})_{\alpha}{}^{\beta} \lambda_{j \lambda} \nabla^{\lambda \rho}{\Phi_{\rho \beta i}\,^{j}}+\frac{19}{320}\Phi_{\rho}\,^{\beta}\,_{i j} (\Gamma_{a})_{\alpha \beta} \lambda^{j}_{\lambda} {W}^{-1} \nabla^{\rho \lambda}{W}+\frac{9}{256}(\Gamma_{a})^{\beta}{}_{\lambda} \lambda_{i \gamma} \nabla^{\gamma \rho}{\nabla^{\lambda}\,_{\alpha}{W_{\beta \rho}}}+\frac{9}{128}(\Gamma_{a})^{\beta}{}_{\lambda} \lambda_{i \gamma} {W}^{-1} \nabla^{\gamma \rho}{W} \nabla^{\lambda}\,_{\alpha}{W_{\beta \rho}}%
+\frac{297}{2560}(\Gamma_{a})_{\alpha \lambda} \lambda_{i \gamma} \nabla^{\gamma \beta}{\nabla^{\lambda \rho}{W_{\beta \rho}}}+\frac{279}{512}(\Gamma_{a})^{\beta}{}_{\lambda} \lambda_{i \gamma} \nabla_{\alpha}\,^{\gamma}{\nabla^{\lambda \rho}{W_{\beta \rho}}}+\frac{639}{5120}(\Gamma_{a})_{\alpha \lambda} \lambda_{i \gamma} {W}^{-1} \nabla^{\gamma \beta}{W} \nabla^{\lambda \rho}{W_{\beta \rho}}+\frac{81}{1280}(\Gamma_{a})_{\alpha}{}^{\beta} \lambda_{i \lambda} \nabla_{\gamma}\,^{\lambda}{\nabla^{\gamma \rho}{W_{\beta \rho}}} - \frac{243}{5120}(\Gamma_{a})_{\alpha}{}^{\beta} \lambda_{i \lambda} {W}^{-1} \nabla_{\gamma}\,^{\lambda}{W} \nabla^{\gamma \rho}{W_{\beta \rho}} - \frac{81}{128}(\Gamma_{a})^{\beta}{}_{\gamma} W^{\lambda \rho} \lambda_{i \lambda} {W}^{-1} \nabla^{\gamma}\,_{\alpha}{F_{\beta \rho}} - \frac{27}{256}(\Gamma_{a})_{\lambda \gamma} W^{\beta}\,_{\rho} \lambda_{i \beta} {W}^{-1} \nabla^{\lambda}\,_{\alpha}{\nabla^{\gamma \rho}{W}} - \frac{27}{64}(\Gamma_{a})^{\beta}{}_{\gamma} W^{\rho}\,_{\lambda} \lambda_{i \alpha} {W}^{-1} \nabla^{\gamma \lambda}{F_{\beta \rho}} - \frac{189}{256}(\Gamma_{a})^{\lambda}{}_{\gamma} W^{\beta}\,_{\rho} \lambda_{i \alpha} \nabla^{\gamma \rho}{W_{\lambda \beta}} - \frac{27}{128}(\Gamma_{a})_{\lambda \gamma} W_{\beta \rho} \lambda_{i \alpha} {W}^{-1} \nabla^{\lambda \beta}{\nabla^{\gamma \rho}{W}}+\frac{9}{64}(\Gamma_{a})^{\beta}{}_{\gamma} W_{\alpha \lambda} \lambda^{\rho}_{i} {W}^{-1} \nabla^{\gamma \lambda}{F_{\beta \rho}}+\frac{9}{128}(\Gamma_{a})_{\rho \lambda} W_{\alpha \beta} \lambda_{i \gamma} {W}^{-1} \nabla^{\rho \beta}{\nabla^{\lambda \gamma}{W}} - \frac{3}{32}(\Gamma_{a})^{\beta \gamma} \lambda^{\rho}_{j} W_{\beta \rho \lambda}\,^{j} {W}^{-1} \nabla_{\alpha}\,^{\lambda}{\lambda_{i \gamma}} - \frac{45}{512}(\Gamma_{a})^{\rho \lambda} \lambda_{j \rho} X^{j}_{\beta} {W}^{-1} \nabla_{\alpha}\,^{\beta}{\lambda_{i \lambda}}+\frac{77}{1024}(\Gamma_{a})^{\beta}{}_{\gamma} \lambda_{i \alpha} \lambda^{\rho}_{j} {W}^{-1} \nabla^{\gamma \lambda}{W_{\beta \rho \lambda}\,^{j}} - \frac{197}{5120}(\Gamma_{a})_{\alpha}{}^{\beta} \lambda_{i \gamma} \lambda^{\rho}_{j} {W}^{-1} \nabla^{\gamma \lambda}{W_{\beta \rho \lambda}\,^{j}}+\frac{11}{256}(\Gamma_{a})^{\beta}{}_{\gamma} \lambda_{j \alpha} \lambda^{\rho}_{i} {W}^{-1} \nabla^{\gamma \lambda}{W_{\beta \rho \lambda}\,^{j}} - \frac{27}{128}(\Gamma_{a})_{\rho \lambda} F_{\alpha \beta} {W}^{-1} \nabla^{\beta \gamma}{\nabla^{\rho \lambda}{\lambda_{i \gamma}}} - \frac{81}{256}(\Gamma_{a})_{\rho \lambda} W_{\alpha \beta} \nabla^{\beta \gamma}{\nabla^{\rho \lambda}{\lambda_{i \gamma}}} - \frac{9}{64}(\Gamma_{a})_{\beta \rho} {W}^{-1} \nabla_{\alpha \gamma}{W} \nabla^{\gamma \lambda}{\nabla^{\beta \rho}{\lambda_{i \lambda}}}%
 - \frac{3}{256}{\rm i} (\Gamma_{a})_{\beta \rho} \lambda_{j \alpha} \lambda^{j}_{\lambda} {W}^{-2} \nabla^{\lambda \gamma}{\nabla^{\beta \rho}{\lambda_{i \gamma}}}+\frac{9}{64}(\Gamma_{a})_{\rho \lambda} \lambda_{i \gamma} {W}^{-1} \nabla^{\gamma \beta}{\nabla^{\rho \lambda}{F_{\alpha \beta}}}+\frac{153}{512}(\Gamma_{a})_{\rho \lambda} \lambda_{i \gamma} {W}^{-1} \nabla^{\gamma \beta}{W} \nabla^{\rho \lambda}{W_{\alpha \beta}}+\frac{27}{256}(\Gamma_{a})_{\rho \lambda} \lambda_{i \gamma} \nabla^{\gamma \beta}{\nabla^{\rho \lambda}{W_{\alpha \beta}}}+\frac{15}{128}(\Gamma_{a})_{\beta \rho} \lambda_{j \lambda} {W}^{-1} \nabla_{\alpha}\,^{\lambda}{\nabla^{\beta \rho}{X_{i}\,^{j}}}+\frac{9}{128}(\Gamma_{a})_{\beta \rho} \lambda_{i \lambda} {W}^{-1} \nabla_{\gamma}\,^{\lambda}{\nabla^{\beta \rho}{\nabla_{\alpha}\,^{\gamma}{W}}} - \frac{81}{1280}(\Gamma_{a})_{\alpha}{}^{\lambda} F^{\beta}\,_{\rho} \lambda_{i \gamma} {W}^{-1} \nabla^{\rho \gamma}{W_{\lambda \beta}}+\frac{9}{2560}(\Gamma_{a})_{\alpha}{}^{\beta} \lambda_{i \lambda} {W}^{-1} \nabla_{\gamma}\,^{\rho}{W} \nabla^{\gamma \lambda}{W_{\beta \rho}}+\frac{81}{640}(\Gamma_{a})_{\alpha}{}^{\lambda} W_{\lambda}\,^{\beta} \lambda_{i \gamma} {W}^{-1} \nabla^{\gamma \rho}{F_{\beta \rho}}+\frac{297}{1280}(\Gamma_{a})_{\alpha}{}^{\beta} W_{\beta \rho} \lambda_{i \lambda} {W}^{-1} \nabla_{\gamma}\,^{\lambda}{\nabla^{\gamma \rho}{W}} - \frac{117}{256}(\Gamma_{a})_{\lambda \gamma} W_{\alpha}\,^{\beta} \lambda^{\rho}_{i} {W}^{-1} \nabla^{\lambda \gamma}{F_{\beta \rho}}+\frac{135}{256}(\Gamma_{a})_{\lambda \gamma} W^{\beta \rho} \lambda_{i \alpha} {W}^{-1} \nabla^{\lambda \gamma}{F_{\beta \rho}} - \frac{27}{512}(\Gamma_{a})_{\rho \lambda} \lambda_{j \alpha} X_{i}^{\beta} {W}^{-1} \nabla^{\rho \lambda}{\lambda^{j}_{\beta}} - \frac{27}{128}(\Gamma_{a})^{\rho \gamma} W_{\rho \lambda} F_{\alpha \beta} {W}^{-1} \nabla^{\lambda \beta}{\lambda_{i \gamma}}+\frac{9}{16}(\Gamma_{a})^{\rho \gamma} W_{\alpha \beta} W_{\rho \lambda} \nabla^{\beta \lambda}{\lambda_{i \gamma}}+\frac{9}{32}(\Gamma_{a})^{\beta \lambda} W_{\beta \rho} {W}^{-1} \nabla_{\alpha \gamma}{W} \nabla^{\gamma \rho}{\lambda_{i \lambda}}+\frac{9}{256}{\rm i} (\Gamma_{a})^{\beta \lambda} W_{\beta \rho} \lambda_{j \alpha} \lambda^{j}_{\gamma} {W}^{-2} \nabla^{\rho \gamma}{\lambda_{i \lambda}}+\frac{9}{64}(\Gamma_{a})^{\rho \beta} W_{\rho \lambda} \lambda_{i \gamma} {W}^{-1} \nabla^{\lambda \gamma}{F_{\alpha \beta}}+\frac{9}{64}(\Gamma_{a})^{\beta}{}_{\lambda} W_{\beta \rho} \lambda_{i \gamma} {W}^{-1} \nabla^{\rho \gamma}{\nabla^{\lambda}\,_{\alpha}{W}}+\frac{27}{64}(\Gamma_{a})_{\rho \lambda} F_{\alpha \beta} {W}^{-1} \nabla^{\rho \beta}{\nabla^{\lambda \gamma}{\lambda_{i \gamma}}}%
+\frac{27}{64}(\Gamma_{a})_{\rho \lambda} W_{\alpha \beta} \nabla^{\rho \beta}{\nabla^{\lambda \gamma}{\lambda_{i \gamma}}}+\frac{9}{32}(\Gamma_{a})_{\beta \rho} {W}^{-1} \nabla_{\alpha \gamma}{W} \nabla^{\beta \gamma}{\nabla^{\rho \lambda}{\lambda_{i \lambda}}}+\frac{3}{128}{\rm i} (\Gamma_{a})_{\beta \rho} \lambda_{j \alpha} \lambda^{j}_{\lambda} {W}^{-2} \nabla^{\beta \lambda}{\nabla^{\rho \gamma}{\lambda_{i \gamma}}} - \frac{9}{32}(\Gamma_{a})_{\rho \lambda} \lambda_{i \gamma} {W}^{-1} \nabla^{\rho \gamma}{\nabla^{\lambda \beta}{F_{\alpha \beta}}} - \frac{3}{64}(\Gamma_{a})_{\beta \rho} \lambda_{j \lambda} {W}^{-1} \nabla^{\beta \lambda}{\nabla^{\rho}\,_{\alpha}{X_{i}\,^{j}}}+\frac{9}{64}(\Gamma_{a})_{\beta \rho} \lambda_{i \lambda} {W}^{-1} \nabla^{\beta \lambda}{\nabla^{\rho}\,_{\gamma}{\nabla_{\alpha}\,^{\gamma}{W}}}+\frac{297}{1280}(\Gamma_{a})_{\alpha \lambda} F^{\beta \rho} \lambda_{i \gamma} {W}^{-1} \nabla^{\lambda \gamma}{W_{\beta \rho}} - \frac{27}{64}(\Gamma_{a})^{\rho}{}_{\lambda} F_{\alpha}\,^{\beta} \lambda_{i \gamma} {W}^{-1} \nabla^{\lambda \gamma}{W_{\rho \beta}}+\frac{27}{128}(\Gamma_{a})^{\beta}{}_{\lambda} \lambda_{i \gamma} {W}^{-1} \nabla_{\alpha}\,^{\rho}{W} \nabla^{\lambda \gamma}{W_{\beta \rho}}+\frac{369}{640}(\Gamma_{a})_{\alpha \lambda} W^{\beta \rho} \lambda_{i \gamma} {W}^{-1} \nabla^{\lambda \gamma}{F_{\beta \rho}} - \frac{51}{10240}(\Gamma_{a})_{\alpha \rho} \lambda_{i \lambda} \lambda^{\beta}_{j} {W}^{-1} \nabla^{\rho \lambda}{X^{j}_{\beta}} - \frac{351}{2048}(\Gamma_{a})^{\beta}{}_{\rho} \lambda_{j \alpha} \lambda_{i \lambda} {W}^{-1} \nabla^{\rho \lambda}{X^{j}_{\beta}}+\frac{69}{5120}(\Gamma_{a})_{\alpha \rho} \lambda^{\beta}_{j} \lambda^{j}_{\lambda} {W}^{-1} \nabla^{\rho \lambda}{X_{i \beta}} - \frac{9}{1024}(\Gamma_{a})^{\beta}{}_{\rho} \lambda_{j \alpha} \lambda^{j}_{\lambda} {W}^{-1} \nabla^{\rho \lambda}{X_{i \beta}} - \frac{9}{32}(\Gamma_{a})^{\lambda}{}_{\gamma} W_{\alpha}\,^{\beta} \lambda_{i \lambda} {W}^{-1} \nabla^{\gamma \rho}{F_{\beta \rho}}+\frac{9}{64}(\Gamma_{a})^{\rho}{}_{\lambda} W_{\alpha \beta} \lambda_{i \rho} {W}^{-1} \nabla^{\lambda}\,_{\gamma}{\nabla^{\gamma \beta}{W}} - \frac{207}{640}(\Gamma_{a})_{\alpha \gamma} W^{\lambda \beta} \lambda_{i \lambda} {W}^{-1} \nabla^{\gamma \rho}{F_{\beta \rho}} - \frac{351}{1280}(\Gamma_{a})_{\alpha \lambda} W^{\beta}\,_{\rho} \lambda_{i \beta} {W}^{-1} \nabla^{\lambda}\,_{\gamma}{\nabla^{\gamma \rho}{W}} - \frac{9}{32}(\Gamma_{a})^{\lambda}{}_{\gamma} W_{\lambda}\,^{\beta} \lambda_{i \alpha} {W}^{-1} \nabla^{\gamma \rho}{F_{\beta \rho}}+\frac{9}{64}(\Gamma_{a})^{\beta}{}_{\lambda} W_{\beta \rho} \lambda_{i \alpha} {W}^{-1} \nabla^{\lambda}\,_{\gamma}{\nabla^{\gamma \rho}{W}}%
+\frac{9}{32}(\Gamma_{a})^{\lambda}{}_{\gamma} W_{\alpha \lambda} \lambda^{\beta}_{i} {W}^{-1} \nabla^{\gamma \rho}{F_{\beta \rho}} - \frac{9}{64}(\Gamma_{a})^{\beta}{}_{\rho} W_{\alpha \beta} \lambda_{i \lambda} {W}^{-1} \nabla^{\rho}\,_{\gamma}{\nabla^{\gamma \lambda}{W}} - \frac{3}{64}(\Gamma_{a})^{\beta}{}_{\rho} X_{i j} {W}^{-1} \nabla_{\alpha \lambda}{\nabla^{\rho \lambda}{\lambda^{j}_{\beta}}} - \frac{3}{32}{\rm i} (\Gamma_{a})^{\beta}{}_{\rho} \lambda_{j \alpha} \lambda_{i \lambda} {W}^{-2} \nabla_{\gamma}\,^{\lambda}{\nabla^{\rho \gamma}{\lambda^{j}_{\beta}}} - \frac{3}{80}(\Gamma_{a})_{\alpha}{}^{\beta} \lambda_{i \gamma} W_{\beta}\,^{\rho}\,_{\lambda j} {W}^{-1} \nabla^{\gamma \lambda}{\lambda^{j}_{\rho}} - \frac{3}{16}(\Gamma_{a})^{\beta}{}_{\gamma} \lambda_{i \alpha} W_{\beta}\,^{\rho}\,_{\lambda j} {W}^{-1} \nabla^{\gamma \lambda}{\lambda^{j}_{\rho}}+\frac{3}{128}(\Gamma_{a})_{\beta \rho} X_{i j} {W}^{-1} \nabla_{\alpha}\,^{\lambda}{\nabla^{\beta \rho}{\lambda^{j}_{\lambda}}} - \frac{3}{64}{\rm i} (\Gamma_{a})_{\beta \rho} \lambda_{j \alpha} \lambda_{i \lambda} {W}^{-2} \nabla^{\lambda \gamma}{\nabla^{\beta \rho}{\lambda^{j}_{\gamma}}} - \frac{9}{128}{\rm i} (\Gamma_{a})^{\beta \lambda} W_{\beta \rho} \lambda_{j \alpha} \lambda_{i \gamma} {W}^{-2} \nabla^{\rho \gamma}{\lambda^{j}_{\lambda}}+\frac{3}{32}{\rm i} (\Gamma_{a})_{\beta \rho} \lambda_{j \alpha} \lambda_{i \lambda} {W}^{-2} \nabla^{\beta \lambda}{\nabla^{\rho \gamma}{\lambda^{j}_{\gamma}}}+\frac{7}{128}{\rm i} \Phi^{\beta \lambda}\,_{i j} (\Gamma_{a})_{\beta}{}^{\rho} \lambda^{j}_{\alpha} \lambda_{k \lambda} \lambda^{k}_{\rho} {W}^{-2} - \frac{63}{1024}(\Gamma_{a})^{\rho \lambda} \lambda_{i \rho} \lambda_{j \lambda} {W}^{-1} \nabla_{\alpha}\,^{\beta}{X^{j}_{\beta}} - \frac{9}{64}(\Gamma_{a})^{\beta \rho} F_{\alpha \beta} \lambda_{i \rho} {W}^{-3} \nabla_{\lambda \gamma}{W} \nabla^{\lambda \gamma}{W} - \frac{3}{64}(\Gamma_{a})_{\alpha}{}^{\beta} X_{i j} \lambda^{j}_{\beta} {W}^{-3} \nabla_{\rho \lambda}{W} \nabla^{\rho \lambda}{W} - \frac{9}{128}{\rm i} (\Gamma_{a})^{\beta \rho} \lambda_{j \alpha} \lambda_{i \beta} \lambda^{j}_{\rho} {W}^{-4} \nabla_{\lambda \gamma}{W} \nabla^{\lambda \gamma}{W} - \frac{3}{64}{\rm i} (\Gamma_{a})^{\beta \rho} \lambda_{i \beta} \lambda_{j \rho} {W}^{-3} \nabla_{\lambda \gamma}{W} \nabla^{\lambda \gamma}{\lambda^{j}_{\alpha}}+\frac{9}{64}(\Gamma_{a})^{\beta \rho} F_{\alpha \beta} {W}^{-2} \nabla_{\lambda \gamma}{W} \nabla^{\lambda \gamma}{\lambda_{i \rho}}+\frac{3}{64}{\rm i} (\Gamma_{a})^{\beta \rho} \lambda_{j \alpha} \lambda^{j}_{\beta} {W}^{-3} \nabla_{\lambda \gamma}{W} \nabla^{\lambda \gamma}{\lambda_{i \rho}} - \frac{3}{128}{\rm i} (\Gamma_{a})^{\beta \rho} \lambda_{j \beta} {W}^{-2} \nabla_{\lambda \gamma}{\lambda^{j}_{\alpha}} \nabla^{\lambda \gamma}{\lambda_{i \rho}}+\frac{9}{256}{\rm i} (\Gamma_{a})^{\lambda \gamma} W^{\beta}\,_{\rho} \lambda_{j \lambda} \lambda^{j}_{\beta} {W}^{-2} \nabla_{\alpha}\,^{\rho}{\lambda_{i \gamma}}%
 - \frac{33}{256}{\rm i} (\Gamma_{a})^{\rho \lambda} W_{\alpha \beta} \lambda_{j \rho} \lambda^{j}_{\gamma} {W}^{-2} \nabla^{\beta \gamma}{\lambda_{i \lambda}} - \frac{9}{64}(\Gamma_{a})^{\rho \beta} \lambda_{i \rho} {W}^{-2} \nabla_{\lambda \gamma}{W} \nabla^{\lambda \gamma}{F_{\alpha \beta}} - \frac{81}{512}(\Gamma_{a})^{\rho \beta} \lambda_{i \rho} {W}^{-1} \nabla_{\lambda \gamma}{W} \nabla^{\lambda \gamma}{W_{\alpha \beta}} - \frac{3}{64}(\Gamma_{a})_{\alpha}{}^{\beta} \lambda_{j \beta} {W}^{-2} \nabla_{\rho \lambda}{W} \nabla^{\rho \lambda}{X_{i}\,^{j}} - \frac{9}{512}(\Gamma_{a})^{\lambda \beta} \lambda_{i \lambda} {W}^{-1} \nabla_{\gamma}\,^{\rho}{W} \nabla_{\alpha}\,^{\gamma}{W_{\beta \rho}}+\frac{9}{512}(\Gamma_{a})^{\rho}{}_{\lambda} \lambda_{i \rho} {W}^{-1} \nabla_{\gamma}\,^{\beta}{W} \nabla^{\lambda \gamma}{W_{\alpha \beta}}+\frac{99}{2560}(\Gamma_{a})_{\alpha}{}^{\lambda} \lambda_{i \lambda} {W}^{-1} \nabla_{\gamma}\,^{\beta}{W} \nabla^{\gamma \rho}{W_{\beta \rho}}+\frac{3}{160}(\Gamma_{a})_{\alpha}{}^{\beta} X_{i j} {W}^{-2} \nabla_{\rho \lambda}{W} \nabla^{\rho \lambda}{\lambda^{j}_{\beta}}+\frac{3}{64}{\rm i} (\Gamma_{a})^{\beta \rho} \lambda_{j \alpha} \lambda_{i \beta} {W}^{-3} \nabla_{\lambda \gamma}{W} \nabla^{\lambda \gamma}{\lambda^{j}_{\rho}}+\frac{3}{128}{\rm i} (\Gamma_{a})^{\beta \rho} \lambda_{i \beta} {W}^{-2} \nabla_{\lambda \gamma}{\lambda_{j \alpha}} \nabla^{\lambda \gamma}{\lambda^{j}_{\rho}} - \frac{3}{128}{\rm i} (\Gamma_{a})^{\beta \rho} \lambda_{j \alpha} {W}^{-2} \nabla_{\lambda \gamma}{\lambda_{i \beta}} \nabla^{\lambda \gamma}{\lambda^{j}_{\rho}} - \frac{9}{64}(\Gamma_{a})^{\beta \rho} {W}^{-1} \nabla_{\lambda \gamma}{F_{\alpha \beta}} \nabla^{\lambda \gamma}{\lambda_{i \rho}} - \frac{3}{160}(\Gamma_{a})_{\alpha}{}^{\beta} {W}^{-1} \nabla_{\rho \lambda}{X_{i j}} \nabla^{\rho \lambda}{\lambda^{j}_{\beta}}+\frac{27}{256}(\Gamma_{a})^{\beta \gamma} W_{\beta}\,^{\rho} W_{\rho \lambda} \nabla_{\alpha}\,^{\lambda}{\lambda_{i \gamma}} - \frac{27}{256}(\Gamma_{a})^{\lambda}{}_{\gamma} W_{\beta \rho} {W}^{-1} \nabla^{\gamma \beta}{W} \nabla_{\alpha}\,^{\rho}{\lambda_{i \lambda}} - \frac{45}{128}(\Gamma_{a})^{\beta \gamma} W_{\alpha \lambda} F_{\beta \rho} {W}^{-1} \nabla^{\lambda \rho}{\lambda_{i \gamma}}+\frac{9}{64}(\Gamma_{a})^{\rho}{}_{\lambda} W_{\alpha \beta} {W}^{-1} \nabla^{\lambda}\,_{\gamma}{W} \nabla^{\gamma \beta}{\lambda_{i \rho}}+\frac{9}{128}(\Gamma_{a})^{\beta \lambda} \nabla_{\alpha \gamma}{W_{\beta \rho}} \nabla^{\gamma \rho}{\lambda_{i \lambda}}+\frac{63}{512}(\Gamma_{a})^{\lambda}{}_{\gamma} \nabla^{\gamma \beta}{W_{\beta \rho}} \nabla_{\alpha}\,^{\rho}{\lambda_{i \lambda}}+\frac{9}{128}(\Gamma_{a})^{\rho}{}_{\lambda} \nabla^{\lambda}\,_{\gamma}{W_{\alpha \beta}} \nabla^{\gamma \beta}{\lambda_{i \rho}}%
+\frac{81}{512}(\Gamma_{a})^{\beta \lambda} \nabla_{\gamma}\,^{\rho}{W_{\beta \rho}} \nabla_{\alpha}\,^{\gamma}{\lambda_{i \lambda}} - \frac{81}{640}(\Gamma_{a})_{\alpha}{}^{\lambda} \nabla_{\gamma}\,^{\beta}{W_{\beta \rho}} \nabla^{\gamma \rho}{\lambda_{i \lambda}} - \frac{3}{128}{\rm i} (\Gamma_{a})^{\beta}{}_{\rho} \lambda^{\lambda}_{i} {W}^{-2} \nabla^{\rho}\,_{\gamma}{\lambda_{j \lambda}} \nabla_{\alpha}\,^{\gamma}{\lambda^{j}_{\beta}}+\frac{3}{128}{\rm i} (\Gamma_{a})_{\beta \rho} \lambda^{\lambda}_{i} {W}^{-2} \nabla^{\beta}\,_{\alpha}{\lambda_{j \gamma}} \nabla^{\rho \gamma}{\lambda^{j}_{\lambda}} - \frac{51}{640}{\rm i} (\Gamma_{a})_{\alpha \beta} \lambda^{\rho}_{i} {W}^{-2} \nabla^{\beta}\,_{\gamma}{\lambda_{j \rho}} \nabla^{\gamma \lambda}{\lambda^{j}_{\lambda}}+\frac{3}{256}{\rm i} (\Gamma_{a})^{\beta}{}_{\lambda} W_{\beta}\,^{\rho} \lambda^{\gamma}_{i} \lambda_{j \rho} {W}^{-2} \nabla^{\lambda}\,_{\alpha}{\lambda^{j}_{\gamma}}+\frac{9}{160}{\rm i} (\Gamma_{a})_{\alpha \lambda} W^{\beta}\,_{\rho} \lambda^{\gamma}_{i} \lambda_{j \beta} {W}^{-2} \nabla^{\lambda \rho}{\lambda^{j}_{\gamma}} - \frac{9}{16}(\Gamma_{a})^{\rho}{}_{\gamma} F_{\alpha}\,^{\beta} F_{\rho \lambda} {W}^{-2} \nabla^{\gamma \lambda}{\lambda_{i \beta}}+\frac{9}{64}(\Gamma_{a})^{\beta}{}_{\lambda} X_{i j} F_{\beta \rho} {W}^{-2} \nabla^{\lambda \rho}{\lambda^{j}_{\alpha}} - \frac{9}{32}(\Gamma_{a})^{\beta}{}_{\lambda} F_{\beta \rho} {W}^{-2} \nabla_{\alpha}\,^{\gamma}{W} \nabla^{\lambda \rho}{\lambda_{i \gamma}} - \frac{3}{16}{\rm i} (\Gamma_{a})^{\beta}{}_{\lambda} F_{\beta \rho} \lambda_{j \alpha} \lambda^{\gamma}_{i} {W}^{-3} \nabla^{\lambda \rho}{\lambda^{j}_{\gamma}}+\frac{9}{32}(\Gamma_{a})^{\beta}{}_{\gamma} F_{\beta \rho} \lambda^{\lambda}_{i} {W}^{-2} \nabla^{\gamma \rho}{F_{\alpha \lambda}} - \frac{9}{64}(\Gamma_{a})^{\beta}{}_{\lambda} F_{\beta \rho} \lambda_{i \gamma} {W}^{-2} \nabla^{\lambda \rho}{\nabla_{\alpha}\,^{\gamma}{W}} - \frac{135}{64}(\Gamma_{a})^{\beta \lambda} W_{\alpha}\,^{\rho} F_{\beta \rho} F_{\lambda}\,^{\gamma} \lambda_{i \gamma} {W}^{-2}+\frac{27}{32}(\Gamma_{a})^{\beta}{}_{\lambda} W_{\alpha}\,^{\rho} F_{\beta \rho} \lambda_{i \gamma} {W}^{-2} \nabla^{\lambda \gamma}{W}+\frac{45}{32}(\Gamma_{a})^{\gamma \beta} W_{\alpha \gamma} F_{\beta}\,^{\rho} F_{\rho}\,^{\lambda} \lambda_{i \lambda} {W}^{-2}+\frac{57}{128}(\Gamma_{a})^{\beta \lambda} F_{\beta}\,^{\rho} \lambda_{i \alpha} \lambda_{j \rho} X^{j}_{\lambda} {W}^{-2} - \frac{99}{320}(\Gamma_{a})_{\alpha}{}^{\beta} F_{\beta}\,^{\rho} \lambda^{\lambda}_{i} \lambda_{j \lambda} X^{j}_{\rho} {W}^{-2}+\frac{9}{32}(\Gamma_{a})_{\rho \lambda} F_{\alpha}\,^{\beta} {W}^{-2} \nabla^{\rho}\,_{\gamma}{W} \nabla^{\lambda \gamma}{\lambda_{i \beta}} - \frac{3}{32}(\Gamma_{a})_{\beta \rho} X_{i j} {W}^{-2} \nabla^{\beta}\,_{\lambda}{W} \nabla^{\rho \lambda}{\lambda^{j}_{\alpha}}%
+\frac{9}{64}(\Gamma_{a})_{\beta \rho} {W}^{-2} \nabla^{\beta}\,_{\gamma}{W} \nabla_{\alpha}\,^{\lambda}{W} \nabla^{\rho \gamma}{\lambda_{i \lambda}}+\frac{3}{32}{\rm i} (\Gamma_{a})_{\beta \rho} \lambda_{j \alpha} \lambda^{\lambda}_{i} {W}^{-3} \nabla^{\beta}\,_{\gamma}{W} \nabla^{\rho \gamma}{\lambda^{j}_{\lambda}}+\frac{3}{128}{\rm i} (\Gamma_{a})_{\beta \rho} \lambda^{\lambda}_{i} {W}^{-2} \nabla^{\beta}\,_{\gamma}{\lambda_{j \alpha}} \nabla^{\rho \gamma}{\lambda^{j}_{\lambda}} - \frac{21}{512}{\rm i} (\Gamma_{a})_{\rho \lambda} W_{\alpha}\,^{\beta} \lambda^{\gamma}_{i} \lambda_{j \beta} {W}^{-2} \nabla^{\rho \lambda}{\lambda^{j}_{\gamma}} - \frac{9}{64}(\Gamma_{a})_{\rho \lambda} \lambda^{\beta}_{i} {W}^{-2} \nabla^{\rho}\,_{\gamma}{W} \nabla^{\lambda \gamma}{F_{\alpha \beta}}+\frac{9}{128}(\Gamma_{a})_{\beta \rho} \lambda_{i \lambda} {W}^{-2} \nabla^{\beta}\,_{\gamma}{W} \nabla^{\rho \gamma}{\nabla_{\alpha}\,^{\lambda}{W}} - \frac{99}{256}(\Gamma_{a})_{\lambda \gamma} W_{\alpha}\,^{\beta} F_{\beta}\,^{\rho} \lambda_{i \rho} {W}^{-2} \nabla^{\lambda \gamma}{W}+\frac{189}{512}(\Gamma_{a})_{\rho \lambda} W_{\alpha \beta} \lambda_{i \gamma} {W}^{-2} \nabla^{\rho \lambda}{W} \nabla^{\beta \gamma}{W} - \frac{27}{128}(\Gamma_{a})_{\alpha \gamma} W^{\beta}\,_{\lambda} F_{\beta}\,^{\rho} \lambda_{i \rho} {W}^{-2} \nabla^{\gamma \lambda}{W}+\frac{81}{320}(\Gamma_{a})_{\alpha \lambda} W_{\beta \rho} \lambda_{i \gamma} {W}^{-2} \nabla^{\lambda \beta}{W} \nabla^{\rho \gamma}{W} - \frac{45}{128}(\Gamma_{a})^{\beta}{}_{\gamma} W_{\alpha \lambda} F_{\beta}\,^{\rho} \lambda_{i \rho} {W}^{-2} \nabla^{\gamma \lambda}{W} - \frac{27}{64}(\Gamma_{a})_{\rho \lambda} W_{\alpha \beta} \lambda_{i \gamma} {W}^{-2} \nabla^{\rho \beta}{W} \nabla^{\lambda \gamma}{W} - \frac{9}{16}(\Gamma_{a})^{\lambda}{}_{\gamma} W_{\alpha \lambda} F^{\beta}\,_{\rho} \lambda_{i \beta} {W}^{-2} \nabla^{\gamma \rho}{W}+\frac{117}{512}(\Gamma_{a})^{\beta}{}_{\rho} \lambda_{i \alpha} \lambda_{j \lambda} X^{j}_{\beta} {W}^{-2} \nabla^{\rho \lambda}{W}+\frac{51}{256}(\Gamma_{a})_{\beta \rho} \lambda^{\lambda}_{i} \lambda_{j \lambda} X^{j}_{\alpha} {W}^{-2} \nabla^{\beta \rho}{W}+\frac{15}{128}(\Gamma_{a})_{\alpha \rho} \lambda^{\lambda}_{i} \lambda_{j \lambda} X^{j}_{\beta} {W}^{-2} \nabla^{\rho \beta}{W}+\frac{63}{128}{\rm i} (\Gamma_{a})^{\beta}{}_{\rho} W_{\alpha \beta} \lambda^{\lambda}_{j} \lambda^{j}_{\gamma} {W}^{-2} \nabla^{\rho \gamma}{\lambda_{i \lambda}} - \frac{9}{128}{\rm i} (\Gamma_{a})_{\beta \rho} \lambda^{\lambda}_{j} {W}^{-2} \nabla^{\beta}\,_{\alpha}{\lambda_{i \lambda}} \nabla^{\rho \gamma}{\lambda^{j}_{\gamma}} - \frac{9}{160}{\rm i} (\Gamma_{a})_{\alpha \beta} \lambda^{\rho}_{j} {W}^{-2} \nabla^{\beta}\,_{\gamma}{\lambda_{i \rho}} \nabla^{\gamma \lambda}{\lambda^{j}_{\lambda}} - \frac{9}{320}{\rm i} (\Gamma_{a})_{\alpha \lambda} W^{\beta}\,_{\rho} \lambda_{j \beta} \lambda^{j \gamma} {W}^{-2} \nabla^{\lambda \rho}{\lambda_{i \gamma}}%
 - \frac{3}{16}{\rm i} (\Gamma_{a})^{\beta}{}_{\lambda} F_{\beta \rho} \lambda_{j \alpha} \lambda^{j \gamma} {W}^{-3} \nabla^{\lambda \rho}{\lambda_{i \gamma}}+\frac{9}{64}(\Gamma_{a})^{\lambda \beta} X_{i j} W_{\alpha \lambda} F_{\beta}\,^{\rho} \lambda^{j}_{\rho} {W}^{-2}+\frac{3}{32}{\rm i} (\Gamma_{a})_{\beta \rho} \lambda_{j \alpha} \lambda^{j \lambda} {W}^{-3} \nabla^{\beta}\,_{\gamma}{W} \nabla^{\rho \gamma}{\lambda_{i \lambda}} - \frac{33}{256}{\rm i} (\Gamma_{a})_{\rho \lambda} W_{\alpha}\,^{\beta} \lambda_{j \beta} \lambda^{j \gamma} {W}^{-2} \nabla^{\rho \lambda}{\lambda_{i \gamma}} - \frac{39}{256}{\rm i} (\Gamma_{a})^{\rho}{}_{\lambda} W_{\alpha \beta} \lambda_{j \rho} \lambda^{j \gamma} {W}^{-2} \nabla^{\lambda \beta}{\lambda_{i \gamma}} - \frac{27}{64}(\Gamma_{a})_{\rho \lambda} X_{i j} W_{\alpha}\,^{\beta} \lambda^{j}_{\beta} {W}^{-2} \nabla^{\rho \lambda}{W} - \frac{153}{320}(\Gamma_{a})_{\alpha \lambda} X_{i j} W^{\beta}\,_{\rho} \lambda^{j}_{\beta} {W}^{-2} \nabla^{\lambda \rho}{W}+\frac{45}{128}(\Gamma_{a})^{\beta}{}_{\rho} X_{i j} W_{\alpha \beta} \lambda^{j}_{\lambda} {W}^{-2} \nabla^{\rho \lambda}{W}+\frac{3}{128}{\rm i} (\Gamma_{a})_{\beta \rho} \lambda_{i \lambda} {W}^{-2} \nabla^{\beta}\,_{\alpha}{\lambda^{\gamma}_{j}} \nabla^{\rho \lambda}{\lambda^{j}_{\gamma}}+\frac{3}{128}{\rm i} (\Gamma_{a})^{\beta}{}_{\rho} \lambda_{i \lambda} {W}^{-2} \nabla^{\rho \lambda}{\lambda_{j \gamma}} \nabla_{\alpha}\,^{\gamma}{\lambda^{j}_{\beta}} - \frac{33}{640}{\rm i} (\Gamma_{a})_{\alpha \beta} \lambda_{i \rho} {W}^{-2} \nabla^{\beta \rho}{\lambda_{j \lambda}} \nabla^{\lambda \gamma}{\lambda^{j}_{\gamma}} - \frac{3}{64}{\rm i} (\Gamma_{a})_{\beta \rho} \lambda_{i \lambda} {W}^{-2} \nabla^{\beta \lambda}{\lambda_{j \alpha}} \nabla^{\rho \gamma}{\lambda^{j}_{\gamma}}+\frac{189}{1280}{\rm i} (\Gamma_{a})_{\alpha \lambda} W^{\beta \rho} \lambda_{i \gamma} \lambda_{j \beta} {W}^{-2} \nabla^{\lambda \gamma}{\lambda^{j}_{\rho}} - \frac{3}{256}{\rm i} (\Gamma_{a})^{\beta}{}_{\lambda} W_{\beta}\,^{\rho} \lambda_{i \gamma} \lambda_{j \rho} {W}^{-2} \nabla^{\lambda \gamma}{\lambda^{j}_{\alpha}} - \frac{9}{16}(\Gamma_{a})^{\rho}{}_{\gamma} F_{\alpha \beta} F_{\rho}\,^{\lambda} {W}^{-2} \nabla^{\gamma \beta}{\lambda_{i \lambda}}+\frac{3}{64}(\Gamma_{a})^{\beta}{}_{\lambda} X_{i j} F_{\beta}\,^{\rho} {W}^{-2} \nabla^{\lambda}\,_{\alpha}{\lambda^{j}_{\rho}} - \frac{9}{32}(\Gamma_{a})^{\beta}{}_{\lambda} F_{\beta}\,^{\rho} {W}^{-2} \nabla_{\alpha \gamma}{W} \nabla^{\lambda \gamma}{\lambda_{i \rho}} - \frac{3}{16}{\rm i} (\Gamma_{a})^{\beta}{}_{\lambda} F_{\beta}\,^{\rho} \lambda_{j \alpha} \lambda_{i \gamma} {W}^{-3} \nabla^{\lambda \gamma}{\lambda^{j}_{\rho}}+\frac{9}{32}(\Gamma_{a})^{\beta}{}_{\lambda} F_{\beta}\,^{\rho} \lambda_{i \gamma} {W}^{-2} \nabla^{\lambda \gamma}{F_{\alpha \rho}} - \frac{9}{64}(\Gamma_{a})^{\beta}{}_{\lambda} F_{\beta \rho} \lambda_{i \gamma} {W}^{-2} \nabla^{\lambda \gamma}{\nabla_{\alpha}\,^{\rho}{W}}%
 - \frac{243}{160}(\Gamma_{a})_{\alpha}{}^{\beta} W^{\lambda \gamma} F_{\beta}\,^{\rho} F_{\lambda \rho} \lambda_{i \gamma} {W}^{-2} - \frac{27}{32}(\Gamma_{a})^{\gamma \beta} W_{\gamma}\,^{\lambda} F_{\beta}\,^{\rho} F_{\lambda \rho} \lambda_{i \alpha} {W}^{-2} - \frac{27}{64}(\Gamma_{a})^{\lambda \beta} W_{\lambda \gamma} F_{\beta \rho} \lambda_{i \alpha} {W}^{-2} \nabla^{\gamma \rho}{W}+\frac{27}{32}(\Gamma_{a})^{\beta \lambda} W_{\alpha}\,^{\gamma} F_{\beta}\,^{\rho} F_{\lambda \rho} \lambda_{i \gamma} {W}^{-2}+\frac{3}{8}(\Gamma_{a})^{\beta \lambda} F_{\beta}\,^{\rho} \lambda^{\gamma}_{i} \lambda_{j \gamma} W_{\alpha \lambda \rho}\,^{j} {W}^{-2} - \frac{129}{256}(\Gamma_{a})_{\alpha}{}^{\beta} F_{\beta}\,^{\rho} \lambda_{i \rho} \lambda^{\lambda}_{j} X^{j}_{\lambda} {W}^{-2}+\frac{477}{1280}(\Gamma_{a})_{\alpha}{}^{\beta} F_{\beta}\,^{\rho} \lambda^{\lambda}_{i} \lambda_{j \rho} X^{j}_{\lambda} {W}^{-2}+\frac{33}{128}{\rm i} (\Gamma_{a})^{\beta}{}_{\rho} \lambda_{j \lambda} {W}^{-2} \nabla^{\rho \lambda}{\lambda_{i \beta}} \nabla_{\alpha}\,^{\gamma}{\lambda^{j}_{\gamma}}+\frac{3}{64}{\rm i} (\Gamma_{a})^{\beta}{}_{\rho} \lambda_{j \lambda} {W}^{-2} \nabla^{\rho \lambda}{\lambda^{j}_{\beta}} \nabla_{\alpha}\,^{\gamma}{\lambda_{i \gamma}}+\frac{39}{128}{\rm i} (\Gamma_{a})^{\rho}{}_{\lambda} W_{\alpha}\,^{\beta} \lambda_{j \beta} \lambda^{j}_{\gamma} {W}^{-2} \nabla^{\lambda \gamma}{\lambda_{i \rho}}+\frac{3}{64}(\Gamma_{a})^{\beta}{}_{\rho} X_{i j} X^{j}\,_{k} {W}^{-2} \nabla^{\rho}\,_{\alpha}{\lambda^{k}_{\beta}}+\frac{3}{64}(\Gamma_{a})^{\beta}{}_{\rho} X_{i j} {W}^{-2} \nabla_{\alpha \lambda}{W} \nabla^{\rho \lambda}{\lambda^{j}_{\beta}}+\frac{3}{32}{\rm i} (\Gamma_{a})^{\beta}{}_{\rho} X_{i j} \lambda_{k \alpha} \lambda^{j}_{\lambda} {W}^{-3} \nabla^{\rho \lambda}{\lambda^{k}_{\beta}} - \frac{9}{64}(\Gamma_{a})^{\beta}{}_{\rho} X_{i j} \lambda^{j}_{\lambda} {W}^{-2} \nabla^{\rho \lambda}{F_{\alpha \beta}}+\frac{9}{128}(\Gamma_{a})_{\beta \rho} X_{i j} \lambda^{j}_{\lambda} {W}^{-2} \nabla^{\beta \lambda}{\nabla^{\rho}\,_{\alpha}{W}} - \frac{9}{64}(\Gamma_{a})^{\lambda \beta} X_{i j} W_{\lambda}\,^{\rho} F_{\beta \rho} \lambda^{j}_{\alpha} {W}^{-2} - \frac{27}{64}(\Gamma_{a})^{\beta}{}_{\lambda} X_{i j} W_{\beta \rho} \lambda^{j}_{\alpha} {W}^{-2} \nabla^{\lambda \rho}{W}+\frac{9}{32}(\Gamma_{a})_{\rho \lambda} F_{\alpha \beta} {W}^{-2} \nabla^{\rho \gamma}{W} \nabla^{\lambda \beta}{\lambda_{i \gamma}}+\frac{9}{64}(\Gamma_{a})_{\beta \rho} {W}^{-2} \nabla^{\beta \lambda}{W} \nabla_{\alpha \gamma}{W} \nabla^{\rho \gamma}{\lambda_{i \lambda}}+\frac{3}{32}{\rm i} (\Gamma_{a})_{\beta \rho} \lambda_{j \alpha} \lambda_{i \lambda} {W}^{-3} \nabla^{\beta \gamma}{W} \nabla^{\rho \lambda}{\lambda^{j}_{\gamma}}%
+\frac{3}{128}{\rm i} (\Gamma_{a})_{\beta \rho} \lambda_{i \lambda} {W}^{-2} \nabla^{\beta \gamma}{\lambda_{j \alpha}} \nabla^{\rho \lambda}{\lambda^{j}_{\gamma}} - \frac{3}{256}{\rm i} (\Gamma_{a})^{\rho}{}_{\lambda} W_{\alpha}\,^{\beta} \lambda_{i \gamma} \lambda_{j \beta} {W}^{-2} \nabla^{\lambda \gamma}{\lambda^{j}_{\rho}} - \frac{9}{64}(\Gamma_{a})_{\rho \lambda} \lambda_{i \gamma} {W}^{-2} \nabla^{\rho \beta}{W} \nabla^{\lambda \gamma}{F_{\alpha \beta}}+\frac{9}{128}(\Gamma_{a})_{\beta \rho} \lambda_{i \lambda} {W}^{-2} \nabla^{\beta}\,_{\gamma}{W} \nabla^{\rho \lambda}{\nabla_{\alpha}\,^{\gamma}{W}}+\frac{27}{40}(\Gamma_{a})_{\alpha \gamma} W^{\beta \lambda} F_{\beta \rho} \lambda_{i \lambda} {W}^{-2} \nabla^{\gamma \rho}{W} - \frac{153}{640}(\Gamma_{a})_{\alpha \lambda} W^{\beta}\,_{\rho} \lambda_{i \beta} {W}^{-2} \nabla^{\lambda}\,_{\gamma}{W} \nabla^{\gamma \rho}{W}+\frac{27}{64}(\Gamma_{a})^{\lambda}{}_{\gamma} W_{\lambda}\,^{\beta} F_{\beta \rho} \lambda_{i \alpha} {W}^{-2} \nabla^{\gamma \rho}{W} - \frac{27}{128}(\Gamma_{a})^{\beta}{}_{\lambda} W_{\beta \rho} \lambda_{i \alpha} {W}^{-2} \nabla^{\lambda}\,_{\gamma}{W} \nabla^{\gamma \rho}{W} - \frac{3}{16}(\Gamma_{a})^{\beta}{}_{\lambda} \lambda^{\gamma}_{i} \lambda_{j \gamma} W_{\alpha \beta \rho}\,^{j} {W}^{-2} \nabla^{\lambda \rho}{W}+\frac{27}{128}(\Gamma_{a})_{\alpha \rho} \lambda_{i \lambda} \lambda^{\beta}_{j} X^{j}_{\beta} {W}^{-2} \nabla^{\rho \lambda}{W}+\frac{135}{512}(\Gamma_{a})_{\rho \lambda} \lambda_{i \alpha} \lambda^{\beta}_{j} X^{j}_{\beta} {W}^{-2} \nabla^{\rho \lambda}{W} - \frac{81}{320}(\Gamma_{a})_{\alpha \rho} \lambda^{\beta}_{i} \lambda_{j \lambda} X^{j}_{\beta} {W}^{-2} \nabla^{\rho \lambda}{W}+\frac{9}{128}{\rm i} (\Gamma_{a})_{\beta \rho} \lambda_{j \lambda} {W}^{-2} \nabla^{\beta \lambda}{\lambda_{i \alpha}} \nabla^{\rho \gamma}{\lambda^{j}_{\gamma}} - \frac{117}{1280}{\rm i} (\Gamma_{a})_{\alpha \lambda} W^{\beta \rho} \lambda_{j \beta} \lambda^{j}_{\gamma} {W}^{-2} \nabla^{\lambda \gamma}{\lambda_{i \rho}} - \frac{3}{16}{\rm i} (\Gamma_{a})^{\beta}{}_{\lambda} F_{\beta}\,^{\rho} \lambda_{j \alpha} \lambda^{j}_{\gamma} {W}^{-3} \nabla^{\lambda \gamma}{\lambda_{i \rho}}+\frac{15}{256}(\Gamma_{a})^{\beta}{}_{\rho} X_{j k} X^{j k} {W}^{-2} \nabla^{\rho}\,_{\alpha}{\lambda_{i \beta}} - \frac{9}{64}{\rm i} (\Gamma_{a})^{\beta}{}_{\rho} X_{j k} \lambda^{j}_{\alpha} \lambda^{k}_{\lambda} {W}^{-3} \nabla^{\rho \lambda}{\lambda_{i \beta}}+\frac{3}{160}(\Gamma_{a})_{\alpha \beta} X_{j k} \lambda^{j}_{\rho} {W}^{-2} \nabla^{\beta \rho}{X_{i}\,^{k}}+\frac{9}{64}(\Gamma_{a})^{\lambda \beta} X_{k j} \lambda^{\rho}_{i} \lambda^{k}_{\lambda} W_{\alpha \beta \rho}\,^{j} {W}^{-2}+\frac{9}{320}(\Gamma_{a})_{\alpha}{}^{\beta} X_{k j} \lambda^{\rho}_{i} \lambda^{k \lambda} W_{\beta \rho \lambda}\,^{j} {W}^{-2}%
+\frac{15}{512}(\Gamma_{a})_{\alpha}{}^{\beta} X_{j k} \lambda^{\rho}_{i} \lambda^{j}_{\rho} X^{k}_{\beta} {W}^{-2}+\frac{3}{32}{\rm i} (\Gamma_{a})_{\beta \rho} \lambda_{j \alpha} \lambda^{j}_{\lambda} {W}^{-3} \nabla^{\beta \gamma}{W} \nabla^{\rho \lambda}{\lambda_{i \gamma}}+\frac{9}{640}(\Gamma_{a})_{\alpha \rho} \lambda^{\beta}_{j} \lambda^{j}_{\lambda} X_{i \beta} {W}^{-2} \nabla^{\rho \lambda}{W}+\frac{9}{16}{\rm i} (\Gamma_{a})^{\beta}{}_{\rho} F_{\alpha \beta} \lambda^{\lambda}_{j} \lambda^{j}_{\gamma} {W}^{-3} \nabla^{\rho \gamma}{\lambda_{i \lambda}}+\frac{3}{80}{\rm i} (\Gamma_{a})_{\alpha \beta} X_{i j} \lambda^{\rho}_{k} \lambda^{k}_{\lambda} {W}^{-3} \nabla^{\beta \lambda}{\lambda^{j}_{\rho}}+\frac{3}{16}{\rm i} (\Gamma_{a})^{\rho}{}_{\lambda} F_{\alpha \beta} \lambda_{i \rho} \lambda^{\gamma}_{j} {W}^{-3} \nabla^{\lambda \beta}{\lambda^{j}_{\gamma}}+\frac{3}{128}{\rm i} (\Gamma_{a})^{\beta}{}_{\rho} X_{j k} \lambda_{i \beta} \lambda^{j \lambda} {W}^{-3} \nabla^{\rho}\,_{\alpha}{\lambda^{k}_{\lambda}}+\frac{3}{32}{\rm i} (\Gamma_{a})^{\beta}{}_{\rho} \lambda_{i \beta} \lambda^{\lambda}_{j} {W}^{-3} \nabla_{\alpha \gamma}{W} \nabla^{\rho \gamma}{\lambda^{j}_{\lambda}}+\frac{3}{16}{\rm i} (\Gamma_{a})^{\rho}{}_{\lambda} F_{\alpha}\,^{\beta} \lambda_{i \rho} \lambda_{j \gamma} {W}^{-3} \nabla^{\lambda \gamma}{\lambda^{j}_{\beta}}+\frac{9}{128}{\rm i} (\Gamma_{a})^{\beta}{}_{\rho} X_{j k} \lambda_{i \beta} \lambda^{j}_{\lambda} {W}^{-3} \nabla^{\rho \lambda}{\lambda^{k}_{\alpha}}+\frac{3}{32}{\rm i} (\Gamma_{a})^{\beta}{}_{\rho} \lambda_{i \beta} \lambda_{j \lambda} {W}^{-3} \nabla_{\alpha}\,^{\gamma}{W} \nabla^{\rho \lambda}{\lambda^{j}_{\gamma}} - \frac{9}{32}(\Gamma_{a})^{\beta}{}_{\rho} \lambda_{j \alpha} \lambda_{i \beta} \lambda^{\lambda}_{k} \lambda^{k}_{\gamma} {W}^{-4} \nabla^{\rho \gamma}{\lambda^{j}_{\lambda}} - \frac{9}{32}{\rm i} (\Gamma_{a})^{\rho}{}_{\lambda} \lambda_{i \rho} \lambda^{\beta}_{j} \lambda^{j}_{\gamma} {W}^{-3} \nabla^{\lambda \gamma}{F_{\alpha \beta}} - \frac{171}{256}{\rm i} (\Gamma_{a})^{\rho}{}_{\lambda} W_{\alpha}\,^{\beta} \lambda_{i \rho} \lambda_{j \beta} \lambda^{j}_{\gamma} {W}^{-3} \nabla^{\lambda \gamma}{W}+\frac{9}{64}{\rm i} (\Gamma_{a})^{\beta}{}_{\rho} \lambda_{i \beta} \lambda_{j \lambda} \lambda^{j}_{\gamma} {W}^{-3} \nabla^{\rho \lambda}{\nabla_{\alpha}\,^{\gamma}{W}} - \frac{273}{128}{\rm i} (\Gamma_{a})^{\beta \gamma} W_{\alpha}\,^{\lambda} F_{\beta}\,^{\rho} \lambda_{i \gamma} \lambda_{j \lambda} \lambda^{j}_{\rho} {W}^{-3}+\frac{99}{64}{\rm i} (\Gamma_{a})^{\lambda \gamma} W_{\alpha \lambda} F^{\beta \rho} \lambda_{i \gamma} \lambda_{j \beta} \lambda^{j}_{\rho} {W}^{-3} - \frac{9}{512}{\rm i} (\Gamma_{a})_{\alpha}{}^{\rho} \lambda_{i \rho} \lambda^{\lambda}_{k} \lambda^{k \beta} \lambda_{j \lambda} X^{j}_{\beta} {W}^{-3} - \frac{3}{16}{\rm i} (\Gamma_{a})^{\rho}{}_{\lambda} F_{\alpha \beta} \lambda_{j \rho} \lambda^{j \gamma} {W}^{-3} \nabla^{\lambda \beta}{\lambda_{i \gamma}}+\frac{3}{128}{\rm i} (\Gamma_{a})^{\beta}{}_{\rho} X_{j k} \lambda^{j}_{\beta} \lambda^{k \lambda} {W}^{-3} \nabla^{\rho}\,_{\alpha}{\lambda_{i \lambda}}%
 - \frac{3}{32}{\rm i} (\Gamma_{a})^{\beta}{}_{\rho} \lambda_{j \beta} \lambda^{j \lambda} {W}^{-3} \nabla_{\alpha \gamma}{W} \nabla^{\rho \gamma}{\lambda_{i \lambda}} - \frac{3}{16}{\rm i} (\Gamma_{a})^{\rho}{}_{\lambda} F_{\alpha}\,^{\beta} \lambda_{j \rho} \lambda^{j}_{\gamma} {W}^{-3} \nabla^{\lambda \gamma}{\lambda_{i \beta}} - \frac{3}{128}{\rm i} (\Gamma_{a})^{\beta}{}_{\rho} X_{j k} \lambda^{j}_{\beta} \lambda^{k}_{\lambda} {W}^{-3} \nabla^{\rho \lambda}{\lambda_{i \alpha}} - \frac{3}{32}{\rm i} (\Gamma_{a})^{\beta}{}_{\rho} \lambda_{j \beta} \lambda^{j}_{\lambda} {W}^{-3} \nabla_{\alpha}\,^{\gamma}{W} \nabla^{\rho \lambda}{\lambda_{i \gamma}} - \frac{9}{32}(\Gamma_{a})^{\beta}{}_{\rho} \lambda_{j \alpha} \lambda^{j}_{\beta} \lambda^{\lambda}_{k} \lambda^{k}_{\gamma} {W}^{-4} \nabla^{\rho \gamma}{\lambda_{i \lambda}}+\frac{3}{32}{\rm i} (\Gamma_{a})^{\beta}{}_{\rho} \lambda_{k \alpha} \lambda^{k}_{\lambda} \lambda_{j \beta} {W}^{-3} \nabla^{\rho \lambda}{X_{i}\,^{j}}+\frac{93}{256}{\rm i} (\Gamma_{a})^{\rho \lambda} X_{i j} W_{\alpha}\,^{\beta} \lambda^{j}_{\rho} \lambda_{k \lambda} \lambda^{k}_{\beta} {W}^{-3}+\frac{237}{1280}{\rm i} (\Gamma_{a})_{\alpha}{}^{\lambda} X_{i j} W^{\beta \rho} \lambda^{j}_{\lambda} \lambda_{k \beta} \lambda^{k}_{\rho} {W}^{-3} - \frac{9}{16}{\rm i} (\Gamma_{a})^{\beta \lambda} X_{i j} W_{\beta}\,^{\rho} \lambda_{k \alpha} \lambda^{j}_{\lambda} \lambda^{k}_{\rho} {W}^{-3} - \frac{3}{32}{\rm i} (\Gamma_{a})^{\lambda \gamma} \lambda^{\beta}_{i} \lambda_{k \lambda} \lambda^{k \rho} \lambda_{j \gamma} W_{\alpha \beta \rho}\,^{j} {W}^{-3} - \frac{3}{80}{\rm i} (\Gamma_{a})_{\alpha}{}^{\gamma} \lambda^{\beta}_{i} \lambda_{j \gamma} \lambda^{\rho}_{k} \lambda^{k \lambda} W_{\beta \rho \lambda}\,^{j} {W}^{-3}+\frac{3}{32}{\rm i} (\Gamma_{a})^{\gamma \beta} \lambda_{k \alpha} \lambda^{\rho}_{i} \lambda^{k \lambda} \lambda_{j \gamma} W_{\beta \rho \lambda}\,^{j} {W}^{-3}+\frac{33}{512}{\rm i} (\Gamma_{a})^{\rho \beta} \lambda_{k \alpha} \lambda^{\lambda}_{i} \lambda^{k}_{\lambda} \lambda_{j \rho} X^{j}_{\beta} {W}^{-3} - \frac{123}{2560}{\rm i} (\Gamma_{a})_{\alpha}{}^{\rho} \lambda^{\lambda}_{i} \lambda_{j \rho} \lambda_{k \lambda} \lambda^{k \beta} X^{j}_{\beta} {W}^{-3} - \frac{39}{512}{\rm i} (\Gamma_{a})^{\beta \rho} \lambda^{\lambda}_{i} \lambda_{k \beta} \lambda^{k}_{\lambda} \lambda_{j \rho} X^{j}_{\alpha} {W}^{-3} - \frac{3}{80}{\rm i} (\Gamma_{a})_{\alpha}{}^{\rho} \lambda_{j \rho} \lambda^{j \lambda} \lambda_{k \lambda} \lambda^{k \beta} X_{i \beta} {W}^{-3}+\frac{27}{256}{\rm i} (\Gamma_{a})^{\rho \beta} \lambda_{j \alpha} \lambda^{j \lambda} \lambda_{k \rho} \lambda^{k}_{\lambda} X_{i \beta} {W}^{-3}+\frac{3}{16}{\rm i} (\Gamma_{a})^{\beta}{}_{\rho} \lambda_{j \alpha} \lambda_{k \beta} \lambda^{k}_{\lambda} {W}^{-3} \nabla^{\rho \lambda}{X_{i}\,^{j}} - \frac{13}{160}{\rm i} \Phi^{\rho \lambda}\,_{i j} (\Gamma_{a})_{\alpha}{}^{\beta} \lambda^{j}_{\rho} \lambda_{k \lambda} \lambda^{k}_{\beta} {W}^{-2} - \frac{9}{64}(\Gamma_{a})^{\lambda \beta} X_{i j} \lambda_{k \lambda} \lambda^{k \rho} W_{\alpha \beta \rho}\,^{j} {W}^{-2}%
+\frac{237}{256}{\rm i} (\Gamma_{a})^{\beta \lambda} X_{i j} W_{\beta}\,^{\rho} \lambda^{j}_{\alpha} \lambda_{k \lambda} \lambda^{k}_{\rho} {W}^{-3}+\frac{3}{8}(\Gamma_{a})^{\gamma \rho} F_{\alpha}\,^{\beta} \lambda_{j \gamma} \lambda^{j \lambda} W_{\rho \beta \lambda i} {W}^{-2} - \frac{3}{32}(\Gamma_{a})^{\gamma \beta} \lambda_{j \gamma} \lambda^{j \rho} W_{\beta \rho \lambda i} {W}^{-2} \nabla_{\alpha}\,^{\lambda}{W} - \frac{9}{32}(\Gamma_{a})^{\beta \rho} F_{\alpha \beta} \lambda^{\lambda}_{j} \lambda^{j \gamma} W_{\rho \lambda \gamma i} {W}^{-2}+\frac{3}{64}(\Gamma_{a})^{\beta}{}_{\gamma} \lambda^{\rho}_{j} \lambda^{j \lambda} W_{\beta \rho \lambda i} {W}^{-2} \nabla^{\gamma}\,_{\alpha}{W} - \frac{9}{64}(\Gamma_{a})^{\lambda \beta} X_{j k} \lambda^{j}_{\lambda} \lambda^{k \rho} W_{\alpha \beta \rho i} {W}^{-2}+\frac{9}{32}{\rm i} (\Gamma_{a})^{\beta \gamma} C_{\alpha \beta}\,^{\rho \lambda} \lambda_{i \rho} \lambda_{j \gamma} \lambda^{j}_{\lambda} {W}^{-2}+\frac{9}{32}{\rm i} (\Gamma_{a})^{\gamma \beta} \lambda_{j \alpha} \lambda^{j \rho} \lambda_{k \gamma} \lambda^{k \lambda} W_{\beta \rho \lambda i} {W}^{-3}+\frac{9}{64}(\Gamma_{a})^{\lambda \rho} F_{\alpha}\,^{\beta} \lambda_{j \lambda} \lambda^{j}_{\beta} X_{i \rho} {W}^{-2}+\frac{9}{512}(\Gamma_{a})^{\rho \beta} \lambda_{j \rho} \lambda^{j}_{\lambda} X_{i \beta} {W}^{-2} \nabla_{\alpha}\,^{\lambda}{W}+\frac{21}{2560}(\Gamma_{a})_{\alpha}{}^{\beta} X_{j k} \lambda^{j \rho} \lambda^{k}_{\rho} X_{i \beta} {W}^{-2}+\frac{69}{256}{\rm i} \Phi_{\alpha}\,^{\beta}\,_{i j} (\Gamma_{a})_{\beta}{}^{\rho} \lambda^{j \lambda} \lambda_{k \rho} \lambda^{k}_{\lambda} {W}^{-2}+\frac{3}{512}{\rm i} (\Gamma_{a})^{\rho \beta} \lambda_{j \alpha} \lambda^{\lambda}_{i} \lambda_{k \rho} \lambda^{k}_{\lambda} X^{j}_{\beta} {W}^{-3}+\frac{9}{32}(\Gamma_{a})^{\beta \rho} F_{\alpha \beta} \lambda^{\lambda}_{i} \lambda^{\gamma}_{j} W_{\rho \lambda \gamma}\,^{j} {W}^{-2} - \frac{3}{64}(\Gamma_{a})^{\beta}{}_{\gamma} \lambda^{\rho}_{i} \lambda^{\lambda}_{j} W_{\beta \rho \lambda}\,^{j} {W}^{-2} \nabla^{\gamma}\,_{\alpha}{W}+\frac{3}{16}{\rm i} (\Gamma_{a})^{\gamma \beta} \lambda_{j \alpha} \lambda^{\rho}_{i} \lambda_{k \gamma} \lambda^{k \lambda} W_{\beta \rho \lambda}\,^{j} {W}^{-3}+\frac{9}{4}{\rm i} (\Gamma_{a})^{\beta \gamma} W^{\rho \lambda} F_{\alpha \beta} \lambda_{i \gamma} \lambda_{j \rho} \lambda^{j}_{\lambda} {W}^{-3} - \frac{81}{64}(\Gamma_{a})^{\lambda \gamma} W^{\beta \rho} \lambda_{j \alpha} \lambda_{i \lambda} \lambda^{j}_{\gamma} \lambda_{k \beta} \lambda^{k}_{\rho} {W}^{-4}+\frac{33}{128}{\rm i} (\Gamma_{a})^{\beta}{}_{\rho} \lambda_{j \beta} {W}^{-2} \nabla^{\rho \lambda}{\lambda_{i \lambda}} \nabla_{\alpha}\,^{\gamma}{\lambda^{j}_{\gamma}}+\frac{9}{32}{\rm i} (\Gamma_{a})^{\beta}{}_{\rho} X_{i j} \lambda_{k \alpha} \lambda^{j}_{\beta} {W}^{-3} \nabla^{\rho \lambda}{\lambda^{k}_{\lambda}}%
 - \frac{93}{1280}(\Gamma_{a})_{\alpha \beta} X_{j k} X^{j k} {W}^{-2} \nabla^{\beta \rho}{\lambda_{i \rho}} - \frac{9}{64}{\rm i} (\Gamma_{a})^{\beta}{}_{\rho} X_{j k} \lambda^{j}_{\alpha} \lambda^{k}_{\beta} {W}^{-3} \nabla^{\rho \lambda}{\lambda_{i \lambda}}+\frac{27}{128}(\Gamma_{a})_{\rho \lambda} F_{\alpha \beta} {W}^{-2} \nabla^{\rho \lambda}{W} \nabla^{\beta \gamma}{\lambda_{i \gamma}}+\frac{3}{64}(\Gamma_{a})_{\beta \rho} X_{i j} {W}^{-2} \nabla^{\beta \rho}{W} \nabla_{\alpha}\,^{\lambda}{\lambda^{j}_{\lambda}}+\frac{9}{128}(\Gamma_{a})_{\beta \rho} {W}^{-2} \nabla^{\beta \rho}{W} \nabla_{\alpha \gamma}{W} \nabla^{\gamma \lambda}{\lambda_{i \lambda}}+\frac{3}{32}{\rm i} (\Gamma_{a})_{\beta \rho} \lambda_{j \alpha} \lambda_{i \lambda} {W}^{-3} \nabla^{\beta \rho}{W} \nabla^{\lambda \gamma}{\lambda^{j}_{\gamma}}+\frac{3}{64}{\rm i} (\Gamma_{a})_{\beta \rho} \lambda_{i \lambda} {W}^{-2} \nabla^{\beta \rho}{\lambda_{j \alpha}} \nabla^{\lambda \gamma}{\lambda^{j}_{\gamma}}+\frac{39}{640}{\rm i} (\Gamma_{a})_{\alpha}{}^{\beta} W_{\beta}\,^{\rho} \lambda_{i \lambda} \lambda_{j \rho} {W}^{-2} \nabla^{\lambda \gamma}{\lambda^{j}_{\gamma}} - \frac{9}{32}(\Gamma_{a})_{\rho \lambda} \lambda_{i \gamma} {W}^{-2} \nabla^{\rho \lambda}{W} \nabla^{\gamma \beta}{F_{\alpha \beta}} - \frac{9}{64}(\Gamma_{a})_{\beta \rho} \lambda_{i \lambda} {W}^{-2} \nabla^{\beta \rho}{W} \nabla_{\gamma}\,^{\lambda}{\nabla_{\alpha}\,^{\gamma}{W}} - \frac{189}{256}(\Gamma_{a})_{\lambda \gamma} W^{\beta \rho} F_{\beta \rho} \lambda_{i \alpha} {W}^{-2} \nabla^{\lambda \gamma}{W}+\frac{189}{256}(\Gamma_{a})_{\lambda \gamma} W^{\beta \rho} F_{\alpha \beta} \lambda_{i \rho} {W}^{-2} \nabla^{\lambda \gamma}{W}+\frac{81}{512}(\Gamma_{a})_{\lambda \gamma} W^{\beta}\,_{\rho} \lambda_{i \beta} {W}^{-2} \nabla^{\lambda \gamma}{W} \nabla_{\alpha}\,^{\rho}{W}+\frac{21}{128}{\rm i} (\Gamma_{a})_{\beta \rho} \lambda_{j \alpha} \lambda^{j}_{\lambda} {W}^{-3} \nabla^{\beta \rho}{W} \nabla^{\lambda \gamma}{\lambda_{i \gamma}} - \frac{9}{256}{\rm i} (\Gamma_{a})_{\beta \rho} \lambda_{j \lambda} {W}^{-2} \nabla^{\beta \rho}{\lambda^{j}_{\alpha}} \nabla^{\lambda \gamma}{\lambda_{i \gamma}} - \frac{21}{320}{\rm i} (\Gamma_{a})_{\alpha}{}^{\beta} W_{\beta}\,^{\rho} \lambda_{j \rho} \lambda^{j}_{\lambda} {W}^{-2} \nabla^{\lambda \gamma}{\lambda_{i \gamma}}+\frac{3}{128}(\Gamma_{a})_{\beta \rho} \lambda_{j \lambda} {W}^{-2} \nabla^{\beta \rho}{W} \nabla_{\alpha}\,^{\lambda}{X_{i}\,^{j}} - \frac{3}{4}{\rm i} (\Gamma_{a})^{\beta \rho} F_{\alpha \beta} \lambda_{j \rho} \lambda^{j}_{\lambda} {W}^{-3} \nabla^{\lambda \gamma}{\lambda_{i \gamma}}+\frac{3}{32}{\rm i} (\Gamma_{a})^{\beta}{}_{\rho} \lambda_{j \beta} \lambda^{j}_{\lambda} {W}^{-3} \nabla^{\rho}\,_{\alpha}{W} \nabla^{\lambda \gamma}{\lambda_{i \gamma}}+\frac{3}{80}{\rm i} (\Gamma_{a})_{\alpha}{}^{\beta} X_{j k} \lambda_{i \beta} \lambda^{j}_{\rho} {W}^{-3} \nabla^{\rho \lambda}{\lambda^{k}_{\lambda}}%
+\frac{9}{32}(\Gamma_{a})^{\beta \rho} \lambda_{j \alpha} \lambda_{i \beta} \lambda_{k \rho} \lambda^{k}_{\lambda} {W}^{-4} \nabla^{\lambda \gamma}{\lambda^{j}_{\gamma}} - \frac{9}{64}{\rm i} (\Gamma_{a})^{\beta \rho} X_{j k} \lambda^{j}_{\beta} \lambda^{k}_{\rho} {W}^{-3} \nabla_{\alpha}\,^{\lambda}{\lambda_{i \lambda}}+\frac{9}{32}(\Gamma_{a})^{\beta \rho} \lambda_{j \alpha} \lambda^{j}_{\beta} \lambda_{k \rho} \lambda^{k}_{\lambda} {W}^{-4} \nabla^{\lambda \gamma}{\lambda_{i \gamma}} - \frac{3}{32}{\rm i} (\Gamma_{a})^{\beta \rho} \lambda_{k \beta} \lambda^{k}_{\lambda} \lambda_{j \rho} {W}^{-3} \nabla_{\alpha}\,^{\lambda}{X_{i}\,^{j}} - \frac{21}{256}(\Gamma_{a})_{\alpha}{}^{\beta} X_{j k} X^{j k} W_{\beta}\,^{\rho} \lambda_{i \rho} {W}^{-2} - \frac{21}{256}{\rm i} (\Gamma_{a})^{\beta \lambda} X_{j k} W_{\beta}\,^{\rho} \lambda^{j}_{\alpha} \lambda_{i \rho} \lambda^{k}_{\lambda} {W}^{-3} - \frac{45}{32}{\rm i} (\Gamma_{a})^{\beta \gamma} W^{\rho \lambda} F_{\alpha \beta} \lambda_{i \rho} \lambda_{j \gamma} \lambda^{j}_{\lambda} {W}^{-3}+\frac{93}{640}{\rm i} (\Gamma_{a})_{\alpha}{}^{\lambda} X_{i j} W^{\beta \rho} \lambda^{j}_{\beta} \lambda_{k \lambda} \lambda^{k}_{\rho} {W}^{-3}+\frac{27}{32}(\Gamma_{a})^{\lambda \gamma} W^{\beta \rho} \lambda_{j \alpha} \lambda_{i \lambda} \lambda^{j}_{\beta} \lambda_{k \gamma} \lambda^{k}_{\rho} {W}^{-4} - \frac{9}{256}{\rm i} (\Gamma_{a})^{\rho \lambda} \lambda_{i \alpha} \lambda_{k \rho} \lambda^{k \beta} \lambda_{j \lambda} X^{j}_{\beta} {W}^{-3}+\frac{9}{80}{\rm i} (\Gamma_{a})_{\alpha}{}^{\lambda} X_{j k} W^{\beta \rho} \lambda_{i \beta} \lambda^{j}_{\lambda} \lambda^{k}_{\rho} {W}^{-3}+\frac{21}{256}{\rm i} (\Gamma_{a})^{\rho \lambda} X_{j k} W_{\alpha}\,^{\beta} \lambda_{i \beta} \lambda^{j}_{\rho} \lambda^{k}_{\lambda} {W}^{-3} - \frac{27}{64}(\Gamma_{a})^{\lambda \gamma} W^{\beta \rho} \lambda_{j \alpha} \lambda_{i \beta} \lambda^{j}_{\lambda} \lambda_{k \gamma} \lambda^{k}_{\rho} {W}^{-4}+\frac{3}{128}{\rm i} (\Gamma_{a})_{\beta \rho} \lambda_{j \lambda} {W}^{-2} \nabla^{\beta \rho}{\lambda^{\gamma}_{i}} \nabla_{\alpha}\,^{\lambda}{\lambda^{j}_{\gamma}}+\frac{3}{128}{\rm i} (\Gamma_{a})_{\beta \rho} \lambda^{\lambda}_{j} {W}^{-2} \nabla^{\beta \rho}{\lambda_{i \gamma}} \nabla_{\alpha}\,^{\gamma}{\lambda^{j}_{\lambda}}+\frac{3}{128}{\rm i} (\Gamma_{a})_{\beta \rho} \lambda_{j \alpha} {W}^{-2} \nabla^{\beta \rho}{\lambda_{i \lambda}} \nabla^{\lambda \gamma}{\lambda^{j}_{\gamma}}+\frac{3}{64}{\rm i} (\Gamma_{a})_{\beta \rho} \lambda_{j \lambda} {W}^{-2} \nabla^{\beta \rho}{\lambda_{i \alpha}} \nabla^{\lambda \gamma}{\lambda^{j}_{\gamma}} - \frac{9}{128}{\rm i} (\Gamma_{a})_{\lambda \gamma} W^{\beta \rho} \lambda_{j \beta} \lambda^{j}_{\rho} {W}^{-2} \nabla^{\lambda \gamma}{\lambda_{i \alpha}} - \frac{9}{64}(\Gamma_{a})_{\lambda \gamma} F_{\alpha}\,^{\beta} F_{\beta}\,^{\rho} {W}^{-2} \nabla^{\lambda \gamma}{\lambda_{i \rho}} - \frac{9}{128}(\Gamma_{a})_{\lambda \gamma} F^{\beta}\,_{\rho} {W}^{-2} \nabla_{\alpha}\,^{\rho}{W} \nabla^{\lambda \gamma}{\lambda_{i \beta}}%
+\frac{3}{32}{\rm i} (\Gamma_{a})_{\lambda \gamma} F^{\beta \rho} \lambda_{j \alpha} \lambda^{j}_{\beta} {W}^{-3} \nabla^{\lambda \gamma}{\lambda_{i \rho}} - \frac{9}{32}(\Gamma_{a})_{\lambda \gamma} F^{\beta \rho} \lambda_{i \beta} {W}^{-2} \nabla^{\lambda \gamma}{F_{\alpha \rho}} - \frac{117}{256}(\Gamma_{a})_{\lambda \gamma} F^{\beta \rho} \lambda_{i \beta} {W}^{-1} \nabla^{\lambda \gamma}{W_{\alpha \rho}} - \frac{3}{16}(\Gamma_{a})_{\rho \lambda} F_{\alpha}\,^{\beta} \lambda_{j \beta} {W}^{-2} \nabla^{\rho \lambda}{X_{i}\,^{j}}+\frac{9}{64}(\Gamma_{a})_{\lambda \gamma} F^{\beta}\,_{\rho} \lambda_{i \beta} {W}^{-2} \nabla^{\lambda \gamma}{\nabla_{\alpha}\,^{\rho}{W}}+\frac{171}{320}(\Gamma_{a})_{\alpha}{}^{\gamma} W_{\gamma}\,^{\beta} F_{\beta}\,^{\rho} F_{\rho}\,^{\lambda} \lambda_{i \lambda} {W}^{-2} - \frac{171}{640}(\Gamma_{a})_{\alpha}{}^{\lambda} W_{\lambda \gamma} F^{\beta}\,_{\rho} \lambda_{i \beta} {W}^{-2} \nabla^{\gamma \rho}{W} - \frac{3}{20}(\Gamma_{a})_{\alpha}{}^{\lambda} F^{\beta \rho} \lambda_{i \beta} \lambda_{j \rho} X^{j}_{\lambda} {W}^{-2}+\frac{39}{1280}(\Gamma_{a})_{\alpha}{}^{\lambda} F^{\beta \rho} \lambda_{j \beta} \lambda^{j}_{\rho} X_{i \lambda} {W}^{-2} - \frac{1}{10}\Phi^{\lambda \beta}\,_{i j} (\Gamma_{a})_{\alpha \lambda} F_{\beta}\,^{\rho} \lambda^{j}_{\rho} {W}^{-1} - \frac{9}{128}(\Gamma_{a})^{\lambda}{}_{\gamma} F^{\beta \rho} \lambda_{i \beta} {W}^{-1} \nabla^{\gamma}\,_{\alpha}{W_{\lambda \rho}}+\frac{27}{128}(\Gamma_{a})^{\lambda}{}_{\gamma} F^{\beta}\,_{\rho} \lambda_{i \beta} {W}^{-1} \nabla^{\gamma \rho}{W_{\alpha \lambda}} - \frac{27}{256}(\Gamma_{a})_{\alpha \gamma} F^{\beta \rho} \lambda_{i \beta} {W}^{-1} \nabla^{\gamma \lambda}{W_{\rho \lambda}}+\frac{3}{64}{\rm i} (\Gamma_{a})^{\beta}{}_{\rho} \lambda^{\lambda}_{j} {W}^{-2} \nabla^{\rho}\,_{\gamma}{\lambda_{i \beta}} \nabla_{\alpha}\,^{\gamma}{\lambda^{j}_{\lambda}}+\frac{3}{64}{\rm i} (\Gamma_{a})^{\beta}{}_{\rho} \lambda_{j \lambda} {W}^{-2} \nabla^{\rho \gamma}{\lambda_{i \beta}} \nabla_{\alpha}\,^{\lambda}{\lambda^{j}_{\gamma}} - \frac{45}{256}{\rm i} (\Gamma_{a})^{\lambda}{}_{\gamma} W^{\beta}\,_{\rho} \lambda_{j \alpha} \lambda^{j}_{\beta} {W}^{-2} \nabla^{\gamma \rho}{\lambda_{i \lambda}} - \frac{3}{32}{\rm i} (\Gamma_{a})^{\beta}{}_{\rho} \lambda_{j \lambda} {W}^{-2} \nabla^{\rho}\,_{\alpha}{\lambda_{i \beta}} \nabla^{\lambda \gamma}{\lambda^{j}_{\gamma}} - \frac{45}{256}{\rm i} (\Gamma_{a})^{\lambda}{}_{\gamma} W^{\beta \rho} \lambda_{j \beta} \lambda^{j}_{\rho} {W}^{-2} \nabla^{\gamma}\,_{\alpha}{\lambda_{i \lambda}}+\frac{9}{16}(\Gamma_{a})^{\lambda}{}_{\gamma} F_{\alpha}\,^{\beta} F_{\beta \rho} {W}^{-2} \nabla^{\gamma \rho}{\lambda_{i \lambda}}+\frac{9}{32}(\Gamma_{a})^{\lambda}{}_{\gamma} F_{\beta \rho} {W}^{-2} \nabla_{\alpha}\,^{\beta}{W} \nabla^{\gamma \rho}{\lambda_{i \lambda}}%
 - \frac{3}{16}{\rm i} (\Gamma_{a})^{\lambda}{}_{\gamma} F^{\beta}\,_{\rho} \lambda_{j \alpha} \lambda^{j}_{\beta} {W}^{-3} \nabla^{\gamma \rho}{\lambda_{i \lambda}}+\frac{9}{32}(\Gamma_{a})^{\lambda}{}_{\gamma} F^{\beta}\,_{\rho} \lambda_{i \beta} {W}^{-2} \nabla^{\gamma \rho}{F_{\alpha \lambda}}+\frac{3}{64}(\Gamma_{a})_{\alpha \lambda} F^{\beta}\,_{\rho} \lambda_{j \beta} {W}^{-2} \nabla^{\lambda \rho}{X_{i}\,^{j}} - \frac{9}{64}(\Gamma_{a})_{\lambda \gamma} F^{\beta}\,_{\rho} \lambda_{i \beta} {W}^{-2} \nabla^{\lambda \rho}{\nabla^{\gamma}\,_{\alpha}{W}}+\frac{243}{64}(\Gamma_{a})^{\gamma \rho} W_{\gamma}\,^{\lambda} F_{\alpha}\,^{\beta} F_{\rho \lambda} \lambda_{i \beta} {W}^{-2} - \frac{27}{128}(\Gamma_{a})^{\rho}{}_{\gamma} W_{\rho \lambda} F_{\alpha}\,^{\beta} \lambda_{i \beta} {W}^{-2} \nabla^{\gamma \lambda}{W}+\frac{3}{64}{\rm i} (\Gamma_{a})^{\beta}{}_{\lambda} W_{\beta}\,^{\rho} \lambda_{i \rho} \lambda^{\gamma}_{j} {W}^{-2} \nabla^{\lambda}\,_{\alpha}{\lambda^{j}_{\gamma}} - \frac{9}{256}{\rm i} (\Gamma_{a})^{\beta \lambda} W_{\beta}\,^{\rho} \lambda_{i \rho} \lambda_{j \gamma} {W}^{-2} \nabla_{\alpha}\,^{\gamma}{\lambda^{j}_{\lambda}} - \frac{111}{1280}{\rm i} (\Gamma_{a})_{\alpha}{}^{\beta} W_{\beta}\,^{\rho} \lambda_{i \rho} \lambda_{j \lambda} {W}^{-2} \nabla^{\lambda \gamma}{\lambda^{j}_{\gamma}}+\frac{2133}{2560}{\rm i} (\Gamma_{a})_{\alpha}{}^{\beta} W_{\beta}\,^{\rho} W^{\lambda \gamma} \lambda_{i \rho} \lambda_{j \lambda} \lambda^{j}_{\gamma} {W}^{-2}+\frac{135}{64}(\Gamma_{a})^{\gamma \rho} W_{\gamma}\,^{\beta} F_{\alpha \beta} F_{\rho}\,^{\lambda} \lambda_{i \lambda} {W}^{-2}+\frac{27}{32}(\Gamma_{a})^{\lambda \rho} W_{\lambda}\,^{\gamma} F_{\alpha}\,^{\beta} F_{\rho \beta} \lambda_{i \gamma} {W}^{-2}+\frac{81}{128}{\rm i} (\Gamma_{a})^{\lambda \beta} W_{\lambda}\,^{\gamma} F_{\beta}\,^{\rho} \lambda_{j \alpha} \lambda_{i \gamma} \lambda^{j}_{\rho} {W}^{-3}+\frac{3}{128}{\rm i} (\Gamma_{a})_{\beta \rho} \lambda_{i \lambda} {W}^{-2} \nabla^{\beta \rho}{\lambda^{\gamma}_{j}} \nabla_{\alpha}\,^{\lambda}{\lambda^{j}_{\gamma}}+\frac{3}{128}{\rm i} (\Gamma_{a})_{\beta \rho} \lambda^{\lambda}_{i} {W}^{-2} \nabla^{\beta \rho}{\lambda_{j \gamma}} \nabla_{\alpha}\,^{\gamma}{\lambda^{j}_{\lambda}}+\frac{3}{64}{\rm i} (\Gamma_{a})_{\rho \lambda} W_{\alpha}\,^{\beta} \lambda_{i \beta} \lambda^{\gamma}_{j} {W}^{-2} \nabla^{\rho \lambda}{\lambda^{j}_{\gamma}} - \frac{27}{512}{\rm i} (\Gamma_{a})_{\rho \lambda} W_{\alpha}\,^{\beta} \lambda^{\gamma}_{i} \lambda_{j \gamma} {W}^{-2} \nabla^{\rho \lambda}{\lambda^{j}_{\beta}} - \frac{3}{64}{\rm i} (\Gamma_{a})_{\beta \rho} \lambda_{i \alpha} {W}^{-2} \nabla^{\beta \rho}{\lambda_{j \lambda}} \nabla^{\lambda \gamma}{\lambda^{j}_{\gamma}}+\frac{9}{64}{\rm i} (\Gamma_{a})_{\lambda \gamma} W^{\beta \rho} \lambda_{i \alpha} \lambda_{j \beta} {W}^{-2} \nabla^{\lambda \gamma}{\lambda^{j}_{\rho}}+\frac{9}{512}{\rm i} (\Gamma_{a})_{\lambda \gamma} W^{\beta \rho} \lambda_{i \beta} \lambda_{j \rho} {W}^{-2} \nabla^{\lambda \gamma}{\lambda^{j}_{\alpha}}%
 - \frac{3}{128}(\Gamma_{a})_{\rho \lambda} X_{i j} F_{\alpha}\,^{\beta} {W}^{-2} \nabla^{\rho \lambda}{\lambda^{j}_{\beta}}+\frac{3}{32}{\rm i} (\Gamma_{a})_{\lambda \gamma} F^{\beta \rho} \lambda_{j \alpha} \lambda_{i \beta} {W}^{-3} \nabla^{\lambda \gamma}{\lambda^{j}_{\rho}}+\frac{3}{64}{\rm i} (\Gamma_{a})^{\beta}{}_{\rho} \lambda^{\lambda}_{i} {W}^{-2} \nabla^{\rho}\,_{\gamma}{\lambda_{j \beta}} \nabla_{\alpha}\,^{\gamma}{\lambda^{j}_{\lambda}}+\frac{3}{64}{\rm i} (\Gamma_{a})^{\beta}{}_{\rho} \lambda_{i \lambda} {W}^{-2} \nabla^{\rho \gamma}{\lambda_{j \beta}} \nabla_{\alpha}\,^{\lambda}{\lambda^{j}_{\gamma}} - \frac{9}{64}{\rm i} (\Gamma_{a})^{\lambda}{}_{\gamma} W^{\beta}\,_{\rho} \lambda_{j \alpha} \lambda_{i \beta} {W}^{-2} \nabla^{\gamma \rho}{\lambda^{j}_{\lambda}}+\frac{3}{64}{\rm i} (\Gamma_{a})^{\beta}{}_{\rho} \lambda_{i \lambda} {W}^{-2} \nabla^{\rho}\,_{\alpha}{\lambda_{j \beta}} \nabla^{\lambda \gamma}{\lambda^{j}_{\gamma}}+\frac{3}{64}{\rm i} (\Gamma_{a})^{\beta}{}_{\rho} \lambda_{i \alpha} {W}^{-2} \nabla^{\rho}\,_{\gamma}{\lambda_{j \beta}} \nabla^{\gamma \lambda}{\lambda^{j}_{\lambda}}+\frac{9}{128}{\rm i} (\Gamma_{a})^{\lambda}{}_{\gamma} W^{\beta \rho} \lambda_{i \beta} \lambda_{j \rho} {W}^{-2} \nabla^{\gamma}\,_{\alpha}{\lambda^{j}_{\lambda}} - \frac{9}{128}{\rm i} (\Gamma_{a})^{\lambda}{}_{\gamma} W^{\beta}\,_{\rho} \lambda_{i \alpha} \lambda_{j \beta} {W}^{-2} \nabla^{\gamma \rho}{\lambda^{j}_{\lambda}} - \frac{3}{16}{\rm i} (\Gamma_{a})^{\lambda}{}_{\gamma} F^{\beta}\,_{\rho} \lambda_{j \alpha} \lambda_{i \beta} {W}^{-3} \nabla^{\gamma \rho}{\lambda^{j}_{\lambda}} - \frac{333}{128}{\rm i} (\Gamma_{a})^{\beta \rho} W_{\alpha \beta} W_{\rho}\,^{\lambda} \lambda^{\gamma}_{i} \lambda_{j \lambda} \lambda^{j}_{\gamma} {W}^{-2} - \frac{513}{640}{\rm i} (\Gamma_{a})_{\alpha}{}^{\beta} W_{\beta}\,^{\rho} W^{\lambda \gamma} \lambda_{i \lambda} \lambda_{j \rho} \lambda^{j}_{\gamma} {W}^{-2} - \frac{63}{128}{\rm i} (\Gamma_{a})^{\lambda \beta} W_{\lambda}\,^{\gamma} F_{\beta}\,^{\rho} \lambda_{j \alpha} \lambda_{i \rho} \lambda^{j}_{\gamma} {W}^{-3} - \frac{45}{256}{\rm i} (\Gamma_{a})^{\beta}{}_{\lambda} W_{\beta}\,^{\rho} \lambda^{\gamma}_{i} \lambda_{j \gamma} {W}^{-2} \nabla^{\lambda}\,_{\alpha}{\lambda^{j}_{\rho}}+\frac{9}{256}{\rm i} (\Gamma_{a})^{\beta \lambda} W_{\beta \rho} \lambda^{\gamma}_{i} \lambda_{j \gamma} {W}^{-2} \nabla_{\alpha}\,^{\rho}{\lambda^{j}_{\lambda}} - \frac{81}{256}{\rm i} (\Gamma_{a})^{\beta}{}_{\rho} W_{\alpha \beta} \lambda^{\lambda}_{i} \lambda_{j \lambda} {W}^{-2} \nabla^{\rho \gamma}{\lambda^{j}_{\gamma}} - \frac{729}{640}{\rm i} (\Gamma_{a})_{\alpha}{}^{\beta} W_{\beta}\,^{\rho} W_{\rho}\,^{\lambda} \lambda^{\gamma}_{i} \lambda_{j \lambda} \lambda^{j}_{\gamma} {W}^{-2}+\frac{135}{64}(\Gamma_{a})^{\lambda \beta} W_{\lambda}\,^{\rho} F_{\beta \rho} \lambda_{i \gamma} {W}^{-2} \nabla_{\alpha}\,^{\gamma}{W}+\frac{207}{128}{\rm i} (\Gamma_{a})^{\lambda \beta} W_{\lambda}\,^{\rho} F_{\beta \rho} \lambda_{j \alpha} \lambda^{\gamma}_{i} \lambda^{j}_{\gamma} {W}^{-3}+\frac{33}{32}{\rm i} (\Gamma_{a})^{\rho}{}_{\lambda} F_{\alpha}\,^{\beta} \lambda_{i \rho} \lambda_{j \beta} {W}^{-3} \nabla^{\lambda \gamma}{\lambda^{j}_{\gamma}}%
 - \frac{141}{64}{\rm i} (\Gamma_{a})^{\rho \gamma} W_{\rho}\,^{\lambda} F_{\alpha}\,^{\beta} \lambda_{i \gamma} \lambda_{j \lambda} \lambda^{j}_{\beta} {W}^{-3}+\frac{3}{16}{\rm i} (\Gamma_{a})^{\beta \lambda} F_{\beta \rho} \lambda_{i \lambda} \lambda^{\gamma}_{j} {W}^{-3} \nabla_{\alpha}\,^{\rho}{\lambda^{j}_{\gamma}}+\frac{3}{16}{\rm i} (\Gamma_{a})^{\beta \lambda} F_{\beta}\,^{\rho} \lambda_{i \lambda} \lambda_{j \gamma} {W}^{-3} \nabla_{\alpha}\,^{\gamma}{\lambda^{j}_{\rho}} - \frac{3}{16}{\rm i} (\Gamma_{a})^{\beta \lambda} F_{\beta \rho} \lambda_{j \alpha} \lambda_{i \lambda} {W}^{-3} \nabla^{\rho \gamma}{\lambda^{j}_{\gamma}}+\frac{9}{4}(\Gamma_{a})^{\beta \rho} F_{\alpha \beta} F_{\rho}\,^{\lambda} F_{\lambda}\,^{\gamma} \lambda_{i \gamma} {W}^{-3}+\frac{3}{80}(\Gamma_{a})_{\alpha}{}^{\beta} X_{i j} F_{\beta}\,^{\rho} F_{\rho}\,^{\lambda} \lambda^{j}_{\lambda} {W}^{-3} - \frac{9}{4}(\Gamma_{a})^{\rho \gamma} F_{\alpha}\,^{\beta} F_{\rho}\,^{\lambda} F_{\beta \lambda} \lambda_{i \gamma} {W}^{-3} - \frac{9}{8}(\Gamma_{a})^{\beta \gamma} F_{\beta}\,^{\rho} F_{\rho \lambda} \lambda_{i \gamma} {W}^{-3} \nabla_{\alpha}\,^{\lambda}{W} - \frac{9}{8}{\rm i} (\Gamma_{a})^{\beta \gamma} F_{\beta}\,^{\rho} F_{\rho}\,^{\lambda} \lambda_{j \alpha} \lambda_{i \gamma} \lambda^{j}_{\lambda} {W}^{-4} - \frac{3}{32}{\rm i} (\Gamma_{a})^{\lambda}{}_{\gamma} F^{\beta \rho} \lambda_{i \beta} \lambda_{j \lambda} {W}^{-3} \nabla^{\gamma}\,_{\alpha}{\lambda^{j}_{\rho}}+\frac{3}{32}{\rm i} (\Gamma_{a})^{\lambda \gamma} F^{\beta}\,_{\rho} \lambda_{i \beta} \lambda_{j \lambda} {W}^{-3} \nabla_{\alpha}\,^{\rho}{\lambda^{j}_{\gamma}} - \frac{9}{160}{\rm i} (\Gamma_{a})_{\alpha}{}^{\lambda} F^{\beta}\,_{\rho} \lambda_{i \beta} \lambda_{j \lambda} {W}^{-3} \nabla^{\rho \gamma}{\lambda^{j}_{\gamma}} - \frac{9}{32}{\rm i} (\Gamma_{a})^{\rho}{}_{\lambda} F_{\alpha}\,^{\beta} \lambda_{i \beta} \lambda_{j \rho} {W}^{-3} \nabla^{\lambda \gamma}{\lambda^{j}_{\gamma}} - \frac{429}{640}{\rm i} (\Gamma_{a})_{\alpha}{}^{\gamma} W^{\beta \lambda} F_{\beta}\,^{\rho} \lambda_{i \rho} \lambda_{j \gamma} \lambda^{j}_{\lambda} {W}^{-3}+\frac{39}{64}{\rm i} (\Gamma_{a})^{\rho \gamma} W_{\rho}\,^{\lambda} F_{\alpha}\,^{\beta} \lambda_{i \beta} \lambda_{j \gamma} \lambda^{j}_{\lambda} {W}^{-3} - \frac{3}{16}{\rm i} (\Gamma_{a})^{\beta \lambda} F_{\beta \rho} \lambda^{\gamma}_{i} \lambda_{j \lambda} {W}^{-3} \nabla_{\alpha}\,^{\rho}{\lambda^{j}_{\gamma}} - \frac{3}{16}{\rm i} (\Gamma_{a})^{\beta \lambda} F_{\beta}\,^{\rho} \lambda_{i \gamma} \lambda_{j \lambda} {W}^{-3} \nabla_{\alpha}\,^{\gamma}{\lambda^{j}_{\rho}}+\frac{117}{128}{\rm i} (\Gamma_{a})^{\beta \lambda} W_{\alpha}\,^{\rho} F_{\beta \rho} \lambda^{\gamma}_{i} \lambda_{j \lambda} \lambda^{j}_{\gamma} {W}^{-3}+\frac{87}{128}{\rm i} (\Gamma_{a})^{\beta \gamma} W_{\alpha}\,^{\lambda} F_{\beta}\,^{\rho} \lambda_{i \lambda} \lambda_{j \gamma} \lambda^{j}_{\rho} {W}^{-3} - \frac{3}{16}{\rm i} (\Gamma_{a})^{\beta \rho} F_{\alpha \beta} \lambda_{i \lambda} \lambda_{j \rho} {W}^{-3} \nabla^{\lambda \gamma}{\lambda^{j}_{\gamma}}%
 - \frac{3}{16}{\rm i} (\Gamma_{a})^{\beta \lambda} F_{\beta \rho} \lambda_{i \alpha} \lambda_{j \lambda} {W}^{-3} \nabla^{\rho \gamma}{\lambda^{j}_{\gamma}} - \frac{9}{32}{\rm i} (\Gamma_{a})^{\beta \gamma} W^{\rho \lambda} F_{\beta \rho} \lambda_{i \alpha} \lambda_{j \gamma} \lambda^{j}_{\lambda} {W}^{-3}+\frac{3}{16}(\Gamma_{a})^{\rho \lambda} X_{i j} F_{\alpha}\,^{\beta} F_{\rho \beta} \lambda^{j}_{\lambda} {W}^{-3}+\frac{9}{8}{\rm i} (\Gamma_{a})^{\beta \gamma} F_{\beta}\,^{\rho} F_{\rho}\,^{\lambda} \lambda_{j \alpha} \lambda_{i \lambda} \lambda^{j}_{\gamma} {W}^{-4} - \frac{45}{64}(\Gamma_{a})^{\beta}{}_{\lambda} F_{\beta \rho} {W}^{-2} \nabla^{\lambda \rho}{W} \nabla_{\alpha}\,^{\gamma}{\lambda_{i \gamma}} - \frac{3}{64}(\Gamma_{a})^{\beta}{}_{\rho} X_{i j} {W}^{-2} \nabla^{\rho}\,_{\lambda}{W} \nabla_{\alpha}\,^{\lambda}{\lambda^{j}_{\beta}} - \frac{9}{16}(\Gamma_{a})^{\beta}{}_{\lambda} X_{i j} F_{\beta \rho} \lambda^{j}_{\alpha} {W}^{-3} \nabla^{\lambda \rho}{W}+\frac{51}{64}{\rm i} (\Gamma_{a})^{\beta \lambda} F_{\beta}\,^{\rho} \lambda_{i \lambda} \lambda_{j \rho} {W}^{-3} \nabla_{\alpha}\,^{\gamma}{\lambda^{j}_{\gamma}} - \frac{15}{128}{\rm i} (\Gamma_{a})^{\beta \rho} X_{j k} \lambda_{i \beta} \lambda^{j}_{\lambda} {W}^{-3} \nabla_{\alpha}\,^{\lambda}{\lambda^{k}_{\rho}} - \frac{141}{128}{\rm i} (\Gamma_{a})^{\beta \rho} X_{j k} W_{\alpha \beta} \lambda_{i \rho} \lambda^{j \lambda} \lambda^{k}_{\lambda} {W}^{-3} - \frac{21}{80}(\Gamma_{a})_{\alpha}{}^{\beta} X_{i j} X^{j}\,_{k} F_{\beta}\,^{\rho} \lambda^{k}_{\rho} {W}^{-3} - \frac{3}{8}(\Gamma_{a})^{\beta}{}_{\lambda} X_{i j} F_{\beta}\,^{\rho} \lambda^{j}_{\rho} {W}^{-3} \nabla^{\lambda}\,_{\alpha}{W} - \frac{9}{16}{\rm i} (\Gamma_{a})^{\beta \lambda} X_{j k} F_{\beta}\,^{\rho} \lambda^{j}_{\alpha} \lambda_{i \lambda} \lambda^{k}_{\rho} {W}^{-4}+\frac{21}{64}{\rm i} (\Gamma_{a})^{\beta \lambda} F_{\beta}\,^{\rho} \lambda_{j \lambda} \lambda^{j}_{\rho} {W}^{-3} \nabla_{\alpha}\,^{\gamma}{\lambda_{i \gamma}} - \frac{75}{256}(\Gamma_{a})^{\beta \lambda} F_{\beta}\,^{\rho} \lambda_{j \lambda} \lambda^{j}_{\rho} X_{i \alpha} {W}^{-2} - \frac{75}{64}{\rm i} (\Gamma_{a})^{\beta \rho} X_{i j} W_{\alpha \beta} \lambda^{j \lambda} \lambda_{k \rho} \lambda^{k}_{\lambda} {W}^{-3} - \frac{45}{128}{\rm i} (\Gamma_{a})^{\beta}{}_{\rho} X_{i j} \lambda^{j}_{\alpha} \lambda_{k \beta} {W}^{-3} \nabla^{\rho \lambda}{\lambda^{k}_{\lambda}} - \frac{21}{16}(\Gamma_{a})^{\beta \rho} X_{i j} X^{j}\,_{k} F_{\alpha \beta} \lambda^{k}_{\rho} {W}^{-3}+\frac{9}{32}(\Gamma_{a})^{\beta \lambda} X_{i j} F_{\beta \rho} \lambda^{j}_{\lambda} {W}^{-3} \nabla_{\alpha}\,^{\rho}{W} - \frac{9}{32}{\rm i} (\Gamma_{a})^{\beta \lambda} X_{i j} F_{\beta}\,^{\rho} \lambda_{k \alpha} \lambda^{j}_{\rho} \lambda^{k}_{\lambda} {W}^{-4}%
 - \frac{81}{128}(\Gamma_{a})^{\beta}{}_{\lambda} W_{\beta \rho} \lambda_{i \gamma} {W}^{-2} \nabla^{\lambda \rho}{W} \nabla_{\alpha}\,^{\gamma}{W}+\frac{9}{256}{\rm i} (\Gamma_{a})^{\beta}{}_{\lambda} W_{\beta \rho} \lambda^{\gamma}_{i} \lambda_{j \gamma} {W}^{-2} \nabla^{\lambda \rho}{\lambda^{j}_{\alpha}}+\frac{27}{64}(\Gamma_{a})^{\rho}{}_{\gamma} W_{\rho}\,^{\lambda} F_{\alpha \beta} \lambda_{i \lambda} {W}^{-2} \nabla^{\gamma \beta}{W} - \frac{81}{64}(\Gamma_{a})^{\rho}{}_{\lambda} W_{\rho}\,^{\beta} F_{\alpha \beta} \lambda_{i \gamma} {W}^{-2} \nabla^{\lambda \gamma}{W} - \frac{27}{128}(\Gamma_{a})^{\beta}{}_{\lambda} W_{\beta \rho} \lambda_{i \gamma} {W}^{-2} \nabla^{\lambda \gamma}{W} \nabla_{\alpha}\,^{\rho}{W}+\frac{63}{256}{\rm i} (\Gamma_{a})^{\beta}{}_{\lambda} W_{\beta}\,^{\rho} \lambda_{j \alpha} \lambda_{i \gamma} \lambda^{j}_{\rho} {W}^{-3} \nabla^{\lambda \gamma}{W} - \frac{81}{256}{\rm i} (\Gamma_{a})^{\beta}{}_{\lambda} W_{\beta}\,^{\rho} \lambda_{j \alpha} \lambda_{i \rho} \lambda^{j}_{\gamma} {W}^{-3} \nabla^{\lambda \gamma}{W} - \frac{3}{64}{\rm i} (\Gamma_{a})^{\beta}{}_{\lambda} W_{\beta}\,^{\rho} \lambda_{i \rho} \lambda_{j \gamma} {W}^{-2} \nabla^{\lambda \gamma}{\lambda^{j}_{\alpha}}+\frac{3}{32}{\rm i} (\Gamma_{a})^{\beta}{}_{\rho} \lambda_{i \beta} \lambda^{\lambda}_{j} {W}^{-3} \nabla^{\rho}\,_{\gamma}{W} \nabla_{\alpha}\,^{\gamma}{\lambda^{j}_{\lambda}}+\frac{3}{32}{\rm i} (\Gamma_{a})^{\beta}{}_{\rho} \lambda_{i \beta} \lambda_{j \lambda} {W}^{-3} \nabla^{\rho \gamma}{W} \nabla_{\alpha}\,^{\lambda}{\lambda^{j}_{\gamma}} - \frac{9}{16}{\rm i} (\Gamma_{a})^{\lambda}{}_{\gamma} W^{\beta}\,_{\rho} \lambda_{j \alpha} \lambda_{i \lambda} \lambda^{j}_{\beta} {W}^{-3} \nabla^{\gamma \rho}{W} - \frac{9}{8}(\Gamma_{a})^{\beta}{}_{\gamma} F_{\alpha \beta} F^{\rho}\,_{\lambda} \lambda_{i \rho} {W}^{-3} \nabla^{\gamma \lambda}{W} - \frac{3}{160}(\Gamma_{a})_{\alpha \lambda} X_{i j} F^{\beta}\,_{\rho} \lambda^{j}_{\beta} {W}^{-3} \nabla^{\lambda \rho}{W} - \frac{9}{8}(\Gamma_{a})^{\lambda}{}_{\gamma} F_{\alpha}\,^{\beta} F_{\beta \rho} \lambda_{i \lambda} {W}^{-3} \nabla^{\gamma \rho}{W} - \frac{9}{16}(\Gamma_{a})^{\lambda}{}_{\gamma} F_{\beta \rho} \lambda_{i \lambda} {W}^{-3} \nabla^{\gamma \beta}{W} \nabla_{\alpha}\,^{\rho}{W} - \frac{9}{16}{\rm i} (\Gamma_{a})^{\lambda}{}_{\gamma} F^{\beta}\,_{\rho} \lambda_{j \alpha} \lambda_{i \lambda} \lambda^{j}_{\beta} {W}^{-4} \nabla^{\gamma \rho}{W} - \frac{3}{32}{\rm i} (\Gamma_{a})^{\beta}{}_{\rho} \lambda^{\lambda}_{i} \lambda_{j \beta} {W}^{-3} \nabla^{\rho}\,_{\gamma}{W} \nabla_{\alpha}\,^{\gamma}{\lambda^{j}_{\lambda}} - \frac{3}{32}{\rm i} (\Gamma_{a})^{\beta}{}_{\rho} \lambda_{i \lambda} \lambda_{j \beta} {W}^{-3} \nabla^{\rho \gamma}{W} \nabla_{\alpha}\,^{\lambda}{\lambda^{j}_{\gamma}}+\frac{9}{32}{\rm i} (\Gamma_{a})^{\lambda}{}_{\gamma} W^{\beta}\,_{\rho} \lambda_{j \alpha} \lambda_{i \beta} \lambda^{j}_{\lambda} {W}^{-3} \nabla^{\gamma \rho}{W}+\frac{81}{256}{\rm i} (\Gamma_{a})^{\rho}{}_{\lambda} W_{\alpha}\,^{\beta} \lambda_{i \beta} \lambda_{j \rho} \lambda^{j}_{\gamma} {W}^{-3} \nabla^{\lambda \gamma}{W}%
 - \frac{3}{32}{\rm i} (\Gamma_{a})^{\beta}{}_{\rho} \lambda_{i \lambda} \lambda_{j \beta} {W}^{-3} \nabla^{\rho}\,_{\alpha}{W} \nabla^{\lambda \gamma}{\lambda^{j}_{\gamma}} - \frac{3}{32}{\rm i} (\Gamma_{a})^{\beta}{}_{\rho} \lambda_{i \alpha} \lambda_{j \beta} {W}^{-3} \nabla^{\rho}\,_{\gamma}{W} \nabla^{\gamma \lambda}{\lambda^{j}_{\lambda}}+\frac{9}{64}{\rm i} (\Gamma_{a})^{\lambda}{}_{\gamma} W^{\beta}\,_{\rho} \lambda_{i \alpha} \lambda_{j \lambda} \lambda^{j}_{\beta} {W}^{-3} \nabla^{\gamma \rho}{W}+\frac{3}{32}(\Gamma_{a})^{\lambda}{}_{\gamma} \lambda^{\beta}_{i} \lambda_{j \lambda} W_{\alpha \beta \rho}\,^{j} {W}^{-2} \nabla^{\gamma \rho}{W} - \frac{9}{32}(\Gamma_{a})^{\rho}{}_{\lambda} X_{i j} F_{\alpha \beta} \lambda^{j}_{\rho} {W}^{-3} \nabla^{\lambda \beta}{W}+\frac{9}{16}{\rm i} (\Gamma_{a})^{\lambda}{}_{\gamma} F^{\beta}\,_{\rho} \lambda_{j \alpha} \lambda_{i \beta} \lambda^{j}_{\lambda} {W}^{-4} \nabla^{\gamma \rho}{W}+\frac{3}{32}{\rm i} (\Gamma_{a})^{\lambda}{}_{\gamma} F^{\beta}\,_{\rho} \lambda_{i \beta} \lambda_{j \lambda} {W}^{-3} \nabla^{\gamma \rho}{\lambda^{j}_{\alpha}}+\frac{15}{128}{\rm i} (\Gamma_{a})^{\beta \gamma} W_{\alpha}\,^{\lambda} F_{\beta}\,^{\rho} \lambda_{i \rho} \lambda_{j \gamma} \lambda^{j}_{\lambda} {W}^{-3} - \frac{45}{64}{\rm i} (\Gamma_{a})^{\beta \lambda} W_{\beta}\,^{\rho} \lambda_{i \lambda} \lambda_{j \rho} \lambda^{j}_{\gamma} {W}^{-3} \nabla_{\alpha}\,^{\gamma}{W}+\frac{27}{16}{\rm i} (\Gamma_{a})^{\rho \beta} W_{\rho}\,^{\lambda} F_{\alpha \beta} \lambda^{\gamma}_{i} \lambda_{j \lambda} \lambda^{j}_{\gamma} {W}^{-3} - \frac{117}{128}{\rm i} (\Gamma_{a})^{\rho \lambda} W_{\rho}\,^{\beta} F_{\alpha \beta} \lambda^{\gamma}_{i} \lambda_{j \lambda} \lambda^{j}_{\gamma} {W}^{-3} - \frac{51}{128}{\rm i} (\Gamma_{a})^{\beta \rho} X_{j k} W_{\alpha \beta} \lambda^{\lambda}_{i} \lambda^{j}_{\rho} \lambda^{k}_{\lambda} {W}^{-3}+\frac{27}{256}{\rm i} (\Gamma_{a})^{\beta \lambda} W_{\beta \rho} \lambda^{\gamma}_{i} \lambda_{j \lambda} \lambda^{j}_{\gamma} {W}^{-3} \nabla_{\alpha}\,^{\rho}{W} - \frac{21}{128}{\rm i} (\Gamma_{a})^{\beta \lambda} X_{j k} W_{\beta}\,^{\rho} \lambda_{i \alpha} \lambda^{j}_{\lambda} \lambda^{k}_{\rho} {W}^{-3}+\frac{9}{32}{\rm i} (\Gamma_{a})^{\beta \lambda} W_{\beta}\,^{\rho} \lambda_{i \gamma} \lambda_{j \lambda} \lambda^{j}_{\rho} {W}^{-3} \nabla_{\alpha}\,^{\gamma}{W}+\frac{153}{128}(\Gamma_{a})^{\beta \lambda} W_{\beta}\,^{\rho} \lambda_{j \alpha} \lambda^{\gamma}_{i} \lambda^{j}_{\lambda} \lambda_{k \rho} \lambda^{k}_{\gamma} {W}^{-4} - \frac{207}{1280}{\rm i} (\Gamma_{a})_{\alpha}{}^{\beta} X_{i j} W_{\beta}\,^{\rho} \lambda^{j \lambda} \lambda_{k \rho} \lambda^{k}_{\lambda} {W}^{-3} - \frac{333}{256}(\Gamma_{a})^{\beta \lambda} W_{\beta}\,^{\rho} \lambda_{j \alpha} \lambda_{i \lambda} \lambda^{j \gamma} \lambda_{k \rho} \lambda^{k}_{\gamma} {W}^{-4} - \frac{9}{32}(\Gamma_{a})^{\beta \rho} \lambda_{i \beta} \lambda_{j \rho} \lambda^{\lambda}_{k} \lambda^{k}_{\gamma} {W}^{-4} \nabla_{\alpha}\,^{\gamma}{\lambda^{j}_{\lambda}} - \frac{315}{256}(\Gamma_{a})^{\rho \lambda} W_{\alpha}\,^{\beta} \lambda_{i \rho} \lambda_{j \lambda} \lambda^{j \gamma} \lambda_{k \beta} \lambda^{k}_{\gamma} {W}^{-4}%
+\frac{9}{32}(\Gamma_{a})^{\beta \rho} \lambda_{j \alpha} \lambda_{i \beta} \lambda^{j}_{\lambda} \lambda_{k \rho} {W}^{-4} \nabla^{\lambda \gamma}{\lambda^{k}_{\gamma}}+\frac{27}{16}{\rm i} (\Gamma_{a})^{\beta \gamma} F_{\alpha \beta} F^{\rho \lambda} \lambda_{i \gamma} \lambda_{j \rho} \lambda^{j}_{\lambda} {W}^{-4}+\frac{9}{80}{\rm i} (\Gamma_{a})_{\alpha}{}^{\lambda} X_{i j} F^{\beta \rho} \lambda^{j}_{\lambda} \lambda_{k \beta} \lambda^{k}_{\rho} {W}^{-4} - \frac{9}{8}{\rm i} (\Gamma_{a})^{\lambda \gamma} F_{\alpha}\,^{\beta} F_{\beta}\,^{\rho} \lambda_{i \lambda} \lambda_{j \gamma} \lambda^{j}_{\rho} {W}^{-4} - \frac{27}{32}{\rm i} (\Gamma_{a})^{\rho \lambda} X_{j k} F_{\alpha}\,^{\beta} \lambda_{i \rho} \lambda^{j}_{\lambda} \lambda^{k}_{\beta} {W}^{-4} - \frac{9}{16}{\rm i} (\Gamma_{a})^{\lambda \gamma} F^{\beta}\,_{\rho} \lambda_{i \lambda} \lambda_{j \gamma} \lambda^{j}_{\beta} {W}^{-4} \nabla_{\alpha}\,^{\rho}{W} - \frac{9}{8}(\Gamma_{a})^{\lambda \gamma} F^{\beta \rho} \lambda_{j \alpha} \lambda_{i \lambda} \lambda^{j}_{\gamma} \lambda_{k \beta} \lambda^{k}_{\rho} {W}^{-5}+\frac{27}{640}{\rm i} (\Gamma_{a})_{\alpha \beta} \lambda_{j \rho} {W}^{-2} \nabla^{\beta \lambda}{\lambda_{i \lambda}} \nabla^{\rho \gamma}{\lambda^{j}_{\gamma}}+\frac{3}{1280}{\rm i} (\Gamma_{a})_{\alpha \lambda} W^{\beta \rho} \lambda_{j \beta} \lambda^{j}_{\rho} {W}^{-2} \nabla^{\lambda \gamma}{\lambda_{i \gamma}} - \frac{9}{32}(\Gamma_{a})^{\rho}{}_{\lambda} F_{\alpha}\,^{\beta} F_{\rho \beta} {W}^{-2} \nabla^{\lambda \gamma}{\lambda_{i \gamma}} - \frac{9}{32}(\Gamma_{a})^{\beta}{}_{\lambda} F_{\beta \rho} {W}^{-2} \nabla_{\alpha}\,^{\rho}{W} \nabla^{\lambda \gamma}{\lambda_{i \gamma}}+\frac{3}{64}{\rm i} (\Gamma_{a})^{\beta}{}_{\lambda} F_{\beta}\,^{\rho} \lambda_{j \alpha} \lambda^{j}_{\rho} {W}^{-3} \nabla^{\lambda \gamma}{\lambda_{i \gamma}} - \frac{9}{32}(\Gamma_{a})^{\beta}{}_{\gamma} F_{\beta}\,^{\rho} \lambda_{i \rho} {W}^{-2} \nabla^{\gamma \lambda}{F_{\alpha \lambda}} - \frac{21}{64}(\Gamma_{a})^{\beta}{}_{\lambda} F_{\beta}\,^{\rho} \lambda_{j \rho} {W}^{-2} \nabla^{\lambda}\,_{\alpha}{X_{i}\,^{j}}+\frac{9}{64}(\Gamma_{a})^{\beta}{}_{\lambda} F_{\beta}\,^{\rho} \lambda_{i \rho} {W}^{-2} \nabla^{\lambda}\,_{\gamma}{\nabla_{\alpha}\,^{\gamma}{W}}+\frac{621}{320}(\Gamma_{a})_{\alpha}{}^{\beta} W^{\lambda \gamma} F_{\beta}\,^{\rho} F_{\lambda \gamma} \lambda_{i \rho} {W}^{-2} - \frac{27}{64}(\Gamma_{a})_{\rho \lambda} F_{\alpha \beta} {W}^{-2} \nabla^{\rho \beta}{W} \nabla^{\lambda \gamma}{\lambda_{i \gamma}}+\frac{9}{32}(\Gamma_{a})_{\beta \rho} {W}^{-2} \nabla^{\beta}\,_{\gamma}{W} \nabla_{\alpha}\,^{\gamma}{W} \nabla^{\rho \lambda}{\lambda_{i \lambda}} - \frac{15}{128}{\rm i} (\Gamma_{a})_{\beta \rho} \lambda_{j \alpha} \lambda^{j}_{\lambda} {W}^{-3} \nabla^{\beta \lambda}{W} \nabla^{\rho \gamma}{\lambda_{i \gamma}}+\frac{3}{64}{\rm i} (\Gamma_{a})_{\beta \rho} \lambda_{j \lambda} {W}^{-2} \nabla^{\beta \lambda}{\lambda^{j}_{\alpha}} \nabla^{\rho \gamma}{\lambda_{i \gamma}}%
+\frac{9}{32}(\Gamma_{a})_{\rho \lambda} \lambda_{i \gamma} {W}^{-2} \nabla^{\rho \gamma}{W} \nabla^{\lambda \beta}{F_{\alpha \beta}}+\frac{3}{16}(\Gamma_{a})_{\beta \rho} \lambda_{j \lambda} {W}^{-2} \nabla^{\beta \lambda}{W} \nabla^{\rho}\,_{\alpha}{X_{i}\,^{j}} - \frac{9}{64}(\Gamma_{a})_{\beta \rho} \lambda_{i \lambda} {W}^{-2} \nabla^{\beta \lambda}{W} \nabla^{\rho}\,_{\gamma}{\nabla_{\alpha}\,^{\gamma}{W}} - \frac{243}{320}(\Gamma_{a})_{\alpha \lambda} W^{\beta \rho} F_{\beta \rho} \lambda_{i \gamma} {W}^{-2} \nabla^{\lambda \gamma}{W}+\frac{3}{64}{\rm i} (\Gamma_{a})^{\beta}{}_{\rho} \lambda_{i \lambda} {W}^{-2} \nabla^{\rho \gamma}{\lambda_{j \gamma}} \nabla_{\alpha}\,^{\lambda}{\lambda^{j}_{\beta}}+\frac{9}{80}{\rm i} (\Gamma_{a})_{\alpha \beta} \lambda_{i \rho} {W}^{-2} \nabla^{\beta \lambda}{\lambda_{j \lambda}} \nabla^{\rho \gamma}{\lambda^{j}_{\gamma}} - \frac{21}{128}{\rm i} (\Gamma_{a})_{\alpha \lambda} W^{\beta \rho} \lambda_{i \beta} \lambda_{j \rho} {W}^{-2} \nabla^{\lambda \gamma}{\lambda^{j}_{\gamma}}+\frac{9}{32}{\rm i} (\Gamma_{a})^{\beta}{}_{\lambda} F_{\beta}\,^{\rho} \lambda_{j \alpha} \lambda_{i \rho} {W}^{-3} \nabla^{\lambda \gamma}{\lambda^{j}_{\gamma}} - \frac{15}{64}{\rm i} (\Gamma_{a})_{\beta \rho} \lambda_{j \alpha} \lambda_{i \lambda} {W}^{-3} \nabla^{\beta \lambda}{W} \nabla^{\rho \gamma}{\lambda^{j}_{\gamma}} - \frac{45}{64}{\rm i} (\Gamma_{a})^{\beta}{}_{\rho} F_{\alpha \beta} \lambda^{\lambda}_{i} \lambda_{j \lambda} {W}^{-3} \nabla^{\rho \gamma}{\lambda^{j}_{\gamma}} - \frac{9}{640}{\rm i} (\Gamma_{a})_{\alpha \beta} X_{i j} \lambda^{j \rho} \lambda_{k \rho} {W}^{-3} \nabla^{\beta \lambda}{\lambda^{k}_{\lambda}} - \frac{15}{128}{\rm i} (\Gamma_{a})_{\beta \rho} \lambda^{\lambda}_{i} \lambda_{j \lambda} {W}^{-3} \nabla^{\beta}\,_{\alpha}{W} \nabla^{\rho \gamma}{\lambda^{j}_{\gamma}}+\frac{3}{4}{\rm i} (\Gamma_{a})^{\beta}{}_{\rho} \lambda_{i \beta} \lambda_{j \lambda} {W}^{-3} \nabla_{\alpha}\,^{\lambda}{W} \nabla^{\rho \gamma}{\lambda^{j}_{\gamma}}+\frac{27}{128}(\Gamma_{a})^{\beta}{}_{\rho} \lambda_{j \alpha} \lambda_{i \beta} \lambda^{j \lambda} \lambda_{k \lambda} {W}^{-4} \nabla^{\rho \gamma}{\lambda^{k}_{\gamma}}+\frac{9}{64}{\rm i} (\Gamma_{a})^{\beta}{}_{\rho} X_{j k} \lambda_{i \alpha} \lambda^{j}_{\beta} {W}^{-3} \nabla^{\rho \lambda}{\lambda^{k}_{\lambda}} - \frac{15}{64}{\rm i} (\Gamma_{a})^{\beta}{}_{\rho} \lambda_{i \lambda} \lambda_{j \beta} {W}^{-3} \nabla_{\alpha}\,^{\lambda}{W} \nabla^{\rho \gamma}{\lambda^{j}_{\gamma}} - \frac{9}{16}(\Gamma_{a})^{\beta}{}_{\rho} \lambda_{j \alpha} \lambda^{\lambda}_{i} \lambda^{j}_{\beta} \lambda_{k \lambda} {W}^{-4} \nabla^{\rho \gamma}{\lambda^{k}_{\gamma}}+\frac{3}{64}{\rm i} (\Gamma_{a})^{\rho}{}_{\lambda} \lambda^{\gamma}_{i} \lambda_{j \rho} \lambda^{j}_{\gamma} {W}^{-3} \nabla^{\lambda \beta}{F_{\alpha \beta}} - \frac{9}{256}{\rm i} (\Gamma_{a})^{\rho}{}_{\lambda} W_{\alpha \beta} \lambda^{\gamma}_{i} \lambda_{j \rho} \lambda^{j}_{\gamma} {W}^{-3} \nabla^{\lambda \beta}{W}+\frac{3}{32}{\rm i} (\Gamma_{a})^{\beta}{}_{\rho} \lambda^{\lambda}_{i} \lambda_{j \beta} \lambda_{k \lambda} {W}^{-3} \nabla^{\rho}\,_{\alpha}{X^{j k}}%
 - \frac{3}{128}{\rm i} (\Gamma_{a})^{\beta}{}_{\rho} \lambda^{\lambda}_{i} \lambda_{j \beta} \lambda^{j}_{\lambda} {W}^{-3} \nabla^{\rho}\,_{\gamma}{\nabla_{\alpha}\,^{\gamma}{W}} - \frac{477}{640}{\rm i} (\Gamma_{a})_{\alpha}{}^{\lambda} W^{\beta \rho} F_{\beta \rho} \lambda^{\gamma}_{i} \lambda_{j \lambda} \lambda^{j}_{\gamma} {W}^{-3} - \frac{153}{1280}{\rm i} (\Gamma_{a})_{\alpha}{}^{\rho} \lambda^{\lambda}_{i} \lambda_{k \rho} \lambda^{k \beta} \lambda_{j \lambda} X^{j}_{\beta} {W}^{-3} - \frac{27}{32}(\Gamma_{a})^{\rho}{}_{\gamma} F_{\alpha}\,^{\beta} \lambda_{i \beta} {W}^{-2} \nabla^{\gamma \lambda}{F_{\rho \lambda}} - \frac{9}{16}(\Gamma_{a})^{\beta}{}_{\lambda} \lambda_{i \gamma} {W}^{-2} \nabla_{\alpha}\,^{\gamma}{W} \nabla^{\lambda \rho}{F_{\beta \rho}} - \frac{15}{64}{\rm i} (\Gamma_{a})^{\beta}{}_{\lambda} \lambda_{j \alpha} \lambda^{\gamma}_{i} \lambda^{j}_{\gamma} {W}^{-3} \nabla^{\lambda \rho}{F_{\beta \rho}} - \frac{3}{64}{\rm i} (\Gamma_{a})^{\beta}{}_{\rho} \lambda^{\lambda}_{i} \lambda_{j \lambda} {W}^{-2} \nabla^{\rho}\,_{\gamma}{\nabla_{\alpha}\,^{\gamma}{\lambda^{j}_{\beta}}} - \frac{3}{64}{\rm i} (\Gamma_{a})_{\beta \rho} \lambda^{\lambda}_{i} \lambda_{j \lambda} {W}^{-2} \nabla^{\beta \gamma}{\nabla^{\rho}\,_{\alpha}{\lambda^{j}_{\gamma}}} - \frac{63}{256}{\rm i} (\Gamma_{a})^{\beta}{}_{\rho} \lambda^{\lambda}_{i} \lambda_{j \lambda} \lambda^{j}_{\gamma} {W}^{-2} \nabla^{\rho \gamma}{W_{\alpha \beta}} - \frac{51}{320}{\rm i} (\Gamma_{a})_{\alpha \beta} \lambda^{\rho}_{i} \lambda_{j \rho} {W}^{-2} \nabla^{\beta}\,_{\gamma}{\nabla^{\gamma \lambda}{\lambda^{j}_{\lambda}}}+\frac{9}{256}{\rm i} (\Gamma_{a})^{\beta}{}_{\lambda} \lambda^{\gamma}_{i} \lambda_{j \gamma} \lambda^{j \rho} {W}^{-2} \nabla^{\lambda}\,_{\alpha}{W_{\beta \rho}} - \frac{27}{512}{\rm i} (\Gamma_{a})_{\alpha \lambda} \lambda^{\gamma}_{i} \lambda_{j \gamma} \lambda^{j \beta} {W}^{-2} \nabla^{\lambda \rho}{W_{\beta \rho}} - \frac{129}{1280}{\rm i} \Phi^{\beta \rho}\,_{j k} (\Gamma_{a})_{\alpha \beta} \lambda^{\lambda}_{i} \lambda^{j}_{\rho} \lambda^{k}_{\lambda} {W}^{-2} - \frac{9}{512}{\rm i} (\Gamma_{a})_{\rho \lambda} \lambda^{\gamma}_{i} \lambda_{j \gamma} \lambda^{j \beta} {W}^{-2} \nabla^{\rho \lambda}{W_{\alpha \beta}}+\frac{9}{128}{\rm i} (\Gamma_{a})_{\alpha}{}^{\beta} \lambda^{\lambda}_{i} \lambda_{j \lambda} \lambda^{j}_{\gamma} {W}^{-2} \nabla^{\gamma \rho}{W_{\beta \rho}} - \frac{15}{128}(\Gamma_{a})_{\beta \rho} \lambda_{j \lambda} {W}^{-2} \nabla_{\alpha}\,^{\lambda}{W} \nabla^{\beta \rho}{X_{i}\,^{j}} - \frac{9}{128}{\rm i} (\Gamma_{a})_{\beta \rho} \lambda_{k \alpha} \lambda^{k \lambda} \lambda_{j \lambda} {W}^{-3} \nabla^{\beta \rho}{X_{i}\,^{j}}+\frac{21}{128}{\rm i} (\Gamma_{a})_{\beta \rho} \lambda^{\lambda}_{i} \lambda_{j \lambda} {W}^{-2} \nabla^{\beta \rho}{\nabla_{\alpha}\,^{\gamma}{\lambda^{j}_{\gamma}}} - \frac{423}{2048}(\Gamma_{a})_{\beta \rho} \lambda^{\lambda}_{i} \lambda_{j \lambda} {W}^{-1} \nabla^{\beta \rho}{X^{j}_{\alpha}}+\frac{19}{256}{\rm i} \Phi^{\beta \rho}\,_{i j} (\Gamma_{a})_{\alpha \beta} \lambda^{j \lambda} \lambda_{k \rho} \lambda^{k}_{\lambda} {W}^{-2}%
+\frac{63}{1280}{\rm i} \Phi^{\beta \rho}\,_{j k} (\Gamma_{a})_{\alpha \beta} \lambda_{i \rho} \lambda^{j \lambda} \lambda^{k}_{\lambda} {W}^{-2}+\frac{27}{64}(\Gamma_{a})_{\rho \lambda} F_{\alpha}\,^{\beta} \lambda_{i \beta} {W}^{-2} \nabla^{\rho}\,_{\gamma}{\nabla^{\lambda \gamma}{W}}+\frac{9}{32}(\Gamma_{a})_{\beta \rho} \lambda_{i \lambda} {W}^{-2} \nabla_{\alpha}\,^{\lambda}{W} \nabla^{\beta}\,_{\gamma}{\nabla^{\rho \gamma}{W}}+\frac{15}{128}{\rm i} (\Gamma_{a})_{\beta \rho} \lambda_{j \alpha} \lambda^{\lambda}_{i} \lambda^{j}_{\lambda} {W}^{-3} \nabla^{\beta}\,_{\gamma}{\nabla^{\rho \gamma}{W}}+\frac{3}{64}{\rm i} (\Gamma_{a})_{\beta \rho} \lambda^{\lambda}_{i} \lambda_{j \lambda} {W}^{-2} \nabla^{\beta}\,_{\gamma}{\nabla^{\rho \gamma}{\lambda^{j}_{\alpha}}} - \frac{15}{256}(\Gamma_{a})_{\beta \rho} X_{j k} \lambda_{i \alpha} {W}^{-2} \nabla^{\beta \rho}{X^{j k}} - \frac{3}{32}{\rm i} (\Gamma_{a})_{\beta \rho} \lambda_{j \alpha} \lambda^{\lambda}_{i} \lambda_{k \lambda} {W}^{-3} \nabla^{\beta \rho}{X^{j k}}+\frac{3}{32}(\Gamma_{a})^{\beta}{}_{\rho} X_{i j} X^{j}\,_{k} \lambda^{k}_{\beta} {W}^{-3} \nabla^{\rho}\,_{\alpha}{W}+\frac{63}{64}{\rm i} (\Gamma_{a})^{\beta \rho} X_{j k} X^{j}\,_{l} \lambda^{k}_{\alpha} \lambda_{i \beta} \lambda^{l}_{\rho} {W}^{-4} - \frac{27}{64}{\rm i} (\Gamma_{a})^{\beta \rho} X_{i j} X^{j}\,_{k} \lambda_{l \alpha} \lambda^{k}_{\beta} \lambda^{l}_{\rho} {W}^{-4}+\frac{9}{32}(\Gamma_{a})_{\beta \rho} {W}^{-2} \nabla^{\beta}\,_{\gamma}{W} \nabla^{\rho \gamma}{W} \nabla_{\alpha}\,^{\lambda}{\lambda_{i \lambda}}+\frac{15}{64}(\Gamma_{a})_{\beta \rho} X_{i j} \lambda^{j}_{\alpha} {W}^{-3} \nabla^{\beta}\,_{\lambda}{W} \nabla^{\rho \lambda}{W}+\frac{81}{128}{\rm i} (\Gamma_{a})^{\beta}{}_{\rho} \lambda_{i \beta} \lambda_{j \lambda} {W}^{-3} \nabla^{\rho \lambda}{W} \nabla_{\alpha}\,^{\gamma}{\lambda^{j}_{\gamma}}+\frac{9}{16}(\Gamma_{a})^{\beta}{}_{\rho} X_{i j} F_{\alpha \beta} \lambda^{j}_{\lambda} {W}^{-3} \nabla^{\rho \lambda}{W}+\frac{21}{160}(\Gamma_{a})_{\alpha \beta} X_{i j} X^{j}\,_{k} \lambda^{k}_{\rho} {W}^{-3} \nabla^{\beta \rho}{W} - \frac{3}{16}(\Gamma_{a})_{\beta \rho} X_{i j} \lambda^{j}_{\lambda} {W}^{-3} \nabla^{\beta}\,_{\alpha}{W} \nabla^{\rho \lambda}{W} - \frac{9}{16}{\rm i} (\Gamma_{a})^{\beta}{}_{\rho} X_{j k} \lambda^{j}_{\alpha} \lambda_{i \beta} \lambda^{k}_{\lambda} {W}^{-4} \nabla^{\rho \lambda}{W}+\frac{9}{128}{\rm i} (\Gamma_{a})^{\beta}{}_{\rho} \lambda_{j \beta} \lambda^{j}_{\lambda} {W}^{-3} \nabla^{\rho \lambda}{W} \nabla_{\alpha}\,^{\gamma}{\lambda_{i \gamma}}+\frac{15}{64}(\Gamma_{a})^{\beta}{}_{\rho} X_{i j} \lambda^{j}_{\beta} {W}^{-3} \nabla^{\rho}\,_{\lambda}{W} \nabla_{\alpha}\,^{\lambda}{W} - \frac{27}{64}{\rm i} (\Gamma_{a})^{\beta}{}_{\rho} X_{i j} \lambda_{k \alpha} \lambda^{j}_{\lambda} \lambda^{k}_{\beta} {W}^{-4} \nabla^{\rho \lambda}{W}%
 - \frac{3}{32}{\rm i} (\Gamma_{a})^{\beta}{}_{\rho} X_{i j} \lambda^{j}_{\lambda} \lambda_{k \beta} {W}^{-3} \nabla^{\rho \lambda}{\lambda^{k}_{\alpha}} - \frac{9}{8}{\rm i} (\Gamma_{a})^{\beta \rho} X_{j k} F_{\alpha \beta} \lambda_{i \rho} \lambda^{j \lambda} \lambda^{k}_{\lambda} {W}^{-4}+\frac{9}{320}{\rm i} (\Gamma_{a})_{\alpha}{}^{\beta} X_{i j} X_{k l} \lambda^{j}_{\beta} \lambda^{k \rho} \lambda^{l}_{\rho} {W}^{-4} - \frac{45}{64}{\rm i} (\Gamma_{a})^{\beta \rho} X_{j k} \lambda_{i \beta} \lambda^{j}_{\rho} \lambda^{k}_{\lambda} {W}^{-4} \nabla_{\alpha}\,^{\lambda}{W}+\frac{9}{8}(\Gamma_{a})^{\beta \rho} X_{j k} \lambda_{l \alpha} \lambda_{i \beta} \lambda^{j \lambda} \lambda^{k}_{\lambda} \lambda^{l}_{\rho} {W}^{-5} - \frac{15}{128}{\rm i} (\Gamma_{a})_{\rho \lambda} F_{\alpha}\,^{\beta} \lambda_{j \beta} \lambda^{j \gamma} {W}^{-3} \nabla^{\rho \lambda}{\lambda_{i \gamma}} - \frac{15}{256}{\rm i} (\Gamma_{a})_{\beta \rho} \lambda^{\lambda}_{j} \lambda^{j}_{\gamma} {W}^{-3} \nabla_{\alpha}\,^{\gamma}{W} \nabla^{\beta \rho}{\lambda_{i \lambda}}+\frac{3}{128}{\rm i} (\Gamma_{a})_{\beta \rho} X_{j k} \lambda^{j}_{\alpha} \lambda^{k \lambda} {W}^{-3} \nabla^{\beta \rho}{\lambda_{i \lambda}} - \frac{3}{64}{\rm i} (\Gamma_{a})_{\beta \rho} X_{j k} \lambda^{j \lambda} \lambda^{k}_{\lambda} {W}^{-3} \nabla^{\beta \rho}{\lambda_{i \alpha}}+\frac{9}{128}(\Gamma_{a})_{\beta \rho} \lambda_{j \alpha} \lambda^{j \lambda} \lambda_{k \lambda} \lambda^{k \gamma} {W}^{-4} \nabla^{\beta \rho}{\lambda_{i \gamma}}+\frac{9}{64}{\rm i} (\Gamma_{a})_{\rho \lambda} \lambda^{\gamma}_{i} \lambda_{j \gamma} \lambda^{j \beta} {W}^{-3} \nabla^{\rho \lambda}{F_{\alpha \beta}} - \frac{9}{128}{\rm i} (\Gamma_{a})_{\rho \lambda} W_{\alpha}\,^{\beta} \lambda^{\gamma}_{i} \lambda_{j \beta} \lambda^{j}_{\gamma} {W}^{-3} \nabla^{\rho \lambda}{W} - \frac{9}{128}{\rm i} (\Gamma_{a})_{\beta \rho} \lambda^{\lambda}_{i} \lambda_{j \lambda} \lambda^{j}_{\gamma} {W}^{-3} \nabla^{\beta \rho}{\nabla_{\alpha}\,^{\gamma}{W}} - \frac{279}{640}{\rm i} (\Gamma_{a})_{\alpha}{}^{\lambda} W_{\lambda}\,^{\beta} F_{\beta}\,^{\rho} \lambda^{\gamma}_{i} \lambda_{j \rho} \lambda^{j}_{\gamma} {W}^{-3}+\frac{279}{1280}{\rm i} (\Gamma_{a})_{\alpha}{}^{\beta} W_{\beta \rho} \lambda^{\lambda}_{i} \lambda_{j \lambda} \lambda^{j}_{\gamma} {W}^{-3} \nabla^{\rho \gamma}{W} - \frac{99}{64}{\rm i} (\Gamma_{a})^{\lambda \beta} W_{\alpha \lambda} F_{\beta}\,^{\rho} \lambda^{\gamma}_{i} \lambda_{j \rho} \lambda^{j}_{\gamma} {W}^{-3}+\frac{207}{2560}{\rm i} (\Gamma_{a})_{\alpha}{}^{\rho} \lambda^{\beta}_{i} \lambda_{k \rho} \lambda^{k \lambda} \lambda_{j \lambda} X^{j}_{\beta} {W}^{-3}+\frac{3}{64}{\rm i} (\Gamma_{a})_{\alpha}{}^{\beta} \lambda^{\rho}_{i} \lambda_{k \rho} \lambda^{k \lambda} \lambda_{j \lambda} X^{j}_{\beta} {W}^{-3} - \frac{51}{2560}{\rm i} (\Gamma_{a})_{\alpha}{}^{\beta} \lambda^{\rho}_{j} \lambda^{j \lambda} \lambda_{k \rho} \lambda^{k}_{\lambda} X_{i \beta} {W}^{-3}+\frac{3}{8}{\rm i} (\Gamma_{a})^{\rho}{}_{\lambda} F_{\alpha}\,^{\beta} \lambda_{j \beta} \lambda^{j}_{\gamma} {W}^{-3} \nabla^{\lambda \gamma}{\lambda_{i \rho}}%
+\frac{3}{16}{\rm i} (\Gamma_{a})^{\beta}{}_{\rho} \lambda_{j \lambda} \lambda^{j}_{\gamma} {W}^{-3} \nabla_{\alpha}\,^{\lambda}{W} \nabla^{\rho \gamma}{\lambda_{i \beta}}+\frac{3}{32}{\rm i} (\Gamma_{a})^{\beta}{}_{\rho} X_{j k} \lambda^{j \lambda} \lambda^{k}_{\lambda} {W}^{-3} \nabla^{\rho}\,_{\alpha}{\lambda_{i \beta}} - \frac{9}{64}(\Gamma_{a})^{\beta}{}_{\rho} \lambda_{j \alpha} \lambda^{j \lambda} \lambda_{k \lambda} \lambda^{k}_{\gamma} {W}^{-4} \nabla^{\rho \gamma}{\lambda_{i \beta}} - \frac{9}{64}{\rm i} (\Gamma_{a})^{\beta}{}_{\rho} \lambda^{\lambda}_{i} \lambda_{j \lambda} \lambda^{j}_{\gamma} {W}^{-3} \nabla^{\rho \gamma}{F_{\alpha \beta}}+\frac{45}{64}{\rm i} (\Gamma_{a})^{\beta}{}_{\rho} W_{\alpha \beta} \lambda^{\lambda}_{i} \lambda_{j \lambda} \lambda^{j}_{\gamma} {W}^{-3} \nabla^{\rho \gamma}{W}+\frac{3}{640}{\rm i} (\Gamma_{a})_{\alpha \beta} \lambda^{\rho}_{k} \lambda^{k}_{\lambda} \lambda_{j \rho} {W}^{-3} \nabla^{\beta \lambda}{X_{i}\,^{j}}+\frac{9}{128}{\rm i} (\Gamma_{a})_{\beta \rho} \lambda^{\lambda}_{i} \lambda_{j \lambda} \lambda^{j}_{\gamma} {W}^{-3} \nabla^{\beta \gamma}{\nabla^{\rho}\,_{\alpha}{W}}+\frac{369}{640}{\rm i} (\Gamma_{a})_{\alpha}{}^{\beta} W^{\rho \lambda} F_{\beta \rho} \lambda^{\gamma}_{i} \lambda_{j \lambda} \lambda^{j}_{\gamma} {W}^{-3} - \frac{81}{1280}{\rm i} (\Gamma_{a})_{\alpha \lambda} W^{\beta}\,_{\rho} \lambda^{\gamma}_{i} \lambda_{j \beta} \lambda^{j}_{\gamma} {W}^{-3} \nabla^{\lambda \rho}{W}+\frac{81}{256}{\rm i} (\Gamma_{a})^{\beta}{}_{\lambda} W_{\beta \rho} \lambda_{j \alpha} \lambda^{\gamma}_{i} \lambda^{j}_{\gamma} {W}^{-3} \nabla^{\lambda \rho}{W}+\frac{3}{8}{\rm i} (\Gamma_{a})^{\rho}{}_{\lambda} F_{\alpha}\,^{\beta} \lambda_{j \rho} \lambda^{j}_{\beta} {W}^{-3} \nabla^{\lambda \gamma}{\lambda_{i \gamma}}+\frac{3}{16}{\rm i} (\Gamma_{a})^{\beta}{}_{\rho} \lambda_{j \beta} \lambda^{j}_{\lambda} {W}^{-3} \nabla_{\alpha}\,^{\lambda}{W} \nabla^{\rho \gamma}{\lambda_{i \gamma}} - \frac{21}{320}{\rm i} (\Gamma_{a})_{\alpha \beta} X_{j k} \lambda^{j \rho} \lambda^{k}_{\rho} {W}^{-3} \nabla^{\beta \lambda}{\lambda_{i \lambda}} - \frac{9}{64}(\Gamma_{a})^{\beta}{}_{\rho} \lambda_{j \alpha} \lambda^{j \lambda} \lambda_{k \beta} \lambda^{k}_{\lambda} {W}^{-4} \nabla^{\rho \gamma}{\lambda_{i \gamma}} - \frac{9}{128}{\rm i} (\Gamma_{a})^{\beta}{}_{\rho} \lambda_{k \beta} \lambda^{k \lambda} \lambda_{j \lambda} {W}^{-3} \nabla^{\rho}\,_{\alpha}{X_{i}\,^{j}}+\frac{15}{512}{\rm i} (\Gamma_{a})^{\rho \beta} \lambda_{i \alpha} \lambda_{k \rho} \lambda^{k \lambda} \lambda_{j \lambda} X^{j}_{\beta} {W}^{-3}+\frac{15}{256}{\rm i} (\Gamma_{a})_{\beta \rho} X_{i j} \lambda_{k \alpha} \lambda^{k \lambda} {W}^{-3} \nabla^{\beta \rho}{\lambda^{j}_{\lambda}}+\frac{3}{64}{\rm i} (\Gamma_{a})_{\rho \lambda} F_{\alpha}\,^{\beta} \lambda_{i \beta} \lambda^{\gamma}_{j} {W}^{-3} \nabla^{\rho \lambda}{\lambda^{j}_{\gamma}} - \frac{15}{128}{\rm i} (\Gamma_{a})_{\beta \rho} X_{j k} \lambda_{i \alpha} \lambda^{j \lambda} {W}^{-3} \nabla^{\beta \rho}{\lambda^{k}_{\lambda}}+\frac{3}{128}{\rm i} (\Gamma_{a})_{\beta \rho} \lambda_{i \lambda} \lambda^{\gamma}_{j} {W}^{-3} \nabla_{\alpha}\,^{\lambda}{W} \nabla^{\beta \rho}{\lambda^{j}_{\gamma}}%
 - \frac{15}{128}{\rm i} (\Gamma_{a})_{\rho \lambda} F_{\alpha}\,^{\beta} \lambda^{\gamma}_{i} \lambda_{j \gamma} {W}^{-3} \nabla^{\rho \lambda}{\lambda^{j}_{\beta}} - \frac{3}{64}{\rm i} (\Gamma_{a})_{\beta \rho} X_{j k} \lambda^{\lambda}_{i} \lambda^{j}_{\lambda} {W}^{-3} \nabla^{\beta \rho}{\lambda^{k}_{\alpha}} - \frac{15}{256}{\rm i} (\Gamma_{a})_{\beta \rho} \lambda^{\lambda}_{i} \lambda_{j \lambda} {W}^{-3} \nabla_{\alpha}\,^{\gamma}{W} \nabla^{\beta \rho}{\lambda^{j}_{\gamma}}+\frac{9}{128}(\Gamma_{a})_{\beta \rho} \lambda_{j \alpha} \lambda^{\lambda}_{i} \lambda_{k \lambda} \lambda^{k \gamma} {W}^{-4} \nabla^{\beta \rho}{\lambda^{j}_{\gamma}}+\frac{357}{2560}{\rm i} (\Gamma_{a})_{\alpha}{}^{\rho} \lambda^{\lambda}_{i} \lambda_{k \rho} \lambda^{k}_{\lambda} \lambda^{\beta}_{j} X^{j}_{\beta} {W}^{-3} - \frac{3}{64}{\rm i} (\Gamma_{a})^{\beta}{}_{\rho} X_{i j} \lambda_{k \alpha} \lambda^{k}_{\lambda} {W}^{-3} \nabla^{\rho \lambda}{\lambda^{j}_{\beta}}+\frac{3}{32}{\rm i} (\Gamma_{a})^{\rho}{}_{\lambda} F_{\alpha \beta} \lambda^{\gamma}_{i} \lambda_{j \gamma} {W}^{-3} \nabla^{\lambda \beta}{\lambda^{j}_{\rho}} - \frac{3}{64}{\rm i} (\Gamma_{a})^{\beta}{}_{\rho} X_{j k} \lambda^{\lambda}_{i} \lambda^{j}_{\lambda} {W}^{-3} \nabla^{\rho}\,_{\alpha}{\lambda^{k}_{\beta}}+\frac{3}{64}{\rm i} (\Gamma_{a})^{\beta}{}_{\rho} \lambda^{\lambda}_{i} \lambda_{j \lambda} {W}^{-3} \nabla_{\alpha \gamma}{W} \nabla^{\rho \gamma}{\lambda^{j}_{\beta}} - \frac{3}{32}{\rm i} (\Gamma_{a})^{\rho}{}_{\lambda} F_{\alpha}\,^{\beta} \lambda_{i \beta} \lambda_{j \gamma} {W}^{-3} \nabla^{\lambda \gamma}{\lambda^{j}_{\rho}} - \frac{21}{256}{\rm i} (\Gamma_{a})^{\rho}{}_{\lambda} W_{\alpha}\,^{\beta} \lambda_{i \beta} \lambda_{j \gamma} {W}^{-2} \nabla^{\lambda \gamma}{\lambda^{j}_{\rho}} - \frac{3}{64}{\rm i} (\Gamma_{a})^{\beta}{}_{\rho} X_{j k} \lambda_{i \alpha} \lambda^{j}_{\lambda} {W}^{-3} \nabla^{\rho \lambda}{\lambda^{k}_{\beta}} - \frac{3}{64}{\rm i} (\Gamma_{a})^{\beta}{}_{\rho} \lambda_{i \lambda} \lambda_{j \gamma} {W}^{-3} \nabla_{\alpha}\,^{\lambda}{W} \nabla^{\rho \gamma}{\lambda^{j}_{\beta}} - \frac{9}{64}(\Gamma_{a})^{\beta}{}_{\rho} \lambda_{j \alpha} \lambda^{\lambda}_{i} \lambda_{k \lambda} \lambda^{k}_{\gamma} {W}^{-4} \nabla^{\rho \gamma}{\lambda^{j}_{\beta}} - \frac{3}{64}{\rm i} (\Gamma_{a})_{\alpha \beta} X_{j k} \lambda^{\rho}_{i} \lambda^{j}_{\rho} {W}^{-3} \nabla^{\beta \lambda}{\lambda^{k}_{\lambda}} - \frac{9}{64}(\Gamma_{a})^{\beta}{}_{\rho} \lambda_{j \alpha} \lambda^{\lambda}_{i} \lambda_{k \beta} \lambda^{k}_{\lambda} {W}^{-4} \nabla^{\rho \gamma}{\lambda^{j}_{\gamma}}+\frac{3}{128}{\rm i} (\Gamma_{a})^{\beta}{}_{\rho} X_{i j} \lambda_{k \beta} \lambda^{k \lambda} {W}^{-3} \nabla^{\rho}\,_{\alpha}{\lambda^{j}_{\lambda}} - \frac{3}{128}{\rm i} (\Gamma_{a})^{\beta \rho} X_{i j} \lambda_{k \beta} \lambda^{k}_{\lambda} {W}^{-3} \nabla_{\alpha}\,^{\lambda}{\lambda^{j}_{\rho}}+\frac{9}{16}{\rm i} (\Gamma_{a})^{\beta \lambda} X_{i j} F_{\beta}\,^{\rho} \lambda^{j}_{\alpha} \lambda_{k \lambda} \lambda^{k}_{\rho} {W}^{-4} - \frac{3}{64}{\rm i} (\Gamma_{a})^{\beta}{}_{\lambda} F_{\beta}\,^{\rho} \lambda^{\gamma}_{i} \lambda_{j \gamma} {W}^{-3} \nabla^{\lambda}\,_{\alpha}{\lambda^{j}_{\rho}}%
+\frac{3}{64}{\rm i} (\Gamma_{a})^{\beta \lambda} F_{\beta \rho} \lambda^{\gamma}_{i} \lambda_{j \gamma} {W}^{-3} \nabla_{\alpha}\,^{\rho}{\lambda^{j}_{\lambda}} - \frac{15}{64}{\rm i} (\Gamma_{a})_{\alpha}{}^{\beta} F_{\beta \rho} \lambda^{\lambda}_{i} \lambda_{j \lambda} {W}^{-3} \nabla^{\rho \gamma}{\lambda^{j}_{\gamma}}+\frac{9}{8}(\Gamma_{a})^{\rho \gamma} F_{\alpha}\,^{\beta} F_{\rho}\,^{\lambda} F_{\gamma \lambda} \lambda_{i \beta} {W}^{-3} - \frac{3}{16}(\Gamma_{a})^{\beta \lambda} X_{i j} F_{\beta}\,^{\rho} F_{\lambda \rho} \lambda^{j}_{\alpha} {W}^{-3}+\frac{9}{16}(\Gamma_{a})^{\beta \lambda} F_{\beta}\,^{\rho} F_{\lambda \rho} \lambda_{i \gamma} {W}^{-3} \nabla_{\alpha}\,^{\gamma}{W}+\frac{9}{16}{\rm i} (\Gamma_{a})^{\beta \lambda} F_{\beta}\,^{\rho} F_{\lambda \rho} \lambda_{j \alpha} \lambda^{\gamma}_{i} \lambda^{j}_{\gamma} {W}^{-4}+\frac{3}{128}{\rm i} (\Gamma_{a})^{\beta}{}_{\rho} \lambda^{\lambda}_{i} \lambda_{j \lambda} {W}^{-3} \nabla^{\rho}\,_{\gamma}{W} \nabla_{\alpha}\,^{\gamma}{\lambda^{j}_{\beta}}+\frac{3}{128}{\rm i} (\Gamma_{a})_{\beta \rho} \lambda^{\lambda}_{i} \lambda_{j \lambda} {W}^{-3} \nabla^{\beta \gamma}{W} \nabla^{\rho}\,_{\alpha}{\lambda^{j}_{\gamma}}+\frac{129}{640}{\rm i} (\Gamma_{a})_{\alpha \beta} \lambda^{\rho}_{i} \lambda_{j \rho} {W}^{-3} \nabla^{\beta}\,_{\gamma}{W} \nabla^{\gamma \lambda}{\lambda^{j}_{\lambda}}+\frac{9}{8}(\Gamma_{a})^{\rho}{}_{\gamma} F_{\alpha}\,^{\beta} F_{\rho \lambda} \lambda_{i \beta} {W}^{-3} \nabla^{\gamma \lambda}{W}+\frac{9}{16}(\Gamma_{a})^{\beta}{}_{\lambda} F_{\beta \rho} \lambda_{i \gamma} {W}^{-3} \nabla^{\lambda \rho}{W} \nabla_{\alpha}\,^{\gamma}{W}+\frac{9}{16}{\rm i} (\Gamma_{a})^{\beta}{}_{\lambda} F_{\beta \rho} \lambda_{j \alpha} \lambda^{\gamma}_{i} \lambda^{j}_{\gamma} {W}^{-4} \nabla^{\lambda \rho}{W}+\frac{3}{64}{\rm i} (\Gamma_{a})^{\beta}{}_{\lambda} F_{\beta \rho} \lambda^{\gamma}_{i} \lambda_{j \gamma} {W}^{-3} \nabla^{\lambda \rho}{\lambda^{j}_{\alpha}}+\frac{15}{64}{\rm i} (\Gamma_{a})^{\beta}{}_{\lambda} F_{\beta}\,^{\rho} \lambda_{i \alpha} \lambda_{j \rho} {W}^{-3} \nabla^{\lambda \gamma}{\lambda^{j}_{\gamma}} - \frac{57}{80}{\rm i} (\Gamma_{a})_{\alpha}{}^{\beta} W^{\lambda \gamma} F_{\beta}\,^{\rho} \lambda_{i \lambda} \lambda_{j \gamma} \lambda^{j}_{\rho} {W}^{-3}+\frac{3}{64}{\rm i} (\Gamma_{a})^{\beta}{}_{\lambda} F_{\beta}\,^{\rho} \lambda_{i \rho} \lambda^{\gamma}_{j} {W}^{-3} \nabla^{\lambda}\,_{\alpha}{\lambda^{j}_{\gamma}} - \frac{3}{64}{\rm i} (\Gamma_{a})^{\beta \lambda} F_{\beta}\,^{\rho} \lambda_{i \rho} \lambda_{j \gamma} {W}^{-3} \nabla_{\alpha}\,^{\gamma}{\lambda^{j}_{\lambda}} - \frac{3}{16}{\rm i} (\Gamma_{a})_{\alpha}{}^{\beta} F_{\beta}\,^{\rho} \lambda_{i \rho} \lambda_{j \lambda} {W}^{-3} \nabla^{\lambda \gamma}{\lambda^{j}_{\gamma}}+\frac{183}{320}{\rm i} (\Gamma_{a})_{\alpha}{}^{\beta} W^{\lambda \gamma} F_{\beta}\,^{\rho} \lambda_{i \rho} \lambda_{j \lambda} \lambda^{j}_{\gamma} {W}^{-3} - \frac{9}{8}(\Gamma_{a})^{\rho \lambda} F_{\alpha}\,^{\beta} F_{\rho \beta} F_{\lambda}\,^{\gamma} \lambda_{i \gamma} {W}^{-3}%
+\frac{9}{16}(\Gamma_{a})^{\beta \lambda} F_{\beta}\,^{\rho} F_{\lambda \gamma} \lambda_{i \rho} {W}^{-3} \nabla_{\alpha}\,^{\gamma}{W}+\frac{9}{16}{\rm i} (\Gamma_{a})^{\beta \lambda} F_{\beta}\,^{\rho} F_{\lambda}\,^{\gamma} \lambda_{j \alpha} \lambda_{i \rho} \lambda^{j}_{\gamma} {W}^{-4} - \frac{15}{128}{\rm i} (\Gamma_{a})_{\beta \rho} \lambda_{i \alpha} \lambda_{j \lambda} {W}^{-3} \nabla^{\beta \lambda}{W} \nabla^{\rho \gamma}{\lambda^{j}_{\gamma}}+\frac{9}{40}{\rm i} (\Gamma_{a})_{\alpha \lambda} W^{\beta \rho} \lambda_{i \beta} \lambda_{j \rho} \lambda^{j}_{\gamma} {W}^{-3} \nabla^{\lambda \gamma}{W}+\frac{9}{16}(\Gamma_{a})^{\rho}{}_{\lambda} F_{\alpha}\,^{\beta} F_{\rho \beta} \lambda_{i \gamma} {W}^{-3} \nabla^{\lambda \gamma}{W}+\frac{9}{32}(\Gamma_{a})^{\beta}{}_{\lambda} F_{\beta \rho} \lambda_{i \gamma} {W}^{-3} \nabla^{\lambda \gamma}{W} \nabla_{\alpha}\,^{\rho}{W}+\frac{9}{16}(\Gamma_{a})^{\rho}{}_{\gamma} F_{\alpha \beta} F_{\rho}\,^{\lambda} \lambda_{i \lambda} {W}^{-3} \nabla^{\gamma \beta}{W} - \frac{9}{32}(\Gamma_{a})^{\beta}{}_{\lambda} F_{\beta}\,^{\rho} \lambda_{i \rho} {W}^{-3} \nabla^{\lambda}\,_{\gamma}{W} \nabla_{\alpha}\,^{\gamma}{W} - \frac{9}{32}{\rm i} (\Gamma_{a})^{\beta}{}_{\lambda} F_{\beta}\,^{\rho} \lambda_{j \alpha} \lambda_{i \rho} \lambda^{j}_{\gamma} {W}^{-4} \nabla^{\lambda \gamma}{W} - \frac{3}{64}{\rm i} (\Gamma_{a})^{\beta}{}_{\lambda} F_{\beta}\,^{\rho} \lambda_{i \rho} \lambda_{j \gamma} {W}^{-3} \nabla^{\lambda \gamma}{\lambda^{j}_{\alpha}} - \frac{3}{128}{\rm i} (\Gamma_{a})_{\beta \rho} \lambda_{i \lambda} \lambda^{\gamma}_{j} {W}^{-3} \nabla^{\beta \lambda}{W} \nabla^{\rho}\,_{\alpha}{\lambda^{j}_{\gamma}} - \frac{3}{128}{\rm i} (\Gamma_{a})^{\beta}{}_{\rho} \lambda_{i \lambda} \lambda_{j \gamma} {W}^{-3} \nabla^{\rho \lambda}{W} \nabla_{\alpha}\,^{\gamma}{\lambda^{j}_{\beta}}+\frac{9}{256}{\rm i} (\Gamma_{a})^{\rho}{}_{\lambda} W_{\alpha}\,^{\beta} \lambda_{i \gamma} \lambda_{j \rho} \lambda^{j}_{\beta} {W}^{-3} \nabla^{\lambda \gamma}{W}+\frac{3}{20}{\rm i} (\Gamma_{a})_{\alpha \beta} \lambda_{i \rho} \lambda_{j \lambda} {W}^{-3} \nabla^{\beta \rho}{W} \nabla^{\lambda \gamma}{\lambda^{j}_{\gamma}} - \frac{9}{64}{\rm i} (\Gamma_{a})_{\alpha \lambda} W^{\beta \rho} \lambda_{i \gamma} \lambda_{j \beta} \lambda^{j}_{\rho} {W}^{-3} \nabla^{\lambda \gamma}{W}+\frac{9}{32}{\rm i} (\Gamma_{a})^{\beta}{}_{\lambda} F_{\beta}\,^{\rho} \lambda_{j \alpha} \lambda_{i \gamma} \lambda^{j}_{\rho} {W}^{-4} \nabla^{\lambda \gamma}{W} - \frac{459}{512}(\Gamma_{a})^{\beta \rho} W_{\alpha \beta} \lambda_{i \rho} \lambda^{\lambda}_{j} \lambda^{j \gamma} \lambda_{k \lambda} \lambda^{k}_{\gamma} {W}^{-4} - \frac{27}{16}{\rm i} (\Gamma_{a})^{\beta \rho} F_{\alpha \beta} F_{\rho}\,^{\lambda} \lambda^{\gamma}_{i} \lambda_{j \lambda} \lambda^{j}_{\gamma} {W}^{-4} - \frac{27}{160}{\rm i} (\Gamma_{a})_{\alpha}{}^{\beta} X_{i j} F_{\beta}\,^{\rho} \lambda^{j \lambda} \lambda_{k \rho} \lambda^{k}_{\lambda} {W}^{-4} - \frac{9}{4}{\rm i} (\Gamma_{a})^{\rho \gamma} F_{\alpha}\,^{\beta} F_{\rho}\,^{\lambda} \lambda_{i \gamma} \lambda_{j \beta} \lambda^{j}_{\lambda} {W}^{-4}%
 - \frac{9}{8}{\rm i} (\Gamma_{a})^{\beta \lambda} F_{\beta}\,^{\rho} \lambda_{i \lambda} \lambda_{j \rho} \lambda^{j}_{\gamma} {W}^{-4} \nabla_{\alpha}\,^{\gamma}{W} - \frac{9}{8}(\Gamma_{a})^{\beta \lambda} F_{\beta}\,^{\rho} \lambda_{j \alpha} \lambda_{i \lambda} \lambda^{j \gamma} \lambda_{k \rho} \lambda^{k}_{\gamma} {W}^{-5} - \frac{9}{128}(\Gamma_{a})^{\beta}{}_{\rho} \lambda^{\lambda}_{i} \lambda_{j \beta} \lambda_{k \lambda} \lambda^{k \gamma} {W}^{-4} \nabla^{\rho}\,_{\alpha}{\lambda^{j}_{\gamma}}+\frac{9}{128}(\Gamma_{a})^{\beta \rho} \lambda^{\lambda}_{i} \lambda_{j \beta} \lambda_{k \lambda} \lambda^{k}_{\gamma} {W}^{-4} \nabla_{\alpha}\,^{\gamma}{\lambda^{j}_{\rho}} - \frac{9}{160}(\Gamma_{a})_{\alpha}{}^{\beta} \lambda^{\rho}_{i} \lambda_{j \beta} \lambda_{k \rho} \lambda^{k}_{\lambda} {W}^{-4} \nabla^{\lambda \gamma}{\lambda^{j}_{\gamma}} - \frac{9}{64}(\Gamma_{a})^{\beta}{}_{\rho} \lambda_{j \alpha} \lambda^{\lambda}_{i} \lambda^{j}_{\lambda} \lambda_{k \beta} {W}^{-4} \nabla^{\rho \gamma}{\lambda^{k}_{\gamma}} - \frac{261}{640}(\Gamma_{a})_{\alpha}{}^{\lambda} W^{\beta \rho} \lambda^{\gamma}_{i} \lambda_{j \lambda} \lambda^{j}_{\beta} \lambda_{k \rho} \lambda^{k}_{\gamma} {W}^{-4} - \frac{9}{32}{\rm i} (\Gamma_{a})^{\beta \lambda} X_{i j} F_{\beta}\,^{\rho} \lambda_{k \alpha} \lambda^{j}_{\lambda} \lambda^{k}_{\rho} {W}^{-4}+\frac{9}{16}{\rm i} (\Gamma_{a})^{\rho \lambda} F_{\alpha}\,^{\beta} F_{\rho \beta} \lambda^{\gamma}_{i} \lambda_{j \lambda} \lambda^{j}_{\gamma} {W}^{-4} - \frac{9}{32}{\rm i} (\Gamma_{a})^{\beta \rho} X_{j k} F_{\alpha \beta} \lambda^{\lambda}_{i} \lambda^{j}_{\rho} \lambda^{k}_{\lambda} {W}^{-4}+\frac{9}{32}{\rm i} (\Gamma_{a})^{\beta \lambda} F_{\beta \rho} \lambda^{\gamma}_{i} \lambda_{j \lambda} \lambda^{j}_{\gamma} {W}^{-4} \nabla_{\alpha}\,^{\rho}{W}+\frac{9}{16}{\rm i} (\Gamma_{a})^{\rho \gamma} F_{\alpha}\,^{\beta} F_{\rho}\,^{\lambda} \lambda_{i \beta} \lambda_{j \gamma} \lambda^{j}_{\lambda} {W}^{-4} - \frac{9}{32}{\rm i} (\Gamma_{a})^{\beta \lambda} X_{j k} F_{\beta}\,^{\rho} \lambda_{i \alpha} \lambda^{j}_{\lambda} \lambda^{k}_{\rho} {W}^{-4}+\frac{9}{32}{\rm i} (\Gamma_{a})^{\beta \lambda} F_{\beta}\,^{\rho} \lambda_{i \gamma} \lambda_{j \lambda} \lambda^{j}_{\rho} {W}^{-4} \nabla_{\alpha}\,^{\gamma}{W}+\frac{9}{8}(\Gamma_{a})^{\beta \lambda} F_{\beta}\,^{\rho} \lambda_{j \alpha} \lambda^{\gamma}_{i} \lambda^{j}_{\lambda} \lambda_{k \rho} \lambda^{k}_{\gamma} {W}^{-5}+\frac{9}{32}{\rm i} (\Gamma_{a})^{\beta}{}_{\rho} X_{i j} \lambda^{j}_{\alpha} \lambda_{k \beta} \lambda^{k}_{\lambda} {W}^{-4} \nabla^{\rho \lambda}{W} - \frac{3}{128}{\rm i} (\Gamma_{a})^{\beta}{}_{\rho} X_{i j} \lambda_{k \beta} \lambda^{k}_{\lambda} {W}^{-3} \nabla^{\rho \lambda}{\lambda^{j}_{\alpha}} - \frac{33}{128}{\rm i} (\Gamma_{a})_{\beta \rho} \lambda^{\lambda}_{i} \lambda_{j \lambda} {W}^{-3} \nabla^{\beta \rho}{W} \nabla_{\alpha}\,^{\gamma}{\lambda^{j}_{\gamma}} - \frac{15}{64}(\Gamma_{a})_{\rho \lambda} X_{i j} F_{\alpha}\,^{\beta} \lambda^{j}_{\beta} {W}^{-3} \nabla^{\rho \lambda}{W}+\frac{15}{64}(\Gamma_{a})_{\beta \rho} X_{i j} X^{j}\,_{k} \lambda^{k}_{\alpha} {W}^{-3} \nabla^{\beta \rho}{W}%
 - \frac{3}{64}(\Gamma_{a})_{\beta \rho} X_{i j} \lambda^{j}_{\lambda} {W}^{-3} \nabla^{\beta \rho}{W} \nabla_{\alpha}\,^{\lambda}{W}+\frac{15}{256}(\Gamma_{a})_{\beta \rho} X_{j k} X^{j k} \lambda_{i \alpha} {W}^{-3} \nabla^{\beta \rho}{W}+\frac{9}{64}{\rm i} (\Gamma_{a})_{\beta \rho} X_{j k} \lambda^{j}_{\alpha} \lambda^{\lambda}_{i} \lambda^{k}_{\lambda} {W}^{-4} \nabla^{\beta \rho}{W}+\frac{63}{1280}{\rm i} (\Gamma_{a})_{\alpha}{}^{\beta} X_{j k} W_{\beta}\,^{\rho} \lambda^{\lambda}_{i} \lambda^{j}_{\rho} \lambda^{k}_{\lambda} {W}^{-3} - \frac{63}{256}{\rm i} (\Gamma_{a})_{\beta \rho} X_{i j} \lambda_{k \alpha} \lambda^{j \lambda} \lambda^{k}_{\lambda} {W}^{-4} \nabla^{\beta \rho}{W} - \frac{9}{256}{\rm i} (\Gamma_{a})_{\beta \rho} X_{i j} \lambda^{j \lambda} \lambda_{k \lambda} {W}^{-3} \nabla^{\beta \rho}{\lambda^{k}_{\alpha}} - \frac{9}{64}{\rm i} (\Gamma_{a})^{\beta \rho} X_{i j} X_{k l} \lambda^{k}_{\alpha} \lambda^{j}_{\beta} \lambda^{l}_{\rho} {W}^{-4} - \frac{9}{8}{\rm i} (\Gamma_{a})^{\beta \rho} X_{i j} F_{\alpha \beta} \lambda^{j \lambda} \lambda_{k \rho} \lambda^{k}_{\lambda} {W}^{-4}+\frac{27}{320}{\rm i} (\Gamma_{a})_{\alpha}{}^{\beta} X_{i j} X^{j}\,_{k} \lambda^{k \rho} \lambda_{l \beta} \lambda^{l}_{\rho} {W}^{-4}+\frac{9}{64}{\rm i} (\Gamma_{a})^{\beta}{}_{\rho} X_{i j} \lambda^{j \lambda} \lambda_{k \beta} \lambda^{k}_{\lambda} {W}^{-4} \nabla^{\rho}\,_{\alpha}{W} - \frac{9}{16}(\Gamma_{a})^{\beta \rho} X_{j k} \lambda^{j}_{\alpha} \lambda_{i \beta} \lambda^{k \lambda} \lambda_{l \rho} \lambda^{l}_{\lambda} {W}^{-5} - \frac{9}{64}(\Gamma_{a})^{\beta \rho} \lambda_{j \beta} \lambda^{j \lambda} \lambda_{k \rho} \lambda^{k}_{\lambda} {W}^{-4} \nabla_{\alpha}\,^{\gamma}{\lambda_{i \gamma}} - \frac{9}{256}(\Gamma_{a})^{\rho \lambda} W_{\alpha}\,^{\beta} \lambda_{i \beta} \lambda_{j \rho} \lambda^{j \gamma} \lambda_{k \lambda} \lambda^{k}_{\gamma} {W}^{-4} - \frac{57}{512}{\rm i} (\Gamma_{a})^{\beta \rho} \lambda_{j \beta} \lambda^{j \lambda} \lambda_{k \rho} \lambda^{k}_{\lambda} X_{i \alpha} {W}^{-3}+\frac{9}{32}{\rm i} (\Gamma_{a})^{\rho \lambda} X_{i j} F_{\alpha}\,^{\beta} \lambda^{j}_{\rho} \lambda_{k \lambda} \lambda^{k}_{\beta} {W}^{-4}+\frac{9}{64}{\rm i} (\Gamma_{a})^{\beta \rho} X_{i j} \lambda^{j}_{\beta} \lambda_{k \rho} \lambda^{k}_{\lambda} {W}^{-4} \nabla_{\alpha}\,^{\lambda}{W}+\frac{9}{320}{\rm i} (\Gamma_{a})_{\alpha}{}^{\beta} X_{i j} X_{k l} \lambda^{j \rho} \lambda^{k}_{\beta} \lambda^{l}_{\rho} {W}^{-4}+\frac{9}{64}{\rm i} (\Gamma_{a})^{\beta \rho} X_{i j} X_{k l} \lambda^{j}_{\alpha} \lambda^{k}_{\beta} \lambda^{l}_{\rho} {W}^{-4} - \frac{9}{16}(\Gamma_{a})^{\beta \rho} X_{i j} \lambda_{k \alpha} \lambda^{j \lambda} \lambda^{k}_{\beta} \lambda_{l \rho} \lambda^{l}_{\lambda} {W}^{-5} - \frac{9}{32}(\Gamma_{a})_{\rho \lambda} F_{\alpha}\,^{\beta} \lambda_{i \beta} {W}^{-3} \nabla^{\rho}\,_{\gamma}{W} \nabla^{\lambda \gamma}{W}%
 - \frac{9}{64}(\Gamma_{a})_{\beta \rho} \lambda_{i \lambda} {W}^{-3} \nabla^{\beta}\,_{\gamma}{W} \nabla^{\rho \gamma}{W} \nabla_{\alpha}\,^{\lambda}{W} - \frac{9}{64}{\rm i} (\Gamma_{a})_{\beta \rho} \lambda_{j \alpha} \lambda^{\lambda}_{i} \lambda^{j}_{\lambda} {W}^{-4} \nabla^{\beta}\,_{\gamma}{W} \nabla^{\rho \gamma}{W} - \frac{3}{128}{\rm i} (\Gamma_{a})_{\beta \rho} \lambda^{\lambda}_{i} \lambda_{j \lambda} {W}^{-3} \nabla^{\beta}\,_{\gamma}{W} \nabla^{\rho \gamma}{\lambda^{j}_{\alpha}}+\frac{9}{32}(\Gamma_{a})_{\rho \lambda} F_{\alpha \beta} \lambda_{i \gamma} {W}^{-3} \nabla^{\rho \beta}{W} \nabla^{\lambda \gamma}{W} - \frac{9}{64}(\Gamma_{a})_{\beta \rho} \lambda_{i \lambda} {W}^{-3} \nabla^{\beta}\,_{\gamma}{W} \nabla^{\rho \lambda}{W} \nabla_{\alpha}\,^{\gamma}{W}+\frac{9}{64}{\rm i} (\Gamma_{a})_{\beta \rho} \lambda_{j \alpha} \lambda_{i \lambda} \lambda^{j}_{\gamma} {W}^{-4} \nabla^{\beta \lambda}{W} \nabla^{\rho \gamma}{W}+\frac{3}{128}{\rm i} (\Gamma_{a})_{\beta \rho} \lambda_{i \lambda} \lambda_{j \gamma} {W}^{-3} \nabla^{\beta \lambda}{W} \nabla^{\rho \gamma}{\lambda^{j}_{\alpha}}+\frac{27}{32}{\rm i} (\Gamma_{a})^{\beta}{}_{\rho} F_{\alpha \beta} \lambda^{\lambda}_{i} \lambda_{j \lambda} \lambda^{j}_{\gamma} {W}^{-4} \nabla^{\rho \gamma}{W}+\frac{27}{320}{\rm i} (\Gamma_{a})_{\alpha \beta} X_{i j} \lambda^{j \rho} \lambda_{k \rho} \lambda^{k}_{\lambda} {W}^{-4} \nabla^{\beta \lambda}{W} - \frac{9}{8}{\rm i} (\Gamma_{a})^{\rho}{}_{\lambda} F_{\alpha}\,^{\beta} \lambda_{i \rho} \lambda_{j \beta} \lambda^{j}_{\gamma} {W}^{-4} \nabla^{\lambda \gamma}{W} - \frac{9}{16}{\rm i} (\Gamma_{a})^{\beta}{}_{\rho} \lambda_{i \beta} \lambda_{j \lambda} \lambda^{j}_{\gamma} {W}^{-4} \nabla^{\rho \lambda}{W} \nabla_{\alpha}\,^{\gamma}{W} - \frac{9}{16}(\Gamma_{a})^{\beta}{}_{\rho} \lambda_{j \alpha} \lambda_{i \beta} \lambda^{j \lambda} \lambda_{k \lambda} \lambda^{k}_{\gamma} {W}^{-5} \nabla^{\rho \gamma}{W} - \frac{9}{64}{\rm i} (\Gamma_{a})^{\beta}{}_{\rho} X_{i j} \lambda_{k \alpha} \lambda^{j}_{\beta} \lambda^{k}_{\lambda} {W}^{-4} \nabla^{\rho \lambda}{W} - \frac{9}{32}{\rm i} (\Gamma_{a})^{\rho}{}_{\lambda} F_{\alpha \beta} \lambda^{\gamma}_{i} \lambda_{j \rho} \lambda^{j}_{\gamma} {W}^{-4} \nabla^{\lambda \beta}{W} - \frac{9}{64}{\rm i} (\Gamma_{a})^{\beta}{}_{\rho} X_{j k} \lambda^{\lambda}_{i} \lambda^{j}_{\beta} \lambda^{k}_{\lambda} {W}^{-4} \nabla^{\rho}\,_{\alpha}{W}+\frac{9}{64}{\rm i} (\Gamma_{a})^{\beta}{}_{\rho} \lambda^{\lambda}_{i} \lambda_{j \beta} \lambda^{j}_{\lambda} {W}^{-4} \nabla^{\rho}\,_{\gamma}{W} \nabla_{\alpha}\,^{\gamma}{W}+\frac{9}{32}{\rm i} (\Gamma_{a})^{\rho}{}_{\lambda} F_{\alpha}\,^{\beta} \lambda_{i \beta} \lambda_{j \rho} \lambda^{j}_{\gamma} {W}^{-4} \nabla^{\lambda \gamma}{W} - \frac{9}{64}{\rm i} (\Gamma_{a})^{\beta}{}_{\rho} X_{j k} \lambda_{i \alpha} \lambda^{j}_{\beta} \lambda^{k}_{\lambda} {W}^{-4} \nabla^{\rho \lambda}{W}+\frac{9}{64}{\rm i} (\Gamma_{a})^{\beta}{}_{\rho} \lambda_{i \lambda} \lambda_{j \beta} \lambda^{j}_{\gamma} {W}^{-4} \nabla^{\rho \gamma}{W} \nabla_{\alpha}\,^{\lambda}{W}+\frac{9}{16}(\Gamma_{a})^{\beta}{}_{\rho} \lambda_{j \alpha} \lambda^{\lambda}_{i} \lambda^{j}_{\beta} \lambda_{k \lambda} \lambda^{k}_{\gamma} {W}^{-5} \nabla^{\rho \gamma}{W}%
+\frac{9}{128}(\Gamma_{a})^{\beta}{}_{\rho} \lambda^{\lambda}_{i} \lambda_{j \beta} \lambda_{k \lambda} \lambda^{k}_{\gamma} {W}^{-4} \nabla^{\rho \gamma}{\lambda^{j}_{\alpha}} - \frac{9}{256}(\Gamma_{a})^{\rho \lambda} W_{\alpha}\,^{\beta} \lambda^{\gamma}_{i} \lambda_{j \rho} \lambda^{j}_{\beta} \lambda_{k \lambda} \lambda^{k}_{\gamma} {W}^{-4} - \frac{27}{32}(\Gamma_{a})^{\beta \rho} F_{\alpha \beta} \lambda_{i \rho} \lambda^{\lambda}_{j} \lambda^{j \gamma} \lambda_{k \lambda} \lambda^{k}_{\gamma} {W}^{-5} - \frac{9}{8}(\Gamma_{a})^{\rho \lambda} F_{\alpha}\,^{\beta} \lambda_{i \rho} \lambda_{j \lambda} \lambda^{j \gamma} \lambda_{k \beta} \lambda^{k}_{\gamma} {W}^{-5} - \frac{9}{16}(\Gamma_{a})^{\beta \rho} X_{j k} \lambda_{l \alpha} \lambda_{i \beta} \lambda^{j}_{\rho} \lambda^{k \lambda} \lambda^{l}_{\lambda} {W}^{-5} - \frac{9}{16}(\Gamma_{a})^{\beta \rho} \lambda_{i \beta} \lambda_{j \rho} \lambda^{j \lambda} \lambda_{k \lambda} \lambda^{k}_{\gamma} {W}^{-5} \nabla_{\alpha}\,^{\gamma}{W} - \frac{45}{64}{\rm i} (\Gamma_{a})^{\beta \rho} \lambda_{j \alpha} \lambda_{i \beta} \lambda^{j}_{\rho} \lambda^{\lambda}_{k} \lambda^{k \gamma} \lambda_{l \lambda} \lambda^{l}_{\gamma} {W}^{-6}+\frac{243}{4096}{\rm i} (\Gamma_{a})_{\rho \lambda} \nabla^{\rho \lambda}{W} \nabla_{\alpha}\,^{\beta}{X_{i \beta}} - \frac{1}{8}(\Gamma_{a})^{\beta \rho} \lambda_{j \beta} \nabla_{\alpha}\,^{\lambda}{\Phi_{\lambda \rho i}\,^{j}} - \frac{117}{1024}(\Gamma_{a})^{\lambda}{}_{\gamma} \lambda_{i \lambda} \nabla_{\alpha}\,^{\beta}{\nabla^{\gamma \rho}{W_{\beta \rho}}} - \frac{81}{1024}(\Gamma_{a})^{\lambda \beta} \lambda_{i \lambda} \nabla_{\alpha \gamma}{\nabla^{\gamma \rho}{W_{\beta \rho}}}+\frac{45}{2048}{\rm i} (\Gamma_{a})^{\beta}{}_{\gamma} W \nabla_{\alpha}\,^{\rho}{\nabla^{\gamma \lambda}{W_{\beta \rho \lambda i}}} - \frac{9}{4096}{\rm i} (\Gamma_{a})^{\beta}{}_{\rho} W \nabla_{\alpha \lambda}{\nabla^{\rho \lambda}{X_{i \beta}}}+\frac{9}{4096}{\rm i} (\Gamma_{a})_{\rho \lambda} W \nabla_{\alpha}\,^{\beta}{\nabla^{\rho \lambda}{X_{i \beta}}}+\frac{7317}{8192}{\rm i} (\Gamma_{a})^{\beta}{}_{\gamma} W W_{\beta}\,^{\rho} \nabla^{\gamma \lambda}{W_{\alpha \rho \lambda i}} - \frac{2931}{4096}{\rm i} (\Gamma_{a})_{\lambda \gamma} W^{\beta \rho} W_{\alpha \beta \rho i} \nabla^{\lambda \gamma}{W}+\frac{3}{64}{\rm i} (\Gamma_{a})^{\gamma \lambda} \lambda_{i \gamma} W_{\alpha}\,^{\beta \rho}\,_{j} W_{\lambda \beta \rho}\,^{j} - \frac{435}{4096}{\rm i} (\Gamma_{a})_{\lambda \gamma} W W^{\beta \rho} \nabla^{\lambda \gamma}{W_{\alpha \beta \rho i}}+\frac{9}{128}(\Gamma_{a})_{\rho \lambda} \lambda^{\beta}_{i} \nabla^{\rho}\,_{\gamma}{\nabla^{\lambda \gamma}{W_{\alpha \beta}}}+\frac{9}{128}(\Gamma_{a})_{\rho \lambda} \lambda_{i \gamma} \nabla^{\rho \beta}{\nabla^{\lambda \gamma}{W_{\alpha \beta}}}%
 - \frac{3}{32}{\rm i} (\Gamma_{a})^{\gamma \lambda} \lambda_{j \gamma} W_{\alpha}\,^{\beta \rho j} W_{\lambda \beta \rho i} - \frac{9}{256}(\Gamma_{a})^{\beta}{}_{\lambda} \lambda^{\rho}_{i} \nabla_{\alpha \gamma}{\nabla^{\lambda \gamma}{W_{\beta \rho}}}+\frac{9}{256}(\Gamma_{a})_{\lambda \gamma} \lambda^{\beta}_{i} \nabla_{\alpha}\,^{\rho}{\nabla^{\lambda \gamma}{W_{\beta \rho}}} - \frac{9}{256}(\Gamma_{a})^{\beta}{}_{\lambda} \lambda_{i \gamma} \nabla_{\alpha}\,^{\rho}{\nabla^{\lambda \gamma}{W_{\beta \rho}}}+\frac{3}{32}{\rm i} (\Gamma_{a})^{\gamma \lambda} \lambda_{j \gamma} W_{\alpha}\,^{\beta \rho}\,_{i} W_{\lambda \beta \rho}\,^{j} - \frac{9}{256}(\Gamma_{a})^{\beta}{}_{\rho} \nabla_{\lambda \gamma}{\nabla^{\lambda \gamma}{\nabla^{\rho}\,_{\alpha}{\lambda_{i \beta}}}} - \frac{9}{512}(\Gamma_{a})_{\beta \rho} \nabla_{\lambda \gamma}{\nabla^{\lambda \gamma}{\nabla^{\beta \rho}{\lambda_{i \alpha}}}} - \frac{27}{512}(\Gamma_{a})^{\rho \beta} \lambda_{i \rho} \nabla_{\lambda \gamma}{\nabla^{\lambda \gamma}{W_{\alpha \beta}}}+\frac{9}{256}(\Gamma_{a})^{\beta \rho} W_{\alpha \beta} \nabla_{\lambda \gamma}{\nabla^{\lambda \gamma}{\lambda_{i \rho}}}+\frac{9}{1280}(\Gamma_{a})_{\alpha \beta} \nabla_{\lambda \gamma}{\nabla^{\lambda \gamma}{\nabla^{\beta \rho}{\lambda_{i \rho}}}}+\frac{27}{128}(\Gamma_{a})^{\lambda}{}_{\gamma} W_{\beta \rho} \nabla^{\gamma \beta}{\nabla_{\alpha}\,^{\rho}{\lambda_{i \lambda}}}+\frac{27}{128}(\Gamma_{a})_{\lambda \gamma} W_{\beta \rho} \nabla^{\lambda \beta}{\nabla^{\gamma \rho}{\lambda_{i \alpha}}}+\frac{297}{256}(\Gamma_{a})^{\beta}{}_{\gamma} W_{\beta}\,^{\rho} W_{\rho \lambda} \nabla^{\gamma \lambda}{\lambda_{i \alpha}} - \frac{45}{128}(\Gamma_{a})^{\beta \lambda} W_{\beta \rho} \nabla_{\gamma}\,^{\rho}{\nabla_{\alpha}\,^{\gamma}{\lambda_{i \lambda}}} - \frac{27}{128}(\Gamma_{a})^{\beta}{}_{\lambda} W_{\beta \rho} \nabla_{\gamma}\,^{\rho}{\nabla^{\lambda \gamma}{\lambda_{i \alpha}}} - \frac{63}{128}(\Gamma_{a})^{\beta \lambda} W_{\beta \rho} W_{\lambda \gamma} \nabla^{\rho \gamma}{\lambda_{i \alpha}}+\frac{9}{64}(\Gamma_{a})^{\beta}{}_{\lambda} W_{\beta \rho} \nabla_{\alpha}\,^{\rho}{\nabla^{\lambda \gamma}{\lambda_{i \gamma}}} - \frac{9}{64}(\Gamma_{a})^{\beta \lambda} W_{\beta}\,^{\rho} W_{\lambda \gamma} \nabla_{\alpha}\,^{\gamma}{\lambda_{i \rho}} - \frac{9}{64}{\rm i} (\Gamma_{a})^{\beta \lambda} W_{\beta \rho} W_{\alpha \lambda \gamma i} \nabla^{\rho \gamma}{W} - \frac{9}{16}{\rm i} (\Gamma_{a})^{\beta}{}_{\gamma} W_{\alpha}\,^{\rho}\,_{\lambda i} \nabla^{\gamma \lambda}{F_{\beta \rho}}%
 - \frac{9}{16}{\rm i} (\Gamma_{a})^{\rho}{}_{\gamma} W_{\rho}\,^{\beta}\,_{\lambda i} \nabla^{\gamma \lambda}{F_{\alpha \beta}} - \frac{9}{16}{\rm i} (\Gamma_{a})^{\lambda \beta} W_{\alpha \lambda \gamma i} \nabla^{\gamma \rho}{F_{\beta \rho}}+\frac{9}{128}{\rm i} (\Gamma_{a})_{\alpha}{}^{\beta} X_{i \lambda} \nabla^{\lambda \rho}{F_{\beta \rho}} - \frac{9}{64}{\rm i} (\Gamma_{a})^{\beta}{}_{\lambda} X_{i}^{\rho} \nabla^{\lambda}\,_{\alpha}{F_{\beta \rho}} - \frac{9}{512}(\Gamma_{a})^{\rho}{}_{\lambda} \lambda_{i \rho} \nabla_{\gamma}\,^{\beta}{\nabla^{\lambda \gamma}{W_{\alpha \beta}}}+\frac{9}{512}(\Gamma_{a})^{\lambda \beta} \lambda_{i \lambda} \nabla_{\gamma}\,^{\rho}{\nabla_{\alpha}\,^{\gamma}{W_{\beta \rho}}} - \frac{117}{5120}(\Gamma_{a})_{\alpha}{}^{\lambda} \lambda_{i \lambda} \nabla_{\gamma}\,^{\beta}{\nabla^{\gamma \rho}{W_{\beta \rho}}} - \frac{117}{1280}{\rm i} (\Gamma_{a})_{\alpha \gamma} F^{\beta \rho} \nabla^{\gamma \lambda}{W_{\beta \rho \lambda i}}+\frac{99}{512}{\rm i} (\Gamma_{a})_{\alpha}{}^{\lambda} F^{\beta}\,_{\rho} \nabla^{\rho \gamma}{W_{\lambda \beta \gamma i}} - \frac{279}{5120}{\rm i} (\Gamma_{a})_{\alpha \lambda} F^{\beta}\,_{\rho} \nabla^{\lambda \rho}{X_{i \beta}} - \frac{369}{1280}{\rm i} (\Gamma_{a})_{\alpha}{}^{\lambda} W_{\lambda}\,^{\gamma} F^{\beta \rho} W_{\gamma \beta \rho i} - \frac{9}{1280}{\rm i} (\Gamma_{a})_{\alpha}{}^{\gamma} W^{\beta \lambda} F_{\beta}\,^{\rho} W_{\gamma \lambda \rho i}+\frac{27}{512}{\rm i} (\Gamma_{a})_{\alpha}{}^{\beta} X_{i \beta} \nabla_{\rho \lambda}{\nabla^{\rho \lambda}{W}}+\frac{207}{4096}{\rm i} (\Gamma_{a})_{\alpha}{}^{\beta} \nabla_{\rho \lambda}{W} \nabla^{\rho \lambda}{X_{i \beta}} - \frac{297}{20480}{\rm i} (\Gamma_{a})_{\alpha}{}^{\beta} W \nabla_{\rho \lambda}{\nabla^{\rho \lambda}{X_{i \beta}}}+\frac{153}{640}(\Gamma_{a})_{\alpha \lambda} W_{\beta \rho} \nabla^{\lambda \beta}{\nabla^{\rho \gamma}{\lambda_{i \gamma}}}+\frac{81}{640}(\Gamma_{a})_{\alpha}{}^{\beta} W_{\beta \rho} \nabla_{\gamma}\,^{\rho}{\nabla^{\gamma \lambda}{\lambda_{i \lambda}}}+\frac{909}{640}(\Gamma_{a})_{\alpha}{}^{\beta} W_{\beta \rho} W^{\lambda}\,_{\gamma} \nabla^{\rho \gamma}{\lambda_{i \lambda}} - \frac{27}{128}(\Gamma_{a})_{\lambda \gamma} W^{\beta}\,_{\rho} \nabla^{\lambda \rho}{\nabla^{\gamma}\,_{\alpha}{\lambda_{i \beta}}} - \frac{27}{128}(\Gamma_{a})^{\beta}{}_{\lambda} W_{\beta \rho} \nabla^{\rho \gamma}{\nabla^{\lambda}\,_{\alpha}{\lambda_{i \gamma}}}%
 - \frac{9}{32}{\rm i} (\Gamma_{a})^{\beta}{}_{\gamma} W_{\beta \rho \lambda i} \nabla^{\gamma \rho}{\nabla_{\alpha}\,^{\lambda}{W}}+\frac{9}{32}{\rm i} (\Gamma_{a})_{\lambda \gamma} W_{\alpha \beta \rho i} \nabla^{\lambda \beta}{\nabla^{\gamma \rho}{W}}+\frac{9}{32}{\rm i} (\Gamma_{a})^{\beta}{}_{\lambda} W_{\alpha \beta \rho i} \nabla_{\gamma}\,^{\rho}{\nabla^{\lambda \gamma}{W}} - \frac{63}{1280}{\rm i} (\Gamma_{a})_{\alpha \rho} X_{i \beta} \nabla_{\lambda}\,^{\beta}{\nabla^{\rho \lambda}{W}}+\frac{9}{128}{\rm i} (\Gamma_{a})_{\rho \lambda} X_{i \beta} \nabla^{\rho}\,_{\alpha}{\nabla^{\lambda \beta}{W}} - \frac{63}{10240}{\rm i} (\Gamma_{a})_{\alpha}{}^{\beta} W \nabla_{\gamma}\,^{\rho}{\nabla^{\gamma \lambda}{W_{\beta \rho \lambda i}}} - \frac{1497}{10240}{\rm i} (\Gamma_{a})_{\alpha}{}^{\beta} \nabla_{\gamma}\,^{\rho}{W} \nabla^{\gamma \lambda}{W_{\beta \rho \lambda i}}+\frac{5727}{20480}{\rm i} (\Gamma_{a})_{\alpha}{}^{\lambda} W^{\beta}\,_{\rho} W_{\lambda \beta \gamma i} \nabla^{\rho \gamma}{W} - \frac{525}{4096}{\rm i} (\Gamma_{a})_{\lambda \gamma} W W_{\alpha}\,^{\beta \rho}\,_{i} \nabla^{\lambda \gamma}{W_{\beta \rho}} - \frac{351}{256}(\Gamma_{a})^{\beta}{}_{\gamma} W_{\beta}\,^{\rho} W_{\rho}\,^{\lambda} \nabla^{\gamma}\,_{\alpha}{\lambda_{i \lambda}} - \frac{567}{20480}{\rm i} (\Gamma_{a})_{\alpha \rho} \nabla_{\lambda}\,^{\beta}{W} \nabla^{\rho \lambda}{X_{i \beta}} - \frac{9}{32}{\rm i} (\Gamma_{a})^{\lambda \gamma} W_{\lambda}\,^{\beta} F_{\beta}\,^{\rho} W_{\alpha \gamma \rho i}+\frac{9}{16}(\Gamma_{a})^{\beta}{}_{\gamma} C_{\alpha \beta}\,^{\rho}\,_{\lambda} \nabla^{\gamma \lambda}{\lambda_{i \rho}} - \frac{9}{32}{\rm i} (\Gamma_{a})^{\beta}{}_{\lambda} W_{\alpha \beta \rho j} \nabla^{\lambda \rho}{X_{i}\,^{j}}+\frac{45}{4096}{\rm i} (\Gamma_{a})_{\beta \rho} W \nabla^{\beta}\,_{\lambda}{\nabla^{\rho \lambda}{X_{i \alpha}}} - \frac{153}{4096}{\rm i} (\Gamma_{a})_{\rho \lambda} W \nabla^{\rho \lambda}{\nabla_{\alpha}\,^{\beta}{X_{i \beta}}}+\frac{9}{1024}{\rm i} (\Gamma_{a})^{\beta}{}_{\lambda} W \nabla_{\gamma}\,^{\rho}{\nabla^{\lambda \gamma}{W_{\alpha \beta \rho i}}}+\frac{81}{4096}{\rm i} (\Gamma_{a})_{\alpha \rho} W \nabla_{\lambda}\,^{\beta}{\nabla^{\rho \lambda}{X_{i \beta}}}+\frac{27}{1024}{\rm i} (\Gamma_{a})^{\beta}{}_{\lambda} \nabla_{\gamma}\,^{\rho}{W} \nabla^{\lambda \gamma}{W_{\alpha \beta \rho i}}+\frac{9}{256}(\Gamma_{a})_{\lambda \gamma} W^{\beta}\,_{\rho} \nabla_{\alpha}\,^{\rho}{\nabla^{\lambda \gamma}{\lambda_{i \beta}}}%
 - \frac{9}{128}(\Gamma_{a})^{\lambda}{}_{\gamma} W_{\beta \rho} \nabla_{\alpha}\,^{\beta}{\nabla^{\gamma \rho}{\lambda_{i \lambda}}}+\frac{63}{512}{\rm i} (\Gamma_{a})^{\lambda}{}_{\gamma} F^{\beta \rho} \nabla^{\gamma}\,_{\alpha}{W_{\lambda \beta \rho i}} - \frac{9}{128}{\rm i} (\Gamma_{a})_{\lambda \gamma} F^{\beta \rho} \nabla^{\lambda \gamma}{W_{\alpha \beta \rho i}} - \frac{153}{512}{\rm i} (\Gamma_{a})^{\lambda}{}_{\gamma} F^{\beta}\,_{\rho} \nabla^{\gamma \rho}{W_{\alpha \lambda \beta i}}+\frac{45}{128}{\rm i} (\Gamma_{a})^{\lambda \beta} X_{i \lambda} \nabla_{\alpha}\,^{\rho}{F_{\beta \rho}}+\frac{45}{256}{\rm i} (\Gamma_{a})^{\beta}{}_{\rho} X_{i \beta} \nabla_{\alpha \lambda}{\nabla^{\rho \lambda}{W}}+\frac{9}{64}(\Gamma_{a})^{\beta}{}_{\lambda} {W}^{-1} \nabla_{\alpha \gamma}{F_{\beta}\,^{\rho}} \nabla^{\lambda \gamma}{\lambda_{i \rho}} - \frac{9}{128}(\Gamma_{a})_{\beta \rho} {W}^{-1} \nabla^{\beta}\,_{\gamma}{\lambda_{i \lambda}} \nabla_{\alpha}\,^{\gamma}{\nabla^{\rho \lambda}{W}}+\frac{9}{128}(\Gamma_{a})^{\beta}{}_{\rho} X_{i j} W_{\alpha \beta} {W}^{-1} \nabla^{\rho \lambda}{\lambda^{j}_{\lambda}} - \frac{3}{80}(\Gamma_{a})_{\alpha \gamma} \lambda^{\beta}_{j} W_{\beta}\,^{\rho}\,_{\lambda i} {W}^{-1} \nabla^{\gamma \lambda}{\lambda^{j}_{\rho}} - \frac{9}{64}(\Gamma_{a})_{\lambda \gamma} \lambda^{\beta}_{j} W_{\alpha \beta}\,^{\rho}\,_{i} {W}^{-1} \nabla^{\lambda \gamma}{\lambda^{j}_{\rho}} - \frac{3}{16}(\Gamma_{a})^{\beta}{}_{\gamma} \lambda_{j \alpha} W_{\beta}\,^{\rho}\,_{\lambda i} {W}^{-1} \nabla^{\gamma \lambda}{\lambda^{j}_{\rho}} - \frac{9}{64}(\Gamma_{a})^{\beta}{}_{\lambda} W_{\beta \rho} {W}^{-1} \nabla^{\lambda \gamma}{W} \nabla_{\alpha}\,^{\rho}{\lambda_{i \gamma}}+\frac{21}{512}(\Gamma_{a})^{\beta \rho} \lambda_{j \lambda} X_{i \beta} {W}^{-1} \nabla_{\alpha}\,^{\lambda}{\lambda^{j}_{\rho}} - \frac{3}{64}(\Gamma_{a})^{\beta}{}_{\gamma} \lambda^{\rho}_{i} W_{\beta \rho}\,^{\lambda}\,_{j} {W}^{-1} \nabla^{\gamma}\,_{\alpha}{\lambda^{j}_{\lambda}}+\frac{9}{64}(\Gamma_{a})^{\beta}{}_{\lambda} {W}^{-1} \nabla_{\alpha}\,^{\gamma}{F_{\beta \rho}} \nabla^{\lambda \rho}{\lambda_{i \gamma}} - \frac{9}{128}(\Gamma_{a})_{\beta \rho} {W}^{-1} \nabla^{\beta}\,_{\gamma}{\lambda_{i \lambda}} \nabla_{\alpha}\,^{\lambda}{\nabla^{\rho \gamma}{W}}+\frac{27}{64}(\Gamma_{a})_{\lambda \gamma} W^{\beta}\,_{\rho} {W}^{-1} \nabla^{\lambda \gamma}{W} \nabla_{\alpha}\,^{\rho}{\lambda_{i \beta}}+\frac{27}{64}(\Gamma_{a})^{\gamma \beta} W^{\rho}\,_{\lambda} \lambda_{i \gamma} {W}^{-1} \nabla_{\alpha}\,^{\lambda}{F_{\beta \rho}}+\frac{27}{128}(\Gamma_{a})^{\lambda}{}_{\gamma} W_{\beta \rho} \lambda_{i \lambda} {W}^{-1} \nabla_{\alpha}\,^{\beta}{\nabla^{\gamma \rho}{W}}%
+\frac{9}{32}(\Gamma_{a})^{\lambda \beta} W_{\lambda \gamma} F_{\beta}\,^{\rho} {W}^{-1} \nabla_{\alpha}\,^{\gamma}{\lambda_{i \rho}} - \frac{45}{128}(\Gamma_{a})^{\lambda \beta} W_{\lambda \gamma} \lambda^{\rho}_{i} {W}^{-1} \nabla_{\alpha}\,^{\gamma}{F_{\beta \rho}} - \frac{45}{256}(\Gamma_{a})^{\beta}{}_{\lambda} W_{\beta \rho} \lambda_{i \gamma} {W}^{-1} \nabla_{\alpha}\,^{\rho}{\nabla^{\lambda \gamma}{W}}+\frac{4641}{10240}(\Gamma_{a})_{\alpha}{}^{\beta} W_{\beta}\,^{\rho} \lambda^{\lambda}_{j} \lambda^{j \gamma} W_{\rho \lambda \gamma i} {W}^{-1} - \frac{489}{10240}(\Gamma_{a})_{\alpha}{}^{\beta} W_{\beta}\,^{\rho} \lambda_{j \rho} \lambda^{j \lambda} X_{i \lambda} {W}^{-1} - \frac{1881}{10240}(\Gamma_{a})_{\alpha}{}^{\beta} W_{\beta}\,^{\rho} \lambda^{\lambda}_{i} \lambda^{\gamma}_{j} W_{\rho \lambda \gamma}\,^{j} {W}^{-1}+\frac{9}{16}(\Gamma_{a})^{\lambda \beta} W_{\lambda}\,^{\gamma} F_{\beta \rho} {W}^{-1} \nabla_{\alpha}\,^{\rho}{\lambda_{i \gamma}} - \frac{9}{32}(\Gamma_{a})^{\beta}{}_{\lambda} W_{\beta}\,^{\rho} {W}^{-1} \nabla^{\lambda}\,_{\gamma}{W} \nabla_{\alpha}\,^{\gamma}{\lambda_{i \rho}} - \frac{267}{1024}(\Gamma_{a})^{\beta \lambda} W_{\beta}\,^{\rho} \lambda_{j \alpha} \lambda^{j}_{\rho} X_{i \lambda} {W}^{-1} - \frac{9}{64}(\Gamma_{a})^{\beta}{}_{\lambda} {W}^{-1} \nabla^{\lambda}\,_{\alpha}{F_{\beta \rho}} \nabla^{\rho \gamma}{\lambda_{i \gamma}}+\frac{9}{128}(\Gamma_{a})_{\beta \rho} {W}^{-1} \nabla_{\gamma}\,^{\lambda}{\lambda_{i \lambda}} \nabla^{\beta}\,_{\alpha}{\nabla^{\rho \gamma}{W}} - \frac{9}{64}(\Gamma_{a})_{\alpha}{}^{\lambda} W_{\lambda}\,^{\beta} F_{\beta \rho} {W}^{-1} \nabla^{\rho \gamma}{\lambda_{i \gamma}}+\frac{81}{1280}(\Gamma_{a})_{\alpha}{}^{\beta} W_{\beta \rho} {W}^{-1} \nabla_{\gamma}\,^{\rho}{W} \nabla^{\gamma \lambda}{\lambda_{i \lambda}}+\frac{123}{2560}(\Gamma_{a})_{\alpha}{}^{\beta} \lambda_{j \rho} X^{j}_{\beta} {W}^{-1} \nabla^{\rho \lambda}{\lambda_{i \lambda}}+\frac{9}{128}{\rm i} (\Gamma_{a})^{\beta \rho} \lambda_{i \beta} {W}^{-2} \nabla_{\alpha \gamma}{\lambda_{j \rho}} \nabla^{\gamma \lambda}{\lambda^{j}_{\lambda}}+\frac{9}{128}(\Gamma_{a})_{\beta \rho} {W}^{-1} \nabla_{\gamma}\,^{\lambda}{\lambda_{i \lambda}} \nabla_{\alpha}\,^{\gamma}{\nabla^{\beta \rho}{W}}+\frac{9}{32}(\Gamma_{a})^{\beta \lambda} {W}^{-1} \nabla_{\gamma}\,^{\rho}{F_{\beta \rho}} \nabla_{\alpha}\,^{\gamma}{\lambda_{i \lambda}}+\frac{81}{64}(\Gamma_{a})^{\beta}{}_{\gamma} W^{\rho \lambda} F_{\beta \rho} {W}^{-1} \nabla^{\gamma}\,_{\alpha}{\lambda_{i \lambda}}+\frac{81}{256}(\Gamma_{a})_{\lambda \gamma} W^{\beta}\,_{\rho} {W}^{-1} \nabla^{\lambda \rho}{W} \nabla^{\gamma}\,_{\alpha}{\lambda_{i \beta}}+\frac{4521}{10240}(\Gamma_{a})_{\alpha}{}^{\lambda} W^{\beta \rho} \lambda_{j \beta} \lambda^{j \gamma} W_{\lambda \rho \gamma i} {W}^{-1}%
 - \frac{993}{10240}(\Gamma_{a})_{\alpha}{}^{\lambda} W^{\beta \rho} \lambda_{j \beta} \lambda^{j}_{\rho} X_{i \lambda} {W}^{-1}+\frac{27}{256}(\Gamma_{a})_{\lambda \gamma} W^{\beta}\,_{\rho} \lambda_{i \beta} {W}^{-1} \nabla_{\alpha}\,^{\rho}{\nabla^{\lambda \gamma}{W}}+\frac{9}{32}(\Gamma_{a})^{\gamma \lambda} F^{\beta}\,_{\rho} \lambda_{i \gamma} {W}^{-2} \nabla_{\alpha}\,^{\rho}{F_{\lambda \beta}}+\frac{9}{64}(\Gamma_{a})^{\lambda}{}_{\gamma} F_{\beta \rho} \lambda_{i \lambda} {W}^{-2} \nabla_{\alpha}\,^{\beta}{\nabla^{\gamma \rho}{W}}+\frac{3}{80}(\Gamma_{a})_{\alpha}{}^{\gamma} F^{\beta \rho} \lambda_{j \gamma} \lambda^{j \lambda} W_{\beta \rho \lambda i} {W}^{-2}+\frac{3}{32}(\Gamma_{a})^{\beta \gamma} F_{\beta}\,^{\rho} \lambda_{j \gamma} \lambda^{j \lambda} W_{\alpha \rho \lambda i} {W}^{-2} - \frac{9}{32}(\Gamma_{a})^{\gamma \lambda} F^{\beta \rho} \lambda_{j \gamma} \lambda^{j}_{\beta} W_{\alpha \lambda \rho i} {W}^{-2} - \frac{33}{1280}(\Gamma_{a})_{\alpha}{}^{\lambda} F^{\beta \rho} \lambda_{j \lambda} \lambda^{j}_{\beta} X_{i \rho} {W}^{-2} - \frac{3}{8}(\Gamma_{a})^{\gamma \lambda} F^{\beta \rho} \lambda_{j \alpha} \lambda_{i \gamma} W_{\lambda \beta \rho}\,^{j} {W}^{-2}+\frac{9}{128}(\Gamma_{a})_{\lambda \gamma} F^{\beta}\,_{\rho} {W}^{-1} \nabla_{\alpha}\,^{\rho}{\nabla^{\lambda \gamma}{\lambda_{i \beta}}} - \frac{9}{64}(\Gamma_{a})^{\lambda}{}_{\gamma} F_{\beta \rho} {W}^{-1} \nabla_{\alpha}\,^{\beta}{\nabla^{\gamma \rho}{\lambda_{i \lambda}}}+\frac{81}{128}(\Gamma_{a})^{\lambda \gamma} W_{\lambda}\,^{\beta} F_{\beta \rho} {W}^{-1} \nabla_{\alpha}\,^{\rho}{\lambda_{i \gamma}}+\frac{9}{64}(\Gamma_{a})^{\beta}{}_{\lambda} F_{\beta \rho} {W}^{-1} \nabla_{\alpha}\,^{\rho}{\nabla^{\lambda \gamma}{\lambda_{i \gamma}}} - \frac{27}{128}(\Gamma_{a})^{\rho}{}_{\lambda} W_{\rho}\,^{\beta} F_{\alpha \beta} {W}^{-1} \nabla^{\lambda \gamma}{\lambda_{i \gamma}}+\frac{45}{128}(\Gamma_{a})_{\alpha}{}^{\gamma} W^{\beta}\,_{\lambda} F_{\beta \rho} {W}^{-1} \nabla^{\lambda \rho}{\lambda_{i \gamma}} - \frac{9}{16}{\rm i} (\Gamma_{a})^{\beta \gamma} F_{\beta}\,^{\rho} F_{\rho}\,^{\lambda} W_{\alpha \gamma \lambda i} {W}^{-1} - \frac{9}{40}{\rm i} (\Gamma_{a})_{\alpha}{}^{\beta} F_{\beta}\,^{\rho} F^{\lambda \gamma} W_{\rho \lambda \gamma i} {W}^{-1} - \frac{9}{64}{\rm i} (\Gamma_{a})^{\beta \lambda} F_{\beta}\,^{\rho} F_{\lambda \rho} X_{i \alpha} {W}^{-1} - \frac{9}{160}{\rm i} (\Gamma_{a})_{\alpha}{}^{\beta} F_{\beta}\,^{\rho} F_{\rho}\,^{\lambda} X_{i \lambda} {W}^{-1} - \frac{63}{320}{\rm i} (\Gamma_{a})_{\alpha}{}^{\lambda} F^{\beta \rho} F_{\beta \rho} X_{i \lambda} {W}^{-1}%
 - \frac{9}{16}\Phi_{\alpha}\,^{\lambda}\,_{i j} (\Gamma_{a})_{\lambda}{}^{\beta} F_{\beta}\,^{\rho} \lambda^{j}_{\rho} {W}^{-1}+\frac{9}{160}{\rm i} (\Gamma_{a})_{\alpha}{}^{\lambda} F^{\beta}\,_{\rho} W_{\lambda \beta \gamma i} {W}^{-1} \nabla^{\rho \gamma}{W}+\frac{9}{40}{\rm i} (\Gamma_{a})_{\alpha \gamma} F^{\beta \rho} W_{\beta \rho \lambda i} {W}^{-1} \nabla^{\gamma \lambda}{W}+\frac{27}{16}{\rm i} (\Gamma_{a})^{\beta}{}_{\gamma} F_{\beta}\,^{\rho} W_{\alpha \rho \lambda i} {W}^{-1} \nabla^{\gamma \lambda}{W} - \frac{9}{32}{\rm i} (\Gamma_{a})^{\beta \lambda} F_{\beta \rho} W_{\alpha \lambda \gamma i} {W}^{-1} \nabla^{\rho \gamma}{W} - \frac{9}{80}{\rm i} (\Gamma_{a})_{\alpha \lambda} F^{\beta}\,_{\rho} X_{i \beta} {W}^{-1} \nabla^{\lambda \rho}{W} - \frac{9}{32}{\rm i} (\Gamma_{a})^{\beta}{}_{\lambda} F_{\beta \rho} X_{i \alpha} {W}^{-1} \nabla^{\lambda \rho}{W}+\frac{9}{32}{\rm i} (\Gamma_{a})_{\alpha}{}^{\beta} F_{\beta \rho} X_{i \lambda} {W}^{-1} \nabla^{\rho \lambda}{W} - \frac{27}{128}(\Gamma_{a})^{\lambda \beta} W_{\lambda \gamma} \lambda_{i \alpha} {W}^{-1} \nabla^{\gamma \rho}{F_{\beta \rho}}+\frac{27}{256}(\Gamma_{a})^{\beta}{}_{\lambda} W_{\beta \rho} \lambda_{i \alpha} {W}^{-1} \nabla_{\gamma}\,^{\rho}{\nabla^{\lambda \gamma}{W}} - \frac{27}{128}(\Gamma_{a})_{\lambda \gamma} W^{\beta \rho} F_{\beta \rho} {W}^{-1} \nabla^{\lambda \gamma}{\lambda_{i \alpha}} - \frac{9}{256}(\Gamma_{a})^{\beta}{}_{\rho} W_{\alpha \beta} \lambda_{i \lambda} {W}^{-1} \nabla_{\gamma}\,^{\lambda}{\nabla^{\rho \gamma}{W}}+\frac{9}{64}(\Gamma_{a})^{\beta}{}_{\lambda} {W}^{-1} \nabla^{\rho \gamma}{F_{\beta \rho}} \nabla^{\lambda}\,_{\alpha}{\lambda_{i \gamma}} - \frac{9}{128}(\Gamma_{a})_{\beta \rho} {W}^{-1} \nabla^{\beta}\,_{\alpha}{\lambda_{i \lambda}} \nabla_{\gamma}\,^{\lambda}{\nabla^{\rho \gamma}{W}}+\frac{9}{64}(\Gamma_{a})^{\beta}{}_{\lambda} {W}^{-1} \nabla_{\gamma}\,^{\rho}{F_{\beta \rho}} \nabla^{\lambda \gamma}{\lambda_{i \alpha}}+\frac{9}{128}(\Gamma_{a})_{\beta \rho} {W}^{-1} \nabla^{\beta}\,_{\lambda}{\lambda_{i \alpha}} \nabla^{\lambda}\,_{\gamma}{\nabla^{\rho \gamma}{W}} - \frac{27}{64}(\Gamma_{a})^{\beta}{}_{\gamma} W^{\rho}\,_{\lambda} F_{\beta \rho} {W}^{-1} \nabla^{\gamma \lambda}{\lambda_{i \alpha}}+\frac{27}{256}(\Gamma_{a})_{\lambda \gamma} W_{\beta \rho} {W}^{-1} \nabla^{\lambda \beta}{W} \nabla^{\gamma \rho}{\lambda_{i \alpha}} - \frac{9}{64}(\Gamma_{a})_{\alpha}{}^{\beta} {W}^{-1} \nabla_{\gamma}\,^{\rho}{F_{\beta \rho}} \nabla^{\gamma \lambda}{\lambda_{i \lambda}}+\frac{27}{640}(\Gamma_{a})_{\alpha \beta} {W}^{-1} \nabla_{\lambda}\,^{\rho}{\lambda_{i \rho}} \nabla_{\gamma}\,^{\lambda}{\nabla^{\beta \gamma}{W}}%
 - \frac{3}{80}{\rm i} (\Gamma_{a})_{\alpha}{}^{\beta} \lambda_{j \beta} {W}^{-2} \nabla_{\gamma}\,^{\rho}{\lambda_{i \rho}} \nabla^{\gamma \lambda}{\lambda^{j}_{\lambda}}+\frac{9}{16}(\Gamma_{a})^{\beta}{}_{\lambda} {W}^{-1} \nabla_{\alpha}\,^{\rho}{F_{\beta \rho}} \nabla^{\lambda \gamma}{\lambda_{i \gamma}} - \frac{9}{32}(\Gamma_{a})_{\beta \rho} {W}^{-1} \nabla^{\beta \lambda}{\lambda_{i \lambda}} \nabla_{\alpha \gamma}{\nabla^{\rho \gamma}{W}}+\frac{27}{128}(\Gamma_{a})_{\alpha}{}^{\beta} W^{\lambda}\,_{\gamma} \lambda_{i \lambda} {W}^{-1} \nabla^{\gamma \rho}{F_{\beta \rho}} - \frac{9}{1280}(\Gamma_{a})_{\alpha \lambda} W^{\beta}\,_{\rho} \lambda_{i \beta} {W}^{-1} \nabla_{\gamma}\,^{\rho}{\nabla^{\lambda \gamma}{W}} - \frac{27}{128}(\Gamma_{a})^{\beta}{}_{\gamma} W^{\rho}\,_{\lambda} \lambda_{i \rho} {W}^{-1} \nabla^{\gamma \lambda}{F_{\alpha \beta}} - \frac{27}{256}(\Gamma_{a})_{\lambda \gamma} W^{\beta}\,_{\rho} \lambda_{i \beta} {W}^{-1} \nabla^{\lambda \rho}{\nabla^{\gamma}\,_{\alpha}{W}} - \frac{9}{32}(\Gamma_{a})^{\lambda \beta} W_{\lambda}\,^{\gamma} \lambda_{i \gamma} {W}^{-1} \nabla_{\alpha}\,^{\rho}{F_{\beta \rho}} - \frac{9}{64}(\Gamma_{a})^{\beta}{}_{\lambda} W_{\beta}\,^{\rho} \lambda_{i \rho} {W}^{-1} \nabla_{\alpha \gamma}{\nabla^{\lambda \gamma}{W}}+\frac{9}{320}(\Gamma_{a})_{\alpha}{}^{\beta} X_{i j} F_{\beta \rho} {W}^{-2} \nabla^{\rho \lambda}{\lambda^{j}_{\lambda}}+\frac{9}{32}(\Gamma_{a})^{\beta}{}_{\lambda} F_{\beta \rho} {W}^{-2} \nabla^{\lambda}\,_{\alpha}{W} \nabla^{\rho \gamma}{\lambda_{i \gamma}} - \frac{9}{32}(\Gamma_{a})^{\beta \lambda} F_{\beta \rho} \lambda_{i \alpha} {W}^{-2} \nabla^{\rho \gamma}{F_{\lambda \gamma}}+\frac{9}{64}(\Gamma_{a})^{\beta}{}_{\lambda} F_{\beta \rho} \lambda_{i \alpha} {W}^{-2} \nabla_{\gamma}\,^{\rho}{\nabla^{\lambda \gamma}{W}} - \frac{27}{64}(\Gamma_{a})^{\beta}{}_{\gamma} W^{\rho}\,_{\lambda} F_{\beta \rho} \lambda_{i \alpha} {W}^{-2} \nabla^{\gamma \lambda}{W} - \frac{45}{256}{\rm i} (\Gamma_{a})^{\beta}{}_{\lambda} W_{\beta}\,^{\rho} \lambda_{j \alpha} \lambda^{j}_{\rho} {W}^{-2} \nabla^{\lambda \gamma}{\lambda_{i \gamma}} - \frac{9}{128}(\Gamma_{a})^{\lambda}{}_{\gamma} F^{\beta \rho} F_{\beta \rho} {W}^{-2} \nabla^{\gamma}\,_{\alpha}{\lambda_{i \lambda}}+\frac{63}{640}(\Gamma_{a})_{\alpha \lambda} F^{\beta \rho} F_{\beta \rho} {W}^{-2} \nabla^{\lambda \gamma}{\lambda_{i \gamma}} - \frac{51}{2560}(\Gamma_{a})_{\alpha}{}^{\rho} X_{j k} \lambda^{j}_{\rho} \lambda^{k \beta} X_{i \beta} {W}^{-2}+\frac{9}{256}{\rm i} (\Gamma_{a})^{\beta}{}_{\lambda} W_{\beta}\,^{\rho} \lambda_{i \alpha} \lambda_{j \rho} {W}^{-2} \nabla^{\lambda \gamma}{\lambda^{j}_{\gamma}}+\frac{9}{64}(\Gamma_{a})^{\beta}{}_{\rho} X_{i j} F_{\alpha \beta} {W}^{-2} \nabla^{\rho \lambda}{\lambda^{j}_{\lambda}}%
 - \frac{3}{64}(\Gamma_{a})_{\beta \rho} X_{j k} X^{j k} {W}^{-2} \nabla^{\beta \rho}{\lambda_{i \alpha}}+\frac{189}{512}(\Gamma_{a})_{\rho \lambda} \lambda_{i \gamma} {W}^{-1} \nabla^{\rho \gamma}{W} \nabla^{\lambda \beta}{W_{\alpha \beta}}+\frac{9}{128}{\rm i} (\Gamma_{a})^{\beta \rho} \lambda_{j \beta} {W}^{-2} \nabla_{\alpha \gamma}{\lambda_{i \rho}} \nabla^{\gamma \lambda}{\lambda^{j}_{\lambda}}+\frac{27}{128}(\Gamma_{a})^{\lambda}{}_{\gamma} W_{\beta \rho} \lambda_{i \lambda} {W}^{-2} \nabla^{\gamma \beta}{W} \nabla_{\alpha}\,^{\rho}{W}+\frac{63}{256}(\Gamma_{a})^{\beta \lambda} W_{\beta \rho} {W}^{-1} \nabla_{\gamma}\,^{\rho}{W} \nabla_{\alpha}\,^{\gamma}{\lambda_{i \lambda}} - \frac{9}{64}(\Gamma_{a})^{\beta}{}_{\rho} {W}^{-1} \nabla_{\alpha \lambda}{\lambda_{i \beta}} \nabla_{\gamma}\,^{\lambda}{\nabla^{\rho \gamma}{W}} - \frac{27}{64}(\Gamma_{a})^{\beta}{}_{\gamma} W^{\rho}\,_{\lambda} F_{\alpha \beta} \lambda_{i \rho} {W}^{-2} \nabla^{\gamma \lambda}{W} - \frac{9}{64}(\Gamma_{a})^{\beta}{}_{\lambda} \lambda_{i \alpha} {W}^{-2} \nabla^{\lambda}\,_{\gamma}{W} \nabla^{\gamma \rho}{F_{\beta \rho}}+\frac{9}{128}(\Gamma_{a})_{\beta \rho} \lambda_{i \alpha} {W}^{-2} \nabla^{\beta}\,_{\lambda}{W} \nabla^{\lambda}\,_{\gamma}{\nabla^{\rho \gamma}{W}}+\frac{9}{128}{\rm i} (\Gamma_{a})^{\beta}{}_{\lambda} W_{\beta}\,^{\rho} \lambda_{j \alpha} \lambda_{i \rho} {W}^{-2} \nabla^{\lambda \gamma}{\lambda^{j}_{\gamma}} - \frac{9}{2048}(\Gamma_{a})^{\beta}{}_{\rho} \lambda_{i \alpha} \lambda_{j \lambda} {W}^{-1} \nabla^{\rho \lambda}{X^{j}_{\beta}} - \frac{201}{2048}(\Gamma_{a})_{\rho \lambda} \lambda_{i \alpha} \lambda^{\beta}_{j} {W}^{-1} \nabla^{\rho \lambda}{X^{j}_{\beta}} - \frac{249}{2048}(\Gamma_{a})_{\rho \lambda} \lambda_{j \alpha} \lambda^{\beta}_{i} {W}^{-1} \nabla^{\rho \lambda}{X^{j}_{\beta}} - \frac{21}{1024}(\Gamma_{a})^{\beta}{}_{\gamma} \lambda_{j \alpha} \lambda^{j \rho} {W}^{-1} \nabla^{\gamma \lambda}{W_{\beta \rho \lambda i}} - \frac{57}{2048}(\Gamma_{a})_{\rho \lambda} \lambda_{j \alpha} \lambda^{j \beta} {W}^{-1} \nabla^{\rho \lambda}{X_{i \beta}}+\frac{9}{16}(\Gamma_{a})^{\beta \gamma} F_{\beta \rho} \lambda_{i \gamma} {W}^{-2} \nabla^{\rho \lambda}{F_{\alpha \lambda}} - \frac{9}{32}(\Gamma_{a})^{\beta}{}_{\lambda} F_{\beta \rho} {W}^{-1} \nabla^{\rho \gamma}{\nabla^{\lambda}\,_{\alpha}{\lambda_{i \gamma}}} - \frac{9}{32}(\Gamma_{a})^{\beta}{}_{\lambda} F_{\beta \rho} {W}^{-1} \nabla_{\gamma}\,^{\rho}{\nabla^{\lambda \gamma}{\lambda_{i \alpha}}} - \frac{9}{32}(\Gamma_{a})^{\lambda \beta} W_{\lambda \gamma} F_{\beta \rho} {W}^{-1} \nabla^{\gamma \rho}{\lambda_{i \alpha}}+\frac{9}{160}(\Gamma_{a})_{\alpha}{}^{\beta} F_{\beta \rho} {W}^{-1} \nabla_{\gamma}\,^{\rho}{\nabla^{\gamma \lambda}{\lambda_{i \lambda}}}%
 - \frac{81}{80}(\Gamma_{a})_{\alpha}{}^{\beta} W^{\lambda}\,_{\gamma} F_{\beta \rho} {W}^{-1} \nabla^{\gamma \rho}{\lambda_{i \lambda}}+\frac{9}{64}(\Gamma_{a})_{\beta \rho} {W}^{-1} \nabla^{\beta}\,_{\gamma}{W} \nabla^{\gamma \lambda}{\nabla^{\rho}\,_{\alpha}{\lambda_{i \lambda}}} - \frac{9}{64}(\Gamma_{a})_{\beta \rho} {W}^{-1} \nabla^{\beta}\,_{\lambda}{W} \nabla^{\lambda}\,_{\gamma}{\nabla^{\rho \gamma}{\lambda_{i \alpha}}} - \frac{9}{64}(\Gamma_{a})^{\beta}{}_{\lambda} W_{\beta \rho} {W}^{-1} \nabla^{\lambda}\,_{\gamma}{W} \nabla^{\gamma \rho}{\lambda_{i \alpha}} - \frac{9}{40}(\Gamma_{a})_{\alpha \lambda} W^{\beta}\,_{\rho} {W}^{-1} \nabla^{\lambda}\,_{\gamma}{W} \nabla^{\gamma \rho}{\lambda_{i \beta}} - \frac{9}{16}{\rm i} (\Gamma_{a})_{\lambda \gamma} F^{\beta \rho} W_{\alpha \beta \rho i} {W}^{-1} \nabla^{\lambda \gamma}{W} - \frac{3}{64}{\rm i} (\Gamma_{a})^{\beta}{}_{\rho} \lambda_{j \beta} \lambda^{j}_{\lambda} {W}^{-2} \nabla^{\lambda \gamma}{\nabla^{\rho}\,_{\alpha}{\lambda_{i \gamma}}} - \frac{3}{64}{\rm i} (\Gamma_{a})^{\beta}{}_{\rho} \lambda_{j \beta} \lambda^{j}_{\lambda} {W}^{-2} \nabla_{\gamma}\,^{\lambda}{\nabla^{\rho \gamma}{\lambda_{i \alpha}}} - \frac{3}{64}{\rm i} (\Gamma_{a})^{\beta \lambda} W_{\beta \rho} \lambda_{j \lambda} \lambda^{j}_{\gamma} {W}^{-2} \nabla^{\rho \gamma}{\lambda_{i \alpha}}+\frac{9}{320}{\rm i} (\Gamma_{a})_{\alpha}{}^{\beta} \lambda_{j \beta} \lambda^{j}_{\rho} {W}^{-2} \nabla_{\gamma}\,^{\rho}{\nabla^{\gamma \lambda}{\lambda_{i \lambda}}} - \frac{147}{640}{\rm i} (\Gamma_{a})_{\alpha}{}^{\lambda} W^{\beta}\,_{\rho} \lambda_{j \lambda} \lambda^{j}_{\gamma} {W}^{-2} \nabla^{\rho \gamma}{\lambda_{i \beta}}+\frac{9}{64}(\Gamma_{a})^{\lambda \beta} \lambda_{j \lambda} \lambda^{j}_{\gamma} W_{\alpha \beta \rho i} {W}^{-2} \nabla^{\gamma \rho}{W}+\frac{3}{64}{\rm i} (\Gamma_{a})^{\beta \lambda} W_{\beta \rho} \lambda_{j \lambda} \lambda^{j \gamma} {W}^{-2} \nabla_{\alpha}\,^{\rho}{\lambda_{i \gamma}}+\frac{9}{128}{\rm i} (\Gamma_{a})^{\beta \lambda} W_{\beta}\,^{\rho} \lambda_{j \lambda} \lambda^{j}_{\gamma} {W}^{-2} \nabla_{\alpha}\,^{\gamma}{\lambda_{i \rho}} - \frac{687}{5120}(\Gamma_{a})_{\alpha}{}^{\beta} \lambda^{\rho}_{j} \lambda^{j}_{\gamma} {W}^{-1} \nabla^{\gamma \lambda}{W_{\beta \rho \lambda i}} - \frac{9}{80}(\Gamma_{a})_{\alpha}{}^{\beta} \lambda_{j \gamma} W_{\beta}\,^{\rho}\,_{\lambda i} {W}^{-1} \nabla^{\gamma \lambda}{\lambda^{j}_{\rho}}+\frac{3}{16}(\Gamma_{a})^{\lambda}{}_{\gamma} \lambda_{j \lambda} W_{\alpha}\,^{\beta}\,_{\rho i} {W}^{-1} \nabla^{\gamma \rho}{\lambda^{j}_{\beta}}+\frac{3}{1024}(\Gamma_{a})_{\alpha \gamma} \lambda^{\beta}_{j} \lambda^{j \rho} {W}^{-1} \nabla^{\gamma \lambda}{W_{\beta \rho \lambda i}}+\frac{57}{10240}(\Gamma_{a})_{\alpha \rho} \lambda^{\beta}_{i} \lambda_{j \lambda} {W}^{-1} \nabla^{\rho \lambda}{X^{j}_{\beta}} - \frac{15}{1024}(\Gamma_{a})_{\alpha \gamma} \lambda^{\beta}_{i} \lambda^{\rho}_{j} {W}^{-1} \nabla^{\gamma \lambda}{W_{\beta \rho \lambda}\,^{j}}%
+\frac{3}{16}(\Gamma_{a})^{\gamma \beta} \lambda_{j \gamma} W_{\beta}\,^{\rho}\,_{\lambda}\,^{j} {W}^{-1} \nabla_{\alpha}\,^{\lambda}{\lambda_{i \rho}}+\frac{3}{128}(\Gamma_{a})^{\beta}{}_{\lambda} \lambda^{\rho}_{j} \lambda^{j}_{\gamma} {W}^{-1} \nabla^{\lambda \gamma}{W_{\alpha \beta \rho i}}+\frac{15}{1024}(\Gamma_{a})^{\beta}{}_{\lambda} \lambda^{\rho}_{i} \lambda_{j \gamma} {W}^{-1} \nabla^{\lambda \gamma}{W_{\alpha \beta \rho}\,^{j}} - \frac{27}{1024}(\Gamma_{a})^{\beta}{}_{\rho} \lambda_{i \lambda} \lambda_{j \beta} {W}^{-1} \nabla^{\rho \lambda}{X^{j}_{\alpha}} - \frac{21}{1024}(\Gamma_{a})^{\beta}{}_{\lambda} \lambda_{i \gamma} \lambda^{\rho}_{j} {W}^{-1} \nabla^{\lambda \gamma}{W_{\alpha \beta \rho}\,^{j}}+\frac{9}{160}{\rm i} (\Gamma_{a})_{\alpha \rho} X_{i \beta} {W}^{-1} \nabla^{\rho}\,_{\lambda}{W} \nabla^{\lambda \beta}{W} - \frac{9}{64}(\Gamma_{a})_{\beta \rho} {W}^{-1} \nabla^{\beta}\,_{\gamma}{W} \nabla_{\alpha}\,^{\gamma}{\nabla^{\rho \lambda}{\lambda_{i \lambda}}}+\frac{9}{32}(\Gamma_{a})^{\beta}{}_{\lambda} \lambda_{i \gamma} {W}^{-1} \nabla_{\alpha}\,^{\gamma}{\nabla^{\lambda \rho}{F_{\beta \rho}}} - \frac{9}{64}(\Gamma_{a})_{\beta \rho} \lambda_{i \lambda} {W}^{-1} \nabla_{\alpha}\,^{\lambda}{\nabla^{\beta}\,_{\gamma}{\nabla^{\rho \gamma}{W}}}+\frac{3}{128}(\Gamma_{a})^{\beta}{}_{\gamma} \lambda^{\rho}_{j} \lambda^{j \lambda} {W}^{-1} \nabla^{\gamma}\,_{\alpha}{W_{\beta \rho \lambda i}} - \frac{3}{512}(\Gamma_{a})^{\beta}{}_{\gamma} \lambda^{\rho}_{i} \lambda^{\lambda}_{j} {W}^{-1} \nabla^{\gamma}\,_{\alpha}{W_{\beta \rho \lambda}\,^{j}}+\frac{9}{32}(\Gamma_{a})^{\beta \lambda} F_{\beta \rho} \lambda_{i \lambda} {W}^{-2} \nabla_{\gamma}\,^{\rho}{\nabla_{\alpha}\,^{\gamma}{W}} - \frac{9}{32}(\Gamma_{a})^{\beta \lambda} F_{\beta \rho} {W}^{-1} \nabla_{\gamma}\,^{\rho}{\nabla_{\alpha}\,^{\gamma}{\lambda_{i \lambda}}}+\frac{9}{64}(\Gamma_{a})^{\beta}{}_{\rho} {W}^{-1} \nabla^{\rho}\,_{\lambda}{W} \nabla^{\lambda}\,_{\gamma}{\nabla_{\alpha}\,^{\gamma}{\lambda_{i \beta}}} - \frac{3}{64}{\rm i} (\Gamma_{a})^{\beta \rho} \lambda_{j \beta} \lambda^{j}_{\lambda} {W}^{-2} \nabla_{\gamma}\,^{\lambda}{\nabla_{\alpha}\,^{\gamma}{\lambda_{i \rho}}} - \frac{15}{64}(\Gamma_{a})^{\lambda}{}_{\gamma} \lambda_{j \lambda} \lambda^{j \beta} W_{\alpha \beta \rho i} {W}^{-2} \nabla^{\gamma \rho}{W}+\frac{9}{512}(\Gamma_{a})^{\beta}{}_{\rho} {W}^{-2} \nabla_{\lambda \gamma}{W} \nabla^{\lambda \gamma}{W} \nabla^{\rho}\,_{\alpha}{\lambda_{i \beta}} - \frac{9}{512}(\Gamma_{a})^{\beta \rho} W_{\alpha \beta} \lambda_{i \rho} {W}^{-2} \nabla_{\lambda \gamma}{W} \nabla^{\lambda \gamma}{W}+\frac{117}{2560}(\Gamma_{a})_{\alpha \beta} {W}^{-2} \nabla_{\lambda \gamma}{W} \nabla^{\lambda \gamma}{W} \nabla^{\beta \rho}{\lambda_{i \rho}}+\frac{9}{256}(\Gamma_{a})^{\beta \rho} W_{\alpha \beta} {W}^{-1} \nabla_{\lambda \gamma}{W} \nabla^{\lambda \gamma}{\lambda_{i \rho}}%
 - \frac{9}{256}(\Gamma_{a})^{\beta}{}_{\rho} {W}^{-1} \nabla_{\lambda \gamma}{W} \nabla^{\lambda \gamma}{\nabla^{\rho}\,_{\alpha}{\lambda_{i \beta}}} - \frac{9}{512}(\Gamma_{a})_{\beta \rho} {W}^{-1} \nabla_{\lambda \gamma}{W} \nabla^{\lambda \gamma}{\nabla^{\beta \rho}{\lambda_{i \alpha}}} - \frac{99}{1280}(\Gamma_{a})_{\alpha \beta} {W}^{-1} \nabla_{\lambda \gamma}{W} \nabla^{\lambda \gamma}{\nabla^{\beta \rho}{\lambda_{i \rho}}} - \frac{9}{64}(\Gamma_{a})^{\beta \lambda} X_{i j} F_{\beta \rho} {W}^{-2} \nabla_{\alpha}\,^{\rho}{\lambda^{j}_{\lambda}} - \frac{9}{32}(\Gamma_{a})^{\beta \lambda} F_{\beta \rho} \lambda^{\gamma}_{i} {W}^{-2} \nabla_{\alpha}\,^{\rho}{F_{\lambda \gamma}} - \frac{9}{64}(\Gamma_{a})^{\beta}{}_{\lambda} F_{\beta \rho} \lambda_{i \gamma} {W}^{-2} \nabla_{\alpha}\,^{\rho}{\nabla^{\lambda \gamma}{W}} - \frac{459}{320}(\Gamma_{a})_{\alpha}{}^{\beta} W^{\rho \lambda} F_{\beta \rho} F_{\lambda}\,^{\gamma} \lambda_{i \gamma} {W}^{-2}+\frac{3}{160}(\Gamma_{a})_{\alpha}{}^{\beta} F_{\beta}\,^{\rho} \lambda^{\lambda}_{j} \lambda^{j \gamma} W_{\rho \lambda \gamma i} {W}^{-2} - \frac{3}{1280}(\Gamma_{a})_{\alpha}{}^{\beta} F_{\beta}\,^{\rho} \lambda_{j \rho} \lambda^{j \lambda} X_{i \lambda} {W}^{-2}+\frac{3}{80}(\Gamma_{a})_{\alpha}{}^{\beta} F_{\beta}\,^{\rho} \lambda^{\lambda}_{i} \lambda^{\gamma}_{j} W_{\rho \lambda \gamma}\,^{j} {W}^{-2} - \frac{33}{256}{\rm i} (\Gamma_{a})^{\beta}{}_{\lambda} W_{\beta}\,^{\rho} \lambda_{j \rho} \lambda^{j \gamma} {W}^{-2} \nabla^{\lambda}\,_{\alpha}{\lambda_{i \gamma}} - \frac{9}{64}(\Gamma_{a})^{\beta}{}_{\lambda} X_{i j} \lambda^{j \rho} {W}^{-2} \nabla^{\lambda}\,_{\alpha}{F_{\beta \rho}}+\frac{9}{128}(\Gamma_{a})_{\beta \rho} X_{i j} \lambda^{j}_{\lambda} {W}^{-2} \nabla^{\beta}\,_{\alpha}{\nabla^{\rho \lambda}{W}} - \frac{9}{64}(\Gamma_{a})^{\beta}{}_{\lambda} \lambda^{\rho}_{i} {W}^{-2} \nabla^{\lambda}\,_{\gamma}{W} \nabla_{\alpha}\,^{\gamma}{F_{\beta \rho}}+\frac{9}{128}(\Gamma_{a})_{\beta \rho} \lambda_{i \lambda} {W}^{-2} \nabla^{\beta}\,_{\gamma}{W} \nabla_{\alpha}\,^{\gamma}{\nabla^{\rho \lambda}{W}} - \frac{27}{320}(\Gamma_{a})_{\alpha \gamma} \lambda^{\beta}_{j} \lambda^{j \rho} W_{\beta \rho \lambda i} {W}^{-2} \nabla^{\gamma \lambda}{W}+\frac{3}{64}(\Gamma_{a})_{\lambda \gamma} \lambda^{\beta}_{j} \lambda^{j \rho} W_{\alpha \beta \rho i} {W}^{-2} \nabla^{\lambda \gamma}{W} - \frac{9}{160}(\Gamma_{a})_{\alpha \gamma} \lambda^{\beta}_{i} \lambda^{\rho}_{j} W_{\beta \rho \lambda}\,^{j} {W}^{-2} \nabla^{\gamma \lambda}{W} - \frac{3}{64}(\Gamma_{a})_{\lambda \gamma} \lambda^{\beta}_{i} \lambda^{\rho}_{j} W_{\alpha \beta \rho}\,^{j} {W}^{-2} \nabla^{\lambda \gamma}{W} - \frac{9}{32}(\Gamma_{a})^{\beta \lambda} F_{\beta}\,^{\rho} \lambda_{i \gamma} {W}^{-2} \nabla_{\alpha}\,^{\gamma}{F_{\lambda \rho}}%
 - \frac{9}{64}(\Gamma_{a})^{\beta}{}_{\lambda} F_{\beta \rho} \lambda_{i \gamma} {W}^{-2} \nabla_{\alpha}\,^{\gamma}{\nabla^{\lambda \rho}{W}} - \frac{3}{16}{\rm i} (\Gamma_{a})^{\beta}{}_{\rho} \lambda_{j \lambda} {W}^{-2} \nabla^{\rho \gamma}{\lambda^{j}_{\gamma}} \nabla_{\alpha}\,^{\lambda}{\lambda_{i \beta}}+\frac{9}{64}{\rm i} (\Gamma_{a})^{\beta \lambda} W_{\beta}\,^{\rho} \lambda_{j \rho} \lambda^{j}_{\gamma} {W}^{-2} \nabla_{\alpha}\,^{\gamma}{\lambda_{i \lambda}}+\frac{9}{128}(\Gamma_{a})_{\beta \rho} X_{i j} \lambda^{j}_{\lambda} {W}^{-2} \nabla_{\alpha}\,^{\lambda}{\nabla^{\beta \rho}{W}} - \frac{9}{64}(\Gamma_{a})^{\beta}{}_{\lambda} \lambda_{i \gamma} {W}^{-2} \nabla^{\lambda \rho}{W} \nabla_{\alpha}\,^{\gamma}{F_{\beta \rho}}+\frac{9}{128}(\Gamma_{a})_{\beta \rho} \lambda_{i \lambda} {W}^{-2} \nabla^{\beta}\,_{\gamma}{W} \nabla_{\alpha}\,^{\lambda}{\nabla^{\rho \gamma}{W}}+\frac{9}{32}{\rm i} (\Gamma_{a})^{\lambda \beta} \lambda_{i \lambda} \lambda^{\rho}_{j} \lambda^{j}_{\gamma} {W}^{-3} \nabla_{\alpha}\,^{\gamma}{F_{\beta \rho}}+\frac{9}{64}{\rm i} (\Gamma_{a})^{\beta}{}_{\rho} \lambda_{i \beta} \lambda_{j \lambda} \lambda^{j}_{\gamma} {W}^{-3} \nabla_{\alpha}\,^{\lambda}{\nabla^{\rho \gamma}{W}}+\frac{9}{160}{\rm i} (\Gamma_{a})_{\alpha}{}^{\gamma} \lambda_{j \gamma} \lambda^{j \beta} \lambda^{\rho}_{k} \lambda^{k \lambda} W_{\beta \rho \lambda i} {W}^{-3}+\frac{9}{256}{\rm i} (\Gamma_{a})_{\beta \rho} \lambda^{\lambda}_{j} \lambda^{j}_{\gamma} {W}^{-2} \nabla_{\alpha}\,^{\gamma}{\nabla^{\beta \rho}{\lambda_{i \lambda}}} - \frac{9}{128}{\rm i} (\Gamma_{a})^{\beta}{}_{\rho} \lambda_{j \lambda} \lambda^{j}_{\gamma} {W}^{-2} \nabla_{\alpha}\,^{\lambda}{\nabla^{\rho \gamma}{\lambda_{i \beta}}} - \frac{9}{128}{\rm i} (\Gamma_{a})^{\beta}{}_{\rho} \lambda_{j \beta} \lambda^{j}_{\lambda} {W}^{-2} \nabla_{\alpha}\,^{\lambda}{\nabla^{\rho \gamma}{\lambda_{i \gamma}}}+\frac{57}{320}{\rm i} (\Gamma_{a})_{\alpha}{}^{\lambda} W^{\beta}\,_{\rho} \lambda_{j \beta} \lambda^{j}_{\gamma} {W}^{-2} \nabla^{\rho \gamma}{\lambda_{i \lambda}} - \frac{3}{80}(\Gamma_{a})_{\alpha}{}^{\lambda} F^{\beta \rho} \lambda_{j \beta} \lambda^{j \gamma} W_{\lambda \rho \gamma i} {W}^{-2}+\frac{33}{256}{\rm i} (\Gamma_{a})^{\beta}{}_{\lambda} W_{\beta}\,^{\rho} \lambda_{j \rho} \lambda^{j}_{\gamma} {W}^{-2} \nabla^{\lambda \gamma}{\lambda_{i \alpha}}+\frac{3}{32}(\Gamma_{a})_{\alpha}{}^{\beta} \lambda^{\rho}_{j} \lambda^{j}_{\gamma} W_{\beta \rho \lambda i} {W}^{-2} \nabla^{\gamma \lambda}{W} - \frac{9}{320}{\rm i} (\Gamma_{a})_{\alpha \beta} \lambda_{j \rho} {W}^{-2} \nabla^{\beta \lambda}{\lambda^{j}_{\lambda}} \nabla^{\rho \gamma}{\lambda_{i \gamma}} - \frac{9}{32}(\Gamma_{a})^{\beta}{}_{\gamma} F_{\beta}\,^{\rho} F_{\rho}\,^{\lambda} {W}^{-2} \nabla^{\gamma}\,_{\alpha}{\lambda_{i \lambda}}+\frac{9}{64}(\Gamma_{a})_{\lambda \gamma} F^{\beta}\,_{\rho} {W}^{-2} \nabla^{\lambda \rho}{W} \nabla^{\gamma}\,_{\alpha}{\lambda_{i \beta}}+\frac{3}{128}{\rm i} (\Gamma_{a})_{\beta \rho} \lambda_{j \lambda} {W}^{-2} \nabla^{\beta}\,_{\alpha}{\lambda^{\gamma}_{i}} \nabla^{\rho \lambda}{\lambda^{j}_{\gamma}}%
+\frac{3}{128}{\rm i} (\Gamma_{a})_{\beta \rho} \lambda^{\lambda}_{j} {W}^{-2} \nabla^{\beta}\,_{\alpha}{\lambda_{i \gamma}} \nabla^{\rho \gamma}{\lambda^{j}_{\lambda}}+\frac{3}{32}{\rm i} (\Gamma_{a})^{\lambda}{}_{\gamma} F^{\beta \rho} \lambda_{j \lambda} \lambda^{j}_{\beta} {W}^{-3} \nabla^{\gamma}\,_{\alpha}{\lambda_{i \rho}}+\frac{9}{32}(\Gamma_{a})^{\beta}{}_{\gamma} F_{\beta}\,^{\rho} F_{\rho \lambda} {W}^{-2} \nabla^{\gamma \lambda}{\lambda_{i \alpha}} - \frac{9}{64}(\Gamma_{a})_{\lambda \gamma} F_{\beta \rho} {W}^{-2} \nabla^{\lambda \beta}{W} \nabla^{\gamma \rho}{\lambda_{i \alpha}}+\frac{3}{128}{\rm i} (\Gamma_{a})_{\beta \rho} \lambda^{\lambda}_{j} {W}^{-2} \nabla^{\beta}\,_{\gamma}{\lambda_{i \alpha}} \nabla^{\rho \gamma}{\lambda^{j}_{\lambda}}+\frac{3}{128}{\rm i} (\Gamma_{a})_{\beta \rho} \lambda_{j \lambda} {W}^{-2} \nabla^{\beta \gamma}{\lambda_{i \alpha}} \nabla^{\rho \lambda}{\lambda^{j}_{\gamma}} - \frac{3}{32}{\rm i} (\Gamma_{a})^{\lambda}{}_{\gamma} F^{\beta}\,_{\rho} \lambda_{j \lambda} \lambda^{j}_{\beta} {W}^{-3} \nabla^{\gamma \rho}{\lambda_{i \alpha}} - \frac{3}{128}{\rm i} (\Gamma_{a})^{\beta}{}_{\lambda} W_{\beta \rho} \lambda_{i \alpha} \lambda^{\gamma}_{j} {W}^{-2} \nabla^{\lambda \rho}{\lambda^{j}_{\gamma}} - \frac{3}{128}{\rm i} (\Gamma_{a})^{\beta}{}_{\lambda} W_{\beta}\,^{\rho} \lambda_{i \alpha} \lambda_{j \gamma} {W}^{-2} \nabla^{\lambda \gamma}{\lambda^{j}_{\rho}}+\frac{3}{32}{\rm i} (\Gamma_{a})^{\lambda \gamma} W_{\lambda}\,^{\beta} F_{\beta}\,^{\rho} \lambda_{i \alpha} \lambda_{j \gamma} \lambda^{j}_{\rho} {W}^{-3} - \frac{99}{1280}{\rm i} (\Gamma_{a})_{\alpha}{}^{\beta} W_{\beta \rho} \lambda^{\lambda}_{i} \lambda_{j \lambda} {W}^{-2} \nabla^{\rho \gamma}{\lambda^{j}_{\gamma}} - \frac{27}{160}(\Gamma_{a})_{\alpha}{}^{\beta} F_{\beta}\,^{\rho} F_{\rho \lambda} {W}^{-2} \nabla^{\lambda \gamma}{\lambda_{i \gamma}}+\frac{9}{80}(\Gamma_{a})_{\alpha \lambda} F_{\beta \rho} {W}^{-2} \nabla^{\lambda \beta}{W} \nabla^{\rho \gamma}{\lambda_{i \gamma}}+\frac{3}{320}{\rm i} (\Gamma_{a})_{\alpha \beta} \lambda^{\rho}_{j} {W}^{-2} \nabla^{\beta}\,_{\gamma}{\lambda^{j}_{\rho}} \nabla^{\gamma \lambda}{\lambda_{i \lambda}} - \frac{3}{80}{\rm i} (\Gamma_{a})_{\alpha \beta} \lambda_{j \rho} {W}^{-2} \nabla^{\beta \rho}{\lambda^{j}_{\lambda}} \nabla^{\lambda \gamma}{\lambda_{i \gamma}}+\frac{3}{64}{\rm i} (\Gamma_{a})_{\alpha}{}^{\lambda} F^{\beta}\,_{\rho} \lambda_{j \lambda} \lambda^{j}_{\beta} {W}^{-3} \nabla^{\rho \gamma}{\lambda_{i \gamma}} - \frac{9}{32}(\Gamma_{a})^{\beta \lambda} F_{\beta}\,^{\rho} F_{\lambda \rho} {W}^{-2} \nabla_{\alpha}\,^{\gamma}{\lambda_{i \gamma}} - \frac{9}{16}(\Gamma_{a})^{\beta \lambda} F_{\beta}\,^{\rho} \lambda_{i \rho} {W}^{-2} \nabla_{\alpha}\,^{\gamma}{F_{\lambda \gamma}}+\frac{27}{128}(\Gamma_{a})^{\lambda \beta} W_{\lambda \gamma} F_{\beta}\,^{\rho} \lambda_{i \rho} {W}^{-2} \nabla_{\alpha}\,^{\gamma}{W} - \frac{9}{32}(\Gamma_{a})^{\beta}{}_{\lambda} F_{\beta}\,^{\rho} \lambda_{i \rho} {W}^{-2} \nabla_{\alpha \gamma}{\nabla^{\lambda \gamma}{W}}%
 - \frac{9}{128}{\rm i} (\Gamma_{a})_{\alpha \lambda} W^{\beta}\,_{\rho} \lambda_{i \beta} \lambda^{\gamma}_{j} {W}^{-2} \nabla^{\lambda \rho}{\lambda^{j}_{\gamma}} - \frac{153}{1280}{\rm i} (\Gamma_{a})_{\alpha \lambda} W^{\beta \rho} \lambda_{i \beta} \lambda_{j \gamma} {W}^{-2} \nabla^{\lambda \gamma}{\lambda^{j}_{\rho}}+\frac{363}{640}{\rm i} (\Gamma_{a})_{\alpha}{}^{\gamma} W^{\beta \lambda} F_{\beta}\,^{\rho} \lambda_{i \lambda} \lambda_{j \gamma} \lambda^{j}_{\rho} {W}^{-3}+\frac{27}{64}(\Gamma_{a})^{\beta}{}_{\gamma} W_{\alpha}\,^{\lambda} F_{\beta \rho} \lambda_{i \lambda} {W}^{-2} \nabla^{\gamma \rho}{W} - \frac{3}{128}(\Gamma_{a})_{\beta \rho} X_{i j} X^{j}\,_{k} {W}^{-2} \nabla^{\beta \rho}{\lambda^{k}_{\alpha}}+\frac{3}{128}{\rm i} (\Gamma_{a})^{\beta}{}_{\rho} \lambda^{\lambda}_{j} {W}^{-2} \nabla^{\rho}\,_{\gamma}{\lambda^{j}_{\lambda}} \nabla_{\alpha}\,^{\gamma}{\lambda_{i \beta}} - \frac{3}{128}{\rm i} (\Gamma_{a})^{\beta}{}_{\rho} \lambda_{j \lambda} {W}^{-2} \nabla^{\rho \lambda}{\lambda^{j}_{\gamma}} \nabla_{\alpha}\,^{\gamma}{\lambda_{i \beta}}+\frac{9}{32}(\Gamma_{a})^{\beta \gamma} F_{\beta}\,^{\rho} F_{\rho \lambda} {W}^{-2} \nabla_{\alpha}\,^{\lambda}{\lambda_{i \gamma}}+\frac{9}{64}(\Gamma_{a})^{\lambda}{}_{\gamma} F_{\beta \rho} {W}^{-2} \nabla^{\gamma \beta}{W} \nabla_{\alpha}\,^{\rho}{\lambda_{i \lambda}} - \frac{3}{32}{\rm i} (\Gamma_{a})^{\lambda \gamma} F^{\beta}\,_{\rho} \lambda_{j \lambda} \lambda^{j}_{\beta} {W}^{-3} \nabla_{\alpha}\,^{\rho}{\lambda_{i \gamma}}+\frac{3}{64}{\rm i} (\Gamma_{a})^{\rho \gamma} W_{\rho}\,^{\lambda} F_{\alpha}\,^{\beta} \lambda_{i \lambda} \lambda_{j \gamma} \lambda^{j}_{\beta} {W}^{-3} - \frac{9}{128}{\rm i} (\Gamma_{a})^{\lambda \beta} W_{\lambda}\,^{\gamma} F_{\beta}\,^{\rho} \lambda_{i \alpha} \lambda_{j \gamma} \lambda^{j}_{\rho} {W}^{-3}+\frac{9}{256}{\rm i} (\Gamma_{a})^{\beta}{}_{\lambda} W_{\beta}\,^{\rho} \lambda_{i \alpha} \lambda_{j \rho} \lambda^{j}_{\gamma} {W}^{-3} \nabla^{\lambda \gamma}{W} - \frac{9}{128}{\rm i} (\Gamma_{a})_{\alpha}{}^{\beta} X_{i j} \lambda_{k \beta} \lambda^{k}_{\rho} {W}^{-3} \nabla^{\rho \lambda}{\lambda^{j}_{\lambda}} - \frac{9}{64}{\rm i} (\Gamma_{a})^{\lambda}{}_{\gamma} F^{\beta \rho} \lambda_{j \beta} \lambda^{j}_{\rho} {W}^{-3} \nabla^{\gamma}\,_{\alpha}{\lambda_{i \lambda}}+\frac{3}{160}{\rm i} (\Gamma_{a})_{\alpha \lambda} F^{\beta \rho} \lambda_{j \beta} \lambda^{j}_{\rho} {W}^{-3} \nabla^{\lambda \gamma}{\lambda_{i \gamma}} - \frac{9}{128}{\rm i} (\Gamma_{a})^{\beta \rho} X_{i j} \lambda^{j}_{\beta} \lambda_{k \rho} {W}^{-3} \nabla_{\alpha}\,^{\lambda}{\lambda^{k}_{\lambda}} - \frac{9}{32}(\Gamma_{a})^{\beta}{}_{\lambda} \lambda_{i \gamma} {W}^{-2} \nabla^{\lambda \gamma}{W} \nabla_{\alpha}\,^{\rho}{F_{\beta \rho}}+\frac{9}{64}(\Gamma_{a})_{\beta \rho} \lambda_{i \lambda} {W}^{-2} \nabla^{\beta \lambda}{W} \nabla_{\alpha \gamma}{\nabla^{\rho \gamma}{W}}+\frac{27}{128}(\Gamma_{a})^{\beta \rho} \lambda_{i \beta} \lambda_{j \rho} \lambda^{j \lambda} \lambda_{k \lambda} {W}^{-4} \nabla_{\alpha}\,^{\gamma}{\lambda^{k}_{\gamma}}%
+\frac{3}{16}{\rm i} (\Gamma_{a})^{\lambda \beta} \lambda^{\gamma}_{i} \lambda_{j \lambda} \lambda^{j}_{\gamma} {W}^{-3} \nabla_{\alpha}\,^{\rho}{F_{\beta \rho}}+\frac{3}{32}{\rm i} (\Gamma_{a})^{\beta}{}_{\rho} \lambda^{\lambda}_{i} \lambda_{j \beta} \lambda^{j}_{\lambda} {W}^{-3} \nabla_{\alpha \gamma}{\nabla^{\rho \gamma}{W}} - \frac{3}{64}{\rm i} (\Gamma_{a})^{\beta}{}_{\rho} \lambda^{\lambda}_{i} \lambda_{j \lambda} {W}^{-2} \nabla_{\alpha \gamma}{\nabla^{\rho \gamma}{\lambda^{j}_{\beta}}}+\frac{3}{128}{\rm i} (\Gamma_{a})_{\beta \rho} \lambda^{\lambda}_{i} \lambda_{j \lambda} {W}^{-2} \nabla_{\alpha}\,^{\gamma}{\nabla^{\beta \rho}{\lambda^{j}_{\gamma}}} - \frac{15}{128}{\rm i} (\Gamma_{a})^{\beta \rho} X_{j k} \lambda^{j}_{\beta} \lambda^{k}_{\lambda} {W}^{-3} \nabla_{\alpha}\,^{\lambda}{\lambda_{i \rho}} - \frac{261}{1280}{\rm i} (\Gamma_{a})_{\alpha}{}^{\beta} X_{j k} W_{\beta}\,^{\rho} \lambda_{i \rho} \lambda^{j \lambda} \lambda^{k}_{\lambda} {W}^{-3} - \frac{3}{16}{\rm i} (\Gamma_{a})^{\beta}{}_{\lambda} F_{\beta}\,^{\rho} \lambda_{j \rho} \lambda^{j \gamma} {W}^{-3} \nabla^{\lambda}\,_{\alpha}{\lambda_{i \gamma}}+\frac{3}{32}{\rm i} (\Gamma_{a})_{\beta \rho} \lambda^{\lambda}_{j} \lambda^{j}_{\gamma} {W}^{-3} \nabla^{\beta \gamma}{W} \nabla^{\rho}\,_{\alpha}{\lambda_{i \lambda}}+\frac{9}{128}(\Gamma_{a})^{\beta}{}_{\rho} \lambda_{j \beta} \lambda^{j \lambda} \lambda_{k \lambda} \lambda^{k \gamma} {W}^{-4} \nabla^{\rho}\,_{\alpha}{\lambda_{i \gamma}}+\frac{3}{16}{\rm i} (\Gamma_{a})^{\beta}{}_{\lambda} F_{\beta}\,^{\rho} \lambda_{j \rho} \lambda^{j}_{\gamma} {W}^{-3} \nabla^{\lambda \gamma}{\lambda_{i \alpha}} - \frac{3}{32}{\rm i} (\Gamma_{a})_{\beta \rho} \lambda_{j \lambda} \lambda^{j}_{\gamma} {W}^{-3} \nabla^{\beta \lambda}{W} \nabla^{\rho \gamma}{\lambda_{i \alpha}} - \frac{9}{128}(\Gamma_{a})^{\beta}{}_{\rho} \lambda_{j \beta} \lambda^{j \lambda} \lambda_{k \lambda} \lambda^{k}_{\gamma} {W}^{-4} \nabla^{\rho \gamma}{\lambda_{i \alpha}}+\frac{9}{128}(\Gamma_{a})^{\beta \lambda} W_{\beta}\,^{\rho} \lambda_{i \alpha} \lambda_{j \lambda} \lambda^{j \gamma} \lambda_{k \rho} \lambda^{k}_{\gamma} {W}^{-4} - \frac{3}{20}{\rm i} (\Gamma_{a})_{\alpha}{}^{\beta} F_{\beta}\,^{\rho} \lambda_{j \rho} \lambda^{j}_{\lambda} {W}^{-3} \nabla^{\lambda \gamma}{\lambda_{i \gamma}}+\frac{3}{40}{\rm i} (\Gamma_{a})_{\alpha \beta} \lambda_{j \rho} \lambda^{j}_{\lambda} {W}^{-3} \nabla^{\beta \rho}{W} \nabla^{\lambda \gamma}{\lambda_{i \gamma}}+\frac{9}{160}(\Gamma_{a})_{\alpha}{}^{\beta} \lambda_{j \beta} \lambda^{j \rho} \lambda_{k \rho} \lambda^{k}_{\lambda} {W}^{-4} \nabla^{\lambda \gamma}{\lambda_{i \gamma}}+\frac{27}{80}(\Gamma_{a})_{\alpha}{}^{\lambda} W^{\beta \rho} \lambda_{i \beta} \lambda_{j \lambda} \lambda^{j \gamma} \lambda_{k \rho} \lambda^{k}_{\gamma} {W}^{-4} - \frac{9}{256}(\Gamma_{a})^{\beta \lambda} W_{\beta}\,^{\rho} \lambda_{j \alpha} \lambda^{\gamma}_{i} \lambda^{j}_{\gamma} \lambda_{k \lambda} \lambda^{k}_{\rho} {W}^{-4}+\frac{3}{16}{\rm i} (\Gamma_{a})^{\beta \lambda} F_{\beta}\,^{\rho} \lambda_{j \rho} \lambda^{j}_{\gamma} {W}^{-3} \nabla_{\alpha}\,^{\gamma}{\lambda_{i \lambda}}+\frac{3}{32}{\rm i} (\Gamma_{a})^{\beta}{}_{\rho} \lambda_{j \lambda} \lambda^{j}_{\gamma} {W}^{-3} \nabla^{\rho \lambda}{W} \nabla_{\alpha}\,^{\gamma}{\lambda_{i \beta}}%
 - \frac{9}{128}(\Gamma_{a})^{\beta \rho} \lambda_{j \beta} \lambda^{j \lambda} \lambda_{k \lambda} \lambda^{k}_{\gamma} {W}^{-4} \nabla_{\alpha}\,^{\gamma}{\lambda_{i \rho}}+\frac{9}{256}(\Gamma_{a})^{\beta \lambda} W_{\beta}\,^{\rho} \lambda_{j \alpha} \lambda_{i \rho} \lambda^{j \gamma} \lambda_{k \lambda} \lambda^{k}_{\gamma} {W}^{-4}+\frac{27}{512}(\Gamma_{a})^{\beta}{}_{\rho} \lambda^{\lambda}_{j} \lambda^{j \gamma} \lambda_{k \lambda} \lambda^{k}_{\gamma} {W}^{-4} \nabla^{\rho}\,_{\alpha}{\lambda_{i \beta}} - \frac{9}{2560}(\Gamma_{a})_{\alpha \beta} \lambda^{\rho}_{j} \lambda^{j \lambda} \lambda_{k \rho} \lambda^{k}_{\lambda} {W}^{-4} \nabla^{\beta \gamma}{\lambda_{i \gamma}}+\frac{9}{640}{\rm i} (\Gamma_{a})_{\alpha}{}^{\beta} \lambda^{\rho}_{j} X_{i}^{\lambda} W_{\beta \rho \lambda}\,^{j}+\frac{333}{640}(\Gamma_{a})_{\alpha \gamma} W^{\beta \rho} W_{\beta \lambda} \nabla^{\gamma \lambda}{\lambda_{i \rho}} - \frac{9}{1280}(\Gamma_{a})_{\alpha}{}^{\gamma} W^{\beta}\,_{\rho} W_{\beta \lambda} \nabla^{\rho \lambda}{\lambda_{i \gamma}} - \frac{117}{2560}(\Gamma_{a})_{\alpha \lambda} \lambda^{\beta}_{i} \nabla_{\gamma}\,^{\rho}{\nabla^{\lambda \gamma}{W_{\beta \rho}}}+\frac{81}{2560}(\Gamma_{a})_{\alpha}{}^{\beta} \lambda^{\rho}_{i} \nabla_{\lambda \gamma}{\nabla^{\lambda \gamma}{W_{\beta \rho}}}+\frac{117}{2560}(\Gamma_{a})_{\alpha}{}^{\beta} \lambda_{i \lambda} \nabla_{\gamma}\,^{\rho}{\nabla^{\gamma \lambda}{W_{\beta \rho}}} - \frac{9}{640}{\rm i} (\Gamma_{a})_{\alpha \lambda} X_{i}^{\beta} \nabla^{\lambda \rho}{F_{\beta \rho}}+\frac{189}{1280}{\rm i} (\Gamma_{a})_{\alpha \rho} X_{i \beta} \nabla^{\rho}\,_{\lambda}{\nabla^{\lambda \beta}{W}}+\frac{3}{640}{\rm i} (\Gamma_{a})_{\alpha}{}^{\beta} \lambda_{j \beta} {W}^{-2} \nabla_{\lambda \gamma}{\lambda^{\rho}_{i}} \nabla^{\lambda \gamma}{\lambda^{j}_{\rho}}+\frac{3}{80}{\rm i} (\Gamma_{a})_{\alpha \beta} \lambda_{j \rho} {W}^{-2} \nabla^{\beta}\,_{\gamma}{\lambda^{j \lambda}} \nabla^{\gamma \rho}{\lambda_{i \lambda}}+\frac{27}{320}(\Gamma_{a})_{\alpha \lambda} {W}^{-1} \nabla^{\lambda}\,_{\gamma}{F^{\beta}\,_{\rho}} \nabla^{\gamma \rho}{\lambda_{i \beta}} - \frac{9}{160}(\Gamma_{a})_{\alpha \lambda} {W}^{-1} \nabla_{\gamma}\,^{\rho}{F^{\beta}\,_{\rho}} \nabla^{\lambda \gamma}{\lambda_{i \beta}} - \frac{27}{1280}(\Gamma_{a})_{\alpha}{}^{\beta} W_{\beta}\,^{\rho} {W}^{-1} \nabla_{\lambda \gamma}{W} \nabla^{\lambda \gamma}{\lambda_{i \rho}}+\frac{9}{640}(\Gamma_{a})_{\alpha \lambda} \nabla^{\lambda}\,_{\gamma}{W^{\beta}\,_{\rho}} \nabla^{\gamma \rho}{\lambda_{i \beta}}+\frac{21}{320}(\Gamma_{a})_{\alpha \beta} {W}^{-1} \nabla_{\lambda}\,^{\rho}{X_{i j}} \nabla^{\beta \lambda}{\lambda^{j}_{\rho}} - \frac{9}{640}(\Gamma_{a})_{\alpha \beta} {W}^{-1} \nabla_{\lambda \gamma}{\lambda_{i \rho}} \nabla^{\lambda \gamma}{\nabla^{\beta \rho}{W}}%
+\frac{9}{320}(\Gamma_{a})_{\alpha \beta} {W}^{-1} \nabla^{\beta}\,_{\lambda}{\lambda_{i \rho}} \nabla^{\lambda}\,_{\gamma}{\nabla^{\gamma \rho}{W}}+\frac{9}{640}(\Gamma_{a})_{\alpha \beta} {W}^{-1} \nabla_{\lambda \gamma}{\lambda_{i \rho}} \nabla^{\beta \lambda}{\nabla^{\gamma \rho}{W}}+\frac{99}{320}(\Gamma_{a})_{\alpha}{}^{\lambda} W_{\lambda \gamma} F^{\beta}\,_{\rho} {W}^{-1} \nabla^{\gamma \rho}{\lambda_{i \beta}} - \frac{27}{320}(\Gamma_{a})_{\alpha}{}^{\beta} W_{\beta \rho} {W}^{-1} \nabla_{\gamma}\,^{\lambda}{W} \nabla^{\gamma \rho}{\lambda_{i \lambda}} - \frac{9}{640}(\Gamma_{a})_{\alpha}{}^{\beta} \nabla_{\gamma}\,^{\lambda}{W_{\beta \rho}} \nabla^{\gamma \rho}{\lambda_{i \lambda}}+\frac{3}{320}{\rm i} (\Gamma_{a})_{\alpha}{}^{\beta} \lambda_{j \beta} {W}^{-2} \nabla_{\gamma}\,^{\lambda}{\lambda_{i \rho}} \nabla^{\gamma \rho}{\lambda^{j}_{\lambda}} - \frac{3}{320}{\rm i} (\Gamma_{a})_{\alpha \beta} \lambda_{j \rho} {W}^{-2} \nabla^{\beta \gamma}{\lambda^{j}_{\lambda}} \nabla^{\rho \lambda}{\lambda_{i \gamma}}+\frac{9}{320}(\Gamma_{a})_{\alpha}{}^{\beta} {W}^{-1} \nabla_{\gamma}\,^{\lambda}{F_{\beta \rho}} \nabla^{\gamma \rho}{\lambda_{i \lambda}} - \frac{27}{320}(\Gamma_{a})_{\alpha \lambda} {W}^{-1} \nabla^{\beta \gamma}{F_{\beta \rho}} \nabla^{\lambda \rho}{\lambda_{i \gamma}}+\frac{9}{640}(\Gamma_{a})_{\alpha \beta} {W}^{-1} \nabla_{\lambda \gamma}{\lambda_{i \rho}} \nabla^{\gamma \rho}{\nabla^{\beta \lambda}{W}} - \frac{27}{640}(\Gamma_{a})_{\alpha \beta} {W}^{-1} \nabla^{\beta}\,_{\lambda}{\lambda_{i \rho}} \nabla_{\gamma}\,^{\rho}{\nabla^{\lambda \gamma}{W}} - \frac{9}{128}(\Gamma_{a})_{\alpha}{}^{\lambda} W_{\beta \rho} {W}^{-1} \nabla_{\gamma}\,^{\beta}{W} \nabla^{\gamma \rho}{\lambda_{i \lambda}}+\frac{3}{128}{\rm i} (\Gamma_{a})_{\alpha}{}^{\lambda} W^{\beta}\,_{\rho} \lambda_{i \lambda} \lambda_{j \gamma} {W}^{-2} \nabla^{\rho \gamma}{\lambda^{j}_{\beta}} - \frac{63}{128}(\Gamma_{a})_{\alpha \gamma} W^{\beta}\,_{\lambda} \lambda^{\rho}_{i} {W}^{-1} \nabla^{\gamma \lambda}{F_{\beta \rho}} - \frac{9}{128}(\Gamma_{a})_{\alpha}{}^{\gamma} W^{\beta}\,_{\lambda} \lambda_{i \gamma} {W}^{-1} \nabla^{\lambda \rho}{F_{\beta \rho}}+\frac{369}{1280}(\Gamma_{a})_{\alpha \lambda} W_{\beta \rho} \lambda_{i \gamma} {W}^{-1} \nabla^{\lambda \beta}{\nabla^{\rho \gamma}{W}} - \frac{477}{1280}(\Gamma_{a})_{\alpha}{}^{\lambda} W_{\beta \rho} \lambda_{i \lambda} {W}^{-1} \nabla_{\gamma}\,^{\beta}{\nabla^{\gamma \rho}{W}}+\frac{27}{320}(\Gamma_{a})_{\alpha}{}^{\lambda} W_{\lambda \gamma} \lambda^{\beta}_{i} {W}^{-1} \nabla^{\gamma \rho}{F_{\beta \rho}}+\frac{99}{640}(\Gamma_{a})_{\alpha}{}^{\beta} W_{\beta \rho} \lambda_{i \lambda} {W}^{-1} \nabla_{\gamma}\,^{\rho}{\nabla^{\gamma \lambda}{W}}+\frac{657}{640}(\Gamma_{a})_{\alpha}{}^{\beta} W_{\beta}\,^{\rho} W_{\rho \lambda} \lambda_{i \gamma} {W}^{-1} \nabla^{\lambda \gamma}{W}%
+\frac{9}{640}{\rm i} (\Gamma_{a})_{\alpha}{}^{\beta} W_{\beta}\,^{\rho} \lambda_{i \lambda} \lambda_{j \gamma} {W}^{-2} \nabla^{\lambda \gamma}{\lambda^{j}_{\rho}} - \frac{3}{160}{\rm i} (\Gamma_{a})_{\alpha}{}^{\beta} \lambda_{j \rho} {W}^{-2} \nabla_{\gamma}\,^{\rho}{\lambda^{j}_{\beta}} \nabla^{\gamma \lambda}{\lambda_{i \lambda}} - \frac{9}{320}(\Gamma_{a})_{\alpha}{}^{\lambda} {W}^{-1} \nabla_{\gamma}\,^{\beta}{F_{\beta \rho}} \nabla^{\gamma \rho}{\lambda_{i \lambda}}+\frac{9}{640}(\Gamma_{a})_{\alpha}{}^{\beta} {W}^{-1} \nabla_{\rho \lambda}{\lambda_{i \beta}} \nabla^{\rho}\,_{\gamma}{\nabla^{\lambda \gamma}{W}} - \frac{33}{320}(\Gamma_{a})_{\alpha \rho} \lambda^{\beta}_{j} X_{i \beta} {W}^{-1} \nabla^{\rho \lambda}{\lambda^{j}_{\lambda}}+\frac{69}{320}{\rm i} (\Gamma_{a})_{\alpha}{}^{\beta} W_{\beta \rho} \lambda^{\lambda}_{j} \lambda^{j}_{\gamma} {W}^{-2} \nabla^{\rho \gamma}{\lambda_{i \lambda}} - \frac{3}{160}{\rm i} (\Gamma_{a})_{\alpha}{}^{\beta} \lambda_{j \rho} {W}^{-2} \nabla_{\gamma}\,^{\rho}{\lambda_{i \beta}} \nabla^{\gamma \lambda}{\lambda^{j}_{\lambda}} - \frac{9}{640}{\rm i} (\Gamma_{a})_{\alpha \beta} \lambda_{i \rho} {W}^{-2} \nabla^{\beta}\,_{\gamma}{\lambda^{\lambda}_{j}} \nabla^{\gamma \rho}{\lambda^{j}_{\lambda}}+\frac{9}{640}{\rm i} (\Gamma_{a})_{\alpha \beta} \lambda_{i \rho} {W}^{-2} \nabla^{\beta \gamma}{\lambda_{j \lambda}} \nabla^{\rho \lambda}{\lambda^{j}_{\gamma}} - \frac{3}{80}{\rm i} (\Gamma_{a})_{\alpha}{}^{\beta} W_{\beta \rho} \lambda_{i \lambda} \lambda^{\gamma}_{j} {W}^{-2} \nabla^{\rho \lambda}{\lambda^{j}_{\gamma}}+\frac{9}{640}{\rm i} (\Gamma_{a})_{\alpha}{}^{\beta} \lambda_{i \rho} {W}^{-2} \nabla_{\gamma}\,^{\rho}{\lambda_{j \beta}} \nabla^{\gamma \lambda}{\lambda^{j}_{\lambda}} - \frac{3}{160}{\rm i} (\Gamma_{a})_{\alpha}{}^{\lambda} W^{\beta}\,_{\rho} \lambda_{i \gamma} \lambda_{j \beta} {W}^{-2} \nabla^{\rho \gamma}{\lambda^{j}_{\lambda}}+\frac{9}{40}(\Gamma_{a})_{\alpha}{}^{\beta} F_{\beta \rho} F^{\lambda}\,_{\gamma} {W}^{-2} \nabla^{\rho \gamma}{\lambda_{i \lambda}}+\frac{3}{40}{\rm i} (\Gamma_{a})_{\alpha}{}^{\lambda} F^{\beta}\,_{\rho} \lambda_{i \gamma} \lambda_{j \lambda} {W}^{-3} \nabla^{\rho \gamma}{\lambda^{j}_{\beta}} - \frac{9}{80}{\rm i} (\Gamma_{a})_{\alpha \lambda} F^{\beta}\,_{\rho} \lambda^{\gamma}_{i} \lambda_{j \gamma} {W}^{-3} \nabla^{\lambda \rho}{\lambda^{j}_{\beta}}+\frac{9}{80}(\Gamma_{a})_{\alpha}{}^{\lambda} F^{\beta}\,_{\rho} \lambda_{i \gamma} {W}^{-2} \nabla^{\rho \gamma}{F_{\lambda \beta}} - \frac{9}{160}(\Gamma_{a})_{\alpha}{}^{\gamma} F^{\beta}\,_{\rho} \lambda_{i \gamma} {W}^{-2} \nabla^{\rho \lambda}{F_{\beta \lambda}}+\frac{27}{64}(\Gamma_{a})_{\alpha}{}^{\lambda} W_{\lambda}\,^{\beta} F_{\beta \rho} \lambda_{i \gamma} {W}^{-2} \nabla^{\rho \gamma}{W}+\frac{27}{320}(\Gamma_{a})_{\alpha}{}^{\gamma} W^{\beta}\,_{\lambda} F_{\beta \rho} \lambda_{i \gamma} {W}^{-2} \nabla^{\lambda \rho}{W}+\frac{27}{1280}(\Gamma_{a})_{\alpha \gamma} F^{\beta}\,_{\rho} \lambda^{\lambda}_{i} {W}^{-1} \nabla^{\gamma \rho}{W_{\beta \lambda}}%
+\frac{9}{160}(\Gamma_{a})_{\alpha \lambda} F_{\beta \rho} \lambda_{i \gamma} {W}^{-2} \nabla^{\beta \gamma}{\nabla^{\lambda \rho}{W}}+\frac{9}{160}(\Gamma_{a})_{\alpha \lambda} F_{\beta \rho} \lambda_{i \gamma} {W}^{-2} \nabla^{\lambda \beta}{\nabla^{\rho \gamma}{W}} - \frac{9}{320}(\Gamma_{a})_{\alpha}{}^{\lambda} F_{\beta \rho} \lambda_{i \lambda} {W}^{-2} \nabla_{\gamma}\,^{\beta}{\nabla^{\gamma \rho}{W}}+\frac{153}{640}(\Gamma_{a})_{\alpha}{}^{\lambda} W_{\lambda}\,^{\gamma} F^{\beta \rho} F_{\beta \rho} \lambda_{i \gamma} {W}^{-2}+\frac{9}{80}(\Gamma_{a})_{\alpha}{}^{\lambda} F^{\beta \rho} \lambda^{\gamma}_{i} \lambda_{j \gamma} W_{\lambda \beta \rho}\,^{j} {W}^{-2}+\frac{3}{80}(\Gamma_{a})_{\alpha}{}^{\gamma} F^{\beta \rho} \lambda_{i \gamma} \lambda^{\lambda}_{j} W_{\beta \rho \lambda}\,^{j} {W}^{-2} - \frac{21}{640}{\rm i} (\Gamma_{a})_{\alpha \beta} \lambda_{j \rho} {W}^{-2} \nabla^{\beta \rho}{\lambda_{i \lambda}} \nabla^{\lambda \gamma}{\lambda^{j}_{\gamma}} - \frac{93}{1280}{\rm i} (\Gamma_{a})_{\alpha}{}^{\lambda} W^{\beta}\,_{\rho} \lambda_{j \lambda} \lambda^{j}_{\beta} {W}^{-2} \nabla^{\rho \gamma}{\lambda_{i \gamma}} - \frac{3}{40}{\rm i} (\Gamma_{a})_{\alpha}{}^{\lambda} F^{\beta}\,_{\rho} \lambda_{j \lambda} \lambda^{j}_{\gamma} {W}^{-3} \nabla^{\rho \gamma}{\lambda_{i \beta}}+\frac{9}{80}(\Gamma_{a})_{\alpha}{}^{\beta} F_{\beta \rho} \lambda_{j \lambda} {W}^{-2} \nabla^{\rho \lambda}{X_{i}\,^{j}} - \frac{3}{40}(\Gamma_{a})_{\alpha}{}^{\lambda} F^{\beta \rho} \lambda^{\gamma}_{i} \lambda_{j \beta} W_{\lambda \rho \gamma}\,^{j} {W}^{-2}+\frac{9}{80}(\Gamma_{a})_{\alpha}{}^{\beta} C_{\beta}\,^{\rho \lambda \gamma} F_{\rho \lambda} \lambda_{i \gamma} {W}^{-1} - \frac{33}{640}{\rm i} (\Gamma_{a})_{\alpha}{}^{\gamma} W^{\beta \lambda} F_{\beta}\,^{\rho} \lambda_{i \gamma} \lambda_{j \lambda} \lambda^{j}_{\rho} {W}^{-3}+\frac{9}{320}(\Gamma_{a})_{\alpha \lambda} {W}^{-1} \nabla^{\lambda \beta}{F_{\beta \rho}} \nabla^{\rho \gamma}{\lambda_{i \gamma}}+\frac{99}{640}(\Gamma_{a})_{\alpha \beta} {W}^{-1} \nabla_{\lambda}\,^{\rho}{\lambda_{i \rho}} \nabla^{\beta}\,_{\gamma}{\nabla^{\gamma \lambda}{W}} - \frac{9}{128}(\Gamma_{a})_{\alpha \beta} {W}^{-1} \nabla^{\beta \rho}{\lambda_{i \rho}} \nabla_{\lambda \gamma}{\nabla^{\lambda \gamma}{W}}+\frac{153}{1280}(\Gamma_{a})_{\alpha}{}^{\beta} W_{\beta}\,^{\rho} \lambda_{i \rho} {W}^{-1} \nabla_{\lambda \gamma}{\nabla^{\lambda \gamma}{W}} - \frac{81}{320}(\Gamma_{a})_{\alpha}{}^{\beta} W^{\rho}\,_{\lambda} F_{\beta \rho} \lambda_{i \gamma} {W}^{-2} \nabla^{\lambda \gamma}{W} - \frac{3}{640}{\rm i} (\Gamma_{a})_{\alpha}{}^{\lambda} W^{\beta}\,_{\rho} \lambda_{i \beta} \lambda_{j \gamma} {W}^{-2} \nabla^{\rho \gamma}{\lambda^{j}_{\lambda}}+\frac{3}{320}{\rm i} (\Gamma_{a})_{\alpha}{}^{\lambda} F^{\beta}\,_{\rho} \lambda_{i \lambda} \lambda_{j \beta} {W}^{-3} \nabla^{\rho \gamma}{\lambda^{j}_{\gamma}}%
+\frac{27}{80}(\Gamma_{a})_{\alpha}{}^{\beta} F_{\beta}\,^{\rho} F^{\lambda \gamma} F_{\lambda \gamma} \lambda_{i \rho} {W}^{-3} - \frac{27}{160}{\rm i} (\Gamma_{a})_{\alpha}{}^{\lambda} F^{\beta \rho} F_{\beta \rho} \lambda^{\gamma}_{i} \lambda_{j \lambda} \lambda^{j}_{\gamma} {W}^{-4} - \frac{3}{320}{\rm i} (\Gamma_{a})_{\alpha}{}^{\beta} X_{j k} \lambda_{i \rho} \lambda^{j}_{\beta} {W}^{-3} \nabla^{\rho \lambda}{\lambda^{k}_{\lambda}} - \frac{3}{40}(\Gamma_{a})_{\alpha \beta} X_{j k} \lambda_{i \rho} {W}^{-2} \nabla^{\beta \rho}{X^{j k}}+\frac{9}{640}(\Gamma_{a})_{\alpha}{}^{\rho} X_{j k} \lambda^{\beta}_{i} \lambda^{j}_{\rho} X^{k}_{\beta} {W}^{-2}+\frac{33}{320}(\Gamma_{a})_{\alpha \lambda} X_{i j} \lambda^{j \beta} {W}^{-2} \nabla^{\lambda \rho}{F_{\beta \rho}} - \frac{27}{320}(\Gamma_{a})_{\alpha}{}^{\beta} X_{i j} \lambda^{j}_{\lambda} {W}^{-1} \nabla^{\lambda \rho}{W_{\beta \rho}}+\frac{15}{128}(\Gamma_{a})_{\alpha \beta} X_{i j} \lambda^{j}_{\rho} {W}^{-2} \nabla^{\beta}\,_{\lambda}{\nabla^{\lambda \rho}{W}}+\frac{3}{128}(\Gamma_{a})_{\alpha}{}^{\beta} X_{i j} \lambda^{j}_{\beta} {W}^{-2} \nabla_{\rho \lambda}{\nabla^{\rho \lambda}{W}}+\frac{3}{80}(\Gamma_{a})_{\alpha}{}^{\beta} X_{i j} {W}^{-1} \nabla_{\rho \lambda}{\nabla^{\rho \lambda}{\lambda^{j}_{\beta}}} - \frac{3}{40}(\Gamma_{a})_{\alpha}{}^{\beta} X_{j k} X^{j k} F_{\beta}\,^{\rho} \lambda_{i \rho} {W}^{-3}+\frac{3}{40}(\Gamma_{a})_{\alpha \beta} X_{j k} X^{j k} \lambda_{i \rho} {W}^{-3} \nabla^{\beta \rho}{W}+\frac{9}{80}{\rm i} (\Gamma_{a})_{\alpha}{}^{\beta} X_{j k} X^{j k} \lambda^{\rho}_{i} \lambda_{l \beta} \lambda^{l}_{\rho} {W}^{-4} - \frac{27}{1280}(\Gamma_{a})_{\alpha}{}^{\rho} \lambda_{j \rho} \lambda^{j}_{\lambda} X_{i \beta} {W}^{-2} \nabla^{\lambda \beta}{W} - \frac{9}{80}(\Gamma_{a})_{\alpha}{}^{\beta} \lambda_{i \rho} \lambda_{j \lambda} X^{j}_{\beta} {W}^{-2} \nabla^{\rho \lambda}{W} - \frac{63}{1280}(\Gamma_{a})_{\alpha}{}^{\rho} \lambda_{i \rho} \lambda_{j \lambda} X^{j}_{\beta} {W}^{-2} \nabla^{\lambda \beta}{W}+\frac{9}{64}(\Gamma_{a})_{\alpha}{}^{\beta} F_{\beta \rho} {W}^{-2} \nabla_{\gamma}\,^{\rho}{W} \nabla^{\gamma \lambda}{\lambda_{i \lambda}}+\frac{27}{320}(\Gamma_{a})_{\alpha \beta} {W}^{-2} \nabla^{\beta}\,_{\lambda}{W} \nabla^{\lambda}\,_{\gamma}{W} \nabla^{\gamma \rho}{\lambda_{i \rho}} - \frac{27}{320}{\rm i} (\Gamma_{a})_{\alpha}{}^{\beta} \lambda_{i \rho} \lambda_{j \beta} {W}^{-3} \nabla_{\gamma}\,^{\rho}{W} \nabla^{\gamma \lambda}{\lambda^{j}_{\lambda}}+\frac{27}{640}{\rm i} (\Gamma_{a})_{\alpha}{}^{\beta} \lambda_{i \beta} \lambda_{j \rho} {W}^{-3} \nabla_{\gamma}\,^{\rho}{W} \nabla^{\gamma \lambda}{\lambda^{j}_{\lambda}}%
 - \frac{9}{64}(\Gamma_{a})_{\alpha}{}^{\beta} \lambda_{i \lambda} {W}^{-2} \nabla_{\gamma}\,^{\lambda}{W} \nabla^{\gamma \rho}{F_{\beta \rho}}+\frac{9}{80}(\Gamma_{a})_{\alpha \lambda} \lambda^{\beta}_{i} {W}^{-2} \nabla^{\lambda}\,_{\gamma}{W} \nabla^{\gamma \rho}{F_{\beta \rho}}+\frac{9}{320}(\Gamma_{a})_{\alpha}{}^{\lambda} \lambda_{i \lambda} {W}^{-2} \nabla_{\gamma}\,^{\beta}{W} \nabla^{\gamma \rho}{F_{\beta \rho}}+\frac{99}{640}(\Gamma_{a})_{\alpha}{}^{\beta} W_{\beta \rho} \lambda_{i \lambda} {W}^{-2} \nabla_{\gamma}\,^{\rho}{W} \nabla^{\gamma \lambda}{W}+\frac{9}{128}(\Gamma_{a})_{\alpha \beta} \lambda_{i \rho} {W}^{-2} \nabla_{\lambda}\,^{\rho}{W} \nabla_{\gamma}\,^{\lambda}{\nabla^{\beta \gamma}{W}} - \frac{9}{160}(\Gamma_{a})_{\alpha \beta} \lambda_{i \rho} {W}^{-2} \nabla^{\beta}\,_{\lambda}{W} \nabla^{\lambda}\,_{\gamma}{\nabla^{\gamma \rho}{W}}+\frac{9}{640}(\Gamma_{a})_{\alpha}{}^{\beta} \lambda_{i \beta} {W}^{-2} \nabla_{\rho \lambda}{W} \nabla^{\rho}\,_{\gamma}{\nabla^{\lambda \gamma}{W}}+\frac{27}{640}{\rm i} (\Gamma_{a})_{\alpha}{}^{\beta} \lambda_{j \beta} \lambda^{j}_{\rho} {W}^{-3} \nabla_{\gamma}\,^{\rho}{W} \nabla^{\gamma \lambda}{\lambda_{i \lambda}}+\frac{9}{320}(\Gamma_{a})_{\alpha \beta} \lambda_{j \rho} {W}^{-2} \nabla_{\lambda}\,^{\rho}{W} \nabla^{\beta \lambda}{X_{i}\,^{j}} - \frac{27}{320}(\Gamma_{a})_{\alpha \beta} \lambda_{j \rho} {W}^{-2} \nabla^{\beta}\,_{\lambda}{W} \nabla^{\lambda \rho}{X_{i}\,^{j}}+\frac{27}{320}(\Gamma_{a})_{\alpha \beta} {W}^{-1} \nabla_{\lambda \gamma}{W} \nabla^{\beta \lambda}{\nabla^{\gamma \rho}{\lambda_{i \rho}}}+\frac{117}{2560}(\Gamma_{a})_{\alpha}{}^{\beta} \lambda^{\rho}_{i} {W}^{-1} \nabla_{\lambda \gamma}{W} \nabla^{\lambda \gamma}{W_{\beta \rho}}+\frac{27}{512}(\Gamma_{a})_{\alpha \lambda} \lambda^{\beta}_{i} {W}^{-1} \nabla_{\gamma}\,^{\rho}{W} \nabla^{\lambda \gamma}{W_{\beta \rho}}+\frac{3}{160}(\Gamma_{a})_{\alpha}{}^{\beta} \lambda^{\rho}_{i} \lambda_{j \gamma} W_{\beta \rho \lambda}\,^{j} {W}^{-2} \nabla^{\gamma \lambda}{W} - \frac{27}{320}(\Gamma_{a})_{\alpha}{}^{\beta} \lambda^{\rho}_{i} \lambda_{j \beta} \lambda^{j}_{\rho} \lambda_{k \lambda} {W}^{-4} \nabla^{\lambda \gamma}{\lambda^{k}_{\gamma}}+\frac{3}{160}{\rm i} (\Gamma_{a})_{\alpha}{}^{\beta} \lambda^{\lambda}_{i} \lambda_{j \lambda} \lambda^{j}_{\gamma} {W}^{-3} \nabla^{\gamma \rho}{F_{\beta \rho}} - \frac{3}{160}{\rm i} (\Gamma_{a})_{\alpha}{}^{\lambda} \lambda^{\beta}_{i} \lambda_{j \lambda} \lambda^{j}_{\gamma} {W}^{-3} \nabla^{\gamma \rho}{F_{\beta \rho}} - \frac{9}{160}{\rm i} (\Gamma_{a})_{\alpha \beta} \lambda^{\rho}_{i} \lambda_{j \rho} \lambda_{k \lambda} {W}^{-3} \nabla^{\beta \lambda}{X^{j k}} - \frac{3}{80}{\rm i} (\Gamma_{a})_{\alpha}{}^{\beta} \lambda_{i \rho} \lambda_{j \beta} \lambda_{k \lambda} {W}^{-3} \nabla^{\rho \lambda}{X^{j k}}+\frac{3}{64}{\rm i} (\Gamma_{a})_{\alpha \beta} \lambda^{\rho}_{i} \lambda_{j \rho} \lambda^{j}_{\lambda} {W}^{-3} \nabla_{\gamma}\,^{\lambda}{\nabla^{\beta \gamma}{W}}%
 - \frac{3}{320}{\rm i} (\Gamma_{a})_{\alpha}{}^{\beta} \lambda_{i \rho} \lambda_{j \beta} \lambda^{j}_{\lambda} {W}^{-3} \nabla_{\gamma}\,^{\lambda}{\nabla^{\gamma \rho}{W}} - \frac{27}{1280}{\rm i} (\Gamma_{a})_{\alpha}{}^{\lambda} W^{\beta}\,_{\rho} \lambda_{i \lambda} \lambda_{j \beta} \lambda^{j}_{\gamma} {W}^{-3} \nabla^{\rho \gamma}{W} - \frac{99}{1280}{\rm i} (\Gamma_{a})_{\alpha}{}^{\lambda} W^{\beta}\,_{\rho} \lambda_{i \gamma} \lambda_{j \lambda} \lambda^{j}_{\beta} {W}^{-3} \nabla^{\rho \gamma}{W} - \frac{27}{160}(\Gamma_{a})_{\alpha}{}^{\beta} F_{\beta \rho} \lambda^{\lambda}_{i} {W}^{-2} \nabla^{\rho \gamma}{F_{\lambda \gamma}}+\frac{27}{160}(\Gamma_{a})_{\alpha}{}^{\beta} F_{\beta}\,^{\rho} \lambda_{i \gamma} {W}^{-2} \nabla^{\gamma \lambda}{F_{\rho \lambda}} - \frac{9}{64}(\Gamma_{a})_{\alpha \lambda} \lambda_{i \gamma} {W}^{-2} \nabla^{\lambda \beta}{W} \nabla^{\gamma \rho}{F_{\beta \rho}}+\frac{3}{40}{\rm i} (\Gamma_{a})_{\alpha}{}^{\lambda} \lambda_{i \gamma} \lambda_{j \lambda} \lambda^{j \beta} {W}^{-3} \nabla^{\gamma \rho}{F_{\beta \rho}} - \frac{3}{320}{\rm i} (\Gamma_{a})_{\alpha \beta} \lambda_{i \rho} \lambda^{\lambda}_{j} {W}^{-2} \nabla_{\gamma}\,^{\rho}{\nabla^{\beta \gamma}{\lambda^{j}_{\lambda}}} - \frac{3}{320}{\rm i} (\Gamma_{a})_{\alpha}{}^{\beta} \lambda_{i \rho} \lambda_{j \lambda} {W}^{-2} \nabla_{\gamma}\,^{\rho}{\nabla^{\gamma \lambda}{\lambda^{j}_{\beta}}}+\frac{3}{320}{\rm i} (\Gamma_{a})_{\alpha \beta} \lambda_{i \rho} \lambda_{j \lambda} {W}^{-2} \nabla^{\rho \gamma}{\nabla^{\beta \lambda}{\lambda^{j}_{\gamma}}} - \frac{3}{40}{\rm i} (\Gamma_{a})_{\alpha \beta} \lambda_{i \rho} \lambda_{j \lambda} {W}^{-2} \nabla^{\beta \rho}{\nabla^{\lambda \gamma}{\lambda^{j}_{\gamma}}}+\frac{3}{64}{\rm i} (\Gamma_{a})_{\alpha \beta} \lambda_{i \rho} \lambda_{j \lambda} {W}^{-2} \nabla^{\rho \lambda}{\nabla^{\beta \gamma}{\lambda^{j}_{\gamma}}}+\frac{87}{2560}{\rm i} (\Gamma_{a})_{\alpha \lambda} \lambda_{i \gamma} \lambda^{\beta}_{j} \lambda^{j \rho} {W}^{-2} \nabla^{\lambda \gamma}{W_{\beta \rho}} - \frac{117}{2560}{\rm i} (\Gamma_{a})_{\alpha}{}^{\beta} \lambda_{i \lambda} \lambda^{\rho}_{j} \lambda^{j}_{\gamma} {W}^{-2} \nabla^{\lambda \gamma}{W_{\beta \rho}}+\frac{93}{2560}{\rm i} (\Gamma_{a})_{\alpha \lambda} \lambda^{\beta}_{i} \lambda^{\rho}_{j} \lambda^{j}_{\gamma} {W}^{-2} \nabla^{\lambda \gamma}{W_{\beta \rho}}+\frac{3}{160}{\rm i} (\Gamma_{a})_{\alpha}{}^{\beta} \lambda_{k \beta} \lambda^{k}_{\rho} \lambda_{j \lambda} {W}^{-3} \nabla^{\rho \lambda}{X_{i}\,^{j}}+\frac{3}{64}{\rm i} (\Gamma_{a})_{\alpha \beta} \lambda_{i \rho} \lambda_{j \lambda} {W}^{-2} \nabla^{\beta \lambda}{\nabla^{\rho \gamma}{\lambda^{j}_{\gamma}}} - \frac{27}{320}(\Gamma_{a})_{\alpha}{}^{\beta} F_{\beta \rho} \lambda_{i \lambda} {W}^{-2} \nabla_{\gamma}\,^{\rho}{\nabla^{\gamma \lambda}{W}} - \frac{9}{80}(\Gamma_{a})_{\alpha \lambda} F^{\beta}\,_{\rho} \lambda_{i \beta} {W}^{-2} \nabla_{\gamma}\,^{\rho}{\nabla^{\lambda \gamma}{W}}+\frac{27}{320}(\Gamma_{a})_{\alpha}{}^{\beta} F_{\beta \rho} \lambda_{i \lambda} {W}^{-2} \nabla_{\gamma}\,^{\lambda}{\nabla^{\gamma \rho}{W}}%
+\frac{9}{128}(\Gamma_{a})_{\alpha \beta} \lambda_{i \rho} {W}^{-2} \nabla^{\beta}\,_{\lambda}{W} \nabla_{\gamma}\,^{\rho}{\nabla^{\lambda \gamma}{W}}+\frac{3}{80}{\rm i} (\Gamma_{a})_{\alpha}{}^{\beta} \lambda_{i \rho} \lambda_{j \beta} \lambda^{j}_{\lambda} {W}^{-3} \nabla_{\gamma}\,^{\rho}{\nabla^{\gamma \lambda}{W}} - \frac{3}{320}{\rm i} (\Gamma_{a})_{\alpha}{}^{\beta} W_{\beta \rho} \lambda^{\lambda}_{i} \lambda_{j \gamma} {W}^{-2} \nabla^{\rho \gamma}{\lambda^{j}_{\lambda}}+\frac{9}{160}(\Gamma_{a})_{\alpha \lambda} F^{\beta}\,_{\rho} {W}^{-1} \nabla_{\gamma}\,^{\rho}{\nabla^{\lambda \gamma}{\lambda_{i \beta}}} - \frac{9}{160}(\Gamma_{a})_{\alpha}{}^{\lambda} F_{\beta \rho} {W}^{-1} \nabla_{\gamma}\,^{\beta}{\nabla^{\gamma \rho}{\lambda_{i \lambda}}} - \frac{9}{320}(\Gamma_{a})_{\alpha \beta} {W}^{-1} \nabla_{\lambda}\,^{\rho}{W} \nabla_{\gamma}\,^{\lambda}{\nabla^{\beta \gamma}{\lambda_{i \rho}}} - \frac{9}{320}(\Gamma_{a})_{\alpha}{}^{\beta} {W}^{-1} \nabla_{\rho \lambda}{W} \nabla^{\rho}\,_{\gamma}{\nabla^{\lambda \gamma}{\lambda_{i \beta}}}+\frac{3}{320}{\rm i} (\Gamma_{a})_{\alpha \beta} \lambda^{\rho}_{j} \lambda^{j}_{\lambda} {W}^{-2} \nabla_{\gamma}\,^{\lambda}{\nabla^{\beta \gamma}{\lambda_{i \rho}}} - \frac{3}{64}{\rm i} (\Gamma_{a})_{\alpha}{}^{\beta} \lambda_{j \rho} \lambda^{j}_{\lambda} {W}^{-2} \nabla_{\gamma}\,^{\rho}{\nabla^{\gamma \lambda}{\lambda_{i \beta}}} - \frac{9}{80}(\Gamma_{a})_{\alpha \beta} \lambda_{i \rho} {W}^{-1} \nabla_{\lambda}\,^{\rho}{\nabla_{\gamma}\,^{\lambda}{\nabla^{\beta \gamma}{W}}} - \frac{9}{160}(\Gamma_{a})_{\alpha \lambda} F_{\beta \rho} {W}^{-1} \nabla^{\beta \gamma}{\nabla^{\lambda \rho}{\lambda_{i \gamma}}} - \frac{9}{80}(\Gamma_{a})_{\alpha \lambda} W_{\beta \rho} \nabla^{\beta \gamma}{\nabla^{\lambda \rho}{\lambda_{i \gamma}}} - \frac{9}{320}(\Gamma_{a})_{\alpha \beta} {W}^{-1} \nabla_{\lambda \gamma}{W} \nabla^{\gamma \rho}{\nabla^{\beta \lambda}{\lambda_{i \rho}}} - \frac{3}{320}{\rm i} (\Gamma_{a})_{\alpha \beta} \lambda_{j \rho} \lambda^{j}_{\lambda} {W}^{-2} \nabla^{\rho \gamma}{\nabla^{\beta \lambda}{\lambda_{i \gamma}}}+\frac{9}{80}(\Gamma_{a})_{\alpha \lambda} \lambda_{i \gamma} {W}^{-1} \nabla^{\gamma \beta}{\nabla^{\lambda \rho}{F_{\beta \rho}}} - \frac{9}{160}(\Gamma_{a})_{\alpha \beta} \lambda_{i \rho} {W}^{-1} \nabla_{\lambda}\,^{\rho}{\nabla^{\beta}\,_{\gamma}{\nabla^{\gamma \lambda}{W}}} - \frac{9}{320}(\Gamma_{a})_{\alpha \beta} \lambda_{i \rho} {W}^{-1} \nabla^{\beta \rho}{\nabla_{\lambda \gamma}{\nabla^{\lambda \gamma}{W}}}+\frac{9}{320}(\Gamma_{a})_{\alpha}{}^{\beta} F_{\beta}\,^{\rho} \lambda_{i \rho} {W}^{-3} \nabla_{\lambda \gamma}{W} \nabla^{\lambda \gamma}{W} - \frac{9}{640}{\rm i} (\Gamma_{a})_{\alpha}{}^{\beta} \lambda^{\rho}_{i} \lambda_{j \beta} \lambda^{j}_{\rho} {W}^{-4} \nabla_{\lambda \gamma}{W} \nabla^{\lambda \gamma}{W} - \frac{3}{640}{\rm i} (\Gamma_{a})_{\alpha}{}^{\beta} \lambda^{\rho}_{i} \lambda_{j \rho} {W}^{-3} \nabla_{\lambda \gamma}{W} \nabla^{\lambda \gamma}{\lambda^{j}_{\beta}}%
+\frac{3}{320}{\rm i} (\Gamma_{a})_{\alpha}{}^{\beta} \lambda^{\rho}_{i} \lambda_{j \beta} {W}^{-3} \nabla_{\lambda \gamma}{W} \nabla^{\lambda \gamma}{\lambda^{j}_{\rho}} - \frac{9}{320}(\Gamma_{a})_{\alpha}{}^{\beta} F_{\beta}\,^{\rho} {W}^{-2} \nabla_{\lambda \gamma}{W} \nabla^{\lambda \gamma}{\lambda_{i \rho}} - \frac{3}{320}{\rm i} (\Gamma_{a})_{\alpha}{}^{\beta} \lambda_{j \beta} \lambda^{j \rho} {W}^{-3} \nabla_{\lambda \gamma}{W} \nabla^{\lambda \gamma}{\lambda_{i \rho}}+\frac{3}{640}{\rm i} (\Gamma_{a})_{\alpha}{}^{\beta} \lambda^{\rho}_{j} {W}^{-2} \nabla_{\lambda \gamma}{\lambda_{i \rho}} \nabla^{\lambda \gamma}{\lambda^{j}_{\beta}} - \frac{9}{80}(\Gamma_{a})_{\alpha \gamma} F^{\beta \rho} F_{\beta \lambda} {W}^{-2} \nabla^{\gamma \lambda}{\lambda_{i \rho}} - \frac{9}{160}(\Gamma_{a})_{\alpha}{}^{\beta} F_{\beta \rho} {W}^{-2} \nabla_{\gamma}\,^{\lambda}{W} \nabla^{\gamma \rho}{\lambda_{i \lambda}} - \frac{3}{160}{\rm i} (\Gamma_{a})_{\alpha \lambda} F^{\beta}\,_{\rho} \lambda^{\gamma}_{i} \lambda_{j \beta} {W}^{-3} \nabla^{\lambda \rho}{\lambda^{j}_{\gamma}} - \frac{9}{80}(\Gamma_{a})_{\alpha \lambda} F^{\beta}\,_{\rho} {W}^{-2} \nabla^{\lambda}\,_{\gamma}{W} \nabla^{\gamma \rho}{\lambda_{i \beta}}+\frac{9}{320}(\Gamma_{a})_{\alpha \lambda} F^{\beta}\,_{\rho} {W}^{-2} \nabla_{\gamma}\,^{\rho}{W} \nabla^{\lambda \gamma}{\lambda_{i \beta}} - \frac{21}{320}(\Gamma_{a})_{\alpha \beta} X_{i j} {W}^{-2} \nabla_{\lambda}\,^{\rho}{W} \nabla^{\beta \lambda}{\lambda^{j}_{\rho}}+\frac{9}{640}(\Gamma_{a})_{\alpha \beta} {W}^{-2} \nabla^{\beta \rho}{W} \nabla_{\lambda \gamma}{W} \nabla^{\lambda \gamma}{\lambda_{i \rho}} - \frac{9}{320}(\Gamma_{a})_{\alpha \beta} {W}^{-2} \nabla^{\beta}\,_{\lambda}{W} \nabla_{\gamma}\,^{\rho}{W} \nabla^{\lambda \gamma}{\lambda_{i \rho}} - \frac{3}{64}{\rm i} (\Gamma_{a})_{\alpha \beta} \lambda^{\rho}_{i} \lambda_{j \lambda} {W}^{-3} \nabla_{\gamma}\,^{\lambda}{W} \nabla^{\beta \gamma}{\lambda^{j}_{\rho}} - \frac{27}{320}(\Gamma_{a})_{\alpha \lambda} \lambda^{\beta}_{i} {W}^{-2} \nabla_{\gamma}\,^{\rho}{W} \nabla^{\lambda \gamma}{F_{\beta \rho}}+\frac{9}{640}(\Gamma_{a})_{\alpha \beta} \lambda_{i \rho} {W}^{-2} \nabla_{\lambda \gamma}{W} \nabla^{\lambda \gamma}{\nabla^{\beta \rho}{W}} - \frac{9}{640}(\Gamma_{a})_{\alpha \beta} \lambda_{i \rho} {W}^{-2} \nabla_{\lambda \gamma}{W} \nabla^{\beta \lambda}{\nabla^{\gamma \rho}{W}}+\frac{3}{80}{\rm i} (\Gamma_{a})_{\alpha \lambda} F^{\beta}\,_{\rho} \lambda_{j \beta} \lambda^{j \gamma} {W}^{-3} \nabla^{\lambda \rho}{\lambda_{i \gamma}}+\frac{3}{20}{\rm i} (\Gamma_{a})_{\alpha}{}^{\beta} F_{\beta \rho} \lambda^{\lambda}_{j} \lambda^{j}_{\gamma} {W}^{-3} \nabla^{\rho \gamma}{\lambda_{i \lambda}} - \frac{3}{40}{\rm i} (\Gamma_{a})_{\alpha \beta} \lambda^{\rho}_{j} \lambda^{j}_{\lambda} {W}^{-3} \nabla^{\beta}\,_{\gamma}{W} \nabla^{\gamma \lambda}{\lambda_{i \rho}}+\frac{3}{640}{\rm i} (\Gamma_{a})_{\alpha}{}^{\beta} \lambda_{i \rho} {W}^{-2} \nabla_{\gamma}\,^{\rho}{\lambda_{j \lambda}} \nabla^{\gamma \lambda}{\lambda^{j}_{\beta}}%
+\frac{3}{20}{\rm i} (\Gamma_{a})_{\alpha \lambda} F^{\beta \rho} \lambda_{i \gamma} \lambda_{j \beta} {W}^{-3} \nabla^{\lambda \gamma}{\lambda^{j}_{\rho}}+\frac{3}{80}{\rm i} (\Gamma_{a})_{\alpha}{}^{\beta} F_{\beta}\,^{\rho} \lambda_{i \lambda} \lambda_{j \gamma} {W}^{-3} \nabla^{\lambda \gamma}{\lambda^{j}_{\rho}}+\frac{9}{80}(\Gamma_{a})_{\alpha \lambda} F^{\beta \rho} \lambda_{i \gamma} {W}^{-2} \nabla^{\lambda \gamma}{F_{\beta \rho}} - \frac{3}{80}{\rm i} (\Gamma_{a})_{\alpha}{}^{\beta} X_{i j} \lambda^{j}_{\rho} \lambda_{k \beta} {W}^{-3} \nabla^{\rho \lambda}{\lambda^{k}_{\lambda}}+\frac{3}{160}{\rm i} (\Gamma_{a})_{\alpha}{}^{\beta} X_{i j} \lambda^{j}_{\rho} \lambda_{k \lambda} {W}^{-3} \nabla^{\rho \lambda}{\lambda^{k}_{\beta}}+\frac{3}{160}{\rm i} (\Gamma_{a})_{\alpha}{}^{\beta} \lambda_{i \rho} \lambda_{j \beta} {W}^{-3} \nabla_{\gamma}\,^{\lambda}{W} \nabla^{\gamma \rho}{\lambda^{j}_{\lambda}}+\frac{3}{160}{\rm i} (\Gamma_{a})_{\alpha \beta} \lambda_{i \rho} \lambda_{j \lambda} {W}^{-3} \nabla^{\lambda \gamma}{W} \nabla^{\beta \rho}{\lambda^{j}_{\gamma}} - \frac{3}{160}{\rm i} (\Gamma_{a})_{\alpha \beta} \lambda_{i \rho} \lambda_{j \lambda} {W}^{-3} \nabla^{\beta \gamma}{W} \nabla^{\rho \lambda}{\lambda^{j}_{\gamma}}+\frac{9}{320}(\Gamma_{a})_{\alpha}{}^{\beta} \lambda_{i \lambda} {W}^{-2} \nabla_{\gamma}\,^{\rho}{W} \nabla^{\gamma \lambda}{F_{\beta \rho}} - \frac{9}{640}(\Gamma_{a})_{\alpha \beta} \lambda_{i \rho} {W}^{-2} \nabla_{\lambda \gamma}{W} \nabla^{\gamma \rho}{\nabla^{\beta \lambda}{W}} - \frac{3}{160}{\rm i} (\Gamma_{a})_{\alpha \lambda} F^{\beta \rho} \lambda_{j \beta} \lambda^{j}_{\gamma} {W}^{-3} \nabla^{\lambda \gamma}{\lambda_{i \rho}}+\frac{3}{160}{\rm i} (\Gamma_{a})_{\alpha \beta} X_{j k} \lambda^{j \rho} \lambda^{k}_{\lambda} {W}^{-3} \nabla^{\beta \lambda}{\lambda_{i \rho}} - \frac{3}{640}{\rm i} (\Gamma_{a})_{\alpha}{}^{\beta} X_{j k} \lambda^{j}_{\rho} \lambda^{k}_{\lambda} {W}^{-3} \nabla^{\rho \lambda}{\lambda_{i \beta}} - \frac{3}{160}{\rm i} (\Gamma_{a})_{\alpha}{}^{\beta} \lambda_{j \beta} \lambda^{j}_{\rho} {W}^{-3} \nabla_{\gamma}\,^{\lambda}{W} \nabla^{\gamma \rho}{\lambda_{i \lambda}} - \frac{3}{80}{\rm i} (\Gamma_{a})_{\alpha}{}^{\beta} F_{\beta \rho} \lambda_{i \lambda} \lambda^{\gamma}_{j} {W}^{-3} \nabla^{\rho \lambda}{\lambda^{j}_{\gamma}} - \frac{3}{64}{\rm i} (\Gamma_{a})_{\alpha \lambda} F^{\beta}\,_{\rho} \lambda_{i \beta} \lambda^{\gamma}_{j} {W}^{-3} \nabla^{\lambda \rho}{\lambda^{j}_{\gamma}} - \frac{69}{640}{\rm i} (\Gamma_{a})_{\alpha \beta} X_{j k} \lambda_{i \rho} \lambda^{j \lambda} {W}^{-3} \nabla^{\beta \rho}{\lambda^{k}_{\lambda}}+\frac{3}{160}{\rm i} (\Gamma_{a})_{\alpha \beta} \lambda_{i \rho} \lambda^{\lambda}_{j} {W}^{-3} \nabla^{\beta}\,_{\gamma}{W} \nabla^{\gamma \rho}{\lambda^{j}_{\lambda}}+\frac{3}{128}{\rm i} (\Gamma_{a})_{\alpha \beta} \lambda_{i \rho} \lambda^{\lambda}_{j} {W}^{-3} \nabla_{\gamma}\,^{\rho}{W} \nabla^{\beta \gamma}{\lambda^{j}_{\lambda}} - \frac{3}{64}{\rm i} (\Gamma_{a})_{\alpha \lambda} F^{\beta \rho} \lambda_{i \beta} \lambda_{j \gamma} {W}^{-3} \nabla^{\lambda \gamma}{\lambda^{j}_{\rho}}%
+\frac{9}{640}{\rm i} (\Gamma_{a})_{\alpha}{}^{\beta} X_{j k} \lambda_{i \rho} \lambda^{j}_{\lambda} {W}^{-3} \nabla^{\rho \lambda}{\lambda^{k}_{\beta}}+\frac{3}{128}{\rm i} (\Gamma_{a})_{\alpha \beta} X_{j k} \lambda^{\rho}_{i} \lambda^{j}_{\lambda} {W}^{-3} \nabla^{\beta \lambda}{\lambda^{k}_{\rho}} - \frac{3}{128}{\rm i} (\Gamma_{a})_{\alpha \beta} \lambda_{i \rho} \lambda_{j \lambda} {W}^{-3} \nabla^{\rho \gamma}{W} \nabla^{\beta \lambda}{\lambda^{j}_{\gamma}}+\frac{9}{160}(\Gamma_{a})_{\alpha}{}^{\beta} \lambda_{i \rho} \lambda_{j \beta} \lambda^{\lambda}_{k} \lambda^{k}_{\gamma} {W}^{-4} \nabla^{\rho \gamma}{\lambda^{j}_{\lambda}}+\frac{9}{160}(\Gamma_{a})_{\alpha \beta} \lambda^{\rho}_{i} \lambda_{j \rho} \lambda^{\lambda}_{k} \lambda^{k}_{\gamma} {W}^{-4} \nabla^{\beta \gamma}{\lambda^{j}_{\lambda}} - \frac{9}{160}{\rm i} (\Gamma_{a})_{\alpha}{}^{\beta} \lambda_{i \lambda} \lambda^{\rho}_{j} \lambda^{j}_{\gamma} {W}^{-3} \nabla^{\lambda \gamma}{F_{\beta \rho}} - \frac{9}{160}{\rm i} (\Gamma_{a})_{\alpha \lambda} \lambda^{\beta}_{i} \lambda^{\rho}_{j} \lambda^{j}_{\gamma} {W}^{-3} \nabla^{\lambda \gamma}{F_{\beta \rho}} - \frac{9}{64}{\rm i} (\Gamma_{a})_{\alpha}{}^{\beta} W_{\beta}\,^{\rho} \lambda_{i \lambda} \lambda_{j \rho} \lambda^{j}_{\gamma} {W}^{-3} \nabla^{\lambda \gamma}{W} - \frac{9}{320}{\rm i} (\Gamma_{a})_{\alpha \beta} \lambda_{i \rho} \lambda_{j \lambda} \lambda^{j}_{\gamma} {W}^{-3} \nabla^{\rho \lambda}{\nabla^{\beta \gamma}{W}} - \frac{9}{320}{\rm i} (\Gamma_{a})_{\alpha \beta} \lambda_{i \rho} \lambda_{j \lambda} \lambda^{j}_{\gamma} {W}^{-3} \nabla^{\beta \lambda}{\nabla^{\rho \gamma}{W}} - \frac{129}{320}{\rm i} (\Gamma_{a})_{\alpha}{}^{\lambda} W_{\lambda}\,^{\gamma} F^{\beta \rho} \lambda_{i \beta} \lambda_{j \gamma} \lambda^{j}_{\rho} {W}^{-3}+\frac{57}{128}{\rm i} (\Gamma_{a})_{\alpha}{}^{\lambda} W_{\lambda}\,^{\gamma} F^{\beta \rho} \lambda_{i \gamma} \lambda_{j \beta} \lambda^{j}_{\rho} {W}^{-3}+\frac{3}{320}{\rm i} (\Gamma_{a})_{\alpha \beta} \lambda^{\rho}_{j} \lambda^{j}_{\lambda} {W}^{-3} \nabla_{\gamma}\,^{\lambda}{W} \nabla^{\beta \gamma}{\lambda_{i \rho}} - \frac{9}{160}(\Gamma_{a})_{\alpha}{}^{\beta} \lambda_{j \beta} \lambda^{j}_{\rho} \lambda^{\lambda}_{k} \lambda^{k}_{\gamma} {W}^{-4} \nabla^{\rho \gamma}{\lambda_{i \lambda}} - \frac{3}{160}{\rm i} (\Gamma_{a})_{\alpha}{}^{\beta} \lambda^{\rho}_{i} \lambda^{\gamma}_{k} \lambda^{k \lambda} \lambda_{j \gamma} W_{\beta \rho \lambda}\,^{j} {W}^{-3} - \frac{3}{160}{\rm i} (\Gamma_{a})_{\alpha}{}^{\gamma} \lambda^{\beta}_{i} \lambda_{k \gamma} \lambda^{k \rho} \lambda^{\lambda}_{j} W_{\beta \rho \lambda}\,^{j} {W}^{-3} - \frac{9}{320}(\Gamma_{a})_{\alpha}{}^{\beta} X_{i j} \lambda^{\rho}_{k} \lambda^{k \lambda} W_{\beta \rho \lambda}\,^{j} {W}^{-2}+\frac{9}{160}{\rm i} (\Gamma_{a})_{\alpha}{}^{\beta} C_{\beta}\,^{\rho \lambda \gamma} \lambda_{i \rho} \lambda_{j \lambda} \lambda^{j}_{\gamma} {W}^{-2}+\frac{333}{1280}(\Gamma_{a})_{\alpha}{}^{\lambda} W^{\beta \rho} \lambda^{\gamma}_{i} \lambda_{j \lambda} \lambda^{j}_{\gamma} \lambda_{k \beta} \lambda^{k}_{\rho} {W}^{-4} - \frac{27}{160}{\rm i} (\Gamma_{a})_{\alpha \beta} \lambda_{i \rho} \lambda_{j \lambda} {W}^{-3} \nabla^{\beta \lambda}{W} \nabla^{\rho \gamma}{\lambda^{j}_{\gamma}}%
+\frac{9}{160}(\Gamma_{a})_{\alpha}{}^{\beta} \lambda^{\rho}_{i} \lambda_{j \beta} \lambda^{j}_{\lambda} \lambda_{k \rho} {W}^{-4} \nabla^{\lambda \gamma}{\lambda^{k}_{\gamma}} - \frac{3}{640}{\rm i} (\Gamma_{a})_{\alpha \beta} \lambda_{j \rho} {W}^{-2} \nabla^{\beta}\,_{\gamma}{\lambda^{\lambda}_{i}} \nabla^{\gamma \rho}{\lambda^{j}_{\lambda}} - \frac{9}{80}(\Gamma_{a})_{\alpha \gamma} F^{\beta \rho} \lambda_{i \beta} {W}^{-2} \nabla^{\gamma \lambda}{F_{\rho \lambda}} - \frac{27}{160}(\Gamma_{a})_{\alpha \lambda} F^{\beta}\,_{\rho} \lambda_{i \beta} {W}^{-2} \nabla^{\lambda}\,_{\gamma}{\nabla^{\gamma \rho}{W}} - \frac{3}{640}{\rm i} (\Gamma_{a})_{\alpha}{}^{\beta} \lambda^{\rho}_{j} {W}^{-2} \nabla_{\lambda \gamma}{\lambda_{i \beta}} \nabla^{\lambda \gamma}{\lambda^{j}_{\rho}}+\frac{3}{640}{\rm i} (\Gamma_{a})_{\alpha \beta} \lambda_{j \rho} {W}^{-2} \nabla^{\beta \gamma}{\lambda_{i \lambda}} \nabla^{\rho \lambda}{\lambda^{j}_{\gamma}} - \frac{3}{640}{\rm i} (\Gamma_{a})_{\alpha}{}^{\beta} \lambda_{j \rho} {W}^{-2} \nabla_{\gamma}\,^{\rho}{\lambda^{j}_{\lambda}} \nabla^{\gamma \lambda}{\lambda_{i \beta}}+\frac{9}{160}(\Gamma_{a})_{\alpha}{}^{\gamma} F^{\beta}\,_{\rho} F_{\beta \lambda} {W}^{-2} \nabla^{\rho \lambda}{\lambda_{i \gamma}}+\frac{9}{320}(\Gamma_{a})_{\alpha \lambda} F_{\beta \rho} {W}^{-2} \nabla^{\beta \gamma}{W} \nabla^{\lambda \rho}{\lambda_{i \gamma}}+\frac{9}{320}(\Gamma_{a})_{\alpha}{}^{\lambda} F_{\beta \rho} {W}^{-2} \nabla_{\gamma}\,^{\beta}{W} \nabla^{\gamma \rho}{\lambda_{i \lambda}}+\frac{9}{160}{\rm i} (\Gamma_{a})_{\alpha}{}^{\lambda} F^{\beta}\,_{\rho} \lambda_{j \beta} \lambda^{j}_{\gamma} {W}^{-3} \nabla^{\rho \gamma}{\lambda_{i \lambda}} - \frac{9}{320}{\rm i} (\Gamma_{a})_{\alpha}{}^{\lambda} F^{\beta}\,_{\rho} \lambda_{i \beta} \lambda_{j \gamma} {W}^{-3} \nabla^{\rho \gamma}{\lambda^{j}_{\lambda}}+\frac{27}{160}{\rm i} (\Gamma_{a})_{\alpha}{}^{\beta} F_{\beta}\,^{\rho} \lambda_{i \lambda} \lambda_{j \rho} {W}^{-3} \nabla^{\lambda \gamma}{\lambda^{j}_{\gamma}} - \frac{51}{320}{\rm i} (\Gamma_{a})_{\alpha \lambda} F^{\beta \rho} \lambda_{i \beta} \lambda_{j \rho} {W}^{-3} \nabla^{\lambda \gamma}{\lambda^{j}_{\gamma}}+\frac{27}{40}(\Gamma_{a})_{\alpha}{}^{\beta} F_{\beta}\,^{\rho} F_{\rho}\,^{\lambda} F_{\lambda}\,^{\gamma} \lambda_{i \gamma} {W}^{-3}+\frac{27}{80}(\Gamma_{a})_{\alpha \gamma} F^{\beta \rho} F_{\beta \lambda} \lambda_{i \rho} {W}^{-3} \nabla^{\gamma \lambda}{W}+\frac{27}{80}(\Gamma_{a})_{\alpha}{}^{\beta} F_{\beta}\,^{\rho} F_{\rho \lambda} \lambda_{i \gamma} {W}^{-3} \nabla^{\lambda \gamma}{W}+\frac{27}{80}{\rm i} (\Gamma_{a})_{\alpha}{}^{\gamma} F^{\beta \rho} F_{\beta}\,^{\lambda} \lambda_{i \rho} \lambda_{j \gamma} \lambda^{j}_{\lambda} {W}^{-4} - \frac{9}{20}{\rm i} (\Gamma_{a})_{\alpha}{}^{\beta} F_{\beta}\,^{\rho} F_{\rho}\,^{\lambda} \lambda^{\gamma}_{i} \lambda_{j \lambda} \lambda^{j}_{\gamma} {W}^{-4}+\frac{27}{160}(\Gamma_{a})_{\alpha}{}^{\beta} X_{i j} F_{\beta \rho} \lambda^{j}_{\lambda} {W}^{-3} \nabla^{\rho \lambda}{W}%
+\frac{9}{80}{\rm i} (\Gamma_{a})_{\alpha}{}^{\lambda} X_{j k} F^{\beta \rho} \lambda_{i \beta} \lambda^{j}_{\lambda} \lambda^{k}_{\rho} {W}^{-4}+\frac{9}{80}{\rm i} (\Gamma_{a})_{\alpha}{}^{\beta} X_{j k} F_{\beta}\,^{\rho} \lambda^{\lambda}_{i} \lambda^{j}_{\rho} \lambda^{k}_{\lambda} {W}^{-4}+\frac{9}{80}{\rm i} (\Gamma_{a})_{\alpha}{}^{\lambda} X_{i j} F^{\beta \rho} \lambda^{j}_{\beta} \lambda_{k \lambda} \lambda^{k}_{\rho} {W}^{-4}+\frac{153}{1280}{\rm i} (\Gamma_{a})_{\alpha}{}^{\lambda} W^{\beta}\,_{\rho} \lambda_{i \beta} \lambda_{j \lambda} \lambda^{j}_{\gamma} {W}^{-3} \nabla^{\rho \gamma}{W} - \frac{27}{80}(\Gamma_{a})_{\alpha}{}^{\beta} F_{\beta \rho} F^{\lambda}\,_{\gamma} \lambda_{i \lambda} {W}^{-3} \nabla^{\rho \gamma}{W} - \frac{27}{160}(\Gamma_{a})_{\alpha \lambda} F_{\beta \rho} \lambda_{i \gamma} {W}^{-3} \nabla^{\lambda \beta}{W} \nabla^{\rho \gamma}{W} - \frac{27}{160}{\rm i} (\Gamma_{a})_{\alpha}{}^{\lambda} F^{\beta}\,_{\rho} \lambda_{i \gamma} \lambda_{j \lambda} \lambda^{j}_{\beta} {W}^{-4} \nabla^{\rho \gamma}{W}+\frac{9}{40}{\rm i} (\Gamma_{a})_{\alpha \lambda} F^{\beta}\,_{\rho} \lambda^{\gamma}_{i} \lambda_{j \beta} \lambda^{j}_{\gamma} {W}^{-4} \nabla^{\lambda \rho}{W}+\frac{27}{160}{\rm i} (\Gamma_{a})_{\alpha}{}^{\lambda} F^{\beta}\,_{\rho} \lambda_{i \beta} \lambda_{j \lambda} \lambda^{j}_{\gamma} {W}^{-4} \nabla^{\rho \gamma}{W}+\frac{153}{640}(\Gamma_{a})_{\alpha}{}^{\beta} W_{\beta}\,^{\rho} \lambda^{\lambda}_{i} \lambda_{j \rho} \lambda^{j \gamma} \lambda_{k \lambda} \lambda^{k}_{\gamma} {W}^{-4}+\frac{9}{20}{\rm i} (\Gamma_{a})_{\alpha}{}^{\beta} F_{\beta}\,^{\rho} F^{\lambda \gamma} \lambda_{i \rho} \lambda_{j \lambda} \lambda^{j}_{\gamma} {W}^{-4}+\frac{9}{40}(\Gamma_{a})_{\alpha}{}^{\lambda} F^{\beta \rho} \lambda^{\gamma}_{i} \lambda_{j \lambda} \lambda^{j}_{\gamma} \lambda_{k \beta} \lambda^{k}_{\rho} {W}^{-5} - \frac{27}{320}(\Gamma_{a})_{\alpha}{}^{\beta} F_{\beta}\,^{\rho} \lambda_{i \rho} {W}^{-2} \nabla_{\lambda \gamma}{\nabla^{\lambda \gamma}{W}} - \frac{9}{160}(\Gamma_{a})_{\alpha \lambda} \lambda_{i \gamma} {W}^{-2} \nabla^{\gamma \beta}{W} \nabla^{\lambda \rho}{F_{\beta \rho}} - \frac{9}{320}(\Gamma_{a})_{\alpha \beta} \lambda_{i \rho} {W}^{-2} \nabla_{\lambda}\,^{\rho}{W} \nabla^{\beta}\,_{\gamma}{\nabla^{\gamma \lambda}{W}}+\frac{9}{160}(\Gamma_{a})_{\alpha \beta} \lambda_{i \rho} {W}^{-2} \nabla^{\beta \rho}{W} \nabla_{\lambda \gamma}{\nabla^{\lambda \gamma}{W}} - \frac{81}{640}{\rm i} (\Gamma_{a})_{\alpha \beta} \lambda_{i \rho} \lambda_{j \lambda} {W}^{-3} \nabla^{\rho \lambda}{W} \nabla^{\beta \gamma}{\lambda^{j}_{\gamma}} - \frac{27}{640}(\Gamma_{a})_{\alpha}{}^{\beta} \lambda_{i \rho} \lambda_{j \beta} \lambda^{j \lambda} \lambda_{k \lambda} {W}^{-4} \nabla^{\rho \gamma}{\lambda^{k}_{\gamma}} - \frac{27}{640}(\Gamma_{a})_{\alpha \beta} \lambda^{\rho}_{i} \lambda_{j \rho} \lambda^{j \lambda} \lambda_{k \lambda} {W}^{-4} \nabla^{\beta \gamma}{\lambda^{k}_{\gamma}}+\frac{3}{40}{\rm i} (\Gamma_{a})_{\alpha \lambda} \lambda^{\gamma}_{i} \lambda_{j \gamma} \lambda^{j \beta} {W}^{-3} \nabla^{\lambda \rho}{F_{\beta \rho}}%
+\frac{3}{40}{\rm i} (\Gamma_{a})_{\alpha \beta} \lambda^{\rho}_{i} \lambda_{j \rho} \lambda^{j}_{\lambda} {W}^{-3} \nabla^{\beta}\,_{\gamma}{\nabla^{\gamma \lambda}{W}}+\frac{3}{128}{\rm i} (\Gamma_{a})_{\alpha}{}^{\beta} \lambda^{\rho}_{i} \lambda_{j \beta} \lambda^{j}_{\rho} {W}^{-3} \nabla_{\lambda \gamma}{\nabla^{\lambda \gamma}{W}}+\frac{3}{320}{\rm i} (\Gamma_{a})_{\alpha}{}^{\beta} \lambda^{\rho}_{i} \lambda_{j \rho} {W}^{-2} \nabla_{\lambda \gamma}{\nabla^{\lambda \gamma}{\lambda^{j}_{\beta}}} - \frac{63}{320}{\rm i} (\Gamma_{a})_{\alpha}{}^{\beta} X_{j k} X^{j}\,_{l} \lambda^{\rho}_{i} \lambda^{k}_{\beta} \lambda^{l}_{\rho} {W}^{-4} - \frac{39}{320}(\Gamma_{a})_{\alpha \beta} X_{i j} \lambda^{j}_{\rho} {W}^{-3} \nabla^{\beta}\,_{\lambda}{W} \nabla^{\lambda \rho}{W}+\frac{9}{80}{\rm i} (\Gamma_{a})_{\alpha}{}^{\beta} X_{j k} \lambda_{i \rho} \lambda^{j}_{\beta} \lambda^{k}_{\lambda} {W}^{-4} \nabla^{\rho \lambda}{W}+\frac{27}{160}{\rm i} (\Gamma_{a})_{\alpha \beta} X_{j k} \lambda^{\rho}_{i} \lambda^{j}_{\rho} \lambda^{k}_{\lambda} {W}^{-4} \nabla^{\beta \lambda}{W} - \frac{9}{320}{\rm i} (\Gamma_{a})_{\alpha}{}^{\beta} X_{j k} \lambda_{i \beta} \lambda^{j}_{\rho} \lambda^{k}_{\lambda} {W}^{-4} \nabla^{\rho \lambda}{W} - \frac{9}{160}{\rm i} (\Gamma_{a})_{\alpha}{}^{\beta} X_{i j} \lambda^{j}_{\rho} \lambda_{k \beta} \lambda^{k}_{\lambda} {W}^{-4} \nabla^{\rho \lambda}{W} - \frac{9}{40}{\rm i} (\Gamma_{a})_{\alpha}{}^{\beta} X_{j k} F_{\beta}\,^{\rho} \lambda_{i \rho} \lambda^{j \lambda} \lambda^{k}_{\lambda} {W}^{-4} - \frac{9}{40}(\Gamma_{a})_{\alpha}{}^{\beta} X_{j k} \lambda^{\rho}_{i} \lambda^{j \lambda} \lambda^{k}_{\lambda} \lambda_{l \beta} \lambda^{l}_{\rho} {W}^{-5}+\frac{3}{160}{\rm i} (\Gamma_{a})_{\alpha \beta} \lambda_{j \rho} \lambda^{j}_{\lambda} {W}^{-3} \nabla^{\rho \gamma}{W} \nabla^{\beta \lambda}{\lambda_{i \gamma}}+\frac{3}{160}{\rm i} (\Gamma_{a})_{\alpha}{}^{\beta} \lambda_{j \rho} \lambda^{j}_{\lambda} {W}^{-3} \nabla_{\gamma}\,^{\rho}{W} \nabla^{\gamma \lambda}{\lambda_{i \beta}} - \frac{9}{640}(\Gamma_{a})_{\alpha}{}^{\beta} \lambda^{\rho}_{j} \lambda^{j}_{\lambda} \lambda_{k \rho} \lambda^{k}_{\gamma} {W}^{-4} \nabla^{\lambda \gamma}{\lambda_{i \beta}} - \frac{9}{320}{\rm i} (\Gamma_{a})_{\alpha \beta} \lambda^{\rho}_{i} \lambda_{j \rho} {W}^{-3} \nabla_{\gamma}\,^{\lambda}{W} \nabla^{\beta \gamma}{\lambda^{j}_{\lambda}} - \frac{9}{640}(\Gamma_{a})_{\alpha \beta} \lambda^{\rho}_{i} \lambda_{j \rho} \lambda^{j \lambda} \lambda_{k \gamma} {W}^{-4} \nabla^{\beta \gamma}{\lambda^{k}_{\lambda}} - \frac{3}{640}{\rm i} (\Gamma_{a})_{\alpha}{}^{\beta} \lambda_{i \rho} \lambda_{j \lambda} {W}^{-3} \nabla_{\gamma}\,^{\rho}{W} \nabla^{\gamma \lambda}{\lambda^{j}_{\beta}} - \frac{9}{640}(\Gamma_{a})_{\alpha \beta} \lambda^{\rho}_{i} \lambda_{j \rho} \lambda^{j}_{\lambda} \lambda^{\gamma}_{k} {W}^{-4} \nabla^{\beta \lambda}{\lambda^{k}_{\gamma}} - \frac{9}{640}(\Gamma_{a})_{\alpha}{}^{\beta} \lambda^{\rho}_{i} \lambda_{j \rho} \lambda^{j}_{\lambda} \lambda_{k \gamma} {W}^{-4} \nabla^{\lambda \gamma}{\lambda^{k}_{\beta}} - \frac{27}{160}(\Gamma_{a})_{\alpha \lambda} F^{\beta \rho} F_{\beta \rho} \lambda_{i \gamma} {W}^{-3} \nabla^{\lambda \gamma}{W}%
 - \frac{9}{160}(\Gamma_{a})_{\alpha}{}^{\beta} F_{\beta \rho} \lambda_{i \lambda} {W}^{-3} \nabla_{\gamma}\,^{\rho}{W} \nabla^{\gamma \lambda}{W}+\frac{9}{40}{\rm i} (\Gamma_{a})_{\alpha}{}^{\beta} F_{\beta \rho} \lambda^{\lambda}_{i} \lambda_{j \lambda} \lambda^{j}_{\gamma} {W}^{-4} \nabla^{\rho \gamma}{W} - \frac{9}{20}{\rm i} (\Gamma_{a})_{\alpha}{}^{\beta} F_{\beta}\,^{\rho} F^{\lambda \gamma} \lambda_{i \lambda} \lambda_{j \rho} \lambda^{j}_{\gamma} {W}^{-4}+\frac{27}{160}(\Gamma_{a})_{\alpha \lambda} F^{\beta}\,_{\rho} \lambda_{i \beta} {W}^{-3} \nabla^{\lambda}\,_{\gamma}{W} \nabla^{\gamma \rho}{W}+\frac{9}{40}{\rm i} (\Gamma_{a})_{\alpha \lambda} F^{\beta \rho} \lambda_{i \beta} \lambda_{j \rho} \lambda^{j}_{\gamma} {W}^{-4} \nabla^{\lambda \gamma}{W} - \frac{9}{40}{\rm i} (\Gamma_{a})_{\alpha \lambda} F^{\beta \rho} \lambda_{i \gamma} \lambda_{j \beta} \lambda^{j}_{\rho} {W}^{-4} \nabla^{\lambda \gamma}{W} - \frac{9}{40}{\rm i} (\Gamma_{a})_{\alpha}{}^{\beta} F_{\beta}\,^{\rho} \lambda_{i \lambda} \lambda_{j \rho} \lambda^{j}_{\gamma} {W}^{-4} \nabla^{\lambda \gamma}{W} - \frac{477}{2560}(\Gamma_{a})_{\alpha}{}^{\beta} W_{\beta}\,^{\rho} \lambda_{i \rho} \lambda^{\lambda}_{j} \lambda^{j \gamma} \lambda_{k \lambda} \lambda^{k}_{\gamma} {W}^{-4}+\frac{9}{40}(\Gamma_{a})_{\alpha}{}^{\lambda} F^{\beta \rho} \lambda_{i \beta} \lambda_{j \lambda} \lambda^{j \gamma} \lambda_{k \rho} \lambda^{k}_{\gamma} {W}^{-5}+\frac{9}{40}(\Gamma_{a})_{\alpha}{}^{\beta} F_{\beta}\,^{\rho} \lambda^{\lambda}_{i} \lambda_{j \rho} \lambda^{j \gamma} \lambda_{k \lambda} \lambda^{k}_{\gamma} {W}^{-5} - \frac{9}{40}(\Gamma_{a})_{\alpha}{}^{\lambda} F^{\beta \rho} \lambda^{\gamma}_{i} \lambda_{j \lambda} \lambda^{j}_{\beta} \lambda_{k \rho} \lambda^{k}_{\gamma} {W}^{-5}+\frac{9}{80}(\Gamma_{a})_{\alpha}{}^{\beta} X_{j k} \lambda^{\rho}_{i} \lambda^{j}_{\beta} \lambda^{k \lambda} \lambda_{l \rho} \lambda^{l}_{\lambda} {W}^{-5}+\frac{9}{80}(\Gamma_{a})_{\alpha}{}^{\beta} X_{j k} \lambda^{\rho}_{i} \lambda^{j}_{\rho} \lambda^{k \lambda} \lambda_{l \beta} \lambda^{l}_{\lambda} {W}^{-5}+\frac{9}{80}(\Gamma_{a})_{\alpha}{}^{\beta} X_{i j} \lambda^{j \rho} \lambda_{k \beta} \lambda^{k \lambda} \lambda_{l \rho} \lambda^{l}_{\lambda} {W}^{-5} - \frac{9}{640}(\Gamma_{a})_{\alpha \beta} \lambda_{i \rho} {W}^{-3} \nabla^{\beta \rho}{W} \nabla_{\lambda \gamma}{W} \nabla^{\lambda \gamma}{W} - \frac{9}{320}(\Gamma_{a})_{\alpha \beta} \lambda_{i \rho} {W}^{-3} \nabla^{\beta}\,_{\lambda}{W} \nabla^{\lambda}\,_{\gamma}{W} \nabla^{\gamma \rho}{W} - \frac{9}{80}{\rm i} (\Gamma_{a})_{\alpha \beta} \lambda^{\rho}_{i} \lambda_{j \rho} \lambda^{j}_{\lambda} {W}^{-4} \nabla^{\beta}\,_{\gamma}{W} \nabla^{\gamma \lambda}{W}+\frac{9}{320}{\rm i} (\Gamma_{a})_{\alpha}{}^{\beta} \lambda_{i \rho} \lambda_{j \beta} \lambda^{j}_{\lambda} {W}^{-4} \nabla_{\gamma}\,^{\rho}{W} \nabla^{\gamma \lambda}{W}+\frac{9}{80}{\rm i} (\Gamma_{a})_{\alpha \beta} \lambda_{i \rho} \lambda_{j \lambda} \lambda^{j}_{\gamma} {W}^{-4} \nabla^{\beta \lambda}{W} \nabla^{\rho \gamma}{W}+\frac{9}{80}(\Gamma_{a})_{\alpha}{}^{\beta} \lambda_{i \rho} \lambda_{j \beta} \lambda^{j \lambda} \lambda_{k \lambda} \lambda^{k}_{\gamma} {W}^{-5} \nabla^{\rho \gamma}{W}%
+\frac{9}{80}(\Gamma_{a})_{\alpha \beta} \lambda^{\rho}_{i} \lambda_{j \rho} \lambda^{j \lambda} \lambda_{k \lambda} \lambda^{k}_{\gamma} {W}^{-5} \nabla^{\beta \gamma}{W} - \frac{9}{80}(\Gamma_{a})_{\alpha}{}^{\beta} \lambda^{\rho}_{i} \lambda_{j \beta} \lambda^{j}_{\lambda} \lambda_{k \rho} \lambda^{k}_{\gamma} {W}^{-5} \nabla^{\lambda \gamma}{W} - \frac{27}{160}(\Gamma_{a})_{\alpha}{}^{\beta} F_{\beta}\,^{\rho} \lambda_{i \rho} \lambda^{\lambda}_{j} \lambda^{j \gamma} \lambda_{k \lambda} \lambda^{k}_{\gamma} {W}^{-5}+\frac{9}{64}{\rm i} (\Gamma_{a})_{\alpha}{}^{\beta} \lambda^{\rho}_{i} \lambda_{j \beta} \lambda^{j}_{\rho} \lambda^{\lambda}_{k} \lambda^{k \gamma} \lambda_{l \lambda} \lambda^{l}_{\gamma} {W}^{-6}+\frac{9}{1280}(\Gamma_{a})_{\alpha}{}^{\beta} W_{\beta}\,^{\rho} \nabla_{\lambda \gamma}{\nabla^{\lambda \gamma}{\lambda_{i \rho}}}+\frac{9}{80}{\rm i} (\Gamma_{a})_{\alpha}{}^{\beta} W_{\beta \rho \lambda i} \nabla_{\gamma}\,^{\rho}{\nabla^{\gamma \lambda}{W}} - \frac{3}{40}(\Gamma_{a})_{\alpha}{}^{\beta} \lambda_{j \gamma} W_{\beta}\,^{\rho}\,_{\lambda}\,^{j} {W}^{-1} \nabla^{\gamma \lambda}{\lambda_{i \rho}} - \frac{9}{2560}(\Gamma_{a})_{\alpha}{}^{\beta} W_{\beta}\,^{\rho} \lambda_{i \rho} {W}^{-2} \nabla_{\lambda \gamma}{W} \nabla^{\lambda \gamma}{W}+\frac{9}{512}(\Gamma_{a})_{\beta \rho} \nabla^{\beta \rho}{\nabla_{\lambda \gamma}{\nabla^{\lambda \gamma}{\lambda_{i \alpha}}}} - \frac{9}{256}(\Gamma_{a})_{\lambda \gamma} \nabla^{\lambda \gamma}{W^{\beta}\,_{\rho}} \nabla_{\alpha}\,^{\rho}{\lambda_{i \beta}} - \frac{9}{256}(\Gamma_{a})_{\lambda \gamma} W^{\beta}\,_{\rho} \nabla^{\lambda \gamma}{\nabla_{\alpha}\,^{\rho}{\lambda_{i \beta}}}+\frac{45}{256}(\Gamma_{a})_{\rho \lambda} W_{\alpha \beta} \nabla^{\rho \lambda}{\nabla^{\beta \gamma}{\lambda_{i \gamma}}} - \frac{9}{64}{\rm i} (\Gamma_{a})_{\lambda \gamma} W_{\alpha}\,^{\beta \rho}\,_{i} \nabla^{\lambda \gamma}{F_{\beta \rho}} - \frac{9}{128}(\Gamma_{a})_{\lambda \gamma} {W}^{-1} \nabla^{\lambda \gamma}{F^{\beta}\,_{\rho}} \nabla_{\alpha}\,^{\rho}{\lambda_{i \beta}}+\frac{9}{128}(\Gamma_{a})_{\lambda \gamma} F^{\beta}\,_{\rho} {W}^{-2} \nabla^{\lambda \gamma}{W} \nabla_{\alpha}\,^{\rho}{\lambda_{i \beta}} - \frac{9}{128}(\Gamma_{a})_{\lambda \gamma} F^{\beta}\,_{\rho} {W}^{-1} \nabla^{\lambda \gamma}{\nabla_{\alpha}\,^{\rho}{\lambda_{i \beta}}}+\frac{81}{256}(\Gamma_{a})_{\lambda \gamma} F^{\beta \rho} \lambda_{i \alpha} {W}^{-1} \nabla^{\lambda \gamma}{W_{\beta \rho}}+\frac{9}{128}(\Gamma_{a})_{\rho \lambda} {W}^{-1} \nabla^{\rho \lambda}{F_{\alpha \beta}} \nabla^{\beta \gamma}{\lambda_{i \gamma}}+\frac{9}{128}(\Gamma_{a})_{\rho \lambda} F_{\alpha \beta} {W}^{-1} \nabla^{\rho \lambda}{\nabla^{\beta \gamma}{\lambda_{i \gamma}}} - \frac{81}{256}(\Gamma_{a})_{\lambda \gamma} F_{\alpha}\,^{\beta} \lambda^{\rho}_{i} {W}^{-1} \nabla^{\lambda \gamma}{W_{\beta \rho}}%
+\frac{9}{128}(\Gamma_{a})_{\lambda \gamma} F^{\beta \rho} \lambda_{i \alpha} {W}^{-2} \nabla^{\lambda \gamma}{F_{\beta \rho}} - \frac{9}{128}(\Gamma_{a})_{\lambda \gamma} F^{\beta \rho} F_{\beta \rho} \lambda_{i \alpha} {W}^{-3} \nabla^{\lambda \gamma}{W} - \frac{9}{128}(\Gamma_{a})_{\beta \rho} {W}^{-1} \nabla_{\gamma}\,^{\lambda}{\lambda_{i \lambda}} \nabla^{\beta \rho}{\nabla_{\alpha}\,^{\gamma}{W}}+\frac{9}{128}(\Gamma_{a})_{\beta \rho} {W}^{-1} \nabla_{\alpha \gamma}{W} \nabla^{\beta \rho}{\nabla^{\gamma \lambda}{\lambda_{i \lambda}}}+\frac{9}{64}{\rm i} (\Gamma_{a})_{\beta \rho} \lambda_{i \alpha} \lambda_{j \lambda} {W}^{-3} \nabla^{\beta \rho}{W} \nabla^{\lambda \gamma}{\lambda^{j}_{\gamma}} - \frac{9}{128}{\rm i} (\Gamma_{a})_{\beta \rho} \lambda_{i \alpha} \lambda_{j \lambda} {W}^{-2} \nabla^{\beta \rho}{\nabla^{\lambda \gamma}{\lambda^{j}_{\gamma}}}+\frac{9}{64}(\Gamma_{a})_{\rho \lambda} \lambda_{i \gamma} {W}^{-1} \nabla^{\rho \lambda}{\nabla^{\gamma \beta}{F_{\alpha \beta}}} - \frac{81}{512}(\Gamma_{a})_{\rho \lambda} W_{\alpha \beta} \lambda_{i \gamma} {W}^{-1} \nabla^{\rho \lambda}{\nabla^{\beta \gamma}{W}}+\frac{9}{128}(\Gamma_{a})_{\beta \rho} {W}^{-1} \nabla_{\alpha}\,^{\lambda}{X_{i j}} \nabla^{\beta \rho}{\lambda^{j}_{\lambda}} - \frac{9}{128}(\Gamma_{a})_{\beta \rho} \lambda_{j \lambda} {W}^{-1} \nabla^{\beta \rho}{\nabla_{\alpha}\,^{\lambda}{X_{i}\,^{j}}}+\frac{9}{128}(\Gamma_{a})_{\beta \rho} \lambda_{i \lambda} {W}^{-1} \nabla^{\beta \rho}{\nabla_{\gamma}\,^{\lambda}{\nabla_{\alpha}\,^{\gamma}{W}}}+\frac{81}{512}(\Gamma_{a})_{\lambda \gamma} \lambda^{\beta}_{i} {W}^{-1} \nabla_{\alpha}\,^{\rho}{W} \nabla^{\lambda \gamma}{W_{\beta \rho}}+\frac{9}{512}(\Gamma_{a})_{\beta \rho} \lambda_{i \alpha} {W}^{-3} \nabla^{\beta \rho}{W} \nabla_{\lambda \gamma}{W} \nabla^{\lambda \gamma}{W} - \frac{9}{512}(\Gamma_{a})_{\beta \rho} \lambda_{i \alpha} {W}^{-2} \nabla_{\lambda \gamma}{W} \nabla^{\beta \rho}{\nabla^{\lambda \gamma}{W}} - \frac{9}{512}(\Gamma_{a})_{\beta \rho} {W}^{-2} \nabla^{\beta \rho}{W} \nabla_{\lambda \gamma}{W} \nabla^{\lambda \gamma}{\lambda_{i \alpha}}+\frac{9}{512}(\Gamma_{a})_{\beta \rho} {W}^{-1} \nabla_{\lambda \gamma}{\lambda_{i \alpha}} \nabla^{\beta \rho}{\nabla^{\lambda \gamma}{W}}+\frac{9}{512}(\Gamma_{a})_{\beta \rho} {W}^{-1} \nabla_{\lambda \gamma}{W} \nabla^{\beta \rho}{\nabla^{\lambda \gamma}{\lambda_{i \alpha}}} - \frac{9}{256}{\rm i} (\Gamma_{a})_{\beta \rho} \lambda^{\lambda}_{j} {W}^{-2} \nabla^{\beta \rho}{\lambda^{j}_{\gamma}} \nabla_{\alpha}\,^{\gamma}{\lambda_{i \lambda}} - \frac{9}{256}{\rm i} (\Gamma_{a})_{\beta \rho} \lambda_{j \lambda} {W}^{-2} \nabla^{\beta \rho}{\lambda^{j \gamma}} \nabla_{\alpha}\,^{\lambda}{\lambda_{i \gamma}}+\frac{9}{128}{\rm i} (\Gamma_{a})_{\beta \rho} \lambda^{\lambda}_{j} \lambda^{j}_{\gamma} {W}^{-3} \nabla^{\beta \rho}{W} \nabla_{\alpha}\,^{\gamma}{\lambda_{i \lambda}}%
 - \frac{9}{256}{\rm i} (\Gamma_{a})_{\beta \rho} \lambda^{\lambda}_{j} \lambda^{j}_{\gamma} {W}^{-2} \nabla^{\beta \rho}{\nabla_{\alpha}\,^{\gamma}{\lambda_{i \lambda}}}+\frac{27}{512}{\rm i} (\Gamma_{a})_{\lambda \gamma} \lambda_{i \alpha} \lambda^{\beta}_{j} \lambda^{j \rho} {W}^{-2} \nabla^{\lambda \gamma}{W_{\beta \rho}} - \frac{27}{256}{\rm i} (\Gamma_{a})_{\lambda \gamma} W^{\beta \rho} \lambda_{i \alpha} \lambda_{j \beta} \lambda^{j}_{\rho} {W}^{-3} \nabla^{\lambda \gamma}{W} - \frac{9}{256}{\rm i} (\Gamma_{a})_{\beta \rho} \lambda_{j \alpha} {W}^{-2} \nabla^{\beta \rho}{\lambda^{j}_{\lambda}} \nabla^{\lambda \gamma}{\lambda_{i \gamma}} - \frac{9}{256}{\rm i} (\Gamma_{a})_{\beta \rho} \lambda_{j \alpha} \lambda^{j}_{\lambda} {W}^{-2} \nabla^{\beta \rho}{\nabla^{\lambda \gamma}{\lambda_{i \gamma}}}+\frac{27}{512}{\rm i} (\Gamma_{a})_{\lambda \gamma} \lambda_{j \alpha} \lambda^{\beta}_{i} \lambda^{j \rho} {W}^{-2} \nabla^{\lambda \gamma}{W_{\beta \rho}} - \frac{27}{256}{\rm i} (\Gamma_{a})_{\lambda \gamma} W^{\beta \rho} \lambda_{j \alpha} \lambda_{i \beta} \lambda^{j}_{\rho} {W}^{-3} \nabla^{\lambda \gamma}{W} - \frac{9}{64}(\Gamma_{a})_{\lambda \gamma} F_{\alpha}\,^{\beta} \lambda^{\rho}_{i} {W}^{-2} \nabla^{\lambda \gamma}{F_{\beta \rho}} - \frac{9}{32}(\Gamma_{a})_{\lambda \gamma} F_{\alpha}\,^{\beta} F_{\beta}\,^{\rho} \lambda_{i \rho} {W}^{-3} \nabla^{\lambda \gamma}{W}+\frac{9}{128}(\Gamma_{a})_{\rho \lambda} X_{i j} \lambda^{j \beta} {W}^{-2} \nabla^{\rho \lambda}{F_{\alpha \beta}} - \frac{9}{128}(\Gamma_{a})_{\lambda \gamma} \lambda^{\beta}_{i} {W}^{-2} \nabla_{\alpha}\,^{\rho}{W} \nabla^{\lambda \gamma}{F_{\beta \rho}} - \frac{9}{64}(\Gamma_{a})_{\lambda \gamma} F^{\beta}\,_{\rho} \lambda_{i \beta} {W}^{-3} \nabla^{\lambda \gamma}{W} \nabla_{\alpha}\,^{\rho}{W}+\frac{9}{128}{\rm i} (\Gamma_{a})_{\lambda \gamma} \lambda_{i \alpha} \lambda^{\beta}_{j} \lambda^{j \rho} {W}^{-3} \nabla^{\lambda \gamma}{F_{\beta \rho}}+\frac{9}{64}{\rm i} (\Gamma_{a})_{\lambda \gamma} F^{\beta \rho} \lambda_{i \alpha} \lambda_{j \beta} {W}^{-3} \nabla^{\lambda \gamma}{\lambda^{j}_{\rho}} - \frac{27}{128}{\rm i} (\Gamma_{a})_{\lambda \gamma} F^{\beta \rho} \lambda_{i \alpha} \lambda_{j \beta} \lambda^{j}_{\rho} {W}^{-4} \nabla^{\lambda \gamma}{W}+\frac{9}{128}{\rm i} (\Gamma_{a})_{\beta \rho} \lambda^{\lambda}_{i} {W}^{-2} \nabla^{\beta \rho}{\lambda_{j \lambda}} \nabla_{\alpha}\,^{\gamma}{\lambda^{j}_{\gamma}}+\frac{9}{128}(\Gamma_{a})_{\beta \rho} X_{i j} \lambda_{k \alpha} {W}^{-2} \nabla^{\beta \rho}{X^{j k}} - \frac{9}{128}(\Gamma_{a})_{\beta \rho} X_{i j} {W}^{-2} \nabla_{\alpha}\,^{\lambda}{W} \nabla^{\beta \rho}{\lambda^{j}_{\lambda}} - \frac{9}{128}(\Gamma_{a})_{\beta \rho} X_{i j} \lambda^{j}_{\lambda} {W}^{-2} \nabla^{\beta \rho}{\nabla_{\alpha}\,^{\lambda}{W}} - \frac{9}{128}{\rm i} (\Gamma_{a})_{\beta \rho} \lambda_{i \alpha} \lambda^{\lambda}_{j} \lambda_{k \lambda} {W}^{-3} \nabla^{\beta \rho}{X^{j k}}%
+\frac{27}{128}{\rm i} (\Gamma_{a})_{\beta \rho} X_{j k} \lambda_{i \alpha} \lambda^{j \lambda} \lambda^{k}_{\lambda} {W}^{-4} \nabla^{\beta \rho}{W} - \frac{9}{128}{\rm i} (\Gamma_{a})_{\rho \lambda} F_{\alpha}\,^{\beta} \lambda^{\gamma}_{i} \lambda_{j \beta} {W}^{-3} \nabla^{\rho \lambda}{\lambda^{j}_{\gamma}}+\frac{27}{128}{\rm i} (\Gamma_{a})_{\rho \lambda} F_{\alpha}\,^{\beta} \lambda^{\gamma}_{i} \lambda_{j \beta} \lambda^{j}_{\gamma} {W}^{-4} \nabla^{\rho \lambda}{W}+\frac{9}{256}{\rm i} (\Gamma_{a})_{\beta \rho} X_{i j} \lambda_{k \alpha} \lambda^{j \lambda} {W}^{-3} \nabla^{\beta \rho}{\lambda^{k}_{\lambda}} - \frac{9}{256}{\rm i} (\Gamma_{a})_{\beta \rho} \lambda^{\lambda}_{i} \lambda_{j \gamma} {W}^{-3} \nabla_{\alpha}\,^{\gamma}{W} \nabla^{\beta \rho}{\lambda^{j}_{\lambda}}+\frac{27}{256}{\rm i} (\Gamma_{a})_{\beta \rho} \lambda^{\lambda}_{i} \lambda_{j \lambda} \lambda^{j}_{\gamma} {W}^{-4} \nabla^{\beta \rho}{W} \nabla_{\alpha}\,^{\gamma}{W} - \frac{27}{256}(\Gamma_{a})_{\beta \rho} \lambda_{i \alpha} \lambda^{\lambda}_{j} \lambda^{j \gamma} \lambda_{k \lambda} {W}^{-4} \nabla^{\beta \rho}{\lambda^{k}_{\gamma}}+\frac{27}{256}(\Gamma_{a})_{\beta \rho} \lambda_{i \alpha} \lambda^{\lambda}_{j} \lambda^{j \gamma} \lambda_{k \lambda} \lambda^{k}_{\gamma} {W}^{-5} \nabla^{\beta \rho}{W}+\frac{9}{640}(\Gamma_{a})_{\alpha \beta} \nabla^{\beta \rho}{\nabla_{\lambda \gamma}{\nabla^{\lambda \gamma}{\lambda_{i \rho}}}}+\frac{9}{320}(\Gamma_{a})_{\alpha \lambda} W^{\beta}\,_{\rho} \nabla^{\lambda}\,_{\gamma}{\nabla^{\gamma \rho}{\lambda_{i \beta}}} - \frac{9}{80}{\rm i} (\Gamma_{a})_{\alpha \gamma} W^{\beta \rho}\,_{\lambda i} \nabla^{\gamma \lambda}{F_{\beta \rho}}+\frac{9}{160}(\Gamma_{a})_{\alpha \lambda} F^{\beta}\,_{\rho} {W}^{-1} \nabla^{\lambda}\,_{\gamma}{\nabla^{\gamma \rho}{\lambda_{i \beta}}}+\frac{9}{160}(\Gamma_{a})_{\alpha \lambda} F_{\beta \rho} {W}^{-1} \nabla^{\lambda \beta}{\nabla^{\rho \gamma}{\lambda_{i \gamma}}}+\frac{9}{80}(\Gamma_{a})_{\alpha \lambda} \lambda_{i \gamma} {W}^{-1} \nabla^{\lambda \beta}{\nabla^{\gamma \rho}{F_{\beta \rho}}}+\frac{9}{160}(\Gamma_{a})_{\alpha \beta} \lambda_{j \rho} {W}^{-1} \nabla^{\beta}\,_{\lambda}{\nabla^{\lambda \rho}{X_{i}\,^{j}}} - \frac{9}{160}(\Gamma_{a})_{\alpha \beta} \lambda_{i \rho} {W}^{-1} \nabla^{\beta}\,_{\lambda}{\nabla_{\gamma}\,^{\rho}{\nabla^{\lambda \gamma}{W}}}+\frac{9}{640}(\Gamma_{a})_{\alpha \beta} {W}^{-1} \nabla_{\lambda \gamma}{W} \nabla^{\beta \rho}{\nabla^{\lambda \gamma}{\lambda_{i \rho}}}+\frac{9}{320}{\rm i} (\Gamma_{a})_{\alpha \beta} \lambda^{\rho}_{j} {W}^{-2} \nabla^{\beta}\,_{\gamma}{\lambda^{j}_{\lambda}} \nabla^{\gamma \lambda}{\lambda_{i \rho}}+\frac{9}{320}{\rm i} (\Gamma_{a})_{\alpha \beta} \lambda^{\rho}_{j} \lambda^{j}_{\lambda} {W}^{-2} \nabla^{\beta}\,_{\gamma}{\nabla^{\gamma \lambda}{\lambda_{i \rho}}} - \frac{9}{80}(\Gamma_{a})_{\alpha \gamma} F^{\beta}\,_{\rho} \lambda^{\lambda}_{i} {W}^{-2} \nabla^{\gamma \rho}{F_{\beta \lambda}}%
+\frac{9}{160}{\rm i} (\Gamma_{a})_{\alpha \lambda} \lambda_{i \gamma} \lambda^{\beta}_{j} \lambda^{j \rho} {W}^{-3} \nabla^{\lambda \gamma}{F_{\beta \rho}} - \frac{9}{160}{\rm i} (\Gamma_{a})_{\alpha \beta} \lambda_{i \rho} \lambda^{\lambda}_{j} \lambda_{k \lambda} {W}^{-3} \nabla^{\beta \rho}{X^{j k}}+\frac{27}{160}{\rm i} (\Gamma_{a})_{\alpha \beta} X_{j k} \lambda_{i \rho} \lambda^{j \lambda} \lambda^{k}_{\lambda} {W}^{-4} \nabla^{\beta \rho}{W} - \frac{9}{320}{\rm i} (\Gamma_{a})_{\alpha \beta} X_{i j} \lambda^{j \rho} \lambda_{k \lambda} {W}^{-3} \nabla^{\beta \lambda}{\lambda^{k}_{\rho}} - \frac{27}{320}(\Gamma_{a})_{\alpha \beta} \lambda_{i \rho} \lambda^{\lambda}_{j} \lambda^{j \gamma} \lambda_{k \lambda} {W}^{-4} \nabla^{\beta \rho}{\lambda^{k}_{\gamma}}+\frac{27}{320}(\Gamma_{a})_{\alpha \beta} \lambda_{i \rho} \lambda^{\lambda}_{j} \lambda^{j \gamma} \lambda_{k \lambda} \lambda^{k}_{\gamma} {W}^{-5} \nabla^{\beta \rho}{W}
\doublespacedmathend
\end{adjustwidth}

\subsubsection{$J^4_{a b, \log}$ Degauged and Gauge Fixed Bosons} \label{J4logWComplete}

\begin{adjustwidth}{0cm}{5cm}
\doublespacedmathbegin
{}\frac{9}{32}W_{\underline{a}}\,^{c} F_{\underline{b} c} Y - \frac{9}{160}\eta_{\underline{a} \underline{b}} W^{c d} F_{c d} Y - \frac{117}{128}W^{d e} W_{\underline{a}}\,^{c} W_{d e} F_{\underline{b} c}+\frac{63}{160}\eta_{\underline{a} \underline{b}} W^{e {e_{1}}} W^{c d} W_{e {e_{1}}} F_{c d} - \frac{687213}{40960}\eta_{\underline{a} \underline{b}} W^{e {e_{1}}} W_{e c} W_{{e_{1}} d} F^{c d}+\frac{873}{64}W^{d c} W_{\underline{a}}\,^{e} W_{e d} F_{\underline{b} c} - \frac{135}{32}W^{e d} W_{\underline{a}}\,^{c} W_{\underline{b} e} F_{c d} - \frac{27}{16}W^{c d} W_{\underline{a}}\,^{e} W_{\underline{b} e} F_{c d}+\frac{853101}{40960}\eta_{\underline{a} \underline{b}} W^{e c} W^{{e_{1}} d} W_{e {e_{1}}} F_{c d}+\frac{27}{256}W_{\underline{a}}\,^{c} W_{\underline{b} c} Y - \frac{27}{1280}\eta_{\underline{a} \underline{b}} W^{c d} W_{c d} Y - \frac{2745}{512}W^{c d} W_{\underline{a}}\,^{e} W_{\underline{b} e} W_{c d}+\frac{1323}{1280}\eta_{\underline{a} \underline{b}} W^{c d} W^{e {e_{1}}} W_{c d} W_{e {e_{1}}} - \frac{3}{5}\Phi_{c d i j} \eta_{\underline{a} \underline{b}} X^{i j} W^{c d}+\frac{39}{32}R_{\underline{a}}\,^{c}\,_{d e} W^{d e} W_{\underline{b} c}+\frac{39}{32}R^{c}\,_{d} W^{d}\,_{\underline{a}} W_{\underline{b} c}+\frac{45}{64}R W_{\underline{a}}\,^{c} W_{\underline{b} c}+\frac{135}{32}R_{\underline{a} c} W^{c d} W_{\underline{b} d} - \frac{39}{160}R_{c d e {e_{1}}} \eta_{\underline{a} \underline{b}} W^{c d} W^{e {e_{1}}}%
+\frac{39}{80}R_{c d} \eta_{\underline{a} \underline{b}} W^{c}\,_{e} W^{d e} - \frac{603}{2560}R \eta_{\underline{a} \underline{b}} W^{c d} W_{c d} - \frac{99}{32}R_{\underline{a}}\,^{c}\,_{\underline{b} d} W^{d e} W_{c e}+\frac{39}{160}R^{c}\,_{d} \eta_{\underline{a} \underline{b}} W^{d e} W_{c e}+\frac{243}{512}R_{\underline{a} \underline{b}} W^{c d} W_{c d}+\frac{99}{512}W_{\underline{a}}\,^{c} W^{d e} W_{\underline{b} c} W_{d e}+\frac{39}{32}R_{\underline{a} c}\,^{d}\,_{e} W^{c e} W_{\underline{b} d} - \frac{39}{160}R_{c d e {e_{1}}} \eta_{\underline{a} \underline{b}} W^{c e} W^{d {e_{1}}}+\frac{3}{10}\Phi_{c d i j} \Phi^{c d i j} \eta_{\underline{a} \underline{b}} - \frac{3}{2}\Phi_{\underline{a}}\,^{c}\,_{i j} \Phi_{\underline{b} c}\,^{i j}+\frac{9}{8}R_{\underline{a}}\,^{c}\,_{d e} W^{d e} F_{\underline{b} c}+\frac{9}{8}R^{c}\,_{d} W^{d}\,_{\underline{a}} F_{\underline{b} c}+\frac{81}{32}R W_{\underline{a}}\,^{c} F_{\underline{b} c}+\frac{9}{4}R_{\underline{a} d} W^{d c} F_{\underline{b} c} - \frac{3}{10}R_{e {e_{1}} c d} \eta_{\underline{a} \underline{b}} W^{e {e_{1}}} F^{c d}+\frac{3}{5}R_{e c} \eta_{\underline{a} \underline{b}} W^{e}\,_{d} F^{c d} - \frac{21}{32}R \eta_{\underline{a} \underline{b}} W^{c d} F_{c d}+\frac{9}{64}W^{d e} W_{\underline{a} d} W^{c}\,_{e} F_{\underline{b} c} - \frac{81}{8}R_{\underline{a}}\,^{c}\,_{\underline{b} e} W^{e d} F_{c d}+\frac{12}{5}R^{c}\,_{e} \eta_{\underline{a} \underline{b}} W^{e d} F_{c d}%
 - \frac{27}{8}R_{\underline{a}}\,^{c} W_{\underline{b}}\,^{d} F_{c d}+\frac{45}{32}R_{\underline{a} \underline{b}} W^{c d} F_{c d}+\frac{81}{128}W_{\underline{a}}\,^{c} W^{d e} W_{d e} F_{\underline{b} c} - \frac{81}{32}W_{\underline{a}}\,^{c} W^{e d} W_{\underline{b} e} F_{c d}+\frac{3}{8}R_{\underline{a}}\,^{e}\,_{c d} W_{\underline{b} e} F^{c d}+\frac{3}{8}R^{d}\,_{c} W_{\underline{b} d} F^{c}\,_{\underline{a}}+\frac{39}{8}R_{\underline{a} c} W_{\underline{b} d} F^{c d}+\frac{9}{8}R_{\underline{a} d}\,^{c}\,_{e} W^{d e} F_{\underline{b} c} - \frac{3}{10}R_{e c {e_{1}} d} \eta_{\underline{a} \underline{b}} W^{e {e_{1}}} F^{c d}+3\Phi_{\underline{a}}\,^{c}\,_{i j} X^{i j} W_{\underline{b} c}+\frac{3}{8}R_{\underline{a} c}\,^{e}\,_{d} W_{\underline{b} e} F^{c d}+\frac{81}{32}W_{\underline{a}}\,^{e} W_{\underline{b} e} F^{c d} F_{c d} - \frac{999}{640}\eta_{\underline{a} \underline{b}} W^{e {e_{1}}} W_{e {e_{1}}} F^{c d} F_{c d} - \frac{27}{8}W_{\underline{a}}\,^{e} W_{\underline{b} c} F^{c d} F_{e d} - \frac{81}{8}W^{e d} W_{\underline{a} e} F_{\underline{b}}\,^{c} F_{c d}+\frac{27}{5}\eta_{\underline{a} \underline{b}} W^{{e_{1}} e} W_{{e_{1}} c} F^{c d} F_{e d}+\frac{675}{128}W^{d e} W_{d e} F_{\underline{a}}\,^{c} F_{\underline{b} c}+\frac{81}{8}W^{e d} W^{c}\,_{e} F_{\underline{a} c} F_{\underline{b} d}+\frac{81}{160}\eta_{\underline{a} \underline{b}} X_{i j} X^{i j} W^{c d} W_{c d} - \frac{81}{32}X_{i j} X^{i j} W_{\underline{a}}\,^{c} W_{\underline{b} c}%
+\frac{9}{4}W_{\underline{a}}\,^{e} F^{c d} F_{\underline{b} e} F_{c d} - \frac{9}{8}\eta_{\underline{a} \underline{b}} W^{e {e_{1}}} F^{c d} F_{e {e_{1}}} F_{c d}+\frac{27}{8}W^{d e} F_{\underline{a}}\,^{c} F_{\underline{b} c} F_{d e}+\frac{9}{4}\eta_{\underline{a} \underline{b}} W^{e {e_{1}}} F^{c d} F_{e c} F_{{e_{1}} d} - \frac{27}{4}W^{d e} F_{\underline{a}}\,^{c} F_{\underline{b} d} F_{c e} - \frac{9}{4}W_{\underline{a} c} F^{c d} F_{\underline{b}}\,^{e} F_{e d} - \frac{9}{4}W_{\underline{a}}\,^{e} F^{c d} F_{\underline{b} c} F_{e d} - \frac{9}{10}\Phi_{c d i j} \eta_{\underline{a} \underline{b}} X^{i j} F^{c d}+\frac{9}{2}\Phi_{\underline{a}}\,^{c}\,_{i j} X^{i j} F_{\underline{b} c}+\frac{3}{2}R_{\underline{a}}\,^{e}\,_{c d} F^{c d} F_{\underline{b} e}+\frac{3}{2}R^{d}\,_{c} F^{c}\,_{\underline{a}} F_{\underline{b} d}+\frac{27}{32}R F_{\underline{a}}\,^{c} F_{\underline{b} c} - \frac{3}{10}R_{c d e {e_{1}}} \eta_{\underline{a} \underline{b}} F^{c d} F^{e {e_{1}}}+\frac{3}{5}R_{c e} \eta_{\underline{a} \underline{b}} F^{c}\,_{d} F^{e d} - \frac{81}{320}R \eta_{\underline{a} \underline{b}} F^{c d} F_{c d} - \frac{27}{4}R_{\underline{a}}\,^{e}\,_{\underline{b} c} F^{c d} F_{e d}+\frac{33}{20}R^{e}\,_{c} \eta_{\underline{a} \underline{b}} F^{c d} F_{e d}+\frac{27}{64}R_{\underline{a} \underline{b}} F^{c d} F_{c d}+\frac{27}{8}W^{e d} W_{\underline{b} e} F_{\underline{a}}\,^{c} F_{d c}+\frac{3}{2}R_{\underline{a} c}\,^{e}\,_{d} F^{c d} F_{\underline{b} e}%
 - \frac{3}{10}R_{c e d {e_{1}}} \eta_{\underline{a} \underline{b}} F^{c d} F^{e {e_{1}}} - \frac{27}{128}F_{\underline{a}}\,^{c} F_{\underline{b} c} Y+\frac{27}{640}\eta_{\underline{a} \underline{b}} F^{c d} F_{c d} Y+\frac{9}{16}F^{c d} F_{\underline{a}}\,^{e} F_{\underline{b} e} F_{c d} - \frac{9}{80}\eta_{\underline{a} \underline{b}} F^{c d} F^{e {e_{1}}} F_{c d} F_{e {e_{1}}} - \frac{9}{80}\eta_{\underline{a} \underline{b}} X_{i j} X^{i j} F^{c d} F_{c d}+\frac{9}{16}X_{i j} X^{i j} F_{\underline{a}}\,^{c} F_{\underline{b} c}+\frac{9}{20}\eta_{\underline{a} \underline{b}} F^{c d} F^{e {e_{1}}} F_{c e} F_{d {e_{1}}} - \frac{9}{4}F^{c d} F_{\underline{a}}\,^{e} F_{\underline{b} c} F_{e d}+\frac{45}{32}\mathcal{D}_{\underline{a}}{\mathcal{D}_{\underline{b}}{W_{c d}}} F^{c d} - \frac{9}{8}\mathcal{D}^{d}{\mathcal{D}_{\underline{a}}{W_{\underline{b}}\,^{c}}} F_{c d} - \frac{9}{2}\mathcal{D}^{d}{\mathcal{D}_{\underline{a}}{W^{c}\,_{d}}} F_{\underline{b} c}+\frac{27}{8}\mathcal{D}_{\underline{a}}{\mathcal{D}^{d}{W_{\underline{b}}\,^{c}}} F_{c d}+\frac{81}{8}\mathcal{D}^{c}{\mathcal{D}^{d}{W_{\underline{a} d}}} F_{\underline{b} c}+\frac{9}{20}\eta_{\underline{a} \underline{b}} \mathcal{D}^{e}{\mathcal{D}^{d}{W^{c}\,_{e}}} F_{c d} - \frac{117}{160}\eta_{\underline{a} \underline{b}} \mathcal{D}_{e}{\mathcal{D}^{e}{W_{c d}}} F^{c d}+\frac{9}{4}\mathcal{D}_{d}{\mathcal{D}^{d}{W_{\underline{a}}\,^{c}}} F_{\underline{b} c} - \frac{27}{8}\mathcal{D}^{d}{\mathcal{D}^{c}{W_{\underline{a} d}}} F_{\underline{b} c} - \frac{9}{4}\eta_{\underline{a} \underline{b}} \mathcal{D}^{d}{\mathcal{D}^{e}{W^{c}\,_{e}}} F_{c d} - \frac{21}{32}R W_{\underline{b} c} F_{\underline{a}}\,^{c}%
 - \frac{57}{8}R^{d c} W_{\underline{a} d} F_{\underline{b} c} - \frac{17055}{16384}\epsilon^{e {e_{1}} {e_{2}} c d} \mathcal{D}_{\underline{a}}{W_{\underline{b} e}} W_{{e_{1}} {e_{2}}} F_{c d}+\frac{19485}{8192}\epsilon_{\underline{a}}\,^{e {e_{1}} {e_{2}} c} \mathcal{D}_{\underline{b}}{W_{e {e_{1}}}} W_{{e_{2}}}\,^{d} F_{c d} - \frac{56421}{32768}\epsilon_{\underline{a}}\,^{d e {e_{1}} {e_{2}}} \mathcal{D}^{c}{W_{d e}} W_{{e_{1}} {e_{2}}} F_{\underline{b} c}+\frac{1647}{16384}\epsilon_{\underline{a} {e_{2}}}\,^{e c d} \mathcal{D}^{{e_{2}}}{W_{e}\,^{{e_{1}}}} W_{\underline{b} {e_{1}}} F_{c d} - \frac{10629}{8192}\epsilon_{\underline{a} {e_{2}}}\,^{{e_{1}} c d} \mathcal{D}^{{e_{2}}}{W_{\underline{b}}\,^{e}} W_{{e_{1}} e} F_{c d} - \frac{1341}{8192}\epsilon^{d e {e_{1}} {e_{2}} c} \mathcal{D}_{\underline{a}}{W_{d e}} W_{{e_{1}} {e_{2}}} F_{\underline{b} c} - \frac{31401}{8192}\epsilon_{\underline{a} {e_{2}}}\,^{d {e_{1}} c} \mathcal{D}^{{e_{2}}}{W_{d}\,^{e}} W_{{e_{1}} e} F_{\underline{b} c}+\frac{37593}{81920}\epsilon^{e {e_{1}} {e_{2}} {e_{3}} c} \eta_{\underline{a} \underline{b}} \mathcal{D}^{d}{W_{e {e_{1}}}} W_{{e_{2}} {e_{3}}} F_{c d}+\frac{6597}{4096}\epsilon^{e {e_{1}} {e_{2}} c d} \mathcal{D}_{\underline{a}}{W_{e {e_{1}}}} W_{\underline{b} {e_{2}}} F_{c d}+\frac{26397}{8192}\epsilon_{\underline{a}}\,^{e {e_{1}} {e_{2}} c} \mathcal{D}_{\underline{b}}{W_{e}\,^{d}} W_{{e_{1}} {e_{2}}} F_{c d} - \frac{18777}{8192}\epsilon_{{e_{2}}}\,^{d e {e_{1}} c} \mathcal{D}^{{e_{2}}}{W_{d e}} W_{\underline{a} {e_{1}}} F_{\underline{b} c} - \frac{7425}{4096}\epsilon_{\underline{a}}\,^{d {e_{1}} {e_{2}} c} \mathcal{D}^{e}{W_{d e}} W_{{e_{1}} {e_{2}}} F_{\underline{b} c}+\frac{1371}{4096}\epsilon_{\underline{a} {e_{2}}}\,^{d e {e_{1}}} \mathcal{D}^{{e_{2}}}{W_{d}\,^{c}} W_{e {e_{1}}} F_{\underline{b} c} - \frac{15027}{8192}\epsilon_{\underline{a}}\,^{{e_{1}} {e_{2}} c d} \mathcal{D}^{e}{W_{\underline{b} e}} W_{{e_{1}} {e_{2}}} F_{c d}+\frac{14661}{16384}\epsilon_{\underline{a} {e_{2}} c}\,^{e {e_{1}}} \mathcal{D}^{{e_{2}}}{W_{e {e_{1}}}} W_{\underline{b} d} F^{c d}+\frac{19971}{16384}\epsilon_{{e_{2}}}\,^{e {e_{1}} c d} \mathcal{D}^{{e_{2}}}{W_{\underline{a} e}} W_{\underline{b} {e_{1}}} F_{c d} - \frac{17799}{16384}\epsilon_{\underline{a} {e_{2}}}\,^{e {e_{1}} c} \mathcal{D}^{{e_{2}}}{W_{\underline{b}}\,^{d}} W_{e {e_{1}}} F_{c d}+\frac{8115}{16384}\epsilon_{{e_{2}}}\,^{d e {e_{1}} c} \mathcal{D}^{{e_{2}}}{W_{\underline{a} d}} W_{e {e_{1}}} F_{\underline{b} c} - \frac{209601}{81920}\epsilon_{{e_{3}}}\,^{e {e_{2}} c d} \eta_{\underline{a} \underline{b}} \mathcal{D}^{{e_{3}}}{W_{e}\,^{{e_{1}}}} W_{{e_{2}} {e_{1}}} F_{c d}%
 - \frac{127521}{81920}\epsilon_{{e_{3}}}\,^{e {e_{1}} {e_{2}} c} \eta_{\underline{a} \underline{b}} \mathcal{D}^{{e_{3}}}{W_{e}\,^{d}} W_{{e_{1}} {e_{2}}} F_{c d}+\frac{129}{2048}\epsilon_{\underline{a} {e_{2}}}\,^{e {e_{1}} c} \mathcal{D}^{{e_{2}}}{W_{e}\,^{d}} W_{\underline{b} {e_{1}}} F_{c d}+\frac{11691}{16384}\epsilon_{\underline{a}}\,^{e {e_{1}} c d} \mathcal{D}^{{e_{2}}}{W_{\underline{b} e}} W_{{e_{1}} {e_{2}}} F_{c d}+\frac{42327}{16384}\epsilon_{\underline{a} d {e_{2}}}\,^{e {e_{1}}} W^{d c} \mathcal{D}^{{e_{2}}}{W_{e {e_{1}}}} F_{\underline{b} c} - \frac{93903}{32768}\epsilon_{\underline{a}}\,^{e {e_{1}} c d} \mathcal{D}^{{e_{2}}}{W_{e {e_{1}}}} W_{\underline{b} {e_{2}}} F_{c d} - \frac{14661}{8192}\epsilon_{e {e_{3}}}\,^{{e_{1}} {e_{2}} c} \eta_{\underline{a} \underline{b}} W^{e d} \mathcal{D}^{{e_{3}}}{W_{{e_{1}} {e_{2}}}} F_{c d}+\frac{19167}{8192}\epsilon_{\underline{a} e {e_{2}}}\,^{{e_{1}} c} W^{e d} \mathcal{D}^{{e_{2}}}{W_{\underline{b} {e_{1}}}} F_{c d}+\frac{7851}{16384}\epsilon_{\underline{a}}\,^{e {e_{2}} c d} \mathcal{D}^{{e_{1}}}{W_{e {e_{1}}}} W_{\underline{b} {e_{2}}} F_{c d}+\frac{243}{256}W^{c d} \mathcal{D}_{\underline{a}}{\mathcal{D}_{\underline{b}}{W_{c d}}} - \frac{423}{64}\mathcal{D}^{d}{\mathcal{D}_{\underline{a}}{W^{c}\,_{d}}} W_{\underline{b} c} - \frac{207}{64}\mathcal{D}^{d}{\mathcal{D}_{\underline{a}}{W_{\underline{b}}\,^{c}}} W_{c d} - \frac{27}{64}\mathcal{D}_{\underline{a}}{\mathcal{D}^{d}{W^{c}\,_{d}}} W_{\underline{b} c}+\frac{189}{64}\mathcal{D}_{\underline{a}}{\mathcal{D}^{d}{W_{\underline{b}}\,^{c}}} W_{c d} - \frac{135}{64}\mathcal{D}^{c}{\mathcal{D}^{d}{W_{\underline{a} c}}} W_{\underline{b} d}+\frac{621}{64}\mathcal{D}^{d}{\mathcal{D}^{c}{W_{\underline{a} c}}} W_{\underline{b} d} - \frac{81}{32}\mathcal{D}_{d}{\mathcal{D}^{d}{W_{\underline{a}}\,^{c}}} W_{\underline{b} c} - \frac{171}{64}\eta_{\underline{a} \underline{b}} \mathcal{D}^{e}{\mathcal{D}^{d}{W^{c}\,_{d}}} W_{c e}+\frac{333}{1280}\eta_{\underline{a} \underline{b}} W^{c d} \mathcal{D}_{e}{\mathcal{D}^{e}{W_{c d}}} - \frac{99}{320}\eta_{\underline{a} \underline{b}} \mathcal{D}^{d}{\mathcal{D}^{e}{W^{c}\,_{d}}} W_{c e} - \frac{57}{16}R^{c d} W_{\underline{a} c} W_{\underline{b} d}%
+\frac{30015}{16384}\epsilon^{c d e {e_{1}} {e_{2}}} \mathcal{D}_{\underline{a}}{W_{c d}} W_{\underline{b} e} W_{{e_{1}} {e_{2}}}+\frac{38763}{8192}\epsilon_{\underline{a}}\,^{c e {e_{1}} {e_{2}}} \mathcal{D}_{\underline{b}}{W_{c}\,^{d}} W_{e {e_{1}}} W_{{e_{2}} d} - \frac{999}{256}\epsilon_{\underline{a}}\,^{c d {e_{1}} {e_{2}}} \mathcal{D}^{e}{W_{c d}} W_{\underline{b} e} W_{{e_{1}} {e_{2}}} - \frac{1053}{2048}\epsilon_{\underline{a} {e_{2}}}\,^{c e {e_{1}}} \mathcal{D}^{{e_{2}}}{W_{c}\,^{d}} W_{\underline{b} d} W_{e {e_{1}}} - \frac{2601}{4096}\epsilon_{\underline{a} {e_{2}}}\,^{d e {e_{1}}} \mathcal{D}^{{e_{2}}}{W_{\underline{b}}\,^{c}} W_{d e} W_{{e_{1}} c} - \frac{117}{1024}\epsilon_{\underline{a} {e_{2}}}\,^{c e {e_{1}}} \mathcal{D}^{{e_{2}}}{W_{c}\,^{d}} W_{\underline{b} e} W_{{e_{1}} d}+\frac{13275}{16384}\epsilon^{c d e {e_{1}} {e_{2}}} \eta_{\underline{a} \underline{b}} \mathcal{D}^{{e_{3}}}{W_{c d}} W_{e {e_{1}}} W_{{e_{2}} {e_{3}}} - \frac{1125}{2048}\epsilon_{\underline{a}}\,^{c e {e_{1}} {e_{2}}} \mathcal{D}^{d}{W_{c d}} W_{\underline{b} e} W_{{e_{1}} {e_{2}}} - \frac{243}{256}\epsilon_{\underline{a}}\,^{d e {e_{1}} {e_{2}}} \mathcal{D}^{c}{W_{\underline{b} c}} W_{d e} W_{{e_{1}} {e_{2}}}+\frac{567}{4096}\epsilon_{{e_{2}}}\,^{c d e {e_{1}}} \mathcal{D}^{{e_{2}}}{W_{\underline{a} c}} W_{\underline{b} d} W_{e {e_{1}}}+\frac{6777}{4096}\epsilon_{\underline{a}}\,^{c d e {e_{1}}} \mathcal{D}^{{e_{2}}}{W_{\underline{b} c}} W_{d e} W_{{e_{1}} {e_{2}}} - \frac{189}{256}\epsilon_{\underline{a}}\,^{c d e {e_{1}}} \mathcal{D}^{{e_{2}}}{W_{c d}} W_{\underline{b} e} W_{{e_{1}} {e_{2}}} - \frac{153}{256}\epsilon_{\underline{a} {e_{2}}}\,^{c d {e_{1}}} \mathcal{D}^{{e_{2}}}{W_{c d}} W_{\underline{b}}\,^{e} W_{{e_{1}} e} - \frac{9}{8}\mathcal{D}^{c}{\mathcal{D}^{d}{R_{\underline{a} c \underline{b} d}}} - \frac{3}{128}\mathcal{D}_{c}{\mathcal{D}^{c}{R}} \eta_{\underline{a} \underline{b}}+\frac{15}{128}\mathcal{D}_{\underline{a}}{\mathcal{D}_{\underline{b}}{R}}+\frac{27}{8}\mathcal{D}^{c}{W_{\underline{a} c}} \mathcal{D}^{d}{W_{\underline{b} d}}+\frac{27}{8}\mathcal{D}^{d}{W_{\underline{a} c}} \mathcal{D}^{c}{W_{\underline{b} d}} - \frac{27}{64}W_{\underline{a} c} \mathcal{D}^{d}{\mathcal{D}^{c}{W_{\underline{b} d}}} - \frac{27}{64}W_{\underline{a} c} \mathcal{D}^{c}{\mathcal{D}^{d}{W_{\underline{b} d}}}%
 - \frac{27}{4}\mathcal{D}^{d}{W_{\underline{b}}\,^{c}} \mathcal{D}_{\underline{a}}{W_{c d}} - \frac{27}{40}\eta_{\underline{a} \underline{b}} \mathcal{D}^{d}{W^{c}\,_{d}} \mathcal{D}^{e}{W_{c e}} - \frac{81}{40}\eta_{\underline{a} \underline{b}} \mathcal{D}^{e}{W^{c}\,_{d}} \mathcal{D}^{d}{W_{c e}} - \frac{27}{8}\mathcal{D}_{d}{W_{\underline{a}}\,^{c}} \mathcal{D}^{d}{W_{\underline{b} c}}+\frac{3}{8}\mathcal{D}^{c}{\mathcal{D}^{e}{R_{c}\,^{d}\,_{e d}}} \eta_{\underline{a} \underline{b}}+\frac{9}{64}\eta_{\underline{a} \underline{b}} W^{c}\,_{d} \mathcal{D}^{d}{\mathcal{D}^{e}{W_{c e}}} - \frac{3}{32}\mathcal{D}_{e}{\mathcal{D}^{e}{R^{c d}\,_{c d}}} \eta_{\underline{a} \underline{b}}+\frac{1053}{1280}\eta_{\underline{a} \underline{b}} \mathcal{D}_{e}{W^{c d}} \mathcal{D}^{e}{W_{c d}} - \frac{3}{8}\mathcal{D}_{\underline{b}}{\mathcal{D}^{d}{R_{\underline{a}}\,^{c}\,_{d c}}} - \frac{3}{16}\mathcal{D}_{\underline{a}}{\mathcal{D}_{\underline{b}}{R^{c}\,_{c}}}+\frac{9}{64}W_{\underline{b}}\,^{c} \mathcal{D}_{\underline{a}}{\mathcal{D}^{d}{W_{c d}}} - \frac{189}{256}\mathcal{D}_{\underline{a}}{W^{c d}} \mathcal{D}_{\underline{b}}{W_{c d}}+\frac{3}{32}\mathcal{D}_{\underline{a}}{\mathcal{D}_{\underline{b}}{R^{c d}\,_{c d}}}+\frac{3}{8}\mathcal{D}_{d}{\mathcal{D}^{d}{R_{\underline{a}}\,^{c}\,_{\underline{b} c}}}+\frac{9}{32}W_{\underline{a}}\,^{c} \mathcal{D}_{d}{\mathcal{D}^{d}{W_{\underline{b} c}}} - \frac{3}{8}\mathcal{D}^{d}{\mathcal{D}_{\underline{b}}{R_{\underline{a}}\,^{c}\,_{d c}}}+\frac{9}{64}W^{c}\,_{d} \mathcal{D}^{d}{\mathcal{D}_{\underline{a}}{W_{\underline{b} c}}}+\frac{9}{64}W_{\underline{b}}\,^{c} \mathcal{D}^{d}{\mathcal{D}_{\underline{a}}{W_{c d}}}+\frac{9}{64}\eta_{\underline{a} \underline{b}} W^{c}\,_{d} \mathcal{D}^{e}{\mathcal{D}^{d}{W_{c e}}}+\frac{9}{64}W^{c}\,_{d} \mathcal{D}_{\underline{a}}{\mathcal{D}^{d}{W_{\underline{b} c}}}%
+\frac{3}{4}R_{\underline{a}}\,^{c}\,_{\underline{b}}\,^{d} R_{c d} - \frac{57}{128}R_{\underline{a} \underline{b}} R - \frac{3}{10}R^{c d} R_{c d} \eta_{\underline{a} \underline{b}}+\frac{69}{640}\eta_{\underline{a} \underline{b}} {R}^{2}+\frac{3}{4}R_{\underline{a}}\,^{c} R_{\underline{b} c} - \frac{3}{32}R R_{\underline{a} \underline{b}}+\frac{9}{32}\epsilon^{d e}\,_{\underline{b}}\,^{{e_{1}} {e_{2}}} R_{\underline{a} c d e} \mathcal{D}^{c}{W_{{e_{1}} {e_{2}}}} - \frac{9}{64}\epsilon^{d e}\,_{\underline{b} {e_{2}}}\,^{{e_{1}}} R_{\underline{a}}\,^{c}\,_{d e} \mathcal{D}^{{e_{2}}}{W_{{e_{1}} c}}+\frac{9}{32}\epsilon^{c d}\,_{\underline{b} {e_{2}}}\,^{{e_{1}}} R_{\underline{a} c d}\,^{e} \mathcal{D}^{{e_{2}}}{W_{{e_{1}} e}} - \frac{3}{16}\epsilon^{d e}\,_{\underline{b}}\,^{{e_{1}} {e_{2}}} \mathcal{D}^{c}{R_{\underline{a} c d e}} W_{{e_{1}} {e_{2}}} - \frac{15}{32}\epsilon^{c}\,_{\underline{b} {e_{1}}}\,^{d e} \mathcal{D}^{{e_{1}}}{R_{\underline{a} c}} W_{d e} - \frac{15}{256}\epsilon_{\underline{a}}\,^{c d {e_{1}} {e_{2}}} W_{c d} \mathcal{D}^{e}{W_{\underline{b} e}} W_{{e_{1}} {e_{2}}} - \frac{15}{256}\epsilon_{\underline{a}}\,^{d e {e_{1}} {e_{2}}} W_{\underline{b} c} W_{d e} \mathcal{D}^{c}{W_{{e_{1}} {e_{2}}}}+\frac{9}{128}\epsilon_{\underline{a}}\,^{c d e {e_{1}}} W_{c d} \mathcal{D}^{{e_{2}}}{W_{\underline{b} e}} W_{{e_{1}} {e_{2}}}+\frac{9}{128}\epsilon_{\underline{a}}\,^{c d e {e_{1}}} W_{\underline{b} c} W_{d e} \mathcal{D}^{{e_{2}}}{W_{{e_{1}} {e_{2}}}} - \frac{3}{32}\epsilon^{c d e}\,_{\underline{a} {e_{2}}} \mathcal{D}^{{e_{2}}}{R_{c d e}\,^{{e_{1}}}} W_{\underline{b} {e_{1}}} - \frac{3}{128}\epsilon_{\underline{a} {e_{2}}}\,^{d e {e_{1}}} W_{\underline{b}}\,^{c} W_{d e} \mathcal{D}^{{e_{2}}}{W_{{e_{1}} c}} - \frac{3}{64}\epsilon_{\underline{a} {e_{2}}}\,^{d e {e_{1}}} W_{\underline{b}}\,^{c} \mathcal{D}^{{e_{2}}}{W_{d c}} W_{e {e_{1}}}+\frac{3}{32}\epsilon^{d e}\,_{\underline{b} {e_{2}}}\,^{{e_{1}}} \mathcal{D}^{{e_{2}}}{R_{\underline{a}}\,^{c}\,_{d e}} W_{{e_{1}} c}+\frac{3}{64}\epsilon_{\underline{a} {e_{2}}}\,^{d e {e_{1}}} W_{\underline{b}}\,^{c} W_{d c} \mathcal{D}^{{e_{2}}}{W_{e {e_{1}}}}%
+\frac{3}{64}\epsilon_{\underline{a} c {e_{2}}}\,^{e {e_{1}}} W^{c d} W_{e d} \mathcal{D}^{{e_{2}}}{W_{\underline{b} {e_{1}}}}+\frac{9}{256}\epsilon_{\underline{a} {e_{2}}}\,^{c d {e_{1}}} W_{\underline{b} c} W_{d}\,^{e} \mathcal{D}^{{e_{2}}}{W_{{e_{1}} e}} - \frac{3}{256}\epsilon_{\underline{a} {e_{2}}}\,^{c d {e_{1}}} W_{\underline{b} c} \mathcal{D}^{{e_{2}}}{W_{d}\,^{e}} W_{{e_{1}} e}+\frac{3}{32}\epsilon^{c d e}\,_{{e_{2}}}\,^{{e_{1}}} \mathcal{D}^{{e_{2}}}{R_{\underline{a} c d e}} W_{\underline{b} {e_{1}}} - \frac{3}{256}\epsilon_{{e_{2}}}\,^{c d e {e_{1}}} W_{\underline{a} c} \mathcal{D}^{{e_{2}}}{W_{\underline{b} d}} W_{e {e_{1}}}+\frac{3}{32}\epsilon^{c d}\,_{\underline{b}}\,^{{e_{1}} {e_{2}}} \mathcal{D}^{e}{R_{\underline{a} c d e}} W_{{e_{1}} {e_{2}}} - \frac{3}{256}\epsilon_{\underline{a} {e_{2}}}\,^{c d {e_{1}}} W_{c d} \mathcal{D}^{{e_{2}}}{W_{\underline{b}}\,^{e}} W_{{e_{1}} e}+\frac{3}{64}\epsilon^{c d e {e_{2}} {e_{3}}} \mathcal{D}^{{e_{1}}}{R_{c d e {e_{1}}}} \eta_{\underline{a} \underline{b}} W_{{e_{2}} {e_{3}}}+\frac{3}{512}\epsilon^{c d e {e_{2}} {e_{3}}} \eta_{\underline{a} \underline{b}} W_{c d} \mathcal{D}^{{e_{1}}}{W_{e {e_{1}}}} W_{{e_{2}} {e_{3}}}+\frac{3}{512}\epsilon^{c d e {e_{2}} {e_{3}}} \eta_{\underline{a} \underline{b}} W_{c d} W_{e {e_{1}}} \mathcal{D}^{{e_{1}}}{W_{{e_{2}} {e_{3}}}} - \frac{3}{512}\epsilon^{c d e {e_{1}} {e_{2}}} \eta_{\underline{a} \underline{b}} W_{c d} \mathcal{D}^{{e_{3}}}{W_{e {e_{1}}}} W_{{e_{2}} {e_{3}}} - \frac{3}{512}\epsilon^{c d e {e_{1}} {e_{2}}} \eta_{\underline{a} \underline{b}} W_{c d} W_{e {e_{1}}} \mathcal{D}^{{e_{3}}}{W_{{e_{2}} {e_{3}}}} - \frac{15}{256}\epsilon_{\underline{a}}\,^{d e {e_{1}} {e_{2}}} W_{\underline{b} c} \mathcal{D}^{c}{W_{d e}} W_{{e_{1}} {e_{2}}} - \frac{15}{256}\epsilon_{\underline{a}}\,^{c d e {e_{1}}} W_{c d} W_{e {e_{1}}} \mathcal{D}^{{e_{2}}}{W_{\underline{b} {e_{2}}}}+\frac{3}{64}\epsilon_{\underline{a}}\,^{c d {e_{1}} {e_{2}}} W_{\underline{b} c} \mathcal{D}^{e}{W_{d e}} W_{{e_{1}} {e_{2}}} - \frac{3}{64}\epsilon_{\underline{a}}\,^{c d e {e_{2}}} W_{c d} W_{e {e_{1}}} \mathcal{D}^{{e_{1}}}{W_{\underline{b} {e_{2}}}}+\frac{3}{256}\epsilon_{{e_{2}}}\,^{c d e {e_{1}}} W_{\underline{a} c} W_{d e} \mathcal{D}^{{e_{2}}}{W_{\underline{b} {e_{1}}}} - \frac{3}{256}\epsilon_{\underline{a} {e_{2}}}\,^{c d e} W_{c d} W_{e}\,^{{e_{1}}} \mathcal{D}^{{e_{2}}}{W_{\underline{b} {e_{1}}}} - \frac{15453}{81920}\epsilon^{c e {e_{1}} {e_{2}} {e_{3}}} \eta_{\underline{a} \underline{b}} \mathcal{D}^{d}{W_{c d}} W_{e {e_{1}}} W_{{e_{2}} {e_{3}}} - \frac{2835}{16384}\epsilon^{c d e {e_{1}} {e_{2}}} \mathcal{D}_{\underline{a}}{W_{\underline{b} c}} W_{d e} W_{{e_{1}} {e_{2}}}%
 - \frac{2835}{8192}\epsilon_{{e_{3}}}\,^{c e {e_{1}} {e_{2}}} \eta_{\underline{a} \underline{b}} \mathcal{D}^{{e_{3}}}{W_{c}\,^{d}} W_{e {e_{1}}} W_{{e_{2}} d}+\frac{27}{2560}\eta_{\underline{a} \underline{b}} \mathcal{D}_{c}{\mathcal{D}^{c}{Y}}+\frac{27}{2560}R \eta_{\underline{a} \underline{b}} Y - \frac{27}{512}\mathcal{D}_{\underline{a}}{\mathcal{D}_{\underline{b}}{Y}} - \frac{27}{512}R_{\underline{a} \underline{b}} Y - \frac{3789}{8192}\epsilon_{\underline{a} {e_{2}}}\,^{e {e_{1}} c} \mathcal{D}^{{e_{2}}}{W_{e {e_{1}}}} W_{\underline{b}}\,^{d} F_{c d}+\frac{3789}{8192}\epsilon_{\underline{a} {e_{2}}}\,^{d e {e_{1}}} \mathcal{D}^{{e_{2}}}{W_{d e}} W_{{e_{1}}}\,^{c} F_{\underline{b} c}+\frac{13491}{8192}\epsilon_{\underline{a}}\,^{e {e_{1}} {e_{2}} c} \mathcal{D}^{d}{W_{\underline{b} e}} W_{{e_{1}} {e_{2}}} F_{c d}+\frac{9489}{16384}\epsilon^{e {e_{2}} {e_{3}} c d} \eta_{\underline{a} \underline{b}} \mathcal{D}^{{e_{1}}}{W_{e {e_{1}}}} W_{{e_{2}} {e_{3}}} F_{c d} - \frac{1143}{2048}\epsilon_{\underline{a} {e_{2}}}\,^{e {e_{1}} c} \mathcal{D}^{{e_{2}}}{W_{\underline{b} e}} W_{{e_{1}}}\,^{d} F_{c d}+\frac{9}{32}\epsilon_{\underline{a}}\,^{e {e_{2}} c d} \mathcal{D}_{\underline{b}}{W_{e}\,^{{e_{1}}}} W_{{e_{2}} {e_{1}}} F_{c d}+\frac{3093}{4096}\epsilon^{e {e_{1}} {e_{2}} c d} \eta_{\underline{a} \underline{b}} \mathcal{D}^{{e_{3}}}{W_{e {e_{1}}}} W_{{e_{2}} {e_{3}}} F_{c d} - \frac{1071}{4096}\epsilon_{\underline{a}}\,^{e {e_{1}} {e_{2}} c} \mathcal{D}^{d}{W_{e {e_{1}}}} W_{\underline{b} {e_{2}}} F_{c d}+\frac{9}{8}\mathcal{D}_{\underline{a}}{W^{c d}} \mathcal{D}_{\underline{b}}{F_{c d}}+\frac{9}{4}\mathcal{D}_{\underline{a}}{W_{\underline{b}}\,^{c}} \mathcal{D}^{d}{F_{c d}} - \frac{9}{8}\mathcal{D}_{\underline{a}}{W^{c}\,_{d}} \mathcal{D}^{d}{F_{\underline{b} c}} - \frac{9}{8}\mathcal{D}^{d}{W_{\underline{b}}\,^{c}} \mathcal{D}_{\underline{a}}{F_{c d}}+\frac{9}{4}\mathcal{D}^{d}{W^{c}\,_{d}} \mathcal{D}_{\underline{a}}{F_{\underline{b} c}}+\frac{27}{4}\mathcal{D}^{d}{W_{\underline{a} d}} \mathcal{D}^{c}{F_{\underline{b} c}} - \frac{9}{8}\eta_{\underline{a} \underline{b}} \mathcal{D}^{d}{W^{c}\,_{e}} \mathcal{D}^{e}{F_{c d}}%
 - \frac{9}{20}\eta_{\underline{a} \underline{b}} \mathcal{D}_{e}{W^{c d}} \mathcal{D}^{e}{F_{c d}}+\frac{9}{8}\mathcal{D}_{d}{W_{\underline{a}}\,^{c}} \mathcal{D}^{d}{F_{\underline{b} c}}+\frac{27}{8}\mathcal{D}^{c}{W_{\underline{a} d}} \mathcal{D}^{d}{F_{\underline{b} c}} - \frac{9}{20}\eta_{\underline{a} \underline{b}} \mathcal{D}^{e}{W^{c}\,_{e}} \mathcal{D}^{d}{F_{c d}}+\frac{27}{32}\epsilon^{e {e_{1}} {e_{2}} c d} W_{\underline{b} e} W_{{e_{1}} {e_{2}}} \mathcal{D}_{\underline{a}}{F_{c d}} - \frac{189}{64}\epsilon_{\underline{a}}\,^{e {e_{1}} {e_{2}} c} W_{e {e_{1}}} W_{{e_{2}}}\,^{d} \mathcal{D}_{\underline{b}}{F_{c d}} - \frac{27}{128}\epsilon_{\underline{a}}\,^{d e {e_{1}} {e_{2}}} W_{d e} W_{{e_{1}} {e_{2}}} \mathcal{D}^{c}{F_{\underline{b} c}} - \frac{27}{32}\epsilon_{{e_{2}}}\,^{d e {e_{1}} c} W_{\underline{a} d} W_{e {e_{1}}} \mathcal{D}^{{e_{2}}}{F_{\underline{b} c}} - \frac{81}{128}\epsilon_{\underline{a}}\,^{d e {e_{1}} c} W_{d e} W_{{e_{1}} {e_{2}}} \mathcal{D}^{{e_{2}}}{F_{\underline{b} c}}+\frac{9}{128}\epsilon_{\underline{a} {e_{2}}}\,^{d e {e_{1}}} W_{d e} W_{{e_{1}}}\,^{c} \mathcal{D}^{{e_{2}}}{F_{\underline{b} c}} - \frac{477}{1280}\epsilon^{e {e_{1}} {e_{2}} {e_{3}} c} \eta_{\underline{a} \underline{b}} W_{e {e_{1}}} W_{{e_{2}} {e_{3}}} \mathcal{D}^{d}{F_{c d}} - \frac{3}{8}\epsilon^{{e_{1}} {e_{2}}}\,_{\underline{b}}\,^{c d} \mathcal{D}^{e}{R_{\underline{a} e {e_{1}} {e_{2}}}} F_{c d} - \frac{15}{16}\epsilon^{e}\,_{\underline{b} {e_{1}}}\,^{c d} \mathcal{D}^{{e_{1}}}{R_{\underline{a} e}} F_{c d} - \frac{15}{128}\epsilon_{\underline{a}}\,^{{e_{1}} {e_{2}} c d} W_{\underline{b} e} \mathcal{D}^{e}{W_{{e_{1}} {e_{2}}}} F_{c d}+\frac{9}{64}\epsilon_{\underline{a}}\,^{e {e_{1}} c d} W_{\underline{b} e} \mathcal{D}^{{e_{2}}}{W_{{e_{1}} {e_{2}}}} F_{c d} - \frac{3}{16}\epsilon^{d e {e_{1}}}\,_{\underline{a} {e_{2}}} \mathcal{D}^{{e_{2}}}{R_{d e {e_{1}}}\,^{c}} F_{\underline{b} c}+\frac{3}{128}\epsilon_{\underline{a} {e_{2}}}\,^{d e {e_{1}}} W_{d e} \mathcal{D}^{{e_{2}}}{W_{{e_{1}}}\,^{c}} F_{\underline{b} c}+\frac{3}{16}\epsilon^{e {e_{1}}}\,_{\underline{b} {e_{2}}}\,^{c} \mathcal{D}^{{e_{2}}}{R_{\underline{a}}\,^{d}\,_{e {e_{1}}}} F_{c d}+\frac{3}{32}\epsilon_{\underline{a} {e_{2}}}\,^{e {e_{1}} c} W_{\underline{b}}\,^{d} \mathcal{D}^{{e_{2}}}{W_{e {e_{1}}}} F_{c d}+\frac{3}{64}\epsilon_{\underline{a} {e_{2}}}\,^{e {e_{1}} c} W_{e {e_{1}}} \mathcal{D}^{{e_{2}}}{W_{\underline{b}}\,^{d}} F_{c d}%
 - \frac{9}{128}\epsilon_{\underline{a} {e_{2}}}\,^{e {e_{1}} c} W_{\underline{b} e} \mathcal{D}^{{e_{2}}}{W_{{e_{1}}}\,^{d}} F_{c d}+\frac{3}{16}\epsilon^{d e {e_{1}}}\,_{{e_{2}}}\,^{c} \mathcal{D}^{{e_{2}}}{R_{\underline{a} d e {e_{1}}}} F_{\underline{b} c}+\frac{3}{128}\epsilon_{{e_{2}}}\,^{d e {e_{1}} c} W_{\underline{a} d} \mathcal{D}^{{e_{2}}}{W_{e {e_{1}}}} F_{\underline{b} c}+\frac{3}{16}\epsilon^{e {e_{1}}}\,_{\underline{b}}\,^{c d} \mathcal{D}^{{e_{2}}}{R_{\underline{a} e {e_{1}} {e_{2}}}} F_{c d} - \frac{9}{128}\epsilon_{\underline{a} {e_{2}}}\,^{{e_{1}} c d} W_{\underline{b}}\,^{e} \mathcal{D}^{{e_{2}}}{W_{{e_{1}} e}} F_{c d}+\frac{3}{32}\epsilon^{e {e_{1}} {e_{2}} c d} \mathcal{D}^{{e_{3}}}{R_{e {e_{1}} {e_{2}} {e_{3}}}} \eta_{\underline{a} \underline{b}} F_{c d}+\frac{3}{256}\epsilon^{e {e_{2}} {e_{3}} c d} \eta_{\underline{a} \underline{b}} W_{e {e_{1}}} \mathcal{D}^{{e_{1}}}{W_{{e_{2}} {e_{3}}}} F_{c d} - \frac{3}{256}\epsilon^{e {e_{1}} {e_{2}} c d} \eta_{\underline{a} \underline{b}} W_{e {e_{1}}} \mathcal{D}^{{e_{3}}}{W_{{e_{2}} {e_{3}}}} F_{c d} - \frac{15}{128}\epsilon_{\underline{a}}\,^{e {e_{1}} c d} W_{e {e_{1}}} \mathcal{D}^{{e_{2}}}{W_{\underline{b} {e_{2}}}} F_{c d} - \frac{3}{32}\epsilon_{\underline{a}}\,^{e {e_{2}} c d} W_{e {e_{1}}} \mathcal{D}^{{e_{1}}}{W_{\underline{b} {e_{2}}}} F_{c d} - \frac{3}{128}\epsilon_{{e_{2}}}\,^{d e {e_{1}} c} W_{d e} \mathcal{D}^{{e_{2}}}{W_{\underline{a} {e_{1}}}} F_{\underline{b} c} - \frac{9}{128}\epsilon_{\underline{a} {e_{2}}}\,^{e c d} W_{e}\,^{{e_{1}}} \mathcal{D}^{{e_{2}}}{W_{\underline{b} {e_{1}}}} F_{c d} - \frac{2637}{8192}\epsilon_{\underline{a}}\,^{d e {e_{1}} c} \mathcal{D}^{{e_{2}}}{W_{d e}} W_{{e_{1}} {e_{2}}} F_{\underline{b} c} - \frac{1845}{8192}\epsilon_{{e_{3}}}\,^{e {e_{1}} {e_{2}} c} \eta_{\underline{a} \underline{b}} \mathcal{D}^{{e_{3}}}{W_{e {e_{1}}}} W_{{e_{2}}}\,^{d} F_{c d}+\frac{27}{4}W_{\underline{a} d} \mathcal{D}^{d}{\mathcal{D}^{c}{F_{\underline{b} c}}} - \frac{315}{128}\epsilon_{\underline{a}}\,^{{e_{1}} {e_{2}} c d} W_{\underline{b} e} W_{{e_{1}} {e_{2}}} \mathcal{D}^{e}{F_{c d}}+\frac{45}{128}\epsilon_{\underline{a} {e_{2}}}\,^{e {e_{1}} c} W_{\underline{b}}\,^{d} W_{e {e_{1}}} \mathcal{D}^{{e_{2}}}{F_{c d}} - \frac{9}{128}\epsilon_{\underline{a}}\,^{e {e_{1}} {e_{2}} c} W_{\underline{b} e} W_{{e_{1}} {e_{2}}} \mathcal{D}^{d}{F_{c d}}+\frac{45}{256}\epsilon^{e {e_{1}} {e_{2}} c d} \eta_{\underline{a} \underline{b}} W_{e {e_{1}}} W_{{e_{2}} {e_{3}}} \mathcal{D}^{{e_{3}}}{F_{c d}} - \frac{63}{128}\epsilon_{\underline{a} {e_{2}}}\,^{{e_{1}} c d} W_{\underline{b}}\,^{e} W_{{e_{1}} e} \mathcal{D}^{{e_{2}}}{F_{c d}}%
 - \frac{63}{128}\epsilon_{\underline{a}}\,^{e {e_{1}} c d} W_{\underline{b} e} W_{{e_{1}} {e_{2}}} \mathcal{D}^{{e_{2}}}{F_{c d}}+\frac{225}{512}\epsilon^{{e_{2}} c d e {e_{1}}} W_{\underline{b} {e_{2}}} \mathcal{D}_{\underline{a}}{F_{c d}} F_{e {e_{1}}}+\frac{459}{256}\epsilon_{\underline{a}}\,^{{e_{1}} {e_{2}} c e} W_{{e_{1}} {e_{2}}} \mathcal{D}_{\underline{b}}{F_{c}\,^{d}} F_{e d} - \frac{153}{512}\epsilon^{{e_{2}} {e_{3}} c e {e_{1}}} \eta_{\underline{a} \underline{b}} W_{{e_{2}} {e_{3}}} \mathcal{D}^{d}{F_{c d}} F_{e {e_{1}}} - \frac{243}{1280}\epsilon_{{e_{3}}}\,^{{e_{2}} c d e} \eta_{\underline{a} \underline{b}} W_{{e_{2}}}\,^{{e_{1}}} \mathcal{D}^{{e_{3}}}{F_{c d}} F_{e {e_{1}}} - \frac{27}{512}\epsilon^{{e_{1}} {e_{2}} c d e} W_{{e_{1}} {e_{2}}} \mathcal{D}_{\underline{a}}{F_{\underline{b} c}} F_{d e}+\frac{45}{256}\epsilon_{\underline{a}}\,^{{e_{2}} c d e} W_{{e_{2}}}\,^{{e_{1}}} \mathcal{D}_{\underline{b}}{F_{c d}} F_{e {e_{1}}}+\frac{81}{512}\epsilon^{{e_{2}} c d e {e_{1}}} \eta_{\underline{a} \underline{b}} W_{{e_{2}} {e_{3}}} \mathcal{D}^{{e_{3}}}{F_{c d}} F_{e {e_{1}}}+\frac{27}{256}\epsilon_{{e_{3}}}\,^{{e_{1}} {e_{2}} c e} \eta_{\underline{a} \underline{b}} W_{{e_{1}} {e_{2}}} \mathcal{D}^{{e_{3}}}{F_{c}\,^{d}} F_{e d} - \frac{81}{64}\epsilon_{\underline{a}}\,^{c d e {e_{1}}} W_{\underline{b} {e_{2}}} \mathcal{D}^{{e_{2}}}{F_{c d}} F_{e {e_{1}}}+\frac{27}{128}\epsilon_{\underline{a} {e_{2}}}\,^{e {e_{1}} c} W_{e {e_{1}}} \mathcal{D}^{{e_{2}}}{F_{c}\,^{d}} F_{\underline{b} d}+\frac{45}{128}\epsilon_{\underline{a} {e_{2}}}\,^{e {e_{1}} d} W_{e {e_{1}}} \mathcal{D}^{{e_{2}}}{F_{\underline{b}}\,^{c}} F_{d c}+\frac{9}{64}\epsilon_{\underline{a} {e_{2}}}\,^{{e_{1}} c e} W_{\underline{b} {e_{1}}} \mathcal{D}^{{e_{2}}}{F_{c}\,^{d}} F_{e d}+\frac{9}{16}\epsilon_{\underline{a} {e_{2}}}\,^{e {e_{1}} c} W_{\underline{b} e} W_{{e_{1}}}\,^{d} \mathcal{D}^{{e_{2}}}{F_{c d}}+\frac{27}{32}\mathcal{D}_{\underline{a}}{F^{c d}} \mathcal{D}_{\underline{b}}{F_{c d}}+\frac{9}{4}\mathcal{D}^{d}{F^{c}\,_{d}} \mathcal{D}_{\underline{a}}{F_{\underline{b} c}}+\frac{9}{4}\mathcal{D}^{d}{F_{\underline{b}}\,^{c}} \mathcal{D}_{\underline{a}}{F_{c d}} - \frac{99}{160}\eta_{\underline{a} \underline{b}} \mathcal{D}_{e}{F^{c d}} \mathcal{D}^{e}{F_{c d}}+\frac{9}{4}\mathcal{D}^{c}{F_{\underline{a} c}} \mathcal{D}^{d}{F_{\underline{b} d}}+\frac{9}{4}\mathcal{D}_{d}{F_{\underline{a}}\,^{c}} \mathcal{D}^{d}{F_{\underline{b} c}}%
+\frac{9}{20}\eta_{\underline{a} \underline{b}} \mathcal{D}^{e}{F^{c}\,_{d}} \mathcal{D}^{d}{F_{c e}}+\frac{3}{8}\epsilon^{e}\,_{\underline{b} {e_{1}}}\,^{c d} R_{\underline{a} e} \mathcal{D}^{{e_{1}}}{F_{c d}} - \frac{9}{8}\epsilon_{\underline{a}}\,^{{e_{1}} {e_{2}} d e} W_{{e_{1}} {e_{2}}} \mathcal{D}^{c}{F_{\underline{b} c}} F_{d e}+\frac{9}{32}\epsilon_{{e_{2}}}\,^{{e_{1}} c d e} W_{\underline{a} {e_{1}}} \mathcal{D}^{{e_{2}}}{F_{\underline{b} c}} F_{d e}+\frac{9}{128}\epsilon_{\underline{a}}\,^{{e_{1}} {e_{2}} c d} W_{{e_{1}} {e_{2}}} \mathcal{D}^{e}{F_{\underline{b} c}} F_{d e} - \frac{9}{128}\epsilon_{\underline{a} {e_{2}}}\,^{c d e} W_{\underline{b}}\,^{{e_{1}}} \mathcal{D}^{{e_{2}}}{F_{c d}} F_{e {e_{1}}}+\frac{9}{64}\epsilon_{\underline{a} {e_{2}}}\,^{{e_{1}} c d} W_{{e_{1}}}\,^{e} \mathcal{D}^{{e_{2}}}{F_{c d}} F_{\underline{b} e} - \frac{117}{64}\epsilon_{\underline{a}}\,^{{e_{1}} {e_{2}} c d} W_{{e_{1}} {e_{2}}} \mathcal{D}^{e}{F_{c d}} F_{\underline{b} e}+\frac{27}{32}\mathcal{D}_{\underline{a}}{X_{i j}} \mathcal{D}_{\underline{b}}{X^{i j}} - \frac{27}{160}\eta_{\underline{a} \underline{b}} \mathcal{D}_{c}{X_{i j}} \mathcal{D}^{c}{X^{i j}}+\frac{45}{64}\epsilon^{{e_{1}} {e_{2}} c d e} W_{{e_{1}} {e_{2}}} \mathcal{D}_{\underline{a}}{F_{c d}} F_{\underline{b} e} - \frac{297}{128}\epsilon_{\underline{a}}\,^{{e_{2}} c e {e_{1}}} W_{{e_{2}}}\,^{d} \mathcal{D}_{\underline{b}}{F_{c d}} F_{e {e_{1}}} - \frac{45}{128}\epsilon_{\underline{a} {e_{2}}}\,^{{e_{1}} d e} W_{{e_{1}}}\,^{c} \mathcal{D}^{{e_{2}}}{F_{\underline{b} c}} F_{d e}+\frac{27}{64}\epsilon_{\underline{a} {e_{2}}}\,^{{e_{1}} c e} W_{{e_{1}}}\,^{d} \mathcal{D}^{{e_{2}}}{F_{c d}} F_{\underline{b} e}+\frac{279}{1280}\epsilon^{{e_{2}} {e_{3}} c d e} \eta_{\underline{a} \underline{b}} W_{{e_{2}} {e_{3}}} \mathcal{D}^{{e_{1}}}{F_{c d}} F_{e {e_{1}}} - \frac{45}{128}\epsilon_{{e_{2}}}\,^{{e_{1}} c d e} W_{\underline{a} {e_{1}}} \mathcal{D}^{{e_{2}}}{F_{c d}} F_{\underline{b} e} - \frac{27}{128}\epsilon_{\underline{a}}\,^{{e_{1}} {e_{2}} c e} W_{{e_{1}} {e_{2}}} \mathcal{D}^{d}{F_{c d}} F_{\underline{b} e} - \frac{27}{64}\epsilon_{\underline{a}}\,^{{e_{2}} c e {e_{1}}} W_{\underline{b} {e_{2}}} \mathcal{D}^{d}{F_{c d}} F_{e {e_{1}}} - \frac{9}{128}\epsilon_{\underline{a}}\,^{{e_{2}} c d e} W_{\underline{b} {e_{2}}} \mathcal{D}^{{e_{1}}}{F_{c d}} F_{e {e_{1}}}+\frac{9}{32}\epsilon^{c d e {e_{1}} {e_{2}}} \mathcal{D}_{\underline{a}}{F_{c d}} F_{\underline{b} e} F_{{e_{1}} {e_{2}}}%
 - \frac{27}{32}\epsilon_{\underline{a}}\,^{c d {e_{1}} {e_{2}}} \mathcal{D}^{e}{F_{c d}} F_{\underline{b} e} F_{{e_{1}} {e_{2}}} - \frac{369}{512}\epsilon^{{e_{2}} c d e {e_{1}}} \mathcal{D}_{\underline{a}}{W_{\underline{b} {e_{2}}}} F_{c d} F_{e {e_{1}}} - \frac{783}{512}\epsilon_{\underline{a}}\,^{{e_{1}} {e_{2}} d e} \mathcal{D}^{c}{W_{{e_{1}} {e_{2}}}} F_{\underline{b} c} F_{d e}+\frac{27}{128}\epsilon_{\underline{a} {e_{2}}}\,^{c d e} \mathcal{D}^{{e_{2}}}{W_{\underline{b}}\,^{{e_{1}}}} F_{c d} F_{e {e_{1}}} - \frac{81}{256}\epsilon_{\underline{a} {e_{2}}}\,^{{e_{1}} d e} \mathcal{D}^{{e_{2}}}{W_{{e_{1}}}\,^{c}} F_{\underline{b} c} F_{d e}+\frac{63}{512}\epsilon^{{e_{1}} {e_{2}} c d e} \mathcal{D}_{\underline{a}}{W_{{e_{1}} {e_{2}}}} F_{\underline{b} c} F_{d e} - \frac{27}{32}\epsilon_{\underline{a} {e_{2}}}\,^{{e_{1}} c d} \mathcal{D}^{{e_{2}}}{W_{{e_{1}}}\,^{e}} F_{\underline{b} c} F_{d e}+\frac{63}{512}\epsilon^{{e_{2}} {e_{3}} c d e} \eta_{\underline{a} \underline{b}} \mathcal{D}^{{e_{1}}}{W_{{e_{2}} {e_{3}}}} F_{c d} F_{e {e_{1}}} - \frac{513}{512}\epsilon_{\underline{a}}\,^{c d e {e_{1}}} \mathcal{D}^{{e_{2}}}{W_{\underline{b} {e_{2}}}} F_{c d} F_{e {e_{1}}}+\frac{27}{256}\epsilon_{\underline{a} {e_{2}} c}\,^{e {e_{1}}} \mathcal{D}^{{e_{2}}}{W_{e {e_{1}}}} F^{c d} F_{\underline{b} d} - \frac{81}{128}\epsilon_{{e_{2}}}\,^{{e_{1}} c d e} \mathcal{D}^{{e_{2}}}{W_{\underline{a} {e_{1}}}} F_{\underline{b} c} F_{d e}+\frac{99}{256}\epsilon_{{e_{3}}}\,^{{e_{2}} c d e} \eta_{\underline{a} \underline{b}} \mathcal{D}^{{e_{3}}}{W_{{e_{2}}}\,^{{e_{1}}}} F_{c d} F_{e {e_{1}}} - \frac{27}{128}\epsilon_{\underline{a}}\,^{{e_{2}} c d e} \mathcal{D}^{{e_{1}}}{W_{\underline{b} {e_{2}}}} F_{c d} F_{e {e_{1}}}+\frac{27}{128}\epsilon_{\underline{a}}\,^{{e_{1}} {e_{2}} c d} \mathcal{D}^{e}{W_{{e_{1}} {e_{2}}}} F_{\underline{b} c} F_{d e}+\frac{27}{4}\mathcal{D}^{d}{\mathcal{D}^{c}{F_{\underline{a} c}}} F_{\underline{b} d}-3R^{c d} F_{\underline{a} c} F_{\underline{b} d}+\frac{81}{128}\epsilon_{{e_{2}}}\,^{e {e_{1}} c d} W_{e {e_{1}}} \mathcal{D}^{{e_{2}}}{F_{\underline{a} c}} F_{\underline{b} d} - \frac{63}{128}\epsilon_{\underline{a}}\,^{{e_{1}} c d e} W_{{e_{1}} {e_{2}}} \mathcal{D}^{{e_{2}}}{F_{\underline{b} c}} F_{d e} - \frac{117}{2560}\epsilon^{{e_{2}} c d e {e_{1}}} \eta_{\underline{a} \underline{b}} \mathcal{D}^{{e_{3}}}{W_{{e_{2}} {e_{3}}}} F_{c d} F_{e {e_{1}}}+\frac{9}{16}\epsilon_{\underline{a}}\,^{{e_{1}} c d e} W_{{e_{1}} {e_{2}}} \mathcal{D}^{{e_{2}}}{F_{c d}} F_{\underline{b} e}%
+\frac{9}{320}R \eta_{\underline{a} \underline{b}} X_{i j} X^{i j}+\frac{9}{160}\eta_{\underline{a} \underline{b}} X_{i j} \mathcal{D}_{c}{\mathcal{D}^{c}{X^{i j}}} - \frac{9}{32}X_{i j} \mathcal{D}_{\underline{a}}{\mathcal{D}_{\underline{b}}{X^{i j}}} - \frac{9}{64}R_{\underline{a} \underline{b}} X_{i j} X^{i j} - \frac{9}{32}\epsilon_{\underline{a}}\,^{c e {e_{1}} {e_{2}}} \mathcal{D}^{d}{F_{c d}} F_{\underline{b} e} F_{{e_{1}} {e_{2}}}+\frac{9}{16}\epsilon_{\underline{a} {e_{2}}}\,^{d e {e_{1}}} \mathcal{D}^{{e_{2}}}{F_{\underline{b}}\,^{c}} F_{d e} F_{{e_{1}} c} - \frac{27}{32}\epsilon_{\underline{a} {e_{2}}}\,^{c e {e_{1}}} \mathcal{D}^{{e_{2}}}{F_{c}\,^{d}} F_{\underline{b} d} F_{e {e_{1}}} - \frac{9}{32}\epsilon_{{e_{2}}}\,^{c d e {e_{1}}} \mathcal{D}^{{e_{2}}}{F_{\underline{a} c}} F_{\underline{b} d} F_{e {e_{1}}} - \frac{9}{16}\epsilon_{\underline{a}}\,^{c d e {e_{1}}} \mathcal{D}^{{e_{2}}}{F_{\underline{b} c}} F_{d e} F_{{e_{1}} {e_{2}}}+\frac{9}{32}\epsilon_{\underline{a}}\,^{c d e {e_{1}}} \mathcal{D}^{{e_{2}}}{F_{c d}} F_{\underline{b} e} F_{{e_{1}} {e_{2}}}+\frac{9}{32}\epsilon_{\underline{a} {e_{2}}}\,^{c d {e_{1}}} \mathcal{D}^{{e_{2}}}{F_{c d}} F_{\underline{b}}\,^{e} F_{{e_{1}} e} - \frac{9}{32}F^{c d} \mathcal{D}_{\underline{a}}{\mathcal{D}_{\underline{b}}{F_{c d}}} - \frac{9}{2}\mathcal{D}^{d}{\mathcal{D}_{\underline{a}}{F_{\underline{b}}\,^{c}}} F_{c d}+\frac{9}{5}\eta_{\underline{a} \underline{b}} \mathcal{D}^{d}{\mathcal{D}^{e}{F^{c}\,_{d}}} F_{c e} - \frac{27}{32}\eta_{\underline{a} \underline{b}} F^{c d} \mathcal{D}_{e}{\mathcal{D}^{e}{F_{c d}}} - \frac{9}{2}\mathcal{D}^{c}{\mathcal{D}^{d}{F_{\underline{a} c}}} F_{\underline{b} d}+\frac{9}{2}\mathcal{D}_{d}{\mathcal{D}^{d}{F_{\underline{a}}\,^{c}}} F_{\underline{b} c}+\frac{9}{2}\mathcal{D}_{\underline{a}}{\mathcal{D}^{d}{F_{\underline{b}}\,^{c}}} F_{c d} - \frac{9}{4}\eta_{\underline{a} \underline{b}} \mathcal{D}^{e}{\mathcal{D}^{d}{F^{c}\,_{d}}} F_{c e} - \frac{99}{640}\epsilon_{{e_{3}}}\,^{{e_{2}} c e {e_{1}}} \eta_{\underline{a} \underline{b}} W_{{e_{2}}}\,^{d} \mathcal{D}^{{e_{3}}}{F_{c d}} F_{e {e_{1}}}%
+\frac{9}{32}W^{c d} \mathcal{D}_{\underline{a}}{\mathcal{D}_{\underline{b}}{F_{c d}}} - \frac{9}{2}W^{c}\,_{d} \mathcal{D}^{d}{\mathcal{D}_{\underline{a}}{F_{\underline{b} c}}}+\frac{9}{5}\eta_{\underline{a} \underline{b}} W^{c}\,_{e} \mathcal{D}^{d}{\mathcal{D}^{e}{F_{c d}}} - \frac{153}{160}\eta_{\underline{a} \underline{b}} W^{c d} \mathcal{D}_{e}{\mathcal{D}^{e}{F_{c d}}} - \frac{9}{2}W_{\underline{a} d} \mathcal{D}^{c}{\mathcal{D}^{d}{F_{\underline{b} c}}}+\frac{9}{2}W_{\underline{a}}\,^{c} \mathcal{D}_{d}{\mathcal{D}^{d}{F_{\underline{b} c}}}+\frac{9}{2}W^{c}\,_{d} \mathcal{D}_{\underline{a}}{\mathcal{D}^{d}{F_{\underline{b} c}}} - \frac{9}{4}\eta_{\underline{a} \underline{b}} W^{c}\,_{e} \mathcal{D}^{e}{\mathcal{D}^{d}{F_{c d}}}+\frac{27}{128}\epsilon_{{e_{3}}}\,^{e {e_{1}} {e_{2}} c} \eta_{\underline{a} \underline{b}} W_{e {e_{1}}} W_{{e_{2}}}\,^{d} \mathcal{D}^{{e_{3}}}{F_{c d}} - \frac{27}{256}\epsilon^{d e {e_{1}} {e_{2}} c} W_{d e} W_{{e_{1}} {e_{2}}} \mathcal{D}_{\underline{a}}{F_{\underline{b} c}}+\frac{9}{256}\epsilon_{\underline{a}}\,^{{e_{2}} c d e} \mathcal{D}_{\underline{b}}{W_{{e_{2}}}\,^{{e_{1}}}} F_{c d} F_{e {e_{1}}} - \frac{9}{32}\epsilon_{\underline{a}}\,^{d e {e_{1}} {e_{2}}} \mathcal{D}^{c}{F_{\underline{b} c}} F_{d e} F_{{e_{1}} {e_{2}}} - \frac{9}{160}\epsilon^{c e {e_{1}} {e_{2}} {e_{3}}} \eta_{\underline{a} \underline{b}} \mathcal{D}^{d}{F_{c d}} F_{e {e_{1}}} F_{{e_{2}} {e_{3}}} - \frac{9}{128}\epsilon_{\underline{a} e {e_{1}}}\,^{c d} \mathcal{D}^{e}{\mathcal{D}_{\underline{b}}{\mathcal{D}^{{e_{1}}}{W_{c d}}}}+\frac{81}{128}\epsilon_{\underline{a} e {e_{1}}}\,^{c d} \mathcal{D}^{e}{\mathcal{D}^{{e_{1}}}{\mathcal{D}_{\underline{b}}{W_{c d}}}}+\frac{27}{64}\epsilon_{\underline{a} d e {e_{1}}}\,^{c} \mathcal{D}^{d}{\mathcal{D}^{e}{\mathcal{D}^{{e_{1}}}{W_{\underline{b} c}}}}+\frac{9}{128}\epsilon_{\underline{a} e {e_{1}}}\,^{c d} \mathcal{D}_{\underline{b}}{\mathcal{D}^{e}{\mathcal{D}^{{e_{1}}}{W_{c d}}}} - \frac{9}{128}\epsilon_{e {e_{1}} {e_{2}}}\,^{c d} \eta_{\underline{a} \underline{b}} \mathcal{D}^{e}{\mathcal{D}^{{e_{1}}}{\mathcal{D}^{{e_{2}}}{W_{c d}}}}+\frac{3}{80}\mathcal{D}_{d}{\mathcal{D}^{d}{R^{c}\,_{c}}} \eta_{\underline{a} \underline{b}}
\doublespacedmathend
\end{adjustwidth}

\subsection{Scalar curvature squared} \label{SupercurrentR2Complete}

\subsubsection{$J^1_{\alpha i, R^2}$}

\begin{adjustwidth}{0cm}{3cm}
\doublespacedmathbegin
- \frac{3}{32}{\rm i} F \boldsymbol{W} \lambda^{i}_{\alpha} {G}^{-1} - \frac{3}{32}{\rm i} F W \boldsymbol{\lambda}^{i}_{\alpha} {G}^{-1}+\frac{3}{16}(\Gamma_{a})_{\alpha}{}^{\beta} W \boldsymbol{W} {G}^{-1} \nabla^{a}{\varphi^{i}_{\beta}} - \frac{9}{64}(\Sigma_{a b})_{\alpha}{}^{\beta} W \boldsymbol{W} W^{a b} {G}^{-1} \varphi^{i}_{\beta} - \frac{27}{64}G^{i}\,_{j} W \boldsymbol{W} X^{j}_{\alpha} {G}^{-1}+\frac{3}{64}F G^{i}\,_{j} W \boldsymbol{W} {G}^{-3} \varphi^{j}_{\alpha} - \frac{3}{64}G_{j k} \boldsymbol{W} \lambda^{i}_{\alpha} {G}^{-3} \varphi^{j \beta} \varphi^{k}_{\beta} - \frac{3}{64}G_{j k} W \boldsymbol{\lambda}^{i}_{\alpha} {G}^{-3} \varphi^{j \beta} \varphi^{k}_{\beta}+\frac{3}{32}{\rm i} W \boldsymbol{W} {G}^{-3} \varphi^{i \beta} \varphi_{j \alpha} \varphi^{j}_{\beta} - \frac{3}{64}\mathcal{H}_{a} G^{i}\,_{j} (\Gamma^{a})_{\alpha}{}^{\beta} W \boldsymbol{W} {G}^{-3} \varphi^{j}_{\beta} - \frac{3}{32}G_{j k} (\Gamma_{a})_{\alpha}{}^{\beta} W \boldsymbol{W} {G}^{-3} \nabla^{a}{G^{i j}} \varphi^{k}_{\beta} - \frac{9}{64}{\rm i} G^{i}\,_{j} G_{k l} W \boldsymbol{W} {G}^{-5} \varphi^{j}_{\alpha} \varphi^{k \beta} \varphi^{l}_{\beta} - \frac{3}{32}G^{i}\,_{j} G_{k l} W \mathbf{X}^{k l} {G}^{-3} \varphi^{j}_{\alpha}+\frac{3}{32}{\rm i} G_{j k} \mathbf{X}^{j k} \lambda^{i}_{\alpha} {G}^{-1}+\frac{3}{16}W \mathbf{X}^{i}\,_{j} {G}^{-1} \varphi^{j}_{\alpha}+\frac{3}{16}{\rm i} G^{i}\,_{j} (\Gamma_{a})_{\alpha}{}^{\beta} W {G}^{-1} \nabla^{a}{\boldsymbol{\lambda}^{j}_{\beta}}+\frac{9}{64}{\rm i} G^{i}\,_{j} (\Sigma_{a b})_{\alpha}{}^{\beta} W W^{a b} \boldsymbol{\lambda}^{j}_{\beta} {G}^{-1} - \frac{3}{32}G^{i}\,_{j} G_{k l} \boldsymbol{W} X^{k l} {G}^{-3} \varphi^{j}_{\alpha}+\frac{3}{16}\boldsymbol{W} X^{i}\,_{j} {G}^{-1} \varphi^{j}_{\alpha}%
+\frac{3}{16}{\rm i} G^{i}\,_{j} (\Gamma_{a})_{\alpha}{}^{\beta} \boldsymbol{W} {G}^{-1} \nabla^{a}{\lambda^{j}_{\beta}}+\frac{9}{64}{\rm i} G^{i}\,_{j} (\Sigma_{a b})_{\alpha}{}^{\beta} \boldsymbol{W} W^{a b} \lambda^{j}_{\beta} {G}^{-1}+\frac{3}{32}{\rm i} G_{j k} X^{j k} \boldsymbol{\lambda}^{i}_{\alpha} {G}^{-1}+\frac{3}{32}{\rm i} G^{i}\,_{j} G_{k l} \lambda^{k \beta} \boldsymbol{\lambda}^{l}_{\beta} {G}^{-3} \varphi^{j}_{\alpha} - \frac{3}{32}{\rm i} \lambda^{i \beta} \boldsymbol{\lambda}_{j \beta} {G}^{-1} \varphi^{j}_{\alpha} - \frac{3}{32}{\rm i} \boldsymbol{\lambda}^{i \beta} \lambda_{j \beta} {G}^{-1} \varphi^{j}_{\alpha}+\frac{3}{32}{\rm i} G^{i}\,_{j} (\Sigma_{a b})_{\alpha}{}^{\beta} F^{a b} \boldsymbol{\lambda}^{j}_{\beta} {G}^{-1} - \frac{3}{32}{\rm i} G_{j k} X^{i j} \boldsymbol{\lambda}^{k}_{\alpha} {G}^{-1}+\frac{3}{32}{\rm i} G^{i}\,_{j} (\Gamma_{a})_{\alpha}{}^{\beta} \boldsymbol{\lambda}^{j}_{\beta} {G}^{-1} \nabla^{a}{W}+\frac{3}{32}{\rm i} G^{i}\,_{j} (\Sigma_{a b})_{\alpha}{}^{\beta} \mathbf{F}^{a b} \lambda^{j}_{\beta} {G}^{-1} - \frac{3}{32}{\rm i} G_{j k} \mathbf{X}^{i j} \lambda^{k}_{\alpha} {G}^{-1}+\frac{3}{32}{\rm i} G^{i}\,_{j} (\Gamma_{a})_{\alpha}{}^{\beta} \lambda^{j}_{\beta} {G}^{-1} \nabla^{a}{\boldsymbol{W}}
\doublespacedmathend
\end{adjustwidth}

\subsubsection{$J^2_{a i j, R^2}$}
\begin{adjustwidth}{0cm}{5cm}
\doublespacedmathbegin
- \frac{3}{16}{\rm i} \boldsymbol{W} \lambda^{\alpha}_{\underline{i}} {G}^{-1} \nabla_{a}{\varphi_{\underline{j} \alpha}}+\frac{3}{8}{\rm i} (\Sigma_{a b})^{\alpha \beta} \boldsymbol{W} \lambda_{\underline{i} \alpha} {G}^{-1} \nabla^{b}{\varphi_{\underline{j} \beta}}+\frac{9}{32}{\rm i} (\Gamma^{b})^{\alpha \beta} \boldsymbol{W} W_{a b} \lambda_{\underline{i} \alpha} {G}^{-1} \varphi_{\underline{j} \beta}+\frac{63}{256}{\rm i} G_{\underline{i} k} (\Gamma_{a})^{\beta \alpha} \boldsymbol{W} \lambda_{\underline{j} \beta} X^{k}_{\alpha} {G}^{-1} - \frac{3}{32}F (\Gamma_{a})^{\alpha \beta} \lambda_{\underline{i} \alpha} \boldsymbol{\lambda}_{\underline{j} \beta} {G}^{-1} - \frac{3}{64}{\rm i} F G_{\underline{i} k} (\Gamma_{a})^{\alpha \beta} \boldsymbol{W} \lambda_{\underline{j} \alpha} {G}^{-3} \varphi^{k}_{\beta} - \frac{3}{16}{\rm i} W \boldsymbol{\lambda}_{\underline{i}}^{\alpha} {G}^{-1} \nabla_{a}{\varphi_{\underline{j} \alpha}}+\frac{3}{8}{\rm i} (\Sigma_{a b})^{\alpha \beta} W \boldsymbol{\lambda}_{\underline{i} \alpha} {G}^{-1} \nabla^{b}{\varphi_{\underline{j} \beta}}+\frac{9}{32}{\rm i} (\Gamma^{b})^{\alpha \beta} W W_{a b} \boldsymbol{\lambda}_{\underline{i} \alpha} {G}^{-1} \varphi_{\underline{j} \beta}+\frac{63}{256}{\rm i} G_{\underline{i} k} (\Gamma_{a})^{\beta \alpha} W \boldsymbol{\lambda}_{\underline{j} \beta} X^{k}_{\alpha} {G}^{-1} - \frac{3}{64}{\rm i} F G_{\underline{i} k} (\Gamma_{a})^{\alpha \beta} W \boldsymbol{\lambda}_{\underline{j} \alpha} {G}^{-3} \varphi^{k}_{\beta} - \frac{171}{512}(\Gamma_{a})^{\alpha \beta} W \boldsymbol{W} X_{\underline{i} \alpha} {G}^{-1} \varphi_{\underline{j} \beta} - \frac{9}{8}{\rm i} G_{\underline{i} \underline{j}} W \boldsymbol{W} {G}^{-1} \nabla^{b}{W_{a b}} - \frac{171}{512}G_{\underline{i} k} G_{\underline{j} l} (\Gamma_{a})^{\alpha \beta} W \boldsymbol{W} X^{k}_{\alpha} {G}^{-3} \varphi^{l}_{\beta}+\frac{3}{32}(\Gamma_{a})^{\alpha \beta} \boldsymbol{W} \lambda_{\underline{i} \alpha} {G}^{-3} \varphi_{\underline{j}}^{\rho} \varphi_{k \beta} \varphi^{k}_{\rho}+\frac{3}{64}{\rm i} G_{k l} (\Gamma_{a})^{\alpha \beta} \lambda_{\underline{i} \alpha} \boldsymbol{\lambda}_{\underline{j} \beta} {G}^{-3} \varphi^{k \rho} \varphi^{l}_{\rho} - \frac{9}{64}G_{\underline{i} k} G_{l m} (\Gamma_{a})^{\alpha \beta} \boldsymbol{W} \lambda_{\underline{j} \alpha} {G}^{-5} \varphi^{k}_{\beta} \varphi^{l \rho} \varphi^{m}_{\rho}+\frac{3}{64}{\rm i} \mathcal{H}_{a} G_{\underline{i} k} \boldsymbol{W} \lambda^{\alpha}_{\underline{j}} {G}^{-3} \varphi^{k}_{\alpha} - \frac{3}{32}{\rm i} \mathcal{H}^{b} G_{\underline{i} k} (\Sigma_{a b})^{\alpha \beta} \boldsymbol{W} \lambda_{\underline{j} \alpha} {G}^{-3} \varphi^{k}_{\beta}%
+\frac{3}{32}{\rm i} G_{k l} \boldsymbol{W} \lambda^{\alpha}_{\underline{i}} {G}^{-3} \nabla_{a}{G_{\underline{j}}\,^{k}} \varphi^{l}_{\alpha} - \frac{3}{16}{\rm i} G_{k l} (\Sigma_{a b})^{\alpha \beta} \boldsymbol{W} \lambda_{\underline{i} \alpha} {G}^{-3} \nabla^{b}{G_{\underline{j}}\,^{k}} \varphi^{l}_{\beta}+\frac{3}{32}(\Gamma_{a})^{\alpha \beta} W \boldsymbol{\lambda}_{\underline{i} \alpha} {G}^{-3} \varphi_{\underline{j}}^{\rho} \varphi_{k \beta} \varphi^{k}_{\rho} - \frac{9}{64}G_{\underline{i} k} G_{l m} (\Gamma_{a})^{\alpha \beta} W \boldsymbol{\lambda}_{\underline{j} \alpha} {G}^{-5} \varphi^{k}_{\beta} \varphi^{l \rho} \varphi^{m}_{\rho}+\frac{3}{64}{\rm i} \mathcal{H}_{a} G_{\underline{i} k} W \boldsymbol{\lambda}_{\underline{j}}^{\alpha} {G}^{-3} \varphi^{k}_{\alpha} - \frac{3}{32}{\rm i} \mathcal{H}^{b} G_{\underline{i} k} (\Sigma_{a b})^{\alpha \beta} W \boldsymbol{\lambda}_{\underline{j} \alpha} {G}^{-3} \varphi^{k}_{\beta}+\frac{3}{32}{\rm i} G_{k l} W \boldsymbol{\lambda}_{\underline{i}}^{\alpha} {G}^{-3} \nabla_{a}{G_{\underline{j}}\,^{k}} \varphi^{l}_{\alpha} - \frac{3}{16}{\rm i} G_{k l} (\Sigma_{a b})^{\alpha \beta} W \boldsymbol{\lambda}_{\underline{i} \alpha} {G}^{-3} \nabla^{b}{G_{\underline{j}}\,^{k}} \varphi^{l}_{\beta} - \frac{9}{32}{\rm i} G_{\underline{i} k} (\Gamma_{a})^{\alpha \beta} W \boldsymbol{W} {G}^{-5} \varphi_{\underline{j}}^{\rho} \varphi^{k}_{\alpha} \varphi_{l \beta} \varphi^{l}_{\rho}+\frac{3}{16}(\Sigma_{a b})^{\alpha \beta} W \boldsymbol{W} {G}^{-3} \nabla^{b}{G_{\underline{i} \underline{j}}} \varphi_{k \alpha} \varphi^{k}_{\beta} - \frac{9}{64}\mathcal{H}_{a} W \boldsymbol{W} {G}^{-3} \varphi_{\underline{i}}^{\alpha} \varphi_{\underline{j} \alpha}+\frac{9}{32}W \boldsymbol{W} {G}^{-3} \nabla_{a}{G_{\underline{i} k}} \varphi_{\underline{j}}^{\alpha} \varphi^{k}_{\alpha}+\frac{3}{16}(\Sigma_{a b})^{\alpha \beta} W \boldsymbol{W} {G}^{-3} \nabla^{b}{G_{\underline{i} k}} \varphi_{\underline{j} \alpha} \varphi^{k}_{\beta} - \frac{9}{64}\mathcal{H}_{a} G_{\underline{i} k} G_{\underline{j} l} W \boldsymbol{W} {G}^{-5} \varphi^{k \alpha} \varphi^{l}_{\alpha} - \frac{9}{32}G_{\underline{i} k} G_{l m} W \boldsymbol{W} {G}^{-5} \nabla_{a}{G_{\underline{j}}\,^{l}} \varphi^{k \alpha} \varphi^{m}_{\alpha}+\frac{9}{16}G_{\underline{i} k} G_{l m} (\Sigma_{a b})^{\alpha \beta} W \boldsymbol{W} {G}^{-5} \nabla^{b}{G_{\underline{j}}\,^{l}} \varphi^{k}_{\alpha} \varphi^{m}_{\beta}+\frac{27}{512}G_{\underline{i} \underline{j}} G_{k l} (\Gamma_{a})^{\alpha \beta} W \boldsymbol{W} X^{k}_{\alpha} {G}^{-3} \varphi^{l}_{\beta} - \frac{9}{128}{\rm i} G_{k l} (\Gamma_{a})^{\alpha \beta} W \boldsymbol{W} {G}^{-5} \varphi_{\underline{i} \alpha} \varphi_{\underline{j} \beta} \varphi^{k \rho} \varphi^{l}_{\rho} - \frac{45}{128}{\rm i} G_{\underline{i} k} G_{\underline{j} l} G_{m n} (\Gamma_{a})^{\alpha \beta} W \boldsymbol{W} {G}^{-7} \varphi^{k}_{\alpha} \varphi^{l}_{\beta} \varphi^{m \rho} \varphi^{n}_{\rho}+\frac{9}{64}\mathcal{H}_{a} G_{\underline{i} \underline{j}} G_{k l} W \boldsymbol{W} {G}^{-5} \varphi^{k \alpha} \varphi^{l}_{\alpha}%
+\frac{9}{32}G_{\underline{i} k} G_{l m} W \boldsymbol{W} {G}^{-5} \nabla_{a}{G_{\underline{j}}\,^{k}} \varphi^{l \alpha} \varphi^{m}_{\alpha} - \frac{3}{64}G_{k l} (\Gamma_{a})^{\alpha \beta} W \mathbf{X}^{k l} {G}^{-3} \varphi_{\underline{i} \alpha} \varphi_{\underline{j} \beta}+\frac{3}{16}G_{\underline{i} k} (\Gamma_{a})^{\alpha \beta} W \mathbf{X}_{\underline{j} l} {G}^{-3} \varphi^{k}_{\alpha} \varphi^{l}_{\beta}+\frac{3}{32}{\rm i} G_{\underline{i} k} G_{l m} (\Gamma_{a})^{\alpha \beta} \mathbf{X}^{l m} \lambda_{\underline{j} \alpha} {G}^{-3} \varphi^{k}_{\beta}+\frac{3}{16}{\rm i} G_{\underline{i} k} G_{\underline{j} l} W {G}^{-3} \nabla_{a}{\boldsymbol{\lambda}^{k \alpha}} \varphi^{l}_{\alpha}+\frac{3}{8}{\rm i} G_{\underline{i} k} G_{\underline{j} l} (\Sigma_{a b})^{\alpha \beta} W {G}^{-3} \nabla^{b}{\boldsymbol{\lambda}^{k}_{\alpha}} \varphi^{l}_{\beta}+\frac{9}{128}{\rm i} \epsilon^{d e}\,_{a}\,^{b c} G_{\underline{i} k} G_{\underline{j} l} (\Sigma_{d e})^{\alpha \beta} W W_{b c} \boldsymbol{\lambda}^{k}_{\alpha} {G}^{-3} \varphi^{l}_{\beta}+\frac{9}{64}{\rm i} G_{\underline{i} k} G_{\underline{j} l} (\Gamma^{b})^{\alpha \beta} W W_{a b} \boldsymbol{\lambda}^{k}_{\alpha} {G}^{-3} \varphi^{l}_{\beta} - \frac{9}{64}G_{\underline{i} k} G_{\underline{j} l} G_{m n} (\Gamma_{a})^{\alpha \beta} W \mathbf{X}^{m n} {G}^{-5} \varphi^{k}_{\alpha} \varphi^{l}_{\beta} - \frac{3}{32}{\rm i} \mathcal{H}_{a} G_{\underline{i} \underline{j}} G_{k l} W \mathbf{X}^{k l} {G}^{-3} - \frac{3}{16}{\rm i} G_{\underline{i} k} G_{l m} W \mathbf{X}^{l m} {G}^{-3} \nabla_{a}{G_{\underline{j}}\,^{k}} - \frac{3}{32}{\rm i} (\Gamma_{a})^{\alpha \beta} \mathbf{X}_{\underline{i} k} \lambda_{\underline{j} \alpha} {G}^{-1} \varphi^{k}_{\beta}+\frac{3}{16}G_{\underline{i} k} \lambda^{\alpha}_{\underline{j}} {G}^{-1} \nabla_{a}{\boldsymbol{\lambda}^{k}_{\alpha}} - \frac{3}{8}G_{\underline{i} k} (\Sigma_{a b})^{\alpha \beta} \lambda_{\underline{j} \alpha} {G}^{-1} \nabla^{b}{\boldsymbol{\lambda}^{k}_{\beta}}+\frac{3}{16}{\rm i} W {G}^{-1} \nabla_{a}{\boldsymbol{\lambda}_{\underline{i}}^{\alpha}} \varphi_{\underline{j} \alpha}+\frac{3}{8}{\rm i} (\Sigma_{a b})^{\alpha \beta} W {G}^{-1} \nabla^{b}{\boldsymbol{\lambda}_{\underline{i} \alpha}} \varphi_{\underline{j} \beta}+\frac{3}{16}{\rm i} \mathcal{H}_{a} W \mathbf{X}_{\underline{i} \underline{j}} {G}^{-1}+\frac{3}{8}{\rm i} W \mathbf{X}_{\underline{i} k} {G}^{-1} \nabla_{a}{G_{\underline{j}}\,^{k}} - \frac{3}{8}{\rm i} G_{\underline{i} \underline{j}} W {G}^{-1} \nabla^{b}{\mathbf{F}_{a b}} - \frac{9}{8}{\rm i} G_{\underline{i} \underline{j}} W W_{a b} {G}^{-1} \nabla^{b}{\boldsymbol{W}}%
+\frac{3}{8}{\rm i} G_{\underline{i} k} W {G}^{-1} \nabla_{a}{\mathbf{X}_{\underline{j}}\,^{k}} - \frac{27}{256}{\rm i} G_{\underline{i} k} (\Gamma_{a})^{\beta \alpha} W \boldsymbol{\lambda}^{k}_{\beta} X_{\underline{j} \alpha} {G}^{-1} - \frac{27}{256}{\rm i} G_{\underline{i} \underline{j}} (\Gamma_{a})^{\beta \alpha} W \boldsymbol{\lambda}_{k \beta} X^{k}_{\alpha} {G}^{-1} - \frac{3}{64}G_{k l} (\Gamma_{a})^{\alpha \beta} \boldsymbol{W} X^{k l} {G}^{-3} \varphi_{\underline{i} \alpha} \varphi_{\underline{j} \beta}+\frac{3}{16}G_{\underline{i} k} (\Gamma_{a})^{\alpha \beta} \boldsymbol{W} X_{\underline{j} l} {G}^{-3} \varphi^{k}_{\alpha} \varphi^{l}_{\beta}+\frac{3}{32}{\rm i} G_{\underline{i} k} G_{l m} (\Gamma_{a})^{\alpha \beta} X^{l m} \boldsymbol{\lambda}_{\underline{j} \alpha} {G}^{-3} \varphi^{k}_{\beta}+\frac{3}{16}{\rm i} G_{\underline{i} k} G_{\underline{j} l} \boldsymbol{W} {G}^{-3} \nabla_{a}{\lambda^{k \alpha}} \varphi^{l}_{\alpha}+\frac{3}{8}{\rm i} G_{\underline{i} k} G_{\underline{j} l} (\Sigma_{a b})^{\alpha \beta} \boldsymbol{W} {G}^{-3} \nabla^{b}{\lambda^{k}_{\alpha}} \varphi^{l}_{\beta}+\frac{9}{128}{\rm i} \epsilon^{d e}\,_{a}\,^{b c} G_{\underline{i} k} G_{\underline{j} l} (\Sigma_{d e})^{\alpha \beta} \boldsymbol{W} W_{b c} \lambda^{k}_{\alpha} {G}^{-3} \varphi^{l}_{\beta}+\frac{9}{64}{\rm i} G_{\underline{i} k} G_{\underline{j} l} (\Gamma^{b})^{\alpha \beta} \boldsymbol{W} W_{a b} \lambda^{k}_{\alpha} {G}^{-3} \varphi^{l}_{\beta} - \frac{9}{64}G_{\underline{i} k} G_{\underline{j} l} G_{m n} (\Gamma_{a})^{\alpha \beta} \boldsymbol{W} X^{m n} {G}^{-5} \varphi^{k}_{\alpha} \varphi^{l}_{\beta} - \frac{3}{32}{\rm i} \mathcal{H}_{a} G_{\underline{i} \underline{j}} G_{k l} \boldsymbol{W} X^{k l} {G}^{-3} - \frac{3}{16}{\rm i} G_{\underline{i} k} G_{l m} \boldsymbol{W} X^{l m} {G}^{-3} \nabla_{a}{G_{\underline{j}}\,^{k}} - \frac{3}{32}{\rm i} (\Gamma_{a})^{\alpha \beta} X_{\underline{i} k} \boldsymbol{\lambda}_{\underline{j} \alpha} {G}^{-1} \varphi^{k}_{\beta}+\frac{3}{16}{\rm i} \boldsymbol{W} {G}^{-1} \nabla_{a}{\lambda^{\alpha}_{\underline{i}}} \varphi_{\underline{j} \alpha}+\frac{3}{8}{\rm i} (\Sigma_{a b})^{\alpha \beta} \boldsymbol{W} {G}^{-1} \nabla^{b}{\lambda_{\underline{i} \alpha}} \varphi_{\underline{j} \beta}+\frac{3}{16}{\rm i} \mathcal{H}_{a} \boldsymbol{W} X_{\underline{i} \underline{j}} {G}^{-1}+\frac{3}{8}{\rm i} \boldsymbol{W} X_{\underline{i} k} {G}^{-1} \nabla_{a}{G_{\underline{j}}\,^{k}}+\frac{3}{16}G_{\underline{i} k} \boldsymbol{\lambda}_{\underline{j}}^{\alpha} {G}^{-1} \nabla_{a}{\lambda^{k}_{\alpha}} - \frac{3}{8}G_{\underline{i} k} (\Sigma_{a b})^{\alpha \beta} \boldsymbol{\lambda}_{\underline{j} \alpha} {G}^{-1} \nabla^{b}{\lambda^{k}_{\beta}}%
 - \frac{3}{8}{\rm i} G_{\underline{i} \underline{j}} \boldsymbol{W} {G}^{-1} \nabla^{b}{F_{a b}} - \frac{9}{8}{\rm i} G_{\underline{i} \underline{j}} \boldsymbol{W} W_{a b} {G}^{-1} \nabla^{b}{W}+\frac{3}{8}{\rm i} G_{\underline{i} k} \boldsymbol{W} {G}^{-1} \nabla_{a}{X_{\underline{j}}\,^{k}} - \frac{27}{256}{\rm i} G_{\underline{i} k} (\Gamma_{a})^{\beta \alpha} \boldsymbol{W} \lambda^{k}_{\beta} X_{\underline{j} \alpha} {G}^{-1} - \frac{27}{256}{\rm i} G_{\underline{i} \underline{j}} (\Gamma_{a})^{\beta \alpha} \boldsymbol{W} \lambda_{k \beta} X^{k}_{\alpha} {G}^{-1}+\frac{3}{64}{\rm i} G_{k l} (\Gamma_{a})^{\alpha \beta} \lambda^{k \rho} \boldsymbol{\lambda}^{l}_{\rho} {G}^{-3} \varphi_{\underline{i} \alpha} \varphi_{\underline{j} \beta} - \frac{3}{32}{\rm i} G_{\underline{i} k} (\Gamma_{a})^{\alpha \beta} \lambda^{\rho}_{\underline{j}} \boldsymbol{\lambda}_{l \rho} {G}^{-3} \varphi^{k}_{\alpha} \varphi^{l}_{\beta} - \frac{3}{32}{\rm i} G_{\underline{i} k} (\Gamma_{a})^{\alpha \beta} \lambda^{\rho}_{l} \boldsymbol{\lambda}_{\underline{j} \rho} {G}^{-3} \varphi^{k}_{\alpha} \varphi^{l}_{\beta}+\frac{3}{64}{\rm i} \epsilon^{d e}\,_{a}\,^{b c} G_{\underline{i} k} G_{\underline{j} l} (\Sigma_{d e})^{\alpha \beta} F_{b c} \boldsymbol{\lambda}^{k}_{\alpha} {G}^{-3} \varphi^{l}_{\beta}+\frac{3}{32}{\rm i} G_{\underline{i} k} G_{\underline{j} l} (\Gamma^{b})^{\alpha \beta} F_{a b} \boldsymbol{\lambda}^{k}_{\alpha} {G}^{-3} \varphi^{l}_{\beta} - \frac{3}{32}{\rm i} G_{\underline{i} k} G_{l m} (\Gamma_{a})^{\alpha \beta} X_{\underline{j}}\,^{l} \boldsymbol{\lambda}^{m}_{\alpha} {G}^{-3} \varphi^{k}_{\beta}+\frac{3}{32}{\rm i} G_{\underline{i} k} G_{\underline{j} l} \boldsymbol{\lambda}^{k \alpha} {G}^{-3} \nabla_{a}{W} \varphi^{l}_{\alpha}+\frac{3}{16}{\rm i} G_{\underline{i} k} G_{\underline{j} l} (\Sigma_{a b})^{\alpha \beta} \boldsymbol{\lambda}^{k}_{\alpha} {G}^{-3} \nabla^{b}{W} \varphi^{l}_{\beta}+\frac{3}{64}{\rm i} \epsilon^{d e}\,_{a}\,^{b c} G_{\underline{i} k} G_{\underline{j} l} (\Sigma_{d e})^{\alpha \beta} \mathbf{F}_{b c} \lambda^{k}_{\alpha} {G}^{-3} \varphi^{l}_{\beta}+\frac{3}{32}{\rm i} G_{\underline{i} k} G_{\underline{j} l} (\Gamma^{b})^{\alpha \beta} \mathbf{F}_{a b} \lambda^{k}_{\alpha} {G}^{-3} \varphi^{l}_{\beta} - \frac{3}{32}{\rm i} G_{\underline{i} k} G_{l m} (\Gamma_{a})^{\alpha \beta} \mathbf{X}_{\underline{j}}\,^{l} \lambda^{m}_{\alpha} {G}^{-3} \varphi^{k}_{\beta}+\frac{3}{32}{\rm i} G_{\underline{i} k} G_{\underline{j} l} \lambda^{k \alpha} {G}^{-3} \nabla_{a}{\boldsymbol{W}} \varphi^{l}_{\alpha}+\frac{3}{16}{\rm i} G_{\underline{i} k} G_{\underline{j} l} (\Sigma_{a b})^{\alpha \beta} \lambda^{k}_{\alpha} {G}^{-3} \nabla^{b}{\boldsymbol{W}} \varphi^{l}_{\beta}+\frac{9}{64}{\rm i} G_{\underline{i} k} G_{\underline{j} l} G_{m n} (\Gamma_{a})^{\alpha \beta} \lambda^{m \rho} \boldsymbol{\lambda}^{n}_{\rho} {G}^{-5} \varphi^{k}_{\alpha} \varphi^{l}_{\beta} - \frac{3}{32}\mathcal{H}_{a} G_{\underline{i} \underline{j}} G_{k l} \lambda^{k \alpha} \boldsymbol{\lambda}^{l}_{\alpha} {G}^{-3}%
 - \frac{3}{16}G_{\underline{i} k} G_{l m} \lambda^{l \alpha} \boldsymbol{\lambda}^{m}_{\alpha} {G}^{-3} \nabla_{a}{G_{\underline{j}}\,^{k}}+\frac{3}{32}{\rm i} (\Gamma_{a})^{\alpha \beta} X_{\underline{i} \underline{j}} \boldsymbol{\lambda}_{k \alpha} {G}^{-1} \varphi^{k}_{\beta}+\frac{3}{64}{\rm i} \epsilon^{d e}\,_{a}\,^{b c} (\Sigma_{d e})^{\alpha \beta} \mathbf{F}_{b c} \lambda_{\underline{i} \alpha} {G}^{-1} \varphi_{\underline{j} \beta}+\frac{3}{32}{\rm i} (\Gamma^{b})^{\alpha \beta} \mathbf{F}_{a b} \lambda_{\underline{i} \alpha} {G}^{-1} \varphi_{\underline{j} \beta}+\frac{3}{32}{\rm i} \lambda^{\alpha}_{\underline{i}} {G}^{-1} \nabla_{a}{\boldsymbol{W}} \varphi_{\underline{j} \alpha}+\frac{3}{16}{\rm i} (\Sigma_{a b})^{\alpha \beta} \lambda_{\underline{i} \alpha} {G}^{-1} \nabla^{b}{\boldsymbol{W}} \varphi_{\underline{j} \beta}+\frac{3}{16}\mathcal{H}_{a} \lambda^{\alpha}_{\underline{i}} \boldsymbol{\lambda}_{\underline{j} \alpha} {G}^{-1}+\frac{3}{16}\lambda^{\alpha}_{\underline{i}} \boldsymbol{\lambda}_{k \alpha} {G}^{-1} \nabla_{a}{G_{\underline{j}}\,^{k}}+\frac{3}{64}{\rm i} \epsilon^{d e}\,_{a}\,^{b c} (\Sigma_{d e})^{\alpha \beta} F_{b c} \boldsymbol{\lambda}_{\underline{i} \alpha} {G}^{-1} \varphi_{\underline{j} \beta}+\frac{3}{32}{\rm i} (\Gamma^{b})^{\alpha \beta} F_{a b} \boldsymbol{\lambda}_{\underline{i} \alpha} {G}^{-1} \varphi_{\underline{j} \beta}+\frac{3}{32}{\rm i} \boldsymbol{\lambda}_{\underline{i}}^{\alpha} {G}^{-1} \nabla_{a}{W} \varphi_{\underline{j} \alpha}+\frac{3}{16}{\rm i} (\Sigma_{a b})^{\alpha \beta} \boldsymbol{\lambda}_{\underline{i} \alpha} {G}^{-1} \nabla^{b}{W} \varphi_{\underline{j} \beta}+\frac{3}{32}{\rm i} (\Gamma_{a})^{\alpha \beta} \mathbf{X}_{\underline{i} \underline{j}} \lambda_{k \alpha} {G}^{-1} \varphi^{k}_{\beta}+\frac{3}{16}\lambda^{\alpha}_{k} \boldsymbol{\lambda}_{\underline{i} \alpha} {G}^{-1} \nabla_{a}{G_{\underline{j}}\,^{k}}+\frac{3}{16}G_{\underline{i} k} \boldsymbol{\lambda}^{k \alpha} {G}^{-1} \nabla_{a}{\lambda_{\underline{j} \alpha}}+\frac{3}{8}G_{\underline{i} k} (\Sigma_{a b})^{\alpha \beta} \boldsymbol{\lambda}^{k}_{\alpha} {G}^{-1} \nabla^{b}{\lambda_{\underline{j} \beta}} - \frac{3}{32}{\rm i} \epsilon_{a}\,^{b c d e} G_{\underline{i} \underline{j}} F_{b c} \mathbf{F}_{d e} {G}^{-1} - \frac{3}{8}{\rm i} G_{\underline{i} \underline{j}} F_{a b} {G}^{-1} \nabla^{b}{\boldsymbol{W}}+\frac{3}{8}{\rm i} G_{\underline{i} k} X_{\underline{j}}\,^{k} {G}^{-1} \nabla_{a}{\boldsymbol{W}} - \frac{3}{8}{\rm i} G_{\underline{i} \underline{j}} \mathbf{F}_{a b} {G}^{-1} \nabla^{b}{W}%
+\frac{3}{8}{\rm i} G_{\underline{i} k} \mathbf{X}_{\underline{j}}\,^{k} {G}^{-1} \nabla_{a}{W}+\frac{3}{16}G_{\underline{i} k} \lambda^{k \alpha} {G}^{-1} \nabla_{a}{\boldsymbol{\lambda}_{\underline{j} \alpha}}+\frac{3}{8}G_{\underline{i} k} (\Sigma_{a b})^{\alpha \beta} \lambda^{k}_{\alpha} {G}^{-1} \nabla^{b}{\boldsymbol{\lambda}_{\underline{j} \beta}}
\doublespacedmathend
\end{adjustwidth}

\subsubsection{$J^2_{a b, R^2}$}

\begin{adjustwidth}{0cm}{5cm}
\doublespacedmathbegin
- \frac{3}{32}{\rm i} \epsilon^{c d}\,_{\hat{a} \hat{b} e} (\Sigma_{c d})^{\alpha \beta} \boldsymbol{W} \lambda_{i \alpha} {G}^{-1} \nabla^{e}{\varphi^{i}_{\beta}}+\frac{3}{16}{\rm i} (\Gamma_{\hat{a}})^{\alpha \beta} \boldsymbol{W} \lambda_{i \alpha} {G}^{-1} \nabla_{\hat{b}}{\varphi^{i}_{\beta}} - \frac{27}{128}{\rm i} \boldsymbol{W} W_{\hat{a} \hat{b}} \lambda^{\alpha}_{i} {G}^{-1} \varphi^{i}_{\alpha}+\frac{27}{256}{\rm i} \epsilon_{\hat{a} \hat{b} e}\,^{c d} (\Gamma^{e})^{\alpha \beta} \boldsymbol{W} W_{c d} \lambda_{i \alpha} {G}^{-1} \varphi^{i}_{\beta} - \frac{9}{32}{\rm i} (\Sigma_{\hat{a} c})^{\alpha \beta} \boldsymbol{W} W^{c}\,_{\hat{b}} \lambda_{i \alpha} {G}^{-1} \varphi^{i}_{\beta}+\frac{9}{64}{\rm i} G_{i j} (\Sigma_{\hat{a} \hat{b}})^{\beta \alpha} \boldsymbol{W} \lambda^{i}_{\beta} X^{j}_{\alpha} {G}^{-1}+\frac{3}{32}F (\Sigma_{\hat{a} \hat{b}})^{\alpha \beta} \lambda_{i \alpha} \boldsymbol{\lambda}^{i}_{\beta} {G}^{-1}+\frac{3}{16}{\rm i} F \boldsymbol{W} F_{\hat{a} \hat{b}} {G}^{-1}+\frac{9}{16}{\rm i} F W \boldsymbol{W} W_{\hat{a} \hat{b}} {G}^{-1} - \frac{3}{64}{\rm i} F G_{i j} (\Sigma_{\hat{a} \hat{b}})^{\alpha \beta} \boldsymbol{W} \lambda^{i}_{\alpha} {G}^{-3} \varphi^{j}_{\beta} - \frac{3}{32}{\rm i} \epsilon^{c d}\,_{\hat{a} \hat{b} e} (\Sigma_{c d})^{\alpha \beta} W \boldsymbol{\lambda}_{i \alpha} {G}^{-1} \nabla^{e}{\varphi^{i}_{\beta}}+\frac{3}{16}{\rm i} (\Gamma_{\hat{a}})^{\alpha \beta} W \boldsymbol{\lambda}_{i \alpha} {G}^{-1} \nabla_{\hat{b}}{\varphi^{i}_{\beta}} - \frac{27}{128}{\rm i} W W_{\hat{a} \hat{b}} \boldsymbol{\lambda}_{i}^{\alpha} {G}^{-1} \varphi^{i}_{\alpha}+\frac{27}{256}{\rm i} \epsilon_{\hat{a} \hat{b} e}\,^{c d} (\Gamma^{e})^{\alpha \beta} W W_{c d} \boldsymbol{\lambda}_{i \alpha} {G}^{-1} \varphi^{i}_{\beta} - \frac{9}{32}{\rm i} (\Sigma_{\hat{a} c})^{\alpha \beta} W W^{c}\,_{\hat{b}} \boldsymbol{\lambda}_{i \alpha} {G}^{-1} \varphi^{i}_{\beta}+\frac{9}{64}{\rm i} G_{i j} (\Sigma_{\hat{a} \hat{b}})^{\beta \alpha} W \boldsymbol{\lambda}^{i}_{\beta} X^{j}_{\alpha} {G}^{-1}+\frac{3}{16}{\rm i} F W \mathbf{F}_{\hat{a} \hat{b}} {G}^{-1} - \frac{3}{64}{\rm i} F G_{i j} (\Sigma_{\hat{a} \hat{b}})^{\alpha \beta} W \boldsymbol{\lambda}^{i}_{\alpha} {G}^{-3} \varphi^{j}_{\beta} - \frac{3}{8}G_{i j} (\Gamma_{\hat{a}})^{\alpha \beta} W \boldsymbol{W} {G}^{-3} \nabla_{\hat{b}}{\varphi^{i}_{\alpha}} \varphi^{j}_{\beta}%
+\frac{3}{8}{\rm i} W \boldsymbol{W} {G}^{-1} \nabla_{\hat{a}}{\mathcal{H}_{\hat{b}}} - \frac{3}{16}W_{\hat{a} \hat{b}}\,^{\alpha}\,_{i} W \boldsymbol{W} {G}^{-1} \varphi^{i}_{\alpha} - \frac{9}{16}(\Sigma_{\hat{a} \hat{b}})^{\alpha \beta} W \boldsymbol{W} X_{i \alpha} {G}^{-1} \varphi^{i}_{\beta}+\frac{3}{8}{\rm i} \Phi_{\hat{a} \hat{b} i j} G^{i j} W \boldsymbol{W} {G}^{-1} - \frac{3}{32}(\Sigma_{\hat{a} \hat{b}})^{\alpha \beta} \boldsymbol{W} \lambda_{i \alpha} {G}^{-3} \varphi^{i \rho} \varphi_{j \beta} \varphi^{j}_{\rho} - \frac{3}{64}{\rm i} G_{i j} (\Sigma_{\hat{a} \hat{b}})^{\alpha \beta} \lambda_{k \alpha} \boldsymbol{\lambda}^{k}_{\beta} {G}^{-3} \varphi^{i \rho} \varphi^{j}_{\rho}+\frac{3}{32}G_{i j} \boldsymbol{W} F_{\hat{a} \hat{b}} {G}^{-3} \varphi^{i \alpha} \varphi^{j}_{\alpha}+\frac{9}{32}G_{i j} W \boldsymbol{W} W_{\hat{a} \hat{b}} {G}^{-3} \varphi^{i \alpha} \varphi^{j}_{\alpha} - \frac{9}{64}G_{i j} G_{k l} (\Sigma_{\hat{a} \hat{b}})^{\alpha \beta} \boldsymbol{W} \lambda^{i}_{\alpha} {G}^{-5} \varphi^{j}_{\beta} \varphi^{k \rho} \varphi^{l}_{\rho} - \frac{3}{128}{\rm i} \epsilon^{c d e}\,_{\hat{a} \hat{b}} \mathcal{H}_{c} G_{i j} (\Sigma_{d e})^{\alpha \beta} \boldsymbol{W} \lambda^{i}_{\alpha} {G}^{-3} \varphi^{j}_{\beta} - \frac{3}{64}{\rm i} \mathcal{H}_{\hat{a}} G_{i j} (\Gamma_{\hat{b}})^{\alpha \beta} \boldsymbol{W} \lambda^{i}_{\alpha} {G}^{-3} \varphi^{j}_{\beta}+\frac{3}{64}{\rm i} \epsilon^{c d}\,_{\hat{a} \hat{b} e} G_{i j} (\Sigma_{c d})^{\alpha \beta} \boldsymbol{W} \lambda_{k \alpha} {G}^{-3} \nabla^{e}{G^{i k}} \varphi^{j}_{\beta} - \frac{3}{32}{\rm i} G_{i j} (\Gamma_{\hat{a}})^{\alpha \beta} \boldsymbol{W} \lambda_{k \alpha} {G}^{-3} \nabla_{\hat{b}}{G^{i k}} \varphi^{j}_{\beta} - \frac{3}{32}(\Sigma_{\hat{a} \hat{b}})^{\alpha \beta} W \boldsymbol{\lambda}_{i \alpha} {G}^{-3} \varphi^{i \rho} \varphi_{j \beta} \varphi^{j}_{\rho}+\frac{3}{32}G_{i j} W \mathbf{F}_{\hat{a} \hat{b}} {G}^{-3} \varphi^{i \alpha} \varphi^{j}_{\alpha} - \frac{9}{64}G_{i j} G_{k l} (\Sigma_{\hat{a} \hat{b}})^{\alpha \beta} W \boldsymbol{\lambda}^{i}_{\alpha} {G}^{-5} \varphi^{j}_{\beta} \varphi^{k \rho} \varphi^{l}_{\rho} - \frac{3}{128}{\rm i} \epsilon^{c d e}\,_{\hat{a} \hat{b}} \mathcal{H}_{c} G_{i j} (\Sigma_{d e})^{\alpha \beta} W \boldsymbol{\lambda}^{i}_{\alpha} {G}^{-3} \varphi^{j}_{\beta} - \frac{3}{64}{\rm i} \mathcal{H}_{\hat{a}} G_{i j} (\Gamma_{\hat{b}})^{\alpha \beta} W \boldsymbol{\lambda}^{i}_{\alpha} {G}^{-3} \varphi^{j}_{\beta}+\frac{3}{64}{\rm i} \epsilon^{c d}\,_{\hat{a} \hat{b} e} G_{i j} (\Sigma_{c d})^{\alpha \beta} W \boldsymbol{\lambda}_{k \alpha} {G}^{-3} \nabla^{e}{G^{i k}} \varphi^{j}_{\beta} - \frac{3}{32}{\rm i} G_{i j} (\Gamma_{\hat{a}})^{\alpha \beta} W \boldsymbol{\lambda}_{k \alpha} {G}^{-3} \nabla_{\hat{b}}{G^{i k}} \varphi^{j}_{\beta}%
 - \frac{9}{32}{\rm i} G_{i j} (\Sigma_{\hat{a} \hat{b}})^{\alpha \beta} W \boldsymbol{W} {G}^{-5} \varphi^{i}_{\alpha} \varphi^{j \rho} \varphi_{k \beta} \varphi^{k}_{\rho}+\frac{3}{32}(\Gamma_{\hat{a}})^{\alpha \beta} W \boldsymbol{W} {G}^{-3} \nabla_{\hat{b}}{G_{i j}} \varphi^{i}_{\alpha} \varphi^{j}_{\beta}+\frac{3}{16}{\rm i} \mathcal{H}_{\hat{a}} G_{i j} W \boldsymbol{W} {G}^{-3} \nabla_{\hat{b}}{G^{i j}}+\frac{9}{32}G_{i j} G_{k l} (\Gamma_{\hat{a}})^{\alpha \beta} W \boldsymbol{W} {G}^{-5} \nabla_{\hat{b}}{G^{i k}} \varphi^{j}_{\alpha} \varphi^{l}_{\beta}+\frac{3}{16}{\rm i} G_{i j} W \boldsymbol{W} {G}^{-3} \nabla_{\hat{a}}{G^{i}\,_{k}} \nabla_{\hat{b}}{G^{j k}}+\frac{9}{64}{\rm i} G_{i j} (\Sigma_{\hat{a} \hat{b}})^{\alpha \beta} W \boldsymbol{W} {G}^{-5} \varphi^{i \rho} \varphi^{j}_{\rho} \varphi_{k \alpha} \varphi^{k}_{\beta} - \frac{3}{16}G_{i j} (\Sigma_{\hat{a} \hat{b}})^{\alpha \beta} W \mathbf{X}^{i}\,_{k} {G}^{-3} \varphi^{j}_{\alpha} \varphi^{k}_{\beta}+\frac{3}{32}{\rm i} G_{i j} G_{k l} (\Sigma_{\hat{a} \hat{b}})^{\alpha \beta} \mathbf{X}^{i j} \lambda^{k}_{\alpha} {G}^{-3} \varphi^{l}_{\beta} - \frac{3}{16}{\rm i} \epsilon^{c d}\,_{\hat{a} \hat{b} e} (\Sigma_{c d})^{\alpha \beta} W {G}^{-1} \nabla^{e}{\boldsymbol{\lambda}_{i \alpha}} \varphi^{i}_{\beta} - \frac{3}{8}{\rm i} (\Gamma_{\hat{a}})^{\alpha \beta} W {G}^{-1} \nabla_{\hat{b}}{\boldsymbol{\lambda}_{i \alpha}} \varphi^{i}_{\beta} - \frac{9}{32}{\rm i} (\Sigma_{\hat{a} \hat{b}})^{\alpha \beta} \mathbf{X}_{i j} \lambda^{i}_{\alpha} {G}^{-1} \varphi^{j}_{\beta} - \frac{3}{16}G_{i j} (\Sigma_{\hat{a} \hat{b}})^{\alpha \beta} \boldsymbol{W} X^{i}\,_{k} {G}^{-3} \varphi^{j}_{\alpha} \varphi^{k}_{\beta}+\frac{3}{32}{\rm i} G_{i j} G_{k l} (\Sigma_{\hat{a} \hat{b}})^{\alpha \beta} X^{i j} \boldsymbol{\lambda}^{k}_{\alpha} {G}^{-3} \varphi^{l}_{\beta} - \frac{3}{16}{\rm i} \epsilon^{c d}\,_{\hat{a} \hat{b} e} (\Sigma_{c d})^{\alpha \beta} \boldsymbol{W} {G}^{-1} \nabla^{e}{\lambda_{i \alpha}} \varphi^{i}_{\beta} - \frac{3}{8}{\rm i} (\Gamma_{\hat{a}})^{\alpha \beta} \boldsymbol{W} {G}^{-1} \nabla_{\hat{b}}{\lambda_{i \alpha}} \varphi^{i}_{\beta} - \frac{9}{32}{\rm i} (\Sigma_{\hat{a} \hat{b}})^{\alpha \beta} X_{i j} \boldsymbol{\lambda}^{i}_{\alpha} {G}^{-1} \varphi^{j}_{\beta}+\frac{3}{32}{\rm i} G_{i j} (\Sigma_{\hat{a} \hat{b}})^{\alpha \beta} \lambda^{i \rho} \boldsymbol{\lambda}_{k \rho} {G}^{-3} \varphi^{j}_{\alpha} \varphi^{k}_{\beta}+\frac{3}{32}{\rm i} G_{i j} (\Sigma_{\hat{a} \hat{b}})^{\alpha \beta} \lambda^{\rho}_{k} \boldsymbol{\lambda}^{i}_{\rho} {G}^{-3} \varphi^{j}_{\alpha} \varphi^{k}_{\beta} - \frac{3}{32}{\rm i} F_{\hat{a} \hat{b}} \boldsymbol{\lambda}_{i}^{\alpha} {G}^{-1} \varphi^{i}_{\alpha}+\frac{3}{64}{\rm i} \epsilon_{\hat{a} \hat{b} e}\,^{c d} (\Gamma^{e})^{\alpha \beta} F_{c d} \boldsymbol{\lambda}_{i \alpha} {G}^{-1} \varphi^{i}_{\beta}%
 - \frac{3}{8}{\rm i} (\Sigma_{\hat{a} c})^{\alpha \beta} F^{c}\,_{\hat{b}} \boldsymbol{\lambda}_{i \alpha} {G}^{-1} \varphi^{i}_{\beta} - \frac{3}{32}{\rm i} G_{i j} G_{k l} (\Sigma_{\hat{a} \hat{b}})^{\alpha \beta} X^{i k} \boldsymbol{\lambda}^{j}_{\alpha} {G}^{-3} \varphi^{l}_{\beta} - \frac{3}{32}{\rm i} \epsilon^{c d}\,_{\hat{a} \hat{b} e} (\Sigma_{c d})^{\alpha \beta} \boldsymbol{\lambda}_{i \alpha} {G}^{-1} \nabla^{e}{W} \varphi^{i}_{\beta} - \frac{3}{16}{\rm i} (\Gamma_{\hat{a}})^{\alpha \beta} \boldsymbol{\lambda}_{i \alpha} {G}^{-1} \nabla_{\hat{b}}{W} \varphi^{i}_{\beta} - \frac{3}{32}{\rm i} \mathbf{F}_{\hat{a} \hat{b}} \lambda^{\alpha}_{i} {G}^{-1} \varphi^{i}_{\alpha}+\frac{3}{64}{\rm i} \epsilon_{\hat{a} \hat{b} e}\,^{c d} (\Gamma^{e})^{\alpha \beta} \mathbf{F}_{c d} \lambda_{i \alpha} {G}^{-1} \varphi^{i}_{\beta} - \frac{3}{8}{\rm i} (\Sigma_{\hat{a} c})^{\alpha \beta} \mathbf{F}^{c}\,_{\hat{b}} \lambda_{i \alpha} {G}^{-1} \varphi^{i}_{\beta} - \frac{3}{32}{\rm i} G_{i j} G_{k l} (\Sigma_{\hat{a} \hat{b}})^{\alpha \beta} \mathbf{X}^{i k} \lambda^{j}_{\alpha} {G}^{-3} \varphi^{l}_{\beta} - \frac{3}{32}{\rm i} \epsilon^{c d}\,_{\hat{a} \hat{b} e} (\Sigma_{c d})^{\alpha \beta} \lambda_{i \alpha} {G}^{-1} \nabla^{e}{\boldsymbol{W}} \varphi^{i}_{\beta} - \frac{3}{16}{\rm i} (\Gamma_{\hat{a}})^{\alpha \beta} \lambda_{i \alpha} {G}^{-1} \nabla_{\hat{b}}{\boldsymbol{W}} \varphi^{i}_{\beta}
\doublespacedmathend
\end{adjustwidth}

\subsubsection{$J^3_{a i \alpha, R^2}$} \label{J3R2Complete}

\begin{adjustwidth}{0cm}{5cm}
\doublespacedmathbegin
{}\frac{3}{16}(\Gamma_{a})^{\beta}{}_{\rho} \boldsymbol{\lambda}_{j \alpha} \lambda^{j}_{\beta} {G}^{-1} \nabla^{\rho \lambda}{\varphi_{i \lambda}}+\frac{9}{8}{\rm i} (\Gamma_{a})^{\beta}{}_{\rho} \boldsymbol{W} F_{\alpha \beta} {G}^{-1} \nabla^{\rho \lambda}{\varphi_{i \lambda}}+\frac{171}{64}{\rm i} (\Gamma_{a})^{\beta}{}_{\rho} W \boldsymbol{W} W_{\alpha \beta} {G}^{-1} \nabla^{\rho \lambda}{\varphi_{i \lambda}}+\frac{9}{64}{\rm i} G_{j k} (\Gamma_{a})^{\beta}{}_{\rho} \boldsymbol{W} \lambda^{j}_{\beta} {G}^{-3} \nabla^{\rho \lambda}{\varphi_{i \lambda}} \varphi^{k}_{\alpha} - \frac{9}{64}(\Gamma_{a})^{\beta}{}_{\rho} \boldsymbol{W} \lambda_{i \beta} {G}^{-1} \nabla^{\rho \lambda}{\mathcal{H}_{\alpha \lambda}} - \frac{3}{32}(\Gamma_{a})^{\beta}{}_{\rho} \boldsymbol{W} \lambda_{j \beta} {G}^{-1} \nabla^{\rho}\,_{\lambda}{\nabla_{\alpha}\,^{\lambda}{G_{i}\,^{j}}} - \frac{63}{64}F (\Gamma_{a})^{\beta \rho} \boldsymbol{W} W_{\alpha \beta} \lambda_{i \rho} {G}^{-1} - \frac{3}{16}{\rm i} (\Gamma_{a})^{\lambda \beta} \boldsymbol{W} \lambda_{j \lambda} W_{\alpha \beta}\,^{\rho j} {G}^{-1} \varphi_{i \rho}+\frac{99}{512}{\rm i} (\Gamma_{a})^{\rho \beta} \boldsymbol{W} \lambda_{j \rho} X^{j}_{\beta} {G}^{-1} \varphi_{i \alpha}+\frac{9}{32}{\rm i} (\Gamma_{a})^{\beta \rho} \boldsymbol{W} \lambda_{j \beta} X^{j}_{\alpha} {G}^{-1} \varphi_{i \rho}+\frac{135}{512}{\rm i} (\Gamma_{a})^{\beta \rho} \boldsymbol{W} \lambda_{i \beta} X_{j \alpha} {G}^{-1} \varphi^{j}_{\rho} - \frac{117}{512}{\rm i} (\Gamma_{a})^{\beta \rho} \boldsymbol{W} \lambda_{j \beta} X_{i \alpha} {G}^{-1} \varphi^{j}_{\rho}+\frac{189}{512}{\rm i} (\Gamma_{a})^{\rho \beta} \boldsymbol{W} \lambda_{i \rho} X_{j \beta} {G}^{-1} \varphi^{j}_{\alpha}+\frac{63}{2560}{\rm i} (\Gamma_{a})_{\alpha}{}^{\rho} \boldsymbol{W} \lambda_{j \rho} X_{i}^{\beta} {G}^{-1} \varphi^{j}_{\beta} - \frac{45}{128}{\rm i} (\Gamma_{a})^{\rho \beta} \boldsymbol{W} \lambda_{j \rho} X_{i \beta} {G}^{-1} \varphi^{j}_{\alpha} - \frac{51}{128}\Phi_{\alpha}\,^{\beta}\,_{j k} G_{i}\,^{j} (\Gamma_{a})_{\beta}{}^{\rho} \boldsymbol{W} \lambda^{k}_{\rho} {G}^{-1} - \frac{45}{512}G_{i j} (\Gamma_{a})^{\rho}{}_{\lambda} \boldsymbol{W} \lambda^{j}_{\rho} {G}^{-1} \nabla^{\lambda \beta}{W_{\alpha \beta}} - \frac{9}{512}G_{i j} (\Gamma_{a})^{\lambda \beta} \boldsymbol{W} \lambda^{j}_{\lambda} {G}^{-1} \nabla_{\alpha}\,^{\rho}{W_{\beta \rho}}+\frac{27}{128}G_{i j} (\Gamma_{a})^{\rho \lambda} \boldsymbol{W} W_{\alpha}\,^{\beta} W_{\rho \beta} \lambda^{j}_{\lambda} {G}^{-1}%
 - \frac{27}{256}G_{i j} (\Gamma_{a})_{\alpha}{}^{\lambda} \boldsymbol{W} W^{\beta \rho} W_{\beta \rho} \lambda^{j}_{\lambda} {G}^{-1} - \frac{9}{32}(\Gamma_{a})^{\beta \lambda} W_{\beta}\,^{\rho} \boldsymbol{\lambda}_{j \alpha} \lambda^{j}_{\lambda} {G}^{-1} \varphi_{i \rho} - \frac{27}{16}{\rm i} (\Gamma_{a})^{\rho \beta} \boldsymbol{W} W_{\rho}\,^{\lambda} F_{\alpha \beta} {G}^{-1} \varphi_{i \lambda}+\frac{189}{64}{\rm i} (\Gamma_{a})^{\beta \rho} W \boldsymbol{W} W_{\alpha \beta} W_{\rho}\,^{\lambda} {G}^{-1} \varphi_{i \lambda}+\frac{3}{32}{\rm i} G_{j k} (\Gamma_{a})^{\beta \lambda} \boldsymbol{W} W_{\beta}\,^{\rho} \lambda^{j}_{\lambda} {G}^{-3} \varphi^{k}_{\alpha} \varphi_{i \rho}+\frac{27}{256}G_{i j} (\Gamma_{a})^{\rho \beta} \boldsymbol{\lambda}_{k \alpha} \lambda^{k}_{\rho} X^{j}_{\beta} {G}^{-1}+\frac{27}{32}{\rm i} G_{i j} (\Gamma_{a})^{\beta \rho} \boldsymbol{W} F_{\alpha \beta} X^{j}_{\rho} {G}^{-1}+\frac{243}{128}{\rm i} G_{i j} (\Gamma_{a})^{\beta \rho} W \boldsymbol{W} W_{\alpha \beta} X^{j}_{\rho} {G}^{-1} - \frac{3}{32}{\rm i} G_{i j} (\Gamma_{a})^{\beta}{}_{\rho} \boldsymbol{W} X^{j}_{\beta} {G}^{-1} \nabla^{\rho}\,_{\alpha}{W}+\frac{45}{256}{\rm i} G_{i j} G_{k l} (\Gamma_{a})^{\rho \beta} \boldsymbol{W} \lambda^{k}_{\rho} X^{j}_{\beta} {G}^{-3} \varphi^{l}_{\alpha}+\frac{3}{16}(\Gamma_{a})^{\beta}{}_{\rho} \boldsymbol{\lambda}_{j \alpha} \lambda_{i \beta} {G}^{-1} \nabla^{\rho \lambda}{\varphi^{j}_{\lambda}}+\frac{39}{160}{\rm i} (\Gamma_{a})_{\alpha \beta} \boldsymbol{W} X_{i j} {G}^{-1} \nabla^{\beta \rho}{\varphi^{j}_{\rho}}+\frac{3}{16}{\rm i} G_{j k} (\Gamma_{a})^{\beta}{}_{\rho} \boldsymbol{W} \lambda_{i \beta} {G}^{-3} \nabla^{\rho \lambda}{\varphi^{j}_{\lambda}} \varphi^{k}_{\alpha}+\frac{3}{8}{\rm i} (\Gamma_{a})^{\lambda \beta} \boldsymbol{W} \lambda_{i \lambda} W_{\alpha \beta}\,^{\rho}\,_{j} {G}^{-1} \varphi^{j}_{\rho}+\frac{99}{128}\Phi_{\alpha}\,^{\beta}\,_{j k} G^{j k} (\Gamma_{a})_{\beta}{}^{\rho} \boldsymbol{W} \lambda_{i \rho} {G}^{-1} - \frac{9}{32}(\Gamma_{a})^{\beta \lambda} W_{\beta}\,^{\rho} \boldsymbol{\lambda}_{j \alpha} \lambda_{i \lambda} {G}^{-1} \varphi^{j}_{\rho} - \frac{3}{32}{\rm i} (\Gamma_{a})_{\alpha}{}^{\beta} \boldsymbol{W} X_{i j} W_{\beta}\,^{\rho} {G}^{-1} \varphi^{j}_{\rho} - \frac{9}{128}{\rm i} G_{j k} (\Gamma_{a})^{\beta \lambda} \boldsymbol{W} W_{\beta}\,^{\rho} \lambda_{i \lambda} {G}^{-3} \varphi^{j}_{\alpha} \varphi^{k}_{\rho} - \frac{51}{256}G_{j k} (\Gamma_{a})^{\rho \beta} \boldsymbol{\lambda}^{j}_{\alpha} \lambda_{i \rho} X^{k}_{\beta} {G}^{-1}+\frac{3}{16}(\Gamma_{a})^{\beta \rho} \lambda_{j \beta} \boldsymbol{\lambda}_{i \rho} {G}^{-1} \nabla_{\alpha}\,^{\lambda}{\varphi^{j}_{\lambda}}%
 - \frac{9}{32}(\Gamma_{a})^{\rho \lambda} W_{\alpha}\,^{\beta} \lambda_{j \rho} \boldsymbol{\lambda}_{i \lambda} {G}^{-1} \varphi^{j}_{\beta}+\frac{9}{64}G_{j k} (\Gamma_{a})^{\beta \rho} \lambda^{j}_{\beta} \boldsymbol{\lambda}_{i \rho} X^{k}_{\alpha} {G}^{-1} - \frac{9}{16}F (\Gamma_{a})^{\beta \rho} F_{\alpha \beta} \boldsymbol{\lambda}_{i \rho} {G}^{-1} - \frac{63}{64}F (\Gamma_{a})^{\beta \rho} W W_{\alpha \beta} \boldsymbol{\lambda}_{i \rho} {G}^{-1} - \frac{9}{16}F (\Gamma_{a})^{\beta \rho} \mathbf{F}_{\alpha \beta} \lambda_{i \rho} {G}^{-1} - \frac{3}{64}F G_{j k} (\Gamma_{a})^{\beta \rho} \lambda^{j}_{\beta} \boldsymbol{\lambda}_{i \rho} {G}^{-3} \varphi^{k}_{\alpha}+\frac{3}{16}(\Gamma_{a})^{\beta \rho} \lambda_{i \beta} \boldsymbol{\lambda}_{j \rho} {G}^{-1} \nabla_{\alpha}\,^{\lambda}{\varphi^{j}_{\lambda}} - \frac{9}{32}(\Gamma_{a})^{\rho \lambda} W_{\alpha}\,^{\beta} \lambda_{i \rho} \boldsymbol{\lambda}_{j \lambda} {G}^{-1} \varphi^{j}_{\beta}+\frac{9}{64}G_{j k} (\Gamma_{a})^{\beta \rho} \lambda_{i \beta} \boldsymbol{\lambda}^{j}_{\rho} X^{k}_{\alpha} {G}^{-1} - \frac{3}{64}F G_{j k} (\Gamma_{a})^{\beta \rho} \lambda_{i \beta} \boldsymbol{\lambda}^{j}_{\rho} {G}^{-3} \varphi^{k}_{\alpha} - \frac{9}{64}{\rm i} G_{i j} (\Gamma_{a})^{\beta \rho} \boldsymbol{W} \lambda_{k \beta} {G}^{-3} \nabla_{\alpha}\,^{\lambda}{\varphi^{k}_{\lambda}} \varphi^{j}_{\rho} - \frac{3}{32}{\rm i} G_{i j} (\Gamma_{a})^{\rho \lambda} \boldsymbol{W} W_{\alpha}\,^{\beta} \lambda_{k \rho} {G}^{-3} \varphi^{j}_{\lambda} \varphi^{k}_{\beta} - \frac{3}{64}F G_{i j} (\Gamma_{a})^{\beta \rho} \boldsymbol{\lambda}_{k \alpha} \lambda^{k}_{\beta} {G}^{-3} \varphi^{j}_{\rho} - \frac{9}{32}{\rm i} F G_{i j} (\Gamma_{a})^{\beta \rho} \boldsymbol{W} F_{\alpha \beta} {G}^{-3} \varphi^{j}_{\rho} - \frac{63}{128}{\rm i} F G_{i j} (\Gamma_{a})^{\beta \rho} W \boldsymbol{W} W_{\alpha \beta} {G}^{-3} \varphi^{j}_{\rho} - \frac{3}{16}{\rm i} G_{j k} (\Gamma_{a})^{\beta \rho} \boldsymbol{W} \lambda_{i \beta} {G}^{-3} \nabla_{\alpha}\,^{\lambda}{\varphi^{j}_{\lambda}} \varphi^{k}_{\rho}+\frac{3}{64}F G_{j k} (\Gamma_{a})^{\beta \rho} \boldsymbol{\lambda}^{j}_{\alpha} \lambda_{i \beta} {G}^{-3} \varphi^{k}_{\rho}+\frac{3}{16}(\Gamma_{a})^{\beta}{}_{\rho} \lambda_{j \alpha} \boldsymbol{\lambda}^{j}_{\beta} {G}^{-1} \nabla^{\rho \lambda}{\varphi_{i \lambda}}+\frac{9}{8}{\rm i} (\Gamma_{a})^{\beta}{}_{\rho} W \mathbf{F}_{\alpha \beta} {G}^{-1} \nabla^{\rho \lambda}{\varphi_{i \lambda}}+\frac{9}{64}{\rm i} G_{j k} (\Gamma_{a})^{\beta}{}_{\rho} W \boldsymbol{\lambda}^{j}_{\beta} {G}^{-3} \nabla^{\rho \lambda}{\varphi_{i \lambda}} \varphi^{k}_{\alpha}%
 - \frac{9}{64}(\Gamma_{a})^{\beta}{}_{\rho} W \boldsymbol{\lambda}_{i \beta} {G}^{-1} \nabla^{\rho \lambda}{\mathcal{H}_{\alpha \lambda}} - \frac{3}{32}(\Gamma_{a})^{\beta}{}_{\rho} W \boldsymbol{\lambda}_{j \beta} {G}^{-1} \nabla^{\rho}\,_{\lambda}{\nabla_{\alpha}\,^{\lambda}{G_{i}\,^{j}}} - \frac{3}{16}{\rm i} (\Gamma_{a})^{\lambda \beta} W \boldsymbol{\lambda}_{j \lambda} W_{\alpha \beta}\,^{\rho j} {G}^{-1} \varphi_{i \rho}+\frac{99}{512}{\rm i} (\Gamma_{a})^{\rho \beta} W \boldsymbol{\lambda}_{j \rho} X^{j}_{\beta} {G}^{-1} \varphi_{i \alpha}+\frac{9}{32}{\rm i} (\Gamma_{a})^{\beta \rho} W \boldsymbol{\lambda}_{j \beta} X^{j}_{\alpha} {G}^{-1} \varphi_{i \rho}+\frac{135}{512}{\rm i} (\Gamma_{a})^{\beta \rho} W \boldsymbol{\lambda}_{i \beta} X_{j \alpha} {G}^{-1} \varphi^{j}_{\rho} - \frac{117}{512}{\rm i} (\Gamma_{a})^{\beta \rho} W \boldsymbol{\lambda}_{j \beta} X_{i \alpha} {G}^{-1} \varphi^{j}_{\rho}+\frac{189}{512}{\rm i} (\Gamma_{a})^{\rho \beta} W \boldsymbol{\lambda}_{i \rho} X_{j \beta} {G}^{-1} \varphi^{j}_{\alpha}+\frac{63}{2560}{\rm i} (\Gamma_{a})_{\alpha}{}^{\rho} W \boldsymbol{\lambda}_{j \rho} X_{i}^{\beta} {G}^{-1} \varphi^{j}_{\beta} - \frac{45}{128}{\rm i} (\Gamma_{a})^{\rho \beta} W \boldsymbol{\lambda}_{j \rho} X_{i \beta} {G}^{-1} \varphi^{j}_{\alpha} - \frac{51}{128}\Phi_{\alpha}\,^{\beta}\,_{j k} G_{i}\,^{j} (\Gamma_{a})_{\beta}{}^{\rho} W \boldsymbol{\lambda}^{k}_{\rho} {G}^{-1} - \frac{45}{512}G_{i j} (\Gamma_{a})^{\rho}{}_{\lambda} W \boldsymbol{\lambda}^{j}_{\rho} {G}^{-1} \nabla^{\lambda \beta}{W_{\alpha \beta}} - \frac{9}{512}G_{i j} (\Gamma_{a})^{\lambda \beta} W \boldsymbol{\lambda}^{j}_{\lambda} {G}^{-1} \nabla_{\alpha}\,^{\rho}{W_{\beta \rho}}+\frac{27}{128}G_{i j} (\Gamma_{a})^{\rho \lambda} W W_{\alpha}\,^{\beta} W_{\rho \beta} \boldsymbol{\lambda}^{j}_{\lambda} {G}^{-1} - \frac{27}{256}G_{i j} (\Gamma_{a})_{\alpha}{}^{\lambda} W W^{\beta \rho} W_{\beta \rho} \boldsymbol{\lambda}^{j}_{\lambda} {G}^{-1} - \frac{9}{32}(\Gamma_{a})^{\beta \lambda} W_{\beta}\,^{\rho} \lambda_{j \alpha} \boldsymbol{\lambda}^{j}_{\lambda} {G}^{-1} \varphi_{i \rho} - \frac{27}{16}{\rm i} (\Gamma_{a})^{\rho \beta} W W_{\rho}\,^{\lambda} \mathbf{F}_{\alpha \beta} {G}^{-1} \varphi_{i \lambda}+\frac{3}{32}{\rm i} G_{j k} (\Gamma_{a})^{\beta \lambda} W W_{\beta}\,^{\rho} \boldsymbol{\lambda}^{j}_{\lambda} {G}^{-3} \varphi^{k}_{\alpha} \varphi_{i \rho}+\frac{27}{256}G_{i j} (\Gamma_{a})^{\rho \beta} \lambda_{k \alpha} \boldsymbol{\lambda}^{k}_{\rho} X^{j}_{\beta} {G}^{-1}+\frac{27}{32}{\rm i} G_{i j} (\Gamma_{a})^{\beta \rho} W \mathbf{F}_{\alpha \beta} X^{j}_{\rho} {G}^{-1}%
 - \frac{3}{32}{\rm i} G_{i j} (\Gamma_{a})^{\beta}{}_{\rho} W X^{j}_{\beta} {G}^{-1} \nabla^{\rho}\,_{\alpha}{\boldsymbol{W}}+\frac{45}{256}{\rm i} G_{i j} G_{k l} (\Gamma_{a})^{\rho \beta} W \boldsymbol{\lambda}^{k}_{\rho} X^{j}_{\beta} {G}^{-3} \varphi^{l}_{\alpha}+\frac{3}{16}(\Gamma_{a})^{\beta}{}_{\rho} \lambda_{j \alpha} \boldsymbol{\lambda}_{i \beta} {G}^{-1} \nabla^{\rho \lambda}{\varphi^{j}_{\lambda}}+\frac{39}{160}{\rm i} (\Gamma_{a})_{\alpha \beta} W \mathbf{X}_{i j} {G}^{-1} \nabla^{\beta \rho}{\varphi^{j}_{\rho}}+\frac{3}{16}{\rm i} G_{j k} (\Gamma_{a})^{\beta}{}_{\rho} W \boldsymbol{\lambda}_{i \beta} {G}^{-3} \nabla^{\rho \lambda}{\varphi^{j}_{\lambda}} \varphi^{k}_{\alpha}+\frac{3}{8}{\rm i} (\Gamma_{a})^{\lambda \beta} W \boldsymbol{\lambda}_{i \lambda} W_{\alpha \beta}\,^{\rho}\,_{j} {G}^{-1} \varphi^{j}_{\rho}+\frac{99}{128}\Phi_{\alpha}\,^{\beta}\,_{j k} G^{j k} (\Gamma_{a})_{\beta}{}^{\rho} W \boldsymbol{\lambda}_{i \rho} {G}^{-1} - \frac{9}{32}(\Gamma_{a})^{\beta \lambda} W_{\beta}\,^{\rho} \lambda_{j \alpha} \boldsymbol{\lambda}_{i \lambda} {G}^{-1} \varphi^{j}_{\rho} - \frac{3}{32}{\rm i} (\Gamma_{a})_{\alpha}{}^{\beta} W \mathbf{X}_{i j} W_{\beta}\,^{\rho} {G}^{-1} \varphi^{j}_{\rho} - \frac{9}{128}{\rm i} G_{j k} (\Gamma_{a})^{\beta \lambda} W W_{\beta}\,^{\rho} \boldsymbol{\lambda}_{i \lambda} {G}^{-3} \varphi^{j}_{\alpha} \varphi^{k}_{\rho} - \frac{51}{256}G_{j k} (\Gamma_{a})^{\rho \beta} \lambda^{j}_{\alpha} \boldsymbol{\lambda}_{i \rho} X^{k}_{\beta} {G}^{-1} - \frac{9}{64}{\rm i} G_{i j} (\Gamma_{a})^{\beta \rho} W \boldsymbol{\lambda}_{k \beta} {G}^{-3} \nabla_{\alpha}\,^{\lambda}{\varphi^{k}_{\lambda}} \varphi^{j}_{\rho} - \frac{3}{32}{\rm i} G_{i j} (\Gamma_{a})^{\rho \lambda} W W_{\alpha}\,^{\beta} \boldsymbol{\lambda}_{k \rho} {G}^{-3} \varphi^{j}_{\lambda} \varphi^{k}_{\beta} - \frac{3}{64}F G_{i j} (\Gamma_{a})^{\beta \rho} \lambda_{k \alpha} \boldsymbol{\lambda}^{k}_{\beta} {G}^{-3} \varphi^{j}_{\rho} - \frac{9}{32}{\rm i} F G_{i j} (\Gamma_{a})^{\beta \rho} W \mathbf{F}_{\alpha \beta} {G}^{-3} \varphi^{j}_{\rho} - \frac{3}{16}{\rm i} G_{j k} (\Gamma_{a})^{\beta \rho} W \boldsymbol{\lambda}_{i \beta} {G}^{-3} \nabla_{\alpha}\,^{\lambda}{\varphi^{j}_{\lambda}} \varphi^{k}_{\rho}+\frac{3}{64}F G_{j k} (\Gamma_{a})^{\beta \rho} \lambda^{j}_{\alpha} \boldsymbol{\lambda}_{i \beta} {G}^{-3} \varphi^{k}_{\rho}+\frac{3}{64}(\Gamma_{a})^{\beta}{}_{\rho} W \boldsymbol{W} {G}^{-3} \nabla^{\rho \lambda}{\varphi_{i \lambda}} \varphi_{j \alpha} \varphi^{j}_{\beta}+\frac{3}{32}(\Gamma_{a})^{\beta}{}_{\rho} W \boldsymbol{W} {G}^{-3} \nabla^{\rho \lambda}{\varphi_{j \lambda}} \varphi_{i \alpha} \varphi^{j}_{\beta} - \frac{3}{32}{\rm i} G_{i j} (\Gamma_{a})^{\beta}{}_{\rho} \boldsymbol{W} \lambda_{k \alpha} {G}^{-3} \nabla^{\rho \lambda}{\varphi^{k}_{\lambda}} \varphi^{j}_{\beta}%
 - \frac{3}{32}{\rm i} G_{i j} (\Gamma_{a})^{\beta}{}_{\rho} W \boldsymbol{\lambda}_{k \alpha} {G}^{-3} \nabla^{\rho \lambda}{\varphi^{k}_{\lambda}} \varphi^{j}_{\beta} - \frac{9}{32}G_{i j} G_{k l} (\Gamma_{a})^{\beta}{}_{\rho} W \boldsymbol{W} {G}^{-5} \nabla^{\rho \lambda}{\varphi^{k}_{\lambda}} \varphi^{l}_{\alpha} \varphi^{j}_{\beta} - \frac{9}{128}{\rm i} G_{i j} (\Gamma_{a})^{\beta}{}_{\rho} W \boldsymbol{W} {G}^{-3} \nabla^{\rho}\,_{\alpha}{F} \varphi^{j}_{\beta} - \frac{9}{128}{\rm i} G_{i j} (\Gamma_{a})^{\beta}{}_{\rho} W \boldsymbol{W} {G}^{-3} \nabla^{\rho \lambda}{\mathcal{H}_{\alpha \lambda}} \varphi^{j}_{\beta} - \frac{63}{256}{\rm i} \mathcal{H}^{\rho \beta} G_{i j} (\Gamma_{a})_{\rho}{}^{\lambda} W \boldsymbol{W} W_{\alpha \beta} {G}^{-3} \varphi^{j}_{\lambda}+\frac{27}{256}{\rm i} \mathcal{H}_{\alpha}\,^{\beta} G_{i j} (\Gamma_{a})^{\rho \lambda} W \boldsymbol{W} W_{\beta \rho} {G}^{-3} \varphi^{j}_{\lambda}+\frac{633}{1024}G_{i j} (\Gamma_{a})^{\beta \rho} W \boldsymbol{W} X_{k \beta} {G}^{-3} \varphi^{k}_{\alpha} \varphi^{j}_{\rho}+\frac{69}{512}G_{i j} (\Gamma_{a})^{\beta \rho} W \boldsymbol{W} X_{k \alpha} {G}^{-3} \varphi^{j}_{\beta} \varphi^{k}_{\rho} - \frac{927}{5120}G_{i j} (\Gamma_{a})_{\alpha}{}^{\rho} W \boldsymbol{W} X_{k}^{\beta} {G}^{-3} \varphi^{j}_{\rho} \varphi^{k}_{\beta}+\frac{21}{320}{\rm i} F G_{i j} (\Gamma_{a})_{\alpha \beta} W \boldsymbol{W} {G}^{-3} \nabla^{\beta \rho}{\varphi^{j}_{\rho}} - \frac{9}{128}{\rm i} \mathcal{H}_{\alpha}\,^{\beta} G_{i j} (\Gamma_{a})_{\beta \rho} W \boldsymbol{W} {G}^{-3} \nabla^{\rho \lambda}{\varphi^{j}_{\lambda}}+\frac{3}{64}{\rm i} G_{i j} (\Gamma_{a})_{\beta \rho} W \boldsymbol{W} {G}^{-3} \nabla^{\beta \lambda}{\varphi_{k \lambda}} \nabla^{\rho}\,_{\alpha}{G^{j k}}+\frac{3}{32}{\rm i} G_{j k} (\Gamma_{a})^{\beta}{}_{\rho} \boldsymbol{W} \lambda^{j}_{\alpha} {G}^{-3} \nabla^{\rho \lambda}{\varphi_{i \lambda}} \varphi^{k}_{\beta}+\frac{3}{32}{\rm i} G_{j k} (\Gamma_{a})^{\beta}{}_{\rho} W \boldsymbol{\lambda}^{j}_{\alpha} {G}^{-3} \nabla^{\rho \lambda}{\varphi_{i \lambda}} \varphi^{k}_{\beta} - \frac{3}{64}{\rm i} G_{j k} (\Gamma_{a})^{\beta}{}_{\rho} W \boldsymbol{W} {G}^{-3} \nabla^{\rho}\,_{\lambda}{\nabla_{\alpha}\,^{\lambda}{G_{i}\,^{j}}} \varphi^{k}_{\beta}+\frac{27}{64}{\rm i} G_{j k} (\Gamma_{a})^{\rho}{}_{\lambda} W \boldsymbol{W} W_{\alpha \beta} {G}^{-3} \nabla^{\lambda \beta}{G_{i}\,^{j}} \varphi^{k}_{\rho}+\frac{3}{128}{\rm i} G_{j k} (\Gamma_{a})^{\beta \lambda} W \boldsymbol{W} W_{\beta \rho} {G}^{-3} \nabla_{\alpha}\,^{\rho}{G_{i}\,^{j}} \varphi^{k}_{\lambda} - \frac{783}{1024}G_{j k} (\Gamma_{a})^{\beta \rho} W \boldsymbol{W} X^{j}_{\beta} {G}^{-3} \varphi_{i \alpha} \varphi^{k}_{\rho} - \frac{159}{512}G_{j k} (\Gamma_{a})^{\beta \rho} W \boldsymbol{W} X^{j}_{\alpha} {G}^{-3} \varphi_{i \beta} \varphi^{k}_{\rho} - \frac{393}{5120}G_{j k} (\Gamma_{a})_{\alpha}{}^{\rho} W \boldsymbol{W} X^{j \beta} {G}^{-3} \varphi_{i \beta} \varphi^{k}_{\rho}%
+\frac{99}{512}G_{j k} (\Gamma_{a})^{\beta \rho} W \boldsymbol{W} X_{i \alpha} {G}^{-3} \varphi^{j}_{\beta} \varphi^{k}_{\rho} - \frac{333}{2560}G_{j k} (\Gamma_{a})_{\alpha}{}^{\rho} W \boldsymbol{W} X_{i}^{\beta} {G}^{-3} \varphi^{j}_{\rho} \varphi^{k}_{\beta}+\frac{243}{512}G_{j k} (\Gamma_{a})^{\beta \rho} W \boldsymbol{W} X_{i \beta} {G}^{-3} \varphi^{j}_{\alpha} \varphi^{k}_{\rho}+\frac{549}{512}{\rm i} (\Gamma_{a})^{\rho}{}_{\lambda} W \boldsymbol{W} {G}^{-1} \nabla^{\lambda \beta}{W_{\alpha \beta}} \varphi_{i \rho} - \frac{621}{512}{\rm i} (\Gamma_{a})^{\beta \lambda} W \boldsymbol{W} {G}^{-1} \nabla_{\alpha}\,^{\rho}{W_{\beta \rho}} \varphi_{i \lambda}+\frac{27}{128}{\rm i} (\Gamma_{a})^{\rho \lambda} W \boldsymbol{W} W_{\alpha}\,^{\beta} W_{\rho \beta} {G}^{-1} \varphi_{i \lambda} - \frac{27}{256}{\rm i} (\Gamma_{a})_{\alpha}{}^{\lambda} W \boldsymbol{W} W^{\beta \rho} W_{\beta \rho} {G}^{-1} \varphi_{i \lambda} - \frac{3}{64}{\rm i} G_{j k} (\Gamma_{a})_{\beta \rho} W \boldsymbol{W} {G}^{-3} \nabla^{\beta \lambda}{\varphi_{i \lambda}} \nabla^{\rho}\,_{\alpha}{G^{j k}} - \frac{3}{16}(\Gamma_{a})_{\beta \rho} \boldsymbol{W} \lambda_{j \alpha} {G}^{-1} \nabla^{\beta}\,_{\lambda}{\nabla^{\rho \lambda}{G_{i}\,^{j}}} - \frac{3}{16}(\Gamma_{a})_{\beta \rho} W \boldsymbol{\lambda}_{j \alpha} {G}^{-1} \nabla^{\beta}\,_{\lambda}{\nabla^{\rho \lambda}{G_{i}\,^{j}}} - \frac{9}{64}{\rm i} G_{j k} (\Gamma_{a})_{\beta \rho} W \boldsymbol{W} {G}^{-3} \nabla^{\beta}\,_{\lambda}{\nabla^{\rho \lambda}{G_{i}\,^{j}}} \varphi^{k}_{\alpha}+\frac{9}{32}{\rm i} (\Gamma_{a})_{\beta \rho} W \boldsymbol{W} {G}^{-1} \nabla^{\beta}\,_{\lambda}{\nabla^{\rho \lambda}{\varphi_{i \alpha}}} - \frac{9}{64}{\rm i} (\Gamma_{a})_{\rho \lambda} W \boldsymbol{W} W_{\alpha}\,^{\beta} {G}^{-1} \nabla^{\rho \lambda}{\varphi_{i \beta}}+\frac{63}{2560}{\rm i} (\Gamma_{a})_{\alpha \lambda} W \boldsymbol{W} {G}^{-1} \nabla^{\lambda \rho}{W^{\beta}\,_{\rho}} \varphi_{i \beta}+\frac{9}{32}{\rm i} (\Gamma_{a})^{\beta}{}_{\lambda} W \boldsymbol{W} {G}^{-1} \nabla^{\lambda}\,_{\alpha}{W_{\beta}\,^{\rho}} \varphi_{i \rho}+\frac{27}{160}{\rm i} (\Gamma_{a})_{\alpha \lambda} W \boldsymbol{W} W^{\beta}\,_{\rho} {G}^{-1} \nabla^{\lambda \rho}{\varphi_{i \beta}} - \frac{45}{64}{\rm i} (\Gamma_{a})^{\rho}{}_{\lambda} W \boldsymbol{W} W_{\alpha \beta} {G}^{-1} \nabla^{\lambda \beta}{\varphi_{i \rho}} - \frac{63}{128}{\rm i} G_{i j} (\Gamma_{a})_{\beta \rho} W \boldsymbol{W} {G}^{-1} \nabla^{\beta \rho}{X^{j}_{\alpha}} - \frac{147}{512}{\rm i} (\Gamma_{a})_{\beta \rho} W \boldsymbol{W} X_{j \alpha} {G}^{-1} \nabla^{\beta \rho}{G_{i}\,^{j}} - \frac{9}{128}{\rm i} G_{i j} (\Gamma_{a})^{\beta}{}_{\rho} W \boldsymbol{W} {G}^{-1} \nabla^{\rho}\,_{\alpha}{X^{j}_{\beta}}%
 - \frac{93}{2560}{\rm i} (\Gamma_{a})_{\alpha \rho} W \boldsymbol{W} X_{j \beta} {G}^{-1} \nabla^{\rho \beta}{G_{i}\,^{j}} - \frac{165}{512}{\rm i} (\Gamma_{a})^{\beta}{}_{\rho} W \boldsymbol{W} X_{j \beta} {G}^{-1} \nabla^{\rho}\,_{\alpha}{G_{i}\,^{j}} - \frac{351}{2560}{\rm i} (\Gamma_{a})_{\alpha}{}^{\beta} W \boldsymbol{W} {G}^{-1} \nabla^{\rho \lambda}{W_{\beta \rho}} \varphi_{i \lambda}+\frac{1107}{512}{\rm i} (\Gamma_{a})^{\beta}{}_{\lambda} W \boldsymbol{W} {G}^{-1} \nabla^{\lambda \rho}{W_{\beta \rho}} \varphi_{i \alpha} - \frac{3}{8}{\rm i} G_{i j} (\Gamma_{a})^{\beta}{}_{\lambda} W \boldsymbol{W} {G}^{-1} \nabla^{\lambda \rho}{W_{\alpha \beta \rho}\,^{j}}+\frac{9}{16}{\rm i} G_{i j} (\Gamma_{a})^{\beta \lambda} W \boldsymbol{W} W_{\beta}\,^{\rho} W_{\alpha \lambda \rho}\,^{j} {G}^{-1}+\frac{63}{512}{\rm i} (\Gamma_{a})^{\beta \rho} \boldsymbol{W} \lambda_{j \alpha} X_{i \beta} {G}^{-1} \varphi^{j}_{\rho}+\frac{63}{512}{\rm i} (\Gamma_{a})^{\beta \rho} W \boldsymbol{\lambda}_{j \alpha} X_{i \beta} {G}^{-1} \varphi^{j}_{\rho} - \frac{111}{256}{\rm i} \Phi_{\alpha}\,^{\beta}\,_{i j} (\Gamma_{a})_{\beta}{}^{\rho} W \boldsymbol{W} {G}^{-1} \varphi^{j}_{\rho} - \frac{27}{64}{\rm i} \mathcal{H}_{\alpha}\,^{\rho} (\Gamma_{a})_{\rho}{}^{\beta} W \boldsymbol{W} X_{i \beta} {G}^{-1}+\frac{45}{512}{\rm i} (\Gamma_{a})^{\beta \rho} \boldsymbol{W} \lambda_{j \alpha} X^{j}_{\beta} {G}^{-1} \varphi_{i \rho}+\frac{45}{512}{\rm i} (\Gamma_{a})^{\beta \rho} W \boldsymbol{\lambda}_{j \alpha} X^{j}_{\beta} {G}^{-1} \varphi_{i \rho}+\frac{537}{1024}G_{j k} (\Gamma_{a})^{\beta \rho} W \boldsymbol{W} X^{j}_{\beta} {G}^{-3} \varphi^{k}_{\alpha} \varphi_{i \rho}+\frac{9}{512}G_{i j} (\Gamma_{a})^{\beta}{}_{\lambda} \boldsymbol{W} \lambda^{j}_{\alpha} {G}^{-1} \nabla^{\lambda \rho}{W_{\beta \rho}}+\frac{9}{512}G_{i j} (\Gamma_{a})^{\beta}{}_{\lambda} W \boldsymbol{\lambda}^{j}_{\alpha} {G}^{-1} \nabla^{\lambda \rho}{W_{\beta \rho}}+\frac{9}{80}{\rm i} G_{i j} (\Gamma_{a})_{\alpha}{}^{\lambda} W \boldsymbol{W} W^{\beta \rho} W_{\lambda \beta \rho}\,^{j} {G}^{-1}+\frac{243}{640}{\rm i} G_{i j} (\Gamma_{a})_{\alpha}{}^{\beta} W \boldsymbol{W} W_{\beta}\,^{\rho} X^{j}_{\rho} {G}^{-1}+\frac{9}{64}G_{i j} (\Gamma_{a})^{\beta \rho} W \boldsymbol{W} X^{j}_{\beta} {G}^{-3} \varphi_{k \alpha} \varphi^{k}_{\rho} - \frac{9}{512}{\rm i} G_{i j} G_{k l} (\Gamma_{a})^{\beta \rho} \boldsymbol{W} \lambda^{k}_{\alpha} X^{l}_{\beta} {G}^{-3} \varphi^{j}_{\rho} - \frac{9}{512}{\rm i} G_{i j} G_{k l} (\Gamma_{a})^{\beta \rho} W \boldsymbol{\lambda}^{k}_{\alpha} X^{l}_{\beta} {G}^{-3} \varphi^{j}_{\rho}%
+\frac{63}{256}{\rm i} \Phi_{\alpha}\,^{\beta}\,_{k l} G_{i j} G^{k l} (\Gamma_{a})_{\beta}{}^{\rho} W \boldsymbol{W} {G}^{-3} \varphi^{j}_{\rho}+\frac{45}{512}{\rm i} G_{i j} G_{k l} (\Gamma_{a})^{\beta}{}_{\rho} W \boldsymbol{W} X^{k}_{\beta} {G}^{-3} \nabla^{\rho}\,_{\alpha}{G^{j l}} - \frac{99}{512}{\rm i} G_{i j} G_{k l} (\Gamma_{a})^{\beta \rho} \boldsymbol{W} \lambda^{k}_{\alpha} X^{j}_{\beta} {G}^{-3} \varphi^{l}_{\rho} - \frac{99}{512}{\rm i} G_{i j} G_{k l} (\Gamma_{a})^{\beta \rho} W \boldsymbol{\lambda}^{k}_{\alpha} X^{j}_{\beta} {G}^{-3} \varphi^{l}_{\rho} - \frac{15}{256}{\rm i} \Phi_{\alpha}\,^{\beta}\,_{j k} G_{i}\,^{j} G^{k}\,_{l} (\Gamma_{a})_{\beta}{}^{\rho} W \boldsymbol{W} {G}^{-3} \varphi^{l}_{\rho} - \frac{9}{512}{\rm i} G_{i j} G_{k l} (\Gamma_{a})^{\beta}{}_{\rho} W \boldsymbol{W} X^{j}_{\beta} {G}^{-3} \nabla^{\rho}\,_{\alpha}{G^{k l}} - \frac{3}{32}{\rm i} (\Gamma_{a})^{\beta \rho} \boldsymbol{\lambda}_{j \alpha} \lambda^{j}_{\beta} {G}^{-3} \varphi_{i}^{\lambda} \varphi_{k \rho} \varphi^{k}_{\lambda}+\frac{9}{16}(\Gamma_{a})^{\beta \rho} \boldsymbol{W} F_{\alpha \beta} {G}^{-3} \varphi_{i}^{\lambda} \varphi_{j \rho} \varphi^{j}_{\lambda}+\frac{27}{16}(\Gamma_{a})^{\beta \rho} W \boldsymbol{W} W_{\alpha \beta} {G}^{-3} \varphi_{i}^{\lambda} \varphi_{j \rho} \varphi^{j}_{\lambda} - \frac{9}{32}G_{j k} (\Gamma_{a})^{\beta \rho} \boldsymbol{W} \lambda^{j}_{\beta} {G}^{-5} \varphi^{k}_{\alpha} \varphi_{i}^{\lambda} \varphi_{l \rho} \varphi^{l}_{\lambda}+\frac{3}{64}{\rm i} \mathcal{H}_{\alpha}\,^{\lambda} (\Gamma_{a})^{\beta \rho} \boldsymbol{W} \lambda_{i \beta} {G}^{-3} \varphi_{j \lambda} \varphi^{j}_{\rho} - \frac{3}{32}{\rm i} (\Gamma_{a})^{\beta \rho} \boldsymbol{W} \lambda_{j \beta} {G}^{-3} \nabla_{\alpha}\,^{\lambda}{G_{i}\,^{j}} \varphi_{k \rho} \varphi^{k}_{\lambda} - \frac{3}{64}{\rm i} \mathcal{H}_{\alpha}\,^{\beta} (\Gamma_{a})_{\beta}{}^{\rho} \boldsymbol{W} \lambda_{j \rho} {G}^{-3} \varphi_{i}^{\lambda} \varphi^{j}_{\lambda}+\frac{3}{64}{\rm i} (\Gamma_{a})^{\beta}{}_{\rho} \boldsymbol{W} \lambda_{j \beta} {G}^{-3} \nabla^{\rho}\,_{\alpha}{G^{j}\,_{k}} \varphi_{i}^{\lambda} \varphi^{k}_{\lambda}+\frac{3}{64}{\rm i} \mathcal{H}_{\alpha}\,^{\lambda} (\Gamma_{a})^{\beta \rho} \boldsymbol{W} \lambda_{j \beta} {G}^{-3} \varphi_{i \lambda} \varphi^{j}_{\rho} - \frac{3}{32}{\rm i} (\Gamma_{a})^{\beta \rho} \boldsymbol{W} \lambda_{j \beta} {G}^{-3} \nabla_{\alpha}\,^{\lambda}{G^{j}\,_{k}} \varphi_{i \lambda} \varphi^{k}_{\rho} - \frac{3}{32}{\rm i} (\Gamma_{a})^{\beta \rho} \boldsymbol{\lambda}_{j \alpha} \lambda_{i \beta} {G}^{-3} \varphi^{j \lambda} \varphi_{k \rho} \varphi^{k}_{\lambda}+\frac{3}{80}(\Gamma_{a})_{\alpha}{}^{\beta} \boldsymbol{W} X_{i j} {G}^{-3} \varphi^{j \rho} \varphi_{k \beta} \varphi^{k}_{\rho} - \frac{9}{32}G_{j k} (\Gamma_{a})^{\beta \rho} \boldsymbol{W} \lambda_{i \beta} {G}^{-5} \varphi^{j}_{\alpha} \varphi^{k \lambda} \varphi_{l \rho} \varphi^{l}_{\lambda} - \frac{3}{32}{\rm i} (\Gamma_{a})^{\beta \rho} \boldsymbol{W} \lambda_{i \beta} {G}^{-3} \nabla_{\alpha}\,^{\lambda}{G_{j k}} \varphi^{j}_{\rho} \varphi^{k}_{\lambda}%
 - \frac{3}{32}{\rm i} (\Gamma_{a})^{\beta \rho} \lambda_{j \beta} \boldsymbol{\lambda}_{i \rho} {G}^{-3} \varphi_{k \alpha} \varphi^{j \lambda} \varphi^{k}_{\lambda}+\frac{9}{32}{\rm i} G_{j k} (\Gamma_{a})^{\beta \rho} F_{\alpha \beta} \boldsymbol{\lambda}_{i \rho} {G}^{-3} \varphi^{j \lambda} \varphi^{k}_{\lambda}+\frac{57}{128}{\rm i} G_{j k} (\Gamma_{a})^{\beta \rho} W W_{\alpha \beta} \boldsymbol{\lambda}_{i \rho} {G}^{-3} \varphi^{j \lambda} \varphi^{k}_{\lambda}+\frac{9}{32}{\rm i} G_{j k} (\Gamma_{a})^{\beta \rho} \mathbf{F}_{\alpha \beta} \lambda_{i \rho} {G}^{-3} \varphi^{j \lambda} \varphi^{k}_{\lambda}+\frac{57}{128}{\rm i} G_{j k} (\Gamma_{a})^{\beta \rho} \boldsymbol{W} W_{\alpha \beta} \lambda_{i \rho} {G}^{-3} \varphi^{j \lambda} \varphi^{k}_{\lambda}+\frac{9}{64}{\rm i} G_{j k} G_{l m} (\Gamma_{a})^{\beta \rho} \lambda^{j}_{\beta} \boldsymbol{\lambda}_{i \rho} {G}^{-5} \varphi^{k}_{\alpha} \varphi^{l \lambda} \varphi^{m}_{\lambda}+\frac{3}{64}\mathcal{H}_{\alpha}\,^{\lambda} G_{j k} (\Gamma_{a})^{\beta \rho} \lambda^{j}_{\beta} \boldsymbol{\lambda}_{i \rho} {G}^{-3} \varphi^{k}_{\lambda} - \frac{3}{32}G_{j k} (\Gamma_{a})^{\beta \rho} \lambda_{l \beta} \boldsymbol{\lambda}_{i \rho} {G}^{-3} \nabla_{\alpha}\,^{\lambda}{G^{j l}} \varphi^{k}_{\lambda} - \frac{3}{32}{\rm i} (\Gamma_{a})^{\beta \rho} \lambda_{i \beta} \boldsymbol{\lambda}_{j \rho} {G}^{-3} \varphi_{k \alpha} \varphi^{j \lambda} \varphi^{k}_{\lambda}+\frac{9}{64}{\rm i} G_{j k} G_{l m} (\Gamma_{a})^{\beta \rho} \lambda_{i \beta} \boldsymbol{\lambda}^{j}_{\rho} {G}^{-5} \varphi^{k}_{\alpha} \varphi^{l \lambda} \varphi^{m}_{\lambda}+\frac{3}{64}\mathcal{H}_{\alpha}\,^{\lambda} G_{j k} (\Gamma_{a})^{\beta \rho} \lambda_{i \beta} \boldsymbol{\lambda}^{j}_{\rho} {G}^{-3} \varphi^{k}_{\lambda} - \frac{3}{32}G_{j k} (\Gamma_{a})^{\beta \rho} \lambda_{i \beta} \boldsymbol{\lambda}_{l \rho} {G}^{-3} \nabla_{\alpha}\,^{\lambda}{G^{j l}} \varphi^{k}_{\lambda}+\frac{9}{64}G_{j k} (\Gamma_{a})^{\beta \rho} \boldsymbol{W} \lambda_{i \beta} {G}^{-5} \varphi_{l \alpha} \varphi^{j \lambda} \varphi^{k}_{\lambda} \varphi^{l}_{\rho} - \frac{9}{64}G_{j k} (\Gamma_{a})^{\beta \rho} \boldsymbol{W} \lambda_{l \beta} {G}^{-5} \varphi_{i \alpha} \varphi^{j \lambda} \varphi^{k}_{\lambda} \varphi^{l}_{\rho} - \frac{9}{32}G_{i j} (\Gamma_{a})^{\beta \rho} \boldsymbol{W} \lambda_{k \beta} {G}^{-5} \varphi_{l \alpha} \varphi^{j}_{\rho} \varphi^{k \lambda} \varphi^{l}_{\lambda}+\frac{9}{64}{\rm i} G_{i j} G_{k l} (\Gamma_{a})^{\beta \rho} \boldsymbol{\lambda}_{m \alpha} \lambda^{m}_{\beta} {G}^{-5} \varphi^{j}_{\rho} \varphi^{k \lambda} \varphi^{l}_{\lambda} - \frac{27}{32}G_{i j} G_{k l} (\Gamma_{a})^{\beta \rho} \boldsymbol{W} F_{\alpha \beta} {G}^{-5} \varphi^{j}_{\rho} \varphi^{k \lambda} \varphi^{l}_{\lambda} - \frac{333}{128}G_{i j} G_{k l} (\Gamma_{a})^{\beta \rho} W \boldsymbol{W} W_{\alpha \beta} {G}^{-5} \varphi^{j}_{\rho} \varphi^{k \lambda} \varphi^{l}_{\lambda}+\frac{45}{64}G_{i j} G_{k l} G_{m n} (\Gamma_{a})^{\beta \rho} \boldsymbol{W} \lambda^{k}_{\beta} {G}^{-7} \varphi^{l}_{\alpha} \varphi^{j}_{\rho} \varphi^{m \lambda} \varphi^{n}_{\lambda} - \frac{9}{128}{\rm i} \mathcal{H}_{\alpha}\,^{\beta} G_{i j} G_{k l} (\Gamma_{a})_{\beta}{}^{\rho} \boldsymbol{W} \lambda^{j}_{\rho} {G}^{-5} \varphi^{k \lambda} \varphi^{l}_{\lambda}%
+\frac{9}{128}{\rm i} G_{i j} G_{k l} (\Gamma_{a})^{\beta}{}_{\rho} \boldsymbol{W} \lambda_{m \beta} {G}^{-5} \nabla^{\rho}\,_{\alpha}{G^{j m}} \varphi^{k \lambda} \varphi^{l}_{\lambda}+\frac{9}{64}{\rm i} \mathcal{H}_{\alpha}\,^{\lambda} G_{i j} G_{k l} (\Gamma_{a})^{\beta \rho} \boldsymbol{W} \lambda^{k}_{\beta} {G}^{-5} \varphi^{j}_{\rho} \varphi^{l}_{\lambda} - \frac{9}{32}{\rm i} G_{i j} G_{k l} (\Gamma_{a})^{\beta \rho} \boldsymbol{W} \lambda_{m \beta} {G}^{-5} \nabla_{\alpha}\,^{\lambda}{G^{k m}} \varphi^{j}_{\rho} \varphi^{l}_{\lambda} - \frac{9}{32}G_{j k} (\Gamma_{a})^{\beta \rho} \boldsymbol{W} \lambda_{i \beta} {G}^{-5} \varphi_{l \alpha} \varphi^{j}_{\rho} \varphi^{k \lambda} \varphi^{l}_{\lambda} - \frac{9}{64}{\rm i} G_{j k} G_{l m} (\Gamma_{a})^{\beta \rho} \boldsymbol{\lambda}^{j}_{\alpha} \lambda_{i \beta} {G}^{-5} \varphi^{k}_{\rho} \varphi^{l \lambda} \varphi^{m}_{\lambda} - \frac{9}{32}{\rm i} G_{j k} G_{l m} (\Gamma_{a})^{\beta \rho} \boldsymbol{W} \lambda_{i \beta} {G}^{-5} \nabla_{\alpha}\,^{\lambda}{G^{j l}} \varphi^{k}_{\rho} \varphi^{m}_{\lambda} - \frac{3}{64}{\rm i} G_{i j} (\Gamma_{a})^{\beta}{}_{\rho} \boldsymbol{W} \lambda_{k \beta} {G}^{-3} \nabla^{\rho}\,_{\alpha}{\varphi^{k \lambda}} \varphi^{j}_{\lambda}+\frac{3}{32}{\rm i} G_{i j} (\Gamma_{a})^{\beta \rho} \boldsymbol{W} \lambda_{k \beta} {G}^{-3} \nabla_{\alpha}\,^{\lambda}{\varphi^{k}_{\rho}} \varphi^{j}_{\lambda}+\frac{3}{32}{\rm i} G_{i j} (\Gamma_{a})^{\beta \lambda} \boldsymbol{W} W_{\beta}\,^{\rho} \lambda_{k \lambda} {G}^{-3} \varphi^{j}_{\alpha} \varphi^{k}_{\rho}+\frac{3}{64}{\rm i} G_{i j} (\Gamma_{a})^{\beta \rho} \boldsymbol{W} W_{\alpha \beta} \lambda_{k \rho} {G}^{-3} \varphi^{j \lambda} \varphi^{k}_{\lambda} - \frac{3}{128}{\rm i} G_{i j} (\Gamma_{a})^{\rho \lambda} \boldsymbol{W} W_{\alpha}\,^{\beta} \lambda_{k \rho} {G}^{-3} \varphi^{j}_{\beta} \varphi^{k}_{\lambda}+\frac{3}{80}{\rm i} G_{i j} (\Gamma_{a})_{\alpha}{}^{\beta} \boldsymbol{W} \lambda_{k \beta} {G}^{-3} \nabla^{\rho \lambda}{\varphi^{k}_{\rho}} \varphi^{j}_{\lambda}+\frac{3}{64}{\rm i} G_{i j} (\Gamma_{a})^{\beta}{}_{\rho} \boldsymbol{W} \lambda_{k \beta} {G}^{-3} \nabla^{\rho \lambda}{\varphi^{k}_{\lambda}} \varphi^{j}_{\alpha}+\frac{3}{64}{\rm i} \mathcal{H}^{\beta \lambda} (\Gamma_{a})_{\beta}{}^{\rho} \boldsymbol{W} \lambda_{i \rho} {G}^{-3} \varphi_{j \alpha} \varphi^{j}_{\lambda}+\frac{3}{64}{\rm i} \mathcal{H}^{\beta \lambda} (\Gamma_{a})_{\beta}{}^{\rho} \boldsymbol{W} \lambda_{j \rho} {G}^{-3} \varphi_{i \alpha} \varphi^{j}_{\lambda}+\frac{3}{64}\mathcal{H}^{\beta \lambda} G_{i j} (\Gamma_{a})_{\beta}{}^{\rho} \boldsymbol{\lambda}_{k \alpha} \lambda^{k}_{\rho} {G}^{-3} \varphi^{j}_{\lambda}+\frac{9}{32}{\rm i} \mathcal{H}^{\rho \lambda} G_{i j} (\Gamma_{a})_{\rho}{}^{\beta} \boldsymbol{W} F_{\alpha \beta} {G}^{-3} \varphi^{j}_{\lambda}+\frac{27}{32}{\rm i} \mathcal{H}^{\rho \lambda} G_{i j} (\Gamma_{a})_{\rho}{}^{\beta} W \boldsymbol{W} W_{\alpha \beta} {G}^{-3} \varphi^{j}_{\lambda} - \frac{9}{64}{\rm i} \mathcal{H}^{\beta \lambda} G_{i j} G_{k l} (\Gamma_{a})_{\beta}{}^{\rho} \boldsymbol{W} \lambda^{k}_{\rho} {G}^{-5} \varphi^{l}_{\alpha} \varphi^{j}_{\lambda} - \frac{3}{128}\mathcal{H}_{\alpha}\,^{\lambda} \mathcal{H}_{\lambda}\,^{\beta} G_{i j} (\Gamma_{a})_{\beta}{}^{\rho} \boldsymbol{W} \lambda^{j}_{\rho} {G}^{-3}%
 - \frac{3}{64}\mathcal{H}_{\lambda}\,^{\beta} G_{i j} (\Gamma_{a})_{\beta}{}^{\rho} \boldsymbol{W} \lambda_{k \rho} {G}^{-3} \nabla^{\lambda}\,_{\alpha}{G^{j k}} - \frac{3}{32}{\rm i} (\Gamma_{a})^{\beta}{}_{\rho} \boldsymbol{W} \lambda_{j \beta} {G}^{-3} \nabla^{\rho \lambda}{G_{i}\,^{j}} \varphi_{k \alpha} \varphi^{k}_{\lambda}+\frac{3}{32}{\rm i} (\Gamma_{a})^{\beta}{}_{\rho} \boldsymbol{W} \lambda_{k \beta} {G}^{-3} \nabla^{\rho \lambda}{G_{i j}} \varphi^{j}_{\alpha} \varphi^{k}_{\lambda} - \frac{3}{32}G_{j k} (\Gamma_{a})^{\beta}{}_{\rho} \boldsymbol{\lambda}_{l \alpha} \lambda^{l}_{\beta} {G}^{-3} \nabla^{\rho \lambda}{G_{i}\,^{j}} \varphi^{k}_{\lambda} - \frac{9}{16}{\rm i} G_{j k} (\Gamma_{a})^{\beta}{}_{\rho} \boldsymbol{W} F_{\alpha \beta} {G}^{-3} \nabla^{\rho \lambda}{G_{i}\,^{j}} \varphi^{k}_{\lambda} - \frac{45}{32}{\rm i} G_{j k} (\Gamma_{a})^{\beta}{}_{\rho} W \boldsymbol{W} W_{\alpha \beta} {G}^{-3} \nabla^{\rho \lambda}{G_{i}\,^{j}} \varphi^{k}_{\lambda}+\frac{9}{32}{\rm i} G_{j k} G_{l m} (\Gamma_{a})^{\beta}{}_{\rho} \boldsymbol{W} \lambda^{j}_{\beta} {G}^{-5} \nabla^{\rho \lambda}{G_{i}\,^{l}} \varphi^{k}_{\alpha} \varphi^{m}_{\lambda} - \frac{15}{64}{\rm i} G_{j k} (\Gamma_{a})^{\beta}{}_{\rho} \boldsymbol{W} \lambda_{i \beta} {G}^{-3} \nabla^{\rho \lambda}{\varphi^{j}_{\alpha}} \varphi^{k}_{\lambda} - \frac{3}{32}{\rm i} G_{j k} (\Gamma_{a})^{\beta}{}_{\rho} \boldsymbol{W} \lambda^{j}_{\beta} {G}^{-3} \nabla^{\rho \lambda}{\varphi_{i \alpha}} \varphi^{k}_{\lambda}+\frac{3}{32}{\rm i} G_{j k} (\Gamma_{a})^{\rho \lambda} \boldsymbol{W} W_{\alpha}\,^{\beta} \lambda^{j}_{\rho} {G}^{-3} \varphi_{i \beta} \varphi^{k}_{\lambda}+\frac{3}{128}{\rm i} G_{j k} (\Gamma_{a})^{\rho \lambda} \boldsymbol{W} W_{\alpha}\,^{\beta} \lambda^{j}_{\rho} {G}^{-3} \varphi_{i \lambda} \varphi^{k}_{\beta}+\frac{3}{64}{\rm i} G_{j k} (\Gamma_{a})^{\beta \rho} \boldsymbol{W} W_{\alpha \beta} \lambda^{j}_{\rho} {G}^{-3} \varphi_{i}^{\lambda} \varphi^{k}_{\lambda}+\frac{9}{512}{\rm i} (\Gamma_{a})_{\alpha}{}^{\rho} \boldsymbol{W} \lambda_{j \rho} X^{j \beta} {G}^{-1} \varphi_{i \beta} - \frac{9}{512}{\rm i} G_{i j} G_{k l} (\Gamma_{a})^{\beta \rho} \boldsymbol{W} \lambda^{k}_{\beta} X^{j}_{\alpha} {G}^{-3} \varphi^{l}_{\rho}+\frac{27}{512}{\rm i} G_{i j} G_{k l} (\Gamma_{a})^{\beta \rho} \boldsymbol{W} \lambda^{j}_{\beta} X^{k}_{\alpha} {G}^{-3} \varphi^{l}_{\rho} - \frac{99}{2560}{\rm i} G_{i j} G_{k l} (\Gamma_{a})_{\alpha}{}^{\rho} \boldsymbol{W} \lambda^{k}_{\rho} X^{j \beta} {G}^{-3} \varphi^{l}_{\beta}+\frac{27}{640}{\rm i} G_{i j} G_{k l} (\Gamma_{a})_{\alpha}{}^{\rho} \boldsymbol{W} \lambda^{j}_{\rho} X^{k \beta} {G}^{-3} \varphi^{l}_{\beta} - \frac{81}{512}{\rm i} G_{i j} G_{k l} (\Gamma_{a})^{\rho \beta} \boldsymbol{W} \lambda^{j}_{\rho} X^{k}_{\beta} {G}^{-3} \varphi^{l}_{\alpha} - \frac{3}{64}\mathcal{H}_{\alpha \lambda} G_{j k} (\Gamma_{a})^{\beta}{}_{\rho} \boldsymbol{W} \lambda^{j}_{\beta} {G}^{-3} \nabla^{\lambda \rho}{G_{i}\,^{k}}+\frac{3}{32}G_{j k} (\Gamma_{a})^{\beta}{}_{\rho} \boldsymbol{W} \lambda_{l \beta} {G}^{-3} \nabla^{\rho}\,_{\lambda}{G_{i}\,^{j}} \nabla_{\alpha}\,^{\lambda}{G^{k l}}%
+\frac{15}{64}{\rm i} G_{j k} (\Gamma_{a})^{\beta \rho} \boldsymbol{W} \lambda_{i \beta} {G}^{-3} \nabla_{\alpha}\,^{\lambda}{\varphi^{j}_{\rho}} \varphi^{k}_{\lambda} - \frac{3}{64}\mathcal{H}^{\beta \lambda} G_{j k} (\Gamma_{a})_{\beta}{}^{\rho} \boldsymbol{\lambda}^{j}_{\alpha} \lambda_{i \rho} {G}^{-3} \varphi^{k}_{\lambda} - \frac{3}{160}{\rm i} \mathcal{H}^{\beta \rho} G_{j k} (\Gamma_{a})_{\alpha \beta} \boldsymbol{W} X_{i}\,^{j} {G}^{-3} \varphi^{k}_{\rho} - \frac{3}{64}\mathcal{H}_{\lambda}\,^{\beta} G_{j k} (\Gamma_{a})_{\beta}{}^{\rho} \boldsymbol{W} \lambda_{i \rho} {G}^{-3} \nabla^{\lambda}\,_{\alpha}{G^{j k}}+\frac{3}{32}{\rm i} (\Gamma_{a})^{\beta}{}_{\rho} \boldsymbol{W} \lambda_{i \beta} {G}^{-3} \nabla^{\rho \lambda}{G_{j k}} \varphi^{j}_{\alpha} \varphi^{k}_{\lambda} - \frac{3}{32}G_{j k} (\Gamma_{a})^{\beta}{}_{\rho} \boldsymbol{\lambda}_{l \alpha} \lambda_{i \beta} {G}^{-3} \nabla^{\rho \lambda}{G^{j l}} \varphi^{k}_{\lambda} - \frac{3}{160}{\rm i} G_{j k} (\Gamma_{a})_{\alpha \beta} \boldsymbol{W} X_{i l} {G}^{-3} \nabla^{\beta \rho}{G^{j l}} \varphi^{k}_{\rho}+\frac{9}{32}{\rm i} G_{j k} G_{l m} (\Gamma_{a})^{\beta}{}_{\rho} \boldsymbol{W} \lambda_{i \beta} {G}^{-5} \nabla^{\rho \lambda}{G^{j l}} \varphi^{k}_{\alpha} \varphi^{m}_{\lambda} - \frac{3}{64}\mathcal{H}_{\alpha \lambda} G_{j k} (\Gamma_{a})^{\beta}{}_{\rho} \boldsymbol{W} \lambda_{i \beta} {G}^{-3} \nabla^{\lambda \rho}{G^{j k}}+\frac{3}{32}G_{j k} (\Gamma_{a})^{\beta}{}_{\rho} \boldsymbol{W} \lambda_{i \beta} {G}^{-3} \nabla^{\rho}\,_{\lambda}{G^{j}\,_{l}} \nabla_{\alpha}\,^{\lambda}{G^{k l}} - \frac{3}{32}{\rm i} (\Gamma_{a})^{\beta \rho} \lambda_{j \alpha} \boldsymbol{\lambda}^{j}_{\beta} {G}^{-3} \varphi_{i}^{\lambda} \varphi_{k \rho} \varphi^{k}_{\lambda}+\frac{9}{16}(\Gamma_{a})^{\beta \rho} W \mathbf{F}_{\alpha \beta} {G}^{-3} \varphi_{i}^{\lambda} \varphi_{j \rho} \varphi^{j}_{\lambda} - \frac{9}{32}G_{j k} (\Gamma_{a})^{\beta \rho} W \boldsymbol{\lambda}^{j}_{\beta} {G}^{-5} \varphi^{k}_{\alpha} \varphi_{i}^{\lambda} \varphi_{l \rho} \varphi^{l}_{\lambda}+\frac{3}{64}{\rm i} \mathcal{H}_{\alpha}\,^{\lambda} (\Gamma_{a})^{\beta \rho} W \boldsymbol{\lambda}_{i \beta} {G}^{-3} \varphi_{j \lambda} \varphi^{j}_{\rho} - \frac{3}{32}{\rm i} (\Gamma_{a})^{\beta \rho} W \boldsymbol{\lambda}_{j \beta} {G}^{-3} \nabla_{\alpha}\,^{\lambda}{G_{i}\,^{j}} \varphi_{k \rho} \varphi^{k}_{\lambda} - \frac{3}{64}{\rm i} \mathcal{H}_{\alpha}\,^{\beta} (\Gamma_{a})_{\beta}{}^{\rho} W \boldsymbol{\lambda}_{j \rho} {G}^{-3} \varphi_{i}^{\lambda} \varphi^{j}_{\lambda}+\frac{3}{64}{\rm i} (\Gamma_{a})^{\beta}{}_{\rho} W \boldsymbol{\lambda}_{j \beta} {G}^{-3} \nabla^{\rho}\,_{\alpha}{G^{j}\,_{k}} \varphi_{i}^{\lambda} \varphi^{k}_{\lambda}+\frac{3}{64}{\rm i} \mathcal{H}_{\alpha}\,^{\lambda} (\Gamma_{a})^{\beta \rho} W \boldsymbol{\lambda}_{j \beta} {G}^{-3} \varphi_{i \lambda} \varphi^{j}_{\rho} - \frac{3}{32}{\rm i} (\Gamma_{a})^{\beta \rho} W \boldsymbol{\lambda}_{j \beta} {G}^{-3} \nabla_{\alpha}\,^{\lambda}{G^{j}\,_{k}} \varphi_{i \lambda} \varphi^{k}_{\rho} - \frac{3}{32}{\rm i} (\Gamma_{a})^{\beta \rho} \lambda_{j \alpha} \boldsymbol{\lambda}_{i \beta} {G}^{-3} \varphi^{j \lambda} \varphi_{k \rho} \varphi^{k}_{\lambda}%
+\frac{3}{80}(\Gamma_{a})_{\alpha}{}^{\beta} W \mathbf{X}_{i j} {G}^{-3} \varphi^{j \rho} \varphi_{k \beta} \varphi^{k}_{\rho} - \frac{9}{32}G_{j k} (\Gamma_{a})^{\beta \rho} W \boldsymbol{\lambda}_{i \beta} {G}^{-5} \varphi^{j}_{\alpha} \varphi^{k \lambda} \varphi_{l \rho} \varphi^{l}_{\lambda} - \frac{3}{32}{\rm i} (\Gamma_{a})^{\beta \rho} W \boldsymbol{\lambda}_{i \beta} {G}^{-3} \nabla_{\alpha}\,^{\lambda}{G_{j k}} \varphi^{j}_{\rho} \varphi^{k}_{\lambda}+\frac{9}{64}G_{j k} (\Gamma_{a})^{\beta \rho} W \boldsymbol{\lambda}_{i \beta} {G}^{-5} \varphi_{l \alpha} \varphi^{j \lambda} \varphi^{k}_{\lambda} \varphi^{l}_{\rho} - \frac{9}{64}G_{j k} (\Gamma_{a})^{\beta \rho} W \boldsymbol{\lambda}_{l \beta} {G}^{-5} \varphi_{i \alpha} \varphi^{j \lambda} \varphi^{k}_{\lambda} \varphi^{l}_{\rho} - \frac{9}{32}G_{i j} (\Gamma_{a})^{\beta \rho} W \boldsymbol{\lambda}_{k \beta} {G}^{-5} \varphi_{l \alpha} \varphi^{j}_{\rho} \varphi^{k \lambda} \varphi^{l}_{\lambda}+\frac{9}{64}{\rm i} G_{i j} G_{k l} (\Gamma_{a})^{\beta \rho} \lambda_{m \alpha} \boldsymbol{\lambda}^{m}_{\beta} {G}^{-5} \varphi^{j}_{\rho} \varphi^{k \lambda} \varphi^{l}_{\lambda} - \frac{27}{32}G_{i j} G_{k l} (\Gamma_{a})^{\beta \rho} W \mathbf{F}_{\alpha \beta} {G}^{-5} \varphi^{j}_{\rho} \varphi^{k \lambda} \varphi^{l}_{\lambda}+\frac{45}{64}G_{i j} G_{k l} G_{m n} (\Gamma_{a})^{\beta \rho} W \boldsymbol{\lambda}^{k}_{\beta} {G}^{-7} \varphi^{l}_{\alpha} \varphi^{j}_{\rho} \varphi^{m \lambda} \varphi^{n}_{\lambda} - \frac{9}{128}{\rm i} \mathcal{H}_{\alpha}\,^{\beta} G_{i j} G_{k l} (\Gamma_{a})_{\beta}{}^{\rho} W \boldsymbol{\lambda}^{j}_{\rho} {G}^{-5} \varphi^{k \lambda} \varphi^{l}_{\lambda}+\frac{9}{128}{\rm i} G_{i j} G_{k l} (\Gamma_{a})^{\beta}{}_{\rho} W \boldsymbol{\lambda}_{m \beta} {G}^{-5} \nabla^{\rho}\,_{\alpha}{G^{j m}} \varphi^{k \lambda} \varphi^{l}_{\lambda}+\frac{9}{64}{\rm i} \mathcal{H}_{\alpha}\,^{\lambda} G_{i j} G_{k l} (\Gamma_{a})^{\beta \rho} W \boldsymbol{\lambda}^{k}_{\beta} {G}^{-5} \varphi^{j}_{\rho} \varphi^{l}_{\lambda} - \frac{9}{32}{\rm i} G_{i j} G_{k l} (\Gamma_{a})^{\beta \rho} W \boldsymbol{\lambda}_{m \beta} {G}^{-5} \nabla_{\alpha}\,^{\lambda}{G^{k m}} \varphi^{j}_{\rho} \varphi^{l}_{\lambda} - \frac{9}{32}G_{j k} (\Gamma_{a})^{\beta \rho} W \boldsymbol{\lambda}_{i \beta} {G}^{-5} \varphi_{l \alpha} \varphi^{j}_{\rho} \varphi^{k \lambda} \varphi^{l}_{\lambda} - \frac{9}{64}{\rm i} G_{j k} G_{l m} (\Gamma_{a})^{\beta \rho} \lambda^{j}_{\alpha} \boldsymbol{\lambda}_{i \beta} {G}^{-5} \varphi^{k}_{\rho} \varphi^{l \lambda} \varphi^{m}_{\lambda} - \frac{9}{32}{\rm i} G_{j k} G_{l m} (\Gamma_{a})^{\beta \rho} W \boldsymbol{\lambda}_{i \beta} {G}^{-5} \nabla_{\alpha}\,^{\lambda}{G^{j l}} \varphi^{k}_{\rho} \varphi^{m}_{\lambda} - \frac{3}{64}{\rm i} G_{i j} (\Gamma_{a})^{\beta}{}_{\rho} W \boldsymbol{\lambda}_{k \beta} {G}^{-3} \nabla^{\rho}\,_{\alpha}{\varphi^{k \lambda}} \varphi^{j}_{\lambda}+\frac{3}{32}{\rm i} G_{i j} (\Gamma_{a})^{\beta \rho} W \boldsymbol{\lambda}_{k \beta} {G}^{-3} \nabla_{\alpha}\,^{\lambda}{\varphi^{k}_{\rho}} \varphi^{j}_{\lambda}+\frac{3}{32}{\rm i} G_{i j} (\Gamma_{a})^{\beta \lambda} W W_{\beta}\,^{\rho} \boldsymbol{\lambda}_{k \lambda} {G}^{-3} \varphi^{j}_{\alpha} \varphi^{k}_{\rho}+\frac{3}{64}{\rm i} G_{i j} (\Gamma_{a})^{\beta \rho} W W_{\alpha \beta} \boldsymbol{\lambda}_{k \rho} {G}^{-3} \varphi^{j \lambda} \varphi^{k}_{\lambda}%
 - \frac{3}{128}{\rm i} G_{i j} (\Gamma_{a})^{\rho \lambda} W W_{\alpha}\,^{\beta} \boldsymbol{\lambda}_{k \rho} {G}^{-3} \varphi^{j}_{\beta} \varphi^{k}_{\lambda}+\frac{3}{80}{\rm i} G_{i j} (\Gamma_{a})_{\alpha}{}^{\beta} W \boldsymbol{\lambda}_{k \beta} {G}^{-3} \nabla^{\rho \lambda}{\varphi^{k}_{\rho}} \varphi^{j}_{\lambda}+\frac{3}{64}{\rm i} G_{i j} (\Gamma_{a})^{\beta}{}_{\rho} W \boldsymbol{\lambda}_{k \beta} {G}^{-3} \nabla^{\rho \lambda}{\varphi^{k}_{\lambda}} \varphi^{j}_{\alpha}+\frac{3}{64}{\rm i} \mathcal{H}^{\beta \lambda} (\Gamma_{a})_{\beta}{}^{\rho} W \boldsymbol{\lambda}_{i \rho} {G}^{-3} \varphi_{j \alpha} \varphi^{j}_{\lambda}+\frac{3}{64}{\rm i} \mathcal{H}^{\beta \lambda} (\Gamma_{a})_{\beta}{}^{\rho} W \boldsymbol{\lambda}_{j \rho} {G}^{-3} \varphi_{i \alpha} \varphi^{j}_{\lambda}+\frac{3}{64}\mathcal{H}^{\beta \lambda} G_{i j} (\Gamma_{a})_{\beta}{}^{\rho} \lambda_{k \alpha} \boldsymbol{\lambda}^{k}_{\rho} {G}^{-3} \varphi^{j}_{\lambda}+\frac{9}{32}{\rm i} \mathcal{H}^{\rho \lambda} G_{i j} (\Gamma_{a})_{\rho}{}^{\beta} W \mathbf{F}_{\alpha \beta} {G}^{-3} \varphi^{j}_{\lambda} - \frac{9}{64}{\rm i} \mathcal{H}^{\beta \lambda} G_{i j} G_{k l} (\Gamma_{a})_{\beta}{}^{\rho} W \boldsymbol{\lambda}^{k}_{\rho} {G}^{-5} \varphi^{l}_{\alpha} \varphi^{j}_{\lambda} - \frac{3}{128}\mathcal{H}_{\alpha}\,^{\lambda} \mathcal{H}_{\lambda}\,^{\beta} G_{i j} (\Gamma_{a})_{\beta}{}^{\rho} W \boldsymbol{\lambda}^{j}_{\rho} {G}^{-3} - \frac{3}{64}\mathcal{H}_{\lambda}\,^{\beta} G_{i j} (\Gamma_{a})_{\beta}{}^{\rho} W \boldsymbol{\lambda}_{k \rho} {G}^{-3} \nabla^{\lambda}\,_{\alpha}{G^{j k}} - \frac{3}{32}{\rm i} (\Gamma_{a})^{\beta}{}_{\rho} W \boldsymbol{\lambda}_{j \beta} {G}^{-3} \nabla^{\rho \lambda}{G_{i}\,^{j}} \varphi_{k \alpha} \varphi^{k}_{\lambda}+\frac{3}{32}{\rm i} (\Gamma_{a})^{\beta}{}_{\rho} W \boldsymbol{\lambda}_{k \beta} {G}^{-3} \nabla^{\rho \lambda}{G_{i j}} \varphi^{j}_{\alpha} \varphi^{k}_{\lambda} - \frac{3}{32}G_{j k} (\Gamma_{a})^{\beta}{}_{\rho} \lambda_{l \alpha} \boldsymbol{\lambda}^{l}_{\beta} {G}^{-3} \nabla^{\rho \lambda}{G_{i}\,^{j}} \varphi^{k}_{\lambda} - \frac{9}{16}{\rm i} G_{j k} (\Gamma_{a})^{\beta}{}_{\rho} W \mathbf{F}_{\alpha \beta} {G}^{-3} \nabla^{\rho \lambda}{G_{i}\,^{j}} \varphi^{k}_{\lambda}+\frac{9}{32}{\rm i} G_{j k} G_{l m} (\Gamma_{a})^{\beta}{}_{\rho} W \boldsymbol{\lambda}^{j}_{\beta} {G}^{-5} \nabla^{\rho \lambda}{G_{i}\,^{l}} \varphi^{k}_{\alpha} \varphi^{m}_{\lambda} - \frac{15}{64}{\rm i} G_{j k} (\Gamma_{a})^{\beta}{}_{\rho} W \boldsymbol{\lambda}_{i \beta} {G}^{-3} \nabla^{\rho \lambda}{\varphi^{j}_{\alpha}} \varphi^{k}_{\lambda} - \frac{3}{32}{\rm i} G_{j k} (\Gamma_{a})^{\beta}{}_{\rho} W \boldsymbol{\lambda}^{j}_{\beta} {G}^{-3} \nabla^{\rho \lambda}{\varphi_{i \alpha}} \varphi^{k}_{\lambda}+\frac{3}{32}{\rm i} G_{j k} (\Gamma_{a})^{\rho \lambda} W W_{\alpha}\,^{\beta} \boldsymbol{\lambda}^{j}_{\rho} {G}^{-3} \varphi_{i \beta} \varphi^{k}_{\lambda}+\frac{3}{128}{\rm i} G_{j k} (\Gamma_{a})^{\rho \lambda} W W_{\alpha}\,^{\beta} \boldsymbol{\lambda}^{j}_{\rho} {G}^{-3} \varphi_{i \lambda} \varphi^{k}_{\beta}+\frac{3}{64}{\rm i} G_{j k} (\Gamma_{a})^{\beta \rho} W W_{\alpha \beta} \boldsymbol{\lambda}^{j}_{\rho} {G}^{-3} \varphi_{i}^{\lambda} \varphi^{k}_{\lambda}%
+\frac{9}{512}{\rm i} (\Gamma_{a})_{\alpha}{}^{\rho} W \boldsymbol{\lambda}_{j \rho} X^{j \beta} {G}^{-1} \varphi_{i \beta} - \frac{9}{512}{\rm i} G_{i j} G_{k l} (\Gamma_{a})^{\beta \rho} W \boldsymbol{\lambda}^{k}_{\beta} X^{j}_{\alpha} {G}^{-3} \varphi^{l}_{\rho}+\frac{27}{512}{\rm i} G_{i j} G_{k l} (\Gamma_{a})^{\beta \rho} W \boldsymbol{\lambda}^{j}_{\beta} X^{k}_{\alpha} {G}^{-3} \varphi^{l}_{\rho} - \frac{99}{2560}{\rm i} G_{i j} G_{k l} (\Gamma_{a})_{\alpha}{}^{\rho} W \boldsymbol{\lambda}^{k}_{\rho} X^{j \beta} {G}^{-3} \varphi^{l}_{\beta}+\frac{27}{640}{\rm i} G_{i j} G_{k l} (\Gamma_{a})_{\alpha}{}^{\rho} W \boldsymbol{\lambda}^{j}_{\rho} X^{k \beta} {G}^{-3} \varphi^{l}_{\beta} - \frac{81}{512}{\rm i} G_{i j} G_{k l} (\Gamma_{a})^{\rho \beta} W \boldsymbol{\lambda}^{j}_{\rho} X^{k}_{\beta} {G}^{-3} \varphi^{l}_{\alpha} - \frac{3}{64}\mathcal{H}_{\alpha \lambda} G_{j k} (\Gamma_{a})^{\beta}{}_{\rho} W \boldsymbol{\lambda}^{j}_{\beta} {G}^{-3} \nabla^{\lambda \rho}{G_{i}\,^{k}}+\frac{3}{32}G_{j k} (\Gamma_{a})^{\beta}{}_{\rho} W \boldsymbol{\lambda}_{l \beta} {G}^{-3} \nabla^{\rho}\,_{\lambda}{G_{i}\,^{j}} \nabla_{\alpha}\,^{\lambda}{G^{k l}}+\frac{15}{64}{\rm i} G_{j k} (\Gamma_{a})^{\beta \rho} W \boldsymbol{\lambda}_{i \beta} {G}^{-3} \nabla_{\alpha}\,^{\lambda}{\varphi^{j}_{\rho}} \varphi^{k}_{\lambda} - \frac{3}{64}\mathcal{H}^{\beta \lambda} G_{j k} (\Gamma_{a})_{\beta}{}^{\rho} \lambda^{j}_{\alpha} \boldsymbol{\lambda}_{i \rho} {G}^{-3} \varphi^{k}_{\lambda} - \frac{3}{160}{\rm i} \mathcal{H}^{\beta \rho} G_{j k} (\Gamma_{a})_{\alpha \beta} W \mathbf{X}_{i}\,^{j} {G}^{-3} \varphi^{k}_{\rho} - \frac{3}{64}\mathcal{H}_{\lambda}\,^{\beta} G_{j k} (\Gamma_{a})_{\beta}{}^{\rho} W \boldsymbol{\lambda}_{i \rho} {G}^{-3} \nabla^{\lambda}\,_{\alpha}{G^{j k}}+\frac{3}{32}{\rm i} (\Gamma_{a})^{\beta}{}_{\rho} W \boldsymbol{\lambda}_{i \beta} {G}^{-3} \nabla^{\rho \lambda}{G_{j k}} \varphi^{j}_{\alpha} \varphi^{k}_{\lambda} - \frac{3}{32}G_{j k} (\Gamma_{a})^{\beta}{}_{\rho} \lambda_{l \alpha} \boldsymbol{\lambda}_{i \beta} {G}^{-3} \nabla^{\rho \lambda}{G^{j l}} \varphi^{k}_{\lambda} - \frac{3}{160}{\rm i} G_{j k} (\Gamma_{a})_{\alpha \beta} W \mathbf{X}_{i l} {G}^{-3} \nabla^{\beta \rho}{G^{j l}} \varphi^{k}_{\rho}+\frac{9}{32}{\rm i} G_{j k} G_{l m} (\Gamma_{a})^{\beta}{}_{\rho} W \boldsymbol{\lambda}_{i \beta} {G}^{-5} \nabla^{\rho \lambda}{G^{j l}} \varphi^{k}_{\alpha} \varphi^{m}_{\lambda} - \frac{3}{64}\mathcal{H}_{\alpha \lambda} G_{j k} (\Gamma_{a})^{\beta}{}_{\rho} W \boldsymbol{\lambda}_{i \beta} {G}^{-3} \nabla^{\lambda \rho}{G^{j k}}+\frac{3}{32}G_{j k} (\Gamma_{a})^{\beta}{}_{\rho} W \boldsymbol{\lambda}_{i \beta} {G}^{-3} \nabla^{\rho}\,_{\lambda}{G^{j}\,_{l}} \nabla_{\alpha}\,^{\lambda}{G^{k l}} - \frac{9}{32}{\rm i} (\Gamma_{a})^{\beta \rho} W \boldsymbol{W} {G}^{-5} \varphi_{j \alpha} \varphi_{i}^{\lambda} \varphi^{j}_{\beta} \varphi_{k \rho} \varphi^{k}_{\lambda}+\frac{9}{32}{\rm i} (\Gamma_{a})^{\beta \rho} W \boldsymbol{W} {G}^{-5} \varphi_{i \alpha} \varphi_{j \beta} \varphi^{j \lambda} \varphi_{k \rho} \varphi^{k}_{\lambda}%
+\frac{9}{32}G_{i j} (\Gamma_{a})^{\beta \rho} \boldsymbol{W} \lambda_{k \alpha} {G}^{-5} \varphi^{j}_{\beta} \varphi^{k \lambda} \varphi_{l \rho} \varphi^{l}_{\lambda}+\frac{9}{32}G_{i j} (\Gamma_{a})^{\beta \rho} W \boldsymbol{\lambda}_{k \alpha} {G}^{-5} \varphi^{j}_{\beta} \varphi^{k \lambda} \varphi_{l \rho} \varphi^{l}_{\lambda}+\frac{45}{32}{\rm i} G_{i j} G_{k l} (\Gamma_{a})^{\beta \rho} W \boldsymbol{W} {G}^{-7} \varphi^{k}_{\alpha} \varphi^{j}_{\beta} \varphi^{l \lambda} \varphi_{m \rho} \varphi^{m}_{\lambda} - \frac{9}{64}F G_{i j} (\Gamma_{a})^{\beta \rho} W \boldsymbol{W} {G}^{-5} \varphi_{k \alpha} \varphi^{j}_{\beta} \varphi^{k}_{\rho}+\frac{9}{64}\mathcal{H}_{\alpha}\,^{\lambda} G_{i j} (\Gamma_{a})^{\beta \rho} W \boldsymbol{W} {G}^{-5} \varphi^{j}_{\beta} \varphi_{k \lambda} \varphi^{k}_{\rho}+\frac{9}{320}F G_{i j} (\Gamma_{a})_{\alpha}{}^{\beta} W \boldsymbol{W} {G}^{-5} \varphi^{j \rho} \varphi_{k \beta} \varphi^{k}_{\rho} - \frac{9}{64}\mathcal{H}_{\alpha}\,^{\beta} G_{i j} (\Gamma_{a})_{\beta}{}^{\rho} W \boldsymbol{W} {G}^{-5} \varphi^{j \lambda} \varphi_{k \rho} \varphi^{k}_{\lambda} - \frac{9}{64}G_{i j} (\Gamma_{a})^{\beta}{}_{\rho} W \boldsymbol{W} {G}^{-5} \nabla^{\rho}\,_{\alpha}{G^{j}\,_{k}} \varphi^{k \lambda} \varphi_{l \beta} \varphi^{l}_{\lambda}+\frac{9}{32}G_{i j} (\Gamma_{a})^{\beta \rho} W \boldsymbol{W} {G}^{-5} \nabla_{\alpha}\,^{\lambda}{G_{k l}} \varphi^{j}_{\beta} \varphi^{k}_{\rho} \varphi^{l}_{\lambda}+\frac{9}{32}G_{j k} (\Gamma_{a})^{\beta \rho} \boldsymbol{W} \lambda^{j}_{\alpha} {G}^{-5} \varphi_{i}^{\lambda} \varphi^{k}_{\beta} \varphi_{l \rho} \varphi^{l}_{\lambda}+\frac{9}{32}G_{j k} (\Gamma_{a})^{\beta \rho} W \boldsymbol{\lambda}^{j}_{\alpha} {G}^{-5} \varphi_{i}^{\lambda} \varphi^{k}_{\beta} \varphi_{l \rho} \varphi^{l}_{\lambda}+\frac{9}{32}G_{j k} (\Gamma_{a})^{\beta \rho} W \boldsymbol{W} {G}^{-5} \nabla_{\alpha}\,^{\lambda}{G_{i}\,^{j}} \varphi^{k}_{\beta} \varphi_{l \rho} \varphi^{l}_{\lambda}+\frac{9}{320}F G_{j k} (\Gamma_{a})_{\alpha}{}^{\beta} W \boldsymbol{W} {G}^{-5} \varphi_{i}^{\rho} \varphi^{j}_{\beta} \varphi^{k}_{\rho} - \frac{9}{64}\mathcal{H}_{\alpha}\,^{\beta} G_{j k} (\Gamma_{a})_{\beta}{}^{\rho} W \boldsymbol{W} {G}^{-5} \varphi_{i}^{\lambda} \varphi^{j}_{\rho} \varphi^{k}_{\lambda}+\frac{9}{64}G_{j k} (\Gamma_{a})^{\beta}{}_{\rho} W \boldsymbol{W} {G}^{-5} \nabla^{\rho}\,_{\alpha}{G^{j}\,_{l}} \varphi_{i}^{\lambda} \varphi^{k}_{\beta} \varphi^{l}_{\lambda}+\frac{9}{64}F G_{j k} (\Gamma_{a})^{\beta \rho} W \boldsymbol{W} {G}^{-5} \varphi_{i \alpha} \varphi^{j}_{\beta} \varphi^{k}_{\rho}+\frac{9}{64}\mathcal{H}_{\alpha}\,^{\lambda} G_{j k} (\Gamma_{a})^{\beta \rho} W \boldsymbol{W} {G}^{-5} \varphi_{i \lambda} \varphi^{j}_{\beta} \varphi^{k}_{\rho} - \frac{9}{32}G_{j k} (\Gamma_{a})^{\beta \rho} W \boldsymbol{W} {G}^{-5} \nabla_{\alpha}\,^{\lambda}{G^{j}\,_{l}} \varphi_{i \lambda} \varphi^{k}_{\beta} \varphi^{l}_{\rho} - \frac{3}{16}{\rm i} (\Gamma_{a})^{\beta}{}_{\rho} \boldsymbol{W} \lambda_{j \alpha} {G}^{-3} \nabla^{\rho \lambda}{G_{i}\,^{j}} \varphi_{k \beta} \varphi^{k}_{\lambda} - \frac{3}{16}{\rm i} (\Gamma_{a})^{\beta}{}_{\rho} W \boldsymbol{\lambda}_{j \alpha} {G}^{-3} \nabla^{\rho \lambda}{G_{i}\,^{j}} \varphi_{k \beta} \varphi^{k}_{\lambda}%
+\frac{9}{16}G_{j k} (\Gamma_{a})^{\beta}{}_{\rho} W \boldsymbol{W} {G}^{-5} \nabla^{\rho \lambda}{G_{i}\,^{j}} \varphi^{k}_{\alpha} \varphi_{l \beta} \varphi^{l}_{\lambda} - \frac{15}{64}(\Gamma_{a})^{\beta}{}_{\rho} W \boldsymbol{W} {G}^{-3} \nabla^{\rho \lambda}{\varphi_{i \alpha}} \varphi_{j \beta} \varphi^{j}_{\lambda}+\frac{21}{320}(\Gamma_{a})_{\alpha}{}^{\lambda} W \boldsymbol{W} W^{\beta \rho} {G}^{-3} \varphi_{i \beta} \varphi_{j \lambda} \varphi^{j}_{\rho} - \frac{57}{128}(\Gamma_{a})^{\beta \lambda} W \boldsymbol{W} W_{\beta}\,^{\rho} {G}^{-3} \varphi_{j \alpha} \varphi_{i \rho} \varphi^{j}_{\lambda} - \frac{27}{128}(\Gamma_{a})^{\rho \lambda} W \boldsymbol{W} W_{\alpha}\,^{\beta} {G}^{-3} \varphi_{i \rho} \varphi_{j \lambda} \varphi^{j}_{\beta}+\frac{9}{320}G_{i j} (\Gamma_{a})_{\alpha}{}^{\rho} W \boldsymbol{W} X^{j \beta} {G}^{-3} \varphi_{k \rho} \varphi^{k}_{\beta}+\frac{3}{128}{\rm i} F (\Gamma_{a})^{\beta}{}_{\rho} W \boldsymbol{W} {G}^{-3} \nabla^{\rho}\,_{\alpha}{G_{i j}} \varphi^{j}_{\beta}+\frac{3}{32}{\rm i} \mathcal{H}_{\alpha \lambda} (\Gamma_{a})^{\beta}{}_{\rho} W \boldsymbol{W} {G}^{-3} \nabla^{\lambda \rho}{G_{i j}} \varphi^{j}_{\beta} - \frac{3}{16}{\rm i} (\Gamma_{a})^{\beta}{}_{\rho} W \boldsymbol{W} {G}^{-3} \nabla^{\rho}\,_{\lambda}{G_{i j}} \nabla_{\alpha}\,^{\lambda}{G^{j}\,_{k}} \varphi^{k}_{\beta}+\frac{21}{640}{\rm i} F (\Gamma_{a})_{\alpha \beta} W \boldsymbol{W} {G}^{-3} \nabla^{\beta \rho}{G_{i j}} \varphi^{j}_{\rho}+\frac{3}{64}{\rm i} \mathcal{H}_{\alpha}\,^{\beta} (\Gamma_{a})_{\beta \rho} W \boldsymbol{W} {G}^{-3} \nabla^{\rho \lambda}{G_{i j}} \varphi^{j}_{\lambda} - \frac{3}{64}{\rm i} (\Gamma_{a})_{\beta \rho} W \boldsymbol{W} {G}^{-3} \nabla^{\beta}\,_{\alpha}{G_{j k}} \nabla^{\rho \lambda}{G_{i}\,^{j}} \varphi^{k}_{\lambda}+\frac{3}{64}(\Gamma_{a})^{\beta}{}_{\rho} W \boldsymbol{W} {G}^{-3} \nabla^{\rho}\,_{\alpha}{\varphi_{j \beta}} \varphi_{i}^{\lambda} \varphi^{j}_{\lambda}+\frac{189}{640}(\Gamma_{a})_{\alpha}{}^{\beta} W \boldsymbol{W} W_{\beta}\,^{\rho} {G}^{-3} \varphi_{i}^{\lambda} \varphi_{j \rho} \varphi^{j}_{\lambda}+\frac{3}{32}(\Gamma_{a})_{\alpha \beta} W \boldsymbol{W} {G}^{-3} \nabla^{\beta \rho}{\varphi_{j \rho}} \varphi_{i}^{\lambda} \varphi^{j}_{\lambda} - \frac{3}{64}{\rm i} \mathcal{H}^{\beta \rho} (\Gamma_{a})_{\beta \rho} \boldsymbol{W} \lambda_{j \alpha} {G}^{-3} \varphi_{i}^{\lambda} \varphi^{j}_{\lambda} - \frac{3}{64}{\rm i} \mathcal{H}^{\beta \rho} (\Gamma_{a})_{\beta \rho} W \boldsymbol{\lambda}_{j \alpha} {G}^{-3} \varphi_{i}^{\lambda} \varphi^{j}_{\lambda}+\frac{9}{64}\mathcal{H}^{\beta \rho} G_{j k} (\Gamma_{a})_{\beta \rho} W \boldsymbol{W} {G}^{-5} \varphi^{j}_{\alpha} \varphi_{i}^{\lambda} \varphi^{k}_{\lambda} - \frac{3}{64}{\rm i} \mathcal{H}^{\beta \rho} (\Gamma_{a})_{\beta \rho} W \boldsymbol{W} {G}^{-3} \nabla_{\alpha}\,^{\lambda}{G_{i j}} \varphi^{j}_{\lambda}+\frac{3}{64}{\rm i} (\Gamma_{a})_{\beta \rho} \boldsymbol{W} \lambda_{k \alpha} {G}^{-3} \nabla^{\beta \rho}{G_{i j}} \varphi^{k \lambda} \varphi^{j}_{\lambda}%
+\frac{3}{64}{\rm i} (\Gamma_{a})_{\beta \rho} W \boldsymbol{\lambda}_{k \alpha} {G}^{-3} \nabla^{\beta \rho}{G_{i j}} \varphi^{k \lambda} \varphi^{j}_{\lambda} - \frac{9}{64}G_{j k} (\Gamma_{a})_{\beta \rho} W \boldsymbol{W} {G}^{-5} \nabla^{\beta \rho}{G_{i l}} \varphi^{j}_{\alpha} \varphi^{k \lambda} \varphi^{l}_{\lambda}+\frac{3}{16}(\Gamma_{a})_{\beta \rho} W \boldsymbol{W} {G}^{-3} \nabla^{\beta \rho}{\varphi_{j \alpha}} \varphi_{i}^{\lambda} \varphi^{j}_{\lambda}+\frac{501}{2560}G_{i j} (\Gamma_{a})_{\alpha}{}^{\beta} W \boldsymbol{W} X_{k \beta} {G}^{-3} \varphi^{j \rho} \varphi^{k}_{\rho} - \frac{87}{2560}G_{j k} (\Gamma_{a})_{\alpha}{}^{\beta} W \boldsymbol{W} X^{j}_{\beta} {G}^{-3} \varphi_{i}^{\rho} \varphi^{k}_{\rho}+\frac{387}{2560}G_{j k} (\Gamma_{a})_{\alpha}{}^{\beta} W \boldsymbol{W} X_{i \beta} {G}^{-3} \varphi^{j \rho} \varphi^{k}_{\rho}+\frac{15}{256}{\rm i} F (\Gamma_{a})_{\beta \rho} W \boldsymbol{W} {G}^{-3} \nabla^{\beta \rho}{G_{i j}} \varphi^{j}_{\alpha} - \frac{3}{256}{\rm i} \mathcal{H}_{\alpha}\,^{\lambda} (\Gamma_{a})_{\beta \rho} W \boldsymbol{W} {G}^{-3} \nabla^{\beta \rho}{G_{i j}} \varphi^{j}_{\lambda}+\frac{3}{64}{\rm i} (\Gamma_{a})_{\beta \rho} W \boldsymbol{W} {G}^{-3} \nabla^{\beta \rho}{G_{i j}} \nabla_{\alpha}\,^{\lambda}{G^{j}\,_{k}} \varphi^{k}_{\lambda}+\frac{3}{64}{\rm i} (\Gamma_{a})_{\beta \rho} \boldsymbol{W} \lambda_{j \alpha} {G}^{-3} \nabla^{\beta \rho}{G^{j}\,_{k}} \varphi_{i}^{\lambda} \varphi^{k}_{\lambda}+\frac{3}{64}{\rm i} (\Gamma_{a})_{\beta \rho} W \boldsymbol{\lambda}_{j \alpha} {G}^{-3} \nabla^{\beta \rho}{G^{j}\,_{k}} \varphi_{i}^{\lambda} \varphi^{k}_{\lambda} - \frac{9}{64}G_{j k} (\Gamma_{a})_{\beta \rho} W \boldsymbol{W} {G}^{-5} \nabla^{\beta \rho}{G^{j}\,_{l}} \varphi^{k}_{\alpha} \varphi_{i}^{\lambda} \varphi^{l}_{\lambda} - \frac{15}{128}{\rm i} (\Gamma_{a})_{\beta \rho} W \boldsymbol{W} {G}^{-3} \nabla^{\beta \rho}{G_{j k}} \nabla_{\alpha}\,^{\lambda}{G_{i}\,^{j}} \varphi^{k}_{\lambda} - \frac{3}{64}{\rm i} (\Gamma_{a})_{\beta \rho} W \boldsymbol{W} {G}^{-3} \nabla^{\beta \rho}{G_{j k}} \nabla_{\alpha}\,^{\lambda}{G^{j k}} \varphi_{i \lambda}+\frac{3}{64}(\Gamma_{a})^{\beta}{}_{\rho} W \boldsymbol{W} {G}^{-3} \nabla^{\rho}\,_{\alpha}{\varphi_{j}^{\lambda}} \varphi_{i \beta} \varphi^{j}_{\lambda}+\frac{3}{32}(\Gamma_{a})^{\beta \rho} W \boldsymbol{W} {G}^{-3} \nabla_{\alpha}\,^{\lambda}{\varphi_{j \beta}} \varphi_{i \rho} \varphi^{j}_{\lambda} - \frac{21}{320}(\Gamma_{a})_{\alpha}{}^{\lambda} W \boldsymbol{W} W^{\beta \rho} {G}^{-3} \varphi_{i \lambda} \varphi_{j \beta} \varphi^{j}_{\rho}+\frac{15}{128}(\Gamma_{a})^{\beta \lambda} W \boldsymbol{W} W_{\beta}\,^{\rho} {G}^{-3} \varphi_{j \alpha} \varphi_{i \lambda} \varphi^{j}_{\rho} - \frac{3}{80}(\Gamma_{a})_{\alpha}{}^{\beta} W \boldsymbol{W} {G}^{-3} \nabla^{\rho \lambda}{\varphi_{j \rho}} \varphi_{i \beta} \varphi^{j}_{\lambda}+\frac{3}{64}(\Gamma_{a})^{\beta}{}_{\rho} W \boldsymbol{W} {G}^{-3} \nabla^{\rho \lambda}{\varphi_{j \lambda}} \varphi^{j}_{\alpha} \varphi_{i \beta}%
+\frac{3}{64}{\rm i} \mathcal{H}^{\beta \lambda} (\Gamma_{a})_{\beta}{}^{\rho} \boldsymbol{W} \lambda_{j \alpha} {G}^{-3} \varphi_{i \rho} \varphi^{j}_{\lambda}+\frac{3}{64}{\rm i} \mathcal{H}^{\beta \lambda} (\Gamma_{a})_{\beta}{}^{\rho} W \boldsymbol{\lambda}_{j \alpha} {G}^{-3} \varphi_{i \rho} \varphi^{j}_{\lambda} - \frac{9}{64}\mathcal{H}^{\beta \lambda} G_{j k} (\Gamma_{a})_{\beta}{}^{\rho} W \boldsymbol{W} {G}^{-5} \varphi^{j}_{\alpha} \varphi_{i \rho} \varphi^{k}_{\lambda} - \frac{3}{128}{\rm i} \mathcal{H}^{\beta \lambda} (\Gamma_{a})_{\beta \rho} W \boldsymbol{W} {G}^{-3} \nabla^{\rho}\,_{\alpha}{G_{i j}} \varphi^{j}_{\lambda} - \frac{3}{32}{\rm i} (\Gamma_{a})^{\beta}{}_{\rho} \boldsymbol{W} \lambda_{k \alpha} {G}^{-3} \nabla^{\rho \lambda}{G_{i j}} \varphi^{k}_{\lambda} \varphi^{j}_{\beta} - \frac{3}{32}{\rm i} (\Gamma_{a})^{\beta}{}_{\rho} W \boldsymbol{\lambda}_{k \alpha} {G}^{-3} \nabla^{\rho \lambda}{G_{i j}} \varphi^{k}_{\lambda} \varphi^{j}_{\beta}+\frac{9}{32}G_{j k} (\Gamma_{a})^{\beta}{}_{\rho} W \boldsymbol{W} {G}^{-5} \nabla^{\rho \lambda}{G_{i l}} \varphi^{j}_{\alpha} \varphi^{k}_{\lambda} \varphi^{l}_{\beta} - \frac{15}{64}(\Gamma_{a})^{\beta}{}_{\rho} W \boldsymbol{W} {G}^{-3} \nabla^{\rho \lambda}{\varphi_{j \alpha}} \varphi_{i \lambda} \varphi^{j}_{\beta} - \frac{21}{64}(\Gamma_{a})^{\beta \lambda} W \boldsymbol{W} W_{\beta}\,^{\rho} {G}^{-3} \varphi_{i \alpha} \varphi_{j \lambda} \varphi^{j}_{\rho}+\frac{231}{5120}G_{i j} (\Gamma_{a})_{\alpha}{}^{\rho} W \boldsymbol{W} X_{k}^{\beta} {G}^{-3} \varphi^{j}_{\beta} \varphi^{k}_{\rho} - \frac{63}{1024}G_{i j} (\Gamma_{a})^{\beta \rho} W \boldsymbol{W} X_{k \beta} {G}^{-3} \varphi^{j}_{\alpha} \varphi^{k}_{\rho} - \frac{15}{64}(\Gamma_{a})^{\beta \rho} W \boldsymbol{W} {G}^{-3} \nabla_{\alpha}\,^{\lambda}{\varphi_{j \beta}} \varphi_{i \lambda} \varphi^{j}_{\rho} - \frac{3}{64}{\rm i} \mathcal{H}^{\beta \lambda} (\Gamma_{a})_{\beta}{}^{\rho} \boldsymbol{W} \lambda_{j \alpha} {G}^{-3} \varphi_{i \lambda} \varphi^{j}_{\rho} - \frac{3}{64}{\rm i} \mathcal{H}^{\beta \lambda} (\Gamma_{a})_{\beta}{}^{\rho} W \boldsymbol{\lambda}_{j \alpha} {G}^{-3} \varphi_{i \lambda} \varphi^{j}_{\rho}+\frac{9}{64}\mathcal{H}^{\beta \lambda} G_{j k} (\Gamma_{a})_{\beta}{}^{\rho} W \boldsymbol{W} {G}^{-5} \varphi^{j}_{\alpha} \varphi_{i \lambda} \varphi^{k}_{\rho}+\frac{3}{64}{\rm i} \mathcal{H}_{\lambda}\,^{\beta} (\Gamma_{a})_{\beta}{}^{\rho} W \boldsymbol{W} {G}^{-3} \nabla^{\lambda}\,_{\alpha}{G_{i j}} \varphi^{j}_{\rho} - \frac{3}{32}{\rm i} (\Gamma_{a})^{\beta}{}_{\rho} \boldsymbol{W} \lambda_{j \alpha} {G}^{-3} \nabla^{\rho \lambda}{G^{j}\,_{k}} \varphi_{i \lambda} \varphi^{k}_{\beta} - \frac{3}{32}{\rm i} (\Gamma_{a})^{\beta}{}_{\rho} W \boldsymbol{\lambda}_{j \alpha} {G}^{-3} \nabla^{\rho \lambda}{G^{j}\,_{k}} \varphi_{i \lambda} \varphi^{k}_{\beta}+\frac{9}{32}G_{j k} (\Gamma_{a})^{\beta}{}_{\rho} W \boldsymbol{W} {G}^{-5} \nabla^{\rho \lambda}{G^{j}\,_{l}} \varphi^{k}_{\alpha} \varphi_{i \lambda} \varphi^{l}_{\beta} - \frac{3}{32}{\rm i} (\Gamma_{a})^{\beta}{}_{\rho} W \boldsymbol{W} {G}^{-3} \nabla^{\rho}\,_{\lambda}{G_{j k}} \nabla_{\alpha}\,^{\lambda}{G_{i}\,^{j}} \varphi^{k}_{\beta}%
 - \frac{3}{64}{\rm i} (\Gamma_{a})_{\beta \rho} W \boldsymbol{W} {G}^{-3} \nabla^{\beta}\,_{\alpha}{G_{j k}} \nabla^{\rho \lambda}{G^{j k}} \varphi_{i \lambda} - \frac{3}{128}(\Gamma_{a})_{\beta \rho} W \boldsymbol{W} {G}^{-3} \nabla^{\beta \rho}{\varphi_{i}^{\lambda}} \varphi_{j \alpha} \varphi^{j}_{\lambda}+\frac{3}{64}{\rm i} G_{j k} (\Gamma_{a})_{\beta \rho} \boldsymbol{W} \lambda^{j}_{\alpha} {G}^{-3} \nabla^{\beta \rho}{\varphi_{i}^{\lambda}} \varphi^{k}_{\lambda}+\frac{3}{64}{\rm i} G_{j k} (\Gamma_{a})_{\beta \rho} W \boldsymbol{\lambda}^{j}_{\alpha} {G}^{-3} \nabla^{\beta \rho}{\varphi_{i}^{\lambda}} \varphi^{k}_{\lambda} - \frac{9}{128}{\rm i} G_{i j} (\Gamma_{a})_{\beta \rho} W \boldsymbol{W} {G}^{-3} \nabla^{\beta \rho}{\mathcal{H}_{\alpha}\,^{\lambda}} \varphi^{j}_{\lambda} - \frac{3}{64}{\rm i} G_{j k} (\Gamma_{a})_{\beta \rho} W \boldsymbol{W} {G}^{-3} \nabla^{\beta \rho}{\nabla_{\alpha}\,^{\lambda}{G_{i}\,^{j}}} \varphi^{k}_{\lambda} - \frac{261}{1280}{\rm i} \mathcal{H}^{\beta \lambda} G_{i j} (\Gamma_{a})_{\alpha}{}^{\rho} W \boldsymbol{W} W_{\beta \rho} {G}^{-3} \varphi^{j}_{\lambda} - \frac{33}{160}{\rm i} G_{j k} (\Gamma_{a})_{\alpha}{}^{\beta} W \boldsymbol{W} W_{\beta \rho} {G}^{-3} \nabla^{\rho \lambda}{G_{i}\,^{j}} \varphi^{k}_{\lambda} - \frac{927}{5120}G_{j k} (\Gamma_{a})_{\alpha}{}^{\rho} W \boldsymbol{W} X^{j \beta} {G}^{-3} \varphi_{i \rho} \varphi^{k}_{\beta}+\frac{243}{640}{\rm i} (\Gamma_{a})_{\alpha}{}^{\beta} W \boldsymbol{W} W_{\beta}\,^{\rho} W_{\rho}\,^{\lambda} {G}^{-1} \varphi_{i \lambda}+\frac{27}{256}{\rm i} (\Gamma_{a})^{\beta \lambda} W \boldsymbol{W} W_{\beta}\,^{\rho} W_{\lambda \rho} {G}^{-1} \varphi_{i \alpha}+\frac{3}{64}{\rm i} G_{j k} (\Gamma_{a})_{\beta \rho} W \boldsymbol{W} {G}^{-3} \nabla^{\beta \rho}{\varphi_{i \lambda}} \nabla_{\alpha}\,^{\lambda}{G^{j k}} - \frac{3}{32}(\Gamma_{a})^{\beta}{}_{\rho} W \boldsymbol{W} {G}^{-3} \nabla^{\rho \lambda}{\varphi_{i \beta}} \varphi_{j \alpha} \varphi^{j}_{\lambda} - \frac{3}{32}{\rm i} G_{j k} (\Gamma_{a})^{\beta}{}_{\rho} \boldsymbol{W} \lambda^{j}_{\alpha} {G}^{-3} \nabla^{\rho \lambda}{\varphi_{i \beta}} \varphi^{k}_{\lambda} - \frac{3}{32}{\rm i} G_{j k} (\Gamma_{a})^{\beta}{}_{\rho} W \boldsymbol{\lambda}^{j}_{\alpha} {G}^{-3} \nabla^{\rho \lambda}{\varphi_{i \beta}} \varphi^{k}_{\lambda}+\frac{27}{640}{\rm i} G_{i j} (\Gamma_{a})_{\alpha \beta} W \boldsymbol{W} {G}^{-3} \nabla^{\beta \rho}{F} \varphi^{j}_{\rho}+\frac{9}{128}{\rm i} G_{i j} (\Gamma_{a})^{\beta}{}_{\rho} W \boldsymbol{W} {G}^{-3} \nabla^{\rho \lambda}{\mathcal{H}_{\alpha \beta}} \varphi^{j}_{\lambda} - \frac{9}{256}{\rm i} \mathcal{H}^{\lambda \beta} G_{i j} (\Gamma_{a})_{\alpha \lambda} W \boldsymbol{W} W_{\beta}\,^{\rho} {G}^{-3} \varphi^{j}_{\rho} - \frac{141}{640}{\rm i} G_{j k} (\Gamma_{a})_{\alpha \lambda} W \boldsymbol{W} W^{\beta}\,_{\rho} {G}^{-3} \nabla^{\lambda \rho}{G_{i}\,^{j}} \varphi^{k}_{\beta} - \frac{27}{256}{\rm i} \mathcal{H}^{\lambda \beta} G_{i j} (\Gamma_{a})_{\lambda}{}^{\rho} W \boldsymbol{W} W_{\beta \rho} {G}^{-3} \varphi^{j}_{\alpha}%
+\frac{27}{256}{\rm i} \mathcal{H}^{\rho \lambda} G_{i j} (\Gamma_{a})_{\rho \lambda} W \boldsymbol{W} W_{\alpha}\,^{\beta} {G}^{-3} \varphi^{j}_{\beta}+\frac{9}{128}{\rm i} G_{j k} (\Gamma_{a})_{\rho \lambda} W \boldsymbol{W} W_{\alpha}\,^{\beta} {G}^{-3} \nabla^{\rho \lambda}{G_{i}\,^{j}} \varphi^{k}_{\beta} - \frac{9}{128}{\rm i} G_{j k} (\Gamma_{a})^{\beta}{}_{\rho} W \boldsymbol{W} {G}^{-3} \nabla^{\rho}\,_{\lambda}{\varphi_{i \beta}} \nabla_{\alpha}\,^{\lambda}{G^{j k}}+\frac{3}{64}(\Gamma_{a})_{\beta \rho} W \boldsymbol{W} {G}^{-3} \nabla^{\beta \rho}{\varphi_{j}^{\lambda}} \varphi_{i \alpha} \varphi^{j}_{\lambda} - \frac{3}{64}{\rm i} G_{i j} (\Gamma_{a})_{\beta \rho} \boldsymbol{W} \lambda_{k \alpha} {G}^{-3} \nabla^{\beta \rho}{\varphi^{k \lambda}} \varphi^{j}_{\lambda} - \frac{3}{64}{\rm i} G_{i j} (\Gamma_{a})_{\beta \rho} W \boldsymbol{\lambda}_{k \alpha} {G}^{-3} \nabla^{\beta \rho}{\varphi^{k \lambda}} \varphi^{j}_{\lambda} - \frac{9}{64}G_{i j} G_{k l} (\Gamma_{a})_{\beta \rho} W \boldsymbol{W} {G}^{-5} \nabla^{\beta \rho}{\varphi^{k \lambda}} \varphi^{l}_{\alpha} \varphi^{j}_{\lambda}+\frac{3}{256}{\rm i} F G_{i j} (\Gamma_{a})_{\beta \rho} W \boldsymbol{W} {G}^{-3} \nabla^{\beta \rho}{\varphi^{j}_{\alpha}} - \frac{15}{256}{\rm i} \mathcal{H}_{\alpha}\,^{\lambda} G_{i j} (\Gamma_{a})_{\beta \rho} W \boldsymbol{W} {G}^{-3} \nabla^{\beta \rho}{\varphi^{j}_{\lambda}}+\frac{3}{64}{\rm i} G_{i j} (\Gamma_{a})_{\beta \rho} W \boldsymbol{W} {G}^{-3} \nabla^{\beta \rho}{\varphi_{k \lambda}} \nabla_{\alpha}\,^{\lambda}{G^{j k}} - \frac{3}{32}(\Gamma_{a})^{\beta}{}_{\rho} W \boldsymbol{W} {G}^{-3} \nabla^{\rho \lambda}{\varphi_{j \beta}} \varphi_{i \alpha} \varphi^{j}_{\lambda}+\frac{3}{32}{\rm i} G_{i j} (\Gamma_{a})^{\beta}{}_{\rho} \boldsymbol{W} \lambda_{k \alpha} {G}^{-3} \nabla^{\rho \lambda}{\varphi^{k}_{\beta}} \varphi^{j}_{\lambda}+\frac{3}{32}{\rm i} G_{i j} (\Gamma_{a})^{\beta}{}_{\rho} W \boldsymbol{\lambda}_{k \alpha} {G}^{-3} \nabla^{\rho \lambda}{\varphi^{k}_{\beta}} \varphi^{j}_{\lambda}+\frac{9}{32}G_{i j} G_{k l} (\Gamma_{a})^{\beta}{}_{\rho} W \boldsymbol{W} {G}^{-5} \nabla^{\rho \lambda}{\varphi^{k}_{\beta}} \varphi^{l}_{\alpha} \varphi^{j}_{\lambda} - \frac{3}{32}{\rm i} F G_{i j} (\Gamma_{a})^{\beta}{}_{\rho} W \boldsymbol{W} {G}^{-3} \nabla^{\rho}\,_{\alpha}{\varphi^{j}_{\beta}}+\frac{3}{64}{\rm i} \mathcal{H}_{\alpha \lambda} G_{i j} (\Gamma_{a})^{\beta}{}_{\rho} W \boldsymbol{W} {G}^{-3} \nabla^{\lambda \rho}{\varphi^{j}_{\beta}} - \frac{9}{128}{\rm i} G_{i j} (\Gamma_{a})^{\beta}{}_{\rho} W \boldsymbol{W} {G}^{-3} \nabla^{\rho}\,_{\lambda}{\varphi_{k \beta}} \nabla_{\alpha}\,^{\lambda}{G^{j k}} - \frac{45}{64}G_{i j} G_{k l} (\Gamma_{a})^{\beta \rho} W \boldsymbol{W} {G}^{-5} \nabla_{\alpha}\,^{\lambda}{\varphi^{k}_{\beta}} \varphi^{j}_{\rho} \varphi^{l}_{\lambda} - \frac{135}{128}G_{i j} G_{k l} (\Gamma_{a})^{\beta \lambda} W \boldsymbol{W} W_{\beta}\,^{\rho} {G}^{-5} \varphi^{k}_{\alpha} \varphi^{j}_{\lambda} \varphi^{l}_{\rho} - \frac{99}{128}G_{i j} G_{k l} (\Gamma_{a})^{\rho \lambda} W \boldsymbol{W} W_{\alpha}\,^{\beta} {G}^{-5} \varphi^{j}_{\rho} \varphi^{k}_{\lambda} \varphi^{l}_{\beta}%
+\frac{9}{64}\mathcal{H}^{\beta \lambda} G_{i j} (\Gamma_{a})_{\beta}{}^{\rho} W \boldsymbol{W} {G}^{-5} \varphi_{k \alpha} \varphi^{j}_{\lambda} \varphi^{k}_{\rho} - \frac{9}{32}\mathcal{H}^{\beta \lambda} G_{j k} (\Gamma_{a})_{\beta}{}^{\rho} W \boldsymbol{W} {G}^{-5} \varphi_{i \alpha} \varphi^{j}_{\lambda} \varphi^{k}_{\rho} - \frac{9}{64}\mathcal{H}^{\beta \lambda} G_{i j} (\Gamma_{a})_{\beta}{}^{\rho} W \boldsymbol{W} {G}^{-5} \varphi_{k \alpha} \varphi^{j}_{\rho} \varphi^{k}_{\lambda} - \frac{9}{64}{\rm i} \mathcal{H}^{\beta \lambda} G_{i j} G_{k l} (\Gamma_{a})_{\beta}{}^{\rho} \boldsymbol{W} \lambda^{k}_{\alpha} {G}^{-5} \varphi^{j}_{\rho} \varphi^{l}_{\lambda} - \frac{9}{64}{\rm i} \mathcal{H}^{\beta \lambda} G_{i j} G_{k l} (\Gamma_{a})_{\beta}{}^{\rho} W \boldsymbol{\lambda}^{k}_{\alpha} {G}^{-5} \varphi^{j}_{\rho} \varphi^{l}_{\lambda} - \frac{9}{128}{\rm i} \mathcal{H}^{\beta \lambda} G_{i j} G_{k l} (\Gamma_{a})_{\beta \rho} W \boldsymbol{W} {G}^{-5} \nabla^{\rho}\,_{\alpha}{G^{j k}} \varphi^{l}_{\lambda} - \frac{9}{64}{\rm i} \mathcal{H}_{\lambda}\,^{\beta} G_{i j} G_{k l} (\Gamma_{a})_{\beta}{}^{\rho} W \boldsymbol{W} {G}^{-5} \nabla^{\lambda}\,_{\alpha}{G^{k l}} \varphi^{j}_{\rho}+\frac{9}{64}G_{i j} G_{k l} (\Gamma_{a})^{\beta}{}_{\rho} W \boldsymbol{W} {G}^{-5} \nabla^{\rho}\,_{\alpha}{\varphi^{k \lambda}} \varphi^{j}_{\lambda} \varphi^{l}_{\beta}+\frac{9}{32}G_{i j} G_{k l} (\Gamma_{a})^{\beta \rho} W \boldsymbol{W} {G}^{-5} \nabla_{\alpha}\,^{\lambda}{\varphi^{k}_{\beta}} \varphi^{j}_{\lambda} \varphi^{l}_{\rho}+\frac{63}{320}G_{i j} G_{k l} (\Gamma_{a})_{\alpha}{}^{\lambda} W \boldsymbol{W} W^{\beta \rho} {G}^{-5} \varphi^{j}_{\beta} \varphi^{k}_{\lambda} \varphi^{l}_{\rho}+\frac{9}{32}G_{i j} G_{k l} (\Gamma_{a})^{\beta \lambda} W \boldsymbol{W} W_{\beta}\,^{\rho} {G}^{-5} \varphi^{j}_{\alpha} \varphi^{k}_{\lambda} \varphi^{l}_{\rho}+\frac{9}{64}G_{i j} G_{k l} (\Gamma_{a})^{\beta \rho} W \boldsymbol{W} W_{\alpha \beta} {G}^{-5} \varphi^{j \lambda} \varphi^{k}_{\rho} \varphi^{l}_{\lambda} - \frac{9}{128}G_{i j} G_{k l} (\Gamma_{a})^{\rho \lambda} W \boldsymbol{W} W_{\alpha}\,^{\beta} {G}^{-5} \varphi^{j}_{\beta} \varphi^{k}_{\rho} \varphi^{l}_{\lambda} - \frac{9}{80}G_{i j} G_{k l} (\Gamma_{a})_{\alpha}{}^{\beta} W \boldsymbol{W} {G}^{-5} \nabla^{\rho \lambda}{\varphi^{k}_{\rho}} \varphi^{j}_{\lambda} \varphi^{l}_{\beta} - \frac{9}{64}G_{i j} G_{k l} (\Gamma_{a})^{\beta}{}_{\rho} W \boldsymbol{W} {G}^{-5} \nabla^{\rho \lambda}{\varphi^{k}_{\lambda}} \varphi^{j}_{\alpha} \varphi^{l}_{\beta}+\frac{9}{64}{\rm i} \mathcal{H}^{\beta \lambda} G_{i j} G_{k l} (\Gamma_{a})_{\beta}{}^{\rho} \boldsymbol{W} \lambda^{k}_{\alpha} {G}^{-5} \varphi^{j}_{\lambda} \varphi^{l}_{\rho}+\frac{9}{64}{\rm i} \mathcal{H}^{\beta \lambda} G_{i j} G_{k l} (\Gamma_{a})_{\beta}{}^{\rho} W \boldsymbol{\lambda}^{k}_{\alpha} {G}^{-5} \varphi^{j}_{\lambda} \varphi^{l}_{\rho} - \frac{9}{64}{\rm i} \mathcal{H}_{\lambda}\,^{\beta} G_{i j} G_{k l} (\Gamma_{a})_{\beta}{}^{\rho} W \boldsymbol{W} {G}^{-5} \nabla^{\lambda}\,_{\alpha}{G^{j k}} \varphi^{l}_{\rho}+\frac{3}{128}{\rm i} \mathcal{H}^{\beta \lambda} G_{i j} (\Gamma_{a})_{\beta \rho} W \boldsymbol{W} {G}^{-3} \nabla^{\rho}\,_{\alpha}{\varphi^{j}_{\lambda}}+\frac{21}{128}{\rm i} \mathcal{H}_{\lambda}\,^{\beta} G_{i j} (\Gamma_{a})_{\beta}{}^{\rho} W \boldsymbol{W} {G}^{-3} \nabla^{\lambda}\,_{\alpha}{\varphi^{j}_{\rho}}%
+\frac{9}{64}{\rm i} \mathcal{H}_{\alpha}\,^{\lambda} G_{i j} (\Gamma_{a})_{\lambda}{}^{\beta} W \boldsymbol{W} W_{\beta}\,^{\rho} {G}^{-3} \varphi^{j}_{\rho}+\frac{3}{128}{\rm i} \mathcal{H}_{\rho}\,^{\beta} G_{i j} (\Gamma_{a})_{\alpha \beta} W \boldsymbol{W} {G}^{-3} \nabla^{\rho \lambda}{\varphi^{j}_{\lambda}}+\frac{3}{128}\mathcal{H}^{\lambda \beta} \mathcal{H}_{\lambda}\,^{\rho} G_{i j} (\Gamma_{a})_{\beta \rho} \boldsymbol{W} \lambda^{j}_{\alpha} {G}^{-3}+\frac{3}{128}\mathcal{H}^{\lambda \beta} \mathcal{H}_{\lambda}\,^{\rho} G_{i j} (\Gamma_{a})_{\beta \rho} W \boldsymbol{\lambda}^{j}_{\alpha} {G}^{-3}+\frac{15}{64}{\rm i} G_{j k} (\Gamma_{a})^{\beta}{}_{\rho} W \boldsymbol{W} {G}^{-3} \nabla_{\alpha \lambda}{\varphi^{j}_{\beta}} \nabla^{\rho \lambda}{G_{i}\,^{k}}+\frac{3}{16}{\rm i} G_{j k} (\Gamma_{a})^{\beta}{}_{\lambda} W \boldsymbol{W} W_{\beta}\,^{\rho} {G}^{-3} \nabla^{\lambda}\,_{\alpha}{G_{i}\,^{j}} \varphi^{k}_{\rho} - \frac{3}{32}{\rm i} G_{j k} (\Gamma_{a})_{\beta \rho} W \boldsymbol{W} {G}^{-3} \nabla^{\beta \lambda}{\varphi^{j}_{\lambda}} \nabla^{\rho}\,_{\alpha}{G_{i}\,^{k}}+\frac{3}{64}{\rm i} \mathcal{H}_{\lambda}\,^{\beta} (\Gamma_{a})_{\beta \rho} W \boldsymbol{W} {G}^{-3} \nabla^{\lambda \rho}{G_{i j}} \varphi^{j}_{\alpha} - \frac{3}{64}\mathcal{H}_{\lambda}\,^{\beta} G_{j k} (\Gamma_{a})_{\beta \rho} \boldsymbol{W} \lambda^{j}_{\alpha} {G}^{-3} \nabla^{\lambda \rho}{G_{i}\,^{k}} - \frac{3}{64}\mathcal{H}_{\lambda}\,^{\beta} G_{j k} (\Gamma_{a})_{\beta \rho} W \boldsymbol{\lambda}^{j}_{\alpha} {G}^{-3} \nabla^{\lambda \rho}{G_{i}\,^{k}} - \frac{15}{128}{\rm i} \mathcal{H}_{\lambda}\,^{\beta} G_{i j} (\Gamma_{a})_{\beta \rho} W \boldsymbol{W} {G}^{-3} \nabla^{\lambda \rho}{\varphi^{j}_{\alpha}} - \frac{27}{128}{\rm i} \mathcal{H}^{\beta \rho} (\Gamma_{a})_{\beta \rho} W \boldsymbol{W} X_{i \alpha} {G}^{-1}+\frac{27}{320}{\rm i} \mathcal{H}^{\rho \beta} (\Gamma_{a})_{\alpha \rho} W \boldsymbol{W} X_{i \beta} {G}^{-1} - \frac{3}{32}{\rm i} G_{i j} (\Gamma_{a})^{\beta}{}_{\rho} W \boldsymbol{W} {G}^{-3} \nabla_{\alpha \lambda}{\varphi_{k \beta}} \nabla^{\rho \lambda}{G^{j k}}+\frac{3}{64}{\rm i} G_{i j} (\Gamma_{a})_{\beta \rho} W \boldsymbol{W} {G}^{-3} \nabla^{\beta}\,_{\alpha}{\varphi_{k \lambda}} \nabla^{\rho \lambda}{G^{j k}}+\frac{3}{64}{\rm i} G_{i j} (\Gamma_{a})^{\beta}{}_{\lambda} W \boldsymbol{W} W_{\beta}\,^{\rho} {G}^{-3} \nabla^{\lambda}\,_{\alpha}{G^{j}\,_{k}} \varphi^{k}_{\rho}+\frac{3}{320}{\rm i} G_{i j} (\Gamma_{a})_{\alpha \lambda} W \boldsymbol{W} W^{\beta}\,_{\rho} {G}^{-3} \nabla^{\lambda \rho}{G^{j}\,_{k}} \varphi^{k}_{\beta}+\frac{3}{64}{\rm i} G_{i j} (\Gamma_{a})_{\rho \lambda} W \boldsymbol{W} W_{\alpha}\,^{\beta} {G}^{-3} \nabla^{\rho \lambda}{G^{j}\,_{k}} \varphi^{k}_{\beta}+\frac{3}{128}{\rm i} G_{i j} (\Gamma_{a})^{\rho}{}_{\lambda} W \boldsymbol{W} W_{\alpha \beta} {G}^{-3} \nabla^{\lambda \beta}{G^{j}\,_{k}} \varphi^{k}_{\rho} - \frac{3}{64}{\rm i} G_{i j} (\Gamma_{a})^{\beta}{}_{\rho} W \boldsymbol{W} W_{\alpha \beta} {G}^{-3} \nabla^{\rho \lambda}{G^{j}\,_{k}} \varphi^{k}_{\lambda}%
 - \frac{21}{640}{\rm i} G_{i j} (\Gamma_{a})_{\alpha \beta} W \boldsymbol{W} {G}^{-3} \nabla_{\rho}\,^{\lambda}{\varphi_{k \lambda}} \nabla^{\beta \rho}{G^{j k}}+\frac{3}{64}\mathcal{H}_{\lambda}\,^{\beta} G_{i j} (\Gamma_{a})_{\beta \rho} \boldsymbol{W} \lambda_{k \alpha} {G}^{-3} \nabla^{\lambda \rho}{G^{j k}}+\frac{3}{64}\mathcal{H}_{\lambda}\,^{\beta} G_{i j} (\Gamma_{a})_{\beta \rho} W \boldsymbol{\lambda}_{k \alpha} {G}^{-3} \nabla^{\lambda \rho}{G^{j k}}+\frac{9}{64}{\rm i} \mathcal{H}_{\lambda}\,^{\beta} G_{i j} G_{k l} (\Gamma_{a})_{\beta \rho} W \boldsymbol{W} {G}^{-5} \nabla^{\lambda \rho}{G^{j k}} \varphi^{l}_{\alpha} - \frac{9}{32}G_{j k} (\Gamma_{a})^{\beta}{}_{\rho} W \boldsymbol{W} {G}^{-5} \nabla^{\rho \lambda}{G_{i}\,^{j}} \varphi_{l \alpha} \varphi^{k}_{\lambda} \varphi^{l}_{\beta}+\frac{9}{32}G_{j k} (\Gamma_{a})^{\beta}{}_{\rho} W \boldsymbol{W} {G}^{-5} \nabla^{\rho \lambda}{G^{j}\,_{l}} \varphi_{i \alpha} \varphi^{k}_{\lambda} \varphi^{l}_{\beta}+\frac{9}{32}G_{i j} (\Gamma_{a})^{\beta}{}_{\rho} W \boldsymbol{W} {G}^{-5} \nabla^{\rho \lambda}{G_{k l}} \varphi^{k}_{\alpha} \varphi^{j}_{\beta} \varphi^{l}_{\lambda} - \frac{9}{32}{\rm i} G_{i j} G_{k l} (\Gamma_{a})^{\beta}{}_{\rho} \boldsymbol{W} \lambda_{m \alpha} {G}^{-5} \nabla^{\rho \lambda}{G^{k m}} \varphi^{j}_{\beta} \varphi^{l}_{\lambda} - \frac{9}{32}{\rm i} G_{i j} G_{k l} (\Gamma_{a})^{\beta}{}_{\rho} W \boldsymbol{\lambda}_{m \alpha} {G}^{-5} \nabla^{\rho \lambda}{G^{k m}} \varphi^{j}_{\beta} \varphi^{l}_{\lambda}+\frac{45}{32}G_{i j} G_{k l} G_{m n} (\Gamma_{a})^{\beta}{}_{\rho} W \boldsymbol{W} {G}^{-7} \nabla^{\rho \lambda}{G^{k m}} \varphi^{l}_{\alpha} \varphi^{j}_{\beta} \varphi^{n}_{\lambda} - \frac{45}{64}G_{i j} G_{k l} (\Gamma_{a})^{\beta}{}_{\rho} W \boldsymbol{W} {G}^{-5} \nabla^{\rho \lambda}{\varphi^{k}_{\alpha}} \varphi^{j}_{\beta} \varphi^{l}_{\lambda}+\frac{9}{640}{\rm i} F G_{i j} G_{k l} (\Gamma_{a})_{\alpha \beta} W \boldsymbol{W} {G}^{-5} \nabla^{\beta \rho}{G^{j k}} \varphi^{l}_{\rho}+\frac{9}{64}{\rm i} \mathcal{H}_{\alpha}\,^{\beta} G_{i j} G_{k l} (\Gamma_{a})_{\beta \rho} W \boldsymbol{W} {G}^{-5} \nabla^{\rho \lambda}{G^{j k}} \varphi^{l}_{\lambda} - \frac{9}{64}{\rm i} G_{i j} G_{k l} (\Gamma_{a})_{\beta \rho} W \boldsymbol{W} {G}^{-5} \nabla^{\beta}\,_{\alpha}{G^{j}\,_{m}} \nabla^{\rho \lambda}{G^{k m}} \varphi^{l}_{\lambda} - \frac{9}{64}{\rm i} \mathcal{H}_{\alpha \lambda} G_{i j} G_{k l} (\Gamma_{a})^{\beta}{}_{\rho} W \boldsymbol{W} {G}^{-5} \nabla^{\lambda \rho}{G^{k l}} \varphi^{j}_{\beta}+\frac{9}{32}{\rm i} G_{i j} G_{k l} (\Gamma_{a})^{\beta}{}_{\rho} W \boldsymbol{W} {G}^{-5} \nabla^{\rho}\,_{\lambda}{G^{k}\,_{m}} \nabla_{\alpha}\,^{\lambda}{G^{l m}} \varphi^{j}_{\beta} - \frac{9}{32}G_{j k} (\Gamma_{a})^{\beta}{}_{\rho} W \boldsymbol{W} {G}^{-5} \nabla^{\rho \lambda}{G_{i}\,^{j}} \varphi_{l \alpha} \varphi^{k}_{\beta} \varphi^{l}_{\lambda}+\frac{9}{32}G_{j k} (\Gamma_{a})^{\beta}{}_{\rho} W \boldsymbol{W} {G}^{-5} \nabla^{\rho \lambda}{G_{i l}} \varphi^{l}_{\alpha} \varphi^{j}_{\beta} \varphi^{k}_{\lambda}+\frac{9}{32}{\rm i} G_{j k} G_{l m} (\Gamma_{a})^{\beta}{}_{\rho} \boldsymbol{W} \lambda^{j}_{\alpha} {G}^{-5} \nabla^{\rho \lambda}{G_{i}\,^{l}} \varphi^{k}_{\beta} \varphi^{m}_{\lambda}+\frac{9}{32}{\rm i} G_{j k} G_{l m} (\Gamma_{a})^{\beta}{}_{\rho} W \boldsymbol{\lambda}^{j}_{\alpha} {G}^{-5} \nabla^{\rho \lambda}{G_{i}\,^{l}} \varphi^{k}_{\beta} \varphi^{m}_{\lambda}%
+\frac{9}{32}{\rm i} G_{j k} G_{l m} (\Gamma_{a})^{\beta}{}_{\rho} W \boldsymbol{W} {G}^{-5} \nabla^{\rho}\,_{\lambda}{G_{i}\,^{j}} \nabla_{\alpha}\,^{\lambda}{G^{k l}} \varphi^{m}_{\beta}+\frac{27}{128}{\rm i} G_{i j} G_{k l} (\Gamma_{a})^{\beta \rho} \boldsymbol{W} \lambda^{j}_{\alpha} X^{k}_{\beta} {G}^{-3} \varphi^{l}_{\rho}+\frac{27}{128}{\rm i} G_{i j} G_{k l} (\Gamma_{a})^{\beta \rho} W \boldsymbol{\lambda}^{j}_{\alpha} X^{k}_{\beta} {G}^{-3} \varphi^{l}_{\rho} - \frac{3}{32}{\rm i} (\Gamma_{a})_{\beta \rho} W \boldsymbol{W} {G}^{-3} \nabla^{\beta}\,_{\lambda}{G_{i j}} \nabla^{\rho \lambda}{G^{j}\,_{k}} \varphi^{k}_{\alpha}+\frac{3}{32}G_{j k} (\Gamma_{a})_{\beta \rho} \boldsymbol{W} \lambda_{l \alpha} {G}^{-3} \nabla^{\beta}\,_{\lambda}{G_{i}\,^{j}} \nabla^{\rho \lambda}{G^{k l}}+\frac{3}{32}G_{j k} (\Gamma_{a})_{\beta \rho} W \boldsymbol{\lambda}_{l \alpha} {G}^{-3} \nabla^{\beta}\,_{\lambda}{G_{i}\,^{j}} \nabla^{\rho \lambda}{G^{k l}}+\frac{9}{32}{\rm i} G_{j k} G_{l m} (\Gamma_{a})_{\beta \rho} W \boldsymbol{W} {G}^{-5} \nabla^{\beta}\,_{\lambda}{G_{i}\,^{j}} \nabla^{\rho \lambda}{G^{k l}} \varphi^{m}_{\alpha} - \frac{3}{32}{\rm i} G_{j k} (\Gamma_{a})_{\beta \rho} W \boldsymbol{W} {G}^{-3} \nabla^{\beta}\,_{\lambda}{\varphi^{j}_{\alpha}} \nabla^{\rho \lambda}{G_{i}\,^{k}} - \frac{3}{32}{\rm i} G_{j k} (\Gamma_{a})_{\beta \rho} W \boldsymbol{W} {G}^{-3} \nabla^{\beta}\,_{\lambda}{\varphi_{i \alpha}} \nabla^{\rho \lambda}{G^{j k}} - \frac{3}{32}{\rm i} G_{j k} (\Gamma_{a})_{\rho \lambda} W \boldsymbol{W} W_{\alpha}\,^{\beta} {G}^{-3} \nabla^{\rho \lambda}{G^{j k}} \varphi_{i \beta} - \frac{3}{160}{\rm i} G_{j k} (\Gamma_{a})_{\alpha \lambda} W \boldsymbol{W} W^{\beta}\,_{\rho} {G}^{-3} \nabla^{\lambda \rho}{G^{j k}} \varphi_{i \beta} - \frac{3}{32}{\rm i} G_{j k} (\Gamma_{a})^{\beta}{}_{\lambda} W \boldsymbol{W} W_{\beta}\,^{\rho} {G}^{-3} \nabla^{\lambda}\,_{\alpha}{G^{j k}} \varphi_{i \rho}+\frac{3}{128}{\rm i} G_{j k} (\Gamma_{a})^{\rho}{}_{\lambda} W \boldsymbol{W} W_{\alpha \beta} {G}^{-3} \nabla^{\lambda \beta}{G^{j k}} \varphi_{i \rho} - \frac{3}{64}{\rm i} G_{j k} (\Gamma_{a})^{\beta}{}_{\rho} W \boldsymbol{W} W_{\alpha \beta} {G}^{-3} \nabla^{\rho \lambda}{G^{j k}} \varphi_{i \lambda} - \frac{9}{512}{\rm i} G_{i j} G_{k l} (\Gamma_{a})_{\beta \rho} W \boldsymbol{W} X^{j}_{\alpha} {G}^{-3} \nabla^{\beta \rho}{G^{k l}}+\frac{45}{512}{\rm i} G_{i j} G_{k l} (\Gamma_{a})_{\beta \rho} W \boldsymbol{W} X^{k}_{\alpha} {G}^{-3} \nabla^{\beta \rho}{G^{j l}} - \frac{9}{2560}{\rm i} G_{i j} G_{k l} (\Gamma_{a})_{\alpha \rho} W \boldsymbol{W} X^{j}_{\beta} {G}^{-3} \nabla^{\rho \beta}{G^{k l}}+\frac{9}{512}{\rm i} G_{i j} G_{k l} (\Gamma_{a})_{\alpha \rho} W \boldsymbol{W} X^{k}_{\beta} {G}^{-3} \nabla^{\rho \beta}{G^{j l}} - \frac{9}{64}{\rm i} G_{j k} (\Gamma_{a})_{\beta \rho} W \boldsymbol{W} {G}^{-3} \nabla^{\beta}\,_{\lambda}{G_{i}\,^{j}} \nabla^{\rho \lambda}{\varphi^{k}_{\alpha}} - \frac{9}{32}{\rm i} (\Gamma_{a})^{\beta \rho} W \boldsymbol{W} {G}^{-5} \varphi_{j \alpha} \varphi_{i \beta} \varphi^{j \lambda} \varphi_{k \rho} \varphi^{k}_{\lambda}%
 - \frac{9}{64}G_{j k} (\Gamma_{a})^{\beta \rho} \boldsymbol{W} \lambda_{l \alpha} {G}^{-5} \varphi_{i \beta} \varphi^{j \lambda} \varphi^{k}_{\lambda} \varphi^{l}_{\rho} - \frac{9}{64}G_{j k} (\Gamma_{a})^{\beta \rho} W \boldsymbol{\lambda}_{l \alpha} {G}^{-5} \varphi_{i \beta} \varphi^{j \lambda} \varphi^{k}_{\lambda} \varphi^{l}_{\rho} - \frac{45}{64}{\rm i} G_{j k} G_{l m} (\Gamma_{a})^{\beta \rho} W \boldsymbol{W} {G}^{-7} \varphi^{j}_{\alpha} \varphi_{i \beta} \varphi^{k}_{\rho} \varphi^{l \lambda} \varphi^{m}_{\lambda}+\frac{9}{320}F G_{j k} (\Gamma_{a})_{\alpha}{}^{\beta} W \boldsymbol{W} {G}^{-5} \varphi_{i \beta} \varphi^{j \rho} \varphi^{k}_{\rho} - \frac{9}{64}\mathcal{H}_{\alpha}\,^{\beta} G_{j k} (\Gamma_{a})_{\beta}{}^{\rho} W \boldsymbol{W} {G}^{-5} \varphi_{i \rho} \varphi^{j \lambda} \varphi^{k}_{\lambda}+\frac{9}{128}G_{j k} (\Gamma_{a})^{\beta}{}_{\rho} W \boldsymbol{W} {G}^{-5} \nabla^{\rho}\,_{\alpha}{G_{i l}} \varphi^{j \lambda} \varphi^{k}_{\lambda} \varphi^{l}_{\beta} - \frac{9}{64}F G_{j k} (\Gamma_{a})^{\beta \rho} W \boldsymbol{W} {G}^{-5} \varphi^{j}_{\alpha} \varphi_{i \beta} \varphi^{k}_{\rho}+\frac{9}{64}\mathcal{H}_{\alpha}\,^{\lambda} G_{j k} (\Gamma_{a})^{\beta \rho} W \boldsymbol{W} {G}^{-5} \varphi_{i \beta} \varphi^{j}_{\lambda} \varphi^{k}_{\rho}+\frac{9}{32}G_{j k} (\Gamma_{a})^{\beta \rho} W \boldsymbol{W} {G}^{-5} \nabla_{\alpha}\,^{\lambda}{G^{j}\,_{l}} \varphi_{i \beta} \varphi^{k}_{\lambda} \varphi^{l}_{\rho} - \frac{135}{64}{\rm i} G_{i j} G_{k l} (\Gamma_{a})^{\beta \rho} W \boldsymbol{W} {G}^{-7} \varphi_{m \alpha} \varphi^{j}_{\beta} \varphi^{k \lambda} \varphi^{l}_{\lambda} \varphi^{m}_{\rho}+\frac{45}{64}{\rm i} G_{j k} G_{l m} (\Gamma_{a})^{\beta \rho} W \boldsymbol{W} {G}^{-7} \varphi_{i \alpha} \varphi^{j}_{\beta} \varphi^{k}_{\rho} \varphi^{l \lambda} \varphi^{m}_{\lambda}+\frac{45}{32}{\rm i} G_{i j} G_{k l} (\Gamma_{a})^{\beta \rho} W \boldsymbol{W} {G}^{-7} \varphi_{m \alpha} \varphi^{j}_{\beta} \varphi^{k}_{\rho} \varphi^{l \lambda} \varphi^{m}_{\lambda}+\frac{45}{64}G_{i j} G_{k l} G_{m n} (\Gamma_{a})^{\beta \rho} \boldsymbol{W} \lambda^{k}_{\alpha} {G}^{-7} \varphi^{j}_{\beta} \varphi^{l}_{\rho} \varphi^{m \lambda} \varphi^{n}_{\lambda}+\frac{45}{64}G_{i j} G_{k l} G_{m n} (\Gamma_{a})^{\beta \rho} W \boldsymbol{\lambda}^{k}_{\alpha} {G}^{-7} \varphi^{j}_{\beta} \varphi^{l}_{\rho} \varphi^{m \lambda} \varphi^{n}_{\lambda} - \frac{45}{128}G_{i j} G_{k l} G_{m n} (\Gamma_{a})^{\beta}{}_{\rho} W \boldsymbol{W} {G}^{-7} \nabla^{\rho}\,_{\alpha}{G^{j k}} \varphi^{l}_{\beta} \varphi^{m \lambda} \varphi^{n}_{\lambda}+\frac{45}{32}G_{i j} G_{k l} G_{m n} (\Gamma_{a})^{\beta \rho} W \boldsymbol{W} {G}^{-7} \nabla_{\alpha}\,^{\lambda}{G^{k m}} \varphi^{j}_{\beta} \varphi^{l}_{\rho} \varphi^{n}_{\lambda}+\frac{9}{128}G_{i j} G_{k l} (\Gamma_{a})^{\beta}{}_{\rho} W \boldsymbol{W} {G}^{-5} \nabla^{\rho}\,_{\alpha}{\varphi^{j}_{\beta}} \varphi^{k \lambda} \varphi^{l}_{\lambda} - \frac{171}{320}G_{i j} G_{k l} (\Gamma_{a})_{\alpha}{}^{\beta} W \boldsymbol{W} W_{\beta}\,^{\rho} {G}^{-5} \varphi^{j}_{\rho} \varphi^{k \lambda} \varphi^{l}_{\lambda}+\frac{117}{640}G_{i j} G_{k l} (\Gamma_{a})_{\alpha \beta} W \boldsymbol{W} {G}^{-5} \nabla^{\beta \rho}{\varphi^{j}_{\rho}} \varphi^{k \lambda} \varphi^{l}_{\lambda} - \frac{9}{64}\mathcal{H}^{\beta \rho} G_{j k} (\Gamma_{a})_{\beta \rho} W \boldsymbol{W} {G}^{-5} \varphi_{i \alpha} \varphi^{j \lambda} \varphi^{k}_{\lambda}%
+\frac{9}{64}\mathcal{H}^{\beta \rho} G_{i j} (\Gamma_{a})_{\beta \rho} W \boldsymbol{W} {G}^{-5} \varphi_{k \alpha} \varphi^{j \lambda} \varphi^{k}_{\lambda} - \frac{9}{128}{\rm i} \mathcal{H}^{\beta \rho} G_{i j} G_{k l} (\Gamma_{a})_{\beta \rho} \boldsymbol{W} \lambda^{j}_{\alpha} {G}^{-5} \varphi^{k \lambda} \varphi^{l}_{\lambda} - \frac{9}{128}{\rm i} \mathcal{H}^{\beta \rho} G_{i j} G_{k l} (\Gamma_{a})_{\beta \rho} W \boldsymbol{\lambda}^{j}_{\alpha} {G}^{-5} \varphi^{k \lambda} \varphi^{l}_{\lambda} - \frac{9}{64}{\rm i} \mathcal{H}^{\beta \rho} G_{i j} G_{k l} (\Gamma_{a})_{\beta \rho} W \boldsymbol{W} {G}^{-5} \nabla_{\alpha}\,^{\lambda}{G^{j k}} \varphi^{l}_{\lambda}+\frac{45}{256}G_{j k} (\Gamma_{a})_{\beta \rho} W \boldsymbol{W} {G}^{-5} \nabla^{\beta \rho}{G_{i l}} \varphi^{l}_{\alpha} \varphi^{j \lambda} \varphi^{k}_{\lambda}+\frac{9}{64}G_{j k} (\Gamma_{a})_{\beta \rho} W \boldsymbol{W} {G}^{-5} \nabla^{\beta \rho}{G_{i}\,^{j}} \varphi_{l \alpha} \varphi^{k \lambda} \varphi^{l}_{\lambda} - \frac{9}{128}{\rm i} G_{j k} G_{l m} (\Gamma_{a})_{\beta \rho} \boldsymbol{W} \lambda^{j}_{\alpha} {G}^{-5} \nabla^{\beta \rho}{G_{i}\,^{k}} \varphi^{l \lambda} \varphi^{m}_{\lambda} - \frac{9}{128}{\rm i} G_{j k} G_{l m} (\Gamma_{a})_{\beta \rho} W \boldsymbol{\lambda}^{j}_{\alpha} {G}^{-5} \nabla^{\beta \rho}{G_{i}\,^{k}} \varphi^{l \lambda} \varphi^{m}_{\lambda}+\frac{9}{32}G_{i j} G_{k l} (\Gamma_{a})_{\beta \rho} W \boldsymbol{W} {G}^{-5} \nabla^{\beta \rho}{\varphi^{j}_{\alpha}} \varphi^{k \lambda} \varphi^{l}_{\lambda} - \frac{9}{64}{\rm i} G_{j k} G_{l m} (\Gamma_{a})_{\beta \rho} W \boldsymbol{W} {G}^{-5} \nabla^{\beta \rho}{G_{i}\,^{j}} \nabla_{\alpha}\,^{\lambda}{G^{k l}} \varphi^{m}_{\lambda}+\frac{9}{64}G_{i j} (\Gamma_{a})_{\beta \rho} W \boldsymbol{W} {G}^{-5} \nabla^{\beta \rho}{G^{j}\,_{k}} \varphi_{l \alpha} \varphi^{k \lambda} \varphi^{l}_{\lambda}+\frac{9}{128}{\rm i} G_{i j} G_{k l} (\Gamma_{a})_{\beta \rho} \boldsymbol{W} \lambda_{m \alpha} {G}^{-5} \nabla^{\beta \rho}{G^{j m}} \varphi^{k \lambda} \varphi^{l}_{\lambda}+\frac{9}{128}{\rm i} G_{i j} G_{k l} (\Gamma_{a})_{\beta \rho} W \boldsymbol{\lambda}_{m \alpha} {G}^{-5} \nabla^{\beta \rho}{G^{j m}} \varphi^{k \lambda} \varphi^{l}_{\lambda} - \frac{45}{128}G_{i j} G_{k l} G_{m n} (\Gamma_{a})_{\beta \rho} W \boldsymbol{W} {G}^{-7} \nabla^{\beta \rho}{G^{j k}} \varphi^{l}_{\alpha} \varphi^{m \lambda} \varphi^{n}_{\lambda}+\frac{9}{128}{\rm i} F G_{i j} G_{k l} (\Gamma_{a})_{\beta \rho} W \boldsymbol{W} {G}^{-5} \nabla^{\beta \rho}{G^{j k}} \varphi^{l}_{\alpha}+\frac{9}{128}{\rm i} \mathcal{H}_{\alpha}\,^{\lambda} G_{i j} G_{k l} (\Gamma_{a})_{\beta \rho} W \boldsymbol{W} {G}^{-5} \nabla^{\beta \rho}{G^{j k}} \varphi^{l}_{\lambda} - \frac{9}{64}{\rm i} G_{i j} G_{k l} (\Gamma_{a})_{\beta \rho} W \boldsymbol{W} {G}^{-5} \nabla^{\beta \rho}{G^{j}\,_{m}} \nabla_{\alpha}\,^{\lambda}{G^{k m}} \varphi^{l}_{\lambda}+\frac{3}{16}(\Gamma_{a})^{\beta \rho} W \mathbf{X}_{j k} {G}^{-3} \varphi^{j}_{\alpha} \varphi_{i \beta} \varphi^{k}_{\rho}+\frac{3}{32}{\rm i} G_{j k} (\Gamma_{a})^{\beta \rho} \mathbf{X}^{j k} \lambda_{l \alpha} {G}^{-3} \varphi_{i \beta} \varphi^{l}_{\rho}+\frac{3}{16}{\rm i} G_{j k} (\Gamma_{a})^{\beta \rho} W {G}^{-3} \nabla_{\alpha}\,^{\lambda}{\boldsymbol{\lambda}^{j}_{\lambda}} \varphi_{i \beta} \varphi^{k}_{\rho}%
+\frac{9}{32}{\rm i} G_{j k} (\Gamma_{a})^{\rho \lambda} W W_{\alpha}\,^{\beta} \boldsymbol{\lambda}^{j}_{\beta} {G}^{-3} \varphi_{i \rho} \varphi^{k}_{\lambda} - \frac{9}{32}G_{j k} G_{l m} (\Gamma_{a})^{\beta \rho} W \mathbf{X}^{j k} {G}^{-5} \varphi^{l}_{\alpha} \varphi_{i \beta} \varphi^{m}_{\rho}+\frac{3}{64}{\rm i} G_{j k} (\Gamma_{a})^{\beta}{}_{\rho} W \mathbf{X}^{j k} {G}^{-3} \nabla^{\rho}\,_{\alpha}{G_{i l}} \varphi^{l}_{\beta} - \frac{3}{16}(\Gamma_{a})^{\beta \rho} W \mathbf{X}_{i j} {G}^{-3} \varphi_{k \alpha} \varphi^{j}_{\beta} \varphi^{k}_{\rho}+\frac{3}{16}{\rm i} G_{j k} (\Gamma_{a})^{\beta \rho} \mathbf{X}_{i l} \lambda^{j}_{\alpha} {G}^{-3} \varphi^{k}_{\beta} \varphi^{l}_{\rho} - \frac{3}{16}{\rm i} G_{i j} (\Gamma_{a})^{\beta \rho} W {G}^{-3} \nabla_{\alpha}\,^{\lambda}{\boldsymbol{\lambda}_{k \lambda}} \varphi^{j}_{\beta} \varphi^{k}_{\rho} - \frac{3}{16}{\rm i} G_{j k} (\Gamma_{a})^{\beta \rho} W {G}^{-3} \nabla_{\alpha}\,^{\lambda}{\boldsymbol{\lambda}_{i \lambda}} \varphi^{j}_{\beta} \varphi^{k}_{\rho} - \frac{9}{32}{\rm i} G_{i j} (\Gamma_{a})^{\rho \lambda} W W_{\alpha}\,^{\beta} \boldsymbol{\lambda}_{k \beta} {G}^{-3} \varphi^{j}_{\rho} \varphi^{k}_{\lambda} - \frac{9}{32}{\rm i} G_{j k} (\Gamma_{a})^{\rho \lambda} W W_{\alpha}\,^{\beta} \boldsymbol{\lambda}_{i \beta} {G}^{-3} \varphi^{j}_{\rho} \varphi^{k}_{\lambda} - \frac{3}{32}{\rm i} \mathcal{H}_{\alpha}\,^{\beta} G_{j k} (\Gamma_{a})_{\beta}{}^{\rho} W \mathbf{X}_{i}\,^{j} {G}^{-3} \varphi^{k}_{\rho} - \frac{3}{32}{\rm i} G_{j k} (\Gamma_{a})^{\beta}{}_{\rho} W \mathbf{X}_{i l} {G}^{-3} \nabla^{\rho}\,_{\alpha}{G^{j l}} \varphi^{k}_{\beta} - \frac{3}{16}(\Gamma_{a})^{\beta \rho} W \mathbf{X}_{j k} {G}^{-3} \varphi_{i \alpha} \varphi^{j}_{\beta} \varphi^{k}_{\rho}+\frac{9}{32}{\rm i} G_{i j} (\Gamma_{a})^{\beta \rho} \mathbf{X}_{k l} \lambda^{k}_{\alpha} {G}^{-3} \varphi^{j}_{\beta} \varphi^{l}_{\rho}+\frac{9}{16}G_{i j} G_{k l} (\Gamma_{a})^{\beta \rho} W \mathbf{X}^{k}\,_{m} {G}^{-5} \varphi^{l}_{\alpha} \varphi^{j}_{\beta} \varphi^{m}_{\rho}+\frac{3}{32}{\rm i} \mathcal{H}_{\alpha}\,^{\beta} G_{i j} (\Gamma_{a})_{\beta}{}^{\rho} W \mathbf{X}^{j}\,_{k} {G}^{-3} \varphi^{k}_{\rho} - \frac{3}{32}{\rm i} G_{i j} (\Gamma_{a})^{\beta}{}_{\rho} W \mathbf{X}_{k l} {G}^{-3} \nabla^{\rho}\,_{\alpha}{G^{j k}} \varphi^{l}_{\beta}+\frac{3}{32}{\rm i} G_{j k} (\Gamma_{a})^{\beta \rho} \mathbf{X}^{j k} \lambda_{i \beta} {G}^{-3} \varphi_{l \alpha} \varphi^{l}_{\rho} - \frac{3}{16}{\rm i} G_{j k} (\Gamma_{a})^{\beta \rho} \mathbf{X}^{j}\,_{l} \lambda_{i \beta} {G}^{-3} \varphi^{l}_{\alpha} \varphi^{k}_{\rho}+\frac{21}{64}(\Gamma_{a})^{\beta \rho} \lambda_{i \beta} {G}^{-1} \nabla_{\alpha}\,^{\lambda}{\boldsymbol{\lambda}_{j \lambda}} \varphi^{j}_{\rho}+\frac{3}{128}(\Gamma_{a})^{\rho \lambda} W_{\alpha}\,^{\beta} \lambda_{i \rho} \boldsymbol{\lambda}_{j \beta} {G}^{-1} \varphi^{j}_{\lambda}%
+\frac{3}{16}{\rm i} G_{i j} G_{k l} (\Gamma_{a})^{\beta \rho} \mathbf{X}^{k l} F_{\alpha \beta} {G}^{-3} \varphi^{j}_{\rho}+\frac{21}{64}{\rm i} G_{i j} G_{k l} (\Gamma_{a})^{\beta \rho} W \mathbf{X}^{k l} W_{\alpha \beta} {G}^{-3} \varphi^{j}_{\rho}+\frac{3}{32}{\rm i} G_{j k} (\Gamma_{a})^{\beta \rho} \mathbf{X}^{j k} \lambda_{l \beta} {G}^{-3} \varphi_{i \alpha} \varphi^{l}_{\rho}+\frac{9}{32}{\rm i} G_{i j} (\Gamma_{a})^{\beta \rho} \mathbf{X}_{k l} \lambda^{k}_{\beta} {G}^{-3} \varphi^{l}_{\alpha} \varphi^{j}_{\rho}+\frac{3}{64}G_{i j} G_{k l} (\Gamma_{a})^{\beta \rho} \lambda^{k}_{\beta} {G}^{-3} \nabla_{\alpha}\,^{\lambda}{\boldsymbol{\lambda}^{l}_{\lambda}} \varphi^{j}_{\rho}+\frac{3}{128}G_{i j} G_{k l} (\Gamma_{a})^{\rho \lambda} W_{\alpha}\,^{\beta} \lambda^{k}_{\rho} \boldsymbol{\lambda}^{l}_{\beta} {G}^{-3} \varphi^{j}_{\lambda} - \frac{9}{32}{\rm i} G_{i j} G_{k l} G_{m n} (\Gamma_{a})^{\beta \rho} \mathbf{X}^{k l} \lambda^{m}_{\beta} {G}^{-5} \varphi^{n}_{\alpha} \varphi^{j}_{\rho} - \frac{3}{64}\mathcal{H}_{\alpha}\,^{\beta} G_{i j} G_{k l} (\Gamma_{a})_{\beta}{}^{\rho} \mathbf{X}^{k l} \lambda^{j}_{\rho} {G}^{-3}+\frac{3}{64}G_{i j} G_{k l} (\Gamma_{a})^{\beta}{}_{\rho} \mathbf{X}^{k l} \lambda_{m \beta} {G}^{-3} \nabla^{\rho}\,_{\alpha}{G^{j m}} - \frac{3}{16}{\rm i} G_{i j} (\Gamma_{a})^{\beta}{}_{\rho} W {G}^{-3} \nabla^{\rho \lambda}{\boldsymbol{\lambda}_{k \lambda}} \varphi^{k}_{\alpha} \varphi^{j}_{\beta}+\frac{3}{8}{\rm i} G_{j k} (\Gamma_{a})^{\beta}{}_{\rho} W {G}^{-3} \nabla^{\rho \lambda}{\boldsymbol{\lambda}^{j}_{\lambda}} \varphi_{i \alpha} \varphi^{k}_{\beta}+\frac{3}{16}{\rm i} G_{i j} (\Gamma_{a})^{\beta}{}_{\rho} W {G}^{-3} \nabla^{\rho \lambda}{\boldsymbol{\lambda}^{j}_{\lambda}} \varphi_{k \alpha} \varphi^{k}_{\beta} - \frac{3}{16}G_{i j} G_{k l} (\Gamma_{a})^{\beta}{}_{\rho} \lambda^{k}_{\alpha} {G}^{-3} \nabla^{\rho \lambda}{\boldsymbol{\lambda}^{j}_{\lambda}} \varphi^{l}_{\beta}+\frac{3}{8}{\rm i} (\Gamma_{a})^{\rho}{}_{\lambda} W {G}^{-1} \nabla^{\lambda \beta}{\mathbf{F}_{\alpha \beta}} \varphi_{i \rho}+\frac{9}{8}{\rm i} (\Gamma_{a})^{\rho}{}_{\lambda} W W_{\alpha \beta} {G}^{-1} \nabla^{\lambda \beta}{\boldsymbol{W}} \varphi_{i \rho} - \frac{3}{32}{\rm i} G_{i j} G_{k l} (\Gamma_{a})^{\beta}{}_{\rho} W {G}^{-3} \nabla^{\rho}\,_{\alpha}{\mathbf{X}^{j k}} \varphi^{l}_{\beta} - \frac{3}{16}{\rm i} (\Gamma_{a})^{\beta}{}_{\rho} W {G}^{-1} \nabla^{\rho}\,_{\lambda}{\nabla_{\alpha}\,^{\lambda}{\boldsymbol{W}}} \varphi_{i \beta} - \frac{3}{16}{\rm i} G_{i j} G_{k l} (\Gamma_{a})^{\beta \rho} W \mathbf{X}^{j k} W_{\alpha \beta} {G}^{-3} \varphi^{l}_{\rho} - \frac{9}{8}{\rm i} (\Gamma_{a})^{\beta \lambda} W W_{\beta \rho} {G}^{-1} \nabla_{\alpha}\,^{\rho}{\boldsymbol{W}} \varphi_{i \lambda} - \frac{81}{2560}{\rm i} G_{i j} G_{k l} (\Gamma_{a})_{\alpha}{}^{\rho} W \boldsymbol{\lambda}^{j \beta} X^{k}_{\beta} {G}^{-3} \varphi^{l}_{\rho}%
+\frac{99}{2560}{\rm i} (\Gamma_{a})_{\alpha}{}^{\rho} W \boldsymbol{\lambda}_{j}^{\beta} X^{j}_{\beta} {G}^{-1} \varphi_{i \rho}+\frac{9}{256}{\rm i} G_{i j} G_{k l} (\Gamma_{a})_{\alpha}{}^{\rho} W \boldsymbol{\lambda}^{k \beta} X^{j}_{\beta} {G}^{-3} \varphi^{l}_{\rho} - \frac{9}{32}{\rm i} G_{i j} (\Gamma_{a})^{\beta \lambda} W W_{\beta}\,^{\rho} \boldsymbol{\lambda}_{k \rho} {G}^{-3} \varphi^{k}_{\alpha} \varphi^{j}_{\lambda}+\frac{9}{16}{\rm i} G_{j k} (\Gamma_{a})^{\beta \lambda} W W_{\beta}\,^{\rho} \boldsymbol{\lambda}^{j}_{\rho} {G}^{-3} \varphi_{i \alpha} \varphi^{k}_{\lambda}+\frac{9}{32}{\rm i} G_{i j} (\Gamma_{a})^{\beta \lambda} W W_{\beta}\,^{\rho} \boldsymbol{\lambda}^{j}_{\rho} {G}^{-3} \varphi_{k \alpha} \varphi^{k}_{\lambda}+\frac{3}{64}G_{i j} G_{k l} (\Gamma_{a})^{\beta}{}_{\rho} \lambda^{k}_{\alpha} {G}^{-3} \nabla^{\rho \lambda}{\boldsymbol{\lambda}^{l}_{\lambda}} \varphi^{j}_{\beta}+\frac{27}{320}F (\Gamma_{a})_{\alpha \beta} W {G}^{-1} \nabla^{\beta \rho}{\boldsymbol{\lambda}_{i \rho}}+\frac{3}{16}\mathcal{H}_{\alpha}\,^{\beta} (\Gamma_{a})_{\beta \rho} W {G}^{-1} \nabla^{\rho \lambda}{\boldsymbol{\lambda}_{i \lambda}} - \frac{3}{32}G_{i j} G_{k l} (\Gamma_{a})_{\beta \rho} W {G}^{-3} \nabla^{\beta \lambda}{\boldsymbol{\lambda}^{k}_{\lambda}} \nabla^{\rho}\,_{\alpha}{G^{j l}} - \frac{63}{320}F (\Gamma_{a})_{\alpha}{}^{\beta} W W_{\beta}\,^{\rho} \boldsymbol{\lambda}_{i \rho} {G}^{-1} - \frac{9}{32}\mathcal{H}_{\alpha}\,^{\lambda} (\Gamma_{a})_{\lambda}{}^{\beta} W W_{\beta}\,^{\rho} \boldsymbol{\lambda}_{i \rho} {G}^{-1}+\frac{9}{64}G_{i j} G_{k l} (\Gamma_{a})^{\beta}{}_{\lambda} W W_{\beta}\,^{\rho} \boldsymbol{\lambda}^{k}_{\rho} {G}^{-3} \nabla^{\lambda}\,_{\alpha}{G^{j l}} - \frac{9}{32}G_{i j} G_{k l} (\Gamma_{a})^{\beta \rho} W \mathbf{X}^{k l} {G}^{-5} \varphi_{m \alpha} \varphi^{j}_{\beta} \varphi^{m}_{\rho}+\frac{9}{32}G_{j k} G_{l m} (\Gamma_{a})^{\beta \rho} W \mathbf{X}^{j k} {G}^{-5} \varphi_{i \alpha} \varphi^{l}_{\beta} \varphi^{m}_{\rho} - \frac{9}{16}G_{i j} G_{k l} (\Gamma_{a})^{\beta \rho} W \mathbf{X}^{k}\,_{m} {G}^{-5} \varphi^{m}_{\alpha} \varphi^{j}_{\beta} \varphi^{l}_{\rho} - \frac{9}{32}{\rm i} G_{i j} G_{k l} G_{m n} (\Gamma_{a})^{\beta \rho} \mathbf{X}^{k l} \lambda^{m}_{\alpha} {G}^{-5} \varphi^{j}_{\beta} \varphi^{n}_{\rho}+\frac{9}{64}{\rm i} G_{i j} G_{k l} G_{m n} (\Gamma_{a})^{\beta}{}_{\rho} W \mathbf{X}^{k l} {G}^{-5} \nabla^{\rho}\,_{\alpha}{G^{j m}} \varphi^{n}_{\beta} - \frac{3}{64}{\rm i} G_{i j} G_{k l} (\Gamma_{a})^{\beta}{}_{\rho} W \mathbf{X}^{k l} {G}^{-3} \nabla^{\rho}\,_{\alpha}{\varphi^{j}_{\beta}}+\frac{21}{320}{\rm i} G_{i j} G_{k l} (\Gamma_{a})_{\alpha}{}^{\beta} W \mathbf{X}^{k l} W_{\beta}\,^{\rho} {G}^{-3} \varphi^{j}_{\rho} - \frac{39}{320}{\rm i} G_{i j} G_{k l} (\Gamma_{a})_{\alpha \beta} W \mathbf{X}^{k l} {G}^{-3} \nabla^{\beta \rho}{\varphi^{j}_{\rho}}%
+\frac{3}{32}{\rm i} \mathcal{H}^{\beta \rho} G_{i j} (\Gamma_{a})_{\beta \rho} W \mathbf{X}^{j}\,_{k} {G}^{-3} \varphi^{k}_{\alpha} - \frac{3}{64}\mathcal{H}^{\beta \rho} G_{i j} G_{k l} (\Gamma_{a})_{\beta \rho} \mathbf{X}^{k l} \lambda^{j}_{\alpha} {G}^{-3} - \frac{3}{16}\mathcal{H}^{\beta \rho} (\Gamma_{a})_{\beta \rho} W {G}^{-1} \nabla_{\alpha}\,^{\lambda}{\boldsymbol{\lambda}_{i \lambda}} - \frac{9}{32}\mathcal{H}^{\rho \lambda} (\Gamma_{a})_{\rho \lambda} W W_{\alpha}\,^{\beta} \boldsymbol{\lambda}_{i \beta} {G}^{-1} - \frac{3}{128}{\rm i} G_{j k} (\Gamma_{a})_{\beta \rho} W \mathbf{X}^{j k} {G}^{-3} \nabla^{\beta \rho}{G_{i l}} \varphi^{l}_{\alpha}+\frac{3}{32}{\rm i} G_{j k} (\Gamma_{a})_{\beta \rho} W \mathbf{X}^{j}\,_{l} {G}^{-3} \nabla^{\beta \rho}{G_{i}\,^{k}} \varphi^{l}_{\alpha} - \frac{3}{64}G_{j k} G_{l m} (\Gamma_{a})_{\beta \rho} \mathbf{X}^{j k} \lambda^{l}_{\alpha} {G}^{-3} \nabla^{\beta \rho}{G_{i}\,^{m}} - \frac{9}{32}(\Gamma_{a})_{\beta \rho} W {G}^{-1} \nabla_{\alpha}\,^{\lambda}{\boldsymbol{\lambda}_{j \lambda}} \nabla^{\beta \rho}{G_{i}\,^{j}} - \frac{27}{128}(\Gamma_{a})_{\rho \lambda} W W_{\alpha}\,^{\beta} \boldsymbol{\lambda}_{j \beta} {G}^{-1} \nabla^{\rho \lambda}{G_{i}\,^{j}} - \frac{3}{16}{\rm i} G_{i j} G_{k l} (\Gamma_{a})_{\beta \rho} W \mathbf{X}^{k l} {G}^{-3} \nabla^{\beta \rho}{\varphi^{j}_{\alpha}} - \frac{3}{32}{\rm i} G_{i j} (\Gamma_{a})_{\beta \rho} W \mathbf{X}_{k l} {G}^{-3} \nabla^{\beta \rho}{G^{j k}} \varphi^{l}_{\alpha}+\frac{3}{64}G_{i j} G_{k l} (\Gamma_{a})_{\beta \rho} \mathbf{X}^{k l} \lambda_{m \alpha} {G}^{-3} \nabla^{\beta \rho}{G^{j m}}+\frac{3}{32}G_{i j} G_{k l} (\Gamma_{a})_{\beta \rho} W {G}^{-3} \nabla_{\alpha}\,^{\lambda}{\boldsymbol{\lambda}^{k}_{\lambda}} \nabla^{\beta \rho}{G^{j l}}+\frac{9}{64}G_{i j} G_{k l} (\Gamma_{a})_{\rho \lambda} W W_{\alpha}\,^{\beta} \boldsymbol{\lambda}^{k}_{\beta} {G}^{-3} \nabla^{\rho \lambda}{G^{j l}}+\frac{9}{64}{\rm i} G_{i j} G_{k l} G_{m n} (\Gamma_{a})_{\beta \rho} W \mathbf{X}^{k l} {G}^{-5} \nabla^{\beta \rho}{G^{j m}} \varphi^{n}_{\alpha} - \frac{15}{64}(\Gamma_{a})^{\beta \rho} \lambda_{j \beta} {G}^{-1} \nabla_{\alpha}\,^{\lambda}{\boldsymbol{\lambda}_{i \lambda}} \varphi^{j}_{\rho} - \frac{3}{64}(\Gamma_{a})^{\rho \lambda} W_{\alpha}\,^{\beta} \lambda_{j \rho} \boldsymbol{\lambda}_{i \beta} {G}^{-1} \varphi^{j}_{\lambda} - \frac{15}{32}{\rm i} (\Gamma_{a})^{\beta \rho} W \mathbf{X}_{i j} W_{\alpha \beta} {G}^{-1} \varphi^{j}_{\rho} - \frac{9}{32}{\rm i} (\Gamma_{a})^{\beta}{}_{\rho} \mathbf{X}_{i j} {G}^{-1} \nabla^{\rho}\,_{\alpha}{W} \varphi^{j}_{\beta}+\frac{3}{32}{\rm i} G_{j k} (\Gamma_{a})^{\beta \rho} \mathbf{X}_{i l} \lambda^{j}_{\beta} {G}^{-3} \varphi^{k}_{\alpha} \varphi^{l}_{\rho}%
+\frac{9}{64}\mathcal{H}_{\alpha}\,^{\beta} (\Gamma_{a})_{\beta}{}^{\rho} \mathbf{X}_{i j} \lambda^{j}_{\rho} {G}^{-1} - \frac{3}{64}(\Gamma_{a})^{\beta}{}_{\rho} \mathbf{X}_{i j} \lambda_{k \beta} {G}^{-1} \nabla^{\rho}\,_{\alpha}{G^{j k}}+\frac{3}{32}{\rm i} G_{j k} (\Gamma_{a})^{\beta \rho} \mathbf{X}^{j}\,_{l} \lambda_{i \beta} {G}^{-3} \varphi^{k}_{\alpha} \varphi^{l}_{\rho}+\frac{3}{64}(\Gamma_{a})^{\beta}{}_{\rho} \mathbf{X}_{j k} \lambda_{i \beta} {G}^{-1} \nabla^{\rho}\,_{\alpha}{G^{j k}} - \frac{9}{16}(\Gamma_{a})^{\beta}{}_{\rho} \lambda_{i \beta} {G}^{-1} \nabla^{\rho \lambda}{\boldsymbol{\lambda}_{j \lambda}} \varphi^{j}_{\alpha}+\frac{9}{32}(\Gamma_{a})^{\beta}{}_{\rho} \lambda_{j \beta} {G}^{-1} \nabla^{\rho \lambda}{\boldsymbol{\lambda}^{j}_{\lambda}} \varphi_{i \alpha} - \frac{9}{32}G_{i j} (\Gamma_{a})^{\beta}{}_{\rho} F_{\alpha \beta} {G}^{-1} \nabla^{\rho \lambda}{\boldsymbol{\lambda}^{j}_{\lambda}} - \frac{27}{32}G_{i j} (\Gamma_{a})^{\beta}{}_{\rho} W W_{\alpha \beta} {G}^{-1} \nabla^{\rho \lambda}{\boldsymbol{\lambda}^{j}_{\lambda}} - \frac{15}{64}G_{i j} (\Gamma_{a})_{\beta \rho} {G}^{-1} \nabla^{\beta}\,_{\alpha}{W} \nabla^{\rho \lambda}{\boldsymbol{\lambda}^{j}_{\lambda}} - \frac{3}{16}G_{i j} G_{k l} (\Gamma_{a})^{\beta}{}_{\rho} \lambda^{k}_{\beta} {G}^{-3} \nabla^{\rho \lambda}{\boldsymbol{\lambda}^{j}_{\lambda}} \varphi^{l}_{\alpha} - \frac{3}{32}G_{i j} (\Gamma_{a})^{\rho}{}_{\lambda} \lambda^{j}_{\rho} {G}^{-1} \nabla^{\lambda \beta}{\mathbf{F}_{\alpha \beta}} - \frac{3}{32}G_{i j} (\Gamma_{a})^{\beta}{}_{\rho} \lambda_{k \beta} {G}^{-1} \nabla^{\rho}\,_{\alpha}{\mathbf{X}^{j k}}+\frac{3}{64}G_{i j} (\Gamma_{a})^{\beta}{}_{\rho} \lambda^{j}_{\beta} {G}^{-1} \nabla^{\rho}\,_{\lambda}{\nabla_{\alpha}\,^{\lambda}{\boldsymbol{W}}}+\frac{9}{64}G_{i j} (\Gamma_{a})^{\beta \lambda} W_{\alpha}\,^{\rho} \mathbf{F}_{\beta \rho} \lambda^{j}_{\lambda} {G}^{-1} - \frac{9}{64}G_{i j} (\Gamma_{a})^{\beta \rho} \mathbf{X}^{j}\,_{k} W_{\alpha \beta} \lambda^{k}_{\rho} {G}^{-1}+\frac{27}{128}G_{i j} (\Gamma_{a})^{\rho}{}_{\lambda} W_{\alpha \beta} \lambda^{j}_{\rho} {G}^{-1} \nabla^{\lambda \beta}{\boldsymbol{W}} - \frac{9}{64}G_{i j} (\Gamma_{a})_{\alpha}{}^{\lambda} W^{\beta \rho} \mathbf{F}_{\beta \rho} \lambda^{j}_{\lambda} {G}^{-1} - \frac{9}{64}G_{i j} (\Gamma_{a})^{\rho \lambda} W_{\rho}\,^{\beta} \mathbf{F}_{\alpha \beta} \lambda^{j}_{\lambda} {G}^{-1} - \frac{9}{128}G_{i j} (\Gamma_{a})^{\beta \lambda} W_{\beta \rho} \lambda^{j}_{\lambda} {G}^{-1} \nabla_{\alpha}\,^{\rho}{\boldsymbol{W}} - \frac{3}{32}G_{i j} (\Gamma_{a})^{\rho \beta} \boldsymbol{\lambda}^{j}_{\alpha} \lambda_{k \rho} X^{k}_{\beta} {G}^{-1}%
 - \frac{15}{128}G_{i j} (\Gamma_{a})^{\beta \rho} \lambda_{k \beta} \boldsymbol{\lambda}^{j}_{\rho} X^{k}_{\alpha} {G}^{-1}+\frac{123}{1280}G_{i j} (\Gamma_{a})_{\alpha}{}^{\rho} \lambda_{k \rho} \boldsymbol{\lambda}^{j \beta} X^{k}_{\beta} {G}^{-1} - \frac{15}{128}G_{i j} (\Gamma_{a})^{\beta \rho} \lambda^{j}_{\beta} \boldsymbol{\lambda}_{k \rho} X^{k}_{\alpha} {G}^{-1}+\frac{57}{640}G_{i j} (\Gamma_{a})_{\alpha}{}^{\rho} \lambda^{j}_{\rho} \boldsymbol{\lambda}_{k}^{\beta} X^{k}_{\beta} {G}^{-1} - \frac{15}{256}G_{i j} (\Gamma_{a})^{\rho \beta} \boldsymbol{\lambda}_{k \alpha} \lambda^{j}_{\rho} X^{k}_{\beta} {G}^{-1}+\frac{27}{1280}G_{i j} (\Gamma_{a})_{\alpha}{}^{\rho} \lambda_{k \rho} \boldsymbol{\lambda}^{k \beta} X^{j}_{\beta} {G}^{-1}+\frac{9}{160}G_{j k} (\Gamma_{a})_{\alpha \beta} X_{i}\,^{j} {G}^{-1} \nabla^{\beta \rho}{\boldsymbol{\lambda}^{k}_{\rho}}+\frac{9}{320}G_{j k} (\Gamma_{a})_{\alpha}{}^{\rho} \lambda_{i \rho} \boldsymbol{\lambda}^{j \beta} X^{k}_{\beta} {G}^{-1}+\frac{9}{32}(\Gamma_{a})^{\beta}{}_{\rho} \lambda_{j \alpha} {G}^{-1} \nabla^{\rho \lambda}{\boldsymbol{\lambda}^{j}_{\lambda}} \varphi_{i \beta} - \frac{3}{16}{\rm i} G_{j k} (\Gamma_{a})^{\beta}{}_{\rho} W {G}^{-3} \nabla^{\rho \lambda}{\boldsymbol{\lambda}^{j}_{\lambda}} \varphi^{k}_{\alpha} \varphi_{i \beta}+\frac{9}{32}(\Gamma_{a})_{\beta \rho} W {G}^{-1} \nabla^{\beta \lambda}{\boldsymbol{\lambda}_{j \lambda}} \nabla^{\rho}\,_{\alpha}{G_{i}\,^{j}} - \frac{9}{32}{\rm i} G_{j k} (\Gamma_{a})^{\beta \lambda} W W_{\beta}\,^{\rho} \boldsymbol{\lambda}^{j}_{\rho} {G}^{-3} \varphi^{k}_{\alpha} \varphi_{i \lambda} - \frac{27}{64}(\Gamma_{a})^{\beta}{}_{\lambda} W W_{\beta}\,^{\rho} \boldsymbol{\lambda}_{j \rho} {G}^{-1} \nabla^{\lambda}\,_{\alpha}{G_{i}\,^{j}}+\frac{3}{64}(\Gamma_{a})^{\beta}{}_{\rho} \lambda_{j \alpha} {G}^{-1} \nabla^{\rho \lambda}{\boldsymbol{\lambda}_{i \lambda}} \varphi^{j}_{\beta} - \frac{3}{16}{\rm i} G_{j k} (\Gamma_{a})^{\beta}{}_{\rho} W {G}^{-3} \nabla^{\rho \lambda}{\boldsymbol{\lambda}_{i \lambda}} \varphi^{j}_{\alpha} \varphi^{k}_{\beta} - \frac{9}{32}{\rm i} (\Gamma_{a})^{\beta}{}_{\rho} W {G}^{-1} \nabla^{\rho}\,_{\alpha}{\mathbf{X}_{i j}} \varphi^{j}_{\beta} - \frac{27}{256}{\rm i} (\Gamma_{a})^{\beta \rho} W \boldsymbol{\lambda}_{i \alpha} X_{j \beta} {G}^{-1} \varphi^{j}_{\rho}+\frac{189}{2560}{\rm i} (\Gamma_{a})_{\alpha}{}^{\rho} W \boldsymbol{\lambda}_{i}^{\beta} X_{j \beta} {G}^{-1} \varphi^{j}_{\rho} - \frac{9}{128}{\rm i} (\Gamma_{a})_{\alpha}{}^{\rho} W \boldsymbol{\lambda}_{j}^{\beta} X_{i \beta} {G}^{-1} \varphi^{j}_{\rho} - \frac{9}{32}{\rm i} G_{j k} (\Gamma_{a})^{\beta \lambda} W W_{\beta}\,^{\rho} \boldsymbol{\lambda}_{i \rho} {G}^{-3} \varphi^{j}_{\alpha} \varphi^{k}_{\lambda}%
+\frac{3}{32}{\rm i} (\Gamma_{a})^{\beta}{}_{\rho} W \mathbf{X}_{i j} {G}^{-1} \nabla^{\rho}\,_{\alpha}{\varphi^{j}_{\beta}}+\frac{9}{64}\mathcal{H}^{\beta \rho} (\Gamma_{a})_{\beta \rho} \mathbf{X}_{i j} \lambda^{j}_{\alpha} {G}^{-1} - \frac{3}{32}{\rm i} \mathcal{H}^{\beta \rho} G_{j k} (\Gamma_{a})_{\beta \rho} W \mathbf{X}_{i}\,^{j} {G}^{-3} \varphi^{k}_{\alpha}+\frac{9}{64}(\Gamma_{a})_{\beta \rho} \mathbf{X}_{k j} \lambda^{k}_{\alpha} {G}^{-1} \nabla^{\beta \rho}{G_{i}\,^{j}} - \frac{3}{32}{\rm i} G_{j k} (\Gamma_{a})_{\beta \rho} W \mathbf{X}^{j}\,_{l} {G}^{-3} \nabla^{\beta \rho}{G_{i}\,^{l}} \varphi^{k}_{\alpha}+\frac{3}{8}{\rm i} (\Gamma_{a})_{\beta \rho} W \mathbf{X}_{i j} {G}^{-1} \nabla^{\beta \rho}{\varphi^{j}_{\alpha}}+\frac{3}{128}(\Gamma_{a})_{\beta \rho} \mathbf{X}_{i j} \lambda_{k \alpha} {G}^{-1} \nabla^{\beta \rho}{G^{j k}} - \frac{3}{32}{\rm i} G_{j k} (\Gamma_{a})_{\beta \rho} W \mathbf{X}_{i l} {G}^{-3} \nabla^{\beta \rho}{G^{j l}} \varphi^{k}_{\alpha}+\frac{3}{4}{\rm i} (\Gamma_{a})^{\rho}{}_{\lambda} W {G}^{-1} \nabla^{\lambda \beta}{\mathbf{F}_{\beta \rho}} \varphi_{i \alpha}+\frac{3}{32}G_{i j} (\Gamma_{a})^{\rho}{}_{\lambda} \lambda^{j}_{\alpha} {G}^{-1} \nabla^{\lambda \beta}{\mathbf{F}_{\beta \rho}}+\frac{3}{32}G_{i j} (\Gamma_{a})^{\beta}{}_{\rho} W {G}^{-1} \nabla^{\rho}\,_{\lambda}{\nabla_{\alpha}\,^{\lambda}{\boldsymbol{\lambda}^{j}_{\beta}}}+\frac{3}{32}G_{i j} (\Gamma_{a})_{\beta \rho} W {G}^{-1} \nabla^{\beta \lambda}{\nabla^{\rho}\,_{\alpha}{\boldsymbol{\lambda}^{j}_{\lambda}}}+\frac{9}{64}G_{i j} (\Gamma_{a})^{\beta}{}_{\lambda} W W_{\beta \rho} {G}^{-1} \nabla^{\lambda \rho}{\boldsymbol{\lambda}^{j}_{\alpha}}+\frac{27}{64}G_{i j} (\Gamma_{a})^{\rho}{}_{\lambda} W W_{\alpha \beta} {G}^{-1} \nabla^{\lambda \beta}{\boldsymbol{\lambda}^{j}_{\rho}}+\frac{9}{128}G_{i j} (\Gamma_{a})^{\beta}{}_{\rho} W \boldsymbol{\lambda}^{j}_{\lambda} {G}^{-1} \nabla^{\rho \lambda}{W_{\alpha \beta}}+\frac{51}{160}G_{i j} (\Gamma_{a})_{\alpha \beta} W {G}^{-1} \nabla^{\beta}\,_{\lambda}{\nabla^{\lambda \rho}{\boldsymbol{\lambda}^{j}_{\rho}}}+\frac{9}{128}G_{i j} (\Gamma_{a})^{\beta}{}_{\lambda} W \boldsymbol{\lambda}^{j \rho} {G}^{-1} \nabla^{\lambda}\,_{\alpha}{W_{\beta \rho}}+\frac{9}{64}G_{i j} (\Gamma_{a})^{\beta}{}_{\lambda} W W_{\beta}\,^{\rho} {G}^{-1} \nabla^{\lambda}\,_{\alpha}{\boldsymbol{\lambda}^{j}_{\rho}}+\frac{1071}{2560}G_{i j} (\Gamma_{a})_{\alpha \lambda} W \boldsymbol{\lambda}^{j \beta} {G}^{-1} \nabla^{\lambda \rho}{W_{\beta \rho}} - \frac{81}{160}G_{i j} (\Gamma_{a})_{\alpha \lambda} W W^{\beta}\,_{\rho} {G}^{-1} \nabla^{\lambda \rho}{\boldsymbol{\lambda}^{j}_{\beta}}%
 - \frac{3}{8}{\rm i} G_{i j} (\Gamma_{a})^{\beta}{}_{\lambda} W W_{\alpha \beta \rho}\,^{j} {G}^{-1} \nabla^{\lambda \rho}{\boldsymbol{W}} - \frac{81}{640}G_{i j} (\Gamma_{a})_{\alpha}{}^{\beta} W W_{\beta}\,^{\rho} W_{\rho}\,^{\lambda} \boldsymbol{\lambda}^{j}_{\lambda} {G}^{-1} - \frac{63}{320}G_{i j} (\Gamma_{a})_{\alpha}{}^{\beta} W W_{\beta \rho} {G}^{-1} \nabla^{\rho \lambda}{\boldsymbol{\lambda}^{j}_{\lambda}}+\frac{9}{64}G_{i j} (\Gamma_{a})^{\beta \lambda} W W_{\beta \rho} {G}^{-1} \nabla_{\alpha}\,^{\rho}{\boldsymbol{\lambda}^{j}_{\lambda}}+\frac{27}{64}G_{i j} (\Gamma_{a})^{\beta \rho} W W_{\alpha \beta} W_{\rho}\,^{\lambda} \boldsymbol{\lambda}^{j}_{\lambda} {G}^{-1}+\frac{27}{256}G_{i j} (\Gamma_{a})^{\beta \lambda} W W_{\beta}\,^{\rho} W_{\lambda \rho} \boldsymbol{\lambda}^{j}_{\alpha} {G}^{-1} - \frac{27}{64}G_{i j} (\Gamma_{a})_{\rho \lambda} W W_{\alpha}\,^{\beta} {G}^{-1} \nabla^{\rho \lambda}{\boldsymbol{\lambda}^{j}_{\beta}} - \frac{51}{640}\Phi^{\beta \rho}\,_{j k} G_{i}\,^{j} (\Gamma_{a})_{\alpha \beta} W \boldsymbol{\lambda}^{k}_{\rho} {G}^{-1}+\frac{81}{128}G_{i j} (\Gamma_{a})_{\rho \lambda} W \boldsymbol{\lambda}^{j \beta} {G}^{-1} \nabla^{\rho \lambda}{W_{\alpha \beta}}+\frac{9}{512}G_{i j} (\Gamma_{a})_{\alpha}{}^{\beta} W \boldsymbol{\lambda}^{j}_{\lambda} {G}^{-1} \nabla^{\lambda \rho}{W_{\beta \rho}}+\frac{9}{4}{\rm i} (\Gamma_{a})^{\beta}{}_{\lambda} W W_{\beta \rho} {G}^{-1} \nabla^{\lambda \rho}{\boldsymbol{W}} \varphi_{i \alpha} - \frac{9}{128}G_{i j} (\Gamma_{a})^{\beta}{}_{\lambda} W_{\beta \rho} \lambda^{j}_{\alpha} {G}^{-1} \nabla^{\lambda \rho}{\boldsymbol{W}} - \frac{159}{320}{\rm i} G_{i j} (\Gamma_{a})_{\alpha \rho} W X^{j}_{\beta} {G}^{-1} \nabla^{\rho \beta}{\boldsymbol{W}} - \frac{9}{64}{\rm i} (\Gamma_{a})_{\beta \rho} W {G}^{-1} \nabla^{\beta \rho}{\mathbf{X}_{i j}} \varphi^{j}_{\alpha}+\frac{3}{64}G_{j k} (\Gamma_{a})_{\beta \rho} \lambda^{j}_{\alpha} {G}^{-1} \nabla^{\beta \rho}{\mathbf{X}_{i}\,^{k}} - \frac{21}{64}G_{i j} (\Gamma_{a})_{\beta \rho} W {G}^{-1} \nabla^{\beta \rho}{\nabla_{\alpha}\,^{\lambda}{\boldsymbol{\lambda}^{j}_{\lambda}}} - \frac{177}{256}{\rm i} G_{i j} (\Gamma_{a})_{\beta \rho} W X^{j}_{\alpha} {G}^{-1} \nabla^{\beta \rho}{\boldsymbol{W}} - \frac{45}{128}\Phi_{\alpha}\,^{\beta}\,_{i j} G^{j}\,_{k} (\Gamma_{a})_{\beta}{}^{\rho} W \boldsymbol{\lambda}^{k}_{\rho} {G}^{-1} - \frac{9}{128}\Phi^{\beta \rho}\,_{i j} G^{j}\,_{k} (\Gamma_{a})_{\alpha \beta} W \boldsymbol{\lambda}^{k}_{\rho} {G}^{-1}+\frac{99}{640}\Phi^{\beta \rho}\,_{j k} G^{j k} (\Gamma_{a})_{\alpha \beta} W \boldsymbol{\lambda}_{i \rho} {G}^{-1}%
 - \frac{3}{8}{\rm i} (\Gamma_{a})_{\beta \rho} W {G}^{-1} \nabla^{\beta}\,_{\lambda}{\nabla^{\rho \lambda}{\boldsymbol{W}}} \varphi_{i \alpha} - \frac{3}{64}G_{i j} (\Gamma_{a})_{\beta \rho} \lambda^{j}_{\alpha} {G}^{-1} \nabla^{\beta}\,_{\lambda}{\nabla^{\rho \lambda}{\boldsymbol{W}}} - \frac{3}{32}G_{i j} (\Gamma_{a})_{\beta \rho} W {G}^{-1} \nabla^{\beta}\,_{\lambda}{\nabla^{\rho \lambda}{\boldsymbol{\lambda}^{j}_{\alpha}}}+\frac{9}{256}G_{j k} (\Gamma_{a})^{\rho \beta} \lambda^{j}_{\alpha} \boldsymbol{\lambda}^{k}_{\rho} X_{i \beta} {G}^{-1} - \frac{3}{32}G_{i j} (\Gamma_{a})^{\rho \beta} \lambda^{j}_{\alpha} \boldsymbol{\lambda}_{k \rho} X^{k}_{\beta} {G}^{-1} - \frac{3}{32}G_{i j} (\Gamma_{a})_{\beta \rho} \lambda_{k \alpha} {G}^{-1} \nabla^{\beta \rho}{\mathbf{X}^{j k}} - \frac{3}{32}{\rm i} G_{i j} G_{k l} (\Gamma_{a})_{\beta \rho} W {G}^{-3} \nabla^{\beta \rho}{\mathbf{X}^{j k}} \varphi^{l}_{\alpha} - \frac{15}{256}G_{i j} (\Gamma_{a})^{\rho \beta} \lambda_{k \alpha} \boldsymbol{\lambda}^{j}_{\rho} X^{k}_{\beta} {G}^{-1}+\frac{3}{16}(\Gamma_{a})^{\beta \rho} \boldsymbol{W} X_{j k} {G}^{-3} \varphi^{j}_{\alpha} \varphi_{i \beta} \varphi^{k}_{\rho}+\frac{3}{32}{\rm i} G_{j k} (\Gamma_{a})^{\beta \rho} X^{j k} \boldsymbol{\lambda}_{l \alpha} {G}^{-3} \varphi_{i \beta} \varphi^{l}_{\rho}+\frac{3}{16}{\rm i} G_{j k} (\Gamma_{a})^{\beta \rho} \boldsymbol{W} {G}^{-3} \nabla_{\alpha}\,^{\lambda}{\lambda^{j}_{\lambda}} \varphi_{i \beta} \varphi^{k}_{\rho}+\frac{9}{32}{\rm i} G_{j k} (\Gamma_{a})^{\rho \lambda} \boldsymbol{W} W_{\alpha}\,^{\beta} \lambda^{j}_{\beta} {G}^{-3} \varphi_{i \rho} \varphi^{k}_{\lambda} - \frac{9}{32}G_{j k} G_{l m} (\Gamma_{a})^{\beta \rho} \boldsymbol{W} X^{j k} {G}^{-5} \varphi^{l}_{\alpha} \varphi_{i \beta} \varphi^{m}_{\rho}+\frac{3}{64}{\rm i} G_{j k} (\Gamma_{a})^{\beta}{}_{\rho} \boldsymbol{W} X^{j k} {G}^{-3} \nabla^{\rho}\,_{\alpha}{G_{i l}} \varphi^{l}_{\beta} - \frac{3}{16}(\Gamma_{a})^{\beta \rho} \boldsymbol{W} X_{i j} {G}^{-3} \varphi_{k \alpha} \varphi^{j}_{\beta} \varphi^{k}_{\rho}+\frac{3}{16}{\rm i} G_{j k} (\Gamma_{a})^{\beta \rho} X_{i l} \boldsymbol{\lambda}^{j}_{\alpha} {G}^{-3} \varphi^{k}_{\beta} \varphi^{l}_{\rho} - \frac{3}{16}{\rm i} G_{i j} (\Gamma_{a})^{\beta \rho} \boldsymbol{W} {G}^{-3} \nabla_{\alpha}\,^{\lambda}{\lambda_{k \lambda}} \varphi^{j}_{\beta} \varphi^{k}_{\rho} - \frac{3}{16}{\rm i} G_{j k} (\Gamma_{a})^{\beta \rho} \boldsymbol{W} {G}^{-3} \nabla_{\alpha}\,^{\lambda}{\lambda_{i \lambda}} \varphi^{j}_{\beta} \varphi^{k}_{\rho} - \frac{9}{32}{\rm i} G_{i j} (\Gamma_{a})^{\rho \lambda} \boldsymbol{W} W_{\alpha}\,^{\beta} \lambda_{k \beta} {G}^{-3} \varphi^{j}_{\rho} \varphi^{k}_{\lambda} - \frac{9}{32}{\rm i} G_{j k} (\Gamma_{a})^{\rho \lambda} \boldsymbol{W} W_{\alpha}\,^{\beta} \lambda_{i \beta} {G}^{-3} \varphi^{j}_{\rho} \varphi^{k}_{\lambda}%
 - \frac{3}{32}{\rm i} \mathcal{H}_{\alpha}\,^{\beta} G_{j k} (\Gamma_{a})_{\beta}{}^{\rho} \boldsymbol{W} X_{i}\,^{j} {G}^{-3} \varphi^{k}_{\rho} - \frac{3}{32}{\rm i} G_{j k} (\Gamma_{a})^{\beta}{}_{\rho} \boldsymbol{W} X_{i l} {G}^{-3} \nabla^{\rho}\,_{\alpha}{G^{j l}} \varphi^{k}_{\beta} - \frac{3}{16}(\Gamma_{a})^{\beta \rho} \boldsymbol{W} X_{j k} {G}^{-3} \varphi_{i \alpha} \varphi^{j}_{\beta} \varphi^{k}_{\rho}+\frac{9}{32}{\rm i} G_{i j} (\Gamma_{a})^{\beta \rho} X_{k l} \boldsymbol{\lambda}^{k}_{\alpha} {G}^{-3} \varphi^{j}_{\beta} \varphi^{l}_{\rho}+\frac{9}{16}G_{i j} G_{k l} (\Gamma_{a})^{\beta \rho} \boldsymbol{W} X^{k}\,_{m} {G}^{-5} \varphi^{l}_{\alpha} \varphi^{j}_{\beta} \varphi^{m}_{\rho}+\frac{3}{32}{\rm i} \mathcal{H}_{\alpha}\,^{\beta} G_{i j} (\Gamma_{a})_{\beta}{}^{\rho} \boldsymbol{W} X^{j}\,_{k} {G}^{-3} \varphi^{k}_{\rho} - \frac{3}{32}{\rm i} G_{i j} (\Gamma_{a})^{\beta}{}_{\rho} \boldsymbol{W} X_{k l} {G}^{-3} \nabla^{\rho}\,_{\alpha}{G^{j k}} \varphi^{l}_{\beta}+\frac{3}{32}{\rm i} G_{j k} (\Gamma_{a})^{\beta \rho} X^{j k} \boldsymbol{\lambda}_{i \beta} {G}^{-3} \varphi_{l \alpha} \varphi^{l}_{\rho} - \frac{3}{16}{\rm i} G_{j k} (\Gamma_{a})^{\beta \rho} X^{j}\,_{l} \boldsymbol{\lambda}_{i \beta} {G}^{-3} \varphi^{l}_{\alpha} \varphi^{k}_{\rho}+\frac{21}{64}(\Gamma_{a})^{\beta \rho} \boldsymbol{\lambda}_{i \beta} {G}^{-1} \nabla_{\alpha}\,^{\lambda}{\lambda_{j \lambda}} \varphi^{j}_{\rho} - \frac{3}{128}(\Gamma_{a})^{\rho \lambda} W_{\alpha}\,^{\beta} \lambda_{j \beta} \boldsymbol{\lambda}_{i \rho} {G}^{-1} \varphi^{j}_{\lambda}+\frac{3}{16}{\rm i} G_{i j} G_{k l} (\Gamma_{a})^{\beta \rho} X^{k l} \mathbf{F}_{\alpha \beta} {G}^{-3} \varphi^{j}_{\rho}+\frac{21}{64}{\rm i} G_{i j} G_{k l} (\Gamma_{a})^{\beta \rho} \boldsymbol{W} X^{k l} W_{\alpha \beta} {G}^{-3} \varphi^{j}_{\rho}+\frac{3}{32}{\rm i} G_{j k} (\Gamma_{a})^{\beta \rho} X^{j k} \boldsymbol{\lambda}_{l \beta} {G}^{-3} \varphi_{i \alpha} \varphi^{l}_{\rho}+\frac{9}{32}{\rm i} G_{i j} (\Gamma_{a})^{\beta \rho} X_{k l} \boldsymbol{\lambda}^{k}_{\beta} {G}^{-3} \varphi^{l}_{\alpha} \varphi^{j}_{\rho}+\frac{3}{64}G_{i j} G_{k l} (\Gamma_{a})^{\beta \rho} \boldsymbol{\lambda}^{k}_{\beta} {G}^{-3} \nabla_{\alpha}\,^{\lambda}{\lambda^{l}_{\lambda}} \varphi^{j}_{\rho} - \frac{3}{128}G_{i j} G_{k l} (\Gamma_{a})^{\rho \lambda} W_{\alpha}\,^{\beta} \lambda^{k}_{\beta} \boldsymbol{\lambda}^{l}_{\rho} {G}^{-3} \varphi^{j}_{\lambda} - \frac{9}{32}{\rm i} G_{i j} G_{k l} G_{m n} (\Gamma_{a})^{\beta \rho} X^{k l} \boldsymbol{\lambda}^{m}_{\beta} {G}^{-5} \varphi^{n}_{\alpha} \varphi^{j}_{\rho} - \frac{3}{64}\mathcal{H}_{\alpha}\,^{\beta} G_{i j} G_{k l} (\Gamma_{a})_{\beta}{}^{\rho} X^{k l} \boldsymbol{\lambda}^{j}_{\rho} {G}^{-3}+\frac{3}{64}G_{i j} G_{k l} (\Gamma_{a})^{\beta}{}_{\rho} X^{k l} \boldsymbol{\lambda}_{m \beta} {G}^{-3} \nabla^{\rho}\,_{\alpha}{G^{j m}}%
 - \frac{3}{16}{\rm i} G_{i j} (\Gamma_{a})^{\beta}{}_{\rho} \boldsymbol{W} {G}^{-3} \nabla^{\rho \lambda}{\lambda_{k \lambda}} \varphi^{k}_{\alpha} \varphi^{j}_{\beta}+\frac{3}{8}{\rm i} G_{j k} (\Gamma_{a})^{\beta}{}_{\rho} \boldsymbol{W} {G}^{-3} \nabla^{\rho \lambda}{\lambda^{j}_{\lambda}} \varphi_{i \alpha} \varphi^{k}_{\beta}+\frac{3}{16}{\rm i} G_{i j} (\Gamma_{a})^{\beta}{}_{\rho} \boldsymbol{W} {G}^{-3} \nabla^{\rho \lambda}{\lambda^{j}_{\lambda}} \varphi_{k \alpha} \varphi^{k}_{\beta} - \frac{3}{16}G_{i j} G_{k l} (\Gamma_{a})^{\beta}{}_{\rho} \boldsymbol{\lambda}^{k}_{\alpha} {G}^{-3} \nabla^{\rho \lambda}{\lambda^{j}_{\lambda}} \varphi^{l}_{\beta}+\frac{3}{8}{\rm i} (\Gamma_{a})^{\rho}{}_{\lambda} \boldsymbol{W} {G}^{-1} \nabla^{\lambda \beta}{F_{\alpha \beta}} \varphi_{i \rho}+\frac{9}{8}{\rm i} (\Gamma_{a})^{\rho}{}_{\lambda} \boldsymbol{W} W_{\alpha \beta} {G}^{-1} \nabla^{\lambda \beta}{W} \varphi_{i \rho} - \frac{3}{32}{\rm i} G_{i j} G_{k l} (\Gamma_{a})^{\beta}{}_{\rho} \boldsymbol{W} {G}^{-3} \nabla^{\rho}\,_{\alpha}{X^{j k}} \varphi^{l}_{\beta} - \frac{3}{16}{\rm i} (\Gamma_{a})^{\beta}{}_{\rho} \boldsymbol{W} {G}^{-1} \nabla^{\rho}\,_{\lambda}{\nabla_{\alpha}\,^{\lambda}{W}} \varphi_{i \beta} - \frac{3}{16}{\rm i} G_{i j} G_{k l} (\Gamma_{a})^{\beta \rho} \boldsymbol{W} X^{j k} W_{\alpha \beta} {G}^{-3} \varphi^{l}_{\rho} - \frac{9}{8}{\rm i} (\Gamma_{a})^{\beta \lambda} \boldsymbol{W} W_{\beta \rho} {G}^{-1} \nabla_{\alpha}\,^{\rho}{W} \varphi_{i \lambda} - \frac{81}{2560}{\rm i} G_{i j} G_{k l} (\Gamma_{a})_{\alpha}{}^{\rho} \boldsymbol{W} \lambda^{j \beta} X^{k}_{\beta} {G}^{-3} \varphi^{l}_{\rho}+\frac{99}{2560}{\rm i} (\Gamma_{a})_{\alpha}{}^{\rho} \boldsymbol{W} \lambda^{\beta}_{j} X^{j}_{\beta} {G}^{-1} \varphi_{i \rho}+\frac{9}{256}{\rm i} G_{i j} G_{k l} (\Gamma_{a})_{\alpha}{}^{\rho} \boldsymbol{W} \lambda^{k \beta} X^{j}_{\beta} {G}^{-3} \varphi^{l}_{\rho} - \frac{9}{32}{\rm i} G_{i j} (\Gamma_{a})^{\beta \lambda} \boldsymbol{W} W_{\beta}\,^{\rho} \lambda_{k \rho} {G}^{-3} \varphi^{k}_{\alpha} \varphi^{j}_{\lambda}+\frac{9}{16}{\rm i} G_{j k} (\Gamma_{a})^{\beta \lambda} \boldsymbol{W} W_{\beta}\,^{\rho} \lambda^{j}_{\rho} {G}^{-3} \varphi_{i \alpha} \varphi^{k}_{\lambda}+\frac{9}{32}{\rm i} G_{i j} (\Gamma_{a})^{\beta \lambda} \boldsymbol{W} W_{\beta}\,^{\rho} \lambda^{j}_{\rho} {G}^{-3} \varphi_{k \alpha} \varphi^{k}_{\lambda}+\frac{3}{64}G_{i j} G_{k l} (\Gamma_{a})^{\beta}{}_{\rho} \boldsymbol{\lambda}^{k}_{\alpha} {G}^{-3} \nabla^{\rho \lambda}{\lambda^{l}_{\lambda}} \varphi^{j}_{\beta}+\frac{27}{320}F (\Gamma_{a})_{\alpha \beta} \boldsymbol{W} {G}^{-1} \nabla^{\beta \rho}{\lambda_{i \rho}}+\frac{3}{16}\mathcal{H}_{\alpha}\,^{\beta} (\Gamma_{a})_{\beta \rho} \boldsymbol{W} {G}^{-1} \nabla^{\rho \lambda}{\lambda_{i \lambda}} - \frac{3}{32}G_{i j} G_{k l} (\Gamma_{a})_{\beta \rho} \boldsymbol{W} {G}^{-3} \nabla^{\beta \lambda}{\lambda^{k}_{\lambda}} \nabla^{\rho}\,_{\alpha}{G^{j l}}%
 - \frac{63}{320}F (\Gamma_{a})_{\alpha}{}^{\beta} \boldsymbol{W} W_{\beta}\,^{\rho} \lambda_{i \rho} {G}^{-1} - \frac{9}{32}\mathcal{H}_{\alpha}\,^{\lambda} (\Gamma_{a})_{\lambda}{}^{\beta} \boldsymbol{W} W_{\beta}\,^{\rho} \lambda_{i \rho} {G}^{-1}+\frac{9}{64}G_{i j} G_{k l} (\Gamma_{a})^{\beta}{}_{\lambda} \boldsymbol{W} W_{\beta}\,^{\rho} \lambda^{k}_{\rho} {G}^{-3} \nabla^{\lambda}\,_{\alpha}{G^{j l}} - \frac{9}{32}G_{i j} G_{k l} (\Gamma_{a})^{\beta \rho} \boldsymbol{W} X^{k l} {G}^{-5} \varphi_{m \alpha} \varphi^{j}_{\beta} \varphi^{m}_{\rho}+\frac{9}{32}G_{j k} G_{l m} (\Gamma_{a})^{\beta \rho} \boldsymbol{W} X^{j k} {G}^{-5} \varphi_{i \alpha} \varphi^{l}_{\beta} \varphi^{m}_{\rho} - \frac{9}{16}G_{i j} G_{k l} (\Gamma_{a})^{\beta \rho} \boldsymbol{W} X^{k}\,_{m} {G}^{-5} \varphi^{m}_{\alpha} \varphi^{j}_{\beta} \varphi^{l}_{\rho} - \frac{9}{32}{\rm i} G_{i j} G_{k l} G_{m n} (\Gamma_{a})^{\beta \rho} X^{k l} \boldsymbol{\lambda}^{m}_{\alpha} {G}^{-5} \varphi^{j}_{\beta} \varphi^{n}_{\rho}+\frac{9}{64}{\rm i} G_{i j} G_{k l} G_{m n} (\Gamma_{a})^{\beta}{}_{\rho} \boldsymbol{W} X^{k l} {G}^{-5} \nabla^{\rho}\,_{\alpha}{G^{j m}} \varphi^{n}_{\beta} - \frac{3}{64}{\rm i} G_{i j} G_{k l} (\Gamma_{a})^{\beta}{}_{\rho} \boldsymbol{W} X^{k l} {G}^{-3} \nabla^{\rho}\,_{\alpha}{\varphi^{j}_{\beta}}+\frac{21}{320}{\rm i} G_{i j} G_{k l} (\Gamma_{a})_{\alpha}{}^{\beta} \boldsymbol{W} X^{k l} W_{\beta}\,^{\rho} {G}^{-3} \varphi^{j}_{\rho} - \frac{39}{320}{\rm i} G_{i j} G_{k l} (\Gamma_{a})_{\alpha \beta} \boldsymbol{W} X^{k l} {G}^{-3} \nabla^{\beta \rho}{\varphi^{j}_{\rho}}+\frac{3}{32}{\rm i} \mathcal{H}^{\beta \rho} G_{i j} (\Gamma_{a})_{\beta \rho} \boldsymbol{W} X^{j}\,_{k} {G}^{-3} \varphi^{k}_{\alpha} - \frac{3}{64}\mathcal{H}^{\beta \rho} G_{i j} G_{k l} (\Gamma_{a})_{\beta \rho} X^{k l} \boldsymbol{\lambda}^{j}_{\alpha} {G}^{-3} - \frac{3}{16}\mathcal{H}^{\beta \rho} (\Gamma_{a})_{\beta \rho} \boldsymbol{W} {G}^{-1} \nabla_{\alpha}\,^{\lambda}{\lambda_{i \lambda}} - \frac{9}{32}\mathcal{H}^{\rho \lambda} (\Gamma_{a})_{\rho \lambda} \boldsymbol{W} W_{\alpha}\,^{\beta} \lambda_{i \beta} {G}^{-1} - \frac{3}{128}{\rm i} G_{j k} (\Gamma_{a})_{\beta \rho} \boldsymbol{W} X^{j k} {G}^{-3} \nabla^{\beta \rho}{G_{i l}} \varphi^{l}_{\alpha}+\frac{3}{32}{\rm i} G_{j k} (\Gamma_{a})_{\beta \rho} \boldsymbol{W} X^{j}\,_{l} {G}^{-3} \nabla^{\beta \rho}{G_{i}\,^{k}} \varphi^{l}_{\alpha} - \frac{3}{64}G_{j k} G_{l m} (\Gamma_{a})_{\beta \rho} X^{j k} \boldsymbol{\lambda}^{l}_{\alpha} {G}^{-3} \nabla^{\beta \rho}{G_{i}\,^{m}} - \frac{9}{32}(\Gamma_{a})_{\beta \rho} \boldsymbol{W} {G}^{-1} \nabla_{\alpha}\,^{\lambda}{\lambda_{j \lambda}} \nabla^{\beta \rho}{G_{i}\,^{j}} - \frac{27}{128}(\Gamma_{a})_{\rho \lambda} \boldsymbol{W} W_{\alpha}\,^{\beta} \lambda_{j \beta} {G}^{-1} \nabla^{\rho \lambda}{G_{i}\,^{j}}%
 - \frac{3}{16}{\rm i} G_{i j} G_{k l} (\Gamma_{a})_{\beta \rho} \boldsymbol{W} X^{k l} {G}^{-3} \nabla^{\beta \rho}{\varphi^{j}_{\alpha}} - \frac{3}{32}{\rm i} G_{i j} (\Gamma_{a})_{\beta \rho} \boldsymbol{W} X_{k l} {G}^{-3} \nabla^{\beta \rho}{G^{j k}} \varphi^{l}_{\alpha}+\frac{3}{64}G_{i j} G_{k l} (\Gamma_{a})_{\beta \rho} X^{k l} \boldsymbol{\lambda}_{m \alpha} {G}^{-3} \nabla^{\beta \rho}{G^{j m}}+\frac{3}{32}G_{i j} G_{k l} (\Gamma_{a})_{\beta \rho} \boldsymbol{W} {G}^{-3} \nabla_{\alpha}\,^{\lambda}{\lambda^{k}_{\lambda}} \nabla^{\beta \rho}{G^{j l}}+\frac{9}{64}G_{i j} G_{k l} (\Gamma_{a})_{\rho \lambda} \boldsymbol{W} W_{\alpha}\,^{\beta} \lambda^{k}_{\beta} {G}^{-3} \nabla^{\rho \lambda}{G^{j l}}+\frac{9}{64}{\rm i} G_{i j} G_{k l} G_{m n} (\Gamma_{a})_{\beta \rho} \boldsymbol{W} X^{k l} {G}^{-5} \nabla^{\beta \rho}{G^{j m}} \varphi^{n}_{\alpha} - \frac{15}{32}{\rm i} (\Gamma_{a})^{\beta \rho} \boldsymbol{W} X_{i j} W_{\alpha \beta} {G}^{-1} \varphi^{j}_{\rho} - \frac{9}{32}{\rm i} (\Gamma_{a})^{\beta}{}_{\rho} X_{i j} {G}^{-1} \nabla^{\rho}\,_{\alpha}{\boldsymbol{W}} \varphi^{j}_{\beta}+\frac{3}{32}{\rm i} G_{j k} (\Gamma_{a})^{\beta \rho} X^{j}\,_{l} \boldsymbol{\lambda}_{i \beta} {G}^{-3} \varphi^{k}_{\alpha} \varphi^{l}_{\rho}+\frac{3}{64}(\Gamma_{a})^{\beta}{}_{\rho} X_{j k} \boldsymbol{\lambda}_{i \beta} {G}^{-1} \nabla^{\rho}\,_{\alpha}{G^{j k}} - \frac{15}{64}(\Gamma_{a})^{\beta \rho} \boldsymbol{\lambda}_{j \beta} {G}^{-1} \nabla_{\alpha}\,^{\lambda}{\lambda_{i \lambda}} \varphi^{j}_{\rho}+\frac{3}{64}(\Gamma_{a})^{\rho \lambda} W_{\alpha}\,^{\beta} \lambda_{i \beta} \boldsymbol{\lambda}_{j \rho} {G}^{-1} \varphi^{j}_{\lambda}+\frac{3}{32}{\rm i} G_{j k} (\Gamma_{a})^{\beta \rho} X_{i l} \boldsymbol{\lambda}^{j}_{\beta} {G}^{-3} \varphi^{k}_{\alpha} \varphi^{l}_{\rho}+\frac{9}{64}\mathcal{H}_{\alpha}\,^{\beta} (\Gamma_{a})_{\beta}{}^{\rho} X_{i j} \boldsymbol{\lambda}^{j}_{\rho} {G}^{-1} - \frac{3}{64}(\Gamma_{a})^{\beta}{}_{\rho} X_{i j} \boldsymbol{\lambda}_{k \beta} {G}^{-1} \nabla^{\rho}\,_{\alpha}{G^{j k}}+\frac{9}{32}(\Gamma_{a})^{\beta}{}_{\rho} \boldsymbol{\lambda}_{j \alpha} {G}^{-1} \nabla^{\rho \lambda}{\lambda^{j}_{\lambda}} \varphi_{i \beta} - \frac{3}{16}{\rm i} G_{j k} (\Gamma_{a})^{\beta}{}_{\rho} \boldsymbol{W} {G}^{-3} \nabla^{\rho \lambda}{\lambda^{j}_{\lambda}} \varphi^{k}_{\alpha} \varphi_{i \beta}+\frac{9}{32}(\Gamma_{a})_{\beta \rho} \boldsymbol{W} {G}^{-1} \nabla^{\beta \lambda}{\lambda_{j \lambda}} \nabla^{\rho}\,_{\alpha}{G_{i}\,^{j}} - \frac{9}{32}{\rm i} G_{j k} (\Gamma_{a})^{\beta \lambda} \boldsymbol{W} W_{\beta}\,^{\rho} \lambda^{j}_{\rho} {G}^{-3} \varphi^{k}_{\alpha} \varphi_{i \lambda} - \frac{27}{64}(\Gamma_{a})^{\beta}{}_{\lambda} \boldsymbol{W} W_{\beta}\,^{\rho} \lambda_{j \rho} {G}^{-1} \nabla^{\lambda}\,_{\alpha}{G_{i}\,^{j}}%
+\frac{3}{64}(\Gamma_{a})^{\beta}{}_{\rho} \boldsymbol{\lambda}_{j \alpha} {G}^{-1} \nabla^{\rho \lambda}{\lambda_{i \lambda}} \varphi^{j}_{\beta} - \frac{3}{16}{\rm i} G_{j k} (\Gamma_{a})^{\beta}{}_{\rho} \boldsymbol{W} {G}^{-3} \nabla^{\rho \lambda}{\lambda_{i \lambda}} \varphi^{j}_{\alpha} \varphi^{k}_{\beta} - \frac{9}{32}{\rm i} (\Gamma_{a})^{\beta}{}_{\rho} \boldsymbol{W} {G}^{-1} \nabla^{\rho}\,_{\alpha}{X_{i j}} \varphi^{j}_{\beta} - \frac{27}{256}{\rm i} (\Gamma_{a})^{\beta \rho} \boldsymbol{W} \lambda_{i \alpha} X_{j \beta} {G}^{-1} \varphi^{j}_{\rho}+\frac{189}{2560}{\rm i} (\Gamma_{a})_{\alpha}{}^{\rho} \boldsymbol{W} \lambda^{\beta}_{i} X_{j \beta} {G}^{-1} \varphi^{j}_{\rho} - \frac{9}{128}{\rm i} (\Gamma_{a})_{\alpha}{}^{\rho} \boldsymbol{W} \lambda^{\beta}_{j} X_{i \beta} {G}^{-1} \varphi^{j}_{\rho} - \frac{9}{32}{\rm i} G_{j k} (\Gamma_{a})^{\beta \lambda} \boldsymbol{W} W_{\beta}\,^{\rho} \lambda_{i \rho} {G}^{-3} \varphi^{j}_{\alpha} \varphi^{k}_{\lambda}+\frac{3}{32}{\rm i} (\Gamma_{a})^{\beta}{}_{\rho} \boldsymbol{W} X_{i j} {G}^{-1} \nabla^{\rho}\,_{\alpha}{\varphi^{j}_{\beta}}+\frac{9}{64}\mathcal{H}^{\beta \rho} (\Gamma_{a})_{\beta \rho} X_{i j} \boldsymbol{\lambda}^{j}_{\alpha} {G}^{-1} - \frac{3}{32}{\rm i} \mathcal{H}^{\beta \rho} G_{j k} (\Gamma_{a})_{\beta \rho} \boldsymbol{W} X_{i}\,^{j} {G}^{-3} \varphi^{k}_{\alpha}+\frac{9}{64}(\Gamma_{a})_{\beta \rho} X_{k j} \boldsymbol{\lambda}^{k}_{\alpha} {G}^{-1} \nabla^{\beta \rho}{G_{i}\,^{j}} - \frac{3}{32}{\rm i} G_{j k} (\Gamma_{a})_{\beta \rho} \boldsymbol{W} X^{j}\,_{l} {G}^{-3} \nabla^{\beta \rho}{G_{i}\,^{l}} \varphi^{k}_{\alpha}+\frac{3}{8}{\rm i} (\Gamma_{a})_{\beta \rho} \boldsymbol{W} X_{i j} {G}^{-1} \nabla^{\beta \rho}{\varphi^{j}_{\alpha}}+\frac{3}{128}(\Gamma_{a})_{\beta \rho} X_{i j} \boldsymbol{\lambda}_{k \alpha} {G}^{-1} \nabla^{\beta \rho}{G^{j k}} - \frac{3}{32}{\rm i} G_{j k} (\Gamma_{a})_{\beta \rho} \boldsymbol{W} X_{i l} {G}^{-3} \nabla^{\beta \rho}{G^{j l}} \varphi^{k}_{\alpha} - \frac{9}{16}(\Gamma_{a})^{\beta}{}_{\rho} \boldsymbol{\lambda}_{i \beta} {G}^{-1} \nabla^{\rho \lambda}{\lambda_{j \lambda}} \varphi^{j}_{\alpha} - \frac{9}{32}G_{i j} (\Gamma_{a})^{\beta}{}_{\rho} \mathbf{F}_{\alpha \beta} {G}^{-1} \nabla^{\rho \lambda}{\lambda^{j}_{\lambda}} - \frac{27}{32}G_{i j} (\Gamma_{a})^{\beta}{}_{\rho} \boldsymbol{W} W_{\alpha \beta} {G}^{-1} \nabla^{\rho \lambda}{\lambda^{j}_{\lambda}}+\frac{9}{160}G_{j k} (\Gamma_{a})_{\alpha \beta} \mathbf{X}_{i}\,^{j} {G}^{-1} \nabla^{\beta \rho}{\lambda^{k}_{\rho}} - \frac{15}{64}G_{i j} (\Gamma_{a})_{\beta \rho} {G}^{-1} \nabla^{\beta}\,_{\alpha}{\boldsymbol{W}} \nabla^{\rho \lambda}{\lambda^{j}_{\lambda}}%
 - \frac{9}{320}G_{j k} (\Gamma_{a})_{\alpha}{}^{\rho} \lambda^{j \beta} \boldsymbol{\lambda}_{i \rho} X^{k}_{\beta} {G}^{-1}+\frac{9}{32}(\Gamma_{a})^{\beta}{}_{\rho} \boldsymbol{\lambda}_{j \beta} {G}^{-1} \nabla^{\rho \lambda}{\lambda^{j}_{\lambda}} \varphi_{i \alpha} - \frac{3}{16}G_{i j} G_{k l} (\Gamma_{a})^{\beta}{}_{\rho} \boldsymbol{\lambda}^{k}_{\beta} {G}^{-3} \nabla^{\rho \lambda}{\lambda^{j}_{\lambda}} \varphi^{l}_{\alpha} - \frac{3}{32}G_{i j} (\Gamma_{a})^{\rho}{}_{\lambda} \boldsymbol{\lambda}^{j}_{\rho} {G}^{-1} \nabla^{\lambda \beta}{F_{\alpha \beta}} - \frac{3}{32}G_{i j} (\Gamma_{a})^{\beta}{}_{\rho} \boldsymbol{\lambda}_{k \beta} {G}^{-1} \nabla^{\rho}\,_{\alpha}{X^{j k}}+\frac{3}{64}G_{i j} (\Gamma_{a})^{\beta}{}_{\rho} \boldsymbol{\lambda}^{j}_{\beta} {G}^{-1} \nabla^{\rho}\,_{\lambda}{\nabla_{\alpha}\,^{\lambda}{W}}+\frac{9}{64}G_{i j} (\Gamma_{a})^{\beta \lambda} W_{\alpha}\,^{\rho} F_{\beta \rho} \boldsymbol{\lambda}^{j}_{\lambda} {G}^{-1} - \frac{9}{64}G_{i j} (\Gamma_{a})^{\beta \rho} X^{j}\,_{k} W_{\alpha \beta} \boldsymbol{\lambda}^{k}_{\rho} {G}^{-1}+\frac{27}{128}G_{i j} (\Gamma_{a})^{\rho}{}_{\lambda} W_{\alpha \beta} \boldsymbol{\lambda}^{j}_{\rho} {G}^{-1} \nabla^{\lambda \beta}{W} - \frac{9}{64}G_{i j} (\Gamma_{a})_{\alpha}{}^{\lambda} W^{\beta \rho} F_{\beta \rho} \boldsymbol{\lambda}^{j}_{\lambda} {G}^{-1} - \frac{9}{64}G_{i j} (\Gamma_{a})^{\rho \lambda} W_{\rho}\,^{\beta} F_{\alpha \beta} \boldsymbol{\lambda}^{j}_{\lambda} {G}^{-1} - \frac{9}{128}G_{i j} (\Gamma_{a})^{\beta \lambda} W_{\beta \rho} \boldsymbol{\lambda}^{j}_{\lambda} {G}^{-1} \nabla_{\alpha}\,^{\rho}{W} - \frac{123}{1280}G_{i j} (\Gamma_{a})_{\alpha}{}^{\rho} \lambda^{j \beta} \boldsymbol{\lambda}_{k \rho} X^{k}_{\beta} {G}^{-1} - \frac{57}{640}G_{i j} (\Gamma_{a})_{\alpha}{}^{\rho} \lambda^{\beta}_{k} \boldsymbol{\lambda}^{j}_{\rho} X^{k}_{\beta} {G}^{-1}+\frac{27}{1280}G_{i j} (\Gamma_{a})_{\alpha}{}^{\rho} \lambda^{\beta}_{k} \boldsymbol{\lambda}^{k}_{\rho} X^{j}_{\beta} {G}^{-1}+\frac{3}{4}{\rm i} (\Gamma_{a})^{\beta}{}_{\lambda} \boldsymbol{W} {G}^{-1} \nabla^{\lambda \rho}{F_{\beta \rho}} \varphi_{i \alpha}+\frac{3}{32}G_{i j} (\Gamma_{a})^{\beta}{}_{\lambda} \boldsymbol{\lambda}^{j}_{\alpha} {G}^{-1} \nabla^{\lambda \rho}{F_{\beta \rho}}+\frac{3}{32}G_{i j} (\Gamma_{a})^{\beta}{}_{\rho} \boldsymbol{W} {G}^{-1} \nabla^{\rho}\,_{\lambda}{\nabla_{\alpha}\,^{\lambda}{\lambda^{j}_{\beta}}}+\frac{3}{32}G_{i j} (\Gamma_{a})_{\beta \rho} \boldsymbol{W} {G}^{-1} \nabla^{\beta \lambda}{\nabla^{\rho}\,_{\alpha}{\lambda^{j}_{\lambda}}}+\frac{9}{64}G_{i j} (\Gamma_{a})^{\beta}{}_{\lambda} \boldsymbol{W} W_{\beta \rho} {G}^{-1} \nabla^{\lambda \rho}{\lambda^{j}_{\alpha}}%
+\frac{27}{64}G_{i j} (\Gamma_{a})^{\rho}{}_{\lambda} \boldsymbol{W} W_{\alpha \beta} {G}^{-1} \nabla^{\lambda \beta}{\lambda^{j}_{\rho}}+\frac{9}{128}G_{i j} (\Gamma_{a})^{\beta}{}_{\rho} \boldsymbol{W} \lambda^{j}_{\lambda} {G}^{-1} \nabla^{\rho \lambda}{W_{\alpha \beta}}+\frac{51}{160}G_{i j} (\Gamma_{a})_{\alpha \beta} \boldsymbol{W} {G}^{-1} \nabla^{\beta}\,_{\lambda}{\nabla^{\lambda \rho}{\lambda^{j}_{\rho}}}+\frac{9}{128}G_{i j} (\Gamma_{a})^{\beta}{}_{\lambda} \boldsymbol{W} \lambda^{j \rho} {G}^{-1} \nabla^{\lambda}\,_{\alpha}{W_{\beta \rho}}+\frac{9}{64}G_{i j} (\Gamma_{a})^{\beta}{}_{\lambda} \boldsymbol{W} W_{\beta}\,^{\rho} {G}^{-1} \nabla^{\lambda}\,_{\alpha}{\lambda^{j}_{\rho}}+\frac{1071}{2560}G_{i j} (\Gamma_{a})_{\alpha \lambda} \boldsymbol{W} \lambda^{j \beta} {G}^{-1} \nabla^{\lambda \rho}{W_{\beta \rho}} - \frac{81}{160}G_{i j} (\Gamma_{a})_{\alpha \lambda} \boldsymbol{W} W^{\beta}\,_{\rho} {G}^{-1} \nabla^{\lambda \rho}{\lambda^{j}_{\beta}} - \frac{3}{8}{\rm i} G_{i j} (\Gamma_{a})^{\beta}{}_{\lambda} \boldsymbol{W} W_{\alpha \beta \rho}\,^{j} {G}^{-1} \nabla^{\lambda \rho}{W} - \frac{81}{640}G_{i j} (\Gamma_{a})_{\alpha}{}^{\beta} \boldsymbol{W} W_{\beta}\,^{\rho} W_{\rho}\,^{\lambda} \lambda^{j}_{\lambda} {G}^{-1} - \frac{63}{320}G_{i j} (\Gamma_{a})_{\alpha}{}^{\beta} \boldsymbol{W} W_{\beta \rho} {G}^{-1} \nabla^{\rho \lambda}{\lambda^{j}_{\lambda}}+\frac{9}{64}G_{i j} (\Gamma_{a})^{\beta \lambda} \boldsymbol{W} W_{\beta \rho} {G}^{-1} \nabla_{\alpha}\,^{\rho}{\lambda^{j}_{\lambda}}+\frac{27}{64}G_{i j} (\Gamma_{a})^{\beta \rho} \boldsymbol{W} W_{\alpha \beta} W_{\rho}\,^{\lambda} \lambda^{j}_{\lambda} {G}^{-1}+\frac{27}{256}G_{i j} (\Gamma_{a})^{\beta \lambda} \boldsymbol{W} W_{\beta}\,^{\rho} W_{\lambda \rho} \lambda^{j}_{\alpha} {G}^{-1} - \frac{27}{64}G_{i j} (\Gamma_{a})_{\rho \lambda} \boldsymbol{W} W_{\alpha}\,^{\beta} {G}^{-1} \nabla^{\rho \lambda}{\lambda^{j}_{\beta}} - \frac{51}{640}\Phi^{\beta \rho}\,_{j k} G_{i}\,^{j} (\Gamma_{a})_{\alpha \beta} \boldsymbol{W} \lambda^{k}_{\rho} {G}^{-1}+\frac{81}{128}G_{i j} (\Gamma_{a})_{\rho \lambda} \boldsymbol{W} \lambda^{j \beta} {G}^{-1} \nabla^{\rho \lambda}{W_{\alpha \beta}}+\frac{9}{512}G_{i j} (\Gamma_{a})_{\alpha}{}^{\beta} \boldsymbol{W} \lambda^{j}_{\lambda} {G}^{-1} \nabla^{\lambda \rho}{W_{\beta \rho}}+\frac{9}{4}{\rm i} (\Gamma_{a})^{\beta}{}_{\lambda} \boldsymbol{W} W_{\beta \rho} {G}^{-1} \nabla^{\lambda \rho}{W} \varphi_{i \alpha} - \frac{9}{128}G_{i j} (\Gamma_{a})^{\beta}{}_{\lambda} W_{\beta \rho} \boldsymbol{\lambda}^{j}_{\alpha} {G}^{-1} \nabla^{\lambda \rho}{W} - \frac{159}{320}{\rm i} G_{i j} (\Gamma_{a})_{\alpha \rho} \boldsymbol{W} X^{j}_{\beta} {G}^{-1} \nabla^{\rho \beta}{W}%
 - \frac{9}{64}{\rm i} (\Gamma_{a})_{\beta \rho} \boldsymbol{W} {G}^{-1} \nabla^{\beta \rho}{X_{i j}} \varphi^{j}_{\alpha}+\frac{3}{64}G_{j k} (\Gamma_{a})_{\beta \rho} \boldsymbol{\lambda}^{j}_{\alpha} {G}^{-1} \nabla^{\beta \rho}{X_{i}\,^{k}} - \frac{21}{64}G_{i j} (\Gamma_{a})_{\beta \rho} \boldsymbol{W} {G}^{-1} \nabla^{\beta \rho}{\nabla_{\alpha}\,^{\lambda}{\lambda^{j}_{\lambda}}} - \frac{177}{256}{\rm i} G_{i j} (\Gamma_{a})_{\beta \rho} \boldsymbol{W} X^{j}_{\alpha} {G}^{-1} \nabla^{\beta \rho}{W} - \frac{45}{128}\Phi_{\alpha}\,^{\beta}\,_{i j} G^{j}\,_{k} (\Gamma_{a})_{\beta}{}^{\rho} \boldsymbol{W} \lambda^{k}_{\rho} {G}^{-1} - \frac{9}{128}\Phi^{\beta \rho}\,_{i j} G^{j}\,_{k} (\Gamma_{a})_{\alpha \beta} \boldsymbol{W} \lambda^{k}_{\rho} {G}^{-1}+\frac{99}{640}\Phi^{\beta \rho}\,_{j k} G^{j k} (\Gamma_{a})_{\alpha \beta} \boldsymbol{W} \lambda_{i \rho} {G}^{-1} - \frac{3}{8}{\rm i} (\Gamma_{a})_{\beta \rho} \boldsymbol{W} {G}^{-1} \nabla^{\beta}\,_{\lambda}{\nabla^{\rho \lambda}{W}} \varphi_{i \alpha} - \frac{3}{64}G_{i j} (\Gamma_{a})_{\beta \rho} \boldsymbol{\lambda}^{j}_{\alpha} {G}^{-1} \nabla^{\beta}\,_{\lambda}{\nabla^{\rho \lambda}{W}} - \frac{3}{32}G_{i j} (\Gamma_{a})_{\beta \rho} \boldsymbol{W} {G}^{-1} \nabla^{\beta}\,_{\lambda}{\nabla^{\rho \lambda}{\lambda^{j}_{\alpha}}}+\frac{9}{256}G_{j k} (\Gamma_{a})^{\rho \beta} \boldsymbol{\lambda}^{j}_{\alpha} \lambda^{k}_{\rho} X_{i \beta} {G}^{-1} - \frac{3}{32}G_{i j} (\Gamma_{a})_{\beta \rho} \boldsymbol{\lambda}_{k \alpha} {G}^{-1} \nabla^{\beta \rho}{X^{j k}} - \frac{3}{32}{\rm i} G_{i j} G_{k l} (\Gamma_{a})_{\beta \rho} \boldsymbol{W} {G}^{-3} \nabla^{\beta \rho}{X^{j k}} \varphi^{l}_{\alpha} - \frac{3}{32}{\rm i} (\Gamma_{a})^{\beta \rho} \lambda^{\lambda}_{j} \boldsymbol{\lambda}_{k \lambda} {G}^{-3} \varphi^{k}_{\alpha} \varphi_{i \beta} \varphi^{j}_{\rho} - \frac{3}{32}{\rm i} (\Gamma_{a})^{\beta \rho} \lambda^{\lambda}_{j} \boldsymbol{\lambda}_{k \lambda} {G}^{-3} \varphi^{j}_{\alpha} \varphi_{i \beta} \varphi^{k}_{\rho}+\frac{3}{16}{\rm i} G_{j k} (\Gamma_{a})^{\rho \lambda} F_{\alpha}\,^{\beta} \boldsymbol{\lambda}^{j}_{\beta} {G}^{-3} \varphi_{i \rho} \varphi^{k}_{\lambda} - \frac{3}{32}{\rm i} G_{j k} (\Gamma_{a})^{\beta \rho} X^{j}\,_{l} \boldsymbol{\lambda}^{k}_{\alpha} {G}^{-3} \varphi_{i \beta} \varphi^{l}_{\rho}+\frac{3}{32}{\rm i} G_{j k} (\Gamma_{a})^{\beta \rho} \boldsymbol{\lambda}^{j}_{\lambda} {G}^{-3} \nabla_{\alpha}\,^{\lambda}{W} \varphi_{i \beta} \varphi^{k}_{\rho}+\frac{3}{16}{\rm i} G_{j k} (\Gamma_{a})^{\rho \lambda} \mathbf{F}_{\alpha}\,^{\beta} \lambda^{j}_{\beta} {G}^{-3} \varphi_{i \rho} \varphi^{k}_{\lambda} - \frac{3}{32}{\rm i} G_{j k} (\Gamma_{a})^{\beta \rho} \mathbf{X}^{j}\,_{l} \lambda^{k}_{\alpha} {G}^{-3} \varphi_{i \beta} \varphi^{l}_{\rho}%
+\frac{3}{32}{\rm i} G_{j k} (\Gamma_{a})^{\beta \rho} \lambda^{j}_{\lambda} {G}^{-3} \nabla_{\alpha}\,^{\lambda}{\boldsymbol{W}} \varphi_{i \beta} \varphi^{k}_{\rho}+\frac{9}{32}{\rm i} G_{j k} G_{l m} (\Gamma_{a})^{\beta \rho} \lambda^{j \lambda} \boldsymbol{\lambda}^{k}_{\lambda} {G}^{-5} \varphi^{l}_{\alpha} \varphi_{i \beta} \varphi^{m}_{\rho}+\frac{3}{64}G_{j k} (\Gamma_{a})^{\beta}{}_{\rho} \lambda^{j \lambda} \boldsymbol{\lambda}^{k}_{\lambda} {G}^{-3} \nabla^{\rho}\,_{\alpha}{G_{i l}} \varphi^{l}_{\beta}+\frac{3}{32}{\rm i} (\Gamma_{a})^{\beta \rho} \lambda^{\lambda}_{i} \boldsymbol{\lambda}_{j \lambda} {G}^{-3} \varphi_{k \alpha} \varphi^{j}_{\beta} \varphi^{k}_{\rho} - \frac{3}{16}{\rm i} G_{i j} (\Gamma_{a})^{\rho \lambda} F_{\alpha}\,^{\beta} \boldsymbol{\lambda}_{k \beta} {G}^{-3} \varphi^{j}_{\rho} \varphi^{k}_{\lambda} - \frac{3}{32}{\rm i} G_{j k} (\Gamma_{a})^{\beta \rho} X_{i}\,^{j} \boldsymbol{\lambda}_{l \alpha} {G}^{-3} \varphi^{k}_{\beta} \varphi^{l}_{\rho} - \frac{3}{32}{\rm i} G_{i j} (\Gamma_{a})^{\beta \rho} \boldsymbol{\lambda}_{k \lambda} {G}^{-3} \nabla_{\alpha}\,^{\lambda}{W} \varphi^{j}_{\beta} \varphi^{k}_{\rho} - \frac{3}{16}{\rm i} G_{j k} (\Gamma_{a})^{\rho \lambda} \mathbf{F}_{\alpha}\,^{\beta} \lambda_{i \beta} {G}^{-3} \varphi^{j}_{\rho} \varphi^{k}_{\lambda} - \frac{3}{32}{\rm i} G_{j k} (\Gamma_{a})^{\beta \rho} \mathbf{X}^{j}\,_{l} \lambda_{i \alpha} {G}^{-3} \varphi^{k}_{\beta} \varphi^{l}_{\rho} - \frac{3}{32}{\rm i} G_{j k} (\Gamma_{a})^{\beta \rho} \lambda_{i \lambda} {G}^{-3} \nabla_{\alpha}\,^{\lambda}{\boldsymbol{W}} \varphi^{j}_{\beta} \varphi^{k}_{\rho} - \frac{3}{320}F G_{j k} (\Gamma_{a})_{\alpha}{}^{\beta} \lambda^{\rho}_{i} \boldsymbol{\lambda}^{j}_{\rho} {G}^{-3} \varphi^{k}_{\beta} - \frac{3}{64}\mathcal{H}_{\alpha}\,^{\beta} G_{j k} (\Gamma_{a})_{\beta}{}^{\rho} \lambda^{\lambda}_{i} \boldsymbol{\lambda}^{j}_{\lambda} {G}^{-3} \varphi^{k}_{\rho} - \frac{3}{64}G_{j k} (\Gamma_{a})^{\beta}{}_{\rho} \lambda^{\lambda}_{i} \boldsymbol{\lambda}_{l \lambda} {G}^{-3} \nabla^{\rho}\,_{\alpha}{G^{j l}} \varphi^{k}_{\beta}+\frac{3}{32}{\rm i} (\Gamma_{a})^{\beta \rho} \lambda^{\lambda}_{j} \boldsymbol{\lambda}_{i \lambda} {G}^{-3} \varphi_{k \alpha} \varphi^{j}_{\beta} \varphi^{k}_{\rho} - \frac{3}{16}{\rm i} G_{j k} (\Gamma_{a})^{\rho \lambda} F_{\alpha}\,^{\beta} \boldsymbol{\lambda}_{i \beta} {G}^{-3} \varphi^{j}_{\rho} \varphi^{k}_{\lambda} - \frac{3}{32}{\rm i} G_{j k} (\Gamma_{a})^{\beta \rho} X^{j}\,_{l} \boldsymbol{\lambda}_{i \alpha} {G}^{-3} \varphi^{k}_{\beta} \varphi^{l}_{\rho} - \frac{3}{32}{\rm i} G_{j k} (\Gamma_{a})^{\beta \rho} \boldsymbol{\lambda}_{i \lambda} {G}^{-3} \nabla_{\alpha}\,^{\lambda}{W} \varphi^{j}_{\beta} \varphi^{k}_{\rho} - \frac{3}{16}{\rm i} G_{i j} (\Gamma_{a})^{\rho \lambda} \mathbf{F}_{\alpha}\,^{\beta} \lambda_{k \beta} {G}^{-3} \varphi^{j}_{\rho} \varphi^{k}_{\lambda} - \frac{3}{32}{\rm i} G_{j k} (\Gamma_{a})^{\beta \rho} \mathbf{X}_{i}\,^{j} \lambda_{l \alpha} {G}^{-3} \varphi^{k}_{\beta} \varphi^{l}_{\rho} - \frac{3}{32}{\rm i} G_{i j} (\Gamma_{a})^{\beta \rho} \lambda_{k \lambda} {G}^{-3} \nabla_{\alpha}\,^{\lambda}{\boldsymbol{W}} \varphi^{j}_{\beta} \varphi^{k}_{\rho}%
 - \frac{3}{320}F G_{j k} (\Gamma_{a})_{\alpha}{}^{\beta} \lambda^{j \rho} \boldsymbol{\lambda}_{i \rho} {G}^{-3} \varphi^{k}_{\beta} - \frac{3}{64}\mathcal{H}_{\alpha}\,^{\beta} G_{j k} (\Gamma_{a})_{\beta}{}^{\rho} \lambda^{j \lambda} \boldsymbol{\lambda}_{i \lambda} {G}^{-3} \varphi^{k}_{\rho} - \frac{3}{64}G_{j k} (\Gamma_{a})^{\beta}{}_{\rho} \lambda^{\lambda}_{l} \boldsymbol{\lambda}_{i \lambda} {G}^{-3} \nabla^{\rho}\,_{\alpha}{G^{j l}} \varphi^{k}_{\beta}+\frac{3}{16}{\rm i} (\Gamma_{a})^{\beta \rho} \lambda^{\lambda}_{j} \boldsymbol{\lambda}_{k \lambda} {G}^{-3} \varphi_{i \alpha} \varphi^{j}_{\beta} \varphi^{k}_{\rho} - \frac{9}{32}{\rm i} G_{i j} G_{k l} (\Gamma_{a})^{\beta \rho} \lambda^{k \lambda} \boldsymbol{\lambda}_{m \lambda} {G}^{-5} \varphi^{l}_{\alpha} \varphi^{j}_{\beta} \varphi^{m}_{\rho}+\frac{3}{64}\mathcal{H}_{\alpha}\,^{\beta} G_{i j} (\Gamma_{a})_{\beta}{}^{\rho} \lambda^{j \lambda} \boldsymbol{\lambda}_{k \lambda} {G}^{-3} \varphi^{k}_{\rho} - \frac{3}{64}G_{i j} (\Gamma_{a})^{\beta}{}_{\rho} \lambda^{\lambda}_{k} \boldsymbol{\lambda}_{l \lambda} {G}^{-3} \nabla^{\rho}\,_{\alpha}{G^{j k}} \varphi^{l}_{\beta} - \frac{9}{32}{\rm i} G_{i j} G_{k l} (\Gamma_{a})^{\beta \rho} \lambda^{\lambda}_{m} \boldsymbol{\lambda}^{k}_{\lambda} {G}^{-5} \varphi^{l}_{\alpha} \varphi^{j}_{\beta} \varphi^{m}_{\rho}+\frac{3}{64}\mathcal{H}_{\alpha}\,^{\beta} G_{i j} (\Gamma_{a})_{\beta}{}^{\rho} \lambda^{\lambda}_{k} \boldsymbol{\lambda}^{j}_{\lambda} {G}^{-3} \varphi^{k}_{\rho} - \frac{3}{64}G_{i j} (\Gamma_{a})^{\beta}{}_{\rho} \lambda^{\lambda}_{l} \boldsymbol{\lambda}_{k \lambda} {G}^{-3} \nabla^{\rho}\,_{\alpha}{G^{j k}} \varphi^{l}_{\beta} - \frac{3}{16}{\rm i} G_{i j} (\Gamma_{a})^{\beta \lambda} F_{\beta}\,^{\rho} \boldsymbol{\lambda}_{k \rho} {G}^{-3} \varphi^{k}_{\alpha} \varphi^{j}_{\lambda}+\frac{3}{8}{\rm i} G_{j k} (\Gamma_{a})^{\beta \lambda} F_{\beta}\,^{\rho} \boldsymbol{\lambda}^{j}_{\rho} {G}^{-3} \varphi_{i \alpha} \varphi^{k}_{\lambda}+\frac{3}{16}{\rm i} G_{i j} (\Gamma_{a})^{\beta \lambda} F_{\beta}\,^{\rho} \boldsymbol{\lambda}^{j}_{\rho} {G}^{-3} \varphi_{k \alpha} \varphi^{k}_{\lambda} - \frac{3}{64}G_{i j} G_{k l} (\Gamma_{a})^{\beta}{}_{\rho} \boldsymbol{\lambda}^{j \lambda} {G}^{-3} \nabla^{\rho}\,_{\alpha}{\lambda^{k}_{\lambda}} \varphi^{l}_{\beta} - \frac{3}{64}G_{i j} G_{k l} (\Gamma_{a})^{\beta \rho} \boldsymbol{\lambda}^{j}_{\lambda} {G}^{-3} \nabla_{\alpha}\,^{\lambda}{\lambda^{k}_{\beta}} \varphi^{l}_{\rho} - \frac{3}{64}G_{i j} G_{k l} (\Gamma_{a})^{\rho \lambda} W_{\alpha}\,^{\beta} \lambda^{k}_{\rho} \boldsymbol{\lambda}^{j}_{\beta} {G}^{-3} \varphi^{l}_{\lambda}+\frac{3}{32}G_{i j} G_{k l} (\Gamma_{a})^{\beta}{}_{\rho} \boldsymbol{\lambda}^{j}_{\alpha} {G}^{-3} \nabla^{\rho \lambda}{\lambda^{k}_{\lambda}} \varphi^{l}_{\beta} - \frac{3}{4}{\rm i} (\Gamma_{a})^{\beta \lambda} \mathbf{F}_{\alpha}\,^{\rho} F_{\beta \rho} {G}^{-1} \varphi_{i \lambda} - \frac{3}{16}{\rm i} G_{i j} G_{k l} (\Gamma_{a})^{\beta \rho} \mathbf{X}^{j k} F_{\alpha \beta} {G}^{-3} \varphi^{l}_{\rho} - \frac{3}{8}{\rm i} (\Gamma_{a})^{\beta \lambda} F_{\beta \rho} {G}^{-1} \nabla_{\alpha}\,^{\rho}{\boldsymbol{W}} \varphi_{i \lambda}%
 - \frac{3}{32}{\rm i} G_{j k} (\Gamma_{a})^{\beta \rho} X_{i}\,^{j} \boldsymbol{\lambda}^{k}_{\beta} {G}^{-3} \varphi_{l \alpha} \varphi^{l}_{\rho}+\frac{3}{32}{\rm i} G_{j k} (\Gamma_{a})^{\beta \rho} X_{i}\,^{j} \boldsymbol{\lambda}_{l \beta} {G}^{-3} \varphi^{l}_{\alpha} \varphi^{k}_{\rho}+\frac{3}{32}{\rm i} G_{j k} (\Gamma_{a})^{\beta \rho} X_{i l} \boldsymbol{\lambda}^{j}_{\beta} {G}^{-3} \varphi^{l}_{\alpha} \varphi^{k}_{\rho} - \frac{3}{16}{\rm i} (\Gamma_{a})^{\beta \rho} X_{i j} \mathbf{F}_{\alpha \beta} {G}^{-1} \varphi^{j}_{\rho} - \frac{3}{32}{\rm i} G_{i j} (\Gamma_{a})^{\beta}{}_{\rho} \boldsymbol{\lambda}_{k \lambda} {G}^{-3} \nabla^{\rho \lambda}{W} \varphi^{k}_{\alpha} \varphi^{j}_{\beta}+\frac{3}{16}{\rm i} G_{j k} (\Gamma_{a})^{\beta}{}_{\rho} \boldsymbol{\lambda}^{j}_{\lambda} {G}^{-3} \nabla^{\rho \lambda}{W} \varphi_{i \alpha} \varphi^{k}_{\beta}+\frac{3}{32}{\rm i} G_{i j} (\Gamma_{a})^{\beta}{}_{\rho} \boldsymbol{\lambda}^{j}_{\lambda} {G}^{-3} \nabla^{\rho \lambda}{W} \varphi_{k \alpha} \varphi^{k}_{\beta}+\frac{3}{8}{\rm i} (\Gamma_{a})^{\rho}{}_{\lambda} \mathbf{F}_{\alpha \beta} {G}^{-1} \nabla^{\lambda \beta}{W} \varphi_{i \rho} - \frac{3}{32}{\rm i} G_{i j} G_{k l} (\Gamma_{a})^{\beta}{}_{\rho} \mathbf{X}^{j k} {G}^{-3} \nabla^{\rho}\,_{\alpha}{W} \varphi^{l}_{\beta} - \frac{3}{16}{\rm i} (\Gamma_{a})^{\beta}{}_{\rho} {G}^{-1} \nabla^{\rho}\,_{\lambda}{W} \nabla_{\alpha}\,^{\lambda}{\boldsymbol{W}} \varphi_{i \beta}+\frac{3}{64}G_{i j} G_{k l} (\Gamma_{a})^{\beta}{}_{\rho} \boldsymbol{\lambda}^{j}_{\lambda} {G}^{-3} \nabla^{\rho \lambda}{\lambda^{k}_{\alpha}} \varphi^{l}_{\beta} - \frac{3}{128}G_{i j} G_{k l} (\Gamma_{a})^{\rho \lambda} W_{\alpha}\,^{\beta} \lambda^{k}_{\beta} \boldsymbol{\lambda}^{j}_{\rho} {G}^{-3} \varphi^{l}_{\lambda}+\frac{9}{64}G_{i j} G_{k l} (\Gamma_{a})^{\beta \rho} W_{\alpha \beta} \lambda^{k \lambda} \boldsymbol{\lambda}^{l}_{\lambda} {G}^{-3} \varphi^{j}_{\rho} - \frac{9}{80}F (\Gamma_{a})_{\alpha}{}^{\beta} F_{\beta}\,^{\rho} \boldsymbol{\lambda}_{i \rho} {G}^{-1} - \frac{3}{16}\mathcal{H}_{\alpha}\,^{\lambda} (\Gamma_{a})_{\lambda}{}^{\beta} F_{\beta}\,^{\rho} \boldsymbol{\lambda}_{i \rho} {G}^{-1}+\frac{3}{32}G_{i j} G_{k l} (\Gamma_{a})^{\beta}{}_{\lambda} F_{\beta}\,^{\rho} \boldsymbol{\lambda}^{k}_{\rho} {G}^{-3} \nabla^{\lambda}\,_{\alpha}{G^{j l}} - \frac{3}{32}{\rm i} G_{j k} (\Gamma_{a})^{\beta \rho} X^{j}\,_{l} \boldsymbol{\lambda}^{k}_{\beta} {G}^{-3} \varphi_{i \alpha} \varphi^{l}_{\rho}+\frac{9}{32}{\rm i} G_{i j} G_{k l} G_{m n} (\Gamma_{a})^{\beta \rho} X^{k m} \boldsymbol{\lambda}^{l}_{\beta} {G}^{-5} \varphi^{n}_{\alpha} \varphi^{j}_{\rho}+\frac{3}{64}\mathcal{H}_{\alpha}\,^{\beta} G_{i j} G_{k l} (\Gamma_{a})_{\beta}{}^{\rho} X^{j k} \boldsymbol{\lambda}^{l}_{\rho} {G}^{-3} - \frac{3}{64}G_{i j} G_{k l} (\Gamma_{a})^{\beta}{}_{\rho} X^{k}\,_{m} \boldsymbol{\lambda}^{l}_{\beta} {G}^{-3} \nabla^{\rho}\,_{\alpha}{G^{j m}}%
+\frac{9}{160}F (\Gamma_{a})_{\alpha \beta} \boldsymbol{\lambda}_{i \rho} {G}^{-1} \nabla^{\beta \rho}{W}+\frac{3}{32}\mathcal{H}_{\alpha}\,^{\beta} (\Gamma_{a})_{\beta \rho} \boldsymbol{\lambda}_{i \lambda} {G}^{-1} \nabla^{\rho \lambda}{W} - \frac{3}{64}G_{i j} G_{k l} (\Gamma_{a})_{\beta \rho} \boldsymbol{\lambda}^{k}_{\lambda} {G}^{-3} \nabla^{\beta \lambda}{W} \nabla^{\rho}\,_{\alpha}{G^{j l}} - \frac{3}{16}{\rm i} G_{i j} (\Gamma_{a})^{\beta \lambda} \mathbf{F}_{\beta}\,^{\rho} \lambda_{k \rho} {G}^{-3} \varphi^{k}_{\alpha} \varphi^{j}_{\lambda}+\frac{3}{8}{\rm i} G_{j k} (\Gamma_{a})^{\beta \lambda} \mathbf{F}_{\beta}\,^{\rho} \lambda^{j}_{\rho} {G}^{-3} \varphi_{i \alpha} \varphi^{k}_{\lambda}+\frac{3}{16}{\rm i} G_{i j} (\Gamma_{a})^{\beta \lambda} \mathbf{F}_{\beta}\,^{\rho} \lambda^{j}_{\rho} {G}^{-3} \varphi_{k \alpha} \varphi^{k}_{\lambda} - \frac{3}{64}G_{i j} G_{k l} (\Gamma_{a})^{\beta}{}_{\rho} \lambda^{j \lambda} {G}^{-3} \nabla^{\rho}\,_{\alpha}{\boldsymbol{\lambda}^{k}_{\lambda}} \varphi^{l}_{\beta} - \frac{3}{64}G_{i j} G_{k l} (\Gamma_{a})^{\beta \rho} \lambda^{j}_{\lambda} {G}^{-3} \nabla_{\alpha}\,^{\lambda}{\boldsymbol{\lambda}^{k}_{\beta}} \varphi^{l}_{\rho}+\frac{3}{64}G_{i j} G_{k l} (\Gamma_{a})^{\rho \lambda} W_{\alpha}\,^{\beta} \lambda^{j}_{\beta} \boldsymbol{\lambda}^{k}_{\rho} {G}^{-3} \varphi^{l}_{\lambda}+\frac{3}{32}G_{i j} G_{k l} (\Gamma_{a})^{\beta}{}_{\rho} \lambda^{j}_{\alpha} {G}^{-3} \nabla^{\rho \lambda}{\boldsymbol{\lambda}^{k}_{\lambda}} \varphi^{l}_{\beta} - \frac{3}{4}{\rm i} (\Gamma_{a})^{\rho \lambda} F_{\alpha}\,^{\beta} \mathbf{F}_{\rho \beta} {G}^{-1} \varphi_{i \lambda} - \frac{3}{16}{\rm i} G_{i j} G_{k l} (\Gamma_{a})^{\beta \rho} X^{j k} \mathbf{F}_{\alpha \beta} {G}^{-3} \varphi^{l}_{\rho} - \frac{3}{8}{\rm i} (\Gamma_{a})^{\rho \lambda} \mathbf{F}_{\beta \rho} {G}^{-1} \nabla_{\alpha}\,^{\beta}{W} \varphi_{i \lambda} - \frac{3}{32}{\rm i} G_{j k} (\Gamma_{a})^{\beta \rho} \mathbf{X}_{i}\,^{j} \lambda^{k}_{\beta} {G}^{-3} \varphi_{l \alpha} \varphi^{l}_{\rho}+\frac{3}{32}{\rm i} G_{j k} (\Gamma_{a})^{\beta \rho} \mathbf{X}_{i}\,^{j} \lambda_{l \beta} {G}^{-3} \varphi^{l}_{\alpha} \varphi^{k}_{\rho}+\frac{3}{32}{\rm i} G_{j k} (\Gamma_{a})^{\beta \rho} \mathbf{X}_{i l} \lambda^{j}_{\beta} {G}^{-3} \varphi^{l}_{\alpha} \varphi^{k}_{\rho} - \frac{3}{16}{\rm i} (\Gamma_{a})^{\beta \rho} \mathbf{X}_{i j} F_{\alpha \beta} {G}^{-1} \varphi^{j}_{\rho} - \frac{3}{32}{\rm i} G_{i j} (\Gamma_{a})^{\beta}{}_{\rho} \lambda_{k \lambda} {G}^{-3} \nabla^{\rho \lambda}{\boldsymbol{W}} \varphi^{k}_{\alpha} \varphi^{j}_{\beta}+\frac{3}{16}{\rm i} G_{j k} (\Gamma_{a})^{\beta}{}_{\rho} \lambda^{j}_{\lambda} {G}^{-3} \nabla^{\rho \lambda}{\boldsymbol{W}} \varphi_{i \alpha} \varphi^{k}_{\beta}+\frac{3}{32}{\rm i} G_{i j} (\Gamma_{a})^{\beta}{}_{\rho} \lambda^{j}_{\lambda} {G}^{-3} \nabla^{\rho \lambda}{\boldsymbol{W}} \varphi_{k \alpha} \varphi^{k}_{\beta}%
+\frac{3}{8}{\rm i} (\Gamma_{a})^{\rho}{}_{\lambda} F_{\alpha \beta} {G}^{-1} \nabla^{\lambda \beta}{\boldsymbol{W}} \varphi_{i \rho} - \frac{3}{32}{\rm i} G_{i j} G_{k l} (\Gamma_{a})^{\beta}{}_{\rho} X^{j k} {G}^{-3} \nabla^{\rho}\,_{\alpha}{\boldsymbol{W}} \varphi^{l}_{\beta}+\frac{3}{16}{\rm i} (\Gamma_{a})^{\beta}{}_{\rho} {G}^{-1} \nabla_{\alpha \lambda}{W} \nabla^{\rho \lambda}{\boldsymbol{W}} \varphi_{i \beta}+\frac{3}{64}G_{i j} G_{k l} (\Gamma_{a})^{\beta}{}_{\rho} \lambda^{j}_{\lambda} {G}^{-3} \nabla^{\rho \lambda}{\boldsymbol{\lambda}^{k}_{\alpha}} \varphi^{l}_{\beta}+\frac{3}{128}G_{i j} G_{k l} (\Gamma_{a})^{\rho \lambda} W_{\alpha}\,^{\beta} \lambda^{j}_{\rho} \boldsymbol{\lambda}^{k}_{\beta} {G}^{-3} \varphi^{l}_{\lambda} - \frac{9}{80}F (\Gamma_{a})_{\alpha}{}^{\beta} \mathbf{F}_{\beta}\,^{\rho} \lambda_{i \rho} {G}^{-1} - \frac{3}{16}\mathcal{H}_{\alpha}\,^{\lambda} (\Gamma_{a})_{\lambda}{}^{\beta} \mathbf{F}_{\beta}\,^{\rho} \lambda_{i \rho} {G}^{-1}+\frac{3}{32}G_{i j} G_{k l} (\Gamma_{a})^{\beta}{}_{\lambda} \mathbf{F}_{\beta}\,^{\rho} \lambda^{k}_{\rho} {G}^{-3} \nabla^{\lambda}\,_{\alpha}{G^{j l}} - \frac{3}{32}{\rm i} G_{j k} (\Gamma_{a})^{\beta \rho} \mathbf{X}^{j}\,_{l} \lambda^{k}_{\beta} {G}^{-3} \varphi_{i \alpha} \varphi^{l}_{\rho}+\frac{9}{32}{\rm i} G_{i j} G_{k l} G_{m n} (\Gamma_{a})^{\beta \rho} \mathbf{X}^{k m} \lambda^{l}_{\beta} {G}^{-5} \varphi^{n}_{\alpha} \varphi^{j}_{\rho}+\frac{3}{64}\mathcal{H}_{\alpha}\,^{\beta} G_{i j} G_{k l} (\Gamma_{a})_{\beta}{}^{\rho} \mathbf{X}^{j k} \lambda^{l}_{\rho} {G}^{-3} - \frac{3}{64}G_{i j} G_{k l} (\Gamma_{a})^{\beta}{}_{\rho} \mathbf{X}^{k}\,_{m} \lambda^{l}_{\beta} {G}^{-3} \nabla^{\rho}\,_{\alpha}{G^{j m}}+\frac{9}{160}F (\Gamma_{a})_{\alpha \beta} \lambda_{i \rho} {G}^{-1} \nabla^{\beta \rho}{\boldsymbol{W}}+\frac{3}{32}\mathcal{H}_{\alpha}\,^{\beta} (\Gamma_{a})_{\beta \rho} \lambda_{i \lambda} {G}^{-1} \nabla^{\rho \lambda}{\boldsymbol{W}} - \frac{3}{64}G_{i j} G_{k l} (\Gamma_{a})_{\beta \rho} \lambda^{k}_{\lambda} {G}^{-3} \nabla^{\beta \lambda}{\boldsymbol{W}} \nabla^{\rho}\,_{\alpha}{G^{j l}}+\frac{9}{32}{\rm i} G_{i j} G_{k l} (\Gamma_{a})^{\beta \rho} \lambda^{k \lambda} \boldsymbol{\lambda}^{l}_{\lambda} {G}^{-5} \varphi_{m \alpha} \varphi^{j}_{\beta} \varphi^{m}_{\rho} - \frac{9}{32}{\rm i} G_{j k} G_{l m} (\Gamma_{a})^{\beta \rho} \lambda^{j \lambda} \boldsymbol{\lambda}^{k}_{\lambda} {G}^{-5} \varphi_{i \alpha} \varphi^{l}_{\beta} \varphi^{m}_{\rho}+\frac{9}{32}{\rm i} G_{i j} G_{k l} (\Gamma_{a})^{\beta \rho} \lambda^{k \lambda} \boldsymbol{\lambda}_{m \lambda} {G}^{-5} \varphi^{m}_{\alpha} \varphi^{j}_{\beta} \varphi^{l}_{\rho}+\frac{9}{32}{\rm i} G_{i j} G_{k l} (\Gamma_{a})^{\beta \rho} \lambda^{\lambda}_{m} \boldsymbol{\lambda}^{k}_{\lambda} {G}^{-5} \varphi^{m}_{\alpha} \varphi^{j}_{\beta} \varphi^{l}_{\rho}+\frac{9}{32}{\rm i} G_{i j} G_{k l} G_{m n} (\Gamma_{a})^{\beta \rho} X^{k m} \boldsymbol{\lambda}^{l}_{\alpha} {G}^{-5} \varphi^{j}_{\beta} \varphi^{n}_{\rho}%
+\frac{9}{32}{\rm i} G_{i j} G_{k l} G_{m n} (\Gamma_{a})^{\beta \rho} \mathbf{X}^{k m} \lambda^{l}_{\alpha} {G}^{-5} \varphi^{j}_{\beta} \varphi^{n}_{\rho}+\frac{9}{64}G_{i j} G_{k l} G_{m n} (\Gamma_{a})^{\beta}{}_{\rho} \lambda^{k \lambda} \boldsymbol{\lambda}^{l}_{\lambda} {G}^{-5} \nabla^{\rho}\,_{\alpha}{G^{j m}} \varphi^{n}_{\beta} - \frac{3}{64}G_{i j} G_{k l} (\Gamma_{a})^{\beta}{}_{\rho} \lambda^{k \lambda} \boldsymbol{\lambda}^{l}_{\lambda} {G}^{-3} \nabla^{\rho}\,_{\alpha}{\varphi^{j}_{\beta}}+\frac{9}{320}G_{i j} G_{k l} (\Gamma_{a})_{\alpha}{}^{\beta} W_{\beta}\,^{\rho} \lambda^{k \lambda} \boldsymbol{\lambda}^{l}_{\lambda} {G}^{-3} \varphi^{j}_{\rho} - \frac{39}{320}G_{i j} G_{k l} (\Gamma_{a})_{\alpha \beta} \lambda^{k \rho} \boldsymbol{\lambda}^{l}_{\rho} {G}^{-3} \nabla^{\beta \lambda}{\varphi^{j}_{\lambda}}+\frac{3}{64}\mathcal{H}^{\beta \rho} G_{i j} (\Gamma_{a})_{\beta \rho} \lambda^{j \lambda} \boldsymbol{\lambda}_{k \lambda} {G}^{-3} \varphi^{k}_{\alpha}+\frac{3}{64}\mathcal{H}^{\beta \rho} G_{i j} (\Gamma_{a})_{\beta \rho} \lambda^{\lambda}_{k} \boldsymbol{\lambda}^{j}_{\lambda} {G}^{-3} \varphi^{k}_{\alpha} - \frac{3}{16}\mathcal{H}^{\rho \lambda} (\Gamma_{a})_{\rho \lambda} F_{\alpha}\,^{\beta} \boldsymbol{\lambda}_{i \beta} {G}^{-1}+\frac{3}{64}\mathcal{H}^{\beta \rho} G_{i j} G_{k l} (\Gamma_{a})_{\beta \rho} X^{j k} \boldsymbol{\lambda}^{l}_{\alpha} {G}^{-3} - \frac{3}{32}\mathcal{H}^{\beta \rho} (\Gamma_{a})_{\beta \rho} \boldsymbol{\lambda}_{i \lambda} {G}^{-1} \nabla_{\alpha}\,^{\lambda}{W} - \frac{3}{16}\mathcal{H}^{\rho \lambda} (\Gamma_{a})_{\rho \lambda} \mathbf{F}_{\alpha}\,^{\beta} \lambda_{i \beta} {G}^{-1}+\frac{3}{64}\mathcal{H}^{\beta \rho} G_{i j} G_{k l} (\Gamma_{a})_{\beta \rho} \mathbf{X}^{j k} \lambda^{l}_{\alpha} {G}^{-3} - \frac{3}{32}\mathcal{H}^{\beta \rho} (\Gamma_{a})_{\beta \rho} \lambda_{i \lambda} {G}^{-1} \nabla_{\alpha}\,^{\lambda}{\boldsymbol{W}} - \frac{3}{128}G_{j k} (\Gamma_{a})_{\beta \rho} \lambda^{j \lambda} \boldsymbol{\lambda}^{k}_{\lambda} {G}^{-3} \nabla^{\beta \rho}{G_{i l}} \varphi^{l}_{\alpha}+\frac{3}{64}G_{j k} (\Gamma_{a})_{\beta \rho} \lambda^{j \lambda} \boldsymbol{\lambda}_{l \lambda} {G}^{-3} \nabla^{\beta \rho}{G_{i}\,^{k}} \varphi^{l}_{\alpha}+\frac{3}{64}G_{j k} (\Gamma_{a})_{\beta \rho} \lambda^{\lambda}_{l} \boldsymbol{\lambda}^{j}_{\lambda} {G}^{-3} \nabla^{\beta \rho}{G_{i}\,^{k}} \varphi^{l}_{\alpha} - \frac{9}{64}(\Gamma_{a})_{\rho \lambda} F_{\alpha}\,^{\beta} \boldsymbol{\lambda}_{j \beta} {G}^{-1} \nabla^{\rho \lambda}{G_{i}\,^{j}}+\frac{3}{64}G_{j k} G_{l m} (\Gamma_{a})_{\beta \rho} X^{j l} \boldsymbol{\lambda}^{k}_{\alpha} {G}^{-3} \nabla^{\beta \rho}{G_{i}\,^{m}} - \frac{9}{128}(\Gamma_{a})_{\beta \rho} \boldsymbol{\lambda}_{j \lambda} {G}^{-1} \nabla_{\alpha}\,^{\lambda}{W} \nabla^{\beta \rho}{G_{i}\,^{j}} - \frac{9}{64}(\Gamma_{a})_{\rho \lambda} \mathbf{F}_{\alpha}\,^{\beta} \lambda_{j \beta} {G}^{-1} \nabla^{\rho \lambda}{G_{i}\,^{j}}%
+\frac{3}{64}G_{j k} G_{l m} (\Gamma_{a})_{\beta \rho} \mathbf{X}^{j l} \lambda^{k}_{\alpha} {G}^{-3} \nabla^{\beta \rho}{G_{i}\,^{m}} - \frac{9}{128}(\Gamma_{a})_{\beta \rho} \lambda_{j \lambda} {G}^{-1} \nabla_{\alpha}\,^{\lambda}{\boldsymbol{W}} \nabla^{\beta \rho}{G_{i}\,^{j}} - \frac{3}{16}G_{i j} G_{k l} (\Gamma_{a})_{\beta \rho} \lambda^{k \lambda} \boldsymbol{\lambda}^{l}_{\lambda} {G}^{-3} \nabla^{\beta \rho}{\varphi^{j}_{\alpha}} - \frac{3}{160}G_{j k} (\Gamma_{a})_{\alpha}{}^{\beta} \lambda^{j \rho} \boldsymbol{\lambda}^{k}_{\rho} X_{i \beta} {G}^{-1} - \frac{3}{64}G_{i j} (\Gamma_{a})_{\beta \rho} \lambda^{\lambda}_{k} \boldsymbol{\lambda}_{l \lambda} {G}^{-3} \nabla^{\beta \rho}{G^{j k}} \varphi^{l}_{\alpha} - \frac{3}{64}G_{i j} (\Gamma_{a})_{\beta \rho} \lambda^{\lambda}_{l} \boldsymbol{\lambda}_{k \lambda} {G}^{-3} \nabla^{\beta \rho}{G^{j k}} \varphi^{l}_{\alpha}+\frac{3}{32}G_{i j} G_{k l} (\Gamma_{a})_{\rho \lambda} F_{\alpha}\,^{\beta} \boldsymbol{\lambda}^{k}_{\beta} {G}^{-3} \nabla^{\rho \lambda}{G^{j l}} - \frac{3}{64}G_{i j} G_{k l} (\Gamma_{a})_{\beta \rho} X^{k}\,_{m} \boldsymbol{\lambda}^{l}_{\alpha} {G}^{-3} \nabla^{\beta \rho}{G^{j m}}+\frac{3}{64}G_{i j} G_{k l} (\Gamma_{a})_{\beta \rho} \boldsymbol{\lambda}^{k}_{\lambda} {G}^{-3} \nabla_{\alpha}\,^{\lambda}{W} \nabla^{\beta \rho}{G^{j l}}+\frac{3}{32}G_{i j} G_{k l} (\Gamma_{a})_{\rho \lambda} \mathbf{F}_{\alpha}\,^{\beta} \lambda^{k}_{\beta} {G}^{-3} \nabla^{\rho \lambda}{G^{j l}} - \frac{3}{64}G_{i j} G_{k l} (\Gamma_{a})_{\beta \rho} \mathbf{X}^{k}\,_{m} \lambda^{l}_{\alpha} {G}^{-3} \nabla^{\beta \rho}{G^{j m}}+\frac{3}{64}G_{i j} G_{k l} (\Gamma_{a})_{\beta \rho} \lambda^{k}_{\lambda} {G}^{-3} \nabla_{\alpha}\,^{\lambda}{\boldsymbol{W}} \nabla^{\beta \rho}{G^{j l}}+\frac{9}{64}G_{i j} G_{k l} G_{m n} (\Gamma_{a})_{\beta \rho} \lambda^{k \lambda} \boldsymbol{\lambda}^{l}_{\lambda} {G}^{-5} \nabla^{\beta \rho}{G^{j m}} \varphi^{n}_{\alpha} - \frac{3}{16}{\rm i} G_{j k} (\Gamma_{a})^{\beta \rho} X_{i}\,^{j} \boldsymbol{\lambda}_{l \beta} {G}^{-3} \varphi^{k}_{\alpha} \varphi^{l}_{\rho} - \frac{3}{64}(\Gamma_{a})^{\beta}{}_{\rho} \lambda^{\lambda}_{j} {G}^{-1} \nabla^{\rho}\,_{\alpha}{\boldsymbol{\lambda}^{j}_{\lambda}} \varphi_{i \beta} - \frac{3}{64}(\Gamma_{a})^{\beta \rho} \lambda_{j \lambda} {G}^{-1} \nabla_{\alpha}\,^{\lambda}{\boldsymbol{\lambda}^{j}_{\beta}} \varphi_{i \rho}+\frac{9}{128}(\Gamma_{a})^{\rho \lambda} W_{\alpha}\,^{\beta} \lambda_{j \beta} \boldsymbol{\lambda}^{j}_{\rho} {G}^{-1} \varphi_{i \lambda}+\frac{3}{80}(\Gamma_{a})_{\alpha}{}^{\beta} \lambda_{j \rho} {G}^{-1} \nabla^{\rho \lambda}{\boldsymbol{\lambda}^{j}_{\lambda}} \varphi_{i \beta} - \frac{3}{16}{\rm i} G_{j k} (\Gamma_{a})^{\beta \lambda} \mathbf{F}_{\beta}\,^{\rho} \lambda^{j}_{\rho} {G}^{-3} \varphi^{k}_{\alpha} \varphi_{i \lambda} - \frac{9}{32}(\Gamma_{a})^{\beta}{}_{\lambda} \mathbf{F}_{\beta}\,^{\rho} \lambda_{j \rho} {G}^{-1} \nabla^{\lambda}\,_{\alpha}{G_{i}\,^{j}}%
 - \frac{3}{32}{\rm i} G_{j k} (\Gamma_{a})^{\beta}{}_{\rho} \lambda^{j}_{\lambda} {G}^{-3} \nabla^{\rho \lambda}{\boldsymbol{W}} \varphi^{k}_{\alpha} \varphi_{i \beta}+\frac{3}{64}(\Gamma_{a})^{\beta}{}_{\rho} \lambda_{j \lambda} {G}^{-1} \nabla^{\rho \lambda}{\boldsymbol{\lambda}^{j}_{\alpha}} \varphi_{i \beta}+\frac{9}{128}(\Gamma_{a})^{\rho \lambda} W_{\alpha}\,^{\beta} \lambda_{j \rho} \boldsymbol{\lambda}^{j}_{\beta} {G}^{-1} \varphi_{i \lambda}+\frac{9}{64}(\Gamma_{a})_{\beta \rho} \lambda_{j \lambda} {G}^{-1} \nabla^{\beta \lambda}{\boldsymbol{W}} \nabla^{\rho}\,_{\alpha}{G_{i}\,^{j}} - \frac{9}{64}(\Gamma_{a})^{\beta \rho} W_{\alpha \beta} \lambda^{\lambda}_{i} \boldsymbol{\lambda}_{j \lambda} {G}^{-1} \varphi^{j}_{\rho} - \frac{15}{64}(\Gamma_{a})^{\beta}{}_{\rho} \lambda_{i \alpha} {G}^{-1} \nabla^{\rho \lambda}{\boldsymbol{\lambda}_{j \lambda}} \varphi^{j}_{\beta} - \frac{3}{16}{\rm i} G_{j k} (\Gamma_{a})^{\beta \lambda} \mathbf{F}_{\beta}\,^{\rho} \lambda_{i \rho} {G}^{-3} \varphi^{j}_{\alpha} \varphi^{k}_{\lambda} - \frac{3}{32}{\rm i} G_{j k} (\Gamma_{a})^{\beta}{}_{\rho} \lambda_{i \lambda} {G}^{-3} \nabla^{\rho \lambda}{\boldsymbol{W}} \varphi^{j}_{\alpha} \varphi^{k}_{\beta}+\frac{3}{64}(\Gamma_{a})^{\beta}{}_{\rho} \lambda^{\lambda}_{j} \boldsymbol{\lambda}_{i \lambda} {G}^{-1} \nabla^{\rho}\,_{\alpha}{\varphi^{j}_{\beta}} - \frac{27}{320}(\Gamma_{a})_{\alpha}{}^{\beta} W_{\beta}\,^{\rho} \lambda^{\lambda}_{j} \boldsymbol{\lambda}_{i \lambda} {G}^{-1} \varphi^{j}_{\rho} - \frac{9}{64}(\Gamma_{a})^{\beta \rho} W_{\alpha \beta} \lambda^{\lambda}_{j} \boldsymbol{\lambda}_{i \lambda} {G}^{-1} \varphi^{j}_{\rho}+\frac{27}{320}(\Gamma_{a})_{\alpha \beta} \lambda^{\rho}_{j} \boldsymbol{\lambda}_{i \rho} {G}^{-1} \nabla^{\beta \lambda}{\varphi^{j}_{\lambda}} - \frac{3}{64}\mathcal{H}^{\beta \rho} G_{j k} (\Gamma_{a})_{\beta \rho} \lambda^{j \lambda} \boldsymbol{\lambda}_{i \lambda} {G}^{-3} \varphi^{k}_{\alpha} - \frac{3}{64}G_{j k} (\Gamma_{a})_{\beta \rho} \lambda^{j \lambda} \boldsymbol{\lambda}_{l \lambda} {G}^{-3} \nabla^{\beta \rho}{G_{i}\,^{l}} \varphi^{k}_{\alpha}+\frac{3}{16}(\Gamma_{a})_{\beta \rho} \lambda^{\lambda}_{i} \boldsymbol{\lambda}_{j \lambda} {G}^{-1} \nabla^{\beta \rho}{\varphi^{j}_{\alpha}} - \frac{27}{320}(\Gamma_{a})_{\alpha}{}^{\beta} W_{\beta}\,^{\rho} \lambda^{\lambda}_{i} \boldsymbol{\lambda}_{j \lambda} {G}^{-1} \varphi^{j}_{\rho}+\frac{27}{320}G_{i j} (\Gamma_{a})_{\alpha}{}^{\beta} \lambda^{\rho}_{k} \boldsymbol{\lambda}^{j}_{\rho} X^{k}_{\beta} {G}^{-1}+\frac{39}{640}G_{j k} (\Gamma_{a})_{\alpha}{}^{\beta} \lambda^{\rho}_{i} \boldsymbol{\lambda}^{j}_{\rho} X^{k}_{\beta} {G}^{-1}+\frac{27}{320}G_{i j} (\Gamma_{a})_{\alpha}{}^{\beta} \lambda^{j \rho} \boldsymbol{\lambda}_{k \rho} X^{k}_{\beta} {G}^{-1}+\frac{3}{64}(\Gamma_{a})^{\beta}{}_{\rho} \lambda^{\lambda}_{i} \boldsymbol{\lambda}_{j \lambda} {G}^{-1} \nabla^{\rho}\,_{\alpha}{\varphi^{j}_{\beta}}%
+\frac{27}{320}(\Gamma_{a})_{\alpha \beta} \lambda^{\rho}_{i} \boldsymbol{\lambda}_{j \rho} {G}^{-1} \nabla^{\beta \lambda}{\varphi^{j}_{\lambda}} - \frac{3}{64}\mathcal{H}^{\beta \rho} G_{j k} (\Gamma_{a})_{\beta \rho} \lambda^{\lambda}_{i} \boldsymbol{\lambda}^{j}_{\lambda} {G}^{-3} \varphi^{k}_{\alpha} - \frac{3}{128}(\Gamma_{a})_{\beta \rho} \mathbf{X}_{j k} \lambda_{i \alpha} {G}^{-1} \nabla^{\beta \rho}{G^{j k}} - \frac{3}{64}G_{j k} (\Gamma_{a})_{\beta \rho} \lambda^{\lambda}_{i} \boldsymbol{\lambda}_{l \lambda} {G}^{-3} \nabla^{\beta \rho}{G^{j l}} \varphi^{k}_{\alpha} - \frac{3}{64}(\Gamma_{a})^{\beta}{}_{\rho} \boldsymbol{\lambda}_{j}^{\lambda} {G}^{-1} \nabla^{\rho}\,_{\alpha}{\lambda^{j}_{\lambda}} \varphi_{i \beta} - \frac{3}{64}(\Gamma_{a})^{\beta \rho} \boldsymbol{\lambda}_{j \lambda} {G}^{-1} \nabla_{\alpha}\,^{\lambda}{\lambda^{j}_{\beta}} \varphi_{i \rho}+\frac{3}{80}(\Gamma_{a})_{\alpha}{}^{\beta} \boldsymbol{\lambda}_{j \lambda} {G}^{-1} \nabla^{\lambda \rho}{\lambda^{j}_{\rho}} \varphi_{i \beta} - \frac{3}{16}{\rm i} G_{j k} (\Gamma_{a})^{\beta \lambda} F_{\beta}\,^{\rho} \boldsymbol{\lambda}^{j}_{\rho} {G}^{-3} \varphi^{k}_{\alpha} \varphi_{i \lambda} - \frac{9}{32}(\Gamma_{a})^{\beta}{}_{\lambda} F_{\beta}\,^{\rho} \boldsymbol{\lambda}_{j \rho} {G}^{-1} \nabla^{\lambda}\,_{\alpha}{G_{i}\,^{j}} - \frac{3}{32}{\rm i} G_{j k} (\Gamma_{a})^{\beta}{}_{\rho} \boldsymbol{\lambda}^{j}_{\lambda} {G}^{-3} \nabla^{\rho \lambda}{W} \varphi^{k}_{\alpha} \varphi_{i \beta}+\frac{3}{64}(\Gamma_{a})^{\beta}{}_{\rho} \boldsymbol{\lambda}_{j \lambda} {G}^{-1} \nabla^{\rho \lambda}{\lambda^{j}_{\alpha}} \varphi_{i \beta}+\frac{9}{64}(\Gamma_{a})_{\beta \rho} \boldsymbol{\lambda}_{j \lambda} {G}^{-1} \nabla^{\beta \lambda}{W} \nabla^{\rho}\,_{\alpha}{G_{i}\,^{j}} - \frac{15}{64}(\Gamma_{a})^{\beta}{}_{\rho} \boldsymbol{\lambda}_{i \alpha} {G}^{-1} \nabla^{\rho \lambda}{\lambda_{j \lambda}} \varphi^{j}_{\beta} - \frac{3}{16}{\rm i} G_{j k} (\Gamma_{a})^{\beta \lambda} F_{\beta}\,^{\rho} \boldsymbol{\lambda}_{i \rho} {G}^{-3} \varphi^{j}_{\alpha} \varphi^{k}_{\lambda} - \frac{3}{32}{\rm i} G_{j k} (\Gamma_{a})^{\beta}{}_{\rho} \boldsymbol{\lambda}_{i \lambda} {G}^{-3} \nabla^{\rho \lambda}{W} \varphi^{j}_{\alpha} \varphi^{k}_{\beta} - \frac{3}{16}{\rm i} G_{j k} (\Gamma_{a})^{\beta \rho} \mathbf{X}_{i}\,^{j} \lambda_{l \beta} {G}^{-3} \varphi^{k}_{\alpha} \varphi^{l}_{\rho} - \frac{3}{64}G_{j k} (\Gamma_{a})_{\beta \rho} \lambda^{\lambda}_{l} \boldsymbol{\lambda}^{j}_{\lambda} {G}^{-3} \nabla^{\beta \rho}{G_{i}\,^{l}} \varphi^{k}_{\alpha}+\frac{3}{16}(\Gamma_{a})_{\beta \rho} \lambda^{\lambda}_{j} \boldsymbol{\lambda}_{i \lambda} {G}^{-1} \nabla^{\beta \rho}{\varphi^{j}_{\alpha}}+\frac{39}{640}G_{j k} (\Gamma_{a})_{\alpha}{}^{\beta} \lambda^{j \rho} \boldsymbol{\lambda}_{i \rho} X^{k}_{\beta} {G}^{-1} - \frac{3}{128}(\Gamma_{a})_{\beta \rho} X_{j k} \boldsymbol{\lambda}_{i \alpha} {G}^{-1} \nabla^{\beta \rho}{G^{j k}}%
 - \frac{3}{64}G_{j k} (\Gamma_{a})_{\beta \rho} \lambda^{\lambda}_{l} \boldsymbol{\lambda}_{i \lambda} {G}^{-3} \nabla^{\beta \rho}{G^{j l}} \varphi^{k}_{\alpha} - \frac{9}{128}(\Gamma_{a})_{\beta \rho} \boldsymbol{\lambda}_{j}^{\lambda} {G}^{-1} \nabla^{\beta \rho}{\lambda_{i \lambda}} \varphi^{j}_{\alpha} - \frac{3}{32}G_{j k} (\Gamma_{a})_{\beta \rho} \mathbf{X}^{j k} {G}^{-1} \nabla^{\beta \rho}{\lambda_{i \alpha}}+\frac{9}{32}G_{i j} (\Gamma_{a})_{\rho \lambda} \boldsymbol{\lambda}^{j \beta} {G}^{-1} \nabla^{\rho \lambda}{F_{\alpha \beta}} - \frac{9}{16}G_{i j} (\Gamma_{a})_{\rho \lambda} W_{\alpha}\,^{\beta} \boldsymbol{\lambda}^{j}_{\beta} {G}^{-1} \nabla^{\rho \lambda}{W} - \frac{9}{64}G_{i j} (\Gamma_{a})_{\beta \rho} \boldsymbol{\lambda}^{j}_{\lambda} {G}^{-1} \nabla^{\beta \rho}{\nabla_{\alpha}\,^{\lambda}{W}} - \frac{27}{320}G_{i j} (\Gamma_{a})_{\alpha}{}^{\lambda} W_{\lambda}\,^{\beta} F_{\beta}\,^{\rho} \boldsymbol{\lambda}^{j}_{\rho} {G}^{-1}+\frac{27}{640}G_{i j} (\Gamma_{a})_{\alpha}{}^{\beta} W_{\beta \rho} \boldsymbol{\lambda}^{j}_{\lambda} {G}^{-1} \nabla^{\rho \lambda}{W}+\frac{9}{32}G_{i j} (\Gamma_{a})^{\lambda \beta} W_{\alpha \lambda} F_{\beta}\,^{\rho} \boldsymbol{\lambda}^{j}_{\rho} {G}^{-1}+\frac{9}{1280}G_{j k} (\Gamma_{a})_{\alpha}{}^{\rho} \lambda^{\beta}_{i} \boldsymbol{\lambda}^{j}_{\rho} X^{k}_{\beta} {G}^{-1}+\frac{9}{32}(\Gamma_{a})^{\beta}{}_{\rho} \boldsymbol{\lambda}_{j \lambda} {G}^{-1} \nabla^{\rho \lambda}{\lambda_{i \beta}} \varphi^{j}_{\alpha}+\frac{3}{64}G_{j k} (\Gamma_{a})^{\beta}{}_{\rho} \mathbf{X}^{j k} {G}^{-1} \nabla^{\rho}\,_{\alpha}{\lambda_{i \beta}} - \frac{9}{32}G_{i j} (\Gamma_{a})^{\beta}{}_{\rho} \boldsymbol{\lambda}^{j}_{\lambda} {G}^{-1} \nabla^{\rho \lambda}{F_{\alpha \beta}} - \frac{9}{32}G_{i j} (\Gamma_{a})^{\beta}{}_{\rho} W_{\alpha \beta} \boldsymbol{\lambda}^{j}_{\lambda} {G}^{-1} \nabla^{\rho \lambda}{W}+\frac{3}{320}G_{j k} (\Gamma_{a})_{\alpha \beta} \boldsymbol{\lambda}^{j}_{\rho} {G}^{-1} \nabla^{\beta \rho}{X_{i}\,^{k}}+\frac{9}{64}G_{i j} (\Gamma_{a})_{\beta \rho} \boldsymbol{\lambda}^{j}_{\lambda} {G}^{-1} \nabla^{\beta \lambda}{\nabla^{\rho}\,_{\alpha}{W}}+\frac{9}{64}G_{i j} (\Gamma_{a})_{\alpha}{}^{\beta} W^{\rho \lambda} F_{\beta \rho} \boldsymbol{\lambda}^{j}_{\lambda} {G}^{-1} - \frac{369}{640}G_{i j} (\Gamma_{a})_{\alpha \lambda} W^{\beta}\,_{\rho} \boldsymbol{\lambda}^{j}_{\beta} {G}^{-1} \nabla^{\lambda \rho}{W}+\frac{9}{64}G_{i j} (\Gamma_{a})^{\lambda \beta} W_{\lambda}\,^{\rho} F_{\beta \rho} \boldsymbol{\lambda}^{j}_{\alpha} {G}^{-1}+\frac{9}{32}(\Gamma_{a})^{\beta}{}_{\rho} \boldsymbol{\lambda}_{j \beta} {G}^{-1} \nabla^{\rho \lambda}{\lambda_{i \lambda}} \varphi^{j}_{\alpha}%
 - \frac{33}{320}G_{j k} (\Gamma_{a})_{\alpha \beta} \mathbf{X}^{j k} {G}^{-1} \nabla^{\beta \rho}{\lambda_{i \rho}}+\frac{3}{64}G_{j k} (\Gamma_{a})^{\beta}{}_{\rho} \boldsymbol{\lambda}^{j}_{\beta} {G}^{-1} \nabla^{\rho}\,_{\alpha}{X_{i}\,^{k}} - \frac{3}{128}G_{j k} (\Gamma_{a})^{\rho \beta} \lambda_{i \alpha} \boldsymbol{\lambda}^{j}_{\rho} X^{k}_{\beta} {G}^{-1} - \frac{3}{64}(\Gamma_{a})_{\beta \rho} \boldsymbol{\lambda}_{j}^{\lambda} {G}^{-1} \nabla^{\beta \rho}{\lambda^{j}_{\lambda}} \varphi_{i \alpha} - \frac{15}{64}G_{i j} (\Gamma_{a})_{\rho \lambda} \mathbf{F}_{\alpha}\,^{\beta} {G}^{-1} \nabla^{\rho \lambda}{\lambda^{j}_{\beta}}+\frac{3}{32}G_{i j} (\Gamma_{a})_{\beta \rho} \mathbf{X}^{j}\,_{k} {G}^{-1} \nabla^{\beta \rho}{\lambda^{k}_{\alpha}} - \frac{15}{128}G_{i j} (\Gamma_{a})_{\beta \rho} {G}^{-1} \nabla_{\alpha}\,^{\lambda}{\boldsymbol{W}} \nabla^{\beta \rho}{\lambda^{j}_{\lambda}} - \frac{3}{64}G_{i j} G_{k l} (\Gamma_{a})_{\beta \rho} \boldsymbol{\lambda}^{j \lambda} {G}^{-3} \nabla^{\beta \rho}{\lambda^{k}_{\lambda}} \varphi^{l}_{\alpha}+\frac{3}{32}(\Gamma_{a})^{\beta}{}_{\rho} \boldsymbol{\lambda}_{j \lambda} {G}^{-1} \nabla^{\rho \lambda}{\lambda^{j}_{\beta}} \varphi_{i \alpha}+\frac{3}{16}G_{i j} (\Gamma_{a})^{\rho}{}_{\lambda} \mathbf{F}_{\alpha \beta} {G}^{-1} \nabla^{\lambda \beta}{\lambda^{j}_{\rho}}+\frac{3}{32}G_{i j} (\Gamma_{a})^{\beta}{}_{\rho} \mathbf{X}^{j}\,_{k} {G}^{-1} \nabla^{\rho}\,_{\alpha}{\lambda^{k}_{\beta}}+\frac{3}{32}G_{i j} (\Gamma_{a})^{\beta}{}_{\rho} {G}^{-1} \nabla_{\alpha \lambda}{\boldsymbol{W}} \nabla^{\rho \lambda}{\lambda^{j}_{\beta}}+\frac{3}{32}G_{i j} G_{k l} (\Gamma_{a})^{\beta}{}_{\rho} \boldsymbol{\lambda}^{j}_{\lambda} {G}^{-3} \nabla^{\rho \lambda}{\lambda^{k}_{\beta}} \varphi^{l}_{\alpha}+\frac{3}{160}G_{i j} (\Gamma_{a})_{\alpha \beta} \mathbf{X}^{j}\,_{k} {G}^{-1} \nabla^{\beta \rho}{\lambda^{k}_{\rho}}+\frac{3}{32}G_{i j} G_{k l} (\Gamma_{a})^{\beta}{}_{\rho} \boldsymbol{\lambda}^{j}_{\beta} {G}^{-3} \nabla^{\rho \lambda}{\lambda^{k}_{\lambda}} \varphi^{l}_{\alpha}+\frac{3}{2}{\rm i} (\Gamma_{a})^{\beta \lambda} F_{\beta}\,^{\rho} \mathbf{F}_{\lambda \rho} {G}^{-1} \varphi_{i \alpha} - \frac{3}{32}G_{i j} (\Gamma_{a})^{\beta}{}_{\lambda} \mathbf{F}_{\beta}\,^{\rho} {G}^{-1} \nabla^{\lambda}\,_{\alpha}{\lambda^{j}_{\rho}}+\frac{3}{32}G_{i j} (\Gamma_{a})^{\rho \lambda} \mathbf{F}_{\beta \rho} {G}^{-1} \nabla_{\alpha}\,^{\beta}{\lambda^{j}_{\lambda}}+\frac{9}{64}G_{i j} (\Gamma_{a})^{\lambda \beta} W_{\lambda}\,^{\rho} \mathbf{F}_{\beta \rho} \lambda^{j}_{\alpha} {G}^{-1}+\frac{9}{32}G_{i j} (\Gamma_{a})^{\lambda \beta} W_{\alpha \lambda} \mathbf{F}_{\beta}\,^{\rho} \lambda^{j}_{\rho} {G}^{-1}%
 - \frac{3}{32}G_{i j} (\Gamma_{a})_{\alpha}{}^{\rho} \mathbf{F}_{\beta \rho} {G}^{-1} \nabla^{\beta \lambda}{\lambda^{j}_{\lambda}}+\frac{9}{64}G_{i j} (\Gamma_{a})_{\alpha}{}^{\beta} W^{\rho \lambda} \mathbf{F}_{\beta \rho} \lambda^{j}_{\lambda} {G}^{-1}+\frac{3}{4}{\rm i} G_{i j} (\Gamma_{a})^{\beta \lambda} W \mathbf{F}_{\beta}\,^{\rho} W_{\alpha \lambda \rho}\,^{j} {G}^{-1} - \frac{3}{32}G_{i j} (\Gamma_{a})^{\beta}{}_{\lambda} F_{\beta}\,^{\rho} {G}^{-1} \nabla^{\lambda}\,_{\alpha}{\boldsymbol{\lambda}^{j}_{\rho}}+\frac{3}{32}G_{i j} (\Gamma_{a})^{\beta \lambda} F_{\beta \rho} {G}^{-1} \nabla_{\alpha}\,^{\rho}{\boldsymbol{\lambda}^{j}_{\lambda}} - \frac{3}{32}G_{i j} (\Gamma_{a})_{\alpha}{}^{\beta} F_{\beta \rho} {G}^{-1} \nabla^{\rho \lambda}{\boldsymbol{\lambda}^{j}_{\lambda}}+\frac{3}{4}{\rm i} G_{i j} (\Gamma_{a})^{\beta \lambda} \boldsymbol{W} F_{\beta}\,^{\rho} W_{\alpha \lambda \rho}\,^{j} {G}^{-1}+\frac{3}{4}{\rm i} (\Gamma_{a})^{\beta}{}_{\lambda} F_{\beta \rho} {G}^{-1} \nabla^{\lambda \rho}{\boldsymbol{W}} \varphi_{i \alpha}+\frac{3}{64}G_{i j} (\Gamma_{a})^{\beta}{}_{\rho} {G}^{-1} \nabla^{\rho}\,_{\lambda}{\boldsymbol{W}} \nabla_{\alpha}\,^{\lambda}{\lambda^{j}_{\beta}}+\frac{3}{64}G_{i j} (\Gamma_{a})_{\beta \rho} {G}^{-1} \nabla^{\beta \lambda}{\boldsymbol{W}} \nabla^{\rho}\,_{\alpha}{\lambda^{j}_{\lambda}} - \frac{9}{32}G_{i j} (\Gamma_{a})^{\beta}{}_{\rho} W_{\alpha \beta} \lambda^{j}_{\lambda} {G}^{-1} \nabla^{\rho \lambda}{\boldsymbol{W}}+\frac{57}{320}G_{i j} (\Gamma_{a})_{\alpha \beta} {G}^{-1} \nabla^{\beta}\,_{\lambda}{\boldsymbol{W}} \nabla^{\lambda \rho}{\lambda^{j}_{\rho}} - \frac{369}{640}G_{i j} (\Gamma_{a})_{\alpha \lambda} W^{\beta}\,_{\rho} \lambda^{j}_{\beta} {G}^{-1} \nabla^{\lambda \rho}{\boldsymbol{W}}+\frac{3}{32}G_{i j} (\Gamma_{a})^{\beta}{}_{\lambda} F_{\beta \rho} {G}^{-1} \nabla^{\lambda \rho}{\boldsymbol{\lambda}^{j}_{\alpha}} - \frac{9}{64}{\rm i} (\Gamma_{a})_{\beta \rho} X_{i j} {G}^{-1} \nabla^{\beta \rho}{\boldsymbol{W}} \varphi^{j}_{\alpha} - \frac{3}{8}G_{i j} (\Gamma_{a})_{\beta \rho} {G}^{-1} \nabla^{\beta \rho}{\boldsymbol{W}} \nabla_{\alpha}\,^{\lambda}{\lambda^{j}_{\lambda}} - \frac{9}{16}G_{i j} (\Gamma_{a})_{\rho \lambda} W_{\alpha}\,^{\beta} \lambda^{j}_{\beta} {G}^{-1} \nabla^{\rho \lambda}{\boldsymbol{W}} - \frac{3}{32}{\rm i} G_{i j} G_{k l} (\Gamma_{a})_{\beta \rho} X^{j k} {G}^{-3} \nabla^{\beta \rho}{\boldsymbol{W}} \varphi^{l}_{\alpha}+\frac{3}{32}G_{i j} (\Gamma_{a})_{\beta \rho} X^{j}\,_{k} {G}^{-1} \nabla^{\beta \rho}{\boldsymbol{\lambda}^{k}_{\alpha}} - \frac{9}{320}G_{i j} (\Gamma_{a})_{\alpha}{}^{\beta} X^{j}\,_{k} W_{\beta}\,^{\rho} \boldsymbol{\lambda}^{k}_{\rho} {G}^{-1}%
+\frac{9}{128}G_{j k} (\Gamma_{a})_{\beta \rho} X_{i}\,^{j} {G}^{-1} \nabla^{\beta \rho}{\boldsymbol{\lambda}^{k}_{\alpha}}+\frac{3}{4}{\rm i} (\Gamma_{a})^{\rho}{}_{\lambda} \mathbf{F}_{\beta \rho} {G}^{-1} \nabla^{\lambda \beta}{W} \varphi_{i \alpha}+\frac{3}{64}G_{i j} (\Gamma_{a})^{\beta}{}_{\rho} {G}^{-1} \nabla^{\rho}\,_{\lambda}{W} \nabla_{\alpha}\,^{\lambda}{\boldsymbol{\lambda}^{j}_{\beta}}+\frac{3}{64}G_{i j} (\Gamma_{a})_{\beta \rho} {G}^{-1} \nabla^{\beta \lambda}{W} \nabla^{\rho}\,_{\alpha}{\boldsymbol{\lambda}^{j}_{\lambda}}+\frac{57}{320}G_{i j} (\Gamma_{a})_{\alpha \beta} {G}^{-1} \nabla^{\beta}\,_{\lambda}{W} \nabla^{\lambda \rho}{\boldsymbol{\lambda}^{j}_{\rho}}+\frac{3}{32}G_{i j} (\Gamma_{a})^{\rho}{}_{\lambda} \mathbf{F}_{\beta \rho} {G}^{-1} \nabla^{\lambda \beta}{\lambda^{j}_{\alpha}} - \frac{9}{64}{\rm i} (\Gamma_{a})_{\beta \rho} \mathbf{X}_{i j} {G}^{-1} \nabla^{\beta \rho}{W} \varphi^{j}_{\alpha} - \frac{3}{8}G_{i j} (\Gamma_{a})_{\beta \rho} {G}^{-1} \nabla^{\beta \rho}{W} \nabla_{\alpha}\,^{\lambda}{\boldsymbol{\lambda}^{j}_{\lambda}}+\frac{9}{128}G_{j k} (\Gamma_{a})_{\beta \rho} \mathbf{X}_{i}\,^{j} {G}^{-1} \nabla^{\beta \rho}{\lambda^{k}_{\alpha}} - \frac{3}{8}{\rm i} (\Gamma_{a})_{\beta \rho} {G}^{-1} \nabla^{\beta}\,_{\lambda}{W} \nabla^{\rho \lambda}{\boldsymbol{W}} \varphi_{i \alpha} - \frac{3}{64}G_{i j} (\Gamma_{a})_{\beta \rho} {G}^{-1} \nabla^{\beta}\,_{\lambda}{\boldsymbol{W}} \nabla^{\rho \lambda}{\lambda^{j}_{\alpha}} - \frac{3}{64}G_{i j} (\Gamma_{a})_{\beta \rho} {G}^{-1} \nabla^{\beta}\,_{\lambda}{W} \nabla^{\rho \lambda}{\boldsymbol{\lambda}^{j}_{\alpha}} - \frac{3}{32}{\rm i} G_{i j} G_{k l} (\Gamma_{a})_{\beta \rho} \mathbf{X}^{j k} {G}^{-3} \nabla^{\beta \rho}{W} \varphi^{l}_{\alpha} - \frac{9}{320}G_{i j} (\Gamma_{a})_{\alpha}{}^{\beta} \mathbf{X}^{j}\,_{k} W_{\beta}\,^{\rho} \lambda^{k}_{\rho} {G}^{-1} - \frac{9}{128}(\Gamma_{a})_{\beta \rho} \lambda^{\lambda}_{j} {G}^{-1} \nabla^{\beta \rho}{\boldsymbol{\lambda}_{i \lambda}} \varphi^{j}_{\alpha} - \frac{3}{32}G_{j k} (\Gamma_{a})_{\beta \rho} X^{j k} {G}^{-1} \nabla^{\beta \rho}{\boldsymbol{\lambda}_{i \alpha}}+\frac{9}{32}G_{i j} (\Gamma_{a})_{\rho \lambda} \lambda^{j \beta} {G}^{-1} \nabla^{\rho \lambda}{\mathbf{F}_{\alpha \beta}} - \frac{9}{64}G_{i j} (\Gamma_{a})_{\beta \rho} \lambda^{j}_{\lambda} {G}^{-1} \nabla^{\beta \rho}{\nabla_{\alpha}\,^{\lambda}{\boldsymbol{W}}} - \frac{27}{320}G_{i j} (\Gamma_{a})_{\alpha}{}^{\lambda} W_{\lambda}\,^{\beta} \mathbf{F}_{\beta}\,^{\rho} \lambda^{j}_{\rho} {G}^{-1}+\frac{27}{640}G_{i j} (\Gamma_{a})_{\alpha}{}^{\beta} W_{\beta \rho} \lambda^{j}_{\lambda} {G}^{-1} \nabla^{\rho \lambda}{\boldsymbol{W}}%
 - \frac{3}{128}G_{j k} (\Gamma_{a})^{\rho \beta} \boldsymbol{\lambda}_{i \alpha} \lambda^{j}_{\rho} X^{k}_{\beta} {G}^{-1} - \frac{9}{1280}G_{j k} (\Gamma_{a})_{\alpha}{}^{\rho} \lambda^{j}_{\rho} \boldsymbol{\lambda}_{i}^{\beta} X^{k}_{\beta} {G}^{-1}+\frac{9}{32}(\Gamma_{a})^{\beta}{}_{\rho} \lambda_{j \lambda} {G}^{-1} \nabla^{\rho \lambda}{\boldsymbol{\lambda}_{i \beta}} \varphi^{j}_{\alpha}+\frac{3}{64}G_{j k} (\Gamma_{a})^{\beta}{}_{\rho} X^{j k} {G}^{-1} \nabla^{\rho}\,_{\alpha}{\boldsymbol{\lambda}_{i \beta}} - \frac{9}{32}G_{i j} (\Gamma_{a})^{\beta}{}_{\rho} \lambda^{j}_{\lambda} {G}^{-1} \nabla^{\rho \lambda}{\mathbf{F}_{\alpha \beta}}+\frac{3}{320}G_{j k} (\Gamma_{a})_{\alpha \beta} \lambda^{j}_{\rho} {G}^{-1} \nabla^{\beta \rho}{\mathbf{X}_{i}\,^{k}}+\frac{9}{64}G_{i j} (\Gamma_{a})_{\beta \rho} \lambda^{j}_{\lambda} {G}^{-1} \nabla^{\beta \lambda}{\nabla^{\rho}\,_{\alpha}{\boldsymbol{W}}}+\frac{9}{32}(\Gamma_{a})^{\beta}{}_{\rho} \lambda_{j \beta} {G}^{-1} \nabla^{\rho \lambda}{\boldsymbol{\lambda}_{i \lambda}} \varphi^{j}_{\alpha} - \frac{33}{320}G_{j k} (\Gamma_{a})_{\alpha \beta} X^{j k} {G}^{-1} \nabla^{\beta \rho}{\boldsymbol{\lambda}_{i \rho}}+\frac{3}{64}G_{j k} (\Gamma_{a})^{\beta}{}_{\rho} \lambda^{j}_{\beta} {G}^{-1} \nabla^{\rho}\,_{\alpha}{\mathbf{X}_{i}\,^{k}} - \frac{3}{64}(\Gamma_{a})_{\beta \rho} \lambda^{\lambda}_{j} {G}^{-1} \nabla^{\beta \rho}{\boldsymbol{\lambda}^{j}_{\lambda}} \varphi_{i \alpha} - \frac{15}{64}G_{i j} (\Gamma_{a})_{\rho \lambda} F_{\alpha}\,^{\beta} {G}^{-1} \nabla^{\rho \lambda}{\boldsymbol{\lambda}^{j}_{\beta}} - \frac{15}{128}G_{i j} (\Gamma_{a})_{\beta \rho} {G}^{-1} \nabla_{\alpha}\,^{\lambda}{W} \nabla^{\beta \rho}{\boldsymbol{\lambda}^{j}_{\lambda}} - \frac{3}{64}G_{i j} G_{k l} (\Gamma_{a})_{\beta \rho} \lambda^{j \lambda} {G}^{-3} \nabla^{\beta \rho}{\boldsymbol{\lambda}^{k}_{\lambda}} \varphi^{l}_{\alpha}+\frac{3}{32}(\Gamma_{a})^{\beta}{}_{\rho} \lambda_{j \lambda} {G}^{-1} \nabla^{\rho \lambda}{\boldsymbol{\lambda}^{j}_{\beta}} \varphi_{i \alpha}+\frac{3}{16}G_{i j} (\Gamma_{a})^{\rho}{}_{\lambda} F_{\alpha \beta} {G}^{-1} \nabla^{\lambda \beta}{\boldsymbol{\lambda}^{j}_{\rho}}+\frac{3}{32}G_{i j} (\Gamma_{a})^{\beta}{}_{\rho} X^{j}\,_{k} {G}^{-1} \nabla^{\rho}\,_{\alpha}{\boldsymbol{\lambda}^{k}_{\beta}}+\frac{3}{32}G_{i j} (\Gamma_{a})^{\beta}{}_{\rho} {G}^{-1} \nabla_{\alpha \lambda}{W} \nabla^{\rho \lambda}{\boldsymbol{\lambda}^{j}_{\beta}}+\frac{3}{32}G_{i j} G_{k l} (\Gamma_{a})^{\beta}{}_{\rho} \lambda^{j}_{\lambda} {G}^{-3} \nabla^{\rho \lambda}{\boldsymbol{\lambda}^{k}_{\beta}} \varphi^{l}_{\alpha}+\frac{3}{160}G_{i j} (\Gamma_{a})_{\alpha \beta} X^{j}\,_{k} {G}^{-1} \nabla^{\beta \rho}{\boldsymbol{\lambda}^{k}_{\rho}}%
+\frac{3}{32}G_{i j} G_{k l} (\Gamma_{a})^{\beta}{}_{\rho} \lambda^{j}_{\beta} {G}^{-3} \nabla^{\rho \lambda}{\boldsymbol{\lambda}^{k}_{\lambda}} \varphi^{l}_{\alpha} - \frac{9}{64}F (\Gamma_{a})^{\beta}{}_{\rho} \boldsymbol{W} {G}^{-1} \nabla^{\rho}\,_{\alpha}{\lambda_{i \beta}} - \frac{9}{64}F (\Gamma_{a})^{\beta}{}_{\rho} W {G}^{-1} \nabla^{\rho}\,_{\alpha}{\boldsymbol{\lambda}_{i \beta}} - \frac{9}{64}(\Gamma_{a})^{\beta \rho} \boldsymbol{W} \lambda_{i \beta} {G}^{-1} \nabla_{\alpha}\,^{\lambda}{\mathcal{H}_{\lambda \rho}}+\frac{3}{32}(\Gamma_{a})^{\beta}{}_{\rho} \boldsymbol{W} \lambda_{j \beta} {G}^{-1} \nabla_{\alpha \lambda}{\nabla^{\rho \lambda}{G_{i}\,^{j}}}+\frac{3}{16}{\rm i} (\Gamma_{a})^{\lambda \beta} \boldsymbol{W} \lambda_{j \lambda} W_{\alpha \beta}\,^{\rho}\,_{i} {G}^{-1} \varphi^{j}_{\rho} - \frac{9}{64}(\Gamma_{a})^{\beta \rho} W \boldsymbol{\lambda}_{i \beta} {G}^{-1} \nabla_{\alpha}\,^{\lambda}{\mathcal{H}_{\lambda \rho}}+\frac{3}{32}(\Gamma_{a})^{\beta}{}_{\rho} W \boldsymbol{\lambda}_{j \beta} {G}^{-1} \nabla_{\alpha \lambda}{\nabla^{\rho \lambda}{G_{i}\,^{j}}}+\frac{3}{16}{\rm i} (\Gamma_{a})^{\lambda \beta} W \boldsymbol{\lambda}_{j \lambda} W_{\alpha \beta}\,^{\rho}\,_{i} {G}^{-1} \varphi^{j}_{\rho}+\frac{9}{64}(\Gamma_{a})^{\beta \rho} W \boldsymbol{W} {G}^{-3} \nabla_{\alpha}\,^{\lambda}{\varphi_{j \lambda}} \varphi_{i \beta} \varphi^{j}_{\rho} - \frac{27}{64}G_{i j} G_{k l} (\Gamma_{a})^{\beta \rho} W \boldsymbol{W} {G}^{-5} \nabla_{\alpha}\,^{\lambda}{\varphi^{k}_{\lambda}} \varphi^{j}_{\beta} \varphi^{l}_{\rho}+\frac{9}{64}{\rm i} G_{i j} (\Gamma_{a})^{\beta \rho} W \boldsymbol{W} {G}^{-3} \nabla_{\alpha}\,^{\lambda}{\mathcal{H}_{\lambda \beta}} \varphi^{j}_{\rho}+\frac{9}{64}{\rm i} G_{j k} (\Gamma_{a})^{\beta}{}_{\rho} W \boldsymbol{W} {G}^{-3} \nabla_{\alpha \lambda}{\nabla^{\rho \lambda}{G_{i}\,^{j}}} \varphi^{k}_{\beta}+\frac{3}{16}G_{j k} (\Gamma_{a})^{\beta \lambda} W \boldsymbol{W} W_{\alpha \beta}\,^{\rho}\,_{i} {G}^{-3} \varphi^{j}_{\lambda} \varphi^{k}_{\rho}+\frac{9}{128}{\rm i} \mathcal{H}^{\beta \rho} G_{i j} (\Gamma_{a})_{\beta \rho} W \boldsymbol{W} {G}^{-3} \nabla_{\alpha}\,^{\lambda}{\varphi^{j}_{\lambda}}+\frac{3}{32}{\rm i} G_{j k} (\Gamma_{a})_{\beta \rho} W \boldsymbol{W} {G}^{-3} \nabla_{\alpha}\,^{\lambda}{\varphi^{j}_{\lambda}} \nabla^{\beta \rho}{G_{i}\,^{k}} - \frac{3}{64}{\rm i} G_{i j} (\Gamma_{a})_{\beta \rho} W \boldsymbol{W} {G}^{-3} \nabla_{\alpha}\,^{\lambda}{\varphi_{k \lambda}} \nabla^{\beta \rho}{G^{j k}} - \frac{9}{32}{\rm i} (\Gamma_{a})_{\beta \rho} W \boldsymbol{W} {G}^{-1} \nabla^{\beta}\,_{\alpha}{\nabla^{\rho \lambda}{\varphi_{i \lambda}}} - \frac{111}{1280}{\rm i} \Phi^{\beta \rho}\,_{i j} (\Gamma_{a})_{\alpha \beta} W \boldsymbol{W} {G}^{-1} \varphi^{j}_{\rho} - \frac{9}{32}{\rm i} (\Gamma_{a})^{\beta}{}_{\rho} W \boldsymbol{W} {G}^{-1} \nabla_{\alpha \lambda}{\nabla^{\rho \lambda}{\varphi_{i \beta}}}%
 - \frac{9}{64}{\rm i} (\Gamma_{a})_{\beta \rho} W \boldsymbol{W} {G}^{-1} \nabla_{\alpha}\,^{\lambda}{\nabla^{\beta \rho}{\varphi_{i \lambda}}} - \frac{3}{32}G_{i j} (\Gamma_{a})^{\beta \lambda} W \boldsymbol{W} W_{\alpha \beta}\,^{\rho}\,_{k} {G}^{-3} \varphi^{j}_{\lambda} \varphi^{k}_{\rho} - \frac{45}{128}{\rm i} G_{i j} (\Gamma_{a})_{\alpha \rho} W \boldsymbol{W} {G}^{-1} \nabla^{\rho \beta}{X^{j}_{\beta}}+\frac{9}{128}{\rm i} G_{j k} (\Gamma_{a})^{\rho \lambda} \boldsymbol{W} W_{\alpha}\,^{\beta} \lambda_{i \rho} {G}^{-3} \varphi^{j}_{\lambda} \varphi^{k}_{\beta}+\frac{9}{128}{\rm i} G_{j k} (\Gamma_{a})^{\rho \lambda} W W_{\alpha}\,^{\beta} \boldsymbol{\lambda}_{i \rho} {G}^{-3} \varphi^{j}_{\lambda} \varphi^{k}_{\beta} - \frac{9}{256}{\rm i} G_{i j} G_{k l} (\Gamma_{a})^{\beta \rho} \boldsymbol{W} \lambda^{k}_{\beta} X^{l}_{\alpha} {G}^{-3} \varphi^{j}_{\rho} - \frac{9}{256}{\rm i} G_{i j} G_{k l} (\Gamma_{a})^{\beta \rho} W \boldsymbol{\lambda}^{k}_{\beta} X^{l}_{\alpha} {G}^{-3} \varphi^{j}_{\rho}+\frac{9}{128}{\rm i} G_{j k} (\Gamma_{a})^{\beta}{}_{\rho} \boldsymbol{W} {G}^{-3} \nabla^{\rho}\,_{\alpha}{\lambda_{i \beta}} \varphi^{j \lambda} \varphi^{k}_{\lambda} - \frac{3}{640}{\rm i} G_{j k} (\Gamma_{a})_{\alpha \beta} \boldsymbol{W} {G}^{-3} \nabla^{\beta \rho}{\lambda_{i \rho}} \varphi^{j \lambda} \varphi^{k}_{\lambda}+\frac{9}{128}{\rm i} G_{j k} (\Gamma_{a})^{\beta}{}_{\rho} W {G}^{-3} \nabla^{\rho}\,_{\alpha}{\boldsymbol{\lambda}_{i \beta}} \varphi^{j \lambda} \varphi^{k}_{\lambda} - \frac{3}{640}{\rm i} G_{j k} (\Gamma_{a})_{\alpha \beta} W {G}^{-3} \nabla^{\beta \rho}{\boldsymbol{\lambda}_{i \rho}} \varphi^{j \lambda} \varphi^{k}_{\lambda}+\frac{3}{64}{\rm i} (\Gamma_{a})^{\beta}{}_{\rho} \boldsymbol{W} \lambda_{k \beta} {G}^{-3} \nabla^{\rho}\,_{\alpha}{G_{i j}} \varphi^{k \lambda} \varphi^{j}_{\lambda}+\frac{3}{64}{\rm i} (\Gamma_{a})^{\beta}{}_{\rho} W \boldsymbol{\lambda}_{k \beta} {G}^{-3} \nabla^{\rho}\,_{\alpha}{G_{i j}} \varphi^{k \lambda} \varphi^{j}_{\lambda} - \frac{9}{64}G_{j k} (\Gamma_{a})^{\beta}{}_{\rho} W \boldsymbol{W} {G}^{-5} \nabla^{\rho}\,_{\alpha}{G_{i l}} \varphi^{j}_{\beta} \varphi^{k \lambda} \varphi^{l}_{\lambda}+\frac{3}{64}{\rm i} (\Gamma_{a})_{\beta \rho} W \boldsymbol{W} {G}^{-3} \nabla^{\beta}\,_{\alpha}{G_{i j}} \nabla^{\rho \lambda}{G^{j}\,_{k}} \varphi^{k}_{\lambda}+\frac{3}{64}(\Gamma_{a})^{\beta}{}_{\rho} W \boldsymbol{W} {G}^{-3} \nabla^{\rho}\,_{\alpha}{\varphi_{i}^{\lambda}} \varphi_{j \beta} \varphi^{j}_{\lambda}+\frac{3}{64}{\rm i} G_{j k} (\Gamma_{a})_{\beta \rho} W \boldsymbol{W} {G}^{-3} \nabla^{\beta}\,_{\alpha}{\nabla^{\rho \lambda}{G_{i}\,^{j}}} \varphi^{k}_{\lambda}+\frac{9}{64}{\rm i} (\Gamma_{a})^{\beta}{}_{\rho} W \boldsymbol{W} {G}^{-1} \nabla^{\rho \lambda}{W_{\alpha \beta}} \varphi_{i \lambda}+\frac{3}{64}{\rm i} G_{j k} (\Gamma_{a})^{\beta}{}_{\rho} \boldsymbol{W} \lambda^{j}_{\beta} {G}^{-3} \nabla^{\rho}\,_{\alpha}{\varphi_{i}^{\lambda}} \varphi^{k}_{\lambda}+\frac{3}{64}{\rm i} G_{j k} (\Gamma_{a})^{\beta}{}_{\rho} W \boldsymbol{\lambda}^{j}_{\beta} {G}^{-3} \nabla^{\rho}\,_{\alpha}{\varphi_{i}^{\lambda}} \varphi^{k}_{\lambda}%
+\frac{3}{64}{\rm i} G_{j k} (\Gamma_{a})_{\beta \rho} W \boldsymbol{W} {G}^{-3} \nabla^{\beta}\,_{\alpha}{\varphi_{i \lambda}} \nabla^{\rho \lambda}{G^{j k}} - \frac{3}{128}{\rm i} G_{j k} (\Gamma_{a})^{\beta \lambda} \boldsymbol{W} W_{\beta}\,^{\rho} \lambda^{j}_{\lambda} {G}^{-3} \varphi_{i \alpha} \varphi^{k}_{\rho} - \frac{3}{128}{\rm i} G_{j k} (\Gamma_{a})^{\beta \lambda} W W_{\beta}\,^{\rho} \boldsymbol{\lambda}^{j}_{\lambda} {G}^{-3} \varphi_{i \alpha} \varphi^{k}_{\rho}+\frac{3}{128}{\rm i} G_{j k} (\Gamma_{a})^{\beta}{}_{\lambda} W \boldsymbol{W} W_{\beta \rho} {G}^{-3} \nabla^{\lambda \rho}{G^{j k}} \varphi_{i \alpha} - \frac{9}{320}(\Gamma_{a})_{\alpha}{}^{\beta} W \boldsymbol{W} {G}^{-3} \nabla^{\rho \lambda}{\varphi_{i \rho}} \varphi_{j \beta} \varphi^{j}_{\lambda}+\frac{27}{320}{\rm i} G_{j k} (\Gamma_{a})_{\alpha \beta} W \boldsymbol{W} {G}^{-3} \nabla_{\rho}\,^{\lambda}{\nabla^{\beta \rho}{G_{i}\,^{j}}} \varphi^{k}_{\lambda} - \frac{3}{80}{\rm i} G_{j k} (\Gamma_{a})_{\alpha}{}^{\beta} \boldsymbol{W} \lambda^{j}_{\beta} {G}^{-3} \nabla^{\rho \lambda}{\varphi_{i \rho}} \varphi^{k}_{\lambda} - \frac{3}{80}{\rm i} G_{j k} (\Gamma_{a})_{\alpha}{}^{\beta} W \boldsymbol{\lambda}^{j}_{\beta} {G}^{-3} \nabla^{\rho \lambda}{\varphi_{i \rho}} \varphi^{k}_{\lambda} - \frac{9}{640}{\rm i} G_{j k} (\Gamma_{a})_{\alpha \beta} W \boldsymbol{W} {G}^{-3} \nabla_{\rho}\,^{\lambda}{\varphi_{i \lambda}} \nabla^{\beta \rho}{G^{j k}} - \frac{3}{64}{\rm i} G_{j k} (\Gamma_{a})^{\beta \rho} \boldsymbol{W} \lambda^{j}_{\beta} {G}^{-3} \nabla_{\alpha}\,^{\lambda}{\varphi_{i \lambda}} \varphi^{k}_{\rho} - \frac{3}{64}{\rm i} G_{j k} (\Gamma_{a})^{\beta \rho} W \boldsymbol{\lambda}^{j}_{\beta} {G}^{-3} \nabla_{\alpha}\,^{\lambda}{\varphi_{i \lambda}} \varphi^{k}_{\rho}+\frac{3}{64}{\rm i} G_{j k} (\Gamma_{a})_{\beta \rho} W \boldsymbol{W} {G}^{-3} \nabla_{\alpha}\,^{\lambda}{\varphi_{i \lambda}} \nabla^{\beta \rho}{G^{j k}} - \frac{9}{64}(\Gamma_{a})^{\beta \rho} W \boldsymbol{W} {G}^{-3} \nabla_{\alpha}\,^{\lambda}{\varphi_{i \beta}} \varphi_{j \rho} \varphi^{j}_{\lambda} - \frac{3}{128}{\rm i} G_{i j} (\Gamma_{a})^{\beta \lambda} \boldsymbol{W} W_{\beta}\,^{\rho} \lambda_{k \lambda} {G}^{-3} \varphi^{k}_{\alpha} \varphi^{j}_{\rho} - \frac{9}{512}{\rm i} G_{i j} G_{k l} (\Gamma_{a})^{\rho \beta} \boldsymbol{W} \lambda^{k}_{\rho} X^{l}_{\beta} {G}^{-3} \varphi^{j}_{\alpha} - \frac{9}{2560}{\rm i} G_{i j} G_{k l} (\Gamma_{a})_{\alpha}{}^{\rho} \boldsymbol{W} \lambda^{k}_{\rho} X^{l \beta} {G}^{-3} \varphi^{j}_{\beta} - \frac{3}{128}{\rm i} G_{i j} (\Gamma_{a})^{\beta \lambda} W W_{\beta}\,^{\rho} \boldsymbol{\lambda}_{k \lambda} {G}^{-3} \varphi^{k}_{\alpha} \varphi^{j}_{\rho} - \frac{9}{512}{\rm i} G_{i j} G_{k l} (\Gamma_{a})^{\rho \beta} W \boldsymbol{\lambda}^{k}_{\rho} X^{l}_{\beta} {G}^{-3} \varphi^{j}_{\alpha} - \frac{9}{2560}{\rm i} G_{i j} G_{k l} (\Gamma_{a})_{\alpha}{}^{\rho} W \boldsymbol{\lambda}^{k}_{\rho} X^{l \beta} {G}^{-3} \varphi^{j}_{\beta}+\frac{9}{128}G_{i j} G_{k l} (\Gamma_{a})^{\beta \lambda} W \boldsymbol{W} W_{\beta}\,^{\rho} {G}^{-5} \varphi^{k}_{\alpha} \varphi^{j}_{\rho} \varphi^{l}_{\lambda}%
+\frac{9}{128}{\rm i} F G_{i j} G_{k l} (\Gamma_{a})^{\beta}{}_{\rho} W \boldsymbol{W} {G}^{-5} \nabla^{\rho}\,_{\alpha}{G^{j k}} \varphi^{l}_{\beta}+\frac{9}{128}{\rm i} G_{i j} (\Gamma_{a})^{\beta \rho} W \boldsymbol{W} {G}^{-3} \nabla_{\alpha}\,^{\lambda}{\mathcal{H}_{\beta \rho}} \varphi^{j}_{\lambda}+\frac{3}{32}{\rm i} G_{j k} (\Gamma_{a})_{\beta \rho} W \boldsymbol{W} {G}^{-3} \nabla_{\alpha}\,^{\lambda}{\nabla^{\beta \rho}{G_{i}\,^{j}}} \varphi^{k}_{\lambda} - \frac{3}{160}G_{i j} (\Gamma_{a})_{\alpha}{}^{\beta} W \boldsymbol{W} W_{\beta}\,^{\rho \lambda}\,_{k} {G}^{-3} \varphi^{j}_{\rho} \varphi^{k}_{\lambda}+\frac{63}{1280}{\rm i} \Phi^{\beta \rho}\,_{k l} G_{i j} G^{k l} (\Gamma_{a})_{\alpha \beta} W \boldsymbol{W} {G}^{-3} \varphi^{j}_{\rho}+\frac{3}{128}{\rm i} G_{i j} (\Gamma_{a})^{\beta}{}_{\lambda} W \boldsymbol{W} W_{\beta \rho} {G}^{-3} \nabla^{\lambda \rho}{G^{j}\,_{k}} \varphi^{k}_{\alpha} - \frac{3}{256}{\rm i} \Phi^{\beta \rho}\,_{j k} G_{i}\,^{j} G^{k}\,_{l} (\Gamma_{a})_{\alpha \beta} W \boldsymbol{W} {G}^{-3} \varphi^{l}_{\rho} - \frac{3}{128}{\rm i} G_{i j} (\Gamma_{a})^{\beta}{}_{\rho} W \boldsymbol{W} {G}^{-3} \nabla_{\alpha \lambda}{G^{j}\,_{k}} \nabla^{\rho \lambda}{\varphi^{k}_{\beta}}+\frac{3}{128}{\rm i} G_{j k} (\Gamma_{a})^{\beta}{}_{\rho} W \boldsymbol{W} {G}^{-3} \nabla_{\alpha \lambda}{G^{j k}} \nabla^{\rho \lambda}{\varphi_{i \beta}} - \frac{9}{64}G_{j k} (\Gamma_{a})^{\beta}{}_{\rho} W \boldsymbol{W} {G}^{-5} \nabla^{\rho}\,_{\alpha}{G_{i}\,^{j}} \varphi^{k \lambda} \varphi_{l \beta} \varphi^{l}_{\lambda} - \frac{9}{128}{\rm i} G_{j k} G_{l m} (\Gamma_{a})^{\beta}{}_{\rho} \boldsymbol{W} \lambda^{j}_{\beta} {G}^{-5} \nabla^{\rho}\,_{\alpha}{G_{i}\,^{k}} \varphi^{l \lambda} \varphi^{m}_{\lambda} - \frac{9}{128}{\rm i} G_{j k} G_{l m} (\Gamma_{a})^{\beta}{}_{\rho} W \boldsymbol{\lambda}^{j}_{\beta} {G}^{-5} \nabla^{\rho}\,_{\alpha}{G_{i}\,^{k}} \varphi^{l \lambda} \varphi^{m}_{\lambda} - \frac{9}{64}{\rm i} G_{j k} G_{l m} (\Gamma_{a})_{\beta \rho} W \boldsymbol{W} {G}^{-5} \nabla^{\beta}\,_{\alpha}{G_{i}\,^{j}} \nabla^{\rho \lambda}{G^{k l}} \varphi^{m}_{\lambda}+\frac{3}{32}{\rm i} G_{j k} (\Gamma_{a})^{\beta}{}_{\rho} W \mathbf{X}^{j}\,_{l} {G}^{-3} \nabla^{\rho}\,_{\alpha}{G_{i}\,^{k}} \varphi^{l}_{\beta} - \frac{3}{64}G_{j k} G_{l m} (\Gamma_{a})^{\beta}{}_{\rho} \mathbf{X}^{j k} \lambda^{l}_{\beta} {G}^{-3} \nabla^{\rho}\,_{\alpha}{G_{i}\,^{m}}+\frac{9}{128}(\Gamma_{a})^{\beta \lambda} W_{\beta}\,^{\rho} \lambda_{j \alpha} \boldsymbol{\lambda}^{j}_{\rho} {G}^{-1} \varphi_{i \lambda}+\frac{3}{128}(\Gamma_{a})^{\beta \lambda} W_{\beta}\,^{\rho} \lambda_{i \alpha} \boldsymbol{\lambda}_{j \rho} {G}^{-1} \varphi^{j}_{\lambda} - \frac{3}{64}G_{i j} G_{k l} (\Gamma_{a})^{\beta \lambda} W_{\beta}\,^{\rho} \lambda^{k}_{\alpha} \boldsymbol{\lambda}^{j}_{\rho} {G}^{-3} \varphi^{l}_{\lambda}+\frac{3}{128}G_{i j} G_{k l} (\Gamma_{a})^{\beta \lambda} W_{\beta}\,^{\rho} \lambda^{k}_{\alpha} \boldsymbol{\lambda}^{l}_{\rho} {G}^{-3} \varphi^{j}_{\lambda}+\frac{9}{64}(\Gamma_{a})^{\beta}{}_{\rho} \mathbf{X}_{k j} \lambda^{k}_{\beta} {G}^{-1} \nabla^{\rho}\,_{\alpha}{G_{i}\,^{j}}%
 - \frac{3}{32}{\rm i} G_{j k} (\Gamma_{a})^{\beta}{}_{\rho} W \mathbf{X}^{j}\,_{l} {G}^{-3} \nabla^{\rho}\,_{\alpha}{G_{i}\,^{l}} \varphi^{k}_{\beta} - \frac{3}{8}{\rm i} (\Gamma_{a})^{\rho \lambda} W {G}^{-1} \nabla_{\alpha}\,^{\beta}{\mathbf{F}_{\beta \rho}} \varphi_{i \lambda}+\frac{3}{16}{\rm i} (\Gamma_{a})^{\beta}{}_{\rho} W {G}^{-1} \nabla_{\alpha \lambda}{\nabla^{\rho \lambda}{\boldsymbol{W}}} \varphi_{i \beta}+\frac{3}{16}G_{i j} (\Gamma_{a})^{\lambda \rho} \lambda^{j}_{\lambda} {G}^{-1} \nabla_{\alpha}\,^{\beta}{\mathbf{F}_{\beta \rho}}+\frac{3}{32}G_{i j} (\Gamma_{a})^{\beta}{}_{\rho} \lambda^{j}_{\beta} {G}^{-1} \nabla_{\alpha \lambda}{\nabla^{\rho \lambda}{\boldsymbol{W}}} - \frac{3}{64}G_{j k} (\Gamma_{a})^{\beta \rho} \lambda^{j}_{\beta} \boldsymbol{\lambda}^{k}_{\rho} X_{i \alpha} {G}^{-1} - \frac{3}{1280}G_{j k} (\Gamma_{a})_{\alpha}{}^{\rho} \lambda^{j}_{\rho} \boldsymbol{\lambda}^{k \beta} X_{i \beta} {G}^{-1}+\frac{3}{32}G_{i j} (\Gamma_{a})^{\beta}{}_{\rho} W {G}^{-1} \nabla_{\alpha \lambda}{\nabla^{\rho \lambda}{\boldsymbol{\lambda}^{j}_{\beta}}} - \frac{3}{64}G_{i j} (\Gamma_{a})_{\beta \rho} W {G}^{-1} \nabla_{\alpha}\,^{\lambda}{\nabla^{\beta \rho}{\boldsymbol{\lambda}^{j}_{\lambda}}}+\frac{3}{32}{\rm i} G_{j k} (\Gamma_{a})^{\beta}{}_{\rho} \boldsymbol{W} X^{j}\,_{l} {G}^{-3} \nabla^{\rho}\,_{\alpha}{G_{i}\,^{k}} \varphi^{l}_{\beta} - \frac{3}{64}G_{j k} G_{l m} (\Gamma_{a})^{\beta}{}_{\rho} X^{j k} \boldsymbol{\lambda}^{l}_{\beta} {G}^{-3} \nabla^{\rho}\,_{\alpha}{G_{i}\,^{m}}+\frac{9}{64}(\Gamma_{a})^{\beta}{}_{\rho} X_{k j} \boldsymbol{\lambda}^{k}_{\beta} {G}^{-1} \nabla^{\rho}\,_{\alpha}{G_{i}\,^{j}} - \frac{3}{32}{\rm i} G_{j k} (\Gamma_{a})^{\beta}{}_{\rho} \boldsymbol{W} X^{j}\,_{l} {G}^{-3} \nabla^{\rho}\,_{\alpha}{G_{i}\,^{l}} \varphi^{k}_{\beta} - \frac{3}{8}{\rm i} (\Gamma_{a})^{\beta \lambda} \boldsymbol{W} {G}^{-1} \nabla_{\alpha}\,^{\rho}{F_{\beta \rho}} \varphi_{i \lambda}+\frac{3}{16}{\rm i} (\Gamma_{a})^{\beta}{}_{\rho} \boldsymbol{W} {G}^{-1} \nabla_{\alpha \lambda}{\nabla^{\rho \lambda}{W}} \varphi_{i \beta}+\frac{3}{16}G_{i j} (\Gamma_{a})^{\lambda \beta} \boldsymbol{\lambda}^{j}_{\lambda} {G}^{-1} \nabla_{\alpha}\,^{\rho}{F_{\beta \rho}}+\frac{3}{32}G_{i j} (\Gamma_{a})^{\beta}{}_{\rho} \boldsymbol{\lambda}^{j}_{\beta} {G}^{-1} \nabla_{\alpha \lambda}{\nabla^{\rho \lambda}{W}}+\frac{3}{1280}G_{j k} (\Gamma_{a})_{\alpha}{}^{\rho} \lambda^{j \beta} \boldsymbol{\lambda}^{k}_{\rho} X_{i \beta} {G}^{-1}+\frac{3}{32}G_{i j} (\Gamma_{a})^{\beta}{}_{\rho} \boldsymbol{W} {G}^{-1} \nabla_{\alpha \lambda}{\nabla^{\rho \lambda}{\lambda^{j}_{\beta}}} - \frac{3}{64}G_{i j} (\Gamma_{a})_{\beta \rho} \boldsymbol{W} {G}^{-1} \nabla_{\alpha}\,^{\lambda}{\nabla^{\beta \rho}{\lambda^{j}_{\lambda}}}%
+\frac{9}{128}(\Gamma_{a})^{\beta \lambda} W_{\beta}\,^{\rho} \boldsymbol{\lambda}_{j \alpha} \lambda^{j}_{\rho} {G}^{-1} \varphi_{i \lambda}+\frac{3}{128}(\Gamma_{a})^{\beta \lambda} W_{\beta}\,^{\rho} \boldsymbol{\lambda}_{i \alpha} \lambda_{j \rho} {G}^{-1} \varphi^{j}_{\lambda} - \frac{3}{64}G_{i j} G_{k l} (\Gamma_{a})^{\beta \lambda} W_{\beta}\,^{\rho} \boldsymbol{\lambda}^{k}_{\alpha} \lambda^{j}_{\rho} {G}^{-3} \varphi^{l}_{\lambda}+\frac{3}{128}G_{i j} G_{k l} (\Gamma_{a})^{\beta \lambda} W_{\beta}\,^{\rho} \boldsymbol{\lambda}^{k}_{\alpha} \lambda^{l}_{\rho} {G}^{-3} \varphi^{j}_{\lambda}+\frac{3}{64}G_{j k} (\Gamma_{a})^{\beta}{}_{\rho} \lambda^{j \lambda} \boldsymbol{\lambda}_{l \lambda} {G}^{-3} \nabla^{\rho}\,_{\alpha}{G_{i}\,^{k}} \varphi^{l}_{\beta}+\frac{3}{64}G_{j k} (\Gamma_{a})^{\beta}{}_{\rho} \lambda^{\lambda}_{l} \boldsymbol{\lambda}^{j}_{\lambda} {G}^{-3} \nabla^{\rho}\,_{\alpha}{G_{i}\,^{k}} \varphi^{l}_{\beta}+\frac{3}{64}G_{j k} G_{l m} (\Gamma_{a})^{\beta}{}_{\rho} X^{j l} \boldsymbol{\lambda}^{k}_{\beta} {G}^{-3} \nabla^{\rho}\,_{\alpha}{G_{i}\,^{m}}+\frac{3}{64}G_{j k} G_{l m} (\Gamma_{a})^{\beta}{}_{\rho} \mathbf{X}^{j l} \lambda^{k}_{\beta} {G}^{-3} \nabla^{\rho}\,_{\alpha}{G_{i}\,^{m}} - \frac{3}{64}G_{j k} (\Gamma_{a})^{\beta}{}_{\rho} \lambda^{j \lambda} \boldsymbol{\lambda}_{l \lambda} {G}^{-3} \nabla^{\rho}\,_{\alpha}{G_{i}\,^{l}} \varphi^{k}_{\beta} - \frac{3}{64}G_{j k} (\Gamma_{a})^{\beta}{}_{\rho} \lambda^{\lambda}_{l} \boldsymbol{\lambda}^{j}_{\lambda} {G}^{-3} \nabla^{\rho}\,_{\alpha}{G_{i}\,^{l}} \varphi^{k}_{\beta} - \frac{9}{64}(\Gamma_{a})^{\beta}{}_{\rho} \boldsymbol{\lambda}_{j}^{\lambda} {G}^{-1} \nabla^{\rho}\,_{\alpha}{\lambda_{i \lambda}} \varphi^{j}_{\beta}+\frac{9}{64}(\Gamma_{a})^{\beta}{}_{\rho} \boldsymbol{\lambda}_{j \lambda} {G}^{-1} \nabla^{\rho \lambda}{\lambda_{i \alpha}} \varphi^{j}_{\beta} - \frac{3}{64}(\Gamma_{a})^{\beta \lambda} W_{\beta}\,^{\rho} \boldsymbol{\lambda}_{j \alpha} \lambda_{i \rho} {G}^{-1} \varphi^{j}_{\lambda}+\frac{3}{128}G_{i j} G_{k l} (\Gamma_{a})^{\beta \lambda} W_{\beta}\,^{\rho} \boldsymbol{\lambda}^{j}_{\alpha} \lambda^{k}_{\rho} {G}^{-3} \varphi^{l}_{\lambda} - \frac{9}{64}(\Gamma_{a})^{\beta \rho} \boldsymbol{\lambda}_{j \lambda} {G}^{-1} \nabla_{\alpha}\,^{\lambda}{\lambda_{i \beta}} \varphi^{j}_{\rho} - \frac{3}{64}(\Gamma_{a})^{\beta \lambda} W_{\beta}\,^{\rho} \lambda_{j \alpha} \boldsymbol{\lambda}_{i \rho} {G}^{-1} \varphi^{j}_{\lambda} - \frac{9}{64}(\Gamma_{a})^{\beta}{}_{\rho} \lambda^{\lambda}_{j} {G}^{-1} \nabla^{\rho}\,_{\alpha}{\boldsymbol{\lambda}_{i \lambda}} \varphi^{j}_{\beta}+\frac{9}{64}(\Gamma_{a})^{\beta}{}_{\rho} \lambda_{j \lambda} {G}^{-1} \nabla^{\rho \lambda}{\boldsymbol{\lambda}_{i \alpha}} \varphi^{j}_{\beta}+\frac{3}{128}G_{i j} G_{k l} (\Gamma_{a})^{\beta \lambda} W_{\beta}\,^{\rho} \lambda^{j}_{\alpha} \boldsymbol{\lambda}^{k}_{\rho} {G}^{-3} \varphi^{l}_{\lambda} - \frac{9}{64}(\Gamma_{a})^{\beta \rho} \lambda_{j \lambda} {G}^{-1} \nabla_{\alpha}\,^{\lambda}{\boldsymbol{\lambda}_{i \beta}} \varphi^{j}_{\rho}%
+\frac{3}{80}(\Gamma_{a})_{\alpha}{}^{\beta} \lambda_{j \rho} \boldsymbol{\lambda}^{j}_{\beta} {G}^{-1} \nabla^{\rho \lambda}{\varphi_{i \lambda}}+\frac{3}{80}(\Gamma_{a})_{\alpha}{}^{\beta} \lambda_{j \beta} \boldsymbol{\lambda}^{j}_{\rho} {G}^{-1} \nabla^{\rho \lambda}{\varphi_{i \lambda}}+\frac{9}{40}{\rm i} (\Gamma_{a})_{\alpha}{}^{\beta} \boldsymbol{W} F_{\beta \rho} {G}^{-1} \nabla^{\rho \lambda}{\varphi_{i \lambda}}+\frac{63}{160}{\rm i} (\Gamma_{a})_{\alpha}{}^{\beta} W \boldsymbol{W} W_{\beta \rho} {G}^{-1} \nabla^{\rho \lambda}{\varphi_{i \lambda}}+\frac{9}{320}{\rm i} G_{j k} (\Gamma_{a})_{\alpha}{}^{\beta} \boldsymbol{W} \lambda^{j}_{\rho} {G}^{-3} \nabla^{\rho \lambda}{\varphi_{i \lambda}} \varphi^{k}_{\beta}+\frac{3}{320}{\rm i} G_{j k} (\Gamma_{a})_{\alpha \beta} \boldsymbol{W} \lambda^{j \rho} {G}^{-3} \nabla^{\beta \lambda}{\varphi_{i \lambda}} \varphi^{k}_{\rho}+\frac{9}{320}(\Gamma_{a})_{\alpha}{}^{\beta} \boldsymbol{W} \lambda_{i \lambda} {G}^{-1} \nabla^{\lambda \rho}{\mathcal{H}_{\rho \beta}}+\frac{9}{320}(\Gamma_{a})_{\alpha \beta} \boldsymbol{W} \lambda^{\lambda}_{i} {G}^{-1} \nabla^{\beta \rho}{\mathcal{H}_{\rho \lambda}} - \frac{3}{160}(\Gamma_{a})_{\alpha \beta} \boldsymbol{W} \lambda_{j \rho} {G}^{-1} \nabla_{\lambda}\,^{\rho}{\nabla^{\beta \lambda}{G_{i}\,^{j}}}+\frac{3}{160}(\Gamma_{a})_{\alpha \beta} \boldsymbol{W} \lambda_{j \rho} {G}^{-1} \nabla^{\beta}\,_{\lambda}{\nabla^{\lambda \rho}{G_{i}\,^{j}}}+\frac{3}{80}(\Gamma_{a})_{\alpha}{}^{\beta} \boldsymbol{W} \lambda_{j \beta} {G}^{-1} \nabla_{\rho \lambda}{\nabla^{\rho \lambda}{G_{i}\,^{j}}} - \frac{3}{80}{\rm i} (\Gamma_{a})_{\alpha}{}^{\beta} \boldsymbol{W} \lambda^{\rho}_{j} W_{\beta \rho}\,^{\lambda j} {G}^{-1} \varphi_{i \lambda}+\frac{9}{160}{\rm i} (\Gamma_{a})_{\alpha}{}^{\beta} \boldsymbol{W} \lambda^{\rho}_{j} X^{j}_{\beta} {G}^{-1} \varphi_{i \rho}+\frac{27}{512}{\rm i} (\Gamma_{a})_{\alpha}{}^{\beta} \boldsymbol{W} \lambda^{\rho}_{i} X_{j \beta} {G}^{-1} \varphi^{j}_{\rho} - \frac{27}{1280}{\rm i} (\Gamma_{a})_{\alpha}{}^{\rho} \boldsymbol{W} \lambda_{i \rho} X_{j}^{\beta} {G}^{-1} \varphi^{j}_{\beta} - \frac{117}{2560}{\rm i} (\Gamma_{a})_{\alpha}{}^{\beta} \boldsymbol{W} \lambda^{\rho}_{j} X_{i \beta} {G}^{-1} \varphi^{j}_{\rho} - \frac{27}{640}(\Gamma_{a})_{\alpha}{}^{\lambda} W^{\beta \rho} \lambda_{j \beta} \boldsymbol{\lambda}^{j}_{\lambda} {G}^{-1} \varphi_{i \rho} - \frac{27}{640}(\Gamma_{a})_{\alpha}{}^{\lambda} W^{\beta \rho} \lambda_{j \lambda} \boldsymbol{\lambda}^{j}_{\beta} {G}^{-1} \varphi_{i \rho} - \frac{27}{80}{\rm i} (\Gamma_{a})_{\alpha}{}^{\beta} \boldsymbol{W} W^{\rho \lambda} F_{\beta \rho} {G}^{-1} \varphi_{i \lambda}+\frac{3}{80}{\rm i} G_{j k} (\Gamma_{a})_{\alpha}{}^{\lambda} \boldsymbol{W} W^{\beta \rho} \lambda^{j}_{\beta} {G}^{-3} \varphi_{i \rho} \varphi^{k}_{\lambda}%
+\frac{3}{160}{\rm i} G_{j k} (\Gamma_{a})_{\alpha}{}^{\beta} \boldsymbol{W} W_{\beta}\,^{\rho} \lambda^{j \lambda} {G}^{-3} \varphi_{i \rho} \varphi^{k}_{\lambda}+\frac{27}{160}{\rm i} G_{i j} (\Gamma_{a})_{\alpha}{}^{\beta} \boldsymbol{W} F_{\beta}\,^{\rho} X^{j}_{\rho} {G}^{-1} - \frac{9}{2560}{\rm i} G_{i j} G_{k l} (\Gamma_{a})_{\alpha}{}^{\beta} \boldsymbol{W} \lambda^{k \rho} X^{j}_{\beta} {G}^{-3} \varphi^{l}_{\rho} - \frac{3}{80}(\Gamma_{a})_{\alpha}{}^{\beta} \lambda_{i \rho} \boldsymbol{\lambda}_{j \beta} {G}^{-1} \nabla^{\rho \lambda}{\varphi^{j}_{\lambda}}+\frac{3}{80}{\rm i} G_{j k} (\Gamma_{a})_{\alpha}{}^{\beta} \boldsymbol{W} \lambda_{i \rho} {G}^{-3} \nabla^{\rho \lambda}{\varphi^{j}_{\lambda}} \varphi^{k}_{\beta}+\frac{3}{80}{\rm i} G_{j k} (\Gamma_{a})_{\alpha \beta} \boldsymbol{W} \lambda^{\rho}_{i} {G}^{-3} \nabla^{\beta \lambda}{\varphi^{j}_{\lambda}} \varphi^{k}_{\rho}+\frac{3}{40}{\rm i} (\Gamma_{a})_{\alpha}{}^{\beta} \boldsymbol{W} \lambda^{\rho}_{i} W_{\beta \rho}\,^{\lambda}\,_{j} {G}^{-1} \varphi^{j}_{\lambda}+\frac{21}{320}(\Gamma_{a})_{\alpha}{}^{\lambda} W^{\beta \rho} \lambda_{i \beta} \boldsymbol{\lambda}_{j \lambda} {G}^{-1} \varphi^{j}_{\rho}+\frac{3}{640}(\Gamma_{a})_{\alpha}{}^{\lambda} W^{\beta \rho} \lambda_{i \lambda} \boldsymbol{\lambda}_{j \beta} {G}^{-1} \varphi^{j}_{\rho} - \frac{9}{128}{\rm i} G_{j k} (\Gamma_{a})_{\alpha}{}^{\lambda} \boldsymbol{W} W^{\beta \rho} \lambda_{i \beta} {G}^{-3} \varphi^{j}_{\lambda} \varphi^{k}_{\rho} - \frac{9}{640}{\rm i} G_{j k} (\Gamma_{a})_{\alpha}{}^{\beta} \boldsymbol{W} W_{\beta}\,^{\rho} \lambda^{\lambda}_{i} {G}^{-3} \varphi^{j}_{\rho} \varphi^{k}_{\lambda}+\frac{3}{80}(\Gamma_{a})_{\alpha}{}^{\beta} \lambda_{j \beta} \boldsymbol{\lambda}_{i \rho} {G}^{-1} \nabla^{\rho \lambda}{\varphi^{j}_{\lambda}} - \frac{21}{320}(\Gamma_{a})_{\alpha}{}^{\lambda} W^{\beta \rho} \lambda_{j \lambda} \boldsymbol{\lambda}_{i \beta} {G}^{-1} \varphi^{j}_{\rho} - \frac{3}{640}(\Gamma_{a})_{\alpha}{}^{\lambda} W^{\beta \rho} \lambda_{j \beta} \boldsymbol{\lambda}_{i \lambda} {G}^{-1} \varphi^{j}_{\rho}+\frac{3}{320}F G_{j k} (\Gamma_{a})_{\alpha}{}^{\beta} \lambda^{j}_{\beta} \boldsymbol{\lambda}_{i}^{\rho} {G}^{-3} \varphi^{k}_{\rho} - \frac{3}{320}F G_{j k} (\Gamma_{a})_{\alpha}{}^{\beta} \lambda^{\rho}_{i} \boldsymbol{\lambda}^{j}_{\beta} {G}^{-3} \varphi^{k}_{\rho}+\frac{9}{320}{\rm i} G_{i j} (\Gamma_{a})_{\alpha \beta} \boldsymbol{W} \lambda^{\rho}_{k} {G}^{-3} \nabla^{\beta \lambda}{\varphi^{k}_{\lambda}} \varphi^{j}_{\rho}+\frac{3}{320}{\rm i} G_{i j} (\Gamma_{a})_{\alpha}{}^{\beta} \boldsymbol{W} \lambda_{k \rho} {G}^{-3} \nabla^{\rho \lambda}{\varphi^{k}_{\lambda}} \varphi^{j}_{\beta} - \frac{3}{160}{\rm i} G_{i j} (\Gamma_{a})_{\alpha}{}^{\beta} \boldsymbol{W} W_{\beta}\,^{\rho} \lambda^{\lambda}_{k} {G}^{-3} \varphi^{j}_{\lambda} \varphi^{k}_{\rho}+\frac{3}{160}{\rm i} G_{i j} (\Gamma_{a})_{\alpha}{}^{\lambda} \boldsymbol{W} W^{\beta \rho} \lambda_{k \beta} {G}^{-3} \varphi^{j}_{\lambda} \varphi^{k}_{\rho}%
 - \frac{3}{320}F G_{i j} (\Gamma_{a})_{\alpha}{}^{\beta} \lambda^{\rho}_{k} \boldsymbol{\lambda}^{k}_{\beta} {G}^{-3} \varphi^{j}_{\rho} - \frac{3}{320}F G_{i j} (\Gamma_{a})_{\alpha}{}^{\beta} \lambda_{k \beta} \boldsymbol{\lambda}^{k \rho} {G}^{-3} \varphi^{j}_{\rho} - \frac{9}{160}{\rm i} F G_{i j} (\Gamma_{a})_{\alpha}{}^{\beta} \boldsymbol{W} F_{\beta}\,^{\rho} {G}^{-3} \varphi^{j}_{\rho} - \frac{63}{640}{\rm i} F G_{i j} (\Gamma_{a})_{\alpha}{}^{\beta} W \boldsymbol{W} W_{\beta}\,^{\rho} {G}^{-3} \varphi^{j}_{\rho}+\frac{9}{40}{\rm i} (\Gamma_{a})_{\alpha}{}^{\rho} W \mathbf{F}_{\beta \rho} {G}^{-1} \nabla^{\beta \lambda}{\varphi_{i \lambda}}+\frac{9}{320}{\rm i} G_{j k} (\Gamma_{a})_{\alpha}{}^{\beta} W \boldsymbol{\lambda}^{j}_{\rho} {G}^{-3} \nabla^{\rho \lambda}{\varphi_{i \lambda}} \varphi^{k}_{\beta}+\frac{3}{320}{\rm i} G_{j k} (\Gamma_{a})_{\alpha \beta} W \boldsymbol{\lambda}^{j \rho} {G}^{-3} \nabla^{\beta \lambda}{\varphi_{i \lambda}} \varphi^{k}_{\rho}+\frac{9}{320}(\Gamma_{a})_{\alpha}{}^{\beta} W \boldsymbol{\lambda}_{i \lambda} {G}^{-1} \nabla^{\lambda \rho}{\mathcal{H}_{\rho \beta}}+\frac{9}{320}(\Gamma_{a})_{\alpha \beta} W \boldsymbol{\lambda}_{i}^{\lambda} {G}^{-1} \nabla^{\beta \rho}{\mathcal{H}_{\rho \lambda}} - \frac{3}{160}(\Gamma_{a})_{\alpha \beta} W \boldsymbol{\lambda}_{j \rho} {G}^{-1} \nabla_{\lambda}\,^{\rho}{\nabla^{\beta \lambda}{G_{i}\,^{j}}}+\frac{3}{160}(\Gamma_{a})_{\alpha \beta} W \boldsymbol{\lambda}_{j \rho} {G}^{-1} \nabla^{\beta}\,_{\lambda}{\nabla^{\lambda \rho}{G_{i}\,^{j}}}+\frac{3}{80}(\Gamma_{a})_{\alpha}{}^{\beta} W \boldsymbol{\lambda}_{j \beta} {G}^{-1} \nabla_{\rho \lambda}{\nabla^{\rho \lambda}{G_{i}\,^{j}}} - \frac{3}{80}{\rm i} (\Gamma_{a})_{\alpha}{}^{\beta} W \boldsymbol{\lambda}_{j}^{\rho} W_{\beta \rho}\,^{\lambda j} {G}^{-1} \varphi_{i \lambda}+\frac{9}{160}{\rm i} (\Gamma_{a})_{\alpha}{}^{\beta} W \boldsymbol{\lambda}_{j}^{\rho} X^{j}_{\beta} {G}^{-1} \varphi_{i \rho}+\frac{27}{512}{\rm i} (\Gamma_{a})_{\alpha}{}^{\beta} W \boldsymbol{\lambda}_{i}^{\rho} X_{j \beta} {G}^{-1} \varphi^{j}_{\rho} - \frac{27}{1280}{\rm i} (\Gamma_{a})_{\alpha}{}^{\rho} W \boldsymbol{\lambda}_{i \rho} X_{j}^{\beta} {G}^{-1} \varphi^{j}_{\beta} - \frac{117}{2560}{\rm i} (\Gamma_{a})_{\alpha}{}^{\beta} W \boldsymbol{\lambda}_{j}^{\rho} X_{i \beta} {G}^{-1} \varphi^{j}_{\rho} - \frac{27}{80}{\rm i} (\Gamma_{a})_{\alpha}{}^{\beta} W W^{\rho \lambda} \mathbf{F}_{\beta \rho} {G}^{-1} \varphi_{i \lambda}+\frac{3}{80}{\rm i} G_{j k} (\Gamma_{a})_{\alpha}{}^{\lambda} W W^{\beta \rho} \boldsymbol{\lambda}^{j}_{\beta} {G}^{-3} \varphi_{i \rho} \varphi^{k}_{\lambda}+\frac{3}{160}{\rm i} G_{j k} (\Gamma_{a})_{\alpha}{}^{\beta} W W_{\beta}\,^{\rho} \boldsymbol{\lambda}^{j \lambda} {G}^{-3} \varphi_{i \rho} \varphi^{k}_{\lambda}%
+\frac{27}{160}{\rm i} G_{i j} (\Gamma_{a})_{\alpha}{}^{\beta} W \mathbf{F}_{\beta}\,^{\rho} X^{j}_{\rho} {G}^{-1} - \frac{9}{2560}{\rm i} G_{i j} G_{k l} (\Gamma_{a})_{\alpha}{}^{\beta} W \boldsymbol{\lambda}^{k \rho} X^{j}_{\beta} {G}^{-3} \varphi^{l}_{\rho}+\frac{3}{80}{\rm i} G_{j k} (\Gamma_{a})_{\alpha}{}^{\beta} W \boldsymbol{\lambda}_{i \rho} {G}^{-3} \nabla^{\rho \lambda}{\varphi^{j}_{\lambda}} \varphi^{k}_{\beta}+\frac{3}{80}{\rm i} G_{j k} (\Gamma_{a})_{\alpha \beta} W \boldsymbol{\lambda}_{i}^{\rho} {G}^{-3} \nabla^{\beta \lambda}{\varphi^{j}_{\lambda}} \varphi^{k}_{\rho}+\frac{3}{40}{\rm i} (\Gamma_{a})_{\alpha}{}^{\beta} W \boldsymbol{\lambda}_{i}^{\rho} W_{\beta \rho}\,^{\lambda}\,_{j} {G}^{-1} \varphi^{j}_{\lambda} - \frac{9}{128}{\rm i} G_{j k} (\Gamma_{a})_{\alpha}{}^{\lambda} W W^{\beta \rho} \boldsymbol{\lambda}_{i \beta} {G}^{-3} \varphi^{j}_{\lambda} \varphi^{k}_{\rho} - \frac{9}{640}{\rm i} G_{j k} (\Gamma_{a})_{\alpha}{}^{\beta} W W_{\beta}\,^{\rho} \boldsymbol{\lambda}_{i}^{\lambda} {G}^{-3} \varphi^{j}_{\rho} \varphi^{k}_{\lambda}+\frac{9}{320}{\rm i} G_{i j} (\Gamma_{a})_{\alpha \beta} W \boldsymbol{\lambda}_{k}^{\rho} {G}^{-3} \nabla^{\beta \lambda}{\varphi^{k}_{\lambda}} \varphi^{j}_{\rho}+\frac{3}{320}{\rm i} G_{i j} (\Gamma_{a})_{\alpha}{}^{\beta} W \boldsymbol{\lambda}_{k \rho} {G}^{-3} \nabla^{\rho \lambda}{\varphi^{k}_{\lambda}} \varphi^{j}_{\beta} - \frac{3}{160}{\rm i} G_{i j} (\Gamma_{a})_{\alpha}{}^{\beta} W W_{\beta}\,^{\rho} \boldsymbol{\lambda}_{k}^{\lambda} {G}^{-3} \varphi^{j}_{\lambda} \varphi^{k}_{\rho}+\frac{3}{160}{\rm i} G_{i j} (\Gamma_{a})_{\alpha}{}^{\lambda} W W^{\beta \rho} \boldsymbol{\lambda}_{k \beta} {G}^{-3} \varphi^{j}_{\lambda} \varphi^{k}_{\rho} - \frac{9}{160}{\rm i} F G_{i j} (\Gamma_{a})_{\alpha}{}^{\beta} W \mathbf{F}_{\beta}\,^{\rho} {G}^{-3} \varphi^{j}_{\rho}+\frac{3}{320}(\Gamma_{a})_{\alpha}{}^{\beta} W \boldsymbol{W} {G}^{-3} \nabla^{\rho \lambda}{\varphi_{j \rho}} \varphi_{i \lambda} \varphi^{j}_{\beta}+\frac{27}{320}G_{i j} G_{k l} (\Gamma_{a})_{\alpha \beta} W \boldsymbol{W} {G}^{-5} \nabla^{\beta \rho}{\varphi^{k}_{\rho}} \varphi^{j \lambda} \varphi^{l}_{\lambda}+\frac{9}{320}G_{i j} G_{k l} (\Gamma_{a})_{\alpha}{}^{\beta} W \boldsymbol{W} {G}^{-5} \nabla^{\rho \lambda}{\varphi^{k}_{\rho}} \varphi^{j}_{\beta} \varphi^{l}_{\lambda} - \frac{9}{160}{\rm i} G_{i j} (\Gamma_{a})_{\alpha}{}^{\beta} W \boldsymbol{W} {G}^{-3} \nabla^{\rho \lambda}{\mathcal{H}_{\rho \beta}} \varphi^{j}_{\lambda} - \frac{9}{160}{\rm i} G_{i j} (\Gamma_{a})_{\alpha \beta} W \boldsymbol{W} {G}^{-3} \nabla^{\beta \rho}{\mathcal{H}_{\rho}\,^{\lambda}} \varphi^{j}_{\lambda}+\frac{9}{320}{\rm i} G_{i j} (\Gamma_{a})_{\alpha}{}^{\beta} W \boldsymbol{W} {G}^{-3} \nabla^{\rho \lambda}{\mathcal{H}_{\rho \lambda}} \varphi^{j}_{\beta}+\frac{3}{40}{\rm i} G_{j k} (\Gamma_{a})_{\alpha \beta} W \boldsymbol{W} {G}^{-3} \nabla^{\beta}\,_{\rho}{\nabla^{\rho \lambda}{G_{i}\,^{j}}} \varphi^{k}_{\lambda}+\frac{3}{64}{\rm i} G_{j k} (\Gamma_{a})_{\alpha}{}^{\beta} W \boldsymbol{W} {G}^{-3} \nabla_{\rho \lambda}{\nabla^{\rho \lambda}{G_{i}\,^{j}}} \varphi^{k}_{\beta}%
 - \frac{9}{160}{\rm i} (\Gamma_{a})_{\alpha}{}^{\beta} W \boldsymbol{W} {G}^{-1} \nabla_{\rho \lambda}{\nabla^{\rho \lambda}{\varphi_{i \beta}}} - \frac{27}{80}{\rm i} (\Gamma_{a})_{\alpha \beta} W \boldsymbol{W} {G}^{-1} \nabla^{\beta}\,_{\rho}{\nabla^{\rho \lambda}{\varphi_{i \lambda}}} - \frac{9}{80}{\rm i} (\Gamma_{a})_{\alpha \beta} W \boldsymbol{W} {G}^{-1} \nabla_{\rho}\,^{\lambda}{\nabla^{\beta \rho}{\varphi_{i \lambda}}} - \frac{9}{1280}{\rm i} G_{i j} G_{k l} (\Gamma_{a})_{\alpha}{}^{\beta} \boldsymbol{W} \lambda^{k \rho} X^{l}_{\beta} {G}^{-3} \varphi^{j}_{\rho} - \frac{9}{2560}{\rm i} G_{i j} G_{k l} (\Gamma_{a})_{\alpha}{}^{\rho} \boldsymbol{W} \lambda^{k \beta} X^{l}_{\beta} {G}^{-3} \varphi^{j}_{\rho} - \frac{9}{1280}{\rm i} G_{i j} G_{k l} (\Gamma_{a})_{\alpha}{}^{\beta} W \boldsymbol{\lambda}^{k \rho} X^{l}_{\beta} {G}^{-3} \varphi^{j}_{\rho} - \frac{9}{2560}{\rm i} G_{i j} G_{k l} (\Gamma_{a})_{\alpha}{}^{\rho} W \boldsymbol{\lambda}^{k \beta} X^{l}_{\beta} {G}^{-3} \varphi^{j}_{\rho} - \frac{3}{160}{\rm i} (\Gamma_{a})_{\alpha}{}^{\beta} \lambda^{\rho}_{j} \boldsymbol{\lambda}^{j}_{\beta} {G}^{-3} \varphi_{i}^{\lambda} \varphi_{k \rho} \varphi^{k}_{\lambda} - \frac{3}{160}{\rm i} (\Gamma_{a})_{\alpha}{}^{\beta} \lambda_{j \beta} \boldsymbol{\lambda}^{j \rho} {G}^{-3} \varphi_{i}^{\lambda} \varphi_{k \rho} \varphi^{k}_{\lambda}+\frac{9}{80}(\Gamma_{a})_{\alpha}{}^{\beta} \boldsymbol{W} F_{\beta}\,^{\rho} {G}^{-3} \varphi_{i}^{\lambda} \varphi_{j \rho} \varphi^{j}_{\lambda}+\frac{9}{160}G_{j k} (\Gamma_{a})_{\alpha}{}^{\beta} \boldsymbol{W} \lambda^{j \rho} {G}^{-5} \varphi_{i}^{\lambda} \varphi^{k}_{\beta} \varphi_{l \rho} \varphi^{l}_{\lambda}+\frac{9}{160}G_{j k} (\Gamma_{a})_{\alpha}{}^{\beta} \boldsymbol{W} \lambda^{j}_{\beta} {G}^{-5} \varphi_{i}^{\rho} \varphi^{k \lambda} \varphi_{l \rho} \varphi^{l}_{\lambda} - \frac{3}{320}{\rm i} \mathcal{H}^{\beta \rho} (\Gamma_{a})_{\alpha \beta} \boldsymbol{W} \lambda^{\lambda}_{i} {G}^{-3} \varphi_{j \rho} \varphi^{j}_{\lambda} - \frac{3}{320}{\rm i} \mathcal{H}^{\rho \lambda} (\Gamma_{a})_{\alpha}{}^{\beta} \boldsymbol{W} \lambda_{i \rho} {G}^{-3} \varphi_{j \lambda} \varphi^{j}_{\beta}+\frac{3}{160}{\rm i} (\Gamma_{a})_{\alpha \beta} \boldsymbol{W} \lambda^{\rho}_{j} {G}^{-3} \nabla^{\beta \lambda}{G_{i}\,^{j}} \varphi_{k \rho} \varphi^{k}_{\lambda} - \frac{3}{160}{\rm i} (\Gamma_{a})_{\alpha}{}^{\beta} \boldsymbol{W} \lambda_{j \rho} {G}^{-3} \nabla^{\rho \lambda}{G_{i}\,^{j}} \varphi_{k \beta} \varphi^{k}_{\lambda} - \frac{3}{320}{\rm i} \mathcal{H}^{\beta \rho} (\Gamma_{a})_{\alpha \beta} \boldsymbol{W} \lambda_{j \rho} {G}^{-3} \varphi_{i}^{\lambda} \varphi^{j}_{\lambda}+\frac{3}{320}{\rm i} (\Gamma_{a})_{\alpha \beta} \boldsymbol{W} \lambda_{j \rho} {G}^{-3} \nabla^{\beta \rho}{G^{j}\,_{k}} \varphi_{i}^{\lambda} \varphi^{k}_{\lambda} - \frac{3}{320}{\rm i} \mathcal{H}^{\beta \rho} (\Gamma_{a})_{\alpha \beta} \boldsymbol{W} \lambda^{\lambda}_{j} {G}^{-3} \varphi_{i \rho} \varphi^{j}_{\lambda} - \frac{3}{160}{\rm i} \mathcal{H}^{\rho \lambda} (\Gamma_{a})_{\alpha}{}^{\beta} \boldsymbol{W} \lambda_{j \beta} {G}^{-3} \varphi_{i \rho} \varphi^{j}_{\lambda}%
+\frac{3}{160}{\rm i} (\Gamma_{a})_{\alpha \beta} \boldsymbol{W} \lambda^{\rho}_{j} {G}^{-3} \nabla^{\beta \lambda}{G^{j}\,_{k}} \varphi_{i \lambda} \varphi^{k}_{\rho}+\frac{3}{160}{\rm i} (\Gamma_{a})_{\alpha}{}^{\beta} \boldsymbol{W} \lambda_{j \beta} {G}^{-3} \nabla^{\rho \lambda}{G^{j}\,_{k}} \varphi_{i \rho} \varphi^{k}_{\lambda}+\frac{3}{160}{\rm i} (\Gamma_{a})_{\alpha}{}^{\beta} \lambda^{\rho}_{i} \boldsymbol{\lambda}_{j \beta} {G}^{-3} \varphi^{j \lambda} \varphi_{k \rho} \varphi^{k}_{\lambda} - \frac{9}{160}G_{j k} (\Gamma_{a})_{\alpha}{}^{\beta} \boldsymbol{W} \lambda^{\rho}_{i} {G}^{-5} \varphi^{j}_{\beta} \varphi^{k \lambda} \varphi_{l \rho} \varphi^{l}_{\lambda} - \frac{9}{160}G_{j k} (\Gamma_{a})_{\alpha}{}^{\beta} \boldsymbol{W} \lambda^{\rho}_{i} {G}^{-5} \varphi^{j}_{\rho} \varphi^{k \lambda} \varphi_{l \beta} \varphi^{l}_{\lambda}+\frac{3}{160}{\rm i} (\Gamma_{a})_{\alpha \beta} \boldsymbol{W} \lambda^{\rho}_{i} {G}^{-3} \nabla^{\beta \lambda}{G_{j k}} \varphi^{j}_{\rho} \varphi^{k}_{\lambda}+\frac{3}{160}{\rm i} (\Gamma_{a})_{\alpha}{}^{\beta} \boldsymbol{W} \lambda_{i \rho} {G}^{-3} \nabla^{\rho \lambda}{G_{j k}} \varphi^{j}_{\beta} \varphi^{k}_{\lambda} - \frac{3}{160}{\rm i} (\Gamma_{a})_{\alpha}{}^{\beta} \lambda_{j \beta} \boldsymbol{\lambda}_{i}^{\rho} {G}^{-3} \varphi^{j \lambda} \varphi_{k \rho} \varphi^{k}_{\lambda}+\frac{3}{160}{\rm i} G_{j k} (\Gamma_{a})_{\alpha}{}^{\beta} F_{\beta}\,^{\rho} \boldsymbol{\lambda}_{i \rho} {G}^{-3} \varphi^{j \lambda} \varphi^{k}_{\lambda}+\frac{21}{640}{\rm i} G_{j k} (\Gamma_{a})_{\alpha}{}^{\beta} W W_{\beta}\,^{\rho} \boldsymbol{\lambda}_{i \rho} {G}^{-3} \varphi^{j \lambda} \varphi^{k}_{\lambda}+\frac{3}{160}{\rm i} G_{j k} (\Gamma_{a})_{\alpha}{}^{\beta} \mathbf{F}_{\beta}\,^{\rho} \lambda_{i \rho} {G}^{-3} \varphi^{j \lambda} \varphi^{k}_{\lambda}+\frac{21}{640}{\rm i} G_{j k} (\Gamma_{a})_{\alpha}{}^{\beta} \boldsymbol{W} W_{\beta}\,^{\rho} \lambda_{i \rho} {G}^{-3} \varphi^{j \lambda} \varphi^{k}_{\lambda}+\frac{9}{320}{\rm i} G_{j k} G_{l m} (\Gamma_{a})_{\alpha}{}^{\beta} \lambda^{j \rho} \boldsymbol{\lambda}_{i \rho} {G}^{-5} \varphi^{k}_{\beta} \varphi^{l \lambda} \varphi^{m}_{\lambda} - \frac{9}{320}{\rm i} G_{j k} G_{l m} (\Gamma_{a})_{\alpha}{}^{\beta} \lambda^{j}_{\beta} \boldsymbol{\lambda}_{i}^{\rho} {G}^{-5} \varphi^{k}_{\rho} \varphi^{l \lambda} \varphi^{m}_{\lambda} - \frac{3}{160}\mathcal{H}^{\beta \rho} G_{j k} (\Gamma_{a})_{\alpha \beta} \lambda^{j \lambda} \boldsymbol{\lambda}_{i \lambda} {G}^{-3} \varphi^{k}_{\rho}+\frac{3}{320}\mathcal{H}^{\rho \lambda} G_{j k} (\Gamma_{a})_{\alpha}{}^{\beta} \lambda^{j}_{\beta} \boldsymbol{\lambda}_{i \rho} {G}^{-3} \varphi^{k}_{\lambda}+\frac{3}{320}G_{j k} (\Gamma_{a})_{\alpha \beta} \lambda^{\rho}_{l} \boldsymbol{\lambda}_{i \rho} {G}^{-3} \nabla^{\beta \lambda}{G^{j l}} \varphi^{k}_{\lambda} - \frac{3}{160}G_{j k} (\Gamma_{a})_{\alpha}{}^{\beta} \lambda_{l \beta} \boldsymbol{\lambda}_{i \rho} {G}^{-3} \nabla^{\rho \lambda}{G^{j l}} \varphi^{k}_{\lambda}+\frac{9}{320}{\rm i} G_{j k} G_{l m} (\Gamma_{a})_{\alpha}{}^{\beta} \lambda^{\rho}_{i} \boldsymbol{\lambda}^{j}_{\rho} {G}^{-5} \varphi^{k}_{\beta} \varphi^{l \lambda} \varphi^{m}_{\lambda}+\frac{9}{320}{\rm i} G_{j k} G_{l m} (\Gamma_{a})_{\alpha}{}^{\beta} \lambda^{\rho}_{i} \boldsymbol{\lambda}^{j}_{\beta} {G}^{-5} \varphi^{k}_{\rho} \varphi^{l \lambda} \varphi^{m}_{\lambda}%
 - \frac{3}{160}\mathcal{H}^{\beta \rho} G_{j k} (\Gamma_{a})_{\alpha \beta} \lambda^{\lambda}_{i} \boldsymbol{\lambda}^{j}_{\lambda} {G}^{-3} \varphi^{k}_{\rho} - \frac{3}{320}\mathcal{H}^{\rho \lambda} G_{j k} (\Gamma_{a})_{\alpha}{}^{\beta} \lambda_{i \rho} \boldsymbol{\lambda}^{j}_{\beta} {G}^{-3} \varphi^{k}_{\lambda}+\frac{3}{320}G_{j k} (\Gamma_{a})_{\alpha \beta} \lambda^{\rho}_{i} \boldsymbol{\lambda}_{l \rho} {G}^{-3} \nabla^{\beta \lambda}{G^{j l}} \varphi^{k}_{\lambda}+\frac{3}{160}G_{j k} (\Gamma_{a})_{\alpha}{}^{\beta} \lambda_{i \rho} \boldsymbol{\lambda}_{l \beta} {G}^{-3} \nabla^{\rho \lambda}{G^{j l}} \varphi^{k}_{\lambda}+\frac{9}{320}G_{j k} (\Gamma_{a})_{\alpha}{}^{\beta} \boldsymbol{W} \lambda^{\rho}_{i} {G}^{-5} \varphi^{j \lambda} \varphi^{k}_{\lambda} \varphi_{l \beta} \varphi^{l}_{\rho} - \frac{9}{320}G_{j k} (\Gamma_{a})_{\alpha}{}^{\beta} \boldsymbol{W} \lambda^{\rho}_{l} {G}^{-5} \varphi_{i \beta} \varphi^{j \lambda} \varphi^{k}_{\lambda} \varphi^{l}_{\rho} - \frac{9}{320}G_{j k} (\Gamma_{a})_{\alpha}{}^{\beta} \boldsymbol{W} \lambda_{l \beta} {G}^{-5} \varphi_{i}^{\rho} \varphi^{j \lambda} \varphi^{k}_{\lambda} \varphi^{l}_{\rho} - \frac{9}{160}G_{i j} (\Gamma_{a})_{\alpha}{}^{\beta} \boldsymbol{W} \lambda^{\rho}_{k} {G}^{-5} \varphi^{j}_{\rho} \varphi^{k \lambda} \varphi_{l \beta} \varphi^{l}_{\lambda}+\frac{9}{160}G_{i j} (\Gamma_{a})_{\alpha}{}^{\beta} \boldsymbol{W} \lambda_{k \beta} {G}^{-5} \varphi^{j \rho} \varphi^{k \lambda} \varphi_{l \rho} \varphi^{l}_{\lambda}+\frac{9}{320}{\rm i} G_{i j} G_{k l} (\Gamma_{a})_{\alpha}{}^{\beta} \lambda^{\rho}_{m} \boldsymbol{\lambda}^{m}_{\beta} {G}^{-5} \varphi^{j}_{\rho} \varphi^{k \lambda} \varphi^{l}_{\lambda}+\frac{9}{320}{\rm i} G_{i j} G_{k l} (\Gamma_{a})_{\alpha}{}^{\beta} \lambda_{m \beta} \boldsymbol{\lambda}^{m \rho} {G}^{-5} \varphi^{j}_{\rho} \varphi^{k \lambda} \varphi^{l}_{\lambda} - \frac{27}{160}G_{i j} G_{k l} (\Gamma_{a})_{\alpha}{}^{\beta} \boldsymbol{W} F_{\beta}\,^{\rho} {G}^{-5} \varphi^{j}_{\rho} \varphi^{k \lambda} \varphi^{l}_{\lambda} - \frac{9}{64}G_{i j} G_{k l} G_{m n} (\Gamma_{a})_{\alpha}{}^{\beta} \boldsymbol{W} \lambda^{k \rho} {G}^{-7} \varphi^{j}_{\rho} \varphi^{l}_{\beta} \varphi^{m \lambda} \varphi^{n}_{\lambda}+\frac{9}{64}G_{i j} G_{k l} G_{m n} (\Gamma_{a})_{\alpha}{}^{\beta} \boldsymbol{W} \lambda^{k}_{\beta} {G}^{-7} \varphi^{j \rho} \varphi^{l}_{\rho} \varphi^{m \lambda} \varphi^{n}_{\lambda} - \frac{9}{640}{\rm i} \mathcal{H}^{\beta \rho} G_{i j} G_{k l} (\Gamma_{a})_{\alpha \beta} \boldsymbol{W} \lambda^{j}_{\rho} {G}^{-5} \varphi^{k \lambda} \varphi^{l}_{\lambda}+\frac{9}{640}{\rm i} G_{i j} G_{k l} (\Gamma_{a})_{\alpha \beta} \boldsymbol{W} \lambda_{m \rho} {G}^{-5} \nabla^{\beta \rho}{G^{j m}} \varphi^{k \lambda} \varphi^{l}_{\lambda} - \frac{9}{320}{\rm i} \mathcal{H}^{\beta \rho} G_{i j} G_{k l} (\Gamma_{a})_{\alpha \beta} \boldsymbol{W} \lambda^{k \lambda} {G}^{-5} \varphi^{j}_{\lambda} \varphi^{l}_{\rho}+\frac{9}{160}{\rm i} \mathcal{H}^{\rho \lambda} G_{i j} G_{k l} (\Gamma_{a})_{\alpha}{}^{\beta} \boldsymbol{W} \lambda^{k}_{\beta} {G}^{-5} \varphi^{j}_{\rho} \varphi^{l}_{\lambda}+\frac{9}{160}{\rm i} G_{i j} G_{k l} (\Gamma_{a})_{\alpha \beta} \boldsymbol{W} \lambda^{\rho}_{m} {G}^{-5} \nabla^{\beta \lambda}{G^{k m}} \varphi^{j}_{\rho} \varphi^{l}_{\lambda} - \frac{9}{160}{\rm i} G_{i j} G_{k l} (\Gamma_{a})_{\alpha}{}^{\beta} \boldsymbol{W} \lambda_{m \beta} {G}^{-5} \nabla^{\rho \lambda}{G^{k m}} \varphi^{j}_{\rho} \varphi^{l}_{\lambda}%
+\frac{9}{160}{\rm i} G_{j k} G_{l m} (\Gamma_{a})_{\alpha \beta} \boldsymbol{W} \lambda^{\rho}_{i} {G}^{-5} \nabla^{\beta \lambda}{G^{j l}} \varphi^{k}_{\rho} \varphi^{m}_{\lambda}+\frac{9}{160}{\rm i} G_{j k} G_{l m} (\Gamma_{a})_{\alpha}{}^{\beta} \boldsymbol{W} \lambda_{i \rho} {G}^{-5} \nabla^{\rho \lambda}{G^{j l}} \varphi^{k}_{\beta} \varphi^{m}_{\lambda} - \frac{3}{320}{\rm i} G_{i j} (\Gamma_{a})_{\alpha \beta} \boldsymbol{W} \lambda_{k \rho} {G}^{-3} \nabla^{\beta \rho}{\varphi^{k \lambda}} \varphi^{j}_{\lambda} - \frac{3}{160}{\rm i} G_{i j} (\Gamma_{a})_{\alpha \beta} \boldsymbol{W} \lambda^{\rho}_{k} {G}^{-3} \nabla^{\beta \lambda}{\varphi^{k}_{\rho}} \varphi^{j}_{\lambda} - \frac{3}{64}{\rm i} G_{i j} (\Gamma_{a})_{\alpha}{}^{\beta} \boldsymbol{W} W_{\beta}\,^{\rho} \lambda_{k \rho} {G}^{-3} \varphi^{j \lambda} \varphi^{k}_{\lambda} - \frac{3}{640}{\rm i} G_{i j} (\Gamma_{a})_{\alpha}{}^{\beta} \boldsymbol{W} W_{\beta}\,^{\rho} \lambda^{\lambda}_{k} {G}^{-3} \varphi^{j}_{\rho} \varphi^{k}_{\lambda}+\frac{39}{640}{\rm i} G_{i j} (\Gamma_{a})_{\alpha}{}^{\lambda} \boldsymbol{W} W^{\beta \rho} \lambda_{k \beta} {G}^{-3} \varphi^{j}_{\rho} \varphi^{k}_{\lambda} - \frac{3}{320}{\rm i} \mathcal{H}^{\rho \lambda} (\Gamma_{a})_{\alpha}{}^{\beta} \boldsymbol{W} \lambda_{j \rho} {G}^{-3} \varphi_{i \beta} \varphi^{j}_{\lambda} - \frac{3}{320}\mathcal{H}^{\rho \lambda} G_{i j} (\Gamma_{a})_{\alpha}{}^{\beta} \lambda_{k \rho} \boldsymbol{\lambda}^{k}_{\beta} {G}^{-3} \varphi^{j}_{\lambda} - \frac{3}{320}\mathcal{H}^{\rho \lambda} G_{i j} (\Gamma_{a})_{\alpha}{}^{\beta} \lambda_{k \beta} \boldsymbol{\lambda}^{k}_{\rho} {G}^{-3} \varphi^{j}_{\lambda} - \frac{9}{160}{\rm i} \mathcal{H}^{\beta \lambda} G_{i j} (\Gamma_{a})_{\alpha}{}^{\rho} \boldsymbol{W} F_{\beta \rho} {G}^{-3} \varphi^{j}_{\lambda} - \frac{9}{320}{\rm i} \mathcal{H}^{\rho \lambda} G_{i j} G_{k l} (\Gamma_{a})_{\alpha}{}^{\beta} \boldsymbol{W} \lambda^{k}_{\rho} {G}^{-5} \varphi^{j}_{\lambda} \varphi^{l}_{\beta}+\frac{3}{640}\mathcal{H}^{\rho \beta} \mathcal{H}_{\rho}\,^{\lambda} G_{i j} (\Gamma_{a})_{\alpha \beta} \boldsymbol{W} \lambda^{j}_{\lambda} {G}^{-3}+\frac{3}{640}\mathcal{H}^{\rho \lambda} \mathcal{H}_{\rho \lambda} G_{i j} (\Gamma_{a})_{\alpha}{}^{\beta} \boldsymbol{W} \lambda^{j}_{\beta} {G}^{-3} - \frac{3}{320}\mathcal{H}_{\rho}\,^{\lambda} G_{i j} (\Gamma_{a})_{\alpha \beta} \boldsymbol{W} \lambda_{k \lambda} {G}^{-3} \nabla^{\rho \beta}{G^{j k}} - \frac{3}{320}\mathcal{H}_{\rho \lambda} G_{i j} (\Gamma_{a})_{\alpha}{}^{\beta} \boldsymbol{W} \lambda_{k \beta} {G}^{-3} \nabla^{\rho \lambda}{G^{j k}} - \frac{3}{160}{\rm i} (\Gamma_{a})_{\alpha}{}^{\beta} \boldsymbol{W} \lambda_{k \rho} {G}^{-3} \nabla^{\rho \lambda}{G_{i j}} \varphi^{k}_{\lambda} \varphi^{j}_{\beta}+\frac{3}{160}{\rm i} (\Gamma_{a})_{\alpha}{}^{\beta} \boldsymbol{W} \lambda_{k \beta} {G}^{-3} \nabla^{\rho \lambda}{G_{i j}} \varphi^{k}_{\rho} \varphi^{j}_{\lambda} - \frac{3}{160}G_{j k} (\Gamma_{a})_{\alpha}{}^{\beta} \lambda_{l \rho} \boldsymbol{\lambda}^{l}_{\beta} {G}^{-3} \nabla^{\rho \lambda}{G_{i}\,^{j}} \varphi^{k}_{\lambda} - \frac{3}{160}G_{j k} (\Gamma_{a})_{\alpha}{}^{\beta} \lambda_{l \beta} \boldsymbol{\lambda}^{l}_{\rho} {G}^{-3} \nabla^{\rho \lambda}{G_{i}\,^{j}} \varphi^{k}_{\lambda}%
 - \frac{9}{80}{\rm i} G_{j k} (\Gamma_{a})_{\alpha}{}^{\beta} \boldsymbol{W} F_{\beta \rho} {G}^{-3} \nabla^{\rho \lambda}{G_{i}\,^{j}} \varphi^{k}_{\lambda}+\frac{9}{160}{\rm i} G_{j k} G_{l m} (\Gamma_{a})_{\alpha}{}^{\beta} \boldsymbol{W} \lambda^{j}_{\rho} {G}^{-5} \nabla^{\rho \lambda}{G_{i}\,^{l}} \varphi^{k}_{\beta} \varphi^{m}_{\lambda}+\frac{9}{160}{\rm i} G_{j k} G_{l m} (\Gamma_{a})_{\alpha}{}^{\beta} \boldsymbol{W} \lambda^{j}_{\beta} {G}^{-5} \nabla^{\rho \lambda}{G_{i}\,^{l}} \varphi^{k}_{\rho} \varphi^{m}_{\lambda} - \frac{3}{64}{\rm i} G_{j k} (\Gamma_{a})_{\alpha}{}^{\beta} \boldsymbol{W} \lambda_{i \rho} {G}^{-3} \nabla^{\rho \lambda}{\varphi^{j}_{\beta}} \varphi^{k}_{\lambda} - \frac{3}{64}{\rm i} G_{j k} (\Gamma_{a})_{\alpha \beta} \boldsymbol{W} \lambda^{\rho}_{i} {G}^{-3} \nabla^{\beta \lambda}{\varphi^{j}_{\rho}} \varphi^{k}_{\lambda} - \frac{3}{160}{\rm i} G_{j k} (\Gamma_{a})_{\alpha}{}^{\beta} \boldsymbol{W} \lambda^{j}_{\rho} {G}^{-3} \nabla^{\rho \lambda}{\varphi_{i \beta}} \varphi^{k}_{\lambda}+\frac{3}{640}{\rm i} G_{j k} (\Gamma_{a})_{\alpha}{}^{\beta} \boldsymbol{W} W_{\beta}\,^{\rho} \lambda^{j \lambda} {G}^{-3} \varphi_{i \lambda} \varphi^{k}_{\rho}+\frac{69}{640}{\rm i} G_{j k} (\Gamma_{a})_{\alpha}{}^{\lambda} \boldsymbol{W} W^{\beta \rho} \lambda^{j}_{\beta} {G}^{-3} \varphi_{i \lambda} \varphi^{k}_{\rho}+\frac{21}{320}{\rm i} G_{j k} (\Gamma_{a})_{\alpha}{}^{\beta} \boldsymbol{W} W_{\beta}\,^{\rho} \lambda^{j}_{\rho} {G}^{-3} \varphi_{i}^{\lambda} \varphi^{k}_{\lambda}+\frac{27}{2560}{\rm i} G_{i j} G_{k l} (\Gamma_{a})_{\alpha}{}^{\beta} \boldsymbol{W} \lambda^{j \rho} X^{k}_{\beta} {G}^{-3} \varphi^{l}_{\rho} - \frac{3}{320}\mathcal{H}_{\rho}\,^{\beta} G_{j k} (\Gamma_{a})_{\alpha \beta} \boldsymbol{W} \lambda^{j}_{\lambda} {G}^{-3} \nabla^{\rho \lambda}{G_{i}\,^{k}}+\frac{3}{320}\mathcal{H}_{\rho \lambda} G_{j k} (\Gamma_{a})_{\alpha}{}^{\beta} \boldsymbol{W} \lambda^{j}_{\beta} {G}^{-3} \nabla^{\rho \lambda}{G_{i}\,^{k}} - \frac{3}{160}G_{j k} (\Gamma_{a})_{\alpha \beta} \boldsymbol{W} \lambda_{l \rho} {G}^{-3} \nabla^{\beta}\,_{\lambda}{G^{j l}} \nabla^{\lambda \rho}{G_{i}\,^{k}} - \frac{3}{160}G_{j k} (\Gamma_{a})_{\alpha}{}^{\beta} \boldsymbol{W} \lambda_{l \beta} {G}^{-3} \nabla_{\rho \lambda}{G_{i}\,^{j}} \nabla^{\rho \lambda}{G^{k l}} - \frac{3}{320}\mathcal{H}_{\rho}\,^{\lambda} G_{j k} (\Gamma_{a})_{\alpha \beta} \boldsymbol{W} \lambda_{i \lambda} {G}^{-3} \nabla^{\rho \beta}{G^{j k}} - \frac{3}{320}\mathcal{H}_{\rho}\,^{\beta} G_{j k} (\Gamma_{a})_{\alpha \beta} \boldsymbol{W} \lambda_{i \lambda} {G}^{-3} \nabla^{\rho \lambda}{G^{j k}}+\frac{3}{160}G_{j k} (\Gamma_{a})_{\alpha \beta} \boldsymbol{W} \lambda_{i \rho} {G}^{-3} \nabla^{\beta}\,_{\lambda}{G^{j}\,_{l}} \nabla^{\lambda \rho}{G^{k l}}+\frac{9}{80}(\Gamma_{a})_{\alpha}{}^{\beta} W \mathbf{F}_{\beta}\,^{\rho} {G}^{-3} \varphi_{i}^{\lambda} \varphi_{j \rho} \varphi^{j}_{\lambda}+\frac{9}{160}G_{j k} (\Gamma_{a})_{\alpha}{}^{\beta} W \boldsymbol{\lambda}^{j \rho} {G}^{-5} \varphi_{i}^{\lambda} \varphi^{k}_{\beta} \varphi_{l \rho} \varphi^{l}_{\lambda}+\frac{9}{160}G_{j k} (\Gamma_{a})_{\alpha}{}^{\beta} W \boldsymbol{\lambda}^{j}_{\beta} {G}^{-5} \varphi_{i}^{\rho} \varphi^{k \lambda} \varphi_{l \rho} \varphi^{l}_{\lambda}%
 - \frac{3}{320}{\rm i} \mathcal{H}^{\beta \rho} (\Gamma_{a})_{\alpha \beta} W \boldsymbol{\lambda}_{i}^{\lambda} {G}^{-3} \varphi_{j \rho} \varphi^{j}_{\lambda} - \frac{3}{320}{\rm i} \mathcal{H}^{\rho \lambda} (\Gamma_{a})_{\alpha}{}^{\beta} W \boldsymbol{\lambda}_{i \rho} {G}^{-3} \varphi_{j \lambda} \varphi^{j}_{\beta}+\frac{3}{160}{\rm i} (\Gamma_{a})_{\alpha \beta} W \boldsymbol{\lambda}_{j}^{\rho} {G}^{-3} \nabla^{\beta \lambda}{G_{i}\,^{j}} \varphi_{k \rho} \varphi^{k}_{\lambda} - \frac{3}{160}{\rm i} (\Gamma_{a})_{\alpha}{}^{\beta} W \boldsymbol{\lambda}_{j \rho} {G}^{-3} \nabla^{\rho \lambda}{G_{i}\,^{j}} \varphi_{k \beta} \varphi^{k}_{\lambda} - \frac{3}{320}{\rm i} \mathcal{H}^{\beta \rho} (\Gamma_{a})_{\alpha \beta} W \boldsymbol{\lambda}_{j \rho} {G}^{-3} \varphi_{i}^{\lambda} \varphi^{j}_{\lambda}+\frac{3}{320}{\rm i} (\Gamma_{a})_{\alpha \beta} W \boldsymbol{\lambda}_{j \rho} {G}^{-3} \nabla^{\beta \rho}{G^{j}\,_{k}} \varphi_{i}^{\lambda} \varphi^{k}_{\lambda} - \frac{3}{320}{\rm i} \mathcal{H}^{\beta \rho} (\Gamma_{a})_{\alpha \beta} W \boldsymbol{\lambda}_{j}^{\lambda} {G}^{-3} \varphi_{i \rho} \varphi^{j}_{\lambda} - \frac{3}{160}{\rm i} \mathcal{H}^{\rho \lambda} (\Gamma_{a})_{\alpha}{}^{\beta} W \boldsymbol{\lambda}_{j \beta} {G}^{-3} \varphi_{i \rho} \varphi^{j}_{\lambda}+\frac{3}{160}{\rm i} (\Gamma_{a})_{\alpha \beta} W \boldsymbol{\lambda}_{j}^{\rho} {G}^{-3} \nabla^{\beta \lambda}{G^{j}\,_{k}} \varphi_{i \lambda} \varphi^{k}_{\rho}+\frac{3}{160}{\rm i} (\Gamma_{a})_{\alpha}{}^{\beta} W \boldsymbol{\lambda}_{j \beta} {G}^{-3} \nabla^{\rho \lambda}{G^{j}\,_{k}} \varphi_{i \rho} \varphi^{k}_{\lambda} - \frac{9}{160}G_{j k} (\Gamma_{a})_{\alpha}{}^{\beta} W \boldsymbol{\lambda}_{i}^{\rho} {G}^{-5} \varphi^{j}_{\beta} \varphi^{k \lambda} \varphi_{l \rho} \varphi^{l}_{\lambda} - \frac{9}{160}G_{j k} (\Gamma_{a})_{\alpha}{}^{\beta} W \boldsymbol{\lambda}_{i}^{\rho} {G}^{-5} \varphi^{j}_{\rho} \varphi^{k \lambda} \varphi_{l \beta} \varphi^{l}_{\lambda}+\frac{3}{160}{\rm i} (\Gamma_{a})_{\alpha \beta} W \boldsymbol{\lambda}_{i}^{\rho} {G}^{-3} \nabla^{\beta \lambda}{G_{j k}} \varphi^{j}_{\rho} \varphi^{k}_{\lambda}+\frac{3}{160}{\rm i} (\Gamma_{a})_{\alpha}{}^{\beta} W \boldsymbol{\lambda}_{i \rho} {G}^{-3} \nabla^{\rho \lambda}{G_{j k}} \varphi^{j}_{\beta} \varphi^{k}_{\lambda}+\frac{9}{320}G_{j k} (\Gamma_{a})_{\alpha}{}^{\beta} W \boldsymbol{\lambda}_{i}^{\rho} {G}^{-5} \varphi^{j \lambda} \varphi^{k}_{\lambda} \varphi_{l \beta} \varphi^{l}_{\rho} - \frac{9}{320}G_{j k} (\Gamma_{a})_{\alpha}{}^{\beta} W \boldsymbol{\lambda}_{l}^{\rho} {G}^{-5} \varphi_{i \beta} \varphi^{j \lambda} \varphi^{k}_{\lambda} \varphi^{l}_{\rho} - \frac{9}{320}G_{j k} (\Gamma_{a})_{\alpha}{}^{\beta} W \boldsymbol{\lambda}_{l \beta} {G}^{-5} \varphi_{i}^{\rho} \varphi^{j \lambda} \varphi^{k}_{\lambda} \varphi^{l}_{\rho} - \frac{9}{160}G_{i j} (\Gamma_{a})_{\alpha}{}^{\beta} W \boldsymbol{\lambda}_{k}^{\rho} {G}^{-5} \varphi^{j}_{\rho} \varphi^{k \lambda} \varphi_{l \beta} \varphi^{l}_{\lambda}+\frac{9}{160}G_{i j} (\Gamma_{a})_{\alpha}{}^{\beta} W \boldsymbol{\lambda}_{k \beta} {G}^{-5} \varphi^{j \rho} \varphi^{k \lambda} \varphi_{l \rho} \varphi^{l}_{\lambda} - \frac{27}{160}G_{i j} G_{k l} (\Gamma_{a})_{\alpha}{}^{\beta} W \mathbf{F}_{\beta}\,^{\rho} {G}^{-5} \varphi^{j}_{\rho} \varphi^{k \lambda} \varphi^{l}_{\lambda}%
 - \frac{9}{64}G_{i j} G_{k l} G_{m n} (\Gamma_{a})_{\alpha}{}^{\beta} W \boldsymbol{\lambda}^{k \rho} {G}^{-7} \varphi^{j}_{\rho} \varphi^{l}_{\beta} \varphi^{m \lambda} \varphi^{n}_{\lambda}+\frac{9}{64}G_{i j} G_{k l} G_{m n} (\Gamma_{a})_{\alpha}{}^{\beta} W \boldsymbol{\lambda}^{k}_{\beta} {G}^{-7} \varphi^{j \rho} \varphi^{l}_{\rho} \varphi^{m \lambda} \varphi^{n}_{\lambda} - \frac{9}{640}{\rm i} \mathcal{H}^{\beta \rho} G_{i j} G_{k l} (\Gamma_{a})_{\alpha \beta} W \boldsymbol{\lambda}^{j}_{\rho} {G}^{-5} \varphi^{k \lambda} \varphi^{l}_{\lambda}+\frac{9}{640}{\rm i} G_{i j} G_{k l} (\Gamma_{a})_{\alpha \beta} W \boldsymbol{\lambda}_{m \rho} {G}^{-5} \nabla^{\beta \rho}{G^{j m}} \varphi^{k \lambda} \varphi^{l}_{\lambda} - \frac{9}{320}{\rm i} \mathcal{H}^{\beta \rho} G_{i j} G_{k l} (\Gamma_{a})_{\alpha \beta} W \boldsymbol{\lambda}^{k \lambda} {G}^{-5} \varphi^{j}_{\lambda} \varphi^{l}_{\rho}+\frac{9}{160}{\rm i} \mathcal{H}^{\rho \lambda} G_{i j} G_{k l} (\Gamma_{a})_{\alpha}{}^{\beta} W \boldsymbol{\lambda}^{k}_{\beta} {G}^{-5} \varphi^{j}_{\rho} \varphi^{l}_{\lambda}+\frac{9}{160}{\rm i} G_{i j} G_{k l} (\Gamma_{a})_{\alpha \beta} W \boldsymbol{\lambda}_{m}^{\rho} {G}^{-5} \nabla^{\beta \lambda}{G^{k m}} \varphi^{j}_{\rho} \varphi^{l}_{\lambda} - \frac{9}{160}{\rm i} G_{i j} G_{k l} (\Gamma_{a})_{\alpha}{}^{\beta} W \boldsymbol{\lambda}_{m \beta} {G}^{-5} \nabla^{\rho \lambda}{G^{k m}} \varphi^{j}_{\rho} \varphi^{l}_{\lambda}+\frac{9}{160}{\rm i} G_{j k} G_{l m} (\Gamma_{a})_{\alpha \beta} W \boldsymbol{\lambda}_{i}^{\rho} {G}^{-5} \nabla^{\beta \lambda}{G^{j l}} \varphi^{k}_{\rho} \varphi^{m}_{\lambda}+\frac{9}{160}{\rm i} G_{j k} G_{l m} (\Gamma_{a})_{\alpha}{}^{\beta} W \boldsymbol{\lambda}_{i \rho} {G}^{-5} \nabla^{\rho \lambda}{G^{j l}} \varphi^{k}_{\beta} \varphi^{m}_{\lambda} - \frac{3}{320}{\rm i} G_{i j} (\Gamma_{a})_{\alpha \beta} W \boldsymbol{\lambda}_{k \rho} {G}^{-3} \nabla^{\beta \rho}{\varphi^{k \lambda}} \varphi^{j}_{\lambda} - \frac{3}{160}{\rm i} G_{i j} (\Gamma_{a})_{\alpha \beta} W \boldsymbol{\lambda}_{k}^{\rho} {G}^{-3} \nabla^{\beta \lambda}{\varphi^{k}_{\rho}} \varphi^{j}_{\lambda} - \frac{3}{64}{\rm i} G_{i j} (\Gamma_{a})_{\alpha}{}^{\beta} W W_{\beta}\,^{\rho} \boldsymbol{\lambda}_{k \rho} {G}^{-3} \varphi^{j \lambda} \varphi^{k}_{\lambda} - \frac{3}{640}{\rm i} G_{i j} (\Gamma_{a})_{\alpha}{}^{\beta} W W_{\beta}\,^{\rho} \boldsymbol{\lambda}_{k}^{\lambda} {G}^{-3} \varphi^{j}_{\rho} \varphi^{k}_{\lambda}+\frac{39}{640}{\rm i} G_{i j} (\Gamma_{a})_{\alpha}{}^{\lambda} W W^{\beta \rho} \boldsymbol{\lambda}_{k \beta} {G}^{-3} \varphi^{j}_{\rho} \varphi^{k}_{\lambda} - \frac{3}{320}{\rm i} \mathcal{H}^{\rho \lambda} (\Gamma_{a})_{\alpha}{}^{\beta} W \boldsymbol{\lambda}_{j \rho} {G}^{-3} \varphi_{i \beta} \varphi^{j}_{\lambda} - \frac{9}{160}{\rm i} \mathcal{H}^{\beta \lambda} G_{i j} (\Gamma_{a})_{\alpha}{}^{\rho} W \mathbf{F}_{\beta \rho} {G}^{-3} \varphi^{j}_{\lambda} - \frac{9}{320}{\rm i} \mathcal{H}^{\rho \lambda} G_{i j} G_{k l} (\Gamma_{a})_{\alpha}{}^{\beta} W \boldsymbol{\lambda}^{k}_{\rho} {G}^{-5} \varphi^{j}_{\lambda} \varphi^{l}_{\beta}+\frac{3}{640}\mathcal{H}^{\rho \beta} \mathcal{H}_{\rho}\,^{\lambda} G_{i j} (\Gamma_{a})_{\alpha \beta} W \boldsymbol{\lambda}^{j}_{\lambda} {G}^{-3}+\frac{3}{640}\mathcal{H}^{\rho \lambda} \mathcal{H}_{\rho \lambda} G_{i j} (\Gamma_{a})_{\alpha}{}^{\beta} W \boldsymbol{\lambda}^{j}_{\beta} {G}^{-3}%
 - \frac{3}{320}\mathcal{H}_{\rho}\,^{\lambda} G_{i j} (\Gamma_{a})_{\alpha \beta} W \boldsymbol{\lambda}_{k \lambda} {G}^{-3} \nabla^{\rho \beta}{G^{j k}} - \frac{3}{320}\mathcal{H}_{\rho \lambda} G_{i j} (\Gamma_{a})_{\alpha}{}^{\beta} W \boldsymbol{\lambda}_{k \beta} {G}^{-3} \nabla^{\rho \lambda}{G^{j k}} - \frac{3}{160}{\rm i} (\Gamma_{a})_{\alpha}{}^{\beta} W \boldsymbol{\lambda}_{k \rho} {G}^{-3} \nabla^{\rho \lambda}{G_{i j}} \varphi^{k}_{\lambda} \varphi^{j}_{\beta}+\frac{3}{160}{\rm i} (\Gamma_{a})_{\alpha}{}^{\beta} W \boldsymbol{\lambda}_{k \beta} {G}^{-3} \nabla^{\rho \lambda}{G_{i j}} \varphi^{k}_{\rho} \varphi^{j}_{\lambda} - \frac{9}{80}{\rm i} G_{j k} (\Gamma_{a})_{\alpha}{}^{\rho} W \mathbf{F}_{\beta \rho} {G}^{-3} \nabla^{\beta \lambda}{G_{i}\,^{j}} \varphi^{k}_{\lambda}+\frac{9}{160}{\rm i} G_{j k} G_{l m} (\Gamma_{a})_{\alpha}{}^{\beta} W \boldsymbol{\lambda}^{j}_{\rho} {G}^{-5} \nabla^{\rho \lambda}{G_{i}\,^{l}} \varphi^{k}_{\beta} \varphi^{m}_{\lambda}+\frac{9}{160}{\rm i} G_{j k} G_{l m} (\Gamma_{a})_{\alpha}{}^{\beta} W \boldsymbol{\lambda}^{j}_{\beta} {G}^{-5} \nabla^{\rho \lambda}{G_{i}\,^{l}} \varphi^{k}_{\rho} \varphi^{m}_{\lambda} - \frac{3}{64}{\rm i} G_{j k} (\Gamma_{a})_{\alpha}{}^{\beta} W \boldsymbol{\lambda}_{i \rho} {G}^{-3} \nabla^{\rho \lambda}{\varphi^{j}_{\beta}} \varphi^{k}_{\lambda} - \frac{3}{64}{\rm i} G_{j k} (\Gamma_{a})_{\alpha \beta} W \boldsymbol{\lambda}_{i}^{\rho} {G}^{-3} \nabla^{\beta \lambda}{\varphi^{j}_{\rho}} \varphi^{k}_{\lambda} - \frac{3}{160}{\rm i} G_{j k} (\Gamma_{a})_{\alpha}{}^{\beta} W \boldsymbol{\lambda}^{j}_{\rho} {G}^{-3} \nabla^{\rho \lambda}{\varphi_{i \beta}} \varphi^{k}_{\lambda}+\frac{3}{640}{\rm i} G_{j k} (\Gamma_{a})_{\alpha}{}^{\beta} W W_{\beta}\,^{\rho} \boldsymbol{\lambda}^{j \lambda} {G}^{-3} \varphi_{i \lambda} \varphi^{k}_{\rho}+\frac{69}{640}{\rm i} G_{j k} (\Gamma_{a})_{\alpha}{}^{\lambda} W W^{\beta \rho} \boldsymbol{\lambda}^{j}_{\beta} {G}^{-3} \varphi_{i \lambda} \varphi^{k}_{\rho}+\frac{21}{320}{\rm i} G_{j k} (\Gamma_{a})_{\alpha}{}^{\beta} W W_{\beta}\,^{\rho} \boldsymbol{\lambda}^{j}_{\rho} {G}^{-3} \varphi_{i}^{\lambda} \varphi^{k}_{\lambda}+\frac{27}{2560}{\rm i} G_{i j} G_{k l} (\Gamma_{a})_{\alpha}{}^{\beta} W \boldsymbol{\lambda}^{j \rho} X^{k}_{\beta} {G}^{-3} \varphi^{l}_{\rho} - \frac{3}{320}\mathcal{H}_{\rho}\,^{\beta} G_{j k} (\Gamma_{a})_{\alpha \beta} W \boldsymbol{\lambda}^{j}_{\lambda} {G}^{-3} \nabla^{\rho \lambda}{G_{i}\,^{k}}+\frac{3}{320}\mathcal{H}_{\rho \lambda} G_{j k} (\Gamma_{a})_{\alpha}{}^{\beta} W \boldsymbol{\lambda}^{j}_{\beta} {G}^{-3} \nabla^{\rho \lambda}{G_{i}\,^{k}} - \frac{3}{160}G_{j k} (\Gamma_{a})_{\alpha \beta} W \boldsymbol{\lambda}_{l \rho} {G}^{-3} \nabla^{\beta}\,_{\lambda}{G^{j l}} \nabla^{\lambda \rho}{G_{i}\,^{k}} - \frac{3}{160}G_{j k} (\Gamma_{a})_{\alpha}{}^{\beta} W \boldsymbol{\lambda}_{l \beta} {G}^{-3} \nabla_{\rho \lambda}{G_{i}\,^{j}} \nabla^{\rho \lambda}{G^{k l}} - \frac{3}{320}\mathcal{H}_{\rho}\,^{\lambda} G_{j k} (\Gamma_{a})_{\alpha \beta} W \boldsymbol{\lambda}_{i \lambda} {G}^{-3} \nabla^{\rho \beta}{G^{j k}} - \frac{3}{320}\mathcal{H}_{\rho}\,^{\beta} G_{j k} (\Gamma_{a})_{\alpha \beta} W \boldsymbol{\lambda}_{i \lambda} {G}^{-3} \nabla^{\rho \lambda}{G^{j k}}%
+\frac{3}{160}G_{j k} (\Gamma_{a})_{\alpha \beta} W \boldsymbol{\lambda}_{i \rho} {G}^{-3} \nabla^{\beta}\,_{\lambda}{G^{j}\,_{l}} \nabla^{\lambda \rho}{G^{k l}}+\frac{9}{80}{\rm i} (\Gamma_{a})_{\alpha}{}^{\beta} W \boldsymbol{W} {G}^{-5} \varphi_{i}^{\rho} \varphi_{j \beta} \varphi^{j \lambda} \varphi_{k \rho} \varphi^{k}_{\lambda}+\frac{9}{160}{\rm i} (\Gamma_{a})_{\alpha}{}^{\beta} W \boldsymbol{W} {G}^{-5} \varphi_{i \beta} \varphi_{j}^{\rho} \varphi^{j \lambda} \varphi_{k \rho} \varphi^{k}_{\lambda} - \frac{9}{32}{\rm i} G_{i j} G_{k l} (\Gamma_{a})_{\alpha}{}^{\beta} W \boldsymbol{W} {G}^{-7} \varphi^{j \rho} \varphi^{k}_{\beta} \varphi^{l \lambda} \varphi_{m \rho} \varphi^{m}_{\lambda} - \frac{9}{32}{\rm i} G_{i j} G_{k l} (\Gamma_{a})_{\alpha}{}^{\beta} W \boldsymbol{W} {G}^{-7} \varphi^{j \rho} \varphi^{k}_{\rho} \varphi^{l \lambda} \varphi_{m \beta} \varphi^{m}_{\lambda} - \frac{9}{160}\mathcal{H}^{\beta \rho} G_{i j} (\Gamma_{a})_{\alpha \beta} W \boldsymbol{W} {G}^{-5} \varphi^{j \lambda} \varphi_{k \rho} \varphi^{k}_{\lambda} - \frac{9}{160}\mathcal{H}^{\rho \lambda} G_{i j} (\Gamma_{a})_{\alpha}{}^{\beta} W \boldsymbol{W} {G}^{-5} \varphi^{j}_{\rho} \varphi_{k \lambda} \varphi^{k}_{\beta} - \frac{9}{320}G_{i j} (\Gamma_{a})_{\alpha \beta} W \boldsymbol{W} {G}^{-5} \nabla^{\beta \rho}{G^{j}\,_{k}} \varphi^{k \lambda} \varphi_{l \lambda} \varphi^{l}_{\rho} - \frac{9}{160}G_{i j} (\Gamma_{a})_{\alpha \beta} W \boldsymbol{W} {G}^{-5} \nabla^{\beta \rho}{G_{k l}} \varphi^{j \lambda} \varphi^{k}_{\lambda} \varphi^{l}_{\rho} - \frac{9}{160}G_{i j} (\Gamma_{a})_{\alpha}{}^{\beta} W \boldsymbol{W} {G}^{-5} \nabla^{\rho \lambda}{G_{k l}} \varphi^{j}_{\rho} \varphi^{k}_{\beta} \varphi^{l}_{\lambda} - \frac{27}{320}G_{j k} (\Gamma_{a})_{\alpha \beta} W \boldsymbol{W} {G}^{-5} \nabla^{\beta \rho}{G_{i}\,^{j}} \varphi^{k \lambda} \varphi_{l \lambda} \varphi^{l}_{\rho} - \frac{9}{160}\mathcal{H}^{\beta \rho} G_{j k} (\Gamma_{a})_{\alpha \beta} W \boldsymbol{W} {G}^{-5} \varphi_{i}^{\lambda} \varphi^{j}_{\rho} \varphi^{k}_{\lambda} - \frac{9}{320}G_{j k} (\Gamma_{a})_{\alpha \beta} W \boldsymbol{W} {G}^{-5} \nabla^{\beta \rho}{G^{j}\,_{l}} \varphi_{i}^{\lambda} \varphi^{k}_{\rho} \varphi^{l}_{\lambda} - \frac{9}{160}\mathcal{H}^{\beta \rho} G_{j k} (\Gamma_{a})_{\alpha \beta} W \boldsymbol{W} {G}^{-5} \varphi_{i \rho} \varphi^{j \lambda} \varphi^{k}_{\lambda}+\frac{9}{160}\mathcal{H}^{\rho \lambda} G_{j k} (\Gamma_{a})_{\alpha}{}^{\beta} W \boldsymbol{W} {G}^{-5} \varphi_{i \rho} \varphi^{j}_{\lambda} \varphi^{k}_{\beta}+\frac{9}{160}G_{j k} (\Gamma_{a})_{\alpha \beta} W \boldsymbol{W} {G}^{-5} \nabla^{\beta \rho}{G^{j}\,_{l}} \varphi_{i \rho} \varphi^{k \lambda} \varphi^{l}_{\lambda}+\frac{9}{160}G_{j k} (\Gamma_{a})_{\alpha}{}^{\beta} W \boldsymbol{W} {G}^{-5} \nabla^{\rho \lambda}{G^{j}\,_{l}} \varphi_{i \rho} \varphi^{k}_{\beta} \varphi^{l}_{\lambda} - \frac{3}{160}(\Gamma_{a})_{\alpha \beta} W \boldsymbol{W} {G}^{-3} \nabla^{\beta \rho}{\varphi_{i}^{\lambda}} \varphi_{j \lambda} \varphi^{j}_{\rho}+\frac{3}{320}{\rm i} \mathcal{H}_{\rho}\,^{\beta} (\Gamma_{a})_{\alpha \beta} W \boldsymbol{W} {G}^{-3} \nabla^{\rho \lambda}{G_{i j}} \varphi^{j}_{\lambda}+\frac{21}{640}{\rm i} \mathcal{H}_{\rho}\,^{\lambda} (\Gamma_{a})_{\alpha \beta} W \boldsymbol{W} {G}^{-3} \nabla^{\rho \beta}{G_{i j}} \varphi^{j}_{\lambda}%
 - \frac{3}{320}{\rm i} \mathcal{H}_{\rho \lambda} (\Gamma_{a})_{\alpha}{}^{\beta} W \boldsymbol{W} {G}^{-3} \nabla^{\rho \lambda}{G_{i j}} \varphi^{j}_{\beta}+\frac{9}{320}{\rm i} (\Gamma_{a})_{\alpha \beta} W \boldsymbol{W} {G}^{-3} \nabla^{\beta}\,_{\rho}{G_{j k}} \nabla^{\rho \lambda}{G_{i}\,^{j}} \varphi^{k}_{\lambda} - \frac{9}{320}{\rm i} (\Gamma_{a})_{\alpha \beta} W \boldsymbol{W} {G}^{-3} \nabla^{\beta}\,_{\rho}{G_{i j}} \nabla^{\rho \lambda}{G^{j}\,_{k}} \varphi^{k}_{\lambda}+\frac{3}{160}{\rm i} (\Gamma_{a})_{\alpha}{}^{\beta} W \boldsymbol{W} {G}^{-3} \nabla_{\rho \lambda}{G_{i j}} \nabla^{\rho \lambda}{G^{j}\,_{k}} \varphi^{k}_{\beta}+\frac{3}{320}{\rm i} (\Gamma_{a})_{\alpha \beta} \boldsymbol{W} \lambda_{k \rho} {G}^{-3} \nabla^{\beta \rho}{G_{i j}} \varphi^{k \lambda} \varphi^{j}_{\lambda}+\frac{3}{320}{\rm i} (\Gamma_{a})_{\alpha \beta} W \boldsymbol{\lambda}_{k \rho} {G}^{-3} \nabla^{\beta \rho}{G_{i j}} \varphi^{k \lambda} \varphi^{j}_{\lambda}+\frac{9}{320}G_{j k} (\Gamma_{a})_{\alpha \beta} W \boldsymbol{W} {G}^{-5} \nabla^{\beta \rho}{G_{i l}} \varphi^{j \lambda} \varphi^{k}_{\rho} \varphi^{l}_{\lambda}+\frac{3}{320}{\rm i} (\Gamma_{a})_{\alpha \beta} W \boldsymbol{W} {G}^{-3} \nabla^{\beta}\,_{\rho}{G_{j k}} \nabla^{\rho \lambda}{G^{j k}} \varphi_{i \lambda}+\frac{9}{160}(\Gamma_{a})_{\alpha \beta} W \boldsymbol{W} {G}^{-3} \nabla^{\beta \rho}{\varphi_{j}^{\lambda}} \varphi_{i \rho} \varphi^{j}_{\lambda} - \frac{3}{40}(\Gamma_{a})_{\alpha \beta} W \boldsymbol{W} {G}^{-3} \nabla^{\beta \rho}{\varphi_{j}^{\lambda}} \varphi_{i \lambda} \varphi^{j}_{\rho}+\frac{3}{64}(\Gamma_{a})_{\alpha}{}^{\beta} W \boldsymbol{W} {G}^{-3} \nabla^{\rho \lambda}{\varphi_{j \beta}} \varphi_{i \rho} \varphi^{j}_{\lambda} - \frac{9}{160}G_{j k} (\Gamma_{a})_{\alpha}{}^{\beta} W \boldsymbol{W} {G}^{-5} \nabla^{\rho \lambda}{G_{i l}} \varphi^{j}_{\beta} \varphi^{k}_{\rho} \varphi^{l}_{\lambda}+\frac{9}{160}G_{j k} (\Gamma_{a})_{\alpha}{}^{\beta} W \boldsymbol{W} {G}^{-5} \nabla^{\rho \lambda}{G_{i l}} \varphi^{j}_{\rho} \varphi^{k}_{\lambda} \varphi^{l}_{\beta}+\frac{3}{320}{\rm i} G_{j k} (\Gamma_{a})_{\alpha \beta} \boldsymbol{W} \lambda^{j}_{\rho} {G}^{-3} \nabla^{\beta \rho}{\varphi_{i}^{\lambda}} \varphi^{k}_{\lambda}+\frac{3}{320}{\rm i} G_{j k} (\Gamma_{a})_{\alpha \beta} W \boldsymbol{\lambda}^{j}_{\rho} {G}^{-3} \nabla^{\beta \rho}{\varphi_{i}^{\lambda}} \varphi^{k}_{\lambda} - \frac{9}{640}{\rm i} G_{i j} (\Gamma_{a})_{\alpha \beta} W \boldsymbol{W} {G}^{-3} \nabla^{\beta \lambda}{\mathcal{H}^{\rho}\,_{\rho}} \varphi^{j}_{\lambda}+\frac{3}{320}{\rm i} G_{j k} (\Gamma_{a})_{\alpha}{}^{\beta} W \boldsymbol{W} {G}^{-3} \nabla_{\rho \lambda}{\varphi_{i \beta}} \nabla^{\rho \lambda}{G^{j k}}+\frac{27}{160}G_{i j} G_{k l} (\Gamma_{a})_{\alpha \beta} W \boldsymbol{W} {G}^{-5} \nabla^{\beta \rho}{\varphi^{k \lambda}} \varphi^{j}_{\lambda} \varphi^{l}_{\rho}+\frac{21}{320}{\rm i} \mathcal{H}_{\rho}\,^{\lambda} G_{i j} (\Gamma_{a})_{\alpha \beta} W \boldsymbol{W} {G}^{-3} \nabla^{\rho \beta}{\varphi^{j}_{\lambda}} - \frac{3}{160}{\rm i} G_{i j} (\Gamma_{a})_{\alpha \beta} W \boldsymbol{W} {G}^{-3} \nabla^{\beta}\,_{\rho}{\varphi_{k \lambda}} \nabla^{\rho \lambda}{G^{j k}}%
 - \frac{9}{40}G_{i j} G_{k l} (\Gamma_{a})_{\alpha \beta} W \boldsymbol{W} {G}^{-5} \nabla^{\beta \rho}{\varphi^{k \lambda}} \varphi^{j}_{\rho} \varphi^{l}_{\lambda} - \frac{9}{64}G_{i j} G_{k l} (\Gamma_{a})_{\alpha}{}^{\beta} W \boldsymbol{W} {G}^{-5} \nabla^{\rho \lambda}{\varphi^{k}_{\beta}} \varphi^{j}_{\rho} \varphi^{l}_{\lambda}+\frac{3}{128}{\rm i} \mathcal{H}_{\rho \lambda} G_{i j} (\Gamma_{a})_{\alpha}{}^{\beta} W \boldsymbol{W} {G}^{-3} \nabla^{\rho \lambda}{\varphi^{j}_{\beta}} - \frac{3}{320}{\rm i} G_{i j} (\Gamma_{a})_{\alpha}{}^{\beta} W \boldsymbol{W} {G}^{-3} \nabla_{\rho \lambda}{\varphi_{k \beta}} \nabla^{\rho \lambda}{G^{j k}}+\frac{117}{640}G_{i j} G_{k l} (\Gamma_{a})_{\alpha}{}^{\beta} W \boldsymbol{W} W_{\beta}\,^{\rho} {G}^{-5} \varphi^{j \lambda} \varphi^{k}_{\rho} \varphi^{l}_{\lambda} - \frac{27}{640}{\rm i} \mathcal{H}_{\rho}\,^{\lambda} G_{i j} G_{k l} (\Gamma_{a})_{\alpha \beta} W \boldsymbol{W} {G}^{-5} \nabla^{\rho \beta}{G^{j k}} \varphi^{l}_{\lambda} - \frac{9}{320}{\rm i} \mathcal{H}_{\rho}\,^{\lambda} G_{i j} G_{k l} (\Gamma_{a})_{\alpha \beta} W \boldsymbol{W} {G}^{-5} \nabla^{\rho \beta}{G^{k l}} \varphi^{j}_{\lambda} - \frac{9}{320}{\rm i} \mathcal{H}_{\rho}\,^{\beta} G_{i j} G_{k l} (\Gamma_{a})_{\alpha \beta} W \boldsymbol{W} {G}^{-5} \nabla^{\rho \lambda}{G^{k l}} \varphi^{j}_{\lambda} - \frac{9}{320}{\rm i} \mathcal{H}_{\rho}\,^{\beta} G_{i j} G_{k l} (\Gamma_{a})_{\alpha \beta} W \boldsymbol{W} {G}^{-5} \nabla^{\rho \lambda}{G^{j k}} \varphi^{l}_{\lambda} - \frac{9}{320}{\rm i} \mathcal{H}_{\rho \lambda} G_{i j} G_{k l} (\Gamma_{a})_{\alpha}{}^{\beta} W \boldsymbol{W} {G}^{-5} \nabla^{\rho \lambda}{G^{j k}} \varphi^{l}_{\beta} - \frac{3}{320}{\rm i} G_{j k} (\Gamma_{a})_{\alpha \beta} W \boldsymbol{W} {G}^{-3} \nabla^{\beta}\,_{\rho}{\varphi^{j}_{\lambda}} \nabla^{\rho \lambda}{G_{i}\,^{k}}+\frac{3}{40}{\rm i} G_{j k} (\Gamma_{a})_{\alpha \beta} W \boldsymbol{W} {G}^{-3} \nabla_{\rho}\,^{\lambda}{\varphi^{j}_{\lambda}} \nabla^{\beta \rho}{G_{i}\,^{k}}+\frac{3}{40}{\rm i} G_{j k} (\Gamma_{a})_{\alpha}{}^{\beta} W \boldsymbol{W} {G}^{-3} \nabla_{\rho \lambda}{\varphi^{j}_{\beta}} \nabla^{\rho \lambda}{G_{i}\,^{k}} - \frac{9}{640}{\rm i} G_{i j} (\Gamma_{a})_{\alpha}{}^{\beta} W \boldsymbol{W} W_{\beta \rho} {G}^{-3} \nabla^{\rho \lambda}{G^{j}\,_{k}} \varphi^{k}_{\lambda} - \frac{9}{160}G_{j k} (\Gamma_{a})_{\alpha}{}^{\beta} W \boldsymbol{W} {G}^{-5} \nabla^{\rho \lambda}{G^{j}\,_{l}} \varphi_{i \beta} \varphi^{k}_{\rho} \varphi^{l}_{\lambda} - \frac{9}{32}G_{i j} G_{k l} G_{m n} (\Gamma_{a})_{\alpha}{}^{\beta} W \boldsymbol{W} {G}^{-7} \nabla^{\rho \lambda}{G^{k m}} \varphi^{j}_{\rho} \varphi^{l}_{\beta} \varphi^{n}_{\lambda} - \frac{9}{32}G_{i j} G_{k l} G_{m n} (\Gamma_{a})_{\alpha \beta} W \boldsymbol{W} {G}^{-7} \nabla^{\beta \rho}{G^{k m}} \varphi^{j \lambda} \varphi^{l}_{\lambda} \varphi^{n}_{\rho}+\frac{9}{320}{\rm i} G_{i j} G_{k l} (\Gamma_{a})_{\alpha \beta} W \boldsymbol{W} {G}^{-5} \nabla^{\beta}\,_{\rho}{G^{j}\,_{m}} \nabla^{\rho \lambda}{G^{k m}} \varphi^{l}_{\lambda}+\frac{9}{160}{\rm i} G_{i j} G_{k l} (\Gamma_{a})_{\alpha \beta} W \boldsymbol{W} {G}^{-5} \nabla^{\beta}\,_{\rho}{G^{k}\,_{m}} \nabla^{\rho \lambda}{G^{l m}} \varphi^{j}_{\lambda} - \frac{9}{160}{\rm i} G_{j k} G_{l m} (\Gamma_{a})_{\alpha \beta} W \boldsymbol{W} {G}^{-5} \nabla^{\beta}\,_{\rho}{G^{j l}} \nabla^{\rho \lambda}{G_{i}\,^{k}} \varphi^{m}_{\lambda}%
+\frac{9}{320}{\rm i} G_{j k} G_{l m} (\Gamma_{a})_{\alpha \beta} W \boldsymbol{W} {G}^{-5} \nabla^{\beta}\,_{\rho}{G_{i}\,^{j}} \nabla^{\rho \lambda}{G^{k l}} \varphi^{m}_{\lambda} - \frac{9}{160}{\rm i} G_{j k} G_{l m} (\Gamma_{a})_{\alpha}{}^{\beta} W \boldsymbol{W} {G}^{-5} \nabla_{\rho \lambda}{G_{i}\,^{j}} \nabla^{\rho \lambda}{G^{k l}} \varphi^{m}_{\beta} - \frac{9}{640}{\rm i} G_{j k} (\Gamma_{a})_{\alpha}{}^{\beta} W \boldsymbol{W} W_{\beta \rho} {G}^{-3} \nabla^{\rho \lambda}{G^{j k}} \varphi_{i \lambda} - \frac{9}{320}{\rm i} G_{j k} (\Gamma_{a})_{\alpha}{}^{\beta} W \boldsymbol{W} {G}^{-3} \nabla_{\rho \lambda}{G_{i}\,^{j}} \nabla^{\rho \lambda}{\varphi^{k}_{\beta}}+\frac{9}{160}{\rm i} G_{j k} (\Gamma_{a})_{\alpha \beta} W \boldsymbol{W} {G}^{-3} \nabla^{\beta}\,_{\rho}{G_{i}\,^{j}} \nabla^{\rho \lambda}{\varphi^{k}_{\lambda}} - \frac{9}{80}{\rm i} G_{j k} (\Gamma_{a})_{\alpha \beta} W \boldsymbol{W} {G}^{-3} \nabla_{\rho}\,^{\lambda}{G_{i}\,^{j}} \nabla^{\beta \rho}{\varphi^{k}_{\lambda}}+\frac{9}{64}{\rm i} G_{j k} G_{l m} (\Gamma_{a})_{\alpha}{}^{\beta} W \boldsymbol{W} {G}^{-7} \varphi_{i}^{\rho} \varphi^{j}_{\beta} \varphi^{k}_{\rho} \varphi^{l \lambda} \varphi^{m}_{\lambda}+\frac{9}{64}{\rm i} G_{j k} G_{l m} (\Gamma_{a})_{\alpha}{}^{\beta} W \boldsymbol{W} {G}^{-7} \varphi_{i \beta} \varphi^{j \rho} \varphi^{k}_{\rho} \varphi^{l \lambda} \varphi^{m}_{\lambda}+\frac{63}{640}G_{j k} (\Gamma_{a})_{\alpha \beta} W \boldsymbol{W} {G}^{-5} \nabla^{\beta \rho}{G_{i l}} \varphi^{j \lambda} \varphi^{k}_{\lambda} \varphi^{l}_{\rho} - \frac{9}{160}\mathcal{H}^{\rho \lambda} G_{j k} (\Gamma_{a})_{\alpha}{}^{\beta} W \boldsymbol{W} {G}^{-5} \varphi_{i \beta} \varphi^{j}_{\rho} \varphi^{k}_{\lambda}+\frac{27}{64}{\rm i} G_{i j} G_{k l} (\Gamma_{a})_{\alpha}{}^{\beta} W \boldsymbol{W} {G}^{-7} \varphi^{j \rho} \varphi^{k \lambda} \varphi^{l}_{\lambda} \varphi_{m \beta} \varphi^{m}_{\rho} - \frac{9}{128}G_{i j} G_{k l} G_{m n} (\Gamma_{a})_{\alpha \beta} W \boldsymbol{W} {G}^{-7} \nabla^{\beta \rho}{G^{j k}} \varphi^{l}_{\rho} \varphi^{m \lambda} \varphi^{n}_{\lambda} - \frac{9}{640}{\rm i} G_{j k} G_{l m} (\Gamma_{a})_{\alpha \beta} \boldsymbol{W} \lambda^{j}_{\rho} {G}^{-5} \nabla^{\beta \rho}{G_{i}\,^{k}} \varphi^{l \lambda} \varphi^{m}_{\lambda} - \frac{9}{640}{\rm i} G_{j k} G_{l m} (\Gamma_{a})_{\alpha \beta} W \boldsymbol{\lambda}^{j}_{\rho} {G}^{-5} \nabla^{\beta \rho}{G_{i}\,^{k}} \varphi^{l \lambda} \varphi^{m}_{\lambda} - \frac{3}{80}(\Gamma_{a})_{\alpha}{}^{\beta} W \mathbf{X}_{j k} {G}^{-3} \varphi_{i}^{\rho} \varphi^{j}_{\beta} \varphi^{k}_{\rho} - \frac{3}{80}(\Gamma_{a})_{\alpha}{}^{\beta} W \mathbf{X}_{j k} {G}^{-3} \varphi_{i \beta} \varphi^{j \rho} \varphi^{k}_{\rho}+\frac{3}{160}{\rm i} G_{j k} (\Gamma_{a})_{\alpha}{}^{\beta} \mathbf{X}^{j k} \lambda_{l \beta} {G}^{-3} \varphi_{i}^{\rho} \varphi^{l}_{\rho}+\frac{3}{160}{\rm i} G_{j k} (\Gamma_{a})_{\alpha}{}^{\beta} \mathbf{X}^{j k} \lambda^{\rho}_{l} {G}^{-3} \varphi_{i \beta} \varphi^{l}_{\rho} - \frac{3}{80}{\rm i} G_{j k} (\Gamma_{a})_{\alpha \beta} W {G}^{-3} \nabla^{\beta \rho}{\boldsymbol{\lambda}^{j}_{\rho}} \varphi_{i}^{\lambda} \varphi^{k}_{\lambda} - \frac{3}{80}{\rm i} G_{j k} (\Gamma_{a})_{\alpha}{}^{\beta} W {G}^{-3} \nabla^{\rho \lambda}{\boldsymbol{\lambda}^{j}_{\rho}} \varphi_{i \lambda} \varphi^{k}_{\beta}%
+\frac{9}{160}G_{j k} G_{l m} (\Gamma_{a})_{\alpha}{}^{\beta} W \mathbf{X}^{j k} {G}^{-5} \varphi_{i}^{\rho} \varphi^{l}_{\beta} \varphi^{m}_{\rho}+\frac{9}{160}G_{j k} G_{l m} (\Gamma_{a})_{\alpha}{}^{\beta} W \mathbf{X}^{j k} {G}^{-5} \varphi_{i \beta} \varphi^{l \rho} \varphi^{m}_{\rho} - \frac{3}{64}{\rm i} G_{j k} (\Gamma_{a})_{\alpha \beta} W \mathbf{X}^{j k} {G}^{-3} \nabla^{\beta \rho}{G_{i l}} \varphi^{l}_{\rho}+\frac{3}{80}{\rm i} G_{j k} (\Gamma_{a})_{\alpha}{}^{\beta} \mathbf{X}_{i l} \lambda^{j}_{\beta} {G}^{-3} \varphi^{k \rho} \varphi^{l}_{\rho}+\frac{3}{160}{\rm i} G_{j k} (\Gamma_{a})_{\alpha}{}^{\beta} \mathbf{X}_{i l} \lambda^{j \rho} {G}^{-3} \varphi^{k}_{\beta} \varphi^{l}_{\rho} - \frac{3}{160}{\rm i} G_{j k} (\Gamma_{a})_{\alpha}{}^{\beta} \mathbf{X}_{i l} \lambda^{j \rho} {G}^{-3} \varphi^{k}_{\rho} \varphi^{l}_{\beta}+\frac{3}{80}{\rm i} G_{i j} (\Gamma_{a})_{\alpha \beta} W {G}^{-3} \nabla^{\beta \rho}{\boldsymbol{\lambda}_{k \rho}} \varphi^{j \lambda} \varphi^{k}_{\lambda} - \frac{3}{80}{\rm i} G_{i j} (\Gamma_{a})_{\alpha}{}^{\beta} W {G}^{-3} \nabla^{\rho \lambda}{\boldsymbol{\lambda}_{k \rho}} \varphi^{j}_{\lambda} \varphi^{k}_{\beta}+\frac{3}{80}{\rm i} G_{j k} (\Gamma_{a})_{\alpha}{}^{\beta} W {G}^{-3} \nabla^{\rho \lambda}{\boldsymbol{\lambda}_{i \rho}} \varphi^{j}_{\beta} \varphi^{k}_{\lambda}+\frac{9}{160}{\rm i} G_{i j} (\Gamma_{a})_{\alpha}{}^{\beta} \mathbf{X}_{k l} \lambda^{k}_{\beta} {G}^{-3} \varphi^{j \rho} \varphi^{l}_{\rho} - \frac{9}{160}{\rm i} G_{i j} (\Gamma_{a})_{\alpha}{}^{\beta} \mathbf{X}_{k l} \lambda^{k \rho} {G}^{-3} \varphi^{j}_{\rho} \varphi^{l}_{\beta} - \frac{9}{80}G_{i j} G_{k l} (\Gamma_{a})_{\alpha}{}^{\beta} W \mathbf{X}^{k}\,_{m} {G}^{-5} \varphi^{j \rho} \varphi^{l}_{\beta} \varphi^{m}_{\rho} - \frac{9}{80}G_{i j} G_{k l} (\Gamma_{a})_{\alpha}{}^{\beta} W \mathbf{X}^{k}\,_{m} {G}^{-5} \varphi^{j \rho} \varphi^{l}_{\rho} \varphi^{m}_{\beta}+\frac{3}{160}{\rm i} \mathcal{H}^{\beta \rho} G_{i j} (\Gamma_{a})_{\alpha \beta} W \mathbf{X}^{j}\,_{k} {G}^{-3} \varphi^{k}_{\rho} - \frac{3}{160}{\rm i} G_{i j} (\Gamma_{a})_{\alpha \beta} W \mathbf{X}_{k l} {G}^{-3} \nabla^{\beta \rho}{G^{j k}} \varphi^{l}_{\rho}+\frac{3}{160}{\rm i} G_{j k} (\Gamma_{a})_{\alpha}{}^{\beta} \mathbf{X}^{j k} \lambda^{\rho}_{i} {G}^{-3} \varphi_{l \beta} \varphi^{l}_{\rho}+\frac{3}{80}{\rm i} G_{j k} (\Gamma_{a})_{\alpha}{}^{\beta} \mathbf{X}^{j}\,_{l} \lambda^{\rho}_{i} {G}^{-3} \varphi^{k}_{\rho} \varphi^{l}_{\beta} - \frac{3}{160}{\rm i} G_{j k} (\Gamma_{a})_{\alpha}{}^{\beta} \mathbf{X}^{j}\,_{l} \lambda_{i \beta} {G}^{-3} \varphi^{k \rho} \varphi^{l}_{\rho}+\frac{3}{160}{\rm i} G_{j k} (\Gamma_{a})_{\alpha}{}^{\beta} \mathbf{X}^{j}\,_{l} \lambda^{\rho}_{i} {G}^{-3} \varphi^{k}_{\beta} \varphi^{l}_{\rho} - \frac{21}{320}(\Gamma_{a})_{\alpha \beta} \lambda^{\rho}_{i} {G}^{-1} \nabla^{\beta \lambda}{\boldsymbol{\lambda}_{j \lambda}} \varphi^{j}_{\rho}%
+\frac{3}{64}(\Gamma_{a})_{\alpha}{}^{\beta} \lambda_{i \beta} {G}^{-1} \nabla^{\rho \lambda}{\boldsymbol{\lambda}_{j \rho}} \varphi^{j}_{\lambda} - \frac{9}{80}(\Gamma_{a})_{\alpha}{}^{\beta} \lambda_{i \rho} {G}^{-1} \nabla^{\rho \lambda}{\boldsymbol{\lambda}_{j \lambda}} \varphi^{j}_{\beta}+\frac{3}{640}(\Gamma_{a})_{\alpha}{}^{\beta} W_{\beta}\,^{\rho} \lambda^{\lambda}_{i} \boldsymbol{\lambda}_{j \rho} {G}^{-1} \varphi^{j}_{\lambda}+\frac{3}{80}{\rm i} G_{i j} G_{k l} (\Gamma_{a})_{\alpha}{}^{\beta} \mathbf{X}^{k l} F_{\beta}\,^{\rho} {G}^{-3} \varphi^{j}_{\rho} - \frac{3}{320}G_{i j} G_{k l} (\Gamma_{a})_{\alpha \beta} \lambda^{k \rho} {G}^{-3} \nabla^{\beta \lambda}{\boldsymbol{\lambda}^{l}_{\lambda}} \varphi^{j}_{\rho} - \frac{3}{320}G_{i j} G_{k l} (\Gamma_{a})_{\alpha}{}^{\beta} \lambda^{k}_{\beta} {G}^{-3} \nabla^{\rho \lambda}{\boldsymbol{\lambda}^{l}_{\rho}} \varphi^{j}_{\lambda}+\frac{3}{640}G_{i j} G_{k l} (\Gamma_{a})_{\alpha}{}^{\beta} W_{\beta}\,^{\rho} \lambda^{k \lambda} \boldsymbol{\lambda}^{l}_{\rho} {G}^{-3} \varphi^{j}_{\lambda}+\frac{3}{640}G_{i j} G_{k l} (\Gamma_{a})_{\alpha}{}^{\lambda} W^{\beta \rho} \lambda^{k}_{\lambda} \boldsymbol{\lambda}^{l}_{\beta} {G}^{-3} \varphi^{j}_{\rho}+\frac{9}{160}{\rm i} G_{i j} G_{k l} G_{m n} (\Gamma_{a})_{\alpha}{}^{\beta} \mathbf{X}^{k l} \lambda^{m \rho} {G}^{-5} \varphi^{j}_{\rho} \varphi^{n}_{\beta} - \frac{9}{160}{\rm i} G_{i j} G_{k l} G_{m n} (\Gamma_{a})_{\alpha}{}^{\beta} \mathbf{X}^{k l} \lambda^{m}_{\beta} {G}^{-5} \varphi^{j \rho} \varphi^{n}_{\rho} - \frac{3}{320}\mathcal{H}^{\beta \rho} G_{i j} G_{k l} (\Gamma_{a})_{\alpha \beta} \mathbf{X}^{k l} \lambda^{j}_{\rho} {G}^{-3}+\frac{3}{320}G_{i j} G_{k l} (\Gamma_{a})_{\alpha \beta} \mathbf{X}^{k l} \lambda_{m \rho} {G}^{-3} \nabla^{\beta \rho}{G^{j m}} - \frac{3}{80}{\rm i} G_{i j} (\Gamma_{a})_{\alpha}{}^{\beta} W {G}^{-3} \nabla^{\rho \lambda}{\boldsymbol{\lambda}^{j}_{\rho}} \varphi_{k \beta} \varphi^{k}_{\lambda}+\frac{3}{80}G_{i j} G_{k l} (\Gamma_{a})_{\alpha}{}^{\beta} \lambda^{k}_{\beta} {G}^{-3} \nabla^{\rho \lambda}{\boldsymbol{\lambda}^{j}_{\rho}} \varphi^{l}_{\lambda} - \frac{3}{80}G_{i j} G_{k l} (\Gamma_{a})_{\alpha}{}^{\beta} \lambda^{k}_{\rho} {G}^{-3} \nabla^{\rho \lambda}{\boldsymbol{\lambda}^{j}_{\lambda}} \varphi^{l}_{\beta} - \frac{3}{40}{\rm i} (\Gamma_{a})_{\alpha}{}^{\rho} W {G}^{-1} \nabla^{\beta \lambda}{\mathbf{F}_{\beta \rho}} \varphi_{i \lambda}+\frac{3}{40}{\rm i} (\Gamma_{a})_{\alpha \lambda} W {G}^{-1} \nabla^{\lambda \beta}{\mathbf{F}_{\beta}\,^{\rho}} \varphi_{i \rho} - \frac{9}{40}{\rm i} (\Gamma_{a})_{\alpha}{}^{\beta} W W_{\beta \rho} {G}^{-1} \nabla^{\rho \lambda}{\boldsymbol{W}} \varphi_{i \lambda}+\frac{9}{160}{\rm i} (\Gamma_{a})_{\alpha \lambda} W W^{\beta}\,_{\rho} {G}^{-1} \nabla^{\lambda \rho}{\boldsymbol{W}} \varphi_{i \beta} - \frac{3}{160}{\rm i} G_{i j} G_{k l} (\Gamma_{a})_{\alpha \beta} W {G}^{-3} \nabla^{\beta \rho}{\mathbf{X}^{j k}} \varphi^{l}_{\rho}%
 - \frac{3}{80}{\rm i} (\Gamma_{a})_{\alpha \beta} W {G}^{-1} \nabla_{\rho}\,^{\lambda}{\nabla^{\beta \rho}{\boldsymbol{W}}} \varphi_{i \lambda}+\frac{3}{80}{\rm i} (\Gamma_{a})_{\alpha \beta} W {G}^{-1} \nabla^{\beta}\,_{\rho}{\nabla^{\rho \lambda}{\boldsymbol{W}}} \varphi_{i \lambda}+\frac{3}{40}{\rm i} (\Gamma_{a})_{\alpha}{}^{\beta} W {G}^{-1} \nabla_{\rho \lambda}{\nabla^{\rho \lambda}{\boldsymbol{W}}} \varphi_{i \beta} - \frac{3}{80}{\rm i} G_{i j} G_{k l} (\Gamma_{a})_{\alpha}{}^{\beta} W \mathbf{X}^{j k} W_{\beta}\,^{\rho} {G}^{-3} \varphi^{l}_{\rho}+\frac{9}{160}{\rm i} G_{i j} (\Gamma_{a})_{\alpha}{}^{\lambda} W W^{\beta \rho} \boldsymbol{\lambda}^{j}_{\beta} {G}^{-3} \varphi_{k \lambda} \varphi^{k}_{\rho} - \frac{3}{80}\mathcal{H}_{\rho}\,^{\beta} (\Gamma_{a})_{\alpha \beta} W {G}^{-1} \nabla^{\rho \lambda}{\boldsymbol{\lambda}_{i \lambda}}+\frac{3}{160}G_{i j} G_{k l} (\Gamma_{a})_{\alpha \beta} W {G}^{-3} \nabla_{\lambda}\,^{\rho}{\boldsymbol{\lambda}^{k}_{\rho}} \nabla^{\beta \lambda}{G^{j l}} - \frac{9}{160}\mathcal{H}^{\lambda \beta} (\Gamma_{a})_{\alpha \lambda} W W_{\beta}\,^{\rho} \boldsymbol{\lambda}_{i \rho} {G}^{-1}+\frac{9}{320}G_{i j} G_{k l} (\Gamma_{a})_{\alpha \lambda} W W^{\beta}\,_{\rho} \boldsymbol{\lambda}^{k}_{\beta} {G}^{-3} \nabla^{\lambda \rho}{G^{j l}}+\frac{9}{160}G_{i j} G_{k l} (\Gamma_{a})_{\alpha}{}^{\beta} W \mathbf{X}^{k l} {G}^{-5} \varphi^{j \rho} \varphi_{m \beta} \varphi^{m}_{\rho}+\frac{9}{320}{\rm i} G_{i j} G_{k l} G_{m n} (\Gamma_{a})_{\alpha \beta} W \mathbf{X}^{k l} {G}^{-5} \nabla^{\beta \rho}{G^{j m}} \varphi^{n}_{\rho}+\frac{3}{160}{\rm i} G_{j k} (\Gamma_{a})_{\alpha \beta} W \mathbf{X}^{j}\,_{l} {G}^{-3} \nabla^{\beta \rho}{G_{i}\,^{k}} \varphi^{l}_{\rho} - \frac{3}{320}G_{j k} G_{l m} (\Gamma_{a})_{\alpha \beta} \mathbf{X}^{j k} \lambda^{l}_{\rho} {G}^{-3} \nabla^{\beta \rho}{G_{i}\,^{m}} - \frac{9}{160}(\Gamma_{a})_{\alpha \beta} W {G}^{-1} \nabla_{\lambda}\,^{\rho}{\boldsymbol{\lambda}_{j \rho}} \nabla^{\beta \lambda}{G_{i}\,^{j}}+\frac{27}{320}(\Gamma_{a})_{\alpha \lambda} W W^{\beta}\,_{\rho} \boldsymbol{\lambda}_{j \beta} {G}^{-1} \nabla^{\lambda \rho}{G_{i}\,^{j}}+\frac{3}{64}(\Gamma_{a})_{\alpha \beta} \lambda^{\rho}_{j} {G}^{-1} \nabla^{\beta \lambda}{\boldsymbol{\lambda}_{i \lambda}} \varphi^{j}_{\rho} - \frac{3}{320}(\Gamma_{a})_{\alpha}{}^{\beta} \lambda_{j \beta} {G}^{-1} \nabla^{\rho \lambda}{\boldsymbol{\lambda}_{i \rho}} \varphi^{j}_{\lambda} - \frac{3}{320}(\Gamma_{a})_{\alpha}{}^{\beta} W_{\beta}\,^{\rho} \lambda^{\lambda}_{j} \boldsymbol{\lambda}_{i \rho} {G}^{-1} \varphi^{j}_{\lambda}+\frac{9}{160}{\rm i} (\Gamma_{a})_{\alpha \beta} \mathbf{X}_{i j} {G}^{-1} \nabla^{\beta \rho}{W} \varphi^{j}_{\rho}+\frac{9}{320}\mathcal{H}^{\beta \rho} (\Gamma_{a})_{\alpha \beta} \mathbf{X}_{i j} \lambda^{j}_{\rho} {G}^{-1}%
+\frac{3}{64}(\Gamma_{a})_{\alpha \beta} \mathbf{X}_{i j} \lambda_{k \rho} {G}^{-1} \nabla^{\beta \rho}{G^{j k}} - \frac{3}{64}(\Gamma_{a})_{\alpha \beta} \mathbf{X}_{j k} \lambda_{i \rho} {G}^{-1} \nabla^{\beta \rho}{G^{j k}} - \frac{9}{160}(\Gamma_{a})_{\alpha}{}^{\beta} \lambda_{j \beta} {G}^{-1} \nabla^{\rho \lambda}{\boldsymbol{\lambda}^{j}_{\rho}} \varphi_{i \lambda}+\frac{3}{80}G_{i j} (\Gamma_{a})_{\alpha}{}^{\rho} \lambda^{j}_{\lambda} {G}^{-1} \nabla^{\lambda \beta}{\mathbf{F}_{\beta \rho}}+\frac{3}{16}G_{i j} (\Gamma_{a})_{\alpha \lambda} \lambda^{j \rho} {G}^{-1} \nabla^{\lambda \beta}{\mathbf{F}_{\beta \rho}} - \frac{3}{160}G_{i j} (\Gamma_{a})_{\alpha \beta} \lambda_{k \rho} {G}^{-1} \nabla^{\beta \rho}{\mathbf{X}^{j k}}+\frac{3}{32}G_{i j} (\Gamma_{a})_{\alpha \beta} \lambda^{j}_{\rho} {G}^{-1} \nabla_{\lambda}\,^{\rho}{\nabla^{\beta \lambda}{\boldsymbol{W}}}+\frac{21}{160}G_{i j} (\Gamma_{a})_{\alpha \beta} \lambda^{j}_{\rho} {G}^{-1} \nabla^{\beta}\,_{\lambda}{\nabla^{\lambda \rho}{\boldsymbol{W}}}+\frac{3}{320}G_{i j} (\Gamma_{a})_{\alpha}{}^{\beta} \lambda^{j}_{\beta} {G}^{-1} \nabla_{\rho \lambda}{\nabla^{\rho \lambda}{\boldsymbol{W}}} - \frac{3}{40}{\rm i} G_{j k} (\Gamma_{a})_{\alpha}{}^{\beta} W {G}^{-3} \nabla^{\rho \lambda}{\boldsymbol{\lambda}^{j}_{\rho}} \varphi_{i \beta} \varphi^{k}_{\lambda}+\frac{9}{160}{\rm i} (\Gamma_{a})_{\alpha \beta} W {G}^{-1} \nabla^{\beta \rho}{\mathbf{X}_{i j}} \varphi^{j}_{\rho}+\frac{9}{320}(\Gamma_{a})_{\alpha \beta} \mathbf{X}_{k j} \lambda^{k}_{\rho} {G}^{-1} \nabla^{\beta \rho}{G_{i}\,^{j}} - \frac{3}{160}{\rm i} G_{j k} (\Gamma_{a})_{\alpha \beta} W \mathbf{X}^{j}\,_{l} {G}^{-3} \nabla^{\beta \rho}{G_{i}\,^{l}} \varphi^{k}_{\rho} - \frac{3}{160}G_{i j} (\Gamma_{a})_{\alpha}{}^{\beta} W {G}^{-1} \nabla_{\rho \lambda}{\nabla^{\rho \lambda}{\boldsymbol{\lambda}^{j}_{\beta}}} - \frac{3}{80}(\Gamma_{a})_{\alpha}{}^{\beta} \boldsymbol{W} X_{j k} {G}^{-3} \varphi_{i}^{\rho} \varphi^{j}_{\beta} \varphi^{k}_{\rho} - \frac{3}{80}(\Gamma_{a})_{\alpha}{}^{\beta} \boldsymbol{W} X_{j k} {G}^{-3} \varphi_{i \beta} \varphi^{j \rho} \varphi^{k}_{\rho}+\frac{3}{160}{\rm i} G_{j k} (\Gamma_{a})_{\alpha}{}^{\beta} X^{j k} \boldsymbol{\lambda}_{l \beta} {G}^{-3} \varphi_{i}^{\rho} \varphi^{l}_{\rho}+\frac{3}{160}{\rm i} G_{j k} (\Gamma_{a})_{\alpha}{}^{\beta} X^{j k} \boldsymbol{\lambda}_{l}^{\rho} {G}^{-3} \varphi_{i \beta} \varphi^{l}_{\rho} - \frac{3}{80}{\rm i} G_{j k} (\Gamma_{a})_{\alpha \beta} \boldsymbol{W} {G}^{-3} \nabla^{\beta \rho}{\lambda^{j}_{\rho}} \varphi_{i}^{\lambda} \varphi^{k}_{\lambda} - \frac{3}{80}{\rm i} G_{j k} (\Gamma_{a})_{\alpha}{}^{\beta} \boldsymbol{W} {G}^{-3} \nabla^{\rho \lambda}{\lambda^{j}_{\rho}} \varphi_{i \lambda} \varphi^{k}_{\beta}%
+\frac{9}{160}G_{j k} G_{l m} (\Gamma_{a})_{\alpha}{}^{\beta} \boldsymbol{W} X^{j k} {G}^{-5} \varphi_{i}^{\rho} \varphi^{l}_{\beta} \varphi^{m}_{\rho}+\frac{9}{160}G_{j k} G_{l m} (\Gamma_{a})_{\alpha}{}^{\beta} \boldsymbol{W} X^{j k} {G}^{-5} \varphi_{i \beta} \varphi^{l \rho} \varphi^{m}_{\rho} - \frac{3}{64}{\rm i} G_{j k} (\Gamma_{a})_{\alpha \beta} \boldsymbol{W} X^{j k} {G}^{-3} \nabla^{\beta \rho}{G_{i l}} \varphi^{l}_{\rho}+\frac{3}{80}{\rm i} G_{j k} (\Gamma_{a})_{\alpha}{}^{\beta} X_{i l} \boldsymbol{\lambda}^{j}_{\beta} {G}^{-3} \varphi^{k \rho} \varphi^{l}_{\rho}+\frac{3}{160}{\rm i} G_{j k} (\Gamma_{a})_{\alpha}{}^{\beta} X_{i l} \boldsymbol{\lambda}^{j \rho} {G}^{-3} \varphi^{k}_{\beta} \varphi^{l}_{\rho} - \frac{3}{160}{\rm i} G_{j k} (\Gamma_{a})_{\alpha}{}^{\beta} X_{i l} \boldsymbol{\lambda}^{j \rho} {G}^{-3} \varphi^{k}_{\rho} \varphi^{l}_{\beta}+\frac{3}{80}{\rm i} G_{i j} (\Gamma_{a})_{\alpha \beta} \boldsymbol{W} {G}^{-3} \nabla^{\beta \rho}{\lambda_{k \rho}} \varphi^{j \lambda} \varphi^{k}_{\lambda} - \frac{3}{80}{\rm i} G_{i j} (\Gamma_{a})_{\alpha}{}^{\beta} \boldsymbol{W} {G}^{-3} \nabla^{\rho \lambda}{\lambda_{k \rho}} \varphi^{j}_{\lambda} \varphi^{k}_{\beta}+\frac{3}{80}{\rm i} G_{j k} (\Gamma_{a})_{\alpha}{}^{\beta} \boldsymbol{W} {G}^{-3} \nabla^{\rho \lambda}{\lambda_{i \rho}} \varphi^{j}_{\beta} \varphi^{k}_{\lambda}+\frac{9}{160}{\rm i} G_{i j} (\Gamma_{a})_{\alpha}{}^{\beta} X_{k l} \boldsymbol{\lambda}^{k}_{\beta} {G}^{-3} \varphi^{j \rho} \varphi^{l}_{\rho} - \frac{9}{160}{\rm i} G_{i j} (\Gamma_{a})_{\alpha}{}^{\beta} X_{k l} \boldsymbol{\lambda}^{k \rho} {G}^{-3} \varphi^{j}_{\rho} \varphi^{l}_{\beta} - \frac{9}{80}G_{i j} G_{k l} (\Gamma_{a})_{\alpha}{}^{\beta} \boldsymbol{W} X^{k}\,_{m} {G}^{-5} \varphi^{j \rho} \varphi^{l}_{\beta} \varphi^{m}_{\rho} - \frac{9}{80}G_{i j} G_{k l} (\Gamma_{a})_{\alpha}{}^{\beta} \boldsymbol{W} X^{k}\,_{m} {G}^{-5} \varphi^{j \rho} \varphi^{l}_{\rho} \varphi^{m}_{\beta}+\frac{3}{160}{\rm i} \mathcal{H}^{\beta \rho} G_{i j} (\Gamma_{a})_{\alpha \beta} \boldsymbol{W} X^{j}\,_{k} {G}^{-3} \varphi^{k}_{\rho} - \frac{3}{160}{\rm i} G_{i j} (\Gamma_{a})_{\alpha \beta} \boldsymbol{W} X_{k l} {G}^{-3} \nabla^{\beta \rho}{G^{j k}} \varphi^{l}_{\rho}+\frac{3}{160}{\rm i} G_{j k} (\Gamma_{a})_{\alpha}{}^{\beta} X^{j k} \boldsymbol{\lambda}_{i}^{\rho} {G}^{-3} \varphi_{l \beta} \varphi^{l}_{\rho}+\frac{3}{80}{\rm i} G_{j k} (\Gamma_{a})_{\alpha}{}^{\beta} X^{j}\,_{l} \boldsymbol{\lambda}_{i}^{\rho} {G}^{-3} \varphi^{k}_{\rho} \varphi^{l}_{\beta} - \frac{3}{160}{\rm i} G_{j k} (\Gamma_{a})_{\alpha}{}^{\beta} X^{j}\,_{l} \boldsymbol{\lambda}_{i \beta} {G}^{-3} \varphi^{k \rho} \varphi^{l}_{\rho}+\frac{3}{160}{\rm i} G_{j k} (\Gamma_{a})_{\alpha}{}^{\beta} X^{j}\,_{l} \boldsymbol{\lambda}_{i}^{\rho} {G}^{-3} \varphi^{k}_{\beta} \varphi^{l}_{\rho} - \frac{21}{320}(\Gamma_{a})_{\alpha \beta} \boldsymbol{\lambda}_{i}^{\lambda} {G}^{-1} \nabla^{\beta \rho}{\lambda_{j \rho}} \varphi^{j}_{\lambda}%
+\frac{3}{64}(\Gamma_{a})_{\alpha}{}^{\beta} \boldsymbol{\lambda}_{i \beta} {G}^{-1} \nabla^{\rho \lambda}{\lambda_{j \rho}} \varphi^{j}_{\lambda} - \frac{9}{80}(\Gamma_{a})_{\alpha}{}^{\beta} \boldsymbol{\lambda}_{i \lambda} {G}^{-1} \nabla^{\lambda \rho}{\lambda_{j \rho}} \varphi^{j}_{\beta} - \frac{3}{640}(\Gamma_{a})_{\alpha}{}^{\beta} W_{\beta}\,^{\rho} \lambda_{j \rho} \boldsymbol{\lambda}_{i}^{\lambda} {G}^{-1} \varphi^{j}_{\lambda}+\frac{3}{80}{\rm i} G_{i j} G_{k l} (\Gamma_{a})_{\alpha}{}^{\beta} X^{k l} \mathbf{F}_{\beta}\,^{\rho} {G}^{-3} \varphi^{j}_{\rho} - \frac{3}{320}G_{i j} G_{k l} (\Gamma_{a})_{\alpha \beta} \boldsymbol{\lambda}^{k \lambda} {G}^{-3} \nabla^{\beta \rho}{\lambda^{l}_{\rho}} \varphi^{j}_{\lambda} - \frac{3}{320}G_{i j} G_{k l} (\Gamma_{a})_{\alpha}{}^{\beta} \boldsymbol{\lambda}^{k}_{\beta} {G}^{-3} \nabla^{\rho \lambda}{\lambda^{l}_{\rho}} \varphi^{j}_{\lambda} - \frac{3}{640}G_{i j} G_{k l} (\Gamma_{a})_{\alpha}{}^{\beta} W_{\beta}\,^{\rho} \lambda^{k}_{\rho} \boldsymbol{\lambda}^{l \lambda} {G}^{-3} \varphi^{j}_{\lambda} - \frac{3}{640}G_{i j} G_{k l} (\Gamma_{a})_{\alpha}{}^{\lambda} W^{\beta \rho} \lambda^{k}_{\beta} \boldsymbol{\lambda}^{l}_{\lambda} {G}^{-3} \varphi^{j}_{\rho}+\frac{9}{160}{\rm i} G_{i j} G_{k l} G_{m n} (\Gamma_{a})_{\alpha}{}^{\beta} X^{k l} \boldsymbol{\lambda}^{m \rho} {G}^{-5} \varphi^{j}_{\rho} \varphi^{n}_{\beta} - \frac{9}{160}{\rm i} G_{i j} G_{k l} G_{m n} (\Gamma_{a})_{\alpha}{}^{\beta} X^{k l} \boldsymbol{\lambda}^{m}_{\beta} {G}^{-5} \varphi^{j \rho} \varphi^{n}_{\rho} - \frac{3}{320}\mathcal{H}^{\beta \rho} G_{i j} G_{k l} (\Gamma_{a})_{\alpha \beta} X^{k l} \boldsymbol{\lambda}^{j}_{\rho} {G}^{-3}+\frac{3}{320}G_{i j} G_{k l} (\Gamma_{a})_{\alpha \beta} X^{k l} \boldsymbol{\lambda}_{m \rho} {G}^{-3} \nabla^{\beta \rho}{G^{j m}} - \frac{3}{80}{\rm i} G_{i j} (\Gamma_{a})_{\alpha}{}^{\beta} \boldsymbol{W} {G}^{-3} \nabla^{\rho \lambda}{\lambda^{j}_{\rho}} \varphi_{k \beta} \varphi^{k}_{\lambda}+\frac{3}{80}G_{i j} G_{k l} (\Gamma_{a})_{\alpha}{}^{\beta} \boldsymbol{\lambda}^{k}_{\beta} {G}^{-3} \nabla^{\rho \lambda}{\lambda^{j}_{\rho}} \varphi^{l}_{\lambda} - \frac{3}{80}G_{i j} G_{k l} (\Gamma_{a})_{\alpha}{}^{\beta} \boldsymbol{\lambda}^{k}_{\lambda} {G}^{-3} \nabla^{\lambda \rho}{\lambda^{j}_{\rho}} \varphi^{l}_{\beta} - \frac{3}{40}{\rm i} (\Gamma_{a})_{\alpha}{}^{\beta} \boldsymbol{W} {G}^{-1} \nabla^{\rho \lambda}{F_{\beta \rho}} \varphi_{i \lambda}+\frac{3}{40}{\rm i} (\Gamma_{a})_{\alpha \lambda} \boldsymbol{W} {G}^{-1} \nabla^{\lambda \rho}{F^{\beta}\,_{\rho}} \varphi_{i \beta} - \frac{9}{40}{\rm i} (\Gamma_{a})_{\alpha}{}^{\beta} \boldsymbol{W} W_{\beta \rho} {G}^{-1} \nabla^{\rho \lambda}{W} \varphi_{i \lambda}+\frac{9}{160}{\rm i} (\Gamma_{a})_{\alpha \lambda} \boldsymbol{W} W^{\beta}\,_{\rho} {G}^{-1} \nabla^{\lambda \rho}{W} \varphi_{i \beta} - \frac{3}{160}{\rm i} G_{i j} G_{k l} (\Gamma_{a})_{\alpha \beta} \boldsymbol{W} {G}^{-3} \nabla^{\beta \rho}{X^{j k}} \varphi^{l}_{\rho}%
 - \frac{3}{80}{\rm i} (\Gamma_{a})_{\alpha \beta} \boldsymbol{W} {G}^{-1} \nabla_{\rho}\,^{\lambda}{\nabla^{\beta \rho}{W}} \varphi_{i \lambda}+\frac{3}{80}{\rm i} (\Gamma_{a})_{\alpha \beta} \boldsymbol{W} {G}^{-1} \nabla^{\beta}\,_{\rho}{\nabla^{\rho \lambda}{W}} \varphi_{i \lambda}+\frac{3}{40}{\rm i} (\Gamma_{a})_{\alpha}{}^{\beta} \boldsymbol{W} {G}^{-1} \nabla_{\rho \lambda}{\nabla^{\rho \lambda}{W}} \varphi_{i \beta} - \frac{3}{80}{\rm i} G_{i j} G_{k l} (\Gamma_{a})_{\alpha}{}^{\beta} \boldsymbol{W} X^{j k} W_{\beta}\,^{\rho} {G}^{-3} \varphi^{l}_{\rho}+\frac{9}{160}{\rm i} G_{i j} (\Gamma_{a})_{\alpha}{}^{\lambda} \boldsymbol{W} W^{\beta \rho} \lambda^{j}_{\beta} {G}^{-3} \varphi_{k \lambda} \varphi^{k}_{\rho} - \frac{3}{80}\mathcal{H}_{\rho}\,^{\beta} (\Gamma_{a})_{\alpha \beta} \boldsymbol{W} {G}^{-1} \nabla^{\rho \lambda}{\lambda_{i \lambda}}+\frac{3}{160}G_{i j} G_{k l} (\Gamma_{a})_{\alpha \beta} \boldsymbol{W} {G}^{-3} \nabla_{\lambda}\,^{\rho}{\lambda^{k}_{\rho}} \nabla^{\beta \lambda}{G^{j l}} - \frac{9}{160}\mathcal{H}^{\lambda \beta} (\Gamma_{a})_{\alpha \lambda} \boldsymbol{W} W_{\beta}\,^{\rho} \lambda_{i \rho} {G}^{-1}+\frac{9}{320}G_{i j} G_{k l} (\Gamma_{a})_{\alpha \lambda} \boldsymbol{W} W^{\beta}\,_{\rho} \lambda^{k}_{\beta} {G}^{-3} \nabla^{\lambda \rho}{G^{j l}}+\frac{9}{160}G_{i j} G_{k l} (\Gamma_{a})_{\alpha}{}^{\beta} \boldsymbol{W} X^{k l} {G}^{-5} \varphi^{j \rho} \varphi_{m \beta} \varphi^{m}_{\rho}+\frac{9}{320}{\rm i} G_{i j} G_{k l} G_{m n} (\Gamma_{a})_{\alpha \beta} \boldsymbol{W} X^{k l} {G}^{-5} \nabla^{\beta \rho}{G^{j m}} \varphi^{n}_{\rho}+\frac{3}{160}{\rm i} G_{j k} (\Gamma_{a})_{\alpha \beta} \boldsymbol{W} X^{j}\,_{l} {G}^{-3} \nabla^{\beta \rho}{G_{i}\,^{k}} \varphi^{l}_{\rho} - \frac{3}{320}G_{j k} G_{l m} (\Gamma_{a})_{\alpha \beta} X^{j k} \boldsymbol{\lambda}^{l}_{\rho} {G}^{-3} \nabla^{\beta \rho}{G_{i}\,^{m}} - \frac{9}{160}(\Gamma_{a})_{\alpha \beta} \boldsymbol{W} {G}^{-1} \nabla_{\lambda}\,^{\rho}{\lambda_{j \rho}} \nabla^{\beta \lambda}{G_{i}\,^{j}}+\frac{27}{320}(\Gamma_{a})_{\alpha \lambda} \boldsymbol{W} W^{\beta}\,_{\rho} \lambda_{j \beta} {G}^{-1} \nabla^{\lambda \rho}{G_{i}\,^{j}}+\frac{9}{160}{\rm i} (\Gamma_{a})_{\alpha \beta} X_{i j} {G}^{-1} \nabla^{\beta \rho}{\boldsymbol{W}} \varphi^{j}_{\rho} - \frac{3}{64}(\Gamma_{a})_{\alpha \beta} X_{j k} \boldsymbol{\lambda}_{i \rho} {G}^{-1} \nabla^{\beta \rho}{G^{j k}}+\frac{3}{64}(\Gamma_{a})_{\alpha \beta} \boldsymbol{\lambda}_{j}^{\lambda} {G}^{-1} \nabla^{\beta \rho}{\lambda_{i \rho}} \varphi^{j}_{\lambda} - \frac{3}{320}(\Gamma_{a})_{\alpha}{}^{\beta} \boldsymbol{\lambda}_{j \beta} {G}^{-1} \nabla^{\rho \lambda}{\lambda_{i \rho}} \varphi^{j}_{\lambda}+\frac{3}{320}(\Gamma_{a})_{\alpha}{}^{\beta} W_{\beta}\,^{\rho} \lambda_{i \rho} \boldsymbol{\lambda}_{j}^{\lambda} {G}^{-1} \varphi^{j}_{\lambda}%
+\frac{9}{320}\mathcal{H}^{\beta \rho} (\Gamma_{a})_{\alpha \beta} X_{i j} \boldsymbol{\lambda}^{j}_{\rho} {G}^{-1}+\frac{3}{64}(\Gamma_{a})_{\alpha \beta} X_{i j} \boldsymbol{\lambda}_{k \rho} {G}^{-1} \nabla^{\beta \rho}{G^{j k}} - \frac{9}{160}(\Gamma_{a})_{\alpha}{}^{\beta} \boldsymbol{\lambda}_{j \beta} {G}^{-1} \nabla^{\rho \lambda}{\lambda^{j}_{\rho}} \varphi_{i \lambda} - \frac{3}{40}{\rm i} G_{j k} (\Gamma_{a})_{\alpha}{}^{\beta} \boldsymbol{W} {G}^{-3} \nabla^{\rho \lambda}{\lambda^{j}_{\rho}} \varphi_{i \beta} \varphi^{k}_{\lambda}+\frac{9}{160}{\rm i} (\Gamma_{a})_{\alpha \beta} \boldsymbol{W} {G}^{-1} \nabla^{\beta \rho}{X_{i j}} \varphi^{j}_{\rho}+\frac{9}{320}(\Gamma_{a})_{\alpha \beta} X_{k j} \boldsymbol{\lambda}^{k}_{\rho} {G}^{-1} \nabla^{\beta \rho}{G_{i}\,^{j}} - \frac{3}{160}{\rm i} G_{j k} (\Gamma_{a})_{\alpha \beta} \boldsymbol{W} X^{j}\,_{l} {G}^{-3} \nabla^{\beta \rho}{G_{i}\,^{l}} \varphi^{k}_{\rho}+\frac{3}{80}G_{i j} (\Gamma_{a})_{\alpha}{}^{\beta} \boldsymbol{\lambda}^{j}_{\lambda} {G}^{-1} \nabla^{\lambda \rho}{F_{\beta \rho}}+\frac{3}{16}G_{i j} (\Gamma_{a})_{\alpha \lambda} \boldsymbol{\lambda}^{j \beta} {G}^{-1} \nabla^{\lambda \rho}{F_{\beta \rho}} - \frac{3}{160}G_{i j} (\Gamma_{a})_{\alpha \beta} \boldsymbol{\lambda}_{k \rho} {G}^{-1} \nabla^{\beta \rho}{X^{j k}}+\frac{3}{32}G_{i j} (\Gamma_{a})_{\alpha \beta} \boldsymbol{\lambda}^{j}_{\rho} {G}^{-1} \nabla_{\lambda}\,^{\rho}{\nabla^{\beta \lambda}{W}}+\frac{21}{160}G_{i j} (\Gamma_{a})_{\alpha \beta} \boldsymbol{\lambda}^{j}_{\rho} {G}^{-1} \nabla^{\beta}\,_{\lambda}{\nabla^{\lambda \rho}{W}}+\frac{3}{320}G_{i j} (\Gamma_{a})_{\alpha}{}^{\beta} \boldsymbol{\lambda}^{j}_{\beta} {G}^{-1} \nabla_{\rho \lambda}{\nabla^{\rho \lambda}{W}} - \frac{3}{160}G_{i j} (\Gamma_{a})_{\alpha}{}^{\beta} \boldsymbol{W} {G}^{-1} \nabla_{\rho \lambda}{\nabla^{\rho \lambda}{\lambda^{j}_{\beta}}} - \frac{3}{160}{\rm i} (\Gamma_{a})_{\alpha}{}^{\beta} \lambda^{\rho}_{j} \boldsymbol{\lambda}_{k \rho} {G}^{-3} \varphi_{i}^{\lambda} \varphi^{j}_{\lambda} \varphi^{k}_{\beta}+\frac{3}{80}{\rm i} (\Gamma_{a})_{\alpha}{}^{\beta} \lambda^{\rho}_{j} \boldsymbol{\lambda}_{k \rho} {G}^{-3} \varphi_{i \beta} \varphi^{j \lambda} \varphi^{k}_{\lambda}+\frac{3}{160}{\rm i} (\Gamma_{a})_{\alpha}{}^{\beta} \lambda^{\rho}_{j} \boldsymbol{\lambda}_{k \rho} {G}^{-3} \varphi_{i}^{\lambda} \varphi^{j}_{\beta} \varphi^{k}_{\lambda}+\frac{3}{80}{\rm i} G_{j k} (\Gamma_{a})_{\alpha}{}^{\beta} F_{\beta}\,^{\rho} \boldsymbol{\lambda}^{j}_{\rho} {G}^{-3} \varphi_{i}^{\lambda} \varphi^{k}_{\lambda}+\frac{3}{40}{\rm i} G_{j k} (\Gamma_{a})_{\alpha}{}^{\lambda} F^{\beta \rho} \boldsymbol{\lambda}^{j}_{\beta} {G}^{-3} \varphi_{i \lambda} \varphi^{k}_{\rho}+\frac{3}{80}{\rm i} G_{j k} (\Gamma_{a})_{\alpha}{}^{\lambda} F^{\beta \rho} \boldsymbol{\lambda}^{j}_{\beta} {G}^{-3} \varphi_{i \rho} \varphi^{k}_{\lambda}%
 - \frac{3}{160}{\rm i} G_{j k} (\Gamma_{a})_{\alpha}{}^{\beta} X^{j}\,_{l} \boldsymbol{\lambda}^{k}_{\beta} {G}^{-3} \varphi_{i}^{\rho} \varphi^{l}_{\rho} - \frac{3}{160}{\rm i} G_{j k} (\Gamma_{a})_{\alpha}{}^{\beta} X^{j}\,_{l} \boldsymbol{\lambda}^{k \rho} {G}^{-3} \varphi_{i \beta} \varphi^{l}_{\rho} - \frac{3}{160}{\rm i} G_{j k} (\Gamma_{a})_{\alpha \beta} \boldsymbol{\lambda}^{j}_{\rho} {G}^{-3} \nabla^{\beta \rho}{W} \varphi_{i}^{\lambda} \varphi^{k}_{\lambda} - \frac{3}{160}{\rm i} G_{j k} (\Gamma_{a})_{\alpha}{}^{\beta} \boldsymbol{\lambda}^{j}_{\rho} {G}^{-3} \nabla^{\rho \lambda}{W} \varphi_{i \lambda} \varphi^{k}_{\beta}+\frac{3}{80}{\rm i} G_{j k} (\Gamma_{a})_{\alpha}{}^{\beta} \mathbf{F}_{\beta}\,^{\rho} \lambda^{j}_{\rho} {G}^{-3} \varphi_{i}^{\lambda} \varphi^{k}_{\lambda}+\frac{3}{40}{\rm i} G_{j k} (\Gamma_{a})_{\alpha}{}^{\lambda} \mathbf{F}^{\beta \rho} \lambda^{j}_{\beta} {G}^{-3} \varphi_{i \lambda} \varphi^{k}_{\rho}+\frac{3}{80}{\rm i} G_{j k} (\Gamma_{a})_{\alpha}{}^{\lambda} \mathbf{F}^{\beta \rho} \lambda^{j}_{\beta} {G}^{-3} \varphi_{i \rho} \varphi^{k}_{\lambda} - \frac{3}{160}{\rm i} G_{j k} (\Gamma_{a})_{\alpha}{}^{\beta} \mathbf{X}^{j}\,_{l} \lambda^{k}_{\beta} {G}^{-3} \varphi_{i}^{\rho} \varphi^{l}_{\rho} - \frac{3}{160}{\rm i} G_{j k} (\Gamma_{a})_{\alpha}{}^{\beta} \mathbf{X}^{j}\,_{l} \lambda^{k \rho} {G}^{-3} \varphi_{i \beta} \varphi^{l}_{\rho} - \frac{3}{160}{\rm i} G_{j k} (\Gamma_{a})_{\alpha \beta} \lambda^{j}_{\rho} {G}^{-3} \nabla^{\beta \rho}{\boldsymbol{W}} \varphi_{i}^{\lambda} \varphi^{k}_{\lambda} - \frac{3}{160}{\rm i} G_{j k} (\Gamma_{a})_{\alpha}{}^{\beta} \lambda^{j}_{\rho} {G}^{-3} \nabla^{\rho \lambda}{\boldsymbol{W}} \varphi_{i \lambda} \varphi^{k}_{\beta} - \frac{9}{160}{\rm i} G_{j k} G_{l m} (\Gamma_{a})_{\alpha}{}^{\beta} \lambda^{j \rho} \boldsymbol{\lambda}^{k}_{\rho} {G}^{-5} \varphi_{i}^{\lambda} \varphi^{l}_{\beta} \varphi^{m}_{\lambda} - \frac{9}{160}{\rm i} G_{j k} G_{l m} (\Gamma_{a})_{\alpha}{}^{\beta} \lambda^{j \rho} \boldsymbol{\lambda}^{k}_{\rho} {G}^{-5} \varphi_{i \beta} \varphi^{l \lambda} \varphi^{m}_{\lambda} - \frac{3}{64}G_{j k} (\Gamma_{a})_{\alpha \beta} \lambda^{j \rho} \boldsymbol{\lambda}^{k}_{\rho} {G}^{-3} \nabla^{\beta \lambda}{G_{i l}} \varphi^{l}_{\lambda} - \frac{3}{80}{\rm i} G_{i j} (\Gamma_{a})_{\alpha}{}^{\beta} F_{\beta}\,^{\rho} \boldsymbol{\lambda}_{k \rho} {G}^{-3} \varphi^{j \lambda} \varphi^{k}_{\lambda}+\frac{3}{80}{\rm i} G_{i j} (\Gamma_{a})_{\alpha}{}^{\lambda} F^{\beta \rho} \boldsymbol{\lambda}_{k \beta} {G}^{-3} \varphi^{j}_{\rho} \varphi^{k}_{\lambda} - \frac{3}{160}{\rm i} G_{j k} (\Gamma_{a})_{\alpha}{}^{\beta} X_{i}\,^{j} \boldsymbol{\lambda}_{l \beta} {G}^{-3} \varphi^{k \rho} \varphi^{l}_{\rho} - \frac{3}{80}{\rm i} G_{j k} (\Gamma_{a})_{\alpha}{}^{\beta} X_{i}\,^{j} \boldsymbol{\lambda}_{l}^{\rho} {G}^{-3} \varphi^{k}_{\beta} \varphi^{l}_{\rho} - \frac{3}{160}{\rm i} G_{j k} (\Gamma_{a})_{\alpha}{}^{\beta} X_{i}\,^{j} \boldsymbol{\lambda}_{l}^{\rho} {G}^{-3} \varphi^{k}_{\rho} \varphi^{l}_{\beta}+\frac{3}{160}{\rm i} G_{i j} (\Gamma_{a})_{\alpha \beta} \boldsymbol{\lambda}_{k \rho} {G}^{-3} \nabla^{\beta \rho}{W} \varphi^{j \lambda} \varphi^{k}_{\lambda}%
 - \frac{3}{160}{\rm i} G_{i j} (\Gamma_{a})_{\alpha}{}^{\beta} \boldsymbol{\lambda}_{k \rho} {G}^{-3} \nabla^{\rho \lambda}{W} \varphi^{j}_{\lambda} \varphi^{k}_{\beta} - \frac{3}{80}{\rm i} G_{j k} (\Gamma_{a})_{\alpha}{}^{\lambda} \mathbf{F}^{\beta \rho} \lambda_{i \beta} {G}^{-3} \varphi^{j}_{\lambda} \varphi^{k}_{\rho} - \frac{3}{320}{\rm i} G_{j k} (\Gamma_{a})_{\alpha \beta} \lambda_{i \rho} {G}^{-3} \nabla^{\beta \rho}{\boldsymbol{W}} \varphi^{j \lambda} \varphi^{k}_{\lambda}+\frac{3}{160}{\rm i} G_{j k} (\Gamma_{a})_{\alpha}{}^{\beta} \lambda_{i \rho} {G}^{-3} \nabla^{\rho \lambda}{\boldsymbol{W}} \varphi^{j}_{\beta} \varphi^{k}_{\lambda} - \frac{3}{80}{\rm i} G_{j k} (\Gamma_{a})_{\alpha}{}^{\lambda} F^{\beta \rho} \boldsymbol{\lambda}_{i \beta} {G}^{-3} \varphi^{j}_{\lambda} \varphi^{k}_{\rho} - \frac{3}{320}{\rm i} G_{j k} (\Gamma_{a})_{\alpha \beta} \boldsymbol{\lambda}_{i \rho} {G}^{-3} \nabla^{\beta \rho}{W} \varphi^{j \lambda} \varphi^{k}_{\lambda}+\frac{3}{160}{\rm i} G_{j k} (\Gamma_{a})_{\alpha}{}^{\beta} \boldsymbol{\lambda}_{i \rho} {G}^{-3} \nabla^{\rho \lambda}{W} \varphi^{j}_{\beta} \varphi^{k}_{\lambda} - \frac{3}{80}{\rm i} G_{i j} (\Gamma_{a})_{\alpha}{}^{\beta} \mathbf{F}_{\beta}\,^{\rho} \lambda_{k \rho} {G}^{-3} \varphi^{j \lambda} \varphi^{k}_{\lambda}+\frac{3}{80}{\rm i} G_{i j} (\Gamma_{a})_{\alpha}{}^{\lambda} \mathbf{F}^{\beta \rho} \lambda_{k \beta} {G}^{-3} \varphi^{j}_{\rho} \varphi^{k}_{\lambda} - \frac{3}{160}{\rm i} G_{j k} (\Gamma_{a})_{\alpha}{}^{\beta} \mathbf{X}_{i}\,^{j} \lambda_{l \beta} {G}^{-3} \varphi^{k \rho} \varphi^{l}_{\rho} - \frac{3}{80}{\rm i} G_{j k} (\Gamma_{a})_{\alpha}{}^{\beta} \mathbf{X}_{i}\,^{j} \lambda^{\rho}_{l} {G}^{-3} \varphi^{k}_{\beta} \varphi^{l}_{\rho} - \frac{3}{160}{\rm i} G_{j k} (\Gamma_{a})_{\alpha}{}^{\beta} \mathbf{X}_{i}\,^{j} \lambda^{\rho}_{l} {G}^{-3} \varphi^{k}_{\rho} \varphi^{l}_{\beta}+\frac{3}{160}{\rm i} G_{i j} (\Gamma_{a})_{\alpha \beta} \lambda_{k \rho} {G}^{-3} \nabla^{\beta \rho}{\boldsymbol{W}} \varphi^{j \lambda} \varphi^{k}_{\lambda} - \frac{3}{160}{\rm i} G_{i j} (\Gamma_{a})_{\alpha}{}^{\beta} \lambda_{k \rho} {G}^{-3} \nabla^{\rho \lambda}{\boldsymbol{W}} \varphi^{j}_{\lambda} \varphi^{k}_{\beta}+\frac{9}{160}{\rm i} G_{i j} G_{k l} (\Gamma_{a})_{\alpha}{}^{\beta} \lambda^{k \rho} \boldsymbol{\lambda}_{m \rho} {G}^{-5} \varphi^{j \lambda} \varphi^{l}_{\beta} \varphi^{m}_{\lambda}+\frac{9}{160}{\rm i} G_{i j} G_{k l} (\Gamma_{a})_{\alpha}{}^{\beta} \lambda^{k \rho} \boldsymbol{\lambda}_{m \rho} {G}^{-5} \varphi^{j \lambda} \varphi^{l}_{\lambda} \varphi^{m}_{\beta}+\frac{3}{320}\mathcal{H}^{\beta \rho} G_{i j} (\Gamma_{a})_{\alpha \beta} \lambda^{j \lambda} \boldsymbol{\lambda}_{k \lambda} {G}^{-3} \varphi^{k}_{\rho} - \frac{3}{320}G_{i j} (\Gamma_{a})_{\alpha \beta} \lambda^{\rho}_{k} \boldsymbol{\lambda}_{l \rho} {G}^{-3} \nabla^{\beta \lambda}{G^{j k}} \varphi^{l}_{\lambda}+\frac{9}{160}{\rm i} G_{i j} G_{k l} (\Gamma_{a})_{\alpha}{}^{\beta} \lambda^{\rho}_{m} \boldsymbol{\lambda}^{k}_{\rho} {G}^{-5} \varphi^{j \lambda} \varphi^{l}_{\beta} \varphi^{m}_{\lambda}+\frac{9}{160}{\rm i} G_{i j} G_{k l} (\Gamma_{a})_{\alpha}{}^{\beta} \lambda^{\rho}_{m} \boldsymbol{\lambda}^{k}_{\rho} {G}^{-5} \varphi^{j \lambda} \varphi^{l}_{\lambda} \varphi^{m}_{\beta}%
+\frac{3}{320}\mathcal{H}^{\beta \rho} G_{i j} (\Gamma_{a})_{\alpha \beta} \lambda^{\lambda}_{k} \boldsymbol{\lambda}^{j}_{\lambda} {G}^{-3} \varphi^{k}_{\rho} - \frac{3}{320}G_{i j} (\Gamma_{a})_{\alpha \beta} \lambda^{\rho}_{l} \boldsymbol{\lambda}_{k \rho} {G}^{-3} \nabla^{\beta \lambda}{G^{j k}} \varphi^{l}_{\lambda}+\frac{3}{80}{\rm i} G_{i j} (\Gamma_{a})_{\alpha}{}^{\lambda} F^{\beta \rho} \boldsymbol{\lambda}^{j}_{\beta} {G}^{-3} \varphi_{k \lambda} \varphi^{k}_{\rho} - \frac{3}{320}G_{i j} G_{k l} (\Gamma_{a})_{\alpha \beta} \boldsymbol{\lambda}^{j \rho} {G}^{-3} \nabla^{\beta \lambda}{\lambda^{k}_{\rho}} \varphi^{l}_{\lambda}+\frac{3}{320}G_{i j} G_{k l} (\Gamma_{a})_{\alpha \beta} \boldsymbol{\lambda}^{j}_{\lambda} {G}^{-3} \nabla^{\beta \lambda}{\lambda^{k \rho}} \varphi^{l}_{\rho} - \frac{3}{320}G_{i j} G_{k l} (\Gamma_{a})_{\alpha}{}^{\beta} \boldsymbol{\lambda}^{j}_{\rho} {G}^{-3} \nabla^{\rho \lambda}{\lambda^{k}_{\beta}} \varphi^{l}_{\lambda} - \frac{3}{320}G_{i j} G_{k l} (\Gamma_{a})_{\alpha}{}^{\beta} W_{\beta}\,^{\rho} \lambda^{k \lambda} \boldsymbol{\lambda}^{j}_{\rho} {G}^{-3} \varphi^{l}_{\lambda} - \frac{3}{320}G_{i j} G_{k l} (\Gamma_{a})_{\alpha}{}^{\lambda} W^{\beta \rho} \lambda^{k}_{\lambda} \boldsymbol{\lambda}^{j}_{\beta} {G}^{-3} \varphi^{l}_{\rho} - \frac{3}{160}G_{i j} G_{k l} (\Gamma_{a})_{\alpha}{}^{\beta} \boldsymbol{\lambda}^{j}_{\beta} {G}^{-3} \nabla^{\rho \lambda}{\lambda^{k}_{\rho}} \varphi^{l}_{\lambda}+\frac{3}{20}{\rm i} (\Gamma_{a})_{\alpha}{}^{\lambda} F^{\beta \rho} \mathbf{F}_{\lambda \beta} {G}^{-1} \varphi_{i \rho} - \frac{3}{20}{\rm i} (\Gamma_{a})_{\alpha}{}^{\beta} F_{\beta}\,^{\rho} \mathbf{F}_{\rho}\,^{\lambda} {G}^{-1} \varphi_{i \lambda}+\frac{3}{10}{\rm i} (\Gamma_{a})_{\alpha}{}^{\lambda} F^{\beta \rho} \mathbf{F}_{\beta \rho} {G}^{-1} \varphi_{i \lambda} - \frac{3}{80}{\rm i} G_{i j} G_{k l} (\Gamma_{a})_{\alpha}{}^{\beta} \mathbf{X}^{j k} F_{\beta}\,^{\rho} {G}^{-3} \varphi^{l}_{\rho}+\frac{3}{40}{\rm i} (\Gamma_{a})_{\alpha \lambda} F^{\beta}\,_{\rho} {G}^{-1} \nabla^{\lambda \rho}{\boldsymbol{W}} \varphi_{i \beta} - \frac{3}{40}{\rm i} (\Gamma_{a})_{\alpha}{}^{\beta} F_{\beta \rho} {G}^{-1} \nabla^{\rho \lambda}{\boldsymbol{W}} \varphi_{i \lambda} - \frac{3}{160}{\rm i} G_{j k} (\Gamma_{a})_{\alpha}{}^{\beta} X_{i}\,^{j} \boldsymbol{\lambda}^{k \rho} {G}^{-3} \varphi_{l \beta} \varphi^{l}_{\rho} - \frac{3}{80}{\rm i} (\Gamma_{a})_{\alpha}{}^{\beta} X_{i j} \mathbf{F}_{\beta}\,^{\rho} {G}^{-1} \varphi^{j}_{\rho} - \frac{3}{160}{\rm i} G_{i j} (\Gamma_{a})_{\alpha}{}^{\beta} \boldsymbol{\lambda}^{j}_{\rho} {G}^{-3} \nabla^{\rho \lambda}{W} \varphi_{k \beta} \varphi^{k}_{\lambda} - \frac{3}{40}{\rm i} (\Gamma_{a})_{\alpha}{}^{\rho} \mathbf{F}_{\beta \rho} {G}^{-1} \nabla^{\beta \lambda}{W} \varphi_{i \lambda}+\frac{3}{40}{\rm i} (\Gamma_{a})_{\alpha \lambda} \mathbf{F}_{\beta}\,^{\rho} {G}^{-1} \nabla^{\lambda \beta}{W} \varphi_{i \rho}%
 - \frac{3}{160}{\rm i} G_{i j} G_{k l} (\Gamma_{a})_{\alpha \beta} \mathbf{X}^{j k} {G}^{-3} \nabla^{\beta \rho}{W} \varphi^{l}_{\rho} - \frac{3}{80}{\rm i} (\Gamma_{a})_{\alpha \beta} {G}^{-1} \nabla_{\rho}\,^{\lambda}{W} \nabla^{\beta \rho}{\boldsymbol{W}} \varphi_{i \lambda}+\frac{3}{80}{\rm i} (\Gamma_{a})_{\alpha \beta} {G}^{-1} \nabla^{\beta}\,_{\rho}{W} \nabla^{\rho \lambda}{\boldsymbol{W}} \varphi_{i \lambda}+\frac{3}{40}{\rm i} (\Gamma_{a})_{\alpha}{}^{\beta} {G}^{-1} \nabla_{\rho \lambda}{W} \nabla^{\rho \lambda}{\boldsymbol{W}} \varphi_{i \beta} - \frac{3}{640}G_{i j} G_{k l} (\Gamma_{a})_{\alpha}{}^{\beta} W_{\beta}\,^{\rho} \lambda^{k}_{\rho} \boldsymbol{\lambda}^{j \lambda} {G}^{-3} \varphi^{l}_{\lambda} - \frac{3}{640}G_{i j} G_{k l} (\Gamma_{a})_{\alpha}{}^{\lambda} W^{\beta \rho} \lambda^{k}_{\beta} \boldsymbol{\lambda}^{j}_{\lambda} {G}^{-3} \varphi^{l}_{\rho} - \frac{3}{80}\mathcal{H}^{\lambda \beta} (\Gamma_{a})_{\alpha \lambda} F_{\beta}\,^{\rho} \boldsymbol{\lambda}_{i \rho} {G}^{-1}+\frac{3}{160}G_{i j} G_{k l} (\Gamma_{a})_{\alpha \lambda} F^{\beta}\,_{\rho} \boldsymbol{\lambda}^{k}_{\beta} {G}^{-3} \nabla^{\lambda \rho}{G^{j l}} - \frac{9}{160}{\rm i} G_{i j} G_{k l} G_{m n} (\Gamma_{a})_{\alpha}{}^{\beta} X^{k m} \boldsymbol{\lambda}^{l \rho} {G}^{-5} \varphi^{j}_{\rho} \varphi^{n}_{\beta}+\frac{9}{160}{\rm i} G_{i j} G_{k l} G_{m n} (\Gamma_{a})_{\alpha}{}^{\beta} X^{k m} \boldsymbol{\lambda}^{l}_{\beta} {G}^{-5} \varphi^{j \rho} \varphi^{n}_{\rho}+\frac{3}{320}\mathcal{H}^{\beta \rho} G_{i j} G_{k l} (\Gamma_{a})_{\alpha \beta} X^{j k} \boldsymbol{\lambda}^{l}_{\rho} {G}^{-3} - \frac{3}{320}G_{i j} G_{k l} (\Gamma_{a})_{\alpha \beta} X^{k}\,_{m} \boldsymbol{\lambda}^{l}_{\rho} {G}^{-3} \nabla^{\beta \rho}{G^{j m}} - \frac{3}{160}\mathcal{H}_{\rho}\,^{\beta} (\Gamma_{a})_{\alpha \beta} \boldsymbol{\lambda}_{i \lambda} {G}^{-1} \nabla^{\rho \lambda}{W}+\frac{3}{320}G_{i j} G_{k l} (\Gamma_{a})_{\alpha \beta} \boldsymbol{\lambda}^{k}_{\rho} {G}^{-3} \nabla_{\lambda}\,^{\rho}{W} \nabla^{\beta \lambda}{G^{j l}}+\frac{3}{80}{\rm i} G_{i j} (\Gamma_{a})_{\alpha}{}^{\lambda} \mathbf{F}^{\beta \rho} \lambda^{j}_{\beta} {G}^{-3} \varphi_{k \lambda} \varphi^{k}_{\rho} - \frac{3}{320}G_{i j} G_{k l} (\Gamma_{a})_{\alpha \beta} \lambda^{j \rho} {G}^{-3} \nabla^{\beta \lambda}{\boldsymbol{\lambda}^{k}_{\rho}} \varphi^{l}_{\lambda}+\frac{3}{320}G_{i j} G_{k l} (\Gamma_{a})_{\alpha \beta} \lambda^{j}_{\rho} {G}^{-3} \nabla^{\beta \rho}{\boldsymbol{\lambda}^{k \lambda}} \varphi^{l}_{\lambda} - \frac{3}{320}G_{i j} G_{k l} (\Gamma_{a})_{\alpha}{}^{\beta} \lambda^{j}_{\rho} {G}^{-3} \nabla^{\rho \lambda}{\boldsymbol{\lambda}^{k}_{\beta}} \varphi^{l}_{\lambda}+\frac{3}{320}G_{i j} G_{k l} (\Gamma_{a})_{\alpha}{}^{\beta} W_{\beta}\,^{\rho} \lambda^{j}_{\rho} \boldsymbol{\lambda}^{k \lambda} {G}^{-3} \varphi^{l}_{\lambda}+\frac{3}{320}G_{i j} G_{k l} (\Gamma_{a})_{\alpha}{}^{\lambda} W^{\beta \rho} \lambda^{j}_{\beta} \boldsymbol{\lambda}^{k}_{\lambda} {G}^{-3} \varphi^{l}_{\rho}%
 - \frac{3}{160}G_{i j} G_{k l} (\Gamma_{a})_{\alpha}{}^{\beta} \lambda^{j}_{\beta} {G}^{-3} \nabla^{\rho \lambda}{\boldsymbol{\lambda}^{k}_{\rho}} \varphi^{l}_{\lambda} - \frac{3}{80}{\rm i} G_{i j} G_{k l} (\Gamma_{a})_{\alpha}{}^{\beta} X^{j k} \mathbf{F}_{\beta}\,^{\rho} {G}^{-3} \varphi^{l}_{\rho} - \frac{3}{160}{\rm i} G_{j k} (\Gamma_{a})_{\alpha}{}^{\beta} \mathbf{X}_{i}\,^{j} \lambda^{k \rho} {G}^{-3} \varphi_{l \beta} \varphi^{l}_{\rho} - \frac{3}{80}{\rm i} (\Gamma_{a})_{\alpha}{}^{\beta} \mathbf{X}_{i j} F_{\beta}\,^{\rho} {G}^{-1} \varphi^{j}_{\rho} - \frac{3}{160}{\rm i} G_{i j} (\Gamma_{a})_{\alpha}{}^{\beta} \lambda^{j}_{\rho} {G}^{-3} \nabla^{\rho \lambda}{\boldsymbol{W}} \varphi_{k \beta} \varphi^{k}_{\lambda} - \frac{3}{160}{\rm i} G_{i j} G_{k l} (\Gamma_{a})_{\alpha \beta} X^{j k} {G}^{-3} \nabla^{\beta \rho}{\boldsymbol{W}} \varphi^{l}_{\rho}+\frac{3}{640}G_{i j} G_{k l} (\Gamma_{a})_{\alpha}{}^{\beta} W_{\beta}\,^{\rho} \lambda^{j \lambda} \boldsymbol{\lambda}^{k}_{\rho} {G}^{-3} \varphi^{l}_{\lambda}+\frac{3}{640}G_{i j} G_{k l} (\Gamma_{a})_{\alpha}{}^{\lambda} W^{\beta \rho} \lambda^{j}_{\lambda} \boldsymbol{\lambda}^{k}_{\beta} {G}^{-3} \varphi^{l}_{\rho} - \frac{3}{80}\mathcal{H}^{\lambda \beta} (\Gamma_{a})_{\alpha \lambda} \mathbf{F}_{\beta}\,^{\rho} \lambda_{i \rho} {G}^{-1}+\frac{3}{160}G_{i j} G_{k l} (\Gamma_{a})_{\alpha \lambda} \mathbf{F}_{\beta}\,^{\rho} \lambda^{k}_{\rho} {G}^{-3} \nabla^{\lambda \beta}{G^{j l}} - \frac{9}{160}{\rm i} G_{i j} G_{k l} G_{m n} (\Gamma_{a})_{\alpha}{}^{\beta} \mathbf{X}^{k m} \lambda^{l \rho} {G}^{-5} \varphi^{j}_{\rho} \varphi^{n}_{\beta}+\frac{9}{160}{\rm i} G_{i j} G_{k l} G_{m n} (\Gamma_{a})_{\alpha}{}^{\beta} \mathbf{X}^{k m} \lambda^{l}_{\beta} {G}^{-5} \varphi^{j \rho} \varphi^{n}_{\rho}+\frac{3}{320}\mathcal{H}^{\beta \rho} G_{i j} G_{k l} (\Gamma_{a})_{\alpha \beta} \mathbf{X}^{j k} \lambda^{l}_{\rho} {G}^{-3} - \frac{3}{320}G_{i j} G_{k l} (\Gamma_{a})_{\alpha \beta} \mathbf{X}^{k}\,_{m} \lambda^{l}_{\rho} {G}^{-3} \nabla^{\beta \rho}{G^{j m}} - \frac{3}{160}\mathcal{H}_{\rho}\,^{\beta} (\Gamma_{a})_{\alpha \beta} \lambda_{i \lambda} {G}^{-1} \nabla^{\rho \lambda}{\boldsymbol{W}}+\frac{3}{320}G_{i j} G_{k l} (\Gamma_{a})_{\alpha \beta} \lambda^{k}_{\rho} {G}^{-3} \nabla_{\lambda}\,^{\rho}{\boldsymbol{W}} \nabla^{\beta \lambda}{G^{j l}} - \frac{9}{160}{\rm i} G_{i j} G_{k l} (\Gamma_{a})_{\alpha}{}^{\beta} \lambda^{k \rho} \boldsymbol{\lambda}^{l}_{\rho} {G}^{-5} \varphi^{j \lambda} \varphi_{m \beta} \varphi^{m}_{\lambda}+\frac{9}{320}G_{i j} G_{k l} G_{m n} (\Gamma_{a})_{\alpha \beta} \lambda^{k \rho} \boldsymbol{\lambda}^{l}_{\rho} {G}^{-5} \nabla^{\beta \lambda}{G^{j m}} \varphi^{n}_{\lambda}+\frac{3}{320}G_{j k} (\Gamma_{a})_{\alpha \beta} \lambda^{j \rho} \boldsymbol{\lambda}_{l \rho} {G}^{-3} \nabla^{\beta \lambda}{G_{i}\,^{k}} \varphi^{l}_{\lambda}+\frac{3}{320}G_{j k} (\Gamma_{a})_{\alpha \beta} \lambda^{\rho}_{l} \boldsymbol{\lambda}^{j}_{\rho} {G}^{-3} \nabla^{\beta \lambda}{G_{i}\,^{k}} \varphi^{l}_{\lambda}%
+\frac{9}{160}(\Gamma_{a})_{\alpha \lambda} F^{\beta}\,_{\rho} \boldsymbol{\lambda}_{j \beta} {G}^{-1} \nabla^{\lambda \rho}{G_{i}\,^{j}}+\frac{3}{320}G_{j k} G_{l m} (\Gamma_{a})_{\alpha \beta} X^{j l} \boldsymbol{\lambda}^{k}_{\rho} {G}^{-3} \nabla^{\beta \rho}{G_{i}\,^{m}}+\frac{9}{320}(\Gamma_{a})_{\alpha \beta} \boldsymbol{\lambda}_{j \rho} {G}^{-1} \nabla_{\lambda}\,^{\rho}{W} \nabla^{\beta \lambda}{G_{i}\,^{j}}+\frac{9}{160}(\Gamma_{a})_{\alpha \lambda} \mathbf{F}_{\beta}\,^{\rho} \lambda_{j \rho} {G}^{-1} \nabla^{\lambda \beta}{G_{i}\,^{j}}+\frac{3}{320}G_{j k} G_{l m} (\Gamma_{a})_{\alpha \beta} \mathbf{X}^{j l} \lambda^{k}_{\rho} {G}^{-3} \nabla^{\beta \rho}{G_{i}\,^{m}}+\frac{9}{320}(\Gamma_{a})_{\alpha \beta} \lambda_{j \rho} {G}^{-1} \nabla_{\lambda}\,^{\rho}{\boldsymbol{W}} \nabla^{\beta \lambda}{G_{i}\,^{j}} - \frac{3}{320}(\Gamma_{a})_{\alpha \beta} \lambda^{\rho}_{j} {G}^{-1} \nabla^{\beta \lambda}{\boldsymbol{\lambda}^{j}_{\rho}} \varphi_{i \lambda}+\frac{3}{320}(\Gamma_{a})_{\alpha \beta} \lambda_{j \rho} {G}^{-1} \nabla^{\beta \rho}{\boldsymbol{\lambda}^{j \lambda}} \varphi_{i \lambda} - \frac{3}{320}(\Gamma_{a})_{\alpha}{}^{\beta} \lambda_{j \rho} {G}^{-1} \nabla^{\rho \lambda}{\boldsymbol{\lambda}^{j}_{\beta}} \varphi_{i \lambda}+\frac{9}{640}(\Gamma_{a})_{\alpha}{}^{\beta} W_{\beta}\,^{\rho} \lambda_{j \rho} \boldsymbol{\lambda}^{j \lambda} {G}^{-1} \varphi_{i \lambda} - \frac{3}{80}{\rm i} G_{j k} (\Gamma_{a})_{\alpha}{}^{\beta} \lambda^{j}_{\rho} {G}^{-3} \nabla^{\rho \lambda}{\boldsymbol{W}} \varphi_{i \beta} \varphi^{k}_{\lambda}+\frac{9}{640}(\Gamma_{a})_{\alpha}{}^{\beta} W_{\beta}\,^{\rho} \lambda^{\lambda}_{j} \boldsymbol{\lambda}^{j}_{\rho} {G}^{-1} \varphi_{i \lambda} - \frac{3}{320}G_{j k} (\Gamma_{a})_{\alpha \beta} \lambda^{j \rho} \boldsymbol{\lambda}_{l \rho} {G}^{-3} \nabla^{\beta \lambda}{G_{i}\,^{l}} \varphi^{k}_{\lambda} - \frac{3}{320}(\Gamma_{a})_{\alpha \beta} \boldsymbol{\lambda}_{j}^{\rho} {G}^{-1} \nabla^{\beta \lambda}{\lambda^{j}_{\rho}} \varphi_{i \lambda}+\frac{3}{320}(\Gamma_{a})_{\alpha \beta} \boldsymbol{\lambda}_{j \lambda} {G}^{-1} \nabla^{\beta \lambda}{\lambda^{j \rho}} \varphi_{i \rho} - \frac{3}{320}(\Gamma_{a})_{\alpha}{}^{\beta} \boldsymbol{\lambda}_{j \rho} {G}^{-1} \nabla^{\rho \lambda}{\lambda^{j}_{\beta}} \varphi_{i \lambda} - \frac{3}{80}{\rm i} G_{j k} (\Gamma_{a})_{\alpha}{}^{\beta} \boldsymbol{\lambda}^{j}_{\rho} {G}^{-3} \nabla^{\rho \lambda}{W} \varphi_{i \beta} \varphi^{k}_{\lambda} - \frac{3}{320}G_{j k} (\Gamma_{a})_{\alpha \beta} \lambda^{\rho}_{l} \boldsymbol{\lambda}^{j}_{\rho} {G}^{-3} \nabla^{\beta \lambda}{G_{i}\,^{l}} \varphi^{k}_{\lambda}+\frac{9}{320}(\Gamma_{a})_{\alpha \beta} \boldsymbol{\lambda}_{j}^{\rho} {G}^{-1} \nabla^{\beta \lambda}{\lambda_{i \rho}} \varphi^{j}_{\lambda}+\frac{9}{320}(\Gamma_{a})_{\alpha \beta} \boldsymbol{\lambda}_{j \lambda} {G}^{-1} \nabla^{\beta \lambda}{\lambda^{\rho}_{i}} \varphi^{j}_{\rho}%
 - \frac{9}{320}(\Gamma_{a})_{\alpha}{}^{\beta} \boldsymbol{\lambda}_{j \rho} {G}^{-1} \nabla^{\rho \lambda}{\lambda_{i \beta}} \varphi^{j}_{\lambda} - \frac{3}{20}G_{i j} (\Gamma_{a})_{\alpha \lambda} \mathbf{F}_{\beta}\,^{\rho} {G}^{-1} \nabla^{\lambda \beta}{\lambda^{j}_{\rho}} - \frac{9}{160}G_{i j} (\Gamma_{a})_{\alpha \beta} {G}^{-1} \nabla_{\lambda}\,^{\rho}{\boldsymbol{W}} \nabla^{\beta \lambda}{\lambda^{j}_{\rho}}+\frac{3}{320}G_{i j} (\Gamma_{a})_{\alpha}{}^{\beta} {G}^{-1} \nabla_{\rho \lambda}{\boldsymbol{W}} \nabla^{\rho \lambda}{\lambda^{j}_{\beta}}+\frac{3}{20}{\rm i} G_{i j} (\Gamma_{a})_{\alpha}{}^{\lambda} W \mathbf{F}^{\beta \rho} W_{\lambda \beta \rho}\,^{j} {G}^{-1} - \frac{3}{20}G_{i j} (\Gamma_{a})_{\alpha \lambda} F^{\beta}\,_{\rho} {G}^{-1} \nabla^{\lambda \rho}{\boldsymbol{\lambda}^{j}_{\beta}}+\frac{3}{20}{\rm i} G_{i j} (\Gamma_{a})_{\alpha}{}^{\lambda} \boldsymbol{W} F^{\beta \rho} W_{\lambda \beta \rho}\,^{j} {G}^{-1} - \frac{9}{160}G_{i j} (\Gamma_{a})_{\alpha \beta} {G}^{-1} \nabla_{\lambda}\,^{\rho}{W} \nabla^{\beta \lambda}{\boldsymbol{\lambda}^{j}_{\rho}}+\frac{3}{320}G_{i j} (\Gamma_{a})_{\alpha}{}^{\beta} {G}^{-1} \nabla_{\rho \lambda}{W} \nabla^{\rho \lambda}{\boldsymbol{\lambda}^{j}_{\beta}}+\frac{9}{320}(\Gamma_{a})_{\alpha \beta} \lambda^{\rho}_{j} {G}^{-1} \nabla^{\beta \lambda}{\boldsymbol{\lambda}_{i \rho}} \varphi^{j}_{\lambda}+\frac{9}{320}(\Gamma_{a})_{\alpha \beta} \lambda_{j \rho} {G}^{-1} \nabla^{\beta \rho}{\boldsymbol{\lambda}_{i}^{\lambda}} \varphi^{j}_{\lambda} - \frac{9}{320}(\Gamma_{a})_{\alpha}{}^{\beta} \lambda_{j \rho} {G}^{-1} \nabla^{\rho \lambda}{\boldsymbol{\lambda}_{i \beta}} \varphi^{j}_{\lambda}+\frac{3}{80}{\rm i} (\Gamma_{a})_{\alpha}{}^{\beta} \boldsymbol{W} \lambda^{\rho}_{j} W_{\beta \rho}\,^{\lambda}\,_{i} {G}^{-1} \varphi^{j}_{\lambda}+\frac{3}{80}{\rm i} (\Gamma_{a})_{\alpha}{}^{\beta} W \boldsymbol{\lambda}_{j}^{\rho} W_{\beta \rho}\,^{\lambda}\,_{i} {G}^{-1} \varphi^{j}_{\lambda}+\frac{3}{640}{\rm i} G_{i j} (\Gamma_{a})_{\alpha \beta} W \boldsymbol{W} {G}^{-3} \nabla^{\beta}\,_{\rho}{G^{j}\,_{k}} \nabla^{\rho \lambda}{\varphi^{k}_{\lambda}} - \frac{3}{320}{\rm i} G_{i j} (\Gamma_{a})_{\alpha \beta} W \boldsymbol{W} {G}^{-3} \nabla_{\rho}\,^{\lambda}{G^{j}\,_{k}} \nabla^{\beta \rho}{\varphi^{k}_{\lambda}} - \frac{3}{320}{\rm i} G_{i j} (\Gamma_{a})_{\alpha}{}^{\beta} W \boldsymbol{W} {G}^{-3} \nabla_{\rho \lambda}{G^{j}\,_{k}} \nabla^{\rho \lambda}{\varphi^{k}_{\beta}} - \frac{3}{640}{\rm i} G_{j k} (\Gamma_{a})_{\alpha \beta} W \boldsymbol{W} {G}^{-3} \nabla^{\beta}\,_{\rho}{G^{j k}} \nabla^{\rho \lambda}{\varphi_{i \lambda}}+\frac{3}{320}{\rm i} G_{j k} (\Gamma_{a})_{\alpha \beta} W \boldsymbol{W} {G}^{-3} \nabla_{\rho}\,^{\lambda}{G^{j k}} \nabla^{\beta \rho}{\varphi_{i \lambda}}+\frac{3}{320}{\rm i} G_{j k} (\Gamma_{a})_{\alpha}{}^{\beta} W \boldsymbol{W} {G}^{-3} \nabla_{\rho \lambda}{G^{j k}} \nabla^{\rho \lambda}{\varphi_{i \beta}}%
+\frac{9}{128}(\Gamma_{a})_{\beta \rho} \boldsymbol{W} \lambda_{i \alpha} {G}^{-1} \nabla^{\beta \rho}{F}+\frac{9}{128}F (\Gamma_{a})_{\beta \rho} \lambda_{i \alpha} {G}^{-1} \nabla^{\beta \rho}{\boldsymbol{W}} - \frac{9}{128}F (\Gamma_{a})_{\beta \rho} \boldsymbol{W} \lambda_{i \alpha} {G}^{-2} \nabla^{\beta \rho}{G}+\frac{9}{128}(\Gamma_{a})_{\beta \rho} W \boldsymbol{\lambda}_{i \alpha} {G}^{-1} \nabla^{\beta \rho}{F}+\frac{9}{128}F (\Gamma_{a})_{\beta \rho} \boldsymbol{\lambda}_{i \alpha} {G}^{-1} \nabla^{\beta \rho}{W} - \frac{9}{128}F (\Gamma_{a})_{\beta \rho} W \boldsymbol{\lambda}_{i \alpha} {G}^{-2} \nabla^{\beta \rho}{G}+\frac{9}{64}{\rm i} (\Gamma_{a})_{\beta \rho} \boldsymbol{W} {G}^{-1} \nabla^{\beta \rho}{W} \nabla_{\alpha}\,^{\lambda}{\varphi_{i \lambda}}+\frac{9}{64}{\rm i} (\Gamma_{a})_{\beta \rho} W {G}^{-1} \nabla^{\beta \rho}{\boldsymbol{W}} \nabla_{\alpha}\,^{\lambda}{\varphi_{i \lambda}} - \frac{9}{64}{\rm i} (\Gamma_{a})_{\beta \rho} W \boldsymbol{W} {G}^{-2} \nabla^{\beta \rho}{G} \nabla_{\alpha}\,^{\lambda}{\varphi_{i \lambda}}+\frac{9}{64}{\rm i} (\Gamma_{a})_{\beta \rho} W \boldsymbol{W} {G}^{-1} \nabla^{\beta \rho}{\nabla_{\alpha}\,^{\lambda}{\varphi_{i \lambda}}} - \frac{27}{128}{\rm i} (\Gamma_{a})_{\rho \lambda} \boldsymbol{W} W_{\alpha}\,^{\beta} {G}^{-1} \nabla^{\rho \lambda}{W} \varphi_{i \beta} - \frac{27}{128}{\rm i} (\Gamma_{a})_{\rho \lambda} W W_{\alpha}\,^{\beta} {G}^{-1} \nabla^{\rho \lambda}{\boldsymbol{W}} \varphi_{i \beta}+\frac{27}{128}{\rm i} (\Gamma_{a})_{\rho \lambda} W \boldsymbol{W} W_{\alpha}\,^{\beta} {G}^{-2} \nabla^{\rho \lambda}{G} \varphi_{i \beta}+\frac{81}{256}{\rm i} G_{i j} (\Gamma_{a})_{\beta \rho} W \boldsymbol{W} X^{j}_{\alpha} {G}^{-2} \nabla^{\beta \rho}{G}+\frac{9}{256}{\rm i} F G_{i j} (\Gamma_{a})_{\beta \rho} \boldsymbol{W} {G}^{-3} \nabla^{\beta \rho}{W} \varphi^{j}_{\alpha}+\frac{9}{256}{\rm i} F G_{i j} (\Gamma_{a})_{\beta \rho} W {G}^{-3} \nabla^{\beta \rho}{\boldsymbol{W}} \varphi^{j}_{\alpha} - \frac{27}{256}{\rm i} F G_{i j} (\Gamma_{a})_{\beta \rho} W \boldsymbol{W} {G}^{-4} \nabla^{\beta \rho}{G} \varphi^{j}_{\alpha} - \frac{9}{256}{\rm i} (\Gamma_{a})_{\beta \rho} \boldsymbol{W} \lambda_{i \alpha} {G}^{-3} \nabla^{\beta \rho}{G_{j k}} \varphi^{j \lambda} \varphi^{k}_{\lambda} - \frac{9}{256}{\rm i} G_{j k} (\Gamma_{a})_{\beta \rho} \lambda_{i \alpha} {G}^{-3} \nabla^{\beta \rho}{\boldsymbol{W}} \varphi^{j \lambda} \varphi^{k}_{\lambda}+\frac{27}{256}{\rm i} G_{j k} (\Gamma_{a})_{\beta \rho} \boldsymbol{W} \lambda_{i \alpha} {G}^{-4} \nabla^{\beta \rho}{G} \varphi^{j \lambda} \varphi^{k}_{\lambda}%
 - \frac{9}{128}{\rm i} G_{j k} (\Gamma_{a})_{\beta \rho} \boldsymbol{W} \lambda_{i \alpha} {G}^{-3} \nabla^{\beta \rho}{\varphi^{j \lambda}} \varphi^{k}_{\lambda} - \frac{9}{256}{\rm i} (\Gamma_{a})_{\beta \rho} W \boldsymbol{\lambda}_{i \alpha} {G}^{-3} \nabla^{\beta \rho}{G_{j k}} \varphi^{j \lambda} \varphi^{k}_{\lambda} - \frac{9}{256}{\rm i} G_{j k} (\Gamma_{a})_{\beta \rho} \boldsymbol{\lambda}_{i \alpha} {G}^{-3} \nabla^{\beta \rho}{W} \varphi^{j \lambda} \varphi^{k}_{\lambda}+\frac{27}{256}{\rm i} G_{j k} (\Gamma_{a})_{\beta \rho} W \boldsymbol{\lambda}_{i \alpha} {G}^{-4} \nabla^{\beta \rho}{G} \varphi^{j \lambda} \varphi^{k}_{\lambda} - \frac{9}{128}{\rm i} G_{j k} (\Gamma_{a})_{\beta \rho} W \boldsymbol{\lambda}_{i \alpha} {G}^{-3} \nabla^{\beta \rho}{\varphi^{j \lambda}} \varphi^{k}_{\lambda}+\frac{9}{128}(\Gamma_{a})_{\beta \rho} \boldsymbol{W} {G}^{-3} \nabla^{\beta \rho}{W} \varphi_{j \alpha} \varphi_{i}^{\lambda} \varphi^{j}_{\lambda}+\frac{9}{128}(\Gamma_{a})_{\beta \rho} W {G}^{-3} \nabla^{\beta \rho}{\boldsymbol{W}} \varphi_{j \alpha} \varphi_{i}^{\lambda} \varphi^{j}_{\lambda} - \frac{27}{128}(\Gamma_{a})_{\beta \rho} W \boldsymbol{W} {G}^{-4} \nabla^{\beta \rho}{G} \varphi_{j \alpha} \varphi_{i}^{\lambda} \varphi^{j}_{\lambda}+\frac{9}{128}(\Gamma_{a})_{\beta \rho} W \boldsymbol{W} {G}^{-3} \nabla^{\beta \rho}{\varphi_{j}^{\lambda}} \varphi^{j}_{\alpha} \varphi_{i \lambda} - \frac{9}{256}{\rm i} \mathcal{H}_{\alpha}\,^{\lambda} G_{i j} (\Gamma_{a})_{\beta \rho} \boldsymbol{W} {G}^{-3} \nabla^{\beta \rho}{W} \varphi^{j}_{\lambda} - \frac{9}{256}{\rm i} \mathcal{H}_{\alpha}\,^{\lambda} G_{i j} (\Gamma_{a})_{\beta \rho} W {G}^{-3} \nabla^{\beta \rho}{\boldsymbol{W}} \varphi^{j}_{\lambda}+\frac{27}{256}{\rm i} \mathcal{H}_{\alpha}\,^{\lambda} G_{i j} (\Gamma_{a})_{\beta \rho} W \boldsymbol{W} {G}^{-4} \nabla^{\beta \rho}{G} \varphi^{j}_{\lambda} - \frac{9}{128}{\rm i} G_{j k} (\Gamma_{a})_{\beta \rho} \boldsymbol{W} {G}^{-3} \nabla^{\beta \rho}{W} \nabla_{\alpha}\,^{\lambda}{G_{i}\,^{j}} \varphi^{k}_{\lambda} - \frac{9}{128}{\rm i} G_{j k} (\Gamma_{a})_{\beta \rho} W {G}^{-3} \nabla^{\beta \rho}{\boldsymbol{W}} \nabla_{\alpha}\,^{\lambda}{G_{i}\,^{j}} \varphi^{k}_{\lambda}+\frac{27}{128}{\rm i} G_{j k} (\Gamma_{a})_{\beta \rho} W \boldsymbol{W} {G}^{-4} \nabla^{\beta \rho}{G} \nabla_{\alpha}\,^{\lambda}{G_{i}\,^{j}} \varphi^{k}_{\lambda} - \frac{9}{128}{\rm i} G_{j k} (\Gamma_{a})_{\beta \rho} W \boldsymbol{W} {G}^{-3} \nabla_{\alpha}\,^{\lambda}{G_{i}\,^{j}} \nabla^{\beta \rho}{\varphi^{k}_{\lambda}}+\frac{27}{256}G_{i j} (\Gamma_{a})_{\beta \rho} W \boldsymbol{W} {G}^{-5} \nabla^{\beta \rho}{G_{k l}} \varphi^{j}_{\alpha} \varphi^{k \lambda} \varphi^{l}_{\lambda}+\frac{27}{256}G_{i j} G_{k l} (\Gamma_{a})_{\beta \rho} \boldsymbol{W} {G}^{-5} \nabla^{\beta \rho}{W} \varphi^{j}_{\alpha} \varphi^{k \lambda} \varphi^{l}_{\lambda}+\frac{27}{256}G_{i j} G_{k l} (\Gamma_{a})_{\beta \rho} W {G}^{-5} \nabla^{\beta \rho}{\boldsymbol{W}} \varphi^{j}_{\alpha} \varphi^{k \lambda} \varphi^{l}_{\lambda} - \frac{135}{256}G_{i j} G_{k l} (\Gamma_{a})_{\beta \rho} W \boldsymbol{W} {G}^{-6} \nabla^{\beta \rho}{G} \varphi^{j}_{\alpha} \varphi^{k \lambda} \varphi^{l}_{\lambda}%
 - \frac{27}{128}G_{i j} G_{k l} (\Gamma_{a})_{\beta \rho} W \boldsymbol{W} {G}^{-5} \nabla^{\beta \rho}{\varphi^{k \lambda}} \varphi^{j}_{\alpha} \varphi^{l}_{\lambda} - \frac{9}{128}{\rm i} G_{i j} (\Gamma_{a})_{\beta \rho} W \mathbf{X}_{k l} {G}^{-3} \nabla^{\beta \rho}{G^{k l}} \varphi^{j}_{\alpha} - \frac{9}{128}{\rm i} G_{i j} G_{k l} (\Gamma_{a})_{\beta \rho} \mathbf{X}^{k l} {G}^{-3} \nabla^{\beta \rho}{W} \varphi^{j}_{\alpha} - \frac{9}{128}{\rm i} G_{i j} G_{k l} (\Gamma_{a})_{\beta \rho} W {G}^{-3} \nabla^{\beta \rho}{\mathbf{X}^{k l}} \varphi^{j}_{\alpha}+\frac{27}{128}{\rm i} G_{i j} G_{k l} (\Gamma_{a})_{\beta \rho} W \mathbf{X}^{k l} {G}^{-4} \nabla^{\beta \rho}{G} \varphi^{j}_{\alpha} - \frac{9}{128}G_{j k} (\Gamma_{a})_{\beta \rho} \lambda_{i \alpha} {G}^{-1} \nabla^{\beta \rho}{\mathbf{X}^{j k}}+\frac{9}{128}G_{j k} (\Gamma_{a})_{\beta \rho} \mathbf{X}^{j k} \lambda_{i \alpha} {G}^{-2} \nabla^{\beta \rho}{G} - \frac{9}{64}{\rm i} (\Gamma_{a})_{\beta \rho} W \mathbf{X}_{i j} {G}^{-2} \nabla^{\beta \rho}{G} \varphi^{j}_{\alpha} - \frac{9}{64}(\Gamma_{a})_{\beta \rho} W {G}^{-1} \nabla^{\beta \rho}{G_{i j}} \nabla_{\alpha}\,^{\lambda}{\boldsymbol{\lambda}^{j}_{\lambda}}+\frac{9}{64}G_{i j} (\Gamma_{a})_{\beta \rho} W {G}^{-2} \nabla^{\beta \rho}{G} \nabla_{\alpha}\,^{\lambda}{\boldsymbol{\lambda}^{j}_{\lambda}}+\frac{27}{128}G_{i j} (\Gamma_{a})_{\rho \lambda} W W_{\alpha}\,^{\beta} \boldsymbol{\lambda}^{j}_{\beta} {G}^{-2} \nabla^{\rho \lambda}{G} - \frac{9}{128}{\rm i} G_{i j} (\Gamma_{a})_{\beta \rho} \boldsymbol{W} X_{k l} {G}^{-3} \nabla^{\beta \rho}{G^{k l}} \varphi^{j}_{\alpha} - \frac{9}{128}{\rm i} G_{i j} G_{k l} (\Gamma_{a})_{\beta \rho} X^{k l} {G}^{-3} \nabla^{\beta \rho}{\boldsymbol{W}} \varphi^{j}_{\alpha} - \frac{9}{128}{\rm i} G_{i j} G_{k l} (\Gamma_{a})_{\beta \rho} \boldsymbol{W} {G}^{-3} \nabla^{\beta \rho}{X^{k l}} \varphi^{j}_{\alpha}+\frac{27}{128}{\rm i} G_{i j} G_{k l} (\Gamma_{a})_{\beta \rho} \boldsymbol{W} X^{k l} {G}^{-4} \nabla^{\beta \rho}{G} \varphi^{j}_{\alpha} - \frac{9}{64}{\rm i} (\Gamma_{a})_{\beta \rho} \boldsymbol{W} X_{i j} {G}^{-2} \nabla^{\beta \rho}{G} \varphi^{j}_{\alpha} - \frac{9}{64}(\Gamma_{a})_{\beta \rho} \boldsymbol{W} {G}^{-1} \nabla^{\beta \rho}{G_{i j}} \nabla_{\alpha}\,^{\lambda}{\lambda^{j}_{\lambda}}+\frac{9}{64}G_{i j} (\Gamma_{a})_{\beta \rho} \boldsymbol{W} {G}^{-2} \nabla^{\beta \rho}{G} \nabla_{\alpha}\,^{\lambda}{\lambda^{j}_{\lambda}}+\frac{27}{128}G_{i j} (\Gamma_{a})_{\rho \lambda} \boldsymbol{W} W_{\alpha}\,^{\beta} \lambda^{j}_{\beta} {G}^{-2} \nabla^{\rho \lambda}{G} - \frac{9}{128}G_{j k} (\Gamma_{a})_{\beta \rho} \boldsymbol{\lambda}_{i \alpha} {G}^{-1} \nabla^{\beta \rho}{X^{j k}}%
+\frac{9}{128}G_{j k} (\Gamma_{a})_{\beta \rho} X^{j k} \boldsymbol{\lambda}_{i \alpha} {G}^{-2} \nabla^{\beta \rho}{G} - \frac{9}{128}G_{i j} (\Gamma_{a})_{\beta \rho} \lambda^{\lambda}_{k} \boldsymbol{\lambda}_{l \lambda} {G}^{-3} \nabla^{\beta \rho}{G^{k l}} \varphi^{j}_{\alpha} - \frac{9}{128}G_{i j} G_{k l} (\Gamma_{a})_{\beta \rho} \boldsymbol{\lambda}^{k \lambda} {G}^{-3} \nabla^{\beta \rho}{\lambda^{l}_{\lambda}} \varphi^{j}_{\alpha} - \frac{9}{128}G_{i j} G_{k l} (\Gamma_{a})_{\beta \rho} \lambda^{k \lambda} {G}^{-3} \nabla^{\beta \rho}{\boldsymbol{\lambda}^{l}_{\lambda}} \varphi^{j}_{\alpha}+\frac{27}{128}G_{i j} G_{k l} (\Gamma_{a})_{\beta \rho} \lambda^{k \lambda} \boldsymbol{\lambda}^{l}_{\lambda} {G}^{-4} \nabla^{\beta \rho}{G} \varphi^{j}_{\alpha}+\frac{9}{128}(\Gamma_{a})_{\beta \rho} \lambda^{\lambda}_{i} {G}^{-1} \nabla^{\beta \rho}{\boldsymbol{\lambda}_{j \lambda}} \varphi^{j}_{\alpha} - \frac{9}{128}(\Gamma_{a})_{\beta \rho} \lambda^{\lambda}_{i} \boldsymbol{\lambda}_{j \lambda} {G}^{-2} \nabla^{\beta \rho}{G} \varphi^{j}_{\alpha}+\frac{9}{128}(\Gamma_{a})_{\beta \rho} \boldsymbol{\lambda}_{i}^{\lambda} {G}^{-1} \nabla^{\beta \rho}{\lambda_{j \lambda}} \varphi^{j}_{\alpha} - \frac{9}{128}(\Gamma_{a})_{\beta \rho} \lambda^{\lambda}_{j} \boldsymbol{\lambda}_{i \lambda} {G}^{-2} \nabla^{\beta \rho}{G} \varphi^{j}_{\alpha}+\frac{9}{64}G_{i j} (\Gamma_{a})_{\rho \lambda} F_{\alpha}\,^{\beta} \boldsymbol{\lambda}^{j}_{\beta} {G}^{-2} \nabla^{\rho \lambda}{G} - \frac{9}{128}G_{j k} (\Gamma_{a})_{\beta \rho} X_{i}\,^{j} \boldsymbol{\lambda}^{k}_{\alpha} {G}^{-2} \nabla^{\beta \rho}{G}+\frac{9}{128}G_{i j} (\Gamma_{a})_{\beta \rho} \boldsymbol{\lambda}^{j}_{\lambda} {G}^{-2} \nabla_{\alpha}\,^{\lambda}{W} \nabla^{\beta \rho}{G}+\frac{9}{64}G_{i j} (\Gamma_{a})_{\rho \lambda} \mathbf{F}_{\alpha}\,^{\beta} \lambda^{j}_{\beta} {G}^{-2} \nabla^{\rho \lambda}{G} - \frac{9}{128}G_{j k} (\Gamma_{a})_{\beta \rho} \mathbf{X}_{i}\,^{j} \lambda^{k}_{\alpha} {G}^{-2} \nabla^{\beta \rho}{G}+\frac{9}{128}G_{i j} (\Gamma_{a})_{\beta \rho} \lambda^{j}_{\lambda} {G}^{-2} \nabla_{\alpha}\,^{\lambda}{\boldsymbol{W}} \nabla^{\beta \rho}{G}+\frac{9}{160}(\Gamma_{a})_{\alpha \beta} \boldsymbol{W} \lambda_{i \rho} {G}^{-1} \nabla^{\beta \rho}{F} - \frac{9}{160}F (\Gamma_{a})_{\alpha \beta} \boldsymbol{W} \lambda_{i \rho} {G}^{-2} \nabla^{\beta \rho}{G}+\frac{9}{160}(\Gamma_{a})_{\alpha \beta} W \boldsymbol{\lambda}_{i \rho} {G}^{-1} \nabla^{\beta \rho}{F} - \frac{9}{160}F (\Gamma_{a})_{\alpha \beta} W \boldsymbol{\lambda}_{i \rho} {G}^{-2} \nabla^{\beta \rho}{G} - \frac{9}{80}{\rm i} (\Gamma_{a})_{\alpha \beta} \boldsymbol{W} {G}^{-1} \nabla^{\beta}\,_{\rho}{W} \nabla^{\rho \lambda}{\varphi_{i \lambda}}%
 - \frac{9}{80}{\rm i} (\Gamma_{a})_{\alpha \beta} W {G}^{-1} \nabla^{\beta}\,_{\rho}{\boldsymbol{W}} \nabla^{\rho \lambda}{\varphi_{i \lambda}}+\frac{9}{80}{\rm i} (\Gamma_{a})_{\alpha \beta} W \boldsymbol{W} {G}^{-2} \nabla^{\beta}\,_{\rho}{G} \nabla^{\rho \lambda}{\varphi_{i \lambda}}+\frac{27}{160}{\rm i} (\Gamma_{a})_{\alpha \lambda} W \boldsymbol{W} W^{\beta}\,_{\rho} {G}^{-2} \nabla^{\lambda \rho}{G} \varphi_{i \beta}+\frac{81}{320}{\rm i} G_{i j} (\Gamma_{a})_{\alpha \rho} W \boldsymbol{W} X^{j}_{\beta} {G}^{-2} \nabla^{\rho \beta}{G}+\frac{9}{320}{\rm i} F G_{i j} (\Gamma_{a})_{\alpha \beta} \boldsymbol{W} {G}^{-3} \nabla^{\beta \rho}{W} \varphi^{j}_{\rho}+\frac{9}{320}{\rm i} F G_{i j} (\Gamma_{a})_{\alpha \beta} W {G}^{-3} \nabla^{\beta \rho}{\boldsymbol{W}} \varphi^{j}_{\rho} - \frac{27}{320}{\rm i} F G_{i j} (\Gamma_{a})_{\alpha \beta} W \boldsymbol{W} {G}^{-4} \nabla^{\beta \rho}{G} \varphi^{j}_{\rho} - \frac{9}{320}{\rm i} (\Gamma_{a})_{\alpha \beta} \boldsymbol{W} \lambda_{i \rho} {G}^{-3} \nabla^{\beta \rho}{G_{j k}} \varphi^{j \lambda} \varphi^{k}_{\lambda}+\frac{27}{320}{\rm i} G_{j k} (\Gamma_{a})_{\alpha \beta} \boldsymbol{W} \lambda_{i \rho} {G}^{-4} \nabla^{\beta \rho}{G} \varphi^{j \lambda} \varphi^{k}_{\lambda} - \frac{9}{160}{\rm i} G_{j k} (\Gamma_{a})_{\alpha \beta} \boldsymbol{W} \lambda_{i \rho} {G}^{-3} \nabla^{\beta \rho}{\varphi^{j \lambda}} \varphi^{k}_{\lambda} - \frac{9}{320}{\rm i} (\Gamma_{a})_{\alpha \beta} W \boldsymbol{\lambda}_{i \rho} {G}^{-3} \nabla^{\beta \rho}{G_{j k}} \varphi^{j \lambda} \varphi^{k}_{\lambda}+\frac{27}{320}{\rm i} G_{j k} (\Gamma_{a})_{\alpha \beta} W \boldsymbol{\lambda}_{i \rho} {G}^{-4} \nabla^{\beta \rho}{G} \varphi^{j \lambda} \varphi^{k}_{\lambda} - \frac{9}{160}{\rm i} G_{j k} (\Gamma_{a})_{\alpha \beta} W \boldsymbol{\lambda}_{i \rho} {G}^{-3} \nabla^{\beta \rho}{\varphi^{j \lambda}} \varphi^{k}_{\lambda} - \frac{9}{160}(\Gamma_{a})_{\alpha \beta} \boldsymbol{W} {G}^{-3} \nabla^{\beta \rho}{W} \varphi_{i}^{\lambda} \varphi_{j \lambda} \varphi^{j}_{\rho} - \frac{9}{160}(\Gamma_{a})_{\alpha \beta} W {G}^{-3} \nabla^{\beta \rho}{\boldsymbol{W}} \varphi_{i}^{\lambda} \varphi_{j \lambda} \varphi^{j}_{\rho}+\frac{27}{160}(\Gamma_{a})_{\alpha \beta} W \boldsymbol{W} {G}^{-4} \nabla^{\beta \rho}{G} \varphi_{i}^{\lambda} \varphi_{j \lambda} \varphi^{j}_{\rho}+\frac{9}{320}{\rm i} \mathcal{H}_{\rho}\,^{\lambda} G_{i j} (\Gamma_{a})_{\alpha \beta} \boldsymbol{W} {G}^{-3} \nabla^{\rho \beta}{W} \varphi^{j}_{\lambda}+\frac{9}{320}{\rm i} \mathcal{H}_{\rho}\,^{\lambda} G_{i j} (\Gamma_{a})_{\alpha \beta} W {G}^{-3} \nabla^{\rho \beta}{\boldsymbol{W}} \varphi^{j}_{\lambda} - \frac{27}{320}{\rm i} \mathcal{H}_{\rho}\,^{\lambda} G_{i j} (\Gamma_{a})_{\alpha \beta} W \boldsymbol{W} {G}^{-4} \nabla^{\rho \beta}{G} \varphi^{j}_{\lambda}+\frac{9}{160}{\rm i} G_{j k} (\Gamma_{a})_{\alpha \beta} \boldsymbol{W} {G}^{-3} \nabla^{\beta}\,_{\rho}{W} \nabla^{\rho \lambda}{G_{i}\,^{j}} \varphi^{k}_{\lambda}%
+\frac{9}{160}{\rm i} G_{j k} (\Gamma_{a})_{\alpha \beta} W {G}^{-3} \nabla^{\beta}\,_{\rho}{\boldsymbol{W}} \nabla^{\rho \lambda}{G_{i}\,^{j}} \varphi^{k}_{\lambda} - \frac{27}{160}{\rm i} G_{j k} (\Gamma_{a})_{\alpha \beta} W \boldsymbol{W} {G}^{-4} \nabla^{\beta}\,_{\rho}{G} \nabla^{\rho \lambda}{G_{i}\,^{j}} \varphi^{k}_{\lambda}+\frac{27}{320}G_{i j} (\Gamma_{a})_{\alpha \beta} W \boldsymbol{W} {G}^{-5} \nabla^{\beta \rho}{G_{k l}} \varphi^{j}_{\rho} \varphi^{k \lambda} \varphi^{l}_{\lambda}+\frac{27}{320}G_{i j} G_{k l} (\Gamma_{a})_{\alpha \beta} \boldsymbol{W} {G}^{-5} \nabla^{\beta \rho}{W} \varphi^{j}_{\rho} \varphi^{k \lambda} \varphi^{l}_{\lambda}+\frac{27}{320}G_{i j} G_{k l} (\Gamma_{a})_{\alpha \beta} W {G}^{-5} \nabla^{\beta \rho}{\boldsymbol{W}} \varphi^{j}_{\rho} \varphi^{k \lambda} \varphi^{l}_{\lambda} - \frac{27}{64}G_{i j} G_{k l} (\Gamma_{a})_{\alpha \beta} W \boldsymbol{W} {G}^{-6} \nabla^{\beta \rho}{G} \varphi^{j}_{\rho} \varphi^{k \lambda} \varphi^{l}_{\lambda} - \frac{9}{160}{\rm i} G_{i j} (\Gamma_{a})_{\alpha \beta} W \mathbf{X}_{k l} {G}^{-3} \nabla^{\beta \rho}{G^{k l}} \varphi^{j}_{\rho} - \frac{9}{160}{\rm i} G_{i j} G_{k l} (\Gamma_{a})_{\alpha \beta} \mathbf{X}^{k l} {G}^{-3} \nabla^{\beta \rho}{W} \varphi^{j}_{\rho} - \frac{9}{160}{\rm i} G_{i j} G_{k l} (\Gamma_{a})_{\alpha \beta} W {G}^{-3} \nabla^{\beta \rho}{\mathbf{X}^{k l}} \varphi^{j}_{\rho}+\frac{27}{160}{\rm i} G_{i j} G_{k l} (\Gamma_{a})_{\alpha \beta} W \mathbf{X}^{k l} {G}^{-4} \nabla^{\beta \rho}{G} \varphi^{j}_{\rho} - \frac{9}{160}G_{j k} (\Gamma_{a})_{\alpha \beta} \lambda_{i \rho} {G}^{-1} \nabla^{\beta \rho}{\mathbf{X}^{j k}}+\frac{9}{160}G_{j k} (\Gamma_{a})_{\alpha \beta} \mathbf{X}^{j k} \lambda_{i \rho} {G}^{-2} \nabla^{\beta \rho}{G} - \frac{9}{80}{\rm i} (\Gamma_{a})_{\alpha \beta} W \mathbf{X}_{i j} {G}^{-2} \nabla^{\beta \rho}{G} \varphi^{j}_{\rho}+\frac{9}{80}(\Gamma_{a})_{\alpha \beta} W {G}^{-1} \nabla^{\beta}\,_{\lambda}{G_{i j}} \nabla^{\lambda \rho}{\boldsymbol{\lambda}^{j}_{\rho}} - \frac{9}{80}G_{i j} (\Gamma_{a})_{\alpha \beta} W {G}^{-2} \nabla^{\beta}\,_{\lambda}{G} \nabla^{\lambda \rho}{\boldsymbol{\lambda}^{j}_{\rho}}+\frac{27}{160}G_{i j} (\Gamma_{a})_{\alpha \lambda} W W^{\beta}\,_{\rho} \boldsymbol{\lambda}^{j}_{\beta} {G}^{-2} \nabla^{\lambda \rho}{G} - \frac{9}{160}{\rm i} G_{i j} (\Gamma_{a})_{\alpha \beta} \boldsymbol{W} X_{k l} {G}^{-3} \nabla^{\beta \rho}{G^{k l}} \varphi^{j}_{\rho} - \frac{9}{160}{\rm i} G_{i j} G_{k l} (\Gamma_{a})_{\alpha \beta} X^{k l} {G}^{-3} \nabla^{\beta \rho}{\boldsymbol{W}} \varphi^{j}_{\rho} - \frac{9}{160}{\rm i} G_{i j} G_{k l} (\Gamma_{a})_{\alpha \beta} \boldsymbol{W} {G}^{-3} \nabla^{\beta \rho}{X^{k l}} \varphi^{j}_{\rho}+\frac{27}{160}{\rm i} G_{i j} G_{k l} (\Gamma_{a})_{\alpha \beta} \boldsymbol{W} X^{k l} {G}^{-4} \nabla^{\beta \rho}{G} \varphi^{j}_{\rho}%
 - \frac{9}{80}{\rm i} (\Gamma_{a})_{\alpha \beta} \boldsymbol{W} X_{i j} {G}^{-2} \nabla^{\beta \rho}{G} \varphi^{j}_{\rho}+\frac{9}{80}(\Gamma_{a})_{\alpha \beta} \boldsymbol{W} {G}^{-1} \nabla^{\beta}\,_{\lambda}{G_{i j}} \nabla^{\lambda \rho}{\lambda^{j}_{\rho}} - \frac{9}{80}G_{i j} (\Gamma_{a})_{\alpha \beta} \boldsymbol{W} {G}^{-2} \nabla^{\beta}\,_{\lambda}{G} \nabla^{\lambda \rho}{\lambda^{j}_{\rho}}+\frac{27}{160}G_{i j} (\Gamma_{a})_{\alpha \lambda} \boldsymbol{W} W^{\beta}\,_{\rho} \lambda^{j}_{\beta} {G}^{-2} \nabla^{\lambda \rho}{G} - \frac{9}{160}G_{j k} (\Gamma_{a})_{\alpha \beta} \boldsymbol{\lambda}_{i \rho} {G}^{-1} \nabla^{\beta \rho}{X^{j k}}+\frac{9}{160}G_{j k} (\Gamma_{a})_{\alpha \beta} X^{j k} \boldsymbol{\lambda}_{i \rho} {G}^{-2} \nabla^{\beta \rho}{G} - \frac{9}{160}G_{i j} (\Gamma_{a})_{\alpha \beta} \lambda^{\rho}_{k} \boldsymbol{\lambda}_{l \rho} {G}^{-3} \nabla^{\beta \lambda}{G^{k l}} \varphi^{j}_{\lambda} - \frac{9}{160}G_{i j} G_{k l} (\Gamma_{a})_{\alpha \beta} \boldsymbol{\lambda}^{k \rho} {G}^{-3} \nabla^{\beta \lambda}{\lambda^{l}_{\rho}} \varphi^{j}_{\lambda} - \frac{9}{160}G_{i j} G_{k l} (\Gamma_{a})_{\alpha \beta} \lambda^{k \rho} {G}^{-3} \nabla^{\beta \lambda}{\boldsymbol{\lambda}^{l}_{\rho}} \varphi^{j}_{\lambda}+\frac{27}{160}G_{i j} G_{k l} (\Gamma_{a})_{\alpha \beta} \lambda^{k \rho} \boldsymbol{\lambda}^{l}_{\rho} {G}^{-4} \nabla^{\beta \lambda}{G} \varphi^{j}_{\lambda}+\frac{9}{160}(\Gamma_{a})_{\alpha \beta} \lambda^{\rho}_{i} {G}^{-1} \nabla^{\beta \lambda}{\boldsymbol{\lambda}_{j \rho}} \varphi^{j}_{\lambda} - \frac{9}{160}(\Gamma_{a})_{\alpha \beta} \lambda^{\rho}_{i} \boldsymbol{\lambda}_{j \rho} {G}^{-2} \nabla^{\beta \lambda}{G} \varphi^{j}_{\lambda}+\frac{9}{160}(\Gamma_{a})_{\alpha \beta} \boldsymbol{\lambda}_{i}^{\rho} {G}^{-1} \nabla^{\beta \lambda}{\lambda_{j \rho}} \varphi^{j}_{\lambda} - \frac{9}{160}(\Gamma_{a})_{\alpha \beta} \lambda^{\rho}_{j} \boldsymbol{\lambda}_{i \rho} {G}^{-2} \nabla^{\beta \lambda}{G} \varphi^{j}_{\lambda}+\frac{9}{80}G_{i j} (\Gamma_{a})_{\alpha \lambda} F^{\beta}\,_{\rho} \boldsymbol{\lambda}^{j}_{\beta} {G}^{-2} \nabla^{\lambda \rho}{G} - \frac{9}{160}G_{j k} (\Gamma_{a})_{\alpha \beta} X_{i}\,^{j} \boldsymbol{\lambda}^{k}_{\rho} {G}^{-2} \nabla^{\beta \rho}{G}+\frac{9}{160}G_{i j} (\Gamma_{a})_{\alpha \beta} \boldsymbol{\lambda}^{j}_{\rho} {G}^{-2} \nabla_{\lambda}\,^{\rho}{W} \nabla^{\beta \lambda}{G}+\frac{9}{80}G_{i j} (\Gamma_{a})_{\alpha \lambda} \mathbf{F}_{\beta}\,^{\rho} \lambda^{j}_{\rho} {G}^{-2} \nabla^{\lambda \beta}{G} - \frac{9}{160}G_{j k} (\Gamma_{a})_{\alpha \beta} \mathbf{X}_{i}\,^{j} \lambda^{k}_{\rho} {G}^{-2} \nabla^{\beta \rho}{G}+\frac{9}{160}G_{i j} (\Gamma_{a})_{\alpha \beta} \lambda^{j}_{\rho} {G}^{-2} \nabla_{\lambda}\,^{\rho}{\boldsymbol{W}} \nabla^{\beta \lambda}{G}
\doublespacedmathend
\end{adjustwidth}
 
\subsubsection{$J^4_{a b, R^2}$ Degauged and Gauge Fixed Bosons} \label{J4R2Complete}

\begin{adjustwidth}{0cm}{5cm}
\doublespacedmathbegin
- \frac{9}{4}\mathcal{D}_{\underline{a}}{\mathcal{H}^{c}} \boldsymbol{W} F_{\underline{b} c} {G}^{-1}+\frac{9}{4}\mathcal{D}^{c}{\mathcal{H}_{\underline{a}}} \boldsymbol{W} F_{\underline{b} c} {G}^{-1}+\frac{9}{10}\mathcal{D}^{c}{\mathcal{H}^{d}} \eta_{\underline{a} \underline{b}} \boldsymbol{W} F_{c d} {G}^{-1} - \frac{9}{2}F \boldsymbol{W} W_{\underline{a}}\,^{c} F_{\underline{b} c} {G}^{-1}+\frac{9}{10}F \eta_{\underline{a} \underline{b}} \boldsymbol{W} W^{c d} F_{c d} {G}^{-1}+\frac{9}{4}\Phi_{\underline{a} c i j} G^{i j} \boldsymbol{W} F^{c}\,_{\underline{b}} {G}^{-1}+\frac{9}{10}\Phi_{c d i j} G^{i j} \eta_{\underline{a} \underline{b}} \boldsymbol{W} F^{c d} {G}^{-1} - \frac{9}{4}\Phi_{\underline{a}}\,^{c}\,_{i j} G^{i j} \boldsymbol{W} F_{\underline{b} c} {G}^{-1} - \frac{9}{2}\mathcal{D}_{\underline{a}}{\mathcal{H}^{c}} \boldsymbol{W} W_{\underline{b} c} {G}^{-1}+\frac{9}{2}\mathcal{D}^{c}{\mathcal{H}_{\underline{a}}} \boldsymbol{W} W_{\underline{b} c} {G}^{-1}+\frac{9}{5}\mathcal{D}^{c}{\mathcal{H}^{d}} \eta_{\underline{a} \underline{b}} \boldsymbol{W} W_{c d} {G}^{-1} - \frac{27}{4}F \boldsymbol{W} W_{\underline{a}}\,^{c} W_{\underline{b} c} {G}^{-1}+\frac{27}{20}F \eta_{\underline{a} \underline{b}} \boldsymbol{W} W^{c d} W_{c d} {G}^{-1}+\frac{9}{2}\Phi_{\underline{a} c i j} G^{i j} \boldsymbol{W} W^{c}\,_{\underline{b}} {G}^{-1}+\frac{9}{5}\Phi_{c d i j} G^{i j} \eta_{\underline{a} \underline{b}} \boldsymbol{W} W^{c d} {G}^{-1} - \frac{9}{2}\Phi_{\underline{a}}\,^{c}\,_{i j} G^{i j} \boldsymbol{W} W_{\underline{b} c} {G}^{-1} - \frac{27}{80}\mathcal{D}_{c}{\mathcal{D}^{c}{G_{i j}}} \eta_{\underline{a} \underline{b}} \boldsymbol{W} X^{i j} {G}^{-1} - \frac{27}{80}R G_{i j} \eta_{\underline{a} \underline{b}} \boldsymbol{W} X^{i j} {G}^{-1} - \frac{27}{40}G_{i j} \eta_{\underline{a} \underline{b}} \boldsymbol{W} X^{i j} W^{c d} W_{c d} {G}^{-1}%
 - \frac{9}{4}F F_{\underline{a}}\,^{c} \mathbf{F}_{\underline{b} c} {G}^{-1}+\frac{9}{16}F \eta_{\underline{a} \underline{b}} F^{c d} \mathbf{F}_{c d} {G}^{-1} - \frac{9}{2}F W_{\underline{a}}\,^{c} \mathbf{F}_{\underline{b} c} {G}^{-1}+\frac{9}{10}F \eta_{\underline{a} \underline{b}} W^{c d} \mathbf{F}_{c d} {G}^{-1} - \frac{9}{4}\mathcal{D}_{\underline{a}}{\mathcal{H}^{c}} \mathbf{F}_{\underline{b} c} {G}^{-1}+\frac{9}{4}\mathcal{D}^{c}{\mathcal{H}_{\underline{a}}} \mathbf{F}_{\underline{b} c} {G}^{-1}+\frac{9}{10}\mathcal{D}^{c}{\mathcal{H}^{d}} \eta_{\underline{a} \underline{b}} \mathbf{F}_{c d} {G}^{-1}+\frac{9}{4}\Phi_{\underline{a} c i j} G^{i j} \mathbf{F}^{c}\,_{\underline{b}} {G}^{-1}+\frac{9}{8}\Phi_{c d i j} G^{i j} \mathbf{F}^{c d} \eta_{\underline{a} \underline{b}} {G}^{-1} - \frac{9}{4}\Phi_{\underline{a}}\,^{c}\,_{i j} G^{i j} \mathbf{F}_{\underline{b} c} {G}^{-1} - \frac{9}{8}\mathcal{D}_{\underline{a}}{F} \mathcal{D}_{\underline{b}}{\boldsymbol{W}} {G}^{-1}+\frac{9}{40}\mathcal{D}_{c}{F} \eta_{\underline{a} \underline{b}} \mathcal{D}^{c}{\boldsymbol{W}} {G}^{-1} - \frac{27}{80}\mathcal{D}_{c}{\mathcal{D}^{c}{G_{i j}}} \eta_{\underline{a} \underline{b}} \mathbf{X}^{i j} {G}^{-1} - \frac{81}{640}R G_{i j} \eta_{\underline{a} \underline{b}} \mathbf{X}^{i j} {G}^{-1} - \frac{27}{40}G_{i j} \eta_{\underline{a} \underline{b}} \mathbf{X}^{i j} W^{c d} W_{c d} {G}^{-1}+\frac{9}{8}F G_{i j} \mathcal{D}_{\underline{a}}{G^{i j}} \mathcal{D}_{\underline{b}}{\boldsymbol{W}} {G}^{-3} - \frac{9}{40}F G_{i j} \mathcal{D}_{c}{G^{i j}} \eta_{\underline{a} \underline{b}} \mathcal{D}^{c}{\boldsymbol{W}} {G}^{-3} - \frac{9}{4}\mathcal{H}_{\underline{a}} G_{i j} \mathcal{D}^{c}{G^{i j}} \boldsymbol{W} W_{\underline{b} c} {G}^{-3}+\frac{9}{4}\mathcal{H}^{c} G_{i j} \mathcal{D}_{\underline{a}}{G^{i j}} \boldsymbol{W} W_{\underline{b} c} {G}^{-3} - \frac{9}{10}\mathcal{H}^{d} G_{i j} \mathcal{D}^{c}{G^{i j}} \eta_{\underline{a} \underline{b}} \boldsymbol{W} W_{c d} {G}^{-3}%
 - \frac{9}{80}R F \eta_{\underline{a} \underline{b}} \boldsymbol{W} {G}^{-1}+\frac{27}{2}\mathcal{H}_{\underline{a}} \boldsymbol{W} \mathcal{D}^{c}{W_{\underline{b} c}} {G}^{-1} - \frac{27}{10}\mathcal{H}^{c} \eta_{\underline{a} \underline{b}} \boldsymbol{W} \mathcal{D}^{d}{W_{c d}} {G}^{-1} - \frac{9}{2}G_{i j} \mathcal{D}_{\underline{a}}{G^{i}\,_{k}} \mathcal{D}^{c}{G^{j k}} \boldsymbol{W} W_{\underline{b} c} {G}^{-3}+\frac{9}{10}G_{i j} \mathcal{D}^{c}{G^{i}\,_{k}} \mathcal{D}^{d}{G^{j k}} \eta_{\underline{a} \underline{b}} \boldsymbol{W} W_{c d} {G}^{-3}+\frac{9}{8}\mathcal{H}^{c} G_{i j} \mathcal{D}_{\underline{a}}{G^{i j}} \boldsymbol{W} F_{\underline{b} c} {G}^{-3} - \frac{9}{8}\mathcal{H}_{\underline{a}} G_{i j} \mathcal{D}^{c}{G^{i j}} \boldsymbol{W} F_{\underline{b} c} {G}^{-3} - \frac{9}{20}\mathcal{H}^{d} G_{i j} \mathcal{D}^{c}{G^{i j}} \eta_{\underline{a} \underline{b}} \boldsymbol{W} F_{c d} {G}^{-3} - \frac{9}{80}\mathcal{H}^{c} \mathcal{H}_{c} G_{i j} \eta_{\underline{a} \underline{b}} \boldsymbol{W} X^{i j} {G}^{-3} - \frac{9}{4}G_{i j} \mathcal{D}_{\underline{a}}{G^{i}\,_{k}} \mathcal{D}^{c}{G^{j k}} \boldsymbol{W} F_{\underline{b} c} {G}^{-3}+\frac{9}{20}G_{i j} \mathcal{D}^{c}{G^{i}\,_{k}} \mathcal{D}^{d}{G^{j k}} \eta_{\underline{a} \underline{b}} \boldsymbol{W} F_{c d} {G}^{-3}+\frac{3}{10}G_{i j} \mathcal{D}_{c}{G^{i}\,_{k}} \mathcal{D}^{c}{G^{j}\,_{l}} \eta_{\underline{a} \underline{b}} \boldsymbol{W} X^{k l} {G}^{-3}+\frac{9}{8}\mathcal{H}^{c} G_{i j} \mathcal{D}_{\underline{a}}{G^{i j}} \mathbf{F}_{\underline{b} c} {G}^{-3} - \frac{9}{8}\mathcal{H}_{\underline{a}} G_{i j} \mathcal{D}^{c}{G^{i j}} \mathbf{F}_{\underline{b} c} {G}^{-3} - \frac{9}{20}\mathcal{H}^{d} G_{i j} \mathcal{D}^{c}{G^{i j}} \eta_{\underline{a} \underline{b}} \mathbf{F}_{c d} {G}^{-3} - \frac{9}{80}\mathcal{H}^{c} \mathcal{H}_{c} G_{i j} \eta_{\underline{a} \underline{b}} \mathbf{X}^{i j} {G}^{-3} - \frac{9}{4}G_{i j} \mathcal{D}_{\underline{a}}{G^{i}\,_{k}} \mathcal{D}^{c}{G^{j k}} \mathbf{F}_{\underline{b} c} {G}^{-3}+\frac{9}{20}G_{i j} \mathcal{D}^{c}{G^{i}\,_{k}} \mathcal{D}^{d}{G^{j k}} \eta_{\underline{a} \underline{b}} \mathbf{F}_{c d} {G}^{-3}+\frac{3}{10}G_{i j} \mathcal{D}_{c}{G^{i}\,_{k}} \mathcal{D}^{c}{G^{j}\,_{l}} \eta_{\underline{a} \underline{b}} \mathbf{X}^{k l} {G}^{-3}+\frac{9}{16}F \mathcal{D}_{\underline{a}}{G_{i j}} \mathcal{D}_{\underline{b}}{G^{i j}} \boldsymbol{W} {G}^{-3}%
 - \frac{9}{80}F \mathcal{D}_{c}{G_{i j}} \mathcal{D}^{c}{G^{i j}} \eta_{\underline{a} \underline{b}} \boldsymbol{W} {G}^{-3} - \frac{9}{80}F G_{i j} G_{k l} \mathcal{D}_{c}{G^{i k}} \mathcal{D}^{c}{G^{j l}} \eta_{\underline{a} \underline{b}} \boldsymbol{W} {G}^{-5} - \frac{9}{16}F G_{i j} G_{k l} \mathcal{D}_{\underline{a}}{G^{i j}} \mathcal{D}_{\underline{b}}{G^{k l}} \boldsymbol{W} {G}^{-5}+\frac{9}{80}F G_{i j} G_{k l} \mathcal{D}_{c}{G^{i j}} \mathcal{D}^{c}{G^{k l}} \eta_{\underline{a} \underline{b}} \boldsymbol{W} {G}^{-5}+\frac{9}{16}F G_{i j} G_{k l} \mathcal{D}_{\underline{a}}{G^{i k}} \mathcal{D}_{\underline{b}}{G^{j l}} \boldsymbol{W} {G}^{-5} - \frac{3}{4}G_{i j} \mathcal{D}_{\underline{a}}{G^{i j}} \mathcal{D}_{\underline{b}}{G_{k l}} \mathbf{X}^{k l} {G}^{-3}+\frac{3}{20}G_{i j} \mathcal{D}_{c}{G^{i j}} \mathcal{D}^{c}{G_{k l}} \eta_{\underline{a} \underline{b}} \mathbf{X}^{k l} {G}^{-3}+\frac{9}{16}\mathcal{H}_{\underline{a}} \mathcal{H}_{\underline{b}} G_{i j} \mathbf{X}^{i j} {G}^{-3} - \frac{3}{2}G_{i j} \mathcal{D}_{\underline{a}}{G^{i}\,_{k}} \mathcal{D}_{\underline{b}}{G^{j}\,_{l}} \mathbf{X}^{k l} {G}^{-3}+\frac{9}{2}\mathcal{H}_{\underline{a}} \mathcal{D}^{c}{\mathbf{F}_{\underline{b} c}} {G}^{-1} - \frac{9}{10}\mathcal{H}^{c} \eta_{\underline{a} \underline{b}} \mathcal{D}^{d}{\mathbf{F}_{c d}} {G}^{-1}+\frac{27}{2}\mathcal{H}_{\underline{a}} \mathcal{D}^{c}{\boldsymbol{W}} W_{\underline{b} c} {G}^{-1} - \frac{27}{10}\mathcal{H}^{c} \eta_{\underline{a} \underline{b}} \mathcal{D}^{d}{\boldsymbol{W}} W_{c d} {G}^{-1} - \frac{3}{2}G_{i j} G_{k l} \mathcal{D}_{\underline{a}}{G^{i k}} \mathcal{D}_{\underline{b}}{\mathbf{X}^{j l}} {G}^{-3}+\frac{3}{10}G_{i j} G_{k l} \mathcal{D}_{c}{G^{i k}} \eta_{\underline{a} \underline{b}} \mathcal{D}^{c}{\mathbf{X}^{j l}} {G}^{-3} - \frac{9}{80}F \eta_{\underline{a} \underline{b}} \mathcal{D}_{c}{\mathcal{D}^{c}{\boldsymbol{W}}} {G}^{-1} - \frac{3}{4}G_{i j} G_{k l} \mathcal{D}_{\underline{a}}{G^{i j}} \mathcal{D}_{\underline{b}}{\mathbf{X}^{k l}} {G}^{-3}+\frac{3}{20}G_{i j} G_{k l} \mathcal{D}_{c}{G^{i j}} \eta_{\underline{a} \underline{b}} \mathcal{D}^{c}{\mathbf{X}^{k l}} {G}^{-3}+\frac{9}{8}G_{i j} G_{k l} G_{m n} \mathcal{D}_{\underline{a}}{G^{i k}} \mathcal{D}_{\underline{b}}{G^{j l}} \mathbf{X}^{m n} {G}^{-5} - \frac{9}{40}G_{i j} G_{k l} G_{m n} \mathcal{D}_{c}{G^{i k}} \mathcal{D}^{c}{G^{j l}} \eta_{\underline{a} \underline{b}} \mathbf{X}^{m n} {G}^{-5}%
+\frac{9}{16}G_{i j} G_{k l} G_{m n} \mathcal{D}_{\underline{a}}{G^{i j}} \mathcal{D}_{\underline{b}}{G^{k l}} \mathbf{X}^{m n} {G}^{-5} - \frac{9}{80}G_{i j} G_{k l} G_{m n} \mathcal{D}_{c}{G^{i j}} \mathcal{D}^{c}{G^{k l}} \eta_{\underline{a} \underline{b}} \mathbf{X}^{m n} {G}^{-5} - \frac{9}{8}G_{i j} G_{k l} \mathcal{D}_{\underline{a}}{\mathcal{D}_{\underline{b}}{G^{i j}}} \mathbf{X}^{k l} {G}^{-3}+\frac{9}{40}G_{i j} G_{k l} \mathcal{D}_{c}{\mathcal{D}^{c}{G^{i j}}} \eta_{\underline{a} \underline{b}} \mathbf{X}^{k l} {G}^{-3}+\frac{27}{16}R_{\underline{a} \underline{b}} G_{i j} \mathbf{X}^{i j} {G}^{-1} - \frac{27}{128}R G_{i j} \mathbf{X}^{i j} \eta_{\underline{a} \underline{b}} {G}^{-1}+\frac{27}{8}G_{i j} \mathbf{X}^{i j} W_{\underline{a}}\,^{c} W_{\underline{b} c} {G}^{-1} - \frac{3}{4}G_{i j} \mathcal{D}_{\underline{a}}{G^{i}\,_{k}} \mathcal{D}_{\underline{b}}{G^{k}\,_{l}} \mathbf{X}^{j l} {G}^{-3} - \frac{21}{8}\mathcal{D}_{\underline{a}}{G_{i j}} \mathcal{D}_{\underline{b}}{\mathbf{X}^{i j}} {G}^{-1} - \frac{27}{80}G_{i j} \eta_{\underline{a} \underline{b}} \mathcal{D}_{c}{\mathcal{D}^{c}{\boldsymbol{W}}} X^{i j} {G}^{-1}+\frac{21}{40}\mathcal{D}_{c}{G_{i j}} \eta_{\underline{a} \underline{b}} \mathcal{D}^{c}{\mathbf{X}^{i j}} {G}^{-1}+\frac{27}{16}\mathcal{D}_{\underline{a}}{\mathcal{D}_{\underline{b}}{G_{i j}}} \mathbf{X}^{i j} {G}^{-1} - \frac{27}{80}G_{i j} \eta_{\underline{a} \underline{b}} \mathcal{D}_{c}{\mathcal{D}^{c}{\mathbf{X}^{i j}}} {G}^{-1}+\frac{27}{16}G_{i j} \mathcal{D}_{\underline{a}}{\mathcal{D}_{\underline{b}}{\mathbf{X}^{i j}}} {G}^{-1} - \frac{3}{4}G_{i j} \mathcal{D}_{\underline{a}}{G^{i j}} \mathcal{D}_{\underline{b}}{G_{k l}} \boldsymbol{W} X^{k l} {G}^{-3}+\frac{3}{20}G_{i j} \mathcal{D}_{c}{G^{i j}} \mathcal{D}^{c}{G_{k l}} \eta_{\underline{a} \underline{b}} \boldsymbol{W} X^{k l} {G}^{-3}+\frac{9}{16}\mathcal{H}_{\underline{a}} \mathcal{H}_{\underline{b}} G_{i j} \boldsymbol{W} X^{i j} {G}^{-3} - \frac{3}{2}G_{i j} \mathcal{D}_{\underline{a}}{G^{i}\,_{k}} \mathcal{D}_{\underline{b}}{G^{j}\,_{l}} \boldsymbol{W} X^{k l} {G}^{-3} - \frac{9}{8}G_{i j} G_{k l} \mathcal{D}_{\underline{a}}{G^{i j}} \mathcal{D}_{\underline{b}}{\boldsymbol{W}} X^{k l} {G}^{-3}+\frac{9}{40}G_{i j} G_{k l} \mathcal{D}_{c}{G^{i j}} \eta_{\underline{a} \underline{b}} \mathcal{D}^{c}{\boldsymbol{W}} X^{k l} {G}^{-3}%
+\frac{9}{2}\mathcal{H}_{\underline{a}} \boldsymbol{W} \mathcal{D}^{c}{F_{\underline{b} c}} {G}^{-1} - \frac{9}{10}\mathcal{H}^{c} \eta_{\underline{a} \underline{b}} \boldsymbol{W} \mathcal{D}^{d}{F_{c d}} {G}^{-1} - \frac{3}{2}G_{i j} G_{k l} \mathcal{D}_{\underline{a}}{G^{i k}} \boldsymbol{W} \mathcal{D}_{\underline{b}}{X^{j l}} {G}^{-3}+\frac{3}{10}G_{i j} G_{k l} \mathcal{D}_{c}{G^{i k}} \eta_{\underline{a} \underline{b}} \boldsymbol{W} \mathcal{D}^{c}{X^{j l}} {G}^{-3} - \frac{3}{4}G_{i j} G_{k l} \mathcal{D}_{\underline{a}}{G^{i j}} \boldsymbol{W} \mathcal{D}_{\underline{b}}{X^{k l}} {G}^{-3}+\frac{3}{20}G_{i j} G_{k l} \mathcal{D}_{c}{G^{i j}} \eta_{\underline{a} \underline{b}} \boldsymbol{W} \mathcal{D}^{c}{X^{k l}} {G}^{-3}+\frac{9}{8}G_{i j} G_{k l} G_{m n} \mathcal{D}_{\underline{a}}{G^{i k}} \mathcal{D}_{\underline{b}}{G^{j l}} \boldsymbol{W} X^{m n} {G}^{-5} - \frac{9}{40}G_{i j} G_{k l} G_{m n} \mathcal{D}_{c}{G^{i k}} \mathcal{D}^{c}{G^{j l}} \eta_{\underline{a} \underline{b}} \boldsymbol{W} X^{m n} {G}^{-5}+\frac{9}{16}G_{i j} G_{k l} G_{m n} \mathcal{D}_{\underline{a}}{G^{i j}} \mathcal{D}_{\underline{b}}{G^{k l}} \boldsymbol{W} X^{m n} {G}^{-5} - \frac{9}{80}G_{i j} G_{k l} G_{m n} \mathcal{D}_{c}{G^{i j}} \mathcal{D}^{c}{G^{k l}} \eta_{\underline{a} \underline{b}} \boldsymbol{W} X^{m n} {G}^{-5} - \frac{9}{8}G_{i j} G_{k l} \mathcal{D}_{\underline{a}}{\mathcal{D}_{\underline{b}}{G^{i j}}} \boldsymbol{W} X^{k l} {G}^{-3}+\frac{9}{40}G_{i j} G_{k l} \mathcal{D}_{c}{\mathcal{D}^{c}{G^{i j}}} \eta_{\underline{a} \underline{b}} \boldsymbol{W} X^{k l} {G}^{-3}+\frac{27}{16}R_{\underline{a} \underline{b}} G_{i j} \boldsymbol{W} X^{i j} {G}^{-1}+\frac{27}{8}G_{i j} \boldsymbol{W} X^{i j} W_{\underline{a}}\,^{c} W_{\underline{b} c} {G}^{-1} - \frac{3}{4}G_{i j} \mathcal{D}_{\underline{a}}{G^{i}\,_{k}} \mathcal{D}_{\underline{b}}{G^{k}\,_{l}} \boldsymbol{W} X^{j l} {G}^{-3} - \frac{21}{8}\mathcal{D}_{\underline{a}}{G_{i j}} \boldsymbol{W} \mathcal{D}_{\underline{b}}{X^{i j}} {G}^{-1} - \frac{9}{4}\mathcal{D}_{\underline{a}}{G_{i j}} \mathcal{D}_{\underline{b}}{\boldsymbol{W}} X^{i j} {G}^{-1}+\frac{9}{20}\mathcal{D}_{c}{G_{i j}} \eta_{\underline{a} \underline{b}} \mathcal{D}^{c}{\boldsymbol{W}} X^{i j} {G}^{-1}+\frac{21}{40}\mathcal{D}_{c}{G_{i j}} \eta_{\underline{a} \underline{b}} \boldsymbol{W} \mathcal{D}^{c}{X^{i j}} {G}^{-1}+\frac{27}{16}\mathcal{D}_{\underline{a}}{\mathcal{D}_{\underline{b}}{G_{i j}}} \boldsymbol{W} X^{i j} {G}^{-1}%
+\frac{27}{8}G_{i j} \mathcal{D}_{\underline{a}}{\boldsymbol{W}} \mathcal{D}_{\underline{b}}{X^{i j}} {G}^{-1} - \frac{27}{40}G_{i j} \eta_{\underline{a} \underline{b}} \mathcal{D}_{c}{\boldsymbol{W}} \mathcal{D}^{c}{X^{i j}} {G}^{-1} - \frac{27}{80}G_{i j} \eta_{\underline{a} \underline{b}} \boldsymbol{W} \mathcal{D}_{c}{\mathcal{D}^{c}{X^{i j}}} {G}^{-1}+\frac{27}{16}G_{i j} \boldsymbol{W} \mathcal{D}_{\underline{a}}{\mathcal{D}_{\underline{b}}{X^{i j}}} {G}^{-1} - \frac{9}{80}F \eta_{\underline{a} \underline{b}} \mathbf{F}^{c d} F_{c d} {G}^{-1}+\frac{9}{8}\epsilon_{\underline{b}}\,^{c d e {e_{1}}} \mathcal{H}_{\underline{a}} F_{c d} \mathbf{F}_{e {e_{1}}} {G}^{-1} - \frac{9}{40}\epsilon^{{e_{2}} c d e {e_{1}}} \mathcal{H}_{{e_{2}}} \eta_{\underline{a} \underline{b}} F_{c d} \mathbf{F}_{e {e_{1}}} {G}^{-1}+\frac{9}{2}\mathcal{H}_{\underline{a}} \mathcal{D}^{c}{\boldsymbol{W}} F_{\underline{b} c} {G}^{-1} - \frac{9}{10}\mathcal{H}^{c} \eta_{\underline{a} \underline{b}} \mathcal{D}^{d}{\boldsymbol{W}} F_{c d} {G}^{-1} - \frac{9}{8}G_{i j} G_{k l} \mathcal{D}_{\underline{a}}{G^{i k}} \mathcal{D}_{\underline{b}}{\boldsymbol{W}} X^{j l} {G}^{-3}+\frac{9}{40}G_{i j} G_{k l} \mathcal{D}_{c}{G^{i k}} \eta_{\underline{a} \underline{b}} \mathcal{D}^{c}{\boldsymbol{W}} X^{j l} {G}^{-3}+\frac{27}{16}G_{i j} \mathcal{D}_{\underline{a}}{\mathcal{D}_{\underline{b}}{\boldsymbol{W}}} X^{i j} {G}^{-1} - \frac{9}{40}\Phi_{c d i j} G^{i j} \eta_{\underline{a} \underline{b}} \mathbf{F}^{c d} {G}^{-1}+\frac{3}{20}G_{i j} \mathcal{D}_{c}{G^{i}\,_{k}} \mathcal{D}^{c}{G^{k}\,_{l}} \eta_{\underline{a} \underline{b}} \mathbf{X}^{j l} {G}^{-3}+\frac{3}{20}G_{i j} \mathcal{D}_{c}{G^{i}\,_{k}} \mathcal{D}^{c}{G^{k}\,_{l}} \eta_{\underline{a} \underline{b}} \boldsymbol{W} X^{j l} {G}^{-3}+\frac{9}{16}R_{\underline{a} \underline{b}} F \boldsymbol{W} {G}^{-1}+\frac{9}{16}F \mathcal{D}_{\underline{a}}{\mathcal{D}_{\underline{b}}{\boldsymbol{W}}} {G}^{-1} - \frac{9}{8}F \mathcal{D}_{\underline{a}}{G} \mathcal{D}_{\underline{b}}{\boldsymbol{W}} {G}^{-2}+\frac{9}{16}\mathcal{D}_{\underline{a}}{\mathcal{D}_{\underline{b}}{F}} \boldsymbol{W} {G}^{-1} - \frac{9}{8}\mathcal{D}_{\underline{a}}{F} \mathcal{D}_{\underline{b}}{G} \boldsymbol{W} {G}^{-2}%
+\frac{9}{8}F \mathcal{D}_{\underline{a}}{G} \mathcal{D}_{\underline{b}}{G} \boldsymbol{W} {G}^{-3} - \frac{9}{16}F \mathcal{D}_{\underline{a}}{\mathcal{D}_{\underline{b}}{G}} \boldsymbol{W} {G}^{-2} - \frac{9}{8}G_{i j} \mathcal{D}_{\underline{a}}{G} \mathcal{D}_{\underline{b}}{G} \mathbf{X}^{i j} {G}^{-3}+\frac{9}{16}G_{i j} \mathcal{D}_{\underline{a}}{\mathcal{D}_{\underline{b}}{G}} \mathbf{X}^{i j} {G}^{-2}+\frac{9}{8}\mathcal{D}_{\underline{a}}{G} \mathcal{D}_{\underline{b}}{G_{i j}} \mathbf{X}^{i j} {G}^{-2}+\frac{9}{8}G_{i j} \mathcal{D}_{\underline{a}}{G} \mathcal{D}_{\underline{b}}{\mathbf{X}^{i j}} {G}^{-2} - \frac{9}{8}G_{i j} \mathcal{D}_{\underline{a}}{G} \mathcal{D}_{\underline{b}}{G} \boldsymbol{W} X^{i j} {G}^{-3}+\frac{9}{16}G_{i j} \mathcal{D}_{\underline{a}}{\mathcal{D}_{\underline{b}}{G}} \boldsymbol{W} X^{i j} {G}^{-2}+\frac{9}{8}\mathcal{D}_{\underline{a}}{G} \mathcal{D}_{\underline{b}}{G_{i j}} \boldsymbol{W} X^{i j} {G}^{-2}+\frac{9}{8}G_{i j} \mathcal{D}_{\underline{a}}{G} \boldsymbol{W} \mathcal{D}_{\underline{b}}{X^{i j}} {G}^{-2}+\frac{9}{8}G_{i j} \mathcal{D}_{\underline{a}}{G} \mathcal{D}_{\underline{b}}{\boldsymbol{W}} X^{i j} {G}^{-2}+\frac{9}{40}F \mathcal{D}_{c}{G} \eta_{\underline{a} \underline{b}} \mathcal{D}^{c}{\boldsymbol{W}} {G}^{-2} - \frac{9}{80}\mathcal{D}_{c}{\mathcal{D}^{c}{F}} \eta_{\underline{a} \underline{b}} \boldsymbol{W} {G}^{-1}+\frac{9}{40}\mathcal{D}_{c}{F} \mathcal{D}^{c}{G} \eta_{\underline{a} \underline{b}} \boldsymbol{W} {G}^{-2} - \frac{9}{40}F \mathcal{D}_{c}{G} \mathcal{D}^{c}{G} \eta_{\underline{a} \underline{b}} \boldsymbol{W} {G}^{-3}+\frac{9}{80}F \mathcal{D}_{c}{\mathcal{D}^{c}{G}} \eta_{\underline{a} \underline{b}} \boldsymbol{W} {G}^{-2}+\frac{9}{40}G_{i j} \mathcal{D}_{c}{G} \mathcal{D}^{c}{G} \eta_{\underline{a} \underline{b}} \mathbf{X}^{i j} {G}^{-3} - \frac{9}{80}G_{i j} \mathcal{D}_{c}{\mathcal{D}^{c}{G}} \eta_{\underline{a} \underline{b}} \mathbf{X}^{i j} {G}^{-2} - \frac{9}{40}\mathcal{D}_{c}{G} \mathcal{D}^{c}{G_{i j}} \eta_{\underline{a} \underline{b}} \mathbf{X}^{i j} {G}^{-2} - \frac{9}{40}G_{i j} \mathcal{D}_{c}{G} \eta_{\underline{a} \underline{b}} \mathcal{D}^{c}{\mathbf{X}^{i j}} {G}^{-2}%
+\frac{9}{40}G_{i j} \mathcal{D}_{c}{G} \mathcal{D}^{c}{G} \eta_{\underline{a} \underline{b}} \boldsymbol{W} X^{i j} {G}^{-3} - \frac{9}{80}G_{i j} \mathcal{D}_{c}{\mathcal{D}^{c}{G}} \eta_{\underline{a} \underline{b}} \boldsymbol{W} X^{i j} {G}^{-2} - \frac{9}{40}\mathcal{D}_{c}{G} \mathcal{D}^{c}{G_{i j}} \eta_{\underline{a} \underline{b}} \boldsymbol{W} X^{i j} {G}^{-2} - \frac{9}{40}G_{i j} \mathcal{D}_{c}{G} \eta_{\underline{a} \underline{b}} \boldsymbol{W} \mathcal{D}^{c}{X^{i j}} {G}^{-2} - \frac{9}{40}G_{i j} \mathcal{D}_{c}{G} \eta_{\underline{a} \underline{b}} \mathcal{D}^{c}{\boldsymbol{W}} X^{i j} {G}^{-2}
\doublespacedmathend
\end{adjustwidth}

\subsubsection{$\lambda^{i}_{\a, R^2}$}

\begin{adjustwidth}{0cm}{3cm}
\doublespacedmathbegin
\frac{3}{8}{\rm i} \mathbf{X}^{i}\,_{j} \lambda^{j \beta} {G}^{-3} \varphi_{k \alpha} \varphi^{k}_{\beta}+\frac{3}{8}{\rm i} \mathbf{X}_{j k} \lambda^{j \beta} {G}^{-3} \varphi^{i}_{\beta} \varphi^{k}_{\alpha} - \frac{3}{8}G^{i}\,_{j} \lambda^{\beta}_{k} {G}^{-3} \nabla_{\alpha}\,^{\rho}{\boldsymbol{\lambda}^{k}_{\rho}} \varphi^{j}_{\beta} - \frac{3}{8}G_{j k} \lambda^{i \beta} {G}^{-3} \nabla_{\alpha}\,^{\rho}{\boldsymbol{\lambda}^{j}_{\rho}} \varphi^{k}_{\beta} - \frac{3}{16}G^{i}\,_{j} W_{\alpha}\,^{\beta} \lambda^{\rho}_{k} \boldsymbol{\lambda}^{k}_{\beta} {G}^{-3} \varphi^{j}_{\rho}+\frac{1}{16}G_{j k} W_{\alpha}\,^{\beta} \lambda^{i \rho} \boldsymbol{\lambda}^{j}_{\beta} {G}^{-3} \varphi^{k}_{\rho} - \frac{3}{32}{\rm i} G^{i}\,_{j} \boldsymbol{W} \lambda^{\beta}_{k} X^{k}_{\alpha} {G}^{-3} \varphi^{j}_{\beta} - \frac{3}{32}{\rm i} G_{j k} \boldsymbol{W} \lambda^{i \beta} X^{j}_{\alpha} {G}^{-3} \varphi^{k}_{\beta} - \frac{1}{4}{\rm i} G_{j k} \mathbf{X}^{i j} F_{\alpha}\,^{\beta} {G}^{-3} \varphi^{k}_{\beta} - \frac{1}{4}{\rm i} G_{j k} W \mathbf{X}^{i j} W_{\alpha}\,^{\beta} {G}^{-3} \varphi^{k}_{\beta} - \frac{3}{8}{\rm i} G_{j k} X^{i}\,_{l} \mathbf{X}^{j l} {G}^{-3} \varphi^{k}_{\alpha} - \frac{5}{8}{\rm i} G_{j k} \mathbf{X}^{i j} {G}^{-3} \nabla_{\alpha}\,^{\beta}{W} \varphi^{k}_{\beta}+\frac{9}{8}{\rm i} G^{i}\,_{j} G_{k l} \mathbf{X}^{k}\,_{m} \lambda^{m \beta} {G}^{-5} \varphi^{j}_{\alpha} \varphi^{l}_{\beta} - \frac{3}{16}F G^{i}\,_{j} \mathbf{X}^{j}\,_{k} \lambda^{k}_{\alpha} {G}^{-3} - \frac{3}{16}\mathcal{H}_{\alpha}\,^{\beta} G^{i}\,_{j} \mathbf{X}^{j}\,_{k} \lambda^{k}_{\beta} {G}^{-3} - \frac{3}{8}G_{j k} \mathbf{X}^{j}\,_{l} \lambda^{l}_{\beta} {G}^{-3} \nabla_{\alpha}\,^{\beta}{G^{i k}}+\frac{1}{2}{\rm i} W {G}^{-3} \nabla^{\beta \rho}{\boldsymbol{\lambda}^{i}_{\beta}} \varphi_{j \alpha} \varphi^{j}_{\rho}+\frac{1}{2}{\rm i} W {G}^{-3} \nabla^{\beta \rho}{\boldsymbol{\lambda}_{j \beta}} \varphi^{i}_{\rho} \varphi^{j}_{\alpha} - \frac{1}{2}G_{j k} \lambda^{i}_{\alpha} {G}^{-3} \nabla^{\beta \rho}{\boldsymbol{\lambda}^{j}_{\beta}} \varphi^{k}_{\rho}%
+\frac{3}{2}{\rm i} G^{i}\,_{j} G_{k l} W {G}^{-5} \nabla^{\beta \rho}{\boldsymbol{\lambda}^{k}_{\beta}} \varphi^{j}_{\alpha} \varphi^{l}_{\rho}-{\rm i} G^{i}\,_{j} W {G}^{-3} \nabla^{\beta \rho}{\mathbf{F}_{\alpha \beta}} \varphi^{j}_{\rho} - \frac{3}{2}{\rm i} G^{i}\,_{j} W W_{\alpha \beta} {G}^{-3} \nabla^{\beta \rho}{\boldsymbol{W}} \varphi^{j}_{\rho} - \frac{3}{2}{\rm i} G^{i}\,_{j} W \boldsymbol{W} {G}^{-3} \nabla^{\beta \rho}{W_{\alpha \beta}} \varphi^{j}_{\rho} - \frac{1}{2}{\rm i} G_{j k} W {G}^{-3} \nabla_{\alpha}\,^{\beta}{\mathbf{X}^{i j}} \varphi^{k}_{\beta}+\frac{1}{2}{\rm i} G^{i}\,_{j} W {G}^{-3} \nabla_{\beta}\,^{\rho}{\nabla_{\alpha}\,^{\beta}{\boldsymbol{W}}} \varphi^{j}_{\rho} - \frac{3}{2}{\rm i} G^{i}\,_{j} W W^{\beta}\,_{\rho} {G}^{-3} \nabla_{\alpha}\,^{\rho}{\boldsymbol{W}} \varphi^{j}_{\beta} - \frac{3}{2}{\rm i} G^{i}\,_{j} W W^{\beta \rho} \mathbf{F}_{\beta \rho} {G}^{-3} \varphi^{j}_{\alpha} - \frac{39}{16}{\rm i} G^{i}\,_{j} W \boldsymbol{W} W^{\beta \rho} W_{\beta \rho} {G}^{-3} \varphi^{j}_{\alpha} - \frac{3}{8}{\rm i} G_{j k} W \boldsymbol{\lambda}^{j}_{\alpha} X^{i \beta} {G}^{-3} \varphi^{k}_{\beta}+\frac{3}{8}{\rm i} G_{j k} W \boldsymbol{\lambda}^{j \beta} X^{i}_{\beta} {G}^{-3} \varphi^{k}_{\alpha} - \frac{3}{32}{\rm i} G^{i}\,_{j} W \boldsymbol{\lambda}_{k}^{\beta} X^{k}_{\alpha} {G}^{-3} \varphi^{j}_{\beta} - \frac{3}{32}{\rm i} G_{j k} W \boldsymbol{\lambda}^{i \beta} X^{j}_{\alpha} {G}^{-3} \varphi^{k}_{\beta} - \frac{3}{8}{\rm i} G^{i}\,_{j} W \boldsymbol{\lambda}_{k \alpha} X^{k \beta} {G}^{-3} \varphi^{j}_{\beta}+\frac{9}{8}{\rm i} G^{i}\,_{j} W \boldsymbol{\lambda}_{k}^{\beta} X^{k}_{\beta} {G}^{-3} \varphi^{j}_{\alpha}+\frac{3}{8}{\rm i} G_{j k} W \boldsymbol{\lambda}^{i}_{\alpha} X^{j \beta} {G}^{-3} \varphi^{k}_{\beta} - \frac{3}{8}{\rm i} G_{j k} W \boldsymbol{\lambda}^{i \beta} X^{j}_{\beta} {G}^{-3} \varphi^{k}_{\alpha} - \frac{3}{2}{\rm i} G^{i}\,_{j} W \boldsymbol{W} {G}^{-3} \nabla_{\alpha}\,^{\rho}{W^{\beta}\,_{\rho}} \varphi^{j}_{\beta} - \frac{1}{4}F G^{i}\,_{j} W {G}^{-3} \nabla_{\alpha}\,^{\beta}{\boldsymbol{\lambda}^{j}_{\beta}} - \frac{1}{4}\mathcal{H}_{\alpha \beta} G^{i}\,_{j} W {G}^{-3} \nabla^{\beta \rho}{\boldsymbol{\lambda}^{j}_{\rho}}%
+\frac{1}{2}G_{j k} W {G}^{-3} \nabla_{\rho}\,^{\beta}{\boldsymbol{\lambda}^{j}_{\beta}} \nabla_{\alpha}\,^{\rho}{G^{i k}} - \frac{3}{4}{\rm i} W W^{\beta \rho} \boldsymbol{\lambda}^{i}_{\beta} {G}^{-3} \varphi_{j \alpha} \varphi^{j}_{\rho} - \frac{3}{4}{\rm i} W W^{\beta \rho} \boldsymbol{\lambda}_{j \beta} {G}^{-3} \varphi^{i}_{\rho} \varphi^{j}_{\alpha} - \frac{9}{4}{\rm i} G^{i}\,_{j} G_{k l} W W^{\beta \rho} \boldsymbol{\lambda}^{k}_{\beta} {G}^{-5} \varphi^{j}_{\alpha} \varphi^{l}_{\rho} - \frac{3}{8}F G^{i}\,_{j} W W_{\alpha}\,^{\beta} \boldsymbol{\lambda}^{j}_{\beta} {G}^{-3}+\frac{3}{8}\mathcal{H}_{\alpha}\,^{\beta} G^{i}\,_{j} W W_{\beta}\,^{\rho} \boldsymbol{\lambda}^{j}_{\rho} {G}^{-3}+\frac{3}{4}G_{j k} W W^{\beta}\,_{\rho} \boldsymbol{\lambda}^{j}_{\beta} {G}^{-3} \nabla_{\alpha}\,^{\rho}{G^{i k}}+\frac{3}{4}W \boldsymbol{W} X^{i \beta} {G}^{-3} \varphi_{j \alpha} \varphi^{j}_{\beta}+\frac{3}{4}W \boldsymbol{W} X_{j}^{\beta} {G}^{-3} \varphi^{i}_{\beta} \varphi^{j}_{\alpha}+\frac{3}{8}{\rm i} G_{j k} \boldsymbol{W} \lambda^{i}_{\alpha} X^{j \beta} {G}^{-3} \varphi^{k}_{\beta}+\frac{3}{32}{\rm i} G^{i}\,_{j} W \boldsymbol{W} Y {G}^{-3} \varphi^{j}_{\alpha}+\frac{9}{4}G^{i}\,_{j} G_{k l} W \boldsymbol{W} X^{k \beta} {G}^{-5} \varphi^{j}_{\alpha} \varphi^{l}_{\beta} - \frac{3}{8}{\rm i} F G^{i}\,_{j} W \boldsymbol{W} X^{j}_{\alpha} {G}^{-3}+\frac{3}{8}{\rm i} \mathcal{H}_{\alpha}\,^{\beta} G^{i}\,_{j} W \boldsymbol{W} X^{j}_{\beta} {G}^{-3}+\frac{3}{4}{\rm i} G_{j k} W \boldsymbol{W} X^{j}_{\beta} {G}^{-3} \nabla_{\alpha}\,^{\beta}{G^{i k}}+\frac{3}{4}G_{j k} W \mathbf{X}^{i}\,_{l} {G}^{-5} \varphi^{l}_{\alpha} \varphi^{j \beta} \varphi^{k}_{\beta}+\frac{3}{4}G_{j k} W \mathbf{X}^{j k} {G}^{-5} \varphi^{i \beta} \varphi_{l \alpha} \varphi^{l}_{\beta}+\frac{3}{8}{\rm i} G_{j k} G_{l m} \mathbf{X}^{j k} \lambda^{i}_{\alpha} {G}^{-5} \varphi^{l \beta} \varphi^{m}_{\beta}+\frac{3}{4}{\rm i} G^{i}\,_{j} G_{k l} W {G}^{-5} \nabla_{\alpha}\,^{\beta}{\boldsymbol{\lambda}^{j}_{\beta}} \varphi^{k \rho} \varphi^{l}_{\rho}+\frac{9}{8}{\rm i} G^{i}\,_{j} G_{k l} W W_{\alpha}\,^{\beta} \boldsymbol{\lambda}^{j}_{\beta} {G}^{-5} \varphi^{k \rho} \varphi^{l}_{\rho}%
 - \frac{9}{8}G^{i}\,_{j} G_{k l} W \boldsymbol{W} X^{j}_{\alpha} {G}^{-5} \varphi^{k \beta} \varphi^{l}_{\beta} - \frac{15}{8}G^{i}\,_{j} G_{k l} G_{m n} W \mathbf{X}^{k l} {G}^{-7} \varphi^{j}_{\alpha} \varphi^{m \beta} \varphi^{n}_{\beta} - \frac{3}{8}{\rm i} F G^{i}\,_{j} G_{k l} W \mathbf{X}^{k l} {G}^{-5} \varphi^{j}_{\alpha}+\frac{3}{8}{\rm i} \mathcal{H}_{\alpha}\,^{\beta} G^{i}\,_{j} G_{k l} W \mathbf{X}^{k l} {G}^{-5} \varphi^{j}_{\beta}+\frac{3}{4}{\rm i} G_{j k} G_{l m} W \mathbf{X}^{j k} {G}^{-5} \nabla_{\alpha}\,^{\beta}{G^{i l}} \varphi^{m}_{\beta} - \frac{1}{4}{\rm i} \mathbf{X}_{j k} \lambda^{i}_{\alpha} {G}^{-3} \varphi^{j \beta} \varphi^{k}_{\beta}+\frac{1}{2}{\rm i} W {G}^{-3} \nabla_{\alpha}\,^{\beta}{\boldsymbol{\lambda}_{j \beta}} \varphi^{i \rho} \varphi^{j}_{\rho}+\frac{3}{4}{\rm i} W W_{\alpha}\,^{\beta} \boldsymbol{\lambda}_{j \beta} {G}^{-3} \varphi^{i \rho} \varphi^{j}_{\rho} - \frac{3}{4}W \boldsymbol{W} X_{j \alpha} {G}^{-3} \varphi^{i \beta} \varphi^{j}_{\beta}+\frac{3}{4}G^{i}\,_{j} W \mathbf{X}_{k l} {G}^{-5} \varphi^{j}_{\alpha} \varphi^{k \beta} \varphi^{l}_{\beta}+\frac{1}{4}{\rm i} F W \mathbf{X}^{i}\,_{j} {G}^{-3} \varphi^{j}_{\alpha} - \frac{1}{4}{\rm i} \mathcal{H}_{\alpha}\,^{\beta} W \mathbf{X}^{i}\,_{j} {G}^{-3} \varphi^{j}_{\beta} - \frac{1}{2}{\rm i} W \mathbf{X}_{j k} {G}^{-3} \nabla_{\alpha}\,^{\beta}{G^{i j}} \varphi^{k}_{\beta} - \frac{1}{2}{\rm i} G_{j k} W \mathbf{X}^{j k} {G}^{-3} \nabla_{\alpha}\,^{\beta}{\varphi^{i}_{\beta}}+{\rm i} G_{j k} W \mathbf{X}^{j k} W_{\alpha}\,^{\beta} {G}^{-3} \varphi^{i}_{\beta} - \frac{5}{16}F G_{j k} \mathbf{X}^{j k} \lambda^{i}_{\alpha} {G}^{-3} - \frac{3}{2}W W_{\alpha \beta} {G}^{-1} \nabla^{\beta \rho}{\boldsymbol{\lambda}^{i}_{\rho}}+X^{i}\,_{j} {G}^{-1} \nabla_{\alpha}\,^{\beta}{\boldsymbol{\lambda}^{j}_{\beta}}+\frac{1}{2}{G}^{-1} \nabla_{\alpha \rho}{W} \nabla^{\rho \beta}{\boldsymbol{\lambda}^{i}_{\beta}} - \frac{1}{2}G^{i}\,_{j} \lambda_{k \beta} {G}^{-3} \nabla^{\beta \rho}{\boldsymbol{\lambda}^{k}_{\rho}} \varphi^{j}_{\alpha}%
+\lambda^{i}_{\rho} {G}^{-1} \nabla^{\rho \beta}{\mathbf{F}_{\alpha \beta}} - \frac{3}{4}W_{\alpha \beta} \lambda^{i}_{\rho} {G}^{-1} \nabla^{\beta \rho}{\boldsymbol{W}} - \frac{1}{2}\lambda_{j \beta} {G}^{-1} \nabla_{\alpha}\,^{\beta}{\mathbf{X}^{i j}}+\frac{1}{2}\lambda^{i}_{\beta} {G}^{-1} \nabla_{\rho}\,^{\beta}{\nabla_{\alpha}\,^{\rho}{\boldsymbol{W}}}+\frac{3}{2}W_{\alpha}\,^{\beta} \mathbf{F}_{\beta}\,^{\rho} \lambda^{i}_{\rho} {G}^{-1}+\frac{9}{4}\boldsymbol{W} W_{\alpha}\,^{\beta} W_{\beta}\,^{\rho} \lambda^{i}_{\rho} {G}^{-1} - \frac{1}{2}\mathbf{X}^{i}\,_{j} W_{\alpha}\,^{\beta} \lambda^{j}_{\beta} {G}^{-1} - \frac{3}{2}W^{\beta}\,_{\rho} \lambda^{i}_{\beta} {G}^{-1} \nabla_{\alpha}\,^{\rho}{\boldsymbol{W}}+\lambda^{\beta}_{j} \boldsymbol{\lambda}^{j \rho} W_{\alpha \beta \rho}\,^{i} {G}^{-1}+\frac{3}{4}\boldsymbol{\lambda}_{j \alpha} \lambda^{j \beta} X^{i}_{\beta} {G}^{-1}+\frac{3}{4}\lambda_{j \alpha} \boldsymbol{\lambda}^{j \beta} X^{i}_{\beta} {G}^{-1} - \frac{3}{8}\lambda^{i \beta} \boldsymbol{\lambda}_{j \beta} X^{j}_{\alpha} {G}^{-1} - \frac{3}{8}\boldsymbol{\lambda}^{i \beta} \lambda_{j \beta} X^{j}_{\alpha} {G}^{-1}+\frac{3}{4}\lambda^{i}_{\alpha} \boldsymbol{\lambda}_{j}^{\beta} X^{j}_{\beta} {G}^{-1}+\frac{3}{4}\lambda^{i \beta} \boldsymbol{\lambda}_{j \alpha} X^{j}_{\beta} {G}^{-1}+\frac{3}{4}\boldsymbol{\lambda}^{i}_{\alpha} \lambda^{\beta}_{j} X^{j}_{\beta} {G}^{-1}+\frac{3}{4}\boldsymbol{\lambda}^{i \beta} \lambda_{j \alpha} X^{j}_{\beta} {G}^{-1}-2\Phi_{\alpha}\,^{\beta i}\,_{j} \boldsymbol{W} \lambda^{j}_{\beta} {G}^{-1}+\frac{3}{2}\boldsymbol{W} \lambda^{i \beta} {G}^{-1} \nabla_{\alpha}\,^{\rho}{W_{\beta \rho}}+\frac{9}{4}W W_{\alpha}\,^{\beta} W_{\beta}\,^{\rho} \boldsymbol{\lambda}^{i}_{\rho} {G}^{-1}%
 - \frac{1}{2}X^{i}\,_{j} W_{\alpha}\,^{\beta} \boldsymbol{\lambda}^{j}_{\beta} {G}^{-1} - \frac{3}{2}W^{\beta}\,_{\rho} \boldsymbol{\lambda}^{i}_{\beta} {G}^{-1} \nabla_{\alpha}\,^{\rho}{W} - \frac{3}{2}G^{i}\,_{j} W^{\beta \rho} \lambda_{k \beta} \boldsymbol{\lambda}^{k}_{\rho} {G}^{-3} \varphi^{j}_{\alpha}+\frac{9}{4}{\rm i} W \boldsymbol{W} W_{\alpha}\,^{\beta} X^{i}_{\beta} {G}^{-1} - \frac{3}{8}{\rm i} \boldsymbol{W} X^{i}\,_{j} X^{j}_{\alpha} {G}^{-1} - \frac{3}{2}{\rm i} \boldsymbol{W} X^{i}_{\beta} {G}^{-1} \nabla_{\alpha}\,^{\beta}{W}+\frac{9}{8}{\rm i} G^{i}\,_{j} \boldsymbol{W} \lambda^{\beta}_{k} X^{k}_{\beta} {G}^{-3} \varphi^{j}_{\alpha}+\mathbf{X}^{i}\,_{j} {G}^{-1} \nabla_{\alpha}\,^{\beta}{\lambda^{j}_{\beta}} - \frac{3}{8}{\rm i} W \mathbf{X}^{i}\,_{j} X^{j}_{\alpha} {G}^{-1}+\frac{3}{4}{\rm i} G^{i}\,_{j} X_{k l} \mathbf{X}^{k l} {G}^{-3} \varphi^{j}_{\alpha} - \frac{1}{4}\lambda^{i}_{\alpha} {G}^{-1} \nabla_{\beta \rho}{\nabla^{\beta \rho}{\boldsymbol{W}}} - \frac{1}{4}{\rm i} G^{i}\,_{j} W {G}^{-3} \nabla_{\beta \rho}{\nabla^{\beta \rho}{\boldsymbol{W}}} \varphi^{j}_{\alpha} - \frac{1}{4}W {G}^{-1} \nabla_{\beta \rho}{\nabla^{\beta \rho}{\boldsymbol{\lambda}^{i}_{\alpha}}}+\frac{3}{2}W \boldsymbol{\lambda}^{i \beta} {G}^{-1} \nabla_{\alpha}\,^{\rho}{W_{\beta \rho}} - \frac{3}{2}W W^{\beta}\,_{\rho} {G}^{-1} \nabla_{\alpha}\,^{\rho}{\boldsymbol{\lambda}^{i}_{\beta}} - \frac{3}{2}{\rm i} W \boldsymbol{W} {G}^{-1} \nabla_{\alpha}\,^{\beta}{X^{i}_{\beta}} - \frac{3}{2}{\rm i} W X^{i}_{\beta} {G}^{-1} \nabla_{\alpha}\,^{\beta}{\boldsymbol{W}}-2\Phi_{\alpha}\,^{\beta i}\,_{j} W \boldsymbol{\lambda}^{j}_{\beta} {G}^{-1}+4{\rm i} W \boldsymbol{W} W^{\beta \rho} W_{\alpha \beta \rho}\,^{i} {G}^{-1}+2{\rm i} W \mathbf{F}^{\beta \rho} W_{\alpha \beta \rho}\,^{i} {G}^{-1}%
+\frac{3}{8}{\rm i} X^{i}\,_{j} \boldsymbol{\lambda}^{j \beta} {G}^{-3} \varphi_{k \alpha} \varphi^{k}_{\beta}+\frac{3}{8}{\rm i} X_{j k} \boldsymbol{\lambda}^{j \beta} {G}^{-3} \varphi^{i}_{\beta} \varphi^{k}_{\alpha} - \frac{3}{8}G^{i}\,_{j} \boldsymbol{\lambda}_{k}^{\rho} {G}^{-3} \nabla_{\alpha}\,^{\beta}{\lambda^{k}_{\beta}} \varphi^{j}_{\rho} - \frac{3}{8}G_{j k} \boldsymbol{\lambda}^{i \rho} {G}^{-3} \nabla_{\alpha}\,^{\beta}{\lambda^{j}_{\beta}} \varphi^{k}_{\rho} - \frac{3}{16}G^{i}\,_{j} W_{\alpha}\,^{\beta} \lambda_{k \beta} \boldsymbol{\lambda}^{k \rho} {G}^{-3} \varphi^{j}_{\rho}+\frac{1}{16}G_{j k} W_{\alpha}\,^{\beta} \boldsymbol{\lambda}^{i \rho} \lambda^{j}_{\beta} {G}^{-3} \varphi^{k}_{\rho} - \frac{1}{4}{\rm i} G_{j k} X^{i j} \mathbf{F}_{\alpha}\,^{\beta} {G}^{-3} \varphi^{k}_{\beta} - \frac{1}{4}{\rm i} G_{j k} \boldsymbol{W} X^{i j} W_{\alpha}\,^{\beta} {G}^{-3} \varphi^{k}_{\beta} - \frac{3}{8}{\rm i} G_{j k} \mathbf{X}^{i}\,_{l} X^{j l} {G}^{-3} \varphi^{k}_{\alpha} - \frac{5}{8}{\rm i} G_{j k} X^{i j} {G}^{-3} \nabla_{\alpha}\,^{\beta}{\boldsymbol{W}} \varphi^{k}_{\beta}+\frac{9}{8}{\rm i} G^{i}\,_{j} G_{k l} X^{k}\,_{m} \boldsymbol{\lambda}^{m \beta} {G}^{-5} \varphi^{j}_{\alpha} \varphi^{l}_{\beta} - \frac{3}{16}F G^{i}\,_{j} X^{j}\,_{k} \boldsymbol{\lambda}^{k}_{\alpha} {G}^{-3} - \frac{3}{16}\mathcal{H}_{\alpha}\,^{\beta} G^{i}\,_{j} X^{j}\,_{k} \boldsymbol{\lambda}^{k}_{\beta} {G}^{-3} - \frac{3}{8}G_{j k} X^{j}\,_{l} \boldsymbol{\lambda}^{l}_{\beta} {G}^{-3} \nabla_{\alpha}\,^{\beta}{G^{i k}}+\frac{1}{2}{\rm i} \boldsymbol{W} {G}^{-3} \nabla^{\beta \rho}{\lambda^{i}_{\beta}} \varphi_{j \alpha} \varphi^{j}_{\rho}+\frac{1}{2}{\rm i} \boldsymbol{W} {G}^{-3} \nabla^{\beta \rho}{\lambda_{j \beta}} \varphi^{i}_{\rho} \varphi^{j}_{\alpha} - \frac{1}{2}G_{j k} \boldsymbol{\lambda}^{i}_{\alpha} {G}^{-3} \nabla^{\beta \rho}{\lambda^{j}_{\beta}} \varphi^{k}_{\rho}+\frac{3}{2}{\rm i} G^{i}\,_{j} G_{k l} \boldsymbol{W} {G}^{-5} \nabla^{\beta \rho}{\lambda^{k}_{\beta}} \varphi^{j}_{\alpha} \varphi^{l}_{\rho}-{\rm i} G^{i}\,_{j} \boldsymbol{W} {G}^{-3} \nabla^{\beta \rho}{F_{\alpha \beta}} \varphi^{j}_{\rho} - \frac{3}{2}{\rm i} G^{i}\,_{j} \boldsymbol{W} W_{\alpha \beta} {G}^{-3} \nabla^{\beta \rho}{W} \varphi^{j}_{\rho}%
 - \frac{1}{2}{\rm i} G_{j k} \boldsymbol{W} {G}^{-3} \nabla_{\alpha}\,^{\beta}{X^{i j}} \varphi^{k}_{\beta}+\frac{1}{2}{\rm i} G^{i}\,_{j} \boldsymbol{W} {G}^{-3} \nabla_{\beta}\,^{\rho}{\nabla_{\alpha}\,^{\beta}{W}} \varphi^{j}_{\rho} - \frac{3}{2}{\rm i} G^{i}\,_{j} \boldsymbol{W} W^{\beta}\,_{\rho} {G}^{-3} \nabla_{\alpha}\,^{\rho}{W} \varphi^{j}_{\beta} - \frac{3}{2}{\rm i} G^{i}\,_{j} \boldsymbol{W} W^{\beta \rho} F_{\beta \rho} {G}^{-3} \varphi^{j}_{\alpha} - \frac{3}{8}{\rm i} G_{j k} \boldsymbol{W} \lambda^{j}_{\alpha} X^{i \beta} {G}^{-3} \varphi^{k}_{\beta}+\frac{3}{8}{\rm i} G_{j k} \boldsymbol{W} \lambda^{j \beta} X^{i}_{\beta} {G}^{-3} \varphi^{k}_{\alpha} - \frac{3}{8}{\rm i} G^{i}\,_{j} \boldsymbol{W} \lambda_{k \alpha} X^{k \beta} {G}^{-3} \varphi^{j}_{\beta} - \frac{3}{8}{\rm i} G_{j k} \boldsymbol{W} \lambda^{i \beta} X^{j}_{\beta} {G}^{-3} \varphi^{k}_{\alpha} - \frac{1}{4}F G^{i}\,_{j} \boldsymbol{W} {G}^{-3} \nabla_{\alpha}\,^{\beta}{\lambda^{j}_{\beta}} - \frac{1}{4}\mathcal{H}_{\alpha \beta} G^{i}\,_{j} \boldsymbol{W} {G}^{-3} \nabla^{\beta \rho}{\lambda^{j}_{\rho}}+\frac{1}{2}G_{j k} \boldsymbol{W} {G}^{-3} \nabla_{\rho}\,^{\beta}{\lambda^{j}_{\beta}} \nabla_{\alpha}\,^{\rho}{G^{i k}} - \frac{3}{4}{\rm i} \boldsymbol{W} W^{\beta \rho} \lambda^{i}_{\beta} {G}^{-3} \varphi_{j \alpha} \varphi^{j}_{\rho} - \frac{3}{4}{\rm i} \boldsymbol{W} W^{\beta \rho} \lambda_{j \beta} {G}^{-3} \varphi^{i}_{\rho} \varphi^{j}_{\alpha} - \frac{9}{4}{\rm i} G^{i}\,_{j} G_{k l} \boldsymbol{W} W^{\beta \rho} \lambda^{k}_{\beta} {G}^{-5} \varphi^{j}_{\alpha} \varphi^{l}_{\rho} - \frac{3}{8}F G^{i}\,_{j} \boldsymbol{W} W_{\alpha}\,^{\beta} \lambda^{j}_{\beta} {G}^{-3}+\frac{3}{8}\mathcal{H}_{\alpha}\,^{\beta} G^{i}\,_{j} \boldsymbol{W} W_{\beta}\,^{\rho} \lambda^{j}_{\rho} {G}^{-3}+\frac{3}{4}G_{j k} \boldsymbol{W} W^{\beta}\,_{\rho} \lambda^{j}_{\beta} {G}^{-3} \nabla_{\alpha}\,^{\rho}{G^{i k}}+\frac{3}{4}G_{j k} \boldsymbol{W} X^{i}\,_{l} {G}^{-5} \varphi^{l}_{\alpha} \varphi^{j \beta} \varphi^{k}_{\beta}+\frac{3}{4}G_{j k} \boldsymbol{W} X^{j k} {G}^{-5} \varphi^{i \beta} \varphi_{l \alpha} \varphi^{l}_{\beta}+\frac{3}{8}{\rm i} G_{j k} G_{l m} X^{j k} \boldsymbol{\lambda}^{i}_{\alpha} {G}^{-5} \varphi^{l \beta} \varphi^{m}_{\beta}%
+\frac{3}{4}{\rm i} G^{i}\,_{j} G_{k l} \boldsymbol{W} {G}^{-5} \nabla_{\alpha}\,^{\beta}{\lambda^{j}_{\beta}} \varphi^{k \rho} \varphi^{l}_{\rho}+\frac{9}{8}{\rm i} G^{i}\,_{j} G_{k l} \boldsymbol{W} W_{\alpha}\,^{\beta} \lambda^{j}_{\beta} {G}^{-5} \varphi^{k \rho} \varphi^{l}_{\rho} - \frac{15}{8}G^{i}\,_{j} G_{k l} G_{m n} \boldsymbol{W} X^{k l} {G}^{-7} \varphi^{j}_{\alpha} \varphi^{m \beta} \varphi^{n}_{\beta} - \frac{3}{8}{\rm i} F G^{i}\,_{j} G_{k l} \boldsymbol{W} X^{k l} {G}^{-5} \varphi^{j}_{\alpha}+\frac{3}{8}{\rm i} \mathcal{H}_{\alpha}\,^{\beta} G^{i}\,_{j} G_{k l} \boldsymbol{W} X^{k l} {G}^{-5} \varphi^{j}_{\beta}+\frac{3}{4}{\rm i} G_{j k} G_{l m} \boldsymbol{W} X^{j k} {G}^{-5} \nabla_{\alpha}\,^{\beta}{G^{i l}} \varphi^{m}_{\beta} - \frac{1}{4}{\rm i} X_{j k} \boldsymbol{\lambda}^{i}_{\alpha} {G}^{-3} \varphi^{j \beta} \varphi^{k}_{\beta}+\frac{1}{2}{\rm i} \boldsymbol{W} {G}^{-3} \nabla_{\alpha}\,^{\beta}{\lambda_{j \beta}} \varphi^{i \rho} \varphi^{j}_{\rho}+\frac{3}{4}{\rm i} \boldsymbol{W} W_{\alpha}\,^{\beta} \lambda_{j \beta} {G}^{-3} \varphi^{i \rho} \varphi^{j}_{\rho}+\frac{3}{4}G^{i}\,_{j} \boldsymbol{W} X_{k l} {G}^{-5} \varphi^{j}_{\alpha} \varphi^{k \beta} \varphi^{l}_{\beta}+\frac{1}{4}{\rm i} F \boldsymbol{W} X^{i}\,_{j} {G}^{-3} \varphi^{j}_{\alpha} - \frac{1}{4}{\rm i} \mathcal{H}_{\alpha}\,^{\beta} \boldsymbol{W} X^{i}\,_{j} {G}^{-3} \varphi^{j}_{\beta} - \frac{1}{2}{\rm i} \boldsymbol{W} X_{j k} {G}^{-3} \nabla_{\alpha}\,^{\beta}{G^{i j}} \varphi^{k}_{\beta} - \frac{1}{2}{\rm i} G_{j k} \boldsymbol{W} X^{j k} {G}^{-3} \nabla_{\alpha}\,^{\beta}{\varphi^{i}_{\beta}}+{\rm i} G_{j k} \boldsymbol{W} X^{j k} W_{\alpha}\,^{\beta} {G}^{-3} \varphi^{i}_{\beta} - \frac{5}{16}F G_{j k} X^{j k} \boldsymbol{\lambda}^{i}_{\alpha} {G}^{-3} - \frac{3}{2}\boldsymbol{W} W_{\alpha \beta} {G}^{-1} \nabla^{\beta \rho}{\lambda^{i}_{\rho}}+\frac{1}{2}{G}^{-1} \nabla_{\alpha \rho}{\boldsymbol{W}} \nabla^{\rho \beta}{\lambda^{i}_{\beta}} - \frac{1}{2}G^{i}\,_{j} \boldsymbol{\lambda}_{k \rho} {G}^{-3} \nabla^{\rho \beta}{\lambda^{k}_{\beta}} \varphi^{j}_{\alpha}+\boldsymbol{\lambda}^{i}_{\rho} {G}^{-1} \nabla^{\rho \beta}{F_{\alpha \beta}}%
 - \frac{3}{4}W_{\alpha \beta} \boldsymbol{\lambda}^{i}_{\rho} {G}^{-1} \nabla^{\beta \rho}{W} - \frac{1}{2}\boldsymbol{\lambda}_{j \beta} {G}^{-1} \nabla_{\alpha}\,^{\beta}{X^{i j}}+\frac{1}{2}\boldsymbol{\lambda}^{i}_{\beta} {G}^{-1} \nabla_{\rho}\,^{\beta}{\nabla_{\alpha}\,^{\rho}{W}}+\frac{3}{2}W_{\alpha}\,^{\beta} F_{\beta}\,^{\rho} \boldsymbol{\lambda}^{i}_{\rho} {G}^{-1} - \frac{1}{4}\boldsymbol{\lambda}^{i}_{\alpha} {G}^{-1} \nabla_{\beta \rho}{\nabla^{\beta \rho}{W}} - \frac{1}{4}{\rm i} G^{i}\,_{j} \boldsymbol{W} {G}^{-3} \nabla_{\beta \rho}{\nabla^{\beta \rho}{W}} \varphi^{j}_{\alpha} - \frac{1}{4}\boldsymbol{W} {G}^{-1} \nabla_{\beta \rho}{\nabla^{\beta \rho}{\lambda^{i}_{\alpha}}} - \frac{3}{2}\boldsymbol{W} W^{\beta}\,_{\rho} {G}^{-1} \nabla_{\alpha}\,^{\rho}{\lambda^{i}_{\beta}}+2{\rm i} \boldsymbol{W} F^{\beta \rho} W_{\alpha \beta \rho}\,^{i} {G}^{-1} - \frac{1}{2}{\rm i} F^{\beta \rho} \boldsymbol{\lambda}^{i}_{\beta} {G}^{-3} \varphi_{j \alpha} \varphi^{j}_{\rho} - \frac{1}{2}{\rm i} F^{\beta \rho} \boldsymbol{\lambda}_{j \beta} {G}^{-3} \varphi^{i}_{\rho} \varphi^{j}_{\alpha}+\frac{1}{4}G_{j k} \boldsymbol{\lambda}^{j}_{\rho} {G}^{-3} \nabla_{\alpha}\,^{\rho}{\lambda^{i \beta}} \varphi^{k}_{\beta} - \frac{1}{4}G_{j k} \boldsymbol{\lambda}^{j \beta} {G}^{-3} \nabla_{\alpha}\,^{\rho}{\lambda^{i}_{\beta}} \varphi^{k}_{\rho}+\frac{1}{4}G_{j k} \boldsymbol{\lambda}^{j}_{\alpha} {G}^{-3} \nabla^{\beta \rho}{\lambda^{i}_{\beta}} \varphi^{k}_{\rho} - \frac{1}{4}G_{j k} \boldsymbol{\lambda}^{j}_{\rho} {G}^{-3} \nabla^{\rho \beta}{\lambda^{i}_{\beta}} \varphi^{k}_{\alpha}+\frac{3}{4}G_{j k} W^{\beta \rho} \lambda^{i}_{\beta} \boldsymbol{\lambda}^{j}_{\rho} {G}^{-3} \varphi^{k}_{\alpha}-{\rm i} G^{i}\,_{j} \mathbf{F}_{\alpha}\,^{\beta} F_{\beta}\,^{\rho} {G}^{-3} \varphi^{j}_{\rho} - \frac{1}{2}{\rm i} G^{i}\,_{j} F^{\beta}\,_{\rho} {G}^{-3} \nabla_{\alpha}\,^{\rho}{\boldsymbol{W}} \varphi^{j}_{\beta} - \frac{3}{2}{\rm i} G^{i}\,_{j} G_{k l} F^{\beta \rho} \boldsymbol{\lambda}^{k}_{\beta} {G}^{-5} \varphi^{j}_{\alpha} \varphi^{l}_{\rho} - \frac{1}{4}F G^{i}\,_{j} F_{\alpha}\,^{\beta} \boldsymbol{\lambda}^{j}_{\beta} {G}^{-3}%
+\frac{1}{4}\mathcal{H}_{\alpha}\,^{\beta} G^{i}\,_{j} F_{\beta}\,^{\rho} \boldsymbol{\lambda}^{j}_{\rho} {G}^{-3}+\frac{1}{2}G_{j k} F^{\beta}\,_{\rho} \boldsymbol{\lambda}^{j}_{\beta} {G}^{-3} \nabla_{\alpha}\,^{\rho}{G^{i k}}+\frac{1}{4}{\rm i} \boldsymbol{\lambda}^{i}_{\beta} {G}^{-3} \nabla^{\beta \rho}{W} \varphi_{j \alpha} \varphi^{j}_{\rho}+\frac{1}{4}{\rm i} \boldsymbol{\lambda}_{j \beta} {G}^{-3} \nabla^{\beta \rho}{W} \varphi^{i}_{\rho} \varphi^{j}_{\alpha} - \frac{1}{2}{\rm i} G^{i}\,_{j} \mathbf{F}_{\alpha \beta} {G}^{-3} \nabla^{\beta \rho}{W} \varphi^{j}_{\rho}+\frac{1}{4}{\rm i} G^{i}\,_{j} {G}^{-3} \nabla_{\beta}\,^{\rho}{W} \nabla_{\alpha}\,^{\beta}{\boldsymbol{W}} \varphi^{j}_{\rho}+\frac{3}{4}{\rm i} G^{i}\,_{j} G_{k l} \boldsymbol{\lambda}^{k}_{\beta} {G}^{-5} \nabla^{\beta \rho}{W} \varphi^{j}_{\alpha} \varphi^{l}_{\rho}+\frac{1}{4}G_{j k} \boldsymbol{\lambda}^{j}_{\beta} {G}^{-3} \nabla^{\beta \rho}{\lambda^{i}_{\alpha}} \varphi^{k}_{\rho}+\frac{1}{8}G_{j k} W_{\alpha}\,^{\beta} \lambda^{i}_{\beta} \boldsymbol{\lambda}^{j \rho} {G}^{-3} \varphi^{k}_{\rho} - \frac{1}{8}F G^{i}\,_{j} \boldsymbol{\lambda}^{j}_{\beta} {G}^{-3} \nabla_{\alpha}\,^{\beta}{W} - \frac{1}{8}\mathcal{H}_{\alpha \beta} G^{i}\,_{j} \boldsymbol{\lambda}^{j}_{\rho} {G}^{-3} \nabla^{\beta \rho}{W}+\frac{1}{4}G_{j k} \boldsymbol{\lambda}^{j}_{\beta} {G}^{-3} \nabla_{\rho}\,^{\beta}{W} \nabla_{\alpha}\,^{\rho}{G^{i k}} - \frac{3}{8}{\rm i} G_{j k} \lambda^{i \beta} \boldsymbol{\lambda}_{l \beta} {G}^{-5} \varphi^{l}_{\alpha} \varphi^{j \rho} \varphi^{k}_{\rho} - \frac{3}{8}{\rm i} G_{j k} \boldsymbol{\lambda}^{i \beta} \lambda_{l \beta} {G}^{-5} \varphi^{l}_{\alpha} \varphi^{j \rho} \varphi^{k}_{\rho} - \frac{3}{4}{\rm i} G_{j k} \lambda^{j \beta} \boldsymbol{\lambda}^{k}_{\beta} {G}^{-5} \varphi^{i \rho} \varphi_{l \alpha} \varphi^{l}_{\rho}+\frac{3}{4}{\rm i} G^{i}\,_{j} G_{k l} F_{\alpha}\,^{\beta} \boldsymbol{\lambda}^{j}_{\beta} {G}^{-5} \varphi^{k \rho} \varphi^{l}_{\rho} - \frac{3}{8}{\rm i} G_{j k} G_{l m} X^{i j} \boldsymbol{\lambda}^{k}_{\alpha} {G}^{-5} \varphi^{l \beta} \varphi^{m}_{\beta}+\frac{3}{8}{\rm i} G^{i}\,_{j} G_{k l} \boldsymbol{\lambda}^{j}_{\beta} {G}^{-5} \nabla_{\alpha}\,^{\beta}{W} \varphi^{k \rho} \varphi^{l}_{\rho}+\frac{3}{4}{\rm i} G^{i}\,_{j} G_{k l} \mathbf{F}_{\alpha}\,^{\beta} \lambda^{j}_{\beta} {G}^{-5} \varphi^{k \rho} \varphi^{l}_{\rho} - \frac{3}{8}{\rm i} G_{j k} G_{l m} \mathbf{X}^{i j} \lambda^{k}_{\alpha} {G}^{-5} \varphi^{l \beta} \varphi^{m}_{\beta}%
+\frac{3}{8}{\rm i} G^{i}\,_{j} G_{k l} \lambda^{j}_{\beta} {G}^{-5} \nabla_{\alpha}\,^{\beta}{\boldsymbol{W}} \varphi^{k \rho} \varphi^{l}_{\rho}+\frac{15}{8}{\rm i} G^{i}\,_{j} G_{k l} G_{m n} \lambda^{k \beta} \boldsymbol{\lambda}^{l}_{\beta} {G}^{-7} \varphi^{j}_{\alpha} \varphi^{m \rho} \varphi^{n}_{\rho} - \frac{3}{8}F G^{i}\,_{j} G_{k l} \lambda^{k \beta} \boldsymbol{\lambda}^{l}_{\beta} {G}^{-5} \varphi^{j}_{\alpha}+\frac{3}{8}\mathcal{H}_{\alpha}\,^{\beta} G^{i}\,_{j} G_{k l} \lambda^{k \rho} \boldsymbol{\lambda}^{l}_{\rho} {G}^{-5} \varphi^{j}_{\beta}+\frac{3}{4}G_{j k} G_{l m} \lambda^{j \beta} \boldsymbol{\lambda}^{k}_{\beta} {G}^{-5} \nabla_{\alpha}\,^{\rho}{G^{i l}} \varphi^{m}_{\rho}+\frac{1}{2}{\rm i} F_{\alpha}\,^{\beta} \boldsymbol{\lambda}_{j \beta} {G}^{-3} \varphi^{i \rho} \varphi^{j}_{\rho}+\frac{1}{4}{\rm i} X^{i}\,_{j} \boldsymbol{\lambda}_{k \alpha} {G}^{-3} \varphi^{j \beta} \varphi^{k}_{\beta}+\frac{1}{4}{\rm i} \boldsymbol{\lambda}_{j \beta} {G}^{-3} \nabla_{\alpha}\,^{\beta}{W} \varphi^{i \rho} \varphi^{j}_{\rho}+\frac{1}{2}{\rm i} \mathbf{F}_{\alpha}\,^{\beta} \lambda_{j \beta} {G}^{-3} \varphi^{i \rho} \varphi^{j}_{\rho}+\frac{1}{4}{\rm i} \mathbf{X}^{i}\,_{j} \lambda_{k \alpha} {G}^{-3} \varphi^{j \beta} \varphi^{k}_{\beta}+\frac{1}{4}{\rm i} \lambda_{j \beta} {G}^{-3} \nabla_{\alpha}\,^{\beta}{\boldsymbol{W}} \varphi^{i \rho} \varphi^{j}_{\rho} - \frac{3}{4}{\rm i} G^{i}\,_{j} \lambda^{\beta}_{k} \boldsymbol{\lambda}_{l \beta} {G}^{-5} \varphi^{j}_{\alpha} \varphi^{k \rho} \varphi^{l}_{\rho}+\frac{1}{8}F \lambda^{i \beta} \boldsymbol{\lambda}_{j \beta} {G}^{-3} \varphi^{j}_{\alpha} - \frac{1}{8}\mathcal{H}_{\alpha}\,^{\beta} \lambda^{i \rho} \boldsymbol{\lambda}_{j \rho} {G}^{-3} \varphi^{j}_{\beta} - \frac{1}{4}\lambda^{\beta}_{j} \boldsymbol{\lambda}_{k \beta} {G}^{-3} \nabla_{\alpha}\,^{\rho}{G^{i j}} \varphi^{k}_{\rho}+\frac{1}{8}F \boldsymbol{\lambda}^{i \beta} \lambda_{j \beta} {G}^{-3} \varphi^{j}_{\alpha} - \frac{1}{8}\mathcal{H}_{\alpha}\,^{\beta} \boldsymbol{\lambda}^{i \rho} \lambda_{j \rho} {G}^{-3} \varphi^{j}_{\beta} - \frac{1}{4}\lambda^{\beta}_{k} \boldsymbol{\lambda}_{j \beta} {G}^{-3} \nabla_{\alpha}\,^{\rho}{G^{i j}} \varphi^{k}_{\rho} - \frac{1}{2}G_{j k} \lambda^{j \beta} \boldsymbol{\lambda}^{k}_{\beta} {G}^{-3} \nabla_{\alpha}\,^{\rho}{\varphi^{i}_{\rho}}+\frac{3}{4}G_{j k} W_{\alpha}\,^{\beta} \lambda^{j \rho} \boldsymbol{\lambda}^{k}_{\rho} {G}^{-3} \varphi^{i}_{\beta}%
+\frac{3}{8}G^{i}\,_{j} G_{k l} \lambda^{k \beta} \boldsymbol{\lambda}^{l}_{\beta} X^{j}_{\alpha} {G}^{-3}+\frac{5}{16}F G_{j k} X^{i j} \boldsymbol{\lambda}^{k}_{\alpha} {G}^{-3} - \frac{1}{4}F G^{i}\,_{j} \mathbf{F}_{\alpha}\,^{\beta} \lambda^{j}_{\beta} {G}^{-3}+\frac{5}{16}F G_{j k} \mathbf{X}^{i j} \lambda^{k}_{\alpha} {G}^{-3} - \frac{1}{8}F G^{i}\,_{j} \lambda^{j}_{\beta} {G}^{-3} \nabla_{\alpha}\,^{\beta}{\boldsymbol{W}}-\mathbf{F}_{\beta}\,^{\rho} {G}^{-1} \nabla_{\alpha}\,^{\beta}{\lambda^{i}_{\rho}}-F^{\beta}\,_{\rho} {G}^{-1} \nabla_{\alpha}\,^{\rho}{\boldsymbol{\lambda}^{i}_{\beta}}-{\rm i} G^{i}\,_{j} F^{\beta \rho} \mathbf{F}_{\beta \rho} {G}^{-3} \varphi^{j}_{\alpha} - \frac{1}{4}{\rm i} G^{i}\,_{j} {G}^{-3} \nabla_{\beta \rho}{W} \nabla^{\beta \rho}{\boldsymbol{W}} \varphi^{j}_{\alpha} - \frac{1}{4}{G}^{-1} \nabla_{\beta \rho}{\boldsymbol{W}} \nabla^{\beta \rho}{\lambda^{i}_{\alpha}} - \frac{1}{4}{G}^{-1} \nabla_{\beta \rho}{W} \nabla^{\beta \rho}{\boldsymbol{\lambda}^{i}_{\alpha}} - \frac{1}{2}{\rm i} \mathbf{F}^{\beta \rho} \lambda^{i}_{\beta} {G}^{-3} \varphi_{j \alpha} \varphi^{j}_{\rho} - \frac{1}{2}{\rm i} \mathbf{F}^{\beta \rho} \lambda_{j \beta} {G}^{-3} \varphi^{i}_{\rho} \varphi^{j}_{\alpha}+\frac{1}{4}G_{j k} \lambda^{j}_{\beta} {G}^{-3} \nabla_{\alpha}\,^{\beta}{\boldsymbol{\lambda}^{i \rho}} \varphi^{k}_{\rho} - \frac{1}{4}G_{j k} \lambda^{j \beta} {G}^{-3} \nabla_{\alpha}\,^{\rho}{\boldsymbol{\lambda}^{i}_{\beta}} \varphi^{k}_{\rho}+\frac{1}{4}G_{j k} \lambda^{j}_{\alpha} {G}^{-3} \nabla^{\beta \rho}{\boldsymbol{\lambda}^{i}_{\beta}} \varphi^{k}_{\rho} - \frac{1}{4}G_{j k} \lambda^{j}_{\beta} {G}^{-3} \nabla^{\beta \rho}{\boldsymbol{\lambda}^{i}_{\rho}} \varphi^{k}_{\alpha}+\frac{3}{4}G_{j k} W^{\beta \rho} \boldsymbol{\lambda}^{i}_{\beta} \lambda^{j}_{\rho} {G}^{-3} \varphi^{k}_{\alpha}-{\rm i} G^{i}\,_{j} F_{\alpha}\,^{\beta} \mathbf{F}_{\beta}\,^{\rho} {G}^{-3} \varphi^{j}_{\rho} - \frac{1}{2}{\rm i} G^{i}\,_{j} \mathbf{F}_{\beta}\,^{\rho} {G}^{-3} \nabla_{\alpha}\,^{\beta}{W} \varphi^{j}_{\rho}%
 - \frac{3}{2}{\rm i} G^{i}\,_{j} G_{k l} \mathbf{F}^{\beta \rho} \lambda^{k}_{\beta} {G}^{-5} \varphi^{j}_{\alpha} \varphi^{l}_{\rho}+\frac{1}{4}\mathcal{H}_{\alpha}\,^{\beta} G^{i}\,_{j} \mathbf{F}_{\beta}\,^{\rho} \lambda^{j}_{\rho} {G}^{-3}+\frac{1}{2}G_{j k} \mathbf{F}_{\beta}\,^{\rho} \lambda^{j}_{\rho} {G}^{-3} \nabla_{\alpha}\,^{\beta}{G^{i k}}+\frac{1}{4}{\rm i} \lambda^{i}_{\beta} {G}^{-3} \nabla^{\beta \rho}{\boldsymbol{W}} \varphi_{j \alpha} \varphi^{j}_{\rho}+\frac{1}{4}{\rm i} \lambda_{j \beta} {G}^{-3} \nabla^{\beta \rho}{\boldsymbol{W}} \varphi^{i}_{\rho} \varphi^{j}_{\alpha} - \frac{1}{2}{\rm i} G^{i}\,_{j} F_{\alpha \beta} {G}^{-3} \nabla^{\beta \rho}{\boldsymbol{W}} \varphi^{j}_{\rho} - \frac{1}{4}{\rm i} G^{i}\,_{j} {G}^{-3} \nabla_{\alpha \beta}{W} \nabla^{\beta \rho}{\boldsymbol{W}} \varphi^{j}_{\rho}+\frac{3}{4}{\rm i} G^{i}\,_{j} G_{k l} \lambda^{k}_{\beta} {G}^{-5} \nabla^{\beta \rho}{\boldsymbol{W}} \varphi^{j}_{\alpha} \varphi^{l}_{\rho}+\frac{1}{4}G_{j k} \lambda^{j}_{\beta} {G}^{-3} \nabla^{\beta \rho}{\boldsymbol{\lambda}^{i}_{\alpha}} \varphi^{k}_{\rho}+\frac{1}{8}G_{j k} W_{\alpha}\,^{\beta} \boldsymbol{\lambda}^{i}_{\beta} \lambda^{j \rho} {G}^{-3} \varphi^{k}_{\rho} - \frac{1}{8}\mathcal{H}_{\alpha \beta} G^{i}\,_{j} \lambda^{j}_{\rho} {G}^{-3} \nabla^{\beta \rho}{\boldsymbol{W}}+\frac{1}{4}G_{j k} \lambda^{j}_{\beta} {G}^{-3} \nabla_{\rho}\,^{\beta}{\boldsymbol{W}} \nabla_{\alpha}\,^{\rho}{G^{i k}}+\frac{5}{8}{\rm i} G_{j k} X^{i}\,_{l} \mathbf{X}^{j k} {G}^{-3} \varphi^{l}_{\alpha}+\frac{5}{8}{\rm i} G_{j k} \mathbf{X}^{i}\,_{l} X^{j k} {G}^{-3} \varphi^{l}_{\alpha} - \frac{1}{2}G^{i}\,_{j} G_{k l} \mathbf{X}^{k l} {G}^{-3} \nabla_{\alpha}\,^{\beta}{\lambda^{j}_{\beta}} - \frac{1}{4}G^{i}\,_{j} G_{k l} \mathbf{X}^{k l} W_{\alpha}\,^{\beta} \lambda^{j}_{\beta} {G}^{-3} - \frac{1}{2}G^{i}\,_{j} G_{k l} X^{k l} {G}^{-3} \nabla_{\alpha}\,^{\beta}{\boldsymbol{\lambda}^{j}_{\beta}} - \frac{1}{4}G^{i}\,_{j} G_{k l} X^{k l} W_{\alpha}\,^{\beta} \boldsymbol{\lambda}^{j}_{\beta} {G}^{-3} - \frac{3}{4}{\rm i} G^{i}\,_{j} G_{k l} G_{m n} X^{k l} \mathbf{X}^{m n} {G}^{-5} \varphi^{j}_{\alpha} - \frac{1}{4}{\rm i} \mathbf{X}^{i}\,_{j} \lambda^{\beta}_{k} {G}^{-3} \varphi^{j}_{\alpha} \varphi^{k}_{\beta}%
+\frac{1}{4}G^{i}\,_{j} \lambda^{\beta}_{k} {G}^{-3} \nabla_{\alpha}\,^{\rho}{\boldsymbol{\lambda}^{j}_{\rho}} \varphi^{k}_{\beta}+\frac{1}{8}G^{i}\,_{j} W_{\alpha}\,^{\beta} \lambda^{\rho}_{k} \boldsymbol{\lambda}^{j}_{\beta} {G}^{-3} \varphi^{k}_{\rho}+\frac{3}{16}{\rm i} G^{i}\,_{j} \boldsymbol{W} \lambda^{\beta}_{k} X^{j}_{\alpha} {G}^{-3} \varphi^{k}_{\beta}+\frac{1}{4}{\rm i} G_{j k} \mathbf{X}^{j k} F_{\alpha}\,^{\beta} {G}^{-3} \varphi^{i}_{\beta}+\frac{1}{8}{\rm i} G_{j k} \mathbf{X}^{j k} {G}^{-3} \nabla_{\alpha}\,^{\beta}{W} \varphi^{i}_{\beta}+\frac{3}{8}{\rm i} G^{i}\,_{j} G_{k l} \mathbf{X}^{k l} \lambda^{\beta}_{m} {G}^{-5} \varphi^{j}_{\alpha} \varphi^{m}_{\beta} - \frac{1}{16}\mathcal{H}_{\alpha}\,^{\beta} G_{j k} \mathbf{X}^{j k} \lambda^{i}_{\beta} {G}^{-3} - \frac{1}{8}G_{j k} \mathbf{X}^{j k} \lambda_{l \beta} {G}^{-3} \nabla_{\alpha}\,^{\beta}{G^{i l}} - \frac{1}{4}{\rm i} X^{i}\,_{j} \boldsymbol{\lambda}_{k}^{\beta} {G}^{-3} \varphi^{j}_{\alpha} \varphi^{k}_{\beta}+\frac{1}{4}G^{i}\,_{j} \boldsymbol{\lambda}_{k}^{\rho} {G}^{-3} \nabla_{\alpha}\,^{\beta}{\lambda^{j}_{\beta}} \varphi^{k}_{\rho} - \frac{1}{8}G^{i}\,_{j} W_{\alpha}\,^{\beta} \lambda^{j}_{\beta} \boldsymbol{\lambda}_{k}^{\rho} {G}^{-3} \varphi^{k}_{\rho}+\frac{3}{16}{\rm i} G^{i}\,_{j} W \boldsymbol{\lambda}_{k}^{\beta} X^{j}_{\alpha} {G}^{-3} \varphi^{k}_{\beta}+\frac{1}{4}{\rm i} G_{j k} X^{j k} \mathbf{F}_{\alpha}\,^{\beta} {G}^{-3} \varphi^{i}_{\beta}+\frac{1}{8}{\rm i} G_{j k} X^{j k} {G}^{-3} \nabla_{\alpha}\,^{\beta}{\boldsymbol{W}} \varphi^{i}_{\beta}+\frac{3}{8}{\rm i} G^{i}\,_{j} G_{k l} X^{k l} \boldsymbol{\lambda}_{m}^{\beta} {G}^{-5} \varphi^{j}_{\alpha} \varphi^{m}_{\beta} - \frac{1}{16}\mathcal{H}_{\alpha}\,^{\beta} G_{j k} X^{j k} \boldsymbol{\lambda}^{i}_{\beta} {G}^{-3} - \frac{1}{8}G_{j k} X^{j k} \boldsymbol{\lambda}_{l \beta} {G}^{-3} \nabla_{\alpha}\,^{\beta}{G^{i l}} - \frac{5}{8}{\rm i} G_{j k} X^{i j} \mathbf{X}^{k}\,_{l} {G}^{-3} \varphi^{l}_{\alpha} - \frac{5}{8}{\rm i} G_{j k} \mathbf{X}^{i j} X^{k}\,_{l} {G}^{-3} \varphi^{l}_{\alpha}+\frac{1}{2}G^{i}\,_{j} G_{k l} \mathbf{X}^{j k} {G}^{-3} \nabla_{\alpha}\,^{\beta}{\lambda^{l}_{\beta}}%
+\frac{1}{4}G^{i}\,_{j} G_{k l} \mathbf{X}^{j k} W_{\alpha}\,^{\beta} \lambda^{l}_{\beta} {G}^{-3}+\frac{3}{8}{\rm i} G^{i}\,_{j} G_{k l} W \mathbf{X}^{j k} X^{l}_{\alpha} {G}^{-3}+\frac{1}{2}G^{i}\,_{j} G_{k l} X^{j k} {G}^{-3} \nabla_{\alpha}\,^{\beta}{\boldsymbol{\lambda}^{l}_{\beta}}+\frac{1}{4}G^{i}\,_{j} G_{k l} X^{j k} W_{\alpha}\,^{\beta} \boldsymbol{\lambda}^{l}_{\beta} {G}^{-3}+\frac{3}{8}{\rm i} G^{i}\,_{j} G_{k l} \boldsymbol{W} X^{j k} X^{l}_{\alpha} {G}^{-3}+\frac{3}{4}{\rm i} G^{i}\,_{j} G_{k l} G_{m n} X^{k m} \mathbf{X}^{l n} {G}^{-5} \varphi^{j}_{\alpha}+\frac{1}{8}{\rm i} X^{i}\,_{j} \boldsymbol{\lambda}_{k}^{\beta} {G}^{-3} \varphi^{k}_{\alpha} \varphi^{j}_{\beta}+\frac{1}{8}{\rm i} X_{j k} \boldsymbol{\lambda}^{i \beta} {G}^{-3} \varphi^{j}_{\alpha} \varphi^{k}_{\beta} - \frac{1}{8}G^{i}\,_{j} \boldsymbol{\lambda}^{j \rho} {G}^{-3} \nabla_{\alpha}\,^{\beta}{\lambda_{k \beta}} \varphi^{k}_{\rho}+\frac{1}{8}G_{j k} \boldsymbol{\lambda}^{j \rho} {G}^{-3} \nabla_{\alpha}\,^{\beta}{\lambda^{k}_{\beta}} \varphi^{i}_{\rho}+\frac{1}{16}G^{i}\,_{j} W_{\alpha}\,^{\beta} \lambda_{k \beta} \boldsymbol{\lambda}^{j \rho} {G}^{-3} \varphi^{k}_{\rho} - \frac{1}{16}G_{j k} W_{\alpha}\,^{\beta} \lambda^{j}_{\beta} \boldsymbol{\lambda}^{k \rho} {G}^{-3} \varphi^{i}_{\rho} - \frac{3}{32}{\rm i} G^{i}\,_{j} W \boldsymbol{\lambda}^{j \beta} X_{k \alpha} {G}^{-3} \varphi^{k}_{\beta}+\frac{3}{32}{\rm i} G_{j k} W \boldsymbol{\lambda}^{j \beta} X^{k}_{\alpha} {G}^{-3} \varphi^{i}_{\beta}+\frac{1}{4}{\rm i} G^{i}\,_{j} X^{j}\,_{k} \mathbf{F}_{\alpha}\,^{\beta} {G}^{-3} \varphi^{k}_{\beta}+\frac{1}{4}{\rm i} G^{i}\,_{j} \boldsymbol{W} X^{j}\,_{k} W_{\alpha}\,^{\beta} {G}^{-3} \varphi^{k}_{\beta}+\frac{1}{8}{\rm i} G^{i}\,_{j} X^{j}\,_{k} {G}^{-3} \nabla_{\alpha}\,^{\beta}{\boldsymbol{W}} \varphi^{k}_{\beta} - \frac{3}{8}{\rm i} G^{i}\,_{j} G_{k l} X^{k}\,_{m} \boldsymbol{\lambda}^{l \beta} {G}^{-5} \varphi^{j}_{\alpha} \varphi^{m}_{\beta}+\frac{1}{16}\mathcal{H}_{\alpha}\,^{\beta} G_{j k} X^{i j} \boldsymbol{\lambda}^{k}_{\beta} {G}^{-3}+\frac{1}{8}G_{j k} X^{j}\,_{l} \boldsymbol{\lambda}^{k}_{\beta} {G}^{-3} \nabla_{\alpha}\,^{\beta}{G^{i l}}%
+\frac{1}{8}{\rm i} \mathbf{X}^{i}\,_{j} \lambda^{\beta}_{k} {G}^{-3} \varphi^{k}_{\alpha} \varphi^{j}_{\beta}+\frac{1}{8}{\rm i} \mathbf{X}_{j k} \lambda^{i \beta} {G}^{-3} \varphi^{j}_{\alpha} \varphi^{k}_{\beta} - \frac{1}{8}G^{i}\,_{j} \lambda^{j \beta} {G}^{-3} \nabla_{\alpha}\,^{\rho}{\boldsymbol{\lambda}_{k \rho}} \varphi^{k}_{\beta}+\frac{1}{8}G_{j k} \lambda^{j \beta} {G}^{-3} \nabla_{\alpha}\,^{\rho}{\boldsymbol{\lambda}^{k}_{\rho}} \varphi^{i}_{\beta} - \frac{1}{16}G^{i}\,_{j} W_{\alpha}\,^{\beta} \lambda^{j \rho} \boldsymbol{\lambda}_{k \beta} {G}^{-3} \varphi^{k}_{\rho}+\frac{1}{16}G_{j k} W_{\alpha}\,^{\beta} \lambda^{j \rho} \boldsymbol{\lambda}^{k}_{\beta} {G}^{-3} \varphi^{i}_{\rho} - \frac{3}{32}{\rm i} G^{i}\,_{j} \boldsymbol{W} \lambda^{j \beta} X_{k \alpha} {G}^{-3} \varphi^{k}_{\beta}+\frac{3}{32}{\rm i} G_{j k} \boldsymbol{W} \lambda^{j \beta} X^{k}_{\alpha} {G}^{-3} \varphi^{i}_{\beta}+\frac{1}{4}{\rm i} G^{i}\,_{j} \mathbf{X}^{j}\,_{k} F_{\alpha}\,^{\beta} {G}^{-3} \varphi^{k}_{\beta}+\frac{1}{4}{\rm i} G^{i}\,_{j} W \mathbf{X}^{j}\,_{k} W_{\alpha}\,^{\beta} {G}^{-3} \varphi^{k}_{\beta}+\frac{1}{8}{\rm i} G^{i}\,_{j} \mathbf{X}^{j}\,_{k} {G}^{-3} \nabla_{\alpha}\,^{\beta}{W} \varphi^{k}_{\beta} - \frac{3}{8}{\rm i} G^{i}\,_{j} G_{k l} \mathbf{X}^{k}\,_{m} \lambda^{l \beta} {G}^{-5} \varphi^{j}_{\alpha} \varphi^{m}_{\beta}+\frac{1}{16}\mathcal{H}_{\alpha}\,^{\beta} G_{j k} \mathbf{X}^{i j} \lambda^{k}_{\beta} {G}^{-3}+\frac{1}{8}G_{j k} \mathbf{X}^{j}\,_{l} \lambda^{k}_{\beta} {G}^{-3} \nabla_{\alpha}\,^{\beta}{G^{i l}}
\doublespacedmathend
\end{adjustwidth}

\subsubsection{$F_{a b, R^2}$}

\begin{adjustwidth}{0cm}{5cm}
\doublespacedmathbegin
- \frac{1}{3}W C_{\hat{a} \hat{b} c d} \mathbf{F}^{c d} {G}^{-1} - \frac{2}{3}W C_{\hat{a} c \hat{b} d} \mathbf{F}^{c d} {G}^{-1} - \frac{1}{3}W C_{c d \hat{a} \hat{b}} \mathbf{F}^{c d} {G}^{-1} - \frac{1}{3}\boldsymbol{W} C_{\hat{a} \hat{b} c d} F^{c d} {G}^{-1} - \frac{2}{3}\boldsymbol{W} C_{\hat{a} c \hat{b} d} F^{c d} {G}^{-1} - \frac{1}{3}\boldsymbol{W} C_{c d \hat{a} \hat{b}} F^{c d} {G}^{-1}-2F_{\hat{b} c} {G}^{-1} \nabla_{\hat{a}}{\nabla^{c}{\boldsymbol{W}}}-2\mathbf{F}_{\hat{b} c} {G}^{-1} \nabla_{\hat{a}}{\nabla^{c}{W}} - \frac{1}{2}W W_{\hat{a} \hat{b}} W_{c d} \mathbf{F}^{c d} {G}^{-1}+\frac{1}{2}W W^{c d} W_{\hat{a} \hat{b}} \mathbf{F}_{c d} {G}^{-1}-W W^{d c} W_{\hat{a} d} \mathbf{F}_{\hat{b} c} {G}^{-1} - \frac{2}{3}W \boldsymbol{W} C_{\hat{a} \hat{b} c d} W^{c d} {G}^{-1} - \frac{4}{3}W \boldsymbol{W} C_{\hat{a} c \hat{b} d} W^{c d} {G}^{-1} - \frac{2}{3}W \boldsymbol{W} C_{c d \hat{a} \hat{b}} W^{c d} {G}^{-1} - \frac{1}{2}\boldsymbol{W} W_{\hat{a} \hat{b}} W_{c d} F^{c d} {G}^{-1}+\frac{1}{2}\boldsymbol{W} W^{c d} W_{\hat{a} \hat{b}} F_{c d} {G}^{-1}-\boldsymbol{W} W^{d c} W_{\hat{a} d} F_{\hat{b} c} {G}^{-1}-6W_{\hat{b} c} {G}^{-1} \nabla_{\hat{a}}{W} \nabla^{c}{\boldsymbol{W}}-6W_{\hat{b} c} {G}^{-1} \nabla^{c}{W} \nabla_{\hat{a}}{\boldsymbol{W}}%
-4W W_{\hat{b} c} {G}^{-1} \nabla^{c}{\nabla_{\hat{a}}{\boldsymbol{W}}}-2W W_{\hat{b} c} {G}^{-1} \nabla_{\hat{a}}{\nabla^{c}{\boldsymbol{W}}}-4\boldsymbol{W} W_{\hat{b} c} {G}^{-1} \nabla^{c}{\nabla_{\hat{a}}{W}}-2\boldsymbol{W} W_{\hat{b} c} {G}^{-1} \nabla_{\hat{a}}{\nabla^{c}{W}}-2W \boldsymbol{W} W^{c d} W_{\hat{a} c} W_{\hat{b} d} {G}^{-1} - \frac{1}{4}X_{i j} \mathbf{X}^{i j} W_{\hat{a} \hat{b}} {G}^{-1}+\Phi_{\hat{a} \hat{b} i j} W \mathbf{X}^{i j} {G}^{-1}+\Phi_{\hat{a} \hat{b} i j} \boldsymbol{W} X^{i j} {G}^{-1}+W {G}^{-1} \nabla_{c}{\nabla^{c}{\mathbf{F}_{\hat{a} \hat{b}}}}+\boldsymbol{W} {G}^{-1} \nabla_{c}{\nabla^{c}{F_{\hat{a} \hat{b}}}}-2{G}^{-1} \nabla_{\hat{a}}{W} \nabla^{c}{\mathbf{F}_{\hat{b} c}}+2{G}^{-1} \nabla_{c}{W} \nabla^{c}{\mathbf{F}_{\hat{a} \hat{b}}}+2{G}^{-1} \nabla^{c}{W} \nabla_{\hat{a}}{\mathbf{F}_{\hat{b} c}}-2{G}^{-1} \nabla_{\hat{a}}{\boldsymbol{W}} \nabla^{c}{F_{\hat{b} c}}+2{G}^{-1} \nabla_{c}{\boldsymbol{W}} \nabla^{c}{F_{\hat{a} \hat{b}}}+2{G}^{-1} \nabla^{c}{\boldsymbol{W}} \nabla_{\hat{a}}{F_{\hat{b} c}}+\frac{1}{2}W_{\hat{a} \hat{b}}\,^{\alpha}\,_{i} X^{i}\,_{j} \boldsymbol{\lambda}^{j}_{\alpha} {G}^{-1}+\frac{1}{2}W_{\hat{a} \hat{b}}\,^{\alpha}\,_{i} \mathbf{X}^{i}\,_{j} \lambda^{j}_{\alpha} {G}^{-1} - \frac{3}{4}F_{\hat{a}}\,^{c} W_{\hat{b} c}\,^{\alpha}\,_{i} \boldsymbol{\lambda}^{i}_{\alpha} {G}^{-1} - \frac{3}{4}\mathbf{F}_{\hat{a}}\,^{c} W_{\hat{b} c}\,^{\alpha}\,_{i} \lambda^{i}_{\alpha} {G}^{-1}%
-4W \boldsymbol{W} {G}^{-1} \nabla^{c}{\nabla_{\hat{a}}{W_{\hat{b} c}}}-2W \boldsymbol{W} {G}^{-1} \nabla_{\hat{a}}{\nabla^{c}{W_{\hat{b} c}}}-6W {G}^{-1} \nabla_{\hat{a}}{\boldsymbol{W}} \nabla^{c}{W_{\hat{b} c}}-6W {G}^{-1} \nabla^{c}{\boldsymbol{W}} \nabla_{\hat{a}}{W_{\hat{b} c}}-6\boldsymbol{W} {G}^{-1} \nabla_{\hat{a}}{W} \nabla^{c}{W_{\hat{b} c}}-6\boldsymbol{W} {G}^{-1} \nabla^{c}{W} \nabla_{\hat{a}}{W_{\hat{b} c}}-W W_{\hat{b}}\,^{d} W_{d c} \mathbf{F}_{\hat{a}}\,^{c} {G}^{-1}-\boldsymbol{W} W_{\hat{b}}\,^{d} W_{d c} F_{\hat{a}}\,^{c} {G}^{-1} - \frac{7}{4}{\rm i} W_{\hat{a} \hat{b}}\,^{\alpha}\,_{i} W \boldsymbol{W} X^{i}_{\alpha} {G}^{-1} - \frac{1}{2}{\rm i} \Phi_{\hat{a} \hat{b} i j} \lambda^{i \alpha} \boldsymbol{\lambda}^{j}_{\alpha} {G}^{-1} - \frac{1}{2}{\rm i} \lambda^{\alpha}_{i} {G}^{-1} \nabla_{\hat{a}}{\nabla_{\hat{b}}{\boldsymbol{\lambda}^{i}_{\alpha}}} - \frac{1}{2}{\rm i} \boldsymbol{\lambda}_{i}^{\alpha} {G}^{-1} \nabla_{\hat{a}}{\nabla_{\hat{b}}{\lambda^{i}_{\alpha}}}+\frac{1}{2}(\Sigma_{\hat{a} c})^{\alpha \beta} F^{c d} W_{\hat{b} d \alpha i} \boldsymbol{\lambda}^{i}_{\beta} {G}^{-1}+\frac{1}{2}(\Sigma_{\hat{a} c})^{\alpha \beta} \mathbf{F}^{c d} W_{\hat{b} d \alpha i} \lambda^{i}_{\beta} {G}^{-1}+\frac{1}{8}(\Sigma_{c d})^{\alpha \beta} F^{c d} W_{\hat{a} \hat{b} \alpha i} \boldsymbol{\lambda}^{i}_{\beta} {G}^{-1}+\frac{1}{8}(\Sigma_{c d})^{\alpha \beta} \mathbf{F}^{c d} W_{\hat{a} \hat{b} \alpha i} \lambda^{i}_{\beta} {G}^{-1}+\frac{11}{128}W W_{\hat{a}}\,^{c} W_{\hat{b} c}\,^{\alpha}\,_{i} \boldsymbol{\lambda}^{i}_{\alpha} {G}^{-1}+\frac{11}{128}\boldsymbol{W} W_{\hat{a}}\,^{c} W_{\hat{b} c}\,^{\alpha}\,_{i} \lambda^{i}_{\alpha} {G}^{-1} - \frac{1}{8}F^{c d} W_{c d}\,^{\alpha}\,_{i} (\Sigma_{\hat{a} \hat{b}})_{\alpha}{}^{\beta} \boldsymbol{\lambda}^{i}_{\beta} {G}^{-1} - \frac{1}{8}\mathbf{F}^{c d} W_{c d}\,^{\alpha}\,_{i} (\Sigma_{\hat{a} \hat{b}})_{\alpha}{}^{\beta} \lambda^{i}_{\beta} {G}^{-1}%
+G_{i j} F_{\hat{b} c} {G}^{-3} \nabla^{c}{\boldsymbol{W}} \nabla_{\hat{a}}{G^{i j}}+G_{i j} \mathbf{F}_{\hat{b} c} {G}^{-3} \nabla^{c}{W} \nabla_{\hat{a}}{G^{i j}} - \frac{1}{2}G_{i j} W \mathbf{X}^{i j} {G}^{-3} \nabla_{\hat{a}}{\mathcal{H}_{\hat{b}}} - \frac{1}{2}G_{i j} \boldsymbol{W} X^{i j} {G}^{-3} \nabla_{\hat{a}}{\mathcal{H}_{\hat{b}}}+\frac{1}{2}\mathcal{H}_{\hat{a}} G_{i j} X^{i j} {G}^{-3} \nabla_{\hat{b}}{\boldsymbol{W}}+\frac{1}{2}\mathcal{H}_{\hat{a}} G_{i j} \mathbf{X}^{i j} {G}^{-3} \nabla_{\hat{b}}{W}+\frac{1}{2}\mathcal{H}_{\hat{a}} G_{i j} W {G}^{-3} \nabla_{\hat{b}}{\mathbf{X}^{i j}}+\frac{1}{2}\mathcal{H}_{\hat{a}} G_{i j} \boldsymbol{W} {G}^{-3} \nabla_{\hat{b}}{X^{i j}}+\frac{1}{2}\mathcal{H}_{\hat{a}} W \mathbf{X}_{i j} {G}^{-3} \nabla_{\hat{b}}{G^{i j}}+\frac{1}{2}\mathcal{H}_{\hat{a}} \boldsymbol{W} X_{i j} {G}^{-3} \nabla_{\hat{b}}{G^{i j}} - \frac{1}{4}\epsilon_{\hat{a} \hat{b}}\,^{e {e_{1}} c} F_{c d} {G}^{-1} \nabla^{d}{\mathbf{F}_{e {e_{1}}}} - \frac{1}{4}\epsilon_{\hat{a} \hat{b}}\,^{c d e} \mathbf{F}_{e {e_{1}}} {G}^{-1} \nabla^{{e_{1}}}{F_{c d}} - \frac{1}{4}\epsilon_{\hat{a} \hat{b}}\,^{e c d} F_{c d} {G}^{-1} \nabla^{{e_{1}}}{\mathbf{F}_{e {e_{1}}}} - \frac{1}{4}\epsilon_{\hat{a} \hat{b}}\,^{c e {e_{1}}} \mathbf{F}_{e {e_{1}}} {G}^{-1} \nabla^{d}{F_{c d}}+\frac{1}{2}\epsilon_{\hat{a} {e_{1}}}\,^{e c d} F_{c d} {G}^{-1} \nabla^{{e_{1}}}{\mathbf{F}_{\hat{b} e}}+\frac{1}{2}\epsilon_{\hat{a} {e_{1}}}\,^{c d e} \mathbf{F}_{d e} {G}^{-1} \nabla^{{e_{1}}}{F_{\hat{b} c}} - \frac{1}{2}\epsilon_{\hat{a} {e_{1}}}\,^{d e c} F_{\hat{b} c} {G}^{-1} \nabla^{{e_{1}}}{\mathbf{F}_{d e}} - \frac{1}{2}\epsilon_{\hat{a} {e_{1}}}\,^{c d e} \mathbf{F}_{\hat{b} e} {G}^{-1} \nabla^{{e_{1}}}{F_{c d}}-2W \boldsymbol{W} W_{\hat{a}}\,^{c} W_{\hat{b}}\,^{d} W_{d c} {G}^{-1}-{\rm i} (\Sigma_{\hat{a} c})^{\alpha \beta} \lambda_{i \alpha} {G}^{-1} \nabla_{\hat{b}}{\nabla^{c}{\boldsymbol{\lambda}^{i}_{\beta}}}%
-{\rm i} (\Sigma_{\hat{a} c})^{\alpha \beta} \lambda_{i \alpha} {G}^{-1} \nabla^{c}{\nabla_{\hat{b}}{\boldsymbol{\lambda}^{i}_{\beta}}}-{\rm i} (\Sigma_{\hat{a} c})^{\alpha \beta} \boldsymbol{\lambda}_{i \alpha} {G}^{-1} \nabla_{\hat{b}}{\nabla^{c}{\lambda^{i}_{\beta}}}-{\rm i} (\Sigma_{\hat{a} c})^{\alpha \beta} \boldsymbol{\lambda}_{i \alpha} {G}^{-1} \nabla^{c}{\nabla_{\hat{b}}{\lambda^{i}_{\beta}}}-2{\rm i} (\Sigma_{\hat{a} c})^{\alpha \beta} {G}^{-1} \nabla^{c}{\lambda_{i \alpha}} \nabla_{\hat{b}}{\boldsymbol{\lambda}^{i}_{\beta}}-2{\rm i} (\Sigma_{\hat{a} c})^{\alpha \beta} {G}^{-1} \nabla_{\hat{b}}{\lambda_{i \alpha}} \nabla^{c}{\boldsymbol{\lambda}^{i}_{\beta}} - \frac{3}{32}{\rm i} X_{i j} \mathbf{F}_{\hat{a} \hat{b}} {G}^{-3} \varphi^{i \alpha} \varphi^{j}_{\alpha} - \frac{3}{32}{\rm i} \mathbf{X}_{i j} F_{\hat{a} \hat{b}} {G}^{-3} \varphi^{i \alpha} \varphi^{j}_{\alpha} - \frac{1}{4}W_{\hat{a} \hat{b}}\,^{\alpha}\,_{i} (\Gamma_{c})_{\alpha}{}^{\beta} W {G}^{-1} \nabla^{c}{\boldsymbol{\lambda}^{i}_{\beta}} - \frac{1}{4}W_{\hat{a} \hat{b}}\,^{\alpha}\,_{i} (\Gamma_{c})_{\alpha}{}^{\beta} \boldsymbol{W} {G}^{-1} \nabla^{c}{\lambda^{i}_{\beta}}+\frac{1}{2}W_{\hat{a} c}\,^{\alpha}\,_{i} (\Gamma^{c})_{\alpha}{}^{\beta} W {G}^{-1} \nabla_{\hat{b}}{\boldsymbol{\lambda}^{i}_{\beta}}+\frac{1}{2}W_{\hat{a} c}\,^{\alpha}\,_{i} (\Gamma^{c})_{\alpha}{}^{\beta} \boldsymbol{W} {G}^{-1} \nabla_{\hat{b}}{\lambda^{i}_{\beta}} - \frac{1}{2}W_{\hat{a} c}\,^{\alpha}\,_{i} (\Gamma_{\hat{b}})_{\alpha}{}^{\beta} W {G}^{-1} \nabla^{c}{\boldsymbol{\lambda}^{i}_{\beta}} - \frac{1}{2}W_{\hat{a} c}\,^{\alpha}\,_{i} (\Gamma_{\hat{b}})_{\alpha}{}^{\beta} \boldsymbol{W} {G}^{-1} \nabla^{c}{\lambda^{i}_{\beta}} - \frac{29}{96}(\Gamma_{\hat{a}})^{\beta \alpha} W \boldsymbol{\lambda}_{i \beta} {G}^{-1} \nabla^{c}{W_{\hat{b} c \alpha}\,^{i}} - \frac{29}{96}(\Gamma_{\hat{a}})^{\beta \alpha} \boldsymbol{W} \lambda_{i \beta} {G}^{-1} \nabla^{c}{W_{\hat{b} c \alpha}\,^{i}}+\frac{15}{32}(\Gamma^{c})^{\beta \alpha} W \boldsymbol{\lambda}_{i \beta} {G}^{-1} \nabla_{\hat{a}}{W_{\hat{b} c \alpha}\,^{i}}+\frac{15}{32}(\Gamma^{c})^{\beta \alpha} \boldsymbol{W} \lambda_{i \beta} {G}^{-1} \nabla_{\hat{a}}{W_{\hat{b} c \alpha}\,^{i}}+\frac{15}{64}(\Gamma_{c})^{\beta \alpha} W \boldsymbol{\lambda}_{i \beta} {G}^{-1} \nabla^{c}{W_{\hat{a} \hat{b} \alpha}\,^{i}}+\frac{15}{64}(\Gamma_{c})^{\beta \alpha} \boldsymbol{W} \lambda_{i \beta} {G}^{-1} \nabla^{c}{W_{\hat{a} \hat{b} \alpha}\,^{i}}+\frac{5}{384}(\Sigma_{\hat{a} \hat{b}})^{\alpha \beta} W W^{c d} W_{c d \alpha i} \boldsymbol{\lambda}^{i}_{\beta} {G}^{-1}%
+\frac{5}{384}(\Sigma_{\hat{a} \hat{b}})^{\alpha \beta} \boldsymbol{W} W^{c d} W_{c d \alpha i} \lambda^{i}_{\beta} {G}^{-1}+\frac{85}{192}(\Sigma_{\hat{a} c})^{\alpha \beta} W W^{c d} W_{\hat{b} d \alpha i} \boldsymbol{\lambda}^{i}_{\beta} {G}^{-1}+\frac{85}{192}(\Sigma_{\hat{a} c})^{\alpha \beta} \boldsymbol{W} W^{c d} W_{\hat{b} d \alpha i} \lambda^{i}_{\beta} {G}^{-1}+\frac{227}{768}(\Sigma_{c d})^{\alpha \beta} W W^{c d} W_{\hat{a} \hat{b} \alpha i} \boldsymbol{\lambda}^{i}_{\beta} {G}^{-1}+\frac{227}{768}(\Sigma_{c d})^{\alpha \beta} \boldsymbol{W} W^{c d} W_{\hat{a} \hat{b} \alpha i} \lambda^{i}_{\beta} {G}^{-1} - \frac{73}{768}W W^{c d} W_{c d}\,^{\alpha}\,_{i} (\Sigma_{\hat{a} \hat{b}})_{\alpha}{}^{\beta} \boldsymbol{\lambda}^{i}_{\beta} {G}^{-1} - \frac{73}{768}\boldsymbol{W} W^{c d} W_{c d}\,^{\alpha}\,_{i} (\Sigma_{\hat{a} \hat{b}})_{\alpha}{}^{\beta} \lambda^{i}_{\beta} {G}^{-1}+3G_{i j} W W_{\hat{b} c} {G}^{-3} \nabla^{c}{\boldsymbol{W}} \nabla_{\hat{a}}{G^{i j}}+3G_{i j} \boldsymbol{W} W_{\hat{b} c} {G}^{-3} \nabla^{c}{W} \nabla_{\hat{a}}{G^{i j}}+\frac{9}{32}(\Sigma_{\hat{a} \hat{b}})^{\beta \alpha} X_{i j} \boldsymbol{\lambda}^{i}_{\beta} X^{j}_{\alpha} {G}^{-1}+\frac{9}{32}(\Sigma_{\hat{a} \hat{b}})^{\beta \alpha} \mathbf{X}_{j i} \lambda^{j}_{\beta} X^{i}_{\alpha} {G}^{-1}+\epsilon_{\hat{a} \hat{b} c d e} {G}^{-1} \nabla^{c}{W} \nabla^{d}{\nabla^{e}{\boldsymbol{W}}}+\epsilon_{\hat{a} \hat{b} c d e} {G}^{-1} \nabla^{c}{\boldsymbol{W}} \nabla^{d}{\nabla^{e}{W}}+\frac{1}{2}{\rm i} G_{i j} \lambda^{i \alpha} \boldsymbol{\lambda}^{j}_{\alpha} {G}^{-3} \nabla_{\hat{a}}{\mathcal{H}_{\hat{b}}} - \frac{1}{2}{\rm i} \mathcal{H}_{\hat{a}} G_{i j} \lambda^{i \alpha} {G}^{-3} \nabla_{\hat{b}}{\boldsymbol{\lambda}^{j}_{\alpha}} - \frac{1}{2}{\rm i} \mathcal{H}_{\hat{a}} G_{i j} \boldsymbol{\lambda}^{i \alpha} {G}^{-3} \nabla_{\hat{b}}{\lambda^{j}_{\alpha}} - \frac{1}{2}{\rm i} \mathcal{H}_{\hat{a}} \lambda^{\alpha}_{i} \boldsymbol{\lambda}_{j \alpha} {G}^{-3} \nabla_{\hat{b}}{G^{i j}} - \frac{3}{32}{\rm i} W \mathbf{X}_{i j} W_{\hat{a} \hat{b}} {G}^{-3} \varphi^{i \alpha} \varphi^{j}_{\alpha} - \frac{3}{32}{\rm i} \boldsymbol{W} X_{i j} W_{\hat{a} \hat{b}} {G}^{-3} \varphi^{i \alpha} \varphi^{j}_{\alpha} - \frac{3}{8}W_{c d}\,^{\alpha}\,_{i} \epsilon^{c d}\,_{\hat{a} \hat{b} e} W {G}^{-1} \nabla^{e}{\boldsymbol{\lambda}^{i}_{\alpha}}%
 - \frac{3}{8}W_{c d}\,^{\alpha}\,_{i} \epsilon^{c d}\,_{\hat{a} \hat{b} e} \boldsymbol{W} {G}^{-1} \nabla^{e}{\lambda^{i}_{\alpha}} - \frac{1}{4}W_{c d}\,^{\alpha}\,_{i} \epsilon^{c d}\,_{\hat{a} \hat{b} e} \lambda^{i}_{\alpha} {G}^{-1} \nabla^{e}{\boldsymbol{W}} - \frac{1}{4}W_{c d}\,^{\alpha}\,_{i} \epsilon^{c d}\,_{\hat{a} \hat{b} e} \boldsymbol{\lambda}^{i}_{\alpha} {G}^{-1} \nabla^{e}{W}+\frac{1}{4}G_{i j} G_{k l} X^{i j} \mathbf{X}^{k l} W_{\hat{a} \hat{b}} {G}^{-3} - \frac{1}{4}G_{i j} G_{k l} X^{i k} \mathbf{X}^{j l} W_{\hat{a} \hat{b}} {G}^{-3} - \frac{1}{2}\Phi_{\hat{a} \hat{b} i j} G^{i j} G_{k l} W \mathbf{X}^{k l} {G}^{-3} - \frac{1}{2}\Phi_{\hat{a} \hat{b} i j} G^{i j} G_{k l} \boldsymbol{W} X^{k l} {G}^{-3} - \frac{1}{2}(\Sigma^{c d})^{\alpha \beta} F_{\hat{a} c} W_{\hat{b} d \alpha i} \boldsymbol{\lambda}^{i}_{\beta} {G}^{-1} - \frac{1}{2}(\Sigma^{c d})^{\alpha \beta} \mathbf{F}_{\hat{a} c} W_{\hat{b} d \alpha i} \lambda^{i}_{\beta} {G}^{-1} - \frac{1}{8}F_{\hat{a} \hat{b}} W^{c d \alpha}\,_{i} (\Sigma_{c d})_{\alpha}{}^{\beta} \boldsymbol{\lambda}^{i}_{\beta} {G}^{-1} - \frac{1}{8}\mathbf{F}_{\hat{a} \hat{b}} W^{c d \alpha}\,_{i} (\Sigma_{c d})_{\alpha}{}^{\beta} \lambda^{i}_{\beta} {G}^{-1}+\frac{1}{2}F_{\hat{a}}\,^{c} W_{c}\,^{d \alpha}\,_{i} (\Sigma_{\hat{b} d})_{\alpha}{}^{\beta} \boldsymbol{\lambda}^{i}_{\beta} {G}^{-1}+\frac{1}{2}\mathbf{F}_{\hat{a}}\,^{c} W_{c}\,^{d \alpha}\,_{i} (\Sigma_{\hat{b} d})_{\alpha}{}^{\beta} \lambda^{i}_{\beta} {G}^{-1} - \frac{9}{32}F (\Sigma_{\hat{a} \hat{b}})^{\alpha \beta} X_{i j} \boldsymbol{\lambda}^{i}_{\alpha} {G}^{-3} \varphi^{j}_{\beta} - \frac{9}{32}F (\Sigma_{\hat{a} \hat{b}})^{\alpha \beta} \mathbf{X}_{i j} \lambda^{i}_{\alpha} {G}^{-3} \varphi^{j}_{\beta} - \frac{5}{8}G_{i j} W {G}^{-3} \nabla_{\hat{a}}{\boldsymbol{\lambda}^{i \alpha}} \nabla_{\hat{b}}{\varphi^{j}_{\alpha}}+\frac{1}{2}G_{i j} W {G}^{-3} \nabla_{\hat{a}}{\nabla_{\hat{b}}{\boldsymbol{\lambda}^{i \alpha}}} \varphi^{j}_{\alpha}+\frac{3}{8}G_{i j} W {G}^{-3} \nabla_{\hat{a}}{\varphi^{i \alpha}} \nabla_{\hat{b}}{\boldsymbol{\lambda}^{j}_{\alpha}} - \frac{5}{8}G_{i j} \boldsymbol{W} {G}^{-3} \nabla_{\hat{a}}{\lambda^{i \alpha}} \nabla_{\hat{b}}{\varphi^{j}_{\alpha}}+\frac{1}{2}G_{i j} \boldsymbol{W} {G}^{-3} \nabla_{\hat{a}}{\nabla_{\hat{b}}{\lambda^{i \alpha}}} \varphi^{j}_{\alpha}%
+\frac{3}{8}G_{i j} \boldsymbol{W} {G}^{-3} \nabla_{\hat{a}}{\varphi^{i \alpha}} \nabla_{\hat{b}}{\lambda^{j}_{\alpha}} - \frac{1}{2}G_{i j} \lambda^{i \alpha} {G}^{-3} \nabla_{\hat{a}}{\boldsymbol{W}} \nabla_{\hat{b}}{\varphi^{j}_{\alpha}} - \frac{1}{2}G_{i j} \boldsymbol{\lambda}^{i \alpha} {G}^{-3} \nabla_{\hat{a}}{W} \nabla_{\hat{b}}{\varphi^{j}_{\alpha}}+\frac{1}{2}G_{i j} {G}^{-3} \nabla_{\hat{a}}{W} \nabla_{\hat{b}}{\boldsymbol{\lambda}^{i \alpha}} \varphi^{j}_{\alpha}+\frac{1}{2}G_{i j} {G}^{-3} \nabla_{\hat{a}}{\boldsymbol{W}} \nabla_{\hat{b}}{\lambda^{i \alpha}} \varphi^{j}_{\alpha} - \frac{1}{2}G_{i j} W {G}^{-3} \nabla_{c}{\mathbf{F}_{\hat{a} \hat{b}}} \nabla^{c}{G^{i j}}-G_{i j} W {G}^{-3} \nabla_{\hat{a}}{\mathbf{F}_{\hat{b} c}} \nabla^{c}{G^{i j}}-G_{i j} W {G}^{-3} \nabla^{c}{\mathbf{F}_{\hat{a} c}} \nabla_{\hat{b}}{G^{i j}} - \frac{1}{2}G_{i j} \boldsymbol{W} {G}^{-3} \nabla_{c}{F_{\hat{a} \hat{b}}} \nabla^{c}{G^{i j}}-G_{i j} \boldsymbol{W} {G}^{-3} \nabla_{\hat{a}}{F_{\hat{b} c}} \nabla^{c}{G^{i j}}-G_{i j} \boldsymbol{W} {G}^{-3} \nabla^{c}{F_{\hat{a} c}} \nabla_{\hat{b}}{G^{i j}}+\frac{3}{64}W_{\hat{a} \hat{b}} \lambda^{\alpha}_{i} \boldsymbol{\lambda}^{i \beta} {G}^{-3} \varphi_{j \alpha} \varphi^{j}_{\beta}+\frac{3}{64}W_{\hat{a} \hat{b}} \lambda^{\alpha}_{i} \boldsymbol{\lambda}_{j}^{\beta} {G}^{-3} \varphi^{i}_{\beta} \varphi^{j}_{\alpha}+\frac{29}{128}G_{i j} X^{i j} W_{\hat{a} \hat{b}} \boldsymbol{\lambda}_{k}^{\alpha} {G}^{-3} \varphi^{k}_{\alpha}+\frac{29}{128}G_{i j} \mathbf{X}^{i j} W_{\hat{a} \hat{b}} \lambda^{\alpha}_{k} {G}^{-3} \varphi^{k}_{\alpha} - \frac{29}{128}G_{i j} X^{i}\,_{k} W_{\hat{a} \hat{b}} \boldsymbol{\lambda}^{j \alpha} {G}^{-3} \varphi^{k}_{\alpha}+\frac{29}{128}G_{i j} X^{i}\,_{k} W_{\hat{a} \hat{b}} \boldsymbol{\lambda}^{k \alpha} {G}^{-3} \varphi^{j}_{\alpha} - \frac{29}{128}G_{i j} \mathbf{X}^{i}\,_{k} W_{\hat{a} \hat{b}} \lambda^{j \alpha} {G}^{-3} \varphi^{k}_{\alpha}+\frac{29}{128}G_{i j} \mathbf{X}^{i}\,_{k} W_{\hat{a} \hat{b}} \lambda^{k \alpha} {G}^{-3} \varphi^{j}_{\alpha}+G_{i j} X^{i}\,_{k} {G}^{-3} \nabla_{\hat{a}}{\boldsymbol{W}} \nabla_{\hat{b}}{G^{j k}}%
+G_{i j} \mathbf{X}^{i}\,_{k} {G}^{-3} \nabla_{\hat{a}}{W} \nabla_{\hat{b}}{G^{j k}}+G_{i j} W {G}^{-3} \nabla_{\hat{a}}{\mathbf{X}^{i}\,_{k}} \nabla_{\hat{b}}{G^{j k}}+G_{i j} \boldsymbol{W} {G}^{-3} \nabla_{\hat{a}}{X^{i}\,_{k}} \nabla_{\hat{b}}{G^{j k}}+\frac{1}{4}\Phi_{\hat{a} \hat{b} i k} G^{i}\,_{j} W \boldsymbol{\lambda}^{k \alpha} {G}^{-3} \varphi^{j}_{\alpha}+\frac{1}{4}\Phi_{\hat{a} \hat{b} i k} G^{i}\,_{j} \boldsymbol{W} \lambda^{k \alpha} {G}^{-3} \varphi^{j}_{\alpha}+\frac{1}{2}W \mathbf{X}_{i k} {G}^{-3} \nabla_{\hat{a}}{G^{i}\,_{j}} \nabla_{\hat{b}}{G^{k j}}+\frac{1}{2}\boldsymbol{W} X_{i k} {G}^{-3} \nabla_{\hat{a}}{G^{i}\,_{j}} \nabla_{\hat{b}}{G^{k j}}+\frac{1}{8}\epsilon_{\hat{a} \hat{b}}\,^{c d e} G_{i j} F_{c d} \mathbf{F}_{e {e_{1}}} {G}^{-3} \nabla^{{e_{1}}}{G^{i j}}+\frac{1}{8}\epsilon_{\hat{a} \hat{b}}\,^{c e {e_{1}}} G_{i j} F_{c d} \mathbf{F}_{e {e_{1}}} {G}^{-3} \nabla^{d}{G^{i j}} - \frac{1}{4}\epsilon_{\hat{a} {e_{1}}}\,^{c d e} G_{i j} F_{\hat{b} c} \mathbf{F}_{d e} {G}^{-3} \nabla^{{e_{1}}}{G^{i j}} - \frac{1}{4}\epsilon_{\hat{a} {e_{1}}}\,^{c d e} G_{i j} F_{c d} \mathbf{F}_{\hat{b} e} {G}^{-3} \nabla^{{e_{1}}}{G^{i j}} - \frac{1}{2}W {G}^{-3} \nabla_{\hat{a}}{\boldsymbol{\lambda}_{i}^{\alpha}} \nabla_{\hat{b}}{G^{i}\,_{j}} \varphi^{j}_{\alpha} - \frac{1}{2}\boldsymbol{W} {G}^{-3} \nabla_{\hat{a}}{\lambda^{\alpha}_{i}} \nabla_{\hat{b}}{G^{i}\,_{j}} \varphi^{j}_{\alpha} - \frac{1}{4}\lambda^{\alpha}_{i} {G}^{-3} \nabla_{\hat{a}}{\boldsymbol{W}} \nabla_{\hat{b}}{G^{i}\,_{j}} \varphi^{j}_{\alpha} - \frac{1}{4}\boldsymbol{\lambda}_{i}^{\alpha} {G}^{-3} \nabla_{\hat{a}}{W} \nabla_{\hat{b}}{G^{i}\,_{j}} \varphi^{j}_{\alpha}+\frac{151}{128}\epsilon_{\hat{a} \hat{b} e}\,^{c d} W \boldsymbol{\lambda}_{i}^{\alpha} {G}^{-1} \nabla^{e}{W_{c d \alpha}\,^{i}}+\frac{151}{128}\epsilon_{\hat{a} \hat{b} e}\,^{c d} \boldsymbol{W} \lambda^{\alpha}_{i} {G}^{-1} \nabla^{e}{W_{c d \alpha}\,^{i}}+\frac{1}{6}{\rm i} (\Sigma^{c d})^{\alpha \beta} C_{\hat{a} \hat{b} c d} \lambda_{i \alpha} \boldsymbol{\lambda}^{i}_{\beta} {G}^{-1}+\frac{1}{3}{\rm i} (\Sigma^{c d})^{\alpha \beta} C_{\hat{a} c \hat{b} d} \lambda_{i \alpha} \boldsymbol{\lambda}^{i}_{\beta} {G}^{-1}+\frac{1}{6}{\rm i} (\Sigma^{c d})^{\alpha \beta} C_{c d \hat{a} \hat{b}} \lambda_{i \alpha} \boldsymbol{\lambda}^{i}_{\beta} {G}^{-1}%
 - \frac{1}{4}{\rm i} G_{i j} W \mathbf{F}_{\hat{a}}\,^{c} W_{\hat{b} c}\,^{\alpha i} {G}^{-3} \varphi^{j}_{\alpha} - \frac{1}{4}{\rm i} G_{i j} \boldsymbol{W} F_{\hat{a}}\,^{c} W_{\hat{b} c}\,^{\alpha i} {G}^{-3} \varphi^{j}_{\alpha} - \frac{1}{4}{\rm i} G_{i j} W_{\hat{a} \hat{b}}\,^{\alpha}\,_{k} W \mathbf{X}^{i j} {G}^{-3} \varphi^{k}_{\alpha} - \frac{1}{4}{\rm i} G_{i j} W_{\hat{a} \hat{b}}\,^{\alpha}\,_{k} \boldsymbol{W} X^{i j} {G}^{-3} \varphi^{k}_{\alpha} - \frac{1}{2}W_{\hat{a} c}\,^{\alpha}\,_{i} \epsilon^{c d e}\,_{\hat{b} {e_{1}}} (\Sigma_{d e})_{\alpha}{}^{\beta} W {G}^{-1} \nabla^{{e_{1}}}{\boldsymbol{\lambda}^{i}_{\beta}} - \frac{1}{2}W_{\hat{a} c}\,^{\alpha}\,_{i} \epsilon^{c d e}\,_{\hat{b} {e_{1}}} (\Sigma_{d e})_{\alpha}{}^{\beta} \boldsymbol{W} {G}^{-1} \nabla^{{e_{1}}}{\lambda^{i}_{\beta}}+\frac{1}{4}W_{c d}\,^{\alpha}\,_{i} \epsilon^{c d e}\,_{\hat{a} \hat{b}} (\Sigma_{e {e_{1}}})_{\alpha}{}^{\beta} W {G}^{-1} \nabla^{{e_{1}}}{\boldsymbol{\lambda}^{i}_{\beta}}+\frac{1}{4}W_{c d}\,^{\alpha}\,_{i} \epsilon^{c d e}\,_{\hat{a} \hat{b}} (\Sigma_{e {e_{1}}})_{\alpha}{}^{\beta} \boldsymbol{W} {G}^{-1} \nabla^{{e_{1}}}{\lambda^{i}_{\beta}} - \frac{7}{192}(\Sigma_{\hat{a}}{}^{\, d})^{\alpha \beta} W W_{\hat{b}}\,^{c} W_{d c \alpha i} \boldsymbol{\lambda}^{i}_{\beta} {G}^{-1} - \frac{7}{192}(\Sigma_{\hat{a}}{}^{\, d})^{\alpha \beta} \boldsymbol{W} W_{\hat{b}}\,^{c} W_{d c \alpha i} \lambda^{i}_{\beta} {G}^{-1}+\frac{1}{192}(\Sigma^{c d})^{\alpha \beta} W W_{\hat{a} \hat{b}} W_{c d \alpha i} \boldsymbol{\lambda}^{i}_{\beta} {G}^{-1} - \frac{85}{192}(\Sigma^{c d})^{\alpha \beta} W W_{\hat{a} c} W_{\hat{b} d \alpha i} \boldsymbol{\lambda}^{i}_{\beta} {G}^{-1}+\frac{1}{192}(\Sigma^{c d})^{\alpha \beta} \boldsymbol{W} W_{\hat{a} \hat{b}} W_{c d \alpha i} \lambda^{i}_{\beta} {G}^{-1} - \frac{85}{192}(\Sigma^{c d})^{\alpha \beta} \boldsymbol{W} W_{\hat{a} c} W_{\hat{b} d \alpha i} \lambda^{i}_{\beta} {G}^{-1} - \frac{73}{768}W W_{\hat{a} \hat{b}} W^{c d \alpha}\,_{i} (\Sigma_{c d})_{\alpha}{}^{\beta} \boldsymbol{\lambda}^{i}_{\beta} {G}^{-1}+\frac{73}{192}W W_{\hat{a}}\,^{c} W_{c}\,^{d \alpha}\,_{i} (\Sigma_{\hat{b} d})_{\alpha}{}^{\beta} \boldsymbol{\lambda}^{i}_{\beta} {G}^{-1} - \frac{73}{768}\boldsymbol{W} W_{\hat{a} \hat{b}} W^{c d \alpha}\,_{i} (\Sigma_{c d})_{\alpha}{}^{\beta} \lambda^{i}_{\beta} {G}^{-1}+\frac{73}{192}\boldsymbol{W} W_{\hat{a}}\,^{c} W_{c}\,^{d \alpha}\,_{i} (\Sigma_{\hat{b} d})_{\alpha}{}^{\beta} \lambda^{i}_{\beta} {G}^{-1}+\frac{1}{2}G_{i j} (\Gamma^{c})^{\alpha \beta} F_{\hat{b} c} \boldsymbol{\lambda}^{i}_{\alpha} {G}^{-3} \nabla_{\hat{a}}{\varphi^{j}_{\beta}}+\frac{1}{2}G_{i j} (\Gamma^{c})^{\alpha \beta} F_{\hat{b} c} {G}^{-3} \nabla_{\hat{a}}{\boldsymbol{\lambda}^{i}_{\alpha}} \varphi^{j}_{\beta}%
+\frac{1}{2}G_{i j} (\Gamma^{c})^{\alpha \beta} \mathbf{F}_{\hat{b} c} \lambda^{i}_{\alpha} {G}^{-3} \nabla_{\hat{a}}{\varphi^{j}_{\beta}}+\frac{1}{2}G_{i j} (\Gamma^{c})^{\alpha \beta} \mathbf{F}_{\hat{b} c} {G}^{-3} \nabla_{\hat{a}}{\lambda^{i}_{\alpha}} \varphi^{j}_{\beta} - \frac{1}{8}G_{i j} (\Sigma_{\hat{a} \hat{b}})^{\alpha \beta} W {G}^{-3} \nabla_{c}{\boldsymbol{\lambda}^{i}_{\alpha}} \nabla^{c}{\varphi^{j}_{\beta}} - \frac{1}{8}G_{i j} (\Sigma_{\hat{a} \hat{b}})^{\alpha \beta} W {G}^{-3} \nabla_{c}{\varphi^{i}_{\alpha}} \nabla^{c}{\boldsymbol{\lambda}^{j}_{\beta}} - \frac{1}{8}G_{i j} (\Sigma_{\hat{a} \hat{b}})^{\alpha \beta} \boldsymbol{W} {G}^{-3} \nabla_{c}{\lambda^{i}_{\alpha}} \nabla^{c}{\varphi^{j}_{\beta}} - \frac{1}{8}G_{i j} (\Sigma_{\hat{a} \hat{b}})^{\alpha \beta} \boldsymbol{W} {G}^{-3} \nabla_{c}{\varphi^{i}_{\alpha}} \nabla^{c}{\lambda^{j}_{\beta}} - \frac{5}{4}G_{i j} (\Sigma_{\hat{a} c})^{\alpha \beta} W {G}^{-3} \nabla^{c}{\boldsymbol{\lambda}^{i}_{\alpha}} \nabla_{\hat{b}}{\varphi^{j}_{\beta}}+\frac{1}{4}G_{i j} (\Sigma_{\hat{a} c})^{\alpha \beta} W {G}^{-3} \nabla_{\hat{b}}{\boldsymbol{\lambda}^{i}_{\alpha}} \nabla^{c}{\varphi^{j}_{\beta}}-G_{i j} (\Sigma_{\hat{a} c})^{\alpha \beta} W {G}^{-3} \nabla_{\hat{b}}{\nabla^{c}{\boldsymbol{\lambda}^{i}_{\alpha}}} \varphi^{j}_{\beta}-G_{i j} (\Sigma_{\hat{a} c})^{\alpha \beta} W {G}^{-3} \nabla^{c}{\nabla_{\hat{b}}{\boldsymbol{\lambda}^{i}_{\alpha}}} \varphi^{j}_{\beta}+\frac{1}{4}G_{i j} (\Sigma_{\hat{a} c})^{\alpha \beta} W {G}^{-3} \nabla^{c}{\varphi^{i}_{\alpha}} \nabla_{\hat{b}}{\boldsymbol{\lambda}^{j}_{\beta}}+\frac{3}{4}G_{i j} (\Sigma_{\hat{a} c})^{\alpha \beta} W {G}^{-3} \nabla_{\hat{b}}{\varphi^{i}_{\alpha}} \nabla^{c}{\boldsymbol{\lambda}^{j}_{\beta}} - \frac{5}{4}G_{i j} (\Sigma_{\hat{a} c})^{\alpha \beta} \boldsymbol{W} {G}^{-3} \nabla^{c}{\lambda^{i}_{\alpha}} \nabla_{\hat{b}}{\varphi^{j}_{\beta}}+\frac{1}{4}G_{i j} (\Sigma_{\hat{a} c})^{\alpha \beta} \boldsymbol{W} {G}^{-3} \nabla_{\hat{b}}{\lambda^{i}_{\alpha}} \nabla^{c}{\varphi^{j}_{\beta}}-G_{i j} (\Sigma_{\hat{a} c})^{\alpha \beta} \boldsymbol{W} {G}^{-3} \nabla_{\hat{b}}{\nabla^{c}{\lambda^{i}_{\alpha}}} \varphi^{j}_{\beta}-G_{i j} (\Sigma_{\hat{a} c})^{\alpha \beta} \boldsymbol{W} {G}^{-3} \nabla^{c}{\nabla_{\hat{b}}{\lambda^{i}_{\alpha}}} \varphi^{j}_{\beta}+\frac{1}{4}G_{i j} (\Sigma_{\hat{a} c})^{\alpha \beta} \boldsymbol{W} {G}^{-3} \nabla^{c}{\varphi^{i}_{\alpha}} \nabla_{\hat{b}}{\lambda^{j}_{\beta}}+\frac{3}{4}G_{i j} (\Sigma_{\hat{a} c})^{\alpha \beta} \boldsymbol{W} {G}^{-3} \nabla_{\hat{b}}{\varphi^{i}_{\alpha}} \nabla^{c}{\lambda^{j}_{\beta}}-G_{i j} (\Sigma_{\hat{a} c})^{\alpha \beta} \lambda^{i}_{\alpha} {G}^{-3} \nabla^{c}{\boldsymbol{W}} \nabla_{\hat{b}}{\varphi^{j}_{\beta}}-G_{i j} (\Sigma_{\hat{a} c})^{\alpha \beta} \lambda^{i}_{\alpha} {G}^{-3} \nabla_{\hat{b}}{\nabla^{c}{\boldsymbol{W}}} \varphi^{j}_{\beta}%
-G_{i j} (\Sigma_{\hat{a} c})^{\alpha \beta} \boldsymbol{\lambda}^{i}_{\alpha} {G}^{-3} \nabla^{c}{W} \nabla_{\hat{b}}{\varphi^{j}_{\beta}}-G_{i j} (\Sigma_{\hat{a} c})^{\alpha \beta} \boldsymbol{\lambda}^{i}_{\alpha} {G}^{-3} \nabla_{\hat{b}}{\nabla^{c}{W}} \varphi^{j}_{\beta}-G_{i j} (\Sigma_{\hat{a} c})^{\alpha \beta} {G}^{-3} \nabla^{c}{W} \nabla_{\hat{b}}{\boldsymbol{\lambda}^{i}_{\alpha}} \varphi^{j}_{\beta}-2G_{i j} (\Sigma_{\hat{a} c})^{\alpha \beta} {G}^{-3} \nabla_{\hat{b}}{W} \nabla^{c}{\boldsymbol{\lambda}^{i}_{\alpha}} \varphi^{j}_{\beta}-G_{i j} (\Sigma_{\hat{a} c})^{\alpha \beta} {G}^{-3} \nabla^{c}{\boldsymbol{W}} \nabla_{\hat{b}}{\lambda^{i}_{\alpha}} \varphi^{j}_{\beta}-2G_{i j} (\Sigma_{\hat{a} c})^{\alpha \beta} {G}^{-3} \nabla_{\hat{b}}{\boldsymbol{W}} \nabla^{c}{\lambda^{i}_{\alpha}} \varphi^{j}_{\beta}-3G_{i j} W \boldsymbol{W} {G}^{-3} \nabla^{c}{W_{\hat{a} c}} \nabla_{\hat{b}}{G^{i j}}+\frac{1}{4}(\Gamma^{c})^{\alpha \beta} F_{\hat{b} c} \boldsymbol{\lambda}_{i \alpha} {G}^{-3} \nabla_{\hat{a}}{G^{i}\,_{j}} \varphi^{j}_{\beta}+\frac{1}{4}(\Gamma^{c})^{\alpha \beta} \mathbf{F}_{\hat{b} c} \lambda_{i \alpha} {G}^{-3} \nabla_{\hat{a}}{G^{i}\,_{j}} \varphi^{j}_{\beta} - \frac{9}{32}\mathcal{H}_{\hat{a}} (\Gamma_{\hat{b}})^{\alpha \beta} X_{i j} \boldsymbol{\lambda}^{i}_{\alpha} {G}^{-3} \varphi^{j}_{\beta} - \frac{9}{32}\mathcal{H}_{\hat{a}} (\Gamma_{\hat{b}})^{\alpha \beta} \mathbf{X}_{i j} \lambda^{i}_{\alpha} {G}^{-3} \varphi^{j}_{\beta}-(\Sigma_{\hat{a} c})^{\alpha \beta} W {G}^{-3} \nabla^{c}{\boldsymbol{\lambda}_{i \alpha}} \nabla_{\hat{b}}{G^{i}\,_{j}} \varphi^{j}_{\beta}-(\Sigma_{\hat{a} c})^{\alpha \beta} \boldsymbol{W} {G}^{-3} \nabla^{c}{\lambda_{i \alpha}} \nabla_{\hat{b}}{G^{i}\,_{j}} \varphi^{j}_{\beta} - \frac{1}{2}(\Sigma_{\hat{a} c})^{\alpha \beta} \lambda_{i \alpha} {G}^{-3} \nabla^{c}{\boldsymbol{W}} \nabla_{\hat{b}}{G^{i}\,_{j}} \varphi^{j}_{\beta} - \frac{1}{2}(\Sigma_{\hat{a} c})^{\alpha \beta} \boldsymbol{\lambda}_{i \alpha} {G}^{-3} \nabla^{c}{W} \nabla_{\hat{b}}{G^{i}\,_{j}} \varphi^{j}_{\beta}+\frac{1}{8}G_{i j} W_{\hat{a} \hat{b}}\,^{\alpha i} \lambda^{\beta}_{k} \boldsymbol{\lambda}^{k}_{\alpha} {G}^{-3} \varphi^{j}_{\beta} - \frac{1}{4}G_{i j} W_{\hat{a} \hat{b}}\,^{\alpha}\,_{k} \lambda^{i \beta} \boldsymbol{\lambda}^{j}_{\beta} {G}^{-3} \varphi^{k}_{\alpha} - \frac{1}{8}\epsilon^{c d e}\,_{\hat{b} {e_{1}}} (\Gamma^{{e_{1}}})^{\alpha \beta} F_{c d} W_{\hat{a} e \alpha i} \boldsymbol{\lambda}^{i}_{\beta} {G}^{-1} - \frac{1}{8}\epsilon^{c d e}\,_{\hat{b} {e_{1}}} (\Gamma^{{e_{1}}})^{\alpha \beta} \mathbf{F}_{c d} W_{\hat{a} e \alpha i} \lambda^{i}_{\beta} {G}^{-1}+\frac{1}{64}\epsilon^{e {e_{1}}}\,_{\hat{a} \hat{b}}\,^{c} (\Sigma_{e {e_{1}}})^{\beta \alpha} W \boldsymbol{\lambda}_{i \beta} {G}^{-1} \nabla^{d}{W_{c d \alpha}\,^{i}}%
+\frac{1}{64}\epsilon^{e {e_{1}}}\,_{\hat{a} \hat{b}}\,^{c} (\Sigma_{e {e_{1}}})^{\beta \alpha} \boldsymbol{W} \lambda_{i \beta} {G}^{-1} \nabla^{d}{W_{c d \alpha}\,^{i}}+\frac{5}{192}\epsilon^{e {e_{1}}}\,_{\hat{a}}\,^{c d} (\Sigma_{e {e_{1}}})^{\beta \alpha} W \boldsymbol{\lambda}_{i \beta} {G}^{-1} \nabla_{\hat{b}}{W_{c d \alpha}\,^{i}}+\frac{5}{192}\epsilon^{e {e_{1}}}\,_{\hat{a}}\,^{c d} (\Sigma_{e {e_{1}}})^{\beta \alpha} \boldsymbol{W} \lambda_{i \beta} {G}^{-1} \nabla_{\hat{b}}{W_{c d \alpha}\,^{i}}+\frac{29}{96}\epsilon^{d e}\,_{\hat{a} {e_{1}}}\,^{c} (\Sigma_{d e})^{\beta \alpha} W \boldsymbol{\lambda}_{i \beta} {G}^{-1} \nabla^{{e_{1}}}{W_{\hat{b} c \alpha}\,^{i}}+\frac{29}{96}\epsilon^{d e}\,_{\hat{a} {e_{1}}}\,^{c} (\Sigma_{d e})^{\beta \alpha} \boldsymbol{W} \lambda_{i \beta} {G}^{-1} \nabla^{{e_{1}}}{W_{\hat{b} c \alpha}\,^{i}}+\frac{11}{48}\epsilon^{e}\,_{\hat{a} \hat{b}}\,^{c d} (\Sigma_{e {e_{1}}})^{\beta \alpha} W \boldsymbol{\lambda}_{i \beta} {G}^{-1} \nabla^{{e_{1}}}{W_{c d \alpha}\,^{i}}+\frac{11}{48}\epsilon^{e}\,_{\hat{a} \hat{b}}\,^{c d} (\Sigma_{e {e_{1}}})^{\beta \alpha} \boldsymbol{W} \lambda_{i \beta} {G}^{-1} \nabla^{{e_{1}}}{W_{c d \alpha}\,^{i}}+\frac{1}{16}\epsilon^{c e {e_{1}}}\,_{\hat{a} \hat{b}} F_{c d} W_{e {e_{1}}}\,^{\alpha}\,_{i} (\Gamma^{d})_{\alpha}{}^{\beta} \boldsymbol{\lambda}^{i}_{\beta} {G}^{-1}+\frac{1}{16}\epsilon^{c e {e_{1}}}\,_{\hat{a} \hat{b}} \mathbf{F}_{c d} W_{e {e_{1}}}\,^{\alpha}\,_{i} (\Gamma^{d})_{\alpha}{}^{\beta} \lambda^{i}_{\beta} {G}^{-1}+\frac{1}{2}{\rm i} \Phi_{\hat{a} \hat{b} i j} G^{i j} G_{k l} \lambda^{k \alpha} \boldsymbol{\lambda}^{l}_{\alpha} {G}^{-3} - \frac{1}{2}{\rm i} G_{i j} (\Sigma_{\hat{a} c})^{\alpha \beta} W \mathbf{F}^{c d} W_{\hat{b} d \alpha}\,^{i} {G}^{-3} \varphi^{j}_{\beta} - \frac{1}{2}{\rm i} G_{i j} (\Sigma_{\hat{a} c})^{\alpha \beta} \boldsymbol{W} F^{c d} W_{\hat{b} d \alpha}\,^{i} {G}^{-3} \varphi^{j}_{\beta} - \frac{1}{8}{\rm i} G_{i j} (\Sigma_{c d})^{\alpha \beta} W \mathbf{F}^{c d} W_{\hat{a} \hat{b} \alpha}\,^{i} {G}^{-3} \varphi^{j}_{\beta} - \frac{1}{8}{\rm i} G_{i j} (\Sigma_{c d})^{\alpha \beta} \boldsymbol{W} F^{c d} W_{\hat{a} \hat{b} \alpha}\,^{i} {G}^{-3} \varphi^{j}_{\beta} - \frac{1}{2}{\rm i} G_{i j} W \boldsymbol{W} W_{\hat{a}}\,^{c} W_{\hat{b} c}\,^{\alpha i} {G}^{-3} \varphi^{j}_{\alpha}+\frac{1}{8}{\rm i} G_{i j} W \mathbf{F}^{c d} W_{c d}\,^{\alpha i} (\Sigma_{\hat{a} \hat{b}})_{\alpha}{}^{\beta} {G}^{-3} \varphi^{j}_{\beta}+\frac{1}{8}{\rm i} G_{i j} \boldsymbol{W} F^{c d} W_{c d}\,^{\alpha i} (\Sigma_{\hat{a} \hat{b}})_{\alpha}{}^{\beta} {G}^{-3} \varphi^{j}_{\beta} - \frac{11}{16}{\rm i} G_{i j} \lambda^{i \alpha} {G}^{-3} \nabla_{\hat{a}}{\boldsymbol{\lambda}_{k \alpha}} \nabla_{\hat{b}}{G^{j k}} - \frac{11}{16}{\rm i} G_{i j} \boldsymbol{\lambda}^{i \alpha} {G}^{-3} \nabla_{\hat{a}}{\lambda_{k \alpha}} \nabla_{\hat{b}}{G^{j k}} - \frac{5}{16}{\rm i} G_{i j} \lambda^{\alpha}_{k} {G}^{-3} \nabla_{\hat{a}}{\boldsymbol{\lambda}^{i}_{\alpha}} \nabla_{\hat{b}}{G^{j k}}%
 - \frac{3}{16}{\rm i} G_{i j} \lambda^{\alpha}_{k} {G}^{-3} \nabla_{\hat{a}}{\boldsymbol{\lambda}^{k}_{\alpha}} \nabla_{\hat{b}}{G^{i j}} - \frac{5}{16}{\rm i} G_{i j} \boldsymbol{\lambda}_{k}^{\alpha} {G}^{-3} \nabla_{\hat{a}}{\lambda^{i}_{\alpha}} \nabla_{\hat{b}}{G^{j k}} - \frac{3}{16}{\rm i} G_{i j} \boldsymbol{\lambda}_{k}^{\alpha} {G}^{-3} \nabla_{\hat{a}}{\lambda^{k}_{\alpha}} \nabla_{\hat{b}}{G^{i j}}+\frac{3}{8}{\rm i} (\Sigma_{\hat{a} \hat{b}})^{\alpha \beta} X_{i j} \mathbf{X}^{i j} {G}^{-3} \varphi_{k \alpha} \varphi^{k}_{\beta} - \frac{3}{4}{\rm i} (\Sigma_{\hat{a} \hat{b}})^{\alpha \beta} X_{i j} \mathbf{X}^{i}\,_{k} {G}^{-3} \varphi^{j}_{\alpha} \varphi^{k}_{\beta} - \frac{1}{2}{\rm i} \lambda^{\alpha}_{i} \boldsymbol{\lambda}_{k \alpha} {G}^{-3} \nabla_{\hat{a}}{G^{i}\,_{j}} \nabla_{\hat{b}}{G^{k j}} - \frac{3}{4}\mathcal{H}_{\hat{a}} G_{i j} G_{k l} W \mathbf{X}^{i j} {G}^{-5} \nabla_{\hat{b}}{G^{k l}} - \frac{3}{4}\mathcal{H}_{\hat{a}} G_{i j} G_{k l} \boldsymbol{W} X^{i j} {G}^{-5} \nabla_{\hat{b}}{G^{k l}} - \frac{1}{256}(\Gamma^{c})^{\alpha \beta} W W_{c d} W_{e {e_{1}} \alpha i} \epsilon_{\hat{a}}\,^{d e {e_{1}}}\,_{\hat{b}} \boldsymbol{\lambda}^{i}_{\beta} {G}^{-1} - \frac{1}{512}(\Gamma_{\hat{a}})^{\alpha \beta} W W_{c d} W_{e {e_{1}} \alpha i} \epsilon^{c d e {e_{1}}}\,_{\hat{b}} \boldsymbol{\lambda}^{i}_{\beta} {G}^{-1} - \frac{1}{256}(\Gamma^{c})^{\alpha \beta} \boldsymbol{W} W_{c d} W_{e {e_{1}} \alpha i} \epsilon_{\hat{a}}\,^{d e {e_{1}}}\,_{\hat{b}} \lambda^{i}_{\beta} {G}^{-1} - \frac{1}{512}(\Gamma_{\hat{a}})^{\alpha \beta} \boldsymbol{W} W_{c d} W_{e {e_{1}} \alpha i} \epsilon^{c d e {e_{1}}}\,_{\hat{b}} \lambda^{i}_{\beta} {G}^{-1} - \frac{1}{256}(\Gamma_{{e_{1}}})^{\alpha \beta} W W_{\hat{b} c} W_{d e \alpha i} \epsilon_{\hat{a}}\,^{{e_{1}} c d e} \boldsymbol{\lambda}^{i}_{\beta} {G}^{-1} - \frac{1}{256}(\Gamma_{{e_{1}}})^{\alpha \beta} \boldsymbol{W} W_{\hat{b} c} W_{d e \alpha i} \epsilon_{\hat{a}}\,^{{e_{1}} c d e} \lambda^{i}_{\beta} {G}^{-1}+\frac{3}{4}G_{i j} (\Gamma^{c})^{\alpha \beta} W_{\hat{b} c} \lambda^{i}_{\alpha} {G}^{-3} \nabla_{\hat{a}}{\boldsymbol{W}} \varphi^{j}_{\beta}+\frac{3}{4}G_{i j} (\Gamma^{c})^{\alpha \beta} W_{\hat{b} c} \boldsymbol{\lambda}^{i}_{\alpha} {G}^{-3} \nabla_{\hat{a}}{W} \varphi^{j}_{\beta}+\frac{3}{4}G_{i j} (\Gamma^{c})^{\alpha \beta} W W_{\hat{b} c} \boldsymbol{\lambda}^{i}_{\alpha} {G}^{-3} \nabla_{\hat{a}}{\varphi^{j}_{\beta}}+\frac{3}{4}G_{i j} (\Gamma^{c})^{\alpha \beta} W W_{\hat{b} c} {G}^{-3} \nabla_{\hat{a}}{\boldsymbol{\lambda}^{i}_{\alpha}} \varphi^{j}_{\beta}+\frac{3}{4}G_{i j} (\Gamma^{c})^{\alpha \beta} \boldsymbol{W} W_{\hat{b} c} \lambda^{i}_{\alpha} {G}^{-3} \nabla_{\hat{a}}{\varphi^{j}_{\beta}}+\frac{3}{4}G_{i j} (\Gamma^{c})^{\alpha \beta} \boldsymbol{W} W_{\hat{b} c} {G}^{-3} \nabla_{\hat{a}}{\lambda^{i}_{\alpha}} \varphi^{j}_{\beta}%
+\frac{3}{8}(\Gamma^{c})^{\alpha \beta} W W_{\hat{b} c} \boldsymbol{\lambda}_{i \alpha} {G}^{-3} \nabla_{\hat{a}}{G^{i}\,_{j}} \varphi^{j}_{\beta}+\frac{3}{8}(\Gamma^{c})^{\alpha \beta} \boldsymbol{W} W_{\hat{b} c} \lambda_{i \alpha} {G}^{-3} \nabla_{\hat{a}}{G^{i}\,_{j}} \varphi^{j}_{\beta} - \frac{1}{256}\epsilon^{c d e}\,_{\hat{b} {e_{1}}} (\Gamma^{{e_{1}}})^{\alpha \beta} W W_{\hat{a} c} W_{d e \alpha i} \boldsymbol{\lambda}^{i}_{\beta} {G}^{-1} - \frac{7}{64}\epsilon^{c d e}\,_{\hat{b} {e_{1}}} (\Gamma^{{e_{1}}})^{\alpha \beta} W W_{c d} W_{\hat{a} e \alpha i} \boldsymbol{\lambda}^{i}_{\beta} {G}^{-1} - \frac{1}{256}\epsilon^{c d e}\,_{\hat{b} {e_{1}}} (\Gamma^{{e_{1}}})^{\alpha \beta} \boldsymbol{W} W_{\hat{a} c} W_{d e \alpha i} \lambda^{i}_{\beta} {G}^{-1} - \frac{7}{64}\epsilon^{c d e}\,_{\hat{b} {e_{1}}} (\Gamma^{{e_{1}}})^{\alpha \beta} \boldsymbol{W} W_{c d} W_{\hat{a} e \alpha i} \lambda^{i}_{\beta} {G}^{-1}+\frac{1}{768}\epsilon_{\hat{a} \hat{b}}\,^{c d e} (\Gamma^{{e_{1}}})^{\alpha \beta} W W_{c d} W_{e {e_{1}} \alpha i} \boldsymbol{\lambda}^{i}_{\beta} {G}^{-1}+\frac{1}{768}\epsilon_{\hat{a} \hat{b}}\,^{c e {e_{1}}} (\Gamma^{d})^{\alpha \beta} W W_{c d} W_{e {e_{1}} \alpha i} \boldsymbol{\lambda}^{i}_{\beta} {G}^{-1}+\frac{1}{768}\epsilon_{\hat{a} \hat{b}}\,^{c d e} (\Gamma^{{e_{1}}})^{\alpha \beta} \boldsymbol{W} W_{c d} W_{e {e_{1}} \alpha i} \lambda^{i}_{\beta} {G}^{-1}+\frac{1}{768}\epsilon_{\hat{a} \hat{b}}\,^{c e {e_{1}}} (\Gamma^{d})^{\alpha \beta} \boldsymbol{W} W_{c d} W_{e {e_{1}} \alpha i} \lambda^{i}_{\beta} {G}^{-1}+\frac{27}{512}\epsilon^{c e {e_{1}}}\,_{\hat{a} \hat{b}} W W_{c d} W_{e {e_{1}}}\,^{\alpha}\,_{i} (\Gamma^{d})_{\alpha}{}^{\beta} \boldsymbol{\lambda}^{i}_{\beta} {G}^{-1}+\frac{27}{512}\epsilon^{c e {e_{1}}}\,_{\hat{a} \hat{b}} \boldsymbol{W} W_{c d} W_{e {e_{1}}}\,^{\alpha}\,_{i} (\Gamma^{d})_{\alpha}{}^{\beta} \lambda^{i}_{\beta} {G}^{-1} - \frac{1}{192}\epsilon_{\hat{a} \hat{b}}\,^{c e {e_{1}}} W W_{c d} W_{e {e_{1}}}\,^{\alpha}\,_{i} (\Gamma^{d})_{\alpha}{}^{\beta} \boldsymbol{\lambda}^{i}_{\beta} {G}^{-1} - \frac{1}{192}\epsilon_{\hat{a} \hat{b}}\,^{c e {e_{1}}} \boldsymbol{W} W_{c d} W_{e {e_{1}}}\,^{\alpha}\,_{i} (\Gamma^{d})_{\alpha}{}^{\beta} \lambda^{i}_{\beta} {G}^{-1}+\frac{3}{32}{\rm i} G_{i j} G_{k l} X^{i j} \mathbf{F}_{\hat{a} \hat{b}} {G}^{-5} \varphi^{k \alpha} \varphi^{l}_{\alpha}+\frac{3}{32}{\rm i} G_{i j} G_{k l} \mathbf{X}^{i j} F_{\hat{a} \hat{b}} {G}^{-5} \varphi^{k \alpha} \varphi^{l}_{\alpha} - \frac{3}{32}{\rm i} G_{i j} G_{k l} X^{i k} \mathbf{F}_{\hat{a} \hat{b}} {G}^{-5} \varphi^{j \alpha} \varphi^{l}_{\alpha} - \frac{3}{32}{\rm i} G_{i j} G_{k l} \mathbf{X}^{i k} F_{\hat{a} \hat{b}} {G}^{-5} \varphi^{j \alpha} \varphi^{l}_{\alpha}+\frac{1}{24}{\rm i} G_{i j} (\Gamma_{\hat{a}})^{\alpha \beta} W \boldsymbol{W} {G}^{-3} \nabla^{c}{W_{\hat{b} c \alpha}\,^{i}} \varphi^{j}_{\beta}+\frac{1}{24}{\rm i} G_{i j} (\Gamma^{c})^{\alpha \beta} W \boldsymbol{W} {G}^{-3} \nabla_{\hat{a}}{W_{\hat{b} c \alpha}\,^{i}} \varphi^{j}_{\beta}%
+\frac{1}{48}{\rm i} G_{i j} (\Gamma_{c})^{\alpha \beta} W \boldsymbol{W} {G}^{-3} \nabla^{c}{W_{\hat{a} \hat{b} \alpha}\,^{i}} \varphi^{j}_{\beta} - \frac{1}{96}{\rm i} G_{i j} (\Sigma_{\hat{a} \hat{b}})^{\alpha \beta} W \boldsymbol{W} W^{c d} W_{c d \alpha}\,^{i} {G}^{-3} \varphi^{j}_{\beta}-{\rm i} G_{i j} (\Sigma_{\hat{a} c})^{\alpha \beta} W \boldsymbol{W} W^{c d} W_{\hat{b} d \alpha}\,^{i} {G}^{-3} \varphi^{j}_{\beta} - \frac{1}{4}{\rm i} G_{i j} (\Sigma_{c d})^{\alpha \beta} W \boldsymbol{W} W^{c d} W_{\hat{a} \hat{b} \alpha}\,^{i} {G}^{-3} \varphi^{j}_{\beta}+\frac{23}{96}{\rm i} G_{i j} W \boldsymbol{W} W^{c d} W_{c d}\,^{\alpha i} (\Sigma_{\hat{a} \hat{b}})_{\alpha}{}^{\beta} {G}^{-3} \varphi^{j}_{\beta} - \frac{7}{16}{\rm i} (\Gamma_{\hat{a}})^{\alpha \beta} X_{i j} {G}^{-3} \nabla_{\hat{b}}{\boldsymbol{W}} \varphi^{i}_{\alpha} \varphi^{j}_{\beta} - \frac{7}{16}{\rm i} (\Gamma_{\hat{a}})^{\alpha \beta} \mathbf{X}_{i j} {G}^{-3} \nabla_{\hat{b}}{W} \varphi^{i}_{\alpha} \varphi^{j}_{\beta}+{\rm i} (\Gamma_{\hat{a}})^{\alpha \beta} W \mathbf{X}_{i j} {G}^{-3} \nabla_{\hat{b}}{\varphi^{i}_{\alpha}} \varphi^{j}_{\beta} - \frac{1}{4}{\rm i} (\Gamma_{\hat{a}})^{\alpha \beta} W {G}^{-3} \nabla_{\hat{b}}{\mathbf{X}_{i j}} \varphi^{i}_{\alpha} \varphi^{j}_{\beta}+{\rm i} (\Gamma_{\hat{a}})^{\alpha \beta} \boldsymbol{W} X_{i j} {G}^{-3} \nabla_{\hat{b}}{\varphi^{i}_{\alpha}} \varphi^{j}_{\beta} - \frac{1}{4}{\rm i} (\Gamma_{\hat{a}})^{\alpha \beta} \boldsymbol{W} {G}^{-3} \nabla_{\hat{b}}{X_{i j}} \varphi^{i}_{\alpha} \varphi^{j}_{\beta} - \frac{3}{8}{\rm i} (\Sigma_{\hat{a} \hat{b}})^{\beta \rho} W \boldsymbol{\lambda}_{i \beta} X^{i \alpha} {G}^{-3} \varphi_{j \rho} \varphi^{j}_{\alpha} - \frac{3}{16}{\rm i} (\Sigma_{\hat{a} \hat{b}})^{\beta \rho} W \boldsymbol{\lambda}_{j \beta} X_{i}^{\alpha} {G}^{-3} \varphi^{j}_{\rho} \varphi^{i}_{\alpha} - \frac{3}{16}{\rm i} (\Sigma_{\hat{a} \hat{b}})^{\beta \rho} W \boldsymbol{\lambda}_{j \beta} X_{i}^{\alpha} {G}^{-3} \varphi^{j}_{\alpha} \varphi^{i}_{\rho}+\frac{3}{64}{\rm i} (\Sigma_{\hat{a} \hat{b}})^{\alpha \beta} W \boldsymbol{\lambda}_{i}^{\rho} X^{i}_{\alpha} {G}^{-3} \varphi_{j \beta} \varphi^{j}_{\rho}+\frac{3}{8}{\rm i} (\Sigma_{\hat{a} \hat{b}})^{\beta \rho} W \boldsymbol{\lambda}_{i}^{\alpha} X^{i}_{\alpha} {G}^{-3} \varphi_{j \beta} \varphi^{j}_{\rho} - \frac{3}{64}{\rm i} (\Sigma_{\hat{a} \hat{b}})^{\alpha \beta} W \boldsymbol{\lambda}_{j}^{\rho} X_{i \alpha} {G}^{-3} \varphi^{j}_{\beta} \varphi^{i}_{\rho}+\frac{3}{32}{\rm i} (\Sigma_{\hat{a} \hat{b}})^{\alpha \beta} W \boldsymbol{\lambda}_{j}^{\rho} X_{i \alpha} {G}^{-3} \varphi^{j}_{\rho} \varphi^{i}_{\beta}+\frac{3}{8}{\rm i} (\Sigma_{\hat{a} \hat{b}})^{\beta \rho} W \boldsymbol{\lambda}_{j}^{\alpha} X_{i \alpha} {G}^{-3} \varphi^{j}_{\beta} \varphi^{i}_{\rho} - \frac{3}{8}{\rm i} (\Sigma_{\hat{a} \hat{b}})^{\beta \rho} \boldsymbol{W} \lambda_{i \beta} X^{i \alpha} {G}^{-3} \varphi_{j \rho} \varphi^{j}_{\alpha}%
 - \frac{3}{16}{\rm i} (\Sigma_{\hat{a} \hat{b}})^{\beta \rho} \boldsymbol{W} \lambda_{j \beta} X_{i}^{\alpha} {G}^{-3} \varphi^{j}_{\rho} \varphi^{i}_{\alpha} - \frac{3}{16}{\rm i} (\Sigma_{\hat{a} \hat{b}})^{\beta \rho} \boldsymbol{W} \lambda_{j \beta} X_{i}^{\alpha} {G}^{-3} \varphi^{j}_{\alpha} \varphi^{i}_{\rho}+\frac{3}{64}{\rm i} (\Sigma_{\hat{a} \hat{b}})^{\alpha \beta} \boldsymbol{W} \lambda^{\rho}_{i} X^{i}_{\alpha} {G}^{-3} \varphi_{j \beta} \varphi^{j}_{\rho}+\frac{3}{8}{\rm i} (\Sigma_{\hat{a} \hat{b}})^{\beta \rho} \boldsymbol{W} \lambda^{\alpha}_{i} X^{i}_{\alpha} {G}^{-3} \varphi_{j \beta} \varphi^{j}_{\rho} - \frac{3}{64}{\rm i} (\Sigma_{\hat{a} \hat{b}})^{\alpha \beta} \boldsymbol{W} \lambda^{\rho}_{j} X_{i \alpha} {G}^{-3} \varphi^{j}_{\beta} \varphi^{i}_{\rho}+\frac{3}{32}{\rm i} (\Sigma_{\hat{a} \hat{b}})^{\alpha \beta} \boldsymbol{W} \lambda^{\rho}_{j} X_{i \alpha} {G}^{-3} \varphi^{j}_{\rho} \varphi^{i}_{\beta}+\frac{3}{8}{\rm i} (\Sigma_{\hat{a} \hat{b}})^{\beta \rho} \boldsymbol{W} \lambda^{\alpha}_{j} X_{i \alpha} {G}^{-3} \varphi^{j}_{\beta} \varphi^{i}_{\rho} - \frac{9}{64}{\rm i} G_{i j} (\Sigma_{\hat{a} \hat{b}})^{\alpha \beta} W \mathbf{X}^{i j} X_{k \alpha} {G}^{-3} \varphi^{k}_{\beta}+\frac{9}{64}{\rm i} G_{i j} (\Sigma_{\hat{a} \hat{b}})^{\alpha \beta} W \mathbf{X}^{i}\,_{k} X^{j}_{\alpha} {G}^{-3} \varphi^{k}_{\beta} - \frac{9}{64}{\rm i} G_{i j} (\Sigma_{\hat{a} \hat{b}})^{\alpha \beta} W \mathbf{X}^{i}\,_{k} X^{k}_{\alpha} {G}^{-3} \varphi^{j}_{\beta} - \frac{9}{64}{\rm i} G_{i j} (\Sigma_{\hat{a} \hat{b}})^{\alpha \beta} \boldsymbol{W} X^{i j} X_{k \alpha} {G}^{-3} \varphi^{k}_{\beta}+\frac{9}{64}{\rm i} G_{i j} (\Sigma_{\hat{a} \hat{b}})^{\alpha \beta} \boldsymbol{W} X^{i}\,_{k} X^{j}_{\alpha} {G}^{-3} \varphi^{k}_{\beta} - \frac{9}{64}{\rm i} G_{i j} (\Sigma_{\hat{a} \hat{b}})^{\alpha \beta} \boldsymbol{W} X^{i}\,_{k} X^{k}_{\alpha} {G}^{-3} \varphi^{j}_{\beta} - \frac{3}{16}{\rm i} G_{i j} (\Sigma_{\hat{a} \hat{b}})^{\alpha \beta} \lambda^{i}_{\alpha} {G}^{-3} \nabla_{c}{\boldsymbol{\lambda}_{k \beta}} \nabla^{c}{G^{j k}} - \frac{3}{16}{\rm i} G_{i j} (\Sigma_{\hat{a} \hat{b}})^{\alpha \beta} \boldsymbol{\lambda}^{i}_{\alpha} {G}^{-3} \nabla_{c}{\lambda_{k \beta}} \nabla^{c}{G^{j k}}+\frac{3}{16}{\rm i} G_{i j} (\Sigma_{\hat{a} \hat{b}})^{\alpha \beta} \lambda_{k \alpha} {G}^{-3} \nabla_{c}{\boldsymbol{\lambda}^{i}_{\beta}} \nabla^{c}{G^{j k}} - \frac{3}{16}{\rm i} G_{i j} (\Sigma_{\hat{a} \hat{b}})^{\alpha \beta} \lambda_{k \alpha} {G}^{-3} \nabla_{c}{\boldsymbol{\lambda}^{k}_{\beta}} \nabla^{c}{G^{i j}}+\frac{3}{16}{\rm i} G_{i j} (\Sigma_{\hat{a} \hat{b}})^{\alpha \beta} \boldsymbol{\lambda}_{k \alpha} {G}^{-3} \nabla_{c}{\lambda^{i}_{\beta}} \nabla^{c}{G^{j k}} - \frac{3}{16}{\rm i} G_{i j} (\Sigma_{\hat{a} \hat{b}})^{\alpha \beta} \boldsymbol{\lambda}_{k \alpha} {G}^{-3} \nabla_{c}{\lambda^{k}_{\beta}} \nabla^{c}{G^{i j}} - \frac{5}{8}{\rm i} G_{i j} (\Sigma_{\hat{a} c})^{\alpha \beta} \lambda^{i}_{\alpha} {G}^{-3} \nabla^{c}{\boldsymbol{\lambda}_{k \beta}} \nabla_{\hat{b}}{G^{j k}}%
+\frac{3}{8}{\rm i} G_{i j} (\Sigma_{\hat{a} c})^{\alpha \beta} \lambda^{i}_{\alpha} {G}^{-3} \nabla_{\hat{b}}{\boldsymbol{\lambda}_{k \beta}} \nabla^{c}{G^{j k}} - \frac{5}{8}{\rm i} G_{i j} (\Sigma_{\hat{a} c})^{\alpha \beta} \boldsymbol{\lambda}^{i}_{\alpha} {G}^{-3} \nabla^{c}{\lambda_{k \beta}} \nabla_{\hat{b}}{G^{j k}}+\frac{3}{8}{\rm i} G_{i j} (\Sigma_{\hat{a} c})^{\alpha \beta} \boldsymbol{\lambda}^{i}_{\alpha} {G}^{-3} \nabla_{\hat{b}}{\lambda_{k \beta}} \nabla^{c}{G^{j k}}+\frac{5}{8}{\rm i} G_{i j} (\Sigma_{\hat{a} c})^{\alpha \beta} \lambda_{k \alpha} {G}^{-3} \nabla^{c}{\boldsymbol{\lambda}^{i}_{\beta}} \nabla_{\hat{b}}{G^{j k}} - \frac{3}{8}{\rm i} G_{i j} (\Sigma_{\hat{a} c})^{\alpha \beta} \lambda_{k \alpha} {G}^{-3} \nabla_{\hat{b}}{\boldsymbol{\lambda}^{i}_{\beta}} \nabla^{c}{G^{j k}}+\frac{3}{8}{\rm i} G_{i j} (\Sigma_{\hat{a} c})^{\alpha \beta} \lambda_{k \alpha} {G}^{-3} \nabla^{c}{\boldsymbol{\lambda}^{k}_{\beta}} \nabla_{\hat{b}}{G^{i j}}+\frac{3}{8}{\rm i} G_{i j} (\Sigma_{\hat{a} c})^{\alpha \beta} \lambda_{k \alpha} {G}^{-3} \nabla_{\hat{b}}{\boldsymbol{\lambda}^{k}_{\beta}} \nabla^{c}{G^{i j}}+\frac{5}{8}{\rm i} G_{i j} (\Sigma_{\hat{a} c})^{\alpha \beta} \boldsymbol{\lambda}_{k \alpha} {G}^{-3} \nabla^{c}{\lambda^{i}_{\beta}} \nabla_{\hat{b}}{G^{j k}} - \frac{3}{8}{\rm i} G_{i j} (\Sigma_{\hat{a} c})^{\alpha \beta} \boldsymbol{\lambda}_{k \alpha} {G}^{-3} \nabla_{\hat{b}}{\lambda^{i}_{\beta}} \nabla^{c}{G^{j k}}+\frac{3}{8}{\rm i} G_{i j} (\Sigma_{\hat{a} c})^{\alpha \beta} \boldsymbol{\lambda}_{k \alpha} {G}^{-3} \nabla^{c}{\lambda^{k}_{\beta}} \nabla_{\hat{b}}{G^{i j}}+\frac{3}{8}{\rm i} G_{i j} (\Sigma_{\hat{a} c})^{\alpha \beta} \boldsymbol{\lambda}_{k \alpha} {G}^{-3} \nabla_{\hat{b}}{\lambda^{k}_{\beta}} \nabla^{c}{G^{i j}}+\frac{3}{16}{\rm i} G_{i j} F_{\hat{a} \hat{b}} \boldsymbol{\lambda}^{i \alpha} {G}^{-5} \varphi^{j \beta} \varphi_{k \alpha} \varphi^{k}_{\beta}+\frac{3}{16}{\rm i} G_{i j} F_{\hat{a} \hat{b}} \boldsymbol{\lambda}_{k}^{\alpha} {G}^{-5} \varphi^{i}_{\alpha} \varphi^{j \beta} \varphi^{k}_{\beta} - \frac{3}{16}{\rm i} G_{i j} F_{\hat{a} \hat{b}} \boldsymbol{\lambda}_{k}^{\alpha} {G}^{-5} \varphi^{i \beta} \varphi^{j}_{\beta} \varphi^{k}_{\alpha}+\frac{3}{16}{\rm i} G_{i j} \mathbf{F}_{\hat{a} \hat{b}} \lambda^{i \alpha} {G}^{-5} \varphi^{j \beta} \varphi_{k \alpha} \varphi^{k}_{\beta}+\frac{3}{16}{\rm i} G_{i j} \mathbf{F}_{\hat{a} \hat{b}} \lambda^{\alpha}_{k} {G}^{-5} \varphi^{i}_{\alpha} \varphi^{j \beta} \varphi^{k}_{\beta} - \frac{3}{16}{\rm i} G_{i j} \mathbf{F}_{\hat{a} \hat{b}} \lambda^{\alpha}_{k} {G}^{-5} \varphi^{i \beta} \varphi^{j}_{\beta} \varphi^{k}_{\alpha} - \frac{1}{32}{\rm i} \epsilon_{\hat{a} \hat{b}}\,^{c d e} (\Gamma^{{e_{1}}})^{\alpha \beta} W_{c d} W_{e {e_{1}}} \lambda_{i \alpha} \boldsymbol{\lambda}^{i}_{\beta} {G}^{-1} - \frac{1}{16}{\rm i} \epsilon_{\hat{a} {e_{1}}}\,^{c d e} (\Gamma^{{e_{1}}})^{\alpha \beta} W_{\hat{b} c} W_{d e} \lambda_{i \alpha} \boldsymbol{\lambda}^{i}_{\beta} {G}^{-1} - \frac{9}{32}G_{i j} G_{k l} (\Sigma_{\hat{a} \hat{b}})^{\beta \alpha} X^{i j} \boldsymbol{\lambda}^{k}_{\beta} X^{l}_{\alpha} {G}^{-3}%
 - \frac{9}{32}G_{i j} G_{k l} (\Sigma_{\hat{a} \hat{b}})^{\beta \alpha} \mathbf{X}^{i j} \lambda^{k}_{\beta} X^{l}_{\alpha} {G}^{-3}+\frac{9}{32}G_{i j} G_{k l} (\Sigma_{\hat{a} \hat{b}})^{\beta \alpha} X^{i k} \boldsymbol{\lambda}^{j}_{\beta} X^{l}_{\alpha} {G}^{-3}+\frac{9}{32}G_{i j} G_{k l} (\Sigma_{\hat{a} \hat{b}})^{\beta \alpha} \mathbf{X}^{i k} \lambda^{j}_{\beta} X^{l}_{\alpha} {G}^{-3} - \frac{1}{256}(\Gamma^{c})^{\alpha \beta} W W_{\hat{b} c} W_{d e \alpha i} \epsilon_{\hat{a}}\,^{d e {e_{1}} {e_{2}}} (\Sigma_{{e_{1}} {e_{2}}})_{\beta}{}^{\rho} \boldsymbol{\lambda}^{i}_{\rho} {G}^{-1} - \frac{1}{256}(\Gamma^{c})^{\alpha \beta} \boldsymbol{W} W_{\hat{b} c} W_{d e \alpha i} \epsilon_{\hat{a}}\,^{d e {e_{1}} {e_{2}}} (\Sigma_{{e_{1}} {e_{2}}})_{\beta}{}^{\rho} \lambda^{i}_{\rho} {G}^{-1} - \frac{1}{4}G_{i j} (\Gamma_{c})^{\alpha \beta} \lambda^{i}_{\alpha} {G}^{-3} \nabla^{c}{\mathbf{F}_{\hat{a} \hat{b}}} \varphi^{j}_{\beta} - \frac{1}{4}G_{i j} (\Gamma_{c})^{\alpha \beta} \boldsymbol{\lambda}^{i}_{\alpha} {G}^{-3} \nabla^{c}{F_{\hat{a} \hat{b}}} \varphi^{j}_{\beta} - \frac{1}{8}(\Gamma_{\hat{a}})^{\alpha \beta} \lambda_{i \alpha} {G}^{-3} \nabla_{\hat{b}}{\boldsymbol{\lambda}^{i \rho}} \varphi_{j \beta} \varphi^{j}_{\rho} - \frac{1}{8}(\Gamma_{\hat{a}})^{\alpha \beta} \lambda_{i \alpha} {G}^{-3} \nabla_{\hat{b}}{\boldsymbol{\lambda}_{j}^{\rho}} \varphi^{i}_{\beta} \varphi^{j}_{\rho} - \frac{1}{8}(\Gamma_{\hat{a}})^{\alpha \beta} \boldsymbol{\lambda}_{i \alpha} {G}^{-3} \nabla_{\hat{b}}{\lambda^{i \rho}} \varphi_{j \beta} \varphi^{j}_{\rho} - \frac{1}{8}(\Gamma_{\hat{a}})^{\alpha \beta} \boldsymbol{\lambda}_{j \alpha} {G}^{-3} \nabla_{\hat{b}}{\lambda^{\rho}_{i}} \varphi^{j}_{\beta} \varphi^{i}_{\rho}+\frac{1}{2}(\Gamma_{\hat{a}})^{\alpha \beta} \lambda^{\rho}_{i} \boldsymbol{\lambda}_{j \rho} {G}^{-3} \nabla_{\hat{b}}{\varphi^{i}_{\alpha}} \varphi^{j}_{\beta}+\frac{1}{2}(\Gamma_{\hat{a}})^{\alpha \beta} \lambda^{\rho}_{i} \boldsymbol{\lambda}_{j \rho} {G}^{-3} \nabla_{\hat{b}}{\varphi^{j}_{\alpha}} \varphi^{i}_{\beta} - \frac{3}{16}(\Gamma_{\hat{a}})^{\alpha \beta} \lambda^{\rho}_{i} {G}^{-3} \nabla_{\hat{b}}{\boldsymbol{\lambda}^{i}_{\alpha}} \varphi_{j \beta} \varphi^{j}_{\rho}+\frac{1}{16}(\Gamma_{\hat{a}})^{\alpha \beta} \lambda^{\rho}_{i} {G}^{-3} \nabla_{\hat{b}}{\boldsymbol{\lambda}_{j \alpha}} \varphi^{i}_{\beta} \varphi^{j}_{\rho} - \frac{1}{4}(\Gamma_{\hat{a}})^{\alpha \beta} \lambda^{\rho}_{i} {G}^{-3} \nabla_{\hat{b}}{\boldsymbol{\lambda}_{j \alpha}} \varphi^{i}_{\rho} \varphi^{j}_{\beta} - \frac{1}{8}(\Gamma_{\hat{a}})^{\alpha \beta} \lambda^{\rho}_{i} {G}^{-3} \nabla_{\hat{b}}{\boldsymbol{\lambda}_{j \rho}} \varphi^{i}_{\alpha} \varphi^{j}_{\beta} - \frac{3}{16}(\Gamma_{\hat{a}})^{\alpha \beta} \boldsymbol{\lambda}_{i}^{\rho} {G}^{-3} \nabla_{\hat{b}}{\lambda^{i}_{\alpha}} \varphi_{j \beta} \varphi^{j}_{\rho}+\frac{1}{16}(\Gamma_{\hat{a}})^{\alpha \beta} \boldsymbol{\lambda}_{j}^{\rho} {G}^{-3} \nabla_{\hat{b}}{\lambda_{i \alpha}} \varphi^{j}_{\beta} \varphi^{i}_{\rho} - \frac{1}{4}(\Gamma_{\hat{a}})^{\alpha \beta} \boldsymbol{\lambda}_{j}^{\rho} {G}^{-3} \nabla_{\hat{b}}{\lambda_{i \alpha}} \varphi^{j}_{\rho} \varphi^{i}_{\beta}%
 - \frac{1}{8}(\Gamma_{\hat{a}})^{\alpha \beta} \boldsymbol{\lambda}_{j}^{\rho} {G}^{-3} \nabla_{\hat{b}}{\lambda_{i \rho}} \varphi^{j}_{\alpha} \varphi^{i}_{\beta} - \frac{3}{32}(\Sigma_{\hat{a}}{}^{\, c})^{\alpha \beta} W_{\hat{b} c} \lambda_{i \alpha} \boldsymbol{\lambda}^{i \rho} {G}^{-3} \varphi_{j \beta} \varphi^{j}_{\rho} - \frac{3}{32}(\Sigma_{\hat{a}}{}^{\, c})^{\alpha \beta} W_{\hat{b} c} \lambda_{i \alpha} \boldsymbol{\lambda}_{j}^{\rho} {G}^{-3} \varphi^{i}_{\rho} \varphi^{j}_{\beta} - \frac{3}{32}(\Sigma_{\hat{a}}{}^{\, c})^{\alpha \beta} W_{\hat{b} c} \lambda^{\rho}_{i} \boldsymbol{\lambda}^{i}_{\alpha} {G}^{-3} \varphi_{j \beta} \varphi^{j}_{\rho} - \frac{3}{32}(\Sigma_{\hat{a}}{}^{\, c})^{\alpha \beta} W_{\hat{b} c} \lambda^{\rho}_{i} \boldsymbol{\lambda}_{j \alpha} {G}^{-3} \varphi^{i}_{\beta} \varphi^{j}_{\rho} - \frac{1}{8}G_{i j} (\Gamma_{\hat{a}})^{\alpha \beta} X^{i j} \boldsymbol{\lambda}_{k \alpha} {G}^{-3} \nabla_{\hat{b}}{\varphi^{k}_{\beta}}+\frac{3}{16}G_{i j} (\Gamma_{\hat{a}})^{\alpha \beta} X^{i j} {G}^{-3} \nabla_{\hat{b}}{\boldsymbol{\lambda}_{k \alpha}} \varphi^{k}_{\beta} - \frac{1}{8}G_{i j} (\Gamma_{\hat{a}})^{\alpha \beta} \mathbf{X}^{i j} \lambda_{k \alpha} {G}^{-3} \nabla_{\hat{b}}{\varphi^{k}_{\beta}}+\frac{3}{16}G_{i j} (\Gamma_{\hat{a}})^{\alpha \beta} \mathbf{X}^{i j} {G}^{-3} \nabla_{\hat{b}}{\lambda_{k \alpha}} \varphi^{k}_{\beta}+\frac{1}{8}G_{i j} (\Gamma_{\hat{a}})^{\alpha \beta} X^{i}\,_{k} \boldsymbol{\lambda}^{j}_{\alpha} {G}^{-3} \nabla_{\hat{b}}{\varphi^{k}_{\beta}}+\frac{3}{8}G_{i j} (\Gamma_{\hat{a}})^{\alpha \beta} X^{i}\,_{k} \boldsymbol{\lambda}^{k}_{\alpha} {G}^{-3} \nabla_{\hat{b}}{\varphi^{j}_{\beta}} - \frac{3}{16}G_{i j} (\Gamma_{\hat{a}})^{\alpha \beta} X^{i}\,_{k} {G}^{-3} \nabla_{\hat{b}}{\boldsymbol{\lambda}^{j}_{\alpha}} \varphi^{k}_{\beta}+\frac{11}{16}G_{i j} (\Gamma_{\hat{a}})^{\alpha \beta} X^{i}\,_{k} {G}^{-3} \nabla_{\hat{b}}{\boldsymbol{\lambda}^{k}_{\alpha}} \varphi^{j}_{\beta}+\frac{1}{8}G_{i j} (\Gamma_{\hat{a}})^{\alpha \beta} \mathbf{X}^{i}\,_{k} \lambda^{j}_{\alpha} {G}^{-3} \nabla_{\hat{b}}{\varphi^{k}_{\beta}}+\frac{3}{8}G_{i j} (\Gamma_{\hat{a}})^{\alpha \beta} \mathbf{X}^{i}\,_{k} \lambda^{k}_{\alpha} {G}^{-3} \nabla_{\hat{b}}{\varphi^{j}_{\beta}} - \frac{3}{16}G_{i j} (\Gamma_{\hat{a}})^{\alpha \beta} \mathbf{X}^{i}\,_{k} {G}^{-3} \nabla_{\hat{b}}{\lambda^{j}_{\alpha}} \varphi^{k}_{\beta}+\frac{11}{16}G_{i j} (\Gamma_{\hat{a}})^{\alpha \beta} \mathbf{X}^{i}\,_{k} {G}^{-3} \nabla_{\hat{b}}{\lambda^{k}_{\alpha}} \varphi^{j}_{\beta} - \frac{1}{2}G_{i j} (\Gamma_{\hat{a}})^{\alpha \beta} \lambda_{k \alpha} {G}^{-3} \nabla_{\hat{b}}{\mathbf{X}^{i k}} \varphi^{j}_{\beta} - \frac{1}{2}G_{i j} (\Gamma_{\hat{a}})^{\alpha \beta} \boldsymbol{\lambda}_{k \alpha} {G}^{-3} \nabla_{\hat{b}}{X^{i k}} \varphi^{j}_{\beta}+\frac{3}{32}G_{i j} (\Sigma_{\hat{a} \hat{b}})^{\beta \alpha} \lambda^{i}_{\beta} \boldsymbol{\lambda}_{k}^{\rho} X^{k}_{\alpha} {G}^{-3} \varphi^{j}_{\rho}%
 - \frac{9}{16}G_{i j} (\Sigma_{\hat{a} \hat{b}})^{\beta \rho} \lambda^{i}_{\beta} \boldsymbol{\lambda}_{k}^{\alpha} X^{k}_{\alpha} {G}^{-3} \varphi^{j}_{\rho}+\frac{3}{32}G_{i j} (\Sigma_{\hat{a} \hat{b}})^{\beta \alpha} \lambda^{i \rho} \boldsymbol{\lambda}_{k \beta} X^{k}_{\alpha} {G}^{-3} \varphi^{j}_{\rho} - \frac{9}{16}G_{i j} (\Sigma_{\hat{a} \hat{b}})^{\beta \rho} \lambda^{i \alpha} \boldsymbol{\lambda}_{k \beta} X^{k}_{\alpha} {G}^{-3} \varphi^{j}_{\rho} - \frac{3}{32}G_{i j} (\Sigma_{\hat{a} \hat{b}})^{\beta \alpha} \lambda_{k \beta} \boldsymbol{\lambda}^{i \rho} X^{k}_{\alpha} {G}^{-3} \varphi^{j}_{\rho}+\frac{9}{16}G_{i j} (\Sigma_{\hat{a} \hat{b}})^{\beta \rho} \lambda_{k \beta} \boldsymbol{\lambda}^{i \alpha} X^{k}_{\alpha} {G}^{-3} \varphi^{j}_{\rho}+\frac{3}{32}G_{i j} (\Sigma_{\hat{a} \hat{b}})^{\beta \alpha} \lambda_{k \beta} \boldsymbol{\lambda}^{k \rho} X^{i}_{\alpha} {G}^{-3} \varphi^{j}_{\rho} - \frac{9}{16}G_{i j} (\Sigma_{\hat{a} \hat{b}})^{\beta \rho} \lambda_{k \beta} \boldsymbol{\lambda}^{k \alpha} X^{i}_{\alpha} {G}^{-3} \varphi^{j}_{\rho} - \frac{3}{32}G_{i j} (\Sigma_{\hat{a} \hat{b}})^{\beta \alpha} \lambda^{\rho}_{k} \boldsymbol{\lambda}^{i}_{\beta} X^{k}_{\alpha} {G}^{-3} \varphi^{j}_{\rho}+\frac{9}{16}G_{i j} (\Sigma_{\hat{a} \hat{b}})^{\beta \rho} \lambda^{\alpha}_{k} \boldsymbol{\lambda}^{i}_{\beta} X^{k}_{\alpha} {G}^{-3} \varphi^{j}_{\rho}+\frac{3}{32}G_{i j} (\Sigma_{\hat{a} \hat{b}})^{\beta \alpha} \lambda^{\rho}_{k} \boldsymbol{\lambda}^{k}_{\beta} X^{i}_{\alpha} {G}^{-3} \varphi^{j}_{\rho} - \frac{9}{16}G_{i j} (\Sigma_{\hat{a} \hat{b}})^{\beta \rho} \lambda^{\alpha}_{k} \boldsymbol{\lambda}^{k}_{\beta} X^{i}_{\alpha} {G}^{-3} \varphi^{j}_{\rho}+\frac{11}{32}G_{i j} (\Sigma_{\hat{a}}{}^{\, c})^{\alpha \beta} X^{i j} W_{\hat{b} c} \boldsymbol{\lambda}_{k \alpha} {G}^{-3} \varphi^{k}_{\beta}+\frac{11}{32}G_{i j} (\Sigma_{\hat{a}}{}^{\, c})^{\alpha \beta} \mathbf{X}^{i j} W_{\hat{b} c} \lambda_{k \alpha} {G}^{-3} \varphi^{k}_{\beta} - \frac{11}{32}G_{i j} (\Sigma_{\hat{a}}{}^{\, c})^{\alpha \beta} X^{i}\,_{k} W_{\hat{b} c} \boldsymbol{\lambda}^{j}_{\alpha} {G}^{-3} \varphi^{k}_{\beta}+\frac{11}{32}G_{i j} (\Sigma_{\hat{a}}{}^{\, c})^{\alpha \beta} X^{i}\,_{k} W_{\hat{b} c} \boldsymbol{\lambda}^{k}_{\alpha} {G}^{-3} \varphi^{j}_{\beta} - \frac{11}{32}G_{i j} (\Sigma_{\hat{a}}{}^{\, c})^{\alpha \beta} \mathbf{X}^{i}\,_{k} W_{\hat{b} c} \lambda^{j}_{\alpha} {G}^{-3} \varphi^{k}_{\beta}+\frac{11}{32}G_{i j} (\Sigma_{\hat{a}}{}^{\, c})^{\alpha \beta} \mathbf{X}^{i}\,_{k} W_{\hat{b} c} \lambda^{k}_{\alpha} {G}^{-3} \varphi^{j}_{\beta}+\frac{3}{16}(\Gamma_{\hat{a}})^{\alpha \beta} X_{k i} \boldsymbol{\lambda}^{k}_{\alpha} {G}^{-3} \nabla_{\hat{b}}{G^{i}\,_{j}} \varphi^{j}_{\beta}+\frac{1}{16}(\Gamma_{\hat{a}})^{\alpha \beta} X_{i j} \boldsymbol{\lambda}_{k \alpha} {G}^{-3} \nabla_{\hat{b}}{G^{i j}} \varphi^{k}_{\beta} - \frac{1}{16}(\Gamma_{\hat{a}})^{\alpha \beta} X_{i k} \boldsymbol{\lambda}_{j \alpha} {G}^{-3} \nabla_{\hat{b}}{G^{i j}} \varphi^{k}_{\beta}%
+\frac{3}{16}(\Gamma_{\hat{a}})^{\alpha \beta} \mathbf{X}_{k i} \lambda^{k}_{\alpha} {G}^{-3} \nabla_{\hat{b}}{G^{i}\,_{j}} \varphi^{j}_{\beta}+\frac{1}{16}(\Gamma_{\hat{a}})^{\alpha \beta} \mathbf{X}_{i j} \lambda_{k \alpha} {G}^{-3} \nabla_{\hat{b}}{G^{i j}} \varphi^{k}_{\beta} - \frac{1}{16}(\Gamma_{\hat{a}})^{\alpha \beta} \mathbf{X}_{i k} \lambda_{j \alpha} {G}^{-3} \nabla_{\hat{b}}{G^{i j}} \varphi^{k}_{\beta}+\Phi_{\hat{a}}\,^{c}\,_{i k} G^{i}\,_{j} (\Sigma_{\hat{b} c})^{\alpha \beta} W \boldsymbol{\lambda}^{k}_{\alpha} {G}^{-3} \varphi^{j}_{\beta}+\Phi_{\hat{a}}\,^{c}\,_{i k} G^{i}\,_{j} (\Sigma_{\hat{b} c})^{\alpha \beta} \boldsymbol{W} \lambda^{k}_{\alpha} {G}^{-3} \varphi^{j}_{\beta} - \frac{1}{8}\epsilon^{e {e_{1}}}\,_{\hat{a}}\,^{c d} G_{i j} (\Sigma_{e {e_{1}}})^{\alpha \beta} F_{c d} \boldsymbol{\lambda}^{i}_{\alpha} {G}^{-3} \nabla_{\hat{b}}{\varphi^{j}_{\beta}} - \frac{1}{8}\epsilon^{e {e_{1}}}\,_{\hat{a}}\,^{c d} G_{i j} (\Sigma_{e {e_{1}}})^{\alpha \beta} F_{c d} {G}^{-3} \nabla_{\hat{b}}{\boldsymbol{\lambda}^{i}_{\alpha}} \varphi^{j}_{\beta} - \frac{1}{8}\epsilon^{e {e_{1}}}\,_{\hat{a}}\,^{c d} G_{i j} (\Sigma_{e {e_{1}}})^{\alpha \beta} \mathbf{F}_{c d} \lambda^{i}_{\alpha} {G}^{-3} \nabla_{\hat{b}}{\varphi^{j}_{\beta}} - \frac{1}{8}\epsilon^{e {e_{1}}}\,_{\hat{a}}\,^{c d} G_{i j} (\Sigma_{e {e_{1}}})^{\alpha \beta} \mathbf{F}_{c d} {G}^{-3} \nabla_{\hat{b}}{\lambda^{i}_{\alpha}} \varphi^{j}_{\beta} - \frac{1}{4}\epsilon^{d e}\,_{\hat{a} {e_{1}}}\,^{c} G_{i j} (\Sigma_{d e})^{\alpha \beta} F_{\hat{b} c} \boldsymbol{\lambda}^{i}_{\alpha} {G}^{-3} \nabla^{{e_{1}}}{\varphi^{j}_{\beta}} - \frac{1}{4}\epsilon^{d e}\,_{\hat{a} {e_{1}}}\,^{c} G_{i j} (\Sigma_{d e})^{\alpha \beta} F_{\hat{b} c} {G}^{-3} \nabla^{{e_{1}}}{\boldsymbol{\lambda}^{i}_{\alpha}} \varphi^{j}_{\beta} - \frac{1}{4}\epsilon^{d e}\,_{\hat{a} {e_{1}}}\,^{c} G_{i j} (\Sigma_{d e})^{\alpha \beta} \mathbf{F}_{\hat{b} c} \lambda^{i}_{\alpha} {G}^{-3} \nabla^{{e_{1}}}{\varphi^{j}_{\beta}} - \frac{1}{4}\epsilon^{d e}\,_{\hat{a} {e_{1}}}\,^{c} G_{i j} (\Sigma_{d e})^{\alpha \beta} \mathbf{F}_{\hat{b} c} {G}^{-3} \nabla^{{e_{1}}}{\lambda^{i}_{\alpha}} \varphi^{j}_{\beta}+\frac{1}{8}\epsilon^{e}\,_{\hat{a} \hat{b}}\,^{c d} G_{i j} (\Sigma_{e {e_{1}}})^{\alpha \beta} F_{c d} \boldsymbol{\lambda}^{i}_{\alpha} {G}^{-3} \nabla^{{e_{1}}}{\varphi^{j}_{\beta}}+\frac{1}{8}\epsilon^{e}\,_{\hat{a} \hat{b}}\,^{c d} G_{i j} (\Sigma_{e {e_{1}}})^{\alpha \beta} F_{c d} {G}^{-3} \nabla^{{e_{1}}}{\boldsymbol{\lambda}^{i}_{\alpha}} \varphi^{j}_{\beta}+\frac{1}{8}\epsilon^{e}\,_{\hat{a} \hat{b}}\,^{c d} G_{i j} (\Sigma_{e {e_{1}}})^{\alpha \beta} \mathbf{F}_{c d} \lambda^{i}_{\alpha} {G}^{-3} \nabla^{{e_{1}}}{\varphi^{j}_{\beta}}+\frac{1}{8}\epsilon^{e}\,_{\hat{a} \hat{b}}\,^{c d} G_{i j} (\Sigma_{e {e_{1}}})^{\alpha \beta} \mathbf{F}_{c d} {G}^{-3} \nabla^{{e_{1}}}{\lambda^{i}_{\alpha}} \varphi^{j}_{\beta} - \frac{1}{2}\epsilon_{\hat{a} \hat{b} c d e} G_{i j} W {G}^{-3} \nabla^{c}{\nabla^{d}{\boldsymbol{W}}} \nabla^{e}{G^{i j}} - \frac{1}{2}\epsilon_{\hat{a} \hat{b} c d e} G_{i j} \boldsymbol{W} {G}^{-3} \nabla^{c}{\nabla^{d}{W}} \nabla^{e}{G^{i j}} - \frac{3}{64}\epsilon^{c d e}\,_{\hat{a} \hat{b}} \mathcal{H}_{c} (\Sigma_{d e})^{\alpha \beta} X_{i j} \boldsymbol{\lambda}^{i}_{\alpha} {G}^{-3} \varphi^{j}_{\beta}%
 - \frac{3}{64}\epsilon^{c d e}\,_{\hat{a} \hat{b}} \mathcal{H}_{c} (\Sigma_{d e})^{\alpha \beta} \mathbf{X}_{i j} \lambda^{i}_{\alpha} {G}^{-3} \varphi^{j}_{\beta} - \frac{1}{16}\epsilon^{e {e_{1}}}\,_{\hat{a}}\,^{c d} (\Sigma_{e {e_{1}}})^{\alpha \beta} F_{c d} \boldsymbol{\lambda}_{i \alpha} {G}^{-3} \nabla_{\hat{b}}{G^{i}\,_{j}} \varphi^{j}_{\beta} - \frac{1}{16}\epsilon^{e {e_{1}}}\,_{\hat{a}}\,^{c d} (\Sigma_{e {e_{1}}})^{\alpha \beta} \mathbf{F}_{c d} \lambda_{i \alpha} {G}^{-3} \nabla_{\hat{b}}{G^{i}\,_{j}} \varphi^{j}_{\beta} - \frac{1}{8}\epsilon^{d e}\,_{\hat{a} {e_{1}}}\,^{c} (\Sigma_{d e})^{\alpha \beta} F_{\hat{b} c} \boldsymbol{\lambda}_{i \alpha} {G}^{-3} \nabla^{{e_{1}}}{G^{i}\,_{j}} \varphi^{j}_{\beta} - \frac{1}{8}\epsilon^{d e}\,_{\hat{a} {e_{1}}}\,^{c} (\Sigma_{d e})^{\alpha \beta} \mathbf{F}_{\hat{b} c} \lambda_{i \alpha} {G}^{-3} \nabla^{{e_{1}}}{G^{i}\,_{j}} \varphi^{j}_{\beta}+\frac{1}{16}\epsilon^{e}\,_{\hat{a} \hat{b}}\,^{c d} (\Sigma_{e {e_{1}}})^{\alpha \beta} F_{c d} \boldsymbol{\lambda}_{i \alpha} {G}^{-3} \nabla^{{e_{1}}}{G^{i}\,_{j}} \varphi^{j}_{\beta}+\frac{1}{16}\epsilon^{e}\,_{\hat{a} \hat{b}}\,^{c d} (\Sigma_{e {e_{1}}})^{\alpha \beta} \mathbf{F}_{c d} \lambda_{i \alpha} {G}^{-3} \nabla^{{e_{1}}}{G^{i}\,_{j}} \varphi^{j}_{\beta}+\frac{5}{48}\epsilon^{e}\,_{\hat{a} \hat{b} {e_{1}}}\,^{c} (\Sigma_{e}{}^{\, d})^{\beta \alpha} W \boldsymbol{\lambda}_{i \beta} {G}^{-1} \nabla^{{e_{1}}}{W_{c d \alpha}\,^{i}}+\frac{5}{48}\epsilon^{e}\,_{\hat{a} \hat{b} {e_{1}}}\,^{c} (\Sigma_{e}{}^{\, d})^{\beta \alpha} \boldsymbol{W} \lambda_{i \beta} {G}^{-1} \nabla^{{e_{1}}}{W_{c d \alpha}\,^{i}}+\frac{3}{32}{\rm i} G_{i j} G_{k l} W \mathbf{X}^{i j} W_{\hat{a} \hat{b}} {G}^{-5} \varphi^{k \alpha} \varphi^{l}_{\alpha} - \frac{3}{32}{\rm i} G_{i j} G_{k l} W \mathbf{X}^{i k} W_{\hat{a} \hat{b}} {G}^{-5} \varphi^{j \alpha} \varphi^{l}_{\alpha}+\frac{3}{32}{\rm i} G_{i j} G_{k l} \boldsymbol{W} X^{i j} W_{\hat{a} \hat{b}} {G}^{-5} \varphi^{k \alpha} \varphi^{l}_{\alpha} - \frac{3}{32}{\rm i} G_{i j} G_{k l} \boldsymbol{W} X^{i k} W_{\hat{a} \hat{b}} {G}^{-5} \varphi^{j \alpha} \varphi^{l}_{\alpha}+\frac{3}{4}{\rm i} \mathcal{H}_{\hat{a}} G_{i j} G_{k l} \lambda^{i \alpha} \boldsymbol{\lambda}^{j}_{\alpha} {G}^{-5} \nabla_{\hat{b}}{G^{k l}} - \frac{3}{2}{\rm i} G_{i j} (\Gamma_{\hat{a}})^{\alpha \beta} W X^{i}_{\alpha} {G}^{-3} \nabla_{\hat{b}}{\boldsymbol{W}} \varphi^{j}_{\beta} - \frac{3}{2}{\rm i} G_{i j} (\Gamma_{\hat{a}})^{\alpha \beta} W \boldsymbol{W} X^{i}_{\alpha} {G}^{-3} \nabla_{\hat{b}}{\varphi^{j}_{\beta}} - \frac{3}{2}{\rm i} G_{i j} (\Gamma_{\hat{a}})^{\alpha \beta} W \boldsymbol{W} {G}^{-3} \nabla_{\hat{b}}{X^{i}_{\alpha}} \varphi^{j}_{\beta} - \frac{3}{2}{\rm i} G_{i j} (\Gamma_{\hat{a}})^{\alpha \beta} \boldsymbol{W} X^{i}_{\alpha} {G}^{-3} \nabla_{\hat{b}}{W} \varphi^{j}_{\beta}+\frac{1}{2}{\rm i} G_{i j} (\Sigma^{c d})^{\alpha \beta} W \mathbf{F}_{\hat{a} c} W_{\hat{b} d \alpha}\,^{i} {G}^{-3} \varphi^{j}_{\beta}+\frac{1}{2}{\rm i} G_{i j} (\Sigma^{c d})^{\alpha \beta} \boldsymbol{W} F_{\hat{a} c} W_{\hat{b} d \alpha}\,^{i} {G}^{-3} \varphi^{j}_{\beta}%
+\frac{1}{8}{\rm i} G_{i j} W \mathbf{F}_{\hat{a} \hat{b}} W^{c d \alpha i} (\Sigma_{c d})_{\alpha}{}^{\beta} {G}^{-3} \varphi^{j}_{\beta} - \frac{1}{2}{\rm i} G_{i j} W \mathbf{F}_{\hat{a}}\,^{c} W_{c}\,^{d \alpha i} (\Sigma_{\hat{b} d})_{\alpha}{}^{\beta} {G}^{-3} \varphi^{j}_{\beta}+\frac{1}{8}{\rm i} G_{i j} \boldsymbol{W} F_{\hat{a} \hat{b}} W^{c d \alpha i} (\Sigma_{c d})_{\alpha}{}^{\beta} {G}^{-3} \varphi^{j}_{\beta} - \frac{1}{2}{\rm i} G_{i j} \boldsymbol{W} F_{\hat{a}}\,^{c} W_{c}\,^{d \alpha i} (\Sigma_{\hat{b} d})_{\alpha}{}^{\beta} {G}^{-3} \varphi^{j}_{\beta} - \frac{3}{4}{\rm i} (\Gamma_{\hat{a}})^{\alpha \beta} W \boldsymbol{W} X_{i \alpha} {G}^{-3} \nabla_{\hat{b}}{G^{i}\,_{j}} \varphi^{j}_{\beta}+\frac{9}{32}{\rm i} G_{i j} W W_{\hat{a} \hat{b}} \boldsymbol{\lambda}^{i \alpha} {G}^{-5} \varphi^{j \beta} \varphi_{k \alpha} \varphi^{k}_{\beta}+\frac{9}{32}{\rm i} G_{i j} W W_{\hat{a} \hat{b}} \boldsymbol{\lambda}_{k}^{\alpha} {G}^{-5} \varphi^{i}_{\alpha} \varphi^{j \beta} \varphi^{k}_{\beta} - \frac{9}{32}{\rm i} G_{i j} W W_{\hat{a} \hat{b}} \boldsymbol{\lambda}_{k}^{\alpha} {G}^{-5} \varphi^{i \beta} \varphi^{j}_{\beta} \varphi^{k}_{\alpha}+\frac{9}{32}{\rm i} G_{i j} \boldsymbol{W} W_{\hat{a} \hat{b}} \lambda^{i \alpha} {G}^{-5} \varphi^{j \beta} \varphi_{k \alpha} \varphi^{k}_{\beta}+\frac{9}{32}{\rm i} G_{i j} \boldsymbol{W} W_{\hat{a} \hat{b}} \lambda^{\alpha}_{k} {G}^{-5} \varphi^{i}_{\alpha} \varphi^{j \beta} \varphi^{k}_{\beta} - \frac{9}{32}{\rm i} G_{i j} \boldsymbol{W} W_{\hat{a} \hat{b}} \lambda^{\alpha}_{k} {G}^{-5} \varphi^{i \beta} \varphi^{j}_{\beta} \varphi^{k}_{\alpha} - \frac{1}{32}{\rm i} \epsilon_{\hat{a} \hat{b} e}\,^{c d} G_{i j} W \boldsymbol{W} {G}^{-3} \nabla^{e}{W_{c d}\,^{\alpha i}} \varphi^{j}_{\alpha}+\frac{1}{4}{\rm i} \epsilon^{d e}\,_{\hat{a} \hat{b} c} \Phi_{d e i j} (\Gamma^{c})^{\alpha \beta} \lambda^{i}_{\alpha} \boldsymbol{\lambda}^{j}_{\beta} {G}^{-1} - \frac{1}{4}{\rm i} \epsilon_{\hat{a} \hat{b} c d e} (\Gamma^{c})^{\alpha \beta} \lambda_{i \alpha} {G}^{-1} \nabla^{d}{\nabla^{e}{\boldsymbol{\lambda}^{i}_{\beta}}} - \frac{1}{4}{\rm i} \epsilon_{\hat{a} \hat{b} c d e} (\Gamma^{c})^{\alpha \beta} \boldsymbol{\lambda}_{i \alpha} {G}^{-1} \nabla^{d}{\nabla^{e}{\lambda^{i}_{\beta}}} - \frac{3}{4}G_{i j} G_{k l} W \mathbf{X}^{i j} {G}^{-5} \nabla_{\hat{a}}{G^{k}\,_{m}} \nabla_{\hat{b}}{G^{l m}} - \frac{3}{4}G_{i j} G_{k l} \boldsymbol{W} X^{i j} {G}^{-5} \nabla_{\hat{a}}{G^{k}\,_{m}} \nabla_{\hat{b}}{G^{l m}}+\frac{9}{32}F G_{i j} G_{k l} (\Sigma_{\hat{a} \hat{b}})^{\alpha \beta} X^{i j} \boldsymbol{\lambda}^{k}_{\alpha} {G}^{-5} \varphi^{l}_{\beta}+\frac{9}{32}F G_{i j} G_{k l} (\Sigma_{\hat{a} \hat{b}})^{\alpha \beta} \mathbf{X}^{i j} \lambda^{k}_{\alpha} {G}^{-5} \varphi^{l}_{\beta} - \frac{9}{32}F G_{i j} G_{k l} (\Sigma_{\hat{a} \hat{b}})^{\alpha \beta} X^{i k} \boldsymbol{\lambda}^{j}_{\alpha} {G}^{-5} \varphi^{l}_{\beta}%
 - \frac{9}{32}F G_{i j} G_{k l} (\Sigma_{\hat{a} \hat{b}})^{\alpha \beta} \mathbf{X}^{i k} \lambda^{j}_{\alpha} {G}^{-5} \varphi^{l}_{\beta}+\frac{3}{64}G_{i j} G_{k l} W_{\hat{a} \hat{b}} \lambda^{i \alpha} \boldsymbol{\lambda}^{j \beta} {G}^{-5} \varphi^{k}_{\alpha} \varphi^{l}_{\beta} - \frac{3}{32}G_{i j} G_{k l} W_{\hat{a} \hat{b}} \lambda^{i \alpha} \boldsymbol{\lambda}^{k \beta} {G}^{-5} \varphi^{j}_{\alpha} \varphi^{l}_{\beta} - \frac{3}{64}G_{i j} G_{k l} W_{\hat{a} \hat{b}} \lambda^{i \alpha} \boldsymbol{\lambda}^{k \beta} {G}^{-5} \varphi^{j}_{\beta} \varphi^{l}_{\alpha}+\frac{3}{2}G_{i j} G_{k l} W {G}^{-5} \nabla_{\hat{a}}{\boldsymbol{\lambda}^{i \alpha}} \nabla_{\hat{b}}{G^{j k}} \varphi^{l}_{\alpha}+\frac{3}{2}G_{i j} G_{k l} \boldsymbol{W} {G}^{-5} \nabla_{\hat{a}}{\lambda^{i \alpha}} \nabla_{\hat{b}}{G^{j k}} \varphi^{l}_{\alpha}+\frac{3}{4}G_{i j} G_{k l} \lambda^{i \alpha} {G}^{-5} \nabla_{\hat{a}}{\boldsymbol{W}} \nabla_{\hat{b}}{G^{j k}} \varphi^{l}_{\alpha}+\frac{3}{4}G_{i j} G_{k l} \boldsymbol{\lambda}^{i \alpha} {G}^{-5} \nabla_{\hat{a}}{W} \nabla_{\hat{b}}{G^{j k}} \varphi^{l}_{\alpha}+\frac{3}{4}G_{i j} (\Gamma^{c})^{\alpha \beta} W \boldsymbol{\lambda}^{i}_{\alpha} {G}^{-3} \nabla_{\hat{a}}{W_{\hat{b} c}} \varphi^{j}_{\beta}+\frac{3}{4}G_{i j} (\Gamma^{c})^{\alpha \beta} \boldsymbol{W} \lambda^{i}_{\alpha} {G}^{-3} \nabla_{\hat{a}}{W_{\hat{b} c}} \varphi^{j}_{\beta}+\frac{3}{8}(\Sigma_{\hat{a} \hat{b}})^{\rho \lambda} (\Gamma_{c})^{\alpha \beta} \lambda_{i \rho} {G}^{-3} \nabla^{c}{\boldsymbol{\lambda}^{i}_{\alpha}} \varphi_{j \lambda} \varphi^{j}_{\beta}+\frac{1}{8}(\Sigma_{\hat{a} \hat{b}})^{\rho \lambda} (\Gamma_{c})^{\alpha \beta} \lambda_{i \rho} {G}^{-3} \nabla^{c}{\boldsymbol{\lambda}_{j \alpha}} \varphi^{i}_{\lambda} \varphi^{j}_{\beta}+\frac{1}{4}(\Sigma_{\hat{a} \hat{b}})^{\rho \lambda} (\Gamma_{c})^{\alpha \beta} \lambda_{i \rho} {G}^{-3} \nabla^{c}{\boldsymbol{\lambda}_{j \alpha}} \varphi^{i}_{\beta} \varphi^{j}_{\lambda}+\frac{3}{8}(\Sigma_{\hat{a} \hat{b}})^{\rho \lambda} (\Gamma_{c})^{\alpha \beta} \boldsymbol{\lambda}_{i \rho} {G}^{-3} \nabla^{c}{\lambda^{i}_{\alpha}} \varphi_{j \lambda} \varphi^{j}_{\beta}+\frac{1}{8}(\Sigma_{\hat{a} \hat{b}})^{\rho \lambda} (\Gamma_{c})^{\alpha \beta} \boldsymbol{\lambda}_{j \rho} {G}^{-3} \nabla^{c}{\lambda_{i \alpha}} \varphi^{j}_{\lambda} \varphi^{i}_{\beta}+\frac{1}{4}(\Sigma_{\hat{a} \hat{b}})^{\rho \lambda} (\Gamma_{c})^{\alpha \beta} \boldsymbol{\lambda}_{j \rho} {G}^{-3} \nabla^{c}{\lambda_{i \alpha}} \varphi^{j}_{\beta} \varphi^{i}_{\lambda}+\frac{1}{8}(\Sigma_{\hat{a} \hat{b}})^{\rho \lambda} (\Gamma_{c})^{\alpha \beta} \lambda_{i \alpha} {G}^{-3} \nabla^{c}{\boldsymbol{\lambda}^{i}_{\rho}} \varphi_{j \lambda} \varphi^{j}_{\beta} - \frac{1}{8}(\Sigma_{\hat{a} \hat{b}})^{\rho \lambda} (\Gamma_{c})^{\alpha \beta} \lambda_{i \alpha} {G}^{-3} \nabla^{c}{\boldsymbol{\lambda}^{i}_{\beta}} \varphi_{j \rho} \varphi^{j}_{\lambda}+\frac{1}{8}(\Sigma_{\hat{a} \hat{b}})^{\rho \lambda} (\Gamma_{c})^{\alpha \beta} \lambda_{i \alpha} {G}^{-3} \nabla^{c}{\boldsymbol{\lambda}_{j \rho}} \varphi^{i}_{\lambda} \varphi^{j}_{\beta} - \frac{1}{8}(\Sigma_{\hat{a} \hat{b}})^{\rho \lambda} (\Gamma_{c})^{\alpha \beta} \lambda_{i \alpha} {G}^{-3} \nabla^{c}{\boldsymbol{\lambda}_{j \beta}} \varphi^{i}_{\rho} \varphi^{j}_{\lambda}%
+\frac{1}{8}(\Sigma_{\hat{a} \hat{b}})^{\rho \lambda} (\Gamma_{c})^{\alpha \beta} \boldsymbol{\lambda}_{i \alpha} {G}^{-3} \nabla^{c}{\lambda^{i}_{\rho}} \varphi_{j \lambda} \varphi^{j}_{\beta} - \frac{1}{8}(\Sigma_{\hat{a} \hat{b}})^{\rho \lambda} (\Gamma_{c})^{\alpha \beta} \boldsymbol{\lambda}_{i \alpha} {G}^{-3} \nabla^{c}{\lambda^{i}_{\beta}} \varphi_{j \rho} \varphi^{j}_{\lambda}+\frac{1}{8}(\Sigma_{\hat{a} \hat{b}})^{\rho \lambda} (\Gamma_{c})^{\alpha \beta} \boldsymbol{\lambda}_{j \alpha} {G}^{-3} \nabla^{c}{\lambda_{i \rho}} \varphi^{j}_{\lambda} \varphi^{i}_{\beta} - \frac{1}{8}(\Sigma_{\hat{a} \hat{b}})^{\rho \lambda} (\Gamma_{c})^{\alpha \beta} \boldsymbol{\lambda}_{j \alpha} {G}^{-3} \nabla^{c}{\lambda_{i \beta}} \varphi^{j}_{\rho} \varphi^{i}_{\lambda} - \frac{3}{8}(\Sigma_{\hat{a} \hat{b}})^{\alpha \beta} (\Sigma_{c d})^{\rho \lambda} W^{c d} \lambda_{i \rho} \boldsymbol{\lambda}^{i}_{\lambda} {G}^{-3} \varphi_{j \alpha} \varphi^{j}_{\beta} - \frac{3}{8}(\Sigma_{\hat{a} \hat{b}})^{\alpha \beta} (\Sigma_{c d})^{\rho \lambda} W^{c d} \lambda_{i \rho} \boldsymbol{\lambda}_{j \lambda} {G}^{-3} \varphi^{i}_{\alpha} \varphi^{j}_{\beta}+\frac{1}{2}G_{i j} W_{\hat{a}}\,^{c \alpha i} (\Sigma_{\hat{b} c})^{\beta \rho} \lambda_{k \beta} \boldsymbol{\lambda}^{k}_{\alpha} {G}^{-3} \varphi^{j}_{\rho} - \frac{9}{64}G_{i j} (\Gamma_{\hat{a}})^{\beta \alpha} W \boldsymbol{\lambda}^{i}_{\beta} X_{k \alpha} {G}^{-3} \nabla_{\hat{b}}{G^{j k}}+\frac{9}{64}G_{i j} (\Gamma_{\hat{a}})^{\beta \alpha} W \boldsymbol{\lambda}_{k \beta} X^{i}_{\alpha} {G}^{-3} \nabla_{\hat{b}}{G^{j k}} - \frac{9}{64}G_{i j} (\Gamma_{\hat{a}})^{\beta \alpha} W \boldsymbol{\lambda}_{k \beta} X^{k}_{\alpha} {G}^{-3} \nabla_{\hat{b}}{G^{i j}} - \frac{9}{64}G_{i j} (\Gamma_{\hat{a}})^{\beta \alpha} \boldsymbol{W} \lambda^{i}_{\beta} X_{k \alpha} {G}^{-3} \nabla_{\hat{b}}{G^{j k}}+\frac{9}{64}G_{i j} (\Gamma_{\hat{a}})^{\beta \alpha} \boldsymbol{W} \lambda_{k \beta} X^{i}_{\alpha} {G}^{-3} \nabla_{\hat{b}}{G^{j k}} - \frac{9}{64}G_{i j} (\Gamma_{\hat{a}})^{\beta \alpha} \boldsymbol{W} \lambda_{k \beta} X^{k}_{\alpha} {G}^{-3} \nabla_{\hat{b}}{G^{i j}} - \frac{3}{4}(\Sigma_{\hat{a} \hat{b}})^{\alpha \beta} W \mathbf{X}_{i j} {G}^{-5} \varphi^{i}_{\alpha} \varphi^{j \rho} \varphi_{k \beta} \varphi^{k}_{\rho}+\frac{3}{8}(\Sigma_{\hat{a} \hat{b}})^{\alpha \beta} W \mathbf{X}_{i j} {G}^{-5} \varphi^{i \rho} \varphi^{j}_{\rho} \varphi_{k \alpha} \varphi^{k}_{\beta} - \frac{3}{4}(\Sigma_{\hat{a} \hat{b}})^{\alpha \beta} \boldsymbol{W} X_{i j} {G}^{-5} \varphi^{i}_{\alpha} \varphi^{j \rho} \varphi_{k \beta} \varphi^{k}_{\rho}+\frac{3}{8}(\Sigma_{\hat{a} \hat{b}})^{\alpha \beta} \boldsymbol{W} X_{i j} {G}^{-5} \varphi^{i \rho} \varphi^{j}_{\rho} \varphi_{k \alpha} \varphi^{k}_{\beta} - \frac{3}{16}\epsilon^{e {e_{1}}}\,_{\hat{a}}\,^{c d} G_{i j} (\Sigma_{e {e_{1}}})^{\alpha \beta} W_{c d} \lambda^{i}_{\alpha} {G}^{-3} \nabla_{\hat{b}}{\boldsymbol{W}} \varphi^{j}_{\beta} - \frac{3}{16}\epsilon^{e {e_{1}}}\,_{\hat{a}}\,^{c d} G_{i j} (\Sigma_{e {e_{1}}})^{\alpha \beta} W_{c d} \boldsymbol{\lambda}^{i}_{\alpha} {G}^{-3} \nabla_{\hat{b}}{W} \varphi^{j}_{\beta} - \frac{3}{16}\epsilon^{e {e_{1}}}\,_{\hat{a}}\,^{c d} G_{i j} (\Sigma_{e {e_{1}}})^{\alpha \beta} W W_{c d} \boldsymbol{\lambda}^{i}_{\alpha} {G}^{-3} \nabla_{\hat{b}}{\varphi^{j}_{\beta}}%
 - \frac{3}{16}\epsilon^{e {e_{1}}}\,_{\hat{a}}\,^{c d} G_{i j} (\Sigma_{e {e_{1}}})^{\alpha \beta} W W_{c d} {G}^{-3} \nabla_{\hat{b}}{\boldsymbol{\lambda}^{i}_{\alpha}} \varphi^{j}_{\beta} - \frac{3}{16}\epsilon^{e {e_{1}}}\,_{\hat{a}}\,^{c d} G_{i j} (\Sigma_{e {e_{1}}})^{\alpha \beta} \boldsymbol{W} W_{c d} \lambda^{i}_{\alpha} {G}^{-3} \nabla_{\hat{b}}{\varphi^{j}_{\beta}} - \frac{3}{16}\epsilon^{e {e_{1}}}\,_{\hat{a}}\,^{c d} G_{i j} (\Sigma_{e {e_{1}}})^{\alpha \beta} \boldsymbol{W} W_{c d} {G}^{-3} \nabla_{\hat{b}}{\lambda^{i}_{\alpha}} \varphi^{j}_{\beta} - \frac{3}{8}\epsilon^{d e}\,_{\hat{a} {e_{1}}}\,^{c} G_{i j} (\Sigma_{d e})^{\alpha \beta} W_{\hat{b} c} \lambda^{i}_{\alpha} {G}^{-3} \nabla^{{e_{1}}}{\boldsymbol{W}} \varphi^{j}_{\beta} - \frac{3}{8}\epsilon^{d e}\,_{\hat{a} {e_{1}}}\,^{c} G_{i j} (\Sigma_{d e})^{\alpha \beta} W_{\hat{b} c} \boldsymbol{\lambda}^{i}_{\alpha} {G}^{-3} \nabla^{{e_{1}}}{W} \varphi^{j}_{\beta} - \frac{3}{8}\epsilon^{d e}\,_{\hat{a} {e_{1}}}\,^{c} G_{i j} (\Sigma_{d e})^{\alpha \beta} W W_{\hat{b} c} \boldsymbol{\lambda}^{i}_{\alpha} {G}^{-3} \nabla^{{e_{1}}}{\varphi^{j}_{\beta}} - \frac{3}{8}\epsilon^{d e}\,_{\hat{a} {e_{1}}}\,^{c} G_{i j} (\Sigma_{d e})^{\alpha \beta} W W_{\hat{b} c} {G}^{-3} \nabla^{{e_{1}}}{\boldsymbol{\lambda}^{i}_{\alpha}} \varphi^{j}_{\beta} - \frac{3}{8}\epsilon^{d e}\,_{\hat{a} {e_{1}}}\,^{c} G_{i j} (\Sigma_{d e})^{\alpha \beta} \boldsymbol{W} W_{\hat{b} c} \lambda^{i}_{\alpha} {G}^{-3} \nabla^{{e_{1}}}{\varphi^{j}_{\beta}} - \frac{3}{8}\epsilon^{d e}\,_{\hat{a} {e_{1}}}\,^{c} G_{i j} (\Sigma_{d e})^{\alpha \beta} \boldsymbol{W} W_{\hat{b} c} {G}^{-3} \nabla^{{e_{1}}}{\lambda^{i}_{\alpha}} \varphi^{j}_{\beta}+\frac{3}{16}\epsilon^{e}\,_{\hat{a} \hat{b}}\,^{c d} G_{i j} (\Sigma_{e {e_{1}}})^{\alpha \beta} W_{c d} \lambda^{i}_{\alpha} {G}^{-3} \nabla^{{e_{1}}}{\boldsymbol{W}} \varphi^{j}_{\beta}+\frac{3}{16}\epsilon^{e}\,_{\hat{a} \hat{b}}\,^{c d} G_{i j} (\Sigma_{e {e_{1}}})^{\alpha \beta} W_{c d} \boldsymbol{\lambda}^{i}_{\alpha} {G}^{-3} \nabla^{{e_{1}}}{W} \varphi^{j}_{\beta}+\frac{3}{16}\epsilon^{e}\,_{\hat{a} \hat{b}}\,^{c d} G_{i j} (\Sigma_{e {e_{1}}})^{\alpha \beta} W W_{c d} \boldsymbol{\lambda}^{i}_{\alpha} {G}^{-3} \nabla^{{e_{1}}}{\varphi^{j}_{\beta}}+\frac{3}{16}\epsilon^{e}\,_{\hat{a} \hat{b}}\,^{c d} G_{i j} (\Sigma_{e {e_{1}}})^{\alpha \beta} W W_{c d} {G}^{-3} \nabla^{{e_{1}}}{\boldsymbol{\lambda}^{i}_{\alpha}} \varphi^{j}_{\beta}+\frac{3}{16}\epsilon^{e}\,_{\hat{a} \hat{b}}\,^{c d} G_{i j} (\Sigma_{e {e_{1}}})^{\alpha \beta} \boldsymbol{W} W_{c d} \lambda^{i}_{\alpha} {G}^{-3} \nabla^{{e_{1}}}{\varphi^{j}_{\beta}}+\frac{3}{16}\epsilon^{e}\,_{\hat{a} \hat{b}}\,^{c d} G_{i j} (\Sigma_{e {e_{1}}})^{\alpha \beta} \boldsymbol{W} W_{c d} {G}^{-3} \nabla^{{e_{1}}}{\lambda^{i}_{\alpha}} \varphi^{j}_{\beta} - \frac{3}{32}\epsilon^{e {e_{1}}}\,_{\hat{a}}\,^{c d} (\Sigma_{e {e_{1}}})^{\alpha \beta} W W_{c d} \boldsymbol{\lambda}_{i \alpha} {G}^{-3} \nabla_{\hat{b}}{G^{i}\,_{j}} \varphi^{j}_{\beta} - \frac{3}{32}\epsilon^{e {e_{1}}}\,_{\hat{a}}\,^{c d} (\Sigma_{e {e_{1}}})^{\alpha \beta} \boldsymbol{W} W_{c d} \lambda_{i \alpha} {G}^{-3} \nabla_{\hat{b}}{G^{i}\,_{j}} \varphi^{j}_{\beta} - \frac{3}{16}\epsilon^{d e}\,_{\hat{a} {e_{1}}}\,^{c} (\Sigma_{d e})^{\alpha \beta} W W_{\hat{b} c} \boldsymbol{\lambda}_{i \alpha} {G}^{-3} \nabla^{{e_{1}}}{G^{i}\,_{j}} \varphi^{j}_{\beta} - \frac{3}{16}\epsilon^{d e}\,_{\hat{a} {e_{1}}}\,^{c} (\Sigma_{d e})^{\alpha \beta} \boldsymbol{W} W_{\hat{b} c} \lambda_{i \alpha} {G}^{-3} \nabla^{{e_{1}}}{G^{i}\,_{j}} \varphi^{j}_{\beta}+\frac{3}{32}\epsilon^{e}\,_{\hat{a} \hat{b}}\,^{c d} (\Sigma_{e {e_{1}}})^{\alpha \beta} W W_{c d} \boldsymbol{\lambda}_{i \alpha} {G}^{-3} \nabla^{{e_{1}}}{G^{i}\,_{j}} \varphi^{j}_{\beta}%
+\frac{3}{32}\epsilon^{e}\,_{\hat{a} \hat{b}}\,^{c d} (\Sigma_{e {e_{1}}})^{\alpha \beta} \boldsymbol{W} W_{c d} \lambda_{i \alpha} {G}^{-3} \nabla^{{e_{1}}}{G^{i}\,_{j}} \varphi^{j}_{\beta}+\frac{1}{128}\epsilon^{e}\,_{\hat{a} \hat{b} {e_{1}} d} (\Gamma^{{e_{1}}})^{\alpha \beta} W W^{c d} W_{e c \alpha i} \boldsymbol{\lambda}^{i}_{\beta} {G}^{-1}+\frac{1}{128}\epsilon^{e}\,_{\hat{a} \hat{b} {e_{1}} d} (\Gamma^{{e_{1}}})^{\alpha \beta} \boldsymbol{W} W^{c d} W_{e c \alpha i} \lambda^{i}_{\beta} {G}^{-1}+\frac{1}{384}\epsilon_{\hat{a} \hat{b} {e_{1}}}\,^{c e} (\Gamma^{{e_{1}}})^{\alpha \beta} W W_{c}\,^{d} W_{e d \alpha i} \boldsymbol{\lambda}^{i}_{\beta} {G}^{-1}+\frac{1}{384}\epsilon_{\hat{a} \hat{b} {e_{1}}}\,^{c e} (\Gamma^{{e_{1}}})^{\alpha \beta} \boldsymbol{W} W_{c}\,^{d} W_{e d \alpha i} \lambda^{i}_{\beta} {G}^{-1}+\frac{1}{32}{\rm i} G_{i j} (\Sigma_{\hat{a}}{}^{\, d})^{\alpha \beta} W \boldsymbol{W} W_{\hat{b}}\,^{c} W_{d c \alpha}\,^{i} {G}^{-3} \varphi^{j}_{\beta}+{\rm i} G_{i j} (\Sigma^{c d})^{\alpha \beta} W \boldsymbol{W} W_{\hat{a} c} W_{\hat{b} d \alpha}\,^{i} {G}^{-3} \varphi^{j}_{\beta} - \frac{1}{192}{\rm i} G_{i j} (\Sigma^{c d})^{\alpha \beta} W \boldsymbol{W} W_{\hat{a} \hat{b}} W_{c d \alpha}\,^{i} {G}^{-3} \varphi^{j}_{\beta}+\frac{23}{96}{\rm i} G_{i j} W \boldsymbol{W} W_{\hat{a} \hat{b}} W^{c d \alpha i} (\Sigma_{c d})_{\alpha}{}^{\beta} {G}^{-3} \varphi^{j}_{\beta} - \frac{23}{24}{\rm i} G_{i j} W \boldsymbol{W} W_{\hat{a}}\,^{c} W_{c}\,^{d \alpha i} (\Sigma_{\hat{b} d})_{\alpha}{}^{\beta} {G}^{-3} \varphi^{j}_{\beta}+\frac{1}{8}{\rm i} \epsilon^{c d e}\,_{\hat{b} {e_{1}}} G_{i j} (\Gamma^{{e_{1}}})^{\alpha \beta} W \mathbf{F}_{c d} W_{\hat{a} e \alpha}\,^{i} {G}^{-3} \varphi^{j}_{\beta}+\frac{1}{8}{\rm i} \epsilon^{c d e}\,_{\hat{b} {e_{1}}} G_{i j} (\Gamma^{{e_{1}}})^{\alpha \beta} \boldsymbol{W} F_{c d} W_{\hat{a} e \alpha}\,^{i} {G}^{-3} \varphi^{j}_{\beta} - \frac{1}{24}{\rm i} \epsilon^{e {e_{1}}}\,_{\hat{a} \hat{b}}\,^{c} G_{i j} (\Sigma_{e {e_{1}}})^{\alpha \beta} W \boldsymbol{W} {G}^{-3} \nabla^{d}{W_{c d \alpha}\,^{i}} \varphi^{j}_{\beta} - \frac{1}{24}{\rm i} \epsilon^{d e}\,_{\hat{a} {e_{1}}}\,^{c} G_{i j} (\Sigma_{d e})^{\alpha \beta} W \boldsymbol{W} {G}^{-3} \nabla^{{e_{1}}}{W_{\hat{b} c \alpha}\,^{i}} \varphi^{j}_{\beta} - \frac{1}{48}{\rm i} \epsilon^{e}\,_{\hat{a} \hat{b}}\,^{c d} G_{i j} (\Sigma_{e {e_{1}}})^{\alpha \beta} W \boldsymbol{W} {G}^{-3} \nabla^{{e_{1}}}{W_{c d \alpha}\,^{i}} \varphi^{j}_{\beta} - \frac{1}{16}{\rm i} \epsilon^{c e {e_{1}}}\,_{\hat{a} \hat{b}} G_{i j} W \mathbf{F}_{c d} W_{e {e_{1}}}\,^{\alpha i} (\Gamma^{d})_{\alpha}{}^{\beta} {G}^{-3} \varphi^{j}_{\beta} - \frac{1}{16}{\rm i} \epsilon^{c e {e_{1}}}\,_{\hat{a} \hat{b}} G_{i j} \boldsymbol{W} F_{c d} W_{e {e_{1}}}\,^{\alpha i} (\Gamma^{d})_{\alpha}{}^{\beta} {G}^{-3} \varphi^{j}_{\beta}+\frac{3}{64}{\rm i} \epsilon_{\hat{a} \hat{b} e}\,^{c d} (\Gamma^{e})^{\alpha \beta} X_{i j} \mathbf{F}_{c d} {G}^{-3} \varphi^{i}_{\alpha} \varphi^{j}_{\beta}+\frac{3}{64}{\rm i} \epsilon_{\hat{a} \hat{b} e}\,^{c d} (\Gamma^{e})^{\alpha \beta} \mathbf{X}_{i j} F_{c d} {G}^{-3} \varphi^{i}_{\alpha} \varphi^{j}_{\beta} - \frac{3}{4}G_{i j} G_{k l} (\Gamma^{c})^{\alpha \beta} F_{\hat{b} c} \boldsymbol{\lambda}^{i}_{\alpha} {G}^{-5} \nabla_{\hat{a}}{G^{j k}} \varphi^{l}_{\beta}%
 - \frac{3}{4}G_{i j} G_{k l} (\Gamma^{c})^{\alpha \beta} \mathbf{F}_{\hat{b} c} \lambda^{i}_{\alpha} {G}^{-5} \nabla_{\hat{a}}{G^{j k}} \varphi^{l}_{\beta}+3G_{i j} G_{k l} (\Sigma_{\hat{a} c})^{\alpha \beta} W {G}^{-5} \nabla^{c}{\boldsymbol{\lambda}^{i}_{\alpha}} \nabla_{\hat{b}}{G^{j k}} \varphi^{l}_{\beta}+3G_{i j} G_{k l} (\Sigma_{\hat{a} c})^{\alpha \beta} \boldsymbol{W} {G}^{-5} \nabla^{c}{\lambda^{i}_{\alpha}} \nabla_{\hat{b}}{G^{j k}} \varphi^{l}_{\beta}+\frac{3}{2}G_{i j} G_{k l} (\Sigma_{\hat{a} c})^{\alpha \beta} \lambda^{i}_{\alpha} {G}^{-5} \nabla^{c}{\boldsymbol{W}} \nabla_{\hat{b}}{G^{j k}} \varphi^{l}_{\beta}+\frac{3}{2}G_{i j} G_{k l} (\Sigma_{\hat{a} c})^{\alpha \beta} \boldsymbol{\lambda}^{i}_{\alpha} {G}^{-5} \nabla^{c}{W} \nabla_{\hat{b}}{G^{j k}} \varphi^{l}_{\beta}+\frac{9}{32}\mathcal{H}_{\hat{a}} G_{i j} G_{k l} (\Gamma_{\hat{b}})^{\alpha \beta} X^{i j} \boldsymbol{\lambda}^{k}_{\alpha} {G}^{-5} \varphi^{l}_{\beta}+\frac{9}{32}\mathcal{H}_{\hat{a}} G_{i j} G_{k l} (\Gamma_{\hat{b}})^{\alpha \beta} \mathbf{X}^{i j} \lambda^{k}_{\alpha} {G}^{-5} \varphi^{l}_{\beta} - \frac{9}{32}\mathcal{H}_{\hat{a}} G_{i j} G_{k l} (\Gamma_{\hat{b}})^{\alpha \beta} X^{i k} \boldsymbol{\lambda}^{j}_{\alpha} {G}^{-5} \varphi^{l}_{\beta} - \frac{9}{32}\mathcal{H}_{\hat{a}} G_{i j} G_{k l} (\Gamma_{\hat{b}})^{\alpha \beta} \mathbf{X}^{i k} \lambda^{j}_{\alpha} {G}^{-5} \varphi^{l}_{\beta}+\frac{9}{8}G_{i j} (\Sigma_{\hat{a} \hat{b}})^{\alpha \beta} W \boldsymbol{W} X^{i}_{\alpha} {G}^{-5} \varphi^{j \rho} \varphi_{k \beta} \varphi^{k}_{\rho} - \frac{9}{8}G_{i j} (\Sigma_{\hat{a} \hat{b}})^{\beta \rho} W \boldsymbol{W} X^{i \alpha} {G}^{-5} \varphi^{j}_{\beta} \varphi_{k \rho} \varphi^{k}_{\alpha} - \frac{9}{8}G_{i j} (\Sigma_{\hat{a} \hat{b}})^{\beta \rho} W \boldsymbol{W} X^{i \alpha} {G}^{-5} \varphi^{j}_{\alpha} \varphi_{k \beta} \varphi^{k}_{\rho}+\frac{9}{8}G_{i j} (\Sigma_{\hat{a} \hat{b}})^{\alpha \beta} W \boldsymbol{W} X_{k \alpha} {G}^{-5} \varphi^{i}_{\beta} \varphi^{j \rho} \varphi^{k}_{\rho} - \frac{9}{8}G_{i j} (\Sigma_{\hat{a} \hat{b}})^{\alpha \beta} W \boldsymbol{W} X_{k \alpha} {G}^{-5} \varphi^{i \rho} \varphi^{j}_{\rho} \varphi^{k}_{\beta} - \frac{9}{8}G_{i j} (\Sigma_{\hat{a} \hat{b}})^{\beta \rho} W \boldsymbol{W} X_{k}^{\alpha} {G}^{-5} \varphi^{i}_{\beta} \varphi^{j}_{\alpha} \varphi^{k}_{\rho}+\frac{3}{16}\epsilon^{c d}\,_{\hat{a} \hat{b} e} G_{i j} (\Sigma_{c d})^{\alpha \beta} X^{i j} \boldsymbol{\lambda}_{k \alpha} {G}^{-3} \nabla^{e}{\varphi^{k}_{\beta}}+\frac{3}{32}\epsilon^{c d}\,_{\hat{a} \hat{b} e} G_{i j} (\Sigma_{c d})^{\alpha \beta} X^{i j} {G}^{-3} \nabla^{e}{\boldsymbol{\lambda}_{k \alpha}} \varphi^{k}_{\beta}+\frac{3}{16}\epsilon^{c d}\,_{\hat{a} \hat{b} e} G_{i j} (\Sigma_{c d})^{\alpha \beta} \mathbf{X}^{i j} \lambda_{k \alpha} {G}^{-3} \nabla^{e}{\varphi^{k}_{\beta}}+\frac{3}{32}\epsilon^{c d}\,_{\hat{a} \hat{b} e} G_{i j} (\Sigma_{c d})^{\alpha \beta} \mathbf{X}^{i j} {G}^{-3} \nabla^{e}{\lambda_{k \alpha}} \varphi^{k}_{\beta} - \frac{3}{16}\epsilon^{c d}\,_{\hat{a} \hat{b} e} G_{i j} (\Sigma_{c d})^{\alpha \beta} X^{i}\,_{k} \boldsymbol{\lambda}^{j}_{\alpha} {G}^{-3} \nabla^{e}{\varphi^{k}_{\beta}}%
+\frac{3}{16}\epsilon^{c d}\,_{\hat{a} \hat{b} e} G_{i j} (\Sigma_{c d})^{\alpha \beta} X^{i}\,_{k} \boldsymbol{\lambda}^{k}_{\alpha} {G}^{-3} \nabla^{e}{\varphi^{j}_{\beta}} - \frac{3}{32}\epsilon^{c d}\,_{\hat{a} \hat{b} e} G_{i j} (\Sigma_{c d})^{\alpha \beta} X^{i}\,_{k} {G}^{-3} \nabla^{e}{\boldsymbol{\lambda}^{j}_{\alpha}} \varphi^{k}_{\beta}+\frac{3}{32}\epsilon^{c d}\,_{\hat{a} \hat{b} e} G_{i j} (\Sigma_{c d})^{\alpha \beta} X^{i}\,_{k} {G}^{-3} \nabla^{e}{\boldsymbol{\lambda}^{k}_{\alpha}} \varphi^{j}_{\beta} - \frac{3}{16}\epsilon^{c d}\,_{\hat{a} \hat{b} e} G_{i j} (\Sigma_{c d})^{\alpha \beta} \mathbf{X}^{i}\,_{k} \lambda^{j}_{\alpha} {G}^{-3} \nabla^{e}{\varphi^{k}_{\beta}}+\frac{3}{16}\epsilon^{c d}\,_{\hat{a} \hat{b} e} G_{i j} (\Sigma_{c d})^{\alpha \beta} \mathbf{X}^{i}\,_{k} \lambda^{k}_{\alpha} {G}^{-3} \nabla^{e}{\varphi^{j}_{\beta}} - \frac{3}{32}\epsilon^{c d}\,_{\hat{a} \hat{b} e} G_{i j} (\Sigma_{c d})^{\alpha \beta} \mathbf{X}^{i}\,_{k} {G}^{-3} \nabla^{e}{\lambda^{j}_{\alpha}} \varphi^{k}_{\beta}+\frac{3}{32}\epsilon^{c d}\,_{\hat{a} \hat{b} e} G_{i j} (\Sigma_{c d})^{\alpha \beta} \mathbf{X}^{i}\,_{k} {G}^{-3} \nabla^{e}{\lambda^{k}_{\alpha}} \varphi^{j}_{\beta}+\frac{3}{32}\epsilon^{c d}\,_{\hat{a} \hat{b} e} (\Sigma_{c d})^{\alpha \beta} X_{k i} \boldsymbol{\lambda}^{k}_{\alpha} {G}^{-3} \nabla^{e}{G^{i}\,_{j}} \varphi^{j}_{\beta} - \frac{3}{32}\epsilon^{c d}\,_{\hat{a} \hat{b} e} (\Sigma_{c d})^{\alpha \beta} X_{i j} \boldsymbol{\lambda}_{k \alpha} {G}^{-3} \nabla^{e}{G^{i j}} \varphi^{k}_{\beta}+\frac{3}{32}\epsilon^{c d}\,_{\hat{a} \hat{b} e} (\Sigma_{c d})^{\alpha \beta} X_{i k} \boldsymbol{\lambda}_{j \alpha} {G}^{-3} \nabla^{e}{G^{i j}} \varphi^{k}_{\beta}+\frac{3}{32}\epsilon^{c d}\,_{\hat{a} \hat{b} e} (\Sigma_{c d})^{\alpha \beta} \mathbf{X}_{k i} \lambda^{k}_{\alpha} {G}^{-3} \nabla^{e}{G^{i}\,_{j}} \varphi^{j}_{\beta} - \frac{3}{32}\epsilon^{c d}\,_{\hat{a} \hat{b} e} (\Sigma_{c d})^{\alpha \beta} \mathbf{X}_{i j} \lambda_{k \alpha} {G}^{-3} \nabla^{e}{G^{i j}} \varphi^{k}_{\beta}+\frac{3}{32}\epsilon^{c d}\,_{\hat{a} \hat{b} e} (\Sigma_{c d})^{\alpha \beta} \mathbf{X}_{i k} \lambda_{j \alpha} {G}^{-3} \nabla^{e}{G^{i j}} \varphi^{k}_{\beta} - \frac{1}{2}\epsilon^{d e}\,_{\hat{a} {e_{1}}}\,^{c} G_{i j} (\Sigma_{d e})^{\alpha \beta} \lambda^{i}_{\alpha} {G}^{-3} \nabla^{{e_{1}}}{\mathbf{F}_{\hat{b} c}} \varphi^{j}_{\beta} - \frac{1}{2}\epsilon^{d e}\,_{\hat{a} {e_{1}}}\,^{c} G_{i j} (\Sigma_{d e})^{\alpha \beta} \boldsymbol{\lambda}^{i}_{\alpha} {G}^{-3} \nabla^{{e_{1}}}{F_{\hat{b} c}} \varphi^{j}_{\beta}+\frac{1}{4}\epsilon^{e}\,_{\hat{a} \hat{b} {e_{1}}}\,^{c} G_{i j} (\Sigma_{e}{}^{\, d})^{\alpha \beta} F_{c d} \boldsymbol{\lambda}^{i}_{\alpha} {G}^{-3} \nabla^{{e_{1}}}{\varphi^{j}_{\beta}}+\frac{1}{4}\epsilon^{e}\,_{\hat{a} \hat{b} {e_{1}}}\,^{c} G_{i j} (\Sigma_{e}{}^{\, d})^{\alpha \beta} F_{c d} {G}^{-3} \nabla^{{e_{1}}}{\boldsymbol{\lambda}^{i}_{\alpha}} \varphi^{j}_{\beta}+\frac{1}{4}\epsilon^{e}\,_{\hat{a} \hat{b} {e_{1}}}\,^{c} G_{i j} (\Sigma_{e}{}^{\, d})^{\alpha \beta} \mathbf{F}_{c d} \lambda^{i}_{\alpha} {G}^{-3} \nabla^{{e_{1}}}{\varphi^{j}_{\beta}}+\frac{1}{4}\epsilon^{e}\,_{\hat{a} \hat{b} {e_{1}}}\,^{c} G_{i j} (\Sigma_{e}{}^{\, d})^{\alpha \beta} \mathbf{F}_{c d} {G}^{-3} \nabla^{{e_{1}}}{\lambda^{i}_{\alpha}} \varphi^{j}_{\beta} - \frac{1}{16}\epsilon^{c d}\,_{\hat{a} \hat{b} e} (\Sigma_{c d})^{\alpha \beta} \lambda_{i \alpha} {G}^{-3} \nabla^{e}{\boldsymbol{\lambda}^{i \rho}} \varphi_{j \beta} \varphi^{j}_{\rho}%
 - \frac{1}{16}\epsilon^{c d}\,_{\hat{a} \hat{b} e} (\Sigma_{c d})^{\alpha \beta} \lambda_{i \alpha} {G}^{-3} \nabla^{e}{\boldsymbol{\lambda}_{j}^{\rho}} \varphi^{i}_{\beta} \varphi^{j}_{\rho} - \frac{1}{16}\epsilon^{c d}\,_{\hat{a} \hat{b} e} (\Sigma_{c d})^{\alpha \beta} \boldsymbol{\lambda}_{i \alpha} {G}^{-3} \nabla^{e}{\lambda^{i \rho}} \varphi_{j \beta} \varphi^{j}_{\rho} - \frac{1}{16}\epsilon^{c d}\,_{\hat{a} \hat{b} e} (\Sigma_{c d})^{\alpha \beta} \boldsymbol{\lambda}_{j \alpha} {G}^{-3} \nabla^{e}{\lambda^{\rho}_{i}} \varphi^{j}_{\beta} \varphi^{i}_{\rho} - \frac{3}{32}\epsilon^{c d}\,_{\hat{a} \hat{b} e} (\Sigma_{c d})^{\alpha \beta} \lambda^{\rho}_{i} {G}^{-3} \nabla^{e}{\boldsymbol{\lambda}^{i}_{\alpha}} \varphi_{j \beta} \varphi^{j}_{\rho}+\frac{1}{16}\epsilon^{c d}\,_{\hat{a} \hat{b} e} (\Sigma_{c d})^{\alpha \beta} \lambda^{\rho}_{i} {G}^{-3} \nabla^{e}{\boldsymbol{\lambda}^{i}_{\rho}} \varphi_{j \alpha} \varphi^{j}_{\beta}+\frac{1}{32}\epsilon^{c d}\,_{\hat{a} \hat{b} e} (\Sigma_{c d})^{\alpha \beta} \lambda^{\rho}_{i} {G}^{-3} \nabla^{e}{\boldsymbol{\lambda}_{j \alpha}} \varphi^{i}_{\beta} \varphi^{j}_{\rho} - \frac{1}{8}\epsilon^{c d}\,_{\hat{a} \hat{b} e} (\Sigma_{c d})^{\alpha \beta} \lambda^{\rho}_{i} {G}^{-3} \nabla^{e}{\boldsymbol{\lambda}_{j \alpha}} \varphi^{i}_{\rho} \varphi^{j}_{\beta}+\frac{1}{16}\epsilon^{c d}\,_{\hat{a} \hat{b} e} (\Sigma_{c d})^{\alpha \beta} \lambda^{\rho}_{i} {G}^{-3} \nabla^{e}{\boldsymbol{\lambda}_{j \rho}} \varphi^{i}_{\alpha} \varphi^{j}_{\beta} - \frac{3}{32}\epsilon^{c d}\,_{\hat{a} \hat{b} e} (\Sigma_{c d})^{\alpha \beta} \boldsymbol{\lambda}_{i}^{\rho} {G}^{-3} \nabla^{e}{\lambda^{i}_{\alpha}} \varphi_{j \beta} \varphi^{j}_{\rho}+\frac{1}{16}\epsilon^{c d}\,_{\hat{a} \hat{b} e} (\Sigma_{c d})^{\alpha \beta} \boldsymbol{\lambda}_{i}^{\rho} {G}^{-3} \nabla^{e}{\lambda^{i}_{\rho}} \varphi_{j \alpha} \varphi^{j}_{\beta}+\frac{1}{32}\epsilon^{c d}\,_{\hat{a} \hat{b} e} (\Sigma_{c d})^{\alpha \beta} \boldsymbol{\lambda}_{j}^{\rho} {G}^{-3} \nabla^{e}{\lambda_{i \alpha}} \varphi^{j}_{\beta} \varphi^{i}_{\rho} - \frac{1}{8}\epsilon^{c d}\,_{\hat{a} \hat{b} e} (\Sigma_{c d})^{\alpha \beta} \boldsymbol{\lambda}_{j}^{\rho} {G}^{-3} \nabla^{e}{\lambda_{i \alpha}} \varphi^{j}_{\rho} \varphi^{i}_{\beta}+\frac{1}{16}\epsilon^{c d}\,_{\hat{a} \hat{b} e} (\Sigma_{c d})^{\alpha \beta} \boldsymbol{\lambda}_{j}^{\rho} {G}^{-3} \nabla^{e}{\lambda_{i \rho}} \varphi^{j}_{\alpha} \varphi^{i}_{\beta}+\frac{1}{8}\epsilon^{e}\,_{\hat{a} \hat{b} {e_{1}}}\,^{c} (\Sigma_{e}{}^{\, d})^{\alpha \beta} F_{c d} \boldsymbol{\lambda}_{i \alpha} {G}^{-3} \nabla^{{e_{1}}}{G^{i}\,_{j}} \varphi^{j}_{\beta}+\frac{1}{8}\epsilon^{e}\,_{\hat{a} \hat{b} {e_{1}}}\,^{c} (\Sigma_{e}{}^{\, d})^{\alpha \beta} \mathbf{F}_{c d} \lambda_{i \alpha} {G}^{-3} \nabla^{{e_{1}}}{G^{i}\,_{j}} \varphi^{j}_{\beta} - \frac{3}{8}{\rm i} G_{i j} G_{k l} (\Sigma_{\hat{a} \hat{b}})^{\alpha \beta} X^{i j} \mathbf{X}^{k l} {G}^{-5} \varphi_{m \alpha} \varphi^{m}_{\beta}+\frac{15}{16}{\rm i} G_{i j} G_{k l} (\Sigma_{\hat{a} \hat{b}})^{\alpha \beta} X^{i j} \mathbf{X}^{k}\,_{m} {G}^{-5} \varphi^{l}_{\alpha} \varphi^{m}_{\beta}+\frac{3}{8}{\rm i} G_{i j} G_{k l} (\Sigma_{\hat{a} \hat{b}})^{\alpha \beta} X^{i k} \mathbf{X}^{j l} {G}^{-5} \varphi_{m \alpha} \varphi^{m}_{\beta} - \frac{15}{16}{\rm i} G_{i j} G_{k l} (\Sigma_{\hat{a} \hat{b}})^{\alpha \beta} X^{i k} \mathbf{X}^{j}\,_{m} {G}^{-5} \varphi^{l}_{\alpha} \varphi^{m}_{\beta} - \frac{15}{16}{\rm i} G_{i j} G_{k l} (\Sigma_{\hat{a} \hat{b}})^{\alpha \beta} X^{i}\,_{m} \mathbf{X}^{j k} {G}^{-5} \varphi^{l}_{\alpha} \varphi^{m}_{\beta}%
+\frac{15}{16}{\rm i} G_{i j} G_{k l} (\Sigma_{\hat{a} \hat{b}})^{\alpha \beta} X^{i}\,_{m} \mathbf{X}^{k l} {G}^{-5} \varphi^{j}_{\alpha} \varphi^{m}_{\beta} - \frac{9}{8}{\rm i} G_{i j} G_{k l} (\Sigma_{\hat{a} \hat{b}})^{\alpha \beta} X^{i}\,_{m} \mathbf{X}^{k m} {G}^{-5} \varphi^{j}_{\alpha} \varphi^{l}_{\beta}+\frac{3}{4}{\rm i} G_{i j} G_{k l} \lambda^{i \alpha} \boldsymbol{\lambda}^{j}_{\alpha} {G}^{-5} \nabla_{\hat{a}}{G^{k}\,_{m}} \nabla_{\hat{b}}{G^{l m}} - \frac{3}{8}{\rm i} G_{i j} (\Sigma_{\hat{a} \hat{b}})^{\alpha \beta} X^{i j} \boldsymbol{\lambda}_{k \alpha} {G}^{-5} \varphi^{k \rho} \varphi_{l \beta} \varphi^{l}_{\rho} - \frac{3}{16}{\rm i} G_{i j} (\Sigma_{\hat{a} \hat{b}})^{\alpha \beta} X^{i j} \boldsymbol{\lambda}_{k}^{\rho} {G}^{-5} \varphi^{k}_{\rho} \varphi_{l \alpha} \varphi^{l}_{\beta} - \frac{3}{8}{\rm i} G_{i j} (\Sigma_{\hat{a} \hat{b}})^{\alpha \beta} \mathbf{X}^{i j} \lambda_{k \alpha} {G}^{-5} \varphi^{k \rho} \varphi_{l \beta} \varphi^{l}_{\rho} - \frac{3}{16}{\rm i} G_{i j} (\Sigma_{\hat{a} \hat{b}})^{\alpha \beta} \mathbf{X}^{i j} \lambda^{\rho}_{k} {G}^{-5} \varphi^{k}_{\rho} \varphi_{l \alpha} \varphi^{l}_{\beta}+\frac{3}{8}{\rm i} G_{i j} (\Sigma_{\hat{a} \hat{b}})^{\alpha \beta} X^{i}\,_{k} \boldsymbol{\lambda}^{j}_{\alpha} {G}^{-5} \varphi^{k \rho} \varphi_{l \beta} \varphi^{l}_{\rho}+\frac{3}{16}{\rm i} G_{i j} (\Sigma_{\hat{a} \hat{b}})^{\alpha \beta} X^{i}\,_{k} \boldsymbol{\lambda}^{j \rho} {G}^{-5} \varphi^{k}_{\rho} \varphi_{l \alpha} \varphi^{l}_{\beta} - \frac{9}{16}{\rm i} G_{i j} (\Sigma_{\hat{a} \hat{b}})^{\alpha \beta} X^{i}\,_{k} \boldsymbol{\lambda}^{k \rho} {G}^{-5} \varphi^{j}_{\alpha} \varphi_{l \beta} \varphi^{l}_{\rho} - \frac{9}{16}{\rm i} G_{i j} (\Sigma_{\hat{a} \hat{b}})^{\alpha \beta} X^{i}\,_{k} \boldsymbol{\lambda}^{k \rho} {G}^{-5} \varphi^{j}_{\rho} \varphi_{l \alpha} \varphi^{l}_{\beta} - \frac{3}{8}{\rm i} G_{i j} (\Sigma_{\hat{a} \hat{b}})^{\alpha \beta} X^{i}\,_{k} \boldsymbol{\lambda}_{l \alpha} {G}^{-5} \varphi^{j}_{\beta} \varphi^{k \rho} \varphi^{l}_{\rho}+\frac{3}{8}{\rm i} G_{i j} (\Sigma_{\hat{a} \hat{b}})^{\alpha \beta} X^{i}\,_{k} \boldsymbol{\lambda}_{l}^{\rho} {G}^{-5} \varphi^{j}_{\alpha} \varphi^{k}_{\beta} \varphi^{l}_{\rho}+\frac{3}{16}{\rm i} G_{i j} (\Sigma_{\hat{a} \hat{b}})^{\alpha \beta} X^{i}\,_{k} \boldsymbol{\lambda}_{l}^{\rho} {G}^{-5} \varphi^{j}_{\alpha} \varphi^{k}_{\rho} \varphi^{l}_{\beta}+\frac{3}{8}{\rm i} G_{i j} (\Sigma_{\hat{a} \hat{b}})^{\alpha \beta} \mathbf{X}^{i}\,_{k} \lambda^{j}_{\alpha} {G}^{-5} \varphi^{k \rho} \varphi_{l \beta} \varphi^{l}_{\rho}+\frac{3}{16}{\rm i} G_{i j} (\Sigma_{\hat{a} \hat{b}})^{\alpha \beta} \mathbf{X}^{i}\,_{k} \lambda^{j \rho} {G}^{-5} \varphi^{k}_{\rho} \varphi_{l \alpha} \varphi^{l}_{\beta} - \frac{9}{16}{\rm i} G_{i j} (\Sigma_{\hat{a} \hat{b}})^{\alpha \beta} \mathbf{X}^{i}\,_{k} \lambda^{k \rho} {G}^{-5} \varphi^{j}_{\alpha} \varphi_{l \beta} \varphi^{l}_{\rho} - \frac{9}{16}{\rm i} G_{i j} (\Sigma_{\hat{a} \hat{b}})^{\alpha \beta} \mathbf{X}^{i}\,_{k} \lambda^{k \rho} {G}^{-5} \varphi^{j}_{\rho} \varphi_{l \alpha} \varphi^{l}_{\beta} - \frac{3}{8}{\rm i} G_{i j} (\Sigma_{\hat{a} \hat{b}})^{\alpha \beta} \mathbf{X}^{i}\,_{k} \lambda_{l \alpha} {G}^{-5} \varphi^{j}_{\beta} \varphi^{k \rho} \varphi^{l}_{\rho}+\frac{3}{8}{\rm i} G_{i j} (\Sigma_{\hat{a} \hat{b}})^{\alpha \beta} \mathbf{X}^{i}\,_{k} \lambda^{\rho}_{l} {G}^{-5} \varphi^{j}_{\alpha} \varphi^{k}_{\beta} \varphi^{l}_{\rho}%
+\frac{3}{16}{\rm i} G_{i j} (\Sigma_{\hat{a} \hat{b}})^{\alpha \beta} \mathbf{X}^{i}\,_{k} \lambda^{\rho}_{l} {G}^{-5} \varphi^{j}_{\alpha} \varphi^{k}_{\rho} \varphi^{l}_{\beta}+\frac{3}{8}{\rm i} G_{i j} (\Sigma_{\hat{a} \hat{b}})^{\alpha \beta} X_{k l} \boldsymbol{\lambda}^{i}_{\alpha} {G}^{-5} \varphi^{j}_{\beta} \varphi^{k \rho} \varphi^{l}_{\rho} - \frac{3}{16}{\rm i} G_{i j} (\Sigma_{\hat{a} \hat{b}})^{\alpha \beta} X_{k l} \boldsymbol{\lambda}^{i \rho} {G}^{-5} \varphi^{j}_{\alpha} \varphi^{k}_{\beta} \varphi^{l}_{\rho}+\frac{9}{16}{\rm i} G_{i j} (\Sigma_{\hat{a} \hat{b}})^{\alpha \beta} X_{k l} \boldsymbol{\lambda}^{k}_{\alpha} {G}^{-5} \varphi^{i \rho} \varphi^{j}_{\rho} \varphi^{l}_{\beta} - \frac{9}{16}{\rm i} G_{i j} (\Sigma_{\hat{a} \hat{b}})^{\alpha \beta} X_{k l} \boldsymbol{\lambda}^{k \rho} {G}^{-5} \varphi^{i}_{\alpha} \varphi^{j}_{\rho} \varphi^{l}_{\beta}+\frac{3}{8}{\rm i} G_{i j} (\Sigma_{\hat{a} \hat{b}})^{\alpha \beta} \mathbf{X}_{k l} \lambda^{i}_{\alpha} {G}^{-5} \varphi^{j}_{\beta} \varphi^{k \rho} \varphi^{l}_{\rho} - \frac{3}{16}{\rm i} G_{i j} (\Sigma_{\hat{a} \hat{b}})^{\alpha \beta} \mathbf{X}_{k l} \lambda^{i \rho} {G}^{-5} \varphi^{j}_{\alpha} \varphi^{k}_{\beta} \varphi^{l}_{\rho}+\frac{9}{16}{\rm i} G_{i j} (\Sigma_{\hat{a} \hat{b}})^{\alpha \beta} \mathbf{X}_{k l} \lambda^{k}_{\alpha} {G}^{-5} \varphi^{i \rho} \varphi^{j}_{\rho} \varphi^{l}_{\beta} - \frac{9}{16}{\rm i} G_{i j} (\Sigma_{\hat{a} \hat{b}})^{\alpha \beta} \mathbf{X}_{k l} \lambda^{k \rho} {G}^{-5} \varphi^{i}_{\alpha} \varphi^{j}_{\rho} \varphi^{l}_{\beta}+\frac{1}{384}{\rm i} G_{i j} (\Gamma^{c})^{\alpha \beta} W \boldsymbol{W} W_{c d} W_{e {e_{1}} \alpha}\,^{i} \epsilon_{\hat{a}}\,^{d e {e_{1}}}\,_{\hat{b}} {G}^{-3} \varphi^{j}_{\beta}+\frac{1}{768}{\rm i} G_{i j} (\Gamma_{\hat{a}})^{\alpha \beta} W \boldsymbol{W} W_{c d} W_{e {e_{1}} \alpha}\,^{i} \epsilon^{c d e {e_{1}}}\,_{\hat{b}} {G}^{-3} \varphi^{j}_{\beta}+\frac{1}{384}{\rm i} G_{i j} (\Gamma_{{e_{1}}})^{\alpha \beta} W \boldsymbol{W} W_{\hat{b} c} W_{d e \alpha}\,^{i} \epsilon_{\hat{a}}\,^{{e_{1}} c d e} {G}^{-3} \varphi^{j}_{\beta} - \frac{3}{4}{\rm i} G_{i j} (\Sigma_{\hat{a}}{}^{\, c})^{\alpha \beta} F_{\hat{b} c} \boldsymbol{\lambda}^{i}_{\alpha} {G}^{-5} \varphi^{j \rho} \varphi_{k \beta} \varphi^{k}_{\rho} - \frac{3}{4}{\rm i} G_{i j} (\Sigma_{\hat{a}}{}^{\, c})^{\alpha \beta} F_{\hat{b} c} \boldsymbol{\lambda}_{k \alpha} {G}^{-5} \varphi^{i}_{\beta} \varphi^{j \rho} \varphi^{k}_{\rho}+\frac{3}{4}{\rm i} G_{i j} (\Sigma_{\hat{a}}{}^{\, c})^{\alpha \beta} F_{\hat{b} c} \boldsymbol{\lambda}_{k \alpha} {G}^{-5} \varphi^{i \rho} \varphi^{j}_{\rho} \varphi^{k}_{\beta} - \frac{3}{4}{\rm i} G_{i j} (\Sigma_{\hat{a}}{}^{\, c})^{\alpha \beta} \mathbf{F}_{\hat{b} c} \lambda^{i}_{\alpha} {G}^{-5} \varphi^{j \rho} \varphi_{k \beta} \varphi^{k}_{\rho} - \frac{3}{4}{\rm i} G_{i j} (\Sigma_{\hat{a}}{}^{\, c})^{\alpha \beta} \mathbf{F}_{\hat{b} c} \lambda_{k \alpha} {G}^{-5} \varphi^{i}_{\beta} \varphi^{j \rho} \varphi^{k}_{\rho}+\frac{3}{4}{\rm i} G_{i j} (\Sigma_{\hat{a}}{}^{\, c})^{\alpha \beta} \mathbf{F}_{\hat{b} c} \lambda_{k \alpha} {G}^{-5} \varphi^{i \rho} \varphi^{j}_{\rho} \varphi^{k}_{\beta}+\frac{3}{8}{\rm i} (\Sigma_{\hat{a} \hat{b}})^{\alpha \beta} \lambda^{\rho}_{i} \boldsymbol{\lambda}_{j \rho} {G}^{-5} \varphi^{i}_{\alpha} \varphi^{j \lambda} \varphi_{k \beta} \varphi^{k}_{\lambda} - \frac{3}{8}{\rm i} (\Sigma_{\hat{a} \hat{b}})^{\alpha \beta} \lambda^{\rho}_{i} \boldsymbol{\lambda}_{j \rho} {G}^{-5} \varphi^{i \lambda} \varphi^{j}_{\alpha} \varphi_{k \beta} \varphi^{k}_{\lambda}%
 - \frac{3}{8}{\rm i} (\Sigma_{\hat{a} \hat{b}})^{\alpha \beta} \lambda^{\rho}_{i} \boldsymbol{\lambda}_{j \rho} {G}^{-5} \varphi^{i \lambda} \varphi^{j}_{\lambda} \varphi_{k \alpha} \varphi^{k}_{\beta}+\frac{3}{32}{\rm i} \epsilon^{d e}\,_{\hat{a} {e_{1}}}\,^{c} G_{i j} (\Sigma_{d e})^{\alpha \beta} W_{\hat{b} c} \lambda^{i}_{\alpha} \boldsymbol{\lambda}_{k \beta} {G}^{-3} \nabla^{{e_{1}}}{G^{j k}} - \frac{3}{32}{\rm i} \epsilon^{d e}\,_{\hat{a} {e_{1}}}\,^{c} G_{i j} (\Sigma_{d e})^{\alpha \beta} W_{\hat{b} c} \lambda_{k \alpha} \boldsymbol{\lambda}^{i}_{\beta} {G}^{-3} \nabla^{{e_{1}}}{G^{j k}}+\frac{3}{32}{\rm i} \epsilon^{d e}\,_{\hat{a} {e_{1}}}\,^{c} G_{i j} (\Sigma_{d e})^{\alpha \beta} W_{\hat{b} c} \lambda_{k \alpha} \boldsymbol{\lambda}^{k}_{\beta} {G}^{-3} \nabla^{{e_{1}}}{G^{i j}} - \frac{3}{64}{\rm i} \epsilon^{e}\,_{\hat{a} \hat{b}}\,^{c d} G_{i j} (\Sigma_{e {e_{1}}})^{\alpha \beta} W_{c d} \lambda^{i}_{\alpha} \boldsymbol{\lambda}_{k \beta} {G}^{-3} \nabla^{{e_{1}}}{G^{j k}}+\frac{3}{64}{\rm i} \epsilon^{e}\,_{\hat{a} \hat{b}}\,^{c d} G_{i j} (\Sigma_{e {e_{1}}})^{\alpha \beta} W_{c d} \lambda_{k \alpha} \boldsymbol{\lambda}^{i}_{\beta} {G}^{-3} \nabla^{{e_{1}}}{G^{j k}} - \frac{3}{64}{\rm i} \epsilon^{e}\,_{\hat{a} \hat{b}}\,^{c d} G_{i j} (\Sigma_{e {e_{1}}})^{\alpha \beta} W_{c d} \lambda_{k \alpha} \boldsymbol{\lambda}^{k}_{\beta} {G}^{-3} \nabla^{{e_{1}}}{G^{i j}}+\frac{1}{384}{\rm i} \epsilon^{c d e}\,_{\hat{b} {e_{1}}} G_{i j} (\Gamma^{{e_{1}}})^{\alpha \beta} W \boldsymbol{W} W_{\hat{a} c} W_{d e \alpha}\,^{i} {G}^{-3} \varphi^{j}_{\beta}+\frac{95}{384}{\rm i} \epsilon^{c d e}\,_{\hat{b} {e_{1}}} G_{i j} (\Gamma^{{e_{1}}})^{\alpha \beta} W \boldsymbol{W} W_{c d} W_{\hat{a} e \alpha}\,^{i} {G}^{-3} \varphi^{j}_{\beta} - \frac{47}{384}{\rm i} \epsilon^{c e {e_{1}}}\,_{\hat{a} \hat{b}} G_{i j} W \boldsymbol{W} W_{c d} W_{e {e_{1}}}\,^{\alpha i} (\Gamma^{d})_{\alpha}{}^{\beta} {G}^{-3} \varphi^{j}_{\beta}+\frac{1}{384}{\rm i} \epsilon_{\hat{a} \hat{b}}\,^{c e {e_{1}}} G_{i j} W \boldsymbol{W} W_{c d} W_{e {e_{1}}}\,^{\alpha i} (\Gamma^{d})_{\alpha}{}^{\beta} {G}^{-3} \varphi^{j}_{\beta}+\frac{3}{64}{\rm i} \epsilon_{\hat{a} \hat{b} e}\,^{c d} (\Gamma^{e})^{\alpha \beta} W \mathbf{X}_{i j} W_{c d} {G}^{-3} \varphi^{i}_{\alpha} \varphi^{j}_{\beta}+\frac{3}{64}{\rm i} \epsilon_{\hat{a} \hat{b} e}\,^{c d} (\Gamma^{e})^{\alpha \beta} \boldsymbol{W} X_{i j} W_{c d} {G}^{-3} \varphi^{i}_{\alpha} \varphi^{j}_{\beta} - \frac{9}{8}G_{i j} G_{k l} (\Gamma^{c})^{\alpha \beta} W W_{\hat{b} c} \boldsymbol{\lambda}^{i}_{\alpha} {G}^{-5} \nabla_{\hat{a}}{G^{j k}} \varphi^{l}_{\beta} - \frac{9}{8}G_{i j} G_{k l} (\Gamma^{c})^{\alpha \beta} \boldsymbol{W} W_{\hat{b} c} \lambda^{i}_{\alpha} {G}^{-5} \nabla_{\hat{a}}{G^{j k}} \varphi^{l}_{\beta} - \frac{21}{256}\epsilon_{\hat{a} \hat{b} e}\,^{c d} G_{i j} (\Gamma^{e})^{\alpha \beta} X^{i j} W_{c d} \boldsymbol{\lambda}_{k \alpha} {G}^{-3} \varphi^{k}_{\beta} - \frac{21}{256}\epsilon_{\hat{a} \hat{b} e}\,^{c d} G_{i j} (\Gamma^{e})^{\alpha \beta} \mathbf{X}^{i j} W_{c d} \lambda_{k \alpha} {G}^{-3} \varphi^{k}_{\beta}+\frac{21}{256}\epsilon_{\hat{a} \hat{b} e}\,^{c d} G_{i j} (\Gamma^{e})^{\alpha \beta} X^{i}\,_{k} W_{c d} \boldsymbol{\lambda}^{j}_{\alpha} {G}^{-3} \varphi^{k}_{\beta} - \frac{21}{256}\epsilon_{\hat{a} \hat{b} e}\,^{c d} G_{i j} (\Gamma^{e})^{\alpha \beta} X^{i}\,_{k} W_{c d} \boldsymbol{\lambda}^{k}_{\alpha} {G}^{-3} \varphi^{j}_{\beta}+\frac{21}{256}\epsilon_{\hat{a} \hat{b} e}\,^{c d} G_{i j} (\Gamma^{e})^{\alpha \beta} \mathbf{X}^{i}\,_{k} W_{c d} \lambda^{j}_{\alpha} {G}^{-3} \varphi^{k}_{\beta}%
 - \frac{21}{256}\epsilon_{\hat{a} \hat{b} e}\,^{c d} G_{i j} (\Gamma^{e})^{\alpha \beta} \mathbf{X}^{i}\,_{k} W_{c d} \lambda^{k}_{\alpha} {G}^{-3} \varphi^{j}_{\beta}+\frac{15}{128}\epsilon^{c d}\,_{\hat{a} \hat{b} e} G_{i j} (\Sigma_{c d})^{\beta \alpha} W \boldsymbol{\lambda}^{i}_{\beta} X_{k \alpha} {G}^{-3} \nabla^{e}{G^{j k}} - \frac{15}{128}\epsilon^{c d}\,_{\hat{a} \hat{b} e} G_{i j} (\Sigma_{c d})^{\beta \alpha} W \boldsymbol{\lambda}_{k \beta} X^{i}_{\alpha} {G}^{-3} \nabla^{e}{G^{j k}}+\frac{15}{128}\epsilon^{c d}\,_{\hat{a} \hat{b} e} G_{i j} (\Sigma_{c d})^{\beta \alpha} W \boldsymbol{\lambda}_{k \beta} X^{k}_{\alpha} {G}^{-3} \nabla^{e}{G^{i j}}+\frac{15}{128}\epsilon^{c d}\,_{\hat{a} \hat{b} e} G_{i j} (\Sigma_{c d})^{\beta \alpha} \boldsymbol{W} \lambda^{i}_{\beta} X_{k \alpha} {G}^{-3} \nabla^{e}{G^{j k}} - \frac{15}{128}\epsilon^{c d}\,_{\hat{a} \hat{b} e} G_{i j} (\Sigma_{c d})^{\beta \alpha} \boldsymbol{W} \lambda_{k \beta} X^{i}_{\alpha} {G}^{-3} \nabla^{e}{G^{j k}}+\frac{15}{128}\epsilon^{c d}\,_{\hat{a} \hat{b} e} G_{i j} (\Sigma_{c d})^{\beta \alpha} \boldsymbol{W} \lambda_{k \beta} X^{k}_{\alpha} {G}^{-3} \nabla^{e}{G^{i j}} - \frac{1}{8}\epsilon^{d e}\,_{\hat{a} \hat{b} c} \Phi_{d e i k} G^{i}\,_{j} (\Gamma^{c})^{\alpha \beta} W \boldsymbol{\lambda}^{k}_{\alpha} {G}^{-3} \varphi^{j}_{\beta} - \frac{1}{8}\epsilon^{d e}\,_{\hat{a} \hat{b} c} \Phi_{d e i k} G^{i}\,_{j} (\Gamma^{c})^{\alpha \beta} \boldsymbol{W} \lambda^{k}_{\alpha} {G}^{-3} \varphi^{j}_{\beta}+\frac{1}{16}\epsilon_{\hat{a} \hat{b} c d e} G_{i j} (\Gamma^{c})^{\alpha \beta} W {G}^{-3} \nabla^{d}{\boldsymbol{\lambda}^{i}_{\alpha}} \nabla^{e}{\varphi^{j}_{\beta}}+\frac{1}{4}\epsilon_{\hat{a} \hat{b} c d e} G_{i j} (\Gamma^{c})^{\alpha \beta} W {G}^{-3} \nabla^{d}{\nabla^{e}{\boldsymbol{\lambda}^{i}_{\alpha}}} \varphi^{j}_{\beta}+\frac{1}{16}\epsilon_{\hat{a} \hat{b} c d e} G_{i j} (\Gamma^{c})^{\alpha \beta} W {G}^{-3} \nabla^{d}{\varphi^{i}_{\alpha}} \nabla^{e}{\boldsymbol{\lambda}^{j}_{\beta}}+\frac{1}{16}\epsilon_{\hat{a} \hat{b} c d e} G_{i j} (\Gamma^{c})^{\alpha \beta} \boldsymbol{W} {G}^{-3} \nabla^{d}{\lambda^{i}_{\alpha}} \nabla^{e}{\varphi^{j}_{\beta}}+\frac{1}{4}\epsilon_{\hat{a} \hat{b} c d e} G_{i j} (\Gamma^{c})^{\alpha \beta} \boldsymbol{W} {G}^{-3} \nabla^{d}{\nabla^{e}{\lambda^{i}_{\alpha}}} \varphi^{j}_{\beta}+\frac{1}{16}\epsilon_{\hat{a} \hat{b} c d e} G_{i j} (\Gamma^{c})^{\alpha \beta} \boldsymbol{W} {G}^{-3} \nabla^{d}{\varphi^{i}_{\alpha}} \nabla^{e}{\lambda^{j}_{\beta}} - \frac{3}{8}\epsilon^{d e}\,_{\hat{a} {e_{1}}}\,^{c} G_{i j} (\Sigma_{d e})^{\alpha \beta} W \boldsymbol{\lambda}^{i}_{\alpha} {G}^{-3} \nabla^{{e_{1}}}{W_{\hat{b} c}} \varphi^{j}_{\beta} - \frac{3}{16}\epsilon^{e {e_{1}}}\,_{\hat{a}}\,^{c d} G_{i j} (\Sigma_{e {e_{1}}})^{\alpha \beta} W \boldsymbol{\lambda}^{i}_{\alpha} {G}^{-3} \nabla_{\hat{b}}{W_{c d}} \varphi^{j}_{\beta} - \frac{3}{8}\epsilon^{d e}\,_{\hat{a} {e_{1}}}\,^{c} G_{i j} (\Sigma_{d e})^{\alpha \beta} \boldsymbol{W} \lambda^{i}_{\alpha} {G}^{-3} \nabla^{{e_{1}}}{W_{\hat{b} c}} \varphi^{j}_{\beta} - \frac{3}{16}\epsilon^{e {e_{1}}}\,_{\hat{a}}\,^{c d} G_{i j} (\Sigma_{e {e_{1}}})^{\alpha \beta} \boldsymbol{W} \lambda^{i}_{\alpha} {G}^{-3} \nabla_{\hat{b}}{W_{c d}} \varphi^{j}_{\beta} - \frac{3}{8}\epsilon^{e}\,_{\hat{a} \hat{b} {e_{1}} d} G_{i j} (\Sigma_{e c})^{\alpha \beta} W \boldsymbol{\lambda}^{i}_{\alpha} {G}^{-3} \nabla^{{e_{1}}}{W^{c d}} \varphi^{j}_{\beta}%
+\frac{3}{16}\epsilon^{e}\,_{\hat{a} \hat{b}}\,^{c d} G_{i j} (\Sigma_{e {e_{1}}})^{\alpha \beta} W \boldsymbol{\lambda}^{i}_{\alpha} {G}^{-3} \nabla^{{e_{1}}}{W_{c d}} \varphi^{j}_{\beta} - \frac{3}{8}\epsilon^{e}\,_{\hat{a} \hat{b} {e_{1}} d} G_{i j} (\Sigma_{e c})^{\alpha \beta} \boldsymbol{W} \lambda^{i}_{\alpha} {G}^{-3} \nabla^{{e_{1}}}{W^{c d}} \varphi^{j}_{\beta}+\frac{3}{16}\epsilon^{e}\,_{\hat{a} \hat{b}}\,^{c d} G_{i j} (\Sigma_{e {e_{1}}})^{\alpha \beta} \boldsymbol{W} \lambda^{i}_{\alpha} {G}^{-3} \nabla^{{e_{1}}}{W_{c d}} \varphi^{j}_{\beta}+\frac{3}{8}\epsilon^{e}\,_{\hat{a} \hat{b} {e_{1}}}\,^{c} G_{i j} (\Sigma_{e}{}^{\, d})^{\alpha \beta} W_{c d} \lambda^{i}_{\alpha} {G}^{-3} \nabla^{{e_{1}}}{\boldsymbol{W}} \varphi^{j}_{\beta}+\frac{3}{8}\epsilon^{e}\,_{\hat{a} \hat{b} {e_{1}}}\,^{c} G_{i j} (\Sigma_{e}{}^{\, d})^{\alpha \beta} W_{c d} \boldsymbol{\lambda}^{i}_{\alpha} {G}^{-3} \nabla^{{e_{1}}}{W} \varphi^{j}_{\beta}+\frac{3}{8}\epsilon^{e}\,_{\hat{a} \hat{b} {e_{1}}}\,^{c} G_{i j} (\Sigma_{e}{}^{\, d})^{\alpha \beta} W W_{c d} \boldsymbol{\lambda}^{i}_{\alpha} {G}^{-3} \nabla^{{e_{1}}}{\varphi^{j}_{\beta}}+\frac{3}{8}\epsilon^{e}\,_{\hat{a} \hat{b} {e_{1}}}\,^{c} G_{i j} (\Sigma_{e}{}^{\, d})^{\alpha \beta} W W_{c d} {G}^{-3} \nabla^{{e_{1}}}{\boldsymbol{\lambda}^{i}_{\alpha}} \varphi^{j}_{\beta}+\frac{3}{8}\epsilon^{e}\,_{\hat{a} \hat{b} {e_{1}}}\,^{c} G_{i j} (\Sigma_{e}{}^{\, d})^{\alpha \beta} \boldsymbol{W} W_{c d} \lambda^{i}_{\alpha} {G}^{-3} \nabla^{{e_{1}}}{\varphi^{j}_{\beta}}+\frac{3}{8}\epsilon^{e}\,_{\hat{a} \hat{b} {e_{1}}}\,^{c} G_{i j} (\Sigma_{e}{}^{\, d})^{\alpha \beta} \boldsymbol{W} W_{c d} {G}^{-3} \nabla^{{e_{1}}}{\lambda^{i}_{\alpha}} \varphi^{j}_{\beta} - \frac{3}{256}\epsilon_{\hat{a} \hat{b} e}\,^{c d} (\Gamma^{e})^{\alpha \beta} W_{c d} \lambda_{i \alpha} \boldsymbol{\lambda}^{i \rho} {G}^{-3} \varphi_{j \beta} \varphi^{j}_{\rho} - \frac{3}{256}\epsilon_{\hat{a} \hat{b} e}\,^{c d} (\Gamma^{e})^{\alpha \beta} W_{c d} \lambda_{i \alpha} \boldsymbol{\lambda}_{j}^{\rho} {G}^{-3} \varphi^{i}_{\rho} \varphi^{j}_{\beta} - \frac{3}{256}\epsilon_{\hat{a} \hat{b} e}\,^{c d} (\Gamma^{e})^{\alpha \beta} W_{c d} \lambda^{\rho}_{i} \boldsymbol{\lambda}^{i}_{\alpha} {G}^{-3} \varphi_{j \beta} \varphi^{j}_{\rho} - \frac{3}{256}\epsilon_{\hat{a} \hat{b} e}\,^{c d} (\Gamma^{e})^{\alpha \beta} W_{c d} \lambda^{\rho}_{i} \boldsymbol{\lambda}_{j \alpha} {G}^{-3} \varphi^{i}_{\beta} \varphi^{j}_{\rho}+\frac{3}{16}\epsilon^{e}\,_{\hat{a} \hat{b} {e_{1}}}\,^{c} (\Sigma_{e}{}^{\, d})^{\alpha \beta} W W_{c d} \boldsymbol{\lambda}_{i \alpha} {G}^{-3} \nabla^{{e_{1}}}{G^{i}\,_{j}} \varphi^{j}_{\beta}+\frac{3}{16}\epsilon^{e}\,_{\hat{a} \hat{b} {e_{1}}}\,^{c} (\Sigma_{e}{}^{\, d})^{\alpha \beta} \boldsymbol{W} W_{c d} \lambda_{i \alpha} {G}^{-3} \nabla^{{e_{1}}}{G^{i}\,_{j}} \varphi^{j}_{\beta}+\frac{3}{16}{\rm i} G_{i j} G_{k l} (\Gamma_{\hat{a}})^{\alpha \beta} X^{i j} {G}^{-5} \nabla_{\hat{b}}{\boldsymbol{W}} \varphi^{k}_{\alpha} \varphi^{l}_{\beta}+\frac{3}{16}{\rm i} G_{i j} G_{k l} (\Gamma_{\hat{a}})^{\alpha \beta} \mathbf{X}^{i j} {G}^{-5} \nabla_{\hat{b}}{W} \varphi^{k}_{\alpha} \varphi^{l}_{\beta} - \frac{15}{16}{\rm i} G_{i j} G_{k l} (\Gamma_{\hat{a}})^{\alpha \beta} X^{i k} {G}^{-5} \nabla_{\hat{b}}{\boldsymbol{W}} \varphi^{j}_{\alpha} \varphi^{l}_{\beta} - \frac{15}{16}{\rm i} G_{i j} G_{k l} (\Gamma_{\hat{a}})^{\alpha \beta} \mathbf{X}^{i k} {G}^{-5} \nabla_{\hat{b}}{W} \varphi^{j}_{\alpha} \varphi^{l}_{\beta} - \frac{3}{2}{\rm i} G_{i j} G_{k l} (\Gamma_{\hat{a}})^{\alpha \beta} W \mathbf{X}^{i j} {G}^{-5} \nabla_{\hat{b}}{\varphi^{k}_{\alpha}} \varphi^{l}_{\beta}%
 - \frac{3}{4}{\rm i} G_{i j} G_{k l} (\Gamma_{\hat{a}})^{\alpha \beta} W {G}^{-5} \nabla_{\hat{b}}{\mathbf{X}^{i k}} \varphi^{j}_{\alpha} \varphi^{l}_{\beta} - \frac{3}{2}{\rm i} G_{i j} G_{k l} (\Gamma_{\hat{a}})^{\alpha \beta} \boldsymbol{W} X^{i j} {G}^{-5} \nabla_{\hat{b}}{\varphi^{k}_{\alpha}} \varphi^{l}_{\beta} - \frac{3}{4}{\rm i} G_{i j} G_{k l} (\Gamma_{\hat{a}})^{\alpha \beta} \boldsymbol{W} {G}^{-5} \nabla_{\hat{b}}{X^{i k}} \varphi^{j}_{\alpha} \varphi^{l}_{\beta}+\frac{9}{16}{\rm i} G_{i j} G_{k l} (\Sigma_{\hat{a} \hat{b}})^{\beta \rho} W \boldsymbol{\lambda}^{i}_{\beta} X^{k \alpha} {G}^{-5} \varphi^{j}_{\rho} \varphi^{l}_{\alpha}+\frac{9}{16}{\rm i} G_{i j} G_{k l} (\Sigma_{\hat{a} \hat{b}})^{\beta \rho} W \boldsymbol{\lambda}^{i}_{\beta} X^{k \alpha} {G}^{-5} \varphi^{j}_{\alpha} \varphi^{l}_{\rho}+\frac{9}{64}{\rm i} G_{i j} G_{k l} (\Sigma_{\hat{a} \hat{b}})^{\alpha \beta} W \boldsymbol{\lambda}^{i \rho} X^{j}_{\alpha} {G}^{-5} \varphi^{k}_{\beta} \varphi^{l}_{\rho} - \frac{9}{64}{\rm i} G_{i j} G_{k l} (\Sigma_{\hat{a} \hat{b}})^{\alpha \beta} W \boldsymbol{\lambda}^{i \rho} X^{k}_{\alpha} {G}^{-5} \varphi^{j}_{\beta} \varphi^{l}_{\rho} - \frac{9}{8}{\rm i} G_{i j} G_{k l} (\Sigma_{\hat{a} \hat{b}})^{\beta \rho} W \boldsymbol{\lambda}^{i \alpha} X^{k}_{\alpha} {G}^{-5} \varphi^{j}_{\beta} \varphi^{l}_{\rho}+\frac{9}{16}{\rm i} G_{i j} G_{k l} (\Sigma_{\hat{a} \hat{b}})^{\beta \rho} \boldsymbol{W} \lambda^{i}_{\beta} X^{k \alpha} {G}^{-5} \varphi^{j}_{\rho} \varphi^{l}_{\alpha}+\frac{9}{16}{\rm i} G_{i j} G_{k l} (\Sigma_{\hat{a} \hat{b}})^{\beta \rho} \boldsymbol{W} \lambda^{i}_{\beta} X^{k \alpha} {G}^{-5} \varphi^{j}_{\alpha} \varphi^{l}_{\rho}+\frac{9}{64}{\rm i} G_{i j} G_{k l} (\Sigma_{\hat{a} \hat{b}})^{\alpha \beta} \boldsymbol{W} \lambda^{i \rho} X^{j}_{\alpha} {G}^{-5} \varphi^{k}_{\beta} \varphi^{l}_{\rho} - \frac{9}{64}{\rm i} G_{i j} G_{k l} (\Sigma_{\hat{a} \hat{b}})^{\alpha \beta} \boldsymbol{W} \lambda^{i \rho} X^{k}_{\alpha} {G}^{-5} \varphi^{j}_{\beta} \varphi^{l}_{\rho} - \frac{9}{8}{\rm i} G_{i j} G_{k l} (\Sigma_{\hat{a} \hat{b}})^{\beta \rho} \boldsymbol{W} \lambda^{i \alpha} X^{k}_{\alpha} {G}^{-5} \varphi^{j}_{\beta} \varphi^{l}_{\rho}+\frac{3}{8}{\rm i} G_{i j} (\Gamma_{\hat{a}})^{\alpha \beta} W \mathbf{X}^{i j} {G}^{-5} \nabla_{\hat{b}}{G_{k l}} \varphi^{k}_{\alpha} \varphi^{l}_{\beta} - \frac{3}{2}{\rm i} G_{i j} (\Gamma_{\hat{a}})^{\alpha \beta} W \mathbf{X}_{k l} {G}^{-5} \nabla_{\hat{b}}{G^{i k}} \varphi^{j}_{\alpha} \varphi^{l}_{\beta}+\frac{3}{8}{\rm i} G_{i j} (\Gamma_{\hat{a}})^{\alpha \beta} \boldsymbol{W} X^{i j} {G}^{-5} \nabla_{\hat{b}}{G_{k l}} \varphi^{k}_{\alpha} \varphi^{l}_{\beta} - \frac{3}{2}{\rm i} G_{i j} (\Gamma_{\hat{a}})^{\alpha \beta} \boldsymbol{W} X_{k l} {G}^{-5} \nabla_{\hat{b}}{G^{i k}} \varphi^{j}_{\alpha} \varphi^{l}_{\beta}+\frac{1}{384}{\rm i} G_{i j} (\Gamma^{c})^{\alpha \beta} W \boldsymbol{W} W_{\hat{b} c} W_{d e \alpha}\,^{i} \epsilon_{\hat{a}}\,^{d e {e_{1}} {e_{2}}} (\Sigma_{{e_{1}} {e_{2}}})_{\beta}{}^{\rho} {G}^{-3} \varphi^{j}_{\rho}+\frac{3}{4}{\rm i} G_{i j} (\Gamma_{\hat{a}})^{\alpha \beta} W {G}^{-5} \nabla_{\hat{b}}{\boldsymbol{\lambda}^{i}_{\alpha}} \varphi^{j \rho} \varphi_{k \beta} \varphi^{k}_{\rho}+\frac{3}{4}{\rm i} G_{i j} (\Gamma_{\hat{a}})^{\alpha \beta} W {G}^{-5} \nabla_{\hat{b}}{\boldsymbol{\lambda}_{k \alpha}} \varphi^{i}_{\beta} \varphi^{j \rho} \varphi^{k}_{\rho}%
 - \frac{3}{4}{\rm i} G_{i j} (\Gamma_{\hat{a}})^{\alpha \beta} W {G}^{-5} \nabla_{\hat{b}}{\boldsymbol{\lambda}_{k \alpha}} \varphi^{i \rho} \varphi^{j}_{\rho} \varphi^{k}_{\beta}+\frac{3}{4}{\rm i} G_{i j} (\Gamma_{\hat{a}})^{\alpha \beta} \boldsymbol{W} {G}^{-5} \nabla_{\hat{b}}{\lambda^{i}_{\alpha}} \varphi^{j \rho} \varphi_{k \beta} \varphi^{k}_{\rho}+\frac{3}{4}{\rm i} G_{i j} (\Gamma_{\hat{a}})^{\alpha \beta} \boldsymbol{W} {G}^{-5} \nabla_{\hat{b}}{\lambda_{k \alpha}} \varphi^{i}_{\beta} \varphi^{j \rho} \varphi^{k}_{\rho} - \frac{3}{4}{\rm i} G_{i j} (\Gamma_{\hat{a}})^{\alpha \beta} \boldsymbol{W} {G}^{-5} \nabla_{\hat{b}}{\lambda_{k \alpha}} \varphi^{i \rho} \varphi^{j}_{\rho} \varphi^{k}_{\beta}+\frac{3}{8}{\rm i} G_{i j} (\Gamma_{\hat{a}})^{\alpha \beta} \lambda^{i}_{\alpha} {G}^{-5} \nabla_{\hat{b}}{\boldsymbol{W}} \varphi^{j \rho} \varphi_{k \beta} \varphi^{k}_{\rho}+\frac{3}{8}{\rm i} G_{i j} (\Gamma_{\hat{a}})^{\alpha \beta} \boldsymbol{\lambda}^{i}_{\alpha} {G}^{-5} \nabla_{\hat{b}}{W} \varphi^{j \rho} \varphi_{k \beta} \varphi^{k}_{\rho}+\frac{3}{8}{\rm i} G_{i j} (\Gamma_{\hat{a}})^{\alpha \beta} \lambda_{k \alpha} {G}^{-5} \nabla_{\hat{b}}{\boldsymbol{W}} \varphi^{i}_{\beta} \varphi^{j \rho} \varphi^{k}_{\rho} - \frac{3}{8}{\rm i} G_{i j} (\Gamma_{\hat{a}})^{\alpha \beta} \lambda_{k \alpha} {G}^{-5} \nabla_{\hat{b}}{\boldsymbol{W}} \varphi^{i \rho} \varphi^{j}_{\rho} \varphi^{k}_{\beta}+\frac{3}{8}{\rm i} G_{i j} (\Gamma_{\hat{a}})^{\alpha \beta} \boldsymbol{\lambda}_{k \alpha} {G}^{-5} \nabla_{\hat{b}}{W} \varphi^{i}_{\beta} \varphi^{j \rho} \varphi^{k}_{\rho} - \frac{3}{8}{\rm i} G_{i j} (\Gamma_{\hat{a}})^{\alpha \beta} \boldsymbol{\lambda}_{k \alpha} {G}^{-5} \nabla_{\hat{b}}{W} \varphi^{i \rho} \varphi^{j}_{\rho} \varphi^{k}_{\beta}+\frac{3}{8}{\rm i} G_{i j} (\Sigma_{\hat{a} \hat{b}})^{\alpha \beta} (\Sigma_{c d})^{\rho \lambda} F^{c d} \boldsymbol{\lambda}^{i}_{\rho} {G}^{-5} \varphi^{j}_{\alpha} \varphi_{k \beta} \varphi^{k}_{\lambda}+\frac{3}{8}{\rm i} G_{i j} (\Sigma_{\hat{a} \hat{b}})^{\alpha \beta} (\Sigma_{c d})^{\rho \lambda} F^{c d} \boldsymbol{\lambda}^{i}_{\rho} {G}^{-5} \varphi^{j}_{\lambda} \varphi_{k \alpha} \varphi^{k}_{\beta}+\frac{3}{8}{\rm i} G_{i j} (\Sigma_{\hat{a} \hat{b}})^{\alpha \beta} (\Sigma_{c d})^{\rho \lambda} F^{c d} \boldsymbol{\lambda}_{k \rho} {G}^{-5} \varphi^{i}_{\alpha} \varphi^{j}_{\lambda} \varphi^{k}_{\beta}+\frac{3}{8}{\rm i} G_{i j} (\Sigma_{\hat{a} \hat{b}})^{\alpha \beta} (\Sigma_{c d})^{\rho \lambda} \mathbf{F}^{c d} \lambda^{i}_{\rho} {G}^{-5} \varphi^{j}_{\alpha} \varphi_{k \beta} \varphi^{k}_{\lambda}+\frac{3}{8}{\rm i} G_{i j} (\Sigma_{\hat{a} \hat{b}})^{\alpha \beta} (\Sigma_{c d})^{\rho \lambda} \mathbf{F}^{c d} \lambda^{i}_{\rho} {G}^{-5} \varphi^{j}_{\lambda} \varphi_{k \alpha} \varphi^{k}_{\beta}+\frac{3}{8}{\rm i} G_{i j} (\Sigma_{\hat{a} \hat{b}})^{\alpha \beta} (\Sigma_{c d})^{\rho \lambda} \mathbf{F}^{c d} \lambda_{k \rho} {G}^{-5} \varphi^{i}_{\alpha} \varphi^{j}_{\lambda} \varphi^{k}_{\beta} - \frac{9}{8}{\rm i} G_{i j} (\Sigma_{\hat{a}}{}^{\, c})^{\alpha \beta} W W_{\hat{b} c} \boldsymbol{\lambda}^{i}_{\alpha} {G}^{-5} \varphi^{j \rho} \varphi_{k \beta} \varphi^{k}_{\rho} - \frac{9}{8}{\rm i} G_{i j} (\Sigma_{\hat{a}}{}^{\, c})^{\alpha \beta} W W_{\hat{b} c} \boldsymbol{\lambda}_{k \alpha} {G}^{-5} \varphi^{i}_{\beta} \varphi^{j \rho} \varphi^{k}_{\rho}+\frac{9}{8}{\rm i} G_{i j} (\Sigma_{\hat{a}}{}^{\, c})^{\alpha \beta} W W_{\hat{b} c} \boldsymbol{\lambda}_{k \alpha} {G}^{-5} \varphi^{i \rho} \varphi^{j}_{\rho} \varphi^{k}_{\beta} - \frac{9}{8}{\rm i} G_{i j} (\Sigma_{\hat{a}}{}^{\, c})^{\alpha \beta} \boldsymbol{W} W_{\hat{b} c} \lambda^{i}_{\alpha} {G}^{-5} \varphi^{j \rho} \varphi_{k \beta} \varphi^{k}_{\rho}%
 - \frac{9}{8}{\rm i} G_{i j} (\Sigma_{\hat{a}}{}^{\, c})^{\alpha \beta} \boldsymbol{W} W_{\hat{b} c} \lambda_{k \alpha} {G}^{-5} \varphi^{i}_{\beta} \varphi^{j \rho} \varphi^{k}_{\rho}+\frac{9}{8}{\rm i} G_{i j} (\Sigma_{\hat{a}}{}^{\, c})^{\alpha \beta} \boldsymbol{W} W_{\hat{b} c} \lambda_{k \alpha} {G}^{-5} \varphi^{i \rho} \varphi^{j}_{\rho} \varphi^{k}_{\beta} - \frac{3}{16}{\rm i} \epsilon^{c d e}\,_{\hat{a} \hat{b}} \mathcal{H}_{c} G_{i j} (\Sigma_{d e})^{\alpha \beta} W \mathbf{X}^{i j} {G}^{-5} \varphi_{k \alpha} \varphi^{k}_{\beta}+\frac{3}{8}{\rm i} \epsilon^{c d e}\,_{\hat{a} \hat{b}} \mathcal{H}_{c} G_{i j} (\Sigma_{d e})^{\alpha \beta} W \mathbf{X}^{i}\,_{k} {G}^{-5} \varphi^{j}_{\alpha} \varphi^{k}_{\beta} - \frac{3}{16}{\rm i} \epsilon^{c d e}\,_{\hat{a} \hat{b}} \mathcal{H}_{c} G_{i j} (\Sigma_{d e})^{\alpha \beta} \boldsymbol{W} X^{i j} {G}^{-5} \varphi_{k \alpha} \varphi^{k}_{\beta}+\frac{3}{8}{\rm i} \epsilon^{c d e}\,_{\hat{a} \hat{b}} \mathcal{H}_{c} G_{i j} (\Sigma_{d e})^{\alpha \beta} \boldsymbol{W} X^{i}\,_{k} {G}^{-5} \varphi^{j}_{\alpha} \varphi^{k}_{\beta}+\frac{3}{16}G_{i j} G_{k l} (\Gamma_{\hat{a}})^{\alpha \beta} X^{i j} \boldsymbol{\lambda}_{m \alpha} {G}^{-5} \nabla_{\hat{b}}{G^{k m}} \varphi^{l}_{\beta}+\frac{3}{16}G_{i j} G_{k l} (\Gamma_{\hat{a}})^{\alpha \beta} \mathbf{X}^{i j} \lambda_{m \alpha} {G}^{-5} \nabla_{\hat{b}}{G^{k m}} \varphi^{l}_{\beta} - \frac{3}{16}G_{i j} G_{k l} (\Gamma_{\hat{a}})^{\alpha \beta} X^{i}\,_{m} \boldsymbol{\lambda}^{j}_{\alpha} {G}^{-5} \nabla_{\hat{b}}{G^{k m}} \varphi^{l}_{\beta} - \frac{9}{16}G_{i j} G_{k l} (\Gamma_{\hat{a}})^{\alpha \beta} X^{i}\,_{m} \boldsymbol{\lambda}^{m}_{\alpha} {G}^{-5} \nabla_{\hat{b}}{G^{j k}} \varphi^{l}_{\beta} - \frac{3}{16}G_{i j} G_{k l} (\Gamma_{\hat{a}})^{\alpha \beta} \mathbf{X}^{i}\,_{m} \lambda^{j}_{\alpha} {G}^{-5} \nabla_{\hat{b}}{G^{k m}} \varphi^{l}_{\beta} - \frac{9}{16}G_{i j} G_{k l} (\Gamma_{\hat{a}})^{\alpha \beta} \mathbf{X}^{i}\,_{m} \lambda^{m}_{\alpha} {G}^{-5} \nabla_{\hat{b}}{G^{j k}} \varphi^{l}_{\beta}+\frac{3}{8}G_{i j} G_{k l} (\Gamma_{\hat{a}})^{\alpha \beta} \lambda^{i}_{\alpha} {G}^{-5} \nabla_{\hat{b}}{\boldsymbol{\lambda}^{k \rho}} \varphi^{j}_{\rho} \varphi^{l}_{\beta}+\frac{3}{8}G_{i j} G_{k l} (\Gamma_{\hat{a}})^{\alpha \beta} \boldsymbol{\lambda}^{i}_{\alpha} {G}^{-5} \nabla_{\hat{b}}{\lambda^{k \rho}} \varphi^{j}_{\rho} \varphi^{l}_{\beta} - \frac{3}{2}G_{i j} G_{k l} (\Gamma_{\hat{a}})^{\alpha \beta} \lambda^{i \rho} \boldsymbol{\lambda}^{j}_{\rho} {G}^{-5} \nabla_{\hat{b}}{\varphi^{k}_{\alpha}} \varphi^{l}_{\beta} - \frac{3}{16}G_{i j} G_{k l} (\Gamma_{\hat{a}})^{\alpha \beta} \lambda^{i \rho} {G}^{-5} \nabla_{\hat{b}}{\boldsymbol{\lambda}^{j}_{\alpha}} \varphi^{k}_{\beta} \varphi^{l}_{\rho}+\frac{9}{16}G_{i j} G_{k l} (\Gamma_{\hat{a}})^{\alpha \beta} \lambda^{i \rho} {G}^{-5} \nabla_{\hat{b}}{\boldsymbol{\lambda}^{k}_{\alpha}} \varphi^{j}_{\beta} \varphi^{l}_{\rho} - \frac{3}{8}G_{i j} G_{k l} (\Gamma_{\hat{a}})^{\alpha \beta} \lambda^{i \rho} {G}^{-5} \nabla_{\hat{b}}{\boldsymbol{\lambda}^{k}_{\rho}} \varphi^{j}_{\alpha} \varphi^{l}_{\beta} - \frac{3}{16}G_{i j} G_{k l} (\Gamma_{\hat{a}})^{\alpha \beta} \boldsymbol{\lambda}^{i \rho} {G}^{-5} \nabla_{\hat{b}}{\lambda^{j}_{\alpha}} \varphi^{k}_{\beta} \varphi^{l}_{\rho}+\frac{9}{16}G_{i j} G_{k l} (\Gamma_{\hat{a}})^{\alpha \beta} \boldsymbol{\lambda}^{i \rho} {G}^{-5} \nabla_{\hat{b}}{\lambda^{k}_{\alpha}} \varphi^{j}_{\beta} \varphi^{l}_{\rho}%
 - \frac{3}{8}G_{i j} G_{k l} (\Gamma_{\hat{a}})^{\alpha \beta} \boldsymbol{\lambda}^{i \rho} {G}^{-5} \nabla_{\hat{b}}{\lambda^{k}_{\rho}} \varphi^{j}_{\alpha} \varphi^{l}_{\beta} - \frac{3}{32}G_{i j} G_{k l} (\Sigma_{\hat{a}}{}^{\, c})^{\alpha \beta} W_{\hat{b} c} \lambda^{i}_{\alpha} \boldsymbol{\lambda}^{j \rho} {G}^{-5} \varphi^{k}_{\beta} \varphi^{l}_{\rho}+\frac{3}{16}G_{i j} G_{k l} (\Sigma_{\hat{a}}{}^{\, c})^{\alpha \beta} W_{\hat{b} c} \lambda^{i}_{\alpha} \boldsymbol{\lambda}^{k \rho} {G}^{-5} \varphi^{j}_{\beta} \varphi^{l}_{\rho}+\frac{3}{32}G_{i j} G_{k l} (\Sigma_{\hat{a}}{}^{\, c})^{\alpha \beta} W_{\hat{b} c} \lambda^{i}_{\alpha} \boldsymbol{\lambda}^{k \rho} {G}^{-5} \varphi^{j}_{\rho} \varphi^{l}_{\beta}+\frac{3}{32}G_{i j} G_{k l} (\Sigma_{\hat{a}}{}^{\, c})^{\alpha \beta} W_{\hat{b} c} \lambda^{i \rho} \boldsymbol{\lambda}^{j}_{\alpha} {G}^{-5} \varphi^{k}_{\beta} \varphi^{l}_{\rho}+\frac{3}{32}G_{i j} G_{k l} (\Sigma_{\hat{a}}{}^{\, c})^{\alpha \beta} W_{\hat{b} c} \lambda^{i \rho} \boldsymbol{\lambda}^{k}_{\alpha} {G}^{-5} \varphi^{j}_{\beta} \varphi^{l}_{\rho}+\frac{3}{16}G_{i j} G_{k l} (\Sigma_{\hat{a}}{}^{\, c})^{\alpha \beta} W_{\hat{b} c} \lambda^{i \rho} \boldsymbol{\lambda}^{k}_{\alpha} {G}^{-5} \varphi^{j}_{\rho} \varphi^{l}_{\beta}+\frac{3}{8}G_{i j} (\Gamma_{\hat{a}})^{\alpha \beta} \lambda^{i \rho} \boldsymbol{\lambda}^{j}_{\rho} {G}^{-5} \nabla_{\hat{b}}{G_{k l}} \varphi^{k}_{\alpha} \varphi^{l}_{\beta} - \frac{3}{4}G_{i j} (\Gamma_{\hat{a}})^{\alpha \beta} \lambda^{\rho}_{k} \boldsymbol{\lambda}_{l \rho} {G}^{-5} \nabla_{\hat{b}}{G^{i k}} \varphi^{j}_{\alpha} \varphi^{l}_{\beta} - \frac{3}{4}G_{i j} (\Gamma_{\hat{a}})^{\alpha \beta} \lambda^{\rho}_{l} \boldsymbol{\lambda}_{k \rho} {G}^{-5} \nabla_{\hat{b}}{G^{i k}} \varphi^{j}_{\alpha} \varphi^{l}_{\beta}+\frac{3}{16}\epsilon^{e {e_{1}}}\,_{\hat{a}}\,^{c d} G_{i j} G_{k l} (\Sigma_{e {e_{1}}})^{\alpha \beta} F_{c d} \boldsymbol{\lambda}^{i}_{\alpha} {G}^{-5} \nabla_{\hat{b}}{G^{j k}} \varphi^{l}_{\beta}+\frac{3}{16}\epsilon^{e {e_{1}}}\,_{\hat{a}}\,^{c d} G_{i j} G_{k l} (\Sigma_{e {e_{1}}})^{\alpha \beta} \mathbf{F}_{c d} \lambda^{i}_{\alpha} {G}^{-5} \nabla_{\hat{b}}{G^{j k}} \varphi^{l}_{\beta}+\frac{3}{8}\epsilon^{d e}\,_{\hat{a} {e_{1}}}\,^{c} G_{i j} G_{k l} (\Sigma_{d e})^{\alpha \beta} F_{\hat{b} c} \boldsymbol{\lambda}^{i}_{\alpha} {G}^{-5} \nabla^{{e_{1}}}{G^{j k}} \varphi^{l}_{\beta}+\frac{3}{8}\epsilon^{d e}\,_{\hat{a} {e_{1}}}\,^{c} G_{i j} G_{k l} (\Sigma_{d e})^{\alpha \beta} \mathbf{F}_{\hat{b} c} \lambda^{i}_{\alpha} {G}^{-5} \nabla^{{e_{1}}}{G^{j k}} \varphi^{l}_{\beta} - \frac{3}{16}\epsilon^{e}\,_{\hat{a} \hat{b}}\,^{c d} G_{i j} G_{k l} (\Sigma_{e {e_{1}}})^{\alpha \beta} F_{c d} \boldsymbol{\lambda}^{i}_{\alpha} {G}^{-5} \nabla^{{e_{1}}}{G^{j k}} \varphi^{l}_{\beta} - \frac{3}{16}\epsilon^{e}\,_{\hat{a} \hat{b}}\,^{c d} G_{i j} G_{k l} (\Sigma_{e {e_{1}}})^{\alpha \beta} \mathbf{F}_{c d} \lambda^{i}_{\alpha} {G}^{-5} \nabla^{{e_{1}}}{G^{j k}} \varphi^{l}_{\beta}+\frac{3}{64}\epsilon^{c d e}\,_{\hat{a} \hat{b}} \mathcal{H}_{c} G_{i j} G_{k l} (\Sigma_{d e})^{\alpha \beta} X^{i j} \boldsymbol{\lambda}^{k}_{\alpha} {G}^{-5} \varphi^{l}_{\beta}+\frac{3}{64}\epsilon^{c d e}\,_{\hat{a} \hat{b}} \mathcal{H}_{c} G_{i j} G_{k l} (\Sigma_{d e})^{\alpha \beta} \mathbf{X}^{i j} \lambda^{k}_{\alpha} {G}^{-5} \varphi^{l}_{\beta} - \frac{3}{64}\epsilon^{c d e}\,_{\hat{a} \hat{b}} \mathcal{H}_{c} G_{i j} G_{k l} (\Sigma_{d e})^{\alpha \beta} X^{i k} \boldsymbol{\lambda}^{j}_{\alpha} {G}^{-5} \varphi^{l}_{\beta} - \frac{3}{64}\epsilon^{c d e}\,_{\hat{a} \hat{b}} \mathcal{H}_{c} G_{i j} G_{k l} (\Sigma_{d e})^{\alpha \beta} \mathbf{X}^{i k} \lambda^{j}_{\alpha} {G}^{-5} \varphi^{l}_{\beta}%
 - \frac{1}{16}\epsilon^{c d}\,_{\hat{a} \hat{b} e} G_{i j} W_{c d}\,^{\alpha i} (\Gamma^{e})^{\beta \rho} \lambda_{k \beta} \boldsymbol{\lambda}^{k}_{\alpha} {G}^{-3} \varphi^{j}_{\rho} - \frac{3}{16}\epsilon^{c d e}\,_{\hat{a} \hat{b}} \mathcal{H}_{c} G_{i j} (\Sigma_{d e})^{\alpha \beta} \lambda^{i \rho} \boldsymbol{\lambda}^{j}_{\rho} {G}^{-5} \varphi_{k \alpha} \varphi^{k}_{\beta}+\frac{3}{16}\epsilon^{c d e}\,_{\hat{a} \hat{b}} \mathcal{H}_{c} G_{i j} (\Sigma_{d e})^{\alpha \beta} \lambda^{i \rho} \boldsymbol{\lambda}_{k \rho} {G}^{-5} \varphi^{j}_{\alpha} \varphi^{k}_{\beta}+\frac{3}{16}\epsilon^{c d e}\,_{\hat{a} \hat{b}} \mathcal{H}_{c} G_{i j} (\Sigma_{d e})^{\alpha \beta} \lambda^{\rho}_{k} \boldsymbol{\lambda}^{i}_{\rho} {G}^{-5} \varphi^{j}_{\alpha} \varphi^{k}_{\beta}+\frac{9}{4}{\rm i} G_{i j} G_{k l} (\Gamma_{\hat{a}})^{\alpha \beta} W \boldsymbol{W} X^{i}_{\alpha} {G}^{-5} \nabla_{\hat{b}}{G^{j k}} \varphi^{l}_{\beta} - \frac{3}{4}{\rm i} G_{i j} (\Sigma_{\hat{a} \hat{b}})^{\rho \lambda} (\Gamma_{c})^{\alpha \beta} W {G}^{-5} \nabla^{c}{\boldsymbol{\lambda}^{i}_{\alpha}} \varphi^{j}_{\rho} \varphi_{k \lambda} \varphi^{k}_{\beta} - \frac{3}{4}{\rm i} G_{i j} (\Sigma_{\hat{a} \hat{b}})^{\rho \lambda} (\Gamma_{c})^{\alpha \beta} W {G}^{-5} \nabla^{c}{\boldsymbol{\lambda}^{i}_{\alpha}} \varphi^{j}_{\beta} \varphi_{k \rho} \varphi^{k}_{\lambda} - \frac{3}{4}{\rm i} G_{i j} (\Sigma_{\hat{a} \hat{b}})^{\rho \lambda} (\Gamma_{c})^{\alpha \beta} W {G}^{-5} \nabla^{c}{\boldsymbol{\lambda}_{k \alpha}} \varphi^{i}_{\rho} \varphi^{j}_{\beta} \varphi^{k}_{\lambda} - \frac{3}{4}{\rm i} G_{i j} (\Sigma_{\hat{a} \hat{b}})^{\rho \lambda} (\Gamma_{c})^{\alpha \beta} \boldsymbol{W} {G}^{-5} \nabla^{c}{\lambda^{i}_{\alpha}} \varphi^{j}_{\rho} \varphi_{k \lambda} \varphi^{k}_{\beta} - \frac{3}{4}{\rm i} G_{i j} (\Sigma_{\hat{a} \hat{b}})^{\rho \lambda} (\Gamma_{c})^{\alpha \beta} \boldsymbol{W} {G}^{-5} \nabla^{c}{\lambda^{i}_{\alpha}} \varphi^{j}_{\beta} \varphi_{k \rho} \varphi^{k}_{\lambda} - \frac{3}{4}{\rm i} G_{i j} (\Sigma_{\hat{a} \hat{b}})^{\rho \lambda} (\Gamma_{c})^{\alpha \beta} \boldsymbol{W} {G}^{-5} \nabla^{c}{\lambda_{k \alpha}} \varphi^{i}_{\rho} \varphi^{j}_{\beta} \varphi^{k}_{\lambda} - \frac{3}{8}{\rm i} G_{i j} (\Sigma_{\hat{a} \hat{b}})^{\rho \lambda} (\Gamma_{c})^{\alpha \beta} \lambda^{i}_{\alpha} {G}^{-5} \nabla^{c}{\boldsymbol{W}} \varphi^{j}_{\rho} \varphi_{k \lambda} \varphi^{k}_{\beta} - \frac{3}{8}{\rm i} G_{i j} (\Sigma_{\hat{a} \hat{b}})^{\rho \lambda} (\Gamma_{c})^{\alpha \beta} \lambda^{i}_{\alpha} {G}^{-5} \nabla^{c}{\boldsymbol{W}} \varphi^{j}_{\beta} \varphi_{k \rho} \varphi^{k}_{\lambda} - \frac{3}{8}{\rm i} G_{i j} (\Sigma_{\hat{a} \hat{b}})^{\rho \lambda} (\Gamma_{c})^{\alpha \beta} \boldsymbol{\lambda}^{i}_{\alpha} {G}^{-5} \nabla^{c}{W} \varphi^{j}_{\rho} \varphi_{k \lambda} \varphi^{k}_{\beta} - \frac{3}{8}{\rm i} G_{i j} (\Sigma_{\hat{a} \hat{b}})^{\rho \lambda} (\Gamma_{c})^{\alpha \beta} \boldsymbol{\lambda}^{i}_{\alpha} {G}^{-5} \nabla^{c}{W} \varphi^{j}_{\beta} \varphi_{k \rho} \varphi^{k}_{\lambda} - \frac{3}{8}{\rm i} G_{i j} (\Sigma_{\hat{a} \hat{b}})^{\rho \lambda} (\Gamma_{c})^{\alpha \beta} \lambda_{k \alpha} {G}^{-5} \nabla^{c}{\boldsymbol{W}} \varphi^{i}_{\rho} \varphi^{j}_{\beta} \varphi^{k}_{\lambda} - \frac{3}{8}{\rm i} G_{i j} (\Sigma_{\hat{a} \hat{b}})^{\rho \lambda} (\Gamma_{c})^{\alpha \beta} \boldsymbol{\lambda}_{k \alpha} {G}^{-5} \nabla^{c}{W} \varphi^{i}_{\rho} \varphi^{j}_{\beta} \varphi^{k}_{\lambda}+\frac{9}{16}{\rm i} G_{i j} (\Sigma_{\hat{a} \hat{b}})^{\alpha \beta} (\Sigma_{c d})^{\rho \lambda} W W^{c d} \boldsymbol{\lambda}^{i}_{\rho} {G}^{-5} \varphi^{j}_{\alpha} \varphi_{k \beta} \varphi^{k}_{\lambda}+\frac{9}{16}{\rm i} G_{i j} (\Sigma_{\hat{a} \hat{b}})^{\alpha \beta} (\Sigma_{c d})^{\rho \lambda} W W^{c d} \boldsymbol{\lambda}^{i}_{\rho} {G}^{-5} \varphi^{j}_{\lambda} \varphi_{k \alpha} \varphi^{k}_{\beta}+\frac{9}{16}{\rm i} G_{i j} (\Sigma_{\hat{a} \hat{b}})^{\alpha \beta} (\Sigma_{c d})^{\rho \lambda} W W^{c d} \boldsymbol{\lambda}_{k \rho} {G}^{-5} \varphi^{i}_{\alpha} \varphi^{j}_{\lambda} \varphi^{k}_{\beta}%
+\frac{9}{16}{\rm i} G_{i j} (\Sigma_{\hat{a} \hat{b}})^{\alpha \beta} (\Sigma_{c d})^{\rho \lambda} \boldsymbol{W} W^{c d} \lambda^{i}_{\rho} {G}^{-5} \varphi^{j}_{\alpha} \varphi_{k \beta} \varphi^{k}_{\lambda}+\frac{9}{16}{\rm i} G_{i j} (\Sigma_{\hat{a} \hat{b}})^{\alpha \beta} (\Sigma_{c d})^{\rho \lambda} \boldsymbol{W} W^{c d} \lambda^{i}_{\rho} {G}^{-5} \varphi^{j}_{\lambda} \varphi_{k \alpha} \varphi^{k}_{\beta}+\frac{9}{16}{\rm i} G_{i j} (\Sigma_{\hat{a} \hat{b}})^{\alpha \beta} (\Sigma_{c d})^{\rho \lambda} \boldsymbol{W} W^{c d} \lambda_{k \rho} {G}^{-5} \varphi^{i}_{\alpha} \varphi^{j}_{\lambda} \varphi^{k}_{\beta} - \frac{3}{32}{\rm i} \epsilon_{\hat{a} \hat{b} c d e} G_{i j} (\Gamma^{c})^{\alpha \beta} \lambda^{i}_{\alpha} {G}^{-3} \nabla^{d}{\boldsymbol{\lambda}_{k \beta}} \nabla^{e}{G^{j k}} - \frac{3}{32}{\rm i} \epsilon_{\hat{a} \hat{b} c d e} G_{i j} (\Gamma^{c})^{\alpha \beta} \boldsymbol{\lambda}^{i}_{\alpha} {G}^{-3} \nabla^{d}{\lambda_{k \beta}} \nabla^{e}{G^{j k}}+\frac{3}{32}{\rm i} \epsilon_{\hat{a} \hat{b} c d e} G_{i j} (\Gamma^{c})^{\alpha \beta} \lambda_{k \alpha} {G}^{-3} \nabla^{d}{\boldsymbol{\lambda}^{i}_{\beta}} \nabla^{e}{G^{j k}} - \frac{3}{32}{\rm i} \epsilon_{\hat{a} \hat{b} c d e} G_{i j} (\Gamma^{c})^{\alpha \beta} \lambda_{k \alpha} {G}^{-3} \nabla^{d}{\boldsymbol{\lambda}^{k}_{\beta}} \nabla^{e}{G^{i j}}+\frac{3}{32}{\rm i} \epsilon_{\hat{a} \hat{b} c d e} G_{i j} (\Gamma^{c})^{\alpha \beta} \boldsymbol{\lambda}_{k \alpha} {G}^{-3} \nabla^{d}{\lambda^{i}_{\beta}} \nabla^{e}{G^{j k}} - \frac{3}{32}{\rm i} \epsilon_{\hat{a} \hat{b} c d e} G_{i j} (\Gamma^{c})^{\alpha \beta} \boldsymbol{\lambda}_{k \alpha} {G}^{-3} \nabla^{d}{\lambda^{k}_{\beta}} \nabla^{e}{G^{i j}} - \frac{1}{192}{\rm i} \epsilon^{e}\,_{\hat{a} \hat{b} {e_{1}} d} G_{i j} (\Gamma^{{e_{1}}})^{\alpha \beta} W \boldsymbol{W} W^{c d} W_{e c \alpha}\,^{i} {G}^{-3} \varphi^{j}_{\beta}+\frac{15}{8}G_{i j} G_{k l} (\Sigma_{\hat{a} \hat{b}})^{\alpha \beta} W \mathbf{X}^{i j} {G}^{-7} \varphi^{k}_{\alpha} \varphi^{l \rho} \varphi_{m \beta} \varphi^{m}_{\rho} - \frac{15}{8}G_{i j} G_{k l} (\Sigma_{\hat{a} \hat{b}})^{\alpha \beta} W \mathbf{X}^{i j} {G}^{-7} \varphi^{k \rho} \varphi^{l}_{\rho} \varphi_{m \alpha} \varphi^{m}_{\beta}+\frac{15}{8}G_{i j} G_{k l} (\Sigma_{\hat{a} \hat{b}})^{\alpha \beta} W \mathbf{X}^{i}\,_{m} {G}^{-7} \varphi^{j}_{\alpha} \varphi^{k \rho} \varphi^{l}_{\rho} \varphi^{m}_{\beta}+\frac{15}{8}G_{i j} G_{k l} (\Sigma_{\hat{a} \hat{b}})^{\alpha \beta} \boldsymbol{W} X^{i j} {G}^{-7} \varphi^{k}_{\alpha} \varphi^{l \rho} \varphi_{m \beta} \varphi^{m}_{\rho} - \frac{15}{8}G_{i j} G_{k l} (\Sigma_{\hat{a} \hat{b}})^{\alpha \beta} \boldsymbol{W} X^{i j} {G}^{-7} \varphi^{k \rho} \varphi^{l}_{\rho} \varphi_{m \alpha} \varphi^{m}_{\beta}+\frac{15}{8}G_{i j} G_{k l} (\Sigma_{\hat{a} \hat{b}})^{\alpha \beta} \boldsymbol{W} X^{i}\,_{m} {G}^{-7} \varphi^{j}_{\alpha} \varphi^{k \rho} \varphi^{l}_{\rho} \varphi^{m}_{\beta} - \frac{3}{4}G_{i j} G_{k l} (\Sigma_{\hat{a} \hat{b}})^{\rho \lambda} (\Gamma_{c})^{\alpha \beta} \lambda^{i}_{\rho} {G}^{-5} \nabla^{c}{\boldsymbol{\lambda}^{k}_{\alpha}} \varphi^{j}_{\lambda} \varphi^{l}_{\beta} - \frac{3}{8}G_{i j} G_{k l} (\Sigma_{\hat{a} \hat{b}})^{\rho \lambda} (\Gamma_{c})^{\alpha \beta} \lambda^{i}_{\rho} {G}^{-5} \nabla^{c}{\boldsymbol{\lambda}^{k}_{\alpha}} \varphi^{j}_{\beta} \varphi^{l}_{\lambda} - \frac{3}{4}G_{i j} G_{k l} (\Sigma_{\hat{a} \hat{b}})^{\rho \lambda} (\Gamma_{c})^{\alpha \beta} \boldsymbol{\lambda}^{i}_{\rho} {G}^{-5} \nabla^{c}{\lambda^{k}_{\alpha}} \varphi^{j}_{\lambda} \varphi^{l}_{\beta} - \frac{3}{8}G_{i j} G_{k l} (\Sigma_{\hat{a} \hat{b}})^{\rho \lambda} (\Gamma_{c})^{\alpha \beta} \boldsymbol{\lambda}^{i}_{\rho} {G}^{-5} \nabla^{c}{\lambda^{k}_{\alpha}} \varphi^{j}_{\beta} \varphi^{l}_{\lambda}%
 - \frac{3}{8}G_{i j} G_{k l} (\Sigma_{\hat{a} \hat{b}})^{\rho \lambda} (\Gamma_{c})^{\alpha \beta} \lambda^{i}_{\alpha} {G}^{-5} \nabla^{c}{\boldsymbol{\lambda}^{k}_{\rho}} \varphi^{j}_{\beta} \varphi^{l}_{\lambda}+\frac{3}{8}G_{i j} G_{k l} (\Sigma_{\hat{a} \hat{b}})^{\rho \lambda} (\Gamma_{c})^{\alpha \beta} \lambda^{i}_{\alpha} {G}^{-5} \nabla^{c}{\boldsymbol{\lambda}^{k}_{\beta}} \varphi^{j}_{\rho} \varphi^{l}_{\lambda} - \frac{3}{8}G_{i j} G_{k l} (\Sigma_{\hat{a} \hat{b}})^{\rho \lambda} (\Gamma_{c})^{\alpha \beta} \boldsymbol{\lambda}^{i}_{\alpha} {G}^{-5} \nabla^{c}{\lambda^{k}_{\rho}} \varphi^{j}_{\beta} \varphi^{l}_{\lambda}+\frac{3}{8}G_{i j} G_{k l} (\Sigma_{\hat{a} \hat{b}})^{\rho \lambda} (\Gamma_{c})^{\alpha \beta} \boldsymbol{\lambda}^{i}_{\alpha} {G}^{-5} \nabla^{c}{\lambda^{k}_{\beta}} \varphi^{j}_{\rho} \varphi^{l}_{\lambda}+\frac{9}{8}G_{i j} G_{k l} (\Sigma_{\hat{a} \hat{b}})^{\alpha \beta} (\Sigma_{c d})^{\rho \lambda} W^{c d} \lambda^{i}_{\rho} \boldsymbol{\lambda}^{k}_{\lambda} {G}^{-5} \varphi^{j}_{\alpha} \varphi^{l}_{\beta}+\frac{9}{32}\epsilon^{e {e_{1}}}\,_{\hat{a}}\,^{c d} G_{i j} G_{k l} (\Sigma_{e {e_{1}}})^{\alpha \beta} W W_{c d} \boldsymbol{\lambda}^{i}_{\alpha} {G}^{-5} \nabla_{\hat{b}}{G^{j k}} \varphi^{l}_{\beta}+\frac{9}{32}\epsilon^{e {e_{1}}}\,_{\hat{a}}\,^{c d} G_{i j} G_{k l} (\Sigma_{e {e_{1}}})^{\alpha \beta} \boldsymbol{W} W_{c d} \lambda^{i}_{\alpha} {G}^{-5} \nabla_{\hat{b}}{G^{j k}} \varphi^{l}_{\beta}+\frac{9}{16}\epsilon^{d e}\,_{\hat{a} {e_{1}}}\,^{c} G_{i j} G_{k l} (\Sigma_{d e})^{\alpha \beta} W W_{\hat{b} c} \boldsymbol{\lambda}^{i}_{\alpha} {G}^{-5} \nabla^{{e_{1}}}{G^{j k}} \varphi^{l}_{\beta}+\frac{9}{16}\epsilon^{d e}\,_{\hat{a} {e_{1}}}\,^{c} G_{i j} G_{k l} (\Sigma_{d e})^{\alpha \beta} \boldsymbol{W} W_{\hat{b} c} \lambda^{i}_{\alpha} {G}^{-5} \nabla^{{e_{1}}}{G^{j k}} \varphi^{l}_{\beta} - \frac{9}{32}\epsilon^{e}\,_{\hat{a} \hat{b}}\,^{c d} G_{i j} G_{k l} (\Sigma_{e {e_{1}}})^{\alpha \beta} W W_{c d} \boldsymbol{\lambda}^{i}_{\alpha} {G}^{-5} \nabla^{{e_{1}}}{G^{j k}} \varphi^{l}_{\beta} - \frac{9}{32}\epsilon^{e}\,_{\hat{a} \hat{b}}\,^{c d} G_{i j} G_{k l} (\Sigma_{e {e_{1}}})^{\alpha \beta} \boldsymbol{W} W_{c d} \lambda^{i}_{\alpha} {G}^{-5} \nabla^{{e_{1}}}{G^{j k}} \varphi^{l}_{\beta} - \frac{3}{64}{\rm i} \epsilon_{\hat{a} \hat{b} e}\,^{c d} G_{i j} G_{k l} (\Gamma^{e})^{\alpha \beta} X^{i j} \mathbf{F}_{c d} {G}^{-5} \varphi^{k}_{\alpha} \varphi^{l}_{\beta} - \frac{3}{64}{\rm i} \epsilon_{\hat{a} \hat{b} e}\,^{c d} G_{i j} G_{k l} (\Gamma^{e})^{\alpha \beta} \mathbf{X}^{i j} F_{c d} {G}^{-5} \varphi^{k}_{\alpha} \varphi^{l}_{\beta}+\frac{3}{64}{\rm i} \epsilon_{\hat{a} \hat{b} e}\,^{c d} G_{i j} G_{k l} (\Gamma^{e})^{\alpha \beta} X^{i k} \mathbf{F}_{c d} {G}^{-5} \varphi^{j}_{\alpha} \varphi^{l}_{\beta}+\frac{3}{64}{\rm i} \epsilon_{\hat{a} \hat{b} e}\,^{c d} G_{i j} G_{k l} (\Gamma^{e})^{\alpha \beta} \mathbf{X}^{i k} F_{c d} {G}^{-5} \varphi^{j}_{\alpha} \varphi^{l}_{\beta} - \frac{3}{32}{\rm i} \epsilon_{\hat{a} \hat{b} e}\,^{c d} G_{i j} (\Gamma^{e})^{\alpha \beta} F_{c d} \boldsymbol{\lambda}^{i}_{\alpha} {G}^{-5} \varphi^{j \rho} \varphi_{k \beta} \varphi^{k}_{\rho} - \frac{3}{32}{\rm i} \epsilon_{\hat{a} \hat{b} e}\,^{c d} G_{i j} (\Gamma^{e})^{\alpha \beta} F_{c d} \boldsymbol{\lambda}_{k \alpha} {G}^{-5} \varphi^{i}_{\beta} \varphi^{j \rho} \varphi^{k}_{\rho}+\frac{3}{32}{\rm i} \epsilon_{\hat{a} \hat{b} e}\,^{c d} G_{i j} (\Gamma^{e})^{\alpha \beta} F_{c d} \boldsymbol{\lambda}_{k \alpha} {G}^{-5} \varphi^{i \rho} \varphi^{j}_{\rho} \varphi^{k}_{\beta} - \frac{3}{32}{\rm i} \epsilon_{\hat{a} \hat{b} e}\,^{c d} G_{i j} (\Gamma^{e})^{\alpha \beta} \mathbf{F}_{c d} \lambda^{i}_{\alpha} {G}^{-5} \varphi^{j \rho} \varphi_{k \beta} \varphi^{k}_{\rho} - \frac{3}{32}{\rm i} \epsilon_{\hat{a} \hat{b} e}\,^{c d} G_{i j} (\Gamma^{e})^{\alpha \beta} \mathbf{F}_{c d} \lambda_{k \alpha} {G}^{-5} \varphi^{i}_{\beta} \varphi^{j \rho} \varphi^{k}_{\rho}%
+\frac{3}{32}{\rm i} \epsilon_{\hat{a} \hat{b} e}\,^{c d} G_{i j} (\Gamma^{e})^{\alpha \beta} \mathbf{F}_{c d} \lambda_{k \alpha} {G}^{-5} \varphi^{i \rho} \varphi^{j}_{\rho} \varphi^{k}_{\beta}+\frac{3}{8}{\rm i} \epsilon^{c d}\,_{\hat{a} \hat{b} e} G_{i j} (\Sigma_{c d})^{\alpha \beta} W {G}^{-5} \nabla^{e}{\boldsymbol{\lambda}^{i}_{\alpha}} \varphi^{j \rho} \varphi_{k \beta} \varphi^{k}_{\rho}+\frac{3}{8}{\rm i} \epsilon^{c d}\,_{\hat{a} \hat{b} e} G_{i j} (\Sigma_{c d})^{\alpha \beta} W {G}^{-5} \nabla^{e}{\boldsymbol{\lambda}_{k \alpha}} \varphi^{i}_{\beta} \varphi^{j \rho} \varphi^{k}_{\rho} - \frac{3}{8}{\rm i} \epsilon^{c d}\,_{\hat{a} \hat{b} e} G_{i j} (\Sigma_{c d})^{\alpha \beta} W {G}^{-5} \nabla^{e}{\boldsymbol{\lambda}_{k \alpha}} \varphi^{i \rho} \varphi^{j}_{\rho} \varphi^{k}_{\beta}+\frac{3}{8}{\rm i} \epsilon^{c d}\,_{\hat{a} \hat{b} e} G_{i j} (\Sigma_{c d})^{\alpha \beta} \boldsymbol{W} {G}^{-5} \nabla^{e}{\lambda^{i}_{\alpha}} \varphi^{j \rho} \varphi_{k \beta} \varphi^{k}_{\rho}+\frac{3}{8}{\rm i} \epsilon^{c d}\,_{\hat{a} \hat{b} e} G_{i j} (\Sigma_{c d})^{\alpha \beta} \boldsymbol{W} {G}^{-5} \nabla^{e}{\lambda_{k \alpha}} \varphi^{i}_{\beta} \varphi^{j \rho} \varphi^{k}_{\rho} - \frac{3}{8}{\rm i} \epsilon^{c d}\,_{\hat{a} \hat{b} e} G_{i j} (\Sigma_{c d})^{\alpha \beta} \boldsymbol{W} {G}^{-5} \nabla^{e}{\lambda_{k \alpha}} \varphi^{i \rho} \varphi^{j}_{\rho} \varphi^{k}_{\beta}+\frac{3}{16}{\rm i} \epsilon^{c d}\,_{\hat{a} \hat{b} e} G_{i j} (\Sigma_{c d})^{\alpha \beta} \lambda^{i}_{\alpha} {G}^{-5} \nabla^{e}{\boldsymbol{W}} \varphi^{j \rho} \varphi_{k \beta} \varphi^{k}_{\rho}+\frac{3}{16}{\rm i} \epsilon^{c d}\,_{\hat{a} \hat{b} e} G_{i j} (\Sigma_{c d})^{\alpha \beta} \boldsymbol{\lambda}^{i}_{\alpha} {G}^{-5} \nabla^{e}{W} \varphi^{j \rho} \varphi_{k \beta} \varphi^{k}_{\rho}+\frac{3}{16}{\rm i} \epsilon^{c d}\,_{\hat{a} \hat{b} e} G_{i j} (\Sigma_{c d})^{\alpha \beta} \lambda_{k \alpha} {G}^{-5} \nabla^{e}{\boldsymbol{W}} \varphi^{i}_{\beta} \varphi^{j \rho} \varphi^{k}_{\rho} - \frac{3}{16}{\rm i} \epsilon^{c d}\,_{\hat{a} \hat{b} e} G_{i j} (\Sigma_{c d})^{\alpha \beta} \lambda_{k \alpha} {G}^{-5} \nabla^{e}{\boldsymbol{W}} \varphi^{i \rho} \varphi^{j}_{\rho} \varphi^{k}_{\beta}+\frac{3}{16}{\rm i} \epsilon^{c d}\,_{\hat{a} \hat{b} e} G_{i j} (\Sigma_{c d})^{\alpha \beta} \boldsymbol{\lambda}_{k \alpha} {G}^{-5} \nabla^{e}{W} \varphi^{i}_{\beta} \varphi^{j \rho} \varphi^{k}_{\rho} - \frac{3}{16}{\rm i} \epsilon^{c d}\,_{\hat{a} \hat{b} e} G_{i j} (\Sigma_{c d})^{\alpha \beta} \boldsymbol{\lambda}_{k \alpha} {G}^{-5} \nabla^{e}{W} \varphi^{i \rho} \varphi^{j}_{\rho} \varphi^{k}_{\beta} - \frac{9}{32}\epsilon^{c d}\,_{\hat{a} \hat{b} e} G_{i j} G_{k l} (\Sigma_{c d})^{\alpha \beta} X^{i j} \boldsymbol{\lambda}_{m \alpha} {G}^{-5} \nabla^{e}{G^{k m}} \varphi^{l}_{\beta} - \frac{9}{32}\epsilon^{c d}\,_{\hat{a} \hat{b} e} G_{i j} G_{k l} (\Sigma_{c d})^{\alpha \beta} \mathbf{X}^{i j} \lambda_{m \alpha} {G}^{-5} \nabla^{e}{G^{k m}} \varphi^{l}_{\beta}+\frac{9}{32}\epsilon^{c d}\,_{\hat{a} \hat{b} e} G_{i j} G_{k l} (\Sigma_{c d})^{\alpha \beta} X^{i}\,_{m} \boldsymbol{\lambda}^{j}_{\alpha} {G}^{-5} \nabla^{e}{G^{k m}} \varphi^{l}_{\beta} - \frac{9}{32}\epsilon^{c d}\,_{\hat{a} \hat{b} e} G_{i j} G_{k l} (\Sigma_{c d})^{\alpha \beta} X^{i}\,_{m} \boldsymbol{\lambda}^{m}_{\alpha} {G}^{-5} \nabla^{e}{G^{j k}} \varphi^{l}_{\beta}+\frac{9}{32}\epsilon^{c d}\,_{\hat{a} \hat{b} e} G_{i j} G_{k l} (\Sigma_{c d})^{\alpha \beta} \mathbf{X}^{i}\,_{m} \lambda^{j}_{\alpha} {G}^{-5} \nabla^{e}{G^{k m}} \varphi^{l}_{\beta} - \frac{9}{32}\epsilon^{c d}\,_{\hat{a} \hat{b} e} G_{i j} G_{k l} (\Sigma_{c d})^{\alpha \beta} \mathbf{X}^{i}\,_{m} \lambda^{m}_{\alpha} {G}^{-5} \nabla^{e}{G^{j k}} \varphi^{l}_{\beta}+\frac{3}{16}\epsilon^{c d}\,_{\hat{a} \hat{b} e} G_{i j} G_{k l} (\Sigma_{c d})^{\alpha \beta} \lambda^{i}_{\alpha} {G}^{-5} \nabla^{e}{\boldsymbol{\lambda}^{k \rho}} \varphi^{j}_{\rho} \varphi^{l}_{\beta}%
+\frac{3}{16}\epsilon^{c d}\,_{\hat{a} \hat{b} e} G_{i j} G_{k l} (\Sigma_{c d})^{\alpha \beta} \boldsymbol{\lambda}^{i}_{\alpha} {G}^{-5} \nabla^{e}{\lambda^{k \rho}} \varphi^{j}_{\rho} \varphi^{l}_{\beta} - \frac{3}{32}\epsilon^{c d}\,_{\hat{a} \hat{b} e} G_{i j} G_{k l} (\Sigma_{c d})^{\alpha \beta} \lambda^{i \rho} {G}^{-5} \nabla^{e}{\boldsymbol{\lambda}^{j}_{\alpha}} \varphi^{k}_{\beta} \varphi^{l}_{\rho}+\frac{9}{32}\epsilon^{c d}\,_{\hat{a} \hat{b} e} G_{i j} G_{k l} (\Sigma_{c d})^{\alpha \beta} \lambda^{i \rho} {G}^{-5} \nabla^{e}{\boldsymbol{\lambda}^{k}_{\alpha}} \varphi^{j}_{\beta} \varphi^{l}_{\rho} - \frac{3}{16}\epsilon^{c d}\,_{\hat{a} \hat{b} e} G_{i j} G_{k l} (\Sigma_{c d})^{\alpha \beta} \lambda^{i \rho} {G}^{-5} \nabla^{e}{\boldsymbol{\lambda}^{k}_{\rho}} \varphi^{j}_{\alpha} \varphi^{l}_{\beta} - \frac{3}{32}\epsilon^{c d}\,_{\hat{a} \hat{b} e} G_{i j} G_{k l} (\Sigma_{c d})^{\alpha \beta} \boldsymbol{\lambda}^{i \rho} {G}^{-5} \nabla^{e}{\lambda^{j}_{\alpha}} \varphi^{k}_{\beta} \varphi^{l}_{\rho}+\frac{9}{32}\epsilon^{c d}\,_{\hat{a} \hat{b} e} G_{i j} G_{k l} (\Sigma_{c d})^{\alpha \beta} \boldsymbol{\lambda}^{i \rho} {G}^{-5} \nabla^{e}{\lambda^{k}_{\alpha}} \varphi^{j}_{\beta} \varphi^{l}_{\rho} - \frac{3}{16}\epsilon^{c d}\,_{\hat{a} \hat{b} e} G_{i j} G_{k l} (\Sigma_{c d})^{\alpha \beta} \boldsymbol{\lambda}^{i \rho} {G}^{-5} \nabla^{e}{\lambda^{k}_{\rho}} \varphi^{j}_{\alpha} \varphi^{l}_{\beta} - \frac{3}{8}\epsilon^{e}\,_{\hat{a} \hat{b} {e_{1}}}\,^{c} G_{i j} G_{k l} (\Sigma_{e}{}^{\, d})^{\alpha \beta} F_{c d} \boldsymbol{\lambda}^{i}_{\alpha} {G}^{-5} \nabla^{{e_{1}}}{G^{j k}} \varphi^{l}_{\beta} - \frac{3}{8}\epsilon^{e}\,_{\hat{a} \hat{b} {e_{1}}}\,^{c} G_{i j} G_{k l} (\Sigma_{e}{}^{\, d})^{\alpha \beta} \mathbf{F}_{c d} \lambda^{i}_{\alpha} {G}^{-5} \nabla^{{e_{1}}}{G^{j k}} \varphi^{l}_{\beta} - \frac{15}{16}{\rm i} G_{i j} G_{k l} G_{m n} (\Sigma_{\hat{a} \hat{b}})^{\alpha \beta} X^{i j} \boldsymbol{\lambda}^{k}_{\alpha} {G}^{-7} \varphi^{l}_{\beta} \varphi^{m \rho} \varphi^{n}_{\rho} - \frac{15}{16}{\rm i} G_{i j} G_{k l} G_{m n} (\Sigma_{\hat{a} \hat{b}})^{\alpha \beta} \mathbf{X}^{i j} \lambda^{k}_{\alpha} {G}^{-7} \varphi^{l}_{\beta} \varphi^{m \rho} \varphi^{n}_{\rho}+\frac{15}{16}{\rm i} G_{i j} G_{k l} G_{m n} (\Sigma_{\hat{a} \hat{b}})^{\alpha \beta} X^{i k} \boldsymbol{\lambda}^{j}_{\alpha} {G}^{-7} \varphi^{l}_{\beta} \varphi^{m \rho} \varphi^{n}_{\rho}+\frac{15}{16}{\rm i} G_{i j} G_{k l} G_{m n} (\Sigma_{\hat{a} \hat{b}})^{\alpha \beta} \mathbf{X}^{i k} \lambda^{j}_{\alpha} {G}^{-7} \varphi^{l}_{\beta} \varphi^{m \rho} \varphi^{n}_{\rho} - \frac{15}{8}{\rm i} G_{i j} G_{k l} (\Sigma_{\hat{a} \hat{b}})^{\alpha \beta} \lambda^{i \rho} \boldsymbol{\lambda}^{j}_{\rho} {G}^{-7} \varphi^{k}_{\alpha} \varphi^{l \lambda} \varphi_{m \beta} \varphi^{m}_{\lambda}+\frac{15}{8}{\rm i} G_{i j} G_{k l} (\Sigma_{\hat{a} \hat{b}})^{\alpha \beta} \lambda^{i \rho} \boldsymbol{\lambda}^{j}_{\rho} {G}^{-7} \varphi^{k \lambda} \varphi^{l}_{\lambda} \varphi_{m \alpha} \varphi^{m}_{\beta} - \frac{15}{16}{\rm i} G_{i j} G_{k l} (\Sigma_{\hat{a} \hat{b}})^{\alpha \beta} \lambda^{i \rho} \boldsymbol{\lambda}_{m \rho} {G}^{-7} \varphi^{j}_{\alpha} \varphi^{k \lambda} \varphi^{l}_{\lambda} \varphi^{m}_{\beta} - \frac{15}{16}{\rm i} G_{i j} G_{k l} (\Sigma_{\hat{a} \hat{b}})^{\alpha \beta} \lambda^{\rho}_{m} \boldsymbol{\lambda}^{i}_{\rho} {G}^{-7} \varphi^{j}_{\alpha} \varphi^{k \lambda} \varphi^{l}_{\lambda} \varphi^{m}_{\beta} - \frac{3}{64}{\rm i} \epsilon_{\hat{a} \hat{b} e}\,^{c d} G_{i j} G_{k l} (\Gamma^{e})^{\alpha \beta} W \mathbf{X}^{i j} W_{c d} {G}^{-5} \varphi^{k}_{\alpha} \varphi^{l}_{\beta}+\frac{3}{64}{\rm i} \epsilon_{\hat{a} \hat{b} e}\,^{c d} G_{i j} G_{k l} (\Gamma^{e})^{\alpha \beta} W \mathbf{X}^{i k} W_{c d} {G}^{-5} \varphi^{j}_{\alpha} \varphi^{l}_{\beta} - \frac{3}{64}{\rm i} \epsilon_{\hat{a} \hat{b} e}\,^{c d} G_{i j} G_{k l} (\Gamma^{e})^{\alpha \beta} \boldsymbol{W} X^{i j} W_{c d} {G}^{-5} \varphi^{k}_{\alpha} \varphi^{l}_{\beta}%
+\frac{3}{64}{\rm i} \epsilon_{\hat{a} \hat{b} e}\,^{c d} G_{i j} G_{k l} (\Gamma^{e})^{\alpha \beta} \boldsymbol{W} X^{i k} W_{c d} {G}^{-5} \varphi^{j}_{\alpha} \varphi^{l}_{\beta} - \frac{9}{64}{\rm i} \epsilon_{\hat{a} \hat{b} e}\,^{c d} G_{i j} (\Gamma^{e})^{\alpha \beta} W W_{c d} \boldsymbol{\lambda}^{i}_{\alpha} {G}^{-5} \varphi^{j \rho} \varphi_{k \beta} \varphi^{k}_{\rho} - \frac{9}{64}{\rm i} \epsilon_{\hat{a} \hat{b} e}\,^{c d} G_{i j} (\Gamma^{e})^{\alpha \beta} W W_{c d} \boldsymbol{\lambda}_{k \alpha} {G}^{-5} \varphi^{i}_{\beta} \varphi^{j \rho} \varphi^{k}_{\rho}+\frac{9}{64}{\rm i} \epsilon_{\hat{a} \hat{b} e}\,^{c d} G_{i j} (\Gamma^{e})^{\alpha \beta} W W_{c d} \boldsymbol{\lambda}_{k \alpha} {G}^{-5} \varphi^{i \rho} \varphi^{j}_{\rho} \varphi^{k}_{\beta} - \frac{9}{64}{\rm i} \epsilon_{\hat{a} \hat{b} e}\,^{c d} G_{i j} (\Gamma^{e})^{\alpha \beta} \boldsymbol{W} W_{c d} \lambda^{i}_{\alpha} {G}^{-5} \varphi^{j \rho} \varphi_{k \beta} \varphi^{k}_{\rho} - \frac{9}{64}{\rm i} \epsilon_{\hat{a} \hat{b} e}\,^{c d} G_{i j} (\Gamma^{e})^{\alpha \beta} \boldsymbol{W} W_{c d} \lambda_{k \alpha} {G}^{-5} \varphi^{i}_{\beta} \varphi^{j \rho} \varphi^{k}_{\rho}+\frac{9}{64}{\rm i} \epsilon_{\hat{a} \hat{b} e}\,^{c d} G_{i j} (\Gamma^{e})^{\alpha \beta} \boldsymbol{W} W_{c d} \lambda_{k \alpha} {G}^{-5} \varphi^{i \rho} \varphi^{j}_{\rho} \varphi^{k}_{\beta} - \frac{3}{256}\epsilon_{\hat{a} \hat{b} e}\,^{c d} G_{i j} G_{k l} (\Gamma^{e})^{\alpha \beta} W_{c d} \lambda^{i}_{\alpha} \boldsymbol{\lambda}^{j \rho} {G}^{-5} \varphi^{k}_{\beta} \varphi^{l}_{\rho}+\frac{3}{128}\epsilon_{\hat{a} \hat{b} e}\,^{c d} G_{i j} G_{k l} (\Gamma^{e})^{\alpha \beta} W_{c d} \lambda^{i}_{\alpha} \boldsymbol{\lambda}^{k \rho} {G}^{-5} \varphi^{j}_{\beta} \varphi^{l}_{\rho}+\frac{3}{256}\epsilon_{\hat{a} \hat{b} e}\,^{c d} G_{i j} G_{k l} (\Gamma^{e})^{\alpha \beta} W_{c d} \lambda^{i}_{\alpha} \boldsymbol{\lambda}^{k \rho} {G}^{-5} \varphi^{j}_{\rho} \varphi^{l}_{\beta}+\frac{3}{256}\epsilon_{\hat{a} \hat{b} e}\,^{c d} G_{i j} G_{k l} (\Gamma^{e})^{\alpha \beta} W_{c d} \lambda^{i \rho} \boldsymbol{\lambda}^{j}_{\alpha} {G}^{-5} \varphi^{k}_{\beta} \varphi^{l}_{\rho}+\frac{3}{256}\epsilon_{\hat{a} \hat{b} e}\,^{c d} G_{i j} G_{k l} (\Gamma^{e})^{\alpha \beta} W_{c d} \lambda^{i \rho} \boldsymbol{\lambda}^{k}_{\alpha} {G}^{-5} \varphi^{j}_{\beta} \varphi^{l}_{\rho}+\frac{3}{128}\epsilon_{\hat{a} \hat{b} e}\,^{c d} G_{i j} G_{k l} (\Gamma^{e})^{\alpha \beta} W_{c d} \lambda^{i \rho} \boldsymbol{\lambda}^{k}_{\alpha} {G}^{-5} \varphi^{j}_{\rho} \varphi^{l}_{\beta} - \frac{9}{16}\epsilon^{e}\,_{\hat{a} \hat{b} {e_{1}}}\,^{c} G_{i j} G_{k l} (\Sigma_{e}{}^{\, d})^{\alpha \beta} W W_{c d} \boldsymbol{\lambda}^{i}_{\alpha} {G}^{-5} \nabla^{{e_{1}}}{G^{j k}} \varphi^{l}_{\beta} - \frac{9}{16}\epsilon^{e}\,_{\hat{a} \hat{b} {e_{1}}}\,^{c} G_{i j} G_{k l} (\Sigma_{e}{}^{\, d})^{\alpha \beta} \boldsymbol{W} W_{c d} \lambda^{i}_{\alpha} {G}^{-5} \nabla^{{e_{1}}}{G^{j k}} \varphi^{l}_{\beta}+\frac{15}{8}{\rm i} G_{i j} G_{k l} G_{m n} (\Gamma_{\hat{a}})^{\alpha \beta} W \mathbf{X}^{i j} {G}^{-7} \nabla_{\hat{b}}{G^{k m}} \varphi^{l}_{\alpha} \varphi^{n}_{\beta}+\frac{15}{8}{\rm i} G_{i j} G_{k l} G_{m n} (\Gamma_{\hat{a}})^{\alpha \beta} \boldsymbol{W} X^{i j} {G}^{-7} \nabla_{\hat{b}}{G^{k m}} \varphi^{l}_{\alpha} \varphi^{n}_{\beta}+\frac{15}{8}G_{i j} G_{k l} G_{m n} (\Gamma_{\hat{a}})^{\alpha \beta} \lambda^{i \rho} \boldsymbol{\lambda}^{j}_{\rho} {G}^{-7} \nabla_{\hat{b}}{G^{k m}} \varphi^{l}_{\alpha} \varphi^{n}_{\beta}
\doublespacedmathend
\end{adjustwidth}

\subsubsection{$X_{i j, R^2}$}

\begin{adjustwidth}{0cm}{5cm}
\doublespacedmathbegin
- \frac{1}{16}(\Gamma_{a})^{\alpha \beta} \lambda^{\rho}_{\underline{i}} {G}^{-3} \nabla^{a}{\boldsymbol{\lambda}_{\underline{j} \alpha}} \varphi_{k \beta} \varphi^{k}_{\rho}+\frac{1}{64}(\Sigma_{a b})^{\alpha \beta} W^{a b} \lambda^{\rho}_{\underline{i}} \boldsymbol{\lambda}_{\underline{j} \alpha} {G}^{-3} \varphi_{k \beta} \varphi^{k}_{\rho}+\frac{3}{64}{\rm i} \boldsymbol{W} \lambda^{\beta}_{\underline{i}} X_{\underline{j}}^{\alpha} {G}^{-3} \varphi_{k \beta} \varphi^{k}_{\alpha}+\frac{1}{16}{\rm i} (\Sigma_{a b})^{\alpha \beta} \mathbf{X}_{\underline{i} \underline{j}} F^{a b} {G}^{-3} \varphi_{k \alpha} \varphi^{k}_{\beta}+\frac{1}{16}{\rm i} (\Sigma_{a b})^{\alpha \beta} W \mathbf{X}_{\underline{i} \underline{j}} W^{a b} {G}^{-3} \varphi_{k \alpha} \varphi^{k}_{\beta} - \frac{9}{16}{\rm i} G_{\underline{i} k} \mathbf{X}_{\underline{j} l} \lambda^{l \alpha} {G}^{-5} \varphi^{k \beta} \varphi_{m \alpha} \varphi^{m}_{\beta} - \frac{15}{32}F \mathbf{X}_{\underline{i} k} \lambda^{k \alpha} {G}^{-3} \varphi_{\underline{j} \alpha}+\frac{3}{32}\mathcal{H}_{a} (\Gamma^{a})^{\alpha \beta} \mathbf{X}_{\underline{i} k} \lambda^{k}_{\alpha} {G}^{-3} \varphi_{\underline{j} \beta} - \frac{3}{16}(\Gamma_{a})^{\alpha \beta} \mathbf{X}_{\underline{i} l} \lambda^{l}_{\alpha} {G}^{-3} \nabla^{a}{G_{\underline{j} k}} \varphi^{k}_{\beta} - \frac{1}{8}(\Gamma_{a})^{\alpha \beta} \lambda^{\rho}_{\underline{i}} {G}^{-3} \nabla^{a}{\boldsymbol{\lambda}_{k \alpha}} \varphi_{\underline{j} \rho} \varphi^{k}_{\beta} - \frac{3}{16}(\Gamma_{a})^{\alpha \beta} \lambda^{\rho}_{k} {G}^{-3} \nabla^{a}{\boldsymbol{\lambda}^{k}_{\alpha}} \varphi_{\underline{i} \beta} \varphi_{\underline{j} \rho}+\frac{3}{64}(\Sigma_{a b})^{\alpha \beta} W^{a b} \lambda^{\rho}_{k} \boldsymbol{\lambda}^{k}_{\alpha} {G}^{-3} \varphi_{\underline{i} \beta} \varphi_{\underline{j} \rho}+\frac{9}{64}{\rm i} \boldsymbol{W} \lambda^{\beta}_{k} X^{k \alpha} {G}^{-3} \varphi_{\underline{i} \beta} \varphi_{\underline{j} \alpha}+\frac{3}{16}{\rm i} X_{\underline{i} k} \mathbf{X}^{k}\,_{l} {G}^{-3} \varphi_{\underline{j}}^{\alpha} \varphi^{l}_{\alpha} - \frac{1}{4}{\rm i} (\Gamma_{a})^{\alpha \beta} \mathbf{X}_{\underline{i} k} {G}^{-3} \nabla^{a}{W} \varphi_{\underline{j} \alpha} \varphi^{k}_{\beta}+\frac{9}{16}{\rm i} G_{\underline{i} k} \mathbf{X}_{l m} \lambda^{l \alpha} {G}^{-5} \varphi_{\underline{j} \alpha} \varphi^{k \beta} \varphi^{m}_{\beta} - \frac{3}{16}(\Gamma_{a})^{\alpha \beta} \mathbf{X}_{k l} \lambda^{k}_{\alpha} {G}^{-3} \nabla^{a}{G_{\underline{i} \underline{j}}} \varphi^{l}_{\beta}+\frac{7}{32}\epsilon^{c d}\,_{e}\,^{a b} G_{\underline{i} k} (\Sigma_{c d})^{\alpha \beta} F_{a b} {G}^{-3} \nabla^{e}{\boldsymbol{\lambda}_{\underline{j} \alpha}} \varphi^{k}_{\beta} - \frac{9}{16}G_{\underline{i} k} (\Gamma^{a})^{\alpha \beta} F_{a b} {G}^{-3} \nabla^{b}{\boldsymbol{\lambda}_{\underline{j} \alpha}} \varphi^{k}_{\beta}%
 - \frac{1}{32}\epsilon^{c d}\,_{e}\,^{a b} G_{\underline{i} k} (\Sigma_{c d})^{\alpha \beta} W W_{a b} {G}^{-3} \nabla^{e}{\boldsymbol{\lambda}_{\underline{j} \alpha}} \varphi^{k}_{\beta} - \frac{25}{16}G_{\underline{i} k} (\Gamma^{a})^{\alpha \beta} W W_{a b} {G}^{-3} \nabla^{b}{\boldsymbol{\lambda}_{\underline{j} \alpha}} \varphi^{k}_{\beta} - \frac{5}{16}G_{\underline{i} k} (\Gamma_{a})^{\alpha \beta} X_{\underline{j} l} {G}^{-3} \nabla^{a}{\boldsymbol{\lambda}^{l}_{\alpha}} \varphi^{k}_{\beta}+\frac{25}{16}G_{\underline{i} k} {G}^{-3} \nabla_{a}{W} \nabla^{a}{\boldsymbol{\lambda}_{\underline{j}}^{\alpha}} \varphi^{k}_{\alpha}+\frac{9}{8}G_{\underline{i} k} (\Sigma_{a b})^{\alpha \beta} {G}^{-3} \nabla^{a}{W} \nabla^{b}{\boldsymbol{\lambda}_{\underline{j} \alpha}} \varphi^{k}_{\beta} - \frac{9}{16}G_{\underline{i} k} G_{\underline{j} l} (\Gamma_{a})^{\alpha \beta} \lambda^{\rho}_{m} {G}^{-5} \nabla^{a}{\boldsymbol{\lambda}^{m}_{\alpha}} \varphi^{k}_{\beta} \varphi^{l}_{\rho}+\frac{3}{4}G_{\underline{i} k} \lambda^{\alpha}_{\underline{j}} {G}^{-3} \nabla_{a}{\nabla^{a}{\boldsymbol{W}}} \varphi^{k}_{\alpha} - \frac{57}{128}G_{\underline{i} k} W^{a b} \mathbf{F}_{a b} \lambda^{\alpha}_{\underline{j}} {G}^{-3} \varphi^{k}_{\alpha} - \frac{17}{32}G_{\underline{i} k} \boldsymbol{W} W^{a b} W_{a b} \lambda^{\alpha}_{\underline{j}} {G}^{-3} \varphi^{k}_{\alpha}+\frac{81}{256}G_{\underline{i} k} \lambda^{\beta}_{l} \boldsymbol{\lambda}^{l \alpha} X_{\underline{j} \alpha} {G}^{-3} \varphi^{k}_{\beta}+\frac{69}{128}G_{\underline{i} k} \lambda^{\beta}_{\underline{j}} \boldsymbol{\lambda}_{l}^{\alpha} X^{l}_{\alpha} {G}^{-3} \varphi^{k}_{\beta} - \frac{57}{256}G_{\underline{i} k} \lambda^{\beta}_{l} \boldsymbol{\lambda}_{\underline{j}}^{\alpha} X^{l}_{\alpha} {G}^{-3} \varphi^{k}_{\beta} - \frac{5}{32}{\rm i} F G_{\underline{i} \underline{j}} (\Gamma_{a})^{\alpha \beta} \lambda_{k \alpha} {G}^{-3} \nabla^{a}{\boldsymbol{\lambda}^{k}_{\beta}}+\frac{3}{32}{\rm i} \mathcal{H}_{a} G_{\underline{i} \underline{j}} \lambda^{\alpha}_{k} {G}^{-3} \nabla^{a}{\boldsymbol{\lambda}^{k}_{\alpha}} - \frac{3}{16}{\rm i} \mathcal{H}^{a} G_{\underline{i} \underline{j}} (\Sigma_{a b})^{\alpha \beta} \lambda_{k \alpha} {G}^{-3} \nabla^{b}{\boldsymbol{\lambda}^{k}_{\beta}}+\frac{3}{16}{\rm i} G_{\underline{i} k} \lambda^{\alpha}_{l} {G}^{-3} \nabla_{a}{\boldsymbol{\lambda}^{l}_{\alpha}} \nabla^{a}{G_{\underline{j}}\,^{k}}+\frac{3}{8}{\rm i} G_{\underline{i} k} (\Sigma_{a b})^{\alpha \beta} \lambda_{l \alpha} {G}^{-3} \nabla^{a}{\boldsymbol{\lambda}^{l}_{\beta}} \nabla^{b}{G_{\underline{j}}\,^{k}} - \frac{5}{16}G_{k l} (\Gamma_{a})^{\alpha \beta} X_{\underline{i} \underline{j}} {G}^{-3} \nabla^{a}{\boldsymbol{\lambda}^{k}_{\alpha}} \varphi^{l}_{\beta} - \frac{9}{16}G_{\underline{i} k} G_{l m} (\Gamma_{a})^{\alpha \beta} \lambda^{\rho}_{\underline{j}} {G}^{-5} \nabla^{a}{\boldsymbol{\lambda}^{l}_{\alpha}} \varphi^{k}_{\beta} \varphi^{m}_{\rho}+\frac{17}{256}G_{k l} \lambda^{\beta}_{\underline{i}} \boldsymbol{\lambda}^{k \alpha} X_{\underline{j} \alpha} {G}^{-3} \varphi^{l}_{\beta}%
 - \frac{9}{256}G_{k l} \lambda^{\beta}_{\underline{i}} \boldsymbol{\lambda}_{\underline{j}}^{\alpha} X^{k}_{\alpha} {G}^{-3} \varphi^{l}_{\beta} - \frac{3}{32}{\rm i} F G_{\underline{i} k} (\Gamma_{a})^{\alpha \beta} \lambda_{\underline{j} \alpha} {G}^{-3} \nabla^{a}{\boldsymbol{\lambda}^{k}_{\beta}}+\frac{5}{32}{\rm i} \mathcal{H}_{a} G_{\underline{i} k} \lambda^{\alpha}_{\underline{j}} {G}^{-3} \nabla^{a}{\boldsymbol{\lambda}^{k}_{\alpha}} - \frac{5}{16}{\rm i} \mathcal{H}^{a} G_{\underline{i} k} (\Sigma_{a b})^{\alpha \beta} \lambda_{\underline{j} \alpha} {G}^{-3} \nabla^{b}{\boldsymbol{\lambda}^{k}_{\beta}}+\frac{7}{16}{\rm i} G_{k l} \lambda^{\alpha}_{\underline{i}} {G}^{-3} \nabla_{a}{\boldsymbol{\lambda}^{k}_{\alpha}} \nabla^{a}{G_{\underline{j}}\,^{l}}+\frac{7}{8}{\rm i} G_{k l} (\Sigma_{a b})^{\alpha \beta} \lambda_{\underline{i} \alpha} {G}^{-3} \nabla^{a}{\boldsymbol{\lambda}^{k}_{\beta}} \nabla^{b}{G_{\underline{j}}\,^{l}} - \frac{57}{128}G_{\underline{i} k} W^{a b} F_{a b} \boldsymbol{\lambda}_{\underline{j}}^{\alpha} {G}^{-3} \varphi^{k}_{\alpha}+\frac{25}{256}\epsilon_{e}\,^{c d a b} G_{\underline{i} k} (\Gamma^{e})^{\alpha \beta} W_{c d} F_{a b} \boldsymbol{\lambda}_{\underline{j} \alpha} {G}^{-3} \varphi^{k}_{\beta}+\frac{23}{32}G_{\underline{i} k} (\Sigma^{a}{}_{\, c})^{\alpha \beta} W^{c b} F_{a b} \boldsymbol{\lambda}_{\underline{j} \alpha} {G}^{-3} \varphi^{k}_{\beta} - \frac{17}{32}G_{\underline{i} k} W W^{a b} W_{a b} \boldsymbol{\lambda}_{\underline{j}}^{\alpha} {G}^{-3} \varphi^{k}_{\alpha}+\frac{37}{256}\epsilon_{e}\,^{a b c d} G_{\underline{i} k} (\Gamma^{e})^{\alpha \beta} W W_{a b} W_{c d} \boldsymbol{\lambda}_{\underline{j} \alpha} {G}^{-3} \varphi^{k}_{\beta} - \frac{11}{64}G_{\underline{i} k} (\Sigma_{a b})^{\alpha \beta} X_{\underline{j} l} W^{a b} \boldsymbol{\lambda}^{l}_{\alpha} {G}^{-3} \varphi^{k}_{\beta}+\frac{25}{128}\epsilon^{c d}\,_{e}\,^{a b} G_{\underline{i} k} (\Sigma_{c d})^{\alpha \beta} W_{a b} \boldsymbol{\lambda}_{\underline{j} \alpha} {G}^{-3} \nabla^{e}{W} \varphi^{k}_{\beta} - \frac{73}{64}G_{\underline{i} k} (\Gamma^{a})^{\alpha \beta} W_{a b} \boldsymbol{\lambda}_{\underline{j} \alpha} {G}^{-3} \nabla^{b}{W} \varphi^{k}_{\beta}+\frac{9}{64}G_{\underline{i} k} G_{\underline{j} l} (\Sigma_{a b})^{\alpha \beta} W^{a b} \lambda^{\rho}_{m} \boldsymbol{\lambda}^{m}_{\alpha} {G}^{-5} \varphi^{k}_{\beta} \varphi^{l}_{\rho} - \frac{21}{64}{\rm i} F G_{\underline{i} \underline{j}} (\Sigma_{a b})^{\alpha \beta} W^{a b} \lambda_{k \alpha} \boldsymbol{\lambda}^{k}_{\beta} {G}^{-3} - \frac{3}{128}{\rm i} \epsilon^{c d e a b} \mathcal{H}_{c} G_{\underline{i} \underline{j}} (\Sigma_{d e})^{\alpha \beta} W_{a b} \lambda_{k \alpha} \boldsymbol{\lambda}^{k}_{\beta} {G}^{-3} - \frac{3}{64}{\rm i} \epsilon^{c d}\,_{e}\,^{a b} G_{\underline{i} k} (\Sigma_{c d})^{\alpha \beta} W_{a b} \lambda_{l \alpha} \boldsymbol{\lambda}^{l}_{\beta} {G}^{-3} \nabla^{e}{G_{\underline{j}}\,^{k}} - \frac{1}{32}(\Sigma_{a b})^{\alpha \beta} W^{a b} \lambda^{\rho}_{\underline{i}} \boldsymbol{\lambda}_{k \alpha} {G}^{-3} \varphi_{\underline{j} \rho} \varphi^{k}_{\beta} - \frac{11}{64}G_{k l} (\Sigma_{a b})^{\alpha \beta} X_{\underline{i} \underline{j}} W^{a b} \boldsymbol{\lambda}^{k}_{\alpha} {G}^{-3} \varphi^{l}_{\beta}%
 - \frac{3}{64}G_{\underline{i} k} G_{l m} (\Sigma_{a b})^{\alpha \beta} W^{a b} \lambda^{\rho}_{\underline{j}} \boldsymbol{\lambda}^{l}_{\alpha} {G}^{-5} \varphi^{k}_{\beta} \varphi^{m}_{\rho} - \frac{3}{64}{\rm i} F G_{\underline{i} k} (\Sigma_{a b})^{\alpha \beta} W^{a b} \lambda_{\underline{j} \alpha} \boldsymbol{\lambda}^{k}_{\beta} {G}^{-3}+\frac{3}{128}{\rm i} \epsilon^{c d e a b} \mathcal{H}_{c} G_{\underline{i} k} (\Sigma_{d e})^{\alpha \beta} W_{a b} \lambda_{\underline{j} \alpha} \boldsymbol{\lambda}^{k}_{\beta} {G}^{-3}+\frac{3}{128}{\rm i} \epsilon^{c d}\,_{e}\,^{a b} G_{k l} (\Sigma_{c d})^{\alpha \beta} W_{a b} \lambda_{\underline{i} \alpha} \boldsymbol{\lambda}^{k}_{\beta} {G}^{-3} \nabla^{e}{G_{\underline{j}}\,^{l}} - \frac{1}{64}{\rm i} G_{k l} (\Gamma^{a})^{\alpha \beta} W_{a b} \lambda_{\underline{i} \alpha} \boldsymbol{\lambda}^{k}_{\beta} {G}^{-3} \nabla^{b}{G_{\underline{j}}\,^{l}}+\frac{3}{64}{\rm i} G_{\underline{i} k} (\Sigma_{a b})^{\alpha \beta} \boldsymbol{W} F^{a b} X_{\underline{j} \alpha} {G}^{-3} \varphi^{k}_{\beta}+\frac{39}{32}{\rm i} G_{\underline{i} k} (\Sigma_{a b})^{\alpha \beta} W \boldsymbol{W} W^{a b} X_{\underline{j} \alpha} {G}^{-3} \varphi^{k}_{\beta} - \frac{51}{64}{\rm i} G_{\underline{i} k} \boldsymbol{W} X_{\underline{j} l} X^{l \alpha} {G}^{-3} \varphi^{k}_{\alpha}+\frac{99}{64}{\rm i} G_{\underline{i} k} (\Gamma_{a})^{\alpha \beta} \boldsymbol{W} X_{\underline{j} \alpha} {G}^{-3} \nabla^{a}{W} \varphi^{k}_{\beta}+\frac{3}{128}G_{\underline{i} k} \boldsymbol{W} Y \lambda^{\alpha}_{\underline{j}} {G}^{-3} \varphi^{k}_{\alpha}+\frac{27}{64}{\rm i} G_{\underline{i} k} G_{\underline{j} l} \boldsymbol{W} \lambda^{\beta}_{m} X^{m \alpha} {G}^{-5} \varphi^{k}_{\beta} \varphi^{l}_{\alpha} - \frac{87}{128}F G_{\underline{i} \underline{j}} \boldsymbol{W} \lambda^{\alpha}_{k} X^{k}_{\alpha} {G}^{-3}+\frac{9}{128}\mathcal{H}_{a} G_{\underline{i} \underline{j}} (\Gamma^{a})^{\beta \alpha} \boldsymbol{W} \lambda_{k \beta} X^{k}_{\alpha} {G}^{-3}+\frac{9}{64}G_{\underline{i} k} (\Gamma_{a})^{\beta \alpha} \boldsymbol{W} \lambda_{l \beta} X^{l}_{\alpha} {G}^{-3} \nabla^{a}{G_{\underline{j}}\,^{k}}+\frac{9}{64}{\rm i} G_{k l} \boldsymbol{W} X_{\underline{i} \underline{j}} X^{k \alpha} {G}^{-3} \varphi^{l}_{\alpha}+\frac{9}{64}{\rm i} G_{\underline{i} k} G_{l m} \boldsymbol{W} \lambda^{\beta}_{\underline{j}} X^{l \alpha} {G}^{-5} \varphi^{k}_{\alpha} \varphi^{m}_{\beta} - \frac{9}{128}F G_{\underline{i} k} \boldsymbol{W} \lambda^{\alpha}_{\underline{j}} X^{k}_{\alpha} {G}^{-3} - \frac{9}{128}\mathcal{H}_{a} G_{\underline{i} k} (\Gamma^{a})^{\beta \alpha} \boldsymbol{W} \lambda_{\underline{j} \beta} X^{k}_{\alpha} {G}^{-3} - \frac{15}{64}G_{k l} (\Gamma_{a})^{\beta \alpha} \boldsymbol{W} \lambda_{\underline{i} \beta} X^{k}_{\alpha} {G}^{-3} \nabla^{a}{G_{\underline{j}}\,^{l}}+\frac{1}{4}G_{k l} (\Gamma_{a})^{\alpha \beta} \mathbf{X}_{\underline{i}}\,^{k} {G}^{-3} \nabla^{a}{\lambda_{\underline{j} \alpha}} \varphi^{l}_{\beta}%
+\frac{5}{16}G_{k l} (\Sigma_{a b})^{\alpha \beta} \mathbf{X}_{\underline{i}}\,^{k} W^{a b} \lambda_{\underline{j} \alpha} {G}^{-3} \varphi^{l}_{\beta}+\frac{3}{16}{\rm i} G_{\underline{i} k} G_{l m} (\Sigma_{a b})^{\alpha \beta} \mathbf{X}_{\underline{j}}\,^{l} F^{a b} {G}^{-5} \varphi^{k}_{\alpha} \varphi^{m}_{\beta} - \frac{7}{8}{\rm i} (\Sigma_{a b})^{\alpha \beta} W \mathbf{X}_{\underline{i} k} W^{a b} {G}^{-3} \varphi_{\underline{j} \alpha} \varphi^{k}_{\beta}+\frac{1}{32}{\rm i} G_{k l} W \mathbf{X}_{\underline{i}}\,^{k} X_{\underline{j}}^{\alpha} {G}^{-3} \varphi^{l}_{\alpha}+\frac{3}{16}{\rm i} G_{\underline{i} k} G_{l m} (\Sigma_{a b})^{\alpha \beta} W \mathbf{X}_{\underline{j}}\,^{l} W^{a b} {G}^{-5} \varphi^{k}_{\alpha} \varphi^{m}_{\beta} - \frac{9}{16}{\rm i} G_{\underline{i} k} G_{l m} X_{\underline{j} n} \mathbf{X}^{l n} {G}^{-5} \varphi^{k \alpha} \varphi^{m}_{\alpha}+\frac{3}{8}F G_{\underline{i} k} X_{\underline{j} l} \mathbf{X}^{k l} {G}^{-3}+\frac{15}{16}{\rm i} G_{\underline{i} k} G_{l m} (\Gamma_{a})^{\alpha \beta} \mathbf{X}_{\underline{j}}\,^{l} {G}^{-5} \nabla^{a}{W} \varphi^{k}_{\alpha} \varphi^{m}_{\beta} - \frac{1}{2}\mathcal{H}_{a} G_{\underline{i} k} \mathbf{X}_{\underline{j}}\,^{k} {G}^{-3} \nabla^{a}{W} - \frac{5}{4}G_{k l} \mathbf{X}_{\underline{i}}\,^{k} {G}^{-3} \nabla_{a}{W} \nabla^{a}{G_{\underline{j}}\,^{l}} - \frac{9}{32}{\rm i} G_{k l} \mathbf{X}^{k}\,_{m} \lambda^{m \alpha} {G}^{-5} \varphi_{\underline{i}}^{\beta} \varphi_{\underline{j} \beta} \varphi^{l}_{\alpha} - \frac{45}{32}{\rm i} G_{\underline{i} k} G_{\underline{j} l} G_{m n} \mathbf{X}^{m}\,_{{i_{1}}} \lambda^{{i_{1}} \alpha} {G}^{-7} \varphi^{k \beta} \varphi^{l}_{\beta} \varphi^{n}_{\alpha}+\frac{9}{16}F G_{\underline{i} \underline{j}} G_{k l} \mathbf{X}^{k}\,_{m} \lambda^{m \alpha} {G}^{-5} \varphi^{l}_{\alpha} - \frac{9}{32}F G_{\underline{i} k} G_{\underline{j} l} \mathbf{X}^{k}\,_{m} \lambda^{m \alpha} {G}^{-5} \varphi^{l}_{\alpha}+\frac{9}{32}\mathcal{H}_{a} G_{\underline{i} k} G_{\underline{j} l} (\Gamma^{a})^{\alpha \beta} \mathbf{X}^{k}\,_{m} \lambda^{m}_{\alpha} {G}^{-5} \varphi^{l}_{\beta}+\frac{9}{16}G_{\underline{i} k} G_{l m} (\Gamma_{a})^{\alpha \beta} \mathbf{X}^{l}\,_{n} \lambda^{n}_{\alpha} {G}^{-5} \nabla^{a}{G_{\underline{j}}\,^{m}} \varphi^{k}_{\beta} - \frac{3}{8}G_{\underline{i} k} (\Gamma_{a})^{\alpha \beta} \mathbf{X}^{k}\,_{l} \lambda^{l}_{\alpha} {G}^{-3} \nabla^{a}{\varphi_{\underline{j} \beta}}+\frac{9}{32}G_{\underline{i} k} (\Sigma_{a b})^{\alpha \beta} \mathbf{X}^{k}\,_{l} W^{a b} \lambda^{l}_{\alpha} {G}^{-3} \varphi_{\underline{j} \beta} - \frac{27}{256}G_{\underline{i} k} G_{\underline{j} l} \mathbf{X}^{k}\,_{m} \lambda^{m \alpha} X^{l}_{\alpha} {G}^{-3}+\frac{385}{256}\mathbf{X}_{\underline{i} k} \lambda^{k \alpha} X_{\underline{j} \alpha} {G}^{-1}%
 - \frac{45}{256}G_{\underline{i} \underline{j}} G_{k l} \mathbf{X}^{k}\,_{m} \lambda^{m \alpha} X^{l}_{\alpha} {G}^{-3} - \frac{3}{4}{\rm i} G_{\underline{i} k} (\Gamma_{a})^{\alpha \beta} W {G}^{-5} \nabla^{a}{\boldsymbol{\lambda}_{\underline{j} \alpha}} \varphi^{k \rho} \varphi_{l \beta} \varphi^{l}_{\rho}+\frac{3}{64}{\rm i} W \boldsymbol{\lambda}_{\underline{i}}^{\beta} X_{\underline{j}}^{\alpha} {G}^{-3} \varphi_{k \beta} \varphi^{k}_{\alpha} - \frac{1}{2}F (\Gamma_{a})^{\alpha \beta} W {G}^{-3} \nabla^{a}{\boldsymbol{\lambda}_{\underline{i} \alpha}} \varphi_{\underline{j} \beta} - \frac{1}{4}\mathcal{H}_{a} W {G}^{-3} \nabla^{a}{\boldsymbol{\lambda}_{\underline{i}}^{\alpha}} \varphi_{\underline{j} \alpha} - \frac{1}{2}\mathcal{H}^{a} (\Sigma_{a b})^{\alpha \beta} W {G}^{-3} \nabla^{b}{\boldsymbol{\lambda}_{\underline{i} \alpha}} \varphi_{\underline{j} \beta}+\frac{1}{4}W {G}^{-3} \nabla_{a}{\boldsymbol{\lambda}_{\underline{i}}^{\alpha}} \nabla^{a}{G_{\underline{j} k}} \varphi^{k}_{\alpha} - \frac{1}{2}(\Sigma_{a b})^{\alpha \beta} W {G}^{-3} \nabla^{a}{\boldsymbol{\lambda}_{\underline{i} \alpha}} \nabla^{b}{G_{\underline{j} k}} \varphi^{k}_{\beta}+\frac{3}{16}(\Gamma_{a})^{\alpha \beta} \lambda^{\rho}_{\underline{i}} {G}^{-3} \nabla^{a}{\boldsymbol{\lambda}_{k \alpha}} \varphi_{\underline{j} \beta} \varphi^{k}_{\rho}+\frac{3}{4}{\rm i} G_{\underline{i} k} (\Gamma_{a})^{\alpha \beta} W {G}^{-5} \nabla^{a}{\boldsymbol{\lambda}_{l \alpha}} \varphi_{\underline{j} \beta} \varphi^{k \rho} \varphi^{l}_{\rho} - \frac{1}{4}{\rm i} (\Gamma^{a})^{\alpha \beta} W {G}^{-3} \nabla^{b}{\mathbf{F}_{a b}} \varphi_{\underline{i} \alpha} \varphi_{\underline{j} \beta} - \frac{3}{4}{\rm i} (\Gamma^{a})^{\alpha \beta} W W_{a b} {G}^{-3} \nabla^{b}{\boldsymbol{W}} \varphi_{\underline{i} \alpha} \varphi_{\underline{j} \beta} - \frac{3}{4}{\rm i} (\Gamma^{a})^{\alpha \beta} W \boldsymbol{W} {G}^{-3} \nabla^{b}{W_{a b}} \varphi_{\underline{i} \alpha} \varphi_{\underline{j} \beta} - \frac{1}{4}{\rm i} (\Gamma_{a})^{\alpha \beta} W {G}^{-3} \nabla^{a}{\mathbf{X}_{\underline{i} k}} \varphi_{\underline{j} \alpha} \varphi^{k}_{\beta} - \frac{1}{2}{\rm i} W {G}^{-3} \nabla_{a}{\nabla^{a}{\boldsymbol{W}}} \varphi_{\underline{i}}^{\alpha} \varphi_{\underline{j} \alpha}+\frac{3}{4}{\rm i} W W^{a b} \mathbf{F}_{a b} {G}^{-3} \varphi_{\underline{i}}^{\alpha} \varphi_{\underline{j} \alpha}+\frac{39}{32}{\rm i} W \boldsymbol{W} W^{a b} W_{a b} {G}^{-3} \varphi_{\underline{i}}^{\alpha} \varphi_{\underline{j} \alpha} - \frac{3}{64}{\rm i} W \boldsymbol{\lambda}_{k}^{\beta} X_{\underline{i}}^{\alpha} {G}^{-3} \varphi_{\underline{j} \beta} \varphi^{k}_{\alpha}+\frac{3}{32}{\rm i} W \boldsymbol{\lambda}_{k}^{\beta} X_{\underline{i}}^{\alpha} {G}^{-3} \varphi_{\underline{j} \alpha} \varphi^{k}_{\beta} - \frac{45}{64}{\rm i} W \boldsymbol{\lambda}_{k}^{\alpha} X_{\underline{i} \alpha} {G}^{-3} \varphi_{\underline{j}}^{\beta} \varphi^{k}_{\beta}%
+\frac{9}{64}{\rm i} W \boldsymbol{\lambda}_{k}^{\beta} X^{k \alpha} {G}^{-3} \varphi_{\underline{i} \beta} \varphi_{\underline{j} \alpha} - \frac{3}{64}{\rm i} W \boldsymbol{\lambda}_{k}^{\alpha} X^{k}_{\alpha} {G}^{-3} \varphi_{\underline{i}}^{\beta} \varphi_{\underline{j} \beta}+\frac{45}{64}{\rm i} W \boldsymbol{\lambda}_{\underline{i}}^{\alpha} X_{k \alpha} {G}^{-3} \varphi_{\underline{j}}^{\beta} \varphi^{k}_{\beta} - \frac{9}{64}{\rm i} W \boldsymbol{\lambda}_{\underline{i}}^{\beta} X_{k}^{\alpha} {G}^{-3} \varphi_{\underline{j} \alpha} \varphi^{k}_{\beta}+\frac{3}{4}G_{\underline{i} k} G_{l m} (\Gamma_{a})^{\alpha \beta} \lambda^{\rho}_{\underline{j}} {G}^{-5} \nabla^{a}{\boldsymbol{\lambda}^{l}_{\alpha}} \varphi^{k}_{\rho} \varphi^{m}_{\beta} - \frac{1}{2}G_{\underline{i} k} (\Gamma^{a})^{\alpha \beta} \lambda_{\underline{j} \alpha} {G}^{-3} \nabla^{b}{\mathbf{F}_{a b}} \varphi^{k}_{\beta}+\frac{25}{128}\epsilon^{c d}\,_{e}\,^{a b} G_{\underline{i} k} (\Sigma_{c d})^{\alpha \beta} W_{a b} \lambda_{\underline{j} \alpha} {G}^{-3} \nabla^{e}{\boldsymbol{W}} \varphi^{k}_{\beta} - \frac{73}{64}G_{\underline{i} k} (\Gamma^{a})^{\alpha \beta} W_{a b} \lambda_{\underline{j} \alpha} {G}^{-3} \nabla^{b}{\boldsymbol{W}} \varphi^{k}_{\beta}+\frac{3}{8}\epsilon^{c d}\,_{e}\,^{a b} G_{\underline{i} k} (\Sigma_{c d})^{\alpha \beta} \boldsymbol{W} \lambda_{\underline{j} \alpha} {G}^{-3} \nabla^{e}{W_{a b}} \varphi^{k}_{\beta} - \frac{3}{4}G_{\underline{i} k} (\Gamma^{a})^{\alpha \beta} \boldsymbol{W} \lambda_{\underline{j} \alpha} {G}^{-3} \nabla^{b}{W_{a b}} \varphi^{k}_{\beta}+\frac{1}{4}G_{k l} (\Gamma_{a})^{\alpha \beta} \lambda_{\underline{i} \alpha} {G}^{-3} \nabla^{a}{\mathbf{X}_{\underline{j}}\,^{k}} \varphi^{l}_{\beta}+\frac{25}{256}\epsilon_{e}\,^{c d a b} G_{\underline{i} k} (\Gamma^{e})^{\alpha \beta} W_{c d} \mathbf{F}_{a b} \lambda_{\underline{j} \alpha} {G}^{-3} \varphi^{k}_{\beta}+\frac{23}{32}G_{\underline{i} k} (\Sigma^{a}{}_{\, c})^{\alpha \beta} W^{c b} \mathbf{F}_{a b} \lambda_{\underline{j} \alpha} {G}^{-3} \varphi^{k}_{\beta}+\frac{37}{256}\epsilon_{e}\,^{a b c d} G_{\underline{i} k} (\Gamma^{e})^{\alpha \beta} \boldsymbol{W} W_{a b} W_{c d} \lambda_{\underline{j} \alpha} {G}^{-3} \varphi^{k}_{\beta}+\frac{1}{4}G_{k l} W^{a b \alpha}\,_{\underline{i}} (\Sigma_{a b})^{\beta \rho} \lambda_{\underline{j} \beta} \boldsymbol{\lambda}^{k}_{\rho} {G}^{-3} \varphi^{l}_{\alpha} - \frac{1}{32}G_{k l} \lambda^{\alpha}_{\underline{i}} \boldsymbol{\lambda}^{k \beta} X_{\underline{j} \alpha} {G}^{-3} \varphi^{l}_{\beta}+\frac{57}{256}G_{\underline{i} k} \lambda^{\alpha}_{\underline{j}} \boldsymbol{\lambda}_{l}^{\beta} X^{l}_{\alpha} {G}^{-3} \varphi^{k}_{\beta}+\frac{9}{256}G_{k l} \lambda^{\alpha}_{\underline{i}} \boldsymbol{\lambda}_{\underline{j}}^{\beta} X^{k}_{\alpha} {G}^{-3} \varphi^{l}_{\beta} - \frac{3}{8}G_{\underline{i} k} \lambda^{\beta}_{\underline{j}} \boldsymbol{\lambda}_{l \beta} X^{l \alpha} {G}^{-3} \varphi^{k}_{\alpha} - \frac{1}{32}\Phi^{a b}\,_{\underline{i} k} G^{k}\,_{l} (\Sigma_{a b})^{\alpha \beta} \boldsymbol{W} \lambda_{\underline{j} \alpha} {G}^{-3} \varphi^{l}_{\beta}%
 - \frac{3}{8}{\rm i} G_{k l} (\Gamma_{a})^{\alpha \beta} W {G}^{-5} \nabla^{a}{\boldsymbol{\lambda}^{k}_{\alpha}} \varphi_{\underline{i}}^{\rho} \varphi_{\underline{j} \rho} \varphi^{l}_{\beta} - \frac{15}{8}{\rm i} G_{\underline{i} k} G_{\underline{j} l} G_{m n} (\Gamma_{a})^{\alpha \beta} W {G}^{-7} \nabla^{a}{\boldsymbol{\lambda}^{m}_{\alpha}} \varphi^{k \rho} \varphi^{l}_{\rho} \varphi^{n}_{\beta} - \frac{3}{4}{\rm i} G_{\underline{i} k} G_{\underline{j} l} (\Gamma^{a})^{\alpha \beta} W {G}^{-5} \nabla^{b}{\mathbf{F}_{a b}} \varphi^{k}_{\alpha} \varphi^{l}_{\beta} - \frac{9}{4}{\rm i} G_{\underline{i} k} G_{\underline{j} l} (\Gamma^{a})^{\alpha \beta} W W_{a b} {G}^{-5} \nabla^{b}{\boldsymbol{W}} \varphi^{k}_{\alpha} \varphi^{l}_{\beta} - \frac{9}{4}{\rm i} G_{\underline{i} k} G_{\underline{j} l} (\Gamma^{a})^{\alpha \beta} W \boldsymbol{W} {G}^{-5} \nabla^{b}{W_{a b}} \varphi^{k}_{\alpha} \varphi^{l}_{\beta}+\frac{3}{4}{\rm i} G_{\underline{i} k} G_{l m} (\Gamma_{a})^{\alpha \beta} W {G}^{-5} \nabla^{a}{\mathbf{X}_{\underline{j}}\,^{l}} \varphi^{k}_{\alpha} \varphi^{m}_{\beta}+\frac{9}{16}{\rm i} G_{\underline{i} k} G_{l m} W \boldsymbol{\lambda}^{l \beta} X_{\underline{j}}^{\alpha} {G}^{-5} \varphi^{k}_{\beta} \varphi^{m}_{\alpha}+\frac{9}{16}{\rm i} G_{\underline{i} k} G_{l m} W \boldsymbol{\lambda}^{l \alpha} X_{\underline{j} \alpha} {G}^{-5} \varphi^{k \beta} \varphi^{m}_{\beta}+\frac{27}{64}{\rm i} G_{\underline{i} k} G_{\underline{j} l} W \boldsymbol{\lambda}_{m}^{\beta} X^{m \alpha} {G}^{-5} \varphi^{k}_{\beta} \varphi^{l}_{\alpha}+\frac{9}{64}{\rm i} G_{\underline{i} k} G_{l m} W \boldsymbol{\lambda}_{\underline{j}}^{\beta} X^{l \alpha} {G}^{-5} \varphi^{k}_{\alpha} \varphi^{m}_{\beta}+\frac{9}{16}{\rm i} G_{\underline{i} k} G_{\underline{j} l} W \boldsymbol{\lambda}_{m}^{\alpha} X^{m}_{\alpha} {G}^{-5} \varphi^{k \beta} \varphi^{l}_{\beta} - \frac{9}{16}{\rm i} G_{\underline{i} k} G_{l m} W \boldsymbol{\lambda}_{\underline{j}}^{\alpha} X^{l}_{\alpha} {G}^{-5} \varphi^{k \beta} \varphi^{m}_{\beta} - \frac{9}{16}{\rm i} G_{\underline{i} k} G_{l m} W \boldsymbol{\lambda}_{\underline{j}}^{\beta} X^{l \alpha} {G}^{-5} \varphi^{k}_{\beta} \varphi^{m}_{\alpha}+\frac{3}{4}F G_{\underline{i} \underline{j}} G_{k l} (\Gamma_{a})^{\alpha \beta} W {G}^{-5} \nabla^{a}{\boldsymbol{\lambda}^{k}_{\alpha}} \varphi^{l}_{\beta} - \frac{3}{4}\mathcal{H}_{a} G_{\underline{i} k} G_{\underline{j} l} W {G}^{-5} \nabla^{a}{\boldsymbol{\lambda}^{k \alpha}} \varphi^{l}_{\alpha} - \frac{3}{2}\mathcal{H}^{a} G_{\underline{i} k} G_{\underline{j} l} (\Sigma_{a b})^{\alpha \beta} W {G}^{-5} \nabla^{b}{\boldsymbol{\lambda}^{k}_{\alpha}} \varphi^{l}_{\beta} - \frac{3}{4}G_{\underline{i} k} G_{l m} W {G}^{-5} \nabla_{a}{\boldsymbol{\lambda}^{l \alpha}} \nabla^{a}{G_{\underline{j}}\,^{m}} \varphi^{k}_{\alpha}+\frac{3}{2}G_{\underline{i} k} G_{l m} (\Sigma_{a b})^{\alpha \beta} W {G}^{-5} \nabla^{a}{\boldsymbol{\lambda}^{l}_{\alpha}} \nabla^{b}{G_{\underline{j}}\,^{m}} \varphi^{k}_{\beta}+G_{\underline{i} k} W {G}^{-3} \nabla_{a}{\nabla^{a}{\boldsymbol{\lambda}_{\underline{j}}^{\alpha}}} \varphi^{k}_{\alpha}+G_{\underline{i} k} (\Sigma_{a b})^{\alpha \beta} W {G}^{-3} \nabla^{a}{\nabla^{b}{\boldsymbol{\lambda}_{\underline{j} \alpha}}} \varphi^{k}_{\beta}%
+\frac{3}{8}\epsilon^{c d}\,_{e}\,^{a b} G_{\underline{i} k} (\Sigma_{c d})^{\alpha \beta} W \boldsymbol{\lambda}_{\underline{j} \alpha} {G}^{-3} \nabla^{e}{W_{a b}} \varphi^{k}_{\beta} - \frac{3}{4}G_{\underline{i} k} (\Gamma^{a})^{\alpha \beta} W \boldsymbol{\lambda}_{\underline{j} \alpha} {G}^{-3} \nabla^{b}{W_{a b}} \varphi^{k}_{\beta}+{\rm i} G_{\underline{i} k} W \boldsymbol{W} W^{a b} W_{a b}\,^{\alpha}\,_{\underline{j}} {G}^{-3} \varphi^{k}_{\alpha}+\frac{1}{2}{\rm i} G_{\underline{i} k} W \mathbf{F}^{a b} W_{a b}\,^{\alpha}\,_{\underline{j}} {G}^{-3} \varphi^{k}_{\alpha}+\frac{3}{64}{\rm i} G_{\underline{i} k} (\Sigma_{a b})^{\alpha \beta} W \mathbf{F}^{a b} X_{\underline{j} \alpha} {G}^{-3} \varphi^{k}_{\beta} - \frac{15}{32}\Phi^{a b}\,_{\underline{i} l} G_{\underline{j} k} (\Sigma_{a b})^{\alpha \beta} W \boldsymbol{\lambda}^{l}_{\alpha} {G}^{-3} \varphi^{k}_{\beta}+\frac{1}{2}\mathcal{H}_{a} G_{\underline{i} \underline{j}} W {G}^{-3} \nabla_{b}{\mathbf{F}^{a b}}-G_{\underline{i} k} W {G}^{-3} \nabla^{a}{\mathbf{F}_{a b}} \nabla^{b}{G_{\underline{j}}\,^{k}}+\frac{99}{64}{\rm i} G_{\underline{i} k} (\Gamma_{a})^{\alpha \beta} W X_{\underline{j} \alpha} {G}^{-3} \nabla^{a}{\boldsymbol{W}} \varphi^{k}_{\beta}+\frac{3}{2}\mathcal{H}^{a} G_{\underline{i} \underline{j}} W W_{a b} {G}^{-3} \nabla^{b}{\boldsymbol{W}}-3G_{\underline{i} k} W W_{a b} {G}^{-3} \nabla^{a}{\boldsymbol{W}} \nabla^{b}{G_{\underline{j}}\,^{k}}+\frac{3}{2}{\rm i} G_{\underline{i} k} (\Gamma_{a})^{\alpha \beta} W \boldsymbol{W} {G}^{-3} \nabla^{a}{X_{\underline{j} \alpha}} \varphi^{k}_{\beta}+\frac{3}{2}\mathcal{H}_{a} G_{\underline{i} \underline{j}} W \boldsymbol{W} {G}^{-3} \nabla_{b}{W^{a b}}-3G_{\underline{i} k} W \boldsymbol{W} {G}^{-3} \nabla^{a}{W_{a b}} \nabla^{b}{G_{\underline{j}}\,^{k}} - \frac{51}{64}{\rm i} G_{\underline{i} k} W \mathbf{X}_{\underline{j} l} X^{l \alpha} {G}^{-3} \varphi^{k}_{\alpha}+\frac{9}{64}{\rm i} G_{k l} W \mathbf{X}_{\underline{i} \underline{j}} X^{k \alpha} {G}^{-3} \varphi^{l}_{\alpha} - \frac{1}{32}\Phi^{a b}\,_{\underline{i} \underline{j}} G_{k l} (\Sigma_{a b})^{\alpha \beta} W \boldsymbol{\lambda}^{k}_{\alpha} {G}^{-3} \varphi^{l}_{\beta} - \frac{1}{32}\Phi^{a b}\,_{\underline{i} k} G^{k}\,_{l} (\Sigma_{a b})^{\alpha \beta} W \boldsymbol{\lambda}_{\underline{j} \alpha} {G}^{-3} \varphi^{l}_{\beta} - \frac{1}{2}\mathcal{H}_{a} G_{\underline{i} k} W {G}^{-3} \nabla^{a}{\mathbf{X}_{\underline{j}}\,^{k}}-G_{k l} W {G}^{-3} \nabla_{a}{\mathbf{X}_{\underline{i}}\,^{k}} \nabla^{a}{G_{\underline{j}}\,^{l}}%
 - \frac{1}{2}F G_{\underline{i} \underline{j}} W {G}^{-3} \nabla_{a}{\nabla^{a}{\boldsymbol{W}}}+\frac{3}{4}F G_{\underline{i} \underline{j}} W W^{a b} \mathbf{F}_{a b} {G}^{-3}+\frac{39}{32}F G_{\underline{i} \underline{j}} W \boldsymbol{W} W^{a b} W_{a b} {G}^{-3}+\frac{9}{128}F G_{\underline{i} k} W \boldsymbol{\lambda}^{k \alpha} X_{\underline{j} \alpha} {G}^{-3}+\frac{9}{128}\mathcal{H}_{a} G_{\underline{i} k} (\Gamma^{a})^{\beta \alpha} W \boldsymbol{\lambda}^{k}_{\beta} X_{\underline{j} \alpha} {G}^{-3}+\frac{3}{16}G_{k l} (\Gamma_{a})^{\beta \alpha} W \boldsymbol{\lambda}^{k}_{\beta} X_{\underline{i} \alpha} {G}^{-3} \nabla^{a}{G_{\underline{j}}\,^{l}}+\frac{3}{128}G_{\underline{i} k} W Y \boldsymbol{\lambda}_{\underline{j}}^{\alpha} {G}^{-3} \varphi^{k}_{\alpha} - \frac{87}{128}F G_{\underline{i} \underline{j}} W \boldsymbol{\lambda}_{k}^{\alpha} X^{k}_{\alpha} {G}^{-3}+\frac{9}{128}\mathcal{H}_{a} G_{\underline{i} \underline{j}} (\Gamma^{a})^{\beta \alpha} W \boldsymbol{\lambda}_{k \beta} X^{k}_{\alpha} {G}^{-3}+\frac{9}{64}G_{\underline{i} k} (\Gamma_{a})^{\beta \alpha} W \boldsymbol{\lambda}_{l \beta} X^{l}_{\alpha} {G}^{-3} \nabla^{a}{G_{\underline{j}}\,^{k}} - \frac{9}{128}F G_{\underline{i} k} W \boldsymbol{\lambda}_{\underline{j}}^{\alpha} X^{k}_{\alpha} {G}^{-3} - \frac{9}{128}\mathcal{H}_{a} G_{\underline{i} k} (\Gamma^{a})^{\beta \alpha} W \boldsymbol{\lambda}_{\underline{j} \beta} X^{k}_{\alpha} {G}^{-3} - \frac{15}{64}G_{k l} (\Gamma_{a})^{\beta \alpha} W \boldsymbol{\lambda}_{\underline{i} \beta} X^{k}_{\alpha} {G}^{-3} \nabla^{a}{G_{\underline{j}}\,^{l}}+\frac{5}{8}G_{\underline{i} k} W {G}^{-3} \nabla_{a}{\varphi_{\underline{j}}^{\alpha}} \nabla^{a}{\boldsymbol{\lambda}^{k}_{\alpha}} - \frac{5}{4}G_{\underline{i} k} (\Sigma_{a b})^{\alpha \beta} W {G}^{-3} \nabla^{a}{\varphi_{\underline{j} \alpha}} \nabla^{b}{\boldsymbol{\lambda}^{k}_{\beta}}+\frac{13}{32}\epsilon^{c d}\,_{e}\,^{a b} G_{\underline{i} k} (\Sigma_{c d})^{\alpha \beta} W W_{a b} {G}^{-3} \nabla^{e}{\boldsymbol{\lambda}^{k}_{\alpha}} \varphi_{\underline{j} \beta}+\frac{13}{16}G_{\underline{i} k} (\Gamma^{a})^{\alpha \beta} W W_{a b} {G}^{-3} \nabla^{b}{\boldsymbol{\lambda}^{k}_{\alpha}} \varphi_{\underline{j} \beta} - \frac{21}{64}G_{\underline{i} k} G_{\underline{j} l} (\Gamma_{a})^{\alpha \beta} W X^{k}_{\alpha} {G}^{-3} \nabla^{a}{\boldsymbol{\lambda}^{l}_{\beta}}+\frac{3}{8}G_{\underline{i} k} W {G}^{-3} \nabla_{a}{\boldsymbol{\lambda}^{k \alpha}} \nabla^{a}{\varphi_{\underline{j} \alpha}} - \frac{3}{4}G_{\underline{i} k} (\Sigma_{a b})^{\alpha \beta} W {G}^{-3} \nabla^{a}{\boldsymbol{\lambda}^{k}_{\alpha}} \nabla^{b}{\varphi_{\underline{j} \beta}}%
+\frac{75}{64}(\Gamma_{a})^{\alpha \beta} W X_{\underline{i} \alpha} {G}^{-1} \nabla^{a}{\boldsymbol{\lambda}_{\underline{j} \beta}} - \frac{27}{64}G_{\underline{i} \underline{j}} G_{k l} (\Gamma_{a})^{\alpha \beta} W X^{k}_{\alpha} {G}^{-3} \nabla^{a}{\boldsymbol{\lambda}^{l}_{\beta}}+\frac{9}{16}{\rm i} G_{\underline{i} k} (\Sigma_{a b})^{\alpha \beta} W W^{a b} \boldsymbol{\lambda}_{\underline{j} \alpha} {G}^{-5} \varphi^{k \rho} \varphi_{l \beta} \varphi^{l}_{\rho}+\frac{3}{8}F (\Sigma_{a b})^{\alpha \beta} W W^{a b} \boldsymbol{\lambda}_{\underline{i} \alpha} {G}^{-3} \varphi_{\underline{j} \beta} - \frac{3}{32}\epsilon^{c d e a b} \mathcal{H}_{c} (\Sigma_{d e})^{\alpha \beta} W W_{a b} \boldsymbol{\lambda}_{\underline{i} \alpha} {G}^{-3} \varphi_{\underline{j} \beta}+\frac{3}{16}\mathcal{H}^{b} (\Gamma^{a})^{\alpha \beta} W W_{a b} \boldsymbol{\lambda}_{\underline{i} \alpha} {G}^{-3} \varphi_{\underline{j} \beta}+\frac{3}{32}\epsilon^{c d}\,_{e}\,^{a b} (\Sigma_{c d})^{\alpha \beta} W W_{a b} \boldsymbol{\lambda}_{\underline{i} \alpha} {G}^{-3} \nabla^{e}{G_{\underline{j} k}} \varphi^{k}_{\beta} - \frac{3}{16}(\Gamma^{a})^{\alpha \beta} W W_{a b} \boldsymbol{\lambda}_{\underline{i} \alpha} {G}^{-3} \nabla^{b}{G_{\underline{j} k}} \varphi^{k}_{\beta}+\frac{1}{64}(\Sigma_{a b})^{\alpha \beta} W^{a b} \lambda^{\rho}_{\underline{i}} \boldsymbol{\lambda}_{k \alpha} {G}^{-3} \varphi_{\underline{j} \beta} \varphi^{k}_{\rho} - \frac{9}{16}{\rm i} G_{\underline{i} k} (\Sigma_{a b})^{\alpha \beta} W W^{a b} \boldsymbol{\lambda}_{l \alpha} {G}^{-5} \varphi_{\underline{j} \beta} \varphi^{k \rho} \varphi^{l}_{\rho}+\frac{9}{32}{\rm i} G_{k l} (\Sigma_{a b})^{\alpha \beta} W W^{a b} \boldsymbol{\lambda}^{k}_{\alpha} {G}^{-5} \varphi_{\underline{i}}^{\rho} \varphi_{\underline{j} \rho} \varphi^{l}_{\beta}+\frac{45}{32}{\rm i} G_{\underline{i} k} G_{\underline{j} l} G_{m n} (\Sigma_{a b})^{\alpha \beta} W W^{a b} \boldsymbol{\lambda}^{m}_{\alpha} {G}^{-7} \varphi^{k \rho} \varphi^{l}_{\rho} \varphi^{n}_{\beta} - \frac{9}{16}F G_{\underline{i} \underline{j}} G_{k l} (\Sigma_{a b})^{\alpha \beta} W W^{a b} \boldsymbol{\lambda}^{k}_{\alpha} {G}^{-5} \varphi^{l}_{\beta} - \frac{9}{32}\epsilon^{c d e a b} \mathcal{H}_{c} G_{\underline{i} k} G_{\underline{j} l} (\Sigma_{d e})^{\alpha \beta} W W_{a b} \boldsymbol{\lambda}^{k}_{\alpha} {G}^{-5} \varphi^{l}_{\beta}+\frac{9}{16}\mathcal{H}^{b} G_{\underline{i} k} G_{\underline{j} l} (\Gamma^{a})^{\alpha \beta} W W_{a b} \boldsymbol{\lambda}^{k}_{\alpha} {G}^{-5} \varphi^{l}_{\beta} - \frac{9}{32}\epsilon^{c d}\,_{e}\,^{a b} G_{\underline{i} k} G_{l m} (\Sigma_{c d})^{\alpha \beta} W W_{a b} \boldsymbol{\lambda}^{l}_{\alpha} {G}^{-5} \nabla^{e}{G_{\underline{j}}\,^{m}} \varphi^{k}_{\beta}+\frac{9}{16}G_{\underline{i} k} G_{l m} (\Gamma^{a})^{\alpha \beta} W W_{a b} \boldsymbol{\lambda}^{l}_{\alpha} {G}^{-5} \nabla^{b}{G_{\underline{j}}\,^{m}} \varphi^{k}_{\beta}+\frac{3}{8}\epsilon^{c d}\,_{e}\,^{a b} G_{\underline{i} k} (\Sigma_{c d})^{\alpha \beta} W W_{a b} \boldsymbol{\lambda}^{k}_{\alpha} {G}^{-3} \nabla^{e}{\varphi_{\underline{j} \beta}} - \frac{3}{4}G_{\underline{i} k} (\Gamma^{a})^{\alpha \beta} W W_{a b} \boldsymbol{\lambda}^{k}_{\alpha} {G}^{-3} \nabla^{b}{\varphi_{\underline{j} \beta}}+\frac{17}{32}G_{\underline{i} k} W W^{a b} W_{a b} \boldsymbol{\lambda}^{k \alpha} {G}^{-3} \varphi_{\underline{j} \alpha}%
 - \frac{37}{256}\epsilon_{e}\,^{a b c d} G_{\underline{i} k} (\Gamma^{e})^{\alpha \beta} W W_{a b} W_{c d} \boldsymbol{\lambda}^{k}_{\alpha} {G}^{-3} \varphi_{\underline{j} \beta}+\frac{87}{256}G_{\underline{i} k} G_{\underline{j} l} (\Sigma_{a b})^{\beta \alpha} W W^{a b} \boldsymbol{\lambda}^{k}_{\beta} X^{l}_{\alpha} {G}^{-3} - \frac{201}{256}(\Sigma_{a b})^{\beta \alpha} W W^{a b} \boldsymbol{\lambda}_{\underline{i} \beta} X_{\underline{j} \alpha} {G}^{-1}+\frac{57}{256}G_{\underline{i} \underline{j}} G_{k l} (\Sigma_{a b})^{\beta \alpha} W W^{a b} \boldsymbol{\lambda}^{k}_{\beta} X^{l}_{\alpha} {G}^{-3} - \frac{9}{8}G_{\underline{i} k} W \boldsymbol{W} X_{\underline{j}}^{\alpha} {G}^{-5} \varphi^{k \beta} \varphi_{l \alpha} \varphi^{l}_{\beta}+\frac{3}{4}{\rm i} F W \boldsymbol{W} X_{\underline{i}}^{\alpha} {G}^{-3} \varphi_{\underline{j} \alpha} - \frac{3}{8}{\rm i} \mathcal{H}_{a} (\Gamma^{a})^{\alpha \beta} W \boldsymbol{W} X_{\underline{i} \alpha} {G}^{-3} \varphi_{\underline{j} \beta}+\frac{3}{8}{\rm i} (\Gamma_{a})^{\alpha \beta} W \boldsymbol{W} X_{\underline{i} \alpha} {G}^{-3} \nabla^{a}{G_{\underline{j} k}} \varphi^{k}_{\beta} - \frac{9}{64}{\rm i} \boldsymbol{W} \lambda^{\beta}_{\underline{i}} X_{k}^{\alpha} {G}^{-3} \varphi_{\underline{j} \alpha} \varphi^{k}_{\beta} - \frac{3}{32}{\rm i} W \boldsymbol{W} Y {G}^{-3} \varphi_{\underline{i}}^{\alpha} \varphi_{\underline{j} \alpha}+\frac{9}{8}G_{\underline{i} k} W \boldsymbol{W} X_{l}^{\alpha} {G}^{-5} \varphi_{\underline{j} \alpha} \varphi^{k \beta} \varphi^{l}_{\beta} - \frac{9}{16}{\rm i} G_{\underline{i} k} G_{l m} \boldsymbol{W} \lambda^{\beta}_{\underline{j}} X^{l \alpha} {G}^{-5} \varphi^{k}_{\beta} \varphi^{m}_{\alpha} - \frac{3}{32}F G_{\underline{i} \underline{j}} W \boldsymbol{W} Y {G}^{-3} - \frac{9}{16}G_{k l} W \boldsymbol{W} X^{k \alpha} {G}^{-5} \varphi_{\underline{i}}^{\beta} \varphi_{\underline{j} \beta} \varphi^{l}_{\alpha} - \frac{45}{16}G_{\underline{i} k} G_{\underline{j} l} G_{m n} W \boldsymbol{W} X^{m \alpha} {G}^{-7} \varphi^{k \beta} \varphi^{l}_{\beta} \varphi^{n}_{\alpha} - \frac{9}{8}{\rm i} F G_{\underline{i} \underline{j}} G_{k l} W \boldsymbol{W} X^{k \alpha} {G}^{-5} \varphi^{l}_{\alpha} - \frac{9}{8}{\rm i} \mathcal{H}_{a} G_{\underline{i} k} G_{\underline{j} l} (\Gamma^{a})^{\alpha \beta} W \boldsymbol{W} X^{k}_{\alpha} {G}^{-5} \varphi^{l}_{\beta} - \frac{9}{8}{\rm i} G_{\underline{i} k} G_{l m} (\Gamma_{a})^{\alpha \beta} W \boldsymbol{W} X^{l}_{\alpha} {G}^{-5} \nabla^{a}{G_{\underline{j}}\,^{m}} \varphi^{k}_{\beta}+\frac{3}{2}{\rm i} G_{\underline{i} k} (\Gamma_{a})^{\alpha \beta} W \boldsymbol{W} X^{k}_{\alpha} {G}^{-3} \nabla^{a}{\varphi_{\underline{j} \beta}} - \frac{39}{32}{\rm i} G_{\underline{i} k} (\Sigma_{a b})^{\alpha \beta} W \boldsymbol{W} W^{a b} X^{k}_{\alpha} {G}^{-3} \varphi_{\underline{j} \beta}%
+\frac{81}{128}{\rm i} G_{\underline{i} k} G_{\underline{j} l} W \boldsymbol{W} X^{k \alpha} X^{l}_{\alpha} {G}^{-3} - \frac{207}{128}{\rm i} W \boldsymbol{W} X_{\underline{i}}^{\alpha} X_{\underline{j} \alpha} {G}^{-1}+\frac{63}{128}{\rm i} G_{\underline{i} \underline{j}} G_{k l} W \boldsymbol{W} X^{k \alpha} X^{l}_{\alpha} {G}^{-3}+\frac{3}{4}W \mathbf{X}_{\underline{i} k} {G}^{-5} \varphi_{\underline{j}}^{\alpha} \varphi^{k \beta} \varphi_{l \alpha} \varphi^{l}_{\beta} - \frac{3}{16}{\rm i} G_{k l} \mathbf{X}_{\underline{i} m} \lambda^{\alpha}_{\underline{j}} {G}^{-5} \varphi^{k \beta} \varphi^{l}_{\beta} \varphi^{m}_{\alpha} - \frac{3}{8}{\rm i} G_{k l} (\Gamma_{a})^{\alpha \beta} W {G}^{-5} \nabla^{a}{\boldsymbol{\lambda}_{\underline{i} \alpha}} \varphi_{\underline{j} \beta} \varphi^{k \rho} \varphi^{l}_{\rho}+\frac{9}{32}{\rm i} G_{k l} (\Sigma_{a b})^{\alpha \beta} W W^{a b} \boldsymbol{\lambda}_{\underline{i} \alpha} {G}^{-5} \varphi_{\underline{j} \beta} \varphi^{k \rho} \varphi^{l}_{\rho} - \frac{9}{16}G_{k l} W \boldsymbol{W} X_{\underline{i}}^{\alpha} {G}^{-5} \varphi_{\underline{j} \alpha} \varphi^{k \beta} \varphi^{l}_{\beta}+\frac{15}{8}G_{\underline{i} k} G_{l m} W \mathbf{X}_{\underline{j} n} {G}^{-7} \varphi^{k \alpha} \varphi^{l \beta} \varphi^{m}_{\beta} \varphi^{n}_{\alpha}+\frac{3}{8}{\rm i} F G_{k l} W \mathbf{X}_{\underline{i} \underline{j}} {G}^{-5} \varphi^{k \alpha} \varphi^{l}_{\alpha}+\frac{3}{4}{\rm i} \mathcal{H}_{a} G_{\underline{i} k} (\Gamma^{a})^{\alpha \beta} W \mathbf{X}_{\underline{j} l} {G}^{-5} \varphi^{k}_{\alpha} \varphi^{l}_{\beta}+\frac{3}{4}{\rm i} G_{k l} (\Gamma_{a})^{\alpha \beta} W \mathbf{X}_{\underline{i} m} {G}^{-5} \nabla^{a}{G_{\underline{j}}\,^{k}} \varphi^{l}_{\alpha} \varphi^{m}_{\beta} - \frac{3}{8}{\rm i} G_{k l} \mathbf{X}^{k l} \lambda^{\alpha}_{\underline{i}} {G}^{-5} \varphi_{\underline{j}}^{\beta} \varphi_{m \alpha} \varphi^{m}_{\beta}+\frac{3}{4}{\rm i} G_{\underline{i} k} (\Gamma_{a})^{\alpha \beta} W {G}^{-5} \nabla^{a}{\boldsymbol{\lambda}^{k}_{\alpha}} \varphi_{\underline{j}}^{\rho} \varphi_{l \beta} \varphi^{l}_{\rho} - \frac{9}{16}{\rm i} G_{\underline{i} k} (\Sigma_{a b})^{\alpha \beta} W W^{a b} \boldsymbol{\lambda}^{k}_{\alpha} {G}^{-5} \varphi_{\underline{j}}^{\rho} \varphi_{l \beta} \varphi^{l}_{\rho}+\frac{9}{8}G_{\underline{i} k} W \boldsymbol{W} X^{k \alpha} {G}^{-5} \varphi_{\underline{j}}^{\beta} \varphi_{l \alpha} \varphi^{l}_{\beta} - \frac{15}{8}G_{\underline{i} k} G_{l m} W \mathbf{X}^{l m} {G}^{-7} \varphi_{\underline{j}}^{\alpha} \varphi^{k \beta} \varphi_{n \alpha} \varphi^{n}_{\beta}+\frac{3}{8}{\rm i} F G_{k l} W \mathbf{X}^{k l} {G}^{-5} \varphi_{\underline{i}}^{\alpha} \varphi_{\underline{j} \alpha} - \frac{3}{16}{\rm i} \mathcal{H}_{a} G_{k l} (\Gamma^{a})^{\alpha \beta} W \mathbf{X}^{k l} {G}^{-5} \varphi_{\underline{i} \alpha} \varphi_{\underline{j} \beta}+\frac{3}{8}{\rm i} G_{k l} (\Gamma_{a})^{\alpha \beta} W \mathbf{X}^{k l} {G}^{-5} \nabla^{a}{G_{\underline{i} m}} \varphi_{\underline{j} \alpha} \varphi^{m}_{\beta}%
 - \frac{3}{8}G_{\underline{i} k} G_{l m} (\Gamma_{a})^{\alpha \beta} \lambda_{\underline{j} \alpha} {G}^{-5} \nabla^{a}{\boldsymbol{\lambda}^{k}_{\beta}} \varphi^{l \rho} \varphi^{m}_{\rho} - \frac{9}{16}G_{\underline{i} k} G_{l m} (\Sigma_{a b})^{\alpha \beta} W^{a b} \lambda_{\underline{j} \alpha} \boldsymbol{\lambda}^{k}_{\beta} {G}^{-5} \varphi^{l \rho} \varphi^{m}_{\rho}+\frac{99}{128}{\rm i} G_{\underline{i} k} G_{l m} \boldsymbol{W} \lambda^{\alpha}_{\underline{j}} X^{k}_{\alpha} {G}^{-5} \varphi^{l \beta} \varphi^{m}_{\beta}+\frac{3}{8}{\rm i} G_{k l} G_{m n} X_{\underline{i} \underline{j}} \mathbf{X}^{k l} {G}^{-5} \varphi^{m \alpha} \varphi^{n}_{\alpha}+\frac{15}{16}{\rm i} G_{\underline{i} k} G_{l m} G_{n {i_{1}}} \mathbf{X}^{l m} \lambda^{\alpha}_{\underline{j}} {G}^{-7} \varphi^{k}_{\alpha} \varphi^{n \beta} \varphi^{{i_{1}}}_{\beta} - \frac{9}{32}F G_{\underline{i} k} G_{l m} \mathbf{X}^{l m} \lambda^{\alpha}_{\underline{j}} {G}^{-5} \varphi^{k}_{\alpha} - \frac{3}{32}\mathcal{H}_{a} G_{\underline{i} k} G_{l m} (\Gamma^{a})^{\alpha \beta} \mathbf{X}^{l m} \lambda_{\underline{j} \alpha} {G}^{-5} \varphi^{k}_{\beta} - \frac{3}{8}G_{k l} G_{m n} (\Gamma_{a})^{\alpha \beta} \mathbf{X}^{k l} \lambda_{\underline{i} \alpha} {G}^{-5} \nabla^{a}{G_{\underline{j}}\,^{m}} \varphi^{n}_{\beta} - \frac{15}{8}{\rm i} G_{\underline{i} k} G_{\underline{j} l} G_{m n} (\Gamma_{a})^{\alpha \beta} W {G}^{-7} \nabla^{a}{\boldsymbol{\lambda}^{k}_{\alpha}} \varphi^{l}_{\beta} \varphi^{m \rho} \varphi^{n}_{\rho}+\frac{3}{4}{\rm i} G_{\underline{i} \underline{j}} G_{k l} W {G}^{-5} \nabla_{a}{\nabla^{a}{\boldsymbol{W}}} \varphi^{k \alpha} \varphi^{l}_{\alpha} - \frac{9}{8}{\rm i} G_{\underline{i} \underline{j}} G_{k l} W W^{a b} \mathbf{F}_{a b} {G}^{-5} \varphi^{k \alpha} \varphi^{l}_{\alpha} - \frac{117}{64}{\rm i} G_{\underline{i} \underline{j}} G_{k l} W \boldsymbol{W} W^{a b} W_{a b} {G}^{-5} \varphi^{k \alpha} \varphi^{l}_{\alpha} - \frac{99}{128}{\rm i} G_{\underline{i} k} G_{l m} W \boldsymbol{\lambda}^{k \alpha} X_{\underline{j} \alpha} {G}^{-5} \varphi^{l \beta} \varphi^{m}_{\beta}+\frac{45}{128}{\rm i} G_{\underline{i} \underline{j}} G_{k l} W \boldsymbol{\lambda}_{m}^{\alpha} X^{m}_{\alpha} {G}^{-5} \varphi^{k \beta} \varphi^{l}_{\beta}+\frac{99}{128}{\rm i} G_{\underline{i} k} G_{l m} W \boldsymbol{\lambda}_{\underline{j}}^{\alpha} X^{k}_{\alpha} {G}^{-5} \varphi^{l \beta} \varphi^{m}_{\beta} - \frac{3}{4}G_{\underline{i} k} G_{l m} W {G}^{-5} \nabla_{a}{\boldsymbol{\lambda}^{k \alpha}} \nabla^{a}{G_{\underline{j}}\,^{l}} \varphi^{m}_{\alpha}+\frac{3}{2}G_{\underline{i} k} G_{l m} (\Sigma_{a b})^{\alpha \beta} W {G}^{-5} \nabla^{a}{\boldsymbol{\lambda}^{k}_{\alpha}} \nabla^{b}{G_{\underline{j}}\,^{l}} \varphi^{m}_{\beta}+\frac{45}{32}{\rm i} G_{\underline{i} k} G_{\underline{j} l} G_{m n} (\Sigma_{a b})^{\alpha \beta} W W^{a b} \boldsymbol{\lambda}^{k}_{\alpha} {G}^{-7} \varphi^{l}_{\beta} \varphi^{m \rho} \varphi^{n}_{\rho} - \frac{9}{32}\epsilon^{c d}\,_{e}\,^{a b} G_{\underline{i} k} G_{l m} (\Sigma_{c d})^{\alpha \beta} W W_{a b} \boldsymbol{\lambda}^{k}_{\alpha} {G}^{-5} \nabla^{e}{G_{\underline{j}}\,^{l}} \varphi^{m}_{\beta}+\frac{9}{16}G_{\underline{i} k} G_{l m} (\Gamma^{a})^{\alpha \beta} W W_{a b} \boldsymbol{\lambda}^{k}_{\alpha} {G}^{-5} \nabla^{b}{G_{\underline{j}}\,^{l}} \varphi^{m}_{\beta}%
+\frac{9}{64}{\rm i} G_{\underline{i} \underline{j}} G_{k l} W \boldsymbol{W} Y {G}^{-5} \varphi^{k \alpha} \varphi^{l}_{\alpha} - \frac{45}{16}G_{\underline{i} k} G_{\underline{j} l} G_{m n} W \boldsymbol{W} X^{k \alpha} {G}^{-7} \varphi^{l}_{\alpha} \varphi^{m \beta} \varphi^{n}_{\beta} - \frac{9}{8}{\rm i} G_{\underline{i} k} G_{l m} (\Gamma_{a})^{\alpha \beta} W \boldsymbol{W} X^{k}_{\alpha} {G}^{-5} \nabla^{a}{G_{\underline{j}}\,^{l}} \varphi^{m}_{\beta} - \frac{15}{32}G_{k l} G_{m n} W \mathbf{X}^{k l} {G}^{-7} \varphi_{\underline{i}}^{\alpha} \varphi_{\underline{j} \alpha} \varphi^{m \beta} \varphi^{n}_{\beta} - \frac{105}{32}G_{\underline{i} k} G_{\underline{j} l} G_{m n} G_{{i_{1}} {i_{2}}} W \mathbf{X}^{m n} {G}^{-9} \varphi^{k \alpha} \varphi^{l}_{\alpha} \varphi^{{i_{1}} \beta} \varphi^{{i_{2}}}_{\beta} - \frac{15}{16}{\rm i} F G_{\underline{i} \underline{j}} G_{k l} G_{m n} W \mathbf{X}^{k l} {G}^{-7} \varphi^{m \alpha} \varphi^{n}_{\alpha} - \frac{15}{16}{\rm i} \mathcal{H}_{a} G_{\underline{i} k} G_{\underline{j} l} G_{m n} (\Gamma^{a})^{\alpha \beta} W \mathbf{X}^{m n} {G}^{-7} \varphi^{k}_{\alpha} \varphi^{l}_{\beta} - \frac{15}{8}{\rm i} G_{\underline{i} k} G_{l m} G_{n {i_{1}}} (\Gamma_{a})^{\alpha \beta} W \mathbf{X}^{l m} {G}^{-7} \nabla^{a}{G_{\underline{j}}\,^{n}} \varphi^{k}_{\alpha} \varphi^{{i_{1}}}_{\beta}+\frac{3}{2}{\rm i} G_{\underline{i} k} G_{l m} (\Gamma_{a})^{\alpha \beta} W \mathbf{X}^{l m} {G}^{-5} \nabla^{a}{\varphi_{\underline{j} \alpha}} \varphi^{k}_{\beta}+\frac{21}{16}{\rm i} G_{\underline{i} k} G_{l m} (\Sigma_{a b})^{\alpha \beta} W \mathbf{X}^{l m} W^{a b} {G}^{-5} \varphi_{\underline{j} \alpha} \varphi^{k}_{\beta}+\frac{27}{128}{\rm i} G_{\underline{i} k} G_{\underline{j} l} G_{m n} W \mathbf{X}^{m n} X^{k \alpha} {G}^{-5} \varphi^{l}_{\alpha}+\frac{3}{16}G_{\underline{i} \underline{j}} G_{k l} W \mathbf{X}^{k l} {F}^{2} {G}^{-5}+\frac{3}{16}\mathcal{H}^{a} \mathcal{H}_{a} G_{\underline{i} \underline{j}} G_{k l} W \mathbf{X}^{k l} {G}^{-5}+\frac{3}{4}\mathcal{H}_{a} G_{\underline{i} k} G_{l m} W \mathbf{X}^{l m} {G}^{-5} \nabla^{a}{G_{\underline{j}}\,^{k}} - \frac{145}{128}{\rm i} G_{k l} W \mathbf{X}^{k l} X_{\underline{i}}^{\alpha} {G}^{-3} \varphi_{\underline{j} \alpha}+\frac{45}{128}{\rm i} G_{\underline{i} \underline{j}} G_{k l} G_{m n} W \mathbf{X}^{k l} X^{m \alpha} {G}^{-5} \varphi^{n}_{\alpha}+\frac{3}{4}G_{k l} G_{m n} W \mathbf{X}^{k l} {G}^{-5} \nabla_{a}{G_{\underline{i}}\,^{m}} \nabla^{a}{G_{\underline{j}}\,^{n}} - \frac{1}{4}(\Gamma_{a})^{\alpha \beta} \lambda_{\underline{i} \alpha} {G}^{-3} \nabla^{a}{\boldsymbol{\lambda}_{k \beta}} \varphi_{\underline{j}}^{\rho} \varphi^{k}_{\rho} - \frac{9}{16}(\Sigma_{a b})^{\alpha \beta} W^{a b} \lambda_{\underline{i} \alpha} \boldsymbol{\lambda}_{k \beta} {G}^{-3} \varphi_{\underline{j}}^{\rho} \varphi^{k}_{\rho}+\frac{45}{64}{\rm i} \boldsymbol{W} \lambda^{\alpha}_{\underline{i}} X_{k \alpha} {G}^{-3} \varphi_{\underline{j}}^{\beta} \varphi^{k}_{\beta}%
 - \frac{1}{16}{\rm i} X_{\underline{i} \underline{j}} \mathbf{X}_{k l} {G}^{-3} \varphi^{k \alpha} \varphi^{l}_{\alpha} - \frac{3}{8}{\rm i} G_{\underline{i} k} \mathbf{X}_{l m} \lambda^{\alpha}_{\underline{j}} {G}^{-5} \varphi^{k}_{\alpha} \varphi^{l \beta} \varphi^{m}_{\beta}+\frac{7}{32}F \mathbf{X}_{\underline{i} k} \lambda^{\alpha}_{\underline{j}} {G}^{-3} \varphi^{k}_{\alpha}+\frac{1}{32}\mathcal{H}_{a} (\Gamma^{a})^{\alpha \beta} \mathbf{X}_{\underline{i} k} \lambda_{\underline{j} \alpha} {G}^{-3} \varphi^{k}_{\beta}+\frac{5}{16}(\Gamma_{a})^{\alpha \beta} \mathbf{X}_{k l} \lambda_{\underline{i} \alpha} {G}^{-3} \nabla^{a}{G_{\underline{j}}\,^{k}} \varphi^{l}_{\beta}+\frac{3}{4}{\rm i} G_{\underline{i} k} (\Gamma_{a})^{\alpha \beta} W {G}^{-5} \nabla^{a}{\boldsymbol{\lambda}_{l \alpha}} \varphi_{\underline{j}}^{\rho} \varphi^{k}_{\beta} \varphi^{l}_{\rho} - \frac{1}{4}W {G}^{-3} \nabla_{a}{\boldsymbol{\lambda}_{k}^{\alpha}} \nabla^{a}{G_{\underline{i}}\,^{k}} \varphi_{\underline{j} \alpha}+\frac{1}{2}(\Sigma_{a b})^{\alpha \beta} W {G}^{-3} \nabla^{a}{\boldsymbol{\lambda}_{k \alpha}} \nabla^{b}{G_{\underline{i}}\,^{k}} \varphi_{\underline{j} \beta} - \frac{9}{16}{\rm i} G_{\underline{i} k} (\Sigma_{a b})^{\alpha \beta} W W^{a b} \boldsymbol{\lambda}_{l \alpha} {G}^{-5} \varphi_{\underline{j}}^{\rho} \varphi^{k}_{\beta} \varphi^{l}_{\rho} - \frac{3}{32}\epsilon^{c d}\,_{e}\,^{a b} (\Sigma_{c d})^{\alpha \beta} W W_{a b} \boldsymbol{\lambda}_{k \alpha} {G}^{-3} \nabla^{e}{G_{\underline{i}}\,^{k}} \varphi_{\underline{j} \beta}+\frac{3}{16}(\Gamma^{a})^{\alpha \beta} W W_{a b} \boldsymbol{\lambda}_{k \alpha} {G}^{-3} \nabla^{b}{G_{\underline{i}}\,^{k}} \varphi_{\underline{j} \beta}+\frac{9}{8}G_{\underline{i} k} W \boldsymbol{W} X_{l}^{\alpha} {G}^{-5} \varphi_{\underline{j}}^{\beta} \varphi^{k}_{\alpha} \varphi^{l}_{\beta} - \frac{3}{8}{\rm i} (\Gamma_{a})^{\alpha \beta} W \boldsymbol{W} X_{k \alpha} {G}^{-3} \nabla^{a}{G_{\underline{i}}\,^{k}} \varphi_{\underline{j} \beta}+\frac{3}{16}W \mathbf{X}_{k l} {G}^{-5} \varphi_{\underline{i}}^{\alpha} \varphi_{\underline{j} \alpha} \varphi^{k \beta} \varphi^{l}_{\beta}+\frac{15}{16}G_{\underline{i} k} G_{\underline{j} l} W \mathbf{X}_{m n} {G}^{-7} \varphi^{k \alpha} \varphi^{l}_{\alpha} \varphi^{m \beta} \varphi^{n}_{\beta}+\frac{3}{8}{\rm i} F G_{\underline{i} \underline{j}} W \mathbf{X}_{k l} {G}^{-5} \varphi^{k \alpha} \varphi^{l}_{\alpha}+\frac{3}{4}{\rm i} G_{\underline{i} k} (\Gamma_{a})^{\alpha \beta} W \mathbf{X}_{l m} {G}^{-5} \nabla^{a}{G_{\underline{j}}\,^{l}} \varphi^{k}_{\alpha} \varphi^{m}_{\beta}-{\rm i} (\Gamma_{a})^{\alpha \beta} W \mathbf{X}_{\underline{i} k} {G}^{-3} \nabla^{a}{\varphi_{\underline{j} \alpha}} \varphi^{k}_{\beta} - \frac{3}{64}{\rm i} G_{\underline{i} k} W \mathbf{X}_{\underline{j} l} X^{k \alpha} {G}^{-3} \varphi^{l}_{\alpha} - \frac{1}{8}W \mathbf{X}_{\underline{i} \underline{j}} {F}^{2} {G}^{-3}%
 - \frac{1}{8}\mathcal{H}^{a} \mathcal{H}_{a} W \mathbf{X}_{\underline{i} \underline{j}} {G}^{-3} - \frac{1}{2}\mathcal{H}_{a} W \mathbf{X}_{\underline{i} k} {G}^{-3} \nabla^{a}{G_{\underline{j}}\,^{k}} - \frac{41}{64}{\rm i} G_{\underline{i} k} W \mathbf{X}^{k}\,_{l} X_{\underline{j}}^{\alpha} {G}^{-3} \varphi^{l}_{\alpha} - \frac{15}{64}{\rm i} G_{\underline{i} \underline{j}} W \mathbf{X}_{k l} X^{k \alpha} {G}^{-3} \varphi^{l}_{\alpha} - \frac{1}{2}W \mathbf{X}_{k l} {G}^{-3} \nabla_{a}{G_{\underline{i}}\,^{k}} \nabla^{a}{G_{\underline{j}}\,^{l}}+\frac{1}{8}G_{k l} (\Gamma_{a})^{\alpha \beta} \mathbf{X}^{k l} \lambda_{\underline{i} \alpha} {G}^{-3} \nabla^{a}{\varphi_{\underline{j} \beta}} - \frac{1}{2}G_{k l} W \mathbf{X}^{k l} {G}^{-3} \nabla_{a}{\nabla^{a}{G_{\underline{i} \underline{j}}}}+\frac{7}{16}G_{\underline{i} \underline{j}} G_{k l} W \mathbf{X}^{k l} W^{a b} W_{a b} {G}^{-3} - \frac{5}{16}G_{k l} (\Sigma_{a b})^{\alpha \beta} \mathbf{X}^{k l} W^{a b} \lambda_{\underline{i} \alpha} {G}^{-3} \varphi_{\underline{j} \beta} - \frac{45}{256}G_{\underline{i} k} G_{l m} \mathbf{X}^{l m} \lambda^{\alpha}_{\underline{j}} X^{k}_{\alpha} {G}^{-3} - \frac{5}{8}F G_{k l} X_{\underline{i} \underline{j}} \mathbf{X}^{k l} {G}^{-3} - \frac{1}{32}{\rm i} \epsilon^{c d}\,_{e}\,^{a b} (\Sigma_{c d})^{\alpha \beta} W_{a b} \lambda_{\underline{i} \alpha} {G}^{-1} \nabla^{e}{\boldsymbol{\lambda}_{\underline{j} \beta}}+\frac{1}{16}{\rm i} (\Gamma^{a})^{\alpha \beta} W_{a b} \lambda_{\underline{i} \alpha} {G}^{-1} \nabla^{b}{\boldsymbol{\lambda}_{\underline{j} \beta}} - \frac{25}{64}W \mathbf{X}_{\underline{i} \underline{j}} W^{a b} W_{a b} {G}^{-1} - \frac{1}{2}W W^{a b} W_{a b}\,^{\alpha}\,_{\underline{i}} \boldsymbol{\lambda}_{\underline{j} \alpha} {G}^{-1} - \frac{9}{4}{\rm i} {G}^{-1} \nabla_{a}{\lambda^{\alpha}_{\underline{i}}} \nabla^{a}{\boldsymbol{\lambda}_{\underline{j} \alpha}}+\frac{1}{2}{\rm i} (\Sigma_{a b})^{\alpha \beta} {G}^{-1} \nabla^{a}{\lambda_{\underline{i} \alpha}} \nabla^{b}{\boldsymbol{\lambda}_{\underline{j} \beta}}+\frac{3}{2}X_{\underline{i} \underline{j}} {G}^{-1} \nabla_{a}{\nabla^{a}{\boldsymbol{W}}} - \frac{1}{2}X_{\underline{i} \underline{j}} W^{a b} \mathbf{F}_{a b} {G}^{-1} - \frac{25}{64}\boldsymbol{W} X_{\underline{i} \underline{j}} W^{a b} W_{a b} {G}^{-1}%
+\frac{385}{256}X_{\underline{i} k} \boldsymbol{\lambda}^{k \alpha} X_{\underline{j} \alpha} {G}^{-1}+\frac{81}{64}X_{\underline{i} \underline{j}} \boldsymbol{\lambda}_{k}^{\alpha} X^{k}_{\alpha} {G}^{-1} - \frac{45}{64}X_{\underline{i} k} \boldsymbol{\lambda}_{\underline{j}}^{\alpha} X^{k}_{\alpha} {G}^{-1}+2{G}^{-1} \nabla_{a}{W} \nabla^{a}{\mathbf{X}_{\underline{i} \underline{j}}}+\frac{63}{128}(\Gamma_{a})^{\beta \alpha} \boldsymbol{\lambda}_{\underline{i} \beta} X_{\underline{j} \alpha} {G}^{-1} \nabla^{a}{W} - \frac{1}{8}(\Gamma_{a})^{\alpha \beta} \lambda_{k \alpha} {G}^{-3} \nabla^{a}{\boldsymbol{\lambda}^{k}_{\beta}} \varphi_{\underline{i}}^{\rho} \varphi_{\underline{j} \rho} - \frac{3}{8}G_{\underline{i} k} G_{\underline{j} l} (\Gamma_{a})^{\alpha \beta} \lambda_{m \alpha} {G}^{-5} \nabla^{a}{\boldsymbol{\lambda}^{m}_{\beta}} \varphi^{k \rho} \varphi^{l}_{\rho}+\frac{1}{4}G_{\underline{i} k} (\Gamma_{a})^{\alpha \beta} \lambda_{l \alpha} {G}^{-3} \nabla^{a}{\mathbf{X}_{\underline{j}}\,^{l}} \varphi^{k}_{\beta} - \frac{11}{64}G_{\underline{i} k} (\Sigma_{a b})^{\alpha \beta} \mathbf{X}_{\underline{j} l} W^{a b} \lambda^{l}_{\alpha} {G}^{-3} \varphi^{k}_{\beta}+\frac{1}{4}G_{\underline{i} k} W^{a b \alpha}\,_{\underline{j}} (\Sigma_{a b})^{\beta \rho} \lambda_{l \beta} \boldsymbol{\lambda}^{l}_{\rho} {G}^{-3} \varphi^{k}_{\alpha}+\frac{81}{256}G_{\underline{i} k} \lambda^{\alpha}_{l} \boldsymbol{\lambda}^{l \beta} X_{\underline{j} \alpha} {G}^{-3} \varphi^{k}_{\beta} - \frac{3}{8}G_{\underline{i} k} \lambda^{\beta}_{l} \boldsymbol{\lambda}_{\underline{j} \beta} X^{l \alpha} {G}^{-3} \varphi^{k}_{\alpha} - \frac{69}{128}G_{\underline{i} k} \lambda^{\alpha}_{l} \boldsymbol{\lambda}_{\underline{j}}^{\beta} X^{l}_{\alpha} {G}^{-3} \varphi^{k}_{\beta} - \frac{15}{32}\Phi^{a b}\,_{\underline{i} l} G_{\underline{j} k} (\Sigma_{a b})^{\alpha \beta} \boldsymbol{W} \lambda^{l}_{\alpha} {G}^{-3} \varphi^{k}_{\beta}-{\rm i} \lambda^{\alpha}_{\underline{i}} {G}^{-1} \nabla_{a}{\nabla^{a}{\boldsymbol{\lambda}_{\underline{j} \alpha}}}+{\rm i} (\Sigma_{a b})^{\alpha \beta} \lambda_{\underline{i} \alpha} {G}^{-1} \nabla^{a}{\nabla^{b}{\boldsymbol{\lambda}_{\underline{j} \beta}}} - \frac{11}{128}{\rm i} W^{a b} W_{a b} \lambda^{\alpha}_{\underline{i}} \boldsymbol{\lambda}_{\underline{j} \alpha} {G}^{-1} - \frac{1}{256}{\rm i} \epsilon_{e}\,^{a b c d} (\Gamma^{e})^{\alpha \beta} W_{a b} W_{c d} \lambda_{\underline{i} \alpha} \boldsymbol{\lambda}_{\underline{j} \beta} {G}^{-1} - \frac{1}{2}\boldsymbol{W} W^{a b} W_{a b}\,^{\alpha}\,_{\underline{i}} \lambda_{\underline{j} \alpha} {G}^{-1} - \frac{1}{2}\mathbf{F}^{a b} W_{a b}\,^{\alpha}\,_{\underline{i}} \lambda_{\underline{j} \alpha} {G}^{-1}%
 - \frac{63}{128}(\Sigma_{a b})^{\beta \alpha} \mathbf{F}^{a b} \lambda_{\underline{i} \beta} X_{\underline{j} \alpha} {G}^{-1} - \frac{7}{16}{\rm i} \Phi^{a b}\,_{\underline{i} k} (\Sigma_{a b})^{\alpha \beta} \lambda_{\underline{j} \alpha} \boldsymbol{\lambda}^{k}_{\beta} {G}^{-1}+\frac{63}{128}(\Gamma_{a})^{\beta \alpha} \lambda_{\underline{i} \beta} X_{\underline{j} \alpha} {G}^{-1} \nabla^{a}{\boldsymbol{W}} - \frac{201}{256}(\Sigma_{a b})^{\beta \alpha} \boldsymbol{W} W^{a b} \lambda_{\underline{i} \beta} X_{\underline{j} \alpha} {G}^{-1} - \frac{45}{64}\mathbf{X}_{\underline{i} k} \lambda^{\alpha}_{\underline{j}} X^{k}_{\alpha} {G}^{-1}+\frac{81}{64}\mathbf{X}_{\underline{i} \underline{j}} \lambda^{\alpha}_{k} X^{k}_{\alpha} {G}^{-1}+\frac{15}{16}{\rm i} \Phi^{a b}\,_{\underline{i} \underline{j}} (\Sigma_{a b})^{\alpha \beta} \lambda_{k \alpha} \boldsymbol{\lambda}^{k}_{\beta} {G}^{-1}+\frac{7}{16}{\rm i} \Phi^{a b}\,_{\underline{i} k} (\Sigma_{a b})^{\alpha \beta} \lambda^{k}_{\alpha} \boldsymbol{\lambda}_{\underline{j} \beta} {G}^{-1} - \frac{1}{2}\mathbf{X}_{\underline{i} \underline{j}} W^{a b} F_{a b} {G}^{-1} - \frac{1}{2}F^{a b} W_{a b}\,^{\alpha}\,_{\underline{i}} \boldsymbol{\lambda}_{\underline{j} \alpha} {G}^{-1}+\frac{3}{32}{\rm i} Y \lambda^{\alpha}_{\underline{i}} \boldsymbol{\lambda}_{\underline{j} \alpha} {G}^{-1} - \frac{63}{128}(\Sigma_{a b})^{\beta \alpha} F^{a b} \boldsymbol{\lambda}_{\underline{i} \beta} X_{\underline{j} \alpha} {G}^{-1} - \frac{1}{32}{\rm i} \epsilon^{c d}\,_{e}\,^{a b} (\Sigma_{c d})^{\alpha \beta} W_{a b} \boldsymbol{\lambda}_{\underline{i} \alpha} {G}^{-1} \nabla^{e}{\lambda_{\underline{j} \beta}}+\frac{1}{16}{\rm i} (\Gamma^{a})^{\alpha \beta} W_{a b} \boldsymbol{\lambda}_{\underline{i} \alpha} {G}^{-1} \nabla^{b}{\lambda_{\underline{j} \beta}} - \frac{3}{16}(\Sigma_{a b})^{\alpha \beta} W^{a b} \lambda_{k \alpha} \boldsymbol{\lambda}^{k}_{\beta} {G}^{-3} \varphi_{\underline{i}}^{\rho} \varphi_{\underline{j} \rho} - \frac{9}{16}G_{\underline{i} k} G_{\underline{j} l} (\Sigma_{a b})^{\alpha \beta} W^{a b} \lambda_{m \alpha} \boldsymbol{\lambda}^{m}_{\beta} {G}^{-5} \varphi^{k \rho} \varphi^{l}_{\rho}+\frac{75}{64}(\Gamma_{a})^{\alpha \beta} \boldsymbol{W} X_{\underline{i} \alpha} {G}^{-1} \nabla^{a}{\lambda_{\underline{j} \beta}} - \frac{3}{64}\boldsymbol{W} X_{\underline{i} \underline{j}} Y {G}^{-1} - \frac{3}{64}{\rm i} \boldsymbol{W} \lambda^{\alpha}_{k} X^{k}_{\alpha} {G}^{-3} \varphi_{\underline{i}}^{\beta} \varphi_{\underline{j} \beta}+\frac{9}{16}{\rm i} G_{\underline{i} k} G_{\underline{j} l} \boldsymbol{W} \lambda^{\alpha}_{m} X^{m}_{\alpha} {G}^{-5} \varphi^{k \beta} \varphi^{l}_{\beta}%
 - \frac{5}{16}G_{\underline{i} k} (\Gamma_{a})^{\alpha \beta} \mathbf{X}_{\underline{j} l} {G}^{-3} \nabla^{a}{\lambda^{l}_{\alpha}} \varphi^{k}_{\beta}+\frac{3}{2}\mathbf{X}_{\underline{i} \underline{j}} {G}^{-1} \nabla_{a}{\nabla^{a}{W}} - \frac{3}{64}W \mathbf{X}_{\underline{i} \underline{j}} Y {G}^{-1}+\frac{3}{16}{\rm i} X_{k l} \mathbf{X}^{k l} {G}^{-3} \varphi_{\underline{i}}^{\alpha} \varphi_{\underline{j} \alpha}+\frac{9}{16}{\rm i} G_{\underline{i} k} G_{\underline{j} l} X_{m n} \mathbf{X}^{m n} {G}^{-5} \varphi^{k \alpha} \varphi^{l}_{\alpha} - \frac{3}{8}F G_{\underline{i} \underline{j}} X_{k l} \mathbf{X}^{k l} {G}^{-3}+W {G}^{-1} \nabla_{a}{\nabla^{a}{\mathbf{X}_{\underline{i} \underline{j}}}} - \frac{1}{16}(\Gamma_{a})^{\alpha \beta} \boldsymbol{\lambda}_{\underline{i}}^{\rho} {G}^{-3} \nabla^{a}{\lambda_{\underline{j} \alpha}} \varphi_{k \beta} \varphi^{k}_{\rho} - \frac{1}{64}(\Sigma_{a b})^{\alpha \beta} W^{a b} \lambda_{\underline{i} \alpha} \boldsymbol{\lambda}_{\underline{j}}^{\rho} {G}^{-3} \varphi_{k \beta} \varphi^{k}_{\rho}+\frac{1}{16}{\rm i} (\Sigma_{a b})^{\alpha \beta} X_{\underline{i} \underline{j}} \mathbf{F}^{a b} {G}^{-3} \varphi_{k \alpha} \varphi^{k}_{\beta}+\frac{1}{16}{\rm i} (\Sigma_{a b})^{\alpha \beta} \boldsymbol{W} X_{\underline{i} \underline{j}} W^{a b} {G}^{-3} \varphi_{k \alpha} \varphi^{k}_{\beta} - \frac{9}{16}{\rm i} G_{\underline{i} k} X_{\underline{j} l} \boldsymbol{\lambda}^{l \alpha} {G}^{-5} \varphi^{k \beta} \varphi_{m \alpha} \varphi^{m}_{\beta} - \frac{15}{32}F X_{\underline{i} k} \boldsymbol{\lambda}^{k \alpha} {G}^{-3} \varphi_{\underline{j} \alpha}+\frac{3}{32}\mathcal{H}_{a} (\Gamma^{a})^{\alpha \beta} X_{\underline{i} k} \boldsymbol{\lambda}^{k}_{\alpha} {G}^{-3} \varphi_{\underline{j} \beta} - \frac{3}{16}(\Gamma_{a})^{\alpha \beta} X_{\underline{i} l} \boldsymbol{\lambda}^{l}_{\alpha} {G}^{-3} \nabla^{a}{G_{\underline{j} k}} \varphi^{k}_{\beta} - \frac{1}{8}(\Gamma_{a})^{\alpha \beta} \boldsymbol{\lambda}_{\underline{i}}^{\rho} {G}^{-3} \nabla^{a}{\lambda_{k \alpha}} \varphi_{\underline{j} \rho} \varphi^{k}_{\beta} - \frac{3}{16}(\Gamma_{a})^{\alpha \beta} \boldsymbol{\lambda}_{k}^{\rho} {G}^{-3} \nabla^{a}{\lambda^{k}_{\alpha}} \varphi_{\underline{i} \beta} \varphi_{\underline{j} \rho}+\frac{3}{64}(\Sigma_{a b})^{\alpha \beta} W^{a b} \lambda_{k \alpha} \boldsymbol{\lambda}^{k \rho} {G}^{-3} \varphi_{\underline{i} \beta} \varphi_{\underline{j} \rho} - \frac{3}{16}{\rm i} X_{k l} \mathbf{X}_{\underline{i}}\,^{k} {G}^{-3} \varphi_{\underline{j}}^{\alpha} \varphi^{l}_{\alpha} - \frac{1}{4}{\rm i} (\Gamma_{a})^{\alpha \beta} X_{\underline{i} k} {G}^{-3} \nabla^{a}{\boldsymbol{W}} \varphi_{\underline{j} \alpha} \varphi^{k}_{\beta}%
+\frac{9}{16}{\rm i} G_{\underline{i} k} X_{l m} \boldsymbol{\lambda}^{l \alpha} {G}^{-5} \varphi_{\underline{j} \alpha} \varphi^{k \beta} \varphi^{m}_{\beta} - \frac{3}{16}(\Gamma_{a})^{\alpha \beta} X_{k l} \boldsymbol{\lambda}^{k}_{\alpha} {G}^{-3} \nabla^{a}{G_{\underline{i} \underline{j}}} \varphi^{l}_{\beta}+\frac{7}{32}\epsilon^{c d}\,_{e}\,^{a b} G_{\underline{i} k} (\Sigma_{c d})^{\alpha \beta} \mathbf{F}_{a b} {G}^{-3} \nabla^{e}{\lambda_{\underline{j} \alpha}} \varphi^{k}_{\beta} - \frac{9}{16}G_{\underline{i} k} (\Gamma^{a})^{\alpha \beta} \mathbf{F}_{a b} {G}^{-3} \nabla^{b}{\lambda_{\underline{j} \alpha}} \varphi^{k}_{\beta} - \frac{1}{32}\epsilon^{c d}\,_{e}\,^{a b} G_{\underline{i} k} (\Sigma_{c d})^{\alpha \beta} \boldsymbol{W} W_{a b} {G}^{-3} \nabla^{e}{\lambda_{\underline{j} \alpha}} \varphi^{k}_{\beta} - \frac{25}{16}G_{\underline{i} k} (\Gamma^{a})^{\alpha \beta} \boldsymbol{W} W_{a b} {G}^{-3} \nabla^{b}{\lambda_{\underline{j} \alpha}} \varphi^{k}_{\beta}+\frac{25}{16}G_{\underline{i} k} {G}^{-3} \nabla_{a}{\boldsymbol{W}} \nabla^{a}{\lambda^{\alpha}_{\underline{j}}} \varphi^{k}_{\alpha}+\frac{9}{8}G_{\underline{i} k} (\Sigma_{a b})^{\alpha \beta} {G}^{-3} \nabla^{a}{\boldsymbol{W}} \nabla^{b}{\lambda_{\underline{j} \alpha}} \varphi^{k}_{\beta} - \frac{9}{16}G_{\underline{i} k} G_{\underline{j} l} (\Gamma_{a})^{\alpha \beta} \boldsymbol{\lambda}_{m}^{\rho} {G}^{-5} \nabla^{a}{\lambda^{m}_{\alpha}} \varphi^{k}_{\beta} \varphi^{l}_{\rho}+\frac{3}{4}G_{\underline{i} k} \boldsymbol{\lambda}_{\underline{j}}^{\alpha} {G}^{-3} \nabla_{a}{\nabla^{a}{W}} \varphi^{k}_{\alpha} - \frac{5}{32}{\rm i} F G_{\underline{i} \underline{j}} (\Gamma_{a})^{\alpha \beta} \boldsymbol{\lambda}_{k \alpha} {G}^{-3} \nabla^{a}{\lambda^{k}_{\beta}}+\frac{3}{32}{\rm i} \mathcal{H}_{a} G_{\underline{i} \underline{j}} \boldsymbol{\lambda}_{k}^{\alpha} {G}^{-3} \nabla^{a}{\lambda^{k}_{\alpha}} - \frac{3}{16}{\rm i} \mathcal{H}^{a} G_{\underline{i} \underline{j}} (\Sigma_{a b})^{\alpha \beta} \boldsymbol{\lambda}_{k \alpha} {G}^{-3} \nabla^{b}{\lambda^{k}_{\beta}}+\frac{3}{16}{\rm i} G_{\underline{i} k} \boldsymbol{\lambda}_{l}^{\alpha} {G}^{-3} \nabla_{a}{\lambda^{l}_{\alpha}} \nabla^{a}{G_{\underline{j}}\,^{k}}+\frac{3}{8}{\rm i} G_{\underline{i} k} (\Sigma_{a b})^{\alpha \beta} \boldsymbol{\lambda}_{l \alpha} {G}^{-3} \nabla^{a}{\lambda^{l}_{\beta}} \nabla^{b}{G_{\underline{j}}\,^{k}} - \frac{5}{16}G_{k l} (\Gamma_{a})^{\alpha \beta} \mathbf{X}_{\underline{i} \underline{j}} {G}^{-3} \nabla^{a}{\lambda^{k}_{\alpha}} \varphi^{l}_{\beta} - \frac{9}{16}G_{\underline{i} k} G_{l m} (\Gamma_{a})^{\alpha \beta} \boldsymbol{\lambda}_{\underline{j}}^{\rho} {G}^{-5} \nabla^{a}{\lambda^{l}_{\alpha}} \varphi^{k}_{\beta} \varphi^{m}_{\rho} - \frac{17}{256}G_{k l} \lambda^{k \alpha} \boldsymbol{\lambda}_{\underline{i}}^{\beta} X_{\underline{j} \alpha} {G}^{-3} \varphi^{l}_{\beta} - \frac{3}{32}{\rm i} F G_{\underline{i} k} (\Gamma_{a})^{\alpha \beta} \boldsymbol{\lambda}_{\underline{j} \alpha} {G}^{-3} \nabla^{a}{\lambda^{k}_{\beta}}+\frac{5}{32}{\rm i} \mathcal{H}_{a} G_{\underline{i} k} \boldsymbol{\lambda}_{\underline{j}}^{\alpha} {G}^{-3} \nabla^{a}{\lambda^{k}_{\alpha}}%
 - \frac{5}{16}{\rm i} \mathcal{H}^{a} G_{\underline{i} k} (\Sigma_{a b})^{\alpha \beta} \boldsymbol{\lambda}_{\underline{j} \alpha} {G}^{-3} \nabla^{b}{\lambda^{k}_{\beta}}+\frac{7}{16}{\rm i} G_{k l} \boldsymbol{\lambda}_{\underline{i}}^{\alpha} {G}^{-3} \nabla_{a}{\lambda^{k}_{\alpha}} \nabla^{a}{G_{\underline{j}}\,^{l}}+\frac{7}{8}{\rm i} G_{k l} (\Sigma_{a b})^{\alpha \beta} \boldsymbol{\lambda}_{\underline{i} \alpha} {G}^{-3} \nabla^{a}{\lambda^{k}_{\beta}} \nabla^{b}{G_{\underline{j}}\,^{l}}+\frac{9}{64}G_{\underline{i} k} G_{\underline{j} l} (\Sigma_{a b})^{\alpha \beta} W^{a b} \lambda_{m \alpha} \boldsymbol{\lambda}^{m \rho} {G}^{-5} \varphi^{k}_{\beta} \varphi^{l}_{\rho}+\frac{1}{32}(\Sigma_{a b})^{\alpha \beta} W^{a b} \lambda_{k \alpha} \boldsymbol{\lambda}_{\underline{i}}^{\rho} {G}^{-3} \varphi_{\underline{j} \rho} \varphi^{k}_{\beta} - \frac{11}{64}G_{k l} (\Sigma_{a b})^{\alpha \beta} \mathbf{X}_{\underline{i} \underline{j}} W^{a b} \lambda^{k}_{\alpha} {G}^{-3} \varphi^{l}_{\beta}+\frac{3}{64}G_{\underline{i} k} G_{l m} (\Sigma_{a b})^{\alpha \beta} W^{a b} \lambda^{l}_{\alpha} \boldsymbol{\lambda}_{\underline{j}}^{\rho} {G}^{-5} \varphi^{k}_{\beta} \varphi^{m}_{\rho}+\frac{3}{64}{\rm i} F G_{\underline{i} k} (\Sigma_{a b})^{\alpha \beta} W^{a b} \lambda^{k}_{\alpha} \boldsymbol{\lambda}_{\underline{j} \beta} {G}^{-3} - \frac{3}{128}{\rm i} \epsilon^{c d e a b} \mathcal{H}_{c} G_{\underline{i} k} (\Sigma_{d e})^{\alpha \beta} W_{a b} \lambda^{k}_{\alpha} \boldsymbol{\lambda}_{\underline{j} \beta} {G}^{-3} - \frac{3}{128}{\rm i} \epsilon^{c d}\,_{e}\,^{a b} G_{k l} (\Sigma_{c d})^{\alpha \beta} W_{a b} \lambda^{k}_{\alpha} \boldsymbol{\lambda}_{\underline{i} \beta} {G}^{-3} \nabla^{e}{G_{\underline{j}}\,^{l}} - \frac{1}{64}{\rm i} G_{k l} (\Gamma^{a})^{\alpha \beta} W_{a b} \lambda^{k}_{\alpha} \boldsymbol{\lambda}_{\underline{i} \beta} {G}^{-3} \nabla^{b}{G_{\underline{j}}\,^{l}}+\frac{1}{4}G_{k l} (\Gamma_{a})^{\alpha \beta} X_{\underline{i}}\,^{k} {G}^{-3} \nabla^{a}{\boldsymbol{\lambda}_{\underline{j} \alpha}} \varphi^{l}_{\beta}+\frac{5}{16}G_{k l} (\Sigma_{a b})^{\alpha \beta} X_{\underline{i}}\,^{k} W^{a b} \boldsymbol{\lambda}_{\underline{j} \alpha} {G}^{-3} \varphi^{l}_{\beta}+\frac{3}{16}{\rm i} G_{\underline{i} k} G_{l m} (\Sigma_{a b})^{\alpha \beta} X_{\underline{j}}\,^{l} \mathbf{F}^{a b} {G}^{-5} \varphi^{k}_{\alpha} \varphi^{m}_{\beta} - \frac{7}{8}{\rm i} (\Sigma_{a b})^{\alpha \beta} \boldsymbol{W} X_{\underline{i} k} W^{a b} {G}^{-3} \varphi_{\underline{j} \alpha} \varphi^{k}_{\beta}+\frac{1}{32}{\rm i} G_{k l} \boldsymbol{W} X_{\underline{i}}\,^{k} X_{\underline{j}}^{\alpha} {G}^{-3} \varphi^{l}_{\alpha}+\frac{3}{16}{\rm i} G_{\underline{i} k} G_{l m} (\Sigma_{a b})^{\alpha \beta} \boldsymbol{W} X_{\underline{j}}\,^{l} W^{a b} {G}^{-5} \varphi^{k}_{\alpha} \varphi^{m}_{\beta}+\frac{9}{16}{\rm i} G_{\underline{i} k} G_{l m} X^{l}\,_{n} \mathbf{X}_{\underline{j}}\,^{n} {G}^{-5} \varphi^{k \alpha} \varphi^{m}_{\alpha} - \frac{3}{8}F G_{\underline{i} k} X^{k}\,_{l} \mathbf{X}_{\underline{j}}\,^{l} {G}^{-3}+\frac{15}{16}{\rm i} G_{\underline{i} k} G_{l m} (\Gamma_{a})^{\alpha \beta} X_{\underline{j}}\,^{l} {G}^{-5} \nabla^{a}{\boldsymbol{W}} \varphi^{k}_{\alpha} \varphi^{m}_{\beta}%
 - \frac{1}{2}\mathcal{H}_{a} G_{\underline{i} k} X_{\underline{j}}\,^{k} {G}^{-3} \nabla^{a}{\boldsymbol{W}} - \frac{5}{4}G_{k l} X_{\underline{i}}\,^{k} {G}^{-3} \nabla_{a}{\boldsymbol{W}} \nabla^{a}{G_{\underline{j}}\,^{l}} - \frac{9}{32}{\rm i} G_{k l} X^{k}\,_{m} \boldsymbol{\lambda}^{m \alpha} {G}^{-5} \varphi_{\underline{i}}^{\beta} \varphi_{\underline{j} \beta} \varphi^{l}_{\alpha} - \frac{45}{32}{\rm i} G_{\underline{i} k} G_{\underline{j} l} G_{m n} X^{m}\,_{{i_{1}}} \boldsymbol{\lambda}^{{i_{1}} \alpha} {G}^{-7} \varphi^{k \beta} \varphi^{l}_{\beta} \varphi^{n}_{\alpha}+\frac{9}{16}F G_{\underline{i} \underline{j}} G_{k l} X^{k}\,_{m} \boldsymbol{\lambda}^{m \alpha} {G}^{-5} \varphi^{l}_{\alpha} - \frac{9}{32}F G_{\underline{i} k} G_{\underline{j} l} X^{k}\,_{m} \boldsymbol{\lambda}^{m \alpha} {G}^{-5} \varphi^{l}_{\alpha}+\frac{9}{32}\mathcal{H}_{a} G_{\underline{i} k} G_{\underline{j} l} (\Gamma^{a})^{\alpha \beta} X^{k}\,_{m} \boldsymbol{\lambda}^{m}_{\alpha} {G}^{-5} \varphi^{l}_{\beta}+\frac{9}{16}G_{\underline{i} k} G_{l m} (\Gamma_{a})^{\alpha \beta} X^{l}\,_{n} \boldsymbol{\lambda}^{n}_{\alpha} {G}^{-5} \nabla^{a}{G_{\underline{j}}\,^{m}} \varphi^{k}_{\beta} - \frac{3}{8}G_{\underline{i} k} (\Gamma_{a})^{\alpha \beta} X^{k}\,_{l} \boldsymbol{\lambda}^{l}_{\alpha} {G}^{-3} \nabla^{a}{\varphi_{\underline{j} \beta}}+\frac{9}{32}G_{\underline{i} k} (\Sigma_{a b})^{\alpha \beta} X^{k}\,_{l} W^{a b} \boldsymbol{\lambda}^{l}_{\alpha} {G}^{-3} \varphi_{\underline{j} \beta} - \frac{27}{256}G_{\underline{i} k} G_{\underline{j} l} X^{k}\,_{m} \boldsymbol{\lambda}^{m \alpha} X^{l}_{\alpha} {G}^{-3} - \frac{45}{256}G_{\underline{i} \underline{j}} G_{k l} X^{k}\,_{m} \boldsymbol{\lambda}^{m \alpha} X^{l}_{\alpha} {G}^{-3} - \frac{3}{4}{\rm i} G_{\underline{i} k} (\Gamma_{a})^{\alpha \beta} \boldsymbol{W} {G}^{-5} \nabla^{a}{\lambda_{\underline{j} \alpha}} \varphi^{k \rho} \varphi_{l \beta} \varphi^{l}_{\rho} - \frac{1}{2}F (\Gamma_{a})^{\alpha \beta} \boldsymbol{W} {G}^{-3} \nabla^{a}{\lambda_{\underline{i} \alpha}} \varphi_{\underline{j} \beta} - \frac{1}{4}\mathcal{H}_{a} \boldsymbol{W} {G}^{-3} \nabla^{a}{\lambda^{\alpha}_{\underline{i}}} \varphi_{\underline{j} \alpha} - \frac{1}{2}\mathcal{H}^{a} (\Sigma_{a b})^{\alpha \beta} \boldsymbol{W} {G}^{-3} \nabla^{b}{\lambda_{\underline{i} \alpha}} \varphi_{\underline{j} \beta}+\frac{1}{4}\boldsymbol{W} {G}^{-3} \nabla_{a}{\lambda^{\alpha}_{\underline{i}}} \nabla^{a}{G_{\underline{j} k}} \varphi^{k}_{\alpha} - \frac{1}{2}(\Sigma_{a b})^{\alpha \beta} \boldsymbol{W} {G}^{-3} \nabla^{a}{\lambda_{\underline{i} \alpha}} \nabla^{b}{G_{\underline{j} k}} \varphi^{k}_{\beta}+\frac{3}{16}(\Gamma_{a})^{\alpha \beta} \boldsymbol{\lambda}_{\underline{i}}^{\rho} {G}^{-3} \nabla^{a}{\lambda_{k \alpha}} \varphi_{\underline{j} \beta} \varphi^{k}_{\rho}+\frac{3}{4}{\rm i} G_{\underline{i} k} (\Gamma_{a})^{\alpha \beta} \boldsymbol{W} {G}^{-5} \nabla^{a}{\lambda_{l \alpha}} \varphi_{\underline{j} \beta} \varphi^{k \rho} \varphi^{l}_{\rho}%
 - \frac{1}{4}{\rm i} (\Gamma^{a})^{\alpha \beta} \boldsymbol{W} {G}^{-3} \nabla^{b}{F_{a b}} \varphi_{\underline{i} \alpha} \varphi_{\underline{j} \beta} - \frac{3}{4}{\rm i} (\Gamma^{a})^{\alpha \beta} \boldsymbol{W} W_{a b} {G}^{-3} \nabla^{b}{W} \varphi_{\underline{i} \alpha} \varphi_{\underline{j} \beta} - \frac{1}{4}{\rm i} (\Gamma_{a})^{\alpha \beta} \boldsymbol{W} {G}^{-3} \nabla^{a}{X_{\underline{i} k}} \varphi_{\underline{j} \alpha} \varphi^{k}_{\beta} - \frac{1}{2}{\rm i} \boldsymbol{W} {G}^{-3} \nabla_{a}{\nabla^{a}{W}} \varphi_{\underline{i}}^{\alpha} \varphi_{\underline{j} \alpha}+\frac{3}{4}{\rm i} \boldsymbol{W} W^{a b} F_{a b} {G}^{-3} \varphi_{\underline{i}}^{\alpha} \varphi_{\underline{j} \alpha} - \frac{3}{64}{\rm i} \boldsymbol{W} \lambda^{\beta}_{k} X_{\underline{i}}^{\alpha} {G}^{-3} \varphi_{\underline{j} \beta} \varphi^{k}_{\alpha}+\frac{3}{32}{\rm i} \boldsymbol{W} \lambda^{\beta}_{k} X_{\underline{i}}^{\alpha} {G}^{-3} \varphi_{\underline{j} \alpha} \varphi^{k}_{\beta} - \frac{45}{64}{\rm i} \boldsymbol{W} \lambda^{\alpha}_{k} X_{\underline{i} \alpha} {G}^{-3} \varphi_{\underline{j}}^{\beta} \varphi^{k}_{\beta}+\frac{3}{4}G_{\underline{i} k} G_{l m} (\Gamma_{a})^{\alpha \beta} \boldsymbol{\lambda}_{\underline{j}}^{\rho} {G}^{-5} \nabla^{a}{\lambda^{l}_{\alpha}} \varphi^{k}_{\rho} \varphi^{m}_{\beta} - \frac{1}{2}G_{\underline{i} k} (\Gamma^{a})^{\alpha \beta} \boldsymbol{\lambda}_{\underline{j} \alpha} {G}^{-3} \nabla^{b}{F_{a b}} \varphi^{k}_{\beta}+\frac{1}{4}G_{k l} (\Gamma_{a})^{\alpha \beta} \boldsymbol{\lambda}_{\underline{i} \alpha} {G}^{-3} \nabla^{a}{X_{\underline{j}}\,^{k}} \varphi^{l}_{\beta} - \frac{1}{4}G_{k l} W^{a b \alpha}\,_{\underline{i}} (\Sigma_{a b})^{\beta \rho} \lambda^{k}_{\beta} \boldsymbol{\lambda}_{\underline{j} \rho} {G}^{-3} \varphi^{l}_{\alpha}+\frac{1}{32}G_{k l} \lambda^{k \beta} \boldsymbol{\lambda}_{\underline{i}}^{\alpha} X_{\underline{j} \alpha} {G}^{-3} \varphi^{l}_{\beta} - \frac{3}{8}{\rm i} G_{k l} (\Gamma_{a})^{\alpha \beta} \boldsymbol{W} {G}^{-5} \nabla^{a}{\lambda^{k}_{\alpha}} \varphi_{\underline{i}}^{\rho} \varphi_{\underline{j} \rho} \varphi^{l}_{\beta} - \frac{15}{8}{\rm i} G_{\underline{i} k} G_{\underline{j} l} G_{m n} (\Gamma_{a})^{\alpha \beta} \boldsymbol{W} {G}^{-7} \nabla^{a}{\lambda^{m}_{\alpha}} \varphi^{k \rho} \varphi^{l}_{\rho} \varphi^{n}_{\beta} - \frac{3}{4}{\rm i} G_{\underline{i} k} G_{\underline{j} l} (\Gamma^{a})^{\alpha \beta} \boldsymbol{W} {G}^{-5} \nabla^{b}{F_{a b}} \varphi^{k}_{\alpha} \varphi^{l}_{\beta} - \frac{9}{4}{\rm i} G_{\underline{i} k} G_{\underline{j} l} (\Gamma^{a})^{\alpha \beta} \boldsymbol{W} W_{a b} {G}^{-5} \nabla^{b}{W} \varphi^{k}_{\alpha} \varphi^{l}_{\beta}+\frac{3}{4}{\rm i} G_{\underline{i} k} G_{l m} (\Gamma_{a})^{\alpha \beta} \boldsymbol{W} {G}^{-5} \nabla^{a}{X_{\underline{j}}\,^{l}} \varphi^{k}_{\alpha} \varphi^{m}_{\beta}+\frac{9}{16}{\rm i} G_{\underline{i} k} G_{l m} \boldsymbol{W} \lambda^{l \beta} X_{\underline{j}}^{\alpha} {G}^{-5} \varphi^{k}_{\beta} \varphi^{m}_{\alpha}+\frac{9}{16}{\rm i} G_{\underline{i} k} G_{l m} \boldsymbol{W} \lambda^{l \alpha} X_{\underline{j} \alpha} {G}^{-5} \varphi^{k \beta} \varphi^{m}_{\beta}%
 - \frac{9}{16}{\rm i} G_{\underline{i} k} G_{l m} \boldsymbol{W} \lambda^{\alpha}_{\underline{j}} X^{l}_{\alpha} {G}^{-5} \varphi^{k \beta} \varphi^{m}_{\beta}+\frac{3}{4}F G_{\underline{i} \underline{j}} G_{k l} (\Gamma_{a})^{\alpha \beta} \boldsymbol{W} {G}^{-5} \nabla^{a}{\lambda^{k}_{\alpha}} \varphi^{l}_{\beta} - \frac{3}{4}\mathcal{H}_{a} G_{\underline{i} k} G_{\underline{j} l} \boldsymbol{W} {G}^{-5} \nabla^{a}{\lambda^{k \alpha}} \varphi^{l}_{\alpha} - \frac{3}{2}\mathcal{H}^{a} G_{\underline{i} k} G_{\underline{j} l} (\Sigma_{a b})^{\alpha \beta} \boldsymbol{W} {G}^{-5} \nabla^{b}{\lambda^{k}_{\alpha}} \varphi^{l}_{\beta} - \frac{3}{4}G_{\underline{i} k} G_{l m} \boldsymbol{W} {G}^{-5} \nabla_{a}{\lambda^{l \alpha}} \nabla^{a}{G_{\underline{j}}\,^{m}} \varphi^{k}_{\alpha}+\frac{3}{2}G_{\underline{i} k} G_{l m} (\Sigma_{a b})^{\alpha \beta} \boldsymbol{W} {G}^{-5} \nabla^{a}{\lambda^{l}_{\alpha}} \nabla^{b}{G_{\underline{j}}\,^{m}} \varphi^{k}_{\beta}+G_{\underline{i} k} \boldsymbol{W} {G}^{-3} \nabla_{a}{\nabla^{a}{\lambda^{\alpha}_{\underline{j}}}} \varphi^{k}_{\alpha}+G_{\underline{i} k} (\Sigma_{a b})^{\alpha \beta} \boldsymbol{W} {G}^{-3} \nabla^{a}{\nabla^{b}{\lambda_{\underline{j} \alpha}}} \varphi^{k}_{\beta}+\frac{1}{2}{\rm i} G_{\underline{i} k} \boldsymbol{W} F^{a b} W_{a b}\,^{\alpha}\,_{\underline{j}} {G}^{-3} \varphi^{k}_{\alpha}+\frac{1}{2}\mathcal{H}_{a} G_{\underline{i} \underline{j}} \boldsymbol{W} {G}^{-3} \nabla_{b}{F^{a b}}-G_{\underline{i} k} \boldsymbol{W} {G}^{-3} \nabla^{a}{F_{a b}} \nabla^{b}{G_{\underline{j}}\,^{k}}+\frac{3}{2}\mathcal{H}^{a} G_{\underline{i} \underline{j}} \boldsymbol{W} W_{a b} {G}^{-3} \nabla^{b}{W}-3G_{\underline{i} k} \boldsymbol{W} W_{a b} {G}^{-3} \nabla^{a}{W} \nabla^{b}{G_{\underline{j}}\,^{k}} - \frac{1}{32}\Phi^{a b}\,_{\underline{i} \underline{j}} G_{k l} (\Sigma_{a b})^{\alpha \beta} \boldsymbol{W} \lambda^{k}_{\alpha} {G}^{-3} \varphi^{l}_{\beta} - \frac{1}{2}\mathcal{H}_{a} G_{\underline{i} k} \boldsymbol{W} {G}^{-3} \nabla^{a}{X_{\underline{j}}\,^{k}}-G_{k l} \boldsymbol{W} {G}^{-3} \nabla_{a}{X_{\underline{i}}\,^{k}} \nabla^{a}{G_{\underline{j}}\,^{l}} - \frac{1}{2}F G_{\underline{i} \underline{j}} \boldsymbol{W} {G}^{-3} \nabla_{a}{\nabla^{a}{W}}+\frac{3}{4}F G_{\underline{i} \underline{j}} \boldsymbol{W} W^{a b} F_{a b} {G}^{-3}+\frac{9}{128}F G_{\underline{i} k} \boldsymbol{W} \lambda^{k \alpha} X_{\underline{j} \alpha} {G}^{-3}+\frac{9}{128}\mathcal{H}_{a} G_{\underline{i} k} (\Gamma^{a})^{\beta \alpha} \boldsymbol{W} \lambda^{k}_{\beta} X_{\underline{j} \alpha} {G}^{-3}%
+\frac{3}{16}G_{k l} (\Gamma_{a})^{\beta \alpha} \boldsymbol{W} \lambda^{k}_{\beta} X_{\underline{i} \alpha} {G}^{-3} \nabla^{a}{G_{\underline{j}}\,^{l}}+\frac{5}{8}G_{\underline{i} k} \boldsymbol{W} {G}^{-3} \nabla_{a}{\varphi_{\underline{j}}^{\alpha}} \nabla^{a}{\lambda^{k}_{\alpha}} - \frac{5}{4}G_{\underline{i} k} (\Sigma_{a b})^{\alpha \beta} \boldsymbol{W} {G}^{-3} \nabla^{a}{\varphi_{\underline{j} \alpha}} \nabla^{b}{\lambda^{k}_{\beta}}+\frac{13}{32}\epsilon^{c d}\,_{e}\,^{a b} G_{\underline{i} k} (\Sigma_{c d})^{\alpha \beta} \boldsymbol{W} W_{a b} {G}^{-3} \nabla^{e}{\lambda^{k}_{\alpha}} \varphi_{\underline{j} \beta}+\frac{13}{16}G_{\underline{i} k} (\Gamma^{a})^{\alpha \beta} \boldsymbol{W} W_{a b} {G}^{-3} \nabla^{b}{\lambda^{k}_{\alpha}} \varphi_{\underline{j} \beta} - \frac{21}{64}G_{\underline{i} k} G_{\underline{j} l} (\Gamma_{a})^{\alpha \beta} \boldsymbol{W} X^{k}_{\alpha} {G}^{-3} \nabla^{a}{\lambda^{l}_{\beta}}+\frac{3}{8}G_{\underline{i} k} \boldsymbol{W} {G}^{-3} \nabla_{a}{\lambda^{k \alpha}} \nabla^{a}{\varphi_{\underline{j} \alpha}} - \frac{3}{4}G_{\underline{i} k} (\Sigma_{a b})^{\alpha \beta} \boldsymbol{W} {G}^{-3} \nabla^{a}{\lambda^{k}_{\alpha}} \nabla^{b}{\varphi_{\underline{j} \beta}} - \frac{27}{64}G_{\underline{i} \underline{j}} G_{k l} (\Gamma_{a})^{\alpha \beta} \boldsymbol{W} X^{k}_{\alpha} {G}^{-3} \nabla^{a}{\lambda^{l}_{\beta}}+\frac{9}{16}{\rm i} G_{\underline{i} k} (\Sigma_{a b})^{\alpha \beta} \boldsymbol{W} W^{a b} \lambda_{\underline{j} \alpha} {G}^{-5} \varphi^{k \rho} \varphi_{l \beta} \varphi^{l}_{\rho}+\frac{3}{8}F (\Sigma_{a b})^{\alpha \beta} \boldsymbol{W} W^{a b} \lambda_{\underline{i} \alpha} {G}^{-3} \varphi_{\underline{j} \beta} - \frac{3}{32}\epsilon^{c d e a b} \mathcal{H}_{c} (\Sigma_{d e})^{\alpha \beta} \boldsymbol{W} W_{a b} \lambda_{\underline{i} \alpha} {G}^{-3} \varphi_{\underline{j} \beta}+\frac{3}{16}\mathcal{H}^{b} (\Gamma^{a})^{\alpha \beta} \boldsymbol{W} W_{a b} \lambda_{\underline{i} \alpha} {G}^{-3} \varphi_{\underline{j} \beta}+\frac{3}{32}\epsilon^{c d}\,_{e}\,^{a b} (\Sigma_{c d})^{\alpha \beta} \boldsymbol{W} W_{a b} \lambda_{\underline{i} \alpha} {G}^{-3} \nabla^{e}{G_{\underline{j} k}} \varphi^{k}_{\beta} - \frac{3}{16}(\Gamma^{a})^{\alpha \beta} \boldsymbol{W} W_{a b} \lambda_{\underline{i} \alpha} {G}^{-3} \nabla^{b}{G_{\underline{j} k}} \varphi^{k}_{\beta} - \frac{1}{64}(\Sigma_{a b})^{\alpha \beta} W^{a b} \lambda_{k \alpha} \boldsymbol{\lambda}_{\underline{i}}^{\rho} {G}^{-3} \varphi_{\underline{j} \beta} \varphi^{k}_{\rho} - \frac{9}{16}{\rm i} G_{\underline{i} k} (\Sigma_{a b})^{\alpha \beta} \boldsymbol{W} W^{a b} \lambda_{l \alpha} {G}^{-5} \varphi_{\underline{j} \beta} \varphi^{k \rho} \varphi^{l}_{\rho}+\frac{9}{32}{\rm i} G_{k l} (\Sigma_{a b})^{\alpha \beta} \boldsymbol{W} W^{a b} \lambda^{k}_{\alpha} {G}^{-5} \varphi_{\underline{i}}^{\rho} \varphi_{\underline{j} \rho} \varphi^{l}_{\beta}+\frac{45}{32}{\rm i} G_{\underline{i} k} G_{\underline{j} l} G_{m n} (\Sigma_{a b})^{\alpha \beta} \boldsymbol{W} W^{a b} \lambda^{m}_{\alpha} {G}^{-7} \varphi^{k \rho} \varphi^{l}_{\rho} \varphi^{n}_{\beta} - \frac{9}{16}F G_{\underline{i} \underline{j}} G_{k l} (\Sigma_{a b})^{\alpha \beta} \boldsymbol{W} W^{a b} \lambda^{k}_{\alpha} {G}^{-5} \varphi^{l}_{\beta}%
 - \frac{9}{32}\epsilon^{c d e a b} \mathcal{H}_{c} G_{\underline{i} k} G_{\underline{j} l} (\Sigma_{d e})^{\alpha \beta} \boldsymbol{W} W_{a b} \lambda^{k}_{\alpha} {G}^{-5} \varphi^{l}_{\beta}+\frac{9}{16}\mathcal{H}^{b} G_{\underline{i} k} G_{\underline{j} l} (\Gamma^{a})^{\alpha \beta} \boldsymbol{W} W_{a b} \lambda^{k}_{\alpha} {G}^{-5} \varphi^{l}_{\beta} - \frac{9}{32}\epsilon^{c d}\,_{e}\,^{a b} G_{\underline{i} k} G_{l m} (\Sigma_{c d})^{\alpha \beta} \boldsymbol{W} W_{a b} \lambda^{l}_{\alpha} {G}^{-5} \nabla^{e}{G_{\underline{j}}\,^{m}} \varphi^{k}_{\beta}+\frac{9}{16}G_{\underline{i} k} G_{l m} (\Gamma^{a})^{\alpha \beta} \boldsymbol{W} W_{a b} \lambda^{l}_{\alpha} {G}^{-5} \nabla^{b}{G_{\underline{j}}\,^{m}} \varphi^{k}_{\beta}+\frac{3}{8}\epsilon^{c d}\,_{e}\,^{a b} G_{\underline{i} k} (\Sigma_{c d})^{\alpha \beta} \boldsymbol{W} W_{a b} \lambda^{k}_{\alpha} {G}^{-3} \nabla^{e}{\varphi_{\underline{j} \beta}} - \frac{3}{4}G_{\underline{i} k} (\Gamma^{a})^{\alpha \beta} \boldsymbol{W} W_{a b} \lambda^{k}_{\alpha} {G}^{-3} \nabla^{b}{\varphi_{\underline{j} \beta}}+\frac{17}{32}G_{\underline{i} k} \boldsymbol{W} W^{a b} W_{a b} \lambda^{k \alpha} {G}^{-3} \varphi_{\underline{j} \alpha} - \frac{37}{256}\epsilon_{e}\,^{a b c d} G_{\underline{i} k} (\Gamma^{e})^{\alpha \beta} \boldsymbol{W} W_{a b} W_{c d} \lambda^{k}_{\alpha} {G}^{-3} \varphi_{\underline{j} \beta}+\frac{87}{256}G_{\underline{i} k} G_{\underline{j} l} (\Sigma_{a b})^{\beta \alpha} \boldsymbol{W} W^{a b} \lambda^{k}_{\beta} X^{l}_{\alpha} {G}^{-3}+\frac{57}{256}G_{\underline{i} \underline{j}} G_{k l} (\Sigma_{a b})^{\beta \alpha} \boldsymbol{W} W^{a b} \lambda^{k}_{\beta} X^{l}_{\alpha} {G}^{-3}+\frac{3}{4}\boldsymbol{W} X_{\underline{i} k} {G}^{-5} \varphi_{\underline{j}}^{\alpha} \varphi^{k \beta} \varphi_{l \alpha} \varphi^{l}_{\beta} - \frac{3}{16}{\rm i} G_{k l} X_{\underline{i} m} \boldsymbol{\lambda}_{\underline{j}}^{\alpha} {G}^{-5} \varphi^{k \beta} \varphi^{l}_{\beta} \varphi^{m}_{\alpha} - \frac{3}{8}{\rm i} G_{k l} (\Gamma_{a})^{\alpha \beta} \boldsymbol{W} {G}^{-5} \nabla^{a}{\lambda_{\underline{i} \alpha}} \varphi_{\underline{j} \beta} \varphi^{k \rho} \varphi^{l}_{\rho}+\frac{9}{32}{\rm i} G_{k l} (\Sigma_{a b})^{\alpha \beta} \boldsymbol{W} W^{a b} \lambda_{\underline{i} \alpha} {G}^{-5} \varphi_{\underline{j} \beta} \varphi^{k \rho} \varphi^{l}_{\rho}+\frac{15}{8}G_{\underline{i} k} G_{l m} \boldsymbol{W} X_{\underline{j} n} {G}^{-7} \varphi^{k \alpha} \varphi^{l \beta} \varphi^{m}_{\beta} \varphi^{n}_{\alpha}+\frac{3}{8}{\rm i} F G_{k l} \boldsymbol{W} X_{\underline{i} \underline{j}} {G}^{-5} \varphi^{k \alpha} \varphi^{l}_{\alpha}+\frac{3}{4}{\rm i} \mathcal{H}_{a} G_{\underline{i} k} (\Gamma^{a})^{\alpha \beta} \boldsymbol{W} X_{\underline{j} l} {G}^{-5} \varphi^{k}_{\alpha} \varphi^{l}_{\beta}+\frac{3}{4}{\rm i} G_{k l} (\Gamma_{a})^{\alpha \beta} \boldsymbol{W} X_{\underline{i} m} {G}^{-5} \nabla^{a}{G_{\underline{j}}\,^{k}} \varphi^{l}_{\alpha} \varphi^{m}_{\beta} - \frac{3}{8}{\rm i} G_{k l} X^{k l} \boldsymbol{\lambda}_{\underline{i}}^{\alpha} {G}^{-5} \varphi_{\underline{j}}^{\beta} \varphi_{m \alpha} \varphi^{m}_{\beta}+\frac{3}{4}{\rm i} G_{\underline{i} k} (\Gamma_{a})^{\alpha \beta} \boldsymbol{W} {G}^{-5} \nabla^{a}{\lambda^{k}_{\alpha}} \varphi_{\underline{j}}^{\rho} \varphi_{l \beta} \varphi^{l}_{\rho}%
 - \frac{9}{16}{\rm i} G_{\underline{i} k} (\Sigma_{a b})^{\alpha \beta} \boldsymbol{W} W^{a b} \lambda^{k}_{\alpha} {G}^{-5} \varphi_{\underline{j}}^{\rho} \varphi_{l \beta} \varphi^{l}_{\rho} - \frac{15}{8}G_{\underline{i} k} G_{l m} \boldsymbol{W} X^{l m} {G}^{-7} \varphi_{\underline{j}}^{\alpha} \varphi^{k \beta} \varphi_{n \alpha} \varphi^{n}_{\beta}+\frac{3}{8}{\rm i} F G_{k l} \boldsymbol{W} X^{k l} {G}^{-5} \varphi_{\underline{i}}^{\alpha} \varphi_{\underline{j} \alpha} - \frac{3}{16}{\rm i} \mathcal{H}_{a} G_{k l} (\Gamma^{a})^{\alpha \beta} \boldsymbol{W} X^{k l} {G}^{-5} \varphi_{\underline{i} \alpha} \varphi_{\underline{j} \beta}+\frac{3}{8}{\rm i} G_{k l} (\Gamma_{a})^{\alpha \beta} \boldsymbol{W} X^{k l} {G}^{-5} \nabla^{a}{G_{\underline{i} m}} \varphi_{\underline{j} \alpha} \varphi^{m}_{\beta} - \frac{3}{8}G_{\underline{i} k} G_{l m} (\Gamma_{a})^{\alpha \beta} \boldsymbol{\lambda}_{\underline{j} \alpha} {G}^{-5} \nabla^{a}{\lambda^{k}_{\beta}} \varphi^{l \rho} \varphi^{m}_{\rho}+\frac{9}{16}G_{\underline{i} k} G_{l m} (\Sigma_{a b})^{\alpha \beta} W^{a b} \lambda^{k}_{\alpha} \boldsymbol{\lambda}_{\underline{j} \beta} {G}^{-5} \varphi^{l \rho} \varphi^{m}_{\rho}+\frac{3}{8}{\rm i} G_{k l} G_{m n} X^{k l} \mathbf{X}_{\underline{i} \underline{j}} {G}^{-5} \varphi^{m \alpha} \varphi^{n}_{\alpha}+\frac{15}{16}{\rm i} G_{\underline{i} k} G_{l m} G_{n {i_{1}}} X^{l m} \boldsymbol{\lambda}_{\underline{j}}^{\alpha} {G}^{-7} \varphi^{k}_{\alpha} \varphi^{n \beta} \varphi^{{i_{1}}}_{\beta} - \frac{9}{32}F G_{\underline{i} k} G_{l m} X^{l m} \boldsymbol{\lambda}_{\underline{j}}^{\alpha} {G}^{-5} \varphi^{k}_{\alpha} - \frac{3}{32}\mathcal{H}_{a} G_{\underline{i} k} G_{l m} (\Gamma^{a})^{\alpha \beta} X^{l m} \boldsymbol{\lambda}_{\underline{j} \alpha} {G}^{-5} \varphi^{k}_{\beta} - \frac{3}{8}G_{k l} G_{m n} (\Gamma_{a})^{\alpha \beta} X^{k l} \boldsymbol{\lambda}_{\underline{i} \alpha} {G}^{-5} \nabla^{a}{G_{\underline{j}}\,^{m}} \varphi^{n}_{\beta} - \frac{15}{8}{\rm i} G_{\underline{i} k} G_{\underline{j} l} G_{m n} (\Gamma_{a})^{\alpha \beta} \boldsymbol{W} {G}^{-7} \nabla^{a}{\lambda^{k}_{\alpha}} \varphi^{l}_{\beta} \varphi^{m \rho} \varphi^{n}_{\rho}+\frac{3}{4}{\rm i} G_{\underline{i} \underline{j}} G_{k l} \boldsymbol{W} {G}^{-5} \nabla_{a}{\nabla^{a}{W}} \varphi^{k \alpha} \varphi^{l}_{\alpha} - \frac{9}{8}{\rm i} G_{\underline{i} \underline{j}} G_{k l} \boldsymbol{W} W^{a b} F_{a b} {G}^{-5} \varphi^{k \alpha} \varphi^{l}_{\alpha} - \frac{99}{128}{\rm i} G_{\underline{i} k} G_{l m} \boldsymbol{W} \lambda^{k \alpha} X_{\underline{j} \alpha} {G}^{-5} \varphi^{l \beta} \varphi^{m}_{\beta}+\frac{45}{128}{\rm i} G_{\underline{i} \underline{j}} G_{k l} \boldsymbol{W} \lambda^{\alpha}_{m} X^{m}_{\alpha} {G}^{-5} \varphi^{k \beta} \varphi^{l}_{\beta} - \frac{3}{4}G_{\underline{i} k} G_{l m} \boldsymbol{W} {G}^{-5} \nabla_{a}{\lambda^{k \alpha}} \nabla^{a}{G_{\underline{j}}\,^{l}} \varphi^{m}_{\alpha}+\frac{3}{2}G_{\underline{i} k} G_{l m} (\Sigma_{a b})^{\alpha \beta} \boldsymbol{W} {G}^{-5} \nabla^{a}{\lambda^{k}_{\alpha}} \nabla^{b}{G_{\underline{j}}\,^{l}} \varphi^{m}_{\beta}+\frac{45}{32}{\rm i} G_{\underline{i} k} G_{\underline{j} l} G_{m n} (\Sigma_{a b})^{\alpha \beta} \boldsymbol{W} W^{a b} \lambda^{k}_{\alpha} {G}^{-7} \varphi^{l}_{\beta} \varphi^{m \rho} \varphi^{n}_{\rho}%
 - \frac{9}{32}\epsilon^{c d}\,_{e}\,^{a b} G_{\underline{i} k} G_{l m} (\Sigma_{c d})^{\alpha \beta} \boldsymbol{W} W_{a b} \lambda^{k}_{\alpha} {G}^{-5} \nabla^{e}{G_{\underline{j}}\,^{l}} \varphi^{m}_{\beta}+\frac{9}{16}G_{\underline{i} k} G_{l m} (\Gamma^{a})^{\alpha \beta} \boldsymbol{W} W_{a b} \lambda^{k}_{\alpha} {G}^{-5} \nabla^{b}{G_{\underline{j}}\,^{l}} \varphi^{m}_{\beta} - \frac{15}{32}G_{k l} G_{m n} \boldsymbol{W} X^{k l} {G}^{-7} \varphi_{\underline{i}}^{\alpha} \varphi_{\underline{j} \alpha} \varphi^{m \beta} \varphi^{n}_{\beta} - \frac{105}{32}G_{\underline{i} k} G_{\underline{j} l} G_{m n} G_{{i_{1}} {i_{2}}} \boldsymbol{W} X^{m n} {G}^{-9} \varphi^{k \alpha} \varphi^{l}_{\alpha} \varphi^{{i_{1}} \beta} \varphi^{{i_{2}}}_{\beta} - \frac{15}{16}{\rm i} F G_{\underline{i} \underline{j}} G_{k l} G_{m n} \boldsymbol{W} X^{k l} {G}^{-7} \varphi^{m \alpha} \varphi^{n}_{\alpha} - \frac{15}{16}{\rm i} \mathcal{H}_{a} G_{\underline{i} k} G_{\underline{j} l} G_{m n} (\Gamma^{a})^{\alpha \beta} \boldsymbol{W} X^{m n} {G}^{-7} \varphi^{k}_{\alpha} \varphi^{l}_{\beta} - \frac{15}{8}{\rm i} G_{\underline{i} k} G_{l m} G_{n {i_{1}}} (\Gamma_{a})^{\alpha \beta} \boldsymbol{W} X^{l m} {G}^{-7} \nabla^{a}{G_{\underline{j}}\,^{n}} \varphi^{k}_{\alpha} \varphi^{{i_{1}}}_{\beta}+\frac{3}{2}{\rm i} G_{\underline{i} k} G_{l m} (\Gamma_{a})^{\alpha \beta} \boldsymbol{W} X^{l m} {G}^{-5} \nabla^{a}{\varphi_{\underline{j} \alpha}} \varphi^{k}_{\beta}+\frac{21}{16}{\rm i} G_{\underline{i} k} G_{l m} (\Sigma_{a b})^{\alpha \beta} \boldsymbol{W} X^{l m} W^{a b} {G}^{-5} \varphi_{\underline{j} \alpha} \varphi^{k}_{\beta}+\frac{27}{128}{\rm i} G_{\underline{i} k} G_{\underline{j} l} G_{m n} \boldsymbol{W} X^{m n} X^{k \alpha} {G}^{-5} \varphi^{l}_{\alpha}+\frac{3}{16}G_{\underline{i} \underline{j}} G_{k l} \boldsymbol{W} X^{k l} {F}^{2} {G}^{-5}+\frac{3}{16}\mathcal{H}^{a} \mathcal{H}_{a} G_{\underline{i} \underline{j}} G_{k l} \boldsymbol{W} X^{k l} {G}^{-5}+\frac{3}{4}\mathcal{H}_{a} G_{\underline{i} k} G_{l m} \boldsymbol{W} X^{l m} {G}^{-5} \nabla^{a}{G_{\underline{j}}\,^{k}} - \frac{145}{128}{\rm i} G_{k l} \boldsymbol{W} X^{k l} X_{\underline{i}}^{\alpha} {G}^{-3} \varphi_{\underline{j} \alpha}+\frac{45}{128}{\rm i} G_{\underline{i} \underline{j}} G_{k l} G_{m n} \boldsymbol{W} X^{k l} X^{m \alpha} {G}^{-5} \varphi^{n}_{\alpha}+\frac{3}{4}G_{k l} G_{m n} \boldsymbol{W} X^{k l} {G}^{-5} \nabla_{a}{G_{\underline{i}}\,^{m}} \nabla^{a}{G_{\underline{j}}\,^{n}} - \frac{1}{4}(\Gamma_{a})^{\alpha \beta} \boldsymbol{\lambda}_{\underline{i} \alpha} {G}^{-3} \nabla^{a}{\lambda_{k \beta}} \varphi_{\underline{j}}^{\rho} \varphi^{k}_{\rho}+\frac{9}{16}(\Sigma_{a b})^{\alpha \beta} W^{a b} \lambda_{k \alpha} \boldsymbol{\lambda}_{\underline{i} \beta} {G}^{-3} \varphi_{\underline{j}}^{\rho} \varphi^{k}_{\rho} - \frac{1}{16}{\rm i} X_{k l} \mathbf{X}_{\underline{i} \underline{j}} {G}^{-3} \varphi^{k \alpha} \varphi^{l}_{\alpha} - \frac{3}{8}{\rm i} G_{\underline{i} k} X_{l m} \boldsymbol{\lambda}_{\underline{j}}^{\alpha} {G}^{-5} \varphi^{k}_{\alpha} \varphi^{l \beta} \varphi^{m}_{\beta}%
+\frac{7}{32}F X_{\underline{i} k} \boldsymbol{\lambda}_{\underline{j}}^{\alpha} {G}^{-3} \varphi^{k}_{\alpha}+\frac{1}{32}\mathcal{H}_{a} (\Gamma^{a})^{\alpha \beta} X_{\underline{i} k} \boldsymbol{\lambda}_{\underline{j} \alpha} {G}^{-3} \varphi^{k}_{\beta}+\frac{5}{16}(\Gamma_{a})^{\alpha \beta} X_{k l} \boldsymbol{\lambda}_{\underline{i} \alpha} {G}^{-3} \nabla^{a}{G_{\underline{j}}\,^{k}} \varphi^{l}_{\beta}+\frac{3}{4}{\rm i} G_{\underline{i} k} (\Gamma_{a})^{\alpha \beta} \boldsymbol{W} {G}^{-5} \nabla^{a}{\lambda_{l \alpha}} \varphi_{\underline{j}}^{\rho} \varphi^{k}_{\beta} \varphi^{l}_{\rho} - \frac{1}{4}\boldsymbol{W} {G}^{-3} \nabla_{a}{\lambda^{\alpha}_{k}} \nabla^{a}{G_{\underline{i}}\,^{k}} \varphi_{\underline{j} \alpha}+\frac{1}{2}(\Sigma_{a b})^{\alpha \beta} \boldsymbol{W} {G}^{-3} \nabla^{a}{\lambda_{k \alpha}} \nabla^{b}{G_{\underline{i}}\,^{k}} \varphi_{\underline{j} \beta} - \frac{9}{16}{\rm i} G_{\underline{i} k} (\Sigma_{a b})^{\alpha \beta} \boldsymbol{W} W^{a b} \lambda_{l \alpha} {G}^{-5} \varphi_{\underline{j}}^{\rho} \varphi^{k}_{\beta} \varphi^{l}_{\rho} - \frac{3}{32}\epsilon^{c d}\,_{e}\,^{a b} (\Sigma_{c d})^{\alpha \beta} \boldsymbol{W} W_{a b} \lambda_{k \alpha} {G}^{-3} \nabla^{e}{G_{\underline{i}}\,^{k}} \varphi_{\underline{j} \beta}+\frac{3}{16}(\Gamma^{a})^{\alpha \beta} \boldsymbol{W} W_{a b} \lambda_{k \alpha} {G}^{-3} \nabla^{b}{G_{\underline{i}}\,^{k}} \varphi_{\underline{j} \beta}+\frac{3}{16}\boldsymbol{W} X_{k l} {G}^{-5} \varphi_{\underline{i}}^{\alpha} \varphi_{\underline{j} \alpha} \varphi^{k \beta} \varphi^{l}_{\beta}+\frac{15}{16}G_{\underline{i} k} G_{\underline{j} l} \boldsymbol{W} X_{m n} {G}^{-7} \varphi^{k \alpha} \varphi^{l}_{\alpha} \varphi^{m \beta} \varphi^{n}_{\beta}+\frac{3}{8}{\rm i} F G_{\underline{i} \underline{j}} \boldsymbol{W} X_{k l} {G}^{-5} \varphi^{k \alpha} \varphi^{l}_{\alpha}+\frac{3}{4}{\rm i} G_{\underline{i} k} (\Gamma_{a})^{\alpha \beta} \boldsymbol{W} X_{l m} {G}^{-5} \nabla^{a}{G_{\underline{j}}\,^{l}} \varphi^{k}_{\alpha} \varphi^{m}_{\beta}-{\rm i} (\Gamma_{a})^{\alpha \beta} \boldsymbol{W} X_{\underline{i} k} {G}^{-3} \nabla^{a}{\varphi_{\underline{j} \alpha}} \varphi^{k}_{\beta} - \frac{3}{64}{\rm i} G_{\underline{i} k} \boldsymbol{W} X_{\underline{j} l} X^{k \alpha} {G}^{-3} \varphi^{l}_{\alpha} - \frac{1}{8}\boldsymbol{W} X_{\underline{i} \underline{j}} {F}^{2} {G}^{-3} - \frac{1}{8}\mathcal{H}^{a} \mathcal{H}_{a} \boldsymbol{W} X_{\underline{i} \underline{j}} {G}^{-3} - \frac{1}{2}\mathcal{H}_{a} \boldsymbol{W} X_{\underline{i} k} {G}^{-3} \nabla^{a}{G_{\underline{j}}\,^{k}} - \frac{41}{64}{\rm i} G_{\underline{i} k} \boldsymbol{W} X^{k}\,_{l} X_{\underline{j}}^{\alpha} {G}^{-3} \varphi^{l}_{\alpha} - \frac{15}{64}{\rm i} G_{\underline{i} \underline{j}} \boldsymbol{W} X_{k l} X^{k \alpha} {G}^{-3} \varphi^{l}_{\alpha}%
 - \frac{1}{2}\boldsymbol{W} X_{k l} {G}^{-3} \nabla_{a}{G_{\underline{i}}\,^{k}} \nabla^{a}{G_{\underline{j}}\,^{l}}+\frac{1}{8}G_{k l} (\Gamma_{a})^{\alpha \beta} X^{k l} \boldsymbol{\lambda}_{\underline{i} \alpha} {G}^{-3} \nabla^{a}{\varphi_{\underline{j} \beta}} - \frac{1}{2}G_{k l} \boldsymbol{W} X^{k l} {G}^{-3} \nabla_{a}{\nabla^{a}{G_{\underline{i} \underline{j}}}}+\frac{7}{16}G_{\underline{i} \underline{j}} G_{k l} \boldsymbol{W} X^{k l} W^{a b} W_{a b} {G}^{-3} - \frac{5}{16}G_{k l} (\Sigma_{a b})^{\alpha \beta} X^{k l} W^{a b} \boldsymbol{\lambda}_{\underline{i} \alpha} {G}^{-3} \varphi_{\underline{j} \beta} - \frac{45}{256}G_{\underline{i} k} G_{l m} X^{l m} \boldsymbol{\lambda}_{\underline{j}}^{\alpha} X^{k}_{\alpha} {G}^{-3} - \frac{5}{8}F G_{k l} X^{k l} \mathbf{X}_{\underline{i} \underline{j}} {G}^{-3}+2{G}^{-1} \nabla_{a}{\boldsymbol{W}} \nabla^{a}{X_{\underline{i} \underline{j}}} - \frac{1}{8}(\Gamma_{a})^{\alpha \beta} \boldsymbol{\lambda}_{k \alpha} {G}^{-3} \nabla^{a}{\lambda^{k}_{\beta}} \varphi_{\underline{i}}^{\rho} \varphi_{\underline{j} \rho} - \frac{3}{8}G_{\underline{i} k} G_{\underline{j} l} (\Gamma_{a})^{\alpha \beta} \boldsymbol{\lambda}_{m \alpha} {G}^{-5} \nabla^{a}{\lambda^{m}_{\beta}} \varphi^{k \rho} \varphi^{l}_{\rho}+\frac{1}{4}G_{\underline{i} k} (\Gamma_{a})^{\alpha \beta} \boldsymbol{\lambda}_{l \alpha} {G}^{-3} \nabla^{a}{X_{\underline{j}}\,^{l}} \varphi^{k}_{\beta}-{\rm i} \boldsymbol{\lambda}_{\underline{i}}^{\alpha} {G}^{-1} \nabla_{a}{\nabla^{a}{\lambda_{\underline{j} \alpha}}}+{\rm i} (\Sigma_{a b})^{\alpha \beta} \boldsymbol{\lambda}_{\underline{i} \alpha} {G}^{-1} \nabla^{a}{\nabla^{b}{\lambda_{\underline{j} \beta}}}+\boldsymbol{W} {G}^{-1} \nabla_{a}{\nabla^{a}{X_{\underline{i} \underline{j}}}}+\frac{3}{8}{\rm i} G_{\underline{i} k} (\Sigma_{a b})^{\alpha \beta} F^{a b} \boldsymbol{\lambda}_{\underline{j} \alpha} {G}^{-5} \varphi^{k \rho} \varphi_{l \beta} \varphi^{l}_{\rho}+\frac{1}{4}F (\Sigma_{a b})^{\alpha \beta} F^{a b} \boldsymbol{\lambda}_{\underline{i} \alpha} {G}^{-3} \varphi_{\underline{j} \beta} - \frac{1}{16}\epsilon^{c d e a b} \mathcal{H}_{c} (\Sigma_{d e})^{\alpha \beta} F_{a b} \boldsymbol{\lambda}_{\underline{i} \alpha} {G}^{-3} \varphi_{\underline{j} \beta}+\frac{1}{8}\mathcal{H}^{b} (\Gamma^{a})^{\alpha \beta} F_{a b} \boldsymbol{\lambda}_{\underline{i} \alpha} {G}^{-3} \varphi_{\underline{j} \beta}+\frac{1}{16}\epsilon^{c d}\,_{e}\,^{a b} (\Sigma_{c d})^{\alpha \beta} F_{a b} \boldsymbol{\lambda}_{\underline{i} \alpha} {G}^{-3} \nabla^{e}{G_{\underline{j} k}} \varphi^{k}_{\beta} - \frac{1}{8}(\Gamma^{a})^{\alpha \beta} F_{a b} \boldsymbol{\lambda}_{\underline{i} \alpha} {G}^{-3} \nabla^{b}{G_{\underline{j} k}} \varphi^{k}_{\beta}%
+\frac{1}{8}(\Gamma_{a})^{\alpha \beta} \boldsymbol{\lambda}_{k \alpha} {G}^{-3} \nabla^{a}{\lambda^{\rho}_{\underline{i}}} \varphi_{\underline{j} \rho} \varphi^{k}_{\beta} - \frac{1}{8}(\Gamma_{a})^{\alpha \beta} \boldsymbol{\lambda}_{k}^{\rho} {G}^{-3} \nabla^{a}{\lambda_{\underline{i} \rho}} \varphi_{\underline{j} \alpha} \varphi^{k}_{\beta} - \frac{3}{8}(\Sigma_{a b})^{\alpha \beta} W^{a b} \lambda^{\rho}_{\underline{i}} \boldsymbol{\lambda}_{k \rho} {G}^{-3} \varphi_{\underline{j} \alpha} \varphi^{k}_{\beta}+\frac{3}{8}(\Gamma_{a})^{\alpha \beta} \boldsymbol{\lambda}_{k \alpha} {G}^{-3} \nabla^{a}{\lambda_{\underline{i} \beta}} \varphi_{\underline{j}}^{\rho} \varphi^{k}_{\rho}+\frac{1}{32}(\Sigma_{a b})^{\alpha \beta} W^{a b} \lambda_{\underline{i} \alpha} \boldsymbol{\lambda}_{k}^{\rho} {G}^{-3} \varphi_{\underline{j} \beta} \varphi^{k}_{\rho}+\frac{1}{4}{\rm i} F^{a b} \mathbf{F}_{a b} {G}^{-3} \varphi_{\underline{i}}^{\alpha} \varphi_{\underline{j} \alpha} - \frac{1}{16}{\rm i} \epsilon_{e}\,^{a b c d} (\Gamma^{e})^{\alpha \beta} F_{a b} \mathbf{F}_{c d} {G}^{-3} \varphi_{\underline{i} \alpha} \varphi_{\underline{j} \beta} - \frac{1}{8}{\rm i} (\Sigma_{a b})^{\alpha \beta} \mathbf{X}_{\underline{i} k} F^{a b} {G}^{-3} \varphi_{\underline{j} \alpha} \varphi^{k}_{\beta} - \frac{1}{4}{\rm i} (\Gamma^{a})^{\alpha \beta} F_{a b} {G}^{-3} \nabla^{b}{\boldsymbol{W}} \varphi_{\underline{i} \alpha} \varphi_{\underline{j} \beta} - \frac{3}{8}{\rm i} G_{\underline{i} k} (\Sigma_{a b})^{\alpha \beta} F^{a b} \boldsymbol{\lambda}_{l \alpha} {G}^{-5} \varphi_{\underline{j} \beta} \varphi^{k \rho} \varphi^{l}_{\rho}+\frac{3}{8}G_{\underline{i} k} G_{l m} (\Gamma_{a})^{\alpha \beta} \boldsymbol{\lambda}^{l}_{\alpha} {G}^{-5} \nabla^{a}{\lambda^{\rho}_{\underline{j}}} \varphi^{k}_{\beta} \varphi^{m}_{\rho} - \frac{1}{4}G_{k l} (\Gamma_{a})^{\alpha \beta} \boldsymbol{\lambda}^{k}_{\alpha} {G}^{-3} \nabla^{a}{X_{\underline{i} \underline{j}}} \varphi^{l}_{\beta}+\frac{3}{32}{\rm i} F G_{\underline{i} k} (\Gamma_{a})^{\alpha \beta} \boldsymbol{\lambda}^{k}_{\alpha} {G}^{-3} \nabla^{a}{\lambda_{\underline{j} \beta}}+\frac{11}{32}{\rm i} \mathcal{H}_{a} G_{\underline{i} k} \boldsymbol{\lambda}^{k \alpha} {G}^{-3} \nabla^{a}{\lambda_{\underline{j} \alpha}}+\frac{5}{16}{\rm i} \mathcal{H}^{a} G_{\underline{i} k} (\Sigma_{a b})^{\alpha \beta} \boldsymbol{\lambda}^{k}_{\alpha} {G}^{-3} \nabla^{b}{\lambda_{\underline{j} \beta}}+\frac{5}{8}{\rm i} G_{k l} \boldsymbol{\lambda}^{k \alpha} {G}^{-3} \nabla_{a}{\lambda_{\underline{i} \alpha}} \nabla^{a}{G_{\underline{j}}\,^{l}} - \frac{3}{4}{\rm i} G_{k l} (\Sigma_{a b})^{\alpha \beta} \boldsymbol{\lambda}^{k}_{\alpha} {G}^{-3} \nabla^{a}{\lambda_{\underline{i} \beta}} \nabla^{b}{G_{\underline{j}}\,^{l}}+\frac{3}{8}G_{\underline{i} k} G_{l m} (\Gamma_{a})^{\alpha \beta} \boldsymbol{\lambda}^{l \rho} {G}^{-5} \nabla^{a}{\lambda_{\underline{j} \rho}} \varphi^{k}_{\alpha} \varphi^{m}_{\beta} - \frac{3}{8}G_{\underline{i} k} G_{l m} (\Gamma_{a})^{\alpha \beta} \boldsymbol{\lambda}^{l \rho} {G}^{-5} \nabla^{a}{\lambda_{\underline{j} \alpha}} \varphi^{k}_{\rho} \varphi^{m}_{\beta} - \frac{3}{8}G_{\underline{i} k} G_{l m} (\Gamma_{a})^{\alpha \beta} \boldsymbol{\lambda}^{l}_{\alpha} {G}^{-5} \nabla^{a}{\lambda_{\underline{j} \beta}} \varphi^{k \rho} \varphi^{m}_{\rho}%
+\frac{9}{16}G_{\underline{i} k} G_{l m} (\Sigma_{a b})^{\alpha \beta} W^{a b} \lambda_{\underline{j} \alpha} \boldsymbol{\lambda}^{l}_{\beta} {G}^{-5} \varphi^{k \rho} \varphi^{m}_{\rho} - \frac{3}{16}{\rm i} \epsilon_{e}\,^{a b c d} G_{\underline{i} k} G_{\underline{j} l} (\Gamma^{e})^{\alpha \beta} F_{a b} \mathbf{F}_{c d} {G}^{-5} \varphi^{k}_{\alpha} \varphi^{l}_{\beta}+\frac{1}{4}F G_{\underline{i} \underline{j}} F^{a b} \mathbf{F}_{a b} {G}^{-3}+\frac{1}{8}\epsilon^{e a b c d} \mathcal{H}_{e} G_{\underline{i} \underline{j}} F_{a b} \mathbf{F}_{c d} {G}^{-3}+\frac{1}{4}\epsilon_{e}\,^{a b c d} G_{\underline{i} k} F_{a b} \mathbf{F}_{c d} {G}^{-3} \nabla^{e}{G_{\underline{j}}\,^{k}} - \frac{3}{4}{\rm i} G_{\underline{i} k} G_{\underline{j} l} (\Gamma^{a})^{\alpha \beta} F_{a b} {G}^{-5} \nabla^{b}{\boldsymbol{W}} \varphi^{k}_{\alpha} \varphi^{l}_{\beta}+\frac{1}{2}\mathcal{H}^{a} G_{\underline{i} \underline{j}} F_{a b} {G}^{-3} \nabla^{b}{\boldsymbol{W}}-G_{\underline{i} k} F_{a b} {G}^{-3} \nabla^{a}{\boldsymbol{W}} \nabla^{b}{G_{\underline{j}}\,^{k}}+\frac{3}{16}{\rm i} G_{k l} (\Sigma_{a b})^{\alpha \beta} F^{a b} \boldsymbol{\lambda}^{k}_{\alpha} {G}^{-5} \varphi_{\underline{i}}^{\rho} \varphi_{\underline{j} \rho} \varphi^{l}_{\beta}+\frac{15}{16}{\rm i} G_{\underline{i} k} G_{\underline{j} l} G_{m n} (\Sigma_{a b})^{\alpha \beta} F^{a b} \boldsymbol{\lambda}^{m}_{\alpha} {G}^{-7} \varphi^{k \rho} \varphi^{l}_{\rho} \varphi^{n}_{\beta} - \frac{3}{8}F G_{\underline{i} \underline{j}} G_{k l} (\Sigma_{a b})^{\alpha \beta} F^{a b} \boldsymbol{\lambda}^{k}_{\alpha} {G}^{-5} \varphi^{l}_{\beta} - \frac{3}{16}\epsilon^{c d e a b} \mathcal{H}_{c} G_{\underline{i} k} G_{\underline{j} l} (\Sigma_{d e})^{\alpha \beta} F_{a b} \boldsymbol{\lambda}^{k}_{\alpha} {G}^{-5} \varphi^{l}_{\beta}+\frac{3}{8}\mathcal{H}^{b} G_{\underline{i} k} G_{\underline{j} l} (\Gamma^{a})^{\alpha \beta} F_{a b} \boldsymbol{\lambda}^{k}_{\alpha} {G}^{-5} \varphi^{l}_{\beta} - \frac{3}{16}\epsilon^{c d}\,_{e}\,^{a b} G_{\underline{i} k} G_{l m} (\Sigma_{c d})^{\alpha \beta} F_{a b} \boldsymbol{\lambda}^{l}_{\alpha} {G}^{-5} \nabla^{e}{G_{\underline{j}}\,^{m}} \varphi^{k}_{\beta}+\frac{3}{8}G_{\underline{i} k} G_{l m} (\Gamma^{a})^{\alpha \beta} F_{a b} \boldsymbol{\lambda}^{l}_{\alpha} {G}^{-5} \nabla^{b}{G_{\underline{j}}\,^{m}} \varphi^{k}_{\beta}+\frac{1}{4}\epsilon^{c d}\,_{e}\,^{a b} G_{\underline{i} k} (\Sigma_{c d})^{\alpha \beta} F_{a b} \boldsymbol{\lambda}^{k}_{\alpha} {G}^{-3} \nabla^{e}{\varphi_{\underline{j} \beta}} - \frac{1}{2}G_{\underline{i} k} (\Gamma^{a})^{\alpha \beta} F_{a b} \boldsymbol{\lambda}^{k}_{\alpha} {G}^{-3} \nabla^{b}{\varphi_{\underline{j} \beta}}+\frac{57}{128}G_{\underline{i} k} W^{a b} F_{a b} \boldsymbol{\lambda}^{k \alpha} {G}^{-3} \varphi_{\underline{j} \alpha} - \frac{25}{256}\epsilon_{e}\,^{c d a b} G_{\underline{i} k} (\Gamma^{e})^{\alpha \beta} W_{c d} F_{a b} \boldsymbol{\lambda}^{k}_{\alpha} {G}^{-3} \varphi_{\underline{j} \beta} - \frac{23}{32}G_{\underline{i} k} (\Sigma^{a}{}_{\, c})^{\alpha \beta} W^{c b} F_{a b} \boldsymbol{\lambda}^{k}_{\alpha} {G}^{-3} \varphi_{\underline{j} \beta}%
+\frac{33}{128}G_{\underline{i} k} G_{\underline{j} l} (\Sigma_{a b})^{\beta \alpha} F^{a b} \boldsymbol{\lambda}^{k}_{\beta} X^{l}_{\alpha} {G}^{-3}+\frac{15}{128}G_{\underline{i} \underline{j}} G_{k l} (\Sigma_{a b})^{\beta \alpha} F^{a b} \boldsymbol{\lambda}^{k}_{\beta} X^{l}_{\alpha} {G}^{-3} - \frac{3}{8}{\rm i} G_{\underline{i} k} (\Gamma_{a})^{\alpha \beta} \boldsymbol{\lambda}_{\underline{j} \alpha} {G}^{-5} \nabla^{a}{W} \varphi^{k \rho} \varphi_{l \beta} \varphi^{l}_{\rho} - \frac{1}{4}F (\Gamma_{a})^{\alpha \beta} \boldsymbol{\lambda}_{\underline{i} \alpha} {G}^{-3} \nabla^{a}{W} \varphi_{\underline{j} \beta} - \frac{1}{8}\mathcal{H}_{a} \boldsymbol{\lambda}_{\underline{i}}^{\alpha} {G}^{-3} \nabla^{a}{W} \varphi_{\underline{j} \alpha} - \frac{1}{4}\mathcal{H}^{a} (\Sigma_{a b})^{\alpha \beta} \boldsymbol{\lambda}_{\underline{i} \alpha} {G}^{-3} \nabla^{b}{W} \varphi_{\underline{j} \beta}+\frac{1}{8}\boldsymbol{\lambda}_{\underline{i}}^{\alpha} {G}^{-3} \nabla_{a}{W} \nabla^{a}{G_{\underline{j} k}} \varphi^{k}_{\alpha} - \frac{1}{4}(\Sigma_{a b})^{\alpha \beta} \boldsymbol{\lambda}_{\underline{i} \alpha} {G}^{-3} \nabla^{a}{W} \nabla^{b}{G_{\underline{j} k}} \varphi^{k}_{\beta} - \frac{1}{4}{\rm i} (\Gamma^{a})^{\alpha \beta} \mathbf{F}_{a b} {G}^{-3} \nabla^{b}{W} \varphi_{\underline{i} \alpha} \varphi_{\underline{j} \beta} - \frac{1}{2}{\rm i} {G}^{-3} \nabla_{a}{W} \nabla^{a}{\boldsymbol{W}} \varphi_{\underline{i}}^{\alpha} \varphi_{\underline{j} \alpha}+\frac{3}{8}{\rm i} G_{\underline{i} k} (\Gamma_{a})^{\alpha \beta} \boldsymbol{\lambda}_{l \alpha} {G}^{-5} \nabla^{a}{W} \varphi_{\underline{j} \beta} \varphi^{k \rho} \varphi^{l}_{\rho} - \frac{1}{8}(\Gamma_{a})^{\alpha \beta} \boldsymbol{\lambda}_{k \alpha} {G}^{-3} \nabla^{a}{\lambda^{\rho}_{\underline{i}}} \varphi_{\underline{j} \beta} \varphi^{k}_{\rho} - \frac{1}{64}(\Sigma_{a b})^{\alpha \beta} W^{a b} \lambda_{\underline{i} \alpha} \boldsymbol{\lambda}_{k}^{\rho} {G}^{-3} \varphi_{\underline{j} \rho} \varphi^{k}_{\beta} - \frac{3}{4}{\rm i} G_{\underline{i} k} G_{\underline{j} l} (\Gamma^{a})^{\alpha \beta} \mathbf{F}_{a b} {G}^{-5} \nabla^{b}{W} \varphi^{k}_{\alpha} \varphi^{l}_{\beta}+\frac{1}{2}\mathcal{H}^{a} G_{\underline{i} \underline{j}} \mathbf{F}_{a b} {G}^{-3} \nabla^{b}{W}-G_{\underline{i} k} \mathbf{F}_{a b} {G}^{-3} \nabla^{a}{W} \nabla^{b}{G_{\underline{j}}\,^{k}} - \frac{1}{2}F G_{\underline{i} \underline{j}} {G}^{-3} \nabla_{a}{W} \nabla^{a}{\boldsymbol{W}} - \frac{3}{16}{\rm i} G_{k l} (\Gamma_{a})^{\alpha \beta} \boldsymbol{\lambda}^{k}_{\alpha} {G}^{-5} \nabla^{a}{W} \varphi_{\underline{i}}^{\rho} \varphi_{\underline{j} \rho} \varphi^{l}_{\beta} - \frac{15}{16}{\rm i} G_{\underline{i} k} G_{\underline{j} l} G_{m n} (\Gamma_{a})^{\alpha \beta} \boldsymbol{\lambda}^{m}_{\alpha} {G}^{-7} \nabla^{a}{W} \varphi^{k \rho} \varphi^{l}_{\rho} \varphi^{n}_{\beta} - \frac{3}{8}G_{\underline{i} k} G_{l m} (\Gamma_{a})^{\alpha \beta} \boldsymbol{\lambda}^{l}_{\alpha} {G}^{-5} \nabla^{a}{\lambda^{\rho}_{\underline{j}}} \varphi^{k}_{\rho} \varphi^{m}_{\beta}%
 - \frac{3}{32}G_{\underline{i} k} G_{l m} (\Sigma_{a b})^{\alpha \beta} W^{a b} \lambda_{\underline{j} \alpha} \boldsymbol{\lambda}^{l \rho} {G}^{-5} \varphi^{k}_{\beta} \varphi^{m}_{\rho}+\frac{3}{8}F G_{\underline{i} \underline{j}} G_{k l} (\Gamma_{a})^{\alpha \beta} \boldsymbol{\lambda}^{k}_{\alpha} {G}^{-5} \nabla^{a}{W} \varphi^{l}_{\beta} - \frac{3}{8}\mathcal{H}_{a} G_{\underline{i} k} G_{\underline{j} l} \boldsymbol{\lambda}^{k \alpha} {G}^{-5} \nabla^{a}{W} \varphi^{l}_{\alpha} - \frac{3}{4}\mathcal{H}^{a} G_{\underline{i} k} G_{\underline{j} l} (\Sigma_{a b})^{\alpha \beta} \boldsymbol{\lambda}^{k}_{\alpha} {G}^{-5} \nabla^{b}{W} \varphi^{l}_{\beta} - \frac{3}{8}G_{\underline{i} k} G_{l m} \boldsymbol{\lambda}^{l \alpha} {G}^{-5} \nabla_{a}{W} \nabla^{a}{G_{\underline{j}}\,^{m}} \varphi^{k}_{\alpha}+\frac{3}{4}G_{\underline{i} k} G_{l m} (\Sigma_{a b})^{\alpha \beta} \boldsymbol{\lambda}^{l}_{\alpha} {G}^{-5} \nabla^{a}{W} \nabla^{b}{G_{\underline{j}}\,^{m}} \varphi^{k}_{\beta}+\frac{1}{2}G_{\underline{i} k} \boldsymbol{\lambda}^{k \alpha} {G}^{-3} \nabla_{a}{W} \nabla^{a}{\varphi_{\underline{j} \alpha}}-G_{\underline{i} k} (\Sigma_{a b})^{\alpha \beta} \boldsymbol{\lambda}^{k}_{\alpha} {G}^{-3} \nabla^{a}{W} \nabla^{b}{\varphi_{\underline{j} \beta}}+\frac{23}{128}\epsilon^{c d}\,_{e}\,^{a b} G_{\underline{i} k} (\Sigma_{c d})^{\alpha \beta} W_{a b} \boldsymbol{\lambda}^{k}_{\alpha} {G}^{-3} \nabla^{e}{W} \varphi_{\underline{j} \beta}+\frac{25}{64}G_{\underline{i} k} (\Gamma^{a})^{\alpha \beta} W_{a b} \boldsymbol{\lambda}^{k}_{\alpha} {G}^{-3} \nabla^{b}{W} \varphi_{\underline{j} \beta} - \frac{33}{128}G_{\underline{i} k} G_{\underline{j} l} (\Gamma_{a})^{\beta \alpha} \boldsymbol{\lambda}^{k}_{\beta} X^{l}_{\alpha} {G}^{-3} \nabla^{a}{W} - \frac{15}{128}G_{\underline{i} \underline{j}} G_{k l} (\Gamma_{a})^{\beta \alpha} \boldsymbol{\lambda}^{k}_{\beta} X^{l}_{\alpha} {G}^{-3} \nabla^{a}{W} - \frac{3}{8}{\rm i} \lambda^{\alpha}_{\underline{i}} \boldsymbol{\lambda}_{k \alpha} {G}^{-5} \varphi_{\underline{j}}^{\beta} \varphi^{k \rho} \varphi_{l \beta} \varphi^{l}_{\rho}+\frac{3}{16}{\rm i} G_{k l} X_{\underline{i} \underline{j}} \boldsymbol{\lambda}_{m}^{\alpha} {G}^{-5} \varphi^{k \beta} \varphi^{l}_{\beta} \varphi^{m}_{\alpha}+\frac{3}{16}{\rm i} G_{k l} (\Sigma_{a b})^{\alpha \beta} \mathbf{F}^{a b} \lambda_{\underline{i} \alpha} {G}^{-5} \varphi_{\underline{j} \beta} \varphi^{k \rho} \varphi^{l}_{\rho} - \frac{3}{16}{\rm i} G_{k l} (\Gamma_{a})^{\alpha \beta} \lambda_{\underline{i} \alpha} {G}^{-5} \nabla^{a}{\boldsymbol{W}} \varphi_{\underline{j} \beta} \varphi^{k \rho} \varphi^{l}_{\rho} - \frac{15}{16}{\rm i} G_{\underline{i} k} G_{l m} \lambda^{\alpha}_{\underline{j}} \boldsymbol{\lambda}_{n \alpha} {G}^{-7} \varphi^{k \beta} \varphi^{l \rho} \varphi^{m}_{\rho} \varphi^{n}_{\beta}+\frac{3}{8}F G_{k l} \lambda^{\alpha}_{\underline{i}} \boldsymbol{\lambda}_{\underline{j} \alpha} {G}^{-5} \varphi^{k \beta} \varphi^{l}_{\beta}+\frac{3}{8}\mathcal{H}_{a} G_{\underline{i} k} (\Gamma^{a})^{\alpha \beta} \lambda^{\rho}_{\underline{j}} \boldsymbol{\lambda}_{l \rho} {G}^{-5} \varphi^{k}_{\alpha} \varphi^{l}_{\beta}+\frac{3}{8}G_{k l} (\Gamma_{a})^{\alpha \beta} \lambda^{\rho}_{\underline{i}} \boldsymbol{\lambda}_{m \rho} {G}^{-5} \nabla^{a}{G_{\underline{j}}\,^{k}} \varphi^{l}_{\alpha} \varphi^{m}_{\beta}%
 - \frac{3}{8}{\rm i} \lambda^{\alpha}_{k} \boldsymbol{\lambda}_{\underline{i} \alpha} {G}^{-5} \varphi_{\underline{j}}^{\beta} \varphi^{k \rho} \varphi_{l \beta} \varphi^{l}_{\rho}+\frac{3}{16}{\rm i} G_{k l} \mathbf{X}_{\underline{i} \underline{j}} \lambda^{\alpha}_{m} {G}^{-5} \varphi^{k \beta} \varphi^{l}_{\beta} \varphi^{m}_{\alpha}+\frac{3}{16}{\rm i} G_{k l} (\Sigma_{a b})^{\alpha \beta} F^{a b} \boldsymbol{\lambda}_{\underline{i} \alpha} {G}^{-5} \varphi_{\underline{j} \beta} \varphi^{k \rho} \varphi^{l}_{\rho} - \frac{3}{16}{\rm i} G_{k l} (\Gamma_{a})^{\alpha \beta} \boldsymbol{\lambda}_{\underline{i} \alpha} {G}^{-5} \nabla^{a}{W} \varphi_{\underline{j} \beta} \varphi^{k \rho} \varphi^{l}_{\rho} - \frac{15}{16}{\rm i} G_{\underline{i} k} G_{l m} \lambda^{\alpha}_{n} \boldsymbol{\lambda}_{\underline{j} \alpha} {G}^{-7} \varphi^{k \beta} \varphi^{l \rho} \varphi^{m}_{\rho} \varphi^{n}_{\beta}+\frac{3}{8}\mathcal{H}_{a} G_{\underline{i} k} (\Gamma^{a})^{\alpha \beta} \lambda^{\rho}_{l} \boldsymbol{\lambda}_{\underline{j} \rho} {G}^{-5} \varphi^{k}_{\alpha} \varphi^{l}_{\beta}+\frac{3}{8}G_{k l} (\Gamma_{a})^{\alpha \beta} \lambda^{\rho}_{m} \boldsymbol{\lambda}_{\underline{i} \rho} {G}^{-5} \nabla^{a}{G_{\underline{j}}\,^{k}} \varphi^{l}_{\alpha} \varphi^{m}_{\beta} - \frac{3}{8}{\rm i} G_{\underline{i} k} (\Sigma_{a b})^{\alpha \beta} F^{a b} \boldsymbol{\lambda}^{k}_{\alpha} {G}^{-5} \varphi_{\underline{j}}^{\rho} \varphi_{l \beta} \varphi^{l}_{\rho}+\frac{3}{8}{\rm i} G_{k l} X_{\underline{i}}\,^{k} \boldsymbol{\lambda}^{l \alpha} {G}^{-5} \varphi_{\underline{j}}^{\beta} \varphi_{m \alpha} \varphi^{m}_{\beta}+\frac{3}{8}{\rm i} G_{\underline{i} k} (\Gamma_{a})^{\alpha \beta} \boldsymbol{\lambda}^{k}_{\alpha} {G}^{-5} \nabla^{a}{W} \varphi_{\underline{j}}^{\rho} \varphi_{l \beta} \varphi^{l}_{\rho} - \frac{3}{8}{\rm i} G_{\underline{i} k} (\Sigma_{a b})^{\alpha \beta} \mathbf{F}^{a b} \lambda^{k}_{\alpha} {G}^{-5} \varphi_{\underline{j}}^{\rho} \varphi_{l \beta} \varphi^{l}_{\rho}+\frac{3}{8}{\rm i} G_{k l} \mathbf{X}_{\underline{i}}\,^{k} \lambda^{l \alpha} {G}^{-5} \varphi_{\underline{j}}^{\beta} \varphi_{m \alpha} \varphi^{m}_{\beta}+\frac{3}{8}{\rm i} G_{\underline{i} k} (\Gamma_{a})^{\alpha \beta} \lambda^{k}_{\alpha} {G}^{-5} \nabla^{a}{\boldsymbol{W}} \varphi_{\underline{j}}^{\rho} \varphi_{l \beta} \varphi^{l}_{\rho}+\frac{15}{8}{\rm i} G_{\underline{i} k} G_{l m} \lambda^{l \alpha} \boldsymbol{\lambda}^{m}_{\alpha} {G}^{-7} \varphi_{\underline{j}}^{\beta} \varphi^{k \rho} \varphi_{n \beta} \varphi^{n}_{\rho}+\frac{3}{8}F G_{k l} \lambda^{k \alpha} \boldsymbol{\lambda}^{l}_{\alpha} {G}^{-5} \varphi_{\underline{i}}^{\beta} \varphi_{\underline{j} \beta} - \frac{3}{16}\mathcal{H}_{a} G_{k l} (\Gamma^{a})^{\alpha \beta} \lambda^{k \rho} \boldsymbol{\lambda}^{l}_{\rho} {G}^{-5} \varphi_{\underline{i} \alpha} \varphi_{\underline{j} \beta}+\frac{3}{8}G_{k l} (\Gamma_{a})^{\alpha \beta} \lambda^{k \rho} \boldsymbol{\lambda}^{l}_{\rho} {G}^{-5} \nabla^{a}{G_{\underline{i} m}} \varphi_{\underline{j} \alpha} \varphi^{m}_{\beta}+\frac{3}{8}G_{\underline{i} k} G_{l m} (\Gamma_{a})^{\alpha \beta} \boldsymbol{\lambda}^{k}_{\alpha} {G}^{-5} \nabla^{a}{\lambda_{\underline{j} \beta}} \varphi^{l \rho} \varphi^{m}_{\rho} - \frac{3}{8}{\rm i} G_{\underline{i} \underline{j}} G_{k l} F^{a b} \mathbf{F}_{a b} {G}^{-5} \varphi^{k \alpha} \varphi^{l}_{\alpha}+\frac{15}{16}{\rm i} G_{\underline{i} k} G_{\underline{j} l} G_{m n} (\Sigma_{a b})^{\alpha \beta} F^{a b} \boldsymbol{\lambda}^{k}_{\alpha} {G}^{-7} \varphi^{l}_{\beta} \varphi^{m \rho} \varphi^{n}_{\rho}%
 - \frac{3}{16}\epsilon^{c d}\,_{e}\,^{a b} G_{\underline{i} k} G_{l m} (\Sigma_{c d})^{\alpha \beta} F_{a b} \boldsymbol{\lambda}^{k}_{\alpha} {G}^{-5} \nabla^{e}{G_{\underline{j}}\,^{l}} \varphi^{m}_{\beta}+\frac{3}{8}G_{\underline{i} k} G_{l m} (\Gamma^{a})^{\alpha \beta} F_{a b} \boldsymbol{\lambda}^{k}_{\alpha} {G}^{-5} \nabla^{b}{G_{\underline{j}}\,^{l}} \varphi^{m}_{\beta} - \frac{3}{4}{\rm i} G_{k l} G_{m n} X_{\underline{i}}\,^{k} \mathbf{X}_{\underline{j}}\,^{l} {G}^{-5} \varphi^{m \alpha} \varphi^{n}_{\alpha} - \frac{15}{16}{\rm i} G_{\underline{i} k} G_{l m} G_{n {i_{1}}} X_{\underline{j}}\,^{l} \boldsymbol{\lambda}^{m \alpha} {G}^{-7} \varphi^{k}_{\alpha} \varphi^{n \beta} \varphi^{{i_{1}}}_{\beta}+\frac{9}{32}F G_{\underline{i} k} G_{l m} X_{\underline{j}}\,^{l} \boldsymbol{\lambda}^{m \alpha} {G}^{-5} \varphi^{k}_{\alpha}+\frac{3}{32}\mathcal{H}_{a} G_{\underline{i} k} G_{l m} (\Gamma^{a})^{\alpha \beta} X_{\underline{j}}\,^{l} \boldsymbol{\lambda}^{m}_{\alpha} {G}^{-5} \varphi^{k}_{\beta}+\frac{3}{8}G_{k l} G_{m n} (\Gamma_{a})^{\alpha \beta} X_{\underline{i}}\,^{k} \boldsymbol{\lambda}^{l}_{\alpha} {G}^{-5} \nabla^{a}{G_{\underline{j}}\,^{m}} \varphi^{n}_{\beta}+\frac{3}{4}{\rm i} G_{\underline{i} \underline{j}} G_{k l} {G}^{-5} \nabla_{a}{W} \nabla^{a}{\boldsymbol{W}} \varphi^{k \alpha} \varphi^{l}_{\alpha} - \frac{15}{16}{\rm i} G_{\underline{i} k} G_{\underline{j} l} G_{m n} (\Gamma_{a})^{\alpha \beta} \boldsymbol{\lambda}^{k}_{\alpha} {G}^{-7} \nabla^{a}{W} \varphi^{l}_{\beta} \varphi^{m \rho} \varphi^{n}_{\rho} - \frac{3}{8}G_{\underline{i} k} G_{l m} \boldsymbol{\lambda}^{k \alpha} {G}^{-5} \nabla_{a}{W} \nabla^{a}{G_{\underline{j}}\,^{l}} \varphi^{m}_{\alpha}+\frac{3}{4}G_{\underline{i} k} G_{l m} (\Sigma_{a b})^{\alpha \beta} \boldsymbol{\lambda}^{k}_{\alpha} {G}^{-5} \nabla^{a}{W} \nabla^{b}{G_{\underline{j}}\,^{l}} \varphi^{m}_{\beta}+\frac{3}{8}G_{\underline{i} k} G_{l m} (\Gamma_{a})^{\alpha \beta} \lambda^{k}_{\alpha} {G}^{-5} \nabla^{a}{\boldsymbol{\lambda}_{\underline{j} \beta}} \varphi^{l \rho} \varphi^{m}_{\rho}+\frac{15}{16}{\rm i} G_{\underline{i} k} G_{\underline{j} l} G_{m n} (\Sigma_{a b})^{\alpha \beta} \mathbf{F}^{a b} \lambda^{k}_{\alpha} {G}^{-7} \varphi^{l}_{\beta} \varphi^{m \rho} \varphi^{n}_{\rho} - \frac{3}{16}\epsilon^{c d e a b} \mathcal{H}_{c} G_{\underline{i} k} G_{\underline{j} l} (\Sigma_{d e})^{\alpha \beta} \mathbf{F}_{a b} \lambda^{k}_{\alpha} {G}^{-5} \varphi^{l}_{\beta}+\frac{3}{8}\mathcal{H}^{b} G_{\underline{i} k} G_{\underline{j} l} (\Gamma^{a})^{\alpha \beta} \mathbf{F}_{a b} \lambda^{k}_{\alpha} {G}^{-5} \varphi^{l}_{\beta} - \frac{3}{16}\epsilon^{c d}\,_{e}\,^{a b} G_{\underline{i} k} G_{l m} (\Sigma_{c d})^{\alpha \beta} \mathbf{F}_{a b} \lambda^{k}_{\alpha} {G}^{-5} \nabla^{e}{G_{\underline{j}}\,^{l}} \varphi^{m}_{\beta}+\frac{3}{8}G_{\underline{i} k} G_{l m} (\Gamma^{a})^{\alpha \beta} \mathbf{F}_{a b} \lambda^{k}_{\alpha} {G}^{-5} \nabla^{b}{G_{\underline{j}}\,^{l}} \varphi^{m}_{\beta} - \frac{15}{16}{\rm i} G_{\underline{i} k} G_{l m} G_{n {i_{1}}} \mathbf{X}_{\underline{j}}\,^{l} \lambda^{m \alpha} {G}^{-7} \varphi^{k}_{\alpha} \varphi^{n \beta} \varphi^{{i_{1}}}_{\beta}+\frac{9}{32}F G_{\underline{i} k} G_{l m} \mathbf{X}_{\underline{j}}\,^{l} \lambda^{m \alpha} {G}^{-5} \varphi^{k}_{\alpha}+\frac{3}{32}\mathcal{H}_{a} G_{\underline{i} k} G_{l m} (\Gamma^{a})^{\alpha \beta} \mathbf{X}_{\underline{j}}\,^{l} \lambda^{m}_{\alpha} {G}^{-5} \varphi^{k}_{\beta}%
+\frac{3}{8}G_{k l} G_{m n} (\Gamma_{a})^{\alpha \beta} \mathbf{X}_{\underline{i}}\,^{k} \lambda^{l}_{\alpha} {G}^{-5} \nabla^{a}{G_{\underline{j}}\,^{m}} \varphi^{n}_{\beta} - \frac{15}{16}{\rm i} G_{\underline{i} k} G_{\underline{j} l} G_{m n} (\Gamma_{a})^{\alpha \beta} \lambda^{k}_{\alpha} {G}^{-7} \nabla^{a}{\boldsymbol{W}} \varphi^{l}_{\beta} \varphi^{m \rho} \varphi^{n}_{\rho} - \frac{3}{8}\mathcal{H}_{a} G_{\underline{i} k} G_{\underline{j} l} \lambda^{k \alpha} {G}^{-5} \nabla^{a}{\boldsymbol{W}} \varphi^{l}_{\alpha} - \frac{3}{4}\mathcal{H}^{a} G_{\underline{i} k} G_{\underline{j} l} (\Sigma_{a b})^{\alpha \beta} \lambda^{k}_{\alpha} {G}^{-5} \nabla^{b}{\boldsymbol{W}} \varphi^{l}_{\beta} - \frac{3}{8}G_{\underline{i} k} G_{l m} \lambda^{k \alpha} {G}^{-5} \nabla_{a}{\boldsymbol{W}} \nabla^{a}{G_{\underline{j}}\,^{l}} \varphi^{m}_{\alpha}+\frac{3}{4}G_{\underline{i} k} G_{l m} (\Sigma_{a b})^{\alpha \beta} \lambda^{k}_{\alpha} {G}^{-5} \nabla^{a}{\boldsymbol{W}} \nabla^{b}{G_{\underline{j}}\,^{l}} \varphi^{m}_{\beta}+\frac{15}{32}{\rm i} G_{k l} G_{m n} \lambda^{k \alpha} \boldsymbol{\lambda}^{l}_{\alpha} {G}^{-7} \varphi_{\underline{i}}^{\beta} \varphi_{\underline{j} \beta} \varphi^{m \rho} \varphi^{n}_{\rho}+\frac{105}{32}{\rm i} G_{\underline{i} k} G_{\underline{j} l} G_{m n} G_{{i_{1}} {i_{2}}} \lambda^{m \alpha} \boldsymbol{\lambda}^{n}_{\alpha} {G}^{-9} \varphi^{k \beta} \varphi^{l}_{\beta} \varphi^{{i_{1}} \rho} \varphi^{{i_{2}}}_{\rho} - \frac{15}{16}F G_{\underline{i} \underline{j}} G_{k l} G_{m n} \lambda^{k \alpha} \boldsymbol{\lambda}^{l}_{\alpha} {G}^{-7} \varphi^{m \beta} \varphi^{n}_{\beta} - \frac{15}{16}\mathcal{H}_{a} G_{\underline{i} k} G_{\underline{j} l} G_{m n} (\Gamma^{a})^{\alpha \beta} \lambda^{m \rho} \boldsymbol{\lambda}^{n}_{\rho} {G}^{-7} \varphi^{k}_{\alpha} \varphi^{l}_{\beta} - \frac{15}{8}G_{\underline{i} k} G_{l m} G_{n {i_{1}}} (\Gamma_{a})^{\alpha \beta} \lambda^{l \rho} \boldsymbol{\lambda}^{m}_{\rho} {G}^{-7} \nabla^{a}{G_{\underline{j}}\,^{n}} \varphi^{k}_{\alpha} \varphi^{{i_{1}}}_{\beta}+\frac{3}{2}G_{\underline{i} k} G_{l m} (\Gamma_{a})^{\alpha \beta} \lambda^{l \rho} \boldsymbol{\lambda}^{m}_{\rho} {G}^{-5} \nabla^{a}{\varphi_{\underline{j} \alpha}} \varphi^{k}_{\beta}+\frac{9}{8}G_{\underline{i} k} G_{l m} (\Sigma_{a b})^{\alpha \beta} W^{a b} \lambda^{l \rho} \boldsymbol{\lambda}^{m}_{\rho} {G}^{-5} \varphi_{\underline{j} \alpha} \varphi^{k}_{\beta}+\frac{99}{128}G_{\underline{i} k} G_{\underline{j} l} G_{m n} \lambda^{m \beta} \boldsymbol{\lambda}^{n}_{\beta} X^{k \alpha} {G}^{-5} \varphi^{l}_{\alpha} - \frac{3}{16}{\rm i} G_{\underline{i} \underline{j}} G_{k l} \lambda^{k \alpha} \boldsymbol{\lambda}^{l}_{\alpha} {F}^{2} {G}^{-5} - \frac{3}{16}{\rm i} \mathcal{H}^{a} \mathcal{H}_{a} G_{\underline{i} \underline{j}} G_{k l} \lambda^{k \alpha} \boldsymbol{\lambda}^{l}_{\alpha} {G}^{-5} - \frac{3}{4}{\rm i} \mathcal{H}_{a} G_{\underline{i} k} G_{l m} \lambda^{l \alpha} \boldsymbol{\lambda}^{m}_{\alpha} {G}^{-5} \nabla^{a}{G_{\underline{j}}\,^{k}} - \frac{93}{128}G_{k l} \lambda^{k \beta} \boldsymbol{\lambda}^{l}_{\beta} X_{\underline{i}}^{\alpha} {G}^{-3} \varphi_{\underline{j} \alpha}+\frac{45}{128}G_{\underline{i} \underline{j}} G_{k l} G_{m n} \lambda^{k \beta} \boldsymbol{\lambda}^{l}_{\beta} X^{m \alpha} {G}^{-5} \varphi^{n}_{\alpha} - \frac{3}{4}{\rm i} G_{k l} G_{m n} \lambda^{k \alpha} \boldsymbol{\lambda}^{l}_{\alpha} {G}^{-5} \nabla_{a}{G_{\underline{i}}\,^{m}} \nabla^{a}{G_{\underline{j}}\,^{n}}%
 - \frac{3}{8}{\rm i} G_{\underline{i} k} (\Sigma_{a b})^{\alpha \beta} F^{a b} \boldsymbol{\lambda}_{l \alpha} {G}^{-5} \varphi_{\underline{j}}^{\rho} \varphi^{k}_{\beta} \varphi^{l}_{\rho} - \frac{1}{16}\epsilon^{c d}\,_{e}\,^{a b} (\Sigma_{c d})^{\alpha \beta} F_{a b} \boldsymbol{\lambda}_{k \alpha} {G}^{-3} \nabla^{e}{G_{\underline{i}}\,^{k}} \varphi_{\underline{j} \beta}+\frac{1}{8}(\Gamma^{a})^{\alpha \beta} F_{a b} \boldsymbol{\lambda}_{k \alpha} {G}^{-3} \nabla^{b}{G_{\underline{i}}\,^{k}} \varphi_{\underline{j} \beta}+\frac{1}{8}{\rm i} X_{\underline{i} k} \mathbf{X}_{\underline{j} l} {G}^{-3} \varphi^{k \alpha} \varphi^{l}_{\alpha}+\frac{3}{8}{\rm i} G_{\underline{i} k} X_{\underline{j} l} \boldsymbol{\lambda}_{m}^{\alpha} {G}^{-5} \varphi^{k}_{\alpha} \varphi^{l \beta} \varphi^{m}_{\beta} - \frac{7}{32}F X_{\underline{i} \underline{j}} \boldsymbol{\lambda}_{k}^{\alpha} {G}^{-3} \varphi^{k}_{\alpha} - \frac{1}{32}\mathcal{H}_{a} (\Gamma^{a})^{\alpha \beta} X_{\underline{i} \underline{j}} \boldsymbol{\lambda}_{k \alpha} {G}^{-3} \varphi^{k}_{\beta} - \frac{1}{16}(\Gamma_{a})^{\alpha \beta} X_{\underline{i} k} \boldsymbol{\lambda}_{l \alpha} {G}^{-3} \nabla^{a}{G_{\underline{j}}\,^{k}} \varphi^{l}_{\beta} - \frac{1}{4}(\Gamma_{a})^{\alpha \beta} X_{\underline{i} l} \boldsymbol{\lambda}_{k \alpha} {G}^{-3} \nabla^{a}{G_{\underline{j}}\,^{k}} \varphi^{l}_{\beta}+\frac{3}{8}{\rm i} G_{\underline{i} k} (\Gamma_{a})^{\alpha \beta} \boldsymbol{\lambda}_{l \alpha} {G}^{-5} \nabla^{a}{W} \varphi_{\underline{j}}^{\rho} \varphi^{k}_{\beta} \varphi^{l}_{\rho} - \frac{1}{8}\boldsymbol{\lambda}_{k}^{\alpha} {G}^{-3} \nabla_{a}{W} \nabla^{a}{G_{\underline{i}}\,^{k}} \varphi_{\underline{j} \alpha}+\frac{1}{4}(\Sigma_{a b})^{\alpha \beta} \boldsymbol{\lambda}_{k \alpha} {G}^{-3} \nabla^{a}{W} \nabla^{b}{G_{\underline{i}}\,^{k}} \varphi_{\underline{j} \beta}+\frac{3}{8}(\Gamma_{a})^{\alpha \beta} \lambda_{k \alpha} {G}^{-3} \nabla^{a}{\boldsymbol{\lambda}_{\underline{i} \beta}} \varphi_{\underline{j}}^{\rho} \varphi^{k}_{\rho} - \frac{3}{8}{\rm i} G_{\underline{i} k} (\Sigma_{a b})^{\alpha \beta} \mathbf{F}^{a b} \lambda_{l \alpha} {G}^{-5} \varphi_{\underline{j}}^{\rho} \varphi^{k}_{\beta} \varphi^{l}_{\rho}+\frac{1}{4}F (\Sigma_{a b})^{\alpha \beta} \mathbf{F}^{a b} \lambda_{\underline{i} \alpha} {G}^{-3} \varphi_{\underline{j} \beta} - \frac{1}{16}\epsilon^{c d e a b} \mathcal{H}_{c} (\Sigma_{d e})^{\alpha \beta} \mathbf{F}_{a b} \lambda_{\underline{i} \alpha} {G}^{-3} \varphi_{\underline{j} \beta}+\frac{1}{8}\mathcal{H}^{b} (\Gamma^{a})^{\alpha \beta} \mathbf{F}_{a b} \lambda_{\underline{i} \alpha} {G}^{-3} \varphi_{\underline{j} \beta} - \frac{1}{16}\epsilon^{c d}\,_{e}\,^{a b} (\Sigma_{c d})^{\alpha \beta} \mathbf{F}_{a b} \lambda_{k \alpha} {G}^{-3} \nabla^{e}{G_{\underline{i}}\,^{k}} \varphi_{\underline{j} \beta}+\frac{1}{8}(\Gamma^{a})^{\alpha \beta} \mathbf{F}_{a b} \lambda_{k \alpha} {G}^{-3} \nabla^{b}{G_{\underline{i}}\,^{k}} \varphi_{\underline{j} \beta}+\frac{3}{8}{\rm i} G_{\underline{i} k} \mathbf{X}_{\underline{j} l} \lambda^{\alpha}_{m} {G}^{-5} \varphi^{k}_{\alpha} \varphi^{l \beta} \varphi^{m}_{\beta}%
 - \frac{7}{32}F \mathbf{X}_{\underline{i} \underline{j}} \lambda^{\alpha}_{k} {G}^{-3} \varphi^{k}_{\alpha} - \frac{1}{32}\mathcal{H}_{a} (\Gamma^{a})^{\alpha \beta} \mathbf{X}_{\underline{i} \underline{j}} \lambda_{k \alpha} {G}^{-3} \varphi^{k}_{\beta} - \frac{1}{16}(\Gamma_{a})^{\alpha \beta} \mathbf{X}_{\underline{i} k} \lambda_{l \alpha} {G}^{-3} \nabla^{a}{G_{\underline{j}}\,^{k}} \varphi^{l}_{\beta} - \frac{1}{4}(\Gamma_{a})^{\alpha \beta} \mathbf{X}_{\underline{i} l} \lambda_{k \alpha} {G}^{-3} \nabla^{a}{G_{\underline{j}}\,^{k}} \varphi^{l}_{\beta}+\frac{3}{8}{\rm i} G_{\underline{i} k} (\Gamma_{a})^{\alpha \beta} \lambda_{l \alpha} {G}^{-5} \nabla^{a}{\boldsymbol{W}} \varphi_{\underline{j}}^{\rho} \varphi^{k}_{\beta} \varphi^{l}_{\rho} - \frac{1}{4}F (\Gamma_{a})^{\alpha \beta} \lambda_{\underline{i} \alpha} {G}^{-3} \nabla^{a}{\boldsymbol{W}} \varphi_{\underline{j} \beta} - \frac{1}{8}\mathcal{H}_{a} \lambda^{\alpha}_{\underline{i}} {G}^{-3} \nabla^{a}{\boldsymbol{W}} \varphi_{\underline{j} \alpha} - \frac{1}{4}\mathcal{H}^{a} (\Sigma_{a b})^{\alpha \beta} \lambda_{\underline{i} \alpha} {G}^{-3} \nabla^{b}{\boldsymbol{W}} \varphi_{\underline{j} \beta} - \frac{1}{8}\lambda^{\alpha}_{k} {G}^{-3} \nabla_{a}{\boldsymbol{W}} \nabla^{a}{G_{\underline{i}}\,^{k}} \varphi_{\underline{j} \alpha}+\frac{1}{4}(\Sigma_{a b})^{\alpha \beta} \lambda_{k \alpha} {G}^{-3} \nabla^{a}{\boldsymbol{W}} \nabla^{b}{G_{\underline{i}}\,^{k}} \varphi_{\underline{j} \beta} - \frac{3}{16}{\rm i} \lambda^{\alpha}_{k} \boldsymbol{\lambda}_{l \alpha} {G}^{-5} \varphi_{\underline{i}}^{\beta} \varphi_{\underline{j} \beta} \varphi^{k \rho} \varphi^{l}_{\rho} - \frac{15}{16}{\rm i} G_{\underline{i} k} G_{\underline{j} l} \lambda^{\alpha}_{m} \boldsymbol{\lambda}_{n \alpha} {G}^{-7} \varphi^{k \beta} \varphi^{l}_{\beta} \varphi^{m \rho} \varphi^{n}_{\rho}+\frac{3}{8}F G_{\underline{i} \underline{j}} \lambda^{\alpha}_{k} \boldsymbol{\lambda}_{l \alpha} {G}^{-5} \varphi^{k \beta} \varphi^{l}_{\beta}+\frac{3}{8}G_{\underline{i} k} (\Gamma_{a})^{\alpha \beta} \lambda^{\rho}_{l} \boldsymbol{\lambda}_{m \rho} {G}^{-5} \nabla^{a}{G_{\underline{j}}\,^{l}} \varphi^{k}_{\alpha} \varphi^{m}_{\beta}+\frac{3}{8}G_{\underline{i} k} (\Gamma_{a})^{\alpha \beta} \lambda^{\rho}_{m} \boldsymbol{\lambda}_{l \rho} {G}^{-5} \nabla^{a}{G_{\underline{j}}\,^{l}} \varphi^{k}_{\alpha} \varphi^{m}_{\beta} - \frac{1}{2}(\Gamma_{a})^{\alpha \beta} \lambda^{\rho}_{\underline{i}} \boldsymbol{\lambda}_{k \rho} {G}^{-3} \nabla^{a}{\varphi_{\underline{j} \alpha}} \varphi^{k}_{\beta} - \frac{33}{128}G_{\underline{i} k} \lambda^{\beta}_{\underline{j}} \boldsymbol{\lambda}_{l \beta} X^{k \alpha} {G}^{-3} \varphi^{l}_{\alpha}+\frac{1}{8}{\rm i} \lambda^{\alpha}_{\underline{i}} \boldsymbol{\lambda}_{\underline{j} \alpha} {F}^{2} {G}^{-3}+\frac{1}{8}{\rm i} \mathcal{H}^{a} \mathcal{H}_{a} \lambda^{\alpha}_{\underline{i}} \boldsymbol{\lambda}_{\underline{j} \alpha} {G}^{-3}+\frac{1}{4}{\rm i} \mathcal{H}_{a} \lambda^{\alpha}_{\underline{i}} \boldsymbol{\lambda}_{k \alpha} {G}^{-3} \nabla^{a}{G_{\underline{j}}\,^{k}}%
 - \frac{15}{128}G_{\underline{i} k} \lambda^{k \beta} \boldsymbol{\lambda}_{l \beta} X_{\underline{j}}^{\alpha} {G}^{-3} \varphi^{l}_{\alpha} - \frac{15}{128}G_{\underline{i} \underline{j}} \lambda^{\beta}_{k} \boldsymbol{\lambda}_{l \beta} X^{k \alpha} {G}^{-3} \varphi^{l}_{\alpha}+\frac{1}{4}{\rm i} \mathcal{H}_{a} \lambda^{\alpha}_{k} \boldsymbol{\lambda}_{\underline{i} \alpha} {G}^{-3} \nabla^{a}{G_{\underline{j}}\,^{k}}+\frac{1}{2}{\rm i} \lambda^{\alpha}_{k} \boldsymbol{\lambda}_{l \alpha} {G}^{-3} \nabla_{a}{G_{\underline{i}}\,^{k}} \nabla^{a}{G_{\underline{j}}\,^{l}} - \frac{1}{2}(\Gamma_{a})^{\alpha \beta} \lambda^{\rho}_{k} \boldsymbol{\lambda}_{\underline{i} \rho} {G}^{-3} \nabla^{a}{\varphi_{\underline{j} \alpha}} \varphi^{k}_{\beta} - \frac{3}{8}(\Sigma_{a b})^{\alpha \beta} W^{a b} \lambda^{\rho}_{k} \boldsymbol{\lambda}_{\underline{i} \rho} {G}^{-3} \varphi_{\underline{j} \alpha} \varphi^{k}_{\beta} - \frac{33}{128}G_{\underline{i} k} \lambda^{\beta}_{l} \boldsymbol{\lambda}_{\underline{j} \beta} X^{k \alpha} {G}^{-3} \varphi^{l}_{\alpha} - \frac{15}{128}G_{\underline{i} k} \lambda^{\beta}_{l} \boldsymbol{\lambda}^{k}_{\beta} X_{\underline{j}}^{\alpha} {G}^{-3} \varphi^{l}_{\alpha} - \frac{15}{128}G_{\underline{i} \underline{j}} \lambda^{\beta}_{l} \boldsymbol{\lambda}_{k \beta} X^{k \alpha} {G}^{-3} \varphi^{l}_{\alpha} - \frac{1}{8}G_{k l} (\Gamma_{a})^{\alpha \beta} X_{\underline{i}}\,^{k} \boldsymbol{\lambda}^{l}_{\alpha} {G}^{-3} \nabla^{a}{\varphi_{\underline{j} \beta}}+\frac{1}{4}\epsilon^{c d}\,_{e}\,^{a b} G_{\underline{i} k} (\Sigma_{c d})^{\alpha \beta} \mathbf{F}_{a b} \lambda^{k}_{\alpha} {G}^{-3} \nabla^{e}{\varphi_{\underline{j} \beta}} - \frac{1}{2}G_{\underline{i} k} (\Gamma^{a})^{\alpha \beta} \mathbf{F}_{a b} \lambda^{k}_{\alpha} {G}^{-3} \nabla^{b}{\varphi_{\underline{j} \beta}} - \frac{1}{8}G_{k l} (\Gamma_{a})^{\alpha \beta} \mathbf{X}_{\underline{i}}\,^{k} \lambda^{l}_{\alpha} {G}^{-3} \nabla^{a}{\varphi_{\underline{j} \beta}}+\frac{1}{2}G_{\underline{i} k} \lambda^{k \alpha} {G}^{-3} \nabla_{a}{\boldsymbol{W}} \nabla^{a}{\varphi_{\underline{j} \alpha}}-G_{\underline{i} k} (\Sigma_{a b})^{\alpha \beta} \lambda^{k}_{\alpha} {G}^{-3} \nabla^{a}{\boldsymbol{W}} \nabla^{b}{\varphi_{\underline{j} \beta}}+\frac{1}{2}{\rm i} G_{k l} \lambda^{k \alpha} \boldsymbol{\lambda}^{l}_{\alpha} {G}^{-3} \nabla_{a}{\nabla^{a}{G_{\underline{i} \underline{j}}}}+\frac{5}{128}{\rm i} G_{\underline{i} \underline{j}} G_{k l} W^{a b} W_{a b} \lambda^{k \alpha} \boldsymbol{\lambda}^{l}_{\alpha} {G}^{-3}+\frac{11}{64}G_{k l} (\Sigma_{a b})^{\alpha \beta} X_{\underline{i}}\,^{k} W^{a b} \boldsymbol{\lambda}^{l}_{\alpha} {G}^{-3} \varphi_{\underline{j} \beta}+\frac{57}{128}G_{\underline{i} k} W^{a b} \mathbf{F}_{a b} \lambda^{k \alpha} {G}^{-3} \varphi_{\underline{j} \alpha} - \frac{25}{256}\epsilon_{e}\,^{c d a b} G_{\underline{i} k} (\Gamma^{e})^{\alpha \beta} W_{c d} \mathbf{F}_{a b} \lambda^{k}_{\alpha} {G}^{-3} \varphi_{\underline{j} \beta}%
 - \frac{23}{32}G_{\underline{i} k} (\Sigma^{a}{}_{\, c})^{\alpha \beta} W^{c b} \mathbf{F}_{a b} \lambda^{k}_{\alpha} {G}^{-3} \varphi_{\underline{j} \beta}+\frac{11}{64}G_{k l} (\Sigma_{a b})^{\alpha \beta} \mathbf{X}_{\underline{i}}\,^{k} W^{a b} \lambda^{l}_{\alpha} {G}^{-3} \varphi_{\underline{j} \beta}+\frac{23}{128}\epsilon^{c d}\,_{e}\,^{a b} G_{\underline{i} k} (\Sigma_{c d})^{\alpha \beta} W_{a b} \lambda^{k}_{\alpha} {G}^{-3} \nabla^{e}{\boldsymbol{W}} \varphi_{\underline{j} \beta}+\frac{25}{64}G_{\underline{i} k} (\Gamma^{a})^{\alpha \beta} W_{a b} \lambda^{k}_{\alpha} {G}^{-3} \nabla^{b}{\boldsymbol{W}} \varphi_{\underline{j} \beta} - \frac{39}{256}G_{\underline{i} k} G_{l m} X_{\underline{j}}\,^{l} \boldsymbol{\lambda}^{m \alpha} X^{k}_{\alpha} {G}^{-3}+\frac{33}{128}G_{\underline{i} k} G_{\underline{j} l} (\Sigma_{a b})^{\beta \alpha} \mathbf{F}^{a b} \lambda^{k}_{\beta} X^{l}_{\alpha} {G}^{-3} - \frac{39}{256}G_{\underline{i} k} G_{l m} \mathbf{X}_{\underline{j}}\,^{l} \lambda^{m \alpha} X^{k}_{\alpha} {G}^{-3} - \frac{33}{128}G_{\underline{i} k} G_{\underline{j} l} (\Gamma_{a})^{\beta \alpha} \lambda^{k}_{\beta} X^{l}_{\alpha} {G}^{-3} \nabla^{a}{\boldsymbol{W}} - \frac{3}{64}{\rm i} G_{\underline{i} \underline{j}} G_{k l} Y \lambda^{k \alpha} \boldsymbol{\lambda}^{l}_{\alpha} {G}^{-3}+\frac{5}{4}F G_{k l} X_{\underline{i}}\,^{k} \mathbf{X}_{\underline{j}}\,^{l} {G}^{-3}+\frac{3}{32}{\rm i} F G_{\underline{i} k} (\Gamma_{a})^{\alpha \beta} \lambda^{k}_{\alpha} {G}^{-3} \nabla^{a}{\boldsymbol{\lambda}_{\underline{j} \beta}}+\frac{3}{8}{\rm i} G_{\underline{i} k} (\Sigma_{a b})^{\alpha \beta} \mathbf{F}^{a b} \lambda_{\underline{j} \alpha} {G}^{-5} \varphi^{k \rho} \varphi_{l \beta} \varphi^{l}_{\rho}+\frac{1}{16}\epsilon^{c d}\,_{e}\,^{a b} (\Sigma_{c d})^{\alpha \beta} \mathbf{F}_{a b} \lambda_{\underline{i} \alpha} {G}^{-3} \nabla^{e}{G_{\underline{j} k}} \varphi^{k}_{\beta} - \frac{1}{8}(\Gamma^{a})^{\alpha \beta} \mathbf{F}_{a b} \lambda_{\underline{i} \alpha} {G}^{-3} \nabla^{b}{G_{\underline{j} k}} \varphi^{k}_{\beta}+\frac{1}{8}(\Gamma_{a})^{\alpha \beta} \lambda_{k \alpha} {G}^{-3} \nabla^{a}{\boldsymbol{\lambda}_{\underline{i}}^{\rho}} \varphi_{\underline{j} \rho} \varphi^{k}_{\beta} - \frac{1}{8}(\Gamma_{a})^{\alpha \beta} \lambda^{\rho}_{k} {G}^{-3} \nabla^{a}{\boldsymbol{\lambda}_{\underline{i} \rho}} \varphi_{\underline{j} \alpha} \varphi^{k}_{\beta} - \frac{1}{32}(\Sigma_{a b})^{\alpha \beta} W^{a b} \lambda^{\rho}_{k} \boldsymbol{\lambda}_{\underline{i} \alpha} {G}^{-3} \varphi_{\underline{j} \beta} \varphi^{k}_{\rho} - \frac{1}{8}{\rm i} (\Sigma_{a b})^{\alpha \beta} X_{\underline{i} k} \mathbf{F}^{a b} {G}^{-3} \varphi_{\underline{j} \alpha} \varphi^{k}_{\beta} - \frac{3}{8}{\rm i} G_{\underline{i} k} (\Sigma_{a b})^{\alpha \beta} \mathbf{F}^{a b} \lambda_{l \alpha} {G}^{-5} \varphi_{\underline{j} \beta} \varphi^{k \rho} \varphi^{l}_{\rho}+\frac{3}{8}G_{\underline{i} k} G_{l m} (\Gamma_{a})^{\alpha \beta} \lambda^{l}_{\alpha} {G}^{-5} \nabla^{a}{\boldsymbol{\lambda}_{\underline{j}}^{\rho}} \varphi^{k}_{\beta} \varphi^{m}_{\rho}%
 - \frac{1}{4}G_{k l} (\Gamma_{a})^{\alpha \beta} \lambda^{k}_{\alpha} {G}^{-3} \nabla^{a}{\mathbf{X}_{\underline{i} \underline{j}}} \varphi^{l}_{\beta}+\frac{11}{32}{\rm i} \mathcal{H}_{a} G_{\underline{i} k} \lambda^{k \alpha} {G}^{-3} \nabla^{a}{\boldsymbol{\lambda}_{\underline{j} \alpha}}+\frac{5}{16}{\rm i} \mathcal{H}^{a} G_{\underline{i} k} (\Sigma_{a b})^{\alpha \beta} \lambda^{k}_{\alpha} {G}^{-3} \nabla^{b}{\boldsymbol{\lambda}_{\underline{j} \beta}}+\frac{5}{8}{\rm i} G_{k l} \lambda^{k \alpha} {G}^{-3} \nabla_{a}{\boldsymbol{\lambda}_{\underline{i} \alpha}} \nabla^{a}{G_{\underline{j}}\,^{l}} - \frac{3}{4}{\rm i} G_{k l} (\Sigma_{a b})^{\alpha \beta} \lambda^{k}_{\alpha} {G}^{-3} \nabla^{a}{\boldsymbol{\lambda}_{\underline{i} \beta}} \nabla^{b}{G_{\underline{j}}\,^{l}}+\frac{3}{8}G_{\underline{i} k} G_{l m} (\Gamma_{a})^{\alpha \beta} \lambda^{l \rho} {G}^{-5} \nabla^{a}{\boldsymbol{\lambda}_{\underline{j} \rho}} \varphi^{k}_{\alpha} \varphi^{m}_{\beta} - \frac{3}{8}G_{\underline{i} k} G_{l m} (\Gamma_{a})^{\alpha \beta} \lambda^{l \rho} {G}^{-5} \nabla^{a}{\boldsymbol{\lambda}_{\underline{j} \alpha}} \varphi^{k}_{\rho} \varphi^{m}_{\beta} - \frac{3}{8}G_{\underline{i} k} G_{l m} (\Gamma_{a})^{\alpha \beta} \lambda^{l}_{\alpha} {G}^{-5} \nabla^{a}{\boldsymbol{\lambda}_{\underline{j} \beta}} \varphi^{k \rho} \varphi^{m}_{\rho} - \frac{9}{16}G_{\underline{i} k} G_{l m} (\Sigma_{a b})^{\alpha \beta} W^{a b} \lambda^{l}_{\alpha} \boldsymbol{\lambda}_{\underline{j} \beta} {G}^{-5} \varphi^{k \rho} \varphi^{m}_{\rho}+\frac{3}{16}{\rm i} G_{k l} (\Sigma_{a b})^{\alpha \beta} \mathbf{F}^{a b} \lambda^{k}_{\alpha} {G}^{-5} \varphi_{\underline{i}}^{\rho} \varphi_{\underline{j} \rho} \varphi^{l}_{\beta}+\frac{15}{16}{\rm i} G_{\underline{i} k} G_{\underline{j} l} G_{m n} (\Sigma_{a b})^{\alpha \beta} \mathbf{F}^{a b} \lambda^{m}_{\alpha} {G}^{-7} \varphi^{k \rho} \varphi^{l}_{\rho} \varphi^{n}_{\beta} - \frac{3}{8}F G_{\underline{i} \underline{j}} G_{k l} (\Sigma_{a b})^{\alpha \beta} \mathbf{F}^{a b} \lambda^{k}_{\alpha} {G}^{-5} \varphi^{l}_{\beta} - \frac{3}{16}\epsilon^{c d}\,_{e}\,^{a b} G_{\underline{i} k} G_{l m} (\Sigma_{c d})^{\alpha \beta} \mathbf{F}_{a b} \lambda^{l}_{\alpha} {G}^{-5} \nabla^{e}{G_{\underline{j}}\,^{m}} \varphi^{k}_{\beta}+\frac{3}{8}G_{\underline{i} k} G_{l m} (\Gamma^{a})^{\alpha \beta} \mathbf{F}_{a b} \lambda^{l}_{\alpha} {G}^{-5} \nabla^{b}{G_{\underline{j}}\,^{m}} \varphi^{k}_{\beta}+\frac{15}{128}G_{\underline{i} \underline{j}} G_{k l} (\Sigma_{a b})^{\beta \alpha} \mathbf{F}^{a b} \lambda^{k}_{\beta} X^{l}_{\alpha} {G}^{-3} - \frac{3}{8}{\rm i} G_{\underline{i} k} (\Gamma_{a})^{\alpha \beta} \lambda_{\underline{j} \alpha} {G}^{-5} \nabla^{a}{\boldsymbol{W}} \varphi^{k \rho} \varphi_{l \beta} \varphi^{l}_{\rho}+\frac{1}{8}\lambda^{\alpha}_{\underline{i}} {G}^{-3} \nabla_{a}{\boldsymbol{W}} \nabla^{a}{G_{\underline{j} k}} \varphi^{k}_{\alpha} - \frac{1}{4}(\Sigma_{a b})^{\alpha \beta} \lambda_{\underline{i} \alpha} {G}^{-3} \nabla^{a}{\boldsymbol{W}} \nabla^{b}{G_{\underline{j} k}} \varphi^{k}_{\beta}+\frac{3}{8}{\rm i} G_{\underline{i} k} (\Gamma_{a})^{\alpha \beta} \lambda_{l \alpha} {G}^{-5} \nabla^{a}{\boldsymbol{W}} \varphi_{\underline{j} \beta} \varphi^{k \rho} \varphi^{l}_{\rho} - \frac{1}{8}(\Gamma_{a})^{\alpha \beta} \lambda_{k \alpha} {G}^{-3} \nabla^{a}{\boldsymbol{\lambda}_{\underline{i}}^{\rho}} \varphi_{\underline{j} \beta} \varphi^{k}_{\rho}%
+\frac{1}{64}(\Sigma_{a b})^{\alpha \beta} W^{a b} \lambda^{\rho}_{k} \boldsymbol{\lambda}_{\underline{i} \alpha} {G}^{-3} \varphi_{\underline{j} \rho} \varphi^{k}_{\beta} - \frac{3}{16}{\rm i} G_{k l} (\Gamma_{a})^{\alpha \beta} \lambda^{k}_{\alpha} {G}^{-5} \nabla^{a}{\boldsymbol{W}} \varphi_{\underline{i}}^{\rho} \varphi_{\underline{j} \rho} \varphi^{l}_{\beta} - \frac{15}{16}{\rm i} G_{\underline{i} k} G_{\underline{j} l} G_{m n} (\Gamma_{a})^{\alpha \beta} \lambda^{m}_{\alpha} {G}^{-7} \nabla^{a}{\boldsymbol{W}} \varphi^{k \rho} \varphi^{l}_{\rho} \varphi^{n}_{\beta} - \frac{3}{8}G_{\underline{i} k} G_{l m} (\Gamma_{a})^{\alpha \beta} \lambda^{l}_{\alpha} {G}^{-5} \nabla^{a}{\boldsymbol{\lambda}_{\underline{j}}^{\rho}} \varphi^{k}_{\rho} \varphi^{m}_{\beta}+\frac{3}{32}G_{\underline{i} k} G_{l m} (\Sigma_{a b})^{\alpha \beta} W^{a b} \lambda^{l \rho} \boldsymbol{\lambda}_{\underline{j} \alpha} {G}^{-5} \varphi^{k}_{\beta} \varphi^{m}_{\rho}+\frac{3}{8}F G_{\underline{i} \underline{j}} G_{k l} (\Gamma_{a})^{\alpha \beta} \lambda^{k}_{\alpha} {G}^{-5} \nabla^{a}{\boldsymbol{W}} \varphi^{l}_{\beta} - \frac{3}{8}G_{\underline{i} k} G_{l m} \lambda^{l \alpha} {G}^{-5} \nabla_{a}{\boldsymbol{W}} \nabla^{a}{G_{\underline{j}}\,^{m}} \varphi^{k}_{\alpha}+\frac{3}{4}G_{\underline{i} k} G_{l m} (\Sigma_{a b})^{\alpha \beta} \lambda^{l}_{\alpha} {G}^{-5} \nabla^{a}{\boldsymbol{W}} \nabla^{b}{G_{\underline{j}}\,^{m}} \varphi^{k}_{\beta} - \frac{15}{128}G_{\underline{i} \underline{j}} G_{k l} (\Gamma_{a})^{\beta \alpha} \lambda^{k}_{\beta} X^{l}_{\alpha} {G}^{-3} \nabla^{a}{\boldsymbol{W}}+\frac{3}{8}G_{k l} (\Gamma_{a})^{\alpha \beta} \mathbf{X}^{k l} {G}^{-3} \nabla^{a}{\lambda_{\underline{i} \alpha}} \varphi_{\underline{j} \beta} - \frac{5}{8}G_{\underline{i} k} (\Gamma_{a})^{\alpha \beta} X_{\underline{j} l} {G}^{-3} \nabla^{a}{\boldsymbol{\lambda}^{k}_{\alpha}} \varphi^{l}_{\beta}+\frac{5}{32}G_{\underline{i} k} (\Sigma_{a b})^{\alpha \beta} X_{\underline{j} l} W^{a b} \boldsymbol{\lambda}^{k}_{\alpha} {G}^{-3} \varphi^{l}_{\beta}+\frac{15}{16}{\rm i} G_{\underline{i} k} G_{l m} X_{\underline{j} n} \mathbf{X}^{l m} {G}^{-5} \varphi^{k \alpha} \varphi^{n}_{\alpha}+\frac{3}{8}G_{k l} (\Gamma_{a})^{\alpha \beta} X^{k l} {G}^{-3} \nabla^{a}{\boldsymbol{\lambda}_{\underline{i} \alpha}} \varphi_{\underline{j} \beta} - \frac{5}{8}G_{\underline{i} k} (\Gamma_{a})^{\alpha \beta} \mathbf{X}_{\underline{j} l} {G}^{-3} \nabla^{a}{\lambda^{k}_{\alpha}} \varphi^{l}_{\beta}+\frac{5}{32}G_{\underline{i} k} (\Sigma_{a b})^{\alpha \beta} \mathbf{X}_{\underline{j} l} W^{a b} \lambda^{k}_{\alpha} {G}^{-3} \varphi^{l}_{\beta}+\frac{15}{16}{\rm i} G_{\underline{i} k} G_{l m} X^{l m} \mathbf{X}_{\underline{j} n} {G}^{-5} \varphi^{k \alpha} \varphi^{n}_{\alpha} - \frac{1}{4}{\rm i} G_{\underline{i} k} G_{\underline{j} l} {G}^{-3} \nabla_{a}{\lambda^{k \alpha}} \nabla^{a}{\boldsymbol{\lambda}^{l}_{\alpha}}+\frac{1}{2}{\rm i} G_{\underline{i} k} G_{\underline{j} l} (\Sigma_{a b})^{\alpha \beta} {G}^{-3} \nabla^{a}{\lambda^{k}_{\alpha}} \nabla^{b}{\boldsymbol{\lambda}^{l}_{\beta}} - \frac{1}{32}{\rm i} \epsilon^{c d}\,_{e}\,^{a b} G_{\underline{i} k} G_{\underline{j} l} (\Sigma_{c d})^{\alpha \beta} W_{a b} \boldsymbol{\lambda}^{k}_{\alpha} {G}^{-3} \nabla^{e}{\lambda^{l}_{\beta}}%
+\frac{1}{16}{\rm i} G_{\underline{i} k} G_{\underline{j} l} (\Gamma^{a})^{\alpha \beta} W_{a b} \boldsymbol{\lambda}^{k}_{\alpha} {G}^{-3} \nabla^{b}{\lambda^{l}_{\beta}}+\frac{3}{4}G_{\underline{i} k} G_{\underline{j} l} G_{m n} (\Gamma_{a})^{\alpha \beta} \mathbf{X}^{m n} {G}^{-5} \nabla^{a}{\lambda^{k}_{\alpha}} \varphi^{l}_{\beta} - \frac{1}{2}G_{\underline{i} \underline{j}} G_{k l} \mathbf{X}^{k l} {G}^{-3} \nabla_{a}{\nabla^{a}{W}}+\frac{1}{2}G_{\underline{i} \underline{j}} G_{k l} \mathbf{X}^{k l} W^{a b} F_{a b} {G}^{-3}+\frac{5}{256}G_{\underline{i} k} G_{l m} \mathbf{X}^{l m} \lambda^{k \alpha} X_{\underline{j} \alpha} {G}^{-3} - \frac{75}{256}G_{\underline{i} \underline{j}} G_{k l} \mathbf{X}^{k l} \lambda^{\alpha}_{m} X^{m}_{\alpha} {G}^{-3} - \frac{1}{32}{\rm i} \epsilon^{c d}\,_{e}\,^{a b} G_{\underline{i} k} G_{\underline{j} l} (\Sigma_{c d})^{\alpha \beta} W_{a b} \lambda^{k}_{\alpha} {G}^{-3} \nabla^{e}{\boldsymbol{\lambda}^{l}_{\beta}}+\frac{1}{16}{\rm i} G_{\underline{i} k} G_{\underline{j} l} (\Gamma^{a})^{\alpha \beta} W_{a b} \lambda^{k}_{\alpha} {G}^{-3} \nabla^{b}{\boldsymbol{\lambda}^{l}_{\beta}}+\frac{1}{128}{\rm i} G_{\underline{i} k} G_{\underline{j} l} W^{a b} W_{a b} \lambda^{k \alpha} \boldsymbol{\lambda}^{l}_{\alpha} {G}^{-3} - \frac{1}{256}{\rm i} \epsilon_{e}\,^{a b c d} G_{\underline{i} k} G_{\underline{j} l} (\Gamma^{e})^{\alpha \beta} W_{a b} W_{c d} \lambda^{k}_{\alpha} \boldsymbol{\lambda}^{l}_{\beta} {G}^{-3} - \frac{3}{16}G_{\underline{i} k} G_{\underline{j} l} G_{m n} (\Sigma_{a b})^{\alpha \beta} \mathbf{X}^{m n} W^{a b} \lambda^{k}_{\alpha} {G}^{-5} \varphi^{l}_{\beta}+\frac{3}{4}G_{\underline{i} k} G_{\underline{j} l} G_{m n} (\Gamma_{a})^{\alpha \beta} X^{m n} {G}^{-5} \nabla^{a}{\boldsymbol{\lambda}^{k}_{\alpha}} \varphi^{l}_{\beta} - \frac{1}{2}G_{\underline{i} \underline{j}} G_{k l} X^{k l} {G}^{-3} \nabla_{a}{\nabla^{a}{\boldsymbol{W}}}+\frac{1}{2}G_{\underline{i} \underline{j}} G_{k l} X^{k l} W^{a b} \mathbf{F}_{a b} {G}^{-3}+\frac{5}{256}G_{\underline{i} k} G_{l m} X^{l m} \boldsymbol{\lambda}^{k \alpha} X_{\underline{j} \alpha} {G}^{-3} - \frac{75}{256}G_{\underline{i} \underline{j}} G_{k l} X^{k l} \boldsymbol{\lambda}_{m}^{\alpha} X^{m}_{\alpha} {G}^{-3} - \frac{3}{16}G_{\underline{i} k} G_{\underline{j} l} G_{m n} (\Sigma_{a b})^{\alpha \beta} X^{m n} W^{a b} \boldsymbol{\lambda}^{k}_{\alpha} {G}^{-5} \varphi^{l}_{\beta} - \frac{3}{16}{\rm i} G_{k l} G_{m n} X^{k l} \mathbf{X}^{m n} {G}^{-5} \varphi_{\underline{i}}^{\alpha} \varphi_{\underline{j} \alpha} - \frac{15}{16}{\rm i} G_{\underline{i} k} G_{\underline{j} l} G_{m n} G_{{i_{1}} {i_{2}}} X^{m n} \mathbf{X}^{{i_{1}} {i_{2}}} {G}^{-7} \varphi^{k \alpha} \varphi^{l}_{\alpha}+\frac{3}{8}F G_{\underline{i} \underline{j}} G_{k l} G_{m n} X^{k l} \mathbf{X}^{m n} {G}^{-5}%
+\frac{3}{8}{\rm i} G_{\underline{i} k} \mathbf{X}_{\underline{j} l} \lambda^{\alpha}_{m} {G}^{-5} \varphi^{k \beta} \varphi^{l}_{\beta} \varphi^{m}_{\alpha}+\frac{1}{32}\epsilon^{c d}\,_{e}\,^{a b} G_{\underline{i} k} (\Sigma_{c d})^{\alpha \beta} F_{a b} {G}^{-3} \nabla^{e}{\boldsymbol{\lambda}^{k}_{\alpha}} \varphi_{\underline{j} \beta}+\frac{1}{16}G_{\underline{i} k} (\Gamma^{a})^{\alpha \beta} F_{a b} {G}^{-3} \nabla^{b}{\boldsymbol{\lambda}^{k}_{\alpha}} \varphi_{\underline{j} \beta} - \frac{1}{16}G_{\underline{i} k} {G}^{-3} \nabla_{a}{W} \nabla^{a}{\boldsymbol{\lambda}^{k \alpha}} \varphi_{\underline{j} \alpha} - \frac{1}{8}G_{\underline{i} k} (\Sigma_{a b})^{\alpha \beta} {G}^{-3} \nabla^{a}{W} \nabla^{b}{\boldsymbol{\lambda}^{k}_{\alpha}} \varphi_{\underline{j} \beta}+\frac{3}{8}G_{\underline{i} k} G_{\underline{j} l} (\Gamma_{a})^{\alpha \beta} \lambda^{\rho}_{m} {G}^{-5} \nabla^{a}{\boldsymbol{\lambda}^{k}_{\alpha}} \varphi^{l}_{\beta} \varphi^{m}_{\rho} - \frac{1}{4}G_{\underline{i} \underline{j}} \lambda^{\alpha}_{k} {G}^{-3} \nabla_{a}{\nabla^{a}{\boldsymbol{W}}} \varphi^{k}_{\alpha}+\frac{33}{128}G_{\underline{i} \underline{j}} W^{a b} \mathbf{F}_{a b} \lambda^{\alpha}_{k} {G}^{-3} \varphi^{k}_{\alpha}+\frac{1}{4}G_{\underline{i} \underline{j}} \boldsymbol{W} W^{a b} W_{a b} \lambda^{\alpha}_{k} {G}^{-3} \varphi^{k}_{\alpha}+\frac{5}{128}G_{\underline{i} k} \lambda^{\beta}_{l} \boldsymbol{\lambda}^{k \alpha} X_{\underline{j} \alpha} {G}^{-3} \varphi^{l}_{\beta} - \frac{15}{128}G_{\underline{i} \underline{j}} \lambda^{\beta}_{l} \boldsymbol{\lambda}_{k}^{\alpha} X^{k}_{\alpha} {G}^{-3} \varphi^{l}_{\beta} - \frac{21}{128}G_{\underline{i} k} \lambda^{\beta}_{l} \boldsymbol{\lambda}_{\underline{j}}^{\alpha} X^{k}_{\alpha} {G}^{-3} \varphi^{l}_{\beta} - \frac{1}{8}{\rm i} G_{\underline{i} k} \lambda^{\alpha}_{l} {G}^{-3} \nabla_{a}{\boldsymbol{\lambda}^{k}_{\alpha}} \nabla^{a}{G_{\underline{j}}\,^{l}} - \frac{1}{4}{\rm i} G_{\underline{i} k} (\Sigma_{a b})^{\alpha \beta} \lambda_{l \alpha} {G}^{-3} \nabla^{a}{\boldsymbol{\lambda}^{k}_{\beta}} \nabla^{b}{G_{\underline{j}}\,^{l}} - \frac{3}{32}G_{\underline{i} k} G_{\underline{j} l} (\Sigma_{a b})^{\alpha \beta} W^{a b} \lambda^{\rho}_{m} \boldsymbol{\lambda}^{k}_{\alpha} {G}^{-5} \varphi^{l}_{\beta} \varphi^{m}_{\rho}+\frac{3}{128}{\rm i} \epsilon^{c d}\,_{e}\,^{a b} G_{\underline{i} k} (\Sigma_{c d})^{\alpha \beta} W_{a b} \lambda_{l \alpha} \boldsymbol{\lambda}^{k}_{\beta} {G}^{-3} \nabla^{e}{G_{\underline{j}}\,^{l}}+\frac{1}{64}{\rm i} G_{\underline{i} k} (\Gamma^{a})^{\alpha \beta} W_{a b} \lambda_{l \alpha} \boldsymbol{\lambda}^{k}_{\beta} {G}^{-3} \nabla^{b}{G_{\underline{j}}\,^{l}} - \frac{3}{64}{\rm i} G_{\underline{i} k} (\Sigma_{a b})^{\alpha \beta} \boldsymbol{W} F^{a b} X^{k}_{\alpha} {G}^{-3} \varphi_{\underline{j} \beta} - \frac{3}{64}{\rm i} G_{\underline{i} k} (\Gamma_{a})^{\alpha \beta} \boldsymbol{W} X^{k}_{\alpha} {G}^{-3} \nabla^{a}{W} \varphi_{\underline{j} \beta} - \frac{3}{128}G_{\underline{i} \underline{j}} \boldsymbol{W} Y \lambda^{\alpha}_{k} {G}^{-3} \varphi^{k}_{\alpha}%
 - \frac{9}{32}{\rm i} G_{\underline{i} k} G_{\underline{j} l} \boldsymbol{W} \lambda^{\beta}_{m} X^{k \alpha} {G}^{-5} \varphi^{l}_{\alpha} \varphi^{m}_{\beta}+\frac{3}{32}G_{\underline{i} k} (\Gamma_{a})^{\beta \alpha} \boldsymbol{W} \lambda_{l \beta} X^{k}_{\alpha} {G}^{-3} \nabla^{a}{G_{\underline{j}}\,^{l}}+\frac{3}{16}{\rm i} G_{\underline{i} k} G_{l m} (\Sigma_{a b})^{\alpha \beta} \mathbf{X}^{l m} F^{a b} {G}^{-5} \varphi_{\underline{j} \alpha} \varphi^{k}_{\beta} - \frac{3}{16}{\rm i} G_{\underline{i} k} G_{l m} (\Gamma_{a})^{\alpha \beta} \mathbf{X}^{l m} {G}^{-5} \nabla^{a}{W} \varphi_{\underline{j} \alpha} \varphi^{k}_{\beta}+\frac{1}{4}G_{k l} \mathbf{X}^{k l} {G}^{-3} \nabla_{a}{W} \nabla^{a}{G_{\underline{i} \underline{j}}} - \frac{3}{32}{\rm i} G_{k l} \mathbf{X}^{k l} \lambda^{\alpha}_{m} {G}^{-5} \varphi_{\underline{i}}^{\beta} \varphi_{\underline{j} \beta} \varphi^{m}_{\alpha} - \frac{15}{32}{\rm i} G_{\underline{i} k} G_{\underline{j} l} G_{m n} \mathbf{X}^{m n} \lambda^{\alpha}_{{i_{1}}} {G}^{-7} \varphi^{k \beta} \varphi^{l}_{\beta} \varphi^{{i_{1}}}_{\alpha}+\frac{3}{16}F G_{\underline{i} \underline{j}} G_{k l} \mathbf{X}^{k l} \lambda^{\alpha}_{m} {G}^{-5} \varphi^{m}_{\alpha}+\frac{3}{16}G_{\underline{i} k} G_{l m} (\Gamma_{a})^{\alpha \beta} \mathbf{X}^{l m} \lambda_{n \alpha} {G}^{-5} \nabla^{a}{G_{\underline{j}}\,^{n}} \varphi^{k}_{\beta}+\frac{3}{8}{\rm i} G_{\underline{i} k} X_{\underline{j} l} \boldsymbol{\lambda}_{m}^{\alpha} {G}^{-5} \varphi^{k \beta} \varphi^{l}_{\beta} \varphi^{m}_{\alpha}+\frac{1}{32}\epsilon^{c d}\,_{e}\,^{a b} G_{\underline{i} k} (\Sigma_{c d})^{\alpha \beta} \mathbf{F}_{a b} {G}^{-3} \nabla^{e}{\lambda^{k}_{\alpha}} \varphi_{\underline{j} \beta}+\frac{1}{16}G_{\underline{i} k} (\Gamma^{a})^{\alpha \beta} \mathbf{F}_{a b} {G}^{-3} \nabla^{b}{\lambda^{k}_{\alpha}} \varphi_{\underline{j} \beta} - \frac{1}{16}G_{\underline{i} k} {G}^{-3} \nabla_{a}{\boldsymbol{W}} \nabla^{a}{\lambda^{k \alpha}} \varphi_{\underline{j} \alpha} - \frac{1}{8}G_{\underline{i} k} (\Sigma_{a b})^{\alpha \beta} {G}^{-3} \nabla^{a}{\boldsymbol{W}} \nabla^{b}{\lambda^{k}_{\alpha}} \varphi_{\underline{j} \beta}+\frac{3}{8}G_{\underline{i} k} G_{\underline{j} l} (\Gamma_{a})^{\alpha \beta} \boldsymbol{\lambda}_{m}^{\rho} {G}^{-5} \nabla^{a}{\lambda^{k}_{\alpha}} \varphi^{l}_{\beta} \varphi^{m}_{\rho} - \frac{1}{4}G_{\underline{i} \underline{j}} \boldsymbol{\lambda}_{k}^{\alpha} {G}^{-3} \nabla_{a}{\nabla^{a}{W}} \varphi^{k}_{\alpha}+\frac{33}{128}G_{\underline{i} \underline{j}} W^{a b} F_{a b} \boldsymbol{\lambda}_{k}^{\alpha} {G}^{-3} \varphi^{k}_{\alpha}+\frac{1}{4}G_{\underline{i} \underline{j}} W W^{a b} W_{a b} \boldsymbol{\lambda}_{k}^{\alpha} {G}^{-3} \varphi^{k}_{\alpha} - \frac{5}{128}G_{\underline{i} k} \lambda^{k \alpha} \boldsymbol{\lambda}_{l}^{\beta} X_{\underline{j} \alpha} {G}^{-3} \varphi^{l}_{\beta}+\frac{15}{128}G_{\underline{i} \underline{j}} \lambda^{\alpha}_{k} \boldsymbol{\lambda}_{l}^{\beta} X^{k}_{\alpha} {G}^{-3} \varphi^{l}_{\beta}%
+\frac{21}{128}G_{\underline{i} k} \lambda^{\alpha}_{\underline{j}} \boldsymbol{\lambda}_{l}^{\beta} X^{k}_{\alpha} {G}^{-3} \varphi^{l}_{\beta} - \frac{1}{8}{\rm i} G_{\underline{i} k} \boldsymbol{\lambda}_{l}^{\alpha} {G}^{-3} \nabla_{a}{\lambda^{k}_{\alpha}} \nabla^{a}{G_{\underline{j}}\,^{l}} - \frac{1}{4}{\rm i} G_{\underline{i} k} (\Sigma_{a b})^{\alpha \beta} \boldsymbol{\lambda}_{l \alpha} {G}^{-3} \nabla^{a}{\lambda^{k}_{\beta}} \nabla^{b}{G_{\underline{j}}\,^{l}}+\frac{3}{32}G_{\underline{i} k} G_{\underline{j} l} (\Sigma_{a b})^{\alpha \beta} W^{a b} \lambda^{k}_{\alpha} \boldsymbol{\lambda}_{m}^{\rho} {G}^{-5} \varphi^{l}_{\beta} \varphi^{m}_{\rho} - \frac{3}{128}{\rm i} \epsilon^{c d}\,_{e}\,^{a b} G_{\underline{i} k} (\Sigma_{c d})^{\alpha \beta} W_{a b} \lambda^{k}_{\alpha} \boldsymbol{\lambda}_{l \beta} {G}^{-3} \nabla^{e}{G_{\underline{j}}\,^{l}}+\frac{1}{64}{\rm i} G_{\underline{i} k} (\Gamma^{a})^{\alpha \beta} W_{a b} \lambda^{k}_{\alpha} \boldsymbol{\lambda}_{l \beta} {G}^{-3} \nabla^{b}{G_{\underline{j}}\,^{l}} - \frac{3}{64}{\rm i} G_{\underline{i} k} (\Sigma_{a b})^{\alpha \beta} W \mathbf{F}^{a b} X^{k}_{\alpha} {G}^{-3} \varphi_{\underline{j} \beta} - \frac{3}{64}{\rm i} G_{\underline{i} k} (\Gamma_{a})^{\alpha \beta} W X^{k}_{\alpha} {G}^{-3} \nabla^{a}{\boldsymbol{W}} \varphi_{\underline{j} \beta} - \frac{3}{128}G_{\underline{i} \underline{j}} W Y \boldsymbol{\lambda}_{k}^{\alpha} {G}^{-3} \varphi^{k}_{\alpha} - \frac{9}{32}{\rm i} G_{\underline{i} k} G_{\underline{j} l} W \boldsymbol{\lambda}_{m}^{\beta} X^{k \alpha} {G}^{-5} \varphi^{l}_{\alpha} \varphi^{m}_{\beta}+\frac{3}{32}G_{\underline{i} k} (\Gamma_{a})^{\beta \alpha} W \boldsymbol{\lambda}_{l \beta} X^{k}_{\alpha} {G}^{-3} \nabla^{a}{G_{\underline{j}}\,^{l}}+\frac{3}{16}{\rm i} G_{\underline{i} k} G_{l m} (\Sigma_{a b})^{\alpha \beta} X^{l m} \mathbf{F}^{a b} {G}^{-5} \varphi_{\underline{j} \alpha} \varphi^{k}_{\beta} - \frac{3}{16}{\rm i} G_{\underline{i} k} G_{l m} (\Gamma_{a})^{\alpha \beta} X^{l m} {G}^{-5} \nabla^{a}{\boldsymbol{W}} \varphi_{\underline{j} \alpha} \varphi^{k}_{\beta}+\frac{1}{4}G_{k l} X^{k l} {G}^{-3} \nabla_{a}{\boldsymbol{W}} \nabla^{a}{G_{\underline{i} \underline{j}}} - \frac{3}{32}{\rm i} G_{k l} X^{k l} \boldsymbol{\lambda}_{m}^{\alpha} {G}^{-5} \varphi_{\underline{i}}^{\beta} \varphi_{\underline{j} \beta} \varphi^{m}_{\alpha} - \frac{15}{32}{\rm i} G_{\underline{i} k} G_{\underline{j} l} G_{m n} X^{m n} \boldsymbol{\lambda}_{{i_{1}}}^{\alpha} {G}^{-7} \varphi^{k \beta} \varphi^{l}_{\beta} \varphi^{{i_{1}}}_{\alpha}+\frac{3}{16}F G_{\underline{i} \underline{j}} G_{k l} X^{k l} \boldsymbol{\lambda}_{m}^{\alpha} {G}^{-5} \varphi^{m}_{\alpha}+\frac{3}{16}G_{\underline{i} k} G_{l m} (\Gamma_{a})^{\alpha \beta} X^{l m} \boldsymbol{\lambda}_{n \alpha} {G}^{-5} \nabla^{a}{G_{\underline{j}}\,^{n}} \varphi^{k}_{\beta}+\frac{3}{8}G_{\underline{i} k} (\Gamma_{a})^{\alpha \beta} \mathbf{X}^{k}\,_{l} {G}^{-3} \nabla^{a}{\lambda_{\underline{j} \alpha}} \varphi^{l}_{\beta} - \frac{7}{32}G_{\underline{i} k} (\Sigma_{a b})^{\alpha \beta} \mathbf{X}^{k}\,_{l} W^{a b} \lambda_{\underline{j} \alpha} {G}^{-3} \varphi^{l}_{\beta}%
+\frac{5}{16}G_{\underline{i} k} (\Gamma_{a})^{\alpha \beta} X_{\underline{j}}\,^{k} {G}^{-3} \nabla^{a}{\boldsymbol{\lambda}_{l \alpha}} \varphi^{l}_{\beta} - \frac{5}{16}G_{k l} (\Gamma_{a})^{\alpha \beta} X_{\underline{i}}\,^{k} {G}^{-3} \nabla^{a}{\boldsymbol{\lambda}^{l}_{\alpha}} \varphi_{\underline{j} \beta} - \frac{5}{64}G_{\underline{i} k} (\Sigma_{a b})^{\alpha \beta} X_{\underline{j}}\,^{k} W^{a b} \boldsymbol{\lambda}_{l \alpha} {G}^{-3} \varphi^{l}_{\beta} - \frac{15}{64}{\rm i} G_{\underline{i} k} \boldsymbol{W} X_{\underline{j}}\,^{k} X_{l}^{\alpha} {G}^{-3} \varphi^{l}_{\alpha}+\frac{15}{64}{\rm i} G_{k l} \boldsymbol{W} X_{\underline{i}}\,^{k} X^{l \alpha} {G}^{-3} \varphi_{\underline{j} \alpha} - \frac{15}{16}{\rm i} G_{\underline{i} k} G_{l m} X_{\underline{j}}\,^{l} \mathbf{X}^{m}\,_{n} {G}^{-5} \varphi^{k \alpha} \varphi^{n}_{\alpha}+\frac{3}{8}G_{\underline{i} k} (\Gamma_{a})^{\alpha \beta} X^{k}\,_{l} {G}^{-3} \nabla^{a}{\boldsymbol{\lambda}_{\underline{j} \alpha}} \varphi^{l}_{\beta} - \frac{7}{32}G_{\underline{i} k} (\Sigma_{a b})^{\alpha \beta} X^{k}\,_{l} W^{a b} \boldsymbol{\lambda}_{\underline{j} \alpha} {G}^{-3} \varphi^{l}_{\beta}+\frac{5}{16}G_{\underline{i} k} (\Gamma_{a})^{\alpha \beta} \mathbf{X}_{\underline{j}}\,^{k} {G}^{-3} \nabla^{a}{\lambda_{l \alpha}} \varphi^{l}_{\beta} - \frac{5}{16}G_{k l} (\Gamma_{a})^{\alpha \beta} \mathbf{X}_{\underline{i}}\,^{k} {G}^{-3} \nabla^{a}{\lambda^{l}_{\alpha}} \varphi_{\underline{j} \beta} - \frac{5}{64}G_{\underline{i} k} (\Sigma_{a b})^{\alpha \beta} \mathbf{X}_{\underline{j}}\,^{k} W^{a b} \lambda_{l \alpha} {G}^{-3} \varphi^{l}_{\beta} - \frac{15}{64}{\rm i} G_{\underline{i} k} W \mathbf{X}_{\underline{j}}\,^{k} X_{l}^{\alpha} {G}^{-3} \varphi^{l}_{\alpha}+\frac{15}{64}{\rm i} G_{k l} W \mathbf{X}_{\underline{i}}\,^{k} X^{l \alpha} {G}^{-3} \varphi_{\underline{j} \alpha} - \frac{15}{16}{\rm i} G_{\underline{i} k} G_{l m} X^{l}\,_{n} \mathbf{X}_{\underline{j}}\,^{m} {G}^{-5} \varphi^{k \alpha} \varphi^{n}_{\alpha}+\frac{1}{4}{\rm i} G_{\underline{i} \underline{j}} G_{k l} {G}^{-3} \nabla_{a}{\lambda^{k \alpha}} \nabla^{a}{\boldsymbol{\lambda}^{l}_{\alpha}} - \frac{1}{2}{\rm i} G_{\underline{i} \underline{j}} G_{k l} (\Sigma_{a b})^{\alpha \beta} {G}^{-3} \nabla^{a}{\lambda^{k}_{\alpha}} \nabla^{b}{\boldsymbol{\lambda}^{l}_{\beta}}+\frac{1}{32}{\rm i} \epsilon^{c d}\,_{e}\,^{a b} G_{\underline{i} \underline{j}} G_{k l} (\Sigma_{c d})^{\alpha \beta} W_{a b} \boldsymbol{\lambda}^{k}_{\alpha} {G}^{-3} \nabla^{e}{\lambda^{l}_{\beta}} - \frac{1}{16}{\rm i} G_{\underline{i} \underline{j}} G_{k l} (\Gamma^{a})^{\alpha \beta} W_{a b} \boldsymbol{\lambda}^{k}_{\alpha} {G}^{-3} \nabla^{b}{\lambda^{l}_{\beta}} - \frac{3}{4}G_{\underline{i} k} G_{\underline{j} l} G_{m n} (\Gamma_{a})^{\alpha \beta} \mathbf{X}^{k m} {G}^{-5} \nabla^{a}{\lambda^{n}_{\alpha}} \varphi^{l}_{\beta}+\frac{1}{2}G_{\underline{i} k} G_{\underline{j} l} \mathbf{X}^{k l} {G}^{-3} \nabla_{a}{\nabla^{a}{W}}%
 - \frac{1}{2}G_{\underline{i} k} G_{\underline{j} l} \mathbf{X}^{k l} W^{a b} F_{a b} {G}^{-3} - \frac{31}{64}G_{\underline{i} k} G_{\underline{j} l} W \mathbf{X}^{k l} W^{a b} W_{a b} {G}^{-3} - \frac{5}{256}G_{\underline{i} k} G_{l m} \mathbf{X}^{k l} \lambda^{m \alpha} X_{\underline{j} \alpha} {G}^{-3}+\frac{15}{64}G_{\underline{i} k} G_{\underline{j} l} \mathbf{X}^{k l} \lambda^{\alpha}_{m} X^{m}_{\alpha} {G}^{-3}+\frac{21}{64}G_{\underline{i} k} G_{l m} \mathbf{X}^{k l} \lambda^{\alpha}_{\underline{j}} X^{m}_{\alpha} {G}^{-3}+\frac{1}{32}{\rm i} \epsilon^{c d}\,_{e}\,^{a b} G_{\underline{i} \underline{j}} G_{k l} (\Sigma_{c d})^{\alpha \beta} W_{a b} \lambda^{k}_{\alpha} {G}^{-3} \nabla^{e}{\boldsymbol{\lambda}^{l}_{\beta}} - \frac{1}{16}{\rm i} G_{\underline{i} \underline{j}} G_{k l} (\Gamma^{a})^{\alpha \beta} W_{a b} \lambda^{k}_{\alpha} {G}^{-3} \nabla^{b}{\boldsymbol{\lambda}^{l}_{\beta}}+\frac{1}{256}{\rm i} \epsilon_{e}\,^{a b c d} G_{\underline{i} \underline{j}} G_{k l} (\Gamma^{e})^{\alpha \beta} W_{a b} W_{c d} \lambda^{k}_{\alpha} \boldsymbol{\lambda}^{l}_{\beta} {G}^{-3}+\frac{3}{16}G_{\underline{i} k} G_{\underline{j} l} G_{m n} (\Sigma_{a b})^{\alpha \beta} \mathbf{X}^{k m} W^{a b} \lambda^{n}_{\alpha} {G}^{-5} \varphi^{l}_{\beta}+\frac{3}{64}G_{\underline{i} k} G_{\underline{j} l} W \mathbf{X}^{k l} Y {G}^{-3}+\frac{9}{16}{\rm i} G_{\underline{i} k} G_{\underline{j} l} G_{m n} W \mathbf{X}^{k m} X^{n \alpha} {G}^{-5} \varphi^{l}_{\alpha} - \frac{3}{4}G_{\underline{i} k} G_{\underline{j} l} G_{m n} (\Gamma_{a})^{\alpha \beta} X^{k m} {G}^{-5} \nabla^{a}{\boldsymbol{\lambda}^{n}_{\alpha}} \varphi^{l}_{\beta}+\frac{1}{2}G_{\underline{i} k} G_{\underline{j} l} X^{k l} {G}^{-3} \nabla_{a}{\nabla^{a}{\boldsymbol{W}}} - \frac{1}{2}G_{\underline{i} k} G_{\underline{j} l} X^{k l} W^{a b} \mathbf{F}_{a b} {G}^{-3} - \frac{31}{64}G_{\underline{i} k} G_{\underline{j} l} \boldsymbol{W} X^{k l} W^{a b} W_{a b} {G}^{-3} - \frac{5}{256}G_{\underline{i} k} G_{l m} X^{k l} \boldsymbol{\lambda}^{m \alpha} X_{\underline{j} \alpha} {G}^{-3}+\frac{15}{64}G_{\underline{i} k} G_{\underline{j} l} X^{k l} \boldsymbol{\lambda}_{m}^{\alpha} X^{m}_{\alpha} {G}^{-3}+\frac{21}{64}G_{\underline{i} k} G_{l m} X^{k l} \boldsymbol{\lambda}_{\underline{j}}^{\alpha} X^{m}_{\alpha} {G}^{-3}+\frac{3}{16}G_{\underline{i} k} G_{\underline{j} l} G_{m n} (\Sigma_{a b})^{\alpha \beta} X^{k m} W^{a b} \boldsymbol{\lambda}^{n}_{\alpha} {G}^{-5} \varphi^{l}_{\beta}+\frac{3}{64}G_{\underline{i} k} G_{\underline{j} l} \boldsymbol{W} X^{k l} Y {G}^{-3}%
+\frac{9}{16}{\rm i} G_{\underline{i} k} G_{\underline{j} l} G_{m n} \boldsymbol{W} X^{k m} X^{n \alpha} {G}^{-5} \varphi^{l}_{\alpha}+\frac{3}{16}{\rm i} G_{k l} G_{m n} X^{k m} \mathbf{X}^{l n} {G}^{-5} \varphi_{\underline{i}}^{\alpha} \varphi_{\underline{j} \alpha}+\frac{15}{16}{\rm i} G_{\underline{i} k} G_{\underline{j} l} G_{m n} G_{{i_{1}} {i_{2}}} X^{m {i_{1}}} \mathbf{X}^{n {i_{2}}} {G}^{-7} \varphi^{k \alpha} \varphi^{l}_{\alpha} - \frac{3}{8}F G_{\underline{i} \underline{j}} G_{k l} G_{m n} X^{k m} \mathbf{X}^{l n} {G}^{-5}+\frac{1}{16}(\Gamma_{a})^{\alpha \beta} \boldsymbol{\lambda}_{k}^{\rho} {G}^{-3} \nabla^{a}{\lambda_{\underline{i} \alpha}} \varphi_{\underline{j} \rho} \varphi^{k}_{\beta}+\frac{3}{16}{\rm i} G_{\underline{i} k} X_{\underline{j} l} \boldsymbol{\lambda}_{m}^{\alpha} {G}^{-5} \varphi^{k \beta} \varphi^{l}_{\alpha} \varphi^{m}_{\beta}+\frac{3}{16}{\rm i} G_{\underline{i} k} X_{l m} \boldsymbol{\lambda}_{\underline{j}}^{\alpha} {G}^{-5} \varphi^{k \beta} \varphi^{l}_{\alpha} \varphi^{m}_{\beta}+\frac{1}{32}\epsilon^{c d}\,_{e}\,^{a b} G_{\underline{i} \underline{j}} (\Sigma_{c d})^{\alpha \beta} \mathbf{F}_{a b} {G}^{-3} \nabla^{e}{\lambda_{k \alpha}} \varphi^{k}_{\beta}+\frac{1}{16}G_{\underline{i} \underline{j}} (\Gamma^{a})^{\alpha \beta} \mathbf{F}_{a b} {G}^{-3} \nabla^{b}{\lambda_{k \alpha}} \varphi^{k}_{\beta}+\frac{1}{32}\epsilon^{c d}\,_{e}\,^{a b} G_{\underline{i} \underline{j}} (\Sigma_{c d})^{\alpha \beta} \boldsymbol{W} W_{a b} {G}^{-3} \nabla^{e}{\lambda_{k \alpha}} \varphi^{k}_{\beta}+\frac{1}{16}G_{\underline{i} \underline{j}} (\Gamma^{a})^{\alpha \beta} \boldsymbol{W} W_{a b} {G}^{-3} \nabla^{b}{\lambda_{k \alpha}} \varphi^{k}_{\beta} - \frac{1}{16}G_{\underline{i} \underline{j}} {G}^{-3} \nabla_{a}{\boldsymbol{W}} \nabla^{a}{\lambda^{\alpha}_{k}} \varphi^{k}_{\alpha} - \frac{1}{8}G_{\underline{i} \underline{j}} (\Sigma_{a b})^{\alpha \beta} {G}^{-3} \nabla^{a}{\boldsymbol{W}} \nabla^{b}{\lambda_{k \alpha}} \varphi^{k}_{\beta} - \frac{3}{16}G_{\underline{i} k} G_{\underline{j} l} (\Gamma_{a})^{\alpha \beta} \boldsymbol{\lambda}^{k \rho} {G}^{-5} \nabla^{a}{\lambda_{m \alpha}} \varphi^{l}_{\beta} \varphi^{m}_{\rho} - \frac{1}{4}G_{\underline{i} k} \boldsymbol{\lambda}^{k \alpha} {G}^{-3} \nabla_{a}{\nabla^{a}{W}} \varphi_{\underline{j} \alpha}+\frac{5}{256}G_{\underline{i} k} \lambda^{\alpha}_{l} \boldsymbol{\lambda}^{k \beta} X_{\underline{j} \alpha} {G}^{-3} \varphi^{l}_{\beta}+\frac{15}{128}G_{\underline{i} k} \lambda^{\alpha}_{l} \boldsymbol{\lambda}^{k \beta} X^{l}_{\alpha} {G}^{-3} \varphi_{\underline{j} \beta} - \frac{21}{256}G_{\underline{i} k} \lambda^{\alpha}_{\underline{j}} \boldsymbol{\lambda}^{k \beta} X_{l \alpha} {G}^{-3} \varphi^{l}_{\beta}+\frac{1}{16}{\rm i} G_{\underline{i} k} \boldsymbol{\lambda}^{k \alpha} {G}^{-3} \nabla_{a}{\lambda_{l \alpha}} \nabla^{a}{G_{\underline{j}}\,^{l}}+\frac{1}{8}{\rm i} G_{\underline{i} k} (\Sigma_{a b})^{\alpha \beta} \boldsymbol{\lambda}^{k}_{\alpha} {G}^{-3} \nabla^{a}{\lambda_{l \beta}} \nabla^{b}{G_{\underline{j}}\,^{l}}%
 - \frac{3}{16}G_{\underline{i} k} G_{l m} (\Gamma_{a})^{\alpha \beta} \boldsymbol{\lambda}^{l \rho} {G}^{-5} \nabla^{a}{\lambda^{m}_{\alpha}} \varphi_{\underline{j} \rho} \varphi^{k}_{\beta} - \frac{5}{256}G_{k l} \lambda^{k \alpha} \boldsymbol{\lambda}^{l \beta} X_{\underline{i} \alpha} {G}^{-3} \varphi_{\underline{j} \beta}+\frac{21}{256}G_{k l} \lambda^{\alpha}_{\underline{i}} \boldsymbol{\lambda}^{k \beta} X^{l}_{\alpha} {G}^{-3} \varphi_{\underline{j} \beta} - \frac{1}{16}{\rm i} G_{k l} \boldsymbol{\lambda}^{k \alpha} {G}^{-3} \nabla_{a}{\lambda^{l}_{\alpha}} \nabla^{a}{G_{\underline{i} \underline{j}}} - \frac{1}{8}{\rm i} G_{k l} (\Sigma_{a b})^{\alpha \beta} \boldsymbol{\lambda}^{k}_{\alpha} {G}^{-3} \nabla^{a}{\lambda^{l}_{\beta}} \nabla^{b}{G_{\underline{i} \underline{j}}} - \frac{1}{256}\epsilon_{e}\,^{c d a b} G_{\underline{i} \underline{j}} (\Gamma^{e})^{\alpha \beta} W_{c d} \mathbf{F}_{a b} \lambda_{k \alpha} {G}^{-3} \varphi^{k}_{\beta}+\frac{1}{32}G_{\underline{i} \underline{j}} (\Sigma^{a}{}_{\, c})^{\alpha \beta} W^{c b} \mathbf{F}_{a b} \lambda_{k \alpha} {G}^{-3} \varphi^{k}_{\beta} - \frac{1}{256}\epsilon_{e}\,^{a b c d} G_{\underline{i} \underline{j}} (\Gamma^{e})^{\alpha \beta} \boldsymbol{W} W_{a b} W_{c d} \lambda_{k \alpha} {G}^{-3} \varphi^{k}_{\beta} - \frac{1}{128}\epsilon^{c d}\,_{e}\,^{a b} G_{\underline{i} \underline{j}} (\Sigma_{c d})^{\alpha \beta} W_{a b} \lambda_{k \alpha} {G}^{-3} \nabla^{e}{\boldsymbol{W}} \varphi^{k}_{\beta}+\frac{1}{64}G_{\underline{i} \underline{j}} (\Gamma^{a})^{\alpha \beta} W_{a b} \lambda_{k \alpha} {G}^{-3} \nabla^{b}{\boldsymbol{W}} \varphi^{k}_{\beta} - \frac{3}{64}G_{\underline{i} k} G_{\underline{j} l} (\Sigma_{a b})^{\alpha \beta} W^{a b} \lambda_{m \alpha} \boldsymbol{\lambda}^{k \rho} {G}^{-5} \varphi^{l}_{\beta} \varphi^{m}_{\rho} - \frac{3}{64}G_{\underline{i} k} G_{l m} (\Sigma_{a b})^{\alpha \beta} W^{a b} \lambda^{l}_{\alpha} \boldsymbol{\lambda}^{m \rho} {G}^{-5} \varphi_{\underline{j} \rho} \varphi^{k}_{\beta}+\frac{1}{32}{\rm i} G_{k l} (\Gamma^{a})^{\alpha \beta} W_{a b} \lambda^{k}_{\alpha} \boldsymbol{\lambda}^{l}_{\beta} {G}^{-3} \nabla^{b}{G_{\underline{i} \underline{j}}} - \frac{3}{64}{\rm i} G_{\underline{i} \underline{j}} (\Sigma_{a b})^{\alpha \beta} W \mathbf{F}^{a b} X_{k \alpha} {G}^{-3} \varphi^{k}_{\beta} - \frac{3}{32}{\rm i} G_{\underline{i} \underline{j}} (\Sigma_{a b})^{\alpha \beta} W \boldsymbol{W} W^{a b} X_{k \alpha} {G}^{-3} \varphi^{k}_{\beta} - \frac{3}{64}{\rm i} G_{\underline{i} \underline{j}} (\Gamma_{a})^{\alpha \beta} W X_{k \alpha} {G}^{-3} \nabla^{a}{\boldsymbol{W}} \varphi^{k}_{\beta} - \frac{3}{128}G_{\underline{i} k} W Y \boldsymbol{\lambda}^{k \alpha} {G}^{-3} \varphi_{\underline{j} \alpha}+\frac{9}{64}{\rm i} G_{\underline{i} k} G_{\underline{j} l} W \boldsymbol{\lambda}^{k \beta} X_{m}^{\alpha} {G}^{-5} \varphi^{l}_{\alpha} \varphi^{m}_{\beta} - \frac{3}{64}G_{\underline{i} k} (\Gamma_{a})^{\beta \alpha} W \boldsymbol{\lambda}^{k}_{\beta} X_{l \alpha} {G}^{-3} \nabla^{a}{G_{\underline{j}}\,^{l}}+\frac{9}{64}{\rm i} G_{\underline{i} k} G_{l m} W \boldsymbol{\lambda}^{l \beta} X^{m \alpha} {G}^{-5} \varphi_{\underline{j} \beta} \varphi^{k}_{\alpha}%
+\frac{3}{64}G_{k l} (\Gamma_{a})^{\beta \alpha} W \boldsymbol{\lambda}^{k}_{\beta} X^{l}_{\alpha} {G}^{-3} \nabla^{a}{G_{\underline{i} \underline{j}}} - \frac{3}{16}{\rm i} G_{\underline{i} k} G_{\underline{j} l} (\Sigma_{a b})^{\alpha \beta} X^{k}\,_{m} \mathbf{F}^{a b} {G}^{-5} \varphi^{l}_{\alpha} \varphi^{m}_{\beta} - \frac{3}{16}{\rm i} G_{\underline{i} k} G_{\underline{j} l} (\Sigma_{a b})^{\alpha \beta} \boldsymbol{W} X^{k}\,_{m} W^{a b} {G}^{-5} \varphi^{l}_{\alpha} \varphi^{m}_{\beta} - \frac{3}{16}{\rm i} G_{\underline{i} k} G_{\underline{j} l} (\Gamma_{a})^{\alpha \beta} X^{k}\,_{m} {G}^{-5} \nabla^{a}{\boldsymbol{W}} \varphi^{l}_{\alpha} \varphi^{m}_{\beta}+\frac{1}{4}G_{\underline{i} k} X^{k}\,_{l} {G}^{-3} \nabla_{a}{\boldsymbol{W}} \nabla^{a}{G_{\underline{j}}\,^{l}}+\frac{3}{32}{\rm i} G_{k l} X^{k}\,_{m} \boldsymbol{\lambda}^{l \alpha} {G}^{-5} \varphi_{\underline{i}}^{\beta} \varphi_{\underline{j} \beta} \varphi^{m}_{\alpha}+\frac{15}{32}{\rm i} G_{\underline{i} k} G_{\underline{j} l} G_{m n} X^{m}\,_{{i_{1}}} \boldsymbol{\lambda}^{n \alpha} {G}^{-7} \varphi^{k \beta} \varphi^{l}_{\beta} \varphi^{{i_{1}}}_{\alpha} - \frac{3}{16}F G_{\underline{i} \underline{j}} G_{k l} X^{k}\,_{m} \boldsymbol{\lambda}^{l \alpha} {G}^{-5} \varphi^{m}_{\alpha} - \frac{3}{16}G_{\underline{i} k} G_{l m} (\Gamma_{a})^{\alpha \beta} X^{l}\,_{n} \boldsymbol{\lambda}^{m}_{\alpha} {G}^{-5} \nabla^{a}{G_{\underline{j}}\,^{n}} \varphi^{k}_{\beta}+\frac{15}{256}G_{\underline{i} \underline{j}} G_{k l} X^{k}\,_{m} \boldsymbol{\lambda}^{l \alpha} X^{m}_{\alpha} {G}^{-3}+\frac{1}{16}(\Gamma_{a})^{\alpha \beta} \lambda^{\rho}_{k} {G}^{-3} \nabla^{a}{\boldsymbol{\lambda}_{\underline{i} \alpha}} \varphi_{\underline{j} \rho} \varphi^{k}_{\beta}+\frac{3}{16}{\rm i} G_{\underline{i} k} \mathbf{X}_{\underline{j} l} \lambda^{\alpha}_{m} {G}^{-5} \varphi^{k \beta} \varphi^{l}_{\alpha} \varphi^{m}_{\beta}+\frac{3}{16}{\rm i} G_{\underline{i} k} \mathbf{X}_{l m} \lambda^{\alpha}_{\underline{j}} {G}^{-5} \varphi^{k \beta} \varphi^{l}_{\alpha} \varphi^{m}_{\beta}+\frac{1}{32}\epsilon^{c d}\,_{e}\,^{a b} G_{\underline{i} \underline{j}} (\Sigma_{c d})^{\alpha \beta} F_{a b} {G}^{-3} \nabla^{e}{\boldsymbol{\lambda}_{k \alpha}} \varphi^{k}_{\beta}+\frac{1}{16}G_{\underline{i} \underline{j}} (\Gamma^{a})^{\alpha \beta} F_{a b} {G}^{-3} \nabla^{b}{\boldsymbol{\lambda}_{k \alpha}} \varphi^{k}_{\beta}+\frac{1}{32}\epsilon^{c d}\,_{e}\,^{a b} G_{\underline{i} \underline{j}} (\Sigma_{c d})^{\alpha \beta} W W_{a b} {G}^{-3} \nabla^{e}{\boldsymbol{\lambda}_{k \alpha}} \varphi^{k}_{\beta}+\frac{1}{16}G_{\underline{i} \underline{j}} (\Gamma^{a})^{\alpha \beta} W W_{a b} {G}^{-3} \nabla^{b}{\boldsymbol{\lambda}_{k \alpha}} \varphi^{k}_{\beta} - \frac{1}{16}G_{\underline{i} \underline{j}} {G}^{-3} \nabla_{a}{W} \nabla^{a}{\boldsymbol{\lambda}_{k}^{\alpha}} \varphi^{k}_{\alpha} - \frac{1}{8}G_{\underline{i} \underline{j}} (\Sigma_{a b})^{\alpha \beta} {G}^{-3} \nabla^{a}{W} \nabla^{b}{\boldsymbol{\lambda}_{k \alpha}} \varphi^{k}_{\beta} - \frac{3}{16}G_{\underline{i} k} G_{\underline{j} l} (\Gamma_{a})^{\alpha \beta} \lambda^{k \rho} {G}^{-5} \nabla^{a}{\boldsymbol{\lambda}_{m \alpha}} \varphi^{l}_{\beta} \varphi^{m}_{\rho}%
 - \frac{1}{4}G_{\underline{i} k} \lambda^{k \alpha} {G}^{-3} \nabla_{a}{\nabla^{a}{\boldsymbol{W}}} \varphi_{\underline{j} \alpha} - \frac{5}{256}G_{\underline{i} k} \lambda^{k \beta} \boldsymbol{\lambda}_{l}^{\alpha} X_{\underline{j} \alpha} {G}^{-3} \varphi^{l}_{\beta} - \frac{15}{128}G_{\underline{i} k} \lambda^{k \beta} \boldsymbol{\lambda}_{l}^{\alpha} X^{l}_{\alpha} {G}^{-3} \varphi_{\underline{j} \beta}+\frac{21}{256}G_{\underline{i} k} \lambda^{k \beta} \boldsymbol{\lambda}_{\underline{j}}^{\alpha} X_{l \alpha} {G}^{-3} \varphi^{l}_{\beta}+\frac{1}{16}{\rm i} G_{\underline{i} k} \lambda^{k \alpha} {G}^{-3} \nabla_{a}{\boldsymbol{\lambda}_{l \alpha}} \nabla^{a}{G_{\underline{j}}\,^{l}}+\frac{1}{8}{\rm i} G_{\underline{i} k} (\Sigma_{a b})^{\alpha \beta} \lambda^{k}_{\alpha} {G}^{-3} \nabla^{a}{\boldsymbol{\lambda}_{l \beta}} \nabla^{b}{G_{\underline{j}}\,^{l}} - \frac{3}{16}G_{\underline{i} k} G_{l m} (\Gamma_{a})^{\alpha \beta} \lambda^{l \rho} {G}^{-5} \nabla^{a}{\boldsymbol{\lambda}^{m}_{\alpha}} \varphi_{\underline{j} \rho} \varphi^{k}_{\beta}+\frac{5}{256}G_{k l} \lambda^{k \beta} \boldsymbol{\lambda}^{l \alpha} X_{\underline{i} \alpha} {G}^{-3} \varphi_{\underline{j} \beta} - \frac{21}{256}G_{k l} \lambda^{k \beta} \boldsymbol{\lambda}_{\underline{i}}^{\alpha} X^{l}_{\alpha} {G}^{-3} \varphi_{\underline{j} \beta} - \frac{1}{16}{\rm i} G_{k l} \lambda^{k \alpha} {G}^{-3} \nabla_{a}{\boldsymbol{\lambda}^{l}_{\alpha}} \nabla^{a}{G_{\underline{i} \underline{j}}} - \frac{1}{8}{\rm i} G_{k l} (\Sigma_{a b})^{\alpha \beta} \lambda^{k}_{\alpha} {G}^{-3} \nabla^{a}{\boldsymbol{\lambda}^{l}_{\beta}} \nabla^{b}{G_{\underline{i} \underline{j}}} - \frac{1}{256}\epsilon_{e}\,^{c d a b} G_{\underline{i} \underline{j}} (\Gamma^{e})^{\alpha \beta} W_{c d} F_{a b} \boldsymbol{\lambda}_{k \alpha} {G}^{-3} \varphi^{k}_{\beta}+\frac{1}{32}G_{\underline{i} \underline{j}} (\Sigma^{a}{}_{\, c})^{\alpha \beta} W^{c b} F_{a b} \boldsymbol{\lambda}_{k \alpha} {G}^{-3} \varphi^{k}_{\beta} - \frac{1}{256}\epsilon_{e}\,^{a b c d} G_{\underline{i} \underline{j}} (\Gamma^{e})^{\alpha \beta} W W_{a b} W_{c d} \boldsymbol{\lambda}_{k \alpha} {G}^{-3} \varphi^{k}_{\beta} - \frac{1}{128}\epsilon^{c d}\,_{e}\,^{a b} G_{\underline{i} \underline{j}} (\Sigma_{c d})^{\alpha \beta} W_{a b} \boldsymbol{\lambda}_{k \alpha} {G}^{-3} \nabla^{e}{W} \varphi^{k}_{\beta}+\frac{1}{64}G_{\underline{i} \underline{j}} (\Gamma^{a})^{\alpha \beta} W_{a b} \boldsymbol{\lambda}_{k \alpha} {G}^{-3} \nabla^{b}{W} \varphi^{k}_{\beta}+\frac{3}{64}G_{\underline{i} k} G_{\underline{j} l} (\Sigma_{a b})^{\alpha \beta} W^{a b} \lambda^{k \rho} \boldsymbol{\lambda}_{m \alpha} {G}^{-5} \varphi^{l}_{\beta} \varphi^{m}_{\rho}+\frac{3}{64}G_{\underline{i} k} G_{l m} (\Sigma_{a b})^{\alpha \beta} W^{a b} \lambda^{l \rho} \boldsymbol{\lambda}^{m}_{\alpha} {G}^{-5} \varphi_{\underline{j} \rho} \varphi^{k}_{\beta} - \frac{3}{64}{\rm i} G_{\underline{i} \underline{j}} (\Sigma_{a b})^{\alpha \beta} \boldsymbol{W} F^{a b} X_{k \alpha} {G}^{-3} \varphi^{k}_{\beta} - \frac{3}{64}{\rm i} G_{\underline{i} \underline{j}} (\Gamma_{a})^{\alpha \beta} \boldsymbol{W} X_{k \alpha} {G}^{-3} \nabla^{a}{W} \varphi^{k}_{\beta}%
 - \frac{3}{128}G_{\underline{i} k} \boldsymbol{W} Y \lambda^{k \alpha} {G}^{-3} \varphi_{\underline{j} \alpha}+\frac{9}{64}{\rm i} G_{\underline{i} k} G_{\underline{j} l} \boldsymbol{W} \lambda^{k \beta} X_{m}^{\alpha} {G}^{-5} \varphi^{l}_{\alpha} \varphi^{m}_{\beta} - \frac{3}{64}G_{\underline{i} k} (\Gamma_{a})^{\beta \alpha} \boldsymbol{W} \lambda^{k}_{\beta} X_{l \alpha} {G}^{-3} \nabla^{a}{G_{\underline{j}}\,^{l}}+\frac{9}{64}{\rm i} G_{\underline{i} k} G_{l m} \boldsymbol{W} \lambda^{l \beta} X^{m \alpha} {G}^{-5} \varphi_{\underline{j} \beta} \varphi^{k}_{\alpha}+\frac{3}{64}G_{k l} (\Gamma_{a})^{\beta \alpha} \boldsymbol{W} \lambda^{k}_{\beta} X^{l}_{\alpha} {G}^{-3} \nabla^{a}{G_{\underline{i} \underline{j}}} - \frac{3}{16}{\rm i} G_{\underline{i} k} G_{\underline{j} l} (\Sigma_{a b})^{\alpha \beta} \mathbf{X}^{k}\,_{m} F^{a b} {G}^{-5} \varphi^{l}_{\alpha} \varphi^{m}_{\beta} - \frac{3}{16}{\rm i} G_{\underline{i} k} G_{\underline{j} l} (\Sigma_{a b})^{\alpha \beta} W \mathbf{X}^{k}\,_{m} W^{a b} {G}^{-5} \varphi^{l}_{\alpha} \varphi^{m}_{\beta} - \frac{3}{16}{\rm i} G_{\underline{i} k} G_{\underline{j} l} (\Gamma_{a})^{\alpha \beta} \mathbf{X}^{k}\,_{m} {G}^{-5} \nabla^{a}{W} \varphi^{l}_{\alpha} \varphi^{m}_{\beta}+\frac{1}{4}G_{\underline{i} k} \mathbf{X}^{k}\,_{l} {G}^{-3} \nabla_{a}{W} \nabla^{a}{G_{\underline{j}}\,^{l}}+\frac{3}{32}{\rm i} G_{k l} \mathbf{X}^{k}\,_{m} \lambda^{l \alpha} {G}^{-5} \varphi_{\underline{i}}^{\beta} \varphi_{\underline{j} \beta} \varphi^{m}_{\alpha}+\frac{15}{32}{\rm i} G_{\underline{i} k} G_{\underline{j} l} G_{m n} \mathbf{X}^{m}\,_{{i_{1}}} \lambda^{n \alpha} {G}^{-7} \varphi^{k \beta} \varphi^{l}_{\beta} \varphi^{{i_{1}}}_{\alpha} - \frac{3}{16}F G_{\underline{i} \underline{j}} G_{k l} \mathbf{X}^{k}\,_{m} \lambda^{l \alpha} {G}^{-5} \varphi^{m}_{\alpha} - \frac{3}{16}G_{\underline{i} k} G_{l m} (\Gamma_{a})^{\alpha \beta} \mathbf{X}^{l}\,_{n} \lambda^{m}_{\alpha} {G}^{-5} \nabla^{a}{G_{\underline{j}}\,^{n}} \varphi^{k}_{\beta}+\frac{15}{256}G_{\underline{i} \underline{j}} G_{k l} \mathbf{X}^{k}\,_{m} \lambda^{l \alpha} X^{m}_{\alpha} {G}^{-3}
 \doublespacedmathend
 \end{adjustwidth}

 \section{Degauged Expressions} \label{DegaugedIdentities}
\subsection{Weyl Multiplet}
\subsubsection{$\nabla^{c} \nabla^{d} W_{a b}$}
\begin{adjustwidth}{0cm}{5cm}
\doublespacedmathbegin
\mathcal{D}^{c}{\nabla^{d}{W_{\hat{a} \hat{b}}}} - \frac{1}{2}\psi^{c}\,_{i}\,^{\alpha} \nabla^{d}{W_{\hat{a} \hat{b} \alpha}\,^{i}} - \frac{1}{2}(\Sigma_{\hat{a} \hat{b}})^{\alpha}{}_{\beta} \psi^{c}\,_{i}\,^{\beta} \nabla^{d}{X^{i}_{\alpha}}+\frac{1}{8}\epsilon^{d {e_{2}} {e_{3}} e {e_{1}}} (\Sigma_{{e_{2}} {e_{3}}})^{\alpha}{}_{\beta} \psi^{c}\,_{i}\,^{\beta} W_{e {e_{1}}} W_{\hat{a} \hat{b} \alpha}\,^{i}+\frac{3}{8}(\Gamma_{e})^{\alpha}{}_{\beta} \psi^{c}\,_{i}\,^{\beta} W^{d e} W_{\hat{a} \hat{b} \alpha}\,^{i} - \frac{3}{16}(\Gamma^{d})^{\alpha}{}_{\beta} \psi^{c}\,_{i}\,^{\beta} W_{\hat{a} \hat{b}} X^{i}_{\alpha} - \frac{1}{32}\epsilon^{d}\,_{\hat{a} \hat{b}}\,^{e {e_{1}}} \psi^{c}\,_{i}\,^{\alpha} W_{e {e_{1}}} X^{i}_{\alpha}+\frac{3}{32}\epsilon^{{e_{2}}}\,_{\hat{a} \hat{b}}\,^{e {e_{1}}} (\Sigma_{{e_{2}}}{}^{\, d})^{\alpha}{}_{\beta} \psi^{c}\,_{i}\,^{\beta} W_{e {e_{1}}} X^{i}_{\alpha} - \frac{3}{16}\epsilon^{d {e_{1}} {e_{2}}}\,_{\hat{a}}\,^{e} (\Sigma_{{e_{1}} {e_{2}}})^{\alpha}{}_{\beta} \psi^{c}\,_{i}\,^{\beta} W_{\hat{b} e} X^{i}_{\alpha} - \frac{3}{8}(\Gamma^{e})^{\alpha}{}_{\beta} \delta^{d}\,_{\hat{a}} \psi^{c}\,_{i}\,^{\beta} W_{\hat{b} e} X^{i}_{\alpha}+\frac{1}{2}(\Gamma_{\hat{a}})^{\alpha}{}_{\beta} \psi^{c}\,_{i}\,^{\beta} W_{\hat{b}}\,^{d} X^{i}_{\alpha}+\frac{1}{16}\epsilon^{d {e_{2}}}\,_{\hat{a} \hat{b}}\,^{e} (\Sigma_{{e_{2}}}{}^{\, {e_{1}}})^{\alpha}{}_{\beta} \psi^{c}\,_{i}\,^{\beta} W_{e {e_{1}}} X^{i}_{\alpha}+\frac{1}{16}\epsilon^{d}\,_{\hat{a} \hat{b}}\,^{{e_{2}} {e_{3}}} (\Sigma_{e {e_{1}}})^{\alpha}{}_{\beta} \psi^{c}\,_{i}\,^{\beta} W^{e {e_{1}}} W_{{e_{2}} {e_{3}} \alpha}\,^{i}+\frac{1}{16}\epsilon_{\hat{a} \hat{b}}\,^{e {e_{1}} {e_{2}}} \psi^{c}\,_{i}\,^{\alpha} W_{e}\,^{d} W_{{e_{1}} {e_{2}} \alpha}\,^{i}+\frac{1}{4}(\Gamma_{\hat{a}})^{\alpha}{}_{\beta} \psi^{c}\,_{i}\,^{\beta} W_{\hat{b}}\,^{e} W^{d}\,_{e \alpha}\,^{i} - \frac{1}{8}(\Gamma^{e})^{\alpha}{}_{\beta} \psi^{c}\,_{i}\,^{\beta} W_{\hat{a} \hat{b}} W_{e}\,^{d}\,_{\alpha}\,^{i}+\frac{1}{8}(\Gamma_{\hat{b}})^{\alpha}{}_{\beta} \delta^{d}\,_{\hat{a}} \psi^{c}\,_{i}\,^{\beta} W^{e {e_{1}}} W_{e {e_{1}} \alpha}\,^{i} - \frac{1}{4}(\Gamma^{{e_{1}}})^{\alpha}{}_{\beta} \delta^{d}\,_{\hat{a}} \psi^{c}\,_{i}\,^{\beta} W_{\hat{b}}\,^{e} W_{{e_{1}} e \alpha}\,^{i}+\frac{1}{8}\epsilon_{\hat{a} \hat{b}}\,^{e {e_{2}} {e_{3}}} (\Sigma^{d {e_{1}}})^{\alpha}{}_{\beta} \psi^{c}\,_{i}\,^{\beta} W_{e {e_{1}}} W_{{e_{2}} {e_{3}} \alpha}\,^{i}%
 - \frac{1}{8}\epsilon^{{e_{3}}}\,_{\hat{a} \hat{b}}\,^{{e_{1}} {e_{2}}} (\Sigma_{{e_{3}} e})^{\alpha}{}_{\beta} \psi^{c}\,_{i}\,^{\beta} W^{d e} W_{{e_{1}} {e_{2}} \alpha}\,^{i} - \frac{1}{4}(\Gamma^{d})^{\alpha}{}_{\beta} \psi^{c}\,_{i}\,^{\beta} W_{\hat{a}}\,^{e} W_{\hat{b} e \alpha}\,^{i} - \frac{1}{8}\epsilon^{d}\,_{\hat{a}}\,^{e {e_{1}} {e_{2}}} \psi^{c}\,_{i}\,^{\alpha} W_{e {e_{1}}} W_{\hat{b} {e_{2}} \alpha}\,^{i}+\frac{1}{4}\epsilon^{{e_{3}}}\,_{\hat{a}}\,^{e {e_{1}} {e_{2}}} (\Sigma_{{e_{3}}}{}^{\, d})^{\alpha}{}_{\beta} \psi^{c}\,_{i}\,^{\beta} W_{e {e_{1}}} W_{\hat{b} {e_{2}} \alpha}\,^{i}+\frac{1}{4}\epsilon^{d {e_{2}} {e_{3}} e {e_{1}}} (\Sigma_{{e_{2}} {e_{3}}})^{\alpha}{}_{\beta} \psi^{c}\,_{i}\,^{\beta} W_{\hat{a} e} W_{\hat{b} {e_{1}} \alpha}\,^{i}+\frac{1}{4}(\Gamma^{e})^{\alpha}{}_{\beta} \psi^{c}\,_{i}\,^{\beta} W_{\hat{a}}\,^{d} W_{\hat{b} e \alpha}\,^{i} - \frac{1}{4}(\Gamma^{e})^{\alpha}{}_{\beta} \psi^{c}\,_{i}\,^{\beta} W_{\hat{a} e} W_{\hat{b}}\,^{d}\,_{\alpha}\,^{i} - \frac{1}{4}\epsilon^{d {e_{2}} {e_{3}}}\,_{\hat{a} e} (\Sigma_{{e_{2}} {e_{3}}})^{\alpha}{}_{\beta} \psi^{c}\,_{i}\,^{\beta} W^{e {e_{1}}} W_{\hat{b} {e_{1}} \alpha}\,^{i} - \frac{1}{4}(\Gamma^{e})^{\alpha}{}_{\beta} \delta^{d}\,_{\hat{a}} \psi^{c}\,_{i}\,^{\beta} W_{e}\,^{{e_{1}}} W_{\hat{b} {e_{1}} \alpha}\,^{i}+\frac{1}{4}(\Gamma_{\hat{a}})^{\alpha}{}_{\beta} \psi^{c}\,_{i}\,^{\beta} W^{d e} W_{\hat{b} e \alpha}\,^{i}+\frac{1}{8}\epsilon^{d}\,_{\hat{a} \hat{b}}\,^{e {e_{2}}} \psi^{c}\,_{i}\,^{\alpha} W_{e}\,^{{e_{1}}} W_{{e_{2}} {e_{1}} \alpha}\,^{i} - \frac{1}{16}\epsilon_{\hat{b}}\,^{e {e_{1}} {e_{2}} {e_{3}}} \delta^{d}\,_{\hat{a}} \psi^{c}\,_{i}\,^{\alpha} W_{e {e_{1}}} W_{{e_{2}} {e_{3}} \alpha}\,^{i}+\frac{1}{16}\epsilon_{\hat{a} \hat{b}}\,^{e {e_{1}} {e_{2}}} \psi^{c}\,_{i}\,^{\alpha} W_{e {e_{1}}} W_{{e_{2}}}\,^{d}\,_{\alpha}\,^{i} - \frac{1}{8}\epsilon^{d}\,_{\hat{a}}\,^{e {e_{1}} {e_{2}}} \psi^{c}\,_{i}\,^{\alpha} W_{\hat{b} e} W_{{e_{1}} {e_{2}} \alpha}\,^{i} - \frac{1}{32}\epsilon^{{e_{2}} {e_{3}}}\,_{\hat{b}}\,^{e {e_{1}}} (\Sigma_{{e_{2}} {e_{3}}})^{\alpha}{}_{\beta} \delta^{d}\,_{\hat{a}} \psi^{c}\,_{i}\,^{\beta} W_{e {e_{1}}} X^{i}_{\alpha} - \frac{1}{2}{\rm i} (\Gamma^{d})^{\alpha}{}_{\beta} \phi^{c}\,_{i}\,^{\beta} W_{\hat{a} \hat{b} \alpha}\,^{i} - \frac{1}{4}{\rm i} \epsilon^{d e {e_{1}}}\,_{\hat{a} \hat{b}} (\Sigma_{e {e_{1}}})^{\alpha}{}_{\beta} \phi^{c}\,_{i}\,^{\beta} X^{i}_{\alpha} - \frac{1}{2}{\rm i} (\Gamma_{\hat{b}})^{\alpha}{}_{\beta} \delta^{d}\,_{\hat{a}} \phi^{c}\,_{i}\,^{\beta} X^{i}_{\alpha}-2W_{\hat{a} \hat{b}} f^{c d}-4\delta^{d}\,_{\hat{a}} W_{\hat{b} e} f^{c e}%
-4W_{\hat{a}}\,^{d} f^{c}\,_{\hat{b}}
\doublespacedmathend
\end{adjustwidth}

\subsubsection{$\nabla^{a} \nabla^{b} X^{i}_{\alpha}$}
\begin{adjustwidth}{0cm}{5cm}
\doublespacedmathbegin
\mathcal{D}^{a}{\nabla^{b}{X^{i}_{\alpha}}} - \frac{1}{3}{\rm i} (\Sigma_{c d})_{\alpha \beta} \psi^{a}\,_{j}\,^{\beta} \nabla^{b}{\Phi^{c d i j}} - \frac{1}{16}{\rm i} \psi^{a i}\,_{\alpha} \nabla^{b}{Y}+\frac{1}{4}{\rm i} \epsilon^{e {e_{1}}}\,_{{e_{2}}}\,^{c d} (\Sigma_{e {e_{1}}})_{\alpha \beta} \psi^{a i \beta} \nabla^{b}{\nabla^{{e_{2}}}{W_{c d}}} - \frac{1}{4}{\rm i} (\Gamma^{c})_{\alpha \beta} \psi^{a i \beta} \nabla^{b}{\nabla^{d}{W_{c d}}} - \frac{5}{32}{\rm i} \epsilon_{{e_{2}}}\,^{e {e_{1}} c d} (\Gamma^{{e_{2}}})_{\alpha \beta} \psi^{a i \beta} W_{c d} \nabla^{b}{W_{e {e_{1}}}} - \frac{1}{2}{\rm i} (\Sigma^{c}{}_{\,e})_{\alpha \beta} \psi^{a i \beta} W_{c d} \nabla^{b}{W^{e d}} - \frac{1}{48}{\rm i} \epsilon^{e {e_{1}} {e_{2}} c d} \Phi_{e {e_{1}}}\,^{i j} (\Sigma_{{e_{2}}}{}^{\, b})_{\alpha \beta} \psi^{a}\,_{j}\,^{\beta} W_{c d}+\frac{7}{24}{\rm i} \epsilon^{b e {e_{1}} {e_{2}} c} \Phi_{e}\,^{d i j} (\Sigma_{{e_{1}} {e_{2}}})_{\alpha \beta} \psi^{a}\,_{j}\,^{\beta} W_{c d} - \frac{1}{4}{\rm i} \Phi_{d c}\,^{i j} (\Gamma^{d})_{\alpha \beta} \psi^{a}\,_{j}\,^{\beta} W^{b c} - \frac{1}{3}{\rm i} \Phi^{b c i j} (\Gamma^{d})_{\alpha \beta} \psi^{a}\,_{j}\,^{\beta} W_{c d}+\frac{1}{128}{\rm i} \epsilon^{b e {e_{1}} c d} (\Sigma_{e {e_{1}}})_{\alpha \beta} \psi^{a i \beta} W_{c d} Y - \frac{1}{32}{\rm i} (\Gamma^{c})_{\alpha \beta} \psi^{a i \beta} W_{c}\,^{b} Y+\frac{7}{256}{\rm i} \epsilon_{{e_{2}}}\,^{e {e_{1}} c d} (\Gamma^{b})_{\alpha \beta} \psi^{a i \beta} W_{c d} \nabla^{{e_{2}}}{W_{e {e_{1}}}} - \frac{1}{2}{\rm i} (\Sigma^{c}{}_{\,e})_{\alpha \beta} \psi^{a i \beta} W_{c d} \nabla^{d}{W^{b e}}+\frac{3}{8}{\rm i} (\Sigma_{d e})_{\alpha \beta} \psi^{a i \beta} W^{b}\,_{c} \nabla^{c}{W^{d e}}+\frac{1}{2}{\rm i} (\Sigma^{c}{}_{\,e})_{\alpha \beta} \psi^{a i \beta} W_{c d} \nabla^{e}{W^{b d}}+\frac{3}{4}{\rm i} (\Sigma_{e d})_{\alpha \beta} \psi^{a i \beta} W^{b}\,_{c} \nabla^{e}{W^{d c}} - \frac{3}{32}{\rm i} \epsilon^{b}\,_{{e_{2}}}\,^{e {e_{1}} c} (\Gamma^{d})_{\alpha \beta} \psi^{a i \beta} W_{c d} \nabla^{{e_{2}}}{W_{e {e_{1}}}}%
+\frac{13}{128}{\rm i} \epsilon_{{e_{1}} {e_{2}}}\,^{d e c} (\Gamma^{{e_{1}}})_{\alpha \beta} \psi^{a i \beta} W_{c}\,^{b} \nabla^{{e_{2}}}{W_{d e}} - \frac{1}{4}{\rm i} (\Sigma_{c d})_{\alpha \beta} \psi^{a i \beta} W^{c d} \nabla_{e}{W^{b e}}+\frac{1}{16}{\rm i} \epsilon^{b}\,_{{e_{2}}}\,^{e c d} (\Gamma^{{e_{2}}})_{\alpha \beta} \psi^{a i \beta} W_{c d} \nabla^{{e_{1}}}{W_{e {e_{1}}}}+\frac{1}{16}{\rm i} \psi^{a i}\,_{\alpha} W^{b}\,_{c} \nabla_{d}{W^{d c}} - \frac{3}{8}{\rm i} (\Sigma_{c d})_{\alpha \beta} \psi^{a i \beta} W^{b c} \nabla_{e}{W^{d e}}+\frac{3}{128}{\rm i} \epsilon^{b {e_{2}} {e_{3}} e {e_{1}}} (\Sigma_{{e_{2}} {e_{3}}})_{\alpha \beta} \psi^{a i \beta} W^{c d} W_{e {e_{1}}} W_{c d}+\frac{5}{64}{\rm i} (\Gamma^{e})_{\alpha \beta} \psi^{a i \beta} W^{c d} W_{e}\,^{b} W_{c d} - \frac{1}{16}{\rm i} \epsilon^{b e {e_{1}} {e_{2}} {e_{3}}} (\Sigma_{c d})_{\alpha \beta} \psi^{a i \beta} W^{c d} W_{e {e_{1}}} W_{{e_{2}} {e_{3}}} - \frac{5}{16}{\rm i} (\Gamma^{e})_{\alpha \beta} \psi^{a i \beta} W^{c d} W_{e c} W^{b}\,_{d} - \frac{5}{32}{\rm i} \epsilon^{c d e {e_{1}} {e_{2}}} (\Sigma^{b {e_{3}}})_{\alpha \beta} \psi^{a i \beta} W_{c d} W_{e {e_{1}}} W_{{e_{2}} {e_{3}}}+\frac{5}{32}{\rm i} \epsilon^{{e_{3}} d e {e_{1}} {e_{2}}} (\Sigma_{{e_{3}} c})_{\alpha \beta} \psi^{a i \beta} W^{b c} W_{d e} W_{{e_{1}} {e_{2}}}+\frac{1}{256}{\rm i} \epsilon^{b e {e_{1}} c d} (\Gamma_{{e_{2}}})_{\alpha \beta} \psi^{a i \beta} W_{c d} \nabla^{{e_{2}}}{W_{e {e_{1}}}} - \frac{3}{64}{\rm i} \epsilon^{b}\,_{{e_{1}} {e_{2}} e}\,^{c} (\Gamma^{{e_{1}}})_{\alpha \beta} \psi^{a i \beta} W_{c d} \nabla^{{e_{2}}}{W^{e d}}+\frac{1}{24}{\rm i} \epsilon^{b e {e_{1}} {e_{2}} c} \Phi_{e {e_{1}}}\,^{i j} (\Sigma_{{e_{2}}}{}^{\, d})_{\alpha \beta} \psi^{a}\,_{j}\,^{\beta} W_{c d}+\frac{1}{48}{\rm i} \epsilon^{e {e_{1}} {e_{2}} c d} \Phi_{e}\,^{b i j} (\Sigma_{{e_{1}} {e_{2}}})_{\alpha \beta} \psi^{a}\,_{j}\,^{\beta} W_{c d}+\frac{1}{128}{\rm i} \epsilon^{b}\,_{{e_{2}}}\,^{e c d} (\Gamma^{{e_{1}}})_{\alpha \beta} \psi^{a i \beta} W_{c d} \nabla^{{e_{2}}}{W_{e {e_{1}}}} - \frac{3}{128}{\rm i} \epsilon^{b}\,_{{e_{2}}}\,^{e {e_{1}} c} (\Gamma^{{e_{2}}})_{\alpha \beta} \psi^{a i \beta} W_{c d} \nabla^{d}{W_{e {e_{1}}}}+\frac{5}{16}{\rm i} (\Gamma^{d})_{\alpha \beta} \psi^{a i \beta} W^{b c} W_{d}\,^{e} W_{e c}+\frac{1}{16}{\rm i} \epsilon_{{e_{1}} {e_{2}} e}\,^{c d} (\Gamma^{{e_{1}}})_{\alpha \beta} \psi^{a i \beta} W_{c d} \nabla^{{e_{2}}}{W^{b e}} - \frac{1}{8}\epsilon^{b e {e_{1}} c d} (\Sigma_{e {e_{1}}})_{\alpha \rho} \psi^{a}\,_{j}\,^{\rho} X^{i \beta} W_{c d \beta}\,^{j}%
+\frac{1}{4}(\Gamma^{c})_{\alpha \rho} \psi^{a}\,_{j}\,^{\rho} X^{i \beta} W_{c}\,^{b}\,_{\beta}\,^{j} - \frac{1}{16}\epsilon^{b e {e_{1}} c d} (\Sigma_{e {e_{1}}})_{\alpha}{}^{\rho} \psi^{a}\,_{j}\,^{\beta} X^{i}_{\rho} W_{c d \beta}\,^{j}+\frac{1}{16}(\Gamma^{b})_{\alpha}{}^{\rho} \psi^{a}\,_{j}\,^{\beta} X^{i}_{\beta} X^{j}_{\rho}+\frac{1}{16}(\Gamma^{b})^{\beta \rho} \psi^{a}\,_{j \alpha} X^{i}_{\beta} X^{j}_{\rho} - \frac{1}{16}(\Gamma^{b})_{\alpha \rho} \psi^{a}\,_{j}\,^{\rho} X^{i \beta} X^{j}_{\beta}+\frac{1}{4}(\Gamma^{b})^{\beta}{}_{\rho} \psi^{a}\,_{j}\,^{\rho} X^{i}_{\beta} X^{j}_{\alpha}+\frac{3}{16}(\Gamma^{b})^{\beta}{}_{\rho} \psi^{a}\,_{j}\,^{\rho} X^{i}_{\alpha} X^{j}_{\beta}+\frac{3}{16}(\Gamma^{b})^{\beta}{}_{\rho} \psi^{a i \rho} X_{j \alpha} X^{j}_{\beta}+\frac{1}{12}{\rm i} \epsilon^{b e {e_{2}} c d} \Phi_{e}\,^{{e_{1}} i j} (\Sigma_{{e_{2}} {e_{1}}})_{\alpha \beta} \psi^{a}\,_{j}\,^{\beta} W_{c d} - \frac{1}{24}{\rm i} \epsilon^{d e {e_{1}} {e_{2}} c} \Phi_{d e}\,^{i j} (\Sigma_{{e_{1}} {e_{2}}})_{\alpha \beta} \psi^{a}\,_{j}\,^{\beta} W_{c}\,^{b} - \frac{1}{2}(\Sigma_{c d})_{\alpha \beta} \phi^{a i \beta} \nabla^{b}{W^{c d}}+\frac{1}{6}\epsilon^{b c d e {e_{1}}} \Phi_{c d}\,^{i j} (\Sigma_{e {e_{1}}})_{\alpha \beta} \phi^{a}\,_{j}\,^{\beta} - \frac{1}{3}\Phi_{c}\,^{b i j} (\Gamma^{c})_{\alpha \beta} \phi^{a}\,_{j}\,^{\beta}+\frac{1}{16}(\Gamma^{b})_{\alpha \beta} \phi^{a i \beta} Y - \frac{1}{4}\phi^{a i}\,_{\alpha} \nabla_{c}{W^{b c}} - \frac{1}{4}\epsilon^{b}\,_{e {e_{1}}}\,^{c d} (\Gamma^{e})_{\alpha \beta} \phi^{a i \beta} \nabla^{{e_{1}}}{W_{c d}} - \frac{1}{2}(\Sigma^{b}{}_{\,c})_{\alpha \beta} \phi^{a i \beta} \nabla_{d}{W^{c d}}-(\Sigma_{d c})_{\alpha \beta} \phi^{a i \beta} \nabla^{d}{W^{b c}} - \frac{1}{16}\epsilon^{b c d e {e_{1}}} \phi^{a i}\,_{\alpha} W_{c d} W_{e {e_{1}}}+\frac{1}{16}\epsilon^{{e_{2}} c d e {e_{1}}} (\Sigma_{{e_{2}}}{}^{\, b})_{\alpha \beta} \phi^{a i \beta} W_{c d} W_{e {e_{1}}}%
+\frac{1}{4}(\Gamma^{d})_{\alpha \beta} \phi^{a i \beta} W^{b c} W_{d c} - \frac{1}{8}\epsilon^{b {e_{2}} c d e} (\Sigma_{{e_{2}}}{}^{\, {e_{1}}})_{\alpha \beta} \phi^{a i \beta} W_{c d} W_{e {e_{1}}} - \frac{1}{16}\epsilon^{{e_{1}} {e_{2}} c d e} (\Sigma_{{e_{1}} {e_{2}}})_{\alpha \beta} \phi^{a i \beta} W_{c d} W_{e}\,^{b}-3X^{i}_{\alpha} f^{a b}+2(\Sigma^{b}{}_{\, c})_{\alpha}{}^{\beta} X^{i}_{\beta} f^{a c}
\doublespacedmathend
\end{adjustwidth}

\subsubsection{$\nabla^{a} \nabla^{b} W_{\alpha \beta \gamma}{}^{i}$}
\begin{adjustwidth}{1em}{5cm}
\doublespacedmathbegin
\mathcal{D}^{a}{\nabla^{b}{W_{\alpha \beta \gamma}\,^{i}}}-{\rm i} \psi^{a i \rho} \nabla^{b}{C_{\alpha \beta \gamma \rho}} - \frac{5}{16}{\rm i} \psi^{a i \rho} W_{\gamma \rho} \nabla^{b}{W_{\alpha \beta}} - \frac{3}{16}{\rm i} \psi^{a i \rho} W_{\alpha \beta} \nabla^{b}{W_{\gamma \rho}} - \frac{5}{16}{\rm i} \psi^{a i \rho} W_{\beta \rho} \nabla^{b}{W_{\alpha \gamma}} - \frac{3}{16}{\rm i} \psi^{a i \rho} W_{\alpha \gamma} \nabla^{b}{W_{\beta \rho}} - \frac{3}{16}{\rm i} \psi^{a i \rho} W_{\beta \gamma} \nabla^{b}{W_{\alpha \rho}} - \frac{5}{16}{\rm i} \psi^{a i \rho} W_{\alpha \rho} \nabla^{b}{W_{\beta \gamma}}+\frac{1}{4}{\rm i} (\Gamma_{c})_{\alpha \rho} \psi^{a i \rho} \nabla^{b}{\nabla^{c}{W_{\beta \gamma}}}+\frac{1}{4}{\rm i} (\Gamma_{c})_{\beta \rho} \psi^{a i \rho} \nabla^{b}{\nabla^{c}{W_{\alpha \gamma}}}+\frac{1}{4}{\rm i} (\Gamma_{c})_{\gamma \rho} \psi^{a i \rho} \nabla^{b}{\nabla^{c}{W_{\alpha \beta}}}+\frac{1}{8}{\rm i} (\Gamma_{c})_{\beta}{}^{\rho} \psi^{a i}\,_{\alpha} \nabla^{b}{\nabla^{c}{W_{\gamma \rho}}}+\frac{1}{8}{\rm i} (\Gamma_{c})_{\gamma}{}^{\rho} \psi^{a i}\,_{\alpha} \nabla^{b}{\nabla^{c}{W_{\beta \rho}}}+\frac{1}{8}{\rm i} (\Gamma_{c})_{\gamma}{}^{\rho} \psi^{a i}\,_{\beta} \nabla^{b}{\nabla^{c}{W_{\alpha \rho}}}+\frac{1}{8}{\rm i} (\Gamma_{c})_{\alpha}{}^{\rho} \psi^{a i}\,_{\beta} \nabla^{b}{\nabla^{c}{W_{\gamma \rho}}}+\frac{1}{8}{\rm i} (\Gamma_{c})_{\alpha}{}^{\rho} \psi^{a i}\,_{\gamma} \nabla^{b}{\nabla^{c}{W_{\beta \rho}}}+\frac{1}{8}{\rm i} (\Gamma_{c})_{\beta}{}^{\rho} \psi^{a i}\,_{\gamma} \nabla^{b}{\nabla^{c}{W_{\alpha \rho}}} - \frac{1}{3}{\rm i} \psi^{a}\,_{j \alpha} \nabla^{b}{\Phi_{\beta \gamma}\,^{i j}} - \frac{1}{3}{\rm i} \psi^{a}\,_{j \beta} \nabla^{b}{\Phi_{\alpha \gamma}\,^{i j}}%
 - \frac{1}{3}{\rm i} \psi^{a}\,_{j \gamma} \nabla^{b}{\Phi_{\alpha \beta}\,^{i j}} - \frac{3}{4}{\rm i} (\Gamma^{b})^{\lambda}{}_{\tau} \psi^{a i \tau} C_{\alpha \beta \gamma}\,^{\rho} W_{\lambda \rho}+\frac{3}{16}{\rm i} (\Gamma^{b})^{\lambda}{}_{\tau} \psi^{a i \tau} W_{\alpha \beta} W_{\gamma}\,^{\rho} W_{\lambda \rho}+\frac{3}{16}{\rm i} (\Gamma^{b})^{\lambda}{}_{\tau} \psi^{a i \tau} W_{\alpha \gamma} W_{\beta}\,^{\rho} W_{\lambda \rho}+\frac{3}{16}{\rm i} (\Gamma^{b})^{\lambda}{}_{\tau} \psi^{a i \tau} W_{\alpha}\,^{\rho} W_{\beta \gamma} W_{\lambda \rho}+\frac{3}{16}{\rm i} (\Gamma^{b})^{\rho}{}_{\tau} (\Gamma_{c})_{\alpha}{}^{\lambda} \psi^{a i \tau} W_{\rho \lambda} \nabla^{c}{W_{\beta \gamma}}+\frac{3}{16}{\rm i} (\Gamma^{b})^{\rho}{}_{\tau} (\Gamma_{c})_{\beta}{}^{\lambda} \psi^{a i \tau} W_{\rho \lambda} \nabla^{c}{W_{\alpha \gamma}}+\frac{3}{16}{\rm i} (\Gamma^{b})^{\rho}{}_{\tau} (\Gamma_{c})_{\gamma}{}^{\lambda} \psi^{a i \tau} W_{\rho \lambda} \nabla^{c}{W_{\alpha \beta}} - \frac{3}{32}{\rm i} (\Gamma^{b})^{\rho}{}_{\tau} (\Gamma_{c})_{\beta}{}^{\lambda} \psi^{a i \tau} W_{\alpha \rho} \nabla^{c}{W_{\gamma \lambda}} - \frac{3}{32}{\rm i} (\Gamma^{b})^{\rho}{}_{\tau} (\Gamma_{c})_{\gamma}{}^{\lambda} \psi^{a i \tau} W_{\alpha \rho} \nabla^{c}{W_{\beta \lambda}} - \frac{3}{32}{\rm i} (\Gamma^{b})^{\rho}{}_{\tau} (\Gamma_{c})_{\gamma}{}^{\lambda} \psi^{a i \tau} W_{\beta \rho} \nabla^{c}{W_{\alpha \lambda}} - \frac{3}{32}{\rm i} (\Gamma^{b})^{\rho}{}_{\tau} (\Gamma_{c})_{\alpha}{}^{\lambda} \psi^{a i \tau} W_{\beta \rho} \nabla^{c}{W_{\gamma \lambda}} - \frac{3}{32}{\rm i} (\Gamma^{b})^{\rho}{}_{\tau} (\Gamma_{c})_{\alpha}{}^{\lambda} \psi^{a i \tau} W_{\gamma \rho} \nabla^{c}{W_{\beta \lambda}} - \frac{3}{32}{\rm i} (\Gamma^{b})^{\rho}{}_{\tau} (\Gamma_{c})_{\beta}{}^{\lambda} \psi^{a i \tau} W_{\gamma \rho} \nabla^{c}{W_{\alpha \lambda}}+\frac{1}{4}{\rm i} \Phi_{\beta \gamma}\,^{i j} (\Gamma^{b})^{\rho}{}_{\lambda} \psi^{a}\,_{j}\,^{\lambda} W_{\alpha \rho}+\frac{1}{4}{\rm i} \Phi_{\alpha \gamma}\,^{i j} (\Gamma^{b})^{\rho}{}_{\lambda} \psi^{a}\,_{j}\,^{\lambda} W_{\beta \rho}+\frac{1}{4}{\rm i} \Phi_{\alpha \beta}\,^{i j} (\Gamma^{b})^{\rho}{}_{\lambda} \psi^{a}\,_{j}\,^{\lambda} W_{\gamma \rho}+\frac{1}{4}{\rm i} (\Gamma^{b})^{\rho \lambda} \psi^{a i \tau} C_{\alpha \beta \gamma \rho} W_{\lambda \tau} - \frac{7}{16}{\rm i} (\Gamma^{b})^{\rho \lambda} \psi^{a i \tau} W_{\alpha \beta} W_{\gamma \rho} W_{\lambda \tau} - \frac{7}{16}{\rm i} (\Gamma^{b})^{\rho \lambda} \psi^{a i \tau} W_{\alpha \gamma} W_{\beta \rho} W_{\lambda \tau}%
 - \frac{7}{16}{\rm i} (\Gamma^{b})^{\rho \lambda} \psi^{a i \tau} W_{\alpha \rho} W_{\beta \gamma} W_{\lambda \tau}+\frac{1}{8}{\rm i} (\Sigma^{b}{}_{\, c})_{\alpha}{}^{\rho} \psi^{a i \lambda} W_{\rho \lambda} \nabla^{c}{W_{\beta \gamma}}+\frac{1}{8}{\rm i} (\Sigma^{b}{}_{\, c})_{\beta}{}^{\rho} \psi^{a i \lambda} W_{\rho \lambda} \nabla^{c}{W_{\alpha \gamma}}+\frac{1}{8}{\rm i} (\Sigma^{b}{}_{\, c})_{\gamma}{}^{\rho} \psi^{a i \lambda} W_{\rho \lambda} \nabla^{c}{W_{\alpha \beta}} - \frac{1}{32}{\rm i} (\Gamma^{b})_{\alpha}{}^{\rho} (\Gamma_{c})_{\beta}{}^{\tau} \psi^{a i \lambda} W_{\rho \lambda} \nabla^{c}{W_{\gamma \tau}} - \frac{1}{32}{\rm i} (\Gamma^{b})_{\alpha}{}^{\rho} (\Gamma_{c})_{\gamma}{}^{\tau} \psi^{a i \lambda} W_{\rho \lambda} \nabla^{c}{W_{\beta \tau}} - \frac{1}{32}{\rm i} (\Gamma^{b})_{\beta}{}^{\rho} (\Gamma_{c})_{\gamma}{}^{\tau} \psi^{a i \lambda} W_{\rho \lambda} \nabla^{c}{W_{\alpha \tau}} - \frac{1}{32}{\rm i} (\Gamma^{b})_{\beta}{}^{\rho} (\Gamma_{c})_{\alpha}{}^{\tau} \psi^{a i \lambda} W_{\rho \lambda} \nabla^{c}{W_{\gamma \tau}} - \frac{1}{32}{\rm i} (\Gamma^{b})_{\gamma}{}^{\rho} (\Gamma_{c})_{\alpha}{}^{\tau} \psi^{a i \lambda} W_{\rho \lambda} \nabla^{c}{W_{\beta \tau}} - \frac{1}{32}{\rm i} (\Gamma^{b})_{\gamma}{}^{\rho} (\Gamma_{c})_{\beta}{}^{\tau} \psi^{a i \lambda} W_{\rho \lambda} \nabla^{c}{W_{\alpha \tau}}+\frac{1}{12}{\rm i} \Phi_{\beta \gamma}\,^{i j} (\Gamma^{b})_{\alpha}{}^{\rho} \psi^{a}\,_{j}\,^{\lambda} W_{\rho \lambda}+\frac{1}{12}{\rm i} \Phi_{\alpha \gamma}\,^{i j} (\Gamma^{b})_{\beta}{}^{\rho} \psi^{a}\,_{j}\,^{\lambda} W_{\rho \lambda}+\frac{1}{12}{\rm i} \Phi_{\alpha \beta}\,^{i j} (\Gamma^{b})_{\gamma}{}^{\rho} \psi^{a}\,_{j}\,^{\lambda} W_{\rho \lambda} - \frac{1}{2}(\Gamma^{b})^{\rho}{}_{\tau} \psi^{a}\,_{j}\,^{\tau} W_{\alpha \rho}\,^{\lambda j} W_{\beta \gamma \lambda}\,^{i} - \frac{1}{2}(\Gamma^{b})^{\lambda}{}_{\tau} \psi^{a}\,_{j}\,^{\tau} W_{\alpha \beta}\,^{\rho i} W_{\gamma \lambda \rho}\,^{j} - \frac{1}{2}(\Gamma^{b})^{\lambda}{}_{\tau} \psi^{a}\,_{j}\,^{\tau} W_{\alpha \gamma}\,^{\rho i} W_{\beta \lambda \rho}\,^{j} - \frac{1}{8}(\Gamma^{b})_{\alpha}{}^{\lambda} \psi^{a}\,_{j}\,^{\tau} W_{\beta \gamma}\,^{\rho i} W_{\lambda \rho \tau}\,^{j}+\frac{1}{8}(\Gamma^{b})^{\rho \tau} \psi^{a}\,_{j}\,^{\lambda} W_{\alpha \rho \lambda}\,^{j} W_{\beta \gamma \tau}\,^{i} - \frac{1}{8}(\Gamma^{b})_{\gamma}{}^{\lambda} \psi^{a}\,_{j}\,^{\tau} W_{\alpha \beta}\,^{\rho i} W_{\lambda \rho \tau}\,^{j}+\frac{1}{8}(\Gamma^{b})^{\rho \lambda} \psi^{a}\,_{j}\,^{\tau} W_{\alpha \beta \rho}\,^{i} W_{\gamma \lambda \tau}\,^{j}%
 - \frac{1}{8}(\Gamma^{b})_{\beta}{}^{\lambda} \psi^{a}\,_{j}\,^{\tau} W_{\alpha \gamma}\,^{\rho i} W_{\lambda \rho \tau}\,^{j}+\frac{1}{8}(\Gamma^{b})^{\rho \lambda} \psi^{a}\,_{j}\,^{\tau} W_{\alpha \gamma \rho}\,^{i} W_{\beta \lambda \tau}\,^{j} - \frac{1}{16}(\Gamma^{b})_{\alpha}{}^{\lambda} \psi^{a}\,_{j}\,^{\rho} X^{j}_{\lambda} W_{\beta \gamma \rho}\,^{i}+\frac{1}{16}(\Gamma^{b})^{\lambda \rho} \psi^{a}\,_{j \alpha} X^{j}_{\lambda} W_{\beta \gamma \rho}\,^{i} - \frac{1}{16}(\Gamma^{b})_{\gamma}{}^{\lambda} \psi^{a}\,_{j}\,^{\rho} X^{j}_{\lambda} W_{\alpha \beta \rho}\,^{i}+\frac{1}{16}(\Gamma^{b})^{\lambda \rho} \psi^{a}\,_{j \gamma} X^{j}_{\lambda} W_{\alpha \beta \rho}\,^{i} - \frac{1}{16}(\Gamma^{b})_{\beta}{}^{\lambda} \psi^{a}\,_{j}\,^{\rho} X^{j}_{\lambda} W_{\alpha \gamma \rho}\,^{i}+\frac{1}{16}(\Gamma^{b})^{\lambda \rho} \psi^{a}\,_{j \beta} X^{j}_{\lambda} W_{\alpha \gamma \rho}\,^{i} - \frac{1}{16}(\Gamma^{b})_{\alpha \lambda} \psi^{a}\,_{j}\,^{\lambda} X^{j \rho} W_{\beta \gamma \rho}\,^{i} - \frac{1}{16}(\Gamma^{b})^{\rho}{}_{\lambda} \psi^{a}\,_{j}\,^{\lambda} X^{j}_{\alpha} W_{\beta \gamma \rho}\,^{i} - \frac{1}{16}(\Gamma^{b})_{\gamma \lambda} \psi^{a}\,_{j}\,^{\lambda} X^{j \rho} W_{\alpha \beta \rho}\,^{i} - \frac{1}{16}(\Gamma^{b})^{\rho}{}_{\lambda} \psi^{a}\,_{j}\,^{\lambda} X^{j}_{\gamma} W_{\alpha \beta \rho}\,^{i} - \frac{1}{16}(\Gamma^{b})_{\beta \lambda} \psi^{a}\,_{j}\,^{\lambda} X^{j \rho} W_{\alpha \gamma \rho}\,^{i} - \frac{1}{16}(\Gamma^{b})^{\rho}{}_{\lambda} \psi^{a}\,_{j}\,^{\lambda} X^{j}_{\beta} W_{\alpha \gamma \rho}\,^{i} - \frac{3}{16}(\Gamma^{b})^{\rho}{}_{\lambda} \psi^{a}\,_{j}\,^{\lambda} X^{j}_{\rho} W_{\alpha \beta \gamma}\,^{i}+\frac{3}{16}(\Gamma^{b})^{\rho}{}_{\lambda} \psi^{a i \lambda} X_{j \rho} W_{\alpha \beta \gamma}\,^{j}+\frac{3}{16}(\Gamma^{b})^{\rho}{}_{\lambda} \psi^{a}\,_{j}\,^{\lambda} X^{i}_{\rho} W_{\alpha \beta \gamma}\,^{j} - \frac{1}{6}{\rm i} \Phi^{\rho}\,_{\lambda}\,^{i j} (\Gamma^{b})_{\alpha \rho} \psi^{a}\,_{j}\,^{\lambda} W_{\beta \gamma} - \frac{1}{6}{\rm i} \Phi^{\rho}\,_{\lambda}\,^{i j} (\Gamma^{b})_{\beta \rho} \psi^{a}\,_{j}\,^{\lambda} W_{\alpha \gamma} - \frac{1}{6}{\rm i} \Phi^{\rho}\,_{\lambda}\,^{i j} (\Gamma^{b})_{\gamma \rho} \psi^{a}\,_{j}\,^{\lambda} W_{\alpha \beta}%
 - \frac{1}{3}{\rm i} \Phi_{\alpha}\,^{\rho i j} (\Gamma^{b})_{\rho \lambda} \psi^{a}\,_{j}\,^{\lambda} W_{\beta \gamma} - \frac{1}{3}{\rm i} \Phi_{\beta}\,^{\rho i j} (\Gamma^{b})_{\rho \lambda} \psi^{a}\,_{j}\,^{\lambda} W_{\alpha \gamma} - \frac{1}{3}{\rm i} \Phi_{\gamma}\,^{\rho i j} (\Gamma^{b})_{\rho \lambda} \psi^{a}\,_{j}\,^{\lambda} W_{\alpha \beta}+\frac{1}{32}{\rm i} (\Gamma^{b})_{\alpha}{}^{\rho} (\Gamma_{c})^{\lambda}{}_{\tau} \psi^{a i \tau} W_{\beta \gamma} \nabla^{c}{W_{\rho \lambda}}+\frac{1}{32}{\rm i} (\Gamma^{b})_{\beta}{}^{\rho} (\Gamma_{c})^{\lambda}{}_{\tau} \psi^{a i \tau} W_{\alpha \gamma} \nabla^{c}{W_{\rho \lambda}}+\frac{1}{32}{\rm i} (\Gamma^{b})_{\gamma}{}^{\rho} (\Gamma_{c})^{\lambda}{}_{\tau} \psi^{a i \tau} W_{\alpha \beta} \nabla^{c}{W_{\rho \lambda}}+\frac{3}{16}{\rm i} (\Sigma^{b}{}_{\, c})^{\rho}{}_{\lambda} \psi^{a i \lambda} W_{\beta \gamma} \nabla^{c}{W_{\alpha \rho}}+\frac{3}{16}{\rm i} (\Sigma^{b}{}_{\, c})^{\rho}{}_{\lambda} \psi^{a i \lambda} W_{\alpha \gamma} \nabla^{c}{W_{\beta \rho}}+\frac{3}{16}{\rm i} (\Sigma^{b}{}_{\, c})^{\rho}{}_{\lambda} \psi^{a i \lambda} W_{\alpha \beta} \nabla^{c}{W_{\gamma \rho}}+\frac{5}{32}{\rm i} (\Gamma^{b})^{\rho}{}_{\tau} (\Gamma_{c})_{\alpha}{}^{\lambda} \psi^{a i \tau} W_{\beta \gamma} \nabla^{c}{W_{\rho \lambda}}+\frac{5}{32}{\rm i} (\Gamma^{b})^{\rho}{}_{\tau} (\Gamma_{c})_{\beta}{}^{\lambda} \psi^{a i \tau} W_{\alpha \gamma} \nabla^{c}{W_{\rho \lambda}}+\frac{5}{32}{\rm i} (\Gamma^{b})^{\rho}{}_{\tau} (\Gamma_{c})_{\gamma}{}^{\lambda} \psi^{a i \tau} W_{\alpha \beta} \nabla^{c}{W_{\rho \lambda}} - \frac{1}{16}{\rm i} (\Sigma^{b}{}_{\, c})_{\alpha}{}^{\rho} \psi^{a i \lambda} W_{\beta \gamma} \nabla^{c}{W_{\rho \lambda}} - \frac{1}{16}{\rm i} (\Sigma^{b}{}_{\, c})_{\beta}{}^{\rho} \psi^{a i \lambda} W_{\alpha \gamma} \nabla^{c}{W_{\rho \lambda}} - \frac{1}{16}{\rm i} (\Sigma^{b}{}_{\, c})_{\gamma}{}^{\rho} \psi^{a i \lambda} W_{\alpha \beta} \nabla^{c}{W_{\rho \lambda}}+\frac{3}{16}{\rm i} (\Sigma^{b}{}_{\, c})^{\rho \lambda} \psi^{a i}\,_{\alpha} W_{\beta \gamma} \nabla^{c}{W_{\rho \lambda}}+\frac{3}{16}{\rm i} (\Sigma^{b}{}_{\, c})^{\rho \lambda} \psi^{a i}\,_{\beta} W_{\alpha \gamma} \nabla^{c}{W_{\rho \lambda}}+\frac{3}{16}{\rm i} (\Sigma^{b}{}_{\, c})^{\rho \lambda} \psi^{a i}\,_{\gamma} W_{\alpha \beta} \nabla^{c}{W_{\rho \lambda}} - \frac{1}{8}{\rm i} (\Gamma^{b})_{\alpha}{}^{\rho} \psi^{a i \tau} W_{\beta \gamma} W_{\rho}\,^{\lambda} W_{\lambda \tau} - \frac{1}{8}{\rm i} (\Gamma^{b})_{\beta}{}^{\rho} \psi^{a i \tau} W_{\alpha \gamma} W_{\rho}\,^{\lambda} W_{\lambda \tau}%
 - \frac{1}{8}{\rm i} (\Gamma^{b})_{\gamma}{}^{\rho} \psi^{a i \tau} W_{\alpha \beta} W_{\rho}\,^{\lambda} W_{\lambda \tau} - \frac{3}{32}{\rm i} (\Gamma^{b})_{\alpha \tau} \psi^{a i \tau} W_{\beta \gamma} W^{\rho \lambda} W_{\rho \lambda} - \frac{3}{32}{\rm i} (\Gamma^{b})_{\beta \tau} \psi^{a i \tau} W_{\alpha \gamma} W^{\rho \lambda} W_{\rho \lambda} - \frac{3}{32}{\rm i} (\Gamma^{b})_{\gamma \tau} \psi^{a i \tau} W_{\alpha \beta} W^{\rho \lambda} W_{\rho \lambda}+\frac{3}{16}{\rm i} (\Gamma^{b})^{\rho \tau} \psi^{a i}\,_{\alpha} W_{\beta \gamma} W_{\rho}\,^{\lambda} W_{\tau \lambda}+\frac{3}{16}{\rm i} (\Gamma^{b})^{\rho \tau} \psi^{a i}\,_{\beta} W_{\alpha \gamma} W_{\rho}\,^{\lambda} W_{\tau \lambda}+\frac{3}{16}{\rm i} (\Gamma^{b})^{\rho \tau} \psi^{a i}\,_{\gamma} W_{\alpha \beta} W_{\rho}\,^{\lambda} W_{\tau \lambda}+\frac{5}{4}\phi^{a i}\,_{\alpha} \nabla^{b}{W_{\beta \gamma}}+\frac{5}{4}\phi^{a i}\,_{\beta} \nabla^{b}{W_{\alpha \gamma}}+\frac{5}{4}\phi^{a i}\,_{\gamma} \nabla^{b}{W_{\alpha \beta}}+(\Gamma^{b})^{\rho}{}_{\lambda} \phi^{a i \lambda} C_{\alpha \beta \gamma \rho}+\frac{1}{2}(\Gamma^{b})^{\rho}{}_{\lambda} \phi^{a i \lambda} W_{\alpha \beta} W_{\gamma \rho}+\frac{1}{2}(\Gamma^{b})^{\rho}{}_{\lambda} \phi^{a i \lambda} W_{\alpha \gamma} W_{\beta \rho}+\frac{1}{2}(\Gamma^{b})^{\rho}{}_{\lambda} \phi^{a i \lambda} W_{\alpha \rho} W_{\beta \gamma}+\frac{1}{2}(\Sigma^{b}{}_{\,c})_{\alpha \rho} \phi^{a i \rho} \nabla^{c}{W_{\beta \gamma}}+\frac{1}{2}(\Sigma^{b}{}_{\,c})_{\beta \rho} \phi^{a i \rho} \nabla^{c}{W_{\alpha \gamma}}+\frac{1}{2}(\Sigma^{b}{}_{\,c})_{\gamma \rho} \phi^{a i \rho} \nabla^{c}{W_{\alpha \beta}} - \frac{1}{8}(\Gamma^{b})_{\alpha \lambda} (\Gamma_{c})_{\beta}{}^{\rho} \phi^{a i \lambda} \nabla^{c}{W_{\gamma \rho}} - \frac{1}{8}(\Gamma^{b})_{\alpha \lambda} (\Gamma_{c})_{\gamma}{}^{\rho} \phi^{a i \lambda} \nabla^{c}{W_{\beta \rho}} - \frac{1}{8}(\Gamma^{b})_{\beta \lambda} (\Gamma_{c})_{\gamma}{}^{\rho} \phi^{a i \lambda} \nabla^{c}{W_{\alpha \rho}}%
 - \frac{1}{8}(\Gamma^{b})_{\beta \lambda} (\Gamma_{c})_{\alpha}{}^{\rho} \phi^{a i \lambda} \nabla^{c}{W_{\gamma \rho}} - \frac{1}{8}(\Gamma^{b})_{\gamma \lambda} (\Gamma_{c})_{\alpha}{}^{\rho} \phi^{a i \lambda} \nabla^{c}{W_{\beta \rho}} - \frac{1}{8}(\Gamma^{b})_{\gamma \lambda} (\Gamma_{c})_{\beta}{}^{\rho} \phi^{a i \lambda} \nabla^{c}{W_{\alpha \rho}}+\frac{1}{3}\Phi_{\beta \gamma}\,^{i j} (\Gamma^{b})_{\alpha \rho} \phi^{a}\,_{j}\,^{\rho}+\frac{1}{3}\Phi_{\alpha \gamma}\,^{i j} (\Gamma^{b})_{\beta \rho} \phi^{a}\,_{j}\,^{\rho}+\frac{1}{3}\Phi_{\alpha \beta}\,^{i j} (\Gamma^{b})_{\gamma \rho} \phi^{a}\,_{j}\,^{\rho}+\frac{1}{4}(\Gamma^{b})_{\alpha}{}^{\rho} \phi^{a i \lambda} W_{\beta \gamma} W_{\rho \lambda}+\frac{1}{4}(\Gamma^{b})_{\beta}{}^{\rho} \phi^{a i \lambda} W_{\alpha \gamma} W_{\rho \lambda}+\frac{1}{4}(\Gamma^{b})_{\gamma}{}^{\rho} \phi^{a i \lambda} W_{\alpha \beta} W_{\rho \lambda}-3W_{\alpha \beta \gamma}\,^{i} f^{a b}+2(\Sigma^{b}{}_{\, c})_{\alpha}{}^{\rho} W_{\beta \gamma \rho}\,^{i} f^{a c}+2(\Sigma^{b}{}_{\, c})_{\gamma}{}^{\rho} W_{\alpha \beta \rho}\,^{i} f^{a c}+2(\Sigma^{b}{}_{\, c})_{\beta}{}^{\rho} W_{\alpha \gamma \rho}\,^{i} f^{a c}
\doublespacedmathend
\end{adjustwidth}
 
\subsubsection{$\nabla^{a} \nabla^{b} Y$}
\begin{adjustwidth}{0cm}{5cm}
\doublespacedmathbegin
\mathcal{D}^{a}{\nabla^{b}{Y}}+4(\Gamma_{c})^{\alpha}{}_{\beta} \psi^{a}\,_{i}\,^{\beta} \nabla^{b}{\nabla^{c}{X^{i}_{\alpha}}} - \frac{5}{2}(\Sigma_{c d})^{\alpha}{}_{\beta} \psi^{a}\,_{i}\,^{\beta} X^{i}_{\alpha} \nabla^{b}{W^{c d}}-(\Sigma_{c d})^{\alpha}{}_{\beta} \psi^{a}\,_{i}\,^{\beta} W^{c d} \nabla^{b}{X^{i}_{\alpha}}+4(\Sigma_{c d})^{\alpha}{}_{\beta} \psi^{a}\,_{i}\,^{\beta} W^{b c} \nabla^{d}{X^{i}_{\alpha}} - \frac{1}{2}\epsilon^{b}\,_{e {e_{1}}}\,^{c d} (\Gamma^{e})^{\alpha}{}_{\beta} \psi^{a}\,_{i}\,^{\beta} W_{c d} \nabla^{{e_{1}}}{X^{i}_{\alpha}}+2\psi^{a}\,_{i}\,^{\alpha} W^{b}\,_{c} \nabla^{c}{X^{i}_{\alpha}}+2(\Sigma^{b c})^{\alpha}{}_{\beta} \psi^{a}\,_{i}\,^{\beta} W_{c d} \nabla^{d}{X^{i}_{\alpha}} - \frac{1}{16}(\Gamma^{b})^{\alpha}{}_{\beta} \psi^{a}\,_{i}\,^{\beta} W^{c d} W_{c d} X^{i}_{\alpha} - \frac{5}{16}\epsilon^{b c d e {e_{1}}} \psi^{a}\,_{i}\,^{\alpha} W_{c d} W_{e {e_{1}}} X^{i}_{\alpha}+\frac{1}{2}\epsilon^{{e_{2}} c d e {e_{1}}} (\Sigma_{{e_{2}}}{}^{\, b})^{\alpha}{}_{\beta} \psi^{a}\,_{i}\,^{\beta} W_{c d} W_{e {e_{1}}} X^{i}_{\alpha} - \frac{3}{2}(\Gamma^{d})^{\alpha}{}_{\beta} \psi^{a}\,_{i}\,^{\beta} W^{b c} W_{d c} X^{i}_{\alpha}+\frac{3}{4}\epsilon^{b {e_{2}} c d e} (\Sigma_{{e_{2}}}{}^{\, {e_{1}}})^{\alpha}{}_{\beta} \psi^{a}\,_{i}\,^{\beta} W_{c d} W_{e {e_{1}}} X^{i}_{\alpha}+\frac{3}{8}\epsilon^{{e_{1}} {e_{2}} c d e} (\Sigma_{{e_{1}} {e_{2}}})^{\alpha}{}_{\beta} \psi^{a}\,_{i}\,^{\beta} W_{c d} W_{e}\,^{b} X^{i}_{\alpha} - \frac{1}{4}(\Gamma^{b})^{\alpha}{}_{\beta} \psi^{a}\,_{i}\,^{\beta} Y X^{i}_{\alpha} - \frac{1}{6}\epsilon^{b c d e {e_{1}}} \Phi_{c d i}\,^{j} (\Sigma_{e {e_{1}}})^{\alpha}{}_{\beta} \psi^{a}\,_{j}\,^{\beta} X^{i}_{\alpha}+\Phi_{c}\,^{b}\,_{i}\,^{j} (\Gamma^{c})^{\alpha}{}_{\beta} \psi^{a}\,_{j}\,^{\beta} X^{i}_{\alpha} - \frac{1}{2}(\Sigma^{b}{}_{\, c})^{\alpha}{}_{\beta} \psi^{a}\,_{i}\,^{\beta} X^{i}_{\alpha} \nabla_{d}{W^{c d}}+\frac{1}{4}\epsilon^{b}\,_{e {e_{1}}}\,^{c d} (\Gamma^{e})^{\alpha}{}_{\beta} \psi^{a}\,_{i}\,^{\beta} X^{i}_{\alpha} \nabla^{{e_{1}}}{W_{c d}}%
 - \frac{3}{4}\psi^{a}\,_{i}\,^{\alpha} X^{i}_{\alpha} \nabla_{c}{W^{b c}}+(\Sigma_{d c})^{\alpha}{}_{\beta} \psi^{a}\,_{i}\,^{\beta} X^{i}_{\alpha} \nabla^{d}{W^{b c}}-8{\rm i} (\Sigma^{b}{}_{\, c})^{\alpha}{}_{\beta} \phi^{a}\,_{i}\,^{\beta} \nabla^{c}{X^{i}_{\alpha}}-{\rm i} \epsilon^{b e {e_{1}} c d} (\Sigma_{e {e_{1}}})^{\alpha}{}_{\beta} \phi^{a}\,_{i}\,^{\beta} W_{c d} X^{i}_{\alpha}+{\rm i} (\Gamma^{c})^{\alpha}{}_{\beta} \phi^{a}\,_{i}\,^{\beta} W_{c}\,^{b} X^{i}_{\alpha}-4Y f^{a b}
  \doublespacedmathend
 \end{adjustwidth}
 
\subsubsection{$\nabla^{c}\Box W_{a b}$}
\begin{adjustwidth}{0cm}{5cm}
\doublespacedmathbegin
{}\mathcal{D}^{c}{\nabla_{d}{\nabla^{d}{W_{\hat{a} \hat{b}}}}}+\frac{3}{32}(\Sigma_{d e})^{\alpha}{}_{\beta} \psi^{c}\,_{i}\,^{\beta} W^{d e} W_{\hat{a}}\,^{{e_{1}}} W_{\hat{b} {e_{1}} \alpha}\,^{i} - \frac{3}{32}(\Sigma_{d e})^{\alpha}{}_{\beta} \psi^{c}\,_{i}\,^{\beta} W^{d e} W_{\hat{b}}\,^{{e_{1}}} W_{\hat{a} {e_{1}} \alpha}\,^{i}+\frac{1}{32}(\Gamma_{d})^{\alpha}{}_{\beta} \psi^{c}\,_{i}\,^{\beta} X^{i}_{\alpha} \nabla^{d}{W_{\hat{a} \hat{b}}}+\frac{1}{8}\Phi^{d e}\,_{i}\,^{j} (\Sigma_{d e})^{\alpha}{}_{\beta} \psi^{c}\,_{j}\,^{\beta} W_{\hat{a} \hat{b} \alpha}\,^{i} - \frac{1}{16}\Phi_{\hat{a} \hat{b} i}\,^{j} \psi^{c}\,_{j}\,^{\alpha} X^{i}_{\alpha}+\frac{1}{32}\epsilon^{e {e_{1}}}\,_{\hat{a} \hat{b} d} \Phi_{e {e_{1}} i}\,^{j} (\Gamma^{d})^{\alpha}{}_{\beta} \psi^{c}\,_{j}\,^{\beta} X^{i}_{\alpha}+\frac{1}{8}\Phi_{\hat{a}}\,^{d}\,_{i}\,^{j} (\Sigma_{\hat{b} d})^{\alpha}{}_{\beta} \psi^{c}\,_{j}\,^{\beta} X^{i}_{\alpha} - \frac{1}{8}\Phi_{\hat{b}}\,^{d}\,_{i}\,^{j} (\Sigma_{\hat{a} d})^{\alpha}{}_{\beta} \psi^{c}\,_{j}\,^{\beta} X^{i}_{\alpha}+\frac{121}{512}\epsilon^{{e_{1}} {e_{2}}}\,_{{e_{3}}}\,^{d e} (\Sigma_{{e_{1}} {e_{2}}})^{\alpha}{}_{\beta} \psi^{c}\,_{i}\,^{\beta} W_{\hat{a} \hat{b} \alpha}\,^{i} \nabla^{{e_{3}}}{W_{d e}} - \frac{127}{256}(\Gamma^{d})^{\alpha}{}_{\beta} \psi^{c}\,_{i}\,^{\beta} W_{\hat{a} \hat{b} \alpha}\,^{i} \nabla^{e}{W_{d e}} - \frac{3}{64}\epsilon_{\hat{a} \hat{b} {e_{1}}}\,^{d e} \psi^{c}\,_{i}\,^{\alpha} X^{i}_{\alpha} \nabla^{{e_{1}}}{W_{d e}}+\frac{15}{128}\epsilon^{{e_{1}}}\,_{\hat{a} \hat{b}}\,^{d e} (\Sigma_{{e_{1}} {e_{2}}})^{\alpha}{}_{\beta} \psi^{c}\,_{i}\,^{\beta} X^{i}_{\alpha} \nabla^{{e_{2}}}{W_{d e}} - \frac{15}{128}\epsilon^{e {e_{1}}}\,_{\hat{b} {e_{2}}}\,^{d} (\Sigma_{e {e_{1}}})^{\alpha}{}_{\beta} \psi^{c}\,_{i}\,^{\beta} X^{i}_{\alpha} \nabla^{{e_{2}}}{W_{\hat{a} d}}+\frac{11}{32}(\Gamma^{d})^{\alpha}{}_{\beta} \psi^{c}\,_{i}\,^{\beta} X^{i}_{\alpha} \nabla_{\hat{b}}{W_{\hat{a} d}} - \frac{25}{64}(\Gamma_{\hat{b}})^{\alpha}{}_{\beta} \psi^{c}\,_{i}\,^{\beta} X^{i}_{\alpha} \nabla^{d}{W_{\hat{a} d}}+\frac{15}{128}\epsilon^{e {e_{1}}}\,_{\hat{a} {e_{2}}}\,^{d} (\Sigma_{e {e_{1}}})^{\alpha}{}_{\beta} \psi^{c}\,_{i}\,^{\beta} X^{i}_{\alpha} \nabla^{{e_{2}}}{W_{\hat{b} d}} - \frac{11}{32}(\Gamma^{d})^{\alpha}{}_{\beta} \psi^{c}\,_{i}\,^{\beta} X^{i}_{\alpha} \nabla_{\hat{a}}{W_{\hat{b} d}}+\frac{25}{64}(\Gamma_{\hat{a}})^{\alpha}{}_{\beta} \psi^{c}\,_{i}\,^{\beta} X^{i}_{\alpha} \nabla^{d}{W_{\hat{b} d}}%
+\frac{61}{512}\epsilon_{\hat{a} \hat{b} {e_{3}}}\,^{d e} (\Sigma_{{e_{1}} {e_{2}}})^{\alpha}{}_{\beta} \psi^{c}\,_{i}\,^{\beta} W_{d e \alpha}\,^{i} \nabla^{{e_{3}}}{W^{{e_{1}} {e_{2}}}} - \frac{3}{512}\epsilon_{\hat{a} \hat{b}}\,^{{e_{1}} d e} \psi^{c}\,_{i}\,^{\alpha} W_{d e \alpha}\,^{i} \nabla^{{e_{2}}}{W_{{e_{1}} {e_{2}}}} - \frac{61}{256}(\Gamma_{\hat{b}})^{\alpha}{}_{\beta} \psi^{c}\,_{i}\,^{\beta} W_{d e \alpha}\,^{i} \nabla^{d}{W_{\hat{a}}\,^{e}}+\frac{61}{256}(\Gamma_{\hat{a}})^{\alpha}{}_{\beta} \psi^{c}\,_{i}\,^{\beta} W_{d e \alpha}\,^{i} \nabla^{d}{W_{\hat{b}}\,^{e}} - \frac{61}{256}(\Gamma^{d})^{\alpha}{}_{\beta} \psi^{c}\,_{i}\,^{\beta} W_{d e \alpha}\,^{i} \nabla^{e}{W_{\hat{a} \hat{b}}}+\frac{61}{512}(\Gamma_{\hat{b}})^{\alpha}{}_{\beta} \psi^{c}\,_{i}\,^{\beta} W_{d e \alpha}\,^{i} \nabla_{\hat{a}}{W^{d e}} - \frac{61}{256}(\Gamma^{d})^{\alpha}{}_{\beta} \psi^{c}\,_{i}\,^{\beta} W_{d e \alpha}\,^{i} \nabla_{\hat{a}}{W_{\hat{b}}\,^{e}} - \frac{61}{512}(\Gamma_{\hat{a}})^{\alpha}{}_{\beta} \psi^{c}\,_{i}\,^{\beta} W_{d e \alpha}\,^{i} \nabla_{\hat{b}}{W^{d e}}+\frac{61}{256}(\Gamma^{d})^{\alpha}{}_{\beta} \psi^{c}\,_{i}\,^{\beta} W_{d e \alpha}\,^{i} \nabla_{\hat{b}}{W_{\hat{a}}\,^{e}} - \frac{67}{256}\epsilon_{\hat{a} \hat{b} {e_{2}}}\,^{d e} (\Sigma_{{e_{3}} {e_{1}}})^{\alpha}{}_{\beta} \psi^{c}\,_{i}\,^{\beta} W_{d e \alpha}\,^{i} \nabla^{{e_{3}}}{W^{{e_{1}} {e_{2}}}}+\frac{67}{256}\epsilon^{{e_{3}}}\,_{\hat{a} \hat{b}}\,^{d e} (\Sigma_{{e_{3}} {e_{1}}})^{\alpha}{}_{\beta} \psi^{c}\,_{i}\,^{\beta} W_{d e \alpha}\,^{i} \nabla_{{e_{2}}}{W^{{e_{1}} {e_{2}}}}+\frac{3}{512}\epsilon_{\hat{b} {e_{2}}}\,^{e {e_{1}} d} \psi^{c}\,_{i}\,^{\alpha} W_{\hat{a} d \alpha}\,^{i} \nabla^{{e_{2}}}{W_{e {e_{1}}}}+\frac{61}{256}(\Gamma_{e})^{\alpha}{}_{\beta} \psi^{c}\,_{i}\,^{\beta} W_{\hat{a} d \alpha}\,^{i} \nabla^{e}{W_{\hat{b}}\,^{d}} - \frac{61}{256}(\Gamma_{\hat{b}})^{\alpha}{}_{\beta} \psi^{c}\,_{i}\,^{\beta} W_{\hat{a} d \alpha}\,^{i} \nabla_{e}{W^{e d}} - \frac{61}{256}(\Gamma^{d})^{\alpha}{}_{\beta} \psi^{c}\,_{i}\,^{\beta} W_{\hat{a} d \alpha}\,^{i} \nabla^{e}{W_{\hat{b} e}} - \frac{3}{128}\epsilon^{{e_{2}}}\,_{\hat{b} {e_{3}} {e_{1}}}\,^{d} (\Sigma_{{e_{2}} e})^{\alpha}{}_{\beta} \psi^{c}\,_{i}\,^{\beta} W_{\hat{a} d \alpha}\,^{i} \nabla^{{e_{3}}}{W^{e {e_{1}}}}+\frac{3}{512}\epsilon^{{e_{2}} {e_{3}} e {e_{1}} d} (\Sigma_{{e_{2}} {e_{3}}})^{\alpha}{}_{\beta} \psi^{c}\,_{i}\,^{\beta} W_{\hat{a} d \alpha}\,^{i} \nabla_{\hat{b}}{W_{e {e_{1}}}}+\frac{67}{256}(\Gamma_{e})^{\alpha}{}_{\beta} \psi^{c}\,_{i}\,^{\beta} W_{\hat{a} d \alpha}\,^{i} \nabla_{\hat{b}}{W^{e d}} - \frac{3}{512}\epsilon^{{e_{2}} {e_{3}}}\,_{\hat{b}}\,^{e {e_{1}}} (\Sigma_{{e_{2}} {e_{3}}})^{\alpha}{}_{\beta} \psi^{c}\,_{i}\,^{\beta} W_{\hat{a} d \alpha}\,^{i} \nabla^{d}{W_{e {e_{1}}}}+\frac{67}{256}(\Gamma^{e})^{\alpha}{}_{\beta} \psi^{c}\,_{i}\,^{\beta} W_{\hat{a} d \alpha}\,^{i} \nabla^{d}{W_{\hat{b} e}}%
 - \frac{3}{512}\epsilon_{\hat{a} {e_{2}}}\,^{e {e_{1}} d} \psi^{c}\,_{i}\,^{\alpha} W_{\hat{b} d \alpha}\,^{i} \nabla^{{e_{2}}}{W_{e {e_{1}}}} - \frac{61}{256}(\Gamma_{e})^{\alpha}{}_{\beta} \psi^{c}\,_{i}\,^{\beta} W_{\hat{b} d \alpha}\,^{i} \nabla^{e}{W_{\hat{a}}\,^{d}}+\frac{61}{256}(\Gamma_{\hat{a}})^{\alpha}{}_{\beta} \psi^{c}\,_{i}\,^{\beta} W_{\hat{b} d \alpha}\,^{i} \nabla_{e}{W^{e d}}+\frac{61}{256}(\Gamma^{d})^{\alpha}{}_{\beta} \psi^{c}\,_{i}\,^{\beta} W_{\hat{b} d \alpha}\,^{i} \nabla^{e}{W_{\hat{a} e}}+\frac{3}{128}\epsilon^{{e_{2}}}\,_{\hat{a} {e_{3}} {e_{1}}}\,^{d} (\Sigma_{{e_{2}} e})^{\alpha}{}_{\beta} \psi^{c}\,_{i}\,^{\beta} W_{\hat{b} d \alpha}\,^{i} \nabla^{{e_{3}}}{W^{e {e_{1}}}} - \frac{3}{512}\epsilon^{{e_{2}} {e_{3}} e {e_{1}} d} (\Sigma_{{e_{2}} {e_{3}}})^{\alpha}{}_{\beta} \psi^{c}\,_{i}\,^{\beta} W_{\hat{b} d \alpha}\,^{i} \nabla_{\hat{a}}{W_{e {e_{1}}}} - \frac{67}{256}(\Gamma_{e})^{\alpha}{}_{\beta} \psi^{c}\,_{i}\,^{\beta} W_{\hat{b} d \alpha}\,^{i} \nabla_{\hat{a}}{W^{e d}}+\frac{3}{512}\epsilon^{{e_{2}} {e_{3}}}\,_{\hat{a}}\,^{e {e_{1}}} (\Sigma_{{e_{2}} {e_{3}}})^{\alpha}{}_{\beta} \psi^{c}\,_{i}\,^{\beta} W_{\hat{b} d \alpha}\,^{i} \nabla^{d}{W_{e {e_{1}}}} - \frac{67}{256}(\Gamma^{e})^{\alpha}{}_{\beta} \psi^{c}\,_{i}\,^{\beta} W_{\hat{b} d \alpha}\,^{i} \nabla^{d}{W_{\hat{a} e}}+\frac{9}{128}\psi^{c}\,_{i}\,^{\alpha} W^{d e} W_{d e} W_{\hat{a} \hat{b} \alpha}\,^{i}+\frac{1}{32}\psi^{c}\,_{i}\,^{\alpha} W^{d e} W_{\hat{a} \hat{b}} W_{d e \alpha}\,^{i}+\frac{1}{256}\epsilon_{{e_{3}}}\,^{d e {e_{1}} {e_{2}}} (\Gamma^{{e_{3}}})^{\alpha}{}_{\beta} \psi^{c}\,_{i}\,^{\beta} W_{d e} W_{{e_{1}} {e_{2}}} W_{\hat{a} \hat{b} \alpha}\,^{i}+\frac{1}{16}(\Sigma^{{e_{1}}}{}_{\, d})^{\alpha}{}_{\beta} \psi^{c}\,_{i}\,^{\beta} W^{d e} W_{{e_{1}} e} W_{\hat{a} \hat{b} \alpha}\,^{i}+\frac{1}{16}\psi^{c}\,_{i}\,^{\alpha} W_{\hat{a}}\,^{d} W_{\hat{b}}\,^{e} W_{d e \alpha}\,^{i} - \frac{1}{64}\epsilon_{\hat{b} {e_{3}}}\,^{e {e_{1}} {e_{2}}} (\Gamma^{{e_{3}}})^{\alpha}{}_{\beta} \psi^{c}\,_{i}\,^{\beta} W_{\hat{a}}\,^{d} W_{e {e_{1}}} W_{{e_{2}} d \alpha}\,^{i}+\frac{1}{16}(\Sigma_{\hat{b} d})^{\alpha}{}_{\beta} \psi^{c}\,_{i}\,^{\beta} W^{d e} W_{\hat{a}}\,^{{e_{1}}} W_{{e_{1}} e \alpha}\,^{i}+\frac{1}{64}\epsilon_{\hat{a} {e_{3}}}\,^{e {e_{1}} {e_{2}}} (\Gamma^{{e_{3}}})^{\alpha}{}_{\beta} \psi^{c}\,_{i}\,^{\beta} W_{\hat{b}}\,^{d} W_{e {e_{1}}} W_{{e_{2}} d \alpha}\,^{i} - \frac{1}{16}(\Sigma_{\hat{a} d})^{\alpha}{}_{\beta} \psi^{c}\,_{i}\,^{\beta} W^{d e} W_{\hat{b}}\,^{{e_{1}}} W_{{e_{1}} e \alpha}\,^{i}+\frac{1}{32}(\Sigma_{\hat{b}}{}^{\, {e_{1}}})^{\alpha}{}_{\beta} \psi^{c}\,_{i}\,^{\beta} W^{d e} W_{\hat{a} {e_{1}}} W_{d e \alpha}\,^{i} - \frac{1}{32}(\Sigma_{\hat{a}}{}^{\, {e_{1}}})^{\alpha}{}_{\beta} \psi^{c}\,_{i}\,^{\beta} W^{d e} W_{\hat{b} {e_{1}}} W_{d e \alpha}\,^{i}%
 - \frac{1}{16}(\Sigma_{\hat{b}}{}^{\, {e_{1}}})^{\alpha}{}_{\beta} \psi^{c}\,_{i}\,^{\beta} W^{d e} W_{\hat{a} d} W_{{e_{1}} e \alpha}\,^{i}+\frac{1}{16}(\Sigma_{\hat{a}}{}^{\, {e_{1}}})^{\alpha}{}_{\beta} \psi^{c}\,_{i}\,^{\beta} W^{d e} W_{\hat{b} d} W_{{e_{1}} e \alpha}\,^{i}+\frac{1}{16}(\Sigma_{\hat{a} \hat{b}})^{\alpha}{}_{\beta} \psi^{c}\,_{i}\,^{\beta} W^{d e} W^{{e_{1}}}\,_{d} W_{{e_{1}} e \alpha}\,^{i} - \frac{1}{128}\epsilon_{\hat{a} \hat{b}}\,^{d {e_{2}} {e_{3}}} (\Gamma^{{e_{1}}})^{\alpha}{}_{\beta} \psi^{c}\,_{i}\,^{\beta} W_{d}\,^{e} W_{{e_{1}} e} W_{{e_{2}} {e_{3}} \alpha}\,^{i} - \frac{1}{16}(\Sigma^{{e_{1}}}{}_{\, d})^{\alpha}{}_{\beta} \psi^{c}\,_{i}\,^{\beta} W^{d e} W_{\hat{b} e} W_{\hat{a} {e_{1}} \alpha}\,^{i}+\frac{1}{128}\epsilon_{\hat{b}}\,^{d e {e_{1}} {e_{3}}} (\Gamma^{{e_{2}}})^{\alpha}{}_{\beta} \psi^{c}\,_{i}\,^{\beta} W_{d e} W_{{e_{1}} {e_{2}}} W_{\hat{a} {e_{3}} \alpha}\,^{i}+\frac{1}{32}\psi^{c}\,_{i}\,^{\alpha} W^{d e} W_{\hat{b} d} W_{\hat{a} e \alpha}\,^{i}+\frac{1}{16}(\Sigma^{{e_{1}}}{}_{\, d})^{\alpha}{}_{\beta} \psi^{c}\,_{i}\,^{\beta} W^{d e} W_{\hat{b} {e_{1}}} W_{\hat{a} e \alpha}\,^{i}+\frac{1}{128}\epsilon_{\hat{b} {e_{3}}}\,^{d e {e_{1}}} (\Gamma^{{e_{3}}})^{\alpha}{}_{\beta} \psi^{c}\,_{i}\,^{\beta} W_{d e} W_{{e_{1}}}\,^{{e_{2}}} W_{\hat{a} {e_{2}} \alpha}\,^{i}+\frac{1}{16}(\Sigma^{{e_{1}}}{}_{\, d})^{\alpha}{}_{\beta} \psi^{c}\,_{i}\,^{\beta} W^{d e} W_{\hat{a} e} W_{\hat{b} {e_{1}} \alpha}\,^{i} - \frac{1}{128}\epsilon_{\hat{a}}\,^{d e {e_{1}} {e_{3}}} (\Gamma^{{e_{2}}})^{\alpha}{}_{\beta} \psi^{c}\,_{i}\,^{\beta} W_{d e} W_{{e_{1}} {e_{2}}} W_{\hat{b} {e_{3}} \alpha}\,^{i} - \frac{1}{32}\psi^{c}\,_{i}\,^{\alpha} W^{d e} W_{\hat{a} d} W_{\hat{b} e \alpha}\,^{i} - \frac{1}{16}(\Sigma^{{e_{1}}}{}_{\, d})^{\alpha}{}_{\beta} \psi^{c}\,_{i}\,^{\beta} W^{d e} W_{\hat{a} {e_{1}}} W_{\hat{b} e \alpha}\,^{i} - \frac{1}{128}\epsilon_{\hat{a} {e_{3}}}\,^{d e {e_{1}}} (\Gamma^{{e_{3}}})^{\alpha}{}_{\beta} \psi^{c}\,_{i}\,^{\beta} W_{d e} W_{{e_{1}}}\,^{{e_{2}}} W_{\hat{b} {e_{2}} \alpha}\,^{i} - \frac{1}{128}\epsilon_{\hat{a} \hat{b}}\,^{d e {e_{3}}} (\Gamma^{{e_{1}}})^{\alpha}{}_{\beta} \psi^{c}\,_{i}\,^{\beta} W_{d e} W_{{e_{1}}}\,^{{e_{2}}} W_{{e_{3}} {e_{2}} \alpha}\,^{i}+\frac{1}{128}\epsilon_{\hat{a} \hat{b}}\,^{d e {e_{1}}} (\Gamma^{{e_{3}}})^{\alpha}{}_{\beta} \psi^{c}\,_{i}\,^{\beta} W_{d e} W_{{e_{1}}}\,^{{e_{2}}} W_{{e_{3}} {e_{2}} \alpha}\,^{i} - \frac{1}{128}\epsilon_{\hat{b}}\,^{d e {e_{2}} {e_{3}}} (\Gamma^{{e_{1}}})^{\alpha}{}_{\beta} \psi^{c}\,_{i}\,^{\beta} W_{\hat{a} d} W_{e {e_{1}}} W_{{e_{2}} {e_{3}} \alpha}\,^{i}+\frac{1}{128}\epsilon_{\hat{a}}\,^{d e {e_{2}} {e_{3}}} (\Gamma^{{e_{1}}})^{\alpha}{}_{\beta} \psi^{c}\,_{i}\,^{\beta} W_{\hat{b} d} W_{e {e_{1}}} W_{{e_{2}} {e_{3}} \alpha}\,^{i} - \frac{1}{2}\psi^{c}\,_{i}\,^{\alpha} \nabla_{d}{\nabla^{d}{W_{\hat{a} \hat{b} \alpha}\,^{i}}} - \frac{1}{2}(\Sigma_{\hat{a} \hat{b}})^{\alpha}{}_{\beta} \psi^{c}\,_{i}\,^{\beta} \nabla_{d}{\nabla^{d}{X^{i}_{\alpha}}}%
+\frac{17}{64}\epsilon^{{e_{1}} {e_{2}}}\,_{{e_{3}}}\,^{d e} (\Sigma_{{e_{1}} {e_{2}}})^{\alpha}{}_{\beta} \psi^{c}\,_{i}\,^{\beta} W_{d e} \nabla^{{e_{3}}}{W_{\hat{a} \hat{b} \alpha}\,^{i}} - \frac{19}{32}(\Gamma^{d})^{\alpha}{}_{\beta} \psi^{c}\,_{i}\,^{\beta} W_{d e} \nabla^{e}{W_{\hat{a} \hat{b} \alpha}\,^{i}} - \frac{1}{8}(\Gamma_{d})^{\alpha}{}_{\beta} \psi^{c}\,_{i}\,^{\beta} W_{\hat{a} \hat{b}} \nabla^{d}{X^{i}_{\alpha}} - \frac{1}{16}\epsilon_{\hat{a} \hat{b} {e_{1}}}\,^{d e} \psi^{c}\,_{i}\,^{\alpha} W_{d e} \nabla^{{e_{1}}}{X^{i}_{\alpha}}+\frac{3}{16}\epsilon^{{e_{1}}}\,_{\hat{a} \hat{b}}\,^{d e} (\Sigma_{{e_{1}} {e_{2}}})^{\alpha}{}_{\beta} \psi^{c}\,_{i}\,^{\beta} W_{d e} \nabla^{{e_{2}}}{X^{i}_{\alpha}} - \frac{3}{16}\epsilon^{e {e_{1}}}\,_{\hat{b} {e_{2}}}\,^{d} (\Sigma_{e {e_{1}}})^{\alpha}{}_{\beta} \psi^{c}\,_{i}\,^{\beta} W_{\hat{a} d} \nabla^{{e_{2}}}{X^{i}_{\alpha}}+\frac{1}{8}(\Gamma^{d})^{\alpha}{}_{\beta} \psi^{c}\,_{i}\,^{\beta} W_{\hat{a} d} \nabla_{\hat{b}}{X^{i}_{\alpha}} - \frac{1}{4}(\Gamma_{\hat{b}})^{\alpha}{}_{\beta} \psi^{c}\,_{i}\,^{\beta} W_{\hat{a} d} \nabla^{d}{X^{i}_{\alpha}}+\frac{3}{16}\epsilon^{e {e_{1}}}\,_{\hat{a} {e_{2}}}\,^{d} (\Sigma_{e {e_{1}}})^{\alpha}{}_{\beta} \psi^{c}\,_{i}\,^{\beta} W_{\hat{b} d} \nabla^{{e_{2}}}{X^{i}_{\alpha}} - \frac{1}{8}(\Gamma^{d})^{\alpha}{}_{\beta} \psi^{c}\,_{i}\,^{\beta} W_{\hat{b} d} \nabla_{\hat{a}}{X^{i}_{\alpha}}+\frac{1}{4}(\Gamma_{\hat{a}})^{\alpha}{}_{\beta} \psi^{c}\,_{i}\,^{\beta} W_{\hat{b} d} \nabla^{d}{X^{i}_{\alpha}}+\frac{3}{64}\epsilon^{{e_{1}}}\,_{\hat{a} \hat{b} {e_{2}} e} (\Sigma_{{e_{1}} d})^{\alpha}{}_{\beta} \psi^{c}\,_{i}\,^{\beta} X^{i}_{\alpha} \nabla^{{e_{2}}}{W^{d e}} - \frac{3}{256}\epsilon^{{e_{1}} {e_{2}}}\,_{\hat{b}}\,^{d e} (\Sigma_{{e_{1}} {e_{2}}})^{\alpha}{}_{\beta} \psi^{c}\,_{i}\,^{\beta} X^{i}_{\alpha} \nabla_{\hat{a}}{W_{d e}}+\frac{3}{256}\epsilon^{{e_{1}} {e_{2}}}\,_{\hat{a}}\,^{d e} (\Sigma_{{e_{1}} {e_{2}}})^{\alpha}{}_{\beta} \psi^{c}\,_{i}\,^{\beta} X^{i}_{\alpha} \nabla_{\hat{b}}{W_{d e}} - \frac{1}{64}\epsilon_{\hat{a} \hat{b} {e_{3}}}\,^{{e_{1}} {e_{2}}} (\Sigma_{d e})^{\alpha}{}_{\beta} \psi^{c}\,_{i}\,^{\beta} W^{d e} \nabla^{{e_{3}}}{W_{{e_{1}} {e_{2}} \alpha}\,^{i}} - \frac{1}{64}\epsilon_{\hat{a} \hat{b}}\,^{d {e_{1}} {e_{2}}} \psi^{c}\,_{i}\,^{\alpha} W_{d e} \nabla^{e}{W_{{e_{1}} {e_{2}} \alpha}\,^{i}}+\frac{5}{64}(\Gamma_{\hat{b}})^{\alpha}{}_{\beta} \psi^{c}\,_{i}\,^{\beta} W^{d e} \nabla_{\hat{a}}{W_{d e \alpha}\,^{i}} - \frac{5}{32}(\Gamma^{e})^{\alpha}{}_{\beta} \psi^{c}\,_{i}\,^{\beta} W_{\hat{b}}\,^{d} \nabla_{\hat{a}}{W_{e d \alpha}\,^{i}}+\frac{5}{32}(\Gamma_{\hat{b}})^{\alpha}{}_{\beta} \psi^{c}\,_{i}\,^{\beta} W_{\hat{a}}\,^{d} \nabla^{e}{W_{d e \alpha}\,^{i}} - \frac{5}{32}(\Gamma^{d})^{\alpha}{}_{\beta} \psi^{c}\,_{i}\,^{\beta} W_{\hat{a} \hat{b}} \nabla^{e}{W_{d e \alpha}\,^{i}}%
 - \frac{5}{32}(\Gamma_{\hat{a}})^{\alpha}{}_{\beta} \psi^{c}\,_{i}\,^{\beta} W_{\hat{b}}\,^{d} \nabla^{e}{W_{d e \alpha}\,^{i}}+\frac{5}{32}(\Gamma^{e})^{\alpha}{}_{\beta} \psi^{c}\,_{i}\,^{\beta} W_{\hat{a}}\,^{d} \nabla_{\hat{b}}{W_{e d \alpha}\,^{i}} - \frac{5}{64}(\Gamma_{\hat{a}})^{\alpha}{}_{\beta} \psi^{c}\,_{i}\,^{\beta} W^{d e} \nabla_{\hat{b}}{W_{d e \alpha}\,^{i}} - \frac{1}{32}\epsilon_{\hat{a} \hat{b}}\,^{d {e_{1}} {e_{2}}} (\Sigma^{e}{}_{\, {e_{3}}})^{\alpha}{}_{\beta} \psi^{c}\,_{i}\,^{\beta} W_{d e} \nabla^{{e_{3}}}{W_{{e_{1}} {e_{2}} \alpha}\,^{i}}+\frac{1}{32}\epsilon^{{e_{3}}}\,_{\hat{a} \hat{b}}\,^{{e_{1}} {e_{2}}} (\Sigma_{{e_{3}}}{}^{\, d})^{\alpha}{}_{\beta} \psi^{c}\,_{i}\,^{\beta} W_{d e} \nabla^{e}{W_{{e_{1}} {e_{2}} \alpha}\,^{i}}+\frac{1}{64}\epsilon_{\hat{b} {e_{2}}}\,^{d e {e_{1}}} \psi^{c}\,_{i}\,^{\alpha} W_{d e} \nabla^{{e_{2}}}{W_{\hat{a} {e_{1}} \alpha}\,^{i}}+\frac{5}{32}(\Gamma_{e})^{\alpha}{}_{\beta} \psi^{c}\,_{i}\,^{\beta} W_{\hat{b}}\,^{d} \nabla^{e}{W_{\hat{a} d \alpha}\,^{i}} - \frac{1}{32}(\Gamma_{\hat{b}})^{\alpha}{}_{\beta} \psi^{c}\,_{i}\,^{\beta} W^{d}\,_{e} \nabla^{e}{W_{\hat{a} d \alpha}\,^{i}} - \frac{5}{32}(\Gamma^{e})^{\alpha}{}_{\beta} \psi^{c}\,_{i}\,^{\beta} W_{\hat{b} d} \nabla^{d}{W_{\hat{a} e \alpha}\,^{i}}+\frac{1}{16}\epsilon^{{e_{2}}}\,_{\hat{b} {e_{3}}}\,^{d {e_{1}}} (\Sigma_{{e_{2}}}{}^{\, e})^{\alpha}{}_{\beta} \psi^{c}\,_{i}\,^{\beta} W_{d e} \nabla^{{e_{3}}}{W_{\hat{a} {e_{1}} \alpha}\,^{i}}+\frac{1}{64}\epsilon^{{e_{2}} {e_{3}} d e {e_{1}}} (\Sigma_{{e_{2}} {e_{3}}})^{\alpha}{}_{\beta} \psi^{c}\,_{i}\,^{\beta} W_{d e} \nabla_{\hat{b}}{W_{\hat{a} {e_{1}} \alpha}\,^{i}}+\frac{1}{32}(\Gamma^{d})^{\alpha}{}_{\beta} \psi^{c}\,_{i}\,^{\beta} W_{d}\,^{e} \nabla_{\hat{b}}{W_{\hat{a} e \alpha}\,^{i}} - \frac{1}{64}\epsilon^{{e_{2}} {e_{3}}}\,_{\hat{b}}\,^{d e} (\Sigma_{{e_{2}} {e_{3}}})^{\alpha}{}_{\beta} \psi^{c}\,_{i}\,^{\beta} W_{d e} \nabla^{{e_{1}}}{W_{\hat{a} {e_{1}} \alpha}\,^{i}}+\frac{7}{32}(\Gamma^{d})^{\alpha}{}_{\beta} \psi^{c}\,_{i}\,^{\beta} W_{\hat{b} d} \nabla^{e}{W_{\hat{a} e \alpha}\,^{i}} - \frac{1}{64}\epsilon_{\hat{a} {e_{2}}}\,^{d e {e_{1}}} \psi^{c}\,_{i}\,^{\alpha} W_{d e} \nabla^{{e_{2}}}{W_{\hat{b} {e_{1}} \alpha}\,^{i}} - \frac{5}{32}(\Gamma_{e})^{\alpha}{}_{\beta} \psi^{c}\,_{i}\,^{\beta} W_{\hat{a}}\,^{d} \nabla^{e}{W_{\hat{b} d \alpha}\,^{i}}+\frac{1}{32}(\Gamma_{\hat{a}})^{\alpha}{}_{\beta} \psi^{c}\,_{i}\,^{\beta} W^{d}\,_{e} \nabla^{e}{W_{\hat{b} d \alpha}\,^{i}}+\frac{5}{32}(\Gamma^{e})^{\alpha}{}_{\beta} \psi^{c}\,_{i}\,^{\beta} W_{\hat{a} d} \nabla^{d}{W_{\hat{b} e \alpha}\,^{i}} - \frac{1}{16}\epsilon^{{e_{2}}}\,_{\hat{a} {e_{3}}}\,^{d {e_{1}}} (\Sigma_{{e_{2}}}{}^{\, e})^{\alpha}{}_{\beta} \psi^{c}\,_{i}\,^{\beta} W_{d e} \nabla^{{e_{3}}}{W_{\hat{b} {e_{1}} \alpha}\,^{i}} - \frac{1}{64}\epsilon^{{e_{2}} {e_{3}} d e {e_{1}}} (\Sigma_{{e_{2}} {e_{3}}})^{\alpha}{}_{\beta} \psi^{c}\,_{i}\,^{\beta} W_{d e} \nabla_{\hat{a}}{W_{\hat{b} {e_{1}} \alpha}\,^{i}}%
 - \frac{1}{32}(\Gamma^{d})^{\alpha}{}_{\beta} \psi^{c}\,_{i}\,^{\beta} W_{d}\,^{e} \nabla_{\hat{a}}{W_{\hat{b} e \alpha}\,^{i}}+\frac{1}{64}\epsilon^{{e_{2}} {e_{3}}}\,_{\hat{a}}\,^{d e} (\Sigma_{{e_{2}} {e_{3}}})^{\alpha}{}_{\beta} \psi^{c}\,_{i}\,^{\beta} W_{d e} \nabla^{{e_{1}}}{W_{\hat{b} {e_{1}} \alpha}\,^{i}} - \frac{7}{32}(\Gamma^{d})^{\alpha}{}_{\beta} \psi^{c}\,_{i}\,^{\beta} W_{\hat{a} d} \nabla^{e}{W_{\hat{b} e \alpha}\,^{i}} - \frac{1}{8}\epsilon^{{e_{1}}}\,_{\hat{a} \hat{b} {e_{2}}}\,^{d} (\Sigma_{{e_{1}}}{}^{\, e})^{\alpha}{}_{\beta} \psi^{c}\,_{i}\,^{\beta} W_{d e} \nabla^{{e_{2}}}{X^{i}_{\alpha}} - \frac{1}{32}\epsilon^{{e_{1}} {e_{2}}}\,_{\hat{b}}\,^{d e} (\Sigma_{{e_{1}} {e_{2}}})^{\alpha}{}_{\beta} \psi^{c}\,_{i}\,^{\beta} W_{d e} \nabla_{\hat{a}}{X^{i}_{\alpha}}+\frac{1}{32}\epsilon^{{e_{1}} {e_{2}}}\,_{\hat{a}}\,^{d e} (\Sigma_{{e_{1}} {e_{2}}})^{\alpha}{}_{\beta} \psi^{c}\,_{i}\,^{\beta} W_{d e} \nabla_{\hat{b}}{X^{i}_{\alpha}}+\frac{3}{32}\epsilon^{{e_{1}} {e_{2}}}\,_{{e_{3}}}\,^{d e} (\Sigma_{{e_{1}} {e_{2}}})^{\alpha}{}_{\beta} \psi^{c}\,_{i}\,^{\beta} W_{\hat{a} \hat{b}} \nabla^{{e_{3}}}{W_{d e \alpha}\,^{i}}+\frac{3}{16}\epsilon^{{e_{1}} {e_{2}}}\,_{\hat{b} d {e_{3}}} (\Sigma_{{e_{1}} {e_{2}}})^{\alpha}{}_{\beta} \psi^{c}\,_{i}\,^{\beta} W^{d e} \nabla^{{e_{3}}}{W_{\hat{a} e \alpha}\,^{i}}+\frac{3}{16}(\Gamma_{d})^{\alpha}{}_{\beta} \psi^{c}\,_{i}\,^{\beta} W^{d e} \nabla_{\hat{b}}{W_{\hat{a} e \alpha}\,^{i}} - \frac{3}{16}(\Gamma_{\hat{b}})^{\alpha}{}_{\beta} \psi^{c}\,_{i}\,^{\beta} W_{d}\,^{e} \nabla^{d}{W_{\hat{a} e \alpha}\,^{i}} - \frac{3}{16}\epsilon^{{e_{1}} {e_{2}}}\,_{\hat{a} d {e_{3}}} (\Sigma_{{e_{1}} {e_{2}}})^{\alpha}{}_{\beta} \psi^{c}\,_{i}\,^{\beta} W^{d e} \nabla^{{e_{3}}}{W_{\hat{b} e \alpha}\,^{i}} - \frac{3}{16}(\Gamma_{d})^{\alpha}{}_{\beta} \psi^{c}\,_{i}\,^{\beta} W^{d e} \nabla_{\hat{a}}{W_{\hat{b} e \alpha}\,^{i}}+\frac{3}{16}(\Gamma_{\hat{a}})^{\alpha}{}_{\beta} \psi^{c}\,_{i}\,^{\beta} W_{d}\,^{e} \nabla^{d}{W_{\hat{b} e \alpha}\,^{i}}+\frac{3}{32}\epsilon^{{e_{1}} {e_{2}}}\,_{\hat{a} \hat{b} {e_{3}}} (\Sigma_{{e_{1}} {e_{2}}})^{\alpha}{}_{\beta} \psi^{c}\,_{i}\,^{\beta} W^{d e} \nabla^{{e_{3}}}{W_{d e \alpha}\,^{i}} - \frac{3}{32}\epsilon_{\hat{a} \hat{b}}\,^{d e {e_{1}}} \psi^{c}\,_{i}\,^{\alpha} W_{d e} \nabla^{{e_{2}}}{W_{{e_{1}} {e_{2}} \alpha}\,^{i}}+\frac{3}{16}\epsilon_{\hat{a} \hat{b}}\,^{d e {e_{1}}} (\Sigma^{{e_{2}}}{}_{\, {e_{3}}})^{\alpha}{}_{\beta} \psi^{c}\,_{i}\,^{\beta} W_{d e} \nabla^{{e_{3}}}{W_{{e_{1}} {e_{2}} \alpha}\,^{i}}+\frac{3}{32}\epsilon_{\hat{b} {e_{2}}}\,^{d e {e_{1}}} \psi^{c}\,_{i}\,^{\alpha} W_{\hat{a} d} \nabla^{{e_{2}}}{W_{e {e_{1}} \alpha}\,^{i}}+\frac{3}{16}\epsilon^{{e_{2}}}\,_{\hat{b}}\,^{d e {e_{1}}} (\Sigma_{{e_{2}} {e_{3}}})^{\alpha}{}_{\beta} \psi^{c}\,_{i}\,^{\beta} W_{\hat{a} d} \nabla^{{e_{3}}}{W_{e {e_{1}} \alpha}\,^{i}}+\frac{3}{16}\epsilon^{{e_{1}} {e_{2}}}\,_{{e_{3}}}\,^{d e} (\Sigma_{{e_{1}} {e_{2}}})^{\alpha}{}_{\beta} \psi^{c}\,_{i}\,^{\beta} W_{\hat{a} d} \nabla^{{e_{3}}}{W_{\hat{b} e \alpha}\,^{i}} - \frac{3}{16}\epsilon^{{e_{1}} {e_{2}}}\,_{\hat{b} {e_{3}}}\,^{e} (\Sigma_{{e_{1}} {e_{2}}})^{\alpha}{}_{\beta} \psi^{c}\,_{i}\,^{\beta} W_{\hat{a}}\,^{d} \nabla^{{e_{3}}}{W_{e d \alpha}\,^{i}}%
 - \frac{3}{32}\epsilon_{\hat{a} {e_{2}}}\,^{d e {e_{1}}} \psi^{c}\,_{i}\,^{\alpha} W_{\hat{b} d} \nabla^{{e_{2}}}{W_{e {e_{1}} \alpha}\,^{i}} - \frac{3}{16}\epsilon^{{e_{2}}}\,_{\hat{a}}\,^{d e {e_{1}}} (\Sigma_{{e_{2}} {e_{3}}})^{\alpha}{}_{\beta} \psi^{c}\,_{i}\,^{\beta} W_{\hat{b} d} \nabla^{{e_{3}}}{W_{e {e_{1}} \alpha}\,^{i}} - \frac{3}{16}\epsilon^{{e_{1}} {e_{2}}}\,_{{e_{3}}}\,^{d e} (\Sigma_{{e_{1}} {e_{2}}})^{\alpha}{}_{\beta} \psi^{c}\,_{i}\,^{\beta} W_{\hat{b} d} \nabla^{{e_{3}}}{W_{\hat{a} e \alpha}\,^{i}}+\frac{3}{16}\epsilon^{{e_{1}} {e_{2}}}\,_{\hat{a} {e_{3}}}\,^{e} (\Sigma_{{e_{1}} {e_{2}}})^{\alpha}{}_{\beta} \psi^{c}\,_{i}\,^{\beta} W_{\hat{b}}\,^{d} \nabla^{{e_{3}}}{W_{e d \alpha}\,^{i}} - \frac{1}{4}\epsilon^{{e_{2}}}\,_{\hat{b}}\,^{e {e_{1}} d} (\Sigma_{{e_{2}} {e_{3}}})^{\alpha}{}_{\beta} \psi^{c}\,_{i}\,^{\beta} W_{\hat{a} d \alpha}\,^{i} \nabla^{{e_{3}}}{W_{e {e_{1}}}} - \frac{1}{4}\epsilon^{{e_{1}} {e_{2}}}\,_{{e_{3}}}\,^{e d} (\Sigma_{{e_{1}} {e_{2}}})^{\alpha}{}_{\beta} \psi^{c}\,_{i}\,^{\beta} W_{\hat{a} d \alpha}\,^{i} \nabla^{{e_{3}}}{W_{\hat{b} e}} - \frac{1}{4}\epsilon^{{e_{1}} {e_{2}}}\,_{\hat{b} {e_{3}} e} (\Sigma_{{e_{1}} {e_{2}}})^{\alpha}{}_{\beta} \psi^{c}\,_{i}\,^{\beta} W_{\hat{a} d \alpha}\,^{i} \nabla^{{e_{3}}}{W^{e d}}+\frac{1}{4}\epsilon^{{e_{2}}}\,_{\hat{a}}\,^{e {e_{1}} d} (\Sigma_{{e_{2}} {e_{3}}})^{\alpha}{}_{\beta} \psi^{c}\,_{i}\,^{\beta} W_{\hat{b} d \alpha}\,^{i} \nabla^{{e_{3}}}{W_{e {e_{1}}}}+\frac{1}{4}\epsilon^{{e_{1}} {e_{2}}}\,_{{e_{3}}}\,^{e d} (\Sigma_{{e_{1}} {e_{2}}})^{\alpha}{}_{\beta} \psi^{c}\,_{i}\,^{\beta} W_{\hat{b} d \alpha}\,^{i} \nabla^{{e_{3}}}{W_{\hat{a} e}}+\frac{1}{4}\epsilon^{{e_{1}} {e_{2}}}\,_{\hat{a} {e_{3}} e} (\Sigma_{{e_{1}} {e_{2}}})^{\alpha}{}_{\beta} \psi^{c}\,_{i}\,^{\beta} W_{\hat{b} d \alpha}\,^{i} \nabla^{{e_{3}}}{W^{e d}} - \frac{1}{8}\epsilon_{\hat{a} \hat{b}}\,^{{e_{1}} {e_{2}} d} \psi^{c}\,_{i}\,^{\alpha} W_{d e \alpha}\,^{i} \nabla^{e}{W_{{e_{1}} {e_{2}}}} - \frac{1}{8}\epsilon_{\hat{b} {e_{2}}}\,^{{e_{1}} d e} \psi^{c}\,_{i}\,^{\alpha} W_{d e \alpha}\,^{i} \nabla^{{e_{2}}}{W_{\hat{a} {e_{1}}}}+\frac{1}{8}\epsilon_{\hat{a} {e_{2}}}\,^{{e_{1}} d e} \psi^{c}\,_{i}\,^{\alpha} W_{d e \alpha}\,^{i} \nabla^{{e_{2}}}{W_{\hat{b} {e_{1}}}}+\frac{9}{128}(\Sigma_{\hat{a} \hat{b}})^{\alpha}{}_{\beta} \psi^{c}\,_{i}\,^{\beta} W^{d e} W_{d e} X^{i}_{\alpha} - \frac{9}{8}{\rm i} (\Sigma_{d e})^{\alpha}{}_{\beta} \phi^{c}\,_{i}\,^{\beta} W^{d e} W_{\hat{a} \hat{b} \alpha}\,^{i}+\frac{9}{16}{\rm i} \phi^{c}\,_{i}\,^{\alpha} W_{\hat{a} \hat{b}} X^{i}_{\alpha} - \frac{9}{32}{\rm i} \epsilon_{\hat{a} \hat{b} {e_{1}}}\,^{d e} (\Gamma^{{e_{1}}})^{\alpha}{}_{\beta} \phi^{c}\,_{i}\,^{\beta} W_{d e} X^{i}_{\alpha} - \frac{9}{8}{\rm i} (\Sigma_{\hat{b}}{}^{\, d})^{\alpha}{}_{\beta} \phi^{c}\,_{i}\,^{\beta} W_{\hat{a} d} X^{i}_{\alpha}+\frac{9}{8}{\rm i} (\Sigma_{\hat{a}}{}^{\, d})^{\alpha}{}_{\beta} \phi^{c}\,_{i}\,^{\beta} W_{\hat{b} d} X^{i}_{\alpha}+\frac{11}{4}{\rm i} \phi^{c}\,_{i}\,^{\alpha} W_{\hat{a}}\,^{d} W_{\hat{b} d \alpha}\,^{i}%
 - \frac{11}{4}{\rm i} \phi^{c}\,_{i}\,^{\alpha} W_{\hat{b}}\,^{d} W_{\hat{a} d \alpha}\,^{i} - \frac{1}{4}{\rm i} (\Gamma_{d})^{\alpha}{}_{\beta} \phi^{c}\,_{i}\,^{\beta} \nabla^{d}{W_{\hat{a} \hat{b} \alpha}\,^{i}} - \frac{1}{8}{\rm i} \epsilon_{\hat{a} \hat{b} {e_{1}}}\,^{d e} \phi^{c}\,_{i}\,^{\alpha} \nabla^{{e_{1}}}{W_{d e \alpha}\,^{i}}+\frac{1}{4}{\rm i} \epsilon^{{e_{1}}}\,_{\hat{a} \hat{b}}\,^{d e} (\Sigma_{{e_{1}} {e_{2}}})^{\alpha}{}_{\beta} \phi^{c}\,_{i}\,^{\beta} \nabla^{{e_{2}}}{W_{d e \alpha}\,^{i}} - \frac{1}{4}{\rm i} \epsilon^{e {e_{1}}}\,_{\hat{b} {e_{2}}}\,^{d} (\Sigma_{e {e_{1}}})^{\alpha}{}_{\beta} \phi^{c}\,_{i}\,^{\beta} \nabla^{{e_{2}}}{W_{\hat{a} d \alpha}\,^{i}}+\frac{1}{4}{\rm i} (\Gamma^{d})^{\alpha}{}_{\beta} \phi^{c}\,_{i}\,^{\beta} \nabla_{\hat{b}}{W_{\hat{a} d \alpha}\,^{i}} - \frac{1}{4}{\rm i} (\Gamma_{\hat{b}})^{\alpha}{}_{\beta} \phi^{c}\,_{i}\,^{\beta} \nabla^{d}{W_{\hat{a} d \alpha}\,^{i}}+\frac{1}{4}{\rm i} \epsilon^{e {e_{1}}}\,_{\hat{a} {e_{2}}}\,^{d} (\Sigma_{e {e_{1}}})^{\alpha}{}_{\beta} \phi^{c}\,_{i}\,^{\beta} \nabla^{{e_{2}}}{W_{\hat{b} d \alpha}\,^{i}} - \frac{1}{4}{\rm i} (\Gamma^{d})^{\alpha}{}_{\beta} \phi^{c}\,_{i}\,^{\beta} \nabla_{\hat{a}}{W_{\hat{b} d \alpha}\,^{i}}+\frac{1}{4}{\rm i} (\Gamma_{\hat{a}})^{\alpha}{}_{\beta} \phi^{c}\,_{i}\,^{\beta} \nabla^{d}{W_{\hat{b} d \alpha}\,^{i}} - \frac{1}{2}{\rm i} \epsilon^{d e}\,_{\hat{a} \hat{b} {e_{1}}} (\Sigma_{d e})^{\alpha}{}_{\beta} \phi^{c}\,_{i}\,^{\beta} \nabla^{{e_{1}}}{X^{i}_{\alpha}} - \frac{1}{2}{\rm i} (\Gamma_{\hat{b}})^{\alpha}{}_{\beta} \phi^{c}\,_{i}\,^{\beta} \nabla_{\hat{a}}{X^{i}_{\alpha}}+\frac{1}{2}{\rm i} (\Gamma_{\hat{a}})^{\alpha}{}_{\beta} \phi^{c}\,_{i}\,^{\beta} \nabla_{\hat{b}}{X^{i}_{\alpha}}+2\nabla^{d}{W_{\hat{a} \hat{b}}} f^{c}\,_{d}-4\nabla_{\hat{a}}{W_{\hat{b}}\,^{d}} f^{c}\,_{d}+4\nabla_{\hat{b}}{W_{\hat{a}}\,^{d}} f^{c}\,_{d}-4\nabla^{d}{W_{\hat{a} d}} f^{c}\,_{\hat{b}}+4\nabla^{d}{W_{\hat{b} d}} f^{c}\,_{\hat{a}}
\doublespacedmathend
\end{adjustwidth}

\subsection{Vector Multiplet} 
\subsubsection{$\nabla^{a} \nabla^{b} W$}
 
\begin{adjustwidth}{0cm}{5cm}
\doublespacedmathbegin
\mathcal{D}^{a}{\nabla^{b}{W}} - \frac{1}{2}{\rm i} \psi^{a}\,_{i}\,^{\alpha} \nabla^{b}{\lambda^{i}_{\alpha}}+\frac{1}{16}{\rm i} \epsilon^{b e {e_{1}} c d} (\Sigma_{e {e_{1}}})^{\alpha}{}_{\beta} \psi^{a}\,_{i}\,^{\beta} W_{c d} \lambda^{i}_{\alpha}+\frac{1}{4}{\rm i} (\Gamma_{c})^{\alpha}{}_{\beta} \psi^{a}\,_{i}\,^{\beta} W^{b c} \lambda^{i}_{\alpha} - \frac{1}{8}(\Gamma^{b})^{\alpha}{}_{\beta} \psi^{a}\,_{i}\,^{\beta} W X^{i}_{\alpha}+\frac{1}{2}(\Gamma^{b})^{\alpha}{}_{\beta} \phi^{a}\,_{i}\,^{\beta} \lambda^{i}_{\alpha}-2W f^{a b}
\doublespacedmathend
\end{adjustwidth}

\subsubsection{$\nabla^{a} \nabla^{b} \lambda^{i}_{\alpha}$}

\begin{adjustwidth}{1em}{5cm}
\doublespacedmathbegin
\mathcal{D}^{a}{\nabla^{b}{\lambda^{i}_{\alpha}}} - \frac{1}{2}(\Sigma_{c d})_{\alpha \beta} \psi^{a i \beta} \nabla^{b}{F^{c d}} - \frac{3}{8}(\Sigma_{c d})_{\alpha \beta} \psi^{a i \beta} W^{c d} \nabla^{b}{W} - \frac{3}{8}(\Sigma_{c d})_{\alpha \beta} \psi^{a i \beta} W \nabla^{b}{W^{c d}} - \frac{1}{2}\psi^{a}\,_{j \alpha} \nabla^{b}{X^{i j}}+\frac{1}{2}(\Gamma_{c})_{\alpha \beta} \psi^{a i \beta} \mathcal{D}^{b}{\nabla^{c}{W}} - \frac{1}{4}{\rm i} (\Gamma_{c})_{\alpha \beta} \psi^{a i \beta} \psi^{b}\,_{j}\,^{\rho} \nabla^{c}{\lambda^{j}_{\rho}}+\frac{1}{16}{\rm i} (\Sigma_{c d})^{\beta}{}_{\rho} \psi^{a i}\,_{\alpha} \psi^{b}\,_{j}\,^{\rho} W^{c d} \lambda^{j}_{\beta} - \frac{3}{16}{\rm i} (\Sigma_{c d})^{\beta}{}_{\rho} \psi^{a i \rho} \psi^{b}\,_{j \alpha} W^{c d} \lambda^{j}_{\beta} - \frac{3}{16}{\rm i} (\Sigma_{c d})_{\alpha}{}^{\beta} \psi^{a i}\,_{\rho} \psi^{b}\,_{j}\,^{\rho} W^{c d} \lambda^{j}_{\beta} - \frac{1}{16}{\rm i} (\Sigma_{c d})_{\beta \rho} \psi^{a i \beta} \psi^{b}\,_{j}\,^{\rho} W^{c d} \lambda^{j}_{\alpha}+\frac{1}{16}{\rm i} (\Sigma_{c d})_{\alpha \beta} \psi^{a i \rho} \psi^{b}\,_{j}\,^{\beta} W^{c d} \lambda^{j}_{\rho} - \frac{1}{16}\psi^{a i}\,_{\alpha} \psi^{b}\,_{j}\,^{\beta} W X^{j}_{\beta}+\frac{1}{8}\psi^{a i \beta} \psi^{b}\,_{j \alpha} W X^{j}_{\beta}+\frac{1}{8}\psi^{a i}\,_{\beta} \psi^{b}\,_{j}\,^{\beta} W X^{j}_{\alpha}+\frac{1}{4}\psi^{a i}\,_{\alpha} \phi^{b}\,_{j}\,^{\beta} \lambda^{j}_{\beta} - \frac{1}{2}\psi^{a i \beta} \phi^{b}\,_{j \alpha} \lambda^{j}_{\beta} - \frac{1}{2}\psi^{a i}\,_{\beta} \phi^{b}\,_{j}\,^{\beta} \lambda^{j}_{\alpha}-(\Gamma^{c})_{\alpha \beta} \psi^{a i \beta} W f^{b}\,_{c}%
 - \frac{1}{16}(\Gamma^{b})_{\alpha \beta} \psi^{a i \beta} W^{c d} F_{c d} - \frac{1}{32}\epsilon^{b e {e_{1}} c d} \psi^{a i}\,_{\alpha} W_{e {e_{1}}} F_{c d}+\frac{3}{32}\epsilon^{{e_{2}} e {e_{1}} c d} (\Sigma_{{e_{2}}}{}^{\, b})_{\alpha \beta} \psi^{a i \beta} W_{e {e_{1}}} F_{c d} - \frac{3}{16}\epsilon^{b {e_{1}} {e_{2}} e c} (\Sigma_{{e_{1}} {e_{2}}})_{\alpha \beta} \psi^{a i \beta} W_{e}\,^{d} F_{c d} - \frac{1}{4}(\Gamma^{c})_{\alpha \beta} \psi^{a i \beta} W^{b d} F_{c d}+\frac{1}{8}(\Gamma^{d})_{\alpha \beta} \psi^{a i \beta} W_{d c} F^{b c} - \frac{7}{64}(\Gamma^{b})_{\alpha \beta} \psi^{a i \beta} W W^{c d} W_{c d}+\frac{1}{64}\epsilon^{b c d e {e_{1}}} \psi^{a i}\,_{\alpha} W W_{c d} W_{e {e_{1}}}+\frac{1}{16}\epsilon^{{e_{2}} c d e {e_{1}}} (\Sigma_{{e_{2}}}{}^{\, b})_{\alpha \beta} \psi^{a i \beta} W W_{c d} W_{e {e_{1}}}+\frac{1}{16}\epsilon^{b e {e_{1}} c d} (\Sigma_{e {e_{1}}})_{\alpha \beta} \psi^{a}\,_{j}\,^{\beta} X^{i j} W_{c d} - \frac{1}{4}(\Gamma^{c})_{\alpha \beta} \psi^{a}\,_{j}\,^{\beta} X^{i j} W_{c}\,^{b}+\frac{1}{2}(\Sigma_{c d})_{\alpha \beta} \psi^{a i \beta} W^{b c} \nabla^{d}{W} - \frac{1}{16}\epsilon^{b}\,_{e {e_{1}}}\,^{c d} (\Gamma^{e})_{\alpha \beta} \psi^{a i \beta} W_{c d} \nabla^{{e_{1}}}{W}+\frac{1}{4}\psi^{a i}\,_{\alpha} W^{b}\,_{c} \nabla^{c}{W}+\frac{1}{4}(\Sigma^{b c})_{\alpha \beta} \psi^{a i \beta} W_{c d} \nabla^{d}{W}+\frac{1}{16}\epsilon^{b {e_{2}} e c d} (\Sigma_{{e_{2}}}{}^{\, {e_{1}}})_{\alpha \beta} \psi^{a i \beta} W_{e {e_{1}}} F_{c d}+\frac{1}{32}\epsilon^{{e_{1}} {e_{2}} d e c} (\Sigma_{{e_{1}} {e_{2}}})_{\alpha \beta} \psi^{a i \beta} W_{d e} F_{c}\,^{b}+\frac{1}{8}(\Gamma^{d})_{\alpha \beta} \psi^{a i \beta} W W^{b c} W_{d c} - \frac{1}{16}\epsilon^{b {e_{2}} c d e} (\Sigma_{{e_{2}}}{}^{\, {e_{1}}})_{\alpha \beta} \psi^{a i \beta} W W_{c d} W_{e {e_{1}}} - \frac{1}{32}\epsilon^{{e_{1}} {e_{2}} c d e} (\Sigma_{{e_{1}} {e_{2}}})_{\alpha \beta} \psi^{a i \beta} W W_{c d} W_{e}\,^{b}%
 - \frac{1}{8}\epsilon^{b e {e_{1}} c d} (\Sigma_{e {e_{1}}})_{\alpha \rho} \psi^{a}\,_{j}\,^{\rho} \lambda^{i \beta} W_{c d \beta}\,^{j}+\frac{1}{4}(\Gamma^{c})_{\alpha \rho} \psi^{a}\,_{j}\,^{\rho} \lambda^{i \beta} W_{c}\,^{b}\,_{\beta}\,^{j} - \frac{1}{16}\epsilon^{b e {e_{1}} c d} (\Sigma_{e {e_{1}}})_{\alpha}{}^{\rho} \psi^{a}\,_{j}\,^{\beta} \lambda^{i}_{\rho} W_{c d \beta}\,^{j}+\frac{1}{16}(\Gamma^{b})_{\alpha}{}^{\beta} \psi^{a}\,_{j}\,^{\rho} \lambda^{i}_{\rho} X^{j}_{\beta}+\frac{1}{16}(\Gamma^{b})^{\rho \beta} \psi^{a}\,_{j \alpha} \lambda^{i}_{\rho} X^{j}_{\beta} - \frac{1}{16}(\Gamma^{b})_{\alpha \rho} \psi^{a}\,_{j}\,^{\rho} \lambda^{i \beta} X^{j}_{\beta}+\frac{1}{16}(\Gamma^{b})^{\beta}{}_{\rho} \psi^{a}\,_{j}\,^{\rho} \lambda^{i}_{\beta} X^{j}_{\alpha}+\frac{3}{16}(\Gamma^{b})^{\beta}{}_{\rho} \psi^{a}\,_{j}\,^{\rho} \lambda^{i}_{\alpha} X^{j}_{\beta}+\frac{3}{16}(\Gamma^{b})^{\beta}{}_{\rho} \psi^{a i \rho} \lambda_{j \alpha} X^{j}_{\beta} - \frac{3}{16}(\Gamma^{b})^{\beta}{}_{\rho} \psi^{a}\,_{j}\,^{\rho} \lambda^{j}_{\alpha} X^{i}_{\beta}+\frac{1}{24}\epsilon^{b c d e {e_{1}}} \Phi_{c d}\,^{i j} (\Sigma_{e {e_{1}}})_{\alpha \beta} \psi^{a}\,_{j}\,^{\beta} W - \frac{1}{4}\Phi_{c}\,^{b i j} (\Gamma^{c})_{\alpha \beta} \psi^{a}\,_{j}\,^{\beta} W - \frac{1}{4}(\Sigma_{d c})_{\alpha \beta} \psi^{a i \beta} W \nabla^{d}{W^{b c}} - \frac{1}{16}\epsilon^{b}\,_{e {e_{1}}}\,^{c d} (\Gamma^{e})_{\alpha \beta} \psi^{a i \beta} W \nabla^{{e_{1}}}{W_{c d}}+\frac{3}{16}\psi^{a i}\,_{\alpha} W \nabla_{c}{W^{b c}}+\frac{1}{8}(\Sigma^{b}{}_{\,c})_{\alpha \beta} \psi^{a i \beta} W \nabla_{d}{W^{c d}} - \frac{3}{2}{\rm i} \phi^{a i}\,_{\alpha} \nabla^{b}{W} - \frac{1}{4}{\rm i} \epsilon^{b e {e_{1}} c d} (\Sigma_{e {e_{1}}})_{\alpha \beta} \phi^{a i \beta} F_{c d}+\frac{1}{2}{\rm i} (\Gamma^{c})_{\alpha \beta} \phi^{a i \beta} F_{c}\,^{b} - \frac{1}{4}{\rm i} \epsilon^{b e {e_{1}} c d} (\Sigma_{e {e_{1}}})_{\alpha \beta} \phi^{a i \beta} W W_{c d}%
+\frac{3}{4}{\rm i} (\Gamma^{c})_{\alpha \beta} \phi^{a i \beta} W W_{c}\,^{b} - \frac{1}{2}{\rm i} (\Gamma^{b})_{\alpha \beta} \phi^{a}\,_{j}\,^{\beta} X^{i j}-{\rm i} (\Sigma^{b}{}_{\,c})_{\alpha \beta} \phi^{a i \beta} \nabla^{c}{W}-3\lambda^{i}_{\alpha} f^{a b}+2(\Sigma^{b}{}_{\, c})_{\alpha}{}^{\beta} \lambda^{i}_{\beta} f^{a c}
\doublespacedmathend
\end{adjustwidth}

\subsubsection{$\nabla^{a} \nabla^{b} X^{i j}$}

\begin{adjustwidth}{1em}{5cm}
\doublespacedmathbegin
\mathcal{D}^{a}{\nabla^{b}{X_{\underline{i} \underline{j}}}}-{\rm i} (\Gamma_{c})^{\alpha}{}_{\beta} \psi^{a \underline{i} \beta} \mathcal{D}^{b}{\nabla^{c}{\lambda_{\underline{j} \alpha}}}+\frac{1}{4}{\rm i} \epsilon^{e {e_{1}}}\,_{{e_{2}}}\,^{c d} (\Sigma_{e {e_{1}}})_{\alpha \beta} \psi^{a \underline{i} \alpha} \psi^{b \underline{j} \beta} \nabla^{{e_{2}}}{F_{c d}}+\frac{1}{2}{\rm i} (\Gamma^{c})_{\alpha \beta} \psi^{a \underline{i} \alpha} \psi^{b \underline{j} \beta} \nabla^{d}{F_{c d}}+\frac{1}{16}{\rm i} \epsilon^{e {e_{1}}}\,_{{e_{2}}}\,^{c d} (\Sigma_{e {e_{1}}})_{\alpha \beta} \psi^{a \underline{i} \alpha} \psi^{b \underline{j} \beta} W_{c d} \nabla^{{e_{2}}}{W}+\frac{7}{8}{\rm i} (\Gamma^{c})_{\alpha \beta} \psi^{a \underline{i} \alpha} \psi^{b \underline{j} \beta} W_{c d} \nabla^{d}{W}+\frac{3}{16}{\rm i} \epsilon^{e {e_{1}}}\,_{{e_{2}}}\,^{c d} (\Sigma_{e {e_{1}}})_{\alpha \beta} \psi^{a \underline{i} \alpha} \psi^{b \underline{j} \beta} W \nabla^{{e_{2}}}{W_{c d}}+\frac{9}{16}{\rm i} (\Gamma^{c})_{\alpha \beta} \psi^{a \underline{i} \alpha} \psi^{b \underline{j} \beta} W \nabla^{d}{W_{c d}}+\frac{1}{2}{\rm i} (\Gamma_{c})_{\alpha \beta} \psi^{a \underline{i} \alpha} \psi^{b}\,_{k}\,^{\beta} \nabla^{c}{X_{\underline{j}}\,^{k}} - \frac{1}{2}{\rm i} \psi^{a \underline{i}}\,_{\alpha} \psi^{b \underline{j} \alpha} \nabla_{c}{\nabla^{c}{W}}-{\rm i} (\Sigma_{c d})_{\alpha \beta} \psi^{a \underline{i} \alpha} \psi^{b \underline{j} \beta} \nabla^{c}{\nabla^{d}{W}}+\frac{7}{16}{\rm i} \psi^{a \underline{i}}\,_{\alpha} \psi^{b \underline{j} \alpha} W^{c d} F_{c d} - \frac{5}{32}{\rm i} \epsilon_{{e_{2}}}\,^{e {e_{1}} c d} (\Gamma^{{e_{2}}})_{\alpha \beta} \psi^{a \underline{i} \alpha} \psi^{b \underline{j} \beta} W_{e {e_{1}}} F_{c d}+\frac{1}{4}{\rm i} (\Sigma^{c}{}_{\,e})_{\alpha \beta} \psi^{a \underline{i} \alpha} \psi^{b \underline{j} \beta} W^{e d} F_{c d}+\frac{27}{64}{\rm i} \psi^{a \underline{i}}\,_{\alpha} \psi^{b \underline{j} \alpha} W W^{c d} W_{c d} - \frac{9}{64}{\rm i} \epsilon_{{e_{2}}}\,^{c d e {e_{1}}} (\Gamma^{{e_{2}}})_{\alpha \beta} \psi^{a \underline{i} \alpha} \psi^{b \underline{j} \beta} W W_{c d} W_{e {e_{1}}} - \frac{1}{8}{\rm i} (\Sigma_{c d})_{\alpha \beta} \psi^{a \underline{i} \alpha} \psi^{b}\,_{k}\,^{\beta} X_{\underline{j}}\,^{k} W^{c d} - \frac{1}{8}{\rm i} (\Sigma^{c d})^{\beta}{}_{\rho} \psi^{a \underline{i} \rho} \psi^{b}\,_{k}\,^{\alpha} \lambda_{\underline{j} \beta} W_{c d \alpha}\,^{k} - \frac{1}{8}{\rm i} \psi^{a \underline{i} \alpha} \psi^{b}\,_{k}\,^{\beta} \lambda_{\underline{j} \beta} X^{k}_{\alpha}%
+\frac{1}{16}{\rm i} \psi^{a \underline{i} \beta} \psi^{b}\,_{k}\,^{\alpha} \lambda_{\underline{j} \beta} X^{k}_{\alpha}+\frac{1}{8}{\rm i} \psi^{a \underline{i}}\,_{\beta} \psi^{b}\,_{k}\,^{\beta} \lambda^{\alpha}_{\underline{j}} X^{k}_{\alpha} - \frac{3}{16}{\rm i} \psi^{a \underline{i} \beta} \psi^{b \underline{j} \alpha} \lambda_{k \beta} X^{k}_{\alpha}+\frac{3}{16}{\rm i} \psi^{a \underline{i} \beta} \psi^{b}\,_{k}\,^{\alpha} \lambda^{k}_{\beta} X_{\underline{j} \alpha} - \frac{3}{8}{\rm i} \psi^{a \underline{i}}\,_{\beta} \psi^{b \underline{j} \beta} \lambda^{\alpha}_{k} X^{k}_{\alpha}+\frac{3}{8}{\rm i} \psi^{a \underline{i} \alpha} \psi^{b \underline{j} \beta} \lambda_{k \beta} X^{k}_{\alpha}+\frac{3}{8}{\rm i} \psi^{a \underline{i}}\,_{\beta} \psi^{b}\,_{k}\,^{\beta} \lambda^{k \alpha} X_{\underline{j} \alpha} - \frac{3}{8}{\rm i} \psi^{a \underline{i} \alpha} \psi^{b}\,_{k}\,^{\beta} \lambda^{k}_{\beta} X_{\underline{j} \alpha} - \frac{1}{4}{\rm i} \Phi^{c d}\,_{\underline{j}}\,^{k} (\Sigma_{c d})_{\alpha \beta} \psi^{a \underline{i} \alpha} \psi^{b}\,_{k}\,^{\beta} W+\frac{1}{2}(\Gamma_{c})_{\alpha \beta} \psi^{a \underline{i} \alpha} \phi^{b \underline{j} \beta} \nabla^{c}{W}+\frac{1}{2}(\Sigma_{c d})_{\alpha \beta} \psi^{a \underline{i} \alpha} \phi^{b \underline{j} \beta} F^{c d} - \frac{5}{2}\psi^{a \underline{i}}\,_{\alpha} \phi^{b}\,_{k}\,^{\alpha} X_{\underline{j}}\,^{k}-{\rm i} (\Gamma^{c})^{\alpha}{}_{\beta} \psi^{a \underline{i} \beta} \lambda_{\underline{j} \alpha} f^{b}\,_{c}+\frac{3}{8}{\rm i} (\Sigma_{c d})^{\alpha}{}_{\beta} \psi^{a \underline{i} \beta} \lambda_{\underline{j} \alpha} \nabla^{b}{W^{c d}} - \frac{3}{4}\psi^{a \underline{i} \alpha} X_{\underline{j} \alpha} \nabla^{b}{W} - \frac{3}{4}\psi^{a \underline{i} \alpha} W \nabla^{b}{X_{\underline{j} \alpha}}-{\rm i} (\Sigma_{c d})^{\alpha}{}_{\beta} \psi^{a \underline{i} \beta} W^{b c} \nabla^{d}{\lambda_{\underline{j} \alpha}}+\frac{1}{8}{\rm i} \epsilon^{b}\,_{e {e_{1}}}\,^{c d} (\Gamma^{e})^{\alpha}{}_{\beta} \psi^{a \underline{i} \beta} W_{c d} \nabla^{{e_{1}}}{\lambda_{\underline{j} \alpha}} - \frac{1}{2}{\rm i} \psi^{a \underline{i} \alpha} W^{b}\,_{c} \nabla^{c}{\lambda_{\underline{j} \alpha}} - \frac{1}{2}{\rm i} (\Sigma^{b c})^{\alpha}{}_{\beta} \psi^{a \underline{i} \beta} W_{c d} \nabla^{d}{\lambda_{\underline{j} \alpha}}%
 - \frac{1}{64}{\rm i} (\Gamma^{b})^{\alpha}{}_{\beta} \psi^{a \underline{i} \beta} W^{c d} W_{c d} \lambda_{\underline{j} \alpha}+\frac{1}{16}{\rm i} \epsilon^{b c d e {e_{1}}} \psi^{a \underline{i} \alpha} W_{c d} W_{e {e_{1}}} \lambda_{\underline{j} \alpha} - \frac{5}{64}{\rm i} \epsilon^{{e_{2}} c d e {e_{1}}} (\Sigma_{{e_{2}}}{}^{\, b})^{\alpha}{}_{\beta} \psi^{a \underline{i} \beta} W_{c d} W_{e {e_{1}}} \lambda_{\underline{j} \alpha}+\frac{3}{32}\epsilon^{b e {e_{1}} c d} (\Sigma_{e {e_{1}}})^{\alpha}{}_{\beta} \psi^{a \underline{i} \beta} W W_{c d} X_{\underline{j} \alpha} - \frac{3}{8}(\Gamma^{c})^{\alpha}{}_{\beta} \psi^{a \underline{i} \beta} W W_{c}\,^{b} X_{\underline{j} \alpha}+\frac{5}{16}{\rm i} (\Gamma^{d})^{\alpha}{}_{\beta} \psi^{a \underline{i} \beta} W^{b c} W_{d c} \lambda_{\underline{j} \alpha} - \frac{5}{32}{\rm i} \epsilon^{b {e_{2}} c d e} (\Sigma_{{e_{2}}}{}^{\, {e_{1}}})^{\alpha}{}_{\beta} \psi^{a \underline{i} \beta} W_{c d} W_{e {e_{1}}} \lambda_{\underline{j} \alpha} - \frac{5}{64}{\rm i} \epsilon^{{e_{1}} {e_{2}} c d e} (\Sigma_{{e_{1}} {e_{2}}})^{\alpha}{}_{\beta} \psi^{a \underline{i} \beta} W_{c d} W_{e}\,^{b} \lambda_{\underline{j} \alpha} - \frac{1}{4}(\Gamma^{b})^{\alpha}{}_{\beta} \psi^{a}\,_{k}\,^{\beta} X_{\underline{i} \underline{j}} X^{k}_{\alpha} - \frac{3}{8}(\Gamma^{b})^{\alpha}{}_{\beta} \psi^{a \underline{i} \beta} X_{\underline{j} k} X^{k}_{\alpha}+\frac{3}{8}(\Gamma^{b})^{\alpha}{}_{\beta} \psi^{a}\,_{k}\,^{\beta} X_{\underline{i}}\,^{k} X_{\underline{j} \alpha}+\frac{1}{24}{\rm i} \epsilon^{b c d e {e_{1}}} \Phi_{c d \underline{i}}\,^{k} (\Sigma_{e {e_{1}}})^{\alpha}{}_{\beta} \psi^{a}\,_{k}\,^{\beta} \lambda_{\underline{j} \alpha} - \frac{1}{4}{\rm i} \Phi_{c}\,^{b}\,_{\underline{i}}\,^{k} (\Gamma^{c})^{\alpha}{}_{\beta} \psi^{a}\,_{k}\,^{\beta} \lambda_{\underline{j} \alpha}+\frac{1}{8}{\rm i} (\Sigma^{b}{}_{\, c})^{\alpha}{}_{\beta} \psi^{a \underline{i} \beta} \lambda_{\underline{j} \alpha} \nabla_{d}{W^{c d}} - \frac{1}{16}{\rm i} \epsilon^{b}\,_{e {e_{1}}}\,^{c d} (\Gamma^{e})^{\alpha}{}_{\beta} \psi^{a \underline{i} \beta} \lambda_{\underline{j} \alpha} \nabla^{{e_{1}}}{W_{c d}}+\frac{3}{16}{\rm i} \psi^{a \underline{i} \alpha} \lambda_{\underline{j} \alpha} \nabla_{c}{W^{b c}} - \frac{1}{4}{\rm i} (\Sigma_{d c})^{\alpha}{}_{\beta} \psi^{a \underline{i} \beta} \lambda_{\underline{j} \alpha} \nabla^{d}{W^{b c}}-2(\Sigma^{b}{}_{\, c})^{\alpha}{}_{\beta} \phi^{a \underline{i} \beta} \nabla^{c}{\lambda_{\underline{j} \alpha}} - \frac{1}{8}\epsilon^{b e {e_{1}} c d} (\Sigma_{e {e_{1}}})^{\alpha}{}_{\beta} \phi^{a \underline{i} \beta} W_{c d} \lambda_{\underline{j} \alpha} - \frac{3}{4}{\rm i} (\Gamma^{b})^{\alpha}{}_{\beta} \phi^{a \underline{i} \beta} W X_{\underline{j} \alpha}%
-4X_{\underline{i} \underline{j}} f^{a b}
\doublespacedmathend
\end{adjustwidth}

\subsubsection{$\nabla^{c} \nabla^{d} F_{a b}$}

\begin{adjustwidth}{1em}{5cm}
\doublespacedmathbegin
\mathcal{D}^{c}{\nabla^{d}{F_{\hat{a} \hat{b}}}}-{\rm i} (\Gamma_{\hat{a}})^{\alpha}{}_{\beta} \psi^{c}\,_{i}\,^{\beta} \nabla^{d}{\nabla_{\hat{b}}{\lambda^{i}_{\alpha}}}+\frac{3}{8}{\rm i} \psi^{c}\,_{i}\,^{\alpha} \lambda^{i}_{\alpha} \nabla^{d}{W_{\hat{a} \hat{b}}}+\frac{1}{2}{\rm i} \psi^{c}\,_{i}\,^{\alpha} W_{\hat{a} \hat{b}} \nabla^{d}{\lambda^{i}_{\alpha}}+\frac{9}{64}{\rm i} \epsilon_{\hat{a} \hat{b} {e_{2}}}\,^{e {e_{1}}} (\Gamma^{{e_{2}}})^{\alpha}{}_{\beta} \psi^{c}\,_{i}\,^{\beta} \lambda^{i}_{\alpha} \nabla^{d}{W_{e {e_{1}}}}+\frac{5}{32}{\rm i} \epsilon_{\hat{a} \hat{b} {e_{2}}}\,^{e {e_{1}}} (\Gamma^{{e_{2}}})^{\alpha}{}_{\beta} \psi^{c}\,_{i}\,^{\beta} W_{e {e_{1}}} \nabla^{d}{\lambda^{i}_{\alpha}}+\frac{1}{2}{\rm i} (\Sigma_{\hat{a}}{}^{\, e})^{\alpha}{}_{\beta} \psi^{c}\,_{i}\,^{\beta} W_{\hat{b} e} \nabla^{d}{\lambda^{i}_{\alpha}}+\frac{1}{2}\psi^{c}\,_{i}\,^{\alpha} W_{\hat{a} \hat{b} \alpha}\,^{i} \nabla^{d}{W}+\frac{1}{2}\psi^{c}\,_{i}\,^{\alpha} W \nabla^{d}{W_{\hat{a} \hat{b} \alpha}\,^{i}} - \frac{3}{32}{\rm i} \epsilon^{d}\,_{\hat{a} \hat{b} {e_{2}}}\,^{e} (\Gamma^{{e_{1}}})^{\alpha}{}_{\beta} \psi^{c}\,_{i}\,^{\beta} W_{e {e_{1}}} \nabla^{{e_{2}}}{\lambda^{i}_{\alpha}}+\frac{3}{32}{\rm i} \epsilon_{\hat{a} \hat{b} {e_{1}} {e_{2}}}\,^{e} (\Gamma^{{e_{1}}})^{\alpha}{}_{\beta} \psi^{c}\,_{i}\,^{\beta} W_{e}\,^{d} \nabla^{{e_{2}}}{\lambda^{i}_{\alpha}} - \frac{1}{4}{\rm i} (\Sigma_{e {e_{1}}})^{\alpha}{}_{\beta} \delta^{d}\,_{\hat{a}} \psi^{c}\,_{i}\,^{\beta} W^{e {e_{1}}} \nabla_{\hat{b}}{\lambda^{i}_{\alpha}} - \frac{1}{2}{\rm i} (\Sigma^{d e})^{\alpha}{}_{\beta} \psi^{c}\,_{i}\,^{\beta} W_{\hat{b} e} \nabla_{\hat{a}}{\lambda^{i}_{\alpha}} - \frac{1}{16}{\rm i} \epsilon^{d}\,_{\hat{a} {e_{2}}}\,^{e {e_{1}}} (\Gamma^{{e_{2}}})^{\alpha}{}_{\beta} \psi^{c}\,_{i}\,^{\beta} W_{e {e_{1}}} \nabla_{\hat{b}}{\lambda^{i}_{\alpha}} - \frac{1}{2}{\rm i} \psi^{c}\,_{i}\,^{\alpha} W_{\hat{b}}\,^{d} \nabla_{\hat{a}}{\lambda^{i}_{\alpha}}+{\rm i} (\Sigma_{\hat{a} e})^{\alpha}{}_{\beta} \psi^{c}\,_{i}\,^{\beta} W^{d e} \nabla_{\hat{b}}{\lambda^{i}_{\alpha}} - \frac{1}{256}{\rm i} \epsilon^{d {e_{2}} {e_{3}} e {e_{1}}} (\Sigma_{{e_{2}} {e_{3}}})^{\alpha}{}_{\beta} \psi^{c}\,_{i}\,^{\beta} W_{\hat{a} \hat{b}} W_{e {e_{1}}} \lambda^{i}_{\alpha}+\frac{1}{16}{\rm i} (\Gamma^{e})^{\alpha}{}_{\beta} \psi^{c}\,_{i}\,^{\beta} W_{\hat{a} \hat{b}} W_{e}\,^{d} \lambda^{i}_{\alpha}+\frac{1}{128}{\rm i} \epsilon^{{e_{3}} {e_{4}} e {e_{1}} {e_{2}}} (\Sigma_{{e_{3}} {e_{4}}})^{\alpha}{}_{\beta} \delta^{d}\,_{\hat{a}} \psi^{c}\,_{i}\,^{\beta} W_{\hat{b} e} W_{{e_{1}} {e_{2}}} \lambda^{i}_{\alpha}%
 - \frac{3}{16}{\rm i} (\Gamma^{{e_{1}}})^{\alpha}{}_{\beta} \delta^{d}\,_{\hat{a}} \psi^{c}\,_{i}\,^{\beta} W_{\hat{b}}\,^{e} W_{{e_{1}} e} \lambda^{i}_{\alpha} - \frac{15}{64}{\rm i} \epsilon^{{e_{2}} {e_{3}}}\,_{\hat{a} e}\,^{{e_{1}}} (\Sigma_{{e_{2}} {e_{3}}})^{\alpha}{}_{\beta} \psi^{c}\,_{i}\,^{\beta} W^{d e} W_{\hat{b} {e_{1}}} \lambda^{i}_{\alpha} - \frac{1}{4}{\rm i} (\Gamma_{\hat{a}})^{\alpha}{}_{\beta} \psi^{c}\,_{i}\,^{\beta} W^{d e} W_{\hat{b} e} \lambda^{i}_{\alpha} - \frac{15}{64}{\rm i} \epsilon^{d {e_{2}} {e_{3}}}\,_{\hat{a} e} (\Sigma_{{e_{2}} {e_{3}}})^{\alpha}{}_{\beta} \psi^{c}\,_{i}\,^{\beta} W^{e {e_{1}}} W_{\hat{b} {e_{1}}} \lambda^{i}_{\alpha}+\frac{15}{64}{\rm i} \epsilon^{{e_{3}} {e_{4}}}\,_{\hat{b} e}\,^{{e_{2}}} (\Sigma_{{e_{3}} {e_{4}}})^{\alpha}{}_{\beta} \delta^{d}\,_{\hat{a}} \psi^{c}\,_{i}\,^{\beta} W^{e {e_{1}}} W_{{e_{2}} {e_{1}}} \lambda^{i}_{\alpha} - \frac{1}{32}{\rm i} (\Gamma_{\hat{b}})^{\alpha}{}_{\beta} \delta^{d}\,_{\hat{a}} \psi^{c}\,_{i}\,^{\beta} W^{e {e_{1}}} W_{e {e_{1}}} \lambda^{i}_{\alpha} - \frac{15}{128}{\rm i} \epsilon^{{e_{2}} {e_{3}}}\,_{\hat{a} \hat{b} e} (\Sigma_{{e_{2}} {e_{3}}})^{\alpha}{}_{\beta} \psi^{c}\,_{i}\,^{\beta} W^{e {e_{1}}} W^{d}\,_{{e_{1}}} \lambda^{i}_{\alpha}+\frac{3}{256}{\rm i} \epsilon^{d {e_{2}} {e_{3}}}\,_{\hat{a} \hat{b}} (\Sigma_{{e_{2}} {e_{3}}})^{\alpha}{}_{\beta} \psi^{c}\,_{i}\,^{\beta} W^{e {e_{1}}} W_{e {e_{1}}} \lambda^{i}_{\alpha} - \frac{23}{256}{\rm i} \epsilon_{\hat{a} \hat{b}}\,^{e {e_{1}} {e_{2}}} \psi^{c}\,_{i}\,^{\alpha} W_{e {e_{1}}} W_{{e_{2}}}\,^{d} \lambda^{i}_{\alpha} - \frac{41}{128}{\rm i} \epsilon^{{e_{3}}}\,_{\hat{a} \hat{b}}\,^{{e_{1}} {e_{2}}} (\Sigma_{{e_{3}} e})^{\alpha}{}_{\beta} \psi^{c}\,_{i}\,^{\beta} W^{d e} W_{{e_{1}} {e_{2}}} \lambda^{i}_{\alpha} - \frac{7}{128}{\rm i} \epsilon^{d}\,_{\hat{a}}\,^{e {e_{1}} {e_{2}}} \psi^{c}\,_{i}\,^{\alpha} W_{\hat{b} e} W_{{e_{1}} {e_{2}}} \lambda^{i}_{\alpha}+\frac{3}{8}{\rm i} \epsilon^{{e_{3}}}\,_{\hat{a}}\,^{e {e_{1}} {e_{2}}} (\Sigma_{{e_{3}}}{}^{\, d})^{\alpha}{}_{\beta} \psi^{c}\,_{i}\,^{\beta} W_{\hat{b} e} W_{{e_{1}} {e_{2}}} \lambda^{i}_{\alpha} - \frac{3}{32}{\rm i} \epsilon^{d {e_{2}} {e_{3}} e {e_{1}}} (\Sigma_{{e_{2}} {e_{3}}})^{\alpha}{}_{\beta} \psi^{c}\,_{i}\,^{\beta} W_{\hat{a} e} W_{\hat{b} {e_{1}}} \lambda^{i}_{\alpha}+\frac{1}{16}{\rm i} (\Gamma^{e})^{\alpha}{}_{\beta} \psi^{c}\,_{i}\,^{\beta} W_{\hat{a} e} W_{\hat{b}}\,^{d} \lambda^{i}_{\alpha}+\frac{3}{32}{\rm i} \epsilon^{d {e_{2}} {e_{3}}}\,_{\hat{a}}\,^{{e_{1}}} (\Sigma_{{e_{2}} {e_{3}}})^{\alpha}{}_{\beta} \psi^{c}\,_{i}\,^{\beta} W_{\hat{b}}\,^{e} W_{{e_{1}} e} \lambda^{i}_{\alpha} - \frac{1}{32}{\rm i} \epsilon^{d}\,_{\hat{a} \hat{b}}\,^{e {e_{1}}} (\Gamma_{{e_{2}}})^{\alpha}{}_{\beta} \psi^{c}\,_{i}\,^{\beta} W_{e {e_{1}}} \nabla^{{e_{2}}}{\lambda^{i}_{\alpha}} - \frac{3}{16}{\rm i} \epsilon^{d}\,_{\hat{a} {e_{1}} {e_{2}}}\,^{e} (\Gamma^{{e_{1}}})^{\alpha}{}_{\beta} \psi^{c}\,_{i}\,^{\beta} W_{\hat{b} e} \nabla^{{e_{2}}}{\lambda^{i}_{\alpha}} - \frac{7}{256}{\rm i} \epsilon^{d}\,_{\hat{a} \hat{b}}\,^{{e_{2}} {e_{3}}} (\Sigma_{e {e_{1}}})^{\alpha}{}_{\beta} \psi^{c}\,_{i}\,^{\beta} W^{e {e_{1}}} W_{{e_{2}} {e_{3}}} \lambda^{i}_{\alpha}+\frac{13}{64}{\rm i} \epsilon_{\hat{a} \hat{b}}\,^{e {e_{1}} {e_{2}}} (\Sigma^{d {e_{3}}})^{\alpha}{}_{\beta} \psi^{c}\,_{i}\,^{\beta} W_{e {e_{1}}} W_{{e_{2}} {e_{3}}} \lambda^{i}_{\alpha} - \frac{1}{16}\epsilon^{d {e_{2}} {e_{3}} e {e_{1}}} (\Sigma_{{e_{2}} {e_{3}}})^{\alpha}{}_{\beta} \psi^{c}\,_{i}\,^{\beta} W W_{e {e_{1}}} W_{\hat{a} \hat{b} \alpha}\,^{i}%
+\frac{1}{4}(\Gamma^{e})^{\alpha}{}_{\beta} \psi^{c}\,_{i}\,^{\beta} W W_{e}\,^{d} W_{\hat{a} \hat{b} \alpha}\,^{i}+\frac{17}{32}{\rm i} \epsilon^{d {e_{3}}}\,_{\hat{a}}\,^{e {e_{1}}} (\Sigma_{{e_{3}}}{}^{\, {e_{2}}})^{\alpha}{}_{\beta} \psi^{c}\,_{i}\,^{\beta} W_{\hat{b} e} W_{{e_{1}} {e_{2}}} \lambda^{i}_{\alpha}+\frac{9}{64}{\rm i} \epsilon^{{e_{2}} {e_{3}}}\,_{\hat{a}}\,^{e {e_{1}}} (\Sigma_{{e_{2}} {e_{3}}})^{\alpha}{}_{\beta} \psi^{c}\,_{i}\,^{\beta} W_{\hat{b}}\,^{d} W_{e {e_{1}}} \lambda^{i}_{\alpha}+\frac{1}{32}{\rm i} \epsilon^{d}\,_{\hat{a} \hat{b} {e_{2}}}\,^{e} (\Gamma^{{e_{2}}})^{\alpha}{}_{\beta} \psi^{c}\,_{i}\,^{\beta} W_{e {e_{1}}} \nabla^{{e_{1}}}{\lambda^{i}_{\alpha}} - \frac{7}{64}{\rm i} \epsilon^{d}\,_{\hat{a} \hat{b}}\,^{e {e_{2}}} (\Sigma^{{e_{1}} {e_{3}}})^{\alpha}{}_{\beta} \psi^{c}\,_{i}\,^{\beta} W_{e {e_{1}}} W_{{e_{2}} {e_{3}}} \lambda^{i}_{\alpha}+\frac{7}{64}{\rm i} \epsilon^{d {e_{3}}}\,_{\hat{a} \hat{b}}\,^{{e_{2}}} (\Sigma_{{e_{3}} e})^{\alpha}{}_{\beta} \psi^{c}\,_{i}\,^{\beta} W^{e {e_{1}}} W_{{e_{2}} {e_{1}}} \lambda^{i}_{\alpha}+\frac{1}{256}{\rm i} \epsilon_{\hat{b}}\,^{e {e_{1}} {e_{2}} {e_{3}}} \delta^{d}\,_{\hat{a}} \psi^{c}\,_{i}\,^{\alpha} W_{e {e_{1}}} W_{{e_{2}} {e_{3}}} \lambda^{i}_{\alpha}+\frac{3}{64}{\rm i} \epsilon^{{e_{4}}}\,_{\hat{b}}\,^{e {e_{1}} {e_{2}}} (\Sigma_{{e_{4}}}{}^{\, {e_{3}}})^{\alpha}{}_{\beta} \delta^{d}\,_{\hat{a}} \psi^{c}\,_{i}\,^{\beta} W_{e {e_{1}}} W_{{e_{2}} {e_{3}}} \lambda^{i}_{\alpha}+\frac{1}{16}\epsilon^{d {e_{2}} {e_{3}} e {e_{1}}} (\Sigma_{{e_{2}} {e_{3}}})^{\alpha}{}_{\beta} \psi^{c}\,_{i}\,^{\beta} F_{e {e_{1}}} W_{\hat{a} \hat{b} \alpha}\,^{i} - \frac{1}{8}(\Gamma^{e})^{\alpha}{}_{\beta} \psi^{c}\,_{i}\,^{\beta} F_{e}\,^{d} W_{\hat{a} \hat{b} \alpha}\,^{i}+\frac{1}{16}\epsilon^{d}\,_{\hat{a} \hat{b}}\,^{{e_{2}} {e_{3}}} (\Sigma_{e {e_{1}}})^{\alpha}{}_{\beta} \psi^{c}\,_{i}\,^{\beta} F^{e {e_{1}}} W_{{e_{2}} {e_{3}} \alpha}\,^{i}+\frac{1}{16}\epsilon_{\hat{a} \hat{b}}\,^{e {e_{1}} {e_{2}}} \psi^{c}\,_{i}\,^{\alpha} F_{e}\,^{d} W_{{e_{1}} {e_{2}} \alpha}\,^{i}+\frac{1}{4}(\Gamma_{\hat{a}})^{\alpha}{}_{\beta} \psi^{c}\,_{i}\,^{\beta} F_{\hat{b}}\,^{e} W^{d}\,_{e \alpha}\,^{i} - \frac{1}{8}(\Gamma^{e})^{\alpha}{}_{\beta} \psi^{c}\,_{i}\,^{\beta} F_{\hat{a} \hat{b}} W_{e}\,^{d}\,_{\alpha}\,^{i}+\frac{1}{8}(\Gamma_{\hat{b}})^{\alpha}{}_{\beta} \delta^{d}\,_{\hat{a}} \psi^{c}\,_{i}\,^{\beta} F^{e {e_{1}}} W_{e {e_{1}} \alpha}\,^{i} - \frac{1}{4}(\Gamma^{{e_{1}}})^{\alpha}{}_{\beta} \delta^{d}\,_{\hat{a}} \psi^{c}\,_{i}\,^{\beta} F_{\hat{b}}\,^{e} W_{{e_{1}} e \alpha}\,^{i}+\frac{1}{8}\epsilon_{\hat{a} \hat{b}}\,^{e {e_{2}} {e_{3}}} (\Sigma^{d {e_{1}}})^{\alpha}{}_{\beta} \psi^{c}\,_{i}\,^{\beta} F_{e {e_{1}}} W_{{e_{2}} {e_{3}} \alpha}\,^{i} - \frac{1}{8}\epsilon^{{e_{3}}}\,_{\hat{a} \hat{b}}\,^{{e_{1}} {e_{2}}} (\Sigma_{{e_{3}} e})^{\alpha}{}_{\beta} \psi^{c}\,_{i}\,^{\beta} F^{d e} W_{{e_{1}} {e_{2}} \alpha}\,^{i} - \frac{1}{4}(\Gamma^{d})^{\alpha}{}_{\beta} \psi^{c}\,_{i}\,^{\beta} F_{\hat{a}}\,^{e} W_{\hat{b} e \alpha}\,^{i} - \frac{1}{8}\epsilon^{d}\,_{\hat{a}}\,^{e {e_{1}} {e_{2}}} \psi^{c}\,_{i}\,^{\alpha} F_{e {e_{1}}} W_{\hat{b} {e_{2}} \alpha}\,^{i}%
+\frac{1}{4}\epsilon^{{e_{3}}}\,_{\hat{a}}\,^{e {e_{1}} {e_{2}}} (\Sigma_{{e_{3}}}{}^{\, d})^{\alpha}{}_{\beta} \psi^{c}\,_{i}\,^{\beta} F_{e {e_{1}}} W_{\hat{b} {e_{2}} \alpha}\,^{i}+\frac{1}{4}\epsilon^{d {e_{2}} {e_{3}} e {e_{1}}} (\Sigma_{{e_{2}} {e_{3}}})^{\alpha}{}_{\beta} \psi^{c}\,_{i}\,^{\beta} F_{\hat{a} e} W_{\hat{b} {e_{1}} \alpha}\,^{i}+\frac{1}{4}(\Gamma^{e})^{\alpha}{}_{\beta} \psi^{c}\,_{i}\,^{\beta} F_{\hat{a}}\,^{d} W_{\hat{b} e \alpha}\,^{i} - \frac{1}{4}(\Gamma^{e})^{\alpha}{}_{\beta} \psi^{c}\,_{i}\,^{\beta} F_{\hat{a} e} W_{\hat{b}}\,^{d}\,_{\alpha}\,^{i} - \frac{1}{4}\epsilon^{d {e_{2}} {e_{3}}}\,_{\hat{a}}\,^{e} (\Sigma_{{e_{2}} {e_{3}}})^{\alpha}{}_{\beta} \psi^{c}\,_{i}\,^{\beta} F_{e}\,^{{e_{1}}} W_{\hat{b} {e_{1}} \alpha}\,^{i} - \frac{1}{4}(\Gamma^{e})^{\alpha}{}_{\beta} \delta^{d}\,_{\hat{a}} \psi^{c}\,_{i}\,^{\beta} F_{e}\,^{{e_{1}}} W_{\hat{b} {e_{1}} \alpha}\,^{i}+\frac{1}{4}(\Gamma_{\hat{a}})^{\alpha}{}_{\beta} \psi^{c}\,_{i}\,^{\beta} F^{d e} W_{\hat{b} e \alpha}\,^{i}+\frac{1}{8}\epsilon^{d}\,_{\hat{a} \hat{b}}\,^{e {e_{2}}} \psi^{c}\,_{i}\,^{\alpha} F_{e}\,^{{e_{1}}} W_{{e_{2}} {e_{1}} \alpha}\,^{i} - \frac{1}{16}\epsilon_{\hat{b}}\,^{e {e_{1}} {e_{2}} {e_{3}}} \delta^{d}\,_{\hat{a}} \psi^{c}\,_{i}\,^{\alpha} F_{e {e_{1}}} W_{{e_{2}} {e_{3}} \alpha}\,^{i}+\frac{1}{16}\epsilon_{\hat{a} \hat{b}}\,^{e {e_{1}} {e_{2}}} \psi^{c}\,_{i}\,^{\alpha} F_{e {e_{1}}} W_{{e_{2}}}\,^{d}\,_{\alpha}\,^{i} - \frac{1}{8}\epsilon^{d}\,_{\hat{a}}\,^{e {e_{1}} {e_{2}}} \psi^{c}\,_{i}\,^{\alpha} F_{\hat{b} e} W_{{e_{1}} {e_{2}} \alpha}\,^{i}+\frac{1}{4}(\Gamma_{\hat{a}})^{\alpha}{}_{\beta} \psi^{c}\,_{i}\,^{\beta} F_{\hat{b}}\,^{d} X^{i}_{\alpha} - \frac{1}{4}(\Gamma^{d})^{\alpha}{}_{\beta} \psi^{c}\,_{i}\,^{\beta} F_{\hat{a} \hat{b}} X^{i}_{\alpha} - \frac{1}{4}(\Gamma^{e})^{\alpha}{}_{\beta} \delta^{d}\,_{\hat{a}} \psi^{c}\,_{i}\,^{\beta} F_{\hat{b} e} X^{i}_{\alpha} - \frac{1}{24}{\rm i} \epsilon^{d e {e_{1}}}\,_{\hat{a} \hat{b}} \Phi_{e {e_{1}} i}\,^{j} \psi^{c}\,_{j}\,^{\alpha} \lambda^{i}_{\alpha} - \frac{1}{2}{\rm i} \Phi_{\hat{a}}\,^{d}\,_{i}\,^{j} (\Gamma_{\hat{b}})^{\alpha}{}_{\beta} \psi^{c}\,_{j}\,^{\beta} \lambda^{i}_{\alpha} - \frac{1}{12}{\rm i} \Phi_{\hat{a} \hat{b} i}\,^{j} (\Gamma^{d})^{\alpha}{}_{\beta} \psi^{c}\,_{j}\,^{\beta} \lambda^{i}_{\alpha} - \frac{1}{6}{\rm i} \epsilon^{d e {e_{2}}}\,_{\hat{a} \hat{b}} \Phi_{e}\,^{{e_{1}}}\,_{i}\,^{j} (\Sigma_{{e_{2}} {e_{1}}})^{\alpha}{}_{\beta} \psi^{c}\,_{j}\,^{\beta} \lambda^{i}_{\alpha} - \frac{1}{12}{\rm i} \epsilon^{e {e_{1}} {e_{2}} {e_{3}}}\,_{\hat{b}} \Phi_{e {e_{1}} i}\,^{j} (\Sigma_{{e_{2}} {e_{3}}})^{\alpha}{}_{\beta} \delta^{d}\,_{\hat{a}} \psi^{c}\,_{j}\,^{\beta} \lambda^{i}_{\alpha}+\frac{1}{6}{\rm i} \Phi_{\hat{a} e i}\,^{j} (\Gamma^{e})^{\alpha}{}_{\beta} \delta^{d}\,_{\hat{b}} \psi^{c}\,_{j}\,^{\beta} \lambda^{i}_{\alpha}%
+\frac{1}{6}{\rm i} \epsilon^{e {e_{1}} {e_{2}}}\,_{\hat{a} \hat{b}} \Phi_{e {e_{1}} i}\,^{j} (\Sigma_{{e_{2}}}{}^{\, d})^{\alpha}{}_{\beta} \psi^{c}\,_{j}\,^{\beta} \lambda^{i}_{\alpha} - \frac{1}{3}{\rm i} \epsilon^{d e {e_{1}} {e_{2}}}\,_{\hat{b}} \Phi_{\hat{a} e i}\,^{j} (\Sigma_{{e_{1}} {e_{2}}})^{\alpha}{}_{\beta} \psi^{c}\,_{j}\,^{\beta} \lambda^{i}_{\alpha} - \frac{1}{64}{\rm i} \epsilon^{d}\,_{\hat{a} \hat{b}}\,^{e {e_{1}}} (\Gamma_{{e_{2}}})^{\alpha}{}_{\beta} \psi^{c}\,_{i}\,^{\beta} \lambda^{i}_{\alpha} \nabla^{{e_{2}}}{W_{e {e_{1}}}}+\frac{1}{4}{\rm i} \psi^{c}\,_{i}\,^{\alpha} \lambda^{i}_{\alpha} \nabla_{\hat{a}}{W^{d}\,_{\hat{b}}} - \frac{1}{2}{\rm i} (\Sigma_{\hat{a} e})^{\alpha}{}_{\beta} \psi^{c}\,_{i}\,^{\beta} \lambda^{i}_{\alpha} \nabla^{e}{W^{d}\,_{\hat{b}}} - \frac{1}{4}{\rm i} (\Sigma^{d}{}_{\, e})^{\alpha}{}_{\beta} \psi^{c}\,_{i}\,^{\beta} \lambda^{i}_{\alpha} \nabla^{e}{W_{\hat{a} \hat{b}}}+\frac{3}{8}{\rm i} (\Sigma_{\hat{a} \hat{b}})^{\alpha}{}_{\beta} \psi^{c}\,_{i}\,^{\beta} \lambda^{i}_{\alpha} \nabla_{e}{W^{d e}}+\frac{1}{2}{\rm i} (\Sigma^{d}{}_{\, e})^{\alpha}{}_{\beta} \psi^{c}\,_{i}\,^{\beta} \lambda^{i}_{\alpha} \nabla_{\hat{a}}{W^{e}\,_{\hat{b}}}+\frac{1}{4}{\rm i} (\Sigma_{\hat{a}}{}^{\, d})^{\alpha}{}_{\beta} \psi^{c}\,_{i}\,^{\beta} \lambda^{i}_{\alpha} \nabla^{e}{W_{\hat{b} e}} - \frac{5}{64}{\rm i} \epsilon^{d}\,_{\hat{a} \hat{b} {e_{2}}}\,^{e} (\Gamma^{{e_{1}}})^{\alpha}{}_{\beta} \psi^{c}\,_{i}\,^{\beta} \lambda^{i}_{\alpha} \nabla^{{e_{2}}}{W_{e {e_{1}}}} - \frac{1}{64}{\rm i} \epsilon^{d}\,_{\hat{a} \hat{b} {e_{2}}}\,^{e} (\Gamma^{{e_{2}}})^{\alpha}{}_{\beta} \psi^{c}\,_{i}\,^{\beta} \lambda^{i}_{\alpha} \nabla^{{e_{1}}}{W_{e {e_{1}}}} - \frac{1}{4}{\rm i} (\Sigma_{e {e_{1}}})^{\alpha}{}_{\beta} \delta^{d}\,_{\hat{a}} \psi^{c}\,_{i}\,^{\beta} \lambda^{i}_{\alpha} \nabla_{\hat{b}}{W^{e {e_{1}}}} - \frac{1}{4}{\rm i} (\Sigma_{\hat{a} e})^{\alpha}{}_{\beta} \delta^{d}\,_{\hat{b}} \psi^{c}\,_{i}\,^{\beta} \lambda^{i}_{\alpha} \nabla_{{e_{1}}}{W^{e {e_{1}}}} - \frac{1}{8}{\rm i} \delta^{d}\,_{\hat{a}} \psi^{c}\,_{i}\,^{\alpha} \lambda^{i}_{\alpha} \nabla^{e}{W_{\hat{b} e}} - \frac{1}{2}{\rm i} (\Sigma_{{e_{1}} e})^{\alpha}{}_{\beta} \delta^{d}\,_{\hat{a}} \psi^{c}\,_{i}\,^{\beta} \lambda^{i}_{\alpha} \nabla^{{e_{1}}}{W^{e}\,_{\hat{b}}}+\frac{1}{2}{\rm i} (\Sigma_{\hat{a} e})^{\alpha}{}_{\beta} \psi^{c}\,_{i}\,^{\beta} \lambda^{i}_{\alpha} \nabla_{\hat{b}}{W^{d e}}+\frac{1}{16}{\rm i} \epsilon_{\hat{a} \hat{b} {e_{2}}}\,^{e {e_{1}}} (\Gamma^{d})^{\alpha}{}_{\beta} \psi^{c}\,_{i}\,^{\beta} \lambda^{i}_{\alpha} \nabla^{{e_{2}}}{W_{e {e_{1}}}}+\frac{3}{32}{\rm i} \epsilon^{d}\,_{\hat{a} {e_{1}} {e_{2}}}\,^{e} (\Gamma^{{e_{1}}})^{\alpha}{}_{\beta} \psi^{c}\,_{i}\,^{\beta} \lambda^{i}_{\alpha} \nabla^{{e_{2}}}{W_{\hat{b} e}} - \frac{5}{64}{\rm i} \epsilon_{\hat{a} \hat{b} {e_{1}} {e_{2}} e} (\Gamma^{{e_{1}}})^{\alpha}{}_{\beta} \psi^{c}\,_{i}\,^{\beta} \lambda^{i}_{\alpha} \nabla^{{e_{2}}}{W^{d e}} - \frac{3}{32}{\rm i} \epsilon^{d}\,_{\hat{a} {e_{2}}}\,^{e {e_{1}}} (\Gamma^{{e_{2}}})^{\alpha}{}_{\beta} \psi^{c}\,_{i}\,^{\beta} \lambda^{i}_{\alpha} \nabla_{\hat{b}}{W_{e {e_{1}}}}%
+\frac{1}{32}{\rm i} \epsilon^{{e_{4}} e {e_{1}} {e_{2}} {e_{3}}} (\Sigma_{\hat{a} {e_{4}}})^{\alpha}{}_{\beta} \delta^{d}\,_{\hat{b}} \psi^{c}\,_{i}\,^{\beta} W_{e {e_{1}}} W_{{e_{2}} {e_{3}}} \lambda^{i}_{\alpha}+\frac{3}{32}{\rm i} \epsilon^{d e {e_{1}} {e_{2}} {e_{3}}} (\Sigma_{\hat{a} \hat{b}})^{\alpha}{}_{\beta} \psi^{c}\,_{i}\,^{\beta} W_{e {e_{1}}} W_{{e_{2}} {e_{3}}} \lambda^{i}_{\alpha}+2(\Sigma_{\hat{a} \hat{b}})^{\alpha}{}_{\beta} \phi^{c}\,_{i}\,^{\beta} \nabla^{d}{\lambda^{i}_{\alpha}}-\delta^{d}\,_{\hat{a}} \phi^{c}\,_{i}\,^{\alpha} \nabla_{\hat{b}}{\lambda^{i}_{\alpha}}+2(\Sigma_{\hat{a}}{}^{\, d})^{\alpha}{}_{\beta} \phi^{c}\,_{i}\,^{\beta} \nabla_{\hat{b}}{\lambda^{i}_{\alpha}} - \frac{1}{2}(\Gamma^{d})^{\alpha}{}_{\beta} \phi^{c}\,_{i}\,^{\beta} W_{\hat{a} \hat{b}} \lambda^{i}_{\alpha}+\frac{1}{8}\epsilon^{d}\,_{\hat{a} \hat{b}}\,^{e {e_{1}}} \phi^{c}\,_{i}\,^{\alpha} W_{e {e_{1}}} \lambda^{i}_{\alpha}+\frac{1}{16}\epsilon^{{e_{2}}}\,_{\hat{a} \hat{b}}\,^{e {e_{1}}} (\Sigma_{{e_{2}}}{}^{\, d})^{\alpha}{}_{\beta} \phi^{c}\,_{i}\,^{\beta} W_{e {e_{1}}} \lambda^{i}_{\alpha} - \frac{1}{8}\epsilon^{d {e_{1}} {e_{2}}}\,_{\hat{a}}\,^{e} (\Sigma_{{e_{1}} {e_{2}}})^{\alpha}{}_{\beta} \phi^{c}\,_{i}\,^{\beta} W_{\hat{b} e} \lambda^{i}_{\alpha} - \frac{1}{4}(\Gamma^{e})^{\alpha}{}_{\beta} \delta^{d}\,_{\hat{a}} \phi^{c}\,_{i}\,^{\beta} W_{\hat{b} e} \lambda^{i}_{\alpha}+\frac{3}{4}(\Gamma_{\hat{a}})^{\alpha}{}_{\beta} \phi^{c}\,_{i}\,^{\beta} W_{\hat{b}}\,^{d} \lambda^{i}_{\alpha} - \frac{1}{8}\epsilon^{d {e_{2}}}\,_{\hat{a} \hat{b}}\,^{e} (\Sigma_{{e_{2}}}{}^{\, {e_{1}}})^{\alpha}{}_{\beta} \phi^{c}\,_{i}\,^{\beta} W_{e {e_{1}}} \lambda^{i}_{\alpha} - \frac{5}{16}\epsilon^{{e_{2}} {e_{3}}}\,_{\hat{b}}\,^{e {e_{1}}} (\Sigma_{{e_{2}} {e_{3}}})^{\alpha}{}_{\beta} \delta^{d}\,_{\hat{a}} \phi^{c}\,_{i}\,^{\beta} W_{e {e_{1}}} \lambda^{i}_{\alpha}+\frac{1}{2}{\rm i} (\Gamma^{d})^{\alpha}{}_{\beta} \phi^{c}\,_{i}\,^{\beta} W W_{\hat{a} \hat{b} \alpha}\,^{i}-4F_{\hat{a} \hat{b}} f^{c d}-4\delta^{d}\,_{\hat{a}} F_{\hat{b} e} f^{c e}-4F_{\hat{a}}\,^{d} f^{c}\,_{\hat{b}}
\doublespacedmathend
\end{adjustwidth}

\subsubsection{$\nabla^{a}\Box W$}
\begin{adjustwidth}{0cm}{5cm}
\doublespacedmathbegin
\mathcal{D}^{a}{\nabla_{b}{\nabla^{b}{W}}} - \frac{1}{2}{\rm i} \psi^{a}\,_{i}\,^{\alpha} \nabla_{b}{\nabla^{b}{\lambda^{i}_{\alpha}}}+\frac{3}{32}{\rm i} \epsilon^{d e}\,_{{e_{1}}}\,^{b c} (\Sigma_{d e})^{\alpha}{}_{\beta} \psi^{a}\,_{i}\,^{\beta} \lambda^{i}_{\alpha} \nabla^{{e_{1}}}{W_{b c}} - \frac{9}{32}{\rm i} (\Gamma^{b})^{\alpha}{}_{\beta} \psi^{a}\,_{i}\,^{\beta} \lambda^{i}_{\alpha} \nabla^{c}{W_{b c}}+\frac{1}{8}{\rm i} \epsilon^{d e}\,_{{e_{1}}}\,^{b c} (\Sigma_{d e})^{\alpha}{}_{\beta} \psi^{a}\,_{i}\,^{\beta} W_{b c} \nabla^{{e_{1}}}{\lambda^{i}_{\alpha}} - \frac{1}{2}{\rm i} (\Gamma^{b})^{\alpha}{}_{\beta} \psi^{a}\,_{i}\,^{\beta} W_{b c} \nabla^{c}{\lambda^{i}_{\alpha}}+\frac{1}{8}(\Gamma_{b})^{\alpha}{}_{\beta} \psi^{a}\,_{i}\,^{\beta} X^{i}_{\alpha} \nabla^{b}{W}+\frac{9}{128}{\rm i} \psi^{a}\,_{i}\,^{\alpha} W^{b c} W_{b c} \lambda^{i}_{\alpha}+\frac{1}{8}{\rm i} \Phi^{b c}\,_{i}\,^{j} (\Sigma_{b c})^{\alpha}{}_{\beta} \psi^{a}\,_{j}\,^{\beta} \lambda^{i}_{\alpha}+\frac{1}{32}\psi^{a}\,_{i}\,^{\alpha} W W^{b c} W_{b c \alpha}\,^{i}+(\Gamma_{b})^{\alpha}{}_{\beta} \phi^{a}\,_{i}\,^{\beta} \nabla^{b}{\lambda^{i}_{\alpha}}+\frac{9}{8}(\Sigma_{b c})^{\alpha}{}_{\beta} \phi^{a}\,_{i}\,^{\beta} W^{b c} \lambda^{i}_{\alpha}+2\nabla^{b}{W} f^{a}\,_{b}
\doublespacedmathend
\end{adjustwidth}

\subsubsection{$\nabla^{a}\Box \lambda^{i}_{\alpha}$}

\begin{adjustwidth}{1em}{5cm}
\doublespacedmathbegin
\mathcal{D}^{a}{\nabla_{b}{\nabla^{b}{\lambda^{i}_{\alpha}}}}+\frac{55}{256}(\Sigma_{b c})_{\alpha \beta} \psi^{a i \beta} W W^{b c} W^{d e} W_{d e} - \frac{19}{32}(\Sigma^{d e})_{\alpha \beta} \psi^{a i \beta} W W^{b c} W_{d b} W_{e c}+\frac{19}{256}\epsilon^{b c d e {e_{1}}} (\Gamma^{{e_{2}}})_{\alpha \beta} \psi^{a i \beta} W W_{b c} W_{d e} W_{{e_{1}} {e_{2}}} - \frac{37}{128}(\Gamma_{d})_{\alpha \beta} \psi^{a i \beta} W^{b c} W_{b c} \nabla^{d}{W}+\frac{1}{32}\epsilon_{{e_{1}}}\,^{b c d e} \psi^{a i}\,_{\alpha} W_{b c} W_{d e} \nabla^{{e_{1}}}{W}+\frac{1}{8}\epsilon^{{e_{1}} b c d e} (\Sigma_{{e_{1}} {e_{2}}})_{\alpha \beta} \psi^{a i \beta} W_{b c} W_{d e} \nabla^{{e_{2}}}{W} - \frac{5}{64}\psi^{a}\,_{j \alpha} W^{b c} \lambda^{i \beta} W_{b c \beta}\,^{j}+\frac{1}{128}\epsilon_{{e_{1}}}\,^{b c d e} (\Gamma^{{e_{1}}})_{\alpha \rho} \psi^{a}\,_{j}\,^{\rho} W_{b c} \lambda^{i \beta} W_{d e \beta}\,^{j}+\frac{1}{16}(\Sigma^{d}{}_{\,b})_{\alpha \rho} \psi^{a}\,_{j}\,^{\rho} W^{b c} \lambda^{i \beta} W_{d c \beta}\,^{j} - \frac{3}{64}(\Sigma_{b c})^{\beta}{}_{\rho} \psi^{a i \rho} W^{b c} \lambda_{j \alpha} X^{j}_{\beta}+\frac{3}{64}(\Sigma_{b c})^{\beta}{}_{\rho} \psi^{a}\,_{j}\,^{\rho} W^{b c} \lambda^{j}_{\alpha} X^{i}_{\beta} - \frac{23}{64}(\Gamma_{d})_{\alpha \beta} \psi^{a i \beta} W W_{b c} \nabla^{d}{W^{b c}} - \frac{1}{128}\epsilon_{{e_{1}}}\,^{d e b c} \psi^{a i}\,_{\alpha} W W_{b c} \nabla^{{e_{1}}}{W_{d e}}+\frac{63}{256}\epsilon^{{e_{1}} d e b c} (\Sigma_{{e_{1}} {e_{2}}})_{\alpha \beta} \psi^{a i \beta} W W_{b c} \nabla^{{e_{2}}}{W_{d e}}+\frac{17}{128}\epsilon^{e {e_{1}}}\,_{{e_{2}} d}\,^{b} (\Sigma_{e {e_{1}}})_{\alpha \beta} \psi^{a i \beta} W W_{b c} \nabla^{{e_{2}}}{W^{d c}}+\frac{1}{32}(\Gamma_{d})_{\alpha \beta} \psi^{a i \beta} W W_{b c} \nabla^{b}{W^{d c}}+\frac{11}{64}(\Gamma^{b})_{\alpha \beta} \psi^{a i \beta} W W_{b c} \nabla_{d}{W^{d c}} - \frac{1}{16}\Phi_{b c}\,^{i j} \psi^{a}\,_{j \alpha} F^{b c}%
+\frac{1}{32}\epsilon^{e {e_{1}}}\,_{d}\,^{b c} \Phi_{e {e_{1}}}\,^{i j} (\Gamma^{d})_{\alpha \beta} \psi^{a}\,_{j}\,^{\beta} F_{b c}+\frac{1}{4}\Phi^{d}\,_{b}\,^{i j} (\Sigma_{d c})_{\alpha \beta} \psi^{a}\,_{j}\,^{\beta} F^{b c} - \frac{1}{32}\Phi_{b c}\,^{i j} \psi^{a}\,_{j \alpha} W W^{b c}+\frac{7}{192}\epsilon^{e {e_{1}}}\,_{d}\,^{b c} \Phi_{e {e_{1}}}\,^{i j} (\Gamma^{d})_{\alpha \beta} \psi^{a}\,_{j}\,^{\beta} W W_{b c}+\frac{11}{24}\Phi^{d}\,_{b}\,^{i j} (\Sigma_{d c})_{\alpha \beta} \psi^{a}\,_{j}\,^{\beta} W W^{b c}+\frac{1}{8}\Phi^{b c}\,_{j}\,^{k} (\Sigma_{b c})_{\alpha \beta} \psi^{a}\,_{k}\,^{\beta} X^{i j}+\frac{1}{48}\epsilon^{b c d e}\,_{{e_{1}}} \Phi_{b c}\,^{i j} (\Sigma_{d e})_{\alpha \beta} \psi^{a}\,_{j}\,^{\beta} \nabla^{{e_{1}}}{W} - \frac{5}{8}\Phi_{b c}\,^{i j} (\Gamma^{b})_{\alpha \beta} \psi^{a}\,_{j}\,^{\beta} \nabla^{c}{W} - \frac{3}{32}(\Gamma_{d})_{\alpha \beta} \psi^{a i \beta} F_{b c} \nabla^{d}{W^{b c}} - \frac{3}{64}\epsilon_{{e_{1}}}\,^{d e b c} \psi^{a i}\,_{\alpha} F_{b c} \nabla^{{e_{1}}}{W_{d e}}+\frac{15}{128}\epsilon^{{e_{1}} d e b c} (\Sigma_{{e_{1}} {e_{2}}})_{\alpha \beta} \psi^{a i \beta} F_{b c} \nabla^{{e_{2}}}{W_{d e}} - \frac{15}{64}\epsilon^{e {e_{1}}}\,_{{e_{2}} d}\,^{b} (\Sigma_{e {e_{1}}})_{\alpha \beta} \psi^{a i \beta} F_{b c} \nabla^{{e_{2}}}{W^{d c}}+\frac{3}{16}(\Gamma_{d})_{\alpha \beta} \psi^{a i \beta} F_{b c} \nabla^{b}{W^{d c}} - \frac{9}{32}(\Gamma^{b})_{\alpha \beta} \psi^{a i \beta} F_{b c} \nabla_{d}{W^{d c}} - \frac{9}{128}\epsilon^{{e_{1}}}\,_{{e_{2}} e}\,^{b c} (\Sigma_{{e_{1}} d})_{\alpha \beta} \psi^{a i \beta} W W_{b c} \nabla^{{e_{2}}}{W^{d e}} - \frac{9}{256}\epsilon^{{e_{1}} {e_{2}} d e b} (\Sigma_{{e_{1}} {e_{2}}})_{\alpha \beta} \psi^{a i \beta} W W_{b c} \nabla^{c}{W_{d e}}+\frac{3}{32}\epsilon^{d e}\,_{{e_{1}}}\,^{b c} (\Sigma_{d e})_{\alpha \beta} \psi^{a}\,_{j}\,^{\beta} X^{i j} \nabla^{{e_{1}}}{W_{b c}} - \frac{9}{32}(\Gamma^{b})_{\alpha \beta} \psi^{a}\,_{j}\,^{\beta} X^{i j} \nabla^{c}{W_{b c}} - \frac{9}{16}(\Sigma_{b c})_{\alpha \beta} \psi^{a i \beta} \nabla_{d}{W} \nabla^{d}{W^{b c}}+\frac{7}{8}(\Sigma_{d b})_{\alpha \beta} \psi^{a i \beta} \nabla_{c}{W} \nabla^{d}{W^{b c}}%
+\frac{1}{32}\epsilon_{d e {e_{1}} b c} (\Gamma^{d})_{\alpha \beta} \psi^{a i \beta} \nabla^{e}{W} \nabla^{{e_{1}}}{W^{b c}}+\frac{3}{32}\psi^{a i}\,_{\alpha} \nabla^{b}{W} \nabla^{c}{W_{b c}}+\frac{13}{16}(\Sigma_{d b})_{\alpha \beta} \psi^{a i \beta} \nabla^{d}{W} \nabla_{c}{W^{b c}}+\frac{9}{128}(\Sigma_{b c})_{\alpha \beta} \psi^{a i \beta} W^{d e} W_{d e} F^{b c}+\frac{9}{128}\psi^{a}\,_{j \alpha} X^{i j} W^{b c} W_{b c} - \frac{3}{64}\psi^{a}\,_{j}\,^{\beta} W^{b c} \lambda^{i}_{\alpha} W_{b c \beta}\,^{j}+\frac{1}{64}\psi^{a}\,_{j}\,^{\beta} W^{b c} \lambda^{i}_{\beta} W_{b c \alpha}\,^{j}+\frac{1}{128}\epsilon_{{e_{1}}}\,^{b c d e} (\Gamma^{{e_{1}}})_{\alpha}{}^{\beta} \psi^{a}\,_{j}\,^{\rho} W_{b c} \lambda^{i}_{\rho} W_{d e \beta}\,^{j}+\frac{1}{16}(\Sigma^{d}{}_{\, b})_{\alpha}{}^{\beta} \psi^{a}\,_{j}\,^{\rho} W^{b c} \lambda^{i}_{\rho} W_{d c \beta}\,^{j} - \frac{1}{2}(\Sigma_{b c})_{\alpha \beta} \psi^{a i \beta} \nabla_{d}{\nabla^{d}{F^{b c}}} - \frac{1}{4}(\Sigma_{b c})_{\alpha \beta} \psi^{a i \beta} W^{b c} \nabla_{d}{\nabla^{d}{W}} - \frac{3}{8}(\Sigma_{b c})_{\alpha \beta} \psi^{a i \beta} W \nabla_{d}{\nabla^{d}{W^{b c}}} - \frac{1}{2}\psi^{a}\,_{j \alpha} \nabla_{b}{\nabla^{b}{X^{i j}}}+\frac{1}{2}(\Gamma_{b})_{\alpha \beta} \psi^{a i \beta} \nabla_{c}{\nabla^{c}{\nabla^{b}{W}}} - \frac{1}{8}(\Gamma_{d})_{\alpha \beta} \psi^{a i \beta} W_{b c} \nabla^{d}{F^{b c}} - \frac{1}{16}\epsilon_{{e_{1}}}\,^{b c d e} \psi^{a i}\,_{\alpha} W_{d e} \nabla^{{e_{1}}}{F_{b c}}+\frac{3}{16}\epsilon^{{e_{1}} b c d e} (\Sigma_{{e_{1}} {e_{2}}})_{\alpha \beta} \psi^{a i \beta} W_{d e} \nabla^{{e_{2}}}{F_{b c}}+\frac{3}{8}\epsilon^{e {e_{1}}}\,_{{e_{2}} b}\,^{d} (\Sigma_{e {e_{1}}})_{\alpha \beta} \psi^{a i \beta} W_{d c} \nabla^{{e_{2}}}{F^{b c}} - \frac{1}{2}(\Gamma_{b})_{\alpha \beta} \psi^{a i \beta} W_{d c} \nabla^{d}{F^{b c}}+\frac{1}{4}(\Gamma^{d})_{\alpha \beta} \psi^{a i \beta} W_{d c} \nabla_{b}{F^{b c}}%
+\frac{1}{8}\epsilon^{d e}\,_{{e_{1}}}\,^{b c} (\Sigma_{d e})_{\alpha \beta} \psi^{a}\,_{j}\,^{\beta} W_{b c} \nabla^{{e_{1}}}{X^{i j}} - \frac{1}{2}(\Gamma^{b})_{\alpha \beta} \psi^{a}\,_{j}\,^{\beta} W_{b c} \nabla^{c}{X^{i j}}-(\Sigma^{b}{}_{\,d})_{\alpha \beta} \psi^{a i \beta} W_{b c} \nabla^{c}{\nabla^{d}{W}}+\frac{1}{8}\epsilon_{d e {e_{1}}}\,^{b c} (\Gamma^{d})_{\alpha \beta} \psi^{a i \beta} W_{b c} \nabla^{e}{\nabla^{{e_{1}}}{W}}+\frac{1}{2}\psi^{a i}\,_{\alpha} W_{b c} \nabla^{b}{\nabla^{c}{W}} - \frac{1}{2}(\Sigma^{b}{}_{\,d})_{\alpha \beta} \psi^{a i \beta} W_{b c} \nabla^{d}{\nabla^{c}{W}}+\frac{3}{64}\epsilon^{{e_{1}}}\,_{{e_{2}} e}\,^{b c} (\Sigma_{{e_{1}} d})_{\alpha \beta} \psi^{a i \beta} F_{b c} \nabla^{{e_{2}}}{W^{d e}}+\frac{3}{128}\epsilon^{{e_{1}} {e_{2}} d e b} (\Sigma_{{e_{1}} {e_{2}}})_{\alpha \beta} \psi^{a i \beta} F_{b c} \nabla^{c}{W_{d e}} - \frac{1}{8}\epsilon^{{e_{1}}}\,_{{e_{2}}}\,^{b c d} (\Sigma_{{e_{1}}}{}^{\, e})_{\alpha \beta} \psi^{a i \beta} W_{d e} \nabla^{{e_{2}}}{F_{b c}}+\frac{1}{16}\epsilon^{{e_{1}} {e_{2}} b d e} (\Sigma_{{e_{1}} {e_{2}}})_{\alpha \beta} \psi^{a i \beta} W_{d e} \nabla^{c}{F_{b c}} - \frac{1}{4}(\Gamma^{b})_{\alpha \beta} \psi^{a i \beta} W_{b}\,^{c} W_{c d} \nabla^{d}{W} - \frac{1}{8}\epsilon^{{e_{1}}}\,_{{e_{2}}}\,^{b c d} (\Sigma_{{e_{1}}}{}^{\, e})_{\alpha \beta} \psi^{a i \beta} W_{b c} W_{d e} \nabla^{{e_{2}}}{W}+\frac{1}{16}\epsilon^{{e_{1}} {e_{2}} b c d} (\Sigma_{{e_{1}} {e_{2}}})_{\alpha \beta} \psi^{a i \beta} W_{b c} W_{d e} \nabla^{e}{W}+\frac{1}{32}\epsilon^{{e_{1}}}\,_{{e_{2}}}\,^{d e b} (\Sigma_{{e_{1}}}{}^{\, c})_{\alpha \beta} \psi^{a i \beta} W W_{b c} \nabla^{{e_{2}}}{W_{d e}} - \frac{1}{64}\epsilon^{{e_{1}} {e_{2}} d b c} (\Sigma_{{e_{1}} {e_{2}}})_{\alpha \beta} \psi^{a i \beta} W W_{b c} \nabla^{e}{W_{d e}} - \frac{3}{16}\epsilon^{d e}\,_{{e_{1}}}\,^{b c} (\Sigma_{d e})_{\alpha \rho} \psi^{a}\,_{j}\,^{\rho} \lambda^{i \beta} \nabla^{{e_{1}}}{W_{b c \beta}\,^{j}}+\frac{3}{8}(\Gamma^{b})_{\alpha \rho} \psi^{a}\,_{j}\,^{\rho} \lambda^{i \beta} \nabla^{c}{W_{b c \beta}\,^{j}} - \frac{1}{4}\epsilon^{d e}\,_{{e_{1}}}\,^{b c} (\Sigma_{d e})_{\alpha \rho} \psi^{a}\,_{j}\,^{\rho} W_{b c}\,^{\beta j} \nabla^{{e_{1}}}{\lambda^{i}_{\beta}}+\frac{1}{2}(\Gamma^{b})_{\alpha \rho} \psi^{a}\,_{j}\,^{\rho} W_{b c}\,^{\beta j} \nabla^{c}{\lambda^{i}_{\beta}}+\frac{1}{8}\epsilon^{d e}\,_{{e_{1}}}\,^{b c} (\Sigma_{d e})_{\alpha}{}^{\rho} \psi^{a}\,_{j}\,^{\beta} W_{b c \beta}\,^{j} \nabla^{{e_{1}}}{\lambda^{i}_{\rho}}%
 - \frac{1}{8}(\Gamma_{b})_{\alpha}{}^{\beta} \psi^{a}\,_{j}\,^{\rho} X^{j}_{\beta} \nabla^{b}{\lambda^{i}_{\rho}}+\frac{1}{8}(\Gamma_{b})^{\beta \rho} \psi^{a}\,_{j \alpha} X^{j}_{\beta} \nabla^{b}{\lambda^{i}_{\rho}} - \frac{1}{8}(\Gamma_{b})_{\alpha \rho} \psi^{a}\,_{j}\,^{\rho} X^{j \beta} \nabla^{b}{\lambda^{i}_{\beta}} - \frac{1}{8}(\Gamma_{b})^{\beta}{}_{\rho} \psi^{a}\,_{j}\,^{\rho} X^{j}_{\alpha} \nabla^{b}{\lambda^{i}_{\beta}}+\frac{3}{16}(\Gamma_{b})^{\beta}{}_{\rho} \psi^{a i \rho} \lambda_{j \alpha} \nabla^{b}{X^{j}_{\beta}}+\frac{3}{8}(\Gamma_{b})^{\beta}{}_{\rho} \psi^{a i \rho} X_{j \beta} \nabla^{b}{\lambda^{j}_{\alpha}} - \frac{3}{16}(\Gamma_{b})^{\beta}{}_{\rho} \psi^{a}\,_{j}\,^{\rho} \lambda^{j}_{\alpha} \nabla^{b}{X^{i}_{\beta}}+\frac{3}{8}(\Gamma_{b})^{\beta}{}_{\rho} \psi^{a}\,_{j}\,^{\rho} X^{i}_{\beta} \nabla^{b}{\lambda^{j}_{\alpha}}+\frac{1}{24}\epsilon^{d e}\,_{{e_{1}} b c} (\Sigma_{d e})_{\alpha \beta} \psi^{a}\,_{j}\,^{\beta} W \nabla^{{e_{1}}}{\Phi^{b c i j}} - \frac{1}{4}(\Gamma_{b})_{\alpha \beta} \psi^{a}\,_{j}\,^{\beta} W \nabla_{c}{\Phi^{b c i j}}+\frac{1}{4}(\Sigma_{d b})_{\alpha \beta} \psi^{a i \beta} W \nabla_{c}{\nabla^{d}{W^{b c}}}+\frac{1}{16}\epsilon_{d e {e_{1}} b c} (\Gamma^{d})_{\alpha \beta} \psi^{a i \beta} W \nabla^{e}{\nabla^{{e_{1}}}{W^{b c}}}+\frac{3}{16}\psi^{a i}\,_{\alpha} W \nabla^{b}{\nabla^{c}{W_{b c}}}+\frac{1}{8}(\Sigma_{d b})_{\alpha \beta} \psi^{a i \beta} W \nabla^{d}{\nabla_{c}{W^{b c}}}+\frac{9}{16}{\rm i} \phi^{a i}\,_{\alpha} W^{b c} F_{b c} - \frac{9}{32}{\rm i} \epsilon_{{e_{1}}}\,^{d e b c} (\Gamma^{{e_{1}}})_{\alpha \beta} \phi^{a i \beta} W_{d e} F_{b c} - \frac{9}{4}{\rm i} (\Sigma^{b}{}_{\,d})_{\alpha \beta} \phi^{a i \beta} W^{d c} F_{b c}+\frac{39}{64}{\rm i} \phi^{a i}\,_{\alpha} W W^{b c} W_{b c} - \frac{11}{64}{\rm i} \epsilon_{{e_{1}}}\,^{b c d e} (\Gamma^{{e_{1}}})_{\alpha \beta} \phi^{a i \beta} W W_{b c} W_{d e} - \frac{9}{8}{\rm i} (\Sigma_{b c})_{\alpha \beta} \phi^{a}\,_{j}\,^{\beta} X^{i j} W^{b c}%
+\frac{1}{16}{\rm i} \epsilon^{d e}\,_{{e_{1}}}\,^{b c} (\Sigma_{d e})_{\alpha \beta} \phi^{a i \beta} W_{b c} \nabla^{{e_{1}}}{W}+\frac{21}{8}{\rm i} (\Gamma^{b})_{\alpha \beta} \phi^{a i \beta} W_{b c} \nabla^{c}{W} - \frac{11}{8}{\rm i} (\Sigma^{b c})_{\alpha}{}^{\rho} \phi^{a}\,_{j}\,^{\beta} \lambda^{i}_{\rho} W_{b c \beta}\,^{j} - \frac{15}{16}{\rm i} \phi^{a i \beta} \lambda_{j \alpha} X^{j}_{\beta}+\frac{15}{16}{\rm i} \phi^{a}\,_{j}\,^{\beta} \lambda^{j}_{\alpha} X^{i}_{\beta} - \frac{3}{4}{\rm i} \Phi^{b c i j} (\Sigma_{b c})_{\alpha \beta} \phi^{a}\,_{j}\,^{\beta} W - \frac{3}{16}{\rm i} \epsilon^{d e}\,_{{e_{1}}}\,^{b c} (\Sigma_{d e})_{\alpha \beta} \phi^{a i \beta} W \nabla^{{e_{1}}}{W_{b c}}+\frac{27}{16}{\rm i} (\Gamma^{b})_{\alpha \beta} \phi^{a i \beta} W \nabla^{c}{W_{b c}}-2{\rm i} \phi^{a i}\,_{\alpha} \nabla_{b}{\nabla^{b}{W}} - \frac{1}{2}{\rm i} \epsilon^{d e}\,_{{e_{1}}}\,^{b c} (\Sigma_{d e})_{\alpha \beta} \phi^{a i \beta} \nabla^{{e_{1}}}{F_{b c}}+{\rm i} (\Gamma^{b})_{\alpha \beta} \phi^{a i \beta} \nabla^{c}{F_{b c}}-{\rm i} (\Gamma_{b})_{\alpha \beta} \phi^{a}\,_{j}\,^{\beta} \nabla^{b}{X^{i j}}-2{\rm i} (\Sigma_{b c})_{\alpha \beta} \phi^{a i \beta} \nabla^{b}{\nabla^{c}{W}}-4(\Sigma_{b c})_{\alpha}{}^{\beta} \nabla^{c}{\lambda^{i}_{\beta}} f^{a b}
\doublespacedmathend
\end{adjustwidth}

\subsubsection{$\nabla^{a}\Box F_{a b}$}

\begin{adjustwidth}{1em}{5cm}
\doublespacedmathbegin
\mathcal{D}^{c}{\nabla_{d}{\nabla^{d}{F_{\hat{a} \hat{b}}}}}+\frac{7}{128}{\rm i} \epsilon^{{e_{1}} {e_{2}}}\,_{\hat{a} \hat{b} {e_{3}}} (\Sigma_{{e_{1}} {e_{2}}})^{\alpha}{}_{\beta} \psi^{c}\,_{i}\,^{\beta} W^{d e} W_{d e} \nabla^{{e_{3}}}{\lambda^{i}_{\alpha}}+\frac{13}{64}{\rm i} (\Gamma_{\hat{a}})^{\alpha}{}_{\beta} \psi^{c}\,_{i}\,^{\beta} W^{d e} W_{d e} \nabla_{\hat{b}}{\lambda^{i}_{\alpha}}+\frac{7}{32}{\rm i} \epsilon_{{e_{3}}}\,^{d e {e_{1}} {e_{2}}} (\Sigma_{\hat{a} \hat{b}})^{\alpha}{}_{\beta} \psi^{c}\,_{i}\,^{\beta} W_{d e} W_{{e_{1}} {e_{2}}} \nabla^{{e_{3}}}{\lambda^{i}_{\alpha}} - \frac{1}{32}{\rm i} \epsilon_{\hat{a}}\,^{d e {e_{1}} {e_{2}}} \psi^{c}\,_{i}\,^{\alpha} W_{d e} W_{{e_{1}} {e_{2}}} \nabla_{\hat{b}}{\lambda^{i}_{\alpha}}+\frac{1}{8}{\rm i} (\Gamma^{d})^{\alpha}{}_{\beta} \psi^{c}\,_{i}\,^{\beta} W_{\hat{a} \hat{b}} W_{d e} \nabla^{e}{\lambda^{i}_{\alpha}}+\frac{1}{8}{\rm i} (\Gamma^{d})^{\alpha}{}_{\beta} \psi^{c}\,_{i}\,^{\beta} W_{\hat{a} d} W_{\hat{b} e} \nabla^{e}{\lambda^{i}_{\alpha}} - \frac{1}{16}{\rm i} \epsilon_{\hat{b}}\,^{d e {e_{1}} {e_{2}}} (\Sigma_{\hat{a} {e_{3}}})^{\alpha}{}_{\beta} \psi^{c}\,_{i}\,^{\beta} W_{d e} W_{{e_{1}} {e_{2}}} \nabla^{{e_{3}}}{\lambda^{i}_{\alpha}}+\frac{1}{16}{\rm i} \epsilon^{{e_{3}} d e {e_{1}} {e_{2}}} (\Sigma_{\hat{a} {e_{3}}})^{\alpha}{}_{\beta} \psi^{c}\,_{i}\,^{\beta} W_{d e} W_{{e_{1}} {e_{2}}} \nabla_{\hat{b}}{\lambda^{i}_{\alpha}} - \frac{55}{256}{\rm i} \psi^{c}\,_{i}\,^{\alpha} W^{d e} W_{\hat{a} \hat{b}} W_{d e} \lambda^{i}_{\alpha}+\frac{19}{256}{\rm i} \epsilon_{{e_{3}}}\,^{d e {e_{1}} {e_{2}}} (\Gamma^{{e_{3}}})^{\alpha}{}_{\beta} \psi^{c}\,_{i}\,^{\beta} W_{\hat{a} \hat{b}} W_{d e} W_{{e_{1}} {e_{2}}} \lambda^{i}_{\alpha}+\frac{57}{64}{\rm i} (\Sigma^{{e_{1}}}{}_{\, d})^{\alpha}{}_{\beta} \psi^{c}\,_{i}\,^{\beta} W^{d e} W_{\hat{a} \hat{b}} W_{{e_{1}} e} \lambda^{i}_{\alpha} - \frac{3}{32}{\rm i} \epsilon_{\hat{a} \hat{b} {e_{3}}}\,^{{e_{1}} {e_{2}}} (\Gamma^{{e_{3}}})^{\alpha}{}_{\beta} \psi^{c}\,_{i}\,^{\beta} W^{d e} W_{{e_{1}} {e_{2}}} W_{d e} \lambda^{i}_{\alpha} - \frac{83}{128}{\rm i} (\Sigma_{\hat{a}}{}^{\, {e_{1}}})^{\alpha}{}_{\beta} \psi^{c}\,_{i}\,^{\beta} W^{d e} W_{\hat{b} {e_{1}}} W_{d e} \lambda^{i}_{\alpha}+\frac{19}{32}{\rm i} \psi^{c}\,_{i}\,^{\alpha} W^{d e} W_{\hat{a} d} W_{\hat{b} e} \lambda^{i}_{\alpha}+\frac{95}{32}{\rm i} (\Sigma_{\hat{a}}{}^{\, {e_{1}}})^{\alpha}{}_{\beta} \psi^{c}\,_{i}\,^{\beta} W^{d e} W_{\hat{b} d} W_{{e_{1}} e} \lambda^{i}_{\alpha} - \frac{19}{64}{\rm i} \epsilon_{\hat{a}}\,^{d e {e_{1}} {e_{2}}} (\Gamma^{{e_{3}}})^{\alpha}{}_{\beta} \psi^{c}\,_{i}\,^{\beta} W_{\hat{b} d} W_{e {e_{1}}} W_{{e_{2}} {e_{3}}} \lambda^{i}_{\alpha} - \frac{19}{64}{\rm i} \epsilon_{{e_{3}}}\,^{d e {e_{1}} {e_{2}}} (\Gamma^{{e_{3}}})^{\alpha}{}_{\beta} \psi^{c}\,_{i}\,^{\beta} W_{\hat{a} d} W_{\hat{b} e} W_{{e_{1}} {e_{2}}} \lambda^{i}_{\alpha} - \frac{19}{64}{\rm i} \epsilon_{\hat{a} {e_{3}}}\,^{e {e_{1}} {e_{2}}} (\Gamma^{{e_{3}}})^{\alpha}{}_{\beta} \psi^{c}\,_{i}\,^{\beta} W_{\hat{b}}\,^{d} W_{e {e_{1}}} W_{{e_{2}} d} \lambda^{i}_{\alpha}%
 - \frac{19}{32}{\rm i} (\Sigma_{\hat{a} d})^{\alpha}{}_{\beta} \psi^{c}\,_{i}\,^{\beta} W^{d e} W_{\hat{b}}\,^{{e_{1}}} W_{{e_{1}} e} \lambda^{i}_{\alpha} - \frac{1}{16}(\Sigma^{d}{}_{\, {e_{1}}})^{\alpha}{}_{\beta} \psi^{c}\,_{i}\,^{\beta} W^{{e_{1}} e} F_{d e} W_{\hat{a} \hat{b} \alpha}\,^{i}+\frac{3}{32}\psi^{c}\,_{i}\,^{\alpha} W^{d e} F_{\hat{a} \hat{b}} W_{d e \alpha}\,^{i} - \frac{1}{256}\epsilon_{{e_{3}}}\,^{d e {e_{1}} {e_{2}}} (\Gamma^{{e_{3}}})^{\alpha}{}_{\beta} \psi^{c}\,_{i}\,^{\beta} W_{d e} F_{\hat{a} \hat{b}} W_{{e_{1}} {e_{2}} \alpha}\,^{i} - \frac{1}{16}(\Sigma^{{e_{1}}}{}_{\, d})^{\alpha}{}_{\beta} \psi^{c}\,_{i}\,^{\beta} W^{d e} F_{\hat{a} \hat{b}} W_{{e_{1}} e \alpha}\,^{i} - \frac{1}{64}\epsilon_{\hat{a} {e_{3}}}\,^{e {e_{1}} {e_{2}}} (\Gamma^{{e_{3}}})^{\alpha}{}_{\beta} \psi^{c}\,_{i}\,^{\beta} W_{e {e_{1}}} F_{\hat{b}}\,^{d} W_{{e_{2}} d \alpha}\,^{i} - \frac{1}{8}(\Sigma^{e {e_{1}}})^{\alpha}{}_{\beta} \psi^{c}\,_{i}\,^{\beta} W_{\hat{a} e} F_{\hat{b}}\,^{d} W_{{e_{1}} d \alpha}\,^{i} - \frac{1}{8}(\Sigma_{\hat{a} e})^{\alpha}{}_{\beta} \psi^{c}\,_{i}\,^{\beta} W^{e {e_{1}}} F_{\hat{b}}\,^{d} W_{d {e_{1}} \alpha}\,^{i} - \frac{1}{256}\epsilon_{\hat{a} \hat{b} {e_{3}}}\,^{{e_{1}} {e_{2}}} (\Gamma^{{e_{3}}})^{\alpha}{}_{\beta} \psi^{c}\,_{i}\,^{\beta} W_{{e_{1}} {e_{2}}} F^{d e} W_{d e \alpha}\,^{i}+\frac{1}{16}(\Sigma_{\hat{a}}{}^{\, {e_{1}}})^{\alpha}{}_{\beta} \psi^{c}\,_{i}\,^{\beta} W_{\hat{b} {e_{1}}} F^{d e} W_{d e \alpha}\,^{i}+\frac{1}{16}(\Sigma^{{e_{1}}}{}_{\, d})^{\alpha}{}_{\beta} \psi^{c}\,_{i}\,^{\beta} W_{\hat{a} \hat{b}} F^{d e} W_{{e_{1}} e \alpha}\,^{i} - \frac{1}{16}(\Sigma^{e {e_{1}}})^{\alpha}{}_{\beta} \psi^{c}\,_{i}\,^{\beta} W_{\hat{a}}\,^{d} F_{\hat{b} d} W_{e {e_{1}} \alpha}\,^{i} - \frac{1}{8}(\Sigma_{\hat{a}}{}^{\, {e_{1}}})^{\alpha}{}_{\beta} \psi^{c}\,_{i}\,^{\beta} W_{\hat{b} d} F^{d e} W_{{e_{1}} e \alpha}\,^{i} - \frac{1}{8}(\Sigma^{d {e_{1}}})^{\alpha}{}_{\beta} \psi^{c}\,_{i}\,^{\beta} W_{\hat{a}}\,^{e} F_{\hat{b} d} W_{{e_{1}} e \alpha}\,^{i}+\frac{1}{8}(\Sigma_{\hat{a} d})^{\alpha}{}_{\beta} \psi^{c}\,_{i}\,^{\beta} W_{\hat{b}}\,^{{e_{1}}} F^{d e} W_{{e_{1}} e \alpha}\,^{i}+\frac{1}{8}(\Sigma_{\hat{a}}{}^{\, {e_{1}}})^{\alpha}{}_{\beta} \psi^{c}\,_{i}\,^{\beta} W^{d e} F_{\hat{b} d} W_{{e_{1}} e \alpha}\,^{i} - \frac{1}{16}(\Sigma_{\hat{a} \hat{b}})^{\alpha}{}_{\beta} \psi^{c}\,_{i}\,^{\beta} W^{e {e_{1}}} F^{d}\,_{e} W_{d {e_{1}} \alpha}\,^{i} - \frac{3}{16}(\Sigma_{\hat{a}}{}^{\, d})^{\alpha}{}_{\beta} \psi^{c}\,_{i}\,^{\beta} W^{e {e_{1}}} F_{\hat{b} d} W_{e {e_{1}} \alpha}\,^{i}+\frac{1}{128}\epsilon_{\hat{a} \hat{b}}\,^{d {e_{2}} {e_{3}}} (\Gamma^{{e_{1}}})^{\alpha}{}_{\beta} \psi^{c}\,_{i}\,^{\beta} W_{{e_{1}}}\,^{e} F_{d e} W_{{e_{2}} {e_{3}} \alpha}\,^{i}+\frac{1}{16}(\Sigma_{d e})^{\alpha}{}_{\beta} \psi^{c}\,_{i}\,^{\beta} W_{\hat{a}}\,^{{e_{1}}} F^{d e} W_{\hat{b} {e_{1}} \alpha}\,^{i}%
+\frac{1}{8}(\Sigma^{{e_{1}}}{}_{\, e})^{\alpha}{}_{\beta} \psi^{c}\,_{i}\,^{\beta} W^{e d} F_{\hat{a} d} W_{\hat{b} {e_{1}} \alpha}\,^{i}+\frac{1}{16}(\Sigma_{e {e_{1}}})^{\alpha}{}_{\beta} \psi^{c}\,_{i}\,^{\beta} W^{e {e_{1}}} F_{\hat{a}}\,^{d} W_{\hat{b} d \alpha}\,^{i}+\frac{1}{16}(\Sigma_{\hat{a}}{}^{\, {e_{1}}})^{\alpha}{}_{\beta} \psi^{c}\,_{i}\,^{\beta} W^{d e} F_{d e} W_{\hat{b} {e_{1}} \alpha}\,^{i} - \frac{1}{8}(\Sigma_{\hat{a} {e_{1}}})^{\alpha}{}_{\beta} \psi^{c}\,_{i}\,^{\beta} W^{{e_{1}} e} F^{d}\,_{e} W_{\hat{b} d \alpha}\,^{i}+\frac{1}{8}(\Sigma^{{e_{1}}}{}_{\, d})^{\alpha}{}_{\beta} \psi^{c}\,_{i}\,^{\beta} W_{\hat{a} e} F^{d e} W_{\hat{b} {e_{1}} \alpha}\,^{i} - \frac{1}{8}(\Sigma^{{e_{1}}}{}_{\, d})^{\alpha}{}_{\beta} \psi^{c}\,_{i}\,^{\beta} W_{\hat{a} {e_{1}}} F^{d e} W_{\hat{b} e \alpha}\,^{i} - \frac{1}{8}(\Sigma^{d}{}_{\, e})^{\alpha}{}_{\beta} \psi^{c}\,_{i}\,^{\beta} W^{e {e_{1}}} F_{\hat{a} d} W_{\hat{b} {e_{1}} \alpha}\,^{i}+\frac{1}{8}(\Sigma_{\hat{a}}{}^{\, d})^{\alpha}{}_{\beta} \psi^{c}\,_{i}\,^{\beta} W^{e {e_{1}}} F_{d e} W_{\hat{b} {e_{1}} \alpha}\,^{i} - \frac{7}{64}{\rm i} \epsilon^{{e_{1}} {e_{2}}}\,_{{e_{3}}}\,^{d e} (\Sigma_{{e_{1}} {e_{2}}})^{\alpha}{}_{\beta} \psi^{c}\,_{i}\,^{\beta} W_{d e} \lambda^{i}_{\alpha} \nabla^{{e_{3}}}{W_{\hat{a} \hat{b}}}+\frac{1}{4}{\rm i} (\Gamma^{d})^{\alpha}{}_{\beta} \psi^{c}\,_{i}\,^{\beta} W_{d e} \lambda^{i}_{\alpha} \nabla^{e}{W_{\hat{a} \hat{b}}} - \frac{7}{64}{\rm i} \epsilon_{\hat{a} \hat{b} {e_{3}}}\,^{{e_{1}} {e_{2}}} (\Sigma_{d e})^{\alpha}{}_{\beta} \psi^{c}\,_{i}\,^{\beta} W^{d e} \lambda^{i}_{\alpha} \nabla^{{e_{3}}}{W_{{e_{1}} {e_{2}}}} - \frac{25}{128}{\rm i} \epsilon_{\hat{a} \hat{b}}\,^{{e_{1}} {e_{2}} d} \psi^{c}\,_{i}\,^{\alpha} W_{d e} \lambda^{i}_{\alpha} \nabla^{e}{W_{{e_{1}} {e_{2}}}} - \frac{9}{32}{\rm i} (\Gamma_{e})^{\alpha}{}_{\beta} \psi^{c}\,_{i}\,^{\beta} W_{\hat{b} d} \lambda^{i}_{\alpha} \nabla_{\hat{a}}{W^{e d}}+\frac{1}{16}{\rm i} (\Gamma_{\hat{a}})^{\alpha}{}_{\beta} \psi^{c}\,_{i}\,^{\beta} W_{d e} \lambda^{i}_{\alpha} \nabla_{\hat{b}}{W^{d e}}+\frac{5}{64}{\rm i} (\Gamma^{d})^{\alpha}{}_{\beta} \psi^{c}\,_{i}\,^{\beta} W_{\hat{a} \hat{b}} \lambda^{i}_{\alpha} \nabla^{e}{W_{d e}} - \frac{1}{64}{\rm i} (\Gamma_{\hat{a}})^{\alpha}{}_{\beta} \psi^{c}\,_{i}\,^{\beta} W_{\hat{b} d} \lambda^{i}_{\alpha} \nabla_{e}{W^{e d}} - \frac{3}{64}{\rm i} \epsilon_{\hat{a} \hat{b}}\,^{{e_{1}} {e_{2}} d} (\Sigma^{e}{}_{\, {e_{3}}})^{\alpha}{}_{\beta} \psi^{c}\,_{i}\,^{\beta} W_{d e} \lambda^{i}_{\alpha} \nabla^{{e_{3}}}{W_{{e_{1}} {e_{2}}}}+\frac{7}{32}{\rm i} \epsilon^{{e_{3}}}\,_{\hat{a} \hat{b}}\,^{{e_{1}} {e_{2}}} (\Sigma_{{e_{3}}}{}^{\, d})^{\alpha}{}_{\beta} \psi^{c}\,_{i}\,^{\beta} W_{d e} \lambda^{i}_{\alpha} \nabla^{e}{W_{{e_{1}} {e_{2}}}}+\frac{3}{64}{\rm i} \epsilon_{\hat{a} {e_{2}}}\,^{{e_{1}} d e} \psi^{c}\,_{i}\,^{\alpha} W_{d e} \lambda^{i}_{\alpha} \nabla^{{e_{2}}}{W_{\hat{b} {e_{1}}}} - \frac{1}{32}{\rm i} (\Gamma^{e})^{\alpha}{}_{\beta} \psi^{c}\,_{i}\,^{\beta} W_{\hat{b} d} \lambda^{i}_{\alpha} \nabla^{d}{W_{\hat{a} e}}%
 - \frac{3}{32}{\rm i} (\Gamma_{e})^{\alpha}{}_{\beta} \psi^{c}\,_{i}\,^{\beta} W_{\hat{b} d} \lambda^{i}_{\alpha} \nabla^{e}{W_{\hat{a}}\,^{d}} - \frac{1}{4}{\rm i} (\Gamma_{\hat{a}})^{\alpha}{}_{\beta} \psi^{c}\,_{i}\,^{\beta} W_{d e} \lambda^{i}_{\alpha} \nabla^{d}{W_{\hat{b}}\,^{e}}+\frac{1}{2}{\rm i} \epsilon^{{e_{2}}}\,_{\hat{a} {e_{3}}}\,^{{e_{1}} d} (\Sigma_{{e_{2}}}{}^{\, e})^{\alpha}{}_{\beta} \psi^{c}\,_{i}\,^{\beta} W_{d e} \lambda^{i}_{\alpha} \nabla^{{e_{3}}}{W_{\hat{b} {e_{1}}}} - \frac{1}{8}{\rm i} \epsilon^{{e_{2}} {e_{3}} {e_{1}} d e} (\Sigma_{{e_{2}} {e_{3}}})^{\alpha}{}_{\beta} \psi^{c}\,_{i}\,^{\beta} W_{d e} \lambda^{i}_{\alpha} \nabla_{\hat{a}}{W_{\hat{b} {e_{1}}}} - \frac{3}{8}{\rm i} (\Gamma^{d})^{\alpha}{}_{\beta} \psi^{c}\,_{i}\,^{\beta} W_{d e} \lambda^{i}_{\alpha} \nabla_{\hat{a}}{W_{\hat{b}}\,^{e}}+\frac{1}{8}{\rm i} \epsilon^{{e_{2}} {e_{3}}}\,_{\hat{a}}\,^{d e} (\Sigma_{{e_{2}} {e_{3}}})^{\alpha}{}_{\beta} \psi^{c}\,_{i}\,^{\beta} W_{d e} \lambda^{i}_{\alpha} \nabla^{{e_{1}}}{W_{\hat{b} {e_{1}}}}+\frac{7}{64}{\rm i} (\Gamma^{d})^{\alpha}{}_{\beta} \psi^{c}\,_{i}\,^{\beta} W_{\hat{b} d} \lambda^{i}_{\alpha} \nabla^{e}{W_{\hat{a} e}}+\frac{5}{64}{\rm i} \epsilon^{{e_{1}} {e_{2}}}\,_{\hat{a} \hat{b} {e_{3}}} (\Sigma_{{e_{1}} {e_{2}}})^{\alpha}{}_{\beta} \psi^{c}\,_{i}\,^{\beta} W_{d e} \lambda^{i}_{\alpha} \nabla^{{e_{3}}}{W^{d e}}+\frac{15}{64}{\rm i} \epsilon_{{e_{3}}}\,^{{e_{1}} {e_{2}} d e} (\Sigma_{\hat{a} \hat{b}})^{\alpha}{}_{\beta} \psi^{c}\,_{i}\,^{\beta} W_{d e} \lambda^{i}_{\alpha} \nabla^{{e_{3}}}{W_{{e_{1}} {e_{2}}}} - \frac{3}{64}{\rm i} \epsilon_{\hat{a}}\,^{{e_{1}} {e_{2}} d e} \psi^{c}\,_{i}\,^{\alpha} W_{d e} \lambda^{i}_{\alpha} \nabla_{\hat{b}}{W_{{e_{1}} {e_{2}}}} - \frac{3}{32}{\rm i} \epsilon_{\hat{b}}\,^{{e_{1}} {e_{2}} d e} (\Sigma_{\hat{a} {e_{3}}})^{\alpha}{}_{\beta} \psi^{c}\,_{i}\,^{\beta} W_{d e} \lambda^{i}_{\alpha} \nabla^{{e_{3}}}{W_{{e_{1}} {e_{2}}}}+\frac{3}{32}{\rm i} \epsilon^{{e_{3}} {e_{1}} {e_{2}} d e} (\Sigma_{\hat{a} {e_{3}}})^{\alpha}{}_{\beta} \psi^{c}\,_{i}\,^{\beta} W_{d e} \lambda^{i}_{\alpha} \nabla_{\hat{b}}{W_{{e_{1}} {e_{2}}}} - \frac{1}{32}{\rm i} \epsilon_{\hat{a} \hat{b} {e_{2}} {e_{1}}}\,^{d} \psi^{c}\,_{i}\,^{\alpha} W_{d e} \lambda^{i}_{\alpha} \nabla^{{e_{2}}}{W^{{e_{1}} e}}+\frac{1}{8}{\rm i} \epsilon^{{e_{2}}}\,_{\hat{b} {e_{3}} {e_{1}}}\,^{d} (\Sigma_{\hat{a} {e_{2}}})^{\alpha}{}_{\beta} \psi^{c}\,_{i}\,^{\beta} W_{d e} \lambda^{i}_{\alpha} \nabla^{{e_{3}}}{W^{{e_{1}} e}} - \frac{1}{32}{\rm i} \epsilon^{{e_{2}} {e_{3}}}\,_{\hat{a} \hat{b}}\,^{d} (\Sigma_{{e_{2}} {e_{3}}})^{\alpha}{}_{\beta} \psi^{c}\,_{i}\,^{\beta} W_{d e} \lambda^{i}_{\alpha} \nabla_{{e_{1}}}{W^{{e_{1}} e}}+\frac{1}{32}{\rm i} \epsilon^{{e_{2}} {e_{3}}}\,_{\hat{a} \hat{b} {e_{1}}} (\Sigma_{{e_{2}} {e_{3}}})^{\alpha}{}_{\beta} \psi^{c}\,_{i}\,^{\beta} W_{d e} \lambda^{i}_{\alpha} \nabla^{d}{W^{{e_{1}} e}} - \frac{1}{8}(\Gamma_{d})^{\alpha}{}_{\beta} \psi^{c}\,_{i}\,^{\beta} X^{i}_{\alpha} \nabla^{d}{F_{\hat{a} \hat{b}}} - \frac{1}{12}{\rm i} \epsilon^{d e}\,_{\hat{a} \hat{b} {e_{1}}} \Phi_{d e i}\,^{j} \psi^{c}\,_{j}\,^{\alpha} \nabla^{{e_{1}}}{\lambda^{i}_{\alpha}} - \frac{1}{6}{\rm i} \Phi_{\hat{a} \hat{b} i}\,^{j} (\Gamma_{d})^{\alpha}{}_{\beta} \psi^{c}\,_{j}\,^{\beta} \nabla^{d}{\lambda^{i}_{\alpha}}-{\rm i} \Phi_{\hat{a} d i}\,^{j} (\Gamma_{\hat{b}})^{\alpha}{}_{\beta} \psi^{c}\,_{j}\,^{\beta} \nabla^{d}{\lambda^{i}_{\alpha}}%
 - \frac{11}{24}{\rm i} \epsilon^{d {e_{1}}}\,_{\hat{a} \hat{b} {e_{2}}} \Phi_{d}\,^{e}\,_{i}\,^{j} (\Sigma_{{e_{1}} e})^{\alpha}{}_{\beta} \psi^{c}\,_{j}\,^{\beta} \nabla^{{e_{2}}}{\lambda^{i}_{\alpha}}+\frac{11}{48}{\rm i} \epsilon^{d e {e_{1}} {e_{2}}}\,_{\hat{a}} \Phi_{d e i}\,^{j} (\Sigma_{{e_{1}} {e_{2}}})^{\alpha}{}_{\beta} \psi^{c}\,_{j}\,^{\beta} \nabla_{\hat{b}}{\lambda^{i}_{\alpha}}+\frac{1}{12}{\rm i} \Phi_{\hat{a} d i}\,^{j} (\Gamma^{d})^{\alpha}{}_{\beta} \psi^{c}\,_{j}\,^{\beta} \nabla_{\hat{b}}{\lambda^{i}_{\alpha}} - \frac{7}{48}{\rm i} \Phi^{d e}\,_{i}\,^{j} (\Sigma_{d e})^{\alpha}{}_{\beta} \psi^{c}\,_{j}\,^{\beta} W_{\hat{a} \hat{b}} \lambda^{i}_{\alpha}+\frac{13}{48}{\rm i} \Phi_{\hat{a}}\,^{d}\,_{i}\,^{j} \psi^{c}\,_{j}\,^{\alpha} W_{\hat{b} d} \lambda^{i}_{\alpha}+\frac{3}{32}{\rm i} \epsilon^{{e_{1}} {e_{2}}}\,_{\hat{a} e}\,^{d} \Phi_{{e_{1}} {e_{2}} i}\,^{j} (\Gamma^{e})^{\alpha}{}_{\beta} \psi^{c}\,_{j}\,^{\beta} W_{\hat{b} d} \lambda^{i}_{\alpha}+\frac{5}{24}{\rm i} \Phi_{\hat{a}}\,^{e}\,_{i}\,^{j} (\Sigma_{e}{}^{\, d})^{\alpha}{}_{\beta} \psi^{c}\,_{j}\,^{\beta} W_{\hat{b} d} \lambda^{i}_{\alpha} - \frac{13}{24}{\rm i} \Phi^{e d}\,_{i}\,^{j} (\Sigma_{\hat{a} e})^{\alpha}{}_{\beta} \psi^{c}\,_{j}\,^{\beta} W_{\hat{b} d} \lambda^{i}_{\alpha}+\frac{13}{48}{\rm i} \epsilon^{d e {e_{1}}}\,_{\hat{a} \hat{b}} \Phi_{d e i}\,^{j} (\Sigma_{{e_{1}} {e_{2}}})^{\alpha}{}_{\beta} \psi^{c}\,_{j}\,^{\beta} \nabla^{{e_{2}}}{\lambda^{i}_{\alpha}} - \frac{19}{24}{\rm i} \epsilon^{d e {e_{1}}}\,_{\hat{b} {e_{2}}} \Phi_{\hat{a} d i}\,^{j} (\Sigma_{e {e_{1}}})^{\alpha}{}_{\beta} \psi^{c}\,_{j}\,^{\beta} \nabla^{{e_{2}}}{\lambda^{i}_{\alpha}} - \frac{1}{8}\Phi^{d e}\,_{i}\,^{j} (\Sigma_{d e})^{\alpha}{}_{\beta} \psi^{c}\,_{j}\,^{\beta} W W_{\hat{a} \hat{b} \alpha}\,^{i} - \frac{1}{32}{\rm i} \epsilon_{\hat{a} \hat{b}}\,^{d e}\,_{{e_{2}}} (\Gamma_{{e_{1}}})^{\alpha}{}_{\beta} \psi^{c}\,_{i}\,^{\beta} \nabla^{{e_{1}}}{W_{d e}} \nabla^{{e_{2}}}{\lambda^{i}_{\alpha}}+\frac{3}{4}{\rm i} \psi^{c}\,_{i}\,^{\alpha} \nabla_{d}{W_{\hat{a} \hat{b}}} \nabla^{d}{\lambda^{i}_{\alpha}}+\frac{1}{2}{\rm i} (\Sigma_{d e})^{\alpha}{}_{\beta} \psi^{c}\,_{i}\,^{\beta} \nabla^{d}{W_{\hat{a} \hat{b}}} \nabla^{e}{\lambda^{i}_{\alpha}} - \frac{1}{2}{\rm i} \psi^{c}\,_{i}\,^{\alpha} \nabla_{\hat{a}}{W_{\hat{b} d}} \nabla^{d}{\lambda^{i}_{\alpha}}+{\rm i} (\Sigma_{\hat{a} e})^{\alpha}{}_{\beta} \psi^{c}\,_{i}\,^{\beta} \nabla^{e}{W_{\hat{b} d}} \nabla^{d}{\lambda^{i}_{\alpha}}-{\rm i} (\Sigma_{d e})^{\alpha}{}_{\beta} \psi^{c}\,_{i}\,^{\beta} \nabla_{\hat{a}}{W^{d}\,_{\hat{b}}} \nabla^{e}{\lambda^{i}_{\alpha}}+\frac{1}{2}{\rm i} (\Sigma_{\hat{a} e})^{\alpha}{}_{\beta} \psi^{c}\,_{i}\,^{\beta} \nabla^{d}{W_{\hat{b} d}} \nabla^{e}{\lambda^{i}_{\alpha}}-{\rm i} (\Sigma_{\hat{a} d})^{\alpha}{}_{\beta} \psi^{c}\,_{i}\,^{\beta} \nabla_{\hat{b}}{W^{d}\,_{e}} \nabla^{e}{\lambda^{i}_{\alpha}} - \frac{3}{4}{\rm i} (\Sigma_{\hat{a} \hat{b}})^{\alpha}{}_{\beta} \psi^{c}\,_{i}\,^{\beta} \nabla^{d}{W_{d e}} \nabla^{e}{\lambda^{i}_{\alpha}}%
 - \frac{5}{128}{\rm i} \epsilon_{\hat{a} \hat{b} {e_{1}}}\,^{d}\,_{{e_{2}}} (\Gamma^{e})^{\alpha}{}_{\beta} \psi^{c}\,_{i}\,^{\beta} \nabla^{{e_{1}}}{W_{d e}} \nabla^{{e_{2}}}{\lambda^{i}_{\alpha}} - \frac{19}{128}{\rm i} \epsilon_{\hat{a} \hat{b} {e_{1}}}\,^{d}\,_{{e_{2}}} (\Gamma^{{e_{1}}})^{\alpha}{}_{\beta} \psi^{c}\,_{i}\,^{\beta} \nabla^{e}{W_{d e}} \nabla^{{e_{2}}}{\lambda^{i}_{\alpha}}+\frac{1}{8}{\rm i} (\Sigma_{d e})^{\alpha}{}_{\beta} \psi^{c}\,_{i}\,^{\beta} \nabla_{\hat{a}}{W^{d e}} \nabla_{\hat{b}}{\lambda^{i}_{\alpha}} - \frac{13}{8}{\rm i} (\Sigma_{\hat{a} d})^{\alpha}{}_{\beta} \psi^{c}\,_{i}\,^{\beta} \nabla_{e}{W^{d e}} \nabla_{\hat{b}}{\lambda^{i}_{\alpha}} - \frac{3}{32}{\rm i} \epsilon_{\hat{a} {e_{1}} {e_{2}}}\,^{d e} (\Gamma^{{e_{1}}})^{\alpha}{}_{\beta} \psi^{c}\,_{i}\,^{\beta} \nabla^{{e_{2}}}{W_{d e}} \nabla_{\hat{b}}{\lambda^{i}_{\alpha}}+\frac{13}{16}{\rm i} \psi^{c}\,_{i}\,^{\alpha} \nabla^{d}{W_{\hat{a} d}} \nabla_{\hat{b}}{\lambda^{i}_{\alpha}}+\frac{1}{4}{\rm i} (\Sigma_{e d})^{\alpha}{}_{\beta} \psi^{c}\,_{i}\,^{\beta} \nabla^{e}{W^{d}\,_{\hat{a}}} \nabla_{\hat{b}}{\lambda^{i}_{\alpha}}+\frac{3}{64}{\rm i} \epsilon^{{e_{1}} {e_{2}}}\,_{{e_{3}}}\,^{d e} (\Sigma_{{e_{1}} {e_{2}}})^{\alpha}{}_{\beta} \psi^{c}\,_{i}\,^{\beta} W_{\hat{a} \hat{b}} \lambda^{i}_{\alpha} \nabla^{{e_{3}}}{W_{d e}}+\frac{1}{16}{\rm i} \epsilon^{{e_{1}} {e_{2}}}\,_{\hat{a} {e_{3}}}\,^{d} (\Sigma_{{e_{1}} {e_{2}}})^{\alpha}{}_{\beta} \psi^{c}\,_{i}\,^{\beta} W_{d e} \lambda^{i}_{\alpha} \nabla^{{e_{3}}}{W_{\hat{b}}\,^{e}} - \frac{13}{128}{\rm i} \epsilon_{\hat{a} \hat{b}}\,^{{e_{1}} d e} \psi^{c}\,_{i}\,^{\alpha} W_{d e} \lambda^{i}_{\alpha} \nabla^{{e_{2}}}{W_{{e_{1}} {e_{2}}}} - \frac{5}{64}{\rm i} \epsilon_{\hat{a} \hat{b} {e_{2}}}\,^{d e} (\Sigma_{{e_{3}} {e_{1}}})^{\alpha}{}_{\beta} \psi^{c}\,_{i}\,^{\beta} W_{d e} \lambda^{i}_{\alpha} \nabla^{{e_{3}}}{W^{{e_{1}} {e_{2}}}}+\frac{1}{16}{\rm i} \epsilon_{\hat{a} {e_{2}}}\,^{e {e_{1}} d} \psi^{c}\,_{i}\,^{\alpha} W_{\hat{b} d} \lambda^{i}_{\alpha} \nabla^{{e_{2}}}{W_{e {e_{1}}}}+\frac{43}{128}{\rm i} \epsilon^{{e_{2}}}\,_{\hat{a}}\,^{e {e_{1}} d} (\Sigma_{{e_{2}} {e_{3}}})^{\alpha}{}_{\beta} \psi^{c}\,_{i}\,^{\beta} W_{\hat{b} d} \lambda^{i}_{\alpha} \nabla^{{e_{3}}}{W_{e {e_{1}}}} - \frac{15}{128}{\rm i} \epsilon^{{e_{1}} {e_{2}}}\,_{{e_{3}}}\,^{e d} (\Sigma_{{e_{1}} {e_{2}}})^{\alpha}{}_{\beta} \psi^{c}\,_{i}\,^{\beta} W_{\hat{b} d} \lambda^{i}_{\alpha} \nabla^{{e_{3}}}{W_{\hat{a} e}}+\frac{73}{128}{\rm i} \epsilon^{{e_{1}} {e_{2}}}\,_{\hat{a} {e_{3}} e} (\Sigma_{{e_{1}} {e_{2}}})^{\alpha}{}_{\beta} \psi^{c}\,_{i}\,^{\beta} W_{\hat{b} d} \lambda^{i}_{\alpha} \nabla^{{e_{3}}}{W^{e d}} - \frac{57}{512}\epsilon^{{e_{1}} {e_{2}}}\,_{{e_{3}}}\,^{d e} (\Sigma_{{e_{1}} {e_{2}}})^{\alpha}{}_{\beta} \psi^{c}\,_{i}\,^{\beta} W W_{\hat{a} \hat{b} \alpha}\,^{i} \nabla^{{e_{3}}}{W_{d e}}+\frac{63}{256}(\Gamma^{d})^{\alpha}{}_{\beta} \psi^{c}\,_{i}\,^{\beta} W W_{\hat{a} \hat{b} \alpha}\,^{i} \nabla^{e}{W_{d e}} - \frac{3}{16}{\rm i} \epsilon_{\hat{a} {e_{1}}}\,^{d e}\,_{{e_{2}}} (\Gamma^{{e_{1}}})^{\alpha}{}_{\beta} \psi^{c}\,_{i}\,^{\beta} \nabla_{\hat{b}}{W_{d e}} \nabla^{{e_{2}}}{\lambda^{i}_{\alpha}}+\frac{21}{64}{\rm i} \epsilon_{\hat{a} \hat{b} {e_{1}}}\,^{d e} (\Gamma^{{e_{1}}})^{\alpha}{}_{\beta} \psi^{c}\,_{i}\,^{\beta} \nabla_{{e_{2}}}{W_{d e}} \nabla^{{e_{2}}}{\lambda^{i}_{\alpha}}+\frac{9}{32}{\rm i} \epsilon^{{e_{3}}}\,_{\hat{a} \hat{b}}\,^{d e} (\Sigma_{{e_{3}} {e_{1}}})^{\alpha}{}_{\beta} \psi^{c}\,_{i}\,^{\beta} W_{d e} \lambda^{i}_{\alpha} \nabla_{{e_{2}}}{W^{{e_{1}} {e_{2}}}}%
+\frac{27}{64}{\rm i} \epsilon^{{e_{2}}}\,_{\hat{a} {e_{3}} {e_{1}}}\,^{d} (\Sigma_{{e_{2}} e})^{\alpha}{}_{\beta} \psi^{c}\,_{i}\,^{\beta} W_{\hat{b} d} \lambda^{i}_{\alpha} \nabla^{{e_{3}}}{W^{e {e_{1}}}} - \frac{27}{256}{\rm i} \epsilon^{{e_{2}} {e_{3}} e {e_{1}} d} (\Sigma_{{e_{2}} {e_{3}}})^{\alpha}{}_{\beta} \psi^{c}\,_{i}\,^{\beta} W_{\hat{b} d} \lambda^{i}_{\alpha} \nabla_{\hat{a}}{W_{e {e_{1}}}}+\frac{27}{256}{\rm i} \epsilon^{{e_{2}} {e_{3}}}\,_{\hat{a}}\,^{e {e_{1}}} (\Sigma_{{e_{2}} {e_{3}}})^{\alpha}{}_{\beta} \psi^{c}\,_{i}\,^{\beta} W_{\hat{b} d} \lambda^{i}_{\alpha} \nabla^{d}{W_{e {e_{1}}}}+\frac{3}{512}\epsilon_{\hat{a} \hat{b} {e_{3}}}\,^{d e} (\Sigma_{{e_{1}} {e_{2}}})^{\alpha}{}_{\beta} \psi^{c}\,_{i}\,^{\beta} W W_{d e \alpha}\,^{i} \nabla^{{e_{3}}}{W^{{e_{1}} {e_{2}}}}+\frac{3}{512}\epsilon_{\hat{a} \hat{b}}\,^{{e_{1}} d e} \psi^{c}\,_{i}\,^{\alpha} W W_{d e \alpha}\,^{i} \nabla^{{e_{2}}}{W_{{e_{1}} {e_{2}}}}+\frac{3}{128}(\Gamma_{\hat{a}})^{\alpha}{}_{\beta} \psi^{c}\,_{i}\,^{\beta} W W_{d e \alpha}\,^{i} \nabla^{d}{W_{\hat{b}}\,^{e}} - \frac{3}{256}(\Gamma^{d})^{\alpha}{}_{\beta} \psi^{c}\,_{i}\,^{\beta} W W_{d e \alpha}\,^{i} \nabla^{e}{W_{\hat{a} \hat{b}}} - \frac{3}{256}(\Gamma_{\hat{a}})^{\alpha}{}_{\beta} \psi^{c}\,_{i}\,^{\beta} W W_{d e \alpha}\,^{i} \nabla_{\hat{b}}{W^{d e}} - \frac{3}{128}(\Gamma^{d})^{\alpha}{}_{\beta} \psi^{c}\,_{i}\,^{\beta} W W_{d e \alpha}\,^{i} \nabla_{\hat{a}}{W_{\hat{b}}\,^{e}}+\frac{3}{256}\epsilon_{\hat{a} \hat{b} {e_{2}}}\,^{d e} (\Sigma_{{e_{3}} {e_{1}}})^{\alpha}{}_{\beta} \psi^{c}\,_{i}\,^{\beta} W W_{d e \alpha}\,^{i} \nabla^{{e_{3}}}{W^{{e_{1}} {e_{2}}}} - \frac{3}{256}\epsilon^{{e_{3}}}\,_{\hat{a} \hat{b}}\,^{d e} (\Sigma_{{e_{3}} {e_{1}}})^{\alpha}{}_{\beta} \psi^{c}\,_{i}\,^{\beta} W W_{d e \alpha}\,^{i} \nabla_{{e_{2}}}{W^{{e_{1}} {e_{2}}}}+\frac{3}{256}\epsilon_{\hat{a} {e_{2}}}\,^{e {e_{1}} d} \psi^{c}\,_{i}\,^{\alpha} W W_{\hat{b} d \alpha}\,^{i} \nabla^{{e_{2}}}{W_{e {e_{1}}}} - \frac{3}{128}(\Gamma_{e})^{\alpha}{}_{\beta} \psi^{c}\,_{i}\,^{\beta} W W_{\hat{b} d \alpha}\,^{i} \nabla^{e}{W_{\hat{a}}\,^{d}}+\frac{3}{128}(\Gamma_{\hat{a}})^{\alpha}{}_{\beta} \psi^{c}\,_{i}\,^{\beta} W W_{\hat{b} d \alpha}\,^{i} \nabla_{e}{W^{e d}}+\frac{3}{128}(\Gamma^{d})^{\alpha}{}_{\beta} \psi^{c}\,_{i}\,^{\beta} W W_{\hat{b} d \alpha}\,^{i} \nabla^{e}{W_{\hat{a} e}} - \frac{3}{64}\epsilon^{{e_{2}}}\,_{\hat{a} {e_{3}} {e_{1}}}\,^{d} (\Sigma_{{e_{2}} e})^{\alpha}{}_{\beta} \psi^{c}\,_{i}\,^{\beta} W W_{\hat{b} d \alpha}\,^{i} \nabla^{{e_{3}}}{W^{e {e_{1}}}}+\frac{3}{256}\epsilon^{{e_{2}} {e_{3}} e {e_{1}} d} (\Sigma_{{e_{2}} {e_{3}}})^{\alpha}{}_{\beta} \psi^{c}\,_{i}\,^{\beta} W W_{\hat{b} d \alpha}\,^{i} \nabla_{\hat{a}}{W_{e {e_{1}}}}+\frac{3}{128}(\Gamma_{e})^{\alpha}{}_{\beta} \psi^{c}\,_{i}\,^{\beta} W W_{\hat{b} d \alpha}\,^{i} \nabla_{\hat{a}}{W^{e d}} - \frac{3}{256}\epsilon^{{e_{2}} {e_{3}}}\,_{\hat{a}}\,^{e {e_{1}}} (\Sigma_{{e_{2}} {e_{3}}})^{\alpha}{}_{\beta} \psi^{c}\,_{i}\,^{\beta} W W_{\hat{b} d \alpha}\,^{i} \nabla^{d}{W_{e {e_{1}}}}+\frac{3}{128}(\Gamma^{e})^{\alpha}{}_{\beta} \psi^{c}\,_{i}\,^{\beta} W W_{\hat{b} d \alpha}\,^{i} \nabla^{d}{W_{\hat{a} e}}%
+\frac{5}{64}{\rm i} \epsilon_{\hat{a} \hat{b} {e_{2}}}\,^{d e} (\Gamma_{{e_{1}}})^{\alpha}{}_{\beta} \psi^{c}\,_{i}\,^{\beta} \nabla^{{e_{2}}}{W_{d e}} \nabla^{{e_{1}}}{\lambda^{i}_{\alpha}}+\frac{27}{64}{\rm i} \epsilon_{\hat{a} e {e_{1}}}\,^{d}\,_{{e_{2}}} (\Gamma^{e})^{\alpha}{}_{\beta} \psi^{c}\,_{i}\,^{\beta} \nabla^{{e_{1}}}{W_{\hat{b} d}} \nabla^{{e_{2}}}{\lambda^{i}_{\alpha}} - \frac{9}{128}\psi^{c}\,_{i}\,^{\alpha} W W^{d e} W_{d e} W_{\hat{a} \hat{b} \alpha}\,^{i} - \frac{3}{256}\epsilon_{\hat{a} \hat{b} {e_{3}}}\,^{d e} (\Gamma^{{e_{3}}})^{\alpha}{}_{\beta} \psi^{c}\,_{i}\,^{\beta} W^{{e_{1}} {e_{2}}} F_{d e} W_{{e_{1}} {e_{2}} \alpha}\,^{i}+\frac{1}{256}\epsilon_{{e_{3}}}\,^{d e {e_{1}} {e_{2}}} (\Gamma^{{e_{3}}})^{\alpha}{}_{\beta} \psi^{c}\,_{i}\,^{\beta} W_{\hat{a} \hat{b}} F_{d e} W_{{e_{1}} {e_{2}} \alpha}\,^{i}+\frac{1}{64}\epsilon_{\hat{a} {e_{3}}}\,^{d e {e_{2}}} (\Gamma^{{e_{3}}})^{\alpha}{}_{\beta} \psi^{c}\,_{i}\,^{\beta} W_{\hat{b}}\,^{{e_{1}}} F_{d e} W_{{e_{2}} {e_{1}} \alpha}\,^{i} - \frac{1}{128}\epsilon_{\hat{a} \hat{b}}\,^{{e_{1}} {e_{2}} {e_{3}}} (\Gamma^{d})^{\alpha}{}_{\beta} \psi^{c}\,_{i}\,^{\beta} W_{{e_{1}}}\,^{e} F_{d e} W_{{e_{2}} {e_{3}} \alpha}\,^{i}-{\rm i} (\Gamma_{\hat{a}})^{\alpha}{}_{\beta} \psi^{c}\,_{i}\,^{\beta} \nabla_{d}{\nabla^{d}{\nabla_{\hat{b}}{\lambda^{i}_{\alpha}}}}+\frac{3}{8}{\rm i} \psi^{c}\,_{i}\,^{\alpha} \lambda^{i}_{\alpha} \nabla_{d}{\nabla^{d}{W_{\hat{a} \hat{b}}}}+\frac{1}{2}{\rm i} \psi^{c}\,_{i}\,^{\alpha} W_{\hat{a} \hat{b}} \nabla_{d}{\nabla^{d}{\lambda^{i}_{\alpha}}}+\frac{9}{64}{\rm i} \epsilon_{\hat{a} \hat{b} {e_{1}}}\,^{d e} (\Gamma^{{e_{1}}})^{\alpha}{}_{\beta} \psi^{c}\,_{i}\,^{\beta} \lambda^{i}_{\alpha} \nabla_{{e_{2}}}{\nabla^{{e_{2}}}{W_{d e}}}+\frac{3}{16}{\rm i} \epsilon_{\hat{a} \hat{b} {e_{1}}}\,^{d e} (\Gamma^{{e_{1}}})^{\alpha}{}_{\beta} \psi^{c}\,_{i}\,^{\beta} W_{d e} \nabla_{{e_{2}}}{\nabla^{{e_{2}}}{\lambda^{i}_{\alpha}}}+\frac{1}{2}{\rm i} (\Sigma_{\hat{a}}{}^{\, d})^{\alpha}{}_{\beta} \psi^{c}\,_{i}\,^{\beta} W_{\hat{b} d} \nabla_{e}{\nabla^{e}{\lambda^{i}_{\alpha}}}+\frac{1}{2}\psi^{c}\,_{i}\,^{\alpha} W_{\hat{a} \hat{b} \alpha}\,^{i} \nabla_{d}{\nabla^{d}{W}}+\psi^{c}\,_{i}\,^{\alpha} \nabla_{d}{W} \nabla^{d}{W_{\hat{a} \hat{b} \alpha}\,^{i}}+\frac{1}{2}\psi^{c}\,_{i}\,^{\alpha} W \nabla_{d}{\nabla^{d}{W_{\hat{a} \hat{b} \alpha}\,^{i}}} - \frac{3}{16}{\rm i} \epsilon_{\hat{a} \hat{b} {e_{1}} {e_{2}}}\,^{d} (\Gamma^{e})^{\alpha}{}_{\beta} \psi^{c}\,_{i}\,^{\beta} W_{d e} \nabla^{{e_{1}}}{\nabla^{{e_{2}}}{\lambda^{i}_{\alpha}}}+\frac{3}{16}{\rm i} \epsilon_{\hat{a} \hat{b} {e_{1}} {e_{2}}}\,^{d} (\Gamma^{{e_{1}}})^{\alpha}{}_{\beta} \psi^{c}\,_{i}\,^{\beta} W_{d e} \nabla^{e}{\nabla^{{e_{2}}}{\lambda^{i}_{\alpha}}} - \frac{1}{2}{\rm i} (\Sigma_{d e})^{\alpha}{}_{\beta} \psi^{c}\,_{i}\,^{\beta} W^{d e} \nabla_{\hat{a}}{\nabla_{\hat{b}}{\lambda^{i}_{\alpha}}}-2{\rm i} (\Sigma_{\hat{a}}{}^{\, d})^{\alpha}{}_{\beta} \psi^{c}\,_{i}\,^{\beta} W_{d e} \nabla^{e}{\nabla_{\hat{b}}{\lambda^{i}_{\alpha}}}%
 - \frac{1}{8}{\rm i} \epsilon_{\hat{a} {e_{1}} {e_{2}}}\,^{d e} (\Gamma^{{e_{1}}})^{\alpha}{}_{\beta} \psi^{c}\,_{i}\,^{\beta} W_{d e} \nabla^{{e_{2}}}{\nabla_{\hat{b}}{\lambda^{i}_{\alpha}}}-{\rm i} \psi^{c}\,_{i}\,^{\alpha} W_{\hat{b} d} \nabla^{d}{\nabla_{\hat{a}}{\lambda^{i}_{\alpha}}}+{\rm i} (\Sigma^{d}{}_{\, e})^{\alpha}{}_{\beta} \psi^{c}\,_{i}\,^{\beta} W_{\hat{b} d} \nabla^{e}{\nabla_{\hat{a}}{\lambda^{i}_{\alpha}}} - \frac{5}{128}{\rm i} \epsilon^{{e_{1}} {e_{2}}}\,_{{e_{3}}}\,^{d e} (\Sigma_{{e_{1}} {e_{2}}})^{\alpha}{}_{\beta} \psi^{c}\,_{i}\,^{\beta} W_{\hat{a} \hat{b}} W_{d e} \nabla^{{e_{3}}}{\lambda^{i}_{\alpha}}+\frac{11}{32}{\rm i} \epsilon^{{e_{2}}}\,_{\hat{a}}\,^{{e_{1}} d e} (\Sigma_{{e_{2}} {e_{3}}})^{\alpha}{}_{\beta} \psi^{c}\,_{i}\,^{\beta} W_{d e} \lambda^{i}_{\alpha} \nabla^{{e_{3}}}{W_{\hat{b} {e_{1}}}}+\frac{9}{32}{\rm i} \epsilon^{{e_{1}} {e_{2}}}\,_{\hat{a} {e_{3}}}\,^{e} (\Sigma_{{e_{1}} {e_{2}}})^{\alpha}{}_{\beta} \psi^{c}\,_{i}\,^{\beta} W_{\hat{b}}\,^{d} W_{e d} \nabla^{{e_{3}}}{\lambda^{i}_{\alpha}} - \frac{3}{8}{\rm i} (\Gamma^{e})^{\alpha}{}_{\beta} \psi^{c}\,_{i}\,^{\beta} W_{\hat{b}}\,^{d} W_{e d} \nabla_{\hat{a}}{\lambda^{i}_{\alpha}}+\frac{1}{2}{\rm i} (\Gamma_{\hat{a}})^{\alpha}{}_{\beta} \psi^{c}\,_{i}\,^{\beta} W_{\hat{b}}\,^{d} W_{d e} \nabla^{e}{\lambda^{i}_{\alpha}} - \frac{13}{64}{\rm i} \epsilon_{\hat{a} \hat{b}}\,^{d e {e_{1}}} \psi^{c}\,_{i}\,^{\alpha} W_{d e} W_{{e_{1}} {e_{2}}} \nabla^{{e_{2}}}{\lambda^{i}_{\alpha}} - \frac{7}{32}{\rm i} \epsilon_{\hat{a} \hat{b}}\,^{d e {e_{1}}} (\Sigma^{{e_{2}}}{}_{\, {e_{3}}})^{\alpha}{}_{\beta} \psi^{c}\,_{i}\,^{\beta} W_{d e} W_{{e_{1}} {e_{2}}} \nabla^{{e_{3}}}{\lambda^{i}_{\alpha}}+\frac{5}{32}{\rm i} \epsilon_{\hat{a} {e_{2}}}\,^{d e {e_{1}}} \psi^{c}\,_{i}\,^{\alpha} W_{\hat{b} d} W_{e {e_{1}}} \nabla^{{e_{2}}}{\lambda^{i}_{\alpha}}+\frac{3}{4}{\rm i} \epsilon^{{e_{2}}}\,_{\hat{a}}\,^{d e {e_{1}}} (\Sigma_{{e_{2}} {e_{3}}})^{\alpha}{}_{\beta} \psi^{c}\,_{i}\,^{\beta} W_{\hat{b} d} W_{e {e_{1}}} \nabla^{{e_{3}}}{\lambda^{i}_{\alpha}} - \frac{3}{16}{\rm i} \epsilon^{{e_{1}} {e_{2}}}\,_{{e_{3}}}\,^{d e} (\Sigma_{{e_{1}} {e_{2}}})^{\alpha}{}_{\beta} \psi^{c}\,_{i}\,^{\beta} W_{\hat{a} d} W_{\hat{b} e} \nabla^{{e_{3}}}{\lambda^{i}_{\alpha}} - \frac{1}{16}{\rm i} \epsilon_{\hat{a} \hat{b} {e_{2}}}\,^{d e} (\Gamma_{{e_{1}}})^{\alpha}{}_{\beta} \psi^{c}\,_{i}\,^{\beta} W_{d e} \nabla^{{e_{2}}}{\nabla^{{e_{1}}}{\lambda^{i}_{\alpha}}} - \frac{3}{8}{\rm i} \epsilon_{\hat{a} e {e_{1}} {e_{2}}}\,^{d} (\Gamma^{e})^{\alpha}{}_{\beta} \psi^{c}\,_{i}\,^{\beta} W_{\hat{b} d} \nabla^{{e_{1}}}{\nabla^{{e_{2}}}{\lambda^{i}_{\alpha}}} - \frac{11}{128}{\rm i} \epsilon_{\hat{a} \hat{b} {e_{3}}}\,^{{e_{1}} {e_{2}}} (\Sigma_{d e})^{\alpha}{}_{\beta} \psi^{c}\,_{i}\,^{\beta} W^{d e} W_{{e_{1}} {e_{2}}} \nabla^{{e_{3}}}{\lambda^{i}_{\alpha}}+\frac{29}{64}{\rm i} \epsilon^{{e_{3}}}\,_{\hat{a} \hat{b}}\,^{d e} (\Sigma_{{e_{3}}}{}^{\, {e_{1}}})^{\alpha}{}_{\beta} \psi^{c}\,_{i}\,^{\beta} W_{d e} W_{{e_{1}} {e_{2}}} \nabla^{{e_{2}}}{\lambda^{i}_{\alpha}} - \frac{11}{64}\epsilon^{{e_{1}} {e_{2}}}\,_{{e_{3}}}\,^{d e} (\Sigma_{{e_{1}} {e_{2}}})^{\alpha}{}_{\beta} \psi^{c}\,_{i}\,^{\beta} W_{d e} W_{\hat{a} \hat{b} \alpha}\,^{i} \nabla^{{e_{3}}}{W}+\frac{13}{32}(\Gamma^{d})^{\alpha}{}_{\beta} \psi^{c}\,_{i}\,^{\beta} W_{d e} W_{\hat{a} \hat{b} \alpha}\,^{i} \nabla^{e}{W} - \frac{11}{64}\epsilon^{{e_{1}} {e_{2}}}\,_{{e_{3}}}\,^{d e} (\Sigma_{{e_{1}} {e_{2}}})^{\alpha}{}_{\beta} \psi^{c}\,_{i}\,^{\beta} W W_{d e} \nabla^{{e_{3}}}{W_{\hat{a} \hat{b} \alpha}\,^{i}}%
+\frac{13}{32}(\Gamma^{d})^{\alpha}{}_{\beta} \psi^{c}\,_{i}\,^{\beta} W W_{d e} \nabla^{e}{W_{\hat{a} \hat{b} \alpha}\,^{i}}+\frac{17}{128}{\rm i} \epsilon_{\hat{a} \hat{b} {e_{1}} {e_{2}}}\,^{d} (\Gamma^{{e_{1}}})^{\alpha}{}_{\beta} \psi^{c}\,_{i}\,^{\beta} \nabla^{{e_{2}}}{W_{d e}} \nabla^{e}{\lambda^{i}_{\alpha}}+\frac{7}{8}{\rm i} \epsilon^{{e_{2}}}\,_{\hat{a} {e_{3}}}\,^{d e} (\Sigma_{{e_{2}}}{}^{\, {e_{1}}})^{\alpha}{}_{\beta} \psi^{c}\,_{i}\,^{\beta} W_{\hat{b} d} W_{e {e_{1}}} \nabla^{{e_{3}}}{\lambda^{i}_{\alpha}} - \frac{7}{32}{\rm i} \epsilon^{{e_{2}} {e_{3}} d e {e_{1}}} (\Sigma_{{e_{2}} {e_{3}}})^{\alpha}{}_{\beta} \psi^{c}\,_{i}\,^{\beta} W_{\hat{b} d} W_{e {e_{1}}} \nabla_{\hat{a}}{\lambda^{i}_{\alpha}}+\frac{7}{32}{\rm i} \epsilon^{{e_{2}} {e_{3}}}\,_{\hat{a}}\,^{e {e_{1}}} (\Sigma_{{e_{2}} {e_{3}}})^{\alpha}{}_{\beta} \psi^{c}\,_{i}\,^{\beta} W_{\hat{b} d} W_{e {e_{1}}} \nabla^{d}{\lambda^{i}_{\alpha}} - \frac{1}{16}{\rm i} \epsilon_{\hat{a} \hat{b} {e_{1}} {e_{2}}}\,^{d} (\Gamma^{{e_{1}}})^{\alpha}{}_{\beta} \psi^{c}\,_{i}\,^{\beta} W_{d e} \nabla^{{e_{2}}}{\nabla^{e}{\lambda^{i}_{\alpha}}}+\frac{1}{64}\epsilon_{\hat{a} \hat{b} {e_{3}}}\,^{{e_{1}} {e_{2}}} (\Sigma_{d e})^{\alpha}{}_{\beta} \psi^{c}\,_{i}\,^{\beta} W^{d e} W_{{e_{1}} {e_{2}} \alpha}\,^{i} \nabla^{{e_{3}}}{W}+\frac{1}{64}\epsilon_{\hat{a} \hat{b}}\,^{d {e_{1}} {e_{2}}} \psi^{c}\,_{i}\,^{\alpha} W_{d e} W_{{e_{1}} {e_{2}} \alpha}\,^{i} \nabla^{e}{W} - \frac{1}{16}(\Gamma_{\hat{a}})^{\alpha}{}_{\beta} \psi^{c}\,_{i}\,^{\beta} W_{\hat{b}}\,^{d} W_{d e \alpha}\,^{i} \nabla^{e}{W} - \frac{1}{32}(\Gamma^{d})^{\alpha}{}_{\beta} \psi^{c}\,_{i}\,^{\beta} W_{\hat{a} \hat{b}} W_{d e \alpha}\,^{i} \nabla^{e}{W} - \frac{1}{32}(\Gamma_{\hat{a}})^{\alpha}{}_{\beta} \psi^{c}\,_{i}\,^{\beta} W^{d e} W_{d e \alpha}\,^{i} \nabla_{\hat{b}}{W} - \frac{1}{16}(\Gamma^{e})^{\alpha}{}_{\beta} \psi^{c}\,_{i}\,^{\beta} W_{\hat{b}}\,^{d} W_{e d \alpha}\,^{i} \nabla_{\hat{a}}{W}+\frac{1}{32}\epsilon_{\hat{a} \hat{b}}\,^{d {e_{1}} {e_{2}}} (\Sigma^{e}{}_{\, {e_{3}}})^{\alpha}{}_{\beta} \psi^{c}\,_{i}\,^{\beta} W_{d e} W_{{e_{1}} {e_{2}} \alpha}\,^{i} \nabla^{{e_{3}}}{W} - \frac{1}{32}\epsilon^{{e_{3}}}\,_{\hat{a} \hat{b}}\,^{{e_{1}} {e_{2}}} (\Sigma_{{e_{3}}}{}^{\, d})^{\alpha}{}_{\beta} \psi^{c}\,_{i}\,^{\beta} W_{d e} W_{{e_{1}} {e_{2}} \alpha}\,^{i} \nabla^{e}{W}+\frac{1}{32}\epsilon_{\hat{a} {e_{2}}}\,^{d e {e_{1}}} \psi^{c}\,_{i}\,^{\alpha} W_{d e} W_{\hat{b} {e_{1}} \alpha}\,^{i} \nabla^{{e_{2}}}{W} - \frac{1}{16}(\Gamma_{e})^{\alpha}{}_{\beta} \psi^{c}\,_{i}\,^{\beta} W_{\hat{a}}\,^{d} W_{\hat{b} d \alpha}\,^{i} \nabla^{e}{W}+\frac{1}{16}(\Gamma^{d})^{\alpha}{}_{\beta} \psi^{c}\,_{i}\,^{\beta} W_{\hat{a} d} W_{\hat{b} e \alpha}\,^{i} \nabla^{e}{W}+\frac{1}{16}(\Gamma^{d})^{\alpha}{}_{\beta} \psi^{c}\,_{i}\,^{\beta} W_{d}\,^{e} W_{\hat{b} e \alpha}\,^{i} \nabla_{\hat{a}}{W}+\frac{1}{16}\epsilon^{{e_{2}}}\,_{\hat{a} {e_{3}}}\,^{d e} (\Sigma_{{e_{2}}}{}^{\, {e_{1}}})^{\alpha}{}_{\beta} \psi^{c}\,_{i}\,^{\beta} W_{d e} W_{\hat{b} {e_{1}} \alpha}\,^{i} \nabla^{{e_{3}}}{W} - \frac{1}{16}\epsilon^{{e_{2}}}\,_{{e_{3}}}\,^{d e {e_{1}}} (\Sigma_{\hat{a} {e_{2}}})^{\alpha}{}_{\beta} \psi^{c}\,_{i}\,^{\beta} W_{d e} W_{\hat{b} {e_{1}} \alpha}\,^{i} \nabla^{{e_{3}}}{W}%
 - \frac{1}{16}\epsilon^{{e_{2}} {e_{3}}}\,_{\hat{a}}\,^{d {e_{1}}} (\Sigma_{{e_{2}} {e_{3}}})^{\alpha}{}_{\beta} \psi^{c}\,_{i}\,^{\beta} W_{d e} W_{\hat{b} {e_{1}} \alpha}\,^{i} \nabla^{e}{W}+\frac{1}{16}(\Gamma^{e})^{\alpha}{}_{\beta} \psi^{c}\,_{i}\,^{\beta} W_{\hat{a} d} W_{\hat{b} e \alpha}\,^{i} \nabla^{d}{W} - \frac{1}{16}(\Gamma_{\hat{a}})^{\alpha}{}_{\beta} \psi^{c}\,_{i}\,^{\beta} W^{d}\,_{e} W_{\hat{b} d \alpha}\,^{i} \nabla^{e}{W}+\frac{1}{64}\epsilon_{\hat{a} \hat{b} {e_{3}}}\,^{{e_{1}} {e_{2}}} (\Sigma_{d e})^{\alpha}{}_{\beta} \psi^{c}\,_{i}\,^{\beta} W W^{d e} \nabla^{{e_{3}}}{W_{{e_{1}} {e_{2}} \alpha}\,^{i}}+\frac{1}{64}\epsilon_{\hat{a} \hat{b}}\,^{d {e_{1}} {e_{2}}} \psi^{c}\,_{i}\,^{\alpha} W W_{d e} \nabla^{e}{W_{{e_{1}} {e_{2}} \alpha}\,^{i}} - \frac{1}{32}(\Gamma_{\hat{a}})^{\alpha}{}_{\beta} \psi^{c}\,_{i}\,^{\beta} W W^{d e} \nabla_{\hat{b}}{W_{d e \alpha}\,^{i}} - \frac{1}{16}(\Gamma^{e})^{\alpha}{}_{\beta} \psi^{c}\,_{i}\,^{\beta} W W_{\hat{b}}\,^{d} \nabla_{\hat{a}}{W_{e d \alpha}\,^{i}} - \frac{1}{16}(\Gamma_{\hat{a}})^{\alpha}{}_{\beta} \psi^{c}\,_{i}\,^{\beta} W W_{\hat{b}}\,^{d} \nabla^{e}{W_{d e \alpha}\,^{i}} - \frac{1}{32}(\Gamma^{d})^{\alpha}{}_{\beta} \psi^{c}\,_{i}\,^{\beta} W W_{\hat{a} \hat{b}} \nabla^{e}{W_{d e \alpha}\,^{i}}+\frac{1}{32}\epsilon_{\hat{a} \hat{b}}\,^{d {e_{1}} {e_{2}}} (\Sigma^{e}{}_{\, {e_{3}}})^{\alpha}{}_{\beta} \psi^{c}\,_{i}\,^{\beta} W W_{d e} \nabla^{{e_{3}}}{W_{{e_{1}} {e_{2}} \alpha}\,^{i}} - \frac{1}{32}\epsilon^{{e_{3}}}\,_{\hat{a} \hat{b}}\,^{{e_{1}} {e_{2}}} (\Sigma_{{e_{3}}}{}^{\, d})^{\alpha}{}_{\beta} \psi^{c}\,_{i}\,^{\beta} W W_{d e} \nabla^{e}{W_{{e_{1}} {e_{2}} \alpha}\,^{i}}+\frac{1}{32}\epsilon_{\hat{a} {e_{2}}}\,^{d e {e_{1}}} \psi^{c}\,_{i}\,^{\alpha} W W_{d e} \nabla^{{e_{2}}}{W_{\hat{b} {e_{1}} \alpha}\,^{i}} - \frac{1}{16}(\Gamma_{e})^{\alpha}{}_{\beta} \psi^{c}\,_{i}\,^{\beta} W W_{\hat{a}}\,^{d} \nabla^{e}{W_{\hat{b} d \alpha}\,^{i}} - \frac{1}{16}(\Gamma_{\hat{a}})^{\alpha}{}_{\beta} \psi^{c}\,_{i}\,^{\beta} W W^{d}\,_{e} \nabla^{e}{W_{\hat{b} d \alpha}\,^{i}}+\frac{1}{16}(\Gamma^{e})^{\alpha}{}_{\beta} \psi^{c}\,_{i}\,^{\beta} W W_{\hat{a} d} \nabla^{d}{W_{\hat{b} e \alpha}\,^{i}}+\frac{1}{8}\epsilon^{{e_{2}}}\,_{\hat{a} {e_{3}}}\,^{d {e_{1}}} (\Sigma_{{e_{2}}}{}^{\, e})^{\alpha}{}_{\beta} \psi^{c}\,_{i}\,^{\beta} W W_{d e} \nabla^{{e_{3}}}{W_{\hat{b} {e_{1}} \alpha}\,^{i}}+\frac{1}{32}\epsilon^{{e_{2}} {e_{3}} d e {e_{1}}} (\Sigma_{{e_{2}} {e_{3}}})^{\alpha}{}_{\beta} \psi^{c}\,_{i}\,^{\beta} W W_{d e} \nabla_{\hat{a}}{W_{\hat{b} {e_{1}} \alpha}\,^{i}}+\frac{1}{16}(\Gamma^{d})^{\alpha}{}_{\beta} \psi^{c}\,_{i}\,^{\beta} W W_{d}\,^{e} \nabla_{\hat{a}}{W_{\hat{b} e \alpha}\,^{i}} - \frac{1}{32}\epsilon^{{e_{2}} {e_{3}}}\,_{\hat{a}}\,^{d e} (\Sigma_{{e_{2}} {e_{3}}})^{\alpha}{}_{\beta} \psi^{c}\,_{i}\,^{\beta} W W_{d e} \nabla^{{e_{1}}}{W_{\hat{b} {e_{1}} \alpha}\,^{i}}+\frac{1}{16}(\Gamma^{d})^{\alpha}{}_{\beta} \psi^{c}\,_{i}\,^{\beta} W W_{\hat{a} d} \nabla^{e}{W_{\hat{b} e \alpha}\,^{i}}%
+\frac{3}{32}\epsilon^{{e_{1}} {e_{2}}}\,_{{e_{3}}}\,^{d e} (\Sigma_{{e_{1}} {e_{2}}})^{\alpha}{}_{\beta} \psi^{c}\,_{i}\,^{\beta} F_{\hat{a} \hat{b}} \nabla^{{e_{3}}}{W_{d e \alpha}\,^{i}} - \frac{3}{16}(\Gamma^{d})^{\alpha}{}_{\beta} \psi^{c}\,_{i}\,^{\beta} F_{\hat{a} \hat{b}} \nabla^{e}{W_{d e \alpha}\,^{i}}+\frac{3}{32}\epsilon^{{e_{1}} {e_{2}}}\,_{{e_{3}}}\,^{d e} (\Sigma_{{e_{1}} {e_{2}}})^{\alpha}{}_{\beta} \psi^{c}\,_{i}\,^{\beta} F_{d e} \nabla^{{e_{3}}}{W_{\hat{a} \hat{b} \alpha}\,^{i}} - \frac{3}{16}(\Gamma^{d})^{\alpha}{}_{\beta} \psi^{c}\,_{i}\,^{\beta} F_{d e} \nabla^{e}{W_{\hat{a} \hat{b} \alpha}\,^{i}} - \frac{3}{8}\epsilon^{{e_{1}} {e_{2}}}\,_{\hat{a} d {e_{3}}} (\Sigma_{{e_{1}} {e_{2}}})^{\alpha}{}_{\beta} \psi^{c}\,_{i}\,^{\beta} F^{d e} \nabla^{{e_{3}}}{W_{\hat{b} e \alpha}\,^{i}} - \frac{3}{8}(\Gamma^{d})^{\alpha}{}_{\beta} \psi^{c}\,_{i}\,^{\beta} F_{d}\,^{e} \nabla_{\hat{a}}{W_{\hat{b} e \alpha}\,^{i}} - \frac{3}{8}(\Gamma_{\hat{a}})^{\alpha}{}_{\beta} \psi^{c}\,_{i}\,^{\beta} F^{d}\,_{e} \nabla^{e}{W_{\hat{b} d \alpha}\,^{i}}+\frac{3}{32}\epsilon^{{e_{1}} {e_{2}}}\,_{\hat{a} \hat{b} {e_{3}}} (\Sigma_{{e_{1}} {e_{2}}})^{\alpha}{}_{\beta} \psi^{c}\,_{i}\,^{\beta} F^{d e} \nabla^{{e_{3}}}{W_{d e \alpha}\,^{i}} - \frac{3}{16}(\Gamma_{\hat{a}})^{\alpha}{}_{\beta} \psi^{c}\,_{i}\,^{\beta} F^{d e} \nabla_{\hat{b}}{W_{d e \alpha}\,^{i}} - \frac{3}{32}\epsilon_{\hat{a} \hat{b}}\,^{d e {e_{1}}} \psi^{c}\,_{i}\,^{\alpha} F_{d e} \nabla^{{e_{2}}}{W_{{e_{1}} {e_{2}} \alpha}\,^{i}}+\frac{3}{16}\epsilon_{\hat{a} \hat{b}}\,^{d e {e_{1}}} (\Sigma^{{e_{2}}}{}_{\, {e_{3}}})^{\alpha}{}_{\beta} \psi^{c}\,_{i}\,^{\beta} F_{d e} \nabla^{{e_{3}}}{W_{{e_{1}} {e_{2}} \alpha}\,^{i}} - \frac{3}{8}(\Gamma_{e})^{\alpha}{}_{\beta} \psi^{c}\,_{i}\,^{\beta} F_{\hat{a}}\,^{d} \nabla^{e}{W_{\hat{b} d \alpha}\,^{i}} - \frac{3}{16}\epsilon_{\hat{a} {e_{2}}}\,^{d e {e_{1}}} \psi^{c}\,_{i}\,^{\alpha} F_{\hat{b} d} \nabla^{{e_{2}}}{W_{e {e_{1}} \alpha}\,^{i}} - \frac{3}{8}\epsilon^{{e_{2}}}\,_{\hat{a}}\,^{d e {e_{1}}} (\Sigma_{{e_{2}} {e_{3}}})^{\alpha}{}_{\beta} \psi^{c}\,_{i}\,^{\beta} F_{\hat{b} d} \nabla^{{e_{3}}}{W_{e {e_{1}} \alpha}\,^{i}}+\frac{3}{8}\epsilon^{{e_{1}} {e_{2}}}\,_{{e_{3}}}\,^{d e} (\Sigma_{{e_{1}} {e_{2}}})^{\alpha}{}_{\beta} \psi^{c}\,_{i}\,^{\beta} F_{\hat{a} d} \nabla^{{e_{3}}}{W_{\hat{b} e \alpha}\,^{i}}+\frac{3}{8}(\Gamma^{e})^{\alpha}{}_{\beta} \psi^{c}\,_{i}\,^{\beta} F_{\hat{a} d} \nabla^{d}{W_{\hat{b} e \alpha}\,^{i}} - \frac{3}{8}(\Gamma^{d})^{\alpha}{}_{\beta} \psi^{c}\,_{i}\,^{\beta} F_{\hat{a} d} \nabla^{e}{W_{\hat{b} e \alpha}\,^{i}}+\frac{3}{8}\epsilon^{{e_{1}} {e_{2}}}\,_{\hat{a} {e_{3}}}\,^{e} (\Sigma_{{e_{1}} {e_{2}}})^{\alpha}{}_{\beta} \psi^{c}\,_{i}\,^{\beta} F_{\hat{b}}\,^{d} \nabla^{{e_{3}}}{W_{e d \alpha}\,^{i}} - \frac{3}{8}(\Gamma^{e})^{\alpha}{}_{\beta} \psi^{c}\,_{i}\,^{\beta} F_{\hat{b}}\,^{d} \nabla_{\hat{a}}{W_{e d \alpha}\,^{i}} - \frac{3}{8}(\Gamma_{\hat{a}})^{\alpha}{}_{\beta} \psi^{c}\,_{i}\,^{\beta} F_{\hat{b}}\,^{d} \nabla^{e}{W_{d e \alpha}\,^{i}}%
+\frac{1}{8}\epsilon^{{e_{1}} {e_{2}}}\,_{{e_{3}}}\,^{d e} (\Sigma_{{e_{1}} {e_{2}}})^{\alpha}{}_{\beta} \psi^{c}\,_{i}\,^{\beta} W_{\hat{a} \hat{b} \alpha}\,^{i} \nabla^{{e_{3}}}{F_{d e}} - \frac{1}{4}(\Gamma^{d})^{\alpha}{}_{\beta} \psi^{c}\,_{i}\,^{\beta} W_{\hat{a} \hat{b} \alpha}\,^{i} \nabla^{e}{F_{d e}}+\frac{1}{8}\epsilon_{\hat{a} \hat{b} {e_{3}}}\,^{{e_{1}} {e_{2}}} (\Sigma_{d e})^{\alpha}{}_{\beta} \psi^{c}\,_{i}\,^{\beta} W_{{e_{1}} {e_{2}} \alpha}\,^{i} \nabla^{{e_{3}}}{F^{d e}}+\frac{1}{2}(\Gamma_{\hat{a}})^{\alpha}{}_{\beta} \psi^{c}\,_{i}\,^{\beta} W_{e d \alpha}\,^{i} \nabla^{e}{F_{\hat{b}}\,^{d}} - \frac{1}{4}(\Gamma^{d})^{\alpha}{}_{\beta} \psi^{c}\,_{i}\,^{\beta} W_{d e \alpha}\,^{i} \nabla^{e}{F_{\hat{a} \hat{b}}} - \frac{1}{4}(\Gamma_{\hat{a}})^{\alpha}{}_{\beta} \psi^{c}\,_{i}\,^{\beta} W_{d e \alpha}\,^{i} \nabla_{\hat{b}}{F^{d e}} - \frac{1}{2}(\Gamma^{e})^{\alpha}{}_{\beta} \psi^{c}\,_{i}\,^{\beta} W_{e d \alpha}\,^{i} \nabla_{\hat{a}}{F_{\hat{b}}\,^{d}} - \frac{1}{4}\epsilon_{\hat{a} \hat{b} e}\,^{{e_{1}} {e_{2}}} (\Sigma_{{e_{3}} d})^{\alpha}{}_{\beta} \psi^{c}\,_{i}\,^{\beta} W_{{e_{1}} {e_{2}} \alpha}\,^{i} \nabla^{{e_{3}}}{F^{d e}}+\frac{1}{4}\epsilon^{{e_{3}}}\,_{\hat{a} \hat{b}}\,^{{e_{1}} {e_{2}}} (\Sigma_{{e_{3}} d})^{\alpha}{}_{\beta} \psi^{c}\,_{i}\,^{\beta} W_{{e_{1}} {e_{2}} \alpha}\,^{i} \nabla_{e}{F^{d e}} - \frac{1}{2}(\Gamma_{e})^{\alpha}{}_{\beta} \psi^{c}\,_{i}\,^{\beta} W_{\hat{b} d \alpha}\,^{i} \nabla^{e}{F_{\hat{a}}\,^{d}}+\frac{1}{2}\epsilon^{{e_{2}}}\,_{\hat{a}}\,^{d e {e_{1}}} (\Sigma_{{e_{2}} {e_{3}}})^{\alpha}{}_{\beta} \psi^{c}\,_{i}\,^{\beta} W_{\hat{b} {e_{1}} \alpha}\,^{i} \nabla^{{e_{3}}}{F_{d e}}+\frac{1}{2}\epsilon^{{e_{1}} {e_{2}}}\,_{{e_{3}}}\,^{d e} (\Sigma_{{e_{1}} {e_{2}}})^{\alpha}{}_{\beta} \psi^{c}\,_{i}\,^{\beta} W_{\hat{b} e \alpha}\,^{i} \nabla^{{e_{3}}}{F_{\hat{a} d}}+\frac{1}{2}(\Gamma^{e})^{\alpha}{}_{\beta} \psi^{c}\,_{i}\,^{\beta} W_{\hat{b} e \alpha}\,^{i} \nabla^{d}{F_{\hat{a} d}} - \frac{1}{2}(\Gamma^{d})^{\alpha}{}_{\beta} \psi^{c}\,_{i}\,^{\beta} W_{\hat{b} e \alpha}\,^{i} \nabla^{e}{F_{\hat{a} d}}+\frac{1}{2}\epsilon^{{e_{1}} {e_{2}}}\,_{\hat{a} {e_{3}} d} (\Sigma_{{e_{1}} {e_{2}}})^{\alpha}{}_{\beta} \psi^{c}\,_{i}\,^{\beta} W_{\hat{b} e \alpha}\,^{i} \nabla^{{e_{3}}}{F^{d e}} - \frac{1}{2}(\Gamma_{d})^{\alpha}{}_{\beta} \psi^{c}\,_{i}\,^{\beta} W_{\hat{b} e \alpha}\,^{i} \nabla_{\hat{a}}{F^{d e}}+\frac{1}{2}(\Gamma_{\hat{a}})^{\alpha}{}_{\beta} \psi^{c}\,_{i}\,^{\beta} W_{\hat{b} e \alpha}\,^{i} \nabla_{d}{F^{d e}} - \frac{1}{8}\epsilon_{\hat{a} \hat{b}}\,^{d e {e_{1}}} \psi^{c}\,_{i}\,^{\alpha} W_{{e_{1}} {e_{2}} \alpha}\,^{i} \nabla^{{e_{2}}}{F_{d e}}+\frac{1}{4}\epsilon_{\hat{a} {e_{2}}}\,^{d e {e_{1}}} \psi^{c}\,_{i}\,^{\alpha} W_{e {e_{1}} \alpha}\,^{i} \nabla^{{e_{2}}}{F_{\hat{b} d}} - \frac{1}{2}(\Gamma^{d})^{\alpha}{}_{\beta} \psi^{c}\,_{i}\,^{\beta} X^{i}_{\alpha} \nabla_{\hat{a}}{F_{\hat{b} d}}%
+\frac{1}{2}(\Gamma_{\hat{a}})^{\alpha}{}_{\beta} \psi^{c}\,_{i}\,^{\beta} X^{i}_{\alpha} \nabla^{d}{F_{\hat{b} d}}+\frac{1}{24}{\rm i} \epsilon_{\hat{a} \hat{b} {e_{1}} d e} \psi^{c}\,_{j}\,^{\alpha} \lambda_{i \alpha} \nabla^{{e_{1}}}{\Phi^{d e i j}}+\frac{1}{12}{\rm i} (\Gamma_{d})^{\alpha}{}_{\beta} \psi^{c}\,_{j}\,^{\beta} \lambda_{i \alpha} \nabla^{d}{\Phi_{\hat{a} \hat{b}}\,^{i j}} - \frac{1}{2}{\rm i} (\Gamma_{\hat{a}})^{\alpha}{}_{\beta} \psi^{c}\,_{j}\,^{\beta} \lambda_{i \alpha} \nabla_{d}{\Phi_{\hat{b}}\,^{d i j}} - \frac{1}{6}{\rm i} \epsilon^{{e_{1}}}\,_{\hat{a} \hat{b} {e_{2}} e} (\Sigma_{{e_{1}} d})^{\alpha}{}_{\beta} \psi^{c}\,_{j}\,^{\beta} \lambda_{i \alpha} \nabla^{{e_{2}}}{\Phi^{d e i j}} - \frac{1}{12}{\rm i} \epsilon^{{e_{1}} {e_{2}}}\,_{\hat{a} d e} (\Sigma_{{e_{1}} {e_{2}}})^{\alpha}{}_{\beta} \psi^{c}\,_{j}\,^{\beta} \lambda_{i \alpha} \nabla_{\hat{b}}{\Phi^{d e i j}}+\frac{1}{6}{\rm i} (\Gamma_{d})^{\alpha}{}_{\beta} \psi^{c}\,_{j}\,^{\beta} \lambda_{i \alpha} \nabla_{\hat{a}}{\Phi_{\hat{b}}\,^{d i j}} - \frac{1}{6}{\rm i} \epsilon^{{e_{1}}}\,_{\hat{a} \hat{b} d e} (\Sigma_{{e_{1}} {e_{2}}})^{\alpha}{}_{\beta} \psi^{c}\,_{j}\,^{\beta} \lambda_{i \alpha} \nabla^{{e_{2}}}{\Phi^{d e i j}} - \frac{1}{3}{\rm i} \epsilon^{e {e_{1}}}\,_{\hat{a} {e_{2}} d} (\Sigma_{e {e_{1}}})^{\alpha}{}_{\beta} \psi^{c}\,_{j}\,^{\beta} \lambda_{i \alpha} \nabla^{{e_{2}}}{\Phi_{\hat{b}}\,^{d i j}} - \frac{1}{64}{\rm i} \epsilon_{\hat{a} \hat{b} {e_{2}}}\,^{d e} (\Gamma_{{e_{1}}})^{\alpha}{}_{\beta} \psi^{c}\,_{i}\,^{\beta} \lambda^{i}_{\alpha} \nabla^{{e_{2}}}{\nabla^{{e_{1}}}{W_{d e}}} - \frac{1}{4}{\rm i} (\Sigma_{d e})^{\alpha}{}_{\beta} \psi^{c}\,_{i}\,^{\beta} \lambda^{i}_{\alpha} \nabla^{d}{\nabla^{e}{W_{\hat{a} \hat{b}}}} - \frac{1}{4}{\rm i} \psi^{c}\,_{i}\,^{\alpha} \lambda^{i}_{\alpha} \nabla^{d}{\nabla_{\hat{a}}{W_{\hat{b} d}}}+\frac{1}{2}{\rm i} (\Sigma_{\hat{a} e})^{\alpha}{}_{\beta} \psi^{c}\,_{i}\,^{\beta} \lambda^{i}_{\alpha} \nabla^{d}{\nabla^{e}{W_{\hat{b} d}}}+\frac{1}{2}{\rm i} (\Sigma_{e d})^{\alpha}{}_{\beta} \psi^{c}\,_{i}\,^{\beta} \lambda^{i}_{\alpha} \nabla^{e}{\nabla_{\hat{a}}{W^{d}\,_{\hat{b}}}}+\frac{1}{4}{\rm i} (\Sigma_{\hat{a} e})^{\alpha}{}_{\beta} \psi^{c}\,_{i}\,^{\beta} \lambda^{i}_{\alpha} \nabla^{e}{\nabla^{d}{W_{\hat{b} d}}}+\frac{3}{8}{\rm i} (\Sigma_{\hat{a} \hat{b}})^{\alpha}{}_{\beta} \psi^{c}\,_{i}\,^{\beta} \lambda^{i}_{\alpha} \nabla^{d}{\nabla^{e}{W_{d e}}} - \frac{5}{64}{\rm i} \epsilon_{\hat{a} \hat{b} {e_{1}} {e_{2}}}\,^{d} (\Gamma^{e})^{\alpha}{}_{\beta} \psi^{c}\,_{i}\,^{\beta} \lambda^{i}_{\alpha} \nabla^{{e_{1}}}{\nabla^{{e_{2}}}{W_{d e}}}+\frac{1}{64}{\rm i} \epsilon_{\hat{a} \hat{b} {e_{1}} {e_{2}}}\,^{d} (\Gamma^{{e_{1}}})^{\alpha}{}_{\beta} \psi^{c}\,_{i}\,^{\beta} \lambda^{i}_{\alpha} \nabla^{{e_{2}}}{\nabla^{e}{W_{d e}}} - \frac{1}{4}{\rm i} (\Sigma_{d e})^{\alpha}{}_{\beta} \psi^{c}\,_{i}\,^{\beta} \lambda^{i}_{\alpha} \nabla_{\hat{a}}{\nabla_{\hat{b}}{W^{d e}}} - \frac{1}{4}{\rm i} (\Sigma_{\hat{a} d})^{\alpha}{}_{\beta} \psi^{c}\,_{i}\,^{\beta} \lambda^{i}_{\alpha} \nabla_{\hat{b}}{\nabla_{e}{W^{d e}}}%
 - \frac{1}{8}{\rm i} \psi^{c}\,_{i}\,^{\alpha} \lambda^{i}_{\alpha} \nabla_{\hat{a}}{\nabla^{d}{W_{\hat{b} d}}} - \frac{1}{2}{\rm i} (\Sigma_{e d})^{\alpha}{}_{\beta} \psi^{c}\,_{i}\,^{\beta} \lambda^{i}_{\alpha} \nabla_{\hat{a}}{\nabla^{e}{W^{d}\,_{\hat{b}}}} - \frac{1}{2}{\rm i} (\Sigma_{\hat{a} d})^{\alpha}{}_{\beta} \psi^{c}\,_{i}\,^{\beta} \lambda^{i}_{\alpha} \nabla_{e}{\nabla_{\hat{b}}{W^{d e}}}+\frac{1}{16}{\rm i} \epsilon_{\hat{a} \hat{b} {e_{2}}}\,^{d e} (\Gamma_{{e_{1}}})^{\alpha}{}_{\beta} \psi^{c}\,_{i}\,^{\beta} \lambda^{i}_{\alpha} \nabla^{{e_{1}}}{\nabla^{{e_{2}}}{W_{d e}}}+\frac{3}{32}{\rm i} \epsilon_{\hat{a} e {e_{1}} {e_{2}}}\,^{d} (\Gamma^{e})^{\alpha}{}_{\beta} \psi^{c}\,_{i}\,^{\beta} \lambda^{i}_{\alpha} \nabla^{{e_{1}}}{\nabla^{{e_{2}}}{W_{\hat{b} d}}}+\frac{5}{64}{\rm i} \epsilon_{\hat{a} \hat{b} {e_{1}} {e_{2}}}\,^{d} (\Gamma^{{e_{1}}})^{\alpha}{}_{\beta} \psi^{c}\,_{i}\,^{\beta} \lambda^{i}_{\alpha} \nabla^{e}{\nabla^{{e_{2}}}{W_{d e}}} - \frac{3}{32}{\rm i} \epsilon_{\hat{a} {e_{1}} {e_{2}}}\,^{d e} (\Gamma^{{e_{1}}})^{\alpha}{}_{\beta} \psi^{c}\,_{i}\,^{\beta} \lambda^{i}_{\alpha} \nabla^{{e_{2}}}{\nabla_{\hat{b}}{W_{d e}}}+\frac{1}{8}{\rm i} \Phi_{d e i}\,^{j} (\Sigma_{\hat{a} \hat{b}})^{\alpha}{}_{\beta} \psi^{c}\,_{j}\,^{\beta} W^{d e} \lambda^{i}_{\alpha}+\frac{1}{24}{\rm i} \Phi_{\hat{a} \hat{b} i}\,^{j} (\Sigma_{d e})^{\alpha}{}_{\beta} \psi^{c}\,_{j}\,^{\beta} W^{d e} \lambda^{i}_{\alpha}+\frac{1}{48}{\rm i} \epsilon^{{e_{1}} {e_{2}}}\,_{\hat{a} \hat{b}}\,^{d} \Phi_{{e_{1}} {e_{2}} i}\,^{j} (\Gamma^{e})^{\alpha}{}_{\beta} \psi^{c}\,_{j}\,^{\beta} W_{d e} \lambda^{i}_{\alpha}+\frac{1}{24}{\rm i} \epsilon^{{e_{2}}}\,_{\hat{b} {e_{1}}}\,^{d e} \Phi_{\hat{a} {e_{2}} i}\,^{j} (\Gamma^{{e_{1}}})^{\alpha}{}_{\beta} \psi^{c}\,_{j}\,^{\beta} W_{d e} \lambda^{i}_{\alpha}+\frac{1}{6}{\rm i} \Phi_{\hat{a} d i}\,^{j} (\Sigma_{\hat{b} e})^{\alpha}{}_{\beta} \psi^{c}\,_{j}\,^{\beta} W^{d e} \lambda^{i}_{\alpha}+\frac{1}{4}\epsilon_{\hat{a} \hat{b} {e_{1}}}\,^{d e} \phi^{c}\,_{i}\,^{\alpha} W_{d e} \nabla^{{e_{1}}}{\lambda^{i}_{\alpha}}-(\Gamma_{d})^{\alpha}{}_{\beta} \phi^{c}\,_{i}\,^{\beta} W_{\hat{a} \hat{b}} \nabla^{d}{\lambda^{i}_{\alpha}}+\frac{3}{2}(\Gamma_{\hat{a}})^{\alpha}{}_{\beta} \phi^{c}\,_{i}\,^{\beta} W_{\hat{b} d} \nabla^{d}{\lambda^{i}_{\alpha}} - \frac{13}{8}\epsilon^{{e_{1}}}\,_{\hat{a} \hat{b} {e_{2}}}\,^{d} (\Sigma_{{e_{1}}}{}^{\, e})^{\alpha}{}_{\beta} \phi^{c}\,_{i}\,^{\beta} W_{d e} \nabla^{{e_{2}}}{\lambda^{i}_{\alpha}}+\frac{13}{16}\epsilon^{{e_{1}} {e_{2}}}\,_{\hat{a}}\,^{d e} (\Sigma_{{e_{1}} {e_{2}}})^{\alpha}{}_{\beta} \phi^{c}\,_{i}\,^{\beta} W_{d e} \nabla_{\hat{b}}{\lambda^{i}_{\alpha}}+\frac{7}{4}(\Gamma^{d})^{\alpha}{}_{\beta} \phi^{c}\,_{i}\,^{\beta} W_{\hat{b} d} \nabla_{\hat{a}}{\lambda^{i}_{\alpha}}+\frac{5}{16}(\Sigma_{d e})^{\alpha}{}_{\beta} \phi^{c}\,_{i}\,^{\beta} W^{d e} W_{\hat{a} \hat{b}} \lambda^{i}_{\alpha} - \frac{27}{8}(\Sigma_{\hat{a} d})^{\alpha}{}_{\beta} \phi^{c}\,_{i}\,^{\beta} W^{d e} W_{\hat{b} e} \lambda^{i}_{\alpha}%
+\frac{15}{32}(\Sigma_{\hat{a} \hat{b}})^{\alpha}{}_{\beta} \phi^{c}\,_{i}\,^{\beta} W^{d e} W_{d e} \lambda^{i}_{\alpha}+\frac{1}{2}\epsilon_{\hat{a} \hat{b}}\,^{d e {e_{1}}} (\Gamma^{{e_{2}}})^{\alpha}{}_{\beta} \phi^{c}\,_{i}\,^{\beta} W_{d e} W_{{e_{1}} {e_{2}}} \lambda^{i}_{\alpha} - \frac{5}{32}\epsilon_{\hat{a} {e_{2}}}\,^{d e {e_{1}}} (\Gamma^{{e_{2}}})^{\alpha}{}_{\beta} \phi^{c}\,_{i}\,^{\beta} W_{\hat{b} d} W_{e {e_{1}}} \lambda^{i}_{\alpha} - \frac{5}{8}(\Sigma^{d e})^{\alpha}{}_{\beta} \phi^{c}\,_{i}\,^{\beta} W_{\hat{a} d} W_{\hat{b} e} \lambda^{i}_{\alpha} - \frac{13}{16}\epsilon^{{e_{1}}}\,_{\hat{a} \hat{b}}\,^{d e} (\Sigma_{{e_{1}} {e_{2}}})^{\alpha}{}_{\beta} \phi^{c}\,_{i}\,^{\beta} W_{d e} \nabla^{{e_{2}}}{\lambda^{i}_{\alpha}}+\frac{17}{8}\epsilon^{e {e_{1}}}\,_{\hat{a} {e_{2}}}\,^{d} (\Sigma_{e {e_{1}}})^{\alpha}{}_{\beta} \phi^{c}\,_{i}\,^{\beta} W_{\hat{b} d} \nabla^{{e_{2}}}{\lambda^{i}_{\alpha}}+\frac{9}{8}{\rm i} (\Sigma_{d e})^{\alpha}{}_{\beta} \phi^{c}\,_{i}\,^{\beta} W W^{d e} W_{\hat{a} \hat{b} \alpha}\,^{i}+\frac{11}{2}{\rm i} \phi^{c}\,_{i}\,^{\alpha} F_{\hat{a}}\,^{d} W_{\hat{b} d \alpha}\,^{i} - \frac{3}{4}\Phi_{\hat{a} \hat{b} i}\,^{j} \phi^{c}\,_{j}\,^{\alpha} \lambda^{i}_{\alpha}+\frac{3}{8}\epsilon^{e {e_{1}}}\,_{\hat{a} \hat{b} d} \Phi_{e {e_{1}} i}\,^{j} (\Gamma^{d})^{\alpha}{}_{\beta} \phi^{c}\,_{j}\,^{\beta} \lambda^{i}_{\alpha}+3\Phi_{\hat{a}}\,^{d}\,_{i}\,^{j} (\Sigma_{\hat{b} d})^{\alpha}{}_{\beta} \phi^{c}\,_{j}\,^{\beta} \lambda^{i}_{\alpha} - \frac{3}{8}(\Gamma_{d})^{\alpha}{}_{\beta} \phi^{c}\,_{i}\,^{\beta} \lambda^{i}_{\alpha} \nabla^{d}{W_{\hat{a} \hat{b}}}+\frac{9}{16}\epsilon_{\hat{a} \hat{b} {e_{1}}}\,^{d e} \phi^{c}\,_{i}\,^{\alpha} \lambda^{i}_{\alpha} \nabla^{{e_{1}}}{W_{d e}} - \frac{15}{32}\epsilon^{{e_{1}}}\,_{\hat{a} \hat{b}}\,^{d e} (\Sigma_{{e_{1}} {e_{2}}})^{\alpha}{}_{\beta} \phi^{c}\,_{i}\,^{\beta} \lambda^{i}_{\alpha} \nabla^{{e_{2}}}{W_{d e}}+\frac{9}{16}\epsilon^{e {e_{1}}}\,_{\hat{a} {e_{2}}}\,^{d} (\Sigma_{e {e_{1}}})^{\alpha}{}_{\beta} \phi^{c}\,_{i}\,^{\beta} \lambda^{i}_{\alpha} \nabla^{{e_{2}}}{W_{\hat{b} d}}+\frac{3}{4}(\Gamma^{d})^{\alpha}{}_{\beta} \phi^{c}\,_{i}\,^{\beta} \lambda^{i}_{\alpha} \nabla_{\hat{a}}{W_{\hat{b} d}}+\frac{15}{8}(\Gamma_{\hat{a}})^{\alpha}{}_{\beta} \phi^{c}\,_{i}\,^{\beta} \lambda^{i}_{\alpha} \nabla^{d}{W_{\hat{b} d}}+\frac{21}{16}\epsilon^{{e_{1}}}\,_{\hat{a} \hat{b} {e_{2}} e} (\Sigma_{{e_{1}} d})^{\alpha}{}_{\beta} \phi^{c}\,_{i}\,^{\beta} \lambda^{i}_{\alpha} \nabla^{{e_{2}}}{W^{d e}}+\frac{21}{32}\epsilon^{{e_{1}} {e_{2}}}\,_{\hat{a}}\,^{d e} (\Sigma_{{e_{1}} {e_{2}}})^{\alpha}{}_{\beta} \phi^{c}\,_{i}\,^{\beta} \lambda^{i}_{\alpha} \nabla_{\hat{b}}{W_{d e}}+2(\Sigma_{\hat{a} \hat{b}})^{\alpha}{}_{\beta} \phi^{c}\,_{i}\,^{\beta} \nabla_{d}{\nabla^{d}{\lambda^{i}_{\alpha}}}%
-2\phi^{c}\,_{i}\,^{\alpha} \nabla_{\hat{a}}{\nabla_{\hat{b}}{\lambda^{i}_{\alpha}}}+4(\Sigma_{\hat{a} d})^{\alpha}{}_{\beta} \phi^{c}\,_{i}\,^{\beta} \nabla^{d}{\nabla_{\hat{b}}{\lambda^{i}_{\alpha}}}+\frac{1}{4}{\rm i} (\Gamma_{d})^{\alpha}{}_{\beta} \phi^{c}\,_{i}\,^{\beta} W_{\hat{a} \hat{b} \alpha}\,^{i} \nabla^{d}{W}+\frac{1}{8}{\rm i} \epsilon_{\hat{a} \hat{b} {e_{1}}}\,^{d e} \phi^{c}\,_{i}\,^{\alpha} W_{d e \alpha}\,^{i} \nabla^{{e_{1}}}{W} - \frac{1}{4}{\rm i} \epsilon^{{e_{1}}}\,_{\hat{a} \hat{b}}\,^{d e} (\Sigma_{{e_{1}} {e_{2}}})^{\alpha}{}_{\beta} \phi^{c}\,_{i}\,^{\beta} W_{d e \alpha}\,^{i} \nabla^{{e_{2}}}{W} - \frac{1}{2}{\rm i} \epsilon^{e {e_{1}}}\,_{\hat{a} {e_{2}}}\,^{d} (\Sigma_{e {e_{1}}})^{\alpha}{}_{\beta} \phi^{c}\,_{i}\,^{\beta} W_{\hat{b} d \alpha}\,^{i} \nabla^{{e_{2}}}{W}+\frac{1}{2}{\rm i} (\Gamma^{d})^{\alpha}{}_{\beta} \phi^{c}\,_{i}\,^{\beta} W_{\hat{b} d \alpha}\,^{i} \nabla_{\hat{a}}{W} - \frac{1}{2}{\rm i} (\Gamma_{\hat{a}})^{\alpha}{}_{\beta} \phi^{c}\,_{i}\,^{\beta} W_{\hat{b} d \alpha}\,^{i} \nabla^{d}{W}+\frac{1}{4}{\rm i} (\Gamma_{d})^{\alpha}{}_{\beta} \phi^{c}\,_{i}\,^{\beta} W \nabla^{d}{W_{\hat{a} \hat{b} \alpha}\,^{i}}+\frac{1}{8}{\rm i} \epsilon_{\hat{a} \hat{b} {e_{1}}}\,^{d e} \phi^{c}\,_{i}\,^{\alpha} W \nabla^{{e_{1}}}{W_{d e \alpha}\,^{i}} - \frac{1}{4}{\rm i} \epsilon^{{e_{1}}}\,_{\hat{a} \hat{b}}\,^{d e} (\Sigma_{{e_{1}} {e_{2}}})^{\alpha}{}_{\beta} \phi^{c}\,_{i}\,^{\beta} W \nabla^{{e_{2}}}{W_{d e \alpha}\,^{i}} - \frac{1}{2}{\rm i} \epsilon^{e {e_{1}}}\,_{\hat{a} {e_{2}}}\,^{d} (\Sigma_{e {e_{1}}})^{\alpha}{}_{\beta} \phi^{c}\,_{i}\,^{\beta} W \nabla^{{e_{2}}}{W_{\hat{b} d \alpha}\,^{i}}+\frac{1}{2}{\rm i} (\Gamma^{d})^{\alpha}{}_{\beta} \phi^{c}\,_{i}\,^{\beta} W \nabla_{\hat{a}}{W_{\hat{b} d \alpha}\,^{i}} - \frac{1}{2}{\rm i} (\Gamma_{\hat{a}})^{\alpha}{}_{\beta} \phi^{c}\,_{i}\,^{\beta} W \nabla^{d}{W_{\hat{b} d \alpha}\,^{i}}-8\nabla_{\hat{a}}{F_{\hat{b}}\,^{d}} f^{c}\,_{d}-8\nabla^{d}{F_{\hat{a} d}} f^{c}\,_{\hat{b}}-2\nabla^{d}{F_{\hat{a} \hat{b}}} f^{c}\,_{d}
\doublespacedmathend
\end{adjustwidth}

\subsubsection{$\Box^2 W$}

\begin{adjustwidth}{0cm}{5cm}
\doublespacedmathbegin
\mathcal{D}_{a}{\nabla^{a}{\nabla_{b}{\nabla^{b}{W}}}} - \frac{1}{2}{\rm i} \psi_{a i}\,^{\alpha} \nabla^{a}{\nabla_{b}{\nabla^{b}{\lambda^{i}_{\alpha}}}}+\frac{3}{32}{\rm i} \epsilon^{c d}\,_{e}\,^{a b} (\Sigma_{c d})^{\alpha}{}_{\beta} \psi_{{e_{1}} i}\,^{\beta} \lambda^{i}_{\alpha} \nabla^{{e_{1}}}{\nabla^{e}{W_{a b}}} - \frac{9}{32}{\rm i} (\Gamma^{a})^{\alpha}{}_{\beta} \psi_{c i}\,^{\beta} \lambda^{i}_{\alpha} \nabla^{c}{\nabla^{b}{W_{a b}}}+\frac{13}{128}{\rm i} \epsilon^{c d}\,_{e}\,^{a b} (\Sigma_{c d})^{\alpha}{}_{\beta} \psi_{{e_{1}} i}\,^{\beta} \nabla^{e}{W_{a b}} \nabla^{{e_{1}}}{\lambda^{i}_{\alpha}} - \frac{11}{32}{\rm i} (\Gamma^{a})^{\alpha}{}_{\beta} \psi_{c i}\,^{\beta} \nabla^{b}{W_{a b}} \nabla^{c}{\lambda^{i}_{\alpha}}+\frac{9}{64}{\rm i} \epsilon^{c d a b}\,_{e} (\Sigma_{c d})^{\alpha}{}_{\beta} \psi_{{e_{1}} i}\,^{\beta} \nabla^{{e_{1}}}{W_{a b}} \nabla^{e}{\lambda^{i}_{\alpha}} - \frac{3}{8}{\rm i} (\Gamma^{a})^{\alpha}{}_{\beta} \psi_{c i}\,^{\beta} \nabla^{c}{W_{a b}} \nabla^{b}{\lambda^{i}_{\alpha}}+\frac{1}{8}{\rm i} \epsilon^{c d}\,_{e}\,^{a b} (\Sigma_{c d})^{\alpha}{}_{\beta} \psi_{{e_{1}} i}\,^{\beta} W_{a b} \nabla^{{e_{1}}}{\nabla^{e}{\lambda^{i}_{\alpha}}} - \frac{1}{2}{\rm i} (\Gamma^{a})^{\alpha}{}_{\beta} \psi_{c i}\,^{\beta} W_{a b} \nabla^{c}{\nabla^{b}{\lambda^{i}_{\alpha}}}+\frac{1}{8}(\Gamma_{a})^{\alpha}{}_{\beta} \psi_{b i}\,^{\beta} X^{i}_{\alpha} \nabla^{b}{\nabla^{a}{W}}+\frac{3}{64}{\rm i} \psi_{c i}\,^{\alpha} W_{a b} \lambda^{i}_{\alpha} \nabla^{c}{W^{a b}}+\frac{57}{2048}{\rm i} \epsilon_{e}\,^{c d a b} (\Gamma^{e})^{\alpha}{}_{\beta} \psi_{{e_{1}} i}\,^{\beta} W_{a b} \lambda^{i}_{\alpha} \nabla^{{e_{1}}}{W_{c d}}+\frac{3}{16}{\rm i} (\Sigma^{a}{}_{\, c})^{\alpha}{}_{\beta} \psi_{d i}\,^{\beta} W_{a b} \lambda^{i}_{\alpha} \nabla^{d}{W^{c b}}+\frac{11}{128}{\rm i} \psi_{c i}\,^{\alpha} W^{a b} W_{a b} \nabla^{c}{\lambda^{i}_{\alpha}} - \frac{1}{8}{\rm i} (\Sigma_{a b})^{\alpha}{}_{\beta} \psi_{c j}\,^{\beta} \lambda_{i \alpha} \nabla^{c}{\Phi^{a b i j}}+\frac{1}{24}{\rm i} \Phi^{a b}\,_{i}\,^{j} (\Sigma_{a b})^{\alpha}{}_{\beta} \psi_{c j}\,^{\beta} \nabla^{c}{\lambda^{i}_{\alpha}}+\frac{1}{32}\psi_{c i}\,^{\alpha} W W_{a b \alpha}\,^{i} \nabla^{c}{W^{a b}}+\frac{1}{32}\psi_{c i}\,^{\alpha} W W^{a b} \nabla^{c}{W_{a b \alpha}\,^{i}}%
+\frac{15}{128}\psi_{c i}\,^{\alpha} W^{a b} W_{a b \alpha}\,^{i} \nabla^{c}{W}+\frac{1}{16}{\rm i} \epsilon^{c d}\,_{e}\,^{a b} (\Sigma_{c d})^{\alpha}{}_{\beta} \psi^{e}\,_{i}\,^{\beta} W_{a b} \nabla_{{e_{1}}}{\nabla^{{e_{1}}}{\lambda^{i}_{\alpha}}} - \frac{1}{4}{\rm i} (\Gamma^{a})^{\alpha}{}_{\beta} \psi^{b}\,_{i}\,^{\beta} W_{a b} \nabla_{c}{\nabla^{c}{\lambda^{i}_{\alpha}}}+\frac{3}{16}{\rm i} \psi_{c i}\,^{\alpha} W_{a b} \lambda^{i}_{\alpha} \nabla^{a}{W^{c b}}+\frac{9}{512}{\rm i} \epsilon_{e {e_{1}}}\,^{c d a} (\Gamma^{e})^{\alpha}{}_{\beta} \psi^{{e_{1}}}\,_{i}\,^{\beta} W_{a b} \lambda^{i}_{\alpha} \nabla^{b}{W_{c d}} - \frac{3}{16}{\rm i} (\Sigma^{a}{}_{\, c})^{\alpha}{}_{\beta} \psi_{d i}\,^{\beta} W_{a b} \lambda^{i}_{\alpha} \nabla^{b}{W^{c d}}+\frac{3}{8}{\rm i} (\Sigma_{d c})^{\alpha}{}_{\beta} \psi^{d}\,_{i}\,^{\beta} W_{a b} \lambda^{i}_{\alpha} \nabla^{a}{W^{c b}}+\frac{9}{64}{\rm i} (\Sigma_{a b})^{\alpha}{}_{\beta} \psi^{c}\,_{i}\,^{\beta} W^{a b} \lambda^{i}_{\alpha} \nabla^{d}{W_{c d}} - \frac{3}{16}{\rm i} (\Sigma_{c d})^{\alpha}{}_{\beta} \psi^{c}\,_{i}\,^{\beta} W_{a b} \lambda^{i}_{\alpha} \nabla^{d}{W^{a b}} - \frac{9}{32}{\rm i} (\Sigma^{a}{}_{\, d})^{\alpha}{}_{\beta} \psi^{d}\,_{i}\,^{\beta} W_{a b} \lambda^{i}_{\alpha} \nabla_{c}{W^{c b}}+\frac{15}{512}{\rm i} \epsilon_{e {e_{1}}}\,^{c a b} (\Gamma^{d})^{\alpha}{}_{\beta} \psi^{e}\,_{i}\,^{\beta} W_{a b} \lambda^{i}_{\alpha} \nabla^{{e_{1}}}{W_{c d}} - \frac{3}{64}{\rm i} (\Sigma_{c d})^{\alpha}{}_{\beta} \psi^{a}\,_{i}\,^{\beta} W_{a b} \lambda^{i}_{\alpha} \nabla^{b}{W^{c d}}+\frac{27}{128}{\rm i} \psi^{a}\,_{i}\,^{\alpha} W_{a b} \lambda^{i}_{\alpha} \nabla_{c}{W^{c b}}+\frac{15}{1024}{\rm i} \epsilon_{e {e_{1}}}\,^{c d a} (\Gamma^{e})^{\alpha}{}_{\beta} \psi^{b}\,_{i}\,^{\beta} W_{a b} \lambda^{i}_{\alpha} \nabla^{{e_{1}}}{W_{c d}}+\frac{9}{64}{\rm i} (\Sigma^{a}{}_{\, c})^{\alpha}{}_{\beta} \psi^{b}\,_{i}\,^{\beta} W_{a b} \lambda^{i}_{\alpha} \nabla_{d}{W^{c d}} - \frac{3}{32}{\rm i} (\Sigma_{d c})^{\alpha}{}_{\beta} \psi^{a}\,_{i}\,^{\beta} W_{a b} \lambda^{i}_{\alpha} \nabla^{d}{W^{c b}} - \frac{1}{128}{\rm i} \epsilon_{e}\,^{a b c d} (\Gamma^{e})^{\alpha}{}_{\beta} \psi_{{e_{1}} i}\,^{\beta} W_{a b} W_{c d} \nabla^{{e_{1}}}{\lambda^{i}_{\alpha}}+\frac{1}{64}{\rm i} (\Sigma^{c}{}_{\, a})^{\alpha}{}_{\beta} \psi_{d i}\,^{\beta} W^{a b} W_{c b} \nabla^{d}{\lambda^{i}_{\alpha}}+\frac{1}{16}{\rm i} \psi^{b}\,_{i}\,^{\alpha} W^{a}\,_{b} W_{a c} \nabla^{c}{\lambda^{i}_{\alpha}}+\frac{1}{128}{\rm i} \epsilon_{e {e_{1}}}\,^{a b c} (\Gamma^{e})^{\alpha}{}_{\beta} \psi^{{e_{1}}}\,_{i}\,^{\beta} W_{a b} W_{c d} \nabla^{d}{\lambda^{i}_{\alpha}}%
+\frac{1}{8}{\rm i} (\Sigma^{a c})^{\alpha}{}_{\beta} \psi^{b}\,_{i}\,^{\beta} W_{a b} W_{c d} \nabla^{d}{\lambda^{i}_{\alpha}} - \frac{3}{8}{\rm i} (\Sigma_{d a})^{\alpha}{}_{\beta} \psi^{d}\,_{i}\,^{\beta} W^{a b} W_{b c} \nabla^{c}{\lambda^{i}_{\alpha}} - \frac{5}{8}{\rm i} (\Sigma_{a d})^{\alpha}{}_{\beta} \psi^{c}\,_{i}\,^{\beta} W^{a b} W_{c b} \nabla^{d}{\lambda^{i}_{\alpha}}+\frac{1}{8}{\rm i} (\Sigma_{a b})^{\alpha}{}_{\beta} \psi^{c}\,_{i}\,^{\beta} W^{a b} W_{c d} \nabla^{d}{\lambda^{i}_{\alpha}}+\frac{1}{32}{\rm i} (\Sigma_{c d})^{\alpha}{}_{\beta} \psi^{c}\,_{i}\,^{\beta} W^{a b} W_{a b} \nabla^{d}{\lambda^{i}_{\alpha}}+\frac{11}{128}{\rm i} \epsilon_{e {e_{1}}}\,^{a b c} (\Gamma^{d})^{\alpha}{}_{\beta} \psi^{e}\,_{i}\,^{\beta} W_{a b} W_{c d} \nabla^{{e_{1}}}{\lambda^{i}_{\alpha}}+\frac{3}{128}{\rm i} \epsilon_{e {e_{1}}}\,^{a b c} (\Gamma^{e})^{\alpha}{}_{\beta} \psi^{d}\,_{i}\,^{\beta} W_{a b} W_{c d} \nabla^{{e_{1}}}{\lambda^{i}_{\alpha}}+\frac{9}{512}{\rm i} \epsilon_{{e_{1}}}\,^{c d a b} (\Gamma_{e})^{\alpha}{}_{\beta} \psi^{e}\,_{i}\,^{\beta} W_{a b} \lambda^{i}_{\alpha} \nabla^{{e_{1}}}{W_{c d}} - \frac{9}{512}{\rm i} \epsilon_{d e {e_{1}} c}\,^{a} (\Gamma^{d})^{\alpha}{}_{\beta} \psi^{e}\,_{i}\,^{\beta} W_{a b} \lambda^{i}_{\alpha} \nabla^{{e_{1}}}{W^{c b}} - \frac{3}{16}{\rm i} (\Sigma^{a}{}_{\, d})^{\alpha}{}_{\beta} \psi_{c i}\,^{\beta} W_{a b} \lambda^{i}_{\alpha} \nabla^{d}{W^{c b}}+\frac{15}{256}{\rm i} \epsilon_{{e_{1}}}\,^{a b c d} (\Gamma_{e})^{\alpha}{}_{\beta} \psi^{{e_{1}}}\,_{i}\,^{\beta} W_{a b} W_{c d} \nabla^{e}{\lambda^{i}_{\alpha}}+\frac{3}{256}{\rm i} \epsilon_{{e_{1}}}\,^{a b c d} (\Gamma_{e})^{\alpha}{}_{\beta} \psi^{e}\,_{i}\,^{\beta} W_{a b} W_{c d} \nabla^{{e_{1}}}{\lambda^{i}_{\alpha}} - \frac{9}{256}{\rm i} \epsilon^{e {e_{1}}}\,_{{e_{2}}}\,^{c d} (\Sigma_{e {e_{1}}})^{\alpha}{}_{\beta} \psi^{{e_{2}}}\,_{i}\,^{\beta} W^{a b} W_{c d} W_{a b} \lambda^{i}_{\alpha}+\frac{9}{512}{\rm i} (\Gamma^{c})^{\alpha}{}_{\beta} \psi^{d}\,_{i}\,^{\beta} W^{a b} W_{c d} W_{a b} \lambda^{i}_{\alpha}+\frac{27}{512}{\rm i} \epsilon_{{e_{2}}}\,^{c d e {e_{1}}} (\Sigma_{a b})^{\alpha}{}_{\beta} \psi^{{e_{2}}}\,_{i}\,^{\beta} W^{a b} W_{c d} W_{e {e_{1}}} \lambda^{i}_{\alpha} - \frac{9}{512}{\rm i} \epsilon^{a b c d e} \psi^{{e_{1}}}\,_{i}\,^{\alpha} W_{a b} W_{c d} W_{e {e_{1}}} \lambda^{i}_{\alpha} - \frac{9}{64}{\rm i} (\Gamma^{c})^{\alpha}{}_{\beta} \psi^{d}\,_{i}\,^{\beta} W^{a b} W_{c a} W_{d b} \lambda^{i}_{\alpha}+\frac{27}{256}{\rm i} \epsilon^{{e_{2}} a b c d} (\Sigma_{{e_{2}}}{}^{\, e})^{\alpha}{}_{\beta} \psi^{{e_{1}}}\,_{i}\,^{\beta} W_{a b} W_{c d} W_{e {e_{1}}} \lambda^{i}_{\alpha} - \frac{27}{256}{\rm i} \epsilon^{a b c d e} (\Sigma^{{e_{1}}}{}_{\, {e_{2}}})^{\alpha}{}_{\beta} \psi^{{e_{2}}}\,_{i}\,^{\beta} W_{a b} W_{c d} W_{e {e_{1}}} \lambda^{i}_{\alpha} - \frac{1}{32}{\rm i} \Phi_{a b i}\,^{j} (\Gamma_{c})^{\alpha}{}_{\beta} \psi^{c}\,_{j}\,^{\beta} W^{a b} \lambda^{i}_{\alpha}%
 - \frac{1}{64}{\rm i} \epsilon^{c d}\,_{e}\,^{a b} \Phi_{c d i}\,^{j} \psi^{e}\,_{j}\,^{\alpha} W_{a b} \lambda^{i}_{\alpha}+\frac{9}{128}{\rm i} \epsilon^{c d e a b} \Phi_{c d i}\,^{j} (\Sigma_{e {e_{1}}})^{\alpha}{}_{\beta} \psi^{{e_{1}}}\,_{j}\,^{\beta} W_{a b} \lambda^{i}_{\alpha}+\frac{15}{64}{\rm i} \epsilon^{c d e}\,_{{e_{1}}}\,^{a} \Phi_{c}\,^{b}\,_{i}\,^{j} (\Sigma_{d e})^{\alpha}{}_{\beta} \psi^{{e_{1}}}\,_{j}\,^{\beta} W_{a b} \lambda^{i}_{\alpha} - \frac{5}{32}{\rm i} \Phi_{c}\,^{a}\,_{i}\,^{j} (\Gamma^{c})^{\alpha}{}_{\beta} \psi^{b}\,_{j}\,^{\beta} W_{a b} \lambda^{i}_{\alpha}+\frac{5}{16}{\rm i} \Phi^{b}\,_{c i}\,^{j} (\Gamma^{a})^{\alpha}{}_{\beta} \psi^{c}\,_{j}\,^{\beta} W_{a b} \lambda^{i}_{\alpha} - \frac{1}{1024}\epsilon^{e {e_{1}}}\,_{{e_{2}}}\,^{c d} (\Sigma_{e {e_{1}}})^{\alpha}{}_{\beta} \psi^{{e_{2}}}\,_{i}\,^{\beta} W W^{a b} W_{c d} W_{a b \alpha}\,^{i}+\frac{3}{512}(\Gamma^{c})^{\alpha}{}_{\beta} \psi^{d}\,_{i}\,^{\beta} W W^{a b} W_{c d} W_{a b \alpha}\,^{i} - \frac{1}{1024}\epsilon_{{e_{2}}}\,^{c d e {e_{1}}} (\Sigma_{a b})^{\alpha}{}_{\beta} \psi^{{e_{2}}}\,_{i}\,^{\beta} W W^{a b} W_{c d} W_{e {e_{1}} \alpha}\,^{i} - \frac{1}{1024}\epsilon^{a b c e {e_{1}}} \psi^{d}\,_{i}\,^{\alpha} W W_{a b} W_{c d} W_{e {e_{1}} \alpha}\,^{i}+\frac{1}{512}(\Gamma^{c})^{\alpha}{}_{\beta} \psi^{d}\,_{i}\,^{\beta} W W^{a b} W_{a b} W_{c d \alpha}\,^{i}+\frac{3}{256}(\Gamma^{d})^{\alpha}{}_{\beta} \psi^{c}\,_{i}\,^{\beta} W W^{a b} W_{c a} W_{d b \alpha}\,^{i} - \frac{1}{128}(\Gamma^{c})^{\alpha}{}_{\beta} \psi^{d}\,_{i}\,^{\beta} W W^{a b} W_{c a} W_{d b \alpha}\,^{i} - \frac{1}{256}\epsilon^{{e_{2}} a b e {e_{1}}} (\Sigma_{{e_{2}}}{}^{\, c})^{\alpha}{}_{\beta} \psi^{d}\,_{i}\,^{\beta} W W_{a b} W_{c d} W_{e {e_{1}} \alpha}\,^{i}+\frac{1}{256}\epsilon^{a b c e {e_{1}}} (\Sigma^{d}{}_{\, {e_{2}}})^{\alpha}{}_{\beta} \psi^{{e_{2}}}\,_{i}\,^{\beta} W W_{a b} W_{c d} W_{e {e_{1}} \alpha}\,^{i} - \frac{1}{256}(\Gamma_{d})^{\alpha}{}_{\beta} \psi^{d}\,_{i}\,^{\beta} W W^{a b} W^{c}\,_{a} W_{c b \alpha}\,^{i}+\frac{1}{256}\epsilon_{{e_{1}}}\,^{a b c e} \psi^{{e_{1}}}\,_{i}\,^{\alpha} W W_{a b} W_{c}\,^{d} W_{e d \alpha}\,^{i} - \frac{3}{512}\epsilon^{{e_{1}} a b c e} (\Sigma_{{e_{1}} {e_{2}}})^{\alpha}{}_{\beta} \psi^{{e_{2}}}\,_{i}\,^{\beta} W W_{a b} W_{c}\,^{d} W_{e d \alpha}\,^{i}+\frac{3}{512}\epsilon^{e {e_{1}}}\,_{{e_{2}}}\,^{c d} (\Sigma_{e {e_{1}}})^{\alpha}{}_{\beta} \psi^{{e_{2}}}\,_{i}\,^{\beta} W W^{a b} W_{c a} W_{d b \alpha}\,^{i} - \frac{3}{512}\epsilon^{e {e_{1}}}\,_{{e_{2}}}\,^{a c} (\Sigma_{e {e_{1}}})^{\alpha}{}_{\beta} \psi^{{e_{2}}}\,_{i}\,^{\beta} W W_{a}\,^{b} W_{c}\,^{d} W_{b d \alpha}\,^{i} - \frac{3}{256}(\Gamma^{a})^{\alpha}{}_{\beta} \psi^{d}\,_{i}\,^{\beta} W W_{a}\,^{b} W^{c}\,_{d} W_{b c \alpha}\,^{i}%
+\frac{3}{2048}{\rm i} \epsilon_{{e_{1}}}\,^{c d a b} (\Gamma_{e})^{\alpha}{}_{\beta} \psi^{{e_{1}}}\,_{i}\,^{\beta} W_{a b} \lambda^{i}_{\alpha} \nabla^{e}{W_{c d}}+\frac{69}{1024}{\rm i} \epsilon_{e {e_{1}}}\,^{c d a} (\Gamma^{b})^{\alpha}{}_{\beta} \psi^{e}\,_{i}\,^{\beta} W_{a b} \lambda^{i}_{\alpha} \nabla^{{e_{1}}}{W_{c d}}+\frac{15}{512}{\rm i} \epsilon_{e {e_{1}}}\,^{c a b} (\Gamma^{e})^{\alpha}{}_{\beta} \psi^{d}\,_{i}\,^{\beta} W_{a b} \lambda^{i}_{\alpha} \nabla^{{e_{1}}}{W_{c d}}+\frac{1}{64}{\rm i} \epsilon^{c d e}\,_{{e_{1}}}\,^{a} \Phi_{c d i}\,^{j} (\Sigma_{e}{}^{\, b})^{\alpha}{}_{\beta} \psi^{{e_{1}}}\,_{j}\,^{\beta} W_{a b} \lambda^{i}_{\alpha} - \frac{1}{128}{\rm i} \epsilon^{c e {e_{1}} a b} \Phi_{c d i}\,^{j} (\Sigma_{e {e_{1}}})^{\alpha}{}_{\beta} \psi^{d}\,_{j}\,^{\beta} W_{a b} \lambda^{i}_{\alpha} - \frac{1}{256}\epsilon^{{e_{1}}}\,_{{e_{2}}}\,^{a c e} (\Sigma_{{e_{1}}}{}^{\, b})^{\alpha}{}_{\beta} \psi^{{e_{2}}}\,_{i}\,^{\beta} W W_{a b} W_{c}\,^{d} W_{e d \alpha}\,^{i} - \frac{1}{1024}\epsilon^{{e_{1}} {e_{2}} a b c} (\Sigma_{{e_{1}} {e_{2}}})^{\alpha}{}_{\beta} \psi^{e}\,_{i}\,^{\beta} W W_{a b} W_{c}\,^{d} W_{d e \alpha}\,^{i}+\frac{1}{1024}\epsilon^{{e_{1}} {e_{2}} a b e} (\Sigma_{{e_{1}} {e_{2}}})^{\alpha}{}_{\beta} \psi^{d}\,_{i}\,^{\beta} W W_{a b} W^{c}\,_{d} W_{e c \alpha}\,^{i} - \frac{3}{8}(\Gamma_{a})^{\alpha}{}_{\beta} \psi^{a}\,_{i}\,^{\beta} X^{i}_{\alpha} \nabla_{b}{\nabla^{b}{W}} - \frac{1}{4}{\rm i} \Phi_{a b i}\,^{j} \psi^{a}\,_{j}\,^{\alpha} \nabla^{b}{\lambda^{i}_{\alpha}} - \frac{1}{24}{\rm i} \epsilon^{b c}\,_{a d e} \Phi_{b c i}\,^{j} (\Gamma^{a})^{\alpha}{}_{\beta} \psi^{d}\,_{j}\,^{\beta} \nabla^{e}{\lambda^{i}_{\alpha}}+\frac{1}{2}{\rm i} \Phi^{a}\,_{b i}\,^{j} (\Sigma_{a c})^{\alpha}{}_{\beta} \psi^{b}\,_{j}\,^{\beta} \nabla^{c}{\lambda^{i}_{\alpha}}+\frac{1}{6}{\rm i} \Phi^{a}\,_{b i}\,^{j} (\Sigma_{a c})^{\alpha}{}_{\beta} \psi^{c}\,_{j}\,^{\beta} \nabla^{b}{\lambda^{i}_{\alpha}} - \frac{3}{32}{\rm i} \epsilon^{c e}\,_{{e_{1}}}\,^{a b} \Phi_{c}\,^{d}\,_{i}\,^{j} (\Sigma_{e d})^{\alpha}{}_{\beta} \psi^{{e_{1}}}\,_{j}\,^{\beta} W_{a b} \lambda^{i}_{\alpha}+\frac{3}{64}{\rm i} \epsilon^{c d e {e_{1}} a} \Phi_{c d i}\,^{j} (\Sigma_{e {e_{1}}})^{\alpha}{}_{\beta} \psi^{b}\,_{j}\,^{\beta} W_{a b} \lambda^{i}_{\alpha} - \frac{1}{8}{\rm i} (\Gamma_{c})^{\alpha}{}_{\beta} \psi^{a}\,_{i}\,^{\beta} \nabla^{c}{W_{a b}} \nabla^{b}{\lambda^{i}_{\alpha}} - \frac{1}{16}{\rm i} \epsilon_{c d}\,^{a b}\,_{e} \psi^{c}\,_{i}\,^{\alpha} \nabla^{d}{W_{a b}} \nabla^{e}{\lambda^{i}_{\alpha}}+\frac{1}{64}{\rm i} \epsilon^{c}\,_{e}\,^{a b}\,_{{e_{1}}} (\Sigma_{c d})^{\alpha}{}_{\beta} \psi^{e}\,_{i}\,^{\beta} \nabla^{d}{W_{a b}} \nabla^{{e_{1}}}{\lambda^{i}_{\alpha}}+\frac{3}{64}{\rm i} \epsilon^{c d}\,_{e}\,^{a}\,_{{e_{1}}} (\Sigma_{c d})^{\alpha}{}_{\beta} \psi^{b}\,_{i}\,^{\beta} \nabla^{e}{W_{a b}} \nabla^{{e_{1}}}{\lambda^{i}_{\alpha}}+\frac{3}{16}{\rm i} (\Gamma_{c})^{\alpha}{}_{\beta} \psi^{a}\,_{i}\,^{\beta} \nabla^{b}{W_{a b}} \nabla^{c}{\lambda^{i}_{\alpha}}%
 - \frac{1}{8}{\rm i} (\Gamma^{a})^{\alpha}{}_{\beta} \psi^{b}\,_{i}\,^{\beta} \nabla_{c}{W_{a b}} \nabla^{c}{\lambda^{i}_{\alpha}}+\frac{1}{16}{\rm i} \epsilon^{c d}\,_{e {e_{1}}}\,^{a} (\Sigma_{c d})^{\alpha}{}_{\beta} \psi^{e}\,_{i}\,^{\beta} \nabla^{{e_{1}}}{W_{a b}} \nabla^{b}{\lambda^{i}_{\alpha}} - \frac{1}{16}{\rm i} (\Gamma_{c})^{\alpha}{}_{\beta} \psi^{c}\,_{i}\,^{\beta} \nabla^{a}{W_{a b}} \nabla^{b}{\lambda^{i}_{\alpha}} - \frac{1}{16}{\rm i} \epsilon^{c}\,_{e {e_{1}}}\,^{a b} (\Sigma_{c d})^{\alpha}{}_{\beta} \psi^{e}\,_{i}\,^{\beta} \nabla^{{e_{1}}}{W_{a b}} \nabla^{d}{\lambda^{i}_{\alpha}}+\frac{3}{64}{\rm i} \epsilon^{c d}\,_{e}\,^{a}\,_{{e_{1}}} (\Sigma_{c d})^{\alpha}{}_{\beta} \psi^{e}\,_{i}\,^{\beta} \nabla^{b}{W_{a b}} \nabla^{{e_{1}}}{\lambda^{i}_{\alpha}}+\frac{27}{1024}{\rm i} \epsilon_{e {e_{1}}}\,^{c a b} (\Gamma^{e})^{\alpha}{}_{\beta} \psi^{{e_{1}}}\,_{i}\,^{\beta} W_{a b} \lambda^{i}_{\alpha} \nabla^{d}{W_{c d}}+\frac{5}{128}{\rm i} \epsilon^{c d}\,_{e}\,^{a b} (\Sigma_{c d})^{\alpha}{}_{\beta} \psi^{e}\,_{i}\,^{\beta} \nabla_{{e_{1}}}{W_{a b}} \nabla^{{e_{1}}}{\lambda^{i}_{\alpha}}+\frac{5}{64}{\rm i} \epsilon^{c}\,_{e}\,^{a b}\,_{{e_{1}}} (\Sigma_{c d})^{\alpha}{}_{\beta} \psi^{d}\,_{i}\,^{\beta} \nabla^{e}{W_{a b}} \nabla^{{e_{1}}}{\lambda^{i}_{\alpha}}+\frac{1}{48}\epsilon^{c d}\,_{e}\,^{a b} (\Sigma_{c d})^{\alpha}{}_{\beta} \psi_{{e_{1}} i}\,^{\beta} \nabla^{e}{W} \nabla^{{e_{1}}}{W_{a b \alpha}\,^{i}} - \frac{1}{12}\epsilon_{c d e}\,^{a b} \psi^{c}\,_{i}\,^{\alpha} \nabla^{d}{W} \nabla^{e}{W_{a b \alpha}\,^{i}} - \frac{1}{24}\epsilon^{c}\,_{e {e_{1}}}\,^{a b} (\Sigma_{c d})^{\alpha}{}_{\beta} \psi^{e}\,_{i}\,^{\beta} \nabla^{d}{W} \nabla^{{e_{1}}}{W_{a b \alpha}\,^{i}}+\frac{1}{12}\epsilon^{c d}\,_{e {e_{1}}}\,^{a} (\Sigma_{c d})^{\alpha}{}_{\beta} \psi^{b}\,_{i}\,^{\beta} \nabla^{e}{W} \nabla^{{e_{1}}}{W_{a b \alpha}\,^{i}}+\frac{1}{24}\epsilon^{c d}\,_{e {e_{1}}}\,^{a} (\Sigma_{c d})^{\alpha}{}_{\beta} \psi^{e}\,_{i}\,^{\beta} \nabla^{{e_{1}}}{W} \nabla^{b}{W_{a b \alpha}\,^{i}}+\frac{1}{6}(\Gamma_{c})^{\alpha}{}_{\beta} \psi^{c}\,_{i}\,^{\beta} \nabla^{a}{W} \nabla^{b}{W_{a b \alpha}\,^{i}} - \frac{1}{48}\epsilon^{c d}\,_{e}\,^{a b} (\Sigma_{c d})^{\alpha}{}_{\beta} \psi_{{e_{1}} i}\,^{\beta} \nabla^{{e_{1}}}{W} \nabla^{e}{W_{a b \alpha}\,^{i}}+\frac{1}{24}\epsilon^{c}\,_{e {e_{1}}}\,^{a b} (\Sigma_{c d})^{\alpha}{}_{\beta} \psi^{e}\,_{i}\,^{\beta} \nabla^{{e_{1}}}{W} \nabla^{d}{W_{a b \alpha}\,^{i}} - \frac{1}{24}\epsilon^{c d}\,_{e {e_{1}}}\,^{a} (\Sigma_{c d})^{\alpha}{}_{\beta} \psi^{e}\,_{i}\,^{\beta} \nabla^{b}{W} \nabla^{{e_{1}}}{W_{a b \alpha}\,^{i}}+\frac{1}{384}\epsilon_{{e_{1}}}\,^{a b c d} (\Gamma_{e})^{\alpha}{}_{\beta} \psi^{{e_{1}}}\,_{i}\,^{\beta} W_{a b} W_{c d \alpha}\,^{i} \nabla^{e}{W} - \frac{7}{192}(\Sigma_{c d})^{\alpha}{}_{\beta} \psi^{c}\,_{i}\,^{\beta} W^{a b} W_{a b \alpha}\,^{i} \nabla^{d}{W} - \frac{11}{64}\psi^{c}\,_{i}\,^{\alpha} W^{a}\,_{b} W_{a c \alpha}\,^{i} \nabla^{b}{W}%
 - \frac{11}{96}(\Sigma_{a d})^{\alpha}{}_{\beta} \psi^{c}\,_{i}\,^{\beta} W^{a b} W_{c b \alpha}\,^{i} \nabla^{d}{W} - \frac{11}{96}(\Sigma^{a c})^{\alpha}{}_{\beta} \psi^{d}\,_{i}\,^{\beta} W_{a b} W_{c d \alpha}\,^{i} \nabla^{b}{W} - \frac{1}{192}(\Sigma_{a b})^{\alpha}{}_{\beta} \psi^{c}\,_{i}\,^{\beta} W^{a b} W_{c d \alpha}\,^{i} \nabla^{d}{W} - \frac{5}{32}(\Sigma^{c}{}_{\, a})^{\alpha}{}_{\beta} \psi_{d i}\,^{\beta} W^{a b} W_{c b \alpha}\,^{i} \nabla^{d}{W}+\frac{7}{96}(\Sigma^{c}{}_{\, d})^{\alpha}{}_{\beta} \psi^{d}\,_{i}\,^{\beta} W^{a}\,_{b} W_{c a \alpha}\,^{i} \nabla^{b}{W} - \frac{19}{96}(\Sigma_{d a})^{\alpha}{}_{\beta} \psi^{d}\,_{i}\,^{\beta} W^{a b} W_{b c \alpha}\,^{i} \nabla^{c}{W}+\frac{1}{192}\epsilon_{e {e_{1}}}\,^{a b c} (\Gamma^{d})^{\alpha}{}_{\beta} \psi^{e}\,_{i}\,^{\beta} W_{a b} W_{c d \alpha}\,^{i} \nabla^{{e_{1}}}{W}+\frac{19}{384}\epsilon_{e {e_{1}}}\,^{a b c} (\Gamma^{e})^{\alpha}{}_{\beta} \psi^{{e_{1}}}\,_{i}\,^{\beta} W_{a b} W_{c d \alpha}\,^{i} \nabla^{d}{W} - \frac{7}{192}(\Sigma^{c d})^{\alpha}{}_{\beta} \psi^{a}\,_{i}\,^{\beta} W_{a b} W_{c d \alpha}\,^{i} \nabla^{b}{W} - \frac{19}{96}(\Sigma^{a c})^{\alpha}{}_{\beta} \psi^{b}\,_{i}\,^{\beta} W_{a b} W_{c d \alpha}\,^{i} \nabla^{d}{W} - \frac{1}{192}\epsilon_{e {e_{1}}}\,^{a c d} (\Gamma^{e})^{\alpha}{}_{\beta} \psi^{b}\,_{i}\,^{\beta} W_{a b} W_{c d \alpha}\,^{i} \nabla^{{e_{1}}}{W} - \frac{11}{64}\psi^{b}\,_{i}\,^{\alpha} W^{a}\,_{b} W_{a c \alpha}\,^{i} \nabla^{c}{W} - \frac{7}{96}(\Sigma^{c}{}_{\, d})^{\alpha}{}_{\beta} \psi^{b}\,_{i}\,^{\beta} W^{a}\,_{b} W_{c a \alpha}\,^{i} \nabla^{d}{W}+\frac{3}{128}\epsilon_{e {e_{1}}}\,^{a c d} (\Gamma^{b})^{\alpha}{}_{\beta} \psi^{e}\,_{i}\,^{\beta} W_{a b} W_{c d \alpha}\,^{i} \nabla^{{e_{1}}}{W}+\frac{3}{128}\epsilon_{e {e_{1}}}\,^{a b c} (\Gamma^{e})^{\alpha}{}_{\beta} \psi^{d}\,_{i}\,^{\beta} W_{a b} W_{c d \alpha}\,^{i} \nabla^{{e_{1}}}{W}+\frac{13}{768}\epsilon_{e}\,^{a b c d} (\Gamma^{e})^{\alpha}{}_{\beta} \psi_{{e_{1}} i}\,^{\beta} W_{a b} W_{c d \alpha}\,^{i} \nabla^{{e_{1}}}{W}+(\Gamma_{a})^{\alpha}{}_{\beta} \phi_{b i}\,^{\beta} \nabla^{b}{\nabla^{a}{\lambda^{i}_{\alpha}}}+\frac{21}{16}(\Sigma_{a b})^{\alpha}{}_{\beta} \phi_{c i}\,^{\beta} \lambda^{i}_{\alpha} \nabla^{c}{W^{a b}}+\frac{11}{8}(\Sigma_{a b})^{\alpha}{}_{\beta} \phi_{c i}\,^{\beta} W^{a b} \nabla^{c}{\lambda^{i}_{\alpha}}+\frac{1}{2}(\Gamma_{a})^{\alpha}{}_{\beta} \phi^{a}\,_{i}\,^{\beta} \nabla_{b}{\nabla^{b}{\lambda^{i}_{\alpha}}}%
 - \frac{9}{32}\phi^{a}\,_{i}\,^{\alpha} \lambda^{i}_{\alpha} \nabla^{b}{W_{a b}}+\frac{3}{32}\epsilon_{c d e}\,^{a b} (\Gamma^{c})^{\alpha}{}_{\beta} \phi^{d}\,_{i}\,^{\beta} \lambda^{i}_{\alpha} \nabla^{e}{W_{a b}}+\frac{3}{8}(\Sigma_{c a})^{\alpha}{}_{\beta} \phi_{b i}\,^{\beta} \lambda^{i}_{\alpha} \nabla^{c}{W^{a b}} - \frac{9}{16}(\Sigma_{c a})^{\alpha}{}_{\beta} \phi^{c}\,_{i}\,^{\beta} \lambda^{i}_{\alpha} \nabla_{b}{W^{a b}} - \frac{3}{4}\phi^{a}\,_{i}\,^{\alpha} W_{a b} \nabla^{b}{\lambda^{i}_{\alpha}}+\frac{1}{8}\epsilon_{c d e}\,^{a b} (\Gamma^{c})^{\alpha}{}_{\beta} \phi^{d}\,_{i}\,^{\beta} W_{a b} \nabla^{e}{\lambda^{i}_{\alpha}}+(\Sigma^{a}{}_{\, c})^{\alpha}{}_{\beta} \phi^{c}\,_{i}\,^{\beta} W_{a b} \nabla^{b}{\lambda^{i}_{\alpha}} - \frac{9}{128}(\Gamma_{c})^{\alpha}{}_{\beta} \phi^{c}\,_{i}\,^{\beta} W^{a b} W_{a b} \lambda^{i}_{\alpha}+\frac{9}{128}\epsilon^{e a b c d} (\Sigma_{e {e_{1}}})^{\alpha}{}_{\beta} \phi^{{e_{1}}}\,_{i}\,^{\beta} W_{a b} W_{c d} \lambda^{i}_{\alpha} - \frac{1}{16}\epsilon^{a b c d}\,_{e} \Phi_{a b i}\,^{j} (\Sigma_{c d})^{\alpha}{}_{\beta} \phi^{e}\,_{j}\,^{\beta} \lambda^{i}_{\alpha}+\frac{1}{8}\Phi_{a b i}\,^{j} (\Gamma^{a})^{\alpha}{}_{\beta} \phi^{b}\,_{j}\,^{\beta} \lambda^{i}_{\alpha}+\frac{1}{128}{\rm i} (\Gamma_{c})^{\alpha}{}_{\beta} \phi^{c}\,_{i}\,^{\beta} W W^{a b} W_{a b \alpha}\,^{i}+\frac{1}{256}{\rm i} \epsilon_{e}\,^{a b c d} \phi^{e}\,_{i}\,^{\alpha} W W_{a b} W_{c d \alpha}\,^{i} - \frac{1}{128}{\rm i} \epsilon^{e a b c d} (\Sigma_{e {e_{1}}})^{\alpha}{}_{\beta} \phi^{{e_{1}}}\,_{i}\,^{\beta} W W_{a b} W_{c d \alpha}\,^{i} - \frac{1}{64}{\rm i} \epsilon^{d e}\,_{{e_{1}}}\,^{a c} (\Sigma_{d e})^{\alpha}{}_{\beta} \phi^{{e_{1}}}\,_{i}\,^{\beta} W W_{a}\,^{b} W_{c b \alpha}\,^{i} - \frac{1}{64}{\rm i} (\Gamma^{a})^{\alpha}{}_{\beta} \phi^{c}\,_{i}\,^{\beta} W W_{a}\,^{b} W_{b c \alpha}\,^{i}+\frac{1}{64}{\rm i} (\Gamma^{c})^{\alpha}{}_{\beta} \phi^{b}\,_{i}\,^{\beta} W W^{a}\,_{b} W_{c a \alpha}\,^{i}+\frac{9}{32}(\Gamma^{a})^{\alpha}{}_{\beta} \phi^{c}\,_{i}\,^{\beta} W_{a}\,^{b} W_{b c} \lambda^{i}_{\alpha} - \frac{9}{64}\epsilon^{e}\,_{{e_{1}}}\,^{a b c} (\Sigma_{e}{}^{\, d})^{\alpha}{}_{\beta} \phi^{{e_{1}}}\,_{i}\,^{\beta} W_{a b} W_{c d} \lambda^{i}_{\alpha}+\frac{9}{128}\epsilon^{e {e_{1}} a b c} (\Sigma_{e {e_{1}}})^{\alpha}{}_{\beta} \phi^{d}\,_{i}\,^{\beta} W_{a b} W_{c d} \lambda^{i}_{\alpha}%
+\frac{1}{4}{\rm i} \phi^{a}\,_{i}\,^{\alpha} W_{a b \alpha}\,^{i} \nabla^{b}{W}+2\nabla^{a}{\nabla^{b}{W}} f_{a b}-6\nabla_{a}{\nabla^{a}{W}} f^{b}\,_{b}
\doublespacedmathend
\end{adjustwidth}

\subsection{Linear Multiplet}

\subsubsection{$\nabla^{a} \nabla^{b} G_{i j}$}

\begin{adjustwidth}{1em}{5cm}
\doublespacedmathbegin
\mathcal{D}_{a}{\nabla_{b}{G_{\underline{i} \underline{j}}}}+\psi_{a \underline{j}}\,^{\alpha} \nabla_{b}{\varphi_{\underline{i} \alpha}} - \frac{1}{8}\epsilon^{e {e_{1}}}\,_{b}\,^{c d} (\Sigma_{e {e_{1}}})^{\alpha}{}_{\beta} \psi_{a \underline{j}}\,^{\beta} W_{c d} \varphi_{\underline{i} \alpha} - \frac{1}{2}(\Gamma^{c})^{\alpha}{}_{\beta} \psi_{a \underline{j}}\,^{\beta} W_{b c} \varphi_{\underline{i} \alpha} - \frac{3}{8}G_{\underline{i} \underline{j}} (\Gamma_{b})^{\alpha}{}_{\beta} \psi_{a k}\,^{\beta} X^{k}_{\alpha} - \frac{3}{8}G_{\underline{i} k} (\Gamma_{b})^{\alpha}{}_{\beta} \psi_{a \underline{j}}\,^{\beta} X^{k}_{\alpha}+\frac{3}{8}G_{\underline{i}}\,^{k} (\Gamma_{b})^{\alpha}{}_{\beta} \psi_{a k}\,^{\beta} X_{\underline{j} \alpha}+{\rm i} (\Gamma_{b})^{\alpha}{}_{\beta} \phi_{a \underline{j}}\,^{\beta} \varphi_{\underline{i} \alpha}-6G_{\underline{i} \underline{j}} f_{a b}
\doublespacedmathend
\end{adjustwidth}

\subsubsection{$\nabla^{a} \nabla^{b} \varphi^{i}_{\alpha}$}

\begin{adjustwidth}{1em}{5cm}
\doublespacedmathbegin
\mathcal{D}^{a}{\nabla^{b}{\varphi^{i}_{\alpha}}}+\frac{1}{4}{\rm i} \psi^{a i}\,_{\alpha} \nabla^{b}{F} - \frac{1}{4}{\rm i} (\Gamma^{c})_{\alpha \beta} \psi^{a i \beta} \nabla^{b}{\mathcal{H}_{c}}+\frac{1}{2}{\rm i} (\Gamma_{c})_{\alpha \beta} \psi^{a}\,_{j}\,^{\beta} \nabla^{b}{\nabla^{c}{G^{i j}}} - \frac{1}{32}{\rm i} \epsilon^{b e {e_{1}} c d} F (\Sigma_{e {e_{1}}})_{\alpha \beta} \psi^{a i \beta} W_{c d}+\frac{1}{8}{\rm i} F (\Gamma^{c})_{\alpha \beta} \psi^{a i \beta} W_{c}\,^{b} - \frac{1}{16}{\rm i} \mathcal{H}^{b} (\Sigma_{c d})_{\alpha \beta} \psi^{a i \beta} W^{c d}+\frac{1}{8}{\rm i} \mathcal{H}_{c} (\Sigma^{b}{}_{\,d})_{\alpha \beta} \psi^{a i \beta} W^{c d} - \frac{1}{32}{\rm i} \epsilon^{b {e_{1}}}\,_{e}\,^{c d} \mathcal{H}_{{e_{1}}} (\Gamma^{e})_{\alpha \beta} \psi^{a i \beta} W_{c d} - \frac{1}{8}{\rm i} \mathcal{H}_{c} \psi^{a i}\,_{\alpha} W^{b c}+\frac{1}{4}{\rm i} \mathcal{H}^{d} (\Sigma_{d c})_{\alpha \beta} \psi^{a i \beta} W^{b c}+\frac{1}{8}{\rm i} (\Sigma_{c d})_{\alpha \beta} \psi^{a}\,_{j}\,^{\beta} W^{c d} \nabla^{b}{G^{i j}}+\frac{1}{2}{\rm i} (\Sigma_{c d})_{\alpha \beta} \psi^{a}\,_{j}\,^{\beta} W^{b c} \nabla^{d}{G^{i j}} - \frac{1}{16}{\rm i} \epsilon^{b}\,_{e {e_{1}}}\,^{c d} (\Gamma^{e})_{\alpha \beta} \psi^{a}\,_{j}\,^{\beta} W_{c d} \nabla^{{e_{1}}}{G^{i j}}+\frac{1}{4}{\rm i} \psi^{a}\,_{j \alpha} W^{b}\,_{c} \nabla^{c}{G^{i j}}+\frac{1}{4}{\rm i} (\Sigma^{b c})_{\alpha \beta} \psi^{a}\,_{j}\,^{\beta} W_{c d} \nabla^{d}{G^{i j}} - \frac{1}{8}\epsilon^{b e {e_{1}} c d} (\Sigma_{e {e_{1}}})_{\alpha \rho} \psi^{a}\,_{j}\,^{\rho} W_{c d}\,^{\beta j} \varphi^{i}_{\beta}+\frac{1}{4}(\Gamma^{c})_{\alpha \rho} \psi^{a}\,_{j}\,^{\rho} W_{c}\,^{b \beta j} \varphi^{i}_{\beta}+\frac{1}{16}\epsilon^{b e {e_{1}} c d} (\Sigma_{e {e_{1}}})_{\alpha}{}^{\rho} \psi^{a}\,_{j}\,^{\beta} W_{c d \beta}\,^{j} \varphi^{i}_{\rho}%
 - \frac{1}{16}(\Gamma^{b})_{\alpha}{}^{\beta} \psi^{a}\,_{j}\,^{\rho} X^{j}_{\beta} \varphi^{i}_{\rho}+\frac{1}{16}(\Gamma^{b})^{\beta \rho} \psi^{a}\,_{j \alpha} X^{j}_{\beta} \varphi^{i}_{\rho} - \frac{1}{16}(\Gamma^{b})_{\alpha \rho} \psi^{a}\,_{j}\,^{\rho} X^{j \beta} \varphi^{i}_{\beta} - \frac{1}{16}(\Gamma^{b})^{\beta}{}_{\rho} \psi^{a}\,_{j}\,^{\rho} X^{j}_{\alpha} \varphi^{i}_{\beta} - \frac{7}{16}(\Gamma^{b})^{\beta}{}_{\rho} \psi^{a}\,_{j}\,^{\rho} X^{j}_{\beta} \varphi^{i}_{\alpha}+\frac{3}{16}(\Gamma^{b})^{\beta}{}_{\rho} \psi^{a i \rho} X_{j \beta} \varphi^{j}_{\alpha}+\frac{3}{16}(\Gamma^{b})^{\beta}{}_{\rho} \psi^{a}\,_{j}\,^{\rho} X^{i}_{\beta} \varphi^{j}_{\alpha}+\frac{1}{8}{\rm i} \epsilon^{b c d e {e_{1}}} \Phi_{c d j}\,^{k} G^{i j} (\Sigma_{e {e_{1}}})_{\alpha \beta} \psi^{a}\,_{k}\,^{\beta} - \frac{3}{4}{\rm i} \Phi_{c}\,^{b}\,_{j}\,^{k} G^{i j} (\Gamma^{c})_{\alpha \beta} \psi^{a}\,_{k}\,^{\beta}+\frac{3}{8}{\rm i} G^{i j} (\Sigma_{c d})_{\alpha \beta} \psi^{a}\,_{j}\,^{\beta} \nabla^{b}{W^{c d}} - \frac{3}{4}{\rm i} G^{i j} (\Sigma_{d c})_{\alpha \beta} \psi^{a}\,_{j}\,^{\beta} \nabla^{d}{W^{b c}} - \frac{3}{16}{\rm i} \epsilon^{b}\,_{e {e_{1}}}\,^{c d} G^{i j} (\Gamma^{e})_{\alpha \beta} \psi^{a}\,_{j}\,^{\beta} \nabla^{{e_{1}}}{W_{c d}}+\frac{9}{16}{\rm i} G^{i j} \psi^{a}\,_{j \alpha} \nabla_{c}{W^{b c}}+\frac{3}{8}{\rm i} G^{i j} (\Sigma^{b}{}_{\,c})_{\alpha \beta} \psi^{a}\,_{j}\,^{\beta} \nabla_{d}{W^{c d}} - \frac{9}{64}{\rm i} G^{i j} (\Gamma^{b})_{\alpha \beta} \psi^{a}\,_{j}\,^{\beta} W^{c d} W_{c d}+\frac{9}{64}{\rm i} \epsilon^{b c d e {e_{1}}} G^{i j} \psi^{a}\,_{j \alpha} W_{c d} W_{e {e_{1}}} - \frac{3}{32}{\rm i} \epsilon^{{e_{2}} c d e {e_{1}}} G^{i j} (\Sigma_{{e_{2}}}{}^{\, b})_{\alpha \beta} \psi^{a}\,_{j}\,^{\beta} W_{c d} W_{e {e_{1}}}+\frac{3}{4}{\rm i} G^{i j} (\Gamma^{d})_{\alpha \beta} \psi^{a}\,_{j}\,^{\beta} W^{b c} W_{d c} - \frac{3}{8}{\rm i} \epsilon^{b {e_{2}} c d e} G^{i j} (\Sigma_{{e_{2}}}{}^{\, {e_{1}}})_{\alpha \beta} \psi^{a}\,_{j}\,^{\beta} W_{c d} W_{e {e_{1}}} - \frac{3}{16}{\rm i} \epsilon^{{e_{1}} {e_{2}} c d e} G^{i j} (\Sigma_{{e_{1}} {e_{2}}})_{\alpha \beta} \psi^{a}\,_{j}\,^{\beta} W_{c d} W_{e}\,^{b}%
+\frac{7}{2}\phi^{a}\,_{j \alpha} \nabla^{b}{G^{i j}} - \frac{1}{4}F (\Gamma^{b})_{\alpha \beta} \phi^{a i \beta} - \frac{1}{4}\mathcal{H}^{b} \phi^{a i}\,_{\alpha} - \frac{1}{2}\mathcal{H}^{c} (\Sigma^{b}{}_{\,c})_{\alpha \beta} \phi^{a i \beta}+(\Sigma^{b}{}_{\,c})_{\alpha \beta} \phi^{a}\,_{j}\,^{\beta} \nabla^{c}{G^{i j}} - \frac{3}{4}G^{i j} (\Gamma^{c})_{\alpha \beta} \phi^{a}\,_{j}\,^{\beta} W_{c}\,^{b}-7\varphi^{i}_{\alpha} f^{a b}+2(\Sigma^{b}{}_{\, c})_{\alpha}{}^{\beta} \varphi^{i}_{\beta} f^{a c}
\doublespacedmathend
\end{adjustwidth}

\subsubsection{$\nabla^{a} \nabla^{b} F$}

\begin{adjustwidth}{1em}{5cm}
\doublespacedmathbegin
\mathcal{D}^{a}{\nabla^{b}{F}}-(\Gamma_{c})^{\alpha}{}_{\beta} \psi^{a}\,_{i}\,^{\beta} \nabla^{b}{\nabla^{c}{\varphi^{i}_{\alpha}}} - \frac{9}{8}(\Sigma_{c d})^{\alpha}{}_{\beta} \psi^{a}\,_{i}\,^{\beta} \nabla^{b}{W^{c d}} \varphi^{i}_{\alpha}-(\Sigma_{c d})^{\alpha}{}_{\beta} \psi^{a}\,_{i}\,^{\beta} W^{c d} \nabla^{b}{\varphi^{i}_{\alpha}}+\frac{3}{4}\psi^{a}\,_{j}\,^{\alpha} X_{i \alpha} \nabla^{b}{G^{i j}} - \frac{3}{4}G_{i}\,^{j} \psi^{a}\,_{j}\,^{\alpha} \nabla^{b}{X^{i}_{\alpha}}-(\Sigma_{c d})^{\alpha}{}_{\beta} \psi^{a}\,_{i}\,^{\beta} W^{b c} \nabla^{d}{\varphi^{i}_{\alpha}}+\frac{1}{8}\epsilon^{b}\,_{e {e_{1}}}\,^{c d} (\Gamma^{e})^{\alpha}{}_{\beta} \psi^{a}\,_{i}\,^{\beta} W_{c d} \nabla^{{e_{1}}}{\varphi^{i}_{\alpha}} - \frac{1}{2}\psi^{a}\,_{i}\,^{\alpha} W^{b}\,_{c} \nabla^{c}{\varphi^{i}_{\alpha}} - \frac{1}{2}(\Sigma^{b c})^{\alpha}{}_{\beta} \psi^{a}\,_{i}\,^{\beta} W_{c d} \nabla^{d}{\varphi^{i}_{\alpha}}+\frac{3}{64}(\Gamma^{b})^{\alpha}{}_{\beta} \psi^{a}\,_{i}\,^{\beta} W^{c d} W_{c d} \varphi^{i}_{\alpha} - \frac{3}{16}\epsilon^{b c d e {e_{1}}} \psi^{a}\,_{i}\,^{\alpha} W_{c d} W_{e {e_{1}}} \varphi^{i}_{\alpha}+\frac{15}{64}\epsilon^{{e_{2}} c d e {e_{1}}} (\Sigma_{{e_{2}}}{}^{\, b})^{\alpha}{}_{\beta} \psi^{a}\,_{i}\,^{\beta} W_{c d} W_{e {e_{1}}} \varphi^{i}_{\alpha}+\frac{3}{32}\epsilon^{b e {e_{1}} c d} G_{i}\,^{j} (\Sigma_{e {e_{1}}})^{\alpha}{}_{\beta} \psi^{a}\,_{j}\,^{\beta} W_{c d} X^{i}_{\alpha} - \frac{3}{8}G_{i}\,^{j} (\Gamma^{c})^{\alpha}{}_{\beta} \psi^{a}\,_{j}\,^{\beta} W_{c}\,^{b} X^{i}_{\alpha} - \frac{15}{16}(\Gamma^{d})^{\alpha}{}_{\beta} \psi^{a}\,_{i}\,^{\beta} W^{b c} W_{d c} \varphi^{i}_{\alpha}+\frac{15}{32}\epsilon^{b {e_{2}} c d e} (\Sigma_{{e_{2}}}{}^{\, {e_{1}}})^{\alpha}{}_{\beta} \psi^{a}\,_{i}\,^{\beta} W_{c d} W_{e {e_{1}}} \varphi^{i}_{\alpha}+\frac{15}{64}\epsilon^{{e_{1}} {e_{2}} c d e} (\Sigma_{{e_{1}} {e_{2}}})^{\alpha}{}_{\beta} \psi^{a}\,_{i}\,^{\beta} W_{c d} W_{e}\,^{b} \varphi^{i}_{\alpha} - \frac{1}{2}F (\Gamma^{b})^{\alpha}{}_{\beta} \psi^{a}\,_{i}\,^{\beta} X^{i}_{\alpha}%
 - \frac{1}{8}\epsilon^{b c d e {e_{1}}} \Phi_{c d i}\,^{j} (\Sigma_{e {e_{1}}})^{\alpha}{}_{\beta} \psi^{a}\,_{j}\,^{\beta} \varphi^{i}_{\alpha}+\frac{3}{4}\Phi_{c}\,^{b}\,_{i}\,^{j} (\Gamma^{c})^{\alpha}{}_{\beta} \psi^{a}\,_{j}\,^{\beta} \varphi^{i}_{\alpha} - \frac{3}{8}(\Sigma^{b}{}_{\, c})^{\alpha}{}_{\beta} \psi^{a}\,_{i}\,^{\beta} \nabla_{d}{W^{c d}} \varphi^{i}_{\alpha}+\frac{3}{16}\epsilon^{b}\,_{e {e_{1}}}\,^{c d} (\Gamma^{e})^{\alpha}{}_{\beta} \psi^{a}\,_{i}\,^{\beta} \nabla^{{e_{1}}}{W_{c d}} \varphi^{i}_{\alpha} - \frac{9}{16}\psi^{a}\,_{i}\,^{\alpha} \nabla_{c}{W^{b c}} \varphi^{i}_{\alpha}+\frac{3}{4}(\Sigma_{d c})^{\alpha}{}_{\beta} \psi^{a}\,_{i}\,^{\beta} \nabla^{d}{W^{b c}} \varphi^{i}_{\alpha}+4{\rm i} \phi^{a}\,_{i}\,^{\alpha} \nabla^{b}{\varphi^{i}_{\alpha}}+2{\rm i} (\Sigma^{b}{}_{\, c})^{\alpha}{}_{\beta} \phi^{a}\,_{i}\,^{\beta} \nabla^{c}{\varphi^{i}_{\alpha}} - \frac{3}{8}{\rm i} \epsilon^{b e {e_{1}} c d} (\Sigma_{e {e_{1}}})^{\alpha}{}_{\beta} \phi^{a}\,_{i}\,^{\beta} W_{c d} \varphi^{i}_{\alpha} - \frac{3}{4}{\rm i} G_{i}\,^{j} (\Gamma^{b})^{\alpha}{}_{\beta} \phi^{a}\,_{j}\,^{\beta} X^{i}_{\alpha}-8F f^{a b}
 \doublespacedmathend
 \end{adjustwidth}
 
\subsubsection{$\nabla^{a} \nabla^{b} \cH_{c}$}

\begin{adjustwidth}{1em}{5cm}
\doublespacedmathbegin
\mathcal{D}^{a}{\nabla^{b}{\mathcal{H}_{c}}}-2(\Sigma_{c d})^{\alpha}{}_{\beta} \psi^{a}\,_{i}\,^{\beta} \nabla^{b}{\nabla^{d}{\varphi^{i}_{\alpha}}}+\frac{3}{16}\epsilon^{{e_{1}} {e_{2}}}\,_{c}\,^{d e} (\Sigma_{{e_{1}} {e_{2}}})^{\alpha}{}_{\beta} \psi^{a}\,_{i}\,^{\beta} \nabla^{b}{W_{d e}} \varphi^{i}_{\alpha}+\frac{3}{4}(\Gamma^{d})^{\alpha}{}_{\beta} \psi^{a}\,_{i}\,^{\beta} \nabla^{b}{W_{c d}} \varphi^{i}_{\alpha}+\frac{9}{32}\epsilon^{{e_{1}} {e_{2}}}\,_{c}\,^{d e} (\Sigma_{{e_{1}} {e_{2}}})^{\alpha}{}_{\beta} \psi^{a}\,_{i}\,^{\beta} W_{d e} \nabla^{b}{\varphi^{i}_{\alpha}}+\frac{1}{2}(\Gamma^{d})^{\alpha}{}_{\beta} \psi^{a}\,_{i}\,^{\beta} W_{c d} \nabla^{b}{\varphi^{i}_{\alpha}} - \frac{1}{4}(\Gamma^{b})^{\alpha}{}_{\beta} \psi^{a}\,_{i}\,^{\beta} W_{c d} \nabla^{d}{\varphi^{i}_{\alpha}} - \frac{1}{8}\epsilon^{b}\,_{c {e_{1}}}\,^{d e} \psi^{a}\,_{i}\,^{\alpha} W_{d e} \nabla^{{e_{1}}}{\varphi^{i}_{\alpha}}+\frac{3}{8}\epsilon^{{e_{1}}}\,_{c {e_{2}}}\,^{d e} (\Sigma_{{e_{1}}}{}^{\, b})^{\alpha}{}_{\beta} \psi^{a}\,_{i}\,^{\beta} W_{d e} \nabla^{{e_{2}}}{\varphi^{i}_{\alpha}}+\frac{5}{16}\epsilon^{b e {e_{1}}}\,_{{e_{2}}}\,^{d} (\Sigma_{e {e_{1}}})^{\alpha}{}_{\beta} \psi^{a}\,_{i}\,^{\beta} W_{c d} \nabla^{{e_{2}}}{\varphi^{i}_{\alpha}} - \frac{1}{2}(\Gamma_{d})^{\alpha}{}_{\beta} \psi^{a}\,_{i}\,^{\beta} W_{c}\,^{b} \nabla^{d}{\varphi^{i}_{\alpha}}+\frac{3}{8}\epsilon^{b {e_{1}} {e_{2}}}\,_{c}\,^{d} (\Sigma_{{e_{1}} {e_{2}}})^{\alpha}{}_{\beta} \psi^{a}\,_{i}\,^{\beta} W_{d e} \nabla^{e}{\varphi^{i}_{\alpha}}+\frac{1}{4}(\Gamma^{d})^{\alpha}{}_{\beta} \delta^{b}\,_{c} \psi^{a}\,_{i}\,^{\beta} W_{d e} \nabla^{e}{\varphi^{i}_{\alpha}} - \frac{1}{2}(\Gamma_{c})^{\alpha}{}_{\beta} \psi^{a}\,_{i}\,^{\beta} W^{b}\,_{d} \nabla^{d}{\varphi^{i}_{\alpha}}+\frac{5}{4}\psi^{a}\,_{i}\,^{\alpha} W^{b d} W_{c d} \varphi^{i}_{\alpha}+\frac{3}{4}(\Sigma_{c}{}^{\, e})^{\alpha}{}_{\beta} \psi^{a}\,_{i}\,^{\beta} W^{b d} W_{e d} \varphi^{i}_{\alpha} - \frac{5}{2}(\Sigma_{c d})^{\alpha}{}_{\beta} \psi^{a}\,_{i}\,^{\beta} W^{d e} W^{b}\,_{e} \varphi^{i}_{\alpha}-(\Sigma^{e}{}_{\, d})^{\alpha}{}_{\beta} \psi^{a}\,_{i}\,^{\beta} W^{b d} W_{c e} \varphi^{i}_{\alpha} - \frac{27}{64}\epsilon_{c {e_{2}}}\,^{d e {e_{1}}} (\Gamma^{{e_{2}}})^{\alpha}{}_{\beta} \psi^{a}\,_{i}\,^{\beta} W_{d e} W_{{e_{1}}}\,^{b} \varphi^{i}_{\alpha}%
 - \frac{1}{4}\delta^{b}\,_{c} \psi^{a}\,_{i}\,^{\alpha} W^{d e} W_{d e} \varphi^{i}_{\alpha}+\frac{1}{2}(\Sigma_{c}{}^{\, b})^{\alpha}{}_{\beta} \psi^{a}\,_{i}\,^{\beta} W^{d e} W_{d e} \varphi^{i}_{\alpha}+\frac{143}{64}(\Sigma^{{e_{1}}}{}_{\, d})^{\alpha}{}_{\beta} \delta^{b}\,_{c} \psi^{a}\,_{i}\,^{\beta} W^{d e} W_{{e_{1}} e} \varphi^{i}_{\alpha}+\frac{3}{8}(\Sigma^{b}{}_{\, d})^{\alpha}{}_{\beta} \psi^{a}\,_{i}\,^{\beta} W^{d e} W_{c e} \varphi^{i}_{\alpha}+\frac{19}{256}\epsilon_{{e_{3}}}\,^{d e {e_{1}} {e_{2}}} (\Gamma^{{e_{3}}})^{\alpha}{}_{\beta} \delta^{b}\,_{c} \psi^{a}\,_{i}\,^{\beta} W_{d e} W_{{e_{1}} {e_{2}}} \varphi^{i}_{\alpha} - \frac{9}{32}\epsilon^{b}\,_{{e_{2}}}\,^{d e {e_{1}}} (\Gamma^{{e_{2}}})^{\alpha}{}_{\beta} \psi^{a}\,_{i}\,^{\beta} W_{c d} W_{e {e_{1}}} \varphi^{i}_{\alpha}+\frac{5}{16}(\Sigma_{d e})^{\alpha}{}_{\beta} \psi^{a}\,_{i}\,^{\beta} W^{d e} W_{c}\,^{b} \varphi^{i}_{\alpha} - \frac{21}{256}\epsilon_{c}\,^{d e {e_{1}} {e_{2}}} (\Gamma^{b})^{\alpha}{}_{\beta} \psi^{a}\,_{i}\,^{\beta} W_{d e} W_{{e_{1}} {e_{2}}} \varphi^{i}_{\alpha} - \frac{19}{128}\epsilon^{b d e {e_{1}} {e_{2}}} (\Gamma_{c})^{\alpha}{}_{\beta} \psi^{a}\,_{i}\,^{\beta} W_{d e} W_{{e_{1}} {e_{2}}} \varphi^{i}_{\alpha}+\frac{1}{32}\epsilon^{b {e_{1}} {e_{2}} d e} (\Sigma_{{e_{1}} {e_{2}}})^{\alpha}{}_{\beta} \psi^{a}\,_{i}\,^{\beta} W_{d e} \nabla_{c}{\varphi^{i}_{\alpha}}+\frac{1}{8}\epsilon^{b {e_{1}}}\,_{c {e_{2}}}\,^{d} (\Sigma_{{e_{1}}}{}^{\, e})^{\alpha}{}_{\beta} \psi^{a}\,_{i}\,^{\beta} W_{d e} \nabla^{{e_{2}}}{\varphi^{i}_{\alpha}} - \frac{1}{16}\epsilon^{{e_{1}} {e_{2}}}\,_{{e_{3}}}\,^{d e} (\Sigma_{{e_{1}} {e_{2}}})^{\alpha}{}_{\beta} \delta^{b}\,_{c} \psi^{a}\,_{i}\,^{\beta} W_{d e} \nabla^{{e_{3}}}{\varphi^{i}_{\alpha}} - \frac{5}{32}\epsilon^{b}\,_{c}\,^{d e {e_{1}}} (\Gamma^{{e_{2}}})^{\alpha}{}_{\beta} \psi^{a}\,_{i}\,^{\beta} W_{d e} W_{{e_{1}} {e_{2}}} \varphi^{i}_{\alpha} - \frac{1}{16}\epsilon^{b {e_{1}}}\,_{c}\,^{d e} (\Sigma_{{e_{1}} {e_{2}}})^{\alpha}{}_{\beta} \psi^{a}\,_{i}\,^{\beta} W_{d e} \nabla^{{e_{2}}}{\varphi^{i}_{\alpha}} - \frac{1}{16}\epsilon^{e {e_{1}}}\,_{c {e_{2}}}\,^{d} (\Sigma_{e {e_{1}}})^{\alpha}{}_{\beta} \psi^{a}\,_{i}\,^{\beta} W_{d}\,^{b} \nabla^{{e_{2}}}{\varphi^{i}_{\alpha}} - \frac{1}{16}\epsilon^{b {e_{1}} {e_{2}} d e} \mathcal{H}_{c} (\Sigma_{{e_{1}} {e_{2}}})^{\alpha}{}_{\beta} \psi^{a}\,_{i}\,^{\beta} W_{d e \alpha}\,^{i}+\frac{1}{8}\mathcal{H}_{c} (\Gamma^{d})^{\alpha}{}_{\beta} \psi^{a}\,_{i}\,^{\beta} W_{d}\,^{b}\,_{\alpha}\,^{i} - \frac{1}{8}\mathcal{H}^{d} (\Gamma^{b})^{\alpha}{}_{\beta} \psi^{a}\,_{i}\,^{\beta} W_{c d \alpha}\,^{i}+\frac{1}{8}\epsilon^{{e_{1}} {e_{2}}}\,_{c}\,^{d e} \mathcal{H}_{{e_{1}}} (\Sigma_{{e_{2}}}{}^{\, b})^{\alpha}{}_{\beta} \psi^{a}\,_{i}\,^{\beta} W_{d e \alpha}\,^{i}+\frac{1}{8}\epsilon^{b e {e_{1}} {e_{2}} d} \mathcal{H}_{e} (\Sigma_{{e_{1}} {e_{2}}})^{\alpha}{}_{\beta} \psi^{a}\,_{i}\,^{\beta} W_{c d \alpha}\,^{i}%
+\frac{1}{8}\mathcal{H}^{b} (\Gamma^{d})^{\alpha}{}_{\beta} \psi^{a}\,_{i}\,^{\beta} W_{c d \alpha}\,^{i} - \frac{1}{8}\mathcal{H}_{d} (\Gamma^{d})^{\alpha}{}_{\beta} \psi^{a}\,_{i}\,^{\beta} W_{c}\,^{b}\,_{\alpha}\,^{i}+\frac{1}{8}\epsilon^{b {e_{1}} {e_{2}}}\,_{c}\,^{d} \mathcal{H}^{e} (\Sigma_{{e_{1}} {e_{2}}})^{\alpha}{}_{\beta} \psi^{a}\,_{i}\,^{\beta} W_{d e \alpha}\,^{i}+\frac{1}{8}\mathcal{H}^{e} (\Gamma^{d})^{\alpha}{}_{\beta} \delta^{b}\,_{c} \psi^{a}\,_{i}\,^{\beta} W_{d e \alpha}\,^{i} - \frac{1}{8}\mathcal{H}^{d} (\Gamma_{c})^{\alpha}{}_{\beta} \psi^{a}\,_{i}\,^{\beta} W^{b}\,_{d \alpha}\,^{i} - \frac{1}{16}\epsilon^{b {e_{1}}}\,_{c}\,^{d e} \mathcal{H}_{{e_{1}}} \psi^{a}\,_{i}\,^{\alpha} W_{d e \alpha}\,^{i} - \frac{1}{8}\mathcal{H}^{b} (\Gamma_{c})^{\alpha}{}_{\beta} \psi^{a}\,_{i}\,^{\beta} X^{i}_{\alpha}+\frac{1}{8}\mathcal{H}_{d} (\Gamma^{d})^{\alpha}{}_{\beta} \delta^{b}\,_{c} \psi^{a}\,_{i}\,^{\beta} X^{i}_{\alpha} - \frac{1}{2}\mathcal{H}_{c} (\Gamma^{b})^{\alpha}{}_{\beta} \psi^{a}\,_{i}\,^{\beta} X^{i}_{\alpha}+\frac{1}{3}\Phi^{d e}\,_{i}\,^{j} (\Sigma_{d e})^{\alpha}{}_{\beta} \delta^{b}\,_{c} \psi^{a}\,_{j}\,^{\beta} \varphi^{i}_{\alpha}-\Phi_{c}\,^{b}\,_{i}\,^{j} \psi^{a}\,_{j}\,^{\alpha} \varphi^{i}_{\alpha}+\frac{1}{6}\epsilon^{b e {e_{1}}}\,_{c d} \Phi_{e {e_{1}} i}\,^{j} (\Gamma^{d})^{\alpha}{}_{\beta} \psi^{a}\,_{j}\,^{\beta} \varphi^{i}_{\alpha} - \frac{2}{3}\Phi_{c}\,^{d}\,_{i}\,^{j} (\Sigma^{b}{}_{\, d})^{\alpha}{}_{\beta} \psi^{a}\,_{j}\,^{\beta} \varphi^{i}_{\alpha}-2\Phi^{b d}\,_{i}\,^{j} (\Sigma_{c d})^{\alpha}{}_{\beta} \psi^{a}\,_{j}\,^{\beta} \varphi^{i}_{\alpha}+\frac{1}{4}(\Gamma^{d})^{\alpha}{}_{\beta} \delta^{b}\,_{c} \psi^{a}\,_{i}\,^{\beta} \nabla^{e}{W_{d e}} \varphi^{i}_{\alpha}+\frac{1}{2}(\Gamma_{d})^{\alpha}{}_{\beta} \psi^{a}\,_{i}\,^{\beta} \nabla^{d}{W^{b}\,_{c}} \varphi^{i}_{\alpha} - \frac{1}{4}\epsilon^{b}\,_{c {e_{1}}}\,^{d e} \psi^{a}\,_{i}\,^{\alpha} \nabla^{{e_{1}}}{W_{d e}} \varphi^{i}_{\alpha}+\frac{1}{16}\epsilon^{b {e_{1}}}\,_{c}\,^{d e} (\Sigma_{{e_{1}} {e_{2}}})^{\alpha}{}_{\beta} \psi^{a}\,_{i}\,^{\beta} \nabla^{{e_{2}}}{W_{d e}} \varphi^{i}_{\alpha}+\frac{1}{4}\epsilon^{b e {e_{1}}}\,_{{e_{2}}}\,^{d} (\Sigma_{e {e_{1}}})^{\alpha}{}_{\beta} \psi^{a}\,_{i}\,^{\beta} \nabla^{{e_{2}}}{W_{c d}} \varphi^{i}_{\alpha} - \frac{1}{4}(\Gamma^{b})^{\alpha}{}_{\beta} \psi^{a}\,_{i}\,^{\beta} \nabla^{d}{W_{c d}} \varphi^{i}_{\alpha}%
+\frac{1}{4}\epsilon^{e {e_{1}}}\,_{c {e_{2}} d} (\Sigma_{e {e_{1}}})^{\alpha}{}_{\beta} \psi^{a}\,_{i}\,^{\beta} \nabla^{{e_{2}}}{W^{b d}} \varphi^{i}_{\alpha} - \frac{1}{2}(\Gamma_{d})^{\alpha}{}_{\beta} \psi^{a}\,_{i}\,^{\beta} \nabla_{c}{W^{b d}} \varphi^{i}_{\alpha} - \frac{3}{4}(\Gamma_{c})^{\alpha}{}_{\beta} \psi^{a}\,_{i}\,^{\beta} \nabla_{d}{W^{b d}} \varphi^{i}_{\alpha}+\frac{3}{16}\epsilon^{b {e_{1}}}\,_{{e_{2}}}\,^{d e} (\Sigma_{c {e_{1}}})^{\alpha}{}_{\beta} \psi^{a}\,_{i}\,^{\beta} \nabla^{{e_{2}}}{W_{d e}} \varphi^{i}_{\alpha}+\frac{1}{8}\epsilon^{b {e_{1}} {e_{2}}}\,_{c}\,^{d} (\Sigma_{{e_{1}} {e_{2}}})^{\alpha}{}_{\beta} \psi^{a}\,_{i}\,^{\beta} \nabla^{e}{W_{d e}} \varphi^{i}_{\alpha} - \frac{3}{16}\epsilon^{b {e_{1}} {e_{2}} d e} (\Sigma_{{e_{1}} {e_{2}}})^{\alpha}{}_{\beta} \psi^{a}\,_{i}\,^{\beta} \nabla_{c}{W_{d e}} \varphi^{i}_{\alpha}+\frac{5}{16}\epsilon^{{e_{1}}}\,_{c {e_{2}}}\,^{d e} (\Sigma_{{e_{1}}}{}^{\, b})^{\alpha}{}_{\beta} \psi^{a}\,_{i}\,^{\beta} \nabla^{{e_{2}}}{W_{d e}} \varphi^{i}_{\alpha}+\frac{1}{8}\epsilon^{b {e_{1}}}\,_{c {e_{2}} e} (\Sigma_{{e_{1}} d})^{\alpha}{}_{\beta} \psi^{a}\,_{i}\,^{\beta} \nabla^{{e_{2}}}{W^{d e}} \varphi^{i}_{\alpha}+5{\rm i} (\Gamma_{c})^{\alpha}{}_{\beta} \phi^{a}\,_{i}\,^{\beta} \nabla^{b}{\varphi^{i}_{\alpha}}-{\rm i} \epsilon^{b d e}\,_{c {e_{1}}} (\Sigma_{d e})^{\alpha}{}_{\beta} \phi^{a}\,_{i}\,^{\beta} \nabla^{{e_{1}}}{\varphi^{i}_{\alpha}}-{\rm i} (\Gamma_{d})^{\alpha}{}_{\beta} \delta^{b}\,_{c} \phi^{a}\,_{i}\,^{\beta} \nabla^{d}{\varphi^{i}_{\alpha}} - \frac{1}{2}{\rm i} (\Sigma_{d e})^{\alpha}{}_{\beta} \delta^{b}\,_{c} \phi^{a}\,_{i}\,^{\beta} W^{d e} \varphi^{i}_{\alpha}+\frac{3}{4}{\rm i} \phi^{a}\,_{i}\,^{\alpha} W_{c}\,^{b} \varphi^{i}_{\alpha} - \frac{1}{4}{\rm i} \epsilon^{b}\,_{c {e_{1}}}\,^{d e} (\Gamma^{{e_{1}}})^{\alpha}{}_{\beta} \phi^{a}\,_{i}\,^{\beta} W_{d e} \varphi^{i}_{\alpha} - \frac{1}{2}{\rm i} (\Sigma^{b d})^{\alpha}{}_{\beta} \phi^{a}\,_{i}\,^{\beta} W_{c d} \varphi^{i}_{\alpha}+3{\rm i} (\Sigma_{c d})^{\alpha}{}_{\beta} \phi^{a}\,_{i}\,^{\beta} W^{b d} \varphi^{i}_{\alpha}-8\mathcal{H}_{c} f^{a b}-2\mathcal{H}^{b} f^{a}\,_{c}+2\mathcal{H}_{d} \delta^{b}\,_{c} f^{a d}
\doublespacedmathend
\end{adjustwidth}

\subsubsection{$\nabla^{a} \Box G_{i j}$}

\begin{adjustwidth}{1em}{5cm}
\doublespacedmathbegin
\mathcal{D}^{a}{\nabla_{b}{\nabla^{b}{G_{\underline{i} \underline{j}}}}}+\frac{3}{32}G_{\underline{i} k} (\Sigma_{b c})^{\alpha}{}_{\beta} \psi^{a}\,_{\underline{j}}\,^{\beta} W^{b c} X^{k}_{\alpha} - \frac{3}{32}G_{\underline{i}}\,^{k} (\Sigma_{b c})^{\alpha}{}_{\beta} \psi^{a}\,_{k}\,^{\beta} W^{b c} X_{\underline{j} \alpha} - \frac{3}{8}(\Gamma_{b})^{\alpha}{}_{\beta} \psi^{a}\,_{k}\,^{\beta} X^{k}_{\alpha} \nabla^{b}{G_{\underline{i} \underline{j}}} - \frac{1}{4}\Phi^{b c}\,_{\underline{i}}\,^{k} (\Sigma_{b c})^{\alpha}{}_{\beta} \psi^{a}\,_{k}\,^{\beta} \varphi_{\underline{j} \alpha} - \frac{3}{16}\epsilon^{d e}\,_{{e_{1}}}\,^{b c} (\Sigma_{d e})^{\alpha}{}_{\beta} \psi^{a}\,_{\underline{j}}\,^{\beta} \nabla^{{e_{1}}}{W_{b c}} \varphi_{\underline{i} \alpha}+\frac{9}{16}(\Gamma^{b})^{\alpha}{}_{\beta} \psi^{a}\,_{\underline{j}}\,^{\beta} \nabla^{c}{W_{b c}} \varphi_{\underline{i} \alpha} - \frac{9}{64}\psi^{a}\,_{\underline{j}}\,^{\alpha} W^{b c} W_{b c} \varphi_{\underline{i} \alpha}+\frac{3}{32}G_{\underline{i} \underline{j}} \psi^{a}\,_{k}\,^{\alpha} W^{b c} W_{b c \alpha}\,^{k}+\psi^{a}\,_{\underline{j}}\,^{\alpha} \nabla_{b}{\nabla^{b}{\varphi_{\underline{i} \alpha}}} - \frac{1}{4}\epsilon^{d e}\,_{{e_{1}}}\,^{b c} (\Sigma_{d e})^{\alpha}{}_{\beta} \psi^{a}\,_{\underline{j}}\,^{\beta} W_{b c} \nabla^{{e_{1}}}{\varphi_{\underline{i} \alpha}}+(\Gamma^{b})^{\alpha}{}_{\beta} \psi^{a}\,_{\underline{j}}\,^{\beta} W_{b c} \nabla^{c}{\varphi_{\underline{i} \alpha}} - \frac{3}{8}G_{\underline{i} k} (\Gamma_{b})^{\alpha}{}_{\beta} \psi^{a}\,_{\underline{j}}\,^{\beta} \nabla^{b}{X^{k}_{\alpha}}+\frac{3}{4}(\Gamma_{b})^{\alpha}{}_{\beta} \psi^{a}\,_{\underline{j}}\,^{\beta} X_{k \alpha} \nabla^{b}{G_{\underline{i}}\,^{k}}+\frac{3}{8}G_{\underline{i}}\,^{k} (\Gamma_{b})^{\alpha}{}_{\beta} \psi^{a}\,_{k}\,^{\beta} \nabla^{b}{X_{\underline{j} \alpha}}+\frac{3}{4}(\Gamma_{b})^{\alpha}{}_{\beta} \psi^{a}\,_{k}\,^{\beta} X_{\underline{i} \alpha} \nabla^{b}{G_{\underline{j}}\,^{k}}+\frac{9}{4}{\rm i} (\Sigma_{b c})^{\alpha}{}_{\beta} \phi^{a}\,_{\underline{j}}\,^{\beta} W^{b c} \varphi_{\underline{i} \alpha}+\frac{15}{8}{\rm i} G_{\underline{i} k} \phi^{a}\,_{\underline{j}}\,^{\alpha} X^{k}_{\alpha} - \frac{15}{8}{\rm i} G_{\underline{i}}\,^{k} \phi^{a}\,_{k}\,^{\alpha} X_{\underline{j} \alpha}%
+2{\rm i} (\Gamma_{b})^{\alpha}{}_{\beta} \phi^{a}\,_{\underline{j}}\,^{\beta} \nabla^{b}{\varphi_{\underline{i} \alpha}}-6\nabla^{b}{G_{\underline{i} \underline{j}}} f^{a}\,_{b}
\doublespacedmathend
\end{adjustwidth}

\begin{footnotesize}

\end{footnotesize}